\documentclass[A4,10pt]{report}

\usepackage[margin=1in,dvips]{geometry}
\usepackage{graphicx,cancel}
\usepackage{graphics}
\usepackage{color, soul}
\usepackage{amsmath, amsthm, slashed}
\usepackage{amssymb}
\usepackage{marvosym}
\usepackage{rotating}
\usepackage{mathrsfs}
\usepackage{upgreek}
\usepackage{epsfig}
\usepackage{verbatim}
\usepackage[affil-it]{authblk}
\usepackage{rotating}
\usepackage{graphicx}
\usepackage{accents}
\usepackage{harmony, slashed}
\usepackage{tikz}
\usepackage{dsfont}
\usepackage{makeidx}

\usepackage{tocloft}

\usepackage{thmtools}
\usepackage{thm-restate}

\usepackage{comment}

\usepackage{enumerate}

\usepackage[colorlinks=true,linkcolor=black,citecolor=blue]{hyperref}

\usepackage{cleveref}

\usepackage{minitoc}

\newtheorem{theorem}{Theorem}[section]
\newtheorem{conjecture}{Conjecture}[section]
\newtheorem*{conjecture*}{Conjecture}
\newtheorem*{theorem*}{Theorem}

\newtheorem{proposition}{Proposition}[section]
\newtheorem{definition}[proposition]{Definition}
\newtheorem{corollary}[proposition]{Corollary}
\newtheorem*{corollary*}{Corollary}
\newtheorem{lemma}[proposition]{Lemma}
\newtheorem{remark}[proposition]{Remark}

\newtheorem{bigishtheorem}{Theorem}[chapter]

\newtheorem{testtheorem}{Theorem}[subsection]
\newtheorem{testcorollary}{Corollary}[subsection]

\newcounter{testcount}[testoutercounter]

\newcommand{\Olin}{\Omega^{-1}\accentset{\scalebox{.6}{\mbox{\tiny (1)}}}{\Omega}}
\newcommand{\Olino}{\accentset{\scalebox{.6}{\mbox{\tiny (1)}}}{\Omega}}
\newcommand{\glinh}{\accentset{\scalebox{.6}{\mbox{\tiny (1)}}}{\hat{\slashed{g}}}}
\newcommand{\glin}{\accentset{\scalebox{.6}{\mbox{\tiny (1)}}}{\slashed{g}}}
  \newcommand{\glinto}{\accentset{\scalebox{.6}{\mbox{\tiny (1)}}}{\sqrt{\slashed{g}}}}
\newcommand{\bmlin}{\accentset{\scalebox{.6}{\mbox{\tiny (1)}}}{b}}
\newcommand{\justglin}{\accentset{\scalebox{.6}{\mbox{\tiny (1)}}}{g}}

\newcommand{\xblin}{\accentset{\scalebox{.6}{\mbox{\tiny (1)}}}{\underline{\hat{\chi}}}}
\newcommand{\xlin}{\accentset{\scalebox{.6}{\mbox{\tiny (1)}}}{{\hat{\chi}}}}
\newcommand{\xli}{\accentset{\scalebox{.6}{\mbox{\tiny (1)}}}{\chi}}
\newcommand{\eblin}{\accentset{\scalebox{.6}{\mbox{\tiny (1)}}}{\underline{\eta}}}
\newcommand{\elin}{\accentset{\scalebox{.6}{\mbox{\tiny (1)}}}{{\eta}}}
\newcommand{\otx}{\accentset{\scalebox{.6}{\mbox{\tiny (1)}}}{\left(\Omega tr \chi\right)}}
\newcommand{\otxb}{\accentset{\scalebox{.6}{\mbox{\tiny (1)}}}{\left(\Omega tr \underline{\chi}\right)}}
\newcommand{\olin}{\accentset{\scalebox{.6}{\mbox{\tiny (1)}}}{\omega}}
\newcommand{\olinb}{\accentset{\scalebox{.6}{\mbox{\tiny (1)}}}{\underline{\omega}}}

\newcommand{\ablin}{\accentset{\scalebox{.6}{\mbox{\tiny (1)}}}{\underline{\alpha}}}
\newcommand{\alin}{\accentset{\scalebox{.6}{\mbox{\tiny (1)}}}{{\alpha}}}
\newcommand{\pblin}{\accentset{\scalebox{.6}{\mbox{\tiny (1)}}}{\underline{\psi}}}

\newcommand{\Pblin}{\accentset{\scalebox{.6}{\mbox{\tiny (1)}}}{\underline{P}}}
\newcommand{\Plin}{\accentset{\scalebox{.6}{\mbox{\tiny (1)}}}{{P}}}

\newcommand{\blin}{\accentset{\scalebox{.6}{\mbox{\tiny (1)}}}{{\beta}}}
\newcommand{\rlin}{\accentset{\scalebox{.6}{\mbox{\tiny (1)}}}{\rho}}
\newcommand{\slin}{\accentset{\scalebox{.6}{\mbox{\tiny (1)}}}{{\sigma}}}
\newcommand{\Klin}{\accentset{\scalebox{.6}{\mbox{\tiny (1)}}}{K}}

\newcommand{\Ylin}{\accentset{\scalebox{.6}{\mbox{\tiny (1)}}}{Y}}

\newcommand{\flin}{\accentset{\scalebox{.6}{\mbox{\tiny (1)}}}{f}}

\newcommand{\gslash}{\slashed{g}}
\newcommand{\Gammaslash}{\slashed{\Gamma}}

\newcommand{\nablaslash}{\slashed{\nabla}}
\newcommand{\Deltaslash}{\slashed{\Delta}}
\newcommand{\Dslash}{\slashed{\mathcal{D}}}
\newcommand{\epsslash}{\slashed{\epsilon}}
\newcommand{\divslash}{\slashed{\mathrm{div}}}
\newcommand{\curlslash}{\slashed{\mathrm{curl}}}

\newcommand{\chibar}{\underline{\chi}}
\newcommand{\chibarhat}{\underline{\hat{\chi}}}
\newcommand{\etabar}{\underline{\eta}}
\newcommand{\mubar}{\underline{\mu}}

\newcommand{\omegabar}{\underline{\omega}}
\newcommand{\alphabar}{\underline{\alpha}}
\newcommand{\betabar}{\underline{\beta}}
\newcommand{\tr}{\mathrm{tr}}

\newcommand{\I}{\mathcal{I}^+}
\newcommand{\Hp}{\mathcal{H}^+}
\newcommand{\Pbar}{\underline{P}}

\newcommand{\psibar}{\underline{\psi}}
\newcommand{\omegahat}{\hat{\omega}}
\newcommand{\omegabarhat}{\underline{\hat{\omega}}}
\newcommand{\Cbar}{\underline{C}}
\newcommand{\Xbar}{\underline{X}}
\newcommand{\kbar}{\underline{k}}

\newcommand{\DcI}{\check{\mathcal{D}}^{\I}}
\newcommand{\DRI}{\mathcal{D}^{\I}}
\newcommand{\DcH}{\check{\mathcal{D}}^{\Hp}}
\newcommand{\DRH}{\mathcal{D}^{\Hp}}

\newcommand{\DRK}{\mathcal{D}^{\mathcal{K}}}
\newcommand{\DREF}{\mathcal{D}^{\mathcal{EF}}}
\newcommand{\CcH}{\check{C}^{\Hp}}
\newcommand{\CcI}{\check{C}^{\I}}
\newcommand{\CbcH}{\underline{\check{C}}^{\Hp}}
\newcommand{\CbcI}{\underline{\check{C}}^{\I}}

\newcommand{\DRHs}{\mathcal{D}^{\Hp}_{r\le s}}

\newcommand{\DRIss}{\mathcal{D}^{\I}_{s \le r\le  \tilde s}}
\newcommand{\DRIs}{\mathcal{D}^{\I}_{r\ge s}}

\newcommand{\fsc}{\mathbf{f}}

\newcommand{\nldq}{\frac{\varepsilon_0^2 + \varepsilon^3}{\tau^2}}

\newcommand{\newsph}[1]{\tilde{#1}}

\DeclareMathAlphabet\mathbfcal{OMS}{cmsy}{b}{n}

\makeatletter \@addtoreset{equation}{section}  \makeatother

\makeindex

\title{The non-linear stability of the Schwarzschild family of black holes}

\date{April 16, 2021}

\author[$\ddag$ $\S$]{Mihalis Dafermos}
\author[$\dag$ $*$]{Gustav Holzegel}
\author[$\ddag$]{Igor Rodnianski}
\author[$\dag$]{Martin Taylor}

\affil[$\dag$]{\small Imperial College London,
Department of Mathematics,
South~Kensington~Campus,~London~SW7~2AZ,~United~Kingdom\vskip.2pc \ }

\affil[$\ddag$]{\small Princeton University, Department of Mathematics, Fine~Hall,~Washington~Road,~Princeton,~NJ~08544,~United~States~of~America\vskip.2pc \ }

\affil[$\S$]{\small University of Cambridge, Department of Pure Mathematics and Mathematical
Statistics, Wilberforce~Road,~Cambridge~CB3~0WA,~United~Kingdom\vskip.2pc \ }

\affil[$*$]{\small Westf\"alische Wilhelms-Universit\"at M\"unster,
Mathematisches~Institut, Einsteinstrasse~62~48149~M\"unster,~Bundesrepublik~Deutschland}

\begin{document}

\pagenumbering{roman}

\maketitle

\begin{abstract}
We prove the  non-linear asymptotic stability of the Schwarzschild family
 as solutions to the Einstein vacuum equations
in the exterior of the black hole region:
general vacuum initial data---with no symmetry assumed---sufficiently close to Schwarzschild data
evolve to a vacuum spacetime which
\begin{enumerate}
\item[(i)] possesses a complete future null infinity $\mathcal{I}^+$ (whose
past $J^-(\mathcal{I}^+)$ is moreover bounded by a regular future complete  event horizon $\mathcal{H}^+$),
\item[(ii)]
remains close to Schwarzschild in its
exterior, and
\item[(iii)] asymptotes  
back to a  member of the Schwarzschild family as an appropriate notion of time goes
to infinity,
\end{enumerate}
provided that the data are themselves constrained to lie on a teleologically constructed
codimension-$3$ ``submanifold'' of  moduli space.
This is the full nonlinear asymptotic  stability of Schwarzschild since
solutions not arising from data lying on this submanifold should by dimensional considerations 
approach a Kerr spacetime with rotation parameter $a\ne0$, i.e.~such solutions cannot satisfy (iii).
The statement is effective, providing quantitative bounds
from explicit initial data quantities, and the global nearness to Schwarzschild at top order
can be 
measured with respect to the same quantity as initial data, i.e.~without loss of derivatives.
The proof employs teleologically normalised double null gauges,
is expressed entirely in physical space and makes essential 
use of the  analysis in our previous study of the \emph{linear} stability of the Kerr family around
Schwarzschild~\cite{holzstabofschw}, as well as
techniques developed over the years to control the non-linearities of the Einstein
equations, in particular in the difficult radiation zone associated to subtle non-linear effects
like Christodoulou memory.  The present work, however, is entirely self-contained.
In view of the recent~\cite{DHRteuk, RitaShlap},
our approach can be applied to the full non-linear asymptotic stability of the subextremal Kerr family.
\end{abstract}

\doparttoc
\dominitoc

\tableofcontents

\cleardoublepage

\pagenumbering{arabic}

\renewcommand{\thesection}{\Roman{section}} 
\setcounter{secnumdepth}{4}

\chapter*{Introduction}
\addcontentsline{toc}{chapter}{Introduction}



The \emph{Schwarzschild metrics}
\begin{equation}
\label{Schwmet}
g_M =-(1-2M/r)dt^2+(1-2M/r)^{-1}dr^2 +r^2(d\theta^2+\sin^2\theta\, d\phi^2)
\index{Schwarzschild background!metric!$g_M$, Schwarzschild metric} \index{Schwarzschild background!parameters!$M$, mass parameter}
\index{Schwarzschild background!metric!$r$,  area radius function}
\end{equation}
give rise to a $1$-parameter family of spherically symmetric, stationary asymptotically
flat
Lorentzian $4$-manifolds $(\mathcal{M},g_M)$, 
parametrised by mass $M\in \mathbb R$, and represent
the most well-known explicit solutions of the 
\emph{Einstein vacuum equations}
\begin{equation}
\label{vaceqhere}
{\rm Ric}[g]=0,
\end{equation}
the equations governing general relativity in the absence of matter.
The metric~\eqref{Schwmet} was originally discovered in local coordinates
by Schwarzschild~\cite{schwarzschild1916}, very soon after Einstein's 
formulation~\cite{Einstein1915}
of~\eqref{vaceqhere}, but
was only later understood by Lema\^itre~\cite{Lemaitre} to admit (in the case $M>0$) a non-trivial
regular extension through an \emph{event horizon}~$\mathcal{H}^+$ \index{Schwarzschild background!sets!$\mathcal{H}^+$,
the event horizon} at~$r=2M$ 
to  what is now known as a \emph{black hole}
region (see also the earlier~\cite{eddingtoncoords}). Whereas freely falling observers may choose to remain in the \emph{exterior} region $r>2M$ for
all proper time (in particular, this region 
possesses a \emph{complete future null infinity} $\mathcal{I}^+$), \index{Schwarzschild background!sets!$\mathcal{I}^+$, future null infinity}
those entering the black hole interior $r< 2M$ encounter  a  ``ferocious singularity'' (see~\cite{misner2017gravitation})
at $r=0$, where curvature blows up and beyond which the metric cannot be
extended---not even merely continuously~\cite{SbierskiCzero}. 
The very notion of black hole
implies, however, that the exterior region remains causally unaffected
by this singularity, and thus the basic properties of this region 
can be studied independently of
the black hole interior.

The most fundamental  question to ask concerning the Schwarzschild metric~\eqref{Schwmet}
is that of its  \emph{nonlinear stability} in the
exterior region
as a solution of~\eqref{vaceqhere}. 
This will be the subject of the present work.

The nonlinear stability  problem is naturally formulated in the context of the \emph{Cauchy
problem}~\cite{Geroch} for~\eqref{vaceqhere}, which associates a unique \emph{maximal Cauchy
development},  solving~\eqref{vaceqhere}, to every \emph{vacuum initial data set}.
We will give a preliminary statement 
of our main result, the {\bf \emph{full non-linear asymptotic stability
of Schwarzschild}}, in {\bf Section~\ref{mainresintint_sec}} below.  This is formulated
as {\bf Theorem~\ref{maintheoremintro}}. Briefly, this theorem says that
for vacuum initial data sets---with no symmetry assumed---sufficiently close to
appropriate Schwarzschild initial data, 
the resulting maximal Cauchy development 
\begin{enumerate} 
\item[(i)] possesses a complete future null infinity $\mathcal{I}^+$ whose
past $J^-(\mathcal{I}^+)$ is bounded to the future 
by a regular future complete event horizon $\mathcal{H}^+$, 
\item[(ii)]
 remains globally close to Schwarzschild~\eqref{Schwmet}  in its exterior and  
\item[(iii)] asymptotes back to a  member of
 the Schwarzschild family as a suitable notion of time goes to infinity,
 \end{enumerate} 
provided that the initial data set itself lies on a codimension-3 ``submanifold'' of the moduli space of vacuum
initial data. Our restriction on data is a necessary
condition for the asymptotic stability statement (iii). 
\begin{figure}
\centering{
\def\svgwidth{13pc}
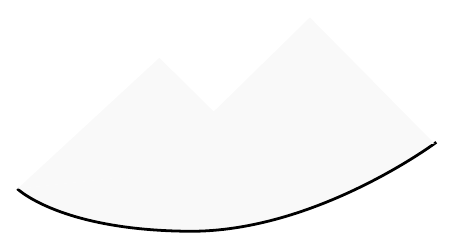}
\caption{Nonlinear asymptotic stability of Schwarzschild: spacetimes satisfying (i)--(iii)}\label{nonlinstabfigure}
\end{figure}
 For as is well known,
the
Schwarzschild family~\eqref{Schwmet} 
is contained as the $a=0$  subcase of a larger family of  stationary 
solutions, \emph{the Kerr family} $g_{a,M}$~\cite{Kerr}, \index{Kerr metric!$g_{a,M}$}
with $aM$ representing  angular momentum.\index{Kerr metric!$a$, Kerr parameter}\index{angular momentum!$aM$, angular momentum of the Kerr metric $g_{a,M}$}
Outside our codimension-$3$ submanifold, one expects solutions to necessarily asymptote
to a Kerr solution with $a\ne 0$, since
the dimension of \emph{linearised Kerr solutions fixing the mass} is equal to $3$
in our parametrisation.

To break
 the general covariance of~\eqref{vaceqhere},
 our proof relies in an essential way on expressing the equations  in 
 \emph{double null gauge}.
 It is natural then to start the problem from characteristic initial data
 (cf.~\cite{rendallchar})
 determined by two null hypersurfaces $C_{\rm out}\cup \underline{C}_{\rm in}$, 
 with $C_{\rm out}$ extending to $\mathcal{I}^+$ and $\underline{C}_{\rm in}$
 entering what will be the black hole region, as
 it follows from~\cite{CK, KN} and Cauchy stability
 that general asymptotically flat  initial data posed on a (spacelike)
  Cauchy hypersurface $\Sigma$ indeed contain such null hypersurfaces in their Cauchy evolution.
  See Figure~\ref{nonlinstabfigure}.
  Our moduli space will thus consist of the set of general such characteristic initial data, without symmetry
  assumptions, in a suitable topology.
  We emphasise that both our asymptotically stable codimension-$3$ submanifold  itself
  and the double null gauges that we shall employ must be constructed ``teleologically'', i.e.~on the basis
  of properties of their future evolution.
  The same applies to the final mass parameter $M$ of the Schwarzschild solution $g_M$ 
 to which our metric $g$ asymptotically settles down.
 (Indeed, in general, the only way to identify data for which (iii) holds, and to moreover determine $M$, is to evolve 
 them towards the future under~\eqref{vaceqhere}!)
 The  event horizon $\mathcal{H}^+$ will represent an asymptotic hypersurface of our double null gauge
 and in particular is also only determined teleologically.  We emphasise that our results give
 stability up to and including $\mathcal{H}^+$, which is moreover shown to be future complete and
 suitably regular; this then determines the boundary of a non-trivial
 black hole region.

 We note that double null gauges have been successfully employed
in recent years for several very different global problems in general relativity~\cite{Chr, lukrodimpulsive, lukweaknull, DafLuk1},
and  their use here means that, despite the complication of the present work,
its basic equations and the general framework for their analysis will be  already familiar to some readers,
and many of the difficulties can be readily understood in 
a far more general context, beyond stability problems. We emphasise however that the present work is 
completely self-contained and will not require previous familiarity with any of the above papers.

It is worth remarking that the global closeness of statement (ii) can be expressed at the top
order energy level with respect to the same quantity that measures a suitable ``initial'' energy quantity,
i.e.~without loss of derivatives. In this sense, we have obtained a true orbital stability statement.
As is well known, however,  the supercriticality of the non-linearities of~\eqref{vaceqhere} 
means that the only path to proving the statements (i) and (ii) is through
a full asymptotic stability statement (iii), and thus one does not expect to be able to obtain
a statement  involving (i)--(ii) alone. See already Section~\ref{orbvsasympt}.

The necessary starting point to proving (iii)  is a robust approach to \emph{linear stability} around the expected
asymptotic state.
This was provided by our
(purely physical space based) method and framework introduced in~\cite{holzstabofschw},
where we proved  the linear stability of 
 the Kerr family around Schwarzschild, in double null gauge.
 In particular, a corollary of~\cite{holzstabofschw} is that the Schwarzschild
 subfamily is itself linearly \emph{asymptotically} stable precisely for data lying on a codimension-$3$ subspace, 
 which, unlike the nonlinear setting, can be identified explicitly, as can the final linearised mass parameter. 
 We note, however, that the necessity
 of teleological normalisation of the double null gauge  is already a key difficulty of linear theory. We 
shall review these results 
in {\bf Section~\ref{linstabintint_sec}}. These in turn can be viewed as the culmination of 
a series of previous
work starting from~\cite{Regge},  using heavily
technology developed 
in the last fifteen years (see~\cite{Sterbenz, DafRod2, BlueSoffer, Mihalisnotes, Holzegelspin2, DafRodnew}) for understanding dispersion of waves outside of black holes.

In addition to the linear theory of~\cite{holzstabofschw}, our results depend on many insights
developed over the years,
starting
from the monumental proof of the stability of Minkowski space~\cite{CK},
for understanding the \emph{nonlinearities} of~\eqref{vaceqhere}, 
 in particular,
in the most difficult radiation zone towards null infinity $\mathcal{I}^+$,
where null structure (cf.~\cite{KlNull}) is paramount.  This full understanding
of the null structure elucidated by the geometric gauge used here
 allows one to understand in our black hole context
subtle non-linear effects associated
to radiation, for instance, Christodoulou memory~\cite{Christmem}.
We shall 
put our work in the context of previous 
non-linear  results for asymptotically flat solutions of~\eqref{vaceqhere} in {\bf Section~\ref{resnonstabintint_sec}}. 
(We shall also compare with the situation when a cosmological constant $\Lambda$ is added to~\eqref{vaceqhere}, in which case there are recent non-linear stability
results if $\Lambda>0$ due to Hintz and Vasy~\cite{hintz2016global}, while in the
$\Lambda<0$ case, all black hole solutions may in fact be unstable (cf.~\cite{holzegel2013decay, moschidis2018newproof}).)

Much previous work on nonlinear stability has considered various symmetric reductions, starting
from work of Christodoulou on the Einstein-scalar field system in spherical symmetry~\cite{maththeory},
followed by~\cite{DRPrice,
dafholz2006nonlinear, holzbiax}, 
and most recently, the impressive work of Klainerman--Szeftel~\cite{klainerman2017global}
for polarised axisymmetric spacetimes, which
is a first work beyond $1+1$ dimensional systems. 
Whereas, as remarked above, the full codimension-$3$ submanifold of moduli space
yielding asymptotic stability (iii) of Schwarzschild, in the absence of symmetry, can only be characterized
teleologically, 
in contrast, in the restrictive setting of symmetry, one can easily identify criteria on initial data which guarantee
that
their only possible end-state within the Kerr family would be Schwarzschild. If these data 
lead to stable evolution, it follows that they 
should then be contained on our submanifold.
Indeed, we will explicitly show,
exploiting the fact that in the vacuum angular momentum cannot radiate to null infinity under axisymmetry,
 that our submanifold does
contain, as an infinite codimension subclass, the set of
all axisymmetric data with vanishing Komar angular momentum.
This is the statement of {\bf Corollary~\ref{axicorollary}}.
Thus in particular, our submanifold contains  also the
polarised axisymmetric data considered recently in~\cite{klainerman2017global}, which
itself is an infinite codimension subfamily  of general axisymmetric data.

The first step of the proof of linear stability of Schwarzschild in~\cite{holzstabofschw} was to
prove decay results
for the decoupled \emph{Teukolsky equation}~\cite{bardeen1973, teukolsky1973},
governing the gauge-invariant part of the perturbations, via a physical
space reinterpretation and novel use of  transformations first introduced by
Chandrasekhar~\cite{Chandraschw}. This has recently been generalised to the very slowly rotating Kerr
case $|a|\ll M$ in~\cite{DHRteuk} (see also~\cite{Ma:2017yui2}) and to the full subextremal range  $|a|<M$ 
by Shlapentokh-Rothman and Teixeira da Costa~\cite{RitaShlap}.
With these more recent developments, the whole approach of this work can in principle be generalised to Kerr, in fact in the
full subextremal range of parameters. 
We shall give a formulation of the full nonlinear stability
of Kerr problem
in {\bf Section~\ref{kerrstatements}}, including a discussion of the extremal case and the relation
with the structure of black hole interiors.

We shall finally give an overview of the proof of Theorem~\ref{maintheoremintro} in
{\bf Section~\ref{firstremarksintint_sec}}.
We will conclude this introduction in {\bf Section~\ref{outline_sec}} with an outline of the work.

\renewcommand{\theequation}{\thesubsection.\arabic{equation}}

\section{The main result: the  nonlinear asymptotic stability of Schwarzschild}
\label{mainresintint_sec}

\addtocontents{toc}{\setcounter{tocdepth}{2}}

We attempt here a first 
rough formulation of our main result, fleshing out the above discussion.

\subsection{Double null gauge}
\label{doublenullgaugeintheintro}
Fundamental to our approach will be expressing  the Einstein
vacuum equations~\eqref{vaceqhere} in a 
\emph{double null gauge},
where the metric takes the form
\begin{equation}
\label{doublenull}
g= -4\Omega^2 du\, dv + \slashed{g}_{CD} (d\theta^C -b^C dv)(d\theta^D-b^D dv).
\index{double null gauge!metric coefficients!$\Omega^2$, conformal metric coefficient}
\index{double null gauge!metric coefficients!$\slashed{g}_{CD}$, angular metric coefficient}
\index{double null gauge!metric coefficients!$b^C$, torsion metric coefficient}
\index{double null gauge!coordinates!$u$, retarded null coordinate}
\index{double null gauge!coordinates!$v$, advanced null coordinate}
\end{equation}
The constant-$u$ and $v$ hypersurfaces are thus null, intersecting
along $2$-surfaces $S_{u,v}$\index{double null gauge!sets!$S_{u,v}$, intersections of constant-$u$ and constant-$v$ hypersurfaces}
with local coordinates $\theta^A$
and induced metric $\slashed{g}_{AB}$ assumed
Riemannian.
Associated to the above is a \emph{normalised double null frame}
\[
e_1, \qquad e_2,\qquad  e_3=\Omega^{-1}\partial_u, \qquad e_4=\Omega^{-1}(\partial_v+b^A\partial_{\theta^A}).
\index{double null gauge!frames!$e_i, i=1,\ldots 4$, normalised double null frame}
\]
The equations reduce to a complicated coupled system for the following
geometric quantities:
\begin{center}
\begin{tabular}{ |c|c| } 
 \hline
 metric coefficients & $\Omega$, \,  $\slashed{g}$, \, $b$  \\ 
 \hline
 connection coefficients & $\chi = g(\nabla_A e_4,e_B)$, \, $\chibar= g(\nabla_A e_3,
e_B)$, \ldots  
\index{double null gauge!connection coefficients!$\chi$} \index{double null gauge!connection coefficients!$\chibar$} 
\\ 
 \hline
 curvature components & $\alpha_{AB}= R(e_A,e_4,e_B,e_4)$, \, $\alphabar_{AB}=R(e_A,e_3,e_B,e_3)$, \ldots
  \index{double null gauge|curvature components!$\alpha$} \index{double null gauge!curvature components!$\alphabar$}
  \\ 
 \hline
\end{tabular}
\end{center}

Some of the resulting equations are displayed below:
\begin{equation}
\label{coupsys}
\begin{tabular}{ |c|c| } 
\hline
transport& $\slashed\nabla_4\slashed{g}=0, \, \slashed\nabla_4\hat\chi+{\rm tr} \chi \, \hat\chi -
\hat\omega \, \hat\chi = \alpha$, \ldots
\index{double null gauge!connection coefficients!$\hat{\chi}$} \index{double null gauge!connection coefficients!$\hat\omega$}
 \\
\hline
elliptic& $\slashed{\rm div} \hat\chi= \slashed\nabla {\rm tr}\chi-\beta +\ldots,$  \ldots
\index{double null gauge!curvature components!$\beta$}
\\
\hline
hyperbolic &$\slashed\nabla_3\alpha+\frac12{\rm tr}\underline\chi\alpha 
+2\omegabarhat\alpha
=-2\slashed{\mathcal{D}}^*_2\beta -3\hat\chi \rho+\ldots,  \,
 \slashed\nabla_4\beta+2{\rm tr}\chi\beta -\hat\omega\beta =
\slashed{\rm div} \alpha + \ldots$, \ldots
\index{double null gauge!connection coefficients!$\omegabarhat$}
\index{double null gauge!curvature components!$\rho$}
\\
\hline
\end{tabular}
\end{equation}
(In the above, $\slashed\nabla_3$ and $\slashed\nabla_4$ denote covariant differential
operators\index{double null gauge!differential operators!$\slashed\nabla_3$, covariant differential operator acting in $e_3$ direction}\index{double null gauge!differential operators!$\slashed\nabla_4$, covariant differential operator acting in $e_4$ direction}
acting in the $e_3$ and $e_4$ directions, whereas $\slashed\nabla$, 
$\slashed{\rm div}$, $\slashed{\mathcal{D}}^*_2$ 
\index{double null gauge!differential operators!$\divslash$, covariant operator acting tangentially on $S_{u,v}$}
\index{double null gauge!differential operators!$\Dslash_2^*$, covariant operator acting tangentially on $S_{u,v}$}
are covariant operators acting tangentially on the spacelike 
surfaces $S_{u,v}$.  The complete equations
will be given in Chapter~\ref{theequationssec}.) We note that the metric and connection coefficients
satisfy \emph{transport} and \emph{elliptic} equations, whereas the essential \emph{hyperbolicity}
of~\eqref{vaceqhere} is captured by the Bianchi identities satisfied by the
curvature components.

The significance of the double null gauge formulation for our work is two-fold: The form~\eqref{coupsys} 
will allow  us to capture
important structure associated to the vacuum equations~\eqref{vaceqhere}, 
both at the linear level  (see Section~\ref{linstabintint_sec}) 
as well as  at the level of the quadratic 
non-linearities (see Section~\ref{resnonstabintint_sec}),
necessary to understand non-linear stability.

As mentioned already above, an 
added benefit of using double null gauge is that the present work can then be seen in a unified context
with a host of other recent works in general relativity where double null gauge has been successfully
employed~\cite{Chr, KN, klailukrod, lukrodimpulsive, lukweaknull, DafLuk1, vacuumscatter}. In particular, the above notations and the form of the above equations are familiar from 
these works.

\subsection{Schwarzschild, Kerr and residual gauge freedom}
\label{SchKerrgaugefreed}
\setcounter{equation}{0}

The Schwarzschild family can itself be written in the form~\eqref{doublenull} with respect
to Eddington--Finkelstein normalised double null coordinates $(u,v)$,
where the exterior is parametrised as $(-\infty,\infty)\times(-\infty,\infty)$ and
\begin{equation}
\label{Schwstandform}
\Omega^2= 1-\frac{2M}r, \qquad \slashed{g} =r^2 \mathring{\gamma}, \qquad b^A=0,
\end{equation}
where $\mathring\gamma$ denotes the standard metric on the unit sphere. 
\index{sphere!metric!$\mathring{\gamma}$, standard metric on $\mathbb S^2$}
The coordinate
$r$ of~\eqref{Schwmet} now appears in the above as a function $r(u,v)$.
The ideal boundary null infinity $\mathcal{I}^+$  can be formally parametrised
as $(-\infty,\infty)\times\{\infty\}\times \mathbb S^2$ and  the parameter $u$ represents a canonical
Bondi time at infinity. In particular, the unboundedness of the range
of $u$ exhibits the ``completeness'' of null infinity.

\begin{figure}
\centering{
\def\svgwidth{10pc}
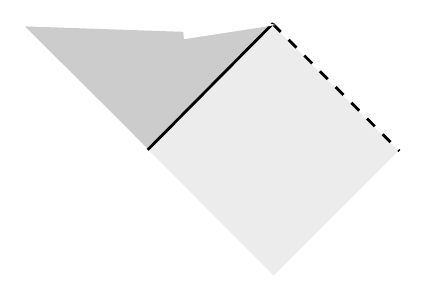}
\caption{The Lemaitre manifold $\mathcal{M}_{\text{Lema\^itre}}$}\label{Lemait}
\end{figure}

The above coordinate system does not cover the \emph{event horizon}.
For this, we can define a new coordinate 
\begin{equation}
\label{transformat}
U(u)=-\Omega^2 (u,0) \frac{r(u,0)}{2M}e^{r(u,0)/2M}= -e^{-u/2M}.
\end{equation}
The transformation~\eqref{transformat} maps
the $u$-range $(-\infty,\infty)$ to the $U$-range $(-\infty,0)$. 
\index{Schwarzschild background!coordinates!$U$, retarded Kruskal null coordinate}
One may easily see that the resulting functions $r(U,v)$ and $\Omega(U,v)$ now extend
by analyticity to positive functions on a region $c(v)> U\ge 0$ 
where $c(v)$  is defined implicitly by the limiting relation
$r(c(v),v)=0$.  We have thus produced
a solution of $(\ref{vaceqhere})$ defined on this larger domain whose underlying manifold
we shall denote 
as $\mathcal{M}_{\text{Lema\^itre}}$. 
The event horizon $\mathcal{H}^+$ is then given by the hypersurface $U=0$ 
and satisfies $r=2M$. It can be characterized as the
boundary in $\mathcal{M}_{\text{Lema\^itre}}$ 
\index{Schwarzschild background!sets!$\mathcal{M}_{\text{Lema\^itre}}$, the Lemaitre  manifold}
of the past of future null infinity $\mathcal{I}^+$,  $J^-(\mathcal{I}^+)$. The region $U> 0$ is
then  what we shall refer to as the \emph{black hole interior}.
See Figure~\ref{Lemait}.

It is convenient
to also tranform $V(v)=e^{v/2M}$, mapping the $v$-range $(-\infty,\infty)$ to the
$V$-range $(0,\infty)$. 
\index{Schwarzschild background!coordinates!$V$, advanced Kruskal null coordinate}
This
transforms the metric coefficient $\Omega^2$ into  
\begin{equation}
\label{SchwsKruskform}
\Omega^2_{\mathcal{K}}(U,V)=\frac{8M^3}r\exp \left(-\frac{r}{2M}\right),
\index{Schwarzschild background!metric!$\Omega^2_{\mathcal{K}}$, metric component in Kruskal coordinates}
\end{equation}
where $r(U,V)$ is implicitly defined by the relation $\left(\frac{r}{2M}-1\right)\exp\left(\frac{r}{2M}\right)=-UV$.
This is the celebrated Kruskal form of the metric. From this form one can see that the metric can in fact be extended
to be defined on an even larger manifold corresponding precisely to the region $UV<1$.  
See already Section~\ref{Kruskalmanifold}.

The above discussion already illustrates a fundamental fact: The double null form~\eqref{doublenull} 
is not
uniquely determined by the metric as~\eqref{Schwstandform} 
and~\eqref{SchwsKruskform} are in particular
both representations of the same metric. 
There is in fact an infinite dimensional family of diffeomorphisms
preserving the form~\eqref{doublenull}. Thus, even after imposing  double null gauge,
finding a particular normalisation
in which the metric coefficients asymptote to~\eqref{Schwstandform} is one of the difficulties of the stability
problem, present---as we shall see in Section~\ref{linstabintint_sec}---already in the context of linear
theory.

As we have  discussed  previously, the Schwarzschild family
appears as a subfamily of the two-parameter Kerr family $g_{a,M}$. 
These latter solutions
are again stationary, but now only axisymmetric, and $aM$\index{angular momentum!$aM$, angular momentum of
the Kerr metric $g_{a,M}$} corresponds to their
total angular momentum. In Boyer--Lindquist coordinates,
the Kerr metric can be written as
\begin{equation}
\label{KerrmetricBL}
g_{a,M} =-\frac{\Delta}{\varrho^2}
(dt-a\sin^2\theta d\phi)^2+\frac{\varrho^2}{\Delta}dr^2
+\varrho^2d\theta^2+\frac{\sin^2\theta}{\varrho^2}(a dt-(r^2+a^2)d\phi)^2,
\end{equation}
where
\begin{equation}
\label{variousdefs}
\Delta =  r^2-2Mr+a^2 ,\qquad \varrho^2 =r^2+a^2\cos^2\theta.
\index{Kerr metric!$\Delta$, function appearing in Kerr metric}
\index{Kerr metric!$\varrho$, function appearing in Kerr metric}
\end{equation}
The Kerr metrics were first expressed globally 
in  double null gauge~\eqref{doublenull} by~\cite{Pretorius}. Note that, given a fixed
underlying double null coordinate system $(u,v,\theta,\phi)$
associated to Schwarzschild, 
it is natural to consider
the ``standard 
Kerr metrics''  as representing a four-dimensional family. In this parametrisation,
there is only one standard Schwarzschild solution for each mass, 
but a three-dimensional family of fixed mass Kerr solutions corresponding to the
group ${\rm SO}(3)$ reflecting the freedom in choosing the axisymmetric
Killing field $\partial_\phi$. 
The codimension of the Schwarzschild family in the  Kerr family is thus 3.

\subsection{First formulation of the main result: Theorem~\ref{maintheoremintro}}
\setcounter{equation}{0}

\begin{figure}
\centering{
\def\svgwidth{10pc}
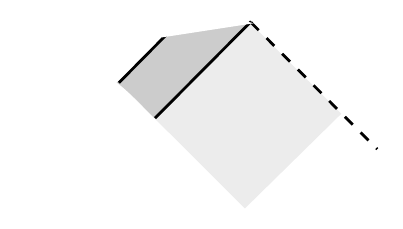}
\caption{Schwarzschild as a Cauchy development of characteristic data}\label{Schwasinit}
\end{figure}

To formulate the stability problem,  let us first restrict to the 
region 
\begin{equation}
\label{restrictoregion}
\mathcal{M}_{\text{Lema\^itre}}\cap \{\delta \ge  U\ge -1\}\cap \{1\le V< \infty\},
\end{equation}
for a $\delta>0$.
See Figure~\ref{Schwasinit}.

We may think of the above region~\eqref{restrictoregion} as 
the unique \emph{maximal future Cauchy development}
of Schwarzschild characteristic initial
data for~\eqref{vaceqhere} posed on the null hypersurfaces
\begin{equation}
\label{initialcones}
C_{\rm out} = \{-1\}\times [1,\infty)\times \mathbb S^2, \qquad 
\underline{C}_{\rm in}=  [-1,\delta] \times \{1\}\times \mathbb S^2.
\index{double null gauge!sets!$C$, outgoing null hypersurface}
\index{double null gauge!sets!$\underline{C}$, ingoing null hypersurface}
\end{equation}
See~\cite{init, Geroch} for the general notion of maximal Cauchy development 
and~\cite{rendallchar} for the characteristic
initial value problem. The significance of taking $\delta>0$ will be clear
momentarily. (Note already that $S=\{\delta\}\times\{1\}\times\mathbb S^2$ 
is a \emph{trapped sphere} in Schwarzschild (see~\cite{Pen1965} for this notion) in view
of the inequalities $\partial_U r(\delta, 1)<0$, $\partial_V r(\delta,1)<0$.)
It is the above data which we will perturb, i.e.~we shall consider characteristic
data sets for~\eqref{coupsys} defined on the  initial hypersurfaces~\eqref{initialcones} which are assumed
suitably
close to Schwarzschild initial data.
We shall give a more detailed discussion of initial data in Section~\ref{initdatasection}.
 In particular, for all data considered,
the cone $C_{\rm out}$ will be future
complete and asymptotically flat, ``terminating'' at null infinity $\mathcal{I}^+$, 
while the terminal sphere
$S$ of $\underline{C}_{\rm in}$ will again be trapped.

The first formulation of our main theorem can then be given as follows: 
\begin{testtheorem}[The nonlinear asymptotic stability of Schwarzschild in full co-dimension]
\label{maintheoremintro}
For all characteristic initial data prescribed on \eqref{initialcones}, 
assumed sufficiently close to Schwarzschild data with mass $M_{\rm init}$
\index{Schwarzschild background!parameters!$M_{\rm init}$, mass parameter associated to initial data}
and lying on a
codimension-3 ``submanifold'' $\mathfrak{M}_{\rm stable}$ of the moduli space $\mathfrak{M}$ of initial data, the maximal Cauchy 
development $\mathcal{M}$
contains \index{spacetime subsets!$\mathcal{M}$, maximal Cauchy development manifold} a region $\mathcal{R}$ 
\index{spacetime subsets!$\mathcal{R}$, subregion of $\mathcal{M}$ for which stability holds}
which can be covered by appropriate (teleologically
normalised) global double null gauges~\eqref{doublenull} 
and which \begin{enumerate}
\item[(i)] possesses a complete future null infinity $\mathcal{I}^+$ 
such that $\mathcal{R}\subset J^-(\mathcal{I}^+)$, and in fact the future boundary
of $\mathcal{R}$ in $\mathcal{M}$ is a regular, future affine complete  
``event horizon'' $\mathcal{H}^+$. Moreover, 
\item[(ii)]
the metric remains close to the Schwarzschild metric with mass $M_{\rm init}$ in $\mathcal{R}$ (moreover, measured
with respect to an energy at the same order as a suitable ``initial'' energy), and
\item[(iii)]
asymptotes, inverse polynomially, to a Schwarzschild metric with mass $M_{\rm final}\approx M_{\rm init}$
\index{Schwarzschild background!parameters!$M_{\rm final}$, final mass parameter}
as $u\to \infty$ and $v\to \infty$, in particular
along $\mathcal{I}^+$ and $\mathcal{H}^+$.
\end{enumerate}
\end{testtheorem}

As we have already remarked, it follows from~\cite{CK, KN} and Cauchy stability arguments that,
were initial data to be posed on an arbitrary asymptotically flat spacelike hypersurface, with
data assumed globally
close (in a suitable sense)
to the induced data on a Cauchy hypersurface of Schwarzschild, then their 
Cauchy evolution would contain
null hypersurfaces $C_{\rm out}$ and $\underline{C}_{\rm in}$ satisfying the assumptions of our 
Theorem~\ref{maintheoremintro}. (See already Remark~\ref{commentaboutpeeling} for further
discussion of this.)
Thus, in the sequel, we will always view the
stability problem as starting from characteristic initial data.

Note that the celebrated \emph{weak cosmic censorship} conjecture (see~\cite{Chrmil}) 
says that for generic
asymptotically flat data, the Cauchy evolution possesses a complete future null infinity $\mathcal{I}^+$.
Thus,  statement (i) can be thought of as showing 
``weak cosmic censorship'' in  a neighbourhood of Schwarzschild, statement (ii) can be thought of as the \emph{orbital stability} of the Schwarzschild exterior,
while statement (iii) represents
the \emph{asymptotic stability} of the Schwarzschild family---all restricted
to  our codimension-$3$  ``submanifold'' $\mathfrak{M}_{\rm stable}$\index{initial data!moduli space!$\mathfrak{M}_{\rm stable}$, asymptotically stable codimension-$3$ ``submanifold'' of $\mathfrak{M}$} 
of data, to be discussed further in Section~\ref{orbvsasympt} below.
We note moreover that (ii) as stated 
is indeed a true orbital stability statement, without loss of derivatives.

We emphasise that stability here is only being proven for the \emph{exterior} region
$J^-(\mathcal{I}^+)$  of
$\mathcal{M}$, up to and including its future boundary $\mathcal{H}^+$, the event horizon.
As is well known, however, one cannot identify this region explicitly from initial data since
$J^-(\mathcal{I}^+)$ (and hence $\mathcal{H}^+$ itself) is defined ``teleologically''. 
It is for this reason that it is essential that the Schwarzschild data extend
slightly beyond the event horizon, i.e.~to
take $\delta>0$ in~\eqref{initialcones}. (Let us note that in view of the fact
that the terminal sphere $S$ of $\underline{C}_{\rm in}$ remains trapped, we can already say
\emph{a priori} that $S\cap  J^-(\mathcal{I}^+)=\emptyset$,\footnote{See~\cite{Chr}. Note that
if our null initial data on $C_{\rm out}\cup\underline{C}_{\rm in}$ 
arose from an asymptotically flat (spacelike) Cauchy hypersurface $\Sigma$, then the presence of $S$ ensures that the maximal Cauchy development of $\Sigma$
is itself future causally geodesically incomplete, by Penrose's celebrated
incompleteness theorem~\cite{Pen1965}. The incompleteness of the Cauchy evolution
of $C_{\rm out}\cup\underline{C}_{\rm in}$ is of course trivial in view of the incompleteness of $\underline{C}_{\rm in}$
itself.}
 and thus one expects
the past of a complete $\mathcal{I}^+$ to be indeed contained in the maximal development
of the data $C_{\rm out}\cup\underline{C}_{\rm in}$, even though $\underline{C}_{\rm in}$ is  itself future 
incomplete.) One can in fact extend the region of stability in the interior of the black hole
all the way to a spacelike hypersurface, foliated by
trapped spheres, complete in one direction. See~\cite{DafLuk2}.
The stability issues ``deep'' in the black hole interior
are of an entirely different nature, however, being intimately related to the question of
\emph{strong cosmic censorship}. See~\cite{DafLuk1} and Section~\ref{kerrstatements} for 
further remarks.

In addition to the region $\mathcal{R}$ being only characterized teleologically,
the double null  gauge determined by the coordinates $(u,v)$ is itself  \emph{normalised} teleologically. In fact, the proof
shall construct \emph{two} distinct double null gauges, corresponding to a near and far region, suitably anchored to one another but with different
normalisations in each case. See already the discussion in 
Section~\ref{two_gauges_intro}. 
An additional statement in the theorem is that
the change of coordinates between these two double null coordinate
systems, as well as
the change of coordinates connecting both these coordinate systems
to the initial data on $C_{\rm out}\cup\underline{C}_{\rm in}$ (or rather, to two
auxiliary double null gauges associated to the data)
are all themselves  controlled
quantitatively from a geometric norm on initial data. See
already the underlined clause in Theorem~\ref{linstabthemintro} for the linear version of
this statement and Section~\ref{bounded_init_intro} for its significance for
the nonlinear proof.

We emphasise that it is the two teleologically normalised double null gauges which our proof employs  that
allows us at the same time to (i) identify the event horizon $\mathcal{H}^+$ and show its regularity 
and global geometric
properties
and (ii) give a  \emph{Bondi foliation}
of null infinity $\mathcal{I}^+$ (see Chapter~17 of~\cite{CK}). One can thus again formulate
the laws of gravitational radiation in our setting,  and our results in particular
control the total displacement of test particles at infinity, involved in the Christodoulou
memory effect~\cite{Christmem} (see already Chapter~\ref{conclusionsection} of the present work).
We note that the final parameter $M_{\rm final}$  has the additional interpretation of  \emph{final Bondi mass}.
Just as with the double null gauges themeselves,  this parameter is again only determined teleologically.

\subsection{The asymptotically stable ``submanifold'' $\mathfrak{M}_{\rm stable}$ and the problem of Kerr}
\label{orbvsasympt}
\setcounter{equation}{0}

The necessity of restricting to an asymptotically stable ``submanifold'' $\mathfrak{M}_{\rm stable}$\index{initial data!moduli space!$\mathfrak{M}_{\rm stable}$, asymptotically stable codimension-$3$ ``submanifold'' of $\mathfrak{M}$} 
arises from our requiring approach
to Schwarzschild in condition (iii) of Theorem~\ref{maintheoremintro}.
For instance,
the induced initial data corresponding to the Kerr solutions themselves
with fixed mass $M$ in standard double null form
evidently do not satisfy (iii) unless one takes $a=0$. 
Moreover, from Section~\ref{SchKerrgaugefreed}, these 
 can be represented as a $3$-dimensional subspace of the moduli
space of initial data, passing through exact  Schwarzschild data.
It is thus natural to expect that the statement of our theorem holds if and only if the data lie on a
``submanifold'' $\mathfrak{M}_{\rm stable}$ of \emph{codimension} $3$. Indeed, this can be understood
in the context of the more general conjectural nonlinear stability of the Kerr family (see
already Section~\ref{kerrstatements}):
initial data outside of  $\mathfrak{M}_{\rm stable}$ should necessarily evolve to a Kerr with $a\ne0$.
Thus, lying on  $\mathfrak{M}_{\rm stable}$ would be a necessary condition for (iii).

We note that, given the necessity of our restriction on data, 
a difficulty of proving Theorem~\ref{maintheoremintro}
is that there is no way to identify in general which initial data satisfy (iii)
other than evolving these data to the future under~\eqref{vaceqhere}. Thus, the asymptotically stable ``submanifold''  
$\mathfrak{M}_{\rm stable}$ cannot
be identified explicitly in the space of initial data, but must also be constructed teleologically, along
with the teleological double null gauges and the parameter $M_{\rm final}$. This is precisely what we do
in the present work. Indeed,
given that the fundamental double null gauges  already have to be constructed teleologically (in fact,
already in linear theory; see 
Section~\ref{linstabintint_sec}), it turns out that 
the additional problem of constructing $\mathfrak{M}_{\rm stable}$ does not in practice present a major conceptual difficulty
(although it gives rise to a number of interesting technical issues; see  Section~\ref{firstremarksintint_sec}).
We note moreover that the codimensionality assumption can be understood very concretely, because
the moduli space of data can be naturally decomposed as a union of disjoint $3$-parameter families (one of which is precisely the one referred to above, i.e.~passing through exact Schwarzschild  
and representing the Kerr family). 
By construction, $\mathfrak{M}_{\rm stable}$
will then contain exactly one data set from each $3$-parameter family (see already Section~\ref{dataintro}).
(In view of this convenient description, we shall not be concerned with the regularity of this set as a subset of moduli space, hence our use of quotations around the word ``submanifold''.)

Though statement (iii) requires our codimensionality assumption, 
statements (i) and (ii) alone should not. (Indeed, according to weak cosmic censorship,
statement (i) is conjecturally true for generic initial data while
statement (ii)  would in particular follow from
the more general stability of Kerr conjecture to be discussed in Section~\ref{kerrstatements}.)
One might wonder then, why not try and prove (i) and (ii) directly in a  neighbourhood of Schwarzschild,
i.e.~without proving the accompanying condition (iii) and the corresponding difficulty of
identifying the ``submanifold'' $\mathfrak{M}_{\rm stable}$ that satisfies this condition?
It is a fundamental limitation of current technology concerning the
analysis of the Einstein vacuum equations~\eqref{vaceqhere}, 
however, that one can only hope to infer ``global'' existence type statements like (i) and
orbital stability type statements like (ii) as a corollary of a much stronger
asymptotic stability statement like (iii). This limitation originates in  the \emph{supercriticality} of 
equations~\eqref{vaceqhere} with respect to natural conserved quantities
like the ADM mass (which fail to be coercive in any case). Thus, it is only
by  identifying the final state to which solutions asymptote as 
perturbations  ``radiate away'' (to null infinity $\mathcal{I}^+$ and---in
the black hole case---to the horizon $\mathcal{H}^+$) that one can hope to control
the total ``backreaction'' of perturbations  and  prove any form of 
nonlinear stability, even simply orbital stability. Moreover, to integrate this backreaction in time
one moreover requires that the decay rate of
these perturbations is sufficiently
fast. We shall see this difficulty
already in our discussion of the non-linear stability of Minkowski space, 
in Section~\ref{nonstabminkintint_sec}.
(We note in contrast, however, that \emph{under sufficient symmetry}, the presence
of a black hole effectively \emph{breaks} the supercriticality of the equations, 
and one can indeed prove orbital stability directly, without a full asymptotic stability; see some results discussed in Section~\ref{undersym_intsec}.)

In view of the above remarks, to study any form of non-linear stability (without symmetry),
one requires quantitative decay estimates  for 
the linearisation of~\eqref{vaceqhere} \emph{around the final state}.
Thus, to show (i) and (ii) in a neighbourhood of Schwarzschild without restriction, one would need to use
an asymptotic linear stability analysis around the very slowly rotating Kerr metrics $|a|\ll M$. 
Carrying this out is indeed in principle now possible (see already the discussion in
Section~\ref{kerrstatements}). The advantage in 
constructing the ``submanifold'' that  satisfies (iii), however, is that we 
only require results concerning  linear stability of the Kerr family
 \emph{around Schwarzschild},
and this is technically slightly easier than the full Kerr problem and moreover already
available from our previous~\cite{holzstabofschw} in a form readily applicable\footnote{readily applicable  
because the energy-based  linear estimates of~\cite{holzstabofschw}
 can be applied directly to the dynamical spacetime itself; see the discussion in Section~\ref{nonstabminkintint_sec}
 in the context of the nonlinear stability of Minkowski space} 
to the nonlinear problem. This important technical simplification compensates for the extra difficulty
of teleologically constructing $\mathfrak{M}_{\rm stable}$.
We turn now to 
a brief review of the linear stability results of~\cite{holzstabofschw}.

\section{The linear stability of Schwarzschild in double null gauge}
\label{linstabintint_sec}

The proof of Theorem~\ref{maintheoremintro} 
will be based on our previous self-contained analysis of linear stability
in~\cite{holzstabofschw}, carried out entirely in physical space. We review this now.

\subsection{The linearised Einstein equations in double null gauge}
As with  our non-linear Theorem~\ref{maintheoremintro},   the problem of \emph{linear}
stability considered in~\cite{holzstabofschw} was formulated in  double null gauge.

That is to say, one fixes a double null coordinate system $(u,v)$ around Schwarzschild,
one embeds the Schwarzschild metric in a $1$-parameter family of solutions
of~\eqref{coupsys} 
and one derives a set of equations 
linearising the system~\eqref{coupsys}. (Note that
the linearisation procedure is covariant with respect to 
change of ambient double null coordinates on Schwarzschild and
can be expressed with respect to Eddington--Finkelstein normalised coordinates
$(u,v)$, provided one considers rescaled quantities at the horizon $\mathcal{H}^+$
so as to be regular.)
This results  
in a closed system for \emph{linearised quantities} $\Olino$, $\glin$, $\xli$, etc.:
\index{linearised theory!metric coefficients!$\Olino$}
\index{linearised theory!metric coefficients!$\glin$}
\index{linearised theory!metric coefficients!$\xli$}
\begin{equation}
\label{coupsysline}
\begin{tabular}{ |c|c| } 
\hline
linearised transport& $\Omega\slashed\nabla_4\left(\frac{\glinto}{\sqrt{\slashed g}}\right)=\otx -\slashed{\rm div}\bmlin,\qquad \slashed\nabla_4\left(\Omega^{-1}\xlin\right)+\Omega^{-1}({\rm tr}\underline{\chi})\xlin =
-\Omega^{-1}\alin$, \ldots
\index{linearised theory!connection coefficients!$\otx$}\index{linearised theory!connection coefficients!$\xlin$} 
\index{linearised theory!curvature components!$\alin$}
 \\
\hline
linearised elliptic& $\slashed{\rm div} \xlin= \frac12\Omega^{-1}\slashed\nabla\otx-\blin +\ldots$, \ldots
\index{linearised theory!curvature components!$\blin$}
 \\
\hline
linearised
hyperbolic &$\slashed\nabla_3\alin+\frac12{\rm tr}\underline\chi\alin +2\underline{\hat\omega}\alin
=-2\slashed{\mathcal{D}}^*_2\blin -3\rho\xlin,  \qquad
 \slashed\nabla_4\blin+2{\rm tr}\chi\blin -\hat\omega\beta =
\slashed{\rm div} \alin$\\
\hline
\end{tabular}
\end{equation}
Here $\Omega$, ${\rm tr}\chi$, (without $(1)$ superscripts) etc., denote Schwarzschild background values,
computable from~\eqref{Schwstandform}. 
The system~\eqref{coupsysline} can be shown to be well-posed,
where, in accordance with the previous section,
initial data for~\eqref{coupsysline} are prescribed on the Schwarzschild
null hypersurfaces 
\begin{equation}
\label{newnullsurfintro}
C_{\rm out}:=
\{0\}\times[0,\infty)\times \mathbb S^2,\qquad C_{\rm in}:=  [0,\infty]\times \{0\}\times \mathbb S^2.
\end{equation}
(Here, the inclusion of $\infty$ in the $u$-range is meant to signify that the data are regular at the horizon
when covariantly changed to Kruskal coordinates.)
Since the equations are linear, one can immediately infer
the existence of a \emph{global} solution on the Schwarzschild
exterior $J^+(C_{\rm out}\cup \underline{C}_{\rm in})\cap (J^-(\mathcal{I}^+)\cup \mathcal{H}^+)=
\{0\le u\le \infty\}\cap \{0\le v< \infty\}$, including the horizon $\mathcal{H}^+$.

The  diffeomorphisms which
preserve the double null form~\eqref{doublenull}, discussed  already in Section~\ref{mainresintint_sec},
give rise in linear theory to an infinite dimensional family
of solutions of~\eqref{coupsysline}, the so-called (residual)
\emph{pure gauge solutions}. An example of such a pure gauge solution
is one generated by an arbitrary smooth function $\flin(v,\theta,\phi)$ \index{linearised theory!pure gauge solutions!$\flin(v,\theta,\phi)$,
smooth function generating pure gauge solution in linear theory}
as follows:
\begin{equation}
\label{puregaugeexamp}
2\Olin=\Omega^{-2}\partial_v(\flin\Omega^2),\qquad \glinh=-4r\slashed{\mathcal{D}}_2^\star\slashed\nabla\flin, \qquad
\ldots
\end{equation}

The $4$-dimensional Kerr family itself, as parametrised in our double null gauge, 
when linearised,
gives rise to an additional $4$-dimensional family of explicit stationary solutions 
of the system~\eqref{coupsysline}, the \emph{linearised Kerr solutions}.
This subspace can in turn be decomposed into a $1$-dimensional
linearised Schwarzschild
family parametrised by $\mathfrak{m}\in \mathbb R$: \index{linearised theory!parameters!$\mathfrak{m}$, linearised Schwarzschild parameter}
\begin{equation}
\label{linschw}
2\Olin= -\mathfrak{m}, \qquad tr_{\slashed{g}}\glin = -2\mathfrak{m}, \qquad
\glinh=0, \qquad \bmlin =0
\end{equation}
and a $3$-dimensional family of \emph{fixed-mass linearised Kerr solutions},
spanned by a basis
\begin{equation}
\label{linker}
\Olin=0, \qquad tr_{\slashed{g}}\glin=0, \qquad \glinh=0, \qquad \bmlin^A=
\frac{4M}r\epsilon^{AB}\partial_BY^{1}_m,
\end{equation}
where $Y^{1}_m$,
$m=-1,0,1$ denote the three standard $\ell=1$ spherical harmonics.
\index{sphere!spherical harmonics!$Y^1_m$, $m=-1,0,1$, $\ell=1$ spherical harmonics}

\subsection{Statement of the linear stability theorem~\cite{holzstabofschw}}
\setcounter{equation}{0}

We may now state 
the main result of~\cite{holzstabofschw}.
\begin{testtheorem}
[Linear stability of the Kerr family around Schwarzschild~\cite{holzstabofschw}]
\label{linstabthemintro}
For all characteristic initial data prescribed on~\eqref{newnullsurfintro},
the arising solution of the linearised Einstein equations~\eqref{coupsysline}
around Schwarzschild remains uniformly bounded in the exterior region 
$\{0\le  u\le  \infty\}\cap\{0\le v< \infty\}$, including the horizon $\mathcal{H}^+$, in
 terms of its initial data. 
Moreover, after adding a pure gauge solution,
\underline{which itself is quantitatively controlled by the data},
the solution approaches inverse polynomially a standard
linearised Kerr metric as $u\to \infty$ and $v\to\infty$,
in particular along $\mathcal{I}^+$ and $\mathcal{H}^+$.

In particular, there is a codimension-$3$ subspace of initial data for which
all solutions approach a linearised Schwarzschild metric~\eqref{linschw}.
\end{testtheorem}

One may already
compare the above with the statement of   Theorem~\ref{maintheoremintro}
in Section~\ref{mainresintint_sec}. We note that the boundedness statement in
Theorem~\ref{linstabthemintro},
at top order energy, is again without loss in derivatives.
We emphasise that the pure gauge solution, referred
to in the statement of Theorem~\ref{linstabthemintro}, which must be added 
to ensure asymptotic stability,
cannot be
 determined explicitly from initial data. This fact is the analogue
of the teleological nature of the normalisation of the double null gauge
of  Theorem~\ref{maintheoremintro}. The underlined statement 
in Theorem~\ref{linstabthemintro} corresponds 
to the property, mentioned already, that the coordinate
representation of the initial data in the future-normalised coordinate
systems of  Theorem~\ref{maintheoremintro} can itself be appropriately bounded. 
(We will describe already in Section~\ref{bounded_init_intro} 
the significance of the underlined statement above
for the proof of non-linear stability.) 
On the other hand,  in contrast to Theorem~\ref{maintheoremintro},
 the ``location'' of the horizon in Theorem~\ref{linstabthemintro},
as well as the standard 
linearised Kerr solution in the span of~\eqref{linker} referred to in the statement, can be 
explicitly determined on the basis of initial data. 
In particular, this allows us to 
  effectively restrict to $\delta=0$ in the definition~\eqref{newnullsurfintro} of
$\underline{C}_{\rm in}$, in comparison with~\eqref{initialcones}, and
the codimension-$3$ subspace
of the last line of the statement of Theorem~\ref{linstabthemintro} is explicit.

\subsection{Proof of linear stability}
\label{proofoflinstabinintro}
\setcounter{equation}{0}

Let us give a brief overview of the proof of
Theorem~\ref{linstabthemintro}, for we will later have to repeat all these arguments in
the more elaborate nonlinear setting.

\subsubsection{The gauge invariant quantities and the Chandrasekhar-type transformation}
\label{ginvquanCha}

The  first advantage of  double null gauge at the linear level is that
it allows one to apply insights from the Newman--Penrose formalism~\cite{newmanpenrose}.
In particular, it was shown already in~\cite{bardeen1973} that the linearised
extremal curvature components $\alin$ and $\ablin$ \index{linearised theory!curvature components!$\ablin$}
decouple from the system
\eqref{coupsysline} and satisfy  wave equations, which in the case of $\alin$ is here
written in tensorial form:
\begin{equation}
\label{teukequationone}
\Omega \nablaslash_4 \Omega \nablaslash_3 (r \Omega^2 \alin)
		+
		\frac{2\Omega^2}{r^2} r^2 \Dslash_2^* \divslash (r\Omega^2 \alin)
	+
		\frac{4}{r} \left( 1 - \frac{3M}{r} \right) \Omega \nablaslash_3 (r \Omega^2 \alin)
		+
		\frac{6M\Omega^2}{r^3} r\Omega^2\alin
=0.
\end{equation}
(If $\alin$ and $\ablin$ are defined with respect to the algebraically
special frame, then an analogous equation can be derived in the Kerr
case, as was shown by Teukolsky~\cite{teukolsky1973}. We will refer
in general to equation~\eqref{teukequationone} as the \emph{Teukolsky
equation}.)  Note that $\Dslash_2^* \divslash = - \frac{1}{2} \Deltaslash  + K$, and thus
the left hand side of~\eqref{teukequationone} indeed
defines a wave operator applied to $\alin$, though with unfamiliar first order terms.
The quantities $\alin$ and $\ablin$ parametrise the 
\emph{gauge-invariant} part of the solution, in the sense that these quantities
vanish in the case of pure gauge solutions like~\eqref{puregaugeexamp}
and standard linearised Kerr solutions like~\eqref{linker},
and conversely,
 any admissible solution with $\alin=\ablin=0$ identically is in fact the sum of
a pure gauge  and a standard linearised Kerr solution.

It turns out, however, that the above equation~\eqref{teukequationone} is difficult
to analyse directly in view of its first order terms.
The approach taken in~\cite{holzstabofschw} was to consider
the higher order quantities
\begin{equation}
\label{Pquantfirstinstance}
\Plin = -\frac12 r^{-3}\Omega^{-1}\slashed\nabla_3(r^2\Omega^{-1}\slashed\nabla_3(r\Omega^2\alin)), \qquad 
\Pblin = -\frac12 r^{-3}\Omega^{-1}\slashed\nabla_4(r^2\Omega^{-1} \slashed\nabla_4
(r\Omega^2\ablin)).
\index{linearised theory!gauge invariant hierarchy!$\Plin$, higher order gauge invariant quantity satisfying Regge--Wheeler equation}
\index{linearised theory!gauge invariant hierarchy!$\Pblin$, higher order gauge invariant quantity satisfying Regge--Wheeler equation}
\end{equation}
These quantities satisfy the so-called \emph{Regge--Wheeler equation},
written below (again in tensorial form):
\begin{equation}
\label{reggewheel}
\Omega \nablaslash_4 \Omega \nablaslash_3 (r^5 \Plin)
		+
		\frac{2\Omega^2}{r^2} r^2 \Dslash_2^* \divslash (r^5 \Plin)
		+
		2\Omega^2 \left( 1 - \frac{3M}{r} \right) r^5 \Plin
		=0. 
\end{equation}
(The Regge--Wheeler equation first appeared in the context of 
the so-called \emph{metric perturbations} approach due to~\cite{Regge}, where it governed,
however,
only ``half'' of the gauge invariant part of the perturbations. It is remarkable
that the very same equation is satisfied by the higher order quantities~\eqref{Pquantfirstinstance}. 
The relations~\eqref{Pquantfirstinstance}
are physical space reformulations of fixed frequency transformations
originally discovered by Chandrasekhar~\cite{Chandraschw}.)

In contrast to the Teukolsky equation~\eqref{teukequationone},
the Regge--Wheeler equation~\eqref{reggewheel} can be analysed in much the same
way as the scalar wave equation 
\begin{equation}
\label{scalarwav}
\Box_g\psi =0,
\end{equation}
which itself  had been the object of much recent work, 
see~\cite{Sterbenz, DafRod2} for
Schwarzschild and~\cite{partiii} for the Kerr case in the full subextremal range 
of parameters $|a|<M$. (For a detailed discussion of~\eqref{scalarwav} 
on Schwarzschild and in particular
the role of the \emph{red-shift} at the event horizon $r=2M$ and
the trapped null geodesics associated to  the \emph{photon sphere} $r=3M$, 
see~\cite{Mihalisnotes}.)
In particular, following previous results for~\eqref{scalarwav},
one can show uniform
boundedness and integrated decay results for~\eqref{reggewheel}.
(See also the treatments in~\cite{BlueSoffer, Holzegelspin2}.)
From this, boundedness and integrated decay for~\eqref{teukequationone}
follow by  integrating the relations~\eqref{Pquantfirstinstance}
\emph{as transport equations}, after suitable multiplication by weighted
quantities. (Note that the boundedness statements at the energy level do not lose derivatives.)
\emph{Inverse polynomial decay} for all quantities $\Plin$, $\alin$, etc.~then follows 
via the $r^p$-weighted
energy hierarchy of~\cite{DafRodnew}.

\subsubsection{Fixing the gauge: initial data normalisation and the linearised
Kerr modes.}
The stability statement of Theorem~\ref{linstabthemintro} 
does not end with showing boundedness and decay
for the gauge invariant quantities above. We must still show  uniform
boundedness and decay to a standard linearised Kerr solution for the remaining
(gauge dependent)
linearised quantities in~\eqref{coupsysline}.

For this, the first task is to normalise the initial data. In the context
of linear theory, this corresponds to adding a pure gauge solution 
(e.g.~\eqref{puregaugeexamp})
such that certain quantities are fixed. 
For instance, one can  arrange so as to fix the quantity
\begin{equation}
\label{horloca}
\otx=0
\end{equation}
at the initial sphere $(u,v)=(\infty,0)$ of $\mathcal{H}^+$.
It turns out that condition~\eqref{horloca} is then preserved
along $\mathcal{H}^+$ by  evolution under~\eqref{coupsysline}.
This can be thought of as ensuring that the event horizon of the perturbed
spacetime coincides with $\mathcal{H}^+$, i.e.~that of
the background Schwarzschild.
In this sense, the ``location'' of the horizon is not teleological in linear theory,
but can be explicitly deduced from the data (and thus one can effectively set the parameter $\delta$ appearing in 
Theorem~\ref{maintheoremintro} to vanish).

Let us note moreover that
in the process of gauge normalisation, the $\ell=0$ and $\ell=1$ spherical harmonic modes
play an anomalous role.
It turns out that solutions supported only on $\ell=0$ and $\ell=1$ can be written
as the sum of a pure
gauge solution (e.g.~\eqref{puregaugeexamp}) and a 
standard linearised Kerr solution (e.g.~\eqref{linschw} or \eqref{linker}). Thus, by
adding a pure gauge solution so as  to fix certain quantities
at the initial sphere $(u,v)=(\infty,0)$ of the horizon $\mathcal{H}^+$,
for instance
\[
\Omega^{-2}\otxb_{\ell=0,1}=0, \qquad  \left(\rlin+\slashed{\rm div}\,\elin\right)_{\ell=1}=0  ,
\index{linearised theory!connection coefficients!$\otxb$}
\index{linearised theory!connection coefficients!$\rlin$}
\index{linearised theory!connection coefficients!$\elin$}
\]
we can arrange so as to be left only with 
a linearised Kerr solution, which will be precisely
that appearing in the statement
of the Theorem~\ref{linstabthemintro}.
Note that, as with the issue of the location of the horizon discussed above, 
this means that that the final
linearised Kerr metric is \underline{not} teleological in linear theory, but
can be \emph{explicitly} deduced from the initial data.

\subsubsection{Estimating the gauge dependent quantities: boundedness}
\label{lineartheoryboundedness}

Having normalised the gauge by adding an appropriate pure gauge solution, we now turn 
to the \emph{boundedness} statement of Theorem~\ref{linstabthemintro}.

The essential point is that one can 
order the remaining linearised quantities hierarchically, 
starting from the outgoing and ingoing shears $\xlin$ and $\xblin$, 
so that the transport equations~\eqref{coupsysline}  contain only the gauge
invariant quantities $\Plin$, $\alin$, $\Pblin$, $\ablin$ estimated already above, or else quantities lower in the hierarchy. Thus the situation resembles the problem of estimating $\alin$, respectively
$\ablin$, from
$\Plin$, respectively $\Pblin$, by integration of the relations~\eqref{Pquantfirstinstance}.
Unlike the case of the gauge invariant quantities, however, as we shall see below, 
this procedure will at this stage only lead to
boundedness.

We can illustrate the issue already with the first step of the hierarchy.
The ingoing shear $\xblin$\index{linearised theory!connection coefficients!$\xblin$} satisfies
 the transport equation
\begin{equation}
\label{transportforlondon}
\slashed\nabla_3(\Omega^{-1}\xblin)+\Omega^{-1}(\rm tr \underline\chi) \xblin=
-\Omega^{-1}\ablin,
\end{equation}
which only contains $\ablin$ on its right hand side.
Already, to obtain boundedness for $\xblin$ upon integration, there is an issue,
as the appropriate integrating factor for~\eqref{transportforlondon}
would require estimating $r^2 \underline{\hat{\chi}}$ initially, which is however generally unbounded along an outgoing cone. 
To resolve this issue, one introduces an auxiliary renormalised quantity 
$\Ylin$ which is essentially a sum of a second order (elliptic) angular operator applied to $\underline{\hat{\chi}}$ and the (decaying) gauge invariant quantity $\pblin:=\frac12r^{-1}\Omega^{-2}\slashed\nabla_4(r\Omega^2\ablin)$, given by
\begin{equation}
\label{Ylindef}
\Ylin= r^2\slashed{\mathcal{D}}^\star_2\slashed{\rm div}(\Omega^{-1}r\xblin)-\Omega^{-1}r^{4}\pblin.
\end{equation}
The quantity $\Ylin$ can be shown to be bounded initially and  
satisfies a transport equation without integrating factor and with integrable right hand side
\begin{equation}
\label{transportforY}
\slashed\nabla_3 \Ylin =\frac12r{\rm tr}\underline\chi\Omega^{-1}r^3\pblin +3M\Omega^{-1}r\ablin .
\end{equation} 

Integration of~\eqref{transportforY} from initial data  will now indeed give rise   to a finite bound for $\Ylin$ (and thus for $\xblin$), but, this bound is dominated at large $v$ values by a flux associated to $\ablin$ which in general will be non-zero.
Thus, the quantity $\Ylin$, and consequently $\xblin$, \emph{will not in general decay} in our gauge.
The solution to the problem described above is to integrate~\eqref{Ylindef}
\emph{backwards}. For this, however,
we  will need to renormalise the gauge teleologically. 

Before turning to teleological normalisation in Section~\ref{forteleologyinintro} below, let us consider also 
the outgoing shear $\xlin$. For this quantity, 
there is a corresponding issue associated with the horizon $\mathcal{H}^+$,
as the transport equation for the regular quantity $\Omega \xlin$ 
\begin{equation}
\label{mustdifferentiatethisone}
\slashed\nabla_4(\Omega \xlin)+(\rm tr \chi)(\Omega \xlin)-2\hat\omega\Omega\xlin =
-\Omega\alin
\end{equation}
 appears to be
``blue-shifted''  when trying to integrate it
forwards in time near $\mathcal{H}^+$, i.e.~the third term on the left hand side
of~\eqref{mustdifferentiatethisone} appears to drive exponential growth! Such exponential growth does
not actually occur, however. In fact, 
exactly along the horizon $\mathcal{H}^+$, it turns out that $\Omega\xlin$ is gauge invariant and fortuitously
can be controlled as part of 
a conserved flux. (See~\cite{Holzegelfluxes} for further elaboration of these conservation
laws.) Given this, in order to estimate $\xlin$ in a neighbourhood of $\mathcal{H}^+$,  it now
suffices to remark that the ``blue-shift'' 
problem of~\eqref{mustdifferentiatethisone} is
cured upon successive commutation by $\Omega^{-1}\slashed\nabla_3$: Commuting
once, one obtains
a ``no-shifted'' quantity, and commuting again, a ``red-shifted'' quantity.
Thus, decay bounds for
higher order $\Omega^{-1}\slashed\nabla_3$ derivatives of $\Omega\xlin$ can be obtained by integrating
the commuted version of~\eqref{mustdifferentiatethisone} forwards in time,
and estimates for $\Omega\xlin$ itself can then be retrieved upon integration along ingoing cones
backwards from the horizon $\mathcal{H}^+$, given the estimate for $\xlin$ on  $\mathcal{H}^+$ itself arising from the flux.

\subsubsection{Estimating the gauge dependent quantities: future-normalisation and decay}
\label{forteleologyinintro}
In the context of the linear theory, teleological normalisation amounts 
to adding a pure gauge solution
related to the  original solution (i.e.~in the ``initial data normalised gauge'') 
by its value \emph{somewhere to the future}, 
so as to normalise certain quantities to vanish there.

It turns out to be sufficient for our purposes here to require that the linearised
metric quantity
$\Olin$ vanish identically \emph{along the entire event horizon $\mathcal{H}^+$:}
\begin{equation}
\label{teleologicalnorm}
\Olin=0.
\end{equation}
Relation~\eqref{teleologicalnorm} can indeed be arranged
by adding a pure gauge solution (at the expense
of relaxing a previous similar normalisation along the initial cone $C_{\rm out}$). 
Let us note that the pure gauge
solution that must be added \emph{is itself quantitatively
bounded from initial data}. This follows from the boundedness statement
just obtained!

One can  now revisit the integration of the transport  equations~\eqref{coupsysline} for various
gauge-dependent quantitites, for instance~\eqref{transportforY} for $\Ylin$ (and thus governing
$\xblin$), by integrating these equations 
backwards, with the condition~\eqref{teleologicalnorm} along
$\mathcal{H}^+$
ensuring that the \emph{future} boundary
terms are controlled. This allows one to indeed inherit decay properties from
those already obtained for the gauge invariant quantities.

Eventually, combining the process again with  the $r^p$-weighted
energy hierarchy of~\cite{DafRodnew},
this allows one to obtain 
inverse polynomial decay for all quantities,
completing the proof of Theorem~\ref{linstabthemintro}.

Let us emphasise that
the fact that the added pure gauge solution which assured~\eqref{teleologicalnorm}
can itself be quantitatively bounded
is fundamental for the nonlinear stability proof to be discussed in Section~\ref{firstremarksintint_sec}.

Finally, let us note   that the necessity of future normalisation is not
new to the Schwarzschild problem but is a  feature of double null gauge,
and more generally, geometric gauges governed by transport 
equations. In particular, as we shall discuss in Section~\ref{nonstabminkintint_sec},
the issue of future normalisation  arises (already at the linear level) in the context of the gauge used in~\cite{CK}
for the stability of
Minkowski space, where one defined  null cones emanating from points on a geodesic.
The fundamental
difference in the black hole case is that since one cannot normalise from such a geodesic, one
must instead also teleologically choose a sphere surrounding the horizon. 
This is exactly one of the conditions accomplished above at the linearised level by our future-normalised gauge of~\cite{holzstabofschw}.
We note already that the nonlinear versions of these will be characterized by analogous equations and constructed by iteration around the linear construction (see already Section~\ref{two_gauges_intro}).

\subsection{The Kerr case}
\setcounter{equation}{0}

In the Kerr case,  the decoupled Teukolsky
equation referred to above
can again be shown to govern the gauge-invariant part of
the perturbations~\cite{wald1978construction}.
To the difficulties of \eqref{teukequationone} discussed previously, however, one must now add
the phenomenon of \emph{superradiance}, which occurs already for~\eqref{scalarwav}.
Thus, even mode stability is highly non-trivial,
shown by~\cite{Whiting, SRT} in the subextremal and~\cite{daCosta:2019muf} in the extremal case.
Recently, the argument of~\cite{holzstabofschw} showing quantitative boundedness and 
decay for the Teukolsky equation has indeed been generalised, first to the
very slowly rotating case $|a|\ll M$ in~\cite{DHRteuk,Ma:2017yui}, and 
then to the full subextremal range $|a|<M$ in~\cite{RitaShlap},
 implementing the first part
of the argument of Section~\ref{proofoflinstabinintro}. 
Here, the boundedness statement at the energy level is proven without loss of 
derivatives and can in particular
be seen to quantify the amplifying strength of superradiance.
With these results, the full linear stability of Kerr in analogy to~\cite{holzstabofschw}
can in principle be obtained, in fact, for the entire subextremal range $|a|<M$.
We will discuss the full \emph{non-linear} stability of Kerr in 
Section~\ref{kerrstatements}.

\subsection{Other approaches}
\setcounter{equation}{0}

The work~\cite{holzstabofschw} was the first complete treatment of linear stability
of Schwarzschild in a well-posed gauge.
Let us note, however, that  double null gauge is not the only such gauge in which
linear stability can be addressed. In particular, in the  recent~\cite{Johnson:2018yci}, 
a version of Theorem~\ref{linstabthemintro} 
has been shown in a ``generalised harmonic gauge''. (For more
on harmonic gauge, see already the discussion in 
Section~\ref{nonstabminkintint_sec}.) See also~\cite{2018arXiv180303881H,
2017arXiv170202843H} and, for some results on 
Kerr, see~\cite{Andersson:2019dwi, Hafner:2019kov}.
(Note that for
the (Schwarzschild) Kerr--de Sitter cases, 
linear stability results are contained in~\cite{hintz2016global}; see 
Section~\ref{instructcomp}
for further discussion.)

\section{Previous results on nonlinear stability}
\label{resnonstabintint_sec}

Before introducing  the main ideas of the proof of Theorem~\ref{maintheoremintro}, 
we describe briefly
previous results relevant to the \emph{non-linear} stability of asymptotically flat solutions of~\eqref{vaceqhere}. 

We begin in Section~\ref{nonstabminkintint_sec}
with a discussion of the non-linear stability of
Minkowski space, followed in Section~\ref{undersym_intsec} 
by a discussion of stability results for non-trivial 
solutions, but under the restrictive assumption of various symmetries. We
review some relevant non-linear model
problems in Section~\ref{nonlinmodel} and finally, we shall discuss 
in Section~\ref{scatteringconstruction} our previous ``scattering theoretic'' construction of
solutions of~\eqref{vaceqhere} representing non-trivial dynamical vacuum black holes. 
(For comparison, we shall comment briefly on results
for $\Lambda\ne 0$ in Section~\ref{instructcomp}.)

\subsection{The nonlinear stability of Minkowski space}
\label{nonstabminkintint_sec}

The study of non-linear stability of solutions of~\eqref{vaceqhere} without symmetry was initiated
with the monumental proof by Christodoulou and Klainerman~\cite{CK} of the stability of Minkowski space, the trivial
solution of~\eqref{vaceqhere} (corresponding to~\eqref{Schwmet} 
in the case $M=0$). Let us discuss briefly the background of this  result.

\subsubsection{Harmonic gauge}
\label{sectionwithnullconddisc}

To understand the main issues, let us begin the discussion from the perspective
of the simplest gauge to consider, namely
 \emph{harmonic gauge}
\begin{equation}
\label{harmgau}
g^{\mu\nu}\Gamma^{\alpha}_{\mu\nu} =0.
\end{equation}
Expressed in the gauge~\eqref{harmgau},
 the  vacuum equations~\eqref{vaceqhere}  reduce
 to a system of quasilinear wave equations
 \begin{equation}
 \label{endok}
 \Box_g g^{\mu\nu} = {Q}^{\mu\nu}(g, \partial g),
 \end{equation}
 where $Q^{\mu\nu}$ is a nonlinear expression quadratic in $\partial g$,
 which when linearised around Minkowski space (see already~\cite{Einsteinwaves}) 
 yield simply
the classical wave equation
\begin{equation}
\label{classwav}
\Box \psi = 0
\end{equation}
for each linearised metric component $\psi={\justglin}^{\mu\nu}$.

Linear stability of Minkowski space in harmonic gauge  follows immediately from classical
boundedness and decay results concerning~\eqref{classwav}.
Nonetheless, proving \emph{non-linear} stability of Minkowski space turned out to be quite difficult! 
To understand why, recall that we have already remarked in Section~\ref{orbvsasympt} on the supercriticality
of~\eqref{vaceqhere} and the resulting necessity of showing not just orbital stability
but the full  \emph{asymptotic stability} of Minkowski space, 
hoping moreover that the
quantitative decay rates back towards the Minkowski metric are indeed sufficiently strong 
so as to ensure
that non-linear terms can be understood as error terms which can be integrated
in time.

For dimensions $4+1$ and higher, one can indeed use 
decay of the linear equation to show relatively easily 
the nonlinear stability of higher dimensional Minkowski space. 
The modern way to do this is via the \emph{vector field method}~\cite{Klainhistory}, 
where energy estimates are applied
\emph{directly} to the nonlinear equation~\eqref{endok} commuted
with weighted commutation vector fields. (Note that one thus does not use the linear
theory proper, but a more robust version of linear estimates that can be applied
to the non-linear equation itself.  This is essential for obtaining a true stability
result without loss of deriatives.)
Sufficient polynomial pointwise decay for
lower order terms can 
be inferred by weighted 
Sobolev estimates, so as to indeed absorb via integration the
non-linear terms. One uses in a fundamental way that the decay of solutions
of~\eqref{classwav} on say $\mathbb R^{4+1}$ 
satisfies $\sup_{x\in \mathbb R^4} |\partial\psi(x,t)| \le Ct^{-\frac32}$,
and the latter decay rate is integrable in $t$.
See~\cite{Loizelet} where this argument for the  non-linear
stability of higher dimensional Minkowski space
is treated in detail.

In the physical case of $3+1$ spacetime dimensions, however, 
solutions to~\eqref{classwav}  decay
along the outgoing null cones  only  as $r^{-1}$.
This slow decay means that the quadratic nature of the non-linearities on the right hand
side of~\eqref{endok}
 is insufficient by itself to
ensure non-linear stability: one must
identify \emph{special structure} in the nonlinear terms. 
This difficulty is already apparent in comparing the following two
model
equations:
\begin{equation}
\label{modelproblems}
\textrm{(i)\ } \Box\psi =(\partial_t\psi)^2, \qquad \textrm{(ii)\ } \Box\psi = -(\partial_t\psi)^2+\sum_{i=1}^3 (\partial_{x_i}\psi)^2.
\end{equation}
John~\cite{fritzblowup} 
showed that (i) of~\eqref{modelproblems} 
exhibits blow up for solutions arising from arbitrary small compactly
supported data, while for (ii), a transformation due to
Nirenberg (see~\cite{klainglobex}) immediately yields global existence. 
The situation was clarified with the formulation of the \emph{null condition} of Klainerman~\cite{KlNull}, which identifies a wide class of ``good'' nonlinearities for which global existence
holds.

\subsubsection{Double null gauge: capturing the null condition for the Einstein equations} 
\label{capturingnullcon}
Unfortunately, as already discussed in~\cite{choqinst}, 
what was later understood as the
null condition is \emph{not} in fact satisfied by the reduced Einstein equations~\eqref{endok} in
harmonic gauge~\eqref{harmgau}. The  approach taken in~\cite{CK} was to completely abandon
harmonic gauge, and introduce   an appropriate geometric
formulation of the equations~\eqref{vaceqhere} in terms of the \emph{structure
relations} associated to a foliation by maximal hypersurfaces and outgoing null cones,
in which an analogue of the  null condition is now captured.
The closely related  double
null gauge formulation used in the present work
 was subsquently developed in~\cite{Chr, KN} and also captures
this null condition. 

The understanding of the  null condition uncovered in the above works can be systematised
with the help of the following schematic notation:
One denotes Ricci coefficients by $\Gamma_p$\index{schematic notation!$\Gamma_p$, schematic notation for Ricci coefficients}, and curvature components by $\mathcal{R}_p$\index{schematic notation!$\mathcal{R}_p$, schematic notation for curvature components},
 where the $p$-subscript\index{schematic notation!$p$, subscript related to $r^p$ weights} denotes  the $r^p$ weighted bound one
 expects to propagate. For instance, the Ricci coefficient $\hat{\underline{\chi}}$ 
 can be denoted as
 $\Gamma_1$, indicating that $r\hat{\underline{\chi}}$ is finite on $\mathcal{I}^+$
 while the curvature component $\underline\alpha$ can be denoted by $\mathcal{R}_1$.
For general products of $\Gamma$ and $\mathcal{R}$,
one can use the $O_p$ notation to denote the total expected $r^p$ weighted bound.
When we do not wish to distinguish between $\Gamma$ and $\mathcal{R}$, we shall use
the symbol $\Phi$ or $\Phi_p$.\index{schematic notation!$\Phi_p$, schematic notation denoting either a Ricci coefficient
or curvature component}

For instance, the  equation for $\hat\chi$ in~\eqref{coupsys} together with~\eqref{eq:chihat3}
can be rewritten in the form 
\begin{equation}
\label{schemevol1}
\slashed\nabla_4(r^{2c[
{\Gamma_p}]}
{\Gamma_p} )
= r^{2c[
{\Gamma_p}]} O_{p+2} ,
\end{equation}
\begin{equation}
\label{schemevol2}
\slashed\nabla_3 
{\Gamma_p}= O_{p} ,
\end{equation}
for $p:=2$ and $c[\Gamma_p]:=1$.
These equations exhibit ``null structure'' in the sense that
the $p$-decay of the quantities on the right hand side is \emph{consistent} upon integration
of~\eqref{schemevol1} and~\eqref{schemevol2} as transport equations.

We note that a similar structure occurs for curvature, but now the equations
are hyperbolic, and must be estimated using additional integration by parts over spheres.
We defer discussion of this to later.

Let us emphasise that the null condition \emph{per se} simply ensures good behaviour \emph{towards}
$\mathcal{I}^+$, i.e.~for $u$ fixed and $v\to\infty$. (Indeed, the issue of the null condition shows up even in 
the semi-global problem of existence for the characteristic initial value problem, all the
way up to $\mathcal{I}^+$,
for small retarded time $u$.) 
To prove the stability of Minkowski space, however,
one requires global decay
of the above quantities, for instance, one requires that the right hand side of 
\eqref{schemevol1} \emph{decay} sufficiently
in the retarded time  $u$ as $u\to\infty$. 
Just like in our present work, obtaining decay for all quantities required teleological normalisations.
We discuss this briefly below.

\subsubsection{The gauge of~\cite{CK} and teleological normalisation}
Though we have based the above discussion of the null condition
on double null gauge,
in the original proof of the stability of Minkowski space~\cite{CK}, 
one in fact used 
a maximal foliation together with an optical function $u$ defining a family
of outgoing null cones.

In this maximal-optical gauge of~\cite{CK}, it is the optical function $u$ 
that must be normalised teleologically.
In practice, two distinct optical functions were used in~\cite{CK}, one normalised
from a timelike geodesic suitable for a ``near'' region and a second
``Bondi'' normalised suitable for the radiation zone. The presence
of two normalisations is thus similar (though
for slightly different reasons---see already Section~\ref{two_gauges_intro}) 
to the situation described after 
Theorem~\ref{maintheoremintro}. 

The first step of the proof of~\cite{CK} was to show that the smallness assumption
expressed geometrically with respect to assumptions on initial data corresponded to 
smallness with respect to energies expressed in a teleologically defined gauge. 
This can be thought of as
the analogue of the underlined clause in the statement of Theorem~\ref{linstabthemintro}.

Let us emphasise that, just as in our discussion in the context
of Section~\ref{linstabintint_sec}, the necessity of teleological normalisation is
in fact a \emph{linear} issue, and would have been present had a full
proof of ``linear stability of  Minkowski space'' 
in the gauge of~\cite{CK} been carried out explicitly.\footnote{In the context of linear stability of Minkowski space,
the entire system of hyperbolic Bianchi equations for curvature
decouple at linear order from the equations 
governing the spin coefficients and can be studied separately,
as opposed to the present setting, where only  the Teukolsky equation satisfied
by $\alin$ and $\ablin$ decouple. Boundedness and decay for this ``spin-2 system'' was indeed proven~\cite{CKasympt}, prior to their~\cite{CK}. 
A full ``linear stability'' in the gauge of~\cite{CK} was never
written out, but can of course be deduced \emph{a posteriori} from the full non-linear stability
proof.}
As discussed already
in Section~\ref{forteleologyinintro},
the fundamental difference between the Minkowski and  Schwarzschild cases is that in the latter, 
due to the presence of the black hole, this normalisation cannot be
 centred upon late points of a timelike geodesic but, rather, upon an appropriate sphere surrounding the horizon.

\subsubsection{Elliptic estimates on spheres}
Let us discuss one additional fundamental element 
of the proof of stability of Minkowski space~\cite{CK}:
In order for the estimates in~\cite{CK} to close, one needed to make use of 
elliptic estimates,
in addition to the hyperbolic estimates and
 integration of  transport equations discussed already above.
We explain this briefly in this section, basing the discussion on the
vacuum equations~\eqref{vaceqhere} expressed in double
null gauge~\eqref{coupsys}, for which similar issues arise.

The equation
\begin{equation}
\label{elliptichere}
\slashed{\rm div} \hat\chi= \slashed\nabla {\rm tr}\chi-\beta +\ldots
\end{equation}
 is an example of an elliptic equation contained in the system~\eqref{coupsys}, 
 corresponding to the celebrated Codazzi relation associated to the sphere (thought of 
 as a codimension-$1$ submanifold of the outoing
 null cone). 
Together with the transport equation
\begin{equation}
\label{goodtransporthere}
\slashed\nabla_4 {\rm tr} \chi  =-\frac12({\rm tr}\chi)^2-\hat\omega{\rm tr}\chi+\ldots ,
\end{equation}
one can use~\eqref{elliptichere} to estimate 
the pair $(\hat\chi, {\rm tr}\chi)$ without loss of derivative, i.e.~at one level
of differentiability more than curvature. The idea is that elliptic
estimates for equation~\eqref{elliptichere} on spheres improve by
one level of differentiability over their right hand side, whereas 
equation~\eqref{goodtransporthere}, which does not improve on its
right hand side, contains 
no curvature.
Remarkably, the  work~\cite{CK} showed
how additional quantities can be 
introduced so as to avoid this this loss of derivatives  for
\emph{all} Ricci coefficients, not just the pair $(\hat\chi, {\rm tr}\chi)$.

In view of the above structure, the authors of~\cite{CK} obtain a true
stability result, where the  stability statement as measured in terms
of top order energies does not lose derivatives. 

Let us note that the remarkable structure described above descends 
from~\eqref{vaceqhere} to the linearised setting
\eqref{coupsysline} considered in Section~\ref{linstabintint_sec}, 
see for instance
the  pair of equations
\[
\slashed{\rm div} \xlin= \frac12\Omega^{-1}\slashed\nabla\otx-\blin +\ldots, \qquad
\Omega\slashed\nabla_4\otx=(-\Omega{\rm tr}\chi+2\omega)\otx+\ldots
\]
contained in the linearised system~\eqref{coupsysline}. This structure can be again exploited to 
indeed estimate all linearised Ricci coefficients at one level of
differentiability higher than the linearised curvature. 
While this was not necessary to close the estimates 
in the context of Theorem~\ref{linstabthemintro}, this 
improvement will be very important
in the non-linear problem in order to close the estimates. (See already
Section~\ref{necessity_sec}.)

\subsubsection{The weak null condition}
We emphasise that  the above type of gauges are
not the only way to capture null structure!
It turned out
that a weaker form of the null condition can be shown to hold for~\eqref{endok} under
harmonic gauge~\eqref{harmgau} by~\cite{LindRodweak}
and with this, an alternative proof
of non-linear stability of Minkowski space
was given in~\cite{LindRod, LindRodAnn}.
For further developments, see also~\cite{Lindblad2017, Hintz:2017xxu, LindbladSchlue, Keir:2018qzh}.

\subsection{Nonlinear results under symmetry}
\label{undersym_intsec}
\setcounter{equation}{0}

The study of 
nonlinear stability of \emph{non-trivial} asympotically flat solutions, like~\eqref{Schwmet} with $M>0$,
has up to now been confined to symmetric
situations. 

By far the simplest  case is where enough symmetry is imposed so as for 
the equations to reduce to a $1+1$-dimensional
system. The only
such  symmetry compatible with asymptotic flatness is spherical symmetry, and this requires
coupling the Einstein equations to appropriate \emph{matter fields} to
evade Birkhoff's theorem~\cite{jebsen1921general}, according to which Schwarzschild
is the \emph{unique} spherically symmetric solution of the vacuum equations~\eqref{vaceqhere}. 
For the case of the \emph{Einstein-scalar field system}
under spherical symmetry, the non-linear asymptotic stability of~\eqref{Schwmet} follows
 from the more general results of~\cite{maththeory, DRPrice}.

 The vacuum equations~\eqref{vaceqhere} in higher space-time dimensions
 admit other symmetries reducing the equations to $1+1$, for instance the triaxial
Bianchi symmetry~\cite{PhysRevLett.95.071102}. The non-linear orbital stability of
the $4+1$-dimensional analogue of~\eqref{Schwmet} under triaxial Bianchi symmetry was proven 
in~\cite{dafholz2006nonlinear} while its full asymptotic stability (in the further
restricted biaxial case)
was proven in~\cite{holzbiax}.  We note that  under any of  the above symmetries,
the presence of the black hole together with the reduction to $1+1$ breaks the
supercriticality of the equations, and thus, in contrast to the situation discussed in Section~\ref{orbvsasympt}, 
orbital stability can be shown \emph{independently of
showing asymptotic stability}. In~\cite{holzbiax}, however, the full strength of
orbital stability was not in fact
used in the proof, to make  clear the relation with the more general strategy
of necessarily proving orbital and
 asymptotic stability at the same time. In particular, the issue of identifying
an approximate final mass parameter in a bootstrap context, via the Hawking mass,
appears already in the work~\cite{holzbiax}.

Beyond reductions to $1+1$, even the symmetric problem is considerably harder.
Non-linear stability of~\eqref{Schwmet} has
been given in the case
of~\eqref{vaceqhere} under polarised axisymmetry (a $2+1$ reduction) in an impressive recent work
by Klainerman--Szeftel~\cite{klainerman2017global}.  Unlike the $1+1$ dimensional problem, 
one must here face already the supercriticality of the equations
and one can  no longer exploit  the special structure of $1+1$ dimensional hyperbolic pde's. Like the present work,
the starting point of~\cite{klainerman2017global} is the linear theory~\cite{holzstabofschw}
discussed in Section~\ref{linstabintint_sec}, though~\cite{klainerman2017global} opts for a different gauge which is still 
however governed by transport equations of geometric quantities associated to foliations. Several of the non-linear difficulties
discussed above already arise in~\cite{klainerman2017global} in simplified form and are addressed in that work in a different framework. We refer the reader to~\cite{klainerman2017global} for more details.

\subsection{Aside:~application of Theorem~\ref{maintheoremintro} to axisymmetry
and the statement of Corollary~\ref{axicorollary}}
\label{asidesecintro}
\setcounter{equation}{0}

It is well known that under the assumption of axisymmetry, vacuum solutions
cannot radiate angular momentum to null infinity. This can be shown using the
conservation~\cite{Wald} of the Komar angular momentum $\mathscr{J}(S)$\index{angular momentum!$\mathscr{J}(S)$, Komar angular momentum associated to $S$} associated to a $2$-surface $S$. This leads 
easily to the following corollary of 
Theorem~\ref{maintheoremintro}.
\begin{testcorollary}[Nonlinear stability of Schwarzschild under axisymmetric 
perturbations with vanishing initial angular momentum]
\label{axicorollary}
All characteristic initial data prescribed on~\eqref{initialcones}, 
assumed sufficiently close to Schwarzschild data with mass $M_{\rm init}$,
which are moreover \underline{axisymmetric} and have \underline{vanishing} Komar angular 
momentum,
are contained in the codimension-$3$ ``submanifold'' $\mathfrak{M}_{\rm stable}$ of Theorem~\ref{maintheoremintro}.
Thus, consequences (i), (ii) and (iii) apply to such data. 
\end{testcorollary}

Thus, even though the full codimension-3 ``submanifold''  $\mathfrak{M}_{\rm stable}$ of
Theorem~\ref{maintheoremintro} can only be characterized teleologically,
the above Corollary identifies a further \emph{infinite}-codimension submanifold
which can indeed be explicitly identified examining only initial data, namely axisymmetric
solutions with vanishing Komar angular momentum. We note that the
polarised axisymmetric solutions of~\cite{klainerman2017global}
in turn themselves form an infinite-codimensional sub-class of the set
of axisymmetric data with vanishing Komar angular momentum.
See already Section~\ref{axisymmtheorem}.

\subsection{Some non-linear model problems}
\label{nonlinmodel}
\setcounter{equation}{0}

We refer briefly here to a series of non-linear model problems which have been studied,
in part motivated by certain of the difficulties of the non-linear stability problem
for Schwarzschild and Kerr.

The simplest such problem is 
the semilinear wave equation
\begin{equation}
\label{semionSchw}
\Box_g\psi = N(\nabla\psi,\nabla\psi)
\end{equation}
on Schwarzschild and more generally on a slowly rotating 
subextremal Kerr, where $N$ is assumed to satisfy Klainerman's 
null condition~\cite{KlNull}.
This problem was originally studied in~\cite{MR3082240}, with a proof based
on a weighted conformal Morawetz multiplier together with weighted commutation
vector fields.
See also~\cite{2016arXiv161000674L} where the results of~\cite{MR3082240} 
are  generalised
to the case of a background metric $g$ suitably decaying in $t$ towards
Schwarzschild and~\cite{oliversterbenz} for a very general framework
of radiating spacetimes.

An interesting model nonlinear problem motivated by the axisymmetric reduction of
the Einstein equations has been studied in~\cite{Ionescu2015}.

Finally, most pertinent for the present work, 
we mention the recent proof of non-linear stability of the trivial solution
for the Maxwell--Born--Infeld system
on Schwarzschild~\cite{Pasqualotto:2017rkh}.
This can be thought of as a quasilinear version of~\eqref{semionSchw} (but with cubic nonlinearities),
which moreover has the additional difficulty that, as with the Einstein
vacuum equations~\eqref{vaceqhere} themselves, 
the system admits non-trivial stationary
solutions. (Note that this system linearises to the usual Maxwell equations
on Schwarzschild, studied previously in~\cite{blueschwarmax, MR3373052, pasqualotto2016spin}.)
To extract the necessary decay, 
the paper~\cite{Pasqualotto:2017rkh} implements an argument based on the so-called ``$r^p$ 
method''~\cite{DafRodnew, Moschnewmeth}, exploiting a hierarchy of $r^p$-weighted
estimates in place of the time-weighted multiplier
and commutation vector fields used in~\cite{MR3082240}.
The most serious non-linear
difficulties can be localised to near null infinity $\mathcal{I}^+$
and near the photon sphere $r=3M$. 
The method of~\cite{Pasqualotto:2017rkh} can
be thought of as a precursor to the arguments described in 
Section~\ref{null_cond_in_intro} and~\ref{photon_sphere_in_intro} below
for understanding the nonlinearities in these  regions in the context
of the present work.

\subsection{A scattering construction of dynamic black holes}
\label{scatteringconstruction}
\setcounter{equation}{0}

A final related non-linear result is the ``scattering construction'' of
non-trivial examples of spacetimes satisfying the
conclusion of Theorem~\ref{maintheoremintro}, i.e.~dynamically settling
down to a Schwarzschild exterior~\cite{vacuumscatter}.
Here, non-trivial means that these examples
are not \emph{identically} Schwarzschild for sufficiently late times.
 Such non-trivial spacetimes
had in fact first been considered in \cite{Holzegelspin2}, where, without
proving  existence, a series of estimates were obtained.

In the  construction of~\cite{vacuumscatter}, 
which in fact produces more generally spacetimes settling down
to Kerr, ``scattering data'' are posed on what will be the event horizon
$\mathcal{H}^+$ and null infinity $\mathcal{I}^+$, and an exterior spacetime $(\mathcal{M},g)$ 
is then proven to exist,
with a global double null coordinate system~\eqref{doublenull},
 admitting $\mathcal{H}^+$ and $\mathcal{I}^+$ as appropriate
boundaries and indeed 
attaining the scattering data. 
Key to the tractability of the problem, however, is requiring that the
scattering data decay \emph{exponentially} to Schwarzschild (or more generally
Kerr) in advanced time $v$ along $\mathcal{H}^+$ and retarded time $u$ along $\mathcal{I}^+$,
at a rate at least as fast as a certain threshold connected to the so-called \emph{surface gravity} of the event horizon.
Note, in contrast, that for generic initial data in  Theorem~\ref{maintheoremintro} (and more generally
in Conjecture~\ref{stabofkerr} to be discussed in Section~\ref{formulationhereinintro}),  
 the solution    is expected to approach Schwarzschild (or more generally Kerr) only
inverse \emph{polynomially}. In fact, in the context of the linear theory,
vanishing of the coefficient of \emph{each} polynomial term in the asymptotic
expansion along $\mathcal{H}^+$ or $\mathcal{I}^+$ 
imposes an \emph{additional} constraint on initial data.  See for 
instance~\cite{aretakisasympt, hintz2020sharp, angelopoulos2021latetime, angelopoulos2021prices}.
Thus, in view of the strong assumption on scattering data,
the solutions constructed in~\cite{vacuumscatter} are expected to be \emph{infinite}
codimension in the moduli space of initial data.

The significance of the exponential decay assumption on scattering data is that all
linear terms in the finite region can be uniformly bounded using nothing other than 
Gronwall's inequality.
In particular, the construction works independently of the validity of the linear stability
of Schwarzschild or Kerr (and indeed, the theorem of~\cite{vacuumscatter} was  
obtained before the linear results described in Section~\ref{linstabintint_sec}). 
On the other hand, in a neighbourhood of null infinity $\mathcal{I}^+$,
the ``null condition'', discussed already in Section~\ref{nonstabminkintint_sec},  is 
relevant, and thus the structure embodied in equations~\eqref{schemevol1}--\eqref{schemevol2} 
must be used already in the proof of~\cite{vacuumscatter}.
We note that in~\cite{vacuumscatter}, the equations
of type~\eqref{schemevol1}  derived
for some of the $
{\Gamma_p}$ quantities 
have only the borderline decay $O_{p+1}$ for their right hand side;
it is important
that these borderline terms can be themselves estimated solely from an equation
of type~\eqref{schemevol2}.

It is interesting finally to remark that there is an essential difficulty in trying to parametrise
solutions in terms of scattering data on $\mathcal{H}^+\cup\mathcal{I}^+$
when the decay along $\mathcal{H}^+$ and along $\mathcal{I}^+$ 
is slower than a certain fixed exponential rate. This has
to do with the red-shift at the horizon $\mathcal{H}^+$, which in the context of backwards
evolution acts as a blue-shift. In particular, one can show already at  the linear level
that the polynomial-decay 
asymptotics of the scattering data on $\mathcal{H}^+$ and $\mathcal{I}^+$ must be 
\emph{correlated}  to all order for the data to have arisen from sufficiently regular 
Cauchy data.  See again~\cite{angelopoulos2021latetime} and~\cite{kerrscattering}. 
Thus, parametrising a full open set of the moduli space of vacuum spacetimes around
Kerr, or even just the finite codimension $\mathfrak{M}_{\rm stable}$  space of Theorem~\ref{maintheoremintro}, 
entirely in terms
of free scattering data on $\mathcal{H}^+\cup\mathcal{I}^+$
appears to be a difficult problem.
For linear scattering theory for gravitational perturbations, 
see the very recent~\cite{masaood2020scattering} for a complete physical space treatment around 
Schwarzschild, and~\cite{daCosta:2019muf, RitaShlap} for fixed frequency statements on Kerr,
up to and including extremality.

\subsection{Comparison with the case of $\Lambda\ne 0$}
\label{instructcomp}
\setcounter{equation}{0}

We compare briefly with the situation when a cosmological constant $\Lambda\ne0$
\index{constants!$\Lambda$, cosmological constant}
is added to the right hand side of the Einstein vacuum equations~\eqref{vaceqhere}.

In the case of $\Lambda>0$, the ground state solution 
analogous to Minkowski space is so-called \emph{de Sitter space}.
The (Schwarzschild) Kerr family of black holes also has an analogue, known 
as (Schwarzschild) Kerr--de Sitter. The analogue of the exterior region considered
in the present work 
is a stationary region bounded by an event and cosmological horizon.

In both the de Sitter and (Schwarzschild) Kerr--de Sitter cases above,
stability of the relevant regions reduces to the domain of development of a compact
Cauchy surface, and moreover one has exponential decay of linear fields. 
Thus, the nonlinear aspects of the stability problem are considerably easier
than the asymptotically flat case, and in particular do not require any special
structure for the non-linearities, like those that were so important in 
Section~\ref{nonstabminkintint_sec}. Nonlinear stability of pure de Sitter space had been obtained
by~\cite{friedrich1986existence}, and recently, for the very slowly rotating Kerr--de Sitter case
by Hintz and Vasy~\cite{hintz2016global}. See~\cite{hintz2016global} for further discussion.

In the case $\Lambda<0$, the ground state solution analogous to
Minkowski space is known as \emph{anti de Sitter space}. Since null infinity
is now timelike, boundary conditions must be imposed there to obtain a well-posed
problem, and the most natural type are so-called reflective conditions. Under
this assumption, anti de Sitter space  has
been conjectured~\cite{eguchiha} 
to be nonlinearly \emph{unstable} already in vacuum,
precisely because of the lack of decay mechanism on such spacetimes. 
In the
case of the Einstein--Vlasov system, this conjecture has recently been 
proven in~\cite{moschidis2018newproof}.
See also~\cite{bizon2011weakly} for numerics and further discussion.

The analogue  for $\Lambda<0$  of the
black hole solutions~\eqref{KerrmetricBL} is the so-called (Schwarzschild) Kerr--anti de Sitter 
family. On these spacetimes, logarithmic decay rates for solutions of the wave
equation have been obtained~\cite{holzegel2013decay}, and 
these have been shown to be sharp in a suitable
sense~\cite{HolzSmulevici}. In view of this very slow decay, 
it remains completely open whether to expect non-linear stability for
these spacetimes for reflective boundary conditions at null infinity. 
See the discussion in~\cite{holzegel2013decay}.

\section{The nonlinear stability of Kerr, extremality and the black hole interior}
\label{kerrstatements}

We include for completeness a statement of the  full non-linear stability
of the Kerr exterior (including a discussion of the extremal case for both vacuum and electrovacuum), as well as the  implications for the \emph{interior} structure of
generic black holes.

\subsection{Formulation of the problem}
\label{formulationhereinintro}

\begin{figure}
\centering{
\def\svgwidth{14pc}
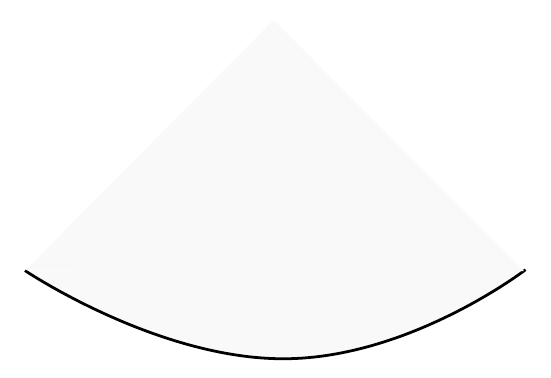}
\caption{Kerr with $0<|a|<M$:  The Cauchy development of characteristic data superimposed
on the Cauchy development of two-ended spacelike data}\label{Kerrfig}
\end{figure}

To compare with our main theorem, we shall again state this problem
in double null gauge. (Recall from our discussion above
that the paper~\cite{Pretorius} exhibits
the Kerr metric itself
 in precisely such 
a gauge (for the full subextremal range $|a|<M$).)
Fixing parameters, we consider then the subregion of Kerr
given as the maximal Cauchy development 
of the union of two
null hypersurfaces $C_{\rm out}\cup \underline{C}_{\rm in}$ of the double null gauge,
where as before $\underline{C}_{\rm in}$ crosses the event horizon,
while $C_{\rm out}$ is future complete and terminates at null infinity.
Note that if $a\ne 0$, the development is depicted
in Figure~\ref{Kerrfig}, superimposed on the larger 
region of Kerr corresponding to the future Cauchy development
of a two-ended asymptotically flat spacelike $\Sigma$.

In marked contrast to Schwarzschild, the above region of Kerr does not terminate at a strongly singular
spacelike boundary corresponding to $r=0$, but rather, can be further extended
beyond a non-trivial null boundary $\mathcal{CH}^+$ known 
as a \emph{Cauchy horizon}~\cite{hawking1967occurrence}. 
These  extensions fail to be 
uniquely determined by initial data,  this pathology
motivating the so-called \emph{strong cosmic censorship conjecture},
according to which, extendibility across Cauchy horizons should
be non-generic.
We shall discuss this  in Section~\ref{implicsec} below.

First we give
the statement of the stability problem for the exterior region:

\begin{conjecture}[Nonlinear stability of the Kerr exterior]
\label{stabofkerr}
For all characteristic data prescribed as above, assumed
sufficiently close to Kerr with parameters $|a_{\rm init}|<M_{\rm init}$,
the maximal Cauchy development $\mathcal{M}$ contains
a region $\mathcal{R}$ which can be covered by appropriate global
double null foliations~\eqref{doublenull} and which (i) possesses
a complete future null infinity $\mathcal{I}^+$ such that
$\mathcal{R}\subset J^-(\mathcal{I}^+)$, and in fact the future boundary
of $\mathcal{R}$ in $\mathcal{M}$ is a regular future affine
complete ``event horizon'' $\mathcal{H}^+$. 
Moreover
(ii) the metric remains close to Kerr in $\mathcal{R}$ and
(iii) asymptotes, inverse polynomially, to a Kerr
metric with parameters $|a_{\rm final}|<M_{\rm final}$ where
\index{Kerr metric!$M_{\rm final}$, final mass parameter}
\index{Kerr metric!$a_{\rm final}$, final rotation parameter}
$a_{\rm final}\approx a_{\rm init}$, 
$M_{\rm final}\approx M_{\rm init}$, as $u\to \infty$
and $v\to \infty$, in particular along $\mathcal{I}^+$ and $\mathcal{H}^+$.

\index{Kerr metric!$M_{\rm init}$, mass parameter associated to initial data}\index{Kerr metric!$a_{\rm init}$, 
rotation parameter associated to initial data}

Moreover, for any given $0<|a_{\rm final}|<M_{\rm final}$, the set of
initial data above attaining these final parameters is codimension-$2$
in the space of all data,  the set
of initial data attaining $a_{\rm final}=0$ for some $M_{\rm final}$
is codimension-$3$,  and the set of initial data attaining $a_{\rm final}=0$
and a fixed $M_{\rm final}$ is codimension-$4$. In particular,
for generic initial data, $a_{\rm final}\ne 0$, while the set
of solutions $a_{\rm final}=0$ corresponds precisely to the solutions
constructed  in Theorem~\ref{maintheoremintro}.
\end{conjecture}

We remark that the difference in dimensionality in the $a_{\rm final}=0$
case has to do with the enhanced symmetry of Schwarzschild
in comparison to Kerr. Let us also note that, as in Theorem~\ref{maintheoremintro},
 a true stability result should
give a top-order orbital stability statement representing (ii) without loss of derivatives.

In view of our discussion and the recent~\cite{DHRteuk, RitaShlap}, 
the path is now open to obtaining 
 Conjecture~\ref{stabofkerr} 
following the approach of the present work, although, at a technical level, the Kerr case introduces several
new complications related to the necessity of applying frequency localisation to deal with 
the issues related to trapping.
In this sense, we view one of the appealing features of having a
 complete, self-contained physical space treatment of the Schwarzschild case 
as in the present work to be that one may understand the essence of the above
conjecture without this additional, largely technical, complication.

\subsection{Extremality and the Aretakis instability}
It is non-trivial even to formulate the extremal analogue $|a_{\rm init}|=M_{\rm init}$
of Conjecture~\ref{stabofkerr}.

Here, already at the linear level, the situation is considerably more complicated. 
In particular, a basic understanding of even the linear scalar wave equation~\eqref{scalarwav}
on an extremal Kerr background (cf.~the discussion
in Section~\ref{ginvquanCha} concerning~\eqref{scalarwav} in the sub-extremal case) 
is not yet available.
One thing that is known, however, is that
all extremal stationary black holes are subject---at the very least---to the
Aretakis instability~\cite{Aretakis2} along their horizon, according to which,
higher order translation invariant derivatives of solutions to the wave equation~\eqref{scalarwav} 
generically blow up polynomially.  
This has been
extended to gravitational perturbations in~\cite{luciettireall, luciettietal}. 
One may view these  derivatives as infinitely growing horizon ``hair'',
whose presence
can moreover be inferred indirectly via measurements at $\mathcal{I}^+$~\cite{horizonhair}.
It is an interesting problem to understand whether the most basic geometric features of these black hole spacetimes
 can still be nonlinearly
stable despite this higher order instability phenomenon associated to their horizons, 
or, rather, whether this growing ``hair'' leads at the nonlinear level to some worse type of blowup, 
for instance the formation
of so-called ``naked singularities'', resulting in a future incomplete $\mathcal{I}^+$ (i.e.~already
violating the analogue of statement (i) of Theorem~\ref{maintheoremintro}).
So far, this question has only been probed numerically for
toy models under spherical symmetry~\cite{whathappens}.

In order to disentangle the Aretakis instability from other difficulties associated to extremal Kerr,
it is natural to first consider  the electrovacuum Reissner--Nordstr\"om metrics (see for instance~\cite{Wald}), a spherically symmetric family
of solutions to the Einstein--Maxwell system with parameters $Q$ and $M$.
(Note that this family contains the Schwarzschild family as the subfamily $Q=0$.)
One expects that
the analogue of Theorem~\ref{maintheoremintro} 
for the sub-extremal case $0 \le Q<M$ of Reissner--Nordstr\"om is a more or less straightforward
adaptation of the results of the present paper,  in view of the recent~\cite{giorgi},
 where a linear stability proof is carried out explicitly
for the subextremal Reissner--Nordstr\"om family, adapting the methods of~\cite{holzstabofschw}.
The interesting case to consider further is thus the extremal case $Q=M$.

To set up the problem,  fix  null hypersurfaces $\underline{C}_{\rm in}\cup C_{\rm out}$ in background extremal Reissner--Nordstr\"om
with parameters $M_{\rm init}=Q_{\rm init}$,
analogous to~\eqref{initialcones}, where the terminal sphere of $\underline{C}_{\rm in}$ lies in the black hole interior. (Note
that, in contrast to the Schwarzschild case, this terminal sphere is no longer trapped.)
Consider now the moduli space $\mathfrak{M}$ of nearby data defined on
$\underline{C}_{\rm in}\cup C_{\rm out}$, suitably normalised. 
Note that we may identify the following  families of initial data corresponding to  explicit
solutions: (a) a $1$-parameter family corresponding to extremal Reissner--Nordstr\"om metrics with charge
$Q=M$: (b) a $1$-parameter family corresponding to  Reissner--Nordstr\"om metrics with fixed
$M=M_{\rm init}$, parametrised by $Q$,  (c) a $3$-parameter family corresponding
to   extremal Kerr--Newman metrics (see~\cite{misner2017gravitation}) 
with charge $Q_{\rm init}$. We note that (b) contains
both subextremal $Q<M_{\rm init}$ and superextremal $Q>M_{\rm init}$ Reissner--Nordstr\"om data on
 $\underline{C}_{\rm in}\cup C_{\rm out}$. The latter lead to spacetimes such that $\underline{C}_{\rm in}\subset
 J^-(\mathcal{I}^+)$, i.e.~spacetimes that fail to form black holes.

For the extremal Reissner--Nordstr\"om family (a) itself, then, 
in view of the above, the best one can hope  is for the existence of a codimension-$4$ asymptotically
stable ``submanifold'' $\mathfrak{M}_{\rm stable}\subset \mathfrak{M}$, where moreover the asymptotic stability
statement is suitably relaxed along $\mathcal{H}^+$ (compared to that of Theorem~\ref{maintheoremintro}),
so as to accommodate the growing horizon ``hair'' associated to the Aretakis instability.
This suggests the following:

\begin{conjecture}
[Asymptotic stability of extremal Reissner--Nordstr\"om but with growing horizon ``hair'']
\label{extremeconjecture}
For all characteristic initial data for the Einstein--Maxwell system prescribed on \eqref{initialcones}, 
assumed sufficiently close to extremal Reissner--Nordstrom data with mass $M_{\rm init}$
and $Q_{\rm init}=M_{\rm init}$
and lying on a
codimension-$4$ ``submanifold'' $\mathfrak{M}_{\rm stable}$ of the moduli space $\mathfrak{M}$ of initial data, the maximal Cauchy 
development $\mathcal{M}$
contains a region $\mathcal{R}$ 
which can be covered by appropriate (teleologically
normalised) global double null gauges~\eqref{doublenull} 
and where the analogues of (i), (ii) and (iii) of Theorem~\ref{maintheoremintro} 
are satisfied
with an extremal Reissner--Nordstr\"om metric 
with parameters $M_{\rm final}=Q_{\rm final}$ in
the place of Schwarzschild. 
Along $\mathcal{H}^+$, however, one has decay to extremal Reissner--Nordstr\"om only
in a
weaker
sense, in particular, for generic data lying on $\mathfrak{M}_{\rm stable}$, 
suitable higher order quantities in the arising solution 
blow up polynomially along $\mathcal{H}^+$ (growing horizon ``hair'').
\end{conjecture}

Given a positive resolution of the above, one would moreover expect 
that the  ``submanifold'' $\mathfrak{M}_{\rm stable}$  itself lies
on a larger codimension-$1$ submanifold $\mathfrak{M}'_{\rm stable}$ of $\mathfrak{M}$ consisting
of data leading to solutions asymptoting
to a very slowly rotating extremal Kerr--Newman,  again with growing horizon hair. Moreover, 
one could hope to prove that this larger
submanifold $\mathfrak{M}'_{\rm stable}$  delimits the boundary
signifying a phase transition between two
very different open regions of  moduli space $\mathfrak{M}$: (1) the set  of data leading to spacetimes
failing to collapse (i.e.~those for which 
$C_{\rm in}\subset J^-(\mathcal{I}^+)$) and (2) the set of data leading to a black hole exterior settling down
to a very slowly rotating subextremal Kerr--Newman.
(Of course, one can already conjecture the analogue of Conjecture~\ref{extremeconjecture} for  extremal Kerr as
a family of the Einstein vacuum equations; we emphasise, however, that the dynamics near this phase
transition in that case may be considerably more complicated!)

In order to prove Conjecture~\ref{extremeconjecture}, one must confront
a fundamental  new difficulty compared to the present work:
In the extremal case, the stabilising mechanism of the red-shift effect, exploited heavily here, degenerates
at $\mathcal{H}^+$. Moreover, 
in view of the expected growing horizon hair, 
it would seem that in order to control the nonlinearities, one must identify and exploit a suitable ``null structure'',
not just near null infinity $\mathcal{I}^+$ as before (cf.~Section~\ref{capturingnullcon}), but now
also in the region near the horizon $\mathcal{H}^+$.  See the recent~\cite{nonlinearextreme} where
this is indeed exploited to show global stability results on a fixed extreme Reissner--Nordstr\"om background  for a 
nonlinear scalar wave equation whose
nonlinearities satisfy the null condition (cf.~equation (ii) of~\eqref{modelproblems}).
We hope that the present work, with its  set-up for proving finite-codimensional stability statements
and 
with one of its teleological gauges normalised at the event horizon $\mathcal{H}^+$, 
may provide a suitable general framework to try to address Conjecture~\ref{extremeconjecture}.

\subsection{Implications for black hole interiors with $a_{\rm final}\ne 0$ and
strong cosmic censorship}
\label{implicsec}
\setcounter{equation}{0}

Returning to the subextremal case,
Conjecture~\ref{stabofkerr} can be applied together with the
following theorem to obtain the $C^0$ stability of the Kerr Cauchy horizon:
\begin{testtheorem}[\cite{DafLuk1}]
\label{withjonathanthemintro}
Consider general characteristic initial data for the
Einstein vacuum equations on $\mathcal{H}^+\cup\underline{C}_{\rm in}$
such that  $\mathcal{H}^+$ is future complete and the
data along $\mathcal{H}^+$ approach that of a sub-extremal rotating Kerr solution (with $0<|a|<M$) 
along its event horizon at a suitable
inverse polynomial rate. Then restricting to a sufficiently short
$\underline{C}_{\rm in}$, the maximal Cauchy development can
be covered by a global double null foliation and can be extended
continuously beyond a non-trivial Cauchy horizon $\mathcal{CH}^+$.

In particular,  for  initial data
as in Conjecture~\ref{stabofkerr}, then as long as $a_{\rm final}\ne 0$
(which would be true generically!), it would follow from the conjecture and the above paragraph
that the  maximal Cauchy development
is extendible beyond a non-trivial Cauchy horizon located in the
black hole interior. In particular, the $C^0$-formulation of strong
cosmic censorship (see~\cite{Chrmil}) is \underline{false}.
\end{testtheorem}

In fact, if one considers   now two-ended Kerr initial data $\Sigma$
as depicted in Figure~\ref{Kerrfig}, then
a further extension of Theorem~\ref{withjonathanthemintro}, see
the upcoming~\cite{DafLuk3}, implies that the \emph{entire} Kerr Penrose diagram
depicted
above is stable, in particular, spacetime is globally
extendible as a $C^0$ metric across a \emph{bifurcate}
null Cauchy horizon such that \emph{all} future inextendible causal geodesics
pass into the extension.

The above result is surprising in view of the presence of a well-known
\emph{blue-shift instability}~\cite{penrose1968battelle} 
associated with the Cauchy horizon,   which provided
 the original evidence for
the  conjecture of strong cosmic censorship. 
The theorem is still compatible, however,
with the possibility that 
for generic initial data, the   boundary $\mathcal{CH}^+$ 
be singular in a weaker sense, specifically, that the metric in particular
fails to be  $H^1_{\rm loc}$ 
in any extension of the maximal Cauchy development.
(This is related to the Christodoulou formulation
of strong cosmic censorship and has been discussed in~\cite{Chr}.)
Proving this is in turn related to 
obtaining a suitable \emph{lower} bound on the rate of approach
to Kerr on $\mathcal{H}^+$ in the statement of Conjecture~\ref{stabofkerr}
for generic initial data. See the discussion in~\cite{DafLuk1}
and~\cite{D2, LukOh2017one, LukOh2017two, maxime} for 
results for a spherically symmetric model.
We also note that here too, the extremal case is exceptional; see~\cite{Gajic:2015csa, gajicluk}.

\subsection{A conjecture for the $a_{\rm final}=0$ case}
\setcounter{equation}{0}

Ironically, it is precisely for data lying on the codimension-3 ``submanifold'' $\mathfrak{M}_{\rm stable}$ 
constructed in Theorem~\ref{maintheoremintro} satisfying
$a_{\rm final}=0$ for which 
there is no general analogue\footnote{Note that, as remarked after the statement of Theorem~\ref{maintheoremintro},  one can still apply~\cite{DafLuk2} in the case $a_{\rm final}=0$ 
to obtain stability up to a suitable spacelike hypersurface foliated by trapped surfaces. In fact, in this case,
using translation invariance, the geometry of Schwarzschild and Cauchy stability, one can then deduce 
relatively easily the statement that the supremum of the Kretschmann curvature (taken over the maximal development
of data $C_{\rm out}\cup \underline{C}_{\rm in}$) is infinite.} of the understanding of the black hole
interior provided by the above theorem. This case is harder from the perspective of the interior
because of the strongly singular nature
of the exact Schwarzschild boundary. In the special
case of polarised axisymmetry studied in~\cite{klainerman2017global}, for which in particular $a_{\rm final}=0$
(cf.~the discussion in Sections~\ref{undersym_intsec} and~\ref{asidesecintro}),
this spacelike singular boundary has very recently been shown to be globally stable~\cite{alexakisfourno2020stable}
in a suitable sense.
This result relies heavily on the polarised assumption, however,
and the precise results proven are not expected to carry over
outside of the symmetry class.
The most basic question one can ask is whether 
$a_{\rm final}=0$ necessarily means that, in contrast to the $a_{\rm final}\ne 0$ case,
there can never exist a Cauchy horizon emanating from ``timelike infinity''.
Thus, it would already
be interesting to prove simply:

\begin{conjecture}
\label{wouldbeinteresting}
For the initial data of Theorem~\ref{maintheoremintro},
the maximal Cauchy development $(\mathcal{M},g)$ will necessarily contain
a TIP whose intersection with
$C_{\rm out}\cup \underline{C}_{\rm in}$ has compact closure.
\end{conjecture}

For the definition of TIPs, see~\cite{forthetips}.
A positive resolution of the above would in particular 
show that the set of vacuum initial data
leading to a TIP whose intersection with spacelike initial data has compact closure, if not open in moduli space,
is at least a set of finite codimension.
(In contrast, the largest class of examples produced so far,
namely the symmetric solutions of~\cite{alexakisfourno2020stable} discussed above, as well as  
previous examples due to~\cite{Fournodavlos2016}, produced by a scattering construction,
 are manifestly of infinite
codimension in  moduli space.)

\section{Brief overview of the proof}
\label{firstremarksintint_sec}

In very broad outline, the proof of Theorem~\ref{maintheoremintro} is an adaptation of the
linear analysis of~\cite{holzstabofschw}---reviewed in 
Section~\ref{linstabintint_sec}---using the insights 
from previous non-linear results---reviewed 
in Section~\ref{resnonstabintint_sec}---to estimate the nonlinear terms
(most specifically the setup of~\cite{vacuumscatter} described in 
Section~\ref{scatteringconstruction}). As in Section~\ref{resnonstabintint_sec}, 
linear analysis is here always understood
in the sense of the modern theory of non-linear wave equations, 
i.e.~applied directly to the non-linear equation. This
is essential for a true non-linear stability result which does not lose derivatives.

In addition, there are various additional specifically nonlinear aspects of the problem, however, 
concerning our modulation scheme, small non-linear corrections to the linear 
teleological determination of the gauge
and other issues. 

The precise statement of Theorem~\ref{maintheoremintro} is given
as Theorem~\ref{thm:main} (see already Section~\ref{maintheoremsec}).
We give a brief overview of the proof in this section.

\subsection{The $3$-parameter families of data and $\mathfrak{M}_{\rm stable}$}
\label{dataintro}
\setcounter{equation}{0}

We decompose the moduli space of characteristic initial data into disjoint $3$-parameter families of initial data
\begin{equation}
\label{threeparamfamintro}
\mathcal{L}^{\varepsilon_0}_{\mathcal{S}_0}:=\{\mathcal{S}_0(\lambda) : \lambda \in  [-c\varepsilon_0, c\varepsilon_0]^3\subset\mathbb R^3\},
\index{initial data!moduli space!$\mathcal{L}^{\varepsilon_0}_{\mathcal{S}_0}$, $3$-parameter family of initial 
data indexed by a reference data set $\mathcal{S}_0$}\index{initial data!seed data!$\mathcal{S}_0(\lambda)$, initial data set derived from 
$\mathcal{S}_0$ associated with parameter $\lambda$}
\end{equation}
each determined by some reference 
data set $\mathcal{S}_0$.
 (Here $c$ and $\varepsilon_0$ are constants and $\mathcal{S}_0(\lambda)$ denotes initial data
 derived from $\mathcal{S}_0$ by adding to one of the seed quantities the three independent $\ell=1$ spherical harmonics
 weighted by the three coefficients $(\lambda_{-1},\lambda_0,\lambda_1)$\index{initial data!parameters!$\lambda=(\lambda_{-1},\lambda_0,\lambda_1)$, modulation vector parameter}\index{angular momentum!$\lambda=(\lambda_{-1},\lambda_0,\lambda_1)$, modulation parameters related to angular momentum of data}  of $\lambda$, and the norm $|\lambda|$ may be thought to be  an approximate measure
 of the norm of the angular momentum of the seed data $\mathcal{S}_0(\lambda)$.) As $\mathcal{S}_0$ is varied in a suitable space, these families
$\mathcal{L}^{\varepsilon_0}_{\mathcal{S}_0}$ 
cover an entire neighbourhood of the moduli space $\mathfrak{M}$ of initial data around Schwarzschild.
(See already Section~\ref{threeparamsection} for a discussion
of initial data.)
We denote by $(\mathcal{M}(\lambda),g(\lambda))$ the maximal Cauchy development
of initial data $\mathcal{S}_0(\lambda)$. (See already Section~\ref{maxcauchysection} for
a general discussion of the maximal Cauchy development.)\index{spacetime subsets!$(\mathcal{M}(\lambda),g(\lambda))$, the maximal Cauchy development of initial data $\mathcal{S}_0(\lambda)$}

The aim is to show that for each $\mathcal{S}_0$, there exists a $\lambda^{\rm final} \in  [-c\varepsilon_0, c\varepsilon_0]^3$ for which $(\mathcal{M}(\lambda^{\rm final}),g(\lambda^{\rm final}))$ asymptotes
to a Schwarzschild metric as described in the statement of Theorem~\ref{thm:main}.
\index{parameters!$\lambda^{\rm final}$, the parameter value for which convergence to Schwarzschild is proven}
The collection of all  $\{\mathcal{S}_0(\lambda^{\rm final})\}$  as one varies $\mathcal{S}_0$ will then form our
codimension-$3$
asymptotically stable ``submanifold'' $\mathfrak{M}_{\rm stable}\subset \mathfrak{M}$.

\subsection{The logic of the proof}
\label{contargintro}
\setcounter{equation}{0}

The logic of the proof (see already Section~\ref{logicoftheproofsection})
proceeds by a continuity argument.
At each step,
governed by a final retarded time $u_f$,\index{bootstrap!parameters! $u_f$, final retarded time associated to bootstrap}
 one defines a subset $\mathfrak{R}(u_f)\subset \mathbb R^3$\index{bootstrap!sets!$\mathfrak{R}(u_f)$, subset of $\lambda$-parameter space depending on $u_f$} of $\lambda$-parameter space,
 and considers, for each $\lambda\in \mathfrak{R}(u_f)$,  
a certain subregion of the spacetime $(\mathcal{M}(\lambda),g(\lambda))$
 together 
with a set of bootstrap assumptions on this region.
The subregion of spacetime
will be described in Section~\ref{two_gauges_intro} below.
We defer further discussion of the role of the
set of parameters  $\mathfrak{R}(u_f)$ to Section~\ref{completingintro}.
(Briefly,  
the set of parameters $\mathfrak{R}(u_f)$ is a closed set in $\mathbb R^3$ 
such that the ``total angular momentum'' of solutions  is less than or equal to $\varepsilon_0 u_f^{-1}$,
with equality holding on the boundary $\partial\mathfrak{R}(u_f)$.
This ``total angular momentum'' is in turn defined as 
the norm of a vector ${\bf J}$ associated to the $\ell=1$ modes of a suitable Ricci coefficient.
The special role of the $\ell=0,1$ modes will be discussed in Section~\ref{lowmodes_sec}.)
At each stage of the bootstrap, the solution is compared to a Schwarzschild
solution whose mass $M_f$ is chosen on the basis of the $\ell=0$ mode of 
a curvature component at time $u_f$.

As usual in  a continuity argument, the statement that the bootstrap assumptions
can be improved is the main difficulty of the proof (see already the statement of
Theorem~\ref{havetoimprovethebootstrap}). 
In Sections~\ref{two_gauges_intro}--\ref{necessity_sec} below, 
we shall   discuss the main difficulties arising in retrieving the
bootstrap assumptions (i.e.~in the proof of Theorem~\ref{havetoimprovethebootstrap}).
These are addressed in the chapters corresponding to Part~\ref{improvingpart} of the work.

We shall discuss issues related to the $\ell=0,1$  modes in Section~\ref{lowmodes_sec},
and then  how one  completes the continuity argument in Section~\ref{completingintro}.
Finally, we shall discuss how one obtains
the asymptotic gauges and  stability  statements in Section~\ref{propsofHandI_sec}.

\subsection{The bootstrap region and the teleological normalisation of the null gauges}
\label{two_gauges_intro}
\setcounter{equation}{0}

As in~\cite{holzstabofschw}, the gauge must be normalised towards \emph{the future} of the bootstrap region. As discussed already, however, for the nonlinear problem it is necessary
to have in fact two separate normalisations   corresponding to a region near the horizon
(the $\mathcal{H}^+$-gauge)
and near infinity (the $\mathcal{I}^+$-gauge).  
(See already Chapter~\ref{almostgaugeandtellychapter} for
details.)

Given a final retarded time $u_f$ and a $\lambda\in \mathfrak{R}(u_f)$, 
the bootstrap region of $(\mathcal{M}(\lambda),g(\lambda))$ 
will be defined as the intersection of the past of a ``late'' outgoing null cone
$C_{u_f}$ and the past of a ``late'' ingoing null cone $\underline{C}_{v_\infty}$, with $v_\infty\to\infty$ as $u_f\to \infty$, intersecting
at the sphere $S_{u_f,v_\infty}$.
In both gauges, defined by double null coordinates $(u_{\mathcal{H}^+}, v_{\mathcal{H}^+})$\index{teleological $\Hp$ gauge!coordinates!$u_{\Hp}$,
retarded null coordinate of the $\Hp$ gauge}\index{teleological $\Hp$ gauge!coordinates!$v_{\Hp}$, advanced null coordinate of the $\Hp$ gauge}
and $(u_{\mathcal{I}^+},v_{\mathcal{I}^+})$\index{teleological $\I$ gauge!coordinates!$u_{\I}$,
retarded null coordinate of the $\I$ gauge}\index{teleological $\I$ gauge!coordinates!$v_{\I}$,
advanced null coordinate of the $\I$ gauge} respectively,
the final outgoing cone $C_{u_f}$ will be a hypersurface of constant
$u_{\mathcal{H}^+}$, respectively $u_{\mathcal{I}^+}$, anchoring the two gauges together.

The $\mathcal{H}^+$-gauge  will be normalised on the ``late'' outgoing null cone
$C_{u_f}$ and an
 initial ingoing cone $\underline{C}^{\mathcal{H}^+}_{v_{-1}}$ defined by
$v_{\mathcal{H}^+}= v_{-1}$. This initial cone does not coincide with the
initial data cone $\underline{C}_{\rm in}$ but remains within a fixed 
distance from $\underline{C}_{\rm in}$ independently
of $u_f$. In the limit as $u_f\to \infty$, the cone $C_{u_f}$ will coincide
with the event horizon $\mathcal{H}^+$.  The gauge will only be considered in a region
$r\le R_2$, where $r$ is a function
of $(u_{\mathcal{H}^+},v_{\mathcal{H}^+})$.

The $\mathcal{I^+}$-gauge will have the cone  $\underline{C}_{v_\infty}$ 
as a $v_{\mathcal{I}^+}=v_\infty$ hypersurface, and the geometry of the cones
will be normalised
on the final ingoing cone $\underline{C}_{v_\infty}$ and the ``initial'' outgoing
cone $C^{\mathcal{I}^+}_{u_{-1}}$ defined by $u_{\mathcal{I}^+}=u_{-1}$.
As before,  this initial cone does not coincide with the initial data cone 
$C_{\rm out}$ but remains within a fixed 
distance from $C_{\rm out}$ independently of $u_f$.
The gauge will only be considered in a region $r\ge  R_{-2}$, with $R_{-2}<R_2$, where $r$ is a function
of $(u_{\mathcal{I}^+},v_{\mathcal{I}^+})$. The two $r$-functions are close and the gauges thus have a nontrivial overlap region.

The normalisations are determined by a series of requirements on the spheres
$S_{u_f,v_{-1}}$ and $S_{u_f,v(R,u_f)}$, for an $R_{-2}<R<R_2$,
and on the cones $C_{u_f}$,  $\underline{C}^{\mathcal{H}^+}_{v_{-1}}$  
in the case of the $\mathcal{H}^+$ gauge
and on the sphere $S_{u_f,v_\infty}$ and the cones $C^{\mathcal{I}^+}_{u_{-1}}$, $\underline{C}_{v_f}$ in the case of
the $\mathcal{I}^+$ gauge, for instance, in the case of $\mathcal{I}^+$ gauge, the requirements
\begin{equation}
\label{forinstancereq1}
\Omega \tr \chi -(\Omega\tr \chi)_{\circ} =0, \qquad \Omega^{-1}\tr \chibar-(\Omega^{-1}\tr\chibar)_{\circ}=0\, {\rm\ on\ } S_{u_f,v_\infty}
\end{equation}
and that 
\begin{equation}
\label{forinstancereq2}
\mu_{\ell \ge1}=0 {\rm\ on\ }\underline{C}_{v_\infty}.\index{double null gauge!connection coefficients!$\mu$, mass aspect
function}
\end{equation}
Here $(\Omega\tr \chi)_{\circ}$\index{Schwarzschild background!connection coefficients!$(\Omega\tr \chi)_{\circ}$}, 
$(\Omega^{-1}\tr\chibar)_{\circ}$\index{Schwarzschild background!connection coefficients!$(\Omega^{-1}\tr\chibar)_{\circ}$} 
denote Schwarzschild background quantities and the Ricci coefficient $\mu$ denotes the mass-aspect function.
Some of the conditions (e.g.~\eqref{forinstancereq2}) 
distingish the behaviour of the $\ell=0$ or $\ell=1$ modes; see already Section~\ref{lowmodes_sec} below.
For issues related to \emph{constructing} a gauge satisfying  conditions 
such as~\eqref{forinstancereq1} and \eqref{forinstancereq2}, see already Section~\ref{completingintro} below.
These normalisations ensure that 
as $(u_f,v_\infty)\to(\infty,\infty)$, the normalisation of the $\mathcal{I}^+$ gauge becomes Bondi, and moreover,
we have vanishing of the quantity $\Sigma_+=\lim_{u_{\mathcal{I}^+}\to\infty} \lim_{v_{\mathcal{I}^+}\to \infty} r^2\hat\chi(u,v)=0$ (see  already Section~\ref{propsofHandI_sec} below).\index{null infinity!$\Sigma_+$, final asymptotic shear}

Note that the two gauge normalisations are non-trivial at the linearised level and
differ in their overlap already in linear theory. 
The future-normalised gauge  used in Theorem~\ref{linstabthemintro}
is closely related to the 
normalisations of the $\mathcal{H}^+$ gauge when it is  anchored as above to the $\mathcal{I}^+$ gauge. 
Though one could have considered the analogue 
of the normalisations specific to the $\mathcal{I}^+$ gauge already in~\cite{holzstabofschw}, 
we did not need this type of gauge in the proof of
Theorem~\ref{linstabthemintro}. Its essential usefulness here lies in the fact that it allows for
better control of the decay towards null infinity $\mathcal{I}^+$, 
important for capturing 
the null condition necessary for controlling the non-linear terms (see already the discussion in~Section~\ref{null_cond_in_intro} below).

\begin{figure}
\centering{
\def\svgwidth{12pc}
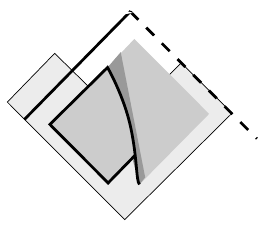}
\caption{The two teleological double null gauges and the two auxiliary double null gauges near data}\label{twogaugesfig}
\end{figure}

The teleological gauges are ``anchored'' to initial data via two auxiliary double null gauges,
one Kruskal-normalised and the other Eddington--Finkelstein normalised,
which cover a fixed region in the vicinity of the initial cones $C_{\rm out}\cup\underline{C}_{\rm in}$,
which
can be constructed by Cauchy stability considerations, together with some local
analysis near $\mathcal{I}^+$ in order to achieve that the latter be
Bondi normalised (see already Section~\ref{localexistencesection}).
(Note that local considerations
imply that one has good estimates in these auxiliary regions, coming from initial data.)
The domains of the four gauges are depicted in Figure~\ref{twogaugesfig}. Note that this depiction is schematic, as each of 
the two teleological null gauges and each of the two auxiliary null gauges  define different foliations of spacetime by spheres. 

\subsection{Boundedness of the initial norm and  the relation of the gauges}
\label{bounded_init_intro}
\setcounter{equation}{0}

Since the two teleological gauges are normalised to the future, one must first show that suitable energies
on the ``initial'' cones
$C^{\mathcal{I}^+}_{u_{-1}}$ and  $\underline{C}^{\mathcal{H}^+}_{v_{-1}}$ can be controlled from initial data.
Here one uses that
these ``initial'' cones lie in the region covered by the two auxiliary double null gauges in which
one already has good estimates. To compare, however,
one must  estimate the diffeomorphism $f$\index{double null gauge!change of gauge!$f$, notation for functions describing diffeomorphisms}
relating teleological to auxiliary gauge in their overlap region.
(See already Chapter~\ref{chap:comparing}.)
As discussed previously,
this is analogous in linear theory
to the underlined part of the statement of Theorem~\ref{linstabthemintro}
(stating that the future normalised pure gauge solution is bounded
by the data). To close our estimates, it is important here that the initial energies  in question 
are ``almost gauge invariant'' (see already Section~\ref{mainestofproofintro}), 
and thus the dependence on the diffeomorphisms will be
at higher order in the smallness parameter $\varepsilon_0$. 
 (For comparison, this corresponds in the proof
of the nonlinear stability of Minkowski space to  the sentence following
formula (10.2.4b) on p.~302 of~\cite{CK}.)

Similarly, one must show that the two future normalisations of double null
gauge  are themselves sufficiently close in their mutual overlap region (see the darker shaded
region in Figure~\ref{twogaugesfig}), by estimating the diffeomorphisms
$f$ relating these two.
 (Again, for comparison, this is in some sense analogous to
 step~2 in the proof
of the nonlinear stability of Minkowski space (see p.~301 and Chapter 15 of~\cite{CK}),
where the exterior optical function is extended to an interior optical function
and appropriate estimates are obtained.)

In general,
the idea for estimating the diffeomorphisms connecting the various
gauges is to use relations expressing
the difference of curvature or Ricci coefficients in the two gauges, in which
derivatives of the gauge functions appear. For instance, one can write
schematically the relation for difference of the curvature component 
$\beta$ expressed in two gauges
\[
\beta-\tilde\beta = \slashed\nabla f+\cdots,
\]
where non-linear terms are omitted and $f$ here denotes one component
of the diffeomorphism connecting the gauges (see already Section~\ref{changeofgaugerelationssection}). 
From the above, estimates on curvature
components like $\beta$
and $\tilde\beta$, to be discussed below,  lead to estimates on $f$.

\subsection{The main estimates: revisiting~\cite{holzstabofschw}}
\label{mainestofproofintro}
\setcounter{equation}{0}

For the main estimates,
the general strategy of the linear stability result~\cite{holzstabofschw} applies, as
described in Section~\ref{linstabintint_sec}. 

In particular, one first considers
the fully non-linear quantities $P$\index{almost gauge invariant hierarchy!$P$, quantity satisfying Regge--Wheeler type equation}, $\underline{P}$\index{almost gauge invariant hierarchy!$\Pbar$, quantity satisfying Regge--Wheeler type equation}, (defined in analogy
with~\eqref{Pquantfirstinstance}
for the linearised $\Plin$ and $\Pblin$)
 which now satisfy equations analogous to~\eqref{reggewheel},
 	\begin{equation}
	\label{Peqnonlinearintro}
	\Omega \nablaslash_4 \Omega \nablaslash_3 (r^5 P)
		+
		\frac{2\Omega^2}{r^2} r^2 \Dslash_2^* \divslash (r^5 P)
		+
		2\Omega^2 \left( 1 - \frac{3M}{r} \right) r^5 P
		= \mathcal{E}[P],
	\end{equation}
	but with  non-linear error terms $\mathcal{E}[P]$ on their right hand sides, 
coupling with other quantities. These error terms can be written in
schematic form as
	\begin{equation}
	\label{RHSschem}
		\mathcal{E}[P]
		=
		\Omega^2
		\sum_{p_1+p_2 \geq 2}
		\sum_{\substack{k_1\leq 3 \\ k_2 \geq 2}}
		r^{-p_1} \cdot (\mathfrak{D}^{k_1} \Phi_{p_2})^{k_2},
\end{equation}
where $\Phi_{p}$ is a schematic notation  (see already
Chapter~\ref{moreprelimchapter}) encompassing
both Ricci coefficients $\Gamma_p$ and curvature components  $\mathcal{R}_p$
(cf.~the discussion in Section~\ref{capturingnullcon} above for a discussion of $p$-weights), but now
 for \emph{differences} of
quantities from their Schwarzschild values (e.g.~the expressions appearing in~\eqref{forinstancereq1}).

An important new technical difficulty that occurs in estimating~\eqref{Peqnonlinearintro} 
is that quantities $P$  and $\underline{P}$ are now only ``almost'' gauge invariant, i.e.~unlike
in~\cite{holzstabofschw}, they now depend on which of the four gauges we are using, i.e.~we have in fact
$P_{\mathcal{H}^+}$, $P_{\mathcal{I}^+}$, etc. This dependence,
however, is  quadratic in the smallness parameter $\varepsilon_0$\index{initial data!parameters!$\varepsilon_0$, smallness parameter appearing in the main theorem}.
Thus, up to quadratic errors quantified already in the bootstrap assumptions, 
one can control the initial data quantities of $P$ evaluated in the teleological
gauge from analogous quantities evaluated in the initial data gauge, and similarly, one can compare
a suitable boundary term (on a timelike boundary $\mathcal{B}$\index{spacetime subsets!$\mathcal{B}$, timelike boundary for energy estimates} in the overlap of the $\mathcal{H}^+$ and
$\mathcal{I}^+$ gauges) which arises when simultaneously applying energy estimates
to~\eqref{Peqnonlinearintro} in the region covered by the two teleological gauges.
With this understanding, one can then repeat the energy estimates as in~\cite{holzstabofschw},
to obtain suitable decay estimates 
for $P$, given estimates on the right hand side~\eqref{RHSschem}.
We emphasise that in accordance with
our comments above, we estimate equation~\eqref{Peqnonlinearintro} directly, viewing
it as a linear equation for $P$, keeping 
the left hand side's dependence on the solution itself. This is essential to prevent
loss of derivatives at top order.

As we shall see, 
the hierarchical
approach to estimates (described in Section~\ref{proofoflinstabinintro} 
in the context
of the linear theory) can be implemented as before, 
to estimate all (differences of) curvature components $\mathcal{R}_p$ (e.g.~$\alpha$, $\rho-\rho_{\circ}$) and
(differences of) Ricci coefficients $\Gamma_p$ (e.g.~$\hat\chi$, $\Omega \tr \chi -(\Omega\tr \chi)_{\circ}$), 
starting from control of
$P$ and $\underline{P}$. (Let us add that in addition to the
transport estimates connecting $\alpha$ to $P$, we also use
the nonlinear analogue of the Teukolsky equation itself~\eqref{teukequationone}
to control certain derivatives
of $\alpha$ which were not explicitly estimated in the linear work~\cite{holzstabofschw}
described in Section~\ref{proofoflinstabinintro}.)
Here, the gauge normalisations, for instance~\eqref{forinstancereq1}--\eqref{forinstancereq2} 
in the $\mathcal{I}^+$ gauge, 
are essential 
to indeed obtain decay of all quantities. See Section~\ref{othernonlinearissues} for a discussion of some specifically
nonlinear issues that arise.

The  estimates thus obtained now allow 
one to absorb the error 
terms arising from~\eqref{RHSschem}
and from similar error terms arising in the other relations, and close the bootstrap.

Let us discuss in more detail the issue of absorbing the errors arising from~\eqref{RHSschem}.
In general, in the context of energy-type estimates, after commutation $N-3$ times by
suitable operators $\mathfrak{D}$\index{double null gauge!differential operators!$\mathfrak{D}$}, the 
$\mathcal{E}[P]$ defined in~\eqref{RHSschem} produces terms which are cubic and higher,
e.g.~terms of the form
\begin{equation}
\label{trilinearintro}
\mathfrak{D}^{N-2}P\cdot \mathfrak{D}^N\Phi \cdot \Phi 
\end{equation}
in  schematic notation,
which must be integrated over spacetime with weights.
The highest order terms $\mathfrak{D}^{N-2}P$ and
$\mathfrak{D}^{N}\Phi$ (an example  of the latter are highest order commuted
curvature terms like
$\mathfrak{D}^N\alpha$),
 must typically be estimated in the energy
norm, whereas  lower order terms $\Phi$  in~\eqref{trilinearintro}
must be taken 
in a higher $L^p$ norm, say $L^\infty$.
The most difficult regions for controlling  these non-linear terms are near null infinity
$\mathcal{I}^+$ 
and near the Schwarzschild photon sphere $r=3M$. 
For it is in these two regions where 
it cannot be ensured that the spacetime integral of the highest
order terms is controlled  by the natural
(weighted) ``integrated local energy decay estimates''.
Thus, in general, to obtain spacetime integrability of the terms~\eqref{trilinearintro}
one must exploit some sort of time-decay for the lower order terms.

We turn to a brief discussion of 
these two regions in Sections~\ref{null_cond_in_intro}
and~\ref{photon_sphere_in_intro} respectively.

\subsection{Estimates near null infinity $\mathcal{I}^+$ and the null condition}
\label{null_cond_in_intro}
\setcounter{equation}{0}

Near null infinity $\mathcal{I}^+$, even local (in $u$) propagation is non-trivial
and requires capturing the so called
``null condition'',  discussed  above in Section~\ref{nonstabminkintint_sec}. 
In the present argument, this enters already at the level of equation~\eqref{Pquantfirstinstance}. 
The null condition can again be captured by 
the subscript-$p$ structure
 in~\eqref{RHSschem} (cf.~the
discussion in  Section~\ref{capturingnullcon}), which ensures
that all nonlinear terms have the correct $r$-decay to propagate.
We emphasise that to infer this structure, we use in an essential way
the Bondi normalisation of the $\mathcal{I}^+$ gauge.

In practice, equation~\eqref{Pquantfirstinstance}  is estimated by applying
the $r^p$ method~\cite{DafRodnew}. (For previous uses of the $r^p$ method for 
nonlinear problems, see~\cite{Yang2013, Pasqualotto:2017rkh, Keir:2018qzh}.)
To globally control the spacetime integral of the terms~\eqref{trilinearintro} arising
from~\eqref{RHSschem},  which are now multiplied by
$r^p$ weights,
one must  in general
exploit decay (in $u$) for the lower order terms. This decay must
match the type of decay which can be obtained from the $r^p$ method itself.
In addition to the null condition, our argument depends on the fortuitous 
absence of certain quantities in~\eqref{RHSschem}, for which
the rate of $u$-decay which can be obtained is very weak.

\subsection{Estimating the non-linear terms near the photon sphere}
\label{photon_sphere_in_intro}
\setcounter{equation}{0}

The region near the Schwarzschild photon 
sphere $r=3M$ presents a similar difficulty to null infinity $\mathcal{I}^+$
because the highest order integrated decay
estimates one obtains from~\eqref{Peqnonlinearintro}
for the quantity $P$ necessarily degenerate.
Thus one again needs to exploit faster decay for the lower order terms in order to
control the part of the spacetime integral arising from~\eqref{RHSschem} 
which is supported near $r=3M$. 

The ``usual'' way of doing this (see for instance~\cite{MR3082240})
is to obtain a pointwise statement  
\begin{equation}
\label{pointwisestate}
|\Phi|\lesssim (u+v)^{-3/2+\delta},
\end{equation}
 for some $\delta<\frac12$
in the region near $r=3M$ for the lower order terms $\Phi$ appearing in~\eqref{RHSschem}.
(Note $u\sim v$ in this region.)
Together with the uniform boundedness
of the highest order energy fluxes (through null hypersurfaces $u=c$ or $v=c$
as appropriate)
 associated to $\mathfrak{D}^{N}\Phi$,
this allows  control of the spacetime integral of terms of the form~\eqref{trilinearintro}, 
restricted to the region near $r=3M$.
Alternatively, however, instead of the pointwise~\eqref{pointwisestate},
it is in fact sufficient to control the 
spacetime integral of
\begin{equation}
\label{averageintro}
u^{2-\delta}\sup_v |\Phi|^2(u,v,\theta^A) +v^{2-\delta}\sup_u|\Phi|^2(u,v,\theta^A)
\end{equation}
for some $\delta>0$, again restricted to the region near $r=3M$. It turns out that one
can 
estimate this spacetime integral of~\eqref{averageintro}  with less assumptions on initial data
than would be required to obtain~\eqref{pointwisestate}. This leads
to important simplifications in the argument.
(See already Section~\ref{seealreadythissubsection}.)

Moreover, the error
terms~\eqref{trilinearintro}
have additional special structure (see already Chapter~\ref{TeukandRegsection}).
For instance, it is useful that certain
terms  do not appear on the right hand
side of~\eqref{RHSschem},
for instance, the term $\nablaslash^{N}\hat\chi$
after $N$ commutations.

\subsection{Other nonlinear issues}
\label{othernonlinearissues}

Let us mention some other nonlinear issues that arise in the estimates of transport equations.
The reader should refer back to the discussion of the linear theory in Section~\ref{lineartheoryboundedness}.

To exploit the nonlinear analogue $Y$\index{double null gauge!connection coefficients!$Y$, quantity used only in the $\mathcal{I}^+$ gauge} of the quantity~\eqref{Ylindef},
necessary to estimate the ingoing shear $r\hat{\underline\chi}$, 
we require additional refinements over the method of~\cite{holzstabofschw}: (i) 
One requires a further renormalisation of the (now non-linear) version of the transport equation~\eqref{transportforY} satisfied by $Y$ to improve the regularity and decay of the right hand side (resulting in the quantity $B$\index{double null gauge!connection coefficients!$B$, quantity used only in the $\mathcal{I}^+$ gauge} , see already~\eqref{Bdef}
in Section~\ref{sec:schematic}). 
(ii) The nonlinear version~\eqref{transportforY} for $Y$ needs in addition to be commuted with $\Omega \slashed\nabla_4$: Decay is then shown for $\Omega \slashed{\nabla}_4 Y$, which in turn, by the gauge conditions on $C^{\mathcal{I}^+}_{v_\infty}$ leads to decay for $Y$ itself.  (iii) A careful analysis of the non-linear errors must be done to keep track of the fact that certain anomalous quantities $\alpha, \beta, \omega-\omega_\circ$ do not appear in the worst terms.  
This is perhaps the most involved part of the estimates for quantities corresponding to 
the $\mathcal{I}^+$ gauge in
 Chapter~\ref{chap:Iestimates}.

Concerning the corresponding issue associated to the ingoing shear $\Omega\hat{\chi}$ and its blue-shifted
transport equation near the horizon $\mathcal{H}^+$,
since the location of the event horizon is not a priori known
we may no longer directly exploit the existence of a conserved flux to obtain a priori bounds on the
horizon itself
as had been done in~\cite{holzstabofschw}.
Instead, we first directly relate
$\Omega\hat{\chi}^{\mathcal{H}^+}$ to $\Omega\hat{\chi}^{\mathcal{I}^+}$
(i.e.~the quantities expressed in the two gauges) on the sphere $S_{u_f,v(R,u_f)}$ and
estimate $\Omega\hat{\chi}$ backwards on the final cone $C_{u_f}$ (for backwards evolution,
the blue-shift is a good sign!). To now estimate $\Omega\hat{\chi}$ is a neighbourhood of
$C_{u_f}$ the ``blue-shift''
problem can be cured, as in~\cite{holzstabofschw}, by successive commutation of the nonlinear analogue of
equation~\eqref{mustdifferentiatethisone} with $\Omega^{-1}\slashed\nabla_3$.
Estimates for $\Omega\hat{\chi}$, and its $\slashed{\nabla}$ derivatives at low orders, are then obtained by integrating
backwards from the final cone $C_{u_f}$, as in~\cite{holzstabofschw}. More care is required, however, in
estimating higher order $\nablaslash$ derivatives of $\Omega \hat{\chi}$ as the above procedure is ``expensive''
in that it ``costs'' two $\Omega^{-1}\slashed{\nabla}_3$ derivatives. (Note that in~\cite{holzstabofschw}
this issue was overcome by exploiting a very special property of the linearised equations.)
To avoid commuting with $\Omega^{-1}\slashed\nabla_3$ too many times, at top order we estimate differences
in place of actual derivatives. Moreover, rather than estimating $\Omega\hat{\chi}$ directly, the
quantity $X=X_1$\index{double null gauge!connection coefficients!$X_1$, quantity used only in the $\mathcal{H}^+$ gauge}  (see already~\eqref{eq:X1def} for its definition), 
is introduced (in complete analogy with the quantity $Y$ discussed above), 
the right hand side of whose equation has
better regularity and decay properties than that of $\Omega \hat{\chi}$.
In fact, to obtain an estimate for $\slashed{\nabla}^N\Omega\hat{\chi}$,  a further
renormalised quantity $X_2$\index{double null gauge!connection coefficients!$X_2$, quantity used only in the $\mathcal{H}^+$ gauge} (see already~\eqref{eq:X2def} for its definition) is considered
(now in analogy with the quantity $B$ discussed above), the right hand side of whose equation has further improved regularity and decay
properties.
Due to the fact that the equation for the difference quotient of $X_2$ is only \emph{noshifted}, unlike the
estimates for $r \hat{\underline{\chi}}$ in the $\mathcal{I}^+$ gauge, the estimates for $\Omega
\hat{\chi}$ in the $\mathcal{H}^+$ gauge grow mildly in $v$ at the highest order.
For more details concerning control of quantities associated to the $\mathcal{H}^+$ gauge, see already
Chapter~\ref{chap:Hestimates}.

\subsection{Top order elliptic estimates for Ricci coefficients}
\label{necessity_sec}
\setcounter{equation}{0}

We have discussed in Section~\ref{nonstabminkintint_sec} 
the elliptic structure of some of the equations
in~\eqref{coupsys} (e.g.~equation~\eqref{elliptichere})
and how this structure can be used to improve the control of the Ricci coefficients
from the point of view of differentiability, allowing for top order estimates
on these coefficients at one degree of differentiability greater than curvature.

In the context of the linear stability of
Theorem~\ref{linstabthemintro}, such top order control was not necessary
for the estimates to close. Thus this control was not part of the statement
of the theorem, although it could be obtained a posteriori.
For the present work, however,
it turns out to indeed be necessary to estimate certain Ricci coefficients at the top order
at the intersection of $C_{u_f}$ and $\underline{C}^{\mathcal{H}^+}_{v_{-1}}$. 
Since the gauge is normalised
to the future,  this in turn requires a top order estimate for quantities
along the whole of $C_{u_f}$ and, in view of the normalisation in the $\mathcal{I}^+$-gauge,
this then requires control of  essentially all Ricci coefficients at top order.
As a result, the top order elliptic estimates as described in Section~\ref{nonstabminkintint_sec}  form an 
integral part of the argument.

\subsection{The $\ell=0,1$ modes}
\label{lowmodes_sec}
\setcounter{equation}{0}

Recall from Section~\ref{proofoflinstabinintro}
that in the context of the linear stability result of Theorem~\ref{linstabthemintro}, 
the projection of a solution
to the $\ell=0$ and $\ell=1$ modes decouples from the
rest of the system and can be written as the sum of a pure gauge solution
and a linearised Kerr solution.
The linearised Kerr metric to which the solution approached
could be read off from the quantities $\rlin_{\ell=0}$, $\slin_{\ell=1,m=-1,0,1}$ 
which, appropriately scaled, were moreover conserved.
At infinity the latter three quantities can be 
related to $(\curlslash\beta)_{\ell=1,m={-1},0,1}$ .

In the present problem,  to  identify
the $\lambda^{\rm final}$ such that $\mathcal{S}_0(\lambda^{\rm final})$ 
corresponds to a
 solution which indeed asymptotes
to a Schwarzschild metric of some mass $M_{\rm final}$, the values of
\begin{equation}
\label{monitor}
-\frac{r^3}2\rho_{\ell=0}, \qquad r^5(\curlslash\beta)_{\ell=1,m={-1},0,1}
\end{equation}
are monitored at late times at the final sphere $S_{u_f,v_\infty}$
for solutions corresponding to general $\lambda$.
The Schwarzschild mass $M_f$ used for taking differences at the $u_f$ stage of
the bootstrap is set equal to $-\frac{r^3}2\rho_{\ell=0}(u_f,v_\infty)$, while the 
vector\index{bootstrap!parameters!${\bf J}(\lambda, u_f)$, vector measuring angular momentum}\index{angular momentum!${\bf J}(\lambda, u_f)$, vector measuring angular momentum at time $u_f$}
\begin{equation}
\label{Jdefherehere}
{\bf J}(\lambda, u_f) = r^5(\curlslash\beta)_{\ell=1,m={-1},0,1}(u_f,v_\infty)
\end{equation}
is used to restrict $\lambda$ to 
a region $\mathfrak{R}(u_f)\subset \mathbb R^3$ of parameter space  determined by
property 
\begin{equation}
\label{imposition}
|{\bf J}(\lambda, u_f)| \le \varepsilon_0 u_f^{-1}.
\end{equation}
The quantity~\eqref{Jdefherehere} will play a fundamental role in completing the
continuity argument discussed in the next section.

 We note  that the $\ell=0,1$ modes above 
 are defined by the projection to the four dimensional kernel space
 of a natural differential operator. (See already Section~\ref{lowmodessection}.)
 
 The fact that, at any given time $u_f$, we are dealing with solutions with potentially
 non-trivial angular momentum~\eqref{Jdefherehere},
 bound only by the linear constraint~\eqref{imposition},
 suggests that, for the sharpest estimates, we should in fact subtract a 
  ``reference fixed-mass
linearised Kerr solution'', with parameters  determined by~\eqref{Jdefherehere}.
This turns out to indeed be necessary in order to improve our bootstrap estimates for the
projections on our $\ell=0,1$ modes. Because these projections no longer exactly decouple,
these lead to various error terms.
In particular, bounds for the derivatives of these projection operators must be included
in our bootstrap assumptions.

\subsection{Completing the continuity argument}
\label{completingintro}
\setcounter{equation}{0}

Once the bootstrap assumptions are improved, one 
aims to show thus that the set $\mathfrak{B}\subset [u^0_f,\infty)$\index{bootstrap!sets!$\mathfrak{B}$, set of allowed
final retarded times $u_f$}  of ``allowed'' 
final retarded times $u_f$ (i.e.~the set $\mathfrak{B}$ of $u_f$ such that the gauges
can be constructed as described and such that the bootstrap assumptions indeed hold)
is a non-empty, open and closed subset of $[u^0_f,\infty)$,
and thus $\mathfrak{B}=[u^0_f,\infty)$.

The standard structure (see already Sections~\ref{hopeitsnotempty},~\ref{opensection} and~\ref{closedsection})
for such an argument is to appeal to Cauchy stability to obtain that $u^0_f\in \mathfrak{B}$ 
and then
to the improvement of the bootstrap assumptions, together
with an appropriate local existence theorem, to show that one can extend the bootstrap region.
In view of our setup, however, 
let us note already
that there are two additional slightly non-standard
features, related to the finite codimensionality nature of the result and the teleological nature
of the gauge: 

1.~Related to the finite codimensionality, 
we have now also a varying set of parameters $\mathfrak{R}(u_f)$ to contend with.
The decay imposed by~\eqref{imposition}, together with faster decay estimates for suitable derivatives
of~\eqref{monitor},
allows us to show the monotonicity statement
\begin{equation}
\label{monotonicity_intro}
\mathfrak{R}(u'_f)\subsetneq \mathfrak{R}(u_f)
\end{equation}
for $u'_f>u_f$ (see already Section~\ref{section:monotonicityR}).
(Note  that we do not have complete freedom in choosing the decay rate~\eqref{imposition},
as the quantities~\eqref{monitor} couple linearly to the other $\ell=0$ and $\ell=1$ modes
and appear in the nonlinear error terms   
coupling to the other modes of the system.)
On the other hand,  we note that by a strengthening of Cauchy stability type arguments, the 
original region $\mathfrak{R}(u_f^0)$ can be seen to be
a closed disc and the map ${\bf J}_0:\mathfrak{R}(u_f^0)\to B$ defined by ${\bf J}_0(\lambda):=(\lambda, u_f^0)$,
mapping to an appropriate closed ball $B\subset \mathbb R^3$,
can be seen to have degree $1$ on $\partial\mathfrak{R}(u_f^0)$ (see already Section~\ref{thedegreeonemap}). 
A topological argument applied to an appropriately defined map ${\bf J}_{u_f}$ associated to~\eqref{Jdefherehere}
can then be used to show
that $\mathfrak{R}(u_f)$  indeed remains non-empty for all $u_f>u_f^0$ (see again
already Section~\ref{section:monotonicityR}).

2.~In view of the teleological normalisation of the gauge (see Section~\ref{two_gauges_intro}
immediately below), enlarging the bootstrap region requires an appeal to an iteration argument around
the analogous linearised construction
in order to select the new ``final'' sphere $S_{u_f',v'_\infty}$ of an enlarged region and achieve
the gauge normalisations such as~\eqref{forinstancereq1}--\eqref{forinstancereq2} 
described previously in Section~\ref{two_gauges_intro}. 
Here again, estimates comparing different gauges (cf.~Section~\ref{bounded_init_intro}) 
are used to ensure the regularity of the relevant
map.

Once it has been established that $\mathfrak{B}=[u_f^0,\infty)$,
one obtains in view of~\eqref{monotonicity_intro} that 
there exists $\lambda^{\rm final}=
(\lambda^{\rm final}_1,\lambda^{\rm final}_2, \lambda^{\rm final}_3)$,
such that $\lambda^{\rm final}\in \mathfrak{R}(u_f)$ for all $u_f \ge  u_f^0$,
and a  solution $(\mathcal{M},g):= (\mathcal{M}(\lambda^{\rm final}), g(\lambda^{\rm final}))$
generated by data $\mathcal{S}:=
\mathcal{S}_0(\lambda^{\rm final})$, with an
$M_f$ tending to the final Schwarzschild mass $M_{\rm final}$ and with ``final total angular momentum''
zero.
We see thus how it is this collapsing of the $3$-parameter family~\eqref{threeparamfamintro} 
which  leads  to the expected
codimension-$3$ statement in the theorem.

\subsection{The properties of null infinity $\mathcal{I}^+$ and the event horizon $\mathcal{H}^+$}
\label{propsofHandI_sec}
\setcounter{equation}{0}

 The asymptotic stability statements for $(\mathcal{M},g)$ defined above
 and the statements about the properties of null infinity $\mathcal{I}^+$ as well as
the existence and regularity of the event horizon $\mathcal{H}^+$
then follow easily via an Arzela--Ascoli argument
by taking the limit of the estimates obtained in the bootstrap regions (see already
Section~\ref{theendoflogic} and Chapter~\ref{conclusionsection} of Part~\ref{conclusionpart}).

The finite gauge normalisations and estimates imply that the 
asymptotic $\mathcal{I}^+$ gauge is ``Bondi-normalised'', meaning that in particular 
one can associate to it  a set $\mathcal{I}^+$ with coordinates $(u,\theta)$,
with $u\in [u_{-1},\infty)$, $\theta\in \mathbb S^2$, and
asymptotic quantities, defined on $\mathcal{I}^+$, familiar from~\cite{BondiBMS, sachsBMS} and~\cite{CK}, for instance
the Bondi news $\Xi$\index{null infinity!$\Xi$, Bondi news}, the asymptotic shear $\Sigma$\index{null infinity! $\Sigma$, asymptotic
shear},  and the rescaled curvature components
${\bf A}$\index{null infinity! ${\bf A}$, rescaled curvature component $\alphabar$}, 
${\bf B}$\index{null infinity! ${\bf B}$, rescaled curvature component $\betabar$}, 
${\bf P}$\index{null infinity! ${\bf P}$, rescaled curvature component $\rho$} 
and ${\bf Q}$\index{null infinity! ${\bf Q}$, rescaled curvature component $\sigma$}, 
satisfying the ``laws of gravitational radiation''.
Given the normalisation, the fact that $u$ exhausts the range 
$\in[u_{-1},\infty)$ corresponds to the (future) completeness of null infinity.
Our teleological normalisation of the gauge imposes in particular the condition that,
defining ${\bf P}^+(\theta) := \lim_{u\to\infty} {\bf P}(u,\theta)$\index{null infinity! ${\bf P}^+$, final value of
${\bf P}$  (related to centre of mass)}, we have
\begin{equation}
\label{centreofmass}
{\bf P}^+_{\ell =0}= M_{\rm final}, \qquad {\bf P}^+_{\ell \ge 1} =0,
\end{equation}
 while
\begin{equation}
\label{BMSbroken}
\Sigma^+:=\lim_{u\to\infty}\Sigma
=0. 
\end{equation}
The condition~\eqref{centreofmass} has the interpretation that the final Schwarzschild black hole
is at rest with respect to our teleologically normalised frames. The condition~\eqref{BMSbroken}, 
in the language of~\cite{newmanpenroseBMS}, is the condition that the final ``cut'' of null infinity is asymptotically
a ``good cut''. Given that our two teleological gauges share a final outgoing cone $C_{u_f}$,
one can think that the normalisation~\eqref{BMSbroken} is a necessary consequence of
the requirement to obtain decay for all quantities at constant $r$.\footnote{This is in distinction
with the proof of the stability of Minkowski space~\cite{CK}, where the supertranslation symmetry is not
broken, corresponding to the fact that the construction of the proof in~\cite{CK} depends on an arbitrary
choice of outgoing cone which is a level surface of the exterior optical function. This is still compatible
with decay of all quantities for fixed $r$ since the ``interior'' region was covered by a different optical functions whose null cones
were not related to the null cones of the exterior optical function. In the exterior region of~\cite{CK}, on the other hand,
the $r$ decay is sufficient to compensate for the lack of decay for some fully rescaled quantities in $u$, so as
for all nonlinearities to still be controllable.} 
 See already Section~\ref{rescaledandgravrad}.

The uniform estimates in the asymptotic $\mathcal{H}^+$ normalised gauge, together with
the presence of a trapped surface in the initial data which provides a barrier, 
allow one to extract a limiting, regular hypersurface $\mathcal{H}^+$. Another appeal to local
existence of the characteristic initial value problem ensures that $\mathcal{H}^+$ is indeed in
the maximal development $(\mathcal{M},g)$ (and thus also a neighbourhood of $\mathcal{I}^+$). 
It is clear moreover that $\mathcal{H}^+$ is the future boundary of $J^-(\mathcal{I}^+)$, when the latter
is interpreted in the obvious way, i.e.~$\mathcal{H}^+$ is indeed the event horizon of the
black hole region of $(\mathcal{M},g)$.
The future completeness of $\mathcal{H}^+$, together with decay along $\mathcal{H}^+$ for various
quantities, follows directly from the estimates. See already Section~\ref{eventsmydear}.

\section{Outline of the work and guide to the reader}
\label{outline_sec}
We end this introduction with an outline of our work and a guide to the reader.

The remainder of the  work is divided into four parts.
In {\bf Part~\ref{preliminlabel}}, we shall give various preliminaries concerning
double null gauge with a background Schwarzschild metric,  the
almost gauge invariant hierarchy satisfying the Regge--Wheeler and Teukolsky equations, teleological gauge normalisations, schematic
notation, and a formalism for change
of double null gauge.
Then, in  {\bf Part~\ref{stateandlogicpart}} we shall introduce the local theory
and relevant setup
necessary  to give 
a detailed statement of the main theorem (stated as
{\bf Theorem~\ref{thm:main}})
 and the logic of its proof. 
 The main analytical content of the proof is contained in {\bf Part~\ref{improvingpart}}
 which corresponds to ``improving the bootstrap assumptions''. The remaining statements
 of the theorem including a necessary existence theorem for the gauge and the final conclusions 
 concerning the event horizon and null infinity are obtained in {\bf Part~\ref{conclusionpart}}.
Each part will be prefaced with a more detailed summary of its contents.

Our work has been written so that it may be read linearly. There are various alternative tracks
through the work, however, that some readers---particularly the impatient reader!---may wish to pursue.  In particular,
the reader anxious to understand the large-scale architecture of the statement and proof of the main theorem
can skip large parts of Part~\ref{preliminlabel} after reading Chapter~\ref{theequationssec}, 
turning directly to Part~\ref{stateandlogicpart} and referring back.
Part~\ref{conclusionpart} can then be read independently of 
Part~\ref{improvingpart}. On the other hand, Part~\ref{improvingpart}
can be understood
somewhat independently of the architecture of the proof of the main theorem. 

More precise directions for these alternative tracks will be given  in the prologues to the various
parts and chapters.

As we have noted above, the double null gauge framework used in this work is common
to a host of different recent works in general relativity~\cite{Chr, KN, klailukrod, lukrodimpulsive, lukweaknull, DafLuk1, vacuumscatter}, as well as providing the framework for our previous complete treatment of the
linear stability problem in~\cite{holzstabofschw}. 
Thus, the reader who has studied previously these works will find familiar
much of the basic setup in Part~\ref{preliminlabel},
as well as the architecture of the proof in Part~\ref{stateandlogicpart} and the general strategy of estimates
in Part~\ref{improvingpart}, if not their precise form.
We have included
helpful comparisons with some of these other works when appropriate. 
Conversely, we hope that the
reader who encounters double null gauge for the first time in the present work will find that the
effort necessary to learn this formalism pays off in making 
the above other works more accessible for future study.

\addcontentsline{toc}{section}{Acknowledgments}
\section*{Acknowledgements}
During the preparation of this work:
MD~acknowledges support through NSF grants DMS-2005464, DMS-1709270 and 
EPSRC grant
EP/K00865X/1.  GH acknowledges support by the Alexander von Humboldt Foundation in the framework of the Alexander von Humboldt Professorship endowed by the Federal Ministry of Education and Research 
and 
ERC Consolidator Grant 772249.  MT~acknowledges support through Royal Society Tata University Research Fellowship URF\textbackslash R1\textbackslash 191409 and ERC Consolidator Grant 772249.    IR~acknowledges support
through NSF grants DMS-2005464,  DMS-1709270 and a Simons Investigator Award.

\renewcommand{\thepart}{\Alph{part}}

\part{Preliminaries}
\label{preliminlabel}

In this part of the work, we will give various preliminaries concerning
the double null gauge, the Schwarzschild background (with various derived concepts),
and the almost gauge invariant hierarchy and teleological gauge normalisations.

\parttoc

We begin in {\bf Chapter~\ref{theequationssec}}
with the formulation of the Einstein vacuum equations in double null gauge
and the Schwarzschild background.

We shall then present in {\bf Chapter~\ref{almostgaugeandtellychapter}} two key concepts 
which will play an important role in our proof: an almost gauge
invariant hierarchy and the teleological gauge normalisations.

In {\bf Chapter~\ref{moreprelimchapter}}, we shall introduce a schematic notation for differences
with Schwarzschild background quantities, and derive the Teukolsky and Regge--Wheeler equations,
exploiting the schematic notation to organise ``error terms''.

Finally, in {\bf Chapter~\ref{bigchangeisgood}}, we shall derive formulas relating change of gauge between
two local double null parametrisations, and discuss how to break the diffeomorphism invariance
of the sphere.
\vskip1pc
\emph{Chapter~\ref{theequationssec} is essential for all other chapters in the work and must be read
first. The reader impatient to proceed to Part~\ref{stateandlogicpart} can skip portions of Chapters~\ref{almostgaugeandtellychapter} and~\ref{bigchangeisgood}
and almost the entirety of Chapter~\ref{moreprelimchapter}, though they will need to refer to these skipped portions to understand the energies appearing in the main theorem. We shall give more precise instructions for such a reader in the preambles
of the various chapters.}

\adjustmtc

\renewcommand{\thesection}{\arabic{chapter}.\arabic{section}} 


\renewcommand{\theequation}{\thesection.\arabic{equation}}

\chapter{The vacuum Einstein equations in double null gauge and the Schwarzschild background}
\label{theequationssec}

\addtocontents{toc}{\setcounter{tocdepth}{1}}

In this chapter we will present the Einstein vacuum equations~\eqref{vaceqhere} 
in double null gauge and associate to such a gauge a Schwarzschild background solution.

\minitoc

We begin in {\bf Section~\ref{geomprelimforvac}} with certain geometric preliminaries
concerning a manifold represented globally by a double null coordinate system.
We shall then give the complete set of  structure equations 
in {\bf Section~\ref{thatsallofthem}} resulting from~\eqref{vaceqhere}.
In {\bf Section~\ref{EFandKrubothsec}}, we introduce the Schwarzschild metric,
expressed in two double null coordinate systems, Eddington--Finkelstein 
and Kruskal
normalised.  This will allow us to associate a Schwarzschild background
to a given general solution of the vacuum equations in double null gauge.
With this, we shall define in {\bf Section~\ref{lowmodessection}} the $\ell=0,1$ modes
and the reference linearised Kerr solutions associated to a solution.

\vskip1pc
\emph{The results and notations of this chapter will be used throughout the work. 
After finishing this chapter, the reader impatient to proceed to Part~\ref{stateandlogicpart} may read Chapter~\ref{thelocaltheorysec} up to and including Section~\ref{initialdatanormsec}.
(We remark in fact
that  Sections~\ref{initdatasection}--\ref{maxcauchysection} of Chapter~\ref{thelocaltheorysec}  only depend on Sections~\ref{geomprelimforvac}--\ref{thatsallofthem} here
and  thus could be read directly after these if the reader prefers.)}

\section{Geometric preliminaries}
\label{geomprelimforvac}

In this section, we give certain geometric preliminaries.

We begin in
{\bf Section~\ref{sphereconcrete}} with some notation concerning the plane
and sphere which will allow us to introduce in
 {\bf Section~\ref{coordsofficial}} the differentiable
structure of the ambient manifold associated to what will be a
double null gauge. We shall consider then the metric and time orientation
in {\bf Section~\ref{timeorientsection}}, interpreting our previously defined
coordinates as double null coordinates.
We shall define associated null frames and differential operators  in {\bf Section~\ref{nullframesopers}}. This
will allow us to define all Ricci coefficients and curvature components 
in {\bf Section~\ref{Riccoefsandcurvsec}}.

The reader may wish to also consult and compare with the references~\cite{Chr, DafLuk1, holzstabofschw} for
equations and notations very similar to those presented here. (We will point out differences arising from our slightly different conventions
wherever appropriate.)

\subsection{The plane $\mathbb R^2$ and the sphere $\mathbb S^2$}
\label{sphereconcrete}

Let us fix some notation concerning the plane $\mathbb R^2$ and the sphere $\mathbb S^2$ which 
will be related to the domains of double null parametrisations.

We will denote the standard coordinates on $\mathbb R^2$ as $(u,v)$\index{double null gauge!coordinates!$u$, retarded null coordinate}\index{double null gauge!coordinates!$v$, advanced null coordinate}
or  $(U,V)$.

We shall consider $\mathbb S^2$\index{sphere!manifold!$\mathbb S^2$, standard sphere in $\mathbb R^3$} concretely, say as a subset $\mathbb S^2\subset \mathbb R^3_{(x,y,z)}$.
Let us fix the north pole $(0,0,1)\in \mathbb S^2$ and let $(\mathring\theta, \mathring\phi)$ denote
standard spherical coordinates\index{sphere!coordinates!$\mathring\theta$, standard spherical coordinate}\index{sphere!coordinates!$\mathring\phi$, standard spherical coordinate} defined by
\begin{equation}
\label{sphcorddef}
\mathring\theta =\cos^{-1}z ,\qquad \mathring \phi =\tan^{-1} (y/x).
\end{equation}

We now let $\mathring{\gamma}$ denote the standard metric on $\mathbb S^2$ induced from
the inclusion\index{sphere!$\mathring{\gamma}$, standard metric on $\mathbb S^2$}
$\mathbb S^2\subset \mathbb R^3$.
In the spherical coordinates defined above, we have
\[
\mathring{\gamma}= d\mathring\theta^2+ \sin^2\mathring \theta\,  d\mathring\phi^2.
\]

Whereas we shall always refer to standard coordinates $(u,v)$ on $\mathbb R^2$, on 
the sphere $\mathbb S^2$, we shall often consider \emph{general} local coordinates. We will denote such
\emph{general} local coordinates on the sphere 
by $\theta^A$,\index{sphere!coordinates!$\theta^A$, general local coordinates}  where $A=1,2$.

We will also sometimes use the notation $\theta\in \mathbb S^2$ to denote a point on the sphere,
with no reference to any particular coordinate system.

\subsection{Coordinates, differentiable structure and $S$-tangent tensors}
\label{coordsofficial}
We fix  some\index{double null gauge!sets!$\mathcal{W}$, subset of $\mathbb R^2$}
\[
\mathcal{W}\subset \mathbb R^2
\]
to be a non-empty open subset. (Later, we shall allow more generally $\mathcal{W}$ to be a
$2$-dimensional
submanifold of $\mathbb R^2$ with piecewise smooth boundary.)

We define now\index{double null gauge!sets!$\mathcal{Z}$, domain of parametrisation}  
\begin{equation}
\label{wemaywriteit}
\mathcal{Z}=\mathcal{W} \times\mathbb S^2
\end{equation}
We note that the standard coordinates $(u,v)$ of $\mathbb R^2$ together with a local coordinate system $\theta^A$, $A=1,2$, on
$\mathbb S^2$, as described in Section~\ref{sphereconcrete}, define a local coordinate system $(u,v,\theta^A)$ on 
$\mathcal{Z}$.

We denote 
\begin{equation}
\label{defSuv}
S_{u,v}=\{(u,v)\}\times \mathbb S^2.\index{double null gauge!sets!$S_{u,v}$, intersections of constant-$u$ and constant-$v$ hypersurfaces}
\end{equation}

An  $S$-tangent $(r,s)$ tensor (or $S$-tensor for short)
will be defined to be a tensor $\phi$ which when expressed
in the coordinate basis defined by the above coordinates takes the form
\[
\phi=\phi^{A_1\ldots A_r}_{B_1\ldots B_s}(u,v,\theta^1,\theta^2)
\frac\partial{\partial \theta^{A_1}}\otimes\cdots\otimes  \frac\partial{\partial \theta^{A_r}}
\otimes
 d\theta^{B_1}\otimes \cdots\otimes d\theta^{B_s} .
\]
In the above, $r$ labels the number of contravariant indices and
$s$ the number of covariant indices. We shall refer as usual 
to  $S$-tangent $(1,0)$ tensors as $S$-tangent vector fields, $(0,1)$ tensors as $S$-tangent
$1$-forms,
and $(0,s)$ tensors as $S$-tangent $s$-covariant  tensors.
Note that the notion of $S$-tensor
does not depend on the choice of local coordinates 
$\theta^A$ on $\mathbb S^2$, and defines for each fixed $(u,v)$ a tensor
on $S_{u,v}$.

\subsection{Metric and time orientation}
\label{timeorientsection}

On $\mathcal{Z}$, let $\Omega^2$\index{double null gauge!metric coefficients! $\Omega^2$, conformal metric coefficient} 
denote a smooth positive function,
let $b$\index{double null gauge!metric coefficients!$b^C$, torsion metric coefficient} denote a smooth $S$-tangent 1-form and let
$\slashed{g}$\index{double null gauge!metric coefficients!$\slashed{g}_{CD}$, angular metric coefficient} denote a smooth $S$-tangent symmetric covariant $2$-tensor
which is assumed to be positive definite (thought of as a tensor on $S_{u,v}$).

(More generally, we will consider such tensors of suitably high regularity so that the operations to be considered
in this Chapter can be defined.)

Under the above assumptions, the expression
\begin{equation}
\label{doublenulllongform}
g=-2\Omega^2(u,v,\theta^A)(du\otimes dv+ dv\otimes du)
+\slashed{g}_{CD}(u,v,\theta^A)(d\theta^C-b^C(u,v,\theta^A)dv)
\otimes (d\theta^D-b^D(u,v,\theta^A)dv)
\end{equation}
defines a smooth Lorentzian metric on $\mathcal{Z}$.
Note that the hypersurfaces 
\begin{equation}
\label{originalconedefs}
C_{u}:=\bigcup_{v: (u,v) \in\mathcal{W}}S_{u,v}, \qquad C_{v}:=\bigcup_{u: (u,v) \in\mathcal{W}}S_{u,v},
\index{double null gauge!sets!$C_u$, ingoing null hypersurface}\index{double null gauge!sets!$C_v$, outgoing null hypersurface}
\end{equation}
of constant $u$ and $v$
respectively, are null with respect to the above metric, while 
$S_{u,v}$ defined in~\eqref{defSuv}
satisfies $S_{u,v}=C_{u}\cap C_{v}$ and 
is spacelike with induced (Riemannian) metric  $\slashed{g}$.
We will denote by $\slashed{g}^{AB}$ the components of the inverse induced
metric $\slashed{g}^{-1}$ on $S_{u,v}$ and we will 
denote by $\slashed\epsilon_{AB}$ the components
of the volume form of $\slashed{g}$.

We may time-orient the above metric~\eqref{doublenulllongform} 
by the timelike vector $\partial_u+\partial_v+b^A\partial_{\theta^A}$\index{double null gauge!time orientation!$\partial_u+
\partial_v +b^A\partial_{\theta^A}$}. This makes $(\mathcal{M},g)$
into a $4$-dimensional spacetime.

Note finally that given an arbitrary $(r,s)$ 
tensor field $\psi$, there is a unique projection
to an $(r,s)$ $S$-tensor $\Pi \psi$\index{double null gauge!projection!$\Pi \phi$} defined by
\[
(\Pi\psi)^{A_1\ldots A_r}_{B_1\ldots B_s}=\psi^{A_1\ldots A_r}_{B_1\ldots B_s},
\qquad \text{all other components vanish}.
\]
This definition is independent of the choice of local coordinates $\theta^A$ on 
$\mathbb S^2$.

\subsection{Null frames and associated differential operators}
\label{nullframesopers}

We define globally 
on $\mathcal{Z}$
the vector fields\index{double null gauge!frames!$e_i, i=1,\ldots 4$, normalised double null frame}
\begin{equation}
\label{nullframedef}
e_3=\Omega^{-1}\partial_u, \qquad 
e_4=\Omega^{-1} \left( \partial_v +b^A\partial_{\theta^A} \right).
\end{equation}
Note that $e_3$ and $e_4$ are future-directed.

Locally, we may complete $e_3$ and $e_4$ to a so-called null frame.
Let $\theta^A$, $A=1,2$ denote local coordinates on $\mathbb S^2$. 
Consider the $S$-tangent vector fields 
\begin{equation}
\label{coordinatevfs}
e_A=\frac{\partial}{\partial\theta^A}
\end{equation}
defined in a local neighbourhood on $\mathcal{M}$.
We shall call the collection $\{e_1, e_2, e_3, e_4\}$ a local null frame.
Note that we have
\[
g(e_3,e_4)=-2,\qquad g(e_3,e_A)=0, \qquad g(e_4,e_B)=0.
\]

We emphasise that $e_1$ and $e_2$ are coordinate vector fields and
thus not in general orthonormal! Recall also that such a frame cannot be defined globally 
on all of $\mathbb S^2$. 

Note also that for covariant $S$-tensors $\phi$, since $e_A$ are defined to be coordinate 
vector fields, then the nontrivial tensor components $\phi_{A_1,\ldots, A_s}$ satisfy
$\phi_{A_1,\ldots, A_s}= \phi(e_{A_1},\ldots, e_{A_s})$.

We will require various differential operators which take $S$-tensors to
$S$-tensors.

We define the projected Lie derivatives  
 $\underline{D}$\index{double null gauge!differential operators!$\underline{D}$, projected Lie derivative in the $e_3$ direction} 
 and $D$\index{double null gauge!differential operators!$D$, projected Lie derivative in the $e_4$ direction}  to act
on an $S$-tensor $\phi$ as the $S$-tensor projection 
of the Lie derivatives $\mathcal{L}_{\Omega e_3}\phi$
and $\mathcal{L}_{\Omega e_4}\phi$, respectively, i.e.
\begin{equation}
\label{projliederivs}
\underline{D}\phi = \Pi ( \mathcal{L}_{\Omega e_3}\phi), \qquad
D\phi=\Pi (\mathcal{L}_{\Omega e_4}\phi).
\end{equation}
Note that the above operators map $S$-tangent $(r,s)$ tensors into $S$-tangent $(r,s)$ tensors.

The operator $\slashed\nabla$ acts on an $S$-tensor as the induced covariant 
derivative on $S_{u,v}$.  Note that this is well defined and raises the order of a tensor.
We use the standard notation $\slashed\nabla_C\phi_{A_1,\ldots, A_s}^{B_1,\ldots, B_r}$
for the components of the tensor $\slashed\nabla\phi$.

The operator $\slashed\nabla_3$ and $\slashed\nabla_4$
acting on $S$-tensors at $p=(u,v,\theta)$
are the $S$-tensor projections
of the covariant derivatives $\nabla_3=\nabla_{e_3}$ and $\nabla_4=\nabla_{e_4}$\index{double null gauge!differential operators!$\slashed\nabla_3$, covariant differential operator acting in $e_3$ direction}\index{double null gauge!differential operators!$\slashed\nabla_4$, covariant differential operator acting in $e_4$ direction} respectively,
i.e.
\begin{equation}
\label{projcovderivdef}
\slashed\nabla_3\phi = \Pi (\nabla_3\phi ), \qquad
\slashed\nabla_4\phi =\Pi (\nabla_4\phi).
\end{equation}
Like their projected Lie derivative analogues, the above 
operators~\eqref{projcovderivdef} preserve the order of $S$-tensors $\phi$.

Finally, for totally symmetric covariant $S$-tensors $\phi$ of rank $s+1$,
we define the covariant rank $s$ $S$-tensors\index{double null gauge!differential operators!$\divslash$, covariant operator acting tangentially on $S_{u,v}$}\index{double null gauge!differential operators!$\curlslash$, covariant operator acting tangentially on $S_{u,v}$}
\[
(\divslash \phi)_{A_1\cdots A_s}:= \slashed{g}^{BC}\nablaslash_B  \phi_{CA_1\cdots A_s}
\]
\[
(\curlslash \phi)_{A_1\cdots A_s}:= \slashed\epsilon^{BC}\nablaslash_B  \phi_{CA_1\cdots A_s}.
\]
For $S$-tangent $1$-forms $\xi$ we will also require the operator\index{double null gauge!differential operators!$\Dslash_2^*$, covariant operator acting tangentially on $S_{u,v}$}
\begin{equation}
\label{Dslash2stardef}
	\Dslash_2^* \xi
	=
	-\frac{1}{2}
	\left(
	\nablaslash \xi
	+
	\nablaslash^T \xi
	-
	\divslash \xi \gslash
	\right),
\end{equation}
where $\nablaslash^T\xi$ denotes the transpose of $\nablaslash \xi$,
\begin{equation}
\label{transposedef}
	(\nablaslash^T \xi)_{AB} = \nablaslash_B \xi_A.
\end{equation}

In addition to the above differential operators, 
we define the following algebraic operations on $S$-tensors.
Let $\vartheta_{AB}$ and $\tilde\vartheta_{AB}$ be symmetric
covariant  $S$-tangent 2-tensors and
$\xi_A$, $\tilde\xi_A$ be covariant $S$-tangent 1-forms.
We define
\index{double null gauge!algebraic operations!$(\vartheta\times\tilde{\vartheta})_{BC}$}\index{double null gauge!algebraic operations!$(\vartheta,\tilde{\vartheta})$}\index{double null gauge!algebraic operations!$(\xi,\tilde{\xi})$}\index{double null gauge!algebraic operations!$(\xi\hat{\otimes}\tilde{\xi})_{AB}$}\index{double null gauge!algebraic operations!$\vartheta\wedge \tilde{\vartheta}$}
\begin{align*}
(\vartheta\times\tilde\vartheta)_{BC}&:=\slashed{g}^{AD}\vartheta_{AB}\tilde\vartheta_{DC}
\\
(\vartheta,\tilde\vartheta)&:=\slashed{g}^{AC}\slashed{g}^{BD}\vartheta_{AB}\tilde\vartheta_{CD}
\\
(\xi,\tilde\xi)&:=\slashed{g}^{AB}\xi_A{\tilde\xi}_B
\\
(\xi\hat{\otimes}\tilde\xi)_{AB}&:=\xi_A\tilde{\xi}_B+\xi_B\tilde\xi_A-\slashed{g}^{CD}
\xi_C\tilde\xi_D\slashed g_{AB}
\\
\vartheta\wedge \tilde\vartheta&: = \slashed\epsilon^{AB}\slashed{g}^{CD}\vartheta_{AC}
\tilde\vartheta_{BD}.
\end{align*}

For totally symmetric covariant $S$-tensors of rank $s+ 2$ we define
\[
(\tr \phi)_{A_1\ldots A_s}:=\slashed{g}^{BC}\phi_{BCA_1\ldots A_s}.\index{double null gauge!algebraic operations!$(\tr \phi)_{A_1\ldots A_s}$}
\]
For $S$-tangent  $1$-forms $\xi_A$ and symmetric covariant $S$-tangent  $2$-tensors $\vartheta_{AB}$
we define the Hodge duals ${}^*\xi_A$ and ${}^*\vartheta_{AB}$ by the expressions
\[
{}^*\xi_A:=\slashed{g}_{AC}\slashed\epsilon^{CB}\xi_B,\qquad
{}^*\vartheta_{AB}:=\slashed{g}_{BD}\slashed\epsilon^{DC}\vartheta_{AC}.
\index{double null gauge!algebraic operations!${}^*\xi_A$, Hodge dual}
\index{double null gauge!algebraic operations!${}^*\vartheta_{AB}$, Hodge dual}
\]
For a (not necessarily symmetric) 2-covariant
$S$-tensor field ${\vartheta}_{AB}$ and an $S$-tangent $1$-form $\xi_A$
we note the musical operations
\[
\vartheta_A^{\sharp \,\, C}=\vartheta_{AB}\slashed{g}^{BC}, \qquad
\xi^{\sharp \,\,C}=\xi_{A}\slashed{g}^{BC},
\index{double null gauge!algebraic operations!$\vartheta_A^{\sharp \,\, C}$}
\index{double null gauge!algebraic operations!$\xi^{\sharp \,\,C}$}
\]
defining $S$-tangent  $(1,1)$ and $(1,0)$ tensors, respectively.

For an $S$-tensor $T$,  we shall often use the coercive expression
\begin{equation}
\label{normforStensors}
	\vert T \vert^2_{\gslash}
	:=
	\gslash^{A_1B_1} \cdots \gslash^{A_kB_k} T_{A_1\ldots A_k} T_{B_1\ldots B_k}.
\end{equation}

\subsection{Ricci coefficients and curvature components}
\label{Riccoefsandcurvsec}

We have already given examples of Ricci coefficients and curvature components in
Section~\ref{doublenullgaugeintheintro}. Here, we give the complete
list, expressed in terms of a local null frame $e_1$, $e_2$, $e_3$, $e_4$.
We use the notation $\nabla_A=\nabla_{e_A}$, where $\nabla$ denotes
the covariant derivative of $g$. 

Note that although the definition is given in terms of a local null frame,
the definitions below determine the  components of
globally defined
covariant $S$-tangent tensors, expressed in  a local coordinate system $\theta^A$.

The complete list of Ricci coefficients\footnote{In comparing formulas with~\cite{DafLuk1},
note that the $\hat\omega$ here corresponds to $-2\omega$ of~\cite{DafLuk1},
and similarly $\underline{\hat\omega}$ here  corresponds to $-2\underline\omega$ 
of~\cite{DafLuk1}.} is given  by:\index{double null gauge!connection coefficients!$\hat\omega$}
\index{double null gauge!connection coefficients!$\omegabarhat$}\index{double null gauge!connection coefficients!$\chi$}\index{double null gauge!connection coefficients!$\chibar$}\index{double null gauge!connection coefficients!$\eta$}\index{double null gauge!connection coefficients!$\etabar$} 
\begin{align*}
\chi_{AB}&:=g(\nabla_A e_4,e_B) ,&  \chibar_{AB}&:= g(\nabla_Ae_3,e_B),
\\
\eta_A&:=-\frac12g(\nabla_3e_A,e_4),& \etabar_A&:=-\frac12g(\nabla_4e_A,e_3),
\\
\hat\omega&:=\frac12g(\nabla_4e_3,e_4),&\hat{\underline{\omega}}&:=\frac12 g(\nabla_3 e_4,e_3).
\end{align*}
We will further decompose
$\chi_{AB}$, resp.~$\chibar_{AB}$ into its trace-free part $\hat\chi_{AB}$, resp.~$\hat{\chibar}_{AB}$, a symmetric trace-free  $S$-tangent $2$-tensor,
and its trace $\tr\chi$, resp.~$\tr\chibar$,
i.e.~we define\index{double null gauge!connection coefficients!$\tr\chi$}\index{double null gauge!connection coefficients!$\tr\chibar$}\index{double null gauge!connection coefficients!$\hat\chi$}\index{double null gauge!connection coefficients!$\hat\chibar$}
\[
\hat\chi_{AB}:=\chi_{AB}-\frac12(\tr \chi)\slashed{g}_{AB} \, ,\qquad 
 \hat\chibar_{AB}= {\chibar}_{AB}-\frac12(\tr \chibar)\slashed{g}_{AB}.
\]

The complete list of curvature components is given  by:\index{double null gauge!curvature components!$\alpha$}\index{double null gauge!curvature components!$\alphabar$}\index{double null gauge!curvature components!$\beta$}\index{double null gauge!curvature components!$\betabar$}\index{double null gauge!curvature components!$\rho$}\index{double null gauge!curvature components!$\sigma$}
\begin{align*}
\alpha_{AB}&:=R(e_A,e_4,e_B,e_4) ,&  \underline\alpha_{AB}&:=R(e_A,e_3,e_B,e_3),\\
\beta_A&:=\frac12R(e_A,e_4,e_3,e_4),& \underline{\beta}_A&:= \frac12 R(e_A,e_3,e_3,e_4),\\
\rho&:=\frac14 R(e_4,e_3,e_4,e_3),&\sigma&:=\frac14{}^*R(e_4,e_3,e_4,e_3).
\end{align*}
Here, $R$ denotes the Riemann curvature tensor of $g$, defined
as usual by
\[
R(W,Z,X,Y)= g(R(X,Y)Z,W)=g(\nabla_{X}\nabla_{Y}Z-
\nabla_{Y}\nabla_{X}Z -  [X,Y]Z  ,W)
\]
and $*$ denotes the Hodge star operation.
Note moreover that by the symmetries of the curvature tensor,
$\alpha$ and $\underline\alpha$ are symmetric.

Define also the mass aspect functions\index{double null gauge!connection coefficients!$\mu$, mass aspect function}\index{double null gauge!connection coefficients!$\mubar$, mass aspect function}\footnote{In comparing formulas with~\cite{Chr},
note that the $\mu$ here corresponds to $-\mu$ of~\cite{Chr},
and similarly $\underline{\mu}$ here  corresponds to $-\underline{\mu}$ 
of~\cite{Chr}.}
\begin{align} \label{massaspectfirstdef}
	\mu
	=
	\slashed{div} \eta + \rho - \frac{1}{2} (\hat{\chi} , \underline{\hat{\chi}}),
	\quad
	\text{and}
	\quad
	\underline{\mu}
	=
	\slashed{div} \underline{\eta} + \rho - \frac{1}{2} (\hat{\chi} , \underline{\hat{\chi}}).
\end{align}

\section{The complete set of equations}
\label{thatsallofthem}

We now assume that the metric $g$ of~\eqref{doublenulllongform} 
satisfies the Einstein vacuum equations
\begin{equation}
\label{Ricciflathere}
{\rm Ric}[g]=0,
\end{equation}
where as usual
\[
{\rm Ric}(X,Y)={\rm tr}( Z\mapsto R(Z,X)Y).
\]
The content of~\eqref{Ricciflathere} is expressed in the null structure
and Bianchi identities satisfied by the Ricci coefficients and curvature components.
Let us note already that $\alpha$ and $\alphabar$ are traceless symmetric $S$-tangent
covariant 2-tensors
under the assumption~\eqref{Ricciflathere}, i.e.
\[
\tr \alpha=0,\qquad \tr\alphabar=0.
\]

We display in {\bf Section~\ref{NSEsec}} the null structure equations (with 
the Ricci coefficients on the left hand side) and in {\bf Section~\ref{Bianchiequationssec}} 
the Bianchi identities
(with the curvature components on the left hand side).
Finally, in {\bf Section~\ref{interchanging}},
we note the different form the equations take if the position of 
the torsion term in~\eqref{doublenulllongform} is interchanged.

\subsection{The null structure equations} 
\label{NSEsec}

We give in this section the null structure equations.

We first note the relation of the projected Lie and covariant derivatives. 
For scalar functions $f$, we clearly have $Df= \Omega \nabla_4f$, $\underline Df= \Omega\nabla_3f$.
For $S$-tangent $1$-forms, we have
\begin{equation}
\label{Dvscovar1froms}
D\xi = \Omega(\slashed\nabla_4\xi +\chi^\sharp\cdot \xi), \qquad
\underline D\xi = \Omega(\slashed\nabla_3\xi+ \chibar^\sharp\cdot\xi).
\end{equation}
For $S$-tangent symmetric  covariant $2$-tensors, we have
\begin{equation}
\label{Dvscovar}
D\vartheta =\Omega(\slashed\nabla_4\vartheta +\chi\times \vartheta+\vartheta\times\chi),
\qquad \underline D\vartheta=\Omega(\slashed\nabla_3\vartheta+\chibar\times\vartheta+\vartheta\times\chibar).
\end{equation}
In particular, in the case where $\vartheta$ is trace-free, these simplify to 
\begin{equation}
\label{Dvscovarsimpl}
D\vartheta =\Omega(\slashed\nabla_4\vartheta+\tr\chi \vartheta),
\qquad \underline D\vartheta=\Omega(\slashed\nabla_3\vartheta+\tr\chibar\vartheta).
\end{equation}

We have the first variational formulae:
\begin{align}
	D \slashed{g}
	=
	2 \Omega \chi
	=
	2 \Omega \hat{\chi} + \Omega \tr \chi \slashed{g}
	\textrm{ \ \ \ \ \ and \ \ \ \ \ }
	\underline{D} \slashed{g}
	=
	2 \Omega \underline{\chi}
	=
	2 \Omega \underline{\hat{\chi}} + \Omega \tr \underline{\chi} \slashed{g} \, .
	\label{eq:firstvariatformula}
\end{align}
These formulae are in fact equivalent to the statement that
 $\slashed{\nabla}_3 \slashed{g} = 0 = \slashed{\nabla}_4\slashed{g}$.

We then have the following set of transport equations for the so-called shear and expansion:
\begin{align}
	\nablaslash_3 \underline{\hat{\chi}} + \tr \chibar \ \hat{\chibar} -\hat{\underline{\omega}} \ \underline{\hat{\chi}}
	= 
	-\underline{\alpha},  
	\ \ \ 
	\nablaslash_4 {\hat{\chi}} + \tr \chi \ \hat{\chi} -\hat{\omega} \ {\hat{\chi}} 
	= - \alpha ,
	\label{eq:chihat4}
\end{align}
\begin{align}
	\nablaslash_3 \tr \chibar  + \frac{1}{2}\left( \tr \chibar \right)^2 - \underline{\hat{\omega}} \tr \chibar
	=
	- \left(\underline{\hat{\chi}} , \underline{\hat{\chi}}\right),
	\ \
	\nablaslash_4 \tr \chi + \frac{1}{2}\left(\tr \chi \right)^2 - \hat{\omega} \tr \chi
	=
	- 
	\left( \hat{\chi}, \hat{\chi} \right).
	\label{eq:Ray}
\end{align}
Note that the last two equations constitute the celebrated Raychaudhuri equations. 

The same quantities satisfy an alternative set of transport equations in the 
conjugate null directions:
\begin{align}
	\nablaslash_3 {\hat{\chi}} + \frac{1}{2} \tr \chibar \ \hat{\chi} +\hat{\underline{\omega}} \ {\hat{\chi}}
	=
	-2 \Dslash_2^* \eta - \frac{1}{2} \tr \chi \ \underline{\hat{\chi}} + \eta \widehat{\otimes} \eta ,
	\label{eq:chihat3}
\end{align}
\begin{align}
	\slashed{\nabla}_4 \underline{\hat{\chi}} + \frac{1}{2} \tr \chi \ \hat{\underline{\chi}} + \hat{\omega} \ \underline{\hat{\chi}} 
	=
	-2 \Dslash_2^* \underline{\eta} - \frac{1}{2} \tr \underline{\chi} \ \hat{\chi} + \underline{\eta} \widehat{\otimes} \underline{\eta},
	\label{eq:chibarhat4}
\end{align}

\begin{align}
	\slashed{\nabla}_3 \tr \chi + \frac{1}{2} \tr \underline{\chi} \tr \chi + \underline{\hat{\omega}} \tr \chi
	=
	- \left(\underline{\hat{\chi}}, {\hat{\chi}}\right) + 2 \left( \eta , \eta\right) + 2\rho +2\divslash \eta \, ,
	\label{eq:trchi3}
\end{align}
\begin{align}
	\slashed{\nabla}_4 \tr \underline{\chi} + \frac{1}{2} \tr \chi  \tr \underline{\chi}+ {\hat{\omega}} \tr \underline{\chi}
	=
	- \left( \underline{\hat{\chi}}, {\hat{\chi}} \right) + 2 \left( \underline{\eta} ,\underline{\eta} \right) + 2\rho + 2\divslash \underline{\eta} \, .
	\label{eq:trchibar4}
\end{align}

We then have the following remaining transport equations:

\begin{align} \label{eq:nabla4eta}
	\slashed{\nabla}_3 \underline{\eta}
	+
	\frac{1}{2} \tr \chibar \etabar
	=
	\frac{1}{2} \tr \chibar \eta
	+
	\hat{\underline{\chi}}^\sharp \cdot \left(\eta - \underline{\eta}\right) + \underline{\beta}
	,
	\ \ \ 
	\slashed{\nabla}_4 {\eta}
	+
	\frac{1}{2} \tr \chi \eta
	=
	\frac{1}{2} \tr \chi \etabar
	-
	\hat{\chi}^\sharp \cdot \left(\eta - \underline{\eta}\right) - \beta,
\end{align}
\begin{align} \label{eq:nabla4etabar}
	\nablaslash_4 \etabar
	+
	\tr \chi \etabar
	=
	\frac{2}{\Omega} \nablaslash ( \Omega \omegahat)
	+
	\beta
	-
	2 \hat{\chi} \cdot \etabar,
\end{align}
\begin{align} \label{eq:nabla3eta}
	\nablaslash_3 \eta
	+
	\tr \chibar \eta
	=
	\frac{2}{\Omega} \nablaslash (\Omega \omegabarhat)
	-
	\betabar
	-
	2 \hat{\chibar} \cdot \eta,
\end{align}
\begin{align}
	\Omega^{-1} \nablaslash_4 \left(\Omega \hat{\underline{\omega}}\right)
	=
	2 \left( \eta, \underline{\eta}\right) - |\eta|^2 - \rho
	\textrm{, \ }
	\Omega^{-1} \nablaslash_3 \left(\Omega \hat{\omega} \right)
	=
	2 \left(\eta, \underline{\eta}\right) - |\underline{\eta}|^2 - \rho,
	\label{eq:omega3omegabar4}
\end{align}
\begin{align} \label{eq:b3}
	\partial_{u} b^A
	=
	2\Omega^2 \left(\eta^A-\underline{\eta}^A\right) .
\end{align}

Finally we have the following tangential equations on $S_{u,v}$:
\begin{align} \label{eq:curletacurletabar}
	\curlslash \eta
	=
	- \frac{1}{2} \hat{\chi} \wedge \underline{\hat{\chi}} + \sigma
	\textrm{ \ \ \ \ \ and \ \ \ \ \ }
	\curlslash \underline{\eta}
	=
	\frac{1}{2} \hat{\chi} \wedge \underline{\hat{\chi}} - \sigma,
\end{align}
\begin{align}
	\divslash \hat{\chi}
	&
	=
	-
	\frac{1}{2} \hat{\chi}^\sharp \cdot \left( \eta - \underline{\eta}\right)
	+
	\frac{1}{4} \tr \chi \left( \eta - \underline{\eta} \right)  + \frac{1}{2} \slashed{\nabla} \tr \chi
	- 
	\beta
	\nonumber
	\\
	&
	=  
	-
	\frac{1}{2} \hat{\chi}^\sharp \cdot \left( \eta - \underline{\eta}\right)
	-
	\frac{1}{2} \tr \chi  \underline{\eta}
	+
	\frac{1}{2 \Omega}\slashed{\nabla} \left( \Omega \tr \chi \right) - \beta,
	\label{eq:Codazzi}
\end{align}
\begin{align}
	\divslash \underline{\hat{\chi}}
	&
	=
	\frac{1}{2} \underline{\hat{\chi}}^\sharp \cdot \left( \eta - \underline{\eta}\right)
	-
	\frac{1}{4} \tr \underline{\chi} \left( \eta - \underline{\eta}\right)
	+
	\frac{1}{2} \slashed{\nabla} \tr \underline{\chi} + \underline{\beta}
	\nonumber
	\\
	&
	=
	\frac{1}{2} \underline{\hat{\chi}}^\sharp \cdot \left( \eta - \underline{\eta}\right) - \frac{1}{2} \tr \chibar \eta
	+
	\frac{1}{2\Omega} \slashed{\nabla} \left(\Omega \tr \underline{\chi}\right) + \underline{\beta}.
	\label{eq:Codazzibar}
\end{align}
The latter two equations represent the  well-known Codazzi equations.

We finally note the Gauss equation\index{double null gauge!connection coefficients!$K$, Gauss curvature of
$S_{u,v}$}
\begin{align} \label{eq:Gauss}
	K
	=
	-\frac{1}{4} \tr \chi \tr \underline{\chi} + \frac{1}{2}\left( \hat{\chi} , \hat{\underline{\chi}} \right) - \rho,
\end{align}
where $K$ denotes the Gauss curvature of $S_{u,v}$,
and the relations
\begin{equation} \label{eq:DlogOmega}
	\eta + \etabar = 2 \nablaslash \log \Omega,
	\qquad
	\omegahat = \nablaslash_4 \log \Omega,
	\qquad
	\omegabarhat = \nablaslash_3 \log \Omega.
\end{equation}

\subsection{The Bianchi identities}
\label{Bianchiequationssec}

To complete our system of equations, we record the  \emph{Bianchi identities}
satisfied by the curvature components:
\begin{align}
	\nablaslash_3 \alpha + \frac{1}{2} \tr \chibar \alpha + 2 \underline{\hat{\omega}} \alpha
	&
	=
	-2 \Dslash_2^* \beta - 3 \hat{\chi} \rho - 3{}^* \hat{\chi} \sigma  + \frac{1}{2} \left(9 \eta - \etabar \right) \hat{\otimes} \beta,
	\label{eq:alpha3}
	\\
	\nablaslash_4 \beta + 2 \tr \chi \beta - \hat{\omega} \beta
	&
	=
	\divslash \alpha + \eta^\sharp \cdot \alpha,
	\label{eq:beta4}
 	\\
	\nablaslash_3 \beta + \tr \chibar \beta + \underline{\hat{\omega}} \beta
	&
	=
	\nablaslash \rho + {}^* \nablaslash \sigma + 3 \eta \rho + 3{}^* \eta \sigma + 2\hat{\chi}^\sharp \cdot \underline{\beta},
	\label{eq:beta3}
 	\\
	\nablaslash_4 \rho  + \frac{3}{2} \tr \chi \rho
	&
	=
	\divslash \beta + \frac{1}{2} \left( \eta + 3\underline{\eta}, \beta \right) - \frac{1}{2} \left(\underline{\hat{\chi}}, \alpha \right),
 	\label{eq:rho4}
	\\
	\nablaslash_4 \sigma + \frac{3}{2} \tr \chi \sigma
	&
	=
	- \curlslash \beta - \frac{1}{2} \left( \eta + 3 \underline{\eta} \right) \wedge \beta  + \frac{1}{2} \underline{\hat{\chi}} \wedge \alpha,
	\label{eq:sigma4}
\end{align}
\begin{align}
	\nablaslash_3 \rho + \frac{3}{2} \tr \chibar \rho
	&
	=
	- \divslash \underline{\beta} - \frac{1}{2} \left(3 \eta + \etabar, \underline{\beta}\right) - \frac{1}{2} \left(\hat{\chi}, \underline{\alpha}\right),
 	\label{eq:rho3}
	\\
	\nablaslash_3 \sigma + \frac{3}{2} \tr \chibar \sigma
	&
	=
	- \curlslash \underline{\beta} 
	-
	\frac{1}{2} \left(3\eta + \etabar \right) \wedge \underline{\beta} - \frac{1}{2} \hat{\chi} \wedge \underline{\alpha},
 	\label{eq:sigma3}
	\\
	\nablaslash_4 \underline{\beta} + \tr \chi  \underline{\beta} + \hat{\omega} \underline{\beta}
	&
	=
	- \nablaslash \rho + {}^* \nablaslash \sigma - 3 \underline{\eta} \rho + 3{}^* \underline{\eta} \sigma
	+
	2 \underline{\hat{\chi}}^\sharp \cdot \beta,
	\label{eq:betabar4}
 	\\
	\nablaslash_3 \underline{\beta} + 2 \tr \chibar  \underline{\beta} - \hat{\underline{\omega}} \underline{\beta}
	&
	=
	- \divslash \underline{\alpha} - \etabar^\sharp \cdot \underline{\alpha},
	\label{eq:betabar3}
 	\\
	\nablaslash_4 \underline{\alpha} + \frac{1}{2}\tr \chi \underline{\alpha} + 2 \hat{\omega} \underline{\alpha}
	&
	=
	2 \Dslash_2^* \underline{\beta} - 3 \underline{\hat{\chi}} \rho + 3{}^* \underline{\hat{\chi}} \sigma
	-
	\frac{1}{2} \left(9 \underline{\eta} - \eta \right) \hat{\otimes} \underline{\beta}.
	\label{eq:alphabar4}
\end{align}

\subsection{Interchanging the position of the torsion term}
\label{interchanging}

We shall also consider double null metrics on $\mathcal{Z}$ in the form
\begin{equation}
\label{doublenulllongforminterchanged}
g=-2\Omega^2(u,v,\theta^A)(du\otimes dv + dv\otimes du)
+\slashed{g}_{CD}(u,v,\theta^A)(d\theta^C-b^C(u,v,\theta^A)du)
\otimes (d\theta^D-b^D(u,v,\theta^A)du),
\end{equation}
i.e.~with the torsion term $b$ multiplying $du$ in place of $dv$.
This just interchanges the role of $u$ and $v$ so that the relevant
formulas can immediately be derived from the formulas above.
Note that the null pair $(\ref{nullframedef})$ is now
replaced by\index{double null gauge!frames!$e_i, i=1,\ldots 4$, normalised double null frame}
\begin{equation}
\label{altnullframedef}
e_3=\Omega^{-1}(\partial_u+b^A\partial_{\theta^A}),\qquad e_4=\Omega^{-1}\partial_v.
\end{equation}
Note that there is an asymmetry in the roles of $e_3$ and $e_4$
in the definition of $\beta$ and $\underline\beta$.
As a result of this, we may repeat all definitions as in Section~\ref{geomprelimforvac} where
we replace $(\ref{doublenulllongform})$ with $(\ref{doublenulllongforminterchanged})$, and
 all equations  in Section~\ref{thatsallofthem} remain valid in precisely their original form,
except for~\eqref{eq:b3}, which becomes
\begin{equation}
\label{newbequation}
	\partial_{v} b^A
	=
	-2\Omega^2 \left(\eta^A-\underline{\eta}^A\right) .
\end{equation}

\section{The Schwarzschild metric in Eddington--Finkelstein and Kruskal normalised double null coordinates}
\label{EFandKrubothsec}

In this section, we shall give two concrete realisations of the Schwarzschild metric with mass $M>0$ expressed
in so-called Eddington--Finkelstein and Kruskal  normalised double null coordinates.
We shall  define these coordinates
in {\bf Sections~\ref{EFmanifold}} and~{\bf\ref{Kruskalmanifold}}, respectively, while in
{\bf Section~\ref{Penrosetype}}  we shall describe
how these coordinate ranges are represented as Penrose diagrams.
In {\bf Section~\ref{diffsofthespheresubsec}}, we shall associate a Schwarzschild
background  with a double null gauge, describing the ambiguity in choosing the diffeomorphism to the
sphere. Finally, in {\bf Section~\ref{compediumparameterssec}} we shall fix a reference parameter $M_{\rm init}$
and collect a number of parameters depending on $M_{\rm init}$ 
which will be used later in this work.

\subsection{Eddington--Finkelstein normalised double null coordinates}
\label{EFmanifold}

In the notation of Section~\ref{coordsofficial}, 
we let $\mathcal{W}_{\mathcal{EF}}:=\mathbb R^2$\index{Schwarzschild background!sets!$\mathcal{W}_{\mathcal{EF}}$, domain of Eddington--Finkelstein double null coordinates} 
with coordinates $u$, $v$. We shall refer to the coordinates $u$ and $v$ as
retarded and advanced Eddington--Finkelstein null coordinates.\index{Schwarzschild background!coordinates!$u$, retarded Eddington--Finkelstein coordinate}\index{Schwarzschild background!coordinates!$v$, advanced Eddington--Finkelstein coordinate}
Let $\ring{\gamma}$ denote the standard metric on $\mathbb S^2$ as defined in Section~\ref{sphereconcrete}.

Let us fix $M>0$.\index{Schwarzschild background!parameters!$M$, mass parameter} 
On the manifold $\mathcal{Z}_{\mathcal{EF}}:=\mathcal{W}_{\mathcal{EF}}\times \mathbb S^2$,\index{Schwarzschild background!sets!$\mathcal{Z}_{\mathcal{EF}}$, underlying Eddington--Finkelstein manifold} we define the metric\index{Schwarzschild background!metric!$g_{\circ, M}$, Schwarzschild metric} 
\begin{equation}
\label{SchwmetricEF}
g_{\circ,M}=-2\Omega^2_{\circ,M}(u,v)(du\otimes dv+dv\otimes du)+r^2_M(u,v)\ring{\gamma}_{AB}d\theta^A\otimes d\theta^B
\end{equation}
where  the function $r_M(u,v)$\index{Schwarzschild background!metric!$r_M$, area radius function} 
is defined implicitly by the relation
\begin{equation}
\label{EFrdef}
	 \left( 1 - \frac{2M}{r_M(u,v)} \right) \frac{r_M(u,v)}{2M}\exp\left(\frac{r_M(u,v)}{2M}\right)=\exp\left(\frac{v-u}{2M}\right),
\end{equation}
and\index{Schwarzschild background!metric!$\Omega^2_{\circ,M}$, metric component}
\begin{equation}
\label{defofOmega}
\Omega^2_{\circ,M} (u,v):= 1 - \frac{2M}{r_M}.
\end{equation}
The expression~\eqref{SchwmetricEF} describes the Schwarzschild
metric with mass $M$, expressed in Eddington--Finkelstein normalised
double null coordinates (see~\cite{eddingtoncoords, finkelsteincoords} for
what is essentially the outgoing null coordinate~$v$).

Note that we may view~\eqref{SchwmetricEF} as being in either the form
$(\ref{doublenulllongform})$ or     in $(\ref{doublenulllongforminterchanged})$,
with the metric coefficient functions given by $(\ref{defofOmega})$, and
\[
b_{\circ, M}=0, \qquad \slashed{g}_{\circ,M} = r^2_M \mathring{\gamma}.
\]

We record here the non-vanishing Ricci coefficients and curvature components
associated to~\eqref{SchwmetricEF}:\index{Schwarzschild background!connection coefficients!$(\Omega \tr \chi)_{\circ,M}$}\index{Schwarzschild background!connection coefficients!$(\Omega \tr \chibar)_{\circ,M}$}\index{Schwarzschild background!connection coefficients!$(\Omega \omegahat)_{\circ,M}$}\index{Schwarzschild background!connection coefficients!$(\Omega \omegabarhat)_{\circ,M}$}
\begin{equation}
\label{recorderhere}
	\Omega^2_{\circ,M} = 1 - \frac{2M}{r_M},
	\quad
	(\Omega \tr \chi)_{\circ,M} = \frac{2\Omega_{\circ,M}^2}{r_M},
	\quad 
	(\Omega \tr \chibar)_{\circ,M} = - \frac{2\Omega_{\circ,M}^2}{r_M},
	\quad
	(\Omega \omegahat)_{\circ,M} = \frac{M}{r^2_M},
	\quad
	(\Omega \omegabarhat)_{\circ,M} = - \frac{M}{r^2_M},
\end{equation}
and\index{Schwarzschild background!curvature components!$\rho_{\circ,M}$}\index{Schwarzschild background!curvature components!$K_{\circ,M}$}
\begin{equation}
\label{recorderhere2}
	\rho_{\circ,M} = - \frac{2M}{r^3_M},
	\quad
	K_{\circ,M} = \frac{1}{r^2_M}.
\end{equation}
The value of the mass $M$ will often be clear from the context, in which case $M$ is omitted from the subscript and we write $\Omega^2_{\circ}$, $r$, etc.

Let us also introduce the following notation:
Given $r'>2M$, we define \index{Schwarzschild background!change of coordinates!$v(r',u)$}
\begin{equation}
\label{vofudefinition}
v(r',u):= v: r_M(u,v)=r'. 
\end{equation}
Note that the above $v$ is unique and for fixed $r'$ this defines a smooth increasing
function of $u$.

\subsection{Kruskal normalised double null coordinates}
\label{Kruskalmanifold}

Now let  $\mathbb R^2_{U,V}$ denote $\mathbb R^2$ with coordinates
$U$ and $V$
and consider  $\mathcal{W}_{\mathcal{K}}\subset \mathbb R^2_{U,V}$\index{Schwarzschild background!sets!$\mathcal{W}_{\mathcal{K}}$, domain of Kruskal normalised double null coordinates}  
to be the region  $UV< 1$. We shall refer to $U$ and $V$ as Kruskal normalised retarded
and advanced coordinates.\index{Schwarzschild background!coordinates!$U$, retarded Kruskal null coordinate}\index{Schwarzschild background!coordinates!$V$, advanced Kruskal null coordinate} 
Let $\ring{\gamma}$ denote the standard metric on $\mathbb S^2$ as defined in Section~\ref{sphereconcrete}.

On the manifold $\mathcal{Z}_{\mathcal{K}}:=\mathcal{W}_{\mathcal{K}}\times \mathbb S^2$\index{Schwarzschild background!sets!$\mathcal{Z}_{\mathcal{K}}$, underlying Kruskal manifold},
we define the metric\index{Schwarzschild background!metric!$g_{\circ, M, \mathcal{K}}$, Schwarzschild metric
in Kruskal coordinates}
\begin{equation}
\label{SchwmetricKruskal}
g_{\circ, M, \mathcal{K}}=-2\Omega^2_{\circ, M, \mathcal{K}}(U,V)(dU\otimes dV+dV\otimes dU) +r^2_{M,\mathcal{K}}(U,V)\ring{\gamma}_{AB}d\theta^A \otimes d\theta^B
\end{equation}
where $r_{M,\mathcal{K}}(U,V)$\index{Schwarzschild background!metric!$r_{M,\mathcal{K}}$, area radius function
of Kruskal coordinates} is defined by the relation
\begin{equation}
\label{Kruskalrdef}
\left(\frac{r_{M,\mathcal{K}}(U,V)}{2M}-1\right)\exp\left(\frac{r_{M,\mathcal{K}}(U,V)}{2M}\right)=-UV
\end{equation}
and\index{Schwarzschild background!metric!$\Omega^2_{\circ, M, \mathcal{K}}$, metric component in Kruskal coordinates}
\begin{equation} \label{eq:Kruskalbackground0}
\Omega^2_{\circ, M, \mathcal{K}}(U,V)=\frac{8M^3}{r_{M,\mathcal{K}}}\exp \left(-\frac{r_{M,\mathcal{K}}}{2M}\right).
\end{equation}

Again, we may view~\eqref{SchwmetricKruskal} as being in either the form
$(\ref{doublenulllongform})$ or     in $(\ref{doublenulllongforminterchanged})$, as $b=0$.

The non-vanishing Ricci coefficients and curvature components associated to~\eqref{SchwmetricKruskal} are
\begin{equation} \label{eq:Kruskalbackground1}
	\Omega \tr \chi_{\circ,M,\mathcal{K}} = \frac{2M}{V} \frac{2}{r_{M,\mathcal{K}}} \left(1-\frac{2M}{r_{M,\mathcal{K}}} \right),
	\quad
	\Omega \tr \chibar_{\circ,M,\mathcal{K}} = - V \frac{8M^2}{r_{M,\mathcal{K}}^2} \exp \left(-\frac{r_{M,\mathcal{K}}}{2M}\right),
	\quad
	\rho_{\circ,M,\mathcal{K}} = - \frac{2M}{r^3_{M,\mathcal{K}}},
\end{equation}
and
\begin{equation} \label{eq:Kruskalbackground2}
	\Omega \omegahat_{\circ,M,\mathcal{K}} = - \frac{1}{2r_{M,\mathcal{K}}} \left( 1+ \frac{2M}{r_{M,\mathcal{K}}} \right) \left(1 - \frac{2M}{r_{M,\mathcal{K}}} \right),
	\quad
	\Omega \omegabarhat_{\circ,M,\mathcal{K}} = V \frac{M}{r_{M,\mathcal{K}}} \left(1 + \frac{2M}{r_{M,\mathcal{K}}} \right) \exp \left(-\frac{r_{M,\mathcal{K}}}{2M}\right).
\end{equation}

As is well known, the expression~\eqref{SchwmetricKruskal} again represents
the Schwarzschild metric with mass $M$, but now in Kruskal normalised double
null coordinates (see~\cite{kruskal}). 
We may identify the manifold $\mathcal{M}_{\rm Lemaitre}$ discussed in Section~\ref{SchKerrgaugefreed}
as the subregion $V>0$. The map\index{Schwarzschild background!change of coordinates!$\iota_M$, map
from Eddington--Finkelstein to Kruskal}
\begin{equation}
\label{iotadef}
\iota_M: \mathcal{W}_{\mathcal{EF}}\to \mathcal{W}_{\mathcal{K}}
\end{equation}
given by
\begin{equation} \label{eq:KruskalEF}
U=-\exp\left(-\frac{u}{2M}\right), \qquad V=\exp\left(\frac{v}{2M}\right), 
\end{equation}
gives rise to an  embedding
$\iota_M\times {id}_{\mathbb S^2\to \mathbb S^2}$ taking $\mathcal{Z}_{\mathcal{EF}}=\mathcal{W}_{\mathcal{EF}}\times \mathbb S^2$
into $\mathcal{Z}_{\mathcal{K}}=\mathcal{W}_{\mathcal{K}}\times \mathbb S^2$ with 
\[
\iota_M(\mathcal{W}_{\mathcal{EF}})=\{U< 0\}\cap\{V> 0\}\subset \mathcal{W}_{\mathcal{K}},
\]
such that 
\[
g_{\circ, M} =(\iota_M\times {id} )^* g_{\circ, M, \mathcal{K}},
\]
where $g_{\circ, M}$ was defined by~\eqref{SchwmetricEF},
in particular
\begin{equation}
\label{obvrel}
r_{M} = \iota_M^* r_{M, \mathcal{K}},
\end{equation}
where $r_M$ was defined by~\eqref{EFrdef}.

In what follows, the use of capital letters $U$ and $V$ will
be sufficient to distinguish the functions $r_{M,\mathcal{K}}(U,V)$ from $r_{M}(u,v)$, and thus, in view
also of $(\ref{obvrel})$, we will drop the $\mathcal{K}$ subscript from $r$ (as well 
as the $M$ subscript, as discussed in Section~\ref{EFrdef}).
Similarly, we will often drop the $M$ subscript as
from the map~\eqref{iotadef}, when the choice of $M$ is implicit.

\subsection{Penrose-type representations}
\label{Penrosetype}

\begin{figure}
\centering{
\def\svgwidth{20pc}
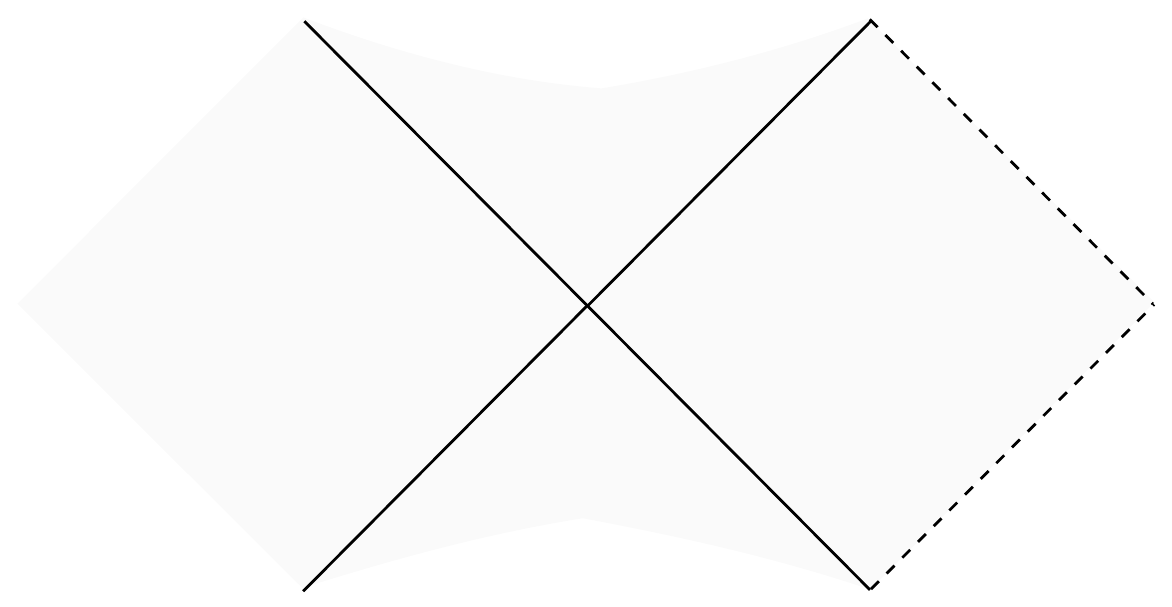}
\caption{Kruskal and Eddington--Finkelstein double null coordinates superimposed on a Penrose diagram}\label{Schwarzschildpicture}
\end{figure}

We shall often depict yet another representation of spacetime, which refers to 
a set of null coordinates which are \emph{globally bounded}. One can obtain such a representation trivially
by suitably rescaling $U$ and $V$. So as to avoid the unnecessary further proliferation of symbols,
we shall not attempt here to name such new coordinates but
we shall often superimpose Kruskal and Eddington--Finkelstein double null coordinate labels on such 
a diagram.
Refer to Figure~\ref{Schwarzschildpicture}.

\subsection{Schwarzschild as a background and diffeomorphisms of the sphere}
\label{diffsofthespheresubsec}

In this work,
we shall use the two alternative double null gauges defined in Sections~\ref{EFmanifold} and~\ref{Kruskalmanifold}
above to define Schwarzschild backgrounds
corresponding to a metric as in Section~\ref{timeorientsection}.

That is to say, consider
$\mathcal{Z}$ as in Section~\ref{coordsofficial} given by~\eqref{wemaywriteit},
with $\mathcal{W}\subset \mathbb R^2$.

Assume that we have a metric $g$
on $\mathcal{Z}$ 
expressed as~\eqref{doublenulllongform} (or alternatively~\eqref{doublenulllongforminterchanged}), satisfying~\eqref{Ricciflathere}.

Since we have $\mathcal{W}\subset \mathcal{W}_{\mathcal{EF}}$, by identifying the
standard coordinates, we may now consider, in addition to $g$, also   
 the Schwarzschild metric $g_{\circ,M}$ with mass $M>0$ defined on the same set
 $\mathcal{Z}$ by the expression~\eqref{SchwmetricEF}.

We shall call $g_{\circ,M}$ the Eddington--Finkelstein  Schwarzschild background defined by
our choice of double null gauge.

(We may alternatively, if $\mathcal{Z}\subset \mathcal{Z}_{\mathcal{K}}$, consider on $\mathcal{Z}$
the Schwarzschild metric defined 
by~\eqref{SchwmetricKruskal}. If we are to do this, we shall denote the coordinates of $\mathcal{Z}$ with
capital $U$, $V$.)

Let us remark already that if we consider a diffeomorphism $\mathcal{Z}_{\mathcal{EF}}\to\mathcal{Z}_{\mathcal{EF}}$ 
of the form
\begin{equation}
\label{diffsoftheform}
{\rm id}\times \psi : \mathcal{W}_{\mathcal{EF}} \times \mathbb S^2\to \mathcal{W}_{\mathcal{EF}} \times \mathbb S^2,
\end{equation}
where $\psi:\mathbb S^2\to \mathbb S^2$ is a diffeomorphism, then this covariantly 
induces a new solution $({\rm id}\times \psi)_* g$ of the vacuum equations on $\mathcal{Z}$,
again 
expressed in the form $(\ref{doublenulllongform})$ (or alternatively $(\ref{doublenulllongforminterchanged})$).  

Considering Schwarzschild as a background, however,  breaks the diffeomorphism invariance,
as $({\rm id}\times\psi)_*g - g_{\circ, M}\ne ({\rm id}\times\psi)_*(g-g_{\circ, M})$ unless $\psi$ is an isometry of 
the standard metric $\mathring\gamma$ on $\mathbb S^2$.
We note already, however, that several of the background functions are invariant to transformation by
such a~\eqref{diffsoftheform}, for instance, the function $r$ and all the quantities~\eqref{recorderhere} and~\eqref{recorderhere2}.

\subsection{The parameter $M_{\rm init}$ and related fixed parameters}
\label{compediumparameterssec}

Let us fix now a parameter 
\begin{equation}
\label{newfixingMinit}
M_{\rm init}>0.
\end{equation}
Associated to $M_{\rm init}$\index{initial data!parameters!$M_{\rm init}$, initial Schwarzschild parameter} will be a smallness parameter\index{initial data!parameters!$\hat\varepsilon_0$, smallness parameter depending on $M_{\rm init}$}
\begin{equation}
\label{smallnessparameterthresh}
\hat\varepsilon_0(M_{\rm init}),
\end{equation}
which will be constrained at various stages within the proof.
We shall consider Schwarzschild metrics of various masses $M$, but
we shall always assume
\begin{equation}
\label{alwaysassumethis}
|M-M_{\rm init}|\lesssim \hat\varepsilon_0.
\end{equation}
Here and in what follows, $A \lesssim B$ for nonnegative $A$ and $B$ denotes $A \leq C B$ where $C>0$ is a constant
depending only on $M_{\rm init}$.

In our proof we shall require a variety of $u$, $v$ and $r$ parameters depending on $M_{\rm init}$.

We first shall require a large value 
\begin{equation}
\label{definitionofRhere}
R:=R(M_{\rm init})>M_{\rm init}\delta^{-2}
\end{equation}
which will moreover be further  constrained later in the paper, specifically
 in (\ref{oneoftheRs2}) and (\ref{oneoftheRs}) and in the proof of Propositions~\ref{prop:l0modes} and~\ref{prop:l1modes}, where each time $R$ is constrained in terms of explicitly computable constants depending only on $M_{\rm init}$. Here 
 \begin{equation}
 \label{deltadefassmallness}
 \delta:=\frac{1}{100}.
 \end{equation}

For a real number $c\in\mathbb R$, let us introduce the notation
\[
R_{c}:= R+cM_{\rm init}.
\]

We shall now require a set of $u$ and $v$ values, together with corresponding sets
of $U$ and $V$ values defined via $\iota_{M_{\rm init}}$.  

Let us define $u_{-2}:=0$, and the corresponding $U_{-2}:=-1$ defined by~\eqref{eq:KruskalEF} with $M=M_{\rm init}$, 
and define $v_{-2}$:=0 such that\index{initial Kruskal gauge!parameters!$V_{-2}$}\index{initial Eddington--Finkelstein gauge!parameters!$u_{-2}$}\index{initial Kruskal gauge!parameters!$U_{-2}$}, and correspondingly $V_{-2}:=1$. 

We choose seven additional parameters\index{teleological $\I$ gauge!parameters!$u_{-1}$}\index{initial Eddington--Finkelstein gauge!parameters!$u_4$}\index{initial Eddington--Finkelstein gauge!parameters!$u_{3}$}\index{teleological $\I$ gauge!parameters!$u_{1}$}\index{teleological $\I$ gauge!parameters!$u_{2}$}\index{bootstrap!parameters!$u^0_f$, ``initial'' final retarded time associated to bootstrap}
\begin{equation}
\label{fourmoreparameters}
u_4>u_3>u_2> u_f^0> u_1> u_0>u_{-1}>u_{-2}=0,
\end{equation} 
and four additional parameters\index{teleological $\Hp$ gauge!parameters!$u_0$}\index{teleological $\Hp$ gauge!parameters!$v_{-1}$}\index{teleological $\Hp$ gauge!parameters!$v_{3}$}\index{initial Eddington--Finkelstein gauge!parameters!$v_{0}$}
\begin{equation}
\label{somevchoices}
v_3>v_2>v_0>v_{-1}>v_{-2}=0,
\end{equation}
all depending only on $M_{\rm init}$,
such that 
\begin{equation}
\label{0minitconstraint}
r_{M_{\rm init}}(u_{-2},v_{0})=r_{M_{\rm init}}(u_{-2},v_{-2})+ M_{\rm init},
\end{equation}
\begin{equation}
\label{34minitconstraint}
r_{M_{\rm init}}(u_0,v_{2})= R_4, \qquad 
r_{M_{\rm init}}(u_f^0,v_3)\ge R_5
 \end{equation}
\begin{equation}
\label{u1constraint}
r_{M_{\rm init}}(u_1,v_{2})= R_3
\end{equation}
and
\begin{equation}
\label{u2constrhere}
r_{M_{\rm init}} (u_3,v_3)\le R_{-2}
\end{equation}
and such that $u_f^0$ is sufficiently large, by a condition to be determined in Section~\ref{thelocaltheorysec} (the condition~\eqref{conditiononuf0} from Remark~\ref{Inlineartheoryremark}, necessary for the validity of Proposition~\ref{assertingdegreeone}) together with the condition 
\begin{equation}
\label{extracondforsob}
u_f^0-2M_{\rm init}\ge u_1.
\end{equation}

(Note that the above constraints
can indeed be satisfied for $R$ and $u_f^0$ arbitrary large. Thus,  we first select $R$ and $u_f^0$ sufficiently
large, and then
the remaining parameters.)

Finally, we shall choose 
\begin{equation}
\label{U4choice}
U_{5}>0 
\end{equation}
such that $r(V_3,U_5)= M_{\rm init}$, where $V_3$ corresponds to $v_3$ by~\eqref{eq:KruskalEF}.

The role of these parameters will become clear when we discuss the boundaries of the domains
of various gauges in Sections~\ref{nullinfgaugesec},~\ref{horizgaugesec} and~\ref{localexistencesection}.
We note already that we shall introduce in Section~\ref{subsubsec:cancelT} an additional
parameter $v_1$\index{teleological $\Hp$ gauge!parameters!$v_{1}=v_1(u_f, \lambda)$}
satisfying $v_0<v_1<v_2$,
but this parameter will depend on an actual solution and will thus be variable.

\section{The $\ell=0, 1$ modes and the reference linearised Kerr solutions}
\label{lowmodessection}

We will consider a subset 
$\mathcal{Z}$ as in Section~\ref{coordsofficial} given by~\eqref{wemaywriteit},
with $\mathcal{W}\subset \mathbb R^2$.
We will assume that we have a metric $g$
on $\mathcal{Z}$ 
expressed as~\eqref{doublenulllongform} (or alternatively~\eqref{doublenulllongforminterchanged}), satisfying~\eqref{Ricciflathere}.

As in Section~\ref{diffsofthespheresubsec}, we shall now consider, in addition to $g$, also   
 the Schwarzschild metric $g_{\circ,M}$ with mass $M>0$ defined on the same set
 $\mathcal{Z}$ by the expression
$(\ref{SchwmetricEF})$.

We define in this section the $\ell=0$ and $\ell=1$ modes and the reference linearised Kerr solutions.
These definitions will all depend on the metric $g$.
We first shall discuss in {\bf Section~\ref{sphintegsec}} integration over spheres. 
We shall define the projections to 
$\ell=0$ and $\ell=1$  spherical harmonics in {\bf Section~\ref{projandthemodessec}}.
(This will require an additional assumption of closeness to the spherical metric.)
We shall finally define the reference linearised Kerr solutions in {\bf Section~\ref{reflinearisedKerrsec}}.

\subsection{Spherical integration and volume form}
\label{sphintegsec}
As usual, let us denote by $r$ the function $r=r_{\circ, M}$  
on $\mathcal{Z}=\mathcal{W}\times\mathbb S^2$  induced by the background Schwarzschild metric $g_{\circ,M}$.

Integrals over spheres $S_{u,v}$ 
will typically involve the volume form $r^{-2} \sqrt{\det \gslash} d \theta^1 d \theta^2$ and so, accordingly, define\index{sphere!norms!$d\theta$, volume form used in integration}
\begin{align} \label{spherevolform}
	d\theta := r^{-2} \sqrt{\det \gslash} d \theta^1 d \theta^2.
\end{align}

We remark already that this volume form will be comparable to that of the round unit sphere metric.

We note however that the volume form changes covariantly under 
diffeomorphisms of the form $(\ref{diffsoftheform})$.
In particular, the integral
\[
\int_{S_{u,v}} f d \theta
\]
is well defined and invariant to diffeomorphism of the form $(\ref{diffsoftheform})$.
We may also define the $L^2$ inner product\index{sphere!norms!$(a,b)$, $L^2$ inner product used}
\begin{equation}
\label{L2inneronspheres}
(a,b) = \int_{S_{u,v}} a b \, d\theta.
\end{equation}

\subsection{Projection onto $\ell=0$ and $\ell=1$ modes and spherical harmonics $Y^\ell_m$}
 \label{projandthemodessec}
 We define in this section the projection onto ``low'' spherical harmonics.
 
\subsubsection{The operator  $\Dslash_2^*\nablaslash$ and the projections for functions and one-forms}
 
Recall the operator $\Dslash_2^*$ acting on $S$-tangent $1$-forms $\xi$ defined by~\eqref{Dslash2stardef}.
Consider now the operator $\Dslash_2^* \nablaslash$. 
This operates on scalar functions.

\begin{proposition}
\label{withclosenesstoroundprop}
Given $(u,v)\in \mathcal{W}$,
suppose $\slashed{g}$ is sufficiently close to the round metric on $S_{u,v}$ in the geometric sense,
i.e.~there exists a diffeomorphism $\psi:\mathbb S^2\to \mathbb S^2$ such that for sufficiently small $\epsilon>0$,
\begin{equation}
\label{closenesstoround}
| r^{-2}\slashed g(u,v,\theta ) -\psi^*\mathring\gamma|<\epsilon.
\end{equation}
Then, thought of as an operator on $C^\infty(S_{u,v})$, we have
\[
\dim {\rm Ker}(\Dslash_2^* \nablaslash) =4,
\]
and the projection $\Pi_{ {\rm Ker}(\Dslash_2^* \nablaslash)}$
defines a continuous map $C^\infty( \mathcal{W}\times\mathbb S^2)\to C^\infty( \mathcal{W}\times\mathbb S^2)$.
\end{proposition}

\begin{remark}
We have formulated the assumption $(\ref{closenesstoround})$ allowing for $\psi$ to 
highlight its geometric nature. In practice, however, we shall always apply this with $\psi$ as the identity.
\end{remark}

Constant functions on $S_{u,v}$ are evidently elements of  ${\rm Ker}(\Dslash_2^* \nablaslash)$.

Under the assumptions of Proposition~\ref{withclosenesstoroundprop}, 
we may decompose
\[
\dim {\rm Ker}(\Dslash_2^* \nablaslash) =  \mathcal{Y}_0 +\mathcal{Y}_1
\]
where\index{sphere!spherical harmonics!$\mathcal{Y}_0$, space of $\ell=0$ spherical harmonics}\index{sphere!spherical harmonics!$\mathcal{Y}_1$, space of $\ell=1$ spherical harmonics}
\[
\mathcal{Y}^0 :=  \mathrm{span} \{ 1\}
\]
and where
\[
	\mathcal{Y}^1 := \mathrm{span} \{ 1\}^{\perp} \subset \mathrm{Ker} (\Dslash_2^* \nablaslash),
\]
where the orthogonal complement is taken in the space Ker$(\Dslash_2^* \nablaslash)$ using the $L^2$ inner product on the sphere $S_{u,v}$ given by $(\ref{L2inneronspheres})$.

For a smooth scalar function $f(u,v,\theta)$, we may now define
\begin{equation}
\label{defellis0projscal}
f_{\ell =0}(u,v,\theta) =\Pi_{\mathcal{Y}^0} f(u,v,\theta),
\end{equation}
\begin{equation}
\label{defellis1projscal}
f_{\ell=1}(u,v,\theta) = \Pi_{\mathcal{Y}^1}  f(u,v,\theta),
\end{equation}
\begin{equation}
\label{defellisgreaterthan1projscal}
f_{\ell\ge 1}(u,v,\theta) =f-f_{\ell=0},
\end{equation}
\begin{equation}
\label{defellisnot1}
f_{\ell\ne1}(u,v,\theta) = f-f_{\ell=1},
\end{equation}
\begin{equation}
\label{defellisgreaterthan2projscal}
f_{\ell\ge 2}(u,v,\theta)=f -\Pi_{{\rm Ker}(\Dslash_2^* \nablaslash)} f= f-f_{\ell=0}-f_{\ell=1},
\end{equation}
where $\Pi$ denotes orthogonal projection with respect to $(\ref{L2inneronspheres})$.\index{sphere!spherical harmonics!$f_{\ell =0}$, projection to $\ell=0$}\index{sphere!spherical harmonics!$f_{\ell =1}$, projection to $\ell=1$}\index{sphere!spherical harmonics!$f_{\ell\ge 2}$, projection to $\ell\ge 2$}
By Propositon~\ref{withclosenesstoroundprop}, these functions are again smooth.

To define projections on $1$-forms, let us first
recall that, for any function $h(u,v,\theta)$, the Hodge dual of the gradient of $h$ is the one form defined by
\[
	r {}^* \nablaslash_A h := \epsslash_{AB} \gslash^{BC} r \nablaslash_C h,
\]
where $\gslash$ is the induced metric on the sphere $S_{u,v}$ and $\epsslash$ is the induced volume form.  Recall the following Hodge decomposition of a $1$-form (which follows from standard elliptic theory):

\begin{proposition} \label{prop:oneformdecomp}
	Under the assumption of Proposition~\ref{withclosenesstoroundprop}, 
	any smooth $S$-tangent $1$-form $\xi$ can be uniquely decomposed as
	\[
		\xi (u,v,\theta) 
		=
		r \nablaslash h_{1,\xi}(u,v,\theta)
		+
		r {}^* \nablaslash h_{2,\xi}(u,v,\theta),
	\]
	where $h_{1,\xi}(u,v,\theta)$, $h_{2,\xi}(u,v,\theta)$ are smooth
	functions such that $(h_{1,\xi})_{\ell=0} = (h_{2,\xi})_{\ell=0} = 0$.
\end{proposition}

Thus, under the assumptions of
Proposition~\ref{withclosenesstoroundprop}, given an $S$-tangent $1$-form $\xi$, define
the smooth $S$-tangent $1$-forms\index{sphere!spherical harmonics!$\xi_{\ell =1}$, projection to $\ell=1$ for $S$ $1$-forms}
\[
	\xi_{\ell=1}(u,v,\theta)
	=
	r \nablaslash (h_{1,\xi})_{\ell=1} (u,v,\theta)
	+
	r {}^* \nablaslash (h_{2,\xi})_{\ell=1} (u,v,\theta),
\]
and\index{sphere!spherical harmonics!$\xi_{\ell\geq 2}$, projection to $\ell\geq2$ for $S$ $1$-forms}
\[
	\xi_{\ell\geq 2}(u,v,\theta)
	=
	r \nablaslash (h_{1,\xi})_{\ell \geq 2} (u,v,\theta)
	+
	r {}^* \nablaslash (h_{2,\xi})_{\ell \geq 2} (u,v,\theta),
\]
where $h_{1,\xi}$ and $h_{2,\xi}$ are as in Proposition \ref{prop:oneformdecomp}.

\begin{remark}
\label{anothernoteoncovar}
We  remark
that  if we consider a diffeomorphism of the form
$(\ref{diffsoftheform})$, the above definitions are covariant in the sense
that $\psi^*(f_{\ell =1})= (\psi^* f)_{\ell=1}$, etc.
\end{remark}

\subsubsection{Round spherical harmonics}
It will be convenient to
introduce actual spherical harmonic functions depending on our metric $g$. 
We first will need to recall the round spherical harmonic functions.

Recall the standard spherical coordinates~\eqref{sphcorddef} on $\mathbb S^2$ defined in 
Section~\ref{sphereconcrete}.

We may lift the standard $\ell=0$ and $\ell=1$ spherical harmonics expressed with respect to this coordinate system
to functions on $\mathcal{W}\times \mathbb S^2$.
Explicitly, we
define the \emph{round $\ell=0$ spherical harmonic} to be\index{sphere!spherical harmonics!$\mathring{Y}^\ell_m$, round spherical harmonics}
\[
	\mathring{Y}^0_0(u,v,\mathring\theta,\mathring\phi) := \sqrt{\frac{1}{4\pi}},
\]
and we define the \emph{round $\ell=1$ spherical harmonics} $\mathring{Y}^1_{m}$, for $m=-1,0,1$, to be
\[
	\mathring{Y}^1_{-1}(u,v,\mathring\theta,\mathring\phi) = \sqrt{\frac{3}{4\pi}} \sin \mathring\theta \sin \mathring\phi,
	\qquad
	\mathring{Y}^1_0(u,v,\mathring\theta,\mathring\phi) = \sqrt{\frac{3}{4\pi}} \cos \mathring\theta,
	\qquad
	\mathring{Y}^1_1(u,v,\mathring\theta,\mathring\phi) = \sqrt{\frac{3}{4\pi}} \sin \mathring\theta \cos \mathring\phi.
\]

\begin{remark}
\label{yetanothernoteoncovar}
We note that, in contrast to~Remark~\ref{anothernoteoncovar},
 the above definitions do \underline{not} change covariantly by 
diffeomorphism of the form
$(\ref{diffsoftheform})$. That is to say, they depend on our concrete realisation of the sphere!
\end{remark}

\subsubsection{A basis for the spaces $\mathcal{Y}^0$ and $\mathcal{Y}^1$ and the coefficients $c_f$, $c_f^m$}
\label{basisforprojspace}

We may now proceed to define  spherical harmonics depending on our metric $g$.
We will always be under the assumption of Proposition~\ref{withclosenesstoroundprop},
but now with the additional restriction that  $\psi=id$.

Define the $\ell=0$ spherical harmonic simply to be the function
\[
	Y^0_0(u,v) := \left(\int_{S^2} r^{-2} \sqrt{\det \slashed{g}} d\theta^1 d \theta^2 \right)^{-\frac{1}{2}}.
\]
For each $(u,v)$, this  manifestly spans the space $\mathcal{Y}^0$.\index{sphere!spherical harmonics!$Y^0_0$, $\ell=0$ spherical harmonic}

Let $\breve{Y}^1_{m}$ denote the projection of the round $\ell=1$ spherical harmonics to the space $\mathcal{Y}$,
\[
	\breve{Y}^1_{m} := \Pi_{\mathcal{Y}^1} \mathring{Y}^1_{m},
	\quad
	\text{ for } m = -1,0,1.
\]
By Proposition~\ref{withclosenesstoroundprop}, these are again smooth.\index{sphere!spherical harmonics!$Y^1_m$, $m=-1,0,1$, $\ell=1$ spherical harmonics}

One easily shows the following
\begin{lemma}
\label{littlelemmahere}
Under the assumption $(\ref{closenesstoround})$, 
with the additional requirement that $\psi={\rm id}$, 
it follows that  $\breve{Y}^1_{m}$ again form a basis for $\mathcal{Y}^1$.
\end{lemma}

Define finally the $\ell=1$ spherical harmonics $Y^1_m$, for $m=-1,0,1$, to be the result of performing the Gram--Schmidt orthonormalisation process to $\breve{Y}^1_{-1}$, $\breve{Y}^1_0$, $\breve{Y}^1_1$.

We have that for each $(u,v)$, 
\[
{\rm span}\{ {Y}^1_{-1}, {Y}^1_0, {Y}^1_1\} = \mathcal{Y}^1,
\]
\[
{\rm span}\{ Y^0_0, {Y}^1_{-1}, {Y}^1_0, {Y}^1_1\} = \mathcal{Y}^0+\mathcal{Y}^1 ={\rm Ker}(\Dslash_2^* \nablaslash).
\]
Thus, for a given smooth scalar function $f:\mathcal{W}\times\mathbb S^2\to \mathbb R$, 
there exist unique smooth functions $c_f, c_f^{-1}, c_f^0, c_f^1:\mathcal{W}\to\mathbb R$\index{sphere!spherical harmonics!$c_f$, $\ell=0$ coefficient of $f$ in spherical harmonic expansion}\index{sphere!spherical harmonics!$c_f^{m}$, $m=-1,0,1$, $\ell=1$ coefficients of $f$ in spherical harmonic expansion} such that
\[
	f_{\ell=0}(u,v) = c_f(u,v) Y^0_0(u,v)
	=
	\left( \int_{S_{u,v}} d\theta \right)^{-1} \int_{S_{u,v}} f(u,v,\theta)d\theta,
\]
and
\begin{equation}
\label{cmfdefishere}
	f_{\ell=1}(u,v,\theta) = \sum_{m=-1}^1 c_f^m (u,v) Y^1_m(u,v,\theta),
\end{equation}
where $f_{\ell=0}$ and $f_{\ell=1}$ were defined in $(\ref{defellis0projscal})$ and 
$(\ref{defellis0projscal})$, respectively.

We have in particular that
\[
	f(u,v,\theta)
	=
	c_f(u,v) Y^0_0(u,v)
	+
	\sum_{m=-1}^1 c_f^m (u,v) Y^1_m(u,v,\theta)
	+
	f_{\ell \geq 2}(u,v,\theta),
\]
where $f_{\ell\geq 2}$ was defined in $(\ref{defellisgreaterthan2projscal})$. Note
that $f_{\ell\geq2}$ satisfies
\[
	\int_{S^2} f_{\ell \geq 2} Y^{\hat\ell}_m r^{-2} \sqrt{\det \gslash} d \theta^1 d \theta^2 = 0,
\]
for $\hat\ell = 0,1$, $\vert m \vert \leq \hat\ell$.

\begin{remark}
\label{yetagainanothernoteoncovar}
In view  of~Remarks~\ref{anothernoteoncovar} and~\ref{yetanothernoteoncovar}, we emphasise that 
that the above definitions of $c_f^m$ (as opposed to the projections $f_{\ell=1}$, etc., themselves) 
are \underline{not} preserved by diffeomorphisms of the form
$(\ref{diffsoftheform})$.
\end{remark}

\subsection{The reference linearised Kerr solutions}
\label{reflinearisedKerrsec}
Finally, we will define in this section a fixed-$M$ linearised Kerr solution on $\mathcal{Z}=\mathcal{W}\times\mathbb S^2$ 
associated to $g$.

Given a 3-vector $(J^{-1},J^0,J^1)\in \mathbb R^3$,\index{angular momentum!$(J^{-1},J^0,J^1)$, angular momentum vector associated to a linearised Kerr solution}
we define\index{Kerr metric!reference linearised Kerr solution!$\sigma_{\rm Kerr}$}\index{Kerr metric!reference linearised Kerr solution!$\etabar_{\rm Kerr}$}\index{Kerr metric!reference linearised Kerr solution!$b_{\rm Kerr}$}\index{Kerr metric!reference linearised Kerr solution!$(\Omega^{-1} \betabar)_{\rm Kerr}$}\index{Kerr metric!reference linearised Kerr solution!$(\Omega \beta)_{\rm Kerr}$}    
\[
	\sigma_{\rm Kerr}
	=
	\sum_{m=-1}^1
	J^m
	\frac{6}{r^4} Y^{\ell=1}_m,
	\qquad
	\eta_{\rm Kerr}
	=
	-
	\etabar_{\rm Kerr}
	=
	\sum_{m=-1}^1
	J^m
	\frac{3}{r^2} {}^* \nablaslash Y^{\ell=1}_m,
	\qquad
	b_{\rm Kerr}
	=
	\sum_{m=-1}^1
	J^m
	\frac{4}{r} {}^* \nablaslash Y^{\ell=1}_m,
\]
\[
	(\Omega^{-1} \betabar)_{\rm Kerr}
	=
	-
	\sum_{m=-1}^1
	J^m
	\frac{3}{r^3} {}^* \nablaslash Y^{\ell=1}_m,
	\qquad
	(\Omega \beta)_{\rm Kerr}
	=
	\Omega_{\circ}^2 \sum_{m=-1}^1
	J^m
	\frac{3}{r^3} {}^* \nablaslash Y^{\ell=1}_m,
\]
where we again assume that  $(\ref{closenesstoround})$ holds with $\psi=id$ so
the spherical harmonics of Section~\ref{basisforprojspace} are indeed well defined on $\mathcal{W}\times\mathbb S^2$.

The above expressions are motivated from~\cite{holzstabofschw}, where they
appeared as solutions to the linearised Einstein vacuum equations (expressed in double null gauge)
around Schwarzschild with mass $M$, corresponding to the fixed-$M$ Kerr family of solutions.

\begin{remark}
\label{stillyetagainanothernoteoncovar}
In view  of~Remark~\ref{yetagainanothernoteoncovar}, we again emphasise that these
definitions are \underline{not} covariant under  diffeomorphisms of the form
$(\ref{diffsoftheform})$.
\end{remark}

\chapter{The almost gauge-invariant hierarchy and the teleological gauge normalisations}
\label{almostgaugeandtellychapter}

In this chapter we shall introduce two essential concepts for our work,
that of the almost gauge-invariant hierarchy and that of teleological gauge normalisations suitable
for black holes.
Both have their origin in~\cite{holzstabofschw}.

\minitoc

In {\bf Section~\ref{AGIQdef}}, we shall introduce the non-linear analogue of the gauge-invariant
hierarchy of~\cite{holzstabofschw}. In the following two sections, we then
introduce  two possible teleological normalisations of a double null gauge,
the so called $\I$ gauge normalisation in {\bf Section~\ref{nullinfgaugesec}}, followed
by the so-called $\Hp$ gauge normalisation in {\bf Section~\ref{horizgaugesec}}.

\vskip1pc
\emph{The constructions of the present chapter are fundamental for the rest of the paper. In principle, the impatient reader anxious to already continue reading Part~\ref{stateandlogicpart} beyond
Section~\ref{initialdatanormsec} could for now
 skip Section~\ref{AGIQdef}, as the gauge invariant quantities only will appear  implicitly in
 the definition of various norms in Sections~\ref{localexistencesection} and~\ref{energiessection} (for which the notations of
 Chapter~\ref{moreprelimchapter} will also be essential; see comments in the prologue to the next chapter). 
Nonetheless, as this is a short section, and given the centrality of this hierarchy to the spirit of the paper, we do not 
recommend so doing! The reader may wish to refer to the discussion 
of~\cite{holzstabofschw} for more background on both the issue of gauge-invariant quantities and of
future gauge normalisation.}

\section{The almost gauge invariant hierarchy: $\alpha$, $\psi$, $P$ and \underline{$\alpha$}, \underline{$\psi$}, \underline{$P$}}
\label{AGIQdef}

We have already remarked in Section~\ref{mainestofproofintro} how  our approach
relies on adapting the gauge-invariant hierarchy of~\cite{holzstabofschw}.
We give the relevant definitions here.

\begin{definition}
\label{defofalmostgaugeinv}
We  consider subsets $\mathcal{Z}$, $\mathcal{W}$
 as in 
Section~\ref{coordsofficial}, 
and we  assume that we have a metric $g$
on $\mathcal{Z}$ 
expressed as~\eqref{doublenulllongform} (or alternatively~\eqref{doublenulllongforminterchanged}), satisfying~\eqref{Ricciflathere}.

Given $M>0$, using the inclusion $\mathcal{W}\subset\mathcal{W}_{\mathcal{EF}}$,
we   consider as in Section~\ref{diffsofthespheresubsec} also  
 the Schwarzschild metric $g_{\circ,M}$ with mass $M$ defined on the same set
 $\mathcal{Z}$ by the expression
$(\ref{SchwmetricEF})$. 
We shall denote $r=r_M$. 

We define the {\bf almost gauge-invariant hierarchy} associated
to $g$ to consist of the following quantities:\index{almost gauge invariant hierarchy!$\alpha$}\index{almost gauge invariant hierarchy!$\psi$}\index{almost gauge invariant hierarchy!$P$, quantity satisfying Regge--Wheeler type equation}\index{almost gauge invariant hierarchy!$\alphabar$}\index{almost gauge invariant hierarchy!$\psibar$}\index{almost gauge invariant hierarchy!$\Pbar$, quantity satisfying Regge--Wheeler type equation}
\begin{align}
\label{gaugeinvariantnobars}
&\alpha,& 	 &\psi := - \frac{1}{2r \Omega^2} \nablaslash_3(r\Omega^2 \alpha),&
 &P := \frac{1}{r^3 \Omega} \nablaslash_3 (r^3 \Omega \psi),
\\
\label{gaugeinvariantbars}
&\alphabar,& 	 &\psibar:= \frac{1}{2r \Omega^2} \nablaslash_4(r\Omega^2 \alphabar),&
	 &\Pbar := - \frac{1}{r^3 \Omega} \nablaslash_4 (r^3 \Omega \psibar).
\end{align}
Moreover, let\index{double null gauge!connection coefficients!$\check{r}$, area
radius defined from $\tr \chi$}\index{Schwarzschild background!metric!$\check{r}$, not a Schwarzschild background
quantity}
\begin{equation}
\label{alternativer}
	\check{r}
	:=
	\frac{2\Omega}{\tr \chi},
\end{equation}
and define also the alternative members\index{almost gauge invariant hierarchy!$\check{\Pbar}$, alternative to
$\Pbar$}\index{almost gauge invariant hierarchy!$\check{\psibar}$, alternative to
$\psibar$} of the hierarchy:
\[
	\check{\psibar} := \frac{1}{2r \Omega^2} \nablaslash_4(\check{r}\Omega^2 \alphabar),
	\quad
	\check{\Pbar} := - \frac{1}{r^3 \Omega} \nablaslash_4 (r^3 \Omega \check{\psibar}).
\]
\end{definition}

In the proof of the main theorem it is useful to consider the quantities $\check{\psibar}$ and $\check{\Pbar}$ when $r$ is large since the error terms in the wave equations that they satisfy have better $r$ behaviour, compared to the error terms in the $\psibar$ and $\Pbar$ equations.  The error terms in the wave equations satisfed by $\psibar$ and $\Pbar$, on the other hand, have better behaviour close to the event horizon.

The above definitions should be compared with the gauge-invariant quantities of linear theory,
defined in~\cite{holzstabofschw},
and discussed in Section~\ref{ginvquanCha}. 
Since the linear theory analogues of~\eqref{gaugeinvariantnobars} and~\eqref{gaugeinvariantbars}
were exactly gauge invariant and
satisfied decoupled equations, it will turn out that
the gauge dependence of~\eqref{gaugeinvariantnobars} and~\eqref{gaugeinvariantbars} 
and their coupling to the other quantities in the full nonlinear
theory, while not trivial, will be at higher order in the smallness parameter $\varepsilon_0$ measuring
distance from Schwarzschild. 
That is to say, if we have two double null parametrisations $i$ and $\tilde{i}$ as in Section~\ref{changeisgood},
then the difference between the quantities corresponding to $i^*g$ and $\tilde{i}^*g$ will be higher
order in $\varepsilon_0$. 

For the generalisations of the  Teukolsky and Regge--Wheeler 
equations satisfied by the gauge invariant hierarchy,
see already~Chapter~\ref{TeukandRegsection}.

\section{The $\I$ gauge}
\label{nullinfgaugesec}

We define in this section what it means for a metric to be expressed in $\I$ gauge.

An $\I$ gauge will be a particular type of double null gauge (in the alternative  form~\eqref{doublenulllongforminterchanged}), characterised by a specific domain $\mathcal{W}$, depending on parameters $u_f$, $M_f$ and $v_{\infty}$, and particular
gauge normalisations on the boundary of $\mathcal{W}$.\index{teleological $\I$ gauge!parameters!$M_f$, reference
Schwarzschild parameter}\index{teleological $\I$ gauge!parameters!$u_f$, final retarded time parameter}  
The nomenclature is motivated by the fact that, when applied later to our maximal developments $(\mathcal{M},g)$,
in the limit where $u_f\to \infty$ and $v_\infty\to \infty$, the ingoing future boundary component of $\mathcal{W}$ will tend to
what will be null infinity $\mathcal{I}^+$, and the normalisations will be such so as
to induce Bondi type behaviour.

\begin{figure}
\centering{
\def\svgwidth{20pc}
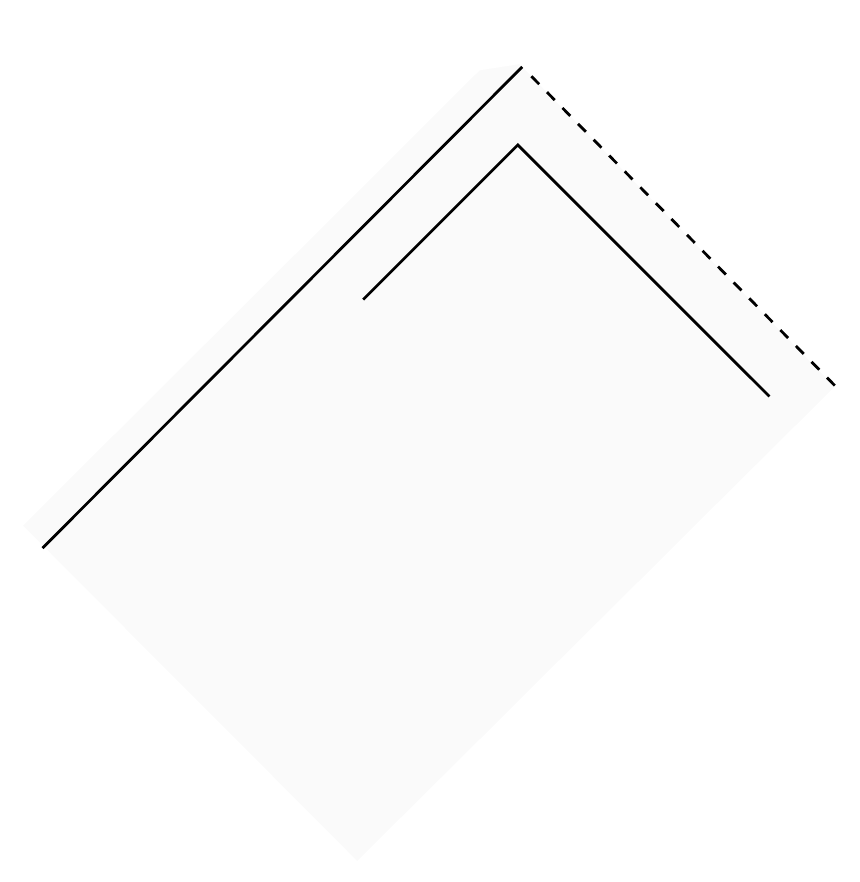}
\caption{The region $\mathcal{W}_{\I}(u_f)$}\label{infplusgaugefig}
\end{figure}

We first introduce the coordinate domain in {\bf Section~\ref{coorddomofI}}.
 Refer already to Figure~\ref{infplusgaugefig}.
The definition of the domain
already refers to a Schwarzschild background metric with mass parameter $M_f$, with which
we can define differences, spherical harmonic projections, etc.
With this, we give the definition of an $\I$ gauge in {\bf Section~\ref{actualdefIpgauge}}.
Finally, we will also define an associated linearised Kerr solution in 
{\bf Section~\ref{linKerrforIp}}.

\subsection{The coordinate domain and Schwarzschild background}
\label{coorddomofI}

Given $M_f>0$ satisfying~\eqref{alwaysassumethis}, let us first recall the Schwarzschild metric
$g_{\circ, M}$ with parameter $M:=M_f$,
 given by $(\ref{SchwmetricEF})$,
defined on the set $\mathcal{Z}_{\mathcal{EF}}=\mathcal{W}_{\mathcal{EF}}\times \mathbb S^2$.
We will denote the resulting 
background quantities as 
\begin{equation}
\label{backgnothere}
r=r_{M_f},  \qquad \Omega_{\circ}^2=\Omega_{\circ, M_f}^2, \, {\rm etc.}
\end{equation}

Recall the parameters $R$
and $u_{-1}$
 from Section~\ref{compediumparameterssec}, and more generally the parameters $R_c$ for $c\in \mathbb R$.

Given now  additional (variable) parameters $u_f \geq u_{-1}$ and $v_{\infty}$
satisfying  $v_{\infty}\ge v(R_4,u)$ for all $u_{-1}\le u\le u_f$ (where $v(R_{c},u)$ is defined by~\eqref{vofudefinition},
and  we note that this definition depends on  $r$ given by~\eqref{backgnothere} and thus on $M_f$),
we may
define $\mathcal{W}_{\I}(u_f, M_f, v_\infty)$\index{teleological $\I$ gauge!sets!$\mathcal{W}_{\I}(u_f, M_f, v_\infty)$, coordinate domain} to be the subset of $\mathbb R^2$ with coordinates
$u$, $v$ given by
\begin{equation}
\label{WI+}
\mathcal{W}_{\I}(u_f, M_f, v_{\infty}) = \{ u_{-1}\le  u \le u_f \}\cap
\{v(R_{-2},u)\le v \le v_\infty\}.
\end{equation}

The domain on which an $\mathcal{I}^+$ gauge is to be defined will then be\index{teleological $\I$ gauge!sets!$\mathcal{Z}_{\I}(u_f)$, domain of parametrisation}
\begin{equation}
\label{thedomainincords}
\mathcal{Z}_{\I}(u_f, M_f, v_{\infty}):=\mathcal{W}_{\mathcal{I}^+} (u_f, M_f, v_{\infty})\times \mathbb S^2.
\end{equation}
We shall typically drop explicit reference to the $M_f$ and $v_{\infty}$ dependence in what follows.

In view of the inclusion
\[
\mathcal{W}_{\mathcal{I}^+}(u_f)\subset \mathcal{W}_{\mathcal{EF}},
\]
we may now also  naturally consider the Schwarzschild metric with parameter $M_f$ as defined
on~\eqref{thedomainincords}, for which  we shall again use the notation~\eqref{backgnothere}.

\subsection{Definition of an $\mathcal{I}^+$ gauge}
\label{actualdefIpgauge}

Let us now assume that we have a metric 
 $g$  on the domain $\mathcal{Z}_{\I}(u_f)$ defined by $(\ref{thedomainincords})$
 in the
form~\eqref{doublenulllongforminterchanged}
satisfying the vacuum equations $(\ref{Ricciflathere})$.
(Recall the alternative form $(\ref{altnullframedef})$ of the null frame
and the resulting equations, described in Section~\ref{interchanging}.)

We may define 
the spherical harmonic projections
as in Section~\ref{lowmodessection},
referring always also to the above Schwarzschild background quantities~\eqref{backgnothere}.
For this we shall require that our metric $g$ satisfies the condition~\eqref{closenesstoround} with $\psi={\rm id}$, i.e.~the
assumptions of both Proposition~\ref{withclosenesstoroundprop} and Lemma~\ref{littlelemmahere}.
We will need one additional quantity, a ``renormalised mass aspect difference'':\index{double null gauge!connection coefficients!$\underline{\tilde{\mu}}$, renormalised mass aspect difference}
\begin{equation}
\label{renormmassaspectdf}
	\underline{\mu}^{\dagger}
	=
	\slashed{div} \underline{\eta} + (\rho -\rho_{\circ}) - \frac{1}{2} \hat{\chi} \cdot \underline{\hat{\chi}} + \frac{\Omega_{\circ}^2}{2r} \left(\frac{tr \underline{\chi}}{\Omega} - \frac{(\Omega \tr \underline{\chi})_\circ}{\Omega^2_\circ}\right).
\end{equation}

We are now ready to give the definition of what it means for a metric to be expressed
in $\mathcal{I}^+$ gauge.

\begin{definition}
\label{Igaugedefinition}
Given $u_f$, $M_f$, $v_{\infty}$ as above,
and defining $\mathcal{W}_{\mathcal{I}^+}(u_f)$ by~\eqref{WI+} and $\mathcal{Z}_{\I}(u_f)$
by~\eqref{thedomainincords},
we say that a metric $g$  in the
form~\eqref{doublenulllongforminterchanged} defined on $\mathcal{Z}_{\I}(u_f)$
and solving the Einstein vacuum equations $(\ref{Ricciflathere})$
is {\bf expressed in $\mathcal{I}^+$ gauge} if,
for all $(u,v)\in \mathcal{W}_{\I}(u_f)$,
the induced metric on the spheres $S_{u,v}$ 
satisfies the roundness condition~\eqref{closenesstoround} with $\psi={\rm id}$, i.e.~the
assumptions of both Proposition~\ref{withclosenesstoroundprop} and Lemma~\ref{littlelemmahere} hold,
and projections to spherical harmonics are thus defined, and
 the following relations hold on the boundary of the domain:
\begin{itemize}
	\item
		 $\frac{1}{2} (r^3 \rho_{\ell =0}) (u_f,v_{\infty})=-M_f $;
	 \item 
		 $b(u,v_\infty,\theta)=0$ for all $u\in [u_{-1},u_f]$, $\theta \in \mathbb S^2$; 
	\item
		$\left(\Omega \tr \chi - (\Omega \tr \chi)_{\circ} \right)_{\ell \neq 1}(u_f,v_{\infty},\theta)=0$ for all $\theta \in \mathbb S^2$;
	\item
		$\left(\divslash \Omega\beta \right)_{\ell = 1} (u_f,v_{\infty},\theta)=0$ for all $\theta \in \mathbb S^2$;
	\item
		$\left(\Omega^{-1} \tr \chibar - (\Omega^{-1} \tr \chibar)_{\circ} \right)(u_f,v_{\infty},\theta)=0$ for all $\theta \in \mathbb S^2$;
	\item
		$\mu_{\ell\geq 1}(u,v_{\infty},\theta)=0$ for all $u \in [u_{-1},u_f]$, $\theta \in
		\mathbb S^2$;
	\item
		$ \underline{\mu}^{\dagger}_{\ell \geq 1}(u_{-1},v,\theta) =0$ for all $v\in [v(R_{-2},u_{-1}),v_{\infty}]$, $\theta \in \mathbb S^2$;
	\item
		$\left(\Omega^2 - \Omega_{\circ}^2 \right)_{\ell=0}(u,v_{\infty})=0$ for all $u \in [u_{-1},u_f]$;
	\item
		$(\Omega \omegahat - (\Omega \omegahat)_{\circ})_{\ell=0} (u_{-1},v)= F(u_{-1}) \frac{\Omega_{\circ}^2}{r^3}(u_{-1},v)$ for all $v \in [v(R_{-2},u_{-1}),v_{\infty}]$, where
		\begin{equation}
		\label{nonlinearcondhere}
			F(u)
			:=
			\frac{1}{2} \int_{u}^{u_f} \int_{\bar{u}}^{u_f}
			r^3 (\Omega \hat{\chi}, \alphabar)_{\ell=0}
			\left(\hat{u},v_\infty\right)
			d\hat{u}
			d\bar{u},
		\end{equation}
\end{itemize}
where the Schwarzschild background is as defined above and  $\underline{\mu}^{\dagger}$ is defined in \eqref{renormmassaspectdf}.
\end{definition}

\begin{remark}
\label{SchwarzschildisIptoo}
Let us remark that  for all $u_f$, $M_f$ and $v_\infty$, the Schwarzschild metric $g_{\circ, M_f}$
 defined by~\eqref{SchwmetricEF},
restricted to~\eqref{thedomainincords},
 is itself expressed in $\mathcal{I}^+$ gauge.
\end{remark}

\begin{remark}
\label{againaboutdiffeos}
Note that because the definition of the Schwarzschild scalar quantities and the projection operations
transform covariantly under  diffeomorphisms of $\mathbb S^2$ (cf.~Remarks~\ref{noteoncovar} 
and~\ref{anothernoteoncovar}), 
it follows that an $\I$ gauge remains such if pulled back by 
the map ${\rm id}\times \psi:\mathcal{W}_{\mathcal{I}^+}(u_f)\times \mathbb S^2 \to
\mathcal{W}_{\mathcal{I}^+}(u_f)\times \mathbb S^2$,  where $\psi$ acts as a diffeomorphism on $\mathbb S^2$.
In particular, we do not really need the assumptions of  Lemma~\ref{littlelemmahere} to hold to be able
to formulate the above definition, but we shall indeed require it in Definition~\ref{assocKerparIplus} below.
\end{remark}

\begin{remark}
All conditions in Definition~\ref{Igaugedefinition} can be directly 
motivated already by their linearised versions, 
except for the final condition~\eqref{nonlinearcondhere}, which is used in Section~\ref{sec:ir01}.
See already Remark~\ref{proplinIpgauge}.
\end{remark}

\subsection{Associated linearised Kerr solution}
\label{linKerrforIp}

Given an $\I$ gauge, we will also define an associated  linearised Kerr solution. 
Recall the $\ell=1$ spherical harmonic functions $Y^1_m$ defined in Section~\ref{basisforprojspace} 
with the help also of the Schwarzschild
background $g_{\circ, M_f}$.
By Proposition~\ref{prop:oneformdecomp}, the curvature component $\Omega \beta$ of a given $\I$ gauge can be decomposed as
\[
	\Omega \beta (u,v,\theta) 
	=
	r \nablaslash h_{1,\Omega \beta}(u,v,\theta)
	+
	r {}^* \nablaslash h_{2,\Omega \beta}(u,v,\theta),
\]
for two functions $h_{1,\Omega \beta}$, $h_{2,\Omega \beta}$.

\begin{definition}
\label{assocKerparIplus}
Given $g$ expressed in $\I$ gauge  with respect to parameters $u_f$, $M_f$ and $v_{\infty}$,
we may define \underline{associated Kerr parameters} $J_{\I}^m$,~\index{teleological $\I$ gauge!parameters!$J_{\I}^m$, $m=-1,0,1$, associated Kerr parameters} for $m=-1,0,1$, by the relation
\[
	(r^4 h_{2,\Omega \beta})_{\ell=1}(u_f,v_{\infty}(u_f),\theta)
	=
	3 \Omega_{\circ,M_f}^2(u_f,v_\infty)
	\sum_{m=-1}^1
	J^m_{\I}
	Y^{1}_m(u_f,v_\infty,\theta).
\]
This then gives rise to an \underline{associated linearised Kerr solution}, which
we shall denote by $\sigma_{\rm Kerr}^{\I}$, $\eta_{\rm Kerr}^{\I}$, etc.,
by the definitions of Section~\ref{reflinearisedKerrsec} with $M=M_f$ and $J^m= J^m_{\I}$ defined above.
We note that in view of Remark~\ref{yetagainanothernoteoncovar}, 
the parameters $J^m_{\I}$ in general change under diffeomorphisms of
the type considered in Remark~\ref{againaboutdiffeos} above.
\end{definition}

Note that, given an $\I$ gauge, the linearised Kerr solution of Definition \ref{assocKerparIplus} satisfies
\begin{align} \label{betal1id}
	\curlslash (\Omega \beta_{\ell=1})(u_f,v_{\infty},\theta)
	=
	\curlslash (\Omega \beta_{\rm Kerr})(u_f,v_{\infty},\theta).
\end{align}

The energies defined in Section~\ref{energiessection} involve the solution minus this corresponding reference linearised Kerr solution and, in the proof of the main theorem, the solution minus the corresponding reference linearised Kerr solution will be shown to satisfy better estimates than the reference linearised Kerr solutions themselves.

\subsection{Aside: The $\mathcal{I}^+$ gauge in linear theory}
\label{linearisedversionoftheIgauge}

The question of existence of an $\mathcal{I}^+$ gauge is already nontrivial in linear theory.
Understanding this  is a pre-requisite for using this gauge in the fully nonlinear theory.
We state already a linearised version of this, which the reader may wish to already try and prove on their own, referring
to~\cite{holzstabofschw} for the relevant notation. 

\begin{proposition}
\label{proplinIpgauge}
Consider a smooth solution $\mathscr{S}$\index{linearised theory!complete solution!$\mathscr{S}$, smooth solution of linearised Einstein
equations in double null gauge} of the linearised Einstein equations in double null gauge
as in~\cite{holzstabofschw} (but with the metric expressed in the alternative double null form~\eqref{doublenulllongforminterchanged}) 
around Schwarzschild with
mass $M_f$, defined on the  domain $\mathcal{W}_{\I}(u_f)\times \mathbb S^2$ of the background
Schwarzschild solution.
Then there exists a pure gauge solution $\mathscr{G}$\index{linearised theory!pure gauge solutions!$\mathscr{G}$, residual smooth pure gauge solution}
 and a linearised Schwarzschild solution $\mathscr{K}_{{\mathfrak m},0,0,0}$\index{linearised theory!linearised Kerr solutions!$\mathscr{K}$, linearised Kerr solution} (see Section~6 of~\cite{holzstabofschw} for this notation; note that the family of linearised Schwarzschild solutions is a 1-dimensional subfamily of the linearised Kerr solutions $\mathscr{K}$) 
so that $\mathscr{S}+\mathscr{G}+\mathscr{K}_{{\mathfrak m},0,0,0}$ satisfies the linearised version
of all requirements of Definition~\ref{Igaugedefinition}, e.g.~$\rlin_{\ell=0}(u_f,v_\infty)=0$, etc.
\end{proposition}

\begin{remark} 
\label{nonuniquenessoflineargauge}
Let us note that the pure gauge solution $\mathscr{G}$ is not unique
as any pure gauge solution generated by a  $1$-parameter family of diffeomorphisms
as in Remark~\ref{againaboutdiffeos} trivially satisfies the (linearised) conditions defining
the gauge. Similarly, pure gauge solutions
generated by the $1$-parameter family generated by the Killing field $\partial_t$
also trivially satisfy these linearised conditions. In essence, in our setting, 
we shall break these symmetries 
 in Section~\ref{anchoredsec} with the help of Proposition~\ref{determiningthesphere} and
 by anchoring the solution to initial data.
\end{remark}

\begin{proof}
See Remark~\ref{linearandnonlinearremarkhere} and the remarks referred to there for instructions
on how to distill this theorem from our later Theorem~\ref{thm:newgauge} of Chapter~\ref{teleoffingchapter}.
\end{proof}

\section{The $\Hp$ gauge}
\label{horizgaugesec}

We define in this section what it means for a metric to be expressed in $\Hp$ gauge. Like
the notion of $\I$ gauge, defined in Section~\ref{nullinfgaugesec},
an $\Hp$ gauge 
 will be a particular type of double null gauge, now however in the  form~\eqref{doublenulllongform}, 
 characterised by a specific domain $\mathcal{Z}$, again depending on
 parameters $M_f$ and $u_f$, and particular
gauge normalisations on its boundary. 
Both the domain and the boundary normalisations will differ from those of the case
of an $\I$ gauge, and the nomenclature is now motivated by the fact that, when 
applied to our maximal Cauchy development $(\mathcal{M},g)$, as $u_f\to \infty$,
the outgoing future boundary of $\mathcal{W}$ will tend to what will be the event horizon $\mathcal{H}^+$,
with suitable normalisations there.

\begin{figure}
\centering{
\def\svgwidth{20pc}
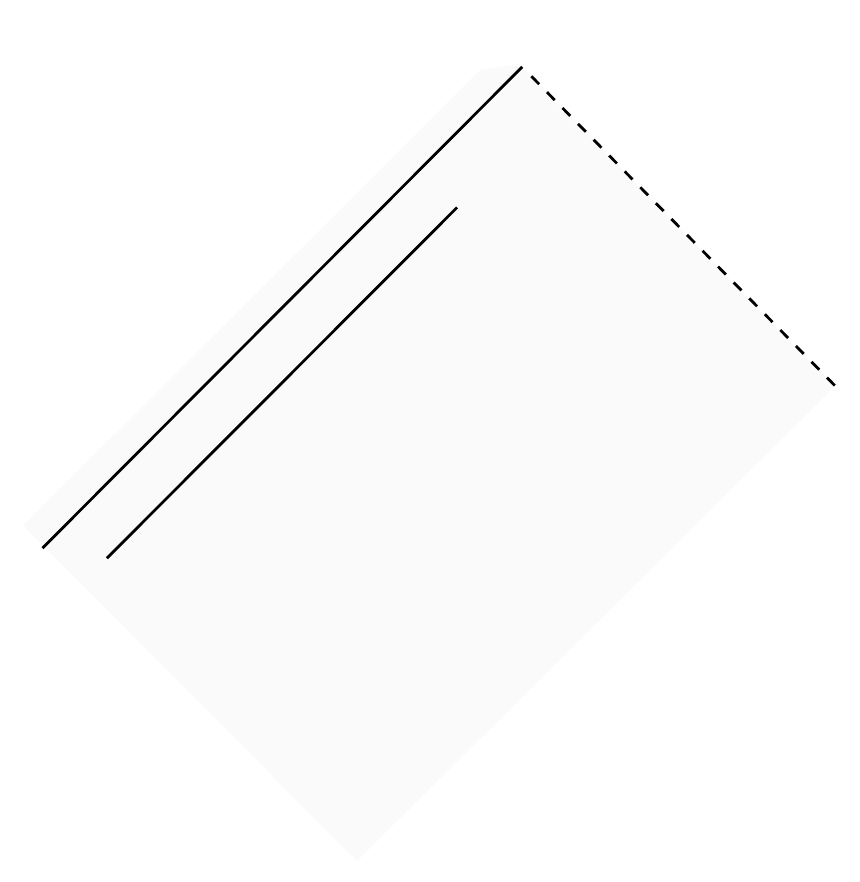}
\caption{The region $\mathcal{W}_{\Hp}(u_f)$}\label{defofhgaugefig}
\end{figure}

We first introduce the coordinate domain in {\bf Section~\ref{coorddomofHp}}, which
again depends on two parameters $u_f$ and $M_f$. 
Refer to Figure~\ref{defofhgaugefig}.
As with the notion of $\I$ gauge, the definition of the domain
already refers to a Schwarzschild background metric with mass parameter $M_f$, with which
we can define differences, spherical harmonic projections, etc. With this,
we give the definition of an $\Hp$ gauge in {\bf Section~\ref{actualdefHpgauge}}.
Again as in the case
of an $\I$ gauge,
we will also define an associated linearised Kerr solution in 
{\bf Section~\ref{linKerrforHp}}.
Finally, we will define an associated Kruskalised gauge in
{\bf Section~\ref{kruskalisedsec}}.

\subsection{The coordinate domain and Schwarzschild background}
\label{coorddomofHp}

As in Section~\ref{coorddomofHp}, given parameter $M_f$,\index{teleological $\Hp$ gauge!parameters!$M_f$
reference Schwarzschild parameter}
satisfying~\eqref{alwaysassumethis},
we will consider  the Schwarzschild metric
$g_{\circ, M}$ with parameter $M:=M_f$,
 given by $(\ref{SchwmetricEF})$,
defined on the set $\mathcal{W}_{\mathcal{EF}}\times \mathbb S^2$
and use the notation $(\ref{backgnothere})$.

Recall the parameters $R$
and $u_{0}$, $v_{-1}$
 from Section~\ref{compediumparameterssec}, and more generally $R_c$ for $c\in \mathbb R$.\index{teleological $\Hp$ gauge!parameters!$u_0$}\index{teleological $\Hp$ gauge!parameters!$v_{-1}$}

Given an additional parameter $u_f>u_0$,\index{teleological $\Hp$ gauge!parameters!$u_f$, final retarded time parameter} 
we define the set $\mathcal{W}_{\Hp}(u_f, M_f)$\index{teleological $\Hp$ gauge!sets!$\mathcal{W}_{\Hp}(u_f, M_f)$, coordinate domain} to  be the subset of $\mathbb R^2$ with coordinates $u, v$ given by
\begin{equation}
\label{WH+}
\mathcal{W}_{\Hp}(u_f, M_f)=\{u_0\le u \le u_f\}
\cap\{v_{-1}\le v \le v(R_2, u)\}.
\end{equation}
where $v(R_2,u)$ is defined by~\eqref{vofudefinition}, 
and where we note again that this
definition depends on $r$ defined above and thus on $M_f$.\index{teleological $\Hp$ gauge!parameters!$R$}
Again, we shall typically drop explicit reference to the $M_f$ dependence in what follows.
We note that by~\eqref{alwaysassumethis}, for $\hat\varepsilon_0$ sufficiently small, we may 
assume that indeed $v_{-1}\le v(R_2, u)$ for all $u\ge u_0$.

The domain on which the $\Hp$ gauge is to be defined will then be
\begin{equation}
\label{actualHpdomain}
\mathcal{W}_{\Hp}(u_f)\times \mathbb S^2.
\end{equation}

Again, in view of the inclusion
\[
\mathcal{W}_{\Hp}\subset \mathcal{W}_{\mathcal{EF}},
\]
we may now also naturally consider the Schwarzschild metric with parameter $M_f$
as defined on~\eqref{thedomainincords}, for which we shall again use the notation~\eqref{backgnothere}.

\subsection{Definition of an $\Hp$ gauge}
\label{actualdefHpgauge}

In analogy to Section~\ref{actualdefHpgauge}, 
let us now assume that we have a metric 
 $g$  on the domain~\eqref{actualHpdomain},
now however in the
form~\eqref{doublenulllongform},
satisfying the vacuum equations~\eqref{Ricciflathere}.

We may define spherical harmonic projections
as in Section~\ref{lowmodessection},
referring always also to the above Schwarzschild background quantities~\eqref{backgnothere}.
For this, we shall again require
the roundness condition~\eqref{closenesstoround} with $\psi={\rm id}$, i.e.~the
assumptions of both Proposition~\ref{withclosenesstoroundprop} and Lemma~\ref{littlelemmahere} hold,
and projections to spherical harmonics are thus defined.

We will need one additional quantity, namely:\index{double null gauge!connection coefficients!$\mu^*$, renormalised mass aspect difference}
\begin{equation}
\label{otherrenormmassaspectdf}
	\mu^* = \divslash \eta + (\rho - \rho_{\circ}) - \frac{3}{2r} \left( \Omega \tr \chi - (\Omega \tr \chi)_{\circ}\right).
\end{equation}

We are now ready to give the definition of what it means for a metric to be expressed in $\Hp$ gauge.

\begin{definition}
\label{Hgaugedefinition}
Given $u_f$, $M_f$ as above,
and defining $\mathcal{W}_{\Hp}(u_f)$ by~\eqref{WH+},
we say that a metric $g$  in the
form~\eqref{doublenulllongform} defined on the domain~\eqref{actualHpdomain}
and solving the Einstein vacuum equations $(\ref{Ricciflathere})$
is {\bf expressed in $\Hp$ gauge} if 
the induced metric on the spheres $S_{u,v}$ satisfy the 
roundness condition~\eqref{closenesstoround} with $\psi={\rm id}$, i.e.~the
assumptions of both Proposition~\ref{withclosenesstoroundprop} and Lemma~\ref{littlelemmahere} hold,
and projections to spherical harmonics are thus defined, and
 the following relations hold on the boundary of the domain:
\begin{itemize}
\item
$b(u_f,v,\theta)=0$ for all $v\in[v_{-1}, v(R_2,u_f)]$ and all $\theta \in\mathbb S^2$;
	\item
		$\mu^*_{\ell \geq 1} (u_f,v(R,u_f),\theta) = 0$ for all  $\theta \in \mathbb S^2$;
	\item
		$\left( \Omega \tr \chi - (\Omega \tr \chi)_{\circ} \right)_{\ell=0} (u_f,v(R,u_f)) = 0$;
	\item
		$\left( \Omega \tr \chibar - (\Omega \tr \chibar)_{\circ} \right)(u_f,v_{-1}, \theta) = 0$ for all $\theta \in \mathbb S^2$;
	\item
		$\Omega(u_f,v,\theta) = \Omega_{\circ} (u_f,v,\theta)$ for all  $v\in[v_{-1}, v(R_2,u_f)]$ and all $\theta \in\mathbb S^2$;
	\item
		$\partial_u \left( r^3 (\divslash \eta)_{\ell\geq 1} + r^3 \rho_{\ell \geq 1} \right) (u,v_{-1},\theta) = 0$ for all
		$u\in [u_0,u_f]$,  $\theta \in \mathbb S^2$;
	\item
		$\Omega(u,v_{-1})_{\ell=0} = \Omega_{\circ} (u,v_{-1})$ for all $u \in[u_0,u_f]$.
\end{itemize}
where
 the Schwarzschild background is as defined above and  $\mu^*$ is defined in~\eqref{otherrenormmassaspectdf}.
\end{definition}

\begin{remark}
\label{SchwarzschildisHptoo}
In analogy with Remark~\ref{SchwarzschildisIptoo}, 
let us remark that  for all $u_f$, $M_f$, the Schwarzschild metric $g_{\circ, M_f}$
 defined by~\eqref{SchwmetricEF},
restricted to domain~\eqref{actualHpdomain},
is itself expressed in~$\Hp$ gauge.
\end{remark}

\begin{remark}
\label{againaboutdiffeosHp}
In analogy with Remark~\ref{againaboutdiffeos}, 
we again note that $\Hp$ gauge remains such if pulled back by 
the map ${\rm id}\times \psi:\mathcal{W}_{\mathcal{I}^+}(u_f)\times \mathbb S^2 \to
\mathcal{W}_{\mathcal{I}^+}(u_f)\times \mathbb S^2$,  where $\psi$ acts as a diffeomorphism on $\mathbb S^2$.
\end{remark}

\subsection{Associated linearised Kerr solution}
\label{linKerrforHp}

As in the case of an $\I$ gauge, 
we will also define an associated  linearised Kerr solution to
a given $\Hp$ gauge.
Recall the $\ell=1$ spherical harmonic functions $Y^1_m$ defined in Section~\ref{basisforprojspace} 
with the help also of the Schwarzschild
background $g_{\circ, M_f}$.
Again, by Proposition~\ref{prop:oneformdecomp} the curvature component $\Omega \beta$ of a given $\Hp$ gauge can be decomposed as
\[
	\Omega \beta (u,v,\theta) 
	=
	r \nablaslash h_{1,\Omega \beta}(u,v,\theta)
	+
	r {}^* \nablaslash h_{2,\Omega \beta}(u,v,\theta),
\]
for two functions $h_{1,\Omega \beta}$, $h_{2,\Omega \beta}$.

\begin{definition} \label{assocKerparHplus}
Given $g$ expressed in  $\Hp$ gauge with respect to parameters $u_f$, $M_f$,
we may define \underline{associated Kerr parameters} $J_{\Hp}^m$,\index{teleological $\Hp$ gauge!parameters!$J_{\Hp}^m$, $m=-1,0,1$, associated Kerr parameters} for $m=-1,0,1$, by the relation
\[
	(r^4 h_{2,\Omega \beta})_{\ell=1}(u_f,v(R,u_f),\theta)
	=
	3 \Omega_{\circ,M_f}^2(u_f,v(R,u_f))
	\sum_{m=-1}^1
	J^m_{\Hp}
	Y^{\ell=1}_m(u_f, v(R,u_f), \theta).
\]
This then gives rise to an \underline{associated linearised Kerr solution}, which
we shall denote by $\sigma_{\rm Kerr}^{\Hp}$, $\eta_{\rm Kerr}^{\Hp}$, etc., given
by the definitions of Section~\ref{reflinearisedKerrsec} with $M=M_f$ and $J^m= J^m_{\Hp}$ defined above.
\end{definition}

Note again that, given an $\Hp$ gauge, the linearised Kerr solution of Definition \ref{assocKerparHplus} satisfies
\[
	\curlslash (\Omega \beta_{\ell=1})(u_f,v(R,u_f),\theta)
	=
	\curlslash (\Omega \beta_{\rm Kerr})(u_f,v(R,u_f),\theta).
\]

\subsection{Associated Kruskalised gauge}
\label{kruskalisedsec}

We define finally the associated Kruskalised $\mathcal{H}^+$ gauge.

Recall the map $\iota_M$ given by~\eqref{iotadef}. Given parameters $M_f$, $u_f$, 
let us define $U_f: = U(u_f)$\index{teleological $\Hp$ gauge!parameters!$U_f$} where $U$ is defined by~\eqref{eq:KruskalEF} with $M:=M_f$, and\index{teleological $\Hp$ gauge!sets!$\mathcal{W}_{\Hp}(U_f)$, Kruskalised coordinate domain}
\[
\mathcal{W}_{\Hp}(U_f) := \iota_{M_f}^{-1}(\mathcal{W}_{\Hp}(u_f)). 
\]
Note that $\iota_{M_f}^{-1}|_{\mathcal{W}_{\Hp}(U_f)}: \mathcal{W}_{\Hp}(U_f) \to
\mathcal{W}_{\Hp}(u_f)$ is well defined.

\begin{definition}
Given $g$ expressed in  $\Hp$ gauge
 with respect to parameters $u_f$, $M_f$, we
define the associated Kruskalised $\mathcal{H}^+$ gauge to be 
given by $(\iota_{M_f}^{-1} \times {\rm id} )^* g$ defined
on 
\begin{equation}
\label{restrictiontothisdomainnow}
\mathcal{W}_{\Hp}(U_f) \times\mathbb S^2.
\end{equation}
\end{definition}

\begin{remark}
\label{SchwarzschildisHptookruskalised}
In view of Remark~\ref{SchwarzschildisHptoo} and the considerations
of Section~\ref{Kruskalmanifold}, 
let us remark that  for all $U_f<0$, $M_f$, the Schwarzschild metric $g_{\circ, M_f, \mathcal{K}}$
 defined by~\eqref{SchwmetricKruskal},
restricted to domain~$\mathcal{W}_{\Hp}(U_f)\times \mathbb S^2$, 
corresponds precisely to the Kruskalised $\mathcal{H}^+$ gauge associated with
$g_{\circ, M_f}$ defined
 on $\mathcal{W}_{\Hp}(U_f)\times \mathbb S^2$.
\end{remark}

\begin{remark}
Note that the values $U_1$  and $V_{-1}$ corresponding to the boundary segments
of $\mathcal{W}_{\mathcal{H}^+}(U_f)$ depend in fact on $M_f$, as opposed to the parameters
of Section~\ref{compediumparameterssec}, which depended only on $M_{\rm init}$.
\end{remark}

\subsection{Aside: The $\mathcal{H}^+$ gauge in linear theory}
\label{linearisedversionoftheHgauge}

We state a counterpart of Proposition~\ref{proplinIpgauge} for the $\Hp$ gauge:
\begin{proposition}
\label{proplinHpgauge}
Consider a smooth solution $\mathscr{S}$ of the linearised Einstein equations in double null gauge
as in~\cite{holzstabofschw} 
around Schwarzschild with
mass $M_f$, defined on the  domain $\mathcal{W}_{\Hp}(u_f)\times \mathbb S^2$ of the background
Schwarzschild solution.
Then there exists a pure gauge solution $\mathscr{G}$ 
so that $\mathscr{S}+\mathscr{G}$ satisfies the linearised version
of all requirements of Definition~\ref{Hgaugedefinition}.
\end{proposition}

\begin{remark} 
In comparison to Remark~\ref{nonuniquenessoflineargauge} concerning the $\I$ gauge,
there is additional non-uniqueness in the statement 
of Proposition~\ref{proplinHpgauge}.  For here, there are additional pure gauge solutions
 $\mathscr{G}$ 
which satisfy the linearised requirements of Definition~\ref{Hgaugedefinition}.
In essence, in our setting, we shall break  this degeneracy in Section~\ref{anchoredsec}
 by anchoring the gauge to the $\I$ gauge.
\end{remark}

\begin{remark}
Let us note that, in the limiting case $u_f=\infty$,
the $\mathcal{H}^+$ gauge, when anchored appropriately to an $\mathcal{I}^+$ gauge as in
Section~\ref{anchoredsec}, is closely
related to the gauge yielding the so-called ``horizon normalised'' solutions considered in~\cite{holzstabofschw}.
\end{remark}

\begin{proof}
See Remark~\ref{sameasbeforebutforthehorizongauge} and the remarks referred to there for instructions
on how to distill this theorem from our later Theorem~\ref{thm:newgauge} of Chapter~\ref{teleoffingchapter}.
\end{proof}

\chapter{Schematic notation and the Teukolsky and Regge--Wheeler equations}
\label{moreprelimchapter}

In this chapter, we will consider a metric in double null gauge  with a background Schwarzschild metric of mass $M$
and we will derive a schematic notation to denote differences. This will allow us to derive in  a compact form
the equations satisfied by the almost gauge invariant quantities.

\minitoc

In {\bf Section~\ref{schemnotsec}}, we will introduce a schematic notation for such differences
and a set of commutation operators.
This will allow us in {\bf Section~\ref{schnotfornonlinearerror}} to introduce a schematic notation for nonlinear
expressions in differences which we shall be able to think of as ``error terms'' in our later estimates.
We shall derive some commutation identities in {\bf Section~\ref{commutetheident}}.
Finally, with the help of the error term notation introduced, we shall derive in {\bf Section~\ref{TeukandRegsection}} 
the nonlinear Teukolsky and Regge--Wheeler equations for the quantities of our gauge invariant hierarchy, 
in a form
which makes clear the relation with the familiar linearised equations from~\cite{holzstabofschw}.

\vskip1pc

\emph{Section~\ref{schemnotsec} is fundamental for understanding the norms appearing later and should
be read before continuing in Part~\ref{stateandlogicpart} beyond
Section~\ref{initialdatanormsec}.
The rest of the chapter may be skipped on a first reading, though it will be necessary for Part~\ref{improvingpart}.
For related schematic notation, see~\cite{vacuumscatter, DafLuk1}.}

\section{Schematic notation  for differences and the commutation  operators}
\label{schemnotsec}

In this section, we will consider throughout a subset 
$\mathcal{Z}$ as in Section~\ref{coordsofficial},
with $\mathcal{W}\subset \mathbb R^2$.
We shall assume moreover that
we  have a metric $g$
on $\mathcal{Z}$
expressed as $(\ref{doublenulllongform})$ (or alternatively $(\ref{doublenulllongforminterchanged})$),
satisfying~\eqref{Ricciflathere}.
By viewing $\mathcal{W}$ as a subset $\mathcal{W}\subset\mathcal{W}_{\mathcal{EF}}$,
we may define a Schwarzschild background as in Section~\ref{diffsofthespheresubsec}, i.e.~we may
define on the same underlying set $\mathcal{Z}$
the Schwarzschild metric $g_{\circ,M}$ with mass $M>0$ as in 
$(\ref{SchwmetricEF})$, 
expressed in Eddington--Finkelstein normalised double null coordinates.

\subsection{Schematic notation for differences $\mathcal{R}_p$, $\Gamma_p$ and $\Phi_p$}
\label{schemfordifssec}
The notation $\mathcal{R}_p$\index{schematic notation!$\mathcal{R}_p$, schematic notation for curvature components} will be schematically used for the following curvature components,
\[
	\mathcal{R}_1 = \{\Omega^{-2}\alphabar\},
	\quad
	\mathcal{R}_2 =\{ \Omega^{-1} \betabar\},
	\quad
	\mathcal{R}_3 =\{ \rho - \rho_{\circ}, \sigma\},
	\quad
	\mathcal{R}_{4} = \{\Omega \beta, \Omega^2 \alpha\},
\]
and the notation $\Gamma_p$\index{schematic notation!$\Gamma_p$, schematic notation for Ricci coefficients} will be used for the following Ricci coefficients:
\begin{align*}
	\Gamma_1
	&=\left\{
	1 - \frac{\Omega^2}{\Omega_{\circ}^2}, \,\,
	1 - \frac{\Omega^2_{\circ}}{\Omega^2}, \,\,
	b, \,\,
	\Omega^{-1} \hat{\chibar},
	\Omega^{-2} \left( \Omega \omegabarhat - (\Omega \omegabarhat)_{\circ} \right)\right\},
	\\
	\Gamma_2
	&=\left\{
	\Omega \tr \chi - (\Omega \tr \chi)_{\circ}, \,\,
	\Omega^{-2} \left( \Omega \tr \chibar - (\Omega \tr \chibar)_{\circ} \right), \,\,
	\Omega\hat{\chi}, \,\,
	\eta, \,\,
	\etabar\right\}, \\
	\Gamma_{\frac{5}{2}}
	&=
	\left\{\Omega \omegahat - (\Omega \omegahat)_{\circ}\right\}.
\end{align*}
(The quantities $1 - \frac{\Omega^2}{\Omega_{\circ}^2}$ and $b$ are metric components, rather than Ricci
 coefficients, though it is convenient to include them with the Ricci coefficients nonetheless.)  

 We may define $\mathcal{R}=\cup_p \mathcal{R}_p$ and $\Gamma=\cup_p \Gamma_p$.
In addition to denoting sets, we will often also use the above notation to denote a general element of the respective set, i.e.~we may write
\[
1 - \frac{\Omega^2}{\Omega_{\circ}^2}=\Gamma_1,
\]
and say informally that  ``$1-\frac{\Omega^2}{\Omega_{\circ}^2}$ is a $\Gamma$'', specifically a $\Gamma_1$,
and write expressions like $\sum_{\Gamma}\|\Gamma\|^2$, where $|\cdot|$ is some norm (where one sees
together the notation being used for the set and for its elements).
  
In the
above definitions, note that  each $\Gamma$ and $\mathcal{R}$  contains an appropriate $\Omega$ weight so that, 
in the proof of the main theorem, it will correspond to a quantity which is in the limit regular on the event horizon.  The $p$-subscript is used to describe the $r$ weights which appear in certain weighted estimates of each quantity in the proof of the main theorem.  See already \eqref{baspw} and \eqref{basiled}.\index{schematic notation!$p$, subscript related to $r^p$ weights}

Since in the present paper it will often not  be necessary to distinguish between curvature components and Ricci coefficients, we will use the schematic notation
$\Phi_p$\index{schematic notation!$\Phi_p$, schematic notation denoting either a Ricci coefficient
or curvature component} to denote either, i.e.
\begin{equation}
\label{defofschemPhi}
	\Phi_p = \mathcal{R}_p \cup  \Gamma_p,\qquad \Phi= \cup_p \Phi_p.
\end{equation}

\begin{remark}
\label{noteoncovar}
Recall from Section~\ref{diffsofthespheresubsec} that
if we consider a diffeomorphism ${\rm id}\times \psi$ 
of the form~\eqref{diffsoftheform},
where $\psi:\mathbb S^2\to \mathbb S^2$ is a diffeomorphism, then this covariantly 
induces a new solution $({\rm id}\times \psi)_* g$ of the vacuum equations, again 
 expressed in the form $(\ref{doublenulllongform})$ (or alternatively $(\ref{doublenulllongforminterchanged})$).  
Because the above quantities $\mathcal{R}_p$, $\Gamma_p$ and $\Phi_p$ involve differences
with Schwarzschild \underline{scalars}, and these are preserved by
such diffeomorphisms (e.g.~$\psi_* \rho_{\circ}=\rho_{\circ}$), it follows that  all these quantities
transform
covariantly under diffeomorphisms of the above form.
\end{remark}

\subsection{The commutation differential operators $\mathfrak{D}$}
\label{diffopsandnonlinerrors}
Let $M$, $\mathcal{Z}$, $g$ and $g_{\circ,M}$ be as in the beginning of the chapter.

Let $\mathfrak{D}$\index{double null gauge!differential operators!$\mathfrak{D}$} denote the collection of differential operators 
$\mathfrak{D} = \{  r \nablaslash, \Omega^{-1} \nablaslash_3, r \Omega \nablaslash_4 \}$.  Given a triple $\kbar = (k_1,k_2,k_3)$ for $k_1,k_2,k_3 \geq 0$, define
\[
	\mathfrak{D}^{\kbar} = (r \nablaslash)^{k_1} (\Omega^{-1} \nablaslash_3)^{k_2} (r \Omega \nablaslash_4)^{k_3},
\]
so that, in particular,
\[
	\mathfrak{D}^{(1,0,0)} = r \nablaslash,
	\quad
	\mathfrak{D}^{(0,1,0)} = \Omega^{-1} \nablaslash_3,
	\quad
	\mathfrak{D}^{(0,0,1)} = r \Omega \nablaslash_4.
\]
Given, moreover, a multi-index $\gamma = (\gamma_1,\ldots,\gamma_k)$, where $\gamma_i \in \{(1,0,0), (0,1,0), (0,0,1)\}$ for each $i=1,\ldots,k$, and given an $S$-tensor $\xi$, define
\[
	\mathfrak{D}^{\gamma} \xi
	:=
	\mathfrak{D}^{\gamma_1} \ldots \mathfrak{D}^{\gamma_k} \xi.
\]
For $\gamma = (\gamma_1,\ldots,\gamma_k)$, define $\vert \gamma \vert = k$.  

Note that, for a given triple $\kbar$ and $S$-tensor $\xi$, in the expression $\mathfrak{D}^{\kbar} \xi$ the $r \Omega \nablaslash_4$ derivatives are always applied first to $\xi$, followed by the $\Omega^{-1} \nablaslash_3$ derivatives, followed, finally, by the $r \nablaslash$ derivatives.  On the other hand, the derivatives $r \nablaslash, \Omega^{-1} \nablaslash_3, r \Omega \nablaslash_4$ can appear in any chosen order, given the correct $\gamma$, in the expression $\mathfrak{D}^{\gamma}$.

\section{Schematic notation for nonlinear error terms}
\label{schnotfornonlinearerror}

In this section, notation is introduced which is used to describe nonlinear error terms throughout the rest of the paper.
Throughout, let $M$, $\mathcal{Z}$, $g$ and $g_{\circ,M}$ be as in Section~\ref{schemnotsec}.

\subsection{Ordering the schematic differences $\Phi$} \label{subsec:orderingdifferences}
Recall the schematic notation $\Phi$ from $(\ref{defofschemPhi})$.  In order to define the nonlinear error notation below, let the above collection of $\Phi$ be given a superscript to order them as follows,
\[
	\Phi^1 = 1 - \frac{\Omega^2}{\Omega_{\circ}^2},
	\quad
	\Phi^2 = 1 - \frac{\Omega^2_{\circ}}{\Omega^2},
	\quad
	\Phi^3 = b,
	\quad
	\Phi^4 = \Omega^{-1} \hat{\chibar},
	\quad
	\Phi^5 = \Omega^{-2} \left( \Omega \omegabarhat - (\Omega \omegabarhat)_{\circ} \right),
	\quad
	\Phi^6 = \Omega^{-2}\alphabar,
\]
\[
	\Phi^7 = \Omega \tr \chi - (\Omega \tr \chi)_{\circ},
	\quad
	\Phi^8 = \Omega^{-2} \left( \Omega \tr \chibar - (\Omega \tr \chibar)_{\circ} \right),
	\quad
	\Phi^9 = \Omega\hat{\chi},
	\quad
	\Phi^{10} = \eta,
	\quad
	\Phi^{11} = \etabar,
	\quad
	\Phi^{12} = \Omega^{-1} \betabar,
\]
\[
	\Phi^{13} = \Omega \omegahat - (\Omega \omegahat)_{\circ},
	\quad
	\Phi^{14} = \rho - \rho_{\circ},
	\quad
	\Phi^{15} = \sigma,
	\quad
	\Phi^{16} = \Omega \beta,
	\quad
	\Phi^{17} = \Omega^2 \alpha.
\]

\subsection{Admissible coefficient functions and schematic notation for traces of $S$-tensors}
\label{subsec:traceindex}

Functions of the following form will appear in the nonlinear error notation introduced below.
Given $N\geq 0$ and a constant $C$, we say that a smooth function\index{schematic notation!$h$, admissible coefficient function} $h : \mathcal{Z} \to \mathbb{R}$ is an \emph{admissible coefficient function} if
\begin{equation} \label{eq:admis}
	\sup_{\mathcal{Z}}
	\sum_{\vert \gamma \vert \leq N}
	\vert \mathfrak{D}^{\gamma} h \vert
	\leq
	C.
\end{equation}
The number of derivatives $N\ge 12$ will be fixed later in the paper, and
the constant $C$ here is understood to be chosen sufficiently large
so that the claimed inequalities later in the paper hold for the $h$'s that naturally
arise. In all such
instances, it will be clear that this largeness can be chosen
 to depend only on the parameter
$M_{\rm init}$ of Section~\ref{compediumparameterssec}.

The following notation, which describes schematically admissible functions multiplying $S$-tensor fields and traces of 
$S$-tensor fields, will be used in defining the schematic notation for nonlinear error terms below.

Consider some $n \geq 0$.  The following objects will act on $(0,n)$ $S$-tensor fields.  A \emph{trace set of order $0$} is defined simply to be an admissible coefficient function $h$, satisfying \eqref{eq:admis}.  Given such an $h$ and 
an $S$-tangent $(0,n)$ 
tensor $\xi$, define $\xi \cdot h$ by pointwise multiplication
\[
	\xi \cdot h : = h \xi.
\]
In schematic expressions we will sometimes write $\xi \cdot_{\gslash} h$ to mean $\xi \cdot h$, even though $\xi \cdot h$ does not involve the metric $\gslash$, in order to not have to distinguish between trace sets of order $0$ and trace sets of order $l\geq 1$, which are defined below.
Suppose now $n \geq 2$.  A \emph{trace set of order $1$} is a collection\index{schematic notation!$I$, trace set} $I = \{ h_{ij} \}_{1\leq i < j \leq n}$, where each $h_{ij}$ is an admissible function, satisfying \eqref{eq:admis}.  Given such an $I$ and 
an $S$-tangent $(0,n)$ tensor $\xi$, define $\xi \cdot_{\gslash} I$ to be the $S$-tangent $(0,n-2)$ tensor
\[
	\big( \xi \cdot_{\gslash} I \big)_{A_1\ldots A_{n-2}}
	=
	\sum_{1 \leq i < j \leq n} h_{ij} \gslash^{BC} \xi_{A_1 \ldots A_{i-1} B A_{i} \ldots A_{j-2} C A_{j-1} \ldots A_{n-2}}.
\]

For example, recall that
\[
	(\Omega \hat{\chi} \times \Omega^{-1} \hat{\chibar})_{A_1A_2}
	=
	\gslash^{BC} \Omega \hat{\chi}_{B A_1}  \Omega^{-1} \hat{\chibar}_{CA_2}
	=
	\gslash^{BC} (\Omega \hat{\chi} \otimes \Omega^{-1} \hat{\chibar})_{BA_1CA_2}.
\]
The tensor field $\Omega \hat{\chi} \times \Omega^{-1} \hat{\chibar}$ can then be written
\[
	\Omega \hat{\chi} \times \Omega^{-1} \hat{\chibar}
	=
	\big( \Omega \hat{\chi} \otimes \Omega^{-1} \hat{\chibar} \big)
	\cdot_{\gslash}
	I,
\]
where $I = \{ h_{ij} \}_{1\leq i < j \leq 4}$, $h_{1\,3} = 1$ and $h_{ij} = 0$ otherwise.

Consider now some $n \geq 2$ and some $1 \leq d \leq \lfloor n/2 \rfloor$.  A \emph{trace set of order $d$} is defined to be a collection $I = \{ I^m \}_{1 \leq m \leq d}$, where each $I^m = \{ h_{ij}^m \}_{1\leq i < j \leq n}$ is a trace set of order $1$.  If $\xi$ is an $S$-tangent $(0,n)$ tensor field and $I = \{ I^m \}_{1 \leq m \leq d}$ is a trace set of order $d\geq 1$, define $\xi \cdot_{\gslash} I$ to be the $S$-tangent $(0,n-2d)$  tensor
\[
	\xi \cdot_{\gslash} I
	=
	\xi \cdot I^1 \cdot_{\gslash} \ldots \cdot_{\gslash} I^d.
\]

For example, recall that
\[
	(\Omega \hat{\chi}, \Omega^{-1} \hat{\chibar})
	=
	\gslash^{AD} \gslash^{BC} \Omega \hat{\chi}_{B A}  \Omega^{-1} \hat{\chibar}_{CD}
	=
	\gslash^{AD} \gslash^{BC} (\Omega \hat{\chi} \otimes \Omega^{-1} \hat{\chibar})_{BACD}.
\]
One can then write
\[
	(\Omega \hat{\chi}, \Omega^{-1} \hat{\chibar})
	=
	\big( \Omega \hat{\chi} \otimes \Omega^{-1} \hat{\chibar} \big)
	\cdot_{\gslash}
	I,
\]
where $I = \{ I^1, I^2 \}$ and $I^1 = \{ h_{ij}^1 \}_{1\leq i < j \leq 4}$, $h_{1\,3}^1 = 1$ and $h_{ij}^1 = 0$ otherwise, is as in the previous example, and $I^2 = \{ h_{1\,2}^2 \}$ where $h_{1\,2}^2 = 1$.

\subsection{The nonlinear error notation $\mathcal{E}^k$ }
\label{sec:nlenotation}

The notation $\mathcal{E}^k$ is used to denote a nonlinear error term involving at most $k$ derivatives of Ricci coefficients and curvature components.

Consider some $k \geq 0$.  Define $\{ (1,0,0), (0,1,0), (0,0,1)\}^0 := \{ (0,0,0)\}$.  Given $i \in \{ 1,\ldots, 17\}$ and $\gamma \in \{ (1,0,0), (0,1,0), (0,0,1)\}^k$, define
\[
	\mathfrak{D}^k \Phi \cdot (i,\gamma)
	:=
	\mathfrak{D}^{\gamma} \Phi^i.
\]

As an example, one can write
\[
	\Omega^{-1} \nablaslash_3 \Omega \hat{\chi}
	=
	\mathfrak{D}^1 \Phi \cdot (9,(0,1,0)),
	\qquad
	\text{and}
	\qquad
	\etabar = \mathfrak{D}^0 \Phi \cdot (11, (0,0,0)).
\]

Consider now some $k \geq 0$, $l \geq 1$ and some
\[
	H = \{ {}^{j,m}H_{k_1\ldots k_{l'}},  {}^jJ_{k_1\ldots k_{l'}} \}_{m=0,\ldots,l',k_1+\ldots + k_{l'} \leq k, l' \geq l, j \geq 1},
\]
where
\[
	{}^{j,m}H_{k_1\ldots k_{l'}}
	\in
	\big\{ (i,\gamma) \mid i \in \{ 1,\ldots, 17\}, \gamma \in \{ (1,0,0), (0,1,0), (0,0,1)\}^{k_m} \big\} \cup \big\{ 0 \big\},
\]
for all $m=0,\ldots,l'$ and ${}^jJ_{k_1\ldots k_{l'}}$ is a trace set of some order $d\geq 0$, as in Section \ref{subsec:traceindex}, for all $j \geq 1$, $k_1+\ldots + k_{l'} \leq k$ and $l' \geq l$.  For such $k \geq 0$, $l \geq 1$ and $H$, define
\[
	\left( \mathfrak{D}^{k} \Phi \right)^{l} \cdot H
	:=
	\sum_{l' \geq l}
	\sum_{k_1+\ldots+k_{l'} \leq k}
	\sum_{j \geq 1}
	\big(
	\mathfrak{D}^{k_1} \Phi \cdot {}^{j,1}H_{k_1\ldots k_{l'}} \otimes \ldots \otimes \mathfrak{D}^{k_{l'}} \Phi \cdot {}^{j,l'}H_{k_1 \ldots k_{l'}}
	\big)
	\cdot_{\gslash}
	{}^jJ_{k_1\ldots k_{l'}}.
\]

\vskip1pc
\noindent\fbox{
    \parbox{6.35in}{
 \emph{Such expressions are only ever considered when sufficiently many of the components of $H$ vanish so that each object appearing in the above summation is an $S$-tensor of the same type, and so that the allowed ranges of $l'\geq l$ and $j \geq 1$ are finite, and so the summation is indeed well defined.}
    }
}
\vskip1pc

For example, one can write
\[
	(r\nablaslash \eta) \otimes \etabar + \Omega \beta \otimes \Omega^{-1} \hat{\chibar} + \Omega^{-2} \alphabar \otimes \eta
	=
	\left( \mathfrak{D}^{1} \Phi \right)^{2} \cdot H,
\]
where ${}^{1,1} H_{1 \, 0} = \{ 10,(1,0,0) \}$, ${}^{1,2} H_{1\,0} = \{ 11, (0,0,0) \}$, ${}^{1,1} H_{0\,0} = \{ 16, (0,0,0) \}$, ${}^{1,2} H_{0\,0} = \{ 4, (0,0,0)\}$, ${}^{2,1} H_{0\,0} = \{ 6, (0,0,0) \}$, ${}^{2,2} H_{0\,0} = \{ 10, (0,0,0)\}$ and ${}^{j,m} H_{k_1\ldots k_{l'}} = 0$ otherwise, and ${}^{j} J_{k_1\ldots k_{l'}} = 0$ are all trace sets of order $0$ (i.\@e.\@ functions satisfying \eqref{eq:admis}) with ${}^{1} J_{1\,0} = {}^{1} J_{0\,0} = {}^{2} J_{0\,0} = 1$, and ${}^{j} J_{k_1\ldots k_{l'}} = 0$ otherwise for all $j,k_1,\ldots,k_{l'}$.

For a given $k\geq 0$, $l \geq 2$ and $H$ as above, define\index{schematic notation!$\mathcal{E}^k(H)$, nonlinear error notation for quantities of up to $k$'th order}
\[
	\mathcal{E}^k(H)
	:=
	\left( \mathfrak{D}^{k} \Phi \right)^{l} \cdot H.
\]
Note that $\mathcal{E}^k(H)$ has the property that, for any $\gamma$,
\[
	\mathfrak{D}^{\gamma} \mathcal{E}^k(H)
	=
	\mathcal{E}^{k+\vert \gamma \vert} (H^{\gamma})
\]
for some $H^{\gamma}$.

We will often abuse notation and write ``$\mathcal{E}^k$'' to mean ``$\mathcal{E}^k(H)$ for some $H$'', and ``$\left( \mathfrak{D}^{k} \Phi \right)^{l}$'' to mean ``$\left( \mathfrak{D}^{k} \Phi \right)^{l} \cdot H$ for some $H$''.

\subsection{The nonlinear error notation $\mathcal{E}^k_p$}
\label{sec:nlepnotation}

The notation $\mathcal{E}^k_p$ is used to denote a nonlinear error term involving at most $k$ derivatives of Ricci coefficients and curvature components which moreover decays, according to the $p$ index notation introduced above, like $r^{-p}$.

Define the sets
\[
	S_1 = \{ 1,\ldots, 6\}, 
	\qquad
	S_2 = \{ 7,\ldots, 12\}, 
	\qquad
	S_{\frac{5}{2}} = \{ 13 \}, 
	\qquad
	S_3 = \{ 14, 15\},
	\qquad
	S_4 = \{ 16, 17\}.
\]
Consider some $k \geq 0$.  Given $i \in \{ 1,\ldots, 17 \}$ and $\gamma \in \{ (1,0,0), (0,1,0), (0,0,1)\}^k$, recall the definition
\[
	\mathfrak{D}^k \Phi \cdot (i,\gamma)
	:=
	\mathfrak{D}^{\gamma} \Phi^i.
\]
Given $p \in \{ 1,2,\frac{5}{2},3,4 \}$, if $i \in S_p$ then we often add a $p$ subscript for emphasis, i.\@e.\@
\[
	\mathfrak{D}^k \Phi_p \cdot (i,\gamma)
	:=
	\mathfrak{D}^{\gamma} \Phi^i
	\qquad
	\text{if }
	i \in S_p.
\]
The $p$ subscript is added to emphasise certain $r$ weights which appear in estimates established for each quantity $\mathfrak{D}^k \Phi_p \cdot (i,\gamma)$ in the proof of the main theorem.

Consider now some $k \geq 0$, $l \geq 1$, $p \geq 0$ and some
\[
	H = \{ {}^{j,m}H^{p_0\ldots p_{l'}}_{k_1\ldots k_{l'}}, {}^j J^{p_0\ldots p_{l'}}_{k_1\ldots k_{l'}} \}_{\substack{
	m=0,\ldots,l',k_1+\ldots + k_{l'} \leq k, 
	\\
	p_0 + \ldots + p_{l'} \geq p, l'\geq l, j \geq 1}}
	,
\]
where
\[
	{}^{j,m}H^{p_0\ldots p_{l'}}_{k_1\ldots k_{l'}}
	\in
	\big\{ (i,\gamma) \mid i \in S_{p_m}, \gamma \in \{ (1,0,0), (0,1,0), (0,0,1)\}^{k_m} \big\} \cup \big\{ 0 \big\},
\]
for all $m=0,\ldots,l'$, and ${}^jJ_{k_1\ldots k_{l'}}^{p_0\ldots p_{l'}}$ is a trace set of some order $d \geq 0$, as in Section \ref{subsec:traceindex}, for all $k_1+\ldots + k_{l'} \leq k$, $p_0+\ldots + p_{l'} \geq p$, $j \geq 1$ and $l' \geq l$.  For such $k \geq 0$, $l \geq 1$ and $H$, define
\[
	\left( \mathfrak{D}^{k} \Phi_p \right)^{l} \cdot H
	:=
	\sum_{l' \geq l}
	\sum_{\substack{
	k_1+\ldots+k_{l'} \leq k
	\\
	p_0 + \ldots+ p_{l'} \geq p
	}}
	\sum_{j \geq 1}
	r^{-p_0}
	\big(
	\mathfrak{D}^{k_1} \Phi_{p_1} \cdot {}^{j,1} H_{k_1\ldots k_{l'}}^{p_0\ldots p_{l'}} \otimes \ldots \otimes \mathfrak{D}^{k_{l'}} \Phi_{p_{l'}} \cdot {}^{j,l'} H_{k_1 \ldots k_{l'}}^{p_0\ldots p_{l'}}
	\big)
	\cdot_{\gslash}
	{}^j J_{k_1\ldots k_{l'}}^{p_0\ldots p_{l'}}.
\]

\vskip1pc
\noindent\fbox{
    \parbox{6.35in}{
 \emph{Such expressions are again only ever considered when sufficiently many of the components of $H$ vanish so that each object appearing in the above summation is an $S$-tensor of the same type, and moreover that the ranges of allowed $l \geq l'$, $j \geq 1$ and $p_0,\ldots, p_{l'} \geq p$ are finite, and so the summation is indeed well defined.}
    }
}
\vskip1pc

For example one can write
\[
	\frac{2}{r} \Omega^{-1} \hat{\chibar} \otimes \eta + \frac{1}{r} \Omega^{-2} \alphabar \otimes \etabar
	=
	\left( \mathfrak{D}^{0} \Phi_4 \right)^{2} \cdot H,
\]
where ${}^{1,1} H_{0\,0}^{1\,1\,2} = \{ 4, (0,0,0)\}$, ${}^{1,2} H_{0\,0}^{1\,1\,2} = \{ 10, (0,0,0) \}$, ${}^{2,1} H_{0\,0}^{1\,1\,2} = \{ 6, (0,0,0) \}$, ${}^{2,2} H_{0\,0}^{1\,1\,2} = \{ 11, (0,0,0) \}$ and ${}^{j,m} H_{k_1 \ldots k_{l'}}^{p_0\ldots p_{l'}} = 0$ otherwise, each ${}^j J_{k_1\ldots k_{l'}}^{p_0\ldots p_{l'}} = 0$ is a trace set of order $0$ with ${}^1 J_{0\,0}^{1\,1\,2} = 2$, ${}^2 J_{0\,0}^{1\,1\,2} = 1$ and ${}^j J_{k_1\ldots k_{l'}}^{p_0\ldots p_{l'}} = 0$ otherwise for all $k_1+\ldots + k_{l'} \leq k$, $p_0+\ldots + p_{l'} \geq p$, $j \geq 1$ and $l' \geq l$.

For a given $k\geq 0$, $l \geq 2$, $p$ and $H$ as above, define\index{schematic notation!$\mathcal{E}^k_p(H)$, nonlinear error notation keeping track of $r^p$ decay}
\[
	\mathcal{E}^k_p(H)
	:=
	\left( \mathfrak{D}^{k} \Phi_p \right)^{l} \cdot H.
\]
Note that the $\mathcal{E}^k_p(H)$ have the property that, for any $\gamma$,
\[
	\mathfrak{D}^{\gamma} \mathcal{E}^k_p(H)
	=
	\mathcal{E}^{k+\vert \gamma \vert}_p (H^{\gamma})
\]
for some $H^{\gamma}$.

We will often abuse notation and write ``$\mathcal{E}^k_p$'' to mean ``$\mathcal{E}^k_p(H)$ for some $H$'', and ``$\left( \mathfrak{D}^{k} \Phi_p \right)^{l}$'' to mean ``$\left( \mathfrak{D}^{k} \Phi_p \right)^{l} \cdot H$ for some $H$''.

\subsection{The nonlinear error notation $\mathcal{E}^{*k}$ and $\mathcal{E}^{*k}_p$} \label{sec:errorstarnot}

In some nonlinear error terms it will be important to keep note of the fact that certain anomalous quantities---which, in the proof of Theorem \ref{thm:main}, will decay at slower rates than other quantities---do not appear.
The notation $\mathcal{E}^{*k}$, therefore, is used to denote a nonlinear error term involving at most $k$ derivatives of Ricci coefficients and curvature components which does not involve the terms
\[
	(r\nablaslash)^k \Omega\hat{\chi},
	\quad
	(r\nablaslash)^k (\Omega \tr \chi - \Omega \tr \chi_{\circ}).
\]
Similarly, the notation $\mathcal{E}^{*k}_p$ is used for nonlinear errors which do not involve the above terms and which moreover decay, according to the $p$ index notation introduced above, like $r^{-p}$.

In order to be more precise, consider again some $k \geq 0$, $l \geq 1$.  Recall that, in the ordering of Section \ref{subsec:orderingdifferences}, $\Phi^7 = \Omega \tr \chi - \Omega \tr \chi_{\circ}$ and $\Phi^9 = \Omega \hat{\chi}$.  Define the set
\[
	\mathcal{P}^*_{k,k}
	=
	\big\{ (i,\gamma) \mid i \in \{ 1,\ldots, 17\}, \gamma \in \{ (1,0,0), (0,1,0), (0,0,1)\}^{k} \big\}
	\smallsetminus
	\big\{ (i,\gamma) \mid i \in \{ 7, 9 \}, \gamma \in (1,0,0)^{k} \big\},
\]
and, for $k' < k$, define
\[
	\mathcal{P}^*_{k,k'}
	=
	\big\{ (i,\gamma) \mid i \in \{ 1,\ldots, 17\}, \gamma \in \{ (1,0,0), (0,1,0), (0,0,1)\}^{k'} \big\}.
\]

Given $l' \geq l$ and $k_1+\ldots + k_{l'} \leq k$, consider some
\[
	H^* = \{ {}^{j,m}H_{k_1\ldots k_{l'}}^*,  {}^jJ_{k_1\ldots k_{l'}} \}_{m=0,\ldots,l',k_1+\ldots + k_{l'} \leq k, l' \geq l, j \geq 1},
\]
as in Section \ref{sec:nlenotation},
where now
\[
	{}^{j,m}H_{k_1\ldots k_{l'}}^*
	\in
	\mathcal{P}^*_{k,k_m}
	\cup \big\{ 0 \big\},
\]
for all $m=0,\ldots,l'$, and again ${}^jJ_{k_1\ldots k_{l'}}$ is a trace set of some order $d\geq 0$, as in Section \ref{subsec:traceindex}, for all $j \geq 1$, $k_1+\ldots + k_{l'} \leq k$ and $l' \geq l$.  For such $k \geq 0$, $l \geq 1$ and $H^*$, define, as in Section \ref{sec:nlenotation},
\[
	\left( \mathfrak{D}^{k} \Phi \right)^{l} \cdot H^*
	:=
	\sum_{l' \geq l}
	\sum_{k_1+\ldots+k_{l'} \leq k}
	\sum_{j \geq 1}
	\big(
	\mathfrak{D}^{k_1} \Phi \cdot {}^{j,1}H_{k_1\ldots k_{l'}}^* \otimes \ldots \otimes \mathfrak{D}^{k_{l'}} \Phi \cdot {}^{j,l'}H_{k_1 \ldots k_{l'}}^*
	\big)
	\cdot_{\gslash}
	{}^jJ_{k_1\ldots k_{l'}}.
\]

For a given $k\geq 0$, $l \geq 2$ and $H^*$ as above, define as before\index{schematic notation!$\mathcal{E}^{*k}(H^*)$, nonlinear error notation for components not involving certain terms}
\[
	\mathcal{E}^{*k}(H^*) := \left( \mathfrak{D}^{k} \Phi \right)^{l} \cdot H^*.
\]
The notation $\mathcal{E}^{*k}_p(H^*)$ is similarly defined\index{schematic notation!$\mathcal{E}^{k^*}_p(H^*)$, nonlinear error notation for components not involving certain terms} in the obvious way.  Note that, for any $\gamma$,
\[
	\mathfrak{D}^{\gamma} \mathcal{E}^{*k}_p(H^*)
	=
	\mathcal{E}^{*,k+\vert \gamma \vert}_p (H^*_{\gamma})
\]
for some $H^*_{\gamma}$.

Again, we will often abuse notation and write ``$\mathcal{E}^{*k}$'' to mean ``$\mathcal{E}^{*k}(H^*)$ for some vector $H^*$ of the above form'', and ``$\mathcal{E}^{*k}_p$'' to mean ``$\mathcal{E}^{*k}_p(H^*)$ for some vector $H^*$ of the above form''.

\subsection{The nonlinear error notation $\check{\mathcal{E}}^{k}$ and $\check{\mathcal{E}}^{k}_p$}
\label{subsec:tildeerrors}

Recall $\check{r}$, defined in Definition \ref{defofalmostgaugeinv}.  The notation $\check{\mathcal{E}}^k$ is used to denote a nonlinear error term involving at most $k$ derivatives of Ricci coefficients and curvature components, which may also include factors of powers of $r^{-1} \check{r}$, and may also contain bad $\Omega$ weights.  Such error terms typically arise only in analysis of the quantities $\check{\psibar}$ and $\check{\Pbar}$ which are used, in the proof of Theorem \ref{havetoimprovethebootstrap}, only in the $\I$ gauge where such bad $\Omega$ weights are irrelevant.  Similarly, $\check{\mathcal{E}}^k_p$ is used to denote such a nonlinear error term which moreover decays, according to the $p$ index notation, like $r^{-p}$.

Consider some $l_1, l_2 \geq 0$ and some $\check{H} = \{ H_{m_1m_2}\}_{\substack{m_1=0,\ldots,l_1 \\ m_2=0,\ldots,l_2}}$, where each $H_{m_1m_2}$ is an array of the form $H$ of Section \ref{sec:nlenotation}.  Define\index{schematic notation!$\check{\mathcal{E}}^{k}(\check{H})$, nonlinear error notation with possible bad metric weights}
\[
	\check{\mathcal{E}}^{k}(\check{H})
	:=
	\sum_{m_1=0}^{l_1} \sum_{m_2=0}^{l_2}
	\Omega^{-m_1}
	\Big( \frac{\check{r}}{r} \Big)^{m_2} \mathcal{E}^k (H_{m_1m_2}).
\]
Similarly define, for $\check{H}$ as in Section \ref{sec:nlepnotation},\index{schematic notation!$\check{\mathcal{E}}^{k}_p(\check{H})$, nonlinear error notation with possible bad metric weights}
\[
	\check{\mathcal{E}}^{k}_p(\check{H})
	:=
	\sum_{m_1=0}^{l_1} \sum_{m_2=0}^{l_2}
	\Omega^{-m_1}
	\Big( \frac{\check{r}}{r} \Big)^{m_2} \mathcal{E}^k_p (H_{m_1m_2}).
\]
Again, such expressions are only ever considered when sufficiently many of the components of $\check{H}$ vanish so that each object appearing in the above summation is an $S$ tensor of the same type, and that the summation is over only finitely many terms.

Note that, for any $\gamma$,
\[
	\mathfrak{D}^{\gamma} \check{\mathcal{E}}^{k}_p(\check{H})
	=
	\check{\mathcal{E}}^{k+\vert \gamma \vert}_p(\check{H}^{\gamma})
\]
for some $\check{H}^{\gamma}$.
 
Again, we will typically abuse notation and write ``$\check{\mathcal{E}}^{k}$'' to mean ``$\check{\mathcal{E}}^{k}(\check{H})$ for some vector $\check{H}$ of the above form'', and ``$\check{\mathcal{E}}^{k}_p$'' to mean ``$\check{\mathcal{E}}^{k}_p(\check{H})$ for some vector $\check{H}$ of the above form''.

\subsection{The nonlinear error notation $\protect\overset{(4)}{\mathcal{E}^k_p}$}
\label{subsec:overset4errors}

In Section \ref{subsec:tildewaveequations},
it will be important to keep track of
error terms which have a gain in decay, which is better than the expected $r^{-1}$, when acted on by $\Omega \nablaslash_4$.  To do so, it is convenient to introduce additional error notation.  Define
\[
	\overset{(4)}{\Phi}{}^1
	=
	\Omega^{-1} \hat{\chibar},
	\quad
	\overset{\tiny{(4)}}{\Phi}{}^2
	=
	\Omega^{-2} \alphabar,
	\quad
	\overset{\tiny{(4)}}{\Phi}{}^3
	=
	\Omega \hat{\chi},
	\quad
	\overset{\tiny{(4)}}{\Phi}{}^4
	=
	\Omega^{-1} \betabar,
	\quad
	\overset{\tiny{(4)}}{\Phi}{}^5
	=
	\rho - \rho_{\circ},
	\quad
	\overset{\tiny{(4)}}{\Phi}{}^6 
	=
	\sigma.
\]
Note that each $\overset{(4)}{\Phi}$ satisfies a null structure or Bianchi equation of the form
\begin{equation} \label{eq:Phi4improvement}
	\Omega\nablaslash_4 (r^p \overset{(4)}{\Phi}_p)
	=
	\frac{1}{r^2}
	\sum_{k =0}^1
	\sum_{\Phi_q}
	c_{k,\Phi_q}
	(r\nablaslash)^k r^{q} \Phi_{q}
	+
	\mathcal{E}^0_2,
\end{equation}
for some admissible functions $c_{k,\Phi_q}$, satisfying \eqref{eq:admis}.

The notation $\overset{(4)}{\mathcal{E}^k_p}$ will be used for nonlinear error terms which involve at most $k$ derivatives of geometric quantities, decay in $r$, according to the $p$ index notation, like $r^{-p}$ and moreover only involve the above $\overset{(4)}{\Phi}$ so that there is a gain of $r^2$ decay when $\nablaslash_4$ acts on the error appropriately (see \eqref{4erroreq}).

Unlike in the previous error notation, general admissible functions are not allowed to appear in the $\overset{(4)}{\mathcal{E}^k_p}$ errors.  In order to define these errors one therefore has to consider trace sets, as in Section \ref{subsec:traceindex}, which do not include any admissible functions.  To be more precise, first define a constant $a$ to be \emph{admissible} if $\vert a \vert \leq C$, where $C$ is as in \eqref{eq:admis}.  
A \emph{trace set of constants of order $0$} is defined simply to be an admissible constant $a$.
Consider now some $n \geq 2$.  A \emph{trace set of constants of order $1$} is a collection $I = \{ a_{ij} \}_{1\leq i < j \leq n}$, where each $a_{ij}$ is an admissible constant.
For $1 \leq d \leq \lfloor n/2 \rfloor$, a \emph{trace set of constants of order $d$} is defined to be a collection $I = \{ I^m \}_{1 \leq m \leq d}$, where each $I^m = \{ a_{ij}^m \}_{1\leq i < j \leq n}$ is a trace set of constants of order $1$.  If $\xi$ is a $(0,n)$ $S$ tensor field and $I$ is a trace set of some order $d\geq 0$, one defines $\xi \cdot_{\gslash} I$ as in Section \ref{subsec:traceindex}.

Define the sets
\[
	\overset{(4)}{S}_1 = \{ 1,2\}, 
	\qquad
	\overset{(4)}{S}_2 = \{ 3, 4 \},
	\qquad
	\overset{(4)}{S}_4 = \{ 5, 6 \}.
\]
Consider some $k \geq 0$.  Given $\gamma \in \{ (1,0,0), (0,1,0), (0,0,1)\}^k$,  $p \in \{ 1,2,4 \}$ and $i \in \overset{(4)}{S}_p$, define
\[
	\mathfrak{D}^k \overset{\tiny{(4)}}{\Phi}_p \cdot (i,\gamma)
	:=
	\mathfrak{D}^{\gamma} \overset{\tiny{(4)}}{\Phi}{}^i
	\qquad
	\text{if }
	i \in \overset{(4)}{S}_p.
\]

Consider now some $k \geq 0$, $l \geq 1$, $p \geq 0$ and some
\[
	H = \{ {}^{j,m}H^{p_0\ldots p_{l'}}_{k_1\ldots k_{l'}}, {}^j J^{p_0\ldots p_{l'}}_{k_1\ldots k_{l'}} \}_{\substack{
	m=0,\ldots,l',k_1+\ldots + k_{l'} \leq k, 
	\\
	p_0 + \ldots + p_{l'} \geq p, l'\geq l, j \geq 1}}
	,
\]
where
\[
	{}^{j,m}H^{p_0\ldots p_{l'}}_{k_1\ldots k_{l'}}
	\in
	\big\{ (i,\gamma) \mid i \in \overset{(4)}{S}_{p_m}, \gamma \in \{ (1,0,0), (0,1,0), (0,0,1)\}^{k_m} \big\} \cup \big\{ 0 \big\},
\]
for all $m=0,\ldots,l'$, and ${}^jJ_{k_1\ldots k_{l'}}^{p_0\ldots p_{l'}}$ is a trace set of constants (defined above) of some order $d \geq 0$, for all $k_1+\ldots + k_{l'} \leq k$, $p_0+\ldots + p_{l'} \geq p$, $j \geq 1$ and $l' \geq l$.  For such $k \geq 0$, $l \geq 1$ and $H$, define
\[
	\big( \mathfrak{D}^{k} \overset{\tiny{(4)}}{\Phi}_p \big)^{l} \cdot H
	:=
	\sum_{l' \geq l}
	\sum_{\substack{
	k_1+\ldots+k_{l'} \leq k
	\\
	p_0 + \ldots+ p_{l'} \geq p
	}}
	\sum_{j \geq 1}
	r^{-p_0}
	\big(
	\mathfrak{D}^{k_1} \overset{\tiny{(4)}}{\Phi}_{p_1} \cdot {}^{j,1} H_{k_1\ldots k_{l'}}^{p_0\ldots p_{l'}} \otimes \ldots \otimes \mathfrak{D}^{k_{l'}} \overset{\tiny{(4)}}{\Phi}_{p_{l'}} \cdot {}^{j,l'} H_{k_1 \ldots k_{l'}}^{p_0\ldots p_{l'}}
	\big)
	\cdot_{\gslash}
	{}^j J_{k_1\ldots k_{l'}}^{p_0\ldots p_{l'}}.
\]

For a given $k\geq 0$, $l \geq 2$, consider some $l_1, l_2 \geq 0$ and some $\check{H} = \{ H_{m_1m_2}\}_{\substack{m_1=0,\ldots,l_1 \\ m_2=0,\ldots,l_2}}$ where each $H_{m_1m_2}$ is an array of the above form $H$.  Define
\begin{equation} \label{4errordef}
	\overset{\tiny{(4)}}{\mathcal{E}^k_p}(\check{H})
	:=
	\sum_{m_1=0}^{l_1} \sum_{m_2=0}^{l_2}
	\Omega^{-m_1}
	\Big( \frac{\check{r}}{r} \Big)^{m_2}
	\big( \mathfrak{D}^{k} \overset{\tiny{(4)}}{\Phi}_p \big)^{l} \cdot H.
\end{equation}
Again, such expressions are only ever considered when sufficiently many of the components of $\check{H}$ vanish so that each object appearing in the above summation is an $S$ tensor of the same type, and that the summation is over only finitely many terms.

Again, we will abuse notation and write ``$\overset{(4)}{\mathcal{E}^k_p}$'' to mean ``$\overset{(4)}{\mathcal{E}^k_p}(\check{H})$ for some vector $\check{H}$''.
In accordance with \eqref{eq:Phi4improvement}, the key property of $\overset{(4)}{\mathcal{E}^k_p}$ errors is summarised in Lemma \ref{lem:nabla4error4}.

\subsection{The nonlinear error notation $\protect\overset{(in)}{\slashed{\mathcal{E}}^{k}_{p}}$ and $\protect\overset{(out)}{\slashed{\mathcal{E}}^{k}_{p}}$} \label{sec:errorterms}
In Chapter \ref{chap:Iestimates},  we are going to derive, in the $\mathcal{I}^+$ gauge, top-order estimates for the Ricci-coefficients, hence we will need to distinguish between Ricci-coefficients and curvature components appearing in the error-terms: Up to $N+1$ derivatives of Ricci-coefficients may appear at top order but only $N$ derivatives of curvature. Moreover, we would like to capture the fact that only certain components appear in certain (ingoing or outgoing) transport equations. To achieve this, we introduce the shorthand notation\index{schematic notation!$T= \Omega tr \chi - (\Omega tr \chi)_\circ$}\index{schematic notation!$\underline{T}= \frac{tr \underline{\chi}}{\Omega} - \frac{tr \underline{\chi}_\circ}{\Omega_\circ}$}\index{schematic notation!$\omega -\omega_\circ = \Omega \hat{\omega} - (\Omega \hat{\omega})_\circ$}\index{schematic notation!$\underline{\omega} -\underline{\omega}_\circ = \Omega \hat{\underline{\omega}} - (\Omega \hat{\underline{\omega}})_\circ$}    
\begin{align} \label{shorthandnot}
T = \Omega tr \chi - (\Omega tr \chi)_\circ,\qquad \underline{T}= \frac{tr \underline{\chi}}{\Omega} - \frac{tr \underline{\chi}_\circ}{\Omega_\circ}, \qquad \omega -\omega_\circ = \Omega \hat{\omega} - (\Omega \hat{\omega})_\circ, \qquad \underline{\omega} -\underline{\omega}_\circ = \Omega \hat{\underline{\omega}} - (\Omega \hat{\underline{\omega}})_\circ.
\end{align}
 and let
\begin{align}
\overset{(in)}{{\Gamma}} &\in \{ \underline{\eta}, \eta, \Omega \hat{\chi}, \Omega^{-1} \underline{\hat{\chi}}, T,\underline{T},1 -\frac{\Omega^2}{\Omega_\circ^2}, 1- \frac{\Omega_\circ^2}{\Omega^2},  \Omega^{-2} \left(\underline{\omega}- \underline{\omega}_\circ \right)\} \nonumber  \\
 \overset{(in)}{\Phi} &\in \{ \underline{\eta}, \eta, \Omega \hat{\chi}, \Omega^{-1} \underline{\hat{\chi}}, T , \underline{T},1 -\frac{\Omega^2}{\Omega_\circ^2}, 1- \frac{\Omega_\circ^2}{\Omega^2},  \Omega^{-2} \left(\underline{\omega}- \underline{\omega}_\circ \right)\}  \cup \{ \Omega^{-2} \underline{\alpha},\Omega^{-1}\underline{\beta}, \rho - \rho_\circ, \sigma \} \nonumber \\
\overset{(out)}{{\Gamma}} &\in \{ \underline{\eta}, \eta, \Omega \hat{\chi}, \Omega^{-1} \underline{\hat{\chi}}, T , \underline{T},1 -\frac{\Omega^2}{\Omega_\circ^2}, 1- \frac{\Omega_\circ^2}{\Omega^2},  \omega-\omega_\circ , b \}\nonumber \\
 \overset{(out)}{\Phi}&\in   \{ \underline{\eta}, \eta, \Omega \hat{\chi}, \Omega^{-1} \underline{\hat{\chi}}, T , \underline{T},1 -\frac{\Omega^2}{\Omega_\circ^2}, 1- \frac{\Omega_\circ^2}{\Omega^2},  \omega-\omega_\circ , b \} \cup \{  \Omega^2{\alpha},\Omega{\beta}, \rho - \rho_\circ, \sigma \}. \nonumber
 \end{align} 
 We also denote by $\overset{(in)}{{\Gamma}_p}$ a $\overset{(in)}{{\Gamma}}$ from the collection $\Gamma_p$ and by $\overset{(in)}{{\Phi}_p}$ a $\overset{(in)}{{\Phi}}$ from the collection $\Phi_p$. The notation $\overset{(in)}{\slashed{\mathcal{E}}^{k}_{p}}$ will be introduced to denote a nonlinear error term involving at most $k$ angular derivatives of Ricci coefficients from $\overset{(in)}{{\Gamma}}$ and at most $k-1$ derivatives of the $\overset{(in)}{\Phi}$ and which decays, according to the $p$ index notation introduced above, like $r^{-p}$.  The analogous definition will be made for $\overset{(out)}{\slashed{\mathcal{E}}^{k}_{p}}$.

Formal definitions can be given using the notation of Section \ref{sec:nlepnotation}. We define for $i \in \{1,...17\}$ the notation $\left[r \slashed{\nabla}\right]^k \Phi_p \cdot (i) := \left[r \slashed{\nabla}\right]^k \Phi^i$. In analogy with Section \ref{sec:nlepnotation}, given $k \geq 0$, $l \geq 1$ and $p\geq 0$, we let 
\begin{align}
\slashed{H} = \left( {}^{j,m}\overset{(in)}{\slashed{H}}{}^{p_0\ldots p_{l'}}_{k_1\ldots k_{l'}} \ , \  {}^j \slashed{J}_{k_1\ldots k_{l'}}^{p_0\ldots p_{l'}} \ , \  {}^{j,m}\overset{(in)}{\slashed{H}}{}^{\tilde{p}_0 \tilde{p}_1\tilde{p}_2}_{top}\ , \  {}^j \slashed{J}_{top}^{\tilde{p}_0 \tilde{p}_1 \tilde{p}_2} \right)
\end{align}
with
\[
	{}^{j,m}\overset{(in)}{\slashed{H}}{}^{p_0\ldots p_{l'}}_{k_1\ldots k_{l'}}
	\in
	\big\{ i \leq 15 \mid i \in S_{p_m} \ \ \textrm{and} \ \  i \notin \{3,13\} \big\} \cup \big\{ 0 \big\} \, , 
\]
\[
 {}^{j,m}\overset{(in)}{\slashed{H}}{}^{\tilde{p}_0 \tilde{p}_1 \tilde{p}_2}_{top}
	\in
	\big\{ i \leq 11 \mid i \in S_{p_m} \ \ \textrm{and} \ \  i \notin \{3,6\} \big\} \cup \big\{ 0 \big\} \, ,
\]
and ${}^j \slashed{J}_{k_1\ldots k_{l'}}^{p_0\ldots p_{l'}}$ and ${}^j \slashed{J}_{top}^{\tilde{p}_0 \tilde{p}_1 \tilde{p}_2}$ trace sets of some order $d, d_{top} \geq 0$, all these being defined for
all $l^\prime \geq l$, $m=0, ..., \ell^\prime$, $k_1+...+k_{l^\prime}\leq k$ and $p_0+...+p_{l^\prime}\geq p$, $\tilde{p}_0+\tilde{p}_1+\tilde{p}_2\geq p$. We finally define
\begin{align} \label{inerrorschematic}
&\left( \left[r \slashed{\nabla}\right]^{k} \overset{(in)}{\Phi_p} \right)^{l} \cdot \slashed{H}
	:= \sum_{\tilde{p}_0+\tilde{p}_1+\tilde{p}_2 \geq p} \sum_{j \geq 1}  r^{-p_0}\big(
	[r \slashed{\nabla}]^{k}{\Phi_{p_1}} \cdot {}^{j,1} \overset{(in)}{\slashed{H}}{}_{top}^{p_0p_1p_2} \otimes  \Phi_{p_{2}} \cdot {}^{j,2} \overset{(in)}{\slashed{H}}{}_{top}^{\tilde{p}_0\tilde{p}_1\tilde{p}_2}
	\big)\cdot_{\gslash}
	{}^j \slashed{J}_{top}^{\tilde{p}_0 \tilde{p}_1 \tilde{p}_2}   \\
&\phantom{XXX} +\sum_{l' \geq l}
	\sum_{\substack{
	k_1+\ldots+k_{l'} \leq k
	\\ k_i \leq k -1 \\
	p_0 + \ldots+ p_{l'} \geq p
	}}
	\sum_{j \geq 1}
	r^{-p_0}
	\big(
	[r \slashed{\nabla}]^{k_1} \Phi_{p_1} \cdot {}^{j,1} \overset{(in)}{\slashed{H}}{}_{k_1\ldots k_{l'}}^{p_0\ldots p_{l'}} \otimes \ldots \otimes [r \slashed{\nabla}]^{k_{l'}} \Phi_{p_{l'}} \cdot {}^{j,l^\prime} \overset{(in)}{\slashed{H}}{}_{k_1 \ldots k_{l'}}^{p_0\ldots p_{l'}}.   
	\big)
	\cdot_{\gslash}
	{}^j \slashed{J}_{k_1\ldots k_{l'}}^{p_0\ldots p_{l'}}. \nonumber 
\end{align}

In other words, top order (i.e.~equal to $k$) derivatives, can only involve Christoffel symbols from $\overset{(in)}{\Gamma}$, not curvature and the terms are necessarily quadratic. Moreover, as is clear from the above notation, in general only angular derivatives can appear. Note that for $k=0$ only the first sum survives and the right hand side equals a sum of products of Christoffel symbols from $\overset{(in)}{\Gamma}$ where each product can also be multiplied with a bounded function.

The same definition is made replacing all superscripts (in) by (out) and setting
\[
	{}^{j,m}\overset{(out)}{\slashed{H}}{}^{p_0\ldots p_{l'}}_{k_1\ldots k_{l'}}
	\in
	\big\{ i \mid i \in S_{p_m} \ \ \textrm{and} \ \  i \notin \{5, 6,12\} \big\} \cup \big\{ 0 \big\} \, , 
\]
\[
 {}^{j,m}\overset{(out)}{\slashed{H}}{}^{p_0p_1p_2}_{top}
	\in
	\big\{ i \leq 13 \mid i \in S_{p_m} \ \ \textrm{and} \ \  i \notin \{5, 6,12\} \big\} \cup \big\{ 0 \big\} \, .
\]
Finally, for a given $k\geq 0$, $l \geq 2$, $p \geq 0$ and given $\slashed{H}$ as above, we define
\begin{align} \label{inerrorschematic2}
	\overset{(in)}{\slashed{\mathcal{E}}}{}^k_p(\slashed{H})
	:= \left( \left[r \slashed{\nabla}\right]^{k} \overset{(in)}{\Phi}_{p} \right)^{l} \cdot \slashed{H}.
\end{align}
The notation $\overset{(out)}{\slashed{\mathcal{E}}^k_p}(\slashed{H})$ is defined replacing (in) by (out).  We will often abuse notation and write ``$\overset{(in)}{\slashed{\mathcal{E}}^k_p}$'' to mean ``$\overset{(in)}{\slashed{\mathcal{E}}^k_p}(\slashed{H})$ for some $\slashed{H}$'' etc. We will also freely use the easily verified schematic identities 
\begin{align} \label{schematicerroradd}
\left[r \slashed{\nabla}\right]^l \overset{(in)}{\slashed{\mathcal{E}}^k_p} = \overset{(in)}{\slashed{\mathcal{E}}^{k+l}_p} \ \ \ \textrm{and} \ \ \ \left[r \slashed{\nabla}\right]^l \overset{(out)}{\slashed{\mathcal{E}}^k_p} = \overset{(out)}{\slashed{\mathcal{E}}^{k+l}_p} \, .
\end{align}

\subsection{The nonlinear error notation  $\protect\overset{(\mathrm{Kerr})}{\slashed{\mathcal{E}}}$} \label{sec:errortermsKerr}
We finally introduce a schematic notation for errors arising from the subtraction of Kerr reference solutions in the Bianchi and null structure equations. This will become relevant only in Section \ref{sec:schematic}. We define
\[
\mathcal{Z}^1_m = (r^2 \slashed{\Delta} +2)Y^{\ell=1}_m \ \ , \ \ \mathcal{Z}^2_m = (r^2 \slashed{\Delta} +2\frac{\Omega_\circ^2}{\Omega^2})Y^{\ell=1}_m \ \ , \ \ \mathcal{Z}^3_m = \Omega^{-1} \slashed{\nabla}_3 Y^{\ell=1}_m \ \ , \ \ \Omega \slashed{\nabla}_4 Y^{\ell=1}_m \ \ ,
\]
\[
\mathcal{Z}^5_m =\Omega^{-1} \slashed{\nabla}_3 (r \cdot {}^* \nablaslash Y^{\ell=1}_m) \ \ , \ \ \mathcal{Z}^6_m =\Omega \slashed{\nabla}_4 (r \cdot {}^* \nablaslash Y^{\ell=1}_m) ,
\]
where we recall the definition of the spherical harmonics (cf.~Section \ref{basisforprojspace}) and observe that $\mathcal{Z}^i_m=0$ holds for Schwarzschild with the standard spherical harmonics. For a given collection of admissible coefficient functions (see (\ref{eq:admis}))  $h_j$ ($j \in \{1,...,17\}$) and $h_{j,m}$ ($j \in \{1, ..., 6\}$, $m \in \{-1,0,1\}$), denoted collectively by $H$, we define the error
\begin{align}
\overset{(\mathrm{Kerr})}{\slashed{\mathcal{E}}}(H) =  \sum_{m=-1}^1 \sum_{j=1}^6 h_{j, m} \cdot a^m_{\I} \cdot \mathcal{Z}_m^j +  \sum_{i \in \{3,10,11,12,15,16\}} h_{i} \left(1-\frac{\Omega_\circ^2}{\Omega^2}\right) r^{p_i} (\Phi^i)_{\mathrm{Kerr}} ,
\end{align}
where $p_3=1$, $p_{10}=p_{11}=p_{12}=2$, $p_{15}=3$ and $p_{16}=4$. Such expressions are only ever considered when sufficiently many of the components of $H$ vanish so that each object appearing in the above summation is an $S$ tensor of the same type. Finally, we will often abuse notation and write ``$\overset{(\mathrm{Kerr})}{\slashed{\mathcal{E}}}$'' to mean ``$\overset{(\mathrm{Kerr})}{\slashed{\mathcal{E}}}(H) $ for some $H$''.

\section{Commutation identities}
\label{commutetheident}
Throughout,
let $M$, $\mathcal{Z}$, $g$ and $g_{\circ,M}$ be as in Section~\ref{schemnotsec}.
The following lemma describes the error terms generated by commuting the various differential operators.

\begin{lemma}[Commutation identities] \label{lem:commutation}
	For any $S$-tangent $(0,k)$ tensor $\xi$,
	\[
		[ \Omega \nablaslash_4, r \nablaslash_B] \xi_{A_1\ldots A_k}
		=
		F_{4 B A_1 \ldots A_k}[\xi],
		\qquad
		[ \Omega \nablaslash_3, r \nablaslash_B] \xi_{A_1\ldots A_k}
		=
		\Omega^2 F_{3 B A_1 \ldots A_k}[\xi],
	\]
	and
	\[
		[ \Omega \nablaslash_3, \Omega \nablaslash_4] \xi_{A_1\ldots A_k}
		=
		\Omega^2 F_{3 4 A_1 \ldots A_k}[\xi],
	\]
	where
	\begin{align*}
		F_{4 B A_1 \ldots A_k}[\xi]
		=
		\
		&
		\frac{1}{2} \left( (\Omega \tr \chi)_{\circ} - \Omega \tr \chi \right) r \nablaslash_B \xi_{A_1 \ldots A_k}
		-
		\Omega {\hat{\chi}_B}^C r \nablaslash_C \xi_{A_1 \ldots A_k}
		\\
		&
		+
		r \Omega \sum_{i=1}^k \left(
		\chi_{A_iB} \etabar^C - {\chi_B}^C \etabar_{A_i} + {}^* \beta_B {\epsslash_{A_i}}^C
		\right)
		\xi_{A_1 \ldots A_{i-1} C A_{i+1} \ldots A_k},
		\\
		F_{3 B A_1 \ldots A_k}[\xi]
		=
		\
		&
		\frac{1}{2} \frac{1}{\Omega^2} \left( (\Omega \tr \chibar)_{\circ} - \Omega \tr \chibar \right) r \nablaslash_B \xi_{A_1 \ldots A_k}
		-
		\Omega^{-1} {\hat{\chibar}_B}^C r \nablaslash_C \xi_{A_1 \ldots A_k}
		\\
		&
		+
		r\Omega^{-1} \sum_{i=1}^k \left(
		\chibar_{A_iB} \eta^C - {\chibar_B}^C \eta_{A_i} + {}^* \betabar_B {\epsslash_{A_i}}^C
		\right)
		\xi_{A_1 \ldots A_{i-1} C A_{i+1} \ldots A_k},
		\\
		F_{3 4 A_1 \ldots A_k}[\xi]
		=
		\
		&
		\left( \eta^B - \etabar^B \right) \nablaslash_B \xi_{A_1 \ldots A_k}
		+
		2 \sum_{i=1}^k \left(
		\etabar_{A_i} \eta^C - \eta_{A_i} \etabar^C + \sigma {\epsslash_{A_i}}^C
		\right)
		\xi_{A_1 \ldots A_{i-1} C A_{i+1} \ldots A_k}.
	\end{align*}
	Finally,
	\begin{equation} \label{eq:angularcommutation}
		[\nablaslash_A,\nablaslash_B] \xi_{C_1\ldots C_k}
		=
		K
		\sum_{i=1}^k
		\gslash_{A C_i} \xi_{C_1\ldots C_{i-1} B C_{i+1}\ldots C_k}
		-
		\gslash_{B C_i} \xi_{C_1\ldots C_{i-1} A C_{i+1}\ldots C_k},
	\end{equation}
	where $K$ is the Gauss curvature of $S_{u,v}$.
\end{lemma}

\begin{proof}
	The proof is standard.  See, for example, Lemma 7.\@3.\@3 of \cite{CK} (note however the difference with the present notation).
\end{proof}

\begin{remark}
	There is an asymmetry in the definitions of $F_3$ and $F_4$.  The reason for this asymmetry is that, whilst $\Omega \nablaslash_4$ is a regular differential operator in the outgoing null direction, in the incoming null direction it is $\Omega^{-1} \nablaslash_3 = \Omega^{-2} \Omega \nablaslash_3$ which is regular.
\end{remark}

Consider some Ricci coefficient or curvature component $\Phi_p$ and some multi index $\gamma$.  In the notation of Section \ref{sec:nlenotation}, the commutation errors take the form
\begin{align}
	F_4[\mathfrak{D}^{\gamma} \Phi_p]
	&
	=
	\mathcal{E}^{\vert \gamma \vert + 1}_{p+2},
	\label{eq:schemcomm4}
	\\
	F_3[\mathfrak{D}^{\gamma} \Phi_p]
	&
	=
	\mathcal{E}^{\vert \gamma \vert + 1}_{p+1},
	\label{eq:schemcomm3}
	\\
	F_{34}[\mathfrak{D}^{\gamma} \Phi_p]
	&
	=
	\mathcal{E}^{\vert \gamma \vert + 1}_{p+3}.
	\label{eq:schemcomm34}
\end{align}

\section{Derivation of equations satisfied by almost gauge invariant quantities}
\label{TeukandRegsection}

Throughout, let $M$, $\mathcal{Z}$, $g$ and $g_{\circ,M}$ be as in Section~\ref{schemnotsec}.
In this section, we will derive the equations satisfied by $\alpha$, $\underline{\alpha}$, $\psi$, $\psibar$, $P$, $\underline{P}$, $\check{\psibar}$, and $\check{\Pbar}$ (see Definition~\ref{defofalmostgaugeinv}).
In particular, we shall exhibit the special structure of the nonlinear terms appearing in these equations.  
Throughout this section, the error notation introduced in Section \ref{schnotfornonlinearerror} is utilised.

Recall the definition~\eqref{Dslash2stardef} of the operator $\Dslash_2^*$ acting on $S$ 1-forms $\xi$ as
\[
	\Dslash_2^* \xi
	=
	-\frac{1}{2}
	\left(
	\nablaslash \xi
	+
	\nablaslash^T \xi
	-
	\divslash \xi \gslash
	\right),
\]
where $\nablaslash^T$ denotes the transpose  of $\nablaslash$ defined by~\eqref{transposedef}.

\subsection{Elliptic relations}

The following proposition gives expressions for $\psi$, $\psibar$, $P$, $\Pbar$ in terms of certain elliptic operators applied to certain Ricci coefficients and curvature components.  It is important to isolate the anomalous error term $\frac{1}{r}(\Omega \omegahat - (\Omega \omegahat)_{\circ}) \cdot \alphabar$ in \eqref{eq:Pbarrhosigma}.

\begin{proposition}[Elliptic relations for $\psi$, $\psibar$, $P$, $\Pbar$] \label{prop:PPbaridentities}
	With $M$, $\mathcal{Z}$, $g$ and $g_{\circ,M}$  as in Section~\ref{schemnotsec}, 
	the quantities $\psi$ and $\psibar$ satisfy
	\begin{align} \label{eq:psipsibar}
		\psi
		=
		\Dslash_2^* \beta
		-
		\frac{3M}{r^3} \hat{\chi}
		+
		\mathcal{E}^0_5,
		\qquad
		\psibar
		=
		\Dslash_2^* \betabar
		+
		\frac{3M}{r^3} \hat{\chibar}
		+
		\mathcal{E}^0_3,
	\end{align}
	and the quantities $P$ and $\Pbar$ satisfy
	\begin{align}
		P
		&
		=
		\Dslash_2^* \left( \nablaslash \rho + {}^* \nablaslash \sigma \right)
		+
		\frac{3M \Omega}{r^4} \left( \hat{\chibar} - \hat{\chi} \right)
		+
		\mathcal{E}^1_5,
		\label{eq:Prhosigma}
		\\
		\Pbar
		&
		=
		\Dslash_2^* \left( \nablaslash \rho - {}^* \nablaslash \sigma \right)
		+
		\frac{3M \Omega}{r^4} \left( \hat{\chibar} - \hat{\chi} \right)
		+
		\frac{1}{r}(\Omega \omegahat - (\Omega \omegahat)_{\circ}) \cdot \alphabar
		+
		\mathcal{E}^1_5.
		\label{eq:Pbarrhosigma}
	\end{align}
\end{proposition}

\begin{proof}
	The expressions for $\psi$ and $\psibar$ follow immediately from the Bianchi equations \eqref{eq:alpha3}, \eqref{eq:alphabar4}.  The expression for $\psi$ implies that
	\begin{align*}
		P
		=
		\frac{1}{r^3\Omega} \nablaslash_3 \left( r^3 \Omega \psi \right)
		=
		\frac{1}{r^3\Omega} \nablaslash_3 \left( r^3 \Omega \Dslash_2^* \beta \right)
		-
		\frac{3M}{r^3\Omega} \nablaslash_3 \left( \Omega \hat{\chi} \right)
		+
		\frac{1}{r^3 \Omega} \nablaslash_3 \left( r^3 \Omega \mathcal{E}^0_5 \right).
	\end{align*}
	The identity \eqref{eq:Prhosigma} for $P$ then follows from the commuted Bianchi equation \eqref{eq:beta3} and the equation \eqref{eq:chihat3}.  The expression for $\Pbar$ is obtained similarly after noting that, in fact,
	\[
		\psibar
		=
		\Dslash_2^* \betabar
		+
		\frac{3M}{r^3} \hat{\chibar}
		-
		\frac{1}{4\Omega} (\Omega \tr \chi - \Omega \tr \chi_{\circ}) \cdot \alphabar
		+
		\mathcal{E}^0_4,
	\]
	and that
	\[
		\Omega \nablaslash_4 (r^2 (\Omega \tr \chi - \Omega \tr \chi_{\circ}) \cdot r \alphabar)
		=
		4 \Omega_{\circ}^2 r^2 (\Omega \omegahat - \Omega \omegahat_{\circ}) \cdot \alphabar
		+
		\mathcal{E}^1_2,
	\]
	which follows from the Raychaudhuri equation, \eqref{eq:Ray}, which can be rewritten
	\begin{equation} \label{eq:Raychaudhuriinschematicformsection3}
		\Omega \nablaslash_4 \left(r^2(
		\Omega \tr \chi - \Omega \tr \chi_{\circ})
		\right)
		=
		2M
		\left(
		\Omega \tr \chi - \Omega \tr \chi_{\circ}
		\right)
		+
		4r \Omega_{\circ}^2 \left(
		\Omega \omegahat - \Omega \omegahat_{\circ}
		\right)
		+
		\mathcal{E}^0_2.
	\end{equation}
\end{proof}

Recall the definition of $\check{r}$
\[
	\check{r}
	:=
	\frac{2\Omega}{\tr \chi},
\]
and the definitions of $\check{\psibar}$ and $\check{\Pbar}$
\[
	\check{\psibar} := \frac{1}{2r \Omega^2} \nablaslash_4(\check{r}\Omega^2 \alphabar),
	\qquad
	\check{\Pbar} := - \frac{1}{r^3 \Omega} \nablaslash_4 (r^3 \Omega \check{\psibar})
\]
introduced in Definition \ref{defofalmostgaugeinv}.
It follows from the Raychaudhuri equation \eqref{eq:Ray} that $\check{r}$ satisfies
\begin{equation} \label{eq:rtilde4}
	\partial_v \check{r}
	=
	\frac{\check{r}}{2} \Omega \tr \chi
	+
	\frac{\check{r}^2}{2 \Omega^2} \left\vert \Omega \hat{\chi} \right\vert^2
	.
\end{equation}
In particular,
\begin{equation} \label{eq:rtildeoverr4}
	\partial_v( r^{-1} \check{r})
	=
	\frac{\check{r}}{2r} ( \Omega \tr \chi - \Omega \tr \chi_{\circ})
	+
	\frac{\check{r}^2}{2r \Omega^2} \left\vert \Omega \hat{\chi} \right\vert^2.
\end{equation}

At various points in this section it is important to keep track of weaker decaying error terms which, in fact, have a gain in decay, which is better than the expected $r^{-1}$, when acted on by $\Omega \nablaslash_4$.  To do so, it is convenient to use error notation of Section \ref{subsec:overset4errors} and the following lemma.

\begin{lemma}[$\nablaslash_4$ derivative of error terms] \label{lem:nabla4error4}
	Each error term of the form $\overset{(4)}{\mathcal{E}^k_p}$ satisfies
	\begin{align} \label{4erroreq}
		\Omega \nablaslash_4 ( r^p \overset{(4)}{\mathcal{E}^k_p}) = \check{\mathcal{E}}^{k+1}_{2}.
	\end{align}
\end{lemma}

\begin{proof}
	The proof is an easy consequence of the property \eqref{eq:Phi4improvement} of the $\overset{(4)}{\Phi}$ geometric quantities, equation \eqref{eq:rtildeoverr4} and equation \eqref{eq:DlogOmega}.
\end{proof}

The following proposition in particular uses the above expression for $\partial_v \check{r}$.

\begin{proposition}[Elliptic relations for $\check{\psibar}$, $\check{\Pbar}$] \label{prop:Pbartildeidentities}
	With $M$, $\mathcal{Z}$, $g$ and $g_{\circ,M}$  as in Section~\ref{schemnotsec},  the quantities $\check{\psibar}$ and $\check{\Pbar}$ satisfy
	\begin{align*}
		\frac{r}{\check{r}} \check{\psibar}
		=
		\Dslash_2^* \betabar
		+
		\frac{3M}{r^3} \hat{\chibar}
		+
		\check{\mathcal{E}}^0_4
		,
		\qquad
		\frac{r}{\check{r}}  \check{\Pbar}
		=
		\Dslash_2^* \left( \nablaslash \rho - {}^* \nablaslash \sigma \right)
		+
		\frac{3M \Omega}{r^4} \left( \hat{\chibar} - \hat{\chi} \right)
		+
		\check{\mathcal{E}}^1_5.
	\end{align*}
\end{proposition}

\begin{proof}
	The proof is similar to that of Proposition \ref{prop:PPbaridentities}, now taking into account \eqref{eq:rtilde4}, which in particular implies that
	\[
		\check{\psibar}
		=
		\frac{\check{r}}{r} \left(
		\Dslash_2^* \betabar
		+
		\frac{3M}{r^3} \hat{\chibar}
		\right)
		+ \frac{1}{4}
		\frac{\check{r}^2}{\Omega^3 r^2} \vert \Omega \hat{\chi} \vert^2 \cdot r \alphabar +\frac{\check{r}}{r} \left(-\frac{3}{2}(\rho-\rho_\circ) \underline{\hat{\chi}} + \frac{3}{2} \sigma {}^* \underline{\hat{\chi}}  -\frac{1}{4}  \left(9 \underline{\eta} - \eta \right) \widehat{\otimes} \underline{\beta} \right) \, .
	\]
	The equation for $\check{\psibar}$ then follows. Using  the definition of $\check{\Pbar}$ from Definition \ref{defofalmostgaugeinv} we deduce
\begin{align} \label{eq:Pbarcheckrefinederrors}
\check{\Pbar} = \frac{\check{r}}{r} \left( \Dslash_2^* \left( \nablaslash \rho - {}^* \nablaslash \sigma \right) 
		+
		\frac{3M \Omega}{r^4} \left( \hat{\chibar} - \hat{\chi} \right)  \right) + \frac{\check{r}}{r} \left( \overset{(4)}{\mathcal{E}^1_5} + k_1 \slashed{\nabla} \hat{\omega} \widehat{\otimes} \underline{\beta} + k_2 ( \Omega \tr \chi - \Omega \tr \chi_{\circ})  \Dslash_2^* \underline{\beta} + \frac{k_3}{r} \underline{\eta} \widehat{\otimes} \underline{\beta} \right) + \check{\mathcal{E}}^1_6,
\end{align}
for some constants $k_1,k_2,k_3$, which implies the result. The refined statement \eqref{eq:Pbarcheckrefinederrors} will be used in Proposition \ref{prop:nabla4pschematic} below.
\end{proof}

Note that, in the proof of Theorem \ref{thm:main}, $\check{\psibar}$ and $\check{\Pbar}$ will only be considered in the $\I$ gauge, where $\check{r}r^{-1}, \Omega^{-1} \sim 1$, and so the presence of the error terms $\check{\mathcal{E}}$, rather than the regular error terms $\mathcal{E}$, in Proposition \ref{prop:Pbartildeidentities} is irrelevant.

The following proposition similarly gives expressions for $\nablaslash_4 ( r^5 P )$, $\nablaslash_4 ( r^5 \check{\Pbar} )$ in terms of certain elliptic operators applied to certain Ricci coefficients and curvature components.  It is important to isolate the anomalous nonlinear error terms involving $\nablaslash(\Omega \omegahat) \hat{\otimes} \betabar$, $\nablaslash_4 r \nablaslash(\Omega \omegahat) \hat{\otimes} \betabar$ and $r^4 (\Omega\hat{\omega} -\Omega\hat{\omega}_{\circ}) \Dslash_2^* \betabar$ in (\ref{eq:nabla4pbartildeschematic}).

\begin{proposition}[Elliptic relations for $\nablaslash_4 ( r^5 P )$, $\nablaslash_4 ( r^5 \check{\Pbar} )$] \label{prop:nabla4pschematic}
	With $M$, $\mathcal{Z}$, $g$ and $g_{\circ,M}$  as in Section~\ref{schemnotsec}, the quantities $P$ and $\check{\Pbar}$ satisfy
	\begin{align}
		\nablaslash_4 \left( r^5 P \right)
		=
		\
		&
		r^5 \Dslash_2^* \nablaslash \divslash \beta
		-
		r^5 \Dslash_2^* \nablaslash \curlslash \beta
		-
		6M \Omega r \Dslash_2^* \etabar
		+
		6M \left( 1 - \frac{3M}{r} \right) \hat{\chi}
		+
		\frac{3M}{\Omega} r^2 \Dslash_2^* \nablaslash \Omega \tr \chi
		+
		3M r \Omega \alpha
		\nonumber
		\\
		&
		+
		\Omega^{-1} \mathcal{E}^2_2,
		\label{eq:nabla4pschematic}
		\\
		 \frac{r}{\check{r}} \nablaslash_4 \left(  r^5 \check{\Pbar} \right)
		=
		\
		&
		r^5 \Dslash_2^* \nablaslash \divslash \beta
		+
		r^5 \Dslash_2^* \nablaslash \curlslash \beta
		-
		6M \Omega r \Dslash_2^* \etabar
		+
		6M \left( 1 - \frac{3M}{r} \right) \hat{\chi}
		+
		\frac{3M}{\Omega} r^2 \Dslash_2^* \nablaslash \Omega \tr \chi
		+
		3M r \Omega \alpha
		\nonumber
		\\
		&
		+
		\check{\mathcal{E}}^2_2
		\label{eq:nabla4pbartildeschematic}
		+
		a_1 r^4 \nablaslash(\Omega \omegahat) \hat{\otimes} \betabar
		+
		a_2 r^4 \nablaslash_4 r \nablaslash(\Omega \omegahat) \hat{\otimes} \betabar
		+
		a_3 r^4 (\Omega\hat{\omega} -\Omega\hat{\omega}_{\circ}) \Dslash_2^* \betabar
		,
	\end{align}
	for some admissible coefficient functions $a_i = a_i(r)$ satisfying \eqref{eq:admis}.  Moreover \eqref{eq:nabla4pbartildeschematic} is also satisfied with $\nablaslash_4 \left(  \frac{r}{\check{r}}  r^5 \check{\Pbar} \right)$ in place of $\frac{r}{\check{r}} \nablaslash_4 \left(  r^5 \check{\Pbar} \right)$ (with a different nonlinear error term $\check{\mathcal{E}}^2_2$ with the same schematic form).
\end{proposition}

\begin{proof}
	Consider first \eqref{eq:nabla4pschematic}.  Revisiting Proposition \ref{prop:PPbaridentities}, note that in fact
	\begin{align*}
		P
		=
		\Dslash_2^* \left( \nablaslash \rho + {}^* \nablaslash \sigma \right)
		+
		\frac{3M \Omega}{r^4} \left( \hat{\chibar} - \hat{\chi} \right)
		+
		\frac{3}{2r^3\Omega} \nablaslash_3 \left( r^3 \Omega \left( (\rho - \rho_{\circ}) \hat{\chi} + \sigma {}^*\hat{\chi} \right) \right)
		+
		2 \Dslash_2^* \left( \hat{\chi} \cdot \betabar \right)
		+
		\mathcal{E}^1_6.
	\end{align*}
	Now equations \eqref{eq:rho4} and \eqref{eq:sigma4} imply that
	\[
		\nablaslash_4 \left( r^5 \Dslash_2^* \nablaslash \rho \right)
		=
		r^5 \Dslash_2^* \nablaslash \divslash \beta
		+
		\frac{3M}{\Omega} r^2 \Dslash_2^* \nablaslash \Omega \tr \chi
		+
		\Omega^{-1} \mathcal{E}^2_2,
		\qquad
		\nablaslash_4 \left( r^5 \Dslash_2^* {}^* \nablaslash \sigma \right)
		=
		-
		r^5 \Dslash_2^* {}^* \nablaslash \curlslash \beta
		+
		\Omega^{-1} \mathcal{E}^2_2.
	\]
	Equations \eqref{eq:chibarhat4} and \eqref{eq:chihat4} imply that
	\[
		\nablaslash_4 \left( r \Omega \hat{\chibar} \right)
		=
		- 2 r \Omega \Dslash_2^* \etabar
		+
		\Omega^2_{\circ} \hat{\chi}
		+
		\Omega \mathcal{E}^0_2,
		\qquad
		\nablaslash_4 \left( r \Omega \hat{\chi} \right)
		=
		-
		\Omega^2_{\circ} \hat{\chi}
		+
		\frac{2M}{r} \hat{\chi}
		-
		r\Omega \alpha
		+
		\Omega^{-1} \mathcal{E}^0_3.
	\]
	Similarly, using also Lemma \ref{lem:commutation},
	\[
		\Omega \nablaslash_4 \left( \frac{r^2}{\Omega} \nablaslash_3 \left( r^3 \Omega (\rho - \rho_{\circ}) \hat{\chi} \right) \right)
		=
		\mathcal{E}^2_2,
		\qquad
		\Omega \nablaslash_4 \left( \frac{r^2}{\Omega} \nablaslash_3 \left( r^3 \Omega \sigma {}^* \hat{\chi} \right) \right)
		=
		\mathcal{E}^2_2,
		\qquad
		\Omega \nablaslash_4 \left( r^5 \Dslash_2^* \left( \hat{\chi} \cdot \betabar \right) \right)
		=
		\mathcal{E}^2_2.
	\]
	The expression \eqref{eq:nabla4pschematic} then follows from the fact that $\Omega \nablaslash_4(r^5 \mathcal{E}^1_6) = \mathcal{E}^2_2$.
	
	The expression \eqref{eq:nabla4pbartildeschematic} is obtained similarly, using now equation \eqref{eq:Pbarcheckrefinederrors}, equation \eqref{eq:rtildeoverr4}, the form \eqref{eq:Raychaudhuriinschematicformsection3} of the Raychaudhuri equation and Lemma~\ref{lem:nabla4error4}.  The statement concerning $\nablaslash_4 \left(  \frac{r}{\check{r}}  r^5 \check{\Pbar} \right)$ follows easily from \eqref{eq:nabla4pbartildeschematic} and equation (\ref{eq:rtildeoverr4}).
\end{proof}

\subsection{Teukolsky equations for $\alpha$ and \underline{$\alpha$}}
In this section, the nonlinear wave equations satisfied by $\alpha$ and \underline{$\alpha$} are derived.  To linear order, both satisfy the \emph{Teukolsky} equation.  See the discussion in Section \ref{proofoflinstabinintro}.  
We will  refer to the nonlinear analogues below  also as Teukolsky equations.

Recall the  $\hat{\otimes}$ product for $S$-tangent 1-forms $\xi$ and $\xi'$, 
\[
	\xi \hat{\otimes} \xi'
	=
	\xi \otimes \xi'
	+
	\xi' \otimes \xi
	-
	(\xi \cdot \xi ') \gslash,
\]
defined previously in Section~\ref{nullframesopers}.
The following product rule for the operator $\Dslash_2^*$ is easily checked.

\begin{lemma}[Product rule for symmetric traceless gradient] \label{lem:D2starproduct}
	For any $S$-tangent $1$-form $\xi$ and any function $f$,
	\[
		\Dslash_2^* (f \xi)
		=
		f \Dslash_2^* \xi
		-
		2 \nablaslash f \hat{\otimes} \xi.
	\]
\end{lemma}

The fact that
\begin{equation} \label{eq:nabla4Omegar}
	\Omega \nablaslash_4 \left( \frac{\Omega^2}{r^2} \right)
	=
	- \frac{2 \Omega^2}{r^2}
	\left(
	\frac{1}{r} \left( 1 - \frac{3M}{r} \right)
	-
	\left( \Omega \omegahat - (\Omega \omegahat)_{\circ} \right)
	\right)
\end{equation}
and
\begin{equation} \label{eq:nabla3Omegar}
	\Omega \nablaslash_3 \left( \frac{\Omega^2}{r^2} \right)
	=
	\frac{2\Omega^2}{r^2}
	\left(
	\frac{1}{r} \left( 1 - \frac{3M}{r} \right)
	+
	\left( \Omega \omegabarhat - (\Omega \omegabarhat)_{\circ} \right)
	\right)
\end{equation}
will be used often in what follows.

\begin{proposition}[Wave equation for $\alpha$] \label{prop:Teukolsky}
	With $M$, $\mathcal{Z}$, $g$ and $g_{\circ,M}$  as in Section~\ref{schemnotsec}, $\alpha$ satisfies the \emph{Teukolsky} equation
	\begin{equation} \label{eq:Teukolsky2}
		\Omega \nablaslash_4 \Omega \nablaslash_3 (r \Omega^2 \alpha)
		+
		\frac{2\Omega^2}{r^2} r^2 \Dslash_2^* \divslash (r\Omega^2 \alpha)
		=
		-
		\frac{4}{r} \left( 1 - \frac{3M}{r} \right) \Omega \nablaslash_3 (r \Omega^2 \alpha)
		-
		\frac{6M\Omega^2}{r^3} r\Omega^2\alpha
		+
		\Omega^2 \mathcal{E}^1_6.
	\end{equation}
\end{proposition}

\begin{proof}
	The Bianchi equation \eqref{eq:alpha3} can be rewritten
	\[
		\Omega \nablaslash_3(r \Omega^2 \alpha)
		=
		-2 \frac{\Omega^4}{r^4} r \Dslash_2^* (r^4 \Omega^{-1} \beta)
		+
		\frac{6M \Omega^2}{r^2} \Omega \hat{\chi} + E_1,
	\]
	where
	\[
		E_1
		=
		\Omega^2 r
		\left(
		-3 (\rho - \rho_{\circ}) \Omega \hat{\chi}
		-
		3\sigma \Omega {}^* \hat{\chi}
		+
		2(\eta + \etabar) \hat{\otimes} (\Omega \beta)
		+
		\frac{1}{2} (9\eta - \etabar) \hat{\otimes} (\Omega \beta)
		-
		\frac{1}{2\Omega^2} (\Omega\tr \chibar - (\Omega \tr \chibar)_{\circ}) \Omega^2 \alpha
		\right),
	\]
	is a nonlinear error term.  Then,
	\begin{align*}
		\Omega \nablaslash_4 \left( \frac{r^4}{\Omega^4} \Omega \nablaslash_3(r \Omega^2 \alpha) \right)
		=
		&
		-2 r \Dslash_2^* \Omega \nablaslash_4(r^4 \Omega^{-1} \beta)
		-
		6M r^2 \alpha
		+
		\Omega \nablaslash_4 \left( \frac{r^4}{\Omega^4} E_1 \right)
		\\
		&
		+
		6M r^2 \Omega^{-2} \left( (\Omega \tr \chi)_{\circ} - \Omega \tr \chi \right) \Omega \hat{\chi}
		-
		2 [\Omega \nablaslash_4,r \Dslash_2^*] (r^4 \Omega^{-1} \beta),
	\end{align*}
	by equation \eqref{eq:chihat4}.  The Bianchi equation \eqref{eq:beta4} can be rewritten
	\[
		\Omega \nablaslash_4 (r^4 \Omega^{-1} \beta)
		=
		r^4 \divslash \alpha
		+
		E_2,
	\]
	where
	\[
		E_2
		=
		\frac{r^4}{\Omega^2}
		\left(
		\eta \cdot \Omega^2 \alpha
		-
		2 (\Omega \tr \chi - (\Omega \tr \chi)_{\circ}) \Omega \beta
		\right).
	\]
	Hence,
	\begin{align*}
		\Omega \nablaslash_4 \left( \frac{r^4}{\Omega^4} \Omega \nablaslash_3(r \Omega^2 \alpha) \right)
		=
		&
		-2 r^5 \Dslash_2^* \divslash \alpha
		-
		6M r^2 \alpha
		-
		2r \Dslash_2^* E_2
		+
		\Omega \nablaslash_4 \left( \frac{r^4}{\Omega^4} E_1 \right)
		\\
		&
		+
		6M r^2 \Omega^{-2} \left( (\Omega \tr \chi)_{\circ} - \Omega \tr \chi \right) \Omega \hat{\chi}
		-
		2 [\Omega \nablaslash_4,r \Dslash_2^*] (r^4 \Omega^{-1} \beta),
	\end{align*}
	which can be rearranged to give
	\begin{align*}
		&
		\Omega \nablaslash_4\Omega \nablaslash_3(r \Omega^2 \alpha)
		+
		2 \frac{\Omega^4}{r} r^2 \Dslash_2^* \divslash \alpha
		=
		-
		\frac{\Omega^4}{r^4} \Omega \nablaslash_4 \left( \frac{r^4}{\Omega^4} \right)
		\Omega \nablaslash_3 (r \Omega^2 \alpha)
		-
		6M \frac{\Omega^4}{r^2} \alpha
		\\
		&
		\qquad \qquad
		+
		\frac{\Omega^4}{r^4}
		\left[
		-
		2r \Dslash_2^* E_2
		+
		\Omega \nablaslash_4 \left( \frac{r^4}{\Omega^4} E_1 \right)
		+
		6M r^2 \Omega^{-2} \left( (\Omega \tr \chi)_{\circ} - \Omega \tr \chi \right) \Omega \hat{\chi}
		-
		2 [\Omega \nablaslash_4,r \Dslash_2^*] (r^4 \Omega^{-1} \beta)
		\right].
	\end{align*}
	The proof follows after noting that, by \eqref{eq:nabla4Omegar},
	\[
		\Omega\nablaslash_4 \left( \frac{r^4}{\Omega^4} \right)
		=
		-2 \frac{r^6}{\Omega^6}
		\Omega\nablaslash_4 \left( \frac{\Omega^2}{r^2} \right)
		=
		-2\frac{r^4}{\Omega^4}
		\left(
		- \frac{2}{r} \left( 1 - \frac{3M}{r} \right)
		+
		2\left( \Omega \omegahat - (\Omega \omegahat)_{\circ} \right)
		\right)
	\]
	and, by Lemma \ref{lem:D2starproduct},
	\begin{align*}
		&
		2 \frac{\Omega^4}{r} r^2 \Dslash_2^* \divslash \alpha
		=
		\frac{2\Omega^2}{r^2} r^2 \Dslash_2^* \divslash (r\Omega^2 \alpha)
		+
		\frac{2\Omega^2}{r^2}
		\left(
		2r^3(\eta + \etabar) \hat{\otimes} \divslash \alpha
		-
		r^3 \Dslash_2^* \left( (\eta + \etabar) \cdot \alpha \right)
		\right)
		,
	\end{align*}
	and checking by inspection that all of the nonlinear error terms have the correct form, using Lemma \ref{lem:commutation} and the schematic expression \eqref{eq:schemcomm4}.  The nontrivial error terms to check are $4 (\Omega \omegahat - (\Omega \omegahat)_{\circ}) \Omega \nablaslash_3(r\Omega^2 \alpha)$, for which the Bianchi equation \eqref{eq:alpha3} is used, and the term $-3\Omega \nablaslash_4(r^5 \Omega^{-2} (\rho - \rho_{\circ}) \Omega \hat{\chi})$ in $\Omega \nablaslash_4 \left( r^4 \Omega^{-4} E_1 \right)$, which is rewritten as
	\[
		-3\Omega \nablaslash_4(r^5 \Omega^{-2} (\rho - \rho_{\circ}) \Omega \hat{\chi})
		=
		-3
		\Omega \nablaslash_4(r^3(\rho - \rho_{\circ})) \cdot r^2 \Omega^{-1} \hat{\chi}
		-
		3 r^3 (\rho-\rho_{\circ}) \Omega \nablaslash_4(r^2 \Omega^{-1} \hat{\chi}),
	\]
	and evaluated using equations \eqref{eq:rho4} and \eqref{eq:chihat4}.  Similarly for the term $-3\Omega \nablaslash_4(r^5 \Omega^{-2} \sigma \Omega {}^* \hat{\chi})$ in the error $\Omega \nablaslash_4 \left( r^4 \Omega^{-4} E_1 \right)$.
\end{proof}

In the proof of the main theorem, the following wave equation for $\alphabar$ will be used only for the  the $\Hp$ gauge and therefore the $r$ behaviour of the error terms are not considered.  For the $\I$ gauge, the 
equation~\eqref{eq:Teukolskybartilde} for $\check{r} \Omega^2 \alphabar$ will be used.  The reason for considering the wave equation for $\check{r} \Omega^2 \alphabar$ in 
the $\I$ gauge is the absence of an error term with bad $r$ behaviour of the form
\[
	\left( \Omega \tr \chi - (\Omega \tr \chi)_{\circ} \right) \Omega \nablaslash_3(r\Omega^{-2} \alphabar)
\]
in equation \eqref{eq:Teukolskybartilde}, which is present in the following equation for $r \Omega^2 \alphabar$.

\begin{proposition}[Wave equation for $\alphabar$]
	With $M$, $\mathcal{Z}$, $g$ and $g_{\circ,M}$  as in Section~\ref{schemnotsec}, $\alphabar$ satisfies the \emph{Teukolsky} equation
	\begin{equation} \label{eq:Teukolskybar}
		\Omega \nablaslash_3 \Omega \nablaslash_4 (r \Omega^2 \alphabar)
		+
		\frac{2\Omega^2}{r^2} r^2\Dslash_2^* \divslash (r\Omega^2 \alphabar)
		=
		\frac{4}{r} \left( 1 - \frac{3M}{r} \right) \Omega \nablaslash_4 (r \Omega^2 \alphabar)
		-
		\frac{6M\Omega^2}{r^3} r\Omega^2\alphabar
		+
		\Omega^6
		\mathcal{E}^{*1}.
	\end{equation}
\end{proposition}

\begin{proof}
	The proof is similar to that of Proposition \ref{prop:Teukolsky}, using now the Bianchi equations \eqref{eq:betabar3}, \eqref{eq:alphabar4}, but is much simpler since the precise $r$ behaviour of the error terms is not recorded.  One sees easily by inspection that the terms $r \nablaslash \Omega \hat{\chi}$ and $r \nablaslash \Omega \tr \chi$ do not appear.
\end{proof}

\subsection{Wave equations for $\psi$ and \underline{$\psi$}}
In this section, the wave equations satisfied by
\[
	\psi
	:=
	- \frac{1}{2r \Omega^2} \nablaslash_3(r\Omega^2 \alpha)
	\quad
	\text{and}
	\quad
	\psibar:= \frac{1}{2r \Omega^2} \nablaslash_4(r\Omega^2 \alphabar)
\]
are derived.  The fact that, for any $p$, $k$ and $\gamma \in \{ (1,0,0), (0,1,0), (0,0,1) \}$,
\begin{equation} \label{eq:derivativeofschematic}
	\mathfrak{D}^{\gamma} \mathcal{E}^k_p = \mathcal{E}^{k+1}_p,
\end{equation}
will be used.

\begin{proposition}[Wave equation for $\psi$] \label{prop:psiwave}
	With $M$, $\mathcal{Z}$, $g$ and $g_{\circ,M}$  as in Section~\ref{schemnotsec},  $\psi$ satisfies the wave equation
	\begin{multline} \label{eq:psiwave}
		\Omega \nablaslash_4 \Omega \nablaslash_3 (r^3 \Omega \psi)
		+
		\frac{2\Omega^2}{r^2} r^2 \Dslash_2^* \divslash (r^3 \Omega \psi)
		\\
		=
		-
		\frac{2}{r} \left( 1 - \frac{3M}{r} \right) \Omega \nablaslash_3 (r^3 \Omega \psi)
		-
		\frac{2\Omega^2}{r^2} \left( 1 - \frac{3M}{r} \right) r^3\Omega \psi
		+
		\frac{3M\Omega^2}{r^2} r \Omega^2 \alpha
		+
		\Omega^2 \mathcal{E}^2_4.
	\end{multline}
\end{proposition}

\begin{proof}
	The Teukolsky equation \eqref{eq:Teukolsky2} can be rewritten
	\[
		-2 \Omega \nablaslash_4 \left( \frac{\Omega^2}{r^2} r^3 \Omega \psi \right)
		+
		\frac{2\Omega^2}{r^2} r^2 \Dslash_2^* \divslash \left( r\Omega \alpha \right)
		=
		\frac{8}{r} \left( 1 - \frac{3M}{r} \right) \frac{\Omega^2}{r^2} r^3 \Omega \psi
		-
		6M \frac{\Omega^2}{r^3} r \Omega^2 \alpha
		+
		\Omega^2 \mathcal{E}^1_6.
	\]
	The expression \eqref{eq:nabla4Omegar} then implies
	\[
		\Omega \nablaslash_4 (r^3 \Omega \psi)
		=
		r^2 \Dslash_2^* \divslash (r\Omega^2 \alpha)
		-
		\frac{2}{r} \left( 1 - \frac{3M}{r} \right) r^3 \Omega \psi
		+
		\frac{3M}{r} r \Omega^2 \alpha
		-
		\frac{r^2}{2\Omega^2} \Omega^2 \mathcal{E}^1_6
		-
		2(\Omega \omegahat - (\Omega \omegahat)_{\circ}) r^3 \Omega \psi,
	\]
	and applying $\Omega \nablaslash_3$ then gives
	\begin{align*}
		&
		\Omega \nablaslash_4 \Omega \nablaslash_3 (r^3 \Omega \psi)
		=
		-2 \Dslash_2^* \divslash \left( \frac{\Omega^2}{r^2} r^3 \Omega \psi \right)
		-
		\frac{2}{r} \left( 1 - \frac{3M}{r} \right) \Omega \nablaslash_3 (r^3 \Omega \psi)
		-
		\frac{2\Omega^2}{r^2} \left( 1 - \frac{3M}{r} \right) r^3 \Omega \psi
		\\
		&
		\quad
		+
		\frac{3M\Omega^2}{r^2} r \Omega^2 \alpha
		-
		\Omega \nablaslash_3 \left(
		\frac{r^2}{2\Omega^2} \Omega^2 \mathcal{E}^1_6
		-
		2(\Omega \omegahat - (\Omega \omegahat)_{\circ}) r^3 \Omega \psi
		\right)
		-
		[\Omega \nablaslash_3,\Omega \nablaslash_4] (r^3 \Omega \psi)
		+
		[ \Omega \nablaslash_3, r^2 \Dslash_2^* \divslash] (r \Omega^2 \alpha)
		\\
		&
		\quad
		-
		\frac{2}{r^2} \left( \Omega_{\circ}^2 - \Omega^2 \right) \left( 1 - \frac{6M}{r} \right) r^3 \Omega \psi
		+
		\frac{3M}{r^2} \left( \Omega_{\circ}^2 - \Omega^2 \right) r \Omega^2 \alpha,
	\end{align*}
	since
	\[
		\Omega \nablaslash_3 \left( \frac{3M}{r} \right)
		=
		\frac{3M}{r^2} \Omega_{\circ}^2,
		\quad
		\text{and}
		\quad
		-
		\Omega\nablaslash_3 \left( \frac{2}{r} \left( 1 - \frac{3M}{r} \right) \right)
		=
		-
		\frac{2\Omega_{\circ}^2}{r^2} \left( 1 - \frac{6M}{r} \right).
	\]
	The proof then follows from checking by inspection that the nonlinear error terms have the correct form.  The expression \eqref{eq:psipsibar} is used, along with \eqref{eq:schemcomm3}, \eqref{eq:schemcomm34} and \eqref{eq:derivativeofschematic}.
\end{proof}

\begin{proposition}[Wave equation for $\psibar$]
	With $M$, $\mathcal{Z}$, $g$ and $g_{\circ,M}$  as in Section~\ref{schemnotsec},  $\psibar$ satisfies the wave equation,
	\begin{multline} \label{eq:psibarwave}
		\Omega \nablaslash_3 \Omega \nablaslash_4 (r^3 \Omega \psibar)
		+
		\frac{2\Omega^2}{r^2} r^2\Dslash_2^* \divslash (r^3 \Omega \psibar)
		\\
		=
		\frac{2}{r} \left( 1 - \frac{3M}{r} \right) \Omega \nablaslash_4 (r^3 \Omega \psibar)
		-
		\frac{2\Omega^2}{r^2} \left( 1 - \frac{3M}{r} \right) r^3 \Omega \psibar
		+
		\frac{3M\Omega^2}{r^2} r \Omega^2 \alphabar
		+
		\Omega^4 \mathcal{E}^{*2}.
	\end{multline}
\end{proposition}

\begin{proof}
	The proof is similar to that of Proposition \ref{prop:psiwave}, using now the Teukolsky equation \eqref{eq:Teukolskybar} and \eqref{eq:nabla3Omegar} instead of \eqref{eq:nabla4Omegar}.  One again easily sees, by inspecting the principal terms, that the terms $(r \nablaslash)^2 \Omega \hat{\chi}$ and $(r \nablaslash)^2 \Omega \tr \chi$ do not appear.
\end{proof}

\subsection{Regge--Wheeler equations for $P$ and \underline{$P$}}
In this section, the wave equations satisfied by
\[
	P := \frac{1}{r^3 \Omega} \nablaslash_3 (r^3 \Omega \psi),
	\quad
	\text{and}
	\quad
	\Pbar := - \frac{1}{r^3 \Omega} \nablaslash_4 (r^3 \Omega \psibar),
\]
are derived.  To linear order, both satisfy the \emph{Regge--Wheeler} equation.  See Section \ref{proofoflinstabinintro}.  We will refer to the  nonlinear analogues below also  as Regge--Wheeler equations.

\begin{proposition}[Wave equation for $P$] \label{prop:Pwave}
	With $M$, $\mathcal{Z}$, $g$ and $g_{\circ,M}$  as in Section~\ref{schemnotsec},  $P$ satisfies the Regge--Wheeler equation,
	\begin{equation*}
		\Omega \nablaslash_4 \Omega \nablaslash_3 (r^5 P)
		+
		\frac{2\Omega^2}{r^2} r^2 \Dslash_2^* \divslash (r^5 P)
		=
		-
		\frac{2\Omega^2}{r^2} \left( 1 - \frac{3M}{r} \right) r^5 P
		+
		\Omega^2 \mathcal{E}^{*3}_2.
	\end{equation*}
\end{proposition}

\begin{proof}
	Equation \eqref{eq:psiwave} can be rewritten as
	\begin{multline*}
		\Omega \nablaslash_4 (\Omega^2 r^3P)
		+
		\frac{2\Omega^2}{r^2} r^2 \Dslash_2^* \divslash (r^3 \Omega \psi)
		\\
		=
		-
		\frac{2}{r} \frac{\Omega^2}{r^2} \left( 1 - \frac{3M}{r} \right) r^5 P
		-
		\frac{2\Omega^2}{r^2} \left( 1 - \frac{3M}{r} \right) r^3\Omega \psi
		+
		\frac{3M\Omega^2}{r^2} r \Omega^2 \alpha
		+
		\Omega^2 \mathcal{E}^2_4.
	\end{multline*}
	Using \eqref{eq:nabla4Omegar}, as in the proof of Proposition \ref{prop:psiwave}, this gives
	\[
		\Omega \nablaslash_4(r^5P)
		+
		2r^2 \Dslash_2^* \divslash (r^3 \Omega \psi)
		=
		-
		2 \left( 1 - \frac{3M}{r} \right) r^3\Omega \psi
		+
		3M r \Omega^2 \alpha
		+
		\frac{r^2}{\Omega^2} \Omega^2 \mathcal{E}^2_4
		-
		2(\Omega \omegahat - (\Omega \omegahat)_{\circ}) r^5P.
	\]
	Note the cancellation which occurs this time.  Applying $\Omega \nablaslash_3$ gives
	\begin{multline*}
		\Omega \nablaslash_4 \Omega \nablaslash_3 (r^5 P)
		+
		\frac{2\Omega^2}{r^2} r^2 \Dslash_2^* \divslash (r^5 P)
		=
		-
		2\Omega^2 \left( 1 - \frac{3M}{r} \right) r^5 P
		\\
		+
		\frac{6M}{r^2} (\Omega_{\circ}^2 - \Omega^2) r^3 \Omega \psi
		-
		2[\Omega \nablaslash_3, r^2 \Dslash_2^* \divslash] (r^3 \Omega \psi)
		+
		\Omega \nablaslash_3 \left(
		\frac{r^2}{\Omega^2} \Omega^2 \mathcal{E}^2_4
		-
		2\left( \Omega \omegahat - (\Omega \omegahat)_{\circ} \right) r^5P
		\right)
	\end{multline*}
	since
	\[
		- \Omega \nablaslash_3 \left( 2 - \frac{6M}{r} \right) = \frac{6M \Omega_{\circ}^2}{r^2}.
	\]
	Again, note the cancellation.  The proof again follows by checking that the nonlinear error terms have the correct form.  The fact that the nonlinear error is of the form $\Omega^2 \mathcal{E}^{*3}_2$, rather than merely of the form $\Omega^2 \mathcal{E}^{3}_2$, is easily seen from the fact that the principal terms all arise from applying $\nablaslash_3$ and $\nablaslash_4$ derivatives to the nonlinear terms in the Bianchi equations \eqref{eq:alpha3} and \eqref{eq:beta4}.
\end{proof}

\begin{proposition}[Wave equation for $\Pbar$]
	With $M$, $\mathcal{Z}$, $g$ and $g_{\circ,M}$  as in Section~\ref{schemnotsec}, $\Pbar$ satisfies the Regge--Wheeler equation,
	\begin{equation*}
		\Omega \nablaslash_4 \Omega \nablaslash_3 (r^5 \Pbar)
		+
		\frac{2\Omega^2}{r^2} r^2 \Dslash_2^* \divslash (r^5 \Pbar)
		=
		-
		\frac{2\Omega^2}{r^2} \left( 1 - \frac{3M}{r} \right) r^5 \Pbar
		+
		\Omega^2 \mathcal{E}^{*3}.
	\end{equation*}
\end{proposition}

\begin{proof}
	The proof is similar to that of Proposition \ref{prop:Pwave}, using now \eqref{eq:nabla3Omegar} and equation \eqref{eq:psibarwave}.
\end{proof}

\subsection{Wave equations for $\check{r} \Omega^2$\underline{$\alpha$}, \underline{$\check{\psi}$}, \underline{$\check{P}$}}
\label{subsec:tildewaveequations}

Recall $\check{r}$, $\check{\psibar}$ and $\check{\Pbar}$ from Definition \ref{defofalmostgaugeinv}.  In this section, 
the wave equations satisfied by $\check{r} \alphabar$, $\check{\psibar}$ and $\check{\Pbar}$ are derived.  Since $\check{\psibar}$ and $\check{\Pbar}$ are only considered, in the proof of Theorem \ref{thm:main}, for the $\I$ gauge, where it will be shown that
$\Omega \sim 1$, it is not necessary to keep track of the $\Omega$ behaviour of error terms in the equations 
for $\check{\psibar}$ and $\check{\Pbar}$, and so only the error notation of Section \ref{subsec:tildeerrors} will be used.

In deriving the wave equation satisfied by $\check{r} \alphabar$, it is important to keep track of weaker decaying error terms which, in fact, have a gain in decay, which is better than the expected $r^{-1}$, when acted on by $\Omega \nablaslash_4$.  To do so, it is convenient to use error notation of Section \ref{subsec:overset4errors}.

In the following proposition it is convenient to isolate the slowest decaying nonlinear error terms.  Moreover, the term $r \Dslash_2^* \left(r \nablaslash \Omega \tr \chi \cdot \alphabar \right)$ involves too many derivatives to be incorporated into the error $\check{\mathcal{E}}^1_3$ and it is important to isolate it as it is, in an appropriate sense, better than the typical error of the form $\check{\mathcal{E}}^2_3$.

\begin{proposition}[Wave equation for $\check{r} \Omega^2 \alphabar$] \label{prop:Teukolskybartilde}
	With $M$, $\mathcal{Z}$, $g$ and $g_{\circ,M}$  as in Section~\ref{schemnotsec}, $\alphabar$ satisfies the Teukolsky equation
	\begin{equation} \label{eq:Teukolskybartilde}
		\Omega \nablaslash_3 \Omega \nablaslash_4 (\check{r} \Omega^2 \alphabar)
		+
		\frac{2\Omega^2}{r^2} r^2\Dslash_2^* \divslash ( \check{r} \Omega^2 \alphabar)
		=
		\frac{4}{r} \left( 1 - \frac{3M}{r} \right) \Omega \nablaslash_4 ( \check{r} \Omega^2 \alphabar)
		-
		\frac{6M\Omega^2}{r^3} \check{r} \Omega^2\alphabar
		+
		\check{\mathcal{E}}[\alphabar],
	\end{equation}
	where the nonlinear error $\check{\mathcal{E}}[\alphabar]$ has the form
	\begin{align*}
		\check{\mathcal{E}}[\alphabar]
		=
		\
		&
		\check{\mathcal{E}}^1_4
		+
		\overset{(4)}{\mathcal{E}^1_3} +
			k_1 r \Dslash_2^* \left(
		r \nablaslash \Omega \tr \chi
		\cdot \alphabar
		\right)
		+
		k_2 r \eta \hat{\otimes} \divslash \alphabar
		+
		k_3 r \etabar \hat{\otimes} \divslash \alphabar
		+
		k_4 r \Dslash_2^* (\eta \cdot \alphabar)
		+
		k_5 r \Dslash_2^* (\etabar \cdot \alphabar)
		\\
		&
		+
		k_6 r^2 \nablaslash \Omega \tr \chi \hat{\otimes} \divslash \alphabar
		+
		k_7 (\Omega \omegabarhat - (\Omega \omegabarhat)_{\circ}) r \Dslash_2^*(\Omega^{-1} \betabar)
		+
		k_8 r \nablaslash(\Omega \omegabarhat) \hat{\otimes} (\Omega^{-1} \betabar)
		\\
		=
		\
		&
		\check{\mathcal{E}}^1_3
		+
		k_1 r \Dslash_2^* \left(
		r \nablaslash \Omega \tr \chi
		\cdot \alphabar
		\right),
	\end{align*}
	for functions $k_i = k_i(r^{-1} \check{r})$, $i=1,\ldots,8$, which each either take the form $k_i=\tilde{k}_i  \frac{\check{r}}{r}$ or $k_i=\tilde{k}_i  \frac{\check{r}^2}{r^2}$, for some constants $\tilde{k}_1,\ldots,\tilde{k}_{8}$.
\end{proposition}

\begin{proof}
	The Bianchi equation \eqref{eq:alphabar4} can be rewritten as
	\[
		\frac{r^4}{\Omega^4} \Omega \nablaslash_4 (\check{r} \Omega^2 \alphabar)
		=
		2 \frac{\check{r}}{r} \cdot r \Dslash_2^* \left( \frac{r^4}{\Omega} \betabar \right)
		+
		6M \frac{\check{r}}{r} \Omega^{-1} r^2\hat{\chibar}
		+
		 \underline{E}_1,
	\]
	where
	\[
		\underline{E}_1
		= \frac{\check{r}}{r} \left( 
		-
		3
		 r \Omega^{-1} r \hat{\chibar} r^3 (\rho - \rho_{\circ})
		-
		3\ r \Omega^{-1} r {}^* \hat{\chibar} r^3 \sigma
		-
		\frac{1}{2} r (13 \etabar + 3 \eta) \hat{\otimes} r^4\betabar \Omega^{-1} \right)
		+
		2 r^4 \Omega^{-4} \vert \Omega \hat{\chi} \vert^2 \check{r}^2 \alphabar.
	\]
	Rewriting the Bianchi equation \eqref{eq:betabar3}
	\[
		\Omega \nablaslash_3 \left( \frac{r^4}{\Omega} \betabar \right)
		=
		-
		r^4 \divslash \alphabar
		+
		\underline{E}_2,
		\quad
		\underline{E}_2
		=
		- \Omega^2 r^4
		\left(
		\etabar \cdot \Omega^{-2} \alphabar
		+
		\Omega^{-2} (\Omega \tr \chibar - (\Omega \tr \chibar)_{\circ}) \Omega^{-1} \betabar
		\right),
	\]
	and the equations \eqref{eq:chihat4}, \eqref{eq:trchi3} as
	\begin{equation} \label{eq:chibarhat3rtilde}
		\Omega \nablaslash_3 \left( \frac{r\check{r}}{\Omega} \hat{\chibar} \right)
		=
		-
		r \check{r} \alphabar
		+
		\underline{E}_3,
	\end{equation}
	with
	\[
		\underline{E}_3
		=
		\frac{r}{\Omega} \hat{\chibar}\left[
		\frac{\check{r}}{2} \left( (\Omega \tr \chibar)_{\circ} - \Omega \tr \chibar \right)
		+
		2\check{r} \left( \Omega \omegabarhat - (\Omega \omegabarhat)_{\circ} \right)
		+
		\frac{2M (\check{r} - r)}{r^3}
		-
		\frac{\check{r}^2}{2} \left( - (\hat{\chibar},\hat{\chi}) + 2 \vert \eta \vert^2 + 2(\rho - \rho_{\circ}) + 2 \divslash \eta \right)
		\right]
	\]
	it follows that
	\[
		\Omega \nablaslash_3 \left( \frac{r^4}{\Omega^4} \Omega \nablaslash_4 (\check{r} \Omega^2 \alphabar) \right)
		=
		-
		2 r^4 \check{r} \Dslash_2^* \divslash \alphabar
		-
		6M r \check{r} \alphabar
		+
		\Omega \nablaslash_3 \left(  \underline{E}_1 \right)
		+
		2 \check{r} \Dslash_2^* \underline{E}_2
		+
		\underline{E}_3
		+
		2[\Omega \nablaslash_3, \check{r} \Dslash_2^*](r^4 \Omega^{-1} \betabar).
	\]
	After rearranging, using \eqref{eq:nabla3Omegar} and the fact that
	\[
		\check{r} \Omega^2 \Dslash_2^* \divslash \alphabar
		=
		\Dslash_2^* \divslash \left( \check{r} \Omega^2 \alphabar \right)
		+
		2 \nablaslash(\check{r} \Omega^2) \hat{\otimes} \divslash \alphabar
		-
		\Dslash_2^* \left( \nablaslash(\check{r} \Omega^2) \cdot \alphabar \right),
	\]
	this gives
	\begin{align*}
		\Omega \nablaslash_3 \Omega \nablaslash_4 (\check{r} \Omega^2 \alphabar)
		+
		\frac{2\Omega^2}{r^2} r^2 \Dslash_2^* \divslash(\check{r} \Omega^2 \alphabar)
		=
		\frac{4}{r} \left( 1 - \frac{3M}{r} \right) \Omega \nablaslash_4 (\check{r} \Omega^2 \alphabar)
		-
		\frac{6M\Omega^4}{r^3} \check{r} \alphabar
		\\
		-
		4 \Omega^4 \nablaslash(\check{r} \Omega^2) \hat{\otimes} \divslash \alphabar
		+
		2 \Omega^4 \Dslash_2^* \left( \nablaslash(\check{r} \Omega^2) \cdot \alphabar \right)
		+
		4 \left( \Omega \omegabarhat - (\Omega \omegabarhat)_{\circ} \right) \Omega \nablaslash_4(\check{r} \Omega^2 \alphabar)
		\\
		+
		\frac{\Omega^4}{r^4} \Omega \nablaslash_3 \left( \underline{E}_1 \right)
		+
		\frac{2 \Omega^4}{r^4} \check{r} \Dslash_2^* \underline{E}_2
		+
		\frac{\Omega^4}{r^4} \underline{E}_3
		+
		\frac{2\Omega^4}{r^4} [\Omega \nablaslash_3, \check{r} \Dslash_2^*](r^4 \Omega^{-1} \betabar).
	\end{align*}
	The proof then follows by inspecting each of the error terms and using the fact that
	\[
		\nablaslash(\check{r} \Omega^2)
		=
		2 \check{r} \Omega^2 (\eta + \etabar)
		-
		\frac{\check{r}^2}{2} \nablaslash(\Omega \tr \chi - (\Omega \tr \chi)_{\circ}).
	\]
	As in Lemma \ref{lem:commutation}, using equation \eqref{eq:trchi3},
	\[
		\frac{2\Omega^4}{r^4} [\Omega \nablaslash_3, \check{r} \Dslash_2^*] (r^4 \Omega^{-1} \betabar)
		=
		4 \Omega^4 \left( \Omega \omegabarhat - (\Omega \omegabarhat)_{\circ} \right) \check{r} \Dslash_2^* (\Omega^{-1} \betabar)
		+
		\check{\mathcal{E}}^1_4
		+
		 \overset{(4)}{\mathcal{E}^1_3}.
	\]
	The most nontrivial terms to check are those in $\frac{\Omega^4}{r^4} \Omega \nablaslash_3 \left( \underline{E}_1 \right)$.  Equations \eqref{eq:rho3} and \eqref{eq:chibarhat3rtilde} are used for the first term after writing
	\[
		-3 \frac{\Omega^4}{r^4} \Omega \nablaslash_3 \left( r^4 \check{r} (\rho - \rho_{\circ}) \Omega^{-1} \hat{\chibar} \right)
		=
		-3 \frac{\Omega^4}{r^4} \Omega \nablaslash_3 \left( r^3 (\rho - \rho_{\circ}) \right) r \check{r} \Omega^{-1} \hat{\chibar}
		+
		-3 \frac{\Omega^4}{r} (\rho - \rho_{\circ}) \Omega \nablaslash_3 \left( r \check{r} \Omega^{-1} \hat{\chibar} \right)
		=
		\check{\mathcal{E}}^1_4
		+
		 \overset{(4)}{\mathcal{E}^1_3}
		.
	\]
	Similarly
	\[
		-3\frac{\Omega^4}{r^4} \Omega \nablaslash_3 \left(r^4 \check{r} \sigma \Omega^{-1} {}^* \hat{\chibar} \right)
		=
		\check{\mathcal{E}}^1_4
		+
		 \overset{(4)}{\mathcal{E}^1_3}.
	\]
	Equations \eqref{eq:nabla3eta} and \eqref{eq:betabar3} are used to show
	\begin{align*}
		-\frac{3}{2} \frac{\Omega^4}{r^4} \Omega \nablaslash_3 \left( r^4 \check{r} \eta \hat{\otimes} \Omega^{-1} \betabar \right)
		=
		&
		-3 \frac{\check{r}}{r} \frac{\Omega^4}{r} \nablaslash(\Omega \omegabarhat) \hat{\otimes} (r^2 \Omega^{-1} \betabar)
		+
		\frac{3}{2}\frac{\check{r}}{r} \Omega^4 r \eta \hat{\otimes} \divslash \alphabar
		+
		\check{\mathcal{E}}^1_4
		+
		 \overset{(4)}{\mathcal{E}^1_3}
		,
	\end{align*}
	and equations \eqref{eq:nabla4eta} and \eqref{eq:betabar3} to show
	\begin{align*}
		- \frac{13}{2} \frac{\Omega^4}{r^4} \Omega \nablaslash_3 \left( r^4 \check{r} \etabar \hat{\otimes} \Omega^{-1} \betabar \right)
		=
		\frac{13}{2} r \Omega^4 \etabar \hat{\otimes} \divslash  \alphabar
		+
		\check{\mathcal{E}}^1_4
		+
		\overset{(4)}{\mathcal{E}^1_3}.
	\end{align*}
	For the final term in $\frac{\Omega^4}{r^4} \Omega \nablaslash_3 \left( \underline{E}_1 \right)$ note that,
	\begin{align*}
		\frac{\Omega^4}{r^4} \Omega \nablaslash_3 \left( \frac{r^4}{\Omega^4} \frac{1}{\Omega \tr \chi} \vert \Omega \hat{\chi} \vert^2 \check{r} \Omega^2 \alphabar \right)
		=
		\frac{\Omega^4}{r^4} \Omega \nablaslash_3 \left( \frac{r^4\check{r}}{2 \Omega^2} \vert \Omega \hat{\chi} \vert^2 \check{r} \Omega^2 \alphabar \right)
		=
		\check{\mathcal{E}}^1_4
		+ \overset{(4)}{\mathcal{E}^1_3}.
	\end{align*}
	
\end{proof}

Again, it is important to isolate the slowest decaying nonlinear error terms in the following proposition.  The additional structure in the error, namely the fact that each term contains at least one factor which is not equal to $\mathfrak{D}^{\gamma} \Omega^2 \alpha$, $\mathfrak{D}^{\gamma} \Omega \beta$ or $\mathfrak{D}^{\gamma} (\Omega \omegahat - \Omega \omegahat_{\circ})$, is used in Chapter \ref{chap:Iestimates} to show that the error in equation \eqref{n4Beq} also has this additional structure, as shown in Proposition \ref{prop:B4errorflux}.

\begin{proposition}[Wave equation for $\check{\psibar}$] \label{prop:psibartildeeq}
	With $M$, $\mathcal{Z}$, $g$ and $g_{\circ,M}$  as in Section~\ref{schemnotsec}, $\check{\psibar}$ satisfies the wave equation,
	\begin{multline} \label{eq:psibartildewave}
		\Omega \nablaslash_3 \Omega \nablaslash_4 (r^3 \Omega \check{\psibar})
		+
		\frac{2\Omega^2}{r^2} r^2\Dslash_2^* \divslash (r^3 \Omega \check{\psibar})
		\\
		=
		\frac{2}{r} \left( 1 - \frac{3M}{r} \right) \Omega \nablaslash_4 (r^3 \Omega \check{\psibar})
		-
		\frac{2\Omega^2}{r^2} \left( 1 - \frac{3M}{r} \right) r^3 \Omega \check{\psibar}
		+
		\frac{3M\Omega^2}{r^2} \check{r} \Omega^2 \alphabar
		+
		\mathcal{E}[\check{\psibar}],
	\end{multline}
	where the nonlinear error $\mathcal{E}[\check{\psibar}]$ has the schematic form,
	\begin{align} 
		\mathcal{E}[\check{\psibar}]
		=
		\
		&
		\check{\mathcal{E}}^2_3
		+
		\overset{(4)}{\mathcal{E}^2_2}
		+
		k_1 \etabar \hat{\otimes} r^2 \divslash \alphabar
		+
		k_2 r\Dslash_2^* (r\etabar \cdot \alphabar)
		+
		k_3 r\nablaslash \Omega \omegahat \hat{\otimes} r^2 \divslash \alphabar
		+
		k_4 r\Dslash_2^*( r\nablaslash \Omega \omegahat \cdot r \alphabar )\nonumber
		\\
		&
		\qquad \quad
		+ k_5 r\nablaslash \Omega \tr \chi \hat{\otimes}  r^2 \divslash \alphabar + k_6 \left(\Omega tr \chi - (\Omega tr \chi)_\circ\right) r^2 \Dslash_2^*  \divslash (r \alphabar)
		\label{psibtex}
		\\
		=
		\
		&
		\check{\mathcal{E}}^2_2
		,
		\nonumber
	\end{align}
	for functions $k_i= k_i(r^{-1} \check{r})$, for $i=1,\ldots,6$, which each take either the form $k_i=\tilde{k}_i  \frac{\check{r}}{r}$ or $k_i=\tilde{k}_i  \frac{\check{r}^2}{r^2}$, for some constants $\tilde{k}_1,\ldots,\tilde{k}_{6}$.  Moreover, each term in the nonlinear error $\mathcal{E}[\check{\psibar}]$ contains at least one factor which is not equal to $\mathfrak{D}^{\gamma} \Omega^2 \alpha$, $\mathfrak{D}^{\gamma} \Omega \beta$ or $\mathfrak{D}^{\gamma} (\Omega \omegahat - \Omega \omegahat_{\circ})$.
\end{proposition}

\begin{proof}
	As in the proof of Proposition \ref{prop:psiwave}, the Teukolsky equation \eqref{eq:Teukolskybartilde} and \eqref{eq:nabla3Omegar} imply that
	\begin{align*}
		&
		\Omega\nablaslash_3 \Omega \nablaslash_4 (r^3 \Omega \check{\psibar}) 
		=
		-
		2r^2 \Dslash_2^* \divslash \left( \frac{\Omega^2}{r^2} r^3 \Omega \check{\psibar} \right)
		+
		\frac{2}{r} \left( 1 - \frac{3M}{r} \right) \Omega \nablaslash_4 (r^3 \Omega \check{\psibar})
		-
		\frac{2\Omega^2}{r^2} \left( 1 - \frac{3M}{r} \right) r^3 \Omega \check{\psibar}
		\\
		&
		\qquad
		+
		\frac{3M\Omega^2}{r^2} \check{r} \Omega^2 \alphabar
		+
		\frac{3M}{r^2} (\Omega_{\circ}^2 - \Omega^2) \check{r} \Omega^2 \alphabar
		-
		\frac{2}{r^2} (\Omega_{\circ}^2 - \Omega^2) \left(1- \frac{6M}{r} \right) r^3 \Omega \check{\psibar}
		-
		[\Omega \nablaslash_4,\Omega\nablaslash_3](r^3 \Omega \check{\psibar})
		\\
		&
		\qquad
		-
		[\Omega \nablaslash_4, r^2 \Dslash_2^* \divslash](\check{r}\Omega^2 \alphabar)
		-
		2 \Omega \nablaslash_4
		\left(
		(\Omega \omegabarhat - (\Omega \omegabarhat)_{\circ}) r^3 \Omega \check{\psibar}
		\right)
		-
		\frac{1}{2} \Omega \nablaslash_4
		\left(
		\frac{r^2}{\Omega^2} \check{\mathcal{E}[\alphabar]}
		\right)
		.
	\end{align*}
	It remains to check the form of the error.  It is straightforward to check by inspection, using also 
	Proposition~\ref{prop:Pbartildeidentities}, that most of the terms have the correct form.  For the penultimate term one also uses the fact that equation \eqref{eq:omega3omegabar4} gives
	\[
		\Omega \nablaslash_4 \left( \Omega \omegabarhat - (\Omega \omegabarhat)_{\circ} \right)
		=
		\Omega^2 \left[
		(\eta,\etabar) - \vert \eta \vert^2 - (\rho - \rho_{\circ})
		\right]
		+
		\rho_{\circ}(\Omega_{\circ}^2 - \Omega^2),
	\]
	so that
	\[
		\Omega \nablaslash_4
		\left(
		(\Omega \omegabarhat - (\Omega \omegabarhat)_{\circ}) r^3 \Omega \check{\psibar}
		\right)
		=
		\Omega \nablaslash_4
		\left(
		(\Omega \omegabarhat - (\Omega \omegabarhat)_{\circ})
		\right)
		r^3 \Omega \check{\psibar}
		+
		(\Omega \omegabarhat - (\Omega \omegabarhat)_{\circ})
		\Omega \nablaslash_4
		\left(
		r^3 \Omega \check{\psibar}
		\right)
		=
		\check{\mathcal{E}}^2_3.
	\]
	For the third from last term, Lemma \ref{lem:commutation} implies that
	\begin{align*}
		2[\Omega \nablaslash_4, r^2 \Dslash_2^* \divslash](\check{r}\Omega^2 \alphabar)
		=
		r \Dslash_2^* \left[
		\left( (\Omega \tr \chi)_{\circ} - \Omega \tr \chi \right) r \divslash(\check{r}\Omega^2 \alphabar)
		\right]
		+
		\left( (\Omega \tr \chi)_{\circ} - \Omega \tr \chi \right) r^2 \Dslash_2^* \divslash(\check{r}\Omega^2 \alphabar)
		+
		\check{\mathcal{E}}^1_3
		+
		\overset{(4)}{\mathcal{E}^1_2}.
	\end{align*}
	The remaining nontrivial terms to check are those in $\Omega \nablaslash_4 \left( \frac{r^2}{\Omega^2} \check{\mathcal{E}}[\alphabar] \right)$, which can be done so in a similar straightforward way by using the relevant null structure and Bianchi equations.  As an example, the first anomalous term in $\check{\mathcal{E}}[\alphabar]$ gives the contribution
	\[
		\Omega \nablaslash_4( r \eta \hat{\otimes} r^2 \divslash \alphabar)
		=
		\Omega_{\circ}^2 \etabar \hat{\otimes} r^2 \divslash \alphabar
		+
		\mathcal{E}^2_3.
	\]
	Lemma \ref{lem:nabla4error4} is used for the terms arising from the $\overset{(4)}{\mathcal{E}^1_3}$ error in $\check{\mathcal{E}}[\alphabar]$ to give
	$
		\Omega \nablaslash_4 ( r^2 \overset{(4)}{\mathcal{E}^1_3})
		=
		\overset{(4)}{\mathcal{E}^2_2} + \check{{\mathcal{E}}}^2_3.
	$
	
	One easily sees that each term in the nonlinear error $\mathcal{E}[\check{\psibar}]$ contains at least one factor which is not equal to $\mathfrak{D}^{\gamma} \Omega^2 \alpha$, $\mathfrak{D}^{\gamma} \Omega \beta$ or $\mathfrak{D}^{\gamma} (\Omega \omegahat - \Omega \omegahat_{\circ})$ by noting the form of the terms in $\check{\mathcal{E}}[\alphabar]$ in Proposition \ref{prop:Teukolskybartilde} and the terms in Lemma \ref{lem:commutation}.
\end{proof}

Finally, the wave equation satisfied by $\check{\Pbar}$ can be derived.  It is convenient to isolate certain anomalous nonlinear error terms.  Accordingly, define
\begin{align} \label{eq:alphabaromegaanomalouserror}
	\mathcal{E}_{\mathrm{anom}}[\check{\Pbar}]
	=
	\
	&
	2k_1 r^4 \nablaslash(\Omega \omegahat) \hat{\otimes} \divslash \alphabar
	+
	2k_2 r^4 \Dslash_2^* \big( \nablaslash(\Omega \omegahat) \cdot \alphabar \big)
	+
	k_3 r^5 \nablaslash \Omega \nablaslash_4(\Omega \omegahat) \hat{\otimes} \divslash \alphabar
	\\
	&
	+
	k_4 r^5 \Dslash_2^* \big( \nablaslash \Omega \nablaslash_4 (\Omega \omegahat) \cdot \alphabar \big)
	+
	4 k_5 r^4 \nablaslash(\Omega \omegahat) \hat{\otimes} \divslash \alphabar
	+
	4k_6 r^5 (\Omega \omegahat - \Omega \omegahat_{\circ}) \Dslash_2^* \divslash \alphabar
	,
	\nonumber
\end{align}
and $k_i = k_i(r^{-1} \check{r})$, for $i=1,\ldots,6$, are as in Proposition \ref{prop:psibartildeeq}.

\begin{proposition}[Wave equation for $\check{\Pbar}$] \label{prop:Pcheckequation}
	With $M$, $\mathcal{Z}$, $g$ and $g_{\circ,M}$  as in Section~\ref{schemnotsec}, $\check{\Pbar}$ satisfies the Regge--Wheeler equation,
	\begin{equation} \label{eq:RWPbartilde}
		\Omega \nablaslash_4 \Omega \nablaslash_3 (r^5 \check{\Pbar})
		+
		\frac{2\Omega^2}{r^2} r^2 \Dslash_2^* \divslash (r^5 \check{\Pbar})
		+
		\frac{2\Omega^2}{r^2} \left( 1 - \frac{3M}{r} \right) r^5 \check{\Pbar}
		=
		\mathcal{E}[\check{\Pbar}],
	\end{equation}
	where the nonlinear error $\mathcal{E}[\check{\Pbar}]$ has the schematic form
	\[
		\mathcal{E}[\check{\Pbar}]
		=
		\check{\mathcal{E}}^3_2
		+
		\mathcal{E}_{\mathrm{anom}}[\check{\Pbar}]
		=
		\check{\mathcal{E}}^3_{\frac{3}{2}},
	\]
	where $\mathcal{E}_{\mathrm{anom}}[\check{\Pbar}]$ is defined in \eqref{eq:alphabaromegaanomalouserror}.
\end{proposition}

\begin{proof}
	As in the proof of Proposition \ref{prop:Pwave}, using now \eqref{eq:nabla3Omegar}, equation \eqref{eq:psibartildewave} implies that
	\begin{multline*}
		\Omega \nablaslash_3 \Omega \nablaslash_4 (r^5 \check{\Pbar})
		=
		-
		2 r^2 \Dslash_2^* \divslash \left( \frac{\Omega^2}{r^2} r^5 \check{\Pbar} \right)
		-
		\frac{2\Omega^2}{r^2} \left( 1 - \frac{3M}{r} \right) r^5 \check{\Pbar}
		+
		\frac{6M}{r^2} \left( \Omega_{\circ}^2 - \Omega^2 \right) r^3 \Omega \check{\psibar}
		\\
		-
		\Omega \nablaslash_4 \left(
		\frac{r^2}{\Omega^2} \mathcal{E} [\check{\psibar}]
		+
		2 (\Omega \omegabarhat - (\Omega \omegabarhat)_{\circ} ) r^5 \check{\Pbar}
		\right)
		-
		[\Omega \nablaslash_4, \Omega \nablaslash_3] (r^5 \check{\Pbar})
		+
		2[\Omega \nablaslash_4, r^2 \Dslash_2^* \divslash] (r^3 \Omega \check{\psibar}).
	\end{multline*}
	Note the cancellations.  It remains to check the structure of the error terms, which is done by inspection.  The most nontrivial terms to check are those in $\Omega \nablaslash_4 \left( \frac{r^2}{\Omega^2} \mathcal{E} [\check{\psibar}] \right)$ for which the equations \eqref{eq:Ray}, \eqref{eq:nabla4etabar}, \eqref{eq:alphabar4} are used, together with Lemma \ref{lem:nabla4error4}.
\end{proof}

\chapter{Change of double null gauge and the diffeomorphism functions}
\label{bigchangeisgood}

The vacuum Lorentzian 
manifold $(\mathcal{M},g)$ of interest in this paper will \underline{not} be covered by a single double null coordinate system.
We will thus consider locally defined double null parametrisations, and being able to understand (and estimate)
the diffeomorphisms relating two such parametrisations will be paramount. This is the goal of the present chapter.
 
 \minitoc

We shall set the notation in {\bf Section~\ref{newdiffysec}} for a local double null parametrisations 
covering a region
of a vacuum spacetime $(\mathcal{M},g)$. We shall then consider  change of double null gauge
in {\bf Section~\ref{changeisgood}},
introducing a formalism for estimating the diffeomorphism functions $f$ to arbitrary order.
We shall then give formulas for how geometric quantities transform under such change of gauge
in {\bf Section~\ref{diffeoidentities}}. 
Finally, in {\bf Section~\ref{determinethesphere}}, we shall give a canonical choice for
breaking the diffeomorphism invariance of the sphere.

\vskip1pc

\emph{The basic notation of Section~\ref{newdiffysec} is fundamental for all of Part~\ref{stateandlogicpart} beyond
Section~\ref{initialdatanormsec}. Section~\ref{changeisgood}
will also be necessary wherever the $f$ notation appears (e.g.~already
in Section~\ref{anchoredsec}), and the operations
defining derivatives of $f$
appear in the energies of Section~\ref{energiessection}.
Section~\ref{diffeoidentities} may be skipped on a first reading, though it will be necessary for Part~\ref{improvingpart}.
The statement of the main proposition of Section~\ref{determinethesphere} is
referred to in Section~\ref{defofanchored}; the proof may also be safely skipped on a first reading.
(We remark here that Section~\ref{determinethesphere} is in fact
independent of the rest of the paper with the exception of the notation of Section~\ref{sphereconcrete}.)}

\section{Local double null gauges in a vacuum spacetime $(\mathcal{M},g)$}
\label{newdiffysec}

Let $(\mathcal{M},g)$ be a smooth time-oriented $4$-dimensional 
Lorentzian manifold satisfying the Einstein vacuum equations $(\ref{Ricciflathere})$.
Consider a
subset as before\index{double null gauge!sets!$\mathcal{Z}$, domain of parametrisation} 
\[
\mathcal{Z} =\mathcal{W}\times\mathbb S^2\subset \mathbb R^2\times \mathbb S^2 
\]
and a
 local parameterisation
\begin{equation}
\label{localformparam}
i: \mathcal{Z} \to (\mathcal{M},g)
\end{equation}
where $i^*g$ satisfies the assumptions of Section~\ref{geomprelimforvac}. 
We will assume that the time-orientation defined in Section~\ref{timeorientsection} 
coincides with that induced from 
$(\mathcal{M},g)$.
We will call such a parametrisation a (locally defined) double null gauge in
$(\mathcal{M},g)$.
In the context of restricting domains, we will may also consider
$\mathcal{Z} \subset \mathbb R^2\times
\mathbb S^2$ to be an arbitrary $4$-dimensional submanfiold  with piecewise smooth boundary.

When working in a single such parametrisation,
we will usually not differentiate our notation between $g$ and $i^*g$,
and, as is standard, we may think of $u,v, \theta^A$  as coordinate functions on $\mathcal{M}$,
i.e.~we may identify these with $u\circ i^{-1}$, etc. Similarly, 
we shall use the notations $C_u$, $C_v$, $S_{u,v}$ etc.,
to denote subsets of $\mathcal{M}$, i.e.~they correspond to $i(C_u)$, $i(C_v)$, etc., 
defined previously.

\begin{remark}
\label{onregofparamrem}
We note that although our underlying manifold $(\mathcal{M},g)$ will be smooth,
as it will be a maximal Cauchy development of smooth data, some gauges which will be defined
via normalisations on null infinity will only have limited regularity. Thus, we shall sometimes refer also to
parametrisations~\eqref{localformparam} which are only $C^k$ for some sufficiently high $k$, in which case
the corresponding geometric quantities will only have finite regularity. We note already that the fundamental
parametrisations of our work, however, i.e.~those corresponding to our $\mathcal{I}^+$ and $\mathcal{H}^+$ gauge
normalised at finite $u_f$ time, will  indeed be smooth.
\end{remark}

\section{Change of double null gauge: the diffeomorphism functions $f$ and their derivatives}
\label{changeisgood}

We shall often cover the same region of $\mathcal{M}$ by multiple local parameterisations of the form 
$(\ref{localformparam})$. For instance, in Chapter~\ref{thelocaltheorysec}, we shall consider the two initial data
gauges~\eqref{initialKgaugehere} and~\eqref{initialEFgaugehere}  (defined in 
Section~\ref{localexistencesection})
and the two  teleologically normalised
$\mathcal{I}^+$ and $\mathcal{H}^+$ gauges~\eqref{fortheIplusgauge} 
and~\eqref{fortheHplusgauge} (defined in Section~\ref{defofanchored}), and the ranges of these parametrisations
will indeed overlap. It will be of fundamental importance to understand how quantities
transform, and in particular, to estimate the diffeomorphisms relating two such parametrisations.

With $(\mathcal{M},g)$ and~\eqref{localformparam} as in Section~\ref{newdiffysec}, 
let us consider then, in addition to a parametrisation~\eqref{localformparam}, a second parametrisation
\begin{equation}
\label{localformparamtwo}
\tilde{i}: \widetilde{\mathcal{Z}} \to \mathcal{M},
\end{equation}
where $\tilde{i}^*g$ again satisfies  the assumptions of Section~\ref{geomprelimforvac}.
We may define
transition functions\index{double null gauge!change of gauge!$F$, notation for functions describing diffeomorphisms} 
\begin{equation}
\label{localformparamcomposition}
F:={i}^{-1}\circ \tilde{i} : \,\, \tilde{i}^{-1}(  i(\mathcal{Z})) \to i^{-1} ( \tilde{i}(\widetilde{\mathcal{Z}})).
\end{equation}

When considering coordinates in two such gauges, it is convenient
to introduce labels on the coordinates so as to distinguish them.
Thus, we may denote by $u$, $v$, $\theta^A$ local coordinates defined by
a parametrisation~\eqref{localformparam} and
$\tilde{u}$, $\tilde{v}$, $\tilde\theta^A$ local coordinates defined by
a second parametrisation~\eqref{localformparamtwo}.
With this notational convention we may consider both $C_u$, etc., and $C_{\tilde{u}}$, etc., as subsets
of $\mathcal{M}$, where the different labelling of the coordinates is sufficient to distinguish these.

\subsection{The diffeomorphism functions $f$}

In coordinates, we may write $F = (F^1,F^2,F^3,F^4)$ in the form\index{double null gauge!change of gauge!$f$, notation for functions describing diffeomorphisms} 
\begin{equation}
\label{expressinglikethis}
	F^A(\widetilde{u}, \widetilde{v}, \widetilde{\theta}) = \widetilde{\theta}^A + f^A(\widetilde{u}, \widetilde{v}, \widetilde{\theta}),
	\quad
	F^3(\widetilde{u}, \widetilde{v}, \widetilde{\theta}) = \widetilde{u} + f^3(\widetilde{u}, \widetilde{v}, \widetilde{\theta}),
	\quad
	F^4(\widetilde{u}, \widetilde{v}, \widetilde{\theta}) = \widetilde{v} + f^4(\widetilde{u}, \widetilde{v}, \widetilde{\theta}),
\end{equation}
where the above expressions define $f^A$, $f^3$ and $f^4$.  
One may think of $f^3$ and $f^4$ as scalar valued functions on $\tilde{i}^{-1}(i(\mathcal{Z}))$.
Their definition does not depend on the choice of local angular coordinates.
The decomposition
of $F^A=\tilde\theta^A+f^A$, on the other hand, is somewhat unnatural, because the linear structure
is heavily dependent on the choice of local coordinates $\theta^A$ and $\tilde\theta^A$. 
Nonetheless, 
 we use this notation to facilitate comparison with~\cite{holzstabofschw}.
In the subsections below, we will define tensorial derivatives of $F$, which we will later
use to estimate the diffeomorphisms.

\subsection{First order derivatives of the diffeomorphism functions}

Consider the set $\widetilde{S}_{\widetilde{u},\widetilde{v}} \subset \widetilde{\mathcal{Z}}$ 
and some point $p \in \widetilde{S}_{\widetilde{u},\widetilde{v}}$.  The image $F(p)$ lies on a unique  $S_{u,v} \subset \mathcal{Z}$.  The restriction of the differential $dF_p$ to $T_p\widetilde{S}_{\widetilde{u},\widetilde{v}}$ can be projected to $T_{F(p)} S_{u,v}$ to give a map which we shall denote\index{double null gauge!change of gauge!$\widetilde{\slashed{d}} \slashed{F}_p$, angular part of first derivatives of diffeomorphism maps}
\begin{equation} \label{eq:dslashFslash}
	\widetilde{\slashed{d}} \slashed{F}_p : T_p\widetilde{S}_{\widetilde{u},\widetilde{v}} \to T_{F(p)} S_{u,v},
\end{equation}
defined by $\widetilde{\slashed{d}} \slashed{F}_p (X) = \Pi_{T_{F(p)} S_{u,v}} \big( d F_p X \big)$, for $X \in T_p\widetilde{S}_{\widetilde{u},\widetilde{v}}$.  
We emphasise that the definition of this map is independent of the choice of local angular coordinates
$\theta^A$ and $\tilde\theta^A$.

In components,
\begin{equation}
\label{thebirthofatensor}
	\widetilde{\slashed{d}} \slashed{F} (\partial_{\widetilde{\theta}^A}, d\theta^B)
	=
	\frac{\partial F^B}{\partial \widetilde{\theta}^A}
	=
	\delta_A^B + \frac{\partial f^B}{\partial \widetilde{\theta}^A}.
\end{equation}
We note that the two summands on the right hand side of above do not have
a natural tensorial meaning individually.

Similarly, define $\partial_{\widetilde{u}} \slashed{F}_p \in T_{F(p)} S_{u,v}$ and $\partial_{\widetilde{v}} \slashed{F}_p \in T_{F(p)} S_{u,v}$ by
\begin{equation}
\label{defofpartuvslashF}
	\partial_{\widetilde{u}} \slashed{F}_p = \Pi_{T_{F(p)} S_{u,v}} \big( d F_p \partial_{\widetilde{u}} \big),
	\qquad
	\partial_{\widetilde{v}} \slashed{F}_p = \Pi_{T_{F(p)} S_{u,v}} \big( d F_p \partial_{\widetilde{v}} \big).
\end{equation}
In components,
\[
	\partial_{\widetilde{u}} \slashed{F} (d\theta^A)
	=
	\frac{\partial F^A}{\partial \widetilde{u}}
	=
	\frac{\partial f^A}{\partial \widetilde{u}},
	\qquad
	\partial_{\widetilde{v}} \slashed{F} (d\theta^A)
	=
	\frac{\partial F^A}{\partial \widetilde{v}}
	=
	\frac{\partial f^A}{\partial \widetilde{v}}.
\]

Define now\index{double null gauge!change of gauge!$\partial_{\widetilde{u}} \slashed{f}$, notation for first derivative of diffeomorphism map}\index{double null gauge!change of gauge!$\partial_{\widetilde{v}} \slashed{f}$, notation for first derivative of diffeomorphism map}
\[
	\partial_{\widetilde{u}} \slashed{f} := \partial_{\widetilde{u}} \slashed{F},
	\qquad
	\partial_{\widetilde{v}} \slashed{f} := \partial_{\widetilde{v}} \slashed{F}.
\]

\subsection{Beyond $S$-tensors: higher order derivatives of the diffeomorphism functions}
\label{beyondSsection}

So as to not overload the notation with restrictions of domain, let us assume without loss
of generality that the transition
map
$F$ of~\eqref{localformparamcomposition}
is in fact a global diffeomorphism $F:\widetilde{\mathcal{Z}}\to \mathcal{Z}$. (Note here
that in this
case, we will not assume that
$\widetilde{\mathcal{Z}}$ and $\mathcal{Z}$ in general factor as products, so it is important
to allow these to be general $4$-dimensional submanifolds of $\mathbb R^2\times \mathbb S^2$ with
boundary, as noted in Section~\ref{newdiffysec}.)

Thought of as defined on $\mathcal{Z}$, then 
$\partial_{\widetilde{u}} \slashed{F}$ and $\partial_{\widetilde{v}} \slashed{F}$
are simply $S$-vector fields.
On the other hand,  $\widetilde{\slashed{d}} \slashed{F}$ is a new kind of tensorial object
which cannot be thought of either as an $S$-tensor or as a $\tilde{S}$-tensor.
Thus, we cannot follow the calculus of Section~\ref{nullframesopers} to define higher order derivatives.

We would like to view both~\eqref{thebirthofatensor} and~\eqref{defofpartuvslashF} 
as tensors on $\widetilde{\mathcal{Z}}$ and to differentiate them with respect to operators which are
connected to the double null structure of $\widetilde{\mathcal{Z}}$.
The purpose of this section is to provide such a calculus.

We first introduce certain vector bundles over $\widetilde{\mathcal{Z}}$
and note that the objects $\partial_{\widetilde{u}} \slashed{f}$, $\partial_{\widetilde{v}} \slashed{f}$, and $\widetilde{\slashed{d}} \slashed{F}$, introduced in the previous section, are
indeed sections of these vector bundles.

Let $TS\subset T\mathcal{Z}$ denote 
the sub-bundle of vectors tangential to the spheres $S_{u,v}$, i.\@e.\@ $TS = \bigcup_{p\in \mathcal{Z}} \{p\} \times T_p S_{u(p),v(p)}$, with the obvious topology and smooth structure.  Let $F^*(TS)$ denote the pull back bundle over
$\widetilde{\mathcal{Z}}$,
\[
	F^*(TS)
	=
	\bigcup_{p \in \widetilde{\mathcal{Z}}}
	\{ p \} \times 
	T_{F(p)}S
	,
\]
where $T_{F(p)}S$ is the fibre of $TS$ over $F(p)$.
Similarly let $T^*\widetilde{S} \subset T^*\mathcal{M}$ denote the sub-bundle of $\widetilde{S}_{\widetilde{u},\widetilde{v}}$ one forms, $T^*\widetilde{S} = \bigcup_{p\in M} \{p\} \times T_p^* \widetilde{S}_{\widetilde{u}(p),\widetilde{v}(p)}$.

Now $\partial_{\widetilde{u}} \slashed{f}, \partial_{\widetilde{v}} \slashed{f} \in \Gamma( F^*(TS))$ and $\widetilde{\slashed{d}} \slashed{F} \in \Gamma \big( T^* \widetilde{S} \otimes F^*(TS) \big)$ and so, in order to apply the operators $\widetilde{\nablaslash}$, $\widetilde{\nablaslash}_3$ and $\widetilde{\nablaslash}_4$ to $\partial_{\widetilde{u}} \slashed{f}$, $\partial_{\widetilde{v}} \slashed{f}$, and $\widetilde{\slashed{d}} \slashed{F}$, we define the operators $\widetilde{\nablaslash}$, $\widetilde{\nablaslash}_3$ and $\widetilde{\nablaslash}_4$ applied to general sections of these vector bundles.

First consider a one form $\xi \in \Gamma( T^*S)$.  Define $\widetilde{\nablaslash}_X \xi$ to be the restriction of the spacetime one form $\nabla_{F_*X} \xi$ to $S_{u,v}$ vectors.  Define moreover $\widetilde{\nablaslash}_{3}\xi$ and $\widetilde{\nablaslash}_{4}\xi$ to be the restriction of $\nabla_{F_*\widetilde{e}_3} \xi$ and $\nabla_{F_*\widetilde{e}_4} \xi$ to $S_{u,v}$ vectors.

Given now a tensor field $T\in \Gamma \big( (T^* \widetilde{S})^s \otimes F^*(TS) \big)$, define $\widetilde{\nablaslash} T\in \Gamma \big( (T^* \widetilde{S})^{s+1} \otimes F^*(TS) \big)$ by
\begin{align*}
	(\widetilde{\nablaslash}_X T) (Y_1,\ldots,Y_s,\xi)
	=
	\
	&
	X\big( T(Y_1,\ldots,Y_s,\xi) \big)
	-
	T(\widetilde{\nablaslash}_XY_1,\ldots,Y_s,\xi)
	-
	\ldots
	-
	T(Y_1,\ldots,\widetilde{\nablaslash}_XY_s,\xi)
	\\
	&
	+
	T(Y_1,\ldots,Y_s,\widetilde{\nablaslash}_{X}\xi)
	,
\end{align*}
for $X,Y_1,\ldots,Y_s\in \Gamma(T \widetilde{S})$ and $\xi \in \Gamma( T^*S)$.
Similarly define
\begin{align*}
	(\widetilde{\nablaslash}_3 T) (Y_1,\ldots,Y_s,\xi)
	=
	\
	&
	\widetilde{e}_3 \big( T(Y_1,\ldots,Y_s,\xi) \big)
	-
	T(\widetilde{\nablaslash}_3 Y_1,\ldots,Y_s,\xi)
	-
	\ldots
	-
	T(Y_1,\ldots,\widetilde{\nablaslash}_3 Y_s,\xi)
	\\
	&
	+
	T(Y_1,\ldots,Y_s, \widetilde{\nablaslash}_{3}\xi)
	,
	\\
	(\widetilde{\nablaslash}_4 T) (Y_1,\ldots,Y_s,\xi)
	=
	\
	&
	\widetilde{e}_4 \big( T(Y_1,\ldots,Y_s,\xi) \big)
	-
	T(\widetilde{\nablaslash}_4 Y_1,\ldots,Y_s,\xi)
	-
	\ldots
	-
	T(Y_1,\ldots,\widetilde{\nablaslash}_4 Y_s,\xi)
	\\
	&
	+
	T(Y_1,\ldots,Y_s, \widetilde{\nablaslash}_{4}\xi)
	.
\end{align*}

Setting, for example, $T = \partial_{\widetilde{u}} \slashed{f}$, the above defines the derivative $\widetilde{\nablaslash} \partial_{\widetilde{u}} \slashed{f} \in \Gamma \big( T^* \widetilde{S} \otimes F^*(TS) \big)$.  Higher order $\widetilde{\nablaslash}$ derivatives are then inductively defined, for $k \geq 2$, by setting $T = \widetilde{\nablaslash}{}^k \partial_{\widetilde{u}} \slashed{f} \in \Gamma \big( (T^* \widetilde{S})^k \otimes F^*(TS) \big)$ in the above and defining
\[
	\widetilde{\nablaslash}{}^{k+1} \partial_{\widetilde{u}} \slashed{f}
	:=
	\widetilde{\nablaslash} \widetilde{\nablaslash}{}^{k} \partial_{\widetilde{u}} \slashed{f}
	\in
	\Gamma \big( (T^* \widetilde{S})^{k+1} \otimes F^*(TS) \big).
\]
One similarly defines general combinations of $\widetilde{\nablaslash}$, $\widetilde{\nablaslash}_3$ and $\widetilde{\nablaslash}_4$ derivatives of $\partial_{\widetilde{u}} \slashed{f}$, and also of $\partial_{\widetilde{v}} \slashed{f}$ and $\widetilde{\slashed{d}} \slashed{F}$.

For $T\in \Gamma \big( (T^* \widetilde{S})^s \otimes F^*(TS) \big)$ and $p \in \widetilde{\mathcal{W}}\times \widetilde{\mathcal{U}}$, we note that the expression
\begin{equation}
\label{normoftensorhere}
	\vert T \vert^2
	:=
	\widetilde{\gslash}^{\widetilde{A_1} \widetilde{B_1}}(p)
	\ldots
	\widetilde{\gslash}^{\widetilde{A_s}\widetilde{B_s}}(p)
	\gslash_{CD}(F(p))
	{T_{\widetilde{A_1}\ldots \widetilde{A_s}}}^C(p) {T_{\widetilde{B_1}\ldots \widetilde{B_s}}}^D(p)
\end{equation}
provides a coercive expression.

\section{Transformation laws for geometric quantities under change of gauge}
\label{diffeoidentities}

In this section, identities are derived for how certain geometric quantities change under a change of gauge relations.  See 
Proposition \ref{prop:metricrelations}, Proposition \ref{prop:Riccirelations} and Proposition \ref{prop:curvaturerelations}.

\subsection{Basic setup: two double null gauges and two Schwarzschild backgrounds with masses $M$ and $\tilde{M}$}
\label{twogaugestwoback}

Throughout, $(\mathcal{M},g)$ and parametrisations~\eqref{localformparam} and~\eqref{localformparamtwo} 
are as in Section~\ref{changeisgood}. As in Section~\ref{beyondSsection}, 
without loss of generality we may assume that
the diffeomorphism $F$ of~\eqref{localformparamcomposition} 
is in fact a global diffeomorphism $F:\widetilde{\mathcal{Z}}\to \mathcal{Z}$.
We may suppress $F$ from the notation, writing~\eqref{expressinglikethis} as 
\begin{equation} \label{eq:xtildexf}
	u = \widetilde{u} + f^3(\widetilde{u}, \widetilde{v}, \widetilde{\theta}^1, \widetilde{\theta}^2),
	\quad
	v = \widetilde{v} + f^4(\widetilde{u}, \widetilde{v}, \widetilde{\theta}^1, \widetilde{\theta}^2),
	\quad
	\theta^A = \widetilde{\theta}^A + f^A(\widetilde{u}, \widetilde{v}, \widetilde{\theta}^1, \widetilde{\theta}^2).
\end{equation}
We may consider~\eqref{eq:xtildexf} as relations on $\widetilde{\mathcal{Z}}$ (or alternatively on the spacetime
region $\tilde{i}(\widetilde{\mathcal{Z}})$) with the usual identifications.

We will typically be interested in the case where parametrisation~\eqref{localformparam} is in the 
form~\eqref{doublenulllongform}
whereas parametrisation~\eqref{localformparamtwo} is in the form~\eqref{doublenulllongforminterchanged}.
In this case, we have the associated null frames
\begin{equation}
\label{onenullframe}
	e_A = \partial_{\theta^A},
	\qquad
	e_3 = \frac{1}{\Omega} \partial_u,
	\qquad
	e_4 = \frac{1}{\Omega} (\partial_v + b),
\end{equation}
and
\begin{equation}
\label{theothernullframe}
	\widetilde{e}_A = \partial_{\widetilde{\theta}^A},
	\qquad
	\widetilde{e}_3 = \frac{1}{\widetilde{\Omega}} (\partial_{\widetilde{u}} + \widetilde{b}),
	\qquad
	\widetilde{e}_4 = \frac{1}{\widetilde{\Omega}} \partial_{\widetilde{v}}.
\end{equation}
Both~\eqref{onenullframe} and~\eqref{theothernullframe}
may be considered as (local) frames on $\widetilde{\mathcal{Z}}$ with the usual identifications, i.e.~where
$e_A$ is identified with $(F_*)^{-1} (e_A)$, etc.

We shall write down identities in Section~\ref{basicidentitiesinthissection} corresponding to the above case.
We shall less frequently consider also the case where~\eqref{localformparamtwo} is  of the form~\eqref{doublenulllongform};
the resulting identities change
in the obvious way.

For given masses $M>0$, $\tilde{M}>0$, we may now associate also Schwarzschild backgrounds to
$i^*g$ and $\tilde{i}^*g$ as in Section~\ref{diffsofthespheresubsec}.  
We will denote the Schwarzschild background corresponding to $\tilde{i}^*g$ with
$\widetilde{\ }$'s and the Schwarzschild background corresponding to $i^*g$ without.
We consider the Schwarzschild scalar quantities defined in both gauges,
denoting the ones corresponding to $\tilde{i}^*g$ with a $\widetilde{\ }$.
Thus, we distinguish $\widetilde{r}_{\tilde{M}}$ from $r_M$.
We will in what follows drop the $M$ and $\tilde{M}$ subscripts.

We will similarly distinguish the operators associated with  $\tilde{i}^*g$ and $i^*g$
by putting $\widetilde{\ }$ on the former.

Again, using the diffeomorphism $F$ of~\eqref{localformparamcomposition}, 
we may consider all such quantities as defined
on $\widetilde{\mathcal{Z}}$.

\subsection{Basic identities}
\label{basicidentitiesinthissection}

With the setup of Section~\ref{twogaugestwoback} and the above conventions, we compute the relations
\begin{align}
	du
	=
	\
	&
	\left( 1 + \frac{\partial f^3}{\partial \widetilde{u}} \right) d \widetilde{u}
	+
	\frac{\partial f^3}{\partial \widetilde{v}} d \widetilde{v}
	+
	\frac{\partial f^3}{\partial \widetilde{\theta}^A} d \widetilde{\theta}^A
	\label{eq:dxdtildex1}
	\\
	dv
	=
	\
	&
	\left( 1 + \frac{\partial f^4}{\partial \widetilde{v}} \right) d \widetilde{v} 
	+
	\frac{\partial f^4}{\partial \widetilde{u}} d \widetilde{u}
	+
	\frac{\partial f^4}{\partial \widetilde{\theta}^A} d \widetilde{\theta}^A
	\label{eq:dxdtildex2}
	\\
	d\theta^A
	=
	\
	&
	\left( \delta^A_B + \frac{\partial f^A}{\partial \widetilde{\theta}^B} \right) d \widetilde{\theta}^B
	+
	\frac{\partial f^A}{\partial \widetilde{u}} d \widetilde{u}
	+
	\frac{\partial f^A}{\partial \widetilde{v}} d \widetilde{v},
	\label{eq:dxdtildex3}
\end{align}
and
\begin{align}
	\partial_{\widetilde{u}}
	=
	\
	&
	\left( 1 + \frac{\partial f^3}{\partial \widetilde{u}} \right) \partial_u
	+
	\frac{\partial f^4}{\partial \widetilde{u}} \partial_v
	+
	\frac{\partial f^A}{\partial \widetilde{u}} \partial_{\theta^A}
	\label{eq:partialxpartialtildex1}
	\\
	\partial_{\widetilde{v}}
	=
	\
	&
	\left( 1 + \frac{\partial f^4}{\partial \widetilde{v}} \right) \partial_v 
	+
	\frac{\partial f^3}{\partial \widetilde{v}} \partial_u
	+
	\frac{\partial f^A}{\partial \widetilde{v}} \partial_{\theta^A}
	\label{eq:partialxpartialtildex2}
	\\
	\partial_{\widetilde{\theta}^A}
	=
	\
	&
	\left( \delta_A^B + \frac{\partial f^B}{\partial \widetilde{\theta}^A} \right) \partial_{\theta^B}
	+
	\frac{\partial f^3}{\partial \widetilde{\theta}^A} \partial_u
	+
	\frac{\partial f^4}{\partial \widetilde{\theta}^A} \partial_v.
	\label{eq:partialxpartialtildex3}
\end{align}
It then follows that the frames~\eqref{onenullframe} and~\eqref{theothernullframe} are related by
\begin{align}
	\widetilde{e}_A
	&
	=
	\left( \delta_A^B + \frac{\partial f^B}{\partial \widetilde{\theta}^A} \right) e_B
	+
	\Omega \frac{\partial f^3}{\partial \widetilde{\theta}^A} e_3
	+
	\Omega \frac{\partial f^4}{\partial \widetilde{\theta}^A} e_4
	-
	\frac{\partial f^4}{\partial \widetilde{\theta}^A} b,
	\label{eq:gaugerelationA}
	\\
	\widetilde{e}_3
	&
	=
	\frac{\Omega}{\widetilde{\Omega}} \left(
	1 + \frac{\partial f^3}{\partial \widetilde{u}}
	\right) e_3
	+
	\frac{\Omega}{\widetilde{\Omega}} \frac{\partial f^4}{\partial \widetilde{u}} e_4
	+
	\frac{1}{\widetilde{\Omega}} \frac{\partial f^A}{\partial \widetilde{u}} e_A
	+
	\frac{1}{\widetilde{\Omega}} \widetilde{b}
	-
	\frac{1}{\widetilde{\Omega}} \frac{\partial f^4}{\partial \widetilde{u}} b,
	\label{eq:gaugerelation3}
	\\
	\widetilde{e}_4
	&
	=
	\frac{\Omega}{\widetilde{\Omega}} \left(
	1 + \frac{\partial f^4}{\partial \widetilde{v}}
	\right) e_4
	+
	\frac{\Omega}{\widetilde{\Omega}} \frac{\partial f^3}{\partial \widetilde{v}} e_3
	+
	\frac{1}{\widetilde{\Omega}} \frac{\partial f^A}{\partial \widetilde{v}} e_A
	-
	\frac{1}{\widetilde{\Omega}} \left( 1 + \frac{\partial f^4}{\partial \widetilde{v}} \right) b.
	\label{eq:gaugerelation4}
\end{align}

Note that
\begin{equation} \label{eq:Riccitab1}
	\nabla_{e_A} e_B
	=
	\Gammaslash_{AB}^C e_C
	+
	\frac{1}{2} \chi_{AB} e_3
	+
	\frac{1}{2} \chibar_{AB} e_4,
\end{equation}
\begin{equation} \label{eq:Riccitab2}
	\nabla_{e_A} e_3 = {\chibar_A}^B e_B + \frac{1}{2} (\eta_A - \etabar_A) e_3,
	\quad 
	\nabla_{e_A} e_4 = {\chi_A}^B e_B + \frac{1}{2} (\etabar_A - \eta_A) e_4,
\end{equation}
\begin{equation} \label{eq:Riccitab3}
	\nabla_{e_3} e_A = {\chibar_A}^B e_B + \eta_A e_3, \quad \nabla_{e_4}e_A = \left[ {\chi_A}^B - e_A(b^B) \right] e_B + \etabar_A e_4,
\end{equation}
\begin{equation} \label{eq:Riccitab4}
	\nabla_{e_3}e_4 = - \omegabarhat e_4 + 2\eta^Ae_A, \quad \nabla_{e_4}e_3 = -\omegahat e_3 + 2\etabar^B e_B,
\end{equation}
\begin{equation} \label{eq:Riccitab5}
	\nabla_{e_3} e_3 = \omegabarhat e_3, \quad \nabla_{e_4} e_4 = \omegahat e_4.
\end{equation}
Similar formulas hold for  $\{\widetilde{e}_{\mu}\}$, with the exception of equations \eqref{eq:Riccitab3} which take the form
\[
	\nabla_{\widetilde{e}_3} \widetilde{e}_A = \left[ {\widetilde{\chibar}_A}^B - \widetilde{e}_A(\widetilde{b}^B) \right] \widetilde{e}_B + \widetilde{\eta}_A \widetilde{e}_3,
	\quad
	\nabla_{\widetilde{e}_4}\widetilde{e}_A = {\widetilde{\chi}_A}^B \widetilde{e}_B + \widetilde{\etabar}_A \widetilde{e}_4.
\]

The null components of the Riemann curvature tensor satisfy
\begin{align}
	R(e_A, e_4, e_B, e_4)
	=
	\alpha (e_A, e_B),
	&
	\qquad
	R(e_A, e_3, e_B, e_3)
	=
	\alphabar (e_A, e_B),
	\label{eq:curvtab1}
	\\
	R(e_A, e_4, e_3, e_4)
	=
	2 \beta (e_A),
	&
	\qquad
	R(e_A, e_3, e_3, e_4)
	=
	2 \betabar (e_A),
	\label{eq:curvtab2}
	\\
	R(e_A,e_4,e_B,e_C)
	=
	-{}^*\beta(e_A) \epsslash(e_B,e_C),
	&
	\qquad
	R(e_A,e_3,e_B,e_C)
	=
	{}^*\betabar(e_A) \epsslash(e_B,e_C),
	\label{eq:curvtab3}
	\\
	R(e_A,e_3,e_B,e_4)
	=
	- \rho \gslash(e_A,e_B) + \sigma \epsslash (e_A,e_B),
	&
	\qquad
	R(e_A,e_B,e_C,e_D)
	=
	\rho \epsslash(e_A,e_B) \epsslash(e_C,e_D),
	\label{eq:curvtab4}
	\\
	R(e_3,e_4,e_3,e_4)
	=
	4 \rho,
	&
	\qquad
	R(e_A,e_B,e_3,e_4)
	=
	2 \sigma \epsslash(e_A,e_B).
	\label{eq:curvtab5}
\end{align}
Similar formulas hold for  $\{\widetilde{e}_{\mu}\}$.

Suppose that the metric $g$ takes the following expressions in each of the two double null coordinate systems
\begin{align*}
	g\vert_{\widetilde{x}=x_0}
	&
	=
	-4 \Omega^2 du dv
	+
	\vert b \vert^2 dv dv
	-
	2 \gslash_{AB} b^A dv d\theta^B
	+
	\gslash_{AB} d \theta^A d \theta^B
	\Big\vert_{x=x_0 + f(x_0)}
	\\
	&
	=
	-4 \widetilde{\Omega}^2 d\widetilde{u} d\widetilde{v}
	+
	\vert \widetilde{b} \vert^2 d\widetilde{u} d\widetilde{u}
	-
	2 \widetilde{\gslash}_{AB} \widetilde{b}^A d\widetilde{u} d\widetilde{\theta}^B
	+
	\widetilde{\gslash}_{AB} d \widetilde{\theta}^A d \widetilde{\theta}^B
	\Big\vert_{\widetilde{x}=x_0}.
\end{align*}
Note the asymmetry between the two expressions in the location of the torsion term $b$.\footnote{The following relations will most commonly be used with the present location of the torsion terms.  Similar relations hold, and will also be used, with alternative locations of the torsion terms.  The following relations all hold with obvious modification, e.\@g.\@ in equations \eqref{eq:metricid4} and \eqref{eq:dthetadv}.}  Expanding the first line, using the relations \eqref{eq:dxdtildex1}--\eqref{eq:dxdtildex3}, and equating with the second line gives the following system of equations for $f$.  The equated coefficients of $d\widetilde{u} d \widetilde{v}$ give
\begin{align}
	-4 \widetilde{\Omega}^2
	=
	\
	&
	-4 \Omega^2
	\left[
	\left( 1 + \frac{\partial f^3}{\partial \widetilde{u}} \right) \left( 1 + \frac{\partial f^4}{\partial \widetilde{v}} \right)
	+
	\frac{\partial f^3}{\partial \widetilde{v}} \frac{\partial f^4}{\partial \widetilde{u}}
	\right]
	+
	2 \left\vert b \right\vert^2 \left( 1 + \frac{\partial f^4}{\partial \widetilde{v}} \right) \frac{\partial f^4}{\partial \widetilde{u}}
	\nonumber
	\\
	&
	-
	2 \gslash_{AB}b^A \left[
	\left( 1 + \frac{\partial f^4}{\partial \widetilde{v}} \right) \frac{\partial f^B}{\partial \widetilde{u}}
	+
	\frac{\partial f^4}{\partial \widetilde{u}} \frac{\partial f^B}{\partial \widetilde{u}}
	\right]
	+
	2 \gslash_{AB} \frac{\partial f^A}{\partial \widetilde{v}} \frac{\partial f^B}{\partial \widetilde{u}}.
	\label{eq:metricid1}
\end{align}
The equated coefficients of $d\widetilde{u} d \widetilde{u}$ give
\begin{align} \label{eq:metricid2}
	\left\vert \widetilde{b} \right\vert^2
	=
	-4 \Omega^2
	\frac{\partial f^4}{\partial \widetilde{u}} \left( 1 + \frac{\partial f^3}{\partial \widetilde{u}} \right)
	+
	\left\vert b \right\vert^2 \left( \frac{\partial f^4}{\partial \widetilde{u}} \right)^2
	-
	2 \gslash_{AB}b^A \frac{\partial f^4}{\partial \widetilde{u}} \frac{\partial f^B}{\partial \widetilde{u}}
	+
	\gslash_{AB} \frac{\partial f^A}{\partial \widetilde{u}} \frac{\partial f^B}{\partial \widetilde{u}}.
\end{align}
The equated coefficients of $d\widetilde{v} d \widetilde{v}$ give
\begin{align} \label{eq:dvdv}
\begin{split}
	0
	=
	-4 \Omega^2
	\frac{\partial f^3}{\partial \widetilde{v}} \left( 1 + \frac{\partial f^4}{\partial \widetilde{v}} \right)
	+
	\left\vert b \right\vert^2 \left(1 + \frac{\partial f^4}{\partial \widetilde{v}} \right)^2
	-
	2 \gslash_{AB}b^A \left( 1 + \frac{\partial f^4}{\partial \widetilde{v}} \right) \frac{\partial f^B}{\partial \widetilde{v}}
	+
	\gslash_{AB}  \frac{\partial f^A}{\partial \widetilde{v}} \frac{\partial f^B}{\partial \widetilde{v}}.
\end{split}
\end{align}
The equated coefficients of $d\widetilde{\theta}^C d \widetilde{u}$ give
\begin{multline} \label{eq:metricid4}
	- 2 \widetilde{\gslash}_{AC} \widetilde{b}^A 
	=
	-4 \Omega^2 \left[
	\left( 1 + \frac{\partial f^3}{\partial \widetilde{u}} \right) \frac{\partial f^4}{\partial \widetilde{\theta}^C}
	+
	\frac{\partial f^3}{\partial \widetilde{\theta}^C} \frac{\partial f^4}{\partial \widetilde{u}}
	\right]
	+
	2 \left\vert b \right\vert^2 \frac{\partial f^4}{\partial \widetilde{u}} \frac{\partial f^4}{\partial \widetilde{\theta}^C}
	\\
	-
	2 \gslash_{AB}b^A \left[
	\left( \delta_C^B + \frac{\partial f^B}{\partial \widetilde{\theta}^C} \right) \frac{\partial f^4}{\partial \widetilde{u}}
	+
	\frac{\partial f^B}{\partial \widetilde{u}} \frac{\partial f^4}{\partial \widetilde{\theta}^C}
	\right]
	+
	2 \gslash_{AB} \left( \delta^A_C + \frac{\partial f^A}{\partial \widetilde{\theta}^C} \right) \frac{\partial f^B}{\partial \widetilde{u}}.
\end{multline}
The equated coefficients of $d\widetilde{\theta}^C d \widetilde{v}$ give
\begin{align} \label{eq:dthetadv}
\begin{split}
	0
	=
	\
	&
	-4 \Omega^2 \left[
	\left( 1 + \frac{\partial f^4}{\partial \widetilde{v}} \right) \frac{\partial f^3}{\partial \widetilde{\theta}^C}
	+
	\frac{\partial f^4}{\partial \widetilde{\theta}^C} \frac{\partial f^3}{\partial \widetilde{v}}
	\right]
	+
	2 \left\vert b \right\vert^2 \left(1 + \frac{\partial f^4}{\partial \widetilde{v}} \right) \frac{\partial f^4}{\partial \widetilde{\theta}^A}
	\\
	&
	-
	2 \gslash_{AB}b^A \left[
	\left( 1 + \frac{\partial f^4}{\partial \widetilde{v}} \right) \left( \delta_C^B + \frac{\partial f^B}{\partial \widetilde{\theta}^C} \right)
	+
	\frac{\partial f^B}{\partial \widetilde{v}} \frac{\partial f^4}{\partial \widetilde{\theta}^C}
	\right]
	+
	2 \gslash_{AB} \left( \delta_C^A + \frac{\partial f^A}{\partial \widetilde{\theta}^C} \right) \frac{\partial f^B}{\partial \widetilde{v}}.
\end{split}
\end{align}
The equated coefficients of $d\widetilde{\theta}^C d \widetilde{\theta}^D$ give
\begin{multline} \label{eq:metricid6}
	\widetilde{\gslash}_{CD}
	=
	-2 \Omega^2 \left(
	\frac{\partial f^3}{\partial \widetilde{\theta}^C} \frac{\partial f^4}{\partial \widetilde{\theta}^D} 
	+
	\frac{\partial f^3}{\partial \widetilde{\theta}^D} \frac{\partial f^4}{\partial \widetilde{\theta}^C}
	\right)
	+
	\left\vert b \right\vert^2 \frac{\partial f^4}{\partial \widetilde{\theta}^C} \frac{\partial f^4}{\partial \widetilde{\theta}^D}
	\\
	-
	\gslash_{AB}b^A \left[
	\frac{\partial f^4}{\partial \widetilde{\theta}^C} \left( \delta_D^B + \frac{\partial f^B}{\partial \widetilde{\theta}^D} \right)
	+
	\frac{\partial f^4}{\partial \widetilde{\theta}^D} \left( \delta_C^B + \frac{\partial f^B}{\partial \widetilde{\theta}^C} \right)
	\right]
	+
	\gslash_{AB} \left( \delta_C^A + \frac{\partial f^A}{\partial \widetilde{\theta}^C} \right) \left( \delta_D^B + \frac{\partial f^B}{\partial \widetilde{\theta}^D} \right).
\end{multline}

Given a $(0,k)$ $S$-tensor $\xi$, the projection $\Pi_{\widetilde{S}} \xi$ is defined to be the restriction of $\xi$ to $\widetilde{S}$ vectors.  The relation \eqref{eq:partialxpartialtildex3} implies that, in components,
\begin{equation} \label{eq:tildeprojection}
	\Pi_{\widetilde{S}} \xi (\widetilde{e}_{A_1},\ldots, \widetilde{e}_{A_k})
	=
	\xi \Big(
	\Big( \delta_{A_1}^{B_1} + \frac{\partial f^{B_1}}{\partial \widetilde{\theta}^{A_1}} \Big) e_{B_1},
	\ldots,
	\Big( \delta_{A_k}^{B_k} + \frac{\partial f^{B_k}}{\partial \widetilde{\theta}^{A_k}} \Big) e_{B_k}
	\Big).
\end{equation}

\subsection{The nonlinear error notation $\mathcal{E}^{1,k}_{\mathfrak{D} \fsc, p}$ and $\mathcal{E}^{2,k}_{\mathfrak{D} \fsc, p}$}
\label{subsec:diffeoerrors}

In this section, we introduce a notation for nonlinear error terms which involve the diffeomorphisms
which relate two gauges.
We will assume throughout that we have the setup of Section~\ref{twogaugestwoback}.

The notation $\widetilde{\mathfrak{D}} \fsc_p$,\index{schematic notation!$\widetilde{\mathfrak{D}} \fsc_p$, schematic notation for first order derivatives of $f$}\index{double null gauge!change of gauge!$\widetilde{\mathfrak{D}} \fsc_p$, schematic notation for first order derivatives of $f$} for $p=-1,0,1$, is used to schematically denote the following first order derivatives
\[
	\widetilde{\mathfrak{D}} \fsc_{-1}
	=
	\widetilde{r\nablaslash} f^4, \widetilde{r \Omega \nablaslash_4} f^4,
	\qquad
	\widetilde{\mathfrak{D}} \fsc_{0}
	=
	\widetilde{\Omega}^{-2} \partial_{\widetilde{u}} \slashed{f}, \widetilde{r} \partial_{\widetilde{v}} \slashed{f},
	\widetilde{r\nablaslash} f^3, \widetilde{\Omega^{-1} \nablaslash_3} f^3, \widetilde{\Omega^{-1} \nablaslash_3} f^4,
	\qquad
	\widetilde{\mathfrak{D}} \fsc_{1}
	=
	\widetilde{r\Omega \nablaslash_4} f^3.
\]

The notation $\mathcal{E}^{1,k}_{\mathfrak{D} \fsc, p}$ will be used to denote a quadratic error term which involves at most $k$ derivatives of $\widetilde{\mathfrak{D}} \fsc$ and $\Phi$, each term involves at least one $\widetilde{\mathfrak{D}} \fsc$ factor, and which decays, according to the $p$ index notation, like $\tilde{r}^{-p}$.  
Similarly, the notation $\mathcal{E}^{2,k}_{\mathfrak{D} \fsc, p}$ will be used to denote a quadratic error term which involves at most $k$ derivatives of $\widetilde{\mathfrak{D}} \fsc$, $\sum_{\vert \gamma \vert =1} \widetilde{\mathfrak{D}}^{\gamma} \widetilde{\mathfrak{D}} \fsc$ and $\Phi$, each term involves at least one $\sum_{\vert \gamma \vert \leq 1} \widetilde{\mathfrak{D}}^{\gamma} \widetilde{\mathfrak{D}} \fsc$ factor, and which decays, according to the $p$ index notation, like $\tilde{r}^{-p}$.

We remark that in the above expressions, the derivatives of $f$ are taken with respect to the
operators of the gauge~\eqref{localformparamtwo}, while the Ricci coefficients and curvature components $\Phi$
are those of the gauge~\eqref{localformparam}. These are the type of terms that will naturally appear later
in the paper.

\subsubsection{Ordering the schematic diffeomorphism components $\widetilde{\mathfrak{D}} \fsc$}

In order to define the error notation more precisely, let the collection $\widetilde{\mathfrak{D}} \fsc$ be ordered as follows
\[
	\widetilde{\mathfrak{D}} \fsc^1 = \widetilde{r\nablaslash} f^4,
	\quad
	\widetilde{\mathfrak{D}} \fsc^2 = \widetilde{r \Omega \nablaslash_4} f^4,
\]
\[
	\widetilde{\mathfrak{D}} \fsc^3 = \widetilde{\Omega}^{-2} \partial_{\widetilde{u}} \slashed{f},
	\quad
	\widetilde{\mathfrak{D}} \fsc^4 = \widetilde{r} \partial_{\widetilde{v}} \slashed{f},
	\quad
	\widetilde{\mathfrak{D}} \fsc^5 = \widetilde{r\nablaslash} f^3,
	\quad
	\widetilde{\mathfrak{D}} \fsc^6 = \widetilde{\Omega^{-1} \nablaslash_3} f^3,
	\quad
	\widetilde{\mathfrak{D}} \fsc^7 = \widetilde{\Omega^{-1} \nablaslash_3} f^4,
	\quad
	\widetilde{\mathfrak{D}} \fsc^8 = \widetilde{r\Omega \nablaslash_4} f^3.
\]

\subsubsection{The notation $\widetilde{\mathfrak{D}}^{k+1} \fsc$ and $\widetilde{\mathfrak{D}}^{k+1} \fsc_p$}

Consider some $k \geq 0$.  Define $\{ (1,0,0), (0,1,0), (0,0,1)\}^0 := \{ (0,0,0)\}$.  Given $i \in \{ 1,\ldots, 8\}$ and $\gamma \in \{ (1,0,0), (0,1,0), (0,0,1)\}^k$, define\index{schematic notation!$\widetilde{\mathfrak{D}}^{k+1} \fsc \cdot (i,\gamma)$, schematic notation for higher order derivatives of $f$}\index{double null gauge!change of gauge!$\widetilde{\mathfrak{D}}^{k+1} \fsc \cdot (i,\gamma)$, schematic notation for higher order derivatives of $f$}
\begin{equation} \label{eq:fschematicderivatives}
	\widetilde{\mathfrak{D}}^{k+1} \fsc \cdot (i,\gamma)
	:=
	\widetilde{\mathfrak{D}}^{\gamma} \widetilde{\mathfrak{D}} \fsc^i.
\end{equation}

As an example, one can write
\[
	\widetilde{\Omega^{-1} \nablaslash_3} \widetilde{r\Omega \nablaslash_4} f^3
	=
	\widetilde{\mathfrak{D}}^2 \fsc \cdot (8,(0,1,0)),
	\qquad
	\text{and}
	\qquad
	\widetilde{r\nablaslash} f^4 = \widetilde{\mathfrak{D}}^1 \fsc \cdot (1, (0,0,0)).
\]

Define now the sets
\[
	Q_{-1} = \{ 1,2\}, 
	\qquad
	Q_0 = \{ 3,\ldots,7\}, 
	\qquad
	Q_1 = \{ 8 \}.
\]
Consider some $k \geq 0$.  Given $p \in \{ -1,0,1 \}$ and $\gamma \in \{ (1,0,0), (0,1,0), (0,0,1)\}^k$, if $i \in Q_p$ then we often add a $p$ subscript in \eqref{eq:fschematicderivatives} for\index{schematic notation!$\widetilde{\mathfrak{D}}^{k+1} \fsc_p \cdot (i,\gamma)$, schematic notation for higher order derivatives of $f$ showing $p$ dependence}\index{double null gauge!change of gauge!$\widetilde{\mathfrak{D}}^{k+1} \fsc_p \cdot (i,\gamma)$, schematic notation for higher order derivatives of $f$ showing $p$ dependence} emphasis, i.\@e.\@
\[
	\widetilde{\mathfrak{D}}^{k+1} \fsc_p \cdot (i,\gamma)
	:=
	\widetilde{\mathfrak{D}}^{\gamma} \widetilde{\mathfrak{D}} \fsc^i
	\qquad
	\text{if }
	i \in Q_p.
\]

\subsubsection{The notation $(\widetilde{\mathfrak{D}}^{k+1} \fsc)^l$}
\label{subsubsec:Dfnotation}

Consider now some $k \geq 0$, $l \geq 1$ and some
\[
	G = \{ {}^{j,m}G_{k_1\ldots k_{l'}},  {}^jL_{k_1\ldots k_{l'}} \}_{m=0,\ldots,l',k_1+\ldots + k_{l'} \leq k, l' \geq l, j \geq 1},
\]
where
\[
	{}^{j,m}G_{k_1\ldots k_{l'}}
	\in
	\big\{ (i,\gamma) \mid i \in \{ 1,\ldots, 8\}, \gamma \in \{ (1,0,0), (0,1,0), (0,0,1)\}^{k_m} \big\} \cup \big\{ 0 \big\},
\]
for all $m=0,\ldots,l'$ and ${}^jL_{k_1\ldots k_{l'}}$ is a trace set of some order $d\geq 0$, as in Section \ref{subsec:traceindex}, for all $j \geq 1$, $k_1+\ldots + k_{l'} \leq k$ and $l' \geq l$.  For such $k \geq 0$, $l \geq 1$ and $G$, define\index{schematic notation!$\left( \widetilde{\mathfrak{D}}^{k+1} \fsc \right)^{l} \cdot G$, schematic notation for products of higher order derivatives of $f$}\index{double null gauge!change of gauge!$\left( \widetilde{\mathfrak{D}}^{k+1} \fsc \right)^{l} \cdot G$, schematic notation for products of higher order derivatives of $f$}
\[
	\left( \widetilde{\mathfrak{D}}^{k+1} \fsc \right)^{l} \cdot G
	:=
	\sum_{l' \geq l}
	\sum_{k_1+\ldots+k_{l'} \leq k}
	\sum_{j \geq 1}
	\big(
	\widetilde{\mathfrak{D}}^{k_1+1} \fsc \cdot {}^{j,1}G_{k_1\ldots k_{l'}} \otimes \ldots \otimes \widetilde{\mathfrak{D}}^{k_{l'}+1} \fsc \cdot {}^{j,l'}G_{k_1 \ldots k_{l'}}
	\big)
	\cdot_{\widetilde{\gslash}}
	{}^jL_{k_1\ldots k_{l'}}.
\]
Such expressions are only ever considered when sufficiently many of the components of $G$ and $L$ vanish so that each object appearing in the above summation is a tensor field of the same type, and so that the allowed ranges of $l_1,l_2$, $p_1,p_2$ and $j \geq 1$ are finite, and so the summation is indeed well defined.

Note that the operation $\cdot_{\widetilde{\gslash}} {}^jL_{k_1\ldots k_{l'}}$ involves multiplying with admissible functions on $\widetilde{\mathcal{Z}}$ or taking traces over the $\ \widetilde{} \ $ indices.
Note also that
\[
	\left( \widetilde{\mathfrak{D}}^{k+1} \fsc \right)^{l} \cdot G
	\in
	\Gamma \big( (T^* \widetilde{S})^s \otimes F^*(TS)^r \big),
\]
for some $r$ and $s$.

\subsubsection{The notation $(\widetilde{\mathfrak{D}}^{k+1} \fsc_p)^l$}

Consider now some $k \geq 0$, $l \geq 1$, $p \geq 0$ and some
\[
	G = \{ {}^{j,m}G^{p_0\ldots p_{l'}}_{k_1\ldots k_{l'}}, {}^j L^{p_0\ldots p_{l'}}_{k_1\ldots k_{l'}} \}_{\substack{
	m=0,\ldots,l',k_1+\ldots + k_{l'} \leq k, 
	\\
	p_0 + \ldots + p_{l'} \geq p, l'\geq l, j \geq 1}}
	,
\]
where
\[
	{}^{j,m}G^{p_0\ldots p_{l'}}_{k_1\ldots k_{l'}}
	\in
	\big\{ (i,\gamma) \mid i \in Q_{p_m}, \gamma \in \{ (1,0,0), (0,1,0), (0,0,1)\}^{k_m} \big\} \cup \big\{ 0 \big\},
\]
for all $m=0,\ldots,l'$, and ${}^j L_{k_1\ldots k_{l'}}$ is a trace set of some order $d \geq 0$, as in Section \ref{subsec:traceindex}, for all $k_1+\ldots + k_{l'} \leq k$, $p_0+\ldots + p_{l'} \geq p$, $j \geq 1$ and $l' \geq l$.  For such $k \geq 0$, $l \geq 1$ and $G$, define\index{schematic notation!$\left( \widetilde{\mathfrak{D}}^{k+1} \fsc_p \right)^{l} \cdot G$, schematic notation for products of higher order derivatives of $f$ with $p$ dependence}\index{double null gauge!change of gauge!$\left( \widetilde{\mathfrak{D}}^{k+1} \fsc_p \right)^{l} \cdot G$, schematic notation for products of higher order derivatives of $f$ with $p$ dependence}
\[
	\left( \widetilde{\mathfrak{D}}^{k+1} \fsc_p \right)^{l} \cdot G
	:=
	\sum_{l' \geq l}
	\sum_{\substack{
	k_1+\ldots+k_{l'} \leq k
	\\
	p_0 + \ldots+ p_{l'} \geq p
	}}
	\sum_{j \geq 1}
	\widetilde{r}^{-p_0}
	\big(
	\widetilde{\mathfrak{D}}^{k_1+1} \fsc_{p_1} \cdot {}^{j,1} G_{k_1\ldots k_{l'}}^{p_0\ldots p_{l'}} \otimes \ldots \otimes \widetilde{\mathfrak{D}}^{k_{l'}+1} \fsc_{p_{l'}} \cdot {}^{j,1} G_{k_1 \ldots k_{l'}}^{p_0\ldots p_{l'}}
	\big)
	\cdot_{\widetilde{\gslash}}
	{}^j L_{k_1\ldots k_{l'}}^{p_0\ldots p_{l'}}.
\]
Again, such expressions will only ever be considered when sufficiently many components of $G$ and $L$ vanish so that the summation is well defined.

\subsubsection{The notation $\mathcal{E}^{1,k}_{\mathfrak{D} \fsc}$ and $\mathcal{E}^{1,k}_{\mathfrak{D} \fsc,p}$}

Given appropriate $H = \{{}^j H^{l_1l_2}_{k_1k_2}\}$ and $G = \{{}^j G^{l_1l_2}_{k_1k_2}\}$, where each ${}^j H^{l_1l_2}_{k_1k_2}$ is as in Section \ref{sec:nlenotation} and each ${}^j G^{l_1l_2}_{k_1k_2}$ is as above, and $K = \{ {}^j K^{l_1l_2}_{k_1k_2} \}$, where each ${}^j K^{l_1l_2}_{k_1k_2}$ is a trace set of some order $d\geq 0$, as in Section \ref{subsec:traceindex}, define\index{schematic notation!$\mathcal{E}^{1,k}_{\mathfrak{D} \fsc}(H,G,K)$, nonlinear error notation associated to $f$}\index{double null gauge!change of gauge!$\mathcal{E}^{1,k}_{\mathfrak{D} \fsc}(H,G,K)$, nonlinear error notation associated to $f$} the $\widetilde{S}$ tensor
\[
	\mathcal{E}^{1,k}_{\mathfrak{D} \fsc}(H,G,K)
	=
	\Pi_{\widetilde{S}}
	\sum_{\substack{k_1+k_2\leq k \\ l_1+l_2 \geq 2, l_2 \geq 1}}
	\sum_{ j \geq 1}
	\Big(
	(\mathfrak{D}^{k_1} \Phi)^{l_1} \cdot {}^jH^{l_1 l_2}_{k_1k_2}
	\otimes
	(\widetilde{\mathfrak{D}}^{k_2+1} \fsc)^{l_2} \cdot {}^jG^{l_1 l_2}_{k_1k_2}
	\Big)
	\cdot_{\gslash} {}^j K^{l_1 l_2}_{k_1k_2},
\]
where $\Pi_{\widetilde{S}}$ denotes the projection to the $\widetilde{S}$ spheres (see \eqref{eq:tildeprojection}).
Note that $l_2\geq 1$ in the summation, so that each term in the nonlinear error $\mathcal{E}^{1,k}_{\mathfrak{D} \fsc}(H,G,K)$ involves at least one factor involving (derivatives of) a diffeomorphism component $\widetilde{\mathfrak{D}} \fsc$.

Similarly, for given $k$, $p$, and for appropriate $H = \{ {}^j H^{l_1l_2}_{\substack{k_1k_2 \\ p_1p_2}} \}$, $G = \{ {}^j G^{l_1l_2}_{\substack{k_1k_2 \\ p_1p_2}}\}$ and $K = \{ {}^j K^{l_1l_2}_{\substack{k_1k_2 \\ p_1p_2}} \}$, define\index{schematic notation!$\mathcal{E}^{1,k}_{\mathfrak{D} \fsc,p}(H,G,K)$, nonlinear error notation associated to $f$ with $p$ dependence}\index{double null gauge!change of gauge!$\mathcal{E}^{1,k}_{\mathfrak{D} \fsc,p}(H,G,K)$, nonlinear error notation associated to $f$ with $p$ dependence}
\[
	\mathcal{E}^{1,k}_{\mathfrak{D} \fsc,p}(H,G,K)
	=
	\Pi_{\widetilde{S}}
	\sum_{\substack{k_1+k_2\leq k \\ l_1+l_2 \geq 2, l_2 \geq 1 \\ p_1+p_2 \geq p}}
	\sum_{j \geq 1}
	\Big(
	(\mathfrak{D}^{k_1} \Phi_{p_1})^{l_1} \cdot {}^j H^{l_1l_2}_{\substack{k_1k_2 \\ p_1p_2}}
	\otimes
	(\widetilde{\mathfrak{D}}^{k_2+1} \fsc_{p_2})^{l_2} \cdot {}^j G^{l_1l_2}_{\substack{k_1k_2 \\ p_1p_2}}
	\Big)
	\cdot_{\gslash}
	{}^j K^{l_1l_2}_{\substack{k_1k_2 \\ p_1p_2}}.
\]

\vskip1pc
\noindent\fbox{%
    \parbox{6.35in}{%
 \emph{Such expressions are only ever considered when sufficiently many of the components of $H$, $G$ and $K$ vanish so that each object appearing in the above summations is a tensor field of the same type, and so that the allowed ranges of $l_1,l_2$, $p_1,p_2$ and $j \geq 1$ are finite, and so each summation is indeed well defined.}
    }%
}
\vskip1pc
Note that, for any $\gamma$,
\[
	\widetilde{\mathfrak{D}}^{\gamma} \mathcal{E}^{1,k}_{\mathfrak{D} \fsc,p}(H,G,K)
	=
	\mathcal{E}^{1,k+\vert \gamma \vert}_{\mathfrak{D} \fsc,p}(H^{\gamma},G^{\gamma},K^{\gamma})
\]
for some $H^{\gamma},G^{\gamma},K^{\gamma}$.

We will typically abuse notation and write ``$\mathcal{E}^{1,k}_{\mathfrak{D} \fsc,p}$'' to mean ``$\mathcal{E}^{1,k}_{\mathfrak{D} \fsc,p}(H,G,K)$ for some $H,G,K$'', etc.

\subsubsection{The notation $\mathcal{E}^{2,k}_{\mathfrak{D} \fsc}$ and $\mathcal{E}^{2,k}_{\mathfrak{D} \fsc,p}$}

The nonlinear error notation $\mathcal{E}^{2,k}_{\mathfrak{D} \fsc}(H,G,K)$ is defined, for appropriate $H,G,K$, by
\[
	\mathcal{E}^{2,k}_{\mathfrak{D} \fsc}(H,G,K)
	=
	\Pi_{\widetilde{S}}
	\sum_{\substack{k_1+k_2\leq k+1, k_1 \leq k \\ l_1+l_2 \geq 2, l_2 \geq 1 }}
	\sum_{j \geq 1}
	\Big(
	(\mathfrak{D}^{k_1} \Phi)^{l_1} \cdot {}^jH^{l_1 l_2}_{k_1k_2}
	\otimes
	(\widetilde{\mathfrak{D}}^{k_2+1} \fsc)^{l_2} \cdot {}^jG^{l_1 l_2}_{k_1k_2}
	\Big)
	\cdot_{\gslash}
	{}^j K^{l_1 l_2}_{k_1k_2},
\]
so that each term in $\mathcal{E}^{2,k}_{\mathfrak{D} \fsc}(H,G,K)$ involves a total of at most $k+1$ derivatives of $\widetilde{\mathfrak{D}} \fsc$ and $\Phi$, but at most $k$ derivatives of $\Phi$.  Similarly, for appropriate $H,G,K$, define
\[
	\mathcal{E}^{2,k}_{\mathfrak{D} \fsc,p}(H,G,K)
	=
	\Pi_{\widetilde{S}}
	\sum_{\substack{k_1+k_2\leq k+1, k_1 \leq k \\ l_1+l_2 \geq 2, l_2 \geq 1 \\ p_1+p_2 \geq p}}
	\sum_{j \geq 1}
	\Big(
	(\mathfrak{D}^{k_1} \Phi_{p_1})^{l_1} \cdot {}^j H^{l_1l_2}_{\substack{k_1k_2 \\ p_1p_2}}
	\otimes
	(\widetilde{\mathfrak{D}}^{k_2+1} \fsc_{p_2})^{l_2} \cdot {}^j G^{l_1l_2}_{\substack{k_1k_2 \\ p_1p_2}}
	\Big)
	\cdot_{\gslash}
	{}^j K^{l_1l_2}_{\substack{k_1k_2 \\ p_1p_2}}.
\]
Here $\Pi_{\widetilde{S}}$ again denotes the projection to the $\widetilde{S}$ spheres (see \eqref{eq:tildeprojection}).
\vskip1pc
\noindent\fbox{%
    \parbox{6.35in}{%
 \emph{Again, such expressions are only ever considered when sufficiently many of the components of $H$, $G$ and $K$ vanish so that each object appearing in the above summations is a tensor field of the same type, and so that the allowed ranges of $l_1,l_2$, $p_1,p_2$ and $j \geq 1$ are finite, and so each summation is indeed well defined.}
    }%
}
\vskip1pc

Note that, for any $\gamma$,
\[
	\widetilde{\mathfrak{D}}^{\gamma} \mathcal{E}^{2,k}_{\mathfrak{D} \fsc,p}(H,G,K)
	=
	\mathcal{E}^{2,k+\vert \gamma \vert}_{\mathfrak{D} \fsc,p}(H^{\gamma},G^{\gamma},K^{\gamma})
\]
for some $H^{\gamma},G^{\gamma},K^{\gamma}$.

Again, we will typically abuse notation and write ``$\mathcal{E}^{2,k}_{\mathfrak{D} \fsc,p}$'' to mean ``$\mathcal{E}^{2,k}_{\mathfrak{D} \fsc,p}(H,G,K)$ for some $H,G,K$'', etc.

\subsection{The change of gauge relations}
\label{changeofgaugerelationssection}

Recall the nonlinear error notation $\mathcal{E}^{1,k}_{\mathfrak{D} \fsc,p}$ and $\mathcal{E}^{2,k}_{\mathfrak{D} \fsc,p}$ from Section \ref{subsec:diffeoerrors}. 

The relations \eqref{eq:metricid1}--\eqref{eq:metricid6} between the metric coefficients can be written in the following schematic form.  Recall the notation $\partial_{\widetilde{u}} \slashed{f}, \partial_{\widetilde{v}} \slashed{f}$ from Section \ref{newdiffysec}.  Since $\partial_{\widetilde{u}} \slashed{f}$ and $\partial_{\widetilde{v}} \slashed{f}$ are $S$-vectors one can also define their $\flat$ with respect to $\gslash$,
\[
	\partial_{\widetilde{u}} \slashed{f}^{\flat}(\partial_{\theta^A})
	:=
	\gslash_{AB} \partial_{\widetilde{u}} f^B,
	\qquad
	\partial_{\widetilde{v}} \slashed{f}^{\flat}(\partial_{\theta^A})
	:=
	\gslash_{AB} \partial_{\widetilde{v}} f^B
	.
\]
Recall also the operator $\Pi_{\widetilde{S}}$, which projects $S$-tensor fields to $\widetilde{S}$-tensor fields (see \eqref{eq:tildeprojection}).  The relations of Propositions \ref{prop:metricrelations}, \ref{prop:Riccirelations} and \ref{prop:curvaturerelations} are relations between $\widetilde{S}$ tensor fields.

\begin{proposition}[Change of gauge relations for metric components] \label{prop:metricrelations}
	With the setup of Section~\ref{twogaugestwoback}, the metric coefficients of the two gauges, viewed as spacetime tensors, satisfy the relations
	\begin{align}
		\Omega^2 \partial_{\widetilde{u}} f^4
		=
		\
		&
		\mathcal{E}^{1,0}_{\mathfrak{D} \fsc,0};
		\label{eq:metriccomp1}
		\\
		\Omega^2 \left( 1 + \partial_{\widetilde{v}} f^4 \right) \partial_{\widetilde{v}} f^3
		=
		\
		&
		\mathcal{E}^{1,0}_{\mathfrak{D} \fsc,2};
		\label{eq:metriccomp2}
		\\
		2 \Omega^2 \widetilde{\nablaslash} f^4
		=
		\
		&
		\Pi_{\widetilde{S}} \partial_{\widetilde{u}} \slashed{f}^{\flat}
		+
		\widetilde{b^{\flat}}
		+
		\mathcal{E}^{1,0}_{\mathfrak{D} \fsc,0};
		\label{eq:metriccomp3}
		\\
		2 \Omega^2 \widetilde{\nablaslash} f^3
		=
		\
		&
		\Pi_{\widetilde{S}} \partial_{\widetilde{v}} \slashed{f}^{\flat}
		-
		\Pi_{\widetilde{S}} b^{\flat}
		+
		\mathcal{E}^{1,0}_{\mathfrak{D} \fsc,1};
		\label{eq:metriccomp4}
	\end{align}
	\begin{align}
		\Pi_{\widetilde{S}} r^2 \gamma
		-
		\widetilde{r^2 \gamma}
		=
		\
		&
		\left( \widetilde{\gslash} - \widetilde{r^2 \gamma} \right)
		-
		\Pi_{\widetilde{S}} \left( \gslash - r^2 \gamma \right)
		+
		\mathcal{E}^{1,0}_{\mathfrak{D} \fsc,1};
		\label{eq:metriccomp5}
		\\
		\Omega^2 \left(
		\partial_{\widetilde{u}} f^3
		+
		\partial_{\widetilde{v}} f^4
		\right)
		+
		\Omega_{\circ,M}^2
		-
		\widetilde{\Omega}_{\circ,\widetilde{M}}^2
		=
		\
		&
		\left(
		\widetilde{\Omega}^2
		-
		\widetilde{\Omega}_{\circ,\widetilde{M}}^2
		\right)
		-
		\left(
		\Omega^2
		-
		\Omega_{\circ,M}^2
		\right)
		+
		\mathcal{E}^{1,0}_{\mathfrak{D} \fsc,0}.
		\label{eq:metriccomp6}
	\end{align}

\end{proposition}

\begin{proof}
	The identities \eqref{eq:metriccomp1}, \eqref{eq:metriccomp2} and \eqref{eq:metriccomp6} are simply the relations \eqref{eq:metricid2}, \eqref{eq:dvdv}, and \eqref{eq:metricid1} respectively written using the nonlinear error notation of Section \ref{subsec:diffeoerrors}.  For \eqref{eq:metriccomp3}, one notes that the relation \eqref{eq:metricid4} can be written as
	\[
		2 \Omega^2 \frac{\partial f^4}{\partial \widetilde{\theta}^C}
		=
		\widetilde{\gslash}_{AC} \widetilde{b}^A
		+
		\gslash_{AB} \left( \delta^A_C + \frac{\partial f^A}{\partial \widetilde{\theta}^C} \right) \frac{\partial f^B}{\partial \widetilde{u}}
		+
		\mathcal{E}^{1,0}_{\mathfrak{D} \fsc,0} (\widetilde{e}_C),
	\]
	from which \eqref{eq:metriccomp3} follows. The relations \eqref{eq:metriccomp4} and \eqref{eq:metriccomp5} follow similarly from \eqref{eq:dthetadv} and \eqref{eq:metriccomp6} respectively.
\end{proof}

Relations between the Ricci coefficients of the two gauges can similarly be obtained.  Note that certain nonlinear error terms have worse $r$ behaviour than the typical terms (such as the term $\partial_{\widetilde{v}} f^4 \widetilde{\nablaslash} \partial_{\widetilde{u}} f^3$ in the relation \eqref{eq:Riccicomp3}), according to the $p$ index notation of Section \ref{subsec:diffeoerrors}, and so are stated explicitly.

\begin{proposition}[Change of gauge relations for Ricci coefficients] \label{prop:Riccirelations}
	With the setup of Section~\ref{twogaugestwoback}, the Ricci coefficients of the two gauges, viewed as spacetime tensors, satisfy the relations
	\begin{align}
		\widetilde{\Omega \hat{\chi}}
		-
		\Pi_{\widetilde{S}} \Omega \hat{\chi}
		=
		\
		&
		-
		2 \Omega^2 \widetilde{\Dslash_2^*} \widetilde{\nablaslash} f^3
		+
		\mathcal{E}^{2,0}_{\mathfrak{D} \fsc,2};
		\label{eq:Riccicomp1}
		\\
		\widetilde{\Omega \hat{\chibar}}
		-
		\Pi_{\widetilde{S}} \Omega \hat{\chibar}
		=
		\
		&
		-
		2 \Omega^2 \widetilde{\Dslash_2^*} \widetilde{\nablaslash} f^4
		+
		\mathcal{E}^{2,0}_{\mathfrak{D} \fsc,1};
		\label{eq:Riccicomp2}
		\\
		\widetilde{\eta}
		-
		\Pi_{\widetilde{S}} \eta
		=
		\
		&
		\left( 1 + \frac{\partial f^4}{\partial \widetilde{v}} \right) \widetilde{\nablaslash} \frac{\partial f^3}{\partial \widetilde{u}}
		+
		(2 \Omega \omegabarhat - \frac{1}{2} \Omega \tr \chibar) \widetilde{\nablaslash} f^3
		+
		\frac{1}{2} \Omega \tr \chi \widetilde{\nablaslash} f^4
		\label{eq:Riccicomp3}
		+
		\mathcal{E}^{2,0}_{\mathfrak{D} \fsc,2};
		\\
		\widetilde{\etabar}
		-
		\Pi_{\widetilde{S}} \etabar
		=
		\
		&
		\left( 1 + \frac{\partial f^3}{\partial \widetilde{u}} \right) \widetilde{\nablaslash} \frac{\partial f^4}{\partial \widetilde{v}}
		+
		(2 \Omega \omegahat - \frac{1}{2} \Omega \tr \chi) \widetilde{\nablaslash} f^4
		+
		\frac{1}{2} \Omega \tr \chibar \widetilde{\nablaslash} f^3
		\label{eq:Riccicomp4}
		+
		\mathcal{E}^{2,0}_{\mathfrak{D} \fsc,2};
	\end{align}
	\begin{align}
		\left(
		\widetilde{\Omega \omegahat} - \widetilde{(\Omega \omegahat)}_{\circ,\widetilde{M}}
		\right)
		-
		\left(
		\Omega \omegahat - (\Omega \omegahat)_{\circ,M}
		\right)
		=
		&
		\
		(\Omega \omegahat)_{\circ,M}
		-
		\widetilde{(\Omega \omegahat)}_{\circ,\widetilde{M}}
		+
		\Omega \omegahat \frac{\partial f^4}{\partial \widetilde{v}}
		+
		\frac{1}{2} \frac{\partial^2 f^4}{\partial \widetilde{v}^2}
		\label{eq:Riccicomp5}
		+
		\mathcal{E}^{2,0}_{\mathfrak{D} \fsc,1};
		\\
		\left(
		\widetilde{\Omega \omegabarhat} - \widetilde{(\Omega \omegabarhat)}_{\circ,\widetilde{M}}
		\right)
		-
		\left(
		\Omega \omegabarhat - (\Omega \omegabarhat)_{\circ,M}
		\right)
		=
		&
		\
		(\Omega \omegabarhat)_{\circ,M}
		-
		\widetilde{(\Omega \omegabarhat)}_{\circ,\widetilde{M}}
		+
		\Omega \omegabarhat \frac{\partial f^3}{\partial \widetilde{u}}
		+
		\frac{1}{2} \frac{\partial^2 f^3}{\partial \widetilde{u}^2}
		\label{eq:Riccicomp6}
		+
		\mathcal{E}^{2,0}_{\mathfrak{D} \fsc,0};
	\end{align}
	and $\Omega \tr \chi$ and $\Omega \tr \chibar$ satisfy
	\begin{align}
		\left( \widetilde{\Omega \tr \chi} - \widetilde{(\Omega \tr \chi)}_{\circ,\widetilde{M}} \right)
		-
		\left( \Omega \tr \chi - (\Omega \tr \chi)_{\circ,M} \right)
		=
		\
		&
		(\Omega \tr \chi)_{\circ,M}
		-
		\widetilde{(\Omega \tr \chi)}_{\circ,\widetilde{M}}
		+
		2 \Omega^2  \widetilde{\Deltaslash} f^3
		+
		\Omega \tr \chi \frac{\partial f^4}{\partial \widetilde{v}}
		\label{eq:Riccicomp7}
		+
		\mathcal{E}^{2,0}_{\mathfrak{D} \fsc,2};
		\\
		\left( \widetilde{\Omega \tr \chibar} - \widetilde{(\Omega \tr \chibar)}_{\circ,\widetilde{M}} \right)
		-
		\left( \Omega \tr \chibar - (\Omega \tr \chibar)_{\circ,M} \right)
		=
		\
		&
		(\Omega \tr \chibar)_{\circ,M}
		-
		\widetilde{(\Omega \tr \chibar)}_{\circ,\widetilde{M}}
		+
		2 \Omega^2 \widetilde{\Deltaslash} f^4
		+
		\Omega \tr \chibar \frac{\partial f^3}{\partial \widetilde{u}}
		\label{eq:Riccicomp8}
		+
		\mathcal{E}^{2,0}_{\mathfrak{D} \fsc,1}.
	\end{align}
\end{proposition}

\begin{proof}
	Consider first \eqref{eq:Riccicomp1} and \eqref{eq:Riccicomp7}.  Writing
	\[
		\widetilde{\Omega \chi}(\widetilde{e}_A,\widetilde{e}_B)
		=
		g(\nabla_{\widetilde{e}_A} \widetilde{\Omega} \widetilde{e}_4,\widetilde{e}_B)
		,
	\]
	expanding $\widetilde{e}_A,\widetilde{e}_B, \widetilde{\Omega} \widetilde{e}_4$ using the relations \eqref{eq:gaugerelationA} and \eqref{eq:gaugerelation4}, and using the metric relation \eqref{eq:metriccomp4}, it follows that
	\[
		\widetilde{\Omega \chi}(\widetilde{e}_A,\widetilde{e}_B)
		=
		(1 + \partial_{\widetilde{v}} f^4 ) \Omega \chi({H_A}^C e_C,{H_B}^D e_D)
		+
		2 \Omega^2 \widetilde{\nablaslash}{}^2_{\partial_{\widetilde{\theta}^A},\partial_{\widetilde{\theta}^B}} f^3
		+
		\mathcal{E}^{2,0}_{\mathfrak{D} \fsc,2}(\widetilde{e}_A,\widetilde{e}_B),
	\]
	where ${H_A}^C = \delta_A^C + \partial_{\widetilde{\theta}^A} f^C$, and so
	\[
		\widetilde{\Omega \chi}
		=
		(1 + \partial_{\widetilde{v}} f^4 ) \Pi_{\widetilde{S}} \Omega \chi
		+
		2 \Omega^2 \widetilde{\nablaslash}{}^2 f^3
		+
		\mathcal{E}^{2,0}_{\mathfrak{D} \fsc,2}.
	\]
	The proof of \eqref{eq:Riccicomp1} and \eqref{eq:Riccicomp7} then follow from decomposing into trace and trace free parts, with respect to $\widetilde{\gslash}$, and using the relation \eqref{eq:metriccomp5}.  The proof of the other relations is similar.
\end{proof}

The following proposition similarly gives relations between the curvature components of the two gauges.

\begin{proposition}[Change of gauge relations for curvature components] \label{prop:curvaturerelations}
	With the setup of Section~\ref{twogaugestwoback}, the curvature components of the two gauges, viewed as spacetime tensors, satisfy the relations
	\begin{align}
		\widetilde{\Omega^2 \alpha}
		-
		\Pi_{\widetilde{S}} \Omega^2 \alpha
		=
		\
		&
		\mathcal{E}^{1,0}_{\mathfrak{D} \fsc,4};
		\label{eq:curvaturecomp1}
		\\
		\widetilde{\Omega^{-2} \alphabar}
		-
		\Pi_{\widetilde{S}} \Omega^{-2} \alphabar
		=
		\
		&
		\mathcal{E}^{1,0}_{\mathfrak{D} \fsc,1};
		\label{eq:curvaturecomp2}
		\\
		\widetilde{\Omega \beta}
		-
		\Pi_{\widetilde{S}} \Omega \beta
		=
		\
		&
		3 \Omega^2 \rho \widetilde{\nablaslash} f^3
		+
		\mathcal{E}^{1,0}_{\mathfrak{D} \fsc,4};
		\label{eq:curvaturecomp3}
		\\
		\widetilde{\Omega^{-1} \betabar}
		-
		\Pi_{\widetilde{S}} \Omega^{-1} \betabar
		=
		\
		&
		-
		3 \rho \widetilde{\nablaslash} f^4
		+
		\mathcal{E}^{1,0}_{\mathfrak{D} \fsc,2};
		\label{eq:curvaturecomp4}
		\\
		(\widetilde{\rho} - \widetilde{\rho}_{\circ,\widetilde{M}})
		-
		(\rho - \rho_{\circ,M})
		=
		\
		&
		\rho_{\circ,M}
		-
		\widetilde{\rho}_{\circ,\widetilde{M}}
		+
		\mathcal{E}^{1,0}_{\mathfrak{D} \fsc,3};
		\label{eq:curvaturecomp5}
		\\
		\widetilde{\sigma}
		-
		\sigma
		=
		\
		&
		\mathcal{E}^{1,0}_{\mathfrak{D} \fsc,3}.
		\label{eq:curvaturecomp6}
	\end{align}
\end{proposition}

\begin{proof}
	Consider first \eqref{eq:curvaturecomp1}.  Using \eqref{eq:curvtab1},
	\[
		\widetilde{\Omega^2\alpha}_{AB}
		=
		R(\widetilde{e}_A, \widetilde{\Omega} \widetilde{e}_4,\widetilde{e}_B,\widetilde{\Omega}\widetilde{e}_4).
	\]
	Expanding $\widetilde{e}_1$, $\widetilde{e}_2$ and $\widetilde{e}_4$ in terms of $e_1,\ldots,e_4$ using the relations \eqref{eq:gaugerelationA}--\eqref{eq:gaugerelation4}, evaluating the resulting curvature components using the relations \eqref{eq:curvtab1}--\eqref{eq:curvtab5}, and checking each of the resulting terms has the correct form, one sees that
	\[
		\widetilde{\Omega^2\alpha}_{AB}
		=
		(\delta_A^C + \partial_{\widetilde{\theta}^A} f^C ) (\delta_B^D + \partial_{\widetilde{\theta}^B} f^D )
		R(e_C, \Omega e_4,e_D, \Omega e_4)
		+
		\mathcal{E}^{1,0}_{\mathfrak{D} \fsc,4} (\widetilde{e}_A, \widetilde{e}_B),
	\]
	and \eqref{eq:curvaturecomp1} follows.
	
	The identities \eqref{eq:curvaturecomp2}--\eqref{eq:curvaturecomp6} follow similarly.  Note that the term $\frac{1}{\widetilde{\Omega}} \left( \partial_{\widetilde{u}}f^A e_A + \widetilde{b} \right)$ in $\widetilde{e}_3$ (see \eqref{eq:gaugerelation3}) can be replaced by $2\frac{\Omega^2}{\widetilde{\Omega}} \widetilde{\nablaslash} f^4$, plus nonlinear terms, using the relation \eqref{eq:metriccomp3}.
\end{proof}

\subsection{Norms and derivatives of projections of $S$-tensors}
\label{subsec:normsderivsStensors}

Recall the setup of Section~\ref{twogaugestwoback} and recall the definition \eqref{normforStensors} of the norm of 
an $S$-tensor.  In this section the $\widetilde{S}$ norms and derivatives of projections of $S$-tensors are related to the corresponding $S$ norms and derivatives.

Proposition \ref{prop:tildenorm} relates the $\gslash$ norm of an $S$-tensor field to the $\widetilde{\gslash}$ norm of its projection to $\widetilde{S}$.  Proposition \ref{prop:tildederivStensor} relates the $\widetilde{\mathfrak{D}}$ derivative of the projection of an $S$-tensor field to an $\widetilde{S}$-tensor field, to the projection of the corresponding $\mathfrak{D}$ derivative of the tensor field.

\begin{proposition}[$\widetilde{\gslash}$ norm of projection to $\widetilde{S}$ of an $S$-tensor field] \label{prop:tildenorm}
	With the setup of Section~\ref{twogaugestwoback}, assuming that the map \eqref{eq:dslashFslash} is a vector space isomorphism, for any multi index $\gamma$ and any $S$-geometric quantity $\Phi$,
	\[
		\vert \mathfrak{D}^{\gamma} \Phi_p \vert_{\gslash}
		=
		\vert \Pi_{\widetilde{S}} \mathfrak{D}^{\gamma} \Phi_p \vert_{\widetilde{\gslash}}
		+
		\mathcal{E}^{1,\vert \gamma \vert}_{\mathfrak{D} \fsc,p}.
	\]
\end{proposition}

\begin{proof}
Let $G_A^B$ denote the components of the inverse of $\big( \delta_{A}^{B} + \frac{\partial f^{B}}{\partial \widetilde{\theta}^{A}} \big)$, so that
\[
	G_A^B \Big( \delta_{B}^{C} + \frac{\partial f^{C}}{\partial \widetilde{\theta}^{B}} \Big) = \delta_A^B.
\]
If $\xi$ and $\xi'$ are two $(0,k)$ $S$-tensors, one can define
\[
	(\Pi_{\widetilde{S}} \gslash) ( \Pi_{\widetilde{S}} \xi, \Pi_{\widetilde{S}} \xi')
	=
	(\Pi_{\widetilde{S}} \gslash)^{A_1 B_1} \ldots (\Pi_{\widetilde{S}} \gslash)^{A_k B_k}
	( \Pi_{\widetilde{S}} \xi)_{A_1\ldots A_k} (\Pi_{\widetilde{S}} \xi')_{B_1\ldots B_k},
	\qquad
	\vert \Pi_{\widetilde{S}} \xi \vert_{\Pi_{\widetilde{S}} \gslash}^2
	=
	(\Pi_{\widetilde{S}} \gslash) ( \Pi_{\widetilde{S}} \xi, \Pi_{\widetilde{S}} \xi),
\]
where
\[
	(\Pi_{\widetilde{S}} \gslash)^{AB}
	=
	G_C^A G_D^B \gslash^{CD},
\]
is the projection of $\gslash^{-1}$ to $\widetilde{S}$.

One easily computes
\[
	(\Pi_{\widetilde{S}} \gslash) ( \Pi_{\widetilde{S}} \xi, \Pi_{\widetilde{S}} \xi')
	=
	\gslash (\xi, \xi'),
	\qquad
	\vert \Pi_{\widetilde{S}} \xi \vert_{\Pi_{\widetilde{S}} \gslash}
	=
	\vert \xi \vert_{\gslash}.
\]
Indeed, if $k=1$,
\[
	(\Pi_{\widetilde{S}} \gslash) ( \Pi_{\widetilde{S}} \xi, \Pi_{\widetilde{S}} \xi')
	=
	\gslash^{-1} (G_A^B d\theta^A, G_C^D d \theta^C)
	\xi
	\Big(
	\Big( \delta_{B}^{E} + \frac{\partial f^{E}}{\partial \widetilde{\theta}^{B}} \Big) e_{E}
	\Big)
	\xi'
	\Big(
	\Big( \delta_{D}^{F} + \frac{\partial f^{F}}{\partial \widetilde{\theta}^{D}} \Big) e_{F}
	\Big)
	=
	\gslash^{AC} \xi_A \xi_C'
	=
	\gslash(\xi, \xi').
\]
Similarly for $k \geq 2$.  The proof then follows from the relation \eqref{eq:metriccomp5}.

\end{proof}

\begin{proposition}[$\widetilde{\mathfrak{D}}$ derivatives of projection to $\widetilde{S}$ of an $S$-tensor field]
	\label{prop:tildederivStensor}
	With the setup of Section~\ref{twogaugestwoback}, for any multi index $\gamma$ and any of the $S$-geometric quantities $\Phi$,
	\begin{align}
		\widetilde{\nablaslash} \Pi_{\widetilde{S}} \mathfrak{D}^{\gamma} \Phi_p
		=
		\
		&
		\Pi_{\widetilde{S}} \nablaslash \mathfrak{D}^{\gamma} \Phi_p
		+
		\mathcal{E}^{1,\vert \gamma \vert + 1}_{\mathfrak{D} \fsc,p+1},
		\label{eq:tildederivStensor1}
		\\
		\widetilde{\Omega} \widetilde{\nablaslash}_3 \Pi_{\widetilde{S}} \mathfrak{D}^{\gamma} \Phi_p
		=
		\
		&
		\Pi_{\widetilde{S}} \Omega \nablaslash_3 \mathfrak{D}^{\gamma} \Phi_p
		+
		\mathcal{E}^{1,\vert \gamma \vert + 1}_{\mathfrak{D} \fsc,p},
		\label{eq:tildederivStensor2}
		\\
		\widetilde{\Omega} \widetilde{\nablaslash}_4 \Pi_{\widetilde{S}} \mathfrak{D}^{\gamma} \Phi_p
		=
		\
		&
		\Pi_{\widetilde{S}} \Omega \nablaslash_4 \mathfrak{D}^{\gamma} \Phi_p
		+
		\mathcal{E}^{1,\vert \gamma \vert + 1}_{\mathfrak{D} \fsc,p+1}.
		\label{eq:tildederivStensor3}
	\end{align}
\end{proposition}

\begin{proof}
	The relation \eqref{eq:gaugerelationA} implies that, for any $S$-tensor $\xi$,
	\[
		\nabla_{\widetilde{e}_B} \xi
		=
		\left( \delta_B^C + \frac{\partial f^C}{\partial \widetilde{\theta}^B} \right) \nabla_{e_C} \xi
		+
		\Omega \frac{\partial f^3}{\partial \widetilde{\theta}^B} \nabla_{e_3} \xi
		+
		\Omega \frac{\partial f^4}{\partial \widetilde{\theta}^B} \nabla_{e_4} \xi
		-
		\frac{\partial f^4}{\partial \widetilde{\theta}^B} b^C \nabla_{e_C} \xi.
	\]
	If $\xi$ is an $S$-tangent $1$-form, the relation \eqref{eq:dxdtildex3} implies that,
	\[
		\Pi_{\widetilde{S}} \xi
		=
		\Big( \delta_{B}^{A} + \frac{\partial f^{A}}{\partial \widetilde{\theta}^{B}} \Big) \xi_A d \widetilde{\theta}^A
		=
		\xi_A \big( d\theta^A - \partial_{\widetilde{u}} f^A d \widetilde{u} - \partial_{\widetilde{v}} f^A d \widetilde{v} \big)
		=
		\xi - \xi\cdot \partial_{\widetilde{u}} \slashed{f} d \widetilde{u} - \xi\cdot \partial_{\widetilde{v}} \slashed{f} d \widetilde{v},
	\]
	and so
	\[
		\widetilde{\nablaslash} \Pi_{\widetilde{S}} \xi
		=
		\Pi_{\widetilde{S}} (
		\nabla \xi)
		-
		\xi\cdot \partial_{\widetilde{u}} \slashed{f} \widetilde{\nablaslash} d \widetilde{u}
		-
		\xi\cdot \partial_{\widetilde{v}} \slashed{f} \widetilde{\nablaslash} d \widetilde{v} 
		.
	\]
	It follows that
	\[
		\widetilde{\nablaslash} \Pi_{\widetilde{S}} \xi
		=
		\Pi_{\widetilde{S}} (\nablaslash \xi)
		+
		\widetilde{\nablaslash} f^3 \otimes \Pi_{\widetilde{S}} (\Omega \nabla_{e_3} \xi) 
		+
		\widetilde{\nablaslash} f^4 \otimes \Pi_{\widetilde{S}} (\Omega \nabla_{e_4} \xi)
		-
		\widetilde{\nablaslash} f^4 \otimes \Pi_{\widetilde{S}} (b\cdot \nablaslash \xi)
		-
		\frac{1}{2} \xi\cdot \partial_{\widetilde{u}} \slashed{f} \widetilde{\Omega}^{-1} \widetilde{\chi}{}^{\sharp}
		-
		\frac{1}{2} \xi\cdot \partial_{\widetilde{v}} \slashed{f} \widetilde{\Omega}^{-1} \widetilde{\chibar}{}^{\sharp}
		.
	\]
	The proof of \eqref{eq:tildederivStensor1} then follows if $\mathfrak{D}^{\gamma} \Phi$ is a one form.  If $\mathfrak{D}^{\gamma} \Phi$ is a higher order tensor the proof of \eqref{eq:tildederivStensor1} is similar, using now the fact that, for any $(0,k)$ $S$-tensor $\xi$,
	\[
		\Pi_{\widetilde{S}} \xi
		=
		\xi_{A_1} \ldots \xi_{A_k}
		\big( d\theta^{A_1} - \partial_{\widetilde{u}} f^{A_1} d \widetilde{u} - \partial_{\widetilde{v}} f^{A_1} d \widetilde{v} \big)
		\ldots
		\big( d\theta^{A_k} - \partial_{\widetilde{u}} f^{A_k} d \widetilde{u} - \partial_{\widetilde{v}} f^{A_k} d \widetilde{v} \big).
	\]
	The proofs of \eqref{eq:tildederivStensor2} and \eqref{eq:tildederivStensor3} are similar, using now the relations \eqref{eq:gaugerelation3} and \eqref{eq:gaugerelation4} respectively.
\end{proof}

\subsection{$\tilde{S}$-tangential operators acting on $S$-tensors}
\label{tangentoperatorssec}

Recall the setup of Section~\ref{twogaugestwoback}.

It is convenient to define the action of the differential operators $\widetilde{\nablaslash}$, $\widetilde{\nablaslash}_3$, $\widetilde{\nablaslash}_4$ associated to the $\widetilde{S}$ double null foliation on $S$-tensors.  
Recall, for any $S$-tangent $1$-form $\xi$, the $\widetilde{\nablaslash}$ derivative $\widetilde{\nablaslash} \xi \in \Gamma \big( (F^{-1})^* T^* \widetilde{S} \otimes T^*S \big)$ defined in Section \ref{beyondSsection}, with $F$ defined by \eqref{localformparamcomposition}.  Similarly, if $\xi \in \Gamma \big( (T^*S)^k \big)$ is an $S$-tangent $(0,k)$ tensor, for any $X\in \Gamma \big( (F^{-1})^* T^* \widetilde{S}\big)$ define $\widetilde{\nablaslash}_X \xi$ to be the restriction of $\nabla_{F_*X} \xi$ to $S$-vectors, so that $\widetilde{\nablaslash} \xi \in \Gamma \big( (F^{-1})^* T^* \widetilde{S} \otimes (T^*S)^k \big)$.  In view of the relation \eqref{eq:gaugerelationA},
\begin{equation} \label{eq:StildederivativeofStensor1}
	\widetilde{\nablaslash}_X \xi
	=
	\nablaslash_{\widetilde{\slashed{d}} \slashed{F}X} \xi
	+
	\widetilde{\nablaslash}_X f^3 \Omega \nablaslash_3 \xi
	+
	\widetilde{\nablaslash}_X f^4 \Omega \nablaslash_4 \xi
	-
	\widetilde{\nablaslash}_X f^4 \nablaslash_b \xi.
\end{equation}
Similarly, define $\widetilde{\nablaslash}_3 \xi$ and $\widetilde{\nablaslash}_4 \xi$ to be the restriction of $\nabla_{F_*\widetilde{e}_3} \xi$ and $\nabla_{F_*\widetilde{e}_4} \xi$ to $S$-vectors respectively.  In view of the relations \eqref{eq:gaugerelation3}, \eqref{eq:gaugerelation4},
\begin{align}
	\widetilde{\nablaslash}_3 \xi
	=
	\
	&
	\frac{\Omega}{\widetilde{\Omega}} \left(
	1 + \frac{\partial f^3}{\partial \widetilde{u}}
	\right) \nablaslash_3 \xi
	+
	\frac{\Omega}{\widetilde{\Omega}} \frac{\partial f^4}{\partial \widetilde{u}} \nablaslash_4 \xi
	+
	\frac{1}{\widetilde{\Omega}} \frac{\partial f^A}{\partial \widetilde{u}} \nablaslash_A \xi
	-
	\frac{1}{\widetilde{\Omega}} \frac{\partial f^4}{\partial \widetilde{u}} \nablaslash_b \xi
	\label{eq:StildederivativeofStensor3}
	\\
	&
	+
	\frac{1}{\widetilde{\Omega}} \nablaslash_{\widetilde{\slashed{d}} \slashed{F} \widetilde{b}} \xi
	+
	\Omega \widetilde{b}( f^3) \nablaslash_3 \xi
	+
	\Omega \widetilde{b}( f^4) \nablaslash_4 \xi
	-
	\widetilde{b}(f^4) \nablaslash_b \xi,
	\nonumber
	\\
	\widetilde{\nablaslash}_4 \xi
	=
	\
	&
	\frac{\Omega}{\widetilde{\Omega}} \left(
	1 + \frac{\partial f^4}{\partial \widetilde{v}}
	\right) \nablaslash_4 \xi
	+
	\frac{\Omega}{\widetilde{\Omega}} \frac{\partial f^3}{\partial \widetilde{v}} \nablaslash_3 \xi
	+
	\frac{1}{\widetilde{\Omega}} \frac{\partial f^A}{\partial \widetilde{v}} \nablaslash_A \xi
	-
	\frac{1}{\widetilde{\Omega}} \left( 1 + \frac{\partial f^4}{\partial \widetilde{v}} \right) \nablaslash_b \xi.
	\label{eq:StildederivativeofStensor4}
\end{align}
Define the coercive expression
\begin{equation} \label{eq:nablastildeofstensor}
	\vert \widetilde{\nablaslash} \xi \vert^2
	:=
	\widetilde{\gslash}^{\widetilde{C} \widetilde{D}} \gslash^{A_1 B_1} \ldots \gslash^{A_k B_k}
	\widetilde{\nablaslash}_{\widetilde{C}} \xi_{A_1\ldots A_k}
	\widetilde{\nablaslash}_{\widetilde{D}} \xi_{B_1\ldots B_k}.
\end{equation}
Note that $\widetilde{\nablaslash}_3 \xi, \widetilde{\nablaslash}_4 \xi \in \Gamma \big( (T^*S)^k \big)$ are $S$-tangent $(0,k)$ tensors and so $\vert \widetilde{\nablaslash}_3 \xi \vert$ and $\vert \widetilde{\nablaslash}_4 \xi \vert$ are defined as usual by \eqref{normforStensors}.

\section{A determination of the sphere diffeomorphism}
\label{determinethesphere}

We have already discussed how some definitions of Section~\ref{lowmodessection} are sensitive to 
diffeomorphisms $\psi$ of the form
$(\ref{diffsoftheform})$ (cf.~Remarks~\ref{yetanothernoteoncovar}, \ref{yetagainanothernoteoncovar}
and~\ref{stillyetagainanothernoteoncovar}). 
Thus, it is important to have a canonical way of choosing the correct
such identification with the standard $\mathbb S^2$ of Section~\ref{sphereconcrete}. 
In this respect, the following proposition will be of use later in the paper. 

\begin{proposition}
\label{determiningthesphere}
Let $(S,h)$\index{sphere!manifold!$S$, manifold diffeomorphic to $\mathbb S^2$}\index{sphere!metric!$h$, metric
close to standard spherical metric} be a compact oriented Riemannian $2$-manifold such that 
\begin{equation}
\label{morethanpinching}
|K-1|\le \epsilon, \qquad |\nabla_h K|\le \epsilon,
\end{equation}
where $K$\index{sphere!metric!$K$, Gauss curvature of $h$} here denotes the Gauss curvature of $h$.
Then for $\epsilon>0$ sufficiently small, it follows that
$S$ is diffeomorphic to the standard sphere $\mathbb S^2$ of Section~\ref{sphereconcrete}, 
and given  $p\in S$, $v\in T_pS$ with $h(v,v)=1$,
we can associate to the triple $((S,h), p, v)$
a ``canonical''
diffeomorphism 
\begin{equation}
\label{itsdiffy}
\psi: \mathbb S^2\to S
\end{equation} 
such that $\psi(1,0,0) = p$, and $\psi_*(0,1,0)=v$,
where we consider $(0,1,0)$ as a vector $(0,1,0)\in T_{(1,0,0)} \mathbb S^2 \subset \mathbb R^3$.

This association is 
canonical in the following sense: If $\xi: \tilde{S} \to S$ is a diffeomorphism 
then $\tilde{\psi}$ associated to the triple $(\xi^* h, \xi^{-1}(p), (\xi_*)^{-1}(v))$ satisfies
\begin{equation}
\label{canonicity}
\psi =\xi \circ\tilde \psi.
\end{equation}

Moreover, the association is equivariant in the following sense: If $(S,h)$ admits a non-trivial
Killing field $X$ such that $X(p)=0$,
then $\partial_{\mathring\phi}$ is a Killing field of $\psi^*h$, where
$\partial_{\mathring\phi}$ is the coordinate vector field on $\mathbb S^2$ defined 
with respect to the canonical coordinates~\eqref{sphcorddef}.

The association depends smoothly on $h$, $p$ and $v$ in the following sense: Let us fix
the manifold $S$ and suppose $h_t$, $p_t$, $v_t$ depend smoothly on a parameter $t$.
Then the resulting map $\psi_t$ given by our association depends smoothly on $t$. 

Moreover,  for all integers $k\ge 3$,  with $\psi$ as above, 
we have the following estimates
\begin{equation}
\label{theestimatesforourspherediff}
\sum_{\tilde{k} =0}^{k+2}
\int_{\mathbb S ^2} |\nabla_{\psi^*h}^{\tilde{k}} (\psi^*h-\mathring{\gamma})|^2_{\psi^*h}  d{\rm Vol}_{\psi^*h}  \le C_k 
\sum_{\tilde{k}=0}^k
\int_{\mathbb S ^2} |\nabla_{\psi^*h}^{\tilde{k}} \psi^*(K-1)|^2_{\psi^*h} d{\rm Vol}_{\psi^*h} .
\end{equation}

Finally, if $(S,h)$ and $(\hat{S},\hat{h})$
are two manifolds, with distinguished points and vectors $p$, $\hat{p}$ and
$v$, $\hat{v}$,
with Gauss curvature $K$ and $\hat{K}$ and with corresponding
canonical diffeomorphisms $\psi$ and $\hat\psi$, we have the estimate
\begin{equation}
\label{theestimatesforourspherediffdifferences}
\sum_{\tilde{k} =0}^{k+2}
\int_{\mathbb S ^2} |\nabla_{\psi^*h}^{\tilde{k}} (\psi^*h-\hat\psi^*{\hat h})|^2_{\psi^*h}  d{\rm Vol}_{\psi^*h}  \le C_k 
\sum_{\tilde{k}=0}^k
\int_{\mathbb S ^2} |\nabla_{\psi^*h}^{\tilde{k}} (\psi^*K-\hat\psi^*\hat{K})|^2_{\psi^*h} d{\rm Vol}_{\psi^*h} .
\end{equation}
If $S=\hat{S}$, $p=\hat{p}$ and $v=\lambda \hat{v}$ for some positive multiple $\lambda$, 
we moreover have the following zeroth and first order
pointwise estimates on the diffeomorphisms $\psi\circ \hat{\psi}^{-1}$, 
which can be expressed geometrically as the statement
\begin{equation}
\label{diffiscloseone}
\sup_{x\in S} {\rm d}^2_{(S,h)}(\psi\circ\hat\psi^{-1}(x),x) \le C \sum_{\tilde k=0}^2\int_{\mathbb S ^2} |\nabla_{\psi^*h}^{\tilde{k}} \psi^* (h-\hat h)|^2_{\psi^*h} d{\rm Vol}_{\psi^*h},
\end{equation}
\begin{equation}
\label{diffisclosetwo}
\sup_{x\in S} {\rm d}^2_{(S,h)} ( d(\psi\circ \hat\psi^{-1})|_x), {d (\rm Id})|_{x}) \le C \sum_{\tilde k=0}^2 \int_{\mathbb S ^2} |\nabla_{\psi^*h}^{\tilde{k}}  \psi^* (h-\hat h)|^2_{\psi^*h} d{\rm Vol}_{\psi^*h}.
\end{equation}
Moreover $\psi\circ \hat\psi^{-1}$ satisfies the higher order estimates
\begin{equation} \label{diffisclosethree}
		\sum_{\tilde k=0}^k
		\int_{\mathbb S^2}
		|\nabla_{\psi^*h}^{\tilde k+1} d (\psi\circ \hat\psi^{-1}) |^2_{\psi^*h}
		d{\rm Vol}_{\psi^*h}
		\le
		C
		\sum_{\tilde k=0}^{k+1} \int_{\mathbb S ^2} |\nabla_{\psi^*h}^{\tilde{k}}  \psi^* (h-\hat h)|^2_{\psi^*h} d{\rm Vol}_{\psi^*h}.
	\end{equation}
In~\eqref{diffiscloseone}, ${\rm d}_{(S,h)}$ 
may be taken to be the Riemannian distance, while in~\eqref{diffisclosetwo}, ${\rm d}_{(S,h)}$ 
simply denotes any appropriate geometric notion of
distance between maps $ T_xS \to T_{\psi\circ\hat\psi^{-1}(x)} S$ and
$T_xS\to T_xS$, defined so as to be
comparable to the differences of the components of the differential maps
expressed in a fixed chart.
In formulas~\eqref{diffiscloseone}--\eqref{diffisclosethree}, the constant $C$ depends on 
\[
\sum_{\tilde k=0}^{k+1} \int_{\mathbb S ^2} |\nabla_{\psi^*h}^{\tilde{k}}  \psi^* h|^2_{\psi^*h} d{\rm Vol}_{\psi^*h}.
\]

\end{proposition}

\begin{remark}
By Gauss--Bonnet, the statement that $S$ is diffeomorphic to the sphere already follows
from the first inequality of~\eqref{morethanpinching} as soon as $\epsilon<1$. 
\end{remark}

\begin{remark}
\label{ifwasoriginallythesphere}
Note that in the case where $(S,h)=(\mathbb S^2,\mathring\gamma)$ defined in Section~\ref{sphereconcrete}, and
$p=(1,0,0)$, $v=(0,1,0)\in T_{(1,0,0)}S \subset \mathbb R^3$, then our $\psi$ as defined below will in fact be the identity map. 
By~\eqref{canonicity}, it then follows that for the same choice $(S,h)=(\mathbb S^2,\mathring{\gamma})$ 
but for a different choice of point
$\tilde{p}$ and vector $\tilde{v}\in T_{\tilde{p}} S$, then the resulting $\tilde\psi : \mathbb S^2\to S=\mathbb S^2$ 
is the unique Euclidean rotation of  $\mathbb S^2$ which brings $(1,0,0)$ to 
$\tilde{p}$ and $(0,1,0)$ to $\tilde{v}$.
\end{remark}

\begin{remark}
Estimate~\eqref{theestimatesforourspherediffdifferences} is the statement that the metrics of $h$ and $\tilde{h}$ 
in their respective canonical coordinates are close if
their curvatures are close as measured in canonical coordinates. Estimates~\eqref{diffiscloseone} and~\eqref{diffisclosetwo}
correspond to the statement that the canonical diffeomorphisms are suitably close to each other, if
the metrics, considered as defined on the same $S$ are sufficiently close, when the anchoring point and vector
is taken to be the same. In practice, estimates~\eqref{diffiscloseone} and~\eqref{diffisclosetwo} will in fact be applied to
metrics $h$ and $\xi^*\hat{h}$ on $S$ where $\xi:S\to \hat{S}$
is a diffeomorphism and $\hat{p}$, $\hat{v}$ are distinguished points and vectors on $\hat{S}$ such that
$\xi^{-1}(\hat{p})=p$, $\xi^{-1}_*\hat{v}=v$.  
\end{remark}

\begin{remark}
\label{equivarianceremark}
The equivariance statement is only relevant for the proof of Corollary~\ref{axicorollary} in Section~\ref{equivariancesection}. 
This will be discussed in Section~\ref{equivariancesection}.
\end{remark}

\begin{proof}
We begin the proof with some additional constructions related to $\mathbb S^2$.

\subsection{Preliminaries on the sphere $\mathbb S^2$}
\label{somestuffforsphere}

Recall the standard sphere $\mathbb S^2$ as described in Section~\ref{sphereconcrete}. 
In addition to the coordinates $\mathring\theta$, $\mathring\phi$ defined by~\eqref{sphcorddef}, 
we may define a standard isothermal coordinate
system $\mathring x$, $\mathring y$ defined by\index{sphere!coordinates!$\mathring x$, standard isothermal coordinate}\index{sphere!coordinates!$\mathring y$, standard isothermal coordinate}
\begin{equation}
\label{isothermalnorthstand}
\mathring x:= \frac{1-\cos  \mathring\theta}{\sin\mathring\theta}\cos\mathring\phi,\qquad
\mathring y:= \frac{1-\cos  \mathring\theta}{\sin\mathring\theta}\sin\mathring\phi.
\end{equation}
Note also the identity
\begin{equation}
\label{identwithtang}
\sqrt{\mathring x^2 +\mathring y^2} = \left|\frac{1-\cos \mathring\theta}{\sin \mathring\theta} \right|=\tan (\mathring\theta/2).
\end{equation}

The above  coordinates satisfy $\Delta_{\mathring \gamma} \mathring x=0$, 
$\Delta_{\mathring\gamma}\mathring y=0$, and $\mathring y$ is a harmonic conjugate of 
$\mathring x$ with respect to the standard metric on the sphere, i.e.
\begin{equation}
\label{harmconjstand}
d\mathring{y} = \mathring{\star}d\mathring{x}.
\end{equation}
Note that $\mathring{y}$ is uniquely determined from $\mathring{x}$ and equation~\eqref{harmconjstand}
together with the additional requirement that $\mathring{y}$ vanish at the ``north pole'' $(0,0,1)$.
Let us note that the standard metric $\mathring\gamma$ in these coordinates takes the form
\begin{equation}
\label{conformalform}
\mathring\gamma = e^{\mathring\lambda(\mathring x,\mathring y)}(d{\mathring x}^2+d{\mathring y}^2)
\end{equation}
where\index{sphere!metric!$\mathring\lambda$, conformal factor in standard isothermal coordinates} 
\begin{equation}
\label{defofmathringlamb}
\mathring\lambda(\mathring x,\mathring y):= \log \frac{4}{(1+{\mathring x}^2+{\mathring y}^2)^2}.
\end{equation}

We may consider an auxiliary set of coordinates associated to the ``south pole'' $(0,0,-1)$. 
We denote these as ${\mathring\theta}'$, ${\mathring\phi}'$.\index{sphere!coordinates!${\mathring\theta}'$, standard southern spherical coordinate}\index{sphere!coordinates!${\mathring\phi}'$, standard southern spherical coordinate}  These are related to $\mathring\theta$, $\mathring\phi$
by the relations
\begin{equation}
\label{relationsonstandsph}
{\mathring\theta}' =\pi-\mathring\theta,\qquad {\mathring\phi}'=\mathring\phi.
\end{equation}
Finally, we may also define an isothermal coordinate system ${\mathring x}'$, ${\mathring y}'$
by replacing $\mathring\theta$ by ${\mathring\theta}'$ and $\mathring\phi$ by
${\mathring\phi}'$ in~\eqref{isothermalnorthstand}.\index{sphere!coordinates!${\mathring x}$, standard southern isothermal coordinate}\index{sphere!coordinates!${\mathring y}'$, standard southern isothermal coordinate}

The two isothermal coordinate systems are themselves related by the transformation law:
\begin{equation}
\label{isothermrelonstandsph}
\mathring{x}' = \frac{\mathring x}{\mathring x^2+\mathring y^2}, \,\,\,
\qquad \mathring{y}'=-\frac{\mathring y}{\mathring x^2+ \mathring y^2}.
\end{equation}

To define now a diffeomorphism $\psi: \mathbb S^2\to S$, it suffices to define open sets
$\mathcal{U}\subset S$, and $\mathcal{U}'\subset S$ with $\mathcal{U}\cup\mathcal{U}'=S$
and to define diffeomorphisms 
\[
\varphi : \mathcal{U}\to  \mathcal{V}\subset \mathbb R^2\, , 
\qquad \varphi' :\mathcal{U}'\to \mathcal{V}'\subset \mathbb R^2
\]
such that  
the transition map 
\begin{equation}
\label{stillintransition}
\varphi'\circ\varphi^{-1}:\mathcal{V} \to \mathcal{V}'
\end{equation}
is given by 
$(\ref{isothermrelonstandsph})$ when we view $\mathcal{V}\subset \mathbb R^2_{\mathring x,\mathring y}$
and $\mathcal{V}'\subset \mathbb R^2_{\mathring x',\mathring y'}$.
(For then $\psi$ can be defined consistently in local coordinates $(\mathring x, \mathring y)$ 
by $\varphi^{-1}$  and in local coordinates $(\mathring x',\mathring y')$ by ${\varphi'}^{-1}$.)
This is what we shall proceed to do.

\subsection{Two geodesic  polar charts on $(S,h)$}
\label{ataleoftwogeodcharts}
We begin with the point $p\in S$ and $v\in T_pS$. By a well known
result of Klingenberg~\cite{klingy}, if $\epsilon$ in~\eqref{morethanpinching} is sufficiently small,
there exists a geodesic polar neighbourhood $S_{p,\frac 78 \pi}\subset S$ around $p$ of, say,
radius $\frac{7}8\pi$. Let us denote the chart by the diffeomorphism
\[
\varphi_{p,\frac{7}8\pi}:  S_{p, \frac 78 \pi} \to \mathbb D_{\frac78\pi}.
\]
Associated to this chart are geodesic polar coordinates which we shall denote as $(\theta,\phi)$.
These are uniquely determined by declaring that $\phi=0$ corresponds to the direction $\partial_v$ and 
the frame $\partial_\theta,\partial_\phi$
is compatible with the orientation where defined.

Let $\gamma(s)$ denote the geodesic in $S$ with initial condition $\gamma(0)=p$,
$\gamma'(0)=v$. This is parametrised by arc length.
We will define an ``antipode'' $q$ as follows:

Note that by the assumption on curvature, for $\epsilon$ sufficiently small,
$\gamma$ will have have a unique conjugate point in the interval $(99\pi/100,101\pi/100)$.
Let $t_0$ correspond to this value and define
$q=\gamma(t_0)$.

Similarly, there  exists a geodesic polar neighbourhood $S_{q,\frac 78 \pi}$ around
$q$ of radius $\frac78\pi$.
Let us denote the chart by
\[
\varphi_{q,\frac{7}8\pi}: S_{q, \frac 78 \pi} \to \mathbb D_{\frac78\pi}.
\]
Again, associated to this chart are geodesic polar coordinates which we shall denote as $(\theta',\phi')$.
These are selected so that $\phi=0$ corresponds to the direction $\gamma(t_0-s)$ and the frame
$\partial_{\theta'},-\partial_{\phi'}$ is compatible with the orientation where defined.

Let us introduce some auxiliary notation: For any $0<\alpha <  \frac78\pi$, we define the sets
\[
S_{p, \alpha} := \{x\in S : {\rm dist}(x,p)<\alpha\},
\]
\[
S_{q, \alpha} := \{x\in S : {\rm dist}(x,q)<\alpha\},
\]
\[
\mathcal{T}_{\alpha} =S_{p,\alpha}\cap S_{q,\alpha},
\]
\[
C_{p,\alpha}= \{x\in S : {\rm dist}(x,p)=\alpha\},
\]
\[
C_{q,\alpha}= \{x\in S : {\rm dist}(x,p)=\alpha\}.
\]
Note that $S_{p,\alpha}\subset S_{p,\frac78 \pi}$ and is itself a geodesic polar neighbourhood.
Similarly, for $S_{q,\alpha}$. The sets $C_{p,\alpha}$ and $C_{q, \alpha}$ are geodesic circles that
generalise the ``lines of constant latitude''.

We recall that in the patches $S_{p, \frac 78 \pi}$ and $S_{q, \frac 78 \pi}$,
the metric $h$ takes the form
\begin{equation}
\label{quiteapatch}
h= d\theta^2 +G(\theta,\phi) d\phi^2 ,\qquad h= d{\theta'}^2 + G'(\theta',\phi')d{\phi'}^2,
\end{equation}
respectively,
where $G$ and $G'$ satisfy the differential equations\index{sphere!metric!$G$, metric function
in geodesic polar coordinates}\index{sphere!metric!$G'$, metric function
in southern geodesic polar coordinates}
\begin{equation}
\label{odeforG}
\sqrt{G}_{\theta\theta} + K\sqrt{G} = 0, \qquad \sqrt{G'}_{\theta\theta} + K\sqrt{G'}=0
\end{equation}
with the initial conditions 
\begin{equation}
\label{initcondsforG}
\sqrt{G}(\theta,0) =0, \sqrt{G}_\theta (\theta, 0)=1, \qquad{\rm and}\qquad
 \sqrt{G'}(\theta',0) =0, \sqrt{G'}_{\theta'} (\theta', 0)=1, 
 \end{equation}
 respectively.
 
Note (as is well known) that if $K=1$ identically, it follows that 
$G=\sin^2\theta$, $G'=\sin^2\theta'$.
In view of the assumptions on $K$, we may quantify the closeness of $G$ and $G'$ to these
values with the following

\begin{lemma} \label{lem:mcclose}
The metric functions $G$ and $G'$ satisfy the following estimates:
\begin{align} \label{mc1}
\sup_{\mathbb{D}_{7/8\pi}} \Big\|\frac{\sqrt{G}-\sin \theta}{\sin^3 \theta} \Big\| + \sup_{\mathbb{D}_{7/8\pi}} \Big\|\frac{\partial_\theta(\sqrt{G}-\sin \theta)}{\sin^2 \theta} \Big\|  + \sup_{\mathbb{D}_{7/8\pi}} \Big\|\frac{\partial_\varphi(\sqrt{G}-\sin \theta)}{\sin^3 \theta} \Big\|  &\leq C \|K-1\|_{C^1 (S)} \\
 \label{mc2}
\sup_{\mathbb{D}_{7/8\pi}} \Big\|\frac{\sqrt{G^\prime}-\sin \theta^\prime}{\sin^3 \theta^\prime} \Big\| + \sup_{\mathbb{D}_{7/8\pi}} \Big\|\frac{\partial_{\theta^\prime}(\sqrt{G^\prime}-\sin \theta^\prime)}{\sin^2 \theta^\prime} \Big\|  + \sup_{\mathbb{D}_{7/8\pi}} \Big\|\frac{\partial_{\varphi^\prime}(\sqrt{G^\prime}-\sin \theta^\prime)}{\sin^3 \theta^\prime} \Big\|  &\leq C \|K-1\|_{C^1 (S)} \, .
\end{align} 
\end{lemma}
\begin{proof}
 From~\eqref{odeforG} and~\eqref{initcondsforG} we derive the ordinary differential equation
\[
(\sqrt{G}-\sin \theta)_{\theta \theta} + (\sqrt{G}-\sin \theta) = -(K-1)\sin \theta - (K-1)(\sqrt{G}-\sin \theta) \, ,
\]
with initial condition $(\sqrt{G}-\sin \theta)(0,0)=0$ and $\partial_\theta (\sqrt{G}-\sin \theta) (0,0)=0$. The solution satisfies
\begin{align}
\sqrt{G}-\sin \theta = &-\sin \theta \int_0^{\theta} d\tilde{\theta} \cos \tilde{\theta} (\sin \tilde{\theta} + \sqrt{G} - \sin \tilde{\theta}) (K-1)(\tilde{\theta},\varphi) \nonumber \\
&+\cos \theta  \int_0^{\theta} d\tilde{\theta} \sin \tilde{\theta} (\sin \tilde{\theta} + \sqrt{G} - \sin \tilde{\theta}) (K-1)(\tilde{\theta},\varphi) \nonumber \, ,
\end{align}
from which the estimates follow. The computation in the south chart is of course analogous. 
\end{proof}

Let us note that the overlap region $\mathcal{T}_{\frac 78 \pi}$ is nonempty.
We have the following estimate for the difference of the coordinates.

\begin{lemma} \label{lem:gpclose}
We have in the overlap region $\mathcal{T}_{\frac{7}{8}\pi}$ the zeroth order estimate
\begin{align}
\| \theta - (\pi - \theta^\prime) \|_{C^0(\mathcal{T}_{\frac{7}{8}\pi})}  + \| \varphi - \varphi^\prime\|_{C^0(\mathcal{T}_{\frac{7}{8}\pi})} \leq C \|K-1\|_{C^1 (S)} \, .
\end{align}
\end{lemma}
\begin{proof}
This follows easily from the estimates of Lemma~\ref{lem:mcclose}.
\end{proof}

In particular, it follows from the above lemma that $\partial S_{p,\frac78 \pi}\subset S_{q, \frac78 \pi}$ and
$\partial S_{q, \frac78 \pi} \subset S_{p, \frac78\pi}$. In particular, $S_{p,\frac78 \pi}\cup S_{q, \frac78 \pi}$
is both open and closed and by connectivity 
we have
\[
S=S_{p,\frac78\pi}\cup S_{q,\frac78\pi}.
\]
More generally, given any $\alpha>\frac12 \pi$, this argument yields
\begin{equation}
\label{makingsure}
S= S_{p,\alpha}\cup S_{q,\alpha} 
\end{equation}
for sufficiently small $\epsilon$.

\subsection{Two isothermal charts on $(S,h)$}

Associated to each of the two geodesic polar neighbourhoods defined above, we may define isothermal charts
in the subregions $S_{p,\frac34 \pi}$ and $S_{q,\frac34 \pi}$ with the following properties:

\begin{lemma}
\label{twoorigisoth}
Let $(S,h)$, $p$, $v$  be as above. 
We may associate two isothermal coordinate systems $(x,y)$ and $(x',y')$, defined\index{sphere!coordinates!$x$, canonical coordinate}\index{sphere!coordinates!$y$, canonical coordinate}
in $S_{p,\frac34 \pi}$ and $S_{q,\frac34 \pi}$, respectively, such that
\begin{equation}
\label{isothermalcondition}
\Delta_hx =0, \Delta_hy=0,\qquad \Delta_hx' =0, \Delta_hy'=0,
\end{equation}
\begin{equation}
\label{harmonicconjcondition}
dy=\star dx,\qquad dy'=\star dx',
\end{equation}
\begin{equation}
\label{vanishingwhereitshould}
x(p)=0, y(p)=0,
\end{equation}
\begin{equation}
\label{isothermalest1}
\left\|x -\frac{1 -\cos\theta}{\sin\theta}\cos \phi\right\|_{H^2(S_{p, \frac34 \pi})} \le C\|K-1\|_{C^1(S)},
\end{equation}
\begin{equation}
\label{isothermalest2}
\left\|y -\frac{1 -\cos\theta}{\sin\theta}\sin \phi\right\|_{H^2(S_{p, \frac34 \pi})} \le C\|K-1\|_{C^1(S)},
\end{equation}
\begin{equation}
\label{isothermalest3}
\left\|x' -\frac{1 -\cos\theta'}{\sin\theta'}\cos \phi'\right\|_{H^2(S_{q, \frac34 \pi})} \le C\|K-1\|_{C^1(S)},
\end{equation}
\begin{equation}
\label{isothermalest4}
\left\|y' -\frac{1 -\cos\theta'}{\sin\theta'}\sin \phi'\right\|_{H^2(S_{q, \frac34 \pi})} \le C\|K-1\|_{C^1(S)}.
\end{equation}
The choice is canonical and, if $(S,h)=(\mathbb S^2,\mathring\gamma)$ with $p=(1,0,0)$ and $v=(0,1,0)$,
then $(x,y)=(\mathring x,\mathring y)$ and $(x',y')=(\mathring x',\mathring y')$.
\end{lemma}

\begin{proof}
Let us first define $x_{\rm prelim}$ to be unique solution of the Dirichlet problem for
\[
\Delta_h x_{\rm prelim} =0
\]
with the boundary condition
\[
x_{\rm prelim}=\frac{1 -\cos(\frac34 \pi)}{\sin(\frac34 \pi)} \cos\phi
\]
on $C_{p, \frac34 \pi}$.

In view of the estimates of Lemma~\ref{lem:mcclose}, it follows
that 
\[
\left\|x_{\rm prelim} - \frac{1 -\cos\theta}{\sin\theta}\cos \phi\right\|_{H^2(S_{p, \frac34 \pi})} \le C\|K-1\|_{C^1(S)}.
\]
In particular, $|x_{\rm prelim}(p)| \le C\|K-1\|_{C^1(S)}$, and defining
$x=x_{\rm prelim}-x_{\rm prelim}(p)$, we have that $x$ satisfies again
$\Delta_h x=0$ and the estimate~\eqref{isothermalest1}.

We define  $y$ to be the harmonic conjugate of $x$
\[
dy:=\star dx,
\]
where $\star$ is the Hodge-star operator associated to the metric $h$,
with the additional condition that $y=0$ at $p$.
Note that this uniquely determines $y$.
The pair $x,y$ clearly satisfy the relations of~\eqref{isothermalcondition},~\eqref{harmonicconjcondition} 
and~\eqref{vanishingwhereitshould}.

We note that $y$ can also be easily seen to satisfy~\eqref{isothermalest2}.

In the case where $(S,h)= (\mathbb S^2, \mathring\gamma)$ and $p=(1,0,0)$, $v=(0,1,0)$, then
$x=\mathring x$, $y=\mathring y$ defined in Section~\ref{somestuffforsphere} above.

In an entirely analogous fashion, we may define isothermal coordinates $x'$ and $y'$
and they clearly satisfy the estimates~\eqref{isothermalest3} and~\eqref{isothermalest4}. 

\end{proof}

Let us denote the diffeomorphisms defined by the above coordinates as
\[
\varphi_{\rm isoth}: S_{p,\frac 34 \pi}\to \mathcal{V}_p \subset \mathbb R^2, \qquad
\varphi_{\rm isoth}': S_{q,\frac 34 \pi}\to \mathcal{V}_q'\subset \mathbb R^2.
\]
These are indeed diffeomorphisms and $\mathcal{V}_p$, $\mathcal{V}_q$ are
bounded open subsets of $\mathbb  R^2$.

We may write the metric $h$ in the coordinates $x$, $y$, as
\begin{equation}
\label{conformformhnorth}
h= e^{\lambda (x,y)} (dx^2+dy^2)
\end{equation}
for a smooth function $\lambda :\mathcal{V}_p\to \mathbb R$.
Similarly,  we have
\begin{equation}
\label{conformformhsouth}
h=e^{\lambda'(x',y')}(dx'{}^2+dy'{}^2)
\end{equation}
for a smooth function $\lambda' :\mathcal{V}_q\to \mathbb R$.

\begin{lemma}
\label{overlaponsphelemma}
We have the following estimates for the  transition functions on the overlap region $\mathcal{T}_{\frac34 \pi}$:
\begin{align}
\label{overlaponsphereformula}
\Big\| x^\prime- \frac{x}{x^2+y^2}\Big\|_{C^0(\mathcal{T}_{\frac{3}{4}\pi})} + \Big\| y^\prime + \frac{y}{x^2+y^2}\Big\|_{C^0(\mathcal{T}_{\frac{3}{4}\pi})}\leq C \|K-1\|_{C^1 (S)} \, .
 \end{align}
\end{lemma}
\begin{proof}
This follows from the exact relation~\eqref{isothermrelonstandsph} in the case of the
transition functions connecting standard isothermal coordinates on the standard sphere, and the estimates
of Lemma~\ref{lem:gpclose} and~\ref{twoorigisoth}.
\end{proof}

Note the following corollary:
\begin{corollary}
\label{corolonchangeofcor}
 \begin{align} 
&\max_{m+n\leq N} \sup_{\varphi_{\rm isoth} (\mathcal{T}_{\frac{11}{16}\pi})} \Big\| (\partial_x)^m (\partial_y)^n \left( x^\prime (x,y)- \frac{x}{x^2+y^2}\right)\Big\| \nonumber \\
\label{maxregc}
&+\max_{m+n\leq N}  \sup_{\varphi_{\rm isoth} (\mathcal{T}_{\frac{11}{16}\pi})} \Big\| (\partial_x)^m (\partial_y)^n \left( y^\prime (x,y)+ \frac{y}{x^2+y^2}\right)\Big\| \leq C_N \|K-1\|_{C^1 (S)} \, .
 \end{align}
\end{corollary}
\begin{proof}
The case $N=0$ follows from Lemma~\ref{overlaponsphelemma}. The higher order estimates for $N\ge 1$ now
follow by elliptic estimates in view of the identities:
\[
\left[\frac{\partial^2}{\partial x^2} +\frac{\partial^2}{\partial y^2} \right] \left( x^\prime (x,y)- \frac{x}{x^2+y^2}\right) =0,
\qquad 
\left[\frac{\partial^2}{\partial x^2} +\frac{\partial^2}{\partial y^2} \right] \left( y^\prime (x,y)- \frac{y}{x^2+y^2}\right) =0.
\]
Note that the identities above follow
since $\triangle_h x=0$, $\triangle_h y=0$, the functions $\frac{x}{x^2+y^2}$ and $\frac{y}{x^2+y^2}$
are harmonic as functions of $(x,y)$, and the fact that the coordinates $x$ and $y$ are isothermal.
\end{proof}

\subsection{Estimates for the metric in isothermal charts}

In this section we prove the following estimates:
\begin{lemma}
Let $\mathring\gamma$ denote the metric on subsets of $\mathbb R^2$ given by the expression~\eqref{conformalform}.
For $N\ge 4$, we have the estimate:
\begin{align}
\label{topchart}
\|  (\varphi_{\rm isoth}^{-1})^* h - \mathring\gamma\|_{H^{N+1}(\varphi_{\rm isoth}(S_{p,\frac{11}{16}\pi}),  (\varphi_{\rm isoth}^{-1})^*h)}  &\leq  C_N \|K-1\|_{H^{N-1}(S)} , \\
\label{bottomchart}
 \| (\varphi_{\rm isoth}'{}^{-1})^*h  -  \mathring{\gamma} \|_{H^{N+1}(\varphi_{\rm isoth}'(S_{q,\frac{11}{16}\pi}),  (\varphi_{\rm isoth}'{}^{-1})^*h)} &\leq  C_N \|K-1\|_{H^{N-1}(S)}.
\end{align}
Here $H^s$ is the geometrically defined Sobolev norm for tensors (defined using covariant derivatives).
\end{lemma}
\begin{proof}
Let us define 
\[
\lambda_0(x,y)= \log\frac{4}{(1+x^2+y^2)^2}, \qquad \lambda'_0=\log\frac{4}{(1+{x'}^2+{y'}^2)^2}.
\]
In view of~\eqref{conformformhnorth},~\eqref{conformformhsouth} 
and~\eqref{conformalform}, it clearly suffices to estimate
\begin{equation}
\label{estimatewithlambda}
\|\lambda - \lambda_0\|_{H^{k+2}(\varphi_{\rm isoth}(S_{p,\frac{11}{16}\pi}) )} \leq C_k \|K-1\|_{H^k(S)} +  C_k  \|K-1\|_{C^1 (S)} \, , 
\end{equation}
\begin{equation}
\label{estimatewithlambdaprime}
\|\lambda^\prime - \lambda^\prime_0\|_{H^{k+2}(\varphi_{\rm isoth}'(S_{q,\frac{11}{16}\pi}))} \leq C_k \|K-1\|_{H^k(S)} +  C_k  \|K-1\|_{C^1 (S)} \, ,
\end{equation}
in view of the fact that for $k\ge 3$ the second term can evidently be dropped on 
the right hand side of~\eqref{estimatewithlambda}
and~\eqref{estimatewithlambdaprime} can evidently be dropped
by a Sobolev inequality (note that a geometric Sobolev inequality with bounded constant holds for metrics satisfying~\eqref{morethanpinching}).
The norm on the left hand side of~\eqref{estimatewithlambda} can~\eqref{estimatewithlambdaprime}
can be understood either as the Euclidean Sobolev norm or as the norm with respect
$(\varphi_{\rm isoth}^{-1})^*h$ or $(\varphi'_{\rm isoth}{}^{-1})^*h$ respectively,
since in view of the estimates on the Christoffel
symbols, these norms are equivalent.

From the change of variable formula relating $h$ in the form~\eqref{quiteapatch} 
and~\eqref{conformformhnorth}, we obtain
that
\begin{align}
\label{changeofvarhere}
\Big|\frac{\partial x}{\partial \theta}\Big|^2 + \Big|\frac{\partial y}{\partial \theta}\Big|^2  = e^{-\lambda(x,y)+\lambda_0(x,y)}e^{-\lambda_0(x,y)}   \, .
\end{align}
Defining $\sigma_x: = x- \frac{1-\cos \theta}{\sin \theta} \cos \varphi$ and $\sigma_y: = y- \frac{1-\cos \theta}{\sin \theta} \sin \varphi$, the left hand side of~\eqref{changeofvarhere} 
can be reexpressed as 
\begin{align}
\frac{1}{4\cos^4(\theta/2)} + \Big|\frac{\partial}{\partial \theta}\sigma_x \Big|^2 + 2 \frac{\partial}{\partial \theta}\sigma_x \frac{\partial}{\partial \theta} \left( \frac{1-\cos \theta}{\sin \theta} \cos \phi \right) + \Big|\frac{\partial}{\partial \theta}\sigma_y \Big|^2 + 2 \frac{\partial}{\partial \theta}\sigma_y \frac{\partial}{\partial \theta} \left( \frac{1-\cos \theta}{\sin \theta} \sin \phi \right) \, .
 \nonumber
\end{align}
We also compute the second factor on the right hand side of~\eqref{changeofvarhere}
\begin{align}
e^{-\lambda_0(x,y)} = \frac{(1+x^2+y^2)^2}{4} = \frac{1}{4} \left(\frac{1}{\cos^2 (\theta/2)} + 2\sigma_x \left(\frac{1-\cos \theta}{\sin \theta} \cos \phi\right) + \sigma_x^2 +  2\sigma_y \left(\frac{1-\cos \theta}{\sin \theta} \sin \phi \right) + \sigma_y^2\right)^2 \nonumber \, .
\end{align}
We conclude using formulas \eqref{isothermalest1},~\eqref{isothermalest2} of Lemma~\ref{twoorigisoth} and
formula~\eqref{overlaponsphereformula} of Lemma~\ref{overlaponsphelemma} that
\begin{equation}
\label{needtoupgrade}
\| e^{\lambda-\lambda_0-1}\|_{L^2(\phi_{\rm isoth}(S_{p,\frac34 \pi}))} \leq C\| K-1\|_{C^1(S)}.
\end{equation}

To obtain~\eqref{estimatewithlambda} as desired we must upgrade~\eqref{needtoupgrade} to higher order. For this,
we note that we have in $\phi_{\rm isoth}(S_{p,\frac34 \pi}))$ the relation
\[
\left[\frac{\partial^2 }{\partial x^2} + \frac{\partial^2 }{\partial y^2} \right] \lambda +2K e^\lambda = 0 \, .
\]
Recalling that $\lambda_0$ satisfies the above relation with $K=1$ we obtain
\begin{equation}
\label{applyintesttothis}
\left[\frac{\partial^2 }{\partial x^2} + \frac{\partial^2 }{\partial y^2} \right] (\lambda - \lambda_0) + 2 \left(e^{\lambda} - e^{\lambda_0} \right) = -2(K-1)e^{\lambda} \, .
\end{equation}
Thus, noting the restriction of the domain, inequality~\eqref{estimatewithlambda} indeed follows
from~\eqref{needtoupgrade} by applying standard Euclidean interior elliptic estimates to~\eqref{applyintesttothis}.

The case of~\eqref{estimatewithlambdaprime} is of course completely analogous.
\end{proof}

\subsection{The modified chart on the southern hemisphere and the completion of the construction}

To define our desired map to the standard sphere $\mathbb S^2$, it suffices to interpolate
between $\left(\frac{x}{x^2+y^2}, \frac{-y}{x^2+y^2}\right)$ and $(x',y')$ via a partition of unity.

\begin{lemma}
\label{modifiedchartlemma}
One can canonically define a third set of coordinates $(\tilde x',\tilde y')$ 
defining a regular chart\index{sphere!coordinates!$\tilde x'$, southern canonical coordinate}\index{sphere!coordinates!$\tilde y'$, southern canonical coordinate}        
\begin{equation}
\label{perturbedregchart}
\tilde\varphi': S_{q,\frac{11}{16}\pi} \to \tilde{\mathcal{V}}_q\subset \mathbb R^2 
\end{equation}
such that the identities
\begin{equation}
\label{transitionisgood}
\tilde{x}' = \frac{x}{x^2+y^2}, \qquad \tilde{y}' =\frac{y}{x^2+y^2}
\end{equation}
hold in $\mathcal{T}_{\frac{17}{32}\pi}$,
and such that the estimate~\eqref{bottomchart} holds with
$\tilde\varphi'$ replacing $\varphi_{\rm isoth}$.
\end{lemma}

\begin{proof}
Let us fix a smooth cutoff function $\chi(\varrho)$ such that $\chi(\varrho)=1$ for $\varrho\ge \tan \frac{ \frac{7}{16}\pi}2$
and $\chi(\varrho)=0$  for  $\varrho\le \tan \frac{ \frac{3}{8}\pi}2$. We define coordinates
\[
\tilde{x}'= \frac{x}{x^2+y^2}\chi (\sqrt{(x')^2+(y')^2}) +(1-\chi (\sqrt{(x')^2+(y')^2}))x',
\]
\[
\tilde{y}'= \frac{-y}{x^2+y^2}\chi (\sqrt{(x')^2+(y')^2}) +(1-\chi (\sqrt{(x')^2+(y')^2}))y'.
\]
Note that the $(x,y)$ coordinates are indeed defined in the region where $\chi(\sqrt{(x')^2+(y')^2})\ne0$.
Using the estimate~\eqref{maxregc} of Corollary~\ref{corolonchangeofcor}, we easily see
that this is indeed a smooth change of coordinates in $S_{q, \frac{11}{16}\pi}$, provided $\epsilon$
in~\eqref{morethanpinching} is sufficiently small. 

Let~\eqref{perturbedregchart} denote the diffeomorphism defined
by these coordinates. In view of the identity~\eqref{identwithtang} and again using the 
estimate~\eqref{maxregc}, it follows that for  $\epsilon$
in~\eqref{morethanpinching} sufficiently small, $\sqrt{(x')^2+(y')^2}\ge \tan \frac{\frac{7}{16}\pi}2$
in  $\mathcal{T}_{\frac{17}{32}\pi}$, and thus  
relation~\eqref{transitionisgood} holds on that region.

Finally, we note that using
\eqref{isothermalest3} and~\eqref{isothermalest4} it is easy to see now that the
estimate~\eqref{bottomchart} holds for   
$\tilde{\varphi}'$ replacing $\varphi$.
\end{proof}

In view of the remarks    at the end of Section~\ref{somestuffforsphere}, then denoting $\varphi:=\varphi_{\rm isoth}|_{S_p,
\frac{17}{32}\pi}$, and requiring that $\epsilon$ in~\eqref{morethanpinching} 
be sufficiently small so that~\eqref{makingsure} holds
with $\alpha=\frac{17}{32}\pi$,
it follows that the  charts
\[
\varphi  : S_{p, \frac{17}{32} \pi  } \to \varphi(S_{p, \frac{17}{32} \pi}) , \qquad  \tilde\varphi': S_{p, \frac{17}{32} \pi  } \to \tilde\varphi'(S_{p, \frac{17}{32} \pi})
\]
indeed determine a well-defined
map $\psi: \mathbb S^2\to S$ in view in particular of the property~\eqref{transitionisgood}
(which corresponds to the transition maps~\eqref{stillintransition} satisfying~\eqref{isothermrelonstandsph}).
The statement~\eqref{theestimatesforourspherediff} 
now follows from~\eqref{topchart} and the analogue of~\eqref{bottomchart} for $\tilde{x}'$ and
$\tilde{y}'$ asserted in Lemma~\ref{modifiedchartlemma}.

\subsection{Equivariance}
\label{equivariancesection}
We note that, in the presence of a Killing field,
the construction is indeed equivariant as claimed in the statement of Proposition~\ref{determiningthesphere}.
For let us assume that
$(S,h)$ admits a nontrivial Killing field $X$ which vanishes at $p$. Note that the geodesics emanating from $p$
are taken to one another under the $1$-parameter group of diffeomorphisms generated by $X$. It follows
that if two such geodesics intersect at another point $\tilde{q}$, then $X$ must vanish at $\tilde{q}$.
Now let $\gamma$, $t_0$ and $q=\gamma(t_0)$ be as in Section~\ref{ataleoftwogeodcharts}. 
It is clear that for $\gamma(t_0)$ to be conjugate
to $p$, it must be that 
all arc-length parameterised geodesics $\tilde\gamma$ emanating from $p$ satisfy
$\tilde\gamma(t_0)=q$ and thus $X(q)=0$. The antipode $q$ is in fact the unique other point of $S$
with this vanishing property.
One clearly sees now that the $G$ and $G'$ of geodesic polar coordinates in both hemispheres are 
manifestly $\phi$, respectively $\phi'$, independent. All subsequent constructions respect this symmetry
and thus the equivariance claim  follows.

\subsection{Estimates for differences in canonical coordinates}
To obtain the geometric estimate~\eqref{theestimatesforourspherediffdifferences} for differences
of two metrics expressed in canonical coordinates,
let us now assume we have two metrics $h$ and $\hat h$ on $S$ as 
in statement.
We essentially repeat the steps of the previous proof,
starting from  the remark that, considering $G$ and $\hat{G}$ as functions on $\mathbb D_{\frac78 \pi}$,
we have
\[
|\sqrt{G}-\sqrt{\hat G} |\le C \|K-\tilde{K}\|_{C^1(\mathbb D_{\frac78 \pi})},
\]
and similarly in the southern hemisphere.

\subsection{Proof of canonicity, smooth dependence and estimates for the diffeomorphisms $\psi\circ \tilde\psi^{-1}$}
We note that the canonicity relation~\eqref{canonicity} follows immediately from the construction
as all coordinate systems are defined geometrically.
It is also clear that all constructions depend smoothly on the metric in the sense described.

To obtain the more quantitative statements~\eqref{diffiscloseone}--\eqref{diffisclosethree}
estimating the diffeomorphisms $\psi\circ \hat\psi^{-1}$,
then denoting $x$, $y$ the final coordinates on the northern hemisphere corresponding
to $\psi$ and $\hat{x}$,  $\hat{y}$ the corresponding coordinates on the northern hemisphere
corresponding to $\hat\psi$, then viewing $\hat{\tilde{x} }  {}'(x,y)$ and  $\hat{\tilde{y} }  {}'(x,y)$ 
on a suitable smaller overlap domain,  
 statement~\eqref{diffiscloseone} is equivalent to
 $C^0$ bounds for $\hat{x} (x,y)-x$ and  $\hat{y} (x,y)-y$, (together with corresponding bounds
 on   $\hat{\tilde{x} }  {}'( \tilde{x}',\tilde{y}')-\hat{\tilde{x}}$ and  $\hat{\tilde{y} }  {}'(\tilde{x}',\tilde{y}')-\hat{\tilde{y}}$ covering the 
 southern hemisphere),
 statement~\eqref{diffisclosetwo} is equivalent to $C^1$ bounds 
 for these functions, while~\eqref{diffisclosethree} is equivalent to $H^{k+2}$ bounds.

Note that because these diffeomorphisms are
based on isothermal coordinates, this easily reduces in coordinates
to showing $k+2$ order interior estimates for differences of
appropriate solutions
of the Laplace equation with respect to  the  metrics $h$ and $\tilde{h}$, which can easily be seen to be bounded
by the quantity on the right hand side of~\eqref{diffisclosethree}

Note finally 
that the statement of Remark~\ref{ifwasoriginallythesphere} follows from the various remarks in the lemmas
concerning the case $(S,h)=(\mathbb S^2,\mathring\gamma)$.

\end{proof}

\part{The main theorem: setup, statement and logic of the proof}
\label{stateandlogicpart}

In this part of the work, we will give the setup of the main theorem,
including its detailed statement and
the complete logic of the proof.

\parttoc

First, in {\bf Chapter~\ref{thelocaltheorysec}}, we shall consider the local theory
for our problem, giving a well-posedness statement for the characteristic initial
value problem which we shall study,  a Cauchy stability type statement which
will allow us to infer the existence of a sufficiently large region of spacetime
covered by initial data gauges,  a decomposition of the set of initial data into $3$-parameter
families, and the anchoring conditions linking teleological normalised gauges
to initial data gauges. This will finally allow us to assert the existence of anchored
teleologically normalised gauges at  time $u_f^0$.

In {\bf Chapter~\ref{maintheoremsec}},  after introducing a variety of energies,
we shall be able to give a complete statement of our main
theorem, stated as {\bf Theorem~\ref{thm:main}}.

Finally, in {\bf Chapter~\ref{logicoftheproofsection}} we shall give the logic of the proof of Theorem~\ref{thm:main}.
The main analytical content will be deferred to subtheorems whose proofs will
appear in Part~\ref{improvingpart} while some of the other statements necessary to complete the proof
will be deferred to subtheorems whose proofs appear in Part~\ref{conclusionpart}.

\vskip1pc
\emph{In general, the chapters of this part will use definitions from Part~\ref{preliminlabel}, but, as discussed
previously, portions can be read
independently. The notations and theorems of Chapter~\ref{thelocaltheorysec} will be fundamental for the rest of the work,
though the proofs may be skipped at a first reading.
The reader who is primarily interested in understanding the large-scale architecture of the proof can skip
some of the detailed definitions of energies in Chapter~\ref{maintheoremsec}.
On the other hand, the reader
more interested in the statement and proof of the bootstrap theorem forming the analytic heart of the argument
can in principle skip the later parts of Chapter~\ref{logicoftheproofsection}.
More detailed instructions to the reader will be provided in the preambles of each chapter.}

\chapter{The local theory}
\label{thelocaltheorysec}

In the present chapter we will develop the local theory associated to the
characteristic initial value problem to be studied in this work. This will also involve
introducing some of the basic setup which will be central to the proof of our main theorem,
including a description of ``seed'' data, a basic well-posedness statement,  the existence of
teleological gauges, anchored to suitable initial data gauges,
and a discussion of
the structure of the moduli space $\mathfrak{M}$ of initial data.

\minitoc

In {\bf Section~\ref{initdatasection}} we shall define  the characteristic initial data
considered, as generated by ``seed data'' $\boldsymbol{\mathcal{S}}$. 
Next, in {\bf Section~\ref{maxcauchysection}} we 
shall give the existence of a maximal future
Cauchy development (Theorem~\ref{maxCauchythe}). 
We note that these two sections depend only on Sections~\ref{geomprelimforvac}--\ref{thatsallofthem}. 

In {\bf Section~\ref{initialdatanormsec}}, 
we shall identify the initial characteristic hypersurfaces with 
analogous hypersurfaces in Schwarzschild, and give a smallness condition on seed  data
guaranteeing that initial data themselves exist globally on the characteristic hypersurfaces.
We shall then introduce in {\bf Section~\ref{globalsmallnessassumptionnorm}} the 
norm on initial data which will define the global smallness condition fundamental  
for our work, determined by a parameter $\varepsilon_0$. 
In {\bf Section~\ref{localexistencesection}}, 
we shall state two theorems ({\bf Theorems~\ref{thm:localKrus}}
and~{\bf \ref{thm:localEF}}) giving that, for data
with small norm, the resulting solution
exists for a sufficiently long  interval so as to contain regions covered
by what we shall term ``the initial data gauges''. This can be thought of as
a more quantitative version of Cauchy stability (using already the null structure exhibited by the equations!), 
allowing existence all the
way up to a ``piece'' of null infinity.

In {\bf Section~\ref{anchoredsec}}, we shall give the anchoring conditions 
connecting two teleologically normalised $\mathcal{I}^+$ and $\mathcal{H}^+$ gauges with the initial data gauges.
We shall address the question of existence of anchored gauges corresponding
to some fixed retarded time parameter $u_f^0$ in
{\bf Section~\ref{existenceofanchored}}.

In {\bf Section~\ref{threeparamsection}}, we shall
discuss the structure of the moduli space of initial data,
foliating it by 
$3$-parameter families $\mathcal{L}_{{\mathcal{S}}_0}$ labelled by some choice of seed data 
${\mathcal{S}}_0\in \mathcal{L}_{{\mathcal{S}}_0}$ and parametrised 
by $\lambda \in \mathbb R^3$. 
This will be essential for the statement of the main theorem in 
Section~\ref{maintheoremsec}.
Finally, in {\bf Section~\ref{thedegreeonemap}},
appealing to the existence of the anchored gauges corresponding
to time $u_f^0$ asserted in Section~\ref{existenceofanchored},
we shall define
the map  
\[
{\bf J}_0:\mathfrak{R}_0\to B_{\varepsilon_0/u_f^0},
\]
defined using the associated Kerr angular momentum parameters of the solution in $\mathcal{I}^+$ gauge,
where $\mathfrak{R}_0\subset  \mathbb R^3$ is a subset of the $\lambda$-parameter space,
which will be key to our topological 
argument in the proof of the main theorem.

\vskip1pc
\emph{This chapter is fundamental for understanding the detailed statement of the main theorem of this work, Theorem~\ref{thm:main} stated in
Chapter~\ref{maintheoremsec}, 
as well as for the large-scale architecture of the proof as outlined in Chapter~\ref{logicoftheproofsection} 
and for the statement of the
main ``bootstrap'' theorem,~\Cref{havetoimprovethebootstrap} stated in Section~\ref{improvingsectionnewdivision}, 
in particular. On a first reading, however, the reader may 
wish to skip some of the details and refer back as necessary. The work~\cite{Chr} provides a useful reference
for setting up data for the characteristic initial value problem.}

\section{Characteristic initial data generated by seed data $\boldsymbol{\mathcal{S}}$}
\label{initdatasection}

In this and the following section, we will set up 
the characteristic initial value problem for the Einstein equations in double
null gauge and state a well-posedness theorem. 
The results here make no reference to a  Schwarzschild background and
only depend on Sections~\ref{geomprelimforvac}--\ref{thatsallofthem}.

Consider arbitrary parameters ${\bf U}_{-2}<{\bf U}_5$, ${\bf V}_{-2}$.\index{initial data!parameters!${\bf U}_{-2}$}\index{initial data!parameters!${\bf V}_{-2}$}\index{initial data!parameters!${\bf U}_{5}$}
Let us define
\[
\boldsymbol{\underline{C}}_{\rm in}= [{\bf U}_{-2},{\bf U}_5] \times\{{\bf V}_{-2}\} \times \mathbb S^2, \qquad
\boldsymbol{C}_{\rm out} = \{{\bf U}_{-2}\}\times [{\bf V}_{-2},\infty) \times \mathbb S^2
\]
and\index{initial data!initial hypersurfaces!$S_{{\bf U}_{-2},{\bf V}_{-2}}$}  
\[
\boldsymbol{S}_{{\bf U}_{-2},{\bf V}_{-2}}=\{ {\bf U}_{-2}\}\times\{ {\bf V}_{-2} \}\times \mathbb S^2.
\]
Note that we can naturally define the notion of $S$-tangent 
tensor on $\boldsymbol{\underline{C}}_{\rm in}\cup \boldsymbol{C}_{\rm out}$.

We define initial data on $\boldsymbol{\underline{C}}_{\rm in}\cup \boldsymbol{C}_{\rm out}$ as follows.\index{initial data!initial surfaces!$\boldsymbol{\underline{C}}_{\rm in}$}\index{initial data!initial surfaces!$\boldsymbol{C}_{\rm out}$}

We prescribe smooth functions $\boldsymbol{\tr\chi}(\theta)$,\index{initial data!seed data!$\boldsymbol{\tr\chi}$}
 $\boldsymbol{\tr\chibar}(\theta)$\index{initial data!seed data!$\boldsymbol{\tr\chibar}$} and a $1$-form
$\boldsymbol{\eta}_A(\theta)$\index{initial data!seed data!$\boldsymbol{\eta}$} 
 on $S_{{\bf U}_{-2},{\bf V}_{-2}}$. 
 
 We further
prescribe smooth positive functions $\boldsymbol{\Omega}^{\rm in}({\bf U}, \theta)$\index{initial data!seed data!$\boldsymbol{\Omega}^{\rm in}$} 
and $\boldsymbol{\Omega}^{\rm out}({\bf V}, \theta)$\index{initial data!seed data!$\boldsymbol{\Omega}^{\rm out}$} 
on $\boldsymbol{\underline{C}}_{\rm in}$ and $\boldsymbol{C}_{\rm out}$ respectively, 
with 
\[
\boldsymbol{\Omega}^{\rm in}({\bf U}_{-2}, \theta)=\boldsymbol{\Omega}^{\rm out}({\bf V}_{-2}, \theta),
\]
and a smooth $S$-tangent vector field
$\boldsymbol{b}^A_{\rm out}({\bf V}, \theta)$\index{initial data!seed data!$\boldsymbol{b}^A_{\rm out}$} 
on $\boldsymbol{C}_{\rm out}$.

We may now define the vector fields
\[
e_3= (\boldsymbol{\Omega}^{\rm in})^{-1}\partial_{\bf U}, \qquad e_4= (\boldsymbol{\Omega}^{\rm out})^{-1}(\partial_{\bf V} +
\boldsymbol{b}^A_{\rm out}\partial_A)
\]
on $\boldsymbol{\underline{C}}_{\rm in}$, $\boldsymbol{C}_{\rm out}$ respectively. 
Note that the operations $\underline{D}$ and $D$ can be defined for covariant $S$-tensors
on $\boldsymbol{\underline{C}}_{\rm in}$
and $\boldsymbol{C}_{\rm out}$ respectively, since the relevant projection does not require the metric.

 Finally we prescribe  smooth
 symmetric positive definite  $S$-tangent covariant $2$-tensors in the angular coordinates
$\boldsymbol{\hat{\gslash}}^{\rm in}_{AB}({\bf U},\theta)$\index{initial data!seed data!$\boldsymbol{\hat{\gslash}}^{\rm in}$}  and 
$\boldsymbol{\hat{\gslash}}^{\rm out}_{AB}({\bf V},\theta)$\index{initial data!seed data!$\boldsymbol{\hat{\gslash}}^{\rm out}$}   
on $\boldsymbol{\underline{C}}_{\rm in}$ and $\boldsymbol{C}_{\rm out}$ respectively, with the property 
that 
\[
\boldsymbol{\hat{\gslash}}^{\rm in}_{{A}{B}}({\bf U}_{-2},\theta) = \boldsymbol{\hat{\gslash}}^{\rm out}_{{A}{B}}({\bf V}_{-2}, \theta)
\]
and that
\begin{equation}
\label{usingthistoo}
(\underline{D} \boldsymbol{\hat{\slashed{\epsilon}}}{}^{\rm in})_{AB} =0, \qquad ({D} \boldsymbol{\hat{\slashed{\epsilon}}}{}^{\rm out})_{AB} =0
\end{equation}
where $\boldsymbol{\hat{\slashed{\epsilon}}}{}^{\rm in}$, 
$\boldsymbol{\hat{\slashed{\epsilon}}}{}^{\rm out}$
 denote the area element of $\boldsymbol{\hat{\gslash}}^{\rm in}_{AB}$, 
$\boldsymbol{\hat{\gslash}}^{\rm out}_{AB}$, respectively.

We refer to this information as the \emph{seed data} and denote this collectively
by the symbol $\boldsymbol{\mathcal{S}}$.\index{initial data!seed data!$\boldsymbol{\mathcal{S}}$, complete
set of seed data quantities}

\begin{proposition}[Existence of the initial data set]
\label{existenceofinitdata}
Consider seed data
\begin{equation}
\label{seeddataforref}
\boldsymbol{\mathcal{S}}=\{
\boldsymbol{\tr \chi}, \boldsymbol{\bf \tr \chibar},
\boldsymbol{\eta}, \boldsymbol{\Omega}^{\rm in}, \boldsymbol{\Omega}^{\rm out},
\boldsymbol{b}^A_{\rm out},
\boldsymbol{\hat{\gslash}}^{\rm in}_{AB}, 
\boldsymbol{\hat{\gslash}}^{\rm out}_{AB}\}
\end{equation}
as defined above.
Then there exists a ${\bf U}_5'$\index{initial data!parameters!${\bf U}_5'$} satisfying 
${\bf U}_{-2}<{\bf U}_5' \le {\bf U}_5$ and 
a ${\bf V}_0'$\index{initial data!parameters!${\bf V}_0'$}  satisfying ${\bf V}_{-2}<{\bf V}_0' \le\infty$
and  unique smooth vector field $\boldsymbol{b}^A_{\rm in}$\index{initial data!derived data!$\boldsymbol{b}^A_{\rm in}$} defined on
$\boldsymbol{\underline{C}}_{\rm in}' =\boldsymbol{\underline{C}}_{\rm in} \cap \{ {\bf U} \le {\bf U}'_4\}$\index{initial data!initial surfaces!$\boldsymbol{\underline{C}}_{\rm in}'$, shorter
hypersurface on which data exist} 
and a smooth    $\boldsymbol{\slashed{g}}_{AB}$\index{initial data!derived data!$\boldsymbol{\slashed{g}}_{AB}$} defined
on $\boldsymbol{\underline{C}}_{\rm in}' \cup \boldsymbol{C}_{\rm out}'$ where $\boldsymbol{C}_{\rm out}'=
\boldsymbol{C}_{\rm out} \cap \{ {\bf V} \le {\bf V}'_0\}$\index{initial data!initial surfaces!$\boldsymbol{C}_{\rm out}'$, shorter
hypersurface on which data exist}, generated by the prescribed data.
Moreover, on $\boldsymbol{\underline{C}}_{\rm in}' \cup \boldsymbol{C}_{\rm out}'$ we may define the operation
$\boldsymbol{\slashed{\nabla}}$,\index{initial data!differential operators!$\boldsymbol{\slashed{\nabla}}$}
we may define $\boldsymbol{\slashed{\nabla}}_3$ on $\boldsymbol{\underline{C}}_{\rm in}'$\index{initial data!differential operators!$\boldsymbol{\slashed{\nabla}}_3$}
and 
$\boldsymbol{\slashed{\nabla}}_4$\index{initial data!differential operators!$\boldsymbol{\slashed{\nabla}}_4$} 
on $\boldsymbol{C}_{\rm out}'$, 
and we may determine ${\mathfrak{D}}^\gamma\vartheta $ applied to 
each of the quantities $\vartheta$ defined in Section~\ref{thatsallofthem} for all $|\gamma|\ge 0$,
yielding definitions for quantities $\boldsymbol{{\mathfrak{D}}^\gamma\vartheta}$\index{initial data!derived quantities!$\boldsymbol{{\mathfrak{D}}^\gamma\vartheta}$}
on $\boldsymbol{\underline{C}'}_{\rm in}\cup \boldsymbol{C'}_{\rm out}$ such that the equations of Section~\ref{thatsallofthem},
and the equations that arise after commutation by  ${\mathfrak{D}}^\gamma$,
are satisfied. We may in fact define
$\boldsymbol\alpha$\index{initial data!derived data!$\boldsymbol\alpha$} on all of $\boldsymbol{C}_{\rm out}$ and 
$\boldsymbol\alphabar$\index{initial data!derived data!$\boldsymbol\alphabar$}
on all of $\boldsymbol{\underline{C}}_{\rm in}$.
\end{proposition}

\begin{proof}
Refer to~\cite{Chr}.

Let us set 
\begin{equation}
\label{definingthemetricfromseed1}
\boldsymbol{\gslash}_{AB}({\bf U}, {\bf V}_{-2}, \theta):=  \phi^{\rm in}({\bf U}, \theta) \boldsymbol{\hat{\gslash}}^{\rm in}_{AB}({\bf U},
\theta)
\end{equation}
\begin{equation}
\label{definingthemetricfromseed2}
\boldsymbol{\gslash}_{AB}({\bf U}_{-2}, {\bf V}, \theta):=  \phi^{\rm out}({\bf V}, \theta)  \boldsymbol{\hat{\gslash}}^{\rm in}_{AB}({\bf V},
\theta)
\end{equation}
for a $\phi^{\rm in}$, $\phi^{\rm out}$ to be determined.
Using~\eqref{usingthistoo}, we 
derive from the relations of Section~\ref{thatsallofthem} the ordinary differential equations
\begin{equation}
\label{pairofodes}
\underline{D}\underline{D}   \phi^{\rm in} +e^{\rm in}\phi^{\rm in} =0, \qquad DD \phi^{\rm out}  +e^{\rm out}\phi^{\rm out} =0
\end{equation}
where\index{initial data!derived data!$e^{\rm in}$}\index{initial data!derived data!$e^{\rm out}$}\index{initial data!derived data!$\phi^{\rm in}$}\index{initial data!derived data!$\phi^{\rm out}$}
\[
e^{\rm in}:= \frac18 (\boldsymbol{\hat\gslash}^{\rm in}{}^{-1})^{AC} (\boldsymbol{\hat\gslash}^{\rm in}{}^{-1})^{BD} \underline
D\boldsymbol{\hat\gslash}_{AB}^{\rm in}
\underline D\boldsymbol{\hat\gslash}_{CD}^{\rm in},
\qquad
e^{\rm out}:= \frac18 (\boldsymbol{\hat\gslash}^{\rm out}{}^{-1})^{AC} (\boldsymbol{\hat\gslash}^{\rm out}{}^{-1})^{BD} D\boldsymbol{\hat\gslash}^{\rm out}_{AB}
D\boldsymbol{\hat\gslash}^{\rm out}_{CD}.
\]

Let us define now $\phi^{\rm in}$, $\phi^{\rm out}$ to be the unique solutions of the~\eqref{pairofodes} with
initial condition 
\[
\phi^{\rm in}({\bf U}_{-2},\theta) =1, \qquad \underline{D}\phi^{\rm in}({\bf U}_{-2}) = \frac12 
\boldsymbol{\Omega}^{\rm in} \boldsymbol{\rm tr \chi},
\qquad {\rm and}\,\,\, \phi^{\rm out}({\bf V}_{-2},\theta) =1, \qquad D\phi^{\rm out}({\bf V}_{-2}) =
\frac12 \boldsymbol{\Omega}^{\rm in} \boldsymbol{\rm tr \chibar} ,
\]
respectively.
These are globally defined on $\boldsymbol{\underline{C}}_{\rm in}$ and 
$\boldsymbol{C}_{\rm out}$ respectively, but may not be positive.
Thus, we do not know yet that the definitions~\eqref{definingthemetricfromseed1}--\eqref{definingthemetricfromseed2}
define metrics. 

Nonetheless, motivated by the relations of Section~\ref{thatsallofthem}, we may already define 
$\boldsymbol{\hat \chi}$  on $\boldsymbol{\underline{C}}_{\rm in}$
and
 $\boldsymbol{\hat \chibar}$
on $\boldsymbol{C}_{\rm out}$ by the expressions
\[
\boldsymbol{\Omega}^{\rm in} \boldsymbol{\hat \chi}_{AB}: = \frac12 (\phi^{\rm in})^2 \underline D\boldsymbol{\hat\gslash}_{AB},
\qquad
\boldsymbol{\Omega}^{\rm out} \boldsymbol{\hat \chi}_{AB}: = \frac12 (\phi^{\rm out})^2 D\boldsymbol{\hat\gslash}_{AB}.
\]

Note that since we may rewrite~\eqref{eq:chihat4} as
$D(\Omega\hat\chi) = -\Omega \alpha$, $\underline D(\Omega \hat\chibar)=-\alphabar$,
we may now already define
$\boldsymbol{\underline\alpha}$ along
$\boldsymbol{\underline{C}}_{\rm in}$ and
 $\boldsymbol\alpha$ along $\boldsymbol{C}_{\rm out}$ as
\begin{equation}
\label{alphaseeddef}
\boldsymbol\Omega^{\rm in}
\boldsymbol\alphabar := \underline D(\frac12 (\phi^{\rm in})^2 \underline D\boldsymbol{\hat\gslash}),
\qquad
\boldsymbol\Omega^{\rm out}
\boldsymbol\alpha: =  D(\frac12 (\phi^{\rm out})^2  D\boldsymbol{\hat\gslash}).
\end{equation}

To define the remaining quantities, we shall require restricting the initial data cones such that
$\phi^{\rm in}$ and $\phi^{\rm out}$ are indeed safely positive.
Let us fix thus  a ${\bf U}'_5$, ${\bf V}'_0$ such that say $\phi^{\rm in}\ge \frac12$, $\phi^{\rm out}\ge \frac12$
for ${\bf U}_{-2}\le {\bf U}\le {\bf U}'_5$ and ${\bf V}_{-2}\le {\bf V}\le {\bf V}_0'$.
This defines the restricted cones $\boldsymbol{\underline{C}'}_{\rm in}$ and $\boldsymbol{C'}_{\rm out}$.

We may define now
$\boldsymbol{\gslash}_{AB}$ from~\eqref{definingthemetricfromseed1},~\eqref{definingthemetricfromseed2}
and these are non-degenerate (Riemannian) metrics on the spheres of constant ${\bf U}$ or ${\bf V}$ respectively.
Thus, we may now define the covariant derivative $\slashed\nabla$ with respect
to $\boldsymbol{\gslash}$, the algebraic relations on 
$S$-tensors  as well as the $S$-tangential differential operators $\slashed{div}$, etc.

Motivated by the relations of Section~\ref{thatsallofthem}, we may also now define
$\boldsymbol{\tr \chi}$, 
 on $\boldsymbol{\underline{C}'}_{\rm in}$
and
$\boldsymbol{\tr \chibar}$
on $\boldsymbol{C'}_{\rm out}$ by the expressions
\[
\boldsymbol{\Omega}^{\rm in} \boldsymbol{\bf \tr \chi}=2(\phi^{\rm in})^{-1}\underline D\phi^{\rm in}\qquad
\boldsymbol{\Omega}^{\rm out} \boldsymbol{\bf \tr \chi}=2(\phi^{\rm out})^{-1}D\phi^{\rm out}.
\]
We may now define $\boldsymbol\chi = \boldsymbol{\hat\chi} +\frac12 \boldsymbol{\rm \tr\chi} \boldsymbol{\gslash}$,
$\boldsymbol\chibar = \boldsymbol{\hat\chibar} +\frac12 \boldsymbol{\rm \tr\chibar} \boldsymbol{\gslash}$.

Note at this point it follows from the relations~\eqref{Dvscovar1froms}--\eqref{Dvscovar} 
that the covariant operators $\slashed\nabla_3$,
$\slashed\nabla_4$ can now be defined along $\boldsymbol{\underline{C}}_{\rm in}'$ 
and $\boldsymbol{C}_{\rm out}'$, respectively.

We note first that we may define $\boldsymbol{\hat\omega}$ and $\boldsymbol{\hat\omegabar}$
along $\boldsymbol{C}_{\rm out}$, $\boldsymbol{\underline{C}}_{\rm in}$ respectively, by the second two relations of~\eqref{eq:DlogOmega}.
To define $\boldsymbol{\eta}$ and $\boldsymbol\etabar$, note
that the first relation of~\eqref{eq:DlogOmega} allows one to
determine $\eta+\underline\eta$ on $\boldsymbol{C}_{\rm out}\cup \boldsymbol{C'}_{\rm in}$. 
In particular, we may use this to define $\boldsymbol\etabar({\bf U}_{-2},{\bf V}_{-2},\theta)$
from $\boldsymbol\eta({\bf U}_{-2},{\bf V}_{-2},\theta)$.

Considering now equations~\eqref{eq:nabla4eta}, we may 
 substitute the $\beta$, respectively $\underline\beta$,
term on the right hand side from~\eqref{eq:Codazzi}, respectively from~\eqref{eq:Codazzibar}. Finally, we may 
substitute the $\etabar$, respectively $\eta$, term by $(\eta+\etabar)-\eta$, respectively $(\eta+\etabar)-\etabar$.
This allows us to integrate~\eqref{eq:nabla4eta} 
as a
linear propagation relation for $\eta$ along $\boldsymbol{C}_{\rm out}$
and $\underline\eta$ along $\boldsymbol{C}_{\rm in}$ with known coefficients
and known initial data at $\boldsymbol{C}_{\rm in}\cap \boldsymbol{C}_{\rm out}$. 
This integration defines our $\boldsymbol\etabar$, $\boldsymbol\eta$
on $\boldsymbol{\underline{C}'}_{\rm in}$ and $\boldsymbol{C'}_{\rm out}$. We may similarly
now integrate~\eqref{eq:nabla4etabar} and \eqref{eq:nabla3eta} to obtain these quantities on
$\boldsymbol{C'}_{\rm out}$ and $\boldsymbol{\underline{C}'}_{\rm in}$.
Thus, we have defined both $\boldsymbol \eta$ and $\boldsymbol\etabar$ on all of 
$\boldsymbol{C'}_{\rm out}\cup \boldsymbol{\underline{C}'}_{\rm in}$. Finally, we may define $\boldsymbol\beta$
as it has been determined.

From~\eqref{eq:b3} we may now determine $b^A$ along $\boldsymbol{\underline{C}'}_{\rm in}$, thus
we define 
\[
\boldsymbol{b}^A_{\rm in}({\bf U}, {\bf V}_{-2}, \theta) :=\boldsymbol{b}^A_{\rm out}({\bf U}_{-2},{\bf V}_{-2},\theta)
+\int_{{\bf U}_{-2}}^{\bf U}   2(\boldsymbol{\Omega}^{\rm in})^2 (\boldsymbol{\eta}^A-\boldsymbol{\etabar}^A).
\]

Note that we may define $\boldsymbol{K}$ on $\boldsymbol{C}_{\rm out}\cup \boldsymbol{\underline{C}'}_{\rm in}$ by the usual expression
for Gauss curvature of the metric $\boldsymbol{\slashed g}$. Using this we may rewrite the
transport equations~\eqref{eq:trchi3} and~\eqref{eq:trchibar4} for ${\rm tr}\chi$ and ${\rm \tr}{\underline\chi}$ in the $
3$ and $4$ directions in terms of known quantities. We may now integrate the transport
equations~\eqref{eq:chihat3} and~\eqref{eq:chibarhat4} 
for $\hat\chi$ and $\hat\chibar$.  This allows us to complete the definition of both $\boldsymbol{\chi}$
and $\boldsymbol\chibar$ to all of $\boldsymbol{\underline{C}}_{\rm in}'\cup \boldsymbol{C}_{\rm out}'$.
Moreover, 
this now determines $\rho$ from 
the Gauss equation~\eqref{eq:Gauss}, whereas $\sigma$ is determined by~\eqref{eq:curletacurletabar}.  
We thus define $\boldsymbol\rho$ and $\boldsymbol\sigma$ according to these
determinations.

Continuing as above, we may inductively determine $\mathfrak{D}^\gamma\vartheta$ for any of the  quantities of Section~\ref{thatsallofthem}
and for any order $|\gamma|$, so that all equations of Section~\ref{thatsallofthem}, commuted by $\mathfrak{D}^\gamma$,
are satisfied along $\boldsymbol{\underline{C}}_{\rm in}'\cup \boldsymbol{C}_{\rm out}'$.

\end{proof}

\begin{remark}
\label{seeddatadiffeo}
We remark already that considering a diffeomorphism $\psi:\mathbb S^2\to \mathbb S^2$,
associated to any seed data set we may define a new data set by replacing all quantities
by their pullback under $\psi^*$. 
\end{remark}

\begin{remark}
We note that the initial data may not exist globally, in particular, the quantities
 $\boldsymbol{\slashed{g}}_{AB}^{\rm in}$ and $\boldsymbol{\slashed{g}}_{AB}^{\rm out}$,
 since if $\phi$ is negative, these no longer define metrics. 
 Global existence of the data on the full $\boldsymbol{\underline{C}}_{\rm in}\cup \boldsymbol{C}_{\rm out}$  will 
 require conditions on
the seed data to be discussed in Section~\ref{initialdatanormsec}.
It is useful, however, that $\boldsymbol\alpha$ and $\boldsymbol\alphabar$ are defined
on $\boldsymbol{C}_{\rm out}$ and $\boldsymbol{\underline{C}}_{\rm in}$ by~\eqref{alphaseeddef},
because this way we may phrase the  conditions
for global existence of data in terms of these quantities.
\end{remark}

\section{The maximal Cauchy development $(\mathcal{M}, g)$}
\label{maxcauchysection}
In the meantime, let us state a general well posedness theorem for
the above characteristic initial value problem.

\begin{theorem}[Existence of the maximal Cauchy evolution~\cite{rendallchar, LukcharIVP, dezornification}]
\label{maxCauchythe}
Consider seed data as in Proposition~\ref{existenceofinitdata} and
$\boldsymbol{\underline{C}}_{\rm in}'\cup \boldsymbol{C}_{\rm out}'$ such that the full set of initial data
exist, as in the statement of the proposition.
Then there exists 
a unique smooth maximal globally hyperbolic future Cauchy evolution $(\mathcal{M},g)$,
with past boundary diffeomorphic to $\boldsymbol{\underline{C}}_{\rm in}'\cup  \boldsymbol{C}_{\rm out}'$, satisfying~\eqref{vaceqhere}
and obtaining the data, in the sense that there exists a globally hyperbolic subset
$\mathcal{W}\times \mathbb{S}^2\subset\mathbb R^2\times \mathbb S^2$ with past boundary
$\boldsymbol{\underline{C}}_{\rm in}'\cup  \boldsymbol{C}_{\rm out}'$, and an embedding  
\begin{equation}
\label{embedintheor}
i:\mathcal{W}\times \mathbb{S}^2\to \mathcal{M}
\end{equation}
such that $i^*(g)$ is of the form~\eqref{doublenulllongform} (and thus satisfies
the 
equations~\eqref{eq:firstvariatformula}--\eqref{eq:alphabar4})
and obtains the prescribed quantities on $\boldsymbol{\underline{C}}_{\rm in}'\cup  \boldsymbol{C}_{\rm out}'$,
and $i|_{\boldsymbol{\underline{C}}_{\rm in}'\cup  \boldsymbol{C}_{\rm out}'}$ 
is a diffeomorphism of $\boldsymbol{\underline{C}}_{\rm in}'\cup \boldsymbol{C}_{\rm out}'$
onto the past boundary
of $(\mathcal{M},g)$.

The spacetime $(\mathcal{M},g)$ is maximal in the sense that, given any other globally hyperbolic $C^k$ Lorentzian manifold
$(\widetilde{\mathcal{M}},\widetilde{g})$, for sufficiently high $k$, satisfying~\eqref{vaceqhere} admitting the data
in the sense of the existence of~\eqref{embedintheor} as above for some map $\tilde{i}:\widetilde{\mathcal{W}}\times
\mathbb{S}^2\to \widetilde{\mathcal{M}}$, then $(\widetilde{\mathcal{M}},\widetilde{g})$ admits a smooth
atlas with respect to which $\widetilde{g}$ is in fact smooth, and
there exists an isometric embedding $j:\widetilde{\mathcal{M}}\to \mathcal{M}$
such that $\tilde{i}\circ j = i$ on $\widetilde{\mathcal{W}}\cap\mathcal{W}\times \mathbb S^2$.

In particular, the following uniqueness statement holds for $i$ in~\eqref{embedintheor}:
If $\mathcal{W}_1$ and $\mathcal{W}_2$ are two open sets satisfying the above in place
of $\mathcal{W}$, with embeddings $i_1$, $i_2$, then $i^*_1(g)=i^*_2(g)$ on $(\mathcal{W}_1\cap\mathcal{W}_2)\times \mathbb S^2$. 
\end{theorem}

\begin{remark} We emphasise here that it is already nontrivial that the maximal future
Cauchy evolution exists 
in a full neighbourhood of $\boldsymbol{\underline{C}}_{\rm in}'\cup  \boldsymbol{C}_{\rm out}'$, as stated in
the above theorem.
This depends in an essential way on the null condition and follows from
the work of~\cite{LukcharIVP}.
\end{remark}

\section{Identification with Schwarzschild and global existence of the data}
\label{initialdatanormsec}

We recall the parameter $M_{\rm init}>0$  fixed in~\eqref{newfixingMinit}.
Recall the Kruskal manifold $\mathcal{W}_{\mathcal{K}}\times\mathbb S^2$
with the Schwarzschild metric~\eqref{SchwmetricKruskal} with mass $M=M_{\rm init}$.\index{initial data!parameters!$M_{\rm init}$, initial Schwarzschild parameter}

\subsection{The initial hypersurfaces and additional normalisations}
We now recall the parameters defined in Section~\ref{compediumparameterssec} associated to $M=M_{\rm init}$.
Let us now identify the ${\bf U}$ with $U$ and ${\bf V}$ with $V$, and thus ${\bf U}_{-2}$, etc., with our chosen 
parameters $U_{-2}$, etc.

For definiteness, in this
context we denote the identified $\boldsymbol{\underline{C}}_{\rm in}$ as $\underline{C}_{\rm in}^{\mathcal{K}}$\index{initial data!initial surfaces!$\underline{C}_{\rm in}^{\mathcal{K}}$} and
$\boldsymbol{C}_{\rm out}$ as $C_{\rm out}^{\mathcal{K}}$.\index{initial data!initial surfaces!$C_{\rm out}^{\mathcal{K}}$} 
We have
\[
\underline{C}^{\mathcal{K}}_{\rm in}: = [U_{-2},U_5]\times\{V_{-2}\}\times \mathbb S^2, \qquad C_{\rm out}^{\mathcal{K}}:=\{U_{-2}\}\times [V_{-2},\infty)\times \mathbb S^2
\]
We thus have that $\underline{C}_{\rm in}^{\mathcal{K}}\cup C_{\rm out}^{\mathcal{K}}
\subset \mathcal{W}_{\mathcal{K}}\times
\mathbb S^{2}$, by our choices 
of $U_{-2}$, $V_{-2}$, $U_5$.
Let us also define\index{initial data!initial surfaces!$C_{\rm out}^{\mathcal{EF}}$} 
\begin{equation}
\label{CEFout}
C_{\rm out}^{\mathcal{EF}}:=(\iota_M\times {\rm id})^{-1}(C_{\rm out})=[v_{-2},\infty)\times \mathbb S^2 \subset \mathcal{W}_{\mathcal{EF}}
\times \mathbb S^2
\end{equation}
and, for any $\underline{C}'{}^{\mathcal{K}}_{\rm in}\subset \underline{C}^{\mathcal{K}}_{\rm in}\cap \{U <0 \}$,
\begin{equation}
\label{CEFin}
\underline{C}'{}_{\rm in}^{\mathcal{EF}}:=(\iota_M\times {\rm id})^{-1}(\underline{C}^{\mathcal{K}_{\rm in}})\subset \mathcal{W}_{\mathcal{EF}}
\end{equation}
where $\iota=\iota_M$ is the map~\eqref{iotadef}.

We consider now seed data $\mathcal{S}=\mathcal{S}^{\mathcal{K}}$
as in Proposition~\ref{existenceofinitdata}, with the following 
additional normalisations:
\begin{equation}
\label{seedrestrictions}
\boldsymbol{b}^A_{out}=0, \qquad
\boldsymbol{\Omega}^{\rm in}= \Omega^{\mathcal{K}}_{\circ}, \qquad
\boldsymbol{\Omega}^{\rm out}=\Omega^{\mathcal{K}}_{\circ}.
\end{equation}
We recall from Proposition~\ref{existenceofinitdata} that
we may \emph{define} $\alpha_{\rm out}^{\mathcal{K}}$\index{initial data!derived data!$\alpha_{\rm out}^{\mathcal{K}}$}
 globally
on  $C_{\rm out}^{\mathcal{K}}$ and $\underline\alpha_{\rm in}^{\mathcal{K}}$\index{initial data!derived data!$\alpha_{\rm in}^{\mathcal{K}}$} on  $\underline{C}_{\rm in}^{\mathcal{K}}$ 
by~\eqref{alphaseeddef}.

\subsection{Associated Eddington--Finkelstein data and covariance}

On~\eqref{CEFout} and any choice of~\eqref{CEFin}, we may define an associated Eddington--Finkelstein
realisation $\mathcal{S}^{\mathcal{EF}}$ 
of the seed data. 
The Eddington--Finkelstein realisation of the data is defined by first setting
\begin{align} \label{seedrestrictions2}
\boldsymbol{b}^A_{out}=0, \qquad
\boldsymbol{\Omega}^{\rm in}= \Omega_\circ^{\mathcal{EF}}, \qquad
\boldsymbol{\Omega}^{\rm out}=\Omega_\circ^{\mathcal{EF}}.
\end{align}
We notice that under this choice $\iota_*(e^{\mathcal{K}}_3)=\frac{\partial{U}}{\partial{u}}e^{\mathcal{EF}}_3$,
$\iota_*(e^{\mathcal{K}}_4)= \frac{\partial{V}}{\partial{v}} e^{\mathcal{EF}}_4$.
We now define\index{initial data!derived data!$\tr\chi_{\mathcal{EF}}$}, $\tr\chibar_{\mathcal{EF}}$\index{initial data!derived data!$\tr\chibar_{\mathcal{EF}}$}
and $\eta_{\mathcal{EF}}$\index{initial data!derived data!$\eta_{\mathcal{EF}}$} 
\[
\tr\chi^{\mathcal{EF}}(u_{-2},v_{-2},\theta)= \tr \chi_{\mathcal{K}}(U_{-2},V_{-2},\theta)
\frac{\partial{v}}{\partial{V}}(U_{-2},V_{-2}),\qquad
\tr\chibar^{\mathcal{EF}}(u_{-2},v_{-2},\theta)= \tr \chi^{\mathcal{K}}(U_{-2},V_{-2},\theta)
\frac{\partial{u}}{\partial{U}}(U_{-2},V_{-2}),
\]
\[
\eta^{\mathcal{EF}}_A =\eta^{\mathcal{K}}_A\frac{\partial{u}}{\partial{U}}(U_{-2},V_{-2}).
\]

Finally, we define ${\hat{\gslash}}^{\rm in}_{\mathcal{EF}}(u,\theta) =
{\hat{\gslash}}^{\rm in}_{\mathcal{K}}(U(u),\theta)$ and 
${\hat{\gslash}}^{\rm in}_{\mathcal{EF}}(u,\theta) =
{\hat{\gslash}}^{\rm in}_{\mathcal{K}}(U(u),\theta)$.
Note that the resulting $\alpha_{\rm out}^{\mathcal{EF}}$\index{initial data!derived data!$\alpha_{\rm out}^{\mathcal{EF}}$}  will satisfy
\[
\alpha_{\rm out}^{\mathcal{EF}}(v,\theta)=  \alpha_{\rm out}^{\mathcal{K}}\left(\frac{\partial{v}}{\partial{V}}\right)^2(V(v),\theta).
\]

Finally, we note that the Gaussian curvature
$K=K^{\mathcal{EF}}=K^{\mathcal{K}}$ on $\underline{C}_{\rm in}\cap C_{\rm out}$  is determined 
by the seed data.\index{initial data!derived data!$K$}

The  uniqueness statement in Theorem~\ref{maxCauchythe} gives us the following
\begin{proposition}
\label{identifiedseeddata}
Fix $M=M_{\rm init}$, consider the seed data $\mathcal{S}^{\mathcal{K}}$
 defined on $\underline{C}_{\rm in}^{\mathcal{K}}\cup C_{\rm out}^{\mathcal{K}}$
 satisfying~\eqref{seedrestrictions}, 
and let $\underline{C}'{}^{\mathcal{K}}_{\rm in}\cup C'{}^{\mathcal{K}}_{\rm out}$, $\mathcal{W}$, $i$,
$(\mathcal{M},g)$
be as in the statement of Theorem~\ref{maxCauchythe}. 
Then,  provided $\underline{C}'{}^{\mathcal{K}}_{\rm in}$ is chosen $\underline{C}'{}^{\mathcal{K}}_{\rm in}\subset \{U <0\}$,
 defining $\mathcal{S}^{\mathcal{EF}}$ as above, 
 and with $\tilde{i}:\widetilde{\mathcal{W}}\times\mathbb S^2\to
(\widetilde{\mathcal{M}},\tilde{g})$ as in the
statement of Theorem~\ref{maxCauchythe} but for data $\mathcal{S}^{\mathcal{EF}}$,
then $\iota^*_M i^* g =\tilde{i}^*\tilde{g}$ on $\iota^{-1}_M(\mathcal{W})\cap \widetilde{\mathcal{W}}$.
\end{proposition}

In view of the above, we may indeed consider the initial data sets induced by
seed data $\mathcal{S}^{\mathcal{K}}$  and corresponding seed data $\mathcal{S}^{\mathcal{EF}}$ as geometrically
the same initial data. In discussing norms, it will be natural to use the $\mathcal{S}^{\mathcal{K}}$
quantities on $\underline{C}_{\rm in}$ while using $\mathcal{S}^{\mathcal{EF}}$ quantities on $C_{\rm out}$
in order to parameterise initial data.

\subsection{Global existence of the full initial data}
We now formulate a smallness assumption guaranteeing the existence
of the full initial data  on all of $\underline{C}_{\rm in}^{\mathcal{K}}\cup C_{\rm out}^{\mathcal{K}}$.

\begin{proposition}
\label{globexistofdata}
Fix $M=M_{\rm init}$, consider seed data $\mathcal{S}^{\mathcal{K}}=\boldsymbol{\mathcal{S}}$ defined 
on $\underline{C}_{\rm in}^{\mathcal{K}}\cup C_{\rm out}^{\mathcal{K}}$
satisfying~\eqref{seedrestrictions}, and define the corresponding $\mathcal{S}^{\mathcal{EF}}$ as above
on $C^{\mathcal{K}}_{\rm out}$. 
Assume in addition the following smallness assumption is satisfied:
\begin{align*}
&\sum_{0\le k\le 3}
\int_{U_{-2}}^{U_5}\int_{\mathbb S^2} |(r\slashed\nabla_\circ)^k\underline\alpha_{\rm in}^{\mathcal{K}}|^2_{\slashed{g}_{\circ}} dU\, d\theta +
\int_{v_{-2}}^{\infty}\int_{\mathbb S^2} r^7 |(r\slashed\nabla_\circ)^k\alpha_{\rm out}^{\mathcal{EF}}|^2_{\slashed{g}_{\circ}} dv\, d\theta
\\
&\qquad+\int_{\mathbb S^2} |(r\slashed\nabla_\circ)^k(\tr\chi_{\mathcal{EF}}-\tr\chi_{\circ,M_{\rm init},\mathcal{EF}})|^2_{\slashed{g}_{\circ}}  +
 |(r\slashed\nabla_\circ)^k(\tr\chibar_{\mathcal{EF}}-\tr\chibar_{\circ,M_{\rm init},\mathcal{EF}})|^2_{\slashed{g}_{\circ}}  \\
 &\qquad\qquad \qquad
  +|(r\slashed\nabla_\circ)^k \hat\chi_{\mathcal{EF}}|^2_{\slashed{g}_{\circ}}
    +|(r\slashed\nabla_\circ)^k \underline{\hat\chi}_{\mathcal{EF}}|^2_{\slashed{g}_{\circ}}
  +|(r\slashed\nabla_\circ)^k\eta_{\mathcal{EF}}|^2_{\slashed{g}_{\circ}} 
 +|(r\slashed\nabla_\circ)^k(K-r^{-2})|^2_{\slashed{g}_{\circ}}  (u_{-2},v_{-2}) d\theta
 \leq \varepsilon_1.
\end{align*}
Then, for sufficiently small $\varepsilon_1=\varepsilon_1(M_{\rm init})$,\index{initial data!parameters!$\varepsilon_1$, smallness parameter ensuring global existence of the data} we may take $U'_5=U_5$, $V'_0=\infty$ in 
Proposition~\ref{existenceofinitdata}, i.e.~the full initial data
exists globally on $\underline{C}_{\rm in}^{\mathcal{K}}\cup C_{\rm out}^{\mathcal{K}}$.

Moreover, $\{U_5\}\times\mathbb S^2$ is a trapped surface in the sense that
$\tr \chi(U_5,\theta)<0$, $\tr\chibar(U_5,\theta)<0$.
\end{proposition}

\begin{proof}
Note that the $\slashed\nabla_\circ$ and the norms are taken with respect to the Schwarzschild metric
as one cannot define globally the metric $\slashed{g}$ before we know that the data exist globally.
Essentially one simply requires that $\phi^{\rm in}$ and $\phi^{\rm out}$
from Proposition~\ref{existenceofinitdata} satisfy globally $\phi^{\rm in}\ge \frac12$,
$\phi^{\rm out}\ge \frac12$. The smallness condition can easily seen to be sufficient
for sufficiently small $\varepsilon_1$ as it ensures weighted pointwise estimates on the quantities
$e^{\rm in}$ and $e^{\rm out}$. The statement concerning the trapped surface follows easily
from examining the first equation of~\eqref{eq:Ray} and equation~\eqref{eq:trchi3}.
\end{proof}

\subsection{Kerr parameters of the data}
\label{Kerpardatasec}

We may associate Kerr parameters to an initial data set  (cf.~Section~\ref{linKerrforIp}).

We first remark that on the sphere $S_{u_{-2},v_{-2}}$ we may define 
 the $\ell=1$ spherical harmonic functions $Y^1_m$ as in Section~\ref{basisforprojspace} 
with the help also of the Schwarzschild
background $g_{\circ, M_{\rm init}}$.
By Proposition~\ref{prop:oneformdecomp} the curvature component $\Omega \beta$
computed from the seed data can be decomposed as
\[
	\Omega \beta (u_{-2},v_{-2},\theta) 
	=
	r \nablaslash h_{1,\Omega \beta}(u_{-2},v_{-2},\theta)
	+
	r {}^* \nablaslash h_{2,\Omega \beta}(u_{-2},v_{-2},\theta),
\]
where ${}^*$ here is defined with respect $\slashed{g}(u_{-2},v_{-2})$.

\begin{definition}
\label{assocKerpardata}
Given seed initial data as above,
we may define \underline{associated Kerr parameters} $J_{{\rm seed}}^m$,~\index{initial data!parameters!$J_{\rm seed}^m$, $m=-1,0,1$, associated Kerr parameters}\index{angular momentum!$(J_{\rm seed}^{-1},J_{\rm seed}^0,J_{\rm seed}^1)$, angular momentum vector associated to data} for $m=-1,0,1$, by the relation
\[
	(r^4 h_{2,\Omega \beta})_{\ell=1}(u_{-2},v_{-2},\theta)
	=
	3 \Omega_{\circ,M_{\rm init}}^2(u_{-2},v_{-2})
	\sum_{m=-1}^1
	J^m_{\rm seed}
	Y^{1}_m(u_{-2},v_{-2},\theta).
\]
We note that in view of Remark~\ref{yetagainanothernoteoncovar}, 
the parameters $J^m_{\rm seed}$ in general change under diffeomorphisms of
the type considered in Remark~\ref{againaboutdiffeos} above.
\end{definition}

\begin{remark}
\label{remarkaboutinitialseed}
We note that the Kerr parameters can be computed from the seed data quantity 
$\eta$ by using the Codazzi equation (\ref{eq:Codazzi}) on the sphere $S_{u_{-2},v_{-2}}$ which yields the relation
\[
\curlslash (\Omega \beta) = \frac{\Omega_\circ^2}{r} \curlslash\, \eta - \curlslash \divslash (\Omega \hat{\chi}) + \curlslash \left(-\frac{1}{2} \left(\Omega tr \chi - (\Omega tr \chi_\circ)\right) \underline{\eta} +\Omega \hat{\chi}^\sharp \underline{\eta} \right) \, .
\]
After projection to $\ell=1$ one obtains the relation $2(h_{2, \Omega \beta})_{\ell=1} \approx  \Omega_\circ^2 \left(\curlslash\, \eta \right)_{\ell=1}$ up to nonlinear terms, and hence
$J^m_{\rm seed}=\frac{1}{6} r^4\left(\curlslash\, \eta\right)_{\ell=1,m}(u_{-2},v_{-2})+O(\varepsilon_1^2)$.
\end{remark}

\subsection{Associated linearised data and linearised solution}
Given 
seed data as in Proposition~\ref{globexistofdata}, 
we may define an associated linearised seed data set.
This gives rise to an associated linearised solution, which we may furthermore renormalise 
at $\mathcal{I}^+$. We summarise this in the following:

\begin{proposition}
\label{linearisedprophere}
Given general
seed data as in Proposition~\ref{globexistofdata}, we may define associated
seed data for the linearised Einstein equations in double null gauge around Schwarzschild
with mass $M_{\rm init}$, restricted to the exterior, as follows:
All  the linearised seed functions are defined by assigning
the linearised quantities their full nonlinear values (in Eddington--Finkelstein representation) minus the
Schwarzschild values, e.g.~$\elin:=\eta_{\mathcal{EF}}$,  $\Olin=0$, etc., 
except for $\glinh_{\rm in}$ and $\glinh_{\rm out}$ which must be constrained to be  
 symmetric traceless $S$-tangent $2$-tensors with respect to the Schwarzschild metric, but which we
 can again choose so that $|\glinh_{\rm in}-\hat\gslash_{\rm in}|\lesssim\varepsilon_1^2$,
 $|\glinh_{\rm out}-\hat\gslash_{\rm out}|\lesssim\varepsilon_1^2$ with respect to  a suitable norm.
 This gives rise to a full data set for the linearised Einstein equations
on $[u_{-2},\infty)\times \{v_{-2}\}\times \mathbb S^2\cup \{u_{-2}\}\times [v_{-2},\infty)\times \mathbb S^2$.

The following hold:
\vskip1pc
\noindent
{\bf (a) Preliminary normalised solution.}
We may associate to this data set
a smooth ``linearised metric''
\begin{eqnarray}
\nonumber
\label{linearisedmetrichere}
\justglin(u,v,\theta) = -2\Olino^2(u,v,\theta) (du\otimes dv+dv\otimes du)  + \glin_{CD}(u,v,\theta)(d\theta^C
\otimes d\theta^D)\\
-\slashed{g}_{\circ,CD}(\bmlin^C(u,v,\theta)du\otimes d\theta^D- \bmlin^D(u,v,\theta)du\otimes d\theta^C)
\end{eqnarray}
defined globally in $[u_{-2},\infty)\times[v_{-2},\infty)\times\mathbb S^2$
attaining the data and solving the linearised Einstein equations in double null gauge. 
We call this the preliminary normalised linearised solution
and we may define associated linearised quantities $\xlin$, etc., as in~\cite{holzstabofschw}.
\vskip1pc
\noindent
{\bf (b) Null-infinity renormalised solution.}
We may now define a pure gauge solution $\mathscr{G}$ so that, upon adding this solution
to~\eqref{linearisedmetrichere}, 
we have that the new associated linearised quantities satisfy:
\begin{equation}
\label{linearisedasymptflat}
\lim_{v\to\infty} r^{-2}\glin(u,v,\theta)=0,
\qquad \lim_{v\to\infty} r^{-2}\Klin(u,v,\theta)=0,\qquad
\lim_{v\to\infty} \Olino(u,v,\theta)=0.
\end{equation}
We call this the null infinity renormalised linearised solution.

\vskip1pc

For both these linearised solutions (a) and (b), we have the following (cf.~Section~\ref{reflinearisedKerrsec}):
 \begin{equation}
 \label{dontchangethem}
	\slashed{curl} \elin_{\ell =1}
	=
	\sum_{m=-1}^1
	J^m_{\rm seed}
	\frac{6}{r^4} \mathring{Y}^{\ell=1}_m,
\qquad
	\slashed{curl} (\Omega_{\circ} \blin)
	=
	\Omega_{\circ}^2 \sum_{m=-1}^1
	J^m_{\rm seed}
	\frac{6}{r^5}  \mathring{Y}^{\ell=1}_m,
\end{equation}
where we recall that $\mathring{Y}$ denote the round spherical harmonics and
the operators above are defined with respect to the Schwarzschild metric with mass $M_{\rm init}$.
\end{proposition}

\begin{proof}
Statement (a) follows directly from the well-posedness statement Theorem~8.1 of~\cite{holzstabofschw}. (Note that,
to apply here~\cite{holzstabofschw},
one must trivially
rewrite the equations~\cite{holzstabofschw} with  the torsion $\bmlin$ term
corresponding to the form~\eqref{doublenulllongforminterchanged} in place of~\eqref{doublenulllongform}.).
For (b), note that the so-called initial data normalisation of Theorem~9.1 of~\cite{holzstabofschw}
already accomplishes the first two relations of~\eqref{linearisedasymptflat}, while
an additional explicit and easily computed pure gauge solution generated by $\flin_3(u, \theta)$ can  be
seen to accomplish $\Olino(u,v,\theta)=0$ along null infinity. The statement~\eqref{dontchangethem} now follows from
Remark~\ref{remarkaboutinitialseed}, the definition of the linearised seed data in the statement of the proposition and
the properties of the propagation of $\ell=1$ modes in linearised theory (see Theorem 9.2 of~\cite{holzstabofschw}). 
\end{proof}

\begin{remark}
\label{notsmoothlineartheory}
Let us note that the above
pure gauge solution $\mathscr{G}$ will not in general be smooth, although, under our assumption on data,
it will be such that all geometric quantities are only $C^k$ for finite $k$. 
This is because, even though our data and thus the
resulting~\eqref{linearisedmetrichere} are smooth,
 the linearised diffeomorphisms realising~\eqref{linearisedasymptflat} (for instance $\flin_3(u,\theta)$)
 depend on the asymptotic behaviour at null infinity.
 Smoothness of this would require weighted estimates to all order, whereas our only assumption here is that
 of Proposition~\ref{globexistofdata}. See also Remark~\ref{notsmoothnonlineargauge}.
\end{remark}

\section{The global smallness assumption and the smallness parameter~$\varepsilon_0$}
\label{globalsmallnessassumptionnorm}
With a criterion ensuring the global existence of initial data,
we now give the smallness assumption on initial data which will appear in the main theorem of this work.

Set $M=M_{\rm init}$ with $M_{\rm init}$ fixed in~\eqref{newfixingMinit}
and consider initial data on $\underline{C}_{\rm in}\cup C_{\rm out}$ generated by
seed data~$\mathcal{S}=\mathcal{S}^{\mathcal{K}}$. Let $\mathcal{S}^{\mathcal{EF}}$ again
denote the corresponding seed data on $C_{\rm out}$.
For general $k$,
we define the following energy quantity\index{energies!initial energies!$\mathbb{E}_{\rm seed}^{k+10}[\mathcal{S}]$, initial energy associated to seed data}\index{initial data!energies!$\mathbb{E}_{\rm seed}^{k+10}[\mathcal{S}]$, initial energy associated to seed data}
\begin{align}
\label{definitionofseennormnow}
&\mathbb{E}_{\rm seed}^{k+10}[\mathcal{S}]:= \sum_{i+j=0}^{k+10}
\int_{U_{-2}}^{U_5}\int_{\mathbb S^2} |(r\slashed\nabla_\circ)^i (\underline{D}^{\mathcal{K}})^j \underline\alpha_{\rm in}^{\mathcal{K}}|^2_{\slashed{g}_{\circ}}  dU\, d\theta  +\sum_{i+j=0}^{k+10}
\int_{v_{-2}}^{\infty}\int_{\mathbb S^2} r^8 |(r\slashed\nabla_\circ)^i (r D^{\mathcal{EF}})^j \alpha_{\rm out}^{\mathcal{EF}}|^2_{\slashed{g}_{\circ}} dv\, d\theta
\\
&+\sum_{i=0}^{k+10} \int_{\mathbb S^2}  \Big\{ |(r\slashed\nabla_\circ)^i(\tr\chi_{\mathcal{EF}}-\tr\chi_{\circ,M_{\rm init},\mathcal{EF}})|^2_{\slashed{g}_{\circ}}  +
 |(r\slashed\nabla_\circ)^i(\tr\chibar_{\mathcal{EF}}-\tr\chibar_{\circ,M_{\rm init},\mathcal{EF}})|^2_{\slashed{g}_{\circ}}   +|(r\slashed\nabla_\circ)^i \hat\chi_{\mathcal{EF}}|^2_{\slashed{g}_{\circ}}\nonumber  \\
    &  \qquad \qquad \qquad  +|(r\slashed\nabla_\circ)^i \underline{\hat\chi}_{\mathcal{EF}}|^2_{\slashed{g}_{\circ}}
  +|(r\slashed\nabla_\circ)^i\eta_{\mathcal{EF}}|^2_{\slashed{g}_{\circ}} 
 +|(r\slashed\nabla_\circ)^i(K-r^{-2})|^2_{\slashed{g}_{\circ}} + | (r\slashed{\nabla}_0)^i \left(\slashed{g}- r^2 \mathring\gamma\right) |^2 \Big\} (u_{-2},v_{-2})\, d\theta \, . \nonumber
\end{align}
Note that for all $k\ge0$, this  quantity is stronger than that appearing in Proposition~\ref{existenceofinitdata} 
both in terms of its
regularity and its $r$ weights.

Let us fix the parameter\index{initial data!parameters!$N$, parameter measuring regularity
required on data} 
\[
N= 12.
\]
This will define the order of differentiability of our smallness assumption.
Consider seed data satisfying\index{initial data!parameters!$\varepsilon_0$, smallness parameter appearing in the main theorem} 
\begin{equation}
\label{smallnessofdata}
	\mathbb{E}^{N+10}_{\rm seed}[\mathcal{S}]\le \varepsilon_0^2,
\end{equation}
and a
$0< \varepsilon_0\le \hat\varepsilon_0(M_{\rm init})$, where
  $\hat\varepsilon_0(M_{\rm init})$ is to be constrained to be sufficiently
small at various points in the work.
\emph{The inequality~\eqref{smallnessofdata} 
will be our fundamental global smallness assumption which we shall require
of all $\mathcal{S}$ whose evolution we shall attempt to estimate.}

More generally, we may set $N\ge 12$, but in this case our constant $\hat\varepsilon$ must
be taken to depend in addition on $N$, i.e.~$\hat\varepsilon(M_{\rm init},N)$.

If follows in particular from~\eqref{smallnessofdata} that if  
\begin{equation}
\label{firstrestriction}
\hat\varepsilon_0\leq \varepsilon_1,
\end{equation} 
then
the assumptions of Proposition~\ref{existenceofinitdata} hold and
the full data exist globally.
For convenience, let us always assume~\eqref{firstrestriction} in what follows.

Recall by Proposition~\ref{existenceofinitdata}
that the operators  $\slashed\nabla$, $\slashed\nabla_4$, $\slashed\nabla_3$, and more generally
 $\mathfrak{D}$ to all order, are defined acting on all the quantities of Section~\ref{thatsallofthem}.
Let us use the convention that these will represent operators
on $\underline{C}_{\rm in}^{\mathcal{K}}$ and on $C_{\rm out}^{\mathcal{EF}}$ as we will express
the norm in terms of $\mathcal{K}$ quantities and $\mathcal{EF}$ quantities, respectively,
on these two initial hypersurfaces.
For quantities defined on their sphere of intersection, we will use the $\mathcal{EF}$ quantities.

With this understanding, under the assumption~\eqref{smallnessofdata}, we may
define, for $k\geq 12$, the following initial data energy,\index{energies!initial energies!$\mathbb{E}^k_0[\mathcal{S}]$, energy associated to full initial data set}\index{initial data!energies!$\mathbb{E}^k_0[\mathcal{S}]$, energy associated
to full initial data set}
\begin{align}
\nonumber
	\mathbb{E}^{k}_0[\mathcal{S}]
	:=
	\
	&
	\sum_{|l| \leq k  }
	\int_{v_{-2}}^{\infty} \int_{S_{u_{-2},v}}
	\left\vert  \mathfrak{D}^l (r^4 \alpha, r^3 \beta, r^2(\rho-\rho_\circ), r^2\sigma, r\underline{\beta}) \right\vert^2
	d \theta\, d v (u_{-2})
	\\
\nonumber
	&
	+
	\sum_{\substack{|l| \leq k }} \int_{U_{-2}}^{U_5} \int_{S_{U,V_{-2}}}
	\left\vert  \mathfrak{D}^l  \left( \beta^{\mathcal{K}}, \rho^{\mathcal{K}}-\rho^{\mathcal{K}}_\circ, \sigma^{\mathcal{K}}, \underline{\beta} ^{\mathcal{K}}, \underline{\alpha}^{\mathcal{K}}\right) \right\vert^2
	d \theta\, dU (V_{-2})
	\\
	&+\sum_{|l| \leq k+1} \sup_{U \in \left[U_{-2},U_5\right]} \int_{S_{U,V_{-2}}} \big|\mathfrak{D}^{l} (\hat{\chi}^{\mathcal{K}}, \underline{\hat{\chi}}^{\mathcal{K}}, tr \chi^{\mathcal{K}} - tr \chi^\mathcal{K}_{\circ}, tr \underline{\chi}^{\mathcal{K}} - tr \underline{\chi}^\mathcal{K}_{\circ}, \eta^{\mathcal{K}}, \underline{\eta}^{\mathcal{K}}, \underline{\hat\omega}^{\mathcal{K}}-\underline{\hat\omega}^{\mathcal{K}}_\circ )|^2\,d\theta  \nonumber \\
	&+ \sum_{|l| \leq k+1} \sup_{v \in \left[v_{-2},\infty\right]}\int_{S_{u_{-2},v}}  \Bigg\{ \Big|\mathfrak{D}^{l} \Big(r^2 \hat{\chi}, r \underline{\hat{\chi}}, r^2 (\Omega tr \chi - \Omega tr \chi_{\circ}), r (\Omega tr \underline{\chi} - \Omega tr \underline{\chi}_{\circ}), r \eta, r^2 \underline{\eta}, \nonumber \\
	 &\qquad \qquad \qquad \qquad \qquad \qquad \qquad r^3 (\hat\omega-\hat\omega_\circ), 1-\frac{\Omega_\circ^2}{\Omega^2}\,\,\Big)\Big|^2 \Bigg\}\,d\theta  +\sum_{j=0}^k \int_{S_{u_{-2},v_{-2}}} \big| \slashed{\nabla}^j \left(\slashed{g}- r^2 \mathring\gamma\right)\big|^2 \,d\theta,\label{datanormtobesmall}
\end{align}
which in general may be infinite.
In the above, we note that we have dropped the $\mathcal{EF}$ labels from the $\mathcal{EF}$ quantities,
but have retained them on the $\mathcal{K}$ quantities.

\begin{proposition}
Under the assumption~\eqref{smallnessofdata}, with the above definitions, 
it follows that  the full initial data satisfy the estimate
\begin{equation}
\label{smallnessofdatacor}
	\mathbb{E}^{N}_0[\mathcal{S}]\lesssim \varepsilon_0^2.
\end{equation}
More generally, for $k\ge 12$, we have the estimate
\[
\mathbb E^k_0[\mathcal{S}] \leq C_k \mathbb E^{k+10}_{\rm seed}[\mathcal{S}],
\]
for a constant $C_k$ depending only on $M_{\rm init}$ and $k$ 
(provided the right hand side is finite).

\end{proposition}

\begin{proof}
See the appendix of~\cite{holzstabofschw} where this construction is carried out explicitly in the linearised setting (the familiar derivative loss being due to the integration of transport equations along the characteristic hypersurfaces).
\end{proof}

Note that for Schwarzschild data of mass $M_{\rm init}$ in its usual form we have
$\mathbb{E}^k_0[\mathcal{S}]=\mathbb{E}^{k+10}_{\rm seed}[\mathcal{S}]=0$.

\begin{remark}
Due to the specific normalisation (\ref{seedrestrictions}), (\ref{seedrestrictions2}) here, the zeroth order bounds on $\hat\omega-\hat\omega_\circ$, $1-\frac{\Omega_\circ^2}{\Omega^2}$ and $\underline{\hat\omega}^{\mathcal{K}}-\underline{\hat\omega}^{\mathcal{K}}_{\circ}$ are in fact trivial and, in view of $\eta + \underline{\eta}=0$, the zeroth order bound on $r \eta$ may be replaced by one on $r^2 \eta$. We have included the quantities because the normalisation  (\ref{seedrestrictions}), (\ref{seedrestrictions2}) is not strictly necessary to construct data satisfying (\ref{smallnessofdata}). 
\end{remark}

\begin{remark}
From the point of view of regularity, the norm (\ref{datanormtobesmall}) is slightly
stronger than what we will need to keep track of: For convenience we have included bounds on $k+1$ derivatives of all Ricci coefficients, although such bounds are only required (and propagated) for some of them. Finally, note that $L^2$ bounds on $k$ derivatives (not all of them in the $4$-direction) of~$\alpha^{\mathcal{K}}$ on the ingoing cone $\underline{C}_{\rm in}$ follow from the Bianchi equations and (\ref{smallnessofdata}). Similarly, weighted $L^2$ bounds on $k$ derivatives (not all of them in the $3$-direction) of $\underline{\alpha}$ follow on the outgoing cone $C_{\rm out}$.  
\end{remark}

\begin{remark}
\label{weakerrratesremark}
Note that the bounds on $(\Omega tr \underline{\chi} - \Omega tr \underline{\chi}_{\circ})$ and $\eta$ 
imposed by  (\ref{smallnessofdata}) are weaker than suggested by the $\Phi_p$ notation. Indeed, 
it is only when the corresponding solution is renormalised with respect to null infinity that 
these quantities will  exhibit $r^{-2}$-decay along outgoing cones.
\end{remark}

\begin{remark}
\label{commentaboutpeeling}
The boundedness~\eqref{smallnessofdata},
together with
standard Sobolev inequalities,
imposes the  pointwise bound
$|r^{9/2}\alpha|\lesssim \varepsilon_0$ and,
through the Bianchi identity~\eqref{eq:beta4} connecting $\beta$ and $\alpha$, the bound
$|r^4\beta| \lesssim \varepsilon_0$.
We see thus that~\eqref{smallnessofdata} is a stronger assumption than that arising from the most general
asymptotically flat initial data considered in~\cite{CK}. We emphasise however that 
one may still recover~\eqref{smallnessofdata} from suitable assumptions on asymptotically
flat spacelike initial data (see for instance~\cite{KN}). We note that this slightly stronger decay, effectively imposed
by the $r^4$ power of $\alpha$ in~\eqref{definitionofseennormnow}, is not essential
to our method but is convenient for our modulation theory used in constructing $\mathfrak{M}_{\rm stable}$.
See already  Remarks~\ref{pointwheredecay} and~\ref{alternativeremark}.
\end{remark}

\begin{remark}
Consider seed data $\mathcal{S}^{\mathcal{K}}$ satisfying~\eqref{smallnessofdata}  for some
$\varepsilon_0$,
but where $\mathbb{E}^N_{\rm seed}[\mathcal{S}]$ is 
defined without the last term in~\eqref{definitionofseennormnow}.
Then, by Proposition~\ref{determiningthesphere} and Remark~\ref{seeddatadiffeo}, 
we may define a new seed data set
related to the original one by pulling back the quantities via a diffeomorphism
$\psi:\mathbb S^2\to \mathbb S^2$, such that~\eqref{smallnessofdata} is satisfied
with $C\varepsilon_0$ in place of $\varepsilon_0$, for some $C$ depending only on $M_{\rm init}$,
provided of course that
$C\varepsilon_0\leq \hat\varepsilon_0$.
\end{remark}

Let us introduce also the notation $\mathbb{E}^k_{0,v_{\infty}}[\mathcal{S}]$\index{initial data!energies!$\mathbb{E}^k_{0,v_\infty}[\mathcal{S}]$, energy of truncated initial data}\index{energies!initial energies!$\mathbb{E}^k_{0,v_\infty}[\mathcal{S}]$, energy of truncated initial data} for~\eqref{datanormtobesmall}
where we replace $v=\infty$ in the limit of
integration with $v=2v_{\infty}$ for a given parameter $v_\infty$.
For smooth initial data, we have trivially by compactness that
\begin{equation}
\label{triviallyforhigher}
\mathbb{E}^k_{0,v_{\infty}}[\mathcal{S}] \le C(k, v_{\infty})
\end{equation}
for some constant depending on the seed data, $k$ and $v_{\infty}$.
This will be useful for considering the propagation of higher regularity
(see already the statement of Theorem~\ref{thehigherordertheorem}).

\begin{remark}
\label{trivialremark}
Let us note that given an embedding~\eqref{embedintheor} for a $\mathcal{W}\times \mathbb S^2$ 
with past boundary $C^{\mathcal{K}}_{\rm out}\cup \underline{C}^{\mathcal{K}}_{\rm in}\cap \mathcal{W}\times \mathbb S^2$
where $(\mathcal{M},g)$ is globally hyperbolic (with past boundary diffeomorphic to 
$C^{\mathcal{K}}_{\rm out}\cup \underline{C}^{\mathcal{K}}_{\rm in}\cap \mathcal{W}\times \mathbb S^2$)
and satisfies the~\eqref{Ricciflathere} 
and where $i^*g$ is expressed in double null 
form~\eqref{doublenulllongform}, 
then if moreover
\begin{equation}
\label{datainthegauge}
\Omega^2(i^*g)|_{C^{\mathcal{K}}_{\rm out}\cup \underline{C}^{\mathcal{K}}_{\rm in}\cap \mathcal{W}\times\mathbb{S}^2 } =\Omega^2_{\circ,M,\mathcal{K}}, 
\qquad  b^A(i^*g)|_{C^{\mathcal{K}}_{\rm out}\cap  \mathcal{W}\times\mathbb{S}^2 }= 0,
\end{equation}
 it follows that
defining the seed data by the quantities corresponding to $i^*g$ on $C_{\rm out}^{\mathcal{K}}\cup
\underline{C}_{\rm in}^{\mathcal{K}}\cap \mathcal{W}\times\mathbb{S}^2 $, then  $(\mathcal{M},g)$ is a Cauchy evolution
corresponding to this seed data.
\end{remark}

\begin{remark}
\label{longertrivialremark}
We have already remarked that for Schwarzschild data of mass $M:=M_{\rm init}$ induced
from~\eqref{SchwmetricKruskal}, 
we have that $\mathbb E^k_0[\mathcal{S}]=\mathbb E^k_{\rm seed}[\mathcal{S}]=0$ 
and thus the smallness condition~\eqref{smallnessofdata}
is trivially satisfied.

Let us also note the following additional
``trivial'' examples of initial data satisfying the smallness condition,
which will all be induced from the Schwarzschild (or more generally Kerr) family itself
via Remark~\ref{trivialremark}.
\begin{enumerate}
\item
\label{firstitemhere}
Consider a diffeomorphism $\psi:\mathbb S^2\to \mathbb S^2$ sufficiently close to the identity and consider the initial data induced as in Remark~\ref{trivialremark} by $i^*g_{\circ, M, \mathcal{K}}$ where 
$i={\rm id}\times \psi:\mathcal{W}\times\mathbb S^2\to (\mathcal{M}_{\rm Kruskal}, g_{\circ, M, \mathcal{K}})$
and $\mathcal{W}=\mathcal{W}_{\mathcal{K}}\cap [U_{-2},U_{5}]\times[V_{-2},\infty)$.
This satisfies the smallness condition~\eqref{smallnessofdata}
but fails to satisfy  $\mathbb E_0^k=0$ 
only on account of the final term in the last integral of $(\ref{datanormtobesmall})$.
\item
\label{seconditemhere}
Consider a sphere $S\subset (\mathcal{M}_{\rm Kruskal}, g_{\circ, M, \mathcal{K}})$ 
in the Schwarzschild solution sufficiently near $S_{U_{-2},V_{-2}}$, and consider
smooth future directed null vector fields  $e_3$, $e_4$ defined along $S$ such that
$g(e_3,e_4)=-2$, and $e_3\perp TS$, $e_4\perp TS$, and such that
$e_3$ and $e_4$ are sufficiently close to the Kruskal vectors $e^{\mathcal{K}}_3$ and $e^{\mathcal{K}}_4$, resp., 
defined previously.
Assume in particular that $S$ is such that the map $S\hookrightarrow \mathcal{W}_{\mathcal{K}}\times \mathbb S^2\xrightarrow{\pi}\mathbb S^2$
is a diffeomorphism $S\to\mathbb S^2$.
For a suitable $\mathcal{W}$ as in Remark~\ref{trivialremark}, we may now define uniquely
a local double null gauge $\mathcal{W}\times \mathbb S^2
\xrightarrow{i}
(\mathcal{M}_{\rm Kruskal}, g_{\circ, M, \mathcal{K}})$ 
with the property that the metric $i^*g_{\circ, M, \mathcal{K}}$ 
takes the double null form~\eqref{doublenulllongform}, $\pi\circ i|_{(U_{-2},V_{-2})\times \mathbb S^2}={\rm id}$,
 such that~\eqref{datainthegauge} 
 are satisfied and such that the vector fields $(i_*)^{-1}(e_3)$ and $(i_*)^{-1}e_4$ 
correspond to~\eqref{nullframedef} on $i^{-1}(S)=S_{U_{-2},V_{-2}}$.
This gives rise to initial data via Remark~\ref{trivialremark} 
satisfying the smallness condition~\eqref{smallnessofdata}.
\item 
\label{thirditemhere}
Given $\tilde{M}\neq M$,  for a suitable $\mathcal{W}$  as in Remark~\ref{trivialremark},
consider the map $i:\mathcal{W}\times\mathbb S^2\to
(\mathcal{M}_{\rm Kruskal},g_{\circ,\tilde{M},\mathcal{K}})$ 
uniquely determined by the conditions
$i_{M,\tilde{M}}\circ\pi_{\mathbb S^2} =
\pi_{\mathbb S^2}\circ i $, $i(U_{-2},V_{-2},\cdot)= (U_{-2},V_{-2},\cdot)$,
$i(U_{5},V_{-2},\cdot)= (U_{5},V_{-2},\cdot)$
and  $i^*g_{\circ,\tilde{M},\mathcal{K}}$ is in double null form~\eqref{doublenulllongform} with
\[
\Omega^2(i^*g_{\circ, \tilde{M},\mathcal{K}})(U_{-2},\cdot,\cdot )=
\Omega^2_{\circ,M,\mathcal{K}}(U_{-2},\cdot, \cdot), \qquad
\Omega^2(i^* g_{\circ, \tilde{M},\mathcal{K}})(\cdot, V_{-2},\cdot)=
\Omega^2_{\circ,M,\mathcal{K}}(\cdot, V_{-2},\cdot).
\]
Note under these assumptions all normalisations~\eqref{datainthegauge} are in fact satisfied.
Then, if $|\tilde{M}-M|$ is sufficiently small,
it follows that the data induced by $i^*g_{\circ,\tilde{M},\mathcal{K}}$ 
satisfy the smallness condition~\eqref{smallnessofdata}.
\item 
\label{fourthitemhere}
Consider (the analogue of the Lemaitre region of) 
a  slowly rotating Kerr solution $(\mathcal{M}_{\rm Kerr},g_{a, M})$ (see~\eqref{KerrmetricBL} for the form of the
metric in local coordinates on the exterior), 
again with $M:=M_{\rm init}$ and with some $a\ne 0$. 
We claim that, for a suitable $\mathcal{W}$, there exists an embedding $i:\mathcal{W}
\times \mathbb S^2 \to (\mathcal{M},g_{a, M})$ satisfying
the assumptions of Remark~\ref{trivialremark}. 
(One can deduce this again via purely geometric considerations
 starting from a sphere $S$ and a null frame $e_3, e_4$ as in example~\ref{seconditemhere} above. Alternatively,
 one can  deduce an explicit form for $i$  starting from the Kerr metric in double null gauge given
in equation (A.43) of~\cite{DafLuk1}, regularising the coordinate across the event horizon,
and rescaling so that the normalisations~\eqref{datainthegauge} hold.) Moreover, 
if $|a|\ll M$, this choice can be made so that
the induced data from $i^* g_{a, M}$
moreover satisfy the smallness condition~\eqref{smallnessofdata}.
\end{enumerate}
We can in fact combine~\ref{firstitemhere}--\ref{fourthitemhere} to construct more complicated
examples  all connected to the explicit Schwarzschild/Kerr family.
\end{remark}

\section{Cauchy stability and the initial data gauges}
\label{localexistencesection}
In this section 
we shall state two results giving that, for data
with small norm, the maximal Cauchy development
contains sufficiently ``large'' regions covered
by what we shall term ``the initial data gauges''.
The initial Kruskal gauge result is given by Theorem~\ref{thm:localKrus} 
in {\bf Section~\ref{initKrusksec}} while the initial Eddington--Finkelstein gauge
is given by 
Theorem~\ref{thm:localEF} in {\bf Section~\ref{initEFsec}}.
The former result is essentially just standard Cauchy stability,
while the latter can be thought of as
a more quantitative version, allowing existence all the
way up to a piece of null infinity. (In  particular, the latter already depends
on the good ``null structure'' satisfied by our equations (cf.~Section~\ref{capturingnullcon}).)

\subsection{The initial Kruskal gauge}
\label{initKrusksec}

The following statement
follows from a standard Cauchy stability argument.

\begin{theorem}[Cauchy stability and the initial Kruskal gauge] 
\label{thm:localKrus}
Set $M=M_{\rm init}$, 
consider initial data as in Proposition~\ref{globexistofdata}   and let $(\mathcal{M},g)$ denote the
maximal Cauchy development given by Theorem~\ref{maxCauchythe}.

Recall the parameter $V_3>V_{-2}$\index{initial Kruskal gauge!parameters!$V_3$} 
from Section~\ref{compediumparameterssec} associated to $v_3$.
Then for sufficiently small $\hat\varepsilon_0$, 
if the initial data satisfy the global smallness assumption~\eqref{smallnessofdata} with $\varepsilon_0\leq\hat\varepsilon_0$,
then, defining $C_{\rm out}'= C_{\rm out}\cap \{V\le V_3\}$, then the
domain $\mathcal{W}$ of Theorem~\ref{maxCauchythe} 
can be chosen to be\index{initial Kruskal gauge!sets!$\mathcal{W}_{\mathcal{K}}(V_3)$, coordinate domain}
\[
\mathcal{W}_{\mathcal{K}}(V_3):=[U_{-2},U_{5}]\times[V_{-2},V_3].
\]
Refer to Figure~\ref{Kruskgaugefig}.
\begin{figure}
\centering{
\def\svgwidth{20pc}
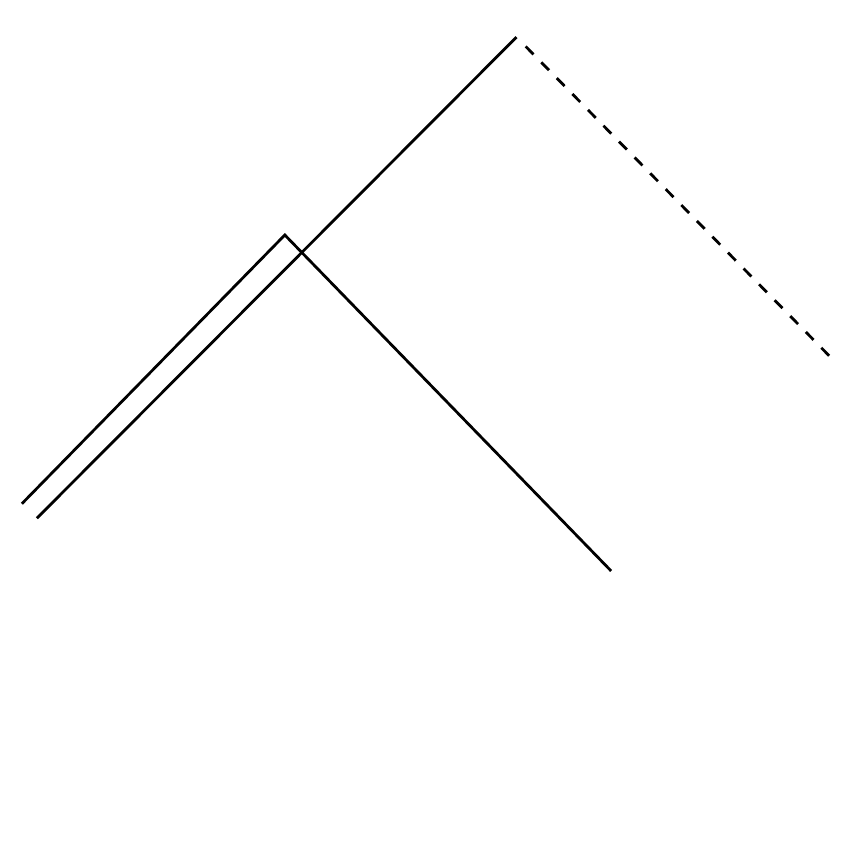}
\caption{The region $\mathcal{W}_{\mathcal{K}}(V_3)$ as a subset of the Kruskal
domain of Schwarzschild with $M=M_{\rm init}$}\label{Kruskgaugefig}
\end{figure}

We will denote the map $(\ref{embedintheor})$ by $i_{\mathcal{K}}$\index{initial Kruskal gauge!double null parametrisation!$i_{\mathcal{K}}$} and the image\index{initial Kruskal gauge!sets!$\DRK(V_3)$, spacetime domain} 
\begin{equation}
\label{DRKdef}
\DRK(V_3):=i_{\mathcal{K}}(\mathcal{W}_{\mathcal{K}}(V_3)\times \mathbb S^2)\subset \mathcal{M}.
\end{equation}

Moreover, the metric $g$ expressed in the above gauge is $\varepsilon_0$-close
to the Schwarzschild metric $(\ref{SchwmetricKruskal})$ expressed in the Kruskal gauge
by making $\varepsilon_0$ sufficiently small, in the following sense:
Noting $\mathcal{W}_{\mathcal{K}}(V_3) \subset \mathcal{W}_{\mathcal{K}}$, then
we may define $g_{\circ, M}$ on $\mathcal{W}_{\mathcal{K}}(V_3)$ by $(\ref{SchwmetricKruskal})$, and the statement is
that $i_{\mathcal{K}}^*g-g_{\circ, M}$ is appropriately controlled by $\varepsilon_0$,  
and in particular the quantities $\Phi$ defined in Section~\ref{schemfordifssec}, 
satisfy
\begin{equation}
\label{willstatelater}
	\mathbb E^N_0[ \Phi_{\mathcal{K}, d} ]  \lesssim \varepsilon_0^2,
	\end{equation}
where the energy $\mathbb E^N_0[ \Phi_{\mathcal{K}, d} ]$ is described in Definition~\ref{defofKnorms} immediately below. 
\end{theorem}

\begin{definition}
\label{defofKnorms}

We will describe here geometrically an energy $\mathbb E^N_0 [ \Phi_{\mathcal{EF}, d} ]$ for
which~\eqref{willstatelater} holds.

Consider any diffeomorphism sphere $\tilde{S} \subset \mathcal{D}^{\mathcal{K}}(V_3)$
which is $C^1$ close to a $S_{U',V'}$ sphere of the gauge~$i_{\mathcal{K}}$, i.e.~such that 
 $\pi_{\mathbb S^2}|_{\tilde{S}}: \tilde{S}\to \mathbb S^2$ is a $C^1$ diffeomorphism and 
such that there exist $(U',V')\in \mathcal{W}_{\mathcal{K}}$ with
\begin{equation}
\label{closetofoliationsphere0}
\sup_{x\in \tilde{S}} |\pi_{\mathcal{W}_{\mathcal{K}}}|_{\tilde{S}} -(U',V') |\lesssim \varepsilon_0,\qquad 
\sup_{x\in \tilde{S}}
|\mathring\nablaslash \pi_{\mathcal{W}_{\mathcal{K}}}|_{\tilde{S}} |\lesssim \varepsilon_0
\end{equation}
where the $|\cdot|$ in the first formula of~\eqref{closetofoliationsphere0} is in the usual sense of distance
in $\mathbb R^2$, while in the second we think of
$\pi_{\mathcal{W}_{\mathcal{K}}}|_{\tilde{S}}$ as a pair of scalar functions and $\mathring\nablaslash$ denotes the
operator on $\tilde{S}$ defined with respect to the pull back of the usual round covariant derivative under 
 $\pi_{\mathbb S^2}|_{\tilde{S}}: \tilde{S}\to \mathbb S^2$ and we interpret finally $|\cdot|$ as the norm on pairs of 
 vectors again defined 
 from the round metric on $\mathbb S^2$.

We consider the past ingoing and outgoing cones $\tilde{C}_{\rm in} \subset  \mathcal{D}^{\mathcal{K}}(V_3)$, 
$ \tilde{C}_{\rm out}\subset \mathcal{D}^{\mathcal{K}}(V_3)$ emanating from $\tilde{S}$. (It follows from~\eqref{closetofoliationsphere0}
that these are again $C^1$ close to  $U=U'$ and  $V=V'$ hypersurfaces of the gauge $i_{\mathcal{K}}$ in the sense
of the analogue of~\eqref{closetofoliationsphere0}.)

We define\index{energies!initial energies!$\mathbb E^N_0[ \Phi_{\mathcal{K}, d} ] $, initial energy associated to the initial Kruskal gauge}
\begin{align}
\mathbb E^N_0[ \Phi_{\mathcal{K}, d} ] 
= \sup_{\tilde{S}  \subset \mathcal{D}^{\mathcal{K}}(V_3)} \Bigg\{ & \phantom{X} \sum_{\Phi, \,|\gamma| \leq N-1} \| (\mathfrak{D}^{\gamma}  \Phi)_{\mathcal{K},d} \|^2_{L^2(\tilde{S})} \nonumber \\
&+ \sum_{\mathcal{R} \setminus \{ \Omega^{-2} \underline{\alpha}\}, \, \vert \gamma \vert \leq N}
		\Vert (\mathfrak{D}^{\gamma} \mathcal{R} )_{\mathcal{K}, \mathrm{d}} \Vert_{\tilde{C}_{\rm out}}^2 	+  \sum_{\mathcal{R} \setminus \{ \Omega^2 \alpha\}\, \vert \gamma \vert \leq N} \Vert (\mathfrak{D}^{\gamma} \mathcal{R} )_{\mathcal{K}, \mathrm{d}}  \Vert_{\tilde{\underline{C}}_{\rm in}}^2 \nonumber \\
	&+\sum_{\vert \gamma \vert \leq N-1}
		\Vert \widetilde{(r \nablaslash)} (\mathfrak{D}^{\gamma} (\Omega^2 \alpha))_{\mathcal{K}, \mathrm{d}} \Vert_{\tilde{\underline{C}}_{\rm in}}^2 + \sum_{\vert \gamma \vert \leq N-1}
		\Vert \widetilde{(\nablaslash_3)} (\mathfrak{D}^{\gamma}  (\Omega^2 \alpha))_{\mathcal{K}, \mathrm{d}} \Vert_{\tilde{\underline{C}}_{\rm in}}^2	\nonumber \\
		\label{initdatanormK}
		&+\sum_{\vert \gamma \vert \leq N-1}
		\Vert \widetilde{(r \nablaslash)} (\mathfrak{D}^{\gamma} (\Omega^{-2} \alpha))_{\mathcal{K}, \mathrm{d}} \Vert_{\tilde{C}_{\rm out}}^2 + \sum_{\vert \gamma \vert \leq N-1}
		\Vert \widetilde{(\nablaslash_4)} (\mathfrak{D}^{\gamma}  (\Omega^{-2} \alpha))_{\mathcal{K}, \mathrm{d}} \Vert_{\tilde{C}_{\rm out}}^2
		\Bigg\},
\end{align}
where the supremum is taken over all $\tilde{S}$ satisfying~\eqref{closetofoliationsphere0},
and the $\tilde{C}_{\rm in}$ and $\tilde{C}_{\rm out}$ are the corresponding null cones defined above.

In the above, the operators  $\widetilde{r \slashed{\nabla}}$, $\widetilde{r \slashed{\nabla}_3}$, $\widetilde{r \slashed{\nabla}_4}$ are appropriate tangential operators  to the cones $\tilde{C}_{\rm in}$, $\tilde{C}_{\rm out}$ (defined in analogy to
Section~\ref{tangentoperatorssec}) and the volume elements are appropriately defined (see
Remark~\ref{analogousprevremark}).

\end{definition}

\begin{remark}
\label{analogousprevremark}
In practice, 
it is only in Propositions~\ref{prop:dHdiff} and~\ref{thm:gidataestimates} where 
we shall apply the  estimate~\eqref{willstatelater} for the quantity~\eqref{initdatanormK}. 
We shall need specifically the control of the expression
 in brackets in~\eqref{initdatanormK} 
 corresponding to a particular $\tilde{S}$ taken to be a sphere of our teleological  $\mathcal{H}^+$ gauge anchored
 to the gauge~$i_{\mathcal{K}}$ by the conditions described in Section~\ref{anchoredsec}. In this case, we 
may represent explicitly the quantities appearing in~\eqref{initdatanormK} using
 diffeomorphism functions relating our teleological $\mathcal{H}^+$ gauge to the gauge~$i_{\mathcal{K}}$
and the pull back measures and 
``mixed tensors" defined in Section \ref{beyondSsection}. For Proposition~\ref{prop:dHdiff}, 
the estimate~\eqref{willstatelater} will be used to control an initial energy controlling the above
diffeomorphism functions. For Proposition~\ref{thm:gidataestimates}, the important point will be that 
the norm~\eqref{initdatanormK} directly controls through the natural energy fluxes for the quantities of the almost gauge invariant hierarchy $\left(\alpha, \psi, P\right)$ and $\left(\underline{\alpha}, \check{\underline{\psi}}, \check{\underline{P}}\right)$.  Thus, the estimate~\eqref{willstatelater} will appear in this work only through the
estimates~\eqref{theestimateshereforfdhp} of Proposition~\ref{prop:dHdiff} and  
estimates~\eqref{thisiswheretheinitialfluxesappearone} and~\eqref{thisiswheretheinitialfluxesappeartwo}
of Proposition \ref{thm:gidataestimates}.
\end{remark}

\begin{remark}
\label{whenwedistinguishkruskaldata}
When we must distinguish between different double null gauges (cf.~the discussion in 
Section~\ref{changeisgood}), 
we shall denote the coordinates on $\DRK(V_3)$
as $(U_{data},V_{data},\theta_{data}^\mathcal{K})$ and refer to these coordinates
as the ``initial Kruskal system''.\index{initial Kruskal gauge!coordinates!$U_{data}$}\index{initial Kruskal gauge!coordinates!$V_{data}$}\index{initial Kruskal gauge!coordinates!$\theta_{data}^{\mathcal{K}}$}
\end{remark}

\begin{proof}
This is essentially a standard Cauchy stability argument applied to the equations of 
Section~\ref{thatsallofthem}.  
The reader unfamiliar with local energy estimates for this system may wish to return to this theorem
after reading the main estimates of this work, which are strictly harder, so as to understand
how the expressions in~\eqref{initdatanormK} naturally appear. 
\end{proof}

\subsection{The initial Eddington--Finkelstein gauge}
\label{initEFsec}

The construction of the initial Eddington--Finkelstein gauge is slightly more involved.

First of all, we need a more quantitative
theorem than Cauchy stability as we would like the gauge to exist up until a fixed
retarded time
$u_{3}$ for all values of $v>v_{0}$, i.e.~we would like to define a ``piece of null infinity $\mathcal{I}^+$''.

Moreover, for this gauge to properly exhibit the asymptotic flatness of the solution (roundness of
spheres and Bondi normalisation) we will need
to re-normalise the gauge ``at'' null infinity $\mathcal{I}^+$, cf.~the conditions~\eqref{linearisedasymptflat}
for the null infinity normalised linearised solution.

Thus, we will obtain the gauge in two stages. We first (a) obtain a preliminary gauge which 
admits the initial data and exists in a large enough region, and then (b) we renormalise it along
the initial outgoing null cone and along null infinity so as to have roundness of the spheres and
Bondi normalisation.

All these statements are summarised in the following theorem:

\begin{theorem}[Existence of the initial Eddington--Finkelstein gauge] \label{thm:localEF}
Set $M=M_{\rm init}$, 
consider initial data as in Proposition~\ref{globexistofdata}   and let $(\mathcal{M},g)$ denote the
maximal Cauchy development given by Theorem~\ref{maxCauchythe}.
\vskip1pc
\noindent
{\bf (a) The preliminary gauge.}
Recall the parameter $u_4>u_{-2}$\index{initial Eddington--Finkelstein gauge!parameters!$u_4$} 
from Section~\ref{compediumparameterssec}.
Then for sufficiently small $\hat\varepsilon_0$, 
if the initial data satisfy the global smallness assumption~\eqref{smallnessofdata} with $\varepsilon_0\le\hat\varepsilon_0$,
then the following is true:

Defining $\underline{C}'{}_{\rm in}^{\mathcal{EF}}=\iota^{-1}_M(C^{\mathcal{K}}_{\rm in}\cap \{U\le U_4\})$, 
then the domain $\mathcal{W}$ of Theorem~\ref{maxCauchythe} applied to  $\mathcal{S}^{\mathcal{EF}}$
restricted to $\underline{C}'{}^{\mathcal{EF}}_{\rm in}\cup C_{\rm out}^{\mathcal{EF}}$
can be chosen to be\index{initial Eddington--Finkelstein gauge!sets!$\mathcal{W}_{\mathcal{EF}}^0(u_3)$, preliminary coordinate domain} 
\[
\mathcal{W}_{\mathcal{EF}}^0(u_4)=[u_{-2},u_4]\times[v_{-2},\infty).
\]

We will denote the map $(\ref{embedintheor})$ by $i^0_{\mathcal{EF}}$\index{initial Eddington--Finkelstein gauge!double null parametrisation!$i^0_{\mathcal{EF}}$, preliminary gauge} and its image by $\mathcal{D}^0_{\mathcal{EF}}(u_4)$.\index{initial Eddington--Finkelstein gauge!sets!$\mathcal{D}^0_{\mathcal{EF}}(u_f)$, spacetime domain of the preliminary gauge}

In the region $\iota_M(\mathcal{W}_{\mathcal{EF}}^0(u_4))\cap \mathcal{W}_{\mathcal{K}}(V_3)$ we have
\begin{equation}
\label{samenesshere}
i_{\mathcal{K}}^*g = (\iota_M^{-1})^* {i^{0*}_{\mathcal{EF}}} g.
\end{equation}

Moreover, the metric $g$ expressed in the above gauge is $\varepsilon_0$-close 
to the Schwarzschild metric $(\ref{SchwmetricEF})$ expressed in the Eddington--Finkelstein
double null gauge:
Noting $\mathcal{W}^0_{\mathcal{EF}}(u_4) \subset \mathcal{W}_{\mathcal{EF}}$, then
we may define $g_{\circ, M}$ on $\mathcal{W}^0_{\mathcal{EF}}(u_4)$ by $(\ref{SchwmetricEF})$, and the statement is
that $i^0{}_{\mathcal{EF}}^*g-g_{\circ, M}$ is
controlled in a suitable norm by $\varepsilon_0$.

In fact,
$i^0{}_{\mathcal{EF}}^*g-g_{\circ, M}-\justglin$ is controllable in a suitable norm by $\varepsilon_0^2$,
where $\justglin$ denotes the initial data normalised linearised solution of Proposition~\ref{linearisedprophere}.

\vskip1pc
\noindent
{\bf (b) The renormalised gauge at $\mathcal{I}^+$.}
Now, recalling also the parameters $u_3$\index{initial Eddington--Finkelstein gauge!parameters!$u_{3}$}  and $v_0$\index{initial Eddington--Finkelstein gauge!parameters!$v_{0}$}  from Section~\ref{compediumparameterssec}
and defining\index{initial Eddington--Finkelstein gauge!sets!$\mathcal{W}_{\mathcal{EF}}(u_3)$, renormalised coordinate domain} 
\[
\mathcal{W}_{\mathcal{EF}}(v_{0},u_3)=[u_{-2},u_3]\times[v_0,\infty),
\]
we may define, for sufficiently small $\varepsilon_0$, a new parameterisation\index{initial Eddington--Finkelstein gauge!double null parametrisation!$i_{\mathcal{EF}}$, renormalised gauge} 
\begin{equation}
\label{herethenewone}
i_{\mathcal{EF}}:\mathcal{W}_{\mathcal{EF}}(v_0,u_3)\times \mathbb S^2\to \mathcal{M}
\end{equation}
such that the following normalisations hold:
\begin{equation}
\label{somenormalisations}
\lim_{v\to \infty} r^{-2}\slashed{g}(u,v,\theta)=\mathring\gamma(\theta),\qquad
\lim_{v\to \infty} r^2K(u,v,\theta )= 1, \qquad \lim_{v\to \infty} \Omega^2(u,v,\theta)=1
\end{equation}
\begin{equation}
\label{normalisationonoutgoing}
\Omega^2(u_{-2},v,\theta)= \Omega^2_{\circ,M,\mathcal{EF}}
\end{equation}
and such that 
\[
i_{\mathcal{EF}}(\mathcal{W}_{\mathcal{EF}}(v_{0},u_3)\times \mathbb S^2)\subset
i^0_{\mathcal{EF}}(\mathcal{W}_{\mathcal{EF}}^0(u_4)\times \mathbb S^2)=\mathcal{D}^0_{\mathcal{EF}}(u_4),
\]
\[
i_{\mathcal{EF}}( \{u_{-2}\}\times[v_0,\infty)\times \mathbb S^2 )\subset i^0_{\mathcal{EF}}({C_{\rm out}}).
\]
and 
\begin{equation}
\label{polesshouldbethesame}
\pi_{\mathbb S^2} \circ (i^0_{\mathcal{EF}})^{-1}\circ  i_{\mathcal{EF}} (\{u_{-2}\} \times [v_0,\infty)\times \{(0,0,1)\}) =(0,0,1)\in \mathbb S^2
\end{equation}
\begin{equation}
\label{meridianstoo}
d (\pi_{\mathbb S^2}\circ (i^0_{\mathcal{EF}})^{-1} \circ  i_{\mathcal{EF}}) |_{(\{u_{-2}\} \times [v_0,\infty)\times \{(0,0,1)\})} (0,1,0)= (0,1+\xi ,0)\in T_{(0,0,1)}\mathbb S^2 \subset \mathbb R^3
\end{equation}
for some real number $|\xi|\lesssim \varepsilon_0$.
Refer to Figure~\ref{initEFgaugefig}.
\begin{figure}
\centering{
\def\svgwidth{20pc}
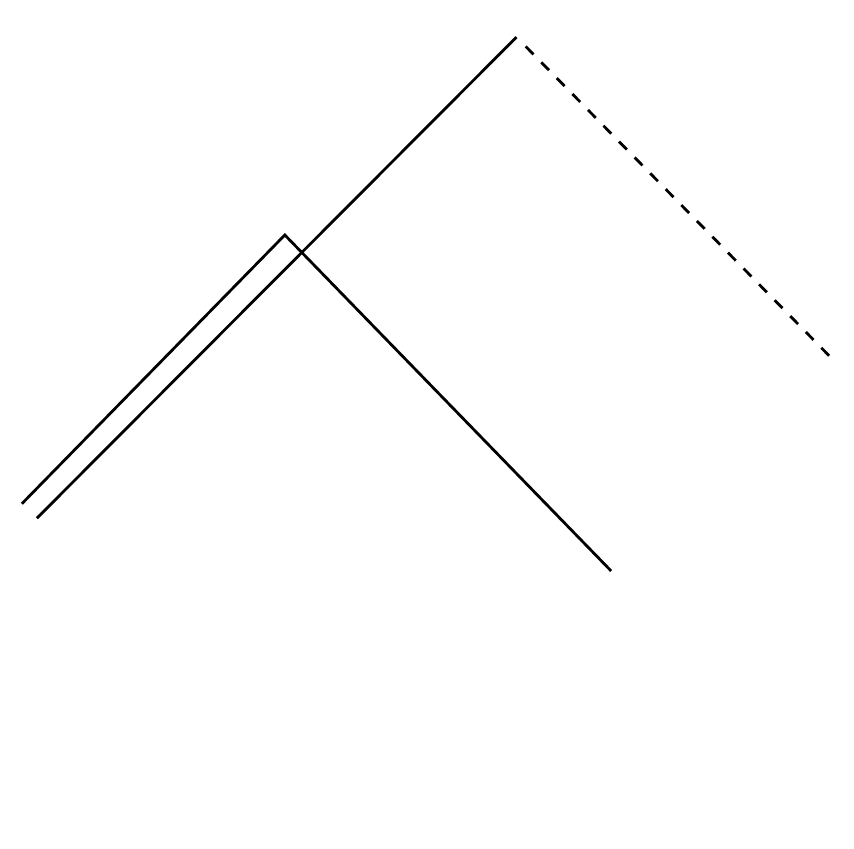}
\caption{The region $\mathcal{W}_{\mathcal{EF}}(u_3)$ as a subset of the Kruskal
domain of Schwarzschild with $M=M_{\rm init}$}\label{initEFgaugefig}
\end{figure}

We will denote the image\index{initial Eddington--Finkelstein gauge!sets!$\DREF(u_2)$, spacetime domain}
\begin{equation}
\label{DREFdef}
\DREF(u_3):=i_{\mathcal{EF}}(\mathcal{W}_{\mathcal{EF}}(u_3)\times \mathbb S^2)\subset \mathcal{M}.
\end{equation}

This parametrisation is close to the parametrisation $i^0_{\mathcal{EF}}$, in particular, in the sense that
\begin{equation}
\label{eq:fddestimates}
|\pi|_{\mathcal{W}^0_{\mathcal{EF}(u_3)}} \circ (i_0^{\mathcal{EF}})^{-1} \circ i^{\mathcal{EF}}(u,v,\theta) - 
(u,v)| \lesssim \varepsilon_0 r_M(u,v).
\end{equation}
(Note, however, the growth in $r$ on the right hand side of~\eqref{eq:fddestimates}.)

As with $i^0{}_{\mathcal{EF}}^*g$, we have that 
 $i_{\mathcal{EF}}^*g-g_{\circ, M}$ is
controlled by $\varepsilon_0$,  
specifically the quantities $\Phi$ defined in Section~\ref{schemfordifssec} satisfy
\begin{align} \label{willstatelater2}
\mathbb E^N_0[ \Phi_{\mathcal{EF}, d} ] \lesssim \varepsilon_0^2,
\end{align}
where the energy $\mathbb E^N_0[ \Phi_{\mathcal{EF}, d} ]$ is described in Definition~\ref{defofEHnorms} immediately below. 

Moreover,  $i_{\mathcal{EF}}^*g-g_{\circ, M}-\justglin$ is in fact controlled by $\varepsilon_0^2$,
where $\justglin$ now denotes the null infinity normalised linearised solution of Proposition~\ref{linearisedprophere},
in particular
\begin{equation}
\label{inparticularherethemodes}
|r^5 \slashed{curl} \Omega\beta_{\ell=1}(u,v,\theta) - \Omega^2 \sum_{m=-1}^1
	J^m_{\rm seed}
	\frac{6}{r^5}  {Y}^{\ell=1}_m(u,v,\theta) | \lesssim \varepsilon_0^2,
\end{equation}
where $Y_{m}^{\ell=1}$ denote the spherical harmonics associated to the parameterisation~\eqref{linearisedasymptflat}
as defined in Section~\ref{basisforprojspace} and $J^m_{\rm seed}$ are defined by Definition~\ref{assocKerpardata}.
\end{theorem}

\begin{definition}
\label{defofEHnorms}
We will describe here geometrically an energy $\mathbb E^N_0 [ \Phi_{\mathcal{EF}, d} ]$ for
which~\eqref{willstatelater2} holds. Compare with Definition~\ref{defofKnorms}.

Consider any diffeomorphism sphere $\tilde{S} \subset \mathcal{D}^{\mathcal{EF}}(u_3)$
which is $C^1$ close to a $S_{u',v'}$ sphere of the gauge~\eqref{herethenewone}, i.e.~such that
 $\pi_{\mathbb S^2}|_{\tilde{S}}: \tilde{S}\to \mathbb S^2$ is a $C^1$ diffeomorphism and 
such that there exist $(u',v')\in \mathcal{W}_{\mathcal{EF}}$  with
\begin{equation}
\label{closetofoliationsphere}
\sup_{x\in \tilde{S}} |v|_{\tilde{S}} -(u',v') |\lesssim \varepsilon_0 r,\qquad \sup_{x\in \tilde{S}} |u|_{\tilde{S}} -(u',v') |\lesssim \varepsilon_0  \qquad 
\sup_{x\in \tilde{S}}
|\mathring\nablaslash \pi_{\mathcal{W}_{\mathcal{EF}}}|_{\tilde{S}} |\lesssim \varepsilon_0
\end{equation}
interpreted as in~\eqref{closetofoliationsphere0}.
(Note the factor of $r$ allowed in the first inequality of~\eqref{closetofoliationsphere}.)

We consider the past ingoing and outgoing cones $\tilde{C}_{\rm in} \subset  \mathcal{D}^{\mathcal{EF}}(u_3)$, $ \tilde{C}_{\rm out}\subset \mathcal{D}^{\mathcal{EF}}(u_3)$ emanating from $\tilde{S}$. (It follows from~\eqref{closetofoliationsphere}
that these are again $C^1$ close to  $u=u'$ and  $v=v'$ hypersurfaces of the gauge~\eqref{herethenewone} in the sense
of the analogue of~\eqref{closetofoliationsphere}.)

We define\index{energies!initial energies!$\mathbb E^N_0[ \Phi_{\mathcal{EF}, d} ]$, initial energy associated to the initial Eddington--Finkelstein gauge}
\begin{align} \label{initdatanormEF}
\mathbb E^N_0[ \Phi_{\mathcal{EF}, d} ]:= \sup_{\tilde{S}  \subset \mathcal{D}^{\mathcal{EF}}(u_3)} \Bigg\{ &\sum_{\Phi, \, |\gamma| \leq N-1} \| (\mathfrak{D}^{\gamma} r^p \Phi_p)_{\mathcal{EF},d} \|^2_{L^2(\tilde{S})} \nonumber \\
&+\sum_{\vert \gamma \vert \leq N}
		\Vert r^{-1} (\mathfrak{D}^{\gamma} (r^5 \alpha, r^4 \beta, r^3 \rho + 2M, r^3 \sigma, r^2 \underline{\beta}))_{\mathcal{EF}, \mathrm{d}} \Vert_{\tilde{C}_{\rm out}}^2  \nonumber \\
		&+\sum_{\vert \gamma \vert \leq N} \Vert  \mathfrak{D}^{\gamma} (r^4 \beta, r^3 \rho + 2M, r^3 \sigma, r^2 \underline{\beta},r \underline{\alpha})_{\mathcal{EF}, \mathrm{d}} \Vert_{\tilde{\underline{C}}_{\rm in}}^2  \nonumber \\
		&+\sum_{\vert \gamma \vert \leq N-1}
		\Vert \widetilde{(r \nablaslash)} (\mathfrak{D}^{\gamma} r^4 \alpha)_{\mathcal{EF}, \mathrm{d}} \Vert_{\tilde{\underline{C}}_{\rm in}}^2 + \sum_{\vert \gamma \vert \leq N-1}
		\Vert \widetilde{(\nablaslash_3)} (\mathfrak{D}^{\gamma} r^5 \alpha)_{\mathcal{EF}, \mathrm{d}} \Vert_{\tilde{\underline{C}}_{\rm in}}^2
		\nonumber \\
		&+\sum_{\vert \gamma \vert \leq N-1} \left[ 
		\Vert r^{-1} \widetilde{(r \nablaslash)}(\mathfrak{D}^{\gamma} r \alphabar)_{\mathcal{EF}, \mathrm{d}} \Vert_{\tilde{C}_{\rm out}}^2
	+\Vert 
		\widetilde{(r \nablaslash_4)}(\mathfrak{D}^{\gamma}  \check{r} \alphabar)_{\mathcal{EF}, \mathrm{d}} \Vert_{\tilde{C}_{\rm out}}^2 \right] \Bigg\},
	\end{align}
where the supremum is taken over all $\tilde{S}$ satisfying~\eqref{closetofoliationsphere},
and the $\tilde{C}_{\rm in}$ and $\tilde{C}_{\rm out}$ are the corresponding null cones defined above.

In the above, the operators  $\widetilde{r \slashed{\nabla}}$, $\widetilde{r \slashed{\nabla}_3}$, $\widetilde{r \slashed{\nabla}_4}$ are appropriate tangential operators to the cones $\tilde{C}_{\rm in}$, $\tilde{C}_{\rm out}$ (defined in analogy to
Section~\ref{tangentoperatorssec}) and the volume elements are appropriately defined (see
Remark~\ref{makeexplicitrem}).
\end{definition}

\begin{remark}
\label{makeexplicitrem}
Analogously with Remark~\ref{analogousprevremark}, 
it is only in Propositions~\ref{prop:dIdiff} and~\ref{thm:gidataestimates} where 
we shall apply the  estimate~\eqref{willstatelater2} for the quantity~\eqref{initdatanormEF}. 
We shall need specifically the control of the expression
 in brackets in~\eqref{initdatanormEF} 
 corresponding to a particular $\tilde{S}$ taken to be a sphere of our teleological $\mathcal{I}^+$ gauge anchored
 to the gauge~\eqref{herethenewone} by the conditions described in Section~\ref{anchoredsec}. In this case, we 
may represent explicitly the quantities appearing in~\eqref{initdatanormEF} using
 diffeomorphism functions relating our teleological $\mathcal{I}^+$ gauge to~\eqref{herethenewone}
and the pull back measures and 
``mixed tensors" defined in Section \ref{beyondSsection}. For Proposition~\ref{prop:dIdiff}, 
the estimate~\eqref{willstatelater2} will be used to control an initial energy controlling the above
diffeomorphism functions. For Proposition~\ref{thm:gidataestimates}, the important point will be that 
the norm~\eqref{initdatanormEF} directly controls through the Bianchi equations energy fluxes for the quantities of the almost gauge invariant hierarchy $\left(\alpha, \psi, P\right)$ and $\left(\underline{\alpha}, \check{\underline{\psi}}, \check{\underline{P}}\right)$.  Thus, the estimate~\eqref{willstatelater2} will appear in this work only through the
estimates~\eqref{theestimateheretoreferto} of Proposition~\ref{prop:dIdiff} and  
estimates~\eqref{thisiswheretheinitialfluxesappearone} and~\eqref{thisiswheretheinitialfluxesappeartwo}
of Proposition \ref{thm:gidataestimates}.
\end{remark}

\begin{remark}
\label{whenwedistinguishEFdata}
When we must distinguish between different double null gauges (cf.~the discussion in 
Section~\ref{changeisgood}), 
we will denote the coordinates on $\DREF(u_3)$
as $(u_{data},v_{data},\theta^{\mathcal{EF}}_{data})$ and refer to these coordinates
as the ``initial Eddington--Finkelstein system''.\index{initial Eddington--Finkelstein gauge!coordinates!$u_{data}$}\index{initial Eddington--Finkelstein gauge!coordinates!$v_{data}$}\index{initial Eddington--Finkelstein gauge!coordinates!$\theta^{\mathcal{EF}}_{data}$}   
\end{remark}

\begin{remark}
We remark that from the Raychaudhuri equation~\eqref{eq:Ray} and the first limiting statement
of~\eqref{somenormalisations}, it follows that $\tr \chi >0$ in the domain of~\eqref{linearisedasymptflat}.
\end{remark}

\begin{remark}
\label{notsmoothnonlineargauge}
We note that while the preliminary gauge $i^0_{\mathcal{EF}}$ is smooth, the 
renormalised gauge~\eqref{herethenewone}, due to its
normalisation on null infinity, has in general only the finite regularity given
from~\eqref{initdatanormEF}  which follows from the smallness of~\eqref{definitionofseennormnow} 
for $k=N$.  (See Remark~\ref{notsmoothlineartheory} for the analogous statement concerning 
linear theory.) For $i^0_{\mathcal{EF}}$ to be smooth, one must require the finiteness (though not smallness!)~of~\eqref{definitionofseennormnow}  for all $k$.
Because it will be convenient to appeal to smoothness in the closedness
argument, we shall circumvent this issue by 
appealing directly to the preliminary gauge $i^0_{\mathcal{EF}}$ in place
of~\eqref{initdatanormEF}. 
See already Theorems~\ref{thehigherordertheorem} and~\ref{thm:closed}.
(These are the only points in this work where the preliminary gauge gauge  $i^0_{\mathcal{EF}}$ will reappear.)
\end{remark}

{\begin{remark}
Conditions~\eqref{polesshouldbethesame} and~\eqref{meridianstoo} are necessary so as to anchor the poles and
standard meridian of the sphere of the two gauges. Without these conditions, relation~\eqref{inparticularherethemodes}
would only be true after a rotation of the vector ${\bf J}_{\rm seed}$.
\end{remark}

\begin{proof}
Compare with the corresponding linear statement, Proposition~\ref{linearisedprophere}.
In contrast to Theorem~\ref{thm:localKrus}, the proof of Theorem~\ref{thm:localEF}, already for (a),  
requires something slightly more quantitative than Cauchy stability,
as one must show that $r$-weighted estimates propagate.  These
estimates are now quite standard and have appeared often in the literature.
For instance, one can infer a proof of (a) directly from~\cite{thetaylorsthesis}.
For~(b), one must moreover show
 the possibility
of achieving~\eqref{somenormalisations} (cf.~the linearised version~\eqref{linearisedasymptflat}) 
and the improved $r$-decay
(see Remark~\ref{weakerrratesremark})  which this leads to, which is itself
incorporated
by the decay inherent in the $p$-notation in the energy of~\eqref{initdatanormEF} of Definition~\ref{defofEHnorms}.  
The estimates required for this  appear implicitly
in a more complicated form as part of the construction
of Chapter~\ref{teleoffingchapter}. We  encourage
the reader unfamiliar with how to accomplish these estimates to provide a detailed  proof after  reading 
Chapter~\ref{teleoffingchapter}.
\end{proof}

\section{The anchoring conditions}
\label{anchoredsec}

In this section we shall define conditions which
will ``anchor'' future normalised $\I$ and $\Hp$ gauges, as defined
in Sections~\ref{nullinfgaugesec} and~\ref{horizgaugesec}, to the maximal Cauchy
development $(\mathcal{M},g)$ with its two initial data gauges defined in Section~\ref{localexistencesection}.
This will uniquely fix the freedom in choosing these gauges and will determine the setup used
in the proof of the main  theorem.

We give the basic setup in {\bf Section~\ref{fourdoublenull}} of the
four local double null parametrisations. We shall then 
review some notational conventions for dealing with multiple double null parametrisations
in {\bf Section~\ref{section:gaugefunctions}}. Finally, we shall give the precise 
 anchoring conditions in {\bf Section~\ref{defofanchored}}.

\subsection{The basic setup: four local double null parametrisations in $(\mathcal{M},g)$}
\label{fourdoublenull}

In this section $(\mathcal{M},g)$ will denote the maximal Cauchy development of
data given by Theorem~\ref{maxCauchythe} applied to data
as in Proposition~\ref{globexistofdata} satisfying the global smallness assumption~\eqref{smallnessofdata}.

In particular, we have
 the two double null parametrisations
\begin{align}
\label{initialKgaugehere}
&i_{\mathcal{K}}:\mathcal{W_{\mathcal{K}}}(V_3) \times\mathbb S^2 \to\DRK(V_3)\subset \mathcal{M}\\
\label{initialEFgaugehere}
&i_{\mathcal{EF}}:\mathcal{W}_{\mathcal{EF}}(u_3)\times \mathbb S^2\to\DREF(u_3)\subset \mathcal{M},
\end{align}
from 
 Theorems~\ref{thm:localKrus} and~\ref{thm:localEF}, respectively.
(Recall that the parameters involved in the definition
of  the above sets, including $u_3$ and $V_3$, are those determined in Section~\ref{compediumparameterssec}.)
 
We now also assume that we are given   additional parameters 
$u_f\ge u_{-1}$ and  $M_f$  satisfying~\eqref{alwaysassumethis}. 
We define\index{teleological $\I$ gauge!parameters!$v_{\infty}$, final advanced time parameter}
\begin{equation} \label{eq:vinfty}
	v_{\infty}(\varepsilon_0,u_f)
	:=
	\varepsilon_0^{-2} (u_f)^{\frac{1}{\delta}},
\end{equation}
where $\delta>0$ is as fixed in Section~\ref{compediumparameterssec}.\index{teleological $\I$ gauge!parameters!$\delta$}\index{teleological $\I$ gauge!parameters!$\epsilon$}
The role of the parameter~\eqref{eq:vinfty} will become clear in Section~\ref{sec:vinfexplain}.
For convenience, we shall require that $\hat\varepsilon_0$ is sufficiently small so that
for any  $M=M_f$ satisfying~\eqref{alwaysassumethis} and $0<\varepsilon_0\le \hat\varepsilon_0$, we have
\begin{equation}
\label{hereanotherrestriction}
v(R_4,u)+M_{\rm init}\le \varepsilon_0^{-2}u^{1/\delta}
\end{equation}
for all $u\ge u_{-1}$, where $v(R_4,u)$ is defined with respect to $M_f$.
Note that by~\eqref{hereanotherrestriction} and the fact that $R_{-2}<R_4$, it follows that
the $v$-range in~\eqref{WI+} is non-empty for all $u_{-1}\le u\le u_f$.

Associated to the above parameters, we assume that we have
two additional parametrisations:\index{teleological $\I$ gauge!double null parametrisation!$i_{\I}$}\index{teleological $\Hp$ gauge!double null parametrisation!$i_{\Hp}$}\index{teleological $\I$ gauge!sets!$\DRI_{u_f}$, spacetime domain}\index{teleological $\Hp$ gauge!sets!$\DRH_{u_f}$, spacetime domain}
\begin{align}
&i_{\I}:\mathcal{W}_{\I}(u_f, M_f, v_\infty)\times \mathbb S^2 \to \DRI_{u_f}\subset \mathcal{M}
\label{fortheIplusgauge}
\\
&i_{\Hp}: \mathcal{W}_{\Hp}(u_f, M_f)
\times\mathbb S^2 \to \DRH_{u_f}\subset \mathcal{M},
\label{fortheHplusgauge}
\end{align}
such that $i_{\I}^*g$ is $\I$ normalised and $i_{\Hp}^*g$ is $\Hp$ normalised, both with respect
to  $u_f$ and $M_f$.

Finally, we note that we may consider a fifth local parametrisation, directly associated to~\eqref{fortheHplusgauge},
namely\index{teleological $\Hp$ gauge!double null parametrisation!$i_{\Hp,\mathcal{K}}$, Kruskalised double null
parametrisation} 
\begin{align}
&i_{\Hp,\mathcal{K}}:
\mathcal{W}_{\Hp}(U_f, M_f)\times \mathbb S^2
  \to \DRH_{u_f}\subset \mathcal{M},
\label{fortheHplusgaugekruskalised}
\end{align}
where $i_{\Hp,\mathcal{K}} = (\iota_{M_f}^{-1} \times {\rm id}) \circ i_{\Hp}$,  and 
thus $i_{\Hp,\mathcal{K}}^*g$ is the metric of the associated 
Kruskalised $\mathcal{H}^+$ gauge to $i_{\Hp}^*g$.

\subsection{Notation for coordinates and diffeomorphism functions} \label{section:gaugefunctions}

As discussed in Section~\ref{newdiffysec}, we may interpret the coordinates  of
the domains of the above parametrisations $(\ref{initialKgaugehere})$--$(\ref{fortheHplusgauge})$
as functions on the images of     $(\ref{initialKgaugehere})$--$(\ref{fortheHplusgauge})$, i.e.~as
functions on
subsets of $\mathcal{M}$. When doing so, however, we shall give the coordinates distinctive
labels. 

\subsubsection{Coordinate labels}
As we have already remarked above (see Remarks~\ref{whenwedistinguishkruskaldata} 
and~\ref{whenwedistinguishEFdata}) we shall refer to the coordinates
of~\eqref{initialKgaugehere} and~\eqref{initialEFgaugehere} as\index{initial Eddington--Finkelstein gauge!coordinates!$u_{data}$}\index{initial Eddington--Finkelstein gauge!coordinates!$v_{data}$}\index{initial Eddington--Finkelstein gauge!coordinates!$\theta^{\mathcal{EF}}_{data}$}\index{initial Kruskal gauge!coordinates!$U_{data}$}\index{initial Kruskal gauge!coordinates!$V_{data}$}\index{initial Kruskal gauge!coordinates!$\theta^{\mathcal{K}}_{data}$}
\[
(U_{data},V_{data},\theta^{\mathcal{K}}_{data}),\qquad (u_{data},v_{data},\theta^{\mathcal{EF}}_{data}),
\]
 respectively.
We shall refer to the coordinates of~\eqref{fortheIplusgauge} and~\eqref{fortheHplusgauge}
as\index{teleological $\I$ gauge!coordinates!$u_{\I}$, retarded null coordinate of the $\I$ gauge}\index{teleological $\I$ gauge!coordinates!$v_{\I}$, advanced null coordinate of the $\I$ gauge}\index{teleological $\I$ gauge!coordinates!$\theta_{\I}$}\index{teleological $\Hp$ gauge!coordinates!$u_{\Hp}$, retarded null coordinate of the $\Hp$ gauge}\index{teleological $\Hp$ gauge!coordinates!$v_{\Hp}$, advanced null coordinate of the $\Hp$ gauge}\index{teleological $\Hp$ gauge!coordinates!$\theta_{\Hp}$}  
\[
(u_{\I},v_{\I}, \theta_{\I}), \qquad (u_{\Hp},v_{\Hp},\theta_{\Hp}),
\]
respectively.
Finally, we shall refer to the coordinates of~\eqref{fortheHplusgaugekruskalised}
as\index{teleological $\Hp$ gauge!coordinates!$U_{\Hp}$, Kruskalised retarded null coordinate of the $\Hp$ gauge}\index{teleological $\Hp$ gauge!coordinates!$V_{\Hp}$, Kruskalised advanced null coordinate of the $\Hp$ gauge} 
\[
(U_{\Hp}, V_{\Hp}, \theta_{\Hp}).
\]
The use of the same symbol $\theta_{\Hp}$ in the $\mathcal{H}^+$ gauge
and its associated Kruskalised gauge will not be confusing as these indeed coincide.

\subsubsection{Labels on cones and spheres}
In order to distinguish the cones and spheres of intersection in both gauges,
we shall typically put the $\Hp$ and $\mathcal{I}^+$ labels on the $C$ and $S$ themselves, i.e.~we denote\index{teleological $\Hp$ gauge!sets!$C_u^{\Hp}$}\index{teleological $\Hp$ gauge!sets!$\Cbar_v^{\Hp}$}
\[
	C_u^{\Hp}
	:=
	\{u_{\Hp} = u \} \cap \DRH,
	\qquad
	\Cbar_v^{\Hp}
	:=
	\{v_{\Hp} = v \} \cap \DRH,
\]
\[
	C_u^{\I}
	:=
	\{u_{\I} = u \} \cap \DRI,
	\qquad
	\Cbar_v^{\I}
	:=
	\{v_{\I} = v \} \cap \DRI
\]
and\index{teleological $\I$ gauge!sets!$C_u^{\I}$}\index{teleological $\I$ gauge!sets!$\Cbar_v^{\I}$}
\index{teleological $\Hp$ gauge!sets!$S^{\Hp}_{u,v}$}\index{teleological $\I$ gauge!sets!$S^{\I}_{u,v}$}
\[
	S^{\Hp}_{u,v}
	:=
	C_u^{\Hp}
	\cap
	\Cbar_v^{\Hp},
	\qquad
	S^{\I}_{u,v}
	:=
	C_u^{\I}
	\cap
	\Cbar_v^{\I}.
\]

\subsubsection{Diffeomorphism functions}
\label{specificdiffeos}

When considering transition diffeomorphisms between our patches we shall follow
the $f$ notation of Section~\ref{changeisgood}.
We summarise this here explicitly in the context of the specific diffeomorphism functions
we shall typically be interested in.

The first pair of such coordinate systems we shall consider is
 the initial data Kruskal gauge $(\ref{initialKgaugehere})$ and a Kruskalised $\mathcal{H}^+$ gauge
 $(\ref{fortheHplusgaugekruskalised})$.
For this, we shall use the notation:\index{double null gauge!change of gauge!$f_{d,\Hp}$}\index{initial Kruskal gauge!change of gauge!$f_{d,\Hp}$, change from teleological $\Hp$ gauge}
\[
	U_{data} = U_{\Hp} + f^3_{d,\Hp}(U_{\Hp}, V_{\Hp}, \theta^1_{\Hp}, \theta^2_{\Hp}),
	\quad
	V_{data} = V_{\Hp} + f^4_{d,\Hp}(U_{\Hp}, V_{\Hp}, \theta^1_{\Hp}, \theta^2_{\Hp}),
\]
\[
	\theta^1_{data} = \theta^1_{\Hp} + f^1_{d,\Hp}(U_{\Hp}, V_{\Hp}, \theta^1_{\Hp}, \theta^2_{\Hp}),
	\quad
	\theta^2_{data} = \theta^2_{\Hp} + f^2_{d,\Hp}(U_{\Hp}, V_{\Hp}, \theta^1_{\Hp}, \theta^2_{\Hp}).
\]

The second pair we shall consider will be the initial Eddington--Finkelstein gauge $(\ref{initialKgaugehere})$
and an $\mathcal{I}^+$ gauge $(\ref{fortheIplusgauge})$.
For this, we shall use the notation:\index{double null gauge!change of gauge!$f_{d,\I}$}\index{initial Eddington--Finkelstein gauge!change of gauge!$f_{d,\I}$, change from teleological $\I$ gauge}
\[
	u_{data} = u_{\I} + f^3_{d,\I}(u_{\I}, v_{\I}, \theta^1_{\I}, \theta^2_{\I}),
	\quad
	v_{data} = v_{\I} + f^4_{d,\I}(u_{\I}, v_{\I}, \theta^1_{\I}, \theta^2_{\I}),
\]
\[
	\theta^1_{data} = \theta^1_{\I} + f^1_{d,\I}(u_{\I}, v_{\I}, \theta^1_{\I}, \theta^2_{\I}),
	\quad
	\theta^2_{data} = \theta^2_{\I} + f^2_{d,\I}(u_{\I}, v_{\I}, \theta^1_{\I}, \theta^2_{\I}).
\]

Finally, the third pair we shall consider will be  an $\mathcal{H}^+$ gauge $(\ref{fortheHplusgauge})$ and an $\mathcal{I}^+$ gauge $(\ref{fortheIplusgauge})$.
For this, we shall use the notation\index{double null gauge!change of gauge!$f_{\Hp,\I}$}\index{teleological $\I$ gauge!change of gauge!$f_{\Hp,\I}$, change to teleological $\Hp$ gauge}\index{teleological $\Hp$ gauge!change of gauge!$f_{\Hp,\I}$, change from teleological $\I$ gauge}
\[
	u_{\Hp} = u_{\I} + f^3_{\Hp,\I}(u_{\I}, v_{\I}, \theta^1_{\I}, \theta^2_{\I}),
	\quad
	v_{\Hp} = v_{\I} + f^4_{\Hp,\I}(u_{\I}, v_{\I}, \theta^1_{\I}, \theta^2_{\I}),
\]
\[
	\theta^1_{\Hp} = \theta^1_{\I} + f^1_{\Hp,\I}(u_{\I}, v_{\I}, \theta^1_{\I}, \theta^2_{\I}),
	\quad
	\theta^2_{\Hp} = \theta^2_{\I} + f^2_{\Hp,\I}(u_{\I}, v_{\I}, \theta^1_{\I}, \theta^2_{\I}).
\]

 In particular, these 
 diffeomorphism functions will already appear in Section~\ref{defofanchored} below,
 where we discuss the anchoring conditions between our gauges.

\subsection{Definition of anchored gauges}
\label{defofanchored}

We may now give a definition of what it means for gauges~\eqref{fortheIplusgauge} and~\eqref{fortheHplusgauge} 
to be anchored in the maximal development
$(\mathcal{M},g)$ of initial data considered in Theorem~\ref{maxCauchythe}.

Let us first introduce some additional notation that will be useful.
Recalling the function $r: =r_{M_f}$ on $\mathcal{W}_{\mathcal{EF}}$ defined by~\eqref{EFrdef}, we may
consider this as a function on $\DRI$ and $\DRH$. We shall denote the resulting functions
$r_{\I}$, $r_{\Hp}$ respectively\index{teleological $\I$ gauge!functions!$r_{\I}$}\index{teleological $\Hp$ gauge!functions!$r_{\Hp}$}, i.e.
\[
r_{\I}=r_{M_f} \circ i^{-1}_{\I}, \qquad r_{\Hp}=r_{M_f} \circ i^{-1}_{\Hp}.
\]
We also recall the parameter $R$ defined in Section~\ref{compediumparameterssec}.

In addition to dropping the explicit dependence on $M_f$, we will often drop the
$u_f$ from the notation, i.e. we shall write $\DRI$ instead of $\DRI_{u_f}$, $\DRH$ instead of $\DRH_{u_f}$, etc.
This will allow us to insert additional labels without the proliferation of symbols.
With this understanding, given $\tilde{s}>s>0$, we may now define the sets\index{teleological $\Hp$ gauge!sets!$\DRHs$}\index{teleological $\I$ gauge!sets!$\DRIs$}\index{teleological $\I$ gauge!sets!$\DRIss$}
\[
	\DRHs: = \DRH \cap \{r_{\Hp}\le s\} ,  
\]
\[
	\DRIs: = \DRI \cap \{r_{\I}\ge s\}, \qquad 
	\DRIss :=  \DRI\cap  \{ s \leq r_{\I} \leq \tilde{s} \}.
\]

Refer to Figure~\ref{defoftheoverlappingds}.
\begin{figure}
\centering{
\def\svgwidth{20pc}
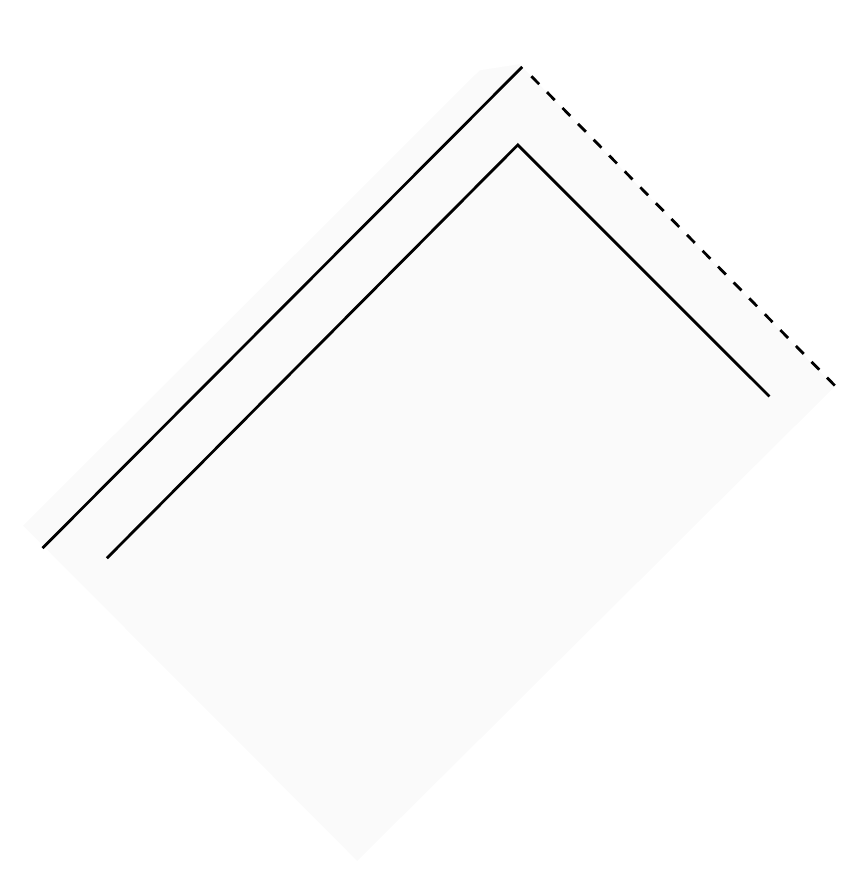}
\caption{The region $\mathcal{D}$ partitioned into the overlapping regions $\mathcal{D}^{\mathcal{H}^+}$ and $\mathcal{D}^{\mathcal{I}^+}$: Beware of reading off incidence relations!}\label{defoftheoverlappingds}
\end{figure}

\begin{definition}
\label{anchoringdef}
Let $(\mathcal{M},g)$ be the maximal development of initial data given by
Theorem~\ref{maxCauchythe} applied to data
as in Proposition~\ref{globexistofdata},
satisfying the global smallness assumption~\eqref{smallnessofdata} with $0<\varepsilon_0\le\hat\varepsilon_0$
for sufficiently small $\hat\varepsilon_0$ 
and let~\eqref{initialKgaugehere} and~\eqref{initialEFgaugehere}  be the initial data gauges
of Theorems~\ref{thm:localKrus} and~\ref{thm:localEF}.

Let us be given parameters $u_f$ and $M_f$, and smooth parametrisations 
\eqref{fortheIplusgauge} and \eqref{fortheHplusgauge} as 
in Section~\ref{fourdoublenull}, i.e.~such that $i_{\I}^* g$ and
$i_{\Hp}^*g$ are expressed in $\mathcal{I}^+$ and $\Hp$ gauge, respectively,
with respect to the parameters $u_f$ and $M_f$.
We say that the gauges are {\bf anchored} in $(\mathcal{M},g)$ with respect to the
parameters $u_f$, $v_\infty=v_\infty(u_f,\varepsilon_0)$ and $M_f$ if the following hold:
\begin{itemize}
\item{\bf Overlap of $\DRH$ and $\DRI$.}
We have that the following inclusions hold
\begin{equation}
{\mathcal{D}}^{\Hp}_{r\ge   R_{-1}}
	\subset
\mathcal{D}^{\I}_{R_{-2}\le r\le R_{2}},
	\label{eq:overlap1}
	\end{equation}
	\begin{equation}
	\mathcal{D}^{\I}_{R_{-2}\le r \le R_{1}} \cap J^+(S^{\mathcal{H}^+}_{u_0,v_{-1}})
	\subset
	\DRH_{r\le R_{2}} ;
	\label{eq:overlap2}
\end{equation}

\item{\bf Common future null cone.}
The final outgoing cones coincide in their common domain,
i.e.~we assume moreover that 
\begin{equation} \label{eq:anchoringdefcommoncone}
C^{\I}_{u_f}\cap \mathcal{D}^{\Hp} = C^{\Hp}_{u_f}\cap \mathcal{D}^{\I};
\end{equation}
note that this implies the statement
\[
f^3_{\Hp,\I}(u_f,v(R,u_f,\theta),\theta) = 0
\]
for all $\theta \in \mathbb S^2$,
where $f^{\mu}_{\Hp,\I}$ is as defined in Section \ref{section:gaugefunctions}.

\item{\bf Overlap with initial data gauges.}
The ``initial hypersurfaces'' defined by the $u_f$-normalised $\I$ gauge and $u_f$-normalised $\Hp$ gauge are contained in 
appropriate regions covered by the initial data gauge, i.e.~they satisfy the inclusions
		\begin{align}
			\bigcup_{v_{-1} \leq v \leq v_2} \Cbar_{v}^{\Hp}
			&
			\subset
			\mathcal{D}^{\mathcal{K}}(V_3),
			\label{eq:overlap3}
			\\
			\bigcup_{u_{-1}\leq u \leq u_2}
			C_{u}^{\I}
			&
			\subset
			\mathcal{D}^{\mathcal{EF}}(u_3);
			\label{eq:overlap5}
		\end{align}
let us assume here  in addition that the  restriction map
\begin{equation}
\label{makethisadiffy}
\pi^{\mathcal{EF}}_{\mathbb{S}^2}|_{S^{\I}_{u_{-1},v_\infty}} S^{\I}_{u_{-1},v_\infty} \to \mathbb S^2
\end{equation}
is a $C^1$-diffeomorphism, where $\pi^{\mathcal{EF}}_{\mathbb{S}^2}:\mathcal{D}^{\mathcal{EF}}\to \mathbb S^2$
denotes $\pi_{\mathbb S^2} \circ (i^{\mathcal{EF}})^{-1}$;
		
\item{\bf Relating $u_f$ with distance from initial data.}
The following diffeomorphism component vanishes:
		\begin{equation} \label{eq:f3f4Iba}
			f^3_{d,\I}(u_{-1}, v_{\infty},\cdot)_{\ell=0}
			=
			0;
		\end{equation}
\item{\bf  Affixing the sphere diffeomorphism of the $\Hp$ gauge to that of the $\I$ gauge.} 
We have the equality of maps
 \begin{equation} \label{eq:anchoringdefaffixingsphere}
\pi_{\mathbb S^2} \circ  i_{\I}^{-1}  |_{S^{\Hp}_{u,v(R,u)}} = \pi_{\mathbb S^2}\circ i_{\Hp}^{-1}|_{S^{\Hp}_{u,v(R,u)}}
 \end{equation}
 as maps $S^{\Hp}_{u,v(R,u)}\to \mathbb S^2$,
  where $\pi_{\mathbb S^2}$ denotes the natural projection to $\mathbb S^2$;
note that in the notation of Section~\ref{specificdiffeos}, this can be written as the statement that 
the following diffeomorphism component vanishes:
\[
f^i_{\mathcal{H}^+,\mathcal{I}^+}(u_f,
v_{\I}(u_f,v_{\Hp}(R,u_{\Hp}), \theta_{\Hp}), \theta_{\I}(u_f,v_{\Hp}(R,u_{\Hp}), \theta_{\Hp})) =0,
\]
for $i=1, 2$, and
 all $\theta_{\mathcal{H}^+}\in \mathbb S^2$;
\item{\bf  Determining the sphere diffeomorphism of the $\I$ gauge and affixing it to initial data.} 
Let us note first that we can distinguish a point $p$ on $S^{\I}_{u_{f},v_\infty}$
and a vector $v\in T_pS^{\I}_{u_{f},v_\infty}$ as follows:

Noting the natural null flow diffeomorphism $j:S^{\I}_{u_{-1},v_\infty}\to S^{\I}_{u_{f},v_\infty}$
associated to the double null gauge, then in view of~\eqref{makethisadiffy} we may 
distinguish 
\begin{equation}
\label{choiceofpandv}
p= j\circ (\pi^{\mathcal{EF}}_{\mathbb S^2}|_{S^{\I}_{u_{-1},v_\infty}})^{-1} (1,0,0),
\qquad
w=  d\left(j\circ  (\pi^{\mathcal{EF}}_{\mathbb S^2}|_{S^{\I}_{u_{-1},v_\infty}})^{-1}\right)|_{(1,0,0)} (0,1,0),
\end{equation}
where we consider here $(1,0,0)\in \mathbb S^2$ and $(0,1,0)\in T_{(1,0,0)}\mathbb S^2$.

Our final condition is thus that the map:
\[
i_{\I}|_{(u_f,v_\infty)\times \mathbb S^2}: (u_f,v_\infty) \times \mathbb S^2\to S^{\I}_{u_{f},v_\infty}\subset \mathcal{M},
\]
thought of as a map $\mathbb S^2\to  S^{\I}_{u_{f},v_\infty}$,
be the unique map  determined by Proposition~\ref{determiningthesphere}, 
with $h=r^{-2} \slashed{g}$, and with 
 $p\in S^{\I}_{u_f,v_\infty}$, $v:=r^{2}g(w,w)^{-1/2}w\in T_p S^{\I}_{u_f,v_\infty}$ given  by~\eqref{choiceofpandv}.

We note finally that~\eqref{choiceofpandv} can be expressed as the following statement.
If $(\theta^1,\theta^2)$ denote local coordinates on $\mathbb S^2$ such that $(\theta^1_0,\theta^2_0)$
corresponds to the point $(1,0,0)$ and the vector $\partial_{\theta^1}(\theta^1_0,\theta^2_0)$ corresponds
to the vector  $(0,1,0)$,
then in the notation of the diffeomorphism functions of Section~\ref{specificdiffeos}, we have
\begin{equation}
\label{makesurewritten}
f^{i}_{d,\mathcal{I}^+} (u_{-1},v_\infty, \theta^1_0, \theta^2_0) = 0 = \frac{\partial f^i_{d,\I}}{\partial \theta^1}(u_{-1},v_\infty, \theta^1_0, \theta^2_0)   ,\qquad i=1,2.
\end{equation}
\end{itemize}
\end{definition}

\begin{remark}
As we shall see later, the above anchoring conditions make the choice of teleological gauges rigid.
Let us note that since~\eqref{fortheIplusgauge} and~\eqref{fortheHplusgauge} will be smooth, but
the parameterisation~\eqref{initialEFgaugehere} is not in general,  the diffeomorphism~\eqref{makethisadiffy} is in general
only of finite regularity.  (In this context, we also note that, since the $\ell=0$ mode is constant on spheres, 
the mild condition~\eqref{eq:f3f4Iba} does 
not impose the  finite regularity
of~\eqref{initialEFgaugehere} onto~\eqref{fortheIplusgauge}.)
\end{remark}

We define\index{spacetime subsets!$\mathcal{D}$}\index{spacetime subsets!$C_{u_f}$}
\[
\mathcal{D}:=\DRH \cup \DRI\subset\mathcal{M},
\]
\[
C_{u_f} = C^{\I}_{u_f} \cup C^{\Hp}_{u_f}.
\]

We note finally the following easy proposition:
\begin{proposition}
\label{lotsofhype}
Consider anchored gauges as in Definition~\ref{anchoringdef}.
Then the set
\[
\mathcal{D}\cup \mathcal{D}^\mathcal{K}\cup\mathcal{D}^\mathcal{EF}
\]
is globally hyperbolic and admits 
$ i_{\mathcal{K}}(\underline{C}_{\rm in}^{\mathcal{K}})\cup
i_{\mathcal{EF}}(C_{\rm out}^{\mathcal{EF}})$
as a (bifurcate null) past Cauchy surface.
\end{proposition}

\section{Existence of anchored teleological gauges at time $u_f^0$}
\label{existenceofanchored}

Let us already state the following existence theorem.\index{bootstrap!parameters!$u^0_f$, ``initial'' final retarded time associated to bootstrap}

\begin{theorem}
\label{existenceofanchoredgaugethe}
Let $(\mathcal{M},g)$ be the maximal development of initial data given by
Theorem~\ref{maxCauchythe} applied to data
as in Proposition~\ref{globexistofdata},
satisfying the global smallness assumption~\eqref{smallnessofdata} with $0<\varepsilon_0\le \hat\varepsilon_0$
for sufficiently small $\hat\varepsilon_0$ 
and let~\eqref{initialKgaugehere} and~\eqref{initialEFgaugehere}  be the initial data gauges
of Theorems~\ref{thm:localKrus} and~\ref{thm:localEF}.

Then, for $u_f^0$ defined in Section~\ref{compediumparameterssec},\index{teleological $\I$ gauge!parameters!$u_f^0$}\index{teleological $\Hp$ gauge!parameters!$u_f^0$} it follows that there exists
an $M_f^0$\index{teleological $\I$ gauge!parameters!$M_f^0$}\index{teleological $\Hp$ gauge!parameters!$M_f^0$} 
satisfying
\begin{equation}
\label{nonumber?}
|M_f^0-M_{\rm init}|\lesssim \varepsilon_0
\end{equation}
and anchored gauges~\eqref{fortheIplusgauge} and~\eqref{fortheHplusgauge}
with respect to parameters $u_f^0$ and $M_f^0$,
in the sense of Definition~\ref{anchoringdef}.

We have moreover the inclusions
\begin{equation}
\label{expinclu}
\mathcal{D}^{\mathcal{H}^+}_{u_f^0} \subset 
\mathcal{D}^{\mathcal{K}}(V_3),\qquad
\mathcal{D}^{\mathcal{I}^+}_{u_f^0}
\subset \mathcal{D}^{\mathcal{EF}}(u_3)
\end{equation}
and the associated Kerr angular momentum parameters given by Definitions~\ref{assocKerparIplus} 
and~\ref{assocKerpardata}
satisfy 
\begin{equation}
\label{closenessofKerrparams}
|J^m_{\rm seed} -J^m_{\mathcal{I}^+}|\lesssim \varepsilon_0^2.
\end{equation}

Moreover, 
the almost gauge invariant quantities of Section~\ref{AGIQdef}, the difference 
quantities of Section~\ref{schemfordifssec},
and the
diffeomorphisms of Section~\ref{section:gaugefunctions} connecting the gauges are all controlled in certain energies
(in particular~\eqref{eq:bamain} and~\eqref{eq:badiffeo} hold for $u_f:=u_f^0$ and with $\lesssim\varepsilon_0^2$
replacing $\le\varepsilon^2$).

Finally, with the understanding that $\varepsilon_0= \sqrt{\varepsilon_0^2}$, 
we may replace $\varepsilon_0^2$ on the right hand side of~\eqref{nonumber?}, \eqref{closenessofKerrparams}
and all the estimates for the energies claimed above by the quantity 
\begin{equation}
\label{replacewiththisquantity}
\mathcal{E}^N_{0}[\Phi_{\mathcal{K},d}]+ \mathbb{E}^N_{0}[\Phi_{\mathcal{EF},d}].
\end{equation}
\end{theorem}

\begin{remark}
The linear version of the existence part of Theorem~\ref{existenceofanchoredgaugethe} is
already given by Proposition~\ref{proplinIpgauge} and~\ref{proplinHpgauge}.
The linear analogue of the  boundedness statement of Theorem~\ref{existenceofanchoredgaugethe} is the statement that the pure gauge solution
$\mathscr{G}$
and linearised Kerr solutions $\mathscr{K}$ are unique and themselves bounded with respect
to the solution $\mathscr{S}$ expressed in the ``initial data normalisation''.
See again~\cite{holzstabofschw}. 
\end{remark}

\begin{proof}
This proof of the existence of the anchored gauges
can be thought of as an easier version of Theorem~\ref{thm:newgauge}, proven
later in this work (see Chapter~\ref{teleoffingchapter}) as part of the continuity argument of the bootstrap. 
See Remark~\ref{rmk:uf0Igaugeexistence0} for precise instructions on how to distill this.
(Briefly, one first constructs the $\mathcal{I}^+$ normalised gauge (cf.~Section~\ref{offing})
 and then the $\mathcal{H}^+$ normalised gauge (cf.~Section~\ref{offing2}).
In view of the expected inclusions~\eqref{expinclu} we can apply our Cauchy stability
statements from Theorems~\ref{thm:localKrus} and~\ref{thm:localEF}.
The construction proceeds by similar iterations to Sections~\ref{offing} and~\ref{offing2},
using the fact that by
Proposition~\ref{proplinIpgauge},
the linearised version of the gauge conditions 
can be readily satisfied.)
The estimate~\eqref{closenessofKerrparams} follows from~\eqref{inparticularherethemodes} 
and the fact that the linearised change of gauge does not alter the quantities~\eqref{dontchangethem}.
Note that to obtain~\eqref{closenessofKerrparams} 
we are using here also the anchoring condition~\eqref{choiceofpandv} determining the sphere diffeomorphism.
(The fact that we may replace $\varepsilon_0$ with~\eqref{replacewiththisquantity} is clear,
because it is from control of the norm~\eqref{replacewiththisquantity} that one constructs the gauges,
and $\varepsilon_0^2$ only entered in the first place
through its appearance in equations~\eqref{willstatelater} and~\eqref{willstatelater2} of
 Theorems~\ref{thm:localKrus} and~\ref{thm:localEF}.)
\end{proof}

\begin{remark}
Implicit in our proof of the above Theorem is a rigidity statement giving
the uniqueness of the above gauges (see again Remark~\ref{rmk:uf0Igaugeexistence0}). 
Let us note in particular that this implies that
were we to consider initial data given by any of the examples~\ref{firstitemhere}--\ref{thirditemhere}
of Remark~\ref{longertrivialremark}, then  $M_f=M, M, \tilde{M}$, respectively,
and our two teleological gauges would give  the standard
Eddington--Finkelstein representation~\eqref{SchwmetricEF} of the Schwarzschild metric
with mass $M_f$, restricted to~\eqref{thedomainincords} or~\eqref{actualHpdomain}
(cf.~Remarks~\ref{SchwarzschildisIptoo} and~\ref{SchwarzschildisHptoo}).
In particular these gauges can be defined
for all $u_f\ge u_f^0$, and the gauge corresponding to $u_f'$ with $u^0_f<u_f'<u_f$ is simply
the restriction of the gauge corresponding to  $u_f$.  (This latter property will  not 
be true for general initial data!) 
In the case of example~\ref{fourthitemhere} of the same remark, it is already not easy to obtain an explicit form of this
gauge, nor is it clear that the gauge can be defined for all sufficiently large $u_f$. 
It will turn out that this latter example of initial data is 
excluded from corresponding to the final $(\mathcal{M},g)$ 
of our main theorem by the codimensionality assumption.
\end{remark}

\section{The $3$-parameter families of initial data $\mathcal{L}_{{\mathcal{S}}_0}$ and the structure 
of the moduli space $\mathfrak{M}$}
\label{threeparamsection}

Given $0<\varepsilon_0\le \hat\varepsilon_0$, we shall denote by
$\mathfrak{M}(M_{\rm init},\varepsilon_0)$\index{initial data!moduli space!$\mathfrak{M}=\mathfrak{M}(M_{\rm init},\varepsilon_0)$, moduli space of data} the set
of all characteristic initial data
as in Proposition~\ref{globexistofdata}  which moreover satisfy
the smallness assumption $(\ref{smallnessofdata})$.
We can think of  $\mathfrak{M}(M_{\rm init},\varepsilon_0)$ as the \emph{moduli space} of characteristic initial data
$\varepsilon_0$-close to the Schwarzschild metric of mass of mass $M_{\rm init}$. When there is
no danger of confusion, we will sometimes simply write $\mathfrak{M}:=\mathfrak{M}(M_{\rm init},\varepsilon_0)$.
Since characteristic initial data are determined by seed data,
we will often refer to an element of $\mathfrak{M}(M_{\rm init},\varepsilon_0)$
by its seed initial data $\mathcal{S}=\mathcal{S}^{\mathcal{K}}$ representing
$(\ref{seeddataforref})$, i.e.~we may write $\mathcal{S}\in \mathfrak{M}(M_{\rm init},\varepsilon_0)$.

Given thus $\mathcal{S}\in \mathfrak{M}(M_{\rm init},\varepsilon_0)$, recall that we denote by
$\mathcal{S}^{\mathcal{EF}}$ the associated realisation of the seed data on $C_{\rm out}$, and
with $\mathcal{EF}$ subscripts the quantities associated to this realisation.
Recall from Section~\ref{Kerpardatasec} 
that the $\ell=1$ spherical harmonics 
$Y^{\ell = 1}_{-1}$, $Y^{\ell = 1}_{0}$, $Y^{\ell = 1}_{1}$ and
the corresponding projection to the $\ell=1$ space
can be 
defined on the sphere $S_{u_{-2},v_{-2}}$ with the help of $g_{\circ, M_{\rm init}}(u_{-2},v_{-2})$
according to Section~\ref{basisforprojspace}.
Using this, we may identify a subset
\[
\mathfrak{M}_0(M_{\rm init}, \varepsilon_0)\subset \mathfrak{M}(M_{\rm init},\varepsilon_0)
\]
consisting of all
seed data\index{initial data!moduli space!$\mathfrak{M}_0(M_{\rm init}, \varepsilon_0)$, set of initial data with $J^m_{\rm seed}=0$} of all seed data
${\mathcal{S}}^{\mathcal{K}}\in \mathfrak{M}(M_{\rm init},\varepsilon_0)$ such that
\[
(\curlslash\, \eta_{\mathcal{EF}})_{\ell =1} =0=(\curlslash\, \eta_{\mathcal{K}})_{\ell=1}
\]
on the sphere $S_{u_{-2},v_{-2}}$.
We shall in general denote seed data contained in $\mathfrak{M}_0$ with 
a $0$ subscript, i.e.~as $\mathcal{S}_0={\mathcal{S}}^{\mathcal{K}}_0$.
In view of Definition~\ref{assocKerpardata} and Remark~\ref{remarkaboutinitialseed}, 
we note that such data satisfy $|J^m_{\rm seed}|\lesssim \varepsilon^2_0$.

Given such seed data $\mathcal{S}_0\in \mathfrak{M}_0(M_{\rm init}, \hat\varepsilon_0)$, 
and real parameters $\lambda_{-1},\lambda_0,\lambda_1$,\index{initial data!parameters!$\lambda_{m}$, $m=-1,0,1$, modulation parameters}
 then, setting\index{angular momentum!$\lambda=(\lambda_{-1},\lambda_0,\lambda_1)$, modulation parameters related to angular momentum of data} 
 \[
 \lambda := (\lambda_{-1},\lambda_0,\lambda_1)\in \mathbb R^3,
 \]
 let us denote by  ${\mathcal{S}}_0(\lambda)$ the seed data
 $(\ref{seeddataforref})$ arising from
 \begin{equation}
 \label{thearisen}
 \eta_{\mathcal{EF}}(\lambda):= \eta_{\mathcal{EF}}
	+ \frac{3}{r^2(u_{-2},v_{-2})} \left(
	\lambda_{-1} {}^* \nablaslash Y^{\ell = 1}_{-1}
	+
	\lambda_0 {}^* \nablaslash Y^{\ell = 1}_{0}
	+
	\lambda_1 {}^* \nablaslash Y^{\ell = 1}_{1} \right),
 \end{equation}
 in place of the $\eta_{\mathcal{EF}}$ of $\mathcal{S}_0$,
 and all other seed data quantities~\eqref{seeddataforref} the same as $\mathcal{S}_0$.
We note here that by Remark~\ref{remarkaboutinitialseed},
\[
|\lambda_m -   J^m_{\rm seed} [\mathcal{S}_0(\lambda)] |\lesssim \varepsilon_0^2 \, .
\]

Given seed data $\mathcal{S}_0\in \mathfrak{M}_0(M_{\rm init}, \varepsilon_0)$, let us denote 
by\index{initial data!moduli space!$\mathcal{L}^{\varepsilon_0}_{\mathcal{S}_0}$, $3$-parameter family of initial data indexed by a reference data set $\mathcal{S}_0$}\index{initial data!seed data!$\mathcal{S}_0(\lambda)$, initial data derived from $\mathcal{S}_0$ associated with parameter $\lambda$} 
$\mathcal{L}^{\varepsilon_0}_{\mathcal{S}_0}$ the $3$-parameter family of initial data
\begin{equation}
\label{familyfirst}
\mathcal{L}^{\varepsilon_0}_{\mathcal{S}_0}=\{\mathcal{S}_0(\lambda) : \lambda \in  [-c\varepsilon_0, c\varepsilon_0]^3\}
\end{equation}
where $c=c(M_{\rm init})>0$ is a constant referred to in Proposition~\ref{theleaves}.
Note that for $\mathcal{S}_0\neq \mathcal{S}'_0$, then
\[
\mathcal{L}^{\varepsilon_0}_{\mathcal{S}_0'}\cap \mathcal{L}^{\varepsilon_0}_{\mathcal{S}_0}=\emptyset.
\]

We have the following:
\begin{proposition}
\label{theleaves}
For $\varepsilon_0$ sufficiently small, there exists a constant $c=c(M_{\rm init})>0$ 
such that the following is true. 
Let $0<\varepsilon_0 \le\hat\varepsilon_0$.
Then 
\begin{equation}
\label{foliationhere}
\mathfrak{M}(M_{\rm init},c^2\varepsilon_0) \subset
\bigcup_{\mathcal{S}_0\in\mathfrak{M}_0(M_{\rm init},c^2\varepsilon_0) } \mathcal{L}^{\varepsilon_0}_{\mathcal{S}_0} \subset  \mathfrak{M}(M_{\rm init},\varepsilon_0).
\end{equation}
In particular, the statement of Theorem~\ref{existenceofanchoredgaugethe} holds for all initial data sets parameterised
by  $\mathcal{L}^{\varepsilon_0}_{{\mathcal{S}}_0}$.
\end{proposition}

 \begin{proof}
Note that the second inclusion follows simply from the triangle inequality, in view of the smoothness of the new term on the right
hand side of~\eqref{definitionofseennormnow}, the restriction of $\lambda$ to $[-c\varepsilon_0,c\varepsilon_0]^3$
and the fact that the 
energy~\eqref{definitionofseennormnow} is expressed
directly in terms of seed data.
 \end{proof}

\begin{remark}
\label{remarkaboutcodim}
The above Proposition  covers $\mathfrak{M}(M_{\rm init},c^2\varepsilon_0)$
by disjoint 3-parameter families, which are themselves contained
in a  $\mathfrak{M}(M_{\rm init}, \varepsilon_0)$. Moreover, 
$\mathfrak{M}_0$ is evidently a codimension-$3$ submanifold of $\mathfrak{M}$ in a natural sense.
One should think of $\mathfrak{M}_0$
as a non-teleological approximation to our actual asymptotically stable ``submanifold''
$\mathfrak{M}_{\rm stable}$.
In our main theorem, we will show that global asymptotic stability holds for
one member of every leaf 
$\mathcal{L}^{\varepsilon_0}_{\mathcal{S}_0}$
of this foliation, corresponding to a $\lambda^{\rm final}(\mathcal{S}_0)$. 
Our actual asymptotically stable ``submanifold'' will then be defined as  union\index{initial data!moduli space!$\mathfrak{M}_{\rm stable}$, asymptotically stable codimension-$3$ ``submanifold'' of $\mathfrak{M}$} 
\[
\mathfrak{M}_{\rm stable}= \bigcup_{\mathcal{S}_0\in \mathfrak{M}_0} \mathcal{S}_0(\lambda^{\rm final})
\]
for some $\lambda^{\rm final}$ which itself depends on $\mathcal{S}_0$. It is in this sense that
our ``submanifold'' can be naturally viewed as codimension $3$. 
See Remark~\ref{remarkafterthetheorem}.
\end{remark}

\section{The homeomorphism  ${\bf J}_0:\mathfrak{R}_0\to B_{\varepsilon_0/u_f^0}$ and its degree 1 property}
\label{thedegreeonemap}

In this section, we shall define a map 
 \[
{\bf J}_0:\mathfrak{R}_0\to B_{\varepsilon_0/u_f^0}
\]
associated to $u_f^0$. 
Here $B_{\varepsilon_0/u_f^0}\subset \mathbb R^3$ denotes the
closed ball of radius $\varepsilon_0/u_f^0$ in $\mathbb R^3$.\index{initial data!sets!$B_{\varepsilon_0/u_f^0}$, closed ball of radius $\varepsilon_0/u_f^0$ in $\mathbb R^3$}\index{bootstrap!sets!$B_{\varepsilon_0/u_f^0}$, closed ball of radius $\varepsilon_0/u_f^0$ in $\mathbb R^3$}

The topological properties
of the extension of this map to later times $u_f\ge u_f^0$ in the context
of the proof of our main theorem will be essential to obtaining our codimension-3 stability result.

\begin{definition}
\label{definitionofmapandunderlset}
For $\hat\varepsilon_0(M_{\rm init})$ chosen sufficiently small (so that in particular  Theorem~\ref{existenceofanchoredgaugethe} applies) and $0<\varepsilon_0\le \hat\varepsilon_0$,
let us fix a leaf $\mathcal{L}_{\mathcal{S}_0}^{\varepsilon_0}$ as in Proposition~\ref{theleaves}. 
For $\lambda \in [-c\varepsilon_0,c\varepsilon_0]^3 $, define\index{initial data!parameters!${\bf J}_0(\lambda)$,  
angular momentum vector at time $u_0^f$}
\begin{equation}
\label{newdefofboldv0}
{\bf J}_0(\lambda): = 	(J^{-1}_{\I},J^{0}_{\I},J^1_{\I}) \in \mathbb R^3
\end{equation}
where $J^m_{\I}$ are the associated Kerr parameters defined in Definition~\ref{assocKerparIplus} 
corresponding
to the anchored $\I$ gauge with respect to parameters $u^0_f$ and $M^0_f(\lambda)$ given by Theorem~\ref{existenceofanchoredgaugethe}
applied to $(\mathcal{M}(\lambda),g(\lambda))$.
We define the set\index{initial data!sets!$\mathfrak{R}_0$, subset of $\lambda$-parameter space at time $u_f^0$} 
\begin{equation}
\label{defofmathfrakR0}
\mathfrak{R}_0=\left\{\lambda \in[-c\varepsilon_0,c\varepsilon_0]^3 : |{\bf J}_0(\lambda)|\le \frac{\varepsilon_0}{u_f^0}\right\}.
\end{equation}
\end{definition}

\begin{proposition}
\label{assertingdegreeone}
For sufficiently small $\hat\varepsilon_0$ and for $0<\varepsilon_0\le \hat\varepsilon_0$,
let $\mathcal{L}_{\mathcal{S}_0}^{\varepsilon_0}$ be as in Definition~\ref{definitionofmapandunderlset}.
Then the set $\mathfrak{R}_0$ defined by~\eqref{defofmathfrakR0} is homeomorphic (in fact
diffeomorphic) to a closed $3$-ball
under the map~\eqref{newdefofboldv0}, i.e.~the map\index{initial data!maps!${\bf J}_0:\mathfrak{R}_0\to B_{\varepsilon_0/u_f^0}$}  
\begin{equation}
\label{isadiffeo}
{\bf J}_0:\mathfrak{R}_0\to B_{\varepsilon_0/u_f^0}
 \end{equation}
 is a diffeomorphism,
 so, in particular,  the restriction
 \begin{equation}
 \label{degreeonestatement}
 {\bf J}_0|_{\partial\mathfrak{R}_0}
 :\partial\mathfrak{R}_0 \to \partial B_{\varepsilon_0/u_f^0}
 \end{equation}
is of degree $1$.
\end{proposition}

\begin{remark}
\label{Inlineartheoryremark}
In view of~\eqref{dontchangethem} of Proposition~\ref{linearisedprophere}, 
it follows that the linear analogue of the map ${\bf J}_0$
is simply the identity map
\[
\lambda \mapsto \lambda \, .
\]
Thus, in linear theory, provided that say
\begin{equation}
\label{conditiononuf0}
u_f^0> 2c^{-1}
\end{equation}
where $c$ is the constant  above,
 then $B_{\varepsilon_0/u^0_f}\subset( -c\varepsilon_0,c\varepsilon)^3 $ and thus
 $\mathfrak R_0 = B_{\varepsilon_0/u^0_f}$,
 whence 
  the analogue of ${\bf J}_0:\mathfrak{R}_0 \to  B_{\varepsilon_0/u_f^0}$ is (trivially)
  a diffeomorphism.
The condition~\eqref{conditiononuf0} thus defines our additional restriction on $u_f^0$
announced in Section~\ref{compediumparameterssec}.
\end{remark}

\begin{remark}
Note that it is the weaker statement that the map~\eqref{degreeonestatement} is degree-$1$ which we
shall propagate in our bootstrap as a property of a $u_f$-dependent
map ${\bf J}_{u_f}$ (see already  Definition~\ref{bootstrapsetdef}),
rather than the stronger statement
that~\eqref{isadiffeo} is in fact a diffeomorphism. In fact, we shall propagate this property by simply
ensuring that our map ${\bf J}_{u_f}$ coincides with ${\bf J}_0$ on $\partial\mathfrak{R}_0$.
\end{remark}

\begin{proof}
In view of Remark~\ref{Inlineartheoryremark}, the result 
is simply an improvement on the statement~\eqref{closenessofKerrparams} and follows again from 
Cauchy stability type arguments.
\end{proof}

\chapter{Final formulation of the main result: Theorem~\ref{thm:main}}
\label{maintheoremsec}

Using the local theory and associated structures defined in the previous chapter,
we may now record in this chapter
the final formulation of our main theorem.

\minitoc

We first will  define in {\bf Section~\ref{energiessection}} a collection of energies controlling
our various quantities,  which
are combined into a set of
master energies,  which 
will then appear explicitly in the
statement of the theorem.  The final formulation of our theorem will appear as 
{\bf Theorem~\ref{thm:main}} 
in {\bf Section~\ref{precisesec}}.
Finally, in {\bf Section~\ref{technicalremarkssec}},
we shall interpret our theorem  as defining a codimension-$3$ asymptotically stable ``submanifold''
$\mathfrak{M}_{\rm stable}$  and conclude with  some technical remarks.

\vskip1pc
\emph{The detailed definitions of the various energies in Section~\ref{energiessection} can be skipped on a first reading.
We note that the reliance of this chapter  on many of the notations from Part~\ref{preliminlabel} (e.g.~the gauge invariant
hierarchy of Section~\ref{AGIQdef}, the calculus of diffeomorphism functions of Chapter~\ref{newdiffysec}, etc.)~is 
only through their presence
in the energies of Section~\ref{energiessection}; thus the present chapter may in fact
be read independently of much of Part~\ref{preliminlabel}, provided that the reader
takes these energies as a black box.}

\section{Energies}
\label{energiessection}

In this section, we shall
fix parameters $u_f$ and $M_f$ and
assume we have a spacetime  as in Definition~\ref{anchoringdef},
i.e.~the maximal Cauchy development $(\mathcal{M},g)$,
satisfying the Einstein vacuum equations $(\ref{Ricciflathere})$, 
equipped with the four
gauges $(\ref{initialKgaugehere})$--$(\ref{fortheHplusgauge})$  defined in 
Section~\ref{fourdoublenull}.
We will proceed to define a variety of energies controlling our basic quantities in both
teleological gauges as well as the diffeomorphism functions. These energies will be combined
to form the master energies appearing explicitly in
the statement of Theorem~\ref{thm:main}.

Many of the estimates shown in the proof of Theorem~\ref{thm:main} amount to stronger statements than the boundedness, stated in Theorem~\ref{thm:main}, of the following energies.  
The reader is referred to Part~\ref{improvingpart} for the sharpest statements.

\subsection{Some auxiliary notation}
\label{notationfortwogauges}

In this section, we introduce various  auxiliary notations concerning
 anchored  gauges which will appear in the definition of energies.
 
We will often typically drop the $u_f$ dependence from the notation.

Recall the definitions of $\DRH$, $\DRI$, $C_u^{\Hp}$, $\Cbar_v^{\Hp}$, $C_u^{\I}$ and $\Cbar_v^{\I}$ from Section \ref{section:gaugefunctions}.  Define also
\[
	\mathcal{D}_{\Hp}^{\I}=\DRH\cap \DRI.
\]
For given $u$, $v$, we will find useful the notations 
\begin{align*}
	\DRH(v)
	&
	:=
	\DRH \cap \{ v_{\Hp} \geq v\},
	&
	\DRI(u)
	&
	:=
	\DRI \cap \{ u_{\I} \geq u\},
	\\
	\DRH(u,v) 
	&
	:=
	\DRH \cap \{ u_{\Hp} \geq u \} \cap \{ v_{\Hp} \geq v\},
	\qquad
	&
	\DRI(u,v)
	&
	:= \DRI \cap \{ u_{\I} \geq u\} \cap \{ v_{\I} \geq v\},
\end{align*}
\[
	\mathcal{D}_{\Hp}^{\I}(u)=\DRH\cap \DRI (u),
\]
and
\[
	C_u^{\Hp}(v) := C_u^{\Hp} \cap \{ v_{\Hp} \geq v\},
	\qquad
	\Cbar_v^{\Hp}(u) := \Cbar_v^{\Hp} \cap \{ u_{\Hp} \geq u\},
\]
\[
	C_u^{\I}(v) := C_u^{\I} \cap \{ v_{\I} \geq v\},
	\qquad
	\Cbar_v^{\I}(u) := \Cbar_v^{\I} \cap \{ u_{\I} \geq u\}.
\]

\subsection{Norms on spheres, cones and spacetime regions}
\label{normsheretoo}

We define the norms\index{energies!norms on spheres!$\Vert \xi \Vert_{S^{\Hp}_{u,v}}^2$} 
\index{energies!norms on spheres!$\Vert \xi \Vert_{S^{\I}_{u,v}}^2$} 
on the spheres, for any $S^{\Hp}$-tensor or $S^{\I}$-tensor $\xi$ by
\[
	\Vert \xi \Vert_{S^{\Hp}_{u,v}}^2
	:=
	\int_{S^{\Hp}_{u,v} }
	\vert \xi \vert^2
	d \theta,
	\qquad
	\Vert \xi \Vert_{S^{\I}_{u,v}}^2
	:=
	\int_{S^{\I}_{u,v} }
	\vert \xi \vert^2
	d \theta,
\]
respectively. 
Recall that the notation $d\theta$ is defined in~\eqref{spherevolform}.  (In particular, the true induced 
volume form is $r^2d\theta$, and, thus, the above integrands are $r^{-2}$ weighted with respect to
the true induced volume form.)

Define the norms on the $\I$ null hypersurfaces,\index{energies!norms on cones!$\Vert \xi \Vert_{C^{\I}_u(v)}^2$}\index{energies!norms on cones!$\Vert \xi \Vert_{\Cbar^{\I}_v(u)}^2$} for any $S^{\I}_{u,v}$ tensor $\xi$,
\[
	\Vert \xi \Vert_{C^{\I}_u(v)}^2
	:=
	\int_v^{v_{\infty}}
	\int_{S^{\I}_{u,v'}}
	\vert \xi \vert^2
	d \theta
	dv',
	\qquad
	\Vert \xi \Vert_{\Cbar^{\I}_v(u)}^2
	:=
	\int_u^{\min \{ u_f , u(R_{-2},v) \}}
	\int_{S^{\I}_{u',v}}
	\vert \xi \vert^2
	\Omega_{\I}^2
	d \theta
	du',
\]
and
\[
	\Vert \xi \Vert_{C^{\I}_u}^2
	:=
	\Vert \xi \Vert_{C^{\I}_u(v(R_{-2},u))}^2,
	\qquad
	\Vert \xi \Vert_{\Cbar^{\I}_v}^2
	:=
	\Vert \xi \Vert_{\Cbar^{\I}_v(u_{-1})}^2.
\]
Similarly define the norms on the $\Hp$ null hypersurfaces,\index{energies!norms on cones!$\Vert \xi \Vert_{C^{\Hp}_u(v)}^2$}\index{energies!norms on cones!$\Vert \xi \Vert_{\Cbar^{\Hp}_v(u)}^2$}  for any $S^{\Hp}$-tensor $\xi$,
\[
	\Vert \xi \Vert_{C^{\Hp}_u(v)}^2
	:=
	\int_{C_u^{\Hp} \cap \{ v_{\Hp} \geq v\} }
	\vert \xi \vert^2
	d \theta dv',
	\qquad
	\Vert \xi \Vert_{\Cbar^{\Hp}_v(u)}^2
	:=
	\int_{\Cbar^{\Hp}_v \cap \{ u_{\Hp} \geq u\}}
	\vert \xi \vert^2
	\Omega_{\Hp}^2
	d \theta d u',
\]
and
\[
	\Vert \xi \Vert_{C^{\Hp}_u}^2
	:=
	\Vert \xi \Vert_{C^{\Hp}_u(v_{-1})}^2,
	\qquad
	\Vert \xi \Vert_{\Cbar^{\Hp}_v}^2
	:=
	\Vert \xi \Vert_{\Cbar^{\Hp}_v(\max \{ u_{-1} , u(R_{2},v) \})}^2.
\]
For the spacetime regions define, for $v_{-1} \leq v \leq v(R_2,u_f)$ and $u_{-1} \leq u \leq u_f$,
\[
\Vert \xi \Vert_{\mathcal{D}^{\Hp}(v)}^2 := \int_{\mathcal{D}^{\Hp}(v)}|\xi|^2 \Omega_{\Hp} ^2 d\theta du dv^\prime  \, ,  \  \ \ \ \ \ \ \ \Vert \xi \Vert_{\mathcal{D}^{\I}(u)}^2  :=  \int_{\mathcal{D}^{\I}(u)}|\xi|^2 \Omega_{\I}^2 d\theta du^\prime dv \, ,
\]
and
\[
	\Vert \xi \Vert_{\mathcal{D}^{\Hp}}^2
	:=
	\Vert \xi \Vert_{\mathcal{D}^{\Hp}(v_{-1})}^2,
	\qquad
	\Vert \xi \Vert_{\mathcal{D}^{\I}}^2
	:=
	\Vert \xi \Vert_{\mathcal{D}^{\I}(u_{-1})}^2,
\]
and finally
\[
	\Vert \xi \Vert_{\mathcal{D}_{\Hp}^{\I}(u)} := \Vert \xi \, \mathds{1}_{\DRH\cap \DRI (u)} \Vert_{\DRI}.
\]

Finally, for any of the above norms $\Vert \cdot \Vert$ we define for any $S_{u,v}$ tensors $\xi_1,\ldots,\xi_k$ and any $\mathfrak{D}^{\gamma}$,
\[
	\Vert \mathfrak{D}^{\gamma} (\xi_1,\ldots,\xi_k) \Vert
	:=
	\Vert \mathfrak{D}^{\gamma} \xi_1 \Vert
	+
	\ldots
	+
	\Vert \mathfrak{D}^{\gamma}\xi_k \Vert.
\]

\subsection{Conventions for the volume form in integrals}
\label{volformconven}

In what follows, we shall use the following convention for integrals, recalling again the
notation $d\theta$  defined in~\eqref{spherevolform}:
\vskip1pc
\noindent\fbox{
    \parbox{6.35in}{
 \emph{If the measure is not spelled out explicitly, integration over an ingoing cone $\underline{C}_v$ will always be with respect to
$du\, d\theta$,
 over an outgoing cone $C_u$ with always be with respect to 
$dv\, d\theta$
 and over a spacetime region will always be with respect to 
 $du\,dv\,d\theta$.}
 }}
 \vskip1pc
 Note that with this convention all $\Omega^2$-weights near (what will be) the horizon and all $r$-weights near infinity will appear explicitly in the integrand.

\subsection{Energies of $P$ and \underline{$P$}}
\label{PandPbarenergydefs}

Recall the quantities $P$, $\underline{P}$ and $\check{\underline{P}}$
appearing in the almost gauge invariant hierarchy of Section~\ref{AGIQdef}, expressed now with respect to the
$\mathcal{H}^+$ and $\mathcal{I}^+$ gauges.

We first define the following rescaled quantities:\index{almost gauge invariant hierarchy!${\Psi}_{\Hp}$, rescaled version of
$P_{\Hp}$}\index{almost gauge invariant hierarchy!${\Psi}_{\I}$, rescaled version of
$P_{\I}$}\index{almost gauge invariant hierarchy!$\underline{\Psi}_{\Hp}$, rescaled version of
$\underline{P}_{\Hp}$}\index{almost gauge invariant hierarchy!$\check{\underline{\Psi}}_{\I}$, rescaled version of
$\check{\underline{P}}_{\I}$}\index{teleological $\I$ gauge!rescaled quantities!${\Psi}_{\I}$, rescaled version of
$P_{\I}$}\index{teleological $\I$ gauge!rescaled quantities!$\check{\underline{\Psi}}_{\I}$, rescaled version of
$\check{\underline{P}}_{\I}$}\index{teleological $\Hp$ gauge!rescaled quantities!${\Psi}_{\Hp}$, rescaled version of
$P_{\Hp}$}\index{teleological $\Hp$ gauge!rescaled quantities!$\underline{\Psi}_{\Hp}$, rescaled version of
$\underline{P}_{\Hp}$}
\begin{align}
{\Psi}_{\Hp} = r^5 P_{\Hp} \ \ \ \ , \ \ \ \  {\Psi}_{\I} = r^5 P_{\I} \ \ \ \ \ \textrm{and} \ \ \ \ \  \underline{\Psi}_{\Hp} = r^5 \underline{P}_{\Hp} \ \ \ \ , \ \ \ \  \check{\underline{\Psi}}_{\I} = r^5 \check{\underline{P}}_{\I}  \, .
\end{align}

We recall the fixed parameters $N \geq 12$ from Section~\ref{globalsmallnessassumptionnorm}
and  $\delta=\frac1{100}$ from Section~\ref{compediumparameterssec}.

We define for $1\leq K \leq N-2$, the energies for $v \geq v_{-1}$\index{energies!energies for $P$ and $\underline{P}$!${\mathbb{E}}^{K} \left[P_{\Hp}\right] \left(v\right)$}
\begin{align} \label{energyPhoz}
&{\mathbb{E}}^{K} \left[P_{\Hp}\right] \left(v\right) := \sup_{\tilde{v} \geq v} \int_{\underline{C}^{\Hp}_{\tilde{v}}} \Omega^2  \sum_{|\underline{k}|=0; k_3\neq K}^{K}  | \mathfrak{D}^{\underline{k}} {\Psi}_{\Hp}|^2 + \sup_{u\leq u_f} \int_{C^{\Hp}_u({v})} \sum_{|\underline{k}|=0; k_1\neq K}^{K}  | \mathfrak{D}^{\underline{k}} \Psi_{\Hp}|^2   \\
& \qquad \ \  +  \int_{\mathcal{D}^{\Hp}\left(v\right)}  \Omega^2 \Bigg\{ \sum_{|\underline{k}|=0}^{K} \left(1-\frac{3M_f}{r}\right)^2 | \mathfrak{D}^{\underline{k}} {\Psi}_{\Hp}|^2  +  \sum_{|\underline{k}|=0}^{K-1}  | \mathfrak{D}^{\underline{k}} {\Psi}_{\Hp}|^2+  | R^\star \mathfrak{D}^{\underline{k}} {\Psi}_{\Hp}|^2  \Bigg\},  \nonumber 
\end{align}
where we have already used the convention of Section~\ref{volformconven} for the volume form,
and for $p \in \{0,1,2\}$, $1 \leq K \leq N-3$ and $\tau \geq u_{-1}$\index{energies!energies for $P$ and $\underline{P}$!${\mathbb{E}}^{K,p} \left[P_{\I}\right] \left(\tau\right)$}
\begin{align} \label{energyPinf}
&{\mathbb{E}}^{K,p} \left[P_{\I}\right] \left(\tau\right) := 
\sup_{\tau \leq u \leq u_f} \int_{C^{\I}_u} \Bigg\{ \sum_{|\underline{k}|=0}^{K-1}  \  r^p | \Omega \slashed{\nabla}_4 \mathfrak{D}^{\underline{k}} \Psi_{\I}|^2 + \frac{1}{r^2} | r \slashed{\nabla} \mathfrak{D}^{\underline{k}} \Psi_{\I}|^2 \Bigg\} \\
& \qquad \qquad \ \   +\sup_{v \leq v_\infty} \int_{\underline{C}^{\I}_{v}(\tau)}  \Bigg\{ \sum_{|\underline{k}|=0}^{K-1}  \   | \Omega \slashed{\nabla}_3 \mathfrak{D}^{\underline{k}} \Psi_{\I}|^2 + \frac{r^p}{r^2} | r \slashed{\nabla} \mathfrak{D}^{\underline{k}} \Psi_{\I}|^2 \Bigg\}  \nonumber \\
& \qquad \qquad \ \  +  \int_{\mathcal{D}^{\I}\left(\tau\right)}  \Bigg\{ \sum_{|\underline{k}|=0}^{K-1}  \   \frac{1}{r^{1+\delta}}  | \Omega \slashed{\nabla}_3 \mathfrak{D}^{\underline{k}} \Psi_{\I}|^2 + r^{p-1-\delta \cdot {\boldsymbol\delta^p_0}} | \Omega \slashed{\nabla}_4 \mathfrak{D}^{\underline{k}} \Psi_{\I}|^2 + \frac{r^p}{r^{3+\delta \cdot {\bf \delta^p_2}}} | r \slashed{\nabla} \mathfrak{D}^{\underline{k}} \Psi_{\I}|^2 \Bigg\} . \nonumber
\end{align}
Furthermore, we define ${\mathbb{E}}^{K} \left[\underline{P}_{\Hp}\right] \left(v\right)$ replacing $\Psi_{\Hp}$ by $\underline{\Psi}_{\Hp}$ on the right hand side of (\ref{energyPhoz}) and ${\mathbb{E}}^{K,p} \left[\check{\underline{P}}_{\I}\right] \left(\tau\right)$ replacing $\Psi_{\I}$ by $\underline{\check{\Psi}}_{\I}$ on the right hand side of (\ref{energyPinf}).

We finally define the master energies (now making the dependence on $u_f$ manifest in the notation)\index{energies!energies for $P$ and $\underline{P}$!$\mathbb{E}_{u_f}^{N-2} [P_{\Hp},P_{\I} ]$}\index{energies!energies for $P$ and $\underline{P}$!$\mathbb{E}_{u_f}^{N-2} [\underline{P}_{\Hp},\check{\underline{P}}_{\I} ]$}
\begin{align} \label{masterPen}
	\mathbb{E}_{u_f}^{N-2} [P_{\Hp},P_{\I} ]
	&:=
	\sum_{s=0,1,2}
	\sup_{u_{-1} \leq \tau \leq u_f} \tau^s \cdot {\mathbb{E}}^{N-2-s,2-s} \left[P_{\I}\right] \left(\tau\right)  + \sum_{s=0,1,2}
	\sup_{v_{-1} \leq v} 
	v^{s} \cdot {\mathbb{E}}^{N-2-s} \left[{P}_{\Hp}\right] \left(v\right)\, , \nonumber \\
	\mathbb{E}_{u_f}^{N-2} [\underline{P}_{\Hp},\check{\underline{P}}_{\I} ]
	&:=
	\sum_{s=0,1,2}
	\sup_{u_{-1} \leq \tau \leq u_f} \tau^s \cdot {\mathbb{E}}^{N-2-s,2-s} \left[\check{\underline{P}}_{\I}\right] \left(\tau\right)  + \sum_{s=0,1,2}
	\sup_{v_{-1} \leq v} 
	v^{s} \cdot {\mathbb{E}}^{N-2-s} \left[\underline{P}_{\Hp}\right] \left(v\right) \, .
\end{align}

\subsection{Energies of $\alpha$ and \underline{$\alpha$}}
\label{twoenergiesalphaandalphabardefs}
Recall  the curvature components $\alpha$ and $\alphabar$ appearing in the almost gauge invariant hierarchy of Section~\ref{AGIQdef}, expressed now with respect to the
$\mathcal{H}^+$ and $\mathcal{I}^+$ gauges.

\subsubsection{Energies of $\alpha$}
We first define the following rescaled quantities:\index{almost gauge invariant hierarchy!${A}_{\Hp}$, rescaled version of $\alpha_{\Hp}$}\index{almost gauge invariant hierarchy!${A}_{\I}$, rescaled version of $\alpha_{\I}$}\index{almost gauge invariant hierarchy!${\Pi}_{\Hp}$, rescaled version of $\psi_{\Hp}$}\index{almost gauge invariant hierarchy!${\Pi}_{\I}$, rescaled version of ${\psi}_{\I}$}\index{teleological $\Hp$ gauge!rescaled quantities!${A}_{\Hp}$, rescaled version of $\alpha_{\Hp}$}\index{teleological $\I$ gauge!rescaled quantities!${A}_{\I}$, rescaled version of $\alpha_{\I}$}\index{teleological $\Hp$ gauge!rescaled quantities!${\Pi}_{\Hp}$, rescaled version of $\psi_{\Hp}$}\index{teleological $\I$ gauge!rescaled quantities!${\Pi}_{\I}$, rescaled version of ${\psi}_{\I}$}
\begin{align}
{A}_{\Hp} = \Omega^2 r \alpha_{\Hp} \ \ \ \ , \ \ \ \  {A}_{\I} = \Omega^2 r \alpha_{\I} \ \ \ \ , \ \ \ \ {\Pi}_{\Hp} = \Omega r^3 \psi_{\Hp} \ \ \ \ , \ \ \ \  {\Pi}_{\I} = \Omega r^3 {\psi}_{\I}
\end{align}

For $1\leq K \leq N$, we define the energies for $v \geq v_{-1}$\index{energies!energies for $\alpha$ and $\underline{\alpha}$!${\mathbb{E}}^{K} \left[{\alpha}_{\Hp}\right] \left(v\right)$}
\begin{align} \label{meah}
&{\mathbb{E}}^{K} \left[{\alpha}_{\Hp}\right] \left(v\right) := \sup_{\tilde{v} \geq v} \int_{\underline{C}^{\Hp}_{\tilde{v}}} \Omega^2  \sum_{|\underline{k}|=0; k_3\neq K}^{K}  | \mathfrak{D}^{\underline{k}} {A}_{\Hp}|^2 + \sup_{u\leq u_f} \int_{C^{\Hp}_u({v})} \sum_{|\underline{k}|=0}^{K}  | \mathfrak{D}^{\underline{k}} A_{\Hp}|^2  \nonumber \\
& \qquad \ \  +  \int_{\mathcal{D}^{\Hp}\left(v\right)}  \Omega^2 \Bigg\{ \sum_{|\underline{k}|=0}^{K} \left(1-\frac{3M_f}{r}\right)^2 | \mathfrak{D}^{\underline{k}} {A}_{\Hp}|^2  +  \sum_{|\underline{k}|=0}^{K-1}  | \mathfrak{D}^{\underline{k}} {A}_{\Hp}|^2+  | R^\star \mathfrak{D}^{\underline{k}} {A}_{\Hp}|^2  \Bigg\} 
\end{align}
and for $p \in \{0,1,2\}$, $1\leq K\leq N$ and $\tau \geq u_{-1}$\index{energies!energies for $\alpha$ and $\underline{\alpha}$!${\mathbb{E}}^{K,p} \left[{\alpha}_{\I}\right] \left(\tau\right)$}
\begin{align}\label{meai}
&{\mathbb{E}}^{K,p} \left[{\alpha}_{\I}\right] \left(\tau\right) := 
 \sup_{\tau \leq u \leq u_f} \int_{C^{\I}_u} \Bigg\{ \sum_{|\underline{k}|=1}^K  \ r^{4+p} | \mathfrak{D}^{\underline{k}} A_{\I}|^2 + \sum_{|\underline{k}|=0}^{K-1}  r^2 | \mathfrak{D}^{\underline{k}} \Pi_{\I}|^2 \Bigg\} \nonumber \\
& \qquad \qquad \ \   +\sup_{v \leq v_\infty} \int_{\underline{C}^{\I}_{v}(\tau)}  \Bigg\{ \sum_{|\underline{k}|=1, k_3 \neq K}^{K}  \ r^{4+p} | \mathfrak{D}^{\underline{k}} A_{\I} |^2 + \sum_{|\underline{k}|=0}^{K-1}  r^{2+p} | \mathfrak{D}^{\underline{k}} \Pi_{\I}|^2 + \sum_{|\underline{k}|=0}^{K-2}  | \mathfrak{D}^{\underline{k}} \Psi_{\I}|^2  \Bigg\} \nonumber \\
& \qquad \qquad \ \  +  \int_{\mathcal{D}^{\I}\left(\tau\right)} \Bigg\{ \sum_{|\underline{k}|=1}^K  \, r^{3+p-\delta} | \mathfrak{D}^{\underline{k}} A_{\I}|^2  + \sum_{|\underline{k}|=0}^{K-1} r^{1+p-\delta} | \mathfrak{D}^{\underline{k}} \Pi_{\I}|^2  +\sum_{|\underline{k}|=0}^{K-2} r^{-1-\delta} | \mathfrak{D}^{\underline{k}} \Psi_{\I}|^2 \Bigg\} .
\end{align}
Now, restoring the dependence on $u_f$, we define the master energy\index{energies!energies for $\alpha$ and $\underline{\alpha}$!$\mathbb{E}_{u_f}^N [\alpha_{\Hp}, \alpha_{\I}]$}
\begin{align} \label{masterenergya}
	\mathbb{E}_{u_f}^N [\alpha_{\Hp}, \alpha_{\I}]
	:=
	\sum_{s=0,1,2}
	\sup_{u_{-1} \leq \tau \leq u_f} \tau^s \cdot {\mathbb{E}}^{N-s,2-s} \left[{\alpha}_{\I}\right] \left(\tau\right)  + \sum_{s=0,1,2}
	\sup_{v_{-1} \leq v} 
	v^{s} \cdot {\mathbb{E}}^{N-s} \left[{\alpha}_{\Hp}\right] \left(v\right).
\end{align}

\subsubsection{Energies of \underline{$\alpha$}}
We define the following rescaled quantities:\index{almost gauge invariant hierarchy!$\underline{A}_{\Hp}$, rescaled version of $\underline{\alpha}_{\Hp}$}\index{almost gauge invariant hierarchy!$\check{\underline{A}}_{\I}$, rescaled version of $\underline{\alpha}_{\I}$}\index{almost gauge invariant hierarchy!$\underline{\Pi}_{\Hp}$, rescaled version of $\underline{\psi}_{\Hp}$}\index{almost gauge invariant hierarchy!$\check{\underline{\Pi}}_{\I}$, rescaled version of $\check{\underline{\psi}}_{\I}$}\index{teleological $\Hp$ gauge!rescaled quantities!$\underline{A}_{\Hp}$, rescaled version of $\underline{\alpha}_{\Hp}$}\index{teleological $\I$ gauge!rescaled quantities!$\check{\underline{A}}_{\I}$, rescaled version of $\underline{\alpha}_{\I}$}\index{teleological $\Hp$ gauge!rescaled quantities!$\underline{\Pi}_{\Hp}$, rescaled version of $\underline{\psi}_{\Hp}$}\index{teleological $\I$ gauge!rescaled quantities!$\check{\underline{\Pi}}_{\I}$, rescaled version of $\check{\underline{\psi}}_{\I}$}
\begin{align}
\underline{A}_{\Hp} = \Omega^2 r \underline{\alpha}_{\Hp} \ \ \ \ , \ \ \ \  \check{\underline{A}}_{\I} = \Omega^2 \check{r} \underline{\alpha}_{\I} \ \ \ \ , \ \ \ \ 
\underline{\Pi}_{\Hp} = \Omega r^3 \underline{\psi}_{\Hp} \ \ \ \ , \ \ \ \  \check{\underline{\Pi}}_{\I} = \Omega r^3 \check{\underline{\psi}}_{\I}.
\end{align}

We define\index{energies!energies for $\alpha$ and $\underline{\alpha}$!${\mathbb{E}}^{K} \left[\underline{\alpha}_{\Hp}\right] \left(v\right)$}\index{energies!energies for $\alpha$ and $\underline{\alpha}$!${\mathbb{E}}^{K} \left[\underline{\alpha}_{\I}\right] \left(\tau\right)$}
\begin{align} \label{abaren1}
&{\mathbb{E}}^{K} \left[\underline{\alpha}_{\Hp}\right] \left(v\right) := \sup_{\tilde{v} \geq v} \int_{\underline{C}^{\Hp}_{\tilde{v}}} \Omega^2  \sum_{|\underline{k}|=0}^{K}  | \mathfrak{D}^{\underline{k}} (\Omega^{-4} \underline{A}_{\Hp})|^2 + \sup_{u\leq u_f} \int_{C^{\Hp}_u({v})} \sum_{|\underline{k}|=0; k_2\neq K}^{K}  | \mathfrak{D}^{\underline{k}} (\Omega^{-4} \underline{A}_{\Hp})|^2   \\
& \qquad \ \  +  \int_{\mathcal{D}^{\Hp}\left(v\right)}  \Omega^2 \Bigg\{ \sum_{|\underline{k}|=0}^{K} \left(1-\frac{3M_f}{r}\right)^2 | \mathfrak{D}^{\underline{k}} (\Omega^{-4} \underline{A}_{\Hp})|^2  +  \sum_{|\underline{k}|=0}^{K-1}  | \mathfrak{D}^{\underline{k}} (\Omega^{-4} \underline{A}_{\Hp})|^2+  | R^\star \mathfrak{D}^{\underline{k}} (\Omega^{-4} \underline{A}_{\Hp})|^2  \Bigg\} \nonumber
\end{align}
\begin{align} \label{abaren2}
{\mathbb{E}}^{K} \left[\underline{\alpha}_{\I}\right] \left(\tau\right) :=  &\sup_{\tau \leq u \leq u_f} \int_{C^{\I}_u} \frac{1}{r^2} \Bigg\{ \sum_{|\underline{k}|=0; k_2\neq K}^{K}  | \mathfrak{D}^{\underline{k}} \check{\underline{A}}_{\I}|^2 + \sum_{|\underline{k}|=0}^{K-1}   | \mathfrak{D}^{\underline{k}} \check{\underline{\Pi}}_{\I}|^2 + \sum_{|\underline{k}|=0}^{K-2} | \mathfrak{D}^{\underline{k}} \check{\underline{\Psi}}_{\I}|^2 \Bigg\} \nonumber \\
&+\sup_{v \leq v_\infty} \int_{\underline{C}^{\I}_{v}(\tau)} \Bigg\{\sum_{|\underline{k}|=0}^{K}  | \mathfrak{D}^{\underline{k}} \check{\underline{A}}_{\I}|^2 + \sum_{|\underline{k}|=0}^{K-1}   | \mathfrak{D}^{\underline{k}} \check{\underline{\Pi}}_{\I}|^2  \Bigg\} \nonumber \\
&+  \int_{\mathcal{D}^{\I}\left(\tau\right)} \frac{1}{r^{1+\delta}}\Bigg\{ \sum_{|\underline{k}|=0}^K | \mathfrak{D}^{\underline{k}} \check{\underline{A}}_{\I}|^2 +\sum_{|\underline{k}|=0}^{K-1} | \mathfrak{D}^{\underline{k}} \check{\underline{\Pi}}_{\I}|^2 + \sum_{k=1}^{K-2} | (R^\star)^k \check{\underline{\Psi}}_{\I}|^2 \Bigg\} 
\end{align}
Now, restoring the dependence on $u_f$, we define the master energy\index{energies!energies for $\alpha$ and $\underline{\alpha}$!$\mathbb{E}_{u_f}^N [\alphabar_{\Hp}, {\alphabar}_{\I}]$}
\begin{align} \label{masterenergyab}
	\mathbb{E}_{u_f}^N [\alphabar_{\Hp}, {\alphabar}_{\I}]
	:=
	\sum_{s=0,1,2}
	\sup_{u_{-1} \leq \tau \leq u_f}
	\tau^{s} \cdot {\mathbb{E}}^{N-s} \left[{\underline{\alpha}}_{\I}\right] \left(\tau\right) + \sum_{s=0,1,2}
	\sup_{v_{-1} \leq v}
	v^{s} \cdot {\mathbb{E}}^{N-s} \left[\underline{\alpha}_{\Hp}\right] \left(v\right) 
\end{align}

\subsection{Energies of $\Phi^{\I}$} \label{sec:energiesig}
Recall the schematic notation from Section \ref{schemnotsec} and the shorthands (\ref{shorthandnot}).
For the energies below we define the following weights which affect the quantities $\omega-\omega_\circ$ and $\beta$ only.\index{schematic notation!$w_s$, $r$-weight} 

For the Ricci coefficients we let $w_s\left(\Gamma_p\right)=1=\tilde{w}_s(\Gamma_p)=1$ for all $\Gamma_p \setminus \{ \omega -\omega_\circ\}$ and $s=0,1,2$ while $w_s \left(\omega -\omega_\circ\right) = r^{1-s}$ and $\tilde{w}_s \left(\omega -\omega_\circ\right) = r^{-s}$. 

For the curvature components, we let $w_s (\mathcal{R}_p) = \tilde{w}  (\mathcal{R}_p)=1$ for all $\mathcal{R}_p \setminus \{\Omega^2 \alpha, \Omega^{-2} \underline{\alpha}, \Omega \beta\}$ and $s=0,1,2$ while for $\beta$ we set $w_s(\Omega \mathcal{\beta})=r^{-s}$ and $\tilde{w}(\Omega \beta)=r^{-1}$.

Finally, we recall  from Section \ref{reflinearisedKerrsec} the Kerr reference solutions. Below we will denote by $(\Gamma_p)_{\mathrm{Kerr}}$ and $(\mathcal{R}_p)_{\mathrm{Kerr}}$ the value of the Kerr reference solution of the relevant quantity $\Gamma_p$ or $\mathcal{R}_p$. In particular, $(\Gamma_p)_{\mathrm{Kerr}}=0$ unless $\Gamma_p \in \{\eta, \underline{\eta}, b\}$ and $(\mathcal{R}_p)_{\mathrm{Kerr}}=0$ unless $\mathcal{R}_p \in \{ \Omega \beta, \Omega^{-1} \underline{\beta}, \sigma\}$.

\subsubsection{Energies on \underline{$C$}$_{v=v_{\infty}}^{\I}$}
We define the following (angular) energies\index{energies!energies for $\Phi^{\I}$!$\slashed{\mathbb{E}}^N_{v_\infty} \left[ \Gamma \right]$, angular energy for $\Gamma$ on cones $\underline{C}_{v=v_{\infty}}^{\I}$}\index{energies!energies for $\Phi^{\I}$!$\slashed{\mathbb{E}}^N_{v_\infty} \left[ \mathcal{R} \right] $, angular energy for $\mathcal{R}$ on cones $\underline{C}_{v=v_{\infty}}^{\I}$} for the Ricci coefficients and curvature on \underline{$C$}$_{v=v_{\infty}}^{\I}$:
\begin{align} \label{estart}
\slashed{\mathbb{E}}^N_{v_\infty} \left[ \Gamma \right] := &\sum_{\Gamma_p \setminus \{\omega-\omega_\circ\}} \sum_{s=0}^2 \sum_{k=0}^{N-s}  \sup_{u \in \left[u_{-1},u_f\right]} u^s \cdot  \| (r\slashed{\nabla})^{k} \left(r^p  \Gamma_p -r^p (\Gamma_p)_{\mathrm{Kerr}}\right) \|^2_{S^2_{u,v_{\infty}}} \nonumber \\
+&\sum_{\Gamma_p \setminus \{\omega-\omega_\circ\}} \sum_{s=0}^2  \sum_{k=0}^{N-s}    \sup_{u \in \left[u_{-1},u_f\right]} u^s \int_{u}^{u_f} d\bar{u} \cdot  \|  (r\slashed{\nabla})^{k}  \left(r^p  \Gamma_p -r^p (\Gamma_p)_{\mathrm{Kerr}}\right) \|^2_{S^2_{\bar{u},v_{\infty}}}
\nonumber \\
+&\int_{u_{-1}}^{u_f} d\bar{u} \cdot  \| (r \slashed{\nabla})^{N+1}  \left( r^2\eta- r^2 \eta_{\mathrm{Kerr}},r^2 \underline{\eta}- r^2\underline{\eta}_{\mathrm{Kerr}},  r^2\hat{\chi}, r \underline{\hat{\chi}}, \underline{\omega} r\right) \|^2_{S^2_{\bar{u},v_{\infty}}} \nonumber \\
+& \sup_{{u} \in \left[u_{-1},u_f\right]} \| (r \slashed{\nabla})^{N+1} \left(r^{5/2} T, r^2 \underline{T}\right) \|^2_{S^2_{{u},v_{\infty}}} + \sum_{i=0}^N \sup_{{u} \in \left[u_{-1},u_f\right]}\|(r \slashed{\nabla})^{i} (r^{3} T) \|^2_{S^2_{{u},v_{\infty}}} \, ,
\end{align}
\begin{align}
\slashed{\mathbb{E}}^N_{v_\infty} \left[ \mathcal{R} \right] := &\sum_{\mathcal{R}_p \setminus \{\Omega^2 \alpha, \Omega^{-2}\underline{\alpha}\}}  \sum_{s=0}^2 \sum_{k=0}^{N-1-s}    \sup_{u \in \left[u_{-1},u_f\right]} u^{\min(\frac{1}{2}+s,2)} \cdot \tilde{w} (\mathcal{R}_p) \cdot \|(r\slashed{\nabla})^{k} \left(r^p  \mathcal{R}_p -r^p (\mathcal{R}_p)_{\mathrm{Kerr}}\right) \|^2_{S^2_{u,v_{\infty}}} \nonumber \\
& \qquad \qquad \ \  \ \ \ \ \  +\sum_{s=0}^2  \sum_{k=0}^{N-1-s}    \sup_{u \in \left[u_{-1},u_f\right]} u^{\frac{s}{2}}  \, \|(r\slashed{\nabla})^{k} \left(r^4 \Omega \beta  -r^4 (\Omega\beta)_{\mathrm{Kerr}}\right) \|^2_{S^2_{u,v_{\infty}}} \nonumber \\
+&\sum_{\mathcal{R}_p \setminus \{\Omega^2 \alpha, \Omega^{-2}\underline{\alpha}\}}  \sum_{s=0}^2 \sum_{k=0}^{N-s}   \sup_{u \in \left[u_{-1},u_f\right]} u^s \int_{u}^{u_f} d\bar{u} \cdot  w_s\left(\mathcal{R}_p\right) \cdot \|  (r\slashed{\nabla})^{k}  \left(r^p  \mathcal{R}_p -r^p (\mathcal{R}_p)_{\mathrm{Kerr}}\right) \|^2_{S^2_{\bar{u},v_{\infty}}} \, . \nonumber
\end{align}

\subsubsection{Energies on $C_{u_{-1}}^{\I}$}
We define the (angular) energies\index{energies!energies for $\Phi^{\I}$!$\slashed{\mathbb{E}}^N_{u_{-1}} \left[ \Gamma \right]$, angular energy for $\Gamma$ on cones $C_{u_{-1}}^{\I}$}\index{energies!energies for $\Phi^{\I}$!$\slashed{\mathbb{E}}^N_{u_{-1}} \left[ \mathcal{R} \right]$, angular energy for $\mathcal{R}$ on cones $C_{u_{-1}}^{\I}$} for the Ricci coefficients and curvature on $C_{u_{-1}}^{\I}$:
\begin{align}
\slashed{\mathbb{E}}^N_{u_{-1}} \left[ \Gamma \right] := &\sum_{\Gamma_p \setminus \{\Omega^{-2} (\Omega\hat{\underline{\omega}}-(\Omega\hat{\underline{\omega}}_\circ))\}} \sum_{k=0}^N  \sup_{v\in \left[v(u_{-1},R_{-2}),v_\infty\right]}  \| (r\slashed{\nabla})^k \left(r^p  \Gamma_p -r^p (\Gamma_p)_{\mathrm{Kerr}}\right) \|^2_{S^2_{u_{-1},v}} \nonumber \\
+&\int_{v(u_{-1},R_{-2})}^{v_\infty} d\bar{v} \frac{1}{r^2} \Big\|  (r \slashed{\nabla})^{N+1} \left( r^2\eta- r^2 \eta_{\mathrm{Kerr}},r^2 \underline{\eta}- r^2\underline{\eta}_{\mathrm{Kerr}},  r^2\hat{\chi}, r \underline{\hat{\chi}}, \omega r^{3}\right) \Big\|^2_{S^2_{u_{-1},\bar{v}}}  \nonumber \\
+& \sum_{k=0}^N \sup_{v\in \left[v(u_{-1},R_{-2}),v_\infty\right]}  
 \| (r \slashed{\nabla})^{k} \left(r \slashed{\nabla} r^{5/2} T, r \slashed{\nabla} r^2 \underline{T}, r^3 T, r^3 (\omega -\omega_\circ)\right) \|^2_{S^2_{u_{-1},v}} \, , 
\end{align}
\begin{align}
\slashed{\mathbb{E}}^N_{u_{-1}} \left[ \mathcal{R} \right] := &\sum_{\mathcal{R}_p \setminus \{\Omega^2 \alpha, \Omega^{-2}\underline{\alpha}\}}  \sum_{k=0}^{N-1}  \sup_{v\in \left[v(u_{-1},R_{-2}),v_\infty\right]}  \| (r \slashed{\nabla})]^k \left(r^p  \mathcal{R}_p -r^p (\mathcal{R}_p)_{\mathrm{Kerr}}\right) \|^2_{S^2_{u_{-1},v}} \nonumber \\
+&\sum_{\mathcal{R}_p \setminus \{\Omega^2 \alpha, \Omega^{-2}\underline{\alpha}\}} \sum_{k=0}^N \int_{v(u_{-1},R_{-2})}^{v_\infty} d\bar{v} \frac{1}{r^2} \|  (r\slashed{\nabla})^k \left(r^p  \mathcal{R}_p -r^p (\mathcal{R}_p)_{\mathrm{Kerr}}\right) \|^2_{S^2_{u_{-1},v}} \, . \nonumber
\end{align}

\subsubsection{Energies in $\mathcal{D}^{\I}$}
\label{Iplusenergiesmastersec}

We define:\index{energies!energies for $\Phi^{\I}$!$\mathbb{E}^N_{\mathcal{D}^{\I}} \left[ \Gamma \right]$, energy for $\Gamma$ in $\mathcal{D}^{\I}$}
\begin{align} \label{egammafull}
\mathbb{E}^N_{\mathcal{D}^{\I}} \left[ \Gamma \right] := &\sum_{\Gamma_p} \sum_{s=0}^2 \sum_{\substack{
	\vert \underline{k} \vert \leq N-s
	\\
	\underline{k}_{\underline{\omega}} \neq (0,N,0) 
	\\
	\underline{k}_{{\omega}} \neq (0,0,N)}} \sup_{\mathcal{D}^{\I}} u^s \cdot w_s(\Gamma_p) \cdot  \| \mathfrak{D}^{\underline{k}} \left(r^p  \Gamma_p -r^p (\Gamma_p)_{\mathrm{Kerr}}\right) \|^2_{S^2_{u,v}} +\sum_{|\underline{k} | \leq N} \sup_{\mathcal{D}^{\I}} \| \mathfrak{D}^{\underline{k}} (r^3 T) \|^2_{S_{u,v}} \nonumber \\
	+& \sup_{\mathcal{D}^{\I}} \|  \left[ r \slashed{\nabla}\right]^{N+1}\left(r^2 T, r^2 \underline{T}, \sqrt{r} b \right) \|^2_{S^2_{u,v}} + \sup_{\mathcal{D}^{\I}} u^{4-2\delta} \| r^2 T_{\ell=0}, r^2 \underline{T}_{\ell=0}, r(\Omega^2-\Omega_\circ^2)_{\ell=0} \|^2_{S^2_{u,v}}\nonumber \\
	+&\sum_{\Gamma_p} \sum_{s=0}^1 \sum_{ |\underline{k}| \leq N-s} \sup_{u \in \left[u_{-1},u_f\right]} u^s \cdot \int_{\mathcal{D}^{\I}(u)}  \frac{d\bar{u}dv}{r^{1+\delta}} \tilde{w}_s\left({\Gamma}_p\right)  \| \mathfrak{D}^{\underline{k}} \left(r^p  {\Gamma}_p -r^p ({\Gamma}_p)_{\mathrm{Kerr}}\right) \|^2_{S^2_{\bar{u},v}} \nonumber \\
	+& \int_{\mathcal{D}^{\I}} dudv  \frac{1}{r^{1+\delta}} \|  (r \slashed{\nabla})^{N+1} \left(r^{2} T, r^2 \hat{\chi}, r^2(\underline{\eta} - \underline{\eta}_{\mathrm{Kerr}}), r^2({\eta} - {\eta}_{\mathrm{Kerr}})  , r(\underline{\omega} - \underline{\omega}_\circ)
	 \right) \|^2_{S^2_{u,v}} \, ,   
\end{align}
where the restriction $\underline{k}_{\underline{\omega}} \neq (0,N,0)$ indicates that the term $[\Omega \slashed{\nabla}_3]^N (r \underline{\omega} - r \underline{\omega}_\circ)$ is excluded from the sum and similarly $\underline{k}_{{\omega}} \neq (0,0,N)$ indicates that the term $[\Omega \slashed{\nabla}_4]^N (r^{\frac{5}{2}} {\omega} - r^{\frac{5}{2}}{\omega}_\circ)$ is excluded,\index{energies!energies for $\Phi^{\I}$!$\mathbb{E}^N_{\mathcal{D}^{\I}} \left[ \mathcal{R} \right]$, energy for $\mathcal{R}$ in $\mathcal{D}^{\I}$}
\begin{align} \label{eRfull}
\mathbb{E}^N_{\mathcal{D}^{\I}} \left[ \mathcal{R} \right] :=	
	&\sum_{\mathcal{R}_p \setminus \{\Omega^2 \alpha, \Omega^{-2}\underline{\alpha}\}}  \sum_{s=0}^2 \sum_{ |\underline{k}| \leq N-1-s}   \sup_{u \in \mathcal{D}^{\I}} u^{\min(\frac{1}{2}+s,2)} \cdot \tilde{w}\left(\mathcal{R}_p\right) \cdot \| \mathfrak{D}^{\underline{k}} \left(r^p  \mathcal{R}_p -r^p (\mathcal{R}_p)_{\mathrm{Kerr}}\right) \|^2_{S^2_{u,v}} 
	\nonumber \\
& \qquad \qquad \ \  \ \ +\sum_{s=0}^2  \sum_{|\underline{k}| \leq N-1-s}    \sup_{u \in \left[u_{-1},u_f\right]} u^{\frac{s}{2}}  \, \| \mathfrak{D}^{\underline{k}} \left(r^4 \Omega \beta  -r^4 (\Omega\beta)_{\mathrm{Kerr}}\right) \|^2_{S^2_{u,v}} \nonumber \\
+&\sum_{\mathcal{R}_p \setminus \{\Omega^2 \alpha, \Omega^{-2}\underline{\alpha}\}}  \sum_{s=0}^1  \sum_{ |\underline{k}| \leq N-s}\sup_{u \in \left[u_{-1},u_f\right]} u^s \cdot \int_{\mathcal{D}^{\I}(u)}  \frac{d\bar{u}dv}{r^{1+\delta}} w_s\left(\mathcal{R}_p\right)  \|  \mathfrak{D}^{\underline{k}} \left(r^p  \mathcal{R}_p -r^p (\mathcal{R}_p)_{\mathrm{Kerr}}\right) \|^2_{S^2_{\bar{u},v}}   \nonumber \\
+&\sum_{\mathcal{R}_p \setminus \{\Omega^2 \alpha, \Omega^{-2}\underline{\alpha}\}}   \sum_{s=0}^2 \sum_{n=0}^{N-s}   \sup_{u \in \left[u_{-1},u_f\right]} u^s \int_{\underline{C}_v(u)} d\bar{u} \cdot  w_s\left(\mathcal{R}_p\right) \|  (r \slashed{\nabla})^n \left(r^p  \mathcal{R}_p -r^p (\mathcal{R}_p)_{\mathrm{Kerr}}\right) \|^2_{S^2_{\bar{u},v}} \nonumber \\
+&\sum_{\mathcal{R}_p \setminus \{\Omega^2 \alpha, \Omega^{-2}\underline{\alpha}\}} \sum_{s=0}^2  \sum_{n=0}^{N-s} \sup_{u \in \left[u_{-1},u_f\right]} u^s \cdot  \int_{C_u} d\bar{v} \frac{1}{r^2} \|(r \slashed{\nabla})^n \left(r^p  \mathcal{R}_p -r^p (\mathcal{R}_p)_{\mathrm{Kerr}}\right) \|^2_{S^2_{u,\bar{v}}} \nonumber \,  \\
+&  \sup_{\mathcal{D}^{\I}} u^{4-2\delta} \| r^3 \rho_{\ell=0}+2M_f\|^2_{S^2_{u,v}} + \sup_v u^{4-2\delta} \| r^5 (\slashed{div} \Omega\beta)_{\ell=1} \|^2_{S^2_{u_f,v}} \, .
\end{align}
We also define the energies\index{energies!energies for $\Phi^{\I}$!$\slashed{\mathbb{E}}^N_{\mathcal{D}^{\I}} \left[ \Gamma \right]$, angular energy for $\Gamma$ in $\mathcal{D}^{\I}$}\index{energies!energies for $\Phi^{\I}$!$\slashed{\mathbb{E}}^N_{\mathcal{D}^{\I}} \left[ \mathcal{R} \right]$, angular energy for $\mathcal{R}$ in $\mathcal{D}^{\I}$} $\slashed{\mathbb{E}}^N_{\mathcal{D}^{\I}} \left[ \Gamma \right]$ and $\slashed{\mathbb{E}}^N_{\mathcal{D}^{\I}} \left[ \mathcal{R} \right] $ to be as above but with the multi-indices $\underline{k}$ restricted to those of the form $\underline{k}=(n,0,0)$, i.e.~only angular derivatives are employed.

We finally define the angular master energy\index{energies!energies for $\Phi^{\I}$!$\slashed{\mathbb{E}}^N_{u_f,\I}$, master angular energy in $\mathcal{D}^{\I}$}
\begin{align} \label{angmasterenergy}
\slashed{\mathbb{E}}^N_{u_f,\I}  := & \slashed{\mathbb{E}}^N_{u_{-1}} \left[ \Gamma \right] + \slashed{\mathbb{E}}^N_{u_{-1}} \left[ \mathcal{R} \right]+  \slashed{\mathbb{E}}^N_{v_{\infty}} \left[ \Gamma \right] + \slashed{\mathbb{E}}^N_{v_\infty} \left[ \mathcal{R} \right] + \slashed{\mathbb{E}}^N_{\mathcal{D}^{\I}} \left[ \Gamma \right] +  \slashed{\mathbb{E}}^N_{\mathcal{D}^{\I}} \left[ \mathcal{R} \right] \\
+& \sup_{\mathcal{D}^{\I}}
		\sum_{s=0}^2 u^s \sum_{k=0}^{N+1-s} \|    \left[ r \slashed{\nabla}\right]^{n}\left(r (\slashed{g} - r^2\gamma) \right) \|^2_{S^2_{u,v}} + \sup_{\mathcal{D}^{\I}}
		\sum_{s=0}^2 u^s \sum_{k=0}^{N+2-s} \sum_{m=-1}^1
		\Vert (r\nablaslash)^{k} (Y^1_m - \mathring{Y}^1_m)^{\I} \Vert_{S_{u,v}^{\I}}^2\,  \nonumber
\end{align}
and the (total) master energy\index{energies!energies for $\Phi^{\I}$!$\mathbb{E}^N_{u_f,\I}$, total master energy in $\mathcal{D}^{\I}$}
\begin{align}
\label{themasterweuseIplus}
\mathbb{E}^N_{u_f,\I}  := & \slashed{\mathbb{E}}^N_{u_f,\I}+ \mathbb{E}^N_{\mathcal{D}^{\I}} \left[ \Gamma \right] + \mathbb{E}^N_{\mathcal{D}^{\I}} \left[ \mathcal{R} \right] 
 + \sup_{\mathcal{D}^{\I}}  \sum_{s=0}^2 u^s \sum_{|\underline{k}| \leq N-s}  \|  \mathfrak{D}^{\underline{k}} \left(r (\slashed{g} - r^2\gamma) \right) \|^2_{S^2_{u,v}} \, .
 \end{align}

\subsubsection{An auxiliary energy for $\omega-\omega_\circ$ and $\underline{\omega}-\underline{\omega}_\circ$ }

We finally define two auxiliary energies on $\omega-\omega_\circ$ and $\underline{\omega}-\underline{\omega}_\circ$ respectively. The motivation is that controlling (1) these auxiliary energies (2) the \emph{angular} master energy $\slashed{\mathbb{E}}^N_{u_f,\I}$ and (3) the energy  $\mathbb{E}^{N}_{u_f}[\alpha_{\Hp}, \alpha_{\I}]+\mathbb{E}^{N}_{u_f}[\alphabar_{\Hp}, \alphabar_{\I}]$ will control the full master energy ${\mathbb{E}}^N_{u_f,\I}$ by exploiting the form of the null structure and Bianchi equations.\footnote{The point is that $\omega$ and $\underline{\omega}$ are the only connection coefficients not satisfying an equation in both the $3$ and the $4$-direction and $\alpha$ and $\underline{\alpha}$ are the only curvature components not satisfying an equation in both the $3$ and the $4$-direction.} 
 See already Theorem~\ref{theo:reducetoangular}.\index{energies!energies for $\Phi^{\I}$!$\mathbb{E}^{N,aux}_{u_f, \mathcal{I}} \left[ \underline{\omega} \right]$, energy for $\underline{\omega}-\underline{\omega}_\circ$}\index{energies!energies for $\Phi^{\I}$!$\mathbb{E}^{N,aux}_{u_f, \mathcal{I}} \left[{\omega} \right]$, energy for $\omega-\omega_\circ$}
 The definitions are as follows:
\begin{align} 
\mathbb{E}^{N,aux}_{u_f, \mathcal{I}} \left[ \underline{\omega} \right] =&\sum_{s=0}^1 \sup_{u \in \left[u_{-1},u_f\right]} \Big\{  u^s \sum_{\substack{k_1+k_3 \leq N-s \\ k_1+k_3 \leq N+1-s \, ,\, k_1\geq 2}} \int_{\mathcal{D}^{\I}(u)} \frac{1}{r^{1+\delta}} \|(\Omega \slashed{\nabla}_3)^{k_3}  (r \slashed{\nabla})^{k_1} (r (\underline{\omega}-\underline{\omega}_\circ)) \|_{S^2_{\bar{u},\bar{v}}}^2 \Big\} \nonumber \\
+&\sum_{s=0}^2 \sup_{\mathcal{D}^{\I}} \Big\{ u^s \sum_{\substack{k_1+k_3 \leq N-s \\ k_1 \neq N-s}} \|(\Omega \slashed{\nabla}_3)^{k_1}  (r \slashed{\nabla})^{k_3} (r (\underline{\omega}-\underline{\omega}_\circ)) \|_{S^2_{{u},{v}}}^2 \Big\} \nonumber \\
+& \sum_{s=0}^2 u^s \sum_{\substack{ k_1+k_3 \leq N-s \\ k_1+k_3 \leq N+1-s \, , \, k_1 \geq 1}} \int_{u}^{u_f} d\bar{u} \|(\Omega \slashed{\nabla}_3)^{k_3}  (r \slashed{\nabla})^{k_1} (r (\underline{\omega}-\underline{\omega}_\circ)) \|_{S^2_{\bar{u}, v_\infty}}^2   \, , \label{angbon}
\end{align}
\begin{align}
\mathbb{E}^{N,aux}_{u_f, \mathcal{I}} \left[{\omega} \right]  =&\sum_{s=0}^1 \sup_{u \in \left[u_{-1},u_f\right]} \Big\{ u^s \sum_{k_4+k_1 \leq N-s} \int_{\mathcal{D}^{\I}(u)} \frac{1}{r^{1+\delta+s}} \| (r \Omega \slashed{\nabla}_4)^{k_4} (r \slashed{\nabla})^{k_1} (r^{5/2} (\omega-\omega_\circ) \|_{S^2_{\bar{u},\bar{v}}}^2 \Big\}  \,  \nonumber
\nonumber \\
+&\sum_{s=0}^2 \sup_{\mathcal{D}^{\I}} \Big\{ u^s \sum_{\substack {k_4+k_1 \leq N-s \\ k_4 \neq N-s}} \| (r \Omega \slashed{\nabla}_4)^{k_4}  (r \slashed{\nabla})^{k_3} (r^{3-s/2} (\omega-\omega_\circ)) \|_{S^2_{{u},{v}}}^2 \Big\} \nonumber \\
+& \int_{v(u_{-1},R_{-2})} d\bar{v} \frac{1}{r^{2}} \| \left[\Omega \slashed{\nabla}_4\right]^{N} (r^{3} (\omega-\omega_\circ)) \|_{S^2_{u_{-1},\bar{v}}}^2 \, .
\end{align}

\subsection{Energies of $\Phi^{\Hp}$}
\label{PhiHpenergysec}

Recall the $\Hp$ linearised Kerr solution of Definition \ref{assocKerparHplus}.  Define\index{schematic notation!$\mathcal{A}_{\mathcal{R}}$, differences with linearised Kerr in $\mathcal{H}^+$ gauge}\index{schematic notation!$\mathcal{A}_{\Gamma}$, differences with linearised Kerr in $\mathcal{H}^+$ gauge}
\[
	\mathcal{A}_{\mathcal{R}}
	=
	\big\{
	(\Omega\beta - \Omega\beta_{\rm Kerr})^{\Hp},
	(\Omega^{-1}\betabar - \Omega^{-1}\betabar_{\rm Kerr})^{\Hp},
	(\rho - \rho_{\circ})^{\Hp},
	(\sigma - \sigma_{\rm Kerr})^{\Hp}
	\big\},
\]
\begin{multline*}
	\mathcal{A}_{\Gamma}
	=
	\big\{
	(\Omega_{\circ}^{-2} \Omega^2 - 1)^{\Hp},
	\Omega^{-1} \hat{\chibar}^{\Hp},
	\Omega^{-2} (\Omega \tr \chibar - \Omega \tr \chibar_{\circ})^{\Hp},
	(\eta - \eta_{\rm Kerr})^{\Hp},
	(\etabar - \etabar_{\rm Kerr})^{\Hp},
	\\
	(\Omega \omegahat - \Omega \omegahat_{\circ})^{\Hp},
	\Omega^{-2} (\Omega \omegabarhat - \Omega \omegabarhat_{\circ})^{\Hp}
	\big\}.
\end{multline*}
Elements of $\mathcal{A}_{\mathcal{R}}$ and $\mathcal{A}_{\Gamma}$ are denoted $\breve{\mathcal{R}}$ and $\breve{\Gamma}$ respectively, with a $\breve{}$ added to emphasise the fact that the linearised Kerr values have been subtracted.  The Ricci coefficients $\Omega\hat{\chi}^{\Hp}$ and $(\Omega \tr \chi - \Omega \tr \chi_{\circ})^{\Hp}$ satisfy weaker estimates than the remaining Ricci coefficients.  Accordingly, define\index{schematic notation!$\mathcal{A}_{\chi}$, differences in $\mathcal{H}^+$ gauge}
\[
	\mathcal{A}_{\chi}
	=
	\{
	\Omega\hat{\chi}^{\Hp}, (\Omega \tr \chi - \Omega \tr \chi_{\circ})^{\Hp}
	\}.
\]
Define the spacetime energy\index{energies!energies for $\Phi^{\Hp}$!$\mathbb{E}^N[\DRH]$, spacetime energy 
in $\DRH$}
\begin{multline*}
	\mathbb{E}^N[\DRH]
	:=
	\sup_{\substack{
	v_{-1} \leq v \leq v(R_{2},u_f)}}
	\sum_{s=0,1,2}
	v^{s}
	\bigg(
	\sum_{\vert \gamma \vert \leq N-s}
	\sum_{\breve{\mathcal{R}} \in \mathcal{A}_{\mathcal{R}}}
	\Vert (1-3M_f/r) \mathfrak{D}^{\gamma} \breve{\mathcal{R}} \Vert^2_{\DRH(v)}
	\\
	+
	\sum_{\vert \gamma \vert \leq N-1-s}
	\sum_{\breve{\mathcal{R}} \in \mathcal{A}_{\mathcal{R}}}
	\Vert \mathfrak{D}^{\gamma} \breve{\mathcal{R}} \Vert^2_{\DRH(v)}
	+
	\sum_{\vert \gamma \vert \leq N-s}
	\sum_{\breve{\Gamma} \in \mathcal{A}_{\Gamma}}
	\Vert \mathfrak{D}^{\gamma} \breve{\Gamma} \Vert^2_{\DRH(v)}
	\bigg),
\end{multline*}
and the null cone energies\index{energies!energies for $\Phi^{\Hp}$!$\mathbb{E}^N[C^{\Hp}]$, null cone energy 
in $C^{\Hp}$}\index{energies!energies for $\Phi^{\Hp}$!$\mathbb{E}^N[\Cbar^{\Hp}]$, null cone energy 
in $\Cbar^{\Hp}$}
\begin{align*}
	\mathbb{E}^N[C^{\Hp}]
	&
	:=
	\sup_{\substack{
	v_{-1} \leq v \leq v(R_2,u_f)}}
	\sum_{s=0,1,2}
	\sum_{\vert \gamma \vert \leq N-s}
	v^{s}
	\sup_{u_{-1} \leq u \leq u_f}
	\sum_{\breve{\Phi} \in \mathcal{A}_{\mathcal{R}} \cup \mathcal{A}_{\Gamma}}
	\Vert \mathfrak{D}^{\gamma} \breve{\Phi} \Vert_{C^{\Hp}_u(v)}^2,
\\
	\mathbb{E}^N[\Cbar^{\Hp}]
	&
	:=
	\sup_{\substack{
	v_{-1} \leq v \leq v(R_2,u_f)}}
	\sum_{s=0,1,2}
	\sum_{\vert \gamma \vert \leq N-s}
	v^{s}
	\sum_{\breve{\Phi} \in \mathcal{A}_{\mathcal{R}} \cup \mathcal{A}_{\Gamma}}
	\Vert \mathfrak{D}^{\gamma} \breve{\Phi} \Vert_{\Cbar^{\Hp}_v}^2,
\end{align*}
along with the energy on spheres\index{energies!energies for $\Phi^{\Hp}$!$\mathbb{E}^N[S^{\Hp}]$, energy on $S^{\Hp}$ spheres}
\[
	\mathbb{E}^N[S^{\Hp}]
	:=
	\sup_{\DRH}
	\sum_{s=0,1,2}
	v^s
	\Big(
	\sum_{\vert \gamma \vert \leq N-1-s}
	\sum_{\breve{\mathcal{R}} \in \mathcal{A}_{\mathcal{R}}}
	\Vert \mathfrak{D}^{\gamma} \breve{\mathcal{R}} \Vert_{S_{u,v}^{\Hp}}^2
	+
	\sum_{\vert \gamma \vert \leq N-s}
	\sum_{\breve{\Gamma} \in \mathcal{A}_{\Gamma}}
	\Vert \mathfrak{D}^{\gamma} \breve{\Gamma} \Vert_{S_{u,v}^{\Hp}}^2
	\Big),
\]
and the energy of $N+1$ derivatives of certain Ricci coefficients on the final hypersurface\index{energies!energies for $\Phi^{\Hp}$!$\mathbb{E}^{N+1}[C_{u_f}^{\Hp}]$, top-order null cone energy on final cone $C_{u_f}^{\Hp}$}
\[
	\mathbb{E}^{N+1}[C_{u_f}^{\Hp}]
	:=
	\sup_{v_{-1} \leq v \leq v(R_2,u)}
	\sum_{s=0,1,2}
	v^{s}
	\sum_{k=0}^{N+1-s}
	\Vert  
	(r\nablaslash)^{k}
	(\Omega \hat{\chi}, \Omega \tr \chi - (\Omega \tr \chi)_{\circ}, \eta - \eta_{\rm Kerr})^{\Hp}
	\Vert_{C^{\Hp}_{u_f}(v)}^2
	.
\]
For the metric components, $\gslash$, $b$ and the mode difference, $Y^1_m - \mathring{Y}^1_m$, define\index{energies!energies for $\Phi^{\Hp}$!$\mathbb{E}^N[g^{\Hp}]$, energy for metric components}
	\begin{align*}
		\mathbb{E}^N[g^{\Hp}]
		:=
		&
		\sup_{\substack{
		u_{0} \leq u \leq u_f
		\\
		v_{-1} \leq v \leq v(R_2,u)
		}}
		\Big(
		v
		\sum_{\vert \gamma \vert \leq N}
		\Vert \mathfrak{D}^{\gamma} (\gslash - r^2 \gamma) \Vert_{S_{u,v}^{\Hp}}^2
		+
		v
		\sum_{k=0}^{N+1} \sum_{m=-1}^1
		\Vert (r\nablaslash)^{k} (Y^1_m - \mathring{Y}^1_m) \Vert_{S_{u,v}^{\Hp}}^2
		\\
		&
		\qquad \qquad \qquad \qquad
		+
		\sum_{s=0,1,2}
		v^s
		\sum_{\vert \gamma \vert \leq N-s}
		\big(
		\Vert \Omega^{-2} \mathfrak{D}^{\gamma} (b - b_{\mathrm{Kerr}}) \Vert_{S_{u,v}}^2
		+
		\Vert \mathfrak{D}^{\gamma} (b - b_{\mathrm{Kerr}}) \Vert_{\Cbar_v}^2
		\\
		&
		\qquad \qquad\qquad \qquad \qquad \qquad
		+
		\Vert \mathfrak{D}^{\gamma} (b - b_{\mathrm{Kerr}}) \Vert_{C_u(v)}^2
		+
		\Vert \mathfrak{D}^{\gamma} (b - b_{\mathrm{Kerr}}) \Vert_{\DRH(v)}^2
		\big)
		\Big)
		.
	\end{align*}
For the quantities $\Omega \hat{\chi}$ and $\Omega \tr \chi - \Omega \tr \chi_{\circ}$ recall that, for $\kbar = (k_1,k_2,k_3)$, $\mathfrak{D}^{\kbar} = (r\nablaslash)^{k_1} (\Omega^{-1} \nablaslash_3)^{k_2} (r\Omega \nablaslash_4)^{k_3}$ and define\index{energies!energies for $\Phi^{\Hp}$!$\mathbb{E}^N[\chi^{\Hp}]$, energy for $\chi^{\Hp}$}
\begin{align*}
	\mathbb{E}^N[\chi^{\Hp}]
	:=
	&
	\sum_{\Gamma \in \mathcal{A}_{\chi}}
	\Big[
	\sup_{\DRH} v^{-\delta}
	\Vert (r\nablaslash)^{N}
	\Gamma
	\Vert_{S^{\Hp}_{u,v}}^2
	+
	\sum_{s=0,1,2}
	\sum_{\substack{ \vert \kbar \vert \leq N-s \\ k_1 \neq N-s}}
	\Big(
	\sup_{\DRH} v^s
	\Vert \mathfrak{D}^{\kbar}
	\Gamma
	\Vert_{S^{\Hp}_{u,v}}^2
	\\
	&
	+
	\sup_{\substack{
	v_{-1} \leq v \leq v(R_2,u_f)}}
	v^{s}
	\big(
	\sup_{u_{-1} \leq u \leq u_f}
	\left\Vert \mathfrak{D}^{\kbar} \Gamma \right\Vert_{C^{\Hp}_u(v)}^2
	+
	\left\Vert \mathfrak{D}^{\kbar} \Gamma \right\Vert_{\Cbar^{\Hp}_v}^2
	+
	\left\Vert \mathfrak{D}^{\kbar} \Gamma \right\Vert^2_{\DRH(v)}
	\big)
	\Big)
	\Big].
\end{align*}

Finally, define\index{energies!energies for $\Phi^{\Hp}$!$\mathbb{E}^N_{u_f,\Hp}$, master energy}
\begin{align}
\nonumber
	\mathbb{E}^N_{u_f,\Hp}
	:=
	\
	&
	\mathbb{E}^N[\DRH]
	+
	\mathbb{E}^N[C^{\Hp}]
	+
	\mathbb{E}^N[\Cbar^{\Hp}]
	+
	\mathbb{E}^N[S^{\Hp}]
	+
	\mathbb{E}^N[\chi^{\Hp}]
	+
	\mathbb{E}^{N+1}[C_{u_f}^{\Hp}]
	+
	\mathbb{E}^N[g^{\Hp}]
	\\
	&
	+
	(u_f)^4 \sum_{m=-1}^1 \vert J_{\Hp}^m - J_{\I}^m \vert^2
	.
	\label{Hplusmasterenergy}
\end{align}
where $J_{\Hp}^m$ are the associated Kerr angular momentum parameters of the $\Hp$ gauge (see Definition \ref{assocKerparHplus}).

\subsection{Pointwise norms of $\Phi^{\mathcal{I}^+}$ and $\Phi^{\mathcal{H}^+}$}
\label{subsec:pointwisenorms}

Define the pointwise norm of the geometric quantities in the $\I$ gauge\index{energies!pointwise norms!$\mathbb P_{u_f}^{N-5}[\Phi^{\mathcal{I}^+}]$, pointwise norm of geometric quantities in $\mathcal{I}^+$ gauge}
\begin{align}
	\mathbb P_{u_f}^{N-5}[\Phi^{\mathcal{I}^+}]
	=
	\
	&
	\sup_{\DRI}
	\sum_{\vert \gamma \vert \leq N-5}
	\Big(
	\vert r^{\frac{9}{2}} \mathfrak{D}^{\gamma} \alpha \vert
	+
	\vert u^{\frac{1}{2}} r^4 \mathfrak{D}^{\gamma} \beta \vert
	+
	\vert r^3 \mathfrak{D}^{\gamma} (\Omega \tr \chi - \Omega \tr \chi_{\circ}) \vert
	+
	\vert r^3 \mathfrak{D}^{\gamma} (\Omega \omegahat - \Omega \omegahat_{\circ}) \vert
	\nonumber
	\\
	&
	+
	u \big(
	\vert r^{\frac{7}{2}} \mathfrak{D}^{\gamma} \alpha \vert
	+
	\vert r \mathfrak{D}^{\gamma} \alphabar \vert
	+
	\vert r^{\frac{7}{2}} \mathfrak{D}^{\gamma} \beta \vert
	+
	\vert r^2 \mathfrak{D}^{\gamma} \betabar \vert
	+
	\vert r^3 \mathfrak{D}^{\gamma} ( \rho - \rho_{\circ}) \vert
	+
	\vert r^3 \mathfrak{D}^{\gamma} \sigma \vert
	\label{eq:Ippointwisenorm}
	\\
	&
	\quad
	+
	\vert r \mathfrak{D}^{\gamma} (\Omega_{\circ}^{-2} \Omega^2 - 1) \vert
	+
	\vert r^2 \mathfrak{D}^{\gamma} (\Omega \tr \chi - \Omega \tr \chi_{\circ}) \vert
	+
	\vert r^2 \mathfrak{D}^{\gamma} (\Omega \tr \chibar - \Omega \tr \chibar_{\circ}) \vert
	+
	\vert r^2 \mathfrak{D}^{\gamma} \hat{\chi} \vert
	+
	\vert r \mathfrak{D}^{\gamma} \hat{\chibar} \vert
	\nonumber
	\\
	&
	\quad
	+
	\vert r^2 \mathfrak{D}^{\gamma} \eta \vert
	+
	\vert r^2 \mathfrak{D}^{\gamma} \etabar \vert
	+
	\vert r^{2} \mathfrak{D}^{\gamma} (\Omega \omegahat - \Omega \omegahat_{\circ}) \vert
	+
	\vert r \mathfrak{D}^{\gamma} (\Omega \omegabarhat - \Omega \omegabarhat_{\circ}) \vert
	+
	\vert r \mathfrak{D}^{\gamma} b \vert
	+
	\vert r \mathfrak{D}^{\gamma} (\gslash - r^2 \gamma) \vert
	\big)
	\Big)
	,
	\nonumber
\end{align}
and of the geometric quantities in the $\Hp$ gauge\index{energies!pointwise norms!$\mathbb P_{u_f}^{N-5}[\Phi^{\mathcal{H}^+}]$, pointwise norm of geometric quantities in $\mathcal{H}^+$ gauge}
\begin{align}
	\mathbb P_{u_f}^{N-5}[\Phi^{\mathcal{H}^+}]
	=
	\
	&
	\sup_{\DRH}
	v
	\sum_{\vert \gamma \vert \leq N-5}
	\big(
	\vert \mathfrak{D}^{\gamma} \Omega^2 \alpha \vert
	+
	\vert \mathfrak{D}^{\gamma} \Omega^{-2} \alphabar \vert
	+
	\vert \mathfrak{D}^{\gamma} \Omega\beta \vert
	+
	\vert \mathfrak{D}^{\gamma} \Omega^{-1}\betabar \vert
	+
	\vert \mathfrak{D}^{\gamma} ( \rho - \rho_{\circ}) \vert
	+
	\vert \mathfrak{D}^{\gamma} \sigma \vert
	\nonumber
	\\
	&
	+
	\vert \mathfrak{D}^{\gamma} (\Omega_{\circ}^{-2} \Omega^2 - 1) \vert
	+
	\vert \mathfrak{D}^{\gamma} (\Omega \tr \chi - \Omega \tr \chi_{\circ}) \vert
	+
	\vert \mathfrak{D}^{\gamma} \Omega^{-2} (\Omega \tr \chibar - \Omega \tr \chibar_{\circ}) \vert
	+
	\vert \mathfrak{D}^{\gamma} \eta \vert
	+
	\vert \mathfrak{D}^{\gamma} \etabar \vert
	\label{eq:Hppointwisenorm}
	\\
	&
	+
	\vert \mathfrak{D}^{\gamma} \Omega \hat{\chi} \vert
	+
	\vert \mathfrak{D}^{\gamma} \Omega^{-1} \hat{\chibar} \vert
	+
	\vert \mathfrak{D}^{\gamma} (\Omega \omegahat - \Omega \omegahat_{\circ}) \vert
	+
	\vert \mathfrak{D}^{\gamma} \Omega^{-2} (\Omega \omegabarhat - \Omega \omegabarhat_{\circ}) \vert
	+
	\vert \mathfrak{D}^{\gamma} \Omega^{-2} b \vert
	+
	\vert \mathfrak{D}^{\gamma} (\gslash - r^2 \gamma) \vert
	\big)
	.
	\nonumber
\end{align}

\subsection{Energies of diffeomorphisms}
\label{energiesofdiffeossec}

Define the energy of the $f_{\Hp,\I}$ diffeomorphisms\index{energies!energies of diffeomorphisms!$\mathbb{E}_{u_f}^N[f_{\Hp,\I}]$, energy associated to $f_{\Hp,\I}$ diffeomorphisms}
\begin{align}
	&
	\mathbb{E}_{u_f}^{N+2}[f_{\Hp,\I}]
	=
	\sup_{\substack{
	u_1 \leq u \leq u_f
	\\
	v(R_{-2},u) \leq v \leq v(R_{3},u)}
	}
	\bigg(
	\sum_{k \leq N-5}
	u^2
	\Big[
	\sum_{\substack{\vert \gamma \vert \leq 2 \\ k + \vert \gamma \vert \geq 1}}
	\Big(
	\Vert
	(r\nablaslash)^{k} \mathfrak{D}^{\gamma} f^3
	\, \mathds{1}
	\Vert_{S_{u,v}^{\I}}^2
	+
	\Vert
	(r\nablaslash)^{k} \mathfrak{D}^{\gamma} f^4
	\, \mathds{1}
	\Vert_{S_{u,v}^{\I}}^2
	\Big)
	\nonumber
	\\
	&
	\qquad \qquad
	+
	\sum_{\vert \gamma \vert \leq 1}
	\Big(
	\Vert
	(r\nablaslash)^{k} \mathfrak{D}^{\gamma} \partial_u \slashed{f}
	\, \mathds{1}
	\Vert_{S_{u,v}^{\I}}^2
	+
	\Vert
	(r\nablaslash)^{k} \mathfrak{D}^{\gamma} \partial_v \slashed{f}
	\, \mathds{1}
	\Vert_{S_{u,v}^{\I}}^2
	\Big)
	\Big]
	\label{EKfHpIpenergy}
	\\
	&
	\qquad \qquad
	+
	u^2
	\vert
	(f^{3} - f^{4})_{\ell =0}
	\, \mathds{1}
	\vert^2
	+
	u^2
	\Vert
	(f^{3} - f^{4})_{\ell =0}
	\Vert^2_{\mathcal{D}_{\Hp}^{\I}(u)}
	+
	u
	\vert
	(f^{3} + f^{4})_{\ell =0}
	\, \mathds{1}
	\vert^2
	\bigg)
	\nonumber
	\\
	&
	\qquad \qquad
	+
	\sum_{s=0}^2\sum_{k=0}^{N-s}
	(u_f)^s
	\Big[
	\sum_{\vert \gamma \vert \leq 2}
	\big(
	\Vert (r \nablaslash)^{k} \mathfrak{D}^{\gamma} f^4 \tilde{\mathds{1}} \Vert^2_{C_{u_f}^{\I}}
	+
	\Vert (r \nablaslash)^{k} \mathfrak{D}^{\gamma} f^3 \tilde{\mathds{1}} \Vert^2_{C_{u_f}^{\I}}
	\big)
	+
	\sum_{\vert \gamma \vert = 0,1}
	\Vert (r \nablaslash)^{k} \mathfrak{D}^{\gamma} \partial_u \slashed{f} \tilde{\mathds{1}} \Vert^2_{C_{u_f}^{\I}}
	\Big],
	\nonumber
\end{align}
where $f = f_{\Hp,\I}$, $\mathds{1} = \mathds{1}_{\DRH\cap \DRI}$, and $\tilde{\mathds{1}} = \mathds{1}_{R_{-1} \leq r \leq R_1}$ so that, for any tensor $\xi$,
\[
	\Vert \xi \mathds{1} \Vert_{C_{u_f}^{\I}}^2
	=
	\int_{v(R_{-1},u_f)}^{v(R_{1},u_f)} \int_{S^2}
	\vert \xi (u_f,v,\theta) \vert^2
	d\theta d v.
\]
Note that the higher order derivatives of $f_{\Hp,\I}$ in the final line of \eqref{EKfHpIpenergy} are only considered on the final hypersurface $u=u_f$.

Define the energy of the $f_{d,\I}$ diffeomorphisms as\index{main theorem!energies!$\mathbb{E}_{u_f}[f_{d,\I}]$, energy measuring diffeomorphisms}\index{energies!energies of diffeomorphisms!$\mathbb{E}_{u_f}[f_{d,\I}]$, energy associated to $f_{d,\I}$ diffeomorphisms}
\begin{align} \label{EfdIpenergy}
	&
	\mathbb{E}_{u_f}[f_{d,\I}]
	:=
	\sup_{\substack{
	u_{-1} \leq u \leq u_{2}
	\\
	v(R_{-2},u) \leq v < v_{\infty}
	}}
	\sum_{k \leq 2}
	\bigg[
	\sum_{\vert \gamma \vert \leq 2}
	\big(
	\Vert (r \nablaslash)^k \mathfrak{D}^{\gamma} f^3 \Vert_{S^{\I}_{u,v}}^2
	+
	r^{-2}
	\Vert (r \nablaslash)^k \mathfrak{D}^{\gamma} f^4 \Vert_{S^{\I}_{u,v}}^2
	\big)
	\\
	&
	+
	\sum_{\vert \gamma \vert \leq 1}
	\big(
	r^2 \Vert (r \nablaslash)^k \mathfrak{D}^{\gamma} r \nablaslash_4 f^3 \Vert_{S^{\I}_{u,v}}^2
	+
	\Vert (r \nablaslash)^k \mathfrak{D}^{\gamma} \nablaslash_3 f^4 \Vert_{S^{\I}_{u,v}}^2
	+
	\Vert (r \nablaslash)^k \mathfrak{D}^{\gamma} \partial_u \slashed{f} \Vert_{S^{\I}_{u,v}}^2
	+
	\Vert (r \nablaslash)^k \mathfrak{D}^{\gamma} \partial_v \slashed{f} \Vert_{S^{\I}_{u,v}}^2
	\big)
	\bigg],
	\nonumber
\end{align}
where $f = f_{d,\I}$.

Finally, define\index{main theorem!energies!$\mathbb{E}_{u_f}[f_{d,\Hp}]$, energy measuring diffeomorphisms}\index{energies!energies of diffeomorphisms!$\mathbb{E}_{u_f}[f_{d,\Hp}]$, energy associated to $f_{d,\Hp}$ diffeomorphisms}
\begin{multline}
	\mathbb{E}_{u_f}[f_{d,\Hp}]
	:=
	\sup_{\substack{
	U_0 \leq U \leq U_f
	\\
	V_{-1} \leq V \leq V_2 \wedge V(R_2,U)
	}}
	\sum_{k \leq 2}
	\bigg[
	\sum_{\vert \gamma \vert \leq 2}
	\big(
	\Vert (r \nablaslash)^k \mathfrak{D}^{\gamma} f^3 \Vert_{S^{\Hp}_{U,V}}^2
	+
	\Vert (r \nablaslash)^k \mathfrak{D}^{\gamma} f^4 \Vert_{S^{\Hp}_{U,V}}^2
	\big)
	\\
	+
	\sum_{\vert \gamma \vert \leq 1}
	\big(
	\Vert (r \nablaslash)^k \mathfrak{D}^{\gamma} \partial_U \slashed{f} \Vert_{S^{\Hp}_{U,V}}^2
	+
	\Vert (r \nablaslash)^k \mathfrak{D}^{\gamma} \partial_V \slashed{f} \Vert_{S^{\Hp}_{U,V}}^2
	\big)
	\bigg],
	\label{EfdHpenergy}
\end{multline}
where $f = f_{d,\Hp}$, $V_2 \wedge V(R_2,U) = \min \{ V_2, V(R_2,U) \}$, and $U_0: = U(u_0)$, where $U$ is defined by~\eqref{eq:KruskalEF} with $M:=M_f$.

\subsection{Pointwise norms of diffeomorphism functions}
\label{subsec:diffeopointwisenorms}

Define the pointwise norm of the diffeomorphism functions\index{energies!pointwise norms!$\mathbb P_{u_f}^{N-7}[f_{\Hp,\I}]$, pointwise norm of diffeomorphisms $f_{\Hp,\I}$}
\begin{align} \label{eq:diffeopointwisenorm1}
	&
	\mathbb P_{u_f}^{N-7}[f_{\Hp,\I}]
	=
	\sup_{\DRH\cap \DRI(u_1)}
	\bigg(
	u
	\vert
	(f^{3} - f^{4})_{\ell =0}
	\vert
	+
	u^{\frac{1}{2}}
	\vert
	(f^{3} + f^{4})_{\ell =0}
	\vert
	\\
	&
	\qquad \qquad \quad
	+
	\sum_{k \leq N-7}
	u
	\Big[
	\sum_{\substack{\vert \gamma \vert \leq 2 \\ k + \vert \gamma \vert \geq 1}}
	\Big(
	\vert
	(r\nablaslash)^{k} \mathfrak{D}^{\gamma} f^3
	\vert
	+
	\vert
	(r\nablaslash)^{k} \mathfrak{D}^{\gamma} f^4
	\vert
	\Big)
	+
	\sum_{\vert \gamma \vert \leq 1}
	\Big(
	\vert
	(r\nablaslash)^{k} \mathfrak{D}^{\gamma} \partial_u \slashed{f}
	\vert
	+
	\vert
	(r\nablaslash)^{k} \mathfrak{D}^{\gamma} \partial_v \slashed{f}
	\vert
	\Big)
	\Big]
	\bigg),
	\nonumber
\end{align}
where $f = f_{\Hp,\I}$,\index{energies!pointwise norms!$\mathbb{P}_{u_f}[f_{d,\I}]$, pointwise norm of diffeomorphisms $f_{d,\I}$}
\begin{multline} \label{eq:diffeopointwisenorm2}
	\mathbb{P}_{u_f}[f_{d,\I}]
	:=
	\sup_{\substack{
	u_{-1} \leq u \leq u_{2}
	\\
	v(R_{-2},u) \leq v < v_{\infty}
	}}
	\bigg[
	\sum_{\vert \gamma \vert \leq 2}
	\big(
	\vert \mathfrak{D}^{\gamma} f^3 \vert
	+
	r^{-1}
	\vert \mathfrak{D}^{\gamma} f^4 \vert
	\big)
	\\
	+
	\sum_{\vert \gamma \vert \leq 1}
	\big(
	r \vert \mathfrak{D}^{\gamma} r \nablaslash_4 f^3 \vert
	+
	\vert \mathfrak{D}^{\gamma} \nablaslash_3 f^4 \vert
	+
	\vert \mathfrak{D}^{\gamma} \partial_u \slashed{f} \vert
	+
	\vert \mathfrak{D}^{\gamma} \partial_v \slashed{f} \vert
	\big)
	\bigg],
\end{multline}
where $f = f_{d,\I}$, and\index{energies!pointwise norms!$\mathbb{P}_{u_f}[f_{d,\Hp}]$, pointwise norm of diffeomorphisms $f_{d,\Hp}$}
\begin{equation} \label{eq:diffeopointwisenorm3}
	\mathbb{P}_{u_f}[f_{d,\Hp}]
	:=
	\sup_{\substack{
	U_0 \leq U \leq U_f
	\\
	V_{-1} \leq V \leq V_2 \wedge V(R_2,U)
	}}
	\bigg[
	\sum_{\vert \gamma \vert \leq 2}
	\big(
	\vert \mathfrak{D}^{\gamma} f^3 \vert
	+
	\vert \mathfrak{D}^{\gamma} f^4 \vert
	\big)
	+
	\sum_{\vert \gamma \vert \leq 1}
	\big(
	\vert \mathfrak{D}^{\gamma} \partial_U \slashed{f} \vert
	+
	\vert \mathfrak{D}^{\gamma} \partial_V \slashed{f} \vert
	\big)
	\bigg],
\end{equation}
where $f = f_{d,\Hp}$, $V_2 \wedge V(R_2,U) = \min \{ V_2, V(R_2,U) \}$, and $U_0: = U(u_0)$, where $U$ is defined by~\eqref{eq:KruskalEF} with $M=M_f$.

\addtocontents{toc}{\setcounter{tocdepth}{2}}

\section{The statement of Theorem~\ref{thm:main}}
\label{precisesec}

Below is the detailed statement of the main theorem (corresponding to Theorem~\ref{maintheoremintro} 
of the introduction):

\begin{bigishtheorem}[Full finite-codimension nonlinear asymptotic stability of Schwarzschild] \label{thm:main}
Let $M_{\rm init}>0$ and associated parameters be fixed as in Section~\ref{compediumparameterssec},\index{main theorem!parameters!$M_{\rm init}$}
with $\hat\varepsilon_0(M_{\rm init})$  sufficiently small.

Given $0<\varepsilon_0\le \hat\varepsilon_0$\index{main theorem!parameters!$\varepsilon_0$}, 
let $\mathcal{S}_0\in \mathfrak{M}(M_{\rm init}, c^2\varepsilon_0)$ be initial data as in 
Proposition~\ref{theleaves}.
Then the  $3$-parameter family $\mathcal{L}^{\varepsilon_0}_{\mathcal{S}_0}$
contains 
an initial data set\index{main theorem!initial data!$\boldsymbol{\mathcal{S}}$}  
\[
\mathcal{S}:=\mathcal{S}(\lambda^{\rm final}) \in \mathcal{L}^{\varepsilon_0}_{\mathcal{S}_0}
\]
such that the following is
true.

Let 
 $(\mathcal{M},g)$\index{main theorem!spacetime sets!$(\mathcal{M},g)$} denote the maximal Cauchy development
 of the initial data $\mathcal{S}$
 as given by Theorem~\ref{maxCauchythe}. Recall that 
 Proposition~\ref{globexistofdata} applies, and moreover
 $(\mathcal{M},g)$
 admits the Kruskal and Eddington--Finkelstein 
 initial data gauges as in Theorems~\ref{thm:localKrus} and~\ref{thm:localEF},
 and Theorem~\ref{existenceofanchoredgaugethe} applies,
so we have in addition anchored (in the sense of Definition~\ref{anchoringdef}) teleologically normalised
$\mathcal{I}^+$ and $\Hp$ gauges~\eqref{fortheIplusgauge} 
  and~\eqref{fortheHplusgauge} corresponding to $u_f^0$ and some parameter $M_f^0$.

Then 
 $(\mathcal{M},g)$ satisfies the following global stability properties:

\renewcommand{\thesubsection}{(\roman{subsection})}

\subsection{Completeness of null infinity $\mathcal{I}^+$ and properties of the event horizon $\mathcal{H}^+$}
 
For all $u_f\ge u_f^0$,\index{main theorem!parameters!$u_f$} there exists an $M_f=M_f(u_f)$\index{main theorem!parameters!$M_f(u_f)$} such that 
 $(\mathcal{M},g)$ 
  admit $\I$ and $\Hp$ normalised gauges~\eqref{fortheIplusgauge} 
  and~\eqref{fortheHplusgauge} anchored in $(\mathcal{M},g)$
  in the sense of Definition~\ref{anchoringdef}.  

Moreover, the $\I$ normalised and $\Hp$ normalised gauges induce 
two limiting double null gauges corresponding to $u_f=\infty$, i.e.~$C^k$ embeddings for a $k\ge 3$
\begin{equation}
\label{limitingembeddinginf}
i_{\I, \infty}: \mathcal{W}_{\mathcal{I}^+}(\infty, M_{\rm final})\times \mathbb S^2\to \mathcal{D}^{\mathcal{I}^+}_{\infty}\subset \mathcal{M}
\end{equation}
\begin{equation}
\label{limitingembeddinghor}
i_{\Hp, \infty}:  \mathcal{W}_{\mathcal{H}^+}(\infty, M_{\rm final})\times \mathbb S^2\to \mathcal{D}^{\mathcal{H}^+}_{\infty}\subset \mathcal{M}
\end{equation}
where the sets $\mathcal{W}_{\I}(\infty, M_{\rm final})$\index{main theorem!coordinate domains!$\mathcal{W}_{\I}(\infty, M_{\rm final})$} and $\mathcal{W}_{\Hp}(\infty, M_{\rm final})$\index{main theorem!coordinate domains!$\mathcal{W}_{\Hp}(\infty, M_{\rm final})$},
are defined as in~\eqref{WI+} and~\eqref{WH+} with $<\infty$ replacing $\le u_f$ and $\le v_{\infty}$, and where
all statements of Definition~\ref{Igaugedefinition} and~\ref{Hgaugedefinition} hold which do not correspond to evaluation
at $u=u_f$ or $v=v_\infty$, where the background Schwarzschild metric is defined with respect to\index{main theorem!parameters!$M_{\rm final}$}
\begin{equation}
\label{Mfinaldefin}
M_{\rm final} := \lim_{u_f\to \infty} M_f (u_f).
\end{equation}
 Moreover, null infinity is complete in the sense of~\cite{Chrmil},  
and the domain of outer
communications 
\[
\mathcal{M} \cap J^-(\mathcal{I}^+)= \mathcal{D}^{\mathcal{EF}}\cup \mathcal{D}^{\mathcal{K}}\cup
\mathcal{D}^{\I}_{\infty}\cup \mathcal{D}^{\Hp}_{\infty} \cap J^-(\mathcal{D}^{\I}_{\infty})
\]
 is bounded to the future in $\mathcal{M}$ by a $C^k$ null hypersurface $\mathcal{H}^+$\index{main theorem!spacetime sets!$\mathcal{H}^+$, the event horizon of $(\mathcal{M},g)$}, for a $k\ge 3$,
to be called the event horizon,
whose null generators are future complete and which can be identified with the
limiting hypersurface $\{u_{\Hp}=\infty\}$ of the limiting $\mathcal{H}^+$ normalised gauge~\eqref{limitingembeddinghor}.

Null infinity can be realised as an ideal boundary corresponding to the limiting $v_{\mathcal{I}^+}=\infty$ hypersurface of
the limiting $\mathcal{I}^+$ gauge~\eqref{limitingembeddinginf} and
the ``laws of gravitational radiation'' hold on $\mathcal{I}^+$\index{main theorem!spacetime sets!$\mathcal{I}^+$, null infinity of $(\mathcal{M},g)$, defined as an asymptotic boundary} as relations 
linking the asymptotic behaviour
of various rescaled quantities~\cite{CK},
and the limiting $\mathcal{I}^+$ gauge satisfies the normalisations\index{main theorem!quantities!$\Sigma_+$}\index{main theorem!quantities!${\bf P}^+_{\ell= 1}$}
\begin{equation}
\label{normalisationsinstatement}
\Sigma_+ =0, \qquad {\bf P}^+_{\ell= 1}=0
\end{equation}
which  express that $u_f=\infty$ is asymptotically a ``good cut'' of $\mathcal{I}^+$
and in centre-of-mass frame.

Given the finiteness of higher order weighted norms on initial data, arbitrary high smoothness of
the asymptotic gauges~\eqref{limitingembeddinginf} and~\eqref{limitingembeddinghor}, 
and thus of the event horizon $\mathcal{H}^+$ itself, follows
and also of the rescaled quantities at $\mathcal{I}^+$. 
In particular, if $(\ref{datanormtobesmall})$ is finite for all $k$, then $\mathcal{H}^+$ is
a smooth hypersurface.

\subsection{Orbital stability}

For all $u_f\ge u_f^0$ above, the masses $M_f(u_f)$ referred to in (i)
remain uniformly close to the initial Schwarzschild
mass $M_{\rm init}$ in the sense that
\begin{equation}
\label{massstaysclose}
|M_f(u_f)-M_{\rm init}|\lesssim  \varepsilon_0,
\end{equation}
the two future normalised gauges corresponding to time $u_f$
remain uniformly close to  the initial data gauges
in the sense that\index{main theorem!energies!$\mathbb{E}_{u_f}[f_{d,\Hp}]$, energy measuring diffeomorphisms}\index{main theorem!energies!$\mathbb{E}_{u_f}[f_{d,\I}]$, energy measuring diffeomorphisms}
\begin{equation}
\label{gaugesstayclose}
\mathbb{E}_{u_f}[f_{d,\Hp}]\lesssim \varepsilon_0^2, \qquad \mathbb{E}_{u_f}[f_{d,\I}]\lesssim \varepsilon_0^2, 
\end{equation}
and the solution in both the $\Hp$ and $\I$ normalised gauges satisfies
\begin{equation}
\label{boundanddecay}
\mathbb{E}^N_{u_f,\Hp} + \mathbb{E}^N_{u_f,\I}
\lesssim \varepsilon_0^2
\end{equation}
and the lower order pointwise estimates 
\begin{equation}
\label{pointwisestatementintheorem}
\mathbb P_{u_f}^{N-5}[\Phi^{\mathcal{I}^+}]+\mathbb P_{u_f}^{N-5}[\Phi^{\mathcal{H}^+}] \lesssim \varepsilon_0.
\end{equation}

Moreover, we may replace $\varepsilon_0^2$ on the right hand side of the above estimates (as well
as all estimates appearing below)
with the quantity
\begin{equation}
\label{initialfluxesfororbital}
\mathbb E^N_0[ \Phi_{\mathcal{K}, d} ]+\mathbb E^N_0[ \Phi_{\mathcal{EF}, d}]
\end{equation}
from~\eqref{initdatanormK} and~\eqref{initdatanormEF}. In view of the 
top order boundedness statement included in the estimates~\eqref{boundanddecay},
the spacetime $(\mathcal{M},g)$ remains uniformly close
to the Schwarzschild metric with mass $M_{\rm init}$ 
in terms of energies at the same level of differentiability as the
``initial'' energy fluxes~\eqref{initialfluxesfororbital}.  (It is this aspect of~\eqref{boundanddecay} which strictly speaking
represents orbital stability.)

\subsection{Asymptotic stability}

For all $u_f\ge u_f^0$ above, the energy and pointwise decay estimates
for the Ricci coefficients and curvature components 
included in~\eqref{boundanddecay} and~\eqref{pointwisestatementintheorem}
can already be considered as a statement of asymptotic stability.

An analogous energy statement 
\begin{equation}
\label{boundanddecayasym}
\mathbb{E}^N_{\infty,\Hp} + \mathbb{E}^N_{\infty,\I}
\lesssim \varepsilon_0^2
\end{equation}
holds for the asymptotically normalised gauges (where the above norms in~\eqref{boundanddecayasym} are defined
by applying the norms of~\eqref{boundanddecay} to the asymptotically normalised gauges~\eqref{limitingembeddinginf} and~\eqref{limitingembeddinghor}, replacing $\le u_f$, $\le v_f$ with $<\infty$,
and including all quantities  which are not evaluated at $u=u_f$ or
$v=v_\infty$).

Similarly,
an
analogous lower order pointwise statement
\begin{equation}
\label{asymptpointwisestatementintheorem}
\mathbb P_{\infty}^{N-5}[\Phi^{\mathcal{I}^+}]+\mathbb P_{\infty}^{N-5}[\Phi^{\mathcal{H}^+}] \lesssim \varepsilon_0
\end{equation}
holds,
where $\mathcal{D}^{\mathcal{I}^+}_{u_f}$ and $\mathcal{D}^{\mathcal{H}^+}_{u_f}$ are
replaced by $\mathcal{D}^{\mathcal{I}^+}_\infty$ and $\mathcal{D}^{\mathcal{H}^+}_\infty$
in formulas~\eqref{eq:Ippointwisenorm} and~\eqref{eq:Hppointwisenorm}, respectively,
and the quantities are taken with respect to the asymptotic gauges~\eqref{limitingembeddinginf} and~\eqref{limitingembeddinghor}.

In this sense, the spacetime $(\mathcal{M},g)$, expressed in the asymptotic $\mathcal{H}^+$ and $\mathcal{I}^+$ gauges,
asymptotically settles down to the 
Schwarzschild metric~\eqref{SchwmetricEF} of mass $M_{\rm final}$ given by~\eqref{Mfinaldefin} in the domain of outer communications.
(Note that by $(\ref{massstaysclose})$, $M_{\rm final}$ satisfies
$|M_{\rm final}-M_{\rm init}|\lesssim \varepsilon_0$.)

Moreover, we have a number of
inverse polynomial decay bounds along $\mathcal{H}^+$  and $\mathcal{I}^+$
for instance the pointwise estimate on $\mathcal{H}^+$
\begin{equation}
\label{pointwiseonhorizon}
|\Omega^2 \alpha_{\mathcal{H}^+} (\infty, v, \cdot ) |\lesssim \varepsilon_0 v^{-1},
\end{equation}
where $\Omega^2 \alpha_{\mathcal{H}^+}(\infty,v,\cdot)$ is defined as the $u\to \infty$ limit of
$\Omega^2 \alpha_{\mathcal{H}^+}$ defined with respect to the asymptotic $\mathcal{H}^+$ gauge~\eqref{limitingembeddinghor},
and the estimates
\begin{equation}
\label{pointwisedecayatinfintity}
0\le M(u) - M_{\rm final} \lesssim \varepsilon_0^2 u^{-2} , \qquad 
|\Sigma(u,\cdot)| \lesssim \varepsilon_0 u^{-1}, \qquad |\Xi (u, \cdot)| \lesssim 
\varepsilon_0 u^{-1}, \qquad |{\bf A} (u, \cdot)| \lesssim \varepsilon_0 u^{-1 },
\end{equation}
where $M(u)$\index{main theorem!quantities!$M(u)$, Bondi mass} denotes the Bondi mass and $\Xi$\index{main theorem!quantities!$\Xi$}, $\Sigma$\index{main theorem!quantities!$\Sigma$} and ${\bf A}$\index{main theorem!quantities!${\bf A}$} are defined in Section~\ref{rescaledquantsec}.
\end{bigishtheorem}

\section{The stable codimension-$3$ ``submanifold'' $\mathfrak{M}_{\rm stable}$ and other remarks on the statement}
\label{technicalremarkssec}

We note a number of remarks:\index{initial data!moduli space!$\mathfrak{M}_{\rm stable}$, asymptotically stable codimension-$3$ ``submanifold'' of $\mathfrak{M}$}

\begin{remark}
\label{remarkafterthetheorem}
The above statement should be compared directly to  the rough statement given in
Theorem~\ref{maintheoremintro}.
Note that by Remark~\ref{remarkaboutcodim},
defining
\[
\mathfrak{M}_{\rm stable}:= \bigcup_{\mathcal{S}_0\in \mathfrak{M}_0} \mathcal{S}_0(\lambda^{\rm final})
\subset \mathfrak{M}(\varepsilon_0, M_{\rm init})
\]
to be union of all data $S(\lambda^{\rm final})$ whose existence follows
from Theorem~\ref{thm:main},
this
can indeed be viewed as defining a ``submanifold'' $\mathfrak{M}_{\rm stable}\subset \mathfrak{M}(\varepsilon_0,M_{\rm init})$
of codimension $3$.
(Here we are primarily interested in the \underline{size} of the set of allowed initial data and not its
regularity as a true submanifold of moduli space. For removing the quotes, see Remark~\ref{smoothnessremark} below.)
Statements (i), (ii) and (iii) above  are elaborations of the corresponding
statements in  Theorem~\ref{maintheoremintro}.
\end{remark}

\begin{remark}
\label{mightrefertowork}
With extra work, one can show that there is in fact a unique
$\mathcal{S} \in \mathcal{L}^{\varepsilon_0}_{\mathcal{S}_0}$ 
satisfying (i), (ii), and (iii). In essence, this follows from our understanding
of the linearised Kerr modes in linear theory~\cite{holzstabofschw}.
Note that this is a weaker statement than the purely teleological statement that, for small $\varepsilon_0$,
$\mathcal{S} \in \mathcal{L}^{\varepsilon_0}_{\mathcal{S}_0}$  is
the unique data set in the family that eventually evolves to Schwarzschild. The latter would of
course follow from a positive resolution to Conjecture~\ref{stabofkerr} for small $|a|\ll M$.
\end{remark}

\begin{remark}
\label{smoothnessremark}
In addition to uniqueness (cf.~Remark~\ref{mightrefertowork}), with more work one can indeed show that 
$\mathfrak{M}_{\rm stable}\subset \mathfrak{M}$
indeed forms a \underline{regular submanifold} of codimension exactly 3.
\end{remark}

\begin{remark}
The inequalities $(\ref{pointwisedecayatinfintity})$ imply that $M_{\rm final}$ can also be identified
as the final Bondi mass, and the total flux of energy radiated to $\mathcal{I}^+$ is bounded (and quadratic
in the smallness parameter $\varepsilon_0$), as measured from retarded time corresponding to $u_f=u_0$
in the asymptotically defined $\mathcal{I}^+$ gauge,
whereas the total   infinitesimal displacement of far-away test masses
given by $|\Sigma_+(\cdot) -\Sigma(u_{-1},\cdot) | =|\Sigma(u_{-1},\cdot)| \lesssim \varepsilon_0$ is also bounded. 
One can show similarly  the flux of linear momentum to $\mathcal{I}^+$ 
to be finite and quadratic in $\varepsilon_0$ (see already~\ref{fluxoflinmom}). 
Let us note finally that had we started from  asymptotically
flat spacelike initial data, then we could define
our asymptotic $\mathcal{I}^+$ gauge for all $u\in (-\infty,\infty)$, and
defining ${\bf P}_-$, $\Sigma_-$, etc.~as the limit $u\to -\infty$, 
we would obtain the relations of~\cite{Christmem} concerning
$\Sigma_+-\Sigma_-$. In particular, note that in this case $|\Sigma_+-\Sigma_-|\lesssim \varepsilon_0^2$.
\end{remark}

\begin{remark}
\label{sizeoflambdafinal}
We remark finally that although one immediately has only the bound
$|\lambda^{\rm final}|\lesssim \varepsilon_0$,
one may show relatively easily that one has  in fact $|\lambda^{\rm final}|\lesssim \varepsilon_0^2$.
\end{remark}

\renewcommand{\thesubsection}{\arabic{chapter}.\arabic{section}.\arabic{subsection}}

\addtocontents{toc}{\setcounter{tocdepth}{1}}

\chapter{The logic of the proof of Theorem~\ref{thm:main}}
\label{logicoftheproofsection}

In the present section, we shall give the logic of the proof of
Theorem \ref{thm:main}, i.e.~the skeletal form of the proof, broken into various 
subtheorems.
The analytical difficulty of the skeleton will then be fleshed out by the proofs of the subtheorems to be given 
over Parts~\ref{improvingpart} and \ref{conclusionpart} of the remainder of this work.

\minitoc

The logic of the proof follows a bootstrap and the first order of business is to 
define the bootstrap set $\mathfrak{B}$ in {\bf Section~\ref{section:bootstrap}}.
The main part of the proof then consists of showing that $\mathfrak{B}$ is nonempty,
open and closed.

We give the nonemptiness statement in {\bf Section~\ref{hopeitsnotempty}}.

The main part of showing openness consists of ``improving the bootstrap assumptions''.
The statement of this bootstrap theorem ({\bf  \Cref{havetoimprovethebootstrap}}) 
will be given in {\bf Section~\ref{improvingsectionnewdivision}} .
(It is precisely this theorem whose proof will form Part~\ref{improvingpart}.)

The 
openness statement itself is then given in
{\bf Section~\ref{opensection}} (depending on an existence theorem for
the teleologically normalised gauges which is deferred to Part~\ref{conclusionpart}).

Higher order estimates are given in {\bf Section~\ref{higherordersec}} (whose proof is also
deferred to Part~\ref{conclusionpart}) followed by
the closedness statement in  {\bf Section~\ref{closedsection}}.

The remaining assertions of  Theorem~\ref{thm:main} are deduced
in {\bf Section~\ref{theendoflogic}}. (Some of these will depend on subtheorems
whose proof is again deferred to Part~\ref{conclusionpart}.)

\vskip1pc
\emph{The reader primarily interested in the large-scale architecture of the proof may read this chapter
without having read Section~\ref{energiessection}, provided they are happy to treat
the energies defined there as black boxes or refer back as necessary for the notation. The
reader primarily interested in~\Cref{havetoimprovethebootstrap} may read up to Section~\ref{improvingsectionnewdivision}
and immediately turn to Part~\ref{improvingpart}. On the other hand,
the reader willing to take~\Cref{havetoimprovethebootstrap}
on faith may turn immediately to Part~\ref{conclusionpart} after reading this chapter.}

\section{The bootstrap set $\mathfrak{B}$}
\label{section:bootstrap}

Let $\mathcal{S}_0\in \mathfrak{M}_0(M_{\rm init},c^2\varepsilon_0)$ 
be as in the statement of Theorem~\ref{thm:main} 
and  let  $\mathcal{L}^{\varepsilon_0}_{\mathcal{S}_0}$
be the three parameter family defined in
$(\ref{familyfirst})$,
parametrised by $\lambda \in[-c\varepsilon_0,c\varepsilon_0]^3$.
 
Let $(\mathcal{M}(\lambda),g(\lambda))$ denote the maximal Cauchy development
corresponding to ${\mathcal{S}}_0(\lambda)$ as given by
Theorem~\ref{maxCauchythe}. 
Recall (cf.~the statement of Theorem~\ref{thm:main}) that 
 Proposition~\ref{globexistofdata} applies, and moreover
 $(\mathcal{M}(\lambda),g(\lambda))$
 admit the Kruskal and Eddington--Finkelstein 
 initial data gauges as in Theorems~\ref{thm:localKrus} and~\ref{thm:localEF}.
Recall also from the statement that Theorem~\ref{existenceofanchoredgaugethe} applies,
so we have in addition anchored (in the sense of Definition~\ref{anchoringdef}) teleologically normalised
$\mathcal{I}^+$ and $\Hp$ gauges~\eqref{fortheIplusgauge} 
 and~\eqref{fortheHplusgauge} corresponding to $u_f^0$ and some parameter $M_f^0$.
Recall finally the set $\mathfrak{R}_0$ defined in Section~\ref{thedegreeonemap}, and the map 
\[
{\bf J}_0:
\mathfrak{R}_0\to B(\varepsilon_0/u_f^0),
\] 
given by~\eqref{newdefofboldv0}, which, by Proposition~\ref{assertingdegreeone}
is a diffeomorphism, and thus in particular
restricts on its boundary to a degree-$1$ map to a $2$-sphere.

Consider the quantity
\begin{equation} \label{eq:bootstrapconstant}
	\varepsilon := \upsilon \varepsilon_0,
\end{equation}
where $\upsilon$ denotes a large constant which will be fixed later.  

\begin{definition}
\label{bootstrapsetdef}
Let $\mathfrak{B}$ denote the set of $\hat{u}_f \in [u^0_f, \infty)$\index{bootstrap!parameters!$\hat{u}_f$}\index{bootstrap!parameters!$u^0_f$, ``initial'' final retarded time associated to bootstrap} such that 
\begin{itemize}
	\item{\bf The $\lambda$-parameter range sets $\mathfrak{R}(u_f)$ and their topology and monotonicity.}
		For all $u_f\in (u^0_f, \hat{u}_f]$\index{bootstrap!parameters!$u_f$, final retarded time associated to bootstrap}, there exists a nonempty closed subset 
		$\mathfrak{R}(u_f)\subset \mathbb [-c\varepsilon_0,c\varepsilon_0]^3\subset \mathbb R^3$\index{bootstrap!sets!$\mathfrak{R}(u_f)$, subset of $\lambda$-parameter space depending on $u_f$}
		with 
			\begin{equation}
	\label{monotonofparam}
			\mathfrak{R}(u''_f) \subset \mathfrak{R}(u'_f), \qquad  	\partial\mathfrak{R}(u'_f) \cap \mathfrak{R}(u''_f) =\emptyset 
			\end{equation}
		for $u''_f>u'_f$, and such that $\mathfrak{R}(u^0_f)$ is homeomorphic to a closed ball;
	\item{\bf The $M_f$ parameter and the existence of anchored
	teleologically normalised  $\mathcal{I}^+$ and $\mathcal{H}^+$ gauges.}
		There exists a  function $M_f: \cup_{u_f\in [u_0,  \hat{u}_f]} \{u_f\}\times \mathfrak{R}(u_f) \to \mathbb R$\index{bootstrap!parameters!$M_f(u_f,\lambda)$} satisfying for all $u_f\in [u_f^0,\hat{u}_f]$, $\lambda\in \mathfrak{R}(u_f)$
		\begin{equation} \label{eq:MMinit}
			\vert M_f(u_f, \lambda) - M_{\rm init} \vert \leq \varepsilon,
		\end{equation}
		and,  anchored (in the sense of Definition~\ref{anchoringdef}) teleologically
		 normalised $\I$  and $\Hp$ gauges~\eqref{fortheIplusgauge} 
 and~\eqref{fortheHplusgauge}  corresponding to parameters $u_f$ and $M_f=M_f(u_f,\lambda)$
		  in $(\mathcal{M}(\lambda),g(\lambda))$;
	\item{\bf Continuous dependence of the anchored teleologically normalised gauges on $u_f$.}
	The function  $M_f(u_f, \lambda)$ is continuous,
	and for fixed $\lambda\in \mathfrak{R}(u_k)$,
	the two teleologically normalised gauges depend continuously on $u_f$,
	in the sense that their domains and all geometric quantities depend continuously on $u_f$.
	\item{\bf Energy estimates for geometric quantities in both teleological gauges.}
		For all $(\mathcal{M}(\lambda),g(\lambda))$ as above,  all $u_f\in [u_f^0,\hat{u}_f]$, and with 
	the above gauges, the estimate\index{bootstrap!energies!$\mathbb{E}^{K-2}_{u_f}[P_{\Hp},\Pbar_{\Hp},P_{\I},\Pbar_{\I}]$}\index{bootstrap!energies!$\mathbb{E}^{K}_{u_f}[\alpha_{\Hp},\alphabar_{\Hp},\alpha_{\I},\alphabar_{\I}]$}\index{bootstrap!energies!$\mathbb{E}^K_{u_f,\Hp}$}\index{bootstrap!energies!$\mathbb{E}^K_{u_f,\I}$}
		\begin{equation} \label{eq:bamain}
			\mathbb{E}^{N-2}_{u_f}[P_{\Hp},\Pbar_{\Hp}]
			+
			\mathbb{E}^{N-2}_{u_f}[P_{\I},\check{\Pbar}_{\I}]
			+
			\mathbb{E}^{N}_{u_f}[\alpha_{\Hp},\alpha_{\I}]+\mathbb{E}^N_{u_f}[\alphabar_{\Hp},\alphabar_{\I}]
			+
			\mathbb{E}^N_{u_f,\Hp}
			+
			\mathbb{E}^N_{u_f,\I}
			\leq
			\varepsilon^2
		\end{equation}
		holds, where the above energies are defined in Section~\ref{PandPbarenergydefs},
		by~\eqref{masterenergya} and~\eqref{masterenergyab} in Section~\ref{twoenergiesalphaandalphabardefs}, by~\eqref{Hplusmasterenergy} in Section~\ref{PhiHpenergysec} and
	by~\eqref{themasterweuseIplus} in	
		 Section~\ref{Iplusenergiesmastersec};
	\item{\bf Energy estimates for the diffeomorphism functions connecting the teleological
	gauges to each other and to initial data.}
		For all $(\mathcal{M}(\lambda),g(\lambda))$ as above,  all $u_f\in [u_f^0,\hat{u}_f]$, and with 
	the above gauges, the estimate\index{bootstrap!energies!$\mathbb{E}_{u_f}[f_{\Hp,\I}]$}\index{bootstrap!energies!$\mathbb{E}_{u_f}[f_{d,\Hp}]$}\index{bootstrap!energies!$\mathbb{E}_{u_f}[f_{d,\I}]$}
		\begin{equation} \label{eq:badiffeo}
			\mathbb{E}_{u_f}^{N+2}[f_{\Hp,\I}]
			+
			\mathbb{E}_{u_f}[f_{d,\Hp}]
			+
			\mathbb{E}_{u_f}[f_{d,\I}]
			\leq
			\varepsilon^2
		\end{equation}
		holds, where the above energies are defined
		by~\eqref{EKfHpIpenergy},~\eqref{EfdHpenergy} and~\eqref{EfdIpenergy}
		 in Section~\ref{energiesofdiffeossec};
	 \item{\bf Zeroth order metric estimates.}
		 For all $(\mathcal{M}(\lambda),g(\lambda))$ as above,  all $u_f\in [u_f^0,\hat{u}_f]$, and with 
	the above gauges, the estimate
		\begin{equation}\label{lowerorderpointwise}
	\sup_{\mathcal{D}^{\Hp}}\left(|r^{-2} \slashed g^{\mathcal{H}^+}-\mathring\gamma|_{\mathring \gamma}
	+|\Omega^{-2}_\circ \Omega^2_{\mathcal{H}^+}-1|\right)
	 + 
		\sup_{\mathcal{D}^{\I}} \left(r|r^{-2}\slashed g^{\mathcal{I}^+} -\mathring\gamma|_{\mathring \gamma} 
		 + |r(\Omega^{-2}_{\circ}\Omega^2_{\mathcal{I}^+}-1)| \right)
		  \leq \sqrt{\varepsilon}
		\end{equation}
		holds;
		\item{\bf Estimate for the angular momentum.}
		For all $(\mathcal{M}(\lambda),g(\lambda))$ as above,  all $u_f\in [u_f^0,\hat{u}_f]$, and with 
	the above gauges, recalling the definition $(\ref{cmfdefishere})$, 
	and defining\index{bootstrap!energies!parameters!${\bf J}(\lambda, u_f)$}\index{angular momentum!${\bf J}(\lambda, u_f)$, vector measuring angular momentum at time $u_f$}
	\begin{equation}
	\label{definofboldv}
	{\bf J}(u_f, \lambda ) =(J^{-1}_{\I},J^{0}_{\I},J^1_{\I}) \in \mathbb R^3
	\end{equation}
where $J^m_{\I}$ are the associated Kerr parameters defined in Definition~\ref{assocKerparIplus} 
corresponding
to the anchored $\I$ gauge  in
 $(\mathcal{M}(\lambda),g(\lambda))$ with parameters $u_f$ and $M_f$ as above, then ${\bf J}:\cup_{u_f\in[u_f^0,\hat{u}_f]}\{u_f\}\times \mathfrak{R}(u_f)\to 
 \mathbb R^3$ is a continuous map and
	the following estimate holds:
		\begin{align}
		|{\bf J} (u_f, \lambda)| \leq \frac{\varepsilon_0}{u_f};
			\label{eq:curletalambda}
		\end{align}
	\item{\bf Norm of angular momentum on the boundary of  the parameter range.}
	For  all $u_f\in [u_f^0,\hat{u}_f]$ and $(\mathcal{M}(\lambda),g(\lambda))$ corresponding to $\lambda\in
	\partial\mathfrak{R}(u_f)$, 
	then 
		\begin{align}
			|{\bf J} (u_f, \lambda)|
			=
			\frac{\varepsilon_0}{u_f};
			\label{eq:curletalao}
		\end{align}
	\item{\bf Degree properties of the map ${\bf J}_{u_f}$.}
	Defining the map\index{bootstrap!maps!${\bf J}_{u_f} : \mathfrak{R}(u^0_f) \to B_{\varepsilon_0/u^0_f}$}
	\begin{equation}
	\label{otherboldvdef}
	{\bf J}_{u_f} : \mathfrak{R}(u^0_f) \to B_{\varepsilon_0/u^0_f}\subset \mathbb R^3
	\end{equation}
	by 
	\begin{equation}
	\label{mapdefinedbythisformula}
	{\bf J}_{u_f} (\lambda) = {\bf J}(u'_f(u_f, \lambda), \lambda) 
	\end{equation}
	where 
	\[
	u'_f(u_f, \lambda):=\sup \{u^0_f\le u'_f \le u_f : \lambda \in \mathfrak{R}(u'_f)\}
	\] 
	then ${\bf J}_{u_f}$ is continuous and
	its restriction satisfies
	\begin{equation}
	\label{restrictiondegree}
	{\bf J}_{u_f}|_{\partial \mathfrak{R}(u^0_f)}= {\bf J}_{u^0_f}|_{\partial\mathfrak{R}(u^0_f)}: \partial \mathfrak{R}(u^0_f) \to \partial B_{\varepsilon_0/u^0_f}
	\end{equation}
	and is thus a degree-$1$ map.  
\end{itemize}
\end{definition}

The essential part of the proof of Theorem \ref{thm:main} will follow from showing that $\mathfrak{B} \neq \emptyset$ and that $\mathfrak{B}$ is open and closed.

\section{Non-emptiness of the bootstrap set $\mathfrak{B}$ }
\label{hopeitsnotempty}

The first step in the proof of Theorem~\ref{thm:main} is to show that $\mathfrak{B} \neq \emptyset$.
This  follows from Theorem~\ref{existenceofanchoredgaugethe} and Proposition~\ref{assertingdegreeone}.

\begin{theorem}[Non-emptiness of the bootstrap set] \label{thm:nonempty}
Let $\mathcal{S}_0$
$\mathcal{L}^{\varepsilon_0}_{\mathcal{S}_0}$,
$(\mathcal{M}(\lambda),g(\lambda))$, $M_f^0(\lambda)$, $\mathfrak{R}_0$ and the map 
${\bf J}_0$ defined by~\eqref{newdefofboldv0}, considered as a map~\eqref{isadiffeo},
be as in the beginning 
of Section~\ref{section:bootstrap}.

Then defining $\mathfrak{R}(u_f^0):=\mathfrak{R}_0$, 
defining $M_f:\{u_f^0\}\times \mathfrak{R}(u_f^0)\to \mathbb R$ by $M_f (u^f_0,\lambda):=M_f^0(\lambda)$,
it follows that, for sufficiently large $\upsilon$ in~\eqref{eq:bootstrapconstant}, all conditions of Definition~\ref{bootstrapsetdef} 
are satisfied by $u_f^0$ with these choices,
i.e.~$u_f^0\in \mathfrak{B}$ and thus $\mathfrak{B}\ne\emptyset$.
\end{theorem}

\begin{proof}
We note that ${\bf J}(\lambda, u_f^0)$ defined by~\eqref{definofboldv} satisfies
${\bf J}(\lambda, u_f^0)= {\bf J}_0$, and thus ${\bf J}_{u_f}$ defined by~\eqref{otherboldvdef} also satisfies
${\bf J}_{u_f}={\bf J}_0$. Thus,~\eqref{eq:curletalambda} holds by definition while~\eqref{eq:curletalao}, the continuity statements in $\lambda$ and the statement that~\eqref{otherboldvdef} is of degree $1$ follow from Proposition~\ref{assertingdegreeone}.
The monotonicity property~\eqref{monotonofparam} is vacuous, while~\eqref{eq:bamain} 
follows from the statement of Theorem~\ref{existenceofanchoredgaugethe}.
The continuous dependence on $u_f$ is also vacuous.
Inequalities \eqref{eq:bamain},~\eqref{eq:badiffeo} 
have already been asserted in
the statement of Theorem~\ref{existenceofanchoredgaugethe}.
\end{proof}

\section{Improving the bootstrap assumptions: the statement of~\Cref{havetoimprovethebootstrap}}
\label{improvingsectionnewdivision}

The main analytical content of this work can be expressed as a theorem stating that
given a $\hat{u}_f\in \mathfrak{B}$, then all quantitative estimates in  Definition~\ref{bootstrapsetdef}
can in fact be improved, i.e.

\begin{restatable}[Improving the bootstrap assumptions]{thmnonum}{improvingthatbootstrap}
\label{havetoimprovethebootstrap}
There exists an $\upsilon=\upsilon(M_{\rm init})$ sufficiently large such that 
for $\hat\varepsilon_0(M_{\rm init})$ sufficiently small, and all
$0<\varepsilon_0\le \hat\varepsilon_0$,
the following is true
with $\varepsilon=\upsilon\varepsilon_0$ defined by~\eqref{eq:bootstrapconstant}.

Let $\mathcal{S}_0$
$\mathcal{L}^{\varepsilon_0}_{\mathcal{S}_0}$,
$(\mathcal{M}(\lambda),g(\lambda))$ be as in the beginning 
of Section~\ref{section:bootstrap}, and let $\mathfrak{B}$ be the set
defined by Definition~\ref{bootstrapsetdef}.

Let $\hat{u}_f\in \mathfrak{B}$. Then for all $u_f\in [u_f^0,\hat{u}_f]$ and all $\lambda \in \mathfrak{R}(u_f)$, 
we have the following improved estimates for the solution $(\mathcal{M}(\lambda), g(\lambda))$:
	\begin{equation} \label{improvementpointwise}
	\sup_{\mathcal{D}^{\Hp}}\left(|r^{-2} \slashed g^{\mathcal{H}^+}-\mathring\gamma|_{\mathring \gamma} + 
	|\Omega^{-2}_\circ \Omega^2_{\mathcal{H}^+}-1|\right)+
		\sup_{\mathcal{D}^{\I}} \left(r|r^{-2}\slashed g^{\mathcal{I}^+} -\mathring\gamma|_{\mathring \gamma} 
		+ |r(\Omega^{-2}_{\circ}\Omega^2_{\mathcal{I}^+}-1)| \right)
		  \leq \frac12\sqrt{\varepsilon},
	\end{equation}	
\begin{equation}
\label{improvement1}
			\vert M_f(u_f, \lambda) - M_{\rm init} \vert \leq \frac12 \varepsilon,
\end{equation}
		\begin{equation} \label{improvement2}
			\mathbb{E}^{N-2}_{u_f}[P_{\Hp},\Pbar_{\Hp}]
			+
			\mathbb{E}^{N-2}_{u_f}[P_{\I},\check{\Pbar}_{\I}]
			+
			\mathbb{E}^{N}_{u_f}[\alpha_{\Hp},\alpha_{\I}]
			+
			\mathbb{E}^{N}_{u_f}[\alphabar_{\Hp},\alphabar_{\I}]
			+
			\mathbb{E}^N_{u_f,\Hp}
			+
			\mathbb{E}^N_{u_f,\I}
			\leq
			\frac12 \varepsilon^2,
		\end{equation}
	\begin{equation} \label{improvement3}
			\mathbb{E}_{u_f}^{N+2}[f_{\Hp,\I}]
			+
			\mathbb{E}_{u_f}[f_{d,\Hp}]
			+
			\mathbb{E}_{u_f}[f_{d,\I}]
			\leq
			\frac12 \varepsilon^2
		\end{equation}
		and the following improved inclusions:
		\begin{align}
		\label{improvedinclus1}
		\mathcal{D}^{\Hp}_{r\ge R_{-1}}&\subset \mathcal{D}_{r\ge R_{-\frac32}}^{\I},\\
		\label{improvedinclus2}
		\mathcal{D}^{\I}_{r \le R_{1}} \cap J^+(S^{\Hp}_{u_0,v_0}) &\subset \mathcal{D}_{r\le R_{\frac32}}^{\Hp}
\end{align}
and	
		\begin{align}
			\bigcup_{v_{-1} \leq v \leq v_2} \Cbar_{v}^{\Hp}
			&
			\subset
			\mathcal{D}^{\mathcal{K}}(V_3) \cap \{ V_{-1} - C\varepsilon \leq V_{data} \leq V_{2}+ C \varepsilon \}
			\cap \{U_0 -C\varepsilon\leq U_{data} \leq C\varepsilon \} ,
			\label{impoverlap3}
			\\
			\bigcup_{u_{-1} \leq u \leq u_2} C_{u}^{\I}
			&
			\subset
			\mathcal{D}^{\mathcal{EF}}(u_3)\cap   \{ u_{-1} - C \varepsilon \leq u_{data} \leq u_{2} + C \varepsilon \},
			\label{impoverlap5}
		\end{align}
		for some constant $C$ depending only on $M_{\rm init}$.
\end{restatable}

\begin{remark}
Here we recall that the energies appearing in~\eqref{improvement2} are defined in Section~\ref{PandPbarenergydefs},
		by~\eqref{masterenergya} and~\eqref{masterenergyab} in Section~\ref{twoenergiesalphaandalphabardefs}, by~\eqref{Hplusmasterenergy} in Section~\ref{PhiHpenergysec} and
	by~\eqref{themasterweuseIplus} in	
		 Sections~\ref{Iplusenergiesmastersec}
		 while the energies appearing in~\eqref{improvement3}
 are defined
		by~\eqref{EKfHpIpenergy},~\eqref{EfdHpenergy} and~\eqref{EfdIpenergy}
		 in Section~\ref{energiesofdiffeossec}. 
		 
	Inequality~\eqref{improvement1} improves~\eqref{eq:MMinit}, 
	inequality~\eqref{improvement2} improves~\eqref{eq:bamain}, 
	inequality~\eqref{improvement3} improves~\eqref{eq:badiffeo} 
	and inequality~\eqref{improvementpointwise} improves~\eqref{lowerorderpointwise}.
	
	The inclusion relations~\eqref{improvedinclus1}--\eqref{improvedinclus2} 
	improve~\eqref{eq:overlap1}--\eqref{eq:overlap2} while
	the inclusion relations~\eqref{impoverlap3}--\eqref{impoverlap5} improve~\eqref{eq:overlap3}--\eqref{eq:overlap5},
	provided that $\hat\varepsilon_0$ is chosen sufficiently small.
\end{remark}

\begin{proof}
The  proof of the above theorem will be the content
of Part~\ref{improvingpart}. See Chapter~\ref{mainestimatessechere}.
\end{proof}

\section{The bootstrap set $\mathfrak{B}$ is open and the statements of Theorems~\ref{thm:newgauge} and~\ref{thm:lambda}} 
\label{opensection}

Having improved the bootstrap assumptions in   Theorem~\ref{havetoimprovethebootstrap},
we now turn to proving that the bootstrap set is open:

\begin{restatable}[The bootstrap set $\mathfrak{B}$ is open]{thm}{justopen}
 \label{thm:justopen}
For $\hat\varepsilon_0(M_{\rm init})$ sufficiently small, let
$0<\varepsilon_0\le \hat\varepsilon_0$, $\mathcal{S}_0$
$\mathcal{L}^{\varepsilon_0}_{\mathcal{S}_0}$,
$(\mathcal{M}(\lambda),g(\lambda))$ be as in the beginning 
of Section~\ref{section:bootstrap}, and let $\mathfrak{B}$ be the set
defined by Definition~\ref{bootstrapsetdef}.

	Then if $\hat{u}_f\in \mathfrak{B}$, then for sufficiently small $\delta_0$, it follows
	that $\hat{u}_f+\delta_0\in\mathfrak{B}$.
\end{restatable}
\begin{proof}

The first step ({\bf Section~\ref{nextstephere}})
is to show that for sufficiently small $\delta$, the spacetimes
$(\mathcal{M}(\lambda), g(\lambda))$ admit anchored  $\I$ and $\Hp$ gauges
with respect to $\hat{u}_f+\delta$ and a suitable $M_f(\hat{u}_f+\delta,\lambda)$, 
continuously depending on $\delta$,
which, in view of Theorem~\ref{havetoimprovethebootstrap}, inherit by continuity
the bootstrap estimates  \eqref{eq:bamain}, \eqref{eq:badiffeo} and~\eqref{lowerorderpointwise}.
This will be the statement of
{\bf Theorem~\ref{thm:newgauge}}, whose proof is deferred to Section~\ref{section:newgauges}.
Finally ({\bf Section~\ref{laststephere}}), 
it is shown that  $\hat{u}_f+\delta_0$  
satisfies the remaining  assumptions 
of Definition~\ref{bootstrapsetdef}, after appropriately defining $\mathfrak{R}(\hat{u}_f+\delta)$ 
for all $0<\delta\le \delta_0$. 
 This will be the statement of
{\bf Theorem~\ref{thm:lambda}}, a statement whose proof is 
deferred to Section~\ref{section:monotonicityR}.
 It follows that indeed, 
$\hat{u}_f+\delta_0$ together with the definition of $\mathfrak{R}(u_f+\delta)$ for $0\le\delta \le \delta_0$ 
satisfy all the conditions 
of Definition~\ref{bootstrapsetdef}. Thus we have indeed $\hat{u}_f+\delta_0
\in\mathfrak{B}$.
This completes now the proof of Theorem~\ref{thm:justopen}.
\end{proof}

\subsection{Existence of anchored $\hat{u}_f+\delta$ normalised gauges: The statement of Theorem~\ref{thm:newgauge}}
\label{nextstephere}

The following  theorem gives that for sufficiently small $\delta_0>0$, there
exist anchored $\hat{u}_f+\delta$ normalised teleological gauges, for all $0\le \delta\le\delta_0$, which inherit the improved
bootstrap estimates.

\begin{restatable}[Existence of anchored $\hat{u}_f+\delta$ normalised gauges]{thmfunnynumbers}{newgauge}
\label{thm:newgauge}
For $\hat\varepsilon_0(M_{\rm init})$ sufficiently small, let
$0<\varepsilon_0\le \hat\varepsilon_0$, $\mathcal{S}_0$
$\mathcal{L}^{\varepsilon_0}_{\mathcal{S}_0}$,
$(\mathcal{M}(\lambda),g(\lambda))$ be as in the beginning 
of Section~\ref{section:bootstrap}, and let $\mathfrak{B}$ be the set
defined by Definition~\ref{bootstrapsetdef}.

	Let $\hat{u}_f\in \mathfrak{B}$. Then there exists a $\delta_0>0$ such that for all $\delta \in [0,\delta_0]$,
	and  all $\lambda\in \mathfrak{R}(\hat{u}_f)$,
	with the gauges as defined above, we have the following:

	There exists  a function
	\begin{equation}
	\label{newMdefhere}
	M_f:[\hat{u}_f,\hat{u}_f+\delta_0]\times \mathfrak{R}(\hat{u}_f)\to \mathbb R
	\end{equation}
	satisfying $(\ref{eq:MMinit})$, and anchored 
	 $\I$ and $\Hp$ normalised gauges with respect to  $\hat{u}_f + \delta$, $v_\infty=v_\infty(\varepsilon_0, \hat{u}_f+\delta)$ and $M_f(\hat{u}_f+\delta,\lambda)$ given by~\eqref{newMdefhere}, defined as in Definition~\ref{anchoringdef}, satisfying
	 \eqref{eq:bamain}, \eqref{eq:badiffeo} and~\eqref{lowerorderpointwise}, with $u_f:=\hat{u}_f+\delta$.
	 
	 The function $M_f$ defined by $(\ref{newMdefhere})$ is
	 continuous and coincides with the previously defined $M_f$ at $\delta=0$. 
	 
	 Finally, one may define the map ${\bf J}(\hat{u}_f+\delta, \lambda)$ by $(\ref{definofboldv})$ as a map 
	 \begin{equation}
	 \label{asamaphere}
	 {\bf J}:[\hat{u}_f,\hat{u}_f+\delta_0]\times\mathfrak{R}(\hat{u}_f)\to \mathbb R^3
	 \end{equation}
	 and $(\ref{asamaphere})$ is a continuous map (i.e.~continuous in both $\lambda$ and $\delta$) and coincides
	 with the previously defined map for $\delta=0$.
\end{restatable}

\begin{proof}
See Section~\ref{section:newgauges}.
\end{proof}

\subsection{Properties of $\mathfrak{R}(\hat{u}_f+\delta)$ and ${\bf J}_{\hat{u}_f+\delta}$:
The statement of Theorem~\ref{thm:lambda}}
\label{laststephere}

Finally, we deduce the existence of $\mathfrak{R}(\hat{u}_f+\delta)$  for all $0\le \delta \le \delta_0$ and infer 
that $\hat{u}_f+\delta_0$ satisfies the remaining hypotheses of Definition~\ref{bootstrapsetdef}.

\begin{restatable}[Definition, non-emptyness and monotonicity properties
of $\mathfrak{R}(\hat{u}_f+\delta)$ and properties of the map 
${\bf J}_{\hat{u}_f+\delta}$]{thmfunnynumbers}{lambdaherehere}
\label{thm:lambda}
	Let
	$\delta_0$, the $\hat{u}_f+\delta$ gauges,
	$(\ref{newMdefhere})$ and  $(\ref{asamaphere})$ be as in Theorem~\ref{thm:newgauge},
	for $0\le \delta<\delta_0$.
	
	Defining the closed set
	\begin{equation}
	\label{herestoadef}
	\mathfrak{R}(\hat{u}_f+\delta):= \left\{ \lambda\in \mathfrak{R}(\hat{u}_f) : |{\bf J}(\hat{u}_f+\delta, \lambda)| \le \frac{\varepsilon_0}{\hat{u}_f+\delta}\right\},
	\end{equation}
	then, after possibly redefining $\delta_0>0$, 
	we have for all $0\le \delta\le \delta_0$ that
	\begin{equation}
	\label{indeednonemptyhere}
	\mathfrak{R}(\hat{u}_f+\delta) \ne\emptyset,
	\end{equation}
	and if
	$0\le \delta'<\delta\le \delta_0$,
	then 
	\begin{equation}
	\label{inclusionstuff}
	\mathfrak{R}(\hat{u}_f+\delta) \subset \mathfrak{R}(\hat{u}_f+\delta'), \qquad \partial\mathfrak{R}(\hat{u}_f+\delta')\cap
	\mathfrak{R}(\hat{u}_f+\delta) =\emptyset.
	\end{equation}
	Moreover, the inequality~\eqref{eq:curletalambda} trivially holds on $\mathfrak{R}(\hat{u}_f+\delta)$ while~\eqref{eq:curletalao} holds on $\partial\mathfrak{R}(\hat{u}_f+\delta)$
 	and the map ${\bf J}_{\hat{u}_f+\delta}$ defined by~\eqref{mapdefinedbythisformula} is continuous and satisfies $(\ref{restrictiondegree})$, thus its
	restriction to the boundary is 
	in particular degree $1$.
\end{restatable}

\begin{proof}
See Section~\ref{section:monotonicityR}.
\end{proof}

\section{Higher order estimates: the statement of Theorem~\ref{thehigherordertheorem}}
\label{higherordersec}

Given the estimates we have obtained in the bootstrap region 
as part of the proof of Theorem~\ref{havetoimprovethebootstrap} for small data, 
we can 
obtain higher order  estimates. 
In particular, for smooth data, this will imply 
pointwise estimates to arbitrary order, depending only on the order and the supremum of ${u}_f$. 
These will be useful for showing closedness in Section~\ref{closedsection}.

\begin{restatable}[Higher order estimates]{thmfunnynumbers}{thehigherordertheoremhere}
\label{thehigherordertheorem}
For $\hat\varepsilon_0(M_{\rm init})$ sufficiently small, let
$0<\varepsilon_0\le \hat\varepsilon_0$, $\mathcal{S}_0$
$\mathcal{L}^{\varepsilon_0}_{\mathcal{S}_0}$,
$(\mathcal{M}(\lambda),g(\lambda))$ be as in the beginning 
of Section~\ref{section:bootstrap}, and let $\mathfrak{B}$ be the set
defined by Definition~\ref{bootstrapsetdef}.

Let $u_f\in \mathfrak{B}$ and let $\mathfrak{R}(u_f)$ be as in  the statement
of Definition~\ref{bootstrapsetdef}.

Then, for all $k\ge N$ there exist constants $C_k$ such that for all $\lambda\in \mathfrak{R}(u_f)$,
we have the following higher order estimates for the solution $(\mathcal{M}(\lambda),g(\lambda))$:
\begin{align}
\label{nonlocalisedstatement}
\nonumber
&\mathbb{E}^{k}_{u_f} [\alpha_{\Hp}, \alpha_{\I}] + \mathbb{E}^{k}_{u_f} [\underline{\alpha}_{\Hp}, \underline{\alpha}_{\I}] +
\mathbb E_{u_f, \mathcal{H}^+}^k
+ \mathbb E_{u_f, \mathcal{I}^+}^k + \mathbb{E}_{u_f}^{k+2}[f_{\Hp,\I}]+\mathbb{P}_{u_f}^{k-5}[\Phi^{\Hp}]+\mathbb{P}_{u_f}^{k-5}[\Phi^{\I}] \\
 &\qquad  \le C_k \mathbb E^k_{0}[\mathcal{S}(\lambda)].
\end{align}
Moreover,
we have in addition the localised estimate
\begin{align}
\nonumber
&\mathbb{E}^{k}_{u_f} [\alpha_{\Hp}, \alpha_{\I}] + \mathbb{E}^{k}_{u_f} [\underline{\alpha}_{\Hp}, \underline{\alpha}_{\I}] +
\mathbb E_{u_f, \mathcal{H}^+}^k
+ \mathbb E_{u_f, \mathcal{I}^+}^k + \mathbb{E}_{u_f}^{k+2}[f_{\Hp,\I}] 
+\mathbb{P}_{u_f}^{k-5}[\Phi^{\Hp}]+\mathbb{P}_{u_f}^{k-5}[\Phi^{\I}]  \\
&\qquad
\lesssim \mathbb E^N_{0}[\mathcal{S}(\lambda)]+ C_k(u_f) \mathbb E^k_{0,v_\infty}[\mathcal{S}(\lambda)],
\label{localisedstatementhere}
\end{align}
where $v_\infty=v_\infty(u_f)$, the
energy $\mathbb E^k_{0,v_\infty}[\mathcal{S}(\lambda)]$ is defined in 
Section~\ref{globalsmallnessassumptionnorm} (appearing in particular in equation~\eqref{triviallyforhigher}) and
 where $C_k(u_f)$ is a constant which depends in addition on $u_f$.
\end{restatable}

\begin{proof}
See Section~\ref{proofofhigherorderestimatessec}.
\end{proof}

\begin{remark}
We will use~\eqref{localisedstatementhere} to show closedness.
We note that~\eqref{localisedstatementhere} does not follow immediately from~\eqref{nonlocalisedstatement}
and the domain of dependence property because  the fact that the initial Eddington--Finkelstein gauge
is normalised at null infinity means that the choice of the teleologically normalised gauges at time $u_f$ 
depend on the full initial data along $C_{\rm out}$, even if the solution itself geometrically can be shown not to.
One expects that~\eqref{localisedstatementhere} is true without the additional $u_f$-dependence, but
due to our crude use of the preliminary initial Eddington--Finkelstein gauge $i^0_{\mathcal{EF}}$ in the proof of Theorem~\ref{localisedstatementhere} 
in  Section~\ref{proofofhigherorderestimatessec} (with bad $r$-dependence of estimates),
we shall not track this.
\end{remark}

\begin{corollary}
\label{smoothestimforclosedness}
Fix an arbitrary constant $u_{\rm max}<\infty$, and let $u_f$ be as in 
Theorem~\ref{thehigherordertheorem} satisfying $u_f\le u_{\rm max}$. 
Then, in view of~\eqref{triviallyforhigher} and statement~\eqref{localisedstatementhere} 
above, it follows that
for all $k$, there exists a constant $D(k, u_{\rm max})$, depending only
on $k$ and $u_{\rm max}$,  such that the $C^k$ norms of $\Omega$, the tensor
$\slashed g_{CD}$  and the vector field $b^D$
in both
teleologically normalised gauges are  bounded by $D(k, u_{\rm max})$.
In particular, with a different labelling of $k$, all geometric quantities are uniformly bounded,
depending only on their order and $u_{\rm max}$.
\end{corollary}

\section{The bootstrap set $\mathfrak{B}$ is closed}
\label{closedsection}

Using the higher order estimates of Theorem~\ref{thehigherordertheorem} and a soft local existence theorem for the
characteristic initial value problem (in the smooth category), the
 bootstrap set $\mathfrak{B}$ can readily be seen to be closed.
For this it will suffice to show the following.

\begin{theorem}[The bootstrap set $\mathfrak{B}$ is closed]\label{thm:closed}
For $\hat\varepsilon_0(M_{\rm init})$ sufficiently small, let
$0<\varepsilon_0\le \hat\varepsilon_0$, $\mathcal{S}_0$
$\mathcal{L}^{\varepsilon_0}_{\mathcal{S}_0}$,
$(\mathcal{M}(\lambda),g(\lambda))$ be as in the beginning 
of Section~\ref{section:bootstrap}, and let $\mathfrak{B}$ be the set
defined by Definition~\ref{bootstrapsetdef}.

Let $u_{f_j}\in \mathfrak{B}$
such that $u_{f_j}\to u_f$ monotonically increasingly. Then
$u_f\in\mathfrak{B}$.
\end{theorem}
\begin{proof}
Define first 
\begin{equation}
\label{letsdefinethisfirsthere}
\mathfrak{R}(u_f):= \bigcap_{u_{f_j}} \mathfrak{R}(u_{f_j}).
\end{equation}
By compactness, we have that $\mathfrak{R}(u_f)\ne\emptyset$.

Consider now $\lambda\in\mathfrak{R}(u_f)$ and consider $(\mathcal{M}(\lambda),g(\lambda))$.
Clearly, $\lambda\in \mathfrak{R}(u_{f_j})$ for  all $u_{f_j}$. Since $u_{f_j}\le u_f$,
by Corollary~\ref{smoothestimforclosedness}, one has  $C^{k+1}$ estimates for all quantities in both
$u_{f_j}$-normalised teleological gauges, independent of $j$. Thus, one can extract a subsequence
converging in all $C^k$ to  limiting $\mathcal{I}^+$ and $\mathcal{H}^+$ normalised metrics
$g_{\mathcal{I}^+}(\lambda)$ and $g_{\mathcal{H}^+}(\lambda)$ 
defined in domains
\begin{equation}
\label{thetwodomainshere}
\mathcal{W}_{\mathcal{I}^+}(u_f)\times\mathbb S^2, \qquad 
\mathcal{W}_{\mathcal{H}^+}(u_f)\times\mathbb S^2, 
\end{equation}
with respect to
\begin{equation}
\label{herealimit}
M_f(u_f,\lambda):= \lim_{j\to\infty} M_f(u_{f_j},\lambda),
\end{equation}
where the existence of this limit is again assured by Arzela--Ascoli.

At this point we have produced normalised metrics satisfying
the assumptions of Sections~\ref{nullinfgaugesec} and~\ref{horizgaugesec}, 
and which can be smoothly attached to another (by applying the limit of the transition functions of the $u_{f_j}$ normalised
gauges), but, we do not yet know that they
are contained in our spacetime $(\mathcal{M}(\lambda),g(\lambda))$,
i.e.~we do not yet know of the existence of the embeddings~\eqref{fortheIplusgauge} and~\eqref{fortheHplusgauge}
defined on the domains~\eqref{thetwodomainshere}.

To argue for this, we first note that, 
again by continuity, 
we may indeed obtain smooth isometric embeddings on the restricted domains
\begin{equation}
\label{restrictedembedone}
\hat{i}_{\mathcal{I}^+} :(\mathcal{W}_{\mathcal{I}^+}(\infty)\times\mathbb S^2\cap \{u<u_f\}\cap \{v<v_\infty(u_f)\}, g_{\mathcal{I}^+}(\lambda)) \to (\mathcal{M}(\lambda), g(\lambda))
\end{equation}
\begin{equation}
\label{restrictedembedtwo}
\hat{i}_{\mathcal{H}^+}: ( \mathcal{W}_{\mathcal{H}^+}(\infty) \times \mathbb S^2
\cap \{u<u_f\} , g_{\mathcal{H}^+}(\lambda)) \to  (\mathcal{M}(\lambda), g(\lambda)),
\end{equation}
by
considering the limit of the embeddings corresponding to $u_{f_j}$.
Let us consider 
the union in $(\mathcal{M}(\lambda), g(\lambda))$ 
of the images of $\hat{i}_{\mathcal{I}^+}$ and $\hat{i}_{\mathcal{H}^+}$
and let us construct a new spacetime with boundary $(\mathcal{M}^{\rm new}_0(\lambda),g^{\rm new}_0(\lambda))$ 
by attaching a future boundary defined from (the  mutually compatible 
in view of our above comments) extensions of $g_{\mathcal{I}^+}$ and $g_{\mathcal{H}^+}$ to $u=u_f$ and
$v=v_\infty(u_f)$, along with parts of the initial data
regions $\mathcal{D}^{\mathcal{K}}$ and $\mathcal{D}_{\mathcal{EF}}^0$ (where
we have here used the preliminary Eddington--Finkelstein gauge  
so as for our solution to be smooth; cf.~Remark~\ref{notsmoothnonlineargauge}). 
Refer to Figure~\ref{forclosedfigure}.
\begin{figure}
\centering{
\def\svgwidth{20pc}
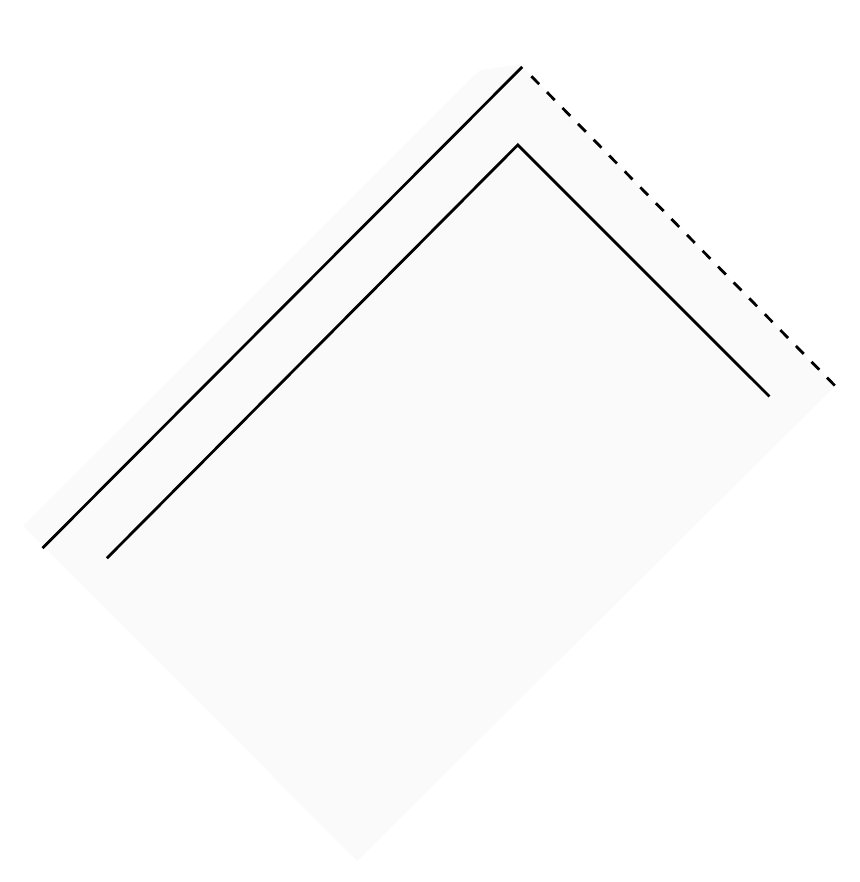}
\caption{The region $\mathcal{D}^0_{\mathcal{EF}}\cup\mathcal{D}^{\mathcal{K}}\cup\mathcal{D}$ and
the extension through local existence}\label{forclosedfigure}
\end{figure}
We may now solve
in the smooth category the obvious characteristic initial value problems~\cite{rendallchar, LukcharIVP} for
the Einstein vacuum equations~\eqref{Ricciflathere}, 
in succession, given by the data on the null future boundary of $\mathcal{M}^{\rm new}_0(\lambda)$  and attach
the maximal solution to $\mathcal{M}^{\rm new}_0$ to form a spacetime $(\mathcal{M}^{\rm new}(\lambda), g^{\rm new}(\lambda))$.
In particular, applying~\cite{LukcharIVP}, $\mathcal{M}^{\rm new}$ contains 
a full neighbourhood of the data including the terminal sphere.
We thus have obtained a smooth vacuum spacetime $\mathcal{M}^{\rm new}$ which is easily seen to be a development
of the original initial data $\mathcal{S}(\lambda)$. It follows then that
this new spacetime is contained in $(\mathcal{M}(\lambda), g(\lambda))$ by the uniqueness
of the maximal Cauchy development as stated in Theorem~\ref{maxCauchythe}. 

Finally
it follows that the embeddings~\eqref{restrictedembedone} and~\eqref{restrictedembedtwo}
can be extended to  embeddings on the entire domains~\eqref{thetwodomainshere} into
$(\mathcal{M}(\lambda), g(\lambda))$.
It is clear again by continuity that these
embeddings correspond to anchored gauges $(\mathcal{M}(\lambda), g(\lambda))$
in the sense of Definition~\ref{anchoringdef}, 
with respect to $u_f$ and $M_f(u_f,\lambda)$ is defined by $(\ref{herealimit})$.

The estimates and relations $(\ref{eq:MMinit})$--$(\ref{eq:badiffeo})$ 
follow by continuity for $u_f$.
The newly defined $\mathfrak{R}(u_f)$ from~\eqref{letsdefinethisfirsthere} 
manifestly inherits the monotonicity properties
$(\ref{monotonofparam})$ with any $u_0\le u'_f<u_f$, whereas,
${\bf J}(u_f, \lambda)$ defined by $(\ref{definofboldv})$  manifestly extends continuously
the previously defined ${\bf J}$, and by continuity
manifestly inherits the properties $(\ref{eq:curletalambda})$ for all $\lambda \in \mathfrak{R}(u_f)$
and $(\ref{eq:curletalao})$ for all $\lambda \in \partial\mathfrak{R}(u_f)$.

Finally, defining 
${\bf J}_{u_f}$ by $(\ref{otherboldvdef})$, by continuous dependence, this map is again continuous
as a map $\mathfrak{R}(u_f^0)\to B_{\varepsilon_0/u_f^0}$ and 
inherits the property $(\ref{restrictiondegree})$ on its boundary restriction.

Thus indeed $u_f\in \mathfrak{B}$.
\end{proof}

\begin{remark}
\label{interprettheopenandclosedness}
It follows now immediately from
Theorem~\ref{thm:nonempty}, Theorem~\ref{thm:justopen} and Theorem~\ref{thm:closed}  
that $\mathfrak{B}= [u_f^0,\infty)$.

\end{remark}

\section{Completing the proof and the statements of Theorems~\ref{limitgaugetwo} and~\ref{limitgaugethree}}
\label{theendoflogic}

\stepcounter{testcount}

The bootstrap having ``closed'' we now obtain relatively easily all assertions
of Theorem~\ref{thm:main}. The only part which will need extra argument (deferred to Chapter~\ref{conclusionsection})
is the precise properties of the limiting gauges and the event horizon $\mathcal{H}^+$.
(These properties will be stated as {\bf Theorems~\ref{limitgaugetwo}} and~{\bf \ref{limitgaugethree}}.)

We first identify the seed data $\mathcal{S}$ in the
statement of the theorem.

Since by Theorem~\ref{thm:nonempty}, Theorem~\ref{thm:justopen} and Theorem~\ref{thm:closed}   
we have
$\mathfrak{B}= [u_f^0,\infty)$ (cf.~Remark~\ref{interprettheopenandclosedness}), it follows that
defining
\[
\mathfrak{R}^{\rm final} = \bigcap_{u_{f_j}} \mathfrak{R}(u_{f_j}) 
\]
for some $u_{f_j}\to \infty$,
we have by compactness and $(\ref{monotonofparam})$  that $\mathfrak{R}^{\rm final} \ne\emptyset$ and thus there exists
an 
\[
\lambda^{\rm final}\in \mathfrak{R}^{\rm final}.
\]
We define
\begin{equation}
\label{definitionofthefinaldata}
\mathcal{S}:=\mathcal{S}_0(\lambda^{\rm final}).
\end{equation}

\begin{remark}
Note that we are not here asserting
$ \mathfrak{R}^{\rm final} = \{\lambda^{\rm final}\}$, nor are we examining the continuous dependence of $\lambda^{\rm final}$ 
on the reference seed $\mathcal{S}_0$. See however
Remarks~\ref{mightrefertowork} and~\ref{smoothnessremark}.
\end{remark}

In {\bf Sections~\ref{proofofisec}},~{\bf \ref{proofofiisec}} and~{\bf \ref{proofofiiisec}} below, we shall obtain
statements (i), (ii) and (iii), respectively, of Therorem~\ref{thm:main}

\subsection{Proof of (i) and the statement of Theorem~\ref{limitgaugetwo}}
\label{proofofisec}

Clearly, since $(\mathcal{M}(\lambda^{\rm final}), g(\lambda^{\rm final}))$ 
satisfies the conditions of Definition~\ref{bootstrapsetdef} for all $u_f\in [u_f^0,\infty)$
the first statement of (i) holds concerning the existence of  anchored  $\mathcal{H}^+$
and $\mathcal{I}^+$ normalised gauges holds with parameters $u_f, M_f(u_f)$, where we define
\[
M_f(u_f):=M_f(u_f,\lambda^{\rm final}).
\]

We remark that we can already deduce from this fact
the statement that null infinity is future complete in the sense of Christodoulou~\cite{Chrmil}.
This is because the generators of the final ingoing cones $\underline{C}_{v_\infty}$ of
the $u_f$-normalised $\mathcal{I}^+$ gauge have affine length  $\sim u_f$
(as measured with respect to that gauge), with the constants in $\sim$ independent
of $u_f$. 
But in view of the estimates on the diffeomorphisms to the initial data gauge
(Theorem~\ref{thm:diffestimates}),
then it follows easily that the affine length of
the generators of $\underline{C}_{v_\infty}$ is again $\sim u_f$ as measured and normalised
from some fixed outgoing cone contained
in the region covered by the Eddington--Finkelstein normalised initial data gauge. 
Thus, completeness in the 
sense of~\cite{Chrmil} follows already since we may take $u_f\to \infty$.

The existence of the induced asymptotic $\I$ and $\Hp$ gauges is given by the following

\begin{theorem}
\label{limitgaugeone}
Let $\lambda^{\rm final}$, $(\mathcal{M}(\lambda^{\rm final}), g(\lambda^{\rm final}))$ be as above.
There exist limiting $\I$ and $\Hp$ gauges, i.e.~$C^k$
embeddings~\eqref{limitingembeddinginf}
and~\eqref{limitingembeddinghor}, for a $k\ge 3$, where
\begin{itemize}
\item
the sets $\mathcal{W}_{\I}(\infty, M_{\rm final})$ and $\mathcal{W}_{\Hp}(\infty, M_{\rm final})$
are defined as in~\eqref{WI+} and~\eqref{WH+} with $<\infty$ replacing $\le u_f$ and $\le v_{\infty}$, 
\item
all statements of Definition~\ref{Igaugedefinition} and~\ref{Hgaugedefinition} hold which do not correspond to evaluation
at $u=u_f$ or $v=v_\infty$, where the background Schwarzschild metric is defined with respect to
$M_{\rm final}$ given by~\eqref{Mfinaldefin}. 
\end{itemize}
The  union $\mathcal{D}^{\Hp}_{\infty} \cup\mathcal{D}^{\I}_{\infty} \cup \mathcal{D}^{\mathcal{EF}}\cup\mathcal{D}^{\mathcal{K}}$ satisfies
\begin{equation}
\label{theunionforever}
\mathcal{D}^{\Hp}_{\infty} \cup\mathcal{D}^{\I}_{\infty} \cup \mathcal{D}^{\mathcal{EF}}\cup\mathcal{D}^{\mathcal{K}}=
\bigcup_{u_f\ge u_f^0} \mathcal{D}^{\Hp}_{u_f}\cup \mathcal{D}^{\I}_{u_f}\cup \mathcal{D}^{\mathcal{K}}\cup\mathcal{D}^{\mathcal{EF}}
 \end{equation}
 and
 is a globally hyperbolic subset of $\mathcal{M}(\lambda^{\rm final})$.

Given the finiteness of the higher order weighted norm $(\ref{datanormtobesmall})$ on initial data, 
arbitrary high smoothness of
the asymptotic gauges can be achieved. 
In particular, if~\eqref{datanormtobesmall} is finite for \underline{all} $k$, then
the embeddings~\eqref{limitingembeddinginf} and~\eqref{limitingembeddinghor} are in fact $C^\infty$.
\end{theorem}

\begin{proof}
Let $\lambda^{\rm final}$ be as given and
$(u_f)_i\to \infty$  be a sequence, and denote $(M_f)_i:= M_f((u_f)_i)$.

Consider first the anchored $\mathcal{I}^+$ gauges with parameters $(u_f)_i$, $(M_f)_i$.
These are defined on sets 
\begin{equation}
\label{hereswheretheyredefined}
\mathcal{W}_{\I}((u_f)_i,(M_f)_i)\times \mathbb S^2.
\end{equation}
These sets ``converge'' to $\mathcal{W}_{\I}(\infty, M_{\rm final})\times \mathbb S^2$
in the sense that for all compact subsets 
\begin{equation}
\label{compactXhere}
\mathcal{X}\subset \mathcal{W}_{\I}(\infty, M_{\rm final})\times\mathbb S^2,
\end{equation}
then $\mathcal{X}\subset \mathcal{W}_{\I}((u_f)_i,(M_f)_i)\times \mathbb S^2$ for all $i$ sufficiently large.

On any compact $\mathcal{X}$ as in~\eqref{compactXhere}, we
have uniform $C^{k+1}$ estimates for a $k\ge 3$
for the metric coefficients $\Omega^2$, $g$, $b$  with respect to all
$(u_f)_i,(M_f)_i$ normalised $\mathcal{I}^+$ gauges containing $\mathcal{X}$.

By exhausting $\mathcal{W}_{\I}(\infty, M_{\rm final})\times \mathbb S^2$ by a sequence of
compact subsets $\mathcal{X}_j$, 
it follows by  Arzela--Ascoli  that there exists a subsequence, which we again denote
as $(u_f)_i$, such that on \emph{all} compact subsets~\eqref{compactXhere}, 
the metric coefficients
$\Omega^2_i$, $g_i$, $b_i$ with respect to the $(u_f)_i, (M_f)_i$ normalised gauges converge
in $C^{k}$ to globally defined metric coefficients 
$\Omega^2$, $g$, $b$ on~\eqref{hereswheretheyredefined}.
This defines a metric expressed in double null gauge
which  satisfies the Einstein vacuum equations~\eqref{Ricciflathere} pointwise,
i.e.~we have obtained a vacuum solution 
to the system on the domain~\eqref{hereswheretheyredefined}.
Note that by continuity this metric indeed satisfies all normalisations in Definition~\ref{Igaugedefinition}
which do not correspond to evaluation
at $u=u_f$ or $v=v_\infty$, where the background Schwarzschild metric is defined with respect to
$M_{\rm final}$ given by~\eqref{Mfinaldefin}.

We now consider the $\mathcal{H}^+$ normalised gauges and similarly extract
via a subsequence
a solution to the system on 
$\mathcal{W}_{\mathcal{H}^+}(\infty, M_{\rm final}) \times \mathbb S^2$.
Again, by continuity this metric indeed satisfies all normalisations in Definition~\ref{Hgaugedefinition}
which do not correspond to evaluation
at $u=u_f$, where as before, the background Schwarzschild metric is defined with respect to
$M_{\rm final}$ given by~\eqref{Mfinaldefin}.

For convenience, we may require that both limiting gauges are in fact
limits of a common subsequence. It follows that there is also a sequence of
diffeomorphisms $f_i$ which connect the gauges, with uniform estimates. Thus,
we may also assume that these diffeomorphisms converge to a limiting diffeomorphism.
Thus, the two limiting gauges are related by a $C^k$ diffeomorphism.

Finally, we also have the  diffeomorphisms  between the gauges and the
initial data.  As we have uniform estimates $C^{k+1}$ on these, we may finally assume that
these also converge to limiting $C^k$ diffeomorphisms. 

The transition functions defined by these diffeomorphisms are all compatible as
they inherit this from the compatibility  relations at finite $(u_f)_i$.
It follows that we may define an abstract
spacetime by gluing together the limiting
teleological metrics on their domains and the metrics corresponding
to the initial data regions $\mathcal{D}^{\mathcal{EF}}$ and $\mathcal{D}^{\mathcal{K}}$ on their
domains via these limiting diffeomorphisms.
This is easily seen (see also below) to yield a globally hyperbolic Lorentzian $C^k$  $4$-manifold  
$(\widetilde{\mathcal{M}},\widetilde{g})$ , which admits
the initial $\underline{C}_{\rm in}\cup C_{\rm out}$ with induced data corresponding to parameter $\lambda^{\rm final}$
as a past bifurcate null Cauchy hypersurface.

It follows from the uniqueness statement as stated in Theorem~\ref{maxCauchythe} 
(in the category of $C^k$ Lorentzian manifolds)
that
$(\widetilde{\mathcal{M}},\widetilde{g})$  indeed embeds into the maximal Cauchy
development 
$(\mathcal{M}(\lambda^{\rm final}),g(\lambda^{\rm final}))$.

Finally, we may define~\eqref{limitingembeddinginf} and~\eqref{limitingembeddinghor} 
to be the restriction of the above embedding
to the domains $\mathcal{W}_{\mathcal{I}^+}(\infty, M_{\rm final})\times \mathbb S^2$ and
$\mathcal{W}_{\mathcal{H}^+}(\infty, M_{\rm final}) \times \mathbb S^2$.
Note that the restriction of the embedding to any $\mathcal{X}$ as in~\eqref{compactXhere}
is easily then seen to be
limit of the embeddings corresponding to the sequence $(u_f)_i$, $(M_f)_i$.
We have then that the image of $(\widetilde{\mathcal{M}},\widetilde{g})$ in $(\mathcal{M},g)$ 
under the aforementioned embedding is precisely   the set on the left hand side 
of~\eqref{theunionforever} and 
the relation~\eqref{theunionforever}  easily follows.
Note that since by Proposition~\ref{lotsofhype}, each set in the big union
on the right hand side of~\eqref{theunionforever} is
globally hyperbolic with a common Cauchy hypersurface, this yields (a posteriori) the global
hyperbolicity of   the left hand side of~\eqref{theunionforever}. 
(We could have thus inferred the global hyperbolicity of $(\widetilde{\mathcal{M}},\widetilde{g})$,
as  needed previously, by interpreting this relation in the abstract manifold before applying the embedding.)

Finally, we note that under the assumption that~\eqref{datanormtobesmall} is finite for sufficient high $k$, 
we may show using~\eqref{nonlocalisedstatement} of Theorem~\ref{thehigherordertheorem}
that the manifold $(\widetilde{\mathcal{M}},\widetilde{g})$ above is a $C^{\tilde{k}}$ Lorentzian manifold
where $\tilde{k}\to \infty$ as $k\to \infty$
and thus the embeddings~\eqref{limitingembeddinginf} and~\eqref{limitingembeddinghor} 
are in fact $C^k$. It follows that  if~\eqref{datanormtobesmall} is finite for all $k$, then the 
embeddings~\eqref{limitingembeddinginf} and~\eqref{limitingembeddinghor} are in fact $C^\infty$.
\end{proof}

The two limiting gauges allow us to define null infinity $\mathcal{I}^+$ and the event horizon
$\mathcal{H}^+$. We summarise the additional properties we can obtain concerning
these asymptotic gauges which are relevant to statement (i).

\begin{restatable}[Properties of $\I$ and $\Hp$]{thmfunnynumbers}{limitgaugetwohere}
\label{limitgaugetwo}
Under the assumptions of Theorem~\ref{limitgaugeone}, 
null infinity can be realised as the 
asymptotic null hypersurface $v=\infty$ of the gauge~\eqref{limitingembeddinginf}, 
denoted now simply as $\mathcal{I}^+$, which is future complete
in the sense that the coordinate $u$ along $\mathcal{I}^+$ is Bondi normalised
and $u\to\infty$ along $\mathcal{I}^+$ towards the future.
The `laws of gravitational radiation' can be formulated along $\mathcal{I}^+$,
and $\mathcal{I}^+$ can be endowed with a frame on which
 the final asymptotic shear 
$\Sigma_+=0$ and final centre of mass normalisation ${\bf P}^+_{\ell = 1}=0$, i.e.~satisfying
$(\ref{normalisationsinstatement})$.

We may characterise the domain
\begin{equation}
\label{domainofout}
\mathcal{D}^{\Hp}_{\infty} \cup\mathcal{D}^{\I}_{\infty} \cup \mathcal{D}^{\mathcal{EF}}\cup\mathcal{D}^{\mathcal{K}}
\cap J^-(\mathcal{D}^{\I}_{\infty})
\end{equation}
as the domain of outer communications of $\mathcal{M}$, and we may write:
\[
J^-(\mathcal{I}^+)\cap \mathcal{M} =\mathcal{D}^{\Hp}_{\infty} \cup\mathcal{D}^{\I}_{\infty} \cup \mathcal{D}^{\mathcal{EF}}\cup\mathcal{D}^{\mathcal{K}}
\cap J^-(\mathcal{D}^{\I}_{\infty}).
\]

On the other hand, the future boundary of~\eqref{domainofout} 
in $\mathcal{M}$ is a future affine complete regular
null hypersurface $\mathcal{H}^+$. 
If $(\ref{datanormtobesmall})$ is finite for a certain $k$, then $\mathcal{H}^+$ is a $C^{\tilde{k}}$ hypersurface,
where $\tilde{k}\to\infty$ as $k\to \infty$.
In particular, if $(\ref{datanormtobesmall})$ is finite for all $k$, then $\mathcal{H}^+$ is a smooth hypersurface.
\end{restatable}

The above  theorem, with more detailed properties, 
will be proven in Chapter~\ref{conclusionsection}.
We note that with the above theorem, all statements in (i) have now been obtained.

\subsection{Proof of (ii)}
\label{proofofiisec}

The energy estimates of (ii) follow from the fact that $(\mathcal{M}(\lambda^{\rm final}), g(\lambda^{\rm final}))$
satisfies the properties of Definition~\ref{bootstrapsetdef} for all $u_f^0\le u_f<\infty$, 
i.e.~\eqref{massstaysclose} follows from \eqref{eq:MMinit},
\eqref{gaugesstayclose} follows from \eqref{eq:badiffeo}
and~\eqref{boundanddecay} follows from~\eqref{eq:bamain}.
Finally, the pointwise estimate~\eqref{pointwisestatementintheorem} follows
from~\eqref{elinfestimates} of Theorem~\ref{thm:sobolevandelinfinity}.

Note that the statement concerning replacing $\varepsilon_0^2$ 
on the right hand side of the estimates by the quantity~\eqref{initialfluxesfororbital} follows immediately (cf.~the analogous statement concerning~\eqref{replacewiththisquantity} in Theorem~\ref{existenceofanchoredgaugethe}), since
for the solution corresponding to $\lambda^{\rm final}$, $\varepsilon_0^2$ only arises in the first place
in the context of bounding~\eqref{initialfluxesfororbital}.

\subsection{Proof of (iii) and the statement of Theorem~\ref{limitgaugethree}}
\label{proofofiiisec}

The decay estimates~\eqref{boundanddecayasym} for 
energies in the asymptotic teleological gauges follow immediately from~\eqref{boundanddecay}
and the way these gauges are defined. Similarly, the pointwise estimates~\eqref{asymptpointwisestatementintheorem} from~\eqref{pointwisestatementintheorem}.

The polynomial decay
estimates $(\ref{pointwiseonhorizon})$ and $(\ref{pointwisedecayatinfintity})$ at $\mathcal{H}^+$ and $\mathcal{I}^+$,
respectively, follow from

\begin{restatable}[Polynomial decay along $\I$ and $\Hp$]{thmfunnynumbers}{limitgaugethreehere}
\label{limitgaugethree}
Under the assumptions of Theorem~\ref{limitgaugeone}, 
we have inverse polynomial decay along $\mathcal{H}^+$  and $\mathcal{I}^+$,
for instance the pointwise estimates 
\begin{equation}
\label{pointwiseonhorizonagain}
|\Omega^2\alpha_{\mathcal{H}^+} (\infty, v, \cdot ) |\lesssim \varepsilon_0 v^{-1},
\end{equation}
where $\Omega^2 \alpha_{\mathcal{H}^+}(\infty,v,\cdot)$ is defined as the $u\to \infty$ limit of
$\Omega^2 \alpha_{\mathcal{H}^+}$ defined with respect to the asymptotic $\mathcal{H}^+$ gauge~\eqref{limitingembeddinghor},
and
\begin{equation}
\label{pointwisedecayatinfintityagain}
0\le M(u) - M_{\rm final} \lesssim \varepsilon_0^2 u^{-2}, \qquad 
|\Sigma(u,\cdot)| \lesssim \varepsilon_0 u^{-1}, \qquad |\Xi (u, \cdot)| \lesssim \varepsilon_0 u^{-1}, \qquad |{\bf A} (u, \cdot)| \lesssim \varepsilon_0 u^{-1},
\end{equation}
where $M(u)$ denotes the Bondi mass and $\Xi$, $\Sigma$ and ${\bf A}$ are defined in Section~\ref{rescaledquantsec}.
\end{restatable}

As with Theorem~\ref{limitgaugetwo}, the above Theorem, together with more detailed properties, 
will be proven in Chapter~\ref{conclusionsection}. We note already, however, that given the
way  $\mathcal{H}^+$ and $\mathcal{I}^+$ are constructed in Theorem~\ref{limitgaugetwo}, 
the result of Theorem~\ref{limitgaugethree} becomes essentially a direct consequence of
the estimates~\eqref{asymptpointwisestatementintheorem}.

\part{Improving the bootstrap assumptions}
\label{improvingpart}

In this part, we shall give the proof of~\Cref{havetoimprovethebootstrap}.
This constitutes the bulk of the analysis of the present work.

\setcounter{parttocdepth}{0}
\parttoc

We begin in {\bf Chapter~\ref{mainestimatessechere}} with the logic of the proof of~\Cref{havetoimprovethebootstrap}.
This will depend on several subtheorems to be proven in the remaining chapters that
constitute this part. These chapters will be outlined in Chapter~\ref{mainestimatessechere}.

\vskip1pc

\emph{This part depends on essentially all of the notation introduced in Part~\ref{preliminlabel}. This part 
can be understood, however, independently of some of the large-scale architecture of the proof
of Theorem~\ref{thm:main}; in particular, this part is independent of Chapter~\ref{logicoftheproofsection} 
beyond Section~\ref{improvingsectionnewdivision}.}

\chapter{The logic of the proof of \Cref{havetoimprovethebootstrap}}
\label{mainestimatessechere}

In this chapter we shall give the logic of the proof of Theorem~\ref{havetoimprovethebootstrap},
first stated in Section~\ref{improvingsectionnewdivision}. The proof will depend on six subtheorems,
which will be proven in the chapters that follow.

For the convenience of the reader, let us first restate the theorem below:

\improvingthatbootstrap*

\begin{proof}

Below, we consider $u_f\in[u_f^0, \hat{u}_f$], with $\hat{u}_f\in \mathfrak{B}$,
the spacetimes  $(\mathcal{M}(\lambda), g(\lambda))$ 
 for all $\lambda \in \mathfrak{R}(u_f)$,
 and 
we shall refer to the anchored $\I$ and $\Hp$ gauges,  corresponding to parameters
$u_f$, $M_f(u_f,\lambda)$,
whose existence is
ensured by Definition~\ref{bootstrapsetdef}.

We shall break up the main part of the 
proof of Theorem~\ref{havetoimprovethebootstrap}  into a collection of subtheorems,
given as Theorems~\ref{thm:sobolevandelinfinity}--\ref{thm:Hestimates}.
\begin{center}
\begin{tabular}{ |c|c|c| } 
\hline{Theorem} & Purpose & Proof  \\
 \hline
Theorem~\ref{thm:sobolevandelinfinity} & Sobolev inequalities, elliptic estimates and basic pointwise bounds & {\bf Chapter~\ref{elliptandcalcchapter}} \\ 
\hline
Theorem~\ref{thm:relatinggauges} & Estimating diffeomorphisms and relating the gauges  & {\bf Chapter~\ref{chap:comparing}} \\ 
\hline
  Theorem~\ref{thm:PPbarestimates} & Estimating almost gauge invariant quantities: $P$ and \underline{$P$}  & {\bf Chapter~\ref{chapter:psiandpsibar}} \\
   \hline
  Theorem~\ref{thm:alphaalphabarestimates} & Estimating  almost gauge invariant quantities: $\alpha$ and \underline{$\alpha$}  & {\bf Chapter~\ref{moreherechapter}}\\
  \hline
   Theorem~\ref{thm:Iestimates} & Estimating geometric quantities in the $\I$ gauge  & {\bf Chapter~\ref{chap:Iestimates}}\\
\hline
   Theorem~\ref{thm:Hestimates} & Estimating geometric quantities in the $\Hp$ gauge &  {\bf Chapter~\ref{chap:Hestimates}}\\
\hline
\end{tabular}
\end{center}

Theorems~\ref{thm:PPbarestimates} and~\ref{thm:alphaalphabarestimates} will rely on a general framework
for estimating inhomogeneous
tensorial wave equations, left to {\bf Chapter~\ref{RWtypechapter}}.

We proceed to discuss briefly these theorems, giving in particular their statements, and show how they
allow us to complete the proof of Theorem~\ref{havetoimprovethebootstrap}.

\vskip1pc
\emph{The remainder of this chapter is essential reading for Part~\ref{improvingpart}. 
The theorems stated below are proven in their logical order in the subsequent chapters, with each
chapter depending on the the previous. One can, however, read various chapters independently. 
Theorems~\ref{thm:sobolevandelinfinity} and~\ref{thm:relatinggauges} of Chapters~\ref{elliptandcalcchapter} 
and~\ref{chap:comparing} are used in all remaining chapters of Part~\ref{improvingpart}, but the reader
can simply refer back to the statements if they want to focus on other aspects of the argument.
In particular, 
Chapters~\ref{RWtypechapter}--\ref{moreherechapter}, concerning estimates for
almost gauge invariant quantities, constitute a cohesive unit, proving Theorems~\ref{thm:PPbarestimates} and~\ref{thm:alphaalphabarestimates} 
while the later Chapters~\ref{chap:Iestimates} and~\ref{chap:Hestimates} can in turn be largely understood
independently of this unit, though they appeal to the statements of Theorems~\ref{thm:PPbarestimates} and~\ref{thm:alphaalphabarestimates}. The order of the estimates follows closely the situation in linear
theory, and thus we recommend the reader refer also back to~\cite{holzstabofschw}.
We note finally that many aspects of the large-scalar architecture of the theorem is
familiar from other works in double null gauge, see for instance~\cite{Chr, DafLuk1}.}

\section{Theorem~\ref{thm:sobolevandelinfinity}: Sobolev inequalities, elliptic estimates and basic pointwise bounds}
\label{elipticstuffetc}

The first step of the proof of Theorem~\ref{havetoimprovethebootstrap} will be 
to obtain basic Sobolev inequalities and elliptic estimates. These  will then allow us to show pointwise
bounds which, as well as being part of the statement of our main theorem, will be important later for controlling
error terms.

 \begin{restatable}[Sobolev inequalities, elliptic estimates and basic pointwise bounds]{thmchap}{sobolevellipticelinf}
\label{thm:sobolevandelinfinity}
Under the assumptions of Theorem~\ref{havetoimprovethebootstrap}, 
then for  all $u_f\in [u_f^0,\hat{u}_f]$ and all $\lambda\in \mathfrak{R}(u_f)$,
with the gauges as defined above,
we have the Sobolev inequality
\begin{equation}
\label{mostbasicSobolev}
		\sup_{\theta \in S_{u,v}} \vert \xi (u,v,\theta) \vert
		\lesssim
		\sum_{k=0}^2 \Vert (r\nablaslash)^k \xi \Vert_{S_{u,v}},
	\end{equation}
for $\xi$ an $S$-tensor field in either the $\mathcal{H}^+$ or $\mathcal{I}^+$ gauges,
and the pointwise estimates
\begin{equation}
\label{elinfestimates}
\mathbb P^{N-5}_{u_f}[\Phi^{\mathcal{H}^+}] +\mathbb P^{N-5}_{u_f}[\Phi^{\mathcal{I}^+}] \lesssim \varepsilon,
\end{equation}
in particular
the  zeroth order estimate 
\begin{equation}\label{hereimprovedbyalot}
	\sup_{\mathcal{D}^{\Hp}}\left(|r^{-2} \slashed g^{\mathcal{H}^+}-\mathring\gamma|_{\mathring \gamma} +
	|\Omega^{-2}_\circ \Omega^2_{\mathcal{H}^+}-1|\right)+ 
		\sup_{\mathcal{D}^{\I}}\left( r|r^{-2}\slashed g^{\mathcal{I}^+} -\mathring\gamma|_{\mathring \gamma} 
		 + |r(\Omega^{-2}_{\circ}\Omega^2_{\mathcal{I}^+}-1)| \right)
		  \lesssim \varepsilon.
		\end{equation}
		
In addition, we have the Sobolev inequalities on cones
given by Propositions~\ref{prop:Sobolevin} and~\ref{prop:Sobolevout},  
the Poincar\'e estimates of Proposition~\ref{prop:Poincare} and the elliptic estimates of
Propositions~\ref{prop:ellipticestimates} and~\ref{prop:divcurl} and the estimates for nonlinear errors associated
to mode projection given by Propositions~\ref{prop:modesdifference},~\ref{prop:almostevalue},~\ref{prop:com0},~\ref{prop:com0oneforms},~\ref{prop:divcurlmodes} and~\ref{prop:Gaussl1}.
\end{restatable}

\begin{remark}
\label{alreadyimprovedheretoo}
We note that estimate~\eqref{hereimprovedbyalot}  already yields  the
improved~\eqref{improvementpointwise} of Theorem~\ref{havetoimprovethebootstrap},
for $\hat{\varepsilon}_0$ sufficiently small.
\end{remark}

\section{Theorem~\ref{thm:relatinggauges}: Estimating diffeomorphisms and relating the gauges}
\label{relatingtheguagessection}

The next step of the proof of Theorem~\ref{havetoimprovethebootstrap} will be to show
that the bootstrap assumptions~\eqref{eq:bamain}  lead to estimates  relating the various gauges.
 
Included in the statement below will be estimates for the diffeomorphisms
 relating the gauges, estimates for the initial data themselves in the teleological
 gauge (recall  Section~\ref{bounded_init_intro} for a discussion of both of these issues)
 as well as estimates for the cancellation of terms arising from different gauges on 
 various domains, including a timelike hypersurface $\mathcal{B}$ (recall Section~\ref{mainestofproofintro} 
 for a discussion of the
 relevance of this issue for the main energy estimates).

 \begin{restatable}[Estimating diffeomorphisms and relating the gauges]{thmchap}{relatinggauges}
\label{thm:relatinggauges}
Under the assumptions of Theorem~\ref{havetoimprovethebootstrap}, 
then  for  all $u_f\in [u_f^0,\hat{u}_f]$ and all $\lambda\in \mathfrak{R}(u_f)$,
with the gauges as defined above,
	the diffeomorphism functions of Section~\ref{specificdiffeos} satisfy the estimates
	\begin{align}
	\label{diffeomorphismest1}
		\mathbb{E}_{u_f}^{N+2}[f_{\Hp,\I}]
		\lesssim
		\
		&
		\mathbb{E}^{N}_{u_f,\Hp}
		+
		\mathbb{E}^{N}_{u_f,\I}
		+ 
		\varepsilon^4,
		\\
	\label{diffeomorphismest2}
		\mathbb{E}_{u_f}[f_{d,\Hp}]
		\lesssim
		\
		&
		\mathbb{E}^{N}_{u_f,\Hp}
		+
		\varepsilon_0^2
		+ 
		\varepsilon^4,
		\\
	\label{diffeomorphismest3}
		\mathbb{E}_{u_f}[f_{d,\I}]
		\lesssim
		\
		&
		\mathbb{E}^{N}_{u_f,\I}
		+
		\varepsilon_0^2
		+ 
		\varepsilon^4,
	\end{align}
	and the mass parameter satisfies the estimate
	\begin{equation}
	\label{hereweimprovethemass}
		\vert M_f(u_f, \lambda) - M_{\rm init} \vert^{2} \lesssim  \mathbb{E}^N_{u_f, \I} + \varepsilon_0^2 + \varepsilon^4,
	\end{equation}
 the  improved inclusions~\eqref{improvedinclus1}--\eqref{improvedinclus2}
and~\eqref{impoverlap3}--\eqref{impoverlap5} are satisfied,
the ``initial'' energies in the teleological gauges are controlled by $\varepsilon_0^2$ according
to the statement of Propositions~\ref{thm:gidataestimates}, the differences
of corresponding quantities in the two teleological gauges satisfy the statements
of Propositions~\ref{thm:cancelT} and~\ref{thm:morecancelnotjustT},
and certain $\Hp$ quantities can be directly estimated from $\I$
quantities according to the statement of Proposition~\ref{thm:inheriting}.
\end{restatable}

 \begin{remark}
We recall here that the energies on the left hand side of~\eqref{diffeomorphismest1},~\eqref{diffeomorphismest2}
and~\eqref{diffeomorphismest3} above are defined 
by~\eqref{EKfHpIpenergy},~\eqref{EfdHpenergy} and~\eqref{EfdIpenergy}, respectively,
in Section~\ref{energiesofdiffeossec} and are the same ones
appearing in formula~\eqref{improvement3} of the statement of Theorem~\ref{havetoimprovethebootstrap}. 
Thus, the above theorem will be key to obtaining the improved statement~\eqref{improvement3}.
On the other hand, the right hand side of~\eqref{diffeomorphismest1},~\eqref{diffeomorphismest2}
and~\eqref{diffeomorphismest3} still contains additional energies 
which must be estimated
at later stages of the argument.
\end{remark}

\begin{remark}
\label{alreadyimprovedinclusionsremark}
We note explicitly that the above statement already contains  the improved
 inclusions~\eqref{improvedinclus1}--\eqref{improvedinclus2}
and~\eqref{impoverlap3}--\eqref{impoverlap5} of Theorem~\ref{havetoimprovethebootstrap} 
concerning the domains.
 \end{remark}

\section{Theorem~\ref{thm:PPbarestimates}: Estimating almost gauge invariant quantities: $P$ and \underline{$P$}}
\label{improvalgainvsechere}

We have arrived at what can be viewed as the main estimates of the proof
(cf.~the discussion in Section~\ref{mainestofproofintro}):
improving
the estimates on the almost gauge invariant quantities, starting with the  quantities $P$ and $\Pbar$,
which are the key to unraveling the almost gauge invariant hierarchy of Section~\ref{AGIQdef}.

Recall from Section~\ref{TeukandRegsection} that the quantities $P$ and $\Pbar$ satisfy Regge--Wheeler equations. 
Wave equation-type estimates lead to the following:
\begin{restatable}[Estimates for $P$ and $\Pbar$]{thmchap}{PPbarestimates}
\label{thm:PPbarestimates}
Under the assumptions of Theorem~\ref{havetoimprovethebootstrap}, 
	then  for  all $u_f\in [u_f^0,\hat{u}_f]$ and all $\lambda\in \mathfrak{R}(u_f)$,
	with the gauges as defined above, then the following holds:

	The quantities $P_{\Hp}$, $P_{\I}$, $\Pbar_{\Hp}$, $\check{\Pbar}_{\I}$ satisfy the estimates
	\[
		\mathbb{E}_{u_f}^{N-2} [P_{\Hp}, P_{\I}] + \mathbb{E}_{u_f}^{N-2} [\Pbar_{\Hp}, \check{\Pbar}_{\I}]
		\lesssim
		\varepsilon_0^2 +  \varepsilon^3.
	\]
\end{restatable}

\begin{proof}
See Chapter~\ref{chapter:psiandpsibar}. 
\end{proof}

\begin{remark}
We recall that the energies appearing in the statement of the theorem are
defined in Section~\ref{PandPbarenergydefs}.
We remark that this theorem depends on 
some of the
cancellation properties included in Theorem~\ref{thm:relatinggauges}.
We recall in particular the discussions of Section~\ref{null_cond_in_intro} 
and~\ref{photon_sphere_in_intro} concerning the special
difficulties near $\mathcal{I}^+$ and near the photon sphere.
\end{remark}

\section{Theorem~\ref{thm:alphaalphabarestimates}: Estimating  almost gauge invariant quantities: $\alpha$ and \underline{$\alpha$}}
Having obtained improved estimates on $P$ and $\Pbar$, we proceed to unlock the rest of the
almost gauge invariant hierarchy of Section~\ref{AGIQdef}, obtaining in the end estimates
for $\alpha$ and $\alphabar$.

This is the content of:
\begin{restatable}[Estimates for $\alpha$ and $\alphabar$]{thmchap}{alphaalphabarestimates}
 \label{thm:alphaalphabarestimates}
Under the assumptions of Theorem~\ref{havetoimprovethebootstrap}, 
	then   for  all $u_f\in [u_f^0,\hat{u}_f]$ and all $\lambda\in \mathfrak{R}(u_f)$,
	with the gauges as defined above, then the quantities $\alpha_{\Hp}$, $\alpha_{\I}$, $\alphabar_{\Hp}$, $\alphabar_{\I}$ satisfy the estimates
	\[
		\mathbb{E}_{u_f}^N [\alpha_{\Hp}, \alpha_{\I}] + \mathbb{E}_{u_f}^N [\alphabar_{\Hp}, \alphabar_{\I}]
		\lesssim
		\varepsilon_0^2 +  \varepsilon^3.
	\]
\end{restatable}
\begin{proof}
See Chapter~\ref{moreherechapter}.
\end{proof}

\begin{remark}
We recall that the energy appearing in the above theorem are defined
by~\eqref{masterenergya} and \eqref{masterenergyab}.
The quantities $\alpha$ and $\alphabar$ are controlled directly from $P$ and $\Pbar$ (just estimated
in Theorem~\ref{thm:PPbarestimates})
using transport estimates as well as by direct use
of wave equation estimates for the Teukolsky equations they satisfy. We will need again to appeal to 
Theorem~\ref{thm:relatinggauges} 
for cancellations on an appropriate boundary. 
\end{remark}

\section{Theorem~\ref{thm:Iestimates}: Estimating geometric quantities in the $\I$ gauge}
\label{improvingipgaugesec}

Having improved the estimates on the almost gauge invariant quantities, 
the next step in the proof of Theorem~\ref{havetoimprovethebootstrap} 
is to improve the estimates on the remaining quantities.
We first consider the $\I$ gauge.

From $\alpha$, $\alphabar$, the gauge conditions and the estimate \eqref{eq:curletalambda}, the geometric quantities in the $\I$ gauge can be controlled. 

\begin{restatable}[Estimates for geometric quantities in the $\I$ gauge]{thmchap}{Iestimates}
\label{thm:Iestimates}
Under the assumptions of Theorem~\ref{havetoimprovethebootstrap}, 
	then  for  all $u_f\in [u_f^0,\hat{u}_f]$ and all $\lambda\in \mathfrak{R}(u_f)$,
	with the gauges as defined above, it follows that the geometric quantities in the $\I$ gauge satisfy the estimates
	\[
		\mathbb{E}^{N}_{u_f,\I}
		\lesssim
		\varepsilon_0^2 +  \varepsilon^3.
	\]
\end{restatable}

\begin{proof}
See Chapter~\ref{chap:Iestimates}.
\end{proof}

\begin{remark}
We recall that the energy  appearing in the above theorem is defined 
by~\eqref{themasterweuseIplus} in	
		 Sections~\ref{Iplusenergiesmastersec}.
As described in Section~\ref{mainestofproofintro}, we shall estimate these via transport equations,
following the method of~\cite{holzstabofschw},
using the estimates obtained for the almost gauge invariant quantities in 
Theorems~\ref{thm:PPbarestimates} and~\ref{thm:alphaalphabarestimates} above,
together with elliptic estimates, in particular to control top order energies of
the Ricci coefficients~(cf.~Section~\ref{necessity_sec}).
The reader may wish to refer back  to some of the specific nonlinear difficulties
 discussed in Section~\ref{othernonlinearissues}.
\end{remark}

\section{Theorem~\ref{thm:Hestimates}: Estimating  geometric quantities in the $\Hp$ gauge}
\label{improvinghpgaugesec}

Having improved the estimates for the geometric quantities in the $\I$ gauge, the next step of
the proof of Theorem~\ref{havetoimprovethebootstrap}  is to improve the estimates for the geometric
quantities in the $\Hp$ gauge.

The geometric quantities in the $\Hp$ gauge can be controlled using the estimates on $\alpha$, $\alphabar$
(from Theorem~\ref{thm:alphaalphabarestimates}), the gauge conditions, 
and the estimates on the geometric quantities in the $\I$ gauge
(from Theorem~\ref{thm:Iestimates}).

\begin{restatable}[Estimates for geometric quantities in the $\Hp$ gauge]{thmchap}{Hestimates}
\label{thm:Hestimates}
	Under the assumptions of Theorem~\ref{havetoimprovethebootstrap}, 
	then   for  all $u_f\in [u_f^0,\hat{u}_f]$ and  all $\lambda\in \mathfrak{R}(u_f)$,
	with the gauges as defined above, it follows that the geometric quantities in the $\Hp$ gauge satisfy the estimates
	\[
		\mathbb{E}^{N}_{u_f,\Hp}
		\lesssim
		\varepsilon_0^2 + \varepsilon^3.
	\]
\end{restatable}

\begin{proof}
See Chapter~\ref{chap:Hestimates}.
\end{proof}

\begin{remark}
We recall that the energy in the above theorem is defined by~\eqref{Hplusmasterenergy} in Section~\ref{PhiHpenergysec}.
As with Theorem~\ref{thm:Iestimates}, we shall estimate these via transport equations, following the method
of~\cite{holzstabofschw}, together with elliptic estimates.
Note that to start the argument,
we shall appeal also to Theorem~\ref{thm:relatinggauges} to directly relate some of the
$\I$ quantities estimated before to some $\Hp$ quantities on $C_{u_f}$.
Again, the reader may wish to refer back  to some of the specific nonlinear difficulties
 discussed in Section~\ref{othernonlinearissues}.
\end{remark}

\section*{Completing the proof of Theorem~\ref{havetoimprovethebootstrap}}
\label{completingnewsectionkeep}

Recall \eqref{eq:bootstrapconstant}.  Theorems~\ref{thm:PPbarestimates}, \ref{thm:alphaalphabarestimates}, \ref{thm:Iestimates} and~\ref{thm:Hestimates} in particular imply that
\[
	\mathbb{E}_{u_f}^{N-2} [P_{\Hp}, P_{\I}] + \mathbb{E}_{u_f}^{N-2} [\Pbar_{\Hp}, \check{\Pbar}_{\I}]
	+
	\mathbb{E}_{u_f}^N [\alpha_{\Hp}, \alpha_{\I}] + \mathbb{E}_{u_f}^N [\alphabar_{\Hp}, \alphabar_{\I}]
	+
	\mathbb{E}^N_{u_f,\Hp}
	+
	\mathbb{E}^N_{u_f,\I}
	\leq
	C(
	 \varepsilon_0^2
	+
	\varepsilon^3),
\]
for some constant $C$, independent of $\upsilon$.  If follows that, if $\upsilon$ is chosen so that $\upsilon^2 \geq 4C$ and if $\hat\varepsilon_0$ is sufficiently small so that $\hat\varepsilon_0 \leq (4 \upsilon C)^{-1}$, the estimate \eqref{eq:bamain} in fact holds with constant $\frac{1}{2}$, i.\@e.\@
\begin{equation} \label{eq:baimproved}
	\mathbb{E}^{N-2}_{u_f}[P_{\Hp},\Pbar_{\Hp}]
	+
	\mathbb{E}^{N-2}_{u_f} [P_{\I},\check{\Pbar}_{\I}]
	+
	\mathbb{E}^{N}_{u_f}[\alpha_{\Hp},\alpha_{\I}]
	+
	\mathbb{E}^{N}_{u_f}[\alphabar_{\Hp},\alphabar_{\I}]
	+
	\mathbb{E}^N_{u_f,\Hp}
	+
	\mathbb{E}^N_{u_f,\I}
	\leq
	\frac{1}{2} \varepsilon^2.
\end{equation}

Theorems~\ref{thm:Iestimates} and~\ref{thm:Hestimates}, together with Theorem~\ref{thm:relatinggauges}, similarly imply that, provided $\upsilon$ is sufficiently large and $\hat\varepsilon_0$ is sufficiently small,
\begin{equation} \label{eq:fbaimproved}
	\mathbb{E}_{u_f}^{N+2}[f_{\Hp,\I}]
	+
	\mathbb{E}_{u_f}[f_{d,\Hp}]
	+
	\mathbb{E}_{u_f}[f_{d,\I}]
	\leq
	\frac{1}{2} \varepsilon^2
\end{equation}
and
\[
\vert M_f(u_f, \lambda) - M_{\rm init} \vert \leq \frac12\varepsilon^2.
\]

With this, in view also  of Remarks~\ref{alreadyimprovedheretoo} and~\ref{alreadyimprovedinclusionsremark},
the proof of Theorem~\ref{havetoimprovethebootstrap} is now complete.
\end{proof}

\chapter{Sobolev inequalities, pointwise bounds and elliptic estimates: the proof of \\ Theorem~\ref{thm:sobolevandelinfinity}}
\label{elliptandcalcchapter}

This section will prove Theorem~\ref{thm:sobolevandelinfinity}, which
yields our basic pointwise estimates, after showing appropriate Sobolev and Poincar\'e inequalities.
The theorem also collects a variety of additional elliptic estimates and  estimates
for mode projection errors.
We restate
the theorem here.

\sobolevellipticelinf*

\minitoc

We shall first prove  in {\bf Section~\ref{sobolev}} the fundamental Sobolev inequality~\eqref{mostbasicSobolev} 
on spheres
and the propositions giving Sobolev  inequalities
on cones, using
only our most basic zeroth order bootstrap assumption~\eqref{lowerorderpointwise}.
We shall then immediately infer the pointwise bounds~\eqref{elinfestimates} 
in {\bf Section~\ref{section:proofofpointwiseestimates}},
already improving in particular~\eqref{lowerorderpointwise} with~\eqref{hereimprovedbyalot}.
We shall then prove the propositions providing Poincar\'e inequalities and 
elliptic estimates in {\bf Section~\ref{section:ellipticestimates}}.
Finally, we shall prove the propositions giving error estimates
arising from mode projections
in {\bf Section~\ref{section:commutingmodeproj}}. 
\vskip1pc

\noindent\fbox{
    \parbox{6.35in}{
We shall assume throughout the assumptions of~\Cref{havetoimprovethebootstrap}. Let us fix an
arbitrary  $u_f\in[u_f^0, \hat{u}_f$], with $\hat{u}_f\in \mathfrak{B}$,
and fix some $\lambda \in \mathfrak{R}(u_f)$.
All propositions below
shall always refer  
to the anchored $\I$ and $\Hp$ gauges in the  spacetime  $(\mathcal{M}(\lambda), g(\lambda))$,  
corresponding to parameters
$u_f$, $M_f(u_f,\lambda)$,
whose existence is
ensured by Definition~\ref{bootstrapsetdef}.
}}
\vskip1pc
\emph{The estimates here will be used throughout the remaining sections of Part~\ref{improvingpart}.
The reader may wish also to refer to~\cite{Chr}. The reader may wish to
skip Section~\ref{section:commutingmodeproj} on a first reading and refer back as necessary.}

\section{Sobolev inequalities}
\label{sobolev}

We begin with the most basic Sobolev inequality~\eqref{mostbasicSobolev}.

\begin{proposition}[Sobolev inequality on spheres] \label{prop:Sobolev}
	In both the $\Hp$ and $\I$ gauges, for any sphere $S_{u,v}$ and any $S_{u,v}$ tensor field $\xi$,
	\[
		\sup_{\theta \in S_{u,v}} \vert \xi (u,v,\theta) \vert
		\lesssim
		\sum_{k=0}^2 \Vert (r\nablaslash)^k \xi \Vert_{S_{u,v}}.
	\]
\end{proposition}

\begin{proof}
	The proof is standard given the closeness to the round sphere provided by
	the bootstrap assumption~\eqref{lowerorderpointwise}.  See, for example, Chapter 5 of~\cite{Chr}.
\end{proof}

Again, using only~\eqref{lowerorderpointwise}, we can also
infer  the following additional Sobolev inequalities on null cones.

\begin{proposition}[Sobolev inequality on incoming cones] \label{prop:Sobolevin}
	In the $\Hp$ gauge, for any $S$-tensor field $\xi$
	\[
		\sup_{\theta \in S_{u,v}} \vert \xi (u,v,\theta) \vert
		\lesssim
		\sum_{\substack{0 \leq k_1+ k_2 \leq 2 \\ 0 \leq k_2 \leq 1 }}
		\left(
		\Vert (r\nablaslash)^{k_1} (\Omega^{-1} \nablaslash_3)^{k_2} \xi \Vert_{\Cbar_{v}^{\Hp}(u)}
		+
		\Vert (r\nablaslash)^{k_1} (\Omega^{-1} \nablaslash_3)^{k_2} \xi \Vert_{\Cbar_{v}^{\Hp}(u_f-M_{\rm init})}
		\right),
	\]
	for all $(u,v)\in \mathcal{W}_{\mathcal{H}^+}(u_f,M_f)$ such that $(u_f-M_{\rm init},v)\in \mathcal{W}_{\mathcal{H}^+}(u_f,M_f)$.
	In the $\I$ gauge, for any $S$-tensor field $\xi$,
	\[
		\sup_{\theta \in S_{u,v}} \vert \xi (u,v,\theta) \vert
		\lesssim
		\sum_{\substack{0 \leq k_1+ k_2 \leq 2 \\ 0 \leq k_2 \leq 1 }}
		\left(
		\Vert (r\nablaslash)^{k_1} (\Omega^{-1} \nablaslash_3)^{k_2} \xi \Vert_{\Cbar_{v}^{\I}(u)}
		+
		\Vert (r\nablaslash)^{k_1} (\Omega^{-1} \nablaslash_3)^{k_2} \xi \Vert_{\Cbar_{v}^{\I}(u_f-M_{\rm init} \wedge u(R_{-1},v) )}
		\right),
	\]
	for all $(u,v)\in \mathcal{W}_{\mathcal{I}^+}(u_f,v_\infty(u_f),M_f)$ such that $(u_f-M_{\rm init}\wedge u(R_{-1},v),v)\in \mathcal{W}_{\mathcal{I}^+}(u_f,M_f)$,
	where $u_f-M_{\rm init}\wedge u(R_{-1},v) := \min \{ u_f-M_{\rm init}, u(R_{-1},v) \}$.
\end{proposition}

\begin{proof}
	The proof is again standard given~\eqref{lowerorderpointwise}.  See, for example, Chapter 10 of \cite{Chr}.  
	Note that it is necessary to include the terms involving the norms $\Vert \cdot \Vert_{\Cbar_{v}^{\Hp}(u_f-M_{\rm init})}$ and $\Vert \cdot \Vert_{\Cbar_{v}^{\I}(u_f-M_{\rm init} \wedge u(R_{-1},v) )}$ in case $u_f - M_{\rm init} \leq u \leq u_f$ and $u(R_{-1},v) \leq u \leq u(R_{-2},v)$ respectively. Note that, by~\eqref{extracondforsob}, 
	we have $u_f-M_{\rm init}\ge u_1$.
\end{proof}

\begin{remark}
Note that the domains of the teleological gauges do not contain sufficiently long incoming cones too close to $u_{\Hp} = u_f$, $v_{\Hp} = v(R_2,u_f)$ and $u_{\I} = u_{-1}$, $v_{\I} = v(R_{-2},u_{-1})$ 
respectively, and so such values of $(u,v)$  have been excluded in the statement above.
\end{remark}

\begin{proposition}[Sobolev inequality on outgoing cones] \label{prop:Sobolevout}
	In the $\Hp$ gauge, for any $S$-tensor field $\xi$ and all  $(u,v)\in \mathcal{W}_{\mathcal{H}^+}(u_f,M_f)$,
	\[
		\sup_{\theta \in S_{u,v}} \vert \xi(u,v,\theta) \vert
		\lesssim
		\sum_{\substack{0 \leq k_1+ k_2 \leq 2 \\ 0 \leq k_2 \leq 1 }}
		\left(
		\Vert (r\nablaslash)^{k_1} (r \Omega \nablaslash_4)^{k_2} \xi \Vert_{C_{u}^{\Hp}(v)}
		+
		\Vert (r\nablaslash)^{k_1} (r \Omega \nablaslash_4)^{k_2} \xi \Vert_{C_{u}^{\Hp}(v(R,u))}
		\right).
	\]
	In the $\I$ gauge, for any function $S$-tensor field $\xi$ and all
	$(u,v)\in \mathcal{W}_{\mathcal{I}^+}(u_f,v_\infty(u_f),M_f)$ ,
	\[
		\sup_{\theta \in S_{u,v}} \vert \xi (u,v,\theta) \vert
		\lesssim
		\sum_{\substack{0 \leq k_1+ k_2 \leq 2 \\ 0 \leq k_2 \leq 1 }}
		\left(
		\Vert (r\nablaslash)^{k_1} (r \Omega \nablaslash_4)^{k_2} \xi \Vert_{C_{u}^{\I}(v)}
		+
		\Vert (r\nablaslash)^{k_1} (r \Omega \nablaslash_4)^{k_2} \xi \Vert_{C_{u}^{\I}(v_{\infty}-M_{\rm init})}
		\right).
	\]
\end{proposition}

\begin{proof}
	Again see, for example, Chapter 10 of \cite{Chr}.  It is again necessary to include the terms involving the norms $\Vert \cdot \Vert_{C_{u}^{\Hp}(v(R,u))}$ and $\Vert \cdot \Vert_{C_{u}^{\I}(v_{\infty}-M_{\rm init})}$ in case $v(R,u) \leq v \leq v(R_1,u)$ and $v_{\infty} - M_{\rm init} \leq v \leq v_{\rm init}$ respectively. Note that by~\eqref{hereanotherrestriction}, 
we have in particular that $(u, v_{\infty}-M_{\rm init})\in  \mathcal{W}_{\mathcal{I}^+}(u_f,v_\infty(u_f),M_f)$ under
our assumptions.
\end{proof}

\section{Pointwise estimates}
\label{section:proofofpointwiseestimates}

We now prove the estimates~\eqref{elinfestimates}.

\begin{proposition} \label{prop:pointwiseboundhereimp}
The pointwise estimates
\begin{equation}
\label{pointwiseboundhereimp}
\mathbb P^{N-5}[\Phi^{\mathcal{H}^+}] +\mathbb P^{N-5}[\Phi^{\mathcal{I}^+}] \lesssim \varepsilon
\end{equation}
hold, in particular, the improved zeroth order bounds
\[
	\sup_{\mathcal{D}^{\Hp}}\left( |r^{-2} \slashed g^{\mathcal{H}^+}-\mathring\gamma|_{\mathring \gamma}
	+|\Omega^{-2}_\circ \Omega^2_{\mathcal{H}^+}-1| \right)
	 + 
		\sup_{\mathcal{D}^{\I}} \left( r|r^{-2}\slashed g^{\mathcal{I}^+} -\mathring\gamma|_{\mathring \gamma} 
		+ |r(\Omega^{-2}_{\circ}\Omega^2_{\mathcal{I}^+}-1)| \right)
		  \lesssim \varepsilon
\]
and the higher order tangential pointwise bounds
\begin{equation}
\label{higherorderpointwiseboundsformetric}
\sum_{k\le N-5}|(r\nablaslash)^k(r^{-2} \slashed g^{\mathcal{H}^+}-\mathring\gamma)|_{\mathring \gamma}
\lesssim\varepsilon,\qquad 
  \sum_{k\le N-5} |(r\nablaslash)^k(r^{-2}\slashed g^{\mathcal{I}^+} -\mathring\gamma)|_{\mathring \gamma} 
		  \lesssim \varepsilon.
\end{equation}
\end{proposition}
\begin{proof}
This is a direct consequence of the Sobolev inequalities and the bootstrap assumption~\eqref{eq:bamain},
on examination of the relation between the integrands of the energies and the expressions in~\eqref{eq:Ippointwisenorm}
and~\eqref{eq:Hppointwisenorm}.
\end{proof}

\section{Poincar\'e inequalities and elliptic estimates}
\label{section:ellipticestimates}

This section concerns 
Poincare inequalities and 
$L^2$ elliptic estimates for certain operators on the spheres $S_{u,v}$ of the $\I$ and $\Hp$ gauges.

We will use here the pointwise bounds~\eqref{higherorderpointwiseboundsformetric} as well as the
bounds
\begin{equation}
\label{followsfrombootstrapformetricandsphharm}
		\sum_{k=0}^N
		\Vert(r\nablaslash)^k (\gslash - r^2 \mathring\gamma) \Vert_{S_{u,v}}^2
	\lesssim\varepsilon^2, \qquad
		\sum_{k=0}^{N+1}
		\Vert (r\nablaslash)^{k} (Y^1_m - \mathring{Y}^1_m) \Vert_{S_{u,v}}^2 \lesssim\varepsilon^2
\end{equation}
which follow in either of the $\mathcal{H}^+$ or $\mathcal{I}^+$ gauges, 
 from the bootstrap estimates $\mathbb{E}^N_{u_f,\mathcal{H}^+}\leq\varepsilon^2$,
$\mathbb{E}^N_{u_f,\mathcal{I}^+}\leq\varepsilon^2$ contained in~\eqref{eq:bamain}, 
in view of the definitions
of Section~\ref{sec:energiesig} and~\ref{PhiHpenergysec}.

We also note the following stronger bound on curvature
\begin{lemma}
We have, in either of the $\mathcal{H}^+$ or $\mathcal{I}^+$ gauges, the bound
\begin{equation}
\label{higherorderintegratedGausscurvaturebound}
\sum_{k=0}^{N-1}
		\Vert (r\nablaslash)^k K \Vert_{S_{u,v}}^2.
	\lesssim\varepsilon^2
\end{equation}
\end{lemma}
\begin{proof}
We apply the operator $r\nablaslash$ to the Gauss equation~\eqref{eq:Gauss} $N-1$ times, using the 
estimates  $\mathbb{E}^N_{u_f,\mathcal{H}^+}\leq\varepsilon^2$ (respectively,
$\mathbb{E}^N_{u_f,\mathcal{I}^+}\leq\varepsilon^2$) and the pointwise bound~\eqref{pointwiseboundhereimp}
to estimate the resulting terms.
\end{proof}

\subsection{Poincar\'e inequalities}

\begin{proposition}[Poincar\'{e} inequalities] \label{prop:Poincare}
	In both the $\Hp$ and $\I$ gauges, for any sphere $S_{u,v}$,
	\begin{align}
		\sqrt{2-\delta} \Vert
		f_{\ell \geq 1}
		\Vert_{S_{u,v}} 
		&
		\leq
		\Vert
		r\nablaslash f
		\Vert_{S_{u,v}}
		\ \  \textrm{for any function $f$ on $S_{u,v}$,}
	\nonumber \\
		\sqrt{1-\delta} \Vert
		\xi
		\Vert_{S_{u,v}}
		&
		\leq
		\Vert
		r\nablaslash \xi
		\Vert_{S_{u,v}} 
		 \ \ \textrm{for any $1$-form $\xi$ on $S_{u,v}$,}
		\nonumber \\
		\sqrt{2-\delta} \Vert
		\xi
		\Vert_{S_{u,v}}
		&
		\leq
		\Vert
		r\nablaslash \xi
		\Vert_{S_{u,v}}
		 \ \ \textrm{for any symmetric traceless tensor $\xi$ on $S_{u,v}$.} \nonumber
	\end{align}

\end{proposition}

\begin{proof}
Note first that
	\[
		f_{\ell = 0}(u,v) = \Big(\int_{S_{u,v}} \sqrt{\det \gslash} d\theta\Big)^{-1} \int_{S_{u,v}} f(u,v,\theta) \sqrt{\det \gslash} d\theta,
	\]
	and so $f_{\ell \geq 1}$ is equal to $f$ minus its average with respect to the sphere $(S_{u,v},\gslash)$. The first inequality is then standard 
	given the control on the induced metric $\gslash$ in each of the $\Hp$ and $\I$ gauges implied by~\eqref{higherorderpointwiseboundsformetric} and the fact that 
	(by a classical result of Lichnerowicz) the lowest positive eigenvalue of $\left(S_{u,v}, \slashed{g}\right)$ is bounded below by  twice the minimum of its Gaussian curvature (whose difference from $1$ is of course bounded by~\eqref{higherorderpointwiseboundsformetric}). The second and third inequalities are proven most easily by establishing them first for the case of the round metric, where they hold with $\delta=0$.
See for instance Proposition 4.4.4 of~\cite{holzstabofschw} for an explicit treatment of the case of symmetric traceless tensors. In a second step one then uses the closeness assumptions~\eqref{higherorderpointwiseboundsformetric} of the induced metric $\slashed{g}$ to estimate (say for symmetric traceless tensors)
\begin{align}
\Vert r\slashed{\nabla} \xi \Vert_{{S_{u,v},\slashed{g}}} \geq (1-C\varepsilon) \Vert r\mathring{\nablaslash} \xi\Vert_{{S_{u,v},\mathring{\slashed{g}}}} - C\epsilon \Vert \xi\Vert_{{S_{u,v},\mathring{\slashed{g}}}} \geq  \sqrt{2(1-C\varepsilon)}  \Vert \xi\Vert_{{S_{u,v},\mathring{\slashed{g}}}} \geq  \sqrt{2(1-C\varepsilon)}  \Vert \xi\Vert_{{S_{u,v},{\slashed{g}}}} \nonumber \, , 
\end{align}
where a ring denotes reference to the round metric and the constant $C$ may be different in each occurrence. 
\end{proof}

\subsection{Elliptic estimates}

The first proposition concerns the operator $\Dslash_2^*$.

\begin{proposition}[Elliptic estimates for symmetric traceless gradient] \label{prop:ellipticestimates}
	Suppose $\varepsilon_0$ is sufficiently small.  In both the $\Hp$ and $\I$ gauges, for any sphere $S_{u,v}$, any function $h$, and any $2 \leq l \leq N+1$,
	\begin{equation} \label{eq:ellipticest3}
		\sum_{k=0}^l \Vert (r \nablaslash)^k h_{\ell \geq 2} \Vert_{S_{u,v}}
		\lesssim
		\sum_{k=0}^{l-2}
		\Vert (r\nablaslash)^k r^2 \Dslash_2^* \nablaslash h \Vert_{S_{u,v}}.
	\end{equation}
	Moreover, for any $S_{u,v}$ one form $\xi$ and any $1\leq l \leq N+1$,
	\begin{equation} \label{eq:ellipticest4}
		\sum_{k=0}^l \Vert (r \nablaslash)^k \xi_{\ell \geq 2} \Vert_{S_{u,v}}
		\lesssim
		\sum_{k=0}^{l-1}
		\Vert (r\nablaslash)^k r \Dslash_2^* \xi \Vert_{S_{u,v}}.
	\end{equation}
\end{proposition}

\begin{proof}
	Let $\mathring{\nablaslash}$ denote the Levi-Civita connection of the round metric $r^2 \mathring{\gamma}$, and let $\mathring{\Dslash}_2^*$ denote the corresponding symmetric traceless gradient operator.  Moreover given a function $h$, let $h_{\mathring{\ell = 0}}$, $h_{\mathring{\ell = 1}}$, $h_{\mathring{\ell \geq 2}}$ denote the projection of $h$ to its $\ell=0$, $\ell = 1$ and $\ell \geq 2$ modes respectively, as defined by the round sphere.  Explicitly, in the notation of Section \ref{projandthemodessec},
	\[
		h_{\mathring{\ell = 0}}
		=
		\frac{1}{\sqrt{4\pi}}
		\int h \mathring{d\theta},
		\qquad
		h_{\mathring{\ell = 1}}
		=
		\sum_{m=-1}^1
		\int h \mathring{Y}^1_m \mathring{d\theta},
		\qquad
		h_{\mathring{\ell \geq 2}}
		=
		h
		-
		h_{\mathring{\ell = 0}}
		-
		h_{\mathring{\ell = 1}},
	\]
	where $\mathring{d\theta} = \sqrt{\det \gamma} d\theta^1 d \theta^2$.  Consider first \eqref{eq:ellipticest3} for $l=2$.  Recall that, using properties of the round spherical harmonics (see, for example, Proposition 4.\@4.\@2 of~\cite{holzstabofschw}), for any smooth function $h$,
	\[
		\sum_{k=0}^2 \int_{S_{u,v}} \vert (r \mathring{\nablaslash})^k h_{\mathring{\ell \geq 2}} \vert^2 \mathring{d\theta}
		\lesssim
		\int_{S_{u,v}} \vert r^2 \mathring{\Dslash}_2^* \mathring{\nablaslash} h \vert^2 \mathring{d\theta}.
	\]
	Now
	\[
		-2 (\Dslash_2^* \nablaslash - \mathring{\Dslash}_2^* \mathring{\nablaslash})h
		=
		(\nablaslash - \mathring{\nablaslash}) \slashed{d} h
		+
		(\nablaslash^T - \mathring{\nablaslash}^T) \slashed{d} h
		-
		(\divslash - \mathring{\divslash}) \slashed{d} h r^2 \mathring{\gamma}
		-
		\Deltaslash h (\gslash - r^2 \mathring{\gamma}),
	\]
	where $\slashed{d}$ is the exterior derivative of the sphere.  Since $((\nablaslash - \mathring{\nablaslash}) \slashed{d} h)_{AB} = (\mathring{\Gammaslash}_{AB}^C - \Gammaslash_{AB}^C) \slashed{d}_C h$, and $\vert \Gammaslash - \mathring{\Gammaslash} \vert \lesssim \varepsilon r^{-2}$, it follows that
	\[
		\vert r^2 (\Dslash_2^* \nablaslash - \mathring{\Dslash}_2^* \mathring{\nablaslash})h \vert
		\lesssim
		\frac{\varepsilon}{r} \sum_{k=0}^2 \vert (r \mathring{\nablaslash})^k h \vert.
	\]
	Similarly $\sum_{k=0}^2 \vert (r \nablaslash)^k h \vert \lesssim \sum_{k=0}^2 \vert (r \mathring{\nablaslash})^k h \vert$.  Let now $h$ be a function with $h_{\ell = 0} = h_{\ell =1} = 0$.  It follows that,
	\[
		\sum_{k=0}^2 \int_{S_{u,v}} \vert (r \nablaslash)^k h_{\mathring{\ell \geq 2}} \vert^2 d \theta
		\lesssim
		\int_{S_{u,v}} \vert r^2 \Dslash_2^* \nablaslash h \vert^2 d \theta
		+
		\frac{\varepsilon}{r} \sum_{k=0}^2 \int_{S_{u,v}} \vert (r \nablaslash)^k h \vert^2 d \theta.
	\]
	Hence, if $\varepsilon_0$ is sufficiently small,
	\begin{align*}
		\sum_{k=0}^2 \Vert (r \nablaslash)^k h \Vert_{S_{u,v}}
		&
		\lesssim
		\Vert r^2 \Dslash_2^* \nablaslash h \Vert_{S_{u,v}}
		+
		\sum_{k=0}^2 \Vert (r \nablaslash)^k (h_{\ell \geq 2} - h_{\mathring{\ell \geq 2}}) \Vert_{S_{u,v}}
		\nonumber
		\\
		&
		\lesssim
		\Vert r^2 \Dslash_2^* \nablaslash h \Vert_{S_{u,v}}
		+
		\Vert h \Vert_{S_{u,v}}
		\Big(
		\sum_{k=0}^2 \sum_{m=-1}^1
		\Vert (r\nablaslash)^{k} (Y^1_m - \mathring{Y}^1_m)\Vert_{S_{u,v}}
		+
		r^{-2} \Vert \gslash - r^2 \mathring{\gamma} \Vert_{S_{u,v}}
		\Big),
	\end{align*}
	since $h_{\ell \geq 2} - h_{\mathring{\ell \geq 2}} = h_{\mathring{\ell =1}} - h_{\ell =1} + h_{\mathring{\ell =0}} - h_{\ell =0}$.  The estimate \eqref{eq:ellipticest3}, for $l=2$, then follows, using the bound~\eqref{followsfrombootstrapformetricandsphharm} for $\gslash - r^2 \mathring{\gamma}$ and $Y^1_m - \mathring{Y}^1_m$, if $\varepsilon_0$ is sufficiently small since, for any function $h$, $\Dslash_2^* \nablaslash h_{\ell \geq 2} = \Dslash_2^* \nablaslash h$.  For $l \geq 3$ the proof of \eqref{eq:ellipticest3} follows similarly using now the fact that, that, for any smooth function $h$,
	\[
		\sum_{k=0}^l \int_{S_{u,v}} \vert (r \mathring{\nablaslash})^k h_{\mathring{\ell \geq 2}} \vert^2 \mathring{d\theta}
		\lesssim
		\sum_{k=0}^{l-2} \int_{S_{u,v}} \vert (r \mathring{\nablaslash})^k r^2 \mathring{\Dslash}_2^* \mathring{\nablaslash} h \vert^2 \mathring{d\theta}.
	\]
	
	The proof of \eqref{eq:ellipticest4} follows similarly using the decomposition of Proposition \ref{prop:oneformdecomp}.
\end{proof}

The next proposition concerns $L^2$ elliptic estimates for div-curl-trace systems.

\begin{proposition}[Elliptic estimates for div-curl-trace systems] \label{prop:divcurl}
	In either the $\Hp$ or the $\I$ gauge, let $\xi$ be an $S$-tangent totally symmetric $(0,j+1)$ tensor satisfying
	\[
		r \divslash \xi = a,
		\qquad
		r \curlslash \xi = b,
		\qquad
		\tr \xi = c,
	\]
	where $\tr \xi := 0$ if $j=0$.  Then, for any $1\leq l \leq N+1$,
	\begin{equation} \label{eq:divcurl1}
		\sum_{k=0}^l \Vert (r\nablaslash)^k \xi \Vert_{S_{u,v}}
		\lesssim
		\sum_{k=0}^{l-1} \left(
		\Vert (r\nablaslash)^k a \Vert_{S_{u,v}}
		+
		\Vert (r\nablaslash)^k b \Vert_{S_{u,v}}
		+
		\Vert (r\nablaslash)^k c \Vert_{S_{u,v}}
		\right).
	\end{equation}
	If $\xi$ is an $S$-tangent symmetric traceless $(0,2)$ tensor such that
	\[
		r \divslash \xi = a,
	\]
	then
	\begin{equation} \label{eq:divcurl2}
		\sum_{k=0}^l \Vert (r\nablaslash)^k \xi \Vert_{S_{u,v}}
		\lesssim
		\sum_{k=0}^{l-1}
		\Vert (r\nablaslash)^k a \Vert_{S_{u,v}}.
	\end{equation}
	Finally, if $\xi$ is an $S$-tangent symmetric traceless $(0,2)$ tensor such that
	\[
		r^2 \divslash \divslash \xi = a,
		\qquad
		r^2 \curlslash \divslash \xi = b,
	\]
	then, for any $2\leq l \leq N+1$,
	\begin{equation} \label{eq:divcurl3}
		\sum_{k=0}^l \Vert (r\nablaslash)^k \xi \Vert_{S_{u,v}}
		\lesssim
		\sum_{k=0}^{l-2} \left(
		\Vert (r\nablaslash)^k a \Vert_{S_{u,v}}
		+
		\Vert (r\nablaslash)^k b \Vert_{S_{u,v}}
		\right).
	\end{equation}
\end{proposition}

\begin{proof}
	The proof, given the bounds~\eqref{followsfrombootstrapformetricandsphharm} and~\eqref{higherorderintegratedGausscurvaturebound}
	on $\gslash - r^2 \mathring{\gamma}$, is standard and so is only sketched.  For more details see, for example, 
	Section 7.\@3 of~\cite{Chr}.
	
	Consider first \eqref{eq:divcurl1} and note the identity
	\[
		\int_{S_{u,v}} \vert r \nablaslash \xi \vert^2 + (j+1) r^2 K \vert \xi \vert^2 d \theta
		=
		\int_{S_{u,v}} \vert a \vert^2 + \vert b \vert^2 + j r^2 K \vert c \vert^2 d\theta.
	\]
	This immediately yields \eqref{eq:divcurl1} for $l=1$.  For $l\geq 2$, the proof follows inductively by relating the higher order symmetrised gradients of $\xi$,
	\[
		\nablaslash^s \xi_{BA_1 \ldots A_{j+1}}
		:=
		\frac{1}{j+2}
		\left(
		\nablaslash_B \xi_{A_1 \ldots A_{j+1}}
		+
		\sum_{i=1}^{j+1}
		\nablaslash_{A_i} \xi_{A_1 \ldots A_{i-1} B A_{i+1} \ldots A_{j+1}}
		\right),
	\]
	to symmetrised gradients of $a$, $b$ and $c$, and using the estimate~\eqref{higherorderintegratedGausscurvaturebound} on derivatives of the Gauss curvature.
	
	The estimate \eqref{eq:divcurl2} follows from \eqref{eq:divcurl1} and the fact that
	\[
		\curlslash \xi = {}^* \divslash \xi,
	\]
	if $\xi$ is an $S$-tangent symmetric traceless $(0,2)$ tensor.
	
	Finally, to show \eqref{eq:divcurl3}, first note that it follows from \eqref{eq:divcurl1} that
	\[
		\sum_{k=0}^{l-1} \Vert (r\nablaslash)^k r \divslash \xi \Vert_{S_{u,v}}
		\lesssim
		\sum_{k=0}^{l-2} \left(
		\Vert (r\nablaslash)^k a \Vert_{S_{u,v}}
		+
		\Vert (r\nablaslash)^k b \Vert_{S_{u,v}}
		\right).
	\]
	The proof then follows from \eqref{eq:divcurl2}.
\end{proof}

\section{Error estimates arising from mode projections}
\label{section:commutingmodeproj}

In this section, we shall give estimates for various nonlinear error terms which arise 
from different ways of applying mode projections.

In contrast to Section~\ref{section:ellipticestimates} where we only used tangential bounds~\eqref{higherorderpointwiseboundsformetric},~\eqref{followsfrombootstrapformetricandsphharm} and~\eqref{higherorderintegratedGausscurvaturebound} 
on spheres $S_{u,v}$, in the present section, we will appeal to 
additional bounds following from the main bootstrap estimate~\eqref{eq:bamain},
including  order bounds on other Ricci coefficients like the expansion $\Omega\tr \chi$ and the shear $\hat\chi$.

\subsection{Estimates for spherical harmonic functions and their eigenvalues}

First, estimates are given for the differences $Y^1_{m} -\mathring{Y}^1_{m}$ of the $\ell = 1$ spherical harmonics with their round counterparts.  

Recall the definitions of Section \ref{projandthemodessec}.  Note that there is a discrepancy between the number of derivatives estimated in the $\Hp$ and $\I$ gauges in the following proposition.  This discrepancy is due to the fact that, in the $\Hp$ gauge, the estimates for $(r\nablaslash)^N \Omega \hat{\chi}$ and $(r\nablaslash)^N \Omega \tr \chi$ grow in $v$.  Though a more optimal statement, in terms of number of derivatives of $\partial_u Y^1_{m}$ and $\partial_u Y^1_{m}$ estimated, can be made in the $\Hp$ gauge, such a statement is not used anywhere and so only non-optimal estimates are stated for simplicity.

\begin{proposition}[Estimates for $Y^1_{m} -\mathring{Y}^1_{m}$] \label{prop:modesdifference}
	In the $\Hp$ gauge, for all $u_{0} \leq u \leq u_f$, $v_{-1} \leq v \leq v(R_2,u)$, the differences $Y^1_{m} -\mathring{Y}^1_{m}$, for $m=-1,0,1$, for any $l \leq N$,
	\begin{equation} \label{eq:modesdifference}
		\sum_{k=0}^{l+1} \Vert (r \nablaslash)^k ( Y^1_{m} -\mathring{Y}^1_{m} ) \Vert_{S_{u,v}^{\Hp}}
		\lesssim
		\sum_{k=0}^{l}
		\Vert (r \nablaslash)^k (\gslash - r^2 \mathring{\gamma}) \Vert_{S_{u,v}^{\Hp}}
		.
	\end{equation}
	and the derivatives $\partial_u Y^1_{m}$ and $\partial_v Y^1_{m}$, for $m=-1,0,1$, satisfy, for $\vert \gamma \vert \leq N-1-s$, $s=0,1,2$, and $k=0,1$,
	\begin{equation} \label{eq:partialuvmodes}
		\Vert (r\nablaslash)^k \mathfrak{D}^{\gamma} \partial_u Y^1_{m} \Vert_{S_{u,v}^{\Hp}}
		+
		\Vert (r\nablaslash)^k \mathfrak{D}^{\gamma}\partial_v Y^1_{m} \Vert_{S_{u,v}^{\Hp}}
		\lesssim
		\frac{\varepsilon}{v^{\frac{s}{2}}}.
	\end{equation}
	In the $\I$ gauge, for all $u_{-1} \leq u \leq u_f$, $v(R_{-2},u) \leq v \leq v_{\infty}$, the differences $Y^1_{m} -\mathring{Y}^1_{m}$, for $m=-1,0,1$, satisfy, for any $l \leq N+1$,
	\begin{equation} \label{eq:modesdifferenceI}
		\sum_{k=0}^{l+1} \Vert (r \nablaslash)^k ( Y^1_{m} -\mathring{Y}^1_{m} ) \Vert_{S_{u,v}^{\I}}
		\lesssim
		\sum_{k=0}^{l}
		\Vert (r \nablaslash)^k (\gslash - r^2 \mathring{\gamma}) \Vert_{S_{u,v}^{\I}}
		.
	\end{equation}
	and the derivatives $\partial_u Y^1_{m}$ and $\partial_v Y^1_{m}$, for $m=-1,0,1$, satisfy, for $\vert \gamma \vert \leq N-1-s$, $s=0,1,2$, and $k = 0,1,2$,
	\begin{equation} \label{eq:partialuvmodesI}
		r \Vert (r\nablaslash)^k \mathfrak{D}^{\gamma} \partial_u Y^1_{m} \Vert_{S_{u,v}^{\I}}
		+
		r^2 \Vert (r\nablaslash)^k \mathfrak{D}^{\gamma} \partial_v Y^1_{m} \Vert_{S_{u,v}^{\I}}
		\lesssim
		\frac{\varepsilon}{u^{\frac{s}{2}}}.
	\end{equation}
\end{proposition}

\begin{proof}
	Consider first \eqref{eq:modesdifference} and \eqref{eq:modesdifferenceI} for $l = 0$.  First note that,  for $m=-1,0,1$,
	\begin{equation} \label{eq:roundmodeslgeq2}
		\sum_{k=0}^2  \Vert (r\nablaslash)^k (\mathring{Y}^1_m)_{\ell \geq 2} \Vert_{S_{u,v}}
		\lesssim
		\Vert r( \Gammaslash - \mathring{\Gammaslash}) \Vert_{S_{u,v}}
		+
		\Vert \gslash - r^2 \mathring{\gamma} \Vert_{S_{u,v}}
		.
	\end{equation}
	Indeed,
	\begin{equation} \label{eq:modesfirstnote}
		- \Dslash_2^* \nablaslash \mathring{Y}^1_{m}
		=
		\Dslash_2^* \nablaslash ( Y^1_{m} -\mathring{Y}^1_{m} )
		=
		(\mathring{\Dslash}_2^* \mathring{\nablaslash} - \Dslash_2^* \nablaslash) \mathring{Y}^1_{m},
	\end{equation}
	where $\mathring{\nablaslash}$ denotes the Levi-Civita connection of the round metric $r^2 \mathring{\gamma}$, and $\mathring{\Dslash}_2^*$ denotes the corresponding symmetric traceless gradient operator.  As in the proof of Proposition \ref{prop:ellipticestimates}, it follows that
	\[
		\vert r^2 (\mathring{\Dslash}_2^* \mathring{\nablaslash} - \Dslash_2^* \nablaslash) \mathring{Y}^1_{m} \vert
		\lesssim
		r \vert \Gammaslash - \mathring{\Gammaslash} \vert
		+
		\vert \gslash - r^2 \mathring{\gamma} \vert,
	\]
	and so \eqref{eq:roundmodeslgeq2} follows from Proposition \ref{prop:ellipticestimates}.  Similarly, the fact that
	$
		\int \mathring{Y}^1_{m} \sqrt{\det \gamma} d\theta^1 d \theta^2 = 0
	$
	implies that
	\[
		\vert (\mathring{Y}^1_{m})_{\ell = 0} \vert \lesssim \Vert \gslash - r^2 \mathring{\gamma} \Vert_{S_{u,v}}.
	\]
	Recall now the notation of Section \ref{projandthemodessec}.  Since $\breve{Y}^1_{m} := \Pi_{\mathcal{Y}^1} \mathring{Y}^1_{m}$ it follows that $(\breve{Y}^1_{m} - \mathring{Y}^1_{m})_{\ell =1} = 0$ for $m=-1,0,1$.  Moreover,
	\begin{equation} \label{eq:roundmodesl=0}
		\big\vert \Vert \mathring{Y}^1_{m} \Vert_{S_{u,v}} - 1 \big\vert
		\lesssim
		\Vert \gslash - r^2 \mathring{\gamma} \Vert_{S_{u,v}},
	\end{equation}
	and, since $\Vert \mathring{Y}^1_{m} \Vert_{S_{u,v}} = \Vert (\mathring{Y}^1_{m})_{\ell=0} \Vert_{S_{u,v}} + \Vert (\mathring{Y}^1_{m})_{\ell =1} \Vert_{S_{u,v}} + \Vert (\mathring{Y}^1_{m})_{\ell \geq 2} \Vert_{S_{u,v}}$ and $\Vert \breve{Y}^1_{m} \Vert_{S_{u,v}} = \Vert (\mathring{Y}^1_{m})_{\ell = 1} \Vert_{S_{u,v}}$, it follows from \eqref{eq:roundmodeslgeq2} and \eqref{eq:roundmodesl=0} that
	\[
		\big\vert \Vert \breve{Y}^1_{m} \Vert_{S_{u,v}} - 1 \big\vert
		\lesssim
		\Vert r( \Gammaslash - \mathring{\Gammaslash}) \Vert_{S_{u,v}}
		+
		\Vert \gslash - r^2 \mathring{\gamma} \Vert_{S_{u,v}}.
	\]
	Now $Y_{-1}^1 = \Vert \breve{Y}^1_{-1} \Vert_{S_{u,v}}^{-1} \breve{Y}^1_{-1}$ and so
	\[
		\sum_{k=0}^2 \Vert (r \nablaslash)^k ( Y^1_{-1} -\mathring{Y}^1_{-1} ) \Vert_{S_{u,v}}
		\leq
		\sum_{k=0}^2 \Vert (r \nablaslash)^k ( \breve{Y}^1_{-1} -\mathring{Y}^1_{-1} ) \Vert_{S_{u,v}}
		+
		\big\vert \Vert \breve{Y}^1_{-1} \Vert_{S_{u,v}}^{-1} - 1 \big\vert
		\sum_{k=0}^2 \Vert (r \nablaslash)^k \breve{Y}^1_{-1} \Vert_{S_{u,v}}.
	\]
	The estimates \eqref{eq:modesdifference} and \eqref{eq:modesdifferenceI} for $m=-1$ then follow.  Similarly for $Y_{0}^1$ and $Y_{1}^1$.  The proof of \eqref{eq:modesdifference} for $1 \leq l \leq N-1-s$ and the proof of \eqref{eq:modesdifferenceI} for $1 \leq l \leq N-s$ are similar.

	The main difficulty in the estimates \eqref{eq:partialuvmodes} and \eqref{eq:partialuvmodesI} are in the $\ell \geq 2$ modes of $\mathfrak{D}^{\gamma} \partial_v Y^1_m$ and $\mathfrak{D}^{\gamma} \partial_u Y^1_m$.  For the $\ell = 0$ modes, for example, one uses the fact that $(Y^1_m)_{\ell = 0}=0$ and so, in the $\I$ gauge for example,
	 \[
	 	0
		=
		\partial_v ((Y^1_m)_{\ell = 0})
		=
		(\partial_v Y^1_m)_{\ell=0}
		+
		\Big( \int_{S_{u,v}} d \theta \Big)^{-1} \int_{S_{u,v}} Y^1_m (\Omega \tr \chi - \Omega \tr \chi_{\circ})_{\ell \geq 1} d\theta,
	\]
	to which one can apply $\mathfrak{D}^{\gamma}$.  Similarly for the $\ell = 1$ modes, using again the fact that $(\breve{Y}^1_{m} - \mathring{Y}^1_{m})_{\ell =1} = 0$.  Consider, therefore, the estimates \eqref{eq:partialuvmodes} and \eqref{eq:partialuvmodesI} for the $\ell \geq 2$ modes.  For the $\partial_v Y^1_m$ estimate of \eqref{eq:partialuvmodes}, consider first the case that $\vert \gamma \vert = 0$.  First note that \eqref{eq:modesfirstnote} implies that 
	\[
		r^2 \Dslash_2^* \nablaslash \partial_v Y^1_{m}
		=
		[r^2 \Dslash_2^* \nablaslash, \Omega\nablaslash_4] ( Y^1_{m} -\mathring{Y}^1_{m} )
		+
		\Omega \nablaslash_4 \left( r^2 (\mathring{\Dslash}_2^* \mathring{\nablaslash} - \Dslash_2^* \nablaslash) \mathring{Y}^1_{m} \right).
	\]
	As in the proof of Proposition \ref{prop:ellipticestimates},
	\[
		-2 r^2 (\mathring{\Dslash}_2^* \mathring{\nablaslash} - \Dslash_2^* \nablaslash) \mathring{Y}^1_{m}
		=
		2 r (\mathring{\Gammaslash} - \Gammaslash) \cdot r \mathring{\nablaslash} \mathring{Y}^1_{m}
		-
		r^2 (\Deltaslash - \mathring{\Deltaslash}) \mathring{Y}^1_{m} \cdot \gslash
		+
		2 (\gslash - r^2 \mathring{\gamma}) \mathring{Y}^1_{m},
	\]	and moreover
	\begin{align}
		\nablaslash_4 (r(\Gammaslash - \mathring{\Gammaslash}))^C_{AB}
		=
		&
		r \nablaslash_A {(\Omega \chi)_B}^C
		+
		r \nablaslash_B {(\Omega \chi)_A}^C
		-
		r \nablaslash^C (\Omega \chi)_{AB}
		\label{eq:chiprincipal}
		\\
		&
		-
		\Omega {\hat{\chi}_A}^Dr(\Gammaslash - \mathring{\Gammaslash})^C_{DB}
		-
		\Omega {\hat{\chi}_B}^Dr(\Gammaslash - \mathring{\Gammaslash})^C_{AD}
		+
		\Omega {\hat{\chi}_D}^Cr(\Gammaslash - \mathring{\Gammaslash})^D_{AB},
		\nonumber
	\end{align}
	and
	\[
		\Omega \nablaslash_4(\gslash - r^2 \mathring{\gamma})_{AB}
		=
		(\Omega \tr \chi - \Omega \tr \chi_{\circ}) r^2 \mathring{\gamma}_{AB}
		+
		\Omega {\hat{\chi}_A}^C r^2 \mathring{\gamma}_{CB}
		+
		\Omega {\hat{\chi}_B}^C r^2 \mathring{\gamma}_{AC}.
	\]
	The proof of the $\partial_v Y^1_m$ estimate of~\eqref{eq:partialuvmodes} then follows from Lemma~\ref{lem:commutation}, the estimate~\eqref{eq:modesdifference}, Proposition \ref{prop:ellipticestimates} and the bootstrap assumption~\eqref{eq:bamain} on the Ricci coefficients and curvature components.  Consider now the $\mathfrak{D}^{\gamma} \partial_v Y^1_m$ estimate of \eqref{eq:partialuvmodes} for $\vert \gamma \vert \neq 0$, and first the case that $\mathfrak{D}^{\gamma} = (r\nablaslash)^k$, for some $k \leq N-1-s$.  Note that the principal terms above are the $\nablaslash (\Omega \chi)$ terms in \eqref{eq:chiprincipal}.  If $k \leq 2$ then proof is as above.  For $3 \leq k \leq N-1-s$ the proof follows similarly by commuting \eqref{eq:modesfirstnote} with $(r\nablaslash)^{k-2} \partial_v$ and using the fact that
	\[
		\sum_{l \leq N-1-s}
		\big(
		\Vert (r\nablaslash)^l \Omega \hat{\chi} \Vert_{S_{u,v}^{\Hp}}
		+
		\Vert (r\nablaslash)^l (\Omega \tr \chi - \Omega \tr \chi_{\circ}) \Vert_{S_{u,v}^{\Hp}}
		\big)
		\lesssim
		\frac{\varepsilon}{v^{\frac{s}{2}}}.
	\]
	For general $\mathfrak{D}^{\gamma}$ with $\vert \gamma \vert \leq N-1-s$, which contains at least one $\Omega^{-1}\nablaslash_3$ or $r \Omega \nablaslash_4$ derivative, the proof is similar (one even controls $\Vert (r\nablaslash)^2 \mathfrak{D}^{\gamma} \partial_v Y^1_m \Vert_{S_{u,v}^{\Hp}}$), commuting \eqref{eq:modesfirstnote} with $\mathfrak{D}^{\gamma}$ and using the fact that, from~\eqref{eq:bamain}, we have from~\eqref{eq:bamain} the estimate
	\[
		\Vert \mathfrak{D}^{\widetilde{\gamma}} \Omega \hat{\chi} \Vert_{S_{u,v}^{\Hp}}
		+
		\Vert \mathfrak{D}^{\widetilde{\gamma}} (\Omega \tr \chi - \Omega \tr \chi_{\circ}) \Vert_{S_{u,v}^{\Hp}}
		\lesssim
		\frac{\varepsilon}{v^{\frac{s}{2}}},
	\]
	for all $\vert \widetilde{\gamma} \vert \leq N-s$ provided $\mathfrak{D}^{\widetilde{\gamma}}$ contains at least one $\Omega^{-1}\nablaslash_3$ or $r \Omega \nablaslash_4$ derivative.
	
	The proof of the $\partial_u Y^1_m$ estimate of \eqref{eq:partialuvmodes} is similar, and is in fact slightly easier since one has better estimates for $\Omega^{-1} \hat{\chibar}$ than for $\Omega \hat{\chi}$, using now the fact that
	\[
		r^2 \Dslash_2^* \nablaslash \partial_u Y^1_{m}
		=
		[r^2 \Dslash_2^* \nablaslash, \Omega\nablaslash_3] ( Y^1_{m} -\mathring{Y}^1_{m} )
		+
		\Omega \nablaslash_3 \left( r^2 (\mathring{\Dslash}_2^* \mathring{\nablaslash} - \Dslash_2^* \nablaslash) \mathring{Y}^1_{m} \right),
	\]
	and
	\begin{align*}
		\nablaslash_3 (r(\Gammaslash - \mathring{\Gammaslash}))^C_{AB}
		=
		&
		r \nablaslash_A {(\Omega \chibar)_B}^C
		+
		r \nablaslash_B {(\Omega \chibar)_A}^C
		-
		r \nablaslash^C (\Omega \chibar)_{AB}
		\\
		&
		-
		\Omega {\hat{\chibar}_A}^Dr(\Gammaslash - \mathring{\Gammaslash})^C_{DB}
		-
		\Omega {\hat{\chibar}_B}^Dr(\Gammaslash - \mathring{\Gammaslash})^C_{AD}
		+
		\Omega {\hat{\chibar}_D}^Cr(\Gammaslash - \mathring{\Gammaslash})^D_{AB},
	\end{align*}
	and
	\[
		\Omega \nablaslash_3(\gslash - r^2 \mathring{\gamma})_{AB}
		=
		(\Omega \tr \chibar - \Omega \tr \chibar_{\circ}) r^2 \mathring{\gamma}_{AB}
		+
		\Omega {\hat{\chibar}_A}^C r^2 \mathring{\gamma}_{CB}
		+
		\Omega {\hat{\chibar}_B}^C r^2 \mathring{\gamma}_{AC}.
	\]

	The proof of \eqref{eq:partialuvmodesI} is similar, and is again slightly easier due to the better estimates for $\hat{\chi}$ in the $\I$ gauge.  One simply commutes \eqref{eq:modesfirstnote} with $\mathfrak{D}^{\gamma} \partial_u$ and $\mathfrak{D}^{\gamma} \partial_v$ and uses the estimates
	\[
		r \Vert \mathfrak{D}^{\widetilde{\gamma}} \Omega \hat{\chibar} \Vert_{S_{u,v}^{\Hp}}
		+
		r^2 \Vert \mathfrak{D}^{\widetilde{\gamma}} (\Omega \tr \chibar - \Omega \tr \chibar_{\circ}) \Vert_{S_{u,v}^{\Hp}}
		+
		r^2 \Vert \mathfrak{D}^{\widetilde{\gamma}} \Omega \hat{\chi} \Vert_{S_{u,v}^{\Hp}}
		+
		r^2 \Vert \mathfrak{D}^{\widetilde{\gamma}} (\Omega \tr \chi - \Omega \tr \chi_{\circ}) \Vert_{S_{u,v}^{\Hp}}
		\lesssim
		\frac{\varepsilon}{u^{\frac{s}{2}}},
	\]
	for any $\vert \widetilde{\gamma} \vert \leq N-s$, contained in~\eqref{eq:bamain}.  Note the better behaviour in $r$ of the commutator $[r^2 \Dslash_2^* \nablaslash, \Omega\nablaslash_4]$, and of $\nablaslash_4(r(\Gammaslash - \mathring{\Gammaslash}))$ and $\nablaslash_4(\gslash - r^2 \mathring{\gamma})$, compared to $[r^2 \Dslash_2^* \nablaslash, \Omega\nablaslash_3]$, $\nablaslash_3(r(\Gammaslash - \mathring{\Gammaslash}))$ and $\nablaslash_3(\gslash - r^2 \mathring{\gamma})$.
\end{proof}

Recall that the round $\ell = 1$ modes, $\mathring{Y}^1_m$, are eigenvalues of the round spherical Laplacian, $r^2 \mathring{\Deltaslash}$, with eigenvalue $-2$.  The following proposition shows that the $\ell =1$ modes $Y^1_m$ are almost eigenvalues of $r^2 \Deltaslash$ with eigenvalue $-2$.

\begin{proposition}[$Y^1_m$ are almost eigenvalues of $\Deltaslash$] \label{prop:almostevalue}
	In the $\Hp$ gauge, for all $u_{0} \leq u \leq u_f$, $v_{-1} \leq v \leq v(R_2,u)$, the $\ell=1$ modes $Y^1_{m}$, for $m=-1,0,1$, satisfy, for any $\vert \gamma \vert \leq N-1-s$, $s=0,1,2$,
	\begin{equation*}
		\Vert \mathfrak{D}^{\gamma} (r^2 \Deltaslash + 2) Y^1_{m} \Vert_{S_{u,v}^{\Hp}}
		\lesssim
		\frac{\varepsilon}{v^{\frac{s}{2}}}.
	\end{equation*}
	In the $\I$ gauge, for all $u_{-1} \leq u \leq u_f$, $v(R_{-2},u) \leq v \leq v_{\infty}$, the $\ell=1$ modes $Y^1_{m}$, for $m=-1,0,1$, satisfy, for any $\vert \gamma \vert \leq N-s$, $s=0,1,2$
	\[
		\Vert \mathfrak{D}^{\gamma} (r^2 \Deltaslash + 2) Y^1_{m} \Vert_{S_{u,v}^{\I}}
		\lesssim
		\frac{\varepsilon}{ru^{\frac{s}{2}}}.
	\]
\end{proposition}

\begin{proof}
	Proposition \ref{prop:modesdifference} in particular implies that
	\[
		\sum_{k \leq 2} \sum_{\vert \gamma \vert \leq N-1-s} \Vert \mathfrak{D}^{\gamma} (r\nablaslash)^k (Y^1_m - \mathring{Y}^1_m)^{\Hp} \Vert_{S_{u,v}^{\Hp}}
		\lesssim
		\frac{\varepsilon}{v^{\frac{s}{2}}},
		\qquad
		\sum_{k \leq 2} \sum_{\vert \gamma \vert \leq N-s} \Vert \mathfrak{D}^{\gamma} (r\nablaslash)^k (Y^1_m - \mathring{Y}^1_m)^{\I} \Vert_{S_{u,v}^{\I}}
		\lesssim
		\frac{\varepsilon}{ru^{\frac{s}{2}}},
	\]
	in the $\Hp$ gauge and $\I$ gauge respectively.  Since $r^2 \mathring{\Deltaslash} \mathring{Y}^1_m = -2 \mathring{Y}^1_m$, it follows that,
	\[
		(r^2 \Deltaslash + 2) Y^1_m
		=
		r^2 \Deltaslash ( Y^1_m - \mathring{Y}^1_m)
		+
		r^2 (\Deltaslash - \mathring{\Deltaslash}) \mathring{Y}^1_m
		+
		2( Y^1_m - \mathring{Y}^1_m),
	\]
	and, after applying $\mathfrak{D}^{\gamma}$, it follows, as in the proof of Proposition \ref{prop:modesdifference}, that
	\[
		\left\vert \mathfrak{D}^{\gamma} (r^2 \Deltaslash + 2) Y^1_m \right\vert
		\lesssim
		\sum_{k\leq 2} \vert \mathfrak{D}^{\gamma} (r \nablaslash)^k ( Y^1_m - \mathring{Y}^1_m) \vert
		+
		r \vert \mathfrak{D}^{\gamma} (\Gammaslash - \mathring{\Gammaslash}) \vert
		+
		\vert \mathfrak{D}^{\gamma} (\gslash - r^2 \mathring{\gamma}) \vert,
	\]
	and the proof follows from Proposition \ref{prop:modesdifference} and the bootstrap assumptions on $\gslash - r^2 \mathring{\gamma}$ contained in~\eqref{eq:bamain} in the two gauges.
\end{proof}

\subsection{Commuting mode projections with derivative operators}
In this section, estimates are given for the nonlinear error terms which arise when projections of functions and 
$S$-tangent $1$-forms to their $\ell = 0$, $\ell =1$ and $\ell \geq 2$ modes are commuted with certain derivative operators.  See Proposition \ref{prop:com0}, Proposition \ref{prop:com0oneforms} and Proposition \ref{prop:divcurlmodes} below.

The following proposition gives estimates for the nonlinear error terms generated by commuting projections of a function to its $\ell=0$, $\ell=1$ and $\ell \geq 2$ modes with the operators $\Omega \nablaslash_3$ and $\Omega \nablaslash_4$.

\begin{proposition}[Estimates for commutator of mode projections with $\nablaslash_3$ and $\nablaslash_4$ derivatives] \label{prop:com0}
	If $f$ is a function in the $\Hp$ gauge then, for any $u_{0} \leq u \leq u_f$, $v_{-1} \leq v \leq v(R_2,u)$ and any $\vert \gamma \vert \leq N-s$, $s=0,1,2$,
	\begin{multline} \label{eq:nabla3modeprojectionho}
		\left\Vert
		\mathfrak{D}^{\gamma} \left(
		(\Omega \nablaslash_3 f)_{\ell=0}
		-
		\Omega \nablaslash_3 (f_{\ell=0})
		\right)
		\right\Vert_{S_{u,v}^{\Hp}}
		+
		\left\Vert
		\mathfrak{D}^{\gamma} \left(
		(\Omega \nablaslash_3 f)_{\ell=1}
		-
		\Omega \nablaslash_3 (f_{\ell=1})
		\right)
		\right\Vert_{S_{u,v}^{\Hp}}
		\\
		+
		\left\Vert
		\mathfrak{D}^{\gamma} \left(
		(\Omega \nablaslash_3 f)_{\ell\geq 2}
		-
		\Omega \nablaslash_3 (f_{\ell \geq 2})
		\right)
		\right\Vert_{S_{u,v}^{\Hp}}
		\lesssim
		\frac{\varepsilon}{v^{\frac{s}{2}}}
		\sum_{\vert \gamma_1 \vert \leq \vert \gamma \vert }
		\Vert \mathfrak{D}^{\gamma_1} f \Vert^2_{S_{u,v}^{\Hp}},
	\end{multline}
	and
	\begin{multline} \label{eq:nabla4modeprojectionho}
		\left\Vert
		\mathfrak{D}^{\gamma} \left(
		(\Omega \nablaslash_4 f)_{\ell=0}
		-
		\Omega \nablaslash_4 (f_{\ell=0})
		\right)
		\right\Vert_{S_{u,v}^{\Hp}}
		+
		\left\Vert
		\mathfrak{D}^{\gamma} \left(
		(\Omega \nablaslash_4 f)_{\ell=1}
		-
		\Omega \nablaslash_4 (f_{\ell=1})
		\right)
		\right\Vert_{S_{u,v}^{\Hp}}
		\\
		+
		\left\Vert
		\mathfrak{D}^{\gamma} \left(
		(\Omega \nablaslash_4 f)_{\ell\geq 2}
		-
		\Omega \nablaslash_4 (f_{\ell \geq 2})
		\right)
		\right\Vert_{S_{u,v}^{\Hp}}
		\lesssim
		\frac{\varepsilon}{v^{\frac{s}{2}}}
		\sum_{\vert \gamma_1 \vert \leq \vert \gamma \vert }
		\Vert \mathfrak{D}^{\gamma_1} f \Vert_{S_{u,v}^{\Hp}}.
	\end{multline}
	Similarly, if $f$ is a function in the $\I$ gauge then, for any $u_{-1} \leq u \leq u_f$, $v(R_{-2},u) \leq v \leq v_{\infty}$ and any $\vert \gamma \vert \leq N-s$, $s=0,1,2$,
	\begin{multline} \label{eq:nabla3modeprojectionhoI}
		\left\Vert
		\mathfrak{D}^{\gamma} \left(
		(\Omega \nablaslash_3 f)_{\ell=0}
		-
		\Omega \nablaslash_3 (f_{\ell=0})
		\right)
		\right\Vert_{S_{u,v}^{\I}}
		+
		\left\Vert
		\mathfrak{D}^{\gamma} \left(
		(\Omega \nablaslash_3 f)_{\ell=1}
		-
		\Omega \nablaslash_3 (f_{\ell=1})
		\right)
		\right\Vert_{S_{u,v}^{\I}}
		\\
		+
		\left\Vert
		\mathfrak{D}^{\gamma} \left(
		(\Omega \nablaslash_3 f)_{\ell\geq 2}
		-
		\Omega \nablaslash_3 (f_{\ell \geq 2})
		\right)
		\right\Vert_{S_{u,v}^{\I}}
		\lesssim
		\frac{\varepsilon}{r u^{\frac{s}{2}}}
		\sum_{\vert \gamma_1 \vert \leq \vert \gamma \vert }
		\Vert \mathfrak{D}^{\gamma_1} f \Vert_{S_{u,v}^{\I}},
	\end{multline}
	and
	\begin{multline} \label{eq:nabla4modeprojectionhoI}
		\left\Vert
		\mathfrak{D}^{\gamma} \left(
		(\Omega \nablaslash_4 f)_{\ell=0}
		-
		\Omega \nablaslash_4 (f_{\ell=0})
		\right)
		\right\Vert_{S_{u,v}^{\I}}
		+
		\left\Vert
		\mathfrak{D}^{\gamma} \left(
		(\Omega \nablaslash_4 f)_{\ell=1}
		-
		\Omega \nablaslash_4 (f_{\ell=1})
		\right)
		\right\Vert_{S_{u,v}^{\I}}
		\\
		+
		\left\Vert
		\mathfrak{D}^{\gamma} \left(
		(\Omega \nablaslash_4 f)_{\ell\geq 2}
		-
		\Omega \nablaslash_4 (f_{\ell \geq 2})
		\right)
		\right\Vert_{S_{u,v}^{\I}}
		\lesssim
		\frac{\varepsilon}{r^2 u^{\frac{s}{2}}}
		\sum_{\vert \gamma_1 \vert \leq \vert \gamma \vert }
		\Vert \mathfrak{D}^{\gamma_1} f \Vert_{S_{u,v}^{\I}}.
	\end{multline}
\end{proposition}

\begin{proof}
	Consider first \eqref{eq:nabla4modeprojectionhoI}, and recall that, in the $\I$ gauge, the double null frame takes the form
	\[
		e_1 = \partial_{\theta^1},
		\quad
		e_2 = \partial_{\theta^2},
		\quad
		e_3 = \frac{1}{\Omega} \partial_{u},
		\quad
		e_4 = \frac{1}{\Omega} (\partial_v + b^A \partial_{\theta^A}).
	\]
	In what follows $\partial_v f_{\ell=0} = \partial_v (f_{\ell=0})$ etc.
	
	First,
	\[
		\partial_v f_{\ell=0}
		=
		\partial_v \bigg( \Big( \int_{S_{u,v}} d \theta \Big)^{-1} \int_{S_{u,v}} f d\theta \bigg)
		=
		(\partial_v f)_{\ell=0}
		+
		\Big( \int_{S_{u,v}} d \theta \Big)^{-1} \int_{S_{u,v}} f (\Omega \tr \chi - \Omega \tr \chi_{\circ})_{\ell \geq 1} d\theta,
	\]
	as $\partial_v(r^{-2} \sqrt{\det \gslash}) = (\Omega \tr \chi - \Omega \tr \chi_{\circ}) r^{-2} \sqrt{\det \gslash}$, and so
	\[
		\left\vert
		(\Omega \nablaslash_4 f)_{\ell=0}
		-
		\Omega \nablaslash_4 (f_{\ell=0})
		\right\vert
		\lesssim
		\Vert f \Vert_{S_{u,v}} \Vert \Omega \tr \chi - \Omega \tr \chi_{\circ} \Vert_{S_{u,v}}.
	\]
	Similarly, for any multi index $\gamma$,
	\[
		\left\vert
		\mathfrak{D}^{\gamma}
		\big(
		(\Omega \nablaslash_4 f)_{\ell=0}
		-
		\Omega \nablaslash_4 (f_{\ell=0})
		\big)
		\right\vert
		\lesssim
		\sum_{k_2+k_3 + l_2 + l_3 \leq \vert \gamma\vert}
		\Vert \mathfrak{D}^{\underline{k}} f \Vert_{S_{u,v}}
		\Vert \mathfrak{D}^{\underline{l}} ( \Omega \tr \chi - \Omega \tr \chi_{\circ}) \Vert_{S_{u,v}}
	\]
	where $\underline{k}=(0,k_2,k_3)$, $\underline{l}=(0,l_2,l_3)$ and $\mathfrak{D}^{\underline{k}} = (\Omega^{-1} \nablaslash_3)^{k_2} (r\Omega \nablaslash_4)^{k_3}$, $\mathfrak{D}^{\underline{l}} = (\Omega^{-1} \nablaslash_3)^{l_2} (r\Omega \nablaslash_4)^{l_3}$, so that, in particular, no angular derivatives of $\Omega \tr \chi - \Omega \tr \chi_{\circ}$ appear.
	
	For $f_{\ell=1}$, note first that,
	\[
		\partial_v f_{\ell=1}
		=
		\sum_{m=-1}^1 \partial_v \Big( \int_{S_{u,v}} f Y^1_m d\theta Y^1_m \Big),
	\]
	and so
	\[
		\partial_v f_{\ell=1}
		=
		(\partial_v f)_{\ell=1}
		+
		\sum_{m=-1}^1 \left[
		\int_{S_{u,v}}
		f \partial_v Y^1_m + f Y^1_m (\Omega \tr \chi - \Omega \tr \chi_{\circ})
		d\theta Y^1_m
		+
		\int_{S_{u,v}} f Y^1_m d\theta \partial_v Y^1_m
		\right],
	\]
	so that
	\[
		\left\vert
		\Omega \nablaslash_4 f_{\ell=1}
		-
		(\Omega \nablaslash_4 f)_{\ell=1}
		\right\vert
		\lesssim
		\Vert f \Vert_{S_{u,v}}
		\Big(
		\Vert \Omega \tr \chi - \Omega \tr \chi_{\circ} \Vert_{S_{u,v}}
		+
		\sum_{m=-1}^1 \left(
		\Vert \partial_v Y^1_m \Vert_{S_{u,v}}
		+
		\vert \partial_v Y^1_m \vert
		\right)
		\Big).
	\]
	Similarly, for any multi index $\gamma$, by Proposition \ref{prop:modesdifference},
	\begin{multline*}
		\left\Vert
		\mathfrak{D}^{\gamma} \big(
		\Omega \nablaslash_4 f_{\ell=1}
		-
		(\Omega \nablaslash_4 f)_{\ell=1}
		\big)
		\right\Vert_{S_{u,v}}
		\\
		\lesssim
		\sum_{\vert \gamma_1\vert + \vert \gamma_2 \vert \leq \vert \gamma \vert}
		\Vert \mathfrak{D}^{\gamma_1} f \Vert_{S_{u,v}}
		\Big(
		\sum_{l_2+l_3 \leq \vert \gamma_2 \vert}
		\Vert \mathfrak{D}^{\underline{l}} (\Omega \tr \chi - \Omega \tr \chi_{\circ}) \Vert_{S_{u,v}}
		+
		\sum_{m=-1}^1
		\Vert \mathfrak{D}^{\gamma_2} \partial_v Y^1_m \Vert_{S_{u,v}}
		\Big),
	\end{multline*}
	where $\underline{l}=(0,l_2,l_3)$ and $\mathfrak{D}^{\underline{l}} = (\Omega^{-1} \nablaslash_3)^{l_2} (r\Omega \nablaslash_4)^{l_3}$, so that no angular derivatives of $\Omega \tr \chi - \Omega \tr \chi_{\circ}$ appear.  The estimate \eqref{eq:nabla4modeprojectionhoI} then follows from the fact that
	\[
		\Omega \nablaslash_4 f_{\ell \geq 2}
		-
		(\Omega \nablaslash_4 f)_{\ell\geq 2}
		=
		(\Omega \nablaslash_4 f)_{\ell=1}
		-
		\Omega \nablaslash_4 f_{\ell=1}
		+
		(\Omega \nablaslash_4 f)_{\ell=0}
		-
		\Omega \nablaslash_4 f_{\ell=0},
	\]
	the estimate for $\Omega \tr \chi - \Omega \tr \chi_{\circ}$ provided directly by the bootstrap estimate~\eqref{eq:bamain} and Proposition \ref{prop:modesdifference}.  The proof of \eqref{eq:nabla3modeprojectionhoI}, and \eqref{eq:nabla3modeprojectionho} and \eqref{eq:nabla4modeprojectionho}, are similar.
\end{proof}

The next proposition similarly gives estimates for the nonlinear error terms generated by commuting projections of an 
$S$-tangent $1$-form to its $\ell=1$ and $\ell \geq 2$ modes with the operators $\Omega \nablaslash_3$ and $\Omega \nablaslash_4$.

\begin{proposition}[Estimates for commutator of mode projections with $\nablaslash_3$ and $\nablaslash_4$ derivatives for $S$-tangent $1$-forms] \label{prop:com0oneforms}
	If $\xi$ is an $S$-tangent $1$-form in the $\Hp$ gauge then, for any $u_0 \leq u \leq u_f$, $v_{-1} \leq v \leq v(R_2,u)$ and any $\vert \gamma \vert \leq N-s$, $s=0,1,2$,
	\begin{equation} \label{eq:nabla3modeprojectionof}
		\left\Vert
		\mathfrak{D}^{\gamma} \left(
		(\Omega \nablaslash_3 \xi)_{\ell=1}
		-
		\Omega \nablaslash_3 (\xi_{\ell=1})
		\right)
		\right\Vert_{S_{u,v}^{\Hp}}
		+
		\left\Vert
		\mathfrak{D}^{\gamma} \left(
		(\Omega \nablaslash_3 \xi)_{\ell\geq 2}
		-
		\Omega \nablaslash_3 (\xi_{\ell \geq 2})
		\right)
		\right\Vert_{S_{u,v}^{\Hp}}
		\lesssim
		\frac{\varepsilon}{v^{\frac{s}{2}}}
		\sum_{\vert \gamma_1 \vert \leq \vert \gamma \vert }
		\Vert \mathfrak{D}^{\gamma_1} \xi \Vert_{S_{u,v}^{\Hp}},
	\end{equation}
	and
	\begin{equation} \label{eq:nabla4modeprojectionof}
		\left\Vert
		\mathfrak{D}^{\gamma} \left(
		(\Omega \nablaslash_4 \xi)_{\ell=1}
		-
		\Omega \nablaslash_4 (\xi_{\ell=1})
		\right)
		\right\Vert_{S_{u,v}^{\Hp}}
		+
		\left\Vert
		\mathfrak{D}^{\gamma} \left(
		(\Omega \nablaslash_4 \xi)_{\ell \geq 2}
		-
		\Omega \nablaslash_4 (\xi_{\ell \geq 2})
		\right)
		\right\Vert_{S_{u,v}^{\Hp}}
		\lesssim
		\frac{\varepsilon}{v^{\frac{s}{2}}}
		\sum_{\vert \gamma_1 \vert \leq \vert \gamma \vert }
		\Vert \mathfrak{D}^{\gamma_1} \xi \Vert_{S_{u,v}^{\Hp}}.
	\end{equation}
	Similarly, if $\xi$ is an $S$-tangent $1$-form in the $\I$ gauge then, for any $u_{-1} \leq u \leq u_f$, $v(R_{-2},u) \leq v \leq v_{\infty}$ and any $\vert \gamma \vert \leq N-s$, $s=0,1,2$,
	\begin{multline} \label{eq:nabla3modeprojectionofI}
		\left\Vert
		\mathfrak{D}^{\gamma} \left(
		(\Omega \nablaslash_3 \xi)_{\ell=1}
		-
		\Omega \nablaslash_3 (\xi_{\ell=1})
		\right)
		\right\Vert_{S_{u,v}^{\I}}
		+
		\left\Vert
		\mathfrak{D}^{\gamma} \left(
		(\Omega \nablaslash_3 \xi)_{\ell \geq 2}
		-
		\Omega \nablaslash_3 (\xi_{\ell \geq 2})
		\right)
		\right\Vert_{S_{u,v}^{\I}}
		\lesssim
		\frac{\varepsilon}{r u^\frac{s}{2}}
		\sum_{\vert \gamma_1 \vert \leq \vert \gamma \vert }
		\Vert \mathfrak{D}^{\gamma_1} \xi \Vert_{S_{u,v}^{\I}},
	\end{multline}
	and
	\begin{multline} \label{eq:nabla4modeprojectionofI}
		\left\Vert
		\mathfrak{D}^{\gamma} \left(
		(\Omega \nablaslash_4 \xi)_{\ell=1}
		-
		\Omega \nablaslash_4 (\xi_{\ell=1})
		\right)
		\right\Vert_{S_{u,v}^{\I}}
		+
		\left\Vert
		\mathfrak{D}^{\gamma} \left(
		(\Omega \nablaslash_4 \xi)_{\ell \geq 2}
		-
		\Omega \nablaslash_4 (\xi_{\ell \geq 2})
		\right)
		\right\Vert_{S_{u,v}^{\I}}
		\lesssim
		\frac{\varepsilon}{r^2 u^\frac{s}{2}}
		\sum_{\vert \gamma_1 \vert \leq \vert \gamma \vert }
		\Vert \mathfrak{D}^{\gamma_1} \xi \Vert_{S_{u,v}^{\I}}.
	\end{multline}
\end{proposition}

\begin{proof}
	Consider first \eqref{eq:nabla3modeprojectionofI} with $\vert \gamma \vert = 0$.  Recall the notation of Proposition \ref{prop:oneformdecomp} and note that
	\[
		\Omega \nablaslash_3 \xi
		=
		\Omega \nablaslash_3 \left( r\nablaslash h_{1,\xi} + r{}^*\nablaslash h_{2,\xi} \right)
		=
		r\nablaslash \Omega \nablaslash_3 h_{1,\xi} + r{}^*\nablaslash \Omega \nablaslash_3 h_{2,\xi}
		+
		[\Omega\nablaslash_3,r\nablaslash] h_{1,\xi}
		+
		[\Omega\nablaslash_3,r{}^*\nablaslash] h_{2,\xi},
	\]
	and so
	\[
		\left\vert
		(\Omega \nablaslash_3 \xi)_{\ell = 1}
		-
		\left(
		r\nablaslash (\Omega \nablaslash_3 h_{1,\xi})_{\ell=1} + r{}^*\nablaslash (\Omega \nablaslash_3 h_{2,\xi})_{\ell=1}
		\right)
		\right\vert
		\lesssim
		\left\vert
		[\Omega\nablaslash_3,r\nablaslash] h_{1,\xi}
		+
		[\Omega\nablaslash_3,r{}^*\nablaslash] h_{2,\xi}
		\right\vert.
	\]
	Proposition \ref{prop:com0} implies that
	\[
		\left\Vert
		r\nablaslash \Omega \nablaslash_3 (h_{1,\xi})_{\ell=1} + r{}^*\nablaslash \Omega \nablaslash_3 (h_{2,\xi})_{\ell=1}
		-
		\left(
		r\nablaslash (\Omega \nablaslash_3 h_{1,\xi})_{\ell=1} + r{}^*\nablaslash (\Omega \nablaslash_3 h_{2,\xi})_{\ell=1}
		\right)
		\right\Vert_{S_{u,v}}
		\\
		\lesssim
		\frac{\varepsilon}{ru} \Vert \xi \Vert_{S_{u,v}},
	\]
	using the Poincar\'{e} inequality, Proposition \ref{prop:Poincare}, and the fact that $\Vert \xi \Vert^2_{S_{u,v}} = \Vert r\nablaslash h_{1,\xi} \Vert^2_{S_{u,v}} + \Vert r{}^*\nablaslash h_{2,\xi} \Vert^2_{S_{u,v}}$.
	Since $\Omega \nablaslash_3 (\xi_{\ell=1}) = \Omega \nablaslash_3 r\nablaslash (h_{1,\xi})_{\ell=1} + \Omega \nablaslash_3 r{}^*\nablaslash (h_{2,\xi})_{\ell=1}$, it therefore follows that
	\[
		\left\Vert
		(\Omega \nablaslash_3 \xi)_{\ell=1}
		-
		\Omega \nablaslash_3 (\xi_{\ell=1})
		\right\Vert_{S_{u,v}}
		\lesssim
		\left\Vert
		[\Omega\nablaslash_3,r\nablaslash] h_{1,\xi}
		\right\Vert_{S_{u,v}}
		+
		\left\Vert
		[\Omega\nablaslash_3,r{}^*\nablaslash] h_{2,\xi}
		\right\Vert_{S_{u,v}}
		+
		\frac{\varepsilon}{ru} \Vert \xi \Vert_{S_{u,v}},
	\]
	and the proof follows from Lemma \ref{lem:commutation}.
	
	The proof when $\vert \gamma \vert \neq 0$ and the proofs of \eqref{eq:nabla3modeprojectionof}, \eqref{eq:nabla4modeprojectionof} and \eqref{eq:nabla4modeprojectionofI} are similar.
\end{proof}

The final proposition of this section gives estimates for the nonlinear error terms generated by commuting projections of an $S$-tangent $1$-form to its $\ell=1$ and $\ell \geq 2$ modes with the operators $\divslash$ and $\curlslash$.

\begin{proposition}[Estimates for commutator of mode projections with $\divslash$ and $\curlslash$] \label{prop:divcurlmodes}
	If $\xi$ is an $S$-tangent $1$-form in the $\Hp$ gauge then, for any $u_0 \leq u \leq u_f$, $v_{-1} \leq v \leq v(R_2,u)$ and any $\vert \gamma \vert \leq N-1-s$, $s=0,1,2$,
	\begin{multline*}
		\Vert \mathfrak{D}^{\gamma} \big( (r \divslash \xi)_{\ell=1} - r \divslash (\xi_{\ell=1}) \big) \Vert_{S_{u,v}}
		+
		\Vert \mathfrak{D}^{\gamma} \big( (r \divslash \xi)_{\ell \geq 2} - r \divslash (\xi_{\ell \geq 2}) \big) \Vert_{S_{u,v}}
		\\
		+
		\Vert \mathfrak{D}^{\gamma} \big( (r \curlslash \xi)_{\ell=1} - r \curlslash (\xi_{\ell=1}) \big) \Vert_{S_{u,v}}
		+
		\Vert \mathfrak{D}^{\gamma} \big( (r \curlslash \xi)_{\ell \geq 2} - r \curlslash (\xi_{\ell \geq 2}) \big) \Vert_{S_{u,v}}
		\lesssim
		\frac{\varepsilon}{v^{\frac{s}{2}}} \sum_{\vert \gamma_1 \vert \leq \vert \gamma \vert} \Vert \mathfrak{D}^{\gamma_1} \xi \Vert_{S_{u,v}}.
	\end{multline*}
	Similarly, if $\xi$ is an $S$-tangent $1$-form in the $\I$ gauge then, for any $u_{-1} \leq u \leq u_f$, $v(R_{-2},u) \leq v \leq v_{\infty}$ and any $\vert \gamma \vert \leq N-s$, $s=0,1,2$,
	\begin{multline*}
		\Vert \mathfrak{D}^{\gamma} \big( (r \divslash \xi)_{\ell=1} - r \divslash (\xi_{\ell=1}) \big) \Vert_{S_{u,v}}
		+
		\Vert \mathfrak{D}^{\gamma} \big( (r \divslash \xi)_{\ell \geq 2} - r \divslash (\xi_{\ell \geq 2}) \big) \Vert_{S_{u,v}}
		\\
		+
		\Vert \mathfrak{D}^{\gamma} \big( (r \curlslash \xi)_{\ell=1} - r \curlslash (\xi_{\ell=1}) \big) \Vert_{S_{u,v}}
		+
		\Vert \mathfrak{D}^{\gamma} \big( (r \curlslash \xi)_{\ell \geq 2} - r \curlslash (\xi_{\ell \geq 2}) \big) \Vert_{S_{u,v}}
		\lesssim
		\frac{\varepsilon}{ru^{\frac{s}{2}}} \sum_{\vert \gamma_1 \vert \leq \vert \gamma \vert} \Vert \mathfrak{D}^{\gamma_1} \xi \Vert_{S_{u,v}}.
	\end{multline*}
\end{proposition}

\begin{proof}
	Recall the decomposition $\xi = r \nablaslash h_{1,\xi} + r {}^* \nablaslash h_{2,\xi}$ and note that, in both the $\Hp$ and $\I$ gauges,
	\[
		(r \divslash \xi)_{\ell =1} = (r^2 \Deltaslash h_{1,\xi})_{\ell =1}
		=
		\sum_{m=-1}^1 \int_{S_{u,v}} r^2 \Deltaslash h_{1,\xi} \cdot Y^1_m d \theta Y^1_m
		=
		\sum_{m=-1}^1 \int_{S_{u,v}} h_{1,\xi} \cdot r^2 \Deltaslash Y^1_m d \theta Y^1_m,
	\]
	and
	\[
		r\divslash (\xi_{\ell=1})
		=
		r^2 \Deltaslash ((h_{1,\xi})_{\ell=1})
		=
		\sum_{m=-1}^1 \int_{S_{u,v}} h_{1,\xi} \cdot Y^1_m d \theta r^2 \Deltaslash Y^1_m.
	\]
	It follows that
	\[
		(r \divslash \xi)_{\ell =1}
		-
		r\divslash (\xi_{\ell=1})
		=
		\sum_{m=-1}^1
		\left(
		\int_{S_{u,v}} h_{1,\xi} \cdot (r^2 \Deltaslash +2) Y^1_m d \theta Y^1_m
		-
		\int_{S_{u,v}} h_{1,\xi} \cdot Y^1_m d \theta (r^2 \Deltaslash +2) Y^1_m
		\right),
	\]
	and Proposition \ref{prop:almostevalue} implies, for example in the $\I$ gauge,
	\[
		\Vert (r \divslash \xi)_{\ell =1}
		-
		r\divslash (\xi_{\ell=1})
		\Vert_{S_{u,v}}
		\lesssim
		\Vert h_{1,\xi} \Vert_{S_{u,v}}
		\Vert (r^2 \Deltaslash + 2) Y^1_m \Vert_{S_{u,v}}
		\lesssim
		\frac{\varepsilon}{ru} \Vert \xi \Vert_{S_{u,v}},
	\]
	by the Poincar\'{e} inequality, Proposition \ref{prop:Poincare}, and the fact that $\Vert \xi \Vert_{S_{u,v}}^2 = \Vert r\nablaslash h_{1,\xi} \Vert_{S_{u,v}}^2 + \Vert r{}^* \nablaslash h_{2,\xi} \Vert_{S_{u,v}}^2$.  The proof of the first part of the two estimates then follows for the case that $\vert \gamma \vert = 0$.  The remainder of the proof is similar after noting that $(\divslash \xi)_{\ell \geq 2} - \divslash (\xi_{\ell \geq 2}) = \divslash (\xi_{\ell=1}) - (\divslash \xi)_{\ell =1}$ and $r \curlslash \xi = - r^2 \Deltaslash h_{2,\xi}$.
\end{proof}

\subsection{$\ell=1$ modes of Gauss curvature}
\label{ellequalsoneofgauss}

The following proposition shows that, in each of the $\Hp$ and $\I$ gauges, the $\ell=1$ modes of the Gauss curvature $K$ vanish to linear order.

\begin{proposition}[$\ell=1$ modes of the Gauss curvature vanish to linear order] \label{prop:Gaussl1}
	The $\ell=1$ modes of the Gauss curvature, in the $\Hp$ and $\I$ gauges, satisfy the estimates
	\[
		\vert (r^2 K^{\I})_{\ell=1} \vert 
		\lesssim \frac{\varepsilon^2}{r^2 u}  \ \ \ \textrm{and} \ \ \ |(r^2 K^{\Hp})_{\ell=1} \vert 
		\lesssim \frac{\varepsilon^2}{v}, \ \ \ \, 
	\]
	in $\mathcal{D}^{\I}$ and $\mathcal{D}^{\Hp}$ respectively.
\end{proposition}

\begin{proof}
Expanding the Gauss curvature $K$ of the metric $\slashed{g}$ in terms of $\slashed{g}-r^2\mathring\gamma$ we obtain the formula
\begin{align} \label{gaussexpand}
r^2 K  -1 =- \frac{1}{4} \left(r^2 \slashed{\Delta} + 2 \right) tr_{\slashed{g}} \left(\slashed{g} - r^2\mathring\gamma\right) + \frac{1}{2} r^2 \divslash\divslash  \widehat{\left(\slashed{g} - r^2\mathring\gamma\right)} + (r^2 K)_{NL} \, ,
\end{align}
where the non-linear error $(r^2 K)_{NL}$ is a finite sum of terms, each of them at least quadratic in $\left(\slashed{g} - r^2\mathring{\gamma}\right)$ and involving two angular derivatives. Equation (\ref{gaussexpand}) is valid both 
the $\I$ gauge and the $\Hp$ gauge.

In the $\I$ gauge, the aforementioned structure of the non-linear error $(r^2 K)_{NL}$ implies that we have
for $n=0,1,\ldots,N-1$ the estimate\footnote{Only $n=0$ is relevant for the current proof but the estimate for higher $n$ will be used later.}
\begin{align} \label{nlgauss}
\| (r \slashed{\nabla})^n (r^2 K)^{\I}_{NL}\|_{S_{u,v}} \lesssim \frac{\varepsilon^2}{r^2 u} \, .
\end{align}
Recalling the definition~\eqref{themasterweuseIplus},
this is a direct consequence of the $L^2$
bounds for the metric components
given by the
estimate $\mathbb E^N_{u_f,\I}\lesssim \varepsilon$ 
(in turn following from the bootstrap assumption~\eqref{eq:bamain}),
together with the pointwise bounds~\eqref{higherorderpointwiseboundsformetric}.

In the $\Hp$ gauge we have similarly for $n=0,1,\ldots,N-2$ the estimate 
\begin{align} \label{nlgausshoz}
\| (r \slashed{\nabla})^n (r^2 K)^{\Hp}_{NL}\|_{S_{u,v}} \lesssim \frac{\varepsilon^2}{v} \, .
\end{align}
We now multiply (\ref{gaussexpand}) by the $Y^{\ell=1}_{m}$ for $m=-1,0,1$ respectively and integrate over $S_{u,v}$. Integrating the angular derivative operators in the linear terms by parts, we
see that the term involving $\slashed{div} \slashed{div}$ vanishes (this is because for $m=-1,0,1$ the $Y^{\ell=1}_{m}$ are in the kernel of $\slashed{\mathcal{D}}_2^\star \slashed{\nabla}$, which is the adjoint of $\slashed{div} \slashed{div}$) while for the term involving $r^2\slashed{\Delta} +2$ we 
can apply Proposition \ref{prop:almostevalue} to include it in the decay term appearing on the right hand side of the estimate. The proof is complete.
\end{proof}

\chapter{Estimating diffeomorphisms and relating the gauges: the proof of Theorem~\ref{thm:relatinggauges}}
\label{chap:comparing}

This section will prove Theorem~\ref{thm:relatinggauges}, which
estimates the diffeomorphisms relating the two teleological gauges to themselves
and initial data, and gives some additional controls of differences of quantities in the two gauges,
as well as   direct estimates for some $\Hp$ quantities in terms of $\I$ quantities. We restate
the theorem here.

\relatinggauges*

\minitoc

We will first 
prove the estimates~\eqref{diffeomorphismest1}--\eqref{diffeomorphismest3} in {\bf Section~\ref{Heretheproofofdiffest}}.
In {\bf Section~\ref{improveinclusionsec}}, we shall obtain the improved
 inclusions~\eqref{improvedinclus1}--\eqref{improvedinclus2}
and~\eqref{impoverlap3}--\eqref{impoverlap5}  for the regions.
In {\bf Section~\ref{subsubsec:cancelT}}, we shall prove Proposition~\ref{thm:cancelT}, 
which concerns cancellations
between corresponding quantities in the two teleological gauges 
on a well-chosen timelike hypersurface $\mathcal{B}$. 
In {\bf Section~\ref{morecancelationssec}}, we shall prove Proposition~\ref{thm:morecancelnotjustT}, 
which concerns such cancellations when
suitably integrated
on null hypersurfaces and spacetime regions.
In {\bf Section~\ref{Heretheproofofgida}}, we shall prove 
Proposition~\ref{thm:gidataestimates}, controlling ``initial'' energies 
expressed in the teleological gauges in terms of $\varepsilon_0$.
Finally, in {\bf Section~\ref{proofofinheritingthem}}, we shall prove Proposition~\ref{thm:inheriting}, which allows
for the direct estimate of certain $\Hp$ quantities from $\I$ quantities.

\vskip1pc

\noindent\fbox{
    \parbox{6.35in}{
We shall assume throughout the assumptions of~\Cref{havetoimprovethebootstrap}. Let us fix an
arbitrary  $u_f\in[u_f^0, \hat{u}_f$], with $\hat{u}_f\in \mathfrak{B}$,
and fix some $\lambda \in \mathfrak{R}(u_f)$.
All propositions below
shall always refer  
to the anchored $\I$ and $\Hp$ gauges in the  spacetime  $(\mathcal{M}(\lambda), g(\lambda))$,  
corresponding to parameters
$u_f$, $M_f(u_f,\lambda)$,
whose existence is
ensured by Definition~\ref{bootstrapsetdef}.
}}

\vskip1pc
\emph{The results of the present chapter depend on those of Chapter~\ref{elliptandcalcchapter}.
The results of Section~\ref{subsubsec:cancelT}--\ref{Heretheproofofgida}  will be used in 
Chapters~\ref{RWtypechapter}--\ref{moreherechapter}
while the results of Section~\ref{proofofinheritingthem}
will be used in Chapter~\ref{chap:Hestimates}.  The reader can, however,
skip any of the proofs here on a first reading and refer back to the statements as necessary.
}

\section{Estimates for the diffeomorphisms functions}
\label{Heretheproofofdiffest}

In this section we shall prove the following proposition, which gives the first part of Theorem~\ref{thm:relatinggauges}
concerning the estimates for
the energies relating to the diffeomorphism functions.

\begin{proposition}[Estimates for the diffeomorphisms] \label{thm:diffestimates}
	The diffeomorphism functions of Section~\ref{specificdiffeos} satisfy the estimates
	\begin{align*}
		\mathbb{E}_{u_f}^{N+2}[f_{\Hp,\I}]
		\lesssim
		\
		&
		\mathbb{E}^{N}_{u_f,\Hp}
		+
		\mathbb{E}^{N}_{u_f,\I}
		+ 
		\varepsilon^4,
		\\
		\mathbb{E}_{u_f}[f_{d,\Hp}]
		\lesssim
		\
		&
		\mathbb{E}^{N}_{u_f,\Hp}
		+
		\varepsilon_0^2
		+ 
		\varepsilon^4,
		\\
		\mathbb{E}_{u_f}[f_{d,\I}]
		\lesssim
		\
		&
		\mathbb{E}^{N}_{u_f,\I}
		+
		\varepsilon_0^2
		+ 
		\varepsilon^4,
	\end{align*}
	and $M_f = M_f(u_f, \lambda)$ satisfies
	\[
		\vert M_f - M_{\rm init} \vert^2
		\lesssim
		\mathbb{E}^{N}_{u_f,\I}
		+
		\varepsilon_0^2
		+ 
		\varepsilon^4.
	\]
\end{proposition}

\begin{remark}
Let us remark already
that the estimates for higher order derivatives of $f_{\Hp,\I}$ are easier to obtain on the final hypersurface 
$C_{u_f}$
than they are elsewhere.  These higher order estimates are used in controlling derivatives of $\hat{\chi}_{\Hp}$ in terms of derivatives of $\Gamma_{\I}$ (see Proposition~\ref{thm:inheriting} below).  It is only on $C_{u_f}$
 that any Ricci coefficients in the $\Hp$ gauge are estimated in terms of Ricci coefficients in the $\I$ gauge (and no Ricci coefficients in the $\I$ gauge are estimated in terms of Ricci coefficients in the $\Hp$ gauge).  Elsewhere it is only the almost gauge invariant quantities in the two different gauges which are compared, for which the lower order estimates of \eqref{eq:badiffeo}, along with estimates on the Ricci coefficients, suffice.
\end{remark}

Proposition \ref{thm:diffestimates} follows from Proposition \ref{prop:HIdiff}, Proposition \ref{prop:fHIuf}, Proposition \ref{prop:dIdiff}, and Proposition \ref{prop:dHdiff} below.
The diffeomorphisms $f_{\Hp,\I}$ are first estimated, followed by $f_{d,\I}$, and finally $f_{d,\Hp}$.

\subsection{Estimates for $f_{\Hp,\I}$}

Recall the pointwise norm $\mathbb P_{u_f}^{N-7}[f_{\Hp,\I}]$ of the $f_{\Hp,\I}$ diffeomorphism functions, defined in \eqref{eq:diffeopointwisenorm1}.  

\begin{proposition} \label{prop:fHpIppointwise}
	The diffeomorphism functions $f_{\Hp,\I}$ satisfy the pointwise estimate
	\begin{equation} \label{eq:fHpIppointwise}
		\mathbb P_{u_f}^{N-7}[f_{\Hp,\I}]
		\lesssim
		\varepsilon.
	\end{equation}
\end{proposition}

\begin{proof}
	The proof is an immediate consequence of the Sobolev inequality, Proposition \ref{prop:Sobolev}, and the bootstrap assumption \eqref{eq:badiffeo} for the energy $\mathbb{E}_{u_f}^{N+2}[f_{\Hp,\I}]$, defined in \eqref{EKfHpIpenergy}.
\end{proof}

Note that the anchoring conditions of Definition \ref{anchoringdef} (in particular the conditions \eqref{eq:anchoringdefcommoncone} and \eqref{eq:anchoringdefaffixingsphere}) imply that
\begin{equation} \label{eq:fHIuf}
	f_{\Hp,\I}^1(u_f,v,\theta) = f_{\Hp,\I}^2(u_f,v,\theta) = f_{\Hp,\I}^3(u_f,v,\theta) = 0,
\end{equation}
for all $v(R_{-2},u_f) \leq v \leq v(R_3,u_f)$ and $\theta \in \mathbb{S}^2$ such that the above diffeormophism functions are defined.

Recall the nonlinear error notation introduced in Section \ref{subsec:diffeoerrors}, and recall the spacetime region $\mathcal{D}_{\Hp}^{\I}(u)=\DRH\cap \DRI (u)$ and the norm
\[
	\Vert \xi \Vert_{\mathcal{D}_{\Hp}^{\I}(u)} := \Vert \xi \, \mathds{1}_{\DRH\cap \DRI (u)} \Vert_{\DRI}.
\]
The error terms in the estimates for $f_{\Hp,\I}$ are controlled as follows.

\begin{lemma}[Estimates for $f_{\Hp,\I}$ nonlinear error terms] \label{lem:fHIerror}
	For any $u_1 \leq u \leq u_f$ and $v(R_{-2},u) \leq v \leq v(R_{3},v)$,
	\[
		\sum_{k \leq N-4} \!\!
		\Vert (r\nablaslash)^k \mathcal{E}^{1,0}_{\mathfrak{D} \fsc} \mathds{1} \Vert_{S_{u,v}^{\I}}^2
		+
		\Vert (r\nablaslash)^k \mathcal{E}^{1,0}_{\mathfrak{D} \fsc} \Vert_{\mathcal{D}_{\Hp}^{\I}(u)}^2
		+ \!
		\sum_{k \leq N-5} \!\!
		\Vert (r\nablaslash)^k \mathcal{E}^{2,0}_{\mathfrak{D} \fsc} \mathds{1} \Vert_{S_{u,v}^{\I}}^2
		+
		\Vert (r\nablaslash)^k \mathcal{E}^{2,0}_{\mathfrak{D} \fsc} \Vert_{\mathcal{D}_{\Hp}^{\I}(u)}^2
		\lesssim
		\frac{\varepsilon^4}{u^4},
	\]
	where $\mathds{1} = \mathds{1}_{\DRH\cap \DRI (u)}$.
\end{lemma}

\begin{proof}
	The proof is an immediate consequence of the pointwise estimate \eqref{eq:fHpIppointwise} and the fact that, for any $u_0 \leq u \leq u_f$ and $v(R_{-2},u) \leq v \leq v(R_{3},v)$, we have from the bootstrap assumption \eqref{eq:badiffeo} the estimates:
	\[
		\Vert (r\nablaslash)^k \mathfrak{D}^{\gamma} \mathfrak{D} \fsc \, \mathds{1} \Vert^2_{S_{u,v}^{\I}}
		\lesssim
		\frac{\varepsilon^2}{u^2}
		\qquad
		\text{for }
		k \leq N-5, \quad \vert \gamma \vert \leq 1,
	\]
	and
	\[
		\Vert \mathfrak{D}^{\gamma} \Phi^{\Hp} \mathds{1} \Vert^2_{S_{u,v}^{\I}}
		+
		\Vert \mathfrak{D}^{\gamma} \Phi^{\Hp} \Vert^2_{\mathcal{D}_{\Hp}^{\I}(u)}
		\lesssim
		\frac{\varepsilon^2}{u^2}
		\qquad
		\text{for }
		\vert \gamma \vert \leq N-3.
	\]
\end{proof}

\begin{lemma}[Estimates for $f_{\Hp,\I}$ nonlinear error terms on the final hypersurface] \label{lem:fHIhoerror}
	On the final hypersurface $u=u_f$,
	\[
		\sum_{s=0,1,2}
		(u_f)^{s+2}
		\sum_{k=0}^{N-s}
		\int_{v(R_{-1},u_f)}^{v(R_1,u_f)}
		\Vert \nablaslash^{k}
		\mathcal{E}^{2,0}_{\mathfrak{D} \fsc}
		\Vert^2_{S_{u_f,v}^{\I}}
		d v
		\lesssim
		\varepsilon^4.
	\]
\end{lemma}

\begin{proof}
	The proof is again a straightforward consequence of the pointwise estimate \eqref{eq:fHpIppointwise} and the fact that
	from the bootstrap assumption~\eqref{eq:badiffeo} we have
	\[
		\sum_{s=0,1,2}
		(u_f)^{s}
		\Big(
		\sum_{k=0}^{N-s}
		\sum_{\vert \gamma \vert \leq 2}
		\int_{v(R_{-1},u_f)}^{v(R_1,u_f)}
		\big\Vert \nablaslash^{k} \mathfrak{D}^{\gamma} f_{\Hp,\I}\big\Vert^2_{S_{u_f,v}^{\I}}
		d v
		+
		\sum_{\vert \gamma \vert \leq N-s}
		\int_{v(R_{-1},u_f)}^{v(R_1,u_f)}
		\big\Vert \mathfrak{D}^{\gamma} \Phi^{\Hp } \big\Vert^2_{S_{u_f,v}^{\I}}
		d v
		\Big)
		\lesssim
		\varepsilon^2.
	\]
\end{proof}

It is possible to estimate more derivatives of $f_{\Hp,\I}$ than are controlled in the following proposition, though such estimates are not required and, for simplicity, are omitted.

Recall the energy $\mathbb{E}_{u_f}^{N+2}[f_{\Hp,\I}]$ from Section \ref{energiesofdiffeossec} and note that the following two propositions in particular imply that
\[
	\mathbb{E}_{u_f}^{N+2}[f_{\Hp,\I}] \lesssim \mathbb{E}^N_{\I}
	+
	\mathbb{E}^N_{\Hp}
	+
	\varepsilon^4,
\]
and thus complete the first part of the proof of Proposition \ref{thm:diffestimates}.

\begin{proposition}[Estimates for $f_{\Hp,\I}$ diffeomorphisms] \label{prop:HIdiff}
	The diffeomorphisms $f = f_{\Hp,\I}$ satisfy the estimates, for $k \leq N-5$ and for any $u_1 \leq u \leq u_f$, $v(R_{-2},u) \leq v \leq v(R_3,u)$,
	\begin{equation} \label{eq:HIdiff1}
		\sum_{\substack{\vert \gamma \vert \leq 2 \\ k + \vert \gamma \vert \geq 1}}
		\Big(
		\Vert
		(r\nablaslash)^{k} \mathfrak{D}^{\gamma} f^3 \, \mathds{1}
		\Vert_{S_{u,v}^{\I}}^2
		+
		\Vert
		(r\nablaslash)^{k} \mathfrak{D}^{\gamma} f^4 \, \mathds{1}
		\Vert_{S_{u,v}^{\I}}^2
		\Big)
		\lesssim
		\frac{
		\mathbb{E}^N_{\I}
		+
		\mathbb{E}^N_{\Hp}
		+
		\varepsilon^4
		}{u^2},
	\end{equation}
	and
	\begin{equation} \label{eq:HIdiff2}
		\sum_{\vert \gamma \vert \leq 1}
		\Big(
		\Vert
		(r\nablaslash)^{k} \mathfrak{D}^{\gamma} \partial_u \slashed{f} \, \mathds{1}
		\Vert_{S_{u,v}^{\I}}^2
		+
		\Vert
		(r\nablaslash)^{k} \mathfrak{D}^{\gamma} \partial_v \slashed{f} \, \mathds{1}
		\Vert_{S_{u,v}^{\I}}^2
		\Big)
		\lesssim
		\frac{
		\mathbb{E}^N_{\I}
		+
		\mathbb{E}^N_{\Hp}
		+
		\varepsilon^4
		}{u^2},
	\end{equation}
	The $\ell=0$ modes of $f^3$ and $f^4$ satisfy,
	\begin{equation} \label{eq:HIdiff4}
		\vert
		(f^{3} - f^{4})_{\ell =0} \, \mathds{1}
		\vert^2
		+
		\Vert
		(f^{3} - f^{4})_{\ell =0}
		\Vert^2_{\mathcal{D}_{\Hp}^{\I}(u)}
		\lesssim
		\frac{
		\mathbb{E}^N_{\I}
		+
		\mathbb{E}^N_{\Hp}
		+
		\varepsilon^4
		}{u^{2}},
		\qquad
		\vert
		(f^{3} + f^{4})_{\ell =0} \, \mathds{1}
		\vert^2
		\lesssim
		\frac{
		\mathbb{E}^N_{\I}
		+
		\mathbb{E}^N_{\Hp}
		+
		\varepsilon^4
		}{u}.
	\end{equation}
\end{proposition}

\begin{proof}
	Consider first $u$, $v$ satisfying $u_1 \leq u \leq u_f$, $v(R_{-2},u) \leq v \leq v(R_1,u)$.  For such $u$, $v$ the diffeomorphism functions $f_{\Hp,\I}(u,v,\cdot)$ are defined on the whole sphere $\mathbb{S}^2$ in view of the inclusion \eqref{eq:overlap2}.  The proof for such $u$, $v$ follows from Propositions \ref{prop:metricrelations}, \ref{prop:Riccirelations} and \ref{prop:curvaturerelations}, with $(u,v,\theta) = (u_{\Hp}, v_{\Hp}, \theta_{\Hp})$ and $(\widetilde{u},\widetilde{v},\widetilde{\theta}) = (u_{\I}, v_{\I}, \theta_{\I})$ and $M=\widetilde{M} = M_f$, using Lemma \ref{lem:fHIerror} to control the nonlinear error terms.  In what follows $f = f_{\Hp,\I}$.  The subscript is omitted for brevity.
	
	Derivatives of the relations in Propositions \ref{prop:metricrelations}, \ref{prop:Riccirelations} and \ref{prop:curvaturerelations} will be taken, and so Proposition \ref{prop:tildederivStensor}, along with Proposition \ref{prop:tildenorm}, will be used throughout the proof.  The different components of the diffeomorphisms are estimated separately in a reductive manner.  Lemma \ref{lem:fHIerror} is used throughout to estimate nonlinear error terms.

\vskip1pc
\noindent \textbf{Estimate for $f^3_{\ell \geq 1}$:} To begin, $(r\nablaslash)^k f^3_{\ell\geq 1}$ is estimated, for $1 \leq k \leq N-3$, by applying $(r\nablaslash)^{k-1}$ to \eqref{eq:curvaturecomp3} and using Proposition \ref{prop:tildederivStensor} to obtain
\[
	3 \Omega^2_{\circ} \rho_{\circ} (\nablaslash^{\I})^{k} f^3
	=
	\big( \nablaslash^{k-1} \Omega \beta \big)^{\I}
	-
	\Pi_{S^{\I}} \big( \nablaslash^{k-1} \Omega \beta \big)^{\Hp}
	+
	(\nablaslash^{\I})^{k-1} \mathcal{E}^{1,0}_{\mathfrak{D} \fsc}.
\]
Proposition \ref{prop:tildenorm} then implies that
\[
	\vert (\nablaslash^{\I})^{k} f^3 \vert_{\gslash_{\I}}
	\lesssim
	\vert \big( \nablaslash^{k-1} \Omega \beta \big)^{\I} \vert_{\gslash_{\I}}
	+
	\vert \big( \nablaslash^{k-1} \Omega \beta \big)^{\Hp} \vert_{\gslash_{\Hp}}
	+
	\vert (\nablaslash^{\I})^{k-1} \mathcal{E}^{1,0}_{\mathfrak{D} \fsc} \vert_{\gslash_{\I}},
\]
from which it follows that, for $k \leq N-3$,
\begin{equation} \label{eq:fHI1}
	\Vert (r\nablaslash)^k f^3_{\ell\geq 1} \Vert^2_{S_{u,v}^{\I}}
	\lesssim
	u^{-2}
	(
	\mathbb{E}^N_{\I}
	+
	\mathbb{E}^N_{\Hp}
	+
	\varepsilon^4),
\end{equation}
where the estimate for $f^3_{\ell\geq 1}$ itself follows from the Poincar\'{e} inequality, Proposition \ref{prop:Poincare}.

\vskip1pc
\noindent \textbf{Estimate for $f^4_{\ell \geq 1}$:} The quantity $(r\nablaslash)^k f^4_{\ell\geq 1}$ is similarly estimated, for $k \leq N-3$, by applying $(r\nablaslash)^{k-1}$ to \eqref{eq:curvaturecomp4} (and using the Poincar\'{e} inequality, Proposition \ref{prop:Poincare}, for $k=0$) from which it follows that, for $k \leq N-3$,
\begin{equation} \label{eq:fHI2}
	\Vert (r\nablaslash)^k f^4_{\ell\geq 1} \Vert^2_{S_{u,v}^{\I}}
	\lesssim
	u^{-2}
	(
	\mathbb{E}^N_{\I}
	+
	\mathbb{E}^N_{\Hp}
	+
	\varepsilon^4).
\end{equation}

\vskip1pc
\noindent \textbf{Estimate for $\nablaslash_3 f^3_{\ell \geq 1}$:}  Next, $(r\nablaslash)^k \nablaslash_3 f^3_{\ell \geq 1}$ is estimated, for $k \leq N-s$, by applying $(r\nablaslash)^k$ to the relation \eqref{eq:Riccicomp3} (or using the Poincar\'{e} inequality, Proposition \ref{prop:Poincare}, for $k=0$) and using the estimates \eqref{eq:fHI1} and \eqref{eq:fHI2} to obtain, for $k \leq N-4$,
\begin{equation} \label{eq:fHI3}
	\Vert (r\nablaslash)^k \nablaslash_3 f^3_{\ell\geq 1} \Vert^2_{S_{u,v}^{\I}}
	\lesssim
	u^{-2}
	(
	\mathbb{E}^N_{\I}
	+
	\mathbb{E}^N_{\Hp}
	+
	\varepsilon^4).
\end{equation}

\vskip1pc
\noindent \textbf{Estimate for $\nablaslash_4 f^4_{\ell \geq 1}$:}  In a similar way, $(r\nablaslash)^k \nablaslash_4 f^4_{\ell \geq 1}$ is estimated, for $k \leq N-4$, using the relation \eqref{eq:Riccicomp4} to obtain
\begin{equation} \label{eq:fHI4}
	\Vert (r\nablaslash)^k \nablaslash_4 f^4_{\ell\geq 1} \Vert^2_{S_{u,v}^{\I}}
	\lesssim
	u^{-2}
	(
	\mathbb{E}^N_{\I}
	+
	\mathbb{E}^N_{\Hp}
	+
	\varepsilon^4).
\end{equation}
(Note that $r$ weights play no role in the present region and so the term $\frac{\partial f^3}{\partial \widetilde{u}} \widetilde{\nablaslash} \frac{\partial f^4}{\partial \widetilde{v}}$ in \eqref{eq:Riccicomp4} can be estimated exactly as the other nonlinear error terms.)

\vskip1pc
\noindent \textbf{Estimates for $\nablaslash_3 f^4$, $(\nablaslash_3)^2 f^4$ and $\nablaslash_4 \nablaslash_3 f^4$:} Next, $(r\nablaslash)^k \nablaslash_3 f^4$ is estimated, for $k \leq N-4$ by applying $(r\nablaslash)^k$ to the metric relation \eqref{eq:metriccomp1} to obtain
\begin{equation} \label{eq:fHI5}
	\Vert (r\nablaslash)^k \nablaslash_3 f^4 \Vert^2_{S_{u,v}^{\I}}
	\lesssim
	u^{-2}
	(
	\mathbb{E}^N_{\I}
	+
	\mathbb{E}^N_{\Hp}
	+
	\varepsilon^4).
\end{equation}
Similarly for $(r\nablaslash)^k (\nablaslash_3)^2 f^4$ and $(r\nablaslash)^k \nablaslash_4 \nablaslash_3 f^4$ for $k \leq N-5$.

\vskip1pc
\noindent \textbf{Estimate for $\nablaslash_4 f^3$, $(\nablaslash_4)^2 f^3$ and $\nablaslash_3\nablaslash_4 f^3$:} Similarly, $(r\nablaslash)^k \nablaslash_4 f^3$ is estimated, for $k \leq N-4$ by applying $(r\nablaslash)^k$ the metric relation \eqref{eq:metriccomp2} to obtain
\begin{equation} \label{eq:fHI6}
	\Vert (r\nablaslash)^k \nablaslash_4 f^3 \Vert^2_{S_{u,v}^{\I}}
	\lesssim
	u^{-2}
	(
	\mathbb{E}^N_{\I}
	+
	\mathbb{E}^N_{\Hp}
	+
	\varepsilon^4).
\end{equation}
Similarly for $(r\nablaslash)^k (\nablaslash_4)^2 f^3$ and $(r\nablaslash)^k \nablaslash_3 \nablaslash_4 f^3$ for $k \leq N-5$.

\vskip1pc
\noindent \textbf{Estimates for $\partial_u \slashed{f}$ and $\nablaslash_3 \partial_u \slashed{f}$:}  The components $(r\nablaslash)^k \partial_u \slashed{f}$, for $k \leq N-4$, are estimated by applying $(r\nablaslash)^k$ to the metric relation \eqref{eq:metriccomp3}.  The bounds \eqref{eq:fHI1} then give
\begin{equation} \label{eq:fHI7}
	\Vert (r\nablaslash)^k \partial_u \slashed{f} \Vert^2_{S_{u,v}^{\I}}
	\lesssim
	u^{-2}
	(
	\mathbb{E}^N_{\I}
	+
	\mathbb{E}^N_{\Hp}
	+
	\varepsilon^4).
\end{equation}
Similarly, $(r\nablaslash)^k \nablaslash_3 \partial_u \slashed{f}$, for $k \leq N-5$, is estimated by applying $(r\nablaslash)^k \nablaslash_3$ to \eqref{eq:metriccomp3}.

\vskip1pc
\noindent \textbf{Estimates for $\partial_v \slashed{f}$, $\nablaslash_3 \partial_v \slashed{f}$ and $\nablaslash_4 \partial_v \slashed{f}$:}  The components $(r\nablaslash)^k \partial_v \slashed{f}$, for $k \leq N-4$, are estimated by applying $(r\nablaslash)^k$ to the metric relation \eqref{eq:metriccomp4}.  The bounds \eqref{eq:fHI2} then give
\begin{equation} \label{eq:fHI8}
	\Vert (r\nablaslash)^k \partial_v \slashed{f} \Vert^2_{S_{u,v}^{\I}}
	\lesssim
	u^{-2}
	(
	\mathbb{E}^N_{\I}
	+
	\mathbb{E}^N_{\Hp}
	+
	\varepsilon^4).
\end{equation}
Similarly for $(r\nablaslash)^k \nablaslash_3 \partial_v \slashed{f}$ and $(r\nablaslash)^k \nablaslash_4 \partial_v \slashed{f}$ for $k \leq N-5$.

\vskip1pc
\noindent \textbf{Estimate for $(f^3-f^4)_{\ell=0}$:}  The component $(f^3 - f^4)_{\ell=0}$ is estimated, by considering the relation \eqref{eq:curvaturecomp5}.  From the fact that,
\begin{equation} \label{eq:rdiffest}
	\left\vert r(x+f(x)) - r(x) + \Omega_{\circ}(x)^2(f^3(x) - f^4(x))  \right\vert
	\lesssim
	\vert f^3 - f^4 \vert^2,
\end{equation}
the first estimate of \eqref{eq:HIdiff4} then follows.

\vskip1pc
\noindent \textbf{Estimate for $(f^3+f^4)_{\ell=0}$:}  For the second estimate of \eqref{eq:HIdiff4}, for $(f^{3} + f^{4})_{\ell =0}$, first note that the first estimate of \eqref{eq:HIdiff4} together with the fact that $f^3=0$ on $\{u_{\I} = u_f\} = \{u_{\Hp} = u_f\}$ (see \eqref{eq:fHIuf}) implies that
		\[
			\left\vert (f^{3} + f^{4})_{\ell =0} (u_f,v) \right\vert^2
			\lesssim
			\frac{
			\mathbb{E}^N_{\I}
			+
			\mathbb{E}^N_{\Hp}
			+
			\varepsilon^2
			}{u_f^{2}},
		\]
		for all $v(R_{-2},u_f) \leq v \leq v(R_{2},u_f)$.  The second estimate of \eqref{eq:HIdiff4} then follows from summing the relations \eqref{eq:metriccomp1}, \eqref{eq:metriccomp2}, \eqref{eq:Riccicomp7} and \eqref{eq:Riccicomp8} to obtain,
\begin{multline*}
	\left\vert (\partial_u + \partial_v)(f^3 + f^4)_{\ell=0} \right\vert
	\lesssim
	\left\vert (f^3 - f^4)_{\ell=0} \right\vert
	+
	\left\vert \left( \Omega \tr \chi - \Omega \tr \chi_{\circ} \right)^{\Hp}_{\ell=0} \right\vert
	+
	\left\vert \left( \Omega \tr \chi - \Omega \tr \chi_{\circ} \right)^{\I}_{\ell=0} \right\vert
	\\
	+
	\left\vert \big( \Omega \tr \chibar - \Omega \tr \chibar_{\circ} \big)^{\Hp}_{\ell=0} \right\vert
	+
	\left\vert \big( \Omega \tr \chibar - \Omega \tr \chibar_{\circ} \big)^{\I}_{\ell=0} \right\vert
	+
	\vert \mathcal{E}^{2,0}_{\mathfrak{D} \fsc} \vert,
\end{multline*}
	and integrating backwards from $\{ u=u_f \}$ along the constant $r_{\I}$, $\theta_{\I}$ curves, using the first estimate of \eqref{eq:HIdiff4}, the fact that
	\[
		\left\vert \left( \Omega \tr \chi - \Omega \tr \chi_{\circ} \right)^{\I}_{\ell=0} \right\vert
		+
		\left\vert \big( \Omega \tr \chibar - \Omega \tr \chibar_{\circ} \big)^{\I}_{\ell=0} \right\vert
		\lesssim
		u^{-\frac{3}{2}} \mathbb{E}^N_{\I},
	\]
	and
	\[
		\sum_{\vert \gamma \vert \leq 3} \Vert \mathfrak{D}^{\gamma} \big( \Omega \tr \chi - \Omega \tr \chi_{\circ} \big)^{\Hp} \Vert_{\DRH_{R_{-1} \leq r \leq R_1} (u)}^2
		\lesssim
		u^{-2} \mathbb{E}^N_{\Hp},
	\]
	so that
	\[
		\int_u^{u_f} \vert \left( \Omega \tr \chi - \Omega \tr \chi_{\circ} \right)^{\Hp}_{\ell=0} (u',u') \vert d u'
		\lesssim
		u^{-\frac{\delta}{2}}
		\sum_{\vert \gamma \vert \leq 3}
		\Vert (u')^{\frac{1+\delta}{2}} \mathfrak{D}^{\gamma} \big( \Omega \tr \chi - \Omega \tr \chi_{\circ} \big)^{\Hp} \Vert_{\DRH_{R_{-1} \leq r \leq R_1}(u)}
		\lesssim
		u^{-\frac{1}{2}} \mathbb{E}^N_{\Hp}.
	\]
	(See Lemma \ref{lem:ILEDtrapping}.)  Similarly for $\big( \Omega \tr \chibar - \Omega \tr \chibar_{\circ} \big)^{\Hp}_{\ell=0}$.

\vskip1pc
\noindent \textbf{Estimates for $\nablaslash_3 f^3_{\ell=0}$ and $\nablaslash_4 f^4_{\ell=0}$:}  The relations \eqref{eq:Riccicomp7} and \eqref{eq:Riccicomp8}, together with the first estimate of \eqref{eq:HIdiff4}, imply that,
\begin{equation} \label{eq:fHI11}
	\left\vert \nablaslash_3 f^3_{\ell=0} \right\vert^2
	+
	\left\vert \nablaslash_4 f^4_{\ell=0} \right\vert^2
	\lesssim
	u^{-2}
	(\mathbb{E}^N_{\I}
	+
	\mathbb{E}^N_{\Hp}
	+
	\varepsilon^4).
\end{equation}

\vskip1pc
\noindent \textbf{Estimates for $(\nablaslash_3)^2 f^3$ and $(\nablaslash_4)^2 f^4$:}  Finally, it follows from the relations \eqref{eq:Riccicomp5} and \eqref{eq:Riccicomp6}, together with the above estimates, that, for $k \leq N-5$,
\begin{equation} \label{eq:fHI12}
	\left\Vert (r\nablaslash)^k (\nablaslash_3)^2 f^3 \right\Vert^2_{S_{u,v}^{\I}}
	+
	\left\Vert (r\nablaslash)^k (\nablaslash_4)^2 f^4 \right\Vert^2_{S_{u,v}^{\I}}
	\lesssim
	u^{-2}
	(\mathbb{E}^N_{\I}
	+
	\mathbb{E}^N_{\Hp}
	+
	\varepsilon^4).
\end{equation}

\vskip1pc
\noindent \textbf{Estimates close to $r_{\Hp} = R_2$:}
The above completes the estimates for $f_{\Hp,\I}$ in the region $R_{-2} \leq r_{\I} \leq R_1$.  Consider now the region $R_{-1} \leq r_{\Hp} \leq R_2$.  Since the diffeomorphism functions $f_{\Hp,\I}$ are not defined on whole spheres close to the boundary of the overlap region, $r_{\Hp} = R_2$, and they are estimated via elliptic estimates, it is more convenient in this region to first estimate the diffeomorphism functions $f_{\I,\Hp}$, defined by \eqref{expressinglikethis} with $F= F_{\I,\Hp} = i^{-1}_{\I} \circ i_{\Hp}$.  Indeed, repeating the above estimates with $f_{\I,\Hp}$ in place of $f_{\Hp,\I}$ and $R_{-1} \leq r_{\Hp} \leq R_2$ in place of $R_{-2} \leq r_{\I} \leq R_1$ (modifying the relations \eqref{eq:metriccomp3}, \eqref{eq:metriccomp4} appropriately in view of the fact that the position of the torsion terms $b_{\Hp}$ and $b_{\I}$ in the respective coordinate expressions of the metric is interchanged), it follows that the estimates \eqref{eq:HIdiff1}--\eqref{eq:HIdiff4} hold for $f=f_{\I,\Hp}$ in the region $R_{-1} \leq r_{\Hp} \leq R_2$.
Now, for appropriate $(u,v,\theta)$, by definition
\[
	u
	=
	u+f^3_{\Hp,\I}(u,v,\theta)
	+
	f^3_{\I,\Hp}(u+f^3_{\Hp,\I}(u,v,\theta),v+f^4_{\Hp,\I}(u,v,\theta), \slashed{F}_{\Hp,\I}(u,v,\theta)),
\]
and so
\begin{equation} \label{eq:HpIpIpHp1}
	f^3_{\Hp,\I}(u,v,\theta)
	=
	-
	f^3_{\I,\Hp}(u+f^3_{\Hp,\I}(u,v,\theta),v+f^4_{\Hp,\I}(u,v,\theta), \slashed{F}_{\Hp,\I}(u,v,\theta)).
\end{equation}
Similarly,
\begin{equation} \label{eq:HpIpIpHp2}
	f^4_{\Hp,\I}(u,v,\theta)
	=
	-
	f^4_{\I,\Hp}(u+f^3_{\Hp,\I}(u,v,\theta),v+f^4_{\Hp,\I}(u,v,\theta), \slashed{F}_{\Hp,\I}(u,v,\theta)),
\end{equation}
and
\begin{equation} \label{eq:HpIpIpHp3}
	\theta^A
	=
	\slashed{F}^A_{\I,\Hp}(u+f^3_{\Hp,\I}(u,v,\theta),v+f^4_{\Hp,\I}(u,v,\theta), \slashed{F}_{\Hp,\I}(u,v,\theta)).
\end{equation}
Applying $\nablaslash^{\Hp}$ to \eqref{eq:HpIpIpHp1} it follows that
\[
	\nablaslash f^3_{\Hp,\I}
	=
	-
	\partial_u f^3_{\I,\Hp} \nablaslash f^3_{\Hp,\I}
	-
	\partial_v f^3_{\I,\Hp} \nablaslash f^4_{\Hp,\I}
	-
	\nablaslash f^3_{\I,\Hp} \cdot \nablaslash \slashed{F}_{\Hp,\I},
\]
where $\nablaslash f^3_{\I,\Hp} \cdot \nablaslash \slashed{F}_{\Hp,\I} (\partial_{\theta^A}) = \partial_{\theta^B} f^3_{\I,\Hp} \partial_{\theta^A} \slashed{F}^B_{\Hp,\I}$.  It then follows from Proposition \ref{prop:tildenorm} that
\[
	\vert \nablaslash f^3_{\Hp,\I} \vert_{\gslash_{\I}}
	=
	\vert \partial_u f^3_{\I,\Hp} \vert \vert \nablaslash f^3_{\Hp,\I} \vert_{\gslash_{\I}}
	+
	\vert \partial_v f^3_{\I,\Hp}\vert \vert \nablaslash f^4_{\Hp,\I}\vert_{\gslash_{\I}}
	+
	\vert \nablaslash f^3_{\I,\Hp} \vert_{\gslash_{\Hp}}
	+
	\varepsilon^2.
\]
Similarly for $\nablaslash_3$ and $\nablaslash_4$ derivatives, for higher order derivatives, and for $f^4_{\Hp,\I}$, $\partial_u\slashed{f}_{\Hp,\I}$ and $\partial_v \slashed{f}_{\Hp,\I}$ using the relations \eqref{eq:HpIpIpHp2} and \eqref{eq:HpIpIpHp3}.

\end{proof}

On the final hypersurface, $u=u_f$, it is important to control extra derivatives of $f_{\Hp,\I}$.  Since $f^3 \equiv 0$ and $f^A \equiv 0$ on $\{u=u_f\}$ (see \eqref{eq:fHIuf}) their derivatives tangential to $\{u=u_f\}$ also all vanish (and, in particular, $\partial_v \slashed{f}$ vanishes on $\{u=u_f\}$).  In the following proposition $\mathds{1} = \mathds{1}_{R_{-1} \leq r \leq R_{1}}$ so that, for any tensor $\xi$,
\[
	\Vert \xi \mathds{1} \Vert_{C_{u_f}^{\I}}^2
	=
	\int_{v(R_{-1},u_f)}^{v(R_1,u_f)} \int_{S^2}
	\vert \xi (u_f,v,\theta) \vert^2
	d\theta d v.
\]

\begin{proposition}[Estimates for extra derivatives of $f_{\Hp,\I}$ on $\{u=u_f\}$] \label{prop:fHIuf}
	On the final hypersurface $\{u=u_f\}$, the diffeomorphisms $f = f_{\Hp,\I}$ satisfy the higher order estimates,
	for $k \leq N-s$, $s=0,1,2$,
	\begin{multline*}
		\sum_{\vert \gamma \vert \leq 2}
		\big(
		\Vert (r \nablaslash)^{k} \mathfrak{D}^{\gamma} f^4 \mathds{1} \Vert^2_{C_{u_f}^{\I}}
		+
		\Vert (r \nablaslash)^{k} \mathfrak{D}^{\gamma} f^3 \mathds{1} \Vert^2_{C_{u_f}^{\I}}
		\big)
		+
		\sum_{\vert \gamma \vert = 0,1}
		\Vert (r \nablaslash)^{k} \mathfrak{D}^{\gamma} \partial_u \slashed{f} \mathds{1} \Vert^2_{C_{u_f}^{\I}}
		\lesssim
		(u_f)^{-s}
		(
		\mathbb{E}^N_{\I}
		+
		\mathbb{E}^N_{\Hp}
		+
		\varepsilon^4)
		,
	\end{multline*}
	where $\mathds{1} = \mathds{1}_{R_{-1} \leq r \leq R_1}$.
\end{proposition}

\begin{proof}
	As in the proof of Proposition \ref{prop:HIdiff}, derivatives of the relations \eqref{eq:metriccomp1}--\eqref{eq:curvaturecomp6} will be taken and Proposition \ref{prop:tildenorm} and Proposition \ref{prop:tildederivStensor} used.  Consider $s=0,1,2$.  Note that, by \eqref{eq:fHIuf}, derivatives of $f^3$ tangential to $\{u=u_f\}$ all vanish.  Lemma \ref{lem:fHIhoerror} is used to estimate the nonlinear error terms.
	
	\vskip1pc
	\noindent \textbf{Estimate for $f^4_{\ell \geq 1}$:} First, $\nablaslash^k f^4_{\ell\geq 2}$ is estimated, for $k \leq N+2-s$, using equation \eqref{eq:Riccicomp2}, the elliptic estimate of Proposition \ref{prop:ellipticestimates}, and the fact that
	\[
		(u_f)^{s}
		\sum_{k=0}^{N-s}
		\Vert (r \nablaslash)^{k} \hat{\chibar}_{\I} \mathds{1} \Vert^2_{C_{u_f}^{\I}}
		\lesssim
		\mathbb{E}^N_{\I},
	\]
	and similarly for $\hat{\chibar}_{\Hp}$, together with Lemma \ref{lem:fHIhoerror} to control the nonlinear error terms.  The $\ell =1$ modes of $f^4$ are controlled using the relation \eqref{eq:curvaturecomp4}, as in the proof of Proposition \ref{prop:HIdiff}.
	
	\vskip1pc
	\noindent \textbf{Estimate for $\partial_v f^4$:} Next, $\nablaslash^k \partial_v f^4$ is estimated, for $1 \leq k \leq N+1-s$, by applying $\nablaslash^{k-1}$ to the relation \eqref{eq:Riccicomp4}, using the above estimate for $\nablaslash^k f^4$ together with the fact that $f^3 \equiv 0$ on $u=u_f$, and controlling the error terms again using Lemma \ref{lem:fHIhoerror}.  The Poincar\'e inequality, Proposition \ref{prop:Poincare}, then gives control of $\partial_v f^4_{\ell \geq 1}$.
	
	\vskip1pc
	\noindent \textbf{Estimates for $\partial_u f^4$, $\partial_u^2 f^4$ and $\partial_u \partial_v f^4$:} The quantity $\nablaslash^k \partial_u f^4$ is estimated, for $0 \leq k \leq N+1-s$, by applying $\nablaslash^k$ to the metric relation \eqref{eq:metriccomp1}.  The quantities $\nablaslash^k \partial_u^2 f^4$ and $\nablaslash^k \partial_u \partial_v f^4$ are similarly estimated, for $0 \leq k \leq N-s$.
	
	\vskip1pc
	\noindent \textbf{Estimates for $\partial_u f^3_{\ell \geq 1}$ and $\partial_v f^4_{\ell \geq 1}$:} Next, $\nablaslash^k \partial_u f^3_{\ell \geq 1}$ is estimated, for $0 \leq k \leq N+1-s$, exactly as $\nablaslash^k \partial_v f^4_{\ell \geq 1}$ is controlled above, using now \eqref{eq:Riccicomp3} instead of \eqref{eq:Riccicomp4}.
	
	\vskip1pc
	\noindent \textbf{Estimates for $\partial_u \slashed{f}$, $\nablaslash_3 \partial_u \slashed{f}$, and $\nablaslash_4 \partial_u \slashed{f}$:} The quantity $\nablaslash^k \partial_u \slashed{f}$ is then controlled, for $0 \leq k \leq N+1-s$, by applying $\nablaslash^k$ to the metric relation \eqref{eq:metriccomp3} and using the above estimate for $\nablaslash^{k+1} f^4$ and the fact that
	\[
		(u_f)^{s}
		\sum_{k=0}^{N+1-s}
		\Vert (r \nablaslash)^{k} b^{\I} \mathds{1} \Vert^2_{C_{u_f}^{\I}}
		\lesssim
		\mathbb{E}^N_{\I}.
	\]
	Similarly for $\nablaslash_3 \partial_u \slashed{f}$ and $\nablaslash_4 \partial_u \slashed{f}$.
	
	\vskip1pc
	\noindent \textbf{Estimates for $f^4_{\ell=0}$, $\partial_u f^3_{\ell=0}$ and $\partial_v f^4_{\ell=0}$:} Now $f^4_{\ell=0}$ is estimated from the relation \eqref{eq:curvaturecomp5}, $\partial_u f^3_{\ell=0}$ is estimated from \eqref{eq:Riccicomp8}, and $\partial_v f^4_{\ell=0}$ from \eqref{eq:Riccicomp7}, using again \eqref{eq:rdiffest} and the fact that $f^3 \equiv 0$ on $u=u_f$.
	
	\vskip1pc
	\noindent \textbf{Estimates for $\partial_v^2 f^4$ and $\partial_u^2 f^3$:} Finally, $\nablaslash^k \partial_v^2 f^4$ and $\nablaslash^k \partial_u^2 f^3$ are estimated, for $0\leq k \leq N-s$, using the above estimates and the relations \eqref{eq:Riccicomp5} and \eqref{eq:Riccicomp6} respectively.
\end{proof}

\subsection{Estimates for $f_{d,\I}$}

The diffeomorphisms $f_{d,\I}$ are estimated in a similar way as $f_{\Hp,\I}$ in the previous section, though it is now necessary to keep track of the $r$ behaviour of the different terms.  First recall the pointwise norm $\mathbb P_{u_f}[f_{d,\I}]$ of the $f_{d,\I}$ diffeomorphism functions, defined in \eqref{eq:diffeopointwisenorm2}.

\begin{proposition} \label{prop:fdIppointwise}
	The diffeomorphism functions $f_{d,\I}$ satisfy the pointwise estimate
	\begin{equation} \label{eq:fdIppointwise}
		\mathbb P_{u_f}[f_{d,\I}]
		\lesssim
		\varepsilon.
	\end{equation}
\end{proposition}

\begin{proof}
	The proof is an immediate consequence of the Sobolev inequality, Proposition \ref{prop:Sobolev}, and the bootstrap assumption \eqref{eq:badiffeo} for the energy $\mathbb{E}_{u_f}[f_{d,\I}]$, defined in \eqref{EfdIpenergy}.
\end{proof}

The following lemma provides estimates for nonlinear error terms which arise in the proof of Proposition \ref{prop:dIdiff} below.

\begin{lemma}[Estimates for $f_{d,\I}$ nonlinear error terms] \label{lem:fdIerror}
	For any $u_{-1} \leq u \leq u_{2}$ and $v(R_{-2},u) \leq v \leq v_{\infty}$,
	\[
		\sum_{k \leq 2}
		\Big(
		\sum_{\vert \gamma\vert \leq 1}
		\Vert (r\nablaslash)^k \mathfrak{D}^{\gamma} \mathcal{E}^{1,0}_{\mathfrak{D} \fsc,p} \Vert_{S_{u,v}^{\I}}^2
		+
		\Vert (r\nablaslash)^k \mathcal{E}^{2,0}_{\mathfrak{D} \fsc,p} \Vert_{S_{u,v}^{\I}}^2
		\Big)
		\lesssim
		\frac{\varepsilon^4}{r^{2p}}.
	\]
\end{lemma}

\begin{proof}
	The proof is an immediate consequence of the fact that, for any $u_{-1} \leq u \leq u_{2}$ and $v(R_{-2},u) \leq v \leq v_{\infty}$,
	\[
		\Vert (r\nablaslash)^k \mathfrak{D}^{\gamma} \mathfrak{D} \fsc_p \Vert^2_{S_{u,v}^{\I}}
		\lesssim
		\frac{\varepsilon^2}{r^{2p}}
		\qquad
		\text{for }
		k \leq 2, \quad \vert \gamma \vert \leq 1,
	\]
	and the geometric quantities of the initial gauge satisfy,
	\[
		\vert \mathfrak{D}^{\gamma} \Phi^{d}_p \vert^2
		\lesssim
		\frac{\varepsilon^2}{r^{2p}}
		\qquad
		\text{for }
		\vert \gamma \vert \leq 4.
	\]
\end{proof}

The diffeomorphisms $f_{d,\I}$ can now be estimated using the change of gauge relations of Chapter \ref{bigchangeisgood} and the estimate \eqref{willstatelater2}.
Recall the definition \eqref{EfdIpenergy} of the energy $\mathbb{E}_{u_f}[f_{d,\I}]$.  It is also possible to estimate higher order derivatives of $f_{d,\I}$, but such estimates are not required and so are not obtained.

\begin{proposition}[Estimates for $f_{d,\I}$ diffeomorphisms and mass difference $M_f - M_{\mathrm{init}}$] \label{prop:dIdiff}
	The diffeomorphisms $f_{d,\I}$ satisfy the estimates,
	\begin{equation}
	\label{theestimateheretoreferto}
		\mathbb{E}_{u_f}[f_{d,\I}]
		\lesssim
		\mathbb{E}^N_{\I}
		+
		\varepsilon_0^2
		+
		\varepsilon^4,
	\end{equation}
	and moreover the mass $M_f$ satisfies,
	\[
		\vert M_f - M_{\mathrm{init}} \vert^2 \lesssim \mathbb{E}^N_{\I} + \varepsilon_0^2 + \varepsilon^4.
	\]
\end{proposition}

\begin{proof}
	The proof follows from Propositions \ref{prop:metricrelations}, \ref{prop:Riccirelations} and \ref{prop:curvaturerelations}, with $(u,v,\theta) = (u_{data}, v_{data}, \theta_{data})$ and $(\widetilde{u},\widetilde{v},\widetilde{\theta}) = (u_{\I}, v_{\I}, \theta_{\I})$, $M= M_{\mathrm{init}}$ and $\widetilde{M} = M_f$.  Derivatives of the relations in Propositions \ref{prop:metricrelations}, \ref{prop:Riccirelations} and \ref{prop:curvaturerelations} will be taken, as in the proof of Proposition \ref{prop:HIdiff}, and Proposition \ref{prop:tildenorm} and Proposition \ref{prop:tildederivStensor} used.  In what follows $f = f_{d,\I}$.  The subscript is omitted for brevity.  Lemma \ref{lem:fdIerror} will be used throughout to estimate the nonlinear error terms which arise, and the estimate \eqref{willstatelater2} will be used to estimate terms in the initial $\mathcal{EF}$ gauge.
	
	Consider some $u_{-1} \leq u \leq u_{2}$ and $v(R_{-2},u) \leq v \leq v_{\infty}$.
	
	\vskip1pc
	\noindent \textbf{Estimate for $f^3_{\ell \geq 1}$:}  To begin, $(r\nablaslash)^k f^3_{\ell \geq 1}$ is estimated for $k \leq 4$, using \eqref{eq:curvaturecomp3} (along with $(r\nablaslash)^{k-1}$ applied to \eqref{eq:curvaturecomp3}) which implies
	\[
		\vert r \nablaslash f^3 \vert_{\gslash^{\I}}
		\lesssim
		\vert r^4 (\Omega \beta)_{data} \vert_{\gslash^{data}}
		+
		\vert r^4 (\Omega \beta)_{\I} \vert_{\gslash^{\I}}
		+
		\vert
		\mathcal{E}^{1,0}_{\mathfrak{D} \fsc,0}
		\vert_{\gslash^{\I}}
	\]
	Similarly for $(r\nablaslash)^k f^3$ for $k = 2,3,4$. The Poincar\'{e} inequality, Proposition \ref{prop:Poincare} then implies
	\begin{equation} \label{eq:fdI1}
		\Vert (r \nablaslash)^k f^3_{\ell\geq 1} \Vert^2_{S_{u,v}^{\I}}
		\lesssim
		\mathbb{E}^N_{\I}
		+
		\varepsilon_0^2
		+
		\varepsilon^4.
	\end{equation}

	\vskip1pc
	\noindent \textbf{Estimate for $f^4_{\ell \geq 1}$:}  Next, $(r\nablaslash)^k f^4_{\ell\geq 1}$ is estimated for $k \leq 4$.  First, to estimate $(r\nablaslash)^k f^4_{\ell\geq 2}$, the relation \eqref{eq:Riccicomp4} implies that
	\[
		\vert r^2 \Dslash_2^* \nablaslash f^4 \vert_{\gslash^{\I}}
		\lesssim
		\vert r^2 \hat{\chibar}_{data} \vert_{\gslash^{data}}
		+
		\vert r^2 \hat{\chibar}_{\I} \vert_{\gslash^{\I}}
		+
		\vert
		\mathcal{E}^{2,0}_{\mathfrak{D} \fsc,-1}
		\vert_{\gslash^{\I}}
	\]
	and similarly after commuting with $r\nablaslash$.  The elliptic estimate, Proposition \ref{prop:ellipticestimates} then implies that
	\[
		\sum_{k\leq 4} \Vert (r \nablaslash)^k f^4_{\ell\geq 2} \Vert^2_{S_{u,v}^{\I}}
		\lesssim
		r^2
		(\mathbb{E}^N_{\I}
		+
		\varepsilon_0^2
		+
		\varepsilon^4).
	\]
	Now the relation \eqref{eq:curvaturecomp5} implies that
	\begin{equation} \label{eq:rhodiffmass}
		(\rho - \rho_{\circ,M_f})^{\I}
		=
		(\rho - \rho_{\circ,M_{\mathrm{init}}})^d
		+
		2M_f \left( \frac{1}{r_d^3} - \frac{1}{r_{\I}^3} \right)
		+
		2\frac{M_{\mathrm{init}}-M_f}{r_d^3}
		+
		\vert
		\mathcal{E}^{1,0}_{\mathfrak{D} \fsc,3}
		\vert.
	\end{equation}
	Projecting to $\ell \geq 1$, using the above estimate \eqref{eq:fdI1}, and the fact that
	\begin{equation} \label{eq:rdrIf}
		\left\vert
		\frac{1}{r_d^3} - \frac{1}{r_{\I}^3}
		-
		\frac{3\Omega_{\circ}^2}{r_d^4}(f^3-f^4)
		-
		\frac{3(r_{M_f} - r_{M_{\mathrm{init}}})}{r_d^4}
		\right\vert
		\lesssim
		\frac{\varepsilon^2}{r_d^3},
	\end{equation}
	and
	\begin{equation} \label{eq:rdrIf2}
		\left\vert
		\frac{r_{M_f} - r_{M_{\mathrm{init}}}}{r_d^4}
		-
		\Big( \frac{M_f}{M_{\mathrm{init}}} - 1 \Big)
		\frac{r_{M_{\mathrm{init}}} - (v_{d} - u_d) \Omega_{\circ,M_{\mathrm{init}}}^2}{r_d^4}
		\right\vert
		\lesssim
		\frac{\varepsilon^2}{r_d^4}
	\end{equation}
	it follows that
	\begin{equation} \label{eq:fdI2}
		\sum_{k\leq 4} \Vert (r \nablaslash)^k f^4_{\ell\geq 1} \Vert^2_{S_{u,v}^{\I}}
		\lesssim
		r^2(
		\mathbb{E}^N_{\I}
		+
		\varepsilon_0^2
		+
		\varepsilon^4).
	\end{equation}

	\vskip1pc
	\noindent \textbf{Estimate for $\nablaslash_3 f^3_{\ell \geq 1}$:}  The quantity $(r\nablaslash)^k \nablaslash_3 f^3_{\ell \geq 1}$ can now be estimated, for $k \leq 3$.  Indeed, it follows from the relation \eqref{eq:Riccicomp3} and the estimates \eqref{eq:fdI1} and \eqref{eq:fdI2} that
	\begin{equation} \label{eq:fdI3}
		\sum_{k\leq 3} \Vert (r \nablaslash)^k \nablaslash_3 f^3_{\ell\geq 1} \Vert^2_{S_{u,v}^{\I}}
		\lesssim
		\mathbb{E}^N_{\I}
		+
		\varepsilon_0^2
		+
		\varepsilon^4.
	\end{equation}

	\vskip1pc
	\noindent \textbf{Estimate for $r \nablaslash_4 f^4_{\ell \geq 1}$:}  Now $(r\nablaslash)^k (r \nablaslash_4) f^4_{\ell \geq 1}$ can now be estimated, for $k \leq 3$, since the relation \eqref{eq:Riccicomp4} and the estimates \eqref{eq:fdI1} and \eqref{eq:fdI2} imply that
	\begin{equation} \label{eq:fdI4}
		\sum_{k\leq 3} \Vert (r \nablaslash)^k (r\nablaslash_4) f^4_{\ell\geq 1} \Vert^2_{S_{u,v}^{\I}}
		\lesssim
		\mathbb{E}^N_{\I}
		+
		\varepsilon_0^2
		+
		\varepsilon^4.
	\end{equation}

	\vskip1pc
	\noindent \textbf{Estimates for $\nablaslash_3 f^4$, $(\nablaslash_3)^2 f^4$ and $r \nablaslash_4 \nablaslash_3 f^4$:}  Next, $(r\nablaslash)^k \nablaslash_3 f^4$ is estimated, for $k \leq 3$ from the metric relation \eqref{eq:metriccomp1}:
	\begin{equation} \label{eq:fdI5}
		\sum_{k\leq 3} \left\Vert (r\nablaslash)^k \nablaslash_3 f^4 \right\Vert^2_{S_{u,v}^{\I}}
		\lesssim
		\mathbb{E}^N_{\I}
		+
		\varepsilon_0^2
		+
		\varepsilon^4.
	\end{equation}
	Similarly for $(r\nablaslash)^k (\nablaslash_3)^2 f^4$ and $(r\nablaslash)^k (r\nablaslash_4) \nablaslash_3 f^4$ for $k \leq 2$.

	\vskip1pc
	\noindent \textbf{Estimates for $r\nablaslash_4 f^3$ and $(r\nablaslash_4)^2 f^3$:}  Similarly, $(r\nablaslash)^k (r\nablaslash_4) f^3$ is estimated, for $k \leq 3$ from the metric relation \eqref{eq:metriccomp2}:
	\begin{equation} \label{eq:fdI6}
		\sum_{k\leq 3} r^2 \left\Vert (r\nablaslash)^k (r \nablaslash_4) f^3 \right\Vert^2_{S_{u,v}^{\I}}
		\lesssim
		\mathbb{E}^N_{\I}
		+
		\varepsilon_0^2
		+
		\varepsilon^4.
	\end{equation}
	Similarly for $(r\nablaslash)^k (r\nablaslash_4)^2 f^3$ for $k \leq 2$.

	\vskip1pc
	\noindent \textbf{Estimates for $\partial_u \slashed{f}$, $\nablaslash_3 \partial_u \slashed{f}$, and $r \nablaslash_4 \partial_u \slashed{f}$:}  The components $(r\nablaslash)^k \partial_u \slashed{f}$, for $k \leq 3$, are estimated from the metric relation \eqref{eq:metriccomp3}, using the bounds \eqref{eq:fdI2}:
	\begin{equation} \label{eq:fdI7}
		\sum_{k\leq 3} \left\Vert (r\nablaslash)^k \partial_u \slashed{f} \right\Vert^2_{S_{u,v}^{\I}}
		\lesssim
		\mathbb{E}^N_{\I}
		+
		\varepsilon_0^2
		+
		\varepsilon^4.
	\end{equation}
	Similarly for $(r\nablaslash)^k \nablaslash_3 \partial_u \slashed{f}$, and $(r\nablaslash)^k (r\nablaslash_4) \partial_u \slashed{f}$ for $k \leq 2$.

	\vskip1pc
	\noindent \textbf{Estimates for $r \partial_v \slashed{f}$, $\nablaslash_3 (r \partial_v \slashed{f})$, and $r \nablaslash_4 (r \partial_v \slashed{f})$:}  The components $(r\nablaslash)^k (r\partial_v \slashed{f})$, for $k \leq 3$, are estimated from the metric relation \eqref{eq:metriccomp4}, using the bounds \eqref{eq:fHI1}:
	\begin{equation} \label{eq:fdI8}
		\sum_{k\leq 3} \left\Vert (r\nablaslash)^k (r\partial_v \slashed{f}) \right\Vert^2_{S_{u,v}^{\I}}
		\lesssim
		\mathbb{E}^N_{\I}
		+
		\varepsilon_0^2
		+
		\varepsilon^4.
	\end{equation}
	Similarly for $(r\nablaslash)^k \nablaslash_3 (r \partial_v \slashed{f})$ and $(r\nablaslash)^k (r \nablaslash_4) (r \partial_v \slashed{f})$ for $k \leq 2$.

	\vskip1pc
	\noindent \textbf{Estimates for $M_f - M_{\mathrm{init}}$, $(\partial_v f^4)_{\ell=0}$, $(\partial_u f^3)_{\ell=0}$, $f^3_{\ell=0}$, $f^4_{\ell=0}$:}  It remains to estimate certain $\ell=0$ modes, along with the difference $M_f - M_{\mathrm{init}}$.  Equation \eqref{eq:metriccomp6} implies that
	\begin{equation} \label{eq:Omegadiffmass}
		\Omega^2_{\circ} \left(
		\partial_{u} f^3
		+
		\partial_{v} f^4
		\right)
		+
		2M_f \left( \frac{1}{r_{\I}} - \frac{1}{r_d} \right)
		+
		2\frac{M_f - M_{\mathrm{init}}}{r_d}
		=
		\left(
		\Omega^2
		-
		\Omega_{\circ,M_f}^2
		\right)^{\I}
		-
		\left(
		\Omega^2
		-
		\Omega_{\circ,M_{\mathrm{init}}}^2
		\right)^d
		+
		\mathcal{E}^{1,0}_{\mathfrak{D} \fsc,0},
	\end{equation}
	equation \eqref{eq:Riccicomp7} implies,
	\begin{multline} \label{eq:trchidiffmass}
		2\frac{r_{\I} - r_d}{r_{\I} r_d} \left( 1 - \frac{4M_f}{r_d} \right)
		+
		4 \frac{M_f - M_{\mathrm{init}}}{r_d^2}
		+
		2 \Omega^2_{\circ} \Deltaslash f^3
		+
		\frac{2 \Omega_{\circ}^2}{r_d} \partial_v f^4
		\\
		=
		\left( \Omega \tr \chi - \Omega \tr \chi_{\circ,M_f} \right)^{\I}
		-
		\left( \Omega \tr \chi - \Omega \tr \chi_{\circ,M_{\mathrm{init}}} \right)^d
		+
		\mathcal{E}^{2,0}_{\mathfrak{D} \fsc,2},
	\end{multline}
	and equation
	\eqref{eq:Riccicomp8} implies,
	\begin{multline} \label{eq:trchibardiffmass}
		- 2\frac{r_{\I} - r_d}{r_{\I} r_d} \left( 1 - \frac{4M_f}{r_d} \right)
		-
		4 \frac{M_f - M_{\mathrm{init}}}{r_d^2}
		+
		2 \Omega^2_{\circ} \Deltaslash f^4
		-
		\frac{2 \Omega_{\circ}^2}{r_d} \partial_u f^3
		\\
		=
		\left( \Omega \tr \chibar - \Omega \tr \chibar_{\circ,M_f} \right)^{\I}
		-
		\left( \Omega \tr \chibar - \Omega \tr \chibar_{\circ,M_{\mathrm{init}}} \right)^d
		+
		\mathcal{E}^{2,0}_{\mathfrak{D} \fsc,1},
	\end{multline}
	with $\Omega^2_{\circ} = (\Omega^2_{\circ,M_{\mathrm{init}}})^d$.  Considering $-\frac{2}{3} \frac{r_d^2}{M_f} \left(1 - \frac{3M_f}{r_d} \right)$ times equation \eqref{eq:rhodiffmass} minus $2/r_d$ times equation \eqref{eq:Omegadiffmass} plus equation \eqref{eq:trchidiffmass} minus equation \eqref{eq:trchibardiffmass} and projecting to $\ell=0$, it follows that,
	\[
		\left\vert M_f - M_{\mathrm{init}} \right\vert^2
		\lesssim
		\mathbb{E}^N_{\I} + \varepsilon_0^2 + \varepsilon^4.
	\]
	Now equation \eqref{eq:rhodiffmass} and \eqref{eq:rdrIf} imply that
	\begin{equation} \label{eq:f3mf4l0}
		\left\vert f^3_{\ell=0} - f^4_{\ell=0} \right\vert^2
		\lesssim
		r_d^2 \left(\mathbb{E}^N_{\I} + \varepsilon_0^2 + \varepsilon^4\right).
	\end{equation}
	Equations \eqref{eq:trchidiffmass} and \eqref{eq:trchibardiffmass} then respectively imply
	\[
		\left\vert (\partial_v f^4)_{\ell=0} \right\vert^2
		\lesssim
		\mathbb{E}^N_{\I} + \varepsilon_0^2 + \varepsilon^4,
		\qquad
		\left\vert (\partial_u f^3)_{\ell=0} \right\vert^2
		\lesssim
		\mathbb{E}^N_{\I} + \varepsilon_0^2 + \varepsilon^4.
	\]
	Finally, the estimate \eqref{eq:fdI6} implies that $\vert \partial_v f^3_{\ell=0} \vert^2 \lesssim r_d^{-4}( \mathbb{E}^K_{\I} +\varepsilon_0^2 + \varepsilon^3)$.  The estimate for $f^3_{\ell=0}$ is estimated by integrating backwards from $u_{\I} = u_0$, $v_{\I} = v_{\infty}$, using the fact that $f^3_{\ell=0}(u_0,v_{\infty}) = 0$, to obtain,
	\[
		\vert f^3_{\ell=0} \vert^2 \lesssim r_d^{-2}( \mathbb{E}^N_{\I} +\varepsilon_0^2 + \varepsilon^4),
		\qquad
		\left\vert f^4_{\ell=0} \right\vert^2
		\lesssim
		r_d^2 ( \mathbb{E}^N_{\I} +\varepsilon_0^2 + \varepsilon^4),
	\]
	where the latter follows from the former and \eqref{eq:f3mf4l0}.
	
	\vskip1pc
	\noindent \textbf{Estimates for $(\nablaslash_3)^2 f^3$ and $(\nablaslash_4)^2 f^4$:}  Finally, the relations \eqref{eq:Riccicomp5} and \eqref{eq:Riccicomp6}, together with the above estimates, then imply that, for $k \leq 2$,
\[
	\left\Vert (r\nablaslash)^k (\nablaslash_3)^2 f^3 \right\Vert^2_{S_{u,v}^{\I}}
	+
	r^{-2} \left\Vert (r\nablaslash)^k (r\nablaslash_4)^2 f^4 \right\Vert^2_{S_{u,v}^{\I}}
	\lesssim
	\mathbb{E}^N_{\I}
	+
	\varepsilon_0^2
	+
	\varepsilon^4.
\]

\end{proof}

\subsection{Estimates for $f_{d,\Hp}$}
\label{subsec:diffeofdH}

Recall the initial Kruskal gauge of Section \ref{initKrusksec}, the Kruskalised $\Hp$ gauge of Section \ref{kruskalisedsec} (see also \eqref{fortheHplusgaugekruskalised}), whose coordinates are denoted $(U_{\Hp}, V_{\Hp},\theta_{\Hp})$, and the diffeomorphisms $f_{d,\Hp}^1,\ldots,f_{d,\Hp}^4$.  Let $\Phi^{\mathcal{K},\Hp}$ denote the geometric quantities of the Kruskalised $\Hp$ gauge of Section \ref{kruskalisedsec}.

The estimates on the geometric quantities of the Eddington--Finkelstein $\Hp$ gauge immediately give estimates for the above geometric quantities of the Kruskalised $\Hp$ gauge.  The quantities in the Kruskalised gauge can, of course, be estimated up to order $N$ in the entire $\Hp$ region, though the following estimate is all that is used in estimating the diffeomorphisms $f_{d,\Hp}$ in Proposition \ref{prop:dHdiff} below.

Recall $U_f$ from Section \ref{kruskalisedsec}, and $U_0: = U(u_0)$, where $U$ is defined by~\eqref{eq:KruskalEF} with $M:=M_f$.

\begin{proposition}[Estimates for $\Hp$ quantities in Kruskalised gauge] \label{prop:HpKruskalisedestimates}
	For any $U_0 \leq U \leq U_f$ and any $V_{-1} \leq V \leq \min \{ V_2, V(R_2,U) \}$, each Ricci coefficient and curvature component of the Kruskalised $\Hp$ gauge (see \eqref{fortheHplusgaugekruskalised}) satisfies
	\[
		\sum_{\vert \gamma \vert \leq 3}
		\big\Vert \mathfrak{D}^{\gamma} \Phi^{\mathcal{K},\Hp} \big\Vert^2_{S_{U,V}^{\Hp}}
		\lesssim
		\mathbb{E}^N_{\Hp}
		+
		\varepsilon^4.
	\]
\end{proposition}

\begin{proof}
	Note that
	\[
		\Omega_{\circ,\mathcal{EF}}^{-2} \partial_u
		=
		\frac{2M_f}{V} \Omega_{\circ,\mathcal{K}}^{-2} \partial_U,
		\qquad
		\partial_v = \frac{V}{2M_f} \partial_V.
	\]
	Consider, for example, $\Omega^{-1} \hat{\chibar}$.  It follows that
	\[
		(\Omega^{-1} \hat{\chibar})_{\mathcal{EF}}
		=
		(\Omega_{\circ}^{-1} \hat{\chibar})_{\mathcal{EF}}
		+
		((1-\Omega_{\circ}^{-1} \Omega) \Omega^{-1} \hat{\chibar})_{\mathcal{EF}}
		=
		\frac{2M_f}{V} (\Omega_{\circ}^{-1} \hat{\chibar})_{\mathcal{K}}
		+
		((1-\Omega_{\circ}^{-1} \Omega) \Omega^{-1} \hat{\chibar})_{\mathcal{EF}},
	\]
	and so
	\[
		\Vert (\Omega^{-1} \hat{\chibar})_{\mathcal{K}} \Vert^2_{S_{U,V}}
		\lesssim
		\mathbb{E}^N_{\Hp}
		+
		\varepsilon^4.
	\]
	Similarly for derivatives and for the other geometric quantities.
\end{proof}

Recall the pointwise norm $\mathbb P_{u_f}[f_{d,\Hp}]$ of the $f_{d,\Hp}$ diffeomorphism functions, defined in \eqref{eq:diffeopointwisenorm3}.

\begin{proposition} \label{prop:fdHppointwise}
	The diffeomorphism functions $f_{d,\Hp}$ satisfy the pointwise estimate
	\begin{equation} \label{eq:fdHppointwise}
		\mathbb P_{u_f}[f_{d,\Hp}]
		\lesssim
		\varepsilon.
	\end{equation}
\end{proposition}

\begin{proof}
	The proof is an immediate consequence of the Sobolev inequality, Proposition \ref{prop:Sobolev}, and the bootstrap assumption \eqref{eq:badiffeo} for the energy $\mathbb{E}_{u_f}[f_{d,\Hp}]$, defined in \eqref{EfdHpenergy}.
\end{proof}

The diffeomorphisms $f_{d,\Hp}$ can now be estimated using the change of gauge relations of Chapter \ref{bigchangeisgood}, Proposition \ref{prop:HpKruskalisedestimates}, and the estimates \eqref{willstatelater}.  Recall the definition \eqref{EfdHpenergy} of the energy $\mathbb{E}_{u_f}[f_{d,\Hp}]$.

\begin{proposition}[Estimates for $f_{d,\Hp}$ diffeomorphisms] \label{prop:dHdiff}
	The diffeomorphisms $f_{d,\Hp}$ satisfy the estimates,
	\begin{equation}
	\label{theestimateshereforfdhp}
		\mathbb{E}_{u_f}[f_{d,\Hp}]
		\lesssim
		\mathbb{E}^N_{\Hp}
		+
		\mathbb{E}^N_{\I}
		+
		\varepsilon_0^2
		+
		\varepsilon^{3}.
	\end{equation}
\end{proposition}

\begin{proof}
	In what follows $f = f_{d,\Hp}$.  The subscript is omitted for brevity.
	
	Consider the relations \eqref{eq:metriccomp1}--\eqref{eq:curvaturecomp6} with $(u,v,\theta) = (U_{data}, V_{data}, \theta_{data})$ and $(\widetilde{u},\widetilde{v},\widetilde{\theta}) = (U_{\Hp}, V_{\Hp}, \theta_{\Hp})$ and $M = M_f$, $\widetilde{M} = M_{\mathrm{init}}$ (modifying the relations \eqref{eq:metriccomp3}, \eqref{eq:metriccomp4} appropriately in view of the position of the torsion terms $b_{\Hp}$ and $b_{data}$ in the respective coordinate expressions of the metric), now with the Schwarzschild background quantities referring to those of the Kruskal Schwarzschild gauge (see \eqref{eq:Kruskalbackground0}--\eqref{eq:Kruskalbackground2}).
	
	The estimates for $(r \nablaslash)^k \mathfrak{D}^{\gamma} f^3_{\ell \geq 1}$ and $(r \nablaslash)^k \mathfrak{D}^{\gamma} f^4_{\ell \geq 1}$, for $k=0,1,2$, $\vert \gamma \vert \leq 2$, and $(r \nablaslash)^k \mathfrak{D}^{\gamma} \partial_U \slashed{f}$ and $(r \nablaslash)^k \mathfrak{D}^{\gamma} \partial_V \slashed{f}$, for $k=0,1,2$, $\vert \gamma \vert \leq 1$, follow as in the proof of Proposition \ref{prop:HIdiff}, using now the estimate~\eqref{willstatelater}.
	
	The estimates for the $\ell = 0$ modes are slightly different to the $\ell = 0$ mode estimates of Proposition \ref{prop:HIdiff} due to the fact that the Kruskal Schwarzschild values differ from the Eddington--Finkelstein Schwarzschild values.
	Recall that $\vert M_{\mathrm{init}} - M_f \vert^2 \lesssim \mathbb{E}^N_{\I} + \varepsilon_0^2 + \varepsilon^3$ (see Proposition \ref{prop:dIdiff}).
	The metric relations \eqref{eq:metriccomp1} and \eqref{eq:metriccomp2} immediately imply
	\begin{equation} \label{eq:Vf3Uf4Krus}
		\vert \partial_V f^3_{\ell = 0} \vert
		+
		\vert \partial_U f^4_{\ell = 0} \vert
		\lesssim
		\varepsilon^2.
	\end{equation}
	Using the expressions \eqref{eq:Kruskalbackground0}--\eqref{eq:Kruskalbackground2} for the Schwarzschild values one easily shows
	\begin{align}
		\Big\vert (\Omega_{\circ}^2)_{\Hp} - (\Omega_{\circ}^2)_{data}
		-
		(r_d - r_{\Hp}) \frac{4M_f}{r} \Big( 1+ \frac{2M_f}{r} \Big) e^{-\frac{r}{2M_f}}
		\Big\vert 
		&
		\lesssim 
		\varepsilon^2,
	\label{eq:SchwdiffKruskal1}
	\\
		\Big\vert
		(\Omega \tr \chibar_{\circ})_{\Hp} - (\Omega \tr \chibar_{\circ})_{data}
		-
		\Big[
		2 M_f f^4 
		+
		(r_{\Hp} - r_d) V \Big( 1 + \frac{4M_f}{r} \Big)
		\Big]
		\frac{4M_f}{r^2} e^{-\frac{r}{2M_f}}
		\Big\vert
		&
		\lesssim
		\varepsilon^2,
	\label{eq:SchwdiffKruskal3}
	\\
		\Big\vert
		\rho_{\Hp} - \rho_{data}
		-
		(r_{\Hp} - r_d) \frac{6M_f}{r^4}
		\Big\vert
		&
		\lesssim
		\varepsilon^2.
	\label{eq:SchwdiffKruskal4}
	\end{align}
	The estimates for $f_{\Hp,\I}$ and $f_{d,\I}$ of Proposition \ref{prop:HIdiff} and Proposition \ref{prop:dIdiff}, along with the estimate \eqref{eq:fddestimates} and the relation \eqref{samenesshere}, imply that
	\begin{equation} \label{eq:f3f4KruskR}
		\vert f^3_{d,\Hp}(U_0,V(R,U_0),\theta) \vert^2
		+
		\vert f^4_{d,\Hp}(U_0,V(R,U_0),\theta) \vert^2
		\lesssim
		\mathbb{E}^N_{\I} + \varepsilon_0^2 + \varepsilon^3.
	\end{equation}
	The relation \eqref{eq:curvaturecomp5}, together with \eqref{eq:SchwdiffKruskal4}, implies that
	\begin{equation} \label{eq:dHdiff1}
		\vert r_{\Hp} - r_{data} \vert^2
		\lesssim
		\mathbb{E}^N_{\Hp}
		+
		\varepsilon_0^2
		+
		\varepsilon^{3},
		\qquad
	\end{equation}
	for $U_0 \leq U \leq U_f$, $V_{-1} \leq V \leq \min\{ V_2 , V(R_2,U) \}$.
	Similarly, the relations \eqref{eq:metriccomp6} and \eqref{eq:Riccicomp8}, together with \eqref{eq:SchwdiffKruskal1} and \eqref{eq:SchwdiffKruskal3}, then imply that
	\begin{equation} \label{eq:dHdiff2}
		\vert \partial_U f^3_{\ell = 0} + \partial_V f^4_{\ell = 0} \vert^2
		+
		\vert 2M_f f^4_{\ell = 0} + \Omega \tr \chibar_{\circ,\mathcal{K}}^{\Hp} \partial_U f^3_{\ell = 0} \vert^2
		\lesssim
		\mathbb{E}^N_{\Hp}
		+
		\varepsilon_0^2
		+
		\varepsilon^{3},
	\end{equation}
	for $U_0 \leq U \leq U_f$, $V_{-1} \leq V \leq \min\{ V_2 , V(R_2,U) \}$.
	In particular,
	\begin{equation} \label{eq:dHdiff3}
		\vert 2M_f f^4_{\ell = 0} - \Omega \tr \chibar_{\circ,\mathcal{K}}^{\Hp} \partial_V f^4_{\ell = 0} \vert^2
		\lesssim
		\mathbb{E}^N_{\Hp}
		+
		\varepsilon_0^2
		+
		\varepsilon^{3},
	\end{equation}
	for $U_0 \leq U \leq U_f$, $V_{-1} \leq V \leq \min\{ V_2 , V(R_2,U) \}$.
	The estimate \eqref{eq:f3f4KruskR} is then used to estimate $f^4_{\ell=0}$ on $C_{U_0}$, and the estimate \eqref{eq:Vf3Uf4Krus} is then used to give
	\begin{equation} \label{eq:dHdiff4}
		\vert f^4_{\ell = 0} \vert^2
		\lesssim
		\mathbb{E}^N_{\Hp}
		+
		\mathbb{E}^N_{\I}
		+
		\varepsilon_0^2
		+
		\varepsilon^{3},
	\end{equation}
	for $U_0 \leq U \leq U_f$, $V_{-1} \leq V \leq \min\{ V_2 , V(R_2,U) \}$.
	Note that, for $r = r_{M_f,\mathcal{K}}$,
	\[
		\partial_U r = - V \frac{4M_f^2}{r} e^{-\frac{r}{2M_f}},
		\qquad
		\partial_V r = \frac{2M_f}{V} \left( 1 - \frac{2M_f}{r} \right),
	\]
	and so
	\[
		\big\vert r (U + f^3, V+f^4)
		-
		r (U, V)
		-
		\big(
		\frac{2M_f}{V} (1- 2M_f/r) f^4
		-
		V \frac{4M_f^2}{r} e^{-\frac{r}{2M_f}} f^3
		\big)
		\big\vert
		\lesssim
		\varepsilon^2.
	\]
	Equations \eqref{eq:dHdiff1}--\eqref{eq:dHdiff4} then combine to give
	\[
		\vert f^3_{\ell = 0} \vert^2
		+
		\vert \partial_U f^3_{\ell = 0} \vert^2
		+
		\vert \partial_V f^4_{\ell = 0} \vert^2
		\lesssim
		\mathbb{E}^N_{\Hp}
		+
		\mathbb{E}^N_{\I}
		+
		\varepsilon_0^2
		+
		\varepsilon^{3},
	\]
	for $U_0 \leq U \leq U_f$, $V_{-1} \leq V \leq \min\{ V_2 , V(R_2,U) \}$, which completes the proof.
	
\end{proof}

\section{Improving the inclusion relations}
\label{improveinclusionsec}

An immediate consequence of Propositions \ref{prop:HIdiff}, \ref{prop:dIdiff} and \ref{prop:dHdiff} is that, if $\hat\varepsilon_0$ is sufficiently small, the inclusions \eqref{eq:overlap1}, \eqref{eq:overlap2}, \eqref{eq:overlap3}, \eqref{eq:overlap5} can
indeed be strengthened to~\eqref{improvedinclus1}--\eqref{improvedinclus2}
and~\eqref{impoverlap3}--\eqref{impoverlap5}.

(These improved inclusions will ensure that the inclusions \eqref{eq:overlap1}, \eqref{eq:overlap2}, \eqref{eq:overlap3}, \eqref{eq:overlap5}, still hold in the $\hat{u}_f+\delta$ gauges, defined in the context
of the proof of Theorem~\ref{thm:newgauge}.)

\section{Cancellations along a timelike hypersurface $\mathcal{B}$}
\label{subsubsec:cancelT}

We first define in {\bf Section~\ref{timelikehype}} 
a timelike hypersurface $\mathcal{B}$ which we will use as a common
boundary for energy estimates in Chapters \ref{chapter:psiandpsibar} and \ref{moreherechapter} 
for almost gauge invariant quantities. Because
these quantities are not exactly gauge invariant, there will be an error term associated to the
difference of the quantities in the two gauges along this $\mathcal{B}$. Controlling this error will be the statement
of Proposition~\ref{thm:cancelT}, given in {\bf Section~\ref{proofofthiscancel}}.
This proposition in turn forms part of the statement of Theorem~\ref{thm:relatinggauges}.

\subsection{The timelike hypersurface $\mathcal{B}_{\tilde{R}}$}
\label{timelikehype}
Recall that we fixed in Section~\ref{compediumparameterssec} a $u_1$\index{teleological $\I$ gauge!parameters!$u_{1}$} satisfying
\begin{equation}
\label{fixinguone}
u_2>u_1>u_0.
\end{equation}

For $u_f \ge u\ge u_1$, $v\ge v (R_{-1}, u_1)$, we may define   the surfaces
\[
	\widehat{S}_{u,v}
	:=
	\{u_{\I} = u\} \cap \{ v_{\Hp} = v \}.
\]
Note that these surfaces are in general \underline{not} spheres of either the $\Hp$ or the $\I$ gauge.

Now, for all $s\in [R_{-1},R_{1}]$,  $u_f\ge u\ge u_1$, it follows from  the pointwise estimate~\eqref{eq:fHpIppointwise} 
 that
$\widehat{S}_{u,v(s,u)}$ is a diffeomorphism sphere, and 
$(\theta^1_{\I}, \theta^2_{\I})$ define coordinates on all $\widehat{S}_{u,v}$.

Finally, for all $s\in [R_{-1},R_{1}]$ we may define:
\[
	\mathcal{B}_s
	:=
	\bigcup_{u=u_1}^{u_f} \widehat{S}_{u,v(s,u)}.
\]

Let us note that by the pointwise estimate \eqref{eq:fHpIppointwise} for the $f_{\Hp,\I}$ diffeomorphism functions
it follows that for fixed $s$, $\mathcal{B}_s$ is a connected timelike hypersurface with boundary
$\widehat{S}_{u_1, v(s, u_1)}\cup \widehat{S}_{u_f, v(s,u_f)}$ and 
$u_{\I}, \theta^1_{\I}, \theta^2_{\I}$ provide coordinates on $\mathcal{B}$.

We finally note that by the the pointwise estimate \eqref{eq:fHpIppointwise} we have in addition:
\begin{equation}
\label{containedinpigeon}
\bigcup_{s\in (R_{-1},R_1)} \mathcal{B}_s \subset \left\{ R_{-\frac32} \le  r(u_{\I},v_{\I})  \le R_{\frac32} \right\}.
\end{equation}

Let $\mathcal{A}_{\Gamma}$ denote the following collection of Ricci coefficients, with their linearised Kerr values subtracted
\begin{align*}
	\mathcal{A}_{\Gamma}
	=
	\
	&
	\Big\{
	1 - \frac{\Omega^2}{\Omega_{\circ}^2},
	\quad
	b
	-
	b_{\mathrm{Kerr}},
	\quad
	\Omega^{-1} \hat{\chibar},
	\quad
	\Omega^{-2} \left( \Omega \omegabarhat - (\Omega \omegabarhat)_{\circ} \right),
	\quad
	\Omega \tr \chi - (\Omega \tr \chi)_{\circ},
	\\
	&
	\quad
	\Omega^{-2} \left( \Omega \tr \chibar - (\Omega \tr \chibar)_{\circ} \right),
	\quad
	\Omega\hat{\chi},
	\quad
	\eta - \eta_{\mathrm{Kerr}},
	\quad
	\etabar - \etabar_{\mathrm{Kerr}},
	\quad
	\Omega \omegahat - (\Omega \omegahat)_{\circ}
	\Big\},
\end{align*}
and let $\mathcal{A}_{\mathcal{R}}$ denote the curvature components $\Omega \beta$, $\Omega^{-1} \betabar$, $\rho - \rho_{\circ}$, $\sigma$, minus their linearised Kerr values,
\[
	\mathcal{A}_{\mathcal{R}}
	=
	\Big\{
	\Omega^{-1} \betabar - \Omega^{-1} \betabar_{\mathrm{Kerr}},
	\quad
	\rho - \rho_{\circ},
	\quad
	\sigma,
	\quad
	\Omega \beta - \Omega \beta_{\mathrm{Kerr}}
	\Big\}.
\]
Elements of $\mathcal{A}_{\mathcal{R}}$ and $\mathcal{A}_{\Gamma}$ are denoted $\breve{\mathcal{R}}$ and $\breve{\Gamma}$ respectively, with a $\breve{}$ added to emphasise the fact that the linearised Kerr values have been subtracted.
Set
\begin{equation} \label{eq:kappadef}
	\kappa(0) = 0,
	\quad
	\kappa(1) = 1-\delta,
	\quad
	\kappa(2) = 2-\delta,
\end{equation}
and define
\begin{align*}
	e_{\delta, N}
	:=
	\
	&
	\sum_{s=0}^1
	u^{\kappa(s)}
	\sum_{\vert \gamma \vert \leq N-s}
	\sum_{\breve{\Gamma} \in \mathcal{A}_{\Gamma}}
	\vert \mathfrak{D}^{\gamma} \breve{\Gamma}^{\I} \vert^2
	+
	\sum_{s=0}^2
	u^{\kappa(s)}
	\Big(
	\sum_{\vert \gamma \vert \leq N-1-s}
	\sum_{\breve{\Gamma} \in \mathcal{A}_{\Gamma}}
	\vert \mathfrak{D}^{\gamma} \breve{\Gamma}^{\Hp} \vert^2
	\\
	&
	+
	\sum_{\vert \gamma \vert \leq N-s}
	\big(
	\sum_{\breve{\mathcal{R}} \in \mathcal{A}_{\mathcal{R}}}
	\vert \mathfrak{D}^{\gamma} \breve{\mathcal{R}}^{\Hp} \vert^2
	+
	\vert \mathfrak{D}^{\gamma} \alpha^{\Hp} \vert^2
	+
	\vert \mathfrak{D}^{\gamma} \alpha^{\I} \vert^2
	+
	\vert \mathfrak{D}^{\gamma} \alphabar^{\Hp} \vert^2
	+
	\vert \mathfrak{D}^{\gamma} \alphabar^{\I} \vert^2
	\big)
	\Big).
\end{align*}

By the mean value theorem (cf.\@ \cite{DafRodnew}) there exists a parameter $\tilde{R} = \tilde{R}(u_f, \lambda)$ 
with the property that $R_{-1} \leq \tilde{R} \leq R_{1}$ and
\begin{equation}
\label{pigeonapplication}
	\int_{\mathcal{B}_{\tilde{R}}} e_{\delta, N} d \theta du
	\leq
	\int_{R_{-1}}^{R_1}\left( \int_{\mathcal{B}_{s}} e_{\delta, N} d \theta du\right) ds.
\end{equation}

Note that the integral on the left hand side of $(\ref{pigeonapplication})$ 
is an integral over the timelike hypersurface $\mathcal{B}_{\tilde{R}}$ and the integral on the right hand side is an integral over a spacetime region around $\mathcal{B}_{\tilde{R}}$.  

As it is precisely this hypersurface $\mathcal{B}_{\tilde{R}}$ which we shall use,
we shall from now on omit the subscript $\tilde{R}$, i.e.~we define
\[
	\mathcal{B} := \mathcal{B}_{\tilde{R}}.
\]
Given $u_1 \leq \tau \leq u_f$, define
\[
	\mathcal{B} (\tau):= \mathcal{B} \cap \{ u \geq \tau \}
	=
	\bigcup_{u=\tau}^{u_f} \widehat{S}_{u,v(\tilde{R},u)}.
\]
We define finally $v_1$\index{teleological $\Hp$ gauge!parameters!$v_{1}=v_1(u_f, \lambda)$} by 
\begin{equation}
\label{v1definition}
v_1=v(\tilde{R}, u_1).
\end{equation}
Refer to Figure~\ref{justTfigure}.
\begin{figure}
\centering{
\def\svgwidth{20pc}
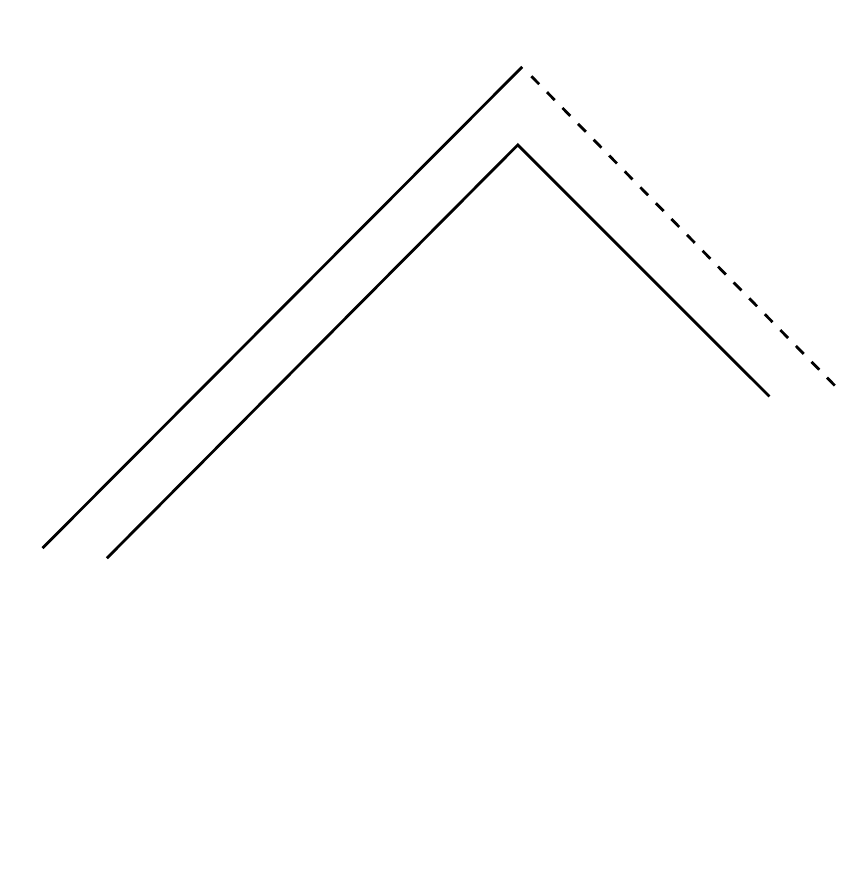}
\caption{The hypersurface $\mathcal{B}$}\label{justTfigure}
\end{figure}
By~\eqref{0minitconstraint},~\eqref{u1constraint} and the pointwise estimate~\eqref{eq:fHpIppointwise}, we have that
\[
v_0<v_1<v_2.
\]

\begin{remark}
Note that $v_1$, as opposed to our other parameters, depends on $u_f$ and $\lambda$.
\end{remark}

Define the $L^2$ norm of a $(0,k)$ tensor $\xi$ on $\mathcal{B}$,\index{energies!norms on timelike hypersurfaces!$\Vert \xi \Vert_{\mathcal{B}}$}\index{energies!norms on timelike hypersurfaces!$\Vert \xi \Vert_{\mathcal{B}}(\tau)$}

\[
	\Vert \xi \Vert_{\mathcal{B}(\tau)}^2
	:=
	\int_{\tau}^{u_f} \int_{\widehat{S}_{\tau',v(\tilde{R},\tau')}} \vert \xi \vert^2 d \mathrm{Vol}_{\mathcal{B}},
\]
where $d \mathrm{Vol}_{\mathcal{B}} = \Omega^2 d \theta \wedge du \wedge dv (n_{\mathcal{B}})$ and $n_{\mathcal{B}}$ is the appropriately oriented unit normal to $\mathcal{B}$.  Define also
\[
	\Vert \xi \Vert_{\mathcal{B}}
	:=
	\Vert \xi \Vert_{\mathcal{B}(u_1)}.
\]

\subsection{Cancellation between almost gauge invariant quantities on the timelike hypersurface $\mathcal{B}$}
\label{proofofthiscancel}
We may now state the result concerning cancellation on $\mathcal{B}$.  If $\xi$ and $\xi'$ are two $S$-tangent $(0,k)$ tensors, their inner product is defined by 
\[
	(\xi,\xi')_{\gslash}
	:=
	\slashed{g}^{A_1B_1} \ldots \slashed{g}^{A_kB_k}
	\xi_{A_1} \ldots \xi_{A_k}
	\xi'_{B_1} \ldots \xi'_{B_k}.
\]

\begin{proposition}[Cancellations between almost gauge invariant quantities on a timelike hypersurface]
 \label{thm:cancelT}
	Recall the hypersurface $\mathcal{B}$ defined in Section~\ref{timelikehype}.
	Given $s=0,1,2$, a smooth admissible coefficient function $h$ as in Section~\ref{subsec:traceindex}, and multi-indices $\gamma_1, \gamma_2$ such that $\vert \gamma_1 \vert \leq N-s$, $\vert \gamma_2 \vert \leq N-s$ and such that $\mathfrak{D}^{\gamma_1} \alpha$ and $\mathfrak{D}^{\gamma_2} \alpha$ are tensor fields of the same order, then,
	\begin{equation} \label{eq:cancelTalpha}
		\sup_{u_1 \leq u \leq u_f} u^s
		\int_{\mathcal{B}(u)}
		\left\vert
		\left( h(r) (\mathfrak{D}^{\gamma_1} \alpha, \mathfrak{D}^{\gamma_2} \alpha)_{\gslash} \right)^{\Hp}
		-
		\left( h(r) (\mathfrak{D}^{\gamma_1} \alpha, \mathfrak{D}^{\gamma_2} \alpha)_{\gslash} \right)^{\I}
		\right\vert
		\lesssim
		\varepsilon^3,
	\end{equation}
	and
	\begin{equation} \label{eq:cancelTalphabar}
		\sup_{u_1 \leq u \leq u_f} u^s
		\int_{\mathcal{B}(u)}
		\left\vert
		\left( h(r) (\mathfrak{D}^{\gamma_1} (r \alphabar), \mathfrak{D}^{\gamma_2} (r \alphabar))_{\gslash} \right)^{\Hp}
		-
		\left( h(r) (\mathfrak{D}^{\gamma_1} (\check{r} \alphabar), \mathfrak{D}^{\gamma_2} (\check{r} \alphabar))_{\gslash} \right)^{\I}
		\right\vert
		\lesssim
		\varepsilon^3.
	\end{equation}
\end{proposition}

Note that Proposition \ref{thm:cancelT} in particular also implies that a cancellation occurs between appropriate derivatives of $P^{\I}$ and $P^{\Hp}$, and also $\Pbar^{\I}$ and $\Pbar^{\Hp}$, on the timelike hypersurface $\mathcal{B}$.

The following three lemmas will be used to estimate nonlinear error terms arising in the proof of Proposition \ref{thm:cancelT}.  First note that, recalling the definition of $\mathcal{B}$ in Section \ref{timelikehype} and the bootstrap assumption \eqref{eq:bamain}, it follows (see the argument of Lemma \ref{lem:ILEDtrapping}) that, for $s=0,1,2$ and $\vert \gamma \vert \leq N-s$,
\[
	\int_{\mathcal{B}} u^{\kappa(s)} 
	\big( \vert \mathfrak{D}^{\gamma} \alpha_{\Hp} \vert^2 + \vert \mathfrak{D}^{\gamma} \alpha_{\I} \vert^2 \big)
	\lesssim
	\int_{\DRI_{R_{-1} \leq r \leq R_1}} u^{\kappa(s)} \big( \vert \mathfrak{D}^{\gamma} \alpha_{\Hp} \vert^2 + \vert \mathfrak{D}^{\gamma} \alpha_{\I} \vert^2 \big)	\lesssim
	\varepsilon^2,
\]
with $\kappa$ defined by \eqref{eq:kappadef}.  It then follows that, for all $u_1 \leq u \leq u_f$,
\[
	\int_{\mathcal{B}(u)} \big( \vert \mathfrak{D}^{\gamma} \alpha_{\Hp} \vert^2 + \vert \mathfrak{D}^{\gamma} \alpha_{\I} \vert^2 \big)
	\lesssim
	\frac{\varepsilon^2}{u^{\kappa(s)}}.
\]
Similarly, the inequalities (in the notation of Section \ref{timelikehype})
\[
	\int_{\DRI_{R_{-1} \leq r \leq R_1}} \Big(
	\sum_{\vert \gamma \vert \leq N-s} 
	\vert \mathfrak{D}^{\gamma} \breve{\mathcal{R}}_{\Hp} \vert^2
	+
	\sum_{\vert \gamma \vert \leq N-1-s}
	\vert \mathfrak{D}^{\gamma} \breve{\Gamma}_{\Hp} \vert^2
	\Big)
	\lesssim
	\frac{\varepsilon^2}{u^s},
\qquad
	\sum_{\vert \gamma \vert \leq N-1}
	\int_{\DRI_{R_{-1} \leq r \leq R_1}}
	\vert \mathfrak{D}^{\gamma} \breve{\Gamma}_{\I} \vert^2
	\lesssim
	\frac{\varepsilon^2}{u},
\]
contained in the bootstrap assumption \eqref{eq:bamain}
imply in particular that
\[
	\sum_{\vert \gamma \vert \leq N-s} \int_{\mathcal{B}(u)}
	\vert \mathfrak{D}^{\gamma} (\mathcal{R} - \mathcal{R}_{\circ})_{\Hp} \vert^2
	\lesssim
	\frac{\varepsilon^2}{u^{\kappa(s)}}
	+
	\frac{\varepsilon^2}{u_f},
	\qquad
	\sum_{s=0}^1
	\sum_{\vert \gamma \vert \leq N-1-s}
	\int_{\mathcal{B}(u)}
	\vert \mathfrak{D}^{\gamma} (\Gamma - \Gamma_{\circ}) \vert^2
	\lesssim
	\frac{\varepsilon^2}{u^{\kappa(s)}},
\]
for each $\Gamma - \Gamma_{\circ} = (\Gamma - \Gamma_{\circ})_{\Hp}$ and $(\Gamma - \Gamma_{\circ})_{\I}$, where $\kappa$ is defined by \eqref{eq:kappadef}.

\begin{lemma}[Cancellation between $r_{\Hp}$ and $r_{\I}$] \label{lem:rHrIerror}
	Given a smooth admissible coefficient function $h(r)$, for any $\vert \gamma \vert \leq N-s$, for $s=0,1,2$,
	\[
		\int_{\mathcal{B}(u)} \vert h(r_{\Hp}) - h(r_{\I}) \vert \vert \mathfrak{D}^{\gamma} \alpha_{\Hp} \vert^2
		\lesssim
		\frac{\varepsilon^{3}}{u^{1+\kappa(s)}},
	\]
	for any $u_1 \leq u \leq u_f$, with $\kappa$ defined by \eqref{eq:kappadef}.  Similarly for $\alpha_{\I}$, $\alphabar_{\Hp}$ and $\alphabar_{\I}$ respectively in place of $\alpha_{\Hp}$.
\end{lemma}

\begin{proof}
	Note that,
	\[
		(r_{\Hp} - r_{\I})\vert_{(u_{\I},v_{\I},\theta_{\I}) = (u,v,\theta)}
		=
		r_{M_f} (u + f^3_{\Hp,\I},v+f^4_{\Hp,\I}) - r_{M_f} (u,v),
	\]
	and so,
	\[
		\int_{\mathcal{B}(u)} \vert h(r_{\Hp}) - h(r_{\I}) \vert \vert \mathfrak{D}^{\gamma} \alpha_{\Hp} \vert^2
		\lesssim
		\int_{\mathcal{B}(u)} \vert f^3 - f^4 \vert \vert \mathfrak{D}^{\gamma} \alpha_{\Hp} \vert^2
		\lesssim
		\frac{\varepsilon}{u} \int_{\mathcal{B}(u)} \vert \mathfrak{D}^{\gamma} \alpha_{\Hp} \vert^2
		\lesssim
		\frac{\varepsilon^{3}}{u^{1+\kappa(s)}}.
	\]
\end{proof}

\begin{lemma}[Cancellation between $r_{\Hp}$ and $\check{r}_{\I}$] \label{lem:rHrIcheckerror}
	For any $\vert \gamma_1 \vert + \vert \gamma_2 \vert \leq N-s$, $\vert \gamma_3 \vert \leq N-s$ for $s=0,1,2$,
	\[
		\int_{\mathcal{B}(u)}
		\vert \mathfrak{D}^{\gamma_1} (r_{\Hp} - \check{r}_{\I}) \vert
		\vert \mathfrak{D}^{\gamma_2} \alphabar_{\Hp} \vert
		\vert \mathfrak{D}^{\gamma_3} \alphabar_{\Hp} \vert
		\lesssim
		\frac{\varepsilon^{3}}{u^{s}},
	\]
	for any $u_1 \leq u \leq u_f$.  Similarly for $\alphabar_{\I}$ in place of $\alpha_{\Hp}$.
\end{lemma}
\begin{proof}
	At most one of $\vert \gamma_1 \vert$ and $\vert \gamma_2 \vert$ can be greater than $N-5$.  Suppose first that $\vert \gamma_1 \vert >N-5$.  Then $\vert \gamma_2 \vert \leq N-5$ and the pointwise estimate \eqref{elinfestimates} implies that
	\[
		\int_{\mathcal{B}(u)}
		\vert \mathfrak{D}^{\gamma_1} (r_{\Hp} - \check{r}_{\I}) \vert
		\vert \mathfrak{D}^{\gamma_2} \alphabar_{\Hp} \vert
		\vert \mathfrak{D}^{\gamma_3} \alphabar_{\Hp} \vert
		\lesssim
		\frac{\varepsilon}{u}
		\Vert \mathfrak{D}^{\gamma_1} (r_{\Hp} - \check{r}_{\I}) \Vert_{\mathcal{B}(u)}
		\Vert \mathfrak{D}^{\gamma_2} \alphabar_{\Hp} \Vert_{\mathcal{B}(u)}.
	\]
	The proof follows the proof of Lemma \ref{lem:rHrIerror} after noting that
	\[
		\check{r} = r + \frac{2\Omega^2}{\Omega \tr \chi} - \frac{2\Omega^2_{\circ}}{\Omega \tr \chi_{\circ}},
	\]
	and that the definition of $\mathcal{B}$ and the bootstrap assumption \eqref{eq:bamain} imply that
	\[
		\sum_{s=0}^1
		\sum_{\vert \gamma \vert \leq N-s}
		\int_{\mathcal{B}(u)}
		\Big(
		\vert \mathfrak{D}^{\gamma} (\Omega \tr \chi - \Omega \tr \chi_{\circ}) \vert^2
		+
		\vert \mathfrak{D}^{\gamma} (\Omega_{\circ}^{-2} \Omega^2 - 1) \vert^2
		\Big)
		\lesssim
		\frac{\varepsilon^2}{u^{\kappa(s)}},
	\]
	where $\kappa$ defined by \eqref{eq:kappadef}.  The proof when $\vert \gamma_1 \vert \leq N-5$ is similar.
\end{proof}

Before proceeding further, it is convenient to introduce additional nonlinear error notation.  Recall the notation of Section \ref{sec:nlenotation}.  Note that if $i \in \{ 6,12,14,15,16,17\}$ and $\gamma \in \{ (1,0,0), (0,1,0), (0,0,1)\}^k$, then $\mathfrak{D}^k \Phi \cdot (i,\gamma)$ is equal to $\mathfrak{D}^{\gamma}$ applied to a \emph{curvature component} (and not a Ricci coefficient).  Accordingly, given $k \geq 0$, $l \geq 1$ and some
\begin{equation} \label{eq:Ferror1}
	H = \{ {}^{j,m}H_{k_1\ldots k_{l'}},  {}^jJ_{k_1\ldots k_{l'}} \}_{m=0,\ldots,l',k_1+\ldots + k_{l'} \leq k, l' \geq l, j \geq 1},
\end{equation}
as in Section \ref{sec:nlenotation}, if
\begin{equation} \label{eq:Ferror2}
	{}^{j,m}H_{k_1\ldots k_{l'}}
	\in
	\big\{ (i,\gamma) \mid i \in \{ 6,12,14,15,16,17\}, \gamma \in \{ (1,0,0), (0,1,0), (0,0,1)\}^{k_m} \big\} \cup \big\{ 0 \big\},
\end{equation}
for all $m=0,\ldots,l'$, $k_1+\ldots + k_{l'} \leq k$, $l' \geq l$, $j \geq 1$, then we denote
\[
	(\mathfrak{D}^k (\mathcal{R} - \mathcal{R}_{\circ}))^l \cdot H
	:=
	(\mathfrak{D}^k \Phi )^l \cdot H.
\]
Similarly, if $H$ is such that
\begin{equation} \label{eq:Ferror3}
	{}^{j,m}H_{k_1\ldots k_{l'}}
	\in
	\big\{ (i,\gamma) \mid i \in \{ 1,\ldots, 5, 7, \ldots, 11, 13 \}, \gamma \in \{ (1,0,0), (0,1,0), (0,0,1)\}^{k_m} \big\} \cup \big\{ 0 \big\},
\end{equation}
for all $m=0,\ldots,l'$, $k_1+\ldots + k_{l'} \leq k$, $l' \geq l$, $j \geq 1$, then $(\mathfrak{D}^k \Phi )^l \cdot H$ involves only \emph{Ricci coefficients} (and no curvature components) and we denote
\[
	(\mathfrak{D}^k (\Gamma - \Gamma_{\circ}))^l \cdot H
	:=
	(\mathfrak{D}^k \Phi )^l \cdot H.
\]

Given $k_1$, $k_2$, $k_3 \geq 0$ and $l_1$, $l_2$, $l_3$, $l_4 \geq 0$, define $\underline{k} = (k_1, k_2,k_3)$, $\underline{l} = (l_1, l_2,l_3, l_4)$ and
\[
	\vert \underline{k} \vert = k_1 + k_2 + k_3,
	\qquad
	\vert \underline{l} \vert = l_1+l_2+l_3+l_4.
\]
Consider some $k \geq 0$, $G = \{ {}^j G^{\underline{k}}_{\underline{l}} \}$ where each ${}^j G^{\underline{k}}_{\underline{l}}$ is as in Section \ref{subsubsec:Dfnotation}, $H = \{ {}^j H^{\underline{k}}_{\underline{l}} \}$ where each ${}^j H^{\underline{k}}_{\underline{l}}$ is of the form \eqref{eq:Ferror1}, \eqref{eq:Ferror2}, $K = \{ {}^j K^{\underline{k}}_{\underline{l}} \}$ and $L = \{ {}^j L^{\underline{k}}_{\underline{l}} \}$ where each ${}^j K^{\underline{k}}_{\underline{l}}$ and ${}^j L^{\underline{k}}_{\underline{l}}$ is of the form \eqref{eq:Ferror1}, \eqref{eq:Ferror3}, and $I = \{ {}^j I^{\underline{k}}_{\underline{l}} \}$ and $J = \{ {}^j J^{\underline{k}}_{\underline{l}} \}$ where each ${}^j I^{\underline{k}}_{\underline{l}}$ and ${}^j J^{\underline{k}}_{\underline{l}}$ is a trace index set of some order $d\geq 0$ as in Section \ref{subsec:traceindex}.
Define $\mathcal{F}^k(G,H,K,L,I,J)$ by,\index{schematic notation!$\mathcal{F}^{k}(G,H,K,L,I,J)$,  non-linear error term notation associated to $f$ and geometric quantities of the two teleological gauges}
\begin{multline} \label{eq:calFerror}
	\mathcal{F}^{k}(G,H,K,L,I,J)
	=
	\Pi_{S^{\I}}
	\sum_{\substack{\vert \underline{l} \vert \geq 2 \\ l_1 \geq 1}}
	\sum_{\substack{\vert \underline{k} \vert \leq k \\ k_2+k_3 \leq k-1 \vee 0}}
	\sum_{j \geq 1}
	\Big(
	(\mathfrak{D} \fsc)^{l_1} \cdot {}^j G^{\underline{k}}_{\underline{l}}
	\otimes
	(\mathfrak{D}^{k_1} (\mathcal{R} - \mathcal{R}_{\circ})_{\Hp})^{l_2} \cdot {}^j H^{\underline{k}}_{\underline{l}}
	\\
	\otimes
	\left( \mathfrak{D}^{k_2} (\Gamma - \Gamma_{\circ})_{\Hp} \right)^{l_3} \cdot {}^j K^{\underline{k}}_{\underline{l}}
	\otimes
	\left( \mathfrak{D}^{k_3} (\Gamma - \Gamma_{\circ})_{\I} \right)^{l_4} \cdot {}^j L^{\underline{k}}_{\underline{l}}
	\Big)
	\cdot_{\gslash_{\Hp}} {}^j I^{\underline{k}}_{\underline{l}}
	\cdot_{\gslash_{\I}} {}^j J^{\underline{k}}_{\underline{l}},
\end{multline}
where $\Pi_{S^{\I}}$ denotes projection to the $S^{\I}$ spheres (see \eqref{eq:tildeprojection}), $\mathfrak{D} \fsc = \mathfrak{D} \fsc_{\Hp,\I}$ and $k-1 \vee 0 = \max \{ k-1,0\}$.  As usual, such expressions are only ever considered when sufficiently many of the elements of $G,H,K,L,I,J$ vanish so that every term in the summation is a tensor of the same type, and so that there are only finitely many terms, so that the summation is well defined.  We typically abuse notation and write ``$\mathcal{F}^k$'' for ``$\mathcal{F}^{k}(G,H,K,L,I,J)$ for some $G,H,K,L,I,J$''.

Note that $\mathcal{F}^k$ is nonlinear, each term involves at least one $\mathfrak{D} \fsc$ factor, and each term involves at most $k-1$ derivatives of $(\Gamma - \Gamma_{\circ})_{\Hp}$ and $(\Gamma - \Gamma_{\circ})_{\I}$.

\begin{lemma}[Error estimate on timelike hypersurface] \label{lem:fHIalphaerror}
	For any $s=0,1,2$ and $k \leq N-s$, $\vert \gamma \vert \leq N-s$, for any $u_1 \leq u \leq u_f$,
	\[
		\int_{\mathcal{B}(u)} \vert \mathcal{F}^{k} \vert_{\gslash^{\I}} \vert \mathfrak{D}^{\gamma} \alpha \vert
		\lesssim
		\frac{\varepsilon^{3}}{u^{\kappa(s) + \frac{1}{2} }}
		,
	\]
	for $\alpha = \alpha_{\Hp}$ and $\alpha_{\I}$, where $\kappa(0)=0$, $\kappa(1) = 1 - \delta$ and $\kappa(2) = 2-\delta$.
\end{lemma}

\begin{proof}
	Recall, for $s=0,1,2$, for $\alpha = \alpha_{\Hp}$ or $\alpha_{\I}$ and for each $\mathcal{R}_{\Hp}$,
	\begin{equation*} 
		\sum_{\vert \gamma \vert \leq N-s} \int_{\mathcal{B}(u)}
		\vert \mathfrak{D}^{\gamma} \alpha \vert^2
		\lesssim
		\frac{\varepsilon^2}{u^{\kappa(s)}},
		\qquad
		\sum_{\vert \gamma \vert \leq N-s} \int_{\mathcal{B}(u)}
		\vert \mathfrak{D}^{\gamma} (\mathcal{R} - \mathcal{R}_{\circ})_{\Hp} \vert^2
		\lesssim
		\frac{\varepsilon^2}{u^{\kappa(s)}}
		+
		\frac{\varepsilon^2}{u_f}.
	\end{equation*}
	Each $(\Gamma - \Gamma_{\circ}) = (\Gamma - \Gamma_{\circ})_{\Hp}$ and $(\Gamma - \Gamma_{\circ})_{\I}$ satisfies,
	\[
		\sum_{s=0}^1
		\sum_{\vert \gamma \vert \leq N-1-s}
		\int_{\mathcal{B}(u)}
		\vert \mathfrak{D}^{\gamma} (\Gamma - \Gamma_{\circ}) \vert^2
		\lesssim
		\frac{\varepsilon^2}{u^{\kappa(s)}}.
	\]
	Recall also that the diffeomorphisms satisfy
	\[
		\vert \mathfrak{D} \fsc \vert
		\lesssim
		\frac{\varepsilon}{u}.
	\]
	See \eqref{eq:fHpIppointwise}.  The proof follows (using Proposition \ref{prop:tildenorm}) from noting that, for $s=0,1,2$, $\vert\gamma _1 \vert \leq N-s$, $\vert \gamma_2 \vert \leq N-s$,
	\[
		\int_{\mathcal{B}(u)} \vert \mathfrak{D} \fsc \vert \vert \mathfrak{D}^{\gamma_1} (\mathcal{R} - \mathcal{R}_{\circ})_{\Hp} \vert \vert \mathfrak{D}^{\gamma_2} \alpha \vert
		\lesssim
		\frac{\varepsilon}{u}
		\Big( \int_{\mathcal{B}(u)} \vert \mathfrak{D}^{\gamma_1} (\mathcal{R} - \mathcal{R}_{\circ})_{\Hp} \vert^2 \Big)^{\frac{1}{2}}
		\Big( \int_{\mathcal{B}(u)} \vert \mathfrak{D}^{\gamma_2} \alpha \vert^2 \Big)^{\frac{1}{2}}
		\lesssim
		\frac{\varepsilon^3}{u^{1+\frac{\kappa(s)}{2}}} \Big( \frac{1}{u^{\kappa(s)}} + \frac{1}{u_f} \Big)^{\frac{1}{2}},
	\]
	and, for $s=0,1,2$, $\vert\gamma _1 \vert \leq N-1-s$, $\vert \gamma_2 \vert \leq N-s$
	\[
		\int_{\mathcal{B}(u)} \vert \mathfrak{D} \fsc \vert \vert \mathfrak{D}^{\gamma_1} (\Gamma - \Gamma_{\circ}) \vert \vert \mathfrak{D}^{\gamma_2} \alpha \vert
		\lesssim
		\frac{\varepsilon}{u}
		\Big( \int_{\mathcal{B}(u)} \vert \mathfrak{D}^{\gamma_1} (\Gamma - \Gamma_{\circ}) \vert^2 \Big)^{\frac{1}{2}}
		\Big( \int_{\mathcal{B}(u)} \vert \mathfrak{D}^{\gamma_2} \alpha \vert^2 \Big)^{\frac{1}{2}}
		\lesssim
		\frac{\varepsilon^3}{u^{1+\frac{\kappa(s) + \kappa(\min\{s,1\})}{2}}},
	\]
	for $\Gamma - \Gamma_{\circ} = (\Gamma - \Gamma_{\circ})_{\Hp}$ and $(\Gamma - \Gamma_{\circ})_{\I}$, and for $\alpha = \alpha_{\Hp}$ and $\alpha_{\I}$.
\end{proof}

The proof of Proposition~\ref{thm:cancelT} can now be given.

\begin{proof}[Proof of Proposition~\ref{thm:cancelT}]
Consider first the estimate \eqref{eq:cancelTalpha} for $\alpha$ and the case $s=2$.

First, suppose $\vert \gamma_1 \vert = \vert \gamma_2 \vert = 0$.  Recall the projection $\Pi_{S^{\I}}$ to the $S^{\I}$ spheres (see \eqref{eq:tildeprojection}) and note that the bootstrap assumption \eqref{eq:badiffeo} implies that $\Pi_{S^{\I}} \gslash^{\Hp}$ is a metric on the $S^{\I}$ spheres.  It follows (see the proof of Proposition \ref{prop:tildenorm}) that
\[
	\vert \alpha_{\Hp} \vert^2_{\gslash^{\Hp}}
	=
	\vert \Pi_{S^{\I}} \alpha_{\Hp} \vert^2_{\Pi_{S^{\I}}\gslash^{\Hp}},
\]
and so
\begin{multline*}
	h(r_{\Hp}) \vert \alpha_{\Hp} \vert^2_{\gslash^{\Hp}} - h(r_{\I}) \vert \alpha_{\I} \vert^2_{\gslash^{\I}}
	=
	\left( h(r_{\Hp}) - h(r_{\I}) \right) \vert \alpha_{\Hp} \vert^2_{\gslash^{\Hp}}
	\\
	+
	h(r_{\I}) \big( \vert \Pi_{S^{\I}} \alpha_{\Hp} \vert^2_{\Pi_{S^{\I}} \gslash^{\Hp}} - \vert \Pi_{S^{\I}} \alpha_{\Hp} \vert^2_{\gslash^{\I}} \big)
	+
	h(r_{\I})
	\left( \Pi_{S^{\I}} \alpha_{\Hp} + \alpha_{\I} , \Pi_{S^{\I}} \alpha_{\Hp} - \alpha_{\I} \right)_{\gslash^{\I}}.
\end{multline*}
Note that the relation \eqref{eq:curvaturecomp1} implies
\[
	\Pi_{S^{\I}} \alpha_{\Hp} - \alpha_{\I}
	=
	\mathcal{E}^{1,0}_{\mathfrak{D} \fsc}.
\]
The proof then follows from Lemma \ref{lem:rHrIerror} and Lemma \ref{lem:fHIalphaerror}.

The proof for $\vert \gamma_1 \vert = \vert \gamma_2 \vert \geq 1$ is similar.  It is helpful to first consider the case $\vert \gamma_1 \vert = \vert \gamma_2 \vert = 1$.  Suppose, for example, that $\gamma_1 = \gamma_2 = (0,0,1)$.  Again,
\begin{multline} \label{eq:nabla4alphaHIdiff}
	h(r_{\Hp}) \vert (r\Omega \nablaslash_4 \alpha)_{\Hp} \vert^2_{\gslash^{\Hp}} - h(r_{\I}) \vert (r\Omega \nablaslash_4 \alpha)_{\I} \vert^2_{\gslash^{\I}}
	\\
	=
	\left( h(r_{\Hp}) - h(r_{\I}) \right) \vert (r\Omega \nablaslash_4 \alpha)_{\Hp} \vert^2_{\gslash^{\Hp}}
	+
	h(r_{\I}) \big( \vert \Pi_{S^{\I}} (r\Omega \nablaslash_4 \alpha)_{\Hp} \vert^2_{\Pi_{S^{\I}} \gslash^{\Hp}} - \vert \Pi_{S^{\I}} (r\Omega \nablaslash_4 \alpha)_{\Hp} \vert^2_{\gslash^{\I}} \big)
	\\
	+
	h(r_{\I})
	\left( \Pi_{S^{\I}} (r\Omega \nablaslash_4 \alpha)_{\Hp} + (r\Omega \nablaslash_4 \alpha)_{\I} , \Pi_{S^{\I}} (r\Omega \nablaslash_4 \alpha)_{\Hp} - (r\Omega \nablaslash_4 \alpha)_{\I} \right)_{\gslash^{\I}}.
\end{multline}
Note that, in both the $\Hp$ and $\I$ gauges,
\[
	\nablaslash_4 \alpha_{AB}
	=
	(\nabla_{e_4} R) (e_A, e_4, e_B, e_4)
	-
	R(e_A, \nabla_{e_4} e_4, e_B, e_4)
	-
	R(e_A, e_4, e_B, \nabla_{e_4} e_4),
\]
and so,
\begin{align*}
	&
	\Pi_{S^{\I}} (\Omega \nablaslash_4 \alpha_{AB})^{\Hp} - (\Omega \nablaslash_4 \alpha_{AB})^{\I}
	=
	(\nabla_{\Omega e_4^{\Hp}} R) ( H_A^C e_C^{\Hp}, e_4^{\Hp}, H_B^D e_D^{\Hp}, e_4^{\Hp})
	-
	(\nabla_{\Omega e_4^{\I}} R) (e_A^{\I}, e_4^{\I}, e_B^{\I}, e_4^{\I})
	\\
	&
	\qquad
	+
	2\left( \Omega \omegahat - (\Omega \omegahat)_{\circ} \right)^{\I}
	R (e_A^{\I}, e_4^{\I}, e_B^{\I}, e_4^{\I})
	-
	2\left( \Omega \omegahat - (\Omega \omegahat)_{\circ} \right)^{\Hp}
	R (H_A^C e_C^{\Hp}, e_4^{\Hp}, H_B^D e_D^{\Hp}, e_4^{\Hp})
	\\
	&
	\qquad
	+
	\frac{2 M_f}{r_{\I}^2}
	R (e_A^{\I}, e_4^{\I}, e_B^{\I}, e_4^{\I})
	-
	\frac{2M_f}{r_{\Hp}^2}
	R (H_A^C e_C^{\Hp}, e_4^{\Hp}, H_B^D e_D^{\Hp}, e_4^{\Hp}),
\end{align*}
where
\[
	H_A^C = \delta_A^C + \frac{\partial f^C_{\Hp,\I}}{\partial \theta^A_{\I}}.
\]
As in the proof of Proposition \ref{prop:curvaturerelations},
\[
	(\nabla_{\Omega e_4^{\Hp}} R) (H_A^C e_C^{\Hp}, e_4^{\Hp}, H_B^D e_D^{\Hp}, e_4^{\Hp})
	-
	(\nabla_{\Omega e_4^{\I}} R) (e_A^{\I}, e_4^{\I}, e_B^{\I}, e_4^{\I})
	\\
	=
	\mathcal{F}^1,
\]
and so it follows from Lemma \ref{lem:fHIalphaerror} that
\[
	\int_{\mathcal{B}(u)}
	\left\vert
	(r\Omega \nablaslash_4 \alpha)_{\Hp}
	\right\vert
	\left\vert
	\left( (r\Omega \nablaslash_4 \alpha)_{\Hp} - (r\Omega \nablaslash_4 \alpha)_{\I} \right)
	\right\vert
	\lesssim
	\frac{\varepsilon^{3}}{u^2}.
\]
The remaining terms in \eqref{eq:nabla4alphaHIdiff} are estimated similarly.

At higher orders, defining,
\[
	\nabla^{(l_1,l_2,l_3)}_{C_1,\ldots,C_{l_1}} := \nabla_{e_{C_1}} \ldots \nabla_{e_{C_{l_1}}} (\nabla_{e_3})^{l_2} (\nabla_{e_4})^{l_3},
\]
it follows that, in both the $\Hp$ and $\I$ gauges,
\[
	\nablaslash_{C_1} \ldots \nablaslash_{C_{k_1}} (\nablaslash_3)^{k_2} (\nablaslash_4)^{k_3} \alpha_{AB}
	=
	\sum_{\substack{ l_1+m_1+n_1 = k_1
	\\
	l_2+m_2+n_2 = k_2
	\\
	l_3+m_3+n_3 = k_3}}
	M_{l_1\ldots n_3}
	(\nabla^{(l_1,l_2,l_3)}_{C_1,\ldots,C_{l_1}} R)
	(e_A,
	\nabla^{(m_1,m_2,m_3)}_{C_1,\ldots,C_{m_1}} e_4,
	e_B,
	\nabla^{(n_1,n_2,n_3)}_{C_1,\ldots,C_{n_1}} e_4),
\]
for some constants $M_{l_1\ldots n_3}$.  The proof then follows similarly from Lemma \ref{lem:fHIalphaerror} since, schematically, for $m_1+m_2+m_3 \geq 1$,
\begin{multline*}
	\nabla^{(m_1,m_2,m_3)}_{C_1,\ldots,C_{m_1}} e_4
	=
	\sum_{\substack{
	j_1 +j_2+j_3 \\
	\leq m_1+m_2+m_3 -1
	\\ l\geq 1}}
	\left( (\nablaslash^{j_1} \nablaslash_3^{j_2} \nablaslash_4^{j_3} \Gamma)^l (e_{C_1}\ldots e_{C_{j_1}}) \right)^A e_A
	\\
	+
	\sum_{\substack{j_1 +j_2+j_3 \\ \leq m_1+m_2+m_3 -1 \\ l\geq 1}}
	(\nablaslash^{j_1} \nablaslash_3^{j_2} \nablaslash_4^{j_3} \Gamma)^l (e_{C_1}\ldots e_{C_{j_1}}) \ e_3
	+
	\sum_{\substack{j_1 +j_2+j_3 \\ \leq m_1+m_2+m_3 -1 \\ l\geq 1}}
	(\nablaslash^{j_1} \nablaslash_3^{j_2} \nablaslash_4^{j_3} \Gamma)^l (e_{C_1}\ldots e_{C_{j_1}}) \ e_4,
\end{multline*}
so that, for any multi-index $\gamma$, schematically,
\begin{equation} \label{eq:alphadifferencehoschematic}
	\Pi_{S^{\I}} \mathfrak{D}^{\gamma} \alpha_{\Hp}
	-
	\mathfrak{D}^{\gamma} \alpha_{\I}
	=
	\mathcal{F}^{\vert \gamma \vert}.
\end{equation}

The cases $s=0,1$ are similar, as are the estimates \eqref{eq:cancelTalphabar} for $\alphabar$ where Lemma \ref{lem:rHrIcheckerror} is also used.

\end{proof}

\section{Cancellations on null hypersurfaces and spacetime regions}
\label{morecancelationssec}

We will also exploit some additional
cancellations in our later estimates that arise from replacing  almost gauge invariant quantities 
in the $\I$ and $\Hp$ gauges.
These are contained in the following proposition, which forms part of the statement
of Theorem~\ref{thm:relatinggauges}.

If $\xi$ is an $S^{\I}$-tangent $(0,k)$ tensor field, recall the definitions of $\nablaslash^{\Hp} \xi$, $\nablaslash^{\Hp}_3 \xi$ and $\nablaslash^{\Hp}_4 \xi$ from Section \ref{tangentoperatorssec}, along with the definitions of $\vert \nablaslash^{\Hp} \xi \vert$, $\vert \nablaslash^{\Hp}_3 \xi \vert$ and $\vert \nablaslash^{\Hp}_4 \xi \vert$.
	
\begin{proposition}[Cancellations between almost gauge invariant quantities on null hypersurfaces and spacetime regions]
\label{thm:morecancelnotjustT}
	Given $s=0,1,2$, $W \in \{ \alpha, \alphabar \}$ and a multi-index $\vert \gamma \vert \leq N-1-s$, for all $u_1 \leq u \leq u_f$, the following estimates on null hypersurfaces hold
	\begin{equation} \label{eq:cancelnullcones1}
		\Vert
		\mathds{1}
		(r\nablaslash)_{\Hp}
		\left(
		(\mathfrak{D}^{\gamma} W)_{\I}
		-
		\Pi (\mathfrak{D}^{\gamma} W)_{\Hp}
		\right)
		\Vert_{C_u^{\Hp}}^2
		+
		\Vert
		\mathds{1}
		(r\Omega\nablaslash_4)_{\Hp}
		\left(
		(\mathfrak{D}^{\gamma} W)_{\I}
		-
		\Pi (\mathfrak{D}^{\gamma} W)_{\Hp}
		\right)
		\Vert_{C_u^{\Hp}}^2
		\lesssim
		\frac{\varepsilon^4}{u^{s+2}},
	\end{equation}
	\begin{equation} \label{eq:cancelnullcones2}
		\Vert
		\mathds{1}
		(r\nablaslash)_{\Hp}
		\left(
		(\mathfrak{D}^{\gamma}W)_{\I}
		-
		\Pi (\mathfrak{D}^{\gamma} W)_{\Hp}
		\right)
		\Vert_{\Cbar_v^{\Hp}}^2
		+
		\Vert
		\mathds{1}
		(\Omega^{-1} \nablaslash_3)_{\Hp}
		\left(
		(\mathfrak{D}^{\gamma} W)_{\I}
		-
		\Pi (\mathfrak{D}^{\gamma} W)_{\Hp}
		\right)
		\Vert_{\Cbar_v^{\Hp}}^2
		\lesssim
		\frac{\varepsilon^4}{v^{s+2}},
	\end{equation}
	for all $v(R_{-3},u_1) \leq v \leq v(R_2,u_f)$, where $\mathds{1} = \mathds{1}_{\{ r_{\I} \geq R_{-2} \}}$, and
	\begin{equation} \label{eq:cancelnullcones3}
		\Vert
		\mathds{1}
		(r\nablaslash)_{\I}
		\left(
		(\mathfrak{D}^{\gamma} W)_{\I}
		-
		\Pi (\mathfrak{D}^{\gamma} W)_{\Hp}
		\right)
		\Vert_{C_u^{\I}}^2
		+
		\Vert
		\mathds{1}
		(r\Omega\nablaslash_4)_{\I}
		\left(
		(\mathfrak{D}^{\gamma} W)_{\I}
		-
		\Pi (\mathfrak{D}^{\gamma} W)_{\Hp}
		\right)
		\Vert_{C_u^{\I}}^2
		\lesssim
		\frac{\varepsilon^4}{u^{s+2}},
	\end{equation}
	\begin{equation} \label{eq:cancelnullcones4}
		\Vert
		\mathds{1}
		(r\nablaslash)_{\I}
		\left(
		(\mathfrak{D}^{\gamma} W)_{\I}
		-
		\Pi (\mathfrak{D}^{\gamma} W)_{\Hp}
		\right)
		\Vert_{\Cbar_v^{\I}}^2
		+
		\Vert
		\mathds{1}
		(\Omega^{-1} \nablaslash_3)_{\I}
		\left(
		(\mathfrak{D}^{\gamma} W)_{\I}
		-
		\Pi (\mathfrak{D}^{\gamma} W)_{\Hp}
		\right)
		\Vert_{\Cbar_v^{\I}}^2
		\lesssim
		\frac{\varepsilon^4}{v^{s+2}},
	\end{equation}
	for all $v(R_{-2},u_1) \leq v \leq v(R_3,u_f)$, where $\mathds{1} = \mathds{1}_{r_{\Hp} \leq R_2}$ and $\Pi = \Pi_{S^{\I}}$.
	
	Moreover, given $s=0,1,2$, $W \in \{ \alpha, \alphabar \}$ and a multi-index $\vert \gamma \vert \leq N-s$, for all $u_1 \leq u \leq u_f$, the following spacetime estimate holds
	\begin{equation} \label{eq:cancelspacetime}
		\Vert
		(\mathfrak{D}^{\gamma} W)_{\I}
		-
		\Pi_{S^{\I}} (\mathfrak{D}^{\gamma} W)_{\Hp}
		\Vert_{\mathcal{D}_{\Hp}^{\I}(u)}^2
		\lesssim
		\frac{\varepsilon^4}{u^{s+1}}.
	\end{equation}
\end{proposition}

The following two lemmas will be used to estimate nonlinear error terms in the proof of Proposition \ref{thm:morecancelnotjustT}.  Recall the nonlinear error notation \eqref{eq:calFerror}.

\begin{lemma}[Spacetime nonlinear error estimate] \label{lem:fHIalphaerrorspacetime}
	For any $s=0,1$, and $k \leq N-s$, for any $u_1 \leq u \leq u_f$,
	\[
		\Vert \mathcal{F}^k \Vert_{\mathcal{D}_{\Hp}^{\I}(u)}^2
		\lesssim
		\frac{\varepsilon^{4}}{u^{s+2}}.
	\]
\end{lemma}

\begin{proof}
	The proof follows as in the proof of Lemma \ref{lem:fHIalphaerror}, now using the fact that each $\Phi = \Phi_{\I}$ or $\Phi_{\Hp}$ and each $\mathfrak{D}\fsc$ satisfies
	\[
		\sum_{s=0}^1 \sum_{\vert \gamma \vert \leq N-s}
		\Vert \mathfrak{D}^{\gamma} \Phi \Vert^2_{\mathcal{D}_{\Hp}^{\I}(u)}
		\lesssim
		\frac{\varepsilon^2}{u^{s}},
		\qquad
		\vert \mathfrak{D} \fsc \vert
		\lesssim
		\frac{\varepsilon}{u},
	\]
	from which it follows that
	\[
		\Vert \mathcal{F}^k \Vert_{\mathcal{D}_{\Hp}^{\I}(u)}^2
		\lesssim
		\frac{\varepsilon^2}{u^2}
		\big(
		\sum_{\vert \gamma \vert \leq k}
		\Vert \mathfrak{D}^{\gamma} \Phi_{\Hp} \Vert_{\mathcal{D}_{\Hp}^{\I}(u)}^2
		+
		\sum_{\mathfrak{D} \fsc}
		\Vert \mathfrak{D} \fsc \Vert_{\mathcal{D}_{\Hp}^{\I}(u)}^2
		\big)
		\lesssim
		\frac{\varepsilon^{4}}{u^{s+2}}.
	\]
\end{proof}

\begin{lemma}[Error estimates on null hypersurfaces] \label{lem:fHIalphaerrorcones}
	For any $s=0,1,2$, and $k\leq N-1-s$, and for any $u_1 \leq u \leq u_f$,
	\[
		\Vert \mathds{1} (r\nablaslash)_{\Hp} \mathcal{F}^k \Vert_{C^{\Hp}_u}^2
		+
		\Vert \mathds{1} (r\Omega\nablaslash_4)_{\Hp} \mathcal{F}^k \Vert_{C^{\Hp}_u}^2
		+
		\Vert \mathds{1} (r\nablaslash)_{\Hp} \mathcal{F}^k \Vert_{\Cbar^{\Hp}_v}^2
		+
		\Vert \mathds{1} (\Omega^{-1} \nablaslash_3)_{\Hp} \mathcal{F}^k \Vert_{\Cbar^{\Hp}_v}^2
		\lesssim
		\frac{\varepsilon^{4}}{u^{s+2}},
	\]
	for all $v(R_{-3},u) \leq v \leq v(R_2,u)$, where $\mathds{1} =  \mathds{1}_{r_{\I} \geq R_{-2}}$, and
	\[
		\Vert \mathds{1} (r\nablaslash)_{\I} \mathcal{F}^k \Vert_{C^{\I}_u}^2
		+
		\Vert \mathds{1} (r\Omega\nablaslash_4)_{\I} \mathcal{F}^k \Vert_{C^{\I}_u}^2
		+
		\Vert \mathds{1} (r\nablaslash)_{\I} \mathcal{F}^k \Vert_{\Cbar^{\I}_v}^2
		+
		\Vert \mathds{1} (\Omega^{-1} \nablaslash_3)_{\I} \mathcal{F}^k \Vert_{\Cbar^{\I}_v}^2
		\lesssim
		\frac{\varepsilon^{4}}{u^{s+2}},
	\]
	for all $v(R_{-2},u) \leq v \leq v(R_3,u)$, where $\mathds{1} =  \mathds{1}_{r_{\Hp} \leq R_2}$.
\end{lemma}

\begin{proof}
	Given an operator $\mathfrak{D}^* \in \{ \mathfrak{D}^{\gamma}_{\Hp}, \mathfrak{D}^{\gamma}_{\I} \mid \vert \gamma \vert =1 \}$, note that
	\[
		\vert \mathfrak{D}^* \mathcal{F}^k \vert
		\lesssim
		\sum_{\vert \gamma\vert \leq 1}\vert \mathfrak{D}^{\gamma} \mathfrak{D} \fsc \vert
		\cdot
		\Big(
		\sum_{\vert \gamma \vert \leq k}
		\vert \mathfrak{D}^* \mathfrak{D}^{\gamma} (\mathcal{R} - \mathcal{R}_{\circ})_{\Hp} \vert
		+
		\sum_{\vert \gamma \vert \leq k}
		\vert \mathfrak{D}^{\gamma} \Phi_{\Hp} \vert
		+
		\sum_{\vert \gamma \vert \leq k}
		\vert \mathfrak{D}^{\gamma} \Phi_{\I} \vert
		+
		\sum_{\vert \gamma \vert \leq 1}
		\sum_{\mathfrak{D} \fsc}
		\vert \mathfrak{D}^{\gamma} \mathfrak{D} \fsc \vert
		\Big).
	\]
	The proof then follows, as in the proof of Lemma \ref{lem:fHIalphaerrorspacetime}, from the fact that
	\begin{multline*}
		\sum_{\vert \gamma \vert \leq N-1-s}
		\Big(
		\Vert \mathds{1} (\mathfrak{D}^{\gamma} \Phi)_{\Hp} \Vert_{C^{\Hp}_u}^2
		+
		\Vert \mathds{1} (r\nablaslash \mathfrak{D}^{\gamma} (\mathcal{R} - \mathcal{R}_{\circ}))_{\Hp} \Vert_{C^{\Hp}_u}^2
		+
		\Vert \mathds{1} (r\Omega\nablaslash_4 \mathfrak{D}^{\gamma} (\mathcal{R} - \mathcal{R}_{\circ}))_{\Hp} \Vert_{C^{\Hp}_u}^2
		\\
		+
		\Vert \mathds{1} (\mathfrak{D}^{\gamma} \Phi)_{\Hp} \Vert_{\Cbar^{\Hp}_v}^2
		+
		\Vert \mathds{1} (r\nablaslash \mathfrak{D}^{\gamma} (\mathcal{R} - \mathcal{R}_{\circ}))_{\Hp} \Vert_{\Cbar^{\Hp}_v}^2
		+
		\Vert \mathds{1} (\Omega^{-1} \nablaslash_3 \mathfrak{D}^{\gamma} (\mathcal{R} - \mathcal{R}_{\circ}))_{\Hp} \Vert_{\Cbar^{\Hp}_v}^2
		\Big)
		\lesssim
		\frac{\varepsilon^2}{u^s},
	\end{multline*}
	for $\mathds{1} =  \mathds{1}_{r_{\I} \geq R_{-2}}$,
	\begin{multline*}
		\sum_{\vert \gamma \vert \leq N-1-s}
		\Big(
		\Vert \mathds{1} (\mathfrak{D}^{\gamma} \Phi)_{\I} \Vert_{C^{\I}_u}^2
		+
		\Vert \mathds{1} (r\nablaslash)^{\I} (\mathfrak{D}^{\gamma} (\mathcal{R} - \mathcal{R}_{\circ}))_{\Hp} \Vert_{C^{\I}_u}^2
		+
		\Vert \mathds{1} (r\Omega\nablaslash_4)^{\I} (\mathfrak{D}^{\gamma} (\mathcal{R} - \mathcal{R}_{\circ}))_{\Hp} \Vert_{C^{\I}_u}^2
		\\
		+
		\Vert \mathds{1} (\mathfrak{D}^{\gamma} \Phi)_{\I} \Vert_{\Cbar^{\I}_v}^2
		+
		\Vert \mathds{1} (r\nablaslash)^{\I} (\mathfrak{D}^{\gamma} (\mathcal{R} - \mathcal{R}_{\circ}))_{\Hp} \Vert_{\Cbar^{\I}_v}^2
		+
		\Vert \mathds{1} (\Omega^{-1} \nablaslash_3)^{\I} (\mathfrak{D}^{\gamma} (\mathcal{R} - \mathcal{R}_{\circ}))_{\Hp} \Vert_{\Cbar^{\I}_v}^2
		\Big)
		\lesssim
		\frac{\varepsilon^2}{u^s},
	\end{multline*}
	for $\mathds{1} =  \mathds{1}_{r_{\Hp} \leq R_2}$, and
	\[
		\sum_{\mathfrak{D} \fsc}
		\sum_{\vert \gamma \vert \leq 1}
		\vert \mathfrak{D}^{\gamma} \mathfrak{D} \fsc \vert
		+
		\sum_{\vert \gamma \vert \leq N-6}
		\big(
		\vert \mathfrak{D}^{\gamma} \Phi_{\Hp} \vert
		+
		\vert \mathfrak{D}^{\gamma} \Phi_{\I} \vert
		\big)
		\lesssim
		\frac{\varepsilon}{u}.
	\]
	Moreover, the geometric quantities of both the $\Hp$ and $\I$ gauges can easily be shown to satisfy the following estimates on the cones of the other gauge at the expense of one derivative
	\[
		\sum_{\vert \gamma \vert \leq N-1-s}
		\big(
		\Vert \mathds{1} (\mathfrak{D}^{\gamma} \Phi)_{\Hp} \Vert_{C^{\I}_u}^2
		+
		\Vert \mathds{1} (\mathfrak{D}^{\gamma} \Phi)_{\Hp} \Vert_{\Cbar^{\I}_v}^2
		\big)
		\lesssim
		\frac{\varepsilon^2}{u^s},
	\]
	for $\mathds{1} = \mathds{1}_{r_{\I} \geq R_{-2}}$, and
	\[
		\sum_{\vert \gamma \vert \leq N-1-s}
		\big(
		\Vert \mathds{1} (\mathfrak{D}^{\gamma} \Phi)_{\I} \Vert_{C^{\Hp}_u}^2
		+
		\Vert \mathds{1} (\mathfrak{D}^{\gamma} \Phi)_{\I} \Vert_{\Cbar^{\Hp}_v}^2
		\big)
		\lesssim
		\frac{\varepsilon^2}{u^s},
	\]
	for $\mathds{1} = \mathds{1}_{r_{\Hp} \leq R_2}$.
\end{proof}

The proof of Proposition \ref{thm:morecancelnotjustT} can now be given.

\begin{proof}[Proof of Proposition \ref{thm:morecancelnotjustT}]
	Consider first the estimates on null hypersurfaces.  As in the proof of Proposition~\ref{thm:cancelT}, schematically
	\[
		(\mathfrak{D}^{\gamma} \alpha)_{\I}
		-
		\Pi_{S^{\I}} (\mathfrak{D}^{\gamma} \alpha)_{\Hp}
		=
		\mathcal{F}^{\vert \gamma \vert}.
	\]
	Similarly for $\alphabar$.  The proof of \eqref{eq:cancelnullcones1}--\eqref{eq:cancelnullcones4} are then an immediate consequence of Lemma \ref{lem:fHIalphaerrorcones}.
	
	The proof of the spacetime estimate \eqref{eq:cancelspacetime} is similar, using now Lemma \ref{lem:fHIalphaerrorspacetime}.
\end{proof}

\section{Estimates for initial energies of $\alpha$ and \underline{$\alpha$}}
\label{Heretheproofofgida}

As discussed already in the introduction in Section~\ref{bounded_init_intro}, 
one aspect of working with teleological normalisations is that it is non-trivial
to estimate ``initial'' energy quantities when expressed in the $\I$ and $\Hp$ gauges.
The next part of the statement of Theorem~\ref{thm:relatinggauges} corresponds to a
bound of the initial energy fluxes of $\alpha$ and $\alphabar$ expressed in the $\I$ and $\Hp$ gauges in
terms of $\varepsilon_0$.

Recall that, for $\underline{k} = (k_1,k_2,k_3)$, $\mathfrak{D}^{\underline{k}} = (r\nablaslash)^{k_1} (\Omega^{-1} \nablaslash_3)^{k_2} (r\Omega \nablaslash_4)^{k_3}$.  Define the following energies of $\alpha$ and $\alphabar$, in the $\Hp$ and $\I$ gauges.  For the $\Hp$ gauge,\index{energies!initial energies!${\mathbb{E}}_0^{N} \left[\alpha_{\Hp}\right]$, initial energy for $\alpha$ in the teleological $\Hp$ gauge}\index{energies!initial energies!${\mathbb{E}}_0^{N} \left[\underline{\alpha}_{\Hp}\right]$, initial energy for $\underline{\alpha}$ in the teleological $\Hp$ gauge}
\begin{align}
	{\mathbb{E}}_0^{N} \left[\alpha_{\Hp}\right]
	:=
	&
	\sup_{v_{-1} \leq v \leq v_{2}} \sum_{|\underline{k}|=0; k_3\neq N}^{N}
	\Vert \mathfrak{D}^{\underline{k}} (\Omega^2 \alpha_{\Hp}) \Vert^2_{\underline{C}^{\Hp}_{v}}
	+
	\sup_{u_{0} \leq u\leq u_f}
	\sum_{|\underline{k}|=0}^{N}
	\Vert \mathfrak{D}^{\underline{k}} (\Omega^2 \alpha_{\Hp}) \mathds{1} \Vert^2_{C^{\Hp}_u}  \, , 
	\nonumber
	\\
	{\mathbb{E}}^{N}_0 \left[\underline{\alpha}_{\Hp}\right]
	:=
	&
	\sup_{v_{-1} \leq v \leq v_{2}}
	\sum_{|\underline{k}|=0}^{N}  
	\Vert \mathfrak{D}^{\underline{k}} (\Omega^{-2} \underline{\alpha}_{\Hp}) \Vert^2_{\underline{C}^{\Hp}_{v}}
	+
	\sup_{u_{0} \leq u\leq u_f}
	\sum_{|\underline{k}|=0; k_2\neq N}^{N}
	\Vert \mathfrak{D}^{\underline{k}} (\Omega^{-2} \underline{\alpha}_{\Hp}) \mathds{1} \Vert^2_{C^{\Hp}_u}
	\, , \nonumber
\end{align}
where $\mathds{1} = \mathds{1}_{v_{-1} \leq v_{\Hp} \leq v_3}$, and for the $\I$ gauge,\index{energies!initial energies!${\mathbb{E}}_0^{N} \left[\alpha_{\I}\right]$, initial energy for $\alpha$ in the teleological $\I$ gauge}\index{energies!initial energies!${\mathbb{E}}_0^{N} \left[\underline{\alpha}_{\I}\right]$, initial energy for $\underline{\alpha}$ in the teleological $\I$ gauge}
\begin{align*}
	&
	{\mathbb{E}}^{N}_0 \left[\alpha_{\I}\right]
	:= 
	\sup_{u_{-1} \leq u \leq u_{2}}
	\sum_{|\underline{k}|=0}^N
	\Vert r^4 \mathfrak{D}^{\underline{k}} \alpha_{\I} \Vert^2_{C^{\I}_u}
	+
	\sup_{v(u_{-1},R_{-2}) \leq v \leq v_{\infty}}
	\sum_{|\underline{k}|=0, k_3 \neq N}^{N}
	\Vert r^4 \mathfrak{D}^{\underline{k}} \alpha_{\I} \mathds{1} \Vert^2_{\underline{C}^{\I}_{v}},
\end{align*}
\begin{align}
	&
	{\mathbb{E}}^{N}_0 \left[\underline{\alpha}_{\I}\right]
	:=
	\sup_{v(u_{-1},R_{-2}) \leq v \leq v_\infty}
	\Big(
	\sum_{|\underline{k}|=0}^{N}
	\Vert \mathfrak{D}^{\underline{k}} (\check{r} \underline{\alpha}_{\I}) \mathds{1} \Vert^2_{\underline{C}^{\I}_{v}}
	+
	\sum_{|\underline{k}|=0}^{N-1}
	\Vert \mathfrak{D}^{\underline{k}} (r^3 \check{\psibar}_{\I}) \mathds{1} \Vert^2_{\underline{C}^{\I}_{v}}
	\Big)
	\nonumber
	\\
	&
	+
	\sup_{u_{-1} \leq u \leq u_{2}}
	\Big(
	\sum_{|\underline{k}|=0; k_2\neq N}^{N}
	\Vert r^{-1} \mathfrak{D}^{\underline{k}} (\check{r} \underline{\alpha}_{\I}) \Vert^2_{C^{\I}_u}
	+
	\sum_{|\underline{k}|=0}^{N-1}
	\Vert r^{-1} \mathfrak{D}^{\underline{k}} (r^3 \check{\psibar}_{\I}) \Vert^2_{C^{\I}_u}
	+
	\sum_{|\underline{k}|=0}^{N-2}
	\Vert r^{-1} \mathfrak{D}^{\underline{k}} (r^5 \check{\Pbar}_{\I}) \Vert^2_{C^{\I}_u}
	\Big)
	\nonumber \, ,
\end{align}
where $\mathds{1} = \mathds{1}_{u_{-1} \leq u \leq u_{2}}$.
These energies should be compared with the energies of Section \ref{twoenergiesalphaandalphabardefs}.
Define also the following energies for $P$ and $\underline{P}$:\index{energies!initial energies!${\mathbb{E}}_0^{N} \left[P_{\Hp}\right]$, initial energy for $P$ in the teleological $\Hp$ gauge}\index{energies!initial energies!${\mathbb{E}}_0^{N} \left[\Pbar_{\Hp}\right]$, initial energy for $\Pbar$ in the teleological $\Hp$ gauge}
\begin{align} 
	{\mathbb{E}}^{N-2}_0 \left[P_{\Hp}\right]
	:=
	&
	\sup_{v_{-1} \leq v \leq v_{2}}
	\sum_{|\underline{k}|=0}^{N-2}
	\Vert \mathfrak{D}^{\underline{k}} (r^5 P_{\Hp}) \Vert^2_{\underline{C}^{\Hp}_{v}}
	+
	\sup_{u_{0} \leq u\leq u_f}
	\sum_{|\underline{k}|=0}^{N-2}
	\Vert \mathfrak{D}^{\underline{k}} (r^5 P_{\Hp}) \mathds{1} \Vert^2_{C^{\Hp}_u},
	\nonumber
	\\
	{\mathbb{E}}^{N-2}_0 \left[\Pbar_{\Hp}\right]
	:=
	&
	\sup_{v_{-1} \leq v \leq v_{2}}
	\sum_{|\underline{k}|=0}^{N-2}
	\Vert \mathfrak{D}^{\underline{k}} (r^5 \Pbar_{\Hp}) \Vert^2_{\underline{C}^{\Hp}_{v}}
	+
	\sup_{u_{0} \leq u\leq u_f}
	\sum_{|\underline{k}|=0}^{N-2}
	\Vert \mathfrak{D}^{\underline{k}} (r^5 \Pbar_{\Hp}) \mathds{1} \Vert^2_{C^{\Hp}_u},
	\nonumber
\end{align}
where $\mathds{1} = \mathds{1}_{v_{-1} \leq v_{\Hp} \leq v_2}$, and\index{energies!initial energies!${\mathbb{E}}_0^{N} \left[P_{\I}\right]$, initial energy for $P$ in the teleological $\I$ gauge}\index{energies!initial energies!${\mathbb{E}}_0^{N} \left[\Pbar_{\I}\right]$, initial energy for $\Pbar$ in the teleological $\I$ gauge}
\begin{align} 
	{\mathbb{E}}^{N-2}_0 \left[P_{\I}\right]
	:= 
	&
	\sup_{u_{-1} \leq u \leq u_{2}}
	\sum_{|\underline{k}|=0}^{N-3}
	\Big(
	\Vert r \slashed{\nabla}_4 \mathfrak{D}^{\underline{k}} (r^5 P_{\I}) \Vert^2_{C^{\I}_u}
	+
	\Vert r^{-1} (r \slashed{\nabla}) \mathfrak{D}^{\underline{k}} (r^5 P_{\I}) \Vert^2_{C^{\I}_u}
	\Big)
	\nonumber
	\\
	& 
	+
	\sup_{v(u_{-1},R_{-2}) \leq v \leq v_\infty}  
	\sum_{|\underline{k}|=0}^{N-3} 
	\Big(
	\Vert \slashed{\nabla}_3 \mathfrak{D}^{\underline{k}} (r^5 P_{\I}) \mathds{1} \Vert^2_{\underline{C}^{\I}_{v}}
	+
	\Vert r \slashed{\nabla} \mathfrak{D}^{\underline{k}} (r^5 P_{\I}) \mathds{1} \Vert^2_{\underline{C}^{\I}_{v}}
	\Big),  \nonumber
	\\
	{\mathbb{E}}^{N-2}_0 \left[\check{\Pbar}_{\I}\right]
	:= 
	&
	\sup_{u_{-1} \leq u \leq u_2}
	\sum_{|\underline{k}|=0}^{N-3}
	\Big(
	\Vert r \slashed{\nabla}_4 \mathfrak{D}^{\underline{k}} (r^5 \check{\Pbar}_{\I}) \Vert^2_{C^{\I}_u}
	+
	\Vert r^{-1} (r \slashed{\nabla}) \mathfrak{D}^{\underline{k}} (r^5 \check{\Pbar}_{\I}) \Vert^2_{C^{\I}_u}
	\Big)
	\nonumber
	\\
	& 
	+
	\sup_{v(u_{-1},R_{-2}) \leq v \leq v_\infty}  
	\sum_{|\underline{k}|=0}^{N-3} 
	\Big(
	\Vert \slashed{\nabla}_3 \mathfrak{D}^{\underline{k}} (r^5 \check{\Pbar}_{\I}) \mathds{1} \Vert^2_{\underline{C}^{\I}_{v}}
	+
	\Vert r \slashed{\nabla} \mathfrak{D}^{\underline{k}} (r^5 \check{\Pbar}_{\I}) \mathds{1} \Vert^2_{\underline{C}^{\I}_{v}}
	\Big),  \nonumber
\end{align}
where $\mathds{1} = \mathds{1}_{u_{-1} \leq u \leq u_{2}}$.  These energies involve integrals over regions which are covered by the initial Kruskal and Eddington--Finkelstein gauges of Theorem \ref{thm:localKrus} and Theorem \ref{thm:localEF} respectively, and can therefore be controlled by comparing $\alpha_{\Hp}$, $\alpha_{\I}$, $\alphabar_{\Hp}$, $\alphabar_{\I}$ and their derivatives to the geometric quantities in the initial gauges, and using the fact that the estimates (\ref{willstatelater}) and (\ref{willstatelater2}) yield appropriate estimates for the geometric quantities in the initial gauges on the relevant $\Hp$ and $\I$ null hypersurfaces.

\begin{proposition}[Estimates for initial energies of $\alpha$ and $\alphabar$] \label{thm:gidataestimates}
	The above initial energies of $\alpha$ and $\alphabar$ satisfy
	\begin{equation}
	\label{thisiswheretheinitialfluxesappearone}
		{\mathbb{E}}_0^{N} \left[{\alpha}_{\Hp}\right] 
		+
		{\mathbb{E}}^{N}_0 \left[{\alpha}_{\I}\right]
		+
		{\mathbb{E}}_0^{N} \left[\underline{\alpha}_{\Hp}\right] 
		+
		{\mathbb{E}}^{N}_0 \left[\underline{\alpha}_{\I}\right] 
		\lesssim
		\varepsilon_0^2
		+
		\varepsilon^4,
	\end{equation}
	and the above initial energies of $P$ and $\Pbar$ satisfy
	\begin{equation}
	\label{thisiswheretheinitialfluxesappeartwo}
		{\mathbb{E}}^{N-2}_0 \left[P_{\Hp}\right]
		+
		{\mathbb{E}}^{N-2}_0 \left[P_{\I}\right]
		+
		{\mathbb{E}}^{N-2}_0 \left[\underline{P}_{\Hp}\right]
		+
		{\mathbb{E}}^{N-2}_0 \left[\check{\Pbar}_{\I}\right]
		\lesssim
		\varepsilon_0^2
		+
		\varepsilon^4.
	\end{equation}	
\end{proposition}

\begin{proof}
	The proof is an immediate consequence of the following estimates for $\alpha = \alpha_{\Hp}$ and $\alphabar = \alphabar_{\Hp}$,
	\begin{align} \label{eq:gidataestimates1}
		\sup_{v_{-1} \leq v \leq v_2}
		\big(
		\sum_{\vert \gamma \vert \leq N}
		\Vert \mathfrak{D}^{\gamma} \Omega^{-2} \alphabar \Vert_{\Cbar_{v_{-1}}^{\Hp}}^2
		+
		\sum_{\vert \gamma \vert \leq N-1}
		\big(
		\Vert \Omega^{-1} \nablaslash_3 \mathfrak{D}^{\gamma} \Omega^2 \alpha \Vert_{\Cbar_{v_{-1}}^{\Hp}}^2
		+
		\Vert r \nablaslash \mathfrak{D}^{\gamma} \Omega^2 \alpha \Vert_{\Cbar_{v_{-1}}^{\Hp}}^2
		\big)
		\big)
		&
		\lesssim
		\varepsilon_0^2
		+
		\varepsilon^4,
	\\
		\sup_{u_0 \leq u \leq u_f}
		\big(
		\sum_{\vert \gamma \vert \leq N}
		\Vert \mathfrak{D}^{\gamma} \Omega^2 \alpha \mathds{1} \Vert_{C_u^{\Hp}}^2
		+
		\sum_{\vert \gamma \vert \leq N-1}
		\big(
		\Vert r \Omega \nablaslash_4 \mathfrak{D}^{\gamma} \Omega^{-2} \alphabar \mathds{1} \Vert_{C_u^{\Hp}}^2
		+
		\Vert r \nablaslash \mathfrak{D}^{\gamma} \Omega^{-2} \alphabar \mathds{1} \Vert_{C_u^{\Hp}}^2
		\big)
		\big)
		&
		\lesssim
		\varepsilon_0^2
		+
		\varepsilon^4,
		\label{eq:gidataestimates2}
	\end{align}
	where $\mathds{1} = \mathds{1}_{v_{-1} \leq v_{\Hp} \leq v_2}$,
	and, for $\alpha = \alpha_{\I}$, $P = P_{\I}$, $\alphabar = \alphabar_{\I}$, $\check{\psibar} = \check{\psibar}_{\I}$, and $\check{\Pbar} = \check{\Pbar}_{\I}$,
	\begin{align}
		\sup_{v(u_{-1},R_{-2}) \leq v \leq v_{\infty}}
		\sum_{|\underline{k}|=0, k_3 \neq N}^{N}
		\Vert r^5 \mathfrak{D}^{\underline{k}} \alpha \mathds{1} \Vert^2_{\underline{C}^{\I}_{v}}
		\lesssim
		\varepsilon_0^2
		+
		\varepsilon^4,
		\label{eq:gidataestimates2a}
	\\
		\sup_{u_{-1} \leq u \leq u_{2}}
		\big(
		\sum_{\vert \gamma \vert \leq N}
		\Vert r^4 \mathfrak{D}^{\gamma} \alpha \Vert_{C_u^{\I}}^2
		+
		\sum_{\vert \gamma \vert \leq N-2}
		\Vert \mathfrak{D}^{\gamma} r \nablaslash_4 r^5 P \Vert_{C_u^{\I}}^2
		\big)
		\lesssim
		\varepsilon_0^2
		+
		\varepsilon^4,
		\label{eq:gidataestimates3}
	\\
		\sup_{v(u_{-1},R_{-2}) \leq v \leq v_\infty}
		\Big(
		\sum_{|\underline{k}|=0}^{N}
		\Vert r \mathfrak{D}^{\underline{k}} \underline{\alpha} \mathds{1} \Vert^2_{\underline{C}^{\I}_{v}}
		+
		\sum_{|\underline{k}|=0}^{N-1}
		\Vert r^3 \mathfrak{D}^{\underline{k}} \check{\psibar} \mathds{1} \Vert^2_{\underline{C}^{\I}_{v}}
		\Big)
		\lesssim
		\varepsilon_0^2
		+
		\varepsilon^4,
		\label{eq:gidataestimates2b}
	\end{align}
	and
	\begin{multline} \label{eq:gidataestimates5}
		\sup_{u_{-1} \leq u \leq u_{2}}
		\big(
		\sum_{\vert \gamma \vert \leq N-1}
		\big(
		\Vert \mathfrak{D}^{\gamma} \alphabar \Vert_{C_u^{\I}}^2
		+
		\Vert (r \nablaslash) \mathfrak{D}^{\gamma} \alphabar \Vert_{C_u^{\I}}^2
		\big)
		+
		\sum_{\vert \gamma \vert \leq N-1}
		\Vert\mathfrak{D}^{\gamma}  r^2 \check{\psibar} \Vert_{C_u^{\I}}^2
		+
		\sum_{\vert \gamma \vert \leq N-2}
		\Vert\mathfrak{D}^{\gamma}  r^4 \check{\Pbar} \Vert_{C_u^{\I}}^2
		\\
		+
		\sum_{\vert \gamma \vert \leq N-3}
		\Vert\mathfrak{D}^{\gamma}  r\nablaslash_4 (r^5 \check{\Pbar}) \Vert_{C_u^{\I}}^2
		\big)
		\lesssim
		\varepsilon_0^2
		+
		\varepsilon^4,
	\end{multline}
	where $\mathds{1} = \mathds{1}_{u_{-1} \leq u \leq u_{2}}$.  The proof of the estimates \eqref{eq:gidataestimates1}--\eqref{eq:gidataestimates5} proceeds by relating $\alpha_{\Hp}$, $\alpha_{\I}$ and their derivatives to the corresponding quantities in the initial Kruskal and Eddington--Finkelstein gauges of Theorem \ref{thm:localKrus} and Theorem \ref{thm:localEF} respectively, and using the fact that the estimates (\ref{willstatelater}) and (\ref{willstatelater2}) yield appropriate estimates for the geometric quantities in the initial gauges on the relevant $\Hp$ and $\I$ null hypersurfaces.
	
	Define the following energy of the geometric quantities of the initial Kruskal gauge of Theorem \ref{thm:localKrus} on the spheres and null cones of the $\Hp$ gauge
	\begin{align}
		\mathbb E^N_{0,i_{\Hp}} [ \Phi_{\mathcal{K}, d}]
		:= 
		\sup_{\substack{
		u_{0} \leq u \leq u_f
		\\
		v_{-1} \leq v \leq v_2
		}} 
		\Bigg(
		&
		\sum_{\Phi, \, |\gamma| \leq N-1} \| (\mathfrak{D}^{\gamma} \Phi)_{\mathcal{K},d} \|^2_{S_{u,v}^{\Hp}}
		\nonumber 
		\\
		&
		+
		\sum_{\vert \gamma \vert \leq N}
		\Big(
		\Vert (\mathfrak{D}^{\gamma} (\alpha, \beta, \rho - \rho_{\circ}, \sigma, \underline{\beta}))_{\mathcal{K}, \mathrm{d}} \mathds{1} \Vert_{C_u^{\Hp}}^2
		+
		\Vert  \mathfrak{D}^{\gamma} (\beta, \rho - \rho_{\circ}, \sigma, \underline{\beta},\underline{\alpha})_{\mathcal{K}, \mathrm{d}} \Vert_{\underline{C}_v^{\Hp}}^2
		\Big)
		\nonumber 
		\\
		&
		+
		\sum_{\vert \gamma \vert \leq N-1}
		\Big(
		\Vert (r \nablaslash)^{\Hp} (\mathfrak{D}^{\gamma} \alpha)_{\mathcal{K}, \mathrm{d}} \Vert_{\underline{C}_v^{\Hp}}^2
		+
		\Vert (\Omega^{-1}\nablaslash_3)^{\Hp} (\mathfrak{D}^{\gamma} \alpha)_{\mathcal{K}, \mathrm{d}} \Vert_{\underline{C}_v^{\Hp}}^2
		\Big)
		\nonumber
		\\
		&
		+
		\sum_{\vert \gamma \vert \leq N-1}
		\Big(
		\Vert (r \nablaslash)^{\Hp}(\mathfrak{D}^{\gamma} \alphabar)_{\mathcal{K}, \mathrm{d}} \mathds{1} \Vert_{C_u^{\Hp}}^2
		+
		\Vert 
		(\Omega \nablaslash_4)^{\Hp} (\mathfrak{D}^{\gamma} \alphabar)_{\mathcal{K}, \mathrm{d}} \mathds{1} \Vert_{C_u^{\Hp}}^2
		\Big)
		\Bigg),
		\nonumber
	\end{align}
	where $\mathds{1} = \mathds{1}_{v_{-1} \leq v_{\Hp} \leq v_2}$, along with the following energy of the geometric quantities of the initial Eddington--Finkelstein gauge of Theorem \ref{thm:localEF} on the spheres and null cones of the $\I$ gauge,  
	\begin{align*}
		\mathbb E^N_{0,i_{\I}} [ \Phi_{\mathcal{EF}, d}]
		:=
		\sup_{\substack{
		u_{-1} \leq u \leq u_2
		\\
		v(R_{-2},u) \leq v \leq v_{\infty}
		}} 
		&
		\Bigg(
		\sum_{\Phi, \, |\gamma| \leq N-1} \| (\mathfrak{D}^{\gamma} r^p \Phi_p)_{\mathcal{EF},d} \|^2_{S_{u,v}^{\I}}
		\\
		&
		+
		\sum_{\vert \gamma \vert \leq N}
		\Vert r^{-1} (\mathfrak{D}^{\gamma} (r^5 \alpha, r^4 \beta, r^3 (\rho - \rho_{\circ}), r^3 \sigma, r^2 \underline{\beta}))_{\mathcal{EF}, \mathrm{d}} \Vert_{C_u^{\I}}^2
		\\
		&
		+
		\sum_{\vert \gamma \vert \leq N}
		\Vert  \mathfrak{D}^{\gamma} (r^4 \beta, r^3 (\rho - \rho_{\circ}), r^3 \sigma, r^2 \underline{\beta},r \underline{\alpha})_{\mathcal{EF}, \mathrm{d}} \mathds{1} \Vert_{\underline{C}_v^{\I}}^2
		\\
		&
		+
		\sum_{\vert \gamma \vert \leq N-1}
		\Big(
		\Vert (r \nablaslash)^{\I} (\mathfrak{D}^{\gamma} r^4 \alpha)_{\mathcal{EF}, \mathrm{d}} \mathds{1} \Vert_{\underline{C}_v^{\I}}^2
		+
		\Vert (\Omega^{-1}\nablaslash_3)^{\I} (\mathfrak{D}^{\gamma} r^5 \alpha)_{\mathcal{EF}, \mathrm{d}} \mathds{1} \Vert_{\underline{C}_v^{\I}}^2
		\Big)
		\\
		&
		+
		\sum_{\vert \gamma \vert \leq N-1}
		\Big(
		\Vert (r \nablaslash)^{\I}(\mathfrak{D}^{\gamma} \alphabar)_{\mathcal{EF}, \mathrm{d}} \Vert_{C_u^{\I}}^2
		+
		\Vert 
		(r \Omega \nablaslash_4)^{\I} (\mathfrak{D}^{\gamma} \check{r} \alphabar)_{\mathcal{EF}, \mathrm{d}} \Vert_{C_u^{\I}}^2
		\Big)
		\Bigg),
	\end{align*}
	where $\mathds{1} = \mathds{1}_{u_{-1} \leq u \leq u_{2}}$.  Here $(r \nablaslash)^{\Hp} (\mathfrak{D}^{\gamma} \alpha)_{\mathcal{K}, \mathrm{d}}$ is defined as in Section \ref{tangentoperatorssec}, and
	\[
		\Vert (r \nablaslash)^{\Hp} (\mathfrak{D}^{\gamma} \alpha)_{\mathcal{K}, \mathrm{d}} \Vert_{\underline{C}_v^{\Hp}}^2
		=
		\int_{\underline{C}_v^{\Hp}}
		\vert (r \nablaslash)^{\Hp} (\mathfrak{D}^{\gamma} \alpha)_{\mathcal{K}, \mathrm{d}} \vert^2
		\Omega^2_{\Hp} d\theta du,
	\]  
	with $\vert (r \nablaslash)^{\Hp} (\mathfrak{D}^{\gamma} \alpha)_{\mathcal{K}, \mathrm{d}} \vert$ defined as in \eqref{eq:nablastildeofstensor}.  Similarly for the other terms involving $\alpha$ and $\alphabar$ in $\mathbb E^N_{0,i_{\I}} [ \Phi_{\mathcal{K}, d}]$ and $\mathbb E^N_{0,i_{\I}} [ \Phi_{\mathcal{EF}, d}]$.  Note that the pointwise control on the $f_{d,\Hp}$ diffeomorphisms in \eqref{eq:fdHppointwise} guarantees that \eqref{closetofoliationsphere0} is satisfied by $\tilde{S} = S^{\Hp}_{u,v}$, $U'=-\exp (-\frac{u}{2M_f} )$, $V' = \exp (\frac{v}{2M_f} )$, for all $u_{0} \leq u \leq u_f$, $v_{-1} \leq v \leq v_2$ and so the fluxes appearing in $\mathbb E^N_{0,i_{\Hp}} [ \Phi_{\mathcal{K}, d}]$ are allowed fluxes in the energy $\mathbb E^N_0[ \Phi_{\mathcal{K}, d}]$, defined in \eqref{initdatanormK}.  Thus
	\begin{equation} \label{eq:initialenergyonHpconesbound}
		\mathbb E^N_{0,i_{\Hp}} [ \Phi_{\mathcal{K}, d}]
		\lesssim
		\mathbb E^N_0[ \Phi_{\mathcal{K}, d}].
	\end{equation}
	Similarly, the pointwise control on the $f_{d,\I}$ diffeomorphisms in \eqref{eq:fdIppointwise} guarantees that \eqref{closetofoliationsphere} is satisfied by $\tilde{S} = S^{\I}_{u,v}$, $u' = u$, $v'=v$ for all  $u_{-1} \leq u \leq u_2$, $v(R_{-2},u) \leq v \leq v_{\infty}$ and thus the fluxes appearing in $\mathbb E^N_{0,i_{\I}} [ \Phi_{\mathcal{EF}, d}]$ are allowed fluxes in the energy $\mathbb E^N_0[ \Phi_{\mathcal{EF}, d}]$, defined in \eqref{initdatanormEF}.  Hence
	\begin{equation} \label{eq:initialenergyonIpconesbound}
		\mathbb E^N_{0,i_{\I}} [ \Phi_{\mathcal{EF}, d}]
		\lesssim
		\mathbb E^N_0[ \Phi_{\mathcal{EF}, d}].
	\end{equation}
	
	Consider first the estimate for $\mathfrak{D}^{\gamma} \Omega^{-2} \alphabar_{\Hp}$, for $\vert \gamma \vert \leq N$, on the incoming cones $\Cbar_v^{\Hp}$ of \eqref{eq:gidataestimates1}.  It suffices to estimate $\alphabar$ and its derivatives in the Kruskalised $\Hp$ gauge (see \eqref{fortheHplusgaugekruskalised} and Section \ref{kruskalisedsec}) since, following Proposition \ref{prop:HpKruskalisedestimates}, for any $v_{-1} \leq v \leq v_3$,
	\[
		\sum_{\vert \gamma \vert \leq N}
		\Vert (\mathfrak{D}^{\gamma} \Omega^{-2} \alphabar)_{\Hp} \Vert_{\Cbar_v^{\Hp}}^2
		\lesssim
		\sum_{\vert \gamma \vert \leq N}
		\Vert (\mathfrak{D}^{\gamma} \alphabar)_{\mathcal{K}, \Hp} \Vert_{\Cbar_v^{\Hp}}^2
		+
		\varepsilon^4.
	\]
	It follows, after relating $(\mathfrak{D}^{\gamma} \alphabar)_{\mathcal{K}, \Hp}$ to covariant derivatives of the Riemann curvature tensor, as in the proof of Proposition~\ref{thm:cancelT} (in fact, the present proof is easier since there is no decay to be captured), that
	\[
		\sum_{\vert \gamma \vert \leq N}
		\Vert (\mathfrak{D}^{\gamma} \alphabar)_{\mathcal{K}, \Hp} \Vert_{\Cbar_v^{\Hp}}^2
		\lesssim
		\sum_{\vert \gamma \vert \leq N}
		\Vert (\mathfrak{D}^{\gamma} \alphabar)_{\mathcal{K}, d} \Vert_{\Cbar_v^{\Hp}}^2
		+
		\big( \mathbb E^N_{0,i_{\Hp}} [ \Phi_{\mathcal{K}, d}] \big)^2
		+
		(\mathbb{E}^N_{\Hp})^2
		+
		\big( \mathbb P_{u_f}[f_{d,\Hp}] \big)^4.
	\]
	Moreover
	\[
		\sum_{\vert \gamma \vert \leq N}
		\Vert (\mathfrak{D}^{\gamma} \alphabar)_{\mathcal{K}, d} \Vert_{\Cbar_v^{\Hp}}^2
		\lesssim
		\mathbb E^N_{0,i_{\Hp}} [ \Phi_{\mathcal{K}, d}],
	\]
	and so, in view of \eqref{eq:initialenergyonHpconesbound}, it then follows from (\ref{willstatelater}), the estimate for $\mathbb{E}^N_{\Hp}$ in the bootstrap assumption \eqref{eq:bamain}, and the pointwise estimate \eqref{eq:fdIppointwise} that
	\[
		\Vert (\mathfrak{D}^{\gamma} \alphabar)_{\mathcal{K}, d} \Vert_{\Cbar_v^{\Hp}}^2
		\lesssim
		\varepsilon_0^2
		+
		\varepsilon^4,
	\]
	thus completing the estimate for $\Omega^{-2} \alphabar_{\Hp}$.
	For $\Omega^{-1} \nablaslash_3 \mathfrak{D}^{\gamma} \Omega^2 \alpha$ one similarly obtains, by relating to covariant derivatives of the Riemann curvature tensor as in the proof of Proposition~\ref{thm:cancelT}, that
	\begin{multline*}
		\sum_{\vert \gamma \vert \leq N-1}
		\Vert (\Omega^{-1} \nablaslash_3 \mathfrak{D}^{\gamma} \Omega^2 \alpha)_{\Hp} \Vert_{\Cbar_v^{\Hp}}^2
		\\
		\lesssim
		\sum_{\vert \gamma \vert \leq N-1}
		\Vert (\Omega^{-1} \nablaslash_3)^{\Hp} (\mathfrak{D}^{\gamma} \alpha)_{\mathcal{K}, d} \Vert_{\Cbar_v^{\Hp}}^2
		+
		\big( \mathbb E^N_{0,i_{\Hp}} [ \Phi_{\mathcal{K}, d}] \big)^2
		+
		(\mathbb{E}^N_{\Hp})^2
		+
		\big( \mathbb P_{u_f}[f_{d,\Hp}] \big)^4
		\lesssim
		\varepsilon_0^2
		+
		\varepsilon^4,
	\end{multline*}
	in view again of \eqref{eq:initialenergyonHpconesbound}, the estimate (\ref{willstatelater}), the estimate for $\mathbb{E}^N_{\Hp}$ in the bootstrap assumption \eqref{eq:bamain}, and the pointwise estimate \eqref{eq:fdIppointwise}.
	The proof of the remaining estimates of \eqref{eq:gidataestimates1} and \eqref{eq:gidataestimates2} are similar.

	The estimates \eqref{eq:gidataestimates2a}--\eqref{eq:gidataestimates5} require slightly more care due to the presence of the $r$ weights.  Consider first the estimate for $r^4 \alpha$ on the cone $C_u^{\I}$, for some $u_{-1} \leq u \leq u_{2}$.  Revisiting the proof of the relation \eqref{eq:curvaturecomp1}, one sees that, in fact,
	\begin{equation} \label{eq:alpharelationbetter}
		\alpha_{\I}
		=
		(1+ \partial_v f^4)^2 \Pi_{S^{\I}} \alpha_{\mathcal{EF},\mathrm{d}}
		+
		\mathcal{E}^{1,0}_{\mathfrak{D} \fsc, 5},
	\end{equation}
	where $f^4 = f^4_{d,\I}$, and so, as in the proof of Proposition \ref{prop:tildenorm}, for any $u_{-1} \leq u \leq u_{2}$,
	\[
		\Vert (r^4 \alpha)_{\I} \Vert_{C_u^{\I}}^2
		\lesssim
		\Vert (r^4 \alpha)_{\mathcal{EF},\mathrm{d}} \Vert_{C_u^{\I}}^2
		+
		\big( \mathbb E^N_{0,i_{\I}}[ \Phi_{\mathcal{EF}, d}] \big)^2
		+
		(\mathbb{E}^N_{\I})^2
		+
		\big( \mathbb P_{u_f}[f_{d,\I}] \big)^4.
	\]
	Since
	\[
		\Vert (r^4 \alpha)_{\mathcal{EF},\mathrm{d}} \Vert_{C_u^{\I}}^2
		\lesssim
		\mathbb E^N_{0,i_{\I}}[ \Phi_{\mathcal{EF}, d}],
	\]
	the estimate (\ref{willstatelater2}), in view of \eqref{eq:initialenergyonIpconesbound}, the estimate for $\mathbb{E}^N_{\I}$ in the bootstrap assumption \eqref{eq:bamain}, and the pointwise control on the $f_{d,\I}$ diffeomorphisms in \eqref{eq:fdIppointwise} then give
	\[
		\Vert (r^4 \alpha)_{\I} \Vert_{C_u^{\I}}^2
		\lesssim
		\varepsilon_0^2
		+
		\varepsilon^4,
	\]
	thus yielding the estimate for the first term in \eqref{eq:gidataestimates3} when $\vert \gamma \vert = 0$.
	Estimates for $\Vert (\mathfrak{D}^{\gamma} r^4 \alpha)_{\I} \Vert_{C_u^{\I}}^2$, for $\vert \gamma \vert \leq N$, are obtained similarly, following the process of the proof of Proposition \ref{thm:cancelT} (see the relation \eqref{eq:alphadifferencehoschematic}).  The estimate \eqref{eq:gidataestimates2a} is similar.
	
	Consider now the second term in \eqref{eq:gidataestimates3}.  For $r \nablaslash_4 ( r^5 P )$, Proposition \ref{prop:nabla4pschematic} and the Codazzi equation~\eqref{eq:Codazzi} imply that
	\begin{align*}
		r \nablaslash_4 ( r^5 P )
		=
		\
		&
		r^6 \Dslash_2^* \nablaslash \divslash \beta
		-
		r^6 \Dslash_2^* \nablaslash \curlslash \beta
		+
		6M_f r \left( 1 - \frac{3M_f}{r} \right) \hat{\chi}
		\\
		&
		+
		3M_f r^3 \Dslash_2^* \divslash \hat{\chi}
		+
		3M_f r^3 \Dslash_2^* \beta
		+
		3M_f r^2 \Omega \alpha
		+
		\Omega^{-1} \mathcal{E}^2_1.
	\end{align*}
	It then follows from the change of gauge relations \eqref{eq:curvaturecomp3}, \eqref{eq:Riccicomp1}, \eqref{eq:alpharelationbetter} and Proposition \ref{prop:tildederivStensor}, after noting that the no linear terms in the diffeomorphism functions appear when the above expression is related to the corresponding expression in the initial $\mathcal{EF}$ gauge, that, for any $u_{-1} \leq u \leq u_{2}$,
	\begin{align*}
		\Vert (r \nablaslash_4 ( r^5 P ))_{\I} \Vert_{C_u^{\I}}^2
		\lesssim
		\
		&
		\sum_{k\leq3}
		\Vert r^3 ((r\nablaslash)^k\beta)_{\mathcal{EF},d} \Vert_{C_u^{\I}}^2
		+
		\sum_{k\leq 2}
		\Vert r ((r\nablaslash)^k\hat{\chi})_{\mathcal{EF},d} \Vert_{C_u^{\I}}^2
		+
		\Vert r^2 \alpha_{\mathcal{EF},d} \Vert_{C_u^{\I}}^2
		\\
		&
		+
		\big( \mathbb E^N_{0,i_{\I}}[ \Phi_{\mathcal{EF}, d}] \big)^2
		+
		(\mathbb{E}^N_{\I})^2
		+
		\big( \mathbb P_{u_f}[f_{d,\I}] \big)^4,
	\end{align*}
	and the estimate for $\Vert (r \nablaslash_4 ( r^5 P ))_{\I} \Vert_{C_u^{\I}}^2$ again follows from (\ref{willstatelater2}), in view of \eqref{eq:initialenergyonIpconesbound}, since
	\[
		\sum_{k\leq3}
		\Vert r^3 ((r\nablaslash)^k\beta)_{\mathcal{EF},d} \Vert_{C_u^{\I}}^2
		+
		\sum_{k\leq 2}
		\Vert r ((r\nablaslash)^k\hat{\chi})_{\mathcal{EF},d} \Vert_{C_u^{\I}}^2
		+
		\Vert r^2 \alpha_{\mathcal{EF},d} \Vert_{C_u^{\I}}^2
		\lesssim
		\mathbb E^N_{0,i_{\I}}[ \Phi_{\mathcal{EF}, d}].
	\]
	Similarly for the higher order derivatives.
	
	The estimates \eqref{eq:gidataestimates2b} and \eqref{eq:gidataestimates5} follow similarly, using now Proposition \ref{prop:Pbartildeidentities} and Proposition \ref{prop:nabla4pschematic}.  For example, for the final estimate of \eqref{eq:gidataestimates5}, Proposition \ref{prop:nabla4pschematic} implies that
	\begin{align*}
		\frac{r^2}{\check{r}} \nablaslash_4 \left(  r^5 \check{\Pbar} \right)
		=
		\
		&
		r^6 \Dslash_2^* \nablaslash \divslash \beta
		+
		r^6 \Dslash_2^* \nablaslash \curlslash \beta
		-
		6M_f \Omega r^2 \Dslash_2^* \etabar
		+
		6M_f r \left( 1 - \frac{3M_f}{r} \right) \hat{\chi}
		+
		\frac{3M_f}{\Omega} r^3 \Dslash_2^* \nablaslash \Omega \tr \chi
		\\
		&
		+
		3M_f r^2 \Omega \alpha
		+
		a_1 r^5 \nablaslash(\Omega \omegahat) \hat{\otimes} \betabar
		+
		a_2 r^5 \nablaslash_4 r \nablaslash(\Omega \omegahat) \hat{\otimes} \betabar
		+
		a_3 r^5 \Omega\hat{\omega}  \Dslash_2^\star \betabar
		+
		\check{\mathcal{E}}^2_1
		,
	\end{align*}
	and so the relations \eqref{eq:curvaturecomp3}, \eqref{eq:Riccicomp1}, \eqref{eq:Riccicomp3}, \eqref{eq:Riccicomp7}, \eqref{eq:alpharelationbetter}, and Proposition \ref{prop:tildederivStensor} imply that, for any $u_{-1} \leq u \leq u_{2}$,
	\begin{multline*}
		\Vert r\nablaslash_4 (r^5 \check{\Pbar}) \Vert_{C_u^{\I}}^2
		\lesssim
		\Vert \frac{r^2}{\check{r}} \nablaslash_4 (r^5 \check{\Pbar}) \Vert_{C_u^{\I}}^2
		\lesssim
		\sum_{k\leq3}
		\Vert r^3 ((r\nablaslash)^k\beta)_{\mathcal{EF},d} \Vert_{C_u^{\I}}^2
		+
		\Vert r (r\nablaslash \eta)_{\mathcal{EF},d} \Vert_{C_u^{\I}}^2
		+
		\Vert r \hat{\chi}_{\mathcal{EF},d} \Vert_{C_u^{\I}}^2
		\\
		+
		\Vert r ((r\nablaslash)^2\Omega \tr \chi)_{\mathcal{EF},d} \Vert_{C_u^{\I}}^2
		+
		\Vert r^2 \alpha_{\mathcal{EF},d} \Vert_{C_u^{\I}}^2
		+
		\mathbb E^N_{0,i_{\I}}[ \Phi_{\mathcal{EF}, d}]^2
		+
		(\mathbb{E}^N_{\I})^2
		+
		\mathbb P_{u_f}[f_{d,\I}]^4
		\lesssim
		\varepsilon_0^2
		+
		\varepsilon^4,
	\end{multline*}
	by (\ref{willstatelater2}) since 
	\begin{multline*}
		\sum_{k\leq3}
		\Vert r^3 ((r\nablaslash)^k\beta)_{\mathcal{EF},d} \Vert_{C_u^{\I}}^2
		+
		\Vert r (r\nablaslash \eta)_{\mathcal{EF},d} \Vert_{C_u^{\I}}^2
		+
		\Vert r \hat{\chi}_{\mathcal{EF},d} \Vert_{C_u^{\I}}^2
		+
		\Vert r ((r\nablaslash)^2\Omega \tr \chi)_{\mathcal{EF},d} \Vert_{C_u^{\I}}^2
		+
		\Vert r^2 \alpha_{\mathcal{EF},d} \Vert_{C_u^{\I}}^2
		\\
		\lesssim 
		\mathbb E^N_{0,i_{\I}}[ \Phi_{\mathcal{EF}, d}].
	\end{multline*}
	Similarly for higher order derivatives of $r\nablaslash_4 (r^5 \check{\Pbar})$.
\end{proof}

\section{Estimates for $\Hp$ quantities from $\I$ quantities on $C_{u_f}$}
\label{proofofinheritingthem}

There is one final statement which we have to show to complete
the proof of Theorem~\ref{thm:relatinggauges}.

Recall that the $\Hp$ and $\I$ gauges share a common final outgoing cone $C_{u_f}$. 
Exploiting this fact and our  estimates on the diffeomorphisms $f_{\Hp,\I}$,
we obtain the following estimates on $\hat{\chi}_{\Hp}$, $\beta^{\Hp}_{\ell = 1}$ and a new quantity
$\Upsilon^{\Hp}_{\ell=0}$ in terms of geometric quantities in the $\I$ gauge.

\begin{proposition}[Estimating $\Hp$ quantities from $\I$ quantities on $C_{u_f}$] \label{thm:inheriting}
	The quantity $\Omega \hat{\chi}_{\Hp}$ satisfies
	\begin{equation} \label{eq:chiHpinheriting}
		\sum_{s=0,1,2} (u_f)^{s} \sum_{k=0}^{N+1-s}
		\int_{v(R_{-1},u_f)}^{v(R_1,u_f)} \int_{S^2}
		\left\vert (r\nablaslash)^k \Omega \hat{\chi}_{\Hp} \right\vert^2
		d\theta dv' (u_f)
		\lesssim
		\mathbb{E}^N_{u_f,\I} + \varepsilon^4,
	\end{equation}
	and the quantities $\divslash \beta^{\Hp}_{\ell = 1}$ and $\Upsilon^{\Hp}_{\ell=0}$ satisfy
	\[
		\left\vert \divslash \beta^{\Hp}_{\ell = 1} (u_f,v(R,u_f)) \right\vert^2
		+
		\left\vert \Upsilon^{\Hp}_{\ell=0} (u_f,v(R,u_f)) \right\vert^2
		\lesssim
		(u_f)^{-4+2\delta}
		\big(
		\mathbb{E}^N_{u_f,\I} + \varepsilon^4
		\big),
	\]
	where\index{double null gauge!connection coefficients!$\Upsilon$, quantity used only in the $\mathcal{H}^+$ gauge}
	\[
		\Upsilon
		=
		\left(1 - \frac{3M_f}{r} \right) (\rho - \rho_{\circ})
		+
		\frac{3M_f}{2r^2} \left( \Omega \tr \chi - (\Omega \tr \chi)_{\circ} \right)
		-
		\frac{3M_f\Omega_{\circ}^2}{2r^2} \Omega^{-2} \left( \Omega \tr \chibar - (\Omega \tr \chibar)_{\circ} \right).
	\]
	Moreover the metric $(\gslash - r^2 \mathring{\gamma})^{\Hp}$ satisfies, for $s=0,1$,
	\begin{align} \label{eq:gslashinheriting}
		\sum_{k \leq N-s}
		\!\!
		\Vert
		(r\nablaslash)^k
		(\gslash - r^2 \mathring{\gamma})^{\Hp}
		\Vert_{S_{u_f,v(R,u_f)}^{\Hp}}^2
		\lesssim
		\frac{\varepsilon^4}{v(R,u_f)^{s+1}}
		+
		\mathbb{E}^N_{u_f,\I}
		+
		\!\!\!
		\sum_{k \leq N-s-2} \!\!
		\Vert (r\nablaslash)^k \hat{\chi}^{\Hp} \Vert_{S_{u_f,v(R,u_f)}^{\Hp}}^2
		\!\!\!\!
		+
		\vert (\rho - \rho_{\circ})^{\Hp} \vert^2
		.
	\end{align}
\end{proposition}

\begin{remark}
We remark that Proposition~\ref{thm:inheriting} will be used in the proof of Theorem~\ref{thm:Hestimates} in Chapter \ref{chap:Hestimates}, after $\mathbb{E}^N_{u_f,\I}$ has been estimated in Theorem~\ref{thm:Iestimates}.  The estimate \eqref{eq:gslashinheriting} will be used in the proof of Theorem~\ref{thm:Hestimates} after $\hat{\chi}^{\Hp}$ and $(\rho - \rho_{\circ})^{\Hp}$ have been estimated (see Proposition \ref{prop:gslashH}).
\end{remark}

In this section the diffeomorphisms $f_{\Hp,\I}$ relating the $\Hp$ and $\I$ gauges are denoted $f$, without the subscript.

In obtaining the estimate \eqref{eq:gslashinheriting} one first shows that
\begin{equation} \label{eq:gslashinheritingprelim}
	\sum_{k \leq N-s}
	\Vert
	(r\nablaslash)^k
	(\gslash - r^2 \gamma)^{\Hp}
	\Vert_{S_{u_f,v(R,u_f)}^{\Hp}}^2
	\lesssim
	\frac{\varepsilon^4}{v(R,u_f)^{s+1}}
	+
	\mathbb{E}^N_{u_f,\I}
	+
	\sum_{k\leq N-s}
	\Vert
	(r\nablaslash)^k
	f^4
	\Vert_{S_{u_f,v(R,u_f)}^{\Hp}}^2.
\end{equation}
(See the proof of Proposition \ref{prop:gslashHpI} below.)  It is more convenient to state the estimate, as in \eqref{eq:gslashinheriting}, after the final term, involving $f^4$, has been estimated by quantities in the $\Hp$ gauge.

\begin{proof}[Proof of Proposition~\ref{thm:inheriting}] The proof will 
be accomplished in the  Sections~\ref{forproofone}--\ref{forproofthree} below.

\subsection{Estimate for $\hat{\chi}^{\Hp}$ from $\hat{\chi}^{\I}$}
\label{forproofone}

When $\hat{\chi}^{\Hp}$ is estimated from $\hat{\chi}^{\I}$ on the final hypersurface $u=u_f$, the following simplification of the identity \eqref{eq:Riccicomp1} is used, which exploits the vanishing of $f^3$ and $f^1,f^2$ on $u=u_f$.

\begin{lemma}[Change of gauge identity for $\hat{\chi}$ on $C_{u_f}$] \label{lem:chihatufsimple}
	The gauge condition $f^3(u_f,v(R,u_f),\cdot) = 0$ implies that, on $\{u_{\Hp} =u_f\} \cap \{u_{\I} =u_f\}$,
	\begin{equation} \label{eq:chihatIchihatH}
		(\Omega \hat{\chi})^{\I}
		=
		\left(
		1 + \frac{\partial f^4}{\partial v_{\I}}
		\right)
		(\Omega \hat{\chi})^{\Hp}.
	\end{equation}
\end{lemma}

\begin{proof}
	Recall \eqref{eq:fHIuf}.
	
	The equalities \eqref{eq:partialxpartialtildex2} and \eqref{eq:partialxpartialtildex3}, together with the fact that $b^{\Hp} = 0$ on $\{u=u_f\}$ (and $\Omega^{\I} e^{\I}_4 = \partial_{v_{\I}}$ everywhere) give
	\[
		e^{\I}_4
		=
		\frac{\Omega^{\Hp}}{\Omega^{\I}} \left( 1 + \frac{\partial f^4}{\partial v_{\I}} \right) e^{\Hp}_4,
		\qquad
		e^{\I}_A
		=
		\partial_{\theta^A_{\I}}
		=
		\partial_{\theta^A_{\Hp}}
		+
		\frac{\partial f^4}{\partial \theta^A_{\I}} \partial_v
		=
		e_A^{\Hp}
		+
		\Omega^{\Hp} \frac{\partial f^4}{\partial \theta^A_{\I}} e_4^{\Hp},
	\]
	on $\{ u_{\Hp} = u_f \} \cap \{ u_{\I} = u_f\}$.  Hence, on $\{ u_{\Hp} = u_f \} \cap \{ u_{\I} = u_f\}$,
	\begin{align*}
		&
		\chi^{\I}_{AB}
		=
		g(\nabla_{e^{\I}_A} e^{\I}_4, e^{\I}_B)
		=
		e^{\I}_A \left( \frac{\Omega^{\Hp}}{\Omega^{\I}} \left( 1 + \frac{\partial f^4}{\partial v_{\I}} \right) \right) g(e_4,e^{\I}_B)
		+
		\frac{\Omega^{\Hp}}{\Omega^{\I}} \left( 1 + \frac{\partial f^4}{\partial v_{\I}} \right) g(\nabla_{e^{\I}_A} e_4, e^{\I}_B)
		\\
		&
		\quad
		=
		\frac{\Omega^{\Hp}}{\Omega^{\I}} \left( 1 + \frac{\partial f^4}{\partial v_{\I}} \right) 
		g \left( \nabla_{e_A + \Omega \frac{\partial f^4}{\partial \widetilde{\theta}^A} e_4} e_4, e_B + \Omega_{\Hp} \frac{\partial f^4}{\partial \theta_{\I}^A} e_4 \right)
		\\
		&
		\quad
		=
		\frac{\Omega^{\Hp}}{\Omega^{\I}} \left( 1 + \frac{\partial f^4}{\partial v_{\I}} \right)
		\bigg[
		g \left( \nabla_{e_A} e_4, e_B \right)
		+
		\Omega^{\Hp} \frac{\partial f^4}{\partial \widetilde{\theta}^A} g \left( \nabla_{e_4} e_4, e_B\right)
		\\
		&
		\quad \qquad \qquad
		+
		\Omega^{\Hp} \frac{\partial f^4}{\partial \widetilde{\theta}^B} g \left( \nabla_{e_A} e_4, e_4 \right)
		+
		\Omega^2_{\Hp} \frac{\partial f^4}{\partial \widetilde{\theta}^A} \frac{\partial f^4}{\partial \widetilde{\theta}^B} g \left( \nabla_{e_4} e_4, e_4\right)
		\bigg]
		\\
		&
		\quad
		=
		\frac{\Omega^{\Hp}}{\Omega^{\I}} \left( 1 + \frac{\partial f^4}{\partial v_{\I}} \right)
		\chi_{AB}^{\Hp},
	\end{align*}
	where $e_A = e_A^{\Hp}$, $e_4 = e_4^{\Hp}$, and the fact that
	\[
		\nabla_{e_A} e_4 = {\chi_A}^B e_C + \etabar_A e_4,
		\quad
		\nabla_{e_4} e_4 = \omegahat e_4,
	\]
	has been used.  Now,
	\[
		\gslash^{\I}_{AB}
		=
		g(e^{\I}_A,e^{\I}_B)
		=
		g\left(
		e_A^{\Hp}
		+
		\Omega^{\Hp} \frac{\partial f^4}{\partial \theta_{\I}^A} e_4^{\Hp},
		e_B^{\Hp}
		+
		\Omega^{\Hp} \frac{\partial f^4}{\partial \theta_{\I}^B} e_4^{\Hp}
		\right)
		=
		g(e_A^{\Hp},e_B^{\Hp})
		=
		\gslash_{AB}^{\Hp},
	\]
	implies that
	\[
		(\tr \chi)^{\I}
		=
		\gslash_{\I}^{AB} \chi^{\I}_{AB}
		=
		\frac{\Omega^{\Hp}}{\Omega^{\I}} \left( 1 + \frac{\partial f^4}{\partial v_{\I}} \right)
		\gslash^{AB}_{\Hp} \chi_{AB}^{\Hp}
		=
		\frac{\Omega^{\Hp}}{\Omega^{\I}} \left( 1 + \frac{\partial f^4}{\partial v_{\I}} \right) \tr \chi^{\Hp},
	\]
	and hence
	\[
		\hat{\chi}^{\I}_{AB}
		=
		\chi^{\I}_{AB}
		-
		\frac{1}{2} (\tr \chi)^{\I} \gslash^{\I}_{AB}
		=
		\frac{\Omega^{\Hp}}{\Omega^{\I}} \left( 1 + \frac{\partial f^4}{\partial v_{\I}} \right)
		\left(
		\chi_{AB}^{\Hp}
		-
		\frac{1}{2} \tr \chi^{\Hp} \gslash_{AB}^{\Hp}
		\right)
		=
		\frac{\Omega^{\Hp}}{\Omega^{\I}} \left( 1 + \frac{\partial f^4}{\partial v_{\I}} \right)
		\hat{\chi}_{AB}^{\Hp}.
	\]
\end{proof}

The estimate \eqref{eq:chiHpinheriting} is now obtained by exploiting Lemma \ref{lem:chihatufsimple}.

\begin{proposition}[Estimate for $\Omega \hat{\chi}_{\Hp}$ on $C_{u_f}$]
	On the final hypersurface $u=u_f$, $\Omega \hat{\chi}_{\Hp}$ satisfies 
	\[
		\sum_{s=0,1,2} (u_f)^{s} \sum_{k=0}^{N+1-s}
		\int_{v(R_{-1},u_f)}^{v(R_1,u_f)} \int_{S^2}
		\left\vert (r\nablaslash)^k \Omega \hat{\chi}_{\Hp} \right\vert^2
		d\theta dv' (u_f)
		\lesssim
		\mathbb{E}^N_{u_f,\I} + \varepsilon^4.
	\]
\end{proposition}

\begin{proof}
	Recall that, for any $S^{\Hp}$-tangent $(0,k)$ tensor $\xi$, on $u=u_f$,
	\[
		\nablaslash^{\I} \xi
		=
		\nablaslash^{\Hp} \xi
		+
		\nablaslash f^4 (\Omega \nablaslash_4)^{\Hp} \xi,
	\]
	and so, for example, the equality \eqref{eq:chihatIchihatH} implies that
	\[
		(\nablaslash \Omega \hat{\chi})^{\I}
		=
		(1+\partial_v f^4) \cdot (\nablaslash \Omega \hat{\chi})^{\Hp}
		+
		(1+\partial_v f^4) \cdot \nablaslash f^4 \cdot (\Omega \nablaslash_4 \Omega \hat{\chi})^{\Hp}
		+
		\nablaslash \partial_v f^4 \cdot (\Omega \hat{\chi})^{\Hp}.
	\]
	Given $s=0,1,2$, applying $\nablaslash^k$ to the equality \eqref{eq:chihatIchihatH}, for any $k \leq N+1-s$, it follows that, in the error notation of Section \ref{schnotfornonlinearerror},
	\[
		\big\vert
		(\nablaslash^k \Omega \hat{\chi})^{\Hp}
		-
		(\nablaslash^k \Omega \hat{\chi})^{\I}
		\big\vert
		\lesssim
		\sum_{\substack{k_1+k_2+\vert \gamma \vert \leq k \\ l_1+l_2+l_3 \geq 2}}
		\vert 
		(\nablaslash^{k_1} f^4)^{l_1}
		\cdot
		(\nablaslash^{k_2} \partial_v f^4)^{l_2}
		\cdot
		(\mathfrak{D}^{\gamma} \hat{\chi})^{l_3}_{\Hp}
		\vert.
	\]
	The proof then follows after controlling the nonlinear terms appropriately.
\end{proof}

\subsection{Estimates for $\ell=0,1$ modes of $\Hp$ quantities from $\I$ quantities}
\label{forprooftwo}

The following estimate for $\beta_{\Hp}$ exploits the fact that $f^3= 0$ on $\{u_{\I}=u_f\} \cap \{u_{\Hp}=u_f\}$.  The estimate for
\[
	\Upsilon^{\Hp}
	=
	\left(1 - \frac{3M_f}{r} \right) (\rho - \rho_{\circ})_{\Hp}
	+
	\frac{3M_f}{2r^2} \left( \Omega \tr \chi - (\Omega \tr \chi)_{\circ} \right)_{\Hp}
	-
	\frac{3M_f\Omega_{\circ}^2}{2r^2} \Omega^{-2} \left( \Omega \tr \chibar - (\Omega \tr \chibar)_{\circ} \right)_{\Hp},
\]
exploits the fact that this quantity, plus a suitable multiple of $(\Omega^2- \Omega^2_{\circ})_{\Hp}$, when projected to $\ell =0$ is almost gauge invariant.

\begin{proposition}[Estimates for $\divslash \beta^{\Hp}_{\ell =1}$ and $\Upsilon_{\ell=0}^{\Hp}$ on $C_{u_f}$]
	The $\ell=1$ modes of $\beta$ and the $\ell=0$ mode of $\Upsilon$ in the $\Hp$ gauge satisfy
	\[
		\left\vert \divslash \beta^{\Hp}_{\ell = 1} (u_f,v(R,u_f)) \right\vert^2
		+
		\left\vert \Upsilon^{\Hp}_{\ell=0} (u_f,v(R,u_f)) \right\vert^2
		\lesssim
		(u_f)^{-4+2\delta}
		\big(
		\mathbb{E}^N_{u_f,\I} + \varepsilon^4
		\big).
	\]
\end{proposition}

\begin{proof}
	The estimate for $\divslash \beta$ follows from applying $\divslash$ to the relation \eqref{eq:curvaturecomp3} and using the fact that $f^3 \equiv 0$ on $\{u_{\Hp} = u_f\} = \{ u_{\I} = u_f\}$.
	
	The estimate for $\Upsilon^{\Hp}_{\ell=0}$ follows from the fact that,
	\[
		\Upsilon_{\ell=0}
		-
		\frac{3M_f}{r^3} \left( \Omega^2 - \Omega_{\circ}^2 \right)_{\ell=0},
	\]
	is almost gauge invariant.  Indeed, the relations \eqref{eq:metriccomp6}, \eqref{eq:Riccicomp7}, \eqref{eq:Riccicomp8} and \eqref{eq:curvaturecomp5} imply that
	\[
		\left\vert
		\left( \Upsilon
		-
		\frac{3M_f}{r^3} \left( \Omega^2 - \Omega_{\circ}^2 \right) \right)^{\I}
		-
		\left( \Upsilon
		-
		\frac{3M_f}{r^3} \left( \Omega^2 - \Omega_{\circ}^2 \right) \right)^{\Hp}
		-
		\frac{3M_f \Omega_{\circ}^2}{r^2} \Deltaslash (f^3-f^4)
		\right\vert
		\lesssim
		\frac{\varepsilon^2}{u^2}.
	\]
	The proof then follows from the fact that $\left( \frac{\Omega_{\circ}^2}{\Omega^2} - 1 \right)^{\Hp}$ vanishes on $u_{\Hp} = u_f$.
\end{proof}

\subsection{Estimate for $(\gslash - r^2 \mathring\gamma)^{\Hp}$ from $(\gslash - r^2 \mathring\gamma)^{\I}$}
\label{forproofthree}

Finally the estimate \eqref{eq:gslashinheriting} is obtained.

\begin{proposition}[Estimates for $(\gslash - r^2 \mathring{\gamma})^{\Hp}$ on $C_{u_f}$] \label{prop:gslashHpI}
	On the final hypersurface $u=u_f$, $(\gslash - r^2 \mathring{\gamma})^{\Hp}$ satisfies, for $s=0,1$,
	\begin{align*}
		\sum_{k \leq N-s}
		\!\!
		\Vert
		(r\nablaslash)^k
		(\gslash - r^2 \mathring{\gamma})^{\Hp}
		\Vert_{S_{u_f,v(R,u_f)}^{\Hp}}^2
		\lesssim
		\frac{\varepsilon^4}{v(R,u_f)^{s+1}}
		+
		\mathbb{E}^N_{u_f,\I}
		+
		\!\!\!
		\sum_{k \leq N-s-2} \!\!
		\Vert (r\nablaslash)^k \hat{\chi}^{\Hp} \Vert_{S_{u_f,v(R,u_f)}^{\Hp}}^2
		\!\!\!\!
		+
		\vert (\rho - \rho_{\circ})^{\Hp} \vert^2
		.
	\end{align*}
\end{proposition}

\begin{proof}
	Recall that, on the final hypersurface $u=u_f$, $f^1 = f^2=f^3 = 0$.
	Using the relation \eqref{eq:metriccomp5} restricted to $\{u=u_f\}$, the fact that
	\[
		\left\vert r_{\Hp} - r_{\I} + \Omega_{\circ}^2(f^3_{\Hp,\I} - f^4_{\Hp,\I})  \right\vert
		\lesssim
		\vert f^3_{\Hp,\I} - f^4_{\Hp,\I} \vert^2,
	\]
	and the fact that, for any $S^{\Hp}$-tangent $(0,k)$ tensor $\xi$, on $u=u_f$,
	\[
		\nablaslash^{\I} \xi
		=
		\nablaslash^{\Hp} \xi
		+
		\nablaslash f^4_{\Hp,\I} (\Omega \nablaslash_4)^{\Hp} \xi,
	\]
	one easily arrives at the estimate \eqref{eq:gslashinheritingprelim}.  The proof then follows from the relation \eqref{eq:curvaturecomp5} restricted to $\{u=u_f\}$ and Lemma \ref{lem:chihatufsimple}.
\end{proof}

This now completes the proof of Proposition~\ref{thm:inheriting},
and thus also of Theorem~\ref{thm:relatinggauges}.
\end{proof}

\chapter{Tensorial wave equations} 
\label{RWtypechapter}

The goal of this chapter is  to provide a unified framework for  estimating 
classes of non-linear $S$-tensorial wave equations which
will encompass the wave equations
of Section~\ref{TeukandRegsection} satisfied by the members of the gauge invariant hierarchy of 
Definition~\ref{defofalmostgaugeinv} and their higher order commutations. 
Specifically, we shall be able to represent these equations in the form
\begin{equation}
\label{generaltypeStensorial}
\frac{1}{2} \left( \Omega \slashed{\nabla}_4 \Omega \slashed{\nabla}_3 + \Omega \slashed{\nabla}_3 \Omega \slashed{\nabla}_4 \right) W+  \frac{\Omega^2}{r^2} r^2 \left(-\slashed{\Delta}+V \right)W  = \mathcal{F}^{lin}[W] + \mathcal{F}^{nlin}[W],
\end{equation}
for $V$ some (non-negative) potential, $\mathcal{F}^{lin}[W]$ a ``linear error" term and $\mathcal{F}^{nlin}[W]$ a ``non-linear error'' term. Depending on the potential, this will be called an $S$-tensorial wave equation of Type 1 or 2.
We shall then derive a set of Propositions giving estimates for equations of the form~\eqref{generaltypeStensorial}
with general right hand side. These will then be applied in the subsequent two chapters in the course
of the proof of Theorems~\ref{thm:PPbarestimates} and~\ref{thm:alphaalphabarestimates}.

\minitoc

The first part of this chapter is purely algebraic.
In {\bf Section~\ref{typeequationsection}},
we shall  introduce the  general formalism for  equations of the type~\eqref{generaltypeStensorial},
specifying in particular the $V$, $\mathcal{F}^{lin}[W]$, $\mathcal{F}^{nlin}[W]$ for the equations satisfied by (weighted versions of) the quantities $\alpha$, $\underline{\alpha}$, $\psi$, $\underline{\psi}$, $P$ and $\underline{P}$, each in their respective gauges. The most important algebraic insight, following from~\cite{holzstabofschw} (cf.~\eqref{reggewheel} in Section~\ref{ginvquanCha}), is that $\mathcal{F}^{lin}\left[r^5 P\right]=\mathcal{F}^{lin}\left[r^5 \underline{P}\right]=0$, in both regions. We then specify the $V$, $\mathcal{F}^{lin}[W]$ and $\mathcal{F}^{nlin}[W]$ arising from commuting tensorial wave equations of various types in  {\bf Section \ref{sec:comrw}}.

The remainder of the chapter concerns the 
analysis of the above tensorial wave equations and will depend on the assumptions of Theorem~\ref{havetoimprovethebootstrap}. 

In {\bf Section \ref{sec:enagi}}, we define various energies (fluxes on null hypersurfaces, integrated decay energies) that will be relevant for the analysis.
{\bf Section \ref{sec:eiledRW}} 
then proves general estimates (energy and integrated decay) for an appropriate pair $(W_{\Hp}, W_{\I})$ satisfying
$S$-tensorial wave equations of type~\eqref{generaltypeStensorial}. 
These estimates arise from summing general multiplier identities applied in the region $\DcH$ (for $W_{\Hp}$) and $\DcI$ (for $W_{\I}$) respectively. Characteristic of all statements is the appearance of a boundary term on the common boundary hypersurface $\mathcal{B}$ that vanishes in linear theory (the contributions from $\DcI$ and $\DcH$ cancelling exactly) and will be estimated later (from Proposition \ref{thm:cancelT}) when specific equations are being considered. Furthermore, the inhomogeneous terms arising from $\mathcal{F}^{lin}$ are kept abstractly in the statements of the propositions. These will again be estimated later when specific $W$'s and equations are considered. 

Finally, {\bf Section \ref{sec:rsagi}} proves redshift multiplier estimates for
$W_{\Hp}$ above and {\bf Section \ref{sec:rpagi}} proves $r^p$-weighted multiplier estimates for
an  $W_{\I}$ above.

\vskip1pc
\emph{As remarked above, Sections~\ref{typeequationsection}--\ref{sec:comrw} lie outside of the proof  of Theorem~\ref{havetoimprovethebootstrap} and can in fact be read immediately
after Chapter~\ref{moreprelimchapter}. 
Starting from Sections~\ref{sec:enagi}, we will assume the assumptions of Theorem~\ref{havetoimprovethebootstrap}.
The results of Sections~\ref{sec:enagi}--\ref{sec:rpagi} 
will depend on Theorems~\ref{thm:sobolevandelinfinity} and~\ref{thm:relatinggauges} 
and will in turn be used in the proofs
of Theorem~\ref{thm:PPbarestimates} and~\ref{thm:alphaalphabarestimates}
of Chapters~\ref{chapter:psiandpsibar} and~\ref{moreherechapter}. Thus, the present chapter together
with the following two form a coherent unit.
The estimates of Sections~\ref{sec:eiledRW},~\ref{sec:rsagi} and~\ref{sec:rpagi} 
build on previous techniques which originate
in the study of the scalar wave equation on Schwarzschild and more generally Kerr spacetimes,
mentioned already in Section~\ref{ginvquanCha}.
The reader can refer to the discussion in Section~2.3 of~\cite{holzstabofschw} for
more context, as well as the lecture notes~\cite{Mihalisnotes} and 
some of the original papers~\cite{BlueSoffer, DafRod2, DafRodnew, partiii}.}

\section{Algebraic preliminaries and definitions} \label{typeequationsection}
Associated to a general double null gauge of the 
form~\eqref{doublenulllongform} or~\eqref{doublenulllongforminterchanged}, endowed
with a Schwarzschild background with mass $M$ as in Section~\ref{diffsofthespheresubsec}, 
we may
define the following operator acting on $S$-tensors:\index{double null gauge!differential operators!$\slashed{\Box}$}
\begin{align}
\slashed{\Box} := \frac{1}{2} \left( \Omega \slashed{\nabla}_4 \Omega \slashed{\nabla}_3 + \Omega \slashed{\nabla}_3 \Omega \slashed{\nabla}_4 \right) +  \frac{\Omega^2}{r^2} r^2 \left(-\slashed{\Delta} \right) = \slashed{\nabla}_T^2 - \slashed{\nabla}_{R^\star}^2 -  \frac{\Omega^2}{r^2} r^2 \slashed{\Delta} \, , 
\end{align}
where\index{double null gauge!differential operators!$T=\frac{1}{2} \left( \Omega e_3 + \Omega e_4 \right)$}\index{double null gauge!differential operators!$R^\star = \frac{1}{2} \left(-\Omega e_3 + \Omega e_4 \right)$}
\begin{equation}
\label{notsogreatnotation}
T=\frac{1}{2} \left( \Omega e_3 + \Omega e_4 \right), \qquad 
R^\star = \frac{1}{2} \left(-\Omega e_3 + \Omega e_4 \right). 
\end{equation}
Define also the potentials\index{double null gauge!functions!$V_I$, potential appearing in Type $1$ wave tensorial wave equations}\index{double null gauge!functions!$V_{II}$, potential appearing in Type $2$ wave tensorial wave equations}
\begin{align}
V_I = \frac{4\Omega^2}{r^2}  - 6M \frac{\Omega^2}{r^3} \ \ \ \textrm{and} \ \ \  V_{II} = \frac{2\Omega^2}{r^2} + 6M \frac{\Omega^2}{r^3} \, .
\end{align}

\begin{remark}
Note that the principal part of $-\slashed{\Box}$ agrees with that of the spacetime wave operator $\Box_g$.
\end{remark}

\begin{definition}[Types of tensorial wave equations] \label{def:tensorwaveclassify}
Consider a double null gauge
of the 
form~\eqref{doublenulllongform} or~\eqref{doublenulllongforminterchanged}, endowed
with a Schwarzschild background with mass $M$ as in Section~\ref{diffsofthespheresubsec}.

We say that an $S$-tensor\footnote{In all applications $W$ will either be a symmetric traceless $S$-tangent $2$-tensor or an $S$-tangent $1$-form.} $W$ 
satisfies a \underline{tensorial wave equation of  Type $1$}\index{schematic notation!$\mathcal{F}^{lin}_1 \left[W\right]$, linear error term associated to tensorial wave equations of Type $1$}\index{schematic notation!$\mathcal{F}^{nlin}_1 \left[W\right]$, non-linear error term associated to tensorial wave equations of Type $1$}
with given expressions $\mathcal{F}^{lin}_1 \left[W \right]$  (``linear errors'') and  $\mathcal{F}^{nlin}_1 \left[W \right]$ (``non-linear errors''), if
\begin{equation} \label{basecase}
\slashed{\Box} W + V_I W = \mathcal{F}^{lin}_{1}\left[W \right] + \mathcal{F}^{nlin}_1 \left[W \right] ,
\end{equation}
 a \underline{tensorial wave equation of  Type $2$}\index{schematic notation!$\mathcal{F}^{lin}_2 \left[W\right]$, linear error term associated to tensorial wave equations of Type $2$}\index{schematic notation!$\mathcal{F}^{nlin}_2 \left[W\right]$, non-linear error term associated to tensorial wave equations of Type $2$} with
given expressions $\mathcal{F}^{lin}_2 \left[W\right]$ and $\mathcal{F}^{nlin}_2 \left[W \right]$, if
\begin{equation} \label{basecase2}
\slashed{\Box} W+ V_{II} W = \mathcal{F}^{lin}_{2}\left[W \right] + \mathcal{F}^{nlin}_2 \left[W\right],
\end{equation}
a  \underline{tensorial wave equation of Type $3_k$}\index{schematic notation!$\mathcal{F}^{lin}_{3_k}\left[W \right]$, linear error term associated to tensorial wave equations of Type $3$}\index{schematic notation!$\mathcal{F}^{nlin}_{3_k}\left[W \right]$, non-linear error term associated to tensorial wave equations of Type $3$} with 
given expressions $\mathcal{F}^{lin}_{3_k}\left[W \right]$ and
$\mathcal{F}^{nlin}_{3_k} \left[W \right]$, if
\begin{equation} \label{redcase}
\slashed{\Box} W + 2k \frac{M}{r^2} \Omega \slashed{\nabla}_3 W = \mathcal{F}^{lin}_{3_k}\left[W \right] + \mathcal{F}^{nlin}_{3_k} \left[W\right]
\end{equation}
and a \underline{tensorial wave equation of Type $4_{k,l}$}\index{schematic notation!$\mathcal{F}^{lin}_{4_{k,l}}\left[W \right]$, linear error term associated to tensorial wave equations of Type $4$}\index{schematic notation!$\mathcal{F}^{nlin}_{4_{k,l}}\left[W \right]$, non-linear error term associated to tensorial wave equations of Type $4$} with
given expressions $\mathcal{F}^{lin}_{4_{k,l}}\left[W \right]$ and 
$\mathcal{F}^{nlin}_{4_{k,l}} \left[W \right]$, if
\begin{equation} \label{rpcase}
\slashed{\Box}  W + \frac{2k}{r} \Omega_\circ^2 \Omega \slashed{\nabla}_4 W + 2l\frac{\Omega^2}{r^2} W = \mathcal{F}^{lin}_{4_{k,l}}\left[W\right] + \mathcal{F}^{nlin}_{4_{k,l}} \left[W\right] .
\end{equation}
\end{definition}

\begin{remark}
We remark already that  while we shall consider Type $1$ and $2$ equations for quantities
of both the $\mathcal{H}^+$ gauge and the $\mathcal{I}^+$ gauge,
we 
shall consider Type $3_{k}$ equations only for the $\mathcal{H}^+$ gauge and Type $4_{k,l}$ equations
only for the $\mathcal{I}^+$ gauge.
\end{remark}

The following propositions reveal the tensorial wave equations satisfied by the quantities in the gauge invariant hierarchy of Definition \ref{defofalmostgaugeinv}. In each case, the relevant double null gauge is that indicated by the subsript 
on the quantitiy, i.e.~$\alpha_{\mathcal{H}^+}$ refers to the $\mathcal{H}^+$ gauge, etc.

\begin{proposition}[Quantities satisfying tensorial wave equation of Type 1 and 2] \label{prop:whythat}
The following quantities satisfy tensorial wave equations of Type 1 or 2, with given linear and nonlinear
error terms:
\begin{itemize}
\item The pairs $\left(A_{\Hp}=r \Omega^2 \alpha_{\Hp}, A_{\I}=r\Omega^2 \alpha_{\I}\right)$ and $\left(\underline{A}_{\Hp}=r \Omega^2 \underline{\alpha}_{\Hp},\check{\underline{A}}_{\I}=\check{r}\Omega^2 \underline{\alpha}_{\I}\right)$ satisfy 
equations of Type $2$ with linear errors given by:
\begin{align}
\mathcal{F}^{lin}_2 \left[A_{\Hp}\right] = \frac{8}{r} \frac{\Omega^2}{r^2} \left(1-\frac{3M}{r}\right) \Pi_{\Hp}  \ \ \ &, \ \ \ \mathcal{F}^{lin}_2 \left[{A}_{\I}\right] =\frac{8}{r} \frac{\Omega^2}{r^2} \left(1-\frac{3M}{r}\right) {\Pi}_{\I}
 \nonumber \\
\mathcal{F}^{lin}_2 \left[\underline{A}_{\Hp}\right] = \frac{8}{r} \frac{\Omega^2}{r^2} \left(1-\frac{3M}{r}\right) \underline{\Pi}_{\Hp}  \ \ \ &, \ \ \ 
\mathcal{F}^{lin}_2 \left[\check{\underline{A}}_{\I}\right] =\frac{8}{r} \frac{\Omega^2}{r^2} \left(1-\frac{3M}{r}\right) \check{\underline{\Pi}}_{\I}
\end{align}
and non-linear errors given by (see Proposition \ref{prop:Teukolskybartilde} for the definition of $k_1$ below):
\begin{align*}
\mathcal{F}^{nlin}_2 \left[A_{\Hp}\right]  = \Omega^2 \mathcal{E}^1 \ \ \ &, \ \ \  \mathcal{F}^{nlin}_2 \left[A_{\I}\right]  =  \mathcal{E}^1_6  \  , \\
   \mathcal{F}^{nlin}_2 \left[\underline{A}_{\Hp}\right] = \Omega^6 (\mathcal{E}^\star)^1 \ \ \ &, \ \ \   \ \mathcal{F}^{nlin}_2 \left[\check{\underline{A}}_{\I}\right] = \check{\mathcal{E}}^1_3 +k_1 r \Dslash_2^* \left(
		r \nablaslash \Omega \tr \chi
		\cdot \alphabar
		\right)  .
\end{align*}
\item The pairs $\left(\Pi_{\Hp}=r^3 \Omega \psi_{\Hp},\Pi_{\I}=r^3\Omega \psi_{\I}\right)$ and $\left(\underline{\Pi}_{\Hp}=r^3 \Omega \underline{\psi}_{\Hp},\check{\underline{\Pi}}_{\I}=r^3\Omega \check{\underline{\psi}}_{\I}\right)$ satisfy
equations of Type $1$ with linear error terms given by
\begin{align}
\mathcal{F}^{lin}_1 \left[ \Pi_{\Hp} \right] &=  \frac{3M}{r^2}\Omega^2 A_{\Hp} - \frac{2}{r} \frac{\Omega^2}{r^2} \left(1-\frac{3M}{r}\right) \Psi_{\Hp} \nonumber \\
 \mathcal{F}^{lin}_1 \left[ \underline{\Pi}_{\Hp} \right] &=  \frac{3M}{r^2}\Omega^2 \underline{A}_{\Hp} - \frac{2}{r} \frac{\Omega^2}{r^2} \left(1-\frac{3M}{r}\right) \underline{\Psi}_{\Hp} \, 
\end{align}
and the same relations replacing all subscripts $\mathcal{H}^+$ by $\mathcal{I}^+$ and putting a check superscript on the underlined-quantities) and non-linear errors given by
\begin{align}
\mathcal{F}^{nlin}_1 \left[\Pi_{\Hp}\right]  = \Omega^2 \mathcal{E}^2  \ , \ \mathcal{F}^{nlin}_1 \left[\Pi_{\I}\right]  =  \mathcal{E}^2_4 \ \  ,  \ \  \mathcal{F}^{nlin}_1 \left[\underline{\Pi}_{\Hp}\right] = \Omega^4 (\mathcal{E}^\star)^2 \ , \ \mathcal{F}^{nlin}_1 \left[\check{\underline{\Pi}}_{\I}\right] = \check{\mathcal{E}}^2_2 \, .
\end{align}
\item The pairs $\left(\Psi_{\Hp},\Psi_{\I}\right)$ and $\left(\underline{\Psi}_{\Hp},\check{\underline{\Psi}}_{\I}\right)$ satisfy
equations of Type 1 with linear errors given by
\begin{align} \label{vgh}
\mathcal{F}^{lin}_1 \left[\Psi_{\Hp}\right] = \mathcal{F}^{lin}_1 \left[\Psi_{\I}\right] =  \mathcal{F}^{lin}_1 \left[\check{\underline{\Psi}}_{\I}\right] = \mathcal{F}^{lin}_1 \left[\underline{\Psi}_{\Hp}\right] = 0 \, .
\end{align}
and non-linear errors given by (see Proposition \ref{prop:Pcheckequation} and in particular (\ref{eq:alphabaromegaanomalouserror}))
\begin{align} \label{nleain}
\mathcal{F}^{nlin}_1 \left[\Psi_{\Hp}\right]  &= \Omega^2 (\mathcal{E}^\star)^3_2 \ , \ \mathcal{F}^{nlin}_1 \left[\Psi_{\I}\right]  = \mathcal{E}^3_2   \\
\mathcal{F}^{nlin}_1 \left[\underline{\Psi}_{\Hp}\right]  &= \Omega^2 (\mathcal{E}^\star)^3_2 \ , \ \mathcal{F}^{nlin}_1 \left[\check{\underline{\Psi}}_{\I}\right]  = \check{\mathcal{E}}^3_2  +
	\mathcal{E}_{\mathrm{anom}}[\check{\Pbar}] \nonumber
\end{align}

\end{itemize}

\end{proposition}

\begin{proof}
This is a rewriting of the equations derived in Chapter \ref{TeukandRegsection}.
\end{proof}

\begin{remark} \label{rem:spestruang}
One may easily check that also $\mathcal{F}^{nlin}_2 \left[A_{\Hp}\right]  = \Omega^2 (\mathcal{E}^\star)^1$ and 
$\mathcal{F}^{nlin}_1 \left[\Pi_{\Hp}\right]  = \Omega^2 (\mathcal{E}^\star)^2$ but we will not need to exploit this additional structure of the error below. We will however exploit  $\mathcal{F}^{nlin}_1 \left[\underline{\Pi}_{\Hp}\right] = \Omega^4 (\mathcal{E}^\star)^2$ and $ \mathcal{F}^{nlin}_2 \left[\underline{A}_{\Hp}\right] = \Omega^6 (\mathcal{E}^\star)^1$ in Section \ref{sec:piabredshift}. (Recall Section \ref{sec:errorstarnot} for the $\mathcal{E}^\star$ notation.)
\end{remark}

\begin{proposition}[Quantities satisfying tensorial wave equations of Type $3_k$] \label{prop:classify3k}
The following quantities satisfy  tensorial wave equations of Type $3_k$ with given linear and non-linear
error terms:
\begin{itemize}
\item The weighted tensor $\Omega^{-4} \underline{A}_{\Hp}$ satisfies an equation of Type $3_2$ with
\begin{align}
\mathcal{F}^{lin}_{3_2} \left[\Omega^{-4} \underline{A}_{\Hp}\right] = \frac{8}{r} \frac{\Omega^2}{r^2} (\Omega^{-2} \underline{\Pi}_{\Hp}) - \frac{8M}{r^3} \Omega^2 (\Omega^{-4} \underline{A}_{\Hp}) \, ,
\end{align}
\begin{align}
\mathcal{F}^{nlin}_{3_2} \left[\Omega^{-4} \underline{A}_{\Hp}\right] = \Omega^2 \mathcal{E}^1 \, .
\end{align}

\item The weighted tensor $\Omega^{-2} \underline{\Pi}_{\Hp}$ satisfies an equation of Type $3_1$ with
\begin{align}
\mathcal{F}^{lin}_{3_1} \left[\Omega^{-2} \underline{\Pi}_{\Hp}\right] &=  \frac{3M}{r^2} \underline{A}_{\Hp} - \frac{2}{r} \frac{\Omega^2}{r^2}  \underline{\Psi}_{\Hp} -2 \frac{2M}{r^3} \Omega^2 \cdot (\Omega^{-2} \underline{\Pi}_{\Hp}) \, ,
\end{align}
\begin{align}
\mathcal{F}^{nlin}_{3_1} \left[\Omega^{-2} \underline{\Pi}_{\Hp}\right] = \Omega^2 \mathcal{E}^2 \, .
\end{align}

\item The tensors $\Psi_{\Hp}$ and $\underline{\Psi}_{\Hp}$ satisfy an equation of Type $3_0$ with
\begin{align} \label{zoer}
\mathcal{F}^{lin}_{3_0} \left[\Psi_{\Hp}\right] = - \frac{4\Omega^2}{r^2} \Psi_{\Hp} + 6M \frac{\Omega^2}{r^3} \Psi_{\Hp} \ \ \ , \ \ \ \mathcal{F}^{lin}_{3_0} \left[\underline{\Psi}_{\Hp}\right] = - \frac{4\Omega^2}{r^2} \underline{\Psi}_{\Hp} + 6M \frac{\Omega^2}{r^3} \underline{\Psi}_{\Hp}
\end{align}
and the non-linear error as in (\ref{nleain}).
\end{itemize}
\end{proposition}

\begin{remark}
The reason for considering $\Omega^{-4} \underline{A}_{\Hp}$ and $\Omega^{-2} \underline{\Pi}_{\Hp}$ is that these quantities are expected to remain uniformly bounded all the way to the event horizon.
\end{remark}

\begin{proof}
The $\Psi, \underline{\Psi}$ case is just a rewriting of the equation of Type 1 from Proposition \ref{prop:whythat}, equation (\ref{vgh}). Now for $W=\underline{\Pi}_{\Hp}$ and $k=1$ as well as for $W=\underline{A}_{\Hp}$ and $k=2$ we note the identity:
\begin{align}
\slashed{\Box} (\Omega^{-2k} W) = \Omega^{-2k} \left( \slashed{\Box} W-k \left(\Omega^{-1} \slashed{\nabla}_3 \Omega^2\right) \Omega \slashed{\nabla}_4 W \right) -2k\omega \Omega \slashed{\nabla}_3 (\Omega^{-2k } W) -2k  \frac{2M}{r^3}  \Omega^2 \cdot \Omega^{-2k}W + \Omega^2 \mathcal{E}^{3-k} \nonumber \, ,
\end{align}
which in turn follows from
\begin{align}
\Omega \slashed{\nabla}_3 \Omega \slashed{\nabla}_4 (\Omega^{-2k} W) &= \Omega \slashed{\nabla}_3 \left(-k \Omega^{-2k}  2\omega W + \Omega^{-2k} \Omega \slashed{\nabla}_4 W  \right) \nonumber \\
&=-2k\omega \Omega \slashed{\nabla}_3 (\Omega^{-2k } W) -2k  \frac{2M}{r^3}  \Omega^2 \cdot \Omega^{-2k}W -k \left(\Omega^{-1} \slashed{\nabla}_3 \Omega^2\right)\Omega^{-2k} \Omega \slashed{\nabla}_4 W \nonumber \\
& \ \ \ +  \left(-2\frac{\Omega^2}{r^2}r^2 \slashed{\mathcal{D}}_2^\star \slashed{div}  (\Omega^{-2k} W) + \Omega^{-2k}\slashed{\Box} W \right)  + \Omega^2 \mathcal{E}^{3-k} \, .
\end{align}
We now apply the above with $k=1$ and $W=\underline{\Pi}_{\Hp}$ as well as with $k=2$ and $W=\underline{A}_{\Hp}$. After inserting the relevant 
 equation of Proposition \ref{prop:whythat} for $\slashed{\Box} \underline{\Pi}_{\Hp}$ and $\slashed{\Box} \underline{A}_{\Hp}$ we obtain the desired form.
\end{proof}

\begin{proposition}[Quantities satisfying tensorial wave equations of Type $4_{k,l}$] \label{prop:classify4kl}
The following quantities satisfy tensorial wave equations of Type $4_{k,l}$ with given linear
and non-linear error terms:
\begin{itemize}
\item The tensors $\Psi_{\I}$ and $\check{\underline{\Psi}}_{\I}$ satisfy equations of
Type $4_{k,l}$ with $k=0$, $\ell=2$ and with
\begin{align}
\mathcal{F}^{lin}_{4_{0,2}} \left[\Psi_{\I}\right] &=  6M\frac{\Omega^2}{r^3}\Psi_{\I} \ \ , \ \ \mathcal{F}^{lin}_{4_{0,2}} \left[\check{\underline{\Psi}}_{\I}\right] =  6M\frac{\Omega^2}{r^3}\check{\underline{\Psi}}_{\I} \, , \nonumber \\
\mathcal{F}^{nlin}_{4_{0,2}} \left[\Psi_{\I}\right] &= {\mathcal{E}}^3_2 \ \ ,  \qquad \qquad \mathcal{F}^{nlin}_{4_{0,2}} \left[\check{\underline{\Psi}}_{\I}\right]   = \check{\mathcal{E}}^3_2   +
	\mathcal{E}_{\mathrm{anom}}[\check{\Pbar}] \nonumber .
\end{align}
\item The tensor $r^2 \Pi_{\I}$ satisfies an equation of Type $4_{k,l}$ a with $k=1$, $\ell=1$ and with
\begin{align}
\mathcal{F}^{lin}_{4_{1,1}} \left[r^2 \Pi_{\I}\right] &=  2M \frac{\Omega^2}{r^2} \Psi_{\I} + \frac{3M \Omega^2}{r^4} (r^4 A_{\I}) +  6M\frac{\Omega^2}{r^3}\left(r^2 \Pi_{\I}\right) \, , \nonumber \\
\mathcal{F}^{nlin}_{4_{1,1}} \left[r^2 \Pi_{\I}\right] &= \mathcal{E}^2_2 \, .
\end{align}
\item The tensor $r^4 A_{\I}$ satisfies an equation of Type $4_{k,l}$ with $k=2$, $\ell=-1$ and with
\begin{align}
\mathcal{F}^{lin}_{4_{2,-1}} \left[r^4 A_{\I}\right] &= -8M \Omega^2\Pi_{\I} - 22M\frac{\Omega_\circ^2}{r^3}\left(r^4 A_{\I}\right) \, ,   \nonumber \\
\mathcal{F}^{nlin}_{4_{2,-1}} \left[r^4 A_{\I}\right] &= \mathcal{E}^1_2 \, .
\end{align}
\end{itemize}
\end{proposition}

\begin{proof}
For $\Psi_{\I}$, this is a rewriting of the equation of Type 1 from Proposition \ref{prop:whythat}, equation (\ref{vgh}). 
For $\Pi_{\I}$, note (dropping subscripts temporarily)
\begin{align}
\Omega \slashed{\nabla}_3 \Omega \slashed{\nabla}_4 \left(r^2 \Pi\right) = \Omega \slashed{\nabla}_3 \left(r^2 \Omega \slashed{\nabla}_4 \Pi +  2 r \Omega_\circ^2 \, \Pi \right) \nonumber \\
= r^2 \left(- 2 \frac{\Omega^2}{r^2} r^2 \slashed{\mathcal{D}}_2^\star \slashed{div} \Pi -2\frac{\Omega^2}{r^2} \Pi +6M \frac{\Omega^2}{r^3}A -\frac{2}{r}  \left(1-\frac{3M}{r}\right) \Omega \slashed{\nabla}_3 \Pi +\frac{3M}{r^2} \Omega^2 A + \mathcal{E}^2_4 \right) \nonumber \\
+2r \Omega_\circ^2 \Omega \slashed{\nabla}_3 \Pi -\frac{2}{r} \Omega_\circ^2 \Omega \slashed{\nabla}_4 \left(r^2 \Pi\right) + 4 \Omega_\circ^4 \Pi -2\Omega_\circ^2 \Pi \nonumber
\end{align}
and collect terms. For $A$, note
\begin{align}
\Omega \slashed{\nabla}_3 \Omega \slashed{\nabla}_4 \left(r^4 A\right) = \Omega \slashed{\nabla}_3 \left(r^4 \Omega \slashed{\nabla}_4 A +  4 r^3 \Omega_\circ^2 \, A \right) \nonumber \\
= r^4 \left(+ \frac{\Omega^2}{r^2} r^2 \slashed{\Delta} A- 2 \frac{\Omega^2}{r^2} A -6M \frac{\Omega^2}{r^3}A -\frac{4}{r}  \left(1-\frac{3M}{r}\right) \Omega \slashed{\nabla}_3 A + \mathcal{E} \left[\alpha\right] + F_{34}[A]\right)  \nonumber \\
+4r^3 \Omega_\circ^2 \Omega \slashed{\nabla}_3 A -\frac{4}{r} \Omega_\circ^2 \Omega \slashed{\nabla}_4 \left(r^4 A\right) + 16r^2 \Omega_\circ^4 A + \left(-12 r^2 + 16Mr\right)\Omega_\circ^2 A \nonumber 
\end{align}
and collect terms.
\end{proof}

\section{The structure of commuted tensorial wave equations} \label{sec:comrw}
The next three propositions capture what structure of the tensorial wave equations is preserved under commutation 
by the operators $\slashed{\nabla}_T$, $r\slashed{div}$, $r^2 \slashed{\Delta}$ and $\slashed{\nabla}_{R^\star}$. 

Before stating them we need to define an additional bit of notation to capture the errors arising from the commutation of an abstract tensorial wave equation.
Recall first from Section \ref{sec:nlepnotation} that given $k_1,p_1 \geq 0$, the notation $\left(\mathfrak{D}^{k_1} \Phi_{p_1}\right)^1 \cdot H$ denotes a term involving a sum of terms (linear, quadratic, cubic, ...)  in the $\Phi_p$ containing at most $k_1$ derivatives in total and that the array ${H}$ encodes the information on exactly what terms appear in the sum. Let $W$ be a symmetric traceless tensor.  Define for $p \geq 0$, $k \geq 2$ the array
\[
\mathscr{H} = \left\{ {}^{j} \bold{H}^{p_1}_{k_1 k_2} \ , \ {}^{j} \tilde{H}^{p_1}_{k_1 k_2}  \ , \  {}^j J^{p_1}_{k_1 k_2} \right\}_{k_1+k_2 \leq k , k_1 \leq k-1 ,  p_1 \geq p, j \geq 1}
\]
where ${}^{j} \bold{H}^{p_1}_{k_1 k_2}$ is a collection of ${H}$ as above,
$
{}^{j} \tilde{H}^{p_1}_{k_1 k_2} \in \{ \gamma \ | \ \gamma \in \{ (1,0,0), (0,1,0), (0,0,1)\}^{k_2} \} \cup \{0 \}
$
and ${}^{j} J^{p_1}_{k_1 k_2}$ is a trace set of some order $d\geq0$. Set $\mathfrak{D}^{k_2} W \cdot (\gamma) := \mathfrak{D}^\gamma W$ for $\gamma \in \{ (1,0,0), (0,1,0), (0,0,1)\}^{k_2}$. We finally define
\begin{align}
\mathfrak{D}^k [\Phi_p, W] \cdot \mathscr{H} =\sum_{\substack{p_1 \geq p \\ k_1+k_2 \leq k \\k_1 \leq k-1}} \sum_{j \geq 1} \left(\left(\mathfrak{D}^{k_1} \Phi_{p_1}\right)^1 \cdot {}^{j} \bold{H}^{p_1}_{k_1 k_2} \otimes \mathfrak{D}^{k_2} W \cdot {}^{j} \tilde{H}^{p_1}_{k_1 k_2}  \right) \cdot_{\slashed{g}} {}^j J^{p_1}_{k_1 k_2} \, .
\end{align}
As usual, such expressions are only ever considered when sufficiently many of the components of $\mathscr{H}$ vanish so that each object appearing in the above summation is an $S$-tensor of the same type and moreover that the range of $p$ and $j$ is finite so that the summation is indeed well-defined. Finally, we note that once $W$ is specified (it will typically be a derivative operator applied to an almost gauge invariant quantity) the above expression can be captured by the familiar $\mathcal{E}^{k^\prime}_{p^\prime}$. For instance $\mathfrak{D}^k [\Phi_p, \left[r \Omega \slashed{\nabla}_4\right]^{m} (r^5 P) ] \cdot \mathscr{H} = \mathcal{E}^{k+m+2}_p$.
\begin{proposition} \label{prop:commutetype1}
Consider a double null gauge
of the 
form~\eqref{doublenulllongform} or~\eqref{doublenulllongforminterchanged}, endowed
with a Schwarzschild background with mass $M$ as in Section~\ref{diffsofthespheresubsec}. Let $W$ be a symmetric traceless $S$-tensor satisfying a tensorial wave equation of Type $1$ or $2$. Then\footnote{We suppress the subscript $1$ or $2$ in $\mathcal{F}^{lin}$ below. It is $1$ or $2$ depending on the type.} 
\begin{itemize}
\item The commuted quantity $\slashed{\nabla}_T W$ satisfies a tensorial wave equation of the same type
with
\begin{align}
\mathcal{F}^{lin}\left[\slashed{\nabla}_T W\right]  &= \slashed{\nabla}_T \left( \mathcal{F}^{lin}\left[W\right] \right) \, , \nonumber \\
\mathcal{F}^{nlin}\left[\slashed{\nabla}_T W\right]  &=  \slashed{\nabla}_T\left( \mathcal{F}^{nlin}\left[W\right] \right) + \mathfrak{D}^2 [\Phi_3, W] \cdot \mathscr{H}.
\end{align}
\item The commuted quantity $r \slashed{div} W$ (which is an $S$-tangent $1$-form) satisfies a tensorial wave equation of the same type with
\begin{align}
\mathcal{F}^{lin}\left[r \slashed{div}  W\right]  &= r \slashed{div}  \mathcal{F}^{lin} \left[W\right]  +3\frac{\Omega^2}{r^2} r \slashed{div} W \, ,  \nonumber \\
\mathcal{F}^{nlin}\left[r \slashed{div}  W\right]  &=  r \slashed{div}  \mathcal{F}^{nlin} \left[W\right]  + \mathfrak{D}^2 [\Phi_2, W] \cdot \mathscr{H}.
\end{align}
\item The commuted quantity $r^2 \slashed{\Delta} W$ satisfies a tensorial wave equation of the same type with
\begin{align}
\mathcal{F}^{lin}_{}\left[r^2 \slashed{\Delta}  W\right]  &= r^2 \slashed{\Delta}  \mathcal{F}^{lin} \left[W\right] \, ,  \nonumber \\
\mathcal{F}^{nlin}_{}\left[r^2 \slashed{\Delta} W\right]  &= r^2 \slashed{\Delta}  \mathcal{F}^{lin} \left[W\right]+ \mathfrak{D}^3 [\Phi_2, W] \cdot \mathscr{H}.
\end{align}
\item The commuted quantity $\slashed{\nabla}_{R^\star} W$ satisfies a tensorial wave equation of the same type with
\begin{align} \label{commuteRstar}
\mathcal{F}^{lin}\left[\slashed{\nabla}_{R^\star} W\right]  &= \slashed{\nabla}_{R^\star} \left( \mathcal{F}^{lin}\left[W\right] \right) - \frac{2}{r} \frac{\Omega^2}{r^2} \left(1-\frac{3M}{r}\right) r^2 \slashed{\Delta} W
+h_{0} \frac{1}{r} \frac{\Omega^2}{r^2} W\, ,  \nonumber \\
\mathcal{F}^{nlin}\left[\slashed{\nabla}_{R^\star}W\right]  &=  \slashed{\nabla}_{R^\star}\left( \mathcal{F}^{nlin}\left[W\right] \right) + \mathfrak{D}^2 [\Phi_3, W] \cdot \mathscr{H}.
\end{align}
\end{itemize}

\end{proposition}

\begin{proof}
The $\slashed{\nabla}_T$ commutation is straightforward using Lemma \ref{lem:commutation}. For the angular commutation we note
\begin{align} \label{divcom}
\slashed{\mathcal{D}}_2 \slashed{\Delta} \xi = \slashed{\mathcal{D}}_2 \left(-2\slashed{\mathcal{D}}_2^\star \slashed{\mathcal{D}}_2 + 2K\right) \xi &= -2 \left(-\frac{1}{2} \slashed{\Delta} - \frac{1}{2}K\right) \slashed{\mathcal{D}}_2 \xi + 2K \slashed{\mathcal{D}}_2 \xi + 2\slashed{\nabla} K\hat{\otimes} \xi  \nonumber \\
&= \slashed{\Delta} \slashed{\mathcal{D}}_2 \xi + 3K  \slashed{\mathcal{D}}_2 \xi + 2\slashed{\nabla} K\hat{\otimes} \xi  \, ,
\end{align}
which yields the result. The $\slashed{\nabla}_{R^\star}$ commutation is again a straightforward commutation using the formulae (\ref{eq:nabla4Omegar}), (\ref{eq:nabla3Omegar}) and Lemma \ref{lem:commutation}.
\end{proof}

\begin{proposition} \label{prop:3kc}
Consider a double null gauge
of the 
form~\eqref{doublenulllongform} or~\eqref{doublenulllongforminterchanged}, endowed
with a Schwarzschild background with mass $M$ as in Section~\ref{diffsofthespheresubsec}. Let $W$ be a symmetric traceless $S$-tensor satisfying a tensorial wave equation of Type $3_k$. Then
\begin{itemize}
\item The commuted quantity $\slashed{\nabla}_{T} W$ satisfies a tensorial wave equation of type $3_{k}$
with
\begin{align}
\mathcal{F}^{lin}_{3_k}\left[\slashed{\nabla}_{T} W\right]  &=\slashed{\nabla}_{T} \left( \mathcal{F}_{3^k}^{lin} \left[W\right] \right) \, , \nonumber \\
\mathcal{F}^{nlin}_{3_k}\left[\slashed{\nabla}_{T}  W\right]  &= \slashed{\nabla}_{T}  \left( \mathcal{F}^{nlin}_{3^k} \left[W\right] \right)  + \mathfrak{D}^2 [\Phi_3, W] \cdot \mathscr{H}\, .
\end{align}
\item The commuted quantity $r \slashed{div} W_{\Hp}$ (which is an $S$-tangent $1$-form) satisfies a tensorial wave equation of type $3_{k}$ with
\begin{align}
\mathcal{F}^{lin}_{3_k}\left[r \slashed{div}  W\right]  &= r \slashed{div}  \mathcal{F}_{3^k}^{lin} \left[W\right]  +3\frac{\Omega^2}{r^2} r \slashed{div} W \, , \nonumber \\
\mathcal{F}^{nlin}_{3_k}\left[r \slashed{div}  W\right]  &= r\slashed{div}  \mathcal{F}_{3^k}^{nlin} \left[W\right]   + \mathfrak{D}^2 [\Phi_2, W] \cdot \mathscr{H}\, .
\end{align}
\item The commuted quantity $r^2 \slashed{\Delta} W$ satisfies a tensorial wave equation of type $3_{k}$ with
\begin{align}
\mathcal{F}^{lin}_{3_k}\left[r^2 \slashed{\Delta}  W\right]  &= r^2 \slashed{\Delta}  \mathcal{F}_{3^k}^{lin} \left[W\right]  \, , \nonumber \\
 \mathcal{F}^{nlin}_{3_k}\left[r^2 \slashed{\Delta} W\right]  &=  r^2 \slashed{\Delta}  \mathcal{F}_{3^k}^{nlin} \left[W\right]  + \mathfrak{D}^3 [\Phi_2, W] \cdot \mathscr{H} \, .
\end{align}
\item The commuted quantity $ \Omega^{-1}\slashed{\nabla}_3 W$ satisfies a tensorial wave equation of type $3_{k+\frac{1}{2}}$with 
\begin{align}
\mathcal{F}^{lin}_{3_k}\left[ \Omega^{-1}\slashed{\nabla}_3 W\right]  &=  \Omega^{-1}\slashed{\nabla}_3  \mathcal{F}_{3^k}^{lin} \left[W\right] + h_3 \Omega^2 \Omega^{-1} \slashed{\nabla}_3 W + h_3 \Omega^2 \slashed{\Delta} W\, , 
\nonumber \\
\mathcal{F}^{nlin}_{3_k} \left[ \Omega^{-1}\slashed{\nabla}_3 W\right]  &= \Omega^{-1}\slashed{\nabla}_3  \mathcal{F}_{3^k}^{nlin} \left[W\right]  + \mathfrak{D}^2 [\Phi_3, W] \cdot \mathscr{H} \label{rsen} \, .
\end{align}
\end{itemize}
\end{proposition}

\begin{remark}
The fact that commutation with $\Omega^{-1} \slashed{\nabla}_3$ leads to a stronger redshift factor ($k+\frac{1}{2}$) is the familiar amplified commuted redshift effect for the wave equation. See~\cite{Mihalisnotes}.
\end{remark}

\begin{remark}
Since these equations are only ever going to be considered in the $\Hp$ gauge, one does not need to keep track of the $r$-weights. We have included them here to unify notation for the error terms.
\end{remark}

\begin{proof}
The $\slashed{\nabla}_T$, $r \slashed{div}$ and $r^2 \slashed{\Delta}$ commutations are analogous to the proof of Proposition \ref{prop:commutetype1}. For $\Omega^{-1} \slashed{\nabla}_3$, write (cf.~Lemma \ref{lem:commutation})
\[
\frac{1}{\Omega^2}\slashed{\Box} = \frac{1}{\Omega} \slashed{\nabla}_3 \Omega \slashed{\nabla}_4 -  \frac{1}{r^2} r^2 \slashed{\Delta} - \frac{1}{2} \Omega^{-2} F_{34} [\cdot] \, 
\]
and commute the operator $\Omega^{-1} \slashed{\nabla}_3$ through, resulting in the commutator
\begin{align}
\left[ \slashed{\Box} , \Omega^{-1} \slashed{\nabla}_3 \right] W \equiv -\frac{2M}{r^2} \Omega^2 \left[ \Omega^{-1} \slashed{\nabla}_3\right]^2 W - \frac{4M}{r^3} \Omega^2 \Omega^{-1} \slashed{\nabla}_3 W -\frac{4}{r^3} \Omega^2 \slashed{\mathcal{D}}_2^\star \slashed{div} W
\end{align}
with $\equiv$ ignoring non-linear errors which are easily seen (using Lemma \ref{lem:commutation}) to be of the form appearing in (\ref{rsen}).
\end{proof}

\begin{proposition} \label{prop:comm4kl}
Consider a double null gauge
of the 
form~\eqref{doublenulllongform} or~\eqref{doublenulllongforminterchanged}, endowed
with a Schwarzschild background with mass $M$ as in Section~\ref{diffsofthespheresubsec}. Let $W$ be a symmetric traceless $S$-tensor satisfying a tensorial wave equation of Type $4_{k,l}$. Then 
\begin{itemize}
\item The commuted quantity $\slashed{\nabla}_{T}  W$ satisfies a tensorial wave equation of type $4_{k,l}$
with
\begin{align}
\mathcal{F}_{4_{k,l}}^{lin}\left[\slashed{\nabla}_{T}  W\right]  &=\slashed{\nabla}_{T}  \left(\mathcal{F}_{4_{k,l}}^{lin} \left[W\right]\right) \, ,  \nonumber \\
\mathcal{F}_{4_{k,l}}^{nlin}\left[\slashed{\nabla}_{T} W\right]  &= \slashed{\nabla}_{T}  \left( \mathcal{F}_{4_{k,l}}^{nlin}\left[W\right] \right)  + \mathfrak{D}^2 [\Phi_3, W] \cdot \mathscr{H} \, .
\end{align}
\item The commuted quantity $r \slashed{div} W_{\I}$  (which is an $S$-tangent $1$-form) satisfies a tensorial wave equation of type $4_{k,l-\frac{3}{2}}$
with 
\begin{align}
\mathcal{F}_{4_{k,l}}^{lin}\left[r \slashed{div}  W\right]  &= r \slashed{div}  \mathcal{F}_{4_{k,l}}^{lin} \left[W\right]  \, ,  \nonumber \\
\mathcal{F}_{4_{k,l}}^{nlin}\left[r \slashed{div}  W\right]  &=r \slashed{div}  \mathcal{F}_{4_{k,l}}^{nlin} \left[W\right]   + \mathfrak{D}^2 [\Phi_2, W] \cdot \mathscr{H}\, .
\end{align}
\item The commuted quantity $r^2 \slashed{\Delta} W$ satisfies a tensorial wave equation of type $4_{k,l}$ with
\begin{align}
\mathcal{F}^{lin}_{4_{k,l}}\left[r^2 \slashed{\Delta}  W\right]  &= r^2 \slashed{\Delta}  \mathcal{F}^{lin} \left[W\right] \, ,  \nonumber \\
\mathcal{F}^{nlin}_{4_{k,l}}\left[r^2 \slashed{\Delta} W\right]  &= r^2 \slashed{\Delta}  \mathcal{F}^{nlin} \left[W\right] + \mathfrak{D}^3 [\Phi_2, W] \cdot \mathscr{H} \, .
\end{align}

\item The commuted quantity $r \Omega\slashed{\nabla}_4 W$ satisfies a tensorial wave equation of type $4_{k+\frac{1}{2}, l-k-1/2}$ with 
\begin{align}
\mathcal{F}^{lin}_{4_{k,l}}\left[r \Omega \slashed{\nabla}_4 W\right]  &= \left( r\Omega \slashed{\nabla}_4 -\Omega_\circ^2 \right) \mathcal{F}_{4_{k,l}}^{lin} \left[W\right] + h_2 r^2 \slashed{\Delta} W + h_3 r \Omega \slashed{\nabla}_4 W +h_2 W  \, , \nonumber \\
\mathcal{F}_{4_{k,l}}^{nlin}\left[r \Omega \slashed{\nabla}_4 W\right]  &=  \left( r\Omega \slashed{\nabla}_4  -\Omega_\circ^2 \right)\mathcal{F}_{4_{k,l}}^{nlin} \left[W\right]  + \mathfrak{D}^2 [\Phi_2, W] \cdot \mathscr{H} \label{vgy} \, .
\end{align}
\end{itemize}

\end{proposition}

\begin{proof}
The $\slashed{\nabla}_T$, $r \slashed{div}$ and $r^2 \slashed{\Delta}$ commutations are analogous to the proof of Proposition \ref{prop:commutetype1}. For $r \Omega \slashed{\nabla}_4$ we first note 
\begin{align}
&r \Omega \slashed{\nabla}_4 \left(\mathcal{F}^{lin}_{4_{k,l}}\left[W\right] + \mathcal{F}^{nlin}_{4_{k,l}}\left[W\right] \right) = r \Omega \slashed{\nabla}_4  \left( \slashed{\Box} + \frac{2k}{r} \Omega_\circ^2 \Omega \slashed{\nabla}_4 W + 2l \frac{\Omega^2}{r^2} W \right) \nonumber \\
& \qquad \qquad \qquad=  \slashed{\Box} \left(r \Omega \slashed{\nabla}_4 W\right) + \frac{2k}{r} \Omega_\circ^2 \Omega \slashed{\nabla}_4 \left(r \Omega \slashed{\nabla}_4 W\right) + 2l \frac{\Omega^2}{r^2} \left(r \Omega \slashed{\nabla}_4 W\right) \nonumber \\
& \qquad \qquad \qquad \ \ + \left[ r \Omega \slashed{\nabla}_4 ,\slashed{\Box} \right] W + \left[r \Omega \slashed{\nabla}_4 , \frac{2k}{r} \Omega_\circ^2 \Omega \slashed{\nabla}_4\right] - 4l \frac{\Omega^2}{r^2} \left(1-\frac{3M}{r}\right)W + \frac{1}{r} \Phi_{5/2} W. \nonumber
\end{align}
To compute the commutator
$
\left[ r\Omega \slashed{\nabla}_4 , \slashed{\Box}\right] 
$
we observe (with $\equiv$ ignoring non-linear terms, wich are easily seen to be of the form claimed)
\begin{align}
r \Omega \slashed{\nabla}_ 4 \left( \Omega \slashed{\nabla}_3 \Omega \slashed{\nabla}_4 - \frac{\Omega^2}{r^2} r^2 \slashed{\Delta} + F_{34}\left[\cdot\right]\right) \equiv r \Omega \slashed{\nabla}_ 3 \left( \Omega \slashed{\nabla}_4 \Omega \slashed{\nabla}_4 \right) - \frac{\Omega^2}{r^2} r^2 \slashed{\Delta} r \Omega \slashed{\nabla}_4 + 2\frac{\Omega^2}{r^2} \left(1-\frac{3M}{r}\right) r^2 \slashed{\Delta} \nonumber \\
\equiv \Omega \slashed{\nabla}_3 \left( r\Omega \slashed{\nabla}_4 \Omega \slashed{\nabla}_4 \right) + \Omega_\circ^2 \Omega \slashed{\nabla}_4 \Omega \slashed{\nabla}_4- \frac{\Omega^2}{r^2} r^2 \slashed{\Delta} r \Omega \slashed{\nabla}_4 + 2\frac{\Omega^2}{r^2} \left(1-\frac{3M}{r}\right) r^2 \slashed{\Delta} \nonumber \\
\equiv \Omega \slashed{\nabla}_3 \left( \Omega \slashed{\nabla}_4 (r\Omega \slashed{\nabla}_4 )- \Omega_\circ^2 \Omega \slashed{\nabla}_4 \right) + \frac{\Omega_\circ^2}{r} \Omega \slashed{\nabla}_4 (r\Omega \slashed{\nabla}_4) - \frac{\Omega_\circ^4}{r} \Omega \slashed{\nabla}_4 - \frac{\Omega^2}{r^2} r^2 \slashed{\Delta} r \Omega \slashed{\nabla}_4 + 2\frac{\Omega^2}{r^2} \left(1-\frac{3M}{r}\right) r^2 \slashed{\Delta} \nonumber \\
\equiv \slashed{\Box} \left(r \Omega \slashed{\nabla}_4 \right) + \frac{\Omega_\circ^2}{r} \Omega \slashed{\nabla}_4 (r\Omega \slashed{\nabla}_4)+ \left(-1+\frac{4M}{r}\right) \frac{\Omega_\circ^2}{r^2} r\Omega \slashed{\nabla}_4-\Omega_\circ^2 \Omega \slashed{\nabla}_3 \Omega \slashed{\nabla}_4 + 2\frac{\Omega^2}{r^2} \left(1-\frac{3M}{r}\right) r^2 \slashed{\Delta} 
\nonumber \\
\equiv \slashed{\Box} \left(r \Omega \slashed{\nabla}_4 \right) + \frac{\Omega_\circ^2}{r} \Omega \slashed{\nabla}_4 (r\Omega \slashed{\nabla}_4)+ \left(-1+2k\right) \frac{\Omega_\circ^2}{r^2} r\Omega \slashed{\nabla}_4   \phantom{XXXX}
\nonumber \\ + h_2 r^2 \slashed{\Delta} + h_2 + h_3 r \Omega \slashed{\nabla}_4 - \Omega_\circ^2 \left(\mathcal{F}^{lin}_{4_{k,l}}\left[\cdot\right] + \mathcal{F}^{nlin}_{4_{k,l}}\left[\cdot\right] \right) \, . \nonumber
\end{align}
To compute the other commutator we note
\begin{align}
r \Omega \slashed{\nabla}_4 \left(\frac{2k}{r} \Omega_\circ^2 \Omega \slashed{\nabla}_4 W\right) = r \Omega \slashed{\nabla}_4 \left(\frac{2k}{r^2} \Omega_\circ^2 r \Omega \slashed{\nabla}_4 W\right) = \frac{2k}{r} \Omega_\circ^2 \Omega \slashed{\nabla}_4 \left(r \Omega \slashed{\nabla}_4 W\right) - \frac{4k}{r^2} \Omega_\circ^2 \left(1-\frac{3M}{r}\right) r \Omega \slashed{\nabla}_4 W \nonumber
\end{align}
Collecting terms now yields the claim.
\end{proof}

\section{Regions and energies for tensorial wave equation estimates} \label{sec:enagi}

We introduce in this section some auxiliary notations for regions convenient for
applying energy type estimates for tensorial wave equations in our setup,
as well as notation for
energy fluxes and spacetime integral quantities that naturally appear in such estimates.

\vskip1pc

\noindent\fbox{
    \parbox{6.35in}{
As in Chapters~\ref{elliptandcalcchapter} and~\ref{chap:comparing}, in the remainder
of this chapter we shall assume throughout the assumptions of~\Cref{havetoimprovethebootstrap}. Let us fix an
arbitrary  $u_f\in[u_f^0, \hat{u}_f$], with $\hat{u}_f\in \mathfrak{B}$,
and fix some $\lambda \in \mathfrak{R}(u_f)$.
All propositions below
shall always refer  
to the anchored $\I$ and $\Hp$ gauges in the  spacetime  $(\mathcal{M}(\lambda), g(\lambda))$,  
corresponding to parameters
$u_f$, $M_f(u_f,\lambda)$,
whose existence is
ensured by Definition~\ref{bootstrapsetdef}.
We shall denote $M=M_f$ throughout the remainder of 
this chapter.}}
\vskip1pc

 The following summary will help the reader familiarise with the notation.
\begin{itemize}
\item Fluxes are generally denoted $\mathbb{F}$ (outgoing cones) and $\underline{\mathbb{F}}$ (ingoing cones), integrated decay energies by $\mathbb{I}$. 
\item A check superscript (i.e.~$\check{\mathbb{I}}$) is used when the cones or spacetimes integrals in the respective spacetime regions are truncated at the timelike hypersurface $\mathcal{B}$ defined in Section~\ref{subsubsec:cancelT}. 
\item A superscript $\diamond$ in the horizon energies denotes an energy that is non-optimal in terms of $\Omega^2$-weights (to be thought of as the horizon-degenerate energy generated by the ultimately Killing field $T$).
\item  A superscript ``deg" in the horizon energies denotes an energy that degenerates near $r=3M$. 
\item A superscript number in the infinity energies denotes an $r$-weight near null infinity.
\item A superscript $\star$ in the horizon region denotes that the energies are restricted to $r\geq 9M_{\rm init}/4$. \\ A superscript $\star$ in the infinity region denotes that the energies are restricted to $r\leq 2R$. 
\end{itemize}

\subsection{The truncated regions}
Recall the hypersurface $\mathcal{B}$ and the values $u_1$, $v_1$
defined in Section~\ref{timelikehype}.
The timelike hypersurface $\mathcal{B}$ partitions the globally hyperbolic domain 
\[
J^+(\underline{C}^{\Hp}_{v_1}) \cap J^-(C_{u_f})
\]
into two connected components whose closures we shall denote by $\DcH$ and $\DcI$, 
characterised by\index{teleological $\I$ gauge!sets!$\DcI$, truncated domain for energy estimates}\index{teleological $\Hp$ gauge!sets!$\DcH$, truncated domain for energy estimates}
\begin{equation}
\label{theCzechs}
\DcH\subset \mathcal{D}^{\Hp}, 
\qquad \DcI \subset \mathcal{D}^{\I}, \qquad \DcH\cap \DcI = \mathcal{B}, \qquad 
 \DcH\cap \DcI = J^+(\underline{C}^{\Hp}_{v_1}) \cap J^-(C_{u_f}).
\end{equation}
Refer to Figure~\ref{truncationfigure}.
\begin{figure}
\centering{
\def\svgwidth{20pc}
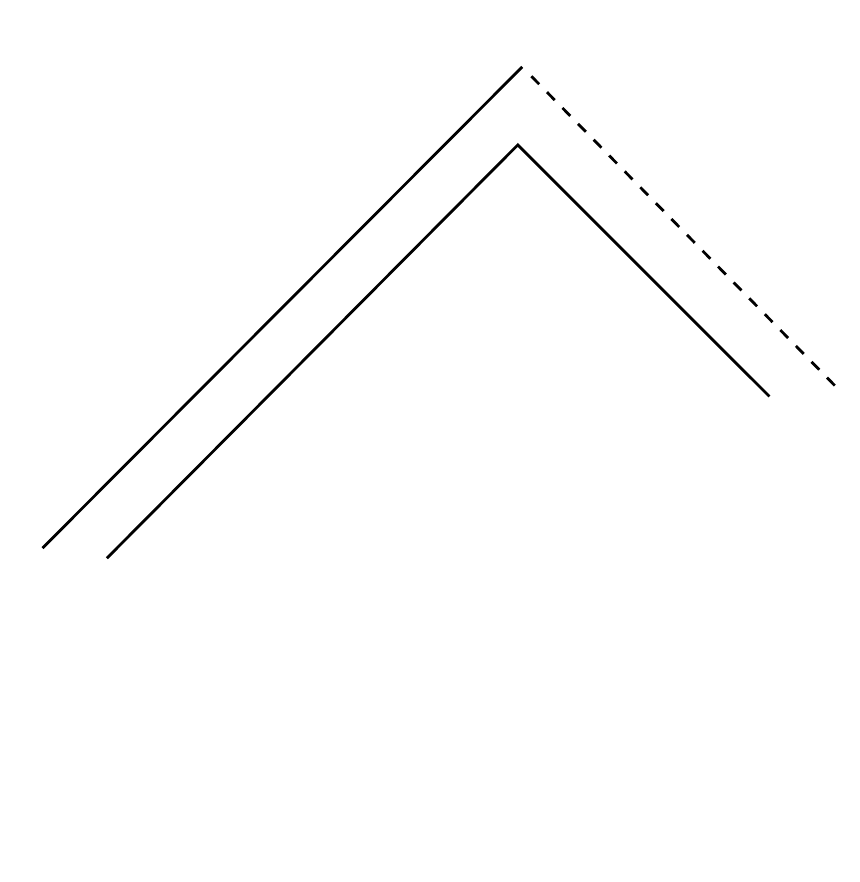}
\caption{The regions $\DcH$ and $\DcI$}\label{truncationfigure}
\end{figure}

More generally we will find useful also the notations, for given $u\ge u_1$, $v\ge v_1$
\begin{align*}
	\DcH(v) 
	&
	:=
	\DcH \cap \{ v_{\Hp} \geq v\},
	&
	\DcI(u)
	&
	:=
	\DcI \cap \{ u_{\I} \geq u\},
	\\
	\DcH(u,v)
	&
	:=
	\DcH \cap \{ u_{\Hp} \geq u \} \cap \{ v_{\Hp} \geq v\},
	\qquad
	&
	\DcI(u,v)
	&
	:=
	\DcI \cap \{ u_{\I} \geq u\} \cap \{ v_{\I} \geq v\} .
\end{align*}

We may now define the truncated cones:\index{teleological $\Hp$ gauge!sets!$\CcH_u(v)$, truncated outgoing null cone for energy estimates}\index{teleological $\Hp$ gauge!sets!$\CbcH_v(u)$, truncated ingoing null cone for energy estimates}\index{teleological $\I$ gauge!sets!$\CcI_u(v)$, truncated outgoing null cone for energy estimates}\index{teleological $\I$ gauge!sets!$\CbcI_v(u)$, truncated ingoing null cone for energy estimates}
For given $u \geq u_1,v\geq v_1$, define the null hypersurfaces of the $\Hp$ gauge truncated at the timelike hypersurface $\mathcal{B}$,
\[
	\CcH_u(v)
	:=
	C^{\Hp}_u(v) \cap \DcH,
	\qquad
	\CbcH_v(u)
	:=
	\Cbar^{\Hp}_v(u) \cap \DcH,
\]
and similarly define the truncated null hypersurfaces in the $\I$ gauge
\[
	\CcI_u(v)
	:=
	C^{\I}_u(v) \cap \DcI,
	\qquad
	\CbcI_v(u)
	:=
	\Cbar^{\I}_v(u) \cap \DcI.
\]
Finally, given a $\tau \geq u_1$ we denote by $v(\tau)$ the number such that $\underline{C}^{\mathcal{H}^+}_{v(\tau)} \cap C_\tau^{\I} \subset \mathcal{B}$.

\subsection{Energies associated to the region $\DcH$}
Recall our conventions from Section~\ref{volformconven} concerning volume forms.

We define the energy fluxes for the truncated cones in $\check{\mathcal{D}}^{\Hp}$ (ending at $\mathcal{B}$, indicated by a check superscript)\index{energies!fluxes!$\check{\mathbb{F}}^\diamond_u \left[W_{\Hp} \right]$, horizon-degenerate energy flux on truncated outgoing null cone}\index{energies!fluxes!$\check{{\mathbb{F}}}_u \left[W_{\Hp} \right] \left(v\right)$,  energy flux on truncated outgoing null cone}\index{energies!fluxes!$\check{\underline{\mathbb{F}}}^\diamond_v \left[W_{\Hp} \right]$, horizon-degenerate energy flux on truncated ingoing null cone}\index{energies!fluxes!$\check{\underline{{\mathbb{F}}}}_v \left[W_{\Hp} \right]$,  energy flux on truncated ingoing null cone}
\begin{align} \label{hre1}
\check{\mathbb{F}}^\diamond_u \left[W_{\Hp} \right] \left(v\right)  &=\int_{\CcH_u(v)}   |\Omega \slashed{\nabla}_4 W_{\Hp}|^2 + \frac{\Omega^2}{r^2} |r\slashed{\nabla}W_{\Hp}|^2 +\frac{\Omega^2}{r^2} | W_{\Hp}|^2  \, ,
\\
\check{{\mathbb{F}}}_u \left[W_{\Hp} \right] \left(v\right) &=\int_{\CcH_u(v)}    |\Omega \slashed{\nabla}_4 W_{\Hp}|^2 + \frac{1}{r^2} |r\slashed{\nabla}W_{\Hp}|^2 +\frac{1}{r^2}| W_{\Hp}|^2  \, , \nonumber
\end{align}
\begin{align} \label{hre2}
\check{\underline{\mathbb{F}}}^\diamond_v \left[W_{\Hp} \right]  &=\int_{\CbcH_v}    |\Omega \slashed{\nabla}_3 W_{\Hp}|^2 + \frac{\Omega^2}{r^2} |r\slashed{\nabla}W_{\Hp}|^2  +  \frac{\Omega^2}{r^2}| W_{\Hp}|^2   \, ,
\\
\check{\underline{{\mathbb{F}}}}_v \left[W_{\Hp} \right]  &=\int_{\CbcH_v}   \Omega^{-2} |\Omega \slashed{\nabla}_3 W_{\Hp}|^2 +  \frac{\Omega^2}{r^2} |r\slashed{\nabla}W_{\Hp}|^2 +\frac{\Omega^2}{r^2} | W_{\Hp}|^2  \, . \nonumber
\end{align}
Note the improvement near the horizon when the $\diamond$ superscript is dropped.

Finally, we define the following auxiliary energy, which contains a $\star$ to indicate it is restricted to the region 
$r \geq 9M_{\rm init}/4$ (i.e.~away from what will be the event horizon):\index{energies!fluxes!$\check{\underline{\mathbb{F}}}^\star_v \left[W_{\Hp} \right]$, energy flux on restricted ingoing null cone}
\begin{align}
\check{\underline{\mathbb{F}}}^\star_v \left[W_{\Hp} \right]  &=\int_{\CbcH_v \cap \{ r\geq 9M_{\rm init}/4\}}    |\Omega \slashed{\nabla}_3 W_{\Hp}|^2 + \frac{\Omega^2}{r^2} |r\slashed{\nabla}W_{\Hp}|^2  +  \frac{\Omega^2}{r^2}| W_{\Hp}|^2 \, .
\end{align}

We furthermore define non-degenerate and degenerate spacetime energies\index{energies!spacetime energies!$\check{\mathbb{I}}^\diamond \left[W_{\Hp}\right] \left(\tau_1,\tau_2\right)$, horizon-degenerate spacetime integral}\index{energies!spacetime energies!$\check{\mathbb{I}} \left[W_{\Hp}\right] \left(\tau_1,\tau_2\right)$, non-degenerate spacetime integral}\index{energies!spacetime energies!$\check{\mathbb{I}}^{\diamond, deg} \left[W_{\Hp}\right] \left(\tau_1,\tau_2\right)$, horizon and trapping-degenerate spacetime integral}\index{energies!spacetime energies!$\check{\mathbb{I}}^{deg} \left[W_{\Hp}\right] \left(\tau_1,\tau_2\right)$, trapping-degenerate spacetime integral}

\begin{align}
\check{\mathbb{I}}^\diamond \left[W_{\Hp}\right] \left(\tau_1,\tau_2\right) &= \int_{\DcH(v(\tau_1)) \setminus \DcH(v(\tau_2))} \frac{ |\Omega \slashed{\nabla}_3 W_{\Hp}|^2}{r^{1+\delta}} + \frac{|\Omega \slashed{\nabla}_4 W_{\Hp}|^2}{r^{1+\delta}} + \frac{ |r \slashed{\nabla} W_{\Hp}|^2}{r^3} + \frac{|W_{\Hp}|^2}{r^3}  \, ,  \\
\check{\mathbb{I}} \left[W_{\Hp}\right] \left(\tau_1,\tau_2\right) &=\check{\mathbb{I}}^\diamond \left[W_{\Hp}\right] \left(\tau_1,\tau_2\right)  +  \int_{\DcH(v(\tau_1)) \setminus \DcH(v(\tau_2))} \frac{1}{r^{1+\delta}} |\Omega^{-1} \slashed{\nabla}_3 W_{\Hp}|^2 ,  \label{ndhe1}
\end{align}
{\small 
\begin{align}
\check{\mathbb{I}}^{\diamond, deg} \left[W_{\Hp}\right] \left(\tau_1,\tau_2\right) &= \int_{\DcH(v(\tau_1)) \setminus \DcH(v(\tau_2))} \frac{ |\slashed{\nabla}_{R^\star} W_{\Hp}|^2}{r^{1+\delta}} + \frac{|W_{\Hp}|^2}{r^3} + \left(1-\frac{3M}{r}\right)^2 \left( \frac{|\slashed{\nabla}_{T}  W_{\Hp}|^2}{r^{1+\delta}} + \frac{ |r\slashed{\nabla} W_{\Hp} |^2}{r^3} \right) \, , \nonumber \\
\check{\mathbb{I}}^{deg} \left[W_{\Hp}\right] \left(\tau_1,\tau_2\right) &=\check{\mathbb{I}}^{\diamond, deg} \left[W_{\Hp}\right] \left(\tau_1,\tau_2\right)  + \int_{\DcH(v(\tau_1)) \setminus \DcH(v(\tau_2))} \frac{1}{r^{1+\delta}}| \Omega^{-1} \slashed{\nabla}_3 W_{\Hp}|^2  \, . \label{dhe1}
\end{align}
}

We also define the auxiliary energy (restricted to $r \geq 9M_{\rm init}/4$)
{\small
\begin{align}
\check{\mathbb{I}}^{\star, deg} \left[W_{\Hp}\right] \left(\tau_1,\tau_2\right) &= \int \frac{ |\slashed{\nabla}_{R^\star} W_{\Hp}|^2}{r^{1+\delta}} + \frac{ |W_{\Hp}|^2}{r^3} + \left(1-\frac{3M}{r}\right)^2 \left( \frac{|\slashed{\nabla}_{T} W_{\Hp}|^2}{r^{1+\delta}} +\frac{ |r\slashed{\nabla} W_{\Hp} |^2}{r^3} \right)  , \nonumber
\end{align}}
\vskip.05pc
\noindent
where the integration is over ${\DcH(v(\tau_1)) \setminus \DcH(v(\tau_2)) \cap \{r \geq 9M_{\rm init}/4\}}$.
For all of the spacetime energies above we agree on the convention that replacing $(\tau_1,\tau_2)$ by just $(\tau_1)$ results in all integrations being over $\DcH(v(\tau_1))$ instead of $\DcH(v(\tau_1)) \setminus \DcH(v(\tau_2))$.

Finally, we define the flux through the timelike hypersurface $\mathcal{B}$ in the horizon region as\index{energies!fluxes!$\mathbb{F}_{\mathcal{B}} \left[W_{\Hp}\right] \left(\tau\right)$, flux through the timelike hypersurface $\mathcal{B}$}
\begin{align}
\mathbb{F}_{\mathcal{B}} \left[W_{\Hp}\right] \left(\tau\right):= \int_{\mathcal{B}(\tau)} |\Omega \slashed{\nabla}_3 W_{\Hp}|^2 + |\Omega \slashed{\nabla}_4 W_{\Hp}|^2 + |r \slashed{\nabla} W_{\Hp}|^2 + |W_{\Hp}|^2 .
\end{align}

\subsection{Energies associated to the region $\DcI$}
We define the following energy fluxes on null cones\index{energies!fluxes!$\check{\mathbb{F}}_u \left[W_{\I} \right]$, energy flux on truncated outgoing null cone}\index{energies!fluxes!$\check{\underline{\mathbb{F}}}_v \left[W_{\I} \right] \left(u\right)$, energy flux on truncated ingoing null cone}\index{energies!fluxes!$\check{\mathbb{F}}^{p}_u \left[W_{\I} \right]$, $p$-weighted energy flux on truncated outgoing null cone}\index{energies!fluxes!$\check{\slashed{\mathbb{F}}}^{p}_u\left[W_{\I} \right]$, $p$-weighted angular energy flux on truncated outgoing null cone}\index{energies!fluxes!$\check{\underline{\slashed{\mathbb{F}}}}^{p}_v \left[W_{\I} \right]$, $p$-weighted angular energy flux on truncated ingoing null cone}
\begin{align}
\check{\mathbb{F}}_u \left[W_{\I} \right]  &=\int_{\CcI_u}  |\Omega \slashed{\nabla}_4 W_{\I}|^2 + \frac{1}{r^2} |r\slashed{\nabla}W_{\I}|^2 + \frac{1}{r^2} | W_{\I}|^2  \, ,
\end{align}
\begin{align}
\check{\underline{\mathbb{F}}}_v \left[W_{\I} \right] \left(u\right) &=\int_{\CbcI_v(u)}  |\Omega \slashed{\nabla}_3 W_{\I}|^2 + \frac{1}{r^2} |r\slashed{\nabla}W_{\I}|^2 + \frac{1}{r^2} | W_{\I}|^2   \, , \nonumber
\end{align}
\begin{align}
\check{\mathbb{F}}^{p}_u \left[W_{\I} \right] =\int_{\CcI_u}    r^p |\Omega \slashed{\nabla}_4 W_{\I}|^2 
\ \ \ \ \textrm{and} \ \ \ \
\check{\slashed{\mathbb{F}}}^{p}_u\left[W_{\I} \right]  =\int_{\CcI_u}    r^{p-2} |r \slashed{\nabla} W_{\I}|^2    \, ,
\end{align}
\begin{align}
\check{\underline{\slashed{\mathbb{F}}}}^{p}_v \left[W_{\I} \right] \left(u\right) &=\int_{\CbcI_v(u)}  \, r^{p-2} |r\slashed{\nabla}W_{\I}|^2   \, . \nonumber
\end{align}

We also define weighted spacetime energies\index{energies!spacetime energies!$\check{\mathbb{I}} \left[W_{\I}\right] \left(\tau_1,\tau_2\right)$, spacetime integral}\index{energies!spacetime energies!$\check{\mathbb{I}}^p \left[W_{\I}\right] \left(\tau_1,\tau_2\right)$, $r^p$ weighted spacetime integral}
\begin{align} \label{infe1}
\check{\mathbb{I}} \left[W_{\I}\right] \left(\tau_1,\tau_2\right) &= \int_{\DcI(\tau_1) \setminus \DcI(\tau_2)} \frac{ \big| \Omega \slashed{\nabla}_4 W_{\I}|^2}{r^{1+
 \delta}} +\frac{ \big| \Omega \slashed{\nabla}_3 W_{\I}|^2}{ r^{1+\delta} }  + \frac{\big| r\slashed{\nabla} W_{\I} \big|^2}{r^{3}} + \frac{|W_{\I}|^2}{r^{3}} \, ,
\end{align}
\begin{align} \label{infe2}
\check{\mathbb{I}}^p \left[W_{\I}\right] \left(\tau_1,\tau_2\right) &= \int_{\DcI(\tau_1) \setminus \DcI(\tau_2)} \left( \frac{ \big| \Omega \slashed{\nabla}_4 W_{\I}|^2}{r^{1+\boldsymbol\delta^p_0}} + \frac{\big| r\slashed{\nabla} W_{\I} \big|^2}{r^{3+\boldsymbol{\delta}^p_2\delta}} + \frac{|W_{\I}|^2}{r^{3+\boldsymbol{\delta}^p_2\delta}} \right) r^p  \, ,
\end{align}
for $p \leq 2$ where $\boldsymbol{\delta}^a_b$ is the Kronecker delta symbol,
and\index{energies!spacetime energies!$\check{\mathbb{I}}^p \left[W_{\I}\right] \left(\tau_1,\tau_2\right)$, $r^p$ weighted angular spacetime integral}
\begin{align}
\check{\slashed{\mathbb{I}}}^p \left[W_{\I}\right] \left(\tau_1,\tau_2\right) &= \int_{\DcI(\tau_1) \setminus \DcI(\tau_2)} r^{p-3} \big| r\slashed{\nabla} W_{\I} \big|^2  \, .
\end{align}

Furthermore, the following auxiliary energies will also be useful. They contain a $\star$ which indicates that these energies are restricted to the region $r \leq 2R$. Specifically
\begin{align}
\check{\mathbb{F}}^{\star}_u \left[W_{\I} \right] &=\int_{\CcI_u\cap\{r\le 2R\}}    \left(  r^p |\Omega \slashed{\nabla}_4 W_{\I}|^2 + \frac{1}{r^2} |r\slashed{\nabla}W_{\I}|^2 + \frac{1}{r^2} | W_{\I}|^2 \right)  \, , 
\end{align}
\begin{align}
\check{\mathbb{I}}^\star \left[W_{\I}\right] \left(\tau_1,\tau_2\right) &= \int_{(\DcI(\tau_1) \setminus \DcI(\tau_2)) \cap \{r\leq 2R\}} \frac{ \big| \Omega \slashed{\nabla}_4 W_{\I}|^2}{r^{1+
 \delta}} +\frac{ \big| \Omega \slashed{\nabla}_3 W_{\I}|^2}{ r^{1+\delta} }  + \frac{\big| r\slashed{\nabla} W_{\I} \big|^2}{r^{3+\delta}} + \frac{|W_{\I}|^2}{r^{3+\delta}}    \, . \nonumber
\end{align}
As before, 
we agree on 
the convention that replacing $(\tau_1,\tau_2)$ by just $(\tau_1)$ in the spacetime energies results in all integrations being over $\DcI(\tau_1)$ instead of $\DcI(\tau_1) \setminus \DcI(\tau_2)$.

Finally, we define the flux through the timelike hypersurface $\mathcal{B}$ in the infinity region:\index{energies!fluxes!$\mathbb{F}_{\mathcal{B}} \left[W_{\I}\right] \left(\tau\right)$, flux through the timelike hypersurface $\mathcal{B}$}
\begin{align} \label{Bfluxnotation}
\mathbb{F}_{\mathcal{B}} \left[W_{\I}\right] \left(\tau\right) := \int_{\mathcal{B}(\tau)} |\Omega \slashed{\nabla}_3 W_{\I}|^2 + |\Omega \slashed{\nabla}_4 W_{\I}|^2 + |r \slashed{\nabla} W_{\I}|^2 + |W_{\I}|^2 .
\end{align}

\section{The  energy estimate and  basic integrated local energy decay} \label{sec:eiledRW}
The next propositions establish good approximate energy conservation and integrated local energy decay estimates for tensorial wave equations
in the region $ \DcI \cup \DcH $ in terms of the energy on the cones $C^{\I}_{u_1}$ and $\underline{C}^{\Hp}_{v(u_1)}$. In the derivation, we will only use the pointwise estimates 
\begin{equation}
\label{useonlyLinfinityestimate}
|r^p \Phi_p | + |\mathfrak{D}(r^p \Phi_p)| \lesssim \varepsilon,
\end{equation} 
valid for the quantities of both guages, in regions $\DcI$ and $\DcH$ respectively, following easily from our general pointwise estimates~\eqref{elinfestimates}. 

\begin{proposition} \label{prop:basecase}
Let $(W_{\Hp},W_{\I})$ be either a pair 
of symmetric traceless $S$-tensors or  of $S$-tangent $1$-forms defined in $\mathcal{D}^{\Hp}$ and $\mathcal{D}^{\I}$ respectively. 
Then, if $(W_{\Hp},W_{\I})$ satisfy tensorial wave equations of Type 1 or Type 2 one has the following estimates:
\begin{enumerate}
\item $T$-Boundedness (non-optimal near horizon): For any $u_1 \leq \tau_1 \leq \tau_2 \leq u_f$,
\begin{align} \label{Tbndb}
\check{\underline{\mathbb{F}}}^\diamond_{v(\tau_2)} [W_{\Hp}] +\check{\mathbb{F}}_{\tau_2} [W_{\I}] + \check{{\mathbb{F}}}^\diamond_{u_f} \left[W_{\Hp} \right] \left(v(\tau_1)\right) + \check{\underline{\mathbb{F}}}_{v_\infty} \left[W_{\I} \right] \left(\tau_1\right)
    \    \lesssim   \ 
 \underline{\check{\mathbb{F}}}^\diamond_{v(\tau_1)} [ W_{\Hp}]
+
\check{\mathbb{F}}_{\tau_1} [W_{\I}]   \nonumber \\    
+\sum_{i=1}^3  \Big|\mathcal{B}_i \left[W_{\Hp}\right] (\tau_1,\tau_2)-  \mathcal{B}_i \left[W_{\I}\right]  (\tau_1,\tau_2)\Big| +\sum_{i=1}^2 \mathcal{G}_i \left[W\right]  (\tau_1,\tau_2)  +  \sum_{i=1}^3 \mathcal{H}_i \left[W\right](\tau_1,\tau_2).
\end{align}

\item  Basic integrated decay (non-optimal near horizon): For any $u_1 \leq \tau \leq u_f$ we have
\begin{align} \label{Tiledb}
\check{\mathbb{I}}^{\diamond, deg} \left[W_{\Hp}\right] \left(v(\tau)\right)  + \check{\mathbb{I}} \left[W_{\I}\right] \left(\tau\right) 
 \  \lesssim \ 
                   \underline{\check{\mathbb{F}}}^\diamond_{v(\tau)} [ W_{\Hp}]
+
\check{\mathbb{F}}_{\tau} [W_{\I}] \nonumber \\
+\sum_{i=1}^3  \Big|\mathcal{B}_i \left[W_{\Hp}\right] (\tau,u_f)-  \mathcal{B}_i \left[W_{\I}\right]   (\tau,u_f)\Big| +\sum_{i=1}^2 \mathcal{G}_i \left[W\right]  (\tau,u_f)  +  \sum_{i=1}^3 \mathcal{H}_i \left[W\right] (\tau,u_f).
\end{align}
\end{enumerate}
Here,  with $\check{\mathcal{D}}^{\Hp} \left(v(\tau_1),v(\tau_2)\right) := \DcH\left(v(\tau_1)\right) \setminus \DcH\left(v(\tau_2)\right)$ and $\check{\mathcal{D}}^{\I} \left(\tau_1,\tau_2\right) := \DcI\left(\tau_1\right) \setminus \DcI\left(\tau_2\right)$, we define
\begin{align} \label{g1t} 
\mathcal{G}_1  \left[W\right] \left(\tau_1,\tau_2\right)= \Big| \int_{\check{\mathcal{D}}^{\Hp} \left(v(\tau_1),v(\tau_2)\right) }  \mathcal{F}^{lin}\left[W_{\Hp}\right] \slashed{\nabla}_{T} W_{\Hp} +  \int_{\check{\mathcal{D}}^{\I} \left(\tau_1,\tau_2\right)}  \mathcal{F}^{lin}\left[W_{\I}\right]  \slashed{\nabla}_{T} W_{\I}\Big| \, , 
\end{align}
\begin{align} \label{g2t}
\mathcal{G}_2  \left[W\right] (\tau_1,\tau_2) = & \Big|   \int_{\check{\mathcal{D}}^{\Hp}\left(v(\tau_1),v(\tau_2)\right)}   \mathcal{F}^{lin}\left[W_{\Hp}\right] \, \slashed{\nabla}_{R^\star} (W_{\Hp}) + \int_{\check{\mathcal{D}}^{\I} \left(\tau_1,\tau_2\right)}  \mathcal{F}^{lin}\left[W_{\I}\right] \, \slashed{\nabla}_{R^\star}W_{\I} \Big|  \\
+& \int_{\check{\mathcal{D}}^{\Hp} \left(v(\tau_1),v(\tau_2)\right) }  |\mathcal{F}^{lin}\left[W_{\Hp}\right]| \, | W_{\Hp} | r^{-1-\delta}+   \int_{\check{\mathcal{D}}^{\I} \left(\tau_1,\tau_2\right)}  |\mathcal{F}^{lin}\left[W_{\I}\right]| \, | W_{\I}| r^{-1-\delta}  \nonumber
\end{align}
and
\begin{align} \label{defh1}
\mathcal{H}_1  \left[W\right]  (\tau_1,\tau_2) =  \int_{\check{\mathcal{D}}^{\Hp} \left(v(\tau_1),v(\tau_2)\right) }  |\mathcal{F}^{nlin}\left[W_{\Hp}\right]| \, |\slashed{\nabla}_{T}  W_{\Hp}| + \int_{\check{\mathcal{D}}^{\I} \left(\tau_1,\tau_2\right)} |\mathcal{F}^{nlin}\left[W_{\I}\right]| \, |\slashed{\nabla}_{T} W_{\I}| \, ,
\end{align}
\begin{align} \label{defh2}
\mathcal{H}_2  \left[W\right]  (\tau_1,\tau_2) = &  \int_{\check{\mathcal{D}}^{\Hp}\left(v(\tau_1),v(\tau_2)\right)}  |\mathcal{F}^{nlin}\left[W_{\Hp}\right]| \, |\slashed{\nabla}_{R^\star} W_{\Hp}| +  \int_{\check{\mathcal{D}}^{\I} \left(\tau_1,\tau_2\right)} |\mathcal{F}^{nlin}\left[W_{\I}\right]| \, |\slashed{\nabla}_{R^\star}W_{\I}|  \\
+&   \int_{\check{\mathcal{D}}^{\Hp} \left(v(\tau_1),v(\tau_2)\right) }  |\mathcal{F}^{nlin}\left[W_{\Hp}\right]| \, |W_{\Hp} | r^{-1-\delta} + \int_{\check{\mathcal{D}}^{\I} \left(\tau_1,\tau_2\right)}  |\mathcal{F}^{nlin}\left[W_{\I}\right]| \, |W_{\I} | r^{-1-\delta} \, , \nonumber
\end{align}
\begin{align} \label{defh3}
\mathcal{H}_3  \left[W\right]  (\tau_1,\tau_2)= & \int_{\check{\mathcal{D}}^{\Hp} \left(v(\tau_1),v(\tau_2)\right) \cap \{ 5M_{init}/2 \leq r \leq 7M_{init}/2\}} \left(| \mathfrak{D} \Phi_p| +   |\Phi_p| \right) \left(|\slashed{\nabla}_{T}  W_{\Hp}|^2 + |\slashed{\nabla} W_{\Hp}|^2\right)
\end{align}
and $\mathcal{B}_i\left[W\right]$ for $i=1,2,3$ are defined in the proof, see (\ref{bTdef}), (\ref{bXdef}) and (\ref{bhdef}).
\end{proposition}

\begin{remark}
To guide the reader's intuition, we briefly discuss the later application of the above estimate to the almost gauge invariant quantities satisfying tensorial wave equations of Type 1 and 2.  The errors $\mathcal{H}_i[W]$ for $i=1,2,3$ are non-linear error terms and will be handled schematically using Section \ref{nonlinearestforwave}. The error terms $\mathcal{B}_i[W]$ are boundary terms on the hypersurface $\mathcal{B}$. They cancel in linear theory and are hence at least cubic. They will be estimated schematically by relating the quantities in the different gauges; see Proposition \ref{thm:cancelT}. No special structure is required. The ``linear'' error terms $\mathcal{G}_1[W]$ and $\mathcal{G}_2[W]$, on the other hand, will be estimated explicitly (which also explains the position of the absolute value bars allowing further integration by parts). Their treatment will depend on a hierarchical structure in the system of tensorial wave equations that we estimate. Note in particular that in
the case of $\Psi$, $\underline{\Psi}$, it follows that the $\mathcal{G}_i[W]$ actually vanish identically.
\end{remark}

\begin{remark}
The terminology and notations: $T$-boundedness, $T$-identity, Morawetz $X$-identity, Lagrangian $h$-identity
that will appear below are used to facilitate comparison with standard notation for the analogous
constructions for the scalar wave equation (see e.g.~\cite{Mihalisnotes}),
where these identities are generated by currents associated to the Lagrangian contracted with
vector fields $T$, $X$, \ldots and auxiliary functions $h$, etc.
Note that in the present work, although we use $X$ as a label, we shall not actually define a vector field $X$!
\end{remark}

\begin{proof}[Proof of Proposition \ref{prop:basecase}]
Below $V$ denotes either $V_I = 2\frac{\Omega^2}{r^2} - \frac{6M\Omega^2}{r^3}$ (Type 1) or $V_{II} = \frac{6M\Omega^2}{r^3}$ (Type 2). Note $V_I \geq \frac{\Omega^2}{2r^2}$ for $r \geq 2M$. 
\vskip1pc
{\bf Step 1: Derivation of multiplier identities.} 
Before we start the proof proper we derive the following multiplier identities.\footnote{The contraction of indices is written below for the case of symmetric traceless tensors. The case of $1$-forms requires trivial modifications and is not spelled out explicitly. Finally, we only explicitly write down the indices involved in the contraction (to a scalar) in case there is ambiguity, e.g.~we write $\slashed{\nabla}_{T} W \slashed{\nabla}_{R^\star} W$ to denote $(\slashed{\nabla}_{T} W)_{AB} (\slashed{\nabla}_{R^\star} W)^{AB}$.} Contracting the tensorial wave equation with $\slashed{\nabla}_{T} W$ produces identities in $\mathcal{D}^{\Hp}$ and $\mathcal{D}^{\I}$ respectively of the form
{\small
\begin{align}
\slashed{\nabla}_{T}   \left( \mathfrak{f}^T_T \left[W_{\mathcal{H}^+}\right]\right)  + \slashed{\nabla}_{R^\star}  \left( \mathfrak{f}^T_{R^\star} \left[W_{\mathcal{H}^+}\right]\right) + \slashed{\nabla}^A \mathfrak{f}^T_A  \left[W_{\Hp}\right]  &= \mathfrak{F}^T  \left[W_{\Hp}\right] + \mathcal{F}^{lin} \left[W_{\Hp}\right] \slashed{\nabla}_{T}  W_{\Hp} + \mathcal{F}^{nlin} \left[W_{\Hp}\right] \slashed{\nabla}_{T}  W_{\Hp}  \, ,  \nonumber \\ 
\slashed{\nabla}_{T}  \left( \mathfrak{f}^T_T \left[W_{\mathcal{I}^+}\right]\right)  + \slashed{\nabla}_{R^\star}  \left( \mathfrak{f}^T_{R^\star} \left[W_{\mathcal{I}^+}\right]\right)  + \slashed{\nabla}^A \mathfrak{f}^T_A  \left[W_{\I}\right] &= \mathfrak{F}^T  \left[W_{\I}\right] + \mathcal{F}^{lin} \left[W_{\I}\right] \slashed{\nabla}_{T}  W_{\Hp} + \mathcal{F}^{nlin} \left[W_{\I}\right] \slashed{\nabla}_{T}  W_{\I}\nonumber \, , 
\end{align} 
}
with\index{multiplier currents!$T$-identity!$\mathfrak{f}^T_T[W]$}\index{multiplier currents!$T$-identity!$\mathfrak{f}^T_{R^\star}[W]$}\index{multiplier currents!$T$-identity!$\mathfrak{f}^T_A[W]$}\index{multiplier currents!$T$-identity!$\mathfrak{F}^T \left[W\right]$}
\begin{align} \label{Tidke}
\mathfrak{f}^T_T \left[W\right] &= \frac{1}{2} \left( |\slashed{\nabla}_T W|^2 + |\slashed{\nabla}_{R^\star} W|^2 + \frac{\Omega^2}{r^2}\left( r^2 |  \slashed{\nabla} W|^2 \right) +V |W|^2 \right)  \, , \nonumber \\
\mathfrak{f}^T_{R^\star} \left[W\right] &= -\slashed{\nabla}_{R^\star} W \slashed{\nabla}_T  W  \, , \nonumber \\
\mathfrak{f}^T_A \left[W\right] &= -\frac{\Omega^2}{r^2} \slashed{\nabla}_A W_{BC}  \slashed{\nabla}_T  W^{BC} \, ,  \nonumber \\
\mathfrak{F}^T \left[W\right] &= +\slashed{\nabla}_{R^\star} W \frac{1}{2} F_{34} \left[W\right]+ \left(K \cdot T\left(\Omega^2 - \Omega_\circ^2\right)+ \Omega^2 T \left(K-K_\circ\right)\right)|W|^2 +\frac{1}{2} \left(T \left(V-V_\circ \right)\right) |W|^2 \, , 
\nonumber \\
&\ \ \  -\Omega^2 \left(\eta + \underline{\eta}\right)^A \slashed{\nabla}_A W_{BC} \slashed{\nabla}_T  W^{BC} + \frac{1}{2} \left(T \left(\Omega^2 - \Omega_\circ^2\right)\right) | \slashed{\nabla} W|^2 +\frac{\Omega^2}{r^2} r \slashed{\nabla}W \left[r \slashed{\nabla}, \slashed{\nabla}_T  \right]W  \, ,
\end{align}
which we shall collectively refer to as the {\bf $T$-identity}.

Contracting the tensorial wave equations with $f \slashed{\nabla}_{R^\star} W + \frac{1}{2} f^\prime W$ where $f$ is a bounded function of $r$ (i.e.~$r=r_{\Hp}$ in $\mathcal{D}^{\Hp}$ and $r=r_{\I}$ in $\mathcal{D}^{\I}$) and a prime denoting a derivative with respect to $R^\star$) produces identities with\index{multiplier currents!Morawetz identity!$\mathfrak{f}^X_T[W]$}\index{multiplier currents!Morawetz identity!$\mathfrak{f}^X_{R^\star}[W]$}\index{multiplier currents!Morawetz identity!$\mathfrak{f}^X_A[W]$}\index{multiplier currents!Morawetz identity!$\mathfrak{f}^X_{bulk} \left[W\right]$}\index{multiplier currents!Morawetz identity!$\mathfrak{F}^X \left[W\right]$}
\begin{align}
&\slashed{\nabla}_T   \left( \mathfrak{f}^X_T \left[W_{\mathcal{H}^+}\right]\right)  + \slashed{\nabla}_{R^\star}  \left( \mathfrak{f}^X_{R^\star} \left[W_{\mathcal{H}^+}\right]\right) + \slashed{\nabla}^A \mathfrak{f}^X_A  \left[W_{\Hp}\right]  + \mathfrak{f}^X_{bulk} \left[W_{\mathcal{H}_+}\right] \nonumber \\
= &\mathfrak{F}^X  \left[W_{\Hp}\right]+ \mathcal{F}^{lin} \left[W_{\Hp}\right]\left( f \slashed{\nabla}_{R^\star} W_{\Hp} + \frac{1}{2} f^\prime W_{\Hp}\right) + \mathcal{F}^{nlin} \left[W_{\Hp}\right] \left( f \slashed{\nabla}_{R^\star} W_{\Hp} + \frac{1}{2} f^\prime W_{\Hp}\right)  \ \ \ \textrm{in $\DRH$}  \, , \nonumber
\end{align}
\begin{align}
&\slashed{\nabla}_T   \left( \mathfrak{f}^X_T \left[W_{\mathcal{I}^+}\right]\right)  + \slashed{\nabla}_{R^\star}  \left( \mathfrak{f}^X_{R^\star} \left[W_{\mathcal{I}^+}\right]\right)  + \slashed{\nabla}^A \mathfrak{f}^X_A  \left[W_{\I}\right] +\mathfrak{f}^X_{bulk} \left[W_{\mathcal{I}^+}\right]  \nonumber \\
= &\mathfrak{F}^X  \left[W_{\I}\right]+ \mathcal{F}^{lin} \left[W_{\I}\right]\left( f \slashed{\nabla}_{R^\star} W_{\I} + \frac{1}{2} f^\prime W_{\I}\right) + \mathcal{F}^{nlin} \left[W_{\I}\right] \left( f \slashed{\nabla}_{R^\star} W_{\I} + \frac{1}{2} f^\prime W_{\I}\right)  \ \ \ \textrm{in $\DRI$} \, ,  \nonumber
\end{align}
with
\begin{align}
\mathfrak{f}^X_T \left[W\right] &= f \slashed{\nabla}_T  W \cdot \slashed{\nabla}_{R^\star} W + \frac{1}{2} f^\prime \slashed{\nabla}_T W \cdot W \, , \nonumber \\
\mathfrak{f}^X_{R^\star} \left[W\right] &= -\frac{1}{2} f \left( |\slashed{\nabla}_T W|^2 + |\slashed{\nabla}_{R^\star} W|^2- \frac{\Omega^2}{r^2} |r \slashed{\nabla} W|^2 \right) + 2 \frac{\Omega^2}{r^2} f |W|^2 + \frac{1}{2} f V |W|^2 - \frac{1}{2}f^\prime \slashed{\nabla}_{R^\star} W \cdot W + \frac{1}{4} f^{\prime \prime} |W|^2\, , \nonumber \\
\mathfrak{f}^X_A \left[W\right] &=  -\Omega^2 \slashed{\nabla}_A W f \slashed{\nabla}_{R^\star} W - \frac{1}{2} \Omega^2 f^\prime \slashed{\nabla}_A W_{BC} W^{BC} \, ,  \nonumber \\
\mathfrak{f}^X_{bulk} \left[W\right]  &=f^\prime |\slashed{\nabla}_{R^\star} W|^2 + r^2 | \slashed{\nabla} W|^2 \frac{\Omega^2}{r^3} \left(1-\frac{3M}{r}\right) f + |W|^2 \left(-f R^\star \left( \frac{V}{2}\right) - \frac{1}{4} f^{\prime \prime \prime} \right) \, , \nonumber \\
\mathfrak{F}^X \left[W\right] &= -\frac{1}{2} \frac{\Omega^2}{r^2} r \slashed{\nabla} W \left[r \slashed{\nabla}, \slashed{\nabla}_{R^\star} \right] W - \Omega^2 \left(\eta + \underline{\eta}\right)^A f \slashed{\nabla}_A W_{BC} \slashed{\nabla}_{R^\star} W^{BC} - 2\frac{\Omega^2}{r^2} \left(r^2 K - 1\right) f W \slashed{\nabla}_{R^\star} W  \nonumber \\
& \  \ \ +\frac{1}{2} (Tf^\prime) W \cdot \slashed{\nabla}_T W - \frac{1}{2} \Omega^2 \left(\eta + \underline{\eta}\right) f^\prime \slashed{\nabla}W \cdot W -\frac{1}{2}\Omega^2 \slashed{\nabla} W \slashed{\nabla} f^\prime \cdot W  \nonumber \\
& \ \ \ + \frac{\Omega^2}{r^2} f r^2 |\slashed{\nabla} W|^2 \left(\omega - \omega_\circ + \underline{\omega} - \underline{\omega}_\circ\right) +|W|^2 \frac{\Omega^2}{r^2} f^\prime \left(K r^2-1\right) \, ,
\end{align}
which we shall collectively refer to as the {\bf Morawetz $X$-identity}.

Finally, contracting the tensorial wave equations with $\frac{1}{2}h W$  produces identities of the form\index{multiplier currents!Lagrangian identity!$\mathfrak{f}^h_T[W]$}\index{multiplier currents!Lagrangian identity!$\mathfrak{f}^h_{R^\star}[W]$}\index{multiplier currents!Lagrangian identity!$\mathfrak{f}^h_A[W]$}\index{multiplier currents!Lagrangian identity!$\mathfrak{f}^h_{bulk}[W]$}\index{multiplier currents!Lagrangian identity!$\mathfrak{F}^h_T[W]$}
\begin{align} \label{lagrangianid}
& \slashed{\nabla}_T   \left( \mathfrak{f}^h_T \left[W_{\mathcal{H}^+}\right]\right)  + \slashed{\nabla}_{R^\star}  \left( \mathfrak{f}^h_{R^\star} \left[W_{\mathcal{H}^+}\right]\right) + \slashed{\nabla}^A \mathfrak{f}^h_A  \left[W_{\Hp}\right]  + \mathfrak{f}^h_{bulk} \left[W_{\mathcal{H}_+}\right] \nonumber \\
= & \ \mathfrak{F}^h  \left[W_{\Hp}\right] + \mathcal{F}^{lin}\left[W_{\Hp}\right] \frac{1}{2} h W_{\Hp} + \mathcal{F}^{nlin} \left[W_{\Hp}\right] \frac{1}{2} h W_{\Hp} \ \ \ \textrm{in $\DRH$}  \, ,
\end{align}
\begin{align}
& \slashed{\nabla}_T   \left( \mathfrak{f}^h_T \left[W_{\mathcal{I}^+}\right]\right)  +\slashed{\nabla}_{ R^\star}  \left( \mathfrak{f}^h_{R^\star} \left[W_{\mathcal{I}^+}\right]\right)  + \slashed{\nabla}^A \mathfrak{f}^h_A  \left[W_{\I}\right] +\mathfrak{f}^h_{bulk} \left[W_{\mathcal{I}^+}\right] \nonumber \\
= & \ \mathfrak{F}^h  \left[W_{\I}\right]+ \mathcal{F}^{lin}\left[W_{\I}\right] \frac{1}{2} h W_{\I} + \mathcal{F}^{nlin} \left[W_{\I}\right] \frac{1}{2} h W_{\I} \ \ \ \textrm{in $\DRI$} \, ,
\end{align}
with
\begin{align} 
\mathfrak{f}^h_T \left[W\right] &=  \frac{1}{2} h \slashed{\nabla}_T  W \cdot W \, , \nonumber \\
\mathfrak{f}^h_{R^\star} \left[W\right] &=  - \frac{1}{2}h \slashed{\nabla}_{R^\star} W \cdot W + \Omega^2 K h |W|^2 + \frac{1}{4} h^{\prime} |W|^2 \, ,   \nonumber  \\
\mathfrak{f}^h_A \left[W\right] &=  - \frac{1}{2} \Omega^2 h \slashed{\nabla}_A W_{BC} \cdot W^{BC} \, , \nonumber \\
\mathfrak{f}^h_{bulk} \left[W\right]  &=\frac{1}{2} h |\slashed{\nabla}_{R^\star} W|^2 - \frac{1}{2} h |\slashed{\nabla}_T  W|^2 + \frac{1}{2} \frac{\Omega^2}{r^2} h r^2 |\slashed{\nabla} W|^2 - \frac{1}{4} h^{\prime \prime} |W|^2 + \frac{1}{2} V h |W|^2 \, , \nonumber \\ 
\mathfrak{F}^h \left[W\right] &= \frac{1}{2} (Th) W \cdot \slashed{\nabla}_T  W - \frac{1}{2} \Omega^2 \left(\eta + \underline{\eta}\right) h \slashed{\nabla}W \cdot W -\frac{1}{2}\Omega^2 \slashed{\nabla} W \slashed{\nabla} h \cdot W \, ,
\end{align}
which we shall collectively refer to as the {\bf Lagrangian $h$-identity}.

\vskip1pc
{\bf Step 2. General strategy and converting projected covariant derivatives into spacetime divergences to apply Stokes' theorem.} With $u_1 \leq \tau \leq u_f$ fixed, we will apply for arbitrary $\tau < u \leq u_f$ successively 
\begin{itemize}
\item The $T$-identity,
\item The Morawetz $X$-identity with $f= \left(1-\frac{3M}{r}\right) \left(1+\frac{M}{r}\right)$,
\item The Lagrangian $h$-identity with $h= -\left(1-\frac{3M}{r}\right)^2 \frac{1}{r^2}$
\item The Morawetz $X$-identity with $f= \left(1-r^{-\delta}\right)\xi$ where $\xi$ is a smooth cut-off function equal to $1$ for $r \geq 2R$ and vanishing for $r \leq 2R_{-1}$
\item The Lagrangian $h$-identity with $h= r^{-1-\delta}\xi$ where $\xi$ is a smooth cut-off function equal to $1$ for $r \geq 2R$ and vanishing for $r \leq 2R_{-1}$.
\end{itemize}
By this we mean multiplying each identity with $\frac{\sqrt{\slashed{g}}}{r^2}$ (recall $r=r_{\I}$ in $\DcI$ and $r=r_{\Hp}$ in the region inside $\DcH$) and integrating 
\begin{itemize}
\item the relevant identity for $W_{\I}$ in the region $\check{\mathcal{D}}^{\I} \left(\tau_1,\tau_2\right)  = \DcI\left(\tau_1\right) \setminus \DcI\left(\tau_2\right)$
\item the relevant identity for $W_{\Hp}$ identity in the region $\check{\mathcal{D}}^{\Hp} \left(v(\tau_1),v(\tau_2)\right) = \DcH(v(\tau_1)) \setminus \DcH(v(\tau_2))$ 
\end{itemize}
and finally \emph{summing} the two identities. 

We then apply Stokes' theorem after using the following Lemma, which converts the projected covariant derivatives appearing in the identities into spacetime divergences.
\begin{lemma} \label{lem:transformder}
For $\mathfrak{f}$ a spacetime function and $\slashed{\mathfrak{f}}$ an $S$-vectorfield we have in $\mathcal{D}^{\Hp}$ the formulae
\begin{align}
\Omega \slashed{\nabla}_3 \mathfrak{f} =  \Omega^2 r^2 \nabla_a \left(\frac{1}{\Omega^2 r^2}(\Omega e_3)^a \mathfrak{f} \right) -  \left( \Omega tr \underline{\chi} - (\Omega tr \underline{\chi})_\circ\right) \mathfrak{f} \, ,
\end{align}
\begin{align}
\Omega \slashed{\nabla}_4 \mathfrak{f} =  \Omega^2 r^2 \nabla_a \left(\frac{1}{\Omega^2 r^2}(\Omega e_4)^a \mathfrak{f} \right) -  \left( \Omega tr {\chi} - (\Omega tr {\chi})_\circ  - b^A \partial_A \log \sqrt{\slashed{g}} \right) \mathfrak{f} \, ,
\end{align}
\begin{align}
\slashed{\nabla}_A \slashed{\mathfrak{f}}^A   = \Omega^2 r^2 \nabla_a \left(\frac{\slashed{\mathfrak{f}}^A}{\Omega^2 r^2} \left(e_A\right)^a\right)   \, .
\end{align}
Similarly, in $\mathcal{D}^{\I}$ (where we recall the metric is of the form (\ref{doublenulllongforminterchanged}) and hence (\ref{altnullframedef}) holds)
\begin{align}
\Omega \slashed{\nabla}_4 \mathfrak{f} =  \Omega^2 r^2 \nabla_a \left(\frac{1}{\Omega^2 r^2}(\Omega e_4)^a \mathfrak{f} \right) -  \left( \Omega tr {\chi} - (\Omega tr {\chi})_\circ\right) \mathfrak{f} \, ,
\end{align}
\begin{align}
\Omega \slashed{\nabla}_3 \mathfrak{f} =  \Omega^2 r^2 \nabla_a \left(\frac{1}{\Omega^2 r^2}(\Omega e_3)^a \mathfrak{f} \right) -  \left( \Omega tr \underline{\chi} - (\Omega tr \underline{\chi})_\circ  - b^A \partial_A \log \sqrt{\slashed{g}} \right) \mathfrak{f} \, ,
\end{align}
\begin{align}
\slashed{\nabla}_A \slashed{\mathfrak{f}}^A   = \Omega^2 r^2 \nabla_a \left(\frac{\slashed{\mathfrak{f}}^A}{\Omega^2 r^2} \left(e_A\right)^a\right)  \, .
\end{align}
\end{lemma}
\begin{proof}
We carry out the proof for the $\mathcal{H}^+$ gauge, the one for the $\mathcal{I}^+$ gauge being entirely analogous. 

We note ($\sqrt{g} = 2\Omega^2 \sqrt{\slashed{g}}$ being the spacetime volume form)
\begin{align}
\Omega \slashed{\nabla}_3 \mathfrak{f} =  \sqrt{g} \nabla_a \left(\frac{1}{\sqrt{g}}(\Omega e_3)^a \mathfrak{f} \right)
\end{align}
and stick in a $\frac{\sqrt{\slashed{g}}}{r^2}$ in the bracket. Analogously for the $4$-direction where we start from
\begin{align}
\Omega \slashed{\nabla}_4 \mathfrak{f} =  \sqrt{g} \nabla_a \left(\frac{1}{\sqrt{g}}(\Omega e_4)^a \mathfrak{f} \right) - \mathfrak{f} \partial_A b^A = \sqrt{g} \nabla_a \left(\frac{1}{\sqrt{g}}(\Omega e_4)^a \mathfrak{f} \right) - \mathfrak{f} \slashed{div} b + \mathfrak{f} b^A \partial_A \log \sqrt{\slashed{g}}
\end{align}
and repeat the computation sticking in a $\frac{\sqrt{\slashed{g}}}{r^2}$ in the bracket. For the last part we compute
\begin{align}
\nabla_a \left(\slashed{\mathfrak{f}}^A \left(e_A\right)^a\right) &= \frac{1}{\sqrt{g}} \partial_a \left(\sqrt{g} \slashed{\mathfrak{f}}^A \left(e_A\right)^a\right) = \frac{1}{\sqrt{\slashed{g}}\Omega^2} \partial_a \left(\sqrt{\slashed{g}} \Omega^2 \slashed{\mathfrak{f}}^A \left(e_A\right)^a\right) =  \frac{1}{\sqrt{\slashed{g}}\Omega^2} \partial_B \left(\sqrt{\slashed{g}} \Omega^2 \slashed{\mathfrak{f}}^B\right) \nonumber \\
&= \frac{1}{\sqrt{\slashed{g}}} \partial_A \left(\sqrt{\slashed{g}}\slashed{\mathfrak{f}}^A\right) + \Omega^2 \left(\eta+\underline{\eta}\right)_B \slashed{\mathfrak{f}}^B = \slashed{\nabla}_A \slashed{\mathfrak{f}}^A + \Omega^2 \left(\eta+\underline{\eta}\right)_B \slashed{\mathfrak{f}}^B  \, .
\end{align}
\end{proof}
Note that indeed Stokes' theorem can be applied to each of the first terms on the right in the Lemma after multiplication by $\frac{\sqrt{\slashed{g}}}{r^2}$ and integration with respect to the volume for $du dv d\theta^1 d\theta^2$. Moreover, note that the second term on the right produces a cubic error term if $\mathfrak{f}$ is quadratic. In summary this produces identities relating various bulk and flux terms as well as a term on the common boundary $\mathcal{B}$.

\vskip1pc
{\bf Step 3: General error estimates.}
We already note the following estimate for the \underline{non-linear error terms} 
arising from the various $\mathfrak{F}$ (recall the definition (\ref{spherevolform})):
\begin{align} \label{hic}
 & \int_{\DcH\left(v(\tau_1),v(\tau_2)\right)} du dv d\theta \left( |\mathfrak{F}^T \left[W_{\Hp}\right]| + |\mathfrak{F}^X \left[W_{\Hp}\right]| + |\mathfrak{F}^h \left[W_{\Hp}\right]| \right) \nonumber \\
+& \int_{\DcI\left(\tau_1,\tau_2\right)} du dv d\theta \left( |\mathfrak{F}^T \left[W_{\I}\right]| + |\mathfrak{F}^X \left[W_{\I}\right]| + |\mathfrak{F}^h \left[W_{\I}\right]| \right) \nonumber \\
& \lesssim  \mathcal{H}_3 \left[W\right] \left(\tau_1,\tau_2\right)+ \varepsilon \left(\mathbb{I}^{\diamond, deg} \left[W_{\Hp}\right] \left(v(\tau_1), v\left(\tau_2\right)\right)  + \mathbb{I} \left[W_{\I}\right] \left(\tau_1,\tau_2\right)\right) \, ,
\end{align}
which follows easily from inserting the pointwise bounds~\eqref{useonlyLinfinityestimate} on the Ricci coefficients (use Lemma \ref{lem:commutation} for the commutator terms before inserting the $L^\infty$-bounds). The non-linear spacetime terms arising after integration from applying the formulae of Lemma \ref{lem:transformder} are also easily seen to be controlled by the right hand side of (\ref{hic}).

\vskip1pc

{\bf  Step 4: Applying the $T$-identity.} Recall $u_1\leq \tau_1 \leq \tau_2 \leq u_f$ has been fixed. Integrating the $T$ identity as described above in the $V$-shaped region produces (after summing) the estimate
\begin{align} \label{tbu}
  \check{\mathbb{F}}_{\tau_2} [W_{\I}] 
+\check{\underline{\mathbb{F}}}^\diamond_{v(\tau_2)} [W_{\Hp}]
                + \check{{\mathbb{F}}}^\diamond_{u_f} \left[W_{\Hp} \right] \left(v(\tau_1)\right) + \check{\underline{\mathbb{F}}}_{v_\infty} \left[W_{\I} \right] \left(\tau_1\right) 
                \lesssim
 \ \  \underline{\check{\mathbb{F}}}^\diamond_{v(\tau_1)} [ W_{\Hp}]  \lesssim
+
\check{\mathbb{F}}_{\tau} [W_{\I}] \nonumber \\
 + \mathcal{G}_1 \left[W\right] \left(\tau_1,\tau_2\right)  + \mathcal{H}_1 \left[W\right] \left(\tau_1,\tau_2\right) + \mathcal{H}_3 \left[W\right] \left(\tau_1,\tau_2\right) + |\mathcal{B}_1 \left[W_{\Hp}\right]\left(\tau_1,\tau_2\right)- \mathcal{B}_1 \left[W_{\I} \right]\left(\tau_1,\tau_2\right)| \nonumber \\
 \ \ + \varepsilon \left(\mathbb{I}^{\diamond, deg} \left[W_{\Hp}\right] \left(v(\tau_1), v\left(\tau_2\right)\right)  + \mathbb{I} \left[W_{\I}\right] \left(\tau_1, \tau_2\right)\right) \, ,
\end{align}
provided we define
\begin{equation} \label{bTdef}
\begin{split}
\mathcal{B}_1 \left[W_{\I}\right] \left(\tau_1,\tau_2\right) &:= \int_{\mathcal{B}(\tau_1,\tau_2)}  \left( \mathfrak{f}_T^T \left[W_{\I}\right] g(T^{\I}, n) + \mathfrak{f}_{R^\star}^T  \left[W_{\I}\right]  g({R^\star}^{\I},n) + \mathfrak{f}_A^T  \left[W_{\I}\right] g(e_A^{\I},n) \right) \\
\mathcal{B}_1 \left[W_{\Hp}\right]  \left(\tau_1,\tau_2\right) &:= \int_{\mathcal{B}(\tau_1,\tau_2)}
 \left( \mathfrak{f}_T^T \left[W_{\Hp}\right] g(T^{\Hp}, n) + \mathfrak{f}_{R^\star}^T  \left[W_{\Hp}\right]  g({R^\star}^{\Hp},n) + f_A^T  \left[W_{\Hp}\right] g(e_A^{\Hp},n) \right) \, .
 \end{split}
\end{equation}
The estimate (\ref{tbu}) arises from the respective boundary terms induced on the hypersurface $\mathcal{B}$, where the integration is with respect to the measure induced on $\mathcal{B}$.

Below we will often add a large multiple (depending only on $M_{\rm init}$) of the estimate~\eqref{tbu}.

\vskip1pc
{\bf Step 5: Applying the Morawetz $X$-identity.} We claim that choosing $f=\left(1-\frac{3M}{r}\right)\left(1+\frac{M}{r}\right)$ in the Morawetz multiplier identity produces 
\begin{align} \label{moty}
 \int_{\DcH\left(v(\tau_1),v(\tau_2)\right)} \left( \frac{1}{r^2} | \slashed{\nabla}_{R^\star} W_{\Hp}|^2 + \frac{\left(1-\frac{3M}{r}\right)^2}{r^3} |\slashed{\nabla} W_{\Hp}|^2 + \frac{1}{r^3}|W_{\Hp}|^2 \right) \nonumber \\
  +  \int_{\DcI (\tau_1,\tau_2)} \left( \frac{1}{r^2} | \slashed{\nabla}_{R^\star} W_{\I}|^2 + \frac{1}{r^3} |\slashed{\nabla} W_{\I}|^2 + \frac{1}{r^3}|W_{\I}|^2 \right)
  \lesssim
                 \underline{\check{\mathbb{F}}}^\diamond_{v(\tau_1)} [ W_{\Hp}]
+
\check{\mathbb{F}}_{\tau_1} [W_{\I}]  \nonumber \\ +  \mathcal{G}_1 \left[W\right] \left(\tau_1,\tau_2\right)+ \mathcal{G}_2 \left[W\right]  \left(\tau_1,\tau_2\right) +  \sum_{i=1}^3 \mathcal{H}_i \left[W\right]\left(\tau_1,\tau_2\right) + \Big|\sum_{i=1}^2  \mathcal{B}_i \left[W_{\Hp}\right]\left(\tau_1,\tau_2\right) -  \mathcal{B}_i \left[W_{\I}\right]\left(\tau_1,\tau_2\right)\Big| \, ,
\end{align}
provided we define
\begin{align} \label{bXdef}
\mathcal{B}_2 \left[W_{\I}\right] &:= \int_{\mathcal{B}(\tau_1,\tau_2)}  \left( \mathfrak{f}_T^X \left[W_{\I}\right] g(T^{\I}, n) + \mathfrak{f}_{R^\star}^X  \left[W_{\I}\right]  g({R^\star}^{\I},n) + \mathfrak{f}_A^X  \left[W_{\I}\right] g(e_A^{\I},n) \right) \nonumber \\
\mathcal{B}_2 \left[W_{\Hp}\right] &:= \int_{\mathcal{B}(\tau_1,\tau_2)}  
 \left( \mathfrak{f}_T^X \left[W_{\Hp}\right] g(T^{\Hp}, n) + \mathfrak{f}_{R^\star}^X  \left[W_{\Hp}\right]  g({R^\star}^{\Hp},n) + \mathfrak{f}_A^X  \left[W_{\Hp}\right] g(e_A^{\Hp},n) \right) \, ,
\end{align}
which arises from the contribution of the boundary terms on the hypersurface $\mathcal{B}$.
Here we have used that 
\begin{itemize}
\item we can control all boundary terms that appear on null hypersurfaces by adding a large multiple of the estimate (\ref{tbu}). 
\item in the region $\DcI\left(\tau_1, \tau_2\right)$ we can use that we have 
\[
-f  \left( \frac{V}{2}\right)^\prime - \frac{1}{4} f^{\prime \prime \prime}  \geq \frac{1-\frac{2M}{r}}{r^3} 
\]
for both $V=V_I$ and $V=V_{II}$.
\item in the region $\DcH(v(\tau_1),v(\tau_2))$ we have for $V_{II}$
\[
-f \left( \frac{V_{II}}{2}\right)^\prime - \frac{1}{4} f^{\prime \prime \prime}  \geq \frac{1-\frac{2M}{r}}{r^3} \, , 
\]
while for $V_I$ we have the above inequality only in $r_{\Hp} \geq 6M_{\rm init}$. However, for $r_{\Hp}\leq 6M_{\rm init}$ we can use in addition the Poincar\'e inequality\footnote{The reason we apply the Poincare inequality only in $r_{\Hp} \leq 6M_{\rm init}$ is to avoid truncated spheres near the hypersurface $\mathcal{B}$.} of Proposition~\ref{prop:Poincare} for a symmetric traceless $S$-tensor $W$,
\begin{align} \label{poincarehere}
\int du dv d\theta r^2 | \slashed{\nabla} W|^2 \frac{\Omega^2}{r^3} \left(1-\frac{3M}{r}\right) f \geq \frac{3}{4} \int du dv d\theta   \frac{\Omega^2}{r^3} \left(1-\frac{3M}{r}\right)^2 \left(1+\frac{M}{r}\right)|W|^2 \, ,
\end{align}
where the integration is over $\DcH(v(\tau_1),v(\tau_2))  \cap \{r_{\Hp} \leq 6M_{\rm init}\}$ on both sides, and the fact that 
\[
-f  \left( \frac{V_I}{2}\right)^\prime - \frac{1}{4} f^{\prime \prime \prime} + \frac{3}{4} \frac{\Omega^2}{r^3} \left(1-\frac{3M}{r}\right)^2 \left(1+\frac{M}{r}\right) \geq \frac{1}{64} \frac{1-\frac{2M}{r}}{r^3} 
\]
holds in $\DcH(v(\tau_1),(\tau_2))  \cap \{r_{\Hp} \leq 6M_{\rm init}\}$.
\end{itemize}

\vskip1pc
{\bf Step 6: Applying the Lagrangian $h$-identity.} To obtain the missing $\slashed{\nabla}_T$ derivatives in (\ref{moty}) we apply the Lagrangian identity with $h=-\left(1-\frac{3M}{r}\right)^2 \frac{1}{r^2}$ and use that we can add a large multiple of the identity (\ref{moty}) and (\ref{tbu}) to produce the identity
\begin{align} \label{iutr}
 \int_{\DcH(v(\tau_1),v(\tau_2))} \left( \frac{1}{r^2} | \slashed{\nabla}_{R^\star} W_{\Hp}|^2 + \frac{\left(1-\frac{3M}{r}\right)^2}{r^3} |\slashed{\nabla} W_{\Hp}|^2 + \frac{\left(1-\frac{3M}{r}\right)^2}{r^2}  |\slashed{\nabla}_T  W_{\Hp}|^2 + \frac{1}{r^3}|W_{\Hp}|^2 \right) \nonumber \\
  +  \int_{\DcI \left(\tau_1,\tau_2\right)} \left( \frac{1}{r^2} | \slashed{\nabla}_{R^\star} W_{\I}|^2 + \frac{1}{r^2} |\slashed{\nabla}_T W_{\I}|^2+ \frac{1}{r^3} |\slashed{\nabla} W_{\I}|^2 + \frac{1}{r^3}|W_{\I}|^2 \right)
 \lesssim
                 \underline{\check{\mathbb{F}}}^\diamond_{v(\tau_1)} [ W_{\Hp}]
+
\check{\mathbb{F}}_{\tau_1} [W_{\I}]  \nonumber \\
+  \sum_{i=1}^2 \mathcal{G}_1 \left[W\right] (\tau_1,\tau_2) +  \sum_{i=1}^3 \mathcal{H}_i \left[W\right](\tau_1,\tau_2) + \Big|\sum_{i=1}^3  \mathcal{B}_i \left[W_{\Hp}\right](\tau_1,\tau_2)  -  \mathcal{B}_i \left[W_{\I}\right](\tau_1,\tau_2)  \Big|\, ,
\end{align}
provided we define
\begin{align} \label{bhdef}
\mathcal{B}_3 \left[W_{\I}\right](\tau_1,\tau_2)   &:= \int_{\mathcal{B}(\tau_1,\tau_2) }  \left( \mathfrak{f}_T^h \left[W_{\I}\right] g(T^{\I}, n) + \mathfrak{f}_{R^\star}^h  \left[W_{\I}\right]  g({R^\star}^{\I},n) + \mathfrak{f}_A^h  \left[W_{\I}\right] g(e_A^{\I},n) \right) \, , \nonumber \\
 \mathcal{B}_3 \left[W_{\Hp}\right] (\tau_1,\tau_2)  &:=\int_{\mathcal{B}(\tau_1,\tau_2) } \left( \mathfrak{f}_T^h \left[W_{\Hp}\right] g(T^{\Hp}, n) + \mathfrak{f}_{R^\star}^h  \left[W_{\Hp}\right]  g({R^\star}^{\Hp},n) + \mathfrak{f}_A^h  \left[W_{\Hp}\right] g(e_A^{\Hp},n) \right)  \, .
\end{align}

{\bf Step 7: Optimising the $r$-weights.}
Except for the $r$-weights at infinity this is already the desired estimate (\ref{Tiled}). To finally optimise the $r$-weights we apply again the Morawetz $X$-identity, this time with $f= \left(1-\frac{1}{r^\delta}\right) \xi$ where $\xi$ is a smooth cut-off function vanishing for $r \leq 2R_{-1}$ and equal to $1$ for $r\geq 2R$ (in particular, no boundary term on $\mathcal{B}$). It is easy to see that
\[
\mathfrak{f}^X_{bulk} \left[W\right] \geq \frac{1}{2}\delta r^{-1-\delta} |\slashed{\nabla}_{R^\star} W|^2 + \frac{1}{r^3} |W|^2
\]
holds for all $ r \geq \tilde{R}$ for some $\tilde{R}$ sufficiently large. 

Since we already control~\eqref{iutr}
and since all boundary terms in the identity remain controlled by the $T$-identity we have obtained the correct $r$-weight for $\slashed{\nabla}_{R^\star} W$ appearing in $\check{\mathbb{I}}\left[W_{\I}\right]$. To obtain the optimal $r$-weight for the $\slashed{\nabla}_T W$ term we apply the Lagrangian $h$-identity with $h=-r^{-1-\delta} \xi$. This finally yields (\ref{Tiledb}) as stated. Coupling (\ref{Tiledb}) to (\ref{tbu}) yields (\ref{Tbndb}). 
\end{proof}

\section{The redshift estimate} \label{sec:rsagi}

We now exploit the redshift (see~\cite{DafRod2, Mihalisnotes}) 
to obtain Proposition \ref{prop:basecase} without the diamond superscripts, i.e.~with the improved weights near the horizon in all energies.

\begin{proposition} \label{prop:redshift}
Let $W_{\Hp}$ be a symmetric traceless $S$-tensor or an $S$-tangent $1$-form satisfying a tensorial wave equation of Type $3_{k}$ with $k\geq 0$. Then, for any $u_1 \leq \tau \leq u_f$ 
the following estimate holds
        \begin{align}  \label{rsdr}
                & \int_{\DcH \left(v(\tau)\right)} |\Omega^{-1} \slashed{\nabla}_3 W_{\Hp}|^2 
                +\sup_{\tau \leq u \leq u_f} \int_{\CbcH_{v(u)}}|\Omega^{-1} \slashed{\nabla}_3 W_{\Hp}|^2  + \int_{\CcH_{u_f} \left(v(\tau)\right) \cap \{r\leq 9M_{\rm init}/4\}} | \slashed{\nabla} W_{\Hp}|^2
                 \\
                &
                \lesssim
                 \int_{\CbcH_{v(\tau)}}|\Omega^{-1} \slashed{\nabla}_3 W_{\Hp}|^2+\sup_{\tau \leq u \leq u_f} \check{\underline{\mathbb{F}}}^\star_{v(u)} [W_{\Hp}] + \mathbb{I}^{\star, deg} \left[W_{\Hp}\right]  \left(v(\tau)\right)   +  \mathcal{G}_4  \left[W_{\Hp}\right] (\tau,u_f) +\mathcal{H}_4  \left[W_{\Hp}\right] (\tau,u_f)\, . \nonumber
\end{align}
Here
\begin{align}
\mathcal{G}_4 \left[W_{\Hp}\right] (\tau,u_f) = \int_{\DcH \left(v(\tau)\right) \cap \{r \leq 5M_{\rm init}/2\}} |\mathcal{F}^{lin}_{3_{k}} \left[W_{\Hp}\right] \cdot   \Omega^{-1} \slashed{\nabla}_3 W_{\Hp} | \, ,
\end{align}
\begin{align} \label{defh4}
\mathcal{H}_4\left[W_{\Hp}\right] (\tau,u_f)=  \int_{\DcH \left(v(\tau)\right)  \cap \{r \leq 5M_{\rm init}/2\}} \Big| \left(\mathcal{F}^{nlin}_{3_{k}} \left[W_{\Hp}\right] - \frac{1}{2} F_{34} \left[W_{\Hp}\right]  \right)  \Omega^{-1} \slashed{\nabla}_3 W_{\Hp}\Big| \, .
\end{align}
\end{proposition}

\begin{remark}
The terms involving a $\star$ on the right hand side are energies supported away from what will be the horizon and controlled from the basic energy and Morawetz estimate, i.e.~Proposition \ref{prop:basecase}.
\end{remark}

\begin{proof}
Let $\xi$ be a radial cut-off function with $\xi=1$ for $ r\leq 9M_{\rm init}/4$ and vanishing for $r \geq 5M_{\rm init}/2$. Note that 
\[
2Mr + 3r^2 \Omega_\circ^2 \geq 2M^2 \ \ \ \textrm{for $r \leq 9M_{\rm init}/4$.}
\]
Contract now the equation of Type $3_k$ by $\frac{r^3}{\Omega^2} \xi \Omega \slashed{\nabla}_3 W_{\Hp}$ to generate the multiplier identity
\begin{align}
\frac{1}{2} \Omega \slashed{\nabla}_4 \left( \xi \frac{r^3}{\Omega^2} |\Omega \slashed{\nabla}_3 W_{\Hp} |^2 \right) + \frac{1}{2} \frac{1}{\Omega^2}  |\Omega \slashed{\nabla}_3 W_{\Hp} |^2 \left( \xi \left(2M \left(k+1\right)r + r^2 \left(\omega - \omega_\circ\right) -3r^2 \Omega_\circ^2\right) - \partial_r \xi r^3 \Omega_\circ^2 \right)  \nonumber \\
+ \frac{1}{2} \Omega \slashed{\nabla}_3 \left(\xi r |r\slashed{\nabla} W_{\Hp}|^2 \right) + \frac{1}{2}  |r\slashed{\nabla} W_{\Hp}|^2 \left(\Omega_\circ^2 \xi + \partial_r \xi \Omega_\circ^2 r \right) 
- \slashed{\nabla}^A \left(r^3 \xi \slashed{\nabla}_A W_{\Hp} \Omega \slashed{\nabla}_3 W_{\Hp}\right) \nonumber \\
=  \mathcal{F}^{lin}_{3_{k}} \left[W_{\Hp}\right] \cdot  \xi r^3 \Omega^{-1} \slashed{\nabla}_3 W_{\Hp} + \mathcal{F}^{nlin}_{3_{k}} \left[W_{\Hp}\right] \cdot  \xi r^3 \Omega^{-1} \slashed{\nabla}_3 W_{\Hp} - \frac{1}{2} F_{34}  \left[W_{\Hp}\right] \xi r^3 \Omega^{-1} \slashed{\nabla}_3 W_{\Hp} \, . \nonumber
\end{align}
Multiply this by $\frac{\sqrt{\slashed{g}}}{r^2}$ and integrate over the region $\DcH \left(v(\tau)\right)$ with respect to $du dv d\theta^1 d\theta^2$ using Lemma~\ref{lem:transformder} and Stokes' theorem. Note that the last term in the second line vanishes and that all spacetime (and boundary) terms have good signs in $r \leq 9M_{\rm init}/4$.
\end{proof}
Revisiting the above proof now integrating the multiplier identity over $\DcH \left(v(\tau)\right) \cap J^-\left(\CcH_{u} \left(v(\tau)\right) \right)$ we easily see
\begin{corollary} \label{cor:rsingfl}
The estimate (\ref{rsdr}) also holds replacing $\int_{\CcH_{u_f} \left(v(\tau)\right) \cap \{r\leq 9/4M_{\rm init}\}} | \slashed{\nabla} W_{\Hp}|^2$ on the left by $\sup_u \int_{\CcH_{u} \left(v(\tau)\right) \cap \{r\leq 9M_{\rm init}/4\}} | \slashed{\nabla} W_{\Hp}|^2$.
\end{corollary}

\section{The $r^p$-weighted hierarchy} \label{sec:rpagi}
In this section we shall derive $r^p$ weighted estimates for tensors $W_{\mathcal{I}^+}$. For the wave
equation, these estimates originate in~\cite{DafRodnew}. 

Since the following Proposition will in particular be used later in the paper to derive  largeness constraints on the constant $R$, we will explicitly denote the additional $R$-dependence in constants in Proposition~\ref{prop:rph} with the notation $C_R$, while
the $\lesssim$ notation in Proposition~\ref{prop:rph} will indicate that the constant can be chosen to depend only on $M_{\rm init}$ independently of the choice of $R$. We shall return to our usual conventions regarding $\lesssim$ in 
Section~\ref{nonlinearestforwave}.

\begin{proposition} \label{prop:rph}
Let $W_{\I}$ be a symmetric traceless $S$-tensor or an $S$-tangent $1$-form satisfying a tensorial wave equation of Type $4_{k,l}$. Then, for any $u_1 \leq \tau \leq u_f$ and $p \in \left[1-4k, 2-\delta\right] \cup \{2\}$, 
the following estimate holds
        \begin{align}  \label{basrpe}
               &  \check{\mathbb{I}}^p \left[W_{\I}\right] \left(\tau, u_f\right) 
                +\sup_{\tau \leq u \leq u_f} \check{\mathbb{F}}^{p}_u [W_{\I}] +  \underline{\check{\slashed{\mathbb{F}}}}^{p}_{v_\infty} [W_{\I}]( \tau ) \nonumber \\   
                 \lesssim &  \ \ \ 
                \check{\mathbb{F}}^{p}_\tau [W_{\I}]  +
                C_R \left[ \sup_{\tau \leq u \leq u_f} \check{\mathbb{F}}^{\star}_u [W_{\I}] + \check{\mathbb{I}}^\star \left[W_{\I}\right] \left(\tau, u_f\right) \right] \nonumber \\
                & + \boxed{ \int_{\DcI\left(\tau\right)} r^{p-3} |W_{\I}|^2 } + \mathcal{G}_{5,p} \left[W_{\I}\right]  \left(\tau,u_f\right)   +\mathcal{H}_{5,p} \left[W_{\I}\right]  \left(\tau,u_f\right)
\end{align}
where the boxed term can be dropped if $l \geq 0$ and also if $
\frac{2l^2}{p+4k-1/2} \leq \frac{3}{8}(2-p)$.\footnote{The latter condition will be exploited in Lemma \ref{barp} for the equation satisfied by $r^4 A_{\I}$.}
Here
\begin{align}
\mathcal{G}_{5,p} \left[W_{\I}\right] \left(\tau,u_f\right) &= \Big| \int_{\DcI\left(\tau\right) \cap \{ r \geq 3R/2\}} \mathcal{F}^{lin}_{4_{k,l}} \left[W_{\I}\right] \cdot  \left(r^p+r^{2-\delta} \boldsymbol\delta^2_p \right)  \Omega \slashed{\nabla}_4 W_{\I} \Big| \, , 
 \\
\label{defh5}
\mathcal{H}_{5,p} \left[W_{\I}\right]  \left(\tau,u_f\right) &=\int_{\DcI\left(\tau\right)  \cap \{r \geq 3R/2\}} |\left(\mathcal{F}^{nlin}_{4_{k,l}} \left[W_{\I}\right] + F_{34} \left[W_{\I}\right] \right)| \cdot  r^p |\Omega \slashed{\nabla}_4 W_{\I}| \, ,
\end{align}
and $C_R$ is a constant depending on $R$.
\end{proposition}

\begin{remark}
The terms in the square bracket are energies supported away from infinity and will  be controlled from an energy and Morawetz estimate, i.e.~Proposition \ref{prop:basecase}. The boxed term will be controlled inductively (and shown not to appear at the lowest order).
\end{remark}

\begin{proof}
Let $\xi$ be a radial cut-off function equal to $1$ for $r \geq 2R$ and equal to zero for $r\leq 3R/2$ and denote the weight $w = 1+\frac{6M}{r}$. Multiply now the equation of type $4_{k,l}$ by $\xi r^p w \Omega \slashed{\nabla}_4 W_{\I}$ to generate the following multiplier identity:
\begin{align} \label{rpid}
&\frac{1}{2} \Omega \slashed{\nabla}_3 \left(\xi r^p w |\Omega \slashed{\nabla}_4 (W_{\I})|^2 \right) + \frac{1}{2} \Omega \slashed{\nabla}_4 \left(\frac{\Omega^2}{r^2} r^p \xi w | r \slashed{\nabla} (W_{\I})|^2 \right) \nonumber \\
& + \frac{1}{2} \left(\xi \left((p+4k) r^{p-1} w- 6M r^{p-2}\right) \Omega_\circ^2+\partial_r \xi r^p w\right) |\Omega \slashed{\nabla}_4 (W_{\I})|^2 \nonumber \\
&+ \frac{1}{2} \left( \xi r^{p-3} \Omega^2 \left( (2-p) +(12-4p)M r^{-1} + M^2 r^{-2} \left(-48+12p\right) )\right) - \partial_r \xi w\Omega^2 r^{p-2} \right)  | r \slashed{\nabla} W_{\I}|^2 \nonumber \\
&\boxed{+ 2l \frac{\Omega^2}{r^2} \xi w r^p W_{\I} \Omega \slashed{\nabla}_4 W_{\I} } -\slashed{div} \left(\frac{\Omega^2}{r^2} r^2 \slashed{\nabla} W_{\I} \xi r^p w \Omega \slashed{\nabla}_4 W_{\I}\right) =  \mathcal{E}_{r^p} \left[W_{\I}\right]  \, , 
\end{align}
where the (at least quadratic) error is
\begin{align} \label{qad}
\mathcal{E}_{r^p} \left[W_{\I}\right] = & ( \mathcal{F}^{lin}_{4_{k,l}} \left[W_{\I}\right] + \mathcal{F}^{nlin}_{4_{k,l}} \left[W_{\I}\right] + \frac{1}{2} F_{34} \left[W_{\I}\right] ) \xi r^p \Omega  \slashed{\nabla}_4 (W_{\I}) -2 \xi r^{p-2} w (\omega - \omega_\circ)  | r \slashed{\nabla} W_{\I}|^2 \nonumber \\
&-  \xi \Omega^2 r^{p-2} (r\slashed{\nabla}(W_{\I})) \left[\Omega \slashed{\nabla}_4 , r \slashed{\nabla}\right]W_{\I}   - \left(\eta +\underline{\eta}\right) \xi r^{p-1}  (r\slashed{\nabla}(W_{\I})) \Omega \slashed{\nabla}_4 (W_{\I}) .
\end{align}

We multiply this identity by $\frac{\sqrt{\slashed{g}}}{r^2}$ and integrate first over $\DcI\left(\tau\right)$ with respect to $du dv d\theta^1 d\theta^2$ using Lemma \ref{lem:transformder} and Stokes' theorem.  Note that the last term on the left then vanishes identically. Ignoring the boxed cross-term for the moment we see that upon integration the first line of (\ref{rpid}) will produce positive future boundary terms as claimed in the proposition. In the second line and third line, the terms involving $\partial_r \xi$ are supported for $r \leq 2R$ only and hence controlled by $\mathbb{I}^\star \left[W_{\I}\right]$. For the other terms in these lines it is not hard to see that the constraint $\frac{M}{R} < \delta$ implies that the following holds  in $r\geq 2R$ for all $p$ with $1-4k \leq p \leq 2-\delta$:
\begin{align}
(p+4k) r^{p-1} w- 6M r^{p-2} \geq \frac{1}{2}r^{p-1} \ \ \  \textrm{and} \ \ \ \  \frac{1}{6}(2-p) +(12-4p)M r^{-1} + M^2 r^{-2} \left(-48+12p\right) ) \geq \frac{\delta}{12}, \nonumber
\end{align}
while for $p=2$ the lower bound in the second estimate changes to $\frac{2M}{r}$.
Therefore, we manifestly control the spacetime energy $\check{\mathbb{I}}^p \left[W_{\I}\right] \left(\tau, u_f\right)$  in the Proposition, at least if $1-4k \leq p \leq 2-\delta$. For $p=2$, we get a weaker estimate in view of the spacetime term containing angular derivatives degenerating for $p=2$. However, we can add to the weaker $p=2$ estimate also the estimate for $p=2-\delta$ to provide control of the angular derivatives as claimed. Turning to the non-linear terms~\eqref{qad}, we see that the first three terms are those appearing in the final estimate while the remaining ones can easily be absorbed on the left using the $L^\infty$ estimates~\eqref{useonlyLinfinityestimate} 
on the Ricci coefficients (use Lemma \ref{lem:commutation} for the commutator terms).

We finally deal with the boxed term. For $l \geq 0$ we use
\begin{align}
 2l \frac{\Omega^2}{r^2} \xi w r^p W_{\I} \Omega \slashed{\nabla}_4 W_{\I} =   l\Omega \slashed{\nabla}_4 \left(  \frac{\Omega^2}{r^2}  |W_{\I}|^2 \xi r^p w\right) \nonumber \\
 +  \left(  \xi r^{p-3} \Omega^2 \left( (2-p) +(12-4p)M r^{-1} + Mr^{-2} \left(-48+12p\right) )\right) - \partial_r \xi w\Omega^2 r^{p-2} \right) |W_{\I}|^2 \nonumber
\end{align}
producing a positive (up to terms supported in $r \leq 2R$ and controlled by $\mathbb{I}^\star \left[W_{\I}\right]$) spacetime term in $r\geq 2R$  and a positive boundary term on $v=v_\infty$.

For $l \leq 0$ we use Cauchy--Schwarz to estimate
\begin{align}
\Big| 2l \frac{\Omega^2}{r^2} \xi w r^p W_{\I} \Omega \slashed{\nabla}_4 W_{\I} \Big|&=  \frac{1}{2} \left(p+4k-1/2\right)r^{p-1} \Omega_\circ^2 w \xi | \Omega \slashed{\nabla}_4 W_{\I}|^2 +  \frac{2l^2}{p+4k-1/2} \frac{\Omega^4}{\Omega_\circ^2}r^{p-3} w \xi | W_{\I}|^2 \, . \nonumber
\end{align}
The first term can be absorbed by the positive term in (\ref{rpid}). The second term is the one appearing on the left hand side. If  $
\frac{2l^2}{p+4k-1/2} \leq \frac{3}{8}(2-p)$ holds we can estimate further 
\begin{align}
 \frac{2l^2}{p+4k-1/2} \frac{\Omega^4}{\Omega_\circ^2}r^{p-3} w \xi | W_{\I}|^2 \leq  \frac{3}{8}  (2-p) \Omega^2 r^{p-3} \xi \frac{16}{15} | r \slashed{\nabla} W_{\I}|^2 =\frac{2}{5} (2-p) \Omega^2 r^{p-3}  | r \slashed{\nabla} W_{\I}|^2 \, , 
\end{align}
where the Poincar\'e inequality of Proposition \ref{prop:Poincare} (for symmetric traceless $S$-tensors and $S$-tangent $1$-forms) has been used. The expression on the right can finally be absorbed   by the positive term in~\eqref{rpid}.

This produces (\ref{basrpe}) except that $\sup_{\tau \leq u \leq u_f} \check{\mathbb{F}}^{p}_u [W_{\I}] $ is replaced by $\check{\mathbb{F}}^{p}_{u_f} [W_{\I}]$ on the left. To finally obtain the boundary term on any constant $u$ hypersurface one repeats the proof integrating now over $\DcI\left(\tau\right) \cap \{u_{\I} \leq u\}$ and using the estimate already obtained.
\end{proof}

Revisiting the above proof now integrating the multiplier identity over the region $\DcI \left(\tau\right) \cap J^-(\CbcI_{v} \left(\tau\right))$ we easily see
\begin{corollary} \label{cor:rsingfl2}
The estimate (\ref{basrpe}) remains true replacing $ \underline{\check{\slashed{\mathbb{F}}}}^{p}_{v_\infty} [W_{\I}]( \tau )$ on the left hand side by \\
$\sup_v \int_{\CbcI_{v} \left(\tau\right) \cap \{r\geq 2R\}} r^{p-2} | r \slashed{\nabla} W_{\I}|^2$.
\end{corollary}

\section{Nonlinear error estimates for wave equations}
\label{nonlinearestforwave}

We recall the non-linear errors $\mathcal{H}_i [W](\tau,u_f)$ (i=1,2,3,4) and $\mathcal{H}_{5,p} [W](\tau,u_f)$ appearing in the energy estimates  of Propositions \ref{prop:basecase}, \ref{prop:redshift}, \ref{prop:rph}. In this section, we prove estimates allowing us to control the resulting nonlinear error terms for the wave equations in Chapters \ref{chapter:psiandpsibar} and \ref{moreherechapter}.  

Recall that $2M <r_0 < r_1 <3M <R$, and recall the nonlinear error notation $\mathcal{E}^k_p$ and $\mathcal{E}^{*k}_p$ defined in Section \ref{sec:nlenotation} and Section \ref{sec:errorstarnot} respectively.  Recall also the following pointwise estimates for the geometric quantities in the $\Hp$ and $\I$ gauges of Proposition \ref{prop:pointwiseboundhereimp},
\begin{equation} \label{pointwiseboundhereimprecalled}
	\mathbb P^{N-5}[\Phi^{\mathcal{H}^+}] +\mathbb P^{N-5}[\Phi^{\mathcal{I}^+}] \lesssim \varepsilon.
\end{equation}
The error terms are estimated separately in different regions.

\subsection{Error terms in the $\Hp$ gauge}
\label{seealreadythissubsection}

The easiest region in which to estimate error terms is the region near the event horizon, $r(u_{\Hp},v_{\Hp}) \leq r_1$.
Recall the set of anomalous quantities
\begin{equation} \label{eq:anomalousquantities}
	\mathcal{A}_s
	=
	\big\{
	(r\nablaslash)^{N-s} \Omega \hat{\chi},
	(r\nablaslash)^{N-s} \left( \Omega \tr \chi - (\Omega \tr \chi)_{\circ} \right)
	\big\},
\end{equation}
for $s=0,1,2$.

\begin{proposition}[Spacetime nonlinear error estimate in the region $r(u_{\Hp},v_{\Hp}) \leq r_1$ of the $\Hp$ gauge]
	\label{prop:wavehorizonerror1}
	For $s=0,1,2$, $0 \leq k \leq N-s$, $0 \leq \vert \gamma \vert \leq N-s$ and $\tau \geq v_{-1}$, for any nonlinear error of the form $\mathcal{E}^{*k}_{\Hp}$ and for any $\Phi_{\Hp}$ such that $\mathfrak{D}^{\gamma} \Phi_{\Hp} \notin \mathcal{A}_s$,
	\[
		\int_{\DRH(\tau)}
		\vert
		\mathfrak{D}^{\gamma} (\Phi^{\Hp} - \Phi^{\Hp}_{\mathrm{Kerr}})
		\cdot 
		\mathcal{E}^{*k}_{\Hp}
		\vert
		\cdot
		\mathds{1}_{\{r_{\Hp} \leq r_1\}}
		\Omega^2 d\theta du dv
		\lesssim
		\frac{\varepsilon^3}{\tau^{\frac{1}{2}+s}}.
	\]
\end{proposition}

\begin{proof}
	A given term in the nonlinear error $\mathcal{E}^{*k}_{\Hp}$ takes the form $\mathfrak{D}^{\gamma_1} \Phi' \cdot \mathfrak{D}^{\gamma_2} \Phi''$ for appropriate geometric quantities $\Phi'$ and $\Phi''$, where $\vert \gamma_1 \vert + \vert \gamma_2 \vert \leq N-s$.  Suppose, without loss of generality, $\vert \gamma_1 \vert \geq \vert \gamma_2 \vert$.  Then $\mathfrak{D}^{\gamma_1} \Phi' \notin \mathcal{A}_s$ and, since $N\geq 12$, it follows that $\vert \gamma_2 \vert \leq N-6$ and so the estimate for $\mathbb{E}^N_{u_f,\Hp}$ in the bootstrap assumption \eqref{eq:bamain} and the pointwise estimate \eqref{pointwiseboundhereimprecalled} imply that
	\[
		\int_{\DRH(\tau)}
		\vert
		\mathfrak{D}^{\gamma_1} (\Phi' - \Phi_{\mathrm{Kerr}}')
		\vert^2
		\cdot
		\mathds{1}_{\{r_{\Hp} \leq r_1\}}
		\Omega^2 d\theta du dv
		\lesssim
		\frac{\varepsilon^2}{\tau^{s}},
		\qquad
		\text{ and }
		\qquad
		\vert \mathfrak{D}^{\gamma_2} \Phi'' \vert \lesssim \frac{\varepsilon}{v},
	\]
	respectively.  It follows that
	\begin{align*}
		&
		\int_{\DRH(\tau)}
		\vert
		\mathfrak{D}^{\gamma} (\Phi - \Phi_{\mathrm{Kerr}})
		\cdot 
		\mathfrak{D}^{\gamma_1} \Phi'
		\cdot
		\mathfrak{D}^{\gamma_2} \Phi''
		\vert
		\cdot
		\mathds{1}_{\{r_{\Hp} \leq r_1\}}
		\Omega^2 d\theta du dv
		\\
		&
		\qquad \qquad
		\lesssim
		\frac{\varepsilon}{\tau}
		\int_{\DRH(\tau)}
		\vert
		\mathfrak{D}^{\gamma} (\Phi - \Phi_{\mathrm{Kerr}})
		\vert
		\vert
		\mathfrak{D}^{\gamma_1} (\Phi' - \Phi_{\mathrm{Kerr}}')
		+
		\mathfrak{D}^{\gamma_1} \Phi_{\mathrm{Kerr}}'
		\vert
		\cdot
		\mathds{1}_{\{r_{\Hp} \leq r_1\}}
		\Omega^2 d\theta du dv
		\\
		&
		\qquad \qquad
		\lesssim
		\frac{\varepsilon}{\tau}
		\Vert
		\mathfrak{D}^{\gamma} (\Phi - \Phi_{\mathrm{Kerr}}) \mathds{1}
		\Vert_{\DRH(\tau)}
		\left(
		\Vert
		\mathfrak{D}^{\gamma_1} (\Phi' - \Phi_{\mathrm{Kerr}}') \mathds{1}
		\Vert_{\DRH(\tau)}
		+
		\Vert
		\mathfrak{D}^{\gamma_1} \Phi_{\mathrm{Kerr}}' \mathds{1}
		\Vert_{\DRH(\tau)}
		\right)
		\\
		&
		\qquad \qquad
		\lesssim
		\frac{\varepsilon}{\tau}
		\frac{\varepsilon}{\tau^{\frac{s}{2}}} \left( \frac{\varepsilon}{\tau^{\frac{s}{2}}} + \frac{\varepsilon}{(u_f)^{\frac{1}{2}}} \right),
	\end{align*}
	since $\Vert \Phi_{\mathrm{Kerr}} \Vert_{\DRH(\tau)}^2 \lesssim \varepsilon^2 (u_f)^{-1}$ for any $\Phi$.  The proof follows.
\end{proof}

The estimates for the error terms in the region $r_1 \leq r(u_{\Hp},v_{\Hp}) \leq R_2$ are more involved than the proof of Proposition \ref{prop:wavehorizonerror1} due to the degeneration of the integrated decay estimates in \eqref{eq:bamain} at $r=3M_f$.

Note that the estimate for $\mathbb{E}^N_{u_f,\Hp}$ in bootstrap assumption \eqref{eq:bamain} implies that each $\Phi^{\Hp}$ satisfies, for all $v \geq v_{-1}$, and for $s=0,1,2$,
\begin{equation} \label{eq:PhiHintegratedlo}
	\sum_{\vert \gamma \vert \leq N-1-s}
	\Vert
	\mathfrak{D}^{\gamma} ( \Phi^{\Hp} - \Phi_{\mathrm{Kerr}}^{\Hp})
	\Vert_{\DRH(v)}^2
	\lesssim
	\frac{\varepsilon^2}{v^{s}},
\end{equation}
and at least one of the two estimates,
\begin{equation} \label{eq:3Mformerlatter}
	\sum_{\vert \gamma \vert \leq N-s}
	\Vert \mathfrak{D}^{\gamma} ( \Phi^{\Hp} - \Phi_{\mathrm{Kerr}}^{\Hp}) \Vert_{\Cbar_v^{\Hp}}^2
	\lesssim
	\frac{\varepsilon^2}{v^s},
	\qquad
	\text{and/or}
	\qquad
	\sum_{\vert \gamma \vert \leq N-s}
	\Vert \mathfrak{D}^{\gamma} ( \Phi^{\Hp} - \Phi_{\mathrm{Kerr}}^{\Hp}) \Vert_{C_u^{\Hp}(v)}^2
	\lesssim
	\frac{\varepsilon^2}{v^s},
\end{equation}
for all $u_0 \leq u \leq u_f$.  Note also that, for each $\Phi^{\Hp}$, the pointwise estimate \eqref{pointwiseboundhereimprecalled} and the estimate \eqref{eq:curletalambda} for the linearised Kerr parameters imply that
\[
	\sum_{\vert \gamma \vert \leq N-6}
	\vert \Phi^{\Hp} \vert
	\lesssim
	\frac{\varepsilon}{v},
	\qquad
	\text{and}
	\qquad
	\vert \Phi^{\Hp}_{\mathrm{Kerr}} \vert \lesssim \frac{\varepsilon}{u_f}.
\]

The following lemma is exploited when estimating nonlinear error terms in the region $r_1 \leq r(u_{\Hp},v_{\Hp}) \leq R_2$ in Proposition \ref{prop:wavehorizonerror2}.

\begin{lemma} \label{lem:ILEDtrapping}
	For all $\Phi^{\Hp}$ and all $\vert \gamma \vert \leq N-4$, $\tau \geq v_{-1}$,
	\begin{equation} \label{eq:ILEDtrappingidentity}
		\Vert
		v^{1 - \delta} \mathfrak{D}^{\gamma} (\Phi^{\Hp} - \Phi^{\Hp}_{\mathrm{Kerr}}) 
		\Vert_{\DRH(\tau)}
		\lesssim
		\varepsilon \tau^{-\delta}.
	\end{equation}
\end{lemma}

\begin{proof}
	The bootstrap assumption \eqref{eq:bamain} in particular implies that
	\begin{equation} \label{eq:PhiHloILED}
		\int_{\DRH(\tau)}
		\left\vert \mathfrak{D}^{\gamma} (\Phi^{\Hp} - \Phi^{\Hp}_{\mathrm{Kerr}}) \right\vert^2
		\Omega^2 d\theta du dv
		\lesssim
		\frac{\varepsilon^2}{\tau^{2}},
	\end{equation}
	since $k \leq N-4$.  To see that \eqref{eq:ILEDtrappingidentity} indeed holds, let $\{ \tau_n \}$ be a dyadic sequence (so that $\tau_{n+1} = 2 \tau_n$, $\tau_0 = v_0$).  The estimate \eqref{eq:PhiHloILED} gives, for any $n$,
	\[
		\int_{\DRH(\tau_n) \smallsetminus \DRH(\tau_{n+1})}
		\left\vert \mathfrak{D}^{\gamma} (\Phi^{\Hp} - \Phi^{\Hp}_{\mathrm{Kerr}}) \right\vert^2
		\Omega^2 d\theta du dv
		\lesssim
		\frac{\varepsilon^2}{(\tau_n)^{2}}.
	\]
	Multiplying by $\tau_{n+1}^{2-2\delta}$ and using the fact that $\tau_{n+1}$ and $v$ are comparable, for any $\tau_n \leq v \leq \tau_{n+1}$, gives
	\[
		\int_{\DRH(\tau_n) \smallsetminus \DRH(\tau_{n+1})}
		v^{2-2\delta} \left\vert \mathfrak{D}^{\gamma} (\Phi^{\Hp} - \Phi^{\Hp}_{\mathrm{Kerr}}) \right\vert^2
		\Omega^2 d\theta du dv
		\lesssim
		\frac{\varepsilon^2}{(\tau_n)^{2\delta}}.
	\]
	Summing over $n \geq l$, and using the fact that $\sum_{n\geq l} \tau_n^{-2\delta} = \tau_l^{-2\delta} \sum_{n \geq l} 2^{-\delta(n-l)} \lesssim \tau_l^{-2\delta}$ gives
	\[
		\int_{\DRH(\tau_n)}
		v^{2-2\delta} \left\vert \mathfrak{D}^{\gamma} (\Phi^{\Hp} - \Phi^{\Hp}_{\mathrm{Kerr}}) \right\vert^2
		\Omega^2 d\theta du dv
		\lesssim
		\frac{\varepsilon^2}{(\tau_n)^{2\delta}},
	\]
	for any $l$, from which \eqref{eq:ILEDtrappingidentity} follows.
\end{proof}

Recall again the anomalous quantities \eqref{eq:anomalousquantities}.

\begin{proposition}[Spacetime nonlinear error estimate in the region $r(u_{\Hp},v_{\Hp}) \geq r_1$ of the $\Hp$ gauge] \label{prop:wavehorizonerror2}
	For $s=0,1,2$, $0 \leq k \leq N-s$, $0 \leq \vert \gamma \vert \leq N-s$ and $\tau \geq v_{-1}$, for any nonlinear error of the form $\mathcal{E}^{*k}_{\Hp}$ and for any $\Phi_{\Hp}$ such that $\mathfrak{D}^{\gamma} \Phi_{\Hp} \notin \mathcal{A}_s$,
	\[
		\int_{\DRH(\tau)}
		\vert
		\mathfrak{D}^{\gamma} (\Phi^{\Hp} - \Phi^{\Hp}_{\mathrm{Kerr}})
		\cdot 
		\mathcal{E}^{*k}_{\Hp}
		\vert
		\cdot
		\mathds{1}_{\{r_{\Hp} \geq r_1\}}
		\Omega^2 d\theta du dv
		\lesssim
		\frac{\varepsilon^3}{\tau^{s}}.
	\]
\end{proposition}

\begin{proof}
	A given term in the nonlinear error $\mathcal{E}^{*k}$ takes the form $\mathfrak{D}^{\gamma_1} \Phi' \cdot \mathfrak{D}^{\gamma_2} \Phi''$, for appropriate geometric quantities $\Phi'$ and $\Phi''$, where $\vert \gamma_1 \vert + \vert \gamma_2 \vert \leq N-s$.  Suppose, without loss of generality, $\vert \gamma_1 \vert \geq \vert \gamma_2 \vert$.  Then $\mathfrak{D}^{\gamma_1} \Phi' \notin \mathcal{A}_s$ and so the bootstrap assumption \eqref{eq:bamain} implies that $\mathfrak{D}^{\gamma_1} (\Phi' - \Phi_{\mathrm{Kerr}}')$ satisfies at least one of the two estimates \eqref{eq:3Mformerlatter}.
	Similarly for $\mathfrak{D}^{\gamma} (\Phi - \Phi_{\mathrm{Kerr}})$.  Moreover, since $N\geq 12$, it follows that $\vert \gamma_2 \vert \leq N-6$ and so $\mathfrak{D}^{\widetilde{\gamma}} (\Phi - \Phi_{\mathrm{Kerr}})$ satisfies the conclusion of Lemma \ref{lem:ILEDtrapping} for all $\vert \widetilde{\gamma} \vert \leq \vert \gamma_2 \vert + 2$.
	
	Now
	\[
		\vert
		\mathfrak{D}^{\gamma} (\Phi - \Phi_{\mathrm{Kerr}})
		\cdot
		\mathfrak{D}^{\gamma_1} \Phi'
		\cdot
		\mathfrak{D}^{\gamma_2} \Phi''
		\vert
		\lesssim
		\vert
		\mathfrak{D}^{\gamma_2} \Phi''
		\vert
		\left(
		\vert
		\mathfrak{D}^{\gamma} (\Phi - \Phi_{\mathrm{Kerr}})
		\vert^2
		+
		\vert
		\mathfrak{D}^{\gamma_1} \Phi'
		\vert^2
		\right).
	\]
	Consider the second term (the first term is simpler to estimate).  Clearly
	\begin{align} \label{eq:3Msecondterm}
		\vert
		\mathfrak{D}^{\gamma_2} \Phi''
		\vert
		\vert
		\mathfrak{D}^{\gamma_1} \Phi'
		\vert^2
		\lesssim
		\
		&
		\vert
		\mathfrak{D}^{\gamma_2} (\Phi'' - \Phi_{\mathrm{Kerr}}'')
		\vert
		\vert
		\mathfrak{D}^{\gamma_1} (\Phi' - \Phi_{\mathrm{Kerr}}')
		\vert^2
		+
		\vert
		\mathfrak{D}^{\gamma_2} \Phi_{\mathrm{Kerr}}''
		\vert
		\vert
		\mathfrak{D}^{\gamma_1} (\Phi' - \Phi_{\mathrm{Kerr}}')
		\vert^2
		\\
		&
		+
		\vert
		\mathfrak{D}^{\gamma_2} (\Phi'' - \Phi_{\mathrm{Kerr}}'')
		\vert
		\vert
		\mathfrak{D}^{\gamma_1} \Phi_{\mathrm{Kerr}}'
		\vert^2
		+
		\vert
		\mathfrak{D}^{\gamma_2} \Phi_{\mathrm{Kerr}}''
		\vert
		\vert
		\mathfrak{D}^{\gamma_1} \Phi_{\mathrm{Kerr}}'
		\vert^2.
		\nonumber
	\end{align}
	Consider the first term and suppose first that the former of \eqref{eq:3Mformerlatter} holds.  By the Sobolev inequality, Proposition \ref{prop:Sobolevin}, and the fact that $\int_{\DRH(\tau)} \Omega^2 d\theta du dv = \int_{\tau}^{v(R_2,u_f)} \int_{\Cbar_v^{\Hp}} \Omega^2 d\theta du dv$,
	\begin{multline*}
		\int_{\DRH(\tau)} \!\!\!
		\vert
		\mathfrak{D}^{\gamma_2} (\Phi'' - \Phi_{\mathrm{Kerr}}'')
		\vert
		\vert
		\mathfrak{D}^{\gamma_1} (\Phi' - \Phi_{\mathrm{Kerr}}')
		\vert^2
		\\
		\lesssim
		\sum_{\vert \widetilde{\gamma} \vert \leq \vert \gamma_2 \vert +2}
		\int_{\tau}^{v(R_2,u_f)}
		\!\!\!
		\Vert
		\mathfrak{D}^{\widetilde{\gamma}} (\Phi'' - \Phi_{\mathrm{Kerr}}'')
		\Vert_{\Cbar_v^{\Hp}}
		\Vert
		\mathfrak{D}^{\gamma_1} (\Phi' - \Phi_{\mathrm{Kerr}}')
		\Vert_{\Cbar_v^{\Hp}}^2
		dv.
	\end{multline*}
	For any $\vert \widetilde{\gamma} \vert \leq \vert \gamma_2 \vert +2$ then, by Cauchy--Schwarz,
	\begin{align*}
		&
		\int_{\tau}^{v(R_2,u_f)}
		\Vert
		\mathfrak{D}^{\widetilde{\gamma}} (\Phi'' - \Phi_{\mathrm{Kerr}}'')
		\Vert_{\Cbar_v^{\Hp}}
		\Vert
		\mathfrak{D}^{\gamma_1} (\Phi' - \Phi_{\mathrm{Kerr}}')
		\Vert_{\Cbar_v^{\Hp}}^2
		dv
		\\
		&
		\qquad
		\qquad
		\lesssim
		\Vert
		v^{1-\delta} \mathfrak{D}^{\widetilde{\gamma}} (\Phi'' - \Phi_{\mathrm{Kerr}}'')
		\Vert_{\DRH(\tau)}
		\left(
		\int_{\tau}^{v(R+2,u_f)}
		v^{-2+2\delta}
		\Vert
		\mathfrak{D}^{\gamma_1} (\Phi' - \Phi_{\mathrm{Kerr}}')
		\Vert_{\Cbar_v^{\Hp}}^4
		dv
		\right)^{\frac{1}{2}}
		\\
		&
		\qquad
		\qquad
		\lesssim
		\frac{\varepsilon}{\tau^{\delta}}
		\left(
		\int_{\tau}^{v(R_2,u_f)}
		v^{-2+2\delta}
		\varepsilon^4 v^{-2s}
		dv
		\right)^{\frac{1}{2}}
		\lesssim
		\frac{\varepsilon^3}{\tau^{s+\frac{1}{2}}},
	\end{align*}
	where \eqref{eq:PhiHintegratedlo}, Lemma \ref{lem:ILEDtrapping} and the first of \eqref{eq:3Mformerlatter} have been used.
	Suppose now that only the latter of \eqref{eq:3Mformerlatter} holds.  Then, similarly, the Sobolev inequality of Proposition \ref{prop:Sobolevout} implies
	\begin{multline*}
		\int_{\DRH(\tau)}
		\vert
		\mathfrak{D}^{\gamma_2} (\Phi'' - \Phi_{\mathrm{Kerr}}'')
		\vert
		\vert
		\mathfrak{D}^{\gamma_1} (\Phi' - \Phi_{\mathrm{Kerr}}')
		\vert^2
		\mathds{1}_{\{r_{\Hp} \geq r_1\}}
		\\
		\lesssim
		\sum_{\vert \widetilde{\gamma} \vert \leq \vert \gamma_2 \vert +2}
		\int_{u(R_2,\tau)}^{u_f}
		\Vert
		v^{1-\delta}
		\mathfrak{D}^{\widetilde{\gamma}} (\Phi'' - \Phi_{\mathrm{Kerr}}'')
		\Omega
		\Vert_{C_u^{\Hp}(\tau)}
		\Vert
		v^{-\frac{1}{2}+\frac{\delta}{2}}
		\mathfrak{D}^{\gamma_1} (\Phi' - \Phi_{\mathrm{Kerr}}')
		\Vert_{C_u^{\Hp}(v(r_0,u))}^2
		du
		\\
		+
		\int_{u(R_2,\tau)}^{u_f}
		\Vert
		v^{1-\delta} \mathfrak{D}^{\widetilde{\gamma}} (\Phi'' - \Phi_{\mathrm{Kerr}}'') \Omega \Vert_{C_{u}^{\Hp}(v(R,u))}
		\Vert
		v^{-\frac{1}{2}+\frac{\delta}{2}}
		\mathfrak{D}^{\gamma_1} (\Phi' - \Phi_{\mathrm{Kerr}}')
		\Vert_{C_u^{\Hp}(v(r_0,u))}^2
		du
		.
	\end{multline*}
	Again, for any $\vert \widetilde{\gamma} \vert \leq \vert \gamma_2 \vert +2$,
	\begin{align*}
		&
		\int_{u(R_2,\tau)}^{u_f}
		\Vert
		v^{1-\delta}
		\mathfrak{D}^{\widetilde{\gamma}} (\Phi'' - \Phi_{\mathrm{Kerr}}'') \Omega
		\Vert_{C_u^{\Hp}(\tau)}
		\Vert
		v^{-\frac{1}{2}+\frac{\delta}{2}}
		\mathfrak{D}^{\gamma_1} (\Phi' - \Phi_{\mathrm{Kerr}}')
		\Vert_{C_u^{\Hp}(v(r_0,u))}^2
		du
		\\
		&
		\qquad
		\lesssim
		\Vert
		v^{1-\delta}
		\mathfrak{D}^{\widetilde{\gamma}} (\Phi'' - \Phi_{\mathrm{Kerr}}'')
		\Vert_{\DRH(\tau)}
		\left(
		\int_{u(R_2,\tau)}^{u_f}
		v(r_0,u)^{-2+2\delta}
		\Vert
		\mathfrak{D}^{\gamma_1} (\Phi' - \Phi_{\mathrm{Kerr}}')
		\Vert_{C_u^{\Hp}(v(r_0,u))}^4
		du
		\right)^{\frac{1}{2}}
		\lesssim
		\frac{\varepsilon^3}{\tau^{s+\frac{1}{2}}}.
	\end{align*}
	The second term on the right hand side of \eqref{eq:3Msecondterm} can be controlled using the fact that
	\[
		\int_{\DRH(\tau)}
		\vert
		\Phi_{\mathrm{Kerr}}''
		\vert
		\vert
		\mathfrak{D}^{\gamma_1} (\Phi' - \Phi_{\mathrm{Kerr}}')
		\vert^2
		\lesssim
		\frac{\varepsilon}{u_f}
		\Vert
		\mathfrak{D}^{\gamma_1} (\Phi' - \Phi_{\mathrm{Kerr}}')
		\Vert_{\DRH(\tau)}^2
		\lesssim
		\frac{\varepsilon^3}{\tau^s},
	\]
	using either of \eqref{eq:3Mformerlatter} when $s=0$, or the integrated decay estimate \eqref{eq:PhiHintegratedlo}
	when $s=1$ or $2$.  The third term on the right hand side of \eqref{eq:3Msecondterm} is estimated similarly.  For the final term one simply uses the fact that
	\[
		\int_{\DRH(\tau)}
		\vert
		\Phi_{\mathrm{Kerr}}''
		\vert
		\vert
		\Phi_{\mathrm{Kerr}}'
		\vert^2
		\lesssim
		\frac{\varepsilon^3}{(u_f)^3} \int_{\DRH(\tau)} \Omega^2 d \theta du dv
		\lesssim
		\frac{\varepsilon^3}{(u_f)^2}.
	\]
\end{proof}

We also note the following estimate whose proof is much easier than that of the previous propositions and hence left to the reader.

\begin{proposition}[Spacetime nonlinear error estimate in the $\Hp$ gauge] \label{prop:wavehorizonerror3}
	For $s=0,1,2$, $0 \leq k \leq N-s$ and $\tau \geq v_{-1}$, for any nonlinear error of the form $\mathcal{E}^{*k}_{\Hp}$ we have
	\[
		\int_{\DRH(\tau)}
		\vert
		\mathcal{E}^{*k}_{\Hp}
		\vert^2
		\Omega^2 d\theta du dv 
		\lesssim
		\frac{\varepsilon^4}{\tau^{1+s}},
	\]
where for $s=1,2$ the estimate also holds for $\mathcal{E}^{k}_{\Hp}$ replacing $\mathcal{E}^{*k}_{\Hp}$.
\end{proposition}

We conclude with an estimate on cones that will be used frequently:
\begin{proposition}[Nonlinear error estimate on null hypersurfaces in the $\Hp$ gauge] \label{prop:wavehorizonerror4}
	For $s=0,1,2$, $0 \leq k \leq N-1-s$ and $\tau \geq v_{-1}$, for any nonlinear error of the form $\mathcal{E}^{k}_{\Hp}$ we have
	\[
		\int_{\underline{C}^{\Hp}_\tau}
		\vert
		\mathcal{E}^{k}_{\Hp}
		\vert^2
		\Omega^2 d\theta du + \int_{{C}^{\Hp}_u(\tau)}
		\vert
		\mathcal{E}^{k}_{\Hp}
		\vert^2
		 d\theta dv 
		\lesssim
		\frac{\varepsilon^4}{\tau^{1+s}}.
	\]
\end{proposition}
\begin{proof}
A given term in the nonlinear error $\mathcal{E}^{k}_{\Hp}$ takes the form $\mathfrak{D}^{\gamma_1} \Phi \cdot \mathfrak{D}^{\gamma_2} \Phi'$ where $\vert \gamma_1 \vert + \vert \gamma_2 \vert \leq N-1-s$.  Suppose, without loss of generality, $\vert \gamma_1 \vert \geq \vert \gamma_2 \vert$. The proof follows from the fact that
$\vert \mathfrak{D}^{\gamma_2} \Phi' \vert \lesssim \frac{\varepsilon}{v}$ and $\vert| \mathfrak{D}^{\gamma_1} \Phi \vert|_{S_{u,v}} \lesssim \frac{\varepsilon}{v^{s/2}}$.
\end{proof}

\subsection{Main error terms in the $\I$ gauge} \label{sec:mainerrorIgauge}

By the pointwise estimate \eqref{pointwiseboundhereimprecalled}, each $\Phi^{\I}_p$ satisfies the pointwise estimates
\begin{align} \label{baspw}
	\sum_{\vert \gamma \vert \leq N-6}
	\vert r^p \mathfrak{D}^{\gamma} \Phi_p^{\I} \vert
	\lesssim
	\frac{\varepsilon}{\sqrt{u}} ,
	\qquad
	\sum_{\vert \gamma \vert \leq N-6}
	\vert r^{p-\frac{1}{2}} \mathfrak{D}^{\gamma} \Phi_p^{\I} \vert
	\lesssim
	\frac{\varepsilon}{u},
\end{align}
and, by the estimate for $\mathbb{E}^N_{u_f,\I}$ in the bootstrap assumption \eqref{eq:bamain}, the integrated decay estimates
\begin{align} \label{basiled}
	\sum_{\vert \gamma \vert \leq N}
	\Vert r^{p-\frac{1}{2} - \frac{\delta}{2}} \mathfrak{D}^{\gamma} \Phi_p^{\I} \Vert^2_{\DRI} \lesssim \varepsilon^2,
\qquad
	\sum_{\vert \gamma \vert \leq N-1}
	\Vert r^{p-1 - \frac{\delta}{2}} \mathfrak{D}^{\gamma} \Phi_p^{\I} \Vert^2_{\DRI(u)} \lesssim \frac{\varepsilon^2}{u}.
\end{align}
Recall the $\check{\mathcal{E}}^k_p$ nonlinear error notation introduced in Section \ref{subsec:tildeerrors}.

\begin{proposition}[Spacetime nonlinear error estimate in the $\I$ gauge] \label{prop:waveIerror1}
	For $s=0,1$, $0 \leq k \leq N-s$, $0 \leq \vert \gamma \vert \leq N-s$ and $\tau \geq u_{-1}$, for any nonlinear error of the form $\mathcal{E}^{k}_2$ and for any $\Phi^{\I}_p$,
	\[
		\int_{\DRI(\tau)}
		r^{-s}
		\vert
		r^p
		\mathfrak{D}^{\gamma} \Phi_p 
		\vert
		\vert
		\mathcal{E}^{k}_{2}
		\vert
		\Omega^2 d\theta du dv 
		\lesssim
		\frac{\varepsilon^3}{\tau^{1+s}}.
	\]
Moreover, the same estimate holds replacing $\mathcal{E}^k_2$ by $\check{\mathcal{E}}^k_2$. 
\end{proposition}

\begin{proof}
	Recall that each term in the error $\mathcal{E}^k_2$ takes the form $r^{-2} r^{p_1} \mathfrak{D}^{\gamma_1}\Phi_{p_1}' \cdot r^{p_2} \mathfrak{D}^{\gamma_2} \Phi_{p_2}''$, for appropriate geometric quantities $\Phi_{p_1}'$ and $\Phi_{p_2}''$, where $\vert \gamma_1 \vert + \vert \gamma_2 \vert \leq N-s$.  Suppose, without loss of generality, $\vert \gamma_1 \vert \geq \vert \gamma_2 \vert$.  
	
	Consider first the case $s=0$.  Then, since $N\geq 12$ it follows that $\vert \gamma_2 \vert \leq N-6$ and so
	\begin{align} \label{kedeco}
		\vert r^{p_2-\frac{1}{2}} \mathfrak{D}^{\gamma_2} \Phi_{p_2}'' \vert \lesssim \frac{\varepsilon}{u},
		\qquad
		\text{and}
		\qquad
		\Vert r^{p_1-\frac{1}{2} - \frac{\delta}{2}} \mathfrak{D}^{\gamma_1} \Phi_{p_1}' \Vert_{\DRI}
		+
		\Vert r^{p-\frac{1}{2} - \frac{\delta}{2}} \mathfrak{D}^{\gamma} \Phi_{p} \Vert_{\DRI}
		\lesssim
		\varepsilon.
	\end{align}
	It follows that
	\[
		\int_{\DRI(\tau)}
		r^{-2}
		\vert
		r^p
		\mathfrak{D}^{\gamma} \Phi_p 
		\vert
		\vert
		r^{p_1} \mathfrak{D}^{\gamma_1}\Phi_{p_1}'
		\vert
		\vert
		r^{p_2} \mathfrak{D}^{\gamma_2} \Phi_{p_2}''
		\vert
		\lesssim
		\varepsilon
		\Vert
		r^{p-\frac{1}{2}-\delta} \mathfrak{D}^{\gamma} \Phi_p
		\Vert_{\DRI(\tau)}
		\Vert
		r^{p_1-\frac{1}{2}-\delta} \mathfrak{D}^{\gamma_1} \Phi_{p_1}''
		\Vert_{\DRI(\tau)}
		\lesssim
		\frac{\varepsilon^3}{\tau}.
	\]
	
	Consider now the case $s=1$.  Since $N\geq 12$, it again follows that $\vert \gamma_2 \vert \leq N-6$ and so,
	\[
		\vert r^{p_2-\frac{1}{2}} \mathfrak{D}^{\gamma_2} \Phi_{p_2}'' \vert \lesssim \frac{\varepsilon}{u},
		\qquad
		\text{and}
		\qquad
		\Vert r^{p_1-1 - \frac{\delta}{2}} \mathfrak{D}^{\gamma_1} \Phi_{p_1}' \Vert^2_{\DRI(\tau)}
		+
		\Vert r^{p-1 - \frac{\delta}{2}} \mathfrak{D}^{\gamma} \Phi_{p} \Vert^2_{\DRI(\tau)}
		\lesssim
		\frac{\varepsilon^2}{\tau}.
	\]
	It follows that
	\begin{multline*}
		\int_{\DRI(\tau)}
		r^{-3}
		\vert
		r^p
		\mathfrak{D}^{\gamma} \Phi_p 
		\vert
		\vert
		r^{p_1} \mathfrak{D}^{\gamma_1}\Phi_{p_1}'
		\vert
		\vert
		r^{p_2} \mathfrak{D}^{\gamma_2} \Phi_{p_2}''
		\vert
		\lesssim
		\frac{\varepsilon}{\tau}
		\Vert
		 r^{p-1-\frac{\delta}{2}} \mathfrak{D}^{\gamma} \Phi_p
		\Vert_{\DRI(\tau)}
		\Vert
		r^{p_1-1-\frac{\delta}{2}} \mathfrak{D}^{\gamma_1} \Phi_{p_1}'
		\Vert_{\DRI(\tau)}
		\lesssim
		\frac{\varepsilon^3}{\tau^{2}}. \nonumber
	\end{multline*}
Finally, the estimate replacing $\mathcal{E}^k_2$ by $\check{\mathcal{E}}^k_2$ follows in exactly the same fashion using in addition that $| \check{r} r^{-1}|\leq 2$ and $|\Omega^{-2l}| \leq C_l$.
\end{proof}

\begin{proposition}[Spacetime nonlinear error estimate in the $\I$ gauge] \label{prop:waveIerror1b}
For $0 \leq k \leq N$ and $\tau \geq u_{-1}$ we have, for $t \in \{0,1\}$,
		\[
		\int_{\DRI(\tau)}
		\left( r^{\frac{3}{2}+t} \vert
		\mathcal{E}^{k}_{2}
		\vert^2 +r^{\frac{3}{2}+t} \vert
		\check{\mathcal{E}}^{k}_{2} 
		\vert^2\right)
		\Omega^2 d\theta du dv 
		\lesssim
		\frac{\varepsilon^4}{\tau^{2-t}}.
	\]
\end{proposition}
\begin{proof}
The proof follows immediately from (\ref{kedeco}) following the proof of the previous proposition.
\end{proof}

\begin{proposition}[Spacetime nonlinear error estimate in the $\I$ gauge] \label{prop:waveIerror2}
	For $s=0,1$, $0 \leq k \leq N-s$, and $\tau \geq u_{-1}$, for any nonlinear error of the form $\mathcal{E}^{k}_2$ and for any $S$-tensor $W$,
	\[
		\int_{\DRI(\tau)}
		\vert
		W
		\vert
		\vert
		\mathcal{E}^{k}_{2}
		\vert
		\Omega^2 d\theta du dv
		\lesssim
		\frac{\varepsilon^2}{\tau^{1+\frac{s}{2}(1-2\delta)}} \Vert r^{-\frac{1}{2} - \frac{\delta}{2}} W\Vert_{\DRI(\tau)}.
	\]
Moreover, the same estimate holds replacing $\mathcal{E}^k_2$ by $\check{\mathcal{E}}^k_2$. 
\end{proposition}

\begin{proof}
	Recall that each term in the error $\mathcal{E}^k_2$ takes the form $r^{-2} r^{p_1} \mathfrak{D}^{\gamma_1}\Phi_{p_1} \cdot r^{p_2} \mathfrak{D}^{\gamma_2} \Phi_{p_2}$ where $\vert \gamma_1 \vert + \vert \gamma_2 \vert \leq N-s$.  Suppose, without loss of generality, $\vert \gamma_1 \vert \geq \vert \gamma_2 \vert$.  Then, since $N\geq 12$ it follows that $\vert \gamma_2 \vert \leq N-6$ and so
	\[
		\vert r^{p_2-\frac{1}{2}} \mathfrak{D}^{\gamma_2} \Phi_{p_2}^{\I} \vert \lesssim \frac{\varepsilon}{u},
		\qquad
		\text{and}
		\qquad
		\Vert r^{p_1-\frac{1}{2} - \frac{\delta}{2}-\frac{s}{2}} \mathfrak{D}^{\gamma_1} \Phi_{p_1}^{\I} \Vert^2_{\DRI(\tau)} \lesssim \frac{\varepsilon^2}{\tau^{s}}.
	\]
	It follows that
	\begin{multline*}
		\int_{\DRI(\tau)}
		r^{-2}
		\vert
		W 
		\vert
		\vert
		r^{p_1} \mathfrak{D}^{\gamma_1}\Phi_{p_1}
		\vert
		\vert
		r^{p_2} \mathfrak{D}^{\gamma_2} \Phi_{p_2}
		\vert
		\\
		\lesssim
		\frac{\varepsilon}{\tau}
		\Vert
		r^{-\frac{1}{2}-\frac{\delta}{2}} W
		\Vert_{\DRI(\tau)}
		\Vert
		r^{p_1-\frac{1}{2}-\frac{\delta}{2}-\frac{1-2\delta}{2}} \mathfrak{D}^{\gamma_1} \Phi_{p_1}
		\Vert_{\DRI(\tau)}
		\lesssim
		\frac{\varepsilon^3}{\tau^{1+ \frac{s}{2}(1-2\delta)}}
		\Vert
		r^{-\frac{1}{2}-\frac{\delta}{2}} W
		\Vert_{\DRI(\tau)}.
	\end{multline*}
\end{proof}

We conclude with an estimate on cones which will be used frequently:
\begin{proposition}[Nonlinear error estimate on null hypersurfaces in the $\I$ gauge] \label{prop:waveIerror1c}
For $0 \leq k \leq N-1$ and $\tau \geq u_{-1}$ we have for $t \in \{0,1\}$:
		\[
		\int_{C^{\I}_\tau}
		\left( r^{-\frac{1}{2}+t} \vert
		 \mathcal{E}^{k}_{1}
		\vert^2 + r^{-\frac{1}{2}+t} \vert
		\check{\mathcal{E}}^{k}_{1} 
		\vert^2\right)
		 dv d\theta + \int_{\underline{C}^{\I}_v(\tau)}
		\left( r^{-\frac{1}{2}+t}  \vert
		 \mathcal{E}^{k}_{1}
		\vert^2 +r^{-\frac{1}{2}+t}  \vert
		\check{\mathcal{E}}^{k}_{1} 
		\vert^2\right)
		 du d\theta
		\lesssim
		\frac{\varepsilon^4}{\tau^{2-t}}.
	\]
\end{proposition}
\begin{proof}
The proof follows immediately from (\ref{kedeco}) and $\| \mathfrak{D}^{N-1} r^p \Phi_p\|_{S_{u,v}} \lesssim \varepsilon$ together with the fact that $r^{-3/2}$ is integrable both in $u$ and $v$.
\end{proof}

\subsection{Anomalous error terms in the $\I$ gauge}
In the equations satisfied by $\check{\Pbar}$ (see Proposition \ref{prop:nabla4pschematic} and Proposition \ref{eq:RWPbartilde}) there are several anomalous error terms which are estimated separately from the others.  Accordingly, given $l$, $k_1$, $k_2$, $k_3$, define
\[
	\mathcal{F}^{k_1k_2k_3} = r^2 (r\nablaslash)^{k_1} (r\Omega \nablaslash_4)^{k_2} (\Omega \omegahat - \Omega \omegahat_{\circ}) \otimes (r\nablaslash)^{k_3} \alphabar,
	\quad
	\mathcal{G}^{l} = r^3 \betabar \otimes (r\Omega \nablaslash_4)^l r \nablaslash (\Omega \omegahat),
	\quad
	\mathring{\mathcal{G}} = r^4 (\Omega\hat{\omega} - \Omega\hat{\omega}_{\circ})  \Dslash_2^\star \betabar.
\]

\begin{proposition}[Anomalous error estimate in the $\I$ gauge] \label{prop:waveIerroranom1}
	For $s=0,1,2$, $0 \leq \vert \gamma \vert \leq N-3-s$, $k_1+k_2+k_3\leq 3$,  $k_3\leq 2$ and $\tau \geq u_{-1}$, and for any $S$-tensor $W$,
	\[
		\int_{\DRI(\tau)}
		\vert
		W
		\vert
		\vert
		\mathfrak{D}^{\gamma}
		\mathcal{F}^{k_1k_2k_3}
		\vert
		\Omega^2 d\theta du dv
		\lesssim
		\frac{\varepsilon^2}{\tau^{\frac{s}{2}}} \Vert r^{-\frac{1}{2} - \frac{\delta}{2}} W\Vert_{\DRI(\tau)}.
	\]
\end{proposition}

\begin{proof}
	Clearly
	\[
		\vert \mathfrak{D}^{\gamma} \mathcal{F}^{k_1k_2k_3} \vert
		\lesssim
		r^2
		\sum_{\substack{\vert \gamma_1\vert + \vert \gamma_2 \vert \leq N-s \\ \vert \gamma_2\vert \leq N-1-s}}
		\vert \mathfrak{D}^{\gamma_1} (\Omega \omegahat - \Omega\omegahat_{\circ})\vert
		\vert \mathfrak{D}^{\gamma_2} \alphabar \vert.
	\]
	Now consider $\gamma_1$, $\gamma_2$ such that $\vert \gamma_1 \vert \leq N-s$, $\vert \gamma_2\vert \leq N-1-s$ and $\vert \gamma_1\vert + \vert \gamma_2 \vert \leq N-s$.  Using the fact that the estimate for $\mathbb{E}^N_{u_f,\I}$ in the bootstrap assumption \eqref{eq:bamain} and the pointwise estimate \eqref{pointwiseboundhereimprecalled} imply that
	\[
	 	\sum_{\vert \gamma \vert \leq N}
		\Vert r^3 \mathfrak{D}^{\gamma} (\Omega \omegahat - \Omega\omegahat_{\circ})\Vert_{S_{u,v}} \lesssim \varepsilon,
		\qquad
		\sum_{\vert \gamma \vert \leq N-5}
		\vert r^3 \mathfrak{D}^{\gamma} (\Omega \omegahat - \Omega\omegahat_{\circ}) \vert \lesssim \varepsilon,
	\]
	respectively, it follows that
	\[
		\int_{\DRI(\tau)}
		\vert
		W
		\vert
		r^2
		\vert \mathfrak{D}^{\gamma_1} (\Omega \omegahat - \Omega\omegahat_{\circ})\vert
		\vert \mathfrak{D}^{\gamma_2} \alphabar \vert
		\Omega^2 d\theta du dv
		\lesssim
		\varepsilon
		\sum_{\vert \gamma_3 \vert \leq N-1-s}
		\Vert
		r^{-\frac{1}{2}-\frac{\delta}{2}} W
		\Vert_{\DRI(\tau)}
		\Vert r^{-\frac{1}{2}+\frac{\delta}{2}} \mathfrak{D}^{\gamma_3} \alphabar \Vert_{\DRI(\tau)}.
	\]
The claim now follows from the fact that $\Vert r^{\frac{1}{2}-\frac{\delta}{2}} \mathfrak{D}^{\gamma_3} \alphabar \Vert_{\DRI(\tau)} \lesssim \varepsilon \tau^{-\frac{s}{2}}$ for $\vert \gamma_3 \vert \leq N-s$ by the estimate for $\mathbb{E}^N_{u_f,\I}$ in the bootstrap assumption \eqref{eq:bamain}.
\end{proof}

\begin{proposition}[Anomalous error estimate in the $\I$ gauge] \label{prop:waveIerroranom2}
	For $s=0,1$, $0 \leq \vert \gamma_1 \vert \leq N-s$, $0 \leq \vert \gamma_2 \vert \leq N-3-s$, $k_1+k_2+k_3\leq 3$, $k_1, k_3\leq 2$, $k_2 \leq 1$ and $\tau \geq u_{-1}$, and for any $\Phi^{\I}_p$,
	\[
		\int_{\DRI(\tau)}
		\vert
		r^p
		\mathfrak{D}^{\gamma_1} \Phi_p^{\I}
		\vert
		\vert
		\mathfrak{D}^{\gamma_2} \mathcal{F}^{k_1k_2k_3}
		\vert
		\Omega^2 d\theta du dv
		\lesssim
		\frac{\varepsilon^3}{\tau^{s}}.
	\]
\end{proposition}

\begin{proof}
	Repeating the proof of Proposition \ref{prop:waveIerroranom1}, we again have
	\[
		\int_{\DRI(\tau)}
		\vert
		r^p
		\mathfrak{D}^{\gamma_1} \Phi_p 
		\vert
		\vert
		\mathfrak{D}^{\gamma_2} \mathcal{F}^{k_1k_2k_3}
		\vert
		\Omega^2 d\theta du dv
		\lesssim
		\varepsilon
		\sum_{\vert \gamma_3 \vert \leq N-1-s}
		\Vert
		r^{p-\frac{3}{2}+\frac{\delta}{2}}
		\mathfrak{D}^{\gamma_1} \Phi_p
		\Vert_{\DRI(\tau)}
		\Vert r^{\frac{1}{2}-\frac{\delta}{2}} \mathfrak{D}^{\gamma_3} \alphabar \Vert_{\DRI(\tau)}.
	\]
The claim follows from the fact that $\Vert r^{\frac{1}{2}-\frac{\delta}{2}} \mathfrak{D}^{\gamma_3} \alphabar \Vert_{\DRI(\tau)} \lesssim \varepsilon \tau^{-\frac{s}{2}}$ and $\Vert r^{p-1-\frac{\delta}{2}} \mathfrak{D}^{\gamma_1} \Phi_p \Vert_{\DRI(\tau)} \lesssim \varepsilon \tau^{-\frac{s}{2}}$.
\end{proof}

\begin{proposition}[Anomalous error estimate in the $\I$ gauge] \label{prop:waveIerroranom3}
	For $s=0,1$, $0 \leq \vert \gamma \vert \leq N-3-s$, $k_1+k_2+k_3\leq 3$, $k_1, k_3 \leq 2$, $k_2\leq  1$, $l=0,1$, and $\tau \geq u_{-1}$,
	\[
		\int_{\DRI(\tau)}
		r^{2}
		\vert
		\mathfrak{D}^{\gamma} \mathcal{G}^l 
		\vert
		\vert
		\mathfrak{D}^{\gamma} \mathcal{F}^{k_1k_2k_3}
		\vert
		\Omega^2 d\theta du dv
		\lesssim
		\frac{\varepsilon^4}{\tau^{s}}.
	\]
The same estimate holds replacing $\mathcal{G}^l$ by $\mathring{\mathcal{G}}$.
\end{proposition}

\begin{proof}
	Clearly
	\[
		\vert \mathfrak{D}^{\gamma} \mathcal{F}^{k_1k_2k_3} \vert
		\lesssim
		r^2
		\!\!\!\!
		\sum_{\substack{\vert \gamma_1\vert + \vert \gamma_2 \vert \leq N-s \\ \vert \gamma_2\vert \leq N-1-s}}
		\!\!\!\!
		\vert \mathfrak{D}^{\gamma_1} (\Omega \omegahat - \Omega\omegahat_{\circ})\vert
		\vert \mathfrak{D}^{\gamma_2} \alphabar \vert,
		\qquad
		\vert \mathfrak{D}^{\gamma} \mathcal{G}^{l} \vert
		\lesssim
		r^2
		\!\!\!\!
		\sum_{\substack{\vert \gamma_3\vert + \vert \gamma_4 \vert \leq N-1-s \\ \vert \gamma_4\vert \leq N-3-s}}
		\!\!\!\!
		\vert \mathfrak{D}^{\gamma_3} (\Omega \omegahat - \Omega\omegahat_{\circ})\vert
		\vert \mathfrak{D}^{\gamma_4} \betabar \vert,
	\]
	and so, using that fact that $\Vert r^3 \mathfrak{D}^{\gamma} (\Omega \omegahat - \Omega\omegahat_{\circ})\Vert_{S_{u,v}} \lesssim \varepsilon$ for all $\vert \gamma \vert \leq N$, together with the pointwise estimate \eqref{pointwiseboundhereimprecalled}, it follows that,
	\[
		\int_{\DRI(\tau)}
		r^{2}
		\vert
		\mathfrak{D}^{\gamma} \mathcal{G}^l 
		\vert
		\vert
		\mathfrak{D}^{\gamma} \mathcal{F}^{k_1k_2k_3}
		\vert
		\Omega^2 d\theta du dv
		\lesssim
		\varepsilon^2
		\sum_{\vert \gamma_1 \vert \leq N-1-s}
		\Vert
		r^{\frac{1}{2} - \frac{\delta}{2}} \mathfrak{D}^{\gamma_1} \alphabar
		\Vert_{\DRI(\tau)}
		\sum_{\vert \gamma_2 \vert \leq N-3-s}
		\Vert
		r^{\frac{1}{2} + \frac{\delta}{2}} \mathfrak{D}^{\gamma_2} \betabar
		\Vert_{\DRI(\tau)},
	\]
	and the claim follows from the fact that $\Vert r^{\frac{1}{2} - \frac{\delta}{2}} \mathfrak{D}^{\gamma} \alphabar \Vert_{\DRI(\tau)} + \Vert r^{\frac{3}{2} - \frac{\delta}{2}} \mathfrak{D}^{\gamma} \betabar \Vert_{\DRI(\tau)} \lesssim \varepsilon \tau^{-\frac{s}{2}}$ for $\vert \gamma \vert \leq N-s$. The claim about $\mathring{\mathcal{G}}$ follows easily by repeating the above proof.
\end{proof}

\begin{proposition}[Anomalous error estimate in the $\I$ gauge] \label{prop:waveIerroranom4}
	For $s=0,1$, $0 \leq k \leq N-s$, $0 \leq \vert \gamma \vert \leq N-3-s$, $l=0,1$ and $\tau \geq u_{-1}$, for any nonlinear error of the form $\mathcal{E}^{k}_2$,
	\[
		\int_{\DRI(\tau)}
		r^{2-s}
		\vert
		\mathfrak{D}^{\gamma} \mathcal{G}^l
		\vert
		\vert
		\mathcal{E}^{k}_{2}
		\vert
		\Omega^2 d\theta du dv + \int_{\DRI(\tau)}
		r^{2-s}
		\vert
		\mathfrak{D}^{\gamma} \mathring{\mathcal{G}}
		\vert
		\vert 
		\mathcal{E}^{k}_{2}
		\vert
		\Omega^2 d\theta du dv
		\lesssim
		\frac{\varepsilon^4}{\tau^{1+s}}.
	\]
	Moreover, the same estimate holds replacing $\mathcal{E}^k_2$ by $\check{\mathcal{E}}^k_2$. 
\end{proposition}

\begin{proof}
The proof follows from respectively applying Proposition \ref{prop:waveIerror2} with $W= r^{1-\frac{s}{2}} \mathfrak{D}^\gamma \mathcal{G}^l$ and $W= r^{1-s/2} \mathfrak{D}^\gamma \mathring{\mathcal{G}}$ and noting that pointwise estimate \eqref{pointwiseboundhereimprecalled} implies
	\[
		\int_{\DRI(\tau)}
		r^{-1-\delta} r^{-1-s}
		r^3 |
		\mathfrak{D}^{\gamma} \mathcal{G}^l |^2
		\Omega^2 d\theta du dv
		 +\int_{\DRI(\tau)}
		r^{-1-\delta} r^{-1-s}
		r^3 |
		\mathfrak{D}^{\gamma} \mathring{\mathcal{G}} |^2
		\Omega^2 d\theta du dv
		\lesssim
		\varepsilon^2 \frac{1}{\tau^2} \, .
	\]
\end{proof}

\section{Some auxiliary estimates for commuted energies} \label{sec:auxcommuted}
When commuting tensorial wave equations for symmetric traceless $S$-tensors we will (as is already clear from 
Section~\ref{sec:comrw}) typically commute with the following modified set of commutation operators:\index{double null gauge!differential operators!$\tilde{\mathfrak{D}}^{\underline{k}}$, modified commutation operator}\index{double null gauge!differential operators!$\tilde{\mathfrak{D}}_{aux}^{\underline{k}}$, auxiliary modified commutation operator}
\begin{align} \label{def:dfraktilde}
\tilde{\mathfrak{D}}^{\underline{k}} = \tilde{\mathfrak{D}}^{(k_1,k_2,k_3)} :=  \left\{
\begin{array}{rl}
\left(\Omega^{-1} \slashed{\nabla}_3\right)^{k_2} \left(r\Omega \slashed{\nabla}_4\right)^{k_3} \left(r^2 \slashed{\Delta}\right)^{k_1/2} &  \text{if } k_1  \ \ \text{even} \\
\left(\Omega^{-1} \slashed{\nabla}_3\right)^{k_2} \left(r\Omega \slashed{\nabla}_4\right)^{k_3}\left(r^2 \slashed{\Delta}\right)^{(k_1-1)/2} r \slashed{div} & \text{if } k_1  \ \ \text{odd}
\end{array} \right.
\end{align}
and
\begin{align}
\tilde{\mathfrak{D}}_{aux}^{\underline{k}} = \tilde{\mathfrak{D}}_{aux}^{(k_1,k_2,k_3)} =  \left\{
\begin{array}{rl}
\left(\slashed{\nabla}_{R^\star}\right)^{k_3}  \left(r^2 \slashed{\Delta}\right)^{k_1/2} \left(\slashed{\nabla}_T\right)^{k_2}  &  \text{if } k_1  \ \ \text{even} \\
 \left(\slashed{\nabla}_{R^\star}\right)^{k_3} \left(r^2 \slashed{\Delta}\right)^{(k_1-1)/2} r \slashed{div} \left(\slashed{\nabla}_T\right)^{k_2}& \text{if } k_1  \ \ \text{odd},
\end{array} \right.
\end{align}
where as usual $|\underline{k}|:=k_1+k_2+k_3$ is the length. 

Note the difference with $\mathfrak{D}^{\underline{k}}$ is that the angular derivatives are $\slashed{div}$ and $\slashed{\Delta}$ (both of which are elliptic operators on the spheres of the double null foliation acting on symmetric traceless tensors) and that derivatives are taken in a slightly different order.

The next proposition asserts that controlling the energy of $\tilde{\mathfrak{D}}^{\underline{k}} W$ is equivalent to controlling that of $\mathfrak{D}^{\underline{k}} W$. Similarly, in a region of $r_{\Hp} \geq 9M_{\rm init}/4$ and $r_{\I} \leq 2R$ respectively, controlling $\tilde{\mathfrak{D}}^{\underline{k}} W$ is equivalent to controlling $\mathfrak{D}^{\underline{k}} W$.

\begin{proposition} \label{sec:ellipticconesI}
Let $W$ denote a symmetric traceless tensor in the $\I$-gauge. We have for $1 \leq K \leq N$ and $u_1 \leq \tau \leq u_f$ the estimates
\begin{align}  \label{sampleelliptic}
\sum_{|\underline{k}|=0, k_2 \neq K}^{K}\int_{\check{C}^{\I}_\tau} |\mathfrak{D}^{\underline{k}} W|^2 \lesssim \sum_{|\underline{k}|=0, k_2 \neq K}^{K}\int_{\check{C}^{\I}_\tau} |\tilde{\mathfrak{D}}^{\underline{k}} W|^2 \, , 
\end{align}
\begin{align}
\sum_{|\underline{k}|=0, k_3 \neq K}^{K}\int_{\check{\underline{C}}^{\I}_v (\tau)} |\mathfrak{D}^{\underline{k}} W|^2 \lesssim \sum_{|\underline{k}|=0, k_3 \neq K}^{K}\int_{\check{\underline{C}}^{\I}_v (\tau)} |\tilde{\mathfrak{D}}^{\underline{k}} W|^2 \, , 
\end{align}
\begin{align}
\sum_{|\underline{k}|=0}^{K}\int_{\check{\mathcal{D}}^{\I} (\tau)} |\mathfrak{D}^{\underline{k}} W|^2 \lesssim \sum_{|\underline{k}|=0}^{K}\int_{\check{\mathcal{D}}^{\I} (\tau)} |\tilde{\mathfrak{D}}^{\underline{k}} W|^2 \, , 
\end{align}
where $\check{\underline{C}}^{\I}_v (\tau)$ is any truncated ingoing cone such that also $\check{\underline{C}}^{\I}_v (\tau, 
\tau+M) \subset \mathcal{D}^{\I}$. 

In addition, for each estimate we can replace $\tilde{\mathfrak{D}}^{\underline{k}}$ by $\tilde{\mathfrak{D}}^{\underline{k}}_{aux}$ on the right provided we restrict the region of integration to $r_{\I} \leq 2R$ in the integrals on the left.
\end{proposition}

\begin{remark}
The condition on the ingoing cone is a technical condition that ensures that the truncated cone is long enough to be bounded by two proper spheres in order to perform the elliptic estimates in the proof. In practise one can always extend an arbitrary ingoing cone in $\check{\mathcal{D}}^{\I}$ slightly to the past if necessary before applying the Proposition. 
\end{remark}

One similarly has for the horizon region:

\begin{proposition} \label{sec:ellipticconesHp}
Let $W$ denote a symmetric traceless tensor in the $\Hp$-gauge. We have for $1 \leq K \leq N$ and  $u_1 \leq \tau \leq u_f$  the estimates
\begin{align}
\sum_{|\underline{k}|=0, k_2 \neq K}^{K}\int_{\check{C}^{\Hp}_{u}(v(\tau))} |\mathfrak{D}^{\underline{k}} W|^2 \lesssim \sum_{|\underline{k}|=0, k_2 \neq K}^{K}\int_{\check{C}^{\Hp}_{u}(v(\tau))} |\tilde{\mathfrak{D}}^{\underline{k}} W|^2 \, , 
\end{align}
\begin{align}
\sum_{|\underline{k}|=0, k_3 \neq K}^{K}\int_{\check{\underline{C}}^{\Hp}_{v (\tau)}} |\mathfrak{D}^{\underline{k}} W|^2 \lesssim \sum_{|\underline{k}|=0, k_3 \neq K}^{K}\int_{\check{\underline{C}}^{\Hp}_{v (\tau)}} |\tilde{\mathfrak{D}}^{\underline{k}} W|^2 \, ,
\end{align}
\begin{align}
\sum_{|\underline{k}|=0}^{K}\int_{\check{\mathcal{D}}^{\Hp} (v(\tau))} |\mathfrak{D}^{\underline{k}} W|^2 \lesssim \sum_{|\underline{k}|=0}^{K}\int_{\check{\mathcal{D}}^{\Hp} (v(\tau))} |\tilde{\mathfrak{D}}^{\underline{k}} W|^2 \, ,
\end{align}
where $\check{C}^{\Hp}_{u}(v(\tau))$ is any outgoing cone which is such that $\check{C}^{\Hp}_{u}(v(\tau), v(\tau)+M) \subset  \check{\mathcal{D}}^{\Hp}$.

In addition, for each estimate we can replace $\tilde{\mathfrak{D}}^{\underline{k}}$ by $\tilde{\mathfrak{D}}^{\underline{k}}_{aux}$ on the right provided we restrict the region of integration to $r_{\Hp} \geq 9M_{\rm init}/4$ in the integrals on the left.
\end{proposition}

\begin{proof}
These estimates are standard, the only complication being the fact that the cones are truncated and hence do not necessarily end in a sphere of the double null foliation. We sketch the proof of (\ref{sampleelliptic}).
One proves the statement successively for $K=1,...,N$. For fixed $K$, one proves the estimate successively for the tuples $\underline{k}=(1,k_2,k_3)$ with $k_2+k_3=K-1$,  $\underline{k}=(2,k_2,k_3)$ with $k_2+k_3=K-2$, until we reach the tuple $\underline{k}=(K,0,0)$. For instance, for the tuple $\underline{k}=(1,k_2,k_3)$ with $k_2+k_3=K-1$ we have
\begin{align}
\int_{\check{C}^{\I}_\tau} | r \slashed{\nabla} \left(\Omega^{-1} \slashed{\nabla}_3\right)^{k_2} \left(r \Omega \slashed{\nabla}_4\right)^{k_3} W|^2 &\lesssim  \int_{\check{C}^{\I}_\tau} | r \slashed{div} \left(\Omega^{-1} \slashed{\nabla}_3\right)^{k_2} \left(r \Omega \slashed{\nabla}_4\right)^{k_3} W|^2 \nonumber \\
& \ \ \ \ + C\varepsilon \int_{\check{C}^{\I}_\tau \cap\{r \leq R_2\}} | \left(\Omega \slashed{\nabla}_4\right) \left(\Omega^{-1} \slashed{\nabla}_3\right)^{k_2} \left(r \Omega \slashed{\nabla}_4\right)^{k_3} W|^2 \nonumber \\
&\lesssim \sum_{|\underline{k}|=0, k_2 \neq K}^{K}\int_{\check{C}^{\I}_{u}(v(\tau))} |\tilde{\mathfrak{D}}^{\underline{k}} W|^2
\end{align}
with the last step following easily from commuting the angular derivative through and using the pointwise bootstrap assumptions ((\ref{useonlyLinfinityestimate}) is sufficient) on the commutator terms. 

To see the first inequality it is sufficient to consider $k_2=k_3=0$. Note first that if the cones were not truncated at a sphere which is (in general) not part of the double null foliation, the first inequality would be true without the $\varepsilon$-term by standard elliptic estimates on the spheres of the cone. 

To reduce the case at hand to standard elliptic estimates on spheres, one re-foliates $\check{C}^{\I}_\tau$ by spheres such that the spheres $\mathcal{B} \cap {\check{C}^{\I}_\tau}$ and the spheres $S_{\tau,v}$ for $v \geq v(\tau,R_2)$ are now part of the new foliation. From $v(s,\theta^\prime)=v^\prime + f(v^\prime,\theta^\prime=\theta)$ one computes from the old frame $\left(\Omega^{-1}e_3, \Omega e_4=\partial_v, e_A=\partial_{\theta^A}\right)$  new frame on $\check{C}^{\I}_\tau \cap \{ r \leq R_2\}$ (see also Proposition \ref{prop:outIconesrelations})
\[
e_A^\prime = e_A + \slashed{\nabla}_A f \cdot \Omega e_4 \ \ \ , \ \ \ e^\prime_4 = \left(1+\partial_{v^\prime} f\right)  \Omega e_4 \ \ \ , \ \ \ e_3^\prime = \frac{1}{1+\partial_{v^\prime} f} \left( \Omega^{-1} e_3 + |\slashed{\nabla}f|^2 \Omega e_4 +2 \slashed{\nabla}^A f e_A\right)
\]
with $e_1^\prime, e_2^\prime$ tangent to the new spheres and with $\|f\|_{C^2(\check{C}^{\I}_\tau)} \lesssim \varepsilon$ by the bootstrap assumptions. Note $\slashed{g}^\prime_{AB} = g(e_A^\prime, e_B^\prime) = g(e_A,e_B)=\slashed{g}_{AB}$. From the symmetric traceless $S_{u,v}$-tensor $W$ we construct a symmetric traceless $S_{u,v^\prime}$-tensor $W^\prime$ through the definition ($\otimes_S$ denoting the symmetric tensor product)
{\small
\[
W^\prime = W - \frac{1}{(1+\partial_{v^\prime} f)^2} g(e^\prime_4, \cdot) \otimes  g(e^\prime_4, \cdot) \left[W (e_A,e_B) \slashed{\nabla}^A f\slashed{\nabla}^B f \right]  + \frac{1}{1+\partial_{v^\prime} f} g(e^\prime_4, \cdot) \otimes_S \slashed{g}^{AC}  g\left(e^\prime_A , \cdot \right) \left[ W (e_C, e_B) \slashed{\nabla}^B  f \right] \, .
\]
}
Indeed, $W^\prime$ is symmetric and once checks $W^\prime(e_A^\prime, e_B^\prime)=W(e_A,e_B)$ and $W^\prime(e_3^\prime, \cdot)=W^\prime(e_4^\prime, \cdot)=0$. Now let $\overset{\prime}{{\slashed{\nabla}}}$ (acting on $S_{u,v^\prime}$-tensors) denote the projection of the covariant derivative to the $S_{u,v^\prime}$-spheres. We have the relation
\begin{align} \label{tensorrel1}
\overset{\prime}{{\slashed{\nabla}}}_{e_C^\prime} W^\prime \left(e^\prime_A,e^\prime_B\right) &= e_C^\prime \left(W^\prime (e^\prime_A,e^\prime_B) \right) -  W^\prime \left(\overset{\prime}{{\slashed{\nabla}}}_{e_C^\prime} e^\prime_A,e^\prime_B\right)  -W^\prime \left(e^\prime_A,\overset{\prime}{{\slashed{\nabla}}}_{e_C^\prime}  e^\prime_B\right)    \nonumber \\ 
&=  \left[e_C + \slashed{\nabla}_C f \cdot \Omega e_4\right]  \left(W (e_A,e_B) \right)  -  W^\prime \left(\overset{\prime}{{\slashed{\nabla}}}_{e_C^\prime} e^\prime_A,e^\prime_B\right)  -W^\prime \left(e^\prime_A,\overset{\prime}{{\slashed{\nabla}}}_{e_C^\prime}  e^\prime_B\right)   \nonumber \\
&=\slashed{\nabla}_{e_C} W \left(e_A,e_B\right) + \varepsilon \cdot \mathcal{O}(W) + \varepsilon \cdot \mathcal{O}(\Omega \slashed{\nabla}_{4} W) \ , 
\end{align}
in particular
\begin{align} \label{tensorrel2}
\overset{\prime}{\slashed{div}} W^\prime \left(e_A^\prime\right)= \slashed{div} W \left(e_A\right)  + \varepsilon \cdot \mathcal{O}(W) + \varepsilon \cdot \mathcal{O}(\slashed{\nabla}_{e_4} W)  \, .
\end{align}
Using these relations we show
\begin{align}
 \int_{\check{C}^{\I}_\tau \cap \{r \leq R_2\}}  | r \slashed{div}  W|^2 &\gtrsim \int_{\check{C}^{\I}_\tau \cap \{r \leq R_2\}}  | r \overset{\prime}{\slashed{div}}  W^\prime |^2 -  \varepsilon | \Omega \slashed{\nabla}_{4}  W|^2 - \varepsilon |W|^2 \\
 &\gtrsim  \int_{\check{C}^{\I}_\tau \cap \{r \leq R_2\}}  | r  \overset{\prime}{\slashed{\nabla}} W^\prime |^2 -  \varepsilon | \Omega \slashed{\nabla}_{4}  W|^2 
\gtrsim \int_{\check{C}^{\I}_\tau \cap \{r \leq R_2\}} | r {\slashed{\nabla}} W |^2 -  \varepsilon | \Omega \slashed{\nabla}_{4}  W|^2 \, . \nonumber
\end{align}
 Here the first and the third inequality follow from (\ref{tensorrel2}) and (\ref{tensorrel1}), while the second one is the standard elliptic estimate on the new spheres.
\end{proof}

\chapter{Estimates for $P$ and  $\protect\underline{P}$: the proof of Theorem~\ref{thm:PPbarestimates}}
\label{chapter:psiandpsibar}
In this section, we prove Theorem~\ref{thm:PPbarestimates}, which we restate here:

\PPbarestimates*

\minitoc

We will give a complete overview of the chapter in {\bf Section~\ref{sec:overviewP}} below, which
will contain the skeleton of the proof, divided into a main subtheorem and various propositions. 
The subsequent sections of the chapter will flesh out this skeleton with the proof of these results. 
We defer a detailed summary 
of the contents of the remaining sections till then.

\vskip1pc
\noindent\fbox{
    \parbox{6.35in}{
As in the previous chapters of Part~\ref{improvingpart},
we shall assume throughout the assumptions of~\Cref{havetoimprovethebootstrap}. Let us fix an
arbitrary  $u_f\in[u_f^0, \hat{u}_f$], with $\hat{u}_f\in \mathfrak{B}$,
and fix some $\lambda \in \mathfrak{R}(u_f)$.
All propositions below
shall always refer  
to the anchored $\I$ and $\Hp$ gauges in the  spacetime  $(\mathcal{M}(\lambda), g(\lambda))$,  
corresponding to parameters
$u_f$, $M_f(u_f,\lambda)$,
whose existence is
ensured by Definition~\ref{bootstrapsetdef}.
We shall denote $M=M_f$ throughout  
this chapter.
}}

\vskip1pc

\emph{In addition to the statements of 
Theorems~\ref{thm:sobolevandelinfinity} and~\ref{thm:relatinggauges}, this chapter depends on the propositions of Chapter~\ref{RWtypechapter}. The reader who is content to understand
the basic structure of the proof of Theorem~\ref{thm:PPbarestimates} may wish to read only Section~\ref{sec:overviewP}.
For the
linear version of Theorem~\ref{thm:PPbarestimates}, the reader
may compare with Theorem~1 in Section~10  of~\cite{holzstabofschw}. }

\section{Overview} \label{sec:overviewP}

The quantities $P$ and $\underline{P}$ are the primal quantities of our almost gauge invariant hierarchy, 
and 
we recall from Proposition \ref{prop:whythat}, that in their various manifestations
$P_{\Hp}$, $P_{\I}$, $\Pbar_{\Hp}$, $\check{\Pbar}_{\I}$, the quantities
 satisfy a \emph{homogeneous up to non-linear error-terms} tensorial wave equation of Type 1.

 We  define the three regions depicted in Figure~\ref{thethreeregions}:
\begin{enumerate}
\item[I.] $\mathcal{D}^{\I}  \cap (\{ u_{\I} \leq u_1\} \cup \{v_{\Hp} \leq v_1 \} ) \subset \mathcal{D}^{\I} \cap \mathcal{D}^{\mathcal{EF}}(u_3)$, the near (Eddington Finkelstein) data region
\item[II.] $\mathcal{D}^{\mathcal{H}^+} \cap (\{v_{\mathcal{H}^+}\le v_1\}\cup \{u_{\mathcal{I}^+}\le u_1\}) \subset \mathcal{D}^{\mathcal{H}^+} \cap \mathcal{D}^{\mathcal{K}}(V_4)$, the near (Kruskal) initial data region
\item[III.]  $ \DcI\left(u_1\right) \cup \DcH\left(v_1=v(u_1)\right)$, the main region.
\end{enumerate}
\begin{figure}
\centering{
\def\svgwidth{20pc}
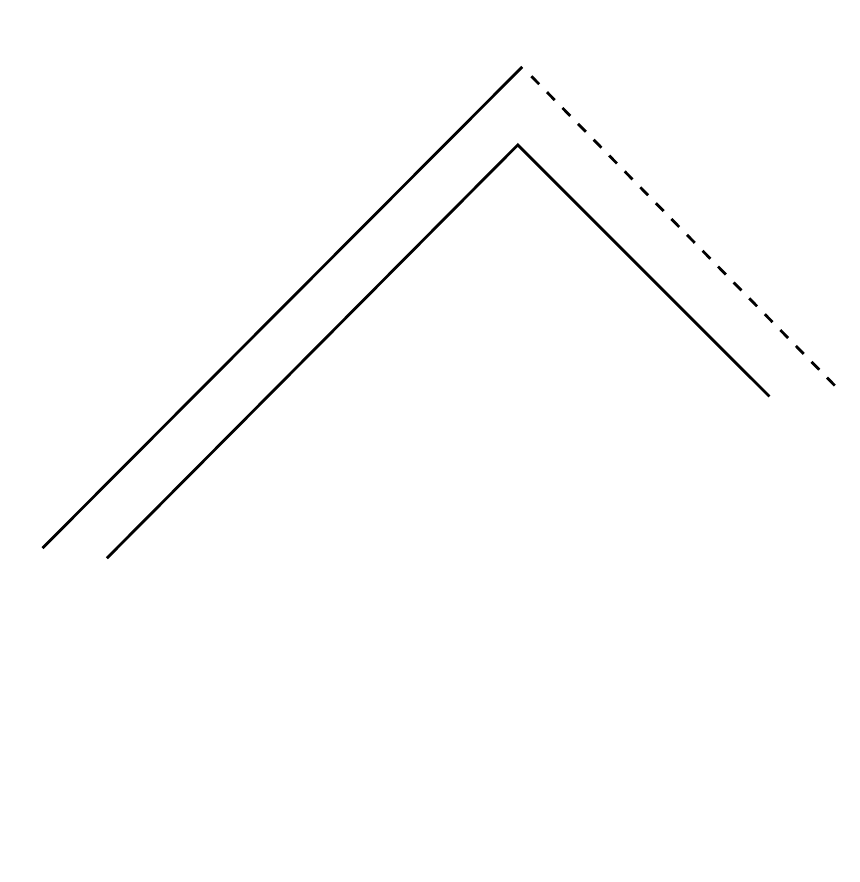}
\caption{The regions I, II and III}\label{thethreeregions}
\end{figure}

Let us note that we may prove Theorem~\ref{thm:PPbarestimates} by obtaining the improved
estimate
separately in the above regions, i.e.
examining the fluxes and spacetime
integrals in
\begin{equation}
\label{presenthereforhere}
\mathbb{E}_{u_f}^{N-2} [P_{\Hp}, P_{\I}] + \mathbb{E}_{u_f}^{N-2} [\Pbar_{\Hp}, \check{\Pbar}_{\I}],
\end{equation}
showing that their restrictions to both the regions I+II and III  are estimated
by $\varepsilon_0^2+\varepsilon^3$.

In the ``main'' region III, we shall estimate these quantities using the wave-equation estimates of
Chapter~\ref{RWtypechapter}. We must first however dispense with regions I+II.

\subsection{Estimates in regions I+II}
\label{estimatinregIplusII}
Obtaining the estimates in regions I+II follows directly from comparison with the respective initial data gauges,
in view of the almost gauge invariant property of the quantities and the bootstrap assumption. 

We have already accomplished this!
In particular, from Proposition \ref{thm:gidataestimates} we have
\begin{align}
\label{estimI+IIhere}
\mathbb{E}_0^{N-2}[P_{\Hp}] + \mathbb{E}_0^{N-2}[\underline{P}_{\Hp}] + \mathbb{E}_0^{N-2}[P_{\I}] + \mathbb{E}_0^{N-2}[\check{\underline{P}}_{\I}] \lesssim \varepsilon_0^2 + \varepsilon^3.
\end{align}
Hence, examining the fluxes present in~\eqref{presenthereforhere},
we see that
all ingoing and outgoing fluxes restricted to regions I+II are included in the left hand side of 
\eqref{estimI+IIhere} and thus are indeed estimated by $\varepsilon_0^2+\varepsilon^3$ as required.

From this one easily concludes that also all the restrictions to regions I+II of the spacetime integrals  
present in~\eqref{presenthereforhere} are similarly estimated by $\varepsilon_0^2+\varepsilon^3$.

\subsection{A preliminary weaker set of estimates in region III}
In view of~Section~\ref{estimatinregIplusII}, we have reduced Theorem \ref{thm:PPbarestimates}
to proving estimates for the fluxes and spacetime integrals present in~\eqref{presenthereforhere}
\emph{restricted to region III}, i.e.~in
$\DcI\left(u_1\right) \cup \DcH\left(v_1\right)$, bounding these by $\varepsilon_0^2 + \varepsilon^3$.

In that regard we will first prove the following weaker set of estimates:

\begin{theorem} \label{theo:Pmainregion2}
The quantities $\Psi_{\Hp}=r^5 P_{\Hp}$, $\Psi_{\I} = r^5 P_{\I}$, $\underline{\Psi}_{\Hp} = r^5\Pbar_{\Hp}$, ${\underline{\Psi}}_{\I} = r^5\check{\Pbar}_{\I}$ satisfy the estimates
\begin{align}
\sum_{s=0,1,2}
	\sup_{u_{1} \leq \tau \leq u_f} \tau^s \cdot {\check{\mathbb{E}}}_{out}^{N-2-s,2-s} \left[P_{\I}\right] \left(\tau\right)  + \sum_{s=0,1,2}
	\sup_{v(u_1) \leq v} 
	v^{s} \cdot {\check{\mathbb{E}}}_{in}^{N-2-s} \left[P_{\Hp}\right] \left(v\right) \lesssim \varepsilon_0^2 + \varepsilon^3\, ,  \label{michelin1} \\ 
	\sum_{s=0,1,2}
	\sup_{u_1 \leq \tau \leq u_f} \tau^s \cdot \check{\mathbb{E}}_{out}^{N-2-s,2-s} \left[\check{\Pbar}_{\I}\right] \left(\tau\right)  + \sum_{s=0,1,2}
	\sup_{v(u_1) \leq v} 
	v^{s} \cdot \check{\mathbb{E}}_{in}^{N-2-s} \left[\underline{P}_{\Hp}\right] \left(v\right)\lesssim   \varepsilon_0^2 + \varepsilon^3  \, , \label{michelin2}
\end{align}
where for $1 \leq K \leq N-2$
\begin{align} 
\check{\mathbb{E}}_{in}^{K} \left[P_{\Hp}\right] \left(v\right) :=& \sup_{\tilde{v} \geq v} \int_{\check{\underline{C}}^{\Hp}_{\tilde{v}}} \Omega^2  \sum_{|\underline{k}|=0; k_3\neq K}^{K}  | \mathfrak{D}^{\underline{k}} {\Psi}_{\Hp}|^2 
 \\
+&\int_{\check{\mathcal{D}}^{\Hp}\left(v\right)}  \Omega^2 \Bigg\{ \sum_{|\underline{k}|=0}^{K} \left(1-\frac{3M_f}{r}\right)^2 | \mathfrak{D}^{\underline{k}} {\Psi}_{\Hp}|^2  +  \sum_{|\underline{k}|=0}^{K-1}  | \mathfrak{D}^{\underline{k}} {\Psi}_{\Hp}|^2+  | R^\star \mathfrak{D}^{\underline{k}} {\Psi}_{\Hp}|^2  \Bigg\}   \nonumber \, , 
\end{align}
\begin{align} 
\check{\mathbb{E}}_{out}^{K,p} \left[P_{\I}\right] \left(\tau\right) &:= 
\sup_{\tau \leq u \leq u_f} \int_{\check{C}^{\I}_u} \Bigg\{ \sum_{|\underline{k}|=0}^{K-1}  \  r^p | \Omega \slashed{\nabla}_4 \mathfrak{D}^{\underline{k}} \Psi_{\I}|^2 + \frac{1}{r^2} | r \slashed{\nabla} \mathfrak{D}^{\underline{k}} \Psi_{\I}|^2 \Bigg\} \\
+&\int_{\check{\mathcal{D}}^{\I}\left(\tau\right)}  \Bigg\{ \sum_{|\underline{k}|=0}^{K-1}  \   \frac{1}{r^{1+\delta}}  | \Omega \slashed{\nabla}_3 \mathfrak{D}^{\underline{k}} \Psi_{\I}|^2 + r^{p-1-\delta \cdot {\boldsymbol\delta^p_0}} | \Omega \slashed{\nabla}_4 \mathfrak{D}^{\underline{k}} \Psi_{\I}|^2 + \frac{r^p}{r^{3+\delta \cdot {\bf \delta^p_2}}} | r \slashed{\nabla} \mathfrak{D}^{\underline{k}} \Psi_{\I}|^2 \Bigg\}  \nonumber
\end{align}
and the same definitions replacing $P_{\Hp}, \Psi_{\Hp}$ by $\underline{P}_{\Hp}, \underline{\Psi}_{\Hp}$ and ${P}_{\I}, \Psi_{\I}$ by $\check{\Pbar}_{\I}, \check{\underline{\Psi}}_{\I}$.

\end{theorem}

Comparing the energies $\check{\mathbb{E}}_{in}^{K} \left[P_{\Hp}\right] \left(v\right)$, $\check{\mathbb{E}}_{out}^{K,p} \left[P_{\I}\right] \left(\tau\right)$ defined in Theorem \ref{theo:Pmainregion2} with the energies defined in (\ref{energyPhoz}), (\ref{energyPinf}) we see that:
\begin{itemize}
\item For $P_{\Hp}$ (and $\underline{P}_{\Hp}$), the energies are restricted to the truncated (at $\mathcal{B}$) ingoing cones $\underline{\check{C}}_v^{\Hp}$ in $\DcH$  and integrated decay estimates on truncated regions $\DcH(\tau)$. However, they do not contain fluxes on outgoing cones in $\DcH$.

\item For $P_{\I}$ (and $\check{\Pbar}_{\I}$), the energies  are restricted to the truncated (at $\mathcal{B}$) outgoing cones $\check{C}_u^{\I}$ in $\DcI$ and integrated decay estimates on truncated regions $\DcI(\tau)$. However, they do not contain fluxes on ingoing cones in $\DcI$.
\end{itemize}
Note already (see Section~\ref{seethisbelownow}) 
that Theorem~\ref{thm:PPbarestimates} will follow from Theorem \ref{theo:Pmainregion2} if we can replace the energies $\check{\mathbb{E}}_{in}^{K} \left[P_{\Hp}\right] \left(v\right)$, $\check{\mathbb{E}}_{out}^{K,p} \left[P_{\I}\right] \left(\tau\right)$ by $\mathbb{E}^{K} \left[P_{\Hp}\right] \left(v\right)$, $\mathbb{E}^{K,p} \left[P_{\I}\right] \left(\tau\right)$ (and similarly for the underlined quantities) respectively in the estimate of Theorem \ref{theo:Pmainregion2}.

The proof of  Theorem \ref{theo:Pmainregion2} proceeds by controlling the left hand side by the ``initial" energies on the cones $\check{C}^{\I}_{u_1}$ and $\check{\underline{C}}^{\Hp}_{v_1}$, namely 
\begin{align} \label{newinitialr3}
 &\int_{\check{C}^{\I}_{u_1}} \sum_{|\underline{k}|=0}^{N-3}  \Big\{   r^2 | \Omega \slashed{\nabla}_4 \mathfrak{D}^{\underline{k}} \Psi_{\I}|^2 + \frac{1}{r^2} | r \slashed{\nabla} \mathfrak{D}^{\underline{k}} \Psi_{\I}|^2+r^2 | \Omega \slashed{\nabla}_4 \mathfrak{D}^{\underline{k}} \underline{\check{\Psi}}_{\I}|^2 + \frac{1}{r^2} | r \slashed{\nabla} \mathfrak{D}^{\underline{k}} \underline{\check{\Psi}}_{\I}|^2 \Big\}  \nonumber \\
 +&\int_{\check{\underline{C}}^{\Hp}_{v_1}} \Omega^2  \sum_{|\underline{k}|=0; k_3\neq N-2}^{N-2} \Big\{ | \mathfrak{D}^{\underline{k}} {\Psi}_{\Hp}|^2+  | \mathfrak{D}^{\underline{k}} \underline{\Psi}_{\Hp}|^2  \Big\}
  \lesssim   \varepsilon_0^2 + \varepsilon^3 \, ,
\end{align}
the inequality being an immediate consequence of (\ref{estimI+IIhere}). We apply the estimates proven in Chapter \ref{typeequationsection} for general tensorial wave equations to obtain boundedness and integrated decay estimates for $\Psi_{\I}$ and $\Psi_{\Hp}$ (and $\check{\underline{\Psi}}_{\I}$ and $\underline{\Psi}_{\Hp}$), as well as $r$-weighted estimates for $\Psi_{\I}$ (and $\check{\underline{\Psi}}_{\I}$) in the infinity region. This is the content of {\bf Sections~\ref{sec:bidp}} and~{\bf \ref{sec:rpwp}}. Here we will use that
\begin{enumerate}
\item  the boundary terms on $\mathcal{B}$ appearing in these estimates cancel up to non-linear terms  which can in turn be controlled by the bootstrap assumptions (see Proposition \ref{thm:cancelT} included as part of the statement of Theorem~\ref{thm:relatinggauges}), and
\item  the non-linear errors are controlled by the bootstrap assumptions (see Section \ref{nonlinearestforwave}).
\end{enumerate}
In {\bf Section \ref{sec:hdp}} we prove higher derivatives estimates. This is based on applying the estimates of Sections \ref{sec:bidp} and \ref{sec:rpwp} to suitably commuted equations. The scheme is as follows:
\begin{enumerate}
\item We first commute the equation for $\Psi$ with the  operator $\slashed{\nabla}_T$ as well as the angular operators $r\slashed{div}$ and $r^2 \slashed{\Delta}$ only. These operators commute up to non-linear terms and (in the case of $r\slashed{div}$) a lower order linear term which can be controlled inductively. We then commute by the operator $\slashed{\nabla}_{R^\star}$ whose commutator can be controlled by the previous estimates.\footnote{Alternatively, one could recover the $R^\star$-derivatives directly from the equation. However, since non-degenerate control near $r=3M$ is required, a further integration by parts argument (near $3M$) would be necessary when deriving the estimates.} We exploit here that the commutator of $\slashed{\Box}$ with $\slashed{\nabla}_{R^\star}$ behaves nicely near trapping (the top order term degenerates linearly at $r=3M$, see the last part of Proposition \ref{prop:commutetype1}). With this, \emph{all} derivatives of $\Psi$ are controlled but with non-optimal weights near the horizon for $\Psi_{\Hp}$ and near null infinity for $\Psi_{\I}$.

\item We improve the weights near the horizon for $\Omega \slashed{\nabla}_3$-derivatives of $\Psi_{\Hp}$ by commuting the equation for $\Psi_{\Hp}$ with a (suitably cut-off) redshift operator $\Omega^{-1} \slashed{\nabla}_3$ and applying Proposition \ref{prop:redshift}. 

\item We improve the weights near null infinity for $\Omega \slashed{\nabla}_4$-derivatives of $\Psi_{\I}$ by commuting the equation for $\Psi_{\I}$ with the (suitably cut-off) operator $r\Omega \slashed{\nabla}_4 $ and applying Proposition \ref{prop:rph}. 
\end{enumerate}

In {\bf Section \ref{sec:decp}}, we run a  pigeonhole argument in the style of~\cite{DafRodnew}, providing inverse polynomial decay estimates for higher derivatives. We finally obtain Proposition~\ref{prop:Psummarydecay} from which 
Theorem~\ref{theo:Pmainregion2} is easily deduced.

\subsection{Completing the proof of Theorem~\ref{thm:PPbarestimates}}
\label{seethisbelownow}

Once Theorem  \ref{theo:Pmainregion2}  has been obtained we use, 
in {\bf Section \ref{sec:missingfluxp}}, the estimates of Theorem \ref{theo:Pmainregion2} and the estimates on the diffeomorphisms relating $\Psi_{\Hp}$ and $\Psi_{\I}$ to deduce the estimates 
on the restriction to region~III of the full fluxes and spacetime integrals contained
in the definition of the energies~\ref{presenthereforhere},
required to obtain
Theorem~\ref{thm:PPbarestimates}: 

We first show that Theorem \ref{theo:Pmainregion2} continues to hold replacing $\DcI$ by $\mathcal{D}^{\I} \cap \left(\DcI\left(u_1\right) \cup \DcH\left(v_1\right)\right)$ and $\DcH$ by $\mathcal{D}^{\Hp}\cap \left(\DcI\left(u_1\right) \cup \DcH\left(v_1\right)\right)$, and 
adding the flux \[
\sup_{u\leq u_f} \int_{\check{C}^{\Hp}_u({v})} \sum_{|\underline{k}|=0; k_1\neq K}^{K}  | \mathfrak{D}^{\underline{k}} \Psi_{\Hp}|^2
\]
to $\check{\mathbb{E}}_{in}^{K} \left[P_{\Hp}\right] \left(v\right)$ and the flux 
\[
\sup_{v \leq v_\infty} \int_{\check{\underline{C}}^{\I}_{v}(\tau)}  \Big\{ \sum_{|\underline{k}|=0}^{K-1}  \   | \Omega \slashed{\nabla}_3 \mathfrak{D}^{\underline{k}} \Psi_{\I}|^2 + \frac{r^p}{r^2} | r \slashed{\nabla} \mathfrak{D}^{\underline{k}} \Psi_{\I}|^2 \Big\}
\]
to $\check{\mathbb{E}}_{out}^{K,p} \left[P_{\I}\right] \left(\tau\right)$. We finally show that truncated cones can be replaced by non-truncated ones.

(For the last step, some localised energy estimates in regions of bounded $r$ (involving null cones that are not part of the respective null-foliation) have to be applied, however, these require no special structure.)

\section{Non-degenerate boundedness and integrated decay} \label{sec:bidp}
\begin{proposition} \label{prop:basicallder}
For any $u_1 \leq \tau \leq u_f$, the pair $({\Psi}_{\I}= r^5 {P}_{\I}$, ${\Psi}_{\Hp}= r^5 {P}_{\Hp})$  satisfies
\begin{enumerate}
\item degenerate boundedness:
\begin{align} \label{Tbndd}
\sup_{\tau \leq u \leq u_f} \check{\underline{\mathbb{F}}}^\diamond_{v(u)} [\Psi_{\Hp}] + \sup_{\tau \leq u \leq u_f} \check{\mathbb{F}}_u [\Psi_{\I}] 
 + \check{{\mathbb{F}}}^\diamond_{u_f} \left[\Psi_{\Hp} \right] \left(v(\tau)\right) 
       \nonumber \\   \lesssim
 \underline{\check{\mathbb{F}}}^\diamond_{v(\tau)} [ \Psi_{\Hp}]
+
\check{\mathbb{F}}_{\tau} [\Psi_{\I}]  
 +  \nldq \, ,
\end{align}

\item boundedness: 
\begin{align} \label{Tbnd}
\sup_{\tau \leq u \leq u_f} \check{\underline{\mathbb{F}}}_{v(u)} [\Psi_{\Hp}] + \sup_{\tau \leq u \leq u_f} \check{\mathbb{F}}_u [\Psi_{\I}] 
 + \check{{\mathbb{F}}}_{u_f} \left[\Psi_{\Hp} \right] \left(v(\tau)\right) + \check{\underline{\mathbb{F}}}_{v_\infty} \left[\Psi_{\I} \right] \left(\tau\right)
\nonumber \\    \lesssim
 \ \  \underline{\check{\mathbb{F}}}_{v(\tau)} [ \Psi_{\Hp}]
+
\check{\mathbb{F}}_{\tau} [\Psi_{\I}]  
 + \nldq \, ,
\end{align}

\item  integrated decay: 
\begin{align} \label{Tiledd}
\check{\mathbb{I}}^{\diamond, deg} \left[\Psi_{\Hp}\right] \left(v(\tau)\right)  + \check{\mathbb{I}} \left[\Psi_{\I}\right] \left(\tau\right) 
 \lesssim
        & \ \      \underline{\check{\mathbb{F}}}^\diamond_{v(\tau)} [ \Psi_{\Hp}]
+
\check{\mathbb{F}}_{\tau} [\Psi_{\I}]  +\nldq \, , 
\end{align}
\begin{align} \label{Tiled}
\check{\mathbb{I}}^{deg} \left[\Psi_{\Hp}\right] \left(v(\tau)\right)  + \check{\mathbb{I}} \left[\Psi_{\I}\right] \left(\tau\right) 
 \lesssim
        & \ \      \underline{\check{\mathbb{F}}}_{v(\tau)} [ \Psi_{\Hp}]
+
\check{\mathbb{F}}_{\tau} [\Psi_{\I}]  +\nldq \, .
\end{align}
\end{enumerate}Moreover, the pair $(\check{\underline{\Psi}}_{\I}= r^5 \check{\underline{P}}_{\I}$, $\underline{\Psi}_{\Hp}= r^5 \underline{P}_{\Hp})$ satisfies the same estimates.
\end{proposition}

\begin{proof}
We first prove (\ref{Tbndd}) and (\ref{Tiledd}). We note that
both pairs $({\Psi}_{\I},{\Psi}_{\Hp})$ and $(\check{\underline{\Psi}}_{\I},\underline{\Psi}_{\Hp})$ satisfy a tensorial wave equation of Type 1 by Proposition \ref{prop:whythat}.\footnote{We focus on the proof for the $({\Psi}_{\I},{\Psi}_{\Hp})$ pair, the underlined pair being entirely analogous.} We can hence apply Proposition \ref{prop:basecase}.  We have $\mathcal{G}_i  \left[
\Psi\right] \left(\tau_1,\tau_2\right)=0$. In view of $\mathcal{F}^{nlin}_1[\Psi]=\mathcal{E}^{3}_2$ we have from Proposition \ref{thm:cancelT} as well as Propositions \ref{prop:wavehorizonerror1}, \ref{prop:wavehorizonerror2}, \ref{prop:waveIerror2} the estimate
\[
\sum_{i=1}^3 \Big| \mathcal{B}_i \left[
\Psi_{\Hp}\right] (\tau_1,\tau_2)-  \mathcal{B}_i \left[\Psi_{\I}\right]  (\tau_1,\tau_2)\Big| + \sum_{i=1}^3 \mathcal{H}_i \left[\Psi\right] (\tau,u_f) \lesssim \nldq \, .
\] 
For $\check{\underline{\Psi}}_{\I}$, we have to use in addition Proposition \ref{prop:waveIerroranom1} in view of the anomalous error term appearing in $\mathcal{F}^{nlin}_1 \left[\check{\underline{\Psi}}_{\I}\right]$; 
see Proposition \ref{prop:whythat} and in particular (\ref{eq:alphabaromegaanomalouserror}). This proves (\ref{Tbndd}) and (\ref{Tiledd}).
 
To obtain (\ref{Tbnd}) and (\ref{Tiled}), we observe first that $\Psi_{\Hp}$ satisfies a tensorial wave equation of type $3_{0}$ by Proposition \ref{prop:classify3k}. Apply now Proposition \ref{prop:redshift}. Note that the energies carrying a $\star$ in (\ref{rsdr}) can be controlled by (\ref{Tbndd}) and (\ref{Tiledd}). The linear error  $\mathcal{G}_4 \left[\Psi_{\Hp}\right]$ can be estimated
\[
\mathcal{G}_4 \left[\Psi_{\Hp}\right] \left(\tau\right) \lesssim \eta \check{\mathbb{I}}^{deg} \left[\Psi_{\Hp}\right] \left(\tau\right) + C_\eta \check{\mathbb{I}}^{\diamond, deg} \left[\Psi_{\Hp}\right]\left(\tau\right)
\]
for any $\eta>0$. The first term on the right can be absorbed on the left of (\ref{rsdr}) and the second is controlled by (\ref{Tiledd}). Finally, by Propositions \ref{prop:wavehorizonerror1}, \ref{prop:wavehorizonerror2} we have $\mathcal{H}_4\left[\Psi_{\Hp}\right]\lesssim \nldq$ for the non-linear error. Combining the resulting estimate with (\ref{Tbndd}) and (\ref{Tiledd}) we deduce (\ref{Tbnd}) and (\ref{Tiled}).
\end{proof}

\section{The $r^p$-weighted estimate} \label{sec:rpwp}

We also have a $p$-weighted estimate near infinity:

\begin{proposition} \label{prop:Pimproved}
For any $u_1 \leq \tau \leq u_f$ and $1 \leq p \leq 2$,
the pair $({\Psi}_{\I}= r^5 P_{\I}$, ${\Psi}_{\Hp}= r^5 {P}_{\Hp})$  satisfies
\begin{align} \label{Pbasic}
                &
        \sup_{\tau \leq u \leq u_f} \check{\underline{\mathbb{F}}}_{v(u)} [\Psi_{\Hp}]  +       \sup_{\tau \leq u \leq u_f} \check{\mathbb{F}}_u [\Psi_{\I}] +
                \sup_{\tau \leq u \leq u_f} \check{\mathbb{F}}^{p}_u [\Psi_{\I}] + \underline{\check{\slashed{\mathbb{F}}}}^{p}_{v_\infty} \left[\Psi_{\I}\right](\tau)
                \nonumber \\
                &
                \qquad \qquad +\check{\mathbb{I}}^{deg} \left[\Psi_{\Hp}\right] \left(v(\tau)\right)  + \check{\mathbb{I}} \left[\Psi_{\I}\right] \left(\tau\right) + \check{\mathbb{I}}^p \left[\Psi_{\I}\right] \left(\tau\right) 
                \nonumber \\
                &
                \qquad \qquad \qquad \qquad
                \lesssim
                \underline{\check{\mathbb{F}}}_{v(\tau)} [ \Psi_{\Hp}]
+
\check{\mathbb{F}}_{\tau} [\Psi_{\I}] +
\check{\mathbb{F}}^p_{\tau} [\Psi_{\I}]  
 +\frac{\varepsilon_0^2 + \varepsilon^3}{\tau} \, .
\end{align}
Moreover, the pair $(\check{\underline{\Psi}}_{\I}= r^5 \underline{\check{P}}_{\I}$, $\underline{\Psi}_{\Hp}= r^5 \underline{P}_{\Hp})$ satisfies the same estimate.
\end{proposition}
\begin{proof}
Note that $\Psi_{\I}$ (and $\check{\underline{\Psi}}_{\I}$) satisfies tensorial wave equation of type $4_{0,2}$ by Proposition \ref{prop:classify4kl}. Hence we can apply Proposition \ref{prop:rph}. Note there is no boxed term and the term in the square brackets of (\ref{basrpe}) can be controlled from Proposition \ref{prop:basicallder}. To control the term 
$\mathcal{G}_{5,p} \left[\Psi_{\I}\right]$ in (\ref{basrpe}) we use Cauchy-Schwarz
\[
\Big| \int_{\DcI\left(\tau\right) \cap \{ r \geq 3/2R\}} \frac{6M\Omega^2}{r^3} \Psi_{\I}\cdot  \left(r^p+r^{2-\delta} \boldsymbol\delta^2_p \right)  \Omega \slashed{\nabla}_4 \Psi_{\I} \Big| \lesssim  \eta \check{\mathbb{I}}^p \left[\Psi_{\I}\right] \left(\tau,  u_f\right) +C_\eta \check{\mathbb{I}} \left[\Psi_{\I}\right] \left(\tau,  u_f\right)
\]
for any $\eta>0$ and $1\leq p \leq 2$. The second term on the right is controlled by (\ref{Tiled}) and the first can, for sufficiently small $\eta$, be absorbed on the left of (\ref{basrpe}). Of course the analogous estimate holds for $\check{\underline{\Psi}}_{\I}$. Finally, we turn to the non-linear error term for $\Psi_{\I}$ and $\check{\underline{\Psi}}_{\I}$ (see Proposition \ref{prop:whythat}):
\begin{align}
\mathcal{H}_{5,p} \left[\check{\underline{\Psi}}_{\I}\right]  \left(\tau,u_f\right) &=\int_{\DcI\left(\tau\right)  \cap \{r \geq 3R/2\}} |\left(\mathcal{F}^{nlin}_{4_{k,l}} \left[\check{\underline{\Psi}}_{\I}\right] + F_{34} \left[[\check{\underline{\Psi}}_{\I}\right] \right)| \,  r^p |\Omega \slashed{\nabla}_4 \check{\underline{\Psi}}_{\I}| \nonumber \\
&=\int_{\DcI\left(\tau\right)  \cap \{r \geq 3R/2\}} \Big| \check{\mathcal{E}}^3_2  + \mathcal{E}_{\mathrm{anom}}[\check{\Pbar}] \Big|r^p |\Omega \slashed{\nabla}_4 \check{\underline{\Psi}}_{\I}|  \, .\nonumber
\end{align}
Inserting the expression for $\slashed{\nabla}_4 \check{\underline{\Psi}}_{\I}$ from Proposition \ref{prop:nabla4pschematic}
we conclude that $\mathcal{H}_{5,p}\left[\check{\underline{\Psi}}_{\I}\right]\lesssim \frac{\varepsilon_0^2 + \varepsilon^3}{\tau}$ after applying the estimates of Propositions \ref{prop:waveIerror1}, \ref{prop:waveIerroranom2}, \ref{prop:waveIerroranom3} and \ref{prop:waveIerroranom4}. The error for $\mathcal{H}_{5,p} \left[{\Psi}_{\I}\right]$ is treated entirely analogously but is strictly easier as no anomalous term appears in $\mathcal{F}^{nlin}_{4_{k,l}} \left[{\Psi}_{\I}\right]$.
\end{proof}

\section{Higher derivative estimates} \label{sec:hdp}

We now state and prove the main decay estimate for higher derivatives, these being the analogues of Propositions \ref{prop:basicallder} and \ref{prop:Pimproved}. We recall the definition $\mathfrak{D}^{\underline{k}}$ from  Section \ref{diffopsandnonlinerrors} and the definitions of $\tilde{\mathfrak{D}}^{\underline{k}}$ and $\tilde{\mathfrak{D}}_{aux}^{\underline{k}}$ from Section \ref{sec:auxcommuted}. 

\subsection{Basic boundedness and integrated decay for higher derivatives}

\begin{proposition} \label{prop:Psummaryall}
For any $u_1 \leq \tau \leq u_f$ and any $1 \leq K\leq N-3$,
the pair $({\Psi}_{\I}= r^5 {{P}}_{\I}$, ${\Psi}_{\Hp}= r^5 {P}_{\Hp})$  satisfies
\begin{align}  \label{thde}
& \sum_{|\underline{k}|\leq K} \left[ \sup_{\tau \leq u \leq u_f} \check{\mathbb{F}}_u [\tilde{\mathfrak{D}}^{\underline{k}} \Psi_{\I}] 
                +  \sup_{\tau \leq u \leq u_f} \check{\underline{\mathbb{F}}}_{v(u)} [\tilde{\mathfrak{D}}^{\underline{k}} \Psi_{\Hp}] 
                +  \check{{\mathbb{F}}}_{u_f} \left[\tilde{\mathfrak{D}}^{\underline{k}} \Psi_{\Hp} \right] \left(v(\tau)\right) + \check{\underline{\mathbb{F}}}_{v_\infty} \left[\tilde{\mathfrak{D}}^{\underline{k}} \Psi_{\I} \right] \left(\tau\right) \right]
                \nonumber \\
                &
                +
                \sum_{|{\underline{k}}|\leq K} \check{\mathbb{I}}^{deg} \left[\tilde{\mathfrak{D}}^{\underline{k}} \Psi_{\Hp}\right] \left(v(\tau)\right) + \sum_{|{\underline{k}}|\leq K-1} \check{\mathbb{I}} \left[\tilde{\mathfrak{D}}^{\underline{k}} \Psi_{\Hp}\right] \left(v(\tau)\right) + \sum_{|{\underline{k}}|\leq K} \check{\mathbb{I}} \left[\tilde{\mathfrak{D}}^{\underline{k}} \Psi_{\I}\right] \left(\tau\right)
                \nonumber \\
                &
                \qquad \qquad
                \lesssim
                \sum_{|{\underline{k}}|\leq K}  \check{\mathbb{F}}_\tau [\tilde{\mathfrak{D}}^{\underline{k}} \Psi_{\I}]    + \sum_{|{\underline{k}}|\leq K}       \check{\underline{\mathbb{F}}}_{v(\tau)} [\tilde{\mathfrak{D}}^{\underline{k}} \Psi_{\Hp}] +\frac{\varepsilon_0^2 + \varepsilon^3}{\tau^{\min(N-3-K,2)}}
                  \, .
\end{align}
Moreover, the pair $(\check{\underline{\Psi}}_{\I}= r^5 \underline{\check{P}}_{\I}$, $\underline{\Psi}_{\Hp}= r^5 \underline{P}_{\Hp})$ satisfies the same estimate.
\end{proposition}

\begin{remark} \label{rem:hadeg}
For $W_{\Hp}$ an $S$-tensor one has of course for any $1\leq K^\prime \leq K$
\begin{align}
\sum_{|\underline{k}|\leq K^\prime-1} \check{\mathbb{I}} \left[\tilde{\mathfrak{D}}^{\underline{k}} \Psi_{\Hp}\right] \left(v(\tau)\right) \lesssim \sum_{|\underline{k}|\leq K^\prime} \check{\mathbb{I}}^{deg} \left[\tilde{\mathfrak{D}}^{\underline{k}} \Psi_{\Hp}\right] \left(v(\tau)\right) \, .
\end{align}
We have  included the non-degenerate term explicitly on the left hand side but in view of the above it suffices to prove the estimate without this term.
\end{remark}

\begin{proof}  The proof will proceed in steps. We first give an overview.
\vskip1pc
\noindent
{\bf Overview.} Note that for $K=0$ we have already proven the above estimate in Proposition \ref{prop:basicallder}. The proof is an induction on $K$. The general strategy is to commute the tensorial wave equations by appropriate derivative operators and observe that they still satisfy the tensorial wave equation up to non-linear errors and a few explicit linear terms (cf.~Section \ref{sec:comrw}). While the non-linear errors can be dealt with schematically (leading to the last term in (\ref{thde})), we will deal with the linear error terms explicitly exploiting a hierarchy in the way one performs the commutation. Finally, in any energy estimate, as before, the boundary terms appearing on $\mathcal{B}$ will always be a non-linear error that can be estimated schematically. The overview is as follows:

\begin{enumerate}
\item Structure of commuted equations and estimating the non-linear errors (Step 0).
\item Prove the estimate (\ref{thde}) with $\tilde{\mathfrak{D}}_{aux}^{\underline{k}}$ instead of $\tilde{\mathfrak{D}}^{\underline{k}}$ (Steps 1 and 2)
\begin{enumerate}
\item Prove it for tuples $\underline{k}=(k_1,k_2,k_3)$ with $k_1$ even.
\item Prove it for tuples $\underline{k}=(k_1,k_2,k_3)$ with $k_1$ odd.
\end{enumerate}
In proving (2a) and (2b), we will first prove (\ref{thde}) for the (weaker horizon) energies containing an additional diamond and then improve using the redshift estimate.
\item Prove the full estimate (\ref{thde}) for all horizon energies on the left (Step 3).
\item Prove the full estimate (\ref{thde}) for all infinity energies on the left (Step 4).
\end{enumerate}

\vskip1pc
\noindent
{\bf Step 0. Structure of commuted equations and non-linear errors.} By Proposition \ref{prop:whythat}, the quantities $\Psi_{\Hp}, \Psi_{\I}, \underline{\Psi}_{\Hp}, \check{\underline{\Psi}}_{\I}$ satisfy an equation of Type 1. Commuting the equations with $\mathfrak{D}^{\underline{k}}_{aux}$ will according to Proposition \ref{prop:commutetype1}  result in a tensorial wave equation of Type 1 with non-linear error given by 
\begin{align}
\mathcal{F}^{nlin} \left[ \tilde{\mathfrak{D}}_{aux}^{\underline{k}} {\Psi}_{\Hp} \right] &= \Omega^2(\mathcal{E}^\star)^{3+|\underline{k}|} \, ,  \\
\mathcal{F}^{nlin} \left[ \tilde{\mathfrak{D}}_{aux}^{\underline{k}} \underline{\Psi}_{\Hp} \right]&= \Omega^2(\mathcal{E}^\star)^{3+|\underline{k}|} \, ,  \\
\mathcal{F}^{nlin} \left[ \tilde{\mathfrak{D}}_{aux}^{\underline{k}} {\Psi}_{\I} \right] &= \mathcal{E}^{3+|\underline{k}|}_2 \, ,  \\ 
\mathcal{F}^{nlin} \left[ \tilde{\mathfrak{D}}_{aux}^{\underline{k}} \check{\underline{\Psi}}_{\I} \right] &= \check{\mathcal{E}}^{3+|\underline{k}|}_2  +  \tilde{\mathfrak{D}}_{aux}^{\underline{k}}  \mathcal{E}_{\mathrm{anom}}[\check{\Pbar}] \, .  \label{Fnlinex}
\end{align}
Similarly, by Proposition \ref{prop:classify3k}, $\Psi_{\Hp}$ and $\underline{\Psi}_{\Hp}$ satisfy an equation of Type $3_0$. Therefore, commuting with $\tilde{\mathfrak{D}}_{\Hp}^{\underline{k}}$ (defined below in (\ref{def:hozdo})) will according to Proposition \ref{prop:3kc} result in an equation of Type $3_0$ with non-linear error given by 
\begin{align}
\mathcal{F}^{nlin} \left[ \tilde{\mathfrak{D}}_{\Hp}^{\underline{k}} {\Psi}_{\Hp} \right] &= \Omega^2(\mathcal{E}^\star)^{3+|\underline{k}|} \, ,  \\
\mathcal{F}^{nlin} \left[ \tilde{\mathfrak{D}}_{\Hp}^{\underline{k}} \underline{\Psi}_{\Hp} \right]&= \Omega^2(\mathcal{E}^\star)^{3+|\underline{k}|} \, .
\end{align}
Finally, by Proposition \ref{prop:classify4kl}, $\Psi_{\I}$ and $\check{\underline{\Psi}}_{\I}$ satisfy an equation of Type $4_{0,2}$. Therefore, commuting with $\tilde{\mathfrak{D}}_{\I}^{\underline{k}}$ (defined below in (\ref{nido})) will according to Proposition \ref{prop:comm4kl} result in an equation of Type $4_{1/2k_3,l}$ (for some $l$) with non-linear error given by 
\begin{align}
\mathcal{F}^{nlin} \left[ \tilde{\mathfrak{D}}_{\I}^{\underline{k}} {\Psi}_{\I} \right] &= \mathcal{E}^{3+|\underline{k}|}_2 \, ,  \\ 
\mathcal{F}^{nlin} \left[ \tilde{\mathfrak{D}}_{\I}^{\underline{k}} \check{\underline{\Psi}}_{\I} \right] &= \check{\mathcal{E}}^{3+|\underline{k}|}_2  +  \tilde{\mathfrak{D}}_{\I}^{\underline{k}}  \mathcal{E}_{\mathrm{anom}}[\check{\Pbar}] \, .  \label{Fnlinex2}
\end{align}

We now recall the errors (\ref{defh1}), (\ref{defh2}), (\ref{defh3}), (\ref{defh4}), (\ref{defh5}) arising in the various energy estimates. We also recall that by the bootstrap assumptions (\ref{eq:bamain}) we have for $s=0,1,2$
\[
\| r^{-\frac{1}{2}-\frac{\delta}{2}} \slashed{\nabla}_{R^\star} \tilde{\mathfrak{D}}^{\underline{k}} \Psi_{\I}\|_{\DcI(\tau)}+\| r^{-\frac{1}{2}-\frac{\delta}{2}} \slashed{\nabla}_T \tilde{\mathfrak{D}}^{\underline{k}} \Psi_{\I}\|_{\DcI(\tau)} + \| r^{-\frac{3}{2}}\tilde{\mathfrak{D}}^{\underline{k}} \Psi_{\I}\|_{\DcI(\tau)} \lesssim \frac{\varepsilon}{\tau^{\min(1,(N-3-|\underline{k}|)/2)}}
\]
and the same estimate replacing $ \tilde{\mathfrak{D}}^{\underline{k}}$ by $\tilde{\mathfrak{D}}_{aux}^{\underline{k}}$ and by $ \tilde{\mathfrak{D}}_{\I}^{\underline{k}}$.

Using Propositions \ref{prop:wavehorizonerror1}, \ref{prop:wavehorizonerror2}  for the errors in $\mathcal{D}^{\mathcal{H}^+}$ and Propositions \ref{prop:waveIerror2}, \ref{prop:waveIerroranom1} for the errors in $\mathcal{D}^{\mathcal{I}^+}$, we conclude\footnote{Note in particular that the integrand in $\mathcal{H}_3  \left[\tilde{\mathfrak{D}}_{aux}^{\underline{k}} \Psi\right] (\tau,u_f)$, (\ref{defh3}), can be rewritten in a form such that Proposition~\ref{prop:wavehorizonerror2} directly applies.}
{\small 
\begin{align}
\sum_{|\underline{k}|\leq K} \left[ \sum_{i=1}^3  \mathcal{H}_i  \left[\tilde{\mathfrak{D}}_{aux}^{\underline{k}} \Psi\right] (\tau,u_f)  + \mathcal{H}_4  \left[\tilde{\mathfrak{D}}_{aux}^{\underline{k}} \Psi\right] (\tau,u_f)  + \mathcal{H}_4  \left[\tilde{\mathfrak{D}}_{\Hp}^{\underline{k}} \Psi\right] (\tau,u_f)  + \mathcal{H}_{5,0}  \left[\tilde{\mathfrak{D}}_{\I}^{\underline{k}} \Psi\right] (\tau,u_f)  \right]\lesssim \frac{\varepsilon_0^2 + \varepsilon^3}{\tau^{\min(N-3-K,2)}}. \nonumber
\end{align}
}
\vskip.1pc
\noindent
Using Proposition \ref{thm:cancelT}, we also conclude
\begin{align}
\sum_{i=1}^3  \sum_{|\underline{k}|\leq K} \Big|   \mathcal{B}_i \left[\tilde{\mathfrak{D}}_{aux}^{\underline{k}} \Psi_{\Hp}\right] (\tau,u_f)-  \mathcal{B}_i \left[\tilde{\mathfrak{D}}_{aux}^{\underline{k}} \Psi_{\I}\right]   (\tau,u_f)\Big| \lesssim  \frac{\varepsilon_0^2 + \varepsilon^3}{\tau^{\min(N-3-K,2)}} 
\end{align}
and the same estimate for the $\left(\underline{\Psi}_{\Hp}, \check{\underline{\Psi}}_{\I}\right)$ pair.

\vskip1pc
\noindent{\bf Step 1. Angular and $T$ derivatives.} We first prove a restricted version of (\ref{thde}) 
\begin{itemize}
\item replacing everywhere $\tilde{\mathfrak{D}}^{\underline{k}}$ by $\tilde{\mathfrak{D}}^{\underline{k}}_{aux}$
\item  restricting all tuples $k$ that appear in the sums to those with $k_3=0$, i.e.~we restrict to derivatives of the form
\begin{enumerate}[a)]
\item $\left(r^2 \slashed{\Delta}\right)^{k_1/2} (\slashed{\nabla}_T)^{k_2} $, \qquad \qquad  \ \ $k_1+k_2 \leq K$; \textrm{ $k_1$ \ \ even,}
\item $ \left(r \slashed{\Delta}\right)^{(k_1-1)/2} \left(r \slashed{div}\right)(\slashed{\nabla}_T)^{k_2} $, \ \  $k_1+k_2 \leq K$;  \textrm{$k_1$ \ \ \ odd.}
\end{enumerate}  
\end{itemize}
For $k_1$ even, the restricted (\ref{thde}) then follows directly: By Proposition \ref{prop:commutetype1}, both $\slashed{\nabla}_T$ and $r^2 \slashed{\Delta}$ commute with the tensorial wave equation up to non-linear terms discussed in Step 0. The commuted equations therefore again satisfy a tensorial wave equation of Type 1 up to non-linear terms and hence repeating the proof of Proposition \ref{prop:basicallder} (i.e.~applying successively Propositions \ref{prop:basecase} and \ref{prop:redshift}) we establish the restricted (\ref{thde}) using the non-linear estimates from Step 0.

For $k_1$ odd, by Proposition \ref{prop:commutetype1} the commuted quantity again satisfies a tensorial wave equation of 
Type~1 up to non-linear terms as above but now with an additional \emph{linear} error term arising from the (angular part of the) commutation. According to Proposition \ref{prop:commutetype1} it has the form
\begin{align}
\mathcal{F}^{lin}_1 \left[ W= \left(r \slashed{\Delta}\right)^{(k_1-1)/2} \left(r \slashed{div}\right) (\slashed{\nabla}_T)^{k_2} \Psi \right] = 3 \frac{\Omega^2}{r^2}  \left(r \slashed{\Delta}\right)^{(k_1-1)/2} \left(r \slashed{div}\right) (\slashed{\nabla}_T)^{k_2}\Psi \, .
\end{align}
We now repeat the proof of Proposition \ref{prop:basicallder}, i.e.~we apply successively Propositions \ref{prop:basecase}, \ref{prop:redshift}. When applying Proposition \ref{prop:basecase} we now need to control in addition the linear errorterms ($u_1 \leq \tau_1 \leq \tau_2 \leq u_f$)
\begin{align} 
\mathcal{G}_1 \left[W\right] \left(\tau_1,\tau_2\right) = \Big| &\int_{\DcH\left(v(\tau_1),v(\tau_2)\right)} 3 \frac{\Omega^2}{r^2}  W_{\Hp} \cdot \slashed{\nabla}_T W_{\Hp} + \int_{\DcI\left(\tau_1,\tau_2\right)} 3 \frac{\Omega^2}{r^2}  W_{\I} \slashed{\nabla}_T W_{\I}  \Big| \nonumber
\end{align}
and (see (\ref{g2t}))
\begin{align} 
\mathcal{G}_2 \left[ W\right] \left(\tau_1,\tau_2\right) = \Big| &\int_{\DcH\left(v(\tau_1),v(\tau_2)\right)} 3 \frac{\Omega^2}{r^2}  W_{\Hp} \cdot \slashed{\nabla}_{R^\star} W_{\Hp} + \int_{\DcI\left(\tau_1,\tau_2\right)} 3 \frac{\Omega^2}{r^2}  W_{\I} \slashed{\nabla}_{R^\star} W_{\I}  \Big|\nonumber \\
+&\int_{\DcH\left(v(\tau_1),v(\tau_2)\right)} 3 \frac{\Omega^2}{r^{3+\delta}} | W_{\Hp} |^2 
+\int_{\DcI\left(\tau_1,\tau_2\right)} 3 \frac{\Omega^2}{r^{3+\delta}}  |W_{\I} |^2 \, ,   \nonumber
\end{align}
where we have denoted and will denote for the rest of this step $W_{\Hp}=\left(r \slashed{\Delta}\right)^{(k_1-1)/2} \left(r \slashed{div}\right) (\slashed{\nabla}_T)^{k_2} \Psi_{\Hp}$ and $W_{\I}=\left(r \slashed{\Delta}\right)^{(k_1-1)/2} \left(r \slashed{div}\right) (\slashed{\nabla}_T)^{k_2} \Psi_{\I}$.
Integrating by parts the $T$ derivative it is easy to see that
\begin{align}
\mathcal{G}_1 \left[W\right] \left(\tau_1,\tau_2\right)
 \lesssim \sum_{\substack{k^\prime_1+k^\prime_2 \leq K-1 \\ k^\prime_3=0, k^\prime_1 \ \textrm{even}}}  \Big\{ &\sup_{\tau_1 \leq u \leq \tau_2} \check{\mathbb{F}}_u [\tilde{\mathfrak{D}}_{aux}^{\underline{k}^\prime} \Psi_{\I}] 
                +  \sup_{\tau_1 \leq u \leq \tau_2} \check{\underline{\mathbb{F}}}_{v(u)} [\tilde{\mathfrak{D}}^{\underline{k}^\prime}_{aux} \Psi_{\Hp}] \nonumber \\
                &+  \check{{\mathbb{F}}}_{u_f} \left[\tilde{\mathfrak{D}}_{aux}^{\underline{k}^\prime} \Psi_{\Hp} \right] \left(v(\tau_1)\right) + \check{\underline{\mathbb{F}}}_{v_\infty} \left[\tilde{\mathfrak{D}}_{aux}^{\underline{k}^\prime} \Psi_{\I} \right] \left(\tau_1\right) \Big\} + \frac{\varepsilon_0^2 + \varepsilon^3}{(\tau_1)^{\min(N-3-K,2)}} \nonumber 
\end{align}
with the last term arising from the boundary term induced on $\mathcal{B}$ (see Proposition  \ref{thm:cancelT}) as well as the non-linear error. Here we used that $| \left(r \slashed{\Delta}\right)^{(k_1-1)/2} \left(r \slashed{div}\right)(\slashed{\nabla}_T)^{k_2} \Psi|^2 \lesssim \sum_{i=0}^{(k_1-1)/2} |r \slashed{\nabla}  \left(r \slashed{\Delta}\right)^{(k_1-1-2i)/2} (\slashed{\nabla}_T)^{k_2} \Psi|^2$ holds up to non-linear errors that can be incorporated into the decay term. Note that the right hand side has already been controlled since we have estimated tuples with $k_1^\prime$ even.

Similarly, integrating the $R^\star$ derivative by parts\footnote{This is necessary only because of the radial decay of this error term. E.g.~in the case $k_1=1$, to control $\int \frac{1}{r^2} r\slashed{div} \Psi \slashed{\nabla}_{R^\star} \slashed{div} \Psi$ one cannot apply Cauchy--Schwarz directly in $\mathcal{D}^{\mathcal{I}^+}$. In $\mathcal{D}^{\mathcal{H}^+}$, one could apply Cauchy--Schwarz with $\gamma$ and (\ref{awgi}). However, conceptually (and to avoid constants depending on $R$) it is easiest to integrate $\slashed{\nabla}_{R^\star}$ by parts in both regions and estimate the non-linear error (boundary) term on $\mathcal{B}$ as usual.} we see that
\begin{align}
\mathcal{G}_2 \left[ W\right] \left(\tau_1,\tau_2\right) \lesssim  
\sum_{\substack{k^\prime_1+k^\prime_2 \leq K-1 \\ k^\prime_3=0, k^\prime_1 \ \textrm{even}}}    \Big\{ \sup_{\tau_1 \leq u \leq \tau_2} \check{\mathbb{F}}_u [\tilde{\mathfrak{D}}_{aux}^{\underline{k}^\prime} \Psi_{\I}] 
                +  \sup_{\tau_1 \leq u \leq \tau_2} \check{\underline{\mathbb{F}}}_{v(u)} [\tilde{\mathfrak{D}}^{\underline{k}^\prime}_{aux} \Psi_{\Hp}]+  \check{{\mathbb{F}}}_{u_f} \left[\tilde{\mathfrak{D}}_{aux}^{\underline{k}^\prime} \Psi_{\Hp} \right] \left(v(\tau_1)\right)  \nonumber \\
       \qquad          + \check{\underline{\mathbb{F}}}_{v_\infty} \left[\tilde{\mathfrak{D}}_{aux}^{\underline{k}^\prime} \Psi_{\I} \right] \left(\tau_1\right) \Big\}  +\int_{\DcH\left(v(\tau_1),v(\tau_2)\right)}  \frac{\Omega^2}{r^{3}} | W_{\Hp} |^2 
+\int_{\DcI\left(\tau_1,\tau_2\right)} \frac{\Omega^2}{r^{3}}  |W_{\I} |^2 + \frac{\varepsilon_0^2 + \varepsilon^3}{(\tau_1)^{\min(N-3-K,2)}}\, . \nonumber
\end{align}
For the spacetime integrals we recall the easily established calculus inequality
\begin{align} \label{awgi}
\int_{\DcH\left(v(\tau_1),v(\tau_2)\right)}  |W|^2 \lesssim \gamma \int_{\DcH\left(v(\tau_1),v(\tau_2)\right)}  |\slashed{\nabla}_{R^\star}  W|^2 + C_\gamma\int_{\DcH\left(v(\tau_1),v(\tau_2)\right)}  \left(1-\frac{3M}{r}\right)^2  | W|^2 \, ,
\end{align}
valid for any $\gamma>0$, which allows us to deduce
\begin{align} \label{adxs}
& \int_{\DcH\left(v(\tau_1),v(\tau_2)\right)}  \frac{\Omega^2}{r^{3}} | W_{\Hp} |^2 
+\int_{\DcH\left(v(\tau_1),v(\tau_2)\right)}  \frac{\Omega^2}{r^{3}}  |W_{\I} |^2  \\
 & \qquad \lesssim  \gamma \cdot   \check{\mathbb{I}}^{deg} \left[ W_{\Hp}\right] \left(v(\tau_1), v(\tau_2)\right) + C_\gamma \sum_{\substack{k^\prime_1+k^\prime_2 \leq K-1 \\ k^\prime_3=0, k^\prime_1 \ \textrm{even}}} \check{\mathbb{I}}^{\diamond, deg} \left[\tilde{\mathfrak{D}}^{\underline{k}^\prime} \Psi_{\Hp}\right] \left(v(\tau_1), v(\tau_2)\right) +\check{\mathbb{I}} \left[\tilde{\mathfrak{D}}^{\underline{k}^\prime} \Psi_{\I}\right] \left(\tau_1, \tau_2\right) . \nonumber 
\end{align}
Since the term multiplying $\gamma$ in the estimate (\ref{adxs}) can be absorbed on the left hand side of (\ref{Tiledb}) for sufficiently small $\gamma$ and the second term has already been estimated when we dealt with $k_1^\prime$ even, we have established the restricted estimate also for $k_1$ odd, however, with all energies appearing carrying an additional diamond superscript (i.e.~the energies being non-optimal near the horizon). To remove that restriction we finally apply Proposition \ref{prop:redshift} to the commuted equation for $W_{\Hp}$, which can also be viewed as a tensorial wave equation of type $3_0$. The additional linear error term $\mathcal{G}_{4}\left[W\right](\tau,u_f)$ is easily controlled using Cauchy--Schwarz in conjunction with the degenerate estimate already established (note the linear error is supported away from trapping). We have proven the restricted version of (\ref{thde}) defined at the beginning of Step 1.

\vskip1pc
\noindent
{\bf Step 2. Including $R^\star$-derivatives.} We next show the desired estimate replacing everywhere $\tilde{\mathfrak{D}}^{\underline{k}}$ by $\tilde{\mathfrak{D}}^{\underline{k}}_{aux}$ but lifting the restriction $k_3=0$. Note that the case $k_3=0$ has been dealt with in Step 1 and that we can again proceed inductively. 

Let $k_3>0$ and $k_1+k_2+k_3 \leq K$. {\bf If $k_1$ is even}, the commuted quantity satisfies a tensorial wave equation of Type 1 with a linear error term $\mathcal{F}_1^{lin}$ determined inductively by Proposition \ref{prop:commutetype1}. Hence we can apply Proposition \ref{prop:basecase}, which requires us to control the errors $\mathcal{G}_1$ and $\mathcal{G}_2$ generated by the linear error term $\mathcal{F}^{lin}$. From (\ref{commuteRstar}) and a simple induction we conclude (denoting $W=(\slashed{\nabla}_{R^\star})^{k_3}  \left(r^2 \slashed{\Delta}\right)^{k_1/2} (\slashed{\nabla}_T)^{k_2} \Psi$ for the rest of this step)
\begin{align}
&\frac{r^2}{\Omega^2} \Big|\mathcal{F}_1^{lin} \left[W\right] \Big|^2  \nonumber \\
& \qquad 
\lesssim  \frac{\Omega^2}{r^3} \Bigg\{ \left(1-\frac{3M}{r}\right)^2 |\left(\slashed{\nabla}_{R^\star}\right)^{k_3-1}   \left(r^2 \slashed{\Delta}\right)^{(k_1+2)/2} (\slashed{\nabla}_T)^{k_2} \Psi|^2 +  |\left(\slashed{\nabla}_{R^\star}\right)^{k_3-1}   \left(r^2 \slashed{\Delta}\right)^{k_1/2} (\slashed{\nabla}_T)^{k_2} \Psi|^2 \Bigg\} \nonumber \\
& \qquad  + \sum_{i=2}^{k_3} \frac{\Omega^2}{r^3} \left( |  \left(\slashed{\nabla}_{R^\star}\right)^{i-2}  \left(r^2 \slashed{\Delta}\right)^{(k_1+2)/2}(\slashed{\nabla}_T)^{k_2} \Psi|^2 +|  \left(\slashed{\nabla}_{R^\star}\right)^{i-2} \left(r^2 \slashed{\Delta}\right)^{k_1/2}(\slashed{\nabla}_T)^{k_2} \Psi|^2 \right)  + |\mathcal{E}^{k_1+k_2+k_3+2}_3|^2,  \nonumber
\end{align}
where the $|\mathcal{E}^{k_1+k_2+k_3+2}_3|^2$ arise from commutation of derivatives. Commuting derivatives further we deduce
\begin{align}
&\frac{r^2}{\Omega^2} \Big|\mathcal{F}_1^{lin} \left[W\right] \Big|^2 \lesssim  \sum_{\tilde{k}_1 \leq k_1 \ \textrm{even}} \frac{\Omega^2}{r^3} \Bigg[ \left(1-\frac{3M}{r}\right)^2 |r \slashed{\nabla}\left(\slashed{\nabla}_{R^\star}\right)^{k_3-1}   \left(r^2 \slashed{\Delta}\right)^{\tilde{k}_1/2}  r\slashed{div} (\slashed{\nabla}_T)^{k_2} \Psi|^2 \nonumber \\
& \qquad \qquad \qquad \qquad \ \ \ + |r \slashed{\nabla}\left(\slashed{\nabla}_{R^\star}\right)^{k_3-1}  \left(r^2 \slashed{\Delta}\right)^{(\tilde{k}_1-2)/2}  r\slashed{div} (\slashed{\nabla}_T)^{k_2} \Psi|^2 \Bigg] \nonumber \\
&\qquad \qquad \qquad \ \ \ + \sum_{i=2}^{k_3} \sum_{\tilde{k}_1 \leq k_1 \ \textrm{even}} \frac{\Omega^2}{r^3} |r \slashed{\nabla}\left(\slashed{\nabla}_{R^\star}\right)^{i-2}  \left(r^2 \slashed{\Delta}\right)^{\tilde{k}_1/2}  \slashed{div} (\slashed{\nabla}_T)^{k_2} \Psi|^2 + |\mathcal{E}^{k_1+k_2+k_3+2}_3|^2 \, .
\end{align}
Since we also have the inequality
\[
|\slashed{\nabla}_T (\slashed{\nabla}_{R^\star})^{k_3}  \left(r^2 \slashed{\Delta}\right)^{k_1/2}(\slashed{\nabla}_T)^{k_2} \Psi|^2 \lesssim |\slashed{\nabla}_{R^\star} (\slashed{\nabla}_{R^\star})^{k_3-1} \left(r^2 \slashed{\Delta}\right)^{k_1/2} (\slashed{\nabla}_T)^{k_2+1} \Psi|^2 + |\mathcal{E}_2^{k_1+k_2+k_3+2}|^2\, , 
\]
it follows from applying Cauchy--Schwarz to the product $\mathcal{F}_1^{lin} \left[W\right] \cdot \slashed{\nabla}_T W$ appearing in $\mathcal{G}_1 \left[W\right]\left(\tau_1,\tau_2\right)$
that
\begin{align}
\mathcal{G}_1  \left[W\right] \left(\tau_1,\tau_2\right) \lesssim &\sum_{\substack{|\underline{k}^\prime| \leq k_1+k_2+k_3 \\ {k}^\prime_3 \leq k_3-1}}  \check{\mathbb{I}}^{\diamond, deg} \left[\tilde{\mathfrak{D}}^{\underline{k}^\prime}_{aux} \Psi_{\Hp}\right] \left(v(\tau_1), v(\tau_2)\right) + \check{\mathbb{I}}   \left[\tilde{\mathfrak{D}}^{\underline{k}^\prime}_{aux} \Psi_{\I}\right]  \left(\tau_1, \tau_2\right) + \frac{\varepsilon_0^2 + \varepsilon^3}{(\tau_1)^{\min(N-3-K,2)}} \,  , \nonumber
\end{align}
thereby completing the induction step for $\mathcal{G}_1$.
Note that we have used Propositions \ref{prop:wavehorizonerror3} and \ref{prop:waveIerror1b} for the non-linear error and that the first term on the right controls also $\sum_{\substack{|\underline{k}^\prime| \leq k_1+k_2+k_3-1 \\ k^\prime_3 \leq k_3-2}}  \check{\mathbb{I}} \left[\tilde{\mathfrak{D}}^{\underline{k}^\prime}_{aux} \Psi_{\Hp}\right] \left(v(\tau_1), v(\tau_2)\right)$  by Remark \ref{rem:hadeg} (the latter term appearing in the application of Cauchy--Schwarz).

For $\mathcal{G}_2$ we obtain similarly 
\begin{align}
\mathcal{G}_2  \left[W\right] \left(\tau_1,\tau_2\right) &\lesssim  C_\gamma \sum_{\substack{|\underline{k}^\prime| \leq k_1+k_2+k_3 \\ k^\prime_3 \leq k_3-1}}  \check{\mathbb{I}}^{\diamond,deg} \left[\tilde{\mathfrak{D}}^{\underline{k}^\prime}_{aux} \Psi_{\Hp}\right] \left(v(\tau_1), v(\tau_2)\right) + \check{\mathbb{I}}   \left[\tilde{\mathfrak{D}}^{\underline{k}^\prime}_{aux} \Psi_{\I}\right]  \left(\tau_1, \tau_2\right)  \nonumber \\
&+\gamma \Bigg\{   \check{\mathbb{I}}^{\diamond, deg}  \left[(\slashed{\nabla}_{R^\star})^{k_3}  \left(r^2 \slashed{\Delta}\right)^{k_1/2} (\slashed{\nabla}_T)^{k_2} \Psi_{\Hp} \right]\left(v(\tau_1), v(\tau_2)\right) \nonumber \\
&\ \ \ \ \ \ \ \ \ \ + \check{\mathbb{I}} \left[(\slashed{\nabla}_{R^\star})^{k_3}  \left(r^2 \slashed{\Delta}\right)^{k_1/2}(\slashed{\nabla}_T)^{k_2} \Psi_{\I} \right]\left(\tau_1, \tau_2\right)\Bigg\}+  \frac{\varepsilon_0^2 + \varepsilon^3}{(\tau_1)^{\min(N-3-K,2)}}   \, . 
\end{align}
For $\gamma$ sufficiently small we can absorb the term in the second line by the left hand side of (\ref{Tiledb}). In summary, we have shown that 
\begin{align} \label{intma}
& \sup_{\tau \leq u \leq u_f} \check{\mathbb{F}}_u [W_{\I}] 
                +  \sup_{\tau \leq u \leq u_f} \check{\underline{\mathbb{F}}}^{\diamond}_{v(u)} [W_{\Hp}] 
                +  \check{{\mathbb{F}}}^{\diamond}_{u_f} \left[W_{\Hp} \right] \left(v(\tau)\right) + \check{\underline{\mathbb{F}}}_{v_\infty} \left[W_{\I} \right] \left(\tau\right)
                +
                \check{\mathbb{I}}^{\diamond, deg} \left[W\right] \left(v(\tau)\right) + \check{\mathbb{I}} \left[W_{\I}\right] \left(\tau\right)
                \nonumber \\
                &
                \qquad \qquad
                \lesssim \check{\mathbb{F}}_\tau [W_{\I}]  + 
                \check{\underline{\mathbb{F}}}^{\diamond}_{v(\tau)} [W_{\Hp}]  +  \sum_{\substack{|\underline{k}^\prime| \leq k_1+k_2+k_3 \\ {k}^\prime_3 \leq k_3-1}}  \Big\{ \check{\mathbb{I}}^{\diamond, deg} \left[\tilde{\mathfrak{D}}^{\underline{k}^\prime}_{aux} \Psi_{\Hp}\right] \left(v(\tau), v(u_f)\right) + \check{\mathbb{I}}   \left[\tilde{\mathfrak{D}}^{\underline{k}^\prime}_{aux} \Psi_{\I}\right]  \left(\tau, u_f\right) \Big\}   \nonumber \\
                & \qquad \qquad \ \ \ +\frac{\varepsilon_0^2 + \varepsilon^3}{\tau^{\min(N-3-K,2)}}
\end{align}
and hence completed the induction step for $k_1$ even. This establishes the desired estimate for $k_1$ even, however, with diamond superscripts on the horizon energies. To remove the latter we finally observe that $W_{\Hp}=(\slashed{\nabla}_{R^\star})^{k_3}  \left(r^2 \slashed{\Delta}\right)^{k_1/2} (\slashed{\nabla}_T)^{k_2} \Psi_{\Hp}$ satisfies a tensorial wave equation of type $3_0$ with 
\[
\mathcal{F}^{lin}_{3_0} \left[W_{\Hp}\right] = -\frac{4\Omega^2}{r^2}W_{\Hp} + 6M \frac{\Omega^2}{r^3} W_{\Hp} + \mathcal{F}_1^{lin}[W_{\Hp}] \, .
\]
We apply Proposition \ref{prop:redshift}. The terms with a $\star$ are easily seen to be controlled by (\ref{intma}) and the linear error $\mathcal{G}_4[W_{\Hp}]$ is controlled using Cauchy--Schwarz in conjunction with (\ref{intma}) noting that the linear error is supported away from $r=3M$. In summary, the estimate (\ref{thde}) has now been established replacing everywhere $\tilde{\mathfrak{D}}^{\underline{k}}$ by $\tilde{\mathfrak{D}}^{\underline{k}}_{aux}$ and restricting to tuples $(k_1,k_2,k_3)$ with $k_1$ even on the left hand side.

Let now $k_3>0$ and $k_1+k_2+k_3 \leq K$ {\bf and $k_1$ odd}. 
We leave the general induction step to the reader but treat here explicitly the key case $k_3=1$. The linear error has the structure (separately for $\Psi_{\Hp}$ and $\Psi_{\I}$)
\begin{align} \label{htd}
\mathcal{F}^{lin} \left[(\slashed{\nabla}_{R^\star})^{k_3=1}  \left(r^2 \slashed{\Delta}\right)^{(k_1-1)/2} r \slashed{div} (\slashed{\nabla}_T)^{k_2} \Psi\right]  = & h_0 \left(1-\frac{3M}{r}\right) \frac{\Omega^2}{r^3} \left(r^2 \slashed{\Delta}\right)^{(k_1+1)/2} r \slashed{div} (\slashed{\nabla}_T)^{k_2} \Psi \nonumber \\
&+h_0 \frac{\Omega^2}{r^3} \left(r^2 \slashed{\Delta}\right)^{(k_1-1)/2} r \slashed{div} (\slashed{\nabla}_T)^{k_2} \Psi \nonumber \\&+ 3 \frac{\Omega^2}{r^2} (\slashed{\nabla}_{R^\star})^{k_3=1}  \left(r^2 \slashed{\Delta}\right)^{(k_1-1)/2} r \slashed{div} (\slashed{\nabla}_T)^{k_2} \Psi \, ,
\end{align}
generating three terms in $\mathcal{G}_1$ and $\mathcal{G}_2$ respectively. 
When applying Proposition \ref{prop:basecase}, the first two can be handled exactly as in the $k_1$ even case using 
Cauchy--Schwarz and noting also the general inequality for a symmetric traceless $S$-tensor $\xi$,
\begin{align}
|r^2 \slashed{\Delta} r \slashed{div} \xi|^2 \lesssim |r \slashed{div} r^2 \slashed{\Delta} \xi|^2 + | r \slashed{div} \xi|^2 \lesssim |r \slashed{\nabla} r^2 \slashed{\Delta} \xi|^2 + | r \slashed{\nabla} \xi|^2 \, .
\end{align}
The term in the third line of (\ref{htd}) generates an integrand in $\mathcal{G}_1$ of the form
\begin{align}
& 3 \frac{\Omega^2}{r^2} (\slashed{\nabla}_{R^\star})^{k_3=1}  \left(r^2 \slashed{\Delta}\right)^{(k_1-1)/2} r \slashed{div} (\slashed{\nabla}_T)^{k_2} \Psi \cdot \slashed{\nabla}_T (\slashed{\nabla}_{R^\star})^{k_3=1}  \left(r^2 \slashed{\Delta}\right)^{(k_1-1)/2} r \slashed{div} (\slashed{\nabla}_T)^{k_2} \Psi \nonumber \\
\lesssim & C_\gamma \frac{\Omega^2}{r^2} |\slashed{\nabla}_{R^\star}  \left(r^2 \slashed{\Delta}\right)^{(k_1-1)/2} r \slashed{div} (\slashed{\nabla}_T)^{k_2} \Psi |^2 + \gamma \cdot \frac{\Omega^2}{r^2}| \slashed{\nabla}_T (\slashed{\nabla}_{R^\star})  \left(r^2 \slashed{\Delta}\right)^{(k_1-1)/2} r \slashed{div} (\slashed{\nabla}_T)^{k_2} \Psi |^2 \nonumber \\
\lesssim & C_\gamma \frac{\Omega^2}{r^2} |\slashed{\nabla}_{R^\star}  \left(r^2 \slashed{\Delta}\right)^{(k_1-1)/2} r \slashed{div} (\slashed{\nabla}_T)^{k_2} \Psi |^2 + \gamma \cdot \frac{\Omega^2}{r^2}| \slashed{\nabla}_{R^\star}  \left(r^2 \slashed{\Delta}\right)^{(k_1-1)/2} r \slashed{div} (\slashed{\nabla}_T)^{k_2+1} \Psi |^2 + |\mathcal{E}_2^{k_1+k_2}|^2 \,. \nonumber
\end{align}
The first term is controlled (non-degenerately at $r=3M$) entirely by Step 1 and the second can be absorbed on the left of (\ref{Tiledb}) for $\gamma$ sufficiently small. The same estimate works for the term $\mathcal{G}_2$ arising from the term in the third line of (\ref{htd}). 

We finally summarise the estimate one obtains for general $k_3\geq 1$:
\begin{align}
&\mathcal{G}_1  \left[(\slashed{\nabla}_{R^\star})^{k_3}  \left(r^2 \slashed{\Delta}\right)^{(k_1-1)/2} r \slashed{div} (\slashed{\nabla}_T)^{k_2} \Psi\right] \left(\tau_1,\tau_2\right)+ \mathcal{G}_2  \left[(\slashed{\nabla}_{R^\star})^{k_3}  \left(r^2 \slashed{\Delta}\right)^{(k_1-1)/2} r \slashed{div} (\slashed{\nabla}_T)^{k_2} \Psi\right] \left(\tau_1,\tau_2\right) \nonumber \\
&\lesssim C_\gamma \sum_{\substack{|\underline{k}^\prime| \leq k_1+k_2+k_3 \\ k^\prime_3 \leq k_3-1}}  \check{\mathbb{I}}^{\diamond, deg} \left[\tilde{\mathfrak{D}}^{\underline{k}^\prime}_{aux} \Psi_{\Hp}\right] \left(v(\tau_1), v(\tau_2)\right) + \check{\mathbb{I}}   \left[\tilde{\mathfrak{D}}^{\underline{k}^\prime}_{aux} \Psi_{\I}\right]  \left(\tau_1, \tau_2\right)+ \frac{\varepsilon_0^2 + \varepsilon^3}{(\tau_1)^{\min(N-3-K,2)}}  \nonumber \\
&+\gamma \  \check{\mathbb{I}}^{\diamond, deg}  \left[(\slashed{\nabla}_{R^\star})^{k_3}  \left(r^2 \slashed{\Delta}\right)^{k_1/2} r \slashed{div} (\slashed{\nabla}_T)^{k_2} \Psi_{\Hp} \right]\left(v(\tau_1), v(\tau_2)\right) \nonumber \\
&+ \gamma \ \check{\mathbb{I}} \left[(\slashed{\nabla}_{R^\star})^{k_3}  \left(r^2 \slashed{\Delta}\right)^{k_1/2} r \slashed{div} (\slashed{\nabla}_T)^{k_2} \Psi_{\I} \right]\left(\tau_1, \tau_2\right) . \nonumber
\end{align}
Taking into account these estimates for the linear errors we easily deduce the analogue of (\ref{intma}) now for $W=(\slashed{\nabla}_{R^\star})^{k_3}  \left(r^2 \slashed{\Delta}\right)^{(k_1-1)/2} r \slashed{div} (\slashed{\nabla}_T)^{k_2}\Psi$, $k_1$ odd on the left. Coupling in the redshift via Proposition~\ref{prop:redshift} proceeds entirely analogously to the $k_1$ even case. This finally produces the desired (\ref{thde}) except that $\tilde{\mathfrak{D}}^{\underline{k}}$ is replaced by $\tilde{\mathfrak{D}}^{\underline{k}}_{aux}$ everywhere.

Note that Propositions \ref{sec:ellipticconesI} and \ref{sec:ellipticconesHp} now imply in particular the estimate (\ref{thde}) if one restricts 
\begin{itemize}
\item all fluxes and spacetime integrals for for $\Psi_{\Hp}$ on the left to $r\geq 9M_{\rm init}/4$
\item all fluxes and spacetime integrals for for $\Psi_{\I}$ on the left to $r\leq 2R$ \, .
\end{itemize}
Therefore, to complete the proof, we only need to improve the estimates for $\Psi_{\Hp}$ (or more precisely $\Omega^{-1} \slashed{\nabla}_3$ derivatives applied to it) in $r \leq 9M_{\rm init}/4$ (Step 3 below) and the estimates for $\Psi_{\I}$ (or more precisely $r \Omega \slashed{\nabla}_4$ derivatives applied to it) in the region $r \geq 2R$ (Step 4 below). 

\vskip1pc
\noindent
{\bf Step 3. Improved control near the horizon for $\Psi_{\Hp}$.}
We introduce another auxiliary set of derivative operators in the horizon region:
\begin{align} \label{def:hozdo}
\tilde{\mathfrak{D}}_{\Hp}^{\underline{k}} = \tilde{\mathfrak{D}}_{\Hp}^{(k_1,k_2,k_3)} :=  \left\{
\begin{array}{rl}
\left(\Omega^{-1} \slashed{\nabla}_3 \right)^{k_3}  \left(r^2 \slashed{\Delta}\right)^{k_1/2} \left(\slashed{\nabla}_T\right)^{k_2}  &  \text{if } k_1  \ \ \text{even} \\
\left(\Omega^{-1} \slashed{\nabla}_3 \right)^{k_3}  \left(r^2 \slashed{\Delta}\right)^{(k_1-1)/2} r \slashed{div} \left(\slashed{\nabla}_T\right)^{k_2}& \text{if } k_1  \ \ \text{odd}.
\end{array} \right.
\end{align}
It is easy to see that if we can establish the estimate (\ref{thde}) for $\tilde{\mathfrak{D}}_{\Hp}^{\underline{k}} \Psi_{\Hp}$ instead of $\tilde{\mathfrak{D}}^{\underline{k}}  \Psi_{\Hp}$, then we have established it for $\tilde{\mathfrak{D}}^{\underline{k}}  \Psi_{\Hp}$ (replace $2\slashed{\nabla}_T=\Omega \slashed{\nabla}_3 + \Omega \slashed{\nabla}_4$, commute, and estimate the non-linear errors via Propositions~\ref{prop:wavehorizonerror3} and~\ref{prop:wavehorizonerror4}). Moreover, we only need to obtain improved estimates for $r\leq 9M_{\rm init}/4$ by the considerations at the end of Step 2. Finally, note that for $k_3=0$ we have already proven the desired estimate in Step 1.

To establish the estimate (\ref{thde}) for $\tilde{\mathfrak{D}}_{\Hp}^{\underline{k}}  \Psi_{\Hp}$, we recall that $\Psi_{\Hp}$ satisfies a tensorial wave equation of Type $3_0$ with (zeroth order) linear error-term given by (\ref{zoer}). Commuting with $\tilde{\mathfrak{D}}_{\Hp}^{\underline{k}}$ for $k_3=1$ produces a tensorial wave equation of Type $3_{1/2}$ whose non-linear error was controlled in Step 0. Applying Proposition~\ref{prop:redshift} with $W_{\Hp}=\tilde{\mathfrak{D}}_{\Hp}^{\underline{k}} \Psi_{\Hp}$ ($k_3=1$) and estimating the linear error by Cauchy--Schwarz (and the estimates from Steps 1 and 2) yields the desired estimate for $k_3=1$ (and $k_1$, $k_2$ arbitrary). Now inductively continue to $k_3=K$. 

\vskip1pc
\noindent
{\bf Step 4. Improved control near infinity for $\Psi_{\I}$ }
We introduce another auxiliary set of derivative operators near infinity:
\begin{align} \label{nido}
\tilde{\mathfrak{D}}_{\I}^{\underline{k}} = \tilde{\mathfrak{D}}_{\I}^{(k_1,k_2,k_3)} :=  \left\{
\begin{array}{rl}
\left(r\Omega \slashed{\nabla}_4 \right)^{k_3}  \left(r^2 \slashed{\Delta}\right)^{k_1/2} \left(\slashed{\nabla}_T\right)^{k_2}  &  \text{if } k_1  \ \ \text{even} \\
\left(r\Omega \slashed{\nabla}_4 \right)^{k_3}  \left(r^2 \slashed{\Delta}\right)^{(k_1-1)/2} r \slashed{div} \left(\slashed{\nabla}_T\right)^{k_2}& \text{if } k_1  \ \ \text{odd}.
\end{array} \right.
\end{align}
It is easy to see that if we can establish the estimate (\ref{thde}) for $\tilde{\mathfrak{D}}_{\I}^{\underline{k}} \Psi_{\I}$ instead of $\tilde{\mathfrak{D}}^{\underline{k}}  \Psi_{\I}$ then we have established it for $\tilde{\mathfrak{D}}^{\underline{k}} \Psi_{\I}$ (replace $2\slashed{\nabla}_T=\Omega \slashed{\nabla}_3 + \Omega \slashed{\nabla}_4$, commute, and estimate the non-linear errors via Propositions \ref{prop:waveIerror1b}, \ref{prop:waveIerror1c}).
Note also that for $k_3=0$ we have already proven the desired estimate in Step~1.

To establish the estimate (\ref{thde}) for $\tilde{\mathfrak{D}}_{\I}^{\underline{k}} \Psi_{\I}$ we proceed inductively, starting with $k_3=1$, i.e.~we prove the desired estimate with the sum on the left restricted to $\sum_{|{\underline{k}}| \leq K, k_3=1}$. 
 By Proposition~\ref{prop:classify4kl}, $\Psi_{\I}$ satisfies a tensorial wave equation of type $4_{0,2}$. Proposition \ref{prop:comm4kl} therefore implies that the tensorial wave equations for $\tilde{\mathfrak{D}}_{\I}^{(k_1,k_2,k_3=1)} \Psi_{\I}$ are of Type $4_{1/2,0}$ ($k_1$ odd) or $4_{1/2,3/2}$ ($k_1$ even). We apply the estimate of Proposition~\ref{prop:rph} for these $W_{\I}=\tilde{\mathfrak{D}}_{\I}^{(k_1,k_2,k_3=1)} \Psi_{\I}$ with $p=0$ and note that the boxed term is absent in view of the values of $k$ and $l$. This leads to the estimate
 {\small
\begin{align} 
 \sum_{k_1+k_2\leq K-1}  \left(         \check{\mathbb{I}}^0 [\tilde{\mathfrak{D}}_{\I}^{(k_1,k_2,k_3=1)} \Psi_{\I}] \left(\tau, u_f\right) 
                +\sup_{\tau \leq u \leq u_f} \check{\mathbb{F}}^{0}_u [\tilde{\mathfrak{D}}_{\I}^{(k_1,k_2,k_3=1)} \Psi_{\I}] +  \underline{\check{\mathbb{F}}}^{0}_{v_\infty} [\tilde{\mathfrak{D}}_{\I}^{(k_1,k_2,k_3=1)} \Psi_{\I}]( \tau )   \right) \nonumber \\
                 \lesssim \textrm{RHS of (\ref{thde})} + 
                \sum_{k_1+k_2\leq K-1}  \Big| \int_{\DcI\left(\tau\right) \cap \{ r \geq 3/2R\}} \mathcal{F}^{lin}_{4_{1/2,l}} [\tilde{\mathfrak{D}}_{\I}^{(k_1,k_2,k_3=1)} \Psi_{\I}] \cdot   \Omega \slashed{\nabla}_4 \left(\tilde{\mathfrak{D}}_{\I}^{(k_1,k_2,k_3=1)} \Psi_{\I}]\right)\Big|. \nonumber
\end{align}
}

We next observe that the left hand side of the above controls the left hand side of~\eqref{thde} 
with the sum restricted to all tuples with $k_3=1$ in view of the estimates
\begin{align} \label{pki}
\sum_{k_1+k_2\leq K-1} \check{\mathbb{F}}_u [\tilde{\mathfrak{D}}_{\I}^{(k_1,k_2,k_3=1)} \Psi_{\I}]  
\lesssim \sum_{k_1+k_2\leq K-1}  \check{\mathbb{F}}^{0}_u [\tilde{\mathfrak{D}}_{\I}^{(k_1,k_2,k_3=1)} \Psi_{\I}]
+ \sum_{k_1+k_2 \leq K}  \check{\mathbb{F}}_u [\tilde{\mathfrak{D}}_{aux}^{(k_1,k_2,k_3=0)} \Psi_{\I}]
\end{align}
and a similar estimate for $\mathbb{I} \left[\tilde{\mathfrak{D}}_{\I}^{(k_1,k_2,k_3=1)} \Psi_{\I}\right]$. The last term in~\eqref{pki} is already controlled from Step~1.

Finally, for the term involving $\mathcal{F}^{lin}_{4_{1/2,l}}$, we note that an application of the Cauchy--Schwarz inequality estimates it for any $\gamma >0$ by 
\[
\gamma \sum_{k_1+k_2\leq K-1}  \check{\mathbb{I}}^0 [\tilde{\mathfrak{D}}_{\I}^{(k_1,k_2,k_3=1)} \Psi_{\I}] \left(\tau, u_f\right) + C_{\gamma} \sum_{k_1+k_2\leq K}  \check{\mathbb{I}} [\tilde{\mathfrak{D}}_{\I}^{(k_1,k_2,k_3=0)} \Psi_{\I}] \left(\tau, u_f\right) \, . 
\]
For $\gamma$ sufficiently small, the first term can be absorbed on the left and the second term is controlled by Step~1. This proves the desired estimate for $\tilde{\mathfrak{D}}_{\I}^{\underline{k}}\Psi_{\I}$ with $\underline{k}$ having $k_3=1$.

It is easy to see we can induct all the way up to $k_3=K$. We leave the details to the reader and note here only the key point. The tensorial wave equations for $\tilde{\mathfrak{D}}_{\I}^{(k_1,k_2,k_3\geq 2)} \Psi_{\I}$ are of type $4_{k_3/2,l}$ for $l$ some (half) integer number depending only on $K$. The exact value of $l$ is irrelevant here because we can now allow for the boxed term to appear in the estimate Proposition \ref{prop:rph}. This is because this term is clearly controlled by $ \check{I}^0_u [ \tilde{\mathfrak{D}}_{\I}^{(k_1,k_2,k_3-1)} \Psi_{\I}]$ (i.e.~the previous step of the induction). The treatment of the inhomogeneous term involving $\mathcal{F}^{lin}_{4_{k_3/2,l}}$ and the non-linear term proceeds completely analogously.  

This concludes the proof for $(\Psi_{\Hp}, \Psi_{\I})$. It is clear that the same steps can be followed for the pair $(\underline{\Psi}_{\Hp}, \check{\underline{\Psi}}_{\I})$ thereby establishing the last claim of the proposition.
\end{proof}

\subsection{The $r^p$ weighted hierarchy for higher derivatives}
\begin{proposition} \label{prop:Psummaryallp}
For any $u_1 \leq \tau \leq u_f$, any $0 \leq K\leq N-3$ and $p \in \{ 1, 2-\delta,2\}$,
the pair $({\Psi}_{\I}= r^5 {{P}}$, ${\Psi}_{\Hp}= r^5 {P})$  satisfies   
\begin{align}  \label{thde2}
& \sum_{|{\underline{k}}|\leq K} \left[ \sup_{\tau \leq u \leq u_f} \check{\underline{\mathbb{F}}}_{v(u)} [\tilde{\mathfrak{D}}^{\underline{k}} \Psi_{\Hp}]  + \sup_{\tau \leq u \leq u_f} \check{\mathbb{F}}_u [\tilde{\mathfrak{D}}^{\underline{k}} \Psi_{\I}]  + \sup_{\tau \leq u \leq u_f} \check{\mathbb{F}}^p_u [\tilde{\mathfrak{D}}^{\underline{k}} \Psi_{\I}] 
                 \right]
                \nonumber \\
                & \sum_{|{\underline{k}}|\leq K} \left[ 
                  \check{{\mathbb{F}}}_{u_f} \left[\tilde{\mathfrak{D}}^{\underline{k}} \Psi_{\Hp} \right] \left(v(\tau)\right) + \check{\underline{\mathbb{F}}}_{v_\infty} \left[\tilde{\mathfrak{D}}^{\underline{k}} \Psi_{\I} \right] \left(\tau\right)+ \check{\underline{\slashed{\mathbb{F}}}}^p_{v_\infty} \left[\tilde{\mathfrak{D}}^{\underline{k}} \Psi_{\I} \right] \left(\tau\right) \right]
                \nonumber \\
                &
                +
                \sum_{|{\underline{k}}|\leq K} \check{\mathbb{I}}^{deg} \left[\tilde{\mathfrak{D}}^{\underline{k}} \Psi_{\Hp}\right] \left(v(\tau)\right) + \sum_{|{\underline{k}}|\leq K-1} \check{\mathbb{I}} \left[\tilde{\mathfrak{D}}^{\underline{k}} \Psi_{\Hp}\right] \left(v(\tau)\right) \nonumber \\
                &
                + \sum_{|{\underline{k}}|\leq K} \check{\mathbb{I}} \left[\tilde{\mathfrak{D}}^{\underline{k}} \Psi_{\I}\right] \left(\tau\right) + \sum_{|{\underline{k}}|\leq K} \check{\mathbb{I}}^p \left[\tilde{\mathfrak{D}}^{\underline{k}} \Psi_{\I}\right] \left(\tau\right)
                \nonumber \\
                &
                \qquad \qquad 
                \lesssim
         \sum_{|{\underline{k}}|\leq K}        \check{\underline{\mathbb{F}}}_{v(\tau)} [\tilde{\mathfrak{D}}^{\underline{k}} \Psi_{\Hp}] +\sum_{|{\underline{k}}|\leq K}  \check{\mathbb{F}}_\tau [\tilde{\mathfrak{D}}^{\underline{k}} \Psi_{\I}] 
         +\sum_{|{\underline{k}}|\leq K}  \check{\mathbb{F}}^p_\tau [\tilde{\mathfrak{D}}^{\underline{k}} \Psi_{\I}]        + \frac{\varepsilon_0^2 + \varepsilon^3}{\tau^{\min(N-3-K,1)}}\, .
\end{align}
Moreover, the pair $(\check{\underline{\Psi}}_{\I}= r^5 \underline{\check{P}}_{\I}$, $\underline{\Psi}_{\Hp}= r^5 \underline{P}_{\Hp})$ satisfies the same estimates.
\end{proposition}

\begin{proof}
Note first that the estimate claimed is an improvement over Proposition \ref{prop:Psummaryall} only in that the weights near infinity for $\Psi_{\I}$ and $\check{\underline{\Psi}}_{\I}$ are improved here.
The proof is  
similar to that of Step 4 of Proposition \ref{prop:Psummaryall} and hence only outlined here. The main idea is to add to the estimate (\ref{thde}) the $r^p$ weighted estimate from Proposition \ref{prop:rph} for the relevant $p$.  Note that we already proved the estimate for $K=0$ in Proposition~\ref{prop:Pimproved}. 

\vskip1pc
\noindent
{\bf Step 0. Non-linear errors.} We first establish the non-linear error estimate
\begin{align} \label{nlpe}
\sum_{|{\underline{k}}|\leq K}  \mathcal{H}_{5,p}  \left[\tilde{\mathfrak{D}}_{\I}^{\underline{k}} \Psi_{\I} \right]  + \sum_{|{\underline{k}}|\leq K}  \mathcal{H}_{5,p}  \left[\tilde{\mathfrak{D}}_{\I}^{\underline{k}} \check{\underline{\Psi}}_{\I} \right]   \lesssim  \frac{\varepsilon_0^2 + \varepsilon^3}{\tau^{\min(N-3-K,1)}}\, .
\end{align}
Focussing on the $\check{\underline{\Psi}}_{\I}$ term (the $\Psi_{\I}$ term being strictly easier as there are no anomalous error terms) we estimate the term
\begin{align}
\mathcal{H}_{5,p} \left[\tilde{\mathfrak{D}}_{\I}^{\underline{k}} \check{\underline{\Psi}}_{\I}\right]  \left(\tau,u_f\right) =\int_{\DcI\left(\tau\right)  \cap \{r \geq 3R/2\}} |\left(\mathcal{F}^{nlin}_{4_{k,l}} \left[\tilde{\mathfrak{D}}_{\I}^{\underline{k}}  \check{\underline{\Psi}}_{\I}\right] + F_{34} \left[\tilde{\mathfrak{D}}_{\I}^{\underline{k}} \check{\underline{\Psi}}_{\I}\right] \right)| \cdot  r^p |\Omega \slashed{\nabla}_4 \tilde{\mathfrak{D}}_{\I}^{\underline{k}}  \check{\underline{\Psi}}_{\I}|  \nonumber
\end{align}
as follows. Wlog $p=2$. We commute the $\Omega \slashed{\nabla}_4$ through on $\check{\underline{\Psi}}_{\I}$ and insert the expression from Proposition~\ref{prop:nabla4pschematic}. Note that errors from commutators (Lemma~\ref{lem:commutation}) either strictly improve regularity of the integrand or its $r$-decay (by at least two powers of $r$) or both. We further note that $\mathcal{F}^{nlin}_{4_{k,l}} \left[\tilde{\mathfrak{D}}_{\I}^{\underline{k}}  \check{\underline{\Psi}}_{\I}\right]$ is of the same form as $\mathcal{F}^{nlin} \left[\tilde{\mathfrak{D}}_{\I}^{\underline{k}}  \check{\underline{\Psi}}_{\I}\right]$ considered in (\ref{Fnlinex2}). Inserting the schematic forms for these terms the error is of the form controlled by Propositions~\ref{prop:waveIerror1},~\ref{prop:waveIerroranom2} and Propositions~\ref{prop:waveIerroranom3},~\ref{prop:waveIerroranom4}. 

\vskip1pc
\noindent
{\bf Step 1. Angular and $T$ derivatives.} We first prove (\ref{thde2}) for any fixed $K \in \{1,2,  ... , N-3\}$ and all $\tilde{\mathfrak{D}}_{\I}^{\underline{k}} \Psi_{\I}$ with $\left(k_1,k_2,k_3=0\right)$, $k_1+k_2=K$. Since such $\tilde{\mathfrak{D}}_{\I}^{\underline{k}} \Psi_{\I}$ satisfy a tensorial wave equation of type $4_{0,1/2}$ or $4_{0,2}$ respectively (depending on $k_2$ even or odd) with $\mathcal{F}^{lin}\left[\tilde{\mathfrak{D}}_{\I}^{\underline{k}} \Psi_{\I} \right]=6M\frac{\Omega^2}{r^3} \tilde{\mathfrak{D}}_{\I}^{\underline{k}} \Psi_{\I}$, we can apply Proposition \ref{prop:rph} (no boxed term in view of $l \geq 0$) for $p \in \{1,2-\delta,2\}$ and estimate the term $\mathcal{G}_{5,p}$ arising from $\mathcal{F}^{lin}$ using the Cauchy--Schwarz inequality and Proposition~\ref{prop:Psummaryall} (just as in the proof of 
Proposition~\ref{prop:Pimproved}). 

\vskip1pc
\noindent
{\bf Step 2. Including $r\slashed{\nabla}_4$-derivatives.} We now prove~\eqref{thde2} for  any fixed $K \in \{1,2,  ... , N-3\}$ and all $\tilde{\mathfrak{D}}_{\I}^{\underline{k}} \Psi_{\I}$ with ${\underline{k}} = \left(k_1,k_2,k_3=1\right)$, $k_1+k_2=K-1$ and then induct all the way to $k_3=K$. Since the main step is going to $k_3=1$, we focus on this. We note that by 
Proposition~\ref{prop:comm4kl} such $\tilde{\mathfrak{D}}_{\I}^{\underline{k}} \Psi_{\I}$ satisfy a tensorial wave equation of type $4_{1/2,l}$ (precise half-integer value of $l$ irrelevant) with linear inhomogeneous term of the form ($h_0$ denoting an admissible coefficient function that can be different in each occurrence)
\begin{align} \label{flio}
\mathcal{F}^{lin}_{4_{1/2,l}} \left[\tilde{\mathfrak{D}}_{\I}^{\underline{k}} \Psi_{\I}\right] = \frac{h_0}{r^3} \tilde{\mathfrak{D}}_{\I}^{\underline{k}} \Psi_{\I} + \frac{h_0}{r^2} r^2 \slashed{\Delta} \tilde{\mathfrak{D}}_{\I}^{(k_1,k_2,k_3=0)} \Psi_{\I} + \frac{h_0}{r^2} \tilde{\mathfrak{D}}_{\I}^{(k_1,k_2,k_3=0)} \Psi_{\I} \, .
\end{align}
We apply Proposition \ref{prop:rph} for this $W_{\I} = \tilde{\mathfrak{D}}_{\I}^k \Psi_{\I}$. The boxed term on the right hand side of (\ref{basrpe}) is controlled for any $p \in \left[1,2\right]$ by
\begin{align}
 \int_{\DcI\left(\tau\right)} r^{p-3} |r \Omega \slashed{\nabla}_4 \tilde{\mathfrak{D}}_{\I}^{(k_1,k_2,k_3=0)}\Psi_{\I}|^2 \lesssim  \check{\mathbb{I}} \left[\tilde{\mathfrak{D}}_{\I}^{(k_1,k_2,k_3=0)} \Psi_{\I}\right] \left(\tau, u_f\right) +  \check{\mathbb{I}}^p \left[\tilde{\mathfrak{D}}_{\I}^{(k_1,k_2,k_3=0)} \Psi_{\I}\right] \left(\tau, u_f\right) \, , \nonumber
\end{align}
with the right hand side being controlled by Step 1. The terms supported in $r\leq 2R$ on the right hand side of (\ref{basrpe}) are controlled already by the estimate (\ref{thde}). The non-linear term is controlled by (\ref{nlpe}). The term
\begin{align}
\mathcal{G}_{5,p} \left[\tilde{\mathfrak{D}}_{\I}^{(k_1,k_2,k_3=1)} \Psi_{\I}\right] = \Big| \int_{\DcI\left(\tau\right) \cap \{ r \geq 3/2R\}} \mathcal{F}^{lin}_{4_{1/2,l}} \left[\tilde{\mathfrak{D}}_{\I}^{\underline{k}} \Psi_{\I}\right] \cdot  \left(r^p+r^{2-\delta} \boldsymbol\delta^2_p \right)   \Omega \slashed{\nabla}_4 \tilde{\mathfrak{D}}_{\I}^k \Psi_{\I} \Big|
\end{align}
arising from (\ref{flio}) on the right hand side of (\ref{basrpe})  is the most difficult. For $p=1$ and $p=2-\delta$ it is again easily seen from the Cauchy--Schwarz inequality that we have for any $\eta>0$
\begin{align}
\mathcal{G}_{5,p} \left[\tilde{\mathfrak{D}}_{\I}^{(k_1,k_2,k_3=1)} \Psi_{\I}\right]  \leq \gamma  \check{\mathbb{I}}^p \left[\tilde{\mathfrak{D}}_{\I}^{(k_1,k_2,k_3=1)} \Psi_{\I}\right] \left(\tau, u_f\right) + C_{\gamma} \check{\mathbb{I}}^p \left[\tilde{\mathfrak{D}}_{\I}^{(k_1,k_2+1,k_3=0)} \Psi_{\I}\right] \left(\tau, u_f\right) + \frac{\varepsilon_0^2 + \varepsilon^3}{\tau} , \nonumber
\end{align}
with the non-linear errors arising from commutations of derivatives. The first term can be absorbed on the left for sufficiently small $\gamma$ and the second term is controlled by Step 1 above.

For $p=2$, this argument still works for the first term appearing in (\ref{flio}) but not for the last two terms. Here we need to control the expression 
\begin{align} \label{savesp}
\int_{\DcI\left(\tau\right) \cap \{ r \geq 3R/2\}}  h_0  r^2 \slashed{\Delta}  \tilde{\mathfrak{D}}_{\I}^{(k_1,k_2,k_3=0)} \Psi_{\I} \cdot \Omega \slashed{\nabla}_4  \tilde{\mathfrak{D}}_{\I}^{(k_1,k_2,k_3=1)} \Psi_{\I}  \, .
\end{align}
Inserting $1= (\chi(r) + (1-\chi(r)))$ into the integrand where $\chi(r)$ is a radial cut-off equal to $1$ for $r\geq 2R$ and zero near $r=3R/2$, we control the integral involving $(1-\chi(r))$, which is supported for $r \leq 2R$, directly from Proposition \ref{prop:Psummaryall}. For the term involving $\chi(r)$ we integrate by parts the angular derivative and the $4$ derivative (using $\left[r \slashed{\nabla}, \Omega \slashed{\nabla}_4\right]$ is non-linear by Lemma \ref{lem:commutation}) resulting in the following estimate. For any $\gamma>0$,
\begin{align}
(\ref{savesp})  \lesssim& \ \  C_\gamma \sum_{k_1^\prime \leq k_1 + 1} \Big\{\check{\mathbb{I}}^{p=2} \left[\tilde{\mathfrak{D}}_{\I}^{(k_1^\prime,k_2,k_3=0)} \Psi_{\I}\right] \left(\tau, u_f\right) + \check{\underline{\slashed{\mathbb{F}}}}^{p=2}_{v_\infty} \left[\tilde{\mathfrak{D}}^{(k_1^\prime,k_2,k_3=0)}\Psi_{\I} \right] \left(\tau\right) \Big\} \nonumber \\
&+ \gamma\check{\underline{\slashed{\mathbb{F}}}}^{p=2}_{v_\infty} \left[\tilde{\mathfrak{D}}^{(k_1,k_2,k_3=1)} \Psi_{\I} \right] \left(\tau\right) + \sum_{|{\underline{k}}|\leq K}  \check{\mathbb{F}}_\tau [\tilde{\mathfrak{D}}^{\underline{k}} \Psi_{\I}]    + \sum_{|{\underline{k}}|\leq K}       \check{\underline{\mathbb{F}}}_{v(\tau)} [\tilde{\mathfrak{D}}^{\underline{k}} \Psi_{\Hp}] + \frac{\varepsilon_0^2 + \varepsilon^3}{\tau^{\min(N-3-K,1)}} . \nonumber
\end{align}
The $\gamma$-term is absorbed on the left of~\eqref{basrpe} and the first line is controlled by the previous estimate from 
Step~1. The last term in (\ref{flio}) is treated entirely analogously. 

Using further commutations it is clear how to induct all the way to $k_3=K$ and the proposition is proven for the pair $\left(\Psi_{\Hp}, \Psi_{\I} \right)$. It is clear that the same steps can be followed for the pair $(\underline{\Psi}_{\Hp}, \check{\underline{\Psi}}_{\I})$ thereby establishing the last claim of the proposition.
\end{proof}

\section{The hierarchy of weighted decay estimates} \label{sec:decp} 
In this section we complete the proof of Theorem \ref{theo:Pmainregion2}.
It is easy to check that the estimates of the following proposition imply the estimates of Theorem \ref{theo:Pmainregion2}.

\begin{proposition} \label{prop:Psummarydecay}
For any $u_1 \leq \tau \leq u_f$,
the pair $({\Psi}_{\I}= r^5 {{P}}$, ${\Psi}_{\Hp}= r^5 {P})$  satisfies the estimates
\begin{align} \label{decay0s}
 \sum_{|\underline{k}|\leq N-3} \left[ \sup_{\tau \leq u \leq u_f} \check{\underline{\mathbb{F}}}_{v(u)} [\tilde{\mathfrak{D}}^{\underline{k}} \Psi_{\Hp}]  + \sup_{\tau \leq u \leq u_f} \check{\mathbb{F}}_u [\tilde{\mathfrak{D}}^{\underline{k}} \Psi_{\I}]  + \sup_{\tau \leq u \leq u_f} \check{\mathbb{F}}^2_u [\tilde{\mathfrak{D}}^{\underline{k}} \Psi_{\I}] 
                 \right] &
                \nonumber \\
                 \sum_{|\underline{k}|\leq N-3} \left[ 
                  \check{{\mathbb{F}}}_{u_f} \left[\tilde{\mathfrak{D}}^{\underline{k}} \Psi_{\Hp} \right] \left(v(\tau)\right) + \check{\underline{\mathbb{F}}}_{v_\infty} \left[\tilde{\mathfrak{D}}^{\underline{k}} \Psi_{\I} \right] \left(\tau\right)+ \check{\underline{\slashed{\mathbb{F}}}}^2_{v_\infty} \left[\tilde{\mathfrak{D}}^{\underline{k}} \Psi_{\I} \right] \left(\tau\right) \right] &
                \nonumber \\
                +
                \sum_{|\underline{k}|\leq N-3} \check{\mathbb{I}}^{deg} \left[\tilde{\mathfrak{D}}^{\underline{k}} \Psi_{\Hp}\right] \left(v(\tau)\right) + \sum_{|\underline{k}|\leq N-4} \check{\mathbb{I}} \left[\tilde{\mathfrak{D}}^{\underline{k}} \Psi_{\Hp}\right] \left(v(\tau)\right)  &\nonumber \\
                + \sum_{|\underline{k}|\leq N-3} \check{\mathbb{I}} \left[\tilde{\mathfrak{D}}^{\underline{k}} \Psi_{\I}\right] \left(\tau\right) + \sum_{|\underline{k}|\leq N-3} \check{\mathbb{I}}^2 \left[\tilde{\mathfrak{D}}^{\underline{k}} \Psi_{\I}\right] \left(\tau\right) &\lesssim \varepsilon_0^2 + \varepsilon^3,
\end{align}
\begin{align} \label{decay1s}
 \sum_{|\underline{k}|\leq N-4} \left[ \sup_{\tau \leq u \leq u_f} \check{\underline{\mathbb{F}}}_{v(u)} [\tilde{\mathfrak{D}}^{\underline{k}} \Psi_{\Hp}]  + \sup_{\tau \leq u \leq u_f} \check{\mathbb{F}}_u [\tilde{\mathfrak{D}}^{\underline{k}} \Psi_{\I}]  + \sup_{\tau \leq u \leq u_f} \check{\mathbb{F}}^1_u [\tilde{\mathfrak{D}}^{\underline{k}} \Psi_{\I}] 
                 \right] &
                \nonumber \\
                 \sum_{|\underline{k}|\leq N-4} \left[ 
                  \check{{\mathbb{F}}}_{u_f} \left[\tilde{\mathfrak{D}}^{\underline{k}} \Psi_{\Hp} \right] \left(v(\tau)\right) + \check{\underline{\mathbb{F}}}_{v_\infty} \left[\tilde{\mathfrak{D}}^{\underline{k}} \Psi_{\I} \right] \left(\tau\right)+ \check{\underline{\slashed{\mathbb{F}}}}^1_{v_\infty} \left[\tilde{\mathfrak{D}}^{\underline{k}} \Psi_{\I} \right] \left(\tau\right) \right] &
                \nonumber \\
                +
                \sum_{|\underline{k}|\leq N-4} \check{\mathbb{I}}^{deg} \left[\tilde{\mathfrak{D}}^{\underline{k}} \Psi_{\Hp}\right] \left(v(\tau)\right) + \sum_{|\underline{k}|\leq N-5} \check{\mathbb{I}} \left[\tilde{\mathfrak{D}}^{\underline{k}} \Psi_{\Hp}\right] \left(v(\tau)\right)  &\nonumber \\
                + \sum_{|\underline{k}|\leq N-4} \check{\mathbb{I}} \left[\tilde{\mathfrak{D}}^{\underline{k}} \Psi_{\I}\right] \left(\tau\right) + \sum_{|\underline{k}|\leq N-4} \check{\mathbb{I}}^1 \left[\tilde{\mathfrak{D}}^{\underline{k}} \Psi_{\I}\right] \left(\tau\right) &\lesssim \frac{\varepsilon_0^2 + \varepsilon^3}{\tau},
\end{align}
\begin{align} \label{decay2s}
 \sum_{|\underline{k}|\leq N-5} \left[ \sup_{\tau \leq u \leq u_f} \check{\underline{\mathbb{F}}}_{v(u)} [\tilde{\mathfrak{D}}^{\underline{k}} \Psi_{\Hp}]  + \sup_{\tau \leq u \leq u_f} \check{\mathbb{F}}_u [\tilde{\mathfrak{D}}^{\underline{k}} \Psi_{\I}] \right]  &
                \nonumber \\
                + \sum_{|\underline{k}|\leq N-5} \left[ 
                  \check{{\mathbb{F}}}_{u_f} \left[\tilde{\mathfrak{D}}^{\underline{k}} \Psi_{\Hp} \right] \left(v(\tau)\right) + \check{\underline{\mathbb{F}}}_{v_\infty} \left[\tilde{\mathfrak{D}}^{\underline{k}} \Psi_{\I} \right] \left(\tau\right) \right] + \sum_{|\underline{k}|\leq N-5} \check{\mathbb{I}} \left[\tilde{\mathfrak{D}}^{\underline{k}} \Psi_{\I}\right] \left(\tau\right) &
                \nonumber \\
                +
                \sum_{|\underline{k}|\leq N-5} \check{\mathbb{I}}^{deg} \left[\tilde{\mathfrak{D}}^{\underline{k}} \Psi_{\Hp}\right] \left(v(\tau)\right) + \sum_{|\underline{k}|\leq N-6} \check{\mathbb{I}} \left[\tilde{\mathfrak{D}}^{\underline{k}} \Psi_{\Hp}\right] \left(v(\tau)\right) 
                 & \lesssim \frac{\varepsilon_0^2 + \varepsilon^3}{\tau^2}.
\end{align}
Moreover, the pair $(\check{\underline{\Psi}}_{\I}= r^5 \underline{\check{P}}_{\I}$, $\underline{\Psi}_{\Hp}= r^5 \underline{P}_{\Hp})$ satisfies the same estimates. 

Finally, all estimates hold replacing $\tilde{\mathfrak{D}}^{\underline{k}}$ by $\mathfrak{D}^{\underline{k}}$.
\end{proposition}

\begin{proof}
Apply the estimate of Proposition~\ref{prop:Psummaryallp} with $\tau=u_1$, $K=N-3$ and $p=2$. This produces (\ref{decay0s}) after using the estimate (\ref{newinitialr3}) for the initial energies. Using the non-degenerate spacetime terms of (\ref{decay0s}) we find a dyadic sequence $\tau_i$ such that
\begin{align}
\sum_{k=0}^{N-4}         \check{\underline{\mathbb{F}}}_{v(\tau_i)} [\tilde{\mathfrak{D}}^{\underline{k}} \Psi_{\Hp}] + \sum_{k=0}^{N-4} \check{\mathbb{F}}_{\tau_i} [\tilde{\mathfrak{D}}^{\underline{k}} \Psi_{\I}]  + \sum_{k=0}^{N-4} \check{\mathbb{F}}^1_{\tau_i} [\tilde{\mathfrak{D}}^{\underline{k}} \Psi_{\I}]
\lesssim \frac{\varepsilon_0^2 + \varepsilon^3}{\tau_i} \, .
\end{align}
Applying the estimate of Proposition~\ref{prop:Psummaryallp} with $\tau=\tau_i$, $K=N-4$ and $p=1$ establishes the estimate~\eqref{decay1s} now for any $\tau$.

From the estimate (\ref{decay1s}) we find another dyadic sequence, also denoted $\tau_i$ along which
\begin{align}
\sum_{k=0}^{N-5}         \check{\underline{\mathbb{F}}}_{v(\tau_i)} [\tilde{\mathfrak{D}}^{\underline{k}} \Psi_{\Hp}] + \sum_{k=0}^{N-5} \check{\mathbb{F}}_{\tau_i} [\tilde{\mathfrak{D}}^{\underline{k}} \Psi_{\I}]  
\lesssim \frac{\varepsilon_0^2 + \varepsilon^3}{\tau_i^2} \, .
\end{align}
Applying now the estimate of Proposition~\ref{prop:Psummaryall} with $\tau=\tau_i$, $K=N-5$ we obtain the estimate (\ref{decay2s}) for any $\tau$. The estimates for $(\check{\underline{\Psi}}_{\I}= r^5 \underline{\check{P}}_{\I}$, $\underline{\Psi}_{\Hp}= r^5 \underline{P}_{\Hp})$ are proven in exactly the same fashion. Finally, the claim regarding $\tilde{\mathfrak{D}}$ follows from Propositions~\ref{sec:ellipticconesI} and~\ref{sec:ellipticconesHp}.
\end{proof}

\section{Completing the proof of Theorem~\ref{thm:PPbarestimates}} \label{sec:missingfluxp}
Having proven Theorem~\ref{theo:Pmainregion2}, we
 complete in this section the proof Theorem~\ref{thm:PPbarestimates}.

We recall once more the master energies~(\ref{masterPen}) and focus on the quantities~$P_{\Hp}, P_{\I}$ 
(the $\underline{P}_{\Hp}, \check{\underline{P}}_{\I}$ case being treated entirely analogously). 
Since, by the considerations of Section~\ref{estimatinregIplusII}, we have reduced the problem
to region III,  it suffices to prove estimates for:
\begin{center}
\begin{tabular}{ |c|c|c| } 
 \hline
 & $P_{\I}$ & $P_{\Hp}$ \\ 
 \hline
outgoing cones & $C^{\I}_{u}$ for $u\geq u_{1}$ & $C^{\Hp}_u(v)$ for $v \geq v_1$ \\ 
\hline
ingoing cones & $\underline{C}^{\I}_v(u)$ for $u \geq u_1$ & $\underline{C}^{\Hp}_{v}$ for $v \geq v_1$ \\ 
\hline
integrated decay regions & $\mathcal{D}^{\I}(u) \cap \{v_{\Hp} \geq v_1\}$ for $u \geq u_1$ & $\mathcal{D}^{\Hp}(v) \cap \{u_{\I} \geq u_1\}$ for $v\geq v_1$ \\
 \hline
\end{tabular}
\end{center}

Note also that in view of Theorem \ref{theo:Pmainregion2}, we have already proved the required estimates for $P_{\I}$ on the truncated cones $\check{C}^{\I}_u$ as well as the regions $\DcI(u)$ for $u\geq u_1$, and the estimates for $P_{\Hp}$ on the truncated cones $\underline{C}_v^{\Hp}$ as well as the regions $\DcH(v)$ for $v\geq v_1$.

\subsection{Integrated decay}
We first focus on the integrated decay. We recall the energies (\ref{dhe1}) and (\ref{ndhe1}) and define the analogous ``unchecked" energies 
\begin{align}
{\mathbb{I}} \left[W_{\Hp}\right] \left(v\right) &= \int_{\mathcal{D}^{\Hp}(v) \cap \{u_{\I} \geq u_1\}} |\Omega^{-1} \slashed{\nabla}_3 W_{\Hp}|^2 + |\Omega \slashed{\nabla}_4 W_{\Hp}|^2 + |r \slashed{\nabla} W_{\Hp}|^2 + |W_{\Hp}|^2  \, \, , 
\end{align}
\begin{align}
\mathbb{I}^{\diamond, deg} \left[W_{\Hp}\right] \left(v\right) &=\int_{\mathcal{D}^{\Hp}(v) \cap \{u_{\I} \geq u_1\}} |\slashed{\nabla}_{R^\star} W_{\Hp}|^2 + |W_{\Hp}|^2 + \left(1-\frac{3M}{r}\right)^2 \left( |\slashed{\nabla}_TW_{\Hp}|^2 + |r\slashed{\nabla} W_{\Hp} |^2 \right) \, , \nonumber \\
\mathbb{I}^{deg} \left[W_{\Hp}\right] \left(v\right) &=\mathbb{I}^{\diamond, deg} \left[W_{\Hp}\right] \left(v\right)  + \int_{\mathcal{D}^{\Hp}(v) \cap \{u_{\I} \geq u_1\}} | \Omega^{-1} \slashed{\nabla}_3 W_{\Hp}|^2  \, .
\end{align}
Similarly, in analogy with~\eqref{infe1} and~\eqref{infe2}, we define
\begin{align}  \label{enfot1}
\mathbb{I} \left[W_{\I}\right] \left(\tau\right) &= \int_{\mathcal{D}^{\I}(\tau) \cap \{v_{\Hp} \geq v_1\}} \frac{ \big| \Omega \slashed{\nabla}_4 W_{\I}|^2}{r^{1+
 \delta}} +\frac{ \big| \Omega \slashed{\nabla}_3 W_{\I}|^2}{ r^{1+\delta} }  + \frac{\big| r\slashed{\nabla} W_{\I} \big|^2}{r^{3}} + \frac{|W_{\I}|^2}{r^{3}} \, ,
\end{align}
\begin{align} \label{enfot2}
\mathbb{I}^p \left[W_{\I}\right] \left(\tau\right) &= \int_{\mathcal{D}^{\I} (\tau) \cap \{v_{\Hp} \geq v_1\}}\left( \frac{ \big| \Omega \slashed{\nabla}_4 W_{\I}|^2}{r^{1+\boldsymbol\delta^p_0}} + \frac{\big| r\slashed{\nabla} W_{\I} \big|^2}{r^{3+\boldsymbol{\delta}^p_2\delta}} + \frac{|W_{\I}|^2}{r^{3+\boldsymbol{\delta}^p_2\delta}} \right) r^p  \, .
\end{align}

\begin{proposition} \label{prop:intdoverlap}
The pair $({\Psi}_{\I}= r^5 P$, ${\Psi}_{\Hp}= r^5 {P})$  satisfies the following estimates
\begin{align} \label{bmp1}
 \sum_{s=0}^2  \sup_{u_{1} \leq \tau \leq u_f}\tau^s \Bigg\{ \sum_{|\underline{k}|\leq N-3-s}{\mathbb{I}} \left[{\mathfrak{D}}^{\underline{k}} \Psi_{\I}\right] \left(\tau\right) + \sum_{|\underline{k}|\leq N-3-s} {\mathbb{I}}^{2-s} \left[{\mathfrak{D}}^{\underline{k}} \Psi_{\I}\right] \left(\tau\right) \Bigg\} \lesssim \varepsilon_0^2 + \varepsilon^3 \, ,
\end{align}
\begin{align} \label{bmp2}
\sum_{s=0}^2  \sup_{v \geq v_1} v^s \Bigg\{ \sum_{|\underline{k}|\leq N-3-s} {\mathbb{I}}^{deg} \left[{\mathfrak{D}}^{\underline{k}} \Psi_{\Hp}\right] \left(v\right) + \sum_{|\underline{k}|\leq N-4-s} {\mathbb{I}} \left[{\mathfrak{D}}^{\underline{k}} \Psi_{\Hp}\right] \left(v\right) \Bigg\} &\lesssim  \varepsilon_0^2 + \varepsilon^3 \, .
\end{align}
Moreover, the quantity $\check{\underline{\Psi}}_{\I}$ in the pair $(\check{\underline{\Psi}}_{\I}= r^5 \underline{\tilde{P}}$, $\underline{\Psi}_{\Hp}= r^5 \underline{P})$ satisfies the same estimates.
\end{proposition}

\begin{proof}
Note that we have proven these estimates already for the $\check{\mathbb{I}}$ energies in Theorem \ref{theo:Pmainregion2}. For $\Psi_{\I}$, it therefore suffices to restrict the integrals in the energies (\ref{enfot1}) and (\ref{enfot2}) appearing in (\ref{bmp1}) to the region $\DcH (v(\tau)) \cap \mathcal{D}^{\I}(\tau)$, which contains ${\mathcal{D}^{\I}(\tau) \cap \{v_{\Hp} \geq v_1\}} \setminus \DcI(\tau)$ and is located far away from $r=3M$. The (non-degenerate) estimates for $\Psi_{\I}$ will then follow if we can replace $\Psi_{\I}$ by $\Psi_{\Hp}$ in these integrals over $\DcH (v(\tau)) \cap \mathcal{D}^{\I}(\tau)$. Indeed, we have $v(\tau) \sim \tau$ and the desired estimates hold for $\Psi_{\Hp}$ in $\DcH(v(\tau))$ by Theorem \ref{theo:Pmainregion2}. To finally replace $\Psi_{\I}$ by $\Psi_{\Hp}$, we apply Proposition~\ref{thm:morecancelnotjustT}, which controls the difference in the region $\DcH (v(\tau)) \cap \mathcal{D}^{\I}(\tau)$ by a decaying error term implicit in  (\ref{bmp1}).

The proof for $\Psi_{\Hp}$ is analogous using again Proposition \ref{thm:morecancelnotjustT} in the overlap region.
\end{proof}
From the definition of the timelike hypersurface $\mathcal{B}$ in Section \ref{timelikehype} we conclude (see (\ref{Bfluxnotation})):
\begin{corollary}  \label{prop:Idc3}
We have in addition the following bounds on the timelike hypersurface $\mathcal{B}$:
\begin{align}
\sup_{u_1 \leq \tau \leq u_f} \sum_{s=0}^2 \tau^s \Bigg\{ \sum_{|\underline{k}|\leq N-3-s} {\mathbb{F}}_{\mathcal{B}(\tau)} [{\mathfrak{D}}^{\underline{k}} \Psi_{\Hp}] + \sum_{|\underline{k}|\leq N-3-s} {\mathbb{F}}_{\mathcal{B}(\tau)} [{\mathfrak{D}}^{\underline{k}} \Psi_{\I}] \Bigg\}    &\lesssim \varepsilon_0^2 + \varepsilon^3 \, .
\end{align}
\end{corollary}

\subsection{Estimates on truncated ingoing cones $\protect\underline{\check C}_v^{\I}$}
\label{gonnabesimilarbelow}
With the improved estimates on the timelike hypersurface $\mathcal{B}$ of Corollary \ref{prop:Idc3}, we can 
obtain estimates on arbitrary \emph{truncated} ingoing cones of the $\mathcal{I}^+$ gauge. 

Indeed, pick such a cone $\check{\underline{C}}_v^{\I} \left(\tau\right)$ with $\tau \geq u_1$ and consider the spacetime region enclosed by $\check{\underline{C}}_v^{\I} \left(\tau\right)$, $\mathcal{B}$, $\check{C}_\tau^{\I}$ and (potentially) $\check{C}^{\I}_{u_f}$. Apply the energy estimate and the $r^p$ weighted estimates in this region using the bounds on the cone $\check{C}_\tau^{\I}$ established in Proposition~\ref{prop:Psummarydecay} and the bounds on the hypersurface $\mathcal{B}$ established in Corollary \ref{prop:Idc3}. This proves in particular the statement of Theorem~\ref{theo:Pmainregion2} for $\Psi_{\I}$ with
 the flux $\sup_{v \leq v_\infty} \int_{\check{\underline{C}}^{\I}_{v}(\tau)}  \Big\{ \sum_{|\underline{k}|=0}^{K-1}  \   | \Omega \slashed{\nabla}_3 \mathfrak{D}^{\underline{k}} \Psi_{\I}|^2 + \frac{r^p}{r^2} | r \slashed{\nabla} \mathfrak{D}^{\underline{k}} \Psi_{\I}|^2 \Big\}$ added to $\check{\mathbb{E}}_{out}^{K,p} \left[P_{\I}\right] \left(\tau\right)$.
Also, the check superscript can be removed for the integrated decay terms in the energies of Theorem~\ref{theo:Pmainregion2} in view of Proposition~\ref{prop:intdoverlap}.

\subsection{Estimates on truncated outgoing cones $\check{\protect\underline{C}}^{\Hp}_v$}
In order obtain estimates on arbitrary \emph{truncated} outgoing cones of the $\mathcal{H}^+$ gauge,
we proceed similarly similarly to Section~\ref{gonnabesimilarbelow} above.

Consider an outgoing cone $\check{C}_u^{\Hp} \left(v\right)$ with $v\geq v_1$ and the region enclosed by $\check{C}_u^{\Hp} \left(v\right)$, $\check{\underline{C}}^{\Hp}_v$ and $\mathcal{B}$. Apply the energy estimate and the redshift estimate in this region using the estimates on the cone $\check{\underline{C}}^{\Hp}_v$ from Proposition \ref{prop:Psummarydecay} and the bounds on the hypersurface $\mathcal{B}$ established in Corollary \ref{prop:Idc3}. This proves in particular the statement of Theorem \ref{theo:Pmainregion2} for $\Psi_{\Hp}$ with the outgoing (truncated) flux
 $\sup_{u\leq u_f} \int_{\check{C}^{\Hp}_u({v})} \sum_{|\underline{k}|=0; k_1\neq K}^{K}  | \mathfrak{D}^{\underline{k}} \Psi_{\Hp}|^2$ added to $\check{\mathbb{E}}_{in}^{K} \left[P_{\Hp}\right] \left(v\right)$. Also, the check superscript can be removed for the integrated decay terms in the energies of Theorem \ref{theo:Pmainregion2} in view of Proposition \ref{prop:intdoverlap}.

\subsection{Estimates on the full cones $C_u^{\mathcal{I}^+}$, $\protect\underline{C}_v^{\mathcal{I}^+}$ 
and $C_u^{\mathcal{H}^+}$, $\protect\underline{C}_v^{\mathcal{H}^+}$} \label{sec:aioci}

To complete the proof of Theorem~\ref{thm:PPbarestimates} for $\Psi_{\I}$, i.e.~to obtain the statement of Theorem \ref{theo:Pmainregion2} for $\Psi_{\I}$ with the check superscripts and the ``in" subscript removed from the energies, it now suffices, in view of the previous estimates, to obtain estimate for the flux on truncated outgoing cones $C^{\I}_u \cap \DcH$ for $u\geq u_1$ and the truncated ingoing ones $\underline{C}^{\I}_v (u) \cap \DcH$ for $u\geq u_1$. This is because the part of the flux on $C^{\I}_u$ to the exterior of $\mathcal{B}$ (which is the flux on $\CcI_u$) has already been estimated, and similarly for $\CbcI_v(u)$. 

\vskip1pc
\noindent
{\bf Step 1: Outgoing cones $C_u^{\mathcal{I}^+}$.} Given a null cone $C_\tau^{\I}$ with  $\tau \geq u_1 + 4M_{\rm init}$ (the ``near data" case $u_1 \leq \tau \leq u_1+4M_{\rm init}$ requiring trivial modifications of the following argument), we consider its past extension into the horizon region, say (for definiteness) up to the timelike hypersurface $r_{\Hp} =R_{-4}=R-4M_{\rm init}$ and denote the resulting cone by $\tilde{C}_\tau$. Note this cone in general does not overlap with an outgoing cone
of the $\mathcal{H}^+$ gauge, but agrees of course with a null cone of the $\mathcal{I}^+$ gauge for $r_{\I}\geq R_{-2} = R-2M_{\rm init}$. We observe that the $v$ coordinate at the intersection sphere satisfies $|v_{\Hp} - \tau + R_{-4}| \leq \frac{1}{10}M_{\rm init}$, which follows from closeness to Schwarzschild and $v-u \approx r$ in view of $R$ satisfying (\ref{definitionofRhere}). 

Next we consider the spacetime region enclosed by the ingoing cone 
$\check{\underline{C}}^{\Hp}_{v=\tau+R-3M_{\rm init}}$ of the $\mathcal{H}^+$ gauge, the hypersurface $\mathcal{B}$ and the hypersurface $\tilde{C}_\tau$. Note that in this region $R_{-4} \leq r_{\Hp} \leq R_1$ 
and in particular $r$-weights and $\Omega^2$-weights can be absorbed into constants in this region. 
Since it is also a finite time region, it suffices to apply a standard energy estimate (e.g.~arising from the vectorfield $T$) to the equation satisfied by $\tilde{\mathfrak{D}}^{\underline{k}} \Psi_{\mathcal{\Hp}}$ in this region. 
The boundary term on $\mathcal{B}$ can be controlled by Proposition \ref{prop:Idc3}. 
The term on $\check{\underline{C}}^{\Hp}_{v=\tau+R-3M_{\rm init}}$ is controlled by Proposition \ref{prop:Psummarydecay}. 
The estimate therefore produces in particular control of all \emph{tangential} derivatives of $\tilde{\mathfrak{D}}^{\underline{k}} \Psi_{\mathcal{\Hp}}$ on ${C}_\tau^{\I} \cap \check{\mathcal{D}}^{\Hp}$. 
With the tangential derivatives of $\tilde{\mathfrak{D}}^{\underline{k}} \Psi_{\Hp}$ controlled on the 
null hypersurface ${C}_\tau^{\I} \cap \check{\mathcal{D}}^{\Hp}$ we can translate $\slashed{\nabla}_{e_4^{\I}} \tilde{\mathfrak{D}}^{\underline{k}} \Psi_{\Hp}$ and $\slashed{\nabla}^{\I}\tilde{\mathfrak{D}}^{\underline{k}} \Psi_{\Hp}$ to $\slashed{\nabla}_{e_4^{\I}}\tilde{\mathfrak{D}}^{\underline{k}} \Psi_{\I}$ and $\slashed{\nabla}^{\I}\tilde{\mathfrak{D}}^{\underline{k}} \Psi_{\I}$ using the estimates of Proposition \ref{thm:morecancelnotjustT} in the overlap region producing the desired flux. [Note that it is important that in the way we carried out the argument, one of the derivatives of $\Psi_{\Hp}$ is already the ``correct" tangential derivative on the cone and does not need to be converted. The expression $\slashed{\nabla}_{e_4^{\I}} \tilde{\mathfrak{D}}^{\underline{k}} \Psi_{\Hp}$ would generally not be controlled at top order on ${C}_\tau^{\I} \cap \check{\mathcal{D}}^{\Hp}$ because it involves derivatives transversal to the cone!]

\vskip1pc
\noindent
{\bf Step 2: Ingoing cones $\underline{C}_v^{\mathcal{I}^+}$.} Consider an arbitrary ingoing cone $\underline{C}^{\I}_v(u) \cap \DcH$ of the infinity gauge. It intersects $\{r_{\I}=R_{-2}\} \cup \{u=u_f\}$ in a sphere of the infinity foliation. The outgoing cone $C$ emanating from that sphere intersects $\mathcal{B}$ and we can hence consider the region enclosed by $C$, the cone $\underline{C}^{\I}_v(u) \cap \DcH$ and $\mathcal{B}$. Doing a backwards energy estimate using the control on $C$ from Step~1 and the estimates of Proposition \ref{prop:Idc3} for the term on $\mathcal{B}$ one obtains the desired estimates.

\vskip1pc

The argument for estimating $\Psi_{\mathcal{H}^+}$ on the cones  $C_u^{\mathcal{H}^+}$, $\underline{C}_v^{\mathcal{H}^+}$
is now completely analogous.

In view of our comments at the beginning of
Section~\ref{sec:missingfluxp}, this now completes the proof of Theorem~\ref{thm:PPbarestimates}.

\chapter{Estimates for $\alpha$ and \underline{$\alpha$}: the proof of Theorem~\ref{thm:alphaalphabarestimates}}
\label{moreherechapter} 
In this chapter we shall prove Theorem~\ref{thm:alphaalphabarestimates}, which
we restate here:

\alphaalphabarestimates*

\minitoc

We shall first prove separate (and slightly weaker) estimates for $\alpha$ and then for $\underline\alpha$. 
These will be the contents of theorems to be stated in {\bf Section~\ref{sec:overviewa}} and~{\bf \ref{sec:overviewab}}, respectively, 
which
will be proven in the four and three sections, respectively immediately following each of them. We defer further discussion of the contents of these
to Sections~\ref{sec:overviewa} and~\ref{sec:overviewab}, respectively.
On the basis of these estimates, the proof of Theorem~\ref{thm:alphaalphabarestimates} will be
completed in~{\bf Section~\ref{sec:teustaro}}.

\vskip1pc
\noindent\fbox{
    \parbox{6.35in}{
As in the previous chapters of Part~\ref{improvingpart},
we shall assume throughout the assumptions of~\Cref{havetoimprovethebootstrap}. Let us fix an
arbitrary  $u_f\in[u_f^0, \hat{u}_f$], with $\hat{u}_f\in \mathfrak{B}$,
and fix some $\lambda \in \mathfrak{R}(u_f)$.
All propositions below
shall always refer  
to the anchored $\I$ and $\Hp$ gauges in the  spacetime  $(\mathcal{M}(\lambda), g(\lambda))$,  
corresponding to parameters
$u_f$, $M_f(u_f,\lambda)$,
whose existence is
ensured by Definition~\ref{bootstrapsetdef}. We shall denote $M=M_f$ throughout  
this chapter.}}
\vskip1pc

\emph{This chapter will depend on all previous chapters of Part~\ref{improvingpart}, and will in particular use 
the propositions of Chapter~\ref{RWtypechapter} as well as the statements of 
Theorems~\ref{thm:sobolevandelinfinity}--\ref{thm:PPbarestimates}. As with 
Chapter~\ref{chapter:psiandpsibar}, the reader who wishes to understand the
structure of the proof without reading the detailed estimates may wish to simply read the overview 
Sections~\ref{sec:overviewa} and~\ref{sec:overviewab}
together with Section~\ref{sec:teustaro}. For the linear version of Theorem~\ref{thm:alphaalphabarestimates}, the reader may
compare with Theorem~2 in Section~11 of~\cite{holzstabofschw}.}

\section{Overview of the estimates on $\alpha$} \label{sec:overviewa}
To prove the estimates on $\alpha_{\I}, \alpha_{\Hp}$ implicit in Theorem \ref{thm:alphaalphabarestimates} we will first prove the following statement:

\begin{theorem} \label{theo:mtheoalphar}
We have the estimates 
\[
	\sum_{s=0,1,2}
	\sup_{u_{-1} \leq \tau \leq u_f} \tau^s \cdot {\mathbb{E}}_\Box^{N-s,2-s} \left[{\alpha}_{\I}\right] \left(\tau\right)  + \sum_{s=0,1,2}
	\sup_{v_{-1} \leq v} 
	v^{s} \cdot {\mathbb{E}}_\Box^{N-s} \left[{\alpha}_{\Hp}\right] \left(v\right)  \lesssim \varepsilon_0^2 + \varepsilon^3 
\]
for the restricted energies (recall $A_{\I}=r\Omega^2 \alpha_{\I}$, $A_{\Hp}=r\Omega^2 \alpha_{\Hp}$ and $\Pi_{\I}=r^3 \Omega \psi_{\I}$, $\Psi_{\I} = r^5 P_{\I}$)
\begin{align}
&{\mathbb{E}}_\Box^{K} \left[{\alpha}_{\Hp}\right] \left(v\right) := \sup_{\tilde{v} \geq v} \int_{\underline{C}^{\Hp}_{\tilde{v}}} \Omega^2  \sum_{|\underline{k}|=0; k_3\neq K}^{K}  | \mathfrak{D}^{\underline{k}} {A}_{\Hp}|^2 + \sup_{u\leq u_f} \int_{C^{\Hp}_u({v})} \sum_{|\underline{k}|=0; \boxed{k_2 \neq K}}^{K}  | \mathfrak{D}^{\underline{k}} A_{\Hp}|^2  \nonumber \\
& \qquad \ \  +  \int_{\mathcal{D}^{\Hp}\left(v\right)}  \Omega^2 \Bigg\{ \sum_{|\underline{k}|=0}^{K} \left(1-\frac{3M}{r}\right)^2 | \mathfrak{D}^{\underline{k}} {A}_{\Hp}|^2  +  \sum_{|\underline{k}|=0}^{K-1}  | \mathfrak{D}^{\underline{k}} {A}_{\Hp}|^2+  | R^\star \mathfrak{D}^{\underline{k}} {A}_{\Hp}|^2  \Bigg\} \nonumber \, ,
\end{align}
\begin{align}
&{\mathbb{E}}^{K,p}_\Box \left[{\alpha}_{\I}\right] \left(\tau\right) := 
 \sup_{\tau \leq u \leq u_f} \int_{C^{\I}_u} \Bigg\{ \sum_{|\underline{k}|=1; \boxed{k_2 \neq K}}^K  \ r^{4+p} | \mathfrak{D}^{\underline{k}} A_{\I}|^2 + \sum_{|\underline{k}|=0; \boxed{k_2 \neq K-1}}^{K-1}  r^2 | \mathfrak{D}^{\underline{k}} \Pi_{\I}|^2 \Bigg\} \nonumber \\
& \qquad \qquad \ \   +\sup_{v \leq v_\infty} \int_{\underline{C}^{\I}_{v}(\tau)}  \Bigg\{ \sum_{|\underline{k}|=1, k_3 \neq K}^{K}  \ r^{4+p} | \mathfrak{D}^{\underline{k}} A_{\I} |^2 + \sum_{|\underline{k}|=0}^{K-1}  r^{2+p} | \mathfrak{D}^{\underline{k}} \Pi_{\I}|^2 + \sum_{|\underline{k}|=0}^{K-2}  | \mathfrak{D}^{\underline{k}} \Psi_{\I}|^2  \Bigg\} \nonumber \\
& \qquad \qquad \ \  +  \int_{\mathcal{D}^{\I}\left(\tau\right)} \Bigg\{ \sum_{|\underline{k}|=1}^K  \, r^{3+p-\delta} | \mathfrak{D}^{\underline{k}} A_{\I}|^2  + \sum_{|\underline{k}|=0}^{K-1} r^{1+p-\delta} | \mathfrak{D}^{\underline{k}} \Pi_{\I}|^2  +\sum_{|\underline{k}|=0}^{K-2} r^{-1-\delta} | \mathfrak{D}^{\underline{k}} \Psi_{\I}|^2 \Bigg\} .
\end{align}
\end{theorem}
Note that the above restricted energies differ from the actual energies defined in (\ref{meah}) and (\ref{meai}) only by the boxed restrictions in the sum. We will finally remove these restrictions in Section \ref{sec:teustaro} (after having proved estimates also on $\underline{\alpha}$), using the Teukolsky--Starobinski identities to deduce control over the missing fluxes concluding the $\mathbb{E}_{u_f}^N [\alpha_{\Hp}, \alpha_{\I}]  \lesssim \varepsilon_0^2 + \varepsilon^3$ part in Theorem \ref{thm:alphaalphabarestimates}.
Observe already at this point that the estimate of Theorem \ref{theo:mtheoalphar} strengthens in particular some of the estimates ($r$-weights) for the derived quantity $\Psi_{\I}=r^5 P_{\I}$ in Theorem \ref{thm:PPbarestimates}.

\begin{remark} \label{rem:equiven}
Note that we can equivalently replace the restrictions ${k}_3 \neq K$, $k_2 \neq K$ and $k_2 \neq K-1$ by $\underline{k}_3 \neq |\underline{k}|$, $k_2 \neq |\underline{k}|$ and $k_2 \neq |\underline{k}|$ in the sums above. This follows by a simple elliptic estimate along cones. 
\end{remark}

To prove Theorem \ref{theo:mtheoalphar} we first recall the three regions introduced at the beginning of Section \ref{sec:overviewP}. Just as for $P$ and $\underline{P}$, in view of Proposition \ref{thm:gidataestimates} it suffices to prove decay estimates for cones and regions contained in the region III, namely
$\DcI\left(u_1\right) \cup \DcH\left(v_1=v(u_1)\right)$, in terms of the weighted energy
\begin{align} \label{initwene}
\int_{\underline{C}^{\Hp}_{v_1}} \Omega^2  \sum_{|\underline{k}|=0; k_3\neq |\underline{k}|}^{N}  | \mathfrak{D}^{\underline{k}} {A}_{\Hp}|^2 +  \int_{C^{\I}_{u_1}} \Bigg\{ \sum_{|\underline{k}|=1; k_2 \neq |\underline{k}|}^N  \ r^{6} | \mathfrak{D}^{\underline{k}} A_{\I}|^2 + \sum_{|\underline{k}|=0; k_2 \neq |\underline{k}|}^{N-1}  r^2 | \mathfrak{D}^{\underline{k}} \Pi_{\I}|^2  \Bigg\}
\end{align}
on $C_{u_1}^{\I} \cup \underline{C}^{\Hp}_{v_1=v(u_1)}$, which is in turn controlled by $\varepsilon_0^2 + \varepsilon^3$ from Proposition \ref{thm:gidataestimates}.

In order to achieve this, we first prove (just as for~$P$ and $\underline{P}$ in Chapter~\ref{chapter:psiandpsibar}) a weaker version of Theorem~\ref{theo:mtheoalphar} which involves only arbitrary truncated (at~$\mathcal{B}$) outgoing cones in the infinity region and arbitrary truncated (at~$\mathcal{B}$) ingoing cones in the horizon region. More precisely, we will prove (recalling Remark~\ref{rem:equiven}):

\begin{theorem} \label{theo:mtheoalphar2}
We have the estimates 
\[
	\sum_{s=0,1,2}
	\sup_{u_{1} \leq \tau \leq u_f} \tau^s \cdot {\check{\mathbb{E}}}_\Box^{N-s,2-s} \left[{\alpha}_{\I}\right] \left(\tau\right)  + \sum_{s=0,1,2}
	\sup_{v \geq v_{1}} 
	v^{s} \cdot {\check{\mathbb{E}}}_\Box^{N-s} \left[{\alpha}_{\Hp}\right] \left(v\right)  \lesssim \varepsilon_0^2 + \varepsilon^3 
\]
for the restricted energies (recall $A_{\I}=r\Omega^2 \alpha_{\I}$, $A_{\Hp}=r\Omega^2 \alpha_{\Hp}$ and $\Pi_{\I}=r^3 \Omega \psi_{\I}$, $\Psi_{\I} = r^5 P_{\I}$)
\begin{align}
&\check{\mathbb{E}}_\Box^{K} \left[{\alpha}_{\Hp}\right] \left(v\right) := \sup_{\tilde{v} \geq v} \int_{\check{\underline{C}}^{\Hp}_{\tilde{v}}} \Omega^2  \sum_{|\underline{k}|=0; k_3\neq |\underline{k}|}^{K}  | \mathfrak{D}^{\underline{k}} {A}_{\Hp}|^2 +\int_{\check{C}^{\Hp}_{u_f}({v})} \sum_{|\underline{k}|=0; {k_2 \neq |\underline{k}|}}^{K}  | \mathfrak{D}^{\underline{k}} A_{\Hp}|^2  \nonumber \\
& \qquad \ \  +  \int_{\check{\mathcal{D}}^{\Hp}\left(v\right)}  \Omega^2 \Bigg\{ \sum_{|\underline{k}|=0}^{K} \left(1-\frac{3M}{r}\right)^2 | \mathfrak{D}^{\underline{k}} {A}_{\Hp}|^2  +  \sum_{|\underline{k}|=0}^{K-1}  | \mathfrak{D}^{\underline{k}} {A}_{\Hp}|^2+  | R^\star \mathfrak{D}^{\underline{k}} {A}_{\Hp}|^2  \Bigg\} \nonumber \, ,
\end{align}
\begin{align}
&\check{\mathbb{E}}^{K,p}_\Box \left[{\alpha}_{\I}\right] \left(\tau\right) := 
 \sup_{\tau \leq u \leq u_f} \int_{\check{C}^{\I}_u} \Bigg\{ \sum_{|\underline{k}|=1; {k_2 \neq |\underline{k}|}}^K  \ r^{4+p} | \mathfrak{D}^{\underline{k}} A_{\I}|^2 + \sum_{|\underline{k}|=0; {k_2 \neq |\underline{k}|}}^{K-1}  r^2 | \mathfrak{D}^{\underline{k}} \Pi_{\I}|^2 \Bigg\} \nonumber \\
& \qquad \qquad \ \   + \int_{\check{\underline{C}}^{\I}_{v_{\infty}}(\tau)}  \Bigg\{ \sum_{|\underline{k}|=1, k_3 \neq |\underline{k}|}^{K}  \ r^{4+p} | \mathfrak{D}^{\underline{k}} A_{\I} |^2 + \sum_{|\underline{k}|=0}^{K-1}  r^{2+p} | \mathfrak{D}^{\underline{k}} \Pi_{\I}|^2 + \sum_{|\underline{k}|=0}^{K-2}  | \mathfrak{D}^{\underline{k}} \Psi_{\I}|^2  \Bigg\} \nonumber \\
& \qquad \qquad \ \  +  \int_{\check{\mathcal{D}}^{\I}\left(\tau\right)} \Bigg\{ \sum_{|\underline{k}|=1}^K  \, r^{3+p-\delta} | \mathfrak{D}^{\underline{k}} A_{\I}|^2  + \sum_{|\underline{k}|=0}^{K-1} r^{1+p-\delta} | \mathfrak{D}^{\underline{k}} \Pi_{\I}|^2  +\sum_{|\underline{k}|=0}^{K-2} r^{-1-\delta} | \mathfrak{D}^{\underline{k}} \Psi_{\I}|^2 \Bigg\} .
\end{align}
\end{theorem}
The proof of Theorem \ref{theo:mtheoalphar2} will extend over {\bf Sections \ref{sec:traporel}--\ref{sec:higherorderw}} and is the main step in proving Theorem~\ref{theo:mtheoalphar}. The full statement of Theorem \ref{theo:mtheoalphar} is then obtained from Theorem \ref{theo:mtheoalphar2} in {\bf Section \ref{sec:missingfluxa}}. 
The latter step (i.e.~removing the check-superscripts, i.e.~adding general ingoing cones in the infinity gauge and general outgoing cones in the horizon gauge as well as considering the non-truncated cones) is straightforward and follows exactly as for $P$, $\underline{P}$ by doing localised energy estimates. 

We end this overview by outlining the proof of Theorem  \ref{theo:mtheoalphar2}.
We first recall the defining relations
\begin{align} \label{recda}
\Omega \slashed{\nabla}_3 A_{\Hp} = (-2)\frac{\Omega^2}{r^2} \Pi_{\Hp}    \ \ \ , \ \ \ \Omega \slashed{\nabla}_3 A_{\I} = (-2)\frac{\Omega^2}{r^2} \Pi_{\I} \, , 
\end{align}
\begin{align} \label{recdpi}
\Omega \slashed{\nabla}_3 \Pi_{\Hp} = \frac{\Omega^2}{r^2} \Psi_{\Hp}    \ \ \ , \ \ \ \Omega \slashed{\nabla}_3 \Pi_{\I} = \frac{\Omega^2}{r^2} \Psi_{\I}  \, ,
\end{align}
as well as the non-linear wave equations (written in ``elliptic" form)
\begin{align} 
 r^2 \slashed{\mathcal{D}}_2^\star \slashed{div} A_{\Hp}  + \frac{3M}{r} A_{\Hp}  &= \Omega \slashed{\nabla}_4 \Pi_{\Hp}  + \frac{2}{r}\left(1- \frac{3M}{r}\right) \Pi_{\Hp}  + \mathcal{E}^1, \label{teueli1} \\ 
 r^2 \slashed{\mathcal{D}}_2^\star \slashed{div} A_{\I} + \frac{3M}{r} A_{\I} &= \Omega \slashed{\nabla}_4 \Pi_{\I} + \frac{2}{r} \left(1- \frac{3M}{r}\right) \Pi_{\I} + \mathcal{E}^1_4 \label{teueli2}
\end{align}
and
\begin{align} 
2 r^2\slashed{\mathcal{D}}_2^\star \slashed{div} \Pi_{\Hp}  + 2 \Pi_{\Hp}  - \frac{6M}{r} \Pi_{\Hp}  &=  -\Omega \slashed{\nabla}_4 \Psi_{\Hp}  + 3MA_{\Hp}  + \mathcal{E}^2_2,  \label{wep1} \\
2 r^2\slashed{\mathcal{D}}_2^\star \slashed{div} \Pi_{\I} + 2 \Pi_{\I} - \frac{6M}{r} \Pi_{\I} &=  -\Omega \slashed{\nabla}_4 \Psi_{\I} + 3MA_{\I} + \mathcal{E}^2_2.  \label{wep2}
\end{align}
Note that we can always replace $r^2\slashed{\mathcal{D}}_2^\star \slashed{div}$ by $-\frac{1}{2} r^2 \slashed{\Delta} + 1$ since the term involving the difference of the Gauss curvature with the round metric can be incorporated into the non-linear error.

The logic in obtaining the required estimates is then as follows:
\begin{enumerate}
\item Interpreting the defining relations (\ref{recda}), (\ref{recdpi}) as \emph{transport equations} we derive estimates on $(\slashed{\nabla}_{R^\star})^k \Pi_{\Hp}$ and $(\slashed{\nabla}_{R^\star})^k A_{\Hp}$ as well as $(\slashed{\nabla}_{R^\star})^k \Pi_{\I}$ and $(\slashed{\nabla}_{R^\star})^k A_{\I}$ for $0\leq k \leq N-2$, i.e.~with a loss of derivatives. These will be useful to control lower order terms. This is the content of Section \ref{sec:traporel}. We remark that from (\ref{recda}), (\ref{recdpi}) we could, by appropriate commutations, derive estimates for \emph{all} derivatives up to order $N-2$, however, we prefer a different argument below.

\item In Section \ref{sec:higherorderw} we consider the \emph{wave equations} satisfied by $(\Pi_{\Hp}, \Pi_{\I})$ and $(A_{\Hp}, A_{\I})$ and prove inductively boundedness and integrated decay estimates for $((\slashed{\nabla}_{R^\star})^k \Pi_{\Hp}, (\slashed{\nabla}_{R^\star})^k\Pi_{\I})$ with $0 \leq k \leq N-1$ and $((\slashed{\nabla}_{R^\star})^k A_{\Hp}, (\slashed{\nabla}_{R^\star})^k A_{\I})$ with $0 \leq k \leq N$. These estimates do not lose derivatives and do not degenerate near $3M$ for $k\geq 1$, see Propositions \ref{prop:pirstar} and \ref{prop:arstar}. With this established we control---with good weights near the horizon but non-optimal weights near infinity---\emph{all} derivatives of $(\Pi_{\Hp}, \Pi_{\I})$ both in integrated decay (Section \ref{sec:susy1}) and fluxes (Section \ref{sec:susy2}). The basic idea here is simple: Angular derivatives are controlled from the relations (\ref{teueli1})--(\ref{wep2}) and previous estimates. For $\Omega \slashed{\nabla}_3$ derivatives one can use (\ref{recdpi}) and the fact that estimates on $(\Psi_{\Hp}, \Psi_{\I})$ have already been established in Theorem~\ref{thm:PPbarestimates}. Finally $\Omega \slashed{\nabla}_4 = \Omega \slashed{\nabla}_3 + 2R^\star$ can be used to reduce $\Omega \slashed{\nabla}_4$ derivatives to $R^\star$ derivatives, which Propositions \ref{prop:pirstar} and \ref{prop:arstar} provide control on.

It remains to optimise the $r$-weights for $\Pi_{\I}$ and $A_{\I}$. This is done in Section \ref{sec:piarpweighted} using estimates for the relevant Bianchi pairs. This finally provides a hierarchy of $r$-weighted estimates, see Proposition \ref{prop:APibp}.  We summarise our weighted estimates concisely in Proposition \ref{prop:sumrp}. This is also a form to which the  pigeonhole principle argument of~\cite{DafRodnew} can then be applied to yield an inverse polynomial decay hierarchy for the weighted energies which provides all of the estimates in Theorem  \ref{theo:mtheoalphar2}. This is the content of
Section~\ref{sec:decaypia}.
\end{enumerate}

\section{Auxiliary transport estimates  at low orders for $\alpha$} \label{sec:traporel}

Exploiting the transport relations we can deduce lower order estimates for $\Pi_{\I}$, $\Pi_{\Hp}$ as well as $A_{\I}$, $A_{\Hp}$.

\begin{proposition} \label{prop:lowordertp}
For any $u_1 \leq \tau \leq u_f$ and $1 \leq K \leq N-2$
the quantities  ${\Pi}_{\I}= r^3 \Omega \psi_{\I}$ and $A_{\I} = r \Omega^2 \alpha_{\I}$  satisfy the estimate
\begin{align}
\sum_{k=1}^{K} \int_{\DcI(\tau)} r^{-1-\delta} |\Pi_{\I}|^2 + r^{1-\delta} |A_{\I}|^2 + r^{1-\delta} \left( | (\slashed{\nabla}_{R^\star})^{k} \Pi_{\I}|^2  +  |(\slashed{\nabla}_{R^\star})^{k} A_{\I}|^2  \right) \nonumber \\
+\sum_{k=1}^{K} \int_{\mathcal{B}(\tau)}  | \Pi_{\I}|^2   + | (\slashed{\nabla}_{R^\star})^{k} \Pi_{\I}|^2 +   |A_{\I}|^2  + | (\slashed{\nabla}_{R^\star})^{k} A_{\I}|^2  \nonumber \\
 \lesssim   \sum_{k=0}^{K} \int_{\check{C}^{\I}_\tau} r^2 | \left(\Omega \slashed{\nabla}_4\right)^k A_{\I}|^2 +  | \left(r\Omega \slashed{\nabla}_4\right)^k \Pi_{\I}|^2 + \frac{\varepsilon_0^2 + \varepsilon^3}{\tau^{\min(2,N-2-K)}}
\end{align}
and the quantities  ${\Pi}_{\Hp}= r^3 \Omega \psi_{\Hp}$ and $A_{\Hp} = r \Omega^2 \alpha_{\Hp}$  satisfy
\begin{align}
\sum_{k=0}^{K} \int_{\DcH(v(\tau))} |  (\slashed{\nabla}_{R^\star})^{k} \Pi_{\Hp}|^2  +  |  (\slashed{\nabla}_{R^\star})^{k} A_{\Hp}|^2  
+\sum_{k=0}^{K} \int_{\mathcal{B}\left(\tau\right)}   |  (\slashed{\nabla}_{R^\star})^{k} \Pi_{\Hp}|^2 + |  (\slashed{\nabla}_{R^\star})^{k} A_{\Hp}|^2 \nonumber \\
 \lesssim   \sum_{k=0}^{K} \int_{\check{C}^{\I}_\tau} r^2 | \left(\Omega \slashed{\nabla}_4\right)^k A_{\I}|^2 +  | \left(r\Omega \slashed{\nabla}_4\right)^k \Pi_{\I}|^2 + \frac{\varepsilon_0^2 + \varepsilon^3}{\tau^{\min(2,N-2-K)}} \, .
\end{align}
Note the right hand side in the two estimates is identical.
\end{proposition}

\begin{proof}
The proof proceeds in several steps.

\vskip1pc
\noindent
{\bf Step 0.} We first show the following auxiliary lemma which implies that it suffices to prove the above estimates with the right hand side replaced by the left hand side of (\ref{auxgoli}).
\begin{lemma} \label{lem:auxrhs}
We have, for any $u_1 \leq \tau \leq u_f$ and $1 \leq K \leq N-2$, the estimate
\begin{align} \label{auxgoli}
\sum_{k=1}^{K} \int_{\check{C}^{\I}_\tau} \, r^{-\delta}  |\Pi_{\I}|^2  + r^{2-\delta} \left( |A_{\I}|^2 +  | (\slashed{\nabla}_{R^\star})^{k} A_{\I}|^2 +  | (\slashed{\nabla}_{R^\star})^k \Pi_{\I}|^2 \right)  \nonumber \\
\lesssim   \sum_{k=0}^{K} \int_{\check{C}^{\I}_\tau} r^2 | \left(\Omega \slashed{\nabla}_4\right)^k A_{\I}|^2 +  | \left(r\Omega \slashed{\nabla}_4\right)^k \Pi_{\I}|^2 + \frac{\varepsilon_0^2 + \varepsilon^3}{\tau^{\min(2,N-2-K)}} \, .
\end{align}
\end{lemma}
\begin{proof}
Use the definition $2R^\star = -\Omega \slashed{\nabla}_3 + \Omega \slashed{\nabla}_4$ and the defining relations (\ref{recda}) and (\ref{recdpi}) in conjunction with the estimates on $\Psi_{\I}$ of Theorem \ref{thm:PPbarestimates}.\footnote{Note that at most the flux of $N-1$ transversal derivatives of $\Psi_{\I}$ can appear.}
\end{proof}

\vskip1pc
\noindent
{\bf Step 1. Proof of the first estimate without the $ (\slashed{\nabla}_{R^\star})^k$ terms}. For technical reasons, namely to avoid elliptic estimates on truncated cones, we will first prove the estimate with the regions replaced by ${\DcI(\tau) \cap \{r\geq R_2\}}$, the hypersurface $\mathcal{B}$ replaced by the hypersurface $\{r= r_{\I}=R_2\} \cap \DcI$ and the cones truncated at the sphere $r=R_2$. From (\ref{wep2}) we have
\begin{align} \label{ibf}
 \int \frac{| \Pi_{\I}|^2 }{r^{1+\delta}}  \leq  \int \frac{|  r^2\slashed{\mathcal{D}}_2^\star \slashed{div}\Pi_{\I}+ \Pi_{\I}|^2 }{r^{1+\delta}} 
\leq \int \frac{|  A_{\I}|^2 }{r^{1+\delta}} + \frac{\varepsilon_0^2 + \varepsilon^3}{\tau^{\min(2,N-2-K)}}\, ,
\end{align}
where the integration is over ${\DcI(\tau) \cap \{r\geq R_2\}}$ with measure $dudvd\theta$ and we have used the fact that $M_{\rm init} R^{-1}< \delta$ by (\ref{definitionofRhere}) in the second inequality. 
From 
(\ref{recda}) we derive
\begin{align} \label{trapo1}
\frac{1}{2} \Omega \slashed{\nabla}_3 (r^s |A_{\I} |^2) + \frac{1}{2} s \cdot r^{s-1} \Omega_\circ^2 |A_{\I}|^2 = \frac{\Omega^2}{r^2} \Pi_{\I} r^s A_{\I} \, .
\end{align}
Applying this with $s=2-\delta$ and integrating over  ${\DcI(\tau) \cap \{r\geq R_2\}}$ with respect to 
$dudvd\theta$ (cf.~Lemma \ref{lem:transformder}) we obtain (after using Cauchy--Schwarz on the right and inserting (\ref{ibf}) using (\ref{definitionofRhere})) the desired estimates for ${A}_{\I}$ on the region ${\DcI(\tau) \cap \{r\geq R_2\}}$ and the hypersurface $r=R_2$ (instead of $\mathcal{B}$).  Revisiting (\ref{ibf}), the estimate for  $\Pi_{\I}$ on ${\DcI(\tau) \cap \{r\geq R_2\}}$ also follows. The missing estimate for $\Pi_{\I}$ on $r=R_2$ (instead of $\mathcal{B}$) is now a consequence of the transport relation (an immediate consequence of (\ref{recdpi}))
\begin{align} \label{trapo2}
\frac{1}{2} \Omega \slashed{\nabla}_3 (r^s |\Pi_{\I} |^2) + \frac{1}{2} s \cdot r^{s-1} \Omega_\circ^2 |\Pi_{\I}|^2 = \frac{\Omega^2}{r^2} \Psi_{\I} r^s \Pi_{\I} \, 
\end{align}
applied with $s=-\delta$ and inserting the previous estimates on $\Pi_{\I}$, (\ref{ibf}), to control the wrong-signed spacetime term on the left. This proves all of the desired estimates for $r\geq R_2$. We can now easily extend them all the way up to $\mathcal{B}$ by integrating (\ref{trapo2}) with $s=+\delta$ over the region $\DcI(\tau) \cap \{r\leq R_2\}$. The boundary term on $r=R_2$ is controlled by the previous step and on the right hand side we can estimate (since we are now in a region with $R_{-2} \leq r \leq R_2$)
\begin{align}
 \frac{\Omega^2}{r^2} \Psi_{\I} r^\delta \Pi_{\I} \leq  \frac{C}{\delta} R^\delta \frac{\Omega^2}{r^3}| \Psi_{\I} |^2 + \frac{1}{4} \delta \cdot r^{\delta-1} \Omega_\circ^2 |\Pi_{\I}|^2+ \frac{\varepsilon_0^2 + \varepsilon^3}{\tau^{\min(2,N-2-K)}}\, .
\end{align}
The first term can be incorporated in the decay term and the second term absorbed on the left. We therefore immediately obtain the desired estimate for $\Pi_{\I}$. Integrating now (\ref{trapo1}) with $r=2-\delta$ over $\DcI(\tau) \cap \{r\leq R_2\}$ and applying Cauchy--Schwarz on the right (using the previous estimate for $\Pi_{\I}$) now yields the desired estimate also for $A_{\I}$.\footnote{The reason for the two-step argument we applied now becomes clear: We cannot apply (\ref{trapo2}) with $s=\delta$ in $r\geq R_2$ without using an $r^p$-weighted norm for $\Psi_{\I}$ with $p\geq \delta$.} 

\vskip1pc
\noindent
{\bf Step 2. Proof of the first estimate including the $ (\slashed{\nabla}_{R^\star})^k$-terms.} We now turn to the commuted versions of the transport equations (\ref{recda}), (\ref{recdpi}):
\[
\Omega \slashed{\nabla}_3  (\slashed{\nabla}_{R^\star})^k \Pi_{\I} = \frac{\Omega^2}{r^2}  (\slashed{\nabla}_{R^\star})^k \Psi_{\I} + \sum_{i=0}^{k-1} \Omega^2 (f_{-3})_i \ (\slashed{\nabla}_{R^\star})^i \Psi_{\I} + \mathcal{E}^{k+2}_3 \, ,
\]
\[
\Omega \slashed{\nabla}_3  (\slashed{\nabla}_{R^\star})^k A_{\I} = \frac{\Omega^2}{r^2}  (\slashed{\nabla}_{R^\star})^k \Pi_{\I} + \sum_{i=0}^{k-1} \Omega^2 (f_{-3})_i  (\slashed{\nabla}_{R^\star})^i \Pi_{\I} + \mathcal{E}^{k+1}_5 \, .
\]
Contracting the first by $\ (\slashed{\nabla}_{R^\star})^k \Pi_{\I} r^{2-\delta}$ and the second by $\ (\slashed{\nabla}_{R^\star})^k A_{\I} r^{2-\delta}$ (stronger weights could be applied for $A_{\I}$) yields the desired estimates after integration over $\DcI(\tau)$ and applying Cauchy--Schwarz on the right.\footnote{The non-linear error satisfies $\int_{\DcI(\tau)} r^{3-\delta} |\mathcal{E}_3^{K+2}|^2 \lesssim \frac{\varepsilon_0^2+ \varepsilon^3}{\tau^{\min(2,N-2-K)}}$ for all $1\leq K \leq N-2$ by Proposition \ref{prop:waveIerror1b}.} The estimate for $ (\slashed{\nabla}_{R^\star})^kA_{\I}$ now follows from the (commuted) transport equation for $A_{\I}$.

\vskip1pc
\noindent
{\bf Step 3. Proof of the second estimate.} This is easier as $r$-weights can be absorbed into the constants. The estimates on $\mathcal{B}$ follow directly from expressing the expression on $\mathcal{B}$ of the first estimate in terms of the horizon frame and using Proposition \ref{thm:cancelT}. The estimates on $\DcH$ are then a direct consequence of applying the transport relations (\ref{trapo1}) with $s>0$ and (\ref{trapo2}), which hold verbatim for $A_{\Hp}$ and $\Pi_{\Hp}$. Note again that in this region $r$-weights can be absorbed into the constant implicit in $\lesssim$.
\end{proof}

\section{Higher order energy estimates for $\alpha$} \label{sec:higherorderw}
In this section, we turn to higher order estimates for $\alpha$.

\subsection{Basic estimates for $\Pi$ and $(\slashed{\nabla}_{R^\star})^k \Pi$}
We begin with a basic estimate arising from commuting the $\Pi$ equations only with $R^\star$ derivatives:

\begin{proposition} \label{prop:pirstar}
For any $u_1 \leq \tau \leq u_f$ and $1\leq K\leq N-2$ the pair $\left(\Pi_{\Hp}=r^3 \Omega \psi_{\Hp},
{\Pi}_{\I}=r^3 \Omega {\psi}_{\I}\right)$ satisfies
\begin{align} \label{golia}
&\sum_{k=0}^K \sup_{\tau \leq u \leq u_f} \check{\underline{\mathbb{F}}}_{v(u)} [ (\slashed{\nabla}_{R^\star})^k\Pi_{\Hp}] + \sum_{k=0}^K \sup_{\tau \leq u \leq u_f} \check{\mathbb{F}}_u [ (\slashed{\nabla}_{R^\star})^k\Pi_{\I}] 
 \\
 &+ \sum_{k=0}^K \check{{\mathbb{F}}}_{u_f} \left[ (\slashed{\nabla}_{R^\star})^k\Pi_{\Hp} \right] \left(v(\tau)\right)+ \sum_{k=0}^K \check{\underline{\mathbb{F}}}_{v_\infty} \left[ (\slashed{\nabla}_{R^\star})^k\Pi_{\I} \right] \left(\tau\right) \nonumber \\
&+\sum_{k=0}^K \check{\mathbb{I}} \left[ (\slashed{\nabla}_{R^\star})^k\Pi_{\Hp}\right] \left(v(\tau)\right)  +\sum_{k=0}^K \check{\mathbb{I}} \left[ (\slashed{\nabla}_{R^\star})^k\Pi_{\I}\right] \left(\tau\right) 
        \nonumber \\
                & \qquad \lesssim
  \sum_{k=0}^K\underline{\check{\mathbb{F}}}_{v(\tau)} [ (\slashed{\nabla}_{R^\star})^k\Pi_{\Hp}]
+
\sum_{k=0}^K\check{\mathbb{F}}_{\tau} [ (\slashed{\nabla}_{R^\star})^k\Pi_{\I}]\nonumber \\
& \qquad \  + \sum_{k=0}^{K} \int_{\check{C}^{\I}_\tau} r^2 | \left(\Omega \slashed{\nabla}_4\right)^k A_{\I}|^2 +  | \left(r\Omega \slashed{\nabla}_4\right)^k \Pi_{\I}|^2 + \frac{\varepsilon_0^2 + \varepsilon^3}{\tau^{\min(2,N-2-K)}}  \, . \nonumber
\end{align}
\end{proposition}

\begin{remark}
Note that we are asserting \underline{non-degenerate} control near $r=3M$ as soon as one commutes at least once with $\slashed{\nabla}_{R^\star}$. In fact, the proof will establish the estimate also for $K=0$ provided one replaces $\check{\mathbb{I}} \left[\Pi_{\Hp}\right] \left(v(\tau)\right)$ by $\check{\mathbb{I}}^{deg} \left[\Pi_{\Hp}\right] \left(v(\tau)\right)$ on the left hand side.
\end{remark}

\begin{proof}
Recall from Proposition \ref{prop:whythat} that $\Pi_{\I}$ and $\Pi_{\Hp}$ satisfy a tensorial wave equation of Type 1 with
\[
\mathcal{F}^{lin} \left[ \Pi_{\Hp} \right] =  \frac{3M}{r^2}\Omega^2 A_{\Hp} - \frac{2}{r} \frac{\Omega^2}{r^2} \left(1-\frac{3M}{r}\right) \Psi_{\Hp} \ \ \ , \ \ \ \mathcal{F}^{lin} \left[ \Pi_{\I} \right] =  \frac{3M}{r^2}\Omega^2 {A}_{\I} - \frac{2}{r} \frac{\Omega^2}{r^2} \left(1-\frac{3M}{r}\right) {\Psi}_{\I} \, .
\]
By a simple induction using Proposition \ref{prop:commutetype1} it is easy to see that $ (\slashed{\nabla}_{R^\star})^k \Pi_{\I}$ and $ (\slashed{\nabla}_{R^\star})^k \Pi_{\Hp}$ satisfy a tensorial wave equation of Type 1 and that the linear error term after $K$ commutations will have the form ($h_0$, $h^k_0$ and $h^{k,i}_{0}$ denoting admissible coefficients functions of $r$ which can be different in different places)
\begin{align} \label{linef}
\mathcal{F}^{lin} \left[  (\slashed{\nabla}_{R^\star})^{K} \Pi_{\I} \right] &=\left(1-\frac{3M}{r}\right) \frac{\Omega^2}{r^3} \left(h_0  (\slashed{\nabla}_{R^\star})^K \Psi_{\I} + h_0 \Omega \slashed{\nabla}_4  (\slashed{\nabla}_{R^\star})^{K-1} \Psi_{\I} \right)  \\
&+ \sum_{k=0}^K  \frac{h^k_{0}}{r^2} \Omega^2  (\slashed{\nabla}_{R^\star})^k A_{\I} + \sum_{i=0}^1  \sum_{k=i}^{K-1} \frac{h^{k,i}_{0}}{r^3} \Omega^2 (\Omega \slashed{\nabla}_4)^i  (\slashed{\nabla}_{R^\star})^{k-i} \Psi_{\I} + \sum_{k=0}^{K-1} \frac{h_{0}^k}{r^3}  (\slashed{\nabla}_{R^\star})^k \Pi_{\I} \nonumber
\end{align}
and with the same schematic form for $\mathcal{F}^{lin} \left[  (\slashed{\nabla}_{R^\star})^{K} \Pi_{\Hp} \right]$. From Propositions \ref{prop:whythat} and \ref{prop:commutetype1} one sees that the non-linear error-term is of the form
\[
\mathcal{F}^{nlin} \left[  (\slashed{\nabla}_{R^\star})^{K} \Pi_{\I} \right] = \Omega^2 \mathcal{E}^{2+K}_{4} \ \ \ , \ \ \ \mathcal{F}^{nlin} \left[  (\slashed{\nabla}_{R^\star})^{K} \Pi_{\Hp} \right] = \Omega^2 (\mathcal{E}^\star)^{K+2} \, .
\]
{\bf Step 1.} We claim that applying the $T$-boundedness estimate and the Morawetz $X$-estimate of Proposition \ref{prop:basecase} to the wave equations for $ (\slashed{\nabla}_{R^\star})^k \Pi_{\Hp}$ and $ (\slashed{\nabla}_{R^\star})^k \Pi_{\I}$ already yields (\ref{golia}) for all $0 \leq K \leq N-2$ ($0$ included), however, 
\begin{itemize}
\item with the horizon fluxes carrying an additional diamond superscript (i.e.~weaker energies on the horizon) and 
\item with $\check{\mathbb{I}} \left[ (\slashed{\nabla}_{R^\star})^k \Pi_{\Hp}\right] \left(v(\tau)\right)$ replaced by $\check{\mathbb{I}}^{\diamond,deg} \left[ (\slashed{\nabla}_{R^\star})^k \Pi_{\Hp}\right] \left(v(\tau)\right)$ on the left. 
\end{itemize}
To verify this claim, we observe that using $T=\Omega \slashed{\nabla}_3 + R^\star$ the linear error $\mathcal{G}_1\left[ (\slashed{\nabla}_{R^\star})^{K} \Pi\right] \left(\tau,u_f\right)$ appearing in Proposition \ref{prop:basecase} can be estimated further by
\begin{align}
&\int_{\DcH(v(\tau))} |\mathcal{F}^{lin} \left[ (\slashed{\nabla}_{R^\star})^{K} \Pi_{\Hp}\right] || \slashed{\nabla}_T  (\slashed{\nabla}_{R^\star})^{K} \Pi_{\Hp}| + \int_{\DcI(\tau)} |\mathcal{F}^{lin} \left[ (\slashed{\nabla}_{R^\star})^{K} \Pi_{\I}\right] || \slashed{\nabla}_T  (\slashed{\nabla}_{R^\star})^{K} \Pi_{\I}| \nonumber \\
&\lesssim \int_{\DcH(v(\tau))} |\mathcal{F}^{lin} \left[ (\slashed{\nabla}_{R^\star})^{K} \Pi_{\Hp}\right]  | \Big|  (\slashed{\nabla}_{R^\star})^{K+1} \Pi_{\Hp} +  (\slashed{\nabla}_{R^\star})^{K} \left(\frac{\Omega^2}{r^2} \Psi_{\Hp} \right) + \Omega^2 (\mathcal{E}^\star)^{K+1} \Big|  \nonumber \\
& \qquad  +\int_{\DcI(\tau)} | \mathcal{F}^{lin} \left[ (\slashed{\nabla}_{R^\star})^{K} \Pi_{\I}\right] | \Big|  (\slashed{\nabla}_{R^\star})^{K+1} \Pi_{\I} +  (\slashed{\nabla}_{R^\star})^{K} \left(\frac{\Omega^2}{r^2} \Psi_{\I} \right) + \Omega^2 \mathcal{E}^{K+1}_4\Big|  \nonumber \\
&\lesssim \gamma \int_{\DcH(v(\tau))} \frac{\Omega^2}{r^{2}} | (\slashed{\nabla}_{R^\star})^{K+1} \Pi_{\Hp}|^2 + \int_{\DcI(\tau)} \frac{\Omega^2}{r^{2}} | (\slashed{\nabla}_{R^\star})^{K+1} \Pi_{\I}|^2 + C_\gamma \left(\textrm{RHS of (\ref{golia})} \right) \nonumber
\end{align}
for any $\gamma>0$. Here we have used: (1) the error estimates of 
Propositions~\ref{prop:wavehorizonerror1},~\ref{prop:wavehorizonerror2}  for the terms involving $(\mathcal{E}^\star)^{K+1}$  in $\mathcal{D}^{\mathcal{H}^+}$ and Proposition~\ref{prop:waveIerror2} for the analogous error in 
$\mathcal{D}^{\mathcal{I}^+}$ and 
(2) Cauchy--Schwarz in the last step, together with the fact that (2a) the lower oder terms in $A_{\Hp}$ and $\Pi_{\Hp}$ (appearing in $|\mathcal{F}^{lin} \left[  (\slashed{\nabla}_{R^\star})^{K} \Pi\right] |^2$) from (\ref{linef}) can be controlled from Proposition \ref{prop:lowordertp} and (2b) the terms involving $\Psi_{\Hp}$ or $\Psi_{\I}$ are controlled from Theorem~\ref{thm:PPbarestimates}.

For the terms $\mathcal{G}_2\left[ (\slashed{\nabla}_{R^\star})^{K} \Pi\right] \left(\tau,u_f\right)$ and $\mathcal{G}_3\left[ (\slashed{\nabla}_{R^\star})^{K} \Pi\right] \left(\tau,u_f\right)$, we note similarly that for any $\gamma>0$, we have
\begin{align}
&\int_{\DcH(v(\tau))} |\mathcal{F}^{lin} \left[ (\slashed{\nabla}_{R^\star})^{K} \Pi_{\Hp}\right]  \slashed{\nabla}_{R^\star}  (\slashed{\nabla}_{R^\star})^{K} \Pi_{\Hp}| + \int_{\DcI(\tau)} |\mathcal{F}^{lin} \left[ (\slashed{\nabla}_{R^\star})^{K} \Pi_{\I}\right]  \slashed{\nabla}_{R^\star}  (\slashed{\nabla}_{R^\star})^{K} \Pi_{\I}| \nonumber \\
+&\int_{\DcH(v(\tau))} |\mathcal{F}^{lin} \left[ (\slashed{\nabla}_{R^\star})^{K} \Pi_{\Hp}\right]  \frac{1}{r^{1+\delta}}   (\slashed{\nabla}_{R^\star})^k \Pi_{\Hp}| + \int_{\DcI(\tau)} |\mathcal{F}^{lin} \left[ (\slashed{\nabla}_{R^\star})^{K} \Pi_{\I}\right] \frac{1}{r^{1+\delta}}   (\slashed{\nabla}_{R^\star})^k \Pi_{\I}| \nonumber \\ 
 &\lesssim \gamma \int_{\DcH(v(\tau))} \frac{\Omega^2}{r^{2}} | (\slashed{\nabla}_{R^\star})^{K+1} \Pi_{\Hp}|^2 + \int_{\DcI(\tau)} \frac{\Omega^2}{r^{2}} | (\slashed{\nabla}_{R^\star})^{K+1} \Pi_{\I}|^2 + C_\gamma \left(\textrm{RHS of (\ref{golia})} \right) \, .\nonumber
\end{align}
Choosing $\gamma$ sufficiently small (depending only on $M_{\rm init}$) we absorb the first terms on the right hand side of (\ref{Tiledb}) and the desired estimate (with the aforementioned restrictions) is proven provided we can control the non-linear errors arising from $\mathcal{H}_i$. For these, we note that,
using Propositions~\ref{prop:wavehorizonerror1},~\ref{prop:wavehorizonerror2}  for the errors in $\mathcal{D}^{\Hp}$ and Proposition~\ref{prop:waveIerror2} for the errors in $\mathcal{D}^{\I}$, we have
\begin{align} \label{nohe2}
\sum_{i=1}^4 \sum_{k\leq K}  \mathcal{H}_i  \left[ (\slashed{\nabla}_{R^\star})^k \Pi\right] \lesssim  \frac{\varepsilon_0^2 + \varepsilon^3}{\tau^{\min(2,N-2-K)}}
\end{align}
and that, using Proposition \ref{thm:cancelT}, we have
\begin{align} 
\sum_{i=1}^3  \sum_{|\underline{k}|\leq K} \Big|   \mathcal{B}_i \left[ (\slashed{\nabla}_{R^\star})^k \Pi\right] (\tau,u_f)-  \mathcal{B}_i \ \left[ (\slashed{\nabla}_{R^\star})^k \Pi\right]   (\tau,u_f)\Big| \lesssim   \frac{\varepsilon_0^2 + \varepsilon^3}{\tau^{\min(2,N-2-K)}} \, .
\end{align}

\vskip1pc
\noindent
{\bf Step 2.} To remove the diamond superscripts, we apply the redshift estimate of Proposition \ref{prop:redshift} to the tensorial wave equation satisfied by $ (\slashed{\nabla}_{R^\star})^k\Pi_{\Hp}$ and $ (\slashed{\nabla}_{R^\star})^k\Pi_{\I}$. The linear errors are easily seen to be controlled by Cauchy--Schwarz and the estimate from Step~1 and the non-linear error is again controlled by (\ref{nohe2}).

\vskip1pc
\noindent
{\bf Step 3.} We have proved (\ref{golia}) except that  $\sum_{k=0}^K\check{\mathbb{I}}^{deg} \left[ (\slashed{\nabla}_{R^\star})^k\Pi_{\Hp}\right] \left(v(\tau)\right)$ appears on the left hand side instead of  $\sum_{k=0}^K\check{\mathbb{I}} \left[ (\slashed{\nabla}_{R^\star})^k\Pi_{\Hp}\right] \left(v(\tau)\right)$. To obtain the non-degenerate control we proceed successively for $K=1,...,N-2$. For $K=1$ we consider the tensorial wave equation for $\slashed{\nabla}_{R^\star} \Pi_{\Hp}$ and integrate over $\DcH$ the Lagrangian identity (\ref{lagrangianid}) with $h$ a radial cut-off function equal to $1$ in $\left[3M-\frac{1}{4}M_{\rm init}, 3M+\frac{1}{4}M_{\rm init}\right]$ and vanishing for $r \leq \frac{5}{2}M_{\rm init}$ and $r \geq \frac{7}{2}M_{\rm init}$. Boundary terms and the term arising from $\mathfrak{F}^h\left[\slashed{\nabla}_{R^\star} \Pi_{\Hp}\right]$ are easily seen to be controlled by the estimate already established. Also, the lower order terms in $\mathfrak{f}^h_{bulk} \left[(\slashed{\nabla}_{R^\star}) \Pi_{\Hp}\right]$ are all directly controlled from $\sum_{k=0}^1\check{\mathbb{I}}^{deg} \left[ (\slashed{\nabla}_{R^\star})^k\Pi_{\Hp}\right] \left(v(\tau)\right)$. The angular derivative term has a good sign and for the wrong signed term involving $\slashed{\nabla}_T (\slashed{\nabla}_{R^\star}) \Pi_{\Hp}$ we observe 
\begin{align}
\slashed{\nabla}_T\slashed{\nabla}_{R^\star} \Pi_{\Hp} = (\slashed{\nabla}_{R^\star} + \Omega \slashed{\nabla}_3) \slashed{\nabla}_{R^\star} \Pi_{\Hp} = (\slashed{\nabla}_{R^\star})^2 \Pi_{\Hp}+ \frac{\Omega^2}{r^2} \slashed{\nabla}_{R^\star} \Psi_{\Hp} -2 \frac{\Omega^2}{r^3} \left(1-\frac{3M}{r}\right) \Psi_{\Hp} 
+\Omega^2 \left( \mathcal{E}^\star\right)^2 \nonumber
\end{align}
with all linear terms on the right already controlled \emph{non-degenerately} near $r=3M$. This gives non-degenerate (near $r=3M$) control on all first derivatives of $\slashed{\nabla}_{R^\star} \Pi$ and hence control on $\check{\mathbb{I}} \left[(\slashed{\nabla}_{R^\star}\Pi_{\Hp}\right] \left(v(\tau)\right)$. By a simple interpolation one also has $\check{\mathbb{I}} \left[\Pi_{\Hp}\right] \left(v(\tau)\right) \lesssim \check{\mathbb{I}} \left[(\slashed{\nabla}_{R^\star}\Pi_{\Hp}\right] \left(v(\tau)\right) + \check{\mathbb{I}}^{deg} \left[\Pi_{\Hp}\right] \left(v(\tau)\right)$. This finishes the proof for $K=1$. For $K=2,3,...,N-2$ we apply successively the same argument, at each order using the Lagrangian multiplier for the wave equation satisfied by $ (\slashed{\nabla}_{R^\star})^K \Pi_{\Hp}$.
\end{proof}

\subsection{Basic estimates for $A$ and $ (\slashed{\nabla}_{R^\star})^k A$}
Completely analogously we prove the above estimate for $A_{\Hp}$ and $A_{\I}$:

\begin{proposition} \label{prop:arstar}
For any $u_1 \leq \tau \leq u_f$ and $1 \leq K\leq N-1$ the pair $\left(A_{\Hp},
A_{\I}\right)$ satisfies
\begin{align}
&\sum_{k=0}^K \sup_{\tau \leq u \leq u_f} \check{\underline{\mathbb{F}}}_{v(u)} [ (\slashed{\nabla}_{R^\star})^kA_{\Hp}] + \sum_{k=0}^K \sup_{\tau \leq u \leq u_f} \check{\mathbb{F}}_u [ (\slashed{\nabla}_{R^\star})^kA_{\I}] 
 \\
 &+ \sum_{k=0}^K \check{{\mathbb{F}}}_{u_f} \left[ (\slashed{\nabla}_{R^\star})^k A_{\Hp} \right] \left(v(\tau)\right)+ \sum_{k=0}^K \check{\underline{\mathbb{F}}}_{v_\infty} \left[ (\slashed{\nabla}_{R^\star})^k A_{\I} \right] \left(\tau\right) \nonumber \\
&+\sum_{k=0}^K \check{\mathbb{I}} \left[ (\slashed{\nabla}_{R^\star})^k A_{\Hp}\right] \left(v(\tau)\right)  +\sum_{k=0}^K \check{\mathbb{I}} \left[ (\slashed{\nabla}_{R^\star})^kA_{\I}\right] \left(\tau\right) 
        \nonumber \\
                & \qquad \lesssim
  \sum_{k=0}^K\underline{\check{\mathbb{F}}}_{v(\tau)} [ (\slashed{\nabla}_{R^\star})^k A_{\Hp}]
+
\sum_{k=0}^K\check{\mathbb{F}}_{\tau} [ (\slashed{\nabla}_{R^\star})^k A_{\I}]
+ \sum_{k=0}^{K-1}\underline{\check{\mathbb{F}}}_{v(\tau)} [ (\slashed{\nabla}_{R^\star})^k\Pi_{\Hp}]
+
\sum_{k=0}^{K-1}\check{\mathbb{F}}_{\tau} [ (\slashed{\nabla}_{R^\star})^k\Pi_{\I}] \nonumber \\
& \qquad \  + \sum_{k=0}^{\max(1,K-1)} \int_{\check{C}^{\I}_\tau} r^2 | \left(\Omega \slashed{\nabla}_4\right)^k A_{\I}|^2 +  | \left(r\Omega \slashed{\nabla}_4\right)^k \Pi_{\I}|^2 + \frac{\varepsilon_0^2 + \varepsilon^3}{\tau^{\min(2,N-2-K)}}  \, .
 \nonumber
\end{align}
\end{proposition}

\begin{proof}
Note that $A_{\Hp}$ and $A_{\I}$ satisfy a tensorial wave equation of Type 2 with 
\begin{align}
\mathcal{F}^{lin}\left[A_{\Hp}\right] = +\frac{8}{r} \frac{\Omega^2}{r^2} \left(1-\frac{3M}{r}\right)\Pi_{\Hp} \ \ \ , \ \ \  \mathcal{F}^{lin}\left[A_{\I}\right] = +\frac{8}{r} \frac{\Omega^2}{r^2} \left(1-\frac{3M}{r}\right)  \Pi_{\I} \, .
\end{align}
Repeating the proof of Proposition \ref{prop:pirstar}, this linear error term and its $R^\star$ commuted analogues are easily controlled using Cauchy--Schwarz and the estimates from Proposition \ref{prop:pirstar}. Note in particular that we obtain $K-1$ in the sums involving $\Pi$ on the right hand side.\footnote{Indeed, for $K=2$ (commuting twice), say, we need to control \emph{two} derivatives of $\Pi$ in the linear error which follow from (the spacetime terms of) $K=1$ in Proposition \ref{prop:pirstar}. Similarly for higher $K$.}
\end{proof}

\subsection{Integrated local energy decay for all derivatives of $\Pi$ and $A$} \label{sec:susy1}

We next conclude that Propositions \ref{prop:pirstar} and \ref{prop:arstar} already provide control over \emph{all} derivatives of $\Pi_{\Hp}$ and $\Pi_{\I}$, however 
 \emph{with non-optimal weights near infinity} (note the $\star$ in the energies below). On the other hand, the weights near the horizon do not have to be improved.

\begin{proposition} \label{prop:piall}
For any $u_1 \leq \tau \leq u_f$ and $1\leq K\leq N-2$ the pair $\left(\Pi_{\Hp}=r^3 \Omega \psi_{\Hp},
{\Pi}_{\I}=r^3 \Omega {\psi}_{\I}\right)$ satisfies
\begin{align} \label{golik}
&\sum_{|\underline{k}|\leq K+1} \check{\mathbb{I}}^{deg} \left[\tilde{\mathfrak{D}}^{\underline{k}} A_{\Hp}\right] \left(v(\tau)\right)  +\sum_{|\underline{k}|\leq K+1} \check{\mathbb{I}}^\star \left[\tilde{\mathfrak{D}}^{\underline{k}} A_{\I}\right] \left(\tau\right) 
    \\
                & \qquad \lesssim
                 \sum_{k=0}^{K+1}\underline{\check{\mathbb{F}}}_{v(\tau)} [ (\slashed{\nabla}_{R^\star})^k A_{\Hp}]
+
\sum_{k=0}^{K+1}\check{\mathbb{F}}_{\tau} [ (\slashed{\nabla}_{R^\star})^k A_{\I}]
+ \sum_{k=0}^{K}\underline{\check{\mathbb{F}}}_{v(\tau)} [ (\slashed{\nabla}_{R^\star})^k\Pi_{\Hp}]
+
\sum_{k=0}^{K}\check{\mathbb{F}}_{\tau} [ (\slashed{\nabla}_{R^\star})^k\Pi_{\I}]\nonumber \\
& \qquad 
+ \sum_{k=0}^K \int_{\check{C}^{\I}_\tau} r^2 | \left(\Omega \slashed{\nabla}_4\right)^k A_{\I}|^2 +  | \left(r\Omega \slashed{\nabla}_4\right)^k \Pi_{\I}|^2 + \frac{\varepsilon_0^2 + \varepsilon^3}{\tau^{\min(2,N-2-K)}} \, .
 \nonumber
 \end{align}
\end{proposition}

\begin{proof}
We only treat the $\DcH$ here since the region $\DcI \left(\tau,u_f\right) \cap \{ r\leq 2R\}$ is treated entirely analogously (but is much easier as there is no potential degeneration near $r=3M$ and weights in $r$ and $\Omega$ are irrelevant). We first prove for $1\leq K \leq N-2$
\begin{align} \label{ghj}
\sum_{|\underline{k}|\leq K} \check{\mathbb{I}}^{deg} \left[\tilde{\mathfrak{D}}^{\underline{k}}\Pi_{\Hp}\right] \left(v(\tau)\right) \nonumber \\
\lesssim
\sum_{k \leq K} \check{\mathbb{I}} \left[ (\slashed{\nabla}_{R^\star})^k \Pi_{\Hp}\right] \left(v(\tau)\right) 
+\sum_{k \leq K-1} \check{\mathbb{I}} \left[ (\slashed{\nabla}_{R^\star})^k A_{\Hp}\right] \left(v(\tau)\right) 
+ \frac{\varepsilon_0^2 + \varepsilon^3}{\tau^{\min(2,N-2-K)}} \, , 
\end{align}
whose right hand side we can estimate by the right hand side of (\ref{golik}) by applying Propositions \ref{prop:pirstar} and \ref{prop:arstar}.
Fixing $K\geq 1$ we look at
\begin{align} \label{hgy}
\slashed{\nabla}_{R^\star} \tilde{\mathfrak{D}}^{(k_1,k_2,k_3)} \Pi_{\Hp}  \ \ \ \textrm{for $k_1+k_2+k_3=K$} \, ,
\end{align}
which we have to control \emph{non-degenerately} near $r=3M$. We first commute the $\left(\frac{1}{\Omega} \slashed{\nabla}_3\right)^{k_1}$ through on the $\Pi_{\Hp}$ and use (\ref{recdpi}) in conjunction with the estimates of Theorem~\ref{thm:PPbarestimates} for the terms involving $\Psi_{\Hp}$ (as well as Proposition \ref{prop:wavehorizonerror3} for the non-linear errors arising from commutation) to obtain
\[
\int_{\DcH(v(\tau))} \Omega^2 |\slashed{\nabla}_{R^\star} \tilde{\mathfrak{D}}^{(k_1,k_2,k_3)} \Pi_{\Hp}|^2 \lesssim  \int_{\DcH(v(\tau))}\Omega^2  | \slashed{\nabla}_{R^\star} \tilde{\mathfrak{D}}^{(k_1,0,k_3)} \Pi_{\Hp} |^2+ \frac{\varepsilon_0^2 + \varepsilon^3}{\tau^{\min(2,N-2-K)}}  .
\]
Similarly, using the relation $\Omega \slashed{\nabla}_4 = \Omega \slashed{\nabla}_3 + 2R^\star$ and commuting the $4$-derivative through we find 
\[
\int_{\DcH(v(\tau))}\Omega^2 |\slashed{\nabla}_{R^\star} \tilde{\mathfrak{D}}^{(k_1,k_2,k_3)} \Pi_{\Hp}|^2 \lesssim  \sum_{i=1}^{k_3+1} \int_{\DcH(v(\tau))} \Omega^2  |  (\slashed{\nabla}_{R^\star})^{i} \tilde{\mathfrak{D}}^{(k_1,0,0)} \Pi_{\Hp} |^2 + \frac{\varepsilon_0^2 + \varepsilon^3}{\tau^{\min(2,N-2-K)}}  .
\]
Now if $k_1=1$ the desired estimate follows directly from Proposition \ref{prop:pirstar}. If $k_1 \geq 2$, we commute all $r^2\slashed{\Delta}$ operators through on $\Pi_{\Hp}$ using repeatedly the relations
(\ref{teueli1}) and (\ref{wep1}) in the form
\begin{align} \label{auxii}
r^2 \slashed{\Delta} \Pi_{\Hp}  = \Omega \slashed{\nabla}_4 \Psi_{\Hp}  +4\Pi_{\Hp} - \frac{6M}{r}\Pi_{\Hp}  - 3MA_{\Hp}  + \mathcal{E}^2 \, , 
\end{align}
\[
 r^2 \slashed{\Delta} A_{\Hp}    = + 2\Pi_{\Hp} + \frac{6M}{r}  A_{\Hp} + -2\Omega 
 \left(\frac{\Omega^2}{r^2}\Psi_{\Hp} - 2\slashed{\nabla}_{R^\star} \Pi_{\Hp}\right) - \frac{4}{r}\left(1- \frac{3M}{r}\right) \Pi_{\Hp}  + \mathcal{E}^1 \, , 
\]
to conclude (using again Proposition \ref{prop:wavehorizonerror3} for the non-linear errors)
\begin{align}
\int_{\DcH(v(\tau))} \Omega^2 |\slashed{\nabla}_{R^\star} \tilde{\mathfrak{D}}^{(k_1,k_2,k_3)} \Pi_{\Hp}|^2 \lesssim  \textrm{RHS of (\ref{ghj})} \, .
\end{align}
The same argument works for general derivatives (which only need to be controlled \emph{degenerately} near $r=3M$) in particular one easily establishes for $k_1+k_2+k_3=K+1$
\begin{align}
\int_{\DcH(v(\tau))} \Omega^2 \left(1-\frac{3M}{r}\right)^2 | \tilde{\mathfrak{D}}^{(k_1,k_2,k_3)} \Pi_{\Hp}|^2 \lesssim  \textrm{RHS of (\ref{ghj})} \, .
\end{align}
The point is that without the $\slashed{\nabla}_{R^\star}$ derivative only the degenerate estimates on $\Psi_{\Hp}$ can be used when inserting relations like (\ref{recdpi}) and (\ref{auxii}). This proves (\ref{ghj}). Next we show
\begin{align} 
\sum_{|\underline{k}|\leq K+1} \check{\mathbb{I}}^{deg} \left[\tilde{\mathfrak{D}}^{\underline{k}}A_{\Hp}\right] \left(v(\tau)\right) \nonumber
\lesssim
\sum_{k \leq K} \check{\mathbb{I}} \left[ (\slashed{\nabla}_{R^\star})^k \Pi_{\Hp}\right] \left(v(\tau)\right) 
+\sum_{k \leq K+1} \check{\mathbb{I}} \left[ (\slashed{\nabla}_{R^\star})^k A_{\Hp}\right] \left(v(\tau)\right) 
+ \frac{\varepsilon_0^2 + \varepsilon^3}{\tau^{\min(2,N-2-K)}} \, . 
\end{align}
The proof of this is entirely analogous to the proof of 
(\ref{ghj}) (in fact easier now that (\ref{ghj})) has already been established) and is therefore left to the reader. 
\end{proof}
The mean value theorem immediately produces the following corollary.

\begin{corollary}\label{cor:Tintermezzoaverage}
 There exists $R_1 \leq \tilde{R} \leq R_2$ such that the right hand side of the Proposition \ref{prop:piall} controls in particular $\sum_{|{\underline{k}} | \leq K+1} \int_{\{r=\tilde{R}\} \cap \DcI(\tau)}  | \tilde{\mathfrak{D}}^{\underline{k}}  {A}_{\I}|^2$.
\end{corollary}
\subsection{Basic fluxes for all derivatives of $\Pi$ and $A$} \label{sec:susy2}
\begin{proposition} \label{prop:piall2}
For any $u_1 \leq \tau \leq u_f$ and $1\leq K\leq N-2$ the pair $\left(\Pi_{\Hp}=r^3 \Omega \psi_{\Hp},
{\Pi}_{\I}=r^3 \Omega {\psi}_{\I}\right)$ satisfies
\begin{align} \label{goli}
&\sum_{|\underline{k}|\leq K+1} \sup_{\tau \leq u \leq u_f} \check{\underline{\mathbb{F}}}_{v(u)} [\tilde{\mathfrak{D}}^{\underline{k}} A_{\Hp}] 
+\sum_{|\underline{k}|\leq K+1} \sup_{\tau \leq u \leq u_f} \check{\mathbb{F}}^\star_u [\tilde{\mathfrak{D}}^{\underline{k}} A_{\I}] +\sum_{|\underline{k}|\leq K+1}  \check{\mathbb{F}}_{u_f} [\tilde{\mathfrak{D}}^{\underline{k}} A_{\Hp}] (v(\tau))
            \\
                & \qquad \lesssim
                 \sum_{k=0}^{K+1}\underline{\check{\mathbb{F}}}_{v(\tau)} [ (\slashed{\nabla}_{R^\star})^k A_{\Hp}]
+
\sum_{k=0}^{K+1}\check{\mathbb{F}}_{\tau} [ (\slashed{\nabla}_{R^\star})^k A_{\I}]
+ \sum_{k=0}^{K}\underline{\check{\mathbb{F}}}_{v(\tau)} [ (\slashed{\nabla}_{R^\star})^k\Pi_{\Hp}]
+
\sum_{k=0}^{K}\check{\mathbb{F}}_{\tau} [ (\slashed{\nabla}_{R^\star})^k\Pi_{\I}]\nonumber \\
& \qquad \ + \sum_{k=0}^K \int_{\check{C}^{\I}_\tau} r^2 | \left(\Omega \slashed{\nabla}_4\right)^k A_{\I}|^2 +  | \left(r\Omega \slashed{\nabla}_4\right)^k \Pi_{\I}|^2 + \frac{\varepsilon_0^2 + \varepsilon^3}{\tau^{\min(2,N-2-K)}}\, .
 \nonumber
 \end{align}
\end{proposition}

\begin{proof}
We prove the estimate for the 
{\bf ingoing boundary term in horizon region}. The proof for the other terms is similar (but easier) and outlined below. We claim that for any $1 \leq K \leq N-2$ we have
\begin{align} \label{detr}
\sum_{|\underline{k}|\leq K} \sup_{\tau \leq u \leq u_f} \check{\underline{\mathbb{F}}}_{v(u)} [\tilde{\mathfrak{D}}^{\underline{k}}\Pi_{\Hp}] +\sum_{|\underline{k}|\leq K+1} \sup_{\tau \leq u \leq u_f} \check{\underline{\mathbb{F}}}_{v(u)} [\tilde{\mathfrak{D}}^{\underline{k}} A_{\Hp}] 
\nonumber \\
\lesssim 
                \sum_{k=0}^K  \sup_{\tau \leq u \leq u_f} \check{\underline{\mathbb{F}}}_{v(u)}  [ (\slashed{\nabla}_{R^\star})^k\Pi_{\Hp}]  + \sum_{k=0}^{K+1}  \sup_{\tau \leq u \leq u_f} \check{\underline{\mathbb{F}}}_{v(u)}  [ (\slashed{\nabla}_{R^\star})^k A_{\Hp}] + \frac{\varepsilon_0^2 + \varepsilon^3}{\tau^{\min(2,N-2-K)}}\, ,
\end{align}
whose right hand side we can estimate by the right hand side of (\ref{goli}) by applying Propositions \ref{prop:pirstar} and \ref{prop:arstar}.\footnote{Note that the first term in (\ref{detr}) is actually redundant since it is manifestly controlled by the second term on the left up to decaying non-linear errors. We have kept it to make the proof more transparent.} To establish this estimate, we will prove for $1\leq K \leq N-2$ and any \emph{fixed} $l=0,1,...,K-1$ the estimate
\begin{align}  \label{wni}
\sum_{|\underline{k}|=K-l} \sup_{\tau \leq u \leq u_f} \check{\underline{\mathbb{F}}}_{v(u)} [\tilde{\mathfrak{D}}^{\underline{k}}  (\slashed{\nabla}_{R^\star})^l \Pi_{\Hp}] \nonumber \\
\lesssim 
                \sum_{|\underline{k}|= K-l-1}  \sup_{\tau \leq u \leq u_f} \check{\underline{\mathbb{F}}}_{v(u)}  [\tilde{\mathfrak{D}}^{\underline{k}} (\slashed{\nabla}_{R^\star})^{l+1} \Pi_{\Hp}] +\sum_{\substack{|\underline{k}|\leq K-l-1 \\ i 
               \leq l }}  \sup_{\tau \leq u \leq u_f} \check{\underline{\mathbb{F}}}_{v(u)} [\tilde{\mathfrak{D}}^{\underline{k}} (\slashed{\nabla}_{R^\star})^i \Pi_{\Hp}] 
          \nonumber \\
  +\sum_{|\underline{k}|\leq K-l-1} \sup_{\tau \leq u \leq u_f} \check{\underline{\mathbb{F}}}_{v(u)} [\tilde{\mathfrak{D}}^{\underline{k}}  (\slashed{\nabla}_{R^\star})^l A_{\Hp}]                
         + \frac{\varepsilon_0^2 + \varepsilon^3}{\tau^{\min(2,N-2-K)}}\, ,
\end{align}
and for $1\leq K \leq N-1$ and any \emph{fixed} $l=0,1,...,K-1$ the estimate
\begin{align}  \label{wni2}
\sum_{|\underline{k}|=K-l} \sup_{\tau \leq u \leq u_f} \check{\underline{\mathbb{F}}}_{v(u)} [\tilde{\mathfrak{D}}^{\underline{k}}  (\slashed{\nabla}_{R^\star})^l A_{\Hp}] \nonumber \\
\lesssim 
                \sum_{|\underline{k}|= K-l-1}  \sup_{\tau \leq u \leq u_f} \check{\underline{\mathbb{F}}}_{v(u)}  [\tilde{\mathfrak{D}}^{\underline{k}} (\slashed{\nabla}_{R^\star})^{l+1} A_{\Hp}] + \sum_{\substack{|\underline{k}|\leq K-l-1 \\ i \leq l }} \sup_{\tau \leq u \leq u_f} \check{\underline{\mathbb{F}}}_{v(u)} [\tilde{\mathfrak{D}}^{\underline{k}} (\slashed{\nabla}_{R^\star})^l A_{\Hp}] \nonumber \\
       + \sum_{|\underline{k}|\leq K-l-1} \sup_{\tau \leq u \leq u_f} \check{\underline{\mathbb{F}}}_{v(u)} [\tilde{\mathfrak{D}}^{\underline{k}}  (\slashed{\nabla}_{R^\star})^l \Pi_{\Hp}]   + \sum_{|\underline{k}| \leq K-l-2} \sup_{\tau \leq u \leq u_f} \check{\underline{\mathbb{F}}}_{v(u)} [\tilde{\mathfrak{D}}^{\underline{k}}  (\slashed{\nabla}_{R^\star})^{l+1} \Pi_{\Hp}]         + \frac{\varepsilon_0^2 + \varepsilon^3}{\tau^{\min(2,N-1-K)}} \, .
\end{align}
Combining (\ref{wni}) and (\ref{wni2}) the estimate (\ref{detr}) follows by a simple induction, which is left to the reader. 

\vskip1pc
\noindent
{\bf Step 1.} We first prove (\ref{wni}) for fixed $1\leq K \leq N-2$ and $l \in \{0,...,K-1\}$.  Since $K-l\geq 1$, at least one of the inequalities $k_1>0$, $k_2>0$, $k_3>0$ must hold for $\underline{k}=(k_1,k_2,k_3)$ in each summand on the left. We prove the inequality for each summand.

If $k_2>0$, we can commute one $\Omega^{-1}\slashed{\nabla}_3$ through\footnote{Observe that up to what will become non-linear terms we have $\Omega^{-1} \slashed{\nabla}_3  (\slashed{\nabla}_{R^\star})^l =  (\slashed{\nabla}_{R^\star})^l \Omega^{-1} \slashed{\nabla}_3 + \sum_{i=1}^k f^i_{2}  (\slashed{\nabla}_{R^\star})^i \Omega^{-1} \slashed{\nabla}_3$.} the expression $\tilde{\mathfrak{D}}^{\underline{k}}  (\slashed{\nabla}_{R^\star})^l \Pi_{\Hp}$ (cf.~Lemma \ref{lem:commutation}) and insert the relation (\ref{recdpi}), $\Omega^{-1} \slashed{\nabla}_3 \Pi_{\Hp} = r^{-2} \Psi_{\Hp}$, to obtain (\ref{wni}) with
the decay term coming from the non-linear commutation errors and applying Theorem~\ref{thm:PPbarestimates} to the terms involving $\Psi_{\Hp}$.

If $k_3>0$ we can assume $k_2=0$ and we write (for $k_1$ even; $k_1$ odd being completely analogous) 
\[
\tilde{\mathfrak{D}}^{\underline{k}}  (\slashed{\nabla}_{R^\star})^l \Pi_{\Hp}= (r\Omega \slashed{\nabla}_4)^{k_3-1} \left(r \Omega\slashed{\nabla}_3 + 2 r \slashed{\nabla}_{R^\star}\right) \left(r^2 \slashed{\Delta})^{k_1/2}\right)  (\slashed{\nabla}_{R^\star})^l \Pi_{\Hp})] \, .
\]
Commuting through the $\Omega^{-1} \slashed{\nabla}_3$ as before as well as commuting the $\slashed{\nabla}_{R^\star}$ through yields (\ref{wni}).

Finally, if $k_1>0$, we can assume in view of the previous that $k_2=0$ and $k_3=0$. We distinguish between $k_1=1$ and $k_1\geq 2$. If $k_1\geq 2$ even, we see from (\ref{wep1}) that 
\begin{align}
(r^2 \slashed{\Delta})^{k_1/2}  (\slashed{\nabla}_{R^\star})^l \Pi_{\Hp} = (r^2 \slashed{\Delta})^{k_1/2-1} (\slashed{\nabla}_{R^\star})^l  \left(\Omega \slashed{\nabla}_4 \Psi_{\Hp} + 4\Pi_{\Hp} - \frac{6M}{r}\Pi_{\Hp} -3M A_{\Hp} \right) +\mathcal{E}^{k_1+l} \nonumber
\end{align}
and similarly for $k_1 \geq 2$ odd (there is an additional term coming from commuting $r^2 \slashed{\Delta}$ and $r \slashed{div}$)
{\small
\begin{align}
(r^2 \slashed{\Delta})^{k_1/2-1/2} r\slashed{div}  (\slashed{\nabla}_{R^\star})^l \Pi_{\Hp} = (r^2 \slashed{\Delta})^{k_1/2-3/2} r \slashed{div}  (\slashed{\nabla}_{R^\star})^l  \left(\Omega \slashed{\nabla}_4 \Psi_{\Hp} + (4-3)\Pi_{\Hp} - \frac{6M}{r}\Pi_{\Hp} -3M A_{\Hp} \right) +\mathcal{E}^{k_1+l} \,. \nonumber
\end{align}
}
In both cases we easily check that (\ref{wni}) holds (using also the error estimates of Proposition \ref{prop:wavehorizonerror4} on cones). In case that $k_1=1$ we need observe that
\begin{align} \label{qws1}
\int_{\CbcH_{v(u)}} \Omega^2 |r\slashed{\nabla} r\slashed{div}  (\slashed{\nabla}_{R^\star})^l \Pi_{
\Hp}|^2 + \int_{\CbcH_{v(u)}} \Omega^2 |\Omega^{-1} \slashed{\nabla}_3 r\slashed{div}  (\slashed{\nabla}_{R^\star})^l \Pi_{
\Hp}|^2 \nonumber \\
 \leq \int_{\CbcH_{v(u)}} \Omega^2 |r^2 \slashed{\mathcal{D}}_2^\star \slashed{div}  (\slashed{\nabla}_{R^\star})^l \Pi_{
\Hp}|^2 + \int_{\CbcH_{v(u)}} \Omega^2 |\Omega^{-1}\slashed{\nabla}_3 \slashed{div}  (\slashed{\nabla}_{R^\star})^l \Pi_{
\Hp}|^2 
\\
\lesssim \int_{\CbcH_{v(u)}} \Omega^2 |r^2 \slashed{\mathcal{D}}_2^\star \slashed{div}  (\slashed{\nabla}_{R^\star})^l \frac{r^2}{\Omega} \slashed{\nabla}_3 A_{\Hp} |^2  + \check{\underline{\mathbb{F}}}_{v(u)} [  (\slashed{\nabla}_{R^\star})^l \Pi_{\Hp}] +  \frac{\varepsilon_0^2+\varepsilon^3}{\tau^{\min(2,N-l-2)}}.  \nonumber
\end{align}
Here we have used an elliptic estimate along truncated cones in the first step and commuted the $\slashed{\nabla}_3$ derivative through in the second term as well as inserted the definition (\ref{recda}) for the first term. We now commute the $r^2 \slashed{\mathcal{D}}_2^\star \slashed{div}$ derivative through on $A_{\Hp}$ in the first term and insert the Teukolsky equation (\ref{teueli1}) where we replace $\Omega \slashed{\nabla}_4 \Pi_{\Hp} = 2\slashed{\nabla}_{R^\star} \Pi_{\Hp} + \frac{\Omega^2}{r^2} \Psi_{\Hp}$ on its right hand side.

\vskip1pc
\noindent
{\bf Step 2.} We now prove (\ref{wni2}) for fixed $1\leq K \leq N-1$ and $l \in \{0,...,K-1\}$.  Since $K-l\geq 1$, at least one of the inequalities $k_1>0$, $k_2>0$, $k_3>0$ must hold for each summand.

If $k_2>0$, we can commute one $\Omega^{-1}\slashed{\nabla}_3$ through the expression $\tilde{\mathfrak{D}}^{\underline{k}}  (\slashed{\nabla}_{R^\star})^l  A_{\Hp}$ and insert the relation $\Omega^{-1} \slashed{\nabla}_3 A_{\Hp} = -2r^{-2} \Pi_{\Hp}$ to obtain (\ref{wni2}) with the decay term coming from the non-linear commutation errors. (Actually, only the second, third and fifth term on the right hand side of  (\ref{wni2}) are needed.)

If $k_3>0$ we can assume $k_2=0$ and write (for $k_1$ even; $k_1$ odd being completely analogous) 
\[
\tilde{\mathfrak{D}}^{\underline{k}}  (\slashed{\nabla}_{R^\star})^l A_{\Hp}= (r\Omega \slashed{\nabla}_4)^{k_3-1} \left(r \Omega\slashed{\nabla}_3 + 2r \slashed{\nabla}_{R^\star}\right) \left(r^2 \slashed{\Delta})^{k_1/2}\right)  (\slashed{\nabla}_{R^\star})^l A_{\Hp}) \, .
\]
Commuting through the $\Omega^{-1} \slashed{\nabla}_3$ as before as well as commuting the $(R^\star)$ through yields (\ref{wni2}). 

Finally, if $k_1>0$, we can assume in view of the previous that $k_2=0$ and $k_3=0$. We distinguish between $k_1=1$ and $k_1\geq 2$. If $k_1\geq 2$ even, we see after inserting (\ref{teueli1}) that 
\begin{align}
(r^2 \slashed{\Delta})^{\frac{k_1}{2}}  (\slashed{\nabla}_{R^\star})^l A_{\Hp} = (r^2 \slashed{\Delta})^{\frac{k_1}{2}-1} (\slashed{\nabla}_{R^\star})^l  \left(  2\slashed{\nabla}_{R^\star} \Pi_{\Hp} - \frac{3M}{r} A_{\Hp} +\frac{\Omega^2}{r^2}\Psi_{\Hp} + \frac{2}{r}\left(1- \frac{3M}{r}\right) \Pi_{\Hp} \right) +\mathcal{E}^{k_1-1+l}_{4} \nonumber
\end{align}
and similarly for $k_1 \geq 2$ odd. In both cases we easily see that (\ref{wni2}) indeed holds. In case that $k_1=1$ we observe that
\begin{align} \label{qws2}
\int_{\CbcH_{v(u)}} \Omega^2 |r\slashed{\nabla} r\slashed{div}  (\slashed{\nabla}_{R^\star})^l A_{
\Hp}|^2 + \int_{\CbcH_{v(u)}} \Omega^2 |\Omega^{-1} \slashed{\nabla}_3 r\slashed{div}  (\slashed{\nabla}_{R^\star})^l A_{
\Hp}|^2 \nonumber \\
 \leq \int_{\CbcH_{v(u)}} \Omega^2 |r^2 \slashed{\mathcal{D}}_2^\star \slashed{div}  (\slashed{\nabla}_{R^\star})^l A_{
\Hp}|^2 + \int_{\CbcH_{v(u)}} \Omega^2 |\Omega^{-1} \slashed{\nabla}_3 \slashed{div}  (\slashed{\nabla}_{R^\star})^l A_{
\Hp}|^2 
\\
\lesssim \check{\underline{\mathbb{F}}}_{v(u)} [  (\slashed{\nabla}_{R^\star})^{l+1} A_{\Hp}]+\check{\underline{\mathbb{F}}}_{v(u)} [  (\slashed{\nabla}_{R^\star})^l A_{\Hp}]+ \check{\underline{\mathbb{F}}}_{v(u)} [  (\slashed{\nabla}_{R^\star})^l \Pi_{\Hp}] +  \frac{\varepsilon^2_0 + \varepsilon^3}{\tau^{\min(2,N-l-2)}}  \, . \nonumber
\end{align}
Here we have used an elliptic estimate along truncated cones in the first step. We then commuted the $\slashed{\nabla}_3$ derivative through for the second term while for the first term we inserted (\ref{teueli1}) in the form
\[
r^2 \slashed{\mathcal{D}}_2^\star \slashed{div} A_{\Hp}    =-  \frac{3M}{r} A_{\Hp} - \frac{1}{2}\frac{r^2}{\Omega^2}  \Omega \slashed{\nabla}_3 \left(-2\frac{\Omega^2}{r^2} \Pi_{\Hp} + \slashed{\nabla}_{R^\star} A_{\Hp} \right)   + \mathcal{E}^1_4 \, .
\]
This finishes the proof for the ingoing flux in the horizon region.  We briefly outline the proof for {\bf the outgoing flux in the infinity region}. Completely analogously to Step~1 (using now the estimates of Proposition~\ref{prop:waveIerror1c} for the non-linear errors) we prove that for any $K \leq N-2$ we have
\begin{align} \label{liv}
\sum_{|\underline{k}|\leq K} \sup_{\tau \leq u \leq u_f} \check{\mathbb{F}}^\star_u [\tilde{\mathfrak{D}}^{\underline{k}}\Pi_{\I}]+\sum_{|\underline{k}|\leq K+1} \sup_{\tau \leq u \leq u_f} \check{\mathbb{F}}^\star_u [\tilde{\mathfrak{D}}^{\underline{k}} A_{\I}] \nonumber \\
\lesssim \sum_{k=0}^K \sup_{\tau \leq u \leq u_f} \check{\mathbb{F}}^\star_u [ (\slashed{\nabla}_{R^\star})^k\Pi_{\I}]  +  \sum_{k=0}^K \sup_{\tau \leq u \leq u_f} \check{\mathbb{F}}^\star_u [ (\slashed{\nabla}_{R^\star})^k A_{\I}]  + \frac{\varepsilon_0^2 + \varepsilon^3}{\tau^{\min(2,N-2-K)}} \, .
\end{align}
Note carefully that all energies are restricted to $r\leq 2R$ so $r$-weights can be absorbed into constants. To the right hand side of (\ref{liv}) we apply Propositions~\ref{prop:pirstar} and~\ref{prop:arstar}. The proof of (\ref{liv}) is in fact a little easier as  now we can directly insert (\ref{teueli2}) and (\ref{wep2}) (when reaching the analogues of (\ref{qws1}), (\ref{qws2})) since these relations produce $\slashed{\nabla}_4$-derivatives which are the ones naturally appearing in the fluxes. Similarly the ingoing flux in the horizon region follows from proving
\begin{align}
\sum_{|\underline{k}|\leq K}  \check{\mathbb{F}}_{u_f} [\tilde{\mathfrak{D}}^{\underline{k}} \Pi_{\Hp}] (v(\tau)) +\sum_{|\underline{k}|\leq K+1}  \check{\mathbb{F}}_{u_f} [\tilde{\mathfrak{D}}^{\underline{k}} A_{\Hp}] (v(\tau)) \nonumber  \\
\lesssim \sum_{k=0}^K  \check{\mathbb{F}}_{u_f} [ (\slashed{\nabla}_{R^\star})^k \Pi_{\Hp}] (v(\tau)) +\sum_{k=0}^{K+1}  \check{\mathbb{F}}_{u_f} [ (\slashed{\nabla}_{R^\star})^k A_{\Hp}] (v(\tau)) + \frac{\varepsilon_0^2 + \varepsilon^3}{\tau^{\min(2,N-2-K)}}
\end{align}
and applying Propositions \ref{prop:pirstar} and \ref{prop:arstar} to the right hand side.
\end{proof}

\subsection{The weighted $r^p$ estimates for $\Pi_{\I}$ and $A_{\I}$} \label{sec:piarpweighted}
In this subsection only we use the following notation (for which we recall (\ref{nido})):\index{double null gauge!differential operators!$\tilde{\mathfrak{D}}_{\nearrow}^{\underline{k}}$, commutators used for $\Pi_{\I}$ and $A_{\I}$}
\begin{align} \label{outgoingderivatives}
\tilde{\mathfrak{D}}_{\nearrow}^{\underline{k}} =\tilde{\mathfrak{D}}_{\I}^{(0,k_2,k_3)} =   \left\{
\begin{array}{rl}
\left(r\Omega \slashed{\nabla}_4 \right)^{k_3}  \left(r^2 \slashed{\Delta}\right)^{k_2/2}   &  \text{if } k_2  \ \ \text{even} \\
\left(r\Omega \slashed{\nabla}_4 \right)^{k_3}  \left(r^2 \slashed{\Delta}\right)^{(k_2-1)/2} r \slashed{div}& \text{if } k_2  \ \ \text{odd}.
\end{array} \right.
\end{align}

\begin{proposition} \label{prop:APibp}
We have, for $3 \leq K \leq N$ and any $u_1 \leq \tau_1 \leq \tau_2 \leq u_f$, the following estimates:
\begin{align} \label{trge}
&\sum_{|\underline{k}|=1}^K \int_{\CcI_{\tau_2}}  \ r^6 | \tilde{\mathfrak{D}}_{\nearrow}^{\underline{k}} A_{\I}|^2 + \sum_{|\underline{k}|=0}^{K-1} \int_{\CbcI_{v_\infty}(\tau_1)}  \ r^4 | \tilde{\mathfrak{D}}_{\nearrow}^{\underline{k}} \Pi_{\I}|^2  \nonumber \\
+ &\sum_{|\underline{k}|=0}^{K} \int_{\DcI\left(\tau_1\right)} \, r^5 | \tilde{\mathfrak{D}}_{\nearrow}^{\underline{k}} A_{\I}|^2  +  \sum_{|\underline{k}|=0}^{K-1}\int_{\DcI\left(\tau_1\right)} r^{3-\delta} | \tilde{\mathfrak{D}}_{\nearrow}^{\underline{k}} \Pi_{\I}|^2 \nonumber \\
\lesssim
& \ \ \sum_{|\underline{k}|=1}^K \int_{\CcI_{\tau_1}}  \ r^6 | \tilde{\mathfrak{D}}_{\nearrow}^{\underline{k}} A_{\I}|^2 + \sum_{|\underline{k}|=0}^{K-1}\int_{\CcI_{\tau_1}} r^2  | \tilde{\mathfrak{D}}_{\nearrow}^{\underline{k}} \Pi_{\I}|^2 \nonumber \\
+& \ \  \sum_{k=0}^{K-1}\underline{\check{\mathbb{F}}}_{v(\tau_1)} [ (\slashed{\nabla}_{R^\star})^k A_{\Hp}] +\sum_{k=0}^{K-2}\underline{\check{\mathbb{F}}}_{v(\tau_1)} [ (\slashed{\nabla}_{R^\star})^k\Pi_{\Hp}]  +\varepsilon^2_0 + \varepsilon^3
\end{align}
and
\begin{align} \label{trgf}
&\sum_{|\underline{k}|=1}^K \int_{\CcI_{\tau_2}}  \ r^5 | \tilde{\mathfrak{D}}_{\nearrow}^{\underline{k}} A_{\I}|^2 + \sum_{|\underline{k}|=0}^{K-1} \int_{\CbcI_{v_\infty}(\tau_1)}  \ r^3 | \tilde{\mathfrak{D}}_{\nearrow}^{\underline{k}} \Pi_{\I}|^2  \nonumber \\
+ &\sum_{|\underline{k}|=1}^K \int_{\DcI\left(\tau_1\right)} \, r^4 | \tilde{\mathfrak{D}}_{\nearrow}^{\underline{k}} A_{\I}|^2  +  \sum_{|\underline{k}|=0}^{K-1} \int_{\DcI\left(\tau_1\right)} r^{2} | \tilde{\mathfrak{D}}_{\nearrow}^{\underline{k}} \Pi_{\I}|^2 \nonumber \\
\lesssim
& \ \ \sum_{|\underline{k}|=1}^K \int_{\CcI_{\tau_1}}  \ r^5 | \tilde{\mathfrak{D}}_{\nearrow}^{\underline{k}} A_{\I}|^2 + \sum_{|\underline{k}|=0}^{K-1}\int_{\CcI_{\tau_1}}  r^2 |\tilde{\mathfrak{D}}_{\nearrow}^{\underline{k}} \Pi_{\I}|^2  \nonumber \\
+ & \ \ \sum_{k=0}^{K-1}\underline{\check{\mathbb{F}}}_{v(\tau_1)} [ (\slashed{\nabla}_{R^\star})^k A_{\Hp}] +\sum_{k=0}^{K-2}\underline{\check{\mathbb{F}}}_{v(\tau_1)} [ (\slashed{\nabla}_{R^\star})^k\Pi_{\Hp}]+\frac{\varepsilon^2_0 + \varepsilon^3}{(\tau_1)^{1-\delta^K_N}},
\end{align}
\begin{align} \label{trg}
& \sum_{|\underline{k}|=1}^K \int_{\CcI_{\tau_2}}  \ r^4 | \tilde{\mathfrak{D}}_{\nearrow}^{\underline{k}} A_{\I}|^2 + \sum_{|\underline{k}|=0}^{K-1} \int_{\CbcI_{v_\infty}(\tau_1)}  \ r^2 |\tilde{\mathfrak{D}}_{\nearrow}^{\underline{k}} \Pi_{\I}|^2  \nonumber \\
+& \sum_{|\underline{k}|=1}^K \int_{\DcI\left(\tau_1\right)} \, r^3 | \tilde{\mathfrak{D}}_{\nearrow}^{\underline{k}} A_{\I}|^2  + \sum_{|\underline{k}|=0}^{K-1} \int_{\DcI\left(\tau_1\right)} r^{1} | \tilde{\mathfrak{D}}_{\nearrow}^{\underline{k}} \Pi_{\I}|^2 \nonumber \\
\lesssim
&\ \  \sum_{|\underline{k}|=1}^K \int_{\CcI_{\tau_1}}  \ r^4 | \tilde{\mathfrak{D}}_{\nearrow}^{\underline{k}} A_{\I}|^2 + \sum_{|\underline{k}|=0}^{K-1}\int_{\CcI_{\tau_1}}  r^2 | \tilde{\mathfrak{D}}_{\nearrow}^{\underline{k}} \Pi_{\I}|^2 \nonumber \\
+ & \ \ \sum_{k=0}^{K-1}\underline{\check{\mathbb{F}}}_{v(\tau_1)} [ (\slashed{\nabla}_{R^\star})^k A_{\Hp}] +\sum_{k=0}^{K-2}\underline{\check{\mathbb{F}}}_{v(\tau_1)} [ (\slashed{\nabla}_{R^\star})^k\Pi_{\Hp}] +\frac{\varepsilon^2_0 + \varepsilon^3}{(\tau_1)^{\min(2,N-K)}}.
\end{align}
\end{proposition}

\begin{proof}
Using the relation $R^\star=\frac{1}{2}(\Omega \slashed{\nabla}_3 + \Omega \slashed{\nabla}_4)$ and inserting (\ref{recda}), (\ref{recdpi}) one first checks that for fixed $3\leq \tilde{K} \leq N$, the right hand side of (\ref{trge})--(\ref{trg})  controls the right hand side of the estimates of Proposition \ref{prop:piall} and \ref{prop:piall2} applied with $K=\tilde{K}-2$. Hence we are free to use the estimates of Propositions \ref{prop:piall} and \ref{prop:piall2} in proving the above.  In particular, it suffices to prove the estimates with all integrals on the left hand sides restricted to $r \geq R_2$ since in the remaining region, the desired estimates already hold by Propositions \ref{prop:piall} and \ref{prop:piall2}.

Key to the proof are the Bianchi pairs
\begin{equation} \label{tff}
\begin{split}
\Omega \slashed{\nabla}_3 (r \slashed{div} A_{\I}) &= - \frac{2 \Omega^2}{r^2} r \slashed{div} \Pi_{\I} + \mathcal{E}^1_4 \, , \\
\Omega \slashed{\nabla}_4 \Pi_{\I} + \left(\frac{2}{r} - \frac{2M}{r^2}\right) \Pi_{\I} &= r^2 \slashed{\mathcal{D}}_2^\star \slashed{div} A_{\I} + \frac{3M}{r} A_{\I} + \mathcal{E}^1_4 \, ,
\end{split}
\end{equation}
which after commutation using the definition
\begin{align} \label{commdef}
W_{\I}^{(k)} = \left\{
\begin{array}{rl}
\left(r \slashed{div} r\slashed{\mathcal{D}}_2^\star \right)^{k/2-1/2} r\slashed{div} W_{\I}    &  \text{if } $k$ \ \textrm{odd}  \\
\left(r\slashed{\mathcal{D}}_2^\star r \slashed{div} \right)^{k/2} W_{\I}  & \text{if } k  \ \textrm{even}
\end{array} \right.
\end{align}
for a symmetric traceless tensor $W_{\I}$ in the $\mathcal{I}^+$ gauge, read, for $k \geq 1$,
\begin{equation} \label{tffc1}
\begin{split}
\Omega \slashed{\nabla}_3 (A^{(k)}_{\I}) &= - \frac{2 \Omega^2}{r^2} r\slashed{div} \Pi^{(k-1)}_{\I} + \mathcal{E}^{k+1}_4\, ,  \\
\Omega \slashed{\nabla}_4 (\Pi^{(k-1)}_{\I}) + \left(\frac{2}{r} - \frac{2M}{r^2}\right) \Pi^{(k-1)}_{\I} &= r \slashed{\mathcal{D}}_2^\star A^{(k)}_{\I} + \frac{3M}{r} A^{(k-1)}_{\I} + \mathcal{E}^{k+1}_4 \, ,
\end{split}
\end{equation}
 if $k$ is odd, and
\begin{equation} \label{tffc2}
\begin{split}
\Omega \slashed{\nabla}_3 (A^{(k)}_{\I}) &= - \frac{2 \Omega^2}{r^2}  r\slashed{\mathcal{D}}_2^\star  \Pi^{(k-1)}_{\I} + \mathcal{E}^{k+1}_4 \, , \\
\Omega \slashed{\nabla}_4 (\Pi^{(k-1)}_{\I}) + \left(\frac{2}{r} - \frac{2M}{r^2}\right) \Pi^{(k-1)}_{\I} &= r \slashed{div} A^{(k)}_{\I} + \frac{3M}{r} A^{(k-1)}_{\I} + \mathcal{E}^{k+1}_4 \, ,
\end{split}
\end{equation}
if $k$ is even.

\vskip1pc
\noindent
{\bf Step 1.} We first prove the desired estimates for angular derivatives, i.e.~replacing $\tilde{\mathfrak{D}}^{\underline{k}}_\nearrow A_{\I}$ by $A^{(|\underline{k}|)}_{\I}$ and $\tilde{\mathfrak{D}}^{\underline{k}}_\nearrow \Pi_{\I}$ by $\Pi^{(|\underline{k}|)}_{\I}$ everywhere. The proof is by induction on $K \geq 1$.\footnote{We will prove the estimates also for $K=1$ and $K=2$ with the term $\sum_{k=0}^{K-1}\underline{\check{\mathbb{F}}}_{v(\tau_1)} [ (\slashed{\nabla}_{R^\star})^k A_{\Hp}] +\sum_{k=0}^{K-2}\underline{\check{\mathbb{F}}}_{v(\tau_1)} [ (\slashed{\nabla}_{R^\star})^k\Pi_{\Hp}]$ on the right replaced by $\sum_{k=0}^{2}\underline{\check{\mathbb{F}}}_{v(\tau_1)} [ (\slashed{\nabla}_{R^\star})^k A_{\Hp}] +\sum_{k=0}^{1}\underline{\check{\mathbb{F}}}_{v(\tau_1)} [ (\slashed{\nabla}_{R^\star})^k\Pi_{\Hp}]$.} Contract the first equation (of the pairs (\ref{tffc1}), (\ref{tffc2}) respectively) by $r^{4+p} A^{(k)}_{\I} \left(1-\frac{\boldsymbol\delta^p_2}{r^\delta}\right)$ and add the second contracted with $2r^{2+p} \Pi^{(k-1)}_{\I} \left(1-\frac{\boldsymbol\delta^p_2}{r^\delta}\right)$ with $p=2,1,0$.\footnote{For $p=2$ the spacetime term for $\Pi^{(k)}_{\I}$ cancels to first order which is the reason for including the $r^{-\delta}$ term in this case.} Then sum over all $k=1,..., K$ and integrate over $\DcI \left( \tau_1,\tau_2 \right) \cap \{r \geq \tilde{R}\}$, where $\tilde{R}$ is the one from Corollary 
\ref{cor:Tintermezzoaverage}. The following observations will then yield the desired (angular restricted) estimate:
\begin{itemize}
\item The terms from the left of (\ref{tffc1}) and (\ref{tffc2}) produce the desired terms on the left hand side of the estimate after an integration by parts except for the boundary term on $r=\tilde{R}$ which is controlled by Corollary \ref{cor:Tintermezzoaverage}.
\item The first term on the right hand side of each of the equations in (\ref{tff}) cancels after an integration by parts up to a cubic error, which for fixed $k$ is of the form (integration being over $\DcI \left( \tau_1,\tau_2 \right) \cap \{r \geq \tilde{R}\}$ with volume form $dudvd\theta$)
\[
\Big| \int \left(\eta + \underline{\eta}\right) \left(1+r^{-\delta}\right) r^{3+p} A^{(k)}_{\I} \cdot \Pi^{(k-1)}_{\I} \Big| \lesssim \varepsilon \int \left( r^{3+p} |A^{(k)}_{\I}|^2 + r^{1+p}|\Pi^{(k-1)}_{\I}|^2\right) \, ,
\]
which can be absorbed on the left hand side. Here we have used $\|r^2(\eta +\underline{\eta})\|_{L^\infty} \lesssim \varepsilon$ following from~\eqref{elinfestimates}.
\item The lower oder linear terms can be estimated for fixed $k$ by
\[
\int \frac{3M}{r}A^{(k-1)}_{\I} \cdot 2r^{2+p} \Pi^{(k-1)}_{\I} \left(1-\frac{\boldsymbol\delta^p_2}{r^\delta}\right) \leq \frac{3M}{R} \int  \left( r^{3+p} |A^{(k-1)}_{\I}|^2 + r^{1+p}|\Pi^{(k-1)}_{\I}|^2\right) .
\]
The $\Pi^{(k-1)}_{\I}$ term can be directly absorbed on the left hand side using (\ref{definitionofRhere}). The $A_{\I}^{(k-1)}$ can be absorbed by the term $A_{\I}^{(k)}$ on the left by a standard elliptic estimate.

\item The non-linear error-term is estimated from Proposition \ref{prop:waveIerror1b} which implies that 
\begin{align}
\int_{\DcI (\tau)} r^6  u^{1+\delta} |\mathcal{E}^N_4|^2 &\lesssim \varepsilon_0^2 + \varepsilon^3    \ \ \ \ \textrm{(used for $p=2$)} \, , \\
\int_{\DcI (\tau)} r^6 |\mathcal{E}^{N-1}_4|^2 &\lesssim \frac{\varepsilon_0^2 + \varepsilon^3}{\tau}  \ \ \ \ \textrm{(used for $p=1$)} \, , \\
\int_{\DcI (\tau)} r^5 |\mathcal{E}^{N-1}_4|^2 &\lesssim \frac{\varepsilon_0^2 + \varepsilon^3}{\tau^2}  \ \ \ \ \textrm{(used for $p=0$)} \, .
\end{align}
\end{itemize}

Note that using standard elliptic estimates on spheres we have obtained the desired estimates for $\mathfrak{D}^{\underline{k}}_\nearrow A_{\I}$ with $\underline{k}$ of the form $(k,0,0)$.

\vskip1pc
\noindent
{\bf Step 2.} We now commute the pairs (\ref{tffc1}), (\ref{tffc2}) for fixed $k$ with $\left(r \Omega \slashed{\nabla}_4\right)^l$ and $l=1,...K-k$ and show the result by an induction on $l$. The commuted equations read, for $k$ odd
\begin{equation} \label{tffc1b}
\begin{split}
\Omega \slashed{\nabla}_3 (\left(r \Omega \slashed{\nabla}_4\right)^l A^{(k)}_{\I}) + (2l-1)\frac{\Omega_\circ^2}{r} (\left(r \Omega \slashed{\nabla}_4\right)^l A^{(k)}_{\I})  &= - \frac{2 \Omega^2}{r^2} r\slashed{div}  (\left(r \Omega \slashed{\nabla}_4\right)^l \Pi^{(k-1)}_{\I}) + \mathcal{F}^{k,l}_{lin}[A_{\I}] + \mathcal{E}^{k+l+1}_4 , \\
\Omega \slashed{\nabla}_4 ((\left(r \Omega \slashed{\nabla}_4\right)^l \Pi^{(k-1)}_{\I}) + \frac{2}{r}  (\left(r \Omega \slashed{\nabla}_4\right)^l \Pi^{(k-1)}_{\I}) &= r \slashed{\mathcal{D}}_2^\star(\left(r \Omega \slashed{\nabla}_4\right)^l A^{(k)}_{\I}) + \mathcal{F}^{k,l}_{lin}[\Pi_{\I}] + \mathcal{E}^{k+1}_4 , \nonumber
\end{split}
\end{equation}
with ($h_0^m$ denoting admissible coefficient functions that can be different in different places)
\[
 \mathcal{F}^{k,l}_{lin}[A_{\I}] = \sum_{m=0}^{l-1} \frac{1}{r}h_0^m (\left(r \Omega \slashed{\nabla}_4\right)^m A^{(k)}_{\I}) + \frac{1}{r^2} h_0^{m} (\left(r \Omega \slashed{\nabla}_4\right)^m \Pi^{(k)}_{\I}) ,
\]
\[
 \mathcal{F}^{k,l}_{lin}[\Pi_{\I}] = \sum_{m=0}^{l-1} h_{0}^{m} (\left(r \Omega \slashed{\nabla}_4\right)^m A^{(k+1)}_{\I}) + \frac{1}{r} {h}_{0}^{m} (\left(r \Omega \slashed{\nabla}_4\right)^m \Pi^{(k-1)}_{\I}) + \sum_{m=0}^l \frac{1}{r} h_0^m \left(r \Omega \slashed{\nabla}_4\right)^m A^{(k-1)}_{\I} \, .
\]
For $k$ even, we interchange $r \slashed{div}$ in the first with $r\slashed{\mathcal{D}}_2^\star$ in the second equation. From the structure of the commuted equations it is clear that we can simply repeat the proof of Step 1 (with the same weights). The linear terms can be dealt with using Cauchy--Schwarz and the estimates from Step~1 (base case) or the previous step of the induction.

Note that with this we have obtained the desired estimates for $\tilde{\mathfrak{D}}^{\underline{k}}_\nearrow A_{\I}$ with $\underline{k}$ of the form $(k,0,l)$ with $k\geq 1$ and all of the desired $\tilde{\mathfrak{D}}^{\underline{k}}_\nearrow \Pi_{\I}$. Therefore, it only remains to prove the estimates for $\tilde{\mathfrak{D}}^{\underline{k}}_\nearrow  A_{\I}$ being $\left(r\Omega \slashed{\nabla}_4\right)^{|\underline{k}|} A_{\I}$. This we turn to in Step 3.

\vskip1pc
\noindent
{\bf Step 3.} To estimate $\left(r\Omega \slashed{\nabla}_4\right)^{|\underline{k}|} A_{\I}$, we apply the $r^p$--hierarchy for the Teukolsky equation:
\begin{lemma} \label{barp}
For any $u_1 \leq \tau_1 \leq \tau_2 \leq u_f$ and any $1\leq K \leq N$ we have for $p \in \{2,1,0\}$ the estimate 
\begin{align} \label{barpest}
\sum_{k=0}^{K-1} \int_{\CcI_{\tau_2}}  \ r^{p-2} | \Omega \slashed{\nabla}_4 \left(r\Omega \slashed{\nabla}_4\right)^k (r^4 A_{\I})|^2 + \sum_{k=0}^{K-1} \int_{\CbcI_{v_{\infty}}(\tau_1)}  \ r^{p-4} | r \slashed{\nabla} \left(r\Omega \slashed{\nabla}_4\right)^k(r^4 A_{\I})|^2  
\nonumber \\
+ \sum_{k=0}^{K-1} \int_{\DcI(\tau_1)} \, r^{p-3} | \Omega \slashed{\nabla}_4 \left(r\Omega \slashed{\nabla}_4\right)^k (r^4 A_{\I})|^2  + r^{p-5}  | r \slashed{\nabla} \left(r\Omega \slashed{\nabla}_4\right)^k(r^4 A_{\I})|^2   
\nonumber \\
\lesssim
\sum_{k=0}^{K} \int_{\CcI_{\tau_1}}   \ r^{p+4} | \left(r\Omega \slashed{\nabla}_4\right)^k A_{\I}|^2
 + \sum_{k=1}^K\int_{\CcI_{\tau_1}} r^2 |\left(r\Omega \slashed{\nabla}_4\right)^{k-1} \Pi_{\I}|^2 +\frac{\varepsilon^2_0 + \varepsilon^3}{(\tau_1)^{\kappa_p}} \nonumber \\
 + \sum_{k=0}^{K-1}\underline{\check{\mathbb{F}}}_{v(\tau_1)} [ (\slashed{\nabla}_{R^\star})^k A_{\Hp}] +\sum_{k=0}^{\max(0,K-2)}\underline{\check{\mathbb{F}}}_{v(\tau_1)} [ (\slashed{\nabla}_{R^\star})^k\Pi_{\Hp}]
\end{align}
where ${\kappa}_2 = 0$, ${\kappa}_1= 1- \delta^K_N$ and ${\kappa}_0 = \min(2,N-K)$.
\end{lemma}
\begin{proof}[Proof of Lemma] 
This will again be proven inductively. We observe from Propositions \ref{prop:classify4kl} and \ref{prop:comm4kl} that $ \left(r\Omega \slashed{\nabla}_4\right)^k (r^4A_{\I})$ satisfies a tensorial wave equation of type $4_{2+k/2,l}$, the precise value of the (half-integer) $l$ being irrelevant, except for $k=0$ where we know by Proposition \ref{prop:classify4kl} that $r^4 A_{\I}$ satisfies a tensorial wave equation of type $4_{2,-1}$. We then apply successively for $K=1,2,...,N$ Proposition \ref{prop:rph} with $p_{\textrm{Proposition \ref{prop:rph}}} = p_{Lemma \ref{barp}}-2$.  

{\bf Non-linear errors.} We first estimate for all $k \leq N-1$ the non-linear error  appearing in Proposition \ref{prop:rph}, $\mathcal{H}_{5,p}\left[\left(r \Omega \slashed{\nabla}_4\right)^k (r^4A_{\I})\right] \left(\tau_1,u_f\right)$. From the schematic notation of Proposition \ref{prop:comm4kl} we deduce
\begin{align} \label{Fnlin4}
\mathcal{F}^{nlin} \left[(r \slashed{\nabla}_4)^{k} (r^4 A_{\I}) \right]  = \mathcal{E}^{k+1}_2 +   \mathcal{E}^{k+1}_1 \left(\textrm{each summand contains $\alpha$}\right) = \mathcal{E}_1^{k+1} \, .
\end{align}
For $p \in \{0,1\}$ we have applying spacetime  Cauchy--Schwarz the estimate 
\begin{align}
\mathcal{H}_{5,p}\left[\left(r \Omega \slashed{\nabla}_4\right)^k (r^4A_{\I}) \right] (\tau_1,u_f) 
\leq   & C_\gamma \int_{\DcI(\tau_1)} r^{p-1}\left( |\mathcal{F}^{nlin} \left[(r \slashed{\nabla}_4)^{k} (r^4 A_{\I}) \right]|^2 + F_{34} \left[(r \slashed{\nabla}_4)^{k} (r^4 A_{\I}) \right]|^2 \right) \nonumber \\
&+\gamma \int_{\DcI(\tau_1)} \, r^{p-3} | \Omega \slashed{\nabla}_4 \left(r\Omega \slashed{\nabla}_4\right)^k (r^4 A_{\I})|^2    \, . 
\end{align}
The second line will be absorbed on the left for sufficiently small $\gamma$ while the first line is estimated by $ \frac{\epsilon^4}{\tau^{2-p}}$ using Lemma \ref{lem:commutation} and Proposition \ref{prop:waveIerror1b}. For $p=2$ the same argument works for the term involving $F_{34}$ in (\ref{defh5}) while for the term involving $\mathcal{F}^{nlin}$ we estimate using 
Cauchy--Schwarz on the cones $C_u^{\I}$
\[
\int_{\DcI(\tau_1)} |\mathcal{F}^{nlin} \left[(r \slashed{\nabla}_4)^{k} (r^4 A_{\I}) \right] | | \Omega \slashed{\nabla}_4 \left(r\Omega \slashed{\nabla}_4\right)^k (r^4 A_{\I})|
\lesssim \varepsilon \sqrt{\int_{\DcI(\tau_1)} u^{1+\delta} |\mathcal{E}_1^{k+1}|^2} \lesssim \varepsilon^3 \, .
\]
Here the last step in a direct consequence of (\ref{baspw}) and (\ref{basiled}).

{\bf The case $K=1$.} This follows directly from the fact that $r^4 A_{\I}$ satisfies a tensorial wave equation of type $4_{2,-1}$ and applying Proposition \ref{prop:rph}. Indeed, we conclude the estimate after 
\begin{itemize}
\item Observing that the boxed term in Proposition \ref{prop:rph} vanishes for $K=1$ since $\frac{2}{p+\frac{15}{2}} \leq \frac{3}{4}$ for $p \in [-2,0]$.
\item Inserting the estimates of Proposition \ref{prop:piall} for the terms in the square bracket in Proposition \ref{prop:rph} and realising that the fluxes appearing on the right hand side of these estimates can be estimated by the right hand side of (\ref{barpest}).

\item Treating the integrand of the linear error term $\mathcal{G}_{5,p}\left[r^4 A_{\I}\right]\left(\tau_1,\tau_2\right)$ as follows. We have for any $\gamma>0$
\[
\mathcal{F}^{lin} \left[r^4 A_{\I}\right] r^{p-2} \Omega \slashed{\nabla}_4 (r^4 A_{\I}) \leq \gamma r^{p-3} | \Omega \slashed{\nabla}_4 (r^4 A_{\I})|^2 + \frac{1}{\gamma} | \mathcal{F}^{lin} \left[r^4 A_{\I}\right] |^2 r^{p-1} \, .
\]
Let $C$ denote the implicit constant in front of $\mathcal{G}_{5,p}\left[r^4 A_{\I}\right]\left(\tau_1,\tau_2\right)$ in~(\ref{basrpe}). Recall from the discussion in Section~\ref{sec:rpagi} that this constant can be chosen to depend only on $M_{\rm init}$ \emph{independently of the choice of $R$.} Choosing now $\gamma=\frac{1}{4C}$ we can absorb the first term above on the left hand side of  (\ref{basrpe}). Furthermore, from Proposition \ref{prop:classify4kl} we compute
\[
\frac{1}{\gamma}| \mathcal{F}^{lin} \left[r^4 A_{\I}\right] |^2 r^{p-1}  \leq \frac{64}{\gamma}\frac{M^2}{R^2} |r^4 A_{\I}|^2 r^{p-5} + \frac{500}{\gamma}\frac{M^{2-\delta}}{R^{2-\delta}} |r^2 \Pi_{\I}|^2 r^{p-3-\delta} \, .
\]
The first term can again be absorbed on the left of (\ref{basrpe}) provided $R$ satisfies 
\begin{align} \label{oneoftheRs2}
\frac{512 C^2 M_{\rm init}^2}{R^2}<1
\end{align}
with $C$ as above.
(Recall that~\eqref{oneoftheRs2} was indeed
one of the constraints announced in Section~\ref{compediumparameterssec} for the choice of $R$.)
The second term is already controlled from Steps~1 and~2.
\item Treating the non-linear error $\mathcal{H}_{5,p}\left[r^4 A_{\I}\right]$ as above.
\end{itemize}
{\bf The case $K>1$.} From Propositions \ref{prop:classify4kl} and \ref{prop:comm4kl} we conclude that $ \left(\Omega \slashed{\nabla}_4\right)^k (r^4A_{\I})$ satisfies a tensorial wave equation of type $4_{2+k/2,l}$, the precise value of the (half-integer) $l$ being irrelevant. We hence again apply Proposition \ref{prop:rph}. The boxed term might no longer vanish but will now be controlled by the estimate proven in the previous step of the induction. For the term $\mathcal{G}_{5,p}\left[(r \slashed{\nabla}_4)^{k-1} (r^4 A_{\I}) \right]$ note (after doing a simple induction using Propositions \ref{prop:classify4kl} and \ref{prop:comm4kl}) that for $2 \leq k \leq K$
\begin{align}
\mathcal{F}^{lin}_{2+\frac{k-1}{2},l} \left[(r \slashed{\nabla}_4)^{k-1} (r^4 A_{\I}) \right] = \sum_{l=0}^{k-2} h_{0}  \Omega \slashed{\nabla}_4(r \Omega \slashed{\nabla}_4)^{l} \Pi _{\I} + \sum_{l=0}^{k-1} \frac{h_{0}}{r^2} \Omega (r \slashed{\nabla}_4)^{l}  (r^4 A_{\I}) + \frac{h_{0}}{r^2} r^2\Pi_{\I} + \frac{h_{0}}{r}  \slashed{\nabla}_4 \Psi_{\I} \nonumber \, .
\end{align}
In particular, for $k=2$
\[
|\mathcal{F}_{4_{2+\frac{k-1}{2},l}}^{lin} \left[r \slashed{\nabla}_4 (r^4 A_{\I})\right] |^2 r^{p-1}  \lesssim r^{p-1} | \slashed{\nabla}_4 \Pi_{\I}|^2 + r^{p-3} |\slashed{\nabla}_4 (r^4 A_{\I})|^2  + r^{p-5} |r^2 \Pi_{\I}|^2 + r^{p-3} |\slashed{\nabla}_4 \Psi_{\I}|^2  \, 
\]
and similarly for the higher $k$. The first and third term on the right can be controlled by the estimates proven in Steps 1 and 2, the second term from the estimate for $K=1$ and the last term by Theorem~\ref{thm:PPbarestimates} for $p \leq 2$. The argument for $k>2$ is analogous and the proof is complete.
\end{proof} 
Combining the estimate of the Lemma with Steps 1 and 2 yields the estimates of the proposition after using standard elliptic estimates on cones.
\end{proof}

We can similarly prove
\begin{proposition} \label{prop:trf}
We have, for $2 \leq K \leq N$ and any $u_1 \leq \tau_1 \leq \tau_2 \leq u_f$, the following estimate:
\begin{align}
\sum_{|\underline{k}|=1}^{K-1}  \int_{\CcI_{\tau_2}}   \ r^2 |\tilde{\mathfrak{D}}_{\nearrow}^{\underline{k}} \Pi_{\I}|^2 + \sum_{|\underline{k}|=0}^{K-2}\int_{\CbcI_{v_\infty}(\tau_1)} \  | \tilde{\mathfrak{D}}_{\nearrow}^{\underline{k}} \Psi_{\I}|^2 \nonumber \\
 + \sum_{|\underline{k}|=1}^{K-1} \int_{\DcI(\tau_1)} \, r | \tilde{\mathfrak{D}}_{\nearrow}^{\underline{k}} \Pi_{\I}|^2  +  \sum_{|\underline{k}|=0}^{K-2} \int_{\DcI(\tau_1)} \, r^{-1-\delta} | \tilde{\mathfrak{D}}_{\nearrow}^{\underline{k}} \Psi_{\I}|^2 \nonumber \\
\lesssim
\sum_{|\underline{k}|=1}^{K-1}  \int_{\CcI_{\tau_1}}   \ r^2 | \tilde{\mathfrak{D}}_{\nearrow}^{\underline{k}} \Pi_{\I}|^2 + \sum_{|\underline{k}|=1}^{K} \int_{\CcI_{\tau_1}}  r^4 | \tilde{\mathfrak{D}}_{\nearrow}^{\underline{k}} A_{\I}|^2 +\frac{\varepsilon_0^2 
 +  \varepsilon^3}{(\tau_1)^{\min(2,N-K)}}  \nonumber \\
+ \sum_{k=0}^{K-1}\underline{\check{\mathbb{F}}}_{v(\tau_1)} [ (\slashed{\nabla}_{R^\star})^k A_{\Hp}] +\sum_{k=0}^{K-2}\underline{\check{\mathbb{F}}}_{v(\tau_1)} [ (\slashed{\nabla}_{R^\star})^k\Pi_{\Hp}] \, .
\end{align}
\end{proposition}

\begin{proof}
The proof is very similar to that of the previous proposition and we shall use terminology from its proof. At the heart of the proof are now the Bianchi pairs
\begin{equation} \label{bpe1}
\begin{split}
\Omega \slashed{\nabla}_3 \left(r^2 \slashed{div} \Pi_{\I} \right)+ \frac{\Omega_\circ^2}{r} \left(r^2 \slashed{div} \Pi_{\I} \right) &= \frac{\Omega^2}{r} r \slashed{div} \Psi_{\I} \boxed{  -
	\Omega^{-1} r{\hat{\chibar}} \cdot r \slashed{div}  \Pi_{\I} 
	+
	r^2 \Omega^{-1} {}^* \betabar \overset{*}{\cdot}  \Pi_{\I} 
	}
	+
 \mathcal{E}^{2}_{3} \, ,
 \\
\Omega \slashed{\nabla}_4 \Psi_{\I} &= -\frac{2}{r} r \slashed{\mathcal{D}}_2^\star \left(r^2 \slashed{div} \Pi_{\I}\right) - 2\Pi_{\I} + \frac{6M}{r} \Pi_{\I} + 3MA_{\I}  + \mathcal{E}^2_2 \, .
\end{split}
\end{equation}
where we have boxed the anomalous error (which would only be $\mathcal{E}^2_2$ in schematic notation).

\vskip1pc
\noindent
{\bf Step 1.} As in the previous proposition, we first prove the estimate for angular derivatives. Hence we look at the angular commuted equations (recall (\ref{commdef})) which for $k\geq 1$ read
\begin{align} \label{bpec}
\Omega \slashed{\nabla}_3 (r \Pi^{(k)}_{\I}) + \frac{\Omega_\circ^2}{r} (r \Pi^{(k)}_{\I})  &= \frac{\Omega^2}{r} r\slashed{div} \Psi^{(k-1)}_{\I}  + \tilde{\mathfrak{D}}^{(k-1,0,0)}  \boxed{  -
	\Omega^{-1} r{\hat{\chibar}} \cdot r \slashed{div}  \Pi_{\I} 
	+
	r^2 \Omega^{-1} {}^* \betabar \overset{*}{\cdot}  \Pi_{\I} 
	} + \mathcal{E}^{k+1}_3\, , \nonumber \\
\Omega \slashed{\nabla}_4 (\Psi^{(k-1)}_{\I})  &= -\frac{2}{r} r \slashed{\mathcal{D}}_2^\star (r \Pi^{(k)}_{\I}) -2 \Pi^{(k-1)}_{\I} +\frac{6M}{r} \Pi^{(k-1)}_{\I} + 3M A^{(k-1)}_{\I} + \mathcal{E}^{k+1}_2 \, ,
\end{align}
 if $k$ is odd and with $r\slashed{div}$ and $r \slashed{\mathcal{D}}_2^\star$ interchanged in the two equations if $k$ is even. To produce the estimate, we contract the first of (\ref{bpec}) by $2 r \left(1+r^{-\delta}\right) r  \Pi^{(k)}_{\I}$ and add the second contracted with
 $\Omega^2 \left(1+r^{-\delta}\right) \Psi^{(k-1)}_{\I}$. We then sum over all $k$ up to $K$ and integrate over $\DcI \left( \tau_1, \tau_2\right) \cap \{r \geq \tilde{R}\}$. This produces the desired estimate after observing
\begin{itemize}
\item The terms on the left produce the desired terms after an integration by parts.
\item The first term on the right hand side of each of (\ref{bpec}) cancels after an integration by parts up to a cubic error 
\[
\left(\eta + \underline{\eta}\right) \left(1+r^{-\delta}\right) r \Pi^{(k)}_{\I} \cdot \Psi^{(k-1)}_{\I} \lesssim \varepsilon \frac{1}{r^2} \left(|r  \Pi^{(k)}_{\I}|^2 + |\Psi^{(k-1)}_{\I}|^2\right) \, ,
\]
which can be absorbed on the left hand side. Here we have used $\|r^2(\eta +\underline{\eta})\|_{L^\infty(\DcI)} \lesssim  \varepsilon$  following from~\eqref{elinfestimates}.
\item All other linear terms on the right hand side, except for the term $-2\Pi^{(k-1)}_{\I}$, can be easily controlled using the third estimate of Proposition \ref{prop:APibp}.
\item For the term $-2\Pi^{(k-1)}_{\I}$, we integrate the relevant term by parts: After using (\ref{recdpi}) we obtain
\begin{align}
\int_{\tau_1}^{\tau_2} d\bar{u} \int_{\CcI_{\bar{u}}} -2\Pi^{(k-1)}_{\I} r^2  \left(1+r^{-\delta}\right) \Omega \slashed{\nabla}_3 \Pi^{(k-1)}_{\I} 
\leq -\int_{\CcI_{\bar{u}}} \,  r^2   \left(1+r^{-\delta}\right) |\Pi^{(k-1)}_{\I}|^2 \Bigg|_{\bar{u}=\tau_1}^{\bar{u}=\tau_2} 
\end{align}
noting that both the spacetime term and the term induced on $r=\tilde{R}$ have favourable (negative) signs. The boundary term on $\tau_2$ also has a good sign and the one on $\tau_1$ is controlled by the right hand side of the desired estimate.

\item For the non-linear term it suffices to observe (after applying Cauchy--Schwarz) that by Proposition~\ref{prop:waveIerror1b}  we have
\begin{align}
\int_{\DcI(\tau_1)} r^{1+\delta} |\mathcal{E}^{K}_2|^2 + \int_{\DcI(\tau_1)} r^{2} |\mathcal{E}^{K}_3|^2 \lesssim
 \frac{\varepsilon^2_0 + \varepsilon^3}{(\tau_1)^{2}}
\end{align}
and for the anomalous term
\begin{align}
\int_{\DcI(\tau_1)} r^2 \Big| \tilde{\mathfrak{D}}^{(K-2,0,0)} ( r{\hat{\chibar}} \cdot r \slashed{div}  \Pi_{\I} )|^2 
	+ r^2 \Big| \tilde{\mathfrak{D}}^{(K-2,0,0)} \left( r^2 {}^* \betabar \overset{*}{\cdot}  \Pi_{\I}\right) \Big|^2  \lesssim
 \frac{\varepsilon^2_0 + \varepsilon^3}{(\tau_1)^{\min(2,N-K)}} \, .
\end{align}
\end{itemize}
\vskip1pc
\noindent
{\bf Step 2.} Proceeds entirely analogously to Step 2 in Proposition \ref{prop:APibp}: Commute the Bianchi pairs (\ref{bpec}) with $\left(r \Omega \slashed{\nabla}_4\right)^l$ and repeat the proof of Step 1. Lower order terms and angular derivative terms can be controlled by the estimates obtained in Step 1. 
\vskip1pc
\noindent {\bf Step 3.} Instead of carrying out the analogue of the proof of Proposition \ref{prop:APibp} (which is also possible), we can directly employ the second equation of (\ref{tff}) and the third estimate of Proposition \ref{prop:APibp} to control the missing flux of $(\Omega \slashed{\nabla}_4)^{K-1} \Pi_{\I}$ on $\CcI\left(\tau_2\right)$.
\end{proof}

 \subsection{Decay estimates for $\Pi$ and $A$: the proof of Theorem \ref{theo:mtheoalphar2}} \label{sec:decaypia}

We first summarise our weighted estimates in the following proposition recalling the energies defined in Theorem \ref{theo:mtheoalphar2}.

\begin{proposition} \label{prop:sumrp}
For any $u_1 \leq \tau \leq u_f$ and $p \in \{0,1,2\}$, we have, for $1 \leq K \leq N$, the estimate
\begin{align} \label{hju}
\check{\mathbb{E}}_\Box^{K,p} \left[{\alpha}_{\I}\right] \left(\tau\right) + \check{\mathbb{E}}_\Box^{K} \left[{\alpha}_{\Hp}\right] \left(v(\tau)\right) \lesssim &  
\int_{\check{\underline{C}}^{\Hp}_{v(\tau)}} \Omega^2  \sum_{|\underline{k}|=0; k_3\neq |\underline{k}|}^{K}  | \mathfrak{D}^{\underline{k}} {A}_{\Hp}|^2 + \frac{\varepsilon^2_0+\varepsilon^3}{\tau^{{\kappa}_p}} \nonumber \\
+ &\int_{\check{C}^{\I}_\tau} \Bigg\{ \sum_{|\underline{k}|=1; {k_2 \neq |\underline{k}|}}^K  \ r^{4+p} | \mathfrak{D}^{\underline{k}} A_{\I}|^2 + \sum_{|\underline{k}|=0; {k_2 \neq |\underline{k}|}}^{K-1}  r^2 | \mathfrak{D}^{\underline{k}} \Pi_{\I}|^2 \Bigg\}  
\end{align}
with ${\kappa}_2 = 0$, $\kappa_1= 1- \delta^K_N$ and $\kappa_0 =\min(2,N-K)$.
\end{proposition}

\begin{proof}
Using Propositions \ref{sec:ellipticconesI} and \ref{sec:ellipticconesHp} one sees that it suffices to prove (\ref{hju}) with  $\mathfrak{D}$ replaced by $\tilde{\mathfrak{D}}$  everywhere in the energies appearing on the left and this is what we will show.

We first note the easily verified estimates
 \begin{align}
  \sum_{k=0}^{K-1}\underline{\check{\mathbb{F}}}_{v(\tau)} [ (\slashed{\nabla}_{R^\star})^k A_{\Hp}] +\sum_{k=0}^{K-2}\underline{\check{\mathbb{F}}}_{v(\tau)} [ (\slashed{\nabla}_{R^\star})^k\Pi_{\Hp}] \lesssim \int_{\CbcH_{v(\tau)}} \Omega^2 \sum_{|\underline{k}|=1, k_3 \neq |\underline{k}|}^{K}  |\mathfrak{D}^{\underline{k}} A_{\Hp}|^2 + \frac{\varepsilon^2_0+\varepsilon^3}{\tau^{\min(2,N-K)}} \, , \nonumber
 \end{align}
 \begin{align}
 \sum_{k=0}^{K-1}\check{\mathbb{F}}_{\tau} [ (\slashed{\nabla}_{R^\star})^k A_{\I}]
+
\sum_{k=0}^{K-2}\check{\mathbb{F}}_{\tau} [ (\slashed{\nabla}_{R^\star})^k\Pi_{\I}] \lesssim & \int_{\check{C}^{\I}_\tau} \Bigg\{ \sum_{|\underline{k}|=1; {k_2 \neq |\underline{k}|}}^K  \ r^{4+0} | \mathfrak{D}^{\underline{k}} A_{\I}|^2 + \sum_{|\underline{k}|=0; {k_2 \neq |\underline{k}|}}^{K-1}  r^2 | \mathfrak{D}^{\underline{k}} \Pi_{\I}|^2 \Bigg\} \nonumber \\
&+ \frac{\varepsilon^2_0+\varepsilon^3}{\tau^{\min(2,N-K)}} \, . \nonumber 
 \end{align}
Together they imply that the right hand side of (\ref{hju}) controls the right hand side (hence the left hand side) of the estimates of Propositions \ref{prop:piall} and \ref{prop:piall2}.
These estimates in turn already  imply that (\ref{hju}) holds for $ \check{\mathbb{E}}_\Box^{K} \left[{\alpha}_{\Hp}\right] \left(v(\tau)\right)$. They also imply (\ref{hju}) for 
$\check{\mathbb{E}}_\Box^{K,p} \left[{\alpha}_{\I}\right] \left(\tau\right)$ provided one restricts all integrals appearing in the definition of that energy $r\leq 2R$.
We now remove that restriction to complete the proof.

We observe that the right hand side of (\ref{hju}) also controls the right hand side of Proposition
\ref{prop:trf} and Proposition \ref{prop:APibp}. We next claim that the estimate of Proposition
\ref{prop:trf} in fact implies
\begin{align}
&\sup_{\tau \leq u \leq u_f} \sum_{|\underline{k}|=1; k_2 \neq |\underline{k}|}^{K-1}  \int_{\CcI_{u}}   \ r^2 | \tilde{\mathfrak{D}}^{\underline{k}} \Pi_{\I}|^2 + \sum_{|\underline{k}|=0}^{K-2}\int_{\CbcI_{v_\infty}(\tau)} \  | \tilde{\mathfrak{D}}^{\underline{k}} \Psi_{\I}|^2 \nonumber \\
&+ \sum_{|\underline{k}|=1}^{K-1} \int_{\DcI(\tau)} \, r^{1-\delta} | \tilde{\mathfrak{D}}^{\underline{k}} \Pi_{\I}|^2  +  \sum_{|\underline{k}|=0}^{K-2} \int_{\DcI(\tau)} \, r^{-1-\delta} | \tilde{\mathfrak{D}}^{\underline{k}} \Psi_{\I}|^2 \lesssim \textrm{RHS of (\ref{hju}) for $p=0$},
\end{align}
that is that the estimate of Proposition~\ref{prop:trf} 
remains true replacing everywhere $\tilde{\mathfrak{D}}_{\nearrow}^{\underline{k}}$ by $\tilde{\mathfrak{D}}^{\underline{k}}$ , provided one accepts an $r^{-\delta}$ in the integrated decay statement 
for $\Pi_{\I}$ and avoids the top-order transversal flux on the cone ($k_2 \neq |\underline{k}|$). 
For the terms involving $\Psi_{\I}$, the claim follows from Theorem~\ref{thm:PPbarestimates}. For the terms involving $\Pi_{\I}$ the claim is readily verified (inductively) by commuting through the $\Omega^{-1} \slashed{\nabla}_3$ of $\tilde{\mathfrak{D}}^{\underline{k}}$ on $\Pi_{\I}$ and inserting the relation (\ref{recdpi}).\footnote{Note the $r^{-\delta}$ loss is necessary here because we do not have $\tau^{-2}$ decay for $ \int_{\DcI(\tau)} \, r^{-1} | \Omega \slashed{\nabla}_4 \Psi_{\I}|^2$; see Proposition \ref{prop:Psummarydecay}.}
By a completely analogous argument (commuting through $\Omega^{-1} \slashed{\nabla}_3$ and using (\ref{recda})), one sees that the estimates of Proposition \ref{prop:APibp} continue to hold replacing $\tilde{\mathfrak{D}}_{\nearrow}^{\underline{k}}$ by $\tilde{\mathfrak{D}}^{\underline{k}}$ everywhere on the left, provided one allows an $r^{-\delta}$ loss in the integrated decay statement for $\Pi_{\I}$ (and omits the top order transversal flux on the cone) in~\eqref{trg}.

Combining these facts one sees that we have shown the desired (\ref{hju}) except for the term 
\[
 \int_{\CbcI_{v_\infty}(\tau)} \sum_{|\underline{k}|=1, k_3 \neq |\underline{k}|}^{K}  \ r^{4+p} | \tilde{\mathfrak{D}}^{\underline{k}} A_{\I} |^2
\]
appearing in $\check{\mathbb{E}}_\Box^{K,p} \left[{\alpha_{\I}}\right] \left(\tau\right)$. Therefore, it suffices to control this term using the estimate (\ref{hju}) with this term in $\check{\mathbb{E}}_\Box^{K,p} \left[{\alpha_{\I}}\right] \left(\tau\right)$ removed.
Now, if one of the derivatives in $\tilde{\mathfrak{D}}^{\underline{k}}$ is a $3$-derivative, we can commute it through, insert (\ref{recda}) and the flux of $\Pi_{\I}$ on $v=v_\infty$ appearing in $\check{\mathbb{E}}^{K,p} \left[{A}\right] \left(\tau\right)$. If one of the derivatives is angular and the remaining ones are $4$-derivatives, the term is already contained in the estimate of Lemma~\ref{barp}. Finally, if there are two angular derivatives in $\underline{k}$, one can use (\ref{teueli2}) and elliptic estimates together with the flux bounds for $\Pi_{\I}$ on $v=v_\infty$ appearing in $\check{\mathbb{E}}_\Box^{K,p} \left[{A}\right] \left(\tau\right)$.
\end{proof}

The following decay estimates are easily seen to imply the statement of Theorem \ref{theo:mtheoalphar2}.

\begin{proposition}  \label{prop:afa}
For any $u_1 \leq \tau \leq u_f$ we have the estimates
\begin{align}
\check{\mathbb{E}}_\Box^{N,2} \left[{\alpha}_{\I}\right] \left(\tau\right) + \check{\mathbb{E}}_\Box^{N} \left[{\alpha}_{\Hp}\right] \left(v(\tau)\right) \lesssim \varepsilon^2_0 + \varepsilon^3 \label{dec1uj} \, , \\
\check{\mathbb{E}}_\Box^{N-1,1} \left[{\alpha}_{\I}\right] \left(\tau\right) + \check{\mathbb{E}}_\Box^{N-1} \left[{\alpha}_{\Hp}\right] \left(v(\tau)\right) \lesssim \frac{ \varepsilon^2_0 + \varepsilon^3}{\tau} \label{dec2uj} \, , \\
\check{\mathbb{E}}_\Box^{N-2,0} \left[{\alpha}_{\I}\right] \left(\tau\right) + \check{\mathbb{E}}_\Box^{N-2} \left[{\alpha}_{\Hp}\right] \left(v(\tau)\right) \lesssim \frac{ \varepsilon^2_0 + \varepsilon^3}{\tau^2} \, .  \label{dec3uj}
\end{align}
\end{proposition}

\begin{proof}
Applying Proposition \ref{prop:sumrp} for $K=N$, $p=2$ and $\tau=u_1$ yields (\ref{dec1uj}) after using that the initial flux (\ref{initwene}) is controlled by $\varepsilon^2_0 + \varepsilon^3$ from Proposition \ref{thm:gidataestimates}. From (\ref{dec1uj}) we extract a dyadic sequence of times $\tau_i$ such that for $\tau \in \{\tau_i\}$
we have
\begin{align} \label{tgb}
&\sum_{|\underline{k}|=1; k_2 \neq |\underline{k}|}^{N-1} \int_{\CcI_{\tau}}  \ r^{5} | \mathfrak{D}^{\underline{k}} A_{\I}|^2 + \sum_{|\underline{k}|=0; k_2 \neq |\underline{k}|}^{N-2} \int_{\CcI_{\tau}} r^2 | \mathfrak{D}^{\underline{k}} \Pi_{\I}|^2 + \int_{\CbcH_{v(\tau_2)}} \Omega^2 \sum_{|\underline{k}|=1, k_3 \neq |\underline{k}|}^{N-1}  |\mathfrak{D}^{\underline{k}} A_{\Hp}|^2 
 \lesssim  \frac{ \varepsilon^2_0 + \varepsilon^3}{\tau} \, .
\end{align}
Applying Proposition \ref{prop:sumrp} with $K=N-1$ and $p=1$ from the slices $\tau_i$ to an arbitrary $\tau$ yields 
(\ref{dec2uj}). From this we extract a (different) dyadic sequence of times $\tau_i$ such that for $\tau \in \{\tau_i\}$ we have
\begin{align} 
&\sum_{|\underline{k}|=1; k_2 \neq |\underline{k}|}^{N-2} \int_{\CcI_{\tau}}  \ r^{4} | \mathfrak{D}^{\underline{k}} A_{\I}|^2 + \sum_{|\underline{k}|=0; k_2 \neq |\underline{k}|}^{N-3} \int_{\CcI_{\tau}} r^2 | \mathfrak{D}^{\underline{k}} \Pi_{\I}|^2 
+ \int_{\CbcH_{v(\tau_2)}} \Omega^2 \sum_{|\underline{k}|=1, k_3 \neq |\underline{k}|}^{N-2}  |\mathfrak{D}^{\underline{k}} A_{\Hp}|^2 \lesssim   \frac{ \varepsilon^2_0 + \varepsilon^3}{\tau^2} \, . \nonumber
\end{align}
Applying Proposition \ref{prop:sumrp} with $K=N-2$ and $p=0$ from the slices $\tau_i$ to an arbitrary $\tau$ yields the
estimate~\eqref{dec3uj}. 
\end{proof}

\begin{remark}
We note that this iteration has in particular provided a \underline{decay} estimate for the flux of $ \int_{\CbcI_{v_\infty}(\tau_1)}  | \mathfrak{D}^{\underline{k}} \Psi_{\I}|^2$ which is much stronger than what we obtained from the $r^p$ method for the tensorial wave equation for $\Psi_{\I}$ itself, where we could only show boundedness. 
\end{remark}

\section{Concluding the main estimates for $\alpha$}
\label{sec:missingfluxa}
In this section we prove Theorem \ref{theo:mtheoalphar} from Theorem \ref{theo:mtheoalphar2}.
To achieve this,
we only need to extend all of our integrated decay estimates and flux estimates to the non-truncated regions and cones. The proof is very similar to that for $P$ and $\underline{P}$ seen in Section \ref{sec:missingfluxp} of the previous chapter. 
\vskip1pc
\noindent
{\bf Step 1.} We observe that the estimates of Theorem \ref{theo:mtheoalphar2} continue to hold if we drop all check superscripts in the \emph{spacetime integrals}. This follows easily from the relations for $A_{\I}$ and $A_{\Hp}$ in the overlap region, i.e.~by applying Proposition \ref{thm:morecancelnotjustT}. As an immediate corollary using the definition of the timelike hypersurface $\mathcal{B}$ we obtain:
\begin{corollary} \label{cor:hypbb}
For all $u_1 \leq \tau \leq u_f$ we have the estimates
\begin{align} 
 \sum_{|{\underline{k}} | \leq N} \int_{\mathcal{B}(u_1)}  | {\mathfrak{D}}^{\underline{k}}  {A}_{\I}|^2 +  | {\mathfrak{D}}^{\underline{k}}  {A}_{\Hp}|^2 &\lesssim \varepsilon^2_0 + \varepsilon^3  \, , \nonumber \\
  \sum_{|{\underline{k}} | \leq N-1} \int_{\mathcal{B}(\tau)}  | {\mathfrak{D}}^{\underline{k}}  {A}_{\I}|^2 +  | {\mathfrak{D}}^{\underline{k}}  {A}_{\Hp}|^2 &\lesssim  \frac{ \varepsilon^2_0 + \varepsilon^3}{\tau} \, , 
  \nonumber \\
  \sum_{|{\underline{k}} | \leq N-2} \int_{\mathcal{B}(\tau)}  |{\mathfrak{D}}^{\underline{k}}  {A}_{\I}|^2 +  | {\mathfrak{D}}^{\underline{k}}  {A}_{\Hp}|^2 &\lesssim  \frac{ \varepsilon^2_0 + \varepsilon^3}{\tau^2} \, .
\end{align}
\end{corollary}

\vskip1pc
\noindent
{\bf Step 2a.} 
We observe that the estimates of Theorem \ref{theo:mtheoalphar2}  continue to hold if we replace the flux
\[
 \int_{\check{\underline{C}}^{\I}_{v_{\infty}}(\tau)}  \Bigg\{ \sum_{|\underline{k}|=1, k_3 \neq |\underline{k}|}^{K}  \ r^{4+p} | \mathfrak{D}^{\underline{k}} A_{\I} |^2 + \sum_{|\underline{k}|=0}^{K-1}  r^{2+p} | \mathfrak{D}^{\underline{k}} \Pi_{\I}|^2 + \sum_{|\underline{k}|=0}^{K-2}  | \mathfrak{D}^{\underline{k}} \Psi_{\I}|^2  \Bigg\}
\]
by the flux over an arbitrary ingoing truncated cone in $\DcI$, i.e. integrating over $\int_{\check{\underline{C}}^{\I}_{v}(\tau)}$ for any cone with $v \leq v_\infty$. Indeed, picking such a cone $\check{\underline{C}}_v^{\I} \left(\tau\right)$ we can consider the spacetime region enclosed by $\check{\underline{C}}_v^{\I} \left(\tau\right)$, $\mathcal{B}$, $\check{C}_\tau^{\I}$ and (potentially) $\check{C}^{\I}_{u_f}$. We reapply the estimates of Propositions \ref{prop:APibp} and \ref{prop:trf} using now the bounds on the cone $\check{C}_\tau^{\I}$ established in \ref{theo:mtheoalphar2} and the bounds on the hypersurface $\mathcal{B}$ established in Corollary~\ref{cor:hypbb}. 

\vskip1pc
\noindent
{\bf Step 2b.}
We observe that the estimates of Theorem \ref{theo:mtheoalphar2}  continue to hold if we replace the flux
\[
\int_{\check{C}^{\Hp}_{u_f}({v})} \sum_{|\underline{k}|=0; {k_2 \neq |\underline{k}|}}^{K}  | \mathfrak{D}^{\underline{k}} A_{\Hp}|^2  
\]
by the flux over an arbitrary outgoing truncated cone in $\DcH$, i.e. integrating over $\check{C}^{\Hp}_{u}({v})$ for any $u \leq u_f$. Indeed picking such a cone $\check{C}_u^{\Hp} \left(v\right)$ we can consider the region enclosed by $\check{C}_u^{\Hp} \left(v\right)$, $\check{\underline{C}}^{\Hp}_v$ and $\mathcal{B}$. We repeat the proof of Propositions \ref{prop:pirstar} and \ref{prop:arstar} in this region using now the estimates of Theorem \ref{theo:mtheoalphar2} on the cone  $\check{\underline{C}}^{\Hp}_v$ and the bounds on the hypersurface $\mathcal{B}$ established in Corollary \ref{cor:hypbb}. One then obtains control over the fluxes ${{\mathbb{F}}}_{u} [\mathfrak{D}^{\underline{k}} A_{\Hp}] (v)$ from that of ${{\mathbb{F}}}_{u} [ (\slashed{\nabla}_{R^\star})^k A_{\Hp}] (v)$ and ${{\mathbb{F}}}_{u} [ (\slashed{\nabla}_{R^\star})^k \Pi_{\Hp}] (v)$ as in the proof of Proposition \ref{prop:piall}.

\vskip1pc
\noindent
{\bf Step 3.} We observe that the estimates of Theorem \ref{theo:mtheoalphar2}  continue to hold if we replace truncated by non-truncated cones in $\DcI \cup \DcH$. This proceeds by doing localised energy estimates (in regions where neither the $r$-weights nor $\Omega$-weights play a role) for the non-linear Teukolsky equation entirely analogously to the case of $P$ and $\underline{P}$ seen in Section \ref{sec:missingfluxp} of the previous chapter. We omit the standard details. Taking into account the remarks at the beginning of Section \ref{theo:mtheoalphar2} about the ``near initial data region" we see that we have proven the estimate of Theorem \ref{theo:mtheoalphar}.

\section{Overview of the estimates on \underline{$\alpha$}} \label{sec:overviewab}

We now turn to the task of obtaining the estimates for $\underline\alpha$  in Theorem~\ref{thm:alphaalphabarestimates}.
This section is closely modelled on Section~\ref{sec:overviewa}.

In complete analogy to the estimate on $\alpha$ we will first prove the following restricted version of the estimates for $\underline{\alpha}_{\I}$, $\underline{\alpha}_{\Hp}$ before concluding the full estimate implied by Theorem~\ref{thm:alphaalphabarestimates}.

\begin{theorem} \label{theo:mtheoalphabr}
We have the estimates 
\[
	\sum_{s=0,1,2}
	\sup_{u_{-1} \leq \tau \leq u_f} \tau^s \cdot {\mathbb{E}}_\Box^{N-s} \left[\underline{\alpha}_{\I}\right] \left(\tau\right)  + \sum_{s=0,1,2}
	\sup_{v_{-1} \leq v} 
	v^{s} \cdot {\mathbb{E}}_\Box^{N-s} \left[\underline{\alpha}_{\Hp}\right] \left(v\right)  \lesssim \varepsilon_0^2 + \varepsilon^3 
\]
for the restricted energies (recall $\check{\underline{A}}_{\I}=\check{r}\Omega^2 \underline{\alpha}_{\I}$, $\underline{A}_{\Hp}=r\Omega^2\underline{\alpha}_{\Hp}$ and $\check{\underline{\Pi}}_{\I}=r^3 \Omega \check{\underline{\psi}}_{\I}$, $\check{\underline{\Psi}}_{\I} = r^5 \check{\underline{P}}_{\I}$)
\begin{align}
&{\mathbb{E}}_\Box^{K} \left[\underline{\alpha}_{\Hp}\right] \left(v\right) := \sup_{\tilde{v} \geq v} \int_{\underline{C}^{\Hp}_{\tilde{v}}} \Omega^2  \sum_{|\underline{k}|=0; \boxed{k_3\neq K}}^{K}  | \mathfrak{D}^{\underline{k}} (\Omega^{-4} \underline{A}_{\Hp})|^2 + \sup_{u\leq u_f} \int_{C^{\Hp}_u({v})} \sum_{|\underline{k}|=0; k_2\neq K}^{K}  | \mathfrak{D}^{\underline{k}} (\Omega^{-4} \underline{A}_{\Hp})|^2  \nonumber \\
& \qquad \ \  +  \int_{\mathcal{D}^{\Hp}\left(v\right)}  \Omega^2 \Bigg\{ \sum_{|\underline{k}|=0}^{K} \left(1-\frac{3M_f}{r}\right)^2 | \mathfrak{D}^{\underline{k}} (\Omega^{-4} \underline{A}_{\Hp})|^2  +  \sum_{|\underline{k}|=0}^{K-1}  | \mathfrak{D}^{\underline{k}} (\Omega^{-4} \underline{A}_{\Hp})|^2+  | R^\star \mathfrak{D}^{\underline{k}} (\Omega^{-4} \underline{A}_{\Hp})|^2  \Bigg\} \, , \nonumber
\end{align}
\begin{align}
{\mathbb{E}}_\Box^{K} \left[\underline{\alpha}_{\I}\right] \left(\tau\right) :=  &\sup_{\tau \leq u \leq u_f} \int_{C^{\I}_u} \frac{1}{r^2} \Bigg\{ \sum_{|\underline{k}|=0; k_2\neq K}^{K}  | \mathfrak{D}^{\underline{k}} \check{\underline{A}}_{\I}|^2 + \sum_{|\underline{k}|=0}^{K-1}   | \mathfrak{D}^{\underline{k}} \check{\underline{\Pi}}_{\I}|^2 + \sum_{|\underline{k}|=0}^{K-2} | \mathfrak{D}^{\underline{k}} \check{\underline{\Psi}}_{\I}|^2 \Bigg\} \nonumber \\
&+\sup_{v \leq v_\infty} \int_{\underline{C}^{\I}_{v}(\tau)} \Bigg\{\sum_{|\underline{k}|=0; \boxed{k_3 \neq K}}^{K}  | \mathfrak{D}^{\underline{k}} \check{\underline{A}}_{\I}|^2 + \sum_{|\underline{k}|=0;  \boxed{k_3 \neq K-1}}^{K-1}   | \mathfrak{D}^{\underline{k}} \check{\underline{\Pi}}_{\I}|^2  \Bigg\} \nonumber \\
&+  \int_{\mathcal{D}^{\I}\left(\tau\right)} \frac{1}{r^{1+\delta}}\Bigg\{ \sum_{|\underline{k}|=0}^K | \mathfrak{D}^{\underline{k}} \check{\underline{A}}_{\I}|^2 +\sum_{|\underline{k}|=0}^{K-1} | \mathfrak{D}^{\underline{k}} \check{\underline{\Pi}}_{\I}|^2 + \sum_{k=1}^{K-2} |  (\slashed{\nabla}_{R^\star})^k \check{\underline{\Psi}}_{\I}|^2 \Bigg\} \, .
\end{align}
\end{theorem}
Note that the above restricted energies differ from the actual energies defined in (\ref{abaren1}) and (\ref{abaren2}) only by the boxed restrictions in the sum. We will finally remove these restrictions in Section \ref{sec:teustaro}, using the 
Teukolsky--Starobinsky identities (in conjunction with the estimates already proven on $\alpha$) to deduce control over the missing fluxes thereby proving Theorem \ref{thm:alphaalphabarestimates}.

To prove Theorem \ref{theo:mtheoalphabr} we again recall the three regions introduced at the beginning of Section \ref{sec:overviewP}. Just as for $P$ and $\underline{P}$, in view of Proposition \ref{thm:gidataestimates} it suffices to prove decay estimates for cones and regions contained in the region $\DcI\left(u_1\right) \cup \DcH\left(v_1=v(u_1)\right)$ in terms of the weighted energy
\begin{align}
\int_{\underline{C}^{\Hp}_{v_1}}  \Omega^2  \sum_{|\underline{k}|=0; k_3\neq N}^{N}  | \mathfrak{D}^{\underline{k}} (\Omega^{-4} \underline{A}_{\Hp})|^2 +  \int_{C^{\I}_{u_1}}\Bigg\{ \sum_{|\underline{k}|=0; k_2\neq N}^{N}  | \mathfrak{D}^{\underline{k}} \check{\underline{A}}_{\I}|^2 + \sum_{|\underline{k}|=0}^{N-1}   | \mathfrak{D}^{\underline{k}} \check{\underline{\Pi}}_{\I}|^2 +\sum_{|\underline{k}|=0}^{N-2} | \mathfrak{D}^{\underline{k}} \check{\underline{\Psi}}_{\I}|^2 \Bigg\}  \nonumber
\end{align}
on $C_{u_1}^{\I} \cup \underline{C}^{\Hp}_{v_1=v(u_1)}$, which is in turn controlled by $\varepsilon_0^2 + \varepsilon^3$ from Proposition \ref{thm:gidataestimates}.

As for $P$ and $\underline{P}$, and for $\alpha$ discussed in Section~\ref{sec:overviewa},
we first prove a weaker version of Theorem \ref{theo:mtheoalphabr} which involves only arbitrary truncated (at $\mathcal{B}$) outgoing cones in the infinity region and arbitrary truncated (at $\mathcal{B}$) ingoing cones in the horizon region. More precisely, we will prove (recall again Remark \ref{rem:equiven}):

\begin{theorem} \label{theo:mtheoalphabr2}
We have the estimates 
\[
	\sum_{s=0,1,2}
	\sup_{u_{1} \leq \tau \leq u_f} \tau^s \cdot {\check{\mathbb{E}}}_\Box^{N-s} \left[\underline{\alpha}_{\I}\right] \left(\tau\right)  + \sum_{s=0,1,2}
	\sup_{v \geq v_{1}} 
	v^{s} \cdot {\check{\mathbb{E}}}_\Box^{N-s} \left[\underline{\alpha}_{\Hp}\right] \left(v\right)  \lesssim \varepsilon_0^2 + \varepsilon^3 
\]
for the restricted energies (recall $\check{\underline{A}}_{\I}=\tilde{r}\Omega^2 \underline{\alpha}_{\I}$, $\underline{A}_{\Hp}=r\Omega^2\underline{\alpha}_{\Hp}$ and $\Pi_{\I}=r^3 \Omega \check{\underline{\Psi}}_{\I}$, $\check{\underline{\Psi}}_{\I} = r^5 \underline{P}_{\I}$)
\begin{align}
&\check{\mathbb{E}}_\Box^{K} \left[\underline{\alpha}_{\Hp}\right] \left(v\right) :=\sup_{\tilde{v} \geq v} \int_{\check{\underline{C}}^{\Hp}_{\tilde{v}}} \Omega^2  \sum_{|\underline{k}|=0; {k_3\neq |\underline{k}|}}^{K}  | \mathfrak{D}^{\underline{k}} (\Omega^{-4} \underline{A}_{\Hp})|^2 +\int_{\check{C}^{\Hp}_{u_f}({v})} \sum_{|\underline{k}|=0; k_2\neq |k|}^{K}  | \mathfrak{D}^{\underline{k}} (\Omega^{-4} \underline{A}_{\Hp})|^2  \nonumber \\
& \qquad \ \  +  \int_{\check{\mathcal{D}}^{\Hp}\left(v\right)}  \Omega^2 \Bigg\{ \sum_{|\underline{k}|=0}^{K} \left(1-\frac{3M_f}{r}\right)^2 | \mathfrak{D}^{\underline{k}} (\Omega^{-4} \underline{A}_{\Hp})|^2  +  \sum_{|\underline{k}|=0}^{K-1}  | \mathfrak{D}^{\underline{k}} (\Omega^{-4} \underline{A}_{\Hp})|^2+  | R^\star \mathfrak{D}^{\underline{k}} (\Omega^{-4} \underline{A}_{\Hp})|^2  \Bigg\} , \nonumber
\end{align}
\begin{align}
\check{\mathbb{E}}_\Box^{K} \left[\underline{\alpha}_{\I}\right] \left(\tau\right) :=  & \sup_{\tau \leq u \leq u_f} \int_{\check{C}^{\I}_u} \frac{1}{r^2} \Bigg\{ \sum_{|\underline{k}|=0; k_2\neq |k|}^{K}  | \mathfrak{D}^{\underline{k}} \check{\underline{A}}_{\I}|^2 + \sum_{|\underline{k}|=0}^{K-1}   | \mathfrak{D}^{\underline{k}} \check{\underline{\Pi}}_{\I}|^2 + \sum_{|\underline{k}|=0}^{K-2} | \mathfrak{D}^{\underline{k}} \check{\underline{\Psi}}_{\I}|^2 \Bigg\} \nonumber \\
& +\int_{\check{\underline{C}}^{\I}_{v_{\infty}}(\tau)}  \Bigg\{\sum_{|\underline{k}|=0; {k_3 \neq |\underline{k}|}}^{K}  | \mathfrak{D}^{\underline{k}} \check{\underline{A}}_{\I}|^2 + \sum_{|\underline{k}|=0;  {k_3 \neq |\underline{k}|}}^{K-1}   | \mathfrak{D}^{\underline{k}}\check{ \underline{\Pi}}_{\I}|^2  \Bigg\} \nonumber \\
&+  \int_{\check{\mathcal{D}}^{\I}\left(\tau\right)} \frac{1}{r^{1+\delta}}\Bigg\{ \sum_{\underline{k}=0}^K | \mathfrak{D}^{|\underline{k}|} \check{\underline{A}}_{\I}|^2 +\sum_{|\underline{k}|=0}^{K-1} | \mathfrak{D}^{\underline{k}} \check{\underline{\Pi}}_{\I}|^2 + \sum_{k=1}^{K-2} |  (\slashed{\nabla}_{R^\star})^{{k}}\check{ \underline{\Psi}}_{\I}|^2 \Bigg\} \, .
\end{align}
\end{theorem}

The proof of Theorem \ref{theo:mtheoalphabr2} will extend over {\bf Sections \ref{sec:traporelb}--\ref{sec:higherorderwb}} and is the main step in proving Theorem~\ref{theo:mtheoalphabr}. The full statement of Theorem \ref{theo:mtheoalphabr} is then obtained from Theorem \ref{theo:mtheoalphabr2} in {\bf Section \ref{sec:missingfluxab}}. The latter step (i.e.~removing the check superscripts from the cones and regions, adding general ingoing cones in the infinity gauge and general outgoing cones in the horizon gauge as well as considering the non-truncated cones) is straightforward and follows exactly as for $P$, $\underline{P}$, and for $\alpha$ discussed in Section~\ref{sec:overviewa}, by doing localised energy estimates. 

We end this overview by outlining the proof of Theorem  \ref{theo:mtheoalphabr2}.
We first recall the defining relations
\begin{align} \label{recdab}
\Omega \slashed{\nabla}_4 (\Omega^{-4} \underline{A}_{\Hp}) + 4\hat{\omega}  (\Omega^{-4} \underline{A}_{\Hp}) = {2}{r^2} \Omega^{-2} \underline{\Pi}_{\Hp}    \ \ \ , \ \ \ \Omega \slashed{\nabla}_4 \check{\underline{A}}_{\I} = 2\frac{\Omega^2}{r^2} \check{\underline{\Pi}}_{\I} \, , 
\end{align}
\begin{align} \label{recdpib}
\Omega \slashed{\nabla}_4 (\Omega^{-2}\underline{\Pi}_{\Hp}) + 2\hat{\omega}  (\Omega^{-2} \underline{\Pi}_{\Hp})   = -\frac{1}{r^2} \underline{\Psi}_{\Hp}    \ \ \ , \ \ \ \Omega \slashed{\nabla}_4 \check{\underline{\Pi}}_{\I} = -\frac{\Omega^2}{r^2} \check{\underline{\Psi}}_{\I}  \, ,
\end{align}
as well as the non-linear wave equations (written in ``elliptic" form) from Proposition \ref{prop:whythat}
\begin{align} 
 r^4 \slashed{\mathcal{D}}_2^\star \slashed{div} (\underline{\Omega}^{-4} \underline{A}_{\Hp})     &= -\frac{1}{\Omega} \slashed{\nabla}_3 \left(\frac{r^2}{\Omega^2}  \underline{\Pi}_{\Hp} \right) -3M r(\underline{\Omega}^{-4} \underline{A}_{\Hp}) + (\mathcal{E}^\star)^1 \, ,
  \label{teueli1b} \\ r^2 \slashed{\mathcal{D}}_2^\star \slashed{div} \check{\underline{A}}_{\I} + \frac{3M}{r}\check{\underline{A}}_{\I} &= -\Omega \slashed{\nabla}_3 \check{\underline{\Pi}}_{\I} +\frac{2}{r}  \left(1- \frac{3M}{r}\right)\check{\underline{\Pi}}_{\I} + \check{\mathcal{E}}^1_1 + \frac{k_1}{r} r \Dslash_2^* \left(
		r \nablaslash r^2 \Omega \tr \chi
		\cdot r\alphabar
		\right)\, ,  \label{teueli2b}
\end{align}
where we explicitly note the anomalous term in the $\check{\underline{A}}_{\I} $ equation and
\begin{align} 
2 r^2\slashed{\mathcal{D}}_2^\star \slashed{div} (\Omega^{-2} \underline{\Pi}_{\Hp})  + 2 \Omega^{-2} \underline{\Pi}_{\Hp}  - \frac{6M}{r} (\Omega^{-2} \underline{\Pi}_{\Hp})  &=  + \frac{1}{\Omega}\slashed{\nabla}_3 \underline{\Psi}_{\Hp}  + 3M \Omega^{-2} \underline{A}_{\Hp}  + (\mathcal{E}^\star)^2 \, ,  \label{wep1b} \\
2 r^2\slashed{\mathcal{D}}_2^\star \slashed{div} \check{\underline{\Pi}}_{\I} + 2 \check{\underline{\Pi}}_{\I} - \frac{6M}{r} \check{\underline{\Pi}}_{\I} &=  +\Omega \slashed{\nabla}_3 \check{\underline{\Psi}}_{\I} + 3M \check{\underline{A}}_{\I} + \check{\mathcal{E}}^2_1  \label{wep2b} \, .
\end{align}
Note that we can always replace $r^2\slashed{\mathcal{D}}_2^\star \slashed{div}$ by $-\frac{1}{2} r^2 \slashed{\Delta} + 1$ since the term involving the difference of the Gauss curvature with the round metric can be incorporated into the non-linear error.

The logic in obtaining the required estimates is then as follows:
\begin{enumerate}
\item Interpreting the defining relations (\ref{recdab}), (\ref{recdpib}) as \emph{transport equations} we derive estimates on $ (\slashed{\nabla}_{R^\star})^k \underline{\Pi}_{\Hp}$ and $ (\slashed{\nabla}_{R^\star})^k \underline{A}_{\Hp}$ as well as $ (\slashed{\nabla}_{R^\star})^k \check{\underline{\Pi}}_{\I}$ and $ (\slashed{\nabla}_{R^\star})^k \check{\underline{A}}_{\I}$ for for $0\leq k \leq N-2$, i.e.~with a loss of derivatives. These will be useful to control lower order terms. This is the content of Section \ref{sec:traporelb}. We remark that from (\ref{recdab}), (\ref{recdpib}) we could derive, by suitable commutation, estimates for \emph{all} derivatives up to order $N-2$, however, we prefer a different argument below. 

\item In Section \ref{sec:higherorderwb}, we consider the \emph{wave equations} satisfied by $(\underline{\Pi}_{\Hp},\check{\underline{\Pi}}_{\I})$ and $(\underline{A}_{\Hp}, \check{\underline{A}}_{\I})$ and prove inductively boundedness and integrated decay estimates for $( (\slashed{\nabla}_{R^\star})^k \underline{\Pi}_{\Hp},  (\slashed{\nabla}_{R^\star})^k\check{\underline{\Pi}}_{\I})$ with $0 \leq k \leq N-1$ and $( (\slashed{\nabla}_{R^\star})^k \underline{A}_{\Hp},  (\slashed{\nabla}_{R^\star})^k \check{\underline{A}}_{\I})$ with $0 \leq k \leq N$. These estimates do not lose derivatives and do not degenerate near $3M$ for $k\geq 1$, see Propositions \ref{prop:pirstarb} and \ref{prop:arstarb}. With this established we control -- with non-optimal weights near the horizon and infinity -- \emph{all} derivatives of $(\underline{\Pi}_{\Hp}, \check{\underline{\Pi}}_{\I})$ and $(\underline{A}_{\Hp}, \check{\underline{A}}_{\I})$ both in integrated decay (Section \ref{sec:susy1b}) and for fluxes (Section \ref{sec:susy1b2}). The basic idea here is simple: Angular derivatives are controlled from the elliptic relations (\ref{teueli1b})--(\ref{wep2b}) and previous estimates. For $\Omega \slashed{\nabla}_4$ derivatives one can use (\ref{recdpib}) and the fact that estimates on $(\underline{\Psi}_{\Hp}, \check{\underline{\Psi}}_{\I})$ have already been established in Theorem~\ref{thm:PPbarestimates}. Finally $\Omega \slashed{\nabla}_3 = \Omega \slashed{\nabla}_4 - 2\slashed{\nabla}_{R^\star}$ can be used to reduce $\Omega \slashed{\nabla}_3$ derivatives to $R^\star$ derivatives, which Proposition \ref{prop:pirstarb} provides control on.

It remains to optimise the weights near infinity for $\check{\underline{\Pi}}_{\I}$ and $\check{\underline{A}}_{\I}$ and near (what will be) the horizon for $\underline{\Pi}_{\Hp}$ and $\underline{A}_{\Hp}$. This is done in Sections \ref{sec:piabinfty} and \ref{sec:piabredshift}  respectively, using estimates for the relevant Bianchi pairs.  The  pigeonhole principle argument of~\cite{DafRodnew} can then be applied to yield an inverse polynomial decay hierarchy for the weighted energies which provides all of the the estimates of Theorem \ref{theo:mtheoalphabr2}. This is the content of Section \ref{sec:decbapi}.

\end{enumerate}

\section{Auxiliary transport estimates at low orders for \underline{$\alpha$}} \label{sec:traporelb}
The reader should compare this section with Section~\ref{sec:traporel}.

Exploiting the transport relations we can deduce lower order estimates for $\check{\underline{\Pi}}_{\I}$, $\underline{\Pi}_{\Hp}$ as well as $\check{\underline{A}}_{\I}$, $\underline{A}_{\Hp}$.

\begin{proposition} \label{prop:lowordertpb}
For any $u_1 \leq \tau \leq u_f$ and $1 \leq K \leq N-2$
the quantities  $\check{\underline{\Pi}}_{\I}= r^3 \Omega \check{\underline{\psi}}_{\I}$ and $\check{\underline{A}}_{\I} = \check{r} \Omega^2 \underline{\alpha}_{\I}$  satisfy the estimate
\begin{align}
\sum_{k=1}^{K} \int_{\DcI(\tau)} \frac{1}{r^{1+\delta}} \left( |\check{\underline{\Pi}}_{\I}|^2 + |\check{\underline{A}}_{\I}|^2 + |  (\slashed{\nabla}_{R^\star})^{k} \check{\underline{\Pi}}_{\I}|^2  +  |  (\slashed{\nabla}_{R^\star})^{k} \check{\underline{A}}_{\I}|^2 \right) \nonumber \\
+\sum_{k=1}^{K} \int_{\mathcal{B}(\tau)}    | \check{\underline{\Pi}}_{\I}|^2   +  |  (\slashed{\nabla}_{R^\star})^{k} \check{\underline{\Pi}}_{\I}|^2 +   |\check{\underline{A}}_{\I}|^2  + |  (\slashed{\nabla}_{R^\star})^{k} \check{\underline{A}}_{\I}|^2  \nonumber \\
\lesssim  \sum_{k=0}^{K-1}  \underline{\check{\mathbb{F}}}_{v(\tau)} [ (\slashed{\nabla}_{R^\star})^k(\Omega^{-2} \underline{\Pi}_{\Hp})] +   \sum_{k=0}^{K-1}  \underline{\check{\mathbb{F}}}_{v(\tau)} [ (\slashed{\nabla}_{R^\star})^k(\Omega^{-4} \underline{A}_{\Hp})] +\frac{\varepsilon_0^2 + \varepsilon^3}{\tau^{\min(2,N-K-2)}} \,  ,\nonumber
\end{align}
and the quantities  $\underline{\Pi}_{\Hp}= r^3 \Omega{\underline{\psi}}_{\Hp}$ and $\underline{A}_{\Hp} = r\Omega^2 \underline{\alpha}_{\Hp}$  satisfy
\begin{align}
\sum_{k=0}^{K} \int_{\DcH(v(\tau))} \Omega^2 \left( |  (\slashed{\nabla}_{R^\star})^{k} (\Omega^{-2} \underline{\Pi}_{\Hp})|^2  +  |  (\slashed{\nabla}_{R^\star})^{k} (\Omega^{-4} \underline{A}_{\Hp})|^2  \right) \nonumber \\
+\sum_{k=0}^{K} \int_{\mathcal{B}\left(\tau\right)}  |  (\slashed{\nabla}_{R^\star})^{k} \underline{\Pi}_{\Hp}|^2 + |  (\slashed{\nabla}_{R^\star})^{k} \underline{A}_{\Hp}|^2\nonumber \\
\lesssim  \sum_{k=0}^{K-1}  \underline{\check{\mathbb{F}}}_{v(\tau)} [ (\slashed{\nabla}_{R^\star})^k(\Omega^{-2} \underline{\Pi}_{\Hp})] +   \sum_{k=0}^{K-1}  \underline{\check{\mathbb{F}}}_{v(\tau)} [ (\slashed{\nabla}_{R^\star})^k(\Omega^{-4} \underline{A}_{\Hp})] +\frac{\varepsilon_0^2 + \varepsilon^3}{\tau^{\min(2,N-K-2)}}\, . \nonumber
\end{align}
\end{proposition}

\begin{proof}
We note that is suffices to prove the estimates with the right hand side being
\[
\sum_{k=1}^K \int_{\underline{\check{C}}^{\Hp}_{v(\tau)}} \Omega^2 \left[  |\Omega^{-2} \underline{\Pi}_{\Hp}|^2  +  \ |\Omega^{-4} \underline{A}_{\Hp}|^2 +  | (\slashed{\nabla}_{R^\star})^{k} (\Omega^{-4}\underline{A}_{\Hp})|^2 +  | (\slashed{\nabla}_{R^\star})^k (\Omega^{-2} \underline{\Pi}_{\Hp})|^2 \right] +\frac{\varepsilon_0^2 + \varepsilon^3}{\tau^{\min(2,N-K-2)}} \, .
 \]
Indeed, it is easy to see (replacing $\Omega \slashed{\nabla}_3=-2\slashed{\nabla}_{R^\star} + \Omega \slashed{\nabla}_4$ and using (\ref{recdab}), (\ref{recdpib})) that the right hand sides appearing in the proposition control the above expression. 

We first prove the second estimate, and this first without the $ (\slashed{\nabla}_{R^\star})^k$ terms. 
From (\ref{recdpib}) we derive
\begin{align} \label{trapo1b}
\frac{1}{2} \Omega \slashed{\nabla}_4 (\Omega^{-2} |\underline{\Pi}_{\Hp} |^2) + \Omega\hat{\omega}\Omega^{-2} |\underline{\Pi}_{\Hp}|^2 = -\frac{1}{r^2} \underline{\Psi}_{\Hp}  \underline{\Pi}_{\Hp} \leq \frac{M}{2r^2} \Omega^{-2} |\underline{\Pi}_{\Hp}|^2 + \frac{2\Omega^2}{Mr^2} |\underline{\Psi}_{\Hp}|^2 \, ,
\end{align}

\begin{align} \label{trapo3b}
\frac{1}{2} \Omega \slashed{\nabla}_4 (\Omega^{-6} |\underline{A}_{\Hp} |^2) + \Omega \hat{\omega} \Omega^{-6} |\underline{A}_{\Hp}|^2 = -\frac{\Omega^{-4}}{r^2} \underline{\Pi}_{\Hp}  \underline{A}_{\Hp} \leq \frac{M}{2r^2} \Omega^{-6} | \underline{A}_{\Hp}|^2 + \frac{2\Omega^{-2}}{Mr^2} |\underline{\Pi}_{\Hp}|^2 \, .
\end{align}
We now integrate (\ref{trapo1b}) over $\DcH(v(\tau))$ (absorbing the first term on the right by the left using $|\Omega\hat{\omega} - Mr^{-2}|\lesssim \varepsilon$) and estimate the term involving $\underline{\Psi}_{\Hp}$ by Theorem \ref{thm:PPbarestimates}. Similarly we integrate (\ref{trapo3b}) over $\DcH(v(\tau))$ (absorbing the first term on the right by the left) and estimate the term involving $\underline{\Pi}_{\Hp}$ by the estimate just obtained. This provides the second estimate of the proposition without the $ (\slashed{\nabla}_{R^\star})^k$-terms. To prove the estimate for the $ (\slashed{\nabla}_{R^\star})^k$-commuted terms one commutes (\ref{recdab}) and (\ref{recdpib}) by $\slashed{\nabla}_{R^\star}$ and
an entirely straightforward induction using the commuted analogues of (\ref{trapo1b}) and (\ref{trapo3b}) finishes the proof of the second estimate of the proposition.

To prove the first estimate (note that in this infinity region weights in $\Omega$ do not play any role), we first convert the estimate for the horizon quantities to the estimate for the infinity quantities on the hypersurface $\mathcal{B}$ using Proposition \ref{thm:cancelT}. This immediately provides control on the terms in the second line of the first estimate. Next we first prove the estimate for the terms appearing in the first line without the $ (\slashed{\nabla}_{R^\star})^k$ terms. To do this we note the easily derived (from (\ref{recdab}) and (\ref{recdpib})) identities for $s>0$
\begin{align} \label{trapo2b}
\frac{1}{2} \Omega \slashed{\nabla}_4 (r^{-s} |\check{\underline{\Pi}}_{\I} |^2) + \frac{1}{2}  \cdot \frac{s}{r^{s+1}}\Omega_\circ^2 |\check{\underline{\Pi}}_{\I}|^2 = -\frac{\Omega^2}{r^2} \check{\underline{\Psi}}_{\I} r^{-s} \check{\underline{\Pi}}_{\I} \lesssim \frac{1}{4} \frac{s}{r^{1+s}} \Omega_\circ^2 |\check{\underline{\Pi}}_{\I}|^2 + \frac{4}{s \cdot r^{3+s}} |\check{\underline{\Psi}}_{\I}|^2  \, ,
\end{align}
\begin{align} \label{trapo4b}
\frac{1}{2} \Omega \slashed{\nabla}_4 ((1+r^{-s}) |\check{\underline{A}}_{\I} |^2) + \frac{1}{2}  \cdot \frac{s}{r^{s+1}}\Omega_\circ^2 |\check{\underline{A}}_{\I}|^2 &= -\frac{\Omega^2}{r^2}\check{ \underline{\Pi}}_{\I} (1+r^{-s}) \check{\underline{A}}_{\I} \nonumber \\
&\lesssim \frac{1}{4} \frac{s}{r^{1+s}} \Omega_\circ^2 |\check{\underline{A}}_{\I}|^2 + \frac{16}{s \cdot r^{3-s}} |\check{\underline{\Pi}}_{\I}|^2  \, .
\end{align}
We now integrate (\ref{trapo2b}) for $s=\delta$ over the region $\DcI(\tau)$ absorbing the first term on the right by the left hand side and using the estimates of Theorem \ref{thm:PPbarestimates}. We then  integrate (\ref{trapo4b}) for $s=\delta$ over the region $\DcI(\tau)$ absorbing the first term on the right by the left hand side and the second term by the estimate just shown on $\check{\underline{\Pi}}_{\I}$.
This yields the desired estimate. Of course the same argument works for the $ (\slashed{\nabla}_{R^\star})^k$ commuted analogue of (\ref{trapo2b}) arising from commuting (\ref{recdab}) and (\ref{recdpib}).
\end{proof}
An easy corollary of the proof (noting that it also produces favourable boundary terms on $v=v_\infty$) is 
\begin{corollary} \label{cor:auxiliaryinfty}
For any $u_1 \leq \tau \leq u_f$ 
the quantities  $\check{\underline{\Pi}}_{\I}= r^3 \Omega \check{\underline{\psi}}_{\I}$ and $\check{\underline{A}}_{\I} = \check{r} \Omega^2 \underline{\alpha}_{\I}$  satisfy the estimate
\begin{align}
\int_{\CbcI_{v_\infty(\tau)}} | \check{\underline{A}}_{\I}|^2 + \frac{1}{r^{\delta}} |\check{\underline{\Pi}}_{\I}|^2 + |\Omega \slashed{\nabla}_3 \check{\underline{A}}_{\I}|^2 \lesssim \underline{\check{\mathbb{F}}}_{v(\tau)} [(\Omega^{-2} \underline{\Pi}_{\Hp})] + \underline{\check{\mathbb{F}}}_{v(\tau)} [(\Omega^{-4} \underline{A}_{\Hp})] +\frac{\varepsilon_0^2 + \varepsilon^3}{\tau^{2}} \, .
\end{align}
\end{corollary}

\section{Higher order energy estimates for \underline{$\alpha$}} \label{sec:higherorderwb}
The reader should compare this section with Section~\ref{sec:traporel}.

\subsection{Basic estimates for $\protect\underline{\Pi}$ and $ (\slashed{\nabla}_{R^\star})^k \protect\underline{\Pi}$}
We begin with a basic estimate arising from commuting the $\underline{\Pi}$ equations only with $R^\star$ derivatives:

\begin{proposition} \label{prop:pirstarb}
For any $u_1 \leq \tau \leq u_f$ and $1\leq K\leq N-2$ the pair $(\underline{\Pi}_{\Hp}=r^3 \Omega {\underline{\psi}}_{\Hp},
\check{\underline{\Pi}}_{\I}=r^3 \Omega \check{\underline{\psi}}_{\I})$ satisfies
\begin{align} \label{goliab}
&\sum_{k=0}^K \sup_{\tau \leq u \leq u_f} \check{\underline{\mathbb{F}}}_{v(u)} [ (\slashed{\nabla}_{R^\star})^k\underline{\Pi}_{\Hp}] + \sum_{k=0}^K \sup_{\tau \leq u \leq u_f} \check{\mathbb{F}}_u [ (\slashed{\nabla}_{R^\star})^k\check{\underline{\Pi}}_{\I}] 
 \\
 &+ \sum_{k=0}^K \check{{\mathbb{F}}}_{u_f} \left[ (\slashed{\nabla}_{R^\star})^k\underline{\Pi}_{\Hp} \right] \left(v(\tau)\right)+ \sum_{k=0}^K \check{\underline{\mathbb{F}}}_{v_\infty} \left[ (\slashed{\nabla}_{R^\star})^k\check{\underline{\Pi}}_{\I} \right] \left(\tau\right) \nonumber \\
&+\sum_{k=0}^K \check{\mathbb{I}} \left[ (\slashed{\nabla}_{R^\star})^k\underline{\Pi}_{\Hp}\right] \left(v(\tau)\right)  +\sum_{k=0}^K \check{\mathbb{I}} \left[ (\slashed{\nabla}_{R^\star})^k\check{\underline{\Pi}}_{\I}\right] \left(\tau\right) 
        \nonumber \\
                & \qquad \lesssim
  \sum_{k=0}^K\underline{\check{\mathbb{F}}}_{v(\tau)} [ (\slashed{\nabla}_{R^\star})^k\underline{\Pi}_{\Hp}]
+
\sum_{k=0}^K\check{\mathbb{F}}_{\tau} [ (\slashed{\nabla}_{R^\star})^k\check{\underline{\Pi}}_{\I}]\nonumber \\
& \qquad \ \ +   \sum_{k=0}^{K-1}  \underline{\check{\mathbb{F}}}_{v(\tau)} [ (\slashed{\nabla}_{R^\star})^k(\Omega^{-2} \underline{\Pi}_{\Hp})] +   \sum_{k=0}^{K-1}  \underline{\check{\mathbb{F}}}_{v(\tau)} [ (\slashed{\nabla}_{R^\star})^k(\Omega^{-4} \underline{A}_{\Hp})]
+\frac{\varepsilon_0^2 + \varepsilon^3}{\tau^{\min(2,N-K-2)}}\, . \nonumber
\end{align}
\end{proposition}

\begin{remark}
Note that we are asserting non-degenerate control near $r=3M$ as soon as one commutes at least once with $\slashed{\nabla}_{R^\star}$. In fact, the proof will establish the above estimate also for $K=0$ provided one replaces $\check{\mathbb{I}} \left[\underline{\Pi}_{\Hp}\right] \left(v(\tau)\right)$ by $\check{\mathbb{I}}^{deg} \left[\underline{\Pi}_{\Hp}\right] \left(v(\tau)\right)$ on the left hand side.
\end{remark}

\begin{proof}
Recall from Proposition \ref{prop:whythat} that $\check{\underline{\Pi}}_{\I}$ and $\underline{\Pi}_{\Hp}$ satisfy a tensorial wave equation of Type 1 with
\[
\mathcal{F}^{lin} \left[\underline{\Pi}_{\Hp} \right] =  \frac{3M}{r^2}\Omega^2 \underline{A}_{\Hp} - \frac{2}{r} \frac{\Omega^2}{r^2} \left(1-\frac{3M}{r}\right) \underline{\Psi}_{\Hp} \ \ \ , \ \ \ \mathcal{F}^{lin} \left[ \check{\underline{\Pi}}_{\I} \right] =  \frac{3M}{r^2}\Omega^2 \check{\underline{A}}_{\I} - \frac{2}{r} \frac{\Omega^2}{r^2} \left(1-\frac{3M}{r}\right)\check{ \underline{\Psi}}_{\I} \, .
\]
By a simple induction using Proposition \ref{prop:commutetype1} it is easy to see that $ (\slashed{\nabla}_{R^\star})^k \check{\underline{\Pi}}_{\I}$ and $ (\slashed{\nabla}_{R^\star})^k \underline{\Pi}_{\Hp}$ satisfy a tensorial wave equation of Type 1 and that the linear error term after $K$ commutations will have the form (cf.~(\ref{linef}) for notation)
\begin{align} \label{linefb}
\mathcal{F}^{lin} \left[  (\slashed{\nabla}_{R^\star})^{K} \check{\underline{\Pi}}_{\I} \right] = &\left(1-\frac{3M}{r}\right) \frac{\Omega^2}{r^3} \left( h_0 (\slashed{\nabla}_{R^\star})^K\check{\underline{\Psi}}_{\I} + h_0\Omega \slashed{\nabla}_3  (\slashed{\nabla}_{R^\star})^{K-1} \check{\underline{\Psi}}_{\I} \right) \\
&+ \sum_{k=0}^K \frac{h_0^k}{r^2} \Omega^2  (\slashed{\nabla}_{R^\star})^k \check{\underline{A}}_{\I} + \sum_{i=0}^1 \sum_{k=i}^{K-1} \frac{h_0^{k,i}}{r^3} \Omega^2 (\Omega \slashed{\nabla}_3 )^i (\slashed{\nabla}_{R^\star})^{k-i} \check{\underline{\Psi}}_{\I} + \sum_{k=0}^{K-1} \frac{h_0^k}{r^3}  (\slashed{\nabla}_{R^\star})^k \check{\underline{\Pi}}_{\I}  \nonumber
\end{align}
and with the same schematic form for $\mathcal{F}^{lin} \left[  (\slashed{\nabla}_{R^\star})^{K} \underline{\Pi}_{\Hp} \right]$. From Propositions \ref{prop:whythat} and \ref{prop:commutetype1} one sees that the non-linear error-term is of the form
\[
\mathcal{F}^{nlin} \left[  (\slashed{\nabla}_{R^\star})^{K} \check{\underline{\Pi}}_{\I} \right] =\check{\mathcal{E}}^{2+K}_{2} \ \ \ , \ \ \ \mathcal{F}^{nlin} \left[  (\slashed{\nabla}_{R^\star})^{K} \underline{\Pi}_{\Hp} \right] = \Omega^4 (\mathcal{E}^\star)^{K+2} \, .
\]

\vskip1pc
\noindent
{\bf Step 1.} We claim that applying the $T$-boundedness estimate and the Morawetz $X$-estimate of Proposition \ref{prop:basecase} to the wave equations for $ (\slashed{\nabla}_{R^\star})^k\underline{\Pi}_{\Hp}$ and $ (\slashed{\nabla}_{R^\star})^k \check{\underline{\Pi}}_{\I}$ already yields the desired estimate for all $0 \leq K \leq N-2$, however, 
\begin{itemize}
\item with the horizon fluxes carrying an additional diamond superscript (i.e.~weaker energies on the horizon) and 
\item with $\check{\mathbb{I}} \left[ (\slashed{\nabla}_{R^\star})^k \underline{\Pi}_{\Hp}\right] \left(v(\tau)\right)$ replaced by $\check{\mathbb{I}}^{\diamond,deg} \left[ (\slashed{\nabla}_{R^\star})^k \underline{\Pi}_{\Hp}\right] \left(v(\tau)\right)$ on the left. 
\end{itemize}
To verify this claim, we observe that using $T=\Omega \slashed{\nabla}_4 - R^\star$ the linear error $\mathcal{G}_1\left[ (\slashed{\nabla}_{R^\star})^{K} \underline{\Pi}\right] \left(\tau,u_f\right)$ can be estimated further by
\begin{align}
&\int_{\DcH(v(\tau))} |\mathcal{F}^{lin} \left[ (\slashed{\nabla}_{R^\star})^{K} \underline{\Pi}_{\Hp}\right]  || \slashed{\nabla}_T (\slashed{\nabla}_{R^\star})^{K} \underline{\Pi}_{\Hp}| + \int_{\DcI(\tau)} |\mathcal{F}^{lin} \left[ (\slashed{\nabla}_{R^\star})^{K} \Pi_{\I}\right] || \slashed{\nabla}_T  (\slashed{\nabla}_{R^\star})^{K} \check{\underline{\Pi}}_{\I}| \nonumber \\
&\lesssim \int_{\DcH(v(\tau))} \Big|\mathcal{F}^{lin} \left[ (\slashed{\nabla}_{R^\star})^{K} \underline{\Pi}_{\Hp}\right] \Big| \Big|  (\slashed{\nabla}_{R^\star})^{K+1} \underline{\Pi}_{\Hp} +  (\slashed{\nabla}_{R^\star})^{K} \left(\frac{\Omega^2}{r^2} \underline{\Psi}_{\Hp} \right)\Big|  \nonumber \\
& \qquad  +\int_{\DcI(\tau)} \Big| \mathcal{F}^{lin} \left[ (\slashed{\nabla}_{R^\star})^{K} \check{\underline{\Pi}}_{\I}\right]  \Big| \Big| (\slashed{\nabla}_{R^\star})^{K+1} \check{\underline{\Pi}}_{\I} +  (\slashed{\nabla}_{R^\star})^{K} \left(\frac{\Omega^2}{r^2} \check{\underline{\Psi}}_{\I} \right)\Big| +\frac{\varepsilon_0^2 + \varepsilon^3}{\tau^{\min(2,N-K-2)}} \nonumber \\
&\lesssim \gamma \int_{\DcH(v(\tau))} \frac{\Omega^2}{r^{2}} | (\slashed{\nabla}_{R^\star})^{K+1} \underline{\Pi}_{\Hp}|^2 + \gamma \int_{\DcI(\tau)} \frac{\Omega^2}{r^{2}} | (\slashed{\nabla}_{R^\star})^{K+1}\check{\underline{\Pi}}_{\I}|^2 + C_\gamma \left(\textrm{RHS of (\ref{goliab})} \right) \nonumber
\end{align}
for any $\gamma>0$, which is obtained in complete analogy to Step~1 below (\ref{linef}). In particular, we have used 
Cauchy--Schwarz and the fact that the lower oder terms in $\underline{A}_{\Hp}$ and $\underline{\Pi}_{\Hp}$ (appearing in $|\mathcal{F}^{lin} \left[  (\slashed{\nabla}_{R^\star})^{K} \underline{\Pi}\right] |^2$) from (\ref{linefb}) can be controlled from Proposition \ref{prop:lowordertpb}.

For the terms $\mathcal{G}_2\left[ (\slashed{\nabla}_{R^\star})^{K} \underline{\Pi}\right] \left(\tau,u_f\right)$ and $\mathcal{G}_3\left[ (\slashed{\nabla}_{R^\star})^{K} \underline{\Pi}\right] \left(\tau,u_f\right)$ we note similarly
\begin{align}
&\int_{\DcH(v(\tau))} |\mathcal{F}^{lin} \left[ (\slashed{\nabla}_{R^\star})^{K} \underline{\Pi}_{\Hp}\right]  R^\star  (\slashed{\nabla}_{R^\star})^{K} \underline{\Pi}_{\Hp}| + \int_{\DcI(\tau)} \mathcal{F}^{lin} |\left[ (\slashed{\nabla}_{R^\star})^{K} \check{\underline{\Pi}}_{\I}\right]  R^\star  (\slashed{\nabla}_{R^\star})^{K} \check{\underline{\Pi}}_{\I}| \nonumber \\
+&\int_{\DcH(v(\tau))} |\mathcal{F}^{lin} \left[ (\slashed{\nabla}_{R^\star})^{K} \underline{\Pi}_{\Hp}\right]  \frac{1}{r^{1+\delta}}   (\slashed{\nabla}_{R^\star})^k \underline{\Pi}_{\Hp}| + \int_{\DcI(\tau)} \mathcal{F}^{lin} \left[ (\slashed{\nabla}_{R^\star})^{K} \check{\underline{\Pi}}_{\I}\right] \frac{1}{r^{1+\delta}}   (\slashed{\nabla}_{R^\star})^k \check{\underline{\Pi}}_{\I}| \nonumber \\ 
 &\lesssim \gamma \int_{\DcH(v(\tau))} \frac{\Omega^2}{r^{2}} | (\slashed{\nabla}_{R^\star})^{K+1}\underline{\Pi}_{\Hp}|^2 + \gamma \int_{\DcI(\tau)} \frac{\Omega^2}{r^{2}} | (\slashed{\nabla}_{R^\star})^{K+1} \check{\underline{\Pi}}_{\I}|^2 + C_\gamma \left(\textrm{RHS of (\ref{goliab})} \right) \, .\nonumber
\end{align}
Choosing $\gamma$ sufficiently small (depending only on $M_{\rm init}$) we absorb the first terms on the right hand side of (\ref{Tiledb}) and the desired estimate (with the aforementioned restrictions) is proven provided we can control the non-linear errors arising from $\mathcal{H}_i$. For these we note that
using Propositions \ref{prop:wavehorizonerror1}, \ref{prop:wavehorizonerror2}  for the errors in the horizon region and Proposition \ref{prop:waveIerror2} for the errors in the infinity region we have
\begin{align} \label{nohe2b}
\sum_{i=1}^4 \sum_{k\leq K}  \mathcal{H}_i  \left[ (\slashed{\nabla}_{R^\star})^k \underline{\Pi}\right] \lesssim \frac{\varepsilon_0^2 + \varepsilon^3}{\tau^{\min(2,N-K-2)}}
\end{align}
and that using Proposition \ref{thm:cancelT} we have
\begin{align} 
\sum_{i=1}^3  \sum_{|\underline{k}|\leq K} \Big|   \mathcal{B}_i \left[ (\slashed{\nabla}_{R^\star})^k \underline{\Pi}_{\Hp}\right] (\tau,u_f)-  \mathcal{B}_i \ \left[ (\slashed{\nabla}_{R^\star})^k \check{\underline{\Pi}}_{\I}\right]   (\tau,u_f)\Big| \lesssim \frac{\varepsilon_0^2 + \varepsilon^3}{\tau^{\min(2,N-K-2)}} \, .
\end{align}

\vskip1pc
\noindent
{\bf Step 2.} To remove the diamond superscripts, we apply the redshift estimate of Proposition \ref{prop:redshift} to the tensorial wave equation satisfied by $ (\slashed{\nabla}_{R^\star})^k\underline{\Pi}_{\Hp}$ and $ (\slashed{\nabla}_{R^\star})^k\check{\underline{\Pi}}_{\I}$. The linear errors are easily seen to be controlled by Cauchy--Schwarz and the estimate from Step 1 and the non-linear error is controlled by (\ref{nohe2b}).

\vskip1pc
\noindent
{\bf Step 3.} We have proved (\ref{goliab}) except that  $\sum_{k=0}^K\check{\mathbb{I}}^{deg} \left[ (\slashed{\nabla}_{R^\star})^k \underline{\Pi}_{\Hp}\right] \left(v(\tau)\right)$ appears on the left hand side instead of  $\sum_{k=0}^K\check{\mathbb{I}} \left[ (\slashed{\nabla}_{R^\star})^k\underline{\Pi}_{\Hp}\right] \left(v(\tau)\right)$. To obtain the non-degenerate control we proceed successively for $K=1,2,..,N-2$. For $K=1$ we consider the tensorial wave equation for $\slashed{\nabla}_{R^\star} \underline{\Pi}_{\Hp}$ and integrate over $\DcH$ the Lagrangian identity (\ref{lagrangianid}) with $h$ a radial cut-off function equal to $1$ in $\left[3M-\frac{1}{4}M_{\rm init}, 3M+\frac{1}{4}M_{\rm init}\right]$ and vanishing for $r \leq \frac{5}{2}M_{\rm init}$ and $r \geq \frac{7}{2}M_{\rm init}$. Boundary terms and the term arising from $\mathfrak{F}^h\left[\slashed{\nabla}_{R^\star}\underline{\Pi}_{\Hp}\right]$ are easily seen to be controlled by the estimate already established. Also, the lower order terms in $\mathfrak{f}^h_{bulk} \left[\slashed{\nabla}_{R^\star} \underline{\Pi}_{\Hp}\right]$ are all directly controlled from $\sum_{k=0}^1\check{\mathbb{I}}^{deg} \left[ (\slashed{\nabla}_{R^\star})^k\underline{\Pi}_{\Hp}\right] \left(v(\tau)\right)$. The angular derivative term has a good sign and for the wrong signed term involving $\slashed{\nabla}_T (\slashed{\nabla}_{R^\star}) \underline{\Pi}_{\Hp}$ we observe
\begin{align}
\slashed{\nabla}_T \slashed{\nabla}_{R^\star} \underline{\Pi}_{\Hp} = (-\slashed{\nabla}_{R^\star} + \Omega \slashed{\nabla}_4)  (\slashed{\nabla}_{R^\star}) \underline{\Pi}_{\Hp} = -\slashed{\nabla}_{R^\star}  \slashed{\nabla}_{R^\star} \underline{\Pi}_{\Hp}- \frac{\Omega^2}{r^2}  (\slashed{\nabla}_{R^\star}) \underline{\Psi}_{\Hp} + \frac{\Omega^2}{r^3}h_0 (\slashed{\nabla}_{R^\star}) \underline{\Pi}_{\Hp} 
+ \Omega^4 (\mathcal{E}^\star)^{2} \nonumber
\end{align}
with all terms on the right already controlled \emph{non-degenerately} near $r=3M$. This gives non-degenerate (near $r=3M$) control on all first derivatives of $\slashed{\nabla}_{R^\star} \underline{\Pi}$ and hence control on $\check{\mathbb{I}} \left[\slashed{\nabla}_{R^\star}\underline{\Pi}_{\Hp}\right] \left(v(\tau)\right)$. By a simple interpolation one also has $\check{\mathbb{I}} \left[\underline{\Pi}_{\Hp}\right] \left(v(\tau)\right) \lesssim \check{\mathbb{I}} \left[\slashed{\nabla}_{R^\star}\underline{\Pi}_{\Hp}\right] \left(v(\tau)\right) + \check{\mathbb{I}}^{deg} \left[\underline{\Pi}_{\Hp}\right] \left(v(\tau)\right)$. This finishes the proof for $K=1$. For $K=2,3,..., N-2$ we apply successively the same argument at each $K$ using the Lagrangian multiplier for the wave equation satisfied by $ (\slashed{\nabla}_{R^\star})^K \underline{\Pi}_{\Hp}$.
\end{proof}

\subsection{Basic estimates for $\protect\underline A$ and $ (\slashed{\nabla}_{R^\star})^k \protect\underline{A}$}
Completely analogously we prove the above estimate for $\underline{A}_{\Hp}$ and $\check{\underline{A}}_{\I}$:

\begin{proposition} \label{prop:arstarb}
For any $u_1 \leq \tau \leq u_f$ and $1 \leq K\leq N-1$ the pair $\left(\underline{A}_{\Hp},
\check{\underline{A}}_{\I}\right)$ satisfies
\begin{align}
&\sum_{k=0}^K \sup_{\tau \leq u \leq u_f} \check{\underline{\mathbb{F}}}_{v(u)} [ (\slashed{\nabla}_{R^\star})^k \underline{A}_{\Hp}] + \sum_{k=0}^K \sup_{\tau \leq u \leq u_f} \check{\mathbb{F}}_u [ (\slashed{\nabla}_{R^\star})^k \check{\underline{A}}_{\I}] 
 \\
 &+ \sum_{k=0}^K \check{{\mathbb{F}}}_{u_f} \left[ (\slashed{\nabla}_{R^\star})^k \underline{A}_{\Hp} \right] \left(v(\tau)\right)+ \sum_{k=0}^K \check{\underline{\mathbb{F}}}_{v_\infty} \left[ (\slashed{\nabla}_{R^\star})^k \check{\underline{A}}_{\I} \right] \left(\tau\right) \nonumber \\
&+\sum_{k=0}^K \check{\mathbb{I}} \left[ (\slashed{\nabla}_{R^\star})^k \underline{A}_{\Hp}\right] \left(v(\tau)\right)  +\sum_{k=0}^K \check{\mathbb{I}} \left[ (\slashed{\nabla}_{R^\star})^k \check{\underline{A}}_{\I}\right] \left(\tau\right) 
        \nonumber \\
                & \qquad \lesssim
  \sum_{k=0}^K\underline{\check{\mathbb{F}}}_{v(\tau)} [ (\slashed{\nabla}_{R^\star})^k \underline{A}_{\Hp}]
+
\sum_{k=0}^K\check{\mathbb{F}}_{\tau} [ (\slashed{\nabla}_{R^\star})^k \check{\underline{A}}_{\I}]
+ \sum_{k=0}^{K-1}\underline{\check{\mathbb{F}}}_{v(\tau)} [ (\slashed{\nabla}_{R^\star})^k \underline{\Pi}_{\Hp}]
+
\sum_{k=0}^{K-1}\check{\mathbb{F}}_{\tau} [ (\slashed{\nabla}_{R^\star})^k \check{\underline{\Pi}}_{\I}]\nonumber \\
& \qquad \ \ +   \sum_{k=0}^{\max(0,K-2)}  \underline{\check{\mathbb{F}}}_{v(\tau)} [ (\slashed{\nabla}_{R^\star})^k(\Omega^{-2} \underline{\Pi}_{\Hp})] +   \sum_{k=0}^{\max(0,K-2)}  \underline{\check{\mathbb{F}}}_{v(\tau)} [ (\slashed{\nabla}_{R^\star})^k(\Omega^{-4} \underline{A}_{\Hp})]
+\frac{C(\epsilon_0)^2 + C^2 \epsilon^4}{\tau^{\kappa(N-K-2)}} \nonumber
\end{align}
\end{proposition}

\begin{proof}
Recall from Proposition \ref{prop:whythat} that $\underline{A}_{\Hp}$ and $\check{\underline{A}}_{\I}$ satisfy a tensorial wave equation of Type 2 with 
\begin{align}
\mathcal{F}^{lin}\left[\underline{A}_{\Hp}\right] = +\frac{8}{r} \frac{\Omega^2}{r^2} \left(1-\frac{3M}{r}\right)\underline{\Pi}_{\Hp} \ \ \ , \ \ \  \mathcal{F}^{lin}\left[\check{\underline{A}}_{\I}\right] = +\frac{8}{r} \frac{\Omega^2}{r^2} \left(1-\frac{3M}{r}\right)  \check{\underline{\Pi}}_{\I} \, .
\end{align}
Repeating the proof of Proposition \ref{prop:pirstarb}, this linear error term and its $R^\star$ commuted analogues are easily controlled using Cauchy--Schwarz and the estimates from Proposition \ref{prop:pirstarb}.\footnote{Note in particular that we obtain $K-1$ in the sums involving $\underline{\Pi}$ on the right hand side. Indeed, for $K=2$ (commuting twice), say, we need to control \emph{two} derivatives of $\underline{\Pi}$ in the linear error which follow from (the spacetime terms of) $K=1$ in Proposition \ref{prop:pirstarb}. Similarly for higher $K$.} The only additional difficulty is concerned with the anomalous non-linear error-term appearing in the Teukolsky equation for $\check{\underline{A}}_{\I}$, see Proposition \ref{prop:whythat} or (\ref{teueli2b}). This corresponds to an additional error term in $\sum_{i=1}^2 \mathcal{H}_i\left[ (\slashed{\nabla}_{R^\star})^k \underline{A}\right]$ of the form:
\begin{align}
\int_{\DcI(\tau)} \Big| (\slashed{\nabla}_{R^\star})^k \left( \frac{1}{r^3} r \Dslash_2^* \left(
		r \nablaslash (r^2 \Omega \tr \chi)
		\cdot r\alphabar \right)\right) \Big| \left(|\slashed{\nabla}_T (\slashed{\nabla}_{R^\star})^k \check{\underline{A}}_{\I}| +| (\slashed{\nabla}_{R^\star})^{k+1} \check{\underline{A}}_{\I}| + \frac{1}{r^{1+\delta}}| (\slashed{\nabla}_{R^\star})^k \check{\underline{A}}_{\I}|\right) \nonumber
\end{align}
for $k=0, ..., N-1$. We can now apply Cauchy--Schwarz (borrowing from the term $\check{\mathbb{I}} \left[ (\slashed{\nabla}_{R^\star})^k \check{\underline{A}}_{\I}\right] \left(\tau\right)$ on the left) reducing the problem to establish the estimate
\begin{align} \label{discussionano}
\sum_{k=0}^{N-1}\int_{\DcI(\tau)} r^2 \Big| (\slashed{\nabla}_{R^\star})^k \left( \frac{1}{r^3} r \Dslash_2^* \left(
		r \nablaslash (r^2 \Omega \tr \chi)
		\cdot r\alphabar \right)\right)  \Big|^2 \lesssim \frac{\varepsilon_0^2 + \varepsilon^3}{\tau^{\min(2,N-K-2)}} \, .
\end{align}
Writing $2\slashed{\nabla}_{R^\star}= -\Omega \slashed{\nabla}_3 + \Omega \slashed{\nabla}_4$ and commuting the derivatives through using the null structure equations for $\Omega tr \chi - (\Omega tr \chi)_\circ$ provides the desired estimate except for the term involving $\left[\Omega \slashed{\nabla}_3\right]^{N-1} \Dslash_2^* \left(
r \nablaslash r^2 \Omega \tr \chi \right)$ which, ignoring terms involving  only $N$ derivatives which are easily controlled,
requires controlling the term $\frac{1}{r^3} r \underline{\alpha} \left[\Omega \slashed{\nabla}_3\right]^{N-3} r^4 \Dslash_2^* 
		 \nablaslash \slashed{div} \slashed{\nabla}  r(\underline{\omega} - \underline{\omega}_\circ)$. However using the bootstrap assumption on the energy defined in (\ref{angbon}) and $|r \underline{\alpha}|^2 \lesssim \frac{\epsilon^2}{\tau^2}$ the estimate also follows for this term.
\end{proof}

\subsection{Integrated local energy decay for all derivatives of $\protect\underline{\Pi}$ and~$\protect\underline{A}$} \label{sec:susy1b}

We next conclude that Propositions \ref{prop:pirstarb} and \ref{prop:arstarb} already provide control over \emph{all} derivatives of $\underline{\Pi}_{\Hp}$ and $\check{\underline{\Pi}}_{\I}$ away from the horizon and infinity (note the $\star$ in the energies below). We start with the integrated decay energies.

\begin{proposition} \label{prop:piallb}
For any $u_1 \leq \tau \leq u_f$ and $1\leq K\leq N-2$, the pair $\left(\underline{\Pi}_{\Hp}=r^3 \Omega \underline{\psi}_{\Hp},
\check{\underline{\Pi}}_{\I}=r^3 \Omega \check{\underline{\psi}}_{\I}\right)$ satisfies
\begin{align} \label{golib}
&\sum_{|\underline{k}|\leq K+1} \check{\mathbb{I}}^{\star, deg} \left[\tilde{\mathfrak{D}}^{\underline{k}} \underline{A}_{\Hp}\right] \left(v(\tau)\right)  +\sum_{|\underline{k}|\leq K+1} \check{\mathbb{I}}^\star\left[\tilde{\mathfrak{D}}^{\underline{k}} \check{\underline{A}}_{\I}\right] \left(\tau\right) +\sum_{|\underline{k}|\leq K} \check{\mathbb{I}}^\star \left[\tilde{\mathfrak{D}}^{\underline{k}}\check{\underline{\Pi}}_{\I}\right] \left(\tau\right) 
         \\
                & \qquad \lesssim
                 \sum_{k=0}^{K+1}\underline{\check{\mathbb{F}}}_{v(\tau)} [ (\slashed{\nabla}_{R^\star})^k \underline{A}_{\Hp}]
+
\sum_{k=0}^{K+1}\check{\mathbb{F}}_{\tau} [ (\slashed{\nabla}_{R^\star})^k \check{\underline{A}}_{\I}]
+ \sum_{k=0}^{K}\underline{\check{\mathbb{F}}}_{v(\tau)} [ (\slashed{\nabla}_{R^\star})^k \underline{\Pi}_{\Hp}]
+
\sum_{k=0}^{K}\check{\mathbb{F}}_{\tau} [ (\slashed{\nabla}_{R^\star})^k \check{\underline{\Pi}}_{\I}]\nonumber \\
& \qquad +   \sum_{k=0}^{K-1}  \underline{\check{\mathbb{F}}}_{v(\tau)} [ (\slashed{\nabla}_{R^\star})^k(\Omega^{-2} \underline{\Pi}_{\Hp})] +   \sum_{k=0}^{K-1}  \underline{\check{\mathbb{F}}}_{v(\tau)} [ (\slashed{\nabla}_{R^\star})^k(\Omega^{-4} \underline{A}_{\Hp})]
+\frac{\varepsilon_0^2 + \varepsilon^3}{\tau^{\min(2,N-K-2)}} \, . 
 \nonumber
 \end{align}
\end{proposition}

\begin{proof}
We first prove
\begin{align} \label{ghjb}
\sum_{|\underline{k}|\leq K} \check{\mathbb{I}}^\star \left[\tilde{\mathfrak{D}}^{\underline{k}}\check{\underline{\Pi}}_{\I}\right] \left(v(\tau)\right)
\lesssim
\sum_{k \leq K} \check{\mathbb{I}} \left[ (\slashed{\nabla}_{R^\star})^k \check{\underline{\Pi}}_{\I}\right] \left(v(\tau)\right) 
+\sum_{k \leq K-1} \check{\mathbb{I}} \left[ (\slashed{\nabla}_{R^\star})^k \check{\underline{A}}_{\I}\right] \left(v(\tau)\right) 
+\frac{\varepsilon_0^2 + \varepsilon^3}{\tau^{\min(2,N-K-2)}}, 
\end{align}
whose right hand side we can estimate by the right hand side of (\ref{golib}) by applying Propositions~\ref{prop:pirstarb} 
and~\ref{prop:arstarb}.
Fixing $K\geq 1$ we look at
\begin{align} \label{hgyb}
 \tilde{\mathfrak{D}}^{(k_1,k_2,k_3)} \check{\underline{\Pi}}_{\I}  \ \ \ \textrm{for $k_1+k_2+k_3=K+1$} \, .
\end{align}
We first use (\ref{recdpib}) for the $\left(r \Omega \slashed{\nabla}_4
\right)^{k_3} \check{\underline{\Pi}}_{\I}$ part in conjunction with the estimates of Theorem \ref{thm:PPbarestimates} for the terms involving $\check{\underline{\Psi}}_{\I}$ to obtain
\[
\int_{\DcI(\tau) \cap \{r \leq 2R\}} | \tilde{\mathfrak{D}}^{(k_1,k_2,k_3)} \check{\underline{\Pi}}_{\I}|^2 \lesssim \int_{\DcI(\tau) \cap \{r \leq 2R\}} | \tilde{\mathfrak{D}}^{(k_1,k_2,0)} \check{\underline{\Pi}}_{\I}|^2 + \frac{\varepsilon_0^2 + \varepsilon^3}{\tau^{\min(2,N-K-2)}} .
\]
Similarly, using the relation $\Omega \slashed{\nabla}_3 = \Omega \slashed{\nabla}_4 -2\slashed{\nabla}_{R^\star}$ and commuting the $3$-derivative through we find 
\[
\int_{\DcI(\tau) \cap \{r \leq 2R\}} | \tilde{\mathfrak{D}}^{(k_1,k_2,k_3)} \check{\underline{\Pi}}_{\I}|^2 \lesssim  \sum_{i=0}^{k_2} \int_{\DcI(\tau) \cap \{r \leq 2R\}} | \tilde{\mathfrak{D}}^{(k_1,0,0)}  (\slashed{\nabla}_{R^\star})^i \check{\underline{\Pi}}_{\I}|^2 +\frac{\varepsilon_0^2 + \varepsilon^3}{\tau^{\min(2,N-K-2)}}.
\]
Now if $k_1=1$ the desired estimate follows directly from Proposition \ref{prop:pirstarb}. If $k_1 \geq 2$ we commute all $r^2\slashed{\Delta}$ operators through on $\check{\underline{\Pi}}_{\I}$ using repeatedly the relations
(\ref{teueli2b}) and (\ref{wep2b}) in the form\footnote{Note we have replaced $\check{\mathcal{E}}^1_1 + \frac{k_1}{r} r \Dslash_2^* \left(
		r \nablaslash r^2 \Omega \tr \chi
		\cdot r\alphabar
		\right) = \check{\mathcal{E}}^2_1$ as we do not need to keep track of the anomalous term when inserting elliptic relations.}
\begin{align} \label{auxiib}
r^2 \slashed{\Delta} \check{\underline{\Pi}}_{\I}  = -\Omega \slashed{\nabla}_3 \check{\underline{\Psi}}_{\I}  +4\check{\underline{\Pi}}_{\I} - \frac{6M}{r}\check{\underline{\Pi}}_{\I}  - 3M \check{\underline{A}}_{\I}  + \check{\mathcal{E}}_1^2
\end{align}
\begin{align} \label{auxiib2}
 r^2 \slashed{\Delta} \check{\underline{A}}_{\I}    = + 2\check{\underline{\Pi}}_{\I} + \frac{6M}{r}  \check{\underline{A}}_{\I} + 
 \left(-\frac{\Omega^2}{r^2}\check{\underline{\Psi}}_{\I} + 2 \slashed{\nabla}_{R^\star} \check{\underline{\Pi}}_{\I}\right) - \frac{4}{r}\left(1- \frac{3M}{r}\right) \check{\underline{\Pi}}_{\I}  + \check{\mathcal{E}}^2_1
\end{align}
to conclude (using Proposition \ref{prop:waveIerror1b} for the non-linear errors)
\begin{align}
\int_{\DcI(\tau) \cap \{r \leq 2R\}}| \tilde{\mathfrak{D}}^{(k_1,k_2,k_3)} \check{\underline{\Pi}}_{\I}|^2 \lesssim  \textrm{RHS of (\ref{ghjb})} \, .
\end{align}
This proves (\ref{ghjb}). Next we show
\begin{align} 
\sum_{|\underline{k}|\leq K+1} \check{\mathbb{I}}^\star \left[\tilde{\mathfrak{D}}^{\underline{k}}\check{\underline{A}}_{\I}\right] \left(\tau\right) \nonumber
\lesssim
\sum_{k \leq K} \check{\mathbb{I}} \left[ (\slashed{\nabla}_{R^\star})^k \check{\underline{\Pi}}_{\I}\right] \left(\tau \right) 
+\sum_{k \leq K+1} \check{\mathbb{I}} \left[ (\slashed{\nabla}_{R^\star})^k \check{\underline{A}}_{\I}\right] \left(\tau\right) 
+\frac{\varepsilon_0^2 + \varepsilon^3}{\tau^{\min(2,N-K-2)}} . 
\end{align}
The proof of this is entirely analogous to the proof of 
(\ref{ghjb}) (note that we can freely use (\ref{ghjb}) now) and is therefore left to the reader. Finally, the estimate 
in $\DcH$ is carried out entirely analogously (see the proof of Proposition \ref{prop:piall} where it is done explicitly). 
\end{proof}

 The following corollaries easily follow from the mean value theorem:
 
 \begin{corollary}\label{cor:Tintermezzobaverage}
 There exists $R_1 \leq \tilde{R}^\prime \leq R_2$ such that the right hand side of Proposition \ref{prop:piallb} controls in particular $\sum_{|{\underline{k}} | \leq K+1} \int_{\DcI(\tau) \cap \{r_{\I}=\tilde{R}^\prime\}} | \tilde{\mathfrak{D}}^{\underline{k}}  \check{\underline{A}}_{\I}|^2$.
\end{corollary}

\begin{corollary} \label{cor:bndhozr}
Suppose $r_{\mathcal{H}}(u_f,v(u_1)) \leq 9M_{\rm init}/4$. Then there exists a $9M_{\rm init}/4<\mathring{R}<5M_{\rm init}/2$ such that
the right hand side of Proposition \ref{prop:piallb}  controls in particular $\sum_{|{\underline{k}} | \leq K+1} \int_{\DcH(v(\tau)) \cap \{ r_{\Hp}=\mathring{R}\}} | \tilde{\mathfrak{D}}^{\underline{k}}  \underline{A}_{\Hp}|^2$.
\end{corollary}

\subsection{Basic fluxes for all derivatives of  $\protect\underline{\Pi}$ and~$\protect\underline{A}$} \label{sec:susy1b2}

We now turn to estimating general fluxes.
\begin{proposition} \label{prop:piallb2}
For any $u_1 \leq \tau \leq u_f$ and $1\leq K\leq N-2$ the pair $\left(\underline{\Pi}_{\Hp}=r^3 \Omega \underline{\psi}_{\Hp},
\check{\underline{\Pi}}_{\I}=r^3 \Omega \check{\underline{\psi}}_{\I}\right)$ satisfies
\begin{align}
&
\sum_{|\underline{k}|\leq K+1} \sup_{\tau \leq u \leq u_f} \check{\underline{\mathbb{F}}}^\star_{v(u)} [\tilde{\mathfrak{D}}^{\underline{k}} \underline{A}_{\Hp}] 
  +\sum_{|\underline{k}|\leq K} \sup_{\tau \leq u \leq u_f} \check{\mathbb{F}}^\star_u [\tilde{\mathfrak{D}}^{\underline{k}}\check{\underline{\Pi}}_{\I}] +\sum_{|\underline{k}|\leq K+1} \sup_{\tau \leq u \leq u_f} \check{\mathbb{F}}^\star_u [\tilde{\mathfrak{D}}^{\underline{k}} \check{\underline{A}}_{\I}]
        \nonumber \\
                & \qquad \lesssim
                 \sum_{k=0}^{K+1}\underline{\check{\mathbb{F}}}_{v(\tau)} [ (\slashed{\nabla}_{R^\star})^k \underline{A}_{\Hp}]
+
\sum_{k=0}^{K+1}\check{\mathbb{F}}_{\tau} [ (\slashed{\nabla}_{R^\star})^k \check{\underline{A}}_{\I}]
+ \sum_{k=0}^{K}\underline{\check{\mathbb{F}}}_{v(\tau)} [ (\slashed{\nabla}_{R^\star})^k \underline{\Pi}_{\Hp}]
+
\sum_{k=0}^{K}\check{\mathbb{F}}_{\tau} [ (\slashed{\nabla}_{R^\star})^k \check{\underline{\Pi}}_{\I}]\nonumber \\
& \qquad +   \sum_{k=0}^{K-1}  \underline{\check{\mathbb{F}}}_{v(\tau)} [ (\slashed{\nabla}_{R^\star})^k(\Omega^{-2} \underline{\Pi}_{\Hp})] +   \sum_{k=0}^{K-1}  \underline{\check{\mathbb{F}}}_{v(\tau)} [ (\slashed{\nabla}_{R^\star})^k(\Omega^{-4} \underline{A}_{\Hp})]
+\frac{\varepsilon_0^2 + \varepsilon^3}{\tau^{\min(2,N-K-2)}} \, . 
 \end{align}

\end{proposition}

\begin{remark}
We omit the proof as it is entirely analogous (but easier) than that of Proposition \ref{prop:piall}. It is easier because here we are restricting the energies to be both away from (what will be) the horizon and from $r \geq 2R$. Hence $r$-weights and $\Omega$-weights are irrelevant in all regions under consideration.
\end{remark}

\subsection{Improving the weights near infinity} \label{sec:piabinfty}
We now obtain $r$-weighted fluxes and integrated decay estimates in $\mathcal{D}^{\I}$.
\begin{proposition} \label{prop:piallbextra}
For any $u_1 \leq \tau \leq u_f$ and $1\leq K\leq N-2$ we have 
\begin{align} \label{golib2}
& \ \   \sum_{|\underline{k}|=0}^{K+1} \sup_{\tau \leq u \leq u_f} \int_{\CcI_u \cap \{r\geq R+2\}} \frac{1}{r^2} |\tilde{\mathfrak{D}}^{\underline{k}} \check{\underline{\Pi}}_{\I} |^2  +\sum_{|\underline{k}|=0}^{K}\sup_{\tau \leq u \leq u_f} \int_{\CcI_u \cap \{r\geq R+2\}}  \frac{1}{r^2} |\tilde{\mathfrak{D}}^{\underline{k}}\check{\underline{\Psi}}_{\I} |^2  \\
+ &\sum_{|\underline{k}|=0; k_3 \neq |\underline{k}|}^{K+2} \int_{\CbcI_{v_\infty}(\tau)} |\tilde{\mathfrak{D}}^{\underline{k}} \check{\underline{A}}_{\I}|^2 + \sum_{|\underline{k}|=0; k_3\neq |\underline{k}|}^{K+1}\int_{\CbcI_{v_\infty}(\tau)} |\tilde{\mathfrak{D}}^{\underline{k}} \check{\underline{\Pi}}_{\I}|^2 \nonumber \\
+ & \ \ \sum_{|\underline{k}|=0}^{K+1}\int_{\DcI(\tau)} \frac{1}{r^{1+\delta}} |\tilde{\mathfrak{D}}^{\underline{k}} \check{\underline{\Pi}}_{\I}|^2 + \sum_{|\underline{k}|=0}^{K+2}\int_{\DcI(\tau)} \frac{1}{r^{1+\delta}} |\tilde{\mathfrak{D}}^{\underline{k}} \check{\underline{A}}_{\I}|^2  \nonumber 
       \\
   &\qquad  \lesssim \int_{\CcI_{\tau}} \frac{1}{r^2} \Bigg\{ \sum_{|\underline{k}|=0; k_2\neq |k|}^{K+2}  | \mathfrak{D}^{\underline{k}} \check{\underline{A}}_{\I}|^2 + \sum_{|\underline{k}|=0}^{K+1}   | \mathfrak{D}^{\underline{k}} \check{\underline{\Pi}}_{\I}|^2  + \sum_{k=0}^{K} |  (\slashed{\nabla}_{R^\star})^k \check{\underline{\Psi}}_{\I}|^2 \Bigg\} \nonumber \\ 
 & \qquad + \int_{\CbcH_{v(\tau)}}  \Omega^2 \sum_{|\underline{k}|=0; k_3\neq |k|}^{K+2} | \mathfrak{D}^{\underline{k}} (\Omega^{-4} \underline{A}_{\Hp})|^2 
+ \frac{\varepsilon_0^2 + \varepsilon^3}{\tau^{\min(2,N-K-2)}} \, . \nonumber
 \end{align}
 Moreover, the restriction $k_3 \neq |\underline{k}|$ can be changed to $k_3 \neq K+2$ in the third sum and to $k_3 \neq K+1$ in the fourth sum.
\end{proposition}

\begin{proof}
Observe that the statement of the restriction in the sums is a simple a posteriori consequence of~\eqref{golib2} and elliptic estimates along the cones using that the tangential derivatives are always included. 

\vskip1pc
\noindent
{\bf Step 0. Simplifying what is to show.} The right hand side of (\ref{golib2}) does control the right hand side of Propositions \ref{prop:piallb} and \ref{prop:piallb2}, hence we are free to use these estimates and Corollary \ref{cor:Tintermezzobaverage}. We also note that it suffices to prove the estimate restricting all sums on the left to tuples $\left(k_1,k_2,0\right)$. Indeed this is clear for the term involving $\check{\underline{\Psi}}_{\I}$ from Theorem \ref{thm:PPbarestimates}. Next, for the expressions involving $\tilde{\mathfrak{D}}^{\underline{k}} \check{\underline{\Pi}}_{\I}$, note that if $\left(k_1,k_2,k_3\right)$ with $k_3 \neq 0$ then we can insert the relation (\ref{recdpib}) which turns the flux into a flux for $\check{\underline{\Psi}}_{\I}$ that has already been controlled in Theorem \ref{thm:PPbarestimates} by the right hand side of (\ref{golib2}). Finally, for the expressions involving $\tilde{\mathfrak{D}}^{\underline{k}} \check{\underline{A}}_{\I}$ note that if $\left(k_1,k_2,k_3\right)$ with $k_3 \neq 0$, then we can insert the relation (\ref{recdab}) and the flux turns into a flux for $\check{\underline{\Pi}}_{\I}$ that has just been controlled.

\vskip1pc
\noindent
{\bf Step 1. Collecting the commuted equations.}
Key to the proof will be the Bianchi pair (\ref{teueli2b}) and (\ref{recdab})
\begin{align} 
 \Omega \slashed{\nabla}_3 \check{\underline{\Pi}}_{\I}  &= -r^2 \slashed{\mathcal{D}}_2^\star \slashed{div} \check{\underline{A}}_{\I} - \frac{3M}{r} \check{\underline{A}}_{\I} +\frac{2}{r}  \left(1- \frac{3M}{r}\right) \check{\underline{\Pi}}_{\I} + \check{\mathcal{E}}^1_1  + \frac{k_1}{r} r \Dslash_2^* \left(
		r \nablaslash r^2 \Omega \tr \chi
		\cdot r\alphabar
		\right) \, , 
 \label{aresc} \\
 \Omega \slashed{\nabla}_4 r\slashed{div} \check{\underline{A}}_{\I} &= 2\frac{\Omega^2}{r^2} r\slashed{div} \check{\underline{\Pi}}_{\I}  + \check{\mathcal{E}}^1_2\, , 
 \end{align}
as well as the pair (\ref{wep2b}) and (\ref{recdpib})
\begin{align}
\Omega \slashed{\nabla}_3 \check{\underline{\Psi}}_{\I} &=  2 r^2\slashed{\mathcal{D}}_2^\star \slashed{div} \check{\underline{\Pi}}_{\I} + 2 \check{\underline{\Pi}}_{\I} -\frac{6M}{r} \check{\underline{\Pi}}_{\I} - 3M \check{\underline{A}}_{\I} + \check{\mathcal{E}}^2_1 \, ,   \\
 \Omega \slashed{\nabla}_4 r\slashed{div} \check{\underline{\Pi}}_{\I} &= -\frac{\Omega^2}{r^2} r\slashed{div} \check{\underline{\Psi}}_{\I} + \check{\mathcal{E}}^2_2 \, .
\end{align}
We first derive commuted versions of these equations, for which we recall the notation (\ref{commdef}). For the first pair we have for $k \geq 1$ and odd, $l \geq 0$
\begin{align} \label{pia3ang}
 \Omega \slashed{\nabla}_3 (\Omega \slashed{\nabla}_3)^l \check{\underline{\Pi}}^{(k-1)}_{\I}  &= -r \slashed{\mathcal{D}}_2^\star (\Omega \slashed{\nabla}_3)^l  \check{\underline{A}}^{(k)}_{\I} + \sum_{i=0}^{l} \frac{h^i_{0}}{r} (\Omega \slashed{\nabla}_3 )^i \check{\underline{A}}^{(k-1)}_{\I} + \sum_{i=0}^{l} \frac{h^i_{0}}{r} (\Omega \slashed{\nabla}_3)^i \check{\underline{\Pi}}^{(k-1)}_{\I}  + \check{\mathcal{E}}^{k+l}_1 \\
 & \qquad \qquad \qquad \qquad \qquad \qquad +  (\Omega \slashed{\nabla}_3)^l\left(r^2 \slashed{\mathcal{D}}_2^\star \slashed{div}\right)^{\frac{k-1}{2}} \frac{h_0}{r} r \Dslash_2^* \left(
		r \nablaslash r^2 \Omega \tr \chi
		\cdot r\alphabar
		\right) \, , \nonumber \\
 \Omega \slashed{\nabla}_4 \left( (\Omega \slashed{\nabla}_3)^l \check{\underline{A}}^{(k)}_{\I}\right) &= 2\frac{\Omega^2}{r^2} r\slashed{div}  (\Omega \slashed{\nabla}_3)^l \check{\underline{\Pi}}^{(k-1)}_{\I}  + \check{\mathcal{E}}^{k+l}_2\nonumber  \, ,
\end{align}
while for $k \geq 1$ and even, $l \geq 0$
\begin{align}  \label{pia3ang2}
 \Omega \slashed{\nabla}_3 (\Omega \slashed{\nabla}_3)^l \check{\underline{\Pi}}^{(k-1)}_{\I}  &= -r \slashed{div} (\Omega \slashed{\nabla}_3)^l  \check{\underline{A}}^{(k)}_{\I} + \sum_{i=0}^{l} \frac{h^i_{0}}{r} (\Omega \slashed{\nabla}_3 )^i \check{\underline{A}}^{(k-1)}_{\I} + \sum_{i=0}^{l} \frac{h^i_{0}}{r} (\Omega \slashed{\nabla}_3)^i \check{\underline{\Pi}}^{(k-1)}_{\I}  +\check{\mathcal{E}}^{k+l}_1 \\
 & \qquad \qquad \qquad \qquad \qquad \qquad+  (\Omega \slashed{\nabla}_3)^l\left(r^2 \slashed{div}\slashed{\mathcal{D}}_2^\star \right)^{\frac{k}{2}} \frac{h_0}{r}  \left(
		r \nablaslash r^2 \Omega \tr \chi
		\cdot r\alphabar
		\right) \, , \nonumber
  \\
 \Omega \slashed{\nabla}_4 \left( (\Omega \slashed{\nabla}_3)^l \check{\underline{A}}^{(k)}_{\I}\right) &= 2\frac{\Omega^2}{r^2} r\slashed{\mathcal{D}}_2^\star (\Omega \slashed{\nabla}_3)^l \check{\underline{\Pi}}^{(k-1)}_{\I}  + \check{\mathcal{E}}^{k+l}_2   \, .
\end{align}
For the second pair we have for $k\geq 1$ and $k$ odd, $l \geq 0$
\begin{align} \label{psipi3ang}
\Omega \slashed{\nabla}_3  (\Omega \slashed{\nabla}_3)^l \check{\underline{\Psi}}^{(k-1)}_{\I} &=  2 r\slashed{\mathcal{D}}_2^\star  (\Omega \slashed{\nabla}_3)^l  \check{\underline{\Pi}}^{(k)}_{\I} + 2  (\Omega \slashed{\nabla}_3)^l \check{\underline{\Pi}}^{(k-1)}_{\I} + \sum_{i=0}^l h_1^i  (\Omega \slashed{\nabla}_3)^l  \check{\underline{\Pi}}^{(k-1)}_{\I} - 3M  (\Omega \slashed{\nabla}_3)^l  \check{\underline{A}}^{(k-1)}_{\I} + \check{\mathcal{E}}^{k+1}_1 \, , \nonumber  \\
 \Omega \slashed{\nabla}_4  (\Omega \slashed{\nabla}_3)^l  \check{\underline{\Pi}}^{(k)}_{\I} &= -\frac{\Omega^2}{r^2} r\slashed{div}  (\Omega \slashed{\nabla}_3)^l  \check{\underline{\Psi}}^{(k-1)}_{\I} + \check{\mathcal{E}}^{k+1}_2 \, ,
\end{align}
while for $k \geq 1$ and even, $l \geq 0$
\begin{align}  \label{psipi3ang2}
\Omega \slashed{\nabla}_3  (\Omega \slashed{\nabla}_3)^l \check{\underline{\Psi}}^{(k-1)}_{\I} &=  2 r\slashed{div} (\Omega \slashed{\nabla}_3)^l  \check{\underline{\Pi}}^{(k)}_{\I} + 2  (\Omega \slashed{\nabla}_3)^l \check{\underline{\Pi}}^{(k-1)}_{\I} + \sum_{i=0}^l h_1^i  (\Omega \slashed{\nabla}_3)^l  \check{\underline{\Pi}}^{(k-1)}_{\I} - 3M  (\Omega \slashed{\nabla}_3)^l  \check{\underline{A}}^{(k-1)}_{\I} + \check{\mathcal{E}}^{k+1}_1, \nonumber  \\
 \Omega \slashed{\nabla}_4  (\Omega \slashed{\nabla}_3)^l  \check{\underline{\Pi}}^{(k)}_{\I} &= -\frac{\Omega^2}{r^2} r\slashed{\mathcal{D}}_2^\star  (\Omega \slashed{\nabla}_3)^l  \check{\underline{\Psi}}^{(k-1)}_{\I} + \check{\mathcal{E}}^{k+1}_2 \, .
\end{align}

\vskip1pc
\noindent
{\bf Step 2. Deriving the estimates for $l=0$.} We first set $l=0$ above to prove the estimate (\ref{golib2}) with the restriction that only angular derivatives can appear in $\tilde{\mathfrak{D}}^{\underline{k}}$. We proceed inductively in $k$. We contract 
the first equation of (\ref{pia3ang}) (and (\ref{pia3ang2}) respectively) with $\frac{2}{r^2} \left(1+M^\delta r^{-\delta}\right) \check{\underline{\Pi}}^{(k-1)}_{\I}$ and the second equation of (\ref{pia3ang}) with $\left(1+M^\delta r^{-\delta}\right) \check{\underline{A}}^{(k-1)}_{\I}$. We then add the relevant equations and integrate over $\DcI(\tau) \setminus  \DcI(u) \cap \{r \geq \tilde{R}^\prime\}$ for arbitrary $\tau \leq u \leq u_f$ with $\tilde{R}^\prime$ the one of Corollary \ref{cor:Tintermezzobaverage} to control boundary terms arising at $r=\tilde{R}^\prime$  in the standard integration by parts. This produces  (for $K=-1,0,\ldots,N-2$)
\begin{align} \label{poty1}
&\int_{\CcI(u) \cap \{r \geq \tilde{R}^\prime\}}\frac{1}{r^2} \sum_{k=0}^{K+1} | \check{\underline{\Pi}}^{(k)}_{\I}|^2 
+ \int_{\CbcI_{v_\infty}(\tau)} \sum_{k=1}^{K+2} | \check{\underline{A}}^{(k)}_{\I}|^2  
 +\int_{\DcI(\tau)\cap \{r \geq \tilde{R}^\prime\}} \frac{M^\delta}{r^{1+\delta}}\sum_{k=1}^{K+2} |\check{\underline{A}}^{(k)}_{\I}|^2 \\
 & \leq  \int_{\DcI(\tau)\cap \{r \geq \tilde{R}^\prime\}} \frac{1}{r^{3}} \Big\{ C_1 \sum_{k=0}^{K+1} |\check{\underline{\Pi}}^{(k)}_{\I}|^2 +  C_2\sum_{k=1}^{K+2} |\check{\underline{A}}^{(k)}_{\I}|^2 \Big\} + \int_{\CcI(\tau)\cap \{r \geq \tilde{R}^\prime\}}\frac{C_3}{r^2}  \sum_{k=0}^{K+1} | \check{\underline{\Pi}}^{(k)}_{\I}|^2 + C_R \left( \textrm{RHS of (\ref{golib})}\right) \nonumber \\
 & \ +  C_4 \int_{\DcI(\tau)\cap \{r \geq \tilde{R}^\prime\}} \sum_{k=0}^{K+1} r^{-2} | \check{\underline{\Pi}}^{(k)}_{\I}| \left( |\check{\mathcal{E}}_1^{k+1}|  +r^{-1} | \mathscr{A}^{[k]} r \Dslash_2^*\left(
		r \nablaslash r^2 \Omega \tr \chi
		\cdot r\alphabar 
		\right) \right) + C_5 \int_{\DcI(\tau)\cap \{r \geq \tilde{R}^\prime\}}\sum_{k=1}^{K+2} | \check{\underline{A}}^{(k)}_{\I}| |\check{\mathcal{E}}_2^{k}|   
  \nonumber
\end{align}
where for $k\geq 0$ we define $\mathscr{A}^{[k]} = \left( r^2 \Dslash_2^*\slashed{div} \right)^{\frac{k}{2}} $ if $k$ is even and $\mathscr{A}^{[k]}= r\slashed{div} \mathscr{A}^{[k-1]}$ if $k$ is odd.\index{double null gauge!differential operators!$\mathscr{A}^{[k]}$, angular operators acting on symmetric traceless $S$-tensors}
 Importantly, the constants $C_1,...,C_5$ above can be chosen \emph{independently of $R$}, 
 i.e.~they only depend on $M_{\rm init}$. 

Similarly, we contract the first equation of (\ref{psipi3ang}) (and (\ref{psipi3ang2}) respectively) with $\frac{1}{r^2}  \left(1+M^\delta r^{-\delta}\right)  \check{\underline{\Psi}}^{(k-1)}_{\I}$ and the second with $2  \left(1+ M^\delta r^{-\delta}\right) \check{\underline{\Pi}}^{(k)}_{\I}$. We then add the relevant equations and integrate over $\DcI(\tau) \setminus  \DcI(u) \cap \{r \geq \tilde{R}^\prime\}$ for arbitrary $\tau \leq u \leq u_f$ with $\tilde{R}^\prime$ the one of Corollary \ref{cor:Tintermezzobaverage}. After an integration by parts this produces  (for $K=0,...,N-2$)
\begin{align} \label{poty2}
&\int_{\CcI(u)\cap \{r \geq \tilde{R}^\prime\}}\frac{1}{r^2} \sum_{k=0}^{K} | \check{\underline{\Psi}}^{(k)}_{\I}|^2 + \int_{\CbcI_{v_\infty}(\tau)} \sum_{k=1}^{K+1} | \check{\underline{\Pi}}^{(k)}_{\I}|^2  
 +\int_{\DcI(\tau)\cap \{r \geq \tilde{R}^\prime\}} \frac{M^\delta}{r^{1+\delta}} \sum_{k=1}^{K+1} | \check{\underline{\Pi}}^{(k)}_{\I}|^2  \nonumber \\
 & \leq C_6 \int_{\DcI(\tau)\cap \{r \geq \tilde{R}^\prime\}} \frac{1}{r^{3+\delta}} \sum_{k=0}^{K} |\check{\underline{\Psi}}^{(k)}_{\I}|^2 +  \int_{\DcI(\tau)\cap \{r \geq \tilde{R}^\prime\}} \Big\{  \frac{C_7}{r^{2}} \sum_{k=0}^{K+1} |\check{\underline{\Pi}}^{(k)}_{\I}|^2 
 +  C_8 \frac{M^\delta}{r^{1+\delta}} \sum_{k=0}^{K} |\check{\underline{A}}^{(k)}_{\I}|^2 \Big\} \nonumber \\
 &+C_9 \int_{\CcI(\tau)\cap \{r \geq \tilde{R}^\prime\}}\frac{1}{r^2}  \sum_{k=0}^{K} | \check{\underline{\Psi}}^{(k)}_{\I}|^2 +C_{10} \int_{\CbcI_{v_\infty}(\tau)\cap \{r \geq \tilde{R}^\prime\}} \sum_{k=0}^{K}  | \check{\underline{A}}^{(k)}_{\I}|^2  + C_R \left( \textrm{RHS of (\ref{golib})} \right) \nonumber \\
 &+C_{11} \int_{\DcI(\tau)\cap \{r \geq \tilde{R}^\prime\}}\sum_{k=0}^{K} r^{-2} | \check{\underline{\Psi}}^{(k)}_{\I}|   |\check{\mathcal{E}}_1^{k+2}|  + 
 C_{12} \int_{\DcI(\tau)\cap \{r \geq \tilde{R}^\prime\}}\sum_{k=1}^{K+1} | \check{\underline{\Pi}}^{(k)}_{\I}|  |\check{\mathcal{E}}_2^{k+2}| 
\end{align}
where again, importantly, the constants $C_6,...C_{12}$ can be chosen independently of $R$. 
Note that the sums for $\check{\underline{\Pi}}^{(k)}_{\I}$ on the left could be extended to $k=0$ by a standard elliptic estimate. We now restrict $R$ to satisfy
\begin{align} \label{oneoftheRs}
\frac{C_1}{M^\delta R^{2-\delta}} + \frac{C_2}{M^\delta R^{2-\delta}} + \frac{C_7}{M^\delta R^{1-\delta}} <\frac{1}{4}  
\end{align}
with the $C_i$ as above. (Recall that~\eqref{oneoftheRs} was indeed of the constraints announced
in Section~\ref{compediumparameterssec} for the choice of $R$.)

Deriving (\ref{poty1}) and (\ref{poty2}) is straightforward, we only discuss here the terms in the derivation of (\ref{poty2}) that require special attention. For the terms
\begin{align}
+ \sum_{k=0}^K  \int_{\DcI(\tau,u) \cap \{r \geq \tilde{R}^\prime\}}   2\check{\underline{\Pi}}^{(k)}_{\I} \frac{1}{r^2}  \left(1+r^{-\delta}\right)  \check{\underline{\Psi}}^{(k)}_{\I} 
\end{align}
one may easily treat the term involving $r^{-\delta}$ with Cauchy--Schwarz and borrowing from the left hand side while for the remaining term one inserts that $\check{\underline{\Psi}}^{(k)}_{\I}  = -\frac{r^2}{\Omega^2} \Omega \slashed{\nabla}_4  \check{\underline{\Pi}}^{(k)}_{\I}+ \check{\mathcal{E}}^{1+k}_0$ and integrates by parts. The spacetime term is then controlled by a term appearing on the right hand side, the boundary term on $v=v_\infty$ has a good sign and the one on $r=\tilde{R}^\prime$ is controlled by Corollary \ref{cor:Tintermezzobaverage}.

For the terms
\begin{align}
+ \sum_{k=0}^K  \int_{\DcI(\tau,u)\cap \{r \geq \tilde{R}^\prime\}}  \left( -3M \check{\underline{A}}^{(k)}_{\I} \right)\frac{1}{r^2}  \left(1+r^{-\delta}\right)  \check{\underline{\Psi}}^{(k)}_{\I} 
\end{align}
one may easily treat the term involving $r^{-\delta}$ with Cauchy--Schwarz. For the remaining term, 
one inserts that $\check{\underline{\Psi}}^{(k)}_{\I}  = -\frac{r^2}{\Omega^2} \Omega \slashed{\nabla}_4  \check{\underline{\Pi}}^{(k-1)}_{\I}+ \mathcal{E}^{2+k}_0$ and integrates by parts. The spacetime term is controlled by terms appearing on the right of (\ref{poty2}), the boundary term on $r=\tilde{R}^\prime$ is controlled by Corollary \ref{cor:Tintermezzobaverage}, while for the boundary term on $v=v_\infty$, we can now estimate for any $\gamma>0$
\[
 \int_{\CbcI_{v_\infty}(\tau)} \sum_{k=1}^{K} | \check{\underline{A}}^{(k)}_{\I}|  | \check{\underline{\Pi}}^{(k)}_{\I}| \leq \gamma \sum_{k=1}^{K}  \int_{\CbcI_{v_\infty}(\tau)}  | \check{\underline{\Pi}}^{(k)}_{\I}|^2 + \frac{1}{\gamma} \int_{\CbcI_{v_\infty}(\tau)}  | \check{\underline{A}}^{(k)}_{\I}|^2 \, .
\]
Choosing $\gamma$ sufficiently small we can absorb the first term on the left. 

We next claim that adding (\ref{poty1}) and (\ref{poty2}) successively for $K=0, \ldots , N-2$  leads immediately to (\ref{golib2}), provided all sums on the left of (\ref{golib2}) are restricted to tuples of the form $(|\underline{k}|,0,0)$. This indeed follows after observing that
\begin{itemize}
\item Theorem~\ref{thm:PPbarestimates} controls all terms involving $\check{\underline{\Psi}}_{\I}$. All other spacetime terms appearing on the right hand side of (\ref{poty1}), (\ref{poty2}) can be directly absorbed on the left using (\ref{oneoftheRs}) except the non-linear errors in the last line and the term involving $ \frac{1}{r^{1+\delta}} \sum_{k=0}^{K} |\check{\underline{A}}^{(k)}_{\I}|^2$ on the left of (\ref{poty2}).
\item  for $K=0$, the term  $\int_{\CbcI_{v_\infty}(\tau)}  | \check{\underline{A}}^{(0)}_{\I}|^2$ on the right hand side of (\ref{poty2}) is controlled by Corollary~\ref{cor:auxiliaryinfty}, while for higher $K$ it is controlled inductively  since (\ref{poty1}) produces for $K=0$ in particular the term $\int_{\CbcI_{v_\infty}(\tau)}  | \check{\underline{A}}^{(1)}_{\I}|^2$. Similarly, for $K=0$ the term $\int_{\DcI(\tau) \cap \{r\geq \tilde{R}^\prime\}}  \frac{1}{r^{1+\delta}} |\check{\underline{A}}^{(0)}_{\I}|^2$ on the right hand side of (\ref{poty2}) is controlled by Proposition \ref{prop:lowordertpb}, while for higher $K$ it is controlled inductively since (\ref{poty1}) produces for $K=0$ in particular the term $\int_{\DcI(\tau) \cap \{r\geq \tilde{R}^\prime\}}  \frac{1}{r^{1+\delta}} |\check{\underline{A}}^{(1)}_{\I}|^2$ on the left hand side.
\item The non-linear errors can be treated using Cauchy--Schwarz and the estimates for $s=0,1,2$,
\begin{align} \label{nljun}
\int_{\mathcal{D}^{\I}(\tau)} r^{1+\delta} |\check{\mathcal{E}}_2^{N-s}|^2 +r^{-1+\delta} |\check{\mathcal{E}}_1^{N-s}|^2  + r^{-3+\delta} | \mathscr{A}^{[N-1-s]} r \Dslash_2^*\left(
		r \nablaslash r^2 \Omega \tr \chi
		\cdot r\alphabar_{\I} 
		\right)|^2 \lesssim \frac{ \varepsilon^4}{\tau^s} \, .
\end{align}
This follows directly from Proposition \ref{prop:waveIerror1b} for the first two terms and is easily established for the third using that the bootstrap assumptions imply $\| r^{-1} \left[r \slashed{\nabla}\right]^{N+1} (r^2 \Omega tr \chi)\|_{\mathcal{D}^{\I}} \lesssim \varepsilon$ and $|r \underline{\alpha}| \lesssim \varepsilon \tau^{-1}$.
\end{itemize}

\vskip1pc
\noindent
{\bf Step 3. Deriving the estimates for $l=1,2,...,N-2$.} We first note that the right hand side of (\ref{golib2}) controls also the following expressions 
\begin{align} \label{lafe}
& \int_{\CbcI_{v_\infty}(\tau)}  | (\Omega \slashed{\nabla}_3)^{K+2} \check{\underline{A}}^{(0)}_{\I}|^2  
 +\int_{\DcI(\tau)} \frac{1}{r^{1+\delta}} |(\Omega \slashed{\nabla}_3)^{K+2} \check{\underline{A}}^{(0)}_{\I}|^2 \nonumber \\
 + &\int_{\CbcI_{v_\infty}(\tau)}  | (\Omega \slashed{\nabla}_3)^{K+1} \check{\underline{\Pi}}^{(0)}_{\I}|^2  
 +\int_{\DcI(\tau)} \frac{1}{r^{1+\delta}} |(\Omega \slashed{\nabla}_3)^{K+1} \check{\underline{\Pi}}^{(0)}_{\I}|^2 \lesssim \textrm{right hand side of (\ref{golib2}) }\, .
\end{align}
This follows immediately from the fluxes and integrated decay estimates of Propositions~\ref{prop:pirstarb} 
and~\ref{prop:arstarb} after inserting the relation $\Omega \slashed{\nabla}_3 = -2R^\star + \Omega \slashed{\nabla}_4$ and the relations (\ref{recdab}) and (\ref{recdpib}) together with Theorem~\ref{thm:PPbarestimates}.

We now repeat the proof of Step 2 for the commuted equations (\ref{pia3ang}), (\ref{psipi3ang}) (and (\ref{pia3ang2}), (\ref{psipi3ang2}) respectively) and fixed $l=1,2,..., N-2$.\footnote{More precisely, for each fixed $l$ we repeat the proof for $K=0,1,\ldots,N-2-l$.} The proof is entirely analogous to Step 2 and therefore left to the reader. We only remark the following important points:
\begin{itemize}
\item Note that this time the boundary term $\int_{\CbcI_{v_\infty}(\tau)}  | (\Omega \slashed{\nabla}_3)^l \check{\underline{A}}^{(0)}_{\I}|^2$ appears for $K=0$ (recall that control on this term corresponded to the ``base case" of the induction in Step 2 and was deduced from Corollary~\ref{cor:auxiliaryinfty}) and $l=1,...,N-2$. Now this term is controlled by (\ref{lafe}). Similarly, the spacetime term $\int_{\DcI(\tau) \cap \{r \geq \tilde{R}^\prime\}} \frac{1}{r^{1+\delta}} |(\Omega \slashed{\nabla}_3)^{l} \check{\underline{A}}^{(0)}_{\I}|^2$ appearing for $K=0$ is also controlled by (\ref{lafe}).

\item For the non-linear terms we now need in addition for $k+l \leq N$ and $k \geq 1$ the estimate
\[
\int_{\mathcal{D}(\tau)} r^{-3} | \left[\Omega \slashed{\nabla}_3\right]^{l} \mathscr{A}^{[k]} \left(
		r \nablaslash r^2 \Omega \tr \chi
		\cdot r\alphabar 
		\right)|^2 \lesssim \frac{\varepsilon_0^2 + \varepsilon^3}{\tau^s} \ \ \ \textrm{for $s=0,1,2$.}
\]
This in turn follows as in the discussion below (\ref{discussionano}). For the top order term (all derivatives falling on $\Omega tr \chi$) we use the null structure equations until it becomes manifest that we can control the term using the bootstrap assumption on the energy defined in (\ref{angbon}) and $|r \underline{\alpha}|^2 \lesssim \frac{\epsilon^2}{\tau^2}$.
\end{itemize}

Having repeated the argument for $l=1,2,..., N-2$, we have now established (\ref{golib2}) except that the sums for $\check{\underline{A}}_{\I}$ (over $|\underline{k}| \leq K+2$) are restricted to tuples with $(k_1,k_2,0)$ with $k_2\leq K$ (i.e.~at least two derivatives have to be angular) and the sums for $\check{\underline{\Pi}}_{\I}$ (over $|\underline{k}| \leq K+1$) are restricted to tuples with $(k_1,k_2,0)$ with $k_2\leq K$ (i.e.~at least one derivative has to be angular). For $\check{\underline{\Pi}}_{\I}$ the tuple $(K+1,0,0)$ is controlled directly from (\ref{golib2}) both on $\check{\underline{C}}_{v_\infty}^{\I}$ and in integrated decay, while for the term
\[
\int_{\CcI(u)}\frac{1}{r^2} \sum_{l=0}^{K+1} | (\Omega \slashed{\nabla}_3)^l \check{\underline{\Pi}}_{\I}|^2
\]
one can insert (\ref{aresc}) and reduce it to a flux already controlled. For $\check{\underline{A}}_{\I}$ one applies 
the $(\Omega \slashed{\nabla}_3)^l$-commuted analogue of (\ref{poty1}) with $K=-1$. The wrong-signed term on the right hand side involving $|(\Omega \slashed{\nabla}_3)^l \check{\underline{\Pi}}^{(k)}_{\I}|^2$ has now already been controlled and (\ref{golib2}) is also shown for $\tilde{\mathfrak{D}}^{(1,K+1,0)} \check{\underline{A}}_{\I}$. Finally, for the tuple $\tilde{\mathfrak{D}}^{(0,K+2,0)} \check{\underline{A}}_{\I}$ the estimate (\ref{golib2}) follows from (\ref{lafe}).
\end{proof}

\begin{corollary} \label{cor:piallbextra}
For any $u_1 \leq \tau \leq u_f$ and $1\leq K\leq N-2$, we also have 
\begin{align} \label{golib2c}
\sum_{|\underline{k}|=0; k_2 \neq |\underline{k}|}^{K+2} \sup_{\tau \leq u \leq u_f} \int_{\CcI_u \cap \{r\geq R+2\}} \frac{1}{r^2} |\tilde{\mathfrak{D}}^{\underline{k}} \check{\underline{A}}_{\I} |^2  \lesssim \textrm{right hand side of (\ref{golib2})} \, .
\end{align}
\end{corollary}
\begin{proof}
We can reduce to tuples with $\underline{k}=(k_1,k_2,0)$ since otherwise using (\ref{recdab}) reduces the expression to a flux involving $\check{\underline{\Pi}}_{\I}$ appearing on the left of (\ref{golib2}). If two or more derivatives are angular, we can insert (\ref{aresc}) and again reduce the expression to a flux involving $\check{\underline{\Pi}}_{\I}$ appearing on the left of (\ref{golib2}). If one derivative is angular we can replace $\Omega \slashed{\nabla}_3 = -2R^\star + \Omega \slashed{\nabla}_4$ and the flux is controlled by the flux on $\underline{A}$ appearing in Proposition \ref{prop:arstarb} and a flux involving $\check{\underline{\Pi}}_{\I}$ appearing on the left of (\ref{golib2}).
\end{proof}

\subsection{The redshift estimates for \underline{${\Pi}$}~and \underline{$A$}} \label{sec:piabredshift}

We now improve the estimates (more specifically the $\Omega$-weights) in $\DcH (v(u_1))  \cap \{r \leq 9M_{\rm init}/4\}$. We can assume wlog that $r_{\Hp}(u_f,v=v(u_1))<\frac{17}{8}M_{\rm init}$ as otherwise we have a uniform lower bound on $\Omega^2$ in all of $\DcH$. Hence $\Omega^2$ weights can be absorbed into constants and the statement below would follow directly from 
 Propositions \ref{prop:piallb} and \ref{prop:piallb2}.

\begin{proposition} \label{prop:redshiftbp1}
For any $u_1\leq \tau<u_f$ and  $1\leq K \leq N-2$, we have
\begin{align} \label{megat}
& \ \ \ \sum_{|\underline{k}|=0, k_3=0}^{K+2} \int_{\CbcH_{v(\tau)}}  \Omega^2 | \tilde{\mathfrak{D}}^{\underline{k}} (\Omega^{-4} \underline{A}_{\Hp})|^2 + \sum_{|\underline{k}|=0, k_3=0}^{K+1}  \int_{\CcH_{u_f}(v(\tau_1))}   | \tilde{\mathfrak{D}}^{\underline{k}} (\Omega^{-2} \underline{\Pi}_{\Hp})|^2   \nonumber \\
  & \ \ \ +\sum_{|\underline{k}|=0}^{K+2}\int_{\DcH\left(v(\tau_1)\right)}  \Omega^2 | \tilde{\mathfrak{D}}^{\underline{k}} (\Omega^{-4} \underline{A}_{\Hp})|^2 +\sum_{|\underline{k}|=0}^{K+1} \int_{\DcH\left(v(\tau_1)\right) }   \Omega^2 | \tilde{\mathfrak{D}}^{\underline{k}} (\Omega^{-2} \underline{\Pi}_{\Hp})|^2  \nonumber \\
&\qquad  \lesssim \int_{\CcI_{\tau}} \frac{1}{r^2} \Bigg\{ \sum_{|\underline{k}|=0; k_2\neq |k|}^{K+2}  | \mathfrak{D}^{\underline{k}} \check{\underline{A}}_{\I}|^2 + \sum_{|\underline{k}|=0}^{K+1}   | \mathfrak{D}^{\underline{k}} \check{\underline{\Pi}}_{\I}|^2  + \sum_{k=0}^{K} |  (\slashed{\nabla}_{R^\star})^k \check{\underline{\Psi}}_{\I}|^2 \Bigg\} \nonumber \\ 
 & \qquad + \int_{\CbcH_{v(\tau)}}  \Omega^2 \sum_{|\underline{k}|=0; k_3\neq |k|}^{K+2} | \mathfrak{D}^{\underline{k}} (\Omega^{-4} \underline{A}_{\Hp})|^2 
+ \frac{\varepsilon_0^2 + \varepsilon^3}{\tau^{\min(2,N-K-2)}} \, . 
\end{align}
\end{proposition}

\begin{proof}
The proof will proceed in steps.

\vskip1pc
\noindent
{\bf Step 0. Simplifying what is to show.}  The right hand side of the estimate claimed controls the right hand side of Propositions \ref{prop:piallb} and \ref{prop:piallb2} hence we are free to use these estimates and Corollary~\ref{cor:bndhozr}. In particular, we only need to prove (\ref{megat}) with all integrals on the left restricted to the region $r \leq \mathring{R}$ because if we restrict all integrals to $r \geq \mathring{R}$, the corresponding estimates already hold by Propositions~\ref{prop:piallb} 
and~\ref{prop:piallb2}. This is what we will do. We also note in advance that it suffices to establish the estimate restricting to tuples $\underline{k}=(k_1,k_2,0)$ also for the integrated decay terms. Indeed, one can use the relations  (\ref{recdpib}) and (\ref{recdab}) and Theorem \ref{thm:PPbarestimates} to a posteriori treat all tuples $\underline{k}=(k_1,k_2,k_3)$.

\vskip1pc
\noindent
{\bf Step 1. Collecting the commuted equations.}
Key to the argument is the Bianchi pair
\begin{align} \label{tffb}
 \Omega \slashed{\nabla}_3 \left(\Omega^{-2} \underline{\Pi}_{\Hp} \right) + \frac{2}{r} \Omega^2 \left(\Omega^{-2} \underline{\Pi}_{\Hp} \right)  &= - \Omega^2 r^2 \slashed{\mathcal{D}}_2^\star \slashed{div} (\underline{\Omega}^{-4} \underline{A}_{\Hp})  -\frac{3M}{r} \Omega^2 (\underline{\Omega}^{-4} \underline{A}_{\Hp})   + \Omega^2(\mathcal{E}^\star)^1 \, ,  \\
 \Omega \slashed{\nabla}_4 (r \slashed{div} \Omega^{-4} \underline{A}_{\Hp}) + \frac{4M}{r^2} (r \slashed{div}  \Omega^{-4} \underline{A}_{\Hp}) &= \frac{2}{r^2}  r \slashed{div} (\Omega^{-2}\underline{\Pi}_{\Hp})  + (\mathcal{E}^\star)^1 \, ,
 \end{align}
 which after commutation with angular derivatives using the definition (\ref{commdef}) reads  for $k \geq 1$
\begin{align} \label{tffbc1}
 \Omega \slashed{\nabla}_3 \left(\Omega^{-2} \underline{\Pi}_{\Hp} \right)^{(k-1)} + \frac{2}{r} \Omega^2 \left(\Omega^{-2} \underline{\Pi}_{\Hp} \right)^{(k-1)}  &= - \Omega^2 r \slashed{\mathcal{D}}_2^\star (\underline{\Omega}^{-4} \underline{A}_{\Hp})^{(k)}  -\frac{3M}{r} \Omega^2 (\underline{\Omega}^{-4} \underline{A}_{\Hp})^{(k-1)}   + \Omega^2(\mathcal{E}^\star)^k\, , \nonumber  \\
 \Omega \slashed{\nabla}_4 ( \Omega^{-4} \underline{A}_{\Hp})^{(k)} + \frac{4M}{r^2} ( \Omega^{-4} \underline{A}_{\Hp})^{(k)} &= \frac{2}{r^2}  r \slashed{div} (\Omega^{-2}\underline{\Pi}_{\Hp})^{(k-1)}  + (\mathcal{E}^\star)^k \, ,
 \end{align}
 if $k$ is odd and
 \begin{align} \label{tffbc2}
 \Omega \slashed{\nabla}_3 \left(\Omega^{-2} \underline{\Pi}_{\Hp} \right)^{(k-1)} + \frac{2}{r} \Omega^2 \left(\Omega^{-2} \underline{\Pi}_{\Hp} \right)^{(k-1)}  &= - \Omega^2 r \slashed{div}(\underline{\Omega}^{-4} \underline{A}_{\Hp})^{(k)}  -\frac{3M}{r} \Omega^2 (\underline{\Omega}^{-4} \underline{A}_{\Hp})^{(k-1)}   + \Omega^2(\mathcal{E}^\star)^k \, , \nonumber  \\
 \Omega \slashed{\nabla}_4 ( \Omega^{-4} \underline{A}_{\Hp})^{(k)} + \frac{4M}{r^2} ( \Omega^{-4} \underline{A}_{\Hp})^{(k)} &= \frac{2}{r^2}  r  \slashed{\mathcal{D}}_2^\star  (\Omega^{-2}\underline{\Pi}_{\Hp})^{(k-1)}  +(\mathcal{E}^\star)^k \, ,
 \end{align}
 if $k$ is even. 
 
 \vskip1pc
 \noindent
 {\bf Step 2. Proving the estimate for tuples $\underline{k}=(\underline{k},0,0)$.} We first prove (\ref{megat}) (with all integrals on the left restricted to $r\geq \mathring{R}$ as mentioned) for angular derivatives, i.e.~with all sums on the left restricted to tuples with $\underline{k}=\left(|\underline{k}|,0,0\right)$. For each $k=1,\ldots,K+2$ we are going to generate an estimate as follows (and then add all of them). We contract the first equation with $2\left(\Omega^{-2} \underline{\Pi}_{\Hp} \right)^{(k-1)}$ and sum it with the second contracted with $r^{2}\Omega^{2}  (\Omega^{-4} \underline{A}_{\Hp})^{(k)}$. We then integrate over $\DcH(v(\tau_1),v(\tau)) \cap \{r \leq \mathring{R}\}$ if $v(\tau) < v(\mathring{R},u_f)$ where $\mathring{R}$ is defined in Corollary \ref{cor:bndhozr}. This yields the desired estimate after observing that 
 \begin{itemize}
 \item The terms on the left hand side produce the good boundary terms and spacetime terms appearing in the proposition noting that $\mathring{R} \leq 5M_{\rm init}/2$. 
 The boundary term on $r=\mathring{R}$ is controlled by Corollary~\ref{cor:bndhozr}.
 
 \item The angular-derivative terms on the left hand side cancel (after an integration by parts of the angular derivative) up to a nonlinear error error-term which is easily controlled.
 
 \item The non-linear terms are easily controlled by Cauchy--Schwarz and that for $s=0,1$ and any $k\leq N-s$ we have by Proposition \ref{prop:wavehorizonerror3}
 \begin{align}
 \int_{D(v(\tau))} \Omega^2  |(\mathcal{E}^\star)^{k}|^2  \lesssim \frac{\varepsilon^4}{\tau^{1+s}} \, . 
 \end{align}
 
\item For the linear ``error", we note 
\begin{align}
&\int_{\DcH(v(\tau_1),v(\tau_2)) \cap \{r \leq \mathring{R}\}} \frac{6M}{r} \Omega^2  ( \Omega^{-4} \underline{A}_{\Hp})^{(k-1)} \left(\Omega^{-2} \underline{\Pi}_{\Hp} \right)^{(k-1)} \nonumber \\
\leq &\int_{\DcH(v(\tau_1),v(\tau_2)) \cap \{r \leq \mathring{R}\}}  \Omega^2 \left(\frac{3}{r}  |\left(\Omega^{-2} \underline{\Pi}_{\Hp} \right)^{(k-1)} |^2+ \frac{3M^2}{r} | (\Omega^{-4} \underline{A}_{\Hp})^{(k-1)}|^2 \right)
\nonumber \\
\leq &\int_{\DcH(v(\tau_1),v(\tau_2)) \cap \{r \leq \mathring{R}\}}  \Omega^2 \left(\frac{3}{r}  |\left(\Omega^{-2} \underline{\Pi}_{\Hp} \right)^{(k-1)} |^2+ 3M | (\Omega^{-4} \underline{A}_{\Hp})^{(k)}|^2 \right)
\end{align}
where we have used that $\frac{M}{r} \leq \frac{3}{4}$ and the elliptic estimate $\int_{S_{u,v}}  | (\Omega^{-4} \underline{A}_{\Hp})^{(k-1)}|^2  \leq \frac{4}{3} \int_{S_{u,v}} | \Omega^{-4} \underline{A}_{\Hp})^{(k)}|^2 $. The right hand side can now be absorbed by the good terms on the left hand side (where they appear with $4$ instead of $3$).
 \end{itemize}
 
 \vskip1pc
 \noindent
{\bf Step 3. Proving the estimate for tuples $(k_1,k_2,0)$ with $k_1\geq 1$.} We next commute (\ref{tffbc1}) and (\ref{tffbc2}) respectively with the redshift vectorfield $\left(\frac{1}{\Omega}\slashed{\nabla}_3\right)^l$.\footnote{Computationally it is much easier to first divide the first of  (\ref{tffbc1}) by $\Omega^{-2}$ and then commute with $\left(\frac{1}{\Omega}\slashed{\nabla}_3\right)^l$ before in the end multiplying by $\Omega^{-2}$ again.} The resulting equations read (restricting to $k$ odd, the even case is treated analogously) 
 \begin{align}
  \Omega \slashed{\nabla}_3 \left[ \left(\frac{1}{\Omega} \slashed{\nabla}_3 \right)^l  \left(\Omega^{-2} \underline{\Pi}_{\Hp} \right)^{(k-1)} \right] + \frac{2}{r} \Omega^2\left[ \left(\frac{1}{\Omega} \slashed{\nabla}_3 \right)^l  \left(\Omega^{-2} \underline{\Pi}_{\Hp} \right)^{(k-1)} \right] 
  = - \Omega^2 r \slashed{\mathcal{D}}_2^\star \left[\left(\frac{1}{\Omega} \slashed{\nabla}_3 \right)^l  (\underline{\Omega}^{-4} \underline{A}_{\Hp})^{(k)}  \right]  \nonumber \\ 
 -\frac{3M}{r} \Omega^2 \left(\frac{1}{\Omega}\slashed{\nabla}_3\right)^l (\underline{\Omega}^{-4} \underline{A}_{\Hp})  + \sum_{i=0}^{l-1} h^i \left(\frac{1}{\Omega} \slashed{\nabla}_3 \right)^i \left(\Omega^{-2} \underline{\Pi}_{\Hp} \right)^{(k-1)}  +   \sum_{i=0}^{l-1} h^i \left(\frac{1}{\Omega} \slashed{\nabla}_3 \right)^i \left(\Omega^{-4} \underline{A}_{\Hp} \right)^{(k-1)} +
  \Omega^2(\mathcal{E}^\star)^{k+l} \nonumber
  \end{align}
\begin{align}
 \Omega \slashed{\nabla}_4 \left[ \left(\frac{1}{\Omega} \slashed{\nabla}_3 \right)^l  ( \Omega^{-4} \underline{A}_{\Hp})^{(k)} \right] + \frac{(4+2l)M}{r^2}\left[ \left(\frac{1}{\Omega} \slashed{\nabla}_3 \right)^l   (\Omega^{-4} \underline{A}_{\Hp})^{(k)} \right] = \frac{2}{r^2}  r \slashed{div}  \left[ \left(\frac{1}{\Omega} \slashed{\nabla}_3 \right)^l  \left(\Omega^{-2} \underline{\Pi}_{\Hp} \right)^{(k-1)} \right] \nonumber \\
   + \sum_{i=0}^{l-1} h^i \left(\frac{1}{\Omega} \slashed{\nabla}_3 \right)^i \left(\Omega^{-2} \underline{A}_{\Hp} \right)^{(k)}  +   \sum_{i=0}^{l-1} h^i \left(\frac{1}{\Omega} \slashed{\nabla}_3 \right)^i \left(\Omega^{-2} \underline{\Pi}_{\Hp} \right)^{(k)} + (\mathcal{E}^\star)^{k+l}  \nonumber
\end{align}
with $h^i$ an admissible coefficient function of $r$ (which may be different in different places).
Since for $l=0$, all estimates have been proven (for all $k$), it is clear that we can repeat the analysis of Step~1 and we proceed inductively in $l$ to obtain (\ref{megat}) with the sums involving $\underline{A}_{\Hp}$ on the right hand side restricted to tuples of the form $\underline{k}=(k_1,k_2,0)$ and $k_1\geq 1$). 

\vskip1pc
\noindent
{\bf Step 4. Proving the estimate for tuples $(0,|\underline{k}|,0)$ with $k_1\geq 1$.}  The only missing bit in the proof of the estimate (\ref{megat}) is to control $\mathfrak{D}^{\underline{k}} \underline{A}_{\Hp}$ for tuples of the form $\underline{k}=(0,|k|,0)$, i.e.~all derivatives being $3$-derivatives. For this we apply the redshift estimates to the Teukolsky equation directly. We recall from Proposition \ref{prop:classify3k} that $\Omega^{-4} \underline{A}_{\Hp}$ satisfies a tensorial wave equation of type $3_2$ and that by Proposition \ref{prop:3kc}, $\left(\frac{1}{\Omega} \slashed{\nabla}_3 \right)^l  \Omega^{-4} \underline{A}_{\Hp}$ satisfies a tensorial wave equation of  type $3_{2+l/2}$ with inhomogeneous term of the form
 \begin{align} \label{acc}
 \mathcal{F}^{lin}_{3_{2+l/2}} \left[ \left(\frac{1}{\Omega} \slashed{\nabla}_3 \right)^l  \Omega^{-4} \underline{A}_{\Hp} \right] =& \sum_{i=0}^l h^i \left(\frac{1}{\Omega} \slashed{\nabla}_3 \right)^i \left(\Omega^{-2} \underline{\Pi}_{\Hp}\right) + \sum_{i=0}^l h^i \left(\frac{1}{\Omega} \slashed{\nabla}_3 \right)^i \left(\Omega^{-4} \underline{A}_{\Hp}\right) \nonumber \\
 &+\sum_{i=0}^{l-1} h^i \left(\frac{1}{\Omega} \slashed{\nabla}_3 \right)^i \slashed{\Delta} \left(\Omega^{-4} \underline{A}_{\Hp}\right) 
 \end{align}
and non-linear error $\mathcal{F}^{nlin}_{3_{2+l/2}} \left[ \left(\frac{1}{\Omega} \slashed{\nabla}_3 \right)^l  \Omega^{-4} \underline{A}_{\Hp} \right] = \Omega^2 (\mathcal{E}^\star)^{1+l}$. We can hence apply Proposition \ref{prop:redshift}. The non-linear error is easily estimated by the last term on the right hand side of (\ref{megat}). For the linear error we apply 
Cauchy--Schwarz and use the estimates already obtained to control  $|\mathcal{F}^{lin}_{3_{2+l/2}}|^2$. This is obvious for the first and the third term on the right hand side of (\ref{acc}). For the second term, note that it involves only $l$ derivatives but that we are estimating $l+1$ derivatives of $\Omega^{-4} \underline{A}_{\Hp}$. The  term can therefore be controlled by an elliptic estimate using that we already proved an estimate for $\left(\frac{1}{\Omega} \slashed{\nabla}_3 \right)^l r\slashed{div} \Omega^{-4} \underline{A}_{\Hp}$ in terms of the right hand side in (\ref{megat}). This proves that
\begin{align} \label{addty}
                +\sup_{\tau \leq u \leq u_f} \int_{\CbcH_{v(u)}} \Big| \left(\frac{1}{\Omega} \slashed{\nabla}_3 \right)^{l+1}  \Omega^{-4} \underline{A}_{\Hp}\Big|^2  + \int_{\CcH_{u_f} \left(v(\tau)\right) \cap \{r\leq 9M_{\rm init}/4\}}  \Big| r\slashed{\nabla} \left(\frac{1}{\Omega} \slashed{\nabla}_3 \right)^{l}  \Omega^{-4} \underline{A}_{\Hp}\Big|^2 \nonumber \\
               + \int_{\DcH \left(v(\tau)\right)} \Big| \left(\frac{1}{\Omega} \slashed{\nabla}_3 \right)^{l+1}  \Omega^{-4} \underline{A}_{\Hp}\Big|^2
                \lesssim \textrm{right hand side of (\ref{megat})} \, ,
\end{align}
thereby completing the proof.
 \end{proof}
 
Completely analogously we prove
 
 \begin{proposition} \label{prop:redshiftbp2}
For any $u_1\leq \tau<u_f$ and $1\leq K \leq N-2$, we have
\begin{align} \label{megat2}
&\sum_{|\underline{k}|=0,  k_3=0}^{K+1} \int_{\CbcH_{v(\tau)}}  \Omega^2 | \tilde{\mathfrak{D}}^{\underline{k}} \Omega^{-2} \underline{\Pi}_{\Hp}|^2  + \sum_{|\underline{k}|=0}^{K}  \int_{\CcH_{u_f}(v(\tau))}   | \tilde{\mathfrak{D}}^{\underline{k}} \underline{\Psi}_{\Hp}|^2 \nonumber \\
&\qquad  \lesssim \int_{\CcI_{\tau}} \frac{1}{r^2} \Bigg\{ \sum_{|\underline{k}|=0; k_2\neq |k|}^{K+2}  | \mathfrak{D}^{\underline{k}} \check{\underline{A}}_{\I}|^2 + \sum_{|\underline{k}|=0}^{K+1}   | \mathfrak{D}^{\underline{k}} \check{\underline{\Pi}}_{\I}|^2  + \sum_{k=0}^{K} |  (\slashed{\nabla}_{R^\star})^k \check{\underline{\Psi}}_{\I}|^2 \Bigg\} \nonumber \\ 
 & \qquad + \int_{\CbcH_{v(\tau)}}  \Omega^2 \sum_{|\underline{k}|=0; k_3\neq |k|}^{K+2} | \mathfrak{D}^{\underline{k}} (\Omega^{-4} \underline{A}_{\Hp})|^2 
+ \frac{\varepsilon_0^2 + \varepsilon^3}{\tau^{\min(2,N-K-2)}} \, . 
\end{align}
\end{proposition}

 \begin{proof}
 Key to the proof is the Bianchi pair
 \begin{align}\label{tffb2}
 \Omega \slashed{\nabla}_3 \underline{\Psi}_{\Hp}   = 2 \Omega^2 r^2\slashed{\mathcal{D}}_2^\star \slashed{div} (\Omega^{-2} \underline{\Pi}_{\Hp})  &+ \left(2 \Omega^2 - \frac{6M}{r}\right) \Omega^2 \Omega^{-2} \underline{\Pi}_{\Hp}  - 3M \Omega^4 \Omega^{-4} \underline{A}_{\Hp} + \Omega^2 (\mathcal{E}^\star)^2 \, , \nonumber \\
 \Omega \slashed{\nabla}_4 (r \slashed{div} \underline{\Pi}_{\Hp} \Omega^{-2}) + \frac{2M}{r^2}(r \slashed{div} \underline{\Pi}_{\Hp} \Omega^{-2})  &= -\frac{1}{r^2} r \slashed{div} \check{\underline{\Psi}}_{\I} + (\mathcal{E}^\star)^2 \, , 
\end{align}
which after commutation with angular derivatives using the definition (\ref{commdef}) reads  for $k \geq 1$
\begin{align}\label{tffb2c1}
&\Omega \slashed{\nabla}_3 \underline{\Psi}^{(k-1)}_{\Hp}   = 2 \Omega^2 r\slashed{\mathcal{D}}_2^\star (\Omega^{-2} \underline{\Pi}_{\Hp})^{(k)} + \left(2 - \frac{6M}{r}\right) \Omega^2 (\Omega^{-2} \underline{\Pi}_{\Hp})^{(k-1)} - 3M \Omega^4 \left(\Omega^{-4} \underline{A}_{\Hp}\right)^{(k-1)} + \Omega^2 (\mathcal{E}^\star)^{k+1} \, ,  \nonumber \\
& \Omega \slashed{\nabla}_4 (\Omega^{-2} \underline{\Pi}_{\Hp} )^{(k)} + \frac{2M}{r^2}(\Omega^{-2} \underline{\Pi}_{\Hp} )^{(k)}  = -\frac{1}{r^2} r \slashed{div} \underline{\Psi}^{(k-1)}_{\I} + (\mathcal{E}^\star)^{k+1}  \, , 
\end{align}
 if $k$ is odd and
 \begin{align}\label{tffb2c2}
&\Omega \slashed{\nabla}_3 \underline{\Psi}^{(k-1)}_{\Hp}   = 2 \Omega^2 r\slashed{div} (\Omega^{-2} \underline{\Pi}_{\Hp})^{(k)} + \left(2 - \frac{6M}{r}\right) \Omega^2 (\Omega^{-2} \underline{\Pi}_{\Hp})^{(k-1)} - 3M \Omega^4 \left(\Omega^{-4} \underline{A}_{\Hp}\right)^{(k-1)} + \Omega^2 (\mathcal{E}^\star)^{k+1},\nonumber \\
&\Omega \slashed{\nabla}_4 (\Omega^{-2} \underline{\Pi}_{\Hp} )^{(k)} + \frac{2M}{r^2}(\Omega^{-2} \underline{\Pi}_{\Hp} )^{(k)}  = -\frac{1}{r^2} r \slashed{\mathcal{D}}_2^\star \underline{\Psi}^{(k-1)}_{\I} + (\mathcal{E}^\star)^{k+1} \, , 
\end{align}
if $k$ is even. Contracting the first equation with $\left( \underline{\Psi}_{\Hp} \right)^{(k-1)}$ and summing it with the second contracted with $2r^{2}\Omega^{2}  (\Omega^{-2} \underline{\Pi}_{\Hp})^{(k)}$ yields after integration over $\DcH(v(\tau_1),v(\tau)) \cap \{r \leq \mathring{R}\}$ if $v(\tau) < v(\mathring{R},u_f)$ (where $\mathring{R}$ is defined in Corollary \ref{cor:bndhozr}) the result for the sums restricted to tuples involving only angular derivatives. Now, if (at least) one of the derivatives of $\mathfrak{D}^{\underline{k}}$  in the first sum of (\ref{megat2}) is a $\Omega^{-1} \slashed{\nabla}_3$ derivative, we can use (\ref{teueli1b}) and the fluxes of Proposition \ref{prop:redshiftbp1} on $\underline{A}_{\Hp}$  to control all fluxes for ${\underline{\Pi}_{\Hp}}$. Similarly, if (at least) one of the $\mathfrak{D}^{\underline{k}}$  in the second sum of (\ref{megat2}) is a $\Omega^{-1} \slashed{\nabla}_3$ derivative, we can use (\ref{wep1b}) and the fluxes of Proposition \ref{prop:redshiftbp1} (as well as the fluxes on $\underline{\Pi}_{\Hp}$ just established) to control the associated flux.  
\end{proof}

\subsection{Decay estimates for \underline{$\Pi$}~and \underline{$A$}: the proof of Theorem \ref{theo:mtheoalphabr2}} \label{sec:decbapi}
We now recall the energies defined in Theorem \ref{theo:mtheoalphabr2} allowing us to concisely summarise the weighted estimates we have proven:
\begin{proposition}
For any $u_1 \leq \tau \leq u_f$, we have, for $3 \leq K \leq N$, the estimate
\begin{align} \label{keyestab}
{\mathbb{E}}_\Box^{K} \left[\underline{\alpha}_{\Hp}\right] \left(v(\tau)\right) + {\mathbb{E}}_\Box^{K} \left[\underline{\alpha}_{\I}\right] \left(\tau\right) \lesssim & \int_{\CcI_{\tau}} \frac{1}{r^2} \Bigg\{ \sum_{|\underline{k}|=0; k_2\neq |k|}^{K}  | \mathfrak{D}^{\underline{k}} \check{\underline{A}}_{\I}|^2 + \sum_{|\underline{k}|=0}^{K-1}   | \mathfrak{D}^{\underline{k}} \check{\underline{\Pi}}_{\I}|^2  + \sum_{k=0}^{K-2} |  (\slashed{\nabla}_{R^\star})^k \check{\underline{\Psi}}_{\I}|^2 \Bigg\} \nonumber \\ 
&+ \int_{\CbcH_{v(\tau)}}  \Omega^2 \sum_{|\underline{k}|=0; k_3\neq |k|}^{K} | \mathfrak{D}^{\underline{k}} (\Omega^{-4} \underline{A}_{\Hp})|^2 
+ \frac{\varepsilon_0^2 + \varepsilon^3}{\tau^{\min(2,N-K)}} \, .
\end{align}
\end{proposition}
\begin{proof}
By Propositions \ref{sec:ellipticconesI} and \ref{sec:ellipticconesHp} it suffices to prove the estimate replacing $\mathfrak{D}^{\underline{k}}$ by $\tilde{\mathfrak{D}}^{\underline{k}}$ everywhere in the energies on the left hand side and this is what we will do.

For  ${\mathbb{E}}_\Box^{K} \left[\underline{\alpha}_{\I}\right] \left(\tau\right)$ this follows immediately from combining the estimates of Theorem \ref{thm:PPbarestimates}, Propositions \ref{prop:piallb} and \ref{prop:piallb2} and Proposition \ref{prop:piallbextra} together with Corollary \ref{cor:piallbextra}. Note in particular that we have, in view of Theorem \ref{thm:PPbarestimates} the estimate
\begin{align}
 \int_{\CcI_{\tau}} \frac{1}{r^2} \sum_{|\underline{k}|=0}^{K-2} | {\mathfrak{D}}^{\underline{k}} \check{\underline{\Psi}}_{\I}|^2 
 \lesssim \int_{\CcI_{\tau}} \frac{1}{r^2} \sum_{k=1}^{K-2} |  (\slashed{\nabla}_{R^\star})^k \check{\underline{\Psi}}_{\I}|^2 + \frac{\varepsilon_0^2 + \varepsilon^3}{\tau^{\min(2,N-K-2)}} \, .
\end{align}
For ${\mathbb{E}}_\Box^{K} \left[\underline{\alpha}_{\Hp}\right] \left(v(\tau)\right)$, the integrated decay terms are already completely included in Proposition \ref{prop:redshiftbp1}. For the fluxes on $v=v(\tau)$ we first note
\begin{align} \label{krak}
 \sup_{\tau \leq u \leq u_f} \int_{\CbcH_{v(u)}} \Omega^2  \sum_{|\underline{k}|=0; k_3\neq |k|}^{K-1}  | \tilde{\mathfrak{D}}^{\underline{k}} (\Omega^{-2} \underline{\Pi}_{\Hp})|^2  \lesssim \textrm{right hand side of (\ref{keyestab}).}
\end{align}
Indeed, if $\underline{k}=(k_1,k_2,0)$ the result is the statement of Proposition  \ref{prop:redshiftbp2}. If 
$\underline{k}=(k_1,k_2,k_3)$ with $k_3\geq 1$ (but $k_3 \neq |\underline{k}|$!) we insert the relation (\ref{recdpib}) and turn it into a flux involving $\underline{\Psi}_{\Hp}$ that is controlled by Theorem \ref{thm:PPbarestimates}. We next claim
\begin{align}
 \sup_{\tau \leq u \leq u_f} \int_{\CbcH_{v(u)}} \Omega^2  \sum_{|\underline{k}|=0; k_3\neq |k|}^{K}  |  \tilde{\mathfrak{D}}^{\underline{k}} (\Omega^{-4} \underline{A}_{\Hp})|^2  \lesssim \textrm{right hand side of (\ref{keyestab}).}
\end{align}
Indeed, if $\underline{k}=(k_1,k_2,0)$ the result is the statement of Proposition  \ref{prop:redshiftbp1}. If 
$\underline{k}=(k_1,k_2,k_3)$ with $k_3\geq 1$ (but $k_3 \neq |\underline{k}|$!) we insert the relation (\ref{recdab}) and turn it into a flux controlled by (\ref{krak}).

Finally, we look at the fluxes on $u=u_f$ appearing in ${\mathbb{E}}_\Box^{K} \left[\underline{\alpha}_{\Hp}\right] \left(v(\tau)\right)$. We first claim
\begin{align} \label{krak2}
 \sum_{|\underline{k}|=0}^{K-1}  \int_{\CcH_{u_f}(v(\tau))}   | \tilde{\mathfrak{D}}^{\underline{k}} (\Omega^{-2} \underline{\Pi}_{\Hp})|^2 \lesssim \textrm{right hand side of (\ref{keyestab}).}
\end{align}
Indeed, if $\underline{k}=(k_1,k_2,0)$ the result is the statement of Proposition  \ref{prop:redshiftbp1}. If 
$\underline{k}=(k_1,k_2,k_3)$ with $k_3\geq 1$ (but $k_3 \neq |\underline{k}|$!) we insert the relation (\ref{recdpib}) and turn it into a flux involving $\underline{\Psi}_{\Hp}$ that is controlled by Proposition \ref{prop:redshiftbp2} and Theorem \ref{thm:PPbarestimates}. 
With (\ref{krak2}) established, we next claim 
\begin{align}
\sum_{|\underline{k}|=0, k_2 \neq |\underline{k}|}^{K}  \int_{\CcH_{u_f}(v(\tau))}   | \tilde{\mathfrak{D}}^{\underline{k}} (\Omega^{-4} \underline{A}_{\Hp})|^2  \lesssim \textrm{right hand side of (\ref{keyestab}).}
\end{align}
It suffices to prove this estimate with the sum starting from $|\underline{k}|=1$ in view of a simple elliptic estimate. If $\underline{k}=\left(k_1,k_2,k_3\right)$ with $k_3 \neq 0$ we can use the relation (\ref{recdab}) to turn the flux into a flux for $\check{\underline{\Pi}}_{\I}$ controlled by (\ref{krak2}). Hence we can restrict to $k_3=0$. If one of the derivatives is angular (and hence the rest $\Omega^{-1} \slashed{\nabla}_3$ derivatives) the flux is controlled by (\ref{addty}). If two derivatives are angular we can insert (\ref{tffb}) and turn it into a flux controlled by (\ref{krak2}).
\end{proof}

The following proposition is easily seen to imply the statement of Theorem \ref{theo:mtheoalphabr2}.

\begin{proposition} \label{prop:decbaredquantities}
For any $u_1 \leq \tau \leq u_f$ we have for $s=0,1,2$ the estimates
\begin{align}
{\mathbb{E}}_\Box^{N-s} \left[\underline{\alpha}_{\Hp}\right] \left(v(\tau)\right) + {\mathbb{E}}_\Box^{N-s} \left[\underline{\alpha}_{\I}\right] \left(\tau\right) &\lesssim \frac{\varepsilon_0^2 + \varepsilon^3}{\tau^s} 
\end{align}
\end{proposition}

\begin{proof}
This follows from applying the  pigeonhole argument of~\cite{DafRodnew} 
to the left hand side of (\ref{keyestab}).
\end{proof}

\section{Concluding the main estimates for \underline{$\alpha$}}  \label{sec:missingfluxab}

In this section we prove Theorem \ref{theo:mtheoalphabr} from Theorem \ref{theo:mtheoalphabr2}.
To achieve this,
we only need to extend all of our integrated decay estimates and flux estimates to the non-truncated regions and cones. The proof is almost identical to that for $\alpha$ seen previously in Section~\ref{sec:missingfluxa}. 

\vskip1pc
\noindent
{\bf Step 1.} We observe that the estimates of Theorem \ref{theo:mtheoalphabr2} continue to hold if we drop all check superscripts in the \emph{spacetime integrals}. This follows easily from the relations for $\check{\underline{A}}_{\I}$ and $\underline{A}_{\Hp}$ in the overlap region, i.e.~by applying Proposition \ref{thm:morecancelnotjustT}. As an immediate corollary using the definition of the timelike hypersurface $\mathcal{B}$ we obtain:
\begin{corollary} \label{cor:hypbb2}
For all $u_1 \leq \tau \leq u_f$ we have for $s=0,1,2$ the estimate
\begin{align} 
 \sum_{|{\underline{k}} | \leq N-s} \int_{\mathcal{B}(\tau)}  | \tilde{\mathfrak{D}}^{\underline{k}}  \check{\underline{A}}_{\I}|^2 +  | \tilde{\mathfrak{D}}^{\underline{k}} \underline{A}_{\Hp}|^2 &\lesssim \frac{\varepsilon^2_0 + \varepsilon^3}{\tau^s}  \, .
\end{align}
\end{corollary}

\vskip1pc
\noindent
{\bf Step 2a.} 
We observe that the estimates of Theorem \ref{theo:mtheoalphabr2}  continue to hold if we replace the flux
\[
 \int_{\check{\underline{C}}^{\I}_{v_{\infty}}(\tau)}  \Bigg\{ \sum_{|\underline{k}|=0, k_3 \neq |\underline{k}|}^{K}  \  | \mathfrak{D}^{\underline{k}} \check{\underline{A}}_{\I} |^2 + \sum_{|\underline{k}|=0; k_3 \neq |\underline{k}|}^{K-1}  | \mathfrak{D}^{\underline{k}} \check{\underline{\Pi}}_{\I}|^2   \Bigg\}
\]
by the flux over an arbitrary ingoing truncated cone in $\DcI$, i.e. integrating over $\int_{\check{\underline{C}}^{\I}_{v}(\tau)}$ for any cone with $v \leq v_\infty$. Indeed, picking such a cone $\check{\underline{C}}_v^{\I} \left(\tau\right)$ we can consider the spacetime region enclosed by $\check{\underline{C}}_v^{\I} \left(\tau\right)$, $\mathcal{B}$, $\check{C}_\tau^{\I}$ and (potentially) $\check{C}^{\I}_{u_f}$. We reapply the estimates of Proposition \ref{prop:piallbextra} (integrating now over $\DcI(\tau) \cap J^-(\check{\underline{C}}_v^{\I} \left(\tau\right))$) using now the bounds on the cone $\check{C}_\tau^{\I}$ established in Theorem \ref{theo:mtheoalphabr2} and the bounds on the hypersurface $\mathcal{B}$ established in Corollary \ref{cor:hypbb2}. 

\vskip1pc
\noindent
{\bf Step 2b.}
We observe that the estimates of Theorem \ref{theo:mtheoalphabr2}  continue to hold if we replace the flux
\[
\int_{\check{C}^{\Hp}_{u_f}({v})} \sum_{|\underline{k}|=0; {k_2 \neq |\underline{k}|}}^{K}  | \mathfrak{D}^{\underline{k}} \underline{A}_{\Hp}|^2  
\]
by the flux over an arbitrary outgoing truncated cone in $\DcH$, i.e. integrating over $\check{C}^{\Hp}_{u}({v})$ for any $u \leq u_f$. Indeed picking such a cone $\check{C}_u^{\Hp} \left(v\right)$ we can consider the region enclosed by $\check{C}_u^{\Hp} \left(v\right)$, $\check{\underline{C}}^{\Hp}_v$ and $\mathcal{B}$. We repeat the proof of Proposition \ref{prop:redshiftbp1}  (integrating now over $\DcH(v(\tau)) \cap J^-(\check{C}^{\Hp}_{u}({v}))$) using the estimates of Theorem \ref{theo:mtheoalphabr2} on the cone  $\check{\underline{C}}^{\Hp}_v$ and the bounds on the hypersurface $\mathcal{B}$ established in Corollary \ref{cor:hypbb2}. Using also the reasoning following (\ref{krak2}) we deduce control of the fluxes on arbitrary truncated outgoing cones.

\vskip1pc
\noindent
{\bf Step 3.} We observe that the estimates of Theorem \ref{theo:mtheoalphabr2}  continue to hold if we replace truncated by non-truncated cones in $\DcI \cup \DcH$. This proceeds by doing localised energy estimates (in regions where neither the $r$-weights nor $\Omega$-weights play a role) for the non-linear Teukolsky equation entirely analogously to the case of $P$ and $\underline{P}$ seen in Chapter \ref{chapter:psiandpsibar}. We omit the standard details.

\section{Completing the proof of Theorem~\ref{thm:alphaalphabarestimates}} \label{sec:teustaro}

To prove Theorem~\ref{thm:alphaalphabarestimates} from Theorems \ref{theo:mtheoalphar} and \ref{theo:mtheoalphabr}, we only need to remove the ``boxed" restrictions in the sums appearing in Theorems \ref{theo:mtheoalphar} and \ref{theo:mtheoalphabr}, i.e.~to obtain control on the top-order fluxes of certain transversal derivatives. This is a direct consequence of certain relations between the gauge invariant hierarchies $(\alpha,\psi,P)$ and $(\underline{\alpha},\underline{\psi}, \underline{P})$, known in linear theory (at the mode decomposed level in the physics literature) as 
the Teukolsky--Starobinsky identities. 

\subsection{Missing transversal fluxes for $\alpha_{\I}, \alpha_{\Hp}$}
Starting from Proposition \ref{prop:PPbaridentities} we derive the identities 
\begin{align}
-\frac{1}{2}  \frac{\Omega^4}{r^4} \left(\frac{r^2}{\Omega^2} \Omega \slashed{\nabla}_3 \right)^2  \Psi_{\Hp} &= \frac{\Omega^4}{r^4} \left(\frac{r^2}{\Omega^2} \Omega \slashed{\nabla}_3 \right)^4  A_{\Hp} = 2r^4 \slashed{\mathcal{D}}_2^\star \slashed{\mathcal{D}}_1^\star \slashed{\mathcal{D}}_1 \slashed{\mathcal{D}}_2 \underline{A}_{\Hp} + 6M (\Omega \slashed{\nabla}_3 + \Omega \slashed{\nabla}_4) \underline{A}_{\Hp}  + \mathcal{E}^3 \nonumber \\ \, , 
-\frac{1}{2}  \frac{\Omega^4}{r^4} \left(\frac{r^2}{\Omega^2} \Omega \slashed{\nabla}_3 \right)^2  \Psi_{\I} &= \frac{\Omega^4}{r^4} \left(\frac{r^2}{\Omega^2} \Omega \slashed{\nabla}_3 \right)^4  A_{\I} = 2r^4 \slashed{\mathcal{D}}_2^\star \slashed{\mathcal{D}}_1^\star \slashed{\mathcal{D}}_1 \slashed{\mathcal{D}}_2 (r \check{r}^{-1} \check{\underline{A}}_{\I}) + 6M (\Omega \slashed{\nabla}_3 + \Omega \slashed{\nabla}_4) \check{\underline{A}}_{\I}  + \check{\mathcal{E}}^3_{0} \nonumber \, .
\end{align}

To obtain the missing fluxes in Theorem \ref{theo:mtheoalphar},  it suffices to prove for $s=0,1,2$ and $\tau \geq u_1$:
\begin{align} \label{fing1}
 \sup_{u\leq u_f} \int_{C^{\Hp}_u({v})} \sum_{k=0}^{N-s}  | \left[\Omega^{-1} \slashed{\nabla}_3 \right]^k A_{\Hp}|^2 \lesssim \frac{\varepsilon^2_0 + \varepsilon^3}{v^s} \, ,
 \end{align}
\begin{align}  \label{fing2}
 \sup_{\tau \leq u \leq u_f} \int_{C^{\I}_u} \Bigg\{ \sum_{k=1}^{N-s}  \ r^{6-s} | \left[\Omega^{-1} \slashed{\nabla}_3 \right]^k A_{\I}|^2 + \sum_{k=0}^{N-1-s}  r^2 | \left[\Omega^{-1} \slashed{\nabla}_3 \right]^k\Pi_{\I}|^2 \Bigg\} \lesssim \frac{\varepsilon^2_0 + \varepsilon^3}{\tau^s} \, .
 \end{align}
 
We can split the sums into a sum for $k<4$ and a sum $k \geq 4$. For the former we can (exploiting that the transversal fluxes are indeed included if we accept the loss of a derivative) estimate 
these fluxes (for fixed $s$) by $\mathbb{E}^{4,2-s}_{\Box}[\alpha_{\I}](\tau)+\mathbb{E}^{4}_{\Box}[\alpha_{\Hp}](\tau)$. For the latter we insert the above identities. For (\ref{fing2}) it is then easy to see that for fixed $s$ the flux is now controlled by ${\mathbb{E}}_\Box^{N-s} \left[\underline{\alpha}_{\I}\right] \left(\tau\right) + {\mathbb{E}}_\Box^{N-s,2-s}[\alpha_{\I}](\tau)+\frac{\varepsilon^2_0 + \varepsilon^3}{\tau^s}$, which is in turn controlled by $\frac{\varepsilon^2_0 + \varepsilon^3}{\tau^s}$ from Theorems \ref{theo:mtheoalphar} and \ref{theo:mtheoalphabr}. Similarly, for the ($k\geq 4$)-part of the sum of (\ref{fing1}) inserting the first identity (note $r$-weights can be absorbed into constants in this region) shows that the flux can be controlled by ${\mathbb{E}}_\Box^{N-s} \left[\underline{\alpha}_{\Hp}\right] \left(v\right) + {\mathbb{E}}_\Box^{N-s,2-s}[\alpha_{\Hp}](v)+\frac{\varepsilon^2_0 + \varepsilon^3}{v^s}$,  which is in turn controlled by $\frac{\varepsilon^2_0 + \varepsilon^3}{v^s}$ from Theorems \ref{theo:mtheoalphar} and \ref{theo:mtheoalphabr}.

\subsection{Missing transversal fluxes for $\protect\underline{\alpha}_{\I}, \protect\underline{\alpha}_{\Hp}$}
Starting from Proposition \ref{prop:nabla4pschematic} we derive the identities 
\begin{align}
-\frac{1}{2}  \frac{\Omega^4}{r^4} \left(\frac{r^2}{\Omega^2} \Omega \slashed{\nabla}_4 \right)^2  \underline{\Psi}_{\Hp}  &= \frac{\Omega^4}{r^4} \left(\frac{r^2}{\Omega^2} \Omega \slashed{\nabla}_4 \right)^4 \underline{A}_{\Hp} = 2r^4 \slashed{\mathcal{D}}_2^\star \slashed{\mathcal{D}}_1^\star \slashed{\mathcal{D}}_1 \slashed{\mathcal{D}}_2 A_{\Hp} - 6M (\Omega \slashed{\nabla}_3 + \Omega \slashed{\nabla}_4) A_{\Hp} + {\mathcal{E}}^3 \, , \nonumber
\end{align}
\begin{align}
-\frac{1}{2}  \frac{\Omega^4}{r^4} \left(\frac{r^2}{\Omega^2} \Omega \slashed{\nabla}_4 \right)^2  \check{\underline{\Psi}}_{\I}  &= \frac{\Omega^4}{r^4} \left(\frac{r^2}{\Omega^2} \Omega \slashed{\nabla}_4 \right)^4 \check{\underline{A}}_{\I} = 2r^4 \slashed{\mathcal{D}}_2^\star \slashed{\mathcal{D}}_1^\star \slashed{\mathcal{D}}_1 \slashed{\mathcal{D}}_2 A_{\I} - 6M (\Omega \slashed{\nabla}_3 + \Omega \slashed{\nabla}_4) A_{\I} + \check{\mathcal{E}}^3_{7/2} \, .\nonumber
\end{align}

Note that both sides of the first identity are regular at $u=u_f$ in view of the schematic identity ($h_i$ a bounded function of $r$ in $\mathcal{D}^{\Hp}$)
\begin{align}
\frac{\Omega^4}{r^4} \left(\frac{r^2}{\Omega^2} \Omega \slashed{\nabla}_4 \right)^4 \underline{A}_{\Hp} = \frac{1}{r^4} \left(r^2 \Omega \slashed{\nabla}_4 \right)^4  (\Omega^{-4} \underline{A}_{\Hp} ) + \sum_{i=0}^3 h_i \left(\Omega \slashed{\nabla}_4 \right)^i  (\Omega^{-4} \underline{A}_{\Hp} )
\end{align}

To obtain the missing fluxes in Theorem \ref{theo:mtheoalphabr},  it suffices to prove for $s=0,1,2$ and $\tau \geq u_1$:
\begin{align} \label{fing1b}
 \sup_{\tilde{v} \geq v_1} \int_{\underline{C}^{\Hp}_{\tilde{v}}} \Omega^2  \sum_{k=0}^{N-s}  | (r \Omega \slashed{\nabla}_4)^k (\Omega^{-4} \underline{A}_{\Hp})|^2 \lesssim \frac{\varepsilon^2_0 + \varepsilon^3}{v^s}
\end{align}
\begin{align} \label{fing2b}
\sup_{v \leq v_\infty} \int_{\underline{C}^{\I}_{v}(\tau)} \Bigg\{\sum_{k=0}^{N-s}  |(r \Omega \slashed{\nabla}_4)^k \check{\underline{A}}_{\I}|^2 + \sum_{k=0}^{N-1-s}   | (r \Omega \slashed{\nabla}_4)^k \check{\underline{\Pi}}_{\I}|^2  \Bigg\} \lesssim \frac{\varepsilon^2_0 + \varepsilon^3}{\tau^s}
\end{align}

For (\ref{fing2b}), using the splitting of sums argument of the previous section we reduce (after inserting (\ref{recdpib})) the problem to establishing the estimate 
$\sup_{v \leq v_\infty} \int_{\underline{C}^{\I}_{v}(\tau)} \sum_{k=2}^{N-2-s}  \frac{1}{r^2}  | (r \Omega \slashed{\nabla}_4)^k \check{\underline{\Psi}}_{\I}|^2 \lesssim \frac{\varepsilon^2_0 + \varepsilon^3}{\tau^s}$. This is easily proven by inserting the second estimate above which yields
\[
\textrm{(\ref{fing2b})}\lesssim {\mathbb{E}}_\Box^{N-s,2-s} \left[{\alpha}_{\I}\right] \left(\tau\right)+ {\mathbb{E}}_\Box^{N-s} \left[\underline{\alpha}_{\I}\right] \left(\tau\right) + \frac{\varepsilon^2_0 + \varepsilon^3}{\tau^s} \lesssim \frac{\varepsilon^2_0 + \varepsilon^3}{\tau^s} \, , 
\]  
with the last step following by applying Theorems \ref{theo:mtheoalphar} and \ref{theo:mtheoalphabr}.
To establish (\ref{fing1b}) one proceeds analogously obtaining 
\[
\textrm{(\ref{fing1b})}\lesssim {\mathbb{E}}_\Box^{N-s} \left[{\alpha}_{\Hp}\right] \left(v\right)+ {\mathbb{E}}_\Box^{N-s} \left[\underline{\alpha}_{\Hp}\right] \left(v\right) + \frac{\varepsilon^2_0 + \varepsilon^3}{v^s} \lesssim \frac{\varepsilon^2_0 + \varepsilon^3}{v^s} \, , 
\]  
with the last step following again by applying Theorems \ref{theo:mtheoalphar} and \ref{theo:mtheoalphabr}.

As mentioned above, Theorems \ref{theo:mtheoalphar} and \ref{theo:mtheoalphabr} and the estimates (\ref{fing1}), (\ref{fing2}), (\ref{fing1b}), (\ref{fing2b}) prove Theorem~\ref{thm:alphaalphabarestimates}.

\subsection{Two corollaries of the proof}
We first record explicitly the following statement, which estimates the horizon quantities on the cones of the infinity gauge. It was actually already obtained in the process of the proof of Step~3 in Sections~\ref{sec:missingfluxa} and~\ref{sec:missingfluxab} respectively. 
\begin{corollary} \label{corollary:HalphaIcones}
We have for $s=0,1,2$ and $\tau \geq u_{-1}$ the estimates
\begin{align}
\sum_{|\underline{k}|=0}^{N-1-s} \int_{\CcI_{\tau} \cap \mathcal{D}^{\Hp}}  | \mathfrak{D}^{\I}\mathfrak{D}_{\Hp}^{\underline{k}} \alpha_{\Hp}|^2 + \sum_{|\underline{k}|=0}^{N-1-s} \int_{\underline{C}^{\I}_{v}(\tau) \cap \mathcal{D}^{\Hp}}  | \mathfrak{D}^{\I}_{\nwarrow} \mathfrak{D}_{\Hp}^{\underline{k}} \alpha_{\Hp}|^2  \lesssim \frac{\varepsilon^2_0 + \varepsilon^3}{\tau^s} \, , 
\end{align}
\begin{align}
\sum_{|\underline{k}|=0}^{N-1-s} \int_{\CcI_{\tau} \cap \mathcal{D}^{\Hp}}  | \mathfrak{D}^{\I}_{\nearrow} \mathfrak{D}_{\Hp}^{\underline{k}} \underline{\alpha}_{\Hp}|^2 + \sum_{|\underline{k}|=0}^{N-1-s} \int_{\underline{C}^{\I}_{v}(\tau) \cap \mathcal{D}^{\Hp}}  | \mathfrak{D}^{\I}\mathfrak{D}_{\Hp}^{\underline{k}} \underline{\alpha}_{\Hp}|^2  \lesssim \frac{\varepsilon^2_0 + \varepsilon^3}{\tau^s} \, .
\end{align}
\end{corollary}
\begin{proof}
Localised energy estimates in the horizon gauge on the infinity null cones produce these estimates with $\mathfrak{D}^{\I}$ replaced by $\mathfrak{D}^{\I}_{\nearrow}$ in the first, and $\mathfrak{D}^{\I}$ replaced by $\mathfrak{D}^{\I}_{\nwarrow}$ in the second estimate. The additional transversal derivatives can be recovered a posteriori as stated using the 
Teukolsky--Starobinski identities as in Section~\ref{sec:teustaro}. 
\end{proof}

The second corollary illustrates the following fact: While working with the weight $\check{r}$ instead of $r$ for the underlined quantities was essential to improve the error in the wave equation for $\check{\underline{P}}$ (as compared to $\underline{P}$) in order to derive optimal estimates,  a posteriori, the estimates for the energies (\ref{abaren2}), (\ref{masterenergyab}) hold verbatim with the ``usual" weight $r$. [We do note in this context, however, that the highest $r$-weighted estimates proven for $\Omega \slashed{\nabla}_4 (r^5 \check{\underline{P}}_{\I})$ may not hold verbatim for $\Omega \slashed{\nabla}_4 (r^5 {\underline{P}}_{\I})$, related to the fact that (\ref{ridico}) below may fail for $i \leq 3$. This fact is irrelevant for future applications.] This conversion is important when one wants to estimate curvature components with the usual $r$-weights from the ($\check{r}$-weighted) almost gauge invariant quantities in Chapter \ref{chap:Iestimates}. See for instance the identities of Proposition \ref{prop:Pbartildeidentities}.

\begin{corollary} \label{cor:replacercheckbyr} 
The estimates of Theorem~\ref{thm:alphaalphabarestimates} remain true if one replaces 
\begin{align}
\check{\underline{A}}_{\I}=\check{r} \Omega^2 \underline{\alpha}_{\I} &\textrm{\ \ by \ \ } \underline{A}_{\I}={r} \Omega^2 \underline{\alpha}_{\I} \nonumber \\
\check{\underline{\Pi}}_{\I}=r^3 \Omega \check{\underline{\psi}}_{\I} &\textrm{\ \ by \ \ } \frac{r}{\check{r}}\check{\underline{\Pi}}_{\I} \nonumber \\
\check{\underline{\Psi}}_{\I}=r^5 \check{\underline{P}}_{\I}  &\textrm{\ \ by \ \ } \frac{r}{\check{r}}\check{\underline{\Psi}}_{\I}
\end{align}
 in the definition of the energies (\ref{abaren2}), (\ref{masterenergyab}). 
\end{corollary}

\begin{proof}
This is an immediate consequence of the definitions of these quantities, the estimate
\begin{align} \label{ridico}
\sup_{\mathcal{D}^{\I}} \sum_{|\underline{k}|+i \leq N; i \leq 2} \| \mathfrak{D}^{\underline{k}} \left(r^2 \Omega \slashed{\nabla}_4\right)^i \left(\frac{\check{r}}{r}-1\right) \|_{S_{u,v}} \lesssim \varepsilon
\end{align}
following easily from (\ref{eq:rtildeoverr4}) and the bootstrap assumptions.
\end{proof}

\chapter{Estimates in the $\I$ gauge: the proof of Theorem \ref{thm:Iestimates}}
\label{chap:Iestimates}

In this chapter we shall prove Theorem~\ref{thm:Iestimates}, which we restate here:

\Iestimates*

\minitoc

 In {\bf Section \ref{sec:schematic}} we derive, after recalling the gauge conditions,  a schematic form of the non-linear null structure and Bianchi equations in the infinity gauge, including a commutation principle governing the behaviour of these equations under angular commutation. 
It is also in this section where the renomalised auxiliary quantity $Y$ is introduced (which, as mentioned in the introduction, will be instrumental in estimating the quantity $\hat{\underline{\chi}}$) and its propagation equation is being derived. 

Non-linear error estimates are collected in  {\bf Section \ref{sec:nlerror}}. 

Finally, Theorem \ref{thm:Iestimates} is 
proven in {\bf Section \ref{sec:transportandelliptic}} via transport and elliptic estimates.

\vskip1pc
\noindent\fbox{
    \parbox{6.35in}{
As in the previous chapters of Part~\ref{improvingpart},
we shall assume throughout the assumptions of~\Cref{havetoimprovethebootstrap}. Let us fix an
arbitrary  $u_f\in[u_f^0, \hat{u}_f$], with $\hat{u}_f\in \mathfrak{B}$,
and fix some $\lambda \in \mathfrak{R}(u_f)$.
In this chapter,  all propositions below
shall always refer  
to the anchored $\I$ gauge in the  spacetime  $(\mathcal{M}(\lambda), g(\lambda))$,  
corresponding to parameters
$u_f$, $M_f(u_f,\lambda)$,
whose existence is
ensured by Definition~\ref{bootstrapsetdef}. 
Thus, we drop the  $\I$ superscripts for geometric quantities without risk of confusion,
writing $\alpha=\alpha_{\I}$, $C_u=C_u^{\I}$, etc.
We shall denote $M=M_f$ throughout  
this chapter.}}

\vskip1pc
\emph{From Part~\ref{improvingpart}, this chapter will only depend on Chapter~\ref{elliptandcalcchapter},~\ref{chapter:psiandpsibar} and~\ref{moreherechapter}. Moreover, Chapters~\ref{elliptandcalcchapter} and~\ref{moreherechapter} are
only appealed to through the estimate in the
statement of Theorems~\ref{thm:PPbarestimates} and~\ref{thm:alphaalphabarestimates}. Thus, the chapter can be read independently
of  Chapter~\ref{chap:comparing}  but also of the entire
gauge-invariant unit Chapters~\ref{RWtypechapter}--\ref{moreherechapter}.}

\emph{
We note that the additional schematic notation of Section~\ref{sec:schematic} will only be used in the present
chapter.  (We note, however, that as Section~\ref{sec:schematic} is exclusively algebraic, it
in fact lies outside of the proof 
of~\Cref{havetoimprovethebootstrap} and may be read immediately after Section~\ref{moreprelimchapter}.)}

\emph{
The reader may wish to compare with the proofs of both Theorems~3 and~4 of~\cite{holzstabofschw} in
Sections~13 and~14, respectively, for  
linear analogues of results proven here, in particular for the role of linearised version $\Ylin$ of quantity $Y$.}

\section{Schematic form of the equations and the quantity $Y$} \label{sec:schematic}

In this section, we provide the schematic form of the null structure and Bianchi equations in the $\I$ gauge. 

We begin in {\bf Section~\ref{sec:gaugerecall}} by recalling the gauge conditions that hold in the infinity gauge as these provide some intuition about writing the equations in the particular form given later.

We then spell out the schematic form of the equations in the $3$- and $4$-directions, in {\bf Section~\ref{Eqin3dirsec}} and
{\bf Section~\ref{Eqin4dirsec}} respectively, followed by the elliptic relations on the spheres of the double null foliation
in {\bf Section~\ref{elliptrelsechere}}. This makes use of the schematic error notation from Sections~\ref{sec:errorterms} 
and~\ref{sec:errortermsKerr}.

A commutation principle for angular commutation is stated in~{\bf Section~\ref{sec:commutationprinciple}}. 
Finally, in {\bf Section~\ref{sec:introduceY}}, the important auxiliary quantity $Y$ (and the related quantity $B$),
discussed already in the introduction Section~\ref{othernonlinearissues}, is introduced and its propagation equation is  derived. 

\subsection{Recalling the gauge conditions} \label{sec:gaugerecall}
We recall from Definition \ref{Igaugedefinition} that the solution in $\mathcal{D}^{\I}$ satisfies:
\begin{enumerate}[(1)]
\item $ r^2 T_{\ell \neq 1}   = r^2 \left(\Omega tr \chi - (\Omega tr \chi)_\circ\right)_{\ell \neq 1} =0$ on $S_{u_f,v_\infty}$ on $S_{u_f,v_\infty}$, \label{choicesphereT}

\item $r^5 \left( \slashed{div} \Omega \beta\right)_{\ell=1} =0 $ on $S_{u_f,v_\infty}$,  \label{betal1gauge}

\item $r^2 \underline{T}  = r^2 \left(\frac{tr \underline{\chi}}{\Omega} - \frac{tr \underline{\chi}_\circ}{\Omega_\circ}\right)= 0$ on $S_{u_f,v_\infty}$, \label{choicesphereTb}

\item $\mu_{\ell \geq 1}=0$ on $\underline{C}_{v_\infty}^{\I}$ \label{mugaugec}  (recall (\ref{massaspectfirstdef}) for the definition of $\mu$), 

\item $\left(\Omega^2 - \Omega_\circ^2 \right)_{\ell=0} =0$ on $\underline{C}_{v_\infty}^{\I}$, \label{zeromodegaugec} 

\item $\left({\underline{\mu}}^\dagger\right)_{\ell \geq 1}=0$ on $C_{u_{-1}}^{\I}$ (see (\ref{conjugatema}) below), \label{skrigaugeoutgoingE}

\item $(\omega- \omega_\circ)_{\ell=0}(u_{-1},v)= F(u_0) \frac{\Omega_\circ^2}{r^3} $ on $C_{u_{-1}}^{\I}$  \label{gaugel0O}, where 
$
F(u) := \frac{1}{2} \int_{u}^{u_f} d\bar{u} \int_{\bar{u}}^{u_f} d\hat{u} r^3 (\Omega \hat{\chi} \underline{\alpha})_{\ell=0} \left(\hat{u},v_\infty\right) \, ,
$

\item $b=0$ on $\underline{C}_{v_\infty}^{\I}$.  \label{gaugebvector} 

\end{enumerate}

We recall also from (\ref{renormmassaspectdf}) that 
\begin{align} \label{conjugatema}
\underline{\mu}^\dagger = \rho +2Mr^{-3} + \slashed{div} \underline{\eta} - \frac{1}{2} \hat{\chi} \underline{\hat{\chi}} + \frac{1}{2r} \Omega_\circ^2 \underline{T} \, ,
\end{align}
and from Definition \ref{Igaugedefinition} and (\ref{eq:curletalambda}), (\ref{betal1id}) respectively
\begin{align}
(\rho r^3)_{\ell=0} =: -2M \ \ \ \ \textrm{and} \ \ \ \ r^5 (\slashed{curl} \Omega \beta)_{\ell=1, m=0,\pm 1 } \in \left[-\frac{\varepsilon_0}{u_f}, \frac{\varepsilon_0}{u_f}\right] \textrm{ \ \ \ at $S_{u_f,v_\infty}$.}
\end{align}
Finally from (\ref{eq:vinfty}) we recall our definition of $v_\infty = \varepsilon_0^{-2} (u_f)^\frac{1}{\delta}$.

 \subsection{Equations in the $3$-direction}
 \label{Eqin3dirsec}
 
 The reader might wish to recall the schematic notation for non-linear error terms defined in  Sections \ref{sec:errorterms} and \ref{sec:errortermsKerr} before reading the next proposition.
\begin{proposition}[Null structure equations in the 3-direction] \label{prop:derivationofequations}
\begin{align}
\Omega \slashed{\nabla}_3 \left(r^2 {\underline{T}} \right) &= - 2 \left(\Omega^2 - \Omega_\circ^2\right)  -r^2 | \underline{\hat{\chi}}|^2 +\Omega^2 \overset{(in)}{\slashed{\mathcal{E}}^{0}_{1}}  \label{Rayni2}
 \\
\Omega \slashed{\nabla}_3 \left(Tr\right)  &= -\Omega_\circ^4 {\underline{T}} + \frac{2}{r}\Omega_\circ^2 \left(\Omega^2 - \Omega_\circ^2\right)   + \boxed{ 2 \Omega^2 r \left(\mu-\rho_\circ\right)} -\frac{4M}{r^2} \left(\Omega^2 - \Omega_\circ^2\right) +\Omega^2 \overset{(in)}{\slashed{\mathcal{E}}^{0}_{3}}  \label{uty}  
 \\
 \Omega  \slashed{\nabla}_3 \left(r^2\Omega^{-1} \underline{\hat{\chi}} \right)   &= - \underline{\alpha} r^2 + \Omega^2 \overset{(in)}{\slashed{\mathcal{E}}^{0}_{0}}  \label{3chib}  
 \\
 \Omega  \slashed{\nabla}_3 \left(r^2\Omega \hat{\chi} \right)   &= \frac{\Omega_\circ^2}{r}  \left(r^2\Omega \hat{\chi} \right)  -2 r^2 \Omega^2 \slashed{\mathcal{D}}_2^\star \eta - \Omega_\circ^4 (r \Omega^{-1} \underline{\hat{\chi}})+ \Omega^2 \overset{(in)}{\slashed{\mathcal{E}}^{0}_{1}} 
 \\
 \Omega \slashed{\nabla}_3 (r \underline{\eta}) &= - \Omega^2 \eta + \Omega r \underline{\beta}  + \Omega^2  \overset{(in)}{\slashed{\mathcal{E}}^{0}_{2}}  \label{etab3dir} 
 \\
 \Omega  \slashed{\nabla}_3 (r^2 {\eta}) &= 2r^2 \slashed{\nabla} (\underline{\omega}-\underline{\omega}_\circ) - \Omega r^2\underline{\beta} + \Omega^2  \overset{(in)}{\slashed{\mathcal{E}}^{0}_{1}} 
 \\
\Omega \slashed{\nabla}_3 \left(\omega - \omega_\circ\right) &= -\Omega^2 \left((\rho - \rho_\circ) - \frac{2M}{r^3}  \left(1- \frac{\Omega_\circ^2}{\Omega^2} \right) \right) + \Omega^2 \overset{(out)}{\slashed{\mathcal{E}}^{0}_{4}} \label{omevol}
\\
 \Omega \slashed{\nabla}_3 \left(1 - \frac{\Omega_\circ^2}{\Omega^2}\right) &= 2(\underline{\omega} - \underline{\omega}_\circ)  - (\underline{\omega} - \underline{\omega}_\circ) \left(1 - \frac{\Omega_\circ^2}{\Omega^2}\right) \label{bigomega3}
\end{align}
Moreover, for the error-term in (\ref{omevol}) we could also replace the superscript (out) by (in).

We finally collect the equation for the renormalised conjugate mass aspect (\ref{conjugatema}):
\begin{align}
\Omega \slashed{\nabla}_3 \left(r^2 \underline{\mu}^\dagger\right) = \frac{2\Omega_\circ^2}{r^2} \left(r^3 \left(\rho - \rho_\circ\right) \right) + \frac{\Omega_\circ^2}{r^2} \left( 1+ \frac{2M}{r} \right)  r^2{\underline{T}} - \frac{1 + \frac{4M}{r} }{r^2} r \left(\Omega^2 -\Omega_\circ^2\right)  - \boxed{\Omega_\circ^2 r \left(\mu-\rho_\circ\right)}+ \Omega^2 \overset{(in)}{\slashed{\mathcal{E}}^{1}_{2}} \, .\label{mumodprop}
\end{align}
\end{proposition}

\begin{proof}
Follows from the equations of Section \ref{NSEsec} and the definitions for the non-linear errors.
\end{proof}

\begin{corollary}
All $\Gamma_p$ except $\Omega^{-2} (\underline{\omega}-\underline{\omega}_\circ)$ and $b$ satisfy a schematic equation of the form
\begin{align} \label{schematicnullstructure3}
\frac{1}{\Omega} \slashed{\nabla}_3 (r^p \Gamma_p) = &  \sum h_{\overset{(in)}{\Phi_{p^\prime}^\prime}} \cdot r^{p^\prime} \overset{(in)}{ \Phi_{p^\prime}^\prime}+\sum_{\Gamma^\prime_{p^\prime} \in \{\eta, \Omega^{-2}(\underline{\omega}-\underline{\omega}_\circ) \}} h_{\Gamma^\prime_{p^\prime}} \cdot r\slashed{\nabla} (r^{p^\prime} \Gamma^\prime_{p^\prime}) + \overset{(in)}{\slashed{\mathcal{E}}^0_{0}} \, ,
\end{align}
with the first sum ranging over all ${\overset{(in)}{\Phi^\prime_{p^\prime}}}$ (of the same tensorial type as the $\Gamma_p$ appearing on the left) and $h_{{\overset{(in)}{\Phi^\prime_{p^\prime}}}}$ and $h_{\Gamma^\prime_{p^\prime}}$ being admissible coefficient functions of $r$ (see (\ref{eq:admis})).  For the same $\Gamma_p$ we also have the following alternative form arising from subtracting the Kerr reference solutions:
\begin{align} \label{schematicnullstructure3withKerr}
\frac{1}{\Omega} \slashed{\nabla}_3 (r^p (\Gamma_p-(\Gamma_p)_{\mathrm{Kerr}})) = &  \sum h_{\overset{(in)}{\Phi_{p^\prime}^\prime}} \cdot r^{p^\prime} (\overset{(in)}{ \Phi_{p^\prime}^\prime}-(\overset{(in)}{ \Phi_{p^\prime}^\prime})_{\mathrm{Kerr}}) \nonumber \\
&+\sum_{\Gamma^\prime_{p^\prime} \in \{\eta, \Omega^{-2}(\underline{\omega}-\underline{\omega}_\circ) \}} h_{\Gamma^\prime_{p^\prime}} \cdot r\slashed{\nabla} (r^{p^\prime} (\Gamma^\prime_{p^\prime}-(\Gamma^\prime_{p^\prime})_{\mathrm{Kerr}}) \
+\overset{(in)}{\slashed{\mathcal{E}}^0_{0}} +  \overset{(Kerr)}{\slashed{\mathcal{E}}} \, .
\end{align}
\end{corollary}

\begin{remark}
The point here is that derivatives of curvature do not appear on the right hand side. The only (potential)  terms containing first order derivatives are linear terms involving $\eta,\Omega^{-2}(\underline{\omega}-\underline{\omega}_\circ)$. Our notation also captures that applying a $3$-derivative to $\Gamma_p$ preserves (or improves) the overall decay in $r$.
\end{remark}

\begin{proof}
Direct inspection of the equations in Proposition \ref{prop:derivationofequations}. For the second schematic form we subtract the Kerr reference solutions. We give one example. For (\ref{etab3dir}) we observe
\[
\frac{1}{\Omega} \slashed{\nabla}_3 (r \underline{\eta}_{\mathrm{Kerr}}) = \frac{1}{\Omega} \slashed{\nabla}_3 (-\frac{1}{r^2} \sum_{m=-1}^1
	a^m
	 r {}^* \nablaslash Y^{\ell=1}_m) = -2\frac{\Omega_\circ^2}{\Omega^2} {\eta}_{\mathrm{Kerr}} -\frac{1}{r} \sum_{m=-1}^1 a^m \frac{1}{\Omega^2}\left( \Omega \slashed{\nabla}_3  r {}^* \nablaslash Y^{\ell=1}_m\right)
\]
and
\[
-  \eta_{\mathrm{Kerr}} + r (\underline{\beta}\Omega^{-1})_{\mathrm{Kerr}} = -2\eta_{\mathrm{Kerr}}
\]
which means we can write
\[
\frac{1}{\Omega}\slashed{\nabla}_3 (r \underline{\eta} - r \underline{\eta}_{\mathrm{Kerr}}) = -\eta + \eta_{\mathrm{Kerr}} + r ((\Omega^{-1} \underline{\beta})-(\Omega^{-1} \underline{\beta})_{\mathrm{Kerr}}) + \overset{(in)}{\slashed{\mathcal{E}}^{0}_{2}}  + \frac{1}{r}  \overset{(Kerr)}{\slashed{\mathcal{E}}}
\]
which after putting in an additional factor of $r$ in the bracket on the left is of the desired form.
\end{proof}

\begin{proposition}[Bianchi equations in the 3-direction] \label{prop:derivationofequations2}
\begin{align}
 \Omega \slashed{\nabla}_3 (\rho r^3 + 2M) &= - r \Omega^2 \slashed{div}  (\Omega^{-1}r^2\underline{\beta}) +3M \hat{\underline{T}} + \Omega^2 {\slashed{\mathcal{E}}^{0}_{0}}  \label{rho3dir}
 \\
  \Omega \slashed{\nabla}_3 (\sigma r^3 ) &= -\Omega^2 r \slashed{curl} (\Omega^{-1} r^2 \underline{\beta}) - \frac{1}{2} \Omega r^3 \hat{\chi} \wedge \underline{\alpha} +  \Omega^2{\slashed{\mathcal{E}}^{0}_{1}}  
 \\
 \Omega \slashed{\nabla}_3 \left(r^2 \Omega \beta\right) &=  \Omega^2 r^2 \nablaslash \rho + \Omega^2 r^2 {}^* \nablaslash \sigma + 3 {\eta} r^2\rho_\circ \Omega^2  + \Omega^2 {\slashed{\mathcal{E}}^{0}_{2}}    
 \\
  \Omega \slashed{\nabla}_3 \left(r^4 \Omega^{-1} \underline{\beta}\right) & = - \Omega^2 r^3 \slashed{div} (\Omega^{-2} r \underline{\alpha}) + \Omega^2 {\slashed{\mathcal{E}}^{0}_{-1}}   \label{betab3dir}
\end{align}
where $\underline{\hat{T}} = \Omega tr \underline{\chi} - (\Omega tr \underline{\chi})_\circ$ and we note the relation $ \hat{\underline{T}} = \Omega^2 \underline{T} -\frac{2}{r} \left(\Omega^2 - \Omega_\circ^2\right)$. 
\end{proposition}

\begin{proof}
Follows from the equations of Section \ref{Bianchiequationssec} and the definitions for the non-linear errors.
\end{proof}

\begin{corollary}
All curvature components $\mathcal{R}_p \in \{ \Omega \beta, \Omega^{-1} \underline{\beta},\rho, \sigma\}$ 
satisfy a schematic equation of the form
\begin{align}
\frac{1}{\Omega} \slashed{\nabla}_3 (r^p \mathcal{R}_p) = & \sum  \sum_{i=0}^1 h_{i,\overset{(in)}{\Phi_{p^\prime}^\prime}} \cdot \left[r \slashed{\nabla}\right]^i \left( r^{p^\prime} \overset{(in)}{ \Phi_{p^\prime}^\prime}\right) + {\slashed{\mathcal{E}}^0_{0}} \, ,
\end{align}
with the first sum ranging over all ${\overset{(in)}{\Phi^\prime_{p^\prime}}}$ and the $h_{i,{\overset{(in)}{\Phi^\prime_{p^\prime}}}}$ admissible coefficient functions of $r$ (see (\ref{eq:admis})). Alternatively, we can write
\begin{align} \label{schematicbianchi3withKerr}
\frac{1}{\Omega} \slashed{\nabla}_3 (r^p (\mathcal{R}_p-(\mathcal{R}_p)_{\mathrm{Kerr}}) = & \sum  \sum_{i=0}^1 h_{i,\overset{(in)}{\Phi_{p^\prime}^\prime}} \cdot \left[r \slashed{\nabla}\right]^i \left( r^{p^\prime} \left( \overset{(in)}{ \Phi_{p^\prime}^\prime} - (\overset{(in)}{ \Phi_{p^\prime}^\prime})_{\mathrm{Kerr}}\right) \right) + \slashed{\mathcal{E}}^0_{0} +  \overset{(Kerr)}{\slashed{\mathcal{E}}} \, .
\end{align}
\end{corollary}

\begin{proof}
The first form can be read off directly from the equations of Proposition \ref{prop:derivationofequations2}. To obtain the second form we subtract the Kerr reference solutions. We again give one example leaving the remaining straightforward computations to the reader. For $\sigma$ we compute
\begin{align}
\Omega^{-1} \slashed{\nabla}_3 (r^3 \sigma_{\mathrm{Kerr}}) &= \frac{3M}{r^2} \sum_{m=-1}^1 a^m \cdot 2 Y^{\ell=1}_m + \frac{6M}{r} \sum_{m=-1}^1 a^m \Omega \slashed{\nabla}_4 Y^{\ell=1}_m
\nonumber \\
r \slashed{curl} (r^2 (\Omega^{-1} \underline{\beta})_{\mathrm{Kerr}}) &= \frac{3M}{r^2} \sum_{m=-1}^1 a^m r^2\slashed{\Delta}  Y^{\ell=1}_m
\end{align}
Adding the two equation we see that the error on the right hand side can be incorporated into $\overset{(Kerr)}{\slashed{\mathcal{E}}}$.
\end{proof}

\subsection{Equations in the $4$-direction}
\label{Eqin4dirsec}

\begin{proposition}[Null structure equations in the 4-direction]
\begin{align}
\Omega \slashed{\nabla}_4 \left(r^2 T \Omega^{-2} \right) &= 4r \left(\omega- \omega_\circ\right)  - |\hat{\chi}|^2 r^2 - \frac{1}{2}T^2 \Omega^{-2} = 4r \left(\omega- \omega_\circ\right) + \Omega^{-2} \overset{(out)}{\mathcal{E}_2^0} \label{T4dir}
 \\
\Omega \slashed{\nabla}_4 \left(\Omega^2 \underline{{T}} r\right) &= +\Omega_\circ^2  T +4 \Omega^2 \left(\omega - \omega_\circ\right) + 2 \Omega^2 r \left(\underline{\mu}-\rho_\circ\right) + \Omega^2 \overset{(out)}{\slashed{\mathcal{E}}^{0}_{3}}   \label{Tb4dirwa} \\
\Omega \slashed{\nabla}_4 \left(\Omega^2 \underline{{T}} r^2\right)  &= +\Omega_\circ^2 r{T}+ 4r \Omega^2 \left(\omega - \omega_\circ\right) + \boxed{2 \Omega^2 r^2 \underline{\mu}^\dagger} + \Omega^2 \overset{(out)}{\slashed{\mathcal{E}}^{0}_{2}}\label{Tb4dir}
\\
 \Omega  \slashed{\nabla}_4 \left(r^2\Omega^{-1} {\hat{\chi}} \right)   &= - {\alpha} r^2 + \Omega^{-2} \overset{(out)}{\slashed{\mathcal{E}}^{0}_{2}} \label{chihat4}
\\
\Omega  \slashed{\nabla}_4 \left(r\Omega \underline{\hat{\chi}} \right)   &= -2\Omega^2 r\slashed{\mathcal{D}}_2^\star \underline{\eta} + \Omega^2 {\hat{\chi}} \Omega + \Omega^2 \overset{(out)}{\slashed{\mathcal{E}}^{0}_{3}}.   \label{chibhat4}
\\
\Omega \slashed{\nabla}_4 (r {\eta}) &=  \Omega^2 \etabar -\Omega r \beta + \overset{(out)}{\slashed{\mathcal{E}}^{0}_{3}} \label{eta4dir} 
\\
\Omega \slashed{\nabla}_4 \left(r \underline{\eta} \right) & = 2r\slashed{\nabla}{\omega} + \Omega r\beta + \overset{(out)}{\slashed{\mathcal{E}}^{0}_{2}}  \label{etab4dir} 
\\
\Omega \slashed{\nabla}_4 \left(\underline{\omega} - \underline{\omega}_\circ\right) &= -\Omega^2 \left((\rho - \rho_\circ) - \frac{2M}{r^3}  \left(1- \frac{\Omega_\circ^2}{\Omega^2} \right) \right) + \Omega^2 \overset{(in)}{\slashed{\mathcal{E}}^{0}_{4}} \, 
\label{om4dir}
\\
 \Omega \slashed{\nabla}_4 \left(1 - \frac{\Omega_\circ^2}{\Omega^2}\right) &= 2({\omega} - {\omega}_\circ)  - ({\omega} - {\omega}_\circ) \left(1 - \frac{\Omega_\circ^2}{\Omega^2}\right). \label{bigomega4}
\end{align}
Moreover, for the errorterm in (\ref{om4dir}) we could also replace the superscript (in) by (out). 

We finally collect the evolution equation of the (linearised) mass aspect (cf.~(\ref{massaspectfirstdef})):
\begin{align}
\Omega \slashed{\nabla}_4 \left(r^2 \left(\rho - \rho_\circ + \slashed{div} \eta\right) \right) &= - 2\Omega_\circ^2 r \left(\rho - \rho_\circ\right) + \frac{3M}{r}T - \frac{1}{2} \Omega_\circ^4 \underline{{T}} + \boxed{\Omega_0^2 r \underline{\mu}^\dagger} + \overset{(out)}{\slashed{\mathcal{E}}^{1}_{2}} \label{mu4dir} \, .
\end{align}
\end{proposition}

\begin{proof}
Follows from the equations of Section \ref{NSEsec} and the definitions for the non-linear errors.
\end{proof}

\begin{corollary}
All $\Gamma_p$ except $(\omega-{\omega}_\circ)$ and $b$ satisfy a schematic equation of the form
\begin{align} \label{schematicnullstructure4}
r \Omega \slashed{\nabla}_4 (r^p \Gamma_p) = &\sum h_{\overset{(out)}{\Phi_{p^\prime}^\prime}} \cdot r^{p^\prime} \overset{(out)}{ \Phi_{p^\prime}^\prime}+\sum_{\Gamma^\prime_{p^\prime} \in \{ \underline{\eta}, {\omega}-{\omega}_\circ \}} h_{\Gamma^\prime_{p^\prime}} \cdot r\slashed{\nabla} (r^{p^\prime} \Gamma^\prime_{p^\prime}) + \overset{(out)}{\slashed{\mathcal{E}}^0_{1}} \, ,
\end{align}
with the first sum ranging over all ${\overset{(out)}{\Phi^\prime_{p^\prime}}}$ (of the same tensorial type as $\Gamma_p$ on the left) and $h_{{\overset{(out)}{\Phi^\prime_{p^\prime}}}}$ and $h_{\Gamma^\prime_{p^\prime}}$ being admissible coefficient functions of $r$ (see (\ref{eq:admis})). Alternatively, we can write
\begin{align} \label{schematicnullstructure4withKerr}
r \Omega \slashed{\nabla}_4 (r^p \Gamma_p-r^p (\Gamma_p)_{\textrm{Kerr}}) = &\sum h_{\overset{(out)}{\Phi_{p^\prime}^\prime}} \left( r^{p^\prime} \overset{(out)}{ \Phi_{p^\prime}^\prime}-r^{p^\prime} (\overset{(out)}{ \Phi_{p^\prime}^\prime})_{\mathrm{Kerr}} \right) \nonumber \\
&+\sum_{\Gamma^\prime_{p^\prime} \in \{ \underline{\eta}, {\omega}-{\omega}_\circ \}} h_{\Gamma^\prime_{p^\prime}} \cdot r\slashed{\nabla} \left(r^{p^\prime} \Gamma^\prime_{p^\prime}-r^{p^\prime} (\Gamma^\prime_{p^\prime})_{\mathrm{Kerr}}\right) + \overset{(out)}{\slashed{\mathcal{E}}^0_{1}} +   \overset{(Kerr)}{\slashed{\mathcal{E}}} \, ,
\end{align}
and also, a bit more refined,
\begin{align} \label{schematicnullstructure4withKerr2}
r \Omega \slashed{\nabla}_4 (r^p \Gamma_p-r^p (\Gamma_p)_{\mathrm{Kerr}}) =& \sum h_{\overset{(in)}{\Phi_{p^\prime}^\prime}} \left( r^{p^\prime} \overset{(in)}{ \Phi_{p^\prime}^\prime}-r^{p^\prime} (\overset{(in)}{ \Phi_{p^\prime}^\prime})_{\mathrm{Kerr}} \right) + h_{\underline{\eta}} \cdot [r\slashed{\nabla}] \left(\underline{\eta}-\underline{\eta}_{\mathrm{Kerr}}\right) r^2   + \overset{(out)}{\slashed{\mathcal{E}}^0_{1}} +   \overset{(Kerr)}{\slashed{\mathcal{E}}}  \nonumber \\
&+ \frac{h_\beta}{r}  (r^4 \beta - r^4 \beta_{\mathrm{Kerr}}) +\frac{h_{\alpha}}{r} (r^4 \alpha)+ \sum_{i=0}^1 \frac{ h_{i, \omega}}{\sqrt{r}} [r\slashed{\nabla}]^i ((\omega-\omega_\circ) r^{\frac{5}{2}}) 
\end{align}
for admissible coefficient functions $h_{\underline{\eta}}$, $h_{\alpha}, h_{\beta}, h_{i, \omega}$ depending only on $r$.
\end{corollary}

\begin{remark} \label{rem:gammastru}
The form (\ref{schematicnullstructure4withKerr2}) makes manifest that the ``anomalous" quantities $\alpha$, $\beta$ and $\omega-\omega_\circ$ can only enter on the right hand side with an additional gain of at least a power of $r^{-1/2}$ (in fact, a full gain of $r^{-1}$ if $\alpha$ or $\beta$ enter). This will ensure that the right hand side decays like $u^{-1}$ at the low orders, which cannot be deduced from the form (\ref{schematicnullstructure4withKerr}).
\end{remark}

\begin{proof}
Analogous to the $3$-direction.
\end{proof}

\begin{proposition}[Bianchi equations in the 4-direction] \label{prop:derivationofequations4}
\begin{align}
 \Omega \slashed{\nabla}_4 (\rho r^3 + 2M) &= + r^3  \slashed{div} \Omega\beta +3M T +  {\slashed{\mathcal{E}}^{0}_{2}}  \label{rho4dir}
 \\
  \Omega \slashed{\nabla}_4 (\sigma r^3 + \sigma_{\mathrm{Kerr}}r^3) &= - r^3  \slashed{curl} (\Omega\beta - (\Omega \beta)_{\mathrm{Kerr}}) +   {\slashed{\mathcal{E}}^{0}_{2}} + \Omega \slashed{\nabla}_4 (\sigma_{\mathrm{Kerr}} r^3) - r^3 \slashed{curl} (\Omega \beta)_{\mathrm{Kerr}} \label{eq:sigmareno}
 \\
 \Omega \slashed{\nabla}_4 \left(r^4 \Omega^{-1} \beta\right) &= r^4 \slashed{div} \alpha + \Omega^{-2} {\slashed{\mathcal{E}}^{0}_{2}}  
 \\
  \Omega \slashed{\nabla}_4 \left(r^2 \Omega \underline{\beta}\right) & = - \Omega^2 r^2 \nablaslash \rho + \Omega^2 r^2 {}^* \nablaslash \sigma - 3 \underline{\eta} r^2\rho_\circ \Omega^2  + \Omega^2 {\slashed{\mathcal{E}}^{0}_{2}} .  \label{betab4dir}
\end{align}
\end{proposition}

\begin{proof}
Follows from the equations of Section \ref{Bianchiequationssec} and the definitions for the non-linear errors.
\end{proof}

\begin{corollary}
All curvature components $\mathcal{R}_p \in \{ \Omega \beta, \Omega^{-1} \underline{\beta},\rho, \sigma\}$ 
satisfy a schematic equation of the form
\begin{align}
r \Omega \slashed{\nabla}_4 (r^p \mathcal{R}_p) = & \sum  \sum_{i=0}^1 h_{i,\overset{(out)}{\Phi_{p^\prime}^\prime}} \cdot \left[r \slashed{\nabla}\right]^i \left( r^{p^\prime} \overset{(out)}{ \Phi_{p^\prime}^\prime}\right) + {\slashed{\mathcal{E}}^0_{1}} \, ,
\end{align}
with the first sum ranging over all ${\overset{(out)}{\Phi^\prime_{p^\prime}}}$ and the $h_{i,{\overset{(out)}{\Phi^\prime_{p^\prime}}}}$ admissible coefficient functions of $r$ (see (\ref{eq:admis})). Alternatively we can write
\begin{align} \label{schematicbianchi4withKerr}
r \Omega \slashed{\nabla}_4 (r^p \mathcal{R}_p - r^p (\mathcal{R}_p)_{\mathrm{Kerr}}) = & \sum  \sum_{i=0}^1 h_{i,\overset{(out)}{\Phi_{p^\prime}^\prime}} \cdot \left[r \slashed{\nabla}\right]^i \left( r^{p^\prime} \overset{(out)}{ \Phi_{p^\prime}^\prime}-r^{p^\prime} (\overset{(out)}{ \Phi_{p^\prime}^\prime})_{\mathrm{Kerr}}\right) + {\slashed{\mathcal{E}}^0_{1}}+ r \cdot   \overset{(Kerr)}{\slashed{\mathcal{E}}}  \, ,
\end{align}
where also $ r \cdot   \overset{(Kerr)}{\slashed{\mathcal{E}}}$ can be replaced by $\overset{(Kerr)}{\slashed{\mathcal{E}}}$ unless $\mathcal{R}_p=\Omega \beta$. 

Finally, for the $\mathcal{R}_p \in \{\Omega^{-1} \underline{\beta},\rho,\sigma\}$, i.e.~the curvature components in $\overset{(in)}{\Phi}$, we can also write
\begin{align} \label{schematicbianchi4withKerr2}
r \Omega \slashed{\nabla}_4 (r^p \mathcal{R}_p - r^p (\mathcal{R}_p)_{\mathrm{Kerr}}) = & \sum  \sum_{i=0}^1 h_{i,\overset{(in)}{\Phi_{p^\prime}^\prime}} \cdot \left[r \slashed{\nabla}\right]^i \left( r^{p^\prime} \overset{(in)}{ \Phi_{p^\prime}^\prime}-r^{p^\prime} (\overset{(in)}{ \Phi_{p^\prime}^\prime})_{\mathrm{Kerr}}\right) + {\slashed{\mathcal{E}}^0_{1}}+  \overset{(Kerr)}{\slashed{\mathcal{E}}} \nonumber \\&+ \frac{h_{\beta}}{r} [r\slashed{\nabla}] (r^4 \beta r^4 - r^4 \beta_{\mathrm{Kerr}})   \, .
\end{align}
\end{corollary}

\begin{remark} \label{rem:Rstru}
The form (\ref{schematicbianchi4withKerr2}) makes manifest that if an $r\Omega \slashed{\nabla}_4$ derivatives falls on a non-anomalous curvature component, then an anomalous component can only appear on the right hand side with an additional gain of $r^{-1}$. See Remark \ref{rem:gammastru}.
\end{remark}

\begin{proof}
Analogous to the $3$-direction. 
\end{proof}

\subsection{Elliptic relations}
\label{elliptrelsechere}

\begin{proposition} \label{prop:derivationofequations5}
We have the following elliptic relations:
\begin{align}
r^3 \slashed{curl} \eta &= -r^3 \slashed{curl} \underline{\eta} = r^3 \sigma - \frac{1}{2} r^3 \hat{\chi} \wedge \hat{\underline{\chi}} \label{curle}  \\
r^3 \slashed{div} \eta & = \boxed{r^3 \mu +2M_f} -r^3\rho -2M_f + \frac{1}{2}r^3 \hat{\chi} \hat{\underline{\chi}} \label{dive} \\
r^3 \slashed{div} \underline{\eta} &= \boxed{r^3 \underline{\mu}^\dagger} -r^3 \rho - 2M_f + \frac{1}{2}r^3 \hat{\chi} \hat{\underline{\chi}} -\frac{1}{2} \Omega_\circ^2 r^2 \underline{T} \label{diveb} \\
2 (r^2\slashed{\Delta} + \Omega_\circ^2) \left( \omega - \omega_\circ\right) &= \boxed{\Omega \slashed{\nabla}_4 (r^2 \underline{\mu}^\dagger)} - 2r^2 \slashed{div} \left(\Omega \beta\right) + (Tr^3)  \left(-\frac{\Omega_\circ^2}{2r^3} - \frac{3M}{r^4}\right) +  \overset{(out)}{\slashed{\mathcal{E}}^{1}_{3}}  \label{ellipticomega}  \\
2r^2 \slashed{\Delta} \underline{\omega} &= \boxed{\frac{1}{r} \Omega \slashed{\nabla}_3 (r^3 {\mu})} + 2r^2 \Omega^2 \slashed{div} \left(\Omega^{-1} \underline{\beta}\right) - (\hat{\underline{T}}r^2) \left(\frac{3M}{r^3}\right)  +\frac{1}{2} \frac{\Omega_\circ^2}{r} r^2|\underline{\hat{\chi}}|^2  + \overset{(in)}{\slashed{\mathcal{E}}^{1}_{2}}   \label{ellipticomegab}
\end{align}
and
\begin{align}
\left[r^2\slashed{\mathcal{D}}_2^\star \slashed{div} +\frac{3M}{r}\right] r^2 \Omega \hat{\chi}  &= -r^4 \Omega \psi - \Omega_\circ^2 r^3 \slashed{\mathcal{D}}_2^\star\underline{\eta} + \frac{1}{2}r^2 \slashed{\mathcal{D}}_2^\star \slashed{\nabla} (r^2 T) +  \overset{(out)}{\slashed{\mathcal{E}}^{1}_{1}}\label{codaz} \\
\left[r^2\slashed{\mathcal{D}}_2^\star \slashed{div} +\frac{3M}{r}\right] r \Omega^{-1}\underline{\hat{\chi}} &= r^3 \Omega^{-1} \frac{r}{\check{r}} \check{\underline{\psi}} + \frac{1}{r} r^3 \slashed{\mathcal{D}}_2^\star \eta + \frac{1}{2r} r^2 \slashed{\mathcal{D}}_2^\star \slashed{\nabla} (r^2 \underline{T})  + \overset{(in)}{\slashed{\mathcal{E}}^{1}_{1}} + \check{\mathcal{E}}^0_1 \, . \label{codazb}
\end{align}
Finally, the Gauss equation
\begin{align} \label{Gauss}
K-K_\circ = -\rho + \rho_\circ + \frac{1}{2} \hat{\chi} \underline{\hat{\chi}} - \frac{1}{2} \frac{\Omega_\circ^2}{r} \underline{T} + \frac{1}{2} \frac{1}{r} T -\frac{1}{4} T \underline{T} = \boxed{- \mu +\rho_\circ} + \slashed{div} \eta - \frac{1}{2} \frac{\Omega_\circ^2}{r} \underline{T} + \frac{1}{2} \frac{1}{r} T -\frac{1}{4} T \underline{T}.
\end{align}
\end{proposition}

\begin{proof}
Direct computation from the Bianchi and null structure equations. For (\ref{codazb}) recall Proposition \ref{prop:Pbartildeidentities}, whose proof also gives that $\check{\mathcal{E}}^0_1$ does not contain the components $\alpha, \beta, \omega$.
\end{proof}

\begin{remark}
The linear terms in boxes will vanish (after projection of the equation to $\ell \geq 1$ or equivalently angular commutation) up to non-linear terms along one of the hypersurfaces defining the gauge, as seen from the gauge conditions (\ref{skrigaugeoutgoingE}) and (\ref{mugaugec}).
\end{remark}

\begin{remark} \label{rem:muexplain}
Note already that after projection to $\ell \geq 1$ the right hand side of (\ref{mumodprop}) decays as $r^{-2}$ suggesting that the quantity $\left(r^3\underline{\mu}^\dagger\right)_{\ell \geq 1}$ is \underline{conserved} along null-infinity. This is the main reason for introducing the renormalised mass aspect (\ref{conjugatema}). See Proposition \ref{prop:mubari}.
\end{remark}

\subsection{Commutation principle} \label{sec:commutationprinciple}
We finally note the following commutation principle, which is tailored to be applied to the schematic transport equations in the previous subsection (and we will do so frequently). The main point here is that the error from commuting the transport operator with angular derivatives can be incorporated into the non-linear error with the correct $r$-weight by Lemma \ref{lem:commutation}.
\begin{lemma} \label{lem:commutationprinciple}
The transport equations (\ref{Rayni2})--(\ref{mumodprop}) and (\ref{T4dir})--(\ref{mu4dir}) are schematically of the form 
\[
\Omega \slashed{\nabla}_3 (A) = B + \overset{(in)}{\slashed{\mathcal{E}}^{k}_{p}} \ \ \ \textrm{and} \ \ \ \Omega \slashed{\nabla}_4 (A^\prime) = B^\prime + \overset{(out)}{\slashed{\mathcal{E}}^{k^\prime}_{p^\prime}}
\]
respectively, with the obvious identifications for $A,B$ and $A^\prime, B^\prime$. Under angular commutation, these equations can be expressed in the following schematic form ($l \geq 0$)
\[
\Omega \slashed{\nabla}_3 ( \left[r\slashed{\nabla}\right]^l (A)) = \left[r\slashed{\nabla}\right]^l B + \overset{(in)}{\slashed{\mathcal{E}}}{}^{k+l}_p  \ \ \ \textrm{and} \ \ \ \ \Omega \slashed{\nabla}_4 ( \left[r\slashed{\nabla}\right]^l (A^\prime)) = \left[r\slashed{\nabla}\right]^l B^\prime + \overset{(out)}{\slashed{\mathcal{E}}}{}^{k^\prime+l}_{p^\prime} \, ,
\]
with the exception of (\ref{omevol}) and (\ref{om4dir}) for which one has 
\[
\Omega \slashed{\nabla}_3 ( \left[r\slashed{\nabla}\right]^l (\omega -\omega_\circ)) = \left[r\slashed{\nabla}\right]^l B + \overset{(out)}{\slashed{\mathcal{E}}}{}^{l}_4 + \slashed{\mathcal{E}}^{l-1}_p   \ \ \ \textrm{and} \ \ \ \ \Omega \slashed{\nabla}_4 ( \left[r\slashed{\nabla}\right]^l (\underline{\omega} - \underline{\omega}_\circ)) = \left[r\slashed{\nabla}\right]^l B^\prime + \overset{(in)}{\slashed{\mathcal{E}}}{}^{l}_{p^\prime} + \slashed{\mathcal{E}}^{l-1}_{p^\prime} \, .
\]
\end{lemma} 
\begin{proof}
Follows directly from the definition of the error, (\ref{schematicerroradd}), and applying inductively Lemma \ref{lem:commutation} for the commutators. The exceptional cases follow from the fact that they involve the $3$-derivative of a quantity not in $\overset{(in)}{\Gamma}$ and the $4$-derivative of a quantity not in $\overset{(out)}{\Gamma}$. Note that by Lemma \ref{lem:commutation} the commutator of a $4$-derivative and an angular derivative produces only terms from $\overset{(in)}{\Gamma}$ except for a lower order curvature term involving $\beta$ which is incorporated into $\slashed{\mathcal{E}}^{l-1}_p$ in the second formula.
\end{proof} 

One similarly has for the Bianchi equations:
\begin{lemma}
The Bianchi equations (\ref{rho3dir})--(\ref{betab3dir}) and  (\ref{rho4dir})--(\ref{betab4dir}) are schematically of the form 
\[
\Omega \slashed{\nabla}_3 (A) = B +{\slashed{\mathcal{E}}^{k}_{p}} \ \ \ \textrm{and} \ \ \ \Omega \slashed{\nabla}_4 (A^\prime) = B^\prime +{\slashed{\mathcal{E}}^{k}_{p^\prime}}
\]
respectively, with the obvious identifications for $A,B$ and $A^\prime, B^\prime$. Under angular commutation, these equations can be expressed in the following schematic form ($l \geq 0$)
\[
\Omega \slashed{\nabla}_3 ( \left[r\slashed{\nabla}\right]^l (A)) = \left[r\slashed{\nabla}\right]^l B + \slashed{\mathcal{E}}{}^{k+l}_p  \ \ \ \textrm{and} \ \ \ \ \Omega \slashed{\nabla}_4 ( \left[r\slashed{\nabla}\right]^l (A^\prime)) = \left[r\slashed{\nabla}\right]^l B^\prime + \slashed{\mathcal{E}}{}^{k+l}_{p^\prime} \, .
\]
\end{lemma}

\subsection{The auxiliary quantities $Y$ and $B$} \label{sec:introduceY}
For this subsection we recall the definitions of the almost gauge invariant quantities $\check{\underline{A}}_{\I}=\check{r} \Omega^2 \underline{\alpha}_{\I}$, $\check{\underline{\Pi}}_{\I} = r^3 \Omega \check{\underline{\psi}}_{\I}$ and $\check{\underline{\Psi}}_{\I}=r^5\check{\underline{P}}_{\I}$, which according to our convention at the beginning of the chapter appear without the $\mathcal{I}^+$ subscript here.  We define the quantity\index{double null gauge!connection coefficients!$Y$, quantity used only in the $\mathcal{I}^+$ gauge}
\begin{align} \label{Ydef}
Y &:= 
-3M r \underline{\hat{\chi}}\Omega^{-1} - r^3 \slashed{\mathcal{D}}_2^\star \underline{\eta} + \frac{1}{2} r^2 \slashed{\mathcal{D}}_2^\star  \slashed{\nabla} (r^2 \underline{T}) \, .
\end{align}
The point of this quantity is that it can be used to prove estimates on $\underline{\hat{\chi}}$ using the following relation.

\begin{lemma} \label{lem:Yrel}
We have the relation
\begin{align} \label{relYgi}
r^2 \mathcal{D}_2^\star \slashed{div} \left(\underline{\hat{\chi}} r \Omega^{-1}\right)  = \frac{Y}{r} +  \frac{r}{\check{r}} \check{\underline{\Pi}} \Omega^{-2} 
+ \overset{(in)}{\slashed{\mathcal{E}}^1_1} +  \check{\mathcal{E}}^0_1\, .
\end{align}
\end{lemma}
\begin{proof}
This is a rewriting of  (\ref{codazb}).
\end{proof}

Using the null structure equations, we next derive a propagation equation for $Y$.

\begin{lemma}
The quantity $Y$ defined in (\ref{Ydef}) satisfies the propagation equation
\begin{align} \label{Yeq}
\Omega \slashed{\nabla}_3 Y = - \frac{r}{\check{r}} \check{\underline{\Pi}} + 3M  \frac{r}{\check{r}} \Omega^{-2} \check{\underline{A}} +\overset{(in)} {\slashed{\mathcal{E}}^1_1}  + \check{\mathcal{E}}^0_1 + \mathcal{E}_{anom} \, ,
\end{align}
where 
\begin{align} \label{anomalous}
\mathcal{E}_{anom}=  - \frac{1}{2} r^2 \slashed{\mathcal{D}}_2^\star  \slashed{\nabla} (r^2 |\underline{\hat{\chi}}|^2) +  \frac{1}{2} \left[\Omega \slashed{\nabla}_3 ,r^2 \slashed{\mathcal{D}}_2^\star  \slashed{\nabla}\right] (r^2 \underline{T})  
\end{align}
and $\check{\mathcal{E}}^0_1$ does not contain the components $\alpha, \beta, \omega$.
\end{lemma}
\begin{proof}
Direct computation starting from (\ref{Ydef}) using Lemma \ref{lem:commutation} and (\ref{etab3dir}), (\ref{3chib}), (\ref{Rayni2}) as well as Proposition \ref{prop:Pbartildeidentities}.
\end{proof}

\begin{remark}
Note that compared with $\overset{(in)} {\slashed{\mathcal{E}}^1_1}  + \check{\mathcal{E}}^0_1$ the isolated anomalous non-linear error $\mathcal{E}_{anom}$ contains both terms with higher derivatives and with less $r$-decay.
\end{remark}
To improve the regularity of the right hand side of (\ref{Yeq}) we first commute the equation with the operator $\mathscr{A}^{[2]}+1$ and renormalise. We use here and below the following definition
for $k\geq 0$ of the (elliptic) angular operators acting on symmetric traceless $S$-tensors\index{double null gauge!differential operators!$\mathscr{A}^{[k]}$, angular operators acting on symmetric traceless $S$-tensors}:
\begin{align}
\mathscr{A}^{[k]} := \left( r^2 \Dslash_2^*\slashed{div} \right)^{\frac{k}{2}}  \textrm{ if $k$ is even \ \ \ \ \  and} \ \ \ \ \  \mathscr{A}^{[k]}:= r\slashed{div} \mathscr{A}^{[k-1]}  \textrm{ \ \ \ if $k$ is odd.}
\end{align}
Note that these operators already appeared in the proof of Proposition \ref{prop:piallbextra}.

More specifically we define the quantity\index{double null gauge!connection coefficients!$B$, quantity used only in the $\mathcal{I}^+$ gauge} 
\begin{align} \label{Bdef}
B :=  \mathscr{A}^{[2]} Y +Y + \frac{1}{2} \frac{r}{\check{r}} \check{\underline{\Psi}} + 3M \frac{r}{\check{r}} \check{\underline{\Pi}} \Omega^{-2} \, ,
\end{align}
for which we can derive the following evolution equation:
\begin{lemma}
The quantity $B$ defined in (\ref{Bdef}) satisfies for $i\geq 0$ the evolution equation
\begin{align} \label{Beq}
\Omega \slashed{\nabla}_3 \left(\mathscr{A}^{[i]} B\right)  = \frac{1}{r}  \frac{r}{\check{r}} h \cdot \mathscr{A}^{[i]} \check{\underline{\Pi}}  + \frac{r}{\check{r}} \tilde{h}\cdot  \mathscr{A}^{[i]} \check{\underline{A}} +  \mathcal{E} \left[\mathscr{A}^{[i]} B\right] 
\end{align}
with $h$ and $\tilde{h}$ admissible coefficient functions of $r$  (see (\ref{eq:admis})) and the non-linear error is given by
\begin{align} \label{errorB1}
\mathcal{E} \left[\mathscr{A}^{[i]} B\right]   = &   \mathscr{A}^{[i]}  \frac{r^2}{\Omega^2}\mathcal{E}[{\check{\underline{\psi}}}]  
+\left(\mathscr{A}^{[i+2]} + \mathscr{A}^{[i]}\right) \mathcal{E}_{anom} +\overset{(in)}{ \slashed{\mathcal{E}}^{3+i}_1}+ \frac{\check{r}}{r} \overset{(in)}{ \slashed{\mathcal{E}}^{3+i}_1} + \check{\mathcal{E}}^{2+i}_1 \, ,
\end{align} 
where moreover the components $\alpha, \beta, \omega$ do not appear in the $\check{\mathcal{E}}^{2+i}_{1}$. See (\ref{psibtex}) for the definition of $\mathcal{E}[{\check{\underline{\psi}}}]$ and (\ref{anomalous}) for that of $\mathcal{E}_{anom}$.
\end{lemma}

\begin{proof}
We prove the identity for $i=0$. The case $i> 0$ will then follow by straightforward commutation. 
We first recall from Propositions \ref{prop:Teukolskybartilde} and \ref{prop:psibartildeeq} the wave equations for the almost gauge invariant quantities
\begin{align} \label{wrf}
 \left(\mathscr{A}^{[2]}+1\right)  \check{\underline{\Pi}}=  +\frac{1}{2} \Omega \slashed{\nabla}_3 \check{\underline{\Psi}} +  \frac{3M}{r} \check{\underline{\Pi}}  + \frac{3M}{2} \check{\underline{A}}+ \frac{r^2}{\Omega^2}\mathcal{E}[{\check{\underline{\psi}}}] - 2(\underline{\omega}-\underline{\omega}_\circ) \check{\underline{\Psi}} \, , 
\end{align}
\begin{align} \label{wrf2}
\left(\mathscr{A}^{[2]}+1\right)\check{\underline{A}}= -\Omega \slashed{\nabla}_3 \check{\underline{\Pi}} + \left(1-\frac{3M}{r}\right)\left(\frac{2}{r}  \check{\underline{\Pi}} + \check{\underline{A}}\right) +\check{\mathcal{E}}^2_1 \, ,
\end{align}
where the components $\alpha, \beta, \omega$ do not appear in the $\check{\mathcal{E}}^2_1$.

Commuting the equation for $Y$, equation (\ref{Yeq}), with the angular operator $\mathscr{A}^{[2]}+1$ yields
\begin{align} 
\Omega \slashed{\nabla}_3 \left(\left(\mathscr{A}^{[2]}+1\right) Y\right) = & -\left(\mathscr{A}^{[2]}+1\right) \frac{r}{\check{r}} \check{\underline{\Pi}} + 3M \left(\mathscr{A}^{[2]}+1\right)  \frac{r}{\check{r}} (\Omega^{-2}\check{\underline{A}}) \nonumber \\
&+ \left(\mathscr{A}^{[2]}+1\right) 
\left(\overset{(in)} {\slashed{\mathcal{E}}^1_1}  + \check{\mathcal{E}}^0_1 + \mathcal{E}_{anom} \right)+ \left[\Omega \slashed{\nabla}_3, \mathscr{A}^{[2]}\right] Y . \nonumber
\end{align}
Commuting the $\mathscr{A}^{[2]}$ past the $\frac{r}{\check{r}}$ and inserting the wave equations for $\check{\underline{A}}$ and $\check{\underline{\Pi}}$ produces 
\begin{align}
\Omega \slashed{\nabla}_3 \left(\mathscr{A}^{[2]}Y + Y + \frac{1}{2}\frac{r}{\check{r}} \check{\underline{\Psi}} +3M\frac{r}{\check{r}} \Omega^{-2} \check{\underline{\Pi}} \right) = \frac{1}{r} h \cdot\check{\underline{\Pi}} + \tilde{h} \cdot \check{\underline{A}} + \mathcal{E}\left[B\right] \, , \, 
\end{align}
from which the result is immediate for $i=0$. Note that the commutator term $\left[\Omega \slashed{\nabla}_3, \mathscr{A}^{[2]}\right] Y$ can indeed be incorporated into the schematic terms after inserting (\ref{Ydef}) for $Y$. To establish the identity for $i> 0$ commute the $i=0$ equation with $\mathscr{A}^{[i]}$ and observe that the commutator term $\left[\Omega \slashed{\nabla}_3, \mathscr{A}^{[i]}\right] B$ can be incorporated into the schematic error terms after inserting (\ref{Bdef}) and (\ref{Ydef}). 
\end{proof}

We finally commute the equation (\ref{Beq}) with $\Omega \slashed{\nabla}_4$.

\begin{lemma}
The quantity $\Omega \slashed{\nabla}_4 B$ satisfies for $i \geq 0$ the evolution equation
\begin{align} \label{n4Beq}
\Omega \slashed{\nabla}_3 \left( \Omega \slashed{\nabla}_4 \mathscr{A}^{[i]} B \right)= \frac{h_1}{r^3} \mathscr{A}^{[i]} \check{\underline{\Psi}} + \frac{h_{2}}{r^2} \mathscr{A}^{[i]} \check{\underline{\Pi}} + \frac{h_3}{r^2}  \mathscr{A}^{[i]} \check{\underline{A} } + \mathcal{E} \left[ \Omega \slashed{\nabla}_4  \mathscr{A}^{[i]} B\right]
\end{align}
with $h_1,h_2,h_3$ admissible coefficient functions of $r$  (see (\ref{eq:admis})) and the non-linear error
\begin{align} \label{nlerror4Bdef}
\mathcal{E} \left[ \Omega \slashed{\nabla}_4  \mathscr{A}^{[i]} B\right] = &-(\eta - \underline{\eta})^C \slashed{\nabla}_C  \mathscr{A}^{[i]} B + \Omega \slashed{\nabla}_4 \left(\mathscr{A}^{[i+2]} + \mathscr{A}^{[i]}\right) \mathcal{E}_{anom}+ \Omega \slashed{\nabla}_4  \left( \mathscr{A}^{[i]} \frac{r^2}{\Omega^2}  \mathcal{E}[{\check{\underline{\psi}}}] \right) \nonumber \\
&+  \check{\mathcal{E}}^{4+i}_{2} \left(\textrm{no $\alpha$, $\beta$, $\omega-\omega_\circ$}\right)+ \check{\mathcal{E}}^{4+i}_{5/2}\left(\textrm{no $\alpha$, $\beta$}\right) +  \check{\mathcal{E}}^{4+i}_{3} \, , 
\end{align}
where we recall the definitions (\ref{psibtex}) and (\ref{anomalous}).
\end{lemma}

\begin{remark}
Note that the first three terms in (\ref{n4Beq}) are \emph{linear} and that the non-linear error involves $i+1$ angular derivatives of $B$.  
\end{remark}

\begin{proof}
We commute (\ref{Beq}) by $\Omega\slashed{\nabla}_4$.  Note that the first term in the error arises from the commutator $\left[\Omega \slashed{\nabla}_3, \Omega \slashed{\nabla}_4 \right] B$ (cf.~Lemma \ref{lem:commutation}) and that the lower order parts of this commutator can be incorporated into the schematic terms. Note also that  $\Omega \slashed{\nabla}_4 \overset{(in)}{ \mathcal{E}^{3+i}_1}$ and $\Omega \slashed{\nabla}_4 \check{\mathcal{E}}^{2+i}_1$ (with $\check{\mathcal{E}}^{2+i}_1$ not containing $\alpha, \beta, \omega-\omega_\circ$) can be incorporated into the last line: Indeed $\alpha$, $\beta$ or $\omega-\omega_\circ$ can only enter through a $4$-equation for $\Phi \setminus \{\alpha, \beta , \omega-\omega_\circ\}$ with an improvement of $r$-decay by two powers if $\alpha$ and $\beta$ enter, and a gain of $r^{-3/2}$ if $\omega - \omega_\circ$ enters (cf.~Remarks \ref{rem:gammastru} and \ref{rem:Rstru}).
\end{proof}

\section{Non-linear error estimates} \label{sec:nlerror}

In this section we collect various error estimates that will be crucial when obtaining estimates from the schematic equations of Section \ref{sec:schematic}.

\subsection{Preliminaries} 

\subsubsection{Bootstrap assumptions for quantities without reference Kerr solution subtracted}

We first note the following important consequence of the bootstrap assumption $\mathbb{E}^K_{u_f,\I} \lesssim \varepsilon^2$. 

\begin{lemma} \label{lem:alsowithout}
The bootstrap assumption $\mathbb{E}^K_{u_f,\I} \lesssim \varepsilon^2$ holds verbatim if in all energies that constitute $\mathbb{E}^K_{u_f,\I}$ (cf.~(\ref{estart})--(\ref{themasterweuseIplus})), the Kerr part is \underline{not} subtracted from the quantities that appear.
\end{lemma}

\begin{proof}
This follows from checking that the Kerr parts individually satisfy all of the bootstrap assumptions, which in turn is a consequence of the definitions of Section \ref{reflinearisedKerrsec} and the estimate $|J^m_{\I}| \lesssim \frac{\varepsilon}{u_f}$ for $m=-1,0,1$.
\end{proof}

The point to make here is that while the bootstrap assumptions hold without the reference Kerr solutions subtracted, we will only be able to \emph{improve} the assumptions \emph{with} the Kerr modes subtracted.

\subsubsection{Basic estimates on spheres for errors $ \overset{(in)}{\slashed{\mathcal{E}}^{n}_{p}}$ and $ \overset{(out)}{\slashed{\mathcal{E}}^{n}_{p}}$}
We next note the following basic estimates for the non-linear errors involving only angular derivatives:
\begin{lemma} \label{lem:basicerrorse}
We have for $n=0,1,...,N+1$ and any sphere $S_{u,v}$ in $\mathcal{D}^{\I}$ the following estimates:
\begin{align}
\| r^p \overset{(in)}{\slashed{\mathcal{E}}^{n}_{p}} \|_{S_{u,v}} \lesssim  \left[\sum_{\Gamma}\sum_{i=0}^n   \| \left[r \slashed{\nabla}\right]^{i} r^{p_1} \overset{(in)}{\Gamma_{p_1}}\|_{S_{u,v}} +  \sum_{\Phi} \sum_{i=0}^{n-1}  \| \left[r \slashed{\nabla}\right]^i r^{p_2} \overset{(in)}{\Phi_{p_2}}\|_{S_{u,v}} \right] \sum_{\Phi} \sum_{j=0}^{N-6} \| \left[r \slashed{\nabla}\right]^{j} r^{p_3} \overset{(in)}{\Phi_{p_3}}\|_{S_{u,v}} \, , \nonumber
\end{align}
\begin{align}
\| r^p \overset{(out)}{\slashed{\mathcal{E}}^{n}_{p}} \|_{S_{u,v}} \lesssim  \left[\sum_{\Gamma}\sum_{i=0}^n   \| \left[r \slashed{\nabla}\right]^{i} r^{p_1} \overset{(out)}{\Gamma_{p_1}}\|_{S_{u,v}} +  \sum_{\Phi} \sum_{i=0}^{n-1}  \| \left[r \slashed{\nabla}\right]^i r^{p_2} \overset{(out)}{\Phi_{p_2}}\|_{S_{u,v}} \right] \sum_{\Phi} \sum_{j=0}^{N-6} \| \left[r \slashed{\nabla}\right]^{j} r^{p_3} \overset{(out)}{\Phi_{p_3}}\|_{S_{u,v}} \nonumber \, .
\end{align}
Moreover, the bootstrap assumption $\slashed{\mathbb{E}}^N_{u_f,\I} \lesssim \varepsilon^2$ implies for $s \in \{0,1\}$ the estimates 
\begin{align}
\sum_{\Phi} \sum_{j=0}^{N-6} \| \left[r \slashed{\nabla}\right]^{j} r^{p_3} \overset{(in)}{\Phi_{p_3}}\|_{S_{u,v}} \lesssim \frac{\varepsilon}{u}  \ \ \ \ \textrm{and} \ \ \  \sum_{\Phi} \sum_{j=0}^{N-6} \frac{1}{r^{\frac{s}{2}}} \| \left[r \slashed{\nabla}\right]^{j} r^{p_3} \overset{(out)}{\Phi_{p_3}}\|_{S_{u,v}} \lesssim \frac{\varepsilon}{u^{\frac{1}{2} + \frac{s}{2}}} \, .
\end{align}
\end{lemma}
\begin{proof}
Recall from (\ref{inerrorschematic}) and (\ref{inerrorschematic2}) that $\overset{(in)}{\slashed{\mathcal{E}}^{n}_{p}}$ denotes a sum of terms which are products of (angular derivatives of) the $\overset{(in)}{\Phi}$. For \emph{quadratic} terms, the estimate follows directly from Cauchy-Schwarz on $S_{u,v}$. For cubic and higher order errors, additional $L^\infty$-Sobolev embedding on $S_{u,v}$ leads to the same estimate (with additional $\varepsilon$ powers) after inserting the bootstrap assumptions. The proof of the second estimate is identical. The third estimate follows directly from the bootstrap assumptions.
\end{proof}

The next lemma looks at $3$- and $4$-derivatives applied to the schematic error.
\begin{lemma} \label{lem:34onerror}
For $n\geq 0$, $m\geq 1$ and $m+n \leq N+1$
\begin{align} \label{awf}
\| \left(\Omega^{-1} \slashed{\nabla}_3\right)^m \overset{(in)}{\slashed{\mathcal{E}}^{n}_{p}} \|_{S_{u,v}} \lesssim & \frac{1}{r^p}  \left[  \sum_{\Phi} \sum_{|\gamma|\leq m+n-1}  \| \mathfrak{D}^\gamma r^{p_2} \overset{(in)}{\Phi_{p_2}}\|_{S_{u,v}} \right] \sum_{\Phi} \sum_{|\gamma|\leq N-6} \| \mathfrak{D}^\gamma r^{p_3} \overset{(in)}{\Phi_{p_3}}\|_{S_{u,v}} \,  \nonumber \\
+&\frac{1}{r^p} \left[  \sum_{i=0}^m \| \left[\Omega^{-1} \slashed{\nabla}_3\right]^{i} \left[r \slashed{\nabla}\right]^{n+i} (r \Omega^{-2} (\underline{\omega}- \underline{\omega}_\circ)) \|_{S_{u,v}} \right] \sum_{\Gamma} \| r^p \overset{(in)}{\Gamma_p} \|_{S_{u,v}} \nonumber \\
+&\frac{1}{r^p} \left[ \sum_{\Gamma \in \{\eta, \underline{\eta}\}} \| \left[r \slashed{\nabla}\right]^{m+n} (r^2 \Gamma) \|_{S_{u,v}}  \right] \sum_{\Gamma} \| r^p \overset{(in)}{\Gamma_p} \|_{S_{u,v}} .
\end{align}
The expressions outside the square brackets are $\lesssim \varepsilon r^{-p} u^{-1}$ by the bootstrap assumptions. In addtion
\begin{align} \label{awf2}
\| \left(r \Omega \slashed{\nabla}_4\right)^m \overset{(out)}{\slashed{\mathcal{E}}^{n}_{p}} \|_{S_{u,v}} \lesssim & \frac{1}{r^p}  \left[  \sum_{\Phi} \sum_{|\gamma|\leq m+n-1}  \| \mathfrak{D}^\gamma r^{p_2} \overset{(out)}{\Phi_{p_2}}\|_{S_{u,v}} \right] \sum_{\Phi} \sum_{|\gamma|\leq N-6} \| \mathfrak{D}^\gamma r^{p_3} \overset{(out)}{\Phi_{p_3}}\|_{S_{u,v}} \,  \nonumber \\
+&\frac{1}{r^p} \left[  \sum_{i=0}^m \| \left[\Omega^{-1} \slashed{\nabla}_4\right]^{i} \left[r \slashed{\nabla}\right]^{n+i} (r^{\frac{5}{2}} ({\omega}- {\omega}_\circ)) \|_{S_{u,v}} \right] \sum_{\Gamma} \| r^p \overset{(out)}{\Gamma_p} \|_{S_{u,v}} \nonumber \\
+&\frac{1}{r^p} \left[ \sum_{\Gamma \in \{\eta, \underline{\eta}\}} \| \left[r \slashed{\nabla}\right]^{m+n} (r^2 \Gamma) \|_{S_{u,v}}  \right] \sum_{\Gamma} \| r^p \overset{(out)}{\Gamma_p} \|_{S_{u,v}} .
\end{align}
The expressions outside the square brackets are $\lesssim \varepsilon r^{-p} u^{-1/2}$ by the bootstrap assumptions. 
\end{lemma}
\begin{proof}
Recall from (\ref{inerrorschematic}) and (\ref{inerrorschematic2}) that $\overset{(in)}{\slashed{\mathcal{E}}^{n}_{p}}$ denotes a sum of terms which are products of (angular derivatives of) the $\overset{(in)}{\Phi}$. Applying $\left(\Omega^{-1} \slashed{\nabla}_3\right)$ to the second term on the right of (\ref{inerrorschematic}) we see that at most $m+n-1$ derivatives appear and hence the resulting expression can be estimated by first term on the right of (\ref{awf}). For the top order term  in the first line of (\ref{inerrorschematic}) (involving $k_1=n$ derivatives) we commute the $\Omega \slashed{\nabla}_3$ through the angular derivatives (the commutator being lower order and controlled by the first term on the right of (\ref{awf})) use the schematic null structure equations (\ref{schematicnullstructure3}) which show that at the top order only $\eta, \underline{\eta}$ and $\underline{\omega}-\underline{\omega}_\circ$ can appear with the derivatives as stated in (\ref{awf}). Note that only top-order \emph{angular} derivatives appear for $\eta$ and $\underline{\eta}$ as there is a schematic equation (\ref{schematicnullstructure3}) available. The claim about the bootstrap assumption is immediate.

The second estimate is proven entirely analogously, now using the schematic equations (\ref{schematicnullstructure4}).
\end{proof}
\subsubsection{Basic estimates on spheres for errors $ \mathcal{E}^{n}_{p}$ and $\overset{(4)}{\mathcal{E}^{n}_{p}}$}
We also recall the error notation $\mathcal{E}^{n}_{p}$, which we recall is essentially a sum of products of the $\Phi$ (involving at most $n$ derivatives from $\mathfrak{D}$) which has total decay $r^{-p}$. We also recall the definition (\ref{4errordef}) of $\overset{(4)}{\mathcal{E}^{n}_{p}}$.

Following the proof of the previous lemma we obtain

\begin{lemma} \label{lem:basicerrorse2}
We have for $n=0,1,...,N$ and any sphere $S_{u,v}$ in $\mathcal{D}^{\I}$ the following estimates:
\begin{align}
\| r^p \overset{(4)}{\mathcal{E}^{n}_{p}} \|_{S_{u,v}} \lesssim  \left[ \sum_{\Phi} \sum_{|k|\leq n}  \| \mathfrak{D}^k r^{p_2} \overset{(4)}{\Phi}_{p_2}\|_{S_{u,v}} \right] \sum_{\Phi} \sum_{|j|\leq N-6} \| \mathfrak{D}^j r^{p_3} \overset{(4)}{\Phi}_{p_3}\|_{S_{u,v}}  \, .
\end{align}
\begin{align} \label{gens}
\| r^p \mathcal{E}^{n}_{p} \|_{S_{u,v}} \lesssim  \left[ \sum_{\Phi} \sum_{|k|\leq n}  \| \mathfrak{D}^k r^{p_2} {\Phi_{p_2}}\|_{S_{u,v}} \right] \sum_{\Phi} \sum_{|j|\leq N-6} \| \mathfrak{D}^j r^{p_3} {\Phi_{p_3}}\|_{S_{u,v}} \, .
\end{align}
Moreover, the bootstrap assumption $\slashed{\mathbb{E}}^N_{u_f,\I} \lesssim \varepsilon^2$ imply for $s \in \{0,1\}$ the estimates 
\begin{align}
\sum_{\Phi} \sum_{|j|\leq N-6} \| \mathfrak{D}^j r^{p_3} \overset{(4)}{\Phi}_{p_3}\|_{S_{u,v}} \lesssim \frac{\varepsilon}{u}  \ \ \ \ \textrm{and} \ \ \  \sum_{\Phi} \sum_{|j|\leq N-6} \frac{1}{r^{\frac{s}{2}}} \| \mathfrak{D}^j r^{p_3} {\Phi_{p_3}}\|_{S_{u,v}}\lesssim \frac{\varepsilon}{u^{\frac{1}{2} + \frac{s}{2}}} \, .
\end{align}
\end{lemma}

\subsection{Error estimates on $\protect\Cbar_{v_\infty}^{\I}$} \label{sec:vinfexplain}
The main idea in proving estimates on $\underline{C}_{v_\infty}^{\I}$ is that any decay in $r$ can be translated into very strong decay in $u$ using our definition $v_\infty = \varepsilon_0^{-2} (u_f)^\frac{1}{\delta} \geq \varepsilon^{-2} u^\frac{1}{\delta}$, which implies in particular
\begin{align} \label{rtou}
\sup_{\underline{C}_{v_\infty}^{\I}} \frac{1}{r(u,v)} \leq \frac{\varepsilon_0^2}{(u_f)^8} \leq  \frac{\varepsilon^2}{u^8}  \, .
\end{align}

We begin with an estimate for \emph{linear} error terms (which are still called error terms because they exhibit additional decay in $r$, which by the above remark can be translated into strong $u$-decay). 

\begin{proposition} \label{prop:linerror}
For $\Gamma_p \in \overset{(in)}{\Gamma} = \{ \hat{\chi} , \underline{\hat{\chi}}, T, \underline{T} , \eta , \underline{\eta}, \Omega^2- \Omega_\circ^2, \underline{\omega}-
\underline{\omega}_\circ\}$ we have for $n=0, ..., N+1$ and $u \in \left[u_{-1},u_f\right]$:
\begin{align}
\left(\int_{u}^{u_f} d\bar{u} \frac{1}{\sqrt{r}}  \| \left[ r\slashed{\nabla}\right]^n \Gamma_p r^p \|_{S_{\bar{u},v_\infty}} \right)^2 \lesssim \frac{\varepsilon^4}{u^4} \ \ \ , \ \ \   \int_{u}^{u_f} d\bar{u} \frac{1}{r}  \| \left[ r\slashed{\nabla}\right]^n \Gamma_p r^p \|_{S_{\bar{u},v_\infty}}^2  \lesssim \frac{\varepsilon^4}{u^4} \, .
\end{align}
For $\mathcal{R}_p \in \{ \underline{\alpha} , \underline{\beta} ,\rho - \rho_\circ, \sigma \}$ 
we have for $n=0, ..., N$:
\begin{align}
\left(\int_{u}^{u_f} d\bar{u} \frac{1}{\sqrt{r}}  \| \left[ r\slashed{\nabla}\right]^n \mathcal{R}_p r^p \|_{S_{\bar{u},v_\infty}}\right)^2 \lesssim \frac{\varepsilon^4}{u^4} \ \ \ , \ \ \   \int_{u}^{u_f} d\bar{u} \frac{1}{r}  \| \left[ r\slashed{\nabla}\right]^n \mathcal{R}_p r^p \|_{S_{\bar{u},v_\infty}}^2  \lesssim \frac{\varepsilon^4}{u^4} \, .
\end{align}
\end{proposition}
\begin{proof}
Note that the bootstrap assumptions imply for the $\Gamma_p$ and $\mathcal{R}_p$ as above the flux estimates\footnote{The bootstrap assumptions are formulated for these quantities with their reference Kerr values subtracted but by Lemma \ref{lem:alsowithout} they also hold without the subtraction.}
\begin{align} \label{kvs}
\sum_{n=0}^{N+1} \int_{u_{-1}}^{u_f} d\bar{u}  \| \left[ r\slashed{\nabla}\right]^{n} \Gamma_p r^p \|^2_{S_{\bar{u},v_\infty}} + \sum_{n=0}^{N} \int_{u_{-1}}^{u_f} d\bar{u}  \| \left[ r\slashed{\nabla}\right]^{n} \mathcal{R}_p r^p \|^2_{S_{\bar{u},v_\infty}}  \lesssim \varepsilon^2  \, .
\end{align}
The desired estimates then follow easily from
 applying Cauchy--Schwarz and using (\ref{rtou}).
\end{proof}

We next provide estimates for \emph{non-linear} error terms on $\underline{C}_{v_\infty}^{\I}$ to be used in the sequel.
\begin{proposition} \label{prop:nlerror}
For all $n=0,...,N+1$ we have for all $u \in \left[u_{-1},u_f\right]$
\begin{align}
\int_{u}^{u_f} d\bar{u} \frac{1}{\sqrt{r}} \| r^p \overset{(in)}{\slashed{\mathcal{E}}^{n}_{p}}\|_{S_{\bar{u},v_\infty}} +\sqrt{\int_{u}^{u_f} d\bar{u}  \frac{1}{r} \| r^p \overset{(in)}{\slashed{\mathcal{E}}^{n}_{p}}\|^2_{S_{\bar{u},v_\infty}}} \lesssim \frac{ \varepsilon^2}{u^2}  \, 
\end{align}
and 
\begin{align} \label{fluxnor}
\int_{u}^{u_f} d\bar{u}  \| r^p \overset{(in)}{\slashed{\mathcal{E}}^{n}_{p}}\|_{S_{\bar{u},v_\infty}} \lesssim \frac{\varepsilon^2}{u} \, .
\end{align}
We also have for $n=0,...,N-2$ the improved estimate
\begin{align} \label{fluxnor2}
\int_{u}^{u_f} d\bar{u}  \| r^p \overset{(in)}{\slashed{\mathcal{E}}^{n}_{p}}\|_{S_{\bar{u},v_\infty}} \, \lesssim \frac{\varepsilon^2}{u^2} \, .
\end{align}
\end{proposition}
\begin{proof}
For the bound (\ref{fluxnor}) we insert the estimate of Lemma \ref{lem:basicerrorse} and apply Cauchy--Schwarz to produce fluxes. Using that the bootstrap assumptions control the flux of up to $N-6$ derivatives of $\overset{(in)}{\Phi}$ by $\frac{\varepsilon^2}{u^2}$ (cf.~Lemma \ref{lem:alsowithout}) and the top-order flux bounds for the quantities appearing in $\overset{(in)}{\Phi}$ and $\overset{(in)}{\Gamma}$, we deduce the result.\footnote{Note that the key here is that all geometric quantities appearing in $ \overset{(in)}{\slashed{\mathcal{E}}^{n}_{p}}$ satisfy a flux estimate on $v=v_\infty$.} The improved bound (\ref{fluxnor2}) follows using that both fluxes now decay like $\frac{\varepsilon^2}{u^2}$. 
\end{proof}

\subsection{Error estimates on {$C$}$_{u_{-1}}^{\I}$}
\begin{proposition} \label{prop:erroroutdata}
For all $n=0,...,N$ and $p \geq 3/2$, we have, for any $v \geq v(u_{-1},R_{-2})$, the estimate
\begin{align}
\int_{v}^\infty d\bar{v}  \|  \overset{(out)}{\slashed{\mathcal{E}}^{n}_{p}} \|_{S_{u_{-1},\bar{v}}} \lesssim \frac{\varepsilon^2}{r^{p-1}(u_{-1},v)} \, .
\end{align}
Also, at the top order\footnote{One loses $3/2$ here if $N$ derivatives of curvature appear because the $r$-weights appearing in the energy flux that has to be used are non-optimal.}, we have, for $p \geq 2$ and any $v \geq v(u_{-1},R_{-2})$,
the estimate
\begin{align}
\int_{v}^\infty d\bar{v}  \| \overset{(out)}{\slashed{\mathcal{E}}^{N+1}_{p}}\|_{S_{u_{-1},\bar{v}}} \lesssim \frac{\varepsilon^2}{r^{p-3/2}(u_{-1},v)} \, .
\end{align}
Finally, for all $n=0,...,N-1$ and $p \geq 0$ we have, for all $v \geq v(u_{-1},R_{-2})$, the estimate
\begin{align}
 \| \overset{(out)}{\slashed{\mathcal{E}}^{n}_{p}} \| _{S_{u_{-1},v}} + \|  \overset{(in)}{\slashed{\mathcal{E}}^{n}_{p}} \| _{S_{u_{-1},v}} \lesssim \frac{\varepsilon^2}{r^p(u_{-1},v)}.
\end{align}
\end{proposition}
\begin{proof}
Follows from structure of the error, Sobolev embedding on spheres and the bootstrap assumptions.
\end{proof}

\subsection{Error estimates in $\mathcal{D}^{\I}$}

\subsubsection{Estimates on $\overset{(in)}{\slashed{\mathcal{E}}^n_p}$ and $\overset{(out)}{\slashed{\mathcal{E}}^n_p}$}

\begin{proposition} \label{prop:errorspacetimeIplus}
 For $s=0,...,N$, we have the estimate
\begin{align} \label{miuy}
\| r^p \overset{(in)}{\slashed{\mathcal{E}}}{}^{N-s}_p\|^2_{S_{u,v}} \lesssim \frac{\varepsilon^4}{u^{\min(4,2+s)}} \, .
\end{align}
For $s=0,1,\ldots ,N$, we have the estimate
\begin{align}
\int_{u}^{u_f} d\bar{u} \int_{v(\bar{u},R_{-2})}^{v_\infty} d\bar{v} \frac{1}{r^{1+\delta}} \| r^p \overset{(in)}{\slashed{\mathcal{E}}}{}^{N-s}_p\|^2_{S_{u,v}} \lesssim \frac{\varepsilon^4}{u^{\min(3,2+s)}} \, .
\end{align}
Also, for the outgoing non-linear error terms we have for $t \in \{0,1\}$ and $s=0,1, 2, \ldots, N$, we have 
\begin{align}
\| r^p \overset{(out)}{\slashed{\mathcal{E}}}{}^{N-s}_p\|^2_{S_{u,v}} \lesssim \frac{\varepsilon^4}{u^{\min(2,1+\frac{s}{2})}}  \ \ \ , \ \ \ \frac{1}{r^{1+t}} \| r^p \overset{(out)}{\slashed{\mathcal{E}}}{}^{N-s}_p\|^2_{S_{u,v}} \lesssim \frac{\varepsilon^4}{u^{\min(3+t,2+\frac{s}{2})}} \, .
\end{align}
For $s=0,1,...,N$ and $t \in \{0,1\}$, we have the estimate
\begin{align}
\int_{u}^{u_f} d\bar{u} \int_{v(\bar{u},R_{-2})}^{v_\infty} d\bar{v} \frac{1}{r^{1+\delta}} \frac{1}{r^t}\| r^p \overset{(out)}{\slashed{\mathcal{E}}}{}^{N-s}_p\|^2_{S_{u,v}} \lesssim \frac{\varepsilon^4}{u^{1+t}} \, .
\end{align}
For $s=2,3,...,N$ and $t \in \{0,1\}$, we have the estimate
\begin{align}
\int_{u}^{u_f} d\bar{u} \int_{v(\bar{u},R_{-2})}^{v_\infty} d\bar{v} \frac{1}{r^{1+\delta}} \frac{1}{r^2}\| r^p \overset{(out)}{\slashed{\mathcal{E}}}{}^{N-s}_p\|^2_{S_{u,v}} \lesssim \frac{\varepsilon^4}{u^{3}} \, .
\end{align}
\end{proposition}

\begin{proof}
The bounds on spheres follow directly from the estimates of Lemma \ref{lem:basicerrorse} and inserting the bootstrap assumptions. For the integrated decay bounds use again the estimates of Lemma \ref{lem:basicerrorse} and in addition the bootstrap assumptions on the remaining spacetime integrals.
\end{proof}

\subsubsection{Estimates on $\mathcal{E}^n_p$, $\check{\mathcal{E}}^n_p$}

\begin{proposition} \label{prop:eebs}
For $s=0,\ldots,N-1$ and $t \in \{0,1\}$, we have
\begin{align} 
\| r^{p-\frac{t}{2}} \mathcal{E}^{N-1-s}_p\|^2_{S_{u,v}} \lesssim \frac{\varepsilon^4}{u^{\min(2+t,1+t+\frac{s}{2})}} \ \ \ \textrm{and}  \ \ \ \| r^{p-1} \mathcal{E}^{N-1-s}_p\|^2_{S_{u,v}} \lesssim \frac{\varepsilon^4}{u^{\min(4,5/2+s)}} \, .
\end{align}
Moreover, assuming that each summand in $\mathcal{E}^{N-1-s}_p$ contains at least one $\overset{(in)}{\Phi}$, we have for $s=0,...,N-2$
\begin{align}
\| r^p \mathcal{E}^{N-2-s}_p\|^2_{S_{u,v}} \lesssim \frac{\varepsilon^4}{u^{\min(3,5/2+s)}}  \, .
\end{align}
Finally, assuming that none of the summands in $\mathcal{E}^{N-1-s}_p$ contains factors from $\{\alpha, \beta, \omega-\omega_\circ\}$, we have for $s=0,...,N-1$
\begin{align}
\| r^p \mathcal{E}^{N-1-s}_p \left(\textrm{no $\alpha$, $\beta$, $\omega-\omega_\circ$}\right)\|^2_{S_{u,v}}+\| r^{p-\frac{1}{2}} \mathcal{E}^{N-1-s}_p \left(\textrm{no $\alpha$, $\beta$}\right)\|^2_{S_{u,v}} \lesssim \frac{\varepsilon^4}{u^{\min(3,5/2+s)}}  \, .
\end{align}
Finally (and more generally), all statements hold verbatim with $\mathcal{E}$ replaced by $\check{\mathcal{E}}$. 
\end{proposition}

\begin{proof}
Follows directly from (\ref{gens}) and the bootstrap assumptions. Note in particular that we have
\begin{align}
\sum_{|\underline{k}|\leq N-1-s} \| \mathfrak{D}^{\underline{k}} (r^p \Phi_p)\|_{S_{u,v}} \lesssim \frac{\varepsilon}{u^{\min(1/4 + s/2,1)}} \ \ \ \textrm{for all $\Phi_p \setminus \{\Omega^2 \alpha, \Omega \beta, \omega-\omega_\circ\}$}
\end{align}
for $s=0,1,2$ by the bootstrap assumptions.
\end{proof}

In terms of integrated decay estimates we have (see also Section \ref{sec:mainerrorIgauge}):
\begin{proposition} \label{prop:intdebs} 
For $s=0,1,...,N$ and $t \in \{0,1\}$
\begin{align}
\int_{u}^{u_f} d\bar{u} \int_{v(\bar{u},R_{-2})}^{v_\infty} d\bar{v} \frac{1}{r^{1+\delta}} \frac{1}{r^t}\| r^p \check{\mathcal{E}}^{N-s}_p\|^2_{S_{u,v}} \lesssim \frac{\varepsilon^4}{u^{1+t}} \, ,
\end{align}
and for $s=0,1,...,N-1$ and $t \in \{0,1\}$
\begin{align}
\int_{u}^{u_f} d\bar{u} \int_{v(\bar{u},R_{-2})}^{v_\infty} d\bar{v} \frac{1}{r^{1+\delta}} \frac{1}{r^{t+1}}\| r^p \check{\mathcal{E}}^{N-1-s}_p\|^2_{S_{u,v}} \lesssim \frac{\varepsilon^4}{u^{2+t}} \, .
\end{align}
\end{proposition}

\begin{proof}
Follows directly from (\ref{gens}) and the bootstrap assumptions.
\end{proof}

\subsubsection{Estimates on $\overset{(Kerr)}{\slashed{\mathcal{E}}}$}

We note the following estimate, which is an immediate consequence of the bootstrap assumptions. Both stronger rates and higher regularity can be proven but the one stated is sufficient for our future applications. 
\begin{proposition} \label{prop:kerrerrorestimate}
We have, for any $|\gamma| \leq N-1$, the estimate
\begin{align}
\| \mathfrak{D}^\gamma \overset{(Kerr)}{\slashed{\mathcal{E}}} \|_{S_{u,v}} \lesssim \frac{\varepsilon^2}{r \cdot u}  \, .
\end{align}
\end{proposition}

\subsection{The error term in the evolution equation of $B$}

In this subsection, we estimate the right hand sides of the propagation equations for the quantities $B$ and $\Omega \slashed{\nabla}_4 B$, see Section~\ref{sec:introduceY}.

Before doing this, 
we collect a straightforward consequence of the bootstrap assumption $\slashed{\mathbb{E}}^N_{u_f,\I}  \leq \varepsilon^2$ in $\mathcal{D}^{\I}$, wich provides top-order fluxes for the shear on arbitrary outgoing and ingoing cones. 

\begin{lemma} \label{lem:shearfluxes}
For fixed $u \in \left[u_{-1}, u_f\right]$, 
we have on any ingoing cone $\underline{C}^{\I}_v(u)$ for $s=0,1,2$ the flux estimates 
\begin{align}
\int_{\underline{C}^{\I}_v(u)} d\bar{u} \|  \left[r \slashed{\nabla}\right]^{N+1} r \underline{\hat{\chi}} \Omega^{-1} \|^2_{S_{\bar{u},v}} +  u^s \int_{\underline{C}^{\I}_v(u)} d\bar{u} \|  \left[r \slashed{\nabla}\right]^{N-s} r \underline{\hat{\chi}} \Omega^{-1} \|^2_{S_{\bar{u},v}}  \lesssim  \varepsilon^2
\end{align}
and on any outgoing cone ${C}^{\I}_u$ the top-order estimate
\begin{align}
\int_{{C}^{\I}_u} d\bar{v} \frac{1}{r^2} \| \left[r \slashed{\nabla}\right]^{N+1} r^2 {\hat{\chi}} \Omega \|^2_{S_{u,\bar{v}}}  \lesssim  \varepsilon^2
\end{align}
\end{lemma}
\begin{proof}
Write the Codazzi equation (\ref{eq:Codazzibar}) in the form
\begin{align}
r^2\slashed{div} (\underline{\hat{\chi}} \Omega^{-1}) &=   r^2(\underline{\beta}\Omega^{-1}-(\Omega^{-1}\underline{\beta})_{\mathrm{Kerr}})-\frac{1}{r} r^2 (\underline{\eta}-\underline{\eta}_{\mathrm{Kerr}}) + \frac{1}{2r} r\slashed{\nabla} r^2\underline{T}- \frac{1}{r}\Omega^{-1} r^2 \underline{\eta} r\underline{\hat{\chi}} + \frac{1}{2r^2} r^2 \underline{T} r^2\underline{\eta} \nonumber \\
&=  r^2(\underline{\beta}\Omega^{-1}-(\Omega^{-1}\underline{\beta})_{\mathrm{Kerr}})-\frac{1}{r} r^2 (\underline{\eta}-\underline{\eta}_{\mathrm{Kerr}}) -\frac{1}{r} r^2 (\underline{\eta}-\underline{\eta}_{\mathrm{Kerr}}) + \frac{1}{2r} r\slashed{\nabla} r^2\underline{T}+ \frac{1}{r}\overset{(in)}{\slashed{\mathcal{E}}^0_0} \, . \nonumber
\end{align}
Apply $N-s$ angular derivatives and use the bootstrap assumptions on the flux on $\underline{\beta}$ and the assumptions on spheres for the Ricci coefficients $T$ and $\eta$. The non-linear error is easily controlled using Proposition \ref{prop:errorspacetimeIplus}. The second estimate follows similarly from the other Codazzi equation (\ref{eq:Codazzi}).
\end{proof}

\begin{proposition} \label{prop:fan}
For $i=0,...,N-3$ and any fixed ingoing cone $\underline{C}^{\I}_v$ in $\mathcal{D}^{\I}$, we have
\begin{align} \label{Berrortop}
\int_{u_0}^u d\bar{u} \Big\| \mathcal{E} \left[ \mathscr{A}^{[i]} B\right] \Big\|_{S_{\bar{u},v}}  \lesssim  \varepsilon^2 \, .
\end{align} 
\end{proposition}

\begin{proof}
We use the defining expression (\ref{errorB1}),
\begin{align}
\mathcal{E} \left[\mathscr{A}^{[i]} B\right]   = &   \mathscr{A}^{[i]}  \frac{r^2}{\Omega^2}\mathcal{E}[{\check{\underline{\psi}}}]  
+\left(\mathscr{A}^{[i+2]} + \mathscr{A}^{[i]}\right) \mathcal{E}_{anom} +\overset{(in)}{ \slashed{\mathcal{E}}^{3+i}_1}+ \frac{\check{r}}{r} \overset{(in)}{ \slashed{\mathcal{E}}^{3+i}_1} + \check{\mathcal{E}}^{2+i}_1 \, ,
\end{align} 
for which we recall also (\ref{psibtex}). We handle the error terms term by term:
\begin{enumerate}
\item We have for all $i \leq N-3$ the estimate $\|\overset{(in)}{ \slashed{\mathcal{E}}^{3+i}_1}\|_{S_{u,v}} \lesssim \frac{\varepsilon^2}{r \sqrt{u}}$ (see Proposition \ref{prop:errorspacetimeIplus}) and similarly $\|\check{\mathcal{E}}^{2+i}_1\|_{S_{u,v}} \lesssim \frac{\varepsilon^2}{r \sqrt{u}}$. Since the right hand side in these estimates is integrable in $u$, the estimate follows for the last three terms of the error. 
\item For the term involving $\mathcal{E}_{anom}$ we recall the definition (\ref{anomalous}). Using the commutator formula of Lemma \ref{lem:commutation} twice we see that besides a number of terms that can be incorporated into the error $\overset{(in)}{ \slashed{\mathcal{E}}^{3+i}_1}$ the desired estimate would follow if products of the form $\left[r \slashed{\nabla}\right]^{k_1}r \underline{\hat{\chi}} \left[ r \slashed{\nabla} \right]^{k_2} r \underline{\hat{\chi}}$,$\left[r \slashed{\nabla}\right]^{k_1}r \underline{T} \left[ r \slashed{\nabla} \right]^{k_2} r \underline{\hat{\chi}}$ and $\left[r \slashed{\nabla}\right]^{k_1}r \underline{T} \left[ r \slashed{\nabla} \right]^{k_2} r \underline{T}$ are integrable along ingoing cones for $k_1+k_2 \leq N+1$. This in turn is an immediate consequence of applying Cauchy--Schwarz and the fact that the bootstrap assumptions and Lemma \ref{lem:shearfluxes} imply a flux estimate for up to $N+1$ angular derivatives of $ r\underline{\hat{\chi}}$ and $r \underline{T}$. 
\item Turning to the term $r^2 \mathcal{E} [\check{\psi}]$ in  (\ref{errorB1})  and the definition (\ref{psibtex}) we first note that the term involving $\left[r \slashed{\nabla}\right]^i r^2 \check{\mathcal{E}}^2_3$ can be handled as in $1.$ Next we have $\| \left[r \slashed{\nabla}\right]^i  r^2 \overset{(4)}{\mathcal{E}^2_2}\| \leq \frac{\varepsilon^2}{u^{5/4}}$ which is also integrable. Finally, we have for $i \leq N-3$
\[
\Big\| \left[  r \slashed{\nabla}\right]^i \left( k_1 \etabar \cdot r^4 \divslash \alphabar
		+
		k_2 r^3 \Dslash_2^* (r\etabar \cdot \alphabar)
		+
		k_3 r^3\nablaslash \Omega \omegahat \cdot r^2 \divslash \alphabar \right) \Big\|_{S_{u,v}} \lesssim \frac{\varepsilon^2}{u^\frac{3}{2}} \, ,
\]
\[
\Big\|  \left[  r \slashed{\nabla}\right]^i \left(   k_4 r^3\Dslash_2^*( r\nablaslash \Omega \omegahat \cdot r \alphabar )
		+
		k_5 r^3\nablaslash \Omega \tr \chi \cdot r^2 \divslash \alphabar\right) \Big\|_{S_{u,v}} \lesssim \frac{\varepsilon^2}{u^\frac{3}{2}} \, ,
\]
as follows easily from Cauchy--Schwarz on $S_{u,v}$ and the bootstrap assumptions on spheres for the quantities involved. 
\end{enumerate}
\end{proof}

For the next proposition, we recall the non-linear error (\ref{nlerror4Bdef}).

\begin{proposition} \label{prop:B4errorflux}
For $i=0,...,N-5$ and any fixed ingoing cone $\underline{C}^{\I}_v$ in $\mathcal{D}^{\I}$ we have
\begin{align} \label{B4errorflux}
\int_{u_{-1}}^u  d\bar{u} \Big\| \mathcal{E} \left[ \Omega \slashed{\nabla}_4  \mathscr{A}^{[i]} B\right] \Big\|_{S_{\bar{u},v}} \lesssim \frac{\varepsilon^2}{v^2} +  \frac{ \varepsilon^2}{r ^{5/4}u} \, .
\end{align}
Finally, for $i=0,...,N-4$ we have
\begin{align} \label{B4errorspacetime}
\int_{\mathcal{D}^{\I}} d\bar{u} d\bar{v} \, r^{3-\delta} \cdot u \Big\| \mathcal{E} \left[\Omega \slashed{\nabla}_4 \mathscr{A}^{[i]} B\right] \Big\|^2_{S_{\bar{u},\bar{v}}} \lesssim  \varepsilon^4 \, .
\end{align}
\end{proposition}

\begin{proof}
{\bf The first estimate.} Recalling the definition (\ref{nlerror4Bdef}), we obtain the first estimate term by term. We have for $i\leq N-5$
\begin{align}
\| (\eta + \underline{\eta}) \slashed{\nabla}  \mathscr{A}^{[i]} B\|_{S_{u,v}} \lesssim \frac{1}{r} \| (\eta + \underline{\eta}) \|_{S_{u,v}} \| r\slashed{\nabla}  \mathscr{A}^{[i]} B\|_{S_{u,v}}  \lesssim \frac{\varepsilon}{r^3 u} \| r\slashed{\nabla}  \mathscr{A}^{[i]} B\|_{S_{u,v}} \lesssim \frac{\varepsilon^2}{r^3 u} \, ,
\end{align}
which is easily seen to integrate to the desired bound. Here the last estimate follows from inserting the definition of $B$ in terms of $Y$, the definition of $Y$ (\ref{Ydef}) and then using the bootstrap assumptions on $S_{u,v}$.

 For the second term in (\ref{nlerror4Bdef}) we recall from the proof of Proposition \ref{prop:fan} that $\left(\mathscr{A}^{[i+2]} + \mathscr{A}^{[i]}\right) \mathcal{E}_{anom}$ is given by terms that can be incorporated into $\overset{(in)}{ \slashed{\mathcal{E}}^{3+i}_1}$ plus a sum of products of the form $\left[r \slashed{\nabla}\right]^{k_1}r \underline{\hat{\chi}} \left[ r \slashed{\nabla} \right]^{k_2} r \underline{\hat{\chi}}$, $\left[r \slashed{\nabla}\right]^{k_1}r \underline{T} \left[ r \slashed{\nabla} \right]^{k_2} r \underline{\hat{\chi}}$ and $\left[r \slashed{\nabla}\right]^{k_1}r \underline{T} \left[ r \slashed{\nabla} \right]^{k_2} r \underline{T}$ with appropriate contractions and $k_1+k_2 \leq i+4$. Since $\Omega \slashed{\nabla}_4 \overset{(in)}{ \mathcal{E}^{3+i}_1} = \check{\mathcal{E}}^{4+i}_{2}+ \check{\mathcal{E}}^{4+i}_{5/2}$ it is clear from the propagation equations for $r^2 \underline{T}$ and $r \underline{\hat{\chi}}$ that it suffices to estimate
 \begin{align}
 \int_{u_{-1}}^{u} d\bar{u} \frac{1}{r^2} \left( \|r \underline{\hat{\chi}}\|_{S_{\bar{u},v}} + \frac{1}{r} \|r^2 \underline{T} \|_{S_{\bar{u},v}} \right) \| \left[r \slashed{\nabla}\right]^{5+i} (\underline{\eta} r^2) \|_{S_{\bar{u},v}} \, .
 \end{align}
 We focus on the term involving $\underline{\hat{\chi}}$, the term involving $\underline{T}$ being much simpler as there is an additional power of $r^{-1}$.
We let $\tilde{u} = \min(u, \frac{1}{2}r(u,v))$ and integrate first
\begin{align}
\int_{u_{-1}}^{\tilde{u}} d\bar{u} \frac{1}{r^2}\|r \underline{\hat{\chi}}\|_{S_{\bar{u},v}} \| \left[r \slashed{\nabla}\right]^{5+i} (\underline{\eta} r^2) \|_{S_{\bar{u},v}}  \lesssim \frac{1}{r^2} \sup_{\mathcal{D}^{\I}} \|  \left[r \slashed{\nabla}\right]^{5+i} (\underline{\eta} r^2) \|_{S_{u,v}} \sqrt{\int_{u_{-1}}^{\tilde{u}} d\bar{u} \| r \underline{\hat{\chi}}\|^2_{S_{\bar{u},v}} \bar{u}^{3/2}} \lesssim \frac{\varepsilon^2}{v^2} \nonumber
\end{align}
and in case that $u > \tilde{u}$ also
\begin{align}
\int_{\tilde{u}}^u d\bar{u} \frac{1}{r^2}\|r \underline{\hat{\chi}}\|_{S_{\bar{u},v}} \| \left[r \slashed{\nabla}\right]^{5+i} (\underline{\eta} r^2) \|_{S_{\bar{u},v}}  \lesssim \sqrt{\int_{\tilde{u}}^u d\bar{u} \frac{1}{r^4(\bar{u},v)} \|  \left[r \slashed{\nabla}\right]^{5+i} (\underline{\eta} r^2) \|^2_{S_{\bar{u},v}}} \sqrt{\int_{\tilde{u}}^u d\bar{u} \| r \underline{\hat{\chi}}\|^2_{S_{\bar{u},v}}} \lesssim \frac{\varepsilon^2}{r^\frac{3}{2} u} \, . \nonumber
\end{align}
Combining the two integrals yields the desired bound for this term.

We turn to the expression $  \check{\mathcal{E}}^{4+i}_{2} \left(\textrm{no $\alpha$, $\beta$, $\omega-\omega_\circ$}\right)+ \check{\mathcal{E}}^{4+i}_{5/2}\left(\textrm{no $\alpha$, $\beta$}\right) +  \check{\mathcal{E}}^{4+i}_{3}$. As a direct consequence of Proposition \ref{prop:eebs}, for $i \leq N-5$
\begin{align} 
 \|  \check{\mathcal{E}}^{4+i}_{2} \left(\textrm{no $\alpha$, $\beta$, $\omega-\omega_\circ$}\right) \|_{S_{u,v}} +  \|  \check{\mathcal{E}}^{4+i}_{5/2} \left(\textrm{no $\alpha$, $\beta$}\right) \|_{S_{u,v}}  + \| \check{\mathcal{E}}^{4+i}_{3} \|_{S_{u,v}}
\lesssim \frac{\varepsilon^2}{r^2 u^\frac{5}{4}} \, ,
\end{align}
which integrates to the desired bound. 
Finally, we recall definition (\ref{psibtex}) according to which we have
\begin{align}
\Omega \slashed{\nabla}_4  \left( \mathscr{A}^{[i]} \frac{r^2}{\Omega^2}  \mathcal{E}[{\check{\underline{\psi}}}] \right) = \check{\mathcal{E}}^{3+i}_2 + \check{\mathcal{E}}^{3+i}_{\frac{5}{2}}+  \Omega \slashed{\nabla}_4 \overset{(4)}{\mathcal{E}}{}^{2+i}_0 + \mathfrak{Y} = \check{\mathcal{E}}^{3+i}_2 + \check{\mathcal{E}}^{3+i}_{\frac{5}{2}} + \mathfrak{Y}
\end{align}
where each summand in $\check{\mathcal{E}}^{3+i}_2$ contains at least one factor of $\overset{(in)}{\Phi}$ and where
\begin{align}
\mathfrak{Y} = \sum_{\substack{k_1+k_2+k_3\leq 3+i \\ k_1\leq 2+i, k_2 \leq 1,k_3\leq 1+i}}
		H_{k_1k_2k_3}
		r(r\nablaslash)^{k_1} (r\Omega \nablaslash_4)^{k_2}(\omega - \omega_\circ)
		\cdot
		(r\nablaslash)^{k_3} \alphabar r
\end{align}
is the anomalous term. Note that the second equality is a consequence of (\ref{4erroreq}). We note that by Proposition \ref{prop:eebs} we have for $i \leq N-5$ the bounds
\begin{align}
\|  \check{\mathcal{E}}^{3+i}_2( \textrm{each summand has a factor of $\overset{(in)}{\Phi}$}) \|_{S_{u,v}} + \| \check{\mathcal{E}}^{3+i}_{\frac{5}{2}} \|_{S_{u,v}}  \lesssim \frac{\varepsilon^2}{r^2 u^{5/4}} \, ,
\end{align}
which integrates to the desired bound. For the anomalous term, we integrate first with $i\leq N-5$
\begin{align}
\int_{u_{-1}}^{\tilde{u}} d\bar{u} \| \mathfrak{Y}\|_{S_{\bar{u},v}} \lesssim \frac{1}{r^2} \sum_{j \leq N-4}\sum_{|k|\leq N-2} \left(  \sup_{\mathcal{D}^{\I}} \|  \mathfrak{D}^k r^3 (\omega - \omega_\circ) \|_{S_{u,v}} \right)\sqrt {\int_{u_{-1}}^{\tilde{u}} d\bar{u} \| \left[r\slashed{\nabla}\right]^j r \underline{\alpha}\|^2_{S_{\bar{u},v}} \bar{u}^{3/2}} \lesssim \frac{\varepsilon^2}{v^2} \nonumber
\end{align}
where we have used Theorem~\ref{thm:alphaalphabarestimates} (and Corollary \ref{cor:replacercheckbyr}) and in case that $u > \tilde{u}$ also
\begin{align}
\int_{\tilde{u}}^u d\bar{u} \| \mathfrak{Y}\|_{S_{\bar{u},v}}  \lesssim  \sum_{j \leq N-4}\sum_{|k|\leq N-2}  \sqrt{\int_{\tilde{u}}^u d\bar{u} \frac{1}{r^4(\bar{u},v)}  \|  \mathfrak{D}^k r^3 (\omega - \omega_\circ) \|_{S_{\bar{u},v}}}
\sqrt{ \int_{u_{-1}}^{\tilde{u}} d\bar{u} \| \left[r\slashed{\nabla}\right]^j r \underline{\alpha}\|^2_{S_{\bar{u},v}}} \lesssim \frac{\varepsilon^2}{r^\frac{3}{2} u} \, . \nonumber
\end{align}
establishing the bound for the anomalous term.

{\bf The second estimate.} This is much easier. We will only estimate the anomalous term leaving the details of the remaining terms to the reader. We have for $i \leq N-4$
\begin{align}
\int_{\mathcal{D}^{\I}} d\bar{u} d\bar{v} \, r^{3-\delta} \cdot u \| \mathfrak{Y} \|^2_{S_{\bar{u},\bar{v}}}  \lesssim \sum_{|k| \leq N-1} \sum_{j \leq N-3} \sup_{\mathcal{D}^{\I}} \|  \mathfrak{D}^k r^3 (\omega - \omega_\circ) \|_{S_{u,v}} \int_{\mathcal{D}^{\I}} d\bar{u} d\bar{v} \frac{\bar{u}}{r^{1+\delta}}  \| \left[r\slashed{\nabla}\right]^j \underline{\alpha} r \|_{S_{\bar{u},\bar{v}}}^2 \lesssim \varepsilon^4 .\nonumber
\end{align}
\end{proof}

\section{The proof of Theorem~\ref{thm:Iestimates}: transport and elliptic estimates} \label{sec:transportandelliptic}

\subsection{Overview of the proof of Theorem \ref{thm:Iestimates}} \label{sec:overviewipg}
We recall that we have already proven Theorem \ref{thm:alphaalphabarestimates}, i.e.~we have obtained improved estimates on the almost gauge invariant quantities $\alpha$ and $\underline{\alpha}$ (note also Corollary \ref{cor:replacercheckbyr}).
The general strategy to obtain the estimates of Theorem \ref{thm:Iestimates} is to estimate connection coefficients and curvature components in the region $\mathcal{D}^{\I}$ using the estimates of Theorem \ref{thm:alphaalphabarestimates} and the gauge conditions on the solution imposed on the null hypersurfaces $\underline{C}^{\I}_{v_\infty}$ and $C^{\I}_{u_{-1}}$, which have been recalled in Section \ref{sec:gaugerecall}.

Our estimates will follow from elliptic estimates on spheres and transport estimates along cones. 
We first state the general form of these estimates which we shall use in {\bf Sections~\ref{sec:basicelliptic}} and {\bf Section~\ref{sec:basictransport}},
respectively.

The actual proof of Theorem \ref{thm:Iestimates} is divided into proving the following three theorems, for which we recall the energies defined in Section \ref{sec:energiesig}:

\begin{theorem}[Improving the angular master energy]  
\label{theo:mtnig2}
We have the estimate
\begin{align}
\slashed{\mathbb{E}}^N_{u_f,\I}  \lesssim \varepsilon_0^2 + \varepsilon^3  \, .
\end{align}
\end{theorem}

\begin{theorem}[Improving the auxiliary energies on $\omega$ and $\underline{\omega}$] 
\label{theo:mtnig3}
We have  the estimate
\begin{align}
\mathbb{E}^{N,aux}_{u_f, \mathcal{I}} \left[ \underline{\omega} \right] + \mathbb{E}^{N,aux}_{u_f, \mathcal{I}} \left[ {\omega} \right] \lesssim \varepsilon_0^2 + \varepsilon^3 \, .
\end{align}
\end{theorem}

\begin{theorem}[Improving the full master energy] 
\label{theo:reducetoangular}
We have the estimate
\begin{align}
\mathbb{E}^N_{u_f,\I} \lesssim \slashed{\mathbb{E}}^N_{u_f,\I} + \mathbb{E}^{N,aux}_{u_f, \mathcal{I}} \left[ \underline{\omega} \right] + \mathbb{E}^{N,aux}_{u_f, \mathcal{I}} \left[ {\omega} \right] +\mathbb{E}^{N}_{u_f}[\alpha_{\Hp},\alpha_{\I}] +\mathbb{E}^{N}_{u_f}[\underline{\alpha}_{\Hp}, \alphabar_{\I}] \, .
\end{align}
\end{theorem}
In view of the fact that we have already proven Theorem \ref{thm:alphaalphabarestimates}, it is clear that Theorems \ref{theo:mtnig2}--\ref{theo:reducetoangular} prove Theorem \ref{thm:Iestimates}. At the heart of the matter is the proof of Theorem \ref{theo:mtnig2}.

\subsubsection{Improving the angular master energy: Proof of Theorem \ref{theo:mtnig2}}

The proof of Theorem \ref{theo:mtnig2} is contained in Sections, \ref{sec:vinfe},~\ref{sec:u0est} 
and~\ref{sec:mre}--\ref{sec:bametric} where we estimate all quantities on $v=v_\infty$, on $u=u_{-1}$ and  
in $\mathcal{D}^{\I}$, respectively. In particular (recall the angular master energy (\ref{angmasterenergy})):
\begin{enumerate}[(1)]
\item In {\bf Section  \ref{sec:vinfe}}, 
we will improve the bootstrap assumptions on $\underline{C}_{v_\infty}^{\I}$, i.e.~we shall prove
\begin{align} \label{target1}
\slashed{\mathbb{E}}^N_{v_{\infty}} \left[ \Gamma \right] + \slashed{\mathbb{E}}^N_{v_\infty} \left[ \mathcal{R} \right] \lesssim \varepsilon_0^2 + \varepsilon^3 \, .
\end{align}

\item In {\bf Section \ref{sec:u0est}}, we will improve the bootstrap assumptions on ${C}_{u_{-1}}^{\I}$, i.e.~we shall prove
\begin{align} \label{target2}
\slashed{\mathbb{E}}^N_{u_{-1}} \left[ \Gamma \right] + \slashed{\mathbb{E}}^N_{u_{-1}} \left[ \mathcal{R} \right]
\lesssim \varepsilon_0^2 + \varepsilon^3 \, . 
\end{align}
\item In {\bf Section \ref{sec:mre}}, we will improve the bootstrap assumptions in $\mathcal{D}^{\I}$, i.e.~we shall prove 
\begin{align} \label{target3}
\slashed{\mathbb{E}}^N_{\mathcal{D}^{\I}} \left[ \Gamma \right] +  \slashed{\mathbb{E}}^N_{\mathcal{D}^{\I}} \left[ \mathcal{R} \right] \lesssim \varepsilon_0^2 + \varepsilon^3 \, .
\end{align}
\item In {\bf Section \ref{sec:bametric}}, we will improve the bootstrap assumptions on the metric, i.e.~we shall prove
\begin{align} \label{target4}
\sup_{\mathcal{D}^{\I}}
		\sum_{s=0}^2 u^s \sum_{k=0}^{N+1-s} \|    (r \slashed{\nabla})^{n}\left(r (\slashed{g} - r^2\gamma) \right) \|^2_{S_{u,v}}   \lesssim \varepsilon_0^2 + \varepsilon^3  \, ,
\end{align}
which by Proposition \ref{prop:modesdifference} implies
\begin{align}
	\sup_{\mathcal{D}^{\I}}
		\sum_{s=0}^2 u^s \sum_{k=0}^{N+2-s} \sum_{m=-1}^1
		\Vert (r\nablaslash)^{k} (Y^1_m - \mathring{Y}^1_m)^{\I} \Vert_{S_{u,v}^{\I}}^2 \, .
\end{align}

\end{enumerate}
This will complete the proof of Theorem \ref{theo:mtnig2}.

The proof will generally proceed by obtaining the estimates (\ref{target1})--(\ref{target4}) separately for each $\Gamma$ and $\mathcal{R}$ appearing in the relevant energies. 
We emphasise the important role of the quantity $Y$ introduced in Section \ref{sec:introduceY} below in estimating the Ricci coefficient $\underline{\hat{\chi}}$ as already discussed in Section \ref{othernonlinearissues} of the introduction.

\subsubsection{Improving the auxiliary energies on $\omega$ and $\underline{\omega}$: Proof of Theorem \ref{theo:mtnig3}}
The proof of Theorem \ref{theo:mtnig3} will be carried out in {\bf Section \ref{sec:proofof3}}.  

This proof is quite straightforward once Theorem~\ref{theo:mtnig2} has been established.
The main idea, for $\underline{\omega}$ say, is to first estimate \emph{all} derivatives of $\underline{\omega}$  elliptically on $\underline{C}_{v_{\infty}}^{\I}$ (using the gauge condition of constant mass aspect) and then in $\mathcal{D}^{\I}$ by commuting the transport equation $\Omega \slashed{\nabla}_4 \underline{\omega}$ with $\left[r \slashed{\nabla}\right]^i\left[\Omega \slashed{\nabla}_3\right]^j$-derivatives exploiting that on the right hand side we can insert the Bianchi or null structure equations to convert $\Omega \slashed{\nabla}_3$ derivatives into (a) angular derivatives (b)  lower order terms and (c) $\Omega \slashed{\nabla}_3$-derivatives of $\underline{\alpha}$, all of which have already been estimated. 

The argument for $\omega$ is completely analogous exchanging the cones $C_{u_{-1}}^{\I}$ and $\underline{C}_{v_{\infty}}^{\I}$ and the $3$-direction by the $4$-direction.

\subsubsection{Improving the full master energy: Proof of Theorem \ref{theo:reducetoangular}  }
The proof of Theorem \ref{theo:reducetoangular} will be carried out in {\bf Section \ref{sec:proofof4}}.

For all quantities in the master energy except the terms involving the metric quantity $b$ and $\slashed{g}-\slashed{g}_\circ$, this follows easily by inductively exploiting the schematic structure of the Bianchi and null structure equations established in Section \ref{sec:schematic}. For $b$ and $\slashed g$ a simple transport argument can be invoked commuting (\ref{transportb}) and (\ref{nabla4g}) with $\mathfrak{D}^\gamma$ to produce the missing estimates on  in the master energy.

\subsection{The basic elliptic estimates} \label{sec:basicelliptic}
The arguments employed in the proof of Theorem \ref{theo:mtnig2} will make use of the following three key propositions which are manifestations of the following general principle: \\

{\bf General Principle: To estimate the angular derivatives of a quantity, it suffices to estimate the $\ell=0$ and $\ell=1$ modes and a suitable \emph{top order} angular derivative operator (whose kernel consists of the $\ell=0,1$ modes) applied to it.} \\

We recall $\mathscr{A}^{[i]}=\left[ r^2 \slashed{\mathcal{D}}_2^\star\slashed{div}\right]^{i/2}$ if $i$ is even and $\mathscr{A}^{[i]}= r\slashed{div} \left[ r^2 \slashed{\mathcal{D}}_2^\star\slashed{div}\right]^{(i-1)/2}$ if $i$ is odd.

\begin{proposition}  \label{prop:symmetrictracelesskeylemma}
For $\xi$ a symmetric traceless $S$-tensor we have for $1\leq n \leq N+1$ the estimate
\begin{align} \label{stth}
\sum_{i=0}^n \| \left[r \slashed{\nabla}\right]^i \xi \|_{S_{u,v}} \lesssim \| \mathscr{A}^{[n]} \xi\|_{S_{u,v}} \, .
\end{align}
\end{proposition}

\begin{proposition}  \label{prop:functionkeylemma}
Let $(f,g)$ be a pair of smooth functions and $f_{\mathrm{Kerr}},g_{\mathrm{Kerr}}$ be functions each supported on $\ell=1$ only. Then we have for $2 \leq n \leq N+1$ 
\begin{align} \label{applyforsigma}
 \sum_{i=0}^n \| [r\slashed{\nabla}]^i (f-f_{\mathrm{Kerr}}, g-g_{\mathrm{Kerr}})\|_{S_{u,v}}   \lesssim  &\phantom{+} \| f_{\ell=0}\|_{S_{u,v}} +  \| f_{\ell=1}-f_{\mathrm{Kerr}}\|_{S_{u,v}}   \\
& + \| g_{\ell=0}\|_{S_{u,v}} +  \| g_{\ell=1}-g_{\mathrm{Kerr}}\|_{S_{u,v}}  + \| \mathscr{A}^{[n-2]} r^2\slashed{\mathcal{D}}_2^\star \slashed{\mathcal{D}}_1^\star (f,g) \|_{S_{u,v}}  \, , \nonumber
\end{align}
where the left hand side denotes the sum $\sum_{i=0}^n \| [r\slashed{\nabla}]^i (f-f_{\mathrm{Kerr}})\|_{S_{u,v}}  + \sum_{i=0}^n \| [r\slashed{\nabla}]^i (g-g_{\mathrm{Kerr}})\|_{S_{u,v}}$.
\end{proposition}

\begin{proposition} \label{prop:oneformkeylemma}
Let $\xi$ be an $S$-tangent $1$-form and $\xi_{\mathrm{Kerr}}$ be a divergence-free  $S$-tangent 
$1$-form supported on $\ell=1$ only. Then we have for $1 \leq n \leq N+1$
\begin{align} \label{applyforbeta}
\sum_{i=0}^n \| \left[r\slashed{\nabla}\right]^i (\xi- \xi_{\mathrm{Kerr}})\|_{S_{u,v}}  \lesssim &  \| (r\slashed{div} \xi)_{\ell=1} \|_{S_{u,v}} + \| \mathscr{A}^{[n-1]} r\slashed{\mathcal{D}}_2^\star \xi\|_{S_{u,v}} 
 \\ 
&+ \| (r\slashed{curl} \xi)_{\ell=1} - r \slashed{curl} \xi_{\mathrm{Kerr}} \|_{S_{u,v}} +\boxed{ \| r \slashed{curl} \xi_{\mathrm{Kerr}} - (r \slashed{curl} \xi_{\mathrm{Kerr}})_{\ell=1} \|_{S_{u,v}} } \, , \nonumber 
\end{align}
and for $3 \leq n \leq N+1$
\begin{align} \label{applyforeta}
\sum_{i=0}^n \| \left[r\slashed{\nabla}\right]^i (\xi- \xi_{\mathrm{Kerr}})\|_{S_{u,v}}  \lesssim &  \| (r\slashed{div} \xi)_{\ell=1} \|_{S_{u,v}} + \| \mathscr{A}^{[n-3]} r^3 \slashed{\mathcal{D}}_2^\star \slashed{\mathcal{D}}_1^\star \slashed{\mathcal{D}}_1 \xi\|_{S_{u,v}}  \\ 
&+ \| (r\slashed{curl} \xi)_{\ell=1} - r \slashed{curl} \xi_{\mathrm{Kerr}} \|_{S_{u,v}} + \boxed{\| r \slashed{curl} \xi_{\mathrm{Kerr}} - (r \slashed{curl} \xi_{\mathrm{Kerr}})_{\ell=1} \|_{S_{u,v}} } \, . \nonumber 
\end{align}
Finally, for the boxed term we have the following estimates (important in applications)
\[
\| r \slashed{curl} \eta_{\mathrm{Kerr}} - (r \slashed{curl} \eta_{\mathrm{Kerr}})_{\ell=1} \|_{S_{u,v}} + \| r \slashed{curl} \underline{\eta}_{\mathrm{Kerr}} - (r \slashed{curl} \underline{\eta}_{\mathrm{Kerr}})_{\ell=1} \|_{S_{u,v}} \lesssim \frac{\varepsilon^2}{r^3 u^2} \, ,
\]
\[
\| r \slashed{curl} (\Omega \beta)_{\mathrm{Kerr}} - (r \slashed{curl}(\Omega \beta)_{\mathrm{Kerr}})_{\ell=1} \|_{S_{u,v}} + \| r \slashed{curl} (\Omega^{-1} \underline{\beta})_{\mathrm{Kerr}} - (r \slashed{curl} (\Omega^{-1} \underline{\beta}_{\mathrm{Kerr}}))_{\ell=1} \|_{S_{u,v}} \lesssim \frac{\varepsilon^2}{r^4 u^2} \, .
\]
\end{proposition}

\begin{remark}
The restriction $ n\leq N+1$ in the above propositions arises from the fact that the proof will require control on up to $N-1$ angular derivatives of the Gauss curvature.
\end{remark}

\begin{proof}[Proof of Proposition \ref{prop:symmetrictracelesskeylemma}]
We first note
\begin{align}
\sum_{i=0}^n \| \left[r \slashed{\nabla}\right]^i \xi \|^2_{S_{u,v}} \lesssim \| \mathscr{A}^{[n]} \xi\|^2_{S_{u,v}} + \sum_{i=0}^{n-1} \| \left[r \slashed{\nabla}\right]^i \xi \|^2_{S_{u,v}} \, .
\end{align}
This follows from first diligently integrating by parts the covariant derivatives on the left to create Laplacians (at the cost of lower order errors involving up to $N-1$ angular derivatives of Gauss curvature, which by (\ref{Gauss}) and the bootstrap assumptions are controlled in $L^2(S_{u,v})$) and then use the relation $r^2 \slashed{\Delta} = -2 \mathscr{A}^{[2]} + 2r^2K$. Secondly, using $r^2 \slashed{\Delta} = -2 \mathscr{A}^{[2]} + 2r^2K$ we can also show that for a symmetric traceless tensor we have both
\begin{align} \label{sttel}
\|  \xi \|^2_{S_{u,v}} \lesssim  \| \mathscr{A}^{[1]} \xi \|^2_{S_{u,v}} \ \ \ \ \textrm{and} \ \ \ \ \| \mathscr{A}^{[1]}  \xi \|^2_{S_{u,v}} \lesssim  \| \mathscr{A}^{[2]} \xi \|^2_{S_{u,v}} \, .
\end{align}
The desired estimate now follows by a simple induction.
\end{proof}

\begin{proof}[Proof of Proposition \ref{prop:functionkeylemma}]
To prove (\ref{applyforsigma}), we first prove that the estimate holds for $n=2$. To achieve this, we observe that it suffices to prove the estimate for $S_{u,v}$ equipped with the round metric $r^2 \mathring{\gamma}$ 
and the projections defined with respect to the honest spherical harmonics. Indeed, all errors arising from this replacement are controlled by the bootstrap assumptions on the metric and the estimates for the difference of the $Y^{\ell=1}_m$ with the spherical harmonics.
For the round metric the estimate follows as in Section~$4.4.2$ of~\cite{holzstabofschw}: Indeed defining $\tilde{f}= f-f_{\mathring{\ell}=1}-(f_{\mathrm{Kerr}})_{\mathring{\ell}=1} -({f})_{\mathring{\ell}=0}$ and $\tilde{g}= g-g_{\mathring{\ell}=1}-(g_{\mathrm{Kerr}})_{\mathring{\ell}=1} -({g})_{\mathring{\ell}=0}$ we can write (in the next formula all operators and volume elements are defined with respect to the round metric)
\begin{align}
\| r^2\slashed{\mathcal{D}}_2^\star \slashed{\mathcal{D}}_1^\star (f,g) \|^2_{S_{u,v}} &= \| r^2\slashed{\mathcal{D}}_2^\star \slashed{\mathcal{D}}_1^\star (\tilde{f},\tilde{g}) \|^2_{S_{u,v}} 
= \int_{S_{u,v}} \frac{1}{2} \slashed{\Delta} \tilde{f} \cdot r^2  \slashed{\Delta} \tilde{f} + \tilde{f} \cdot r^2 \slashed{\Delta} \tilde{g}+\frac{1}{2} \slashed{\Delta} \tilde{g} \cdot r^2  \slashed{\Delta} \tilde{g} + \tilde{g} \cdot r^2 \slashed{\Delta} \tilde{f} \, . \nonumber
\end{align}
Expanding into spherical harmonics of the round sphere and using that $\tilde{f}$, $\tilde{g}$ are supported on $\mathring{\ell} \geq 2$ as well as the orthogonality properties of the spherical harmonics now leads to the desired bound for $n=2$.

The proof of $n>2$ proceeds in analogy to the proof of Proposition \ref{prop:symmetrictracelesskeylemma}. We use that for any pair of functions $(f,g)$ and $n\geq 3$
\begin{align}
\sum_{i=0}^n \| \left[r \slashed{\nabla}\right]^i (f,g) \|^2_{S_{u,v}} \lesssim \| \mathscr{A}^{[n-2]} r^2\slashed{\mathcal{D}}_2^\star \slashed{\mathcal{D}}_1^\star(f,g)\|^2_{S_{u,v}} + \sum_{i=0}^{n-1} \| \left[r \slashed{\nabla}\right]^i (f,g) \|^2_{S_{u,v}} \, ,
\end{align}
which follows once more by diligently integrating by parts using in particular the relation $2\slashed{\mathcal{D}}_1\slashed{\mathcal{D}}_2 \slashed{\mathcal{D}}_2^\star \slashed{\mathcal{D}}_1^\star = \slashed{\Delta} \slashed{\Delta}-2\slashed{\mathcal{D}}_1 (K\slashed{\mathcal{D}}_1^\star)$. Invoking also (\ref{sttel}) and the fact that the estimate has been shown for $n=2$, we conclude the proof of (\ref{applyforsigma}).
\end{proof}

\begin{proof}[Proof of Proposition \ref{prop:oneformkeylemma}]
We turn to the proof of (\ref{applyforbeta}). We claim we can reduce the proof to the case of functions using that there is a unique pair of functions $(f,g)$, each with vanishing spherical mean, such that $\xi-\xi_{\mathrm{Kerr}} = \slashed{\mathcal{D}}_1^\star (f,g)$ holds. The estimate (\ref{applyforsigma}) applied with $f_{\mathrm{Kerr}}=g_{\mathrm{Kerr}}=0$ yields
\begin{align}
 \sum_{i=0}^n \| [r\slashed{\nabla}]^i( (f,g) \|_{S_{u,v}} &\lesssim \ \| f_{\ell=1}\|_{S_{u,v}} +\| g_{\ell=1}\|_{S_{u,v}} +  \| \mathscr{A}^{[n-2]} r^2\slashed{\mathcal{D}}_2^\star \slashed{\mathcal{D}}_1^\star (f,g) \|_{S_{u,v}}  \, \nonumber \\
 &\lesssim \ \| (r^2\slashed{\Delta} f)_{\ell=1}\|_{S_{u,v}} +\| (r^2\slashed{\Delta} g)_{\ell=1}\|_{S_{u,v}} +  \| \mathscr{A}^{[n-2]} r^2\slashed{\mathcal{D}}_2^\star \slashed{\mathcal{D}}_1^\star (f,g) \|_{S_{u,v}}  \nonumber \\
 &\lesssim   \| (r\slashed{div} \xi)_{\ell=1}\|_{S_{u,v}}  + \| (r\slashed{curl} (\xi-\xi_{\mathrm{Kerr}}))_{\ell=1}\|_{S_{u,v}}+  \| \mathscr{A}^{[n-2]} r^2\slashed{\mathcal{D}}_2^\star \slashed{\mathcal{D}}_1^\star (f,g) \|_{S_{u,v}}  \nonumber \, ,
\end{align}
from which the estimate (\ref{applyforbeta}) follows. The estimate (\ref{applyforeta}) can be reduced to (\ref{applyforbeta}) after noting the relation
\begin{align} \label{angularconversionformula}
r^3 \slashed{\mathcal{D}}_2^\star \slashed{\mathcal{D}}_1^\star \slashed{\mathcal{D}}_1 \xi = 2 \mathscr{A}^{[2]} r  \slashed{\mathcal{D}}_2^\star \xi + 2  \slashed{\mathcal{D}}_2^\star (K \xi ) \, .
\end{align}
Finally, the bounds on the boxed term follows readily from the definition of the reference Kerr solution and the fact that the angular momentum parameter satisfies $|J^m_{\I}| \lesssim \varepsilon$.
\end{proof}

\subsection{The basic transport estimates} \label{sec:basictransport}
We rely on the following basic transport estimates. 
\begin{lemma} \label{intlemskri}
Let $Q$ be an $S$-tensor and $\Omega \slashed{\nabla}_3 Q = \xi$ hold along $\underline{C}_{v}^{\I}$. Let $u_+ \geq u_-$ such that $\{u_\pm\} \times S_{u_\pm,v} \subset \underline{C}_{v}^{\I}$. Then
\begin{align}
\| Q \|_{S_{u_\pm,v}} \lesssim \| Q \|_{S_{u_\mp,v}} + \int^{u_+}_{u_{-}} d\bar{u} \, \| \xi\|_{S_{\bar{u},v}} \, .
\end{align}
Similarly, let $Q$ be an $S$-tensor and $\Omega \slashed{\nabla}_4 Q = \xi$ hold along ${C}_{u}^{\I}$. Let $v_+ \geq v_-$ such that $\{v_\pm\} \times S_{u,v_\pm} \subset {C}_{u}^{\I}$. Then
\begin{align}
\| Q \|_{S_{u,v_\pm}} \lesssim \| Q \|_{S_{u,v_\mp}} + \int^{v_+}_{v_{-}} d\bar{v} \, \| \xi\|_{S_{u,\bar{v}}} \, .
\end{align}
\end{lemma}

\subsection{Estimates for the angular master energy for quantities on $\protect\Cbar_{v_{\infty}}^{\I}$} \label{sec:vinfe}
The objective of this section is to prove the estimate (\ref{target1}).

\subsubsection{Estimating \underline{$\hat{\chi}$} and \underline{$\beta$}}

\begin{proposition} \label{lem:chibarfluxes}
On $\underline{C}_{v_\infty}^{\I}$ we have for all $u \in \left[u_{-1},u_f\right]$ and $s=0,1,2$ 
\begin{align}  \label{skrifluxest1}
\sum_{i=0}^{N+1-s} \int_{u}^{u_f} du \|  \left[r \slashed{\nabla}\right]^{i} (r \underline{\hat{\chi}} \Omega^{-1}) \|_{S_{u,v_{\infty}}}^2 \lesssim  \frac{(\varepsilon_0)^2+ \varepsilon^3}{u^{s}} \, .
\end{align}
On the spheres we have
 for $s=0,1,2$
\begin{align}  \label{skrisphereest1}
\sum_{i=0}^{N-s} \|\left[r \slashed{\nabla}\right]^{i} (r \underline{\hat{\chi}}\Omega^{-1})\|^2_{S_{u,\infty}} \lesssim\frac{(\varepsilon_0)^2+ \varepsilon^3}{u^{\min(\frac{1}{2}+s,2)}}  \, .
\end{align}
\end{proposition}
\begin{proof}
The flux estimate follows  directly from the relation (\ref{codazb}) together with the bounds on the flux of $\check{\underline{\Pi}}=r^3\check{\underline{\psi}} \Omega$ in Theorem~\ref{thm:alphaalphabarestimates}  and Proposition \ref{prop:linerror} for the linear lower order (in $r$) terms, as well as Propositions \ref{prop:nlerror} and \ref{prop:eebs} for the non-linear errors (which have additional decay in $r$).

For the bound on spheres recall the null structure equation (\ref{3chib}), which after invoking the flux estimates on $\underline{\alpha}$ of Theorem~\ref{thm:alphaalphabarestimates}  (see also Corollary \ref{cor:replacercheckbyr}) leads to the bound
\begin{align} \label{fvb}
\sum_{i=0}^{N-s} \int_{u}^{u_f} d\bar{u} \| \Omega \slashed{\nabla}_3 \left[r \slashed{\nabla}\right]^{i}  (r \underline{\hat{\chi}} \Omega^{-1}) \|_{S_{\bar{u},v_{\infty}}}^2 \lesssim  \frac{(\varepsilon_0)^2+ \varepsilon^3}{u^{s}} \, 
\end{align}
for $s=0, 1,2$. Now from (\ref{skrifluxest1}) we easily extract a dyadic sequence of spheres $S_{u_i,v_\infty}$ on which (\ref{skrisphereest1}) already holds. The bound can then be extended to any sphere $S_{u,v_\infty}$ using the fundamental theorem of calculus in conjunction with the flux bounds (\ref{fvb}) and (\ref{skrifluxest1}).
\end{proof}

\begin{corollary} \label{cor:betabinf}
On $\underline{C}_{v_\infty}^{\I}$ we have for all $u \in \left[u_{-1},u_f\right]$ and $s=0,1,2$ 
\begin{align} 
\sum_{i=0}^{N-s} \int_{u}^{u_f} du \|  \left[r \slashed{\nabla}\right]^{i} \left( r^2\Omega^{-1}  \underline{\beta}-r^2 (\Omega^{-1}  \underline{\beta})_{\mathrm{Kerr}} \right)  \|_{S_{u,v_{\infty}}}^2 \lesssim  \frac{(\varepsilon_0)^2+ \varepsilon^3}{u^{s}} \, .
\end{align}
On the spheres we have
\begin{align}
\sum_{i=0}^{N-1-s}  \|\left[ r \slashed{\nabla}\right]^{i} \left( r^2\Omega^{-1}  \underline{\beta}-r^2 (\Omega^{-1}  \underline{\beta})_{\mathrm{Kerr}} \right)  \|^2_{S_{u,\infty}} \lesssim\frac{(\varepsilon_0)^2+ \varepsilon^3}{u^{\min(\frac{1}{2}+s,2)}} \, .
\end{align} 
Moreover, the same estimates hold without subtracting the Kerr reference solution.
\end{corollary}

\begin{proof}
The last claim is immediate from $r^2(\Omega^{-1}\underline{\beta})_{\mathrm{Kerr}} \sim r^{-2}$ manifestly satisfying the above estimates in view of (\ref{rtou}). It hence suffices to show the estimate without the Kerr reference solutions subtracted and this  follows directly from the Codazzi equation (\ref{eq:Codazzibar}) using that $\Omega^{-1}\underline{\beta} = \slashed{div}(\Omega^{-1} \underline{\hat{\chi}})$ holds up to terms of order $\frac{1}{r}$, see Propositions \ref{prop:linerror} and \ref{prop:nlerror}. 
\end{proof}

\subsubsection{Estimating \underline{$T$}} 
\begin{proposition} \label{lem:trxbflux}
We have for $\underline{T} = \frac{tr \chi}{\Omega} - \frac{tr \underline{\chi}_\circ}{\Omega_\circ}$ on $\underline{C}_{v_\infty}^{\I}$ for all $u \in \left[u_{-1},u_f\right]$ and $s=0,1,2$ 
\begin{align} \label{trxbflux1top}
\sum_{i=0}^{N+1-s} \int_{u}^{u_f} d\bar{u} \| \left[r\slashed{\nabla}\right]^{i} r^2 \underline{T} \|_{S_{\bar{u},v_\infty}}^2 \lesssim \frac{ \varepsilon^4}{u^{s+1}} 
\end{align}
and
\begin{align} \label{trxbsptop}
\sum_{i=0}^{N+1-s} \| \left[r\slashed{\nabla}\right]^{i}r^2 
 \underline{T}  \|^2_{S_{u,\infty}}\lesssim \frac{ \varepsilon^4}{u^{s+2}}  \, .
\end{align}
\end{proposition}
\begin{proof}
One first proves (\ref{trxbsptop}) by integrating the (angular commuted) Raychaudhuri equation (\ref{Rayni2}) from the sphere $S_{u_f,v_\infty}$ where $\underline{T}=0$ by the gauge condition (\ref{choicesphereTb}). For the right hand side one uses the flux estimates of Proposition \ref{lem:chibarfluxes} for the $\left[r \slashed{\nabla}\right]^n |\underline{\hat{\chi}}|^2$ term on the right hand side, the error estimate of Proposition \ref{prop:nlerror} for the other non-linear terms and that 
\[
\sum_{i=0}^{N+1} \int^{u_f}_u d\bar{u} \| \left[ r\slashed{\nabla}\right]^i (\Omega^2 - \Omega_\circ^2)\|_{S_{\bar{u},v_\infty}}  \lesssim \frac{\ \varepsilon^4}{u^{2}} \, ,
\, 
\]
the latter exploiting that this term is lower oder in $r$ and hence Proposition \ref{prop:linerror} applies. This proves (\ref{trxbsptop}) and the flux estimate (\ref{trxbflux1top}) then follows directly by integration. 
\end{proof}

\subsubsection{Estimating \underline{${\mu}$}$^\dagger$}

\begin{proposition} \label{prop:mubari} 
On $\underline{C}_{v_\infty}^{\I}$ we have for all $u \in \left[u_{-1},u_f\right]$
\begin{align}
\sum_{n=1}^N \| \left[ r \slashed{\nabla}\right]^n (r^3\underline{\mu}^\dagger) \|^2_{S_{u,v_{\infty}}} \lesssim \frac{\epsilon^4}{u^4} \, ,
\end{align}
\end{proposition}
\begin{proof}
Commute equation (\ref{mumodprop}) with $n$ angular derivatives. By the commutation principle 
(Section~\ref{sec:commutationprinciple}), the error from $n$ commutations can be incorporated into $\overset{(in)}{\slashed{\mathcal{E}}^{n+1}_2}$ and we have $\slashed{\nabla} \mu =0$ on the right hand side by gauge condition (\ref{mugaugec}).
We then integrate the resulting equation from the sphere $S_{u_{-1},v_\infty}$ where $\slashed{\nabla} \underline{\mu}^\dagger=0$ holds by gauge condition (\ref{skrigaugeoutgoingE}) and estimate the right hand side using Propositions \ref{prop:nlerror} and \ref{prop:linerror}. Since we can pull out an overall factor of $r^{-\frac{1}{2}}$ on the right we obtain the first estimate after using (\ref{rtou}). 
\end{proof}

\subsubsection{Estimating the $\ell=1$ and $\ell=0$ modes} \label{sec:ir01}

We turn to estimating the $\ell=0$ and $\ell=1$ modes of all quantities. We prove 
\begin{proposition} \label{prop:l01coll}
On $\underline{C}_{v_\infty}^{\I}$ we have for all $u \in \left[u_{-1},u_f\right]$ the following estimates for the $\ell=0,1$ modes of the Ricci coefficients:
\begin{align} 
 \| r \underline{\omega}_{\ell=0} - r\underline{\omega}_\circ \|_{S_{\bar{u},v_\infty}}  &\lesssim \frac{\varepsilon^2}{u^4} , \label{ombbas}  \\
 \| r^4 (\slashed{curl} \underline{\eta})_{\ell=1} -  r^4 \slashed{curl} \underline{\eta}_{\mathrm{Kerr}} \|_{S_{u,v_\infty}}   + \| r^4 (\slashed{curl} {\eta})_{\ell=1} -  r^4 \slashed{curl}{\eta}_{\mathrm{Kerr}} \|_{S_{u,v_\infty}}  &\lesssim \frac{\varepsilon^2}{u^{3/2}}  \label{etabcurl1} \\
 \| r^4 (\slashed{div} \underline{\eta})_{\ell=1} \|_{S_{u,v_\infty}}   &\lesssim \varepsilon^2  \label{etabdiv1} \\
  \| r^3 (\slashed{div} {\eta})_{\ell=1} \|_{S_{u,v_\infty}}&\lesssim \frac{\varepsilon^2}{u^2} \label{etadiv1} \\
\|r^2 \underline{T}_{\ell=1} \|_{S_{\bar{u},v_\infty}} + \|r^2 \underline{T}_{\ell=0} \|_{S_{\bar{u},v_\infty}} &\lesssim \frac{\varepsilon^2}{u^2} \label{Tb01} \\
\|r^3 T_{\ell=1} \|_{S_{\bar{u},v_\infty}} + \|r^3 T_{\ell=0} \|_{S_{\bar{u},v_\infty}} &\lesssim \varepsilon^2 \label{Tl1}.
\end{align}
Moreover, on $\underline{C}_{v_\infty}^{\I}$ we have for all $u \in \left[u_{-1},u_f\right]$ the following estimates for the $\ell=0,1$ modes of the curvature components:
\begin{align}
 \| r^3 \left(\slashed{div} \underline{\beta} \Omega^{-1} \right)_{\ell=1}\|_{S_{u,v_\infty}} + \| r^3 \left(\slashed{curl} \underline{\beta}\Omega^{-1} \right)_{\ell=1} - r^3 \slashed{curl} \left(\underline{\beta}\Omega^{-1} \right)_{\mathrm{Kerr}} \|_{S_{u,v_\infty}}  &\lesssim \frac{\varepsilon^2}{u^4}  \, , \label{betabvinf} \\
 \|r^3 \rho_{\ell=1} \|_{S_{\bar{u},v_\infty}}  + \|r^3 \sigma_{\ell=1}  -r^3 \sigma_{\mathrm{Kerr}} \|_{S_{\bar{u},v_\infty}}  &\lesssim \frac{\varepsilon^2}{u^2} , \label{rhosigmal1} \\
 \| r^3 \rho_{\ell=0} +2M \|_{S_{{u},v_\infty}} + \|r^3 \sigma_{\ell=0} \|_{S_{\bar{u},v_\infty}} &\lesssim \frac{\varepsilon^2}{u^2}  \label{rhosigma0} \\
 \| r^5 (\slashed{curl} \beta \Omega)_{\ell=1} - r^5 \slashed{curl} (\Omega \beta)_{\mathrm{Kerr}}  \|_{S_{u,v_\infty}} &\lesssim \frac{\varepsilon^2}{u^{3/2}} \label{cbl1} \\
 \|r^5 \left(\slashed{div} \beta \Omega \right)_{\ell=1} \|_{S_{\bar{u},v_\infty}} &\lesssim \frac{\varepsilon^2}{u}  \, . \label{dbl1}
\end{align}
Finally, on the sphere $S_{u_f,v_\infty}$ the estimates (\ref{etabcurl1}) and (\ref{cbl1}) hold with the improved rate $\frac{\epsilon^2}{u^{2}}$ on the right hand side.
\end{proposition}

\begin{remark}
Note that by our choice of $v_\infty$ we have
\[
 \| r^3 \slashed{curl} (\Omega^{-1} \underline{\beta})_{\mathrm{Kerr}} \|_{S_{u,v_\infty}}  + \| r^3 \sigma_{\mathrm{Kerr}}  \|_{S_{u,v_\infty}}  \lesssim \frac{\varepsilon^2}{u^4} \, 
\]
so the above estimates for $\underline{\beta}$ and $\sigma$ hold verbatim without their Kerr modes subtracted. 
\end{remark}

\begin{remark}
Estimates on $\underline{\omega}_{\ell=1}$ will be obtained in the context of estimating all $\ell\geq 1$ modes of $\underline{\omega}$ in Section \ref{sec:ombei}. 
\end{remark} 

\begin{proof}
{\bf Step 1: Preliminary estimates.}
Note the estimate (\ref{Tb01}) is already implied by Proposition~\ref{lem:trxbflux}. 
We first recall that from Proposition \ref{prop:Gaussl1} we have for $u \in \left[u_{-1},u_f\right]$
\begin{align} \label{betterl1gauss}
\|r^4 \left(K-K_\circ\right)_{\ell=1}\|_{S_{u,v_\infty}} \lesssim \frac{\varepsilon^2}{u}\ \textrm{ \ and by (\ref{rtou}) in particular \  } \|r^3 \left(K-K_\circ\right)_{\ell=1} \|_{S_{u,v_\infty}}  \lesssim \frac{\varepsilon^2}{u^2}.
\end{align}
Invoking the Gauss equation (\ref{Gauss}) on $S_{u,v_\infty}$ together with the improved estimate (\ref{trxbsptop}) for $\underline{T}$ and the bootstrap assumption $\|T r^3\|_{S_{u,v_\infty}}  \leq \varepsilon$ (which using (\ref{rtou}) implies $\| T r^2\|_{S_{u,v_\infty}}  \lesssim \varepsilon^2 u^{-2}$) we conclude 
\begin{align} \label{rvu}
\| r^3 \left(\rho - \rho_\circ\right)_{\ell=1}\|_{S_{u,v_\infty}}  \lesssim \frac{\varepsilon^2}{u^2}
\end{align}
for $u \in \left[u_{-1},u_f\right]$, which is already the first part of (\ref{rhosigmal1}).
Using Proposition \ref{prop:mubari} and $\mu_{\ell=1}=0$ on $v=v_\infty$ as well as  the improved estimate (\ref{trxbsptop}) for $\underline{T}$ we deduce
\begin{align} \label{rvu2}
\| r^3 \left( \slashed{div} \left(\eta + \underline{\eta}\right) \right)_{\ell=1}\|_{S_{u,v_\infty}}  \lesssim \frac{\varepsilon^2}{u^2} \ \ \ \textrm{and hence} \ \ \ \| r \left(\Omega^2 - \Omega_\circ^2\right)_{\ell=1}\|_{S_{u,v_\infty}}  \lesssim \frac{\varepsilon^2}{u^2}  \, 
\end{align}
for $u \in \left[u_{-1},u_f\right]$. Here the last step follows from (\ref{eq:DlogOmega}) and applying Proposition \ref{prop:almostevalue}. We finally revisit (\ref{mumodprop}) projected to $\ell=1$:
\begin{align}
\Omega \slashed{\nabla}_3 \left(r^2 \underline{\mu}^\dagger_{\ell=1}\right) = \frac{2\Omega_\circ^2}{r^2} \left(r^3 \left(\rho - \rho_\circ\right)_{\ell=1} \right) + \frac{\Omega_\circ^2}{r^2} \left( 1+ \frac{2M}{r} \right)  r^2{\underline{T}}_{\ell=1} - \frac{1 + \frac{4M}{r} }{r^2} r \left(\Omega^2 -\Omega_\circ^2\right)_{\ell=1} \nonumber \\
+ \left(\Omega^2 \overset{(in)}{\slashed{\mathcal{E}}^{1}_{2}}\right)_{\ell=1} + \Omega \slashed{\nabla}_3 \left(r^2 \underline{\mu}^\dagger_{\ell=1}\right) - \left( \Omega \slashed{\nabla}_3 \left(r^2 \underline{\mu}^\dagger\right)\right)_{\ell=1} \, ,
\end{align} 
from which we conclude after integration for all $u \in \left[u_{-1},u_f\right]$ using Proposition \ref{prop:com0}  the estimate
\begin{align}
\| r^4 \underline{\mu}^\dagger_{\ell=1}\|_{S_{u,v_\infty}} \lesssim \varepsilon^2 \, .
\end{align}
{\bf Step 2: Estimates on $S_{u_f,v_\infty}$.} Note that $\underline{T}=0$ on this sphere. From the Gauss equation and the definition of $\underline{\mu}^\dagger_{\ell=1}$ we conclude after using (\ref{betterl1gauss})
\[
\| r^4 \left(\slashed{div} \underline{\eta} \right)_{\ell=1} + \frac{1}{2} r^3 T_{\ell=1} \|_{S_{u_f,v_\infty}}   \lesssim \varepsilon^2  \, .
\]
From the Gauss equation and the definition of ${\mu}_{\ell=1}$ (which vanishes on $v=v_\infty$) we conclude after using (\ref{betterl1gauss})
\[
\| r^4 \left(\slashed{div} {\eta} \right)_{\ell=1} + \frac{1}{2} r^3 T_{\ell=1} \|_{S_{u_f,v_\infty}}   \lesssim \varepsilon^2  \, .
\]
From the Codazzi equation (\ref{eq:Codazzi}) evaluated on the sphere $S_{u_f,v_\infty}$ we obtain using the gauge condition (\ref{betal1gauge})
\[
\| r^4 \left(\slashed{div} \underline{\eta} \right)_{\ell=1} +  r^3 T_{\ell=1} \|_{S_{u_f,v_\infty}}   \lesssim \varepsilon^2  \, .
\]
From the previous three estimates we conclude 
\begin{align} \label{betterbetal1}
\| r^4 \left(\slashed{div} \underline{\eta} \right)_{\ell=1} \|_{S_{u_f,v_\infty}} +  \| r^4 \left(\slashed{div} {\eta} \right)_{\ell=1} \|_{S_{u_f,v_\infty}} + \| r^3 T_{\ell=1} \|_{S_{u_f,v_\infty}}    \lesssim \varepsilon^2 \, .
\end{align}
Using the Codazzi equation (\ref{eq:Codazzibar}) projected to $\ell=1$ on the sphere $S_{u_f,v_\infty}$,
\[
r^4 (\slashed{div} \Omega^{-1} \underline{\beta})_{\ell=1} = \frac{1}{2} (\Omega^{-1} tr \underline{\chi})_\circ r^4 (\slashed{div} \underline{\eta})_{\ell=1}+ (\overset{(in)}{\slashed{\mathcal{E}}^{1}_{0}} )_{\ell=1}\, ,
\]
we conclude after using (\ref{miuy}) for the error
\begin{align} \label{divbetabfinals}
\| r^4 \left(\slashed{div} \Omega^{-1} \underline{\beta} \right)_{\ell=1}\|_{S_{u_f,v_\infty}}  \lesssim \frac{\varepsilon^2}{u^2} \ \textrm{ \ \ and hence \  \  }  \| r^3 \left(\slashed{div} \Omega^{-1}\underline{\beta} \right)_{\ell=1}\|_{S_{u_f,v_\infty}} \lesssim \frac{\varepsilon^2}{u^4} \, .
\end{align}
From the $\slashed{curl}$-part of the Codazzi equation (\ref{eq:Codazzi}) we have
\begin{align} \label{codazu}
\| r^5 \left(\slashed{curl} (\beta \Omega)\right)_{\ell=1}  + r^4 \Omega_\circ^2 \left(\slashed{curl} \underline{\eta}\right)_{\ell=1}  \|_{S_{u_f,v_\infty}} \lesssim \frac{\varepsilon^2}{u^2} \, 
\end{align}
and hence in particular using the triangle inequality and the definition of the Kerr modes 
\begin{align} \label{kmeta01}
 \|  r^4 \left(\slashed{curl} \underline{\eta}\right)_{\ell=1}  - r^4 \slashed{curl} \underline{\eta}_{\mathrm{Kerr}}  \|_{S_{u_f,v_\infty}}+\|  r^4 \left(\slashed{curl} \eta\right)_{\ell=1}  - r^4 \slashed{curl} \eta_{\mathrm{Kerr}}  \|_{S_{u_f,v_\infty}}  \lesssim \frac{\varepsilon^2}{u^2} \, ,
\end{align}
where the bound on $\eta$ follows from $\slashed{curl} \eta = - \slashed{curl} \underline{\eta}$.
We conclude from (\ref{eq:curletacurletabar}) also
\begin{align} \label{datv}
\| r^3 (\sigma_{\ell=1}-\sigma_{\mathrm{Kerr}})\|_{S_{u_f,v_\infty}} \lesssim \frac{\varepsilon^2}{u^2}  \ \ \ \ \textrm{as well as} \ \ \ \| r^4 (\slashed{curl} \Omega^{-1} \underline{\beta})_{\ell=1}\|_{S_{u_f,v_\infty}} \lesssim \frac{\varepsilon^2}{u^2}  \ \ \ \, ,
\end{align}
with the second bound following from applying $\slashed{curl}$ to the Codazzi equation  (\ref{eq:Codazzibar}) and projecting to $\ell=1$. This already proves the last statement of the proposition regarding the improved decay on the sphere $S_{u_f,v_\infty}$.

{\bf Step 3: Estimates for $\ell=0$ on $\underline{C}_{v_\infty}^{\I}$.}
The projected $\slashed{\nabla}_3$-propagation equation for $(\rho r^3+2M)_{\ell=0}$ can be written
\begin{align} \label{rhoprop0}
\Omega \slashed{\nabla}_3 \left( r^3 \rho_{\ell=0} + 2M + f(u) \right) = \frac{3M\Omega^2}{r^2} r^2\underline{T}_{\ell=0} +  (\overset{(in)}{\slashed{\mathcal{E}}^{1}_{1}})_{\ell=0} + \Omega \slashed{\nabla}_3 (r^3 \rho_{\ell=0} + 2M) - \left(\Omega \slashed{\nabla}_3 (r^3 \rho + 2M)\right)_{\ell=0}
\end{align}
where we defined (recall gauge condition (\ref{gaugel0O})) 
\begin{align} \label{fudef}
f(u):=F^\prime(u) = \frac{1}{2} \int_{u}^{u_f} d\hat{u} r^3 (\Omega \hat{\chi} \underline{\alpha})_{\ell=0} \left(\hat{u},v_\infty\right) \, .
\end{align}
We integrate (\ref{rhoprop0}) from $S_{u_f,v_\infty}$ where the quantity in brackets on the left is zero. For the error we observe using the fact that $1 \leq \frac{r(u,v_\infty)}{r(u_f,v_\infty)} \lesssim 1$ and (\ref{fluxnor2}) the estimate
\[
\int_{u}^{u_f} d \bar{u} \| \overset{(in)}{\slashed{\mathcal{E}}^{1}_{1}} \|_{S_{\bar{u},v_{\infty}}} \lesssim \frac{1}{r} \int_{u}^{u_f} d \bar{u} \| r \overset{(in)}{\slashed{\mathcal{E}}^{1}_{1}} \|_{S_{\bar{u},v_{\infty}}} \lesssim \frac{\varepsilon^2}{r u^2} \, .
\]
The same estimate holds for the commutator of the derivative and projection in (\ref{rhoprop0}) in view of $
		\left\vert
		(\Omega \nablaslash_3 (\rho r^3 + 2M))_{\ell=0}
		-
		\Omega \nablaslash_3 ((\rho r^3 + 2M)_{\ell=0})
		\right\vert
		\lesssim
		\Vert \rho r^3 + 2M \Vert_{S_{u,v}} \Vert \Omega \tr \underline{\chi} - \Omega \tr \underline{\chi}_{\circ} \Vert_{S_{u,v}}$.\footnote{Applying the weaker statement  (\ref{eq:nabla3modeprojectionhoI}) produces (\ref{rhol0improved}) with $u^{-1}$ instead of $u^{-2}$ which is sufficient in applications.}
Combining these facts we are lead to the estimate
\begin{align} \label{rhol0improved}
\| (r^3 \rho + 2M)_{\ell=0} + f(u) \|_{S_{{u},v_\infty}} \lesssim \frac{\varepsilon^2}{r u^2} \ \ \ \  \textrm{for all $u \in \left[u_{-1},u_f\right]$.} 
\end{align}
From this one easily deduces (using the bootstrap assumptions on the fluxes for $\hat{\chi}$ and the estimates on $\underline{\alpha}$) the estimate for $\rho$ of (\ref{rhosigma0}). From the equation $\slashed{curl} \eta = -\frac{1}{2} \hat{\chi} \wedge \hat{\underline{\chi}} + \sigma$ projected to $\ell=0$ one deduces the $\sigma$ part of (\ref{rhosigma0}). 

We next estimate $T_{\ell=0}$. For this we derive the projected propagation equation
\[
\Omega \slashed{\nabla}_3 \left(T_{\ell=0} r\right)  = -\Omega_\circ^4 \underline{T}_{\ell=0}  - (\Omega^2 r \hat{\chi} \hat{\underline{\chi}})_{\ell=0} + (\Omega^2 \overset{(in)}{\slashed{\mathcal{E}}^{0}_{3}})_{\ell=0} \, .
\]
We conclude the $\ell=0$ part of (\ref{Tl1}) after integration from $S_{u_f,v_\infty}$ (where $T_{\ell=0}=0$ vanishes by the gauge condition (\ref{choicesphereT})) using Proposition \ref{lem:trxbflux}. The $\ell=0$ part of (\ref{Tb01}) then immediately follows using (\ref{rtou}). We finally turn to $(\underline{\omega} -  \underline{\omega}_\circ)_{\ell=0}$. We project the equation (\ref{bigomega3})
to $\ell=0$ and use the gauge condition (\ref{zeromodegaugec}) to deduce
\begin{align} \label{ombpre}
2 (\underline{\omega} - \underline{\omega}_\circ)_{\ell=0} = & \frac{1}{\Omega_\circ^{2}} \Big[\left(\Omega \slashed{\nabla}_3 (\Omega^2 - \Omega_\circ^2)\right)_{\ell=0} - \Omega \slashed{\nabla}_3   (\Omega^2 - \Omega_\circ^2)_{\ell=0} \Big] -\frac{2}{\Omega_\circ^{2}} \left((\Omega^2 - \Omega_\circ^2)(\underline{\omega} - \underline{\omega}_\circ)\right)_{\ell=0} \, .
\end{align}
We conclude from (\ref{eq:nabla3modeprojectionhoI}) and the bootstrap assumptions that
\begin{align} \label{ompredetail}
\| (\underline{\omega} - \underline{\omega}_\circ)_{\ell=0} \| \lesssim \frac{\varepsilon^2}{r^2 u} \, .
\end{align} 
The $\ell=0$ part of the estimate (\ref{ombbas}) follows immediately using (\ref{rtou}).
We close this step noting that from the definition of $\underline{\mu}^\dagger$ and the estimates established on the $\ell=0$ modes we easily deduce
\begin{align} \label{mubu}
|r^3 \left( \underline{\mu}^\dagger\right)_{\ell=0}| \lesssim \frac{\varepsilon^2}{u^2}  \, .
\end{align}
{\bf Step 4: Estimates for $\ell=1$ on $\underline{C}_{v_\infty}^{\I}$.}
We start from the commuted and projected Bianchi equation 
\begin{align} \label{3bbh}
\Omega \slashed{\nabla}_3 \left(r^5 (\slashed{div} \Omega^{-1} \underline{\beta})_{\ell=1}\right) = r\overset{(in)}{\slashed{\mathcal{E}}^{1}_{0}} + \left[\Omega \slashed{\nabla}_3 \left(r^5 (\slashed{div} \Omega^{-1} \underline{\beta})_{\ell=1}\right) - \left( \Omega \slashed{\nabla}_3 \left(r^5 (\slashed{div} \Omega^{-1} \underline{\beta}\right) \right))_{\ell=1} \right]  \, ,
\end{align}
where we observe that the right hand side grows linearly in $r$ and has the additional structure (Proposition \ref{prop:com0}, (\ref{eq:betabar3}) and Lemma \ref{lem:commutation}) that the non-linearity contains at least one factor of $\underline{\beta}$ or $\underline{\alpha}$, both of which are integrable in $u$ using the flux bounds on these quantities. Integrating (\ref{3bbh}) using (\ref{divbetabfinals}) for the initial term we can hence deduce
$
\| r^4 \left(\slashed{div} \Omega^{-1} \underline{\beta} \right)_{\ell=1}\|_{S_{u,v_\infty}}  \lesssim\varepsilon^2  \, ,
$
and similarly, from the propagation  for the $\slashed{curl}$-part using (\ref{datv}) we find 
the estimate $
\| r^4 \left(\slashed{curl} \Omega^{-1}\underline{\beta} \right)_{\ell=1}\|_{S_{u,v_\infty}}  \lesssim\varepsilon^2  \, .
$
Noting also $\|r^4 \slashed{curl} (\Omega^{-1}\underline{\beta})_{\mathrm{Kerr}}(u_f,v_\infty,\theta)\|_{S_{u,v_\infty}} \lesssim \frac{\varepsilon^2}{u^4}$)
we obtain (\ref{betabvinf}) using (\ref{rtou}).

Turning to $(\rho, \sigma)$ we recall that the $\rho$-part was already obtained in Step 1. For $\sigma_{\ell=1}$ we derive 
\begin{align}
\Omega \slashed{\nabla}_3 \left(r^3 \sigma_{\ell=1} \right) &= r^3 \Omega^2 (\slashed{curl} \Omega^{-1}\underline{\beta})_{\ell=1}  -\frac{1}{2} \left( \Omega^2 r^2 \Omega \hat{\chi} \wedge r\Omega^{-2} \underline{\alpha} \right)_{\ell=1}  + \left( {\slashed{\mathcal{E}}^{0}_{1}} \right)_{\ell=1} +\Omega \slashed{\nabla}_3 \left(r^3 \sigma_{\ell=1} \right) -  \left( \Omega \slashed{\nabla}_3 \left(r^3 \sigma \right)\right)_{\ell=1} \nonumber \, .
\end{align}
Using Proposition \ref{prop:com0} we conclude the $\sigma$-part of  the estimate (\ref{rhosigmal1}) using the estimates just shown on $\underline{\beta}$ and (\ref{fluxnor2}). Note that one needs to use the flux estimates for $\hat{\chi}$ and $\underline{\alpha}$ to obtain the rate of $u^{-2}$ claimed.

From $\mu_{\ell=1}=0$ and the definition of $\mu$ (cf.~(\ref{massaspectfirstdef})) we deduce (\ref{etadiv1}) using (\ref{rvu}). 

We turn to $\underline{\eta}$:
\begin{align} \label{etabpropa}
\Omega \slashed{\nabla}_3 \left(r^3 \underline{\eta}\right) = -2\Omega_\circ^2 \left(\eta + \underline{\eta}\right)r^2 + \Omega_\circ r^3 \slashed{div}  \underline{\hat{\chi}} - \frac{1}{2} \Omega^2 r \slashed{\nabla} (\underline{T} r^2)  + \overset{(in)}{\slashed{\mathcal{E}}^{0}_{0}} \, ,
\end{align}
where $\overset{(in)}{\slashed{\mathcal{E}}^{0}_{0}}$ satisfies $\int_{u}^{u_f} d\bar{u} \| \overset{(in)}{\slashed{\mathcal{E}}^{0}_{0}}\|_{S_{\bar{u},v_\infty}} \lesssim \frac{\varepsilon^2}{u^2}$ from (\ref{fluxnor2}). Commuting and projecting to $\ell=1$ we find
\begin{align}
\Omega \slashed{\nabla}_3 \left(r^4 (\slashed{curl} \underline{\eta})_{\ell=1} - r^4 \slashed{curl} \underline{\eta}_{\mathrm{Kerr}}\right) &= + \overset{(in)}{\slashed{\mathcal{E}}^{1}_{0}} + \left[
\Omega \slashed{\nabla}_3 \left(r^4 (\slashed{curl} \underline{\eta})_{\ell=1} \right)- \left(\Omega \slashed{\nabla}_3 (r^4 \slashed{curl} \underline{\eta})\right)_{\ell=1} \right] \nonumber \\
& \ \ -3M \sum_{m=-1}^1 a^m_{\I}  \Omega \slashed{\nabla}_3 r^2 \slashed{\Delta} Y^{\ell=1}_n   \, ,
\label{mals}
\end{align}
where $\overset{(in)}{\slashed{\mathcal{E}}^{1}_{0}}$ satisfies $\int_{u}^{u_f} d\bar{u} \| \overset{(in)}{\slashed{\mathcal{E}}^{1}_{0}}\|_{S_{\bar{u},v_\infty}} \lesssim \frac{\varepsilon^2}{u^2}$ from (\ref{fluxnor2}). The commutator term in square brackets satisfies the same estimate with $\frac{\varepsilon^2}{u^{3/2}}$ (use  (\ref{eq:nabla3modeprojectionhoI}) and the flux estimate for $\slashed{curl} \underline{\eta}$).\footnote{This could be improved to $u^{-2}$ by improving the estimate (\ref{eq:nabla3modeprojectionhoI}) to a flux estimate but we will not require or use this. This is the origin of the small loss in decay in the $\slashed{curl}$ estimates.} The last term in (\ref{mals}) is order $\varepsilon^2 u_f^{-1} r^{-1}$ by Proposition \ref{prop:modesdifference} and hence easily integrated using (\ref{rtou}). We conclude  using (\ref{codazu})  for all $u \in \left[u_{-1},u_f\right]$ the estimate (\ref{etabcurl1}), first for $\underline{\eta}$ and in view of $\slashed{curl} \eta =-\slashed{curl} \underline{\eta}$ also for $\eta$. Commuting now (\ref{etabpropa}) with $r\slashed{div}$ and projecting to $\ell=1$, the right hand side is integrable using and the estimates of Proposition \ref{lem:trxbflux} for $\underline{T}$ and (\ref{rvu2}) for $\slashed{div} \left(\eta + \underline{\eta}\right)$ and we deduce (\ref{etabdiv1}).

We turn to estimating $\beta$. From the Codazzi equation (\ref{eq:Codazzi}) we deduce
\begin{align} \label{curlcodaz}
r^5 \slashed{curl} \slashed{div} \Omega \hat{\chi} = -\Omega_\circ^2 r^4 \slashed{curl} \underline{\eta} - r^5 \slashed{curl} (\beta \Omega) + \overset{(in)}{\slashed{\mathcal{E}}^{1}_{0}} \, , 
\end{align}
which after projection to $\ell=1$ allows one to conclude (using (\ref{miuy}) for the non-linear term)
\begin{align} \label{hiad}
\|  r^4  \Omega_\circ^2 (\slashed{curl} \underline{\eta})_{\ell=1} + r^5 (\slashed{curl} \beta \Omega)_{\ell=1} \|_{S_{u,v_\infty}} \lesssim \frac{\varepsilon^2}{u^2} \, .
\end{align}
Note that (\ref{hiad}) actually holds for any sphere $S_{u,v}$ in $\mathcal{D}^{\I}$.
Using the estimate (\ref{etabcurl1}), the triangle inequality and the definition of the Kerr modes we conclude (\ref{cbl1}). Turning to the propagation equation 
\begin{align}
\Omega \slashed{\nabla}_3 \left(r^5 (\slashed{div}\beta \Omega)_{\ell=1}\right) &= \Omega_\circ^2 r^5 \slashed{\Delta} \rho_{\ell=1} + 3 \rho_0 \Omega_\circ^2 r^5 (\slashed{div} \eta)_{\ell=1} + (\overset{(in)}{\slashed{\mathcal{E}}^{1}_{0}})_{\ell=0} + (\slashed{\mathcal{E}}^1_1)_{\ell=0} \nonumber \\
&+ \Omega \slashed{\nabla}_3 \left(r^5 (\slashed{div}\beta \Omega)_{\ell=1}\right)  - \left(\Omega \slashed{\nabla}_3 \left(r^5 (\slashed{div}\beta \Omega)\right)\right)_{\ell=1} -  \Omega_\circ^2 r^5 \slashed{\Delta} \rho_{\ell=1}  +  \Omega_\circ^2 r^5 (\slashed{\Delta} \rho)_{\ell=1} \, ,  \nonumber
\end{align}
we easily derive using Propositions  \ref{prop:nlerror} as well as \ref{prop:com0} and \ref{prop:almostevalue}, the estimate (\ref{dbl1}).

Finally, we estimate the $\ell=1$ modes of $T$. Using the Codazzi equation 
\begin{align}
r^5 \slashed{div}  \slashed{div} \Omega \hat{\chi} &= -\Omega_\circ^2 r^4 \slashed{div} \underline{\eta} - r^5 \slashed{div} (\beta \Omega) + \frac{1}{2} r^5 \slashed{\Delta} T + \overset{(in)}{\slashed{\mathcal{E}}^{1}_{0}}
\end{align}
projected to $\ell=1$ we conclude the $\ell=1$ part of (\ref{Tl1}). The $\ell=1$ part of the bound (\ref{Tb01}) then immediately follows using (\ref{rtou}).
\end{proof}

From the proof we explicitly note the following corollary:
\begin{corollary} \label{cor:extra01}
With $f(u)$ defined in (\ref{fudef}) we also have the estimate (\ref{rhol0improved}) for $\rho_{\ell=0}$. 
\end{corollary}

\subsubsection{Estimating $\rho$ and $\sigma$}

\begin{proposition} \label{lem:rhoflux}
On $\underline{C}_{v_\infty}^{\I}$ we have for all $u \in \left[u_{-1},u_f\right]$ and $s=0,1,2$ 
\begin{align}  \label{bimpro}
\sum_{i=0}^{N-s}  \int^{u_f}_u  d\bar{u}  \| \left[r\slashed{\nabla}\right]^{i} (r^3\rho + 2M_f, r^3\sigma) \|_{S_{\bar{u},v_\infty}}^2 \lesssim \frac{(\varepsilon_0)^2+\varepsilon^3}{u^s} \, \ \  \, .
\end{align}
We also have on spheres
\begin{align} \label{rhoimp}
 \sum_{i=0}^{N-1-s}  \|  \left[r\slashed{\nabla}\right]^{i}   (r^3\rho + 2M_f, r^3\sigma) \|^2_{S_{u,v_\infty}} \lesssim \frac{(\varepsilon_0)^2+ \varepsilon^3}{u^{\min(s+\frac{1}{2},2)}}  \, .
\end{align}
\end{proposition}

\begin{proof}
We first note that it suffices to prove the above estimates with the sums starting from $i=2$ and replacing $\left[r \slashed{\nabla}\right]^i$ by $\mathscr{A}^{[i-2]} r^2 \slashed{\mathcal{D}}_2^\star\slashed{\mathcal{D}}_1^\star$. This follows from Proposition \ref{prop:functionkeylemma}  after noting that the projections $ (r^3\rho + 2M, r^3\sigma)_{\ell=1}$ and $ (r^3\rho + 2M, r^3\sigma)_{\ell=0}$ already manifestly satisfy the above estimates from the results of the previous section; see (\ref{rhosigmal1}), (\ref{rhosigma0}).

To prove the latter, we recall that we have the following (already improved) flux estimates for $\Psi=r^5 P$:
\begin{align}  \label{helpi}
\sum_{i=0}^{N-2-s} \int^{u_f}_u  d\bar{u}  \|\mathscr{A}^{[i]}  \Psi \|_{S_{\bar{u},v_\infty}}^2 \lesssim \frac{(\varepsilon_0)^2+ \varepsilon^3}{u^s} \, 
\end{align}
and 
\begin{align} \label{helpi2}
\sum_{i=0}^{N-3-s}  \int^{u_f}_u  d\bar{u}   \|\slashed{\nabla}_3 \mathscr{A}^{[i]}  \Psi \|_{S_{\bar{u},v_\infty}}^2 \lesssim \frac{(\varepsilon_0)^2+ \varepsilon^3}{u^s} \, .
\end{align}
The first estimate of the proposition now follows from the flux estimate on $\Psi = r^5 P$ of Proposition \ref{prop:afa}, the relations of Proposition \ref{prop:PPbaridentities} and (\ref{skrifluxest1}). To obtain the estimates on spheres we will first deduce  
\begin{align} \label{titu}
\sum_{i=0}^{N-3-s} \| \mathscr{A}^{[i]} \Psi \|^2_{S_{u,v_\infty}} & \lesssim \frac{(\varepsilon_0)^2+\varepsilon^3}{u^{\max(s+\frac{1}{2},2)}}  \, ,
\end{align}
after which the desired estimates follows immediately 
from  the relations of Proposition \ref{prop:PPbaridentities} and (\ref{skrisphereest1}). To establish (\ref{titu}) we extract from (\ref{helpi}) a dyadic sequence of good spheres with additional decay and integrate from these spheres using 
Cauchy--Schwarz and the flux bounds for the $3$-derivatives (\ref{helpi2}). 
\end{proof}

\subsubsection{Estimating \underline{$\eta$} and $\eta$}
We finally exploit Proposition \ref{prop:mubari} regarding the modified mass aspect function $\underline{\mu}^\dagger$ which allows us to deduce decay estimates on $\underline{\eta}$.
\begin{proposition} \label{prop:etbv}
On $\underline{C}_{v_\infty}^{\I}$ we have for all $u \in \left[u_{-1},u_f\right]$ and $s=0,1,2$ 
\begin{align} \label{etbflux}
\sum_{i=0}^{N+1-s} \int^{u_f}_u d\bar{u}   \|\left[r\slashed{\nabla}\right]^{i}   r^2( \underline{\eta} - \underline{\eta}_{\mathrm{Kerr}})  \|_{S_{\bar{u},v_{\infty}}}^2  \lesssim \frac{(\varepsilon_0)^2+\varepsilon^3}{u^s}   \, .
\end{align}
On spheres we have 
 \begin{align} \label{etbsphere}
\sum_{i=0}^{N-s}   \| \left[r\slashed{\nabla}\right]^{i}   r^2( \underline{\eta} - \underline{\eta}_{\mathrm{Kerr}})  \|_{S_{u,v_\infty}}^2   \lesssim \frac{(\varepsilon_0)^2+ \varepsilon^3}{u^s}     \, .
  \end{align}
 Finally, (\ref{etbflux}) and (\ref{etbsphere}) also hold if one replaces $( \underline{\eta} - \underline{\eta}_{\mathrm{Kerr}})$ by $\underline{\eta}$ itself.
\end{proposition}

\begin{proof}
In view of Proposition \ref{prop:oneformkeylemma} and the estimates on the $\ell=1$ modes of $\underline{\eta}$ in Proposition \ref{prop:l01coll} it suffices to prove these estimate for the sums starting from $i=3$ and with $\left[r \slashed{\nabla}\right]^i r^3( \underline{\eta} - \underline{\eta}_{\mathrm{Kerr}}) $ replaced by $\mathscr{A}^{[i-3]}  r^2 \slashed{\mathcal{D}}_2^\star\slashed{\mathcal{D}}_1^\star \slashed{\mathcal{D}}_1 (r^3 \underline{\eta})$. 
Now, from the estimate of Proposition \ref{prop:mubari} and equation (\ref{curle}) we have
\begin{align}
 \int_{u}^{u_f} d\bar{u} \Big\|   \mathscr{A}^{[i-2]}  r^2 \slashed{\mathcal{D}}_2^\star\slashed{\mathcal{D}}_1^\star  \left( r^3 \slashed{div} \underline{\eta} + r^3 \rho + 2M_f + \frac{1}{2} r^3 \hat{\chi} \underline{\hat{\chi}} + r^2 \underline{T}, r^3\slashed{curl} \underline{\eta} + r^3 \sigma - \frac{1}{2} r^3 \hat{\chi} \wedge \underline{\hat{\chi}} \right) \Big\|^2_{S_{\bar{u},v_\infty}} \lesssim \frac{\varepsilon^4}{u^3} \, . \nonumber
\end{align}
We now use Proposition \ref{lem:rhoflux} to control the flux on $\left(r^3 \rho + 2M_f,\sigma\right)$ and Proposition \ref{lem:trxbflux} to control the flux on $\underline{T}$. It is easy to see that the bootstrap assumptions and Sobolev embedding implies for $n\leq N$
\begin{align} \label{bht}
\int_{u}^{u_f} d\bar{u} \|  \mathscr{A}^{[n-2]} r^2 \slashed{\mathcal{D}}_2^\star\slashed{\mathcal{D}}_1^\star  (r^3 \hat{\chi} \underline{\hat{\chi}}) \|_{S_{\bar{u},v_\infty}}^2 
\lesssim \  \frac{\varepsilon^4}{u^{2}} \, ,
\end{align}
and similarly for the product $\hat{\chi} \wedge \underline{\hat{\chi}}$.
This establishes (\ref{etbflux}). The estimate on the spheres is similar. Here we ``lose a derivative" because at the level of $N$ derivatives we only have the \emph{flux} of $\rho$ at our disposal.  For the final claim note that $\underline{\eta}_{\mathrm{Kerr}}$ itself satisfies the estimate in view of $\underline{\eta}_{\mathrm{Kerr}} \sim r^{-3}$ and (\ref{rtou}).
\end{proof}

\begin{proposition} \label{prop:etv}
Proposition \ref{prop:etbv} holds verbatim replacing $\underline{\eta}$ by $\eta$.
\end{proposition}
\begin{proof}
The proof is identical to that of Proposition \ref{prop:etbv} now starting from the fact that by 
gauge condition (\ref{mugaugec}) and  (\ref{curle}) we have
 $ \| \mathscr{A}^{[i-2]}  r^2 \slashed{\mathcal{D}}_2^\star\slashed{\mathcal{D}}_1^\star  \left( r^3 \mu, r^3\slashed{curl} {\eta} - r^3 \sigma + \frac{1}{2} r^3 \hat{\chi} \wedge \underline{\hat{\chi}} \right)\|^2_{S_{u,v_\infty}} = 0$ along $v=v_\infty$.
\end{proof}

\subsubsection{Estimating $\hat{\chi}$ and $\beta$}
Before we estimate $\hat{\chi}$, we deduce the following estimates:
\begin{proposition} \label{prop:consequencesIgauge}
On $\underline{C}_{v_\infty}^{\I}$ we have for all $u \in \left[u_{-1},u_f\right]$ the following bounds.
For $1\leq n \leq N$
\begin{align} \label{fluxchiiseta}
 \int_{u}^{u_f} d\bar{u} \| \mathscr{A}^{[n-1]}r \slashed{\mathcal{D}}_2^\star (r^2  \underline{\eta} -r^3 \slashed{div} \Omega \hat{\chi}) \|^2_{S_{\bar{u},v_\infty}} \lesssim \frac{\varepsilon^4}{u^4}  \, ,
\end{align}
and for $2\leq n \leq N$
\begin{align} \label{spherechiiseta}
 \|  \mathscr{A}^{[n-2]} r\slashed{\mathcal{D}}_2^\star  (r^2 \underline{\eta} -r^3 \slashed{div} \Omega \hat{\chi}) \|^2_{S_{{u},v_\infty}} \lesssim \frac{\varepsilon^4}{u^4}   \, .
\end{align}
\end{proposition}

\begin{proof}
This is a direct consequence of the Codazzi equation (\ref{codazb}) evaluated on $\underline{C}_{v_\infty}^{\I}$ using Propositions \ref{prop:nlerror} and \ref{prop:linerror} and recalling (\ref{rtou}).
\end{proof}

From (\ref{fluxchiiseta}) and (\ref{spherechiiseta}) and Proposition \ref{prop:etbv} we can now deduce applying  Proposition \ref{prop:symmetrictracelesskeylemma} the following corollary.

\begin{corollary} \label{cor:bimp}
On $\underline{C}_{v_\infty}^{\I}$ we have for all $u \in \left[u_{-1},u_f\right]$ and $s=0,1$
\begin{align} \label{chifluxby}
\sum_{i=0}^{N+1-s} \int^{u_f}_u d\bar{u}  \|\left[r\slashed{\nabla}\right]^{i}   \left(r^3 \Omega \hat{\chi}\right) \|_{S_{\bar{u},v_\infty}}^2 \lesssim \frac{(\varepsilon_0)^2+ \varepsilon^3}{u^{1+s}}\, .
\end{align} 
On spheres, we have
\begin{align}  \label{chisphereby}
\sum_{i=0}^{N}   \|\left[r\slashed{\nabla}\right]^{i}   \left(r^3 \Omega \hat{\chi}\right) \|_{S_{u,v_\infty}}^2 \lesssim \frac{(\varepsilon_0)^2+ \varepsilon^3}{u}    \, 
  \end{align}
  and
  \begin{align}  \label{chisphereby2}
\sum_{i=0}^{N-2}  \|\left[r\slashed{\nabla}\right]^{i}   \left(r^3\Omega \hat{\chi}\right) \|_{S_{u,v_\infty}}^2  \lesssim \frac{(\varepsilon_0)^2+ \varepsilon^3}{u^2}   \, .
  \end{align}
\end{corollary}

From the relation of Proposition \ref{prop:PPbaridentities} and the estimates on $\psi$ of Theorem \ref{thm:alphaalphabarestimates} we deduce:

\begin{corollary} \label{cor:bimpa}
On $\underline{C}_{v_\infty}^{\I}$ we have for all $u \in \left[u_{-1},u_f\right]$ 
\begin{align} \label{betaflux}
\sum_{i=1}^{N} \int^{u_f}_u d\bar{u}  \|\left[r\slashed{\nabla}\right]^{i} r^{4}(\Omega \beta - (\Omega\beta)_{\mathrm{Kerr}}) \|_{S_{\bar{u},v_\infty}}^2 \lesssim (\varepsilon_0)^2+ \varepsilon^3\, .
\end{align} 
On spheres we have for $s=0,1,2$
 \begin{align}  \label{betapshere}
\sum_{i=1}^{N-s-1} \| \left[r\slashed{\nabla}\right]^{i} r^{4}(\Omega \beta - (\Omega\beta)_{\mathrm{Kerr}}) \|_{S_{u,v_\infty}}^2 \lesssim \frac{(\varepsilon_0)^2+ \varepsilon^3}{u^\frac{s}{2}}    \, .
  \end{align}
 Finally, by (\ref{rtou}) both estimates hold with $\frac{(\varepsilon_0)^2+ \varepsilon^3}{u^4}$ on the right provided we replace $r^4$ by $r^{7/2}$ on the left.
\end{corollary}

\begin{proof}
In view of Proposition~\ref{prop:oneformkeylemma} and the estimates on the $\ell=1$ modes of $\Omega{\beta}$ in 
Proposition~\ref{prop:l01coll}, it suffices to prove these estimates replacing $ \| \left[r\slashed{\nabla}\right]^{i} r^{4-s/2}(\Omega \beta -(\Omega \beta)_{\mathrm{Kerr}}) \|$ by $ \|  \mathscr{A}^{[i-1]} r^{4-s/2} r \slashed{\mathcal{D}}_2^\star \Omega \beta \|$.

For the latter, we first note that the estimates of Theorem \ref{thm:alphaalphabarestimates} imply that for $s=0,1,2$, we have
\[
\sum_{i=1}^{N} \int^{u_f}_u d\bar{u}  \|\mathscr{A}^{[i]} r^5 \psi \Omega \|_{S_{\bar{u},v_\infty}}^2 \lesssim (\varepsilon_0)^2+ \varepsilon^3 \ \ \ \textrm{and} \ \ \ \sum_{i=1}^{N-s-1} \| \mathscr{A}^{[i]} r^{5} \psi \Omega \|_{S_{u,v_\infty}}^2 \lesssim \frac{(\varepsilon_0)^2+ \varepsilon^3}{u^{\frac{s}{2}}} \, .
\]
From the relation of Proposition~\ref{prop:PPbaridentities} and the estimates of Corollary~\ref{cor:bimp} 
and Proposition~\ref{prop:eebs}, the desired estimates now follow.
\end{proof}

\subsubsection{Estimating $T=\Omega tr \chi- (\Omega tr \chi)_\circ$} 
\begin{proposition} \label{prop:bimp2}
On $\underline{C}_{v_\infty}^{\I}$ we have for all $u \in \left[u_{-1},u_f\right]$ 
\begin{align} \label{Timp}
\sum_{i=0}^{N+1}  \| \left[r \slashed{\nabla}\right]^{i} r^{5/2} T\|_{S_{u,v_\infty}}^2 \lesssim \frac{\varepsilon^3}{u^4} \, .
 \end{align}
 For $n \leq N$ we have
 \begin{align} \label{Timp2}
 \sum_{i=0}^N \|\left[r \slashed{\nabla}\right]^{n} r^{3} T\|_{S_{u,v_\infty}}^2\lesssim \varepsilon_0^2 + \varepsilon^3 \, ,
 \end{align}
 hence in particular
 \begin{align} \label{Timp3}
 \sum_{i=0}^N \| \left[r \slashed{\nabla}\right]^{n} r^{2} T\|_{S_{u,v_\infty}}^2\lesssim \frac{(\varepsilon_0)^2 +\varepsilon^3}{u^4} \, .
 \end{align}
\end{proposition}
\begin{proof}
In view of Proposition \ref{prop:functionkeylemma} and the estimates on the $\ell=0,1$ modes of $T$ in Proposition \ref{prop:l01coll} it suffices to prove these estimates with the sums starting from $i=2$ and with $\left[r \slashed{\nabla}\right]^i$ replaced by $\mathscr{A}^{[i-2]} r^2 \slashed{\mathcal{D}}_2^\star\slashed{\nabla}$. 

The estimate (\ref{Timp}) then follows by commuting  the equation (\ref{uty}) with $\mathscr{A}^{[i-2]}  r^2 \slashed{\mathcal{D}}_2^\star\slashed{\nabla}$ (using that the boxed term then vanishes by the gauge condition (\ref{mugaugec}), $\mu_{\ell \geq 1}=0$, and integrating. To show (\ref{Timp2}) we proceed as before but now observe that for all $2 \leq i \leq N$ we have
\[
\int_{u_{-1}}^{u_f}  | \mathscr{A}^{[i-2]}  r^2 \slashed{\mathcal{D}}_2^\star\slashed{\nabla} \underline{T} r^2|^2 u^{1+\delta} du  + \int_{u_{-1}}^{u_f} r^4 |\mathscr{A}^{[i-2]}r \slashed{\mathcal{D}}_2^\star ( \eta + \underline{\eta})|^2 u^{1+\delta} du \lesssim \varepsilon^4
\]
following from Propositions \ref{prop:etbv} and \ref{prop:etv} as well as the estimate (\ref{trxbsptop}) with $s=1$. The last estimate is immediate from (\ref{rtou}).
\end{proof}

\subsubsection{Estimating \underline{$\omega$}} \label{sec:ombei}

\begin{proposition} \label{prop:ombe}
On $\underline{C}_{v_\infty}^{\I}$ we have for all $u \in \left[u_{-1},u_f\right]$ and $s=0,1,2$
\begin{align} \label{ombfluximp}
\sum_{i=0}^{N+1-s} \int^{u_f}_u d\bar{u}   \|\left[r\slashed{\nabla}\right]^{i}   \left(r \underline{\omega} -r \underline{\omega}_\circ \right) \|_{S_{\bar{u},v_\infty}}^2 \lesssim \frac{(\varepsilon_0)^2+ \varepsilon^3}{u^{s}} \, .
\end{align}
We also have
\begin{align} \label{ombimp}
\sum_{i=0}^{N-s}  \|\left[r\slashed{\nabla}\right]^{n}   \left(r\underline{\omega} - r \underline{\omega}_\circ \right)\|_{S_{u,v_\infty}}^2  \lesssim \frac{(\varepsilon_0)^2+ \varepsilon^3}{u^{s}} \, .
\end{align}
\end{proposition}

\begin{proof}
Note that by Proposition \ref{prop:l01coll} the $\ell=0$ modes of $\underline{\omega}$ already manifestly satisfy the estimates claimed.
Projecting (\ref{ellipticomegab}) to $\ell \geq 1$ we obtain 
\begin{align} \label{eop}
2r^2 \slashed{\Delta} \underline{\omega}_{\ell \geq 1} = & + 2r^2 \Omega_\circ^2 (\slashed{div} \left(\Omega^{-1} \underline{\beta}\right))_{\ell \geq 1} + \Omega_{\circ}^2 (\underline{T}r^2)_{\ell \geq 1} \left(\frac{3M_f}{r^3}\right)  + \frac{1}{2r} r^2 \left( |r \underline{\hat{\chi}}|^2\right)_{\ell \geq 1}+ \left(\overset{(in)}{\slashed{\mathcal{E}}^{1}_{2}}\right)_{\ell \geq 1} \nonumber \\
&+\frac{1}{r}  \Big[ \left(\Omega \slashed{\nabla}_3 (r^3 \mu+2M)\right)_{\ell \geq 1} - \Omega \slashed{\nabla}_3 (r^3 \mu+2M)_{\ell \geq 1}\Big]  + 2 r^2  \left[ \slashed{\Delta} \underline{\omega}_{\ell \geq 1} -  \left(\slashed{\Delta} \underline{\omega}\right)_{\ell \geq 1}  \right] \, .
\end{align}
The estimates now follow from (angular commutation and) Corollary \ref{cor:betabinf} and Proposition \ref{lem:trxbflux} as well as Proposition \ref{prop:divcurlmodes} (with $\xi = \slashed{\nabla}_A  \underline{\omega}$) and (\ref{eq:nabla3modeprojectionhoI}) in conjunction with Proposition \ref{prop:nlerror}.
\end{proof}

\subsubsection{Conclusions}
We now combine all our estimates to establish that 
we have shown (\ref{target1}). We recall the energies $\slashed{\mathbb{E}}^N_{v_\infty} \left[\Gamma\right]$ and $\slashed{\mathbb{E}}^N_{v_\infty} \left[\mathcal{R}\right]$. They contain sums over all Ricci-coefficients $\Gamma_p$ and curvature components $\mathcal{R}_p$. The following list shows where the estimate is established for each individual quantity:
\begin{enumerate}
\item Corollary \ref{cor:betabinf} shows it for $\Omega^{-1} \underline{\beta}$. Proposition \ref{lem:rhoflux} shows it for ${\rho}, {\sigma}$. \\
Corollary \ref{cor:bimpa} shows it for $\Omega{\beta}$.
\item Proposition \ref{lem:chibarfluxes} shows it for $\Omega^{-1}\underline{\hat{\chi}}$. 
\item Proposition \ref{lem:trxbflux} shows it for $\underline{T}$. 
\item Propositions \ref{prop:etbv} and \ref{prop:etv} show it for $\underline{\eta}$ and $\eta$. 
\item Corollary \ref{cor:bimp}  shows it for $\Omega\hat{\chi}$. 
\item Proposition \ref{prop:bimp2}  shows it for ${T}$.
\item Proposition \ref{prop:ombe} shows it for $\underline{\omega}-\underline{\omega}_\circ$ and it is easy to see the estimates hold verbatim for $\Omega^{-2}\left(\underline{\omega}-\underline{\omega}_\circ\right)$.
\item The estimate on $(r (\Omega^2 - \Omega_\circ^2)$ is easily obtained from the gauge condition $(r (\Omega^2 - \Omega_\circ^2)_{\ell=0} = 0$ on $C^{\I}_{v_\infty}$, the relation $ \slashed{\nabla} (\Omega^2 - \Omega_\circ^2) = \Omega^2 \left(\eta + \underline{\eta}\right)$ and the estimates on $\eta$ and $\underline{\eta}$ in Propositions \ref{prop:etbv} and \ref{prop:etv}.
\end{enumerate}

\subsubsection{Estimating the quantity $Y$}

We finally deduce an estimate on the quantity $Y$ (both flux and estimates on spheres) which follows directly from the definition (\ref{Ydef}) and the previous bounds on $\underline{\eta}$, $\underline{\hat{\chi}}$ and $\underline{T}$.
\begin{corollary} \label{cor:auxYonskri}
On $\underline{C}_{v_\infty}^{\I}$ we have for all $u \in \left[u_{-1},u_f\right]$ and $s=0,1,2$
\begin{align} \label{fluxYi}
\sum_{i=0}^{N-s} \int_{u}^{u_f} d\bar{u} \|  \left[r \slashed{\nabla}\right]^{i} Y\|_{S_{\bar{u},v_\infty}}^2 \lesssim \frac{(\varepsilon_0)^2+\varepsilon^3}{u^{s+1}}
\end{align}
while on spheres
\begin{align} \label{sphereYi}
\sum_{i=0}^{N-1-s} \|  \left[r \slashed{\nabla}\right]^{i} Y\|_{S_{u,v_\infty}}^2 \lesssim \frac{(\varepsilon_0)^2+ \varepsilon^3}{u^{s}}  \, .
\end{align}
\end{corollary}

\subsection{Estimates for the angular master energy for quantities on $C_{u_{-1}}^{\I}$} \label{sec:u0est}

The objective of this section is to prove the estimate (\ref{target2}).

\subsubsection{The $\ell=0$ and $\ell=1$ modes}
\begin{proposition} \label{prop:datal01}
For the Ricci coefficients along $C_{u_{-1}}^{\I}$ we have for all $v \in \left[v(u_{-1},R_{-2}),v_\infty\right]$
\begin{align}
\|r^3 {\omega}_{\ell=1} \|_{S_{u_{-1},v}} + \|r^3 {\omega}_{\ell=0} \|_{S_{u_{-1},v}} + \|r^2 (\Omega^2- \Omega_\circ^2)_{\ell=0} \|_{S_{u_{-1},v}} + \|r^2 (\Omega^2- \Omega_\circ^2)_{\ell=0} \|_{S_{u_{-1},v}} &\lesssim \varepsilon^2 \, ,  \label{angom} \\
\|r^{2} \underline{T}_{\ell=1}\|_{S_{u_{-1},v}} + \|r^{2} \underline{T}_{\ell=0}\|_{S_{u_{-1},v}} + \|r^{3} T_{\ell=1}\|_{S_{u_{-1},v}} + \|r^{3} T_{\ell=0}\|_{S_{u_{-1},v}}   &\lesssim  \varepsilon^2 \, , \label{Tzme} \\
\|r^{3}  (\slashed{div} \eta)_{\ell=1}\|_{S_{u_{-1},v}} + \|( r^4 (\slashed{curl} \eta)_{\ell=1} - r^4 \slashed{curl} \eta_{\mathrm{Kerr}} \|_{S_{u_{-1},v}}  &\lesssim   \varepsilon^2 \, , \label{etadatal1} \\
\|r^{3}  (\slashed{div} \underline{\eta})_{\ell=1}\|_{S_{u_{-1},v}} + \|(r^4 ( \slashed{curl} \underline{\eta})_{\ell=1} - r^4 \slashed{curl} \underline{\eta}_{\mathrm{Kerr}} \|_{S_{u_{-1},v}}  &\lesssim   \varepsilon^2 \label{etabdatal1} \, .
\end{align}
For the curvature components along $C_{u_{-1}}^{\I}$ we have for all $v \in \left[v(u_{-1},R_{-2}),v_\infty\right]$
\begin{align}
 \|r^{5}  (\slashed{div} \Omega \beta)_{\ell=1}\|_{S_{u_{-1},v}} + \|( r^5 ( \slashed{curl} \beta \Omega)_{\ell=1} - r^5 (\slashed{curl} (\beta \Omega)_{\mathrm{Kerr}}) |_{S_{u_{-1},v}}  & \lesssim  \varepsilon^2 \, ,  \label{betal1} \\
 \|r^{3}\rho_{\ell=1}\|_{S_{u_{-1},v}} + \|r^{3}\rho_{\ell=0} + 2M\|_{S_{u_{-1},v}} + \|r^{3}\sigma_{\ell=1}- r^3 \sigma_{\mathrm{Kerr}} \|_{S_{u_{-1},v}} + \|r^{3}\sigma_{\ell=0}\|_{S_{u_{-1},v}}  & \lesssim  \varepsilon^2 \, , \label{rhosigmadatal01} \\
\|r^{3}(\slashed{div} \Omega^{-1} \underline{\beta})_{\ell=1}\|_{S_{u_{-1},v}} +  \|r^{3}(\slashed{curl} (\Omega^{-1}\underline{\beta}))_{\ell=1} + r^3 \slashed{curl} (\Omega^{-1}\underline{\beta})_{\mathrm{Kerr}}\|_{S_{u_{-1},v}}& \lesssim   \varepsilon^2 \label{betabdatal01} \, .
\end{align}
\end{proposition}

\begin{proof}
{\bf Step 1. We first prove all the estimates for $\ell=0$.} The estimate for $\omega_{\ell=0}$ is immediate from gauge condition (\ref{gaugel0O}). The one on $\left(\Omega^2-\Omega_\circ^2\right)_{\ell=0}$ follows from projecting the equation
\[
2({\omega} - {\omega}_\circ)  +  2\left(\frac{\Omega^2-\Omega_\circ^2}{\Omega_\circ^2}\right)({\omega} - {\omega}_\circ) = \Omega \slashed{\nabla}_4 \left(\frac{\Omega^2 - \Omega_\circ^2}{\Omega_\circ^2}\right)  \, 
\] 
to $\ell=0$ and integrating from $S_{u_{-1},v_\infty}$ (where $\left(\Omega^2-\Omega_\circ^2\right)_{\ell=0}=0$ by gauge condition (\ref{zeromodegaugec})). Note here the estimate for the commutator 
\[
\Big| \left(\Omega \slashed{\nabla}_4 (\Omega^2 - \Omega_\circ^2)\right)_{\ell=0} - \Omega \slashed{\nabla}_4   (\Omega^2 - \Omega_\circ^2)_{\ell=0} \Big| \lesssim \frac{\varepsilon^2}{r^3}
\]
following directly from (\ref{eq:nabla4modeprojectionhoI}). 
Projecting (\ref{T4dir}) to $\ell=0$ 
and integrating from $S_{u_{-1},v_\infty}$ we obtain after using the estimate on $(\omega-\omega_\circ)_{\ell=0}$ and the estimates on the non-linear terms (including (\ref{eq:nabla4modeprojectionhoI}) to deal with the commutator term from the projection) the estimate for $T_{\ell=0}$. Projecting the $\rho$-equation to $\ell=0$ we obtain
\[
\Omega \slashed{\nabla}_4 (\rho r^3 + 2M)_{\ell=0} = +3M T_{\ell=0} +  (\overset{(out)}{\slashed{\mathcal{E}}^{0}_{2}})_{\ell=0} - \left[ \left(\Omega \slashed{\nabla}_4 (\rho r^3 + 2M)\right)_{\ell=0} - \Omega \slashed{\nabla}_4 (\rho r^3 + 2M)_{\ell=0} \right]
\]
which we integrate from $S_{u_{-1},v_\infty}$. Using Proposition \ref{prop:l01coll} for the initial term and the previous estimates as well as (\ref{eq:nabla4modeprojectionhoI}), we obtain the desired estimate for $\rho_{\ell=0}$. For $\sigma_{\ell=0}$ we can directly project (\ref{curle}) to $\ell=0$.
Finally, for $\underline{T}_{\ell=0}$ we similarly project (\ref{Tb4dirwa}) to $\ell=0$ and use previous estimate on the right hand side including the fact that $|r^3\underline{\mu}_{\ell=0}|\lesssim \varepsilon^2$ by the previous estimate on $\rho_{\ell=0}$.

{\bf Step 2. We now prove all the estimates for $\ell=1$.}
Commuting the Bianchi equation for $r^4 \beta$ with $r\slashed{\mathcal{D}}_1$ and projecting to $\ell=1$, we derive
\begin{align} \label{notagainfrom4}
\Omega \slashed{\nabla}_4 \left(r^5 ( \slashed{div} \Omega^{-1}\beta)_{\ell=1} \right) =  (\overset{(out)}{\slashed{\mathcal{E}}^{1}_{2}})_{\ell=1} - \left[ \left(\Omega \slashed{\nabla}_4  \left(r^5  \slashed{div} (\Omega^{-1} \beta)\right) \right)_{\ell=1} - \Omega \slashed{\nabla}_4 \left(r^5 \slashed{div} (\Omega^{-1} \beta)\right)_{\ell=1} \right] \, ,
\end{align}
\begin{align}
\Omega \slashed{\nabla}_4 \left(r^5  (\slashed{curl} \Omega^{-1} \beta)_{\ell=1} - r^5 \Omega_\circ^{-2} (\slashed{curl} (\Omega \beta)_{\mathrm{Kerr}}) \right) &= + 3M \sum_{m=-1}^1 a^m_{\I}  \Omega \slashed{\nabla}_4 \left( r^2 \slashed{\Delta} Y^{\ell=1}_m\right) +
(\overset{(out)}{\slashed{\mathcal{E}}^{1}_{2}})_{\ell=1}  \\
& - \left[ \left(\Omega \slashed{\nabla}_4  \left(r^5  \slashed{curl} (\Omega^{-1} \beta)\right) \right)_{\ell=1} - \Omega \slashed{\nabla}_4 \left(r^5   \slashed{curl} (\Omega^{-1} \beta)\right)_{\ell=1} \right]\, . \nonumber
\end{align}
We note that the right hand sides are integrable and order $\varepsilon^2$. We integrate and using the global estimate
\[ 
\| r^5 (\slashed{curl} \Omega^{-1} \beta)_{\ell=1}  - r^5 \Omega_\circ^{-2}(\slashed{curl} \Omega \beta)_{\ell=1} \|_{S_{u,v}} \lesssim \varepsilon^2
\]
as well as the estimates of Proposition \ref{prop:l01coll} for the initial term on $S_{u_{-1},v_\infty}$, we deduce the estimate (\ref{betal1}).

We now turn to estimate $T_{\ell=1}$ and $\omega_{\ell=1}$ which will have to be proven simultaneously. We first compute from (\ref{ellipticomega})
\begin{align} \label{eop2}
2\left( r^2 \slashed{\Delta} + 1 -\frac{2M}{r}\right) {\omega}_{\ell=1} =  &- 2r^2 \left(\slashed{div} \left(\Omega \beta\right)\right)_{\ell=1} + T_{\ell=1}  \left(-\frac{1}{2} - \frac{2M}{r}\right) +  (\overset{(out)}{\slashed{\mathcal{E}}^{1}_{3}} )_{\ell=1} \nonumber \\
&+  \Big[ \left(\Omega \slashed{\nabla}_4 (r^2 \underline{\mu}^\dagger)\right)_{\ell = 1} - \Omega \slashed{\nabla}_4 (r^2 \underline{\mu}^\dagger)_{\ell = 1}\Big]  + 2 r^2  \left[ \slashed{\Delta} {\omega}_{\ell = 1} -  \left(\slashed{\Delta} {\omega}\right)_{\ell = 1}  \right]  .
\end{align}
For future reference we note that (\ref{eop2}) holds verbatim replacing $\ell=1$ by $\ell \geq 1$. With the estimate on $\left(\slashed{div} \left(\Omega \beta\right)\right)_{\ell=1}$ already shown we deduce using Proposition \ref{prop:almostevalue}
\begin{align} \label{tomrel}
\Big\| 4 \omega_{\ell=1}- \frac{1+\frac{4M}{r}}{1+\frac{2M}{r}} (Tr)_{\ell=1} \frac{1}{r} \Big\|_{S_{u_{-1},v}} \lesssim \frac{\varepsilon^2}{r^3} \, .
\end{align}
Projecting (\ref{T4dir}) to $\ell=1$ one obtains after inserting (\ref{tomrel}) 
an ODE for the quantity $(T_{\ell=1} \Omega_\circ^{-2} r)$ whose right hand side is order $\varepsilon^2 r^{-3}$. The desired estimate on $r^3T_{\ell=1}$ follows by integrating this ODE
and using the estimate of Proposition \ref{prop:l01coll} for the initial term. The estimate on $r^3 \omega_{\ell=1}$ now follows from (\ref{tomrel}). 

Projecting the Bianchi equation (\ref{rho4dir}) to $\ell=1$ one obtains the estimate for $\rho_{\ell=1}$ by the usual integration. Similarly, projecting (\ref{eq:sigmareno}) to $\ell=1$ and integrating one obtains the estimate on $\sigma_{\ell=1}$. Projecting (\ref{Tb4dir}) to $\ell=1$ using that $\left(\underline{\mu}^\dagger\right)_{\ell=1}=0$ one obtains the desired estimate for $\underline{T}_{\ell=1}$.

Estimates on the $\ell=1$ modes of $\slashed{div} \underline{\eta}$ and $\slashed{curl} \underline{\eta}$ now follow from $\left(\underline{\mu}^\dagger\right)_{\ell=1}=0$ and the elliptic equation $-\left(\slashed{curl} \underline{\eta}\right)_{\ell=1} - \slashed{curl}\underline{\eta}_{\mathrm{Kerr}} = \sigma - \sigma_{\mathrm{Kerr}} - \frac{1}{2} \hat{\chi} \wedge \hat{\underline{\chi}} - ( \slashed{curl}\underline{\eta}_{\mathrm{Kerr}}  -\sigma_{\mathrm{Kerr}})$ respectively. 

For the bounds on $(\slashed{div} \eta)_{\ell=1}$ we recall $
\Omega \slashed{\nabla}_4 \left(r \slashed{div} r {\eta}\right)	
	=
	+\Omega_\circ^2 r\slashed{div} \etabar
	-r^2 \slashed{div} \Omega\beta + \overset{(in)}{\slashed{\mathcal{E}}^1_2}$ and project to $\ell=1$.
Integrating then provides the estimates on $(\slashed{div} {\eta})_{\ell=1}$. The relation $\slashed{curl} \eta = -\slashed{curl} \underline{\eta}$ the analogous estimate for the curl. 

Finally, from the Codazzi equation (\ref{eq:Codazzibar}) commuted with $r \slashed{div}$ and $r \slashed{curl}$ respectively we obtain after  projection to $\ell=1$ and inserting previous bounds on $\underline{T}$ and $\underline{\eta}$ the bound (\ref{betabdatal01}).
\end{proof}

\subsubsection{Estimating $\beta$}
We note the following auxiliary estimate
\begin{align} \label{trilow}
\sup_v \| \left[r\slashed{\nabla}\right]^{N} r^2 \Omega \hat{\chi} \|_{S_{u_{-1},v}} \lesssim \varepsilon_0 + \varepsilon^{\frac{3}{2}} \, , 
\end{align}
which follows easily from (\ref{chihat4}), the estimates on $\alpha$ from Theorem \ref{theo:mtheoalphar} and the estimates on the error from Proposition \ref{prop:erroroutdata}. From the relation of Proposition \ref{prop:PPbaridentities} and the estimates on $\psi$ of Theorem \ref{thm:alphaalphabarestimates} as well as (\ref{trilow}):
\begin{align} \label{betcom}
\int_{v(u_{-1},R_{-2})}^{v_\infty} d\bar{v} \frac{1}{r^2} \| \mathscr{A}^{[N-1]} r \slashed{\mathcal{D}}_2^\star (r^4 \beta \Omega) \|_{S_{u_{-1},\bar{v}}}^2 + \sup_v  \| \mathscr{A}^{[N-1]} r \slashed{\mathcal{D}}_2^\star (r^4 \beta \Omega) \|_{S_{u_{-1},v}}^2 \lesssim \varepsilon_0^2 + \varepsilon^3 \, .
\end{align}
Note that in view of the estimate (\ref{betal1}) and Proposition \ref{prop:oneformkeylemma} {\bf we have shown the estimate (\ref{target2}) for all terms containing $\beta$.}

\subsubsection{Estimating $T$ and $\omega-\omega_\circ$}
Next we look at the coupled system of transport and elliptic equations  (\ref{T4dir}) and (\ref{ellipticomega}) for $T$ and $\omega - \omega_\circ$. We first claim that the structure of these equations implies that if we can show the lowest order estimate\footnote{Note we already proved this estimate for $\ell=1$ and $\ell=0$ in Proposition \ref{prop:datal01}.}
\begin{align} \label{locT}
\sup_v \| r^3 T_{\ell \geq 2} \|_{S_{u_{-1},v}}^2 \lesssim   \varepsilon_0^2 + \varepsilon^3 \, 
\end{align}
for $r^3 T$, then it follows that 
\begin{align} \label{Tofinal1}
\sup_v \| \left[r \slashed{\nabla}\right]^{N} (r^3 T) \|_{S_{u_{-1},v}}^2 +\sup_v \| \left[r \slashed{\nabla}\right]^{N} (r^3 (\omega- \omega_\circ) \|_{S_{u_{-1},v}}^2  \lesssim \varepsilon_0^2 + \varepsilon^3
\end{align}
and
\begin{align} \label{Tofinal2}
\sup_v \| \left[r \slashed{\nabla}\right]^{N+1} (r^{5/2} T) \|_{S_{u_{-1},v}}^2 +  \int_{v(u_{-1},R_{-2})}^{v_\infty} d\bar{v} \frac{1}{r^2}\| \left[r \slashed{\nabla}\right]^{N+1} (r^3 (\omega- \omega_\circ) \|_{S_{u_{-1},\bar{v}}}^2   \lesssim   \varepsilon_0^2 + \varepsilon^3 \, .
\end{align}
To see this, note that (\ref{locT}) immediately implies---using the standard elliptic estimate for (\ref{ellipticomega}) (use the $\ell \geq 1$ version of (\ref{eop2}))---control of two angular derivatives of $r^3(\omega-\omega_\circ)$. This means that we can commute (\ref{T4dir}) twice with angular derivatives and upgrade (\ref{locT}) to two angular derivatives of $r^3 T$. This can be iterated: Commuting now (\ref{ellipticomega}) (or better the $\ell \geq 1$ version of (\ref{eop2}))  with $r^2 \slashed{\Delta}$ we estimate four angular derivatives of $r^3(\omega -\omega_\circ)$ allowing to estimate four angular derivatives of $r^3 T$ etc. For the top derivatives of $\omega - \omega_\circ$ we only control the flux as it relies on the top order flux of $\beta$ improved in (\ref{betcom}) and this leads to a loss in $r$-power for $T$ because one needs to use the flux estimate for $\omega-\omega_\circ$ when integrating the $N+1$-angular commuted transport equation (\ref{T4dir}).

To finally establish (\ref{locT}) we first note that by comparison with the round sphere and the properties of spherical harmonics
\begin{align}
6 \|(\omega - \omega_\circ)_{\ell \geq 2} \|^2_{(S_{u_{-1},v}, \slashed{g})} &\leq 6 \|(\omega - \omega_\circ)_{\mathring{\ell} \geq 2} \|^2_{(S_{u_{-1},v}, \mathring{\slashed{g})}} + \varepsilon^4  \leq  \|- (\omega - \omega_\circ)_{\mathring{\ell} \geq 2} \mathring{\slashed{\Delta}} (\omega - \omega_\circ)_{\mathring{\ell} \geq 2} \|_{(S_{u_{-1},v}, \mathring{\slashed{g})}} + \varepsilon^4 \nonumber \\ 
&\leq \| - (\omega - \omega_\circ)_{\ell \geq 2} \slashed{\Delta} (\omega - \omega_\circ)_{\ell \geq 2} \|_{(S_{u_{-1},v}, \slashed{g})} +\varepsilon^4 \, ,
\end{align}
which follows easily from the bootstrap assumptions on the closeness of the metric and the spherical harmonics to their round analogues.

This implies that the projection of (\ref{ellipticomega}) to $\ell \geq 2$ leads to the estimate
\begin{align} \label{omtr}
 \|r^3(\omega - \omega_\circ)_{\ell \geq 2} \|_{S_{u_{-1},v}} \leq \frac{1}{5} \sup_v \| r^3 T_{\ell \geq 2} \|_{S_{u_{-1},v}} +  \varepsilon_0 + \varepsilon^{\frac{3}{2}} \, .
\end{align}
Projecting now the transport equation (\ref{T4dir}) to $\ell \geq 2$ and integrating yields (\ref{locT}) after inserting the estimate (\ref{omtr}) on the right hand side (and exploiting that $r \geq R_{-2}$ is large in the region under consideration, see (\ref{definitionofRhere})). With this (\ref{Tofinal1}) and (\ref{Tofinal2}) are proven and {\bf we have shown the estimate (\ref{target2}) for all terms containing $T$ and $\omega - \omega_\circ$.}

\subsubsection{Estimating \underline{$T$}}
Commuting (\ref{Tb4dir}) with angular derivatives we see (using that $\underline{\mu}^\dagger_{\ell \geq 2}=0$ on $\underline{C}^{\I}_{u_{-1}}$) that
\[
\Omega \slashed{\nabla}_4 \left(\mathscr{A}^{[N-1]} r^2 \slashed{\mathcal{D}}_2^\star \slashed{\nabla} \Omega^2 \underline{{T}} r^2\right)  = \frac{\Omega_\circ^2}{r^2} (\mathscr{A}^{[N-1]} r^2 \slashed{\mathcal{D}}_2^\star \slashed{\nabla} r^3 {T}) + \frac{4 \Omega_\circ^2}{r^2}  \mathscr{A}^{[N-1]} r^2 \slashed{\mathcal{D}}_2^\star \slashed{\nabla} r^3 \left(\omega - \omega_\circ\right)  +\overset{(out)}{\slashed{\mathcal{E}}^{N+1}_{2}} \, .
\]
We integrate and use the estimates (\ref{Tofinal1}) and (\ref{Tofinal2}) on the right hand side. Finally, we conclude using Proposition \ref{prop:functionkeylemma} with the fact that the estimate has already been shown for the $\ell=0$ and $\ell=1$ projections by Proposition \ref{prop:datal01} the estimate
\begin{align} \label{auxTb}
\sup_v \| \left[r \slashed{\nabla}\right]^{N+1} (r^2 \underline{T}) \|_{S_{u_{-1},v}}^2 \lesssim \varepsilon_0^2 + \varepsilon^3 \, .
\end{align}
With this {\bf we have shown the estimate (\ref{target2}) for all terms containing $\underline{T}$.}

\subsubsection{Estimating $\rho$, $\sigma$ and \underline{$\eta$}}
Integrating the Bianchi equations for $\rho$ and $\sigma$ as transport equations we find
\begin{align} \label{auxrhos}
\sup_v \| \mathscr{A}^{[N-3]} r^2 \slashed{\mathcal{D}}_2^\star  \slashed{\mathcal{D}}_1^\star (r^3\rho + 2M, \sigma) \|_{S_{u_{-1},v}}^2   &\lesssim  \varepsilon_0^2 + \varepsilon^3\, .
\end{align}
From the elliptic equations (\ref{diveb}) and (\ref{curle}) we have using the gauge condition $\left(\underline{\mu}^\dagger\right)_{\ell \geq 1} = 0$,
\begin{align}
r^5\slashed{\mathcal{D}}_2^\star \slashed{\mathcal{D}}_1^\star \left(\slashed{div} \underline{\eta}, \slashed{curl} \underline{\eta}\right) = r^2 \slashed{\mathcal{D}}_2^\star \slashed{\mathcal{D}}_1^\star \left(-r^3 \rho - 2M_f + \frac{1}{2}r^3 \hat{\chi} \hat{\underline{\chi}} -\frac{1}{2} \Omega_\circ^2 r^2 \underline{T},  r^3 \sigma - \frac{1}{2} r^3 \hat{\chi} \wedge \hat{\underline{\chi}} \right), \nonumber 
\end{align}
from which we easily deduce using (\ref{auxrhos}) and (\ref{auxTb})
\begin{align}
\sup_v \| \mathscr{A}^{[N-3]} \slashed{\mathcal{D}}_2^\star \slashed{\mathcal{D}}_1^\star \slashed{\mathcal{D}}_1\underline{\eta} r^2  \|_{S_{u_{-1},v}}^2  &\lesssim \varepsilon_0^2 + \varepsilon^3 \, .
\end{align}
Commuting the equation (\ref{chibhat4}) we first deduce the lower order estimate
\begin{align}
\sup_v \|\mathscr{A}^{[N-1]} r \Omega^{-1}\underline{\hat{\chi}}  \|_{S_{u_{-1},v}}^2  &\lesssim   \varepsilon_0^2 + \varepsilon^3 \, .
\end{align}
Using this and (\ref{trilow}) we obtain immediately from the flux estimates on $\Psi = r^5 P$ of Proposition \ref{prop:afa} and the relations of Proposition \ref{prop:PPbaridentities} the top order flux estimate
\begin{align}
\int_{v(u_{-1},R_{-2})}^{v_\infty} d\bar{v} \frac{1}{r^2} \| \mathscr{A}^{[N-2]} r^2 \slashed{\mathcal{D}}_2^\star \slashed{\mathcal{D}}_1^\star \left(r^3 \rho + 2M_f, r^3 \sigma\right)\|^2_{S_{u_{-1},\bar{v}}} \lesssim \varepsilon_0^2 + \varepsilon^3 
\end{align}
and revisiting the elliptic relation for $\underline{\eta}$ above also the flux estimate
\begin{align}
\int_{v(u_{-1},R_{-2})}^{v_\infty} d\bar{v} \frac{1}{r^2} \|  \mathscr{A}^{[N-2]} r^3 \slashed{\mathcal{D}}_2^\star \slashed{\mathcal{D}}_1^\star \slashed{\mathcal{D}}_1\underline{\eta} r^2  \|_{S_{u_{-1},\bar{v}}}^2  &\lesssim \varepsilon_0^2 + \varepsilon^3  \, .
\end{align}
In view of Propositions \ref{prop:functionkeylemma}, \ref{prop:oneformkeylemma} and the estimates on the $\ell=0,1$ modes in Proposition \ref{prop:datal01} now {\bf we have shown the estimate (\ref{target2}) for all terms containing $\rho-\rho_\circ, \sigma, \underline{\eta}$.}

\subsubsection{Estimating $\eta$}
Commuting (\ref{mu4dir}) with $\mathscr{A}^{[N-2]}r^2\slashed{\mathcal{D}}_2^\star \slashed{\nabla}$ and using the gauge condition (\ref{skrigaugeoutgoingE}) on the right hand side (which makes the $\underline{\mu}^\dagger$-term vanish) and the fact that $\mu_{\ell \geq 1} =0$ on $S_{u_f, v_\infty}$ we deduce after integration bounds on $N$ derivatives of the linearised mass aspect, which easily translate (using also the commuted  (\ref{curle})) into the following bounds:
\begin{align}
\sup_v \| \mathscr{A}^{[N-3]}r^2 \slashed{\mathcal{D}}_2^\star \slashed{\mathcal{D}}_1^\star \left( r\slashed{div}
  {\eta} r^2, 0 \right)  \|_{S_{u_{-1},v}}^2  &\lesssim   \varepsilon_0^2 + \varepsilon^3 \, 
\end{align}
and
\begin{align}
\int_{v(u_{-1},R_{-2})}^{v_\infty} d\bar{v} \frac{1}{r^2} \|\mathscr{A}^{[N-2]}r^2 \slashed{\mathcal{D}}_2^\star \slashed{\mathcal{D}}_1^\star \left( r\slashed{div}
  {\eta} r^2, 0 \right)  \|_{S_{u_{-1},\bar{v}}}^2  &\lesssim   \varepsilon_0^2 + \varepsilon^3 \, .
\end{align}
We now note that the estimates remain valid if we replace $0$ by $r^3 \slashed{curl} \eta$.
In view of Proposition \ref{prop:oneformkeylemma} and the estimates on the $\ell=0,1$ modes in Proposition \ref{prop:datal01} now {\bf we have shown the estimate (\ref{target2}) for all terms containing $\eta$.}

\subsubsection{Estimating \underline{$\beta$}, \underline{$\hat{\chi}$} and $\hat{\chi}$ }
From the estimates for $\underline{\psi}$ we deduce
\begin{align}
\sup_v \| \mathscr{A}^{[N-2]}r\slashed{\mathcal{D}}_2^\star (\Omega^{-1} \underline{\beta}r^2)   \|_{S_{u_{-1},v}}^2  &\lesssim    \varepsilon_0^2 + \varepsilon^3 \,
\end{align}
and
\begin{align}
\int_{v(u_{-1},R_{-2})}^{v_\infty} d\bar{v} \frac{1}{r^2} \| \mathscr{A}^{[N-1]}r\slashed{\mathcal{D}}_2^\star (\Omega^{-1} \underline{\beta} r^2)   \|_{S_{u_{-1},\bar{v}}}^2  &\lesssim   \varepsilon_0^2 + \varepsilon^3 \, .
\end{align}
The Codazzi equation (\ref{codazb}) (cf.~(\ref{eq:Codazzi})) and previous estimates now lead to
\begin{align}
\sup_v \| \mathscr{A}^{[N]} (\Omega^{-1} \underline{\hat{\chi}}r  ) \|_{S_{u_{-1},v}}^2 +\int_{v(u_{-1},R_{-2})}^{v_\infty} d\bar{v} \frac{1}{r^2} \|\mathscr{A}^{[N+1]} (\Omega^{-1} \underline{\hat{\chi}}r )  \|_{S_{u_{-1},\bar{v}}}^2
 &\lesssim   \varepsilon_0^2 + \varepsilon^3 \, .
\end{align}
Similarly, the Codazzi equation (\ref{codaz}) (cf.~(\ref{eq:Codazzibar})) and previous estimates lead to
\begin{align}
 \sup_v \| \mathscr{A}^{[N]} (\Omega {\hat{\chi}}r^2)   \|_{S_{u_{-1},v}}^2 + \int_{v(u_{-1},R_{-2})}^{v_\infty} d\bar{v} \frac{1}{r^2}  \| \mathscr{A}^{[N+1]} (\Omega {\hat{\chi}}r^2 )  \|_{S_{u_{-1},\bar{v}}}^2  &\lesssim    \varepsilon_0^2 + \varepsilon^3\, \, .
\end{align}
With this, after invoking Proposition \ref{prop:symmetrictracelesskeylemma}, {\bf we have shown the estimate (\ref{target2}) for all terms containing $\underline{\beta}$, $\underline{\hat{\chi}}$ and $\hat{\chi}$.}

This completes the proof of (\ref{target2}).

\subsubsection{Estimating $Y$}
We collect some estimates concerning the quantity $Y$ introduced in (\ref{Ydef}). 
\begin{corollary}
The following estimates hold on $\underline{C}^{\I}_{u_{-1}}$.
\begin{align}
 \sup_{v \geq v(u_{-1},R_{-2})} \|  \mathscr{A}^{[N-1]} Y\|_{S_{u_{-1},v}}^2 \lesssim    \varepsilon_0^2 + \varepsilon^3 \, , 
\end{align}
\begin{align} \label{sphbY}
 \sup_{v \geq v(u_{-1},R_{-2})}  \| \mathscr{A}^{[N-2]} r^2 \Omega \slashed{\nabla}_4 Y\|_{S_{u_{-1},v}}^2 \lesssim    \varepsilon_0^2 + \varepsilon^3 \, , 
\end{align}
\begin{align} \label{fluxYu0}
 \int_{v(u_{-1},R_{-2})}^{v_\infty} \| r^{3-\delta} \mathscr{A}^{[N-2]} \slashed{\nabla}_4 Y\|_{S_{u_{-1},v}}^2 dv \lesssim   \varepsilon_0^2 + \varepsilon^3 \, .
\end{align}
\end{corollary}
\begin{proof}
The first estimate is a direct consequence of the expression (\ref{Ydef}) and the estimates already proven. For the second estimate one computes $\Omega \slashed{\nabla}_4 Y$ from the definition (\ref{Ydef}) using the relevant null structure equations for $\underline{\hat{\chi}}$, $\underline{\eta}$ and $\underline{T}$ after commuting $\Omega \slashed{\nabla}_4$ through. Note that when inserting (\ref{Tb4dir}) one does not produce linear terms with $N+1$ derivatives because gauge condition (\ref{skrigaugeoutgoingE}) implies $\slashed{\mathcal{D}}_2^\star  \slashed{\nabla} \underline{\mu}^\dagger=0$. All linear terms involve only $N-1$ derivatives of curvature and $N$ derivatives of Ricci-coefficients. The only non-linear term with $N+1$ derivatives involves $N+1$ derivatives of $\underline{T}$, which we control on spheres. Therefore, one deduces the second estimate invoking the bounds proven \emph{on spheres}. The third estimate follows from integrating the second. 
\end{proof}

\subsection{Estimates for the angular master energy for quantities in $\mathcal{D}^{\I}$} \label{sec:mre}

The purpose of this Section is to prove (\ref{target3}). We begin by estimating the $\ell=0$ and $\ell=1$ modes of all quantities in Sections \ref{sec:l0inD} and \ref{sec:l1inD}.

\subsubsection{The $\ell=0$ modes} \label{sec:l0inD}
For the $\ell=0$ modes we prove a strong decay estimate. 
\begin{proposition} \label{prop:l0modes}
We have for any $(u,v,\theta) \in \mathcal{D}^{\I}$ the following estimates 
\begin{align}
 |r^2 T_{\ell=0}| +  | r^2(\omega-\omega_\circ)_{\ell=0} | &+ | r(\Omega^2-\Omega^2_\circ)_{\ell=0}| + |r^3\rho_{\ell=0}+2M_f|  \lesssim   \frac{ \varepsilon^2}{u^{2-\delta}} \label{bmef}
\end{align}
and
\begin{align} 
|r^2 \underline{T}_{\ell=0} | +  |r (\underline{\omega} - \underline{\omega}_\circ)_{\ell=0}|  \lesssim \frac{ \varepsilon^2}{u^{2-\delta}}. \label{bme}
\end{align}
In addition,
\begin{align}
| r^3 (\omega-\omega_\circ)_{\ell=0} |  \lesssim \frac{ \varepsilon^2}{u^{1-\delta}} \, . \label{bmee}
 \end{align}
\end{proposition}
\begin{proof}
\emph{The proof exploits that  $R \gg M_f$ in $\mathcal{D}^{\I}$
and will put an explicit constraint on $R$ depending only on $M_{f}$. Hence for the duration of the proof, 
we will denote dependences on $R$ explicitly, while the implicit constants in
the notation $\lesssim$  can here be chosen  \underline{independently
of the choice of $R$.}}

{\bf Step 1. We first establish the above estimates in the region $\mathcal{A} := \mathcal{D}^{\I} \cap \{r \geq \frac{1}{2}u\}$}. By our choice of parameters $R_{-2}>\frac{1}{2}u_{-1}$ so that the hypersurface $r=\frac{1}{2}u$ does not intersect $u=u_{-1}$ in $\mathcal{D}^{\I}$.

For the proof, we let for fixed $(u,v) \in \mathcal{W}_{\I}(u_f)\cap \{r \geq \frac{1}{2}u\}$
\[
\mathcal{A}^\star(u,v):=\{ (\bar{u},\bar{v},\theta) \in \mathcal{A} \ \ \textrm{with $\bar{u} \leq u$ and $\bar{v} \geq v$}\}
\]
and note that $\sup_{\mathcal{A}^\star(u,v)} \frac{1}{r} = \frac{1}{r(u,v)}$. 

Integrating the projected Bianchi equation for $\rho$,
\begin{align} \label{birho0}
\Omega \slashed{\nabla}_4 (r^3 (\rho - \rho_\circ)_{\ell=0} + f(u) ) =  3M_f T_{\ell=0} + (\overset{(out)}{\slashed{\mathcal{E}}^{1}_{2}})_{\ell=0} + \Omega \slashed{\nabla}_4 (r^3 (\rho - \rho_\circ)_{\ell=0} ) -\left(\Omega \slashed{\nabla}_4 (r^3 (\rho - \rho_\circ)) \right)_{\ell=0}  
\end{align}
 backwards from $v=v_{\infty}$, we have for any $(u,v,\theta) \in \mathcal{A}$
\begin{align} \label{hit}
|r^3 (\rho -\rho_\circ)_{\ell=0} \left(u,v\right) + f(u)| \lesssim   \frac{\varepsilon^2}{r u} + \frac{3M_f}{r u^{2-\delta}}  \sup_{\mathcal{A}^\star(u,v)} |T_{\ell=0} r^2 u^{2-\delta}| \, ,
\end{align}
with the $\frac{\varepsilon^2}{r u}$ term on the right arising from $|\overset{(out)}{\slashed{\mathcal{E}}^{1}_{2}}| \lesssim \frac{\varepsilon^2}{r^2 u}$ (cf.~Proposition \ref{prop:errorspacetimeIplus}), Proposition \ref{prop:com0} and (\ref{rhol0improved}) for the term on $v=v_\infty$.

We now integrate the projected $\slashed{\nabla}_3 \omega$ equation forwards from $(u_{-1},v,\theta)$ to $(u,v,\theta) \in \mathcal{A}$. From
\begin{align} \label{omprop0}
\Omega \slashed{\nabla}_3 \left((\omega - \omega_\circ)_{\ell=0} - \frac{\Omega_\circ^2}{r^3} F(u) \right) 
= &-\frac{\Omega_\circ^2}{r^3} \left(r^3 (\rho - \rho_\circ)_{\ell=0} + f(u)\right) + \frac{2M_f}{r^3}  \left(\Omega^2- \Omega_\circ^2 \right)_{\ell=0} - \frac{\Omega_\circ^2}{r^4} \left(3-\frac{4M_f}{r}\right) F(u) \nonumber \\
&+( \overset{(in)}{\slashed{\mathcal{E}}^{0}_{4}})_{\ell=0}+ \Omega \slashed{\nabla}_3 (\omega - \omega_\circ)_{\ell=0} - \left( \Omega \slashed{\nabla}_3 (\omega - \omega_\circ)\right)_{\ell=0}  
\end{align}
we derive 
\begin{align} \label{oml0}
\Big|(\omega-\omega_\circ)_{\ell=0} \left(u,v\right) - \frac{\Omega_\circ^2}{r^3} F(u) \Big| \lesssim \frac{ \varepsilon^2}{r^{4-\delta}} + \frac{ \sup_{\mathcal{A}^\star} | r \left(\Omega^2 -\Omega_\circ^2\right)_{\ell=0}| }{r^3}
+ \frac{ \sup_{\mathcal{A}^\star(u,v)} |T_{\ell=0} r^2 u^{2-\delta}|}{r^4} \, 
\end{align}
for all $(u,v, \theta) \in \mathcal{A}$, where we have used 
\begin{itemize}
\item that the quantity in brackets on the left of (\ref{omprop0}) vanishes on $u=u_{-1}$ by the gauge condition (\ref{gaugel0O}), 
\item (\ref{hit}) and $|\overset{(in)}{\slashed{\mathcal{E}}^{0}_{4}}| \lesssim \frac{\varepsilon^2}{r^4 u}$ (cf. Proposition \ref{prop:errorspacetimeIplus}) as well as $F(u) \lesssim \frac{ \varepsilon^2}{u}$.
\item Proposition \ref{prop:com0} which implies that $| \Omega \slashed{\nabla}_3 (\omega - \omega_\circ)_{\ell=0} - \left( \Omega \slashed{\nabla}_3 (\omega - \omega_\circ)\right)_{\ell=0}  |\lesssim \frac{ \varepsilon^2}{r^4 u}$.
\end{itemize}
Integrating backwards from $v=v_{\infty}$ the equation 
\begin{align} \label{ominf}
\Omega \slashed{\nabla}_4 \left(\Omega^2 -\Omega_\circ^2\right)_{\ell=0} &= 2\Omega_\circ^2 (\omega - \omega_\circ)_{\ell=0} -\frac{2M_f}{r^3} \left(\Omega^2 -\Omega_\circ^2\right)_{\ell=0} \nonumber \\
&+ 2  \left[ \left(\Omega^2 -\Omega_\circ^2\right) \left(\omega-\omega_\circ\right)\right]_{\ell=0} + \Omega \slashed{\nabla}_4 \left(\Omega^2 -\Omega_\circ^2\right)_{\ell=0} - \left(\Omega \slashed{\nabla}_4 \left(\Omega^2 -\Omega_\circ^2\right)\right)_{\ell=0} \, ,
\end{align}
we deduce after inserting (\ref{oml0}) and using Proposition \ref{prop:com0} and the fact that the initial term on $v=v_\infty$ vanishes by the gauge condition (\ref{zeromodegaugec}), that we have
\begin{align}
 \Big| \left(\Omega^2 -\Omega_\circ^2\right)_{\ell=0}(u,v) \Big| \lesssim  \frac{\varepsilon^2}{r^2 u^{1-\delta}}  + \frac{1}{r^2}  \sup_{\mathcal{A}^\star} | r \left(\Omega^2 -\Omega_\circ^2\right)_{\ell=0}|  + \frac{1}{r^3} \sup_{\mathcal{A}^\star(u,v)} |T_{\ell=0} r^2 u^{2-\delta}|
\end{align}
and hence in particular
\begin{align} \label{muehe1}
\sup_{\mathcal{A}^\star(u,v)} \Big|r \left(\Omega^2 -\Omega_\circ^2\right)_{\ell=0}(u,v) \Big| \lesssim  \frac{\varepsilon^2}{r u^{1-\delta}}  + \frac{1}{r^2} \sup_{\mathcal{A}^\star(u,v)} |T_{\ell=0} r^2 u^{2-\delta}|\, ,
\end{align}
which allows to simplify (\ref{oml0}) to 
\begin{align} \label{oml0p}
\Big|(\omega-\omega_\circ)_{\ell=0} \left(u,v\right) - \frac{\Omega_\circ^2}{r^3} F(u) \Big| \lesssim \frac{\varepsilon^2}{r^{4-\delta}} 
+ \frac{1}{r^4} \sup_{\mathcal{A}^\star(u,v)} |T_{\ell=0} r^2 u^{2-\delta}|\, 
\end{align}
and
\begin{align} \label{muehe2}
|(\omega-\omega_\circ)_{\ell=0} \left(u,v\right)| \lesssim \frac{ \varepsilon^2}{r^{3-\delta}u} 
+ \frac{1}{r^4} \sup_{\mathcal{A}^\star(u,v)} |T_{\ell=0} r^2 u^{2-\delta}|\, .
\end{align}
Note that since we are in the region $\mathcal{A}$, we could replace $r^{3-\delta}$ by $v^{3-\delta}$, something that will be convenient to do below.
We finally project (\ref{T4dir}) to $\ell=0$,
\begin{align} \label{T4dirl0}
\Omega \slashed{\nabla}_4 ((\Omega^{-2} r^2 T)_{\ell=0}) = r (\omega-\omega_\circ)_{\ell=0} + (\overset{(out)}{\slashed{\mathcal{E}}^{0}_{2}})_{\ell=0} + \Omega \slashed{\nabla}_4 ((\Omega^{-2} r^2 T)_{\ell=0})  - \left( \Omega \slashed{\nabla}_4 ((\Omega^{-2} r^2 T )\right)_{\ell=0} \, ,
\end{align}
and integrate backwards from $v=v_\infty$. Inserting the previous estimates and using (\ref{eq:nabla4modeprojectionhoI}) we conclude 
\begin{align} \label{inhd}
|r^2 T_{\ell=0} \left(u,v\right)| \lesssim \frac{ \varepsilon^2}{u^2}  +  \frac{ \varepsilon^2}{r^{1-\delta}u} + \frac{1}{r^2} \sup_{\mathcal{A}^\star(u,v)} |T_{\ell=0} r^2 u^{2-\delta}|
\end{align}
for all $(u,v,\theta) \in \mathcal{A}$. Multiplying by $u^{2-\delta}$ and using that $|u r^{-1}| \leq 2$ in $\mathcal{A}$, we finally conclude
\begin{align} \label{r3Tl0b}
\sup_{\mathcal{A}^\star(u,v)} |r^2 T_{\ell=0} u^{2-\delta}| \lesssim \varepsilon^2 ,
\end{align}
provided $R^{-\delta} C < \frac{1}{2}$ where $C$ is the constant implicit in (\ref{inhd}), which can easily be computed explicitly. 
Since $(u,v)$ was arbitrary we have this estimate in all of $\mathcal{A}$. The estimates (\ref{bmef}) now follow in $\mathcal{A}$ by revisiting (\ref{muehe1}), (\ref{muehe2}) and (\ref{hit}). The estimate (\ref{bmee}) follows from (\ref{oml0p}). To show (\ref{bme}) in $\mathcal{A}$ we integrate 
\begin{align} \label{ombl0}
\Omega \slashed{\nabla}_4 \left((\underline{\omega} - \underline{\omega}_\circ)_{\ell=0}\right) = &-\Omega_\circ^2 (\rho - \rho_\circ)_{\ell=0} + \frac{2M_f}{r^3}  \left(\Omega^2- \Omega_\circ^2 \right)_{\ell=0}  \nonumber \\
 &+ \overset{(out)}{\slashed{\mathcal{E}}^{0}_{4}}+ \Omega \slashed{\nabla}_4 (\underline{\omega} - \underline{\omega}_\circ)_{\ell=0} - \left( \Omega \slashed{\nabla}_4(\underline{\omega} - \underline{\omega}_\circ)\right)_{\ell=0} 
\end{align}
 from $v=v_\infty$ (where we have $|(\underline{\omega} - \underline{\omega}_\circ)_{\ell=0}| \leq \frac{\varepsilon^2}{r^2 u}$ by (\ref{ompredetail})) to conclude the bound claimed. For $\underline{T}$ we integrate the projected equation (\ref{Tb4dirwa}),
 \begin{align} \label{Tbarl0}
 \Omega \slashed{\nabla}_4 \left(\Omega^2 \underline{{T}} r\right)_{\ell=0} = &\Omega_\circ^2  T_{\ell=0} +4 \Omega_\circ^2\left(\omega - \omega_\circ\right)_{\ell=0}  +  \frac{2 \Omega_\circ^2}{r^2}  \left(r^3 (\rho - \rho_\circ)_{\ell=0} - \frac{1}{2} (r^3 \hat{\chi} \underline{\hat{\chi}})_{\ell=0}\right) + (\overset{(out)}{\slashed{\mathcal{E}}^{0}_{3}})_{\ell=0} \nonumber \\
 &+\Omega \slashed{\nabla}_4 \left(\Omega^2 \underline{{T}} r\right)_{\ell=0}  -  \left(\Omega \slashed{\nabla}_4 \left(\Omega^2 \underline{{T}} r\right)\right)_{\ell=0} ,
 \end{align}
 to obtain
 \[
 \left(\Omega^2 \underline{{T}} r\right)_{\ell=0} (u,v)  \lesssim \frac{ \varepsilon^2}{r u^{2-\delta}} \, ,
 \]
 from which the estimate claimed for $\underline{T}$ follows easily. 
 
{\bf Step 2. We now establish the estimates in the region $\mathcal{B}= \mathcal{D}^{\I} \cap \{ r \leq \frac{1}{2} u\}$.} \\
Note that in $\mathcal{B}$ we have $u\leq v \leq 2u$. Note also that from $R_{-2} > \frac{1}{2} u_{-1}$, the past ingoing cone through any sphere in $\mathcal{B}$ intersects $\{r=\frac{1}{2}u\}$ (and so does the future outgoing cone). For the proof, it is useful to define for $(u,v) \in \mathcal{W}_{\I}(u_f)\cap \{ r \leq \frac{1}{2} u\}$ the set 
\[
\mathcal{B}^\star(u,v):=\{ (\bar{u},\bar{v},\theta) \in \mathcal{B} \ \ \textrm{with $\bar{u} \leq u$ and $\bar{v} \geq v$}\}.
\]
Note that $\inf_{(u^\prime, v^\prime) \in \mathcal{B}^\star(u,v)} u^\prime \geq \frac{1}{3}u$ so the $u$ value in $\mathcal{B}^\star(u,v)$ is everywhere comparable to $u$ itself.

Integrating (\ref{birho0}) (now dropping the $f(u)$ on the left) from $r=\frac{1}{2}u$ (where $r \sim u \sim v$) yields
\[
|r^3 \rho_{\ell=0} (u,v) + 2M| \lesssim   \frac{ \varepsilon^2}{u^{2-\delta}} + \frac{1}{R}  \sup_{\mathcal{B}^\star(u,v)} |T_{\ell=0} r^{2}| \ \ \ \ \textrm{ for all $(u,v,\theta) \in \mathcal{B}$} \, ,
\]
where we have used $|\overset{(out)}{\slashed{\mathcal{E}}^{1}_{2}}| \lesssim \frac{\varepsilon^2}{u^{2-\delta} r^{1+\delta}}$ from Proposition \ref{prop:errorspacetimeIplus}. Integrating the $\omega- \omega_\circ$ equation (\ref{omprop0}) (again with all terms involving $F(u)$ and $f(u)$ removed on both sides) forwards  from $\{r=\frac{1}{2} u\}$ yields
\[
|(\omega-\omega_\circ)_{\ell=0} \left(u,v\right)| \lesssim \frac{\varepsilon^2}{v^{3-\delta} u}  +  \frac{\varepsilon^2}{r^2 u^{2}}  + \frac{1}{r^3} \sup_{\mathcal{B}^\star(u,v)} \Big| r \left(\Omega^2 -\Omega_\circ^2\right)_{\ell=0}\Big| 
+ \frac{1}{r^2 R}   \sup_{\mathcal{B}^\star(u,v)} |T_{\ell=0} r^{2}| \, ,
\]
where we have used $| \overset{(in)}{\slashed{\mathcal{E}}^{0}_{4}}+ \Omega \slashed{\nabla}_3 (\omega - \omega_\circ)_{\ell=0} - \left( \Omega \slashed{\nabla}_3 (\omega - \omega_\circ)\right)_{\ell=0} | \lesssim \frac{ \varepsilon^2}{r^2 u^2}$ and the first term arises from the initial term on $r=\frac{1}{2}u$.\footnote{Note that when integrating $u^{-\gamma}$ along the constant $v$ hypersurface in $\mathcal{B}$ we call pull out $u^{-\gamma}$ from the integral because in $\mathcal{B}$, the infimum and the supremum of $u$ along that hypersurface are comparable. Moreover, note that $r$ is decreasing towards the future along that hypersurface.}
Integrating (\ref{ominf}) we find
\begin{align}
\sup_{\mathcal{B}^\star(u,v)}  \Big| r\left(\Omega^2 -\Omega_\circ^2\right)_{\ell=0}(u,v) \Big| \lesssim  \frac{\varepsilon^2}{u^{2-\delta}}  + \frac{1}{R}   \sup_{\mathcal{B}^\star(u,v)} |T_{\ell=0} r^2|\, ,
\end{align}
and hence
\[
|(\omega-\omega_\circ)_{\ell=0} \left(u,v\right)| \lesssim  \frac{ \varepsilon^2}{v^{3-\delta} u}  +  \frac{ \varepsilon^2}{r^2 u^{2}} 
+ \frac{1}{r^2 R}   \sup_{\mathcal{B}^\star(u,v)} |T_{\ell=0} r^{2}| \, .
\]
Integrating backwards (\ref{T4dirl0}) from $r=\frac{1}{2}u$ we deduce
\begin{align} \label{inhd2}
|r^2 T_{\ell=0} \left(u,v\right)| \lesssim \frac{\varepsilon^2}{u^{2-\delta}} +   \frac{1}{r}  \sup_{\mathcal{B}^\star(u,v)} |T_{\ell=0} r^2| \ \ \ \ \textrm{ for all $(u,v,\theta) \in \mathcal{B}$,}
\end{align}
and hence, provided $R^{-1} C<\frac{1}{2}$ with $C$ the constant implicit in (\ref{inhd2}),
\begin{align}
\sup_{\mathcal{B}^\star(u,v)} |r^2 T_{\ell=0} \left(u,v\right)| \lesssim \frac{\varepsilon^2}{u^{2-\delta}} .
\end{align}
Since $(u,v)$ was arbitrary, we have shown the estimate in all of $\mathcal{B}$ and hence in all of $\mathcal{D}^{\I}$. The estimates for $(\omega- \omega_\circ)_{\ell=0}$, $(\Omega^2- \Omega_\circ)_{\ell=0}$ and $r^3 \rho_{\ell=0}+2M$ now follow as before by revisiting the conditional estimates of Step 2. Integrating (\ref{ombl0}) and (\ref{Tbarl0}) from $r=\frac{1}{2}u$ using the previous estimates establishes also the bounds for $\underline{T}$ and $\underline{\omega}-\underline{\omega}_\circ$. This ends the proof.
\end{proof}

\subsubsection{The $\ell=1$ modes} \label{sec:l1inD}
We now turn to the $\ell=1$ modes. Key is an improved estimate for $(\slashed{curl} \eta)_{\ell=1}$. For all other (independent) quantities we confine ourselves with the standard $u^{-1}$ decay, see Proposition \ref{prop:l1modes} below.

\begin{proposition} \label{prop:l1modesspecial}
We have for any $(u,v,\theta) \in \mathcal{D}^{\I}$ the estimates
\begin{align}
 \|r^5 \slashed{curl} \left( (\Omega\beta)_{\ell=1} - (\Omega \beta)_{\mathrm{Kerr}}) \right)\|_{S_{u,v}} + \|r^3 \slashed{curl} \left( (\Omega^{-1}\underline{\beta})_{\ell=1} - (\Omega^{-1} \underline{\beta})_{\mathrm{Kerr}} \right)\|_{S_{u,v}} &\lesssim \frac{\varepsilon^2}{u^{3/2}}   \\
 \|r^3 \left(\sigma_{\ell=1} -\sigma_{\mathrm{Kerr}} \right)\|_{S_{u,v}} &\lesssim \frac{\varepsilon^2}{u^2} 
\end{align} 
and
\begin{align}
\|r^3 \slashed{curl} (\eta_{\ell=1} - \eta_{\mathrm{Kerr}}) \|_{S_{u,v}}  +  \|r^3 \slashed{curl} (\underline{\eta}_{\ell=1}- \underline{\eta}_{\mathrm{Kerr}}) \|_{S_{u,v}}  \lesssim \frac{ \varepsilon^2}{u^{3/2}}.
\end{align} 
Finally, on $u=u_f$ all estimates hold with the improved rate $\varepsilon^2 (u_f)^{-2}$.
\end{proposition}

\begin{proof}
We recall from (\ref{notagainfrom4}) the projected equations
\begin{align} 
\Omega \slashed{\nabla}_4 \left(r^5 ( \slashed{div} \Omega^{-1}\beta)_{\ell=1} \right) =  (\overset{(out)}{\slashed{\mathcal{E}}^{1}_{2}})_{\ell=1} - \left[ \left(\Omega \slashed{\nabla}_4  \left(r^5  \slashed{div} (\Omega^{-1} \beta)\right) \right)_{\ell=1} - \Omega \slashed{\nabla}_4 \left(r^5 \slashed{div} (\Omega^{-1} \beta)\right)_{\ell=1} \right] \, ,
\end{align}
\begin{align}
\Omega \slashed{\nabla}_4 \left(r^5  (\slashed{curl} \Omega^{-1} \beta)_{\ell=1} - r^5 \Omega_\circ^{-2} (\slashed{curl} (\Omega \beta)_{\mathrm{Kerr}}) \right) &= + 3M_f \sum_{m=-1}^1 a^m_{\I}  \Omega \slashed{\nabla}_4 \left( r^2 \slashed{\Delta} Y^{\ell=1}_m\right) +
(\overset{(out)}{\slashed{\mathcal{E}}^{1}_{2}} )_{\ell=1} \\
& - \left[ \left(\Omega \slashed{\nabla}_4  \left(r^5  \slashed{curl} (\Omega^{-1} \beta)\right) \right)_{\ell=1} - \Omega \slashed{\nabla}_4 \left(r^5   \slashed{curl} (\Omega^{-1} \beta)\right)_{\ell=1} \right]\, . \nonumber
\end{align}
In fact, we need to be slightly more precise about the terms appearing in $\overset{(out)}{\slashed{\mathcal{E}}^{1}_{2}}$. It is easy to check that the terms in $\overset{(out)}{\slashed{\mathcal{E}}^{1}_{2}}$ are either actually $\overset{(out)}{\slashed{\mathcal{E}}^{1}_{5/2}}$ or each summand in the remaining terms will contain at least one factor from $\overset{(in)}{\Phi_p}$.
This special structure leads to the estimate $\|\overset{(out)}{\slashed{\mathcal{E}}^{1}_{2}}\|_{S_{u,v}} \lesssim \varepsilon^2 r^{-3/2} u^{-2}$ (while without it only $\|\overset{(out)}{\slashed{\mathcal{E}}^{1}_{2}}\|_{S_{u,v}} \lesssim \varepsilon^2 r^{-3/2} u^{-3/2}$ would hold). Using also the estimate of Proposition \ref{prop:com0} we conclude that the entire right hand sides of the above two equations satisfy $\|RHS\|_{S_{u,v}} \lesssim  \varepsilon^2 r^{-3/2} u^{-2}$. In view of the estimates on $v=v_\infty$ for the initial term (\ref{cbl1}), (\ref{dbl1}) we therefore conclude the estimates
\begin{align} \label{bypro}
\| r^5 (\slashed{div} \Omega \beta)_{\ell=1}\|_{S_{u,v}} \lesssim   \frac{ \varepsilon^2}{u} 
 \ \ \ \ \textrm{and} \ \ \ \ 
\| r^5 (\slashed{curl} \Omega \beta)_{\ell=1} - r^5 \slashed{curl} (\Omega \beta)_{\mathrm{Kerr}} \|_{S_{u,v}} \lesssim   \frac{\varepsilon^2}{u^{3/2}} \, , 
\end{align}
with an improvement to $\frac{\varepsilon^2}{u^{2}}$ on $u=u_f$ for the latter estimate, in view of the improvement  of the estimate for the initial term (\ref{cbl1}) in Proposition \ref{prop:l01coll}.
The $\slashed{curl}$-part of the Codazzi equation (\ref{eq:Codazzi}) projected to $\ell=1$ as well as $\slashed{curl} (\eta + \underline{\eta})=0$ now provide the bounds claimed on $\slashed{curl} \eta$ and $\slashed{curl} \underline{\eta}$. The equation $\sigma = \slashed{curl} \eta + \frac{1}{2} \hat{\chi} \wedge \hat{\underline{\chi}}$ provides the bound on $\sigma$ and the other Codazzi equation the bound on $\slashed{curl}\Omega^{-1} \underline{\beta}$.
\end{proof}

\begin{proposition} \label{prop:l1modes}
We have for any $(u,v,\theta) \in \mathcal{D}^{\I}$ the following estimates 
\begin{align} \label{l1s1}
 \|r^2 T_{\ell=1}\|_{S_{u,v}} +  \| r^2(\omega-\omega_\circ)_{\ell=1} \|_{S_{u,v}}  + &\| r(\Omega^2-\Omega^2_\circ)_{\ell=1}\|_{S_{u,v}}  \nonumber \\
 +& \|r^3\rho_{\ell=1}\|_{S_{u,v}}  + \| r^5 (\slashed{div} \Omega \beta)_{\ell=1}\|_{S_{u,v}}  \lesssim   \frac{\varepsilon^2}{u} 
\end{align}
and
\begin{align} \label{l1s2}
& \|r^2 \underline{T}_{\ell=1} \|_{S_{u,v}}  +  \|r (\underline{\omega} - \underline{\omega}_\circ)_{\ell=1}\|_{S_{u,v}}  \nonumber \\
+& \|r^3 (\slashed{div} \eta)_{\ell=1}\|_{S_{u,v}}  +  \|r^3 (\slashed{div} \underline{\eta})_{\ell=1}\|_{S_{u,v}}  +  \|r^3 (\slashed{div} \Omega^{-1}\underline{\beta})_{\ell=1}\|_{S_{u,v}}  \lesssim \frac{ \varepsilon^2}{u}.
\end{align}
In addition,
\begin{align} \label{l1s3}
\| r^3 (\omega-\omega_\circ)_{\ell=1} \|_{S_{u,v}}   \lesssim  \varepsilon^2 \, . 
 \end{align}
 \end{proposition}

\begin{proof}
Note the bound on $\beta$ was already obtained in (\ref{bypro}).
The proof proceeds very similarly to Proposition \ref{prop:l0modes}. We use the notation of the regions etc.~from that proof (in particular $\lesssim$ does not involve constants that depend on $R$) and streamline the argument, which is easier here as we are not aiming for improved decay.

{\bf Step 1. Estimates in $\mathcal{A}$.} The analogue of the  (now to $\ell=1$) projected propagation equation contains an additional term $r^3 (\slashed{div}\Omega \beta)_{\ell=1}$ on the right hand side, for which we have just proven the estimate (\ref{bypro}). Therefore, we have
\begin{align}  \label{wayout}
\|r^3 (\rho -\rho_\circ)_{\ell=1} \|_{S_{u,v}} \lesssim   \frac{ \varepsilon^2}{u^2} + \frac{1}{r u}  \sup_{(u^\prime, v^\prime) \in \mathcal{A}^\star(u,v)} \|T_{\ell=1} r^2 u\|_{S_{u^\prime,v^\prime}} \, ,
\end{align}
for all $(u,v,\theta) \in \mathcal{A}$. Integrating the projected $\Omega \slashed{\nabla}_3 (\omega-\omega_\circ)_{\ell=1}$ equation yields or all $(u,v,\theta) \in \mathcal{A}$
\begin{align}
\|(\omega-\omega_\circ)_{\ell=1}\|_{S_{u,v}} \lesssim \frac{ \varepsilon^2}{r^{3}} + \frac{ \sup_{ \mathcal{A}^\star(u,v)} \| u \cdot r \left(\Omega^2 -\Omega_\circ^2\right)_{\ell=1}\|_{S_{u^\prime,v^\prime}} }{r^{4-\delta}}
+ \frac{ \sup_{\mathcal{A}^\star(u,v)} \|T_{\ell=1} r^2 u\|_{S_{u^\prime,v^\prime}}}{r^{4-\delta}} \, ,
\end{align}
where we write shorthand $\sup_{\mathcal{A}^\star(u,v)}$ for $\sup_{(u^\prime,v^\prime) \in \mathcal{A}^\star(u,v)}$.
Integrating $\left(\Omega^2 -\Omega_\circ^2\right)_{\ell=1}$ backwards using (\ref{rvu2}) for the initial term gives
\begin{align}
\| \left(\Omega^2 -\Omega_\circ^2\right)_{\ell=1} \|_{S_{u^\prime,v^\prime}} \lesssim  \frac{\varepsilon^2}{r u}  + \frac{1}{r^{3-\delta}}  \sup_{\mathcal{A}^\star (u,v)} \| u \cdot r \left(\Omega^2 -\Omega_\circ^2\right)_{\ell=1}\|_{S_{u^\prime,v^\prime}}+ \frac{1}{r^{3-\delta}} \sup_{\mathcal{A}^\star(u,v)} \|T_{\ell=1} r^2 u\|_{S_{u^\prime,v^\prime}} \nonumber
\end{align}
and hence in particular
\begin{align}
\sup_{\mathcal{A}^\star(u,v)} \| u \cdot r \left(\Omega^2 -\Omega_\circ^2\right)_{\ell=1}\|_{S_{u^\prime,v^\prime}}  \lesssim  \varepsilon^2 + \frac{1}{r^{1-\delta}} \sup_{\mathcal{A}^\star(u,v)} \|T_{\ell=0} r^2 u\|_{S_{u^\prime,v^\prime}} \, ,
\end{align}
and
\begin{align}
\|(\omega-\omega_\circ)_{\ell=1}\|_{S_{u,v}} \lesssim \frac{ \varepsilon^2}{r^{3}}
+ \frac{ 1}{r^{4-\delta}} \sup_{\mathcal{A}^\star(u,v)} \|T_{\ell=1} r^2 u\|_{S_{u^\prime,v^\prime}}\, 
\end{align}
holds in $\mathcal{A}$. Integrating backwards the $\ell=1$ analogue of (\ref{T4dirl0}) yields (using (\ref{Tl1}) for the initial term)
\begin{align} \label{inhd3}
\|r^2 T_{\ell=1} \|_{S_{u,v}} \lesssim \frac{ \varepsilon^2}{r} + \frac{1}{r^{2-\delta}} \sup_{\mathcal{A}^\star(u,v)} \|T_{\ell=1} r^2 u\|_{S_{u^\prime,v^\prime}} 
\end{align}
for all $(u,v,\theta) \in \mathcal{A}$. Multiplying by $u$ and using that $|u r^{-1}| \leq 2$ in $\mathcal{A}$ we conclude
\begin{align}
\sup_{\mathcal{A}^\star(u,v)} \|r^2 T_{\ell=1} u \|_{S_{u^\prime,v^\prime}}  \lesssim  \varepsilon^2 ,
\end{align}
provided $R^{-2+\delta} C<\frac{1}{2}$ where $C$ is the constant implicit in (\ref{inhd3}).
Since $(u,v)$ was arbitrary we have this estimate in all of $\mathcal{A}$. Revisiting the estimates shows (\ref{l1s1}) and (\ref{l1s3}) in $\mathcal{A}$. Applying $r^4 \slashed{div}$ to the Codazzi equation (\ref{eq:Codazzi}) and projecting to $\ell=1$ immediately establishes the estimate for $r^3 \slashed{div} \underline{\eta}_{\ell=1}$ after using that $\|r^4 \slashed{\Delta} T_{\ell=1}\|_{S_{u,v}}  \lesssim \frac{\varepsilon^2}{u}$ by (\ref{r3Tl0b}) and Proposition \ref{prop:almostevalue}. We next integrate
\begin{align} \label{propl1eta}
\Omega \slashed{\nabla}_4 \left(r^2 (\slashed{div} \eta)_{\ell=1} \right) =  \frac{\Omega_\circ^2}{r^2} r^3 (\slashed{div} \underline{\eta})_{\ell=1}- r^2 (\slashed{div} \Omega \beta)_{\ell=1} + (\overset{(out)}{\slashed{\mathcal{E}}^{1}_{3}})_{\ell=0} + \Omega \slashed{\nabla}_4 \left(r^2 (\slashed{div} \eta)_{\ell=1} \right) - \left(\Omega \slashed{\nabla}_4 \left(r^2 \slashed{div} \eta\right)\right)_{\ell=1} 
\end{align}
from infinity (using (\ref{etadiv1})) to establish $\|(r^3 \slashed{div} \eta)_{\ell=1}\|_{S_{u,v}}  \lesssim \frac{\varepsilon^2}{u}$. The estimates for the $\ell=1$ part of $\underline{T}$, $\underline{\omega}-\underline{\omega}_\circ$ and $\slashed{div} \Omega^{-1}\underline{\beta}$ easily follow from their projected propagation equations in the $4$-direction and the estimates already shown.

{\bf Step 2. Estimates in $\mathcal{B}$.}

Integrating the $\ell=1$ analogue of (\ref{birho0}) from $r=\frac{1}{2}u$ (where $r \sim u \sim v$) and using (\ref{wayout}) for the initial term yields
\[
\|r^3 \rho_{\ell=1} (u,v) \|_{S_{u,v}}  \lesssim   \frac{ \varepsilon^2}{u^2} + \frac{ \varepsilon^2}{r u}  + \frac{1}{R}  \sup_{\mathcal{B}^\star(u,v)} \|T_{\ell=1} r^{2}\|_{S_{u^\prime,v^\prime}}   \ \ \ \ \textrm{ for all $(u,v,\theta) \in \mathcal{B}$} \, .
\]
Integrating the projected $\Omega \slashed{\nabla}_3 (\omega-\omega_\circ)_{\ell=1}$ equation forwards  from $\{r=\frac{1}{2} u\}$ yields
\[
\|(\omega-\omega_\circ)_{\ell=1}\|_{S_{u,v}}  \lesssim \frac{ \varepsilon^2}{v^3}+\frac{ \varepsilon^2}{r^3 u}  + \frac{1}{r^3} \sup_{\mathcal{B}^\star(u,v)} \| r \left(\Omega^2 -\Omega_\circ^2\right)_{\ell=1}\|_{S_{u^\prime,v^\prime}} 
+ \frac{1}{r^2 R}   \sup_{\mathcal{B}^\star(u,v)} \|T_{\ell=1} r^{2}\|_{S_{u^\prime,v^\prime}}   \, .
\]
Integrating the $\ell=1$ analogue of (\ref{ominf}) we find
\begin{align}
\sup_{\mathcal{B}^\star(u,v)}  \| r\left(\Omega^2 -\Omega_\circ^2\right)_{\ell=1}\|_{S_{u^\prime,v^\prime}}   \lesssim  \frac{ \varepsilon^2}{u}  + \frac{1}{R}   \sup_{\mathcal{B}^\star(u,v)} \|T_{\ell=1} r^2\|_{S_{u^\prime,v^\prime}} , 
\end{align}
and hence
\[
\|(\omega-\omega_\circ)_{\ell=1}\|_{S_{u,v}}  \lesssim \frac{ \varepsilon^2}{v^3}+\frac{ \varepsilon^2}{r^3 u}
+ \frac{1}{r^2 R}   \sup_{\mathcal{B}^\star(u,v)} \|T_{\ell=1} r^{2}\|_{S_{u^\prime,v^\prime}}   \, .
\]
Integrating backwards the $\ell=1$ analogue of (\ref{T4dirl0}) from $r=\frac{1}{2}u$ we deduce
\begin{align} \label{inhd4}
\|r^2 T_{\ell=1}\|_{S_{u,v}}  \lesssim \frac{\varepsilon^2}{u} +   \frac{1}{r}  \sup_{\mathcal{B}^\star(u,v)} \|T_{\ell=1} r^2\|_{S_{u^\prime,v^\prime}}   \ \ \ \ \textrm{ for all $(u,v,\theta) \in \mathcal{B}$,}
\end{align}
and hence, provided $R^{-1} C<\frac{1}{2}$ with $C$ the constant implicit in (\ref{inhd4}),
\begin{align}
\sup_{\mathcal{B}^\star(u,v)} \|r^2 T_{\ell=1} \|_{S_{u^\prime,v^\prime}}   \lesssim \frac{\varepsilon^2}{u} .
\end{align}
Since $(u,v)$ was arbitrary, we have shown the estimate in all of $\mathcal{B}$ and hence in all of $\mathcal{D}^{\I}$. The estimates for $(\omega- \omega_\circ)_{\ell=1}$, $(\Omega^2- \Omega_\circ)_{\ell=1}$ and $r^3 \rho_{\ell=1}+2M$ now follow as before by revisiting the conditional estimates of Step 2. The estimate for $(\slashed{div} \underline{\eta})_{\ell=1}$ again follows from Codazzi and the one for $(\slashed{div} {\eta})_{\ell=1}$ again straight from the propagation equation (\ref{propl1eta}).

Integrating the $\ell=1$ analogue of (\ref{ombl0}) and (\ref{Tbarl0}) from $r=\frac{1}{2}u$ using the previous estimates establishes also the bounds for $\underline{T}$ and $\underline{\omega}-\underline{\omega}_\circ$. The estimate for $\slashed{div} \Omega^{-1}\underline{\beta}$ easily follows again from their propagation equations in the $4$-direction. This ends the proof.
\end{proof}

From Proposition \ref{prop:l0modes} and (\ref{bypro}) we conclude in particular (referring back to the terms in the second line of the energy (\ref{egammafull}) and in the last line of the energy (\ref{eRfull})):
\begin{center}
$\boxed{\textrm{The estimate (\ref{target3}) has been shown for all terms containing explicitly the $\ell=0,1$ modes.}}$
\end{center}

\subsubsection{Estimating $\hat{\chi}$}
\begin{proposition} \label{prop:chistart}
For any $u \in \left[u_{-1},u_f\right]$ we have for $s=0,1$ the integrated decay estimates
\begin{align} \label{chiest}
\int_u^{u_f} d\bar{u} \int_{v(\bar{u},R_{-2})}^{v_{\infty}} d\bar{v}  \frac{1}{r^{1+\delta}}\|\mathscr{A}^{[N+1-s]} {\hat{\chi}}{\Omega} r^2\|_{S_{\bar{u},\bar{v}}}^2 
\lesssim 
 \frac{(\varepsilon_0)^2+\varepsilon^3}{u^{s}}  \, .
\end{align}
Along any ingoing cone $\underline{C}^{\I}_v(u) \in \mathcal{D}^{\I}$ we have for $s=0,1$
\begin{align} \label{chiestcone}
\int_{\underline{C}^{\I}_v(u)} d\bar{u} \| \mathscr{A}^{[N+1-s]} {\hat{\chi}}{\Omega} r^2\|^2_{S_{\bar{u},v}} 
\lesssim 
 \frac{(\varepsilon_0)^2+\varepsilon^3}{u^{s}}  \, .
\end{align}
On spheres we have for $s=0,1,2$
\begin{align} \label{chiest2}
 \big\| \mathscr{A}^{[N-s]} {\hat{\chi}}{\Omega} r^2 \big\|^2_{S_{u,v}} 
\lesssim 
  \frac{(\varepsilon_0)^2+\varepsilon^3}{u^{\min(\frac{1}{2}+s,2)}}   \, .
\end{align}
Along any outgoing cone we have the flux estimate
\begin{align} \label{chiest4}
\int_{v\left(u,R_{-2}\right)}^\infty d\bar{v} \, \frac{1}{r^2} \|\mathscr{A}^{[N+1]} {\hat{\chi}}{\Omega} r^2\|_{S_{u,\bar{v}}}^2 \lesssim 
(\varepsilon_0)^2+\varepsilon^3   \, .
\end{align}
In particular, in view of Proposition \ref{prop:symmetrictracelesskeylemma}
\begin{center}
$\boxed{\textrm{The estimate (\ref{target3}) has been shown for all terms containing $ \hat{\chi}$.}}$
\end{center}
\end{proposition}

\begin{remark}
As the proof will reveal, the estimates (\ref{chiest}) and (\ref{chiestcone}) also hold for $s=2$ provided the decay rate on the right is changed to $2-2\delta$.
\end{remark}

\begin{proof}
We first note that it suffices to establish all estimates with $\frac{\hat{\chi}}{\Omega}$ in place of $\hat{\chi}\Omega$. This follows easily from the estimates on the non-linear errors caused by the difference of these quantities, cf.~Proposition \ref{prop:errorspacetimeIplus}. 

\noindent{\bf Step 1: Renormalised quantity and propagation equation.} We introduce the renormalised quantity
\begin{align} \label{Zdef}
Z := -\frac{3M}{r} r^2 \hat{\chi} \Omega^{-1} - r^3 \slashed{\mathcal{D}}_2^\star\underline{\eta} + \frac{1}{2}r^2 \slashed{\mathcal{D}}_2^\star \slashed{\nabla} ( \Omega^{-2} r^2 T)=r^2 \slashed{\mathcal{D}}_2^\star \slashed{div} \left({\hat{\chi}} r^2 \Omega^{-1}\right) + r \Omega^{-2} \Pi + \Omega^{-2} \overset{(out)}{\slashed{\mathcal{E}}^{1}_{1}}
\end{align}
with the second equality following from (\ref{codaz}) recalling $\Pi=r^3 \Omega \psi$. Clearly, in view of the second equality, estimates on $Z$ will translate into estimates on $\hat{\chi}$ using estimates on the almost gauge invariant quantity $\Pi$. We compute (using the null structure equations and noting that the linear term in $\omega-\omega_\circ$ appearing from  (\ref{T4dir}) and (\ref{eq:nabla4etabar}) cancels) the schematic propagation equation
\begin{align} 
\Omega \slashed{\nabla}_4 Z = -\Pi + 3M A \Omega^{-2} +\sum_{\substack{k_1+k_2 \leq 2 \\ \Phi_{p_1}, \Phi_{p_2} \in \{{\hat{\chi}} ,{T}\} }} h_{k_1, k_2, \Phi_{p_1},\Phi_{p_2}}\Omega^{-2} r^2 [r\slashed{\nabla}]^{k_1} \Phi_{p_1} [r\slashed{\nabla}]^{k_2} \Phi_{p_2}  +\Omega^{-2} \overset{(out)}{\slashed{\mathcal{E}}^{1}_{2}} \, .
\end{align}
Here implicit in the sum involving the admissible coefficient functions $ h_{k_1, k_2, \Phi_{p_1},\Phi_{p_2}}$ are various contractions with the metric $\slashed{g}$ (to produce a symmetric traceless tensor) that we do not keep track of.
Commuting yields
\begin{align}  \label{zcomm}
\Omega \slashed{\nabla}_4 \left(\mathscr{A}^{[N-1-s]} Z \right) = -\mathscr{A}^{[N-1-s]} \Pi + \frac{3M}{\Omega^2} \mathscr{A}^{[N-1-s]} A +\sum_{\substack{k_1+k_2 \leq N+1-s \\ \Phi_{p_1}, \Phi_{p_2} \in \{{\hat{\chi}} ,{T}\} }} \Omega^{-2} r^2 [r\slashed{\nabla}]^{k_1} \Phi_{p_1} [r\slashed{\nabla}]^{k_2} \Phi_{p_2} +  \Omega^{-2} \overset{(out)}{\slashed{\mathcal{E}}^{N-s}_{2}} \, ,
\end{align}
where we recall $A=r \Omega^2 \alpha$ and we have further abused notation and suppressed the admissible coefficient functions multiplying each term in the sum.

\noindent{\bf Step 2: Estimates for the right hand side and boundary terms} 
\begin{enumerate}[(1)]
\item (Error estimates)
Denoting the last two terms in (\ref{zcomm}) momentarily by ``$NL_s$" we have for any $s$
\begin{align}
\int_{v}^{v_\infty} d\bar{v} \|NL_s \|_{S_{u,\bar{v}}}^2 r^{1+\delta} \lesssim \frac{\varepsilon^3}{u^{2-\delta^s_0}} \ \ \ , \ \ \ \int_u^{u_f} d\bar{u} \int_{v(\bar{u},R_{-2})}^{v_{\infty}} d\bar{v}  r^{1+\delta} \|NL_s \|_{S_{\bar{u},\bar{v}}}^2 \lesssim \frac{\varepsilon^3}{u^2} \,.
\end{align}
This follows easily from the error estimates of Proposition \ref{prop:errorspacetimeIplus} and the flux estimate of Lemma \ref{lem:shearfluxes} for the top order derivative term in the sum. For the spacetime integral one uses in addition the bootstrap assumption on $N+1$ angular derivatives of $\hat{\chi}$ and $T$.
\item (Almost gauge invariant quantities)
From Theorem~\ref{thm:alphaalphabarestimates}, we  have, for any $u \in \left[u_{-1},u_f\right]$ and $s=0,1,2$, the flux estimates
\begin{align}
\int_{v}^{v_\infty} d\bar{v} \| \mathscr{A}^{[N-s]} A\|_{S_{u,\bar{v}}}^2 r^{1+\delta} \lesssim \frac{(\varepsilon_0)^2+\varepsilon^4}{u^s} \ \ \ , \ \ \  \int_{v}^{v_\infty} d\bar{v} \| \mathscr{A}^{[N-1-s]} \Pi\|_{S_{u,\bar{v}}}^2 r^{1+\delta} \lesssim \frac{(\varepsilon_0)^2+\varepsilon^3}{u^s} \nonumber
\end{align}
as well as the spacetime estimates
\begin{align}
\int_u^{u_f} d\bar{u} \int_{v(\bar{u},R_{-2})}^{v_{\infty}} d\bar{v}  r^{1+\delta} \|\mathscr{A}^{[N-s]} A\|_{S_{\bar{u},\bar{v}}}^2 &\lesssim \frac{(\varepsilon_0)^2+\varepsilon^3}{u^s} \ \ \ \textrm{for $s=0,1,2$,} \nonumber \\
 \int_u^{u_f} d\bar{u} \int_{v(\bar{u},R_{-2})}^{v_{\infty}} d\bar{v}  r^{1+\delta} \| \mathscr{A}^{[N-1-s]} \Pi\|_{S_{\bar{u},\bar{v}}}^2 &\lesssim \frac{(\varepsilon_0)^2+\varepsilon^3}{u^{s}} \ \ \ \textrm{for $s=0,1$,} \nonumber \\
  \int_u^{u_f} d\bar{u} \int_{v(\bar{u},R_{-2})}^{v_{\infty}} d\bar{v}  r^{1+\delta} \|\mathscr{A}^{[N-3]} \Pi\|_{S_{\bar{u},\bar{v}}}^2 &\lesssim \frac{(\varepsilon_0)^2+\varepsilon^3}{u^{2-2\delta}}  \, .\end{align}
\item (boundary terms on $\underline{C}_{v_\infty}^{\I}$)
From estimates of Propositions \ref{lem:chibarfluxes}, \ref{lem:trxbflux} and \ref{prop:etbv}, we have on $\underline{C}_{v_\infty}^{\I}$ the estimates
\begin{align}
\| \mathscr{A}^{[N-1-s]}Z\|^2_{S_{u,v_\infty}} \lesssim \frac{(\varepsilon_0)^2+\varepsilon^3}{u^s} \ \ \ , \ \ \ \int_{u}^{u_f} d\bar{u} \|\mathscr{A}^{[N-1-s]}Z\|^2_{S_{\bar{u},v_\infty}} \lesssim \frac{(\varepsilon_0)^2+\varepsilon^3}{u^s} \, .
\end{align}
\end{enumerate}
{\bf Step 3: Completing the proof.} Using (1)--(3) above, the desired estimates follow:  We contract (\ref{zcomm}) with $(1-r^{-\delta})\mathscr{A}^{[N-1-s]} Z$ and integrate successively over the following regions
\begin{enumerate}
\item $\mathcal{D}^{\I}(u)$ (to obtain the spacetime integral) 
\item $\mathcal{D}^{\I}(u) \cap J^+(\underline{C}^{\I}_v(u))$ (to produce the boundary term on the ingoing cone $\underline{C}^{\I}_v(u)$) \, .
\end{enumerate}
After an integration by parts with the boundary term controlled by (3) we produce a spacetime term with a favourable sign on the left of the form 
\[
\delta \int_u^{u_f} d\bar{u} \int_{v(\bar{u},R_{-2})}^{v_{\infty}} d\bar{v}  r^{-1-\delta} \| \mathscr{A}^{[N-1-s]}  Z\|_{S_{\bar{u},\bar{v}}}^2 \, .
\]
Applying Cauchy--Schwarz to the terms borrowing from this spacetime term and using the estimates from (1) and (2) above yields for $s=0,1$ the estimate
\begin{align} \label{Z1a}
\int_{\underline{C}^{\I}_v(u)} d\bar{u} \Big\|\mathscr{A}^{[N-1-s]}  Z\Big\|^2_{S_{\bar{u},v}} +  \int_u^{u_f} d\bar{u} \int_{v(\bar{u},R_{-2})}^{v_{\infty}} d\bar{v}  r^{-1-\delta} \| \mathscr{A}^{[N-1-s]}  Z\|_{S_{\bar{u},\bar{v}}}^2 \lesssim \frac{(\varepsilon_0)^2+\varepsilon^3}{u^s} \, .
\end{align}
The estimate also holds for $s=2$ provided the decay rate is changed to $2-2\delta$.

Following the same argument but integrating now only over a fixed cone $C_u^{\I}$ we also obtain for $s=0,1,2$ 
\begin{align} \label{Z1b}
 \Big\|\mathscr{A}^{[N-1-s]}  Z\Big\|^2_{S_{{u},v}} + \int_{v(u,R_{-2})}^{v_{\infty}} d\bar{v}  r^{-1-\delta} \| \mathscr{A}^{[N-1-s]}  Z\|_{S_{u,\bar{v}}}^2 \lesssim \frac{(\varepsilon_0)^2+\varepsilon^3}{u^s} \, .
\end{align}
Using the relation (\ref{Zdef}), the above estimates on $Z$ readily translate into estimates for $r^2 \Omega^{-1}\hat{\chi}$ using the estimates on the almost gauge invariant quantity $\Pi$ from Theorem~\ref{thm:alphaalphabarestimates}.
\end{proof}

We conclude with an $L^1$-estimate for $\hat{\chi}$ along the outgoing cones $C_u^{\I}$ which will be useful when estimating the metric components later.
\begin{corollary}
For any $u \in \left[u_{-1},u_f\right]$ and $v > v(u,R_{-2})$ we have for $s=0,1,2$ the estimate
\begin{align} \label{l1chi}
\int_{v}^{v_\infty} d\bar{v} \| \left[ r\slashed{\nabla} \right]^{N+1-s} (\Omega \hat{\chi})\|_{S_{u,\bar{v}}}  \lesssim \frac{\varepsilon_0+\varepsilon^{\frac{3}{2}}}{r \cdot u^{s/2}} \, .
\end{align}
\end{corollary}
\begin{proof}
Note that we have 
\begin{align}
\int_{v}^{v_\infty} d\bar{v} \| \mathscr{A}^{[N-1-s]} r^2 \slashed{\mathcal{D}}_2^\star \slashed{div} (\Omega \hat{\chi})\|_{S_{u,\bar{v}}}=
\int_{v}^{v_\infty} d\bar{v} \|  \mathscr{A}^{[N-1-s]}  \frac{1}{r^2} (\Omega^2 Z - r  \Pi -  \overset{(out)}{\slashed{\mathcal{E}}^{1}_{1}} )\|_{S_{u,\bar{v}}} \lesssim \frac{\varepsilon_0+\varepsilon^{\frac{3}{2}}}{r \cdot u^{s/2}} \, , \nonumber
\end{align}
where the equality follows from (\ref{Zdef}) and the estimate from Proposition  \ref{prop:errorspacetimeIplus} on the non-linear error, the estimate (\ref{Z1b}) on $Z$ and using Cauchy--Schwarz on the term involving $\Pi$ so that its flux appearing in Theorem~\ref{thm:alphaalphabarestimates} can be exploited. Applying Proposition \ref{prop:symmetrictracelesskeylemma} to the left hand side the result follows.
\end{proof}

\subsubsection{Estimating $\beta$} \label{sec:betainD}
Using that the bootstrap estimates have been improved for $\hat{\chi}$ we can improve them for $\beta$ using the control of the $\ell=1$ modes from Propositions \ref{prop:l1modesspecial} and \ref{prop:l1modes} and the almost gauge invariant quantity $\psi$.

\begin{proposition} \label{prop:allbetaimproved}
$\phantom{X}$
\begin{center}
$\boxed{\textrm{The estimate (\ref{target3}) has been shown for all terms containing $\beta$.}}$
\end{center}
\end{proposition}

\begin{proof}
Note that using the estimates of Propositions \ref{prop:l1modesspecial} and \ref{prop:l1modes} on the $\ell=1$ mode of $\beta$ it suffices in view of Proposition \ref{prop:oneformkeylemma} to obtain estimates for $\mathscr{A}^{[N-1-s]}\Dslash_2^*  \beta$. To obtain the latter we recall from the proof of Proposition \ref{prop:Teukolsky} the relation
\[
\Dslash_2^* \Omega \beta =  \Pi  r^{-3} +\frac{3M}{r^3} \Omega \hat{\chi} 
+\frac{1}{2} \left(-3 (\rho - \rho_{\circ}) \Omega \hat{\chi}-3\sigma \Omega  {}^* \hat{\chi}\right) + \mathcal{E}^0_6 \, .
\]
We commute this with $\mathscr{A}^{[N-1-s]}$ and use the estimates of Proposition \ref{prop:chistart} on $\hat{\chi}$ and  the estimates on the almost gauge invariant quantity $\Pi$ from Theorem~\ref{thm:alphaalphabarestimates} for the linear terms. The non-linear term $\mathcal{E}^0_6$ is controlled using Propositions \ref{prop:eebs} and \ref{prop:intdebs}. For the other non-linear terms note in particular that we have  ingoing flux estimates for $\rho$ and $\sigma$ from the bootstrap assumptions and also such estimates for $\Omega\hat{\chi}$ from (\ref{chiestcone}).  
\end{proof}

\subsubsection{Estimating \underline{$\hat{\chi}$}} \label{sec:chibare}

The quantity $\underline{\hat{\chi}}$ will be estimated from the quantity $B$ which in turn will be estimated from its propagation equation in the $3$-direction (\ref{Beq}). 

\begin{proposition} \label{prop:Ytopspheres}
We have in $\mathcal{D}^{\I}$ the estimates
\begin{align} \label{Btopest}
\| \mathscr{A}^{[N-3]} B \|_{S_{u,v}} + \| \mathscr{A}^{[N-1]} Y \|_{S_{u,v}} \lesssim  \varepsilon_0 +  \varepsilon^{\frac{3}{2}} \, .
\end{align}
\end{proposition}

\begin{proof}
The estimate for $B$ follows directly from (\ref{Beq}) noting the non-linear error estimate (\ref{Berrortop}). For the most difficult linear error term on the right hand side of (\ref{Beq}) note that by Cauchy--Schwarz
\begin{align}
\int_{u_{-1}}^u d\bar{u} \|  \mathscr{A}^{[N-3]}  \check{\underline{A}}\|_{S_{\bar{u},v}} \lesssim \sqrt{\int_{u_{-1}}^u d\bar{u} \|   \mathscr{A}^{[N-3]} \check{\underline{A}}\|^2_{S_{\bar{u},v}} \bar{u}^{1+\delta}} \lesssim \varepsilon_0 + \varepsilon^{\frac{3}{2}} \, ,
\end{align}
with the last estimate following from Theorem~\ref{thm:alphaalphabarestimates}.
Once the estimate holds for $B$, the relation (\ref{Bdef}) implies it for $Y$ using the estimates on $\check{\underline{\Psi}}$ and $\check{\underline{\psi}}$ from Theorem~\ref{thm:alphaalphabarestimates} and Corollary \ref{cor:replacercheckbyr}.
\end{proof}

\begin{corollary} \label{cor:topxb}
We have in $\mathcal{D}^{\I}$ the estimates
\begin{align}
\sup_{\mathcal{D}^{\I}} \|  \mathscr{A}^{[N]} r \Omega^{-1} \underline{\hat{\chi}} \|_{S_{u,v}} +\sqrt{\int_{u_{-1}}^{u_f} d\bar{u} \|  \mathscr{A}^{[N+1]} r \Omega^{-1} \underline{\hat{\chi}} \|^2_{S_{\bar{u},v}}} &\lesssim \varepsilon_0 + \varepsilon^{\frac{3}{2}} \, .
\end{align}
\end{corollary}

\begin{proof}
This follows directly the relation (\ref{relYgi}), and estimates on $\check{\underline{\psi}}$ of Theorem~\ref{thm:alphaalphabarestimates}. For the flux estimate start from the fact that we control $\int d\bar{u} \frac{1}{r^2}\|\mathscr{A}^{[N-1]} Y\|^2_{S_{\bar{u},v}}$ from (\ref{Btopest}) and (\ref{relYgi}).
\end{proof}

For $\Omega \slashed{\nabla}_4 B$ we have estimates on spheres
\begin{proposition}
We have in $\mathcal{D}^{\I}$ the estimates
\begin{align}
\label{n4Btopest}
\| \Omega \slashed{\nabla}_4 \mathscr{A}^{[N-5]} B \|_{S_{u,v}} &\lesssim \frac{\varepsilon_0 + \varepsilon^{\frac{3}{2}}}{v^2}+  \frac{\varepsilon_0 + \varepsilon^{\frac{3}{2}}}{r ^{5/4}u} \, .
\end{align}
The same estimate holds for any order of the derivatives $\Omega \slashed{\nabla}_4$ and $\mathscr{A}^{[N-5]}$ in the above.
\end{proposition}

\begin{proof}
Note first that on $u=u_{-1}$ we have by (\ref{sphbY}) the bound
\[
\| \Omega \slashed{\nabla}_4  \mathscr{A}^{[N-5]} B \|_{S_{u_{-1},v}} \lesssim  \frac{\varepsilon_0 + \varepsilon^{\frac{3}{2}}}{v^2} \, .
\]
Integrating (\ref{n4Beq}) along a hypersurface of fixed constant $v$, we use that by Theorem~\ref{thm:alphaalphabarestimates} for $i=1,...,N-5$
\begin{align}
\int_{u_{-1}}^{u} d\bar{u} \frac{1}{r^3} \| \mathscr{A}^{[i]} \check{\underline{\Psi}}\|_{S_{\bar{u},v}} + \frac{1}{r^2}\| \mathscr{A}^{[i]} \check{\underline{\Pi} }\|_{S_{\bar{u},v}} + \frac{1}{r^2} \| \check{\mathcal{A}}^{[i]} \underline{A} \|_{S_{\bar{u},v}}  \lesssim  \frac{\varepsilon_0 + \varepsilon^{\frac{3}{2}}}{v^2} +  \frac{\varepsilon_0 + \varepsilon^{\frac{3}{2}}}{r ^{5/4}u} \, .
\end{align}
Combining this with (\ref{B4errorflux}) the estimate follows.
For the interchange of derivatives, note that the commutator terms that arise can be controlled by (\ref{Btopest}) and the $u$-decay of the lower order terms arising in the commutator formula of Lemma \ref{lem:commutation}. 
\end{proof}

\begin{corollary} \label{cor:Yi}
We have
\begin{align}
\|  \mathscr{A}^{[N-5]} B \|_{S_{u,v}} + \|  \mathscr{A}^{[N-3]}Y \|_{S_{u,v}}   &\lesssim \frac{\varepsilon_0 + \varepsilon^{\frac{3}{2}}}{u} 
\end{align} 
and for $s=0,1$
\begin{align}
\|  \mathscr{A}^{[N-1-s]} r \Omega^{-1}\underline{\hat{\chi}} \|_{S_{u,v}} &\lesssim \frac{\varepsilon_0 + \varepsilon^{\frac{3}{2}}}{u^{\frac{s+1}{2}}}.
\end{align}
\end{corollary}
\begin{proof}
For the bound on $B$ we first observe that it holds for $\|  \mathscr{A}^{[N-5]} B \|_{S_{u,v_\infty}}$ by Corollary \ref{cor:auxYonskri}, the relation (\ref{Bdef}) and Theorem~\ref{thm:alphaalphabarestimates} (Theorem \ref{thm:PPbarestimates} is also sufficient, the estimate on spheres is easily obtained from Sobolev embedding and the control of the fluxes on $v=v_\infty$). We then integrate backwards along constant $u$ using the estimate (\ref{n4Btopest}) to finish the proof. 

To obtain the bound on $Y$, we first prove it for $Y$ replaced by $\frac{Y}{r}$. In this case, the bound follows directly from the relation (\ref{Bdef}) and use Theorem~\ref{thm:alphaalphabarestimates}.\footnote{The reason we can a priori only establish the weaker estimate is that only know $\frac{1}{r} \| \check{\underline{\Psi}}\|_{S_{u,v}}  \lesssim  \frac{\varepsilon_0 + \varepsilon^{\frac{3}{2}}}{u}$ from Theorems \ref{thm:PPbarestimates} and \ref{thm:alphaalphabarestimates}. After having proven the estimate for $r \underline{\hat{\chi}}$, we also have $\|\check{\underline{\Psi}}\|_{S_{u,v}}  \lesssim  \frac{\varepsilon_0 + \varepsilon^{\frac{3}{2}}}{u}$ and hence the stronger estimate for $Y$.} With this weaker bound, the bound on $r \Omega^{-1}\underline{\hat{\chi}}$ already follows from (\ref{relYgi}). Finally, to get the estimate claimed for $Y$, note that by Propositions \ref{prop:PPbaridentities} and \ref{prop:Pbartildeidentities} we have $\|  \mathscr{A}^{[N-5]} \frac{r}{\check{r}} \check{\underline{\Psi} }\| \lesssim \|  \mathscr{A}^{[N-5]}  \Psi \| + \|  \mathscr{A}^{[N-5]} r \Omega^{-1} \underline{\hat{\chi}}\| + \frac{1}{r} \|  \mathscr{A}^{[N-5]}  r^2 \Omega \hat{\chi}\| +\frac{\varepsilon_0 + \varepsilon^{\frac{3}{2}}}{u} \lesssim \frac{\varepsilon_0 + \varepsilon^{\frac{3}{2}}}{u}$, the last inequality following in view of what we have already proven (in particular on $r \underline{\hat{\chi}}$). Now revisit the relation (\ref{Bdef}) with the improved estimate on $\|  \mathscr{A}^{[N-5]}\frac{r}{\check{r}} \check{\underline{\Psi}} \|$ to deduce the estimate for $Y$.
\end{proof}

We next prove an integrated decay estimate for $\slashed{\nabla}_4 B$ which will then produce an integrated decay estimate for $B$ via a Hardy inequality, see Corollary \ref{cor:BHardy}.

\begin{proposition} \label{prop:n4B}
We have the top-order estimate 
\begin{align} \label{imolfinish}
\sup_{u \in \left[u_{-1},u_f\right]} \int_{v(u,R_{-2})}^{v_{\infty}} d\bar{v} & \ r^{2+\delta} \| \Omega \slashed{\nabla}_4  \mathscr{A}^{[N-4]} B \|_{S_{\bar{u},\bar{v}}}^2  
\nonumber \\
\int_{u_{-1}}^{u_f} d\bar{u} \int_{v(\bar{u},R_{-2})}^{v_{\infty}} d\bar{v} & \ r^{1+\delta}  \|\Omega \slashed{\nabla}_4   \mathscr{A}^{[N-4]} B \|_{S_{\bar{u},\bar{v}}}^2 \lesssim  \varepsilon_0^2 + \varepsilon^3
\end{align}
and the lower order integrated decay estimate
\begin{align}
\int_{u}^{u_f} d\bar{u} \int_{v(\bar{u},R_{-2})}^{v_{\infty}} d\bar{v} & \ r^{1-\delta}  \|  \Omega \slashed{\nabla}_4  \mathscr{A}^{[N-5]} B \|_{S_{\bar{u},\bar{v}}}^2 \lesssim  \frac{\varepsilon_0^2 + \varepsilon^3}{u}
\end{align}
Finally, both estimates hold interchanging the order of $\Omega \slashed{\nabla}_4$ and the angular operator $\mathscr{A}^{[N-4]}$ (and $\mathscr{A}^{[N-5]}$ respectively) in the above.
\end{proposition}

\begin{proof}
The second estimate is a direct consequence of (\ref{n4Btopest}) using that $v>u$ in $\mathcal{D}^{\I}$. 

For the first estimate we derive from (\ref{n4Beq}) upon contraction with $r^{2+\delta} \Omega \slashed{\nabla}_4 \mathscr{A}^{[N-4]} B$ and integration over $\mathcal{D}^{\I}(u)$ the estimate
\begin{align} \label{imol}
\sup_{\left[u_{-1},u_f\right]} \int_{v(\bar{u},R_{-2})}^{v_{\infty}} d\bar{v} & \ r^{2+\delta} \|\Omega \slashed{\nabla}_4 \mathscr{A}^{[N-4]} B \|_{S_{\bar{u},\bar{v}}}^2 
\nonumber \\
+ \int_{u_{-1}}^{u_f} d\bar{u} \int_{v(\bar{u},R_{-2})}^{v_{\infty}} d\bar{v} & \ r^{1+\delta}  \|\Omega \slashed{\nabla}_4 \mathscr{A}^{[N-4]} B \|_{S_{\bar{u},\bar{v}}}^2 \lesssim \varepsilon_0^2 + \varepsilon^3 \, ,
\end{align}
where we have ignored the favourable boundary term on $r=R_{-2}$.
To prove (\ref{imol}), we first note the estimates for the error-term (for any $\gamma>0$)
\begin{align}
\int_{\mathcal{D}^{\I} \cap \{ r \geq \frac{1}{2}u \}} dudv d\theta \mathcal{E} \left[ \Omega \slashed{\nabla}_4  \mathscr{A}^{[N-4]} B\right] r^{2+\delta} \Omega \slashed{\nabla}_4\mathcal{A} ^{[N-4]} B \nonumber \\
\lesssim C_\gamma \int_{\mathcal{D}^{\I} \cap \{ r \geq \frac{1}{2}u\}} dudv d\theta u^{1+\delta} r^{2+\delta} |\mathcal{E} \left[ \Omega \slashed{\nabla}_4  \mathscr{A}^{[N-4]} B\right]|^2 + \gamma \sup_{\left[u_{-1},u\right]} \int_{v(\bar{u},R_{-2})}^{v_{\infty}} d\bar{v} \ r^{2+\delta} \| \Omega \slashed{\nabla}_4 \mathscr{A}^{[N-4]} B \|_{S_{\bar{u},\bar{v}}}^2 \nonumber \, 
\end{align}
and
\begin{align}
\int_{\mathcal{D}^{\I} \cap \{ r \leq \frac{1}{2}u\}} dudv d\theta \mathcal{E} \left[ \Omega \slashed{\nabla}_4  \mathscr{A}^{[N-4]} B\right] r^{2+\delta} \Omega \slashed{\nabla}_4\mathcal{A} ^{[N-4]} B \nonumber \\
\lesssim C_\gamma \int_{\mathcal{D}^{\I} \cap \{ r \leq \frac{1}{2}u\}} dudv d\theta r^{3+\delta} |\mathcal{E} \left[ \Omega \slashed{\nabla}_4  \mathscr{A}^{[N-4]} B\right]|^2 + \gamma \int_{u_{-1}}^{u} d\bar{u} \int_{v(\bar{u},R_{-2})}^{v_{\infty}} d\bar{v} \ r^{1+\delta}  \|\Omega \slashed{\nabla}_4 \mathscr{A}^{[N-4]} B \|_{S_{\bar{u},\bar{v}}}^2 \nonumber \, .
\end{align}
The first term in both estimates is easily controlled by (\ref{B4errorspacetime}) while the second one is absorbed on the left for sufficiently small $\gamma$. The ``linear" terms 
\begin{align}
\int_{\mathcal{D}^{\I}} dudv d\theta  \left( h_{3}  \mathscr{A}^{[N-4]} \check{\underline{\Psi}} + h_{2} \mathscr{A}^{[N-4]} \check{\underline{\Pi}} + h_{2}  \mathscr{A}^{[N-4]} \check{\underline{A}}  \right) r^{2+\delta} \Omega \slashed{\nabla}_4\mathscr{A}^{[N-4]} B \, 
\end{align}
can be handled similarly
using again Cauchy--Schwarz and the estimates of Theorem~\ref{thm:alphaalphabarestimates} for the almost gauge invariant quantities. For the term on the data $u=u_{-1}$ we use the estimate (\ref{fluxYu0}) in conjunction with (\ref{Bdef}) and Theorem~\ref{thm:alphaalphabarestimates}. This proves the desired estimate. The claim about the interchange of derivatives follows easily from the commutator formula, the lower order bootstrap assumptions on spheres and Proposition \ref{prop:Ytopspheres}.
\end{proof}

\begin{corollary} \label{cor:BHardy}
We have for $s=0,1$
\begin{align}  \label{imcimol}
\int_{u}^{u_f} d\bar{u} \int_{v(\bar{u},R_{-2})}^{v_{\infty}} d\bar{v} \ r^{-1-\delta}  \|\mathscr{A}^{[N-4-s]} B \|_{S_{\bar{u},\bar{v}}}^2 \lesssim \frac{\varepsilon_0^2 + \varepsilon^3}{u^s}, \\
\int_{u}^{u_f} d\bar{u} \int_{v(\bar{u},R_{-2})}^{v_{\infty}} d\bar{v} \ r^{-1-\delta}  \| |\mathscr{A}^{[N-2-s]}Y \|_{S_{\bar{u},\bar{v}}}^2 \lesssim \frac{\varepsilon_0^2 + \varepsilon^3}{u^s}.
\end{align}
\end{corollary}
\begin{proof}
Note that for $s=1$ both estimates follow directly from Corollary \ref{cor:Yi}. For $B$ and $s=0$ the result follows from the following Hardy inequality: Let $f$ be a smooth symmetric traceless tensor on $S_{u,v}$. Then we have for any $u_{-1} \leq u \leq u_f$ and any fixed $R_{-2} \leq \hat{R} \leq R$
the inequality
\begin{align} 
\int_{u}^{u_f} d\bar{u} \int_{v(\bar{u},\hat{R})}^{v_\infty}d\bar{v} d\theta \frac{1}{r^{1+\delta}} \|   f \|^2_{S_{\bar{u},\bar{v}}} + \int_{u}^{u_f} du d\theta \| f\|^2_{S_{u,v\left(u,\hat{R}\right)}}
\nonumber \\ \lesssim  \int_{u}^{u_f} d\bar{u} \int_{v(\bar{u},\hat{R})}^{v_\infty}d\bar{v} d\theta \, r^{1+\delta} \| \Omega \slashed{\nabla}_4 f\|^2_{S_{\bar{u},\bar{v}}}  + \int^{u_f}_{u} \int_{S} d\bar{u} d\theta \|  f\|^2_{S_{\bar{u},v_\infty}} \, .
\end{align}
Applying the inequality with $f=\mathscr{A}^{[N-4]} B$ and using that the last term on $v=v_\infty$ is controlled by Corollary \ref{cor:auxYonskri}, the relation (\ref{Bdef}) and Theorem~\ref{thm:alphaalphabarestimates}, the result follows. For the estimate on $Y$ use (\ref{Bdef}) and the estimates of Theorem~\ref{thm:alphaalphabarestimates}. 
\end{proof}

\begin{corollary} \label{cor:imcimol}
We have for $s=0,1$ the integrated decay estimate
\begin{align}
\int_{u}^{u_f} d\bar{u} \int_{v(\bar{u},R_{-2})}^{v_{\infty}} d\bar{v} \ r^{-1-\delta} \Omega^2  \||\mathscr{A}^{[N-s]} r \Omega^{-1} \underline{\hat{\chi}} \|_{S_{\bar{u},\bar{v}}}^2 \lesssim \frac{\varepsilon_0^2 + \varepsilon^3}{u^s} \, .
\end{align}
\end{corollary}
\begin{proof}
Use (\ref{relYgi}) in conjunction with the estimate on $Y$ from Corollary \ref{cor:BHardy} and the estimates for $\underline{\Pi}$ from Theorem~\ref{thm:alphaalphabarestimates}.
\end{proof}

\begin{corollary} \label{cor:fits}
We have for $u \in \left[u_{-1},u_f\right]$ along any cone $\underline{C}_v^{\I}(u) \in \mathcal{D}^{\I}(u)$ the flux estimate
\begin{align}
\int_{u}^{u_f} d\bar{u} \||\mathscr{A}^{[N-1]} (r \Omega^{-1} \underline{\hat{\chi}}) \|_{S_{\bar{u},v}}^2  
\lesssim \frac{\varepsilon_0^2 + \varepsilon^3}{u^2} \, .
\end{align}
\end{corollary}
\begin{proof}
From Corollary \ref{cor:Yi} we have
\[
\int_{u}^{u_f} d\bar{u} \frac{1}{r^2} \||\mathscr{A}^{[N-3]} Y \|_{S_{\bar{u},v}}^2 \lesssim\frac{\varepsilon_0^2 + \varepsilon^3}{u^2}  \, .
\]
Now use (\ref{relYgi})  as well as the flux bound on $\check{\underline{\Pi}}$  from Theorem~\ref{thm:alphaalphabarestimates} to establish the desired flux estimate. 
\end{proof}

\begin{corollary}
We have for $u \in \left[u_{-1},u_f\right]$ along any cone $\underline{C}_v^{\I}(u) \in \mathcal{D}^{\I}(u)$   the flux estimate
\begin{align}
\int_{u}^{u_f} du \| \mathscr{A}^{[N-5]} \check{\underline{\Psi}} \|_{S_{u,v}}^2  
\lesssim \frac{ \varepsilon_0^2 + \varepsilon^3}{u^{2}} \, .
\end{align}
and the integrated decay estimate
\begin{align} \label{iledbetterPb}
\int_{u}^{u_f} d\bar{u} \int_{v(\bar{u},R_{-2})}^{v_{\infty}} d\bar{v} \ r^{-1-\delta} \|  \mathscr{A}^{[N-4]} \check{\underline{\Psi}} \|_{S_{\bar{u},\bar{v}}}^2 \lesssim \frac{\varepsilon_0^2 + \varepsilon^3}{u}
\end{align}
\end{corollary}
\begin{proof}
Use that $\|  \mathscr{A}^{[N-5]} \frac{r}{\check{r}} \check{\underline{\Psi} }\| \lesssim \|  \mathscr{A}^{[N-5]}  \Psi \| + \|  \mathscr{A}^{[N-5]} r \Omega^{-1} \underline{\hat{\chi}}\| + \frac{1}{r} \|  \mathscr{A}^{[N-5]}  r^2 \Omega \hat{\chi}\| +\frac{\varepsilon_0 + \varepsilon^{\frac{3}{2}}}{u} \lesssim \frac{\varepsilon_0 + \varepsilon^{\frac{3}{2}}}{u}$ as in 
 the proof of Corollary \ref{cor:Yi} 
\end{proof}

We finally estimate also the \emph{ingoing} flux of $N+1$ derivatives of $\underline{\hat{\chi}}$:
\begin{corollary} \label{cor:fits2}
We have for any $C^{\I}_u \subset \mathcal{D}^{\I}$
\begin{align}
\int_{C^{\I}_u} d\bar{v} \frac{1}{r^2} \|\mathscr{A}^{[N+1]} (r \Omega^{-1} \underline{\hat{\chi}}) \|_{S_{u,v}}^2   \lesssim\varepsilon_0^2 + \varepsilon^3 \, .
\end{align}
\end{corollary}
\begin{proof}
Use (\ref{Btopest}), which implies $\int_{C_v} \frac{1}{r^2} \|\mathscr{A}^{[N-1]} Y \|^2_{S_{u,v}} \lesssim \varepsilon_0^2 + \varepsilon^3$. Convert back using (\ref{relYgi}) and the estimates  on $\check{\underline{\Pi}}$ from Theorem~\ref{thm:alphaalphabarestimates} and Corollary \ref{cor:replacercheckbyr}.
\end{proof} 
Collecting the estimates of Corollaries \ref{cor:topxb}, \ref{cor:Yi}, \ref{cor:imcimol} and \ref{cor:fits}, \ref{cor:fits2} we conclude applying Proposition \ref{prop:symmetrictracelesskeylemma}:
\begin{center}
$\boxed{\textrm{The estimate (\ref{target3}) has been shown for all terms containing $\underline{\hat{\chi}}$.}}$
\end{center}

\subsubsection{Estimating \underline{$\beta$}} \label{sec:betabimprove}
\begin{proposition} \label{prop:allbetabarimproved}
$\phantom{X}$
\begin{center}
$\boxed{\textrm{The estimate (\ref{target3}) has been shown for all terms containing $\underline{\beta}$.}}$
\end{center}
\end{proposition}

\begin{proof}
With the bootstrap estimates having been improved for $\underline{\hat{\chi}}$ the proof is entirely analogous to that of Proposition \ref{prop:allbetaimproved}. In particular, control of the $\ell=1$ modes for $\underline{\beta}$ from Propositions \ref{prop:l1modesspecial} and \ref{prop:l1modes} implies (by Proposition \ref{prop:oneformkeylemma}) that it suffices to obtain estimates for $\mathscr{A}^{[N-1-s]} \Dslash_2^* \betabar$. These are obtained using  the relation of Proposition  \ref{prop:Pbartildeidentities} and the the estimates on the almost gauge invariant quantity $\check{\underline{\Pi}}=r^3 \Omega \check{\underline{\psi}}$ from Theorem~\ref{thm:alphaalphabarestimates} (and Corollary \ref{cor:replacercheckbyr}). Propositions \ref{prop:eebs} and \ref{prop:intdebs} for the non-linear error term  yields the result. 
\end{proof}

\subsubsection{Estimating $\rho, \sigma$}
\begin{proposition} \label{prop:allrhosigmaimproved}
$\phantom{X}$
\begin{center}
$\boxed{\textrm{The estimate (\ref{target3}) has been shown for all terms containing $(\rho-\rho_\circ,\sigma)$.}}$
\end{center}
\end{proposition}

\begin{proof}
We control of the $\ell=0,1$ modes for $\rho-\rho_\circ$ and $\sigma-\sigma_{\mathrm{Kerr}}$ from Propositions \ref{prop:l1modesspecial} and \ref{prop:l1modes}, which implies by Proposition \ref{prop:functionkeylemma} that it suffices to obtain estimates for $\mathscr{A}^{[N-2-s]} \Dslash_2^* \left( \nablaslash \rho + {}^* \nablaslash \sigma \right)$. The latter are obtained from the relation
\begin{align}
P
		=
		\Dslash_2^* \left( \nablaslash \rho + {}^* \nablaslash \sigma \right)
		+
		\frac{3M \Omega}{r^4} \left( \hat{\chibar} - \hat{\chi} \right)
		+
		\frac{3}{2r^3\Omega} \nablaslash_3 \left( r^3 \Omega \left( (\rho - \rho_{\circ}) \hat{\chi} + \sigma {}^*\hat{\chi} \right) \right)
		+
		2 \Dslash_2^* \left( \hat{\chi} \cdot \betabar \right)
		+
		\mathcal{E}^1_6.
\end{align}
using previous estimates and Propositions \ref{prop:eebs} and \ref{prop:intdebs} for the non-linear error terms. For the most difficult ingoing flux note that all non-linear terms are either $\mathcal{E}^1_6$ or involve a product of quantities from the collection  $\hat{\chi}, \underline{\hat{\chi}}, \rho, \sigma, \underline{\beta}$ for each of which we have ingoing flux estimates available
\end{proof}

\subsubsection{Estimating $r^2 \Omega \slashed{\nabla}_4 (r$\underline{$\hat{\chi}$})}

Returning now to Proposition \ref{prop:n4B} we convert our estimates for $\Omega \slashed{\nabla}_4 B$ into estimate for $\Omega \slashed{\nabla}_4 Y$ and eventually $\Omega \slashed{\nabla}_4 (r\underline{\hat{\chi}})$ using the definitions of $B$ and $Y$. Key in all proofs is the relation
\begin{align} \label{n4Yrel}
r^2 \Omega \slashed{\nabla}_4 \left(r^2\slashed{\mathcal{D}}_2^\star \slashed{div} (r \underline{\hat{\chi}}\Omega^{-1})\right)  =r \Omega \slashed{\nabla}_4 Y- \Omega_\circ^2 Y + h \frac{r}{\check{r}} \check{\underline{\Psi}} + \tilde{h} \frac{r}{\check{r}} \check{\underline{\Psi}} + \overset{(out)}{\slashed{\mathcal{E}}^{2}_{0}}+\check{\slashed{\mathcal{E}}}^{1}_{0} \, ,
\end{align}
with $h, \tilde{h}$ admissible coefficient functions. This follows from Lemma \ref{lem:Yrel} and (\ref{eq:rtildeoverr4}). Note that using the Bianchi and null structure equations we have that $r^2 \Omega \slashed{\nabla}_4 \overset{(in)}{\slashed{\mathcal{E}}^{1}_{1}}=\overset{(out)}{\slashed{\mathcal{E}}^{2}_{0}}+{\slashed{\mathcal{E}}^{1}_{0}}$ holds for the $\overset{(in)}{\slashed{\mathcal{E}}^{1}_{1}}$ appearing in Lemma \ref{lem:Yrel}.

We begin with a Lemma estimating $\Omega \slashed{\nabla}_4 Y$ for $\Omega \slashed{\nabla}_4 B$ on spheres:

\begin{lemma}
We have 
\begin{align} \label{n4Yspherestop}
\| r\Omega \slashed{\nabla}_4 \left[ r \slashed{\nabla}\right]^{N-2} Y \|_{S_{u,v}} \lesssim \varepsilon_0+\varepsilon^{\frac{3}{2}}
\end{align}
and for $s=0,1$ 
\begin{align}  \label{n4Yspheres}
\|  r\Omega \slashed{\nabla}_4 \left[ r \slashed{\nabla}\right]^{N-3-s} Y \|_{S_{u,v}} &\lesssim \frac{ \varepsilon_0 +  \varepsilon^{\frac{3}{2}}}{u^{\frac{1+s}{2}}} 
\end{align}
\begin{align}  \label{n4Yfluxout}
\int_{v}^{v_\infty} d\bar{v} \frac{1}{r^2} \|  r\Omega \slashed{\nabla}_4 \left[ r \slashed{\nabla}\right]^{N-3} Y \|^2_{S_{{u},\bar{v}}} &\lesssim \frac{ (\varepsilon_0)^2 +  \varepsilon^3}{u^{2}} 
\end{align}
\end{lemma}
\begin{proof}
This follows from the estimate (\ref{n4Btopest}), the definition of $B$ (\ref{Bdef}) and the estimates on $\check{\underline{\psi}}$ and $\check{\underline{\Psi}}$ on spheres following from Theorem~\ref{thm:alphaalphabarestimates}, in particular that for $s=0,1$
\begin{align} \label{pio}
 \|   \mathscr{A}^{[N-5-s]} r \slashed{\nabla}_4 \left(\frac{r}{\check{r}} \check{\underline{\Psi}}\right) \|_{S_{u,v}}   \lesssim \frac{\varepsilon_0 +  \varepsilon^{\frac{3}{2}}}{u^{\frac{1+s}{2}}}  \ \ \ \textrm{and} \ \ \ \|  \mathscr{A}^{[N-4]} r \slashed{\nabla}_4 \left( \frac{r}{\check{r}}\check{\underline{\Psi}}\right) \|_{S_{u,v}}   \lesssim \varepsilon_0 +  \varepsilon^{\frac{3}{2}} \, .
\end{align}
To prove (\ref{pio}) recall that it holds for $\Psi$ in place of $\underline{\Psi}$ by Proposition \ref{prop:afa}. Then use Proposition \ref{prop:nabla4pschematic} to derive the relation 
\[
r\Omega \slashed{\nabla}_4 \left(\frac{r}{\check{r}} \underline{\Psi} \right) = r\Omega \slashed{\nabla}_4  \Psi + \Omega r^6 \slashed{\mathcal{D}}_2^\star  \slashed{\mathcal{D}}_1^\star  \left(0, \slashed{curl} \beta \right) + \check{\mathcal{E}}^2_{\frac{1}{2}} \, .
\]
The claim (\ref{pio}) now follows after angular commutation and using that the bootstrap assumptions for $\beta$ have been improved in Section \ref{sec:betainD} already. 
\end{proof}

\begin{proposition} \label{prop:co2}
We have the following estimates on spheres  for $s=0,1,2$
\begin{align}
\| r^2\Omega \slashed{\nabla}_4 \mathscr{A}^{[N-1-s]} (r \Omega^{-1} \hat{\underline{\chi}}) \|_{S_{u,v}} \lesssim \frac{ \varepsilon_0 + \varepsilon^{\frac{3}{2}}}{u^{\min(\frac{1}{4}+\frac{s}{2},1)}} \, .
\end{align}
\end{proposition}

\begin{proof}
The estimate follows from the identity (\ref{n4Yrel}) using (\ref{n4Yspheres}) and the estimates of Corollary~\ref{cor:Yi}, in particular the improved bound $\| \mathscr{A}^{[N-5]}  \check{\underline{\Psi}} \| \lesssim \frac{\varepsilon_0 + \varepsilon^2}{u}$ established in the proof of Corollary~\ref{cor:Yi}. 
\end{proof}

\begin{proposition} \label{prop:po3}
We have for any $u_{-1}\leq u \leq u_f$ the estimates
\begin{align} \label{imolfinish3}
 \int_{u_{-1}}^{u_f} d\bar{u} \int_{v(\bar{u},R_{-2})}^{v_{\infty}} d\bar{v}  \ r^{-1-\delta} \|  r^2 \Omega \slashed{\nabla}_4 \mathscr{A}^{[N]} (r \Omega^{-1} \underline{\hat{\chi}}) \|_{S_{u,\bar{v}}}^2
&\lesssim\varepsilon_0^2 + \varepsilon^3 \, , \nonumber \\
 \int_{u}^{u_f} d\bar{u} \int_{v(\bar{u},R_{-2})}^{v_{\infty}} d\bar{v}  \ r^{-1-\delta} \|  r^2 \Omega \slashed{\nabla}_4  \mathscr{A}^{[N-2]} (r \Omega^{-1} \underline{\hat{\chi}}) \|_{S_{u,\bar{v}}}^2 &\lesssim \frac{\varepsilon_0^2 +  \varepsilon^3}{u} \, .
\end{align}
We also have the flux estimates
\begin{align}
\sup_{\bar{u} \in \left[u,u_f\right]} \int_{v(\bar{u},R_{-2})}^{v_{\infty}} d\bar{v}  \ r^{-2} \| r^2 \Omega \slashed{\nabla}_4 \mathscr{A}^{[N]}(r  \Omega^{-1} \underline{\hat{\chi}}) \|_{S_{\bar{u},\bar{v}}}^2  &\lesssim\varepsilon_0^2 + \varepsilon^3 \, , \nonumber \\
\sup_{\bar{u} \in \left[u,u_f\right]} \int_{v(\bar{u},R_{-2})}^{v_{\infty}} d\bar{v}  \ r^{-2} \| r^2 \Omega \slashed{\nabla}_4 \mathscr{A}^{[N-2]} (r \Omega^{-1} \underline{\hat{\chi}}) \|_{S_{\bar{u},\bar{v}}}^2   &\lesssim \frac{( \varepsilon_0)^2 +  \varepsilon^3}{u} \, .
\end{align}
\end{proposition}

\begin{proof}
From Proposition \ref{prop:n4B} and the relation  (\ref{Bdef})  we have (using the estimates on $\Omega \slashed{\nabla}_4 \check{\underline{\Psi}}$ from Theorem~\ref{thm:alphaalphabarestimates} which lead to the small loss in $r$-decay)
\begin{align} \label{imolfinish2}
&\sup_{\bar{u} \in \left[u,u_f\right]} \int_{v(\bar{u},R_{-2})}^{v_{\infty}} d\bar{v} \ r^{2} \|  \Omega \slashed{\nabla}_4 \mathscr{A}^{[N-2]}Y \|_{S_{\bar{u},\bar{v}}}^2 \nonumber \\
+ &\int_{u}^{u_f} d\bar{u} \int_{v(\bar{u},R_{-2})}^{v_{\infty}} d\bar{v} \ r^{1}  \|  \Omega \slashed{\nabla}_4  \mathscr{A}^{[N-2]} Y \|_{S_{\bar{u},\bar{v}}}^2 \lesssim \varepsilon_0^2 + \varepsilon^3 \, .
\end{align}
Using (\ref{imolfinish2}),  (\ref{n4Yrel}) and Corollary \ref{cor:BHardy} as well as the estimates on $\check{\underline{\psi}}$, $\check{\underline{\Psi}}$ from Theorem~\ref{thm:alphaalphabarestimates} we conclude the first and third estimate. For the second we consider the spacetime integral of the identity (\ref{n4Yrel}) and use the estimate (\ref{n4Yspheres}) for $s=1$ on spheres as well as the decay estimate for $Y$ of Corollary \ref{cor:BHardy}. The fourth estimate is already immediate from Proposition \ref{prop:co2}. 
\end{proof}

\subsubsection{Estimating \underline{$\eta$}} \label{sec:etabaux}

\begin{proposition} \label{prop:etabILED}
We have for any $u_{-1} \leq u \leq u_f$ the integrated decay estimates
\begin{align}  \label{ebtopflux}
 \int_{u}^{u_f} d\bar{u} \int_{v(\bar{u},R_{-2})}^{v_{\infty}} d\bar{v} & \  r^{-1-\delta} \|   \mathscr{A}^{[N]} r \slashed{\mathcal{D}}_2^\star (r^2 \underline{\eta}) \|_{S_{u,\bar{v}}}^2 
\lesssim\varepsilon_0^2 + \varepsilon^3   \, , \nonumber \\
\int_{u}^{u_f} d\bar{u} \int_{v(\bar{u},R_{-2})}^{v_{\infty}} d\bar{v} & \  r^{-1-\delta} \|   \mathscr{A}^{[N-2]} r \slashed{\mathcal{D}}_2^\star (r^2 \underline{\eta}) \|_{S_{u,\bar{v}}}^2 
\lesssim \frac{( \varepsilon_0)^2 +  \varepsilon^3}{u}   \, .
\end{align}
We also have the flux estimates
\begin{align} \label{etabtopflux}
\sup_{\bar{u} \in \left[u,u_f\right]} \int_{v(\bar{u},R_{-2})}^{v_{\infty}} d\bar{v} & \ r^{-2} \|  \mathscr{A}^{[N]} r \slashed{\mathcal{D}}_2^\star (r^2 \underline{\eta}) \|_{S_{\bar{u},\bar{v}}}^2 \lesssim\varepsilon_0^2 + \varepsilon^3 \, , \nonumber \\
\sup_{\bar{u} \in \left[u,u_f\right]} \int_{v(\bar{u},R_{-2})}^{v_{\infty}} d\bar{v} & \ r^{-2} \|  \mathscr{A}^{[N-2]}r \slashed{\mathcal{D}}_2^\star (r^2 \underline{\eta}) \|_{S_{\bar{u},\bar{v}}}^2 \lesssim \frac{( \varepsilon_0)^2 +  \varepsilon^3}{u}  \, .
\end{align}
On spheres we have for $s=0,1,2$
\begin{align} \label{seba1}
\|   \mathscr{A}^{[N-1-s]} r \slashed{\mathcal{D}}_2^\star (r^2 \underline{\eta}) \|_{S_{u,v}} \lesssim \frac{\varepsilon_0 +  \varepsilon^{\frac{3}{2}}}{u^{\min(\frac{s}{2}+\frac{1}{4},1)}}
\end{align}
\end{proposition}

\begin{proof}
Follows directly from the angular commuted propagation equation (\ref{eq:chibarhat4}) for $\slashed{\nabla}_4 (r\underline{\hat{\chi}})$ combining the estimates of Proposition  \ref{prop:chistart} with Proposition \ref{prop:po3} (in the spacetime case) and the estimates (\ref{chiest2}) and Proposition \ref{prop:co2} (in the case of spheres). 
\end{proof}

Applying as usual Proposition \ref{prop:oneformkeylemma} in conjunction with the estimates on the $\ell=0,1$ modes from Propositions \ref{prop:l1modesspecial} and \ref{prop:l1modes} we conclude
\begin{center}
$\boxed{\textrm{The estimate (\ref{target3}) has been shown for all terms containing $\underline{\eta}$.}}$
\end{center}

\subsubsection{Estimating {$\eta$}} \label{sec:etaaux}

\begin{proposition} \label{prop:etafluxNiled}
We have for any $u_{-1} \leq u \leq u_f$ the integrated decay estimates
\begin{align} 
\int_{u}^{u_f} d\bar{u} \int_{v(\bar{u},R_{-2})}^{v_{\infty}} d\bar{v} \  r^{-1-\delta} \|  \mathscr{A}^{[N]} r \slashed{\mathcal{D}}_2^\star (r^2 {\eta}) \|_{S_{\bar{u},\bar{v}}}^2 
\lesssim \varepsilon_0^2 + \varepsilon^3 \, , \\
\int_{u}^{u_f} d\bar{u}\int_{v(\bar{u},R_{-2})}^{v_{\infty}} d\bar{v} \  r^{-1-\delta} \|  \mathscr{A}^{[N-2]} r \slashed{\mathcal{D}}_2^\star (r^2 {\eta}) \|_{S_{\bar{u},\bar{v}}}^2 
\lesssim \frac{\varepsilon_0^2 + \varepsilon^3}{u}
\end{align}
On spheres we have  for any $u_{-1} \leq u \leq u_f$ and $s=0,1,2$
\begin{align}  \label{seba2}
\|  \mathscr{A}^{[N-1-s]} r \slashed{\mathcal{D}}_2^\star (r^2 {\eta}) \|^2_{S_{u,v}} \lesssim \frac{( \varepsilon_0)^2 +  \varepsilon^3}{u^{\min(s+\frac{1}{2},2)}} \, .
\end{align}
In particular,
\begin{center}
$\boxed{\textrm{The estimate (\ref{target3}) has been shown for all terms containing $\eta$.}}$
\end{center}
Finally, for the outgoing fluxes we also have the top order estimate
\begin{align} \label{seba3}
\int_{{C}^{\I}_u} d\bar{v} \frac{1}{r^2} \Big\| \left[r \slashed{\nabla}\right]^{N+1}  (r^2(\eta-\eta_{\mathrm{Kerr}}))   \Big\|_{S_{u,\bar{v}}}^2 \lesssim \varepsilon_0^2 + \varepsilon^3 \, .
\end{align}
\end{proposition}

\begin{proof}
To estimate $\eta$, we derive from (\ref{eta4dir}) its commuted version 
\begin{align}
\Omega \slashed{\nabla}_4 (\mathscr{A}^{[n]} r^2 \slashed{\mathcal{D}}^\star_2 {\eta}) =  \Omega_\circ^2 \mathscr{A}^{[n]} r  \slashed{\mathcal{D}}^\star_2 \etabar -\Omega r^2 \mathscr{A}^{[n]}  \slashed{\mathcal{D}}^\star_2 \Omega \beta  +  \overset{(out)}{\slashed{\mathcal{E}}^{n+1}_{3}} \, .
\end{align}
Contracting with $ r^{2-\delta}\mathscr{A}^{[n]} r^2 \slashed{\mathcal{D}}^\star_2 {\eta}$ for appropriate $n$ and integrating either over $\mathcal{D}^{\I}(u)$, 
or $C_u^{\I}$ yields after applying Cauchy-Schwarz on the right hand side using the improved estimates for $\underline{\eta}$ and $\beta$ the second and third estimate. For the top order estimate (\ref{seba3}) we derive 
\begin{align} \label{etanore}
\Omega \slashed{\nabla}_4 \left(\frac{r^3 \rho + 2M}{r} + r^2 \slashed{div} \eta \right) = +\frac{\Omega_\circ^2}{r^2} \left(r^3 \slashed{div} \underline{\eta} - (r^3 \rho + 2M) - \frac{1}{2} r^3 \hat{\chi} \underline{\hat{\chi}}\right) +  \overset{(out)}{\slashed{\mathcal{E}}^{1}_{3}} \, ,  
\end{align}
and note that the only top order (=first) derivatives that can appear in the non-linear error-term are angular derivatives of $\eta$, $\underline{\eta}$ and $T$ and $\Omega\hat{\chi}$.  For $\underline{\eta}$, $T$ and $\Omega\hat{\chi}$ we already have an outgoing flux bound for $N+1$ derivatives: For $T$ and $\hat{\chi}$ from the bootstrap assumptions and Lemma \ref{lem:shearfluxes} respectively, for $\underline{\eta}$ from Proposition \ref{prop:etabILED} (and Proposition \ref{prop:oneformkeylemma}). For (\ref{etanore}) we also note that quantity in brackets on the left is equal to $\frac{1}{r}(\mu + \frac{1}{2}r^3 \hat{\chi} \underline{\hat{\chi}})$ which on $v=v_\infty$ for $\ell \geq 1$ reduces to $ \frac{1}{2r}r^3( \hat{\chi} \underline{\hat{\chi}})_{\ell \geq 1}$.

Therefore, commuting (\ref{etanore}) with $\mathscr{A}^{[N-2]}r^2 \slashed{\mathcal{D}}_2^\star \slashed{\nabla}$ and integrating along a cone $C_u^{\I}$ from the sphere $S_{u,v_\infty}$ (where clearly $\|\frac{1}{2r} \mathscr{A}^{[N-2]}r^2 \slashed{\mathcal{D}}_2^\star \slashed{\nabla} r^3 \hat{\chi} \underline{\hat{\chi}}\|_{S_{u,v_\infty}} \lesssim \frac{\varepsilon^2}{r \cdot u}$ by the bootstrap assumptions) we obtain using the flux estimate for $\underline{\eta}$ of Proposition \ref{prop:etabILED} and the flux estimate for $\rho$ of Proposition \ref{prop:allrhosigmaimproved} 
\begin{align} \label{slre}
\sup_{C_u^{\I}} \Big\|\mathscr{A}^{[N-2]}r^2 \slashed{\mathcal{D}}_2^\star \slashed{\nabla}\left( r^3 \rho + 2M + r^3 \slashed{div} \eta \right)\Big\|_{S_{u,v}} \lesssim \varepsilon_0 + \varepsilon^{\frac{3}{2}} + \varepsilon \sqrt{ \int_{C_u^{\I}} \frac{1}{r^2} \| \left[r\slashed{\nabla}\right]^{N+1} (\eta - \eta_{\mathrm{Kerr}}) \|^2} \, , 
\end{align}
where the last term arises from the non-linear error-term involving $N+1$ derivatives of $\eta$.\footnote{Note that we can write $\eta=\eta-\eta_{\mathrm{Kerr}} + \eta_{\mathrm{Kerr}}$ in the non-linear error-term and estimate the term involving $\eta_{\mathrm{Kerr}}\sim \frac{\varepsilon}{r^3 u_f}$ (see Section \ref{reflinearisedKerrsec}) separately exploiting that we can estimate $N+2$ derivatives of the $Y^{\ell=1}_m$ from Proposition \ref{prop:modesdifference}.}

We also have directly from (\ref{curle}) and the bootstrap assumptions
\begin{align}
\sup_{C_u^{\I}}  \Big\| \mathscr{A}^{[N-2]}r^2 \slashed{\mathcal{D}}_2^\star {}^\star \slashed{\nabla}\left( r^3 \sigma - r^3 \slashed{curl} \eta \right)\Big\|_{S_{u,v}} \lesssim \varepsilon_0 + \varepsilon^{\frac{3}{2}} \, .
 \end{align}
Using the top order fluxes on $\rho$ and $\sigma$ from Proposition \ref{prop:allrhosigmaimproved} we deduce
\begin{align} 
\int_{C_u^{\I}} \frac{1}{r^2} \|  \mathscr{A}^{[N-2]}r^3 \slashed{\mathcal{D}}_2^\star \slashed{\mathcal{D}}_1^\star \slashed{\mathcal{D}}_1\eta \|^2 \lesssim \varepsilon_0^2 + \varepsilon^3 + \varepsilon^2  \int_{C_u^{\I}} \frac{1}{r^2} \| \left[r\slashed{\nabla}\right]^{N+1} (\eta - \eta_{\mathrm{Kerr}}) \|^2 \, .
\end{align}
Using Proposition \ref{prop:oneformkeylemma} on the term on the left (and the estimates on the $\ell=1$ modes of $\eta$ from Propositions~\ref{prop:l1modesspecial} and~\ref{prop:l1modes}) we can absorb the last term on the right and deduce the desired (\ref{seba3}).

Finally, for the top order integrated decay estimate we commute (\ref{etanore}) with $\mathscr{A}^{[N-2]}r^2 \slashed{\mathcal{D}}_2^\star \slashed{\nabla}$ contract with $r^{2-\delta} \mathscr{A}^{[N-2]}r^2 \slashed{\mathcal{D}}_2^\star \slashed{\nabla}\left( r^3 \rho + 2M + r^3 \slashed{div} \eta\right)$ and integrate over $\mathcal{D}^{\I}(u_{-1})$.
\end{proof}

\subsubsection{Estimating $\Omega^2-\Omega_\circ^2$} 
Using the estimates for the $\ell=0$ and $\ell=1$ modes from Proposition  \ref{prop:l0modes} and \ref{prop:l1modes} respectively, we conclude using Proposition \ref{prop:functionkeylemma} together with the relation $\Omega^2(\eta + \etabar) = \nablaslash (\Omega^2-\Omega_\circ^2)$ from (\ref{eq:DlogOmega}) and the estimates on $\underline{\eta}$ and $\eta$ obtained in the previous sections:
\begin{center}
$\boxed{\textrm{The estimate (\ref{target3}) has been shown for all terms containing $\Omega^2-\Omega_\circ^2$.}}$
\end{center}

\subsubsection{Estimating $\omega$}

\begin{proposition} \label{prop:omauxspheres}
We have on spheres for any $u_{-1} \leq u \leq u_f$ and $s=0,1,2$
\begin{align} \label{omfstr}
\sum_{i=0}^{N-s} \|   \left[ r \slashed{\nabla}\right]^{i}  r^{3-\frac{s}{2}} ({\omega}-\omega_\circ) \|_{S_{u,v}}  \lesssim \frac{ \varepsilon_0 +  \varepsilon^{\frac{3}{2}}}{u^{\frac{s}{2}}} \, .
\end{align}
We also have the integrated decay estimates
\begin{align} 
\sum_{i=0}^{N} \int_{u}^{u_f} d\bar{u} \int_{v(\bar{u},R_{-2})}^{v_{\infty}}  d\bar{v} \  r^{-1-\delta} \|  \left[r \slashed{\nabla}\right]^{i} r^{5/2} ({\omega}-\omega_\circ) \|_{S_{\bar{u},\bar{v}}}^2 
\lesssim \varepsilon_0^2 + \varepsilon^3  \, , \\
\sum_{i=0}^{N-1} \int_{u}^{u_f} d\bar{u} \int_{v(\bar{u},R_{-2})}^{v_{\infty}}  d\bar{v} \  r^{-1-\delta} \|  \left[r \slashed{\nabla}\right]^{i} r^{2} ({\omega}-\omega_\circ) \|_{S_{\bar{u},\bar{v}}}^2 
\lesssim \frac{\varepsilon_0^2 + \varepsilon^3}{u} \, 
\end{align}
and the flux estimate
\begin{align} 
\sum_{i=0}^{N}  \sup_{\bar{u} \in \left[u,u_f\right]} \int_{v(\bar{u},R_{-2})}^{v_{\infty}} d\bar{v} \ r^{-\delta} \|  \left[r \slashed{\nabla}\right]^{i} (\omega-\omega_\circ) r^{2} \|_{S_{u,\bar{v}}}^2 \lesssim \varepsilon_0^2 + \varepsilon^3 \, .
\end{align}

\end{proposition}

\begin{proof}
For the proof observe that for a function $f$ and $N+1 \geq n \geq 2$ we have the elliptic estimate  
\begin{align} \label{easyelliptic}
\sum_{i=0}^n \| \left[r\slashed{\nabla}\right]^i f\|^2_{S_{u,v}} \lesssim \sum_{i=0}^{n-2} \| \left[r\slashed{\nabla}\right]^i r^2 \slashed{\Delta} f\|^2_{S_{u,v}} + \|f_{\ell=0}\|^2_{S_{u,v}} \, .
\end{align}
We will first prove that for any $u_{-1} \leq u \leq u_f$ and $s=0,1$ 
\begin{align} \label{idaho}
\sum_{i=0}^{N-2-s}\|  \left[ r \slashed{\nabla}\right]^{i} r^{3-s}  (r^{2}\slashed{\Delta} {\omega} +r^2 \slashed{div} (\beta\Omega))\|_{S_{u,v}}  \lesssim \frac{ \varepsilon_0 +  \varepsilon^3}{u^{s}} \, .
\end{align}
From this, the desired estimates on spheres follow using that the bootstrap estimates on $\beta$ have already been improved in Section \ref{sec:betainD} and that the $\ell=0$ mode of $\omega-\omega_{\ell=0}$ is controlled from Proposition \ref{prop:l0modes}. To prove (\ref{idaho}), we first derive for $h_0$ an admissible coefficient function (different in different places)
\begin{align} \label{omcomu}
\Omega \slashed{\nabla}_3 \left(r^2 \slashed{\Delta} \omega + r^2 \slashed{div} ( \beta \Omega) \right) =& h_0 r \slashed{div} \beta \Omega +h_0 \frac{1}{r} \slashed{div} \eta - h_0 \frac{1}{r} \slashed{div} \underline{\eta}+ \overset{(in)}{\slashed{\mathcal{E}}^{2}_{4}} + \left[\Omega \slashed{\nabla}_3, r^2 \slashed{\Delta}\right] \left(\omega - \omega_\circ\right) \, .
\end{align}
Commute (\ref{omcomu}) with $\left[r \slashed{\nabla}\right]^{i}$. Integrating the resulting transport equation it is easy to see that we have for all $i \leq N-2$ the estimate\footnote{We could in fact estimate $N+1$ derivatives.}
\begin{align} \label{idaho1}
\|   \left[ r \slashed{\nabla}\right]^{i}   (r^{2}\slashed{\Delta} {\omega} +r^2 \slashed{div} (\beta\Omega))\|_{S_{u,v}}  \lesssim \frac{\varepsilon_0 + \varepsilon^{\frac{3}{2}}}{r^3} \, ,
\end{align}
which already proves the desired estimate for $s=0$ and also for $s=1$ in the subregion $\mathcal{D} \cap \{r \geq \frac{1}{2}u\}$. For $s=1$ and the region $\mathcal{D} \cap \{r \leq \frac{1}{2}u\}$ we integrate from $r=\frac{1}{2}u$ where the estimate already holds to prove
\begin{align} \label{idaho2}
\|   \left[ r \slashed{\nabla}\right]^{i}   (r^{2}\slashed{\Delta} {\omega} +r^2 \slashed{div} (\beta\Omega))\|_{S_{u,v}}  \lesssim \frac{\varepsilon_0 + \varepsilon^{\frac{3}{2}}}{r^2 u} + \frac{\varepsilon_0 + \varepsilon^{\frac{3}{2}}}{r^3 u}  \, 
\end{align}
for all $i \leq N-3$. Here we have used the (improved in Section \ref{sec:betainD}) flux estimates for the $\beta$ term and the (already improved) estimates on spheres for the $\eta$ and $\underline{\eta}$ term as well as estimating the non-linear error on spheres by $\frac{\epsilon^2}{r^4 u}$.

The second integrated decay and the flux estimate now follows directly from integrating (\ref{idaho}) and converting back to $\omega$ using that the bootstrap assumptions on $\beta$ have already been improved. 
For the first integrated decay estimate we return to (\ref{omcomu}), commute with $\left[r \slashed{\nabla}\right]^{N-2}$ and contract with $r^{5-\delta} \left[r \slashed{\nabla}\right]^{N-2} (r^{2}\slashed{\Delta} {\omega} +r^2 \slashed{div} (\beta\Omega))$. Integration over $\mathcal{D}^{\I}$ will then produce the desired integrated decay estimate (and the boundary term) for $(r^{2}\slashed{\Delta} {\omega} +r^2 \slashed{div} (\beta\Omega))$. This is easily converted to $\omega-\omega_\circ$ using the improved estimates on $\beta$ and the estimates on the $\ell=0$ modes of $\omega-\omega_\circ$ from Proposition \ref{prop:l1modes}.
\end{proof}

\begin{center}
$\boxed{\textrm{The estimate (\ref{target3}) has been shown for all terms containing $\omega-\omega_\circ$.}}$
\end{center}

\subsubsection{Estimating \underline{$\omega$}}
For $\underline{\omega}$ we need to prove an integrated decay estimate for $N+1$ derivatives, the reason being its appearance in the anomalous error-term of the top order commuted Teukolsky equation for $\underline{\alpha}$.

\begin{proposition} \label{prop:ombspheres}
For $u_{-1} \leq u \leq u_f$ we have for $s=0,1,2$
\begin{align} \label{ombspheres}
\sum_{i=0}^{N-s} \|  \left[ r \slashed{\nabla}\right]^{i}  (r \Omega^{-2} ( \underline{\omega} - \underline{\omega}_\circ)) \|_{S_{u,v}} \lesssim \frac{\varepsilon_0 +  \varepsilon^{\frac{3}{2}}}{u^{\frac{s}{2}}} \, .
\end{align}
Also, for $u_{-1} \leq u \leq u_f$ we have for $s=0,1,2$ the integrated decay estimates
\begin{align}  \label{iledomb}
\sum_{i=0}^{N+1-s} \int_{u}^{u_f} d\bar{u} \int_{v(\bar{u},R_{-2})}^{v_{\infty}} d\bar{v} \ r^{-1-\delta}  \|  (r \Omega^{-2} ( \underline{\omega} - \underline{\omega}_\circ))  \|_{S_{\bar{u},\bar{v}}}^2 
\lesssim \frac{\varepsilon_0^2 + \varepsilon^3}{u^{\min(1,S-s)}}   \, .
\end{align}
\end{proposition}

\begin{proof}
It clearly suffices to prove the estimates without the factor of $\Omega^{-2}$. 
For the proof, recall (\ref{easyelliptic}). Recalling Lemma \ref{lem:commutationprinciple} we derive the propagation equation
\begin{align} \label{ombcomu}
\Omega \slashed{\nabla}_4 \left(r^2 \slashed{\Delta} (\underline{\omega}-\underline{\omega}_\circ) - \Omega^2 r \slashed{div} (r \underline{\beta} \Omega^{-1}) \right) = & \Omega_\circ^4 r \slashed{div} ( \underline{\beta} \Omega^{-1}) -\frac{4M}{r} \Omega_\circ^2 \slashed{div} \underline{\eta} + \frac{2M}{r} \Omega_\circ^2 \slashed{div} \eta  + \overset{(in)}{\slashed{\mathcal{E}}^{2}_{3}}  + \slashed{\mathcal{E}}^1_1 \, .
\end{align}
We observe in particular that $\overset{(in)}{\slashed{\mathcal{E}}^{2}_{3}}$ in (\ref{ombcomu}) does not contain second (angular) derivatives of $\underline{\hat{\chi}}$, $\underline{T}$ or $\omega-\omega_\circ$. 
To prove (\ref{ombspheres}), commute (\ref{ombcomu}) with $\left[r \slashed{\nabla}\right]^{i}$ for $i=0,1,...,N-2$. 
We then integrate from the sphere $S_{u,v_\infty}$ along the outgoing cone $C_u^{\I}$ and use: 
\begin{itemize}
\item The fact that (\ref{target1}) has already been established to control the boundary term on the sphere $S_{u,v_\infty}$.
\item Cauchy--Schwarz and Proposition \ref{prop:errorspacetimeIplus} for the non-linear error on the right hand side
\item The already improved estimates for $\slashed{div} \underline{\beta}\Omega^{-1}$ (Section \ref{sec:betabimprove}, note $\slashed{div} (\underline{\beta}\Omega^{-1})_{\mathrm{Kerr}}=0$),  $\slashed{div} \underline{\eta}$ (Section \ref{sec:etabaux}) and  $\slashed{div} {\eta}$ (Section \ref{sec:etaaux}) \emph{on spheres} on the right hand side.
\end{itemize}
Finally, use once more the improved estimates on $\slashed{div} \underline{\beta}$ on spheres (Section \ref{sec:betabimprove}) to convert the resulting estimate into the desired estimates for $\underline{\omega}$ itself. 

For (\ref{iledomb}) we commute (\ref{ombcomu}) with $\left[r \slashed{\nabla}\right]^{i}$ for $i=0,1,...N-1$, contract the resulting equation with $r^{2-\delta}\left[r \slashed{\nabla}\right]^{i} \left(r^2 \slashed{\Delta} \underline{\omega} - \Omega^2 r \slashed{div} (r \underline{\beta} \Omega^{-1}) \right) $ and integrate over $\mathcal{D}^{\I}(u)$. Use 
Corollary \ref{cor:betabinf} and the estimate (\ref{ombfluximp}) for the flux term appearing on the hypersurface $\underline{C}_{v_\infty}^{\I}$. Use Cauchy--Schwarz on the right hand side, the already improved integrated decay estimates for $\slashed{div} \underline{\beta}\Omega^{-1}$,  $\slashed{div} \underline{\eta}$ and  $\slashed{div} {\eta}$. For the non-linear error 
use the fact that $\overset{(in)}{\slashed{\mathcal{E}}^{N+1}_{3}}$ is equal to ${\slashed{\mathcal{E}}^{N}_{3}}$ (controlled by Proposition \ref{prop:errorspacetimeIplus}) except for terms involving $N+1$ angular derivatives of connections coefficients for which we do have an integrated decay estimate available. 
Finally, use once more the integrated decay estimate on $\slashed{div} \underline{\beta}\Omega^{-1}$ to convert the resulting estimate into an estimate for $\underline{\omega}$ itself. 
\end{proof}
We conclude
\begin{center}
$\boxed{\textrm{The estimate (\ref{target3}) has been shown for all terms containing $\underline{\omega}-\underline{\omega}_\circ$.}}$
\end{center}

\subsubsection{Estimating $T$ } \label{sec:Test}
We recall now (\ref{Zdef}) written in the form
\begin{align} \label{sreo}
 \frac{1}{2}r^2 \slashed{\mathcal{D}}_2^\star \slashed{\nabla} (\Omega^{-2} r^2 T) = Z +\frac{3M}{r} r^2 \hat{\chi} \Omega^{-1} + \Omega r^3 \slashed{\mathcal{D}}_2^\star\underline{\eta} \, .
\end{align}

\begin{proposition} \label{prop:TfluxNiled}
We have for any $u_{-1}\leq u \leq u_f$ and $s=0,1$ the estimates
\begin{align} 
\int_{u}^{u_f} d\bar{u} \int_{v(\bar{u},R_{-2})}^{v_{\infty}} d\bar{v} \  r^{-1-\delta}  \|  \mathscr{A}^{[N-1-s]} r^2 \slashed{\mathcal{D}}_2^\star \slashed{\nabla} (r^2 {T}) \|_{S_{u,\bar{v}}}^2 
\lesssim \frac{\varepsilon_0^2 + \varepsilon^3}{u^s}   \, ,
\end{align}
On spheres we have for $s=0,1,2$
\begin{align} \label{finitos}
\|  \mathscr{A}^{[N-1-s]} r^2 \slashed{\mathcal{D}}_2^\star \slashed{\nabla}  r^2 {T} \|_{S_{u,v}} \lesssim \frac{\varepsilon_0 +  \varepsilon^{\frac{3}{2}}}{u^{\frac{s}{2}}} \, 
\end{align}
and
\begin{align} \label{Tglobale}
\sum_{i=0}^N \| \left[r \slashed{\nabla}\right]^i (r^3 T) \|_{S_{u,v}} \lesssim \varepsilon_0 +  \varepsilon^{\frac{3}{2}} \, .
\end{align}
\end{proposition}
\begin{proof}
Commuting (\ref{sreo}) by $\mathscr{A}^{[n]}$ for appropriate $n$ we obtain the first two estimates using the previous estimates of Propositions \ref{prop:etabILED}, \ref{prop:etabILED}, \ref{prop:chistart} and the estimates  (\ref{Z1a}) and (\ref{Z1b}) on $Z$. For the estimate (\ref{Tglobale}) we integrate the $[r\slashed{\nabla}]^i$--commuted (\ref{T4dir})  backwards from $v=v_\infty$ and use (\ref{Timp2}) for the term on $v=v_\infty$ and the estimate (\ref{omfstr}) on the right hand side of  (\ref{T4dir}).
\end{proof}

We conclude after applying Proposition \ref{prop:functionkeylemma} with the estimates on the $\ell=0,1$ modes from Propositions \ref{prop:l0modes}, \ref{prop:l1modes}:
\begin{center}
$\boxed{\textrm{The estimate (\ref{target3}) has been shown for all terms containing $T$.}}$
\end{center}

\subsubsection{Estimating \underline{$T$}}
We recall (\ref{Ydef}), now written in the form
\begin{align}  \label{Yexpagain}
  \frac{1}{2} r^2 \slashed{\mathcal{D}}_2^\star  \slashed{\nabla} r^2\underline{T} =Y + 3M r \underline{\hat{\chi}}\Omega^{-1} + r^3 \slashed{\mathcal{D}}_2^\star \underline{\eta}.
\end{align}

Using the estimates for the quantities on the right hand side (see Section \ref{sec:chibare} as well as Propositions \ref{prop:etabILED}) we easily deduce the following two propositions.

\begin{proposition} \label{prop:TbarfluxNiled}
We have for any $u_{-1}\leq u \leq u_f$ and $s=0,1$ the estimates
\begin{align} 
\int_{u}^{u_f} d\bar{u} \int_{v(\bar{u},R_{-2})}^{v_{\infty}} d\bar{v} \  r^{-1-\delta}  \| \mathscr{A}^{[N-2-s]} r^2 \slashed{\mathcal{D}}_2^\star \slashed{\nabla} (r^2 {\underline{T}}) \|_{S_{u,\bar{v}}}^2 
\lesssim \frac{\varepsilon_0^2 + \varepsilon^3}{u^s}   \, .
\end{align}
On spheres we have
\begin{align}
\| \mathscr{A}^{[N-1-s]} r^2 \slashed{\mathcal{D}}_2^\star \slashed{\nabla}  r^2 \underline{T} \|_{S_{u,v}} \lesssim \frac{\varepsilon_0 +  \varepsilon^{\frac{3}{2}}}{u^{\frac{s}{2}}} \, .
\end{align}
\end{proposition}
\begin{proof}
For the estimate on spheres commute (\ref{Yexpagain}) by $\mathscr{A}^{[N-1-s]}$ and use previous estimates. For the integrated decay estimate commute by  $\mathscr{A}^{[N-1-s]}$ and use previous integrated decay estimates.
\end{proof}
We conclude after applying Proposition \ref{prop:functionkeylemma} with the estimates on the $\ell=0,1$ modes from Propositions \ref{prop:l0modes}, \ref{prop:l1modes}:
\begin{center}
$\boxed{\textrm{The estimate (\ref{target3}) has been shown for all terms containing $\underline{T}$.}}$
\end{center}

\subsubsection{Estimating $b$}
\begin{proposition} \label{prop:b}
We have for any $u_{-1} \leq u \leq u_f$ and $s=0,1,2$ 
\begin{align} 
\|\left[ r \slashed{\nabla}\right]^{N+1} (b-b_{\mathrm{Kerr}}) \sqrt{r}\|_{S_{u,v}}  \lesssim \varepsilon_0 +  \varepsilon^{\frac{3}{2}},
\end{align}
\begin{align}
\sum_{i=0}^{N-s} \|  \left[ r \slashed{\nabla}\right]^{i} (b-b_{\mathrm{Kerr}}) r\|_{S_{u,v}} \lesssim \frac{ \varepsilon_0 +  \varepsilon^{\frac{3}{2}}}{u^{\min(\frac{1}{4}+\frac{s}{2},1)}}.
\end{align}
In addition, we have for $s=0,1$ the integrated decay estimate
\begin{align}  \label{biled}
\sum_{i=0}^{N-s} \int_{u}^{u_f} d\bar{u}\int_{v(\bar{u},R_{-2})}^{v_{\infty}}  d\bar{v} \  r^{-1-\delta}  \|  \left[ r \slashed{\nabla}\right]^{i} (b-b_{\mathrm{Kerr}}) r\|_{S_{\bar{u},\bar{v}}}^2 
\lesssim \frac{\varepsilon_0^2 + \varepsilon^3}{u^s} \, ,
\end{align}
\end{proposition}

\begin{proof}
We have the transport equation
\begin{align} \label{transportb}
\Omega \slashed{\nabla}_4 \left(\frac{1}{r} (b-b_{\mathrm{Kerr}})\right) = 2 \frac{\Omega_\circ^2}{r}  \left((\eta - \eta_{\mathrm{Kerr}}) - (\underline{\eta}-\underline{\eta}_{\mathrm{Kerr}})\right)+\overset{(out)}{\slashed{\mathcal{E}}^0_4 }\, ,
\end{align}
where one easily sees that the only connection coefficients appearing in $\overset{(out)}{\slashed{\mathcal{E}}^0_4 }$ are $\eta, \underline{\eta}, \hat{\chi}, T, b$.

To obtain the second estimate we first commute (\ref{transportb}) with $\left[r \slashed{\nabla}\right]^{i}$ for $i=0,\ldots,N-s$ and the integrate the resulting equation along the cone $C_u^{\I}$. For the linear terms on the right hand side we can use the estimates for $\eta$ and $\underline{\eta}$ implicit in (\ref{target3}) which have already been improved at the end of Sections \ref{sec:etabaux} and \ref{sec:etaaux} respectively. 

To obtain the first estimate we commute with $\left[r \slashed{\nabla}\right]^{N+1}$ and integrate using again the estimates for $\eta$ and $\underline{\eta}$ implicit in (\ref{target3}) which have already been improved at the end of Sections \ref{sec:etabaux} and \ref{sec:etaaux} respectively. For the non-linear term we now note that for $N+1$ angular derivatives of $\eta, \underline{\eta}, \hat{\chi}, T, b$ we have an outgoing flux estimate or an estimate on spheres available\footnote{For $b$ and $T$ by the bootstrap assumptions on spheres, for $\eta$ by (\ref{seba3}), for $\underline{\eta}$ by (\ref{etabtopflux}) and for $\hat{\chi}$ by (\ref{chiest4})).}, which allow one to show using Lemma \ref{lem:basicerrorse}
\[
\int_v^{v_\infty} d\bar{v} \int_{S_{u,\bar{v}}} \| \left[r\slashed{\nabla}\right]^{N+1} \overset{(out)}{\slashed{\mathcal{E}}^0_4}\|_{S_{u,\bar{v}}} \lesssim \frac{\varepsilon^2}{r^2 u} \ \ \ \textrm{for the particular $\overset{(out)}{\slashed{\mathcal{E}}^0_4 }$ appearing in (\ref{transportb}).}
\] 
Since the error arising from the (iteration of) commutators on the left can contain at most $N+1$ angular derivatives of $(b-b_{\mathrm{Kerr}})$, the first estimate follows. Note the loss in $r$-weight for the top-order estimate arising from having to use (weaker) outgoing flux estimates for $\eta, \underline{\eta}$ instead of estimates on spheres, which are not available at top order.

To obtain (\ref{biled}) one commutes (\ref{transportb}) with $\left[r \slashed{\nabla}\right]^{i}$ for $i=0,1,\ldots,N-s$ 
an contracts the resulting equation with $r^{3-\delta} \frac{1}{r}\left[r \slashed{\nabla}\right]^{i} (b-b_{\mathrm{Kerr}})$ and integrates over $\mathcal{D}^{\I}(u)$.  After standard integration by parts the boundary term on $v=v_\infty$ is always controlled from the gauge condition (\ref{gaugebvector})
and the fact that $\| \left[r \slashed{\nabla}\right]^{i} b_{\mathrm{Kerr}}\|_{S_{u,v_\infty}} \lesssim \frac{a}{r^2} \lesssim \frac{\varepsilon^3}{r u^4}$ in view of (\ref{rtou}). Finally, on the right hand side, we apply Cauchy--Schwarz and invoke the (improved) integrated decay estimates on $\eta$ and $\underline{\eta}$ implicit in (\ref{target3}), which have already been established at the end of Sections \ref{sec:etabaux} and \ref{sec:etaaux} respectively. The non-linear error-term is controlled using the integrated decay estimates of Proposition \ref{prop:errorspacetimeIplus}.
\end{proof}

This finishes the proof of Theorem \ref{theo:mtnig2}.

\subsection{Estimates for the angular master energy for the metric components} \label{sec:bametric}
The objective of this section is to prove (\ref{target4}).

From the Gauss equation (\ref{Gauss}) we first deduce in view of the previous estimates (more precisely the estimate (\ref{target1}) from Theorem \ref{theo:mtnig2}) the following estimate on the Gauss curvature:
\begin{align}
\| \left[r \slashed{\nabla}\right]^{N-1-s} r (r^2 K-1) \|_{S_{u,v_\infty}} \lesssim \frac{\varepsilon_0+\varepsilon^{\frac{3}{2}}}{u^{\min(1,\frac{s}{2}+\frac{1}{4})}} \ \ \ \textrm{for $s=0,1,\ldots,N-1$.}
\end{align}

\subsubsection{Proof of (\ref{target4}) on the final sphere $S_{u_f,v_\infty}$} \label{subsec:finalspheremetricestimate}
Restricting now to the final sphere $S_{u_f,v_\infty}$ we recall the estimate (\ref{theestimatesforourspherediff}) which implies the following estimate on the metric components in their respective charts:
\begin{align} \label{eq:finalspheremetricestimate}
\|   \left[ r \slashed{\nabla}\right]^{N+1-s}\left(r (\slashed{g} - r^2\mathring\gamma) \right) \|_{S_{u_f,v_\infty}} \lesssim \frac{\varepsilon_0+\varepsilon^{\frac{3}{2}}}{(u_f)^{\min(1,\frac{s}{2}+\frac{1}{4})}} \ \ \ \textrm{for $s=0,1,\ldots ,N+1$.} 
\end{align}
This establishes (\ref{target4}) on the final sphere.

\subsubsection{Proof of (\ref{target4}) on the final ingoing cone}
To estimate the metric components along the cone $\underline{C}^{\I}_{v_\infty}$ we first derive the following renormalised (tensorial) transport equation.\footnote{The general equation has an additional term $ \slashed{\nabla}_A b_B + \slashed{\nabla}_B b_A$ on the right hand side which vanishes by the gauge condition (\ref{gaugebvector}) on the cone.}
\begin{align}
\Omega \slashed{\nabla}_3 \left(\slashed{g} - r^2 \mathring{\gamma}\right) = & \ r^2 \mathring\gamma \left(\Omega tr \underline{\chi} - (\Omega tr \underline{\chi})_\circ\right) + 2\Omega \underline{\hat{\chi}} 
- \left( \Omega \underline{\hat{\chi}} \times  \left(\slashed{g} - r^2 \mathring{\gamma}\right)\right) - \left(   \left(\slashed{g} - r^2 {\mathring\gamma}\right) \times \Omega  \underline{\hat{\chi}} \right) \, .
\end{align}
We could integrate this directly but this would lead to a loss in the estimates because $\underline{\hat{\chi}}$ decays only like $r^{-1} u^{-1}$. Therefore we first deduce an equation for the trace:
\begin{align} \label{traceme}
\Omega \slashed{\nabla}_3 \left( tr_{\slashed{g}} \left(\slashed{g} - r^2 {\mathring\gamma}\right) \right) = 2 \underline{\hat{T}} - \left(\textrm{$\underline{\hat{T}}$ or $\Omega \underline{\hat{\chi}}$}\right)\left(\slashed{g} - r^2 {\mathring\gamma}\right) \, ,
\end{align}
where the last (non-linear) term denotes a finite sum of arbitrary contractions between the tensors involved. Note that the right hand side of the equation for the trace decays like $r^{-2} u^{-1}$. 
Integrating (\ref{traceme}) using Lemma \ref{intlemskri} we therefore deduce\footnote{Recall again $1\leq \frac{r(u,v_\infty)}{r(u_f,v)} \lesssim \frac{v_\infty-u}{v_\infty - u_f} \lesssim 1$ in view of our definition of $v_\infty$.} using the (now improved) estimates on $\underline{\hat{T}}$ the estimate
\begin{align}
\|    \left[ r \slashed{\nabla}\right]^{N+1-s}\left(r \cdot tr_{\slashed{g}} (\slashed{g} - r^2\mathring\gamma) \right) \|_{S_{u,v_\infty}} \lesssim \frac{\varepsilon_0+\varepsilon^{\frac{3}{2}}}{u^{\min(1,\frac{s}{2}+\frac{1}{4})}} \ \ \ \textrm{for $s=0,1,...,N+1$.} 
\end{align}
Using this estimate on the trace we can revisit the relation (\ref{gaussexpand}) and using also the estimate (\ref{nlgauss}) deduce
\begin{align} \label{traceskri1}
\|    \left[ r \slashed{\nabla}\right]^{N-1-s}\left(r \cdot r^2 \slashed{div} \slashed{div} (\widehat{\slashed{g} - r^2\mathring\gamma}) \right) \|_{S_{u,v_\infty}} \lesssim \frac{\varepsilon_0+\varepsilon^{\frac{3}{2}}}{u^{\min(1,\frac{s}{2}+\frac{1}{4})}} \ \ \ \textrm{for $s=0,1,...,N-1$.} 
\end{align}
To estimate the $\slashed{curl} \slashed{div} (\widehat{\slashed{g} - r^2\mathring\gamma})$-part we note 
the propagation equation equation for the trace-free part,
\begin{align} \label{tracefreeskri2a}
\Omega \slashed{\nabla}_3 \left( \widehat{\slashed{g} - r^2 \mathring{\gamma}} \right) = 2 \Omega \underline{\hat{\chi}} - \left(\textrm{$\underline{\hat{T}}$ or $\Omega \underline{\hat{\chi}}$}\right)\left(\slashed{g} - r^2 \mathring{\gamma}\right) \, .
\end{align}
From this we can derive an improved equation for the renormalised quantity
\begin{align} 
\Omega \slashed{\nabla}_3 \left( r^2 \slashed{curl}\slashed{div} \left( \widehat{\slashed{g} - r^2 \mathring{\gamma}}\right) + 2r^2 \sigma \right) = &+\frac{\Omega_\circ^2}{r^2}\sigma r^3 + \frac{\Omega_\circ^2}{r^2} r \slashed{curl} (\eta r^2)  - \frac{1}{r} \Omega \hat{\chi} \wedge \underline{\alpha} + \mathcal{E}^1_2 \nonumber \\
& -  r^2 \slashed{curl}\slashed{div} \left( \left(\textrm{$\underline{\hat{T}}$ or $\Omega \underline{\hat{\chi}}$}\right)\left(\slashed{g} - r^2 {\mathring\gamma}\right) \right) \, .
\end{align}
Integrating and using the estimates (\ref{target1}) from Theorem \ref{theo:mtnig2} for the right hand side we deduce
\begin{align} \label{tracefreeskri2b}
\|    \left[ r \slashed{\nabla}\right]^{N-1-s}\left(r \cdot r^2 \slashed{curl} \slashed{div} (\widehat{\slashed{g} - r^2\mathring\gamma}) \right) \|_{S_{u,v_\infty}} \lesssim \frac{\varepsilon_0+\varepsilon^{\frac{3}{2}}}{u^{\min(1,\frac{s}{2}+\frac{1}{4})}} \ \ \ \textrm{for $s=0,1,...,N-1$.} 
\end{align}
Combining (\ref{traceskri1}), (\ref{tracefreeskri2a}) and (\ref{tracefreeskri2b}) we have established 
\begin{align} \label{summarygv}
\|   \left[ r \slashed{\nabla}\right]^{N+1-s}\left(r (\slashed{g} - r^2\mathring\gamma) \right) \|_{S_{u,v_\infty}} \lesssim \frac{\varepsilon_0+\varepsilon^{\frac{3}{2}}}{u^{\min(1,\frac{s}{2}+\frac{1}{4})}} \ \ \ \textrm{for $s=0,1,...,N+1$} 
\end{align}
and hence (a stronger version of) the estimate (\ref{target4}) along the cone $\underline{C}^{\I}_{v_\infty}$.

\subsubsection{Proof of (\ref{target4}) in all of $\mathcal{D}^{\I}$}
Note first the propagation equation along each outgoing cone $C_u^{\I}$ with $u_{-1} \leq u \leq u_f$,  
\begin{align} \label{nabla4g}
\Omega \slashed{\nabla}_4 \left(\slashed{g} - r^2 \mathring{\gamma}\right) = & \ r^2 \mathring{\gamma} \left(\Omega tr {\chi} - (\Omega tr {\chi})_\circ\right) + 2\Omega {\hat{\chi}} 
- \left( \Omega {\hat{\chi}} \times  \left(\slashed{g} - r^2 \mathring{\gamma}\right)\right) - \left(   \left(\slashed{g} - r^2 \mathring{\gamma}\right) \times \Omega  {\hat{\chi}} \right) \, .
\end{align}
We commute with up to $N+1$ angular derivatives
and integrate backwards from $S_{u,v_\infty}$.
Using (\ref{summarygv}) for the initial data term and the estimate (\ref{finitos}) for $T =\Omega tr {\chi} - (\Omega tr {\chi})_\circ$ and the estimate (\ref{l1chi}) on $\hat{\chi}$  we finally deduce (\ref{target4}) as desired.

\subsection{Estimates for the auxiliary energies for $\omega$ and \underline{$\omega$}} \label{sec:proofof3}

The objective of this section is to prove Theorem \ref{theo:mtnig3}. In this regard, we prove (\ref{omap}) in Section \ref{sec:omball} and (\ref{omap2}) in Section \ref{sec:omall}.

\subsubsection{Estimates on all derivatives of $\underline{\omega}$} \label{sec:omball}

We first estimate $\left[r \slashed{\nabla}\right]^i \left[\slashed{\nabla}_3 \right]^j   \left(r \underline{\omega} -r\underline{\omega}_\circ\right)$ derivatives on $\underline{C}_{v_\infty}^{\I}$.

\begin{proposition} \label{prop:fluxallominf}
On $\underline{C}_{v_\infty}^{\I}$ we have for all $u \in \left[u_{-1},u_f\right]$ and $s=0,1,2$
\begin{align} 
&\sum_{i+j =0; i \geq 1}^{N+1-s} \int^{u_f}_u d\bar{u}   \| \left[r \slashed{\nabla}\right]^{i} \left[ \Omega \slashed{\nabla}_3 \right]^j   \left(r \Omega^{-2} ( \underline{\omega} -r\underline{\omega}_\circ)\right) \|_{S_{\bar{u},v_\infty}}^2 \nonumber \\ 
&\qquad+\sum_{i+j =0}^{N-s} \int^{u_f}_u d\bar{u}   \|\left[r \slashed{\nabla}\right]^{i} \left[ \Omega \slashed{\nabla}_3 \right]^j    \left(r \Omega^{-2} ( \underline{\omega} -r\underline{\omega}_\circ)\right) \|_{S_{\bar{u},v_\infty}}^2 \lesssim \frac{(\varepsilon_0)^2+\varepsilon^3}{u^{s}} \, . \label{jr1}
\end{align}
On spheres we have for $s=0,1,2$
\begin{align} 
&\sum_{i+j =0 ; i \geq 1}^{N-s} \| \left[r \slashed{\nabla}\right]^{i} \left[ \Omega \slashed{\nabla}_3 \right]^j    \left(r \Omega^{-2} ( \underline{\omega} -r\underline{\omega}_\circ)\right) \|_{S_{u,v_\infty}}^2 \nonumber \\
&\qquad+\sum_{i+j =0}^{N-1-s} \|\left[r \slashed{\nabla}\right]^{i} \left[ \Omega \slashed{\nabla}_3 \right]^j    \left(r \Omega^{-2} ( \underline{\omega} -r\underline{\omega}_\circ)\right) \|_{S_{u,v_\infty}}^2 \lesssim  \frac{(\varepsilon_0)^2+ \varepsilon^3}{u^{\min(\frac{1}{2}+s, 2)}} . 
\label{jr2}
\end{align}
\end{proposition}

\begin{remark} \label{rem:imd}
Note that the above proposition already establishes the required estimate in Theorem~\ref{theo:mtnig3} for the term on $v=v_\infty$ appearing in $\mathbb{E}^{N,aux}_{u_f, \mathcal{I}} \left[ \underline{\omega} \right]$ (see (\ref{angbon})) as well as the estimates on spheres restricted to $v=v_\infty$.
\end{remark}

\begin{proof}
Note these estimates have already been shown for $j=0$ in Proposition \ref{prop:ombe}.
Note also that we can equivalently prove the estimates without the $\Omega^{-2}$ factor, which is what we are going to do. 

We will prove the estimates separately for $ \left(r \underline{\omega} -r\underline{\omega}_\circ\right)_{\ell=0}$ and $ \left(r \underline{\omega} -r\underline{\omega}_\circ\right)_{\ell\geq 1}$. For the former, we recall the projected equation (\ref{ombpre}) and deduce in view of our definition of $v_\infty$, using the estimates of Proposition~\ref{prop:com0} for derivatives of the commutator of projections and $\Omega \slashed{\nabla}_3$ the following estimates for the $\ell=0$ mode. For $u \in \left[u_{-1},u_f\right]$, we have
\begin{align}
 \sum_{i=0}^{N-1} \|\left[\Omega \slashed{\nabla}_3\right]^i (r \underline{\omega}- r \underline{\omega}_\circ)_{\ell=0} \|_{S_{u,v_\infty}} +  \int_{u}^{u_f} d\bar{u} \|\left[\Omega \slashed{\nabla}_3\right]^{N} (r \underline{\omega} - r\underline{\omega}_\circ)_{\ell=0} \|_{S_{\bar{u},v_\infty}}  &\lesssim \frac{\varepsilon^2}{u^4} \, .  \label{ombhigh} 
\end{align}
The estimates (\ref{ombhigh}) already imply that the estimates claimed in the proposition hold for $ \left(r \underline{\omega} -r\underline{\omega}_\circ\right)_{\ell=0}$ (instead of the full $ \left(r \underline{\omega} -r\underline{\omega}_\circ\right)$). From this we conclude that it suffices to prove both estimates for the respective \emph{first} sum appearing in (\ref{jr1}) and (\ref{jr2}). This is trivial for $i\geq 1$ and follows for $i=0$ from the Poincar\'e inequality after noting that $\| \left( (\Omega \slashed{\nabla}_3)^k \underline{\omega}_{\ell \geq 1} \right)_{\ell=0} \|_{S_{u,v}}$ is non-linear and estimated by Proposition~\ref{prop:com0}.

We turn to estimating the respective first sums in (\ref{jr1}) and (\ref{jr2}), restricting first further to $i \geq 2$ in these sums.
In this case we commute (\ref{eop}) with $\left[r \slashed{\nabla}\right]^{i} \left[ \Omega \slashed{\nabla}_3 \right]^j$ (with $i+j \leq N-2$ for the top order estimate) and induct on $j$.
More precisely, we can commute the $\Omega \slashed{\nabla}_3$ derivatives with the $\ell \geq 1$ projections (estimating the commutator terms repeatedly by Proposition \ref{prop:com0} and insert the Bianchi and null structure equation in the $3$-direction for all \emph{linear} terms on the right hand side converting $3$-derivatives into angular derivatives which have already been controlled in Theorem \ref{theo:mtnig2}. The only quantities for which there exists no equation converting a $3$-derivative to an angular derivative are $\underline{\omega}-\underline{\omega}_\circ$ and $\underline{\alpha}$. However, the latter has been controlled in Theorem~\ref{thm:alphaalphabarestimates} while the former is easily controlled inductively as it is necessarily lower order in terms of derivatives (and decays stronger in $r$).

We now treat the case $i=1$. We define unique functions $b_1,b_2$ with vanishing mean satisfying
$\underline{\beta} = r \slashed{\nabla} b_1 + r {}^\star \slashed{\nabla} b_2$. We can then write (derive the formula first for $j=0$ where the commutator term in the last line vanishes)
\begin{align}
2r^2 \slashed{\Delta} \left[\Omega \slashed{\nabla}_3 \right]^j \left(\underline{\omega} - b_1 \cdot r \right)_{\ell \geq 1} =  &\left[\Omega \slashed{\nabla}_3\right]^j \Big\{ (\underline{T}r^2)_{\ell \geq 1} \left(\frac{3M}{r^3}\right)   + \frac{1}{2r} r^2 \left( |\underline{\hat{\chi}}|^2\right)_{\ell \geq 1}+ (\overset{(in)}{\slashed{\mathcal{E}}^{1}_{2}})_{\ell \geq 1} \Big\}  \\
+ & \left[\Omega \slashed{\nabla}_3 \right]^j \Big\{ \frac{1}{r}  \Big[ \left(\Omega \slashed{\nabla}_3 (r^3 \mu+2M)\right)_{\ell \geq 1} - \Omega \slashed{\nabla}_3 (r^3 \mu+2M)_{\ell \geq 1}\Big] \Big\} \nonumber \\
+ &\left[\Omega \slashed{\nabla}_3 \right]^j \Big\{  2 r^2  \left[ \slashed{\Delta} (\underline{\omega}-b_1 \cdot r)_{\ell \geq 1} -  \left(\slashed{\Delta} (\underline{\omega}-b_1\cdot r)\right)_{\ell \geq 1}  \right] \Big\} \nonumber \\
+ & 2\left[r^2 \slashed{\Delta}, \left[\Omega \slashed{\nabla}_3\right]^j \right]\left(\underline{\omega} - b_1 \cdot r \right)_{\ell \geq 1} \, .
\end{align}
Denoting the entire right hand side momentarily by $\mathcal{F}^j$ we have
by the standard elliptic estimate 
\begin{align}
\|r \slashed{\nabla} \left[\Omega \slashed{\nabla}_3 \right]^j \underline{\omega}_{\ell \geq 1} \|_{S_{u,v_\infty}} &\lesssim \| r\slashed{\nabla} \left[\Omega \slashed{\nabla}_3 \right]^j (r (b_1)_{\ell \geq 1} \|_{S_{u,v_\infty}}  + \|r \mathcal{F}^j \|_{S_{u,v_\infty}} \nonumber \\
&\lesssim \| \left[\Omega \slashed{\nabla}_3 \right]^j r^2 \underline{\beta} \|_{S_{u.v_\infty}}  + r \| \mathcal{F}^j \|_{S_{u,v_\infty}} + \frac{\varepsilon^2}{u^2} \, . \nonumber
\end{align}
Here we have used that the commutator is lower order in $r$ and the definition of $v_\infty$ to exploit decay in the non-linear term.

To estimate $\mathcal{F}^j$ we can again commute $\Omega \slashed{\nabla}_3$-derivative with the $l \geq 1$ projections on the right hand side (estimating the errors by Propositions \ref{prop:com0} and \ref{prop:divcurlmodes}) and insert the $\Omega \slashed{\nabla}_3$-Bianchi and null structure equations for all terms (to convert $\Omega \slashed{\nabla}_3$-derivatives into angular derivatives and lower oder terms controlled by Theorem \ref{theo:mtnig2}) and treat the lower order linear terms in $\left[\Omega \slashed{\nabla}_3\right]^{\tilde{j}}(\underline{\omega}-\underline{\omega}_\circ)$ that appear (and for which there is no equation to insert) inductively. Note that for $j=0$ no linear terms in $\underline{\omega}-\underline{\omega}_\circ$ are present. We conclude
\[
\| \mathcal{F}^{j-1} \|_{S_{u,v_\infty}} \lesssim \frac{1}{r} \sum_{i=0}^{j-1} \| \left[\Omega \slashed{\nabla}_3 \right]^i \underline{\omega}_{\ell \geq 1} \|_{S_{u,v_\infty}} +  \frac{\varepsilon_0+ \varepsilon^{\frac{3}{2}}}{u \cdot (r(u,v))} \ \ \ \textrm{for $1 \leq j \leq N$} \, , 
\]
\[
\int_u^{u_f} d\bar{u} r^2 \| \mathcal{F}^{j} \|^2_{S_{\bar{u},v_\infty}} \lesssim \int_u^{u_f} d\bar{u} \sum_{i=0}^{j-1} \| \left[\Omega \slashed{\nabla}_3 \right]^i \underline{\omega}_{\ell \geq 1} \|^2_{S_{u,v_\infty}} +  \frac{(\varepsilon_0)^2+ \varepsilon^3}{u^2} \ \ \ \textrm{for $0 \leq j \leq N$} \, .
\]
The first term is always controlled by the previous step of the induction. Combining the estimates, the estimates for the first sum in (\ref{jr1}) and (\ref{jr2}) respectively now follow.
\end{proof}

We now turn to proving
\begin{align} \label{omap}
\mathbb{E}^{N,aux}_{u_f, \mathcal{I}} \left[ \underline{\omega} \right]  \lesssim (\varepsilon_\circ)^2 + \varepsilon^3 \, .
\end{align}
where we recall (\ref{angbon}) and Remark \ref{rem:imd}. We first prove the estimate for the energy $\mathbb{E}^{N,aux}_{u_f, \mathcal{I}} \left[ \underline{\omega} \right]$ defined in (\ref{angbon}) with the sum $\sum_{\substack{k_1+k_3 \leq N-s \\ k_1+k_3 \leq N+1-s \, \wedge \, k_3\geq 2}}$ replaced by $\sum_{\substack{k_1+k_3 \leq N-s}}$ (we refer to this as \emph{the retricted (\ref{omap}) below}).
For this it suffices to recall the propagation equation (\ref{om4dir}) and commute it (using Lemma \ref{lem:commutationprinciple}), repeatedly inserting the Bianchi and null structure equations on the right hand side. This yields the following schematic form
\begin{align} \label{struco}
\Omega \slashed{\nabla}_4 \left( \left[\Omega \slashed{\nabla}_3\right]^{k_3} \left[r \slashed{\nabla}\right]^{k_1} \left(\underline{\omega} - \underline{\omega}_\circ\right)\right) = &\sum_{\Phi_p} \sum_{k \leq k_1+k_3} \frac{1}{r^3} h_0 \left[r\slashed{\nabla}\right]^k (r^p({\Phi_p} -(\Phi_p)_{\mathrm{Kerr}})) \\
&+ \sum_{\substack{j_1 \leq k_1 \\ j_3 < k_3}} \frac{1}{r^3} h_0 \left[\Omega \slashed{\nabla}_3\right]^{j_3}\left[r\slashed{\nabla}\right]^{j_1}  (\underline{\omega}- \underline{\omega}_\circ)
\nonumber \\
+&\sum_{\substack{j_1+j_3 \leq k_1+k_3 \\ j_3 < k_3-1}} \frac{1}{r^3} h_0 \left[r\slashed{\nabla}\right]^{j_1} \left[\Omega \slashed{\nabla}_3\right]^{j_3} (\underline{\alpha} r)+ \left[\Omega \slashed{\nabla}_3 \right]^{k_3} \overset{(in)}{\slashed{\mathcal{E}}}{}^{k_1}_{3} + \left[\Omega \slashed{\nabla}_3 \right]^{k_3}{\slashed{\mathcal{E}}}^{k_1-1}_{3} .\nonumber
\end{align}
A simple induction on $k_3$ (and the usual procedure to derive integrated decay estimates and estimates on spheres) now proves (\ref{omap}) with the aforementioned restriction on the sums appearing on the left. Indeed, we can use that
\begin{itemize}
\item The first term on the right of (\ref{struco}) is controlled by the estimate of Theorem \ref{theo:mtnig2}. Use the flux for the curvature components and estimates on spheres for connection coefficients. 
\item The second term on the right of (\ref{struco}) is controlled inductively.
\item The third term on the right of (\ref{struco}) is controlled by Theorem \ref{thm:alphaalphabarestimates}.
\item The non-linear error term $\left[\Omega \slashed{\nabla}_3 \right]^{k_3} \overset{(in)}{\slashed{\mathcal{E}}}{}^{k_1}_{3} + \left[\Omega \slashed{\nabla}_3 \right]^{k_3}{\slashed{\mathcal{E}}}^{k_1-1}_{3}$ involves only up to $N$ derivatives of the connection coefficients and $N-1$ derivatives of curvature components and is hence controlled by the bootstrap assumptions
on spheres and Lemma \ref{lem:34onerror}. 
\end{itemize}
To obtain (\ref{omap}) also without the restriction on the sum on the left, we need to derive  integrated decay estimates for $N+1-s$ derivatives with two of them being angular. For this we commute (\ref{ombcomu}) with $\left[\Omega \slashed{\nabla}_3\right]^{k_3} \left[r \slashed{\nabla}\right]^{k_1}$ for $k_1+k_3 \leq N-1-s$ and (as before) repeatedly insert null-structure and Bianchi equations on the right hand side to convert the $3$-derivatives into angular ones.\footnote{Note that the order the derivatives are taken in is irrelevant as the commutator is always non-linear with improved decay and regularity.} Then one contracts the resulting equation with $r^{2-\delta} \left[\Omega \slashed{\nabla}_3\right]^{k_3} \left[r \slashed{\nabla}\right]^{k_1}  \left( r^2 \slashed{\Delta} (\underline{\omega}-\underline{\omega}_\circ) - r\Omega^2 \slashed{div} (r \underline{\beta} \Omega^{-1})\right)$, applies Cauchy-Schwarz on the right and invokes the estimates of Theorem \ref{theo:mtnig2} as well as the restricted (\ref{omap}) for the linear terms 
and the estimates of Lemma \ref{lem:34onerror} and Proposition \ref{prop:intdebs} for the non-linear error-terms. In the end, one converts the resulting integrated decay estimate for $\left[\Omega \slashed{\nabla}_3\right]^{k_3} \left[r \slashed{\nabla}\right]^{k_1}  \left( r^2 \slashed{\Delta} (\underline{\omega}-\underline{\omega}_\circ) - r\Omega^2 \slashed{div} (r \underline{\beta} \Omega^{-1})\right)$ where $k_1+k_3 \leq N-1-s$ into one for $\left[\Omega \slashed{\nabla}_3\right]^{k_3} \left[r \slashed{\nabla}\right]^{k_1+2}  (\underline{\omega}-\underline{\omega}_\circ)$ itself (by inserting again the Bianchi equation for $\left[\Omega \slashed{\nabla}_3\right]^{k_3} \left( r \slashed{div} (r \underline{\beta} \Omega^{-1})\right)$ and using the estimates of Theorem \ref{theo:mtnig2} as well as the restricted (\ref{omap}) proven before.

\subsubsection{Estimates on $\omega$} \label{sec:omall}

We begin by estimating $\left[\slashed{\nabla}_4 \right]^j   \left({\omega} -{\omega}_\circ\right)$ derivatives on $C_{u_{-1}}^{\I}$:
\begin{proposition} \label{prop:fluxallomu}
On $C_{u_{-1}}^{\I}$ we have 
\begin{align} 
&\sum_{i+j=0; i \geq 1}^{N+1} \int_{v(u_{-1},R_{-2})}^{v_\infty} d\bar{v} \frac{1}{r^2}   \|\left[r \slashed{\nabla}\right]^{i} \left[ \Omega \slashed{\nabla}_4 \right]^j   \left(r^3 {\omega} -r^3{\omega}_\circ\right) \|_{S_{u_{-1},\bar{v}}}^2 \nonumber \\
&\qquad +\sum_{i+j =0}^{N} \int_{v(u_{-1},R_{-2})}^{v_\infty} d\bar{v} \frac{1}{r^2}   \|\left[r \slashed{\nabla}\right]^{i} \left[ \Omega \slashed{\nabla}_4 \right]^j   \left(r^3 {\omega} -r^3{\omega}_\circ\right) \|_{S_{u_{-1},\bar{v}}}^2 \lesssim (\varepsilon_0)^2+\varepsilon^3 \, .
\end{align}
On spheres we have for all $v \in \left[v(u_{-1},R_{-2}),v_\infty\right]$
\begin{align} 
& \sum_{i+j =0 ; i \geq 1}^{N} \| \left[r \slashed{\nabla}\right]^{i} \left[ \Omega \slashed{\nabla}_4 \right]^j   \left(r^3 {\omega} -r^3{\omega}_\circ\right) \|_{S_{u_{-1},v}}^2 \nonumber \\
&\qquad +\sum_{i+j =0}^{N-1} \|\left[r \slashed{\nabla}\right]^{i} \left[ \Omega \slashed{\nabla}_4 \right]^j   \left(r^3 {\omega} -r^3{\omega}_\circ\right) \|_{S_{u_{-1},v}}^2 
 \lesssim   (\varepsilon_0)^2+\varepsilon^3 \, .  
\end{align}
\end{proposition}

\begin{proof}
Analogous to Proposition \ref{prop:fluxallominf} hence omitted: For the $\ell=0$ modes the estimate is immediate from gauge condition (\ref{gaugel0O}). For $\ell \geq 1$ one uses the relation (\ref{eop2}) which holds verbatim replacing $\ell=1$ by $\ell \geq 1$ and proceeds as in Proposition \ref{prop:fluxallominf}.
\end{proof}

We finally prove
\begin{align} \label{omap2}
\mathbb{E}^{N,aux}_{u_f, \mathcal{I}} \left[{\omega} \right]  \lesssim (\varepsilon_\circ)^2 + \varepsilon^3 \, .
\end{align}
In this case, the global estimates are a little more involved than for $\underline{\omega} - \underline{\omega}_\circ$ as we have to integrate forwards. On the other hand, we do not have to show an integrated decay estimate for $N+1$ derivatives (although we could).

The previous proposition in conjunction with following proposition easily imply
(\ref{omap2}). 

\begin{proposition} \label{prop:omauxspheresall}
We have on spheres for any $u_{-1} \leq u \leq u_f$ and $s=0,1,2$, $k_4+k_1 \leq N-s, k_4\neq N$
\begin{align} \label{om4ons}
\|   \left[\Omega \slashed{\nabla}_4\right]^{k_4} \left[ r \slashed{\nabla}\right]^{k_1}  r^{3-\frac{s}{2}} ({\omega}-\omega_\circ) \|^2_{S_{u,v}}  \lesssim \frac{ (\varepsilon_0)^2 +  \varepsilon^3}{u^{s}} \, .
\end{align}
We also have the integrated decay estimates for $k_4+k_1 \leq N$
\begin{align} 
\int_{u}^{u_f} d\bar{u} \int_{v(R_{-2},\bar{u})}^{v_{\infty}} d\bar{v} \  r^{-1-\delta} \| \left[r \Omega \slashed{\nabla}_4\right]^{k_4} \left[ r \slashed{\nabla}\right]^{k_1}  r^{5/2} ({\omega}-\omega_\circ) \|_{S_{\bar{u},\bar{v}}}^2 
\lesssim \varepsilon_0^2 + \varepsilon^3  \, ,
\end{align}
and for $k_4+k_1 \leq N-1$
\begin{align}
\int_{u}^{u_f} d\bar{u} \int_{v(R_{-2},\bar{u})}^{v_{\infty}} d\bar{v} \  r^{-1-\delta} \| \left[r \Omega \slashed{\nabla}_4\right]^{k_4} \left[ r \slashed{\nabla}\right]^{k_1}  r^{2} ({\omega}-\omega_\circ) \|_{S_{\bar{u},\bar{v}}}^2 
\lesssim \frac{\varepsilon_0^2 + \varepsilon^3}{u} \, .
\end{align}
\end{proposition}

\begin{proof}
The proof is very similar to that of Proposition \ref{prop:omauxspheres}. In particular, for $k_4=0$, Proposition \ref{prop:omauxspheres} already established the desired bounds. 

{\bf Step 1.} We establish (\ref{om4ons}) for $k_4=1$. Commuting (\ref{omcomu}) by $\left[r\slashed{\nabla}\right]^{N-3-s}\left[\Omega \slashed{\nabla}_4\right]$ yields
\begin{align}
\Omega \slashed{\nabla}_3 \left( \left[r\slashed{\nabla}\right]^{N-3-s}\left[\Omega \slashed{\nabla}_4\right] \left(r^2 \slashed{\Delta} \omega + r^2 \slashed{div} ( \beta \Omega)\right) \right) = \left[ \Omega \slashed{\nabla}_3 , \left[r\slashed{\nabla}\right]^{N-3-s}\left[\Omega \slashed{\nabla}_4\right]  \right] \left(r^2\slashed{\Delta} \omega + r^2 \slashed{div} ( \beta \Omega)\right) \nonumber \\
+ \left[r\slashed{\nabla}\right]^{N-3-s}\left[\Omega \slashed{\nabla}_4\right] \Big( h_0 r \slashed{div} \beta \Omega +h_0 \frac{1}{r} \slashed{div} \eta - h_0 \frac{1}{r} \slashed{div} \underline{\eta}+ \overset{(in)}{\slashed{\mathcal{E}}^{2}_{4}} + \left[\Omega \slashed{\nabla}_3, r^2 \slashed{\Delta}\right] \left(\omega - \omega_\circ\right)\Big) \, . \nonumber
\end{align}
We now insert all the (available) null structure and Bianchi equations in the $4$-direction on the right hand side gaining (at least) a power of $\frac{1}{r}$ in decay. With this, all (linear) terms on the right hand side can then either be controlled by Theorem  \ref{theo:mtnig2} or Theorem \ref{thm:alphaalphabarestimates}. The non-linear terms are easily seen to be handled by the bootstrap assumptions. Proceeding as in the proof of Proposition \ref{prop:omauxspheres} therefore yields
 \begin{align}
\|  r^{3-\frac{s}{2}}  \left[r\slashed{\nabla}\right]^{N-3-s}\left[r \Omega \slashed{\nabla}_4\right] (r^{2}\slashed{\Delta} {\omega} +r^2 \slashed{div} (\beta\Omega))\|_{S_{u,v}}  \lesssim \frac{ \varepsilon_0 +  \varepsilon^{\frac{3}{2}}}{u^{s}}\, .
\end{align}
Using the fact that
\[
\| r^{3-s}  \left[r\slashed{\nabla}\right]^{N-3-s}\left[r \Omega \slashed{\nabla}_4\right] r^2 \slashed{div} (\beta\Omega)\|_{S_{u,v}}  \lesssim \frac{ \varepsilon_0 +  \varepsilon^{\frac{3}{2}}}{u^{\min(\frac{1}{4}+\frac{s}{2},1)}} \, 
\]
is easily seen to hold after inserting the Bianchi equation for $\beta$ and using Theorem  \ref{theo:mtnig2} and Theorem \ref{thm:alphaalphabarestimates}, the result for $\omega -\omega_\circ$ on spheres follows for $k_4=1$. 

{\bf Step 2.} We next establish (\ref{om4ons}) for the case $k_4 \geq 2$. Commuting the propagation equation
\[
\Omega \slashed{\nabla}_3 \left({\omega} - {\omega}_\circ\right) = -\Omega^2 (\rho - \rho_\circ) + \frac{2M}{r^3}  \left(\Omega^2- \Omega_\circ^2 \right) + 2 \Omega^2 \left(\eta, \underline{\eta}\right) - \Omega^2 |\underline{\eta}|^2 
\]
with $ \left[ r \slashed{\nabla}\right]^{k_1} \left[\Omega \slashed{\nabla}_4\right]^{k_4} $ we first find for $k_4 \geq 1$, $k_1 \leq N-k_4-s$ schematically
\begin{align}
\Omega \slashed{\nabla}_3 \left(  \left[ r \slashed{\nabla}\right]^{k_1} \left[\Omega \slashed{\nabla}_4\right]^{k_4}  \left({\omega} - {\omega}_\circ\right) \right) = & +\frac{h_0}{r^{3+k_4}} \left(  \left[ r \slashed{\nabla}\right]^{k_1}  \left(\Omega^2 - \Omega_\circ^2\right) +  \left[ r \slashed{\nabla}\right]^{k_1} (\rho - \rho_\circ)r^3 \right) 
\nonumber \\
& +\frac{h_{0}}{r^3}  \left[ r \slashed{\nabla}\right]^{k_1} \left[\Omega \slashed{\nabla}_4\right]^{k_4-1}   (\omega-\omega_\circ)  
+ \frac{h_0}{r^{4+k_4}} \left[ r \slashed{\nabla}\right]^{k_1+1} r^3(\Omega \beta - (\Omega \beta)_{\mathrm{Kerr}}) \nonumber \\
&+\frac{h_0}{r^{5+k_4-2}}  \left[ r \slashed{\nabla}\right]^{k_1+2} \left[r \Omega \slashed{\nabla}_4\right]^{k_4-2}   r^3 \alpha \nonumber \\
&+ \left[ r \slashed{\nabla}\right]^{k_1} \left[\Omega \slashed{\nabla}_4\right]^{k_4} \overset{(in)}{\slashed{\mathcal{E}}^{0}_{4}} + \left[ \Omega \slashed{\nabla}_3 ,  \left[ r \slashed{\nabla}\right]^{k_1} \left[\Omega \slashed{\nabla}_4\right]^{k_4} \right]  \left({\omega} - {\omega}_\circ\right)  \, . \label{schemalong}
\end{align}
Assuming $k_4 \geq 2$ and integrating we use
\begin{itemize}
\item the estimates of Proposition \ref{prop:fluxallomu} for the data term on $u=u_{-1}$ (this is where $k_4 \neq N$ is used)
\item the estimates of Theorem \ref{thm:alphaalphabarestimates} for the term involving $\alpha$ 
\item the estimates of Theorem \ref{theo:mtnig2} for the linear terms (this is the only place $k_4=2$ is used\footnote{Note that the improved ingoing curvature flux bound $\int_{u_0}^u d\bar{u} \bar{u}^{2-\delta} \| \left[ r \slashed{\nabla}\right]^{N-2} (\rho-\rho_\circ)\|^2$ is crucial here to recover the full $r$-decay. This is also why the argument does not work directly for $k_4=1$: The estimate does not hold for $N-1$ derivatives.})
\item the bootstrap assumptions on the non-linear terms (note they can involve at most $N-s$ derivatives of Ricci coefficients or $N-1$ derivatives of curvature and decay strongly in $r$) 
\end{itemize}
yields after a simple induction on $k_4$ (using the estimate for $k_4=1$ from Step 1) for $k_4+k_1 \leq N-s$ and $k_4\geq 2$ the estimate
\begin{align}
\| r^{-1} \cdot  r^{3-\frac{s}{2}}  \left[r\slashed{\nabla}\right]^{k_1}\left[r \Omega \slashed{\nabla}_4\right]^{k_4} \left({\omega}-\omega_\circ\right) \|_{S_{u,v}}  \lesssim \frac{ \varepsilon_0 +  \varepsilon^2}{u^{s/2}}\, ,
\end{align}
which is the desired estimate on spheres. 

{\bf Step 3. } We now prove the integrated decay estimate, which can be done inductively in $k_4$ without distinguishing the case $k_4=1$. We integrate  (\ref{schemalong}) contracted with $r^{2k_4+5-\delta}  \left[ r \slashed{\nabla}\right]^{k_1} \left[\Omega \slashed{\nabla}_4\right]^{k_4}  \left({\omega} - {\omega}_\circ\right)$ 
over spacetime to obtain (inducting on $k_4$ using that for $k_4=0$ the estimate has already been proven) the boundedness estimate
\begin{align} 
\sum_{k_4+k_1 \leq N} \int_{u_{-1}}^{u_f} d\bar{u} \int_{v(R_{-2},\bar{u})}^{v_{\infty}} d\bar{v} \  r^{-1-\delta} \| \left[r \Omega \slashed{\nabla}_4\right]^{k_4} \left[ r \slashed{\nabla}\right]^{k_1}  r^{5/2} ({\omega}-\omega_\circ) \|_{S_{\bar{u},\bar{v}}}^2 
\lesssim \varepsilon_0^2 + \varepsilon^3  \, .
\end{align} 
One may now extract a dyadic sequence of constant $u$ hypersurfaces such that 
\begin{align}
\sum_{k_4+k_1 \leq N} \int_{v(R_{-2},u_i)}^{v_{\infty}} d\bar{v} \  r^{-\delta} \| \left[r \Omega \slashed{\nabla}_4\right]^{k_4} \left[ r \slashed{\nabla}\right]^{k_1}  r^{2} ({\omega}-\omega_\circ) \|_{S_{\bar{u},\bar{v}}}^2 
\lesssim \frac{\varepsilon_0^2 + \varepsilon^3}{u_i} \, .
\end{align}
For $k_4+k_1 \leq N-1$, therefore, integrating (\ref{schemalong}) now contracted with $r^{2k_4+4-\delta}\left[ r \slashed{\nabla}\right]^{k_1} \left[\Omega \slashed{\nabla}_4\right]^{k_4}  \left({\omega} - {\omega}_\circ\right)$ over $\mathcal{D}^{\I}\left(u_i, u_f\right)$, i.e.~now starting from a ``good slice", yields also the last estimate of the proposition.
\end{proof}

\subsection{Estimates the full master energy}  \label{sec:proofof4} 
We now prove the estimate of Theorem  \ref{theo:reducetoangular}. 
Inspecting the master energies in Section \ref{Iplusenergiesmastersec}, we see that 
our task is to estimate up to $N$ arbitrary derivatives of Ricci coefficients and curvature (both in integrated decay and on spheres). Note that we can assume that not all derivatives are angular, as these have already been estimated.

{\bf Step 1.} We show the estimate of Theorem  \ref{theo:reducetoangular} excluding the metric quantities $b$ and $\slashed{g}-\slashed{g}_\circ$ in the energy on the left hand side and excluding the estimate for $\sum_{|\underline{k} | \leq N} \sup_{\mathcal{D}^{\I}} \| \mathfrak{D}^{\underline{k}} (r^3 T) \|^2_{S_{u,v}}$. This follows from repeatedly applying the schematic equations collected in
(\ref{schematicnullstructure3withKerr}), (\ref{schematicnullstructure4withKerr}) and (\ref{schematicbianchi3withKerr}), (\ref{schematicbianchi4withKerr}), as these will provide expressions for $1\leq n \leq N$ derivatives of any Ricci coefficient $\mathfrak{D}^{\gamma} (r^p\Gamma_p-r^p(\Gamma_p)_{\mathrm{Kerr}})$ 
(except $\omega - \omega_\circ$ and $\underline{\omega}-\underline{\omega}_\circ$, which however have already been estimated in Theorem \ref{sec:proofof3}), where $1\leq |\gamma|\leq n$ (and $\gamma \neq (0,0,n)$, i.e.~not all derivatives angular), in terms of 
\begin{itemize}
\item up to $n$ angular derivatives of Ricci-coefficients
\item up to $n$ derivatives of $\omega$ and $\underline{\omega}$ (at least one of them angular if $N$ derivatives appear) 
\item up to $n-1$ angular derivatives of curvature
\item non-linear errors involving at most $n$ derivatives of $\Gamma$'s and $n-1$ derivatives of $\mathcal{R}$'s.
\end{itemize}
Moreover, if $\alpha$, $\Omega \beta - (\Omega \beta)_{\mathrm{Kerr}}$ or $\omega-\omega_\circ$ appears on the right hand side, then it does so with an additional weight of at least $r^{-1/2}$, cf.~Remarks \ref{rem:gammastru} and \ref{rem:Rstru}. 

Similarly, \ref{schematicnullstructure3withKerr}), (\ref{schematicnullstructure4withKerr}) and (\ref{schematicbianchi3withKerr}), (\ref{schematicbianchi4withKerr}) imply expressions for $n$ derivatives of any curvature component (except $\alpha$ and $\underline{\alpha}$ itself, which however have already been estimated), i.e.~$\mathfrak{D}^{\gamma} (r^p \mathcal{R}_p-r^p (\mathcal{R}_p)_{\mathrm{Kerr}})$, where $1\leq |\gamma|\leq n$ (and $\gamma \neq (0,0,n)$, i.e.~not all derivatives angular) in terms of 
\begin{itemize}
\item up to $n$ angular derivatives of Ricci coefficients and curvature components
\item up to $n$ derivatives of $\omega$ and $\underline{\omega}$ (at least one of them angular if $n$ derivatives appear)
\item non-linear errors involving at most $n$ derivatives of the $\Gamma$'s and $n$ derivatives of the $\mathcal{R}$'s.
\end{itemize}
Moreover, if $\alpha$, $\Omega \beta - (\Omega \beta)_{\mathrm{Kerr}}$ or $\omega-\omega_\circ$ appears on the right hand side, then it does so with an additional weight of at least $r^{-1/2}$, cf.~Remarks \ref{rem:gammastru} and \ref{rem:Rstru}. 

The integrated decay estimates and the estimates on spheres follow for all $\Gamma \setminus \{b , \omega-\omega_\circ, \underline{\omega}-\underline{\omega}_\circ\}$ and all $\mathcal{R} \setminus \{ \alpha, \underline{\alpha} \}$ directly from these observations.  

{\bf Step 2.} The estimate 
\[
\sum_{|\underline{k} | \leq N} \sup_{\mathcal{D}^{\I}} \| \mathfrak{D}^{\underline{k}} (r^3 T) \|_{S_{u,v}} \lesssim \varepsilon_0 + \varepsilon^{\frac{3}{2}}
\]
now follows by differentiating equations (\ref{T4dir}) and (\ref{uty}). (Reall again that we have proven the estimate for $\underline{k}=(|\underline{k}|,0,0)$ in the angular master energy already.) 

{\bf Step 3.} We prove the estimate of Theorem~\ref{theo:reducetoangular} for the terms involving $b$ in the energy on the left. This follows from commuting the transport equation (\ref{transportb}) with derivatives of the form $\left[\Omega^{-1} \slashed{\nabla}_3\right]^i \left[ r\slashed{\nabla}\right]^j$ for $1 \leq i+j = n \leq N$.\footnote{Note that a tuple of derivatives involving one $\Omega \slashed{\nabla}_4$ can of course be estimated directly from the equation.} The non-linear error on the right involves at most $n$ derivatives of the Ricci-coefficients and the linear term involve up to $n\leq N$ derivatives of $\eta-\eta_{\mathrm{Kerr}}$ and $\underline{\eta}-\underline{\eta}_{\mathrm{Kerr}}$ which have been estimated in Step 1. Integrating the commuted transport equation therefore yields the desired estimates.

{\bf Step 4.} We prove the estimate of Theorem  \ref{theo:reducetoangular} for the terms involving the metric difference $\slashed{g}-\slashed{g}_\circ$. This follows from the transport equation (\ref{nabla4g}) and repeating the argument of Step 3.

\chapter{Estimates in the $\Hp$ gauge: the proof of Theorem \ref{thm:Hestimates}}

\label{chap:Hestimates}

In this chapter we shall prove Theorem~\ref{thm:Hestimates}, which we restate here:

\Hestimates*

\minitoc

Section {\bf Section~\ref{section:Hprelim}} concerns certain preliminaries, including estimates for general \emph{redshifted} and \emph{blueshifted} transport equations and the derivation of an equation satisfied by the quantity $X$, which is later used to estimate $\Omega^{-1} \hat{\chi}$ in what is perhaps the most involved part of the proof of Theorem~\ref{thm:Hestimates}.  See already the discussion in Section \ref{othernonlinearissues}.  Section {\bf Section~\ref{nonlinestforH}} involves estimates for certain nonlinear error terms.  These preliminaries and error estimates are used in {\bf Section~\ref{section:Hmainestimates}}, which concerns transport and elliptic estimates for the geometric quantities in the $\Hp$ gauge and constitutes the main part of the proof of Theorem~\ref{thm:Hestimates}.

\vskip1pc
\noindent\fbox{
    \parbox{6.35in}{
As in the previous chapters of Part~\ref{improvingpart},
we shall assume throughout the assumptions of~\Cref{havetoimprovethebootstrap}. Let us fix an
arbitrary  $u_f\in[u_f^0, \hat{u}_f$], with $\hat{u}_f\in \mathfrak{B}$,
and fix some $\lambda \in \mathfrak{R}(u_f)$.
In this chapter, unless explictly stated otherwise, all propositions below
shall always refer  
to the anchored $\Hp$ gauge in the  spacetime  $(\mathcal{M}(\lambda), g(\lambda))$,  
corresponding to parameters
$u_f$, $M_f(u_f,\lambda)$,
whose existence is
ensured by Definition~\ref{bootstrapsetdef}. 
Thus, we drop the  $\Hp$ superscripts for geometric quantities without risk of confusion,
writing $\alpha=\alpha_{\Hp}$, $C_u=C_u^{\Hp}$, etc.
We shall denote $M=M_f$ throughout  
this chapter.}}
\vskip1pc

\emph{This Chapter will depend on results from all previous chapters of  Part~\ref{improvingpart}. 
Note, however, that from Theorem~\ref{thm:relatinggauges} of Chapter~\ref{chap:comparing}, this chapter will only use
the statement of Proposition~\ref{thm:inheriting}.
Chapters~\ref{elliptandcalcchapter} and~\ref{moreherechapter} are
only appealed to through the estimate in the
statement of Theorems~\ref{thm:PPbarestimates} and~\ref{thm:alphaalphabarestimates}, and
Chapter~\ref{chap:Iestimates} only through the statement of Theorem~\ref{thm:Iestimates}. Thus, if the reader
is willing to refer back for these statements, 
the chapter can be read independently of the rest of Part~\ref{improvingpart}.}

\emph{As with the previous chapter, the reader may again wish to compare with the proofs of both Theorems~3 and~4 of~\cite{holzstabofschw} in
Sections~13 and~14, respectively, for  
linear analogues of results proven here.}

\section{General transport estimates for redshifted and blueshifted equations, the quantity $X$, and other preliminaries}
\label{section:Hprelim}

This section contains some preliminaries which will be used in Section \ref{section:Hmainestimates} to estimate the Ricci coefficients and curvature components.  {\bf Section~\ref{subsec:Htransportestimates}} concerns some basic transport estimates for $S$-tensors, including estimates for \emph{redshifted} and \emph{blueshifted} transport equations in the outgoing null direction, which will be used in Section~\ref{section:Hmainestimates} along with the elliptic estimates of Section \ref{section:ellipticestimates}.  {\bf Section \ref{section:Hequations}} involves the derivation of equations satisfied by certain renormalised quantities, including the quantity $X$ discussed in Section \ref{othernonlinearissues}.

\subsection{General transport estimates}
\label{subsec:Htransportestimates}

The general transport estimates of this section are used in Section \ref{section:Hmainestimates} in order to estimate the Ricci coefficients and curvature components.

\subsubsection{$L^2$ transport estimates in the $\nablaslash_4$ direction}
\label{subsubsec:nabla4transport}

A transport equation in the $\nablaslash_4$ direction for an $S$ tensor $\xi$ of the form
\begin{equation} \label{eq:shifted}
	\Omega \nablaslash_4 \xi + a (\Omega \omegahat)_{\circ} \xi= F,
\end{equation}
for some inhomogeneous term $F$ and some $a \in \mathbb{R}$, is referred to as \emph{redshifted}, \emph{noshifted} or \emph{blueshifted} according to whether the sign of $a$ is positive, zero or negative respectively.

The following lemma provides an estimate for quantities which satisfy a blueshifted transport equation.

\begin{lemma}[Estimates for blueshifted transport equation] \label{lem:nabla4bluexi}
	Consider some $u_0 \leq u \leq u_f$ and $v_{-1} \leq v \leq v(R,u)$.  Let $\xi$ be an $S$-tensor field such that
	\begin{equation*}
		\Omega \nablaslash_4 \xi - l (\Omega \omegahat) \xi = F,
	\end{equation*}
	for some tensor field $F$ and some $l \geq 2$.  Then,
	\[
		\Vert \xi \Vert_{S_{u,v}}^2
		+
		\Vert \xi \Vert_{C_u(v)}^2
		\lesssim
		\Omega(u,v)^{l}
		\Vert \xi \Vert_{S_{u,v(R,u)}}^2
		+
		\Vert F \Vert_{C_{u}(v)}^2
		\lesssim
		\Vert \xi \Vert_{S_{u,v(R,u)}}^2
		+
		\Vert F \Vert_{C_{u}(v)}^2.
	\]
\end{lemma}

\begin{proof}
	Since $\partial_v \log \Omega = \Omega \omegahat$, it follows that
	\[
		\Omega \nablaslash_4 (\Omega^{-\frac{l}{2}}\xi) - \frac{l}{2} (\Omega \omegahat) \Omega^{-\frac{l}{2}} \xi
		=
		\Omega^{-\frac{l}{2}} F.
	\]
	The proof follows from contracting with $\Omega^{\frac{l}{2}}\xi$ to give
	\[
		\partial_v (\Omega^{-l}\vert \xi \vert^2) - \frac{l}{2} \Omega \omegahat_{\circ} \Omega^{-l} \vert \xi \vert^2 = 2 \Omega^{-l} \xi \cdot F.
	\]
	Integrating over $S_{u,v'}$ and from $v$ to $v(R,u)$ one obtains
	\begin{align*}
		\Vert \Omega^{-\frac{l}{2}} \xi \Vert_{S_{u,v}}^2
		+
		\Vert \Omega^{-\frac{l}{2}} \xi \Vert_{C_{u}(v)}^2
		&
		\lesssim
		\Vert \Omega^{-\frac{l}{2}} \xi \Vert_{S_{u,v(R,u)}}^2
		+
		\int_{v}^{v(R,u)} \int_{S_{u,v'}} \Omega^{-l} \vert \xi \cdot F \vert dv'
		\\
		&
		\lesssim
		\Vert \Omega^{-\frac{l}{2}} \xi \Vert_{S_{u,v(R,u)}}^2
		+
		\lambda
		\Vert \Omega^{-\frac{l}{2}} \xi \Vert^2_{C_{u}(v)}
		+
		\lambda^{-1}
		\Vert \Omega^{-\frac{l}{2}} F \Vert^2_{C_{u}(v)},
	\end{align*}
	for any $\lambda>0$.  The proof follows from taking $\lambda$ suitably small and multiplying by $\Omega(u,v)^l$.
\end{proof}

The next lemma provides an estimate for quantities which satisfy a redshifted transport equation.

\begin{lemma}[Estimates for redshifted transport equation] \label{lem:nabla4redxi}
	Let $\xi$ be an $S$-tensor field such that
	\begin{equation} \label{eq:nabla4redxi}
		\Omega \nablaslash_4 \xi + l (\Omega \omegahat) \xi= F,
	\end{equation}
	for some $l \geq 2$ and some tensor field $F$.  Then, for any $\kappa \geq 0$, and any $2M_f < r_* \leq R_2$,
	\begin{equation} \label{eq:nabla4redxispheres}
		\Vert \xi \Vert_{S_{u,v}}^2
		+
		\Vert \xi \mathds{1} \Vert_{C_u^{\Hp}(v)}^2
		\lesssim
		\frac{1}{v^{\kappa}}
		\Vert \xi \Vert_{S_{u,v_{-1}}}^2
		+
		\Vert F \mathds{1} \Vert_{C_u^{\Hp}(v_{-1} \vee v/2)}^2,
	\end{equation}
	and
	\begin{equation} \label{eq:nabla4redxispacetime}
		\Vert \xi \mathds{1} \Vert_{\Cbar^{\Hp}_v}^2
		+
		\Vert \xi \mathds{1} \Vert_{\DRH(v)}^2
		\lesssim
		\frac{1}{v^{\kappa}}
		\Vert \xi \mathds{1} \Vert_{\Cbar^{\Hp}_{v_{-1}}}^2
		+
		\Vert F \mathds{1} \Vert_{\DRH(v_{-1} \vee v/2)}^2,
	\end{equation}
	for all $u_0 \leq u \leq u_f$, $v_{-1} \leq v \leq v(r_*,u)$, where $v_{-1} \vee v/2 := \max \{ v_{-1}, v/2 \}$ and $\mathds{1} = \mathds{1}_{r\leq r_*}$.
\end{lemma}

\begin{proof}
	After contracting \eqref{eq:nabla4redxi} with $\xi$ and integrating between any $v_{-1} \leq v_1<v_2 \leq v(r_*,u)$, it follows that,
	\begin{equation} \label{eq:nabla4redxi2}
		\Vert \xi \Vert_{S_{u,v_2}}^2
		+
		\int_{v_1}^{v_2} \Vert \xi \Vert_{S_{u,v}}^2 dv
		\lesssim
		\Vert \xi \Vert_{S_{u,v_1}}^2
		+
		\int_{v_1}^{v_2} \Vert F \Vert_{S_{u,v}}^2 dv.
	\end{equation}
	Let $\{ \widetilde{v}_n\}$ be an appropriate dyadic sequence.  By the pigeonhole principle, for each $n$ there exists $v_n \in [\widetilde{v}_n, \widetilde{v}_{n+1})$ such that
	\[
		\Vert \xi \Vert_{S_{u,v_n}}^2
		=
		\frac{1}{\widetilde{v}_{n+1} - \widetilde{v}_n}
		\int_{\widetilde{v}_{n}}^{\widetilde{v}_{n+1}}
		\Vert \xi \Vert_{S_{u,v}}^2
		dv
		\lesssim
		\frac{1}{v_n}
		\Big(
		\Vert \xi \Vert_{S_{u,v_{-1}}}^2
		+
		\int_{\widetilde{v}_n}^{v(r_*,u)} \Vert F \Vert_{S_{u,v}}^2 dv
		\Big),
	\]
	where the inequality follows from \eqref{eq:nabla4redxi2}.  Returning to \eqref{eq:nabla4redxi2}, it follows that
	\[
		\Vert \xi \Vert_{S_{u,v}}^2
		+
		\int_{v}^{v(r_*,u)} \Vert \xi \Vert_{S_{u,v}}^2 dv
		\lesssim
		\frac{1}{v}
		\Vert \xi \Vert_{S_{u,v_{-1}}}^2
		+
		\int_{\lambda v}^{v(r_*,u)} \Vert F \Vert_{S_{u,v}}^2 dv,
	\]
	for some appropriate $\lambda <1$ (depending on exactly how the original dyadic sequence was chosen).  The estimate \eqref{eq:nabla4redxispheres} follows by iterating the above argument $\left\lfloor \kappa -1 \right\rfloor$ more times.  The estimate \eqref{eq:nabla4redxispacetime} follows from replacing $\xi$ with $\Omega \xi$ in \eqref{eq:nabla4redxispheres} and integrating from $u$ to $u_f$.
\end{proof}

\subsubsection{$L^2$ transport estimates in the $\nablaslash_3$ direction}

The following lemma provides an estimate for quantities which satisfy transport equations in the $\nablaslash_3$ direction.

\begin{lemma}[Estimates for transport equation in $\nablaslash_3$ direction] \label{lem:nabla3xi}
	Let $\xi$ be an $S$-tangent tensor field satisfying
	\begin{equation} \label{eq:nabla3xi}
		\Omega \nablaslash_3 \xi = \Omega^2 F,
	\end{equation}
	for some tensor field $F$.  Then, for all $u_0 \leq u \leq u_f$, $v_{-1} \leq v \leq v(R_2,u)$
	\[
		\Vert \xi \Vert^2_{S_{u,v}}
		+
		\Vert \xi \Vert^2_{\Cbar_v(u)}
		\lesssim
		\Vert \xi \Vert^2_{S_{u_f,v}}
		+
		\Vert F \Vert^2_{\Cbar_v(u)},
		\quad
		\text{and}
		\quad
		\Vert \xi \Vert^2_{C_u(v)}
		+
		\Vert \xi \Vert^2_{\DRH(v)}
		\lesssim
		\Vert \xi \Vert^2_{C_{u_f}(v)}
		+
		\Vert F \Vert^2_{\DRH(v)}.
	\]
\end{lemma}

\begin{proof}
	Renormalising \eqref{eq:nabla3xi} with $r^{-1}$ and contracting with $\xi$ gives,
	\[
		\partial_u ( \vert r^{-1} \xi \vert^2)
		-
		2 r^{-1} \Omega_{\circ}^2 \vert r^{-1} \xi \vert^2
		=
		2 r^{-2} \Omega^2 \xi \cdot F,
	\]
	and so, integrating over the cone $\Cbar_{v}(u)$ gives
	\begin{align*}
		\Vert \xi \Vert^2_{S_{u,v}}
		+
		\Vert \xi \Vert^2_{\Cbar_v(u)} 
		\leq
		\Vert \xi \Vert^2_{S_{u_f,v}}
		+
		\Vert F \Vert^2_{\Cbar_v(u)} .
	\end{align*}
	The second estimate follows similarly by integrating over the spacetime region $\DRH(v)$.
\end{proof}

\subsubsection{$L^2$ transport estimates for difference quotients}
\label{subsubsec:L2transportestimatesH}

Recall that any $S$-tangent $(0,k)$ tensor field $\xi$ can be uniquely decomposed as
\[
	\xi = \xi_{A_1\ldots A_k} d\theta^{A_1} \otimes \ldots \otimes d\theta^{A_k}.
\]
For such $\xi$ define the difference
\[
	\xi(u,v,\theta) - \xi(u_f,v,\theta)
	:=
	\big(
	\xi(u,v,\theta)_{A_1\ldots A_k} - \xi(u_f,v,\theta)_{A_1\ldots A_k}
	\big)
	d\theta^{A_1} \otimes \ldots \otimes d\theta^{A_k},
\]
(note that this definition is independent of the choice of local coordinates $\theta^A$ on 
$\mathbb S^2$) and define the difference quotient
\[
	D_{u_f} \xi (u,v,\theta) := \frac{\xi(u,v,\theta) - \xi(u_f,v,\theta)}{\Omega_{\circ}(u,v,\theta)^2}.
\]

This section contains transport estimates concerning the above difference quotients of $S_{u,v}$ tensor fields.  The following lemma gives an estimate for the difference quotient of a noshifted equation (recall the nomenclature of Section \ref{subsubsec:nabla4transport}).

\begin{lemma}[Difference quotient estimate for noshifted transport equation] \label{lem:xinoshiftdifference}
	If $\xi$ is an $S$-tangent $(0,k)$  tensor which satisfies
	\begin{equation} \label{eq:xinoshift}
		\Omega \nablaslash_4 \xi
		=
		F,
	\end{equation}
	then, for any $u_0 \leq u \leq u_f$, $v_{-1} \leq v \leq v(R_2,u)$,
	and any $\kappa \geq 0$,
	\[
		\Vert D_{u_f} \xi \Vert_{S_{u,v}}^2
		+
		\Vert D_{u_f} \xi \Vert_{C_{u}^{\Hp}(v)}^2
		\lesssim
		\frac{1}{v^{\kappa}}
		\Vert D_{u_f} \xi \Vert_{S_{u,v_{-1}}}^2
		+
		\Vert D_{u_f} F \Vert_{C_{u}^{\Hp}(v_{-1} \vee v/2)}^2,
	\]
	and
	\[
		\Vert D_{u_f} \xi \Vert_{\Cbar^{\Hp}_v}^2
		+
		\Vert D_{u_f} \xi \Vert_{\DRH(v)}^2
		\lesssim
		\frac{1}{v^{\kappa}}
		\Vert D_{u_f} \xi \Vert_{\Cbar^{\Hp}_{v_{-1}}}^2
		+
		\Vert D_{u_f} F \Vert_{\DRH(v_{-1} \vee v/2)}^2,
	\]
	where $v_{-1} \vee v/2 := \max \{ v_{-1}, v/2 \}$.
\end{lemma}

\begin{proof}
	The equation \eqref{eq:xinoshift} implies that
	\[
		\Omega \nablaslash_4
		D_{u_f} \xi
		+
		2(\Omega \omegahat)_{\circ} D_{u_f} \xi
		=
		D_{u_f} F.
	\]
	The proof then follows from Lemma \ref{lem:nabla4redxi}.
\end{proof}

The next lemma in particular shows that the difference quotient $D_{u_f} \xi$ can be appropriately controlled by the derivative $\Omega^{-1} \nablaslash_3 \xi$.

\begin{lemma}[Estimate for difference by $\nablaslash_3$ derivative] \label{lem:nabla3differenceestimate}
	Consider some $u_0 \leq u \leq u_f$, $v_{-1} \leq v \leq v(R_2,u)$ and an $S$-tangent tensor field $\xi$.  Then
	\begin{equation} \label{eq:nabla3differenceestimate1}
		\vert \xi(u,v,\theta) - \xi(u_f,v,\theta) \vert
		\lesssim
		\int_{u}^{u_f}
		(\vert \Omega^{-1} \nablaslash_3 \xi \vert + \vert \xi \vert)
		(u',v,\theta)
		\Omega^2 du'.
	\end{equation}
	In particular, it follows that
	\begin{equation} \label{eq:nabla3differenceestimate2}
		\Vert
		D_{u_f} \xi
		\Vert^2_{S_{u,v}}
		\leq
		\sup_{u \leq u' \leq u_f}
		\big(
		\Vert \Omega^{-1} \nablaslash_3 \xi \Vert^2_{S_{u',v}}
		+
		\Vert \xi \Vert^2_{S_{u',v}}
		\big).
	\end{equation}
\end{lemma}

\begin{proof}
	For simplicity, suppose $\xi$ is an $S$-tangent $1$-form.  Write
	\[
		\xi(u,v,\theta) - \xi(u_f,v,\theta)
		=
		(\xi(u,v,\theta)_A - \xi(u_f,v,\theta)_A) d \theta^A
		=
		\int_{u}^{u_f} \partial_u \big( \xi(u',v,\theta)_A \big) du' d \theta^A.
	\]
	The inequality \eqref{eq:nabla3differenceestimate1} follows from the fact that
	\[
		\partial_u (\xi_A)
		=
		\Omega \nablaslash_3 \xi_A
		+
		\Omega {\chibar_A}^B \xi_B.
	\]
	The inequality \eqref{eq:nabla3differenceestimate2} is then immediate from the fact that $\Omega_{\circ}(u,v)^{-2} \int_{u}^{u_f}\Omega^2 du' \lesssim 1$.  The proof for more general $S$-tangent tensor fields is similar.
\end{proof}

The next lemma concerns particular blueshifted transport equations which become noshifted after one commutation with $\Omega^{-1} \nablaslash_3$.

\begin{lemma}[Estimates for blueshifted transport equations which are noshifted after one commutation] \label{lem:noshifteddifferencequotient}
	Let $\xi$ be an $S$-tangent tensor field satisfying
	\[
		\Omega \nablaslash_4 \Omega^{-1} \nablaslash_3 \xi = h(r) \xi + F,
	\]
	for some $S$-tangent tensor field $F$ and some smooth function $h$.  For any $u_0 \leq u \leq u_f$, $v_{-1} \leq v \leq v(R_2,u)$ and any $\kappa \geq 0$,
	\begin{align} \label{eq:noshifteddifferencequotient1}
		&
		\sum_{l=0,1}
		\Big(
		\Vert (\Omega^{-1} \nablaslash_3)^l \xi \Vert_{S_{u,v}}^2
		+
		\Vert (\Omega^{-1} \nablaslash_3)^l \xi \Vert^2_{C_{u}(v)}
		\Big)
		\\
		&
		\quad
		\lesssim
		v^{-\kappa} \Vert D_{u_f} \Omega^{-1} \nablaslash_3 \xi \Vert_{S_{u,v_{-1}}}^2
		+
		\Vert D_{u_f} F \Vert_{C_{u}(v_{-1} \vee \frac{v}{2})}^2
		+
		\sum_{l=0,1}
		\Big(
		\Vert (\Omega^{-1} \nablaslash_3)^l \xi \Vert_{S_{u_f,v}}^2
		+
		\Vert (\Omega^{-1} \nablaslash_3)^l \xi \Vert^2_{C_{u_f}(v_{-1} \vee \frac{v}{2})}
		\Big),
		\nonumber
	\end{align}
	and
	\begin{align} \label{eq:noshifteddifferencequotient2}
		&
		\sum_{l=0,1}
		\Big(
		\Vert (\Omega^{-1} \nablaslash_3)^l \xi \Vert^2_{\Cbar_{v}(u)}
		+
		\Vert (\Omega^{-1} \nablaslash_3)^l \xi \Vert^2_{\DRH(v)}
		\Big)
		\\
		&
		\quad
		\lesssim
		v^{-\kappa} \Vert D_{u_f} \Omega^{-1} \nablaslash_3 \xi \Vert_{\Cbar_{v_{-1}}}^2
		+
		\Vert D_{u_f} F \Vert_{\DRH(v_{-1} \vee \frac{v}{2})}^2
		+
		\sum_{l=0,1}
		\Big(
		\Vert (\Omega^{-1} \nablaslash_3)^l \xi \Vert_{S_{u_f,v}}^2
		+
		\Vert (\Omega^{-1} \nablaslash_3)^l \xi \Vert^2_{C_{u_f}(v_{-1} \vee \frac{v}{2})}
		\Big),
		\nonumber
	\end{align}
	where $v_{-1} \vee v/2 := \max \{ v_{-1}, v/2 \}$.
\end{lemma}

\begin{proof}
	The difference quotient $D_{u_f} \Omega^{-1} \nablaslash_3 \xi$ satisfies
	\begin{equation} \label{eq:noshifteddifferencequotient}
		\Omega\nablaslash_4 D_{u_f} \Omega^{-1} \nablaslash_3 \xi
		+
		2 (\Omega \omegahat)_{\circ} D_{u_f} \Omega^{-1} \nablaslash_3 \xi
		=
		h(r) D_{u_f} \xi
		+
		\xi (u_f,v) D_{u_f} h
		+
		D_{u_f} F.
	\end{equation}
	Given $v_{-1} \leq v_1 < v_2 \leq v(R_2,u)$, after contracting \eqref{eq:noshifteddifferencequotient} with $D_{u_f} \Omega^{-1} \nablaslash_3 \xi$ and integrating it follows that
	\begin{align*}
		\Vert D_{u_f} \Omega^{-1} \nablaslash_3 \xi \Vert_{S_{u,v_2}}^2
		+
		\int_{v_1}^{v_2} \Vert D_{u_f} \Omega^{-1} \nablaslash_3 \xi \Vert^2_{S_{u,v}} dv
		\lesssim
		\
		&
		\Vert D_{u_f} \Omega^{-1} \nablaslash_3 \xi \Vert_{S_{u,v_1}}^2
		\\
		&
		+
		\int_{v_1}^{v_2}
		\Vert \xi \Vert_{S_{u_f,v}}^2
		+
		\Vert D_{u_f} F \Vert_{S_{u,v}}^2
		+
		\Vert D_{u_f} \xi \Vert_{S_{u,v}}^2
		dv.
	\end{align*}
	Now, by Lemma \ref{lem:nabla3differenceestimate}
	\[
		\vert D_{u_f} \xi (u,v) \vert
		\leq
		\Omega_{\circ}(u,v)^{-2} \int_u^{u_f}
		\big(
		\vert \Omega^{-1} \nablaslash_3 \xi (u',v) \vert
		+
		\vert \xi (u',v) \vert
		\big)
		\Omega^2
		du',
	\]
	and so
	\begin{align*}
		\int_{v_1}^{v_2}
		\Vert D_{u_f} \xi \Vert_{S_{u,v}}^2
		dv
		&
		\lesssim
		\int_{v_1}^{v_2} \int_u^{u_f}
		\big(
		\Vert \Omega^{-1} \nablaslash_3 \xi \Vert_{S_{u',v}}^2
		+
		\Vert \xi \Vert_{S_{u',v}}^2
		\big) \Omega_{\circ}^2 du'
		\Omega_{\circ}(u,v)^{-4} \int_u^{u_f} \Omega(u',v)^2 d u' dv
		\\
		&
		\lesssim
		\int_u^{u_f} \int_{v_1}^{v_2}
		\Vert \Omega^{-1} \nablaslash_3 \xi \Vert_{S_{u',v}}^2
		+
		\Vert \xi \Vert_{S_{u',v}}^2
		dv
		\frac{\Omega_{\circ}(u',v_2)^2}{\Omega_{\circ}(u,v_2)^{2}} d u'.
	\end{align*}
	Now since,
	\[
		\Vert \Omega^{-1} \nablaslash_3 \xi \Vert^2_{S_{u,v}}
		\lesssim
		\Vert \Omega^{-1} \nablaslash_3 \xi \Vert^2_{S_{u_f,v}}
		+
		\Vert D_{u_f} \Omega^{-1} \nablaslash_3 \xi \Vert^2_{S_{u,v}}
	\]
	and
	\[
		\Vert \xi \Vert^2_{S_{u,v}}
		\lesssim
		\Vert \xi \Vert^2_{S_{u_f,v}}
		+
		\int_u^{u_f}
		\Vert \Omega^{-1} \nablaslash_3 \xi \Vert^2_{S_{u',v}}
		\Omega^2
		du',
	\]
	it follows that
	\begin{multline*}
		\Vert \xi \Vert^2_{S_{u,v}}
		+
		\Vert \Omega^{-1} \nablaslash_3 \xi \Vert^2_{S_{u,v}}
		+
		\int_{v_1}^{v_2}
		\Vert \xi \Vert^2_{S_{u,v}}
		+
		\Vert \Omega^{-1} \nablaslash_3 \xi \Vert^2_{S_{u,v}}
		dv
		\lesssim
		\Vert \xi \Vert^2_{S_{u_f,v}}
		+
		\Vert \Omega^{-1} \nablaslash_3 \xi \Vert^2_{S_{u_f,v}}
		\\
		+
		\int_{v_1}^{v_2}
		\Vert \xi \Vert^2_{S_{u_f,v}}
		+
		\Vert \Omega^{-1} \nablaslash_3 \xi \Vert^2_{S_{u_f,v}}
		dv
		+
		\int_u^{u_f}
		\big(
		\Vert D_{u_f} \Omega^{-1} \nablaslash_3 \xi \Vert_{S_{u',v_2}}^2
		+
		\int_{v_1}^{v_2} \Vert D_{u_f} \Omega^{-1} \nablaslash_3 \xi \Vert^2_{S_{u',v}} dv
		\big)
		\Omega^2
		du',
	\end{multline*}
	and the Gr\"{o}nwall inequality implies
	\begin{multline*}
		\int_{v_1}^{v_2}
		\Vert \xi \Vert^2_{S_{u,v}}
		+
		\Vert \Omega^{-1} \nablaslash_3 \xi \Vert^2_{S_{u,v}}
		+
		\Vert D_{u_f} \Omega^{-1} \nablaslash_3 \xi \Vert^2_{S_{u,v}}
		dv
		\\
		\lesssim
		\Vert D_{u_f} \Omega^{-1} \nablaslash_3 \xi \Vert_{S_{u,v_1}}^2
		+
		\int_{v_1}^{v_2}
		\Vert \xi \Vert_{S_{u_f,v}}^2
		+
		\Vert \Omega^{-1} \nablaslash_3 \xi \Vert^2_{S_{u_f,v}}
		+
		\Vert D_{u_f} F \Vert_{S_{u,v}}^2
		dv
	\end{multline*}
	and
	\begin{multline*}
		\Vert \xi \Vert^2_{S_{u,v_2}}
		+
		\Vert \Omega^{-1} \nablaslash_3 \xi \Vert_{S_{u,v_2}}^2
		+
		\Vert D_{u_f} \Omega^{-1} \nablaslash_3 \xi \Vert^2_{S_{u,v_2}}
		\\
		\lesssim
		\Vert D_{u_f} \Omega^{-1} \nablaslash_3 \xi \Vert_{S_{u,v_1}}^2
		+
		\Vert \xi \Vert^2_{S_{u_f,v_2}}
		+
		\Vert \Omega^{-1} \nablaslash_3 \xi \Vert_{S_{u_f,v_2}}^2
		+
		\int_{v_1}^{v_2}
		\Vert \xi \Vert_{S_{u_f,v}}^2
		+
		\Vert D_{u_f} F \Vert_{S_{u,v}}^2
		dv.
	\end{multline*}
	The proof of \eqref{eq:noshifteddifferencequotient1} then follows from a pigeonhole argument, as in the proof of Lemma \ref{lem:nabla4redxi}.  The proof of \eqref{eq:noshifteddifferencequotient2}, this time using the fact that the difference quotient satisfies
	\[
		\Omega\nablaslash_4 \Omega_{\circ} D_{u_f} \Omega^{-1} \nablaslash_3 \xi
		+
		(\Omega \omegahat)_{\circ} \Omega_{\circ} D_{u_f} \Omega^{-1} \nablaslash_3 \xi
		=
		h(r) \Omega_{\circ} D_{u_f} \xi
		+
		\Omega_{\circ} \xi (u_f,v) D_{u_f} h
		+
		\Omega_{\circ} D_{u_f} F.
	\]
\end{proof}

The next lemma similarly concerns particular blueshifted transport equations which become noshifted after one commutation with $\Omega^{-1} \nablaslash_3$.  Unlike Lemma \ref{lem:noshifteddifferencequotient}, the following lemma involves commuting with $D_{u_f}$, rather than commuting with $\Omega^{-1} \nablaslash_3$.

\begin{lemma}[Estimates for difference quotient for blueshifted transport equations which are noshifted after one commutation] \label{lem:xidifference}
	If $\xi$ is an $S$-tangent $(0,k)$  tensor which satisfies
	\begin{equation} \label{eq:xidifference}
		\Omega \nablaslash_4 \xi - 2(\Omega \omegahat)_{\circ} \xi
		=
		F,
	\end{equation}
	then, for any $u_0 \leq u \leq u_f$, $v_{-1} \leq v \leq v(R_2,u)$,
	\[
		\left\Vert
		D_{u_f} \xi
		\right\Vert^2_{S_{u,v}}
		\lesssim
		\left\Vert
		D_{u_f} \xi
		\right\Vert^2_{S_{u,v_{-1}}}
		+
		\int_{v_{-1}}^v
		(v')^{1+\frac{\delta}{2}}
		\left\Vert
		\xi
		\right\Vert^2_{S_{u_f,v'}}
		dv'
		+
		\int_{v_{-1}}^v
		(v')^{1+\frac{\delta}{2}}
		\Vert
		D_{u_f} F
		\Vert^2_{S_{u,v'}}
		dv'.
	\]
\end{lemma}

\begin{proof}
	The equation \eqref{eq:xidifference} implies that
	\[
		\Omega \nablaslash_4
		D_{u_f} \xi
		=
		\hat{F},
		\qquad
		\text{where},
		\quad
		\hat{F}(u,v)
		=
		2 \xi(u_f,v) D_{u_f} (\Omega \omegahat)_{\circ}(u,v)
		+
		D_{u_f} F(u,v),
	\]
	Hence
	\[
		\partial_v \Vert D_{u_f} \xi \Vert^2_{S_{u,v}}
		=
		2 \int_{S^2} D_{u_f} \xi \cdot \hat{F} d \theta,
	\]
	and,
	\begin{align*}
		\Vert D_{u_f} \xi \Vert^2_{S_{u,v}}
		&
		\leq
		\Vert D_{u_f} \xi \Vert^2_{S_{u,v_{-1}}}
		+
		2
		\int_{v_{-1}}^v 
		\Vert D_{u_f} \xi \Vert_{S_{u,v'}}
		\Vert \hat{F} \Vert_{S_{u,v'}}
		dv'
		\\
		&
		\leq
		\Vert D_{u_f} \xi \Vert^2_{S_{u,v_{-1}}}
		+
		2 \sup_{v_{-1} \leq v' \leq v}
		\Vert D_{u_f} \xi \Vert_{S_{u,v'}}
		\left( \int_{v_{-1}}^v (v')^{-1-\frac{\delta}{2}} d v' \right)^{\frac{1}{2}}
		\left( \int_{v_{-1}}^v (v')^{1+\frac{\delta}{2}} \Vert \hat{F} \Vert_{S_{u,v'}} dv' \right)^{\frac{1}{2}}
		dv'.
	\end{align*}
	The result then follows after dividing by
	$
		\sup_{v_{-1} \leq v' \leq v}
		\Vert D_{u_f} \xi \Vert_{S_{u,v'}},
	$
	taking the supremum over $v_{-1} \leq v' \leq v$ and noting that
	\[
		\frac{r(u,v) - r(u_f,v)}{\Omega_{\circ}(u,v)^2}
		=
		\frac{r(u,v) - r(u_f,v)}{
		\frac{2M(r(u,v) - r(u_f,v))}{r(u_f,v) r(u,v)}
		+
		\Omega_{\circ}(u_f,v)^2
		}
		\leq
		\frac{r(u_f,v) r(u,v)}{2M_f},
	\]
	and so
	\[
		\frac{(\Omega \omegahat)_{\circ}(u,v) - (\Omega \omegahat)_{\circ}(u_f,v)}{\Omega_{\circ}(u,v)^2}
		\lesssim
		1.
	\]
\end{proof}

\subsection{Equations for commuted and renormalised quantities}
\label{section:Hequations}

In this section several equations, which are used in the later sections, are collected.

When estimating $\hat{\chi}$ it is convenient to consider the renormalised quantity\index{double null gauge!connection coefficients!$X_1$, quantity used only in the $\mathcal{H}^+$ gauge}
\begin{equation} \label{eq:X1def}
	X
	=
	X_1
	=
	r\Dslash_2^* r \divslash (r^2 \Omega \hat{\chi})
	-
	\frac{r^3}{2} \Omega^{-1} \nablaslash_3(r\Omega^2 \alpha),
\end{equation}
for which, unlike $\hat{\chi}$, $\alpha$ does not appear as a top order quantity on the right hand side of its evolution equation in the outgoing direction.

Recall the error notation of Section \ref{sec:nlenotation}.

\begin{proposition}[Equation for $X_1$] \label{prop:X1}
	The quantity $X_1$ satisfies
	\[
		\Omega \nablaslash_4 X_1
		-
		2 (\Omega \omegahat)_{\circ} X_1
		=
		\mathcal{L}[X_1]
		+
		\mathcal{E}[X_1],
	\]
	where the linear term $\mathcal{L}[X_1]$ takes the form
	\[
		\mathcal{L}[X_1]
		=
		\frac{r^2 \Omega_{\circ}^2}{2} \Omega^{-1} \nablaslash_3(r \Omega^2 \alpha)
		+
		3M_f r \Omega^2 \alpha,
	\]
	and the nonlinear error $\mathcal{E}[X_1]$ has the schematic form,
	\begin{align*}
		\mathcal{E}[X_1]
		=
		\mathcal{E}^1
		+
		(H^1 \cdot \Phi) \cdot (r \nablaslash)^2 \Omega \hat{\chi}
		+
		(H^2 \cdot\Phi) \cdot
		(r\nablaslash)^2 \Omega \omegahat.
	\end{align*}
	for some vectors of admissible coefficient functions (see \eqref{eq:admis}) $H^j = \{ H_{i}^j \}_{i=1,\ldots,17}$, $j=1,2$ and $H^{k_1k_2} = \{ H_{i}^{k_1k_2} \}_{i=1,\ldots,17}$.
\end{proposition}

\begin{proof}
	Equation \eqref{eq:chihat4} can be written
	\[
		\Omega \nablaslash_4 (r^2 \Omega \hat{\chi})
		-
		2 (\Omega \omegahat)_{\circ} r^2 \Omega\hat{\chi}
		=
		-
		r^2 \Omega^2 \alpha
		+
		E_4[r^2 \Omega \hat{\chi}],
		\quad
		E_4[r^2 \Omega \hat{\chi}]
		=
		2 (\Omega \omegahat - (\Omega \omegahat)_{\circ}) r^2 \Omega \hat{\chi}
		-
		(\Omega \tr \chi - (\Omega \tr \chi)_{\circ}) r^2 \Omega \hat{\chi}.
	\]
	Equation \eqref{eq:chihat4} and equation \eqref{eq:Teukolsky2} then imply that,
	\begin{align*}
		\Omega \nablaslash_4 X_1
		=
		\
		&
		r^2 \Dslash_2^* \divslash \left( 2(\Omega\omegahat)_{\circ} r^2 \Omega \hat{\chi}
		-
		r^2 \Omega^2 \alpha \right)
		-
		\Omega \nablaslash_4 \left( \frac{r^3}{2\Omega^2} \right) \Omega \nablaslash_3 (r\Omega^2 \alpha)
		-
		\frac{r^3}{2\Omega^2} \Omega\nablaslash_4 \Omega\nablaslash_3 (r\Omega^2 \alpha)
		\\
		&
		+
		[\Omega \nablaslash_4, r^2 \Dslash_2^* \divslash](r^2 \Omega \hat{\chi})
		+
		r^2 \Dslash_2^* \divslash E_4[r^2 \Omega \hat{\chi}]
		\\
		=
		\
		&
		2(\Omega \omegahat)_{\circ} X_1
		+
		\left[
		2(\Omega \omegahat)_{\circ}
		\frac{r^3}{2 \Omega^2}
		-
		\Omega \nablaslash_4 \left( \frac{r^3}{2\Omega^2} \right)
		+
		\frac{2r^2}{\Omega^2} \left( 1 - \frac{3M_f}{r} \right)
		\right]
		\Omega \nablaslash_3 (r \Omega^2 \alpha)
		+
		3M_f r \Omega^2 \alpha
		\\
		&
		-
		\frac{r^3}{2} \mathcal{E}^1
		+
		[\Omega \nablaslash_4, r^2 \Dslash_2^* \divslash](r^2 \Omega \hat{\chi})
		+
		r^2 \Dslash_2^* \divslash E_4[r^2 \Omega \hat{\chi}].
	\end{align*}
	The proof follows from the fact that
	\[
		\Omega \nablaslash_4 \left( \frac{r^3}{2\Omega^2} \right)
		=
		\frac{3r^2}{2\Omega^2} \left( 1 - \frac{2M_f}{r} \right)
		-
		\frac{M_fr}{\Omega^2}
		-
		\frac{r^3}{\Omega^2} \left( \Omega \omegahat - (\Omega \omegahat)_{\circ} \right),
	\]
	and so
	\[
		2(\Omega \omegahat)_{\circ}
		\frac{r^3}{2 \Omega^2}
		-
		\Omega \nablaslash_4 \left( \frac{r^3}{2\Omega^2} \right)
		+
		\frac{2r^2}{\Omega^2} \left( 1 - \frac{3M_f}{r} \right)
		=
		\frac{r^2}{2} \frac{\Omega_{\circ}^2}{\Omega^2}
		+
		\frac{r^3}{\Omega^2} \left( \Omega \omegahat - (\Omega \omegahat)_{\circ} \right).
	\]
	It follows from inspecting the form of the error (using also Lemma \ref{lem:commutation}) that the only second order terms involve two angular derivatives of $\Omega \hat{\chi}$, $\Omega \omegahat$ or $\Omega \tr \chi$.  The terms involving $\nablaslash \Omega \tr \chi$ can be replaced by terms involving $\divslash \Omega \hat{\chi}$, plus zeroth order terms, using the Codazzi equation \eqref{eq:Codazzi}.
\end{proof}

Proposition \ref{prop:X1} shows that $X_1$ satisfies a transport equation which contains linear terms involving only one derivative of $\alpha$ on the right hand side (instead of the ``expected'' two derivatives).  Define\index{double null gauge!connection coefficients!$X_2$, quantity used only in the $\mathcal{H}^+$ gauge}
\begin{equation} \label{eq:X2def}
	X_2
	=
	r\Dslash_2^* r \divslash X_1
	+
	\frac{r^4 \Omega^2}{4} (\Omega^{-1} \nablaslash_3)^2 (r \Omega^2 \alpha).
\end{equation}
The following proposition shows that $X_2$ satisfies a transport equation such that the linear terms on the right hand side involve only two derivatives of $\alpha$, instead of the ``expected'' four derivatives.  This ``gap'' of two derivatives will be exploited when estimating $\Omega \hat{\chi}$ in Section \ref{section:Hestimateslgeq2}.

\begin{proposition}[Equation for $X_2$] \label{prop:X2}
	The quantity $X_2$ satisfies
	\[
		\Omega \nablaslash_4 X_2
		-
		2 (\Omega \omegahat)_{\circ} X_2
		=
		\mathcal{L}[X_2]
		+
		\mathcal{E}[X_2],
	\]
	where the linear term $\mathcal{L}[X_2]$ takes the form
	\[
		\mathcal{L}[X_2]
		=
		\sum_{k_1+k_2 \leq 2} H^{k_1k_2} \cdot (\Omega^{-1} \nablaslash_3)^{k_1} (r \nablaslash)^{k_2} (r \Omega^2 \alpha)
	\]
	and the nonlinear error $\mathcal{E}[X_2]$ has the schematic form
	\begin{align*}
		\mathcal{E}[X_2]
		=
		\mathcal{E}^3
		+
		(H^1 \cdot \Phi) \cdot (r \nablaslash)^4 \Omega \hat{\chi}
		+
		(H^2 \cdot\Phi) \cdot
		(r\nablaslash)^4 \Omega \omegahat,
	\end{align*}
	for some vectors of admissible coefficient functions (see \eqref{eq:admis}) $H^j = \{ H_{i}^j \}_{i=1,\ldots,17}$, $j=1,2$ and $H^{k_1k_2} = \{ H_{i}^{k_1k_2} \}_{i=1,\ldots,17}$.
\end{proposition}

\begin{proof}
	The proof is a direct computation using Proposition \ref{prop:X1}, Proposition \ref{prop:Teukolsky} and the commutation formulae of Lemma \ref{lem:commutation}.  Explicitly one obtains,
	\begin{align*}
		\Omega \nablaslash_4 X_2
		-
		2 (\Omega \omegahat)_{\circ} X_2
		=
		&
		-
		\frac{r^2\Omega^2}{2} (\Omega^{-1} \nablaslash_3)^2 (r \Omega^2 \alpha)
		+
		(3M_f-r^2 \Omega^2) r^2 \Dslash_2^* \divslash (r\Omega^2 \alpha)
		\\
		&
		-
		r^4 \Omega \nablaslash_3 \left(
		\frac{1}{r}\left(1 - \frac{3M_f}{r} \right) \Omega^{-1} \nablaslash_3 (r\Omega^2 \alpha)
		\right)
		-
		\frac{r^4}{2} \Omega \nablaslash_3 \left( \frac{3M_f}{r^3} r \Omega^2 \alpha \right)
		\\
		&
		+
		r^3 \Omega^2 \Omega \nablaslash_3 \Omega^{-1} \nablaslash_3(r \Omega^2 \alpha)
		-
		\frac{r^4}{2} \Omega \nablaslash_3 \left( \frac{M_f}{r^2} \Omega^{-1} \nablaslash_3 (r\Omega^2 \alpha) \right)
		+
		\mathcal{E}[X_2],
	\end{align*}
	where the error $\mathcal{E}[X_2]$ has the above schematic form.
\end{proof}

Define
\begin{equation} \label{eq:Xbar1}
	\Xbar_1
	=
	r^2 \Dslash_2^* \divslash (r^2 \Omega \hat{\chibar})
	-
	\frac{r^3}{2 \Omega^2} \Omega \nablaslash_4 (r \Omega^2 \alphabar).
\end{equation}
Similar to the $\nablaslash_4$ equation satisfied by $X_1$, $\Xbar_1$ satisfies a transport equation in the $\nablaslash_3$ direction such that the linear terms on the right hand side involve only first order derivatives of $\alphabar$.

\begin{proposition}[Equation for $\Xbar_1$] \label{prop:Xbar1}
	The quantity $\Xbar_1$ satisfies
	\begin{align*}
		\Omega \nablaslash_3 (\Omega^{-2} \Xbar_1)
		=
		-
		\frac{r^2}{2\Omega^2} \Omega \nablaslash_4 (r \Omega^2 \alphabar)
		+
		3M_f r \alphabar
		+
		\mathcal{E}^2.
	\end{align*}
\end{proposition}

\begin{proof}
	The proof is a direct computation and is similar to the proof of Proposition \ref{prop:X1}, using now equations \eqref{eq:chihat4} and \eqref{eq:Teukolskybar}.
\end{proof}

Recall the quantity $\mu^* = \divslash \eta + (\rho - \rho_{\circ}) - \frac{3}{2r} \left( \Omega \tr \chi - (\Omega \tr \chi)_{\circ}\right)$, which satisfies the following equation which is \emph{noshifted}, but has only nonlinear terms on the right hand side.  

\begin{proposition}[Equation for $\mu^*$] \label{prop:nabla4mustaruf}
	On the final hypersurface $\{ u=u_f\}$, the quantity $\mu^*$ satisfies
	\[
		\partial_v (r^3 \mu^*)
		=
		\mathcal{E}[r^3 \mu^*].
	\]
	and the nonlinear error $\mathcal{E}[r^3 \mu^*]$ has the schematic form
	\begin{align*}
		\mathcal{E}[r^3 \mu^*]
		=
		\mathcal{E}^0
		+
		(H^1 \cdot \Phi) \cdot r \nablaslash \Omega \hat{\chi}
		+
		(H^2 \cdot\Phi) \cdot
		r\nablaslash \eta
	\end{align*}
	for some vectors of admissible coefficient functions (see \eqref{eq:admis}) $H^j = \{ H_{i}^j \}_{i=1,\ldots,17}$, $j=1,2$.
\end{proposition}

\begin{proof}
	First note that the equations \eqref{eq:nabla4eta}, \eqref{eq:rho4}, together with \eqref{eq:ufetaetabar}, imply that, on $\{u=u_f\}$,
	\begin{align*}
		\partial_v (r^3 \divslash \eta)
		=
		&
		- r^3 \divslash (\Omega \beta)
		+
		[\Omega \nablaslash_4, r \divslash] r^2 \eta
		-
		2 r^3 \divslash \left( \Omega \hat{\chi} \cdot \eta \right)
		+
		\mathcal{E}^0,
		\\
		\partial_v (r^3\rho)
		=
		&
		-
		\frac{3r^3}{2} \rho_{\circ} \left( \Omega \tr \chi - (\Omega \tr \chi)_{\circ} \right)
		+
		r^3 \Omega \divslash \beta
		+
		\mathcal{E}^0 
	\end{align*}
	and the Raychaudhuri equation \eqref{eq:Ray} can be written in renormalised form as
	\[
		\partial_v \big( r^2 \left( \Omega \tr \chi - (\Omega \tr \chi)_{\circ} \right) \big)
		=
		2 r^2 (\Omega \omegahat)_{\circ}
		\big(
		\Omega \tr \chi - (\Omega \tr \chi)_{\circ}
		\big)
		+
		\mathcal{E}^0,
	\]
	on $\{u=u_f\}$.  The result then follows after noting the cancellations in the linear terms.
\end{proof}

The following proposition contains the commuted form of equation \eqref{eq:chihat4}.  Note that, in the language of Section \ref{subsubsec:nabla4transport}, the equation is \emph{blueshifted}, \emph{noshifted} or \emph{redshifted} depending on whether it is commuted with zero, one, or two or more $\Omega^{-1} \nablaslash_3$ derivatives respectively.

\begin{proposition}[Commuted equation for $\Omega \hat{\chi}$] \label{prop:chihatHschematic}
	For any $k_1$, $k_2$, $\Omega \hat{\chi}$ satisfies the commuted equation
	\[
		\Omega \nablaslash_4 \left(
		(r \nablaslash)^{k_1} (\Omega^{-1} \nablaslash_3)^{k_2} r^2 \Omega \hat{\chi}
		\right)
		+
		2 (k_2 -1) \Omega \omegahat
		(r \nablaslash)^{k_1} (\Omega^{-1} \nablaslash_3)^{k_2} r^2 \Omega \hat{\chi}
		=
		\mathcal{L}_{k_1k_2} [\Omega \hat{\chi} ]
		+
		\mathcal{E}^{k_1+k_2},
	\]
	where
	\[
		\vert \mathcal{L}_{k_1k_2} [\Omega \hat{\chi} ] \vert
		\lesssim
		\sum_{l_1 + l_2 =0}^{k_1-1}
		\vert (r \nablaslash)^{l_1} (\Omega^{-1} \nablaslash_3)^{l_2} \Omega \hat{\chi} \vert
		+
		\sum_{l_1+l_2=0}^{k_1+k_2}
		\vert (r \nablaslash)^{l_1} (\Omega^{-1} \nablaslash_3)^{l_2} \Omega^2 \alpha \vert.
	\]
	If $k_2 \geq k_1+1$ then the one can replace $\mathcal{E}^{k_1+k_2}$ with $\mathcal{E}^{*k_1+k_2}$.
\end{proposition}

\begin{proof}
	The proof follows by commuting equation \eqref{eq:chihat4}, using Lemma \ref{lem:commutation}.
\end{proof}

Finally, the following proposition collects equations which will be used to estimate $\sigma$.

\begin{lemma}[Equations satisfied by $\sigma$] \label{lem:sigma33sigma44}
	The following schematic equations are satisfied by $\sigma$,
	\begin{equation} \label{eq:sigma44}
		\Omega \nablaslash_4 \left( r^2 \Omega \nablaslash_4 (r^3 \sigma) \right)
		=
		-
		r^5 \curlslash \divslash ( \Omega^2 \alpha)
		+
		2 \Omega \omegahat r^2 \Omega \nablaslash_4(r^3 \sigma)
		+
		\mathcal{E}^1,
	\end{equation}
	and,
	\begin{equation} \label{eq:sigma33}
		\Omega^{-1} \nablaslash_3 \left( r^2 \Omega^{-1} \nablaslash_3(r^3 \sigma) \right)
		=
		r^5 \curlslash \divslash (\Omega^{-2} \alphabar)
		+
		\mathcal{E}^1.
	\end{equation}
\end{lemma}

\begin{proof}
	The Bianchi equation \eqref{eq:sigma4} implies that
	\[
		\Omega \nablaslash_4 (r^3 \sigma)
		=
		- r^3 \curlslash (\Omega \beta) + E_4[r^3\sigma],
	\]
	where
	\[
		E_4[r^3\sigma]
		=
		\frac{r^3}{2} {}^* (\eta + \etabar) \cdot \Omega \beta
		-
		\frac{r^3}{2} \left( \eta + 3 \underline{\eta} \right) \wedge \Omega \beta
		+
		\frac{r^3}{2} \Omega^{-1} \underline{\hat{\chi}} \wedge \Omega^2 \alpha.
	\]
	Similarly, the Bianchi equation \eqref{eq:beta4} implies
	\[
		\Omega \nablaslash_4 \left( \frac{r^4}{\Omega} \beta \right)
		=
		r^4 \divslash \alpha
		+
		E_4 \left[ \frac{r^4}{\Omega} \beta \right],
		\qquad
		E_4 \left[ \frac{r^4}{\Omega} \beta \right]
		=
		\Omega^{-2} r^4 \left(
		\eta^{\sharp} \cdot \alpha
		-
		2(\Omega \tr \chi - (\Omega \tr \chi)_{\circ}) \Omega \beta
		\right),
	\]
	which then gives,
	\begin{multline*}
		\Omega \nablaslash_4 \left( r^5 \curlslash \Omega \beta \right)
		=
		\Omega^2 r^5 \curlslash \divslash \alpha
		-
		2 \Omega \omegahat r^2 \Omega \nablaslash_4 (r^3\sigma)
		+
		2 \Omega \omegahat r^2 E_4[r^3 \sigma]
		+
		\Omega^2 \Omega \nablaslash_4 \left( \frac{r^5}{\Omega^2} {}^* (\eta + \etabar) \cdot \Omega \beta \right)
		\\
		+
		\Omega^2 r \curlslash E_4 \left[ \frac{r^4}{\Omega} \beta\right]
		-
		\Omega^2 [ r \curlslash, \Omega \nablaslash_4] \left( \frac{r^4}{\Omega} \beta \right).
	\end{multline*}
	Equation \eqref{eq:sigma44} then follows.  Equation \eqref{eq:sigma33} follows similarly, now using equations \eqref{eq:sigma3} and \eqref{eq:betabar3}.
\end{proof}

\section{Non-linear error estimates}
\label{nonlinestforH}

The results of this section will be used to control the nonlinear error terms which arise when estimating the geometric quantities in Section \ref{section:Hmainestimates}.  Recall again the pointwise estimates for the geometric quantities in the $\Hp$ gauge of Proposition \ref{prop:pointwiseboundhereimp}, which take the form
\[
	\mathbb P^{N-5}_{u_f}[\Phi^{\mathcal{H}^+}] 
	\lesssim
	\varepsilon
\]
with the pointwise norm $\mathbb P^{N-5}_{u_f}[\Phi^{\mathcal{H}^+}]$ defined in \eqref{eq:Hppointwisenorm}.  Recall also the estimate for the energy $\mathbb{E}^N_{u_f,\Hp}$, defined in \eqref{Hplusmasterenergy}, in the bootstrap assumption \eqref{eq:bamain}, which takes the form
\[
	\mathbb{E}^N_{u_f,\Hp}
	\lesssim
	\varepsilon^2.
\]
These estimates will be used throughout this section.

\subsection{Spacetime error estimates}

The first nonlinear error estimate concerns nonlinear errors estimated in the spacetime $L^2$ norm.

\begin{proposition}[Spacetime nonlinear error estimate] \label{prop:spacetimeerrorH1}
	For $s=0,1,2$, if $k \leq N-s$, then, for any $v_{-1} \leq v \leq v(R_2,u_f)$,
	\[
		\Vert
		\mathcal{E}^k
		\Vert_{\DRH(v)}^2
		\lesssim
		\frac{\varepsilon^4}{v^{\kappa(s)}},
	\]
	where $\kappa(0) = 1-\delta$, $\kappa(1) = 2$ and $\kappa(2) = 3$.
\end{proposition}

\begin{proof}
	Consider a nonlinear term of the form $\mathfrak{D}^{\gamma_1} \Phi^{(1)} \cdot \mathfrak{D}^{\gamma_2} \Phi^{(2)}$ for some $\vert \gamma_1 \vert + \vert \gamma_2 \vert \leq N-s$ (the cubic and higher order terms are easier to estimate).  Without loss of generality assume $\vert \gamma_2 \vert \geq \vert \gamma_1 \vert$.  Suppose first that $\mathfrak{D}^{\gamma_2} \Phi^{(2)} \neq (r\nablaslash)^{N-s} \Omega \hat{\chi}$ and $\mathfrak{D}^{\gamma_2} \Phi^{(2)} \neq (r\nablaslash)^{N-s} \Omega \tr \chi$.  Then
	\[
		\Vert \mathfrak{D}^{\gamma_2}( \Phi^{(2)} - \Phi^{(2)}_{\mathrm{Kerr}}) \Vert_{\DRH(v)}^2
		\lesssim
		\frac{\varepsilon^2}{v^s}.
	\]
	Now, since $\vert \mathfrak{D}^{\gamma_1} \Phi^{(1)} \vert \lesssim \varepsilon v^{-1}$,
	\[
		\Vert \mathfrak{D}^{\gamma_1} \Phi^{(1)} \cdot \mathfrak{D}^{\gamma_2} \Phi^{(2)} \Vert_{\DRH(v)}^2
		\lesssim
		\frac{\varepsilon^2}{v^2}
		\Big(
		\Vert \Phi^{(2)}_{\mathrm{Kerr}} \Vert_{\DRH(v)}^2
		+
		\Vert \mathfrak{D}^{\gamma_2}( \Phi^{(2)} - \Phi^{(2)}_{\mathrm{Kerr}}) \Vert_{\DRH(v)}^2
		\Big),
	\]
	and the result follows from the fact that $\Vert \Phi^{(2)}_{\mathrm{Kerr}} \Vert_{\DRH(v)}^2 \lesssim \varepsilon^2 v^{-1}$.
	
	Suppose now that $\mathfrak{D}^{\gamma_2} \Phi^{(2)} = (r\nablaslash)^{N-s} \Omega \hat{\chi}$ or $(r\nablaslash)^{N-s} \Omega \tr \chi$.  Consider first the case that $s=0$.  Then
	\[
		\Vert \mathfrak{D}^{\gamma_2} \Phi^{(2)} \Vert^2_{S_{u,v}}
		\lesssim
		\varepsilon^2 v^{\delta},
	\]
	and
	\[
		\Vert \mathfrak{D}^{\gamma_1} \Phi^{(1)} \cdot \mathfrak{D}^{\gamma_2} \Phi^{(2)} \Vert_{\DRH(v)}^2
		\lesssim
		\int_{\DRH(v)} \varepsilon^4 (v')^{\delta-2} \Omega^2_{\circ} du dv
		\lesssim
		\varepsilon^4 v^{\delta-1}.
	\]
	The cases $s=1$ and $s=2$ are similar, using the fact that
	\[
		\Vert \mathfrak{D}^{\gamma_2} \Phi^{(2)} \Vert^2_{S_{u,v}}
		\lesssim
		\varepsilon^2,
		\text{ if } s=1,
		\qquad
		\Vert \mathfrak{D}^{\gamma_2} \Phi^{(2)} \Vert^2_{S_{u,v}}
		\lesssim
		\varepsilon^2 v^{-1},
		\text{ if } s=2.
	\]
\end{proof}

\subsection{Error estimates on spheres}

In this section nonlinear error terms are estimate on spheres.

\begin{proposition}[Nonlinear error estimate on spheres] \label{prop:sphereserrorH}
	For $s=0,1,2$, if $k \leq N-1-s$ then, for any $u_0 \leq u \leq u_f$, $v_{-1} \leq v \leq v(R_2,u)$,
	\[
		\left\Vert
		\mathcal{E}^k
		\right\Vert^2_{S_{u,v}^{\Hp}}
		\lesssim
		\frac{\varepsilon^4}{v^{2+s}}.
	\]
\end{proposition}

\begin{proof}
	The proof is an immediate consequence of the fact that each $\Phi$ satisfies, for $s=0,1,2$,
	\[
		\sum_{\vert \gamma \vert \leq N-1-s}
		\Vert \mathfrak{D}^{\gamma} \Phi \Vert_{S_{u,v}}^2
		\lesssim
		\frac{\varepsilon^2}{v^s}.
	\]
\end{proof}

An extra derivative can be controlled if the nonlinear error involves only Ricci coefficients, and no curvature components.

\begin{proposition}[Nonlinear error estimate on spheres] \label{prop:sphereserrorGammaH}
	For $s=0,1,2$, if $\vert \gamma \vert \leq N-s$ then, for any $\Gamma$ and $\widetilde{\Gamma}$ and any $u_0 \leq u \leq u_f$, $v_{-1} \leq v \leq v(R_2,u)$,
	\[
		\left\Vert
		\mathfrak{D}^{\gamma} \big( \Gamma \cdot \widetilde{\Gamma} \big)
		\right\Vert_{S_{u,v}^{\Hp}}^2
		\lesssim
		\frac{\varepsilon^4}{v^{1+\kappa(s)}},
	\]
	where $\kappa(0) = 1-\delta$, $\kappa(1) = 2$ and $\kappa(2) = 3$.
\end{proposition}

\begin{proof}
	The proof is an immediate consequence of the fact that, for $s=0,1,2$, for each $\Gamma$,
	\[
		\sum_{\vert \gamma \vert \leq N-s}
		\Vert \mathfrak{D}^{\gamma} \Gamma \Vert_{S_{u,v}^{\Hp}}^2
		\lesssim
		\frac{\varepsilon^2}{v^{\kappa(s) - 1}}.
	\]
\end{proof}

\subsection{Error estimates on null hypersurfaces}

In this section, nonlinear error terms are estimated in $L^2$ on null hypersurfaces.  The first proposition concerns outgoing null hypersurfaces.

\begin{proposition}[Nonlinear error estimate on outgoing null hypersurfaces] \label{prop:errorout}
	Given $s=0,1,2$ and $k \leq N-s$, let $\mathcal{E}^k$ be a nonlinear error which does not involve the top order quantity $(\Omega^{-1} \nablaslash_3)^k \Omega^{-2} \alphabar$.  Then, for any $u_0 \leq u \leq u_f$, $v_{-1} \leq v \leq v(R_2,u)$,
	\[
		\Vert \mathcal{E}^k \Vert_{C_u^{\Hp}(v)}^2 \lesssim \frac{\varepsilon^4}{v^{\kappa(s)}},
	\]
	where $\kappa(0) = 1-\delta$, $\kappa(1) = 2$ and $\kappa(2) = 3$.
\end{proposition}

\begin{proof}
	The proof is similar to that of Proposition \ref{prop:spacetimeerrorH1} since, for all $\vert \gamma \vert \leq N-s$ and $\mathfrak{D}^{\gamma} \Phi \neq (\Omega^{-1} \nablaslash_3)^{N-s} \Omega^{-2} \alphabar$, $(r\nablaslash)^{N-s} \Omega \hat{\chi}$ or $(r\nablaslash)^{N-s} \Omega \tr \chi$,
	\[
		\Vert
		\mathfrak{D}^{\gamma} ( \Phi - \Phi_{\mathrm{Kerr}})
		\Vert_{C_{u}^{\Hp}(v)}^2
		\lesssim
		\frac{\varepsilon^2}{v^s}.
	\]
	Terms of the form $\mathfrak{D}^{\gamma_1} \Phi^{(1)} \cdot \mathfrak{D}^{\gamma_2} \Phi^{(2)}$ for $\mathfrak{D}^{\gamma_2} \Phi^{(2)} =(r\nablaslash)^{N-s} \Omega \hat{\chi}$ or $(r\nablaslash)^{N-s} \Omega \tr \chi$ are again estimated separately, exactly as in the proof of Proposition \ref{prop:spacetimeerrorH1}.
\end{proof}

The next proposition concerns nonlinear error terms on incoming null hypersurfaces.

\begin{proposition}[Nonlinear error estimate on incoming null hypersurfaces] \label{prop:errorin}
	Given $s=0,1,2$ and $k \leq N-s$, let $\mathcal{E}^k$ be a nonlinear error which does not involve the top order quantity $(r \Omega \nablaslash_4)^k \Omega^2 \alpha$.  Then, for any $v_{-1} \leq v \leq v(R_2,u_f)$,
	\[
		\Vert \mathcal{E}^k \Vert_{\Cbar_v^{\Hp}}^2 \lesssim \frac{\varepsilon^4}{v^{\kappa(s)}},
	\]
	where $\kappa(0) = 1-\delta$, $\kappa(1) = 2$ and $\kappa(2) = 3$.
\end{proposition}

\begin{proof}
	The proof is again similar to that of Proposition \ref{prop:spacetimeerrorH1} since, for all $\vert \gamma \vert \leq N-s$ and $\mathfrak{D}^{\gamma} \Phi \neq (\Omega^{-1} \nablaslash_3)^{N-s} \Omega^{-2} \alphabar$, $(r\nablaslash)^{N-s} \Omega \hat{\chi}$ or $(r\nablaslash)^{N-s} \Omega \tr \chi$,
	\[
		\Vert
		\mathfrak{D}^{\gamma} ( \Phi - \Phi_{\mathrm{Kerr}})
		\Vert_{\Cbar_{v}^{\Hp}}^2
		\lesssim
		\frac{\varepsilon^2}{v^s}.
	\]
	Terms of the form $\mathfrak{D}^{\gamma_1} \Phi^{(1)} \cdot \mathfrak{D}^{\gamma_2} \Phi^{(2)}$ for $\mathfrak{D}^{\gamma_2} \Phi^{(2)} =(r\nablaslash)^{N-s} \Omega \hat{\chi}$ or $(r\nablaslash)^{N-s} \Omega \tr \chi$ are again estimated separately, exactly as in the proof of Proposition \ref{prop:spacetimeerrorH1}.
\end{proof}

On the final hypersurface $\{u_{\Hp} =u_f\}$, nonlinear error terms involving even higher order derivatives of certain Ricci coefficients can be controlled.  Such estimates are used in Section \ref{subsec:ufH}.

\begin{proposition}[Nonlinear error estimate on the final outgoing hypersurfaces]  \label{prop:nonlinerroruf}
	On the final hypersurface $\{u=u_f\}$, for $s=0,1$, for any $\Phi$, any $\xi = \Omega \hat{\chi}, \Omega \tr \chi - (\Omega \tr \chi)_{\circ}, \eta - \eta_{\mathrm{Kerr}}$, any $k \leq N+1-s$, and any $v_{-1} \leq v \leq v(R_2,u_f)$,
	\[
		\Big(
		\int_{C_u(v)}
		\vert \Phi \cdot (r\nablaslash)^{k} \xi \vert
		d \theta d v'
		\Big)^2
		\lesssim
		\frac{\varepsilon^4}{v^{2+s}}.
	\]
\end{proposition}

\begin{proof}
	The proof is an immediate consequence of the Cauchy--Schwarz inequality, the fact that $\vert \Phi \vert \lesssim \varepsilon v^{-1}$ for all $\Phi$, and
	\[
		\sum_{\vert \gamma \vert \leq N+1-s}
		\Vert
		(r\nablaslash)^{k} \xi
		\Vert_{C_{u_f}^{\Hp}(v)}^2
		\lesssim
		\frac{\varepsilon^2}{v^s},
	\]
	for $\xi = \Omega \hat{\chi}, \Omega \tr \chi - (\Omega \tr \chi)_{\circ}, \eta - \eta_{\mathrm{Kerr}}$ and $s=0,1,2$.
\end{proof}

The following proposition exploits the fact that nonlinear error terms have faster decay if they do not contain quadratic terms for which both factors are non-vanishing in the reference linearised Kerr solution.  Note that such terms for which one of the factors is $\Omega \beta$ are allowed in the following proposition since, in the reference linearised Kerr solution, $\Omega \beta_{\mathrm{Kerr}}$ behaves like $\Omega^2$.

\begin{proposition}[Nonlinear error estimate on outgoing null hypersurfaces] \label{prop:newnoshifterrorestimate}
	Given $s=0,1,2$ and $k \leq N-1-s$, let $\mathcal{E}^{k}$ be a nonlinear error which moreover does not contain any quadratic terms of the form
	\[
		\mathfrak{D}^{\gamma_1} \Phi^{(1)} \cdot \mathfrak{D}^{\gamma_2} \Phi^{(2)},
		\qquad
		\Phi^{(1)}, \Phi^{(2)}
		\in
		\{ \eta, \etabar, \sigma, \Omega^{-1} \betabar \}.
	\]
	Then, for any $u_0 \leq u \leq u_f$, $v_{-1} \leq v \leq v(R_2,u)$,
	\begin{equation} \label{eq:newnoshifterrorestimate}
		\Big(
		\int_{C_u(v)}
		\vert \mathcal{E}^k \vert
		d \theta d v'
		\Big)^2
		\lesssim
		\frac{\varepsilon^4}{v^{1+s}},
		\qquad
		\text{and}
		\qquad
		\Vert \mathcal{E}^k \Vert_{C_u^{\Hp}(v)}^2 \lesssim \frac{\varepsilon^4}{v^{2+s}}.
	\end{equation}
\end{proposition}

\begin{proof}
	Consider a quadratic term $\mathfrak{D}^{\gamma_1} \Phi^{(1)} \cdot \mathfrak{D}^{\gamma_2} \Phi^{(2)}$ and suppose that $\Phi^{(1)}_{\mathrm{Kerr}} = 0$.  Then
	\[
		\Vert
		\mathfrak{D}^{\gamma_1} \Phi^{(1)}
		\Vert_{C_u(v)}^2
		\lesssim
		\varepsilon^2 v^{-s},
		\qquad
		\Vert
		\mathfrak{D}^{\gamma_2} (\Phi^{(2)} - \Phi^{(2)}_{\mathrm{Kerr}})
		\Vert_{C_u(v)}^2
		\lesssim
		\varepsilon^2 v^{-s},
	\]
	for $\vert \gamma_1 \vert \leq N-1-s$ and $\vert \gamma_2 \vert \leq N-1-s$ respectively.  Hence, by Cauchy--Schwarz,
	\begin{align*}
		\Big(
		\int_{C_u(v)}
		\vert \mathfrak{D}^{\gamma_1} \Phi^{(1)} \cdot \mathfrak{D}^{\gamma_2} \Phi^{(2)} \vert
		d \theta d v'
		\Big)^2
		&
		\lesssim
		\Big(
		\int_{C_u(v)}
		\frac{\varepsilon}{u_f} \vert \mathfrak{D}^{\gamma_1} \Phi^{(1)} \vert
		+
		\vert \mathfrak{D}^{\gamma_1} \Phi^{(1)} \cdot \mathfrak{D}^{\gamma_2} (\Phi^{(2)} - \Phi^{(2)}_{\mathrm{Kerr}}) \vert
		d \theta d v'
		\Big)^2
		\\
		&
		\lesssim
		\Big(
		\frac{\varepsilon^2}{u_f}
		+
		\Vert \mathfrak{D}^{\gamma_2} (\Phi^{(2)} - \Phi^{(2)}_{\mathrm{Kerr}}) \Vert_{C_u(v)}^2
		\Big)
		\Vert \mathfrak{D}^{\gamma_1} \Phi^{(1)} \Vert_{C_u(v)}^2
		\lesssim
		\frac{\varepsilon^4}{v^{s+1}}.
	\end{align*}
	For the remaining case, when $\Phi^{(1)} = \Omega \beta$, one estimates the term arising from $\mathfrak{D}^{\gamma_1} (\Phi^{(1)} - \Phi^{(1)}_{\mathrm{Kerr}})$ exactly as above.  Now $\vert \Phi^{(1)}_{\mathrm{Kerr}} \vert \lesssim \Omega_{\circ}^2 \varepsilon (u_f)^{-1}$ and so
	\begin{align*}
		\Big(
		\int_{C_u(v)}
		\vert \Phi^{(1)}_{\mathrm{Kerr}} \cdot \mathfrak{D}^{\gamma_2} \Phi^{(2)} \vert
		d \theta d v'
		\Big)^2
		&
		\lesssim
		\Big(
		\int_{C_u(v)}
		\frac{\varepsilon \Omega_{\circ}^2}{u_f} 
		\big(
		\frac{\varepsilon}{u_f} 
		+
		\vert \mathfrak{D}^{\gamma_2} (\Phi^{(2)} - \Phi^{(2)}_{\mathrm{Kerr}}) \vert
		\big)
		d \theta d v'
		\Big)^2
		\\
		&
		\lesssim
		\Big(
		\frac{\varepsilon^2}{u_f}
		+
		\Vert \mathfrak{D}^{\gamma_2} (\Phi^{(2)} - \Phi^{(2)}_{\mathrm{Kerr}}) \Vert_{C_u(v)}^2
		\Big)
		\frac{\varepsilon^2}{(u_f)^2}
		\Vert \Omega_{\circ}^2 \Vert_{C_u(v)}^2
		\lesssim
		\frac{\varepsilon^4}{v^{s+1}}.
	\end{align*}
	The first of \eqref{eq:newnoshifterrorestimate} then follows.  The proof of the second is similar.
\end{proof}

\subsection{Difference quotient nonlinear error estimates for $\Omega \omegahat - (\Omega \omegahat)_{\circ}$}

The term $\Omega \omegahat - (\Omega \omegahat)_{\circ}$ appears at top order as an error in the $\nablaslash_4 X_2$ equation (see Section \ref{section:Hequations}) and so the following will be used in estimating the difference quotient $D_{u_f} X_2$.

\begin{proposition}[Difference quotient nonlinear error estimates for $\Omega \omegahat - (\Omega \omegahat)_{\circ}$] \label{prop:omegahatdifferenceH2}
	For any $k \leq N-s$, for $s=0,1,2$, and for any $\Phi$, the difference quotient $D_{u_f} \big(
		\Phi \cdot (r\nablaslash)^k (\Omega \omegahat - (\Omega \omegahat)_{\circ})
		\big)$ satisfies, for all $u_0 \leq u \leq u_f$, $v_{-1} \leq v \leq v(R_2,u)$,
	\[
		\Vert
		D_{u_f} \big(
		\Phi \cdot (r\nablaslash)^k (\Omega \omegahat - (\Omega \omegahat)_{\circ})
		\big)
		\Vert_{C_u(v)}^2
		+
		\Vert
		D_{u_f} \big(
		\Phi \cdot (r\nablaslash)^k (\Omega \omegahat - (\Omega \omegahat)_{\circ})
		\big)
		\Vert_{\DRH(v)}^2
		\lesssim
		\frac{\varepsilon^4}{v^{2+s}}.
	\]
\end{proposition}

\begin{proof}
	Note first that
	\begin{multline*}
		D_{u_f} \left( \Phi \cdot (r\nablaslash)^k (\Omega \omegahat - (\Omega \omegahat)_{\circ}) \right) (u,v)
		\\
		=
		D_{u_f}\Phi(u,v) \cdot (r\nablaslash)^k (\Omega \omegahat - (\Omega \omegahat)_{\circ}) (u_f,v)
		+
		D_{u_f} (r\nablaslash)^k (\Omega \omegahat - (\Omega \omegahat)_{\circ}) (u,v)
		\cdot
		\Phi(u,v).
	\end{multline*}
	Equation \eqref{eq:omega3omegabar4} and Lemma \ref{lem:commutation} imply that
	\[
		\Omega^{-1} \nablaslash_3 (r\nablaslash)^k (\Omega \omegahat - (\Omega \omegahat)_{\circ})
		=
		(r\nablaslash)^k (\rho - \rho_{\circ})
		+
		\mathcal{E}^{*k},
	\]
	and so it follows from Proposition \ref{lem:nabla3differenceestimate} that
	\[
		\Vert
		D_{u_f} \big(
		\Phi \cdot (r\nablaslash)^k (\Omega \omegahat - (\Omega \omegahat)_{\circ})
		\big)
		\Vert_{C_u(v)}^2
		\lesssim
		\sup_{u \leq u' \leq u_f}
		\int_v^{v(R_2,u)} \int_{S_{u',v'}}
		\vert
		\mathcal{E}^{*k}
		\vert^2
		d\theta d v'
		\lesssim
		\frac{\varepsilon^4}{v^{2+s}}.
	\]
	The estimate on $\DRH(v)$ is similar, using now the fact that the estimate \eqref{eq:nabla3differenceestimate1} implies that, denoting 
	\[
		W_k = (r\nablaslash)^k (\Omega \omegahat - (\Omega \omegahat)_{\circ})
	\]
	\begin{align*}
		\Vert 
		\Phi \cdot D_{u_f} W_k
		\Vert_{\DRH(v)}^2
		&
		\lesssim
		\int_u^{u_f} \int_v^{v(R_2,u')} \int_{S_{u',v'}}
		\vert \Phi \vert^2 \Omega^{-4}
		\Big(
		\int_{u'}^{u_f}
		(
		\vert \Omega^{-1} \nablaslash_3 W_k \vert
		+
		\vert W_k \vert
		)
		\Omega^2 du''
		\Big)^2
		\Omega^2 d\theta dv' du'
		\\
		&
		\lesssim
		\frac{\varepsilon^2}{v^2}
		\int_u^{u_f} \int_v^{v(R_2,u')} \int_{S_{u',v'}}
		\int_{u'}^{u_f}
		(
		\vert \Omega^{-1} \nablaslash_3 W_k \vert^2
		+
		\vert W_k \vert^2
		)
		\Omega^2 du''
		d\theta dv' du'
		\\
		&
		\lesssim
		\frac{\varepsilon^2}{v^2}
		\sup_{u \leq u' \leq u_f}(
		\Vert \Omega^{-1} \nablaslash_3 W_k \Vert^2_{C_{u'}(v)}
		+
		\Vert W_k \Vert^2_{C_{u'}(v)}
		),
	\end{align*}
	where the fact that
	\[
		\int_u^{u_f} \int_{u'}^{u_f} \Omega(u'',v')^2 du'' du' \lesssim 1,
	\]
	for all $v'$ has been used.  The term $D_{u_f}\Phi \cdot (r\nablaslash)^k (\Omega \omegahat - (\Omega \omegahat)_{\circ})$ is estimated similarly.
\end{proof}

\section{The proof of Theorem~\ref{thm:Hestimates}: transport and elliptic estimates}
\label{section:Hmainestimates}

This section concerns the main part of the proof of Theorem~\ref{thm:Hestimates}.
In Section \ref{subsec:proofofHestimates} the proof of Theorem~\ref{thm:Hestimates} is outlined and reduced to the main statements of the following sections.  
Some of the geometric quantities are \emph{almost gauge invariant} and so can be estimated directly, using only elliptic estimates, without directly exploiting any of the defining properties of the $\Hp$ gauge, from Theorem~\ref{thm:alphaalphabarestimates}.  These quantities are estimated in Section \ref{section:gaugeinvariantH}.  The most involved part of the proof of Theorem~\ref{thm:Hestimates} concerns the estimates for the $\ell \geq 2$ modes of the geometric quantities.  The $\ell \geq 2$ modes are first estimated, via transport and elliptic estimates, on the hypersurfaces $C_{u_f}$ and $\Cbar_{v_{-1}}$ in Section \ref{section:finalinitialH}, and then in the spacetime region $\DRH$ in Section \ref{section:Hestimateslgeq2}.  The $\ell =0,1$ modes of the geometric quantities are then estimated on the hypersurfaces $C_{u_f}$ and $\Cbar_{v_{-1}}$ in Section \ref{subsec:Hell0ell1ufv1}, and then in the spacetime region $\DRH$ in Section \ref{subsec:Hell0ell1}.  Section \ref{subsec:Hmetric} concerns estimates for the metric components.

\subsection{Outline of the proof of Theorem~\ref{thm:Hestimates}}
\label{subsec:proofofHestimates}

In this section the proof of Theorem~\ref{thm:Hestimates} is reduced to Theorems \ref{prop:finalHenergy}, \ref{prop:ellgeq2Henergy}, \ref{prop:ell1Henergy}, \ref{prop:ell0Henergy}, and \ref{prop:metricHenergy}, whose proofs are given in Sections \ref{section:gaugeinvariantH}--\ref{subsec:Hmetric}.

Recall the energy $\mathbb{E}^N_{u_f,\Hp}$ from Section \ref{energiessection},
\begin{align*}
	\mathbb{E}^N_{u_f,\Hp}
	:=
	\
	&
	\mathbb{E}^N[\DRH]
	+
	\mathbb{E}^N[C^{\Hp}]
	+
	\mathbb{E}^N[\Cbar^{\Hp}]
	+
	\mathbb{E}^N[S^{\Hp}]
	+
	\mathbb{E}^N[\chi^{\Hp}]
	+
	\mathbb{E}^{N+1}[C_{u_f}^{\Hp}]
	+
	\mathbb{E}^N[g^{\Hp}]
	\\
	&
	+
	(u_f)^4 \sum_{m=-1}^1 \vert J^m_{\Hp} - J^m_{\I} \vert^2
	.
\end{align*}
The $\ell \geq 2$, $\ell = 1$ and $\ell=0$ modes of the geometric quantities are estimated separately.  Accordingly, recall
\[
	\mathcal{A}_{\mathcal{R}}
	=
	\big\{
	(\Omega\beta - \Omega\beta_{\rm Kerr})^{\Hp},
	(\Omega^{-1}\betabar - \Omega^{-1}\betabar_{\rm Kerr})^{\Hp},
	(\rho - \rho_{\circ})^{\Hp},
	(\sigma - \sigma_{\rm Kerr})^{\Hp}
	\big\},
\]
\begin{multline*}
	\mathcal{A}_{\Gamma}
	=
	\big\{
	(\Omega_{\circ}^{-2} \Omega^2 - 1)^{\Hp},
	\Omega^{-1} \hat{\chibar}^{\Hp},
	\Omega^{-2} (\Omega \tr \chibar - \Omega \tr \chibar_{\circ})^{\Hp},
	(\eta - \eta_{\rm Kerr})^{\Hp},
	(\etabar - \etabar_{\rm Kerr})^{\Hp},
	\\
	(\Omega \omegahat - \Omega \omegahat_{\circ})^{\Hp},
	\Omega^{-2} (\Omega \omegabarhat - \Omega \omegabarhat)^{\Hp}
	\big\}.
\end{multline*}
\[
	\mathcal{A}_{\chi}
	=
	\{
	\Omega\hat{\chi}^{\Hp}, (\Omega \tr \chi - \Omega \tr \chi_{\circ})^{\Hp}
	\},
\]
and recall that elements of $\mathcal{A}_{\mathcal{R}}$ and $\mathcal{A}_{\Gamma}$ are denoted $\breve{\mathcal{R}}$ and $\breve{\Gamma}$ respectively, with a $\breve{}$ added to emphasise the fact that the linearised Kerr values have been subtracted.
Define the spacetime energies
\begin{align}
	\mathbb{E}^N_{\ell \geq 2}[\DRH]
	:=
	\
	&
	\sup_{\substack{
	v_{-1} \leq v \leq v(R_2,u_f)}}
	\sum_{s=0,1,2}
	v^{s}
	\bigg(
	\sum_{\vert \gamma \vert \leq N-s}
	\sum_{\breve{\mathcal{R}} \in \mathcal{A}_{\mathcal{R}}}
	\Vert (1-3M_f/r) \mathfrak{D}^{\gamma} \breve{\mathcal{R}}_{\ell \geq 2} \Vert^2_{\DRH(v)}
	\label{eq:spacetimeHenergy1}
	\\
	&
	+
	\sum_{\vert \gamma \vert \leq N-1-s}
	\sum_{\breve{\mathcal{R}} \in \mathcal{A}_{\mathcal{R}}}
	\Vert \mathfrak{D}^{\gamma} \breve{\mathcal{R}}_{\ell \geq 2} \Vert^2_{\DRH(v)}
	+
	\sum_{\vert \gamma \vert \leq N-s}
	\sum_{\breve{\Gamma} \in \mathcal{A}_{\Gamma}}
	\Vert \mathfrak{D}^{\gamma} \breve{\Gamma}_{\ell \geq 2} \Vert^2_{\DRH(v)}
	\bigg),
	\nonumber
	\\
	\mathbb{E}^N_{\ell = 1}[\DRH]
	:=
	\
	&
	\sup_{\substack{
	v_{-1} \leq v \leq v(R_2,u_f)}}
	\sum_{s=0,1,2}
	v^{s}
	\sum_{\vert \gamma \vert \leq N-s}
	\sum_{\breve{\Phi} \in \mathcal{A}_{\mathcal{R}} \cup \mathcal{A}_{\Gamma} \cup \mathcal{A}_{\chi}}
	\Vert \mathfrak{D}^{\gamma} \breve{\Phi}_{\ell = 1} \Vert^2_{\DRH(v)}
	\label{eq:spacetimeHenergy2}
	\\
	\mathbb{E}^N_{\ell = 0}[\DRH]
	:=
	\
	&
	\sup_{\substack{
	v_{-1} \leq v \leq v(R_2,u_f)}}
	\sum_{s=0,1,2}
	v^{s}
	\sum_{\vert \gamma \vert \leq N-s}
	\sum_{\breve{\Phi} \in \mathcal{A}_{\mathcal{R}} \cup \mathcal{A}_{\Gamma} \cup \mathcal{A}_{\chi}}
	\Vert \mathfrak{D}^{\gamma} \breve{\Phi}_{\ell = 0} \Vert^2_{\DRH(v)},
	\label{eq:spacetimeHenergy3}
\end{align}
and the null cone energies
\begin{align}
	\mathbb{E}^N_{\ell \geq 2}[C^{\Hp}]
	&
	:=
	\sup_{\substack{
	v_{-1} \leq v \leq v(R_2,u_f)}}
	\sum_{s=0,1,2}
	\sum_{\vert \gamma \vert \leq N-s}
	v^{s}
	\sup_{u_{-1} \leq u \leq u_f}
	\sum_{\breve{\Phi} \in \mathcal{A}_{\mathcal{R}} \cup \mathcal{A}_{\Gamma}}
	\Vert \mathfrak{D}^{\gamma} \breve{\Phi}_{\ell \geq 2} \Vert_{C^{\Hp}_u(v)}^2,
\label{eq:outgoingHenergy1}
\\
	\mathbb{E}^N_{\ell = 1}[C^{\Hp}]
	&
	:=
	\sup_{\substack{
	v_{-1} \leq v \leq v(R_2,u_f)}}
	\sum_{s=0,1,2}
	\sum_{\vert \gamma \vert \leq N-s}
	v^{s}
	\sup_{u_{-1} \leq u \leq u_f}
	\sum_{\breve{\Phi} \in \mathcal{A}_{\mathcal{R}} \cup \mathcal{A}_{\Gamma} \cup \mathcal{A}_{\chi}}
	\Vert \mathfrak{D}^{\gamma} \breve{\Phi}_{\ell = 1} \Vert_{C^{\Hp}_u(v)}^2,
\label{eq:outgoingHenergy2}
\\
	\mathbb{E}^N_{\ell =0}[C^{\Hp}]
	&
	:=
	\sup_{\substack{
	v_{-1} \leq v \leq v(R_2,u_f)}}
	\sum_{s=0,1,2}
	\sum_{\vert \gamma \vert \leq N-s}
	v^{s}
	\sup_{u_{-1} \leq u \leq u_f}
	\sum_{\breve{\Phi} \in \mathcal{A}_{\mathcal{R}} \cup \mathcal{A}_{\Gamma} \cup \mathcal{A}_{\chi}}
	\Vert \mathfrak{D}^{\gamma} \breve{\Phi}_{\ell = 0} \Vert_{C^{\Hp}_u(v)}^2,
\label{eq:outgoingHenergy3}
\end{align}
and
\begin{align}
	\mathbb{E}^N_{\ell \geq 2}[\Cbar^{\Hp}]
	&
	:=
	\sup_{\substack{
	v_{-1} \leq v \leq v(R_2,u_f)}}
	\sum_{s=0,1,2}
	\sum_{\vert \gamma \vert \leq N-s}
	v^{s}
	\sum_{\breve{\Phi} \in \mathcal{A}_{\mathcal{R}} \cup \mathcal{A}_{\Gamma}}
	\Vert \mathfrak{D}^{\gamma} \breve{\Phi}_{\ell \geq 2} \Vert_{\Cbar^{\Hp}_v}^2,
	\label{eq:incomingHenergy1}
	\\
	\mathbb{E}^N_{\ell = 1}[\Cbar^{\Hp}]
	&
	:=
	\sup_{\substack{
	v_{-1} \leq v \leq v(R_2,u_f)}}
	\sum_{s=0,1,2}
	\sum_{\vert \gamma \vert \leq N-s}
	v^{s}
	\sum_{\breve{\Phi} \in \mathcal{A}_{\mathcal{R}} \cup \mathcal{A}_{\Gamma}}
	\Vert \mathfrak{D}^{\gamma} \breve{\Phi}_{\ell = 1} \Vert_{\Cbar^{\Hp}_v}^2,
	\label{eq:incomingHenergy2}
	\\
	\mathbb{E}^N_{\ell = 0}[\Cbar^{\Hp}]
	&
	:=
	\sup_{\substack{
	v_{-1} \leq v \leq v(R_2,u_f)}}
	\sum_{s=0,1,2}
	\sum_{\vert \gamma \vert \leq N-s}
	v^{s}
	\sum_{\breve{\Phi} \in \mathcal{A}_{\mathcal{R}} \cup \mathcal{A}_{\Gamma}}
	\Vert \mathfrak{D}^{\gamma} \breve{\Phi}_{\ell = 0} \Vert_{\Cbar^{\Hp}_v}^2,
	\label{eq:incomingHenergy3}
\end{align}
the energies on spheres
\begin{align}
	\mathbb{E}^N_{\ell \geq 2}[S^{\Hp}]
	:=
	\
	&
	\sup_{\DRH}
	\sum_{s=0,1,2}
	v^s
	\Big(
	\sum_{\vert \gamma \vert \leq N-1-s}
	\sum_{\breve{\mathcal{R}} \in \mathcal{A}_{\mathcal{R}}}
	\Vert \mathfrak{D}^{\gamma} \breve{\mathcal{R}}_{\ell \geq 2} \Vert_{S_{u,v}^{\Hp}}^2
	+
	\sum_{\vert \gamma \vert \leq N-s}
	\sum_{\breve{\Gamma} \in \mathcal{A}_{\Gamma}}
	\Vert \mathfrak{D}^{\gamma} \breve{\Gamma}_{\ell \geq 2} \Vert_{S_{u,v}^{\Hp}}^2
	\Big),
	\label{eq:spheresHenergy1}
	\\
	\mathbb{E}^N_{\ell = 1}[S^{\Hp}]
	:=
	\
	&
	\sup_{\DRH}
	\sum_{s=0,1,2}
	v^s
	\sum_{\vert \gamma \vert \leq N-s}
	\sum_{\breve{\Phi} \in \mathcal{A}_{\mathcal{R}} \cup \mathcal{A}_{\Gamma} \cup \mathcal{A}_{\chi}}
	\Vert \mathfrak{D}^{\gamma} \breve{\Phi}_{\ell = 1} \Vert_{S_{u,v}^{\Hp}}^2,
	\label{eq:spheresHenergy2}
	\\
	\mathbb{E}^N_{\ell = 0}[S^{\Hp}]
	:=
	\
	&
	\sup_{\DRH}
	\sum_{s=0,1,2}
	v^s
	\sum_{\vert \gamma \vert \leq N-s}
	\sum_{\breve{\Phi} \in \mathcal{A}_{\mathcal{R}} \cup \mathcal{A}_{\Gamma} \cup \mathcal{A}_{\chi}}
	\Vert \mathfrak{D}^{\gamma} \breve{\Phi}_{\ell = 0} \Vert_{S_{u,v}^{\Hp}}^2,
	\label{eq:spheresHenergy3}
\end{align}
and the energy of $N+1$ derivatives of certain Ricci coefficients through the final hypersurface
\begin{equation} \label{eq:finalHenergy}
	\mathbb{E}^{N+1}_{\ell \geq 2} [C_{u_f}^{\Hp}]
	:=
	\sup_{v_{-1} \leq v \leq v(R_2,u)}
	\sum_{s=0,1,2}
	v^{s}
	\sum_{k=0}^{N+1-s}
	\Vert  
	(r\nablaslash)^{k}
	(\Omega \hat{\chi}_{\ell \geq 2}, (\Omega \tr \chi - \Omega \tr \chi_{\circ})_{\ell \geq 2}, (\eta - \eta_{\rm Kerr})_{\ell \geq 2})^{\Hp}
	\Vert_{C^{\Hp}_{u_f}(v)}^2
	.
\end{equation}
For the quantities $\Omega \hat{\chi}$ and $\Omega \tr \chi - \Omega \tr \chi_{\circ}$ recall again that, for $\kbar = (k_1,k_2,k_3)$, $\mathfrak{D}^{\kbar} = (r\nablaslash)^{k_1} (\Omega^{-1} \nablaslash_3)^{k_2} (r\Omega \nablaslash_4)^{k_3}$ and define
\begin{align} \label{eq:anomalousHenergy}
	\mathbb{E}^N_{\ell \geq 2}[\chi^{\Hp}]
	:=
	&
	\sum_{\Gamma \in \mathcal{A}_{\chi}}
	\Big[
	\sup_{\DRH} v^{-\delta}
	\Vert (r\nablaslash)^{N}
	\Gamma_{\ell \geq 2}
	\Vert_{S^{\Hp}_{u,v}}^2
	+
	\sum_{s=0,1,2}
	\sum_{\substack{ \vert \kbar \vert \leq N-s \\ k_1 \neq N-s}}
	\Big(
	\sup_{\DRH} v^s
	\Vert \mathfrak{D}^{\kbar}
	\Gamma_{\ell \geq 2}
	\Vert_{S^{\Hp}_{u,v}}^2
	\\
	&
	+
	\sup_{\substack{
	v_{-1} \leq v \leq v(R_2,u_f)}}
	v^{s}
	\big(
	\sup_{u_{-1} \leq u \leq u_f}
	\left\Vert \mathfrak{D}^{\kbar} \Gamma_{\ell \geq 2} \right\Vert_{C^{\Hp}_u(v)}^2
	+
	\left\Vert \mathfrak{D}^{\kbar} \Gamma_{\ell \geq 2} \right\Vert_{\Cbar^{\Hp}_v}^2
	+
	\left\Vert \mathfrak{D}^{\kbar} \Gamma_{\ell \geq 2} \right\Vert^2_{\DRH(v)}
	\big)
	\Big)
	\Big].
	\nonumber
\end{align}
Here $\xi_{\ell = 1} = \xi_{\ell =0} = 0$ if $\xi$ is a symmetric trace free $(0,2)$ $S$ tensor.

\begin{proof}[Proof of Theorem~\ref{thm:Hestimates}]
	The estimates for $\mathbb{E}^N[\DRH]$, $\mathbb{E}^N[C^{\Hp}]$, $\mathbb{E}^N[\Cbar^{\Hp}]$, $\mathbb{E}^N[S^{\Hp}]$, follow from Theorems \ref{prop:ellgeq2Henergy}, \ref{prop:ell1Henergy} and \ref{prop:ell0Henergy} (for the $\ell \geq 2$, $\ell = 1$ and $\ell = 0$ modes respectively) and the fact that
	\[
		\mathbb{E}^N[\DRH]
		\lesssim
		\mathbb{E}^N_{\ell \geq 2} [\DRH]
		+
		\mathbb{E}^N_{\ell = 1} [\DRH]
		+
		\mathbb{E}^N_{\ell = 0}[\DRH],
	\]
	etc.  The estimate for $\mathbb{E}^{N+1} [C_{u_f}^{\Hp}]$ then follows from the estimate for $\mathbb{E}^{N+1}_{\ell \geq 2} [C_{u_f}^{\Hp}]$ of Theorem \ref{prop:finalHenergy} and the fact that
	\[
		\mathbb{E}^{N+1} [C_{u_f}^{\Hp}]
		\lesssim
		\mathbb{E}^{N+1}_{\ell \geq 2} [C_{u_f}^{\Hp}]
		+
		\mathbb{E}^{N+1}_{\ell = 1} [C^{\Hp}]
		+
		\mathbb{E}^{N+1}_{\ell = 0} [C^{\Hp}],
	\]
	and the estimate for $\mathbb{E}^N[\chi^{\Hp}]$ follows from the estimate for $\mathbb{E}^N_{\ell \geq 2}[\chi^{\Hp}]$ of Theorem \ref{prop:ellgeq2Henergy} and the fact that
	\[
		\mathbb{E}^N[\chi^{\Hp}]
		\lesssim
		\mathbb{E}^N_{\ell \geq 2}[\chi^{\Hp}]
		+
		\mathbb{E}^N_{\ell = 0,1}[\DRH]
		+
		\mathbb{E}^N_{\ell =0,1}[C^{\Hp}]
		+
		\mathbb{E}^N_{\ell =0,1}[\Cbar^{\Hp}]
		+
		\mathbb{E}^N_{\ell =0,1}[S^{\Hp}],
	\]
	where
	\[
		\mathbb{E}^N_{\ell = 0,1}[\DRH] = \mathbb{E}^N_{\ell = 0}[\DRH] + \mathbb{E}^N_{\ell = 1}[\DRH],
	\]
	etc.  The estimate for $\mathbb{E}^N[g^{\Hp}]$ follows from Theorem \ref{prop:metricHenergy} and the estimate for the Kerr parameters $J^m_{\Hp} - J^m_{\I}$ follows from Theorem \ref{prop:finalHenergy}.
\end{proof}

\subsection{Estimates for almost gauge invariant quantities in $\DRH$}
\label{section:gaugeinvariantH}

The quantities $\sigma_{\ell \geq 2}$, $\curlslash \eta_{\ell \geq 2}$, $\curlslash \etabar_{\ell \geq 2}$, $\curlslash \beta_{\ell \geq 2}$, $\curlslash \betabar_{\ell \geq 2}$, $\curlslash \divslash \hat{\chi}$, $\curlslash \divslash \hat{\chibar}$ are all almost gauge invariant and so can be directly estimated in terms of $\alpha$, $\alphabar$ and higher order derivatives, without directly exploiting the defining conditions of the $\Hp$ gauge.  The estimates of this section will be used in Section \ref{section:finalinitialH} and Section \ref{section:Hestimateslgeq2}.

Given an $S_{u,v}$ tensor $\xi$ and $u_0 \leq u \leq u_f$, $v_{-1} \leq v \leq v(R_2,u)$, define the norm
\begin{align} \label{eq:Hcombinednorm}
	\Vert \xi \Vert_{\DRH(v), \Cbar_v, C_u(v), S_{u,v}}^2
	:=
	&
	\Vert R^* \xi \Vert_{\DRH(v)}^2
	+
	\Vert \xi \Vert_{\DRH(v)}^2
	+
	\Vert \xi \Vert_{S_{u,v}}^2
	\\
	&
	+
	\sum_{\vert \gamma \vert \leq 1}
	\Big(
	\Vert (1-3M_f/r) \mathfrak{D}^{\gamma} \xi \Vert_{\DRH(v)}^2
	+
	\Vert \mathfrak{D}^{\gamma} \xi \Vert_{\Cbar_v}^2
	+
	\Vert \mathfrak{D}^{\gamma} \xi \Vert_{C_u(v)}^2
	\Big),
	\nonumber
\end{align}
where $R^* = \Omega \nablaslash_4 - \Omega \nablaslash_3$.  Note that the norm $\Vert \xi \Vert_{\DRH(v), \Cbar_v, C_u(v), S_{u,v}}^2$ involves first order derivatives of $\xi$.

\begin{proposition}[Estimate for $\sigma_{\ell \geq 2}$ in $\DRH$] \label{prop:sigmaH}
	For any $u_0 \leq u \leq u_f$, $v \geq v_{-1}$, for $s=0,1,2$ and for $0 \leq \vert \gamma \vert \leq N-1-s$,
	\[
		\Vert \mathfrak{D}^{\gamma} \sigma_{\ell\geq 2} \Vert_{\DRH(v), \Cbar_v, C_u(v), S_{u,v}}^2
		\lesssim
		\frac{
		\varepsilon_0^2
		+
		\varepsilon^3
		}
		{v^{s}},
	\]
	where the norm $\Vert \cdot \Vert_{\DRH(v), \Cbar_v, C_u(v), S_{u,v}}$ is defined in \eqref{eq:Hcombinednorm}.
\end{proposition}

\begin{proof}
	Consider first the case that $\mathfrak{D}^{\gamma} = \mathfrak{D}^{\widetilde{\gamma}} (r\nablaslash)^k$ for some $\vert \widetilde{\gamma} \vert \leq N-4-s$ and $k \leq 3$.  Proposition \ref{prop:PPbaridentities} implies that
	\[
		\Dslash_2^* {}^* \nablaslash \sigma
		=
		\frac{1}{2} (P - \Pbar)
		+
		\mathcal{E}^1,
	\]
	and the result then follows from Proposition \ref{prop:ellipticestimates} and Theorem \ref{thm:PPbarestimates}, using Propositions \ref{prop:spacetimeerrorH1}, \ref{prop:sphereserrorH}, \ref{prop:errorout}, \ref{prop:errorin} to control the error terms.
	
	If, now, $\mathfrak{D}^{\gamma} = \mathfrak{D}^{\widetilde{\gamma}} (r \Omega \nablaslash_4)^{k}$ for some $\vert \widetilde{\gamma} \vert \leq N-3-s$ and $k \leq 2$ then the proof follows inductively from Lemma \ref{lem:sigma33sigma44} and Theorem \ref{thm:alphaalphabarestimates}, again using Propositions \ref{prop:spacetimeerrorH1}, \ref{prop:sphereserrorH}, \ref{prop:errorout}, \ref{prop:errorin} to control the error terms (it is readily checked that the error terms have the correct form).  Similarly if $\mathfrak{D}^{\gamma} = \mathfrak{D}^{\widetilde{\gamma}} (\Omega^{-1} \nablaslash_3)^{k}$ for some $\vert \widetilde{\gamma} \vert \leq N-3-s$ and $k \leq 2$.  The remaining cases follow easily from commuting the previous cases.
\end{proof}

\begin{proposition}[Estimates for $\curlslash \eta_{\ell \geq 2}$ and $\curlslash \etabar_{\ell \geq 2}$ in $\DRH$] \label{prop:curletaetabarH}
	For any $u_0 \leq u \leq u_f$, $v \geq v_{-1}$, for $s=0,1,2$ and for $0 \leq \vert \gamma \vert \leq N-1-s$,
	\[
		\Vert \mathfrak{D}^{\gamma} \curlslash \eta_{\ell\geq 2} \Vert_{\DRH(v), \Cbar_v, C_u(v), S_{u,v}}^2
		+
		\Vert \mathfrak{D}^{\gamma} \curlslash \etabar_{\ell\geq 2} \Vert_{\DRH(v), \Cbar_v, C_u(v), S_{u,v}}^2
		\lesssim
		\frac{
		\varepsilon_0^2
		+
		\varepsilon^3
		}
		{v^{s}},
	\]
	where the norm $\Vert \cdot \Vert_{\DRH(v), \Cbar_v, C_u(v), S_{u,v}}$ is defined in \eqref{eq:Hcombinednorm}.
\end{proposition}

\begin{proof}
	The proof follows from the equations \eqref{eq:curletacurletabar} and Proposition \ref{prop:sigmaH}, using Propositions \ref{prop:spacetimeerrorH1}, \ref{prop:sphereserrorH}, \ref{prop:errorout}, \ref{prop:errorin} to control the nonlinear error terms.
\end{proof}

\begin{proposition}[Estimates for $\curlslash \Omega \beta_{\ell \geq 2}$ and $\curlslash \Omega^{-1} \betabar_{\ell \geq 2}$ in $\DRH$] \label{prop:curlbetabetabarH}
	For any $u_0 \leq u \leq u_f$, $v \geq v_{-1}$, for $s=0,1,2$ and for $0 \leq \vert \gamma \vert \leq N-2-s$,
	\[
		\Vert \mathfrak{D}^{\gamma} \curlslash \Omega \beta_{\ell\geq 2} \Vert_{\DRH(v), \Cbar_v, C_u(v), S_{u,v}}^2
		+
		\Vert \mathfrak{D}^{\gamma} \curlslash \Omega^{-1} \betabar_{\ell\geq 2} \Vert_{\DRH(v), \Cbar_v, C_u(v), S_{u,v}}^2
		\lesssim
		\frac{
		\varepsilon_0^2
		+
		\varepsilon^3
		}
		{v^{s}},
	\]
	where the norm $\Vert \cdot \Vert_{\DRH(v), \Cbar_v, C_u(v), S_{u,v}}$ is defined in \eqref{eq:Hcombinednorm}.
\end{proposition}

\begin{proof}
	Since $\curlslash \eta = - \curlslash \etabar$, equation \eqref{eq:nabla4eta} implies that
	\[
		r^3 \curlslash (\Omega \beta)
		=
		\Omega \nablaslash_4 (r^3 \curlslash \eta)
		+
		\mathcal{E}^1.
	\]
	The proof for $\curlslash \beta$ then follows from Proposition \ref{prop:curletaetabarH} and Propositions \ref{prop:spacetimeerrorH1}, \ref{prop:sphereserrorH}, \ref{prop:errorout}, \ref{prop:errorin} (after noting that $\alpha$ and $\alphabar$ do not appear as top order terms in the error).  The proof for $\curlslash \betabar$ is similar, using now
	\[
		r^3 \curlslash (\Omega^{-1} \betabar)
		=
		\Omega^{-1} \nablaslash_3 (r^3 \curlslash \etabar)
		+
		\mathcal{E}^1.
	\]
\end{proof}

\begin{proposition}[Estimates for $\curlslash \divslash \Omega \hat{\chi}$ and $\curlslash \divslash \Omega^{-1} \hat{\chibar}$ in $\DRH$] \label{prop:curlchihatchibarhatH}
	For any $u_0 \leq u \leq u_f$, $v \geq v_{-1}$, for $s=0,1,2$ and for $0 \leq \vert \gamma \vert \leq N-2-s$, for $s=0,1,2$,
	\[
		\Vert \mathfrak{D}^{\gamma} \curlslash \divslash \Omega \hat{\chi} \Vert_{\DRH(v), \Cbar_v, C_u(v), S_{u,v}}^2
		+
		\Vert \mathfrak{D}^{\gamma} \curlslash \divslash \Omega^{-1} \hat{\chibar} \Vert_{\DRH(v), \Cbar_v, C_u(v), S_{u,v}}^2
		\lesssim
		\frac{
		\varepsilon_0^2
		+
		\varepsilon^3
		}
		{v^{s}},
	\]
	where the norm $\Vert \cdot \Vert_{\DRH(v), \Cbar_v, C_u(v), S_{u,v}}$ is defined in \eqref{eq:Hcombinednorm}.
\end{proposition}

\begin{proof}
	Applying $\curlslash$ to the Codazzi equation \eqref{eq:Codazzi} gives
	\[
		\curlslash \divslash (\Omega \hat{\chi})
		=
		-
		\curlslash \Omega \beta
		-
		\frac{1}{2} (\Omega \tr \chi)_{\circ} \curlslash \etabar
		+
		\mathcal{E}^1.
	\]
	The proof then follows from Proposition \ref{prop:curletaetabarH}, Proposition \ref{prop:curlbetabetabarH}, using Propositions \ref{prop:spacetimeerrorH1}, \ref{prop:sphereserrorH}, \ref{prop:errorout}, \ref{prop:errorin} (after noting that $\alpha$ and $\alphabar$ do not appear as top order terms in the error) to control the nonlinear error terms.  The proof for $\curlslash \divslash \hat{\chibar}$ follows similarly, using now the Codazzi equation \eqref{eq:Codazzibar}.
\end{proof}

\subsection{Estimates for quantities on $C_{u_f}$ and $\protect\Cbar_{v_{-1}}$: the $\ell \geq 2$ modes}
\label{section:finalinitialH}

In this section the defining conditions of the $\Hp$ gauge (see Definition \ref{Hgaugedefinition} or \eqref{eq:Hgauge1}--\eqref{eq:Hgauge6} below) are exploited, together with the estimates on $\alpha_{\Hp}$ and $\alphabar_{\Hp}$ of Theorem \ref{thm:alphaalphabarestimates} and the estimates on $\hat{\chi}_{\I}$ of Theorem \ref{thm:Iestimates}, in order to control the $\ell \geq 2$ modes of the geometric quantities of the $\Hp$ gauge, first on the final outgoing hypersurface $u_{\Hp} = u_f$ (see Section \ref{subsec:ufH}), and then on the initial incoming hypersurface $v_{\Hp} = v_{-1}$ (see Section \ref{subsec:v0H}).  The main goal of this section is to obtain results which will be exploited in the proof of Theorem \ref{prop:ellgeq2Henergy} in Section \ref{section:Hestimateslgeq2}.  As a consequence, the following is also shown.  Recall the energy \eqref{eq:finalHenergy}.

\begin{theorem}[Improving bootstrap assumptions for $\mathbb{E}^{N+1}_{\ell \geq 2} \lbrack C_{u_f}^{\Hp} \rbrack$ and angular momentum parameters] \label{prop:finalHenergy}
	The energy $\mathbb{E}^{N+1}_{\ell \geq 2}[C_{u_f}^{\Hp}]$ satisfies
	\begin{equation} \label{eq:finalHenergy1}
		\mathbb{E}^{N+1}_{\ell \geq 2}[C_{u_f}^{\Hp}]
		\lesssim
		\varepsilon_0^2
		+
		\varepsilon^3,
	\end{equation}
	and the Kerr angular momentum parameters satisfy the estimate, for $m=-1,0,1$,
	\begin{equation} \label{eq:finalHenergy2}
		\vert J^m_{\Hp} - J^m_{\I} \vert \lesssim \frac{\varepsilon^2}{(u_f)^{2}}.
	\end{equation}
\end{theorem}

\begin{proof}
	The proof of \eqref{eq:finalHenergy1} is a direct consequence of Propositions \ref{prop:chihatuf}, \ref{prop:trchiuf2} and \ref{prop:etabarufR} below.  The proof of \eqref{eq:finalHenergy2} follows from Proposition \ref{prop:amestimate}.
\end{proof}

Recall the quantity $\mu^*$,
\[
	\mu^* = \divslash \eta + \rho - \rho_{\circ} - \frac{3}{2r} \left( \Omega \tr \chi - (\Omega \tr \chi)_{\circ}\right).
\]
Recall that the geometric quantities in the $\Hp$ gauge satisfy
\begin{align}
	b(u_f,v,\theta) = 0
	&
	\qquad
	\text{on } C_{u_f}
	\\
	\label{eq:Hgauge1}
	\Omega(u_f,v,\theta) = \Omega_{\circ} (u_f,v) 
	&
	\qquad
	\text{on } C_{u_f}
	\\
	\label{eq:Hgauge2}
	\partial_u \left( r^3 (\divslash \eta)_{\ell\geq 1} + r^3 \rho_{\ell \geq 1} \right) (u,v_{-1},\theta) = 0
	&
	\qquad \text{for all }
	u_0 \leq u \leq u_f, \theta \in S^2;
	\\
	\label{eq:Hgauge8}
	\Omega_{\ell=0}(u,v_{-1}) = \Omega_{\circ} (u,v_{-1}) 
	&
	\qquad \text{for all }u_0 \leq u \leq u_f;
	\\
	\label{eq:Hgauge3}
	f^3_{\Hp,\I}(u_f, v(R,u_f), \theta) = 0
	&
	\qquad \text{for all } \theta \in S^2;
	\\
	\label{eq:Hgauge4}
	\mu^*_{\ell \geq 1} (u_f,v(R,u_f),\theta) = 0
	&
	\qquad \text{for all } \theta \in S^2;
	\\
	\label{eq:Hgauge4A}
	\left( \Omega \tr \chi - (\Omega \tr \chi)_{\circ} \right)_{\ell=0}(u_f,v(R,u_f),\theta) = 0
	&
	\qquad \text{for all } \theta \in S^2;
	\\
	\label{eq:Hgauge5}
	\divslash \divslash \chibarhat(u_f,v_{-1}, \theta) = 0
	&
	\qquad \text{for all } \theta \in S^2;
	\\
	\label{eq:Hgauge6}
	\left( \Omega \tr \chibar - (\Omega \tr \chibar)_{\circ} \right)_{\ell=0,1}(u_f,v_{-1}, \theta) = 0
	&
	\qquad \text{for all } \theta \in S^2.
\end{align}
The gauge condition \eqref{eq:Hgauge1} involving $\Omega$ in particular implies that
\[
	\Omega \omegahat = (\Omega \omegahat)_{\circ}
	\qquad
	\text{on }
	C_{u_f},
\]
and
\begin{equation} \label{eq:ufetaetabar}
	\eta = - \etabar
	\qquad
	\text{on }
	C_{u_f}.
\end{equation}

Recall that $2M_f < r_0 < 3M_f < R$, and $r_0$ will be chosen so that $r_0 - 2M_f$ is suitably small.

\subsubsection{Estimates for quantities on  the final hypersurface $C_{u_f}$}
\label{subsec:ufH}

Recall that a transport equation in the $\nablaslash_4$ direction of the form \eqref{eq:shifted} is referred to as \emph{redshifted}, \emph{noshifted} or \emph{blueshifted} according to whether the sign of $a$ in \eqref{eq:shifted} is positive, zero or negative respectively.  Recall also, see Proposition \ref{prop:chihatHschematic}, that $\Omega \hat{\chi}$ satisfies a blueshifted transport equation, $\Omega^{-1} \nablaslash_3 \Omega \hat{\chi}$ satisfies a noshifted transport equation, and higher order $\Omega^{-1} \nablaslash_3$ derivatives of $\Omega \hat{\chi}$ satisfy redshifted transport equations.

One can view the main goal of this section as estimating $\Omega \hat{\chi}$ and its derivatives on the final hypersurface $C_{u_f}$.  The estimates are accordingly obtained differently depending whether they involve zero, one, or two or more $\Omega^{-1} \nablaslash_3$ derivatives and hence the three cases are considered separately.  See Propositions \ref{prop:chihatuf}, \ref{prop:nabla3chihatufR} and \ref{prop:nabla3hochihatufR} respectively.  Estimates for other Ricci coefficients and curvature components are also obtained in the process.

\subsubsection*{Estimates for Kerr parameters}

First, however, the following estimate is obtained on the difference between the Kerr parameters $J^m_{\Hp}-J^m_{\I}$ (recall Definition \ref{assocKerparIplus} and Definition \ref{assocKerparHplus}).

\begin{proposition}[Estimate for $\Hp$ Kerr parameters] \label{prop:amestimate}
	The Kerr angular momentum parameters satisfy the estimate, for $m=-1,0,1$,
	\[
		\vert J^m_{\Hp} - J^m_{\I} \vert \lesssim \frac{\varepsilon^2}{(u_f)^{2}}.
	\]
\end{proposition}

\begin{proof}
	Setting $x=x_{\Hp}$ and $\widetilde{x} = x_{\I}$ in \eqref{eq:curvaturecomp3} and applying $\curlslash_{\I}$, it follows from Lemma \ref{lem:fHIhoerror} that, on $\{u_{\I} = u_f\}$,
	\[
		\vert
		(\curlslash \Omega \beta)^{\I} - (\curlslash \Omega \beta)^{\Hp}
		\vert
		\lesssim
		\frac{\varepsilon^2}{(u_f)^2}.
	\]
	The proof then follows from Proposition \ref{prop:l1modesspecial}, which in particular implies that, on $\{u_{\I} = u_f\}$,
	\[
		\vert (r^5 \curlslash \Omega \beta)^{\I}_{\ell=1}(u_f,v)
		-
		r^5 \curlslash \Omega \beta^{\I}_{\mathrm{Kerr}}
		\vert
		\lesssim \frac{\varepsilon^2}{(u_f)^{2}}.
	\]
\end{proof}

\subsubsection*{Estimates for $\Omega \hat{\chi}$ and $\Omega \beta$}

First the \emph{blueshifted} quantities $\Omega \hat{\chi}$ and $\Omega \beta$ are estimated on $C_{u_f}$.

To begin, the gauge condition \eqref{eq:Hgauge3} is exploited (in the form of Proposition \ref{thm:inheriting}) along with the estimates on $\hat{\chi}^{\I}$ obtained in Theorem \ref{thm:Iestimates} to give the following estimate for $\Omega \hat{\chi}^{\Hp}$ on $C_{u_f}$ around $S_{u_f,v(R,u_f)}$.

\begin{proposition}[Estimate for $\Omega \hat{\chi}$ around $S_{u_f,v(R,u_f)}$] \label{prop:chihatufdata}
	For $k \leq N+1-s$, for $s=0,1,2$,
	\[
		\int_{v(R_{-1},u_f)}^{v(R_2,u_f)} \int_{S^2} \left\vert
		(r\nablaslash)^k \hat{\chi}_{\Hp}
		\right\vert^2
		d \theta dv
		(u_f)
		\lesssim
		\frac{
		\varepsilon_0^2
		+
		\varepsilon^3
		}
		{v(R,u_f)^{s}}.
	\]
\end{proposition}

\begin{proof}
	The proof is an immediate consequence of Proposition \ref{thm:inheriting} and Theorem \ref{thm:Iestimates}.
\end{proof}

In order to estimate $\Omega\hat{\chi}$ on the entire hypersurface $C_{u_f}$, it is convenient to first consider the renormalised quantity $X_1$ (see Section \ref{section:Hequations}).

\begin{proposition}[Estimate for $X_1$ on $C_{u_f}$] \label{prop:X1uf}
	On the final hypersurface $u=u_f$, for $k \leq N-1-s$, for $s=0,1,2$, for any $v_{-1} \leq v \leq v(R_2,u_f)$,
	\[
		\sum_{k \leq N-2-s}
		\Vert (r\nablaslash)^k X_1 \Vert^2_{S_{u_f,v}}
		+
		\sum_{k \leq N-1-s}
		\Vert (r\nablaslash)^k X_1 \Vert^2_{C_{u_f}(v)}
		\lesssim
		\frac{
		\varepsilon_0^2
		+
		\varepsilon^3
		}
		{{v}^s}.
	\]
\end{proposition}

\begin{proof}
	Let $\phi$ be a smooth cut off function such that $\phi(v(R_2,u_f)) = 0$, and $\phi(v) = 1$ for all $v\leq v(R_{-1},u_f)$.  It follows from Proposition \ref{prop:X1} that $(r\nablaslash)^kX_1$ satisfies
	\begin{multline*}
		\left\vert
		\Omega \nablaslash_4
		\left( \phi (r\nablaslash)^k X_1 \right)
		-
		\frac{2M_f}{r} \phi (r\nablaslash)^k X_1
		\right\vert
		\\
		\lesssim
		\left\vert \phi' \cdot (r\nablaslash)^k X_1 \right\vert
		+
		\sum_{\substack{ k_1\leq k
		\\
		l \leq1}}
		\left\vert
		\phi \cdot (r\nablaslash)^{k_1} (\Omega^{-1} \nablaslash_3)^l \Omega^2 \alpha
		\right\vert
		+
		\sum_{k_1\leq k} \left\vert \phi \cdot \mathcal{E}^1 \right\vert
		+
		\phi \vert (r\nablaslash)^{k+2} \Omega \hat{\chi} \vert
		\sum \vert \Phi \vert  .
	\end{multline*}
	The result follows from Lemma \ref{lem:nabla4bluexi} (with $v_2 = v(R_2,u_f)$) using the vanishing of $\phi(v(R_{2},u_f))$, the fact that $\phi'$ is supported in $v(R_{-1},u_f) \leq v \leq v(R_2,u_f)$, Proposition~\ref{prop:chihatufdata} 
	and Theorem~\ref{thm:alphaalphabarestimates}, using Propositions~\ref{prop:sphereserrorH},~\ref{prop:errorout}, and~\ref{prop:nonlinerroruf} to control the nonlinear error terms.
\end{proof}

Proposition \ref{prop:X1uf} yields the following estimate for $\Omega \hat{\chi}$.

\begin{proposition}[Estimate for $\Omega \hat{\chi}$ on $C_{u_f}$] \label{prop:chihatuf}
	On the final hypersurface $u=u_f$, for $s=0,1,2$, for any $v_{-1} \leq v \leq v(R_2,u_f)$,
	\[
		\sum_{k \leq N-s}
		\Vert
		(r\nablaslash)^k (\Omega \hat{\chi})
		\Vert_{S_{u_f,v}}^2
		\lesssim
		\frac{
		\varepsilon_0^2
		+
		\varepsilon^3
		}
		{{v}^s},
		\qquad
		\sum_{k \leq N+1-s}
		\Vert
		(r\nablaslash)^k (\Omega \hat{\chi})
		\Vert_{C_{u_f}(v)}^2
		\lesssim
		\frac{
		\varepsilon_0^2
		+
		\varepsilon^3
		}
		{{v}^s}.
	\]
\end{proposition}

\begin{proof}
	The proof follows from Proposition \ref{prop:X1uf}, Propositions \ref{prop:ellipticestimates} and \ref{prop:divcurl}, and Theorem \ref{thm:alphaalphabarestimates}.
\end{proof}

The following estimate for $\Omega \beta$ follows easily from Proposition \ref{prop:chihatuf}.

\begin{proposition}[Estimate for $\Omega \beta_{\ell \geq 2}$ on $C_{u_f}$] \label{prop:betauf}
	On the final hypersurface $u=u_f$, for $s=0,1,2$, for any $v_{-1} \leq v \leq v(R_2,u_f)$,
	\[
		\sum_{k \leq N-1-s}
		\Vert
		(r\nablaslash)^k (\Omega \beta_{\ell\geq 2})
		\Vert_{S_{u_f,v}}^2
		\lesssim
		\frac{
		\varepsilon_0^2
		+
		\varepsilon^3
		}
		{{v}^s},
		\qquad
		\sum_{k \leq N-s}
		\Vert
		(r\nablaslash)^k (\Omega \beta_{\ell\geq 2})
		\Vert_{C_{u_f}(v)}^2
		\lesssim
		\frac{
		\varepsilon_0^2
		+
		\varepsilon^3
		}
		{{v}^s}.
	\]
\end{proposition}

\begin{proof}
	The Bianchi equation \eqref{eq:alpha3} can be rewritten,
	\[
		\Dslash_2^* \Omega \beta
		=
		-
		\frac{1}{2r \Omega} \nablaslash_3(r \Omega^2 \alpha)
		+
		\frac{3M}{r^3} \Omega \hat{\chi}
		+
		\mathcal{E}^0.
	\]
	The result then follows from Proposition \ref{prop:ellipticestimates}, Theorem \ref{thm:alphaalphabarestimates} and Proposition \ref{prop:chihatuf}, using also Propositions \ref{prop:sphereserrorH} and \ref{prop:errorout} to control the nonlinear error terms.
\end{proof}

\subsubsection*{Estimates for $\Omega^{-1} \nablaslash_3 \Omega \hat{\chi}$, $\eta$, $\Omega \tr \chi - \Omega \tr \chi_{\circ}$ and $\Omega^{-1}$\underline{$\hat{\chi}$}}

The outgoing propagation equation for the quantity $\Omega^{-1} \nablaslash_3 \Omega \hat{\chi}$ is \emph{noshifted} and so it is convenient to first consider the quantity $\mu^*$ which is also \emph{noshifted}, but its outgoing propagation equation contains only nonlinear terms on the right hand side.  See Proposition \ref{prop:nabla4mustaruf}.

\begin{proposition}[Estimate for $\mu^*_{\ell \geq 1}$ on $C_{u_f}$] \label{prop:Xuf}
	On the final hypersurface $u=u_f$, for $k \leq N-s$, for $s=0,1,2$, for any $v_{-1} \leq v \leq v(R_2,u_f)$,
	\begin{equation} \label{eq:mustar}
		\Vert (r\nablaslash)^k \mu^*_{\ell \geq 1} \Vert_{S_{u_f,v}}^2
		\lesssim
		\frac{\varepsilon^4}{v^{s+1}},
		\qquad
		\Vert (r\nablaslash)^k \mu^*_{\ell \geq 1} \Vert_{C_{u_f}(v)}^2
		\lesssim
		\frac{\varepsilon^4}{v^{s}}.
	\end{equation}
\end{proposition}

\begin{proof}
	By Proposition \ref{prop:nabla4mustaruf} and Lemma \ref{lem:commutation}, for $k \leq N-s$, it follows that each nonlinear term in the equation for $\Omega \nablaslash_4 (r\nablaslash)^k \mu^*$ involves at least one $\Phi$ for which $\Phi_{\mathrm{Kerr}} = 0$.  Hence, by Proposition \ref{prop:com0},
	\begin{multline} \label{eq:mustar1}
		\Vert \Omega \nablaslash_4 (r\nablaslash)^k \mu^*_{\ell \geq 1} \Vert_{S_{u_f,v}}^2
		\lesssim
		\sum_{k_1+k_2\leq k}
		\Vert
		(r\nablaslash)^{k_1} \Phi \cdot (r\nablaslash)^{k_2} (\Phi - \Phi_{\mathrm{Kerr}})
		\Vert_{S_{u_f,v}}^2
		\\
		+
		\sum_{\Phi, \xi}
		\Vert
		\Phi \cdot (r\nablaslash)^{k+1} \xi
		\Vert_{S_{u_f,v}}^2
		+
		\frac{\varepsilon^2}{v^s} \sum_{l \leq k}
		\Vert
		(r\nablaslash)^l \mu^*
		\Vert_{S_{u_f,v}}^2,
	\end{multline}
	for $\xi = \Omega \hat{\chi}, \Omega \tr \chi - \Omega \tr \chi_{\circ}, \eta - \eta_{\mathrm{Kerr}}$.  Consider $k_1+k_2 \leq k$.  Since $\vert \Phi_{\mathrm{Kerr}} \vert \lesssim \varepsilon (u_f)^{-1}$ for all $\Phi$,
	\[
		\Big(
		\int_v^{v(R_2,u_f)} \!\!\! \int_{S_{u_f,v'}} \!\!
		\vert
		\Phi_{\mathrm{Kerr}} \cdot (r\nablaslash)^{k_2} (\Phi - \Phi_{\mathrm{Kerr}})
		\vert
		d\theta d v'
		\Big)^2
		\lesssim
		\int_v^{v(R_2,u_f)} \!\!\!
		\frac{\varepsilon^2}{(u_f)^2} d v'
		\Vert (r\nablaslash)^{k_2} (\Phi - \Phi_{\mathrm{Kerr}}) \Vert_{C_{u_f}(v)}^2
		\lesssim
		\frac{\varepsilon^4}{v^{s+1}},
	\]
	and
	\begin{multline*}
		\Big(
		\int_v^{v(R_2,u_f)} \!\!\! \int_{S_{u_f,v'}} \!\!
		\vert
		(r\nablaslash)^{k_1} (\Phi - \Phi_{\mathrm{Kerr}})
		\cdot
		(r\nablaslash)^{k_2} (\Phi - \Phi_{\mathrm{Kerr}})
		\vert
		d\theta d v'
		\Big)^2
		\\
		\lesssim
		\Vert (r\nablaslash)^{k_1} (\Phi - \Phi_{\mathrm{Kerr}}) \Vert_{C_{u_f}(v)}^2
		\Vert (r\nablaslash)^{k_2} (\Phi - \Phi_{\mathrm{Kerr}}) \Vert_{C_{u_f}(v)}^2
		\lesssim
		\frac{\varepsilon^4}{v^{s+2}}.
	\end{multline*}
	The second term on the right hand side of \eqref{eq:mustar1} is estimated in Proposition \ref{prop:nonlinerroruf}, and an estimate for the third term follows immediately from the bootstrap assumption \eqref{eq:bamain}.
	The first of \eqref{eq:mustar} then follows from
	the fact that $\vert (r\nablaslash)^k \mu^*_{\ell \geq 1} (v) \vert \leq \vert (r\nablaslash)^k \mu^*_{\ell \geq 1} (v(R,u_f)) \vert + \int_{v}^{v(R,u_f)} \vert \Omega \nablaslash_4 (r\nablaslash)^k \mu^*_{\ell \geq 1} (v') \vert dv'$
	and the gauge condition \eqref{eq:Hgauge4}.  The second of \eqref{eq:mustar} follows from integrating the first.
\end{proof}

Recall that $r_0 >2M_f$ will be chosen to be close to $2M_f$ later.  The following provides control over $\Omega \tr \chi_{\ell \geq 2}$ in terms of $\Omega_{\circ}^2$ times $\eta$.  The $\Omega_{\circ}^2$ factor will be later be used as a smallness factor in the region $r \leq r_0$.

In what follows, if $a,b \in \mathbb{R}$, then $a \vee b := \max\{ a,b\}$.

\begin{proposition}[Preliminary estimate for $\Omega \tr \chi_{\ell \geq 2}$ on $C_{u_f}\cap \{ r \leq r_0 \}$] \label{prop:trchiuf}
	On the final hypersurface $u=u_f$, for any $v_{-1} \leq v \leq v(R_2,u_f)$, for $0 \leq k \leq N+1-s$, for $s=0,1,2$,
	\[
		\Vert
		(r\nablaslash)^k (\Omega \tr \chi)_{\ell \geq 2}
		\Vert_{S_{u_f,v}}^2
		\lesssim
		\frac{\varepsilon_0^2+\varepsilon^3}{v^s}
		+
		\Vert
		\Omega_{\circ}^2
		(r\nablaslash)^{(k-1)\vee 0} \eta_{\ell \geq 2}
		\Vert_{S_{u_f,v}}^2,
	\]
	\[
		\Vert
		\mathds{1}
		(r\nablaslash)^k (\Omega \tr \chi)_{\ell \geq 2}
		\Vert_{C_{u_f}(v)}^2
		\lesssim
		\frac{\varepsilon_0^2+\varepsilon^3}{v^s}
		+
		\Vert
		\mathds{1}
		\Omega_{\circ}^2
		(r\nablaslash)^{(k-1)\vee 0} \eta_{\ell \geq 2}
		\Vert_{C_{u_f}(v)}^2.
	\]
	where $\mathds{1} = \mathds{1}_{r\leq r_0}$.
\end{proposition}

\begin{proof}
	The Codazzi equation \eqref{eq:Codazzi} implies that
	\[
		\Dslash_2^* \nablaslash (\Omega \tr \chi)
		=
		2 \Dslash_2^* \big(
		\divslash (\Omega \hat{\chi})
		+
		\Omega \beta
		+
		\frac{1}{2} \Omega \tr \chi_{\circ} \etabar
		+
		\mathcal{E}^0
		\big)
		=
		2X_1 + \frac{2\Omega_{\circ}^2}{r} \Dslash_2^* \etabar
		+
		\mathcal{E}^1,
	\]
	by the Bianchi equation \eqref{eq:alpha3}.  The proof then follows from the elliptic estimate of Proposition \ref{prop:ellipticestimates}, using the estimates of Proposition \ref{prop:X1uf} and Propositions \ref{prop:sphereserrorGammaH} and \ref{prop:errorout} to control the nonlinear error terms, after recalling that $\eta = - \etabar$ on $\{u=u_f\}$.
\end{proof}

The following two propositions similarly control $\rho$ and $\eta$ in terms of $\Omega_{\circ}^2$ times $\Omega^{-1} \hat{\chibar}$.  Again, the $\Omega_{\circ}^2$ factor will be used as a smallness factor in the region $r\leq r_0$.

\begin{proposition}[Preliminary estimate for $\rho_{\ell \geq 2}$ on $C_{u_f}$] \label{prop:rhouf}
	On the final hypersurface $u=u_f$, for $0 \leq k \leq N-s$, for $s=0,1,2$, for any $v_{-1} \leq v \leq v(r_0,u_f)$,
	\[
		\Vert
		\mathds{1}
		(r\nablaslash)^k \rho_{\ell \geq 2}
		\Vert_{C_{u_f}(v)}^2
		\lesssim
		\frac{\varepsilon_0^2+\varepsilon^3}{v^s}
		+
		\Vert
		\mathds{1}
		\Omega_{\circ}^2
		(r\nablaslash)^{(k-2)\vee 0} (\Omega^{-1} \hat{\chibar})
		\Vert_{C_{u_f}(v)}^2,
	\]
	where $\mathds{1} = \mathds{1}_{r\leq r_0}$, and, for $0 \leq k \leq N-1-s$, for $s=0,1,2$, for any $v_{-1} \leq v \leq v(R_2,u_f)$,
	\[
		\Vert
		(r\nablaslash)^k \rho_{\ell \geq 2}
		\Vert_{S_{u_f,v}}^2
		\lesssim
		\frac{\varepsilon_0^2+\varepsilon^3}{v^s}
		+
		\Vert
		\Omega_{\circ}^2
		((r\nablaslash)^{(k-2)\vee 0} (\Omega^{-1} \hat{\chibar})
		\Vert_{S_{u_f,v}}^2.
	\]
\end{proposition}

\begin{proof}
	The proof follows from the equality \eqref{eq:Prhosigma}, together with the elliptic estimate Proposition \ref{prop:ellipticestimates}, Theorem \ref{thm:PPbarestimates} and Propositions \ref{prop:sigmaH} and \ref{prop:chihatuf}, using Propositions \ref{prop:sphereserrorH} and \ref{prop:errorout} to control the nonlinear error terms.
\end{proof}

\begin{proposition}[Preliminary estimate for $\eta_{\ell \geq 2}$ on $C_{u_f}$] \label{prop:etauf}
	On the final hypersurface $u=u_f$, if $r_0 - 2M_f$ is sufficiently small then, for $0 \leq k \leq N+1-s$, for $s=0,1,2$, for any $v_{-1} \leq v \leq v(r_0,u_f)$,
	\[
		\Vert
		(r\nablaslash)^k \eta_{\ell \geq 2}
		\Vert_{S_{u_f,v}}^2
		\lesssim
		\frac{\varepsilon_0^2+\varepsilon^3}{v^s}
		+
		\Vert
		\Omega_{\circ}^2
		((r\nablaslash)^{(k-3)\vee 0} (\Omega^{-1} \hat{\chibar})
		\Vert_{S_{u_f,v}}^2,
	\]
	and, for $0\leq k \leq N+1-s$, for $s=0,1,2$, with $\mathds{1} = \mathds{1}_{r\leq r_0}$,
	\[
		\Vert
		\mathds{1}
		(r\nablaslash)^k \eta_{\ell \geq 2}
		\Vert_{C_{u_f}(v)}^2
		\lesssim
		\frac{\varepsilon_0^2+\varepsilon^3}{v^s}
		+
		\Vert
		\mathds{1}
		\Omega_{\circ}^2
		(r\nablaslash)^{(k-3)\vee 0} (\Omega^{-1} \hat{\chibar})
		\Vert_{C_{u_f}(v)}^2.
	\]
\end{proposition}

\begin{proof}
	Recall that
	\[
		\divslash \eta
		=
		\mu^*
		-
		(\rho - \rho_{\circ})
		+
		\frac{3}{2r} \left( \Omega \tr \chi - (\Omega \tr \chi)_{\circ}\right).
	\]
	The proof follows immediately from Propositions \ref{prop:Xuf}, \ref{prop:trchiuf}, \ref{prop:rhouf}, along with Proposition \ref{prop:curletaetabarH} and the elliptic estimate of Proposition \ref{prop:divcurl}, using also Proposition \ref{prop:divcurlmodes}, provided $r_0-2M_f$, and hence $\Omega_{\circ}$, is sufficiently small.
\end{proof}

The next propositions exploit the factors of $\Omega_{\circ}^2$ in the above to control $\Omega^{-1} \hat{\chibar}$, $\rho_{\ell \geq 2}$, $\eta_{\ell \geq 2}$ and $\Omega \tr \chi_{\ell \geq 2}$.

\begin{proposition}[Estimate for $\Omega^{-1} \hat{\chibar}$ on $C_{u_f}\cap \{ r \leq r_0 \}$] \label{prop:chibarhatuf}
	On the final hypersurface $u=u_f$, provided $r_0 - 2M$ is sufficiently small, for $k \leq N-s$, for $s=0,1,2$, for any $v_{-1} \leq v \leq v(r_0,u_f)$, with $\mathds{1} = \mathds{1}_{r\leq r_0}$,
	\[
		\Vert
		(r\nablaslash)^k (\Omega^{-1} \hat{\chibar})
		\Vert_{S_{u_f,v}}^2
		+
		\Vert
		\mathds{1}
		(r\nablaslash)^k (\Omega^{-1} \hat{\chibar})
		\Vert_{C_{u_f}(v)}^2
		\lesssim
		\frac{\varepsilon_0^2+\varepsilon^3}{v^s}.
	\]
\end{proposition}

\begin{proof}
	The gauge condition \eqref{eq:Hgauge5}, Proposition \ref{prop:curlchihatchibarhatH} and Proposition \ref{prop:divcurl} imply that,
	\[
		\Vert (r\nablaslash)^k (\Omega^{-1} \hat{\chibar}) \Vert^2_{S_{u_f,v_{-1}}}
		\lesssim
		\varepsilon_0^2
		+
		\varepsilon^3.
	\]
	The equation \eqref{eq:chibarhat4} and Lemma \ref{lem:commutation} imply that, for any $k \geq 0$, on $\{u=u_f\}$,
	\[
		\Omega \nablaslash_4((r\nablaslash)^k r \Omega^{-1} \chibarhat)
		+
		\frac{2M_f}{r^2} (r\nablaslash)^k (r \Omega^{-1} \chibarhat)
		=
		-2r (r\nablaslash)^k \Dslash_2^* \etabar
		+
		(r\nablaslash)^k (\Omega \hat{\chi})
		+
		(r\nablaslash)^k \mathcal{E}^0.
	\]
	The proof then follows that of Lemma \ref{lem:nabla4redxi}.  Proposition \ref{prop:chihatuf} and Proposition \ref{prop:etauf}, together with \eqref{eq:ufetaetabar}, are used to control the linear error terms and Proposition \ref{prop:errorout} to control the nonlinear error terms.  The $\int_{v}^{v(r_0,u_f)} \Omega_{\circ}^2 \int_{S^2} \left\vert (r\nablaslash)^{(k-2)\vee 0} (\Omega^{-1} \hat{\chibar}) \right\vert^2 d \theta dv' (u_f)$ term arising from Proposition \ref{prop:etauf} is absorbed on the left by taking $r_0-2M$ (and hence $\Omega_{\circ}$ in $r \geq r_0$) sufficiently small.
\end{proof}

\begin{proposition}[Estimate for $\eta_{\ell \geq 2}$ on $C_{u_f}$] \label{prop:etabarufRlor}
	On the final hypersurface $u=u_f$, for $s=0,1,2$, for any $v_{-1} \leq v \leq v(R_2,u_f)$,
	\[
		\sum_{k \leq N-s}
		\Big(
		\Vert
		(r\nablaslash)^k \eta_{\ell \geq 2}
		\Vert_{S_{u_f,v}}^2
		+
		\Vert
		(r\nablaslash)^k \eta_{\ell \geq 2}
		\Vert_{C_{u_f}(v)}^2
		\Big)
		\lesssim
		\frac{\varepsilon_0^2+\varepsilon^3}{v^s}.
	\]
\end{proposition}

\begin{proof}
	Proposition \ref{prop:etauf} and Proposition \ref{prop:chibarhatuf} imply that, for $v_{-1} \leq v\leq v(r_0,u_f)$,
	\[
		\sum_{k \leq N-s}
		\Big(
		\Vert
		(r\nablaslash)^k \eta_{\ell \geq 2}
		\Vert_{S_{u_f,v}}^2
		+
		\Vert
		\mathds{1}
		(r\nablaslash)^k \eta_{\ell \geq 2}
		\Vert_{C_{u_f}(v)}^2
		\Big)
		\lesssim
		\frac{\varepsilon_0^2+\varepsilon^3}{v^s},
	\]
	where $\mathds{1} = \mathds{1}_{r\leq r_0}$.  For $v(r_0,u_f) \leq v \leq v(R_{2},u_f)$, the equation \eqref{eq:nabla4eta} can be renormalised with a bad $\Omega$ weight to introduce a \emph{redshift term}
	\[
		\Omega \slashed{\nabla}_4 \left( \Omega^{-2} r^2\eta \right)
		+
		2 \Omega^{-2} (\Omega \omegahat)_{\circ} r^2 \eta
		=
		- \Omega^{-2}
		\left(
		2 r^2 \Omega \beta
		-
		2r^2 \Omega \hat{\chi} \cdot \eta
		-
		r^2(\Omega \tr \chi - (\Omega \tr \chi)_{\circ}) \eta
		\right),
	\]
	where \eqref{eq:ufetaetabar} has also been used.  The proof then follows from Lemma \ref{lem:nabla4redxi}, Proposition \ref{prop:betauf}, and Proposition \ref{prop:com0oneforms}, using Proposition \ref{prop:errorout} to control the error terms, since $\Omega \sim 1$ in the region $r \geq r_0$.
\end{proof}

\begin{proposition}[Estimate for $\rho_{\ell \geq 2}$ on $C_{u_f}$] \label{prop:angularrhoufR}
	On the final hypersurface $u=u_f$, for $s=0,1,2$, for any $v_{-1} \leq v \leq v(R_2,u_f)$,
	\[
		\sum_{k \leq N-1-s}
		\Vert
		(r\nablaslash)^k \rho_{\ell \geq 2}
		\Vert_{S_{u_f,v}}^2
		+
		\sum_{k \leq N-s}
		\Vert
		(r\nablaslash)^k \rho_{\ell \geq 2}
		\Vert_{C_{u_f}(v)}^2
		\lesssim
		\frac{\varepsilon_0^2+\varepsilon^3}{v^s}.
	\]
\end{proposition}

\begin{proof}
	First note that
	\begin{equation} \label{eq:angularrhoufRchibarhat}
		\sum_{k \leq N-1-s}
		\Big(
		\Vert
		(r\nablaslash)^k \Omega^{-1} \hat{\chibar}
		\Vert_{S_{u_f,v}}^2
		+
		\Vert
		(r\nablaslash)^k \Omega^{-1} \hat{\chibar}
		\Vert_{C_{u_f}(v)}^2
		\Big)
		\lesssim
		\frac{\varepsilon_0^2+\varepsilon^3}{v^s}.
	\end{equation}
	Indeed, for $v_{-1} \leq v \leq v(r_0,u_f)$ \eqref{eq:angularrhoufRchibarhat} follows from Proposition \ref{prop:chibarhatuf}.  For $v \geq v(r_0,u_f)$, the proof follows by integrating the equation \eqref{eq:chibarhat4} forwards, using Lemma \ref{lem:nabla4redxi}.  Proposition \ref{prop:chihatuf} and Proposition \ref{prop:etabarufRlor} are used to control the linear error terms.  The nonlinear error terms are controlled by Propositions \ref{prop:errorout}.
	
	The proof of the Proposition then follows from the relation \eqref{eq:Prhosigma}, Theorem \ref{thm:PPbarestimates}, Proposition \ref{prop:sigmaH} and the elliptic estimate of Proposition \ref{prop:ellipticestimates}.
\end{proof}

The sharper part of the following estimate will be used in the proof of Proposition \ref{prop:gslashH}.

\begin{proposition}[Estimate for $\Omega \tr \chi_{\ell \geq 2}$ on $C_{u_f}$] \label{prop:trchiuf2}
	On the final hypersurface $u=u_f$, for $s=0,1,2$, for any $v_{-1} \leq v \leq v(R_2,u_f)$,
	\[
		\sum_{k \leq N+1-s}
		\Vert
		(r\nablaslash)^k \Omega \tr \chi_{\ell \geq 2}
		\Vert_{S_{u_f,v}}^2
		+
		\Vert
		(r\nablaslash)^k \Omega \tr \chi_{\ell \geq 2}
		\Vert_{C_{u_f}(v)}^2
		\lesssim
		\Omega(u_f,v)^2
		\frac{
		\varepsilon_0^2
		+
		\varepsilon^3
		}
		{{v(R,u_f)}^{s}}
		+
		\frac{
		\varepsilon_0^2
		+
		\varepsilon^3
		}
		{{v}^{s+2}}
		\lesssim
		\frac{
		\varepsilon_0^2
		+
		\varepsilon^3
		}
		{{v}^{s}}.
	\]
\end{proposition}

\begin{proof}
	Consider first the cruder estimate.  It follows from Proposition \ref{prop:trchiuf} and Proposition \ref{prop:etabarufRlor} that
	\[
		\sum_{k \leq N+1-s}
		\Big(
		\Vert
		(r\nablaslash)^k (\Omega \tr \chi)_{\ell \geq 2}
		\Vert_{S_{u_f,v}}^2
		+
		\Vert
		\mathds{1}
		(r\nablaslash)^k (\Omega \tr \chi)_{\ell \geq 2}
		\Vert_{C_{u_f}(v)}^2
		\Big)
		\lesssim
		\frac{\varepsilon_0^2+\varepsilon^3}{v^s},
	\]
	where $\mathds{1} = \mathds{1}_{r\leq r_0}$.  For $v(r_0,u_f) \leq v \leq v(R_{2},u_f)$, the equation
	\begin{equation} \label{eq:trchiuf2raychaudhuri}
		\partial_v \big( r^2(\Omega \tr \chi - \Omega \tr \chi_{\circ}) \big)
		-
		2 \Omega \omegahat_{\circ} r^2(\Omega \tr \chi - \Omega \tr \chi_{\circ})
		=
		- \frac{1}{2} (\Omega \tr \chi - \Omega \tr \chi_{\circ})^2
		-
		\vert \Omega \hat{\chi} \vert^2,
	\end{equation}
	can be renormalised with a bad $\Omega$ weight to introduce a \emph{redshift term} and $(r\nablaslash)^k (\Omega \tr \chi - \Omega \tr \chi_{\circ})$ can be estimated as $\eta_{\ell \geq 2}$ is estimated in the proof of Proposition \ref{prop:etabarufRlor}, using now the fact that
	\begin{equation} \label{eq:trchiuf2trchihatchierrors}
		\Vert (r\nablaslash)^k (\Omega \tr \chi - \Omega \tr \chi_{\circ})^2 \Vert_{C_{u_f}(v)}^2
		+
		\Vert (r\nablaslash)^k \vert \Omega \hat{\chi} \vert^2 \Vert_{C_{u_f}(v)}^2
		\lesssim
		\frac{\varepsilon^4}{v^{s+2}},
	\end{equation}
	for $k \leq N+1-s$.
	
	For the sharper estimate note that, on $C_{u_f}$,  the equation \eqref{eq:trchiuf2raychaudhuri} for $\Omega \tr \chi - \Omega \tr \chi_{\circ}$ is \emph{blueshifted}, and only has nonlinear terms on the right hand side.
	The proof then follows from Proposition \ref{lem:nabla4bluexi}, together with \eqref{eq:trchiuf2trchihatchierrors}, where the cruder estimate, which has already been established, is used to control $\Vert (r\nablaslash)^k \Omega \tr \chi_{\ell \geq 2} \Vert_{S_{u_f,v_{-1}}}^2$.
\end{proof}

An additional derivative of $\eta_{\ell \geq 2}$ can now be estimated on $C_{u_f}$.

\begin{proposition}[Estimate for $\eta_{\ell \geq 2}$ on $C_{u_f}$] \label{prop:etabarufR}
	On the final hypersurface $u=u_f$, for $s=0,1,2$, for any $v_{-1} \leq v \leq v(R_2,u_f)$,
	\[
		\sum_{k \leq N-s}
		\Vert
		(r\nablaslash)^k \eta_{\ell \geq 2}
		\Vert_{S_{u_f,v}}^2
		+
		\sum_{k \leq N+1-s}
		\Vert
		(r\nablaslash)^k \eta_{\ell \geq 2}
		\Vert_{C_{u_f}(v)}^2
		\lesssim
		\frac{\varepsilon_0^2+\varepsilon^3}{v^s}.
	\]
\end{proposition}

\begin{proof}
	Recall that
	\[
		\divslash \eta
		=
		\mu^*
		-
		(\rho - \rho_{\circ})
		+
		\frac{3}{2r} \left( \Omega \tr \chi - (\Omega \tr \chi)_{\circ}\right).
	\]
	The proof then follows from the elliptic estimate of Proposition \ref{prop:divcurl} along with Propositions \ref{prop:curletaetabarH}, \ref{prop:Xuf}, \ref{prop:angularrhoufR} and \ref{prop:trchiuf2}.
\end{proof}

\begin{proposition}[Estimate for $\Omega^{-1} \hat{\chibar}$ on $C_{u_f}$] \label{prop:chibarhatufR}
	On the final hypersurface $u=u_f$, for $k \leq N-s$, for $s=0,1,2$, for any $v_{-1} \leq v \leq v(R_2,u_f)$,
	\[
		\Vert
		(r\nablaslash)^k (\Omega^{-1} \hat{\chibar})
		\Vert_{S_{u_f,v}}^2
		+
		\Vert
		(r\nablaslash)^k (\Omega^{-1} \hat{\chibar})
		\Vert_{C_{u_f}(v)}^2
		\lesssim
		\frac{\varepsilon_0^2+\varepsilon^3}{v^s}.
	\]
\end{proposition}

\begin{proof}
	For $v_{-1} \leq v \leq v(r_0,u_f)$ the statement reduces to Proposition \ref{prop:chibarhatuf}.  For $v \geq v(r_0,u_f)$, the proof follows by integrating the equation \eqref{eq:chibarhat4} forwards, using Lemma \ref{lem:nabla4redxi}.  Proposition \ref{prop:chihatuf} and Proposition \ref{prop:etabarufR} are used to control the linear error terms.  The nonlinear error terms are controlled by Propositions \ref{prop:errorout}.
\end{proof}

\begin{proposition}[Estimate for $\Omega^{-1} \hat{\chibar}$ on $C_{u_f}$] \label{prop:chibarhatnabla4uf}
	On the final hypersurface $u=u_f$, for $k_1+k_2 \leq N-s$, for $s=0,1,2$, for any $v_{-1} \leq v \leq v(R_2,u_f)$,
	\[
		\Vert
		(\Omega \nablaslash_4)^{k_1} (r\nablaslash)^{k_2} (\Omega^{-1} \hat{\chibar})
		\Vert_{S_{u_f,v}}^2
		+
		\Vert
		(\Omega \nablaslash_4)^{k_1} (r\nablaslash)^{k_2} (\Omega^{-1} \hat{\chibar})
		\Vert_{C_{u_f}(v)}^2
		\lesssim
		\frac{\varepsilon_0^2+\varepsilon^3}{v^s}.
	\]
\end{proposition}

\begin{proof}
	When $k_1=0$ the proposition reduces to Proposition \ref{prop:chibarhatufR}.  When $k_1=1$ the proof follows, by equation \eqref{eq:chibarhat4} and \eqref{eq:ufetaetabar}, from Proposition \ref{prop:etabarufR}, Proposition \ref{prop:chihatuf} and Proposition \ref{prop:errorout}.  For $k_1 \geq 2$ the proof follows inductively, using again equation \eqref{eq:chihat4} and also equations \eqref{eq:nabla4eta}, \eqref{eq:beta4} and the $\nablaslash_4 \hat{\chi}$ equation from \eqref{eq:chihat4}, along with Theorem \ref{thm:alphaalphabarestimates}, Proposition \ref{prop:betauf} and Propositions \ref{prop:sphereserrorH} and \ref{prop:errorout}.
\end{proof}

One $\Omega^{-1} \nablaslash_3$ derivative of $\hat{\chi}$ can now be controlled on $C_{u_f}$ using the equation \eqref{eq:chihat3}.

\begin{proposition}[Estimate for $\Omega^{-1} \nablaslash_3( \Omega \hat{\chi})$ on $C_{u_f}$] \label{prop:nabla3chihatufR}
	On the final hypersurface $u=u_f$, for $k \leq N-1-s$, for $s=0,1$, for any $v_{-1} \leq v \leq v(R_2,u_f)$,
	\[
		\Vert
		(r\nablaslash)^k \Omega^{-1} \nablaslash_3( \Omega \hat{\chi})
		\Vert_{S_{u_f,v}}^2
		+
		\Vert
		(r\nablaslash)^k \Omega^{-1} \nablaslash_3( \Omega \hat{\chi})
		\Vert_{C_{u_f}(v)}^2
		\lesssim
		\frac{\varepsilon_0^2+\varepsilon^3}{v^s}.
	\]
\end{proposition}

\begin{proof}
	The proof follows from equation \eqref{eq:chihat3}, which implies that
	\[
		\Omega^{-1} \nablaslash_3(r \Omega \hat{\chi})
		=
		-2r \Dslash_2^* \eta - \Omega \hat{\chibar}
		+
		\mathcal{E}^0,
	\]
	along with Proposition \ref{prop:etabarufR} and Proposition \ref{prop:chibarhatufR}, using Propositions \ref{prop:sphereserrorH} and \ref{prop:errorout} to control the nonlinear error terms.
\end{proof}

\subsubsection*{Estimates for higher order $\Omega^{-1} \nablaslash_3$ derivatives of $\Omega \hat{\chi}$}

In order to control higher order $\Omega^{-1} \nablaslash_3$ derivatives of $\hat{\chi}$ on $\{u=u_f\}$, it is first necessary to exploit the gauge condition \eqref{eq:Hgauge2} to control them on the sphere $S_{u_f,v_{-1}}$.

\begin{proposition}[Estimate for $\Omega^{-1}\nablaslash_3$ derivatives of $\rho_{\ell \geq 2}$ on $S_{u_f,v_{-1}}$] \label{prop:rhoufv0}
	On the sphere $S_{u_f,v_{-1}}$, for $k_1+k_2 \leq N-1$,
	\[
		\Vert
		(\Omega^{-1}\nablaslash_3)^{k_1} (r\nablaslash)^{k_2} \rho_{\ell \geq 2}
		\Vert_{S_{u_f,v_{-1}}}
		\lesssim
		\varepsilon_0^2+\varepsilon^3
		+
		\sum_{l_1+l_2 \leq k_1+k_2}
		\Vert
		(\Omega^{-1} \nablaslash_3)^{l_1} (r\nablaslash)^{l_2} ( \Omega \hat{\chi})
		\Vert_{S_{u_f,v_{-1}}}.
	\]
\end{proposition}

\begin{proof}
	The proof follows from the relation \eqref{eq:Prhosigma}, along with Theorem \ref{thm:PPbarestimates}, Proposition \ref{prop:sigmaH}, Proposition \ref{prop:chibarhatufR}, the equation \eqref{eq:chihat4} and Theorem \ref{thm:alphaalphabarestimates}.  The nonlinear error terms are controlled using Proposition \ref{prop:sphereserrorH}.
\end{proof}

\begin{proposition}[Estimate for $\Omega^{-1}\nablaslash_3$ derivatives of $\eta_{\ell \geq 2}$ on $S_{u_f,v_{-1}}$] \label{prop:etaufv0}
	On the sphere $S_{u_f,v_{-1}}$, for $k_1+k_2 \leq N-1$,
	\[
		\Vert
		(\Omega^{-1}\nablaslash_3)^{k_1} (r\nablaslash)^{k_2+1} \eta_{\ell \geq 2}
		\Vert_{S_{u_f,v_{-1}}}
		\lesssim
		\varepsilon_0^2+\varepsilon^3
		+
		\sum_{l_1+l_2 \leq k_1+k_2}
		\Vert
		(\Omega^{-1} \nablaslash_3)^{l_1} (r\nablaslash)^{l_2} ( \Omega \hat{\chi})
		\Vert_{S_{u_f,v_{-1}}}.
	\]
\end{proposition}

\begin{proof}
	If $k_1 = 0$ then the proof follows from Proposition \ref{prop:etabarufR}.  The proof, for $k_1 \geq 1$, follows from the gauge condition \eqref{eq:Hgauge2} along with Proposition \ref{prop:rhoufv0} and Proposition \ref{prop:curletaetabarH}, using Proposition \ref{prop:divcurl} and Proposition \ref{prop:divcurlmodes}.
\end{proof}

\begin{proposition}[Estimate for $\Omega^{-1}\nablaslash_3$ derivatives of $\Omega \hat{\chi}$ on $S_{u_f,v_{-1}}$] \label{prop:nabla3chihatufv0}
	On the sphere $S_{u_f,v_{-1}}$, for $k_1+k_2 \leq N$,
	\[
		\Vert
		(\Omega^{-1}\nablaslash_3)^{k_1} (r\nablaslash)^{k_2} (\Omega \hat{\chi})
		\Vert_{S_{u_f,v_{-1}}}
		\lesssim
		\varepsilon_0^2+\varepsilon^3.
	\]
\end{proposition}

\begin{proof}
	The result for $k_1=0$ and $k_1=1$ follows from Proposition \ref{prop:chihatuf} and Proposition \ref{prop:nabla3chihatufR}. Applying $\Omega^{-1} \nablaslash_3$ to the appropriately normalised equation \eqref{eq:chihat3} gives, schematically,
	\[
		(\Omega^{-1} \nablaslash_3)^2 (r \Omega \hat{\chi})
		=
		-2 \Omega^{-1} \nablaslash_3( r \Dslash_2^* \eta)
		-
		\Omega^{-1} \nablaslash_3 (\Omega^2 \cdot \Omega^{-1} \hat{\chibar})
		+
		\mathcal{E}^1,
	\]
	and so the result for $k_1=2$ follows from Proposition \ref{prop:nabla3chihatufR}, Proposition \ref{prop:etaufv0}, the equation \eqref{eq:chihat4}, Theorem \ref{thm:alphaalphabarestimates} and Proposition \ref{prop:sphereserrorH}.  A straightforward inductive argument on $k_1$ completes the proof.
\end{proof}

With this control of higher order $\Omega^{-1} \nablaslash_3$ derivatives of $\Omega \hat{\chi}$ on the initial sphere, $S_{u_f,v_{-1}}$, the \emph{redshifted} equations satisfied by these higher order derivatives can now be integrated forwards to control them on the entire final hypersurface $\{u=u_f\}$.

\begin{proposition}[Estimate for $\Omega^{-1}\nablaslash_3$ derivatives of $\Omega \hat{\chi}$ on $C_{u_f}$] \label{prop:nabla3hochihatufR}
	On the final hypersurface $u=u_f$, for $k_1+k_2 \leq N-s$, for $s=0,1,2$, for any $v_{-1} \leq v \leq v(R_2,u_f)$,
	\[
		\Vert
		(\Omega^{-1} \nablaslash_3)^{k_1} (r\nablaslash)^{k_2} ( \Omega \hat{\chi})
		\Vert_{S_{u_f,v}}^2
		+
		\Vert
		(\Omega^{-1} \nablaslash_3)^{k_1} (r\nablaslash)^{k_2} ( \Omega \hat{\chi})
		\Vert_{C_{u_f}(v)}^2
		\lesssim
		\frac{\varepsilon_0^2+\varepsilon^3}{v^s}.
	\]
\end{proposition}

\begin{proof}
	For $k_1=0$ and $k_1=1$ the proposition reduces to Proposition \ref{prop:chihatuf} and Proposition \ref{prop:nabla3chihatufR} respectively.  For $k_1 \geq 2$ the proof uses the fact that the twice (and higher order) $\Omega^{-1} \nablaslash_3$ commuted equation \eqref{eq:chihat4} is redshifted.  See Proposition \ref{prop:chihatHschematic}.  The proof then follows from Lemma \ref{lem:nabla4redxi}, Proposition \ref{prop:nabla3chihatufv0}, Theorem \ref{thm:alphaalphabarestimates} and Propositions \ref{prop:errorout}.
\end{proof}

The following estimate on $\Omega \nablaslash_4$ derivatives of $\eta_{\ell \geq 2}$ will be used in the proof of Proposition \ref{prop:nabla4etabarH}.

\begin{proposition}[Estimate for $\Omega \nablaslash_4$ derivatives of $\eta_{\ell \geq 2}$ on $C_{u_f}$]  \label{prop:etabar4Huf}
	On the final hypersurface $u=u_f$, for $k \leq N-s$, for $s=0,1,2$, for any $v_{-1} \leq v \leq v(R_2,u_f)$,
	\[
		\Vert
		(\Omega \nablaslash_4)^k \eta_{\ell \geq 2}
		\Vert_{S_{u_f,v}}^2
		+
		\Vert
		(\Omega \nablaslash_4)^k \eta_{\ell \geq 2}
		\Vert_{C_{u_f}(v)}^2
		\lesssim
		\frac{\varepsilon_0^2+\varepsilon^3}{v^s}.
	\]
\end{proposition}

\begin{proof}
	The case that $k=0$ follows from Proposition \ref{prop:etabarufR}.  Equation \eqref{eq:nabla4etabar} on $u=u_f$ takes the schematic form,
	\[
		\Omega \nablaslash_4 (r^2 \etabar)
		=
		r^2 \Omega \beta
		+
		\mathcal{E}^0.
	\]
	The proof, for $k \geq 1$, then follows from applying $(\Omega \nablaslash_4)^{k-1}$, using equation \eqref{eq:beta4} and Theorem \ref{thm:alphaalphabarestimates}.
\end{proof}

Finally, for Proposition \ref{prop:trchibarv0} below, it is convenient to control, for $k \leq N$, $(r \nablaslash)^k \Omega \tr \chibar_{\ell \geq 2}$ at the sphere $S_{u_f,v_{-1}}$.

\begin{proposition}[Estimate for $\Omega \tr \chibar_{\ell \geq 2}$ on $C_{u_f}$] \label{prop:trchibarufR}
	On the final hypersurface $u=u_f$, for $k \leq N-s$, for $s=0,1,2$, for any $v_{-1} \leq v \leq v(R_2,u_f)$,
	\[
		\Vert
		(r\nablaslash)^k \Omega^{-2} (\Omega \tr \chibar)_{\ell \geq 2}
		\Vert_{S_{u_f,v}}^2
		+
		\Vert
		(r\nablaslash)^k \Omega^{-2} (\Omega \tr \chibar)_{\ell \geq 2}
		\Vert_{C_{u_f}(v)}^2
		\lesssim
		\frac{\varepsilon_0^2+\varepsilon^3}{v^s}.
	\]
\end{proposition}

\begin{proof}
	The Codazzi equation \eqref{eq:Codazzibar} can be rewritten schematically as
	\[
		\frac{1}{2} \slashed{\nabla} \left( \Omega^{-2} \left(
		\Omega \tr \underline{\chi} - (\Omega \tr \chi)_{\circ}
		\right) \right)
		=
		-
		\divslash \Omega^{-1} \underline{\hat{\chi}}
		-
		\frac{1}{r} \eta
		+
		\Omega^{-1} \underline{\beta}
		+
		\mathcal{E}^0.
	\]
	The proof then follows after applying $(r\nablaslash)^{k-1}$.  The resulting term involving $\Omega^{-1} \hat{\chibar}$ is controlled using Proposition \ref{prop:chibarhatufR}, the term involving $\eta$ is controlled using Proposition \ref{prop:etabarufR} and the term involving $\Omega^{-1} \betabar$ is controlled using the relation \eqref{eq:psipsibar}, Theorem \ref{thm:alphaalphabarestimates} and Proposition \ref{prop:chibarhatufR}.  The nonlinear terms are controlled using Propositions \ref{prop:sphereserrorH} and \ref{prop:errorout}.
\end{proof}

\subsubsection{Estimates for quantities on the initial incoming hypersurface $\Cbar_{v_{-1}}$}
\label{subsec:v0H}

In this section the curvature components and Ricci coefficients are estimated on the incoming null hypersurface $\Cbar_{v_{-1}}$.

\subsubsection*{Estimates for $\Omega \hat{\chi}$, $\Omega^{-1} \hat{\chibar}$, $\Omega^{-1} \betabar_{\ell \geq 2}$, $\eta_{\ell \geq 2}$, $\etabar_{\ell \geq 2}$}

To begin, the quantities $\Omega \hat{\chi}$, $\Omega^{-1} \hat{\chibar}$, $\Omega^{-1} \betabar_{\ell \geq 2}$, $\eta_{\ell \geq 2}$, $\etabar_{\ell \geq 2}$ are estimated on $\Cbar_{v_{-1}}$.  First, the gauge condition \eqref{eq:Hgauge2} is exploited to estimate $(\divslash \eta + \rho)_{\ell\geq 2}$.

\begin{proposition}[Estimate for $(\divslash \eta + \rho)_{\ell\geq 2}$ on $\Cbar_{v_{-1}}$] \label{prop:divetarhov0}
	On the initial hypersurface $v=v_{-1}$, for $k \leq N$, for any $u_0 \leq u \leq u_f$,
	\[
		\Vert
		(r\nablaslash)^{k} ( \divslash \eta + \rho)_{\ell\geq 2}
		\Vert_{S_{u,v_{-1}}}^2
		+
		\Vert
		(r\nablaslash)^{k} ( \divslash \eta + \rho)_{\ell\geq 2}
		\Vert_{\Cbar_{v_{-1}}}^2
		\lesssim
		\varepsilon_0^2+\varepsilon^3.
	\]
\end{proposition}

\begin{proof}
	The gauge condition \eqref{eq:Hgauge2} gives
	\[
		\Omega \nablaslash_3 \left( r^3 \rho_{\ell \geq 2} + r^3 \divslash \eta_{\ell \geq 2} \right)
		=
		0
	\]
	on $\{ v=v_{-1}\}$.  The proof then follows from Lemma \ref{lem:nabla3xi}, after inductively commuting with $(r\nablaslash)^k$, using Lemma \ref{lem:commutation}, and Proposition \ref{prop:Xuf}, Proposition \ref{prop:trchiuf} and Proposition \ref{prop:etabarufR} to control the initial condition, and Proposition \ref{prop:errorin} to control the nonlinear error terms resulting from commutation.
\end{proof}

The equation \eqref{eq:chihat4} can be immediately used, along with the estimates on $\Omega^{-2} \alphabar$ of Theorem \ref{thm:alphaalphabarestimates}, to estimate $\Omega^{-1} \hat{\chibar}$.

\begin{proposition}[Estimate for $\Omega^{-1} \hat{\chibar}$ on $\Cbar_{v_{-1}}$] \label{prop:chibarhatv0}
	On the initial hypersurface $v=v_{-1}$, for $k \leq N$, for any $u_0 \leq u \leq u_f$,
	\[
		\Vert
		(r\nablaslash)^{k} ( \Omega^{-1} \hat{\chibar})
		\Vert_{S_{u,v_{-1}}}^2
		+
		\Vert
		(r\nablaslash)^{k} ( \Omega^{-1} \hat{\chibar})
		\Vert_{\Cbar_{v_{-1}}}^2
		\lesssim
		\varepsilon_0^2+\varepsilon^3.
	\]
\end{proposition}

\begin{proof}
	The proof follows from Lemma \ref{lem:nabla3xi} and the equation \eqref{eq:chihat4}, using Theorem \ref{thm:alphaalphabarestimates}, Lemma \ref{lem:commutation} and Proposition \ref{prop:chibarhatufR} to control the initial condition, and Proposition \ref{prop:errorin} to control the nonlinear error terms.
\end{proof}

The $\Omega^{-1} \nablaslash_3$ derivatives of $\Omega^{-1} \hat{\chibar}$ can be similarly estimated.

\begin{proposition}[Estimate for $\Omega^{-1} \nablaslash_3$ derivatives of $\Omega^{-1} \hat{\chibar}$ on $\Cbar_{v_{-1}}$] \label{prop:chibarhatnabla3v0}
	On the initial hypersurface $v=v_{-1}$, for $k_1 + k_2 \leq N$, for any $u_0 \leq u \leq u_f$,
	\[
		\Vert
		(\Omega^{-1} \nablaslash_3)^{k_1} (r\nablaslash)^{k_2} ( \Omega^{-1} \hat{\chibar})
		\Vert_{S_{u,v_{-1}}}^2
		+
		\Vert
		(\Omega^{-1} \nablaslash_3)^{k_1} (r\nablaslash)^{k_2} ( \Omega^{-1} \hat{\chibar})
		\Vert_{\Cbar_{v_{-1}}}^2
		\lesssim
		\varepsilon_0^2+\varepsilon^3.
	\]
\end{proposition}

\begin{proof}
	When $k_1 = 0$ the proposition reduces to Proposition \ref{prop:chibarhatv0}.  For $k_1 \geq 1$ the proof follows inductively from equation \eqref{eq:chihat4}, Lemma \ref{lem:commutation}, using Theorem \ref{thm:alphaalphabarestimates} to control the linear error terms involving $\alphabar$ and Proposition \ref{prop:sphereserrorH} and \ref{prop:errorin} to control the nonlinear error terms.
\end{proof}

Proposition \ref{prop:divetarhov0} and Proposition \ref{prop:chibarhatv0} can now be used to give the following estimates for $\Omega \hat{\chi}$.

\begin{proposition}[Estimate for $\Omega \hat{\chi}$ on $\Cbar_{v_{-1}}$] \label{prop:chihatv0}
	On the initial hypersurface $v=v_{-1}$, for $k \leq N$, for any $u_0 \leq u \leq u_f$,
	\[
		\Vert
		(r\nablaslash)^{k} ( \Omega \hat{\chi})
		\Vert_{S_{u,v_{-1}}}^2
		+
		\Vert
		(r\nablaslash)^{k} ( \Omega \hat{\chi})
		\Vert_{\Cbar_{v_{-1}}}^2
		\lesssim
		\varepsilon_0^2+\varepsilon^3.
	\]
\end{proposition}

\begin{proof}
	Lemma \ref{lem:nabla3xi}, Proposition \ref{prop:chihatuf}, equation \eqref{eq:chihat3} and Proposition \ref{prop:chibarhatv0} imply that,
	\[
		\Vert
		(r\nablaslash)^{k} ( \Omega \hat{\chi})
		\Vert_{S_{u,v_{-1}}}^2
		+
		\Vert
		(r\nablaslash)^{k} ( \Omega \hat{\chi})
		\Vert_{\Cbar_{v_{-1}}}^2
		\lesssim
		\Vert
		(r\nablaslash)^{k+1} \eta_{\ell \geq 2}
		\Vert_{\Cbar_{v_{-1}}}^2
		+
		\varepsilon_0^2+\varepsilon^3
		\lesssim
		\Vert
		(r\nablaslash)^{k} \rho_{\ell \geq 2}
		\Vert_{\Cbar_{v_{-1}}}^2
		+
		\varepsilon_0^2+\varepsilon^3,
	\]
	where the second inequality follows from Proposition \ref{prop:divetarhov0} and Proposition \ref{prop:curletaetabarH}.  Equation \eqref{eq:Prhosigma}, Theorem \ref{thm:PPbarestimates}, Proposition \ref{prop:sigmaH} and Proposition \ref{prop:chibarhatv0} imply that, for $k\geq 2$,
	\[
		\Vert
		(r\nablaslash)^{k} \rho_{\ell \geq 2}
		\Vert_{\Cbar_{v_{-1}}}^2
		\lesssim
		\Vert
		(r\nablaslash)^{(k-2) \vee 0} \hat{\chi}
		\Vert_{\Cbar_{v_{-1}}}^2
		+
		\varepsilon_0^2+\varepsilon^3,
	\]
	where $a\vee b = \max\{ a,b\}$.  The proof then follows from the Gr\"{o}nwall inequality.
\end{proof}

Proposition \ref{prop:chihatv0} yields the following estimates for $\Omega^{-1} \betabar_{\ell \geq 2}$.

\begin{proposition}[Estimate for $\Omega^{-1} \nablaslash_3$ derivatives of $\Omega^{-1} \betabar_{\ell \geq 2}$ on $\Cbar_{v_{-1}}$]  \label{prop:betabarnabla3v0}
	On the initial hypersurface $v=v_{-1}$, for $k_1 + k_2 \leq N$, for any $u_0 \leq u \leq u_f$,
	\[
		\Vert
		(\Omega^{-1} \nablaslash_3)^{k_1} (r\nablaslash)^{k_2} ( \Omega^{-1} \betabar)_{\ell \geq 2}
		\Vert_{\Cbar_{v_{-1}}}^2
		\lesssim
		\varepsilon_0^2+\varepsilon^3.
	\]
\end{proposition}

\begin{proof}
	If $k_1=0$ then the proof follows from equation \eqref{eq:psipsibar}, Lemma \ref{prop:ellipticestimates}, Theorem \ref{thm:alphaalphabarestimates}, Proposition \ref{prop:chibarhatnabla3v0} and Proposition \ref{prop:errorin}, after recalling that $2r \Omega^2 \psibar = \nablaslash_4(r\Omega^2 \alphabar)$.  If $k_1 \geq 1$ then the proof follows from equation \eqref{eq:betabar3}, Theorem \ref{thm:alphaalphabarestimates} and Proposition \ref{prop:errorin}.
\end{proof}

The following estimates for $\eta_{\ell \geq 2}$ and $\etabar_{\ell \geq 2}$ can now be obtained.

\begin{proposition}[Estimates for $\Omega^{-1} \nablaslash_3$ derivatives of $\eta_{\ell \geq 2}$ and $\etabar_{\ell \geq 2}$ on $\Cbar_{v_{-1}}$]  \label{prop:etav0}
	On the initial hypersurface $v=v_{-1}$, for $k_1+k_2 \leq N$, for any $u_0 \leq u \leq u_f$,
	\begin{align*}
		\Vert
		(r\nablaslash)^{k_1} (\Omega^{-1} \nablaslash_3)^{k_2} \eta_{\ell \geq 2}
		\Vert_{S_{u,v_{-1}}}^2
		+
		\Vert
		(r\nablaslash)^{k_1} (\Omega^{-1} \nablaslash_3)^{k_2} \eta_{\ell \geq 2}
		\Vert_{\Cbar_{v_{-1}}}^2
		&
		\lesssim
		\varepsilon_0^2+\varepsilon^3,
		\\
		\Vert
		(r\nablaslash)^{k_1+1} (\Omega^{-1} \nablaslash_3)^{k_2} \eta_{\ell \geq 2}
		\Vert_{\Cbar_{v_{-1}}}^2
		&
		\lesssim
		\varepsilon_0^2+\varepsilon^3.
	\end{align*}
\end{proposition}

\begin{proof}
	If $k_2 = 0$ then, by Proposition \ref{prop:divcurl},
	\[
		\Vert (r\nablaslash)^{k_1} \eta_{\ell \geq 2} \Vert^2_{S_{u,v_{-1}}}
		\lesssim
		\Vert (r\nablaslash)^{k_1-1} r\divslash \eta_{\ell \geq 2} \Vert^2_{S_{u,v_{-1}}}
		+
		\Vert (r\nablaslash)^{k_1-1 } r \curlslash \eta_{\ell \geq 2} \Vert^2_{S_{u,v_{-1}}},
	\]
	where $k_1-1$ can be replaced by $0$ if $k_1=0$, and the proof follows from Proposition \ref{prop:curletaetabarH} for the $\curlslash$ part, and Proposition \ref{prop:divetarhov0}, equation \eqref{eq:Prhosigma}, Theorem \ref{thm:PPbarestimates}, Proposition \ref{prop:sigmaH}, Proposition \ref{prop:chihatv0} and Proposition \ref{prop:chibarhatv0} for the $\divslash$ part, using Propositions \ref{prop:sphereserrorH} and \ref{prop:errorin} to control nonlinear error terms.
	
	If $k_2 \geq 1$, by Proposition \ref{prop:divcurl} and Lemma \ref{lem:commutation},
	\begin{multline*}
		\Vert (r\nablaslash)^{k_1} (\Omega^{-1} \nablaslash_3)^{k_2} \eta_{\ell \geq 2} \Vert^2_{S_{u,v_{-1}}}
		\lesssim
		\Vert (r\nablaslash)^{k_1-1} r\divslash (\Omega^{-1} \nablaslash_3)^{k_2} \eta_{\ell \geq 2} \Vert^2_{S_{u,v_{-1}}}
		+
		\Vert (r\nablaslash)^{k_1-1} r \curlslash (\Omega^{-1} \nablaslash_3)^{k_2} \eta_{\ell \geq 2} \Vert^2_{S_{u,v_{-1}}}
		\\
		\lesssim
		\Vert (r\nablaslash)^{k_1-1} (\Omega^{-1} \nablaslash_3)^{k_2} r\divslash \eta_{\ell \geq 2} \Vert^2_{S_{u,v_{-1}}}
		+
		\Vert (r\nablaslash)^{k_1-1} (\Omega^{-1} \nablaslash_3)^{k_2} r \curlslash \eta_{\ell \geq 2} \Vert^2_{S_{u,v_{-1}}}
		d \theta
		+
		\varepsilon_0^2+\varepsilon^3,
	\end{multline*}
	where $k_1-1$ can be replaced by $0$ if $k_1=0$.  Now,
	\[
		(r\nablaslash)^{k_1-1} (\Omega^{-1} \nablaslash_3)^{k_2} r\divslash \eta
		=
		(r\nablaslash)^{k_1-1} (\Omega^{-1} \nablaslash_3)^{k_2-1} \Omega^{-2} \partial_u ( r\divslash \eta),
	\]
	and
	\begin{align*}
		\partial_u (r \divslash \eta)
		=
		-\frac{1}{r^2} \partial_u(r^3 \rho)
		+
		\partial_u \left( \frac{1}{r^2} \right) r^3 \divslash \eta
		=
		r \Omega \divslash \betabar
		-
		\frac{3M_f}{r^2} (\Omega \tr \chibar - \Omega \tr \chibar_{\circ})
		+
		2 \Omega^2 \divslash \eta
		+
		\Omega^2 \mathcal{E}^0,
	\end{align*}
	by equation \eqref{eq:rho3}.  Similarly,
	\[
		(r\nablaslash)^{k_1-1} (\Omega^{-1} \nablaslash_3)^{k_2} r\curlslash \eta
		=
		(r\nablaslash)^{k_1-1} (\Omega^{-1} \nablaslash_3)^{k_2-1} \Omega^{-2} \partial_u ( r\curlslash \eta),
	\]
	and,
	\[
		\partial_u (r \curlslash \eta)
		=
		-\frac{1}{r^2} \partial_u \left( r^3 \sigma - \frac{r^3}{2} \hat{\chi} \wedge \hat{\chibar} \right)
		+
		\partial_u \left( \frac{1}{r^2} \right) r^3 \curlslash \eta
		=
		r \Omega \curlslash \betabar
		+
		2 \Omega^2 \curlslash \eta
		+
		\Omega^2 \mathcal{E}^1,
	\]
	by equation \eqref{eq:curletacurletabar} and equation \eqref{eq:sigma3}.  The proof then follows from Proposition \ref{prop:chibarhatnabla3v0} and the Codazzi equation \eqref{eq:Codazzibar}, Proposition \ref{prop:betabarnabla3v0}, Proposition \ref{prop:sphereserrorH}, Proposition \ref{prop:errorin} and an induction argument.
\end{proof}

Finally, the following estimates for $\Omega^{-1} \nablaslash_3$ derivatives of $\Omega \hat{\chi}$ are obtained.

\begin{proposition}[Estimate for $\Omega^{-1} \nablaslash_3$ derivatives of $\Omega \hat{\chi}$ on $\Cbar_{v_{-1}}$]  \label{prop:chihatnabla3v0}
	On the initial hypersurface $v=v_{-1}$, for $k_1+k_2 \leq N$, for any $u_0 \leq u \leq u_f$,
	\[
		\Vert
		(r\nablaslash)^{k_1} (\Omega^{-1} \nablaslash_3)^{k_2} (\Omega \hat{\chi})
		\Vert_{S_{u,v_{-1}}}^2
		+
		\Vert
		(r\nablaslash)^{k_1} (\Omega^{-1} \nablaslash_3)^{k_2} (\Omega \hat{\chi})
		\Vert_{\Cbar_{v_{-1}}}^2
		\lesssim
		\varepsilon_0^2+\varepsilon^3.
	\]
\end{proposition}

\begin{proof}
	When $k_2=0$ the proposition reduces to Proposition \ref{prop:chihatv0}.  For $k_2\geq 1$, the proof follows inductively from Proposition \ref{prop:chibarhatnabla3v0} and Proposition \ref{prop:etav0} after applying $(r\nablaslash)^{k_1} (\Omega^{-1} \nablaslash_3)^{k_2-1}$ to equation \eqref{eq:chihat3}, which can be schematically rewritten as,
	\[
		\Omega^{-1} \nablaslash_3 (\Omega \hat{\chi})
		=
		-2 \Dslash_2^\star \eta
		-
		\frac{\Omega^2}{r} \Omega^{-1} \underline{\hat{\chi}}
		+
		\frac{1}{r} \Omega \hat{\chi}
		+
		\mathcal{E}^0.
	\]
\end{proof}

\subsubsection*{Estimates for difference quotients}

Recall the quantity $X_2$ from Section \ref{section:Hequations}.  It is necessary to control the difference quotients
\[
	D_{u_f} X_2(u,v_{-1})
	\qquad
	\text{and} 
	\qquad
	D_{u_f} \big( \Omega^{-1} \nablaslash_3 (\Omega \hat{\chi}) \big) (u,v_{-1}),
\]
along with their higher order angular derivatives, on the initial incoming hypersurface.  See Section \ref{subsubsec:L2transportestimatesH} for the definition of $D_{u_f} \xi$ for any $S$-tangent $(0,k)$ tensor $\xi$.  The main results of this Section are Proposition \ref{prop:X2diffv0} and Proposition \ref{prop:nabla3chihatdiffv0}.  Other Ricci coefficients and difference quotients are estimated on $\Cbar_{v_{-1}}$ in the process.

To estimate $D_{u_f} X_2$, it is convenient to first control the corresponding difference quotient for $(\Omega \tr \chi)_{\ell \geq 2}$.  In order to use the propagation equation for $(\Omega \tr \chi)_{\ell \geq 2}$ in the $\nablaslash_3$ direction, it is first necessary to control the $(\Omega \omegabarhat)_{\ell \geq 2}$ term which appears linearly on its right hand side.

\begin{proposition}[Estimate for $\Omega^{-2} (\Omega \omegabarhat)_{\ell \geq 2}$ on $\Cbar_{v_{-1}}$]  \label{prop:omegabarv0}
	On the initial hypersurface $v=v_{-1}$, for $k \leq N$, for any $u_0 \leq u \leq u_f$,
	\[
		\Vert
		(r\nablaslash)^{k} \Omega^{-2} (\Omega \omegabarhat)_{\ell \geq 2}
		\Vert_{S_{u,v_{-1}}}^2
		+
		\Vert
		(r\nablaslash)^{k} \Omega^{-2} (\Omega \omegabarhat)_{\ell \geq 2}
		\Vert_{\Cbar_{v_{-1}}}^2
		\lesssim
		\varepsilon_0^2+\varepsilon^3.
	\]
\end{proposition}

\begin{proof}
	The proof is a direct consequence of equation \eqref{eq:nabla3eta}, Proposition \ref{prop:betabarnabla3v0} and Proposition \ref{prop:etav0}, using Propositions \ref{prop:sphereserrorH} and \ref{prop:errorin} to control the nonlinear error terms.
\end{proof}

The following estimates for $\Omega_{\circ}^{-2} \Omega^2_{\ell \geq 2}$ are similarly obtained.

\begin{proposition}[Estimate for $\Omega_{\circ}^{-2} \Omega^2_{\ell \geq 2}$ on $\Cbar_{v_{-1}}$] \label{prop:Omegav0}
	On the initial hypersurface $v=v_{-1}$, for $k \leq N$, for any $u_0 \leq u \leq u_f$,
	\[
		\Big\Vert
		(r\nablaslash)^{k} \Big(1 - \frac{\Omega^2}{\Omega_{\circ}^2} \Big)_{\ell \geq 2}
		\Big\Vert_{S_{u,v_{-1}}}^2
		+
		\Big\Vert
		(r\nablaslash)^{k} \Big(1 - \frac{\Omega^2}{\Omega_{\circ}^2} \Big)_{\ell \geq 2}
		\Big\Vert_{\Cbar_{v_{-1}}}^2
		\lesssim
		\varepsilon_0^2+\varepsilon^3.
	\]
\end{proposition}

\begin{proof}
	Equation \eqref{eq:DlogOmega} implies
	\[
		\Omega \nablaslash_3 \log \left( \frac{\Omega}{\Omega_{\circ}} \right)
		=
		\Omega \omegabarhat - (\Omega \omegabarhat)_{\circ}.
	\]
	Lemma \ref{lem:commutation}, Lemma \ref{lem:nabla3xi}, Proposition \ref{prop:omegabarv0} and Proposition \ref{prop:com0} then imply that, using the gauge condition \eqref{eq:Hgauge1},
	\[
		\Big\Vert
		(r\nablaslash)^{k} \log \Big( \frac{\Omega}{\Omega_{\circ}} \Big)_{\ell \geq 2}
		\Big\Vert_{S_{u,v_{-1}}}^2
		+
		\Big\Vert
		(r\nablaslash)^{k} \log \Big( \frac{\Omega}{\Omega_{\circ}} \Big)_{\ell \geq 2}
		\Big\Vert_{C_{v_{-1}}}^2
		\lesssim
		\varepsilon_0^2+\varepsilon^3.
	\]
	where Proposition \ref{prop:errorin} is used to control the error terms arising from commutation, from which the proof follows.
\end{proof}

Proposition \ref{prop:omegabarv0} is now used to give the following estimates on $\Omega_{\circ}^{-2} (\Omega \tr \chibar)_{\ell \geq 2}$.

\begin{proposition}[Estimate for $\Omega_{\circ}^{-2} (\Omega \tr \chibar)_{\ell \geq 2}$ on $\Cbar_{v_{-1}}$] \label{prop:trchibarv0}
	On the initial hypersurface $v=v_{-1}$, for $k \leq N$, for any $u_0 \leq u \leq u_f$,
	\[
		\Vert
		(r\nablaslash)^{k} \Omega^{-2} (\Omega \tr \chibar)_{\ell \geq 2}
		\Vert_{S_{u,v_{-1}}}^2
		+
		\Vert
		(r\nablaslash)^{k} \Omega^{-2} (\Omega \tr \chibar)_{\ell \geq 2}
		\Vert_{\Cbar_{v_{-1}}}^2
		\lesssim
		\varepsilon_0^2+\varepsilon^3.
	\]
\end{proposition}

\begin{proof}
	The equation \eqref{eq:Ray} can be rewritten schematically as
	\[
		\Omega \nablaslash_3 \Big( \frac{r^2}{\Omega^2} \left( \Omega \tr \chibar - (\Omega \tr \chibar)_{\circ} \right) \Big)
		=
		\Omega^2 \Big[
		-4r \Omega^{-2} \left( \Omega \omegabarhat - (\Omega \omegabarhat)_{\circ} \right)
		+
		\mathcal{E}^0
		\Big].
	\]
	The proof then follows from Proposition \ref{prop:com0}, Lemma \ref{lem:nabla3xi}, Proposition \ref{prop:trchibarufR} and Proposition \ref{prop:omegabarv0}.  The nonlinear terms are controlled using Proposition \ref{prop:errorin}.
\end{proof}

Recall again, given an $S_{u,v}$ tensor $\xi$, the difference quotient $D_{u_f} \xi$ is defined by
\[
	(D_{u_f} \xi) (u,v,\theta) := \frac{\xi (u,v,\theta) - \xi (u_f,v,\theta)}{\Omega_{\circ}(u,v)^2}.
\]
(See Section \ref{subsubsec:L2transportestimatesH} for the definition of the difference $\xi (u,v,\theta) - \xi (u_f,v,\theta)$.)  The difference quotient $D_{u_f} (\Omega \tr \chi)_{\ell \geq 2}$ can now be estimated.

\begin{proposition}[Estimate for $D_{u_f} (\Omega \tr \chi)_{\ell \geq 2}$ on $\Cbar_{v_{-1}}$] \label{prop:trchidiffv0}
	On the initial hypersurface $v=v_{-1}$, for $k \leq N$, for any $u_0 \leq u \leq u_f$, the difference quotient satisfies
	\[
		\Vert
		D_{u_f} \big( (r\nablaslash)^{k} (\Omega \tr \chi)_{\ell \geq 2} \big)
		\Vert_{S_{u,v_{-1}}}^2
		+
		\Vert
		D_{u_f} \big( (r\nablaslash)^{k} (\Omega \tr \chi)_{\ell \geq 2} \big)
		\Vert_{\Cbar_{v_{-1}}}^2
		\lesssim
		\varepsilon_0^2+\varepsilon^3.
	\]
\end{proposition}

\begin{proof}
	Equation \eqref{eq:trchi3} can be schematically rewritten as
	\[
		\Omega^{-1} \nablaslash_3
		\left( r \left(
		\Omega \tr \chi - (\Omega \tr \chi)_{\circ}
		\right)
		\right)
		=
		2 r \left( \divslash \eta + (\rho - \rho_{\circ}) \right)
		-
		\left(
		\Omega \tr \chibar - (\Omega \tr \chibar)_{\circ}
		\right)
		-
		\frac{4M}{r^2} \left( 1 - \frac{\Omega_{\circ}^2}{\Omega^2} \right)
		+
		\mathcal{E}^0,
	\]
	(see equation \eqref{eq:trchi3H}). The result then follows from Proposition \ref{prop:com0}, Lemma \ref{lem:nabla3differenceestimate}, together with Proposition \ref{prop:divetarhov0}, Proposition \ref{prop:Omegav0} and Proposition \ref{prop:trchibarv0}.  The nonlinear error terms are controlled using Proposition \ref{prop:errorin}.
\end{proof}

The difference quotient $D_{u_f} (X_2)_{\ell \geq 2}$ can now be similarly estimated.

\begin{proposition}[Estimate for $D_{u_f} (X_2)_{\ell \geq 2}$ on $\Cbar_{v_{-1}}$] \label{prop:X2diffv0}
	On the initial hypersurface $v=v_{-1}$, for $k \leq N-4$, for any $u_0 \leq u \leq u_f$,the difference quotient satisfies
	\[
		\Vert
		D_{u_f} \big( (r\nablaslash)^{k} X_2 \big)
		\Vert_{S_{u,v_{-1}}}^2
		+
		\Vert
		D_{u_f} \big( (r\nablaslash)^{k} X_2 \big)
		\Vert_{\Cbar_{v_{-1}}}^2
		\lesssim
		\varepsilon_0^2+\varepsilon^3.
	\]
\end{proposition}

\begin{proof}
	Equation \eqref{eq:alpha3} and the Codazzi equation \eqref{eq:Codazzi} imply that,
	\begin{align*}
		X_2
		&
		=
		r^3 \Dslash_2^* \divslash \Dslash_2^*
		\left(
		r^3 \divslash \Omega \hat{\chi}
		+
		r^3 \Omega \beta
		-
		3M_fr \Omega \hat{\chi}
		+
		\mathcal{E}^0
		\right)
		+
		\frac{1}{2} (3r^3M_f - 2r^2M_f^2 - r^4) (\Omega^{-1} \nablaslash_3)^2 (r \Omega^2 \alpha)
		\\
		&
		\!
		=
		r^3 \Dslash_2^* \divslash \Dslash_2^*
		\! \left(
		\frac{r^3}{2} \nablaslash (\Omega \tr \chi)
		-
		\Omega^2 r^2 \etabar
		-
		3M_fr \Omega \hat{\chi}
		+
		\mathcal{E}^0
		\right)
		\! +
		\frac{1}{2} (3r^3M_f - 2r^2M_f^2 - r^4) (\Omega^{-1} \nablaslash_3)^2 (r \Omega^2 \alpha).
	\end{align*}
	The proof follows from Proposition \ref{prop:trchidiffv0} by applying $(r \nablaslash)^k$, subtracting $(r \nablaslash)^k X_2 (u_f,v_{-1})$ and dividing by $\Omega_{\circ}(u,v_{-1})^2$.  The differences of the other linear terms can be estimated as in the proof of Proposition \ref{prop:trchidiffv0}, now using equations \eqref{eq:alpha3}, \eqref{eq:chihat3}, \eqref{eq:nabla4eta} and the above estimates.  The differences of the nonlinear error terms are controlled by Lemma \ref{lem:nabla3differenceestimate} and Proposition \ref{prop:errorin}.
\end{proof}

The difference quotient $D_{u_f} \Omega^{-1} \nablaslash_3 (\Omega \tr \chi)_{\ell \geq 2}$ is estimated using equation \eqref{eq:trchi3}.

\begin{proposition}[Estimate for $D_{u_f} \Omega^{-1} \nablaslash_3 (\Omega \tr \chi)_{\ell \geq 2}$ on $\Cbar_{v_{-1}}$] \label{prop:nabla3trchidiffv0}
	On the initial hypersurface $v=v_{-1}$, for $k \leq N$, for any $u_0 \leq u \leq u_f$, the difference quotient satisfies
	\[
		\Vert
		D_{u_f} \big( \Omega^{-1} \nablaslash_3 (r\nablaslash)^{k} (\Omega \tr \chi)_{\ell \geq 2} \big)
		\Vert_{S_{u,v_{-1}}}^2
		+
		\Vert
		D_{u_f} \big( \Omega^{-1} \nablaslash_3 (r\nablaslash)^{k} (\Omega \tr \chi)_{\ell \geq 2} \big)
		\Vert_{\Cbar_{v_{-1}}}^2
		\lesssim
		\varepsilon_0^2+\varepsilon^3.
	\]
\end{proposition}

\begin{proof}
	Recall again equation \eqref{eq:trchi3},
	\[
		\Omega^{-1} \nablaslash_3
		\left( r \left(
		\Omega \tr \chi - (\Omega \tr \chi)_{\circ}
		\right)
		\right)
		=
		2 r \left( \divslash \eta + (\rho - \rho_{\circ}) \right)
		-
		\left(
		\Omega \tr \chibar - (\Omega \tr \chibar)_{\circ}
		\right)
		-
		\frac{4M_f}{r^2} \left( 1 - \frac{\Omega_{\circ}^2}{\Omega^2} \right)
		+
		\mathcal{E}^0.
	\]
	The proof again follows by commuting with $(r \nablaslash)^k$, projecting to the $\ell \geq 2$ modes, subtracting the value at $S_{u_f,v_{-1}}$, $\Omega^{-1} \nablaslash_3 (r\nablaslash)^{k} (\Omega \tr \chi)_{\ell \geq 2}(u_f,v_{-1})$, and dividing by $\Omega_{\circ}(u',v_{-1})^2$.  The differences of the linear terms are controlled, as in the proof of Proposition \ref{prop:trchidiffv0}, using now the gauge condition \eqref{eq:Hgauge2}, equation \eqref{eq:Ray} and \eqref{eq:DlogOmega}.  The differences of the nonlinear error terms are again controlled using Lemma \ref{lem:nabla3differenceestimate} and Proposition \ref{prop:errorin}.
\end{proof}

Finally, the difference quotient $D_{u_f} \Omega^{-1} \nablaslash_3 X_2$ is estimated.

\begin{proposition}[Estimate for $D_{u_f} \Omega^{-1} \nablaslash_3 X_2$ on $\Cbar_{v_{-1}}$] \label{prop:nabla3chihatdiffv0}
	On the initial hypersurface $v=v_{-1}$, for $k \leq N-5$, for any $u_0 \leq u \leq u_f$,
	the difference quotient satisfies
	\[
		\Vert
		D_{u_f} \big( \Omega^{-1} \nablaslash_3 (r\nablaslash)^{k} X_2 \big)
		\Vert_{S_{u,v_{-1}}}^2
		+
		\Vert
		D_{u_f} \big( \Omega^{-1} \nablaslash_3 (r\nablaslash)^{k} X_2 \big)
		\Vert_{\Cbar_{v_{-1}}}^2
		\lesssim
		\varepsilon_0^2+\varepsilon^3.
	\]
\end{proposition}

\begin{proof}
	As in the proof of Proposition \ref{prop:X2diffv0}, the proof follows from the Codazzi equation
	\[
		X_2
		=
		r^3 \Dslash_2^* \divslash \Dslash_2^*
		\Big(
		\frac{r^3}{2} \nablaslash (\Omega \tr \chi)
		-
		\Omega^2 r^2 \etabar
		-
		3M_f r \Omega \hat{\chi}
		+
		\mathcal{E}^0
		\Big)
		+
		\frac{1}{2} (3r^3M_f - 2r^2M_f^2 - r^4) (\Omega^{-1} \nablaslash_3)^2 (r \Omega^2 \alpha),
	\]
	after applying $\Omega^{-1} \nablaslash_3 (r \nablaslash)^k$ and using now Proposition \ref{prop:nabla3trchidiffv0}.
\end{proof}

\subsubsection*{Estimate for $\Omega^{-1} \nablaslash_3$ derivatives of $\eta_{\ell \geq 2}$ and $\Omega^{-2} ( \Omega \omegabarhat - (\Omega \omegabarhat)_{\circ} )_{\ell \geq 2}$}

In this section $\Omega^{-1} \nablaslash_3$ derivatives of $\eta_{\ell \geq 2}$ and $\Omega^{-2} ( \Omega \omegabarhat - (\Omega \omegabarhat)_{\circ} )_{\ell \geq 2}$ are estimated.

First, the estimates of the following proposition, for $\Omega^{-1} \nablaslash_3$ derivatives of $\eta_{\ell \geq 2}$, will be used in the proof of Proposition \ref{prop:etaH}.

\begin{proposition}[Estimate for $\Omega^{-1} \nablaslash_3$ derivatives of $\eta_{\ell \geq 2}$ on $\Cbar_{v_{-1}}$] \label{prop:DetaHv0}
	On the initial hypersurface $v=v_{-1}$, for $k \leq N+1$,
	\[
		\Vert
		(\Omega^{-1} \nablaslash_3)^k \eta_{\ell \geq 2}
		\Vert_{\Cbar_{v_{-1}}}^2
		\lesssim
		\varepsilon_0^2+\varepsilon^3.
	\]
\end{proposition}

\begin{proof}
	First note that the result for $k \leq N$ follows from Proposition \ref{prop:etav0}.
	
	Now, equation \eqref{eq:nabla3eta} and Propositions \ref{prop:etav0} and \ref{prop:betabarnabla3v0} imply that, for any $l=0,1$, $k \leq N-1$,
	\[
		\Vert (r\nablaslash)^l (\Omega^{-1} \nablaslash_3)^k \Omega^{-2} (\Omega \omegahat - \Omega \omegahat_{\circ}) \Vert_{S_{u,v_{-1}}}^2
		\lesssim
		\varepsilon_0^2 + \varepsilon^3,
	\]
	and hence the Raychaudhuri equation \eqref{eq:Ray} implies that, for any $k \leq N$,
	\[
		\Vert (\Omega^{-1} \nablaslash_3)^k \Omega^{-2} (\Omega \tr \chibar - \Omega \tr \chibar_{\circ}) \Vert_{S_{u,v_{-1}}}^2
		\lesssim
		\varepsilon_0^2 + \varepsilon^3.
	\]
	Revisiting the proof of Proposition \ref{prop:etav0}, note that $\eta$ satisfies
	\begin{align*}
		&
		\Vert r\divslash \left(
		(\Omega^{-1} \nablaslash_3)^{N+1} r^2 \eta_{\ell \geq 2}
		-
		(\Omega^{-1} \nablaslash_3)^{N} r^2 \Omega^{-1} \betabar_{\ell \geq 2}
		\right) \Vert_{S_{u,v_{-1}}}^2
		\\
		&
		\qquad
		\lesssim
		\varepsilon^2 \Vert (\Omega^{-1} \nablaslash_3)^{N+1}  \eta_{\ell \geq 2} \Vert_{S_{u,v_{-1}}}^2
		\!\!\!
		+
		\varepsilon^2 \Vert (\Omega^{-1} \nablaslash_3)^{N}(r\nablaslash) \eta_{\ell \geq 2} \Vert_{S_{u,v_{-1}}}^2
		\!\!\!
		+
		\Vert (\Omega^{-1} \nablaslash_3)^{N} \Omega^{-2} (\Omega \tr \chibar - \Omega \tr \chibar_{\circ}) \Vert_{S_{u,v_{-1}}}^2
		\\
		&
		\qquad \quad
		+
		\Vert \mathcal{E}^N \Vert_{S_{u,v_{-1}}}^2
		\lesssim
		\varepsilon_0^2 + \varepsilon^3
		+
		\varepsilon^2 \Vert (\Omega^{-1} \nablaslash_3)^{N+1}  \eta_{\ell \geq 2} \Vert_{S_{u,v_{-1}}}^2,
	\end{align*}
	by Proposition \ref{prop:etav0} and the above, and similarly,
	\[
		\Vert r\curlslash \left(
		(\Omega^{-1} \nablaslash_3)^{N+1} r^2 \eta_{\ell \geq 2}
		+
		(\Omega^{-1} \nablaslash_3)^{N} r^2 \Omega^{-1} \betabar_{\ell \geq 2}
		\right) \Vert_{S_{u,v_{-1}}}^2
		\lesssim
		\varepsilon_0^2 + \varepsilon^3
		+
		\varepsilon^2 \Vert (\Omega^{-1} \nablaslash_3)^{N+1}  \eta_{\ell \geq 2} \Vert_{S_{u,v_{-1}}}^2.
	\]
	Writing
	\[
		(\Omega^{-1} \nablaslash_3)^{N+1} r^2 \eta_{\ell \geq 2}
		=
		r\nablaslash h_1 + r {}^* \nablaslash h_2,
		\qquad
		(\Omega^{-1} \nablaslash_3)^{N} r^2 \Omega^{-1} \betabar_{\ell \geq 2}
		=
		r \nablaslash g_1 + r {}^* \nablaslash g_2,
	\]
	for some functions $h_1,h_2,g_1,g_2$, it follows from standard elliptic theory that
	\[
		\sum_{l=0,1,2}
		\Vert (r\nablaslash)^l (h_1 - g_1) \Vert_{S_{u,v_{-1}}}^2
		+
		\Vert (r\nablaslash)^l (h_2 + g_2) \Vert_{S_{u,v_{-1}}}^2
		\lesssim
		\varepsilon_0^2 + \varepsilon^3
		+
		\varepsilon^2 \Vert (\Omega^{-1} \nablaslash_3)^{N+1}  \eta_{\ell \geq 2} \Vert_{S_{u,v_{-1}}}^2,
	\]
	and hence
	\begin{multline*}
		\Vert (\Omega^{-1} \nablaslash_3)^{N+1} r^2 \eta_{\ell \geq 2} \Vert_{S_{u,v_{-1}}}^2
		=
		\Vert r \nablaslash h_1 \Vert_{S_{u,v_{-1}}}^2
		+
		\Vert r \nablaslash h_2 \Vert_{S_{u,v_{-1}}}^2
		\lesssim
		\varepsilon_0^2 + \varepsilon^3
		+
		\varepsilon^2 \Vert (\Omega^{-1} \nablaslash_3)^{N+1}  \eta_{\ell \geq 2} \Vert_{S_{u,v_{-1}}}^2
		\\
		+
		\Vert r \nablaslash g_1 \Vert_{S_{u,v_{-1}}}^2
		+
		\Vert r \nablaslash g_2 \Vert_{S_{u,v_{-1}}}^2
		=
		\varepsilon_0^2 + \varepsilon^3
		+
		\varepsilon^2 \Vert (\Omega^{-1} \nablaslash_3)^{N+1}  \eta_{\ell \geq 2} \Vert_{S_{u,v_{-1}}}^2
		+
		\Vert (\Omega^{-1} \nablaslash_3)^{N} r^2 \Omega^{-1} \betabar_{\ell \geq 2} \Vert_{S_{u,v_{-1}}}^2.
	\end{multline*}
	The proof then follows from Proposition \ref{prop:betabarnabla3v0} after integrating in $u$ and taking $\varepsilon$ sufficiently small.
\end{proof}

Finally, the following proposition, on estimates for $\Omega^{-1} \nablaslash_3$ derivatives of $\Omega^{-2} ( \Omega \omegabarhat - (\Omega \omegabarhat)_{\circ} )_{\ell \geq 2}$, will be used in the proof of Proposition \ref{prop:omegabarhatH}.

\begin{proposition}[Estimate for $\Omega^{-1} \nablaslash_3$ derivatives of $\Omega^{-2} ( \Omega \omegabarhat - (\Omega \omegabarhat)_{\circ} )_{\ell \geq 2}$ on $\Cbar_{v_{-1}}$] \label{prop:DomegabarHv0}
	On the initial hypersurface $v=v_{-1}$, for $k \leq N$, $l=0,1$,
	\[
		\Vert
		(r\nablaslash)^l (\Omega^{-1} \nablaslash_3)^k \left( \Omega^{-2} \left( \Omega \omegabarhat - (\Omega \omegabarhat)_{\circ} \right) \right)_{\ell \geq 2}
		\Vert_{\Cbar_{v_{-1}}}^2
		\lesssim
		\varepsilon_0^2+\varepsilon^3.
	\]
\end{proposition}

\begin{proof}
	The proof follows from projecting equation \eqref{eq:nabla3eta} to $\ell \geq 2$ and commuting with $(\Omega^{-1} \nablaslash_3)^k$ and using Proposition \ref{prop:betabarnabla3v0}, Proposition \ref{prop:com0oneforms} and Proposition \ref{prop:DetaHv0}, and the Poincar\'{e} inequality, Proposition \ref{prop:Poincare}.  The nonlinear error terms are controlled using Propostion \ref{prop:errorin}.
\end{proof}

\subsection{Estimates for quantities in $\DRH$: the $\ell \geq 2$ modes}
\label{section:Hestimateslgeq2}

The goal of this section is to estimate the $\ell \geq 2$ modes of the geometric quantities of the $\Hp$ gauge using the estimates for the corresponding quantities on $C_{u_f}$ and $\Cbar_{v_{-1}}$ obtained in the previous section.  The main result of this section is the following proposition.  Recall the energies \eqref{eq:spacetimeHenergy1}, \eqref{eq:outgoingHenergy1}, \eqref{eq:incomingHenergy1}, \eqref{eq:spheresHenergy1}, \eqref{eq:anomalousHenergy}.

\begin{theorem}[Improving bootstrap assumptions for energies of $\ell \geq 2$ modes] \label{prop:ellgeq2Henergy}
	The energies 
	defined in \eqref{eq:spacetimeHenergy1}, \eqref{eq:outgoingHenergy1}, \eqref{eq:incomingHenergy1}, \eqref{eq:spheresHenergy1}, \eqref{eq:anomalousHenergy}
	satisfy,
	\[
		\mathbb{E}^N_{\ell \geq 2}[\DRH]
		+
		\mathbb{E}^{N+1}_{\ell \geq 2} [C_{u_f}^{\Hp}]
		+
		\mathbb{E}^N_{\ell \geq 2}[\Cbar^{\Hp}]
		+
		\mathbb{E}^N_{\ell \geq 2}[S^{\Hp}]
		+
		\mathbb{E}^N_{\ell \geq 2}[\chi^{\Hp}]
		\lesssim
		\varepsilon_0^2
		+
		\varepsilon^3.
	\]
\end{theorem}

\begin{proof}
	The proof is a direct consequence of the estimate for $\sigma_{\ell \geq 2}$ of Proposition \ref{prop:sigmaH}, the estimates for $\Omega^{-1} \hat{\chibar}$ of Propositions \ref{prop:chibarhatH} and \ref{prop:chibarhatH2}, the estimates for $\Omega \hat{\chi}$ of Propositions \ref{prop:chihatHtop} and \ref{prop:chihatHn3aspacetime}, the estimates for $\rho_{\ell \geq 2}$ of Proposition \ref{prop:rhoH1}, the estimate for $\Omega \beta_{\ell \geq 2}$ of Proposition \ref{prop:nabla3betatopH}, the estimate for $\Omega^{-1} \betabar_{\ell \geq 2}$ of Proposition \ref{prop:nabla4betabartopH}, the estimate for $\eta_{\ell \geq 2}$ of Proposition \ref{prop:nabla4etaH}, the estimate for $\etabar_{\ell \geq 2}$ of Proposition \ref{prop:nabla4etabarH}, the estimate for $\Omega_{\ell \geq 2}$ of Proposition \ref{prop:OmegaH}, the estimate for $\Omega\omegahat_{\ell \geq 2}$ of Proposition \ref{prop:omegahatH}, the estimate for $\Omega^{-2}(\Omega \omegabarhat)_{\ell \geq 2}$ of Proposition \ref{prop:omegabarhatH}, the estimates for $\Omega \tr \chi_{\ell \geq 2}$ of Propositions \ref{prop:trchiH3} and \ref{prop:trchiH2}, and the estimate for $\Omega^{-2} (\Omega \tr \chibar)_{\ell \geq 2}$ of Proposition \ref{prop:trchibarH1}.
\end{proof}

The remainder of this section involves the proof of the propositions which constitute the proof of Proposition \ref{prop:ellgeq2Henergy}.  First $\Omega^{-1} \hat{\chibar}$ is estimated in Section \ref{subsubsec:Hchihatbar}.  The main difficulties are in estimating $\Omega \hat{\chi}$.  Such estimates are obtained in Section \ref{subsubsec:Hchihat}.  In Section \ref{subsubsec:Hcurvature} the curvature components $\Omega \beta_{\ell \geq 2}$, $\Omega^{-1} \betabar_{\ell \geq 2}$ and $\rho_{\ell \geq 2}$ are estimated.  The quantities $\eta_{\ell \geq 2}$ and $\etabar_{\ell \geq 2}$ are estimated in Section \ref{subsec:etaetabarH}, $\Omega_{\circ}^{-2} \Omega^2_{\ell \geq 2}$ is estimated in Section \ref{subsec:OmegaH}, $(\Omega \omegahat - (\Omega \omegahat)_{\circ})_{\ell \geq 2}$ and $\Omega^{-2} (\Omega \omegabarhat - (\Omega \omegabarhat)_{\circ})_{\ell \geq 2}$
are estimated in Section \ref{subsubsec:omegahatomegabarhatH}, and $(\Omega \tr \chi - \Omega \tr \chi_{\circ})_{\ell \geq 2}$ and $\Omega^{-2}(\Omega \tr \chibar - \Omega \tr \chibar_{\circ})_{\ell \geq 2}$ are estimated in Section \ref{subsubsec:trchitrchibarH}.

\subsubsection{Estimates for $\Omega^{-1}$\underline{$\hat{\chi}$}}
\label{subsubsec:Hchihatbar}

To begin, $\Omega^{-1} \hat{\chibar}$ can immediately be estimated by integrating backwards from the final hypersurface $\{u=u_f\}$.

\begin{proposition}[Estimates for $\Omega^{-1} \hat{\chibar}$ on $C_u$ and $\DRH$] \label{prop:chibarhatH}
	For $\vert \gamma \vert \leq N-s$, for $s=0,1,2$, for any $v_{-1} \leq v \leq v(R_2,u_f)$, and any $u_0 \leq u \leq u_f$,
	\[
		\Vert
		\mathfrak{D}^{\gamma} (\Omega^{-1} \hat{\chibar})
		\Vert_{C_{u}(v)}^2
		+
		\Vert
		\mathfrak{D}^{\gamma} (\Omega^{-1} \hat{\chibar})
		\Vert_{\DRH(v)}^2
		\lesssim
		\frac{\varepsilon_0^2+\varepsilon^3}{v^s}.
	\]
\end{proposition}

\begin{proof}
	Recall the quantity $\Xbar_1$ (see equation \eqref{eq:Xbar1}) and the $\nablaslash_3$ equation it satisfies (see Proposition \ref{prop:Xbar1}).  Since $\Omega^{-2} \alphabar$ only appears at lower orders on the right hand side of the equation (and so the estimates of Theorem \ref{thm:alphaalphabarestimates} do not degenerate at $r=3M_f$), Lemma \ref{lem:nabla3xi} and Theorem \ref{thm:alphaalphabarestimates} imply that
	\[
		\sum_{k\leq N-2-s}
		\Vert
		(r\nablaslash)^k (\Omega^{-2} \Xbar_1)
		\Vert_{C_{u}(v)}^2
		+
		\Vert
		(r\nablaslash)^k (\Omega^{-2} \Xbar_1)
		\Vert_{\DRH(v)}^2
		\lesssim
		\frac{\varepsilon_0^2+\varepsilon^3}{v^s},
	\]
	where Proposition \ref{prop:spacetimeerrorH1} is used to control the nonlinear error terms.  It follows from Theorem \ref{thm:alphaalphabarestimates} and Proposition \ref{prop:ellipticestimates} that the result holds in the case that $\mathfrak{D}^{\gamma} = (r\nablaslash)^{k}$.
	
	Consider now the case that $\mathfrak{D}^{\gamma} = (\Omega^{-1} \nablaslash_3)^{k_1} (r \nablaslash)^{k_2}$, for some $k_1 +k_2 = \vert \gamma\vert$.  The equation \eqref{eq:chihat4} can schematically be written
	\begin{equation} \label{eq:chibarhat3schematic}
		\Omega \nablaslash_3 (\Omega^{-1} \hat{\chibar})
		-
		\frac{2\Omega^2}{r} \Omega^{-1} \hat{\chibar}
		=
		-
		\Omega^2 \cdot \Omega^{-2} \alphabar
		+
		\Omega^2 \mathcal{E}^0.
	\end{equation}
	Given $l_1+l_2 \leq N-1$, it follows from commuting the equation \eqref{eq:chibarhat3schematic} that
	\begin{align*}
		\left\vert (\Omega^{-1} \nablaslash_3)^{l_1+1} (r\nablaslash)^{l_2} \Omega^{-1} \hat{\chibar} \right\vert
		\lesssim
		&
		\sum_{k_1 + k_2 \leq l_1+l_2}\left(
		\left\vert (\Omega^{-1} \nablaslash_3)^{l_1} (r\nablaslash)^{l_2} \Omega^{-1} \hat{\chibar} \right\vert
		+
		\left\vert (\Omega^{-1} \nablaslash_3)^{l_1} (r\nablaslash)^{l_2} \Omega^{-2} \hat{\alphabar} \right\vert \right)
		+
		\vert
		\mathcal{E}^{l_1+l_2}
		\vert,
	\end{align*}
	and the result in the case that $\mathfrak{D}^{\gamma} = (\Omega^{-1} \nablaslash_3)^{k_1} (r \nablaslash)^{k_2}$ follows by a straightforward induction argument by integrating the above inequality and using Theorem \ref{thm:alphaalphabarestimates}.
	
	Consider now $\mathfrak{D}^{\gamma}$ of the form $\mathfrak{D}^{\gamma} = (r \Omega \nablaslash_4)^{k_1} (\Omega^{-1} \nablaslash_3)^{k_2} (r \nablaslash)^{k_3}$, for some $k_1+k_2 \leq N$ with $k_1 \geq 1$.  Given such $k_1$, $k_2$, define
	\[
		\xi := (\Omega\nablaslash_4 - \Omega\nablaslash_3) (\Omega \nablaslash_4)^{k_1-1} (\Omega^{-1} \nablaslash_3)^{k_2} (r \nablaslash)^{k_3} \Omega^{-1} \hat{\chibar}.
	\]
	By Lemma \ref{lem:commutation} and equation \eqref{eq:chibarhat3schematic} it follows that that,
	\[
		\vert
		\Omega^{-1} \nablaslash_3 \xi
		-
		\frac{2}{r} \xi
		\vert
		\lesssim
		\vert
		(\Omega\nablaslash_4 - \Omega\nablaslash_3) (\Omega \nablaslash_4)^{k_1-1} (\Omega^{-1} \nablaslash_3)^{k_2} (r\nablaslash)^{k_2} \Omega^{-2} \alphabar
		\vert
		+
		\vert \mathcal{E}^{k_1+k_2+k_3} \vert,
	\]
	and Lemma \ref{lem:nabla3xi}, Proposition \ref{prop:chibarhatnabla4uf} and Theorem \ref{thm:alphaalphabarestimates} then imply that
	\[
		\Vert
		\xi
		\Vert_{C_{u}(v)}^2
		+
		\Vert
		\xi
		\Vert_{\DRH(v)}^2
		\lesssim
		\frac{\varepsilon_0^2+\varepsilon^3}{v^s}.
	\]
	since the integrated decay estimate for $(\Omega \nablaslash_4 - \Omega \nablaslash_3) \Omega^{-2}\alphabar$ does not degenerate at $r=3M_f$.  The result then inductively follows for the case that $\mathfrak{D}^{\gamma} = (r\Omega\nablaslash_4)^{k_1} (\Omega^{-1} \nablaslash_3)^{k_2} (r \nablaslash)^{k_3}$ by combining with the previous estimates.  The result for general $\mathfrak{D}^{\gamma}$ then follows from Lemma \ref{lem:commutation}.
\end{proof}

The following estimates for $\Omega^{-1} \hat{\chibar}$ on $S_{u,v}$ and $\Cbar_v$ are obtained similarly.

\begin{proposition}[Estimates for $\Omega^{-1} \hat{\chibar}$ on $S_{u,v}$ and $\Cbar_v$] \label{prop:chibarhatH2}
	For $k \leq N-s$, for $s=0,1,2$, for any $v_{-1} \leq v \leq v(R_2,u_f)$, and any $\max \{ u_0, u(R_2,v)\} \leq u \leq u_f$,
	\[
		\Vert
		\mathfrak{D}^k (\Omega^{-1} \hat{\chibar})
		\Vert_{S_{u,v}}^2
		+
		\Vert
		\mathfrak{D}^k (\Omega^{-1} \hat{\chibar})
		\Vert_{\Cbar_{v}}^2
		\lesssim
		\frac{\varepsilon_0^2+\varepsilon^3}{v^s}.
	\]
\end{proposition}

\begin{proof}
	The proof is similar to the proof of Proposition \ref{prop:chibarhatH}, though is much simpler since the degeneration of the estimates at $r=3M_f$ is not an issue, using now Lemma \ref{lem:nabla3xi}.
\end{proof}

\subsubsection{Estimates for $\Omega\hat{\chi}$}
\label{subsubsec:Hchihat}

Recall that a transport equation in the $\nablaslash_4$ direction of the form \eqref{eq:shifted} is referred to as \emph{redshifted}, \emph{noshifted} or \emph{blueshifted} according to whether the sign of $a$ in \eqref{eq:shifted} is positive, zero or negative respectively.  Recall also, see Proposition \ref{prop:chihatHschematic}, that $\Omega \hat{\chi}$ satisfies a blueshifted transport equation, $\Omega^{-1} \nablaslash_3 \Omega \hat{\chi}$ satisfies a noshifted transport equation, and higher order $\Omega^{-1} \nablaslash_3$ derivatives of $\Omega \hat{\chi}$ satisfy redshifted transport equations.  In Section \ref{subsec:ufH}, the derivatives of $\Omega \hat{\chi}$ were accordingly estimated on the final hypersurface $C_{u_f}$ differently depending whether they involve zero, one, or two or more $\Omega^{-1} \nablaslash_3$ derivatives.  In this section the derivatives of $\Omega \hat{\chi}$ are similarly estimated in $\DRH$ differently depending whether they involve zero, one, or two or more $\Omega^{-1} \nablaslash_3$ derivatives.

\subsubsection*{Estimates for $\Omega \hat{\chi}$ and angular derivatives up to order $N$}

To estimate angular derivatives of $\Omega \hat{\chi}$, $(r \nablaslash)^k \Omega \hat{\chi}$ for $k \leq N$, rather than using the equation \eqref{eq:chihat4} directly, which is blueshifted, it is convenient to first estimate the difference quotient
\[
	D_{u_f} \big( (r \nablaslash)^k \Omega \hat{\chi} \big),
\]
(recall the definition from Section \ref{subsubsec:L2transportestimatesH}).  Further, it is convenient to consider the renormalised quantities $X_1$ and $X_2$, introduced in Section \ref{section:Hequations}, which take the form
\begin{equation} \label{eq:X2X1def}
	X_1
	=
	r\Dslash_2^* r \divslash (r^2 \Omega \hat{\chi})
	-
	\frac{r^3}{2} \Omega^{-1} \nablaslash_3(r\Omega^2 \alpha),
	\qquad
	X_2
	=
	r\Dslash_2^* r \divslash X_1
	+
	\frac{r^4 \Omega^2}{4} (\Omega^{-1} \nablaslash_3)^2 (r \Omega^2 \alpha).
\end{equation}
Note that term term involving $\alpha$ on the right hand side of the equation in the $\nablaslash_4$ direction satisfied by $X_2$ (see Proposition \ref{prop:X2}), involves only two derivatives, in contrast to the commuted equation \eqref{eq:chihat4} for $(r\nablaslash)^4\Omega \hat{\chi}$ which involves four derivatives of $\Omega^2 \alpha$.

The estimates of Proposition \ref{prop:chihatHtop} are most relevant for the top order quantity $(r\nablaslash)^N \Omega \hat{\chi}$.  A better estimate for lower order derivatives will be obtained in Proposition \ref{prop:chihatnabla3Htop}.  First, estimates for the difference quotient $D_{u_f}(\Omega \hat{\chi})$ are obtained.

\begin{proposition}[Estimate for $D_{u_f}(\Omega \hat{\chi})$ on $S_{u,v}$] \label{prop:chihatdifferenceHtop}
	For $k \leq N$, for any $v_{-1} \leq v \leq v(R_2,u_f)$, and any $\max \{ u_0, u(R_2,v)\} \leq u \leq u_f$,
	\[
		\Vert
		D_{u_f} \big( (r \nablaslash)^k \Omega \hat{\chi} \big)
		\Vert_{S_{u,v}}^2
		\lesssim
		v^{\delta}(\varepsilon_0^2 + \varepsilon^3)
		+
		\sum_{\substack{
		l_1+l_2 \leq k-1
		\\
		l_2 \leq 2
		}}
		\Vert
		D_{u_f} \big( (r \nablaslash)^{l_1} (\Omega^{-1} \nablaslash_3)^{l_2} (\Omega^2 \alpha) \big)
		\Vert_{S_{u,v}}^2.
	\]
\end{proposition}

\begin{proof}
	The proof proceeds by considering the quantity $X_2$.  See equation \eqref{eq:X2X1def} above.  Suppose $k \geq 4$ (the case $k \leq 4$ will follow from Propositions \ref{prop:ellipticestimates} and \ref{prop:divcurl}).  By Proposition \ref{prop:X2} and Lemma \ref{lem:commutation},
	\[
		\Omega \nablaslash_4 (r\nablaslash)^{k-4} X_2
		-
		2 (\Omega \omegahat)_{\circ} (r\nablaslash)^{k-4} X_2
		=
		\mathcal{L}[(r\nablaslash)^{k-4}X_2]
		+
		\mathcal{E}[(r\nablaslash)^{k-4}X_2],
	\]
	where the linear term $\mathcal{L}[(r\nablaslash)^{k-4}X_2]$ takes the form
	\[
		\mathcal{L}[(r\nablaslash)^{k-4} X_2]
		=
		\sum_{\substack{
		l_1+l_2 \leq k-2
		\\
		l_2 \leq 2
		}}
		H^{l_1l_2} \cdot (r \nablaslash)^{l_1} (\Omega^{-1} \nablaslash_3)^{l_2} (r \Omega^2 \alpha)
	\]
	and the nonlinear error $\mathcal{E}[(r\nablaslash)^{k-4} X_2]$ has the schematic form,
	\[
		\mathcal{E}[(r\nablaslash)^{k-4} X_2]
		=
		\mathcal{E}^{k-1}
		+
		(H^{k,1} \cdot \Phi) \cdot (r \nablaslash)^k (\Omega \hat{\chi})
		+
		(H^{k,2} \cdot\Phi) \cdot
		(r\nablaslash)^k \Omega \omegahat,
	\]
	for some vectors of admissible coefficient functions (see \eqref{eq:admis}) $H^{l_1l_2}$, $H^{k,1}$, $H^{k,2}$.

	The proof will follow from Lemma \ref{lem:xidifference}.  First note that the initial condition, at $v=v_{-1}$, is controlled by Proposition \ref{prop:X2diffv0}.  Now, to control the term on $\{u=u_f\}$ arising from applying Lemma \ref{lem:xidifference}, Proposition \ref{prop:X1uf} gives
	\[
		\Vert
		(r\nablaslash)^{k-2} X_1
		\Vert_{C_{u_f}(v)}^2
		\lesssim
		\frac{\varepsilon_0^2
		+
		\varepsilon^3}{v},
	\]
	for all $v_{-1} \leq v \leq v(R_2,u_f)$ and so, for any $v_{-1} \leq v \leq v(R_2,u_f)$,
	\[
		\int_{v_{-1}}^v
		(v')^{1-\frac{\delta}{2}}
		\left\Vert (r\nablaslash)^{k-2} X_1 \right\Vert^2_{S_{u_f,v'}}
		dv'
		\lesssim
		\varepsilon_0^2
		+
		\varepsilon^3,
		\quad
		\text{hence,}
		\quad
		\int_{v_{-1}}^v
		(v')^{1+ \frac{\delta}{2}}
		\left\Vert (r\nablaslash)^{k-2} X_1 \right\Vert^2_{S_{u_f,v'}}
		dv'
		\\
		\lesssim
		v^{\delta} \left(
		\varepsilon_0^2
		+
		\varepsilon^3 \right).
	\]
	Theorem \ref{thm:alphaalphabarestimates} then implies that,
	\[
		\int_{v_{-1}}^v
		(v')^{1+\frac{\delta}{2}}
		\left\Vert (r\nablaslash)^{k-4} X_2 \right\Vert^2_{S_{u_f,v'}}
		dv'
		\lesssim
		v^{\delta} \left(
		\varepsilon_0^2
		+
		\varepsilon^3 \right).
	\]
	For the remaining terms arising from applying Lemma \ref{lem:xidifference} first note that
	Lemma \ref{lem:nabla3differenceestimate}
	implies that, for $\mathfrak{D}^{\gamma} = (r \nablaslash)^{l_1} (\Omega^{-1} \nablaslash_3)^{l_2}$ for some $l_1+l_2 \leq k-2$,
	\[
		\int_{v_{-1}}^v (v')^{1+\frac{\delta}{2}}
		\Vert
		D_{u_f} \mathfrak{D}^{\gamma} \alpha
		\Vert_{S_{u,v'}}^2
		dv'
		\lesssim
		\sup_{u\leq u' \leq u_f}
		\int_{v_{-1}}^v
		(v')^{1+\frac{\delta}{2}}
		(
		\Vert
		\Omega^{-1} \nablaslash_3 \mathfrak{D}^{\gamma} \alpha
		\Vert_{S_{u,v'}}^2
		+
		\Vert
		\mathfrak{D}^{\gamma} \alpha
		\Vert_{S_{u,v'}}^2
		) dv'
		\\
		\lesssim
		v^{\delta}(\varepsilon_0^2
		+
		\varepsilon^3),
	\]
	by Theorem \ref{thm:alphaalphabarestimates}.
	
	The nonlinear error terms arising from the first collection of nonlinear terms in $\mathcal{E}[(r\nablaslash)^{k-4} X_2]$ are estimated using Lemma \ref{lem:nabla3differenceestimate} and Proposition \ref{prop:spacetimeerrorH1}.

	For the second collection of nonlinear terms in $\mathcal{E}[(r\nablaslash)^{k-4} X_2]$, write
	\[
		D_{u_f} \left( \Phi \cdot (r\nablaslash)^k (\Omega \hat{\chi}) \right) (u,v)
		=
		D_{u_f}\Phi(u,v) \cdot (r\nablaslash)^k (\Omega \hat{\chi}) (u_f,v)
		+
		D_{u_f} (r\nablaslash)^k (\Omega \hat{\chi}) (u,v)
		\cdot
		\Phi(u,v).
	\]
	The first resulting term can be estimated exactly as the first collection of nonlinear terms are above.  For the second,
	\begin{align*}
		\int_{v_{-1}}^v (v')^{1+\frac{\delta}{2}}
		\!\!
		\int_{S^2}
		\vert
		D_{u_f} \big( (r\nablaslash)^k \Omega \hat{\chi} \big)
		\vert^2
		\left\vert \Phi \right\vert^2
		d\theta dv'
		\lesssim
		\sum_{k \leq 2}
		\int_{v_{-1}}^v (v')^{1+\frac{\delta}{2}}
		\!\!
		\int_{S^2}
		\left\vert (r\nablaslash)^k \Phi \right\vert^2
		d\theta
		\int_{S^2}
		\vert
		D_{u_f} \big( (r\nablaslash)^k \Omega \hat{\chi} \big)
		\vert^2
		d\theta
		dv'
		,
	\end{align*}
	by the Sobolev inequality, Proposition \ref{prop:Sobolev}.
	The nonlinear error terms arising from the third collection of nonlinear terms in $\mathcal{E}[(r\nablaslash)^{k-4} X_2]$ are estimated in Proposition \ref{prop:omegahatdifferenceH2}.
	
	Lemma \ref{lem:xidifference} then implies that
	\[
		\Vert
		D_{u_f} \big( (r \nablaslash)^{k-4} X_2 \big)
		\Vert_{S_{u,v}}^2
		\lesssim
		v^{\delta}(\varepsilon_0^2 + \varepsilon^3)
		+
		\sum_{\Phi, k \leq 2}
		\int_{v_{-1}}^v (v')^{1+\frac{\delta}{2}}
		\int_{S^2}
		\left\vert (r\nablaslash)^k \Phi \right\vert^2
		d\theta
		\int_{S^2}
		\vert
		D_{u_f} \big( (r\nablaslash)^k (\Omega \hat{\chi}) \big)
		\vert^2
		d\theta
		dv'
		.
	\]
	
	Using the definition of $X_2$ (see equation \eqref{eq:X2X1def}), it follows that,
	\begin{multline*}
		\Vert
		D_{u_f} \big( (r\nablaslash)^{k-4} (r^2 \Dslash_2^* \divslash)^2 (r^2 \Omega \hat{\chi}) \big)
		\Vert_{S_{u,v}}^2
		\lesssim
		v^{\delta} \left(
		\varepsilon_0^2
		+
		\varepsilon^3 \right)
		+
		\sum_{\substack{
		l_1+l_2 \leq k-1
		\\
		l_2 \leq 2
		}}
		\Vert
		D_{u_f} \big( (r \nablaslash)^{l_1} (\Omega^{-1} \nablaslash_3)^{l_2} (\Omega^2 \alpha) \big)
		\Vert_{S_{u,v}}^2
		\\
		+
		\sum_{\Phi, k \leq 2}
		\int_{v_{-1}}^v (v')^{1+\frac{\delta}{2}}
		\int_{S^2}
		\left\vert (r\nablaslash)^k \Phi \right\vert^2
		d\theta
		\int_{S^2}
		\vert
		D_{u_f} \big( (r\nablaslash)^k (\Omega \hat{\chi}) \big)
		\vert^2
		d\theta
		dv'
		.
	\end{multline*}
	The result then follows from the Gr\"{o}nwall inequality and Propositions \ref{prop:ellipticestimates} and \ref{prop:divcurl} since
	\[
		\int_{v_{-1}}^v (v')^{1+\frac{\delta}{2}}
		\int_{S^2}
		\left\vert (r\nablaslash)^k \Phi \right\vert^2
		d\theta
		dv'
		\lesssim
		\varepsilon^2
	\]
	for all $\Phi$ and $k \leq 2$.
\end{proof}

The following proposition provides weak control over $(r \nablaslash)^k \Omega \hat{\chi}$ for $k \leq N$.  Better estimates for lower order angular derivatives of $\Omega \hat{\chi}$ are obtained in Proposition \ref{prop:chihatnabla3Htop} and Proposition \ref{prop:chihatHn3aspacetime} below.

\begin{proposition}[Estimate for $\Omega \hat{\chi}$ on $S_{u,v}$ and $\Cbar_v$] \label{prop:chihatHtop}
	For $k \leq N$, for any $v_{-1} \leq v \leq v(R_2,u_f)$, and any $\max \{ u_0, u(R_2,v)\} \leq u \leq u_f$,
	\[
		\Vert
		(r\nablaslash)^k (\Omega \hat{\chi})
		\Vert_{S_{u,v}}^2
		+
		\Vert
		(r\nablaslash)^k (\Omega \hat{\chi})
		\Vert_{\Cbar_{v}}^2
		\lesssim
		v^{\delta} (\varepsilon_0^2+\varepsilon^3).
	\]
\end{proposition}

\begin{proof}
	From Proposition \ref{prop:chihatdifferenceHtop} it follows that
	\begin{align*}
		&
		\Vert (r\nablaslash)^k (\Omega \hat{\chi}) \Vert_{S_{u,v}}^2
		\leq
		\Omega_{\circ}(u,v)^2
		\Vert D_{u_f} \big( (r\nablaslash)^k (\Omega \hat{\chi}) \big) \Vert_{S_{u,v}}^2
		+
		\Vert (r\nablaslash)^k (\Omega \hat{\chi}) \Vert_{S_{u_f,v}}^2
		\\
		&
		\qquad \qquad
		\lesssim
		v^{\delta} (\varepsilon_0^2+\varepsilon^3)
		+
		\sum_{\vert \gamma \vert \leq k-1}
		\Omega_{\circ}(u,v)^2
		\Vert D_{u_f} \big( \mathfrak{D}^{\gamma} (\Omega^2 \alpha) \big) \Vert_{S_{u,v}}^2
		+
		\Vert (r\nablaslash)^k (\Omega \hat{\chi}) \Vert_{S_{u_f,v}}^2
		\lesssim
		v^{\delta} (\varepsilon_0^2+\varepsilon^3),
	\end{align*}
	by Theorem \ref{thm:alphaalphabarestimates} and Proposition \ref{prop:nabla3hochihatufR}.
\end{proof}

\subsubsection*{Estimates for $\Omega \hat{\chi}$, $\Omega^{-1} \nablaslash_3 (\Omega \hat{\chi})$ and angular derivatives up to order $N-1$}

The $\nablaslash_4$ equation for $\Omega^{-1} \nablaslash_3 (\Omega \hat{\chi})$ is \emph{noshifted} (see the $k_1=1$ case of Proposition \ref{prop:chihatHschematic}) and so, in order to obtain the best estimates for $\Omega^{-1} \nablaslash_3 (\Omega \hat{\chi})$ and its angular derivatives up to order $N-1$, it is again convenient to consider the difference quotient $D_{u_f} \big( \Omega^{-1} \nablaslash_3 (\Omega \hat{\chi}) \big)$.  As in the proof of Proposition \ref{prop:chihatdifferenceHtop}, the renormalised quantity $X_2$ is again considered.

\begin{proposition}[Estimate for $\Omega^{-1} \nablaslash_3 (\Omega \hat{\chi})$] \label{prop:chihatnabla3Htop}
	For $s=0,1,2$ and $k_1=0,1$, $k_2 \leq N-1-s$, for any $v_{-1} \leq v \leq v(R_2,u_f)$, and any $\max \{ u_0, u(R_2,v)\} \leq u \leq u_f$, if $\varepsilon$ is sufficiently small,
	\[
		\Vert
		(\Omega^{-1} \nablaslash_3)^{k_1} (r\nablaslash)^{k_2} (\Omega \hat{\chi})
		\Vert_{S_{u,v}}^2
		+
		\Vert
		(\Omega^{-1} \nablaslash_3)^{k_1} (r\nablaslash)^{k_2} (\Omega \hat{\chi})
		\Vert_{C_u(v)}^2
		\lesssim
		\frac{\varepsilon_0^2+\varepsilon^3}{v^s},
	\]
	and
	\[
		\Vert
		(\Omega^{-1} \nablaslash_3)^{k_1} (r\nablaslash)^{k_2} (\Omega \hat{\chi})
		\Vert_{\Cbar_v}^2
		+
		\Vert
		(\Omega^{-1} \nablaslash_3)^{k_1} (r\nablaslash)^{k_2} (\Omega \hat{\chi})
		\Vert_{\DRH(v)}^2
		\lesssim
		\frac{\varepsilon_0^2+\varepsilon^3}{v^s}.
	\]
\end{proposition}

\begin{proof}
	Consider some $l \leq N-5-s$.  Proposition \ref{prop:X2} and Lemma \ref{lem:commutation} imply that $\Omega^{-1} \nablaslash_3 (r\nablaslash)^l X_2$ satisfies the \emph{noshifted} equation
	\[
		\Omega \nablaslash_4 \Omega^{-1} \nablaslash_3 (r\nablaslash)^l X_2
		=
		- 2 \rho_{\circ} (r\nablaslash)^l X_2
		+
		\mathcal{L}[\Omega^{-1} \nablaslash_3 (r\nablaslash)^l X_2]
		+
		\mathcal{E}[\Omega^{-1} \nablaslash_3 (r\nablaslash)^l X_2],
	\]
	where the linear term $\mathcal{L}[\Omega^{-1} \nablaslash_3 (r\nablaslash)^l X_2]$ takes the form
	\[
		\mathcal{L}[\Omega^{-1} \nablaslash_3 (r\nablaslash)^l X_2]
		=
		\sum_{\substack{
		l_1 + l_2 \leq l+3
		\\
		l_2 \leq 3
		}}
		H^{l_1l_2} \cdot (r \nablaslash)^{l_1} (\Omega^{-1} \nablaslash_3)^{l_2} (r \Omega^2 \alpha)
	\]
	and the nonlinear error $\mathcal{E}[\Omega^{-1} \nablaslash_3 (r\nablaslash)^l X_2]$ has the schematic form,
	\[
		\mathcal{E}[\Omega^{-1} \nablaslash_3 (r\nablaslash)^l X_2]
		=
		\mathcal{E}^{l+4}
		+
		(H^{k,1} \cdot \Phi) \cdot (r \nablaslash)^{l+5} (\Omega \hat{\chi})
		+
		(H^{k,1} \cdot \Phi) \cdot \Omega^{-1} \nablaslash_3 (r \nablaslash)^{l+4} (\Omega \hat{\chi})
		+
		(H^{k,3} \cdot\Phi) \cdot
		(r\nablaslash)^{l+5} \Omega \omegahat,
	\]
	for some vectors of admissible coefficient functions (see \eqref{eq:admis}) $H^{l_1l_2}$, $H^{k,1}$, $H^{k,2}$, $H^{k,3}$, where the second equation of \eqref{eq:omega3omegabar4} has been used.

	The proof then follows that of Lemma \ref{lem:noshifteddifferencequotient}.
	Since, for any $\Phi$,
	\begin{multline*}
		D_{u_f} \big( \Phi \cdot \Omega^{-1} \nablaslash_3 (r\nablaslash)^{l+4} (\Omega \hat{\chi}) \big) (u,v)
		\\
		=
		D_{u_f} \Phi (u,v) \cdot \Omega^{-1} \nablaslash_3 (r\nablaslash)^{l+4} (\Omega \hat{\chi})(u_f,v)
		+
		D_{u_f} \big( \Omega^{-1} \nablaslash_3 (r\nablaslash)^{l+4} (\Omega \hat{\chi}) \big) (u,v)
		\cdot
		\Phi (u_f,v),
	\end{multline*}
	it follows that, for any $v_{-1} \leq v_1 < v_2 \leq v(R_2,u)$,
	\begin{multline*}
		\Vert
		D_{u_f} \Omega^{-1} \nablaslash_3 (r\nablaslash)^l X_2
		\Vert_{S_{u,v_2}}^2
		+
		\int_{v_1}^{v_2}
		\Vert
		D_{u_f} \Omega^{-1} \nablaslash_3 (r\nablaslash)^l X_2
		\Vert_{S_{u,v}}^2
		dv
		\lesssim
		\Vert
		D_{u_f} \Omega^{-1} \nablaslash_3 (r\nablaslash)^l X_2
		\Vert_{S_{u,v_1}}^2
		\\
		+
		\int_{v_1}^{v_2}
		\Vert
		(r\nablaslash)^l X_2
		\Vert_{S_{u_f,v}}^2
		+
		\Vert
		D_{u_f} (r\nablaslash)^l X_2
		\Vert_{S_{u,v}}^2
		+
		\Vert
		D_{u_f} \mathcal{L}
		\Vert_{S_{u,v}}^2
		+
		\Vert
		\hat{\mathcal{E}}
		\Vert_{S_{u,v}}^2
		+
		\varepsilon^2
		\Vert
		D_{u_f} \Omega^{-1} \nablaslash_3 (r\nablaslash)^{l+4} \Omega \hat{\chi}
		\Vert_{S_{u,v}}^2
		dv,
	\end{multline*}
	where $\mathcal{L} = \mathcal{L}[\Omega^{-1} \nablaslash_3 (r\nablaslash)^l X_2]$ and 
	\begin{align*}
		\hat{\mathcal{E}} (u,v)
		=
		\
		&
		D_{u_f}\mathcal{E}^{l+4} (u,v)
		+
		D_{u_f} \Big(
		(H^{k,1} \cdot \Phi) \cdot (r \nablaslash)^{l+5} (\Omega \hat{\chi})
		+
		(H^{k,3} \cdot\Phi) \cdot
		(r\nablaslash)^{l+5} \Omega \omegahat
		\Big) (u,v)
		\\
		&
		+
		\big( D_{u_f}(H^{k,1} \cdot \Phi) \big) (u,v)
		\cdot
		\Omega^{-1} \nablaslash_3 (r \nablaslash)^{l+4} (\Omega \hat{\chi})(u_f,v)
		.
	\end{align*}
	Note that the term $D_{u_f} \big( \Omega^{-1} \nablaslash_3 (r\nablaslash)^{l+4} (\Omega \hat{\chi}) \big) (u,v) \cdot \Phi (u_f,v)$ in $D_{u_f} \mathcal{E}[\Omega^{-1} \nablaslash_3 (r\nablaslash)^l X_2]$ is considered separately.  Now, since
	\[
		D_{u_f} \Omega^{-1} \nablaslash_3 (r^2 \Dslash_2^* \divslash)^2 (r\nablaslash)^l \Omega \hat{\chi}
		=
		(r^2 \Dslash_2^* \divslash)^2 D_{u_f} \Omega^{-1} \nablaslash_3 (r\nablaslash)^l \Omega \hat{\chi}
		+
		D_{u_f} [\Omega^{-1} \nablaslash_3, (r^2 \Dslash_2^* \divslash)^2] (r\nablaslash)^l \Omega \hat{\chi},
	\]
	it follows form Proposition \ref{prop:ellipticestimates} and Proposition \ref{prop:divcurl} that
	\begin{multline*}
		\sum_{l' \leq l+4}
		\Vert D_{u_f} \Omega^{-1} \nablaslash_3 (r\nablaslash)^{l'} \Omega \hat{\chi} \Vert_{S_{u,v}}^2
		\lesssim
		\Vert
		D_{u_f} \Omega^{-1} \nablaslash_3 (r^2 \Dslash_2^* \divslash)^2 (r\nablaslash)^l \Omega \hat{\chi} 
		\Vert_{S_{u,v}}^2
		\\
		+
		\frac{\varepsilon^2}{v^2}
		\sum_{\substack{l_1 +l_2 \leq l+2 \\ l_1 \leq 1}}
		\Vert (\Omega^{-1} \nablaslash_3)^{l_1} (r\nablaslash)^{l_2} \Omega \hat{\chi} \Vert_{S_{u,v}}^2,
	\end{multline*}
	and hence, by Theorem \ref{thm:alphaalphabarestimates}, if $\varepsilon$ is sufficiently small, for $l \leq l' \leq l+4$,
	\begin{multline*}
		\sum_{l' \leq l+4}
		\Vert
		D_{u_f} \Omega^{-1} \nablaslash_3 (r\nablaslash)^{l'} \Omega \hat{\chi}
		\Vert_{S_{u,v_2}}^2
		+
		\int_{v_1}^{v_2}
		\Vert
		D_{u_f} \Omega^{-1} \nablaslash_3 (r\nablaslash)^{l'} \Omega \hat{\chi}
		\Vert_{S_{u,v}}^2
		dv
		\lesssim
		\sum_{l' \leq l+4}
		\Vert
		D_{u_f} \Omega^{-1} \nablaslash_3 (r\nablaslash)^{l'} \Omega \hat{\chi}
		\Vert_{S_{u,v_1}}^2
		\\
		+
		\int_{v_1}^{v_2}
		\sum_{l' \leq l}
		\Big(
		\Vert
		(r\nablaslash)^{l'} X_2
		\Vert_{S_{u_f,v}}^2
		+
		\Vert
		D_{u_f} (r\nablaslash)^{l'} \Omega \hat{\chi}
		\Vert_{S_{u,v}}^2
		\Big)
		+
		\Vert
		D_{u_f} \mathcal{L}
		\Vert_{S_{u,v}}^2
		+
		\Vert
		\hat{\mathcal{E}}
		\Vert_{S_{u,v}}^2
		dv
		+
		\frac{\varepsilon_0^2 + \varepsilon^3}{v_1^s}
		.
	\end{multline*}
	It then follows, as in the proof of Lemma \ref{lem:noshifteddifferencequotient}, using Propositions \ref{prop:nabla3hochihatufR} and \ref{prop:nabla3chihatdiffv0}, that, for $l' \leq l+4$ and $i = 0,1$,
	\begin{align*}
		&
		\Vert (\Omega^{-1} \nablaslash_3)^i (r\nablaslash)^{l'} \Omega \hat{\chi} \Vert_{S_{u,v}}^2
		+
		\Vert (\Omega^{-1} \nablaslash_3)^i (r\nablaslash)^{l'} \Omega \hat{\chi} \Vert^2_{C_{u}(v)}
		+
		\Vert (\Omega^{-1} \nablaslash_3)^i (r\nablaslash)^{l'} \Omega \hat{\chi} \Vert^2_{\Cbar_{v}}
		\\
		&
		+
		\Vert (\Omega^{-1} \nablaslash_3)^i (r\nablaslash)^{l'} \Omega \hat{\chi} \Vert^2_{\DRH(v)}
		\\
		&
		\qquad \qquad
		\lesssim
		\frac{\varepsilon_0^2+\varepsilon^3}{v^s}
		+
		\Vert D_{u_f} \mathcal{L} \Vert_{C_{u}(v_{-1} \vee v/2)}^2
		+
		\Vert \hat{\mathcal{E}} \Vert_{C_{u}(v_{-1} \vee v/2)}^2
		+
		\Vert D_{u_f} \mathcal{L} \Vert_{\DRH(v_{-1} \vee v/2)}^2
		+
		\Vert \hat{\mathcal{E}} \Vert_{\DRH(v_{-1} \vee v/2)}^2
		,
	\end{align*}
	where $v_{-1} \vee v/2 := \max \{ v_{-1}, v/2 \}$.  To control the first $D_{u_f} \mathcal{L}$ term note that, by Lemma \ref{lem:nabla3differenceestimate}, for $\mathfrak{D}^{\gamma} = (r \nablaslash)^{l_1} (\Omega^{-1} \nablaslash_3)^{l_2}$ with $l_1+l_1\leq l-2$,
	\[
		\left\Vert
		D_{u_f} \big(
		\mathfrak{D}^{\gamma} ( \Omega^2 \alpha) \big)
		\right\Vert_{C_u(v_{-1} \vee v/2)}^2
		\lesssim
		\sup_{u\leq u'\leq u_f}
		\sum_{k=0,1}
		\left\Vert
		(\Omega^{-1} \nablaslash_3)^k
		\mathfrak{D}^{\gamma} ( \Omega^2 \alpha) 
		\right\Vert_{C_{u'}(v_{-1} \vee v/2)}^2
		\lesssim
		\frac{\varepsilon_0^2+\varepsilon^3}{v^s},
	\]
	by Theorem \ref{thm:alphaalphabarestimates}.  The term $\Vert D_{u_f} \mathcal{L} \Vert_{\DRH(v_{-1} \vee v/2)}^2$ is controlled similarly.  For the first nonlinear term in $\hat{\mathcal{E}}$, Lemma \ref{lem:nabla3differenceestimate} and Proposition \ref{prop:spacetimeerrorH1} imply that
	\[
		\left\Vert
		D_{u_f}
		\mathcal{E}^{l+4}
		\right\Vert_{C_u(v_{-1} \vee v/2)}^2
		\lesssim
		\frac{\varepsilon^4}{v^s}.
	\]
	The nonlinear term in $\hat{\mathcal{E}}$ arising from the nonlinear term involving $(r\nablaslash)^{l+5} (\Omega \omegahat - \Omega \omegahat_{\circ})$ in $\mathcal{E}$ is controlled by Proposition \ref{prop:omegahatdifferenceH2}.  For the nonlinear terms involving $\Omega \hat{\chi}$, for any $\Phi$ write
	\[
		D_{u_f} \big( \Phi \cdot (r\nablaslash)^{l+5} (\Omega \hat{\chi}) \big) (u,v)
		=
		D_{u_f} \Phi (u,v) \cdot (r\nablaslash)^{l+5} (\Omega \hat{\chi}) (u_f,v)
		+
		D_{u_f} \big(
		(r\nablaslash)^{l+5} (\Omega \hat{\chi}) \big) (u,v)
		\cdot
		\Phi(u,v).
	\]
	For the first term it follows exactly as above, using Lemma \ref{lem:nabla3differenceestimate} and Proposition \ref{prop:spacetimeerrorH1}, that
	\[
		\left\Vert
		(r\nablaslash)^{l+5} (\Omega \hat{\chi})
		\cdot
		D_{u_f} \Phi
		\right\Vert_{C_u(v_{-1} \vee v/2)}^2
		\lesssim
		\frac{\varepsilon^4}{v^s}.
	\]
	If $l \leq N-6$ then the second term can be estimated similarly.  For the remaining case, Proposition \ref{prop:chihatdifferenceHtop} implies that,
	\begin{multline*}
		\left\Vert
		\Phi \cdot D_{u_f} \big( (r\nablaslash)^{N} (\Omega \hat{\chi}) \big)
		\right\Vert^2_{C_u(v_{-1} \vee \frac{v}{2})}
		\lesssim
		\int_{v_{-1} \vee v/2}^{v(R_2,u)}
		\frac{\varepsilon^4}{(v')^{2-\delta}}
		dv'
		\\
		+
		\sum_{\substack{
		l_1+l_2 \leq N-1
		\\
		l_2 \leq 2
		}}
		\frac{\varepsilon^2}{v^2}
		\left\Vert
		D_{u_f} \big( (r \nablaslash)^{l_1} (\Omega^{-1} \nablaslash_3)^{l_2} (\Omega^2 \alpha) \big)
		\right\Vert^2_{C_u(v_{-1} \vee \frac{v}{2})}
		\lesssim
		\frac{\varepsilon^4}{v^{1-\delta}},
	\end{multline*}
	where the last inequality follows from Lemma \ref{lem:nabla3differenceestimate} and Theorem \ref{thm:alphaalphabarestimates}.  The remaining nonlinear term in $\hat{\mathcal{E}}$ is estimated similarly, and the Proposition follows.

\end{proof}

\subsubsection*{All remaining derivatives of $\Omega \hat{\chi}$}

In this section the remaining derivatives of $\Omega \hat{\chi}$ are estimated.  The top order angular derivatives of $\Omega \hat{\chi}$ only satisfy the weaker estimates of Propositions \ref{prop:chihatHtop} above and hence are excluded from the results of this section.

Recall that, given $k_1,k_2,k_3 \geq 0$, $\mathfrak{D}^{(k_1,k_2,k_3)} = (r\nablaslash)^{k_1} (\Omega^{-1} \nablaslash_3)^{k_2} (r\Omega \nablaslash_4)^{k_3}$.

\begin{proposition}[Estimate for remaining derivatives of $\Omega \hat{\chi}$ in $\DRH$] \label{prop:chihatHn3aspacetime}
	For $s=0,1,2$ and $k_1+k_2+k_3 \leq N-s$, $k_1 \leq N-1-s$, for any $v_{-1} \leq v \leq v(R_2,u_f)$, and any $\max \{ u_0, u(R_2,v)\} \leq u \leq u_f$,
	\[
		\Vert
		\mathfrak{D}^{\kbar} (\Omega \hat{\chi})
		\Vert_{S_{u,v}}^2
		+
		\Vert
		\mathfrak{D}^{\kbar} (\Omega \hat{\chi})
		\Vert_{C_u(v)}^2
		+
		\Vert
		\mathfrak{D}^{\kbar} (\Omega \hat{\chi})
		\Vert_{\Cbar_v}^2
		+
		\Vert
		\mathfrak{D}^{\kbar} (\Omega \hat{\chi})
		\Vert_{\DRH(v)}^2
		\lesssim
		\frac{\varepsilon_0^2+\varepsilon^3}{v^s},
	\]
	where $\kbar = (k_1,k_2,k_3)$.
\end{proposition}

\begin{proof}
	Consider first the region $2M_f \leq r \leq r_0$, and suppose first that $k_3=0$.  The proof proceeds by induction on $k_2$.  Note that, when $k_2=0$ and $k_2=1$, the proposition reduces to Proposition \ref{prop:chihatnabla3Htop}.  Suppose then that the result holds for some $1\leq k_2 \leq N-1-s$.  Proposition \ref{prop:chihatHschematic} and Lemma \ref{lem:nabla4redxi} then imply that
	\begin{multline*}
		\Vert (r\nablaslash)^{k_1} (\Omega^{-1} \nablaslash_3)^{k_2+1} \Omega \hat{\chi} \mathds{1} \Vert_{\Cbar^{\Hp}_v}^2
		+
		\Vert (r\nablaslash)^{k_1} (\Omega^{-1} \nablaslash_3)^{k_2+1} \Omega \hat{\chi} \mathds{1} \Vert_{\DRH(v)}^2
		\lesssim
		\frac{1}{v^{s}}
		\Vert (r\nablaslash)^{k_1} (\Omega^{-1} \nablaslash_3)^{k_2+1} \Omega \hat{\chi} \mathds{1} \Vert_{\Cbar^{\Hp}_{v_{-1}}}^2
		\\
		+
		\sum_{l_1+l_2 =0}^{k_2} \Vert (r\nablaslash)^{l_1} (\Omega^{-1} \nablaslash_3)^{l_2} \Omega \hat{\chi} \mathds{1} \Vert_{\DRH(v_{-1} \vee v/2)}^2
		+
		\sum_{l_1+l_2 =0}^{k_1+ k_2+1} \Vert (r\nablaslash)^{l_1} (\Omega^{-1} \nablaslash_3)^{l_2} \Omega^2 \alpha \mathds{1} \Vert_{\DRH(v_{-1} \vee v/2)}^2
		+
		\frac{\varepsilon^4}{v^s}
		,
	\end{multline*}
	where $\mathds{1} = \mathds{1}_{r\leq r_0}$ and the nonlinear terms are controlled using Propositions \ref{prop:spacetimeerrorH1} and \ref{prop:errorout}.  It then follows from Proposition \ref{prop:chihatnabla3v0}, Theorem \ref{thm:alphaalphabarestimates} and the inductive hypothesis that
	\[
		\Vert (r\nablaslash)^{k_1} (\Omega^{-1} \nablaslash_3)^{k_2+1} \Omega \hat{\chi} \mathds{1} \Vert_{\Cbar^{\Hp}_v}^2
		+
		\Vert (r\nablaslash)^{k_1} (\Omega^{-1} \nablaslash_3)^{k_2+1} \Omega \hat{\chi} \mathds{1} \Vert_{\DRH(v)}^2
		\lesssim
		\frac{\varepsilon_0^2+\varepsilon^3}{v^{s}}.
	\]
	Similarly for $\Vert (r\nablaslash)^{k_1} (\Omega^{-1} \nablaslash_3)^{k_2+1} \Omega \hat{\chi} \Vert_{S_{u,v}}^2$ and $\Vert (r\nablaslash)^{k_1} (\Omega^{-1} \nablaslash_3)^{k_2+1} \Omega \hat{\chi} \mathds{1} \Vert_{C_u^{\Hp}(v)}^2$.  For the case that $k_3 \geq 1$ it then inductively follows from applying $(r\nablaslash)^{k_1} (\Omega^{-1} \nablaslash_3)^{k_2} (\Omega \nablaslash_4)^{k_3-1}$ to (the appropriately normalised) equation \eqref{eq:chihat4} that
	\[
		\Vert \mathfrak{D}^{\kbar} \Omega \hat{\chi} \Vert_{S_{u,v}}^2
		+
		\Vert \mathfrak{D}^{\kbar} \Omega \hat{\chi} \mathds{1} \Vert_{C_u^{\Hp}(v)}^2
		+
		\Vert \mathfrak{D}^{\kbar} \Omega \hat{\chi} \mathds{1} \Vert_{\Cbar^{\Hp}_v}^2
		+
		\Vert \mathfrak{D}^{\kbar} \Omega \hat{\chi} \mathds{1} \Vert_{\DRH(v)}^2
		\lesssim
		\frac{\varepsilon_0^2+\varepsilon^3}{v^{s}},
	\]
	with $\kbar = (k_1,k_2,k_3)$.
	
	The region $r\geq r_0$ is simpler since $\Omega \sim 1$ and hence the redshift effect plays no role.  One does, however, have to take care with the degeneration of the integrated decay estimates for $\alpha$ at $r=3M_f$.  Consider again first the case that $k_3=0$.  Note again that, when $k_2=0$ and $k_2=1$, the proposition reduces to  Proposition \ref{prop:chihatnabla3Htop}.  Let $\tau$ be a smooth cut off function such that $\tau(u,v) = 1$ for $r(u,v) \geq r_0$ and $\tau(u,v) = 0$ for $r(u,v) \leq \frac{r_0-2M_f}{2}$.  Given $k_2 \geq 1$, from equation \eqref{eq:chihat4} it follows, by Lemma \ref{lem:commutation}, that
	\begin{multline*}
		\big\vert
		\Omega\nablaslash_4
		(
		\Omega^{-4}
		\tau \cdot R^* (\Omega \nablaslash_3)^{k_2-1} (r\nablaslash)^{k_1} \Omega \hat{\chi}
		)
		+
		2 \Omega \omegahat
		\Omega^{-4}
		\tau \cdot R^* (\Omega \nablaslash_3)^{k_2-1} (r\nablaslash)^{k_1} \Omega \hat{\chi}
		\big\vert
		\\
		\lesssim
		\vert
		\partial_v \tau
		\cdot
		R^* (\Omega \nablaslash_3)^{k_2-1} (r\nablaslash)^{k_1} \Omega \hat{\chi}
		\vert
		+
		\vert
		\tau \cdot
		R^* (\Omega \nablaslash_3)^{k_2-1} (r\nablaslash)^{k_1} \Omega^2 \alpha
		\vert
		+
		\vert \mathcal{E}^{* k_1+k_2} \vert,
	\end{multline*}
	where $R^* = \Omega \nablaslash_4 - \Omega \nablaslash_3$.  Lemma \ref{eq:nabla4redxi} (now with $r_* =R_2$), Theorem \ref{thm:alphaalphabarestimates} and the above estimates in $r \leq r_0$ then imply that
	\[
		\Vert
		R^* (\Omega \nablaslash_3)^{k_2-1} (r\nablaslash)^{k_1} \Omega \hat{\chi}
		\Vert_{\Cbar^{\Hp}_v}^2
		+
		\Vert 
		R^* (\Omega \nablaslash_3)^{k_2-1} (r\nablaslash)^{k_1} \Omega \hat{\chi}
		\Vert_{\DRH(v)}^2
		\lesssim
		\frac{\varepsilon_0^2+\varepsilon^3}{v^{s}},
	\]
	where Proposition \ref{prop:spacetimeerrorH1} is used to control the nonlinear error terms.  Similarly for $\Vert R^* (\Omega \nablaslash_3)^{k_2-1} (r\nablaslash)^{k_1} \Omega \hat{\chi} \Vert_{S_{u,v}}^2$ and $\Vert R^* (\Omega \nablaslash_3)^{k_2-1} (r\nablaslash)^{k_1} \Omega \hat{\chi} \Vert_{C_u^{\Hp}(v)}^2$.  The estimates then hold for $(\Omega \nablaslash_3)^{k_2} (r\nablaslash)^{k_1} \Omega \hat{\chi}$ after inserting equation \eqref{eq:chihat4}.  Combining with the above estimates in $r\leq r_0$ then gives the proposition in the case that $k_3 = 0$.  The case $k_3\geq 1$ then again follows inductively by applying $(r\nablaslash)^{k_1} (\Omega^{-1} \nablaslash_3)^{k_2} (\Omega \nablaslash_4)^{k_3-1}$ to equation \eqref{eq:chihat4}.
\end{proof}

\subsubsection{Estimates for curvature components}
\label{subsubsec:Hcurvature}

In this section most of the remaining curvature components are estimated (though there are some derivatives of $\Omega \beta_{\ell \geq 2}$ and $\Omega^{-1} \betabar_{\ell \geq 2}$ which are estimated in Section \ref{subsec:etaetabarH} after $\eta_{\ell \geq 2}$ and $\etabar_{\ell \geq 2}$ are estimated).

Recall, for an $S$-tensor $\xi$ and $u_0 \leq u \leq u_f$, $v \geq v_{-1}$, the norm
\begin{align*}
	\Vert \xi \Vert_{\DRH(v), \Cbar_v, C_u(v), S_{u,v}}^2
	:=
	&
	\Vert R^* \xi \Vert_{\DRH(v)}^2
	+
	\Vert \xi \Vert_{\DRH(v)}^2
	+
	\Vert \xi \Vert_{S_{u,v}}^2
	\\
	&
	+
	\sum_{\vert \gamma \vert \leq 1}
	\left(
	\Vert (1-3M_f/r) \mathfrak{D}^{\gamma} \xi \Vert_{\DRH(v)}^2
	+
	\Vert \mathfrak{D}^{\gamma} \xi \Vert_{\Cbar_v}^2
	+
	\Vert \mathfrak{D}^{\gamma} \xi \Vert_{C_u(v)}^2
	\right),
\end{align*}
where $R^* = \Omega \nablaslash_4 - \Omega \nablaslash_3$.  Note that the norm $\Vert \xi \Vert_{\DRH(v), \Cbar_v, C_u(v), S_{u,v}}^2$ involves first order derivatives of $\xi$.

\subsubsection*{Estimates for $\Omega \beta_{\ell \geq 2}$ and $\Omega^{-1}$\underline{$\beta$}$_{\ell \geq 2}$}

The following two propositions control most derivatives of $\beta$ and $\betabar$.  The exceptions are $(\Omega^{-1} \nablaslash_3)^N (\Omega \beta)$ and $(\Omega \nablaslash_4)^N (\Omega^{-1} \betabar)$, which will be controlled in Section \ref{subsec:etaetabarH} below after all derivatives of $\rho_{\ell \geq 2}$ and $\eta_{\ell \geq 2}$ and $\etabar_{\ell \geq 2}$ have been controlled.  Recall again that, given $k_1,k_2,k_3 \geq 0$, $\mathfrak{D}^{(k_1,k_2,k_3)} = (r\nablaslash)^{k_1} (\Omega^{-1} \nablaslash_3)^{k_2} (r\Omega \nablaslash_4)^{k_3}$.

\begin{proposition}[Estimate for $\Omega \beta_{\ell \geq 2}$ in $\DRH$] \label{prop:betaH}
	For $s=0,1,2$ and $\kbar=(k_1,k_2,k_3)$ with $k_1 + k_2 +k_3 \leq N-1-s$, $k_2 \leq N-2-s$, for any $u_0 \leq u \leq u_f$ and any $v_{-1} \leq v \leq v(R_2,u)$,
	\[
		\Vert \mathfrak{D}^{\kbar} (\Omega \beta)_{\ell \geq 2} \Vert_{\DRH(v), \Cbar_v, C_u(v), S_{u,v}}^2
		\lesssim
		\frac{\varepsilon_0^2 + \varepsilon^3}{v^s}.
	\]
\end{proposition}

\begin{proof}
	Consider a multi index $\kbar$ such that $\vert \kbar \vert \leq N-2-s$.  Equation \eqref{eq:alpha3} takes the schematic form
	\[
		-2 \Dslash_2^* \Omega \beta
		=
		\Omega^{-1} \nablaslash_3 (r \Omega^2 \alpha)
		-
		\frac{6M_f}{r^3} \Omega \hat{\chi}
		+
		\mathcal{E}^0,
	\]
	and so it follows from Lemma \ref{prop:ellipticestimates}, Theorem \ref{thm:alphaalphabarestimates} and Proposition \ref{prop:chihatHn3aspacetime} that
	\[
		\sum_{l=0,1} \Vert (r\nablaslash)^l \mathfrak{D}^{\kbar} (\Omega \beta)_{\ell \geq 2} \Vert_{\DRH(v), \Cbar_v, C_u(v), S_{u,v}}^2
		\lesssim
		\frac{\varepsilon_0^2 + \varepsilon^3}{v^s},
	\]
	using Propositions \ref{prop:spacetimeerrorH1}, \ref{prop:sphereserrorH}, \ref{prop:errorout} and \ref{prop:errorin} to control the nonlinear error terms.  Similarly, it follows from equation \eqref{eq:beta4} that
	\[
		\Vert \mathfrak{D}^{\kbar} r\Omega \nablaslash_4 (\Omega \beta)_{\ell \geq 2} \Vert_{\DRH(v), \Cbar_v, C_u(v), S_{u,v}}^2
		\lesssim
		\frac{\varepsilon_0^2 + \varepsilon^3}{v^s}.
	\]
\end{proof}

\begin{proposition}[Estimate for $\Omega^{-1} \betabar_{\ell \geq 2}$ in $\DRH$] \label{prop:betabarH}
	For $s=0,1,2$ and $\kbar=(k_1,k_2,k_3)$ with $k_1 + k_2 +k_3 \leq N-1-s$, $k_3 \leq N-2-s$, for any $u_0 \leq u \leq u_f$ and any $v_{-1} \leq v \leq v(R_2,u)$,
	\[
		\Vert \mathfrak{D}^{\kbar} (\Omega^{-1} \betabar)_{\ell \geq 2} \Vert_{\DRH(v), \Cbar_v, C_u(v), S_{u,v}}^2
		\lesssim
		\frac{\varepsilon_0^2 + \varepsilon^3}{v^s}.
	\]
\end{proposition}

\begin{proof}
	The proof is similar to that of Proposition \ref{prop:betaH}, using now equation \eqref{eq:alphabar4} and equation \eqref{eq:betabar3}.
\end{proof}

\subsubsection*{Estimates for $\rho_{\ell \geq 2}$}

The following proposition provides control over all derivatives of $\rho$ up to order $N$.

\begin{proposition}[Estimate for $\rho_{\ell \geq 2}$ in $\DRH$] \label{prop:rhoH1}
	For $s=0,1,2$ and $\vert \gamma \vert \leq N-1-s$, for any $u_0 \leq u \leq u_f$ and any $v_{-1} \leq v \leq v(R_2,u)$,
	\[
		\Vert \mathfrak{D}^{\gamma} \rho_{\ell \geq 2} \Vert_{\DRH(v), \Cbar_v, C_u(v), S_{u,v}}^2
		\lesssim
		\frac{\varepsilon_0^2 + \varepsilon^3}{v^s}.
	\]
\end{proposition}

\begin{proof}
	Consider first the case $\mathfrak{D}^{\gamma} = (r\nablaslash)^k$ for some $k \leq N-1-s$.  After commuting the equality \eqref{eq:Prhosigma} by $(r\nablaslash)^{k-2}$ (or not commuting if $k \leq 1$) and using Theorem \ref{thm:PPbarestimates} and Propositions \ref{prop:sigmaH}, \ref{prop:chibarhatH}, \ref{prop:chibarhatH2} and \ref{prop:chihatHn3aspacetime} to control the linear terms, and Propositions \ref{prop:spacetimeerrorH1}, \ref{prop:sphereserrorH}, \ref{prop:errorout} and \ref{prop:errorin} to control the nonlinear error terms, it follows from Proposition \ref{prop:ellipticestimates} that
	\[
		\sum_{l \leq k}
		\Vert (r\nablaslash)^l \rho_{\ell \geq 2} \Vert_{\DRH(v), \Cbar_v, C_u(v), S_{u,v}}^2
		\lesssim
		\frac{\varepsilon_0^2 + \varepsilon^3}{v^s}.
	\]
	Moreover, for $\vert \gamma \vert \leq N-2-s$, the estimates
	\[
		\Vert \mathfrak{D}^{\gamma} \Omega \nablaslash_4 \rho_{\ell \geq 2} \Vert_{\DRH(v), \Cbar_v, C_u(v), S_{u,v}}^2
		+
		\Vert \mathfrak{D}^{\gamma} \Omega^{-1} \nablaslash_3 \rho_{\ell \geq 2} \Vert_{\DRH(v), \Cbar_v, C_u(v), S_{u,v}}^2
		\lesssim
		\frac{\varepsilon_0^2 + \varepsilon^3}{v^s},
	\]
	follow from applying $\mathfrak{D}^{\gamma}$ to equations \eqref{eq:rho4} and \eqref{eq:rho3} respectively.  The proposition then follows.
\end{proof}

\subsubsection{Estimates for $\eta_{\ell \geq 2}$ and \underline{$\eta$}$_{\ell \geq 2}$}
\label{subsec:etaetabarH}

In this section $\eta_{\ell \geq 2}$ and \underline{$\eta$}$_{\ell \geq 2}$ are estimated.

\subsubsection*{Estimates for $\eta_{\ell \geq 2}$}
Most of the derivatives of $\eta$ can be estimated directly using the equation \eqref{eq:chihat3}, which can be rewritten schematically as,
\begin{equation} \label{eq:Dslashetaschematic}
	\Dslash_2^* \eta
	=
	-
	\frac{1}{2} \Omega^{-1} \nablaslash_3 (\Omega \hat{\chi})
	+
	\frac{1}{2r} \Omega \hat{\chi}
	-
	\frac{\Omega_{\circ}^2}{2r} \Omega^{-1} \hat{\chibar}
	+
	\mathcal{E}^0.
\end{equation}

\begin{proposition}[Estimate for $\nablaslash \eta_{\ell \geq 2}$ in $\DRH$] \label{prop:etaH}
	For $s=0,1,2$ and $\vert \gamma \vert \leq N-1-s$, for any $u_0 \leq u \leq u_f$ and any $v_{-1} \leq v \leq v(R_2,u)$,
	\[
		\sum_{l=0,1}
		\Vert (r\nablaslash)^l \mathfrak{D}^{\gamma} \eta_{\ell \geq 2} \Vert_{S_{u,v}}^2
		+
		\Vert (r\nablaslash)^l \mathfrak{D}^{\gamma} \eta_{\ell \geq 2} \Vert_{C_u(v)}^2
		+
		\Vert (r\nablaslash)^l \mathfrak{D}^{\gamma} \eta_{\ell \geq 2} \Vert_{\Cbar_v}^2
		+
		\Vert (r\nablaslash)^l \mathfrak{D}^{\gamma} \eta_{\ell \geq 2} \Vert_{\DRH(v)}^2
		\lesssim
		\frac{\varepsilon_0^2 + \varepsilon^3}{v^s}.
	\]
\end{proposition}

\begin{proof}
	After commuting equation \eqref{eq:Dslashetaschematic} with $\mathfrak{D}^{\gamma}$, the proof follows from Proposition \ref{prop:ellipticestimates}, Propositions \ref{prop:chibarhatH}, \ref{prop:chibarhatH2} and \ref{prop:chihatHn3aspacetime}, and using Propositions \ref{prop:spacetimeerrorH1}, \ref{prop:sphereserrorH}, \ref{prop:errorout} and \ref{prop:errorin} to control the nonlinear error terms.
\end{proof}

The remaining derivatives of $\eta$ are estimated using the equation \eqref{eq:nabla4eta} and so it is first necessary to estimate $\etabar$ and the remaining derivatives of $\Omega \beta$.

\subsubsection*{Estimates for \underline{$\eta$}$_{\ell \geq 2}$}

Derivatives of $\nablaslash \etabar$ will be controlled using the equation \eqref{eq:chibarhat4}, which can be written schematically as,
\begin{equation} \label{eq:Dslashetabarschematic}
	\Dslash_2^* \etabar
	=
	- \frac{1}{2} \Omega \nablaslash_4 (\Omega^{-1} \hat{\chibar})
	-
	\left( \frac{1}{4} (\Omega \tr \chi)_{\circ} + (\Omega \omegahat)_{\circ} \right) \Omega^{-1} \hat{\chibar}
	+
	\frac{1}{2r} \Omega \hat{\chi}
	+
	\mathcal{E}^0.
\end{equation}

\begin{proposition}[Estimate for $\nablaslash \etabar_{\ell \geq 2}$ in $\DRH$] \label{prop:etabarH}
	For $s=0,1,2$ and $\vert \gamma \vert \leq N-1-s$, for any $u_0 \leq u \leq u_f$ and any $v_{-1} \leq v \leq v(R_2,u)$,
	\[
		\sum_{l=0,1}
		\Vert (r\nablaslash)^l \mathfrak{D}^{\gamma} \etabar_{\ell \geq 2} \Vert_{S_{u,v}}^2
		+
		\Vert (r\nablaslash)^l \mathfrak{D}^{\gamma} \etabar_{\ell \geq 2} \Vert_{C_u(v)}^2
		+
		\Vert (r\nablaslash)^l \mathfrak{D}^{\gamma} \etabar_{\ell \geq 2} \Vert_{\Cbar_v}^2
		+
		\Vert (r\nablaslash)^l \mathfrak{D}^{\gamma} \etabar_{\ell \geq 2} \Vert_{\DRH(v)}^2
		\lesssim
		\frac{\varepsilon_0^2 + \varepsilon^3}{v^s}.
	\]
\end{proposition}

\begin{proof}
	The proof again follows from Proposition \ref{prop:ellipticestimates}, Propositions \ref{prop:chibarhatH}, \ref{prop:chibarhatH2} and \ref{prop:chihatHn3aspacetime}, now commuting equation \eqref{eq:Dslashetabarschematic} with $\mathfrak{D}^{\gamma}$.
\end{proof}

The remaining derivatives of $\etabar$ are estimated using the equation \eqref{eq:nabla4eta}.  The case when one of the derivatives is $\Omega^{-1} \nablaslash_3$ can be estimated already.

\begin{proposition}[Estimate for $\Omega^{-1} \nablaslash_3 \etabar_{\ell \geq 2}$ in $\DRH$] \label{prop:nabla3etabarH}
	For $s=0,1,2$ and $\vert \gamma \vert \leq N-1-s$, for any $u_0 \leq u \leq u_f$ and any $v_{-1} \leq v \leq v(R_2,u)$,
	\[
		\Vert \mathfrak{D}^{\gamma} \Omega^{-1} \nablaslash_3 \etabar_{\ell \geq 2} \Vert_{S_{u,v}}^2
		+
		\Vert \mathfrak{D}^{\gamma} \Omega^{-1} \nablaslash_3 \etabar_{\ell \geq 2} \Vert_{C_u(v)}^2
		+
		\Vert \mathfrak{D}^{\gamma} \Omega^{-1} \nablaslash_3 \etabar_{\ell \geq 2} \Vert_{\Cbar_v}^2
		+
		\Vert \mathfrak{D}^{\gamma} \Omega^{-1} \nablaslash_3 \etabar_{\ell \geq 2} \Vert_{\DRH(v)}^2
		\lesssim
		\frac{\varepsilon_0^2 + \varepsilon^3}{v^s}.
	\]
\end{proposition}

\begin{proof}
	The proof follows from projecting \eqref{eq:nabla4eta} to $\ell \geq 2$, applying $(\Omega^{-1} \nablaslash_3)^{N-1}$ and using Proposition~\ref{prop:com0oneforms}.  The linear terms involving $\beta$ can be estimated using Proposition \ref{prop:betaH}.  The linear terms involving $\eta$ can be estimated using Proposition \ref{prop:etaH}.
\end{proof}

\subsubsection*{Remaining derivatives of $\Omega\beta_{\ell \geq 2}$ and $\Omega^{-1}$\underline{$\beta$}$_{\ell \geq 2}$}

Before estimating the remaining derivatives of $\eta$, it is first necessary (in particular for the $(\Omega^{-1} \nablaslash_3)^N$ derivative) to control the remaining $(\Omega^{-1} \nablaslash_3)^{N-s}$ derivatives of $\Omega \beta$, which it is now possible to do using Proposition~\ref{prop:etaH} and the estimates for $\rho$ and $\sigma$.

\begin{proposition}[Estimate for $\Omega \beta_{\ell \geq 2}$ in $\DRH$] \label{prop:nabla3betatopH}
	For $s=0,1,2$ and $\vert \gamma \vert \leq N-1-s$, for any $u_0 \leq u \leq u_f$ and any $v_{-1} \leq v \leq v(R_2,u)$,
	\[
		\Vert \mathfrak{D}^{\gamma} (\Omega \beta)_{\ell \geq 2} \Vert_{\DRH(v), \Cbar_v, C_u(v), S_{u,v}}^2
		\lesssim
		\frac{\varepsilon_0^2 + \varepsilon^3}{v^s}.
	\]
\end{proposition}

\begin{proof}
	In the case that $\mathfrak{D}^{\gamma} = (\Omega^{-1} \nablaslash_3)^k$, for some $1\leq k \leq N-1-s$, the proof follows from projecting equation \eqref{eq:beta3}, which can be schematically rewritten as
	\[
		\Omega^{-1} \nablaslash_3 (\Omega \beta)
		=
		\frac{2}{r} \Omega \beta
		+
		\nablaslash \rho
		+
		{}^* \nablaslash \sigma
		-
		\frac{6M_f}{r^3} \eta
		+
		\mathcal{E}^0,
	\]
	to $\ell \geq 2$, applying $(\Omega^{-1} \nablaslash_3)^{k-1}$ and using Proposition \ref{prop:com0oneforms}, and using Propositions \ref{prop:betaH}, \ref{prop:rhoH1}, \ref{prop:sigmaH} and \ref{prop:etaH} to control the linear terms.  The nonlinear terms are controlled using Propositions \ref{prop:spacetimeerrorH1}, \ref{prop:sphereserrorH}, \ref{prop:errorout} and \ref{prop:errorin}.  In the remaining cases the proposition reduces to Proposition \ref{prop:betaH}.
\end{proof}

In order to estimate the remaining derivatives of $\etabar$, it is similarly necessary to control the $(\Omega \nablaslash_4)^{N-s}$ derivatives of $\Omega^{-1} \betabar$, which it is now possible to do using Proposition \ref{prop:etabarH}.

\begin{proposition}[Estimate for $\Omega^{-1} \betabar_{\ell \geq 2}$ in $\DRH$] \label{prop:nabla4betabartopH}
	For $s=0,1,2$ and $\vert \gamma \vert \leq N-1-s$, for any $u_0 \leq u \leq u_f$ and any $v_{-1} \leq v \leq v(R_2,u)$,
	\[
		\Vert \mathfrak{D}^{\gamma} (\Omega^{-1} \betabar)_{\ell \geq 2} \Vert_{\DRH(v), \Cbar_v, C_u(v), S_{u,v}}^2
		\lesssim
		\frac{\varepsilon_0^2 + \varepsilon^3}{v^s}.
	\]
\end{proposition}

\begin{proof}
	The proof is similar to that of Proposition \ref{prop:nabla3betatopH}, using now equation \eqref{eq:betabar4}, which can be schematically written
	\[
		\Omega \nablaslash_4 (\Omega^{-1} \betabar)
		=
		-
		\left(
		(\Omega \tr \chi)_{\circ}
		+
		2 (\Omega \omegahat)_{\circ}
		\right)
		\Omega^{-1} \betabar
		-
		\nablaslash \rho
		+
		{}^* \nablaslash \sigma
		+
		\frac{6M_f}{r^3} \etabar
		+
		\mathcal{E}^0.
	\]
\end{proof}

\subsubsection*{Remaining derivatives of $\eta_{\ell \geq 2}$ and \underline{$\eta$}$_{\ell \geq 2}$}

The remaining derivatives of $\eta$ are estimated using the equation \eqref{eq:nabla4eta}, which can schematically be written,
\begin{equation} \label{eq:nabla4etaHschematic}
	\Omega \nablaslash_4 (r \eta)
	=
	\Omega^2 \etabar
	-
	r \Omega \beta
	+
	\mathcal{E}^0.
\end{equation}

\begin{proposition}[Estimate for $\eta_{\ell \geq 2}$ in $\DRH$] \label{prop:nabla4etaH}
	For $s=0,1,2$ and $\vert \gamma \vert \leq N-s$, for any $u_0 \leq u \leq u_f$ and any $v_{-1} \leq v \leq v(R_2,u)$,
	\[
		\Vert \mathfrak{D}^{\gamma} \eta_{\ell \geq 2} \Vert_{S_{u,v}}^2
		+
		\Vert \mathfrak{D}^{\gamma} \eta_{\ell \geq 2} \Vert_{C_u(v)}^2
		+
		\Vert \mathfrak{D}^{\gamma} \eta_{\ell \geq 2} \Vert_{\Cbar_v}^2
		+
		\Vert \mathfrak{D}^{\gamma} \eta_{\ell \geq 2} \Vert_{\DRH(v)}^2
		\lesssim
		\frac{\varepsilon_0^2 + \varepsilon^3}{v^s}.
	\]
\end{proposition}

\begin{proof}
	First, the case that $\mathfrak{D}^{\gamma} = \mathfrak{D}^{\widetilde{\gamma}} r\Omega \nablaslash_4$, for some $\vert \widetilde{\gamma} \vert \leq \vert \gamma \vert -1$ follows from projecting equation \eqref{eq:nabla4etaHschematic} to $\ell \geq 2$, applying $\mathfrak{D}^{\widetilde{\gamma}}$ and using Proposition \ref{prop:com0oneforms}, Proposition \ref{prop:etabarH} and Proposition \ref{prop:nabla3betatopH}.
	
	Consider now the case that $\mathfrak{D}^{\gamma} = (\Omega^{-1} \nablaslash_3)^{k}$ for some $k \leq N-s$.  The proof is similar to that of Proposition \ref{prop:chihatHn3aspacetime} in which $\Omega \hat{\chi}$ is estimated.  The equation \eqref{eq:nabla4etaHschematic} is first projected to $\ell \geq 2$, commuted with $(\Omega^{-1} \nablaslash_3)^k$ and (noting that the resulting equation is redshifted) Lemma \ref{eq:nabla4redxi} is used, along with Propositions \ref{prop:nabla3etabarH} and \ref{prop:nabla3betatopH} and Proposition \ref{prop:DetaHv0} to control the term on $v=v_{-1}$, to estimate $(\Omega^{-1} \nablaslash_3)^k \eta$ in the region $r \leq r_0$, where the integrated decay estimate for $\beta$ does not degenerate.  The equation \eqref{eq:nabla4etaHschematic} is then commuted with $R^* (\Omega^{-1} \nablaslash_3)^{k-1}$ and Lemma \ref{eq:nabla4redxi} is again used, along with the fact that the integrated decay estimate for $R^* \Omega \beta$ of Proposition \ref{prop:nabla3betatopH} does not degenerate at $r=3M_f$, to estimate $R^* (\Omega^{-1} \nablaslash_3)^{k-1} \eta$ in $\DRH$.
	
	The remaining cases follow from Proposition \ref{prop:etaH}.
\end{proof}

The remaining derivatives of $\etabar$ are estimated by commuting the equation \eqref{eq:nabla4eta}.

\begin{proposition}[Estimate for $\etabar_{\ell \geq 2}$ in $\DRH$] \label{prop:nabla4etabarH}
	For $s=0,1,2$ and $\vert \gamma \vert \leq N-s$, for any $u_0 \leq u \leq u_f$ and any $v_{-1} \leq v \leq v(R_2,u)$,
	\[
		\Vert \mathfrak{D}^{\gamma} \etabar_{\ell \geq 2} \Vert_{S_{u,v}}^2
		+
		\Vert \mathfrak{D}^{\gamma} \etabar_{\ell \geq 2} \Vert_{C_u(v)}^2
		+
		\Vert \mathfrak{D}^{\gamma} \etabar_{\ell \geq 2} \Vert_{\Cbar_v}^2
		+
		\Vert \mathfrak{D}^{\gamma} \etabar_{\ell \geq 2} \Vert_{\DRH(v)}^2
		\lesssim
		\frac{\varepsilon_0^2 + \varepsilon^3}{v^s}.
	\]
\end{proposition}

\begin{proof}
	Suppose $\mathfrak{D}^{\gamma} = (r\Omega \nablaslash_4)^k$ for some $1\leq k \leq N-s$.  The proof follows from considering \eqref{eq:nabla4eta}, which can be schematically written
	\[
		\Omega \nablaslash_3 (r \etabar)
		=
		-
		\Omega^2 \eta
		+
		\Omega^2
		r \Omega^{-1} \betabar
		+
		\Omega^2 \mathcal{E}^0,
	\]
	projecting to $\ell \geq 2$ and commuting with $(\Omega \nablaslash_4)^{k-1} R^*$, and using Lemma \ref{lem:nabla3xi} and Proposition \ref{prop:com0oneforms}.  The term at $u=u_f$ is controlled by Proposition \ref{prop:etabar4Huf}, and the linear terms are controlled by Proposition \ref{prop:nabla4betabartopH} and Proposition \ref{prop:nabla4etaH}.  The result then follows from Proposition \ref{prop:nabla3etabarH}.  The remaining cases follow from Proposition \ref{prop:etabarH} and Proposition \ref{prop:nabla3etabarH}.
\end{proof}

\subsubsection{Estimates for $\Omega_{\circ}^{-2}\Omega^2_{\ell \geq 2}$}
\label{subsec:OmegaH}

Estimates for $\Omega^2$ can easily be obtained from the estimates for $\eta$ and $\etabar$ using the equation
\begin{equation} \label{eq:nablaslashOmegaH}
	\nablaslash \log \Omega^2 = \eta + \etabar.
\end{equation}

\begin{proposition}[Estimate for $\Omega_{\circ}^{-2}\Omega^2_{\ell \geq 2}$ in $\DRH$] \label{prop:OmegaH}
	For $s=0,1,2$ and $\vert \gamma \vert \leq N-s$, for any $u_0 \leq u \leq u_f$ and any $v_{-1} \leq v \leq v(R_2,u)$,
	\[
		\Big\Vert \mathfrak{D}^{\gamma} (\Omega_{\circ}^{-2}\Omega^2_{\ell \geq 2}) \Big\Vert_{S_{u,v}}^2
		+
		\Big\Vert \mathfrak{D}^{\gamma} (\Omega_{\circ}^{-2}\Omega^2_{\ell \geq 2}) \Big\Vert_{C_u(v)}^2
		+
		\Big\Vert \mathfrak{D}^{\gamma} (\Omega_{\circ}^{-2}\Omega^2_{\ell \geq 2}) \Big\Vert_{\Cbar_v}^2
		+
		\Big\Vert \mathfrak{D}^{\gamma} (\Omega_{\circ}^{-2}\Omega^2_{\ell \geq 2}) \Big\Vert_{\DRH(v)}^2
		\lesssim
		\frac{\varepsilon_0^2 + \varepsilon^3}{v^s}.
	\]
\end{proposition}

\begin{proof}
	The equation \eqref{eq:nablaslashOmegaH} implies that
	\begin{equation} \label{eq:nablaslashOmegaH2}
		\nablaslash \log \left( \frac{\Omega^2}{\Omega_{\circ}^2} \right)
		=
		\eta + \etabar.
	\end{equation}
	The proof follows from Proposition \ref{prop:etaH} and Proposition \ref{prop:etabarH} and the Poincar\'{e} inequality, Proposition \ref{prop:Poincare}, after projecting to $\ell \geq 2$ and commuting with $\mathfrak{D}^{\gamma}$.
\end{proof}

Since $\Omega^2$ is a metric component, it is also possible to estimate more than $N$ of its derivatives.  Such estimates will not be obtained here.

\subsubsection{Estimates for $(\Omega \omegahat - (\Omega \omegahat)_{\circ})_{\ell \geq 2}$ and $\Omega^{-2} (\Omega \omegabarhat - (\Omega \omegabarhat)_{\circ})_{\ell \geq 2}$}
\label{subsubsec:omegahatomegabarhatH}

The propagation equation \eqref{eq:omega3omegabar4} for $\Omega \omegahat$ can be schematically rewritten as,
\begin{equation} \label{eq:omegahatH}
	\partial_u \left( \Omega \omegahat - (\Omega \omegahat)_{\circ} \right)
	=
	- \Omega^2 (\rho - \rho_{\circ})
	+
	\frac{2M_f}{r^3} \left( \Omega^2 - \Omega_{\circ}^2 \right)
	+
	\Omega^2 \mathcal{E}^0,
\end{equation}
and equation \eqref{eq:nabla4etabar} can be schematically rewritten
\begin{equation} \label{eq:nablaomegahatH}
	2 \nablaslash ( \Omega \omegahat)
	=
	\Omega \nablaslash_4 \etabar
	+
	2 \frac{\Omega^2}{r} \etabar
	-
	\Omega \beta
	+
	\mathcal{E}^0.
\end{equation}

\begin{proposition}[Estimate for $(\Omega \omegahat)_{\ell \geq 2}$ in $\DRH$] \label{prop:omegahatH}
	For $s=0,1,2$ and $\vert \gamma \vert \leq N-s$, for any $u_0 \leq u \leq u_f$ and any $v_{-1} \leq v \leq v(R_2,u)$,
	\[
		\Vert \mathfrak{D}^{\gamma} (\Omega \omegahat)_{\ell \geq 2} \Vert_{S_{u,v}}^2
		+
		\Vert \mathfrak{D}^{\gamma} (\Omega \omegahat)_{\ell \geq 2} \Vert_{C_u(v)}^2
		+
		\Vert \mathfrak{D}^{\gamma} (\Omega \omegahat)_{\ell \geq 2} \Vert_{\Cbar_v}^2
		+
		\Vert \mathfrak{D}^{\gamma} (\Omega \omegahat)_{\ell \geq 2} \Vert_{\DRH(v)}^2
		\lesssim
		\frac{\varepsilon_0^2 + \varepsilon^3}{v^s}.
	\]
\end{proposition}

\begin{proof}
	Consider first the case that $\mathfrak{D}^{\gamma} = \mathfrak{D}^{\widetilde{\gamma}} r \nablaslash$ for some $\vert \widetilde{\gamma} \vert = \vert \gamma \vert -1$.  The proof follows from projecting equation \eqref{eq:nablaomegahatH} to $\ell \geq 2$, applying $\mathfrak{D}^{\widetilde{\gamma}}$, using Proposition \ref{prop:nabla3betatopH} and Proposition \ref{prop:nabla4etabarH} to control the linear terms, and Propositions \ref{prop:spacetimeerrorH1}, \ref{prop:sphereserrorH}, \ref{prop:errorout} and \ref{prop:errorin} to control the nonlinear terms.  Next, if $\mathfrak{D}^{\gamma} = \mathfrak{D}^{\widetilde{\gamma}} \Omega^{-1} \nablaslash_3$ for some $\vert \widetilde{\gamma} \vert = \vert \gamma \vert -1$, the proof follows from dividing equation \eqref{eq:omegahatH} by $\Omega^2$, projecting to $\ell \geq 2$ and using Proposition \ref{prop:com0}, applying $\mathfrak{D}^{\widetilde{\gamma}}$ and using Proposition \ref{prop:rhoH1} and Proposition \ref{prop:OmegaH}.  Finally, suppose $\mathfrak{D}^{\gamma} = (r\Omega \nablaslash_4)^k$ for some $1 \leq k \leq N-s$.  The proof then follows from projecting \eqref{eq:omegahatH} to $\ell \geq 2$, commuting with $R^* (\Omega \nablaslash_4)^{k-1}$ and using Lemma \ref{lem:nabla3xi} and Proposition \ref{prop:com0}.  The linear terms are controlled using Proposition \ref{prop:rhoH1} and Proposition \ref{prop:OmegaH}.  The proof then follows from the previous cases.
\end{proof}

The propagation equation \eqref{eq:omega3omegabar4} for $\omegabarhat$ can be rewritten as,
\begin{equation} \label{eq:omegabarhatH}
	\partial_v \left(
	\Omega^{-2}
	\left(
	\Omega \omegabarhat - (\Omega \omegabarhat)_{\circ}
	\right)
	\right)
	+
	2 \Omega \omegahat
	\Omega^{-2}
	\left(
	\Omega \omegabarhat - (\Omega \omegabarhat)_{\circ}
	\right)
	=
	-(\rho - \rho_{\circ})
	+
	\frac{2M_f}{r^3} \left( 1 - \frac{\Omega_{\circ}^2}{\Omega^2} \right)
	+
	\mathcal{E}^0.
\end{equation}
Equation \eqref{eq:nabla3eta} can be schematically written,
\begin{equation} \label{eq:omegabarhatnablaH}
	2 \nablaslash (\Omega^{-2} (\Omega \omegabarhat - \Omega\omegahat_{\circ}))
	=
	\Omega^{-1} \nablaslash_3 \eta
	-
	\frac{2}{r} \eta
	+
	\Omega^{-1} \betabar
	+
	\mathcal{E}^0.
\end{equation}

In the following, $(\omegabarhat)_{*, \ell \geq 2}$ will denote the quantity
\[
	(\omegabarhat)_{*, \ell \geq 2}
	:=
	\left(
	\Omega^{-2}
	\left(
	\Omega \omegabarhat - \Omega \omegabarhat_{\circ}
	\right)
	\right)_{\ell \geq 2}.
\]

\begin{proposition}[Estimate for $\left( \Omega^{-2} \left( \Omega \omegabarhat - \Omega \omegabarhat_{\circ} \right) \right)_{\ell \geq 2}$ in $\DRH$] \label{prop:omegabarhatH}
	For $s=0,1,2$ and $\vert \gamma \vert \leq N-s$, for any $u_0 \leq u \leq u_f$ and any $v_{-1} \leq v \leq v(R_2,u)$,
	\[
		\Vert \mathfrak{D}^{\gamma} (\omegabarhat)_{*, \ell \geq 2} \Vert_{S_{u,v}}^2
		+
		\Vert \mathfrak{D}^{\gamma} (\omegabarhat)_{*, \ell \geq 2} \Vert_{C_u(v)}^2
		+
		\Vert \mathfrak{D}^{\gamma} (\omegabarhat)_{*, \ell \geq 2} \Vert_{\Cbar_v}^2
		+
		\Vert \mathfrak{D}^{\gamma} (\omegabarhat)_{*, \ell \geq 2} \Vert_{\DRH(v)}^2
		\lesssim
		\frac{\varepsilon_0^2 + \varepsilon^3}{v^s}.
	\]
\end{proposition}

\begin{proof}
	In the cases that $\mathfrak{D}^{\gamma} = \mathfrak{D}^{\widetilde{\gamma}} r\nablaslash$ and $\mathfrak{D}^{\gamma} = \mathfrak{D}^{\widetilde{\gamma}} r\Omega \nablaslash_4$ for some $\vert \widetilde{\gamma} \vert = \vert \gamma \vert -1$ the proof follows from respectively projecting \eqref{eq:omegabarhatnablaH} and \eqref{eq:omegabarhatH} to $\ell \geq 2$ and applying $\mathfrak{D}^{\widetilde{\gamma}}$, using Propositions \ref{prop:com0}, \ref{prop:rhoH1}, \ref{prop:nabla3betatopH}, \ref{prop:nabla4etaH}, and \ref{prop:OmegaH}.  In the case that $\mathfrak{D}^{\gamma} = (\Omega^{-1} \nablaslash_3)^k$ for some $1\leq k \leq N-s$ the proof follows the approach of Proposition \ref{prop:chihatHn3aspacetime}.  First equation \eqref{eq:omegabarhatH} is inductively commuted with $(\Omega^{-1} \nablaslash_3)^k$ for all $1\leq k \leq N-s$, and Lemma \ref{lem:nabla4redxi} is used with $r_* = r_0$ to control $(\Omega^{-1} \nablaslash_3)^k (\omegabarhat)_{*, \ell \geq 2}$ in the region $r \leq r_0$.  Then equation \eqref{eq:omegabarhatH} is commuted with $R^* (\Omega \nablaslash_3)^{k-1}$ and Lemma \ref{lem:nabla4redxi} is used again to control $(\Omega^{-1} \nablaslash_3)^k (\omegabarhat)_{*, \ell \geq 2}$ in the region $r_0 \leq r \leq R_2$.  Proposition \ref{prop:rhoH1} and Proposition \ref{prop:OmegaH} are used to control the linear error terms, and Propositions \ref{prop:spacetimeerrorH1} and \ref{prop:errorout} to control the nonlinear terms.  The terms on the initial hypersurface $v=v_{-1}$ are controlled by Proposition \ref{prop:DomegabarHv0}.
\end{proof}

\subsubsection{Estimates for $(\Omega \tr \chi - \Omega \tr \chi_{\circ})_{\ell \geq 2}$ and $(\Omega^{-2}(\Omega \tr \chibar - \Omega \tr \chibar_{\circ}))_{\ell \geq 2}$}
\label{subsubsec:trchitrchibarH}

In this section, the quantities $(\Omega \tr \chi - \Omega \tr \chi_{\circ})_{\ell \geq 2}$ and $(\Omega^{-2}(\Omega \tr \chibar - \Omega \tr \chibar_{\circ}))_{\ell \geq 2}$ are estimated.

\subsubsection*{Equations}
The equations satisfied by $\tr \chi$ and $\tr \chibar$ are more convenient in the following renormalised forms.
The Codazzi equation \eqref{eq:Codazzi} can be rewritten,
\begin{align}
	\nablaslash \left(
	\Omega \tr \chi - (\Omega \tr \chi)_{\circ}
	\right)
	&
	=
	2 \divslash (\Omega \hat{\chi})
	+
	(\Omega \tr \chi)_{\circ} \etabar
	+
	2\Omega \beta
	+
	2 \Omega \hat{\chi} \cdot \eta
	+
	\left(
	\Omega \tr \chi - (\Omega \tr \chi)_{\circ}
	\right) \etabar
	\nonumber
	\\
	&
	=
	2 \divslash (\Omega \hat{\chi})
	+
	(\Omega \tr \chi)_{\circ} \etabar
	+
	2\Omega \beta
	+
	\mathcal{E}^0.
	\label{eq:CodazziH}
\end{align}

The Raychaudhuri equation for $\tr \chi$, \eqref{eq:Ray}, can be rewritten
\begin{align}
	\Omega \nablaslash_4 \left(
	\Omega \tr \chi - (\Omega \tr \chi)_{\circ}
	\right)
	=
	\
	&
	-
	\left(
	(\Omega \tr \chi)_{\circ}
	-
	2(\Omega \omegahat)_{\circ}
	\right)
	\left(
	\Omega \tr \chi - (\Omega \tr \chi)_{\circ}
	\right)
	+
	\frac{4\Omega_{\circ}^2}{r} \left(
	\Omega \omegahat - (\Omega \omegahat)_{\circ}
	\right)
	\nonumber
	\\
	&
	-
	\frac{1}{2}
	\left(
	\Omega \tr \chi - (\Omega \tr \chi)_{\circ}
	\right)^2
	+
	2
	\left(
	\Omega \omegahat - (\Omega \omegahat)_{\circ}
	\right)
	\left(
	\Omega \tr \chi - (\Omega \tr \chi)_{\circ}
	\right)
	-
	\vert \Omega \hat{\chi} \vert^2
	\nonumber
	\\
	=
	\
	&
	-
	\left(
	(\Omega \tr \chi)_{\circ}
	-
	2(\Omega \omegahat)_{\circ}
	\right)
	\left(
	\Omega \tr \chi - (\Omega \tr \chi)_{\circ}
	\right)
	+
	\frac{4\Omega_{\circ}^2}{r} \left(
	\Omega \omegahat - (\Omega \omegahat)_{\circ}
	\right)
	+
	\mathcal{E}^0.
	\label{eq:trchi4H}
\end{align}
The equation \eqref{eq:trchi3} can be rewritten,
\begin{align}
	\Omega^{-1} \nablaslash_3
	\left(
	\Omega \tr \chi - (\Omega \tr \chi)_{\circ}
	\right)
	=
	\
	&
	2\divslash \eta
	+
	\frac{\Omega_{\circ}^2}{r \Omega^2}
	\left(
	\Omega \tr \chi - (\Omega \tr \chi)_{\circ}
	\right)
	-\frac{\Omega_{\circ}^2}{r \Omega^2}
	\left(
	\Omega \tr \chibar - (\Omega \tr \chibar)_{\circ}
	\right)
	+
	2 \frac{\Omega_{\circ}^2}{\Omega^2} (\rho - \rho_{\circ})
	\nonumber
	\\
	&
	-
	\frac{4M_f}{r^3} \left( 1 - \frac{\Omega_{\circ}^2}{\Omega^2} \right)
	-
	\frac{1}{2\Omega^2}
	\left(
	\Omega \tr \chibar - (\Omega \tr \chibar)_{\circ}
	\right)
	\left(
	\Omega \tr \chi - (\Omega \tr \chi)_{\circ}
	\right)
	\nonumber
	\\
	&
	+
	2
	\left( 1 - \frac{\Omega_{\circ}^2}{\Omega^2} \right)
	(\rho - \rho_{\circ})
	-
	(\Omega^{-1} \hat{\chibar}) \cdot (\Omega \hat{\chi})
	+
	2 \eta \cdot \eta
	\nonumber
	\\
	=
	\
	&
	2\divslash \eta
	+
	\frac{\Omega_{\circ}^2}{r \Omega^2}
	\left(
	\Omega \tr \chi - (\Omega \tr \chi)_{\circ}
	\right)
	-\frac{\Omega_{\circ}^2}{r \Omega^2}
	\left(
	\Omega \tr \chibar - (\Omega \tr \chibar)_{\circ}
	\right)
	+
	2 \frac{\Omega_{\circ}^2}{\Omega^2} (\rho - \rho_{\circ})
	\nonumber
	\\
	&
	-
	\frac{4M_f}{r^3} \left( 1 - \frac{\Omega_{\circ}^2}{\Omega^2} \right)
	+
	\mathcal{E}^0.
	\label{eq:trchi3H}
\end{align}

The Codazzi equation \eqref{eq:Codazzibar} can be rewritten,
\begin{align}
	\nablaslash 
	\left(
	\Omega^{-2}
	\left(
	\Omega \tr \chibar - (\Omega \tr \chibar)_{\circ}
	\right)
	\right)
	&
	=
	2 \divslash (\Omega^{-1} \hat{\chibar})
	-
	\frac{2}{r} \frac{\Omega_{\circ}^2}{\Omega^2} \eta
	-
	2 \Omega^{-1} \betabar
	+
	2 \Omega^{-1} \hat{\chibar} \cdot ( \eta + 2 \etabar)
	-
	\Omega^{-2} \left(
	\Omega \tr \chibar - (\Omega \tr \chibar)_{\circ}
	\right) \etabar
	\nonumber
	\\
	&
	=
	2 \divslash (\Omega^{-1} \hat{\chibar})
	-
	\frac{2}{r} \eta
	-
	2 \Omega^{-1} \betabar
	+
	\mathcal{E}^0.
	\label{eq:CodazzibarH}
\end{align}
The Raychaudhuri equation for $\tr \chibar$, \eqref{eq:Ray}, can be rewritten
\begin{align}
	\Omega^{-1} \nablaslash_3
	\left( \Omega^{-2}
	\left(
	\Omega \tr \chibar - (\Omega \tr \chibar)_{\circ}
	\right)
	\right)
	=
	\
	&
	\frac{2}{r}
	\frac{\Omega_{\circ}^2}{\Omega^2}
	\Omega^{-2}
	\left(
	\Omega \tr \chibar - (\Omega \tr \chibar)_{\circ}
	\right)
	-
	\frac{4}{r} \frac{\Omega_{\circ}^2}{\Omega^2} \Omega^{-2} \left(
	\Omega \omegabarhat - (\Omega \omegabarhat)_{\circ}
	\right)
	\nonumber
	\\
	&
	-
	\frac{1}{2}
	\left( \Omega^{-2}
	\left(
	\Omega \tr \chibar - (\Omega \tr \chibar)_{\circ}
	\right)
	\right)^2
	-
	\vert \Omega^{-1} \hat{\chibar} \vert^2
	\nonumber
	\\
	=
	\
	&
	\frac{2}{r}
	\Omega^{-2}
	\left(
	\Omega \tr \chibar - (\Omega \tr \chibar)_{\circ}
	\right)
	-
	\frac{4}{r} \Omega^{-2} \left(
	\Omega \omegabarhat - (\Omega \omegabarhat)_{\circ}
	\right)
	+
	\mathcal{E}^0.
	\label{eq:trchibar3H}
\end{align}
The equation \eqref{eq:trchibar4} can be rewritten,
\begin{align}
	\Omega \nablaslash_4
	\left( \Omega^{-2}
	\left(
	\Omega \tr \chibar - (\Omega \tr \chibar)_{\circ}
	\right)
	\right)
	=
	\
	&
	2\divslash \etabar
	-
	\left( 2(\Omega \omegahat)_{\circ} + \frac{1}{2} (\Omega \tr \chi)_{\circ} \right)
	\Omega^{-2}
	\left(
	\Omega \tr \chibar - (\Omega \tr \chibar)_{\circ}
	\right)
	\nonumber
	\\
	&
	+
	\frac{\Omega_{\circ}^2}{r \Omega^2}
	\left(
	\Omega \tr \chi - (\Omega \tr \chi)_{\circ}
	\right)
	+
	2 \frac{\Omega_{\circ}^2}{\Omega^2} (\rho - \rho_{\circ})
	-
	\frac{4M_f}{r^3} \left( 1 - \frac{\Omega_{\circ}^2}{\Omega^2} \right)
	-
	(\Omega^{-1} \hat{\chibar}) \cdot (\Omega \hat{\chi})
	\nonumber
	\\
	&
	-
	\frac{1}{2\Omega^2}
	\left(
	\Omega \tr \chibar - (\Omega \tr \chibar)_{\circ}
	\right)
	\left(
	\Omega \tr \chi - (\Omega \tr \chi)_{\circ}
	\right)
	+
	2
	\left( 1 - \frac{\Omega_{\circ}^2}{\Omega^2} \right)
	(\rho - \rho_{\circ})
	+
	2 \etabar \cdot \etabar
	\nonumber
	\\
	=
	\
	&
	2\divslash \etabar
	-
	\left( 2(\Omega \omegahat)_{\circ} + \frac{1}{2} (\Omega \tr \chi)_{\circ} \right)
	\Omega^{-2}
	\left(
	\Omega \tr \chibar - (\Omega \tr \chibar)_{\circ}
	\right)
	+
	\frac{1}{r}
	\left(
	\Omega \tr \chi - (\Omega \tr \chi)_{\circ}
	\right)
	\nonumber
	\\
	&
	+
	2(\rho - \rho_{\circ})
	-
	\frac{4M_f}{r^3} \left( 1 - \frac{\Omega_{\circ}^2}{\Omega^2} \right)
	+
	\mathcal{E}^0.
	\label{eq:trchibar4H}
\end{align}

\subsubsection*{Estimates for $\Omega \tr \chi_{\ell \geq 2}$}

Most of the derivatives of $(\Omega \tr \chi)_{\ell \geq 2}$ can immediately be estimated.  The quantity $(r\nablaslash)^N (\Omega \tr \chi)_{\ell \geq 2}$ only satisfies the weaker estimate of Proposition \ref{prop:trchiH2} below.  The quantity $(\Omega^{-1} \nablaslash_3)^N (\Omega \tr \chi)_{\ell \geq 2}$ is estimated in Proposition \ref{prop:trchiH3}, after derivatives of $(\Omega \tr \chibar)_{\ell \geq 2}$ have been estimated.

Recall again that $\mathfrak{D}^{(k_1,k_2,k_3)} = (r\nablaslash)^{k_1} (\Omega^{-1} \nablaslash_3)^{k_2} (r\Omega \nablaslash_4)^{k_3}$.

\begin{proposition}[Estimate for $\Omega \tr \chi_{\ell \geq 2}$ in $\DRH$] \label{prop:trchiH1}
	For $s=0,1,2$ and $k_1+k_2+k_3 \leq N-s$, $k_1 +k_2 \leq N-1-s$, for any $v_{-1} \leq v \leq v(R_2,u_f)$, and any $\max \{ u_0, u(R_2,v)\} \leq u \leq u_f$,
	\[
		\Vert
		\mathfrak{D}^{\kbar} (\Omega \tr \chi)_{\ell \geq 2}
		\Vert_{S_{u,v}}^2
		+
		\Vert
		\mathfrak{D}^{\kbar} (\Omega \tr \chi)_{\ell \geq 2}
		\Vert_{C_u(v)}^2
		+
		\Vert
		\mathfrak{D}^{\kbar} (\Omega \tr \chi)_{\ell \geq 2}
		\Vert_{\Cbar_v}^2
		+
		\Vert
		\mathfrak{D}^{\kbar} (\Omega \tr \chi)_{\ell \geq 2}
		\Vert_{\DRH(v)}^2
		\lesssim
		\frac{\varepsilon_0^2+\varepsilon^3}{v^s},
	\]
	where $\kbar = (k_1,k_2,k_3)$.
\end{proposition}

\begin{proof}
	Suppose first that $k_1+k_2+k_3 \leq N-1-s$.  The proof follows from projecting equation \eqref{eq:CodazziH} to $\ell \geq 2$, commuting with $(r\nablaslash)^{k_1-1} (\Omega^{-1} \nablaslash_3)^{k_2} (r\Omega \nablaslash_4)^{k_3}$ (or with $(\Omega^{-1} \nablaslash_3)^{k_2} (r\Omega \nablaslash_4)^{k_3}$ if $k_1=0$) and using Propositions \ref{prop:chihatHn3aspacetime}, \ref{prop:nabla3betatopH} and \ref{prop:nabla4etabarH}, and Propositions \ref{prop:spacetimeerrorH1}, \ref{prop:sphereserrorH}, \ref{prop:errorout} and \ref{prop:errorin} to control the nonlinear terms.  The case that $\mathfrak{D}^{\kbar} = (r\Omega \nablaslash_4)^{k_3}$ for some $1 \leq k_3 \leq N-s$ follows inductively from applying $(\Omega \nablaslash_4)^{k_3-1}$ to equation \eqref{eq:trchi4H} (after projecting to $\ell \geq 2$ and using Proposition \ref{prop:com0}) and using Proposition \ref{prop:omegahatH}.
\end{proof}

The top order quantity $(r\nablaslash)^N (\Omega \tr \chi)_{\ell \geq 2}$ satisfies the following weaker estimate.

\begin{proposition}[Estimate for $(r\nablaslash)^N (\Omega \tr \chi)_{\ell \geq 2}$ in $\DRH$] \label{prop:trchiH2}
	For any $k \leq N$, $u_0 \leq u \leq u_f$ and any $v_{-1} \leq v \leq v(R_2,u)$,
	\begin{align*}
		\Vert
		(r\nablaslash)^{k} (\Omega \tr \chi)_{\ell \geq 2}
		\Vert_{S_{u,v}}^2
		+
		\Vert
		(r\nablaslash)^{k} (\Omega \tr \chi)_{\ell \geq 2}
		\Vert_{\Cbar_v}^2
		\lesssim
		v^{\delta}
		(
		\varepsilon_0^2
		+
		\varepsilon^3).
	\end{align*}
\end{proposition}

\begin{proof}
	The proof follows from Propositions \ref{prop:chihatHtop}, \ref{prop:nabla3betatopH} and \ref{prop:nabla4etabarH} after applying $(r\nablaslash)^{k-1}$ to equation \eqref{eq:CodazziH}.
\end{proof}

\subsubsection*{Estimates for $(\Omega^{-2}(\Omega \tr \chibar - (\Omega \tr \chibar)_{\circ}))_{\ell \geq 2}$}

In the following, $(\tr \chibar)_{*, \ell \geq 2}$ will denote the quantity
\[
	(\tr \chibar)_{*, \ell \geq 2}
	:=
	\left(
	\Omega^{-2}
	\left(
	\Omega \tr \chibar - (\Omega \tr \chibar)_{\circ}
	\right)
	\right)_{\ell \geq 2}.
\]

\begin{proposition}[Estimate for $(\Omega^{-2}(\Omega \tr \chibar - (\Omega \tr \chibar)_{\circ}))_{\ell \geq 2}$ in $\DRH$] \label{prop:trchibarH1}
	For $s=0,1,2$ and $\vert \gamma \vert \leq N-s$, for any $v_{-1} \leq v \leq v(R_2,u_f)$, and any $\max \{ u_0, u(R_2,v)\} \leq u \leq u_f$,
	\[
		\Vert
		\mathfrak{D}^{\gamma} (\tr \chibar)_{*, \ell \geq 2}
		\Vert_{S_{u,v}}^2
		+
		\Vert
		\mathfrak{D}^{\gamma} (\tr \chibar)_{*, \ell \geq 2}
		\Vert_{C_u(v)}^2
		+
		\Vert
		\mathfrak{D}^{\gamma} (\tr \chibar)_{*, \ell \geq 2}
		\Vert_{\Cbar_v}^2
		+
		\Vert
		\mathfrak{D}^{\gamma} (\tr \chibar)_{*, \ell \geq 2}
		\Vert_{\DRH(v)}^2
		\lesssim
		\frac{\varepsilon_0^2+\varepsilon^3}{v^s}.
	\]
\end{proposition}

\begin{proof}
	The proof for $\gamma = 0$ follows from equation \eqref{eq:CodazzibarH}, the Poincar\'{e} inequality, Proposition \ref{prop:Poincare}, and Propositions \ref{prop:chibarhatH}, \ref{prop:chibarhatH2}, \ref{prop:nabla4betabartopH} and \ref{prop:nabla4etaH}.  The proof for $\vert \gamma \vert \geq 1$ follows from inductively considering some $\widetilde{\gamma}$ with $\vert \widetilde{\gamma} \vert = \vert \gamma \vert -1$ and applying $\mathfrak{D}^{\widetilde{\gamma}}$ to either equation \eqref{eq:CodazzibarH}, \eqref{eq:trchibar3H} or \eqref{eq:trchibar4H} and using Propositions \ref{prop:chibarhatH}, \ref{prop:chibarhatH2}, \ref{prop:rhoH1}, \ref{prop:nabla4betabartopH}, \ref{prop:nabla4etaH}, \ref{prop:nabla4etabarH}, \ref{prop:OmegaH}, \ref{prop:omegabarhatH} and \ref{prop:trchiH1}.
\end{proof}

\subsubsection*{Estimates for remaining derivatives of $(\Omega \tr \chi)_{\ell \geq 2}$}

The remaining derivatives of $(\Omega \tr \chi)_{\ell \geq 2}$ can now be estimated.

\begin{proposition}[Estimate for $\Omega \tr \chi_{\ell \geq 2}$ in $\DRH$] \label{prop:trchiH3}
	For $s=0,1,2$ and $k_1+k_2+k_3 \leq N-s$, $k_1 \leq N-1-s$, for any $v_{-1} \leq v \leq v(R_2,u_f)$, and any $\max \{ u_0, u(R_2,v)\} \leq u \leq u_f$,
	\[
		\Vert
		\mathfrak{D}^{\kbar} (\Omega \tr \chi)_{\ell \geq 2}
		\Vert_{S_{u,v}}^2
		+
		\Vert
		\mathfrak{D}^{\kbar} (\Omega \tr \chi)_{\ell \geq 2}
		\Vert_{C_u(v)}^2
		+
		\Vert
		\mathfrak{D}^{\kbar} (\Omega \tr \chi)_{\ell \geq 2}
		\Vert_{\Cbar_v}^2
		+
		\Vert
		\mathfrak{D}^{\kbar} (\Omega \tr \chi)_{\ell \geq 2}
		\Vert_{\DRH(v)}^2
		\lesssim
		\frac{\varepsilon_0^2+\varepsilon^3}{v^s},
	\]
	where $\kbar = (k_1,k_2,k_3)$.
\end{proposition}

\begin{proof}
	By Proposition \ref{prop:trchiH1} the only remaining case to show is when $\mathfrak{D}^{\kbar} = (\Omega^{-1} \nablaslash_3)^{k_2}$ for some $1 \leq k_2 \leq N-s$.  The proof then follows from Propositions \ref{prop:rhoH1}, \ref{prop:nabla4etaH}, \ref{prop:OmegaH}, \ref{prop:trchiH1} and \ref{prop:trchibarH1} after projecting equation \eqref{eq:trchi3H} to $\ell \geq 2$, applying $(\Omega^{-1} \nablaslash_3)^{k_2-1}$ and using Proposition \ref{prop:com0}.
\end{proof}

\subsection{Estimates for quantities on $C_{u_f}$ and $\protect\Cbar_{v_{-1}}$: the $\ell =0,1$ modes}
\label{subsec:Hell0ell1ufv1}

In this section, the defining conditions of the $\Hp$ gauge are exploited to estimate the $\ell = 0,1$ modes of the geometric quantities on $C_{u_f}$ and $\Cbar_{v_{-1}}$.
Recall that, restricted to their $\ell=0,1$ modes, the geometric quantities in the $\Hp$ gauge satisfy the gauge conditions
\begin{align}
	\label{eq:Hgaugelm0}
	b_{\ell = 0,1}(u_f,v,\theta) = 0
	&
	\qquad
	\text{on } C_{u_f};
	\\
	\label{eq:Hgaugelm1}
	\left(\Omega(u_f,v,\theta) - \Omega_{\circ} (u_f,v)\right)_{\ell = 0,1} = 0
	&
	\qquad
	\text{on } C_{u_f};
	\\
	\label{eq:Hgaugelm2}
	\partial_u \left( r^3 (\divslash \eta)_{\ell = 1} + r^3 \rho_{\ell = 1} \right) (u,v_{-1},\theta) = 0
	&
	\qquad \text{for all }
	u_0 \leq u \leq u_f, \theta \in S^2;
	\\
	\label{eq:Hgaugelm8}
	\left(\Omega(u,v_{-1}) - \Omega_{\circ} (u,v_{-1}) \right)_{\ell=0} = 0
	&
	\qquad \text{for all }u_0 \leq u \leq u_f;
	\\
	\label{eq:Hgaugelm3}
	(f^3_{\Hp,\I})_{\ell=0,1}(u_f, v(R,u_f), \theta) = 0
	&
	\qquad \text{for all } \theta \in S^2;
	\\
	\label{eq:Hgaugelm4}
	\mu^*_{\ell =1} (u_f,v(R,u_f),\theta) = 0
	&
	\qquad \text{for all } \theta \in S^2;
	\\
	\label{eq:Hgaugelm4A}
	\left( \Omega \tr \chi - (\Omega \tr \chi)_{\circ} \right)_{\ell=0} (u_f,v(R,u_f),\theta) = 0
	&
	\qquad \text{for all } \theta \in S^2;
	\\
	\label{eq:Hgaugelm5}
	\left( \Omega \tr \chibar - (\Omega \tr \chibar)_{\circ} \right)_{\ell=0,1}(u_f,v_{-1}, \theta) = 0
	&
	\qquad \text{for all } \theta \in S^2;
\end{align}
where
\[
	\mu^* = \divslash \eta + (\rho - \rho_{\circ}) - \frac{3}{2r} \left( \Omega \tr \chi - (\Omega \tr \chi)_{\circ}\right).
\]

Recall the reference linearised Kerr solution from Section \ref{reflinearisedKerrsec} and Section \ref{linKerrforHp}.  The estimates of the following sections will be shown to hold for the solution minus this reference linearised Kerr solution, as the linearised Kerr solution itself does not satisfy the estimates.
 
The fact that. if $\xi$ is an $S$-tangent  symmetric trace free $(0,2)$ tensor, 
\[
	(\divslash \divslash \xi)_{\ell=0,1}
	=
	(\curlslash \divslash \xi)_{\ell=0,1}
	=
	0
\]
will be used throughout this section.

\subsubsection{Estimates for $\ell=1$ modes of quantities on the final hypersurface $C_{u_f}$}
\label{subsec:finall1uf}
In this section the $\ell = 1$ modes are estimated on $C_{u_f}$.
Note that that the gauge condition \eqref{eq:Hgaugelm3} implies, by Theorem \ref{thm:inheriting} and Theorem \ref{thm:Iestimates}, that $\divslash (\Omega \beta)_{\ell=1}$ satisfies the estimate
\begin{equation} \label{eq:betaufdata}
	\left\vert
	\divslash (\Omega \beta)_{\ell=1}
	(u_f,v(R,u_f))
	\right\vert^2
	\lesssim
	\frac{\varepsilon_0^2
	+
	\varepsilon^3}{v(R,u_f)^{4-2\delta}}
	,
\end{equation}
and moreover, by the definition of the $\Hp$ linearised Kerr solution,
\begin{equation} \label{eq:curlbetaufdata}
	\curlslash \big( (\Omega \beta)_{\ell=1}
	-
	\Omega \beta_{\mathrm{Kerr}} 
	\big)
	(u_f,v(R,u_f),\theta)
	=
	0
	\qquad
	\text{for all }
	\theta \in S_{u_f,v(R,u_f)}.
\end{equation}

\begin{proposition}[Estimate for $(\Omega \beta)_{\ell=1} - \Omega \beta_{\mathrm{Kerr}}$ on $C_{u_f}$] \label{prop:divbetal1uf}
	On the final hypersurface $u=u_f$, for any $v_{-1} \leq v \leq v(R_2,u_f)$, for $k \leq N-s$, $s=0,1,2$,
	\begin{equation*}
		\left\Vert (\Omega \nablaslash_4)^k (\Omega \beta - \Omega \beta_{\mathrm{Kerr}})_{\ell=1} \right\Vert^2_{S_{u_f,v}}
		+
		\left\Vert (\Omega \nablaslash_4)^k (\Omega \beta - \Omega \beta_{\mathrm{Kerr}})_{\ell=1} \right\Vert^2_{C_{u_f}(v)}
		\lesssim
		\frac{\varepsilon_0^2
		+
		\varepsilon^3}{v^{2-2\delta +s}}.
	\end{equation*}
	Moreover, for $k \leq N-1-s$, $s=0,1,2$,
	\begin{equation*}
		\left\Vert (\Omega \nablaslash_4)^k (\divslash \Omega \beta)_{\ell=1} \right\Vert^2_{S_{u_f,v}}
		+
		\left\Vert (\Omega \nablaslash_4)^k (\divslash \Omega \beta)_{\ell=1} \right\Vert^2_{C_{u_f}(v)}
		\lesssim
		\frac{\varepsilon_0^2
		+
		\varepsilon^3}{v^{2-2\delta +s}}.
	\end{equation*}
\end{proposition}

\begin{proof}
	The Bianchi equation \eqref{eq:beta4} implies
	\begin{equation} \label{eq:beta4l1schem}
		\Omega \nablaslash_4 \left( r^4 \big( \Omega \beta - \Omega \beta_{\mathrm{Kerr}} \big) \right)_{\ell=1}
		-
		2 (\Omega \omegahat)_{\circ}
		r^4 \big( (\Omega \beta)_{\ell=1} - \Omega \beta_{\mathrm{Kerr}} \big)
		=
		\mathcal{E}^0_{\ell=1},
	\end{equation}
	where the error term has the correct structure to use Proposition \ref{prop:newnoshifterrorestimate}.  The proof of the first estimate, for $k=0$, then follows from Lemma \ref{lem:nabla4bluexi}, the ``final condition'' \eqref{eq:betaufdata} and \eqref{eq:curlbetaufdata}, and Proposition \ref{prop:newnoshifterrorestimate}, using also Proposition \ref{prop:com0}.  For $k\geq 1$ the proof follows inductively from applying $(\Omega \nablaslash_4)^{k-1}$ to \eqref{eq:beta4l1schem}.  The second estimate, for $(\divslash \Omega \beta)_{\ell=1}$, is similar.
\end{proof}

\begin{proposition}[Equations for $\curlslash \eta_{\ell=1}$] \label{prop:curletal1eqnschematic}
	The quantity $\curlslash \eta_{\ell=1}$ satisfies the equations
	\begin{equation} \label{eq:curletal1schem}
		\Omega \nablaslash_3 (r^4 \curlslash \eta_{\ell=1})
		=
		\Omega^2 \mathcal{E}^1,
		\qquad
		\Omega \nablaslash_4 (r^4 \curlslash \eta_{\ell=1})
		=
		\mathcal{E}^1,
	\end{equation}
	where the nonlinear errors contain no quadratic terms which are both non-vanishing in the reference linearised Kerr solution, i.\@e.\@ no terms of the form
	\[
		\mathfrak{D}^{l_1} \Phi^{(1)} \cdot \mathfrak{D}^{l_2} \Phi^{(2)},
		\qquad
		\Phi^{(1)}, \Phi^{(2)}
		\in
		\{ \eta, \etabar, \sigma, \Omega^{-1} \betabar, \Omega \beta \}.
	\]
\end{proposition}

\begin{proof}
	From equation \eqref{eq:nabla4etabar} it follows that
	\[
		\Omega \nablaslash_4 (r^3 \curlslash \eta) = r^3 \curlslash (\Omega \beta) + \mathcal{E}^1.
	\]
	The Codazzi equation equation \eqref{eq:Codazzi} implies that
	\[
		\frac{(\Omega \tr \chi)_{\circ}}{2} \curlslash \eta_{\ell =1}
		=
		-
		\curlslash (\Omega \beta)_{\ell =1}
		+
		\mathcal{E}^1,
	\]
	and the equation for $\Omega \nablaslash_4 (r^4 \curlslash \eta_{\ell =1})$ follows after noting that $\curlslash \eta = - \curlslash \etabar$.  The $\Omega \nablaslash_3$ equation is obtained similarly.
\end{proof}

\begin{proposition}[Estimate for $\curlslash \eta_{\ell=1} - \curlslash \eta_{\mathrm{Kerr}}$ on $C_{u_f}$] \label{prop:curletal1uf}
	On the final hypersurface $u=u_f$, for any $v_{-1} \leq v \leq v(R_2,u_f)$, for $k \leq N-s$, $s=0,1,2$,
	\begin{equation*}
		\left\Vert (\Omega \nablaslash_4)^k \left( 
		\curlslash \eta_{\ell=1}
		-
		\curlslash \eta_{\mathrm{Kerr}}
		\right) \right\Vert^2_{S_{u_f,v}}
		\lesssim
		\frac{\varepsilon^2_0 + \varepsilon^3}{v^{2-2\delta +s}}.
	\end{equation*}
\end{proposition}

\begin{proof}
	Consider first the case $k =0$.  The Codazzi equation \eqref{eq:Codazzi} projected to $\ell =1$ together with \eqref{eq:curlbetaufdata} implies that, on $C_{u_f}$.
	\[
		\vert
		\curlslash \eta_{\ell=1}
		-
		\curlslash \eta_{\mathrm{Kerr}}
		\vert^2
		\lesssim
		\frac{\varepsilon^2_0 + \varepsilon^3}{v^{4-2\delta}}.
	\]
	For $k \geq 1$ the proof follows from Proposition \ref{prop:divbetal1uf} after applying $(\Omega \nablaslash_4)^{k-1}$ to the latter of \eqref{eq:curletal1schem} and using the error estimate of Proposition \ref{prop:sphereserrorH}.
\end{proof}

\begin{proposition}[Estimate for $\sigma_{\ell=1} - \sigma_{\mathrm{Kerr}}$ on $C_{u_f}$] \label{prop:sigmal1uf}
	On the final hypersurface $u=u_f$, for any $v_{-1} \leq v \leq v(R_2,u_f)$, for $k \leq N-s$, $s=0,1,2$,
	\begin{equation*}
		\left\Vert (\Omega \nablaslash_4)^k \left( 
		\sigma_{\ell=1}
		-
		\sigma_{\mathrm{Kerr}}
		\right) \right\Vert^2_{S_{u_f,v}}
		\lesssim
		\frac{\varepsilon^2_0 + \varepsilon^3}{v^{2-2\delta + s}}.
	\end{equation*}
\end{proposition}

\begin{proof}
	The proof follows from Proposition \ref{prop:curletal1uf} and equation \eqref{eq:curletacurletabar}.
\end{proof}

Recall that $\mu^*_{\ell =1}$ was estimated on $C_{u_f}$ in Proposition \ref{prop:Xuf}.

\begin{proposition}[Estimate for $\mu^*_{\ell =1}$ on $C_{u_f}$] \label{prop:Xl1uf}
	On the final hypersurface $u=u_f$, for any $v_{-1} \leq v \leq v(R_2,u_f)$, for $k \leq N-s$, $s=0,1,2$,
	\begin{equation*}
		\left\Vert (\Omega \nablaslash_4)^k
		\mu^*_{\ell =1}
		\right\Vert^2_{S_{u_f,v}}
		\lesssim
		\frac{\varepsilon^4}{v^{s+1}}.
	\end{equation*}
\end{proposition}

\begin{proof}
	See Proposition \ref{prop:Xuf}.
\end{proof}

\begin{proposition}[Preliminary estimate for $(\Omega \tr \chi)_{\ell =1}$ on $C_{u_f}$] \label{prop:trchir0l1uf}
	On the final hypersurface $u=u_f$, for any $v_{-1} \leq v \leq v(R_2,u_f)$,
	\[
		\left\Vert (\Omega \tr \chi)_{\ell =1} \right\Vert^2_{S_{u_f,v}}
		\lesssim
		\frac{
		\varepsilon_0^2
		+
		\varepsilon^3
		}
		{{v}^{4-2\delta}}
		+
		\Omega_{\circ}^2
		\left\Vert (\divslash \eta)_{\ell =1} \right\Vert^2_{S_{u_f,v}}
		,
	\]
	and, for any $v_{-1} \leq v \leq v(r_0,u_f)$,
	\[
		\int_v^{v(r_0,u_f)}
		\int_{S^2}
		\left\vert (\Omega \tr \chi)_{\ell =1} \right\vert^2
		d\theta dv
		(u_f)
		\lesssim
		\frac{
		\varepsilon_0^2
		+
		\varepsilon^3
		}
		{{v}^{4-2\delta}}
		+
		\int_v^{v(r_0,u_f)}
		\Omega_{\circ}^2
		\int_{S^2} \left\vert (\divslash \eta)_{\ell = 1} \right\vert^2 d \theta
		dv'
		(u_f).
	\]
\end{proposition}

\begin{proof}
	The Codazzi equation \eqref{eq:Codazzi} implies
	\[
		(\Deltaslash \Omega \tr \chi)_{\ell=1}
		=
		2 (\divslash \Omega \beta)_{\ell=1}
		+
		\frac{2\Omega_{\circ}^2}{r} (\divslash \etabar)_{\ell=1}
		+
		\mathcal{E}^1_{\ell =1}.
	\]
	Proposition \ref{prop:divcurlmodes} implies that
	\[
		\vert (\Deltaslash \Omega \tr \chi)_{\ell=1} - \Deltaslash (\Omega \tr \chi_{\ell=1}) \vert
		\lesssim
		\frac{\varepsilon^2}{v^2}
	\]
	The proof then follows from Proposition \ref{prop:divbetal1uf}, Proposition \ref{prop:newnoshifterrorestimate} and the fact that $\eta = - \etabar$ on $C_{u_f}$.
\end{proof}

\begin{proposition}[Preliminary estimate for $\rho_{\ell =1}$ on $C_{u_f}$] \label{prop:rhor0l1uf}
	On the final hypersurface $u=u_f$, for any $v_{-1} \leq v \leq v(R_2,u_f)$,
	\[
		\left\Vert \rho_{\ell =1} \right\Vert^2_{S_{u_f,v}}
		\lesssim
		\frac{
		\varepsilon_0^2
		+
		\varepsilon^3
		}
		{{v}^{4-2\delta}}
		+
		\Omega_{\circ}^2
		\left(
		\left\Vert (\divslash \eta)_{\ell = 1} \right\Vert^2_{S_{u_f,v}}
		+
		\left\Vert \Omega^{-2} (\Omega \tr \chibar)_{\ell = 1} \right\Vert^2_{S_{u_f,v}}
		\right)
		,
	\]
	and, for any $v_{-1} \leq v \leq v(r_0,u_f)$,
	\[
		\int_v^{v(r_0,u_f)}
		\int_{S^2}
		\left\vert \rho_{\ell =1} \right\vert^2
		d\theta dv
		(u_f)
		\lesssim
		\frac{
		\varepsilon_0^2
		+
		\varepsilon^3
		}
		{{v}^{4-2\delta}}
		+
		\int_v^{v(r_0,u_f)}
		\Omega_{\circ}^2
		\int_{S^2} \left\vert (\divslash \eta)_{\ell = 1} \right\vert^2
		+
		\left\vert \Omega^{-2} (\Omega \tr \chibar)_{\ell = 1} \right\vert^2
		d \theta
		dv'
		(u_f).
	\]
\end{proposition}

\begin{proof}
	The Gauss equation \eqref{eq:Gauss} can be schematically written,
	\[
		\rho - \rho_{\circ}
		=
		\frac{\Omega_{\circ}^2}{r^2} \left( \frac{\Omega_{\circ}^2}{\Omega^2} - 1 \right)
		+
		\frac{1}{2r} \left( \Omega \tr \chi - (\Omega \tr \chi)_{\circ} \right)
		-
		\frac{\Omega_{\circ}^2}{2r} \Omega^{-2} \left( \Omega \tr \chibar - (\Omega \tr \chibar)_{\circ} \right)
		-
		(K - K_{\circ})
		+
		\mathcal{E}^0,
	\]
	and so Proposition \ref{prop:Gaussl1} implies that,
	\[
		\rho_{\ell=1}
		=
		\frac{1}{2r} \left( \Omega \tr \chi - (\Omega \tr \chi)_{\circ} \right)_{\ell=1}
		-
		\frac{\Omega_{\circ}^2}{2r} \Omega^{-2} \left( \Omega \tr \chibar - (\Omega \tr \chibar)_{\circ} \right)_{\ell=1}
		+
		\mathcal{E}^0,
	\]
	since $\Omega = \Omega_{\circ}$ on $\{u=u_f\}$.  The proof then follows from Proposition \ref{prop:trchir0l1uf}.
\end{proof}

\begin{proposition}[Preliminary estimate for $(\divslash \eta)_{\ell =1}$ on $C_{u_f}$] \label{prop:divetar0l1uf}
	On the final hypersurface $u=u_f$, if $r_0$ is sufficiently small then, for any $v_{-1} \leq v \leq v(r_0,u_f)$,
	\[
		\left\Vert (\divslash \eta)_{\ell =1}
		\right\Vert^2_{S_{u_f,v}}
		\lesssim
		\frac{
		\varepsilon_0^2
		+
		\varepsilon^3
		}
		{{v}^3}
		+
		\Omega_{\circ}^2
		\left\Vert \Omega^{-2} (\Omega \tr \chibar)_{\ell = 1} \right\Vert^2_{S_{u_f,v}}
		.
	\]
\end{proposition}

\begin{proof}
	The proof follows from Propositions \ref{prop:Xl1uf}, \ref{prop:trchir0l1uf} and \ref{prop:rhor0l1uf} after taking $r_0$, and hence $\Omega_{\circ}$ in $r\leq r_0$, sufficiently small.
\end{proof}

In what follows the equation \eqref{eq:betabar4} for $\betabar$ will be used, which can be schematically written,
\begin{equation} \label{eq:betabar4l1schem}
	\Omega \nablaslash_4 (r^2 \Omega^{-1} \betabar)
	+
	2 r^2 \Omega \omegahat \Omega^{-1} \betabar
	=
	-r^2 \nablaslash \rho
	+
	r^2 {}^* \nablaslash \sigma
	+
	\frac{6M_f}{r} \etabar
	+
	\mathcal{E}^0.
\end{equation}

\begin{proposition}[Preliminary estimate for $(\divslash \Omega^{-1} \betabar)_{\ell =1}$ on $C_{u_f}$] \label{prop:divbetabarl1ufr0}
	On the final hypersurface $u=u_f$ for $r \leq r_0$, i.\@e.\@ for any $v_{-1} \leq v \leq v(r_0,u_f)$,
	\begin{multline*}
		\int_{S^2} \left\vert (\divslash \Omega^{-1} \betabar)_{\ell=1} \right\vert^2 d \theta
		(u_f,v)
		+
		\int_v^{v(r_0,u_f)}
		\int_{S^2}
		\left\vert (\divslash \Omega^{-1} \betabar)_{\ell=1} \right\vert^2
		d\theta dv'
		(u_f)
		\\
		\lesssim
		\frac{
		\varepsilon_0^2
		+
		\varepsilon^3
		}
		{{v}^3}
		+
		\int_v^{v(r_0,u_f)}
		\Omega_{\circ}^2
		\int_{S^2}
		\left\vert \Omega^{-2} (\Omega \tr \chibar)_{\ell = 1} \right\vert^2
		d \theta
		dv'
		(u_f).
	\end{multline*}
\end{proposition}

\begin{proof}
	First note that, by the Codazzi equation,
	\[
		(\divslash \Omega^{-1} \betabar)_{\ell=1}
		=
		-
		\frac{1}{r} (\divslash \eta)_{\ell =1}
		-
		\Deltaslash \big( \Omega^{-2} ( \Omega \tr \chibar - \Omega \tr \chibar_{\circ} ) \big)_{\ell=1}
		+
		\mathcal{E}^1_{\ell=1},
	\]
	on $\{u=u_f\}$ and Proposition \ref{prop:divetar0l1uf} and the gauge condition \eqref{eq:Hgaugelm5}, it follows that, on the sphere $S_{u_f,v_{-1}}$,
	\[
		\left\Vert (\divslash \Omega^{-1} \betabar)_{\ell =1}
		\right\Vert^2_{S_{u_f,v_{-1}}}
		\lesssim
		\varepsilon_0^2
		+
		\varepsilon^3
		.
	\]
	The equation \eqref{eq:betabar4l1schem} implies,
	\[
		\Omega \nablaslash_4 \left(r^3 ( \divslash \Omega^{-1} \betabar)_{\ell=1} \right)
		+
		2 r^3 \Omega \omegahat_{\circ} ( \divslash \Omega^{-1} \betabar)_{\ell=1}
		=
		-
		r^3 (\Deltaslash \rho)_{\ell=1}
		+
		6M_f (\divslash \etabar)_{\ell=1}
		+
		\mathcal{E}^1_{\ell=1}.
	\]
	The result then follows from Lemma \ref{lem:nabla4redxi}, Proposition \ref{prop:divcurlmodes} and Propositions \ref{prop:rhor0l1uf}, \ref{prop:divetar0l1uf}.
\end{proof}

\begin{proposition}[Preliminary estimate for $\Omega^{-2} (\Omega \tr \chibar)_{\ell=1}$ on $C_{u_f}$] \label{prop:trchibarl1ufr0}
	On the final hypersurface $u=u_f$ for $r \leq r_0$, i.\@e.\@ for any $v_{-1} \leq v \leq v(r_0,u_f)$,
	\[
		\int_{S^2} \left\vert \Omega^{-2} (\Omega \tr \chibar)_{\ell=1} \right\vert^2 d \theta
		(u_f,v)
		+
		\int_v^{v(r_0,u_f)}
		\int_{S^2}
		\left\vert \Omega^{-2} (\Omega \tr \chibar)_{\ell=1} \right\vert^2
		d\theta dv'
		(u_f)
		\lesssim
		\frac{
		\varepsilon_0^2
		+
		\varepsilon^3
		}
		{{v}^3}.
	\]
\end{proposition}

\begin{proof}
	The Codazzi equation \eqref{eq:Codazzibar} implies that
	\[
		\big( \Deltaslash \Omega^{-2} (\Omega \tr \chibar) \big)_{\ell=1}
		=
		- \frac{2}{r} (\divslash \eta)_{\ell=1}
		-
		2 (\divslash \Omega^{-1} \betabar)_{\ell=1}
		+
		\mathcal{E}^1_{\ell =1}.
	\]
	Proposition \ref{prop:divcurlmodes} implies that
	\[
		\big\vert \big( \Deltaslash \Omega^{-2} (\Omega \tr \chibar) \big)_{\ell=1} - \Deltaslash \Omega^{-2} (\Omega \tr \chibar)_{\ell=1} \big\vert
		\lesssim
		\frac{\varepsilon^2}{v^2},
	\]
	and so the result follows from Propositions \ref{prop:divetar0l1uf} and \ref{prop:divbetabarl1ufr0} after taking $r_0-2M_f$, and hence $\Omega_{\circ}^2$ in $\{ r \leq r_0\}$, sufficiently small.
\end{proof}

\begin{proposition}[Estimate for $\Omega \tr \chi_{\ell=1}$ on $C_{u_f}$] \label{prop:trchil1uf}
	On the final hypersurface $u=u_f$, for any $v_{-1} \leq v \leq v(R_2,u_f)$, for $k\leq N-s$, $s=0,1,2$,
	\[
		\left\Vert (\Omega \nablaslash_4)^{k} (\Omega \tr \chi)_{\ell =1} \right\Vert^2_{S_{u_f,v}}
		+
		\left\Vert (\Omega \nablaslash_4)^{k} (\Omega \tr \chi)_{\ell =1} \right\Vert^2_{C_{u_f}(v)}
		\lesssim
		\frac{
		\varepsilon_0^2
		+
		\varepsilon^3
		}
		{{v}^{s+1}}
		.
	\]
\end{proposition}

\begin{proof}
	First consider the case $k=0$.  By Propositions \ref{prop:trchir0l1uf}, \ref{prop:divetar0l1uf} and \ref{prop:trchibarl1ufr0} it suffices to consider the case $v \geq v(r_0,u_f)$.  For such $v$, the redshift plays no role and the Raychaudhuri equation \eqref{eq:Ray} can be normalised with a bad $\Omega$ weight to give
	\begin{equation} \label{eq:trchi4l1schem}
		\Omega\nablaslash_4 \left( r^2 \Omega^{-4} \left( \Omega \tr \chi \right) \right)
		+
		2 r^2 (\Omega \omegahat)_{\circ} \Omega^{-4} (\Omega\tr \chi)
		=
		\Omega^{-4} \mathcal{E}^0,
	\end{equation}
	since $\Omega \omegahat = (\Omega \omegahat)_{\circ}$ on $\{u=u_f\}$.  The result then follows from Lemma \ref{lem:nabla4redxi} after projecting to $\ell =1$ and using Proposition \ref{prop:com0}.  For $k\geq 1$ the proof follows inductively from applying $(\Omega \nablaslash_4)^{k-1}$ to \eqref{eq:trchi4l1schem}, again after projecting to $\ell =1$ and using Proposition \ref{prop:com0}.
\end{proof}

\begin{proposition}[Estimate for $\divslash \eta_{\ell = 1}$ on $C_{u_f}$] \label{prop:divetal1uf}
	On the final hypersurface $u=u_f$, for any $v_{-1} \leq v \leq v(R_2,u_f)$, for $k\leq N-s$, $s=0,1,2$,
	\[
		\left\Vert (\Omega \nablaslash_4)^{k} (\divslash \eta)_{\ell =1}
		\right\Vert^2_{S_{u_f,v}}
		\lesssim
		\frac{
		\varepsilon_0^2
		+
		\varepsilon^3
		}
		{{v}^{s+1}},
		\qquad
		\left\Vert (\Omega \nablaslash_4)^{k} (\divslash \eta)_{\ell =1}
		\right\Vert^2_{C_{u_f}(v)}
		\lesssim
		\frac{
		\varepsilon_0^2
		+
		\varepsilon^3
		}
		{{v}^{s}}
		.
	\]
\end{proposition}

\begin{proof}
	First suppose $k=0$.  Again, by Propositions \ref{prop:divetar0l1uf} and \ref{prop:trchibarl1ufr0} it suffices to assume $v\geq v(r_0,u_f)$.  Equation \eqref{eq:nabla4eta} implies that,
	\begin{equation} \label{eq:divetal1ufschematic}
		\Omega \nablaslash_4 (\Omega^{-2} r^3 \divslash \eta)
		+
		2 \Omega \omegahat \Omega^{-2} r^3 \divslash \eta
		=
		-
		\Omega^{-2} r^3 \divslash \Omega \beta
		+
		\Omega^{-2} \mathcal{E}^1,
	\end{equation}
	since $\eta = - \etabar$ on $\{u=u_f\}$.  The proof then follows from Lemma \ref{lem:nabla4redxi} and Proposition \ref{prop:divbetal1uf}, since the $\Omega$ weights are irrelevant in the region $r_0 \leq r \leq R_2$, after projecting to $\ell =1$ and using Proposition \ref{prop:com0}.  For $k \geq 1$ the proof follows inductively from Proposition \ref{prop:divbetal1uf} after applying $(\Omega \nablaslash_4)^{k-1}$ to the equation~\eqref{eq:divetal1ufschematic}.
\end{proof}

\begin{proposition}[Estimate for $\eta_{\ell = 1} - \eta_{\mathrm{Kerr}}$ on $C_{u_f}$] \label{prop:etal1uf}
	On the final hypersurface $u=u_f$, for any $v_{-1} \leq v \leq v(R_2,u_f)$, for $k\leq N-s$, $s=0,1,2$,
	\[
		\left\Vert (\Omega \nablaslash_4)^{k} \left( \eta_{\ell =1}
		-
		\eta_{\mathrm{Kerr}} \right)
		\right\Vert^2_{S_{u_f,v}}
		\lesssim
		\frac{
		\varepsilon_0^2
		+
		\varepsilon^3
		}
		{{v}^{s+1}},
		\qquad
		\left\Vert (\Omega \nablaslash_4)^{k} \left( \eta_{\ell =1}
		-
		\eta_{\mathrm{Kerr}} \right)
		\right\Vert^2_{C_{u_f}(v)}
		\lesssim
		\frac{
		\varepsilon_0^2
		+
		\varepsilon^3
		}
		{{v}^{s}}
		.
	\]
\end{proposition}

\begin{proof}
	The result for $k=0$ follows from Proposition \ref{prop:divcurl} and Propositions \ref{prop:curletal1uf} and \ref{prop:divetal1uf}.  For $k \geq 1$ the proof follows inductively from Proposition \ref{prop:divbetal1uf} after applying $(\Omega \nablaslash_4)^{k-1}$ to equation \eqref{eq:nabla4eta} which can be schematically rewritten,
	\[
		\Omega \nablaslash_4 (r^2 \eta)_{\ell=1}
		=
		-
		r^2 (\Omega \beta)_{\ell=1}
		+
		\mathcal{E}^0,
	\]
	and using Proposition \ref{prop:com0oneforms}.
\end{proof}

\begin{proposition}[Estimate for $\rho_{\ell = 1}$ on $C_{u_f}$] \label{prop:rhol1uf}
	On the final hypersurface $u=u_f$, for any $v_{-1} \leq v \leq v(R_2,u_f)$, for $k\leq N-s$, $s=0,1,2$,
	\[
		\left\Vert 
		(\Omega \nablaslash_4)^{k} \rho_{\ell =1}
		\right\Vert^2_{S_{u_f,v}}
		\lesssim
		\frac{
		\varepsilon_0^2
		+
		\varepsilon^3
		}
		{{v}^{s+1}},
		\qquad
		\left\Vert 
		(\Omega \nablaslash_4)^{k} \rho_{\ell =1}
		\right\Vert^2_{C_{u_f}(v)}
		\lesssim
		\frac{
		\varepsilon_0^2
		+
		\varepsilon^3
		}
		{{v}^{s}}
		.
	\]
\end{proposition}

\begin{proof}
	The proof follows from Propositions \ref{prop:Xl1uf}, \ref{prop:trchil1uf} and \ref{prop:divetal1uf}.
\end{proof}

\begin{proposition}[Estimate for $(\Omega^{-1} \betabar)_{\ell=1} - \Omega^{-1} \betabar_{\mathrm{Kerr}}$ on $C_{u_f}$] \label{prop:divbetabarl1uf}
	On the final hypersurface $u=u_f$, for any $v_{-1} \leq v \leq v(R_2,u_f)$, for $k\leq N-s$, $s=0,1,2$,
	\[
		\left\Vert
		(\Omega \nablaslash_4)^{k} \left(
		(\Omega^{-1} \betabar)_{\ell=1}
		-
		\Omega^{-1} \betabar_{\mathrm{Kerr}}
		\right)
		\right\Vert^2_{S_{u_f,v}}
		+
		\left\Vert 
		(\Omega \nablaslash_4)^{k} \left(
		(\Omega^{-1} \betabar)_{\ell=1}
		-
		\Omega^{-1} \betabar_{\mathrm{Kerr}}
		\right)
		\right\Vert^2_{C_{u_f}(v)}
		\lesssim
		\frac{
		\varepsilon_0^2
		+
		\varepsilon^3
		}
		{{v}^{s}}
		.
	\]
\end{proposition}

\begin{proof}
	First suppose $k=0$.  The Codazzi equation \eqref{eq:Codazzibar} gives
	\[
		(\Omega^{-1} \betabar)_{\ell=1}
		=
		-
		\frac{1}{r} \eta_{\ell =1}
		-
		\nablaslash \left( \Omega^{-2} \left( \Omega \tr \chibar - (\Omega \tr \chibar)_{\circ}\right)_{\ell=1} \right)
		+
		\mathcal{E}^0,
	\]
	and hence, on the initial sphere $S_{u_f,v_{-1}}$,
	\[
		\Vert (\Omega^{-1} \betabar)_{\ell =1} 
		-
		\Omega^{-1} \betabar_{\mathrm{Kerr}} \Vert_{S_{u_f,v_{-1}}}^2
		\lesssim
		\varepsilon_0^2 + \varepsilon^3
	\]
	by Proposition \ref{prop:etal1uf} and the gauge condition \eqref{eq:Hgaugelm5}.  The proof then follows from Lemma \ref{lem:nabla4redxi} and Propositions \ref{prop:sigmal1uf}, \ref{prop:etal1uf} and \ref{prop:rhol1uf}, using the equation \eqref{eq:betabar4l1schem} projected to $\ell =1$.  For $k\geq 1$ the proof follows inductively from applying $(\Omega \nablaslash_4)^{k-1}$ directly to the equation \eqref{eq:betabar4l1schem} projected to $\ell =1$ and using Proposition \ref{prop:com0oneforms}.
\end{proof}

\begin{proposition}[Estimate for $\Omega^{-1} \nablaslash_3 (\Omega \beta - \Omega \beta_{\mathrm{Kerr}})_{\ell=1}$ on $C_{u_f}$] \label{prop:nabla3betal1uf}
	On the final hypersurface $u=u_f$, for any $v_{-1} \leq v \leq v(R_2,u_f)$,
	\[
		\left\Vert 
		\Omega^{-1} \nablaslash_3 (\Omega \beta - \Omega \beta_{\mathrm{Kerr}})_{\ell=1}
		\right\Vert^2_{S_{u_f,v}}
		\lesssim
		\frac{
		\varepsilon_0^2
		+
		\varepsilon^3
		}
		{{v}^{s+1}},
		\qquad
		\left\Vert 
		\Omega^{-1} \nablaslash_3 (\Omega \beta - \Omega \beta_{\mathrm{Kerr}})_{\ell=1}
		\right\Vert^2_{C_{u_f}(v)}
		\lesssim
		\frac{
		\varepsilon_0^2
		+
		\varepsilon^3
		}
		{{v}^{s}}
		.
	\]
\end{proposition}

\begin{proof}
	The proof follows from the Bianchi equation \eqref{eq:beta3}, which can be schematically written,
	\begin{align*}
		\Omega^{-1} \nablaslash_3 \left( r^2 (\Omega \beta - \Omega \beta_{\mathrm{Kerr}}) \right)_{\ell=1}
		=
		r^2 \nablaslash \rho_{\ell=1}
		+
		r^2 {}^* \nablaslash \left( \sigma_{\ell=1} - \sigma_{\mathrm{Kerr}} \right)
		-
		\frac{6M_f}{r} \left( \eta_{\ell=1} - \eta_{\mathrm{Kerr}} \right)
		+
		\mathcal{E}^0_{\ell=1},
	\end{align*}
	and Propositions \ref{prop:sigmal1uf}, \ref{prop:etal1uf} and \ref{prop:rhol1uf}.
\end{proof}

\subsubsection{Estimates for $\ell=1$ modes of quantities on the initial hypersurface $\Cbar_{v_{-1}}$}

In this section the $\ell = 1$ modes of quantities on $\Cbar_{v_{-1}}$ are estimated.  Since the estimates of this section do not involve \emph{decay}, it is not necessary to subtract the $\Hp$ linearised Kerr solution in this section.

\begin{proposition}[Estimate for $(\Omega^{-1} \betabar)_{\ell=1}$ on $\Cbar_{v_{-1}}$] \label{prop:divbetabarl1v0}
	On the initial hypersurface $v=v_{-1}$, for any $u_0 \leq u \leq u_f$, for $k\leq N$,
	\[
		\left\Vert (\Omega^{-1} \nablaslash_3)^k (\Omega^{-1} \betabar)_{\ell=1} \right\Vert^2_{S_{u,v_{-1}}}
		+
		\left\Vert (\Omega^{-1} \nablaslash_3)^k (\Omega^{-1} \betabar)_{\ell=1} \right\Vert^2_{\Cbar_{v_{-1}}}
		\lesssim
		\varepsilon_0^2
		+
		\varepsilon^3
		.
	\]
\end{proposition}

\begin{proof}
	By equation \eqref{eq:betabar3},
	\begin{equation} \label{eq:betabar3l1schem}
		\Omega \nablaslash_3 \left( r^4 (\Omega^{-1} \betabar) \right)_{\ell=1}
		=
		\Omega^2
		\mathcal{E}^0_{\ell=1}.
	\end{equation}
	The proof for $k=0$ then follows from Lemma \ref{lem:nabla3xi} and Proposition \ref{prop:com0oneforms}, using Proposition \ref{prop:divbetabarl1uf} to control the final condition.  For $k\geq 1$ the proof follows inductively by applying $(\Omega^{-1} \nablaslash_3)^{k-1}$ to \eqref{eq:betabar3l1schem}.
\end{proof}

\begin{proposition}[Estimate for $\sigma_{\ell=1}$ on $\Cbar_{v_{-1}}$] \label{prop:sigmal1v0}
	On the initial hypersurface $v=v_{-1}$, for any $u_0 \leq u \leq u_f$, for $k\leq N$,
	\[
		\left\Vert (\Omega^{-1} \nablaslash_3)^k \sigma_{\ell=1} \right\Vert^2_{S_{u,v_{-1}}}
		+
		\left\Vert (\Omega^{-1} \nablaslash_3)^k \sigma_{\ell=1} \right\Vert^2_{\Cbar_{v_{-1}}}
		\lesssim
		\varepsilon_0^2
		+
		\varepsilon^3
		.
	\]
\end{proposition}

\begin{proof}
	Equations \eqref{eq:sigma3}, \eqref{eq:curletacurletabar} and \eqref{eq:Codazzibar} imply that
	\[
		\Omega\nablaslash_3 \left( r^4 \sigma \right)_{\ell=1}
		=
		\Omega^2
		\mathcal{E}^1_{\ell=1}.
	\]
	The proof when $k=0$ then follows from Lemma \ref{lem:nabla3xi} and Proposition \ref{prop:com0oneforms}, using Proposition \ref{prop:sigmal1uf} to control the final condition.  For $k \geq1$, the proof follows inductively from Proposition \ref{prop:divbetabarl1v0} by applying $(\Omega^{-1} \nablaslash_3)^{k-1}$ to \eqref{eq:sigma3}, which can be written,
	\[
		\Omega^{-1} \nablaslash_3 \left( r^3 \sigma \right)_{\ell=1}
		=
		-
		r^3 (\curlslash \Omega^{-1} \betabar)_{\ell=1}
		+
		\mathcal{E}^0_{\ell=1}.
	\]
\end{proof}

\begin{proposition}[Estimate for $\eta_{\ell=1}$ on $\Cbar_{v_{-1}}$] \label{prop:divetal1v0}
	On the initial hypersurface $v=v_{-1}$, for any $u_0 \leq u \leq u_f$, for $k\leq N$,
	\[
		\left\Vert (\Omega^{-1} \nablaslash_3)^k \eta_{\ell=1} \right\Vert^2_{S_{u,v_{-1}}}
		+
		\left\Vert (\Omega^{-1} \nablaslash_3)^k \eta_{\ell=1} \right\Vert^2_{\Cbar_{v_{-1}}}
		\lesssim
		\varepsilon_0^2
		+
		\varepsilon^3
		.
	\]
\end{proposition}

\begin{proof}
	Since $\nablaslash (\Omega \omegabarhat)_{\ell=1} = 0$ on $\{v =v_{-1}\}$, equation \eqref{eq:nabla3eta} implies
	\begin{equation} \label{eq:eta3l1schem}
		\Omega\nablaslash_3 \left( r^2 \eta \right)_{\ell=1}
		=
		-
		\Omega^2 r^2 (\Omega^{-1} \betabar)_{\ell=1}
		+
		\Omega^2
		\mathcal{E}^0_{\ell=1}.
	\end{equation}
	The proof when $k=0$ then follows from Lemma \ref{lem:nabla3xi}, Proposition \ref{prop:com0oneforms} and Proposition \ref{prop:divbetabarl1v0}, using Proposition \ref{prop:etal1uf} to control the final condition.  When $k\geq 1$, the proof follows inductively from Proposition \ref{prop:divbetabarl1v0} by dividing \eqref{eq:eta3l1schem} by $\Omega^2$ and applying $(\Omega^{-1} \nablaslash_3)^{k-1}$.
\end{proof}

\begin{proposition}[Estimate for $\Omega^{-2} ( \Omega \tr \chibar)_{\ell=1}$ on $\Cbar_{v_{-1}}$] \label{prop:trchibarl1v0}
	On the initial hypersurface $v=v_{-1}$, for any $u_0 \leq u \leq u_f$, for $k\leq N$,
	\[
		\left\Vert (\Omega^{-1} \nablaslash_3)^k \Omega^{-2} ( \Omega \tr \chibar)_{\ell=1} \right\Vert^2_{S_{u,v_{-1}}}
		+
		\left\Vert (\Omega^{-1} \nablaslash_3)^k \Omega^{-2} ( \Omega \tr \chibar)_{\ell=1} \right\Vert^2_{\Cbar_{v_{-1}}}
		\lesssim
		\varepsilon_0^2
		+
		\varepsilon^3
		.
	\]
\end{proposition}

\begin{proof}
	By the Raychaudhuri equation \eqref{eq:Ray} and the gauge condition \eqref{eq:Hgaugelm8},
	\begin{equation} \label{eq:trchibar3l1schem}
		\Omega \nablaslash_3 \left( r^2 \Omega^{-2} (\Omega \tr \chibar) \right)_{\ell=1}
		=
		\Omega^2
		\mathcal{E}^0_{\ell=1},
	\end{equation}
	and the proof, when $k=0$, follows from Lemma \ref{lem:nabla3xi} and the gauge condition \eqref{eq:Hgaugelm5}.  For $k \geq 1$ the proof follows from applying $(\Omega^{-1} \nablaslash_3)^{k-1}$ to \eqref{eq:trchibar3l1schem} and using Proposition \ref{prop:com0}.
\end{proof}

\begin{proposition}[Estimate for $\rho_{\ell=1}$ on $\Cbar_{v_{-1}}$] \label{prop:rhol1v0}
	On the initial hypersurface $v=v_{-1}$, for any $u_0 \leq u \leq u_f$, for $k\leq N$,
	\[
		\left\Vert (\Omega^{-1} \nablaslash_3)^k \rho_{\ell=1} \right\Vert^2_{S_{u,v_{-1}}}
		+
		\left\Vert (\Omega^{-1} \nablaslash_3)^k \rho_{\ell=1} \right\Vert^2_{\Cbar_{v_{-1}}}
		\lesssim
		\varepsilon_0^2
		+
		\varepsilon^3
		.
	\]
\end{proposition}

\begin{proof}
	The equation \eqref{eq:rho3} gives,
	\[
		\Omega \nablaslash_3 (r^3 \rho)_{\ell=1}
		=
		3M \Omega^2 \Omega^{-2} (\Omega \tr \chibar)_{\ell=1}
		-
		\Omega^2 r^3 (\divslash \Omega^{-1} \betabar)_{\ell=1}
		+
		\Omega^2
		\mathcal{E}^1_{\ell=1},
	\]
	and the result follows from Lemma \ref{lem:nabla3xi}, Proposition \ref{prop:com0}, and Propositions \ref{prop:divbetabarl1v0} and \ref{prop:trchibarl1v0}.
\end{proof}

\begin{proposition}[Estimate for $(\Omega \beta)_{\ell=1}$ on $\Cbar_{v_{-1}}$] \label{prop:divbetal1v0}
	On the initial hypersurface $v=v_{-1}$, for any $u_0 \leq u \leq u_f$, for $k\leq N$,
	\[
		\left\Vert (\Omega^{-1} \nablaslash_3)^k (\Omega \beta)_{\ell=1} \right\Vert^2_{S_{u,v_{-1}}}
		+
		\left\Vert (\Omega^{-1} \nablaslash_3)^k (\Omega \beta)_{\ell=1} \right\Vert^2_{\Cbar_{v_{-1}}}
		\lesssim
		\varepsilon_0^2
		+
		\varepsilon^3
		.
	\]
\end{proposition}

\begin{proof}
	Equation \eqref{eq:beta3} implies
	\[
		\Omega \nablaslash_3 \left( r^3 (\divslash \Omega \beta) \right)_{\ell=1}
		=
		\Omega^2 r^3 (\Deltaslash \rho)_{\ell=1}
		-
		6M \Omega^2 (\divslash \eta)_{\ell=1}
		+
		\Omega^2
		\mathcal{E}^0_{\ell=1},
	\]
	and the proof then follows from Lemma \ref{lem:nabla3xi}, Proposition \ref{prop:com0}, and Propositions \ref{prop:divetal1v0} and \ref{prop:rhol1v0}.
\end{proof}

\subsubsection{Estimates for $\ell=0$ modes of quantities on the final hypersurface $C_{u_f}$}
\label{subsec:finall0uf}

This section concerns estimates for the $\ell =0$ modes of the Ricci coefficients and 
curvature components on~$C_{u_f}$.

Recall the quantity\index{double null gauge!connection coefficients!$\Upsilon$, quantity used only in the $\mathcal{H}^+$ gauge}
\[
	\Upsilon
	=
	\left(1 - \frac{3M_f}{r} \right) (\rho - \rho_{\circ})
	+
	\frac{3M_f}{2r^2} \left( \Omega \tr \chi - (\Omega \tr \chi)_{\circ} \right)
	-
	\frac{3M_f\Omega_{\circ}^2}{2r^2} \Omega^{-2} \left( \Omega \tr \chibar - (\Omega \tr \chibar)_{\circ} \right).
\]
Proposition \ref{thm:inheriting} and Theorem \ref{thm:Iestimates} imply that, at $S_{u_f,v(R,u_f)}$,
\begin{equation} \label{eq:Upsilonl0inherited}
	\left\vert
	\Upsilon_{\ell=0} (u_f,v(R,u_f))
	\right\vert^2
	\lesssim
	\frac{
	\varepsilon_0^2
	+
	\varepsilon^3}{v(R,u_f)^{4-2\delta}}.
\end{equation}

\begin{proposition}[Estimate for $(\Omega \tr \chi - (\Omega \tr \chi)_{\circ})_{\ell =0}$ on $C_{u_f}$] \label{prop:trchil0uf}
	On the final hypersurface $u=u_f$, for any $v_{-1} \leq v \leq v(R_2,u_f)$, for $k \leq N-s$, $s=0,1,2$,
	\begin{equation*}
		\left\Vert (\Omega \nablaslash_4)^k
		(\Omega \tr \chi - (\Omega \tr \chi)_{\circ})_{\ell =0}
		\right\Vert^2_{S_{u_f,v}}
		+
		\left\Vert (\Omega \nablaslash_4)^k
		(\Omega \tr \chi - (\Omega \tr \chi)_{\circ})_{\ell =0}
		\right\Vert^2_{C_{u_f}(v)}
		\lesssim
		\frac{\varepsilon_0^2
		+
		\varepsilon^3}{v^{2+s}}.
	\end{equation*}
\end{proposition}

\begin{proof}
	The gauge condition \eqref{eq:Hgaugelm1} and the equation \eqref{eq:Ray} (see \eqref{eq:trchi4H}) imply that $(\Omega \tr \chi - (\Omega \tr \chi)_{\circ})_{\ell =0}$ satisfies the blueshifted equation
	\begin{equation} \label{eq:trchil0ufschematic}
		\Omega \nablaslash_4
		\left(
		r^2
		(\Omega \tr \chi - (\Omega \tr \chi)_{\circ})
		\right)_{\ell =0}
		-
		2(\Omega \omegahat)_{\circ}
		r^2
		(\Omega \tr \chi - (\Omega \tr \chi)_{\circ})_{\ell =0}
		=
		\mathcal{E}^0_{\ell =0}.
	\end{equation}
	The proof for $k =0$ follows from Lemma \ref{lem:nabla4bluexi} and Proposition \ref{prop:com0}, using the gauge condition \eqref{eq:Hgaugelm4A}.  The nonlinear error terms are controlled by Proposition \ref{prop:errorout}.  For $k\geq 1$ the proof follows inductively from applying $(\Omega \nablaslash_4)^{k-1}$ directly to the equation \eqref{eq:trchil0ufschematic}.
\end{proof}

\begin{proposition}[Estimate for $\Upsilon_{\ell =0}$ on $C_{u_f}$] \label{prop:Upsilonuf}
	On the final hypersurface $u=u_f$, for any $v_{-1} \leq v \leq v(R_2,u_f)$, for $k \leq N-s$, $s=0,1,2$,
	\begin{equation*}
		\left\Vert (\Omega \nablaslash_4)^k
		\Upsilon_{\ell =0}
		\right\Vert^2_{S_{u_f,v}}
		\lesssim
		\frac{\varepsilon^4}{v^{s+1}}.
	\end{equation*}
\end{proposition}

\begin{proof}
	The equations \eqref{eq:trchi4H}, \eqref{eq:trchibar4H} and \eqref{eq:rho4}, along with the gauge condition \eqref{eq:Hgaugelm1}, imply that $\Upsilon_{\ell=0}$ on $\{u=u_f\}$ satisfies the noshifted equation
	\[
		\partial_v (r^3 \Upsilon)_{\ell=0}
		=
		\mathcal{E}^0_{\ell=0}
		-
		3M_f r \Omega_{\circ}^2 \vert \etabar \vert^2,
	\]
	where the nonlinear error $\mathcal{E}^0$ has the correct structure to apply Proposition \ref{prop:newnoshifterrorestimate}.  The proof then follows as in Proposition \ref{prop:Xuf}, using also the fact that
	\[
		\Big(
		\int_{C_u(v)}
		\Omega_{\circ}^2 \vert \etabar \vert^2
		d \theta d v'
		\Big)^2
		\lesssim
		\frac{\varepsilon^4}{v^{4}}.
	\]
\end{proof}

\begin{proposition}[Preliminary estimate for $\Omega^{-2} (\Omega \tr \chibar - (\Omega \tr \chibar)_{\circ})_{\ell =0}$ on $C_{u_f}$] \label{prop:trchibarl0ufr0}
	On the final hypersurface $u=u_f$ for $r \leq r_0$, i.\@e.\@ for any $v_{-1} \leq v \leq v(r_0,u_f)$, if $r_0-2M_f$ is sufficiently small,
	\[
		\left\Vert
		\Omega^{-2} (\Omega \tr \chibar - (\Omega \tr \chibar)_{\circ})_{\ell =0}
		\right\Vert^2_{S_{u_f,v}}
		+
		\int_v^{v(r_0,u_f)}
		\left\Vert
		\Omega^{-2} (\Omega \tr \chibar - (\Omega \tr \chibar)_{\circ})_{\ell =0}
		\right\Vert^2_{S_{u_f,v'}}
		dv'
		\lesssim
		\frac{
		\varepsilon_0^2
		+
		\varepsilon^3
		}
		{{v}^2}.
	\]
\end{proposition}

\begin{proof}
	On $u=u_f$, $\Omega \tr \chibar_{\ell=0}$ satisfies the redshifted equation
	\begin{equation} \label{eq:trchibarl0}
		\Omega \nablaslash_4
		\Big( \frac{r}{\Omega^{2}}
		(
		\Omega \tr \chibar - (\Omega \tr \chibar)_{\circ}
		)
		\Big)_{\ell=0}
		+
		2(\Omega \omegahat)_{\circ} 
		\frac{r}{\Omega^{2}}
		\left(
		\Omega \tr \chibar - (\Omega \tr \chibar)_{\circ}
		\right)_{\ell=0}
		=
		\left(
		\Omega \tr \chi - (\Omega \tr \chi)_{\circ}
		\right)_{\ell=0}
		+
		2(\rho - \rho_{\circ})_{\ell=0}
		+
		\mathcal{E}^0_{\ell=0}.
	\end{equation}
	The proof then follows, after taking $r_0-2M_f$, and hence $\Omega_{\circ}^2$ in $r\leq r_0$, sufficiently small, from Lemma \ref{eq:nabla4redxi}, Proposition \ref{prop:com0} and Propositions \ref{prop:trchil0uf} and \ref{prop:Upsilonuf} after noting that, in $r\leq r_0$,
	\[
		\vert (\rho - \rho_{\circ})_{\ell =0} \vert
		\lesssim
		\vert
		(\Omega \tr \chi - (\Omega \tr \chi)_{\circ})_{\ell =0}
		\vert
		+
		\vert
		\Upsilon_{\ell=0}
		\vert
		+
		\Omega_{\circ}^2
		\vert
		\Omega^{-2} (\Omega \tr \chibar - (\Omega \tr \chibar)_{\circ})_{\ell =0}
		\vert.
	\]
\end{proof}

\begin{proposition}[Estimate for $(\rho - \rho_{\circ})_{\ell =0}$ on $C_{u_f}$] \label{prop:rhol0uf}
	On the final hypersurface $u=u_f$, for any $v_{-1} \leq v \leq v(R_2,u_f)$, for $k \leq N-s$, $s=0,1,2$,
	\begin{equation*}
		\left\Vert (\Omega \nablaslash_4)^k
		(\rho - \rho_{\circ})_{\ell =0}
		\right\Vert^2_{S_{u_f,v}}
		+
		\left\Vert (\Omega \nablaslash_4)^k
		(\rho - \rho_{\circ})_{\ell =0}
		\right\Vert^2_{C_{u_f}(v)}
		\lesssim
		\frac{\varepsilon_0^2
		+
		\varepsilon^3}{v^{s}}.
	\end{equation*}
\end{proposition}

\begin{proof}
	Suppose first that $k=0$.  In the region $r\leq r_0$ the proof follows from Propositions \ref{prop:trchil0uf}, \ref{prop:Upsilonuf} and \ref{prop:trchibarl0ufr0}, and the fact that, in $r\leq r_0$,
	\[
		\vert (\rho - \rho_{\circ})_{\ell =0} \vert
		\lesssim
		\vert
		(\Omega \tr \chi - (\Omega \tr \chi)_{\circ})_{\ell =0}
		\vert
		+
		\vert
		\Upsilon_{\ell=0}
		\vert
		+
		\Omega_{\circ}^2
		\vert
		\Omega^{-2} (\Omega \tr \chibar - (\Omega \tr \chibar)_{\circ})_{\ell =0}
		\vert.
	\]
	In the region $r_0 \leq r \leq R_2$, the redshift plays no role and so the Bianchi equation \eqref{eq:rho4} can be normalised with a bad $\Omega$ weight to give
	\[
		\Omega \nablaslash_4 (\Omega^{-2} r^3 (\rho - \rho_{\circ}))_{\ell=0}
		+
		2(\Omega\omegahat)_{\circ} \Omega^{-2} r^3 (\rho - \rho_{\circ})_{\ell=0}
		=
		3M_f \Omega^{-2} (\Omega \tr \chi - \Omega \tr \chi_{\circ})
		+
		\Omega^{-2} \mathcal{E}^0_{\ell=0}
		.
	\]
	The proof then follows from Lemma \ref{eq:nabla4redxi}, Proposition \ref{prop:com0} and Proposition \ref{prop:trchil0uf}.  For $k \geq 1$ the proof follows from Proposition \ref{prop:trchil0uf} after applying $(\Omega\nablaslash_4)^{k-1}$ to \eqref{eq:rho4}.
\end{proof}

\begin{proposition}[Estimate for $\Omega^{-2} (\Omega \tr \chibar - (\Omega \tr \chibar)_{\circ})_{\ell =0}$ on $C_{u_f}$] \label{prop:trchibarl0uf}
	On the final hypersurface $u=u_f$, for any $v_{-1} \leq v \leq v(R_2,u_f)$, for $k \leq N-s$, $s=0,1,2$,
	\begin{equation*}
		\left\Vert (\Omega \nablaslash_4)^k
		\Omega^{-2} (\Omega \tr \chibar - (\Omega \tr \chibar)_{\circ})_{\ell =0}
		\right\Vert^2_{S_{u_f,v}}
		+
		\left\Vert (\Omega \nablaslash_4)^k
		\Omega^{-2} (\Omega \tr \chibar - (\Omega \tr \chibar)_{\circ})_{\ell =0}
		\right\Vert^2_{C_{u_f}(v)}
		\lesssim
		\frac{\varepsilon_0^2
		+
		\varepsilon^3}{v^s}.
	\end{equation*}
\end{proposition}

\begin{proof}
	For $k=0$, the proof follows as in Proposition \ref{prop:trchibarl0ufr0}, using now Proposition \ref{prop:rhol0uf}.  For $k \geq 1$ the proof follows from applying $(\Omega \nablaslash_4)^{k-1}$ to equation \eqref{eq:trchibarl0}, and using Proposition \ref{prop:trchil0uf} and Proposition \ref{prop:rhol0uf}.
\end{proof}

\begin{proposition}[Estimate for $\Omega^{-1} \nablaslash_3 (\Omega \tr \chi - (\Omega \tr \chi)_{\circ})_{\ell =0}$ on $C_{u_f}$] \label{prop:trchi3l0uf}
	On the final hypersurface $u=u_f$, for any $v_{-1} \leq v \leq v(R_2,u_f)$,
	\begin{equation*}
		\left\Vert \Omega^{-1} \nablaslash_3
		(\Omega \tr \chi - (\Omega \tr \chi)_{\circ})_{\ell =0}
		\right\Vert^2_{S_{u_f,v}}
		+
		\left\Vert \Omega^{-1} \nablaslash_3
		(\Omega \tr \chi - (\Omega \tr \chi)_{\circ})_{\ell =0}
		\right\Vert^2_{C_{u_f}(v)}
		\lesssim
		\frac{\varepsilon_0^2
		+
		\varepsilon^3}{v^2}.
	\end{equation*}
\end{proposition}

\begin{proof}
	The proof follows directly from Propositions \ref{prop:trchil0uf}, \ref{prop:rhol0uf} and \ref{prop:trchibarl0uf} using equation \eqref{eq:trchi3} which, after projecting to $\ell =0$, on $u=u_f$ takes the form
	\[
		\Omega^{-1} \nablaslash_3
		\left(
		\Omega \tr \chi - (\Omega \tr \chi)_{\circ}
		\right)_{\ell=0}
		=
		r^{-1}
		\left(
		\Omega \tr \chi - (\Omega \tr \chi)_{\circ}
		\right)_{\ell=0}
		-
		r^{-1}
		\left(
		\Omega \tr \chibar - (\Omega \tr \chibar)_{\circ}
		\right)_{\ell=0}
		+
		2 (\rho - \rho_{\circ})_{\ell=0}
		+
		\mathcal{E}^0_{\ell = 0}.
	\]
\end{proof}

\subsubsection{Estimates for $\ell=0$ modes of quantities on the initial hypersurface $\Cbar_{v_{-1}}$}

This section concerns estimates for the $\ell =0$ modes of the Ricci coefficients and curvature components on $\Cbar_{v_{-1}}$.

\begin{proposition}[Estimate for $\Omega^{-2} (\Omega \tr \chibar - (\Omega \tr \chibar)_{\circ})_{\ell=0}$ on $\Cbar_{v_{-1}}$] \label{prop:trchibarl0v0}
	On the initial hypersurface $v=v_{-1}$, for any $u_0 \leq u \leq u_f$, for $k\leq N$,
	\[
		\left\Vert (\Omega^{-1} \nablaslash_3)^k
		\Omega^{-2} (\Omega \tr \chibar - (\Omega \tr \chibar)_{\circ})_{\ell=0}
		\right\Vert^2_{S_{u,v_{-1}}}
		+
		\left\Vert
		(\Omega^{-1} \nablaslash_3)^k
		\Omega^{-2} (\Omega \tr \chibar - (\Omega \tr \chibar)_{\circ})_{\ell=0}
		\right\Vert^2_{\Cbar_{v_{-1}}}
		\lesssim
		\varepsilon_0^2
		+
		\varepsilon^3
		.
	\]
\end{proposition}

\begin{proof}
	The gauge condition \eqref{eq:Hgaugelm8} and the equation \eqref{eq:Ray} (see \eqref{eq:trchibar3H}) imply that, on $\{v=v_{-1}\}$, $(\Omega \tr \chibar - (\Omega \tr \chibar)_{\circ})_{\ell =0}$ satisfies
	\begin{equation} \label{eq:trchibarl0v0schematic}
		\Omega \nablaslash_3
		\left(
		r^2 \Omega^{-2}
		(\Omega \tr \chibar - (\Omega \tr \chibar)_{\circ})
		\right)_{\ell =0}
		=
		\Omega^2 \mathcal{E}^0_{\ell =0}.
	\end{equation}
	The proof for $k =0$ follows from Lemma \ref{lem:nabla3xi}, Proposition \ref{prop:com0} and the gauge condition \eqref{eq:Hgaugelm5}.  For $k\geq 1$ the proof follows inductively after dividing the equation \eqref{eq:trchibarl0v0schematic} by $\Omega^2$ and applying $(\Omega^{-1} \nablaslash_3)^{k-1}$.
\end{proof}

\begin{proposition}[Estimate for $(\rho - \rho_{\circ})_{\ell=0}$ on $\Cbar_{v_{-1}}$] \label{prop:rhol0v0}
	On the initial hypersurface $v=v_{-1}$, for any $u_0 \leq u \leq u_f$, for $k\leq N$,
	\[
		\left\Vert (\Omega^{-1} \nablaslash_3)^k
		(\rho - \rho_{\circ})_{\ell=0}
		\right\Vert^2_{S_{u,v_{-1}}}
		+
		\left\Vert
		(\Omega^{-1} \nablaslash_3)^k
		(\rho - \rho_{\circ})_{\ell=0}
		\right\Vert^2_{\Cbar_{v_{-1}}}
		\lesssim
		\varepsilon_0^2
		+
		\varepsilon^3
		.
	\]
\end{proposition}

\begin{proof}
	The equation \eqref{eq:rho3} implies
	\begin{equation} \label{eq:rhol0v0schematic}
		\Omega \nablaslash_3 \left( r^3 (\rho - \rho_{\circ}) \right)_{\ell =0}
		=
		-
		\frac{3}{2} \Omega^2 \rho_{\circ} \Omega^{-2} \left( \Omega \tr \chibar - (\Omega \tr \chibar)_{\circ} \right)_{\ell =0}
		+
		\Omega^2
		\mathcal{E}^0_{\ell =0},
	\end{equation}
	and the proof for $k=0$ follows from Lemma \ref{lem:nabla3xi}, Proposition \ref{prop:com0}, Proposition \ref{prop:trchibarl0v0} and Proposition \ref{prop:rhol0uf}.  For $k\geq 1$ the proof follows inductively after dividing the equation \eqref{eq:rhol0v0schematic} by $\Omega^2$ and applying $(\Omega^{-1} \nablaslash_3)^{k-1}$.
\end{proof}

\begin{proposition}[Estimate for $(\Omega \tr \chi - (\Omega \tr \chi)_{\circ})_{\ell=0}$ on $\Cbar_{v_{-1}}$]  \label{prop:trchil0v0}
	On the initial hypersurface $v=v_{-1}$, for any $u_0 \leq u \leq u_f$, for $k\leq N$,
	\[
		\left\Vert (\Omega^{-1} \nablaslash_3)^k
		(\Omega \tr \chi - (\Omega \tr \chi)_{\circ})_{\ell=0}
		\right\Vert^2_{S_{u,v_{-1}}}
		+
		\left\Vert
		(\Omega^{-1} \nablaslash_3)^k
		(\Omega \tr \chi - (\Omega \tr \chi)_{\circ})_{\ell=0}
		\right\Vert^2_{\Cbar_{v_{-1}}}
		\lesssim
		\varepsilon_0^2
		+
		\varepsilon^3
		.
	\]
\end{proposition}

\begin{proof}
	Equation \eqref{eq:trchi3} (see also \eqref{eq:trchi3H}) implies
	\begin{equation} \label{eq:trchil0v0schematic}
		\Omega \nablaslash_3 \left(
		r (\Omega \tr \chi - (\Omega \tr \chi)_{\circ})
		\right)_{\ell=0}
		=
		-
		\Omega^2 (\Omega \tr \chibar - (\Omega \tr \chibar)_{\circ})_{\ell=0}
		+
		2 \Omega^2 r (\rho - \rho_{\circ})_{\ell=0}
		+
		\Omega^2
		\mathcal{E}^0_{\ell=0},
	\end{equation}
	and the proof for $k=0$ follows from Lemma \ref{lem:nabla3xi}, Proposition \ref{prop:com0}, Proposition \ref{prop:trchibarl0v0}, Proposition \ref{prop:rhol0v0} and Proposition \ref{prop:trchil0uf}.  For $k\geq 1$ the proof follows inductively after dividing the equation \eqref{eq:trchil0v0schematic} by $\Omega^2$ and applying $(\Omega^{-1} \nablaslash_3)^{k-1}$.
\end{proof}

\subsection{Estimates for quantities in $\DRH$: the $\ell =0,1$ modes}
\label{subsec:Hell0ell1}

In this section the estimates for the $\ell= 0,1$ modes of the geometric quantities on $C_{u_f}$ and $\Cbar_{v_{-1}}$ obtained in Section \ref{subsec:Hell0ell1ufv1} are exploited to estimate the $\ell= 0,1$ modes of the geometric quantities in $\DRH$.  The main results of this section are Proposition \ref{prop:ell1Henergy} and Proposition \ref{prop:ell0Henergy}.

\subsubsection{Estimates for $\ell =1$ modes in $\DRH$}

The main result of this section is the following proposition.

\begin{theorem}[Improving bootstrap assumptions for energies of $\ell = 1$ modes] \label{prop:ell1Henergy}
	The energies defined in \eqref{eq:spacetimeHenergy2}, \eqref{eq:outgoingHenergy2}, \eqref{eq:incomingHenergy2}, \eqref{eq:spheresHenergy2}
	satisfy,
	\[
		\mathbb{E}^N_{\ell = 1}[\DRH]
		+
		\mathbb{E}^N_{\ell =1}[C^{\Hp}]
		+
		\mathbb{E}^N_{\ell =1}[\Cbar^{\Hp}]
		+
		\mathbb{E}^N_{\ell =1}[S^{\Hp}]
		\lesssim
		\varepsilon_0^2
		+
		\varepsilon^3.
	\]
\end{theorem}

\begin{proof}
	The proof is a direct consequence of the estimate for $\Omega^{-1} \betabar_{\ell = 1}$ of Proposition \ref{prop:betabarHl1}, the estimate for $\Omega \beta_{\ell = 1}$ of Proposition \ref{prop:betaHl1}, the estimate for $\sigma_{\ell = 1}$ of Proposition \ref{prop:sigmal1H}, the estimate for $\rho_{\ell = 1}$ of Proposition \ref{prop:rhol1H}, the estimate for $\Omega_{\ell =1}$ of Proposition \ref{prop:OmegaHl1}, the estimates for $(\Omega \omegahat)_{\ell = 1}$ and $\Omega^{-2} (\Omega \omegabarhat)_{\ell =1}$ of Proposition \ref{prop:omegahatomegabarhatHl1}, the estimates for $\eta_{\ell =1}$ and $\etabar_{\ell = 1}$ of Proposition \ref{prop:etaetabarl1H}, and the estimates for $(\Omega \tr \chi)_{\ell =1}$ and $\Omega^{-2} (\Omega \tr \chibar)_{\ell = 1}$ of Proposition \ref{prop:trchitrchibarl1H}.
\end{proof}

Recall that, for $k_1, k_2 \geq 0$, $\mathfrak{D}^{(0,k_1,k_2)} = (\Omega^{-1} \nablaslash_3)^{k_1} (r\Omega \nablaslash_4)^{k_2}$.

\subsubsection*{Estimates for $(\Omega^{-1} \betabar)_{\ell=1}$}
To begin, $\Omega^{-1} \betabar$ can immediately be estimated by integrating the Bianchi equation \eqref{eq:betabar3} backwards from $u=u_f$.  Compare with Proposition \ref{prop:chibarhatH} and Proposition \ref{prop:chibarhatH2}.

\begin{proposition}[Estimate for $(\Omega^{-1} \betabar)_{\ell=1}$ in $\DRH$] \label{prop:betabarHl1}
	Define $\underline{B} = (\Omega^{-1} \betabar)_{\ell=1} - \Omega^{-1} \betabar_{\mathrm{Kerr}}$.  For $\kbar = (0,k_1,k_2)$, with $k_1+k_2 \leq N-s$, for $s=0,1,2$, for any $v_{-1} \leq v \leq v(R_2,u_f)$, and any $\max \{ u_0, u(R_2,v)\} \leq u \leq u_f$,
	\[
		\Vert \mathfrak{D}^{\kbar} \underline{B} \Vert_{S_{u,v}}^2
		+
		\Vert \mathfrak{D}^{\kbar} \underline{B} \Vert_{C_{u}(v)}^2
		+
		\Vert \mathfrak{D}^{\kbar} \underline{B} \Vert_{\Cbar_v}^2
		+
		\Vert \mathfrak{D}^{\kbar} \underline{B} \Vert_{\DRH(v)}^2
		\lesssim
		\frac{
		\varepsilon_0^2
		+
		\varepsilon^3
		}
		{{v}^{s}}.
	\]
\end{proposition}

\begin{proof}
	The equation \eqref{eq:betabar4}, after projecting to $\ell=1$, can schematically be written
	\[
		\Omega \nablaslash_3 (
		\Omega^{-1} \betabar
		-
		\Omega^{-1} \betabar_{\mathrm{Kerr}}
		)_{\ell=1}
		-
		\frac{4\Omega^2}{r} (
		\Omega^{-1} \betabar
		-
		\Omega^{-1} \betabar_{\mathrm{Kerr}}
		)_{\ell=1}
		=
		\Omega^2 \mathcal{E}^0_{\ell=1}.
	\]
	The proof then inductively follows, after commuting the equation, from Lemma \ref{lem:nabla3xi} and Proposition \ref{prop:divbetabarl1uf}, using Proposition \ref{prop:com0} and Propositions \ref{prop:spacetimeerrorH1} and \ref{prop:errorin} to control the nonlinear error terms.
\end{proof}

\subsubsection*{Estimates for $(\Omega \beta)_{\ell=1}$}

In order to estimate $\Omega \beta$, since the Bianchi equation \eqref{eq:beta4} is blueshifted (recall the nomenclature of Section \ref{subsubsec:nabla4transport}) it is first commuted twice with $\Omega^{-1} \nablaslash_3$ so that the resulting equation satisfied by $(\Omega^{-1} \nablaslash_3)^2 \Omega \beta$ is redshifted.

By Lemma \ref{lem:commutation}, for any $k \geq 0$, $(\Omega \beta)_{\ell=1}$ satisfies the commuted equation
\begin{multline} \label{eq:beta4hoschematic}
	\Omega \nablaslash_4 \left(
	(\Omega^{-1} \nablaslash_3)^{k} r^4 \Omega \beta
	\right)_{\ell=1}
	+
	2 (k -1) \Omega \omegahat
	(\Omega^{-1} \nablaslash_3)^{k} r^4 (\Omega \beta)_{\ell=1}
	=
	\sum_{l =0}^{k-1}
	H_l \cdot (\Omega^{-1} \nablaslash_3)^{l} (\Omega \beta)_{\ell=1}
	+
	\mathcal{E}^{k}_{\ell=1}.
\end{multline}
for some admissible coefficient functions (see \eqref{eq:admis}) $H_1,\ldots,H_{k-1}$.

Compare the following proposition with Proposition \ref{prop:chihatnabla3Htop}.

\begin{proposition}[Estimate for $(\Omega \beta)_{\ell=1}$ in $\DRH$] \label{prop:betaHl1}
	Define $B = (\Omega \beta)_{\ell=1} - \Omega \beta_{\mathrm{Kerr}}$.  For $\kbar = (0,k_1,k_2)$, with $k_1+k_2 \leq N-s$, for $s=0,1,2$, for any $v_{-1} \leq v \leq v(R_2,u_f)$, and any $\max \{ u_0, u(R_2,v)\} \leq u \leq u_f$,
	\[
		\Vert \mathfrak{D}^{\kbar} B \Vert_{S_{u,v}}^2
		+
		\Vert \mathfrak{D}^{\kbar} B \Vert_{C_{u}(v)}^2
		+
		\Vert \mathfrak{D}^{\kbar} B \Vert_{\Cbar_v}^2
		+
		\Vert \mathfrak{D}^{\kbar} B \Vert_{\DRH(v)}^2
		\lesssim
		\frac{
		\varepsilon_0^2
		+
		\varepsilon^3
		}
		{{v}^{s}}.
	\]
\end{proposition}

\begin{proof}
	Consider first the case $k_2=0$.  The proof, for $k_1=0,1$ follows from equation \eqref{eq:beta4hoschematic} and Lemma \ref{lem:noshifteddifferencequotient}, using Lemma \ref{lem:nabla3differenceestimate} and Proposition \ref{prop:errorout} to control the nonlinear error term
	\[
		\Vert D_{u_f} \mathcal{E}^{1} \Vert_{C_{u}(v)}^2
		\lesssim
		\frac{
		\varepsilon_0^2
		+
		\varepsilon^3
		}
		{{v}^{2}},
	\]
	and Propositions \ref{prop:divbetabarl1uf}, \ref{prop:nabla3betal1uf} and \ref{prop:divbetal1v0}.  For $k_1 \geq 2$ the proof proceeds by induction, using Lemma \ref{lem:nabla4redxi} and the fact that equation \eqref{eq:beta4hoschematic} is redshifted, using Propositions \ref{prop:spacetimeerrorH1} and \ref{prop:errorout} to control the nonlinear error terms.  For $k_2 \geq 1$ the proof proceeds inductively by applying $(\Omega^{-1} \nablaslash_3)^{k_1} (\Omega \nablaslash_4)^{k_2-1}$ to equation \eqref{eq:beta4hoschematic}.
\end{proof}

\subsubsection*{Estimates for $\sigma_{\ell=1}$}

The Bianchi equation \eqref{eq:sigma3}, the $\curlslash$ of the Codazzi equation \eqref{eq:Codazzibar} projected to $\ell=1$ and \eqref{eq:curletacurletabar} imply that $\sigma_{\ell=1}$ satisfies
\begin{equation} \label{eq:sigmal13}
	\Omega \nablaslash_3 (\sigma - \sigma_{\mathrm{Kerr}})_{\ell=1}
	-
	4 \frac{\Omega_{\circ}^2}{r} (\sigma_{\ell=1} - \sigma_{\mathrm{Kerr}})
	=
	\Omega^2 \mathcal{E}^1_{\ell=1}.
\end{equation}
Similarly, equations \eqref{eq:sigma4} and \eqref{eq:Codazzi} imply that
\begin{equation} \label{eq:sigmal14}
	\Omega \nablaslash_4 (\sigma - \sigma_{\mathrm{Kerr}})_{\ell=1}
	+
	4 \frac{\Omega_{\circ}^2}{r} (\sigma_{\ell=1} - \sigma_{\mathrm{Kerr}})
	=
	\mathcal{E}^1_{\ell=1}.
\end{equation}
These equations are used to obtain the following estimates for $\sigma_{\ell = 1}$.

\begin{proposition}[Estimate for $\sigma_{\ell=1}$ in $\DRH$] \label{prop:sigmal1H}
	Define $S = \sigma_{\ell=1} - \sigma_{\mathrm{Kerr}}$.  For $\kbar = (0,k_1,k_2)$, with $k_1+k_2 \leq N-s$, for $s=0,1,2$, for any $v_{-1} \leq v \leq v(R_2,u_f)$, and any $\max \{ u_0, u(R_2,v)\} \leq u \leq u_f$,
	\[
		\Vert \mathfrak{D}^{\kbar} S \Vert_{S_{u,v}}^2
		+
		\Vert \mathfrak{D}^{\kbar} S \Vert_{C_{u}(v)}^2
		+
		\Vert \mathfrak{D}^{\kbar} S \Vert_{\Cbar_v}^2
		+
		\Vert \mathfrak{D}^{\kbar} S \Vert_{\DRH(v)}^2
		\lesssim
		\frac{
		\varepsilon_0^2
		+
		\varepsilon^3
		}
		{{v}^{s}}.
	\]
\end{proposition}

\begin{proof}
	Consider first the case that $k_2=0$.  The result for $k_1=0$ follows, using equation \eqref{eq:sigmal13} from Lemma \ref{lem:nabla3xi}, Proposition \ref{prop:com0}, Proposition \ref{prop:sigmal1uf} and Propositions \ref{prop:spacetimeerrorH1} and \ref{prop:errorin}.  For $k_1 \geq 1$ the proof follows from dividing \eqref{eq:sigmal13} by $\Omega^2$ and applying $(\Omega^{-1} \nablaslash_3)^{k_1-1}$.  For $k_2 \geq 1$ the proof follows from applying $(\Omega^{-1} \nablaslash_3)^{k_1} (\Omega \nablaslash_4)^{k_2}$ to \eqref{eq:sigmal14}.
\end{proof}

\subsubsection*{Estimates for $\eta_{\ell=1}$ and \underline{$\eta$}$_{\ell=1}$}

The quantities $\eta_{\ell =1}$ and $\etabar_{\ell = 1}$ and most of their derivatives can now be estimated.  Define
\[
	E = \eta_{\ell=1} - \eta_{\mathrm{Kerr}},
	\qquad
	\underline{E} = \etabar_{\ell=1} - \etabar_{\mathrm{Kerr}}.
\]

First a preliminary estimate for the difference $\eta_{\ell=1} - \etabar_{\ell =1}$ is obtained.

\begin{proposition}[Estimate for $\eta_{\ell=1} - \etabar_{\ell =1}$ in $\DRH$] \label{prop:etaminusetabarl1H}
	For $\kbar = (0,k_1,k_2)$, with $k_1+k_2 \leq N-1-s$, for $s=0,1,2$, for any $v_{-1} \leq v \leq v(R_2,u_f)$, and any $\max \{ u_0, u(R_2,v)\} \leq u \leq u_f$,
	\[
		\Vert \mathfrak{D}^{\kbar} ( E - \underline{E}) \Vert_{S_{u,v}}^2
		+
		\Vert \mathfrak{D}^{\kbar} ( E - \underline{E}) \Vert_{C_{u}(v)}^2
		+
		\Vert \mathfrak{D}^{\kbar} ( E - \underline{E}) \Vert_{\Cbar_v}^2
		+
		\Vert \mathfrak{D}^{\kbar} ( E - \underline{E}) \Vert_{\DRH(v)}^2
		\lesssim
		\frac{
		\varepsilon_0^2
		+
		\varepsilon^3
		}
		{{v}^{s}}.
	\]
\end{proposition}

\begin{proof}
	Adding together equations \eqref{eq:beta3} and \eqref{eq:betabar4} gives
	\begin{multline*}
		\frac{6M_f}{r} (\eta - \eta_{\mathrm{Kerr}} - \etabar + \etabar_{\mathrm{Kerr}})
		=
		2r^2 {}^*\nablaslash (\sigma - \sigma_{\mathrm{Kerr}})
		-
		\Omega^{-1} \nablaslash_3(r^2 (\Omega \beta - \Omega \beta_{\mathrm{Kerr}}))
		\\
		-
		\Omega \nablaslash_4 (r^2 (\Omega^{-1} \betabar - \Omega^{-1} \betabar_{\mathrm{Kerr}}))
		-
		2 r^2 \Omega \omegahat (\Omega^{-1} \betabar - \Omega^{-1} \betabar_{\mathrm{Kerr}})
		+
		\mathcal{E}^0.
	\end{multline*}
	The proof the follows from Propositions \ref{prop:betabarHl1}, \ref{prop:betaHl1} and \ref{prop:sigmal1H} after projecting to $\ell=1$ and applying $(\Omega \nablaslash_4)^{k_1} (\Omega^{-1} \nablaslash_3)^{k_2}$.
\end{proof}

Most of the derivatives of $\etabar_{\ell =1}$ can now be estimated.

\begin{proposition}[Estimate for $\etabar_{\ell=1}$ in $\DRH$] \label{prop:etabarl1loH}
	For $\kbar = (0,k_1,k_2)$, with $k_1+k_2 \leq N-s$, $k_2 \leq N-1-s$, for $s=0,1,2$, for any $v_{-1} \leq v \leq v(R_2,u_f)$, and any $\max \{ u_0, u(R_2,v)\} \leq u \leq u_f$,
	\[
		\Vert \mathfrak{D}^{\kbar} \underline{E} \Vert_{S_{u,v}}^2
		+
		\Vert \mathfrak{D}^{\kbar} \underline{E} \Vert_{C_{u}(v)}^2
		+
		\Vert \mathfrak{D}^{\kbar} \underline{E} \Vert_{\Cbar_v}^2
		+
		\Vert \mathfrak{D}^{\kbar} \underline{E} \Vert_{\DRH(v)}^2
		\lesssim
		\frac{
		\varepsilon_0^2
		+
		\varepsilon^3
		}
		{{v}^{s}}.
	\]
\end{proposition}

\begin{proof}
	Equation \eqref{eq:nabla4eta} implies,
	\[
		\Omega \nablaslash_3 (r^{-1} (\etabar - \etabar_{\mathrm{Kerr}}))
		-
		\frac{\Omega_{\circ}^2}{r} r^{-1} (\etabar - \etabar_{\mathrm{Kerr}})
		=
		\frac{\Omega^2}{r}
		\left(
		r^{-1} (\etabar - \etabar_{\mathrm{Kerr}} - \eta + \eta_{\mathrm{Kerr}})
		+
		\Omega^{-1} \betabar
		-
		\Omega^{-1} \betabar_{\mathrm{Kerr}}
		+
		\mathcal{E}^0
		\right).
	\]
	Consider first the case that $k_2=0$.  The proof follows inductively, after projecting the above equation to $\ell=1$ and using Proposition \ref{prop:com0oneforms}, from Lemma \ref{lem:nabla3xi} and Propositions \ref{prop:etal1uf}, \ref{prop:betabarHl1},  \ref{prop:etaminusetabarl1H} after commuting with $(\Omega \nablaslash_4)^{k_1}$.  For $k_2 \geq 1$, the proof follows inductively from Propositions \ref{prop:betabarHl1} and \ref{prop:etaminusetabarl1H} after projecting the above equation to $\ell=1$, dividing by $\Omega^2$, and applying $(\Omega \nablaslash_4)^{k_1} (\Omega^{-1} \nablaslash_3)^{k_2-1}$.
\end{proof}

Most of the derivatives of $\eta_{\ell = 1}$ are now estimated.  The remaining derivatives of $\eta_{\ell = 1}$ and $\etabar_{\ell = 1}$ are estimated in Proposition \ref{prop:etaetabarl1H} below.

\begin{proposition}[Estimate for $\eta_{\ell=1}$ in $\DRH$] \label{prop:etal1loH}
	For $\kbar = (0,k_1,k_2)$, with $k_1+k_2 \leq N-1-s$, for $s=0,1,2$, for any $v_{-1} \leq v \leq v(R_2,u_f)$, and any $\max \{ u_0, u(R_2,v)\} \leq u \leq u_f$,
	\[
		\Vert \mathfrak{D}^{\kbar} E \Vert_{S_{u,v}}^2
		+
		\Vert \mathfrak{D}^{\kbar} E \Vert_{C_{u}(v)}^2
		+
		\Vert \mathfrak{D}^{\kbar} E \Vert_{\Cbar_v}^2
		+
		\Vert \mathfrak{D}^{\kbar} E \Vert_{\DRH(v)}^2
		\lesssim
		\frac{
		\varepsilon_0^2
		+
		\varepsilon^3
		}
		{{v}^{s}}.
	\]
\end{proposition}

\begin{proof}
	The proof is an immediate consequence of Proposition \ref{prop:etaminusetabarl1H} and Proposition \ref{prop:etabarl1loH}.
\end{proof}

\subsubsection*{Estimates for $(\Omega \tr \chi)_{\ell=1}$ and $\Omega^{-2}(\Omega \tr$\underline{$\chi$}$)_{\ell=1}$}

The Codazzi equations now yield the following estimates for $(\Omega \tr \chi)_{\ell=1}$ and $\Omega^{-2}(\Omega \tr$\underline{$\chi$}$)_{\ell=1}$.

\begin{proposition}[Estimate for $(\Omega \tr \chi)_{\ell=1}$ in $\DRH$] \label{prop:trchil1loH}
	For $\kbar = (0,k_1,k_2)$, with $k_1+k_2 \leq N-1-s$, for $s=0,1,2$, for any $v_{-1} \leq v \leq v(R_2,u_f)$, and any $\max \{ u_0, u(R_2,v)\} \leq u \leq u_f$,
	\[
		\Vert \mathfrak{D}^{\kbar} (\Omega \tr \chi)_{\ell=1} \Vert_{S_{u,v}}^2
		+
		\Vert \mathfrak{D}^{\kbar} (\Omega \tr \chi)_{\ell=1} \Vert_{C_{u}(v)}^2
		+
		\Vert \mathfrak{D}^{\kbar} (\Omega \tr \chi)_{\ell=1} \Vert_{\Cbar_v}^2
		+
		\Vert \mathfrak{D}^{\kbar} (\Omega \tr \chi)_{\ell=1} \Vert_{\DRH(v)}^2
		\lesssim
		\frac{
		\varepsilon_0^2
		+
		\varepsilon^3
		}
		{{v}^{s}}.
	\]
\end{proposition}

\begin{proof}
	The proof follows from the Codazzi equation \eqref{eq:Codazzi} projected to $\ell=1$ and Propositions \ref{prop:betaHl1} and \ref{prop:etal1loH}.
\end{proof}

\begin{proposition}[Estimate for $\Omega^{-2} (\Omega \tr \chibar)_{\ell=1}$ in $\DRH$] \label{prop:trchibarl1loH}
	For $\kbar = (0,k_1,k_2)$, with $k_1+k_2 \leq N-1-s$, for $s=0,1,2$, for any $v_{-1} \leq v \leq v(R_2,u_f)$, and any $\max \{ u_0, u(R_2,v)\} \leq u \leq u_f$,
	\begin{multline*}
		\Vert \mathfrak{D}^{\kbar} \big( \Omega^{-2} (\Omega \tr \chibar)_{\ell=1} \big) \Vert_{S_{u,v}}^2
		+
		\Vert \mathfrak{D}^{\kbar} \big( \Omega^{-2} (\Omega \tr \chibar)_{\ell=1} \big) \Vert_{C_{u}(v)}^2
		+
		\Vert \mathfrak{D}^{\kbar} \big( \Omega^{-2} (\Omega \tr \chibar)_{\ell=1} \big) \Vert_{\Cbar_v}^2
		\\
		+
		\Vert \mathfrak{D}^{\kbar} \big( \Omega^{-2} (\Omega \tr \chibar)_{\ell=1} \big) \Vert_{\DRH(v)}^2
		\lesssim
		\frac{
		\varepsilon_0^2
		+
		\varepsilon^3
		}
		{{v}^{s}}.
	\end{multline*}
\end{proposition}

\begin{proof}
	The proof follows from the Codazzi equation \eqref{eq:Codazzibar} projected to $\ell=1$ and Propositions \ref{prop:betabarHl1} and \ref{prop:etabarl1loH}.
\end{proof}

\subsubsection*{Estimates for $\rho_{\ell=1}$}

Using the Bianchi equations \eqref{eq:rho4} and \eqref{eq:rho3}, $\rho_{\ell =1}$ can now be estimated.

\begin{proposition}[Estimate for $\rho_{\ell=1}$ in $\DRH$] \label{prop:rhol1H}
	For $\kbar = (0,k_1,k_2)$, with $k_1+k_2 \leq N-s$, for $s=0,1,2$, for any $v_{-1} \leq v \leq v(R_2,u_f)$, and any $\max \{ u_0, u(R_2,v)\} \leq u \leq u_f$,
	\[
		\Vert \mathfrak{D}^{\kbar} \rho_{\ell=1} \Vert_{S_{u,v}}^2
		+
		\Vert \mathfrak{D}^{\kbar} \rho_{\ell=1} \Vert_{C_{u}(v)}^2
		+
		\Vert \mathfrak{D}^{\kbar} \rho_{\ell=1} \Vert_{\Cbar_v}^2
		+
		\Vert \mathfrak{D}^{\kbar} \rho_{\ell=1} \Vert_{\DRH(v)}^2
		\lesssim
		\frac{
		\varepsilon_0^2
		+
		\varepsilon^3
		}
		{{v}^{s}}.
	\]
\end{proposition}

\begin{proof}
	The equation \eqref{eq:rho3} gives,
	\[
		\Omega \nablaslash_3 (r^3\rho)_{\ell=1}
		=
		3M_f \left( \Omega \tr \chibar \right)_{\ell=1}
		-
		r^3 \Omega^2 (\divslash \Omega^{-1}\betabar)_{\ell=1}
		+
		\mathcal{E}^0_{\ell=1},
	\]
	and the result for $k_1=k_2=0$ follows from Lemma \ref{lem:nabla3xi}, Proposition \ref{prop:com0}, Proposition \ref{prop:rhol1uf} and Propositions \ref{prop:betabarHl1} and \ref{prop:trchibarl1loH}.  The result for $k_1 \geq 1$, $k_2=0$ follows inductively from dividing the above equation by $\Omega^2$ and applying $(\Omega^{-1} \nablaslash_3)^{k_1-1}$.  For $k_2 \geq 0$ the proof follows similarly from applying $(\Omega^{-1} \nablaslash_3)^{k_1} (\Omega \nablaslash_4)^{k_2-1}$ to the projection of equation \eqref{eq:rho4} and using Propositions \ref{prop:betaHl1} and \ref{prop:trchil1loH}.
\end{proof}

\subsubsection*{Estimates for $(\Omega \omegahat)_{\ell=1}$ and $\Omega^{-2} (\Omega$\underline{$\omegahat$}$)_{\ell=1}$}

The quantities $(\Omega_{\circ}^{-2} \Omega^2)_{\ell =1}$, $(\Omega \omegahat)_{\ell=1}$ and $\Omega^{-2} (\Omega$\underline{$\omegahat$}$)_{\ell=1}$ can now be estimated.

\begin{proposition}[Estimate for $(\Omega_{\circ}^{-2} \Omega^2)_{\ell =1}$ in $\DRH$] \label{prop:OmegaHl1}
	For any $v_{-1} \leq v \leq v(R_2,u_f)$, and any $\max \{ u_0, u(R_2,v)\} \leq u \leq u_f$,
	\[
		\Vert ( \Omega_{\circ}^{-2} \Omega^2 )_{\ell=1} \Vert_{S_{u,v}}^2
		+
		\Vert ( \Omega_{\circ}^{-2} \Omega^2 )_{\ell=1} \Vert_{C_{u}(v)}^2
		+
		\Vert ( \Omega_{\circ}^{-2} \Omega^2 )_{\ell=1} \Vert_{\Cbar_v}^2
		+
		\Vert ( \Omega_{\circ}^{-2} \Omega^2 )_{\ell=1} \Vert_{\DRH(v)}^2
		\lesssim
		\frac{
		\varepsilon_0^2
		+
		\varepsilon^3
		}
		{{v}^{s}}.
	\]
\end{proposition}

\begin{proof}
	The proof follows from Propositions \ref{prop:etabarl1loH} and \ref{prop:etal1loH}, Proposition \ref{prop:Poincare} and the fact that
	\[
		\nablaslash \log \left( \frac{\Omega^2}{\Omega_{\circ}^2} \right)
		=
		\eta + \etabar.
	\]
\end{proof}

\begin{proposition}[Estimates for $(\Omega \omegahat)_{\ell=1}$ and $\Omega^{-2} (\Omega \omegabarhat)_{\ell=1}$ in $\DRH$] \label{prop:omegahatomegabarhatHl1}
	For $\kbar = (0,k_1,k_2)$, with $k_1+k_2 \leq N-s$, for $s=0,1,2$, for any $v_{-1} \leq v \leq v(R_2,u_f)$, and any $\max \{ u_0, u(R_2,v)\} \leq u \leq u_f$,
	\begin{equation} \label{eq:omegahatHl1}
		\Vert \mathfrak{D}^{\kbar} (\Omega \omegahat)_{\ell=1} \Vert_{S_{u,v}}^2
		+
		\Vert \mathfrak{D}^{\kbar} (\Omega \omegahat)_{\ell=1} \Vert_{C_{u}(v)}^2
		+
		\Vert \mathfrak{D}^{\kbar} (\Omega \omegahat)_{\ell=1} \Vert_{\Cbar_v}^2
		+
		\Vert \mathfrak{D}^{\kbar} (\Omega \omegahat)_{\ell=1} \Vert_{\DRH(v)}^2
		\lesssim
		\frac{
		\varepsilon_0^2
		+
		\varepsilon^3
		}
		{{v}^{s}},
	\end{equation}
	and
	\begin{equation} \label{eq:omegabarhatHl1}
		\Vert \mathfrak{D}^{\kbar} \Omega^{-2} (\Omega \omegabarhat)_{\ell=1} \Vert_{S_{u,v}}^2
		+
		\Vert \mathfrak{D}^{\kbar} \Omega^{-2} (\Omega \omegabarhat)_{\ell=1} \Vert_{C_{u}(v)}^2
		+
		\Vert \mathfrak{D}^{\kbar} \Omega^{-2} (\Omega \omegabarhat)_{\ell=1} \Vert_{\Cbar_v}^2
		+
		\Vert \mathfrak{D}^{\kbar} \Omega^{-2} (\Omega \omegabarhat)_{\ell=1} \Vert_{\DRH(v)}^2
		\lesssim
		\frac{
		\varepsilon_0^2
		+
		\varepsilon^3
		}
		{{v}^{s}}.
	\end{equation}
\end{proposition}

\begin{proof}
	Consider first the estimate \eqref{eq:omegahatHl1} in the case that $k_1=k_2=0$.  Equation \eqref{eq:omega3omegabar4} implies that,
	\[
		\Omega \nablaslash_3 \left( r^{-1} (\Omega \omegahat) \right)_{\ell=1}
		-
		\frac{\Omega_{\circ}^2}{r^2} (\Omega \omegahat)_{\ell=1}
		=
		\frac{\Omega^2}{r} \left(
		-
		\rho_{\circ} ( \Omega_{\circ}^{-2} \Omega^2 )_{\ell=1}
		-
		\rho_{\ell=1}
		+
		\mathcal{E}^0_{\ell=1}
		\right),
	\]
	and the proof follows from Lemma \ref{lem:nabla3xi}, Proposition \ref{prop:com0}, and Propositions \ref{prop:rhol1H} and \ref{prop:OmegaHl1} after recalling that $(\Omega \omegahat)_{\ell=1}=0$ on $\{u=u_f\}$.  Consider now the estimate \eqref{eq:omegabarhatHl1} in the case that $k_1=k_2=0$.  Equation \eqref{eq:omega3omegabar4} implies,
	\[
		\Omega \nablaslash_4( \Omega^{-2} (\Omega \omegabarhat) )_{\ell=1}
		+
		2 (\Omega \omegahat)_{\circ} \Omega^{-2} (\Omega \omegabarhat)_{\ell=1}
		=
		-
		\rho_{\circ} ( \Omega_{\circ}^{-2} \Omega^2 )_{\ell=1}
		-
		\rho_{\ell=1}
		+
		\mathcal{E}^0_{\ell=1},
	\]
	and the proof follows from Lemma \ref{lem:nabla4redxi}, Proposition \ref{prop:com0}, and Propositions \ref{prop:rhol1H} and \ref{prop:OmegaHl1}, after recalling that $(\Omega \omegabarhat)_{\ell=1}=0$ on $\{ v=v_{-1}\}$.
	
	For $k_1+k_2 \geq 1$, the proofs follow inductively by commuting the equations with either $(\Omega^{-1} \nablaslash_3)^{k_1} (\Omega \nablaslash_4)^{k_2-1}$ or $(\Omega^{-1} \nablaslash_3)^{k_1-1} (\Omega \nablaslash_4)^{k_2}$ and using the fact that
	\[
		\Omega^{-1} \nablaslash_3 \left( \Omega_{\circ}^{-2} \Omega^2 \right)
		=
		2 \Omega^{-2} \left( \Omega \omegabarhat - (\Omega \omegabarhat)_{\circ} \right)
		+
		2 \left( \Omega_{\circ}^{-2} \Omega^2 - 1 \right) \Omega^{-2} \left( \Omega \omegabarhat - (\Omega \omegabarhat)_{\circ} \right),
	\]
	and
	\[
		\Omega \nablaslash_4 \left( \Omega_{\circ}^{-2} \Omega^2 \right)
		=
		2 \left( \Omega \omegahat - (\Omega \omegahat)_{\circ} \right)
		+
		2 \left( \Omega_{\circ}^{-2} \Omega^2 - 1 \right) \left( \Omega \omegahat - (\Omega \omegahat)_{\circ} \right).
	\]
\end{proof}

\subsubsection*{Estimates for top order derivatives of $\eta_{\ell=1}$, \underline{$\eta$}$_{\ell=1}$, $(\Omega \tr \chi)_{\ell=1}$ and $\Omega^{-2}(\Omega \tr$\underline{$\chi$}$)_{\ell=1}$}

The following proposition gives estimates for all $\Omega^{-1} \nablaslash_3$ and $\Omega \nablaslash_4$ derivatives of $\eta_{\ell=1}$ and $\etabar_{\ell=1}$ up to order $N$, meaning that the estimates of Proposition \ref{prop:etabarl1loH} and Proposition \ref{prop:etal1loH} in fact hold for all $k_1+k_2 \leq N-s$, $s=0,1,2$.

\begin{proposition}[Estimates for $\eta_{\ell=1}$ and $\etabar_{\ell=1}$ in $\DRH$] \label{prop:etaetabarl1H}
	For $\kbar = (0,k_1,k_2)$, with $k_1+k_2 \leq N-s$, for $s=0,1,2$, for any $v_{-1} \leq v \leq v(R_2,u_f)$, and any $\max \{ u_0, u(R_2,v)\} \leq u \leq u_f$,
	\[
		\Vert \mathfrak{D}^{\kbar} \underline{E} \Vert_{S_{u,v}}^2
		+
		\Vert \mathfrak{D}^{\kbar} \underline{E} \Vert_{C_{u}(v)}^2
		+
		\Vert \mathfrak{D}^{\kbar} \underline{E} \Vert_{\Cbar_v}^2
		+
		\Vert \mathfrak{D}^{\kbar} \underline{E} \Vert_{\DRH(v)}^2
		\lesssim
		\frac{
		\varepsilon_0^2
		+
		\varepsilon^3
		}
		{{v}^{s}},
	\]
	and
	\[
		\Vert \mathfrak{D}^{\kbar} E \Vert_{S_{u,v}}^2
		+
		\Vert \mathfrak{D}^{\kbar} E \Vert_{C_{u}(v)}^2
		+
		\Vert \mathfrak{D}^{\kbar} E \Vert_{\Cbar_v}^2
		+
		\Vert \mathfrak{D}^{\kbar} E \Vert_{\DRH(v)}^2
		\lesssim
		\frac{
		\varepsilon_0^2
		+
		\varepsilon^3
		}
		{{v}^{s}},
	\]
	where $E = \eta_{\ell=1} - \eta_{\mathrm{Kerr}}$ and $\underline{E} = \etabar_{\ell=1} - \etabar_{\mathrm{Kerr}}$.
\end{proposition}

\begin{proof}
	When $k_1+k_2 \leq N-1-s$ the proposition reduces to Proposition \ref{prop:etabarl1loH} and Proposition \ref{prop:etal1loH}.  For the remaining cases the proofs follow from commuting equations \eqref{eq:nabla4eta}, \eqref{eq:nabla4etabar}, \eqref{eq:nabla3eta} with either $(\Omega^{-1} \nablaslash_3)^{k_1} (\Omega \nablaslash_4)^{k_2-1}$ or $(\Omega^{-1} \nablaslash_3)^{k_1-1}(\Omega \nablaslash_4)^{k_1}$ and using Propositions \ref{prop:betabarHl1}, \ref{prop:betaHl1} and \ref{prop:omegahatomegabarhatHl1}.
\end{proof}

The following proposition gives estimates for all $\Omega^{-1} \nablaslash_3$ and $\Omega \nablaslash_4$ derivatives of $(\Omega \tr \chi)_{\ell=1}$ and $\Omega^{-2} (\Omega \tr \chibar)_{\ell=1}$ up to order $N$, meaning that the estimates of Proposition \ref{prop:trchil1loH} and Proposition \ref{prop:trchibarl1loH} in fact hold for all $k_1+k_2 \leq N-s$, $s=0,1,2$.

\begin{proposition}[Estimate for $(\Omega \tr \chi)_{\ell=1}$ and $\Omega^{-2} (\Omega \tr \chibar)_{\ell =1}$ in $\DRH$] \label{prop:trchitrchibarl1H}
	For $\kbar = (0,k_1,k_2)$, with $k_1+k_2 \leq N-s$, for $s=0,1,2$, for any $v_{-1} \leq v \leq v(R_2,u_f)$, and any $\max \{ u_0, u(R_2,v)\} \leq u \leq u_f$,
	\[
		\Vert \mathfrak{D}^{\kbar} (\Omega \tr \chi)_{\ell=1} \Vert_{S_{u,v}}^2
		+
		\Vert \mathfrak{D}^{\kbar} (\Omega \tr \chi)_{\ell=1} \Vert_{C_{u}(v)}^2
		+
		\Vert \mathfrak{D}^{\kbar} (\Omega \tr \chi)_{\ell=1} \Vert_{\Cbar_v}^2
		+
		\Vert \mathfrak{D}^{\kbar} (\Omega \tr \chi)_{\ell=1} \Vert_{\DRH(v)}^2
		\lesssim
		\frac{
		\varepsilon_0^2
		+
		\varepsilon^3
		}
		{{v}^{s}},
	\]
	and
	\begin{multline*}
		\Vert \mathfrak{D}^{\kbar} \big( \Omega^{-2} (\Omega \tr \chibar)_{\ell=1} \big) \Vert_{S_{u,v}}^2
		+
		\Vert \mathfrak{D}^{\kbar} \big( \Omega^{-2} (\Omega \tr \chibar)_{\ell=1} \big) \Vert_{C_{u}(v)}^2
		+
		\Vert \mathfrak{D}^{\kbar} \big( \Omega^{-2} (\Omega \tr \chibar)_{\ell=1} \big) \Vert_{\Cbar_v}^2
		\\
		+
		\Vert \mathfrak{D}^{\kbar} \big( \Omega^{-2} (\Omega \tr \chibar)_{\ell=1} \big) \Vert_{\DRH(v)}^2
		\lesssim
		\frac{
		\varepsilon_0^2
		+
		\varepsilon^3
		}
		{{v}^{s}}.
	\end{multline*}
\end{proposition}

\begin{proof}
	When $k_1+k_2 \leq N-1-s$ the proposition reduces to Proposition \ref{prop:trchil1loH} and Proposition \ref{prop:trchibarl1loH}.  For the remaining cases the proofs follow from commuting equations \eqref{eq:Ray}, \eqref{eq:trchi3}, \eqref{eq:trchibar4} with either $(\Omega^{-1} \nablaslash_3)^{k_1} (\Omega \nablaslash_4)^{k_2-1}$ or $(\Omega^{-1} \nablaslash_3)^{k_1-1} (\Omega \nablaslash_4)^{k_2}$ and using Propositions \ref{prop:rhol1H}, \ref{prop:etaetabarl1H} and \ref{prop:omegahatomegabarhatHl1}.
\end{proof}

\subsubsection{Estimates for $\ell =0$ modes in $\DRH$}

This section concerns estimates for the $\ell =0$ modes of the Ricci coefficients and curvature components on $\DRH$.  The main result is the following Proposition.

\begin{theorem}[Improving bootstrap assumptions for energies of $\ell = 0$ modes] \label{prop:ell0Henergy}
	The energies defined in \eqref{eq:spacetimeHenergy3}, \eqref{eq:outgoingHenergy3}, \eqref{eq:incomingHenergy3}, \eqref{eq:spheresHenergy3}
	satisfy,
	\[
		\mathbb{E}^N_{\ell = 0}[\DRH]
		+
		\mathbb{E}^N_{\ell =0}[C^{\Hp}]
		+
		\mathbb{E}^N_{\ell =0}[\Cbar^{\Hp}]
		+
		\mathbb{E}^N_{\ell =0}[S^{\Hp}]
		\lesssim
		\varepsilon_0^2
		+
		\varepsilon^3.
	\]
\end{theorem}

\begin{proof}
	The proof is a direct consequence of Proposition \ref{prop:l0modesR} below.
\end{proof}

The quantities are estimated separately in the region $r\leq r_0$, where the smallness of $\Omega^2_{\circ}$ can be exploited, and the region $r_0 \leq r \leq R_2$, where a softer argument is sufficient.

\subsubsection*{Estimates for $\ell = 0$ modes in the region $r\leq r_0$}

The goal of this section is to estimate the $\ell=0$ modes of the geometric quantities in the region $r\leq r_0$ by exploiting the smallness of $r_0$.  The main difficulty is in estimating the lower order derivatives.  See Propositions \ref{prop:trchil0Homega1}--\ref{prop:trchibarl0Htrchibar}.  Once the lower order derivatives have been estimated the higher order derivatives are estimated in Proposition \ref{prop:l0modesr0}.

By Lemma \ref{lem:commutation} and the Raychaudhuri equation \eqref{eq:Ray} (see also \eqref{eq:trchi4H}), for any $k \geq 1$, $(\Omega \tr \chi - (\Omega \tr \chi)_{\circ})_{\ell=0}$ satisfies the commuted equation
\begin{multline} \label{eq:trchil04hoschematic}
	\Omega \nablaslash_4 \left(
	(\Omega^{-1} \nablaslash_3)^{k} r^2 (\Omega \tr \chi - (\Omega \tr \chi)_{\circ})_{\ell=0}
	\right)
	+
	2 (k -1) \Omega \omegahat
	(\Omega^{-1} \nablaslash_3)^{k} r^2 (\Omega \tr \chi - (\Omega \tr \chi)_{\circ})_{\ell=0}
	\\
	=
	\Omega_{\circ}^2
	\sum_{l =0}^k
	H^1_l \cdot
	(\Omega^{-1} \nablaslash_3)^l
	\left(
	\Omega \omegahat - (\Omega \omegahat)_{\circ}
	\right)_{\ell=0}
	+
	\sum_{l =0}^{k-1}
	H^2_l \cdot (\Omega^{-1} \nablaslash_3)^{l} (\Omega \tr \chi - (\Omega \tr \chi)_{\circ})_{\ell=0}
	+
	\mathcal{E}^{*k},
\end{multline}
for some admissible coefficient functions (see \eqref{eq:admis}) $H^1_0,\ldots,H^1_k, H^2_0,\ldots,H^2_{k-1}$.
Note that the terms involving $(\Omega \omegahat - (\Omega \omegahat)_{\circ})_{\ell=0}$ in \eqref{eq:trchil04hoschematic} involve a smallness factor of $\Omega_{\circ}^2$.

Define
\begin{multline*}
	F_2[(\Omega \omegahat - (\Omega \omegahat)_{\circ})_{\ell=0}](v)
	:=
	(r_0-2M_f)
	\sup_{u(r_0,v) \leq u \leq u_f}
	\sum_{l=0}^2
	\Big(
	\frac{1}{v^2}
	\int_{v_{-1}}^{v(r_0,u)}
	\left\Vert (\Omega^{-1} \nablaslash_3)^l
	(\Omega \omegahat - (\Omega \omegahat)_{\circ})_{\ell=0}
	\right\Vert^2_{S_{u,v'}} dv'
	\\
	+
	\frac{1}{v}
	\int_{\frac{v}{4} \vee v_{-1}}^{v(r_0,u)}
	\left\Vert (\Omega^{-1} \nablaslash_3)^l
	(\Omega \omegahat - (\Omega \omegahat)_{\circ})_{\ell=0}
	\right\Vert^2_{S_{u,v'}} dv'
	+
	\int_{v}^{v(r_0,u)} 
	\left\Vert (\Omega^{-1} \nablaslash_3)^l
	(\Omega \omegahat - (\Omega \omegahat)_{\circ})_{\ell=0}
	\right\Vert^2_{S_{u,v'}} dv'
	\Big),
\end{multline*}
where $\frac{v}{4} \vee v_{-1} = \max\{ v_{-1},v/4\}$.  Note the smallness factor of $r_0 -2M_f$.  This quantity arises in Propositions \ref{prop:trchil0Homega1} and \ref{prop:rhol0Homega} below, before $(\Omega \omegahat - (\Omega \omegahat)_{\circ})_{\ell=0}$ itself is estimated.  The quantity is then eventually, in Proposition \ref{prop:omegal0Htrchibar}, absorbed by exploiting the smallness factor.  Similarly define
\begin{multline*}
	F_1[\Omega^{-2} (\Omega \tr \chibar - (\Omega \tr \chibar)_{\circ})_{\ell=0}](v)
	:=
	(r_0-2M_f)
	\sup_{u(r_0,v) \leq u \leq u_f}
	\sum_{l=0}^1
	\Big(
	\frac{1}{v^2}
	\int_{v_{-1}}^{v(r_0,u)}
	\left\Vert (\Omega^{-1} \nablaslash_3)^l
	\underline{T}_{\ell =0}
	\right\Vert^2_{S_{u,v'}} dv'
	\\
	+
	\frac{1}{v}
	\int_{\frac{v}{4} \vee v_{-1}}^{v(r_0,u)}
	\left\Vert (\Omega^{-1} \nablaslash_3)^l
	\underline{T}_{\ell =0}
	\right\Vert^2_{S_{u,v'}} dv'
	+
	\int_{v}^{v(r_0,u)}
	\left\Vert (\Omega^{-1} \nablaslash_3)^l
	\underline{T}_{\ell =0}
	\right\Vert^2_{S_{u,v'}} dv'
	\Big),
\end{multline*}
where $\underline{T}_{\ell =0} = \Omega^{-2} (\Omega \tr \chibar - (\Omega \tr \chibar)_{\circ})_{\ell=0}$.

\begin{proposition}[Estimate for $(\Omega \tr \chi - (\Omega \tr \chi)_{\circ})_{\ell=0}$ for $r\leq r_0$] \label{prop:trchil0Homega1}
	For any $u(r_0,v_{-1}) \leq u \leq u_f$ and any $v_{-1} \leq v \leq v(r_0,u)$, for $k = 0,1,2$, provided $r_0$ is sufficiently small,
	\begin{multline*}
		\Vert (\Omega^{-1} \nablaslash_3)^k T_{\ell=0} \Vert_{S_{u,v}}^2
		+
		\Vert (\Omega^{-1} \nablaslash_3)^k T_{\ell=0} \mathds{1} \Vert_{C_u(v)}^2
		+
		\Vert (\Omega^{-1} \nablaslash_3)^k T_{\ell=0} \mathds{1} \Vert_{\Cbar_v}^2
		+
		\Vert (\Omega^{-1} \nablaslash_3)^k T_{\ell=0} \mathds{1} \Vert_{\DRH(v)}^2
		\\
		\lesssim
		\frac{
		\varepsilon_0^2
		+
		\varepsilon^3
		}
		{v^2}
		+
		F_2[(\Omega \omegahat - (\Omega \omegahat)_{\circ})_{\ell=0}](v),
	\end{multline*}
	where $T = \Omega \tr \chi - (\Omega \tr \chi)_{\circ}$ and $\mathds{1} = \mathds{1}_{r\leq r_0}$.
\end{proposition}

\begin{proof}
	Consider some $v_{-1} \leq v_1 < v_2 \leq v(r_0,u)$.  Integrating backwards from $u=u_f$, Propositions \ref{prop:trchil0uf} and \ref{prop:trchi3l0uf} imply
	\begin{multline*}
		\int_{v_1}^{v_2}
		\Vert (\Omega \tr \chi - (\Omega \tr \chi)_{\circ})_{\ell=0} \Vert_{S_{u,v}}^2
		+
		\Vert \Omega^{-1} \nablaslash_3 (\Omega \tr \chi - (\Omega \tr \chi)_{\circ})_{\ell=0} \Vert_{S_{u,v}}^2
		dv
		\\
		\lesssim
		\frac{
		\varepsilon_0^2
		+
		\varepsilon^3
		}
		{v^2_1}
		+
		(r_0 - 2M_f)^2
		\int_{v_1}^{v_2}
		\Vert (\Omega^{-1} \nablaslash_3)^2 (\Omega \tr \chi - (\Omega \tr \chi)_{\circ})_{\ell=0} \Vert_{\Cbar_{v}}^2
		dv.
	\end{multline*}
	Now by equation \eqref{eq:trchil04hoschematic} (with $k=2$) it follows, from Proposition \ref{prop:com0} and Proposition \ref{prop:trchil0v0} that,
	\begin{align*}
		&
		\Vert (\Omega^{-1} \nablaslash_3)^2 (\Omega \tr \chi - (\Omega \tr \chi)_{\circ})_{\ell=0} \Vert_{S_{u,v_1}}^2
		+
		\int_{v_1}^{v_2} \Vert (\Omega^{-1} \nablaslash_3)^2 (\Omega \tr \chi - (\Omega \tr \chi)_{\circ})_{\ell=0} \Vert_{S_{u,v}}^2 dv
		\\
		&
		\qquad
		\lesssim
		\Vert (\Omega^{-1} \nablaslash_3)^2 (\Omega \tr \chi - (\Omega \tr \chi)_{\circ})_{\ell=0} \Vert_{S_{u,v_2}}^2
		+
		\frac{\varepsilon_0^2
		+
		\varepsilon^3}{v^2_1}
		\\
		&
		\qquad \quad
		+
		\int_{v_1}^{v_2}
		\sum_{l=0}^1
		\Vert (\Omega^{-1} \nablaslash_3)^l (\Omega \tr \chi - (\Omega \tr \chi)_{\circ})_{\ell=0} \Vert_{S_{u,v}}^2
		+
		\sum_{l=0}^2
		\Vert (\Omega^{-1} \nablaslash_3)^l (\Omega \omegahat - (\Omega \omegahat)_{\circ})_{\ell=0} \mathds{1} \Vert_{S_{u,v}}^2
		dv,
	\end{align*}
	and so
	\begin{align*}
		&
		\sum_{k=0}^2
		\Vert (\Omega^{-1} \nablaslash_3)^k (\Omega \tr \chi - (\Omega \tr \chi)_{\circ})_{\ell=0} \Vert_{S_{u,v_1}}^2
		+
		\sum_{k=0}^2
		\int_{v_1}^{v_2} \Vert (\Omega^{-1} \nablaslash_3)^k (\Omega \tr \chi - (\Omega \tr \chi)_{\circ})_{\ell=0} \Vert_{S_{u,v}}^2 dv
		\\
		&
		\qquad
		\lesssim
		\Vert (\Omega^{-1} \nablaslash_3)^2 (\Omega \tr \chi - (\Omega \tr \chi)_{\circ})_{\ell=0} \Vert_{S_{u,v_2}}^2
		+
		\frac{\varepsilon_0^2
		+
		\varepsilon^3}{v^2_1}
		+
		\sum_{l=0}^2
		\int_{v_1}^{v_2}
		\Vert (\Omega^{-1} \nablaslash_3)^l (\Omega \omegahat - (\Omega \omegahat)_{\circ})_{\ell=0} \Vert_{S_{u,v}}^2
		dv.
	\end{align*}
	The proof then follows as in that of Lemma \ref{lem:nabla4redxi}.
\end{proof}

\begin{proposition}[Estimate for $(\rho - \rho_{\circ})_{\ell=0}$ for $r\leq r_0$] \label{prop:rhol0Homega}
	For any $u(r_0,v_{-1}) \leq u \leq u_f$ and any $v_{-1} \leq v \leq v(r_0,u)$, for $k = 0,1,2$, provided $r_0$ is sufficiently small,
	\begin{multline*}
		\Vert (\Omega^{-1} \nablaslash_3)^k (\rho - \rho_{\circ})_{\ell=0} \Vert_{S_{u,v}}^2
		+
		\Vert (\Omega^{-1} \nablaslash_3)^k (\rho - \rho_{\circ})_{\ell=0} \mathds{1} \Vert_{C_u(v)}^2
		+
		\Vert (\Omega^{-1} \nablaslash_3)^k (\rho - \rho_{\circ})_{\ell=0} \mathds{1} \Vert_{\Cbar_v}^2
		\\
		+
		\Vert (\Omega^{-1} \nablaslash_3)^k (\rho - \rho_{\circ})_{\ell=0} \mathds{1} \Vert_{\DRH(v)}^2
		\lesssim
		\frac{
		\varepsilon_0^2
		+
		\varepsilon^3
		}
		{v^2}
		+
		F_2[(\Omega \omegahat - (\Omega \omegahat)_{\circ})_{\ell=0}](v),
	\end{multline*}
	where $\mathds{1} = \mathds{1}_{r\leq r_0}$.
\end{proposition}

\begin{proof}
	The proof follows, as in the proof of Proposition \ref{prop:trchil0Homega1}, using now equation \eqref{eq:rho4}, which implies,
	\[
		\Omega \nablaslash_4 (r^3 (\rho - \rho_{\circ}))
		=
		r^3 \divslash \beta
		+
		3M_f \left( \Omega \tr \chi - (\Omega \tr \chi)_{\circ} \right)
		+
		\mathcal{E}^0,
	\]
	after projecting to $\ell = 0$, using Proposition \ref{prop:com0} and the fact that $(\divslash \beta)_{\ell = 0}$, by Proposition \ref{prop:trchil0Homega1} and Proposition \ref{prop:rhol0v0}.
\end{proof}

Equations \eqref{eq:trchi3} (see also \eqref{eq:trchi3H}) and \eqref{eq:omega3omegabar4} imply
\begin{multline} \label{eq:trchiomegal0schematic}
	\Omega \nablaslash_3 \Big(
	(\Omega \tr \chi - (\Omega \tr \chi)_{\circ} )
	+
	(\Omega \omegahat - (\Omega \omegahat)_{\circ})
	\Big)_{\ell=0}
	\\
	=
	\frac{\Omega^2}{r} (\Omega \tr \chi - (\Omega \tr \chi)_{\circ} )_{\ell=0}
	-
	\frac{\Omega^4}{r}
	\Omega^{-2}
	(\Omega \tr \chibar - (\Omega \tr \chibar)_{\circ} )_{\ell=0}
	+
	\Omega^2 \mathcal{E}^0_{\ell=0}.
\end{multline}
Note the smallness factor of $\Omega^2$ accompanying the $\Omega^{-2} (\Omega \tr \chibar - (\Omega \tr \chibar)_{\circ} )_{\ell=0}$ term.

\begin{proposition}[Estimate for $(\Omega \omegahat - (\Omega \omegahat)_{\circ})_{\ell=0}$ for $r\leq r_0$] \label{prop:omegal0Htrchibar}
	For any $u(r_0,v_{-1}) \leq u \leq u_f$ and any $v_{-1} \leq v \leq v(r_0,u)$, for $k = 0,1,2$, provided $r_0$ is sufficiently small,
	\begin{multline} \label{eq:omegal0Htrchibar1}
		\Vert (\Omega^{-1} \nablaslash_3)^k
		(\Omega \omegahat - (\Omega \omegahat)_{\circ})_{\ell=0}
		\mathds{1} \Vert_{C_u(v)}^2
		+
		\Vert (\Omega^{-1} \nablaslash_3)^k
		(\Omega \omegahat - (\Omega \omegahat)_{\circ})_{\ell=0}
		\mathds{1} \Vert_{\DRH(v)}^2
		\\
		\lesssim
		\frac{
		\varepsilon_0^2
		+
		\varepsilon^3
		}
		{v^2}
		+
		(r_0-2M_f)
		\sum_{l=0}^1
		\sup_{u_0 \leq u \leq u_f}
		\Vert (\Omega^{-1} \nablaslash_3)^l
		\Omega^{-2} (\Omega \tr \chibar - (\Omega \tr \chibar)_{\circ})_{\ell=0}
		\mathds{1} \Vert_{C_u(v)}^2
		,
	\end{multline}
	and
	\begin{multline} \label{eq:omegal0Htrchibar2}
		\Vert (\Omega^{-1} \nablaslash_3)^k 
		(\Omega \omegahat - (\Omega \omegahat)_{\circ})_{\ell=0}
		\Vert_{S_{u,v}}^2
		+
		\Vert (\Omega^{-1} \nablaslash_3)^k
		(\Omega \omegahat - (\Omega \omegahat)_{\circ})_{\ell=0}
		\mathds{1} \Vert_{\Cbar_v}^2
		\\
		\lesssim
		\frac{
		\varepsilon_0^2
		+
		\varepsilon^3
		}
		{v^2}
		+
		(r_0-2M_f)
		\sum_{l=0}^1
		\sup_{u_0 \leq u' \leq u_f}
		\Vert (\Omega^{-1} \nablaslash_3)^l
		\Omega^{-2} (\Omega \tr \chibar - (\Omega \tr \chibar)_{\circ})_{\ell=0}
		\mathds{1} \Vert_{S_{u',v}}^2
		,
	\end{multline}
	where $\mathds{1} = \mathds{1}_{r\leq r_0}$.
\end{proposition}

\begin{proof}
	Setting
	\[
		\xi
		=
		(\Omega \tr \chi - (\Omega \tr \chi)_{\circ} )_{\ell=0}
		+
		(\Omega \omegahat - (\Omega \omegahat)_{\circ})_{\ell=0},
	\]
	it follows as in Lemma \ref{lem:nabla3xi}, using equation \eqref{eq:trchiomegal0schematic} (or rather the corresponding equation for $r^{-1} \xi$) that
	\begin{multline*}
		\Vert 
		\xi
		\mathds{1} \Vert_{C_u(v)}^2
		+
		\Vert
		\xi
		\mathds{1} \Vert_{\DRH(v)}^2
		\lesssim
		\frac{
		\varepsilon_0^2
		+
		\varepsilon^3
		}
		{v^2}
		+
		F_2[(\Omega \omegahat - (\Omega \omegahat)_{\circ})_{\ell=0}](v)
		\\
		+
		(r_0-2M_f)
		\sup_{u_0 \leq u \leq u_f}
		\Vert
		\Omega^{-2} (\Omega \tr \chibar - (\Omega \tr \chibar)_{\circ})_{\ell=0}
		\mathds{1} \Vert_{C_u(v)}^2
		,
	\end{multline*}
	by Proposition \ref{prop:trchil0uf}, Proposition \ref{prop:trchil0Homega1} and the fact that $(\Omega \omegahat - (\Omega \omegahat)_{\circ})_{\ell=0}=0$ on $u=u_f$.  It moreover follows from the equation \eqref{eq:trchiomegal0schematic} directly, and from applying $\Omega^{-1} \nablaslash_3$ to \eqref{eq:trchiomegal0schematic} that
	\begin{multline} \label{eq:omegal0Htrchibarestimate}
		\sum_{k=0}^2
		\Big(
		\Vert 
		(\Omega^{-1} \nablaslash_3)^k \xi
		\mathds{1} \Vert_{C_u(v)}^2
		+
		\Vert
		(\Omega^{-1} \nablaslash_3)^k \xi
		\mathds{1} \Vert_{\DRH(v)}^2
		\Big)
		\lesssim
		\frac{
		\varepsilon_0^2
		+
		\varepsilon^3
		}
		{v^2}
		+
		F_2[(\Omega \omegahat - (\Omega \omegahat)_{\circ})_{\ell=0}](v)
		\\
		+
		(r_0-2M_f)
		\sum_{l=0}^1
		\sup_{u_0 \leq u \leq u_f}
		\Vert
		(\Omega^{-1} \nablaslash_3)^l
		\Omega^{-2} (\Omega \tr \chibar - (\Omega \tr \chibar)_{\circ})_{\ell=0}
		\mathds{1} \Vert_{C_u(v)}^2,
	\end{multline}
	and, by Proposition \ref{prop:trchil0Homega1}, the estimate \eqref{eq:omegal0Htrchibarestimate} holds with $(\Omega \omegahat - (\Omega \omegahat)_{\circ})_{\ell=0}$ in place of $\xi$.  The proof of \eqref{eq:omegal0Htrchibar1} then follows by taking $r_0$ sufficiently small and exploiting the $\Omega_{\circ}^2$ smallness factor in $F_2[(\Omega \omegahat - (\Omega \omegahat)_{\circ})_{\ell=0}]$.  The proof of \eqref{eq:omegal0Htrchibar2} follows similarly.
\end{proof}

Now $(\Omega_{\circ}^{-2}\Omega^2 - 1)_{\ell=0}$ can be estimated in the region $r\leq r_0$.

\begin{proposition}[Estimate for $(\Omega_{\circ}^{-2}\Omega^2 - 1)_{\ell=0}$ for $r\leq r_0$] \label{prop:Omegal0Htrchibar}
	For any $u(r_0,v_{-1}) \leq u \leq u_f$ and any $v_{-1} \leq v \leq v(r_0,u)$, for $k = 0,1$, provided $r_0$ is sufficiently small,
	\begin{multline*}
		\Vert (\Omega^{-1} \nablaslash_3)^k \Big( \frac{\Omega^2}{\Omega_{\circ}^2} - 1 \Big)_{\ell=0} \Vert_{S_{u,v}}^2
		+
		\Vert (\Omega^{-1} \nablaslash_3)^k \Big( \frac{\Omega^2}{\Omega_{\circ}^2} - 1 \Big)_{\ell=0} \mathds{1} \Vert_{C_u(v)}^2
		+
		\Vert (\Omega^{-1} \nablaslash_3)^k \Big( \frac{\Omega^2}{\Omega_{\circ}^2} - 1 \Big)_{\ell=0} \mathds{1} \Vert_{\Cbar_v}^2
		\\
		+
		\Vert (\Omega^{-1} \nablaslash_3)^k \Big( \frac{\Omega^2}{\Omega_{\circ}^2} - 1 \Big)_{\ell=0} \mathds{1} \Vert_{\DRH(v)}^2
		\lesssim
		\frac{
		\varepsilon_0^2
		+
		\varepsilon^3
		}
		{v^2}
		+
		F_1[\Omega^{-2} (\Omega \tr \chibar - (\Omega \tr \chibar)_{\circ})_{\ell=0}](v),
	\end{multline*}
	where $\mathds{1} = \mathds{1}_{r\leq r_0}$.
\end{proposition}

\begin{proof}
	The proof is similar to that of Proposition \ref{prop:trchil0Homega1}, using now Proposition \ref{prop:omegal0Htrchibar} and equation \eqref{eq:DlogOmega}, which implies that,
	\[
		\Omega \nablaslash_4 ( \Omega_{\circ}^{-2} \Omega^2 - 1 )_{\ell=0}
		=
		2 \left( \Omega \omegahat - (\Omega \omegahat)_{\circ} \right)_{\ell=0}
		+
		\mathcal{E}^0_{\ell=0},
	\]
	along with the gauge conditions \eqref{eq:Hgaugelm1} and \eqref{eq:Hgaugelm8}.
\end{proof}

Similarly, $\Omega^{-2} (\Omega \tr \chibar - (\Omega \tr \chibar)_{\circ})_{\ell=0}$ can be estimated for $r\leq r_0$.

\begin{proposition}[Estimate for $\Omega^{-2} (\Omega \tr \chibar - (\Omega \tr \chibar)_{\circ})_{\ell=0}$ for $r\leq r_0$] \label{prop:trchibarl0Htrchibar}
	For any $u(r_0,v_{-1}) \leq u \leq u_f$ and any $v_{-1} \leq v \leq v(r_0,u)$, for $k = 0,1$, provided $r_0$ is sufficiently small,
	\begin{multline*}
		\Vert (\Omega^{-1} \nablaslash_3)^k
		\Omega^{-2} (\Omega \tr \chibar - (\Omega \tr \chibar)_{\circ})_{\ell=0}
		\Vert_{S_{u,v}}^2
		+
		\Vert (\Omega^{-1} \nablaslash_3)^k
		\Omega^{-2} (\Omega \tr \chibar - (\Omega \tr \chibar)_{\circ})_{\ell=0}
		\mathds{1} \Vert_{C_u(v)}^2
		\\
		+
		\Vert (\Omega^{-1} \nablaslash_3)^k
		\Omega^{-2} (\Omega \tr \chibar - (\Omega \tr \chibar)_{\circ})_{\ell=0}
		\mathds{1} \Vert_{\Cbar_v}^2
		+
		\Vert (\Omega^{-1} \nablaslash_3)^k
		\Omega^{-2} (\Omega \tr \chibar - (\Omega \tr \chibar)_{\circ})_{\ell=0}
		\mathds{1} \Vert_{\DRH(v)}^2
		\lesssim
		\frac{
		\varepsilon_0^2
		+
		\varepsilon^3
		}
		{v^2},
	\end{multline*}
	where $\mathds{1} = \mathds{1}_{r\leq r_0}$.
\end{proposition}

\begin{proof}
	Equation \eqref{eq:Ray} (see also equation \eqref{eq:trchibar3H}) implies
	\begin{equation} \label{eq:trchibarl0schematic}
		\Omega^{-1} \nablaslash_3
		\left( \Omega^{-2}
		\left(
		\Omega \tr \chibar - (\Omega \tr \chibar)_{\circ}
		\right)_{\ell=0}
		\right)
		-
		\frac{2}{r}
		\Omega^{-2}
		\left(
		\Omega \tr \chibar - (\Omega \tr \chibar)_{\circ}
		\right)_{\ell=0}
		=
		-
		\frac{4}{r} \Omega^{-2} \left(
		\Omega \omegabarhat - (\Omega \omegabarhat)_{\circ}
		\right)_{\ell=0}
		+
		\mathcal{E}^0.
	\end{equation}
	Using Lemma \ref{lem:nabla3xi} and Proposition \ref{prop:trchibarl0uf} for $k=0$, and the equation \eqref{eq:trchibarl0schematic} directly for $k=1$, it follows from Proposition \ref{prop:Omegal0Htrchibar} that
	\begin{multline*}
		\Vert (\Omega^{-1} \nablaslash_3)^k
		\Omega^{-2} (\Omega \tr \chibar - (\Omega \tr \chibar)_{\circ})_{\ell=0}
		\mathds{1} \Vert_{C_u(v)}^2
		\\
		+
		\Vert (\Omega^{-1} \nablaslash_3)^k
		\Omega^{-2} (\Omega \tr \chibar - (\Omega \tr \chibar)_{\circ})_{\ell=0}
		\mathds{1} \Vert_{\DRH(v)}^2
		\lesssim
		\frac{
		\varepsilon_0^2
		+
		\varepsilon^3
		}
		{v^2}
		+
		F_1[\Omega^{-2} (\Omega \tr \chibar - (\Omega \tr \chibar)_{\circ})_{\ell=0}](v),
	\end{multline*}
	since
	\[
		2 \Omega^{-2} \left( \Omega \omegabarhat - (\Omega \omegabarhat)_{\circ} \right)_{\ell=0}
		=
		\Omega^{-1} \nablaslash_3
		( \Omega_{\circ}^{-2}\Omega^2 - 1)_{\ell=0}
		+
		\mathcal{E}^0.
	\]
	The estimate for the $\Vert \cdot \Vert_{C_u(v)}$ and $\Vert \cdot \Vert_{\DRH(v)}$ norms follow after taking $r_0$ sufficiently small.  The estimate for the $\Vert \cdot \Vert_{\Cbar_v}$ and $\Vert \cdot \Vert_{S_{u,v}}$ norms are similar.
\end{proof}

Estimates for the higher order derivatives of the $\ell = 0$ modes of the quantities now easily follow.

\begin{proposition}[Estimate for higher order derivatives in $r\leq r_0$] \label{prop:l0modesr0}
	For
	\begin{align*}
		\Phi_{\ell=0}
		=
		\
		&
		(\Omega \tr \chi - (\Omega \tr \chi)_{\circ})_{\ell=0}
		,
		\Omega^{-2} (\Omega \tr \chibar - (\Omega \tr \chibar)_{\circ})_{\ell=0}
		,
		\left( \frac{\Omega^2}{\Omega_{\circ}^2} - 1 \right)_{\ell=0}
		,
		\left( \Omega \omegahat - (\Omega \omegahat)_{\circ} \right)_{\ell=0}
		,
		\\
		&
		\Omega^{-2} \left( \Omega \omegabarhat - (\Omega \omegabarhat)_{\circ} \right)_{\ell=0}
		,
		(\rho - \rho_{\circ})_{\ell=0}
		,
		\sigma_{\ell=0},
	\end{align*}
	for any $u(r_0,v_{-1}) \leq u \leq u_f$ and any $v_{-1} \leq v \leq v(r_0,u)$, for $\kbar = (0,k_1,k_2)$ with $k_1+k_2 \leq N-s$ and $s=0,1,2$,
	\[
		\Vert \mathfrak{D}^{\kbar}
		\Phi_{\ell=0}
		\Vert_{S_{u,v}}^2
		+
		\Vert \mathfrak{D}^{\kbar}
		\Phi_{\ell=0}
		\mathds{1} \Vert_{C_u(v)}^2
		+
		\Vert \mathfrak{D}^{\kbar}
		\Phi_{\ell=0}
		\mathds{1} \Vert_{\Cbar_v}^2
		+
		\Vert \mathfrak{D}^{\kbar}
		\Phi_{\ell=0}
		\mathds{1} \Vert_{\DRH(v)}^2
		\lesssim
		\frac{
		\varepsilon_0^2
		+
		\varepsilon^3
		}
		{v^{s}},
	\]
	where $\mathds{1} = \mathds{1}_{r\leq r_0}$.
\end{proposition}

\begin{proof}
	For $\Phi_{\ell=0} = \sigma_{\ell=0}$, the equation \eqref{eq:curletacurletabar} projected to $\ell=0$ takes the schematic form
	\[
		\sigma_{\ell=0}
		=
		\mathcal{E}^0_{\ell=0},
	\]
	and proof follows from applying $(\Omega^{-1} \nablaslash_3)^{k_1} (\Omega \nablaslash_4)^{k_2}$.
	
	For the remaining quantities, consider first the case that $k_2=0$.  The proof for $\Phi_{\ell=0} = (\Omega \tr \chi - (\Omega \tr \chi)_{\circ})_{\ell=0}, \left( \Omega \omegahat - (\Omega \omegahat)_{\circ} \right)_{\ell=0}, (\rho - \rho_{\circ})_{\ell=0}$ for $k_1\leq 2, k_2=0$ follows from Propositions \ref{prop:trchil0Homega1}--\ref{prop:trchibarl0Htrchibar}, as does the proof for $\Phi_{\ell=0} = \Omega^{-2} (\Omega \tr \chibar - (\Omega \tr \chibar)_{\circ})_{\ell=0}, \left( \frac{\Omega^2}{\Omega_{\circ}^2} - 1 \right)_{\ell=0}$ for $k_1\leq 1, k_2=0$.  For $k_1 = 0$, the proof for $\Omega^{-2} \left( \Omega \omegabarhat - (\Omega \omegabarhat)_{\circ} \right)_{\ell=0}$ follows from Proposition \ref{prop:Omegal0Htrchibar} and the fact that
	\[
		2 \Omega^{-2} \left( \Omega \omegabarhat - (\Omega \omegabarhat)_{\circ} \right)_{\ell=0}
		=
		\Omega^{-1} \nablaslash_3
		( \Omega_{\circ}^{-2}\Omega^2 - 1)_{\ell=0}
		+
		\mathcal{E}^0_{\ell=0}.
	\]
	The proof then follows inductively after applying $(\Omega^{-1} \nablaslash_3)^{k_1-1}$ to equations \eqref{eq:Ray}, \eqref{eq:trchi3}, \eqref{eq:omega3omegabar4}, \eqref{eq:rho3} projected to $\ell=0$, and commuting equation \eqref{eq:omega3omegabar4}, which takes the schematic form
	\[
		\partial_v \left(
		\Omega^{-2}
		\left(
		\Omega \omegabarhat - (\Omega \omegabarhat)_{\circ}
		\right)
		\right)_{\ell=0}
		+
		2 \Omega \omegahat
		\Omega^{-2}
		\left(
		\Omega \omegabarhat - (\Omega \omegabarhat)_{\circ}
		\right)_{\ell=0}
		=
		-(\rho - \rho_{\circ})_{\ell=0}
		+
		\frac{2M_f}{r^3} \left( 1 - \frac{\Omega_{\circ}^2}{\Omega^2} \right)_{\ell=0}
		+
		\mathcal{E}^0_{\ell=0},
	\]
	with $(\Omega^{-1} \nablaslash_3)^{k_1}$ and using Lemma \ref{lem:nabla4redxi}.
	
	For $k_2 \geq 1$, the proof again follows inductively from applying $(\Omega^{-1} \nablaslash_3)^{k_1} (\Omega \nablaslash_4)^{k_2-1}$ to the equations \eqref{eq:Ray}, \eqref{eq:trchibar4}, \eqref{eq:omega3omegabar4}, \eqref{eq:rho4} projected to $\ell=0$ and commuting equation \eqref{eq:omega3omegabar4} with $(\Omega^{-1} \nablaslash_3)^{k_1} (\Omega \nablaslash_4)^{k_2}$ and using Lemma \ref{lem:nabla3xi}.
\end{proof}

\subsubsection*{Estimates for $\ell = 0$ modes in the region $r_0 \leq r \leq R_2$}

The results of the previous section, namely Proposition \ref{prop:l0modesr0}, can be extended to the region $r_0 \leq r \leq R_2$.  The smallness of $\Omega_{\circ}^2$ can no longer be exploited, however the proof of Proposition \ref{prop:l0modesR} below is much simpler than that of Proposition \ref{prop:l0modesr0} since, for each $u$, the difference $v(R_2,u) - v(r_0,u)$ is uniformly bounded independently of $u$.

\begin{proposition}[Estimates for $\ell =0$ modes in $\DRH$] \label{prop:l0modesR}
	For
	\begin{align*}
		\Phi_{\ell=0}
		=
		\
		&
		(\Omega \tr \chi - (\Omega \tr \chi)_{\circ})_{\ell=0}
		,
		\Omega^{-2} (\Omega \tr \chibar - (\Omega \tr \chibar)_{\circ})_{\ell=0}
		,
		\left( \frac{\Omega^2}{\Omega_{\circ}^2} - 1 \right)_{\ell=0}
		,
		\left( \Omega \omegahat - (\Omega \omegahat)_{\circ} \right)_{\ell=0}
		,
		\\
		&
		\Omega^{-2} \left( \Omega \omegabarhat - (\Omega \omegabarhat)_{\circ} \right)_{\ell=0}
		,
		(\rho - \rho_{\circ})_{\ell=0}
		,
		\sigma_{\ell=0},
	\end{align*}
	for any $u_0 \leq u \leq u_f$ and any $\max\{ v_{-1}, v(r_0,u)\} \leq v \leq v(R_2,u)$, for $\kbar = (0,k_1,k_2)$ with $k_1+k_2 \leq N-s$ and $s=0,1,2$,
	\[
		\Vert \mathfrak{D}^{\kbar}
		\Phi_{\ell=0}
		\Vert_{S_{u,v}}^2
		+
		\Vert \mathfrak{D}^{\kbar}
		\Phi_{\ell=0}
		\Vert_{C_u(v)}^2
		+
		\Vert \mathfrak{D}^{\kbar}
		\Phi_{\ell=0}
		\Vert_{\Cbar_v}^2
		+
		\Vert \mathfrak{D}^{\kbar}
		\Phi_{\ell=0}
		\Vert_{\DRH(v)}^2
		\lesssim
		\frac{
		\varepsilon_0^2
		+
		\varepsilon^3
		}
		{v^{s}}.
	\]
\end{proposition}

\begin{proof}
	Define
	\begin{align*}
		\overset{(3)}{\Phi}_{\ell=0}
		=
		(\Omega \tr \chi - (\Omega \tr \chi)_{\circ})_{\ell=0}
		,
		\Omega^{-2} (\Omega \tr \chibar - (\Omega \tr \chibar)_{\circ})_{\ell=0}
		,
		\left( \frac{\Omega^2}{\Omega_{\circ}^2} - 1 \right)_{\ell=0}
		,
		\left( \Omega \omegahat - (\Omega \omegahat)_{\circ} \right)_{\ell=0}
		,
		(\rho - \rho_{\circ})_{\ell=0}
		,
		\sigma_{\ell=0},
	\end{align*}
	and
	\begin{align*}
		\overset{(4)}{\Phi}_{\ell=0}
		=
		\
		&
		(\Omega \tr \chi - (\Omega \tr \chi)_{\circ})_{\ell=0}
		,
		\Omega^{-2} (\Omega \tr \chibar - (\Omega \tr \chibar)_{\circ})_{\ell=0}
		,
		\left( \frac{\Omega^2}{\Omega_{\circ}^2} - 1 \right)_{\ell=0}
		,
		\\
		&
		\Omega^{-2} \left( \Omega \omegabarhat - (\Omega \omegabarhat)_{\circ} \right)_{\ell=0}
		,
		(\rho - \rho_{\circ})_{\ell=0}
		,
		\sigma_{\ell=0}.
	\end{align*}
	For each $\overset{(3)}{\Phi}_{\ell=0}$ and $\overset{(4)}{\Phi}_{\ell=0}$, by the equations \eqref{eq:Ray}, \eqref{eq:trchibar4}, \eqref{eq:trchi3}, \eqref{eq:omega3omegabar4}, \eqref{eq:DlogOmega}, \eqref{eq:rho4}, \eqref{eq:sigma4}, \eqref{eq:rho3}, \eqref{eq:sigma3} there is some function $h_{\overset{(3)}{\Phi}}(r)$ and $h_{\overset{(4)}{\Phi}}(r)$ respectively, both comparable to $1$ in the region $r_0 \leq r \leq R_2$, such that
	\[
		\Omega \nablaslash_3 \left( h_{\overset{(3)}{\Phi}}(r) \overset{(3)}{\Phi}_{\ell=0} \right)
		=
		\sum_{\overset{(4)}{\Phi}}
		h \cdot \overset{(4)}{\Phi}_{\ell=0}
		+
		\Omega^2 \mathcal{E}^0,
		\quad
		\text{and}
		\quad
		\Omega \nablaslash_4 \left( h_{\overset{(4)}{\Phi}}(r) \overset{(4)}{\Phi}_{\ell=0} \right)
		=
		\sum_{\overset{(3)}{\Phi}}
		h \cdot \overset{(3)}{\Phi}_{\ell=0}
		+
		\mathcal{E}^0.
	\]
	Consider first the case that $k_1=k_2 =0$.  For any $\lambda >0$, the first equations can be renormalised to give
	\[
		\Omega \nablaslash_3 \left( r^{-\lambda} h_{\overset{(3)}{\Phi}}(r) \overset{(3)}{\Phi}_{\ell=0} \right)
		-
		\frac{\lambda \Omega_{\circ}^2}{r} r^{-\lambda} h_{\overset{(3)}{\Phi}}(r) \overset{(3)}{\Phi}_{\ell=0}
		=
		r^{-\lambda}
		\sum_{\overset{(4)}{\Phi}}
		h \cdot \overset{(4)}{\Phi}_{\ell=0},
	\]
	and so, after contracting with $r^{-\lambda} h_{\overset{(3)}{\Phi}}(r) \overset{(3)}{\Phi}_{\ell=0}$,
	\begin{equation} \label{eq:Phi3l0estimate}
		\bigg\vert
		\partial_u \Big(
		\Big\vert r^{-\lambda} h_{\overset{(3)}{\Phi}}(r) \overset{(3)}{\Phi}_{\ell=0} \Big\vert^2
		\Big)
		-
		\frac{\lambda \Omega_{\circ}^2}{r}
		\Big\vert r^{-\lambda} h_{\overset{(3)}{\Phi}}(r) \overset{(3)}{\Phi}_{\ell=0} \Big\vert^2
		\bigg\vert
		\lesssim
		\sum_{\overset{(4)}{\Phi}}
		\Big\vert r^{-\lambda} \overset{(4)}{\Phi}_{\ell=0} \Big\vert^2
		+
		\vert \mathcal{E}^0 \vert^2,
	\end{equation}
	and similarly,
	\begin{equation} \label{eq:Phi4l0estimate}
		\bigg\vert
		\partial_v \Big(
		\Big\vert r^{-\lambda} h_{\overset{(4)}{\Phi}}(r) \overset{(4)}{\Phi}_{\ell=0} \Big\vert^2
		\Big)
		+
		\frac{\lambda \Omega_{\circ}^2}{r}
		\Big\vert r^{-\lambda} h_{\overset{(4)}{\Phi}}(r) \overset{(4)}{\Phi}_{\ell=0} \Big\vert^2
		\bigg\vert
		\lesssim
		\sum_{\overset{(3)}{\Phi}}
		\Big\vert r^{-\lambda} \overset{(3)}{\Phi}_{\ell=0} \Big\vert^2
		+
		\vert \mathcal{E}^0 \vert^2,
	\end{equation}
	where the constant implicit in $\lesssim$ is independent of $\lambda$.  Integrating over the spacetime region $u \leq u' \leq u_f$ and any $v \leq v' \leq v(R_{2},u)$ and summing, Propositions \ref{prop:trchil0uf}--\ref{prop:trchil0v0} give
	\begin{align*}
		\lambda
		\sum_{\Phi} \left\Vert 
		r^{- \lambda} \Phi_{\ell=0}
		\right\Vert^2_{\DRH(v)}
		+
		\sum_{\overset{(4)}{\Phi}}
		\big\Vert 
		r^{-\lambda} \overset{(4)}{\Phi}_{\ell=0}
		\big\Vert^2_{\Cbar_v}
		+
		\sum_{\overset{(3)}{\Phi}}
		\big\Vert 
		r^{-\lambda} \overset{(3)}{\Phi}_{\ell=0}
		\big\Vert^2_{C_u(v)}
		\lesssim
		C_{\lambda} \frac{
		\varepsilon_0^2
		+
		\varepsilon^3
		}
		{v^{2}}
		+
		\sum_{\Phi}
		\big\Vert 
		r^{- \lambda} \Phi_{\ell=0}
		\big\Vert^2_{\DRH(v)}
		,
	\end{align*}
	where Proposition \ref{prop:l0modesr0} is used, after averaging, to estimate the boundary terms on $r=r_0$.  After taking $\lambda$ sufficiently large, it follows that,
	\begin{align*}
		\sum_{\Phi} \left\Vert 
		\Phi_{\ell=0}
		\right\Vert^2_{\DRH(v)}
		+
		\sum_{\overset{(4)}{\Phi}}
		\big\Vert 
		\overset{(4)}{\Phi}_{\ell=0}
		\big\Vert^2_{\Cbar_v}
		+
		\sum_{\overset{(3)}{\Phi}}
		\big\Vert 
		\overset{(3)}{\Phi}_{\ell=0}
		\big\Vert^2_{C_u(v)}
		\lesssim
		\frac{
		\varepsilon_0^2
		+
		\varepsilon^3
		}
		{v^{2}}
		,
	\end{align*}
	Integrating the estimates \eqref{eq:Phi3l0estimate} over $u \leq u' \leq u_f$ and integrating \eqref{eq:Phi4l0estimate} over $v_{-1} \leq v' \leq v$ it follows that, for each $\overset{(3)}{\Phi}_{\ell=0}$ and $\overset{(4)}{\Phi}_{\ell=0}$ respectively,
	\[
		\big\Vert 
		\overset{(3)}{\Phi}_{\ell=0}
		\big\Vert^2_{S_{u,v}}
		+
		\big\Vert 
		\overset{(4)}{\Phi}_{\ell=0}
		\big\Vert^2_{S_{u,v}}
		\lesssim
		\frac{
		\varepsilon_0^2
		+
		\varepsilon^3
		}
		{v^{2}},
	\]
	which completes the proof for $k_1=k_2=0$ (note that, for each $u_0 \leq u \leq u_f$, $\vert v(R_2,u) - v(r_0,u) \vert \lesssim 1$ and so the estimates on the outgoing null hypersurfaces follow from the estimates on spheres).
	For $k_1 +k_2 \geq 1$ the proof follows inductively from commuting equations \eqref{eq:omega3omegabar4} to control $\Omega \omegahat - (\Omega \omegahat)_{\circ}$ and $\Omega^{-2} (\Omega \omegahat - (\Omega \omegahat)_{\circ})$, and using the equations directly to control the remaining quantities, as in the proof of Proposition \ref{prop:l0modesr0}.
\end{proof}

\subsection{Estimates for metric components in $\DRH$}
\label{subsec:Hmetric}

The main result of this section is the following estimate for the metric components.

\begin{theorem}[Improving bootstrap assumptions for energy of metric components] \label{prop:metricHenergy}
	The energy for the metric components, defined in Section \ref{PhiHpenergysec},
	satisfies,
	\[
		\mathbb{E}^N[g^{\Hp}]
		\lesssim
		\varepsilon_0^2
		+
		\varepsilon^3.
	\]
\end{theorem}

\begin{proof}
	The proof is a direct consequence of the estimates for $b$ of Proposition \ref{prop:bH}, the estimates for $\gslash - r^2 \mathring\gamma$ of Proposition \ref{prop:gslashH}, and the estimates for the mode differences $Y^1_m - \mathring{Y}^1_m$ of Proposition \ref{prop:modesdiffH}.
\end{proof}

First, the metric component $b$ is estimated.

\begin{proposition}[Estimate for $b - b_{\mathrm{Kerr}}$ in $\DRH$] \label{prop:bH}
	For $s=0,1,2$ and $\vert \gamma \vert \leq N-s$, for any $u_0 \leq u \leq u_f$ and any $v_{-1} \leq v \leq v(R_2,u)$,
	\[
		\Vert \Omega^{-2} \mathfrak{D}^{\gamma} (b - b_{\mathrm{Kerr}}) \Vert_{S_{u,v}}^2
		+
		\Vert \mathfrak{D}^{\gamma} (b - b_{\mathrm{Kerr}}) \Vert_{\Cbar_v}^2
		\lesssim
		\frac{\varepsilon_0^2 + \varepsilon^3}{v^s},
	\]
	and
	\[
		\Vert \mathfrak{D}^{\gamma} (b - b_{\mathrm{Kerr}}) \Vert_{C_u(v)}^2
		+
		\Vert \mathfrak{D}^{\gamma} (b - b_{\mathrm{Kerr}}) \Vert_{\DRH(v)}^2
		\lesssim
		\frac{\varepsilon_0^2 + \varepsilon^3}{v^s}.
	\]
\end{proposition}

\begin{proof}
	Equation \eqref{eq:b3} implies that
	\[
		\partial_u ( b^A - b_{\mathrm{Kerr}}^A)
		=
		2\Omega^2 \left(\eta^A-\underline{\eta}^A\right)
		-
		2\Omega^2_{\circ} ( \eta_{\mathrm{Kerr}}^A-\underline{\eta}_{\mathrm{Kerr}}^A )
	\]
	The proof then follows from Lemma \ref{lem:nabla3xi}, Propositions \ref{prop:nabla4etaH}, \ref{prop:nabla4etabarH} and \ref{prop:etaetabarl1H}, and the fact that $b \equiv 0$ on $u=u_f$.  The $\Omega^{-2}$ factor in the first estimate can indeed be included by using the fact that, for example,
	\[
		\Vert \mathfrak{D}^{\gamma}(\eta - \eta_{\mathrm{Kerr}}) \Vert^2_{\Cbar_v(u)}
		\lesssim
		\sup_{u\leq u'\leq u_f} \Vert \mathfrak{D}^{\gamma}( \eta - \eta_{\mathrm{Kerr}}) \Vert^2_{S_{u',v}}
		\int_u^{u_f} \Omega(u',v)^2 du'
		\lesssim
		\Omega(u,v)^2 \frac{\varepsilon_0^2 + \varepsilon^3}{v^s}.
	\]
\end{proof}

Next, the difference between the induced metric on $S_{u,v}$ and the round metric, $\gslash - r^2 \gamma$, is estimated.

\begin{proposition}[Estimate for $\gslash - r^2 \mathring\gamma$ in $\DRH$] \label{prop:gslashH}
	For any $\vert \gamma \vert \leq N$, for any $u_0 \leq u \leq u_f$ and any $v_{-1} \leq v \leq v(R_2,u)$,
	\[
		\Vert \mathfrak{D}^{\gamma} (\gslash - r^2 \mathring\gamma) \Vert_{S_{u,v}}^2
		+
		\Vert \mathfrak{D}^{\gamma} (\gslash - r^2 \mathring\gamma) \Vert_{\Cbar_v}^2
		\lesssim
		\frac{\varepsilon_0^2 + \varepsilon^3}{v}.
	\]
\end{proposition}

\begin{proof}
	Proposition \ref{thm:inheriting}, Theorem \ref{thm:Iestimates} and Propositions \ref{prop:chihatHn3aspacetime}, \ref{prop:rhoH1}, \ref{prop:rhol1H}, \ref{prop:l0modesR} imply that, on the sphere $S_{u_f,v(R,u_f)}$, for $k \leq N$,
	\[
		\Vert (r\nablaslash)^k (\gslash -  r^2 \mathring\gamma)\Vert_{S_{u_f,v(R,u_f)}}^2
		\lesssim
		v(R,u_f)^{-1}( \varepsilon_0^2 + \varepsilon^3). 
	\]
	Now, on $u=u_f$,
	\[
		\Omega \nablaslash_4( \gslash - r^2 \mathring\gamma)
		=
		(\Omega \tr \chi - \Omega \tr \chi_{\circ}) r^2 \mathring\gamma
		+
		\Omega \hat{\chi}\times r^2 \mathring\gamma
		+
		r^2 \mathring\gamma \times \Omega \hat{\chi},
	\]
	(recalling the notation of Section \ref{nullframesopers}) and so, in the notation of Lemma \ref{lem:commutation},
	\[
		\Omega \nablaslash_4( \tr (\gslash - r^2 \mathring\gamma))
		=
		2(\Omega \tr \chi - \Omega \tr \chi_{\circ})
		+
		\tr
		\Big(
		(\Omega \tr \chi - \Omega \tr \chi_{\circ}) (r^2 \mathring\gamma - \gslash)
		+
		\Omega \hat{\chi}\times (r^2 \mathring\gamma - \gslash)
		+
		(r^2 \mathring\gamma - \gslash) \times \Omega \hat{\chi}
		\Big),
	\]
	\begin{multline*}
		\Omega \nablaslash_4 \big( 
		r^2 \divslash \divslash (\gslash - r^2 \mathring\gamma) + 2 r^2 \rho
		\big)
		=
		r^2 \divslash \divslash \Big(
		(\Omega \tr \chi - \Omega \tr \chi_{\circ}) r^2 \mathring\gamma
		-
		\Omega \hat{\chi} \times (\gslash - r^2 \mathring\gamma)
		+
		(\gslash - r^2 \mathring\gamma) \times \Omega \hat{\chi}
		\Big)
		\\
		-
		2 r \Omega_{\circ}^2 \divslash \etabar
		+
		r^2 \Deltaslash \Omega \tr \chi
		-
		r^2 \Omega \tr \chi \rho
		+
		\mathcal{E}^1
		+
		\tr F_4 [r\divslash (\gslash - r^2 \mathring\gamma)]
		+
		r\divslash \tr F_4 [\gslash - r^2 \mathring\gamma],
	\end{multline*}
	and
	\begin{multline*}
		\Omega \nablaslash_4 \big( 
		r^2 \curlslash \divslash (\gslash - r^2 \mathring\gamma) - 2 r^2 \sigma
		\big)
		=
		r^2 \curlslash \divslash \Big(
		(\Omega \tr \chi - \Omega \tr \chi_{\circ}) r^2 \mathring\gamma
		-
		\Omega \hat{\chi} \times (\gslash - r^2 \mathring\gamma)
		+
		(\gslash - r^2 \mathring\gamma) \times \Omega \hat{\chi}
		\Big)
		\\
		-
		2 r \Omega_{\circ}^2 \curlslash \etabar
		+
		r^2 \Omega \tr \chi \sigma
		+
		\mathcal{E}^1
		+
		\tr F_4 [r\curlslash (\gslash - r^2 \mathring\gamma)]
		+
		r\curlslash \tr F_4 [\gslash - r^2 \mathring\gamma].
	\end{multline*}
	Note that, for any $\xi$,
	\[
		\Vert \xi \Vert_{L^1(S_{u_f,v})}
		\lesssim
		\Vert \xi \Vert_{L^1(S_{u_f,v(R,u_f)})}
		+
		\int_v^{v(R_2,u_f)}
		\Vert \Omega \nablaslash_4 \xi \Vert_{L^1(S_{u_f,v'})}
		dv'.
	\]
	Proposition \ref{prop:trchil1uf}, Proposition \ref{prop:trchil0uf} and the sharper estimate of Proposition \ref{prop:trchiuf2} imply that,
	\[
		\sum_{k \leq N}
		\Big(
		\int_v^{v(R_2,u_f)}
		\Vert (r\nablaslash)^k (\Omega \tr \chi - \Omega \tr \chi_{\circ}) \Vert_{L^1(S_{u_f,v'})}
		dv'
		\Big)^2
		\lesssim
		\frac{\varepsilon_0^2 + \varepsilon^3}{v(R,u_f)} \Big(
		\int_v^{v(R_2,u_f)}
		\Omega
		dv'
		\Big)^2
		+
		\frac{\varepsilon_0^2 + \varepsilon^3}{v}
		\lesssim
		\frac{\varepsilon_0^2 + \varepsilon^3}{v},
	\]
	and so it follows that
	\[
		\sum_{k \leq N} \Vert (r\nablaslash)^k \tr (\gslash - r^2 \mathring\gamma) \Vert_{S_{u_f,v}}^2
		\lesssim
		\frac{\varepsilon_0^2 + \varepsilon^3}{v}.
	\]
	
	Note now that, for any $\xi$ and any $\delta >0$,
	\[
		\Vert \xi \Vert_{S_{u_f,v}}^2
		\lesssim
		\Vert \xi \Vert_{S_{u_f,v(R,u_f)}}^2
		+
		v^{-\delta}
		\int_v^{v(R_2,u_f)}
		(v')^{1+\delta}
		\Vert \Omega \nablaslash_4 \xi \Vert_{S_{u_f,v'}}^2
		dv'.
	\]
	Propositions \ref{prop:etabarufR}, \ref{prop:etal1uf}, \ref{prop:sigmaH}, \ref{prop:sigmal1H}, \ref{prop:angularrhoufR}, \ref{prop:rhol1uf} and \ref{prop:l0modesR} imply that
	\[
		\int_v^{v(R_2,u_f)}
		(v')^{1+\delta}
		\big(
		\sum_{k \leq N-1}
		\Vert (r\nablaslash)^k \etabar \Vert_{S_{u_f,v'}}^2
		+
		\sum_{k \leq N-2}
		\big(
		\Vert (r\nablaslash)^k \rho \Vert_{S_{u_f,v'}}^2
		+
		\Vert (r\nablaslash)^k \sigma \Vert_{S_{u_f,v'}}^2
		\big)
		\big)
		dv'
		\lesssim
		\frac{\varepsilon_0^2 + \varepsilon^3}{v^{1-\delta}},
	\]
	and so
	\[
		\sum_{k \leq N-2}
		\Big(
		\Big\Vert (r\nablaslash)^k \big( r^2 \divslash \divslash (\gslash - r^2 \mathring\gamma) +2 r^2 \rho \big) \Big\Vert_{S_{u_f,v}}^2
		+
		\Big\Vert (r\nablaslash)^k \big( r^2 \curlslash \divslash (\gslash - r^2 \mathring\gamma) - 2 r^2 \sigma \big) \Big\Vert_{S_{u_f,v}}^2
		\Big)
		\lesssim
		\frac{\varepsilon_0^2 + \varepsilon^3}{v}.
	\]
	It then follows from Propositions~\ref{prop:ellipticestimates} and~\ref{prop:divcurl}, and 
	Propositions~\ref{prop:sigmaH},~\ref{prop:sigmal1H},~\ref{prop:angularrhoufR},~\ref{prop:rhol1uf} and~\ref{prop:l0modesR} that, on $C_{u_f}$,
	\[
		\sum_{k \leq N}
		\Vert (r\nablaslash)^k (\gslash - r^2 \mathring\gamma) \Vert_{S_{u_f,v}}^2
		\lesssim
		\frac{\varepsilon_0^2 + \varepsilon^3}{v}.
	\]

	The proof, when $\mathfrak{D}^{\gamma} = (r\nablaslash)^k$, then follows from commuting the equation
	\begin{equation} \label{eq:gslash3}
		\Omega \nablaslash_3( \gslash - r^2 \mathring\gamma)
		=
		(\Omega \tr \chibar - \Omega \tr \chibar_{\circ}) r^2 \mathring\gamma
		+
		\Omega \hat{\chibar} \times r^2 \mathring\gamma
		+
		r^2 \mathring\gamma \times \Omega \hat{\chibar},
	\end{equation}
	and integrating backwards from $C_{u_f}$, using Lemma \ref{lem:nabla3xi} and Propositions \ref{prop:chibarhatH2}, \ref{prop:trchibarH1}, \ref{prop:trchitrchibarl1H} and \ref{prop:l0modesR}.  The proof for $\mathfrak{D}^{\gamma}$ otherwise follows from directly using the equations \eqref{eq:gslash3} and
	\[
		\Omega \nablaslash_4( \gslash - r^2 \mathring\gamma)_{AB}
		=
		(\Omega \tr \chi - \Omega \tr \chi_{\circ}) r^2 \mathring\gamma_{AB}
		+
		\Omega {\hat{\chi}_A}^C \cdot r^2 \mathring\gamma_{CB}
		+
		\Omega {\hat{\chi}_B}^C \cdot r^2 \mathring\gamma_{CA}
		-
		e_A(b^C) r^2 \mathring\gamma_{CB}
		-
		e_B(b^C) r^2 \mathring\gamma_{AC}.
	\]
\end{proof}

Proposition \ref{prop:gslashH} and Proposition \ref{prop:modesdifference} immediately yield the following estimate on $Y^1_m - \mathring{Y}^1_m$.

\begin{proposition}[Estimates for $Y^1_{m} -\mathring{Y}^1_{m}$] \label{prop:modesdiffH}
	For all $u_{-1} \leq u \leq u_f$, $v_{-1} \leq v \leq v(R_2,u)$, the differences $Y^1_{m} -\mathring{Y}^1_{m}$ satisfy, for $m=-1,0,1$,
	\[
		\sum_{k=0}^{N+1} \Vert (r \nablaslash)^k ( Y^1_{m} -\mathring{Y}^1_{m} ) \Vert_{S_{u,v}}
		\lesssim
		\frac{\varepsilon_0^2 + \varepsilon^3}{v}
		.
	\]
\end{proposition}

Proposition \ref{prop:modesdiffH} completes the proof of Proposition \ref{prop:metricHenergy} and hence of Theorem \ref{thm:Hestimates}.

\part{Finishing the proof of the main theorem}
\label{conclusionpart}

In this final part of the paper, 
we shall obtain the final subtheorems necessary to complete the proof 
of Theorem~\ref{thm:main}.

\setcounter{parttocdepth}{1}
\parttoc

In {\bf Chapter~\ref{teleoffingchapter}}, we shall at last address the issue of the existence
of the teleological gauges necessary to complete the openness argument. 
In {\bf Chapter~\ref{conclusionsection}}, we shall 
obtain the final subtheorems of the proof of Theorem~\ref{thm:main}, concerning
the existence of the limiting gauges, and the properties
of the event horizon $\mathcal{H}^+$ and future null infinity $\mathcal{I}^+$.

\vskip1pc

\emph{This part may in principle be read independently of Part~\ref{improvingpart}, though some of the proofs will refer back
to arguments from there. Chapters~\ref{teleoffingchapter} and~\ref{conclusionsection} 
may be read completely independently of one another.}

\chapter{The construction of the teleological gauges and higher order estimates}
\label{teleoffingchapter}

This chapter is concerned with the proofs of Theorem~\ref{thm:newgauge}, Theorem~\ref{thm:lambda} and 
Theorem~\ref{thehigherordertheorem}, first stated in Sections~\ref{opensection} and~\ref{higherordersec}.

\setcounter{minitocdepth}{3}

\minitoc

In {\bf Section~\ref{section:newgauges}}, we  give the proof  of Theorem~\ref{thm:newgauge}, on the existence of $\hat{u}_f+\delta$ normalised gauges. 
We then give in {\bf Section~\ref{section:monotonicityR}} the proof of Theorem~\ref{thm:lambda}, concerning the monotonicity properties of the set $\mathfrak{R}(\hat{u}_f+\delta)$.
Finally, in {\bf Section~\ref{proofofhigherorderestimatessec}}, we give 
the proof of Theorem~\ref{thehigherordertheorem}, concerning estimates for higher order derivatives.
\vskip1pc

\emph{The three sections of this chapter may be read independently, although Section~\ref{section:monotonicityR} 
will refer to some estimates from Section~\ref{section:newgauges}. For the construction of
Section~\ref{section:newgauges}, the reader should compare with the
construction of the future normalised pure gauge solutions in~\cite{holzstabofschw} through solving of elliptic equations on 
spheres and transport equations. Here, a very analogous linear analysis must be supplemented with
a nonlinear iteration. For the convenience of the reader, we will explicitly distill the analogous purely linear statements
in the language of~\cite{holzstabofschw} in Remarks~\ref{linearandnonlinearremarkhere} and~\ref{sameasbeforebutforthehorizongauge}. These will in particular prove Propositions~\ref{proplinIpgauge}
and~\ref{proplinHpgauge} from Chapter~\ref{almostgaugeandtellychapter}. We will also explain how to distill the proof of Theorem~\ref{existenceofanchoredgaugethe} of Chapter~\ref{thelocaltheorysec} in 
Remarks~\ref{rmk:uf0Igaugeexistence0} and~\ref{theotheronexiststoo}.}

\section{Existence of anchored \texorpdfstring{$\hat{u}_f+\delta$}{TEXT} normalised gauges: the proof of Theorem \ref{thm:newgauge}}
\label{section:newgauges}

The main result of this section is Theorem \ref{thm:newgauge}, on the existence of the teleological gauges, which we restate here.

\newgauge*

The proof relies first on, for any given $\hat{u}_f$ in the bootstrap set $\mathfrak{B}$, extending the $\hat{u}_f$-normalised $\I$ and $\Hp$ gauges.  In {\bf Section~\ref{firststephere}} the definition of such extended gauges is given, and a theorem on the existence of such extensions is stated.  In {\bf Section~\ref{offing}}, the part of the proof of Theorem \ref{thm:newgauge} concerning the existence of $\hat{u}_f + \delta$ normalised $\I$ gauges is given, along with the properties of the functions $M_f$ and ${\bf J}$.  In {\bf Section~\ref{offing2}} the proof of the existence of $\hat{u}_f + \delta$ normalised $\Hp$ gauges is given, thus completing the proof of Theorem \ref{thm:newgauge}.  (Note that $\delta$ will here be used as a variable and thus in this section $\delta$ no longer denotes \eqref{deltadefassmallness}.)

\subsection{Extended $\hat{u}_f$ normalised gauges}

\label{firststephere}

This section is concerned with Theorem \ref{thm:extendedgauges}, stated below, which guarantees that the $\hat{u}_f$-normalised $\I$ and $\Hp$ gauges can be extended for some small time.  First, definitions of appropriate extensions of the $\hat{u}_f$-normalised $\I$ and $\Hp$ gauges are stated.

Recall the definition of the set $\mathcal{W}_{\mathcal{I}^+}$ from Chapter \ref{almostgaugeandtellychapter} (see \eqref{WI+}).  
For general $\tilde{u}_f$ and $\varepsilon_0$, let us define 
$v_\infty(\tilde{u}_f,\varepsilon_0)=\varepsilon_0^{-2}(\tilde{u}_f)^{100}$. 
Given $u_f$, $\epsilon_0$, and $M_f$, 
for $\delta_1\geq 0$, define\index{teleological $\I$ gauge!sets!$\mathcal{W}_{\I}(u_f,\delta_1)$, extended coordinates domain}
\begin{equation} \label{WI+ext}
	\mathcal{W}_{\I}(u_f,\delta_1)
	=
	\{ u_{-1} - \delta_1 \le  u \le u_f + \delta_1 \} \cap \{v(R_{-2},u)\le v \le v_\infty(u_f+\delta_1)  \},
\end{equation}
and define the corresponding domain\index{teleological $\I$ gauge!sets!$\mathcal{Z}_{\I}(u_f,\delta_1)$, extended domain of parametrisation}
\begin{equation} \label{actualIdomainext}
	\mathcal{Z}_{\I}(u_f,\delta_1):=\mathcal{W}_{\mathcal{I}^+} (u_f,\delta_1)\times \mathbb S^2.
\end{equation}
(Here $v(R_{-2},u)$ is defined with respect to $M_f$. These sets depend in addition on our choice of $M_f$  and $\varepsilon_0$ but we suppress this from the notation.)
Note that $\mathcal{W}_{\mathcal{I}^+}(u_f,0) = \mathcal{W}_{\mathcal{I}^+}(u_f)$ and $\mathcal{Z}_{\mathcal{I}^+}(u_f,0) = \mathcal{Z}_{\mathcal{I}^+}(u_f)$.
The following definition extends Definition~\ref{Igaugedefinition} of a $u_f$-normalised $\I$ gauge to allow for a larger domain $\mathcal{W}_{\mathcal{I}^+}(u_f,\delta_1)$ with $\delta_1>0$.

\begin{definition} \label{def:Igaugeextension}
	Given $u_f$, $M_f$, $\varepsilon_0$ and $\delta_1 \geq 0$, let $\mathcal{W}_{\mathcal{I}^+}(u_f,\delta_1)$ and $\mathcal{Z}_{\mathcal{I}^+}(u_f,\delta_1)$ be defined by~\eqref{WI+ext} and \eqref{actualIdomainext} respectively.  We say that a metric $g$  in the form~\eqref{doublenulllongforminterchanged} defined on the domain $\mathcal{Z}_{\mathcal{I}^+}(u_f,\delta_1)$ and solving the Einstein vacuum equations $(\ref{Ricciflathere})$
is {\bf expressed in $\delta_1$ extended $u_f$ normalised $\mathcal{I}^+$ gauge} if 
for all $(u,v)\in \mathcal{W}_{\I}(u_f,\delta_1)$,
the induced metric on the spheres $S_{u,v}$, 
satisfies the roundness condition~\eqref{closenesstoround} with $\psi={\rm id}$, i.e.~the
assumptions of both Proposition~\ref{withclosenesstoroundprop} and Lemma~\ref{littlelemmahere} hold,
and projections to spherical harmonics are thus defined, and
 the following relations hold:
\begin{itemize}
	\item
		$\frac{1}{2} (r^3 \rho_{\ell =0}) (u_f,v_{\infty}(u_f))=-M_f $;
	\item 
		 $b(u,v_\infty,\theta)=0$ for all $u\in [u_{-1} - \delta_1,u_f + \delta_1]$, $\theta \in \mathbb S^2$; 
	\item
		$\left(\Omega \tr \chi - (\Omega \tr \chi)_{\circ} \right)_{\ell \neq 1}(u_f,v_{\infty}(u_f),\theta)=0$ for all $\theta \in \mathbb S^2$;
	\item
		$\left(\divslash \Omega\beta \right)_{\ell = 1} (u_f,v_{\infty}(u_f),\theta)=0$ for all $\theta \in \mathbb S^2$;
	\item
		$\left(\Omega^{-1} \tr \chibar - (\Omega^{-1} \tr \chibar)_{\circ} \right)(u_f,v_{\infty}(u_f),\theta)=0$ for all $\theta \in \mathbb S^2$;
	\item
		$\mu_{\ell\geq 1}(u,v_{\infty}(u_f),\theta)=0$ for all $u \in [u_{-1} - \delta_1,u_f+ \delta_1]$, $\theta \in
		\mathbb S^2$;
	\item
		$ \underline{\mu}^{\dagger}_{\ell \geq 1}(u_{-1},v,\theta) =0$ for all $v\in [v(R_{-2},u_{-1}),v_{\infty}(u_f+ \delta_1)]$, $\theta \in \mathbb S^2$;
	\item
		$\left(\Omega^2 - \Omega_{\circ}^2 \right)_{\ell=0}(u,v_{\infty}(u_f))=0$ for all $u \in [u_{-1} - \delta_1,u_f+ \delta_1]$;
	\item
		$(\Omega \omegahat - (\Omega \omegahat)_{\circ})_{\ell=0} (u_{-1},v)= F(u_{-1}) \frac{\Omega_{\circ}^2}{r^3}(u_{-1},v)$ for all $v \in [v(R_{-2},u_{-1}),v_{\infty}(u_f+ \delta_1)]$, where
		\[
			F(u)
			:=
			\frac{1}{2} \int_{u}^{u_f} \int_{\bar{u}}^{u_f}
			r^3 (\Omega \hat{\chi}, \alphabar)_{\ell=0}
			\left(\hat{u},v_\infty(u_f)\right)
			d\hat{u}
			d\bar{u},
		\]
\end{itemize}
where $\underline{\mu}^{\dagger}$ is defined in \eqref{renormmassaspectdf}.
\end{definition}

Note that Definition \ref{def:Igaugeextension} coincides with Definition \ref{Igaugedefinition} if $\delta_1=0$.

The following definition similarly extends Definition \ref{Hgaugedefinition} of a $u_f$-normalised $\Hp$ gauge to allow for a larger domain.  Recall the definition of $\mathcal{W}_{\Hp}$ from Chapter \ref{almostgaugeandtellychapter} (see \eqref{WH+}).  Given $u_f$, $M_f$, for $\delta_1\geq 0$ define\index{teleological $\Hp$ gauge!sets!$\mathcal{W}_{\Hp}(u_f,\delta_1)$, extended coordinates domain}
\begin{equation} \label{WH+ext}
	\mathcal{W}_{\Hp}(u_f,\delta_1)
	=
	\{u_0\le u \le u_f + \delta_1 \} \cap \{v_{-1} - \delta_1 \le v \le v(R_2 + \delta_1, u)\},
\end{equation}
and the corresponding domain\index{teleological $\Hp$ gauge!sets!$\mathcal{Z}_{\Hp}(u_f,\delta_1)$, extended coordinates domain}
\begin{equation} \label{actualHpdomainext}
	\mathcal{Z}_{\mathcal{H}^+}(u_f,\delta_1)
	=
	\mathcal{W}_{\Hp}(u_f,\delta_1)\times \mathbb S^2.
\end{equation}
Again, these domains depend also on $M_f$ to define $v(R_2+\delta_1,u)$ but we omit this dependence
from the notation.

\begin{definition} \label{def:Hgaugeextension}
	Given $u_f$, $M_f$ and $\delta_1 \geq 0$, 
	let $\mathcal{W}_{\mathcal{H}^+}(u_f,\delta_1)$ and $\mathcal{Z}_{\mathcal{H}^+}(u_f,\delta_1)$ be defined by~\eqref{WH+ext} and \eqref{actualHpdomainext} respectively.
We say that a metric $g$  in the
form~\eqref{doublenulllongform} defined on the domain $\mathcal{Z}_{\mathcal{H}^+}(u_f,\delta_1)$
and solving the Einstein vacuum equations $(\ref{Ricciflathere})$
is {\bf expressed in $\delta_1$ extended $u_f$ normalised $\Hp$ gauge} if 
for all $(u,v)\in \mathcal{W}_{\Hp}(u_f,\delta_1)$,
the induced metric on the spheres $S_{u,v}$, 
satisfies the roundness condition~\eqref{closenesstoround} with $\psi={\rm id}$, i.e.~the
assumptions of both Proposition~\ref{withclosenesstoroundprop} and Lemma~\ref{littlelemmahere} hold,
and projections to spherical harmonics are thus defined, and
 the following relations hold:
\begin{itemize}
\item
$b(u_f,v,\theta)=0$ for all $v\in[v_{-1} - \delta_1, v(R_2,u_f+\delta_1)]$,  $\theta \in \mathbb S^2$;
	\item
		$\mu^*_{\ell \geq 1} (u_f,v(R,u_f),\theta) = 0$ for all  $\theta \in \mathbb S^2$;
	\item
		$\left( \Omega \tr \chi - (\Omega \tr \chi)_{\circ} \right)_{\ell=0} (u_f,v(R,u_f)) = 0$;
	\item
		$\left( \Omega \tr \chibar - (\Omega \tr \chibar)_{\circ} \right)(u_f,v_{-1}, \theta) = 0$ for all $\theta \in S^2$;
	\item
		$\Omega(u_f,v,\theta) = \Omega_{\circ} (u_f,v,\theta)$ for all  $v\in[v_{-1} - \delta_1, v(R_2,u_f+\delta_1)]$, $\theta \in \mathbb S^2$;
	\item
		$\partial_u \left( r^3 (\divslash \eta)_{\ell\geq 1} + r^3 \rho_{\ell \geq 1} \right) (u,v_{-1},\theta) = 0$ for all
		$u\in [u_0,u_f+\delta_1]$,  $\theta \in \mathbb S^2$;
	\item
		$\Omega(u,v_{-1})_{\ell=0} = \Omega_{\circ} (u,v_{-1})$ for all $u \in[u_0,u_f+\delta_1]$,
\end{itemize}
where $\mu^*$ is defined in~\eqref{otherrenormmassaspectdf}.
\end{definition}

The following theorem guarantees that the $\hat{u}_f$-normalised $\I$ and $\Hp$ gauges can be indeed extended for some small time $\delta_1$.

\begin{theorem}[Extension of the $\hat{u}_f$ normalised gauges]
\label{thm:extendedgauges}
	Under the assumptions of Theorem~\ref{thm:justopen}, for all $\lambda\in \mathfrak{R}(\hat{u}_f)$, the following holds:

	There exists some small $\delta_1>0$ and $\delta_1$ extensions of the $\hat{u}_f$-normalised $\I$ and $\Hp$ gauges, in the sense of Definition \ref{def:Igaugeextension} and Definition \ref{def:Hgaugeextension} respectively, with $\varepsilon_0$ as in the statement of Theorem~\ref{thm:justopen} and $M_f=M_f(\hat{u}_f, \lambda)$ given by Definition~\ref{bootstrapsetdef}, 
	satisfying the anchoring conditions of Definition~\ref{anchoringdef},
	where, in place of the inclusions \eqref{eq:overlap1} and \eqref{eq:overlap2}, the domains $\DRH = \mathcal{D}^{\Hp}_{\hat{u}_f + \delta_1}$ and $\DRI = \DRI_{\hat{u}_f + \delta_1}$ of the $\delta_1$ extensions of the $\hat{u}_f$-normalised $\I$ and $\Hp$ gauges satisfy the inclusions,
	\begin{align}
		\label{impoverlapext1}
		\mathcal{D}^{\Hp}_{r\ge R_{-1}}&\subset \mathcal{D}_{r\ge R_{-\frac74}}^{\I},
		\\
		\label{impoverlapext2}
		\mathcal{D}^{\I}_{r \le R_{1}} \cap J^+(S^{\Hp}_{u_0,v_0}) &\subset \mathcal{D}_{r\le R_{\frac74}}^{\Hp}.
	\end{align}
	Moreover, the null cones of the $\delta_1$ extensions of the $\hat{u}_f$-normalised $\I$ and $\Hp$ gauges satisfy
	\begin{align}
		\bigcup_{v_{-1} \leq v \leq v_2} \Cbar_{v}^{\Hp}
		&
		\subset
		\mathcal{D}^{\mathcal{K}}(V_3) \cap \{ V_{-1} - 2C\varepsilon \leq V_{data} \leq V_{-1}+ 2C \varepsilon \}
			\cap \{U_0 -2C\varepsilon\leq U_{data} \leq 2C\varepsilon \} ,
		\label{impoverlapext3}
		\\
		\bigcup_{u_{-1} \leq u \leq u_2} C_{u}^{\I}
		&
		\subset
		\mathcal{D}^{\mathcal{EF}}(u_3)\cap   \{ u_{-1} - 2C \varepsilon \leq u_{data} \leq u_{2} + 2C \varepsilon \},
		\label{impoverlapext5}
	\end{align}
	where the constant $C$ is as in Theorem \ref{havetoimprovethebootstrap}, and the geometric quantities of the $\hat{u}_f$ 
	normalised $\I$ gauge and the $\hat{u}_f$ normalised $\Hp$ gauge satisfy
	\begin{equation}\label{eq:lowerorderpointwiseext}
		\sup_{\mathcal{D}^{\Hp}_{\hat{u}_f+\delta_1}}
		|r^{-2} \slashed g^{\mathcal{H}^+}-\mathring\gamma|_{\mathring \gamma}
		+ 
		\sup_{\mathcal{D}^{\I}_{\hat{u}_f+\delta_1}}
		r
		|r^{-2}\slashed g^{\mathcal{I}^+} -\mathring\gamma|_{\mathring \gamma} 
		+
		\sup_{\mathcal{D}^{\Hp}_{\hat{u}_f+\delta_1}}
		|\Omega^{-2}_\circ \Omega^2_{\mathcal{H}^+}-1|
		+
		\sup_{\mathcal{D}^{\I}_{\hat{u}_f+\delta_1}}
		|r(\Omega^{-2}_{\circ}\Omega^2_{\mathcal{I}^+}-1)| 
		\leq
		\frac{3}{4} \sqrt{\varepsilon},
	\end{equation}
	\begin{multline} \label{improvement2ext}
		\mathbb{E}^{N-2}_{\hat{u}_f+\delta_1}[P_{\Hp},\Pbar_{\Hp}]
		+
		\mathbb{E}^{N-2}_{\hat{u}_f+\delta_1}[P_{\I},\check{\Pbar}_{\I}]
		+
		\mathbb{E}^{N}_{\hat{u}_f+\delta_1}[\alpha_{\Hp},\alpha_{\I}]
		+
		\mathbb{E}^{N}_{\hat{u}_f+\delta_1}[\alphabar_{\Hp},\alphabar_{\I}]
		\\
		+
		\mathbb{E}^N_{\hat{u}_f+\delta_1,\Hp}
		+
		\mathbb{E}^N_{\hat{u}_f+\delta_1,\I}
		\leq
		\frac34 \varepsilon^2,
	\end{multline}
	\begin{equation} \label{improvement3ext}
			\mathbb{E}^{N+2}_{\hat{u}_f+\delta_1}[f_{\Hp,\I}]
			+
			\mathbb{E}_{\hat{u}_f+\delta_1}[f_{d,\Hp}]
			+
			\mathbb{E}_{\hat{u}_f+\delta_1}[f_{d,\I}]
			\leq
			2 \varepsilon^2.
	\end{equation}
	Note that the inclusions \eqref{impoverlapext1}--\eqref{impoverlapext5} guarantee that all of the diffeomorphism functions in \eqref{improvement3ext} are indeed well defined.
	
\end{theorem}

\begin{proof}[Proof]
	The estimates \eqref{improvement2}, \eqref{improvement3} guarantee that the cones $C_{u_{-1}}^{\I}$, $\Cbar_{v_{\infty}}^{\I}$, $C_{\hat{u}_f}^{\Hp}$, and $\Cbar_{v_{-1}}^{\Hp}$ can be extended as regular cones for some small time $\delta_1$ and, following Propositions \ref{prop:foliationIincomingcone}, \ref{prop:foliationIoutgoingcone}, \ref{prop:foliationHincomingcone}, \ref{prop:foliationHoutgoingcone} below, the relevant gauge conditions concerning the foliations of these cones by spheres can be extended.  By compactness of the domains $W_{\I}(\hat{u}_f,\delta_1)$ and $W_{\Hp}(\hat{u}_f,\delta_1)$ these extended foliations of the extended cones $C_{u_{-1}}^{\I}$, $\Cbar_{v_{\infty}}^{\I}$, $C_{\hat{u}_f}^{\Hp}$, and $\Cbar_{v_{-1}}^{\Hp}$ define regular extensions of the spacetime double null parametrisations $i_{\I} \colon W_{\I}(\hat{u}_f,\delta_1) \to \mathcal{M}$ and $i_{\Hp} \colon W_{\Hp}(\hat{u}_f,\delta_1) \to \mathcal{M}$.  The definitions \eqref{fourmoreparameters}, \eqref{somevchoices} and \eqref{U4choice} of $u_3$, $V_3$ and $U_{5}$ guarantee that the initial data regions $\mathcal{D}^{\mathcal{EF}}(u_3)$ and $\mathcal{D}^{\mathcal{K}}(V_3)$ are large enough so that the improved inclusions \eqref{impoverlap3}--\eqref{impoverlap5} imply that the inclusions \eqref{impoverlapext3}--\eqref{impoverlapext5} hold.  The inclusions \eqref{impoverlapext1}, \eqref{impoverlapext2} similarly follow from the improved inclusions \eqref{improvedinclus1}, \eqref{improvedinclus2}.  The inequalities \eqref{eq:lowerorderpointwiseext}, \eqref{improvement2ext} and \eqref{improvement3ext} follow from the compactness of the domains $W_{\I}(\hat{u}_f)$ and $W_{\Hp}(\hat{u}_f)$, Theorem \ref{havetoimprovethebootstrap}, and the fact that the norms depend continuously on $\delta_1$.
	
\end{proof}

\subsection{Existence of $\hat{u}_f+\delta$ normalised $\I$ gauge}
\label{offing}

In this section, and in Section \ref{offing2}, the proof of Theorem \ref{thm:newgauge} is given.  This section concerns the following theorem on the existence of $\hat{u}_f+\delta$ normalised $\I$ gauges.

\begin{theorem}[Existence of $\hat{u}_f+\delta$ normalised $\I$ gauge] \label{thm:Igaugeexistence}
	There exists $\delta_0>0$ and a function
	\[
		M_f:[\hat{u}_f,\hat{u}_f+\delta_0]\times \mathfrak{R}(\hat{u}_f)\to \mathbb R
	\]
	satisfying $(\ref{eq:MMinit})$
	such that, if $\hat{\varepsilon}_0$ is sufficiently small, for all $\delta \in [0,\delta_0]$ and all $\lambda\in \mathfrak{R}(\hat{u}_f)$, there exists a smooth $\hat{u}_f + \delta$ normalised $\I$ gauge with respect to $M_f(\hat{u}_f+\delta,\lambda)$
	\[
		i_{\delta} \colon \mathcal{Z}_{\I}(\hat{u}_f + \delta) \to \mathcal{M},
	\]
	where $\mathcal{Z}_{\I}(\hat{u}_f + \delta)$ is defined in \eqref{thedomainincords}, satisfying the anchoring conditions \eqref{eq:f3f4Iba} and \eqref{makesurewritten} and the overlap relation \eqref{eq:overlap5}, and such that the map
	\begin{equation} \label{eq:Igaugefinalsphereiden}
		i_{\delta}|_{(\hat{u}_{f} + \delta,v_\infty)\times \mathbb S^2}: (\hat{u}_{f} + \delta,v_\infty) \times \mathbb S^2\to S_{\hat{u}_{f} + \delta,v_\infty}\subset \mathcal{M},
	\end{equation}
	thought of as a map $\mathbb S^2 \to S_{\hat{u}_{f} + \delta,v_\infty}$, is the unique map determined by Proposition~\ref{determiningthesphere} with $h=r^{-2} \slashed{g}$ (and $p \in S_{\hat{u}_{f} + \delta,v_\infty}$ and $v \in T_p S_{\hat{u}_{f} + \delta,v_\infty}$ chosen so that \eqref{makesurewritten} holds).  The energies $\mathbb{E}^{N-2}_{\hat{u}_f+\delta}[P_{\I},\check{\Pbar}_{\I}]$, $\mathbb{E}^{N}_{\hat{u}_f+\delta}[\alpha_{\I},\alphabar_{\I}]$, $\mathbb{E}^N_{\hat{u}_f+\delta,\I}$ of the geometric quantities of the $\hat{u}_f + \delta$ normalised $\I$ gauge with respect to $M_f(\hat{u}_f+\delta,\lambda)$, and the associated diffeomorphism energies $\mathbb{E}_{\hat{u}_f+\delta}[f_{d,\I}]$, depend continuously on $\delta$.  The function $M_f$ satisfies
	\begin{equation} \label{eq:Igaugeexistencemass}
		\vert M_f(\hat{u}_f+\delta, \lambda) - M_f(\hat{u}_f,\lambda) \vert
		\lesssim
		\frac{\varepsilon \delta}{\hat{u}_f}.
	\end{equation}
	If $\delta_1$ is sufficiently small, then each $\hat{u}_f + \delta$ normalised $\I$ gauge can be extended to a smooth $\delta_1$ extended $\hat{u}_f + \delta$ normalised $\I$ gauge
	\[
		i_{\delta} \colon \mathcal{Z}_{\I}(\hat{u}_f + \delta, \delta_1) \to \mathcal{M},
	\]
	in the sense of Definition \ref{def:Igaugeextension}.  For any $0 \leq \delta' < \delta \leq \delta_0$, define\index{double null gauge!change of gauge!$F_{\delta',\delta}$}
	\[
		F_{\delta',\delta}
		=
		i_{\delta'}^{-1} \circ i_{\delta}
		\colon
		\mathcal{Z}_{\I}(\hat{u}_f + \delta)
		\to
		i_{\delta'}^{-1} ( i_{\delta}(\mathcal{Z}_{\I}(\hat{u}_f + \delta', \delta_1))),
	\]
	and define $f^3_{\delta',\delta}$, $f^4_{\delta',\delta}$, $\partial_u \slashed{f}_{\delta',\delta}$, $\partial_v \slashed{f}_{\delta',\delta}$ in terms of $F_{\delta',\delta}$ by \eqref{expressinglikethis}.
	These diffeomorphism functions satisfy, for all $u_{-1} \leq u \leq \hat{u}_f+\delta$, the estimates
	\begin{multline} \label{eq:diffeodeltadeltaprime1}
		\sum_{\vert \gamma \vert \leq 4}
		\Big(
		\Vert \mathfrak{D}^{\gamma} \partial_u \slashed{f}_{\delta',\delta} \Vert_{S_{u,v_{\infty}}}
		+
		\Vert \mathfrak{D}^{\gamma} \partial_v \slashed{f}_{\delta',\delta} \Vert_{S_{u,v_{\infty}}}
		+
		\Vert \mathfrak{D}^{\gamma} \partial_u f^4_{\delta',\delta} \Vert_{S_{u,v_{\infty}}}
		+
		r \Vert \mathfrak{D}^{\gamma} \partial_v f^3_{\delta',\delta} \Vert_{S_{u,v_{\infty}}}
		\Big)
		\\
		+
		\sum_{\vert \gamma \vert \leq 5}
		\Big(
		\Vert \mathfrak{D}^{\gamma} f^3_{\delta',\delta} \Vert_{S_{u,v_{\infty}}}
		+
		r^{-1} \Vert \mathfrak{D}^{\gamma} f^4_{\delta',\delta} \Vert_{S_{u,v_{\infty}}}
		\Big)
		\lesssim
		\frac{\varepsilon (\delta - \delta')}{\hat{u}_f+\delta'},
	\end{multline}
	and, for all $\hat{u}_f+\delta' \leq u \leq \hat{u}_f+\delta$,
	\begin{equation} \label{eq:diffeodeltadeltaprime2}
		\sup_{\theta \in \mathbb{S}^2} \mathrm{dist}_{\gamma}(\theta,\slashed{F}_{\delta',\delta}(u,v_{\infty},\theta))
		\lesssim
		\frac{\varepsilon (\delta - \delta')}{\hat{u}_f +\delta'}
		,
	\end{equation}
	where $v_{\infty}=v_{\infty}(\hat{u}_f+\delta)$ and $\slashed{F}_{\delta',\delta}(u,v_{\infty},\theta) = \pi_{\mathbb{S}^2} \circ F_{\delta',\delta}(u,v_{\infty},\theta)$, where $\pi_{\mathbb{S}^2}$ is the projection to the $\mathbb{S}^2$ argument.
	
	Finally, one may extend the map ${\bf J}(\hat{u}_f+\delta,\lambda)$ defined by $(\ref{definofboldv})$ as a map 
	 \begin{equation} \label{asamaphereext}
	 {\bf J}: [\hat{u}_f,\hat{u}_f+\delta_0] \times \mathfrak{R}(\hat{u}_f) \to \mathbb R^3.
	 \end{equation}
	 The map $(\ref{asamaphereext})$ is continuous in both $\lambda$ and $\delta$ and coincides
	 with the previously defined map for $\delta=0$.
\end{theorem}

\begin{remark}
\label{manifestdestiny}
	It is manifest from the proof of Theorem \ref{thm:Igaugeexistence} that, for fixed $\lambda\in \mathfrak{R}(\hat{u}_f)$ and fixed $0 \leq \delta \leq \delta_0$, the mass $M_f(\hat{u}_f+\delta,\lambda)$ and the $\hat{u}_f + \delta$ normalised $\I$ gauge with respect to $M_f(\hat{u}_f+\delta,\lambda)$ are unique amongst all such nearby masses and gauges, satisfying the anchoring conditions \eqref{eq:f3f4Iba} and \eqref{makesurewritten}, such that the map \eqref{eq:Igaugefinalsphereiden},
	thought of as a map $\mathbb S^2 \to S_{\hat{u}_{f} + \delta,v_\infty}$, is the unique map determined by Proposition~\ref{determiningthesphere} with $h=r^{-2} \slashed{g}$ (and $p$ and $v$ chosen so that \eqref{makesurewritten} holds).
\end{remark}

\begin{remark}
	For $0 < \delta \leq \delta_0$ the proof of Theorem \ref{thm:Igaugeexistence} involves proving suitable a priori estimates for the diffeomorphism functions $f_{0,\delta}$.  Weaker estimates than \eqref{eq:diffeodeltadeltaprime1} and \eqref{eq:diffeodeltadeltaprime2} suffice for the proof, where the $\delta$ factor provides sufficient smallness and the decay factor of $(\hat{u}_f)^{-1}$ is not necessary.  The stronger estimates, with the decay factor of $(\hat{u}_f)^{-1}$, and the more general estimates for $\delta' \neq 0$, are used in the proof of Theorem \ref{thm:lambda}.
\end{remark}

\begin{remark}
\label{linearandnonlinearremarkhere}
	For any $u_f$, the existence of the analogue of a $u_f$ normalised $\I$ gauge in linear theory, namely the proof of Proposition \ref{proplinIpgauge}, can be inferred from (a considerable simplification of) the proof of Theorem \ref{thm:Igaugeexistence}.  We will explicitly distill this proof in Remarks \ref{rmk:linearIgaugefoliations} and \ref{rmk:linearIgaugesphere} below.
\end{remark}

\begin{remark} \label{rmk:uf0Igaugeexistence0}
	Recall the double null parametrisation \eqref{herethenewone} of Theorem \ref{thm:localEF}, and let $u_f^0$ be as defined in Section~\ref{compediumparameterssec}.
	The proof of Theorem \ref{thm:Igaugeexistence} can readily be adapted to give the existence of a mass $M_f(u_f^0,\lambda)$ and a $u_f^0$ normalised $\I$ gauge with respect to $M_f(u_f^0,\lambda)$, satisfying the anchoring conditions \eqref{eq:f3f4Iba} and \eqref{makesurewritten}, and such that the map \eqref{eq:Igaugefinalsphereiden}, thought of as a map $\mathbb S^2 \to S_{\hat{u}_{f} + \delta,v_\infty}$, is the unique map determined by Proposition~\ref{determiningthesphere} with $h=r^{-2} \slashed{g}$ (and $p$ and $v$ chosen so that \eqref{makesurewritten} holds) (cf.\@ the proof of Theorem \ref{existenceofanchoredgaugethe}) by replacing the role of the $\delta_1$ extended $\hat{u}_f$ normalised $\I$ gauge of Theorem \ref{thm:extendedgauges} with the double null parametrisation \eqref{herethenewone} of Theorem \ref{thm:localEF}, and replacing any invocation of the smallness of $\delta_0$ with the smallness of $\hat{\varepsilon}_0$.  In fact, the existence of $u_f^0$ normalised $\I$ gauge is much simpler than the proof of Theorem \ref{thm:Igaugeexistence} in view of the fact that $\hat{\varepsilon}_0$ can be chosen to be small with respect to $u_f^0$.  Moreover, it is implicit from the construction that this mass and gauge are unique amongst all such nearby masses and gauges (cf.\@ Remark \ref{manifestdestiny}).  See Remarks \ref{rmk:uf0Igaugeexistence1} and \ref{rmk:uf0Igaugeexistence2} below for further discussion.
\end{remark}

The proof of Theorem \ref{thm:Igaugeexistence} is divided into two parts.  For a fixed $\delta \in [0,\delta_0]$, the first part of the proof involves showing that, for any sphere suitably close to the sphere $S_{\hat{u}_f+\delta, v_{\infty}(\hat{u}_f+\delta)}$ of the extended $\hat{u}_f$ normalised gauge and for any mass $M$ suitably close to $M_f$, there exist certain foliations of the corresponding null hypersurfaces defined by the sphere.  The foliations attain a subset of the defining conditions of the $\hat{u}_f+\delta$ normalised $\I$ gauge with mass $M$.  The existence of such foliations is addressed in {\bf Section~\ref{subsec:Ifoliationsexistence}}.  See Proposition \ref{prop:Iconesfoliations}.  The second part of the proof of Theorem \ref{thm:Igaugeexistence} involves finding an appropriate sphere and mass such that, in the double null foliation arising from the foliations of the resulting cones of Proposition \ref{prop:Iconesfoliations}, the remaining defining conditions of the $\hat{u}_f+\delta$ normalised $\I$ gauge are attained.  The existence of such a sphere and mass is addressed in {\bf Section~\ref{subsec:Isphereexistence}}, along with the properties of the map \eqref{asamaphereext} and the diffeomorphsims $F_{\delta',\delta}$.

\subsubsection{Foliations of null hypersurfaces}
\label{subsec:Ifoliationsexistence}

Throughout this section $\lambda\in \mathfrak{R}(\hat{u}_f)$ is fixed and the dependence of quantities on $\lambda$ is suppressed.
Recall that the boundary components of the causal future and causal past of any spacelike $2$-sphere are null hypersurfaces.
Suppose $S$ is any sphere suitably close to $S_{\hat{u}_f+\delta, v_{\infty}(\hat{u}_f+\delta)}$.  Let $\newsph{\underline{C}}_{v_{\infty}(\hat{u}_f+\delta)}$ denote the incoming component of the boundary of the causal past of $S$.  Suppose a foliation of $\newsph{\underline{C}}_{v_{\infty}(\hat{u}_f+\delta)}$ by spheres is given, described by the level hypersurfaces of a function $\newsph{u}$.  This foliation defines a sphere $\newsph{S}$, the sphere of the foliation at parameter time $u_{-1}$.  Let $\newsph{C}_{u_{-1}}$ denote the outgoing component of the boundary of the causal past of $\newsph{S}$.  Any foliation of $\newsph{C}_{u_{-1}}$ by spheres then defines a spacetime double null foliation by considering the outgoing components of the boundaries of the causal past of the spheres of the foliation of $\newsph{\underline{C}}_{v_{\infty}(\hat{u}_f+\delta)}$, and the incoming components of the boundaries of the causal future of the spheres of the foliation of $\newsph{C}_{u_{-1}}$.  Under suitable smallness conditions, this spacetime double null foliation will be regular in the image of the domain of the extended $\hat{u}_f$ normalised $\I$ gauge, $i_{\I}(\mathcal{Z}_{\I}(\hat{u}_f,\delta_1))$.
The first step in the proof of Theorem \ref{thm:Igaugeexistence} is the following proposition which states that, given any such sphere suitably close to $S_{\hat{u}_f+\delta, v_{\infty}(\hat{u}_f+\delta)}$, and any mass $M$ suitably close to $M_f = M_f(\hat{u}_f)$, there exist foliations of the two corresponding null hypersurfaces defined by this sphere such that, in the corresponding spacetime double null foliation, a subset of the defining properties of the $\hat{u}_f+\delta$ normalised $\I$ gauge are satisfied with mass $M$.  Recall that
\[
	\underline{\mu}^{\dagger}_M
	=
	\slashed{div} \underline{\eta}
	+
	\rho - \rho_{\circ,M}
	-
	\frac{1}{2} \hat{\chi} \cdot \underline{\hat{\chi}}
	+
	\frac{\Omega_{\circ,M}^2}{2r_M}
	\left(
	\frac{tr \underline{\chi}}{\Omega} - \frac{(\Omega \tr \underline{\chi})_{\circ,M}}{\Omega^2_{\circ,M}}
	\right).
\]

\begin{proposition}[Foliation of cones corresponding to any sphere close to $S_{\hat{u}_f+\delta, v_{\infty}(\hat{u}_f+\delta)}$] \label{prop:Iconesfoliations}
	Let $S$ be a sphere defined, in the $(u_{\I}, v_{\I},\theta_{\I})$ coordinate system of the extended $\hat{u}_f$ normalised $\I$ gauge, by smooth functions $s^3,s^4 \colon \mathbb{S}^2 \to \mathbb{R}$,
	\[
		S = \{ i_{\I} (\hat{u}_f+\delta + s^3(\theta), v_{\infty}(\hat{u}_f+\delta) + s^4(\theta),\theta) \mid \theta \in \mathbb{S}^2\}.
	\]
	Suppose that $s^3$ and $s^4$ satisfy
	\[
		\sum_{k \leq 5}
		\big(
		\Vert (r \nablaslash)^k s^3 \Vert_{\mathbb{S}^2}
		+
		\Vert (r \nablaslash)^k s^4 \Vert_{\mathbb{S}^2}
		\big)
		\leq
		\tau,
	\]
	for some $\tau$ sufficiently small, so that $S$ is appropriately close to the sphere $S_{\hat{u}_f+\delta, v_{\infty}(\hat{u}_f+\delta)}$.  Let also $M>0$ denote any mass which is close to the mass $M_f = M_f(\hat{u}_f)$ in the sense that
	\[
		\vert M - M_f \vert \leq \tau.
	\]
	Then, if $\delta_0$ and $\tau$ are sufficiently small with respect to $\hat{u}_f$ and $\hat{\varepsilon}_0$ is sufficiently small (independent of $\hat{u}_f$), there exists a smooth foliation of the past incoming cone, $\newsph{\underline{C}}_{v_{\infty}(\hat{u}_f+\delta)}$, of $S$ and a smooth foliation of the past outgoing cone, $\newsph{C}_{u_{-1}}$, of the sphere $\newsph{S}_{u_{-1},v_{\infty}(\hat{u}_f+\delta)}$ (defined to be the sphere at parameter time $u_{-1} - (\hat{u}_f+\delta)$ in the past of $S$ along $\newsph{\underline{C}}_{v_{\infty}(\hat{u}_f+\delta)}$) such that the geometric quantities of the associated spacetime double null foliation satisfy
	\begin{align}
		\mu_{\ell\geq 1}(u,v_{\infty}(\hat{u}_f+\delta),\theta)
		&
		=
		0
		\quad
		\text{for all } u \in [u_{-1},\hat{u}_f+\delta], \theta \in
		\mathbb{S}^2;
		\label{eq:Ifoliation1}
		\\
		\underline{\mu}^{\dagger}_{\ell \geq 1}(u_{-1},v,\theta)
		&
		=
		0 
		\quad
		\text{for all }
		v\in [v(R_{-2},u_{-1}),v_{\infty}(\hat{u}_f+\delta)], \theta \in \mathbb{S}^2;
		\label{eq:Ifoliation2}
		\\
		\left(\Omega^2 - \Omega_{\circ,M}^2 \right)_{\ell=0}(u,v_{\infty}(\hat{u}_f+\delta))
		&
		=
		0
		\quad
		\text{for all }
		u \in [u_{-1},\hat{u}_f+\delta];
		\label{eq:Ifoliation3}
		\\
		\left(\Omega \tr \chi - (\Omega \tr \chi)_{\circ,M} \right)_{\ell = 0} (\hat{u}_f+\delta, v_{\infty}(\hat{u}_f+\delta),\theta)
		&
		=
		0
		\quad
		\text{for all }
		\theta \in \mathbb{S}^2;
		\label{eq:Ifoliation4}
	\end{align}
	and
	\begin{align}
		(\Omega \omegahat - (\Omega \omegahat)_{\circ,M})_{\ell=0} (u_{-1},v)
		=
		F(u_{-1}) \frac{\Omega_{\circ,M}^2}{r^3}(u_{-1},v)
		\quad
		\text{for all }
		v \in [v(R_{-2},u_{-1}),v_{\infty}(\hat{u}_f+\delta)],
		\label{eq:Ifoliation5}
	\end{align}
	where
	\[
		F(u)
		:=
		\frac{1}{2} \int_{u}^{\hat{u}_f+\delta} \int_{\bar{u}}^{\hat{u}_f+\delta}
		r^3 (\Omega \hat{\chi}, \alphabar)_{\ell=0}
		\left(\hat{u},v_\infty\right)
		d\hat{u}
		d\bar{u}.
	\]
\end{proposition}

Instead of showing Proposition \ref{prop:Iconesfoliations} directly, it will first be shown that, for any sphere close to $S_{u_{-1},v_{\infty}(\hat{u}_f+\delta)}$ (rather than the sphere $S_{\hat{u}_f+\delta,v_{\infty}(\hat{u}_f+\delta)}$), the desired foliations of its corresponding cones can be found.  It will then afterwards be shown that any sphere close to $S_{\hat{u}_f+\delta,v_{\infty}(\hat{u}_f+\delta)}$ can be attained by an appropriate choice of sphere close to $S_{u_{-1},v_{\infty}(\hat{u}_f+\delta)}$.  It will therefore be supposed that $\newsph{S}$ is a sphere defined, in the $(u_{\I}, v_{\I},\theta_{\I})$ coordinate system, by smooth functions $j^3$, $j^4$,
\begin{equation} \label{eq:newjsphere}
	\newsph{S} = \{ i_{\I} (u_{-1} + j^3(\theta), v_{\infty}(\hat{u}_f+\delta) + j^4(\theta),\theta) \mid \theta \in \mathbb{S}^2\}.
\end{equation}
The functions $j^3$ and $j^4$ will be assumed to satisfy
\[
	\sum_{k \leq 5}
	\big(
	\Vert (r \nablaslash)^k j^3 \Vert_{\mathbb{S}^2}
	+
	\Vert (r \nablaslash)^k j^4 \Vert_{\mathbb{S}^2}
	\big)
	\leq
	\tau,
\]
so that $\newsph{S}$ is appropriately close to the sphere $S_{u_{-1}, v_{\infty}(\hat{u}_f+\delta)}$.  For such an $\newsph{S}$, let $\newsph{\Cbar}_{v_{\infty}(\hat{u}_f+\delta)}$ be the incoming component of the causal future of $\newsph{S}$ and let $\newsph{C}_{u_{-1}}$ be the outgoing component of the causal past of $\newsph{S}$.  Note that, provided $\tau$ is suitably small, the sphere $\newsph{S}$ is contained in the image of the domain of the extended $\hat{u}_f$ normalised $\I$ gauge, $\newsph{S} \subset i_{\I}(\mathcal{Z}_{\I}(\hat{u}_f,\delta_1))$, and the cones $\newsph{\Cbar}_{v_{\infty}(\hat{u}_f+\delta)}$ and $\newsph{C}_{u_{-1}}$ are arbitrarily close to the cones $\Cbar_{v_{\infty}(\hat{u}_f+\delta)}$ and $C_{u_{-1}}$ respectively of the extended $\hat{u}_f$ normalised $\I$ gauge.  In particular, $\newsph{\Cbar}_{v_{\infty}(\hat{u}_f+\delta)}$ and $\newsph{C}_{u_{-1}}$, when restricted to $i_{\I}(\mathcal{Z}_{\I}(\hat{u}_f,\delta_1))$, are regular null hypersurfaces.  The cones $\newsph{\Cbar}_{v_{\infty}(\hat{u}_f+\delta)}$ and $\newsph{C}_{u_{-1}}$ will each be foliated in such a way that the relevant regions of the cones also live in the region $i_{\I}(\mathcal{Z}_{\I}(\hat{u}_f,\delta_1))$ covered by the extended $\hat{u}_f$ normalised $\I$ gauge.

\subsubsection*{Foliations of the incoming null hypersurface $\newsph{\Cbar}_{v_{\infty}(\hat{u}_f+\delta)}$}

Consider an incoming null hypersurface $\Cbar$ in $(\mathcal{M},g)$ foliated by spheres described by the level hypersurfaces of a function $u:\Cbar \to \mathbb{R}$.  Consider the null generator $\underline{L}$ of $\Cbar$ satisfying $\underline{L}(u) = 1$.  There exists a unique \emph{conjugate} null vector field $L'$ on $\Cbar$ satisfying
\[
	g(\underline{L},L') = -2.
\]

If $\Cbar$ is an incoming null hypersurface of a given spacetime double null foliation, and its foliation by spheres coincides with that obtained by intersecting $\Cbar$ with the outgoing null hypersurfaces of the spacetime double null foliation, then the vectors of the double null frame $e_3$, $e_4$ are related to $\underline{L}$, $L'$ by
\[
	\underline{L} = \Omega e_3,
	\qquad
	L' = \Omega^{-1} e_4.
\]
It follows that many of the Ricci coefficients and curvature components of the spacetime double null foliation can be expressed entirely in terms of $\underline{L}$ and $L'$.  For example, if $X$ is a vector field tangent to the level hypersurfaces of $u$,
\begin{equation} \label{eq:foliationIincomingcone1}
	\eta (X) = -\frac{1}{2} g(\nabla_{\underline{L}} X, L'),
	\qquad
	\rho = \frac{1}{4} R(L',\underline{L},L',\underline{L}).
\end{equation}
Though the same is not true of Ricci coefficients such as $\omegabarhat$ and $\hat{\chi}$, certain $\Omega$ rescaled quantities can also be expressed entirely in terms of $\underline{L}$ and $L'$.  For example, if $X$ and $Y$ are tangential to the level hypersurfaces of $u$,
\begin{equation} \label{eq:foliationIincomingcone2}
	\Omega \omegabarhat
	=
	\frac{1}{4} g(\nabla_{\underline{L}} L',\underline{L}),
	\quad
	\Omega^{-1} \chi(X,Y)
	=
	g(\nabla_{X} L',Y),
	\quad
	\Omega \chibar(X,Y)
	=
	g(\nabla_{X} \underline{L},Y),
\end{equation}
\begin{equation} \label{eq:foliationIincomingcone3}
	\Omega^{-1} \hat{\chi}(X,Y)
	=
	\Omega^{-1} \chi(X,Y)
	-
	\Omega^{-1} \tr \chi \gslash(X,Y),
	\quad
	\Omega \hat{\chibar}(X,Y)
	=
	\Omega \chibar(X,Y)
	-
	\Omega \tr \chibar \gslash(X,Y),
\end{equation}
and
\begin{equation} \label{eq:foliationIincomingcone4}
	\Omega^2\alphabar(X,Y)
	=
	R(X,\underline{L},Y,\underline{L}),
\end{equation}

The following proposition relates certain geometric quantities associated to two different foliations of the incoming null hypersurface $\newsph{\underline{C}}_{v_{\infty}(\hat{u}_f+\delta)}$.

\begin{proposition}[Change of foliation relations on incoming cones] \label{prop:inIconesrelations}
	For a given sphere $\newsph{S}$ as in \eqref{eq:newjsphere}, let $u$ denote the restriction to $\newsph{\underline{C}}_{v_{\infty}(\hat{u}_f+\delta)}$ of the $u$ coordinate of the extended $\hat{u}_f$ normalised gauge.  Consider another foliation of $\newsph{\underline{C}}_{v_{\infty}(\hat{u}_f+\delta)}$ by spheres, described by the level hypersurfaces of a function $\widetilde{u}$.  Suppose $u$ and $\widetilde{u}$ are related by
	\begin{equation} \label{eq:inIconesuutilde}
		u = \widetilde{u} + f^3(\widetilde{u},\theta),
	\end{equation}
	for some function $f^3:[u_{-1},\hat{u}_f+\delta_1] \times \mathbb{S}^2 \to \mathbb{R}$.  Then the corresponding geometric quantities $\mu$ and $\Omega \omegabarhat$ are related by
	\[
		\widetilde{\Deltaslash} \log (1+\partial_{\widetilde{u}} f^3(\widetilde{u},\theta))
		=
		\widetilde{\mu}(\widetilde{u},\theta)
		-
		\mu(\widetilde{u} + f^3,\theta)
		+
		F_{\mu} (\widetilde{\nablaslash} f^3, \widetilde{\nablaslash}{}^2 f^3,\partial_{\widetilde{u}} f^3, \widetilde{\nablaslash} \partial_{\widetilde{u}} f^3),
	\]
	and
	\[
		\frac{1}{2} \partial_{\widetilde{u}} \log(1+\partial_{\widetilde{u}} f^3)(\widetilde{u},\theta)
		=
		\widetilde{\Omega \omegabarhat}(\widetilde{u},\theta)
		-
		(1+\partial_{\widetilde{u}} f^3(\widetilde{u},\theta)) \Omega \omegabarhat(\widetilde{u} + f^3,\theta).
	\]
	Here $\widetilde{\nablaslash}$ denotes the Levi-Civita connection of the induced metric on the level hypersurfaces of $\widetilde{u}$, and $F_{\mu}$ is a function which satisfies
	\begin{equation} \label{eq:inIconesrelationsFmuestimate}
		r^2 \vert F_{\mu} (\widetilde{\nablaslash} f^3, \widetilde{\nablaslash}{}^2 f^3,\partial_{\widetilde{u}} f^3, \widetilde{\nablaslash} \partial_{\widetilde{u}} f^3) \vert
		\lesssim
		\big(
		\vert \widetilde{r \nablaslash} f^3 \vert + \vert (\widetilde{r \nablaslash})^2 f^3 \vert+\vert \partial_{\widetilde{u}} f^3 \vert + \vert \widetilde{r \nablaslash} \partial_{\widetilde{u}} f^3 \vert
		+
		\frac{1}{r}
		\big)
		\vert \widetilde{r \nablaslash} f^3 \vert
		+
		\vert (\widetilde{r \nablaslash})^2 f^3 \vert
		.
	\end{equation}
	In particular, if $\vert \widetilde{r \nablaslash} f^3 \vert + \vert (\widetilde{r \nablaslash})^2 f^3 \vert+\vert \partial_{\widetilde{u}} f^3 \vert + \vert \widetilde{r \nablaslash} \partial_{\widetilde{u}} f^3 \vert \lesssim 1$, then $F_{\mu}$ satisfies
	\[
		r^2 \vert F_{\mu} (\widetilde{\nablaslash} f^3, \widetilde{\nablaslash}{}^2 f^3,\partial_{\widetilde{u}} f^3, \widetilde{\nablaslash} \partial_{\widetilde{u}} f^3) \vert
		\lesssim
		\vert \widetilde{r \nablaslash} f^3 \vert
		+
		\vert (\widetilde{r \nablaslash})^2 f^3 \vert
		.
	\]
\end{proposition}

\begin{proof}
	Let $\underline{L}$ and $\widetilde{\underline{L}}$ denote the null generators of $\newsph{\underline{C}}_{v_{\infty}(\hat{u}_f+\delta)}$ which satisfy
	\[
		\underline{L}(u) 
		=
		1,
		\qquad 
		\widetilde{\underline{L}}(\widetilde{u})
		=
		1,
	\]
	respectively.  Clearly $\widetilde{\underline{L}} = a \underline{L}$ for some $a$ and, by applying to the relation \eqref{eq:inIconesuutilde}, it follows that
	\[
		\widetilde{\underline{L}}
		=
		(1+\partial_{\widetilde{u}} f^3) \underline{L}.
	\]
	Now defining $e_A = \partial_{\theta^A}$ in the $(u,\theta)$ coordinate system for $\newsph{\underline{C}}_{v_{\infty}(\hat{u}_f+\delta)}$, and $\widetilde{e}_A = \widetilde{\partial}_{\theta^A}$ in the $(\widetilde{u},\theta)$ coordinate system, it follows again from \eqref{eq:inIconesuutilde} that
	\[
		\widetilde{e}_A
		=
		e_A
		+
		\partial_{\theta^A} f^3 \underline{L}.
	\]
	In particular, $\widetilde{\gslash}_{AB} = g(\widetilde{e}_A,\widetilde{e}_B) = g(e_A,e_B) = \gslash_{AB}$.  If now $L'$ and $\widetilde{L}'$ denote the conjugate null vectors to $\underline{L}$ and $\widetilde{\underline{L}}$ respectively, satisfying $g(L',\underline{L}) = g (\widetilde{L}',\widetilde{\underline{L}}) = -2$, it follows, after writing $\widetilde{L}' = a L' + b \underline{L} + c^A e_A$ and using the fact that $g(\widetilde{L}', \widetilde{L}') = 0$, $g(\widetilde{L}',\widetilde{\underline{L}}) = -2$ and $g(\widetilde{L}',\widetilde{e}_A) = 0$ to compute $a$, $b$ and $c^A$, that
	\[
		\widetilde{L}'
		=
		(1+\partial_{\widetilde{u}} f^3)^{-1}
		\big[
		L'
		+
		\vert \widetilde{\nablaslash} f^3 \vert^2
		\underline{L}
		+
		2
		\gslash^{AB}
		\partial_{\theta^A} f^3 e_B
		\big],
	\]
	where $\vert \widetilde{\nablaslash} f^3 \vert^2 = \gslash^{AB} \partial_{\theta^A} f^3 \partial_{\theta^B} f^3$.
	
	From the above relations, it follows that
	\begin{align*}
		-2 \widetilde{\eta} (\widetilde{e}_A)
		&
		=
		g(\nabla_{\widetilde{\underline{L}}} \widetilde{e}_A, \widetilde{L}')
		\\
		&
		=
		(1+\partial_{\widetilde{u}} f^3)^{-1}
		\partial_{\widetilde{u}} \partial_{\theta^A} f^3 g(\underline{L},L')
		+
		g(\nabla_{\underline{L}} e_A + \partial_{\theta^A} f^3 \nabla_{\underline{L}} \underline{L},
		L' + \vert \widetilde{\nablaslash} f^3 \vert^2 \underline{L} + 2 \gslash^{BC} \partial_{\theta^B} f^3 e_C)
		\\
		&
		=
		-2 (1+\partial_{\widetilde{u}} f^3)^{-1}
		\partial_{\widetilde{u}} \partial_{\theta^A} f^3
		-2 \eta(e_A)
		+
		\Omega {\chibar_A}^B \partial_{\theta^B} f^3
		-
		4 \Omega \omegabarhat \partial_{\theta^A} f^3.
	\end{align*}
	Similarly,
	\[
		\widetilde{\rho}
		-
		\rho
		=
		\widetilde{\nablaslash} f^3 \otimes \widetilde{\nablaslash} f^3 \cdot \Omega^2 \alphabar
		-
		2 \widetilde{\nablaslash} f^3 \cdot \Omega \betabar,
	\]
	where
	\[
		\widetilde{\nablaslash} f^3 \otimes \widetilde{\nablaslash} f^3 \cdot \Omega^2 \alphabar
		=
		\gslash^{AB} \gslash^{CD} \partial_{\theta^A} f^3 \partial_{\theta^C} f^3 \Omega^2 \alphabar_{BD},
		\qquad
		\widetilde{\nablaslash} f^3 \cdot \Omega \betabar
		=
		\gslash^{AB} \partial_{\theta^A} f^3 \Omega \betabar_B.
	\]
	Moreover,
	\begin{align*}
		&
		\widetilde{\Omega \chibar} (\widetilde{e}_A, \widetilde{e}_B)
		=
		(1+\partial_{\widetilde{u}} f^3) \Omega \chibar (e_A,e_B),
		\\
		&
		\widetilde{\Omega^{-1} \chi} (\widetilde{e}_A, \widetilde{e}_B)
		=
		(1+\partial_{\widetilde{u}} f^3)^{-1} 
		\big[
		2 \widetilde{\nablaslash}{}^2_{A,B} f^3
		+
		\Omega^{-1} \chi (e_A,e_B)
		+
		2(\eta_A \partial_{\theta^B}f^3 + \eta_B \partial_{\theta^A}f^3)
		+
		4 \Omega \omegabarhat \partial_{\theta^A} f^3 \partial_{\theta^B} f^3
		\\
		&
		+
		\vert \widetilde{\nablaslash} f^3\vert^2 \Omega \chibar_{AB}
		-
		2\partial_{\theta^C} f^3 (\Omega \chibar_A^C \partial_{\theta^B} f^3 + \Omega \chibar_A^C \partial_{\theta^B} f^3)
		-
		2(\partial_{\theta^A} \log(1+\partial_{\widetilde{u}} f^3) \partial_{\theta^B} f^3
		+
		\partial_{\theta^B} \log(1+\partial_{\widetilde{u}} f^3) \partial_{\theta^A} f^3
		)
		\big].
	\end{align*}
	After recalling that $\mu = \slashed{div} \eta + \rho - \frac{1}{2} \hat{\chi} \cdot \underline{\hat{\chi}}$, these relations combine to give the expression for $\widetilde{\mu} - \mu$.  The expression involving $\widetilde{\Omega \omegabarhat}$ can be computed similarly, using the fact that
	\[
		\Omega \omegabarhat = \frac{1}{4} g(\nabla_{\underline{L}} L', \underline{L}).
	\]
	We finally note also that
	\[
		\widetilde{\Omega^2 \alphabar}(\widetilde{e}_A, \widetilde{e}_B)
		=
		(1+\partial_{\widetilde{u}} f^3)^2 \Omega^2 \alphabar(e_A,e_B).
	\]
\end{proof}

Rather than foliate the incoming cone $\newsph{\Cbar}_{v_{\infty}(\hat{u}_f+\delta)}$ directly by the condition that $(\Omega^2 - \Omega_{\circ}^2)_{\ell=0} = 0$, which is not a condition which can be expressed entirely in terms of the vectors $\underline{L}$ and $L'$, it is first shown that the incoming cone can be foliated by certain conditions which involve only the above quantities $\Omega \omegabarhat$ and $\Omega \tr \chibar$, which can be expressed entirely in terms of $\underline{L}$ and $L'$ (see \eqref{eq:foliationIincomingcone2} and \eqref{eq:foliationIincomingcone3}).  The foliation achieving the condition $(\Omega^2 - \Omega_{\circ}^2)_{\ell=0} = 0$ will be shown to exist simultaneously with the foliation of $\newsph{C}_{u_{-1}}$ achieving the conditions \eqref{eq:Ifoliation2} and \eqref{eq:Ifoliation5} in the proof of Proposition \ref{prop:Iconesfoliations} below.

In order to motivate the foliations constructed in Proposition \ref{prop:foliationIincomingcone} below, note first that, in a given spacetime double null foliation,
\begin{equation} \label{eq:OmegacalFbar}
	\partial_u \big( \Omega_{\circ}^{-1} \Omega_{\ell=0} - 1 \big)
	=
	\Omega \omegabarhat_{\ell=0}
	-
	\Omega \omegabarhat_{\circ}
	-
	\underline{\mathcal{F}},
\end{equation}
where
\[
	\underline{\mathcal{F}}
	=
	-
	\big( \Omega_{\circ}^{-1} \Omega_{\ell=0} - 1 \big)
	\big( \Omega \omegabarhat_{\ell=0}
	-
	\Omega \omegabarhat_{\circ} \big)
	-
	\big(
	\Omega_{\circ}^{-1} \Omega_{\ell \geq 1} (\Omega \omegabarhat)_{\ell \geq 1}
	\big)_{\ell =0}
	-
	\big( \Omega_{\circ}^{-1} \Omega (\Omega \tr \chibar)_{\ell \geq 1} \big)_{\ell=0}.
\]
Equation \eqref{eq:DlogOmega} implies that,
\begin{equation} \label{eq:Omegaeomegabar}
	\Omega_{\circ}^{-1} \Omega (u,v_{\infty},\theta)
	=
	\Omega_{\circ}^{-1} \Omega (u_{-1},v_{\infty},\theta)
	e^{\int_{u_{-1}}^u (\Omega \omegabarhat - \Omega \omegabarhat_{\circ})(u',v_{\infty},\theta) du'},
\end{equation}
and so $\underline{\mathcal{F}}$ can be viewed as a function of $u$, for any given $\Omega \omegabarhat - \Omega \omegabarhat_{\circ}$, $\Omega \tr \chibar - \Omega \tr \chibar_{\circ}$, and $\Omega_{\circ}^{-1} \Omega (u_{-1},v_{\infty},\cdot)$, i.\@e.\@, abusing notation,
\[
	\underline{\mathcal{F}}
	=
	\underline{\mathcal{F}}
	( \Omega \omegabarhat
	-
	\Omega \omegabarhat_{\circ}
	,
	\Omega \tr \chibar - \Omega \tr \chibar_{\circ}
	,
	\Omega_{\circ}^{-1} \Omega (u_{-1},v_{\infty},\cdot))
	: [u_{-1}, \hat{u}_f+\delta] \to \mathbb{R}
	.
\]

\begin{proposition}[Foliations of the incoming cone $\newsph{\underline{C}}_{v_{\infty}(\hat{u}_f+\delta)}$] \label{prop:foliationIincomingcone}
	Let $\newsph{S}$ be a sphere defined, in the coordinate system of the extended $\hat{u}_f$ normalised $\I$ gauge, by smooth functions $j^3$, $j^4$,
	\[
		\newsph{S} = \{ i_{\I} (u_{-1} + j^3(\theta), v_{\infty}(\hat{u}_f+\delta) + j^4(\theta),\theta) \mid \theta \in \mathbb{S}^2\}.
	\]
	such that $j^3$ and $j^4$ satisfy
	\[
		\sum_{k \leq 5}
		\big(
		\Vert (r \nablaslash)^k j^3 \Vert_{\newsph{S}}
		+
		\Vert (r \nablaslash)^k j^4 \Vert_{\newsph{S}}
		\big)
		\leq
		\tau.
	\]
	Consider also a mass $M$ satisfying the smallness assumption of Proposition \ref{prop:Iconesfoliations}, a constant $A\in \mathbb{R}$, and a smooth function $q : \mathbb{S}^2 \to \mathbb{R}$ satisfying
	\[
		\vert A \vert 
		+
		\sup_{\theta \in \mathbb{S}^2} \vert q(\theta) \vert
		\leq
		\tau.
	\]
	Then, provided $\tau$ and $\delta_0$ are sufficiently small, there exists a smooth foliation of the incoming cone, $\newsph{\underline{C}}_{v_{\infty}(\hat{u}_f+\delta)}$, of $\newsph{S}$ such that, on $\newsph{\underline{C}}_{v_{\infty}(\hat{u}_f+\delta)}$, the geometric quantities $\mu$ and $\Omega \omegabarhat$ of this foliation satisfy
	\[
		\mu_{\ell \geq 1}(\widetilde{u},\theta) = 0,
		\qquad
		(\Omega \omegabarhat - \Omega \omegabarhat_{\circ,M})_{\ell = 0}(\widetilde{u})
		=
		\underline{\mathcal{F}}
		[
		\Omega \omegabarhat - \Omega \omegabarhat_{\circ,M}
		,
		\Omega \tr \chibar - \Omega \tr \chibar_{\circ,M}
		,
		q
		]
		(\widetilde{u}),
	\]
	for all $\widetilde{u} \in [u_{-1}, \hat{u}_f+\delta]$, $\theta \in \mathbb{S}^2$, and,
	\[
		\partial_{\widetilde{u}} f^3_{\ell = 0} (u_{-1})
		=
		A,
	\]
	where $f^3$ is the diffeomorphism which relates this foliation of $\newsph{\underline{C}}_{v_{\infty}(\hat{u}_f+\delta)}$ to the foliation of $\newsph{\underline{C}}_{v_{\infty}(\hat{u}_f+\delta)}$ obtained by intersecting with the outgoing cones $C_u$, for $u_{-1} \leq u \leq \hat{u}_f + \delta_1$, of the extended $\hat{u}_f$ normalised gauge, so that $u = \widetilde{u} + f^3(\widetilde{u},\theta)$.
\end{proposition}

\begin{proof}
	Let $\mu$, $\Omega \omegabarhat$, $\Omega \tr \chibar$ denote the geometric quantities of the foliation of $\newsph{\underline{C}}_{v_{\infty}(\hat{u}_f+\delta)}$ obtained by intersecting $\newsph{\underline{C}}_{v_{\infty}(\hat{u}_f+\delta)}$ with the outgoing hypersurfaces of the extended $\hat{u}_f$ normalised gauge.  Note that, by continuity and compactness, these quantities are sufficiently close to their values in the extended $\hat{u}_f$ normalised gauge if $\tau$ and $\delta$ are sufficiently small.
	By Proposition \ref{prop:inIconesrelations}, the goal is to construct a function $f^3:[u_{-1},\hat{u}_f+\delta] \times \mathbb{S}^2 \to \mathbb{R}$ satisfying
	\[
		\widetilde{\Deltaslash} \log (1+\partial_{\widetilde{u}} f^3(\widetilde{u},\theta))
		=
		\Big[
		-
		\mu(\widetilde{u} + f^3,\theta)
		+
		F_{\mu} (\widetilde{\nablaslash} f^3, \widetilde{\nablaslash}{}^2 f^3,\partial_{\widetilde{u}} f^3, \widetilde{\nablaslash} \partial_{\widetilde{u}} f^3)
		\Big]_{\widetilde{\ell \geq 1}},
	\]
	and
	\[
		\frac{1}{2}
		\big[
		\partial_{\widetilde{u}} \log(1+\partial_{\widetilde{u}} f^3)(\widetilde{u},\cdot)
		\big]_{\widetilde{\ell =0}}
		=
		\mathring{\underline{\mathcal{F}}}(\widetilde{u})
		+
		\big[
		\Omega \omegabarhat_{\circ,M}(\widetilde{u})
		-
		(1+\partial_{\widetilde{u}} f^3(\widetilde{u},\cdot)) \Omega \omegabarhat(\widetilde{u} + f^3(\widetilde{u},\cdot),\cdot)
		\big]_{\widetilde{\ell =0}},
	\]
	along with the initial conditions,
	\[
		f^3(u_{-1},\cdot) = j^3,
		\qquad
		\partial_{\widetilde{u}} f^3_{\ell = 0} (u_{-1})
		=
		A,
	\]
	where
	\[
		\mathring{\underline{\mathcal{F}}}
		=
		\underline{\mathcal{F}}
		( 
		\Omega\omegabarhat_{\circ,M}
		+
		\frac{1}{2} \partial_{\widetilde{u}} \log(1+\partial_{\widetilde{u}} f^3)
		+
		(1+\partial_{\widetilde{u}} f^3)\Omega \omegabarhat_{f^3}
		,
		\Omega \tr \chibar_{\circ,M}
		+
		(1+\partial_{\widetilde{u}} f^3)
		\Omega \tr \chibar_{f^3}
		,
		q)
		,
	\]
	and
	\[
		\Omega \omegabarhat_{f^3}(\widetilde{u},\theta)
		=
		\Omega \omegabarhat( \widetilde{u} + f^3(\widetilde{u},\theta),\theta),
		\qquad
		\Omega \tr \chibar_{f^3}(\widetilde{u},\theta)
		=
		\Omega \tr \chibar( \widetilde{u} + f^3(\widetilde{u},\theta),\theta).
	\]
	Here $\widetilde{\ell \geq 1}$ and $\widetilde{\ell =0}$ denote the mode projections with respect to the spheres defined as the level hypersurfaces of $\widetilde{u}$, where $u = \widetilde{u} + f^3(\widetilde{u},\theta)$.  Define, therefore, $f_{[0]} = 0$ and, for $n \geq 1$, iterates $f_{[n]}$ as solutions of
	\begin{equation} \label{eq:vinftyfoliationiterates1}
		\Deltaslash_{[n-1]} \log (1+\partial_{u} f^3_{[n]}(u,\theta))
		=
		\Big[
		-
		\mu(u + f^3_{[n-1]},\theta)
		+
		F_{\mu} (\nablaslash f^3_{[n-1]}, \nablaslash^2 f^3_{[n-1]},\partial_{u} f^3_{[n-1]}, \nablaslash \partial_{u} f^3_{[n-1]})
		\Big]_{(\ell \geq 1)_{[n-1]}},
	\end{equation}
	and
	\begin{equation} \label{eq:vinftyfoliationiterates2}
		\frac{1}{2}
		\big[
		\partial_{u} \log(1+\partial_{u} f^3_{[n]})(u,\cdot)
		\big]_{(\ell =0)_{[n-1]}}
		=
		\underline{\mathcal{F}}_{[n-1]}(u)
		+
		\big[
		\Omega \omegabarhat_{\circ,M}
		-
		(1+\partial_{u} f^3_{[n-1]}(u,\cdot)) \Omega \omegabarhat(u + f^3_{[n-1]}(u,\cdot),\cdot)
		\big]_{(\ell =0)_{[n-1]}},
	\end{equation}
	along with the initial conditions,
	\[
		f^3_{[n]}(u_{-1},\cdot) = j^3,
		\qquad
		\partial_{\widetilde{u}} (f^3_{[n]})_{(\ell =0)_{[n-1]}} (u_{-1})
		=
		A,
	\]
	where
	\[
		\underline{\mathcal{F}}_{[n-1]}
		=
		\underline{\mathcal{F}}
		( 
		\Omega \omegabarhat_{\circ,M}
		+
		\frac{1}{2} \partial_{u} \log(1+\partial_{u} f^3_{[n-1]})
		+
		(1+\partial_{u} f^3_{[n-1]})\Omega \omegabarhat_{f^3_{[n-1]}}
		,
		\Omega \tr \chibar_{\circ,M}
		+
		(1+\partial_{u} f^3_{[n-1]})
		\Omega \tr \chibar_{f^3_{[n-1]}}
		,
		q)
		,
	\]
	where $(\ell \geq 1)_{[n-1]}$ and $(\ell =0)_{[n-1]}$ denote the mode projections with respect to the spheres defined as the level hypersurfaces of $u_{[n-1]}$, where $u_{[n-1]}$ is defined implicitly by $u = u_{[n-1]} + f^3_{[n-1]}(u_{[n-1]},\theta)$.
	
	Define
	\[
		g_{[n]} = \log (1+ \partial_u f^3_{[n]}),
	\]
	so that
	\[
		f^3_{[n]}(u,\theta)
		=
		j^3(\theta)
		+
		\int_{u_{-1}}^u e^{g_{[n]}(u',\theta)} - 1 du'.
	\]
	
	Consider some $U> u_{-1}$.  It follows, by induction, that, provided $U - u_{-1}$ is sufficiently small,
	\begin{equation} \label{eq:f3vinftyinduction}
		\sup_{u_{-1} \leq u \leq U}
		\sum_{k \leq 5}
		\Vert (r\nablaslash)^k f^3_{[n]} \Vert_{S_{u,v_{\infty}}}
		\leq
		2 \tau,
		\qquad
		\sup_{u_{-1} \leq u \leq U}
		\sum_{k \leq 5}
		\Vert (r\nablaslash)^k \partial_u f^3_{[n]} \Vert_{S_{u,v_{\infty}}}
		\leq
		C_1 \tau,
	\end{equation}
	for all $n$, for some constant $C_1$ to be chosen.  Indeed, suppose that \eqref{eq:f3vinftyinduction} holds for some $n \geq 1$.  A standard elliptic estimate for \eqref{eq:vinftyfoliationiterates1} and transport estimate for \eqref{eq:vinftyfoliationiterates2} gives, for all $u_{-1} \leq u \leq U$,
	\begin{align*}
		\sup_{u_{-1} \leq u \leq U}
		\sum_{k \leq 5} \Vert (r\nablaslash)^k g_{[n+1]} \Vert_{S_{u,v_{\infty}}}
		\lesssim
		&
		\sup_{u_{-1} \leq u \leq U}
		\sum_{k \leq 5} \Vert (r\nablaslash)^k f^3_{[n]} \Vert_{S_{u,v_{\infty}}}
		+
		(U - u_{-1})
		\sup_{u_{-1} \leq u \leq U}
		\sum_{k \leq 4} \Vert (r\nablaslash)^k \partial_u f^3_{[n]} \Vert_{S_{u,v_{\infty}}}
		\\
		&
		+
		\vert M - M_f \vert
		+
		\vert \log( 1+A) \vert
		+
		(U - u_{-1}) \sup_{\theta \in \mathbb{S}^2} \vert q(\theta) \vert.
	\end{align*}
	Now
	\[
		\sup_{u_{-1} \leq u \leq U} \sum_{k \leq 5} \Vert (r\nablaslash)^k \partial_u f^3_{[n+1]} \Vert_{S_{u,v_{\infty}}}
		\lesssim
		\sup_{u_{-1} \leq u \leq U}
		\sum_{k \leq 5} \Vert (r\nablaslash)^k g_{[n+1]} \Vert_{S_{u,v_{\infty}}},
	\]
	and
	\[
		\sup_{u_{-1} \leq u \leq U} \sum_{k \leq 5} \Vert (r\nablaslash)^k f^3_{[n+1]} \Vert_{S_{u,v_{\infty}}}
		\leq
		\sum_{k \leq 5} \Vert (r\nablaslash)^k j^3 \Vert_{S_{u_{-1},v_{\infty}}}
		+
		C
		(U - u_{-1})
		\sup_{u_{-1} \leq u \leq U}
		\sum_{k \leq 5} \Vert (r\nablaslash)^k g_{[n+1]} \Vert_{S_{u,v_{\infty}}},
	\]
	and hence \eqref{eq:f3vinftyinduction} holds with $n+1$ in place of $n$ if $U - u_{-1}$ is sufficiently small, and $C_1$ is chosen to be sufficiently large.
	
	Similarly, considering now the differences of the systems \eqref{eq:vinftyfoliationiterates1} and \eqref{eq:vinftyfoliationiterates2} for $n+1$ and $n$ respectively, and using the estimates \eqref{eq:f3vinftyinduction}, it follows from \eqref{eq:f3vinftyinduction} that
	\begin{align*}
		\sup_{u_{-1} \leq u \leq U}
		\sum_{k \leq 5}
		\Vert (r\nablaslash)^k(g_{[n+1]} - g_{[n]}) \Vert_{S_{u,v_{\infty}}}
		\lesssim
		&
		\sup_{u_{-1} \leq u \leq U}
		\sum_{k \leq 5} \Vert (r\nablaslash)^k( f^3_{[n]} - f^3_{[n-1]} ) \Vert_{S_{u,v_{\infty}}}
		\\
		&
		+
		(\tau + (U - u_{-1}))
		\sup_{u_{-1} \leq u \leq U}
		\sum_{k \leq 4} \Vert (r\nablaslash)^k ( \partial_u f^3_{[n]} - \partial_u f^3_{[n-1]} ) \Vert_{S_{u,v_{\infty}}},
	\end{align*}
	and hence
	\begin{align*}
		\sup_{u_{-1} \leq u \leq U}
		\sum_{k \leq 5} \Vert (r\nablaslash)^k ( \partial_u f^3_{[n+1]} - \partial_u f^3_{[n]} ) \Vert_{S_{u,v_{\infty}}}
		\lesssim
		&
		\sup_{u_{-1} \leq u \leq U}
		\sum_{k \leq 5} \Vert (r\nablaslash)^k( f^3_{[n]} - f^3_{[n-1]} ) \Vert_{S_{u,v_{\infty}}}
		\\
		&
		+
		(\tau + (U - u_{-1}))
		\sup_{u_{-1} \leq u \leq U}
		\sum_{k \leq 4} \Vert (r\nablaslash)^k ( \partial_u f^3_{[n]} - \partial_u f^3_{[n-1]} ) \Vert_{S_{u,v_{\infty}}},
	\end{align*}
	and
	\begin{multline*}
		\sup_{u_{-1} \leq u \leq U}
		\sum_{k \leq 5} \Vert (r\nablaslash)^k( f^3_{[n+1]} - f^3_{[n]} ) \Vert_{S_{u,v_{\infty}}}
		\lesssim
		(U - u_{-1})
		\sup_{u_{-1} \leq u \leq U}
		\sum_{k \leq 5} \Vert (r\nablaslash)^k( f^3_{[n]} - f^3_{[n-1]} ) \Vert_{S_{u,v_{\infty}}}
		\\
		+
		(U - u_{-1}) (\tau + (U - u_{-1}))
		\sup_{u_{-1} \leq u \leq U}
		\sum_{k \leq 4} \Vert (r\nablaslash)^k ( \partial_u f^3_{[n]} - \partial_u f^3_{[n-1]} ) \Vert_{S_{u,v_{\infty}}},
	\end{multline*}
	from which it follows that
	\begin{align*}
		\sup_{u_{-1} \leq u \leq U}
		\sum_{k \leq 5} \Vert (r\nablaslash)^k( f^3_{[n+1]} - f^3_{[n]} ) \Vert_{S_{u,v_{\infty}}}
		\lesssim
		(U - u_{-1})
		\sup_{u_{-1} \leq u \leq U}
		\sum_{k \leq 5} \Vert (r\nablaslash)^k( f^3_{[n]} - f^3_{[n-1]} ) \Vert_{S_{u,v_{\infty}}}.
	\end{align*}
	Hence, if $U - u_{-1}$ is sufficiently small, the map which takes $f^3_{[n]}$ to $f^3_{[n+1]}$ is a contraction and the sequence $\{f^3_{[n]}\}$ converges to a limit $f^3$ on $[u_{-1},U]$ satisfying the estimate \eqref{eq:f3vinftyinduction}.
	
	The proof then follows from dividing the interval $[u_{-1},\hat{u}_f+\delta]$ into subintervals of size $U-u_{-1}$ and repeating the proof, using now the estimate \eqref{eq:f3vinftyinduction} for $f^3$ in place of the assumptions on $j^3$ and $A$.  The desired smallness holds if $\tau$ is sufficiently small with respect to $\hat{u}_f+\delta$.  The smoothness of the foliation follows from the smoothness of the ambient spacetime, the smoothness of the $\hat{u}_f$ normalised $\I$ gauge, the smoothness of the functions $j^3$, $j^4$ and $q$, and thus the smoothness of $f^3$.
\end{proof}

\subsubsection*{Foliations of the outgoing null hypersurface $\newsph{C}_{u_{-1}}$}

Consider now an outgoing null hypersurface $C$ in $(\mathcal{M},g)$ foliated by spheres described by the level hypersurfaces of a function $v:C \to \mathbb{R}$.  Consider the null generator $L$ of $C$ satisfying $L (v) = 1$.  There exists a unique \emph{conjugate} null vector field $\underline{L}'$ on $C$ satisfying
\[
	g(L,\underline{L}') = -2.
\]

If $C$ is an outgoing null hypersurface of a given spacetime double null foliation, and its foliation by spheres coincides with that obtained by intersecting $C$ with the incoming null hypersurfaces of the spacetime double null foliation, then the vectors of the double null frame $e_3$, $e_4$ are related to $L$, $\underline{L}'$ by
\[
	\underline{L}' = \Omega^{-1} e_3,
	\qquad
	L = \Omega e_4.
\]
Again, many of the Ricci coefficients and curvature components of the spacetime double null foliation can be expressed entirely in terms of $L$ and $\underline{L}'$.  For example, if $X$ is a vector field tangent to the level hypersurfaces of $v$,
\[
	\etabar (X) = -\frac{1}{2} g(\nabla_{L} X, \underline{L}'),
	\qquad
	\rho = \frac{1}{4} R(L,\underline{L}',L,\underline{L}').
\]
Moreover, certain $\Omega$ rescaled quantities can again also be expressed entirely in terms of $L$ and $\underline{L}'$.  For example, if $X$ and $Y$ are tangential to the level hypersurfaces of $v$,
\[
	\Omega \omegahat
	=
	\frac{1}{4} g(\nabla_{L} \underline{L}',L),
	\quad
	\Omega \chi(X,Y)
	=
	g(\nabla_{X} L,Y),
	\quad
	\Omega^{-1} \chibar(X,Y)
	=
	g(\nabla_{X} \underline{L}',Y),
\]
and
\[
	\Omega \hat{\chi}(X,Y)
	=
	\Omega \chi(X,Y)
	-
	\Omega \tr \chi \gslash(X,Y),
	\quad
	\Omega^{-1} \hat{\chibar}(X,Y)
	=
	\Omega^{-1} \chibar(X,Y)
	-
	\Omega^{-1} \tr \chibar \gslash(X,Y).
\]

The following proposition relates certain geometric quantities associated to two different foliations of the outgoing null hypersurface $\newsph{C}_{u_{-1}}$.

\begin{proposition}[Change of foliation relations on outgoing cones] \label{prop:outIconesrelations}
	For a given sphere $\newsph{S}$ as in \eqref{eq:newjsphere}, let $v$ denote the restriction to $\newsph{C}_{u_{-1}}$ of the $v$ coordinate of the extended $\hat{u}_f$ normalised gauge.  Consider another foliation of $\newsph{C}_{u_{-1}}$ by spheres, described by the level hypersurfaces of a function $\widetilde{v}$, and a mass $M$.  Suppose $v$ and $\widetilde{v}$ are related by
	\[
		v = \widetilde{v} + f^4(\widetilde{v},\theta),
	\]
	for some function $f^4:[v(R_{-2},u_{-1}),v_{\infty}(\hat{u}_f+\delta)] \times \mathbb{S}^2 \to \mathbb{R}$.  Then the corresponding geometric quantities $\mubar^{\dagger}$ and $\Omega \omegahat$ are related by
	\begin{align*}
		\widetilde{\Deltaslash} \log (1+\partial_{\widetilde{v}} f^4(\widetilde{v},\theta))
		=
		&
		\widetilde{\mubar}^{\dagger}_{M}(\widetilde{v},\theta)
		-
		\mubar^{\dagger}_{M_f}(\widetilde{v} + f^4,\theta)
		+
		\frac{\Omega_{\circ}^2}{2r} \Omega^{-1} \tr \chibar (\widetilde{v} + f^4,\theta)
		\big( (1+ \partial_{\widetilde{v}} f^4(\widetilde{v},\theta))^{-1} - 1 \big)
		\\
		&
		+
		F_{\mubar^{\dagger}} (f^4,\widetilde{\nablaslash} f^4, \widetilde{\nablaslash}{}^2 f^4,\partial_{\widetilde{v}} f^4, \widetilde{\nablaslash} \partial_{\widetilde{v}} f^4,M),
	\end{align*}
	for a function $F_{\mubar^{\dagger}}$ which, provided $\vert f^4 \vert + \vert \widetilde{r \nablaslash} f^4 \vert + \vert (\widetilde{r \nablaslash})^2 f^4 \vert+\vert \partial_{\widetilde{v}} f^4 \vert + \vert \widetilde{r \nablaslash} \partial_{\widetilde{v}} f^4 \vert \lesssim 1$, satisfies
	\[
		r^2 \vert F_{\mubar^{\dagger}} (f^4,\widetilde{\nablaslash} f^4, \widetilde{\nablaslash}{}^2 f^4,\partial_{\widetilde{v}} f^4, \widetilde{\nablaslash} \partial_{\widetilde{v}} f^4) \vert
		\lesssim
		\vert f^4 \vert
		+
		\vert \widetilde{r \nablaslash} f^4 \vert
		+
		\vert (\widetilde{r \nablaslash})^2 f^4 \vert
		+
		\vert M - M_f \vert
		,
	\]
	and
	\[
		\frac{1}{2} \partial_{\widetilde{v}} \log(1+\partial_{\widetilde{v}} f^4)(\widetilde{v},\theta)
		=
		\widetilde{\Omega \omegahat}(\widetilde{v},\theta)
		-
		(1+\partial_{\widetilde{v}} f^4) \Omega \omegahat(\widetilde{v} + f^4,\theta).
	\]
	Here $\widetilde{\nablaslash}$ denotes the Levi-Civita connection of the induced metric on the level hypersurfaces of $\widetilde{v}$.
\end{proposition}

\begin{proof}
	Let $L$ and $\underline{L}$ denote the null generators of $C$ which satisfy
	\[
		L(v) 
		=
		1,
		\qquad 
		\widetilde{L}(\widetilde{v})
		=
		1,
	\]
	respectively, and let $\underline{L}'$ and $\widetilde{\underline{L}}'$ denote the conjugate null vectors to $L$ and $\widetilde{L}$ respectively, satisfying $g(L,\underline{L}') = g (\widetilde{L},\widetilde{\underline{L}}') = -2$.  Let $e_A = \partial_{\theta^A}$ in the $(v,\theta)$ coordinate system for $C$, and $\widetilde{e}_A = \widetilde{\partial}_{\theta^A}$ in the $(\widetilde{v},\theta)$ coordinate system.  It follows, as in the proof of Proposition \ref{prop:inIconesrelations}, that
	\begin{align*}
		\widetilde{L}
		&
		=
		(1+\partial_{\widetilde{v}} f^4) L,
		\\
		\widetilde{\underline{L}}'
		&
		=
		(1+\partial_{\widetilde{v}} f^4)^{-1}
		\big[
		\underline{L}'
		+
		\vert \widetilde{\nablaslash} f^4 \vert^2
		L
		+
		2
		\gslash^{AB}
		\partial_{\theta^A} f^4 e_B
		\big],
		\\
		\widetilde{e}_A
		&
		=
		e_A
		+
		\partial_{\theta^A} f^4 L,
	\end{align*}
	where $\vert \widetilde{\nablaslash} f^4 \vert^2 = \gslash^{AB} \partial_{\theta^A} f^4 \partial_{\theta^B} f^3$.
	
	As in the proof of Proposition \ref{prop:inIconesrelations} one computes
	\begin{align*}
		\widetilde{\etabar} (\widetilde{e}_A)
		&
		=
		(1+\partial_{\widetilde{v}} f^4)^{-1}
		\partial_{\widetilde{v}} \partial_{\theta^A} f^4
		+
		\etabar(e_A)
		-
		\frac{1}{2}
		\Omega {\chi_A}^B \partial_{\theta^B} f^4
		+
		2 \Omega \omegahat \partial_{\theta^A} f^4,
		\\
		\widetilde{\rho}
		&
		=
		\rho
		+
		\widetilde{\nablaslash} f^4 \otimes \widetilde{\nablaslash} f^4 \cdot \Omega^2 \alpha
		-
		2 \widetilde{\nablaslash} f^4 \cdot \Omega \beta,
		\\
		\widetilde{\Omega \chi} (\widetilde{e}_A, \widetilde{e}_B)
		&
		=
		(1+\partial_{\widetilde{v}} f^4) \Omega \chi (e_A,e_B),
	\end{align*}
	and
	\begin{align*}
		&
		\widetilde{\Omega^{-1} \chibar} (\widetilde{e}_A, \widetilde{e}_B)
		=
		(1+\partial_{\widetilde{v}} f^4)^{-1} 
		\big[
		2 \widetilde{\nablaslash}{}^2_{A,B} f^4
		+
		\Omega^{-1} \chibar (e_A,e_B)
		+
		2(\etabar_A \partial_{\theta^B}f^4 + \etabar_B \partial_{\theta^A}f^4)
		+
		4 \Omega \omegahat \partial_{\theta^A} f^4 \partial_{\theta^B} f^4
		\\
		&
		+
		\vert \widetilde{\nablaslash} f^4\vert^2 \Omega \chi_{AB}
		-
		2\partial_{\theta^C} f^4 (\Omega \chi_A^C \partial_{\theta^B} f^4 + \Omega \chi_B^C \partial_{\theta^A} f^4)
		-
		2(\partial_{\theta^A} \log(1+\partial_{\widetilde{v}} f^4) \partial_{\theta^B} f^4
		+
		\partial_{\theta^B} \log(1+\partial_{\widetilde{v}} f^4) \partial_{\theta^A} f^4
		)
		\big].
	\end{align*}
	The latter in particular implies that
	\[
		\widetilde{\Omega^{-1} \tr \chibar}
		=
		(1+\partial_{\widetilde{v}} f^4)^{-1}
		\big[
		\Omega^{-1} \tr \chi
		+
		2 \widetilde{\Deltaslash} f^4
		+
		4 \etabar \cdot \widetilde{\nablaslash} f^4
		+
		(4 \Omega \omegahat + \Omega \tr \chi) \vert \widetilde{\nablaslash} f^4 \vert^2
		-
		4 \Omega \chi \cdot \widetilde{\nablaslash} f^4 \otimes \widetilde{\nablaslash} f^4
		-
		4
		\widetilde{\nablaslash} \log(1+ \partial_{\widetilde{v}} f^4) \cdot \widetilde{\nablaslash} f^4
		\big].
	\]
	The expression involving $\widetilde{\Omega \omegahat}$ can be computed similarly, using the fact that,
	\[
		\Omega \omegahat = \frac{1}{4} g(\nabla_{L} \underline{L}', L).
	\]
	The proof then follows from the fact that
	\[
		\underline{\mu}^{\dagger}_M
		=
		\slashed{div} \underline{\eta}
		+
		\rho - \rho_{\circ,M}
		-
		\frac{1}{2} \hat{\chi} \cdot \underline{\hat{\chi}}
		+
		\frac{\Omega_{\circ,M}^2}{2r_M}
		\left(
		\frac{tr \underline{\chi}}{\Omega} - \frac{(\Omega \tr \underline{\chi})_{\circ,M}}{\Omega^2_{\circ,M}}
		\right).
	\]
	and
	\[
		\vert
		\rho_{\circ,M}(\widetilde{v},\theta)
		-
		\rho_{\circ,M_f}(\widetilde{v} + f^4,\theta)
		\vert
		+
		\Big\vert
		\frac{1}{r_M}(\Omega \tr \underline{\chi})_{\circ,M}(\widetilde{v},\theta)
		-
		\frac{1}{r_{M_f}} (\Omega \tr \underline{\chi})_{\circ,M_f}(\widetilde{v} + f^4,\theta)
		\Big\vert
		\lesssim
		\frac{\vert M - M_f \vert + \vert f^4 \vert}{r_{M_f}^2}.
	\]
\end{proof}

The following proposition shows that certain foliations of the cone $\newsph{C}_{u_{-1}}$ can be attained.

\begin{proposition}[Foliations of the outgoing cone $\newsph{C}_{u_{-1}}$] \label{prop:foliationIoutgoingcone}
	Let $\newsph{S}$ be a sphere defined, in the $(u_{\I}, v_{\I},\theta_{\I})$ coordinate system, by smooth functions $j^3$, $j^4$,
	\[
		\newsph{S} = \{ i_{\I} (u_{-1} + j^3(\theta), v_{\infty}(\hat{u}_f+\delta) + j^4(\theta),\theta) \mid \theta \in \mathbb{S}^2 \}.
	\]
	such that $j^3$ and $j^4$ satisfy
	\[
		\sum_{k \leq 5}
		\big(
		\Vert (r \nablaslash)^k j^3 \Vert_{\newsph{S}}
		+
		\Vert (r \nablaslash)^k j^4 \Vert_{\newsph{S}}
		\big)
		\leq
		\tau.
	\]
	Consider also a mass $M$ satisfying the smallness assumption of Proposition \ref{prop:Iconesfoliations}, a constant $B \in \mathbb{R}$, and a smooth function $\mathcal{F}:[v(R_{-2},u_{-1}), v_{\infty}(\hat{u}_f+\delta)] \to \mathbb{R}$ satisfying
	\[
		\vert B \vert 
		+
		\sup_{v(R_{-2},u_{-1}) \leq v \leq v_{\infty}(\hat{u}_f+\delta)}
		\vert \mathcal{F}(v) \vert
		\leq
		\tau.
	\]
	Then, provided $\tau$ and $\delta_0$ are sufficiently small, there exists a smooth foliation of the past outgoing cone, $\newsph{C}_{u_{-1}}$, of $\newsph{S}$ such that, on $\newsph{C}_{u_{-1}}$,
	\[
		\mubar^{\dagger}_{\ell \geq 1}(\widetilde{v},\theta) = 0,
		\qquad
		(\Omega \omegahat - \Omega \omegahat_{\circ,M})_{\ell = 0}(\widetilde{v})
		=
		\mathcal{F}(\widetilde{v}),
	\]
	for all $\widetilde{v} \in [v(R_{-2},u_{-1}), v_{\infty}(\hat{u}_f+\delta)]$, $\theta \in \mathbb{S}^2$.  Moreover,
	\[
		\partial_{\widetilde{v}} f^4_{\ell = 0} (u_{-1},v_{\infty}(\hat{u}_f+\delta))
		=
		B,
	\]
	where $f^4$ is the diffeomorphism which relates this foliation of $\newsph{C}_{u_{-1}}$ to the foliation of $\newsph{C}_{u_{-1}}$ obtained by intersecting with the incoming cones $\Cbar_v$, for $v(R_{-2},u_{-1}) \leq v \leq v_{\infty}(\hat{u}_f+\delta)$, of the extended $\hat{u}_f$ normalised gauge, so that $v = \widetilde{v} + f^3(\widetilde{v},\theta)$.
\end{proposition}

\begin{proof}
	The proof is similar to that of Proposition \ref{prop:foliationIincomingcone}.  By Proposition \ref{prop:outIconesrelations}, the goal is to construct a function $f^4:[v(R_{-2},u_{-1}),v_{\infty}(\hat{u}_f+\delta)] \times \mathbb{S}^2 \to \mathbb{R}$ satisfying
	\[
		\Big[
		\widetilde{\Deltaslash} \log (1+\partial_{\widetilde{v}} f^4(\widetilde{v},\theta))
		+
		\frac{1}{r^2} \log(1+\partial_{\widetilde{v}} f^4(\widetilde{v},\theta))
		\Big]_{\widetilde{\ell \geq 1}}
		=
		\Big[
		-
		\mubar^{\dagger}(\widetilde{v} + f^4,\theta)
		+
		F_{\mubar^{\dagger}}' (f^4,\widetilde{\nablaslash} f^4, \widetilde{\nablaslash}{}^2 f^4,\partial_{\widetilde{v}} f^4, \widetilde{\nablaslash} \partial_{\widetilde{v}} f^4)
		\Big]_{\widetilde{\ell \geq 1}},
	\]
	and
	\[
		\frac{1}{2}
		\big[
		\partial_{\widetilde{v}} \log(1+\partial_{\widetilde{v}} f^4)(\widetilde{v},\cdot)
		\big]_{\widetilde{\ell =0}}
		=
		\mathcal{F}(\widetilde{v})
		+
		\big[
		\Omega \omegahat_{\circ}(\widetilde{v})
		-
		(1+\partial_{\widetilde{v}} f^4(\widetilde{v},\cdot)) \Omega \omegahat(\widetilde{v} + f^4(\widetilde{v},\cdot),\cdot)
		\big]_{\widetilde{\ell =0}},
	\]
	along with the \emph{final} conditions,
	\[
		f^4(v_{\infty}(\hat{u}_f+\delta),\cdot) = j^4,
		\qquad
		\partial_{\widetilde{v}} f^4_{\ell = 0} (v_{\infty}(\hat{u}_f+\delta))
		=
		B,
	\]
	where
	\begin{multline*}
		F_{\mubar^{\dagger}}' (f^4,\widetilde{\nablaslash} f^4, \widetilde{\nablaslash}{}^2 f^4,\partial_{\widetilde{v}} f^4, \widetilde{\nablaslash} \partial_{\widetilde{v}} f^4)
		=
		F_{\mubar^{\dagger}} (f^4,\widetilde{\nablaslash} f^4, \widetilde{\nablaslash}{}^2 f^4,\partial_{\widetilde{v}} f^4, \widetilde{\nablaslash} \partial_{\widetilde{v}} f^4)
		\\
		+
		\frac{\Omega_{\circ}^2}{2r} \Omega^{-1} \tr \chibar (\widetilde{v} + f^4,\theta)
		\big( (1+ \partial_{\widetilde{v}} f^4(\widetilde{v},\theta))^{-1} - 1 \big)
		+
		\frac{1}{r^2} \log(1+\partial_{\widetilde{v}} f^4(\widetilde{v},\theta)),
	\end{multline*}
	and $\mubar^{\dagger}$ and $\Omega \omegahat$ denote the quantities of the gauge obtained by intersecting $\newsph{C}_{u_{-1}}$ with the incoming hypersurfaces of the extended $\hat{u}_f$ normalised gauge (which, again, are arbitrarily close to their values of the extended $\hat{u}_f$ normalised gauge provided $\tau$ and $\delta$ are sufficiently small).  Define, therefore, $f_{[0]} = 0$ and, for $n \geq 1$, iterates $f_{[n]}$ as solutions of
	\begin{multline} \label{eq:u0foliationiterates1}
		\Big[
		\Deltaslash_{[n-1]} \log (1+\partial_{v} f^4_{[n]}(v,\theta))
		+
		\frac{1}{r^2} \log(1+\partial_{v} f^4_{[n]}(v,\theta))
		\Big]_{(\ell \geq 1)_{[n-1]}}
		=
		\Big[
		-
		\mubar^{\dagger}(v + f^4_{[n-1]},\theta)
		\\
		+
		F_{\mubar^{\dagger}}' (f^4_{[n-1]},\nablaslash f^4_{[n-1]}, \nablaslash^2 f^4_{[n-1]},\partial_{v} f^4_{[n-1]}, \nablaslash \partial_{v} f^4_{[n-1]})
		\Big]_{(\ell \geq 1)_{[n-1]}},
	\end{multline}
	and
	\begin{equation} \label{eq:u0foliationiterates2}
		\frac{1}{2}
		\big[
		\partial_{v} \log(1+\partial_{v} f^4_{[n]})(v,\cdot)
		\big]_{(\ell =0)_{[n-1]}}
		=
		\mathcal{F}(v)
		+
		\big[
		\Omega \omegahat_{\circ}(v)
		-
		(1+\partial_{v} f^4_{[n-1]}(v,\cdot)) \Omega \omegahat(v + f^4_{[n-1]}(v,\cdot),\cdot)
		\big]_{(\ell =0)_{[n-1]}},
	\end{equation}
	along with the initial conditions,
	\[
		f^4_{[n]}(v_{\infty}(\hat{u}_f+\delta),\cdot) = j^4,
		\qquad
		\partial_{\widetilde{v}} (f^4_{[n]})_{(\ell =0)_{[n-1]}} (v_{\infty}(\hat{u}_f+\delta))
		=
		B,
	\]
	where $(\ell \geq 1)_{[n-1]}$ and $(\ell =0)_{[n-1]}$ denote the mode projections with respect to the spheres defined as the level hypersurfaces of $v_{[n-1]}$, where $v_{[n-1]}$ is defined implicitly by $v = v_{[n-1]} + f^4_{[n-1]}(v_{[n-1]},\theta)$.
	
	Consider some $V < v_{\infty}(\hat{u}_f+\delta)$.  As in the proof of Proposition \ref{prop:foliationIincomingcone} it follows by induction, using an elliptic estimate for \eqref{eq:u0foliationiterates1} and transport estimate for \eqref{eq:u0foliationiterates2}, that, provided $v_{\infty}(\hat{u}_f+\delta) - V$ is sufficiently small,
	\begin{equation} \label{eq:f4u0induction}
		\sup_{V \leq v \leq v_{\infty}(\hat{u}_f+\delta)}
		\sum_{k \leq 5}
		\Vert (r\nablaslash)^k f^4_{[n]} \Vert_{S_{u_{-1},v}}
		\lesssim
		\tau,
		\qquad
		\sup_{V \leq v \leq v_{\infty}(\hat{u}_f+\delta)}
		\sum_{k \leq 5}
		\Vert (r\nablaslash)^k \partial_v f^4_{[n]} \Vert_{S_{u_{-1},v}}
		\lesssim
		\tau,
	\end{equation}
	for all $n$, provided $\tau$, $v_{\infty}(\hat{u}_f+\delta) - V$ and $\varepsilon$ are sufficiently small, and moreover,
	\begin{align*}
		\sup_{V \leq v \leq v_{\infty}(\hat{u}_f+\delta)}
		\sum_{k \leq 5} \Vert (r\nablaslash)^k( f^4_{[n+1]} - f^4_{[n]} ) \Vert_{S_{{u_{-1},v}}}
		\lesssim
		(v_{\infty} - V)
		\sup_{V \leq v \leq v_{\infty}(\hat{u}_f+\delta)}
		\sum_{k \leq 5} \Vert (r\nablaslash)^k( f^4_{[n]} - f^4_{[n-1]} ) \Vert_{S_{u_{-1},v}}.
	\end{align*}
	Hence, if $v_{\infty}(\hat{u}_f+\delta) - V$ is sufficiently small, the map which takes $f^4_{[n]}$ to $f^4_{[n+1]}$ is a contraction and the sequence $\{f^4_{[n]}\}$ converges to a limit $f^4$ on $[V,v_{\infty}(\hat{u}_f+\delta)]$ satisfying the estimate \eqref{eq:f4u0induction}.
	
	As in the proof of Proposition \ref{prop:foliationIincomingcone}, the proof follows from dividing the interval $[v(R_{-2},u_{-1}), v_{\infty}(\hat{u}_f+\delta)]$ into subintervals of size $v_{\infty} - V$ and repeating the proof, using now the estimate \eqref{eq:f4u0induction} for $f^4$ in place of the assumptions on $j^4$ and $B$.  The desired smallness holds if $\tau$ is sufficiently small with respect to $\hat{u}_f+\delta$.
\end{proof}

\subsubsection*{The proof of Proposition \ref{prop:Iconesfoliations}: foliation of cones of any sphere close to $S_{\hat{u}_f+\delta, v_{\infty}(\hat{u}_f+\delta)}$}

The proof of Proposition \ref{prop:Iconesfoliations} can now be given.

\begin{proof}[Proof of Proposition \ref{prop:Iconesfoliations}]
	\underline{\textbf{Overview:}}
	It will first be shown that, for any sphere close to $S_{u_{-1},v_{\infty}(\hat{u}_f+\delta)}$ (rather than the sphere $S_{\hat{u}_f+\delta,v_{\infty}(\hat{u}_f+\delta)}$), the desired foliations of its corresponding cones can be found.  Suppose therefore that $\newsph{S}$ is a sphere defined, in the $(u_{\I}, v_{\I},\theta_{\I})$ coordinate system, by smooth functions $j^3$, $j^4$,
	\begin{equation} \label{eq:initialIjsphere}
		\newsph{S} = \{ i_{\I} (u_{-1} + j^3(\theta), v_{\infty}(\hat{u}_f+\delta) + j^4(\theta),\theta) \mid \theta \in \mathbb{S}^2\},
	\end{equation}
	where $j^3$ and $j^4$ satisfy
	\begin{equation} \label{eq:initialIjsphere2}
		\sum_{k \leq 5}
		\big(
		\Vert (r \nablaslash)^k j^3 \Vert_{S_{u_{-1},v_{\infty}}}
		+
		\Vert (r \nablaslash)^k j^4 \Vert_{S_{u_{-1},v_{\infty}}}
		\big)
		\leq
		\tau,
	\end{equation}
	so that $\newsph{S}$ is appropriately close to the sphere $S_{u_{-1}, v_{\infty}(\hat{u}_f+\delta)}$.
	
	Let $\newsph{\Cbar}_{v_{\infty}(\hat{u}_f+\delta)}$ denote the incoming null hypersurface of the sphere $\newsph{S}$ and let $\newsph{C}_{u_{-1}}$ denote the outgoing null hypersurface of $\newsph{S}$.  Let $\newsph{S}$ be identified with the standard sphere so that the $\mathbb{S}^2$ part of the diffeomorphisms relating the new double null foliation to the extended $\hat{u}_f$ normalised gauge satisfy $f^1 = f^2 =0$ on $\newsph{S}$.  It then follows that
	\[
		f^4(u_{-1},v_{\infty}(\hat{u}_f+\delta),\theta) = j^4(\theta),
		\qquad
		f^3(u_{-1},v_{\infty}(\hat{u}_f+\delta),\theta) = j^3(\theta),
	\]
	for all $\theta \in \mathbb{S}^2$.
	
	Throughout the proof $v_{\infty} = v_{\infty}(\hat{u}_f+\delta)$.
	
	The goal is to choose $A$, $B$, $\mathcal{F}$, $q$ appropriately so that the corresponding foliations of $\newsph{\Cbar}_{v_{\infty}(\hat{u}_f+\delta)}$ and $\newsph{C}_{u_{-1}}$ of Proposition \ref{prop:foliationIincomingcone} and Proposition \ref{prop:foliationIoutgoingcone} give rise to a resulting spacetime double null foliation with the desired properties.  The constants $A$ and $B$ will be chosen to achieve, in the resulting spacetime double null foliation,
	\begin{equation} \label{eq:wanttoachieve1}
		( \Omega_{\circ,M}^{-2} \Omega^2 - 1)_{\ell = 0}(u_{-1},v_{\infty}) = 0,
	\end{equation}
	and
	\begin{equation} \label{eq:wanttoachieve2}
		(\Omega \tr \chi - \Omega \tr \chi_{\circ,M})_{\ell=0} (\hat{u}_f+\delta,v_{\infty}) = 0.
	\end{equation}
	The function $q$ will be chosen so that, in the resulting spacetime double null foliation, the function $\underline{\mathcal{F}}$ of Proposition \ref{prop:foliationIincomingcone} satisfies
	\begin{equation} \label{eq:wanttoachieve3}
		\underline{\mathcal{F}}
		=
		\partial_u \big( \big( \Omega^2 - \Omega_{\circ,M}^{2} \big)_{\ell=0}\big)
		-
		(\Omega \omegabarhat - \Omega \omegabarhat_{\circ,M})_{\ell=0}.
	\end{equation}
	The conditions \eqref{eq:wanttoachieve1} and \eqref{eq:wanttoachieve3} will then imply that
	\[
		( \Omega^2 - \Omega_{\circ,M}^{2})_{\ell = 0}(u,v_{\infty}) = 0,
	\]
	for all $u_{-1} \leq u \leq \hat{u}_f+\delta$ (see equation \eqref{eq:OmegacalFbar}).
	Finally, $\mathcal{F}$ will be chosen so that, in the resulting spacetime double null foliation,
	\begin{equation} \label{eq:wanttoachieve4}
		\mathcal{F}(v)
		=
		\frac{\Omega_{\circ,M}^2}{r^3}(u_{-1},v)
		\frac{1}{2} \int_{u_{-1}}^{\hat{u}_f+\delta} \int_{\bar{u}}^{\hat{u}_f+\delta}
		r^3 (\Omega^{-1} \hat{\chi}, \Omega^2 \alphabar)_{\ell=0}
		\left(\hat{u}, v_{\infty} \right)
		d\hat{u}
		d\bar{u},
	\end{equation}
	as desired.
	
	The desired $A$ and $B$ will be obtained as solutions of two nonlinear equations.
	The change of spacetime gauge relations \eqref{eq:metriccomp1}--\eqref{eq:metriccomp6} give a nonlinear equation for $A$ and $B$, involving $j^3$, $j^4$ and quantities in the extended $\hat{u}_f$ normalised gauge, which guarantees that the condition \eqref{eq:wanttoachieve1} will be achieved in the resulting spacetime double null foliation.  The corresponding relation required to achieve \eqref{eq:wanttoachieve2} is more complicated, however, due to the fact that \eqref{eq:wanttoachieve2} is a condition at $(\hat{u}_f+\delta, v_{\infty})$, rather than at $(u_{-1},v_{\infty})$.  To arrive at such an equation then, consider some given spacetime double null foliation.  Equation \eqref{eq:trchi3} can be used to relate $(\Omega \tr \chi - \Omega \tr \chi_{\circ})_{\ell=0} (\hat{u}_f+\delta,v_{\infty})$ to $(\Omega \tr \chi - \Omega \tr \chi_{\circ})_{\ell=0} (u_{-1},v_{\infty})$.  The right hand side of the equation involves quantities such as $\rho- \rho_{\circ}$, $\Omega^{-1} \hat{\chi}$, etc., which can be related, by Proposition \ref{prop:inIconesrelations}, to the corresponding quantities in the extended $\hat{u}_f$ normalised gauge, $f^3$, $\partial_u f^3$, and also the quantity $\Omega_{\circ}^{-1} \Omega$.  This latter quantity can be re-expressed, using equation \eqref{eq:DlogOmega}, in terms of $\Omega \omegabarhat - \Omega \omegabarhat_{\circ}$, which can again be expressed in terms of $f^3$ by Proposition \ref{prop:inIconesrelations}, and $\Omega_{\circ}^{-1} \Omega(u_{-1},v_{\infty})$ (see equation \eqref{eq:Omegaeomegabar}).  Using again the change of spacetime gauge relations \eqref{eq:metriccomp1}--\eqref{eq:metriccomp6}, $\Omega_{\circ}^{-1} \Omega(u_{-1},v_{\infty})$ can then be expressed in terms of $A$, $B$, $j^3$, $j^4$ and quantities in the extended $\hat{u}_f$ normalised gauge.  
	
	Similarly, in order to achieve the condition \eqref{eq:wanttoachieve3} the fact that $\Omega_{\circ}^{-1} \Omega(u_{-1},v_{\infty})$ can be expressed in terms of $A$, $B$, $j^3$, $j^4$ and quantities in the extended $\hat{u}_f$ normalised gauge is used to define $q$ appropriately.

	The result is a closed system for $A$ and $B$ which involves the $f^3(u)$ and $\partial_u f^3(u)$, for all $u_{-1} \leq u \leq \hat{u}_f+\delta$, arising from the foliation of Proposition \ref{prop:foliationIincomingcone}.  Sequences of iterates $\{ A_{[n]} \}$ and $\{ B_{[n]}\}$ are defined and are shown to be bounded, uniformly in $n$, from which it follows that there exist convergent subsequences.  One then considers the foliation of $\newsph{\Cbar}_{v_{\infty}(\hat{u}_f+\delta)}$ defined by the limits and Proposition \ref{prop:foliationIincomingcone}.  Given this foliation of $\newsph{\Cbar}_{v_{\infty}(\hat{u}_f+\delta)}$, the quantities in \eqref{eq:wanttoachieve4} are all defined and $\newsph{C}_{u_{-1}}$ can be foliated, according to Proposition \ref{prop:foliationIoutgoingcone}, with this $\mathcal{F}$.

	\noindent \underline{\textbf{Equations:}}
	Suppose that, for some given $A$, $q$, the cone $\newsph{\Cbar}_{v_{\infty}(\hat{u}_f+\delta)}$ is foliated according to Proposition \ref{prop:foliationIincomingcone} and, for some given $B$, $\mathcal{F}$, the outgoing hypersurface $\newsph{C}_{u_{-1}}$, is foliated according to Proposition \ref{prop:foliationIoutgoingcone}.  Suppose that these foliations are described as the level hypersurfaces of functions $\newsph{u}$ and $\newsph{v}$ respectively.  The resulting spacetime double null foliation, obtained by considering the outgoing null cones associated to the spheres of constant $\newsph{u}$ in $\newsph{\Cbar}_{v_{\infty}(\hat{u}_f+\delta)}$ and the incoming null cones associated to the spheres of constant $\newsph{v}$ in $\newsph{C}_{u_{-1}}$, is related to the extended $\hat{u}_f$ normalised double null foliation by functions $f^3$, $f^4$ satisfying
	\begin{equation} \label{eq:initialIjsphere3}
		u = \newsph{u} + f^3(\newsph{u},\newsph{v},\newsph{\theta}),
		\qquad
		v = \newsph{v} + f^4(\newsph{u},\newsph{v},\newsph{\theta}),
		\qquad
		\theta^A = \newsph{\theta}^A + f^A(\newsph{u},\newsph{v},\newsph{\theta}),
	\end{equation}
	where $f^3(\cdot,v_{\infty}(\hat{u}_f+\delta),\cdot)$ agrees with $f^3$ of Proposition \ref{prop:foliationIincomingcone}, $f^4(u_{-1},\cdot,\cdot)$ agrees with $f^4$ of Proposition \ref{prop:foliationIoutgoingcone}, and $f^1$, $f^2$ are defined so that
	\[
		f^A(u_{-1},v_{\infty}(\hat{u}_f+\delta),\cdot) \equiv 0,
		\qquad
		A = 1,2,
	\]
	and the spacetime metric takes the standard double null form.  In particular,
	\[
		f^3(u_{-1},v_{\infty}(\hat{u}_f+\delta),\cdot) = j^3,
		\qquad
		f^4(u_{-1},v_{\infty}(\hat{u}_f+\delta),\cdot) = j^4.
	\]
	The change of spacetime gauge relations \eqref{eq:metriccomp6} and \eqref{eq:Riccicomp7} then imply that
	\begin{multline}
		\label{eq:IfoliationsA}
		(\newsph{\Omega \tr \chi} - \newsph{\Omega \tr \chi}_{\circ,M})_{\ell =0}(u_{-1},v_{\infty})
		-
		\frac{2}{r}( \newsph{\Omega}^2 - \newsph{\Omega}_{\circ,M}^2)_{\ell = 0}(u_{-1},v_{\infty})
		=
		-\frac{2\Omega^2}{r} A
		+
		\frac{2\Omega^2}{r^2} (j^3-j^4)_{\ell = 0}
		\\
		+
		(\Omega \tr \chi - \Omega \tr \chi_{\circ,M})(u_{-1}+j^3,v_{\infty}+j^4)_{\ell =0}
		-
		\frac{2}{r}( \Omega^2 - \Omega_{\circ,M}^2)(u_{-1}+j^3,v_{\infty}+j^4)_{\ell =0}
		+
		\mathfrak{E}^1,
	\end{multline}
	and
	\begin{equation} \label{eq:IfoliationsAB}
		( \newsph{\Omega}_{\circ,M}^{-2} \newsph{\Omega}^2 - 1)_{\ell = 0}(u_{-1},v_{\infty})
		=
		A + B
		-
		\frac{2M}{r^2} (j^3 - j^4)_{\ell=0}
		+
		( \Omega_{\circ,M}^{-2} \Omega^2 - 1)(u_{-1} + j^3,v_{\infty}+j^4)_{\ell = 0}
		+
		\mathfrak{E}^2,
	\end{equation}
	where $v_{\infty} = v_{\infty}(\hat{u}_f+\delta)$ and $\mathfrak{E}^1$, $\mathfrak{E}^2$ are nonlinearities, as determined by the relations of Proposition \ref{prop:metricrelations} and Proposition \ref{prop:Riccirelations}, which both take the schematic form
	\begin{align*}
		\mathfrak{E}^1
		&
		=
		\Big(
		\sum_{\substack{k \leq1 \\ \vert \gamma_1 \vert, \vert \gamma_2 \vert \leq 1}}
		\Pi_{\newsph{S}} \Phi \cdot (r\newsph{\nablaslash})^k \newsph{\mathfrak{D}}^{\gamma_1} f(x_{-1})
		+
		\newsph{\mathfrak{D}}^{\gamma_1} f \cdot (r\newsph{\nablaslash})^k \newsph{\mathfrak{D}}^{\gamma_2} f(x_{-1}) \Big)_{\ell=0},
	\\
		\mathfrak{E}^2
		&
		=
		\Big(
		\sum_{\vert \gamma_1 \vert, \vert \gamma_2 \vert \leq 1}
		\Pi_{\newsph{S}} \Phi \cdot \newsph{\mathfrak{D}}^{\gamma_1} f(x_{-1})
		+
		\newsph{\mathfrak{D}}^{\gamma_1} f \cdot \newsph{\mathfrak{D}}^{\gamma_2} f(x_{-1}) \Big)_{\ell=0},
	\end{align*}
	with $x_{-1} = (u_{-1},v_{\infty}(\hat{u}_f+\delta),\cdot)$,
	and $\newsph{\Omega}$ and $\newsph{\Omega \tr \chi}$ denote the values of $\Omega$ and $\Omega \tr \chi$ in the resulting spacetime double null foliation.
	
	Note that equation \eqref{eq:trchi3} can be rewritten
	\begin{align*}
		\Omega^{-1} \nablaslash_3
		\left(
		\Omega \tr \chi - (\Omega \tr \chi)_{\circ}
		\right)
		=
		\
		&
		2\divslash \eta
		+
		\frac{\Omega_{\circ}^2}{r \Omega^2}
		\left(
		\Omega \tr \chi - (\Omega \tr \chi)_{\circ}
		\right)
		-\frac{\Omega_{\circ}^2}{r \Omega^2}
		\left(
		\Omega \tr \chibar - (\Omega \tr \chibar)_{\circ}
		\right)
		+
		2 \frac{\Omega_{\circ}^2}{\Omega^2} (\rho - \rho_{\circ})
		\\
		&
		-
		\frac{4M}{r^3} \left( 1 - \frac{\Omega_{\circ}^2}{\Omega^2} \right)
		-
		\frac{1}{2\Omega^2}
		\left(
		\Omega \tr \chibar - (\Omega \tr \chibar)_{\circ}
		\right)
		\left(
		\Omega \tr \chi - (\Omega \tr \chi)_{\circ}
		\right)
		\\
		&
		+
		2
		\left( 1 - \frac{\Omega_{\circ}^2}{\Omega^2} \right)
		(\rho - \rho_{\circ})
		-
		(\Omega^{-1} \hat{\chibar}) \cdot (\Omega \hat{\chi})
		+
		2 \eta \cdot \eta,
	\end{align*}
	and so,
	\[
		(\newsph{\Omega \tr \chi} - \newsph{\Omega \tr \chi}_{\circ})(u_{-1},v_{\infty})
		=
		(\newsph{\Omega \tr \chi} - \newsph{\Omega \tr \chi}_{\circ})(\hat{u}_f+\delta,v_{\infty})
		+
		\mathcal{G}_0(\newsph{\Omega}_{\circ}^{-1} \newsph{\Omega}, \newsph{\Omega \tr \chibar} - \newsph{\Omega \tr \chibar}_{\circ}, \newsph{\Omega^{-1} \hat{\chi}}, \newsph{\Omega \hat{\chibar}},\newsph{\rho},\newsph{\eta})
		,
	\]
	for an appropriate function $\mathcal{G}_0$, which depends on the values of $\newsph{\Omega}_{\circ}^{-1} \newsph{\Omega}(u,v_{\infty})$, $(\newsph{\Omega \tr \chibar} - \newsph{\Omega \tr \chibar}_{\circ})(u,v_{\infty})$, etc.\@ for all $u_{-1} \leq u \leq \hat{u}_f+\delta$.  Recall equation \eqref{eq:Omegaeomegabar} and note that, by Proposition \ref{prop:inIconesrelations}, the quantities $\newsph{\Omega \omegabarhat} - \newsph{\Omega \omegabarhat}_{\circ}$, $\newsph{\Omega \tr \chibar} - \newsph{\Omega \tr \chibar}_{\circ}$, $\newsph{\Omega^{-1} \hat{\chi}}$, $\newsph{\Omega \hat{\chibar}}$, $\newsph{\rho}$, $\newsph{\eta}$ can be related to the corresponding quantities in the extended $\hat{u}_f$ normalised gauge and $f^3$ and its derivatives.  It follows then that
	\begin{multline*}
		(\newsph{\Omega \tr \chi} - \newsph{\Omega \tr \chi}_{\circ,M})(u_{-1},v_{\infty})
		=
		(\newsph{\Omega \tr \chi} - \newsph{\Omega \tr \chi}_{\circ,M})(\hat{u}_f+\delta,v_{\infty})
		\\
		+
		\mathcal{G}(\newsph{\Omega}_{\circ,M}^{-1} \newsph{\Omega}(u_{-1},v_{\infty}),
		f^3, \nablaslash f^3, \nablaslash^2 f^3,\partial_u f^3, \nablaslash \partial_u f^3, \nablaslash^2 \partial_u f^3, M)
		,
	\end{multline*}
	for some appropriate $\mathcal{G}$, which depends on $\Omega \omegabarhat - \Omega \omegabarhat_{\circ}$, $\Omega \tr \chibar - \Omega \tr \chibar_{\circ}$, etc.\@ in the extended $\hat{u}_f$ normalised gauge, and satisfies
	\begin{multline} \label{eq:foliationsG1}
		v_{\infty}^2 \vert \mathcal{G} \vert
		\lesssim
		\frac{\hat{u}_f}{v_{\infty}}
		\sup_{u_{-1} \leq u \leq \hat{u}_f+\delta}
		\sum_{k \leq 2}
		(\vert (r \nablaslash)^k f^3 \vert + \vert (r \nablaslash)^k \partial_u f^3 \vert)
		+
		\hat{u}_f
		\sup_{u_{-1} \leq u \leq \hat{u}_f+\delta}
		\sum_{k \leq 2}
		(\vert (r \nablaslash)^k f^3 \vert^2 + \vert (r \nablaslash)^k \partial_u f^3 \vert^2)
		\\
		+
		\frac{\hat{u}_f}{v_{\infty}} \vert \newsph{\Omega}_{\circ}^{-1} \newsph{\Omega} (u_{-1},v_{\infty}) \vert
		+
		\frac{\hat{u}_f}{v_{\infty}} \vert M - M_f\vert
		+
		\frac{\hat{u}_f \varepsilon}{v_{\infty}}
		.
	\end{multline}
	Moreover, the change of spacetime gauge relation \eqref{eq:metriccomp6} implies that
	\begin{multline} \label{eq:foliationsG1A}
		\vert \newsph{\Omega}_{\circ,M}^{-1} \newsph{\Omega} (u_{-1},v_{\infty}) - 1\vert
		\lesssim
		\sum_{k\leq 1}
		(
		\vert (r\nablaslash)^k j^3 \vert
		+
		\vert (r\nablaslash)^k j^3 \vert)
		+
		\vert \partial_u f^3 (u_{-1},v_{\infty}) \vert
		+
		\vert \partial_v f^3 (u_{-1},v_{\infty}) \vert
		\\
		+
		\vert \partial_v f^4 (u_{-1},v_{\infty}) \vert
		+
		\vert \partial_u f^4 (u_{-1},v_{\infty}) \vert
		+
		\vert M-M_f\vert
		+
		\varepsilon.
	\end{multline}
	The relations \eqref{eq:metriccomp1}--\eqref{eq:metriccomp6} moreover imply that, if $\tau$ and $\delta_0$ are sufficiently small,
	\begin{align}
		&
		\sum_{k \leq 4}
		\big(
		\Vert (r \nablaslash)^k \partial_u f^3 \Vert_{S_{u_{-1},v_{\infty}}}
		+
		\Vert (r \nablaslash)^k \partial_v f^3 \Vert_{S_{u_{-1},v_{\infty}}}
		+
		\Vert (r \nablaslash)^k \partial_u f^4 \Vert_{S_{u_{-1},v_{\infty}}}
		+
		\Vert (r \nablaslash)^k \partial_v f^4 \Vert_{S_{u_{-1},v_{\infty}}}
		\big)
		\nonumber
		\\
		&
		\qquad
		\lesssim
		\sum_{k \leq 5}
		\big(
		\Vert (r \nablaslash)^k j^3 \Vert_{S_{u_{-1},v_{\infty}}}
		+
		\Vert (r \nablaslash)^k j^4 \Vert_{S_{u_{-1},v_{\infty}}}
		\big)
		+
		\vert A \vert
		+
		\vert B \vert,
		\label{eq:foliationsG2}
	\end{align}
	since $f^A(u_{-1},v_{\infty},\cdot) = 0$ and $\newsph{b}(u_{-1},v_{\infty},\cdot) = 0$.

	The goal then is to find $A$ and $B$ such that
	\begin{multline*}
		\mathcal{G}(\newsph{\Omega}_{\circ,M}^{-1} \newsph{\Omega}(u_{-1},v_{\infty}),
		f^3, \nablaslash f^3, \nablaslash^2 f^3,\partial_u f^3, \nablaslash \partial_u f^3, \nablaslash^2 \partial_u f^3,M)
		=
		-\frac{2\Omega^2}{r} A
		+
		\frac{2\Omega^2}{r^2} (j^3-j^4)_{\ell = 0}
		\\
		+
		(\Omega \tr \chi - \Omega \tr \chi_{\circ,M})(u_{-1}+j^3,v_{\infty}+j^4)_{\ell =0}
		-
		\frac{2}{r}( \Omega^2 - \Omega_{\circ,M}^2)(u_{-1}+j^3,v_{\infty}+j^4)_{\ell =0}
		+
		\mathfrak{E}^1,
	\end{multline*}
	and
	\begin{equation*}
		0
		=
		A + B
		-
		\frac{2M}{r^2} (j^3 - j^4)_{\ell=0}
		+
		( \Omega_{\circ,M}^{-2} \Omega^2 - 1)(u_{-1} + j^3,v_{\infty}+j^4)_{\ell = 0}
		+
		\mathfrak{E}^2,
	\end{equation*}
	where $f^3$ denotes the change of foliation of the cone arising from Proposition \ref{prop:foliationIincomingcone} with $\partial_u f^3_{\ell = 0} (u_{-1},v_{\infty}) = A$ and
	\[
		q
		=
		\newsph{\Omega}_{\circ,M}^{-1} \newsph{\Omega} (u_{-1},v_{\infty}),
	\]
	where $\newsph{\Omega}_{\circ,M}^{-1} \newsph{\Omega}(u_{-1},v_{\infty})$ can be computed from $A$, $B$, $j^3$, $j^4$ from the relations \eqref{eq:metriccomp1}--\eqref{eq:metriccomp6}.

	\noindent \underline{\textbf{Definition of iterates:}}
	In order to begin the iteration, define first
	\[
		A_{[1]} = q = 0.
	\]
	Consider the foliation of $\newsph{\Cbar}_{v_{\infty}}$ arising from Proposition \ref{prop:foliationIincomingcone} with $A = A_{[1]} = 0$ and $q = q_{[1]} = 0$.  This foliation of $\newsph{\Cbar}_{v_{\infty}}$, described by the level hypersurfaces of a function $u_{[1]}$, is related to $u$ by a function $f^3_{[1]}$ by
	\[
		u = u_{[1]} + f^3_{[1]}(u_{[1]},\theta).
	\]
	The corresponding geometric quantities of this foliation of $\newsph{\Cbar}_{v_{\infty}}$, defined by \eqref{eq:foliationIincomingcone1}--\eqref{eq:foliationIincomingcone4}, are denoted $\eta_{[1]}$, $\Omega\omegabarhat_{[1]}$ etc.
	
	Define also
	\[
		B_{[1]} = 0,
		\qquad
		\mathcal{F}_{[1]}(v)
		=
		\frac{\Omega_{\circ,M}^2}{r^3}(u_{-1},v)
		F_{[1]}(u_{-1}),
	\]
	\[
		F_{[1]}(u_{-1})
		=
		\frac{1}{2} \int_{u_{-1}}^{\hat{u}_f+\delta} \int_{\bar{u}}^{\hat{u}_f+\delta}
		r^3 (\Omega^{-1} \hat{\chi}_{[1]}, \Omega^2 \alphabar_{[1]})_{\ell=0}
		\left(\hat{u}, v_{\infty}(\hat{u}_f+\delta) \right)
		d\hat{u}
		d\bar{u},
	\]
	and consider the foliation of $\newsph{C}_{u_{-1}}$ arising from Proposition \ref{prop:foliationIoutgoingcone} with $B = B_{[1]} = 0$ and $\mathcal{F} = \mathcal{F}_{[1]}$.  This foliation of $\newsph{C}_{u_{-1}}$, described by the level hypersurfaces of a function $v_{[1]}$, is related to $v$ by a function $f^4_{[1]}$ by
	\[
		v = v_{[1]} + f^4_{[1]}(v_{[1]},\theta).
	\]
	
	Similarly, for $n \geq 2$, inductively define
	$\mathfrak{E}^1_{[n-1]}$ and $\mathfrak{E}^2_{[n-1]}$ by replacing $f$ with $f_{[n-1]}$ in the definition of $\mathfrak{E}^1$ and $\mathfrak{E}^2$ respectively.  Schematically, 
	\begin{align*}
		\mathfrak{E}^1_{[n-1]}
		&
		=
		\Big(
		\sum_{\substack{k \leq1 \\ \vert \gamma_1 \vert, \vert \gamma_2 \vert \leq 1}}
		\Pi_{\newsph{S}} \Phi \cdot ((r \nablaslash)^k \mathfrak{D}^{\gamma_1} f)_{[n-1]} (x_{-1})
		+
		(\mathfrak{D}^{\gamma_1} f)_{[n-1]} \cdot ((r \nablaslash)^k \mathfrak{D}^{\gamma_2} f)_{[n-1]}(x_{-1}) \Big)_{\ell=0}.
	\\
		\mathfrak{E}^2_{[n-1]}
		&
		=
		\Big(
		\sum_{\vert \gamma_1 \vert, \vert \gamma_2 \vert \leq 1}
		\Pi_{\newsph{S}} \Phi \cdot (\mathfrak{D}^{\gamma_1} f)_{[n-1]} (x_{-1})
		+
		(\mathfrak{D}^{\gamma_1} f)_{[n-1]} \cdot (\mathfrak{D}^{\gamma_2} f)_{[n-1]}(x_{-1}) \Big)_{\ell=0}.
	\end{align*}
	Define moreover,
	\[
		\mathfrak{F}^1
		=
		\frac{1}{r} (j^3-j^4)_{\ell = 0}
		+
		\frac{r}{2\Omega^2_{\circ,M}}
		(\Omega \tr \chi - \Omega \tr \chi_{\circ,M})(u_{-1}+j^3,v_{\infty}+j^4)_{\ell =0}
		-
		( \Omega_{\circ,M}^{-2} \Omega^2 - 1)(u_{-1}+j^3,v_{\infty}+j^4)_{\ell =0},
	\]
	\[
		\mathfrak{F}^2
		=
		\frac{2M}{r^2} (j^3 - j^4)_{\ell=0}
		-
		( \Omega_{\circ,M}^{-2} \Omega^2 - 1)(u_{-1} + j^3,v_{\infty}+j^4)_{\ell =0},
	\]
	\[
		A_{[n]} = \mathfrak{F}^1 + \frac{r}{2\Omega^2_{\circ,M}} \big( \mathcal{G}_{[n-1]} + \mathfrak{E}^1_{[n-1]} \big),
		\qquad
		q_{[n]}
		=
		\Omega_{\circ,M}^{-1} \Omega_{[n-1]} (u_{-1},v_{\infty})
		,
	\]
	where
	\[
		\mathcal{G}_{[n-1]}
		=
		\mathcal{G}(\Omega_{\circ,M}^{-1} \Omega_{[n-1]}(u_{-1},v_{\infty}),
		f^3_{[n-1]}, \nablaslash f^3_{[n-1]}, \nablaslash^2 f^3_{[n-1]},\partial_u f^3_{[n-1]}, \nablaslash \partial_u f^3_{[n-1]}, \nablaslash^2 \partial_u f^3_{[n-1]},M).
	\]
	Consider the foliation of $\newsph{\Cbar}_{v_{\infty}}$ arising from Proposition \ref{prop:foliationIincomingcone} with $A = A_{[n]}$ and $q = q_{[n]}$.  This foliation of $\newsph{\Cbar}_{v_{\infty}}$, described by the level hypersurfaces of a function $u_{[n]}$, is related to $u$ by a function $f^3_{[n]}$ by
	\[
		u = u_{[n]} + f^3_{[n]}(u_{[n]},\theta).
	\]
	The corresponding geometric quantities of this foliation of $\newsph{\Cbar}_{v_{\infty}}$, defined by \eqref{eq:foliationIincomingcone1}--\eqref{eq:foliationIincomingcone4}, are denoted $\eta_{[n]}$, $\Omega\omegabarhat_{[n]}$ etc.
	
	Define now
	\[
		B_{[n]}
		=
		\mathfrak{F}^2 - \mathfrak{E}^2_{[n-1]} - A_{[n]},
		\qquad
		\mathcal{F}_{[n]}(v)
		=
		\frac{\Omega_{\circ,M}^2}{r^3}(u_{-1},v)
		F_{[n]}(u_{-1}),
	\]
	\[
		F_{[n]}(u_{-1})
		=
		\frac{1}{2} \int_{u_{-1}}^{\hat{u}_f+\delta} \int_{\bar{u}}^{\hat{u}_f+\delta}
		r^3 (\Omega^{-1} \hat{\chi}_{[n]}, \Omega^2 \alphabar_{[n]})_{\ell=0}
		\left(\hat{u}, v_{\infty}(\hat{u}_f+\delta) \right)
		d\hat{u}
		d\bar{u}.
	\]
	Consider the foliation of $\newsph{C}_{u_{-1}}$ arising from Proposition \ref{prop:foliationIoutgoingcone} with $B = B_{[n]}$ and $\mathcal{F} = \mathcal{F}_{[n]}$.  This foliation of $\newsph{C}_{u_{-1}}$, described by the level hypersurfaces of a function $v_{[n]}$, is related to $v$ by a function $f^4_{[n]}$ by
	\[
		v = v_{[n]} + f^4_{[n]}(v_{[n]},\theta).
	\]
	
	\noindent \underline{\textbf{Estimates and existence of limits:}}
	It will be shown, by induction, that
	\[
		r^2 \vert \mathcal{G}_{[n]} \vert
		\leq
		\tau
		+
		\varepsilon,
		\qquad
		r^2 \vert
		\mathfrak{E}^1_{[n]}
		\vert
		+
		r
		\vert
		\mathfrak{E}^2_{[n]}
		\vert
		\leq
		\tau
		+
		\frac{\varepsilon}{v_{\infty}},
	\]
	for all $n$.  The estimate trivially holds for $n=0$.  Assume the estimate is true for some $n-1$.
	
	By definition,
	\[
		\vert A_{[n]} \vert
		\lesssim
		\vert
		\mathfrak{F}^1
		\vert
		+
		\frac{r}{2\Omega^2_{\circ}}
		\big(
		\vert
		\mathcal{G}_{[n-1]}
		\vert
		+
		\vert
		\mathfrak{E}^1_{[n-1]}
		\vert
		\big),
		\qquad
		\vert B_{[n]} \vert
		\lesssim
		\vert
		\mathfrak{F}^2
		\vert
		+
		\vert \mathfrak{E}^2_{[n-1]} \vert
		+
		\vert
		\mathfrak{F}^1
		\vert
		+
		\frac{r}{2\Omega^2_{\circ}}
		\big(
		\vert
		\mathcal{G}_{[n-1]}
		\vert
		+
		\vert
		\mathfrak{E}^1_{[n-1]}
		\vert
		\big),
	\]
	and so it follows, from the inductive hypothesis, that
	\begin{equation} \label{eq:AnBnbound}
		r \vert A_{[n]} \vert
		+
		r \vert B_{[n]} \vert
		\lesssim
		\tau
		+
		\frac{\varepsilon}{v_{\infty}}.
	\end{equation}

	Proposition \ref{prop:inIconesrelations} implies that,
	\begin{equation} \label{eq:foliationsf3n1}
		\Deltaslash_{[n]} \log (1+\partial_{u} f^3_{[n]}(u,\theta))
		=
		\big[
		-
		\mu(u + f^3_{[n]},\theta)
		+
		F_{\mu} (\nablaslash f^3_{[n]}, \nablaslash^2 f^3_{[n]},\partial_{u} f^3_{[n]}, \nablaslash \partial_{u} f^3_{[n]})
		\big]_{(\ell \geq 1)_{[n]}},
	\end{equation}
	where $F_{\mu}$ is as in Proposition \ref{prop:inIconesrelations}, and
	\begin{equation} \label{eq:foliationsf3n2}
		\frac{1}{2} \big[ \partial_{\widetilde{u}} \log(1+\partial_{u} f^3_{[n]}) \big]_{(\ell =0)_{[n]}}(u)
		=
		\underline{\mathcal{F}}_{[n-1]}(u)
		+
		\Omega \omegabarhat_{\circ,M}(u)
		-
		\big[
		(1+\partial_{u} f^3_{[n]} (u,\cdot))
		\Omega \omegabarhat (u + f^3_{[n]}(u,\cdot),\cdot)
		\big]_{(\ell =0)_{[n]}},
	\end{equation}
	where $\Deltaslash_{[n]}$ denotes the Laplacian, $(\ell \geq 1)_{[n]}$ and $(\ell =0)_{[n]}$ denote the appropriate mode projections, of the spheres defined by the level hypersurfaces of $u_{[n]}$ in $\newsph{\Cbar}_{v_{\infty}(\hat{u}_f+\delta)}$, and
	\[
		\underline{\mathcal{F}}_{[n-1]}
		=
		\underline{\mathcal{F}}
		( (\Omega \omegabarhat
		-
		\Omega \omegabarhat_{\circ,M})_{[n]}
		,
		(\Omega \tr \chibar - \Omega \tr \chibar_{\circ,M})_{[n]}
		,
		\Omega_{\circ,M}^{-1} \Omega_{[n-1]} (u_{-1},v_{\infty},\cdot)).
	\]
	
	The estimate \eqref{eq:foliationsG2} implies that, if $\tau$ is sufficiently small,
	\begin{align*}
		&
		\sum_{k \leq 4}
		\big(
		\Vert (r \nablaslash)^k \partial_u f^3_{[n]} \Vert_{S_{u_{-1},v_{\infty}}}
		+
		\Vert (r \nablaslash)^k \partial_v f^3_{[n]} \Vert_{S_{u_{-1},v_{\infty}}}
		+
		\Vert (r \nablaslash)^k \partial_u f^4_{[n]} \Vert_{S_{u_{-1},v_{\infty}}}
		+
		\Vert (r \nablaslash)^k \partial_v f^4_{[n]} \Vert_{S_{u_{-1},v_{\infty}}}
		\big)
		\\
		&
		\qquad
		\lesssim
		\sum_{k \leq 5}
		\big(
		\Vert (r \nablaslash)^k j^3 \Vert_{S_{u_{-1},v_{\infty}}}
		+
		\Vert (r \nablaslash)^k j^4 \Vert_{S_{u_{-1},v_{\infty}}}
		\big)
		+
		\vert A_{[n]} \vert
		+
		\vert B_{[n]} \vert,
	\end{align*}
	from which it follows, using also \eqref{eq:AnBnbound}, that
	\[
		\vert
		\mathfrak{E}^1_{[n]}
		\vert
		\lesssim
		\frac{\varepsilon}{r^2}( \tau + \varepsilon v_{\infty}^{-1}) + \tau^2,
		\qquad
		\vert
		\mathfrak{E}^2_{[n]}
		\vert
		\lesssim
		\frac{\varepsilon}{r} ( \tau + \varepsilon v_{\infty}^{-1}) + \tau^2.
	\]
	Recalling \eqref{eq:inIconesrelationsFmuestimate}, an estimate for the system \eqref{eq:foliationsf3n1}--\eqref{eq:foliationsf3n2} gives
	\begin{multline*}
		\sup_{u_{-1} \leq u \leq \hat{u}_f+\delta}
		\sum_{k \leq 4}
		\big(
		\Vert (r\nablaslash)^k f^3_{[n]} \Vert_{S_{u,v_{\infty}}}
		+
		\Vert (r\nablaslash)^k f^3_{[n]} \Vert_{S_{u,v_{\infty}}}
		\big)
		\\
		\lesssim
		\sum_{k \leq 5}
		\big(
		\Vert (r \nablaslash)^k j^3 \Vert_{S_{u_{-1},v_{\infty}}}
		+
		\Vert (r \nablaslash)^k j^4 \Vert_{S_{u_{-1},v_{\infty}}}
		\big)
		+
		\vert A_{[n]} \vert
		+
		\vert B_{[n]} \vert
		+
		\vert M - M_f \vert
		.
	\end{multline*}
	The estimates \eqref{eq:foliationsG1} and \eqref{eq:foliationsG1A} then combine to give, provided $\hat{\varepsilon}_0$ and $\tau$ are sufficiently small,
	\[
		v_{\infty}^2 \vert \mathcal{G}_{[n]} \vert
		\lesssim
		\frac{\hat{u}_f}{v_{\infty}}
		\Big(
		\sum_{k \leq 5}
		\big(
		\Vert (r \nablaslash)^k j^3 \Vert_{S_{u_{-1},v_{\infty}}}
		+
		\Vert (r \nablaslash)^k j^4 \Vert_{S_{u_{-1},v_{\infty}}}
		\big)
		+
		\vert A_{[n]} \vert
		+
		\vert B_{[n]} \vert
		+
		\vert M - M_f \vert
		+
		\varepsilon
		\Big).
	\]
	The induction is completed after inserting \eqref{eq:AnBnbound} and taking $\tau$ to be suitably small.
	
	From \eqref{eq:AnBnbound} it follows that there exists a subsequences of $\{ A_{[n]}\}$ and $\{B_{[n]} \}$ which converge to limits $A$ and $B$ respectively, from which the desired foliations are constructed.

	\noindent \textbf{\underline{Achieving any final sphere:}}  It remains to show that, for any \emph{final} sphere $S$ defined, in the extended $\hat{u}_f$ normalised coordinate system, by functions $s^3$, $s^4$,
	\[
		S = \{ i_{\I} (\hat{u}_f+\delta + s^3(\theta), v_{\infty}(\hat{u}_f+\delta) + s^4(\theta),\theta) \mid \theta \in \mathbb{S}^2 \},
	\]
	which satisfy
	\[
		\sum_{k \leq 5}
		\big(
		\Vert (r \nablaslash)^k s^3 \Vert_{\mathbb{S}^2}
		+
		\Vert (r \nablaslash)^k s^4 \Vert_{\mathbb{S}^2}
		\big)
		\leq
		\tau,
	\]
	the desired foliations of the corresponding null hypersurfaces exist.
	This fact is established by showing that any such functions $s^3$, $s^4$ are attained by suitable functions $j^3$, $j^4$ as above.  More precisely, this fact is established by showing that, for any such functions $s^3$, $s^4$, there exist functions $j^3$, $j^4$ such that, in the above double foliation corresponding to the initial sphere \eqref{eq:initialIjsphere} described by $j^3$, $j^4$ satisfying \eqref{eq:initialIjsphere2}, the functions $f^3$ and $f^4$ which relate this new double null foliation to the extended $\hat{u}_f$ normalised double null foliation, \eqref{eq:initialIjsphere3}, satisfy
	\[
		f^3(\hat{u}_f+\delta, v_{\infty}(\hat{u}_f+\delta),\theta) = s^3(\theta),
		\qquad
		f^4(\hat{u}_f+\delta, v_{\infty}(\hat{u}_f+\delta),\theta) = s^4(\theta).
	\]
	Indeed, for any such $j^3$, $j^4$ as above, one writes, with $v_{\infty} = v_{\infty}(\hat{u}_f+\delta)$,
	\[
		f^3(\hat{u}_f+\delta, v_{\infty},\theta)
		=
		j^3(\theta)
		+
		\int_{u_{-1}}^{\hat{u}_f+\delta} \partial_u f^3(u, v_{\infty},\theta) du,
		\quad
		f^4(\hat{u}_f+\delta, v_{\infty},\theta)
		=
		j^4(\theta)
		+
		\int_{u_{-1}}^{\hat{u}_f+\delta} \partial_u f^4(u, v_{\infty},\theta) du.
	\]
	The result then follows, after taking $\tau$ and $\hat{\varepsilon}_0$ suitably small, from defining suitable iterates $j^3_{[n]}$ and $j^4_{[n]}$ and using the fact that if $j^3_{[n]}$ and $j^4_{[n]}$ satisfy \eqref{eq:initialIjsphere2} then the corresponding diffeomorphism functions $f^3_{[n]}$ and $f^4_{[n]}$ satisfy, 
	\[
		\int_{u_{-1}}^{\hat{u}_f+\delta} \vert \partial_u f^3_{[n]}(u, v_{\infty}) \vert du
		+
		\int_{u_{-1}}^{\hat{u}_f+\delta} \vert \partial_u f^4_{[n]}(u, v_{\infty})\vert du
		\lesssim
		\hat{u}_f
		\Big(
		\frac{\tau}{v_{\infty}}
		+
		\tau^2
		\Big).
	\]
\end{proof}

\begin{remark} \label{rmk:linearIgaugefoliations}
	In the linear setting of Proposition \ref{proplinIpgauge}, one can distill from the proof of 
	Proposition~\ref{prop:Iconesfoliations} an analogous linear statement, namely that,  for any 
	solution $\tilde{\mathscr{S}}$ of the linearised equations around Schwarzschild of mass $M_f$,
	there exist functions $f^3(u,\theta)$, $f^4(v,\theta)$, with $f^3(u_f,\cdot) =0$, $f^4(v_{\infty},\cdot)= 0$, which generate a pure gauge solution (see Section 6 of~\cite{holzstabofschw}), denoted $\mathscr{G}_{\mathrm{foliations}}$, such that the solution 
	$\tilde{\mathscr{S}}+\mathscr{G}_{\mathrm{foliations}}$ 
	of the linearised equations 
	satisfies the linearised analogues of \eqref{eq:Ifoliation1}--\eqref{eq:Ifoliation5}.  Indeed, in the notation of~\cite{holzstabofschw}, by the linearised analogues of the relations of Proposition \ref{prop:inIconesrelations}, Proposition \ref{prop:outIconesrelations} and the the relations \eqref{eq:metriccomp6} and \eqref{eq:Riccicomp7}, it suffices to find $f^3(u,\theta)$ satisfying the equations
	\begin{align*}
		\Deltaslash \partial_u f^3_{\ell \geq 1}
		+
		\frac{1}{r} \left( 1 - \frac{4M_f}{r} \right) \Deltaslash f^3_{\ell \geq 1}
		-
		\frac{6M_f \Omega^2}{r^4} f^3_{\ell \geq 1}
		&
		=
		-
		\big(
		\divslash \elin
		+
		\rlin
		\big)_{\ell \geq 1}^{\tilde{\mathscr{S}}} (\cdot, v_{\infty},\cdot),
		\\
		\frac{1}{2} \partial_u^2 f^3_{\ell =0}
		-
		\frac{M_f}{r^2} \partial_u f^3_{\ell = 0}
		-
		\frac{2M_f\Omega^2}{r^3} f^3_{\ell = 0}
		&
		=
		-
		\olinb_{\ell =0}^{\tilde{\mathscr{S}}} (\cdot,v_{\infty})
		,
	\end{align*}
	and $f^4(v,\theta)$ satisfying the equations
	\begin{multline*}
		\Deltaslash \partial_v f^4_{\ell \geq 1}
		+
		\frac{\Omega^2}{r^2} \partial_v f^4_{\ell \geq 1}
		+
		\frac{2M_f}{r^2} \Deltaslash f^4_{\ell \geq 1}
		+
		\frac{\Omega^2}{r^3} \left( 1 + \frac{4M_f}{r} \right) f^4_{\ell \geq 1}
		\\
		=
		-
		\big(
		\divslash \eblin
		+
		\rlin
		+
		\frac{1}{2r}  \otxb
		+
		\frac{\Omega^2}{r^2} 2 \Olin
		\big)_{\ell \geq 1}^{\tilde{\mathscr{S}}} (u_f,\cdot,\cdot),
	\end{multline*}
	\begin{align*}
		\frac{1}{2} \partial_v^2 f^4_{\ell =0}
		+
		\frac{M_f}{r^2} \partial_v f^4_{\ell = 0}
		-
		\frac{2M_f\Omega^2}{r^3} f^4_{\ell = 0}
		=
		-
		\olin_{\ell =0}^{\tilde{\mathscr{S}}} (u_f,\cdot),
	\end{align*}
	along with the final conditions
	\begin{align*}
		f^3(u_f,\cdot) = f^4(v_{\infty},\cdot)
		&
		= 0,
	\\
		\partial_u f^3_{\ell=0} (u_f) + \partial_v f^4_{\ell=0}(v_{\infty})
		&
		=
		-
		2 \Olin_{\ell =0}^{\tilde{\mathscr{S}}} (u_f,v_{\infty}),
		\\
		\partial_u f^3_{\ell=0} (u_f)
		&
		=
		\big(
		\frac{r}{2\Omega^2} \otx - 2 \Olin
		\big)_{\ell = 0}^{\tilde{\mathscr{S}}} (u_f,v_{\infty}),
	\end{align*}
	(the latter two being the linear analogues of \eqref{eq:IfoliationsA} and \eqref{eq:IfoliationsAB}) where $r= r_{M_f}$ is defined by \eqref{EFrdef}, $\Deltaslash$ now denotes the Laplacian of the round sphere of radius $r$, and $\ell \geq 1$ now denote projections with respect to the spherical harmonics of the round sphere, etc.  In this simplified linear setting, the proofs of Propositions \ref{prop:foliationIincomingcone}, \ref{prop:foliationIoutgoingcone}, and \ref{prop:Iconesfoliations} reduce to showing the existence of functions $f^3(u,\theta)$ and $f^4(v,\theta)$ satisfying these linear equations.
\end{remark}

\begin{remark} \label{rmk:uf0Igaugeexistence1}
	Recall the double null parametrisation \eqref{herethenewone} of Theorem \ref{thm:localEF}, and let $u_f^0$ be as defined in Section~\ref{compediumparameterssec}.
	The proof of Proposition \ref{prop:Iconesfoliations} can easily be adapted to yield the statement that, for any sphere,
	\begin{equation} \label{eq:uf0Igaugeexistencesphere}
		S = \{ i_{\mathcal{EF}} (u_f^0 + s^3(\theta), v_{\infty}(u_f^0) + s^4(\theta),\theta) \mid \theta \in \mathbb{S}^2\},
	\end{equation}
	defined by smooth functions $s^3,s^4 \colon \mathbb{S}^2 \to \mathbb{R}$, and any mass $M>0$, satisfying
	\[
		\sum_{k \leq 5}
		\big(
		\Vert (r \nablaslash)^k s^3 \Vert_{\mathbb{S}^2}
		+
		\Vert (r \nablaslash)^k s^4 \Vert_{\mathbb{S}^2}
		\big)
		+
		\vert M - M_{\mathrm{init}} \vert
		\leq
		\tau,
	\]
	for some $\tau$ sufficiently small, then, if $\hat{\varepsilon}_0$ is sufficiently small, there exists a smooth foliation of the past incoming cone, $\newsph{\underline{C}}_{v_{\infty}(u_f^0)}$, of $S$ and a smooth foliation of the past outgoing cone, $\newsph{C}_{u_{-1}}$, of the sphere $\newsph{S}_{u_{-1},v_{\infty}(u_f^0)}$ (defined to be the sphere at parameter time $u_{-1} - u_f^0$ in the past of $S$ along $\newsph{\underline{C}}_{v_{\infty}(u_f^0)}$) such that the geometric quantities of the associated spacetime double null foliation satisfy \eqref{eq:Ifoliation1}--\eqref{eq:Ifoliation5} with $u_f^0$ in place of $\hat{u}_f+\delta$.  Indeed, the proof is adapted by replacing the role of the $\delta_1$ extended $\hat{u}_f$ normalised $\I$ gauge of Theorem \ref{thm:extendedgauges} with the double null parametrisation \eqref{herethenewone} of Theorem \ref{thm:localEF}, and exploiting only the smallness of $\hat{\varepsilon}_0$ (which may depend on $u_f^0$) in place of the smallness of $\delta_0$.
\end{remark}

\subsubsection{Existence of a new good sphere and mass: the proof of Theorem \ref{thm:Igaugeexistence}}
\label{subsec:Isphereexistence}

The remaining defining conditions of the $\hat{u}_f+\delta$ normalised $\I$ gauge, namely,
\begin{align}
	\left(\Omega \tr \chi - (\Omega \tr \chi)_{\circ,M_{\hat{u}_f+\delta}} \right)_{\ell \geq 2}(\hat{u}_f+\delta, v_{\infty}(\hat{u}_f+\delta),\theta)
	&
	=
	0
	\text{ for all } \theta \in \mathbb{S}^2;
	\label{eq:Iextragauge1}
	\\
	\left( \slashed{div} \Omega \beta \right)_{\ell=1}
	(\hat{u}_f+\delta, v_{\infty}(\hat{u}_f+\delta),\theta)
	&
	=
	0
	\text{ for all } \theta \in \mathbb{S}^2;
	\label{eq:Iextragauge1a}
	\\
	\left(\Omega^{-1} \tr \chibar - (\Omega^{-1} \tr \chibar)_{\circ,M_{\hat{u}_f+\delta}} \right) (\hat{u}_f+\delta, v_{\infty}(\hat{u}_f+\delta),\theta)
	&
	=
	0
	\text{ for all }
	\theta \in \mathbb{S}^2;
	\label{eq:Iextragauge2}
	\\
	\frac{1}{2} (r^3 \rho_{\ell =0}) (\hat{u}_f+\delta, v_{\infty}(\hat{u}_f+\delta))
	&
	=
	-M_{\hat{u}_f+\delta};
	\label{eq:Iextragauge4}
	\\
	b(u,v_\infty(\hat{u}_f+\delta),\theta)
	&
	=
	0
	\text{ for all }
	u\in [u_{-1},\hat{u}_f+\delta], \theta \in \mathbb{S}^2,
	\label{eq:Iextragauge5}
\end{align}
for some $M_{\hat{u}_f+\delta} = M_f(\hat{u}_f+\delta,\lambda)$, along with the anchoring conditions of Definition \ref{anchoringdef} (using the notation of Definition \ref{anchoringdef})
\begin{align}
	f^3_{d,\mathcal{I}+}(u_{-1},v_{\infty}(\hat{u}_f+\delta))_{\ell=0} 
	&
	=
	0,
	\label{eq:Iextragauge3}
	\\
	f^{A}_{d,\mathcal{I}+} (u_{-1},v_{\infty}(\hat{u}_f+\delta), \theta^1_0, \theta^2_0) 
	&
	=
	0,\qquad A=1,2,
	\label{eq:Iextragauge6a}
	\\
	\frac{\partial f^A_{d,\I}}{\partial \theta^1}(u_{-1},v_{\infty}(\hat{u}_f+\delta), \theta^1_0, \theta^2_0)
	&
	=
	0,\qquad A=1,2,
	\label{eq:Iextragauge6}
\end{align}
will be achieved by iteratively solving for a suitable sphere $\mathring{S}$ and considering the resulting spacetime double null foliation generated by Proposition \ref{prop:Iconesfoliations}.

Note that the inclusion \eqref{eq:overlap5} is trivially satisfied, provided $\delta_0$ is sufficiently small, by continuity and the improved bootstrap assumptions of Theorem \ref{havetoimprovethebootstrap}.

\begin{proof}[Proof of Theorem \ref{thm:Igaugeexistence}]
	Let $\lambda\in \mathfrak{R}(\hat{u}_f)$ be fixed.
	Note that any two smooth functions $h^3, h^4 : \mathbb{S}^2 \to \mathbb{R}$ and a diffeomorphism $\slashed{H} \colon \mathbb{S}^2 \to \mathbb{S}^2$ define a \emph{new final sphere} given (in the coordinate system of the extended $\hat{u}_f$ normalised gauge) by
	\begin{equation} \label{eq:Igaugeexistencenewsphere}
		\mathring{S}
		=
		\{ i_{\I}(\hat{u}_f+\delta + h^3(\theta), v_{\infty}(\hat{u}_f+\delta) + h^4(\theta),\slashed{H}(\theta)) \mid \theta \in \mathbb{S}^2 \}.
	\end{equation}
	Recall the domain \eqref{thedomainincords}.  Any such sphere $\mathring{S}$ (provided $h^3\circ \slashed{H}^{-1}$ and $h^4\circ \slashed{H}^{-1}$ are sufficiently small) defines, by Proposition \ref{prop:Iconesfoliations} (with $s^3 = h^3\circ \slashed{H}^{-1}$, $s^4 = h^4\circ \slashed{H}^{-1}$), a unique double null parametrisation
	\[
		\newsph{i} \colon \mathcal{Z}_{\I}(\hat{u}_f + \delta) \to \mathcal{M},
	\]
	such that the associated geometric quantities satisfy the gauge conditions of Proposition \ref{prop:Iconesfoliations}, \eqref{eq:Iextragauge5} holds, the image of $\mathbb{S}^2$ under $\newsph{i} (\hat{u}_f+\delta, v_{\infty}(\hat{u}_f+\delta), \cdot)$ is the sphere $\mathring{S}$, and
	\[
		\newsph{i} (\hat{u}_f+\delta, v_{\infty}(\hat{u}_f+\delta), \cdot) = \psi^{-1},
	\]
	where $\psi \colon \mathring{S} \to \mathbb{S}^2$ is the canonical diffeomorphism of Proposition \ref{determiningthesphere} arising from the normalised induced metric $r^{-2} g\vert_{\mathring{S}}$ and $p \in \mathring{S}$, $v\in T_p \mathring{S}$ chosen so that the diffeomorphisms relating this double null parametrisation to the initial Eddington--Finkelstein gauge $(\ref{initialEFgaugehere})$ satisfies \eqref{eq:Iextragauge6a}, \eqref{eq:Iextragauge6}.
	Note that, if $\delta$, $h^3\circ \slashed{H}^{-1}$ and $h^4\circ \slashed{H}^{-1}$ are sufficiently small then $\newsph{i}(\mathcal{Z}_{\I}(\hat{u}_f + \delta)) \subset i(\mathcal{Z}_{\mathcal{I}^+}(\hat{u}_f,\delta_1))$ (recalling \eqref{actualIdomainext} and Theorem \ref{thm:extendedgauges}) where $i$ (see \eqref{localformparam}) is the local parametrisation of the extended $\hat{u}_f$ normalised $\I$ gauge, and, by the compactness of the domain $\mathcal{Z}_{\I}(\hat{u}_f + \delta)$, the parametrisation $\newsph{i}$ can be made arbitrarily close to the parametrisation $i$.  It follows that the inclusion \eqref{impoverlapext5} implies that the cones $\newsph{C}_{u}^{\I}$ of the parametrisation $\newsph{i}$ satisfy
	\[
		\bigcup_{u_{-1} \leq u \leq u_2} \newsph{C}_{u}^{\I}
		\subset
		\mathcal{D}^{\mathcal{EF}}(u_3)\cap   \{ u_{-1} - 3C \varepsilon \leq u_{data} \leq u_{2} + 3C \varepsilon \},
	\]
	and in particular the diffeomophism functions appearing in \eqref{eq:Iextragauge5}, \eqref{eq:Iextragauge6a}, \eqref{eq:Iextragauge6} are well defined.
	The goal is to find appropriate $h^3,h^4, \slashed{H}$ so that this double null foliation satisfies  \eqref{eq:Iextragauge1}--\eqref{eq:Iextragauge4} and \eqref{eq:Iextragauge3}.

	Using the expressions \eqref{eq:Riccicomp1}, \eqref{eq:Riccicomp2}, \eqref{eq:Riccicomp3}, \eqref{eq:Riccicomp4}, \eqref{eq:Riccicomp7}, \eqref{eq:Riccicomp8}, \eqref{eq:curvaturecomp5} for how geometric quantities change under a change of gauge, along with the gauge conditions \eqref{eq:Ifoliation1}, \eqref{eq:Ifoliation2}, \eqref{eq:Ifoliation3} of Proposition \ref{prop:Iconesfoliations} (with $M = M_{\hat{u}_f+\delta}$), it follows that the goal is to find $M_{\hat{u}_f+\delta}$ and functions $h^3, h^4 \colon \mathbb{S}^2 \to \mathbb{R}$ and a diffeomorphism $\slashed{H} \colon \mathbb{S}^2 \to \mathbb{S}^2$ such that
	\begin{align}
		&
		\Big[
		2 \newsph{\Deltaslash} \newsph{\Deltaslash} h^3(\theta)
		+
		\frac{2}{r^2} \Big( 3 - \frac{8M_f}{r} \Big) \newsph{\Deltaslash} h^3(\theta)
		-
		\frac{2\Omega_{\circ,M_f}^2}{r^2} \newsph{\Deltaslash} h^4(\theta)
		+
		\mathfrak{a}^1(M_f,M_{\hat{u}_f+\delta},r,\newsph{r})
		\Big]_{\ell \geq 2}
		\label{eq:newIgauge1}
		\\
		&
		\qquad
		=
		\Big[
		-
		\Omega_{\circ,M_f}^{-2}
		\Deltaslash \Omega \tr \chi (x_{\delta}+f(x_{\delta}))
		-
		r^{-2}
		(\Omega \tr \chi - \Omega \tr \chi_{\circ,M_f})(x_{\delta}+f(x_{\delta}))
		-
		2 r^{-1} \mathfrak{m}
		+
		\mathfrak{E}^1
		\Big]_{\ell \geq 2},
		\nonumber
		\\
		&
		\Big[
		\frac{6M_f}{r^3} \Big( 1 - \frac{2M_f}{r} \Big) \newsph{\Deltaslash} h^3(\theta)
		\Big]_{\ell =1}
		=
		\Big[
		\divslash \Omega \beta (x_{\delta}+f(x_{\delta}))
		+
		\mathfrak{E}^2
		\Big]_{\ell =1},
		\label{eq:newIgauge1a}
		\\
		&
		\Big[
		2 \newsph{\Deltaslash} \newsph{\Deltaslash} h^4(\theta)
		+
		\frac{4}{r^2} \Big( 1 - \frac{3M_f}{r} \Big) \newsph{\Deltaslash} h^4(\theta)
		+
		\mathfrak{a}^3(M_f,M_{\hat{u}_f+\delta},r,\newsph{r})
		\Big]_{\ell \geq 1}
		\label{eq:newIgauge2}
		\\
		&
		\qquad
		=
		\Big[
		-
		\Deltaslash \Omega^{-1} \tr \chibar (x_{\delta}+f(x_{\delta}))
		-
		\frac{2}{r} \mathfrak{m}
		-
		\frac{\Omega_{\circ,M_f}^2}{r^{2}}
		\big(
		\Omega^{-1} \tr \chibar
		-
		\Omega^{-1} \tr \chibar_{\circ,M_f}
		\big)(x_{\delta}+f(x_{\delta}))
		+
		\mathfrak{E}^3
		\Big]_{\ell \geq 1},
		\nonumber
		\\
		&
		\Big[
		\mathfrak{a}^4(M_f,M_{\hat{u}_f+\delta},r,\newsph{r})
		\Big]_{\ell = 0}
		\label{eq:newIgauge3}
		\\
		&
		\qquad
		=
		\Big[
		-
		\big(
		\Omega^{-1} \tr \chibar
		-
		\Omega_{\circ,M_f}^{-2} (\Omega \tr \chibar)_{\circ,M_f}
		\big)
		(x_{\delta}+f(x_{\delta}))
		+
		\Omega_{\circ,M_f}^{-2} 
		\big( \Omega \tr \chi - \Omega \tr \chi_{\circ,M_f} \big) (x_{\delta}+f(x_{\delta}))
		+
		\mathfrak{E}^4
		\Big]_{\ell = 0},
		\nonumber
		\\
		&
		h^3_{\ell=0}(\hat{u}_f+\delta,v_{\infty}(\hat{u}_f+\delta))
		=
		\int_{u_{-1}}^{\hat{u}_f+\delta}
		\partial_u \big( f^3_{\ell=0} \big) (u,v_{\infty}(\hat{u}_f+\delta))
		du,
		\label{eq:newIgauge4}
		\\
		&
		\frac{1}{2} (\newsph{r}^3 \newsph{\rho}_{\ell =0}) (\hat{u}_f+\delta, v_{\infty}(\hat{u}_f+\delta))=-M_{\hat{u}_f+\delta},
		\label{eq:newIgauge5}
	\end{align}
	where $\newsph{\Deltaslash}$ is the Laplacian of the new sphere $\mathring{S}$, and the mode projections refer to the modes of the new sphere $\mathring{S}$, and such that, if $\psi \colon \mathring{S} \to \mathbb{S}^2$ denotes the canonical diffeomorphism of Proposition \ref{determiningthesphere} arising from the sphere defined by \eqref{eq:Igaugeexistencenewsphere} with the normalised induced metric $r^{-2} g\vert_{\mathring{S}}$ and $p \in \mathring{S}$, $v\in T_p \mathring{S}$ chosen so that \eqref{eq:Iextragauge6a}, \eqref{eq:Iextragauge6} are satisfied, then 
	\begin{equation} \label{eq:psiiIgaugeh3h4}
		\psi \circ i (\hat{u}_f+\delta + h^3(\theta), v_{\infty}(\hat{u}_f+\delta) + h^4(\theta),\slashed{H}(\theta))
		=
		\theta,
	\end{equation}
	for all $\theta \in \mathbb{S}^2$, where $i$ (see \eqref{localformparam}) is the local parametrisation of the extended $\hat{u}_f$ normalised $\I$ gauge.
	Here $x_{\delta}=x_{\delta}(\theta)$ denotes the point $x_{\delta}=(\hat{u}_f+\delta,v_{\infty}(\hat{u}_f+\delta),\theta)$,
	\[
		x_{\delta}+f(x_{\delta})
		=
		(\hat{u}_f+\delta + f^3(\hat{u}_f+\delta,v_{\infty},\theta)
		,
		v_{\infty}
		+
		f^4(\hat{u}_f+\delta,v_{\infty},\theta),\slashed{F}(\hat{u}_f+\delta,v_{\infty},\theta)),
	\]
	where $v_{\infty} = v_{\infty}(\hat{u}_f+\delta)$, and $\mathfrak{E}^1$, $\mathfrak{E}^2$, $\mathfrak{E}^3$ are nonlinearities determined by the appropriate nonlinearities in the relations \eqref{eq:Riccicomp1}, \eqref{eq:Riccicomp2}, \eqref{eq:Riccicomp3}, \eqref{eq:Riccicomp4}, \eqref{eq:Riccicomp7}, \eqref{eq:Riccicomp8}, \eqref{eq:curvaturecomp5}.  The functions $f^1$, $f^2$, $f^3$, $f^4$ denote the diffeomorphisms relating the new double null foliation defined by the sphere $\mathring{S}$ 
	(whose double null coordinates are denoted $(\newsph{u}, \newsph{v}, \newsph{\theta})$), to the extended $\hat{u}_f$ normalised $\I$ double null foliation, so that
	\begin{equation} \label{eq:newIdiffeomorphisms}
		u = \newsph{u} + f^3(\newsph{u}, \newsph{v}, \newsph{\theta}),
		\quad
		v = \newsph{v} + f^4(\newsph{u}, \newsph{v}, \newsph{\theta}),
		\quad
		\theta^A 
		=
		\slashed{F}^A(\newsph{u}, \newsph{v},\newsph{\theta})
		= \newsph{\theta}^A + f^A(\newsph{u}, \newsph{v}, \newsph{\theta}).
	\end{equation}
	Note that the condition \eqref{eq:psiiIgaugeh3h4} guarantees that
	\[
		h^3(\theta)
		=
		f^3(\hat{u}_f+\delta,v_{\infty},\theta),
		\qquad
		h^4(\theta)
		=
		f^4(\hat{u}_f+\delta,v_{\infty},\theta),
		\qquad
		\slashed{H}(\theta)
		=
		\slashed{F}(\hat{u}_f+\delta,v_{\infty},\theta),
	\]
	with $v_{\infty} = v_{\infty}(\hat{u}_f+\delta)$.
	The quantity $\mathfrak{m}$ is defined by
	\[
		\mathfrak{m}
		=
		\newsph{\Deltaslash} \partial_v f^4 + \frac{\Omega^2_{\circ,M_f}}{r^2} \partial_v f^4
		+
		\frac{2M_f}{r^2} \newsph{\Deltaslash} h^4
		-
		\frac{\Omega^2_{\circ,M_f}}{r} \newsph{\Deltaslash} h^3
		+
		\mathfrak{a}(M_f,M_{\hat{u}_f+\delta},r,\newsph{r})
		,
	\]
	where $\mathfrak{a}$ is a smooth function, determined from the relations \eqref{eq:Riccicomp1}, \eqref{eq:Riccicomp2}, \eqref{eq:Riccicomp4}, \eqref{eq:Riccicomp8} and \eqref{eq:curvaturecomp5}, such that
	\[
		\vert
		\mathfrak{a}(M_f,M_{\hat{u}_f+\delta},r,\newsph{r})
		+
		\frac{\Omega^2_{\circ,M_f}}{r^3} \left( 1 + \frac{4M_f}{r} \right) (h^3 - h^4)
		\vert
		\lesssim
		\vert h \vert^2
		+
		\vert M_f - M_{\hat{u}_f+\delta} \vert^2.
	\]
	Note that $\newsph{\mubar}^{\dagger}(x_{\delta}) = \mubar^{\dagger}(x_{\delta}+f(x_{\delta})) + \mathfrak{m}(x_{\delta}) + \mathcal{O}(\varepsilon^2)$.  The functions $\mathfrak{a}^i$ are smooth for each $i$ and satisfy
	\begin{align*}
		\big\vert
		\mathfrak{a}^1(M_f,M_{\hat{u}_f+\delta},r,\newsph{r})
		-
		\frac{4\Omega_{\circ,M_f}^2}{r^4} (h^3(\theta) - h^4(\theta))
		\big\vert
		&
		\lesssim
		\vert h \vert^2
		+
		\vert M_f - M_{\hat{u}_f+\delta} \vert^2,
		\\
		\big\vert
		\mathfrak{a}^3(M_f,M_{\hat{u}_f+\delta},r,\newsph{r})
		-
		\frac{12M_f \Omega_{\circ,M_f}^2}{r^5} (h^3(\theta) - h^4(\theta))
		\big\vert
		&
		\lesssim
		\vert h \vert^2
		+
		\vert M_f - M_{\hat{u}_f+\delta} \vert^2,
		\\
		\big\vert
		\mathfrak{a}^4(M_f,M_{\hat{u}_f+\delta},r,\newsph{r})
		+
		\frac{4}{r^2} \Big( 1 - \frac{3M_f}{r} \Big) (h^3(\theta) - h^4(\theta))
		\big\vert
		&
		\lesssim
		\vert h \vert^2
		+
		\vert M_f - M_{\hat{u}_f+\delta} \vert.
	\end{align*}
	The function $\newsph{r}$ is the $r_M$ (see \eqref{EFrdef}) associated to the $\newsph{i}$ double null parametrisation and mass $M=M_{\hat{u}_f+\delta}$, and $r$ is the $r_M$ associated to the extended $\hat{u}_f$ normalised $\I$ double null gauge, with mass $M=M_f$.  In particular, in the coordinates associated to the $\newsph{i}$ double null parametrisation,
	\[
		\newsph{r}(u,v,\theta) = r_{M_{\hat{u}_f+\delta}}(u,v),
		\qquad
		r(u,v,\theta) = r_{M_f}(u+f^3(u,v,\theta),v+f^4(u,v,\theta)).
	\]
	Schematically, the nonlinear terms take the form, for $i=1,2,3,4$,
	\begin{equation} \label{eq:IgaugecalEerrorsdef}
		\mathfrak{E}^i(\theta)
		=
		\sum_{\vert \gamma_1 \vert, \vert \gamma_2 \vert \leq 4}
		\Pi_{\newsph{S}} \Phi_{M_f} \cdot \mathfrak{D}^{\gamma_1} f (x_{\delta})
		+
		\mathfrak{D}^{\gamma_1} f \cdot \mathfrak{D}^{\gamma_2} f(x_{\delta}).
	\end{equation}
	The quantities $\Deltaslash \Omega \tr \chi$, $\mubar^{\dagger}$, $\Phi$ etc.\@ refer to the quantities in the extended $\hat{u}_f$ normalised gauge.

	Indeed, given a solution to the system \eqref{eq:newIgauge1}--\eqref{eq:newIgauge5}, it follows from the expressions \eqref{eq:metriccomp1}--\eqref{eq:curvaturecomp6} for how geometric quantities change under a change of gauge, together with the fact that the gauge conditions \eqref{eq:Ifoliation1}, \eqref{eq:Ifoliation2}, \eqref{eq:Ifoliation3} of Proposition \ref{prop:Iconesfoliations} already hold, that the gauge conditions \eqref{eq:Iextragauge1}--\eqref{eq:Iextragauge4} hold in the spacetime double null foliation obtained from Proposition \ref{prop:Iconesfoliations} using the new sphere defined by the functions $s^3 = h^3\circ \slashed{H}^{-1}$, $s^4 = h^4 \circ \slashed{H}^{-1}$.
	
	Consider the following sequence of iterates.  Let $h_{[0]} = \slashed{H}_{[0]} = f_{[0]} = 0$, $\rho_{[0]} = \rho$.  For $n \geq 1$, $M_{[n]}$, $h_{[n]}$ and $f_{[n]}$ are defined inductively as follows.  Given $\rho_{[n-1]}$, let
	\[
		M_{[n-1]} := - \frac{1}{2} (r^3 \rho_{\ell =0})_{[n-1]} (\hat{u}_f+\delta, v_{\infty}(\hat{u}_f+\delta)),
	\]
	and, given $h_{[n-1]}$ and $f_{[n-1]}$, consider the equations,
	\begin{align}
		&
		\Big[
		2 \Deltaslash_{[n-1]} \Deltaslash_{[n-1]} h^3_{[n]}(\theta)
		+
		\frac{2}{r^2_{[n-1]}} \Big( 3 - \frac{8M_f}{r_{[n-1]}} \Big) \Deltaslash_{[n-1]} h^3_{[n]}(\theta)
		-
		\frac{2}{r^2_{[n-1]}}\Big(1 - \frac{2M_f}{r_{[n-1]}} \Big) \Deltaslash_{[n-1]} h^4_{[n]}(\theta)
		\nonumber
		\\
		&
		+
		\frac{4}{r^4_{[n-1]}}\Big(1 - \frac{2M_f}{r_{[n-1]}} \Big) (h^3_{[n]}(\theta) - h^4_{[n]}(\theta))
		\Big]_{(\ell \geq 2)_{[n-1]}}
		\label{eq:nIgauge1}
		=
		\Big[
		\mathcal{G}^1_{[n-1]}
		+
		\mathfrak{E}^1_{[n-1]}
		+
		A^1_{[n-1]}
		\Big]_{(\ell \geq 2)_{[n-1]}},
		\\
		&
		\Big[
		\frac{6M_f}{r^3_{[n-1]}} \Big( 1 - \frac{2M_f}{r_{[n-1]}} \Big) \Deltaslash_{[n-1]} h^3_{[n]}(\theta)
		\Big]_{(\ell =1)_{[n-1]}}
		=
		\Big[
		\mathcal{G}^2_{[n-1]}
		+
		\mathfrak{E}^2_{[n-1]}
		+
		A^2_{[n-1]}
		\Big]_{(\ell =1)_{[n-1]}},
		\label{eq:nIgauge1a}
		\\
		&
		2 \Deltaslash_{[n-1]} \Deltaslash_{[n-1]} h^4_{[n]}(\theta)
		+
		\frac{4}{r^2_{[n-1]}} \Big( 1 - \frac{3M_f}{r_{[n-1]}} \Big) \Deltaslash_{[n-1]} h^4_{[n]}(\theta)
		+
		\frac{12M_f}{r^5_{[n-1]}} \Big( 1 - \frac{2M_f}{r_{[n-1]}} \Big)
		(h^3_{[n]}(\theta) - h^4_{[n]}(\theta))
		\nonumber
		\\
		&
		\qquad
		=
		\Big[
		\label{eq:nIgauge2}
		\mathcal{G}^3_{[n-1]}
		+
		\mathfrak{E}^3_{[n-1]}
		+
		A^3_{[n-1]}
		\Big]_{(\ell \geq 1)_{[n-1]}},
	\end{align}
	\begin{align}
		&
		-
		\frac{4}{r^2_{[n-1]}} \Big( 1 - \frac{3M_f}{r_{[n-1]}} \Big)
		(h^3_{[n]}(\theta) - h^4_{[n]}(\theta))
		\label{eq:nIgauge3}
		=
		\Big[
		\mathcal{G}^4_{[n-1]}
		+
		\mathfrak{E}^4_{[n-1]}
		+
		A^4_{[n-1]}
		\Big]_{(\ell = 0)_{[n-1]}},
		\\
		&
		(h^3_{\ell=0})_{[n]}(\hat{u}_f+\delta,v_{\infty}(\hat{u}_f+\delta))
		=
		\int_{u_{-1}}^{\hat{u}_f+\delta}
		\partial_u \big( f^3_{\ell=0} \big)_{[n]} (u,v_{\infty}(\hat{u}_f+\delta))
		du,
		\label{eq:nIgauge4}
	\end{align}
	where, $\Deltaslash = \Deltaslash_{[n-1]}$ is the Laplacian of the sphere defined by $h_{[n-1]}$ (see \eqref{eq:Snhsphere} below), the notation $(\ell \geq 1)_{[n-1]}$ is used for the projection to $\ell \geq 1$ associated the spheres of the double foliation defined by $h_{[n-1]}$,  and, schematically,
\[
		\mathfrak{E}^i_{[n-1]}(\theta)
		=
		\sum_{\vert \gamma_1 \vert, \vert \gamma_2 \vert \leq 4}
		\Pi_{S_{[n-1]}} \Phi_{M_f} \cdot \mathfrak{D}^{\gamma_1} f_{[n-1]} (x_{\delta}) 
		+
		\mathfrak{D}^{\gamma_1} f_{[n-1]} \cdot \mathfrak{D}^{\gamma_2} f_{[n-1]}(x_{\delta}).
	\]
Again,
\begin{equation} \label{eq:xdeltaplusfn}
	x_{\delta}+f_{[n-1]}(x_{\delta})
	=
	(\hat{u}_f+\delta + f^3_{[n-1]}(\hat{u}_f+\delta,v_{\infty},\theta)
	,
	v_{\infty}
	+
	f^4_{[n-1]}(\hat{u}_f+\delta,v_{\infty},\theta),\slashed{F}_{[n-1]}(\hat{u}_f+\delta,v_{\infty},\theta)),
\end{equation}
where $v_{\infty} = v_{\infty}(\hat{u}_f+\delta)$.
The functions $\mathcal{G}^i_{[n-1]}$ are defined by
\begin{align*}
	\mathcal{G}^1_{[n-1]}
	=
	&
	-
	\Omega_{\circ,M_f}^{-2}
	\Deltaslash \Omega \tr \chi (x_{\delta}+f_{[n-1]}(x_{\delta}))
	-
	r^{-2}
	(\Omega \tr \chi - \Omega \tr \chi_{\circ})(x_{\delta}+f_{[n-1]}(x_{\delta}))
	-
	2 r^{-1} \mathfrak{m}_{[n-1]}(x_{\delta}),
	\\
	\mathcal{G}^2_{[n-1]}
	=
	&
	\divslash \Omega \beta (x_{\delta}+f_{[n-1]}(x_{\delta})),
	\\
	\mathcal{G}^3_{[n-1]}
	=
	&
	-
	\Deltaslash \Omega^{-1} \tr \chibar (x_{\delta}+f_{[n-1]}(x_{\delta}))
	-
	\Omega_{\circ,M_f}^2 r^{-2}
	\big(
	\Omega^{-1} \tr \chibar
	-
	\Omega^{-1} \tr \chibar_{\circ,M_f}
	\big)(x_{\delta}+f_{[n-1]}(x_{\delta}))
	-
	2 r^{-1} \mathfrak{m}_{[n-1]}(x_{\delta}),
	\\
	\mathcal{G}^4_{[n-1]}
	=
	&
	-
	\big(
	\Omega^{-1} \tr \chibar
	-
	\Omega_{\circ,M_f}^{-2} (\Omega \tr \chibar)_{\circ,M_f}
	\big)
	(x_{\delta}+f_{[n-1]}(x_{\delta}))
	+
	\Omega_{\circ,M_f}^{-2} 
	\big( \Omega \tr \chi - \Omega \tr \chi_{\circ,M_f} \big) (x_{\delta}+f_{[n-1]}(x_{\delta})),
\end{align*}
where
\[
	\mathfrak{m}_{[n-1]}
	=
	\Deltaslash_{[n-1]} \partial_v f^4_{[n-1]} + \frac{\Omega^2_{\circ,M_f}}{r^2} \partial_v f^4_{[n-1]}
	+
	\frac{2M_f}{r^2} \Deltaslash_{[n-1]} h^4_{[n-1]}
	-
	\frac{\Omega^2_{\circ,M_f}}{r} \Deltaslash_{[n-1]} h^3_{[n-1]}
	+
	\mathfrak{a}(M_f,M_{[n-1]},r,r_{[n-1]}),
\]
and the functions $A^i_{[n-1]}$ are defined by
\begin{align*}
	A^1_{[n-1]}
	=
	&
	\Big[
	\frac{2}{r^2_{[n-1]}} \Big( 3 - \frac{8M_f}{r_{[n-1]}} \Big)
	-
	\frac{2}{r^2} \Big( 3 - \frac{8M_f}{r} \Big)
	\Big]
	\Deltaslash_{[n-1]} h^3_{[n-1]}(\theta)
	+
	\frac{4}{r^4_{[n-1]}}\big(1 - \frac{2M_f}{r_{[n-1]}} \big)
	(h^3_{[n-1]}(\theta) - h^4_{[n-1]}(\theta))
	\\
	&
	-
	\Big[
	\frac{2}{r^2_{[n-1]}}\big(1 - \frac{2M_f}{r_{[n-1]}} \big)
	-
	\frac{2}{r^2}\Big(1 - \frac{2M_f}{r} \Big)
	\Big]
	\Deltaslash_{[n-1]} h^4_{[n-1]}(\theta)
	-
	\mathfrak{a}^1(M_f,M_{[n-1]},r,r_{[n-1]}),
	\\
	A^2_{[n-1]}
	=
	&
	\Big[
	\frac{6M_f}{r^3_{[n-1]}} \Big( 1 - \frac{2M_f}{r_{[n-1]}} \Big)
	-
	\frac{6M_f}{r^3} \Big( 1 - \frac{2M_f}{r} \Big)
	\Big]
	\Deltaslash_{[n-1]} h^3_{[n-1]}(\theta),
	\\
	A^3_{[n-1]}
	=
	&
	\Big[
	\frac{4}{r^2_{[n-1]}} \Big( 1 - \frac{3M_f}{r_{[n-1]}} \Big)
	-
	\frac{4}{r^2} \Big( 1 - \frac{3M_f}{r} \Big)
	\Big] \Deltaslash_{[n-1]} h^4_{[n-1]}(\theta)
	\\
	&
	+
	\frac{12M_f}{r^5_{[n-1]}} \Big( 1 - \frac{2M_f}{r_{[n-1]}} \Big) (h^3_{[n-1]}(\theta) - h^4_{[n-1]}(\theta))
	-
	\mathfrak{a}^3(M_f,M_{[n-1]},r,r_{[n-1]}),
	\\
	A^4_{[n-1]}
	=
	&
	-
	\mathfrak{a}^4(M_f,M_{[n-1]},r,r_{[n-1]})
	+
	\frac{4}{r^2_{[n-1]}} \Big( 1 - \frac{3M_f}{r_{[n-1]}} \Big) (h^3_{[n-1]}(\theta) - h^4_{[n-1]}(\theta)),
\end{align*}
$r_{[n-1]}$ is the $r_M$ (see \eqref{EFrdef}) associated to the $i_{[n-1]}$ double null parametrisation and mass $M=M_{[n-1]}$, and $r$ is the $r_M$ associated to the extended $\hat{u}_f$ normalised $\I$ double null gauge, with mass $M=M_f$, so that, in the coordinates associated to the $i_{[n-1]}$ double null parametrisation,
\begin{align*}
	r_{[n-1]}(\hat{u}_f+\delta,v_{\infty},\theta)
	&
	=
	r_{M_{[n-1]}}(\hat{u}_f+\delta,v_{\infty}),
	\\
	r(\hat{u}_f+\delta,v_{\infty},\theta)
	&
	=
	r_{M_f}(\hat{u}_f+\delta+f^3_{[n-1]}(\hat{u}_f+\delta,v_{\infty},\theta),v+f^4_{[n-1]}(\hat{u}_f+\delta,v_{\infty},\theta))
	,
\end{align*}
with $v_{\infty} = v_{\infty}(\hat{u}_f+\delta)$.

It follows from standard elliptic theory that there exist unique solutions $h^3_{[n]}$, $h^4_{[n]}$ of the linear elliptic system \eqref{eq:nIgauge1}, \eqref{eq:nIgauge1a}, \eqref{eq:nIgauge2} which are supported on $(\ell \geq 1)_{[n-1]}$.  Moreover the unique solutions $h^3_{[n]}$, $h^4_{[n]}$ of \eqref{eq:nIgauge3}, \eqref{eq:nIgauge4} are supported on $(\ell =0)_{[n-1]}$.  Define, therefore
\[
	h^3_{[n]}
	=
	(h^3_{[n]})_{(\ell =0)_{[n-1]}}
	+
	(h^3_{[n]})_{(\ell \geq 1)_{[n-1]}},
	\qquad
	h^4_{[n]}
	=
	(h^4_{[n]})_{(\ell =0)_{[n-1]}}
	+
	(h^4_{[n]})_{(\ell \geq 1)_{[n-1]}},
\]
where the functions $(h^3_{[n]})_{(\ell \geq 1)_{[n-1]}}$, $(h^4_{[n]})_{(\ell \geq 1)_{[n-1]}}$ are the unique solutions of the system \eqref{eq:nIgauge1}, \eqref{eq:nIgauge1a}, \eqref{eq:nIgauge2} and the functions $(h^3_{[n]})_{(\ell =0)_{[n-1]}}$, $(h^4_{[n]})_{(\ell =0)_{[n-1]}}$ are the unique solutions of \eqref{eq:nIgauge3}, \eqref{eq:nIgauge4}.

Such solutions $h^3_{[n]}$, $h^4_{[n]}$, together with the ``previous'' spherical diffeomorphism $\slashed{H}_{[n-1]}$, define an ``$n$-th'' sphere $S_{[n]}$, expressed in the $(u_{\I}, v_{\I},\theta_{\I})$ coordinate system of the extended $\hat{u}_f$ normalised $\I$ gauge by
	\begin{equation} \label{eq:Snhsphere}
		S_{[n]}
		=
		\{ i_{\I} (\hat{u}_f+\delta + h^3_{[n]}(\theta), v_{\infty}(\hat{u}_f+\delta) + h^4_{[n]}(\theta), \slashed{H}_{[n-1]}(\theta)) \mid \theta \in \mathbb{S}^2\}.
	\end{equation}
	Each such sphere $S_{[n]}$ defines (provided $h^3_{[n]}\circ \slashed{H}^{-1}_{[n-1]}$ and $h^4_{[n]}\circ \slashed{H}^{-1}_{[n-1]}$ are sufficiently small), by Proposition \ref{prop:Iconesfoliations} (with $s^3 = h^3_{[n]}\circ \slashed{H}^{-1}_{[n-1]}$, $s^4 = h^4_{[n]}\circ \slashed{H}^{-1}_{[n-1]}$), a unique double null parametrisation
	\[
		i_{[n]} \colon \mathcal{Z}_{\I}(\hat{u}_f + \delta) \to \mathcal{M},
	\]
	such that the associated geometric quantities satisfy the gauge conditions of Proposition \ref{prop:Iconesfoliations}, \eqref{eq:Iextragauge5} holds, the image of $\mathbb{S}^2$ under $i_{[n]} (\hat{u}_f+\delta, v_{\infty}(\hat{u}_f+\delta), \cdot)$ is the sphere $S_{[n]}$, and
	\[
		i_{[n]} (\hat{u}_f+\delta, v_{\infty}(\hat{u}_f+\delta), \cdot) = \psi^{-1}_{[n]},
	\]
	where $\psi_{[n]} \colon S_{[n]} \to \mathbb{S}^2$ is the canonical diffeomorphism of Proposition \ref{determiningthesphere} arising from the normalised induced metric $r^{-2}_{[n]} g\vert_{S_{[n]}}$ and $p \in S_{[n]}$, $v\in T_p S_{[n]}$ chosen so that the diffeomorphisms relating this double null parametrisation to the initial Eddington--Finkelstein gauge $(\ref{initialEFgaugehere})$ satisfies, using the notation of Definition \ref{anchoringdef},
	\[
		(f_{[n]})^{A}_{d,\mathcal{I}+} (u_{-1},v_{\infty}(\hat{u}_f+\delta), \theta^1_0, \theta^2_0)
		=
		\partial_{\theta^1} (f_{[n]})^A_{d,\I} (u_{-1},v_{\infty}(\hat{u}_f+\delta), \theta^1_0, \theta^2_0)
		=
		0,
		\qquad A=1,2.
	\]
	 
	Let $(u_{[n]}, v_{[n]}, \theta_{[n]})$ denote the coordinates of the local representation which define this $n$-th double null foliation, and define $f_{[n]}$ to be the diffeomorphism functions which relate this ``$n$-th'' double null foliation to the extended $\hat{u}_f$ normalised double null foliation,
\[
	u = u_{[n]} + f^3_{[n]}(u_{[n]}, v_{[n]}, \theta_{[n]}),
	\quad
	v = v_{[n]} + f^4_{[n]}(u_{[n]}, v_{[n]}, \theta_{[n]}),
\]
\[
	\theta^A
	=
	\slashed{F}^A_{[n]} (u_{[n]}, v_{[n]}, \theta_{[n]})
	=
	\theta^A_{[n]} + f^A_{[n]}(u_{[n]}, v_{[n]}, \theta_{[n]}),
\]
and define $\rho_{[n]}$ etc.\@ to be the geometric quantities of this ``$n$-th'' double null foliation (schematically denoted $\Phi_{[n]}$).  Define also the diffeomorphism $F_{[n]}$ by
\[
	(u,v,\theta)
	=
	F_{[n]}(u_{[n]}, v_{[n]}, \theta_{[n]}).
\]
	Let $\slashed{H}_{[n]}$ be defined by the relation 
	\[
		\psi_{[n]} \circ i (\hat{u}_f+\delta + h^3_{[n]}(\theta), v_{\infty}(\hat{u}_f+\delta) + h^4_{[n]}(\theta),\slashed{H}_{[n-1]}(\theta))
		=
		\slashed{H}_{[n]}^{-1} \circ \slashed{H}_{[n-1]} (\theta),
	\]
	for all $\theta \in \mathbb{S}^2$, where $i$ (see \eqref{localformparam}) is the local parametrisation of the extended $\hat{u}_f$ normalised $\I$ gauge.
	It follows that
	\[
		(i_{[n]})^{-1} \circ i (\hat{u}_f+\delta + h^3_{[n]}(\theta), v_{\infty}(\hat{u}_f+\delta) + h^4_{[n]}(\theta),\slashed{H}_{[n-1]}(\theta))
		=
		(\hat{u}_f+\delta, v_{\infty}(\hat{u}_f+\delta), \slashed{H}_{[n]}^{-1} \circ \slashed{H}_{[n-1]} (\theta)),
	\]
	hence
	\begin{equation} \label{eq:Igaugef3h3H}
		f^3_{[n]}(\hat{u}_f+\delta,v_{\infty},\theta)
		=
		h^3_{[n]} \circ \slashed{H}_{[n-1]}^{-1} \circ \slashed{H}_{[n]} (\theta)
		,
		\qquad
		f^4_{[n]}(\hat{u}_f+\delta,v_{\infty},\theta)
		=
		h^4_{[n]} \circ \slashed{H}_{[n-1]}^{-1} \circ \slashed{H}_{[n]} (\theta),
	\end{equation}
	and
	\[
		\slashed{H}_{[n]}(\theta)
		=
		\slashed{F}_{[n]}(\hat{u}_f+\delta,v_{\infty}, \theta),
	\]
	for all $\theta \in \mathbb{S}^2$, with $v_{\infty} = v_{\infty}(\hat{u}_f+\delta)$.

In order to show that the sequence $\{h_{[n]}\}$ converges it is first necessary to estimate the iterates $f_{[n]}$ and their derivatives.

In what follows $x$ will be used to denote a value $x = (u,v,\theta)$.  For fixed $(u,v)$ the geometric quantities $\Phi(u,v)$ and $\Phi_{[n]}(u,v)$ can be viewed, via the identifications \eqref{localformparam} and \eqref{localformparamtwo}, as tensor fields on the sphere $\mathbb{S}^2$.  When the quantities $\Phi$ and $\Phi_{[n]}$ are compared, they will be sometimes be identified using the values of their respective coordinate functions, i.\@e.\@ $\Phi(u,v)$ and $\Phi_{[n]}(u,v)$ will be identified, in this way, as tensor fields on $\mathbb{S}^2$, in which case we write $\Phi(x) - \Phi_{[n]}(x)$.  They will also be considered at the same spacetime point, as in the relations of Propositions \ref{prop:metricrelations}, \ref{prop:Riccirelations} and \ref{prop:curvaturerelations}, in which case we write $\Pi_{S_{[n]}}\Phi - \Phi_{[n]}$, where the projection $\Pi_{S_{[n]}}$ is as in \eqref{eq:tildeprojection}.  For example, for $\Phi = \rho$, the difference $(\Pi_{S_{[n]}} \rho - \rho_{[n]} )$ at the spacetime point $(u_{[n]},v_{[n]},\theta_{[n]}) = (u,v,\theta)$ takes the form
\[
	(\Pi_{S_{[n]}} \rho - \rho_{[n]} )(u,v,\theta)
	=
	\rho(F_{[n]}(u,v,\theta))
	-
	\rho_{[n]}(u,v,\theta)
	=
	\rho(x+f_{[n]}(x))
	-
	\rho_{[n]}(x),
\]
and for $\Phi = \alpha$, the difference takes the form
\[
	(\Pi_{S_{[n]}} \alpha - \alpha_{[n]} )(u,v,\theta)
	=
	\Big(
	\alpha(F_{[n]}(u,v,\theta))_{CD} \partial_{\theta_{[n]}^A} F^C_{[n]} \partial_{\theta_{[n]}^B} F^D_{[n]}
	-
	\alpha_{[n]}(u,v,\theta)_{AB}
	\Big)
	d\theta_{[n]}^A d \theta_{[n]}^B.
\]
The difference $\alpha(u,v,\theta) - \alpha_{[n]}(u,v,\theta)$, on the other hand, takes the form
\[
	\alpha(u,v,\theta) - \alpha_{[n]}(u,v,\theta)
	=
	\Big( \alpha(u,v,\theta)_{AB} - \alpha_{[n]}(u,v,\theta)_{AB} \Big) d\theta_{[n]}^A d \theta_{[n]}^B.
\]
Similarly, for fixed $(u,v)$, $\alpha$ and $\Pi_{S_{[n]}} \alpha$ can be identified as tensor fields on $\mathbb{S}^2$ and we write $\alpha(u,v,\theta) - \Pi_{S_{[n]}} \alpha(u,v,\theta)$ to mean
\begin{equation} \label{eq:alphadifferencetomean}
	\alpha(u,v,\theta) - \Pi_{S_{[n]}} \alpha(u,v,\theta)
	=
	\Big( \alpha(u,v,\theta)_{AB} - 
	\alpha(F_{[n]}(u,v,\theta))_{CD} \partial_{\theta_{[n]}^A} F^C_{[n]} \partial_{\theta_{[n]}^B} F^D_{[n]}
	\Big) d\theta_{[n]}^A d \theta_{[n]}^B.
\end{equation}

For some given $u_{-1} \leq u \leq \hat{u}_f+\delta$ and $k \geq 1$, define the norm on diffeomorphisms $f$,
\begin{align} \label{eq:iteratesdiffeonorms}
	\sum_{\vert \gamma \vert \leq k} \Vert \mathfrak{D}^{\gamma} f \Vert_{S_{u,v_{\infty}}}
	=
	\
	&
	\sum_{\vert \gamma \vert \leq k}
	\Big(
	\Vert \mathfrak{D}^{\gamma} f^3 \Vert_{S_{u,v_{\infty}}}
	+
	r^{-1} \Vert \mathfrak{D}^{\gamma} f^4 \Vert_{S_{u,v_{\infty}}}
	\Big)
	\\
	&
	+
	\sum_{\vert \gamma \vert \leq k-1}
	\Big(
	\Vert \mathfrak{D}^{\gamma} \partial_u \slashed{f} \Vert_{S_{u,v_{\infty}}}
	+
	\Vert \mathfrak{D}^{\gamma} \partial_v \slashed{f} \Vert_{S_{u,v_{\infty}}}
	+
	\Vert \mathfrak{D}^{\gamma} \partial_u f^4 \Vert_{S_{u,v_{\infty}}}
	+
	r \Vert \mathfrak{D}^{\gamma} \partial_v f^3 \Vert_{S_{u,v_{\infty}}}
	\Big),
	\nonumber
\end{align}
with, for example,
\[
	\Vert \mathfrak{D}^{\gamma} f^3 \Vert_{S_{u,v_{\infty}}}^2
	=
	\int_{\mathbb{S}^2} \vert \mathfrak{D}^{\gamma} f^3 (u,v_{\infty},\theta) \vert^2_{r^2 \gamma} \sqrt{\det \gamma(\theta)} d \theta.
\]
Define also $\sum_{\vert \gamma \vert \leq k} \Vert \mathfrak{D}^{\gamma} f_{\ell \geq 1} \Vert_{S_{u,v_{\infty}}}$ by replacing $f^3$ and $f^4$ with $f^3_{\ell \geq 1}$ and $f^4_{\ell \geq 1}$ respectively in \eqref{eq:iteratesdiffeonorms}.

The following will be used in the proof of Lemma \ref{lem:fnI}.

\begin{lemma}[Estimates for differences between $S$ tensors with pullback by $\slashed{F}_{[n]}$] \label{lem:mvt}
	Let $v_{\infty} = v_{\infty}(\hat{u}_f+\delta)$, and recall the norm \eqref{eq:iteratesdiffeonorms}.  Suppose that, in each coordinate chart $U \subset \mathbb{S}^2$, with associated coordinates $(\theta^1,\theta^2)$,
	\[
		\sum_{k_1+k_2\leq 5}
		\sum_{A=1,2}
		\Vert \partial_{\theta^1}^{k_1} \partial_{\theta^2}^{k_2} (\slashed{H}_{[n]}^A(\theta) - \theta^A) \mathds{1}_U \Vert_{\mathbb{S}^2}
		+
		\sup_{u_{-1} \leq u \leq \hat{u}_f+\delta}
		\sum_{\vert \gamma \vert \leq 5} \Vert \mathfrak{D}^{\gamma} f_{[n]} \Vert_{S_{u,v_{\infty}}}
		\lesssim
		1.
	\]
	Then, for any $u_{-1} \leq u \leq \hat{u}_f + \delta$,
	\begin{equation} \label{eq:diffiscloseiteratesone}
		|h(u,v_{\infty},\theta) - h(u,v_{\infty},\slashed{F}_{[n]}(u,v_{\infty},\theta))|
		\lesssim
		\sup_{S_{u,v_{\infty}}} |r\nablaslash h |_{\gslash_{[n]}}
		\Big(
		\frac{\varepsilon \delta}{r}
		+
		\sup_{u_{-1} \leq u \leq \hat{u}_f+\delta}
		\sum_{\vert \gamma \vert \leq 3} \Vert \mathfrak{D}^{\gamma} f_{[n]} \Vert_{S_{u,v_{\infty}}}
		\Big)
		,
	\end{equation}
	for all functions $h$, and
	\begin{equation} \label{eq:diffiscloseiteratestwo}
		\big\vert \big( \omega - (\slashed{F}_{[n]}(u,v_{\infty},\cdot))^*\omega \big) (u,v_{\infty},\theta) \big\vert_{\gslash_{[n]}}
		\lesssim
		\sup_{S_{u,v_{\infty}}}(|\omega|_{\gslash_{[n]}}
		+
		|r\nablaslash\omega|_{\gslash_{[n]}})
		\Big(
		\frac{\varepsilon \delta}{r}
		+
		\sup_{u_{-1} \leq u \leq \hat{u}_f+\delta}
		\sum_{\vert \gamma \vert \leq 3} \Vert \mathfrak{D}^{\gamma} f_{[n]} \Vert_{S_{u,v_{\infty}}}
		\Big).
	\end{equation}
	for any $S$-tangent $(0,k)$ tensor $\omega$.  Moreover,
	\begin{equation} \label{eq:diffiscloseiteratesthree}
		\sum_{k \leq 3} \Vert (r\nablaslash)^{k+2} \slashed{H}_{[n]} \Vert_{\mathbb{S}^2}
		\lesssim
		\frac{\varepsilon \delta}{r}
		+
		\sup_{u_{-1} \leq u \leq \hat{u}_f+\delta}
		\sum_{\vert \gamma \vert \leq 5} \Vert \mathfrak{D}^{\gamma} f_{[n]} \Vert_{S_{u,v_{\infty}}}
		.
	\end{equation}
	
	Similarly,
	\begin{multline} \label{eq:diffiscloseiteratesdifferenceone}
		|
		h(\hat{u}_f+\delta,v_{\infty},\slashed{F}_{[n+1]}(\hat{u}_f+\delta,v_{\infty},\theta))
		-
		h(u,v_{\infty},\slashed{F}_{[n]}(u,v_{\infty},\theta))
		|
		\\
		\lesssim
		\sup_{S_{u,v_{\infty}}} |r\nablaslash h |_{\gslash_{[n]}}
		\sup_{u_{-1} \leq u \leq \hat{u}_f+\delta}
		\sum_{\vert \gamma \vert \leq 3} \Vert \mathfrak{D}^{\gamma} f_{[n+1]} - \mathfrak{D}^{\gamma} f_{[n]} \Vert_{S_{u,v_{\infty}}}
		,
	\end{multline}
	for all functions $h$, and
	\begin{multline} \label{eq:diffiscloseiteratesdifferencetwo}
		\big\vert \big( (\slashed{F}_{[n+1]}(u,v_{\infty},\cdot))^* \omega - (\slashed{F}_{[n]}(u,v_{\infty},\cdot))^*\omega \big) (u,v_{\infty},\theta) \big\vert_{\gslash_{[n]}}
		\\
		\lesssim
		\sup_{S_{u,v_{\infty}}}(|\omega|_{\gslash_{[n]}}
		+
		|r\nablaslash\omega|_{\gslash_{[n]}})
		\sup_{u_{-1} \leq u \leq \hat{u}_f+\delta}
		\sum_{\vert \gamma \vert \leq 3}
		\Vert \mathfrak{D}^{\gamma} f_{[n+1]} - \mathfrak{D}^{\gamma} f_{[n]} \Vert_{S_{u,v_{\infty}}}
		.
	\end{multline}
	for any $S$-tangent $(0,k)$ tensor $\omega$.
	Finally,
	\begin{equation} \label{eq:diffiscloseiteratesdifferencethree}
		\sum_{k \leq 3} \Vert (r\nablaslash)^{k+2} \slashed{H}_{[n]}^{-1} \circ \slashed{H}_{[n+1]} \Vert_{\mathbb{S}^2}
		\lesssim
		\sup_{u_{-1} \leq u \leq \hat{u}_f+\delta}
		\sum_{\vert \gamma \vert \leq 5}
		\Vert \mathfrak{D}^{\gamma} f_{[n+1]} - \mathfrak{D}^{\gamma} f_{[n]} \Vert_{S_{u,v_{\infty}}}
		.
	\end{equation}
\end{lemma}

\begin{proof}
	Let $(\theta^1,\theta^2)$ be the local coordinates of some coordinate chart on $\mathbb{S}^2$, and let $i$ denote the parametrisation of the extended $\hat{u}_f$ normalised $\I$ gauge.  In what follows the identification
	\begin{equation} \label{eq:mvtsphereiden1}
		\pi_{\mathbb{S}^2} \circ i^{-1} \colon S^{[n]}_{\hat{u}_{f} + \delta,v_\infty} \to \mathbb{S}^2,
	\end{equation}
	\begin{equation} \label{eq:mvtsphereiden2}
		i(\hat{u}_f+\delta + h^3_{[n]} \circ \slashed{H}_{[n-1]}^{-1}(\theta)
		,
		v_{\infty} + h^4_{[n]} \circ \slashed{H}_{[n-1]}^{-1}(\theta)
		,
		\theta)
		\mapsto
		\theta
	\end{equation}
	will be used, where $\pi_{\mathbb{S}^2}$ denotes the projection onto the $\mathbb{S}^2$ argument.  Consider Proposition \ref{determiningthesphere}, in particular equations \eqref{diffiscloseone}, \eqref{diffisclosetwo} and \eqref{diffisclosethree}, with $S = \mathbb{S}^2$, $p=(1,0,0)$, $v=(0,1,0)\in T_p \mathbb{S}^2$, $h = i^*( r^{-2} \gslash)(\hat{u}_f,v_{\infty}(\hat{u}_f),\cdot)$ and $\hat{h} = R^* \mathfrak{g}$ where
	\[
		\mathfrak{g}
		=
		r_{[n]}(\hat{u}_f+\delta,v_{\infty})^{-2} (i^*g)
		(\hat{u}_f+\delta + h^3_{[n]} \circ \slashed{H}_{[n-1]}^{-1}(\cdot)
		,
		v_{\infty} + h^4_{[n]} \circ \slashed{H}_{[n-1]}^{-1}(\cdot)
		,
		\cdot),
	\]
	is the ($r_{[n]}^{-2}$ normalised) induced metric on the sphere $S^{[n]}_{\hat{u}_f+\delta,v_{\infty}}$ in the parametrisation of the extended $\hat{u}_f$ normalised $\I$ gauge (i.\@e.\@ viewed as a metric on $\mathbb{S}^2$ via the identification \eqref{eq:mvtsphereiden1}, \eqref{eq:mvtsphereiden2}), $v_{\infty} = v_{\infty}(\hat{u}_f+\delta)$ and $R\colon \mathbb{S}^2 \to \mathbb{S}^2$ is the unique rotation which satisfies
	\begin{align*}
		R(1,0,0)
		&
		=
		\pi_{\mathbb{S}^2} \circ i^{-1} \circ j_{[n]}\circ (\pi^{\mathcal{EF}}_{\mathbb S^2}|_{S^{[n]}_{u_{-1},v_\infty}})^{-1} (1,0,0),
		\\
		R_* (0,1,0)
		&
		=
		(\pi_{\mathbb{S}^2})_* \circ (i^{-1})_* \circ (j_{[n]})_*\circ (\pi^{\mathcal{EF}}_{\mathbb S^2}|_{S^{[n]}_{u_{-1},v_\infty}})^{-1}_* (0,L^{-1},0),
	\end{align*}
	where $j_{[n]}:S^{[n]}_{u_{-1},v_\infty}\to S^{[n]}_{\hat{u}_{f} + \delta,v_\infty}$ is the natural null flow diffeomorphism of the $n$-th double null foliation (see \eqref{choiceofpandv}) and
	\[
		L
		=
		\big\vert (\pi_{\mathbb{S}^2})_* \circ (i^{-1})_* \circ (j_{[n]})_*\circ (\pi^{\mathcal{EF}}_{\mathbb S^2}|_{S^{[n]}_{u_{-1},v_\infty}})^{-1}_* (0,1,0) \big\vert_{\mathring{\gamma}}.
	\]
	Now, in order to estimate the right hand sides of equations \eqref{diffiscloseone}, \eqref{diffisclosetwo} and \eqref{diffisclosethree}, for $k = 0,1,2$, in view of the $\mathbb{S}^2$ identification of the extended $\hat{u}_f$ normalised $\I$ double null parametrisation,
	\begin{align} \label{eq:rotationest1}
		&
		\Vert \nabla_{\psi^* h}^k \psi^* (h - \hat{h}) \Vert_{\mathbb{S}^2}
		=
		\Vert (r\nablaslash)^k (r^{-2}\gslash(\hat{u}_f,v_{\infty}(\hat{u}_f),\cdot) - R^* \mathfrak{g}) \Vert_{\mathbb{S}^2}
		\\
		&
		\lesssim
		\Vert (r\nablaslash)^k (r^{-2}\gslash(\hat{u}_f,v_{\infty}(\hat{u}_f),\cdot) - R^*(r^{-2}\gslash)(\hat{u}_f,v_{\infty}(\hat{u}_f),\cdot)) \Vert_{\mathbb{S}^2}
		+
		\Vert (r\nablaslash)^k (R^*(r^{-2}\gslash)(\hat{u}_f,v_{\infty}(\hat{u}_f),\cdot)) - R^* \mathfrak{g}) \Vert_{\mathbb{S}^2}.
		\nonumber
	\end{align}
	It follows from the relation \eqref{eq:metricid6} that
	\begin{equation} \label{eq:rotationest2}
		\Vert (r\nablaslash)^k (r^{-2}\gslash(\hat{u}_f,v_{\infty}(\hat{u}_f),\cdot) - \mathfrak{g}) \Vert_{\mathbb{S}^2}
		\lesssim
		\frac{\varepsilon \delta}{r}
		+
		r^{-1}
		\sum_{\tilde{k} \leq k+1}
		\big(
		\Vert (r\nablaslash)^{\tilde{k}} (h^3_{[n]} \circ \slashed{H}_{[n-1]}^{-1}) \Vert_{\mathbb{S}^2}
		+
		\Vert (r\nablaslash)^{\tilde{k}} (h^4_{[n]} \circ \slashed{H}_{[n-1]}^{-1}) \Vert_{\mathbb{S}^2}
		\big).
	\end{equation}
	Now, for any $(0,k)$ tensor field $\omega$ on $\mathbb{S}^2$,
	\begin{align} \label{eq:rotationMVT}
		\sum_{l \leq k} \Vert \mathring{\nabla}^l(\omega - R^* \omega) \Vert_{\mathbb{S}^2}
		\lesssim
		\sum_{l \leq k+1 } \Vert \mathring{\nabla}^l \omega \Vert_{\mathbb{S}^2}
		\sum_{A=1,2}
		\big(
		\vert R(x_{[n]})^{A} - x_{[n]}^{A} \vert
		+
		\vert ( R_* \partial_{\theta^1} \vert_{R(x_{[n]})})^{A} - (\partial_{\theta^1} \vert_{x_{[n]}})^{A} \vert
		\big),
	\end{align}
	where
	\[
		x_{[n]} = (\hat{u}_f+\delta + f^3_{[n]}(\hat{u}_f+\delta,v_{\infty},\slashed{H}_{[n]}^{-1} (\theta_0))
		,
		v_{\infty}
		+
		f^4_{[n]}(\hat{u}_f+\delta,v_{\infty},\slashed{H}_{[n]}^{-1} (\theta_0))
		, \theta_0),
	\]
	which, recall, is identified with the point $\theta_0 \in \mathbb{S}^2$ via \eqref{eq:mvtsphereiden1}, \eqref{eq:mvtsphereiden2}.
	Before proceeding to estimate the terms on the right hand side of \eqref{eq:rotationMVT}, first note that the diffeomorphism $j_{[n]}$ has the property that, for a point $q = i_{\I} (u(q),v(q),\theta(q)) \in S^{[n]}_{u_{-1},v_{\infty}}$, its image under $j_{[n]}$ takes the form $j_{[n]}(q) = i_{\I} (u(j_{[n]}q),v(j_{[n]}q),\theta(j_{[n]}q)) \in S^{[n]}_{\hat{u}_{f}+\delta,v_{\infty}}$ with coordinates satisfying
	\begin{align}
		\vert u(j_{[n]} q) -(\hat{u}_f+\delta) - u(q) + u_{-1} \vert
		&
		\lesssim
		\sup_{\theta \in \mathbb{S}^2}
		\big(
		\vert f^3_{[n]}(\hat{u}_f+\delta,v_{\infty},\theta) \vert
		+
		\vert f^3_{[n]}(u_{-1},v_{\infty},\theta) \vert
		\big),
		\label{eq:jncoordsdiff1}
		\\
		\vert v(j_{[n]} q) - v(q) \vert
		&
		\lesssim
		\sup_{\theta \in \mathbb{S}^2}
		\big(
		\vert f^4_{[n]}(\hat{u}_f+\delta,v_{\infty},\theta) \vert
		+
		\vert f^4_{[n]}(u_{-1},v_{\infty},\theta) \vert
		\big),
		\label{eq:jncoordsdiff2}
		\\
		\sum_{A=1,2} \vert \theta^A (j_{[n]} q) - \theta^A(q) \vert
		&
		\lesssim
		\sup_{\theta \in \mathbb{S}^2}
		\int_{u_{-1}}^{\hat{u}_f+\delta}
		r^{-1}
		\vert \partial_u \slashed{f}_{[n]} (u,v_{\infty}) \vert
		du
		.
		\label{eq:jncoordsdiff3}
	\end{align}
	Moreover, if a vector $w \in T_q S^{[n]}_{u_{-1},v_{\infty}}$ takes the form $w = w^A \partial_{\theta^A} + w^u \partial_u + w^v \partial_v$ with respect to the coordinate frame of the extended $\hat{u}_f$ normalised $\I$ gauge, then its image $(j_{[n]})_* w \in T_{j_{[n]}(q)} S^{[n]}_{\hat{u}_f+\delta,v_{\infty}}$ takes the form $(j_{[n]})_* w = ((j_{[n]})_* w)^A \partial_{\theta^A} + ((j_{[n]})_* w)^u \partial_u + ((j_{[n]})_* w)^v \partial_v$, where the components satisfy
	\begin{align} \label{eq:jnstarcoordsdiff1}
		\vert ((j_{[n]})_* w)^u - w^u \vert
		&
		\lesssim
		\sup_{\theta \in \mathbb{S}^2}
		\big(
		\vert
		w^A \partial_{\theta^A} (\slashed{H}^{-1}_{[n]})^B(\theta)
		\partial_{\theta^B} f^3_{[n]}(u_{-1},v_{\infty},\theta)
		\vert
		+
		\vert
		w^A \partial_{\theta^A} (\slashed{H}^{-1}_{[n]})^B(\theta)
		\partial_{\theta^B} f^3_{[n]}(\hat{u}_f+\delta,v_{\infty},\theta)
		\vert
		\big)
		\\
		\vert ((j_{[n]})_* w)^v - w^v \vert
		&
		\lesssim
		\sup_{\theta \in \mathbb{S}^2}
		\big(
		\vert
		w^A \partial_{\theta^A} (\slashed{H}^{-1}_{[n]})^B(\theta)
		\partial_{\theta^B} f^4_{[n]}(u_{-1},v_{\infty},\theta)
		\vert
		+
		\vert
		w^A \partial_{\theta^A} (\slashed{H}^{-1}_{[n]})^B(\theta)
		\partial_{\theta^B} f^4_{[n]}(\hat{u}_f+\delta,v_{\infty},\theta)
		\vert
		\big)
		\label{eq:jnstarcoordsdiff2}
		\\
		\vert
		((j_{[n]})_* w)^A - w^A
		\vert
		&
		\lesssim
		\vert
		w^B \partial_{\theta^B} (\slashed{H}^{-1}_{[n]})^C(\theta)
		\vert
		\int_{u_{-1}}^{\hat{u}_f+\delta}
		\vert \partial_{\theta^C} \partial_u \slashed{f}_{[n]}^A (u,v_{\infty}) \vert
		du
		.
		\label{eq:jnstarcoordsdiff3}
	\end{align}
	Indeed, \eqref{eq:jncoordsdiff1}--\eqref{eq:jncoordsdiff3} follow from the fact that any such $q$ takes the form
	\[
		q = i_{\I}(u_{-1} + f^3_{[n]}(u_{-1},v_{\infty},\slashed{H}_{[n]}^{-1} (\theta))
		,
		v_{\infty}
		+
		f^4_{[n]}(u_{-1},v_{\infty},\slashed{H}_{[n]}^{-1} (\theta))
		, 
		\slashed{F}_{[n]}(u_{-1},v_{\infty},\slashed{H}_{[n]}^{-1} (\theta))),
	\]
	for some $\theta\in \mathbb{S}^2$ and then
	\begin{multline*}
		j_{[n]} \circ i_{\I}(u_{-1} + f^3_{[n]}(u_{-1},v_{\infty},\slashed{H}_{[n]}^{-1} (\theta))
		,
		v_{\infty}
		+
		f^4_{[n]}(u_{-1},v_{\infty},\slashed{H}_{[n]}^{-1} (\theta))
		,
		\slashed{F}_{[n]}(u_{-1},v_{\infty},\slashed{H}_{[n]}^{-1} (\theta)))
		\\
		=
		i_{\I} (\hat{u}_f+\delta + f^3_{[n]}(\hat{u}_f+\delta,v_{\infty},\slashed{H}_{[n]}^{-1} (\theta))
		,
		v_{\infty}
		+
		f^4_{[n]}(\hat{u}_f+\delta,v_{\infty},\slashed{H}_{[n]}^{-1} (\theta))
		, \theta),
	\end{multline*}
	along with the fact that
	\[
		\vert
		\slashed{F}_{[n]}^A(u_{-1},v_{\infty},\slashed{H}_{[n]}^{-1} (\theta))
		-
		\theta^A
		\vert
		\lesssim
		\int_{u_{-1}}^{\hat{u}_f+\delta}
		\vert \partial_u \slashed{f}^A_{[n]} (u,v_{\infty},\slashed{H}_{[n]}^{-1} (\theta)) \vert du,
	\]
	and
	\[
		\vert \partial_u f^A_{[n]} \vert
		\lesssim
		\vert \partial_u \slashed{f}_{[n]} \vert \vert d \theta^A \vert_{\gslash}
		\lesssim
		r^{-1} \vert \partial_u \slashed{f}_{[n]} \vert.
	\]
	Similarly, \eqref{eq:jnstarcoordsdiff1} and \eqref{eq:jnstarcoordsdiff2} follow from the fact that any $w \in T_q S^{[n]}_{u_{-1},v_{\infty}}$ can be written
	\begin{multline*}
		w = w^A \Big(
		\partial_{\theta^B} f^3_{[n]}(u_{-1},v_{\infty},\slashed{H}_{[n]}^{-1}(\theta)) \partial_{\theta^A} (\slashed{H}_{[n]}^{-1})^B(\theta) \partial_u
		+
		\partial_{\theta^B} f^4_{[n]}(u_{-1},v_{\infty},\slashed{H}_{[n]}^{-1}(\theta)) \partial_{\theta^A} (\slashed{H}_{[n]}^{-1})^B(\theta) \partial_v
		\\
		+
		\partial_{\theta^B} \slashed{F}^C_{[n]}(u_{-1},v_{\infty},\slashed{H}_{[n]}^{-1}(\theta)) \partial_{\theta^A} (\slashed{H}_{[n]}^{-1})^B(\theta)
		\partial_{\theta^C}
		\Big),
	\end{multline*}
	and so $(j_{[n]})_* w$ takes the form
	\begin{multline*}
		(j_{[n]})_* w
		=
		w^A \Big(
		\partial_{\theta^B} f^3_{[n]}(\hat{u}_f+\delta,v_{\infty},\slashed{H}_{[n]}^{-1}(\theta)) \partial_{\theta^A} (\slashed{H}_{[n]}^{-1})^B(\theta) \partial_u
		\\
		+
		\partial_{\theta^B} f^4_{[n]}(\hat{u}_f+\delta,v_{\infty},\slashed{H}_{[n]}^{-1}(\theta)) \partial_{\theta^A} (\slashed{H}_{[n]}^{-1})^B(\theta) \partial_v
		+
		\partial_{\theta^A}
		\Big).
	\end{multline*}
	
	The terms on the right hand side of \eqref{eq:rotationMVT} are estimated as follows.  The rotation $R$ has the property that (recalling the identification \eqref{eq:mvtsphereiden1}, \eqref{eq:mvtsphereiden2} of $S^{[n]}_{\hat{u}_f+\delta,v_{\infty}}$ with $\mathbb{S}^2$)
	\[
		R
		(\hat{u}_f+\delta + f^3_{[n]}(\hat{u}_f+\delta,v_{\infty},\slashed{H}_{[n]}^{-1} (\theta_0))
		,
		v_{\infty}
		+
		f^4_{[n]}(\hat{u}_f+\delta,v_{\infty},\slashed{H}_{[n]}^{-1} (\theta_0))
		, \theta_0)
		=
		j_{[n]}(u_*,v_*,\slashed{F}_{\I,d}(u_*,v_*,\theta_0)),
	\]
	and
	\[
		R_* \partial_{\theta^1}
		=
		L^{-1}
		(j_{[n]})_*( \Pi_{S^{[n]}_{u_{-1},v_{\infty}}} \partial_{\theta^1_{\mathcal{EF}}})
	\]
	for some $u_*$, $v_*$ satisfying
	\begin{equation} \label{eq:defofustarvstarthetastar}
		u_* = u_{-1} + f^3_{[n]}(u_{-1},v_{\infty},\theta_*),
		\qquad
		v_* = v_{\infty} + f^4_{[n]}(u_{-1},v_{\infty},\theta_*),
	\end{equation}
	and $\theta_*$ satisfying
	\[
		\slashed{F}_{\I,d}(u_*,v_*,\theta_0)
		=
		\slashed{F}_{[n]} (u_{-1} ,v_{\infty},\theta_*)
		,
	\]
	so that indeed
	\[
		i(u_*,v_*,\slashed{F}_{\I,d}(u_*,v_*,\theta_0))
		\in
		S^{[n]}_{u_{-1} ,v_{\infty}},
	\]
	where $\slashed{F}_{\I,d} = \pi_{\mathbb{S}^2} \circ i_{\I}^{-1} \circ i_{\mathcal{EF}}$ and $i_{\mathcal{EF}}$ is the initial Eddington--Finkelstein parametrisation \eqref{initialEFgaugehere}, $i_{\I}$ is the parametrisation of the extended $\hat{u}_f$ normalised $\I$ gauge, and $\pi_{\mathbb{S}^2}$ denotes the projection onto the $\mathbb{S}^2$ argument.  
	To estimate $\vert R(x_{[n]})^{A} - x_{[n]}^{A} \vert$, note that \eqref{eq:jncoordsdiff3} implies that
	\[
		\big\vert
		\big(
		R(x_{[n]})^A - x_{[n]}^A
		\big)
		-
		\big(
		\slashed{F}_{\I,d}(u_*,v_*, \theta_0)^A
		-
		\theta_0^A
		\big)
		\big\vert
		\lesssim
		\sup_{\theta \in \mathbb{S}^2}
		\int_{u_{-1}}^{\hat{u}_f+\delta}
		r^{-1}
		\vert \partial_u \slashed{f}_{[n]} (u,v_{\infty}) \vert
		du.
	\]
	Recall now that $\slashed{F}_{\I,d}(u_{-1},v_{\infty}(\hat{u}_f), \theta_0) = \theta_0$ and so
	\begin{align*}
		\vert
		\slashed{F}_{\I,d}(u_*,v_*, \theta_0)^A
		-
		\theta_0^A
		\vert
		&
		\lesssim
		\int_{u_{-1}}^{u_*}
		\vert \partial_u \slashed{f}^A_{\I,d} (u,v_*,\theta_0) \vert
		du
		+
		\int_{v_{\infty}(\hat{u}_f)}^{v_*}
		\vert \partial_v \slashed{f}^A_{\I,d} (u_{-1},v,\theta_0) \vert
		dv
		\\
		&
		\lesssim
		\frac{1}{r}
		\big(
		\varepsilon \delta
		+
		\varepsilon
		\sup_{\theta \in \mathbb{S}^2}
		\big(
		\vert f^3_{[n]}(u_{-1},v_{\infty}(\hat{u}_f+\delta),\theta) \vert
		+
		\vert f^4_{[n]}(u_{-1},v_{\infty}(\hat{u}_f+\delta),\theta) \vert
		\big)
		\big),
	\end{align*}
	by \eqref{eq:defofustarvstarthetastar}, since $\vert v_{\infty}(\hat{u}_f) - v_{\infty}(\hat{u}_f+\delta) \vert \lesssim \delta$.  Hence, for $A=1,2$,
	\begin{equation} \label{eq:rotationest5}
		\big\vert
		R(x_{[n]})^A - x_{[n]}^A
		\big\vert
		\lesssim
		\frac{\varepsilon \delta}{r}
		+
		r^{-1}
		\sup_{\theta \in \mathbb{S}^2}
		\big(
		\vert f^3_{[n]}(u_{-1},v_{\infty}(\hat{u}_f+\delta)) \vert
		+
		\vert f^4_{[n]}(u_{-1},v_{\infty}(\hat{u}_f+\delta)) \vert
		+
		\int_{u_{-1}}^{\hat{u}_f+\delta}
		\vert \partial_u \slashed{f}_{[n]} (u,v_{\infty}) \vert
		du
		\big).
	\end{equation}
	
	Note now that
	\[
		\Pi_{S^{[n]}_{u_{-1},v_{\infty}}} \partial_{\theta^1_{\mathcal{EF}}}
		=
		\partial_{\theta^1_{\mathcal{EF}}}
		+
		\frac{1}{2} g( \partial_{\theta^1_{\mathcal{EF}}}, e_4^{[n]}) e_3^{[n]}
		+
		\frac{1}{2} g(\partial_{\theta^1_{\mathcal{EF}}}, e_3^{[n]}) e_4^{[n]}.
	\]
	With respect to the double null frame of the extended $\I$ gauge,
	\begin{align*}
		\partial_{\theta^1_{\mathcal{EF}}}
		=
		\
		&
		\partial_{{\theta}^1} F_{\I,d}^A e_A
		+
		\Omega \partial_{{\theta}^1} f^3_{\I,d} e_3
		+
		\Omega \partial_{{\theta}^1} f^4_{\I,d} e_4
		-
		\partial_{{\theta}^1} f^3_{\I,d} b,
		\\
		e_3^{[n]}
		=
		\
		&
		\Omega \left(
		1 + \partial_u f^3_{[n]}
		\right) e_3
		+
		\Omega \partial_u f^4_{[n]} e_4
		+
		\partial_u f^A_{[n]} e_A
		-
		\left( 1 + \partial_u f^3_{[n]} \right) b,
		\\
		e_4^{[n]}
		=
		\
		&
		\Omega \left(
		1 + \partial_v f^4_{[n]}
		\right) e_4
		+
		\Omega \partial_v f^3_{[n]} e_3
		+
		\partial_v f^A_{[n]} e_A
		-
		\partial_v f^3_{[n]} b
		,
	\end{align*}
	since $b_{[n]} \equiv 0$ and $\Omega_{[n]} \equiv 1$ on $S^{[n]}_{u_{-1},v_{\infty}}$,
	and so it follows that
	\[
		\Pi_{S^{[n]}_{u_{-1},v_{\infty}}} \partial_{\theta^1_{\mathcal{EF}}}
		=
		\mathcal{U}^A \partial_{\theta^A} + \mathcal{U}^u \partial_u + \mathcal{U}^v \partial_v,
	\]
	where the components satisfy
	\begin{align}
		&
		\sum_{A=1,2} \vert \mathcal{U}^A(x_0) - \partial_{{\theta}^1} \slashed{F}_{\I,d}^A(u_*,v_*,\theta_0) \vert
		+
		\vert \mathcal{U}^u(x_0) - \gslash_{AB}(x_0) \partial_v \slashed{F}^A_{[n]}(u_{-1},v_{\infty},\slashed{H}_{[n]}^{-1}(\theta)) \partial_{\theta^1} \slashed{F}^B_{\I,d}(u_*,v_*,\theta_0) \vert
		\nonumber
		\\
		&
		\qquad \qquad
		+
		\vert \mathcal{U}^v(x_0) - \gslash_{AB}(x_0) \partial_u \slashed{F}^A_{[n]}(u_{-1},v_{\infty},\slashed{H}_{[n]}^{-1}(\theta)) \partial_{\theta^1} \slashed{F}^B_{\I,d}(u_*,v_*,\theta_0) \vert
		\nonumber
		\\
		&
		\lesssim
		\sup_{\theta \in \mathbb{S}^2}
		\big(
		\vert \partial_u \slashed{f}_{[n]} (u_{-1},v_{\infty}) \vert
		+
		\vert \partial_v \slashed{f}_{[n]} (u_{-1},v_{\infty}) \vert
		+
		\vert \partial_u f^3_{[n]} (u_{-1},v_{\infty}) \vert
		+
		\vert \partial_v f^3_{[n]} (u_{-1},v_{\infty}) \vert
		+
		\vert \partial_u f^4_{[n]} (u_{-1},v_{\infty}) \vert
		\nonumber
		\\
		&
		\qquad \qquad
		+
		\vert \partial_v f^4_{[n]} (u_{-1},v_{\infty}) \vert
		\big),
		\label{eq:componentsofcalU}
	\end{align}
	using the relation \eqref{eq:metricid1} and the estimates \eqref{improvement3ext}, which easily yield appropriate estimates for the $f_{\I,d}$ diffeomorphism functions, where
	\[
		x_0
		=
		(u_*,v_*,\slashed{F}_{\I,d}(u_*,v_*,\theta_0)).
	\]
	It follows from \eqref{eq:jnstarcoordsdiff3} and \eqref{eq:componentsofcalU} that, for $A=1,2$,
	\begin{align*}
		&
		\big\vert
		\big(
		( R_* \partial_{\theta^1} \vert_{R(x_{[n]})})^A - (\partial_{\theta^1} \vert_{x_{[n]}})^A
		\big)
		-
		\big(
		\partial_{{\theta}^1} \slashed{F}_{\I,d}^A(u_*,v_*,\theta_0) - \delta_1^A
		\big)
		\big\vert
		\\
		&
		\qquad
		\lesssim
		\sup_{\theta \in \mathbb{S}^2}
		\big(
		\int_{u_{-1}}^{\hat{u}_f+\delta}
		r^{-1} \vert r\nablaslash \partial_u \slashed{f}_{[n]} (u,v_{\infty}) \vert
		du
		+
		\vert \partial_u \slashed{f}_{[n]} (u_{-1},v_{\infty}) \vert
		+
		\vert \partial_v \slashed{f}_{[n]} (u_{-1},v_{\infty}) \vert
		+
		\vert \partial_u f^3_{[n]} (u_{-1},v_{\infty}) \vert
		\\
		&
		\qquad \quad
		+
		\vert \partial_v f^3_{[n]} (u_{-1},v_{\infty}) \vert
		+
		\vert \partial_u f^4_{[n]} (u_{-1},v_{\infty}) \vert
		+
		\vert \partial_v f^4_{[n]} (u_{-1},v_{\infty}) \vert
		\big)
		.
	\end{align*}
	Now recall that
	\[
		\partial_{{\theta}^1} \slashed{F}_{\I,d}^A(u_{-1},v_{\infty}(\hat{u}_f),\theta_0) = \delta_1^A,
	\]
	and so
	\[
		\vert \partial_{{\theta}^1} \slashed{F}_{\I,d}^A(u_*,v_*,\theta_0) - \delta_1^A \vert
		\lesssim
		\frac{1}{r}
		\big(
		\varepsilon \delta
		+
		\varepsilon
		\sup_{\theta \in \mathbb{S}^2}
		\big(
		\vert f^3_{[n]}(u_{-1},v_{\infty}(\hat{u}_f+\delta),\theta) \vert
		+
		\vert f^4_{[n]}(u_{-1},v_{\infty}(\hat{u}_f+\delta),\theta) \vert
		\big)
		\big).
	\]
	Hence
	\begin{align}
		&
		\big\vert
		( R_* \partial_{\theta^1} \vert_{R(x_{[n]})})^A - (\partial_{\theta^1} \vert_{x_{[n]}})^A
		\big\vert
		\lesssim
		\frac{\varepsilon \delta}{r}
		+
		\varepsilon
		\sup_{\theta \in \mathbb{S}^2}
		\big(
		r^{-1}
		\big(
		\vert f^3_{[n]}(u_{-1},v_{\infty}(\hat{u}_f+\delta),\theta) \vert
		+
		\vert f^4_{[n]}(u_{-1},v_{\infty}(\hat{u}_f+\delta),\theta) \vert
		\big)
		\nonumber
		\\
		&
		\qquad \quad
		+
		\int_{u_{-1}}^{\hat{u}_f+\delta}
		r^{-1} \vert r\nablaslash \partial_u \slashed{f}_{[n]} (u,v_{\infty}) \vert
		+
		\vert \partial_u \slashed{f}_{[n]} (u_{-1},v_{\infty}) \vert
		+
		\vert \partial_v \slashed{f}_{[n]} (u_{-1},v_{\infty}) \vert
		+
		\vert \partial_u f^3_{[n]} (u_{-1},v_{\infty}) \vert
		\nonumber
		\\
		&
		\qquad \quad
		+
		\vert \partial_v f^3_{[n]} (u_{-1},v_{\infty}) \vert
		+
		\vert \partial_u f^4_{[n]} (u_{-1},v_{\infty}) \vert
		+
		\vert \partial_v f^4_{[n]} (u_{-1},v_{\infty}) \vert
		\big)
		.
		\label{eq:rotationest6}
	\end{align}
	The estimates \eqref{eq:rotationest1}, \eqref{eq:rotationest2}, \eqref{eq:rotationMVT}, \eqref{eq:rotationest5}, \eqref{eq:rotationest6}, together with the Sobolev inequality (Proposition \ref{prop:Sobolev}) and the fact that $\hat{u}_f r(\hat{u}_f,v_{\infty})^{-1} \leq 1$, then imply that, for any $2\leq k \leq 4$,
	\begin{equation} \label{eq:finallymetricdifference}
		\sum_{l\leq k}
		\Vert \nabla_{\psi^* h}^l \psi^* (h - \hat{h}) \Vert_{\mathbb{S}^2}
		\lesssim
		\frac{\varepsilon \delta}{r}
		+
		\sup_{u_{-1} \leq u \leq \hat{u}_f+\delta}
		\sum_{\vert \gamma \vert \leq k+1} \Vert \mathfrak{D}^{\gamma} f_{[n]} \Vert_{S_{u,v_{\infty}}}.
	\end{equation}

	Equation \eqref{diffiscloseone} and the estimate \eqref{eq:finallymetricdifference} imply that
	\begin{equation} \label{eq:diffiscloseiteratescoords}
		\vert \slashed{F}^A_{[n]}(\hat{u}_f + \delta,v_{\infty},\theta) - \theta^A \vert
		\lesssim
		\frac{\varepsilon \delta}{r}
		+
		\sup_{u_{-1} \leq u \leq \hat{u}_f+\delta}
		\sum_{\vert \gamma \vert \leq 3} \Vert \mathfrak{D}^{\gamma} f_{[n]} \Vert_{S_{u,v_{\infty}}},
	\end{equation}
	for $A=1,2$.  Hence, for any $u_{-1} \leq u \leq \hat{u}_f + \delta$,
	\[
		\vert \slashed{F}^A_{[n]}(u,v_{\infty},\theta) - \theta^A \vert
		\lesssim
		\frac{\varepsilon \delta}{r}
		+
		\sup_{u_{-1} \leq u \leq \hat{u}_f+\delta}
		\sum_{\vert \gamma \vert \leq 3} \Vert \mathfrak{D}^{\gamma} f_{[n]} \Vert_{S_{u,v_{\infty}}}
		+
		\int_u^{\hat{u}_f + \delta} \vert \partial_u f^A_{[n]} (u',v_{\infty},\theta) \vert du'.
	\]
	Now \eqref{eq:diffiscloseiteratesone} follows from the fact that
	\[
		|h(u,v_{\infty},\theta) - h(u,v_{\infty},\slashed{F}_{[n]}(u,v_{\infty},\theta))|
		\lesssim
		\big( \sup_{S_{u,v_{\infty}}} | r\nablaslash h |_{\gslash_{[n]}} \big)
		\sum_{A=1}^2\vert \slashed{F}^A_{[n]}(u ,v_{\infty},\theta) - \theta^A \vert,
	\]
	and
	\[
		\vert \partial_u f^A_{[n]} \vert
		\lesssim
		\vert \partial_u \slashed{f}_{[n]} \vert \vert d \theta^A \vert_{\gslash}
		\lesssim
		r^{-1} \vert \partial_u \slashed{f}_{[n]} \vert.
	\]
	For \eqref{eq:diffiscloseiteratestwo} note that, in the $(\theta^1,\theta^2)$ coordinate chart,
	\begin{multline*}
		\big( \omega - (\slashed{F}_{[n]}(u,v_{\infty},\cdot))^*\omega \big) (u,v_{\infty},\theta)
		\\
		=
		\Big( \omega_{A_1\ldots A_k}(u,v_{\infty},\theta)
		- 
		\omega_{B_1\ldots B_k}(u,v_{\infty},\slashed{F}_{[n]}(u,v,\theta))
		\partial_{\theta_{[n]}^{A_1}} \slashed{F}^{B_1}_{[n]} \ldots \partial_{\theta_{[n]}^{A_k}} \slashed{F}^{B_k}_{[n]}
		\Big) d\theta_{[n]}^{A_1} \ldots d\theta_{[n]}^{A_k}.
	\end{multline*}
	Proposition \ref{determiningthesphere}, in particular equation \eqref{diffisclosetwo}, and the estimate \eqref{eq:finallymetricdifference} imply that
	\[
		\vert \partial_{\theta^B} \slashed{F}^A_{[n]}(\hat{u}_f + \delta,v_{\infty},\theta) - \delta^A_B \vert
		\lesssim
		\frac{\varepsilon \delta}{r}
		+
		\sup_{u_{-1} \leq u \leq \hat{u}_f+\delta}
		\sum_{\vert \gamma \vert \leq 3} \Vert \mathfrak{D}^{\gamma} f_{[n]} \Vert_{S_{u,v_{\infty}}},
	\]
	for $A,B=1,2$.  Hence, for any $u_{-1} \leq u \leq \hat{u}_f + \delta$,
	\begin{align*}
		\vert \partial_{\theta^B} \slashed{F}^A_{[n]}(u,v_{\infty},\theta) - \delta^A_B \vert
		\lesssim
		\frac{\varepsilon \delta}{r}
		+
		\sup_{u_{-1} \leq u \leq \hat{u}_f+\delta}
		\sum_{\vert \gamma \vert \leq 3} \Vert \mathfrak{D}^{\gamma} f_{[n]} \Vert_{S_{u,v_{\infty}}}
		+
		\int_u^{\hat{u}_f + \delta} \vert \partial_{\theta^B} \partial_u f^A_{[n]} (u',v_{\infty},\theta) \vert du'
		.
	\end{align*}
	Moreover,
	\begin{align*}
		&
		\big\vert \big( \omega_{A_1\ldots A_k}(u,v_{\infty},\theta)
		- 
		\omega_{A_1\ldots A_k}(u,v_{\infty},\slashed{F}_{[n]}(u,v,\theta))
		\big) d\theta_{[n]}^{A_1} \ldots d\theta_{[n]}^{A_k} 
		\big\vert
		\\
		&
		\qquad \qquad
		\lesssim
		r^k \big( \sup_{S_{u,v_{\infty}}} | r\nablaslash \omega |_{\gslash_{[n]}} \big)
		\sum_{A=1}^2 \vert \slashed{F}^A_{[n]}(u ,v_{\infty},\theta) - \theta^A \vert
		\vert d\theta_{[n]}^{A_1}\vert_{\gslash_{[n]}} \ldots \vert d\theta_{[n]}^{A_k}\vert_{\gslash_{[n]}}
		\\
		&
		\qquad \qquad
		\lesssim
		\big( \sup_{S_{u,v_{\infty}}} | r\nablaslash \omega |_{\gslash_{[n]}} \big)
		\sum_{A=1}^2 \vert \slashed{F}^A_{[n]}(u ,v_{\infty},\theta) - \theta^A \vert.
	\end{align*}
	It follows that
	\begin{align*}
		\big\vert
		\big( \omega - (\slashed{F}_{[n]}(u,v_{\infty},\cdot))^*\omega \big) (u,v_{\infty},\theta)
		\big\vert
		\lesssim
		&
		\int_u^{\hat{u}_f + \delta} \sum_{A,B=1}^2 \vert \partial_{\theta^B} \partial_u f^A_{[n]} (u',v_{\infty},\theta) \vert du'
		\\
		&
		+
		\frac{\varepsilon \delta}{r}
		+
		\sup_{u_{-1} \leq u \leq \hat{u}_f+\delta}
		\sum_{\vert \gamma \vert \leq 3} \Vert \mathfrak{D}^{\gamma} f_{[n]} \Vert_{S_{u,v_{\infty}}}.
	\end{align*}
	The proof of \eqref{eq:diffiscloseiteratestwo} then follows from the fact that
	\[
		\vert \partial_{\theta^B} \partial_u f^A_{[n]} \vert
		\lesssim
		\vert \nablaslash \partial_u \slashed{f}_{[n]} \vert.
	\]
	The estimate \eqref{eq:diffiscloseiteratesthree} similarly follows from \eqref{diffisclosethree} and \eqref{eq:finallymetricdifference}.
	
	For the estimates \eqref{eq:diffiscloseiteratesdifferenceone}, \eqref{eq:diffiscloseiteratesdifferencetwo} and \eqref{eq:diffiscloseiteratesdifferencethree}, consider Proposition \ref{determiningthesphere}, in particular equations \eqref{diffiscloseone}, \eqref{diffisclosetwo} and \eqref{diffisclosethree}, with $S = \mathbb{S}^2$, $p=(1,0,0)$, $v=(0,1,0)\in T_p \mathbb{S}^2$,
	\[
		h
		=
		r_{[n]}(\hat{u}_f+\delta,v_{\infty})^{-2}
		(i^*g)
		(\hat{u}_f+\delta + h^3_{[n]} \circ \slashed{H}_{[n-1]}^{-1} \circ \slashed{H}_{[n]} (\cdot)
		,
		v_{\infty} + h^4_{[n]} \circ \slashed{H}_{[n-1]}^{-1} \circ \slashed{H}_{[n]}(\cdot)
		,
		\slashed{H}_{[n]}( \cdot))
	\]
	\[
		\hat{h}
		=
		r_{[n+1]}(\hat{u}_f+\delta,v_{\infty})^{-2} ( R^*_{[n+1]} i^*g)
		(\hat{u}_f+\delta + h^3_{[n+1]} \circ \slashed{H}_{[n]}^{-1}(\cdot)
		,
		v_{\infty} + h^4_{[n+1]} \circ \slashed{H}_{[n]}^{-1}(\cdot)
		,
		\cdot),
	\]
	where $R_{[n+1]} \colon \mathbb{S}^2 \to \mathbb{S}^2$ is the unique rotations which satisfies
	\begin{align*}
		R_{[n+1]}(1,0,0)
		&
		=
		\pi_{\mathbb{S}^2} \circ i_{[n]}^{-1} \circ j_{[n+1]}\circ (\pi^{\mathcal{EF}}_{\mathbb S^2}|_{S^{\I}_{u_{-1},v_\infty}})^{-1} (1,0,0),
		\\
		(R_{[n+1]})_* (0,1,0)
		&
		=
		(\pi_{\mathbb{S}^2})_* \circ (i_{[n]}^{-1})_* \circ (j_{[n+1]})_*\circ (\pi^{\mathcal{EF}}_{\mathbb S^2}|_{S^{\I}_{u_{-1},v_\infty}})^{-1}_* (0,1,0).
	\end{align*}
	As in the proof of \eqref{eq:finallymetricdifference}, it follows that, for any $2\leq k \leq 4$,
	\begin{equation} \label{eq:finallymetricdifference2}
		\sum_{l\leq k}
		\Vert \nabla_{\psi^* h}^l \psi^* (h - \hat{h}) \Vert_{\mathbb{S}^2}
		\lesssim
		\sup_{u_{-1} \leq u \leq \hat{u}_f+\delta}
		\sum_{\vert \gamma \vert \leq k+1} \Vert \mathfrak{D}^{\gamma} f_{[n+1]} - \mathfrak{D}^{\gamma} f_{[n]} \Vert_{S_{u,v_{\infty}}}.
	\end{equation}
	
	For \eqref{eq:diffiscloseiteratesdifferenceone}, Proposition \ref{determiningthesphere}, in particular equation \eqref{diffiscloseone}, and \eqref{eq:finallymetricdifference2} imply that
	\begin{equation} \label{eq:diffeomorphismsconverging}
		\vert \slashed{F}^A_{[n+1]}(\hat{u}_f + \delta,v_{\infty},\theta) - \slashed{F}^A_{[n]}(\hat{u}_f + \delta,v_{\infty},\theta) \vert
		\lesssim
		\sup_{u_{-1} \leq u \leq \hat{u}_f+\delta}
		\sum_{\vert \gamma \vert \leq 3} \Vert \mathfrak{D}^{\gamma} f_{[n+1]} - \mathfrak{D}^{\gamma} f_{[n]} \Vert_{S_{u,v_{\infty}}},
	\end{equation}
	for $A=1,2$.  The estimate \eqref{eq:diffiscloseiteratesdifferenceone} then follows as in the proof of \eqref{eq:diffiscloseiteratesone}, using now the fact that
	\[
		|h(u,v_{\infty},\slashed{F}_{[n+1]}(u,v_{\infty},\theta)) - h(u,v_{\infty},\slashed{F}_{[n]}(u,v_{\infty},\theta))|
		\lesssim
		\big( \sup_{S_{u,v_{\infty}}} | r\nablaslash h |_{\gslash_{[n]}} \big)
		\sum_{A=1}^2\vert \slashed{F}^A_{[n+1]}(u ,v_{\infty},\theta) - \slashed{F}^A_{[n]}(u ,v_{\infty},\theta) \vert.
	\]
	Similarly for \eqref{eq:diffiscloseiteratesdifferencetwo}, using the fact that Proposition \ref{determiningthesphere}, in particular equation \eqref{diffisclosetwo}, implies that
	\begin{equation} \label{eq:diffeomorphismsconverging2}
		\vert
		\partial_{\theta^B} \slashed{F}^A_{[n+1]}(\hat{u}_f + \delta,v_{\infty},\theta)
		-
		\partial_{\theta^B} \slashed{F}^A_{[n]}(\hat{u}_f + \delta,v_{\infty},\theta)
		\vert
		\lesssim
		\sup_{u_{-1} \leq u \leq \hat{u}_f+\delta}
		\sum_{\vert \gamma \vert \leq 3} \Vert \mathfrak{D}^{\gamma} f_{[n+1]} - \mathfrak{D}^{\gamma} f_{[n]} \Vert_{S_{u,v_{\infty}}},
	\end{equation}
	for $A,B=1,2$.
	The estimate \eqref{eq:diffiscloseiteratesdifferencethree} follows similarly from \eqref{diffisclosethree}.
\end{proof}

Estimates for the iterates $f_{[n]}$ can now be given.

\begin{lemma}[Estimates for iterates] \label{lem:fnI}
	Provided $\hat{\varepsilon}_0$ is sufficiently small (independent of $\hat{u}_f$) and $\delta_0$ is sufficiently small (with respect to $\hat{u}_f$), for all $n \geq 1$ the diffeomorphisms $f_{[n]}$ satisfy, for all $u_{-1} \leq u \leq \hat{u}_f+\delta$, the estimates
	\[
		\sum_{\vert \gamma \vert \leq 5} \Vert \mathfrak{D}^{\gamma} f_{[n]} \Vert_{S_{u,v_{\infty}}}
		+
		\vert M_{[n]} - M_f \vert
		\lesssim
		\frac{\varepsilon \delta}{\hat{u}_f},
	\]
	where $v_{\infty}=v_{\infty}(\hat{u}_f+\delta)$ and the norm $\Vert \mathfrak{D}^{\gamma} f_{[n]} \Vert_{S_{u,v_{\infty}}}$ is defined in \eqref{eq:iteratesdiffeonorms}.
\end{lemma}

\begin{proof}
The diffeomorphisms $\mathfrak{D}^{\gamma} f_{[n]}$ are estimated inductively.  Suppose $n\geq1$ is such that
\[
	\sup_{u_{-1} \leq u \leq \hat{u}_f+\delta}
	\sum_{\vert \gamma \vert \leq 5}
	\Vert \mathfrak{D}^{\gamma} f_{[n-1]} \Vert_{S_{u,v_{\infty}}}
	+
	\vert M_{[n-1]} - M_f \vert
	\leq
	\frac{C_1 \varepsilon \delta}{\hat{u}_f},
\]
for some appropriate $C_1$, to be chosen, where the norm $\sum_{\vert \gamma \vert \leq 5} \Vert \mathfrak{D}^{\gamma} f_{[n-1]} \Vert_{S_{u,v_{\infty}}}$ is defined in \eqref{eq:iteratesdiffeonorms}.  Under this assumption, the estimates for $f_{[n]}$ and $M_{[n]}$ are divided into several steps.

Define
\begin{align*}
			&
	\mathcal{S}_{[n]}
	=
	\sum_{k \leq 3}
	\Big[
	r^2
	\big\Vert
	\big[
	(r\nablaslash)^k (\Omega \tr \chi - \Omega \tr \chi_{\circ,M_{[n]}})_{\ell \neq 1}^{[n]}
	-
	(r\nablaslash)^k(\Omega \tr \chi - \Omega \tr \chi_{\circ,M_f})_{\ell \neq 1}
	\big] (\hat{u}_f+\delta,v_{\infty})
	\big\Vert_{S_{\hat{u}_f+\delta,v_{\infty}}}
	\\
	&
	\qquad
	+
	r^2
	\big\Vert
	\big[
	(r\nablaslash)^k(\Omega^{-1} \tr \chibar - \Omega^{-1} \tr \chibar_{\circ,M_{[n]}})_{[n]}
	-
	(r\nablaslash)^k(\Omega^{-1} \tr \chibar - \Omega^{-1} \tr \chibar_{\circ,M_f})
	\big] (\hat{u}_f+\delta,v_{\infty})
	\big\Vert_{S_{\hat{u}_f+\delta,v_{\infty}}} \Big]
	\\
	&
	\qquad
	+
	r^5
	\Vert
	\big[
	(\divslash \Omega \beta)_{\ell = 1}^{[n]}
	-
	(\divslash \Omega \beta)_{\ell = 1}
	\big] (\hat{u}_f+\delta,v_{\infty})
	\Vert_{S_{\hat{u}_f+\delta,v_{\infty}}}
\end{align*}
to be the sum of the difference between the values of $(\Omega \tr \chi - \Omega \tr \chi_{\circ,M_{[n]}})_{\ell \neq 1}^{[n]} (\hat{u}_f+\delta,v_{\infty})$ and $(\Omega \tr \chi - \Omega \tr \chi_{\circ,M_f})_{\ell \neq 1} (\hat{u}_f+\delta,v_{\infty})$, and $(\Omega^{-1} \tr \chibar - \Omega^{-1} \tr \chibar_{\circ,M_{[n]}})_{[n]}(\hat{u}_f+\delta,v_{\infty})$ and $(\Omega^{-1} \tr \chibar - \Omega^{-1} \tr \chibar_{\circ,M_f}) (\hat{u}_f+\delta,v_{\infty})$, along with the differences of their angular derivatives up to order $3$, and the difference between the $\ell =1$ modes $(\divslash \Omega \beta)_{\ell = 1}^{[n]}(\hat{u}_f+\delta,v_{\infty})$ and $(\divslash \Omega \beta)_{\ell = 1} (\hat{u}_f+\delta,v_{\infty})$.

\noindent \underline{\textbf{Estimates for Ricci coefficients of iterates $\Phi_{[n]}$:}}
The first step is to obtain the following estimates for the Ricci coefficients of the $n$-th gauge, schematically denoted $\Phi_{[n]}$, along with non-sharp estimates for the diffeomorphisms $f_{[n]}$ and their derivatives, and the mass difference $M_{[n]} - M_f$:
\begin{equation} \label{eq:PhinfnInonsharp}
	\sum_{\vert \gamma \vert \leq 4}
	\Vert r^p \mathfrak{D}^{\gamma} \Phi^{[n]}_p \Vert_{S_{u,v_{\infty}(\hat{u}_f+\delta)}}
	+
	\sum_{\vert \gamma \vert \leq 6}
	\Vert \mathfrak{D}^{\gamma} f_{[n]} \Vert_{S_{u,v_{\infty}(\hat{u}_f+\delta)}}
	+
	\vert M_{[n]} - M_f \vert
	\lesssim
	\varepsilon,
\end{equation}
for all $n \geq 1$.  Since this is similar to, but less involved than, the main part of the proof, the estimates are only sketched.  The estimates for $f_{[n]}$ and $M_{[n]} - M_f$ are non-sharp, and much easier to obtain than the sharp estimates, as they do not exploit the fact that the differences $\Phi_{[n]}(x) - \Phi(x)$ can be estimated by $\delta$, but only that $\Phi_{[n]}(x)$ and $\Phi(x)$ can individually be estimated by $\varepsilon$.

First one estimates $\alpha_{[n]}(u,v_{\infty}(\hat{u}_f+\delta))$ and $\alphabar_{[n]}(u,v_{\infty}(\hat{u}_f+\delta))$ and their derivatives up to order $4$, for $u_{-1} \leq u \leq \hat{u}_f+\delta$.  The estimates are obtained, as in the proof of Proposition \ref{thm:gidataestimates}, in terms of nonlinearities involving $\Phi$ and the diffeomorphisms $f_{[n]}$, and linear terms involving $\Pi_{S_{[n]}} \alpha$ and $\Pi_{S_{[n]}} \alphabar$ and their derivatives up to order $4$ (see, for example, equation \eqref{eq:curvaturecomp2}).  One then uses, for example, the fact that
\[
	\sum_{k \leq 4} r \Vert \Pi_{S_{[n]}} (r\nablaslash)^k \alphabar \Vert_{S_{u,v_{\infty}}}
	\lesssim
	\varepsilon \Big(
	1
	+
	\sup_{u_{-1} \leq u \leq \hat{u}_f+\delta}
	\sum_{\vert \gamma \vert \leq 3} \Vert \mathfrak{D}^{\gamma} f_{[n]} \Vert_{S_{u,v_{\infty}}}
	\Big)
	,
\]
using the estimates on $\alphabar$ and its derivatives up to order $6$ (see the expression \eqref{eq:alphadifferencetomean}) and Lemma \ref{lem:mvt}.

Next one estimates, as in the proof of Proposition \ref{prop:dIdiff}, the diffeomorphisms $f_{[n]}(u,v_{\infty}(\hat{u}_f+\delta))$ and their derivatives, along with $M_{[n]} - M_f$, in terms of $\Phi_{[n]}$, $\Phi$ and their derivatives.  One then revisits the estimates of Chapter \ref{chap:Iestimates} (or rather the part of Chapter \ref{chap:Iestimates} involving the estimates on the hypersurface $v=v_{\infty}$) and estimates $\Phi_{[n]}(u,v_{\infty}(\hat{u}_f+\delta))$ and their derivatives up to order $4$ in terms of $\alpha_{[n]}(u,v_{\infty}(\hat{u}_f+\delta))$, $\alphabar_{[n]}(u,v_{\infty}(\hat{u}_f+\delta))$, and their derivatives, and
\[
	\sum_{k \leq 4}
	r^2
	\Vert
	(r\nablaslash)^k (\Omega \tr \chi - \Omega \tr \chi_{\circ,M_{[n]}})_{[n]}
	\Vert_{S_{\hat{u}_f+\delta,v_{\infty}}}
	+
	r^2
	\Vert
	(r\nablaslash)^k(\Omega^{-1} \tr \chibar - \Omega^{-1} \tr \chibar_{\circ,M_{[n]}})_{[n]}
	\Vert_{S_{\hat{u}_f+\delta,v_{\infty}}}.
\]
A simple induction argument, using the change of gauge relations \eqref{eq:Riccicomp1}, \eqref{eq:Riccicomp2}, \eqref{eq:Riccicomp3}, \eqref{eq:Riccicomp4}, \eqref{eq:Riccicomp7}, \eqref{eq:Riccicomp8}, \eqref{eq:curvaturecomp5} and the equations \eqref{eq:nIgauge1}--\eqref{eq:nIgauge4}, then completes the proof of the estimate \eqref{eq:PhinfnInonsharp}.

\noindent \underline{\textbf{Estimates for differences of Ricci coefficients $\Phi_{[n]} - \Phi$:}}  Now the differences between the $n$-th Ricci coefficients and curvature components and those of the extended $\hat{u}_f$ normalised $\I$ gauge, $\Phi_{[n]}(x) - \Phi(x)$, \emph{identified by the values of their respective coordinate functions}, are estimated.  

		Considering the equations for the differences $\Phi_{[n]}(x) - \Phi(x)$, revisiting the estimates of Chapter \ref{chap:Iestimates}, and using the estimates \eqref{eq:PhinfnInonsharp} for $\Phi_{[n]}$, it follows that, if $v_{\infty}$ is sufficiently large, for any $u_{-1} \leq u \leq \hat{u}_f+\delta$,
		\begin{align}
			&
			\sum_{\vert \gamma \vert \leq 3} r^p \Vert \mathfrak{D}^{\gamma} \Phi_p^{[n]}(x) - \mathfrak{D}^{\gamma} \Phi_p(x) \Vert_{S_{u,v_{\infty}}}
			\!\!\!\!
			+
			\Vert
			(r^p \Phi_p^{[n]})_{\ell=0}(x) - (r^p \Phi_p)_{\ell=0}(x)
			\Vert_{L^1(\Cbar_{v_{\infty}})}
			\nonumber
			\lesssim
			\mathcal{S}_{[n]}
			+
			\frac{\varepsilon \delta}{\hat{u}_f}
			+
			\varepsilon \frac{\hat{u}_f}{v_{\infty}} \vert M_{[n]} - M_f \vert
			\\
			&
			+
			\sum_{\vert \gamma \vert \leq 2}
			\Vert r^5 \mathfrak{D}^{\gamma} \nablaslash_3 \alpha_{[n]}(x) - r^5 \mathfrak{D}^{\gamma} \nablaslash_3 \alpha(x) \Vert_{S, \Cbar_{v_{\infty}}}
			\label{eq:Ifestimate1}
			+
			\sum_{\substack{
			\vert \gamma \vert + k \leq 3
			\\
			k \leq 2
			}}
			\Vert  \mathfrak{D}^{\gamma} (r^2 \nablaslash_4)^k r \alphabar_{[n]}(x)
			-
			\mathfrak{D}^{\gamma} (r^2 \nablaslash_4)^k r \alphabar (x) \Vert_{S, \Cbar_{v_{\infty}}},
		\end{align}
		where
		\[
			\Vert F \Vert_{L^1(\Cbar_{v_{\infty}})}
			=
			\int_{u_{-1}}^{\hat{u}_f} \int_{S_{u,v_{\infty}(\hat{u}_f+\delta)}} \vert F \vert d \theta du,
			\qquad
			\Vert F \Vert_{S, \Cbar_{v_{\infty}}}
			=
			\sup_{u_{-1} \leq u \leq \hat{u}_f+\delta}
			\Vert F \Vert_{S_{u,v_{\infty}(\hat{u}_f+\delta)}}
			+
			\Vert F \Vert_{\Cbar_{v_{\infty}(\hat{u}_f+\delta)}}.
		\]
		Clearly also, by the Gauss equation \eqref{eq:Gauss} and the estimates for $K$,
		\begin{equation} \label{eq:GaussminusGaussn}
			\big\Vert  (r^{2} K)(\hat{u}_f,v_{\infty}(\hat{u}_f),\cdot) - (r^{2} K)_{[n]}(\hat{u}_f + \delta,v_{\infty},\cdot) \big\Vert_{\mathbb{S}^2}
			\lesssim
			\sum_{\vert \gamma \vert \leq 3} r^p \Vert \mathfrak{D}^{\gamma} \Phi_p^{[n]}(x) - \mathfrak{D}^{\gamma} \Phi_p(x) \Vert_{S_{\hat{u}_f + \delta,v_{\infty}}}
			+
			\frac{\varepsilon \delta}{\hat{u}_f}.
		\end{equation}

		The difference between the metrics, $\gslash(x) - \gslash_{[n]}(x)$, are similarly be estimated.  The estimate \eqref{theestimatesforourspherediffdifferences} implies that
		\begin{multline*}
			\vert r^{-2} \gslash (\hat{u}_f,v_{\infty}(\hat{u}_f),\theta)
			-
			(r^{-2} \gslash)_{[n]} (\hat{u}_f + \delta,v_{\infty},\theta)
			\vert_{\gamma}
			\lesssim
			\Vert r^2 K (\hat{u}_f,v_{\infty}(\hat{u}_f),\cdot)
			-
			(r^{2} K)_{[n]}(\hat{u}_f + \delta,v_{\infty},\cdot)
			\Vert_{\mathbb{S}^2}
			\\
			\lesssim
			\Vert r^2 K (\hat{u}_f + \delta,v_{\infty},\cdot)
			-
			(r^{2}K)_{[n]}(\hat{u}_f + \delta,v_{\infty},\cdot)
			\Vert_{\mathbb{S}^2}
			+
			\frac{\varepsilon \delta}{\hat{u}_f},
		\end{multline*}
		from which it follows that
		\begin{multline} \label{eq:gslashgslashnufdeltaestimate}
			\vert \gslash (\hat{u}_f + \delta,v_{\infty},\theta)
			-
			\gslash_{[n]}(\hat{u}_f + \delta,v_{\infty},\theta)
			\vert_{\gslash}
			\lesssim
			\frac{\varepsilon \delta}{r \hat{u}_f}
			+
			\frac{\vert f^3_{[n]}(\hat{u}_f + \delta,v_{\infty},\theta) \vert + \vert f^4_{[n]}(\hat{u}_f + \delta,v_{\infty},\theta) \vert}{r}
			\\
			+
			\vert M_{[n]} - M_f \vert
			+
			r
			\Vert K (\hat{u}_f + \delta,v_{\infty},\cdot)
			-
			K_{[n]}(\hat{u}_f + \delta,v_{\infty},\cdot)
			\Vert_{S_{\hat{u}_f + \delta,v_{\infty}}}
			,
		\end{multline}
		with $v_{\infty} = v_{\infty}(\hat{u}_f+\delta)$.  The difference $\gslash(x) - \gslash_{[n]}(x)$ is then estimated as in Chapter \ref{chap:Iestimates}, using this estimate in place of where the estimate \eqref{theestimatesforourspherediff} is used (see Section \ref{subsec:finalspheremetricestimate}).

\noindent \underline{\textbf{Estimates for differences $\alpha_{[n]} - \alpha$ and $\alphabar_{[n]} - \alphabar$:}} 
Next, the differences $\alpha_{[n]}(x) - \alpha(x)$ and $\alphabar_{[n]}(x) - \alphabar(x)$ are estimated.  The relation \eqref{eq:curvaturecomp2} implies that, schematically,
		\[
			r (\alphabar_{[n]} - \Pi_{S_{[n]}} \alphabar)
			=
			\Pi_{S_{[n]}} r^p \Phi_p \cdot \mathfrak{D} f_{[n]}
			+
			(\mathfrak{D} f_{[n]})^2,
		\]
		and so by the Sobolev inequality,
		\[
			\Vert r \alphabar_{[n]} - \Pi_{S_{[n]}} r \alphabar \Vert_{S, \Cbar_{v_{\infty}}}
			\lesssim
			\varepsilon \sup_{u_{-1} \leq u \leq \hat{u}_f+\delta}
			\sum_{\vert \gamma \vert \leq 1} \Vert \mathfrak{D}^{\gamma} f_{[n]} \Vert_{S_{u,v_{\infty}}}
			+
			\big( \sum_{\vert \gamma \vert \leq 1} \Vert \mathfrak{D}^{\gamma} f_{[n]} \Vert_{S, \Cbar_{v_{\infty}}} \big)^2.
		\]
		where the norm $\sum_{\vert \gamma \vert \leq 1} \Vert \mathfrak{D}^{\gamma} f_{[n]} \Vert_{S_{u,v_{\infty}}}$ is defined in \eqref{eq:iteratesdiffeonorms}, and the norm $\sum_{\vert \gamma \vert \leq 1} \Vert \mathfrak{D}^{\gamma} f_{[n]} \Vert_{S, \Cbar_{v_{\infty}}}$ is similarly defined, in the obvious way.
		Moreover (recall \eqref{eq:alphadifferencetomean}),
		\[
			\Vert \Pi_{S_{[n]}} r \alphabar(x) - r \alphabar(x) \Vert_{S, \Cbar_{v_{\infty}}}
			\lesssim
			\varepsilon
			\Big(
			\frac{\delta}{r}
			+
			\sup_{u_{-1} \leq u \leq \hat{u}_f+\delta}
		\sum_{\vert \gamma \vert \leq 3} \Vert \mathfrak{D}^{\gamma} f_{[n]} \Vert_{S_{u,v_{\infty}}}
			\Big),
		\]
		where Lemma \ref{lem:mvt} has been used.  Similarly for higher order derivatives and for $\alpha$ (note that the estimates for derivatives with better $r$ weights, e.\@g.\@ for $r^2 \nablaslash_4 r \alphabar_{[n]} - \Pi_{S_{[n]}} r^2 \nablaslash_4 r \alphabar$, follow as in the proof of Proposition \ref{thm:gidataestimates}).  It therefore follows, using \eqref{eq:GaussminusGaussn}, that
		\begin{multline} \label{eq:Ifestimate2}
			\sum_{\vert \gamma \vert \leq 2}
			\Vert r^5 \mathfrak{D}^{\gamma} \nablaslash_3 \alpha_{[n]}(x) - r^5 \mathfrak{D}^{\gamma} \nablaslash_3 \alpha(x) \Vert_{S, \Cbar_{v_{\infty}}}
			+
			\sum_{\substack{
			\vert \gamma \vert + k \leq 3
			\\
			k \leq 2
			}}
			\Vert  \mathfrak{D}^{\gamma} (r^2 \nablaslash_4)^k (r \alphabar_{[n]})(x)
			-
			\mathfrak{D}^{\gamma} (r^2 \nablaslash_4)^k (r \alphabar)(x) \Vert_{S, \Cbar_{v_{\infty}}}
			\\
			\lesssim
			\frac{\varepsilon^2 \delta}{\hat{u}_f}
			+
			\varepsilon
			\sum_{\vert \gamma \vert \leq 3} r^p \Vert \mathfrak{D}^{\gamma} \Phi_p^{[n]}(x) - \mathfrak{D}^{\gamma} \Phi_p(x) \Vert_{S_{u,v_{\infty}}}
			+
			\varepsilon \sup_{u_{-1} \leq u \leq \hat{u}_f+\delta}
			\sum_{\vert \gamma \vert \leq 5}
			\Vert \mathfrak{D}^{\gamma} f_{[n]} \Vert_{S_{u,v_{\infty}}}
			+
			\big(
			\sum_{\vert \gamma \vert \leq 4}
			\Vert \mathfrak{D}^{\gamma} f_{[n]} \Vert_{S, \Cbar_{v_{\infty}}}
			\big)^2.
		\end{multline}

\noindent \underline{\textbf{Estimates for diffeomorphisms of iterates $f_{[n]}$:}}  Consider now the entire system \eqref{eq:metriccomp1}--\eqref{eq:curvaturecomp6}, each equation of which takes the schematic form
		\begin{equation} \label{eq:ellipticfIschematic}
			\mathfrak{D}^2 f_{[n]}
			=
			\Phi_{[n]} - \Pi_{S_{[n]}} \Phi
			+
			\sum_{1\leq \vert \gamma_1 \vert, \vert \gamma_2 \vert \leq 2}
			\big(
			\Pi_{S_{[n]}} \Phi \cdot \mathfrak{D}^{\gamma_1} f_{[n]}
			+
			\mathfrak{D}^{\gamma_1} f_{[n]} \cdot \mathfrak{D}^{\gamma_2} f_{[n]}
			\big).
		\end{equation}
		Exploiting the reductive structure of the left hand sides, as in the proof of Proposition \ref{prop:dIdiff}, and using the fact that, for each $\Phi$,
		\[
			r^p \Vert \mathfrak{D}^{\gamma} \Phi_p(x) - \Pi_{S_{[n]}} \mathfrak{D}^{\gamma} \Phi_p(x) \Vert_{S_{u,v_{\infty}}}
			\lesssim
			\sum_{\vert \tilde{\gamma} \vert \leq \vert \gamma \vert +1}
			\sup_{S_{u,v_{\infty}}}
			\vert \mathfrak{D}^{\tilde{\gamma}} \Phi_p \vert
			\Big(
			\frac{\varepsilon \delta}{r}
			+
			\sup_{u_{-1} \leq u \leq \hat{u}_f+\delta}
		\sum_{\vert \gamma \vert \leq 3} \Vert \mathfrak{D}^{\gamma} f_{[n]} \Vert_{S_{u,v_{\infty}}}
			\Big),
		\]
		by Lemma \ref{lem:mvt}, it follows that the $\ell \geq 1$ modes of $\mathfrak{D}^{\gamma} f_{[n]}$ satisfy, for $\vert \gamma \vert \leq 5$,
		\begin{equation} \label{eq:Ifestimate3}
			\Vert \mathfrak{D}^{\gamma} f^{[n]}_{\ell \geq 1} \Vert_{S_{u,v_{\infty}}}
			\lesssim
			\sum_{\vert \widetilde{\gamma} \vert \leq \vert \gamma \vert -2}
			r^p
			\Vert \mathfrak{D}^{\widetilde{\gamma}} \Phi^{[n]}_p(x) - \mathfrak{D}^{\widetilde{\gamma}} \Phi_p(x) \Vert_{S_{u,v_{\infty}}}
			+
			\varepsilon \sum_{\vert \widetilde{\gamma} \vert \leq \vert \gamma \vert} \Vert \mathfrak{D}^{\widetilde{\gamma}} f_{[n]} \Vert_{S_{u,v_{\infty}}}
			+
			\frac{\varepsilon^2 \delta}{\hat{u}_f}
			,
		\end{equation}
		where \eqref{eq:GaussminusGaussn} has been used.  For the $\ell =0$ modes, a similar argument to that of Proposition \ref{prop:dIdiff} gives
		\begin{multline} \label{eq:Ifestimate3a}
			\vert (f^3_{[n]} - f^4_{[n]})_{\ell=0} \vert
			+
			\sum_{1\leq \vert \gamma \vert \leq 2} \vert (\mathfrak{D}^{\gamma} f_{[n]})_{\ell =0} \vert
			\lesssim
			r^p \vert (\Phi_p^{[n]}(x) - \Phi_p(x))_{\ell =0} \vert
			\\
			+
			r^p \vert \Phi_p \vert \sum_{\vert \gamma \vert \leq 2} \vert \mathfrak{D}^{\gamma} f_{[n]} \vert
			+
			( \sum_{\vert \gamma \vert \leq 2} \vert \mathfrak{D}^{\gamma} f_{[n]} \vert )^2
			+
			\frac{\varepsilon^2 \delta}{\hat{u}_f},
		\end{multline}
		and, using the analogue of \eqref{eq:rhodiffmass}, \eqref{eq:rdrIf}, \eqref{eq:rdrIf2} for $r$ and $r_{[n]}$, along with the fact that $M_{[n]} + \frac{1}{2} (r^3\rho_{\ell = 0})(\hat{u}_f+\delta,v_{\infty}(\hat{u}_f+\delta))_{[n]} = 0$ and $\vert M_f  + \frac{1}{2} (r^3 \rho_{\ell = 0})(\hat{u}_f+\delta,v_{\infty}(\hat{u}_f+\delta)) \vert \lesssim \varepsilon \delta (\hat{u}_f)^{-1}$, 
		\begin{equation} \label{eq:Ifestimate3b}
			\vert M_{[n]} - M_f \vert
			\lesssim
			\frac{\varepsilon \delta}{\hat{u}_f}
			+
			\frac{1}{r}
			r^p \vert (\Phi_p^{[n]}(x) - \Phi_p(x))_{\ell =0} \vert
			+
			r^p \vert \Phi_p \vert \sum_{\vert \gamma \vert \leq 2} \vert \mathfrak{D}^{\gamma} f_{[n]} \vert
			+
			( \sum_{\vert \gamma \vert \leq 2} \vert \mathfrak{D}^{\gamma} f_{[n]} \vert )^2,
		\end{equation}
		and moreover, equation \eqref{eq:nIgauge4} implies
		\begin{multline*}
			\vert (f^3_{\ell=0})_{[n]}(u,v_{\infty}) \vert
			\leq
			\int_{u_{-1}}^{u}
			\vert \partial_u \big( f^3_{\ell=0} \big)_{[n]} (u',v_{\infty}) \vert
			du'
			\\
			\lesssim
			\int_{u_{-1}}^{u}
			r^p \vert (\Phi_p^{[n]}(x) - \Phi_p(x))_{\ell =0} \vert
			+
			r^p \vert \Phi_p \vert \sum_{\vert \gamma \vert \leq 2} \vert \mathfrak{D}^{\gamma} f_{[n]} \vert
			+
			( \sum_{\vert \gamma \vert \leq 2} \vert \mathfrak{D}^{\gamma} f_{[n]} \vert )^2
			du'
			+
			\frac{\varepsilon^2 \delta}{\hat{u}_f}
			.
		\end{multline*}

\noindent \underline{\textbf{Estimate for $\mathcal{S}_{[n]}$:}} Consider now $\mathcal{S}_{[n]}$.  The change of gauge relations \eqref{eq:Riccicomp1}, \eqref{eq:Riccicomp2}, \eqref{eq:Riccicomp4}, \eqref{eq:Riccicomp7}, \eqref{eq:Riccicomp8}, \eqref{eq:curvaturecomp5}, the expressions \eqref{eq:Igaugef3h3H}, and equation \eqref{eq:nIgauge1} imply that
		\begin{align*}
			&
			\Big[
			\Omega_{\circ,M_f}^{-2}
			\Deltaslash \Omega \tr \chi_{[n]} (x_{\delta})
			+
			r^{-2}
			(\Omega \tr \chi - \Omega \tr \chi_{\circ})_{[n]}(x_{\delta})
			\Big]_{(\ell \geq 2)_{[n-1]}}
			=
			\Big[
			\Omega_{\circ,M_f}^{-2}
			\Deltaslash \Omega \tr \chi (x_{f_{[n]}})
			+
			r^{-2}
			(\Omega \tr \chi - \Omega \tr \chi_{\circ}) (x_{f_{[n]}})
			\\
			&
			\quad
			+
			2 \Deltaslash \Deltaslash \big( h^3_{[n]}(\theta_{[n]}) \big)
			+
			\frac{2}{r^2} \Big( 3 - \frac{8M_f}{r} \Big) \Deltaslash \big( h^3_{[n]}(\theta_{[n]}) \big)
			-
			\frac{2\Omega_{\circ,M_f}^2}{r^2} \Deltaslash \big( h^4_{[n]}(\theta_{[n]}) \big)
			\\
			&
			\quad
			+
			\frac{4\Omega_{\circ,M_f}^2}{r^4} (h^3_{[n]}(\theta_{[n]}) - h^4_{[n]}(\theta_{[n]}))
			+
			\frac{2}{r} \big( \mubar^{\dagger}_{[n]}(x) - \mubar^{\dagger} (x_{f_{[n]}}) \big)
			-
			\mathfrak{E}^1_{[n]}(\theta)
			-
			A^1_{[n]}(\theta)
			\Big]_{(\ell \geq 2)_{[n-1]}}
			\\
			&
			=
			\Big[
			\Omega_{\circ,M_f}^{-2}
			\Deltaslash \Omega \tr \chi (x_{f_{[n]}})
			+
			r^{-2}
			(\Omega \tr \chi - \Omega \tr \chi_{\circ}) (x_{f_{[n]}})
			-
			\Omega_{\circ,M_f}^{-2}
			\Deltaslash \Omega \tr \chi (x_{[n-1]})
			-
			r^{-2}
			(\Omega \tr \chi - \Omega \tr \chi_{\circ}) (x_{[n-1]})
			\\
			&
			\quad
			+
			\frac{2}{r} \big( \mubar^{\dagger}_{[n]}(x) - \mubar^{\dagger} (x_{f_{[n]}}) \big)
			-
			\frac{2}{r} \big( \mubar^{\dagger}_{[n-1]}(x) - \mubar^{\dagger} (x_{[n-1]}) \big)
			-
			\mathfrak{E}^1_{[n]}(\theta)
			+
			\mathfrak{E}^1_{[n-1]}(\theta_{[n]})
			\\
			&
			\quad
			-
			A^1_{[n]}(\theta)
			+
			A^1_{[n-1]}(\theta_{[n]})
			+
			\mathfrak{H}_{[n]}(\theta)
			\Big]_{(\ell \geq 2)_{[n-1]}},
		\end{align*}
		where $x_{\delta} = x_{\delta}(\theta) = (\hat{u}_f+\delta,v_{\infty}(\hat{u}_f+\delta),\theta)$,
		\[
			\theta_{[n]} = \slashed{H}_{[n-1]}^{-1} \circ \slashed{H}_{[n]} (\theta),
		\]
		$x_{f_{[n]}} = x_{\delta} + f_{[n]}(x_{\delta})$ is as defined in \eqref{eq:xdeltaplusfn}, and
		\[
			x_{[n-1]}
			= 
			(\hat{u}_f+\delta + f^3_{[n-1]}(\hat{u}_f+\delta,v_{\infty}, \slashed{H}_{[n-1]}^{-1} \circ \slashed{H}_{[n]} (\theta))
			,
			v_{\infty} + f^4_{[n-1]}(\hat{u}_f+\delta,v_{\infty},\slashed{H}_{[n-1]}^{-1} \circ \slashed{H}_{[n]} (\theta))
			,
			\slashed{H}_{[n]} (\theta)),
		\]
		and $\mathfrak{H}_{[n]}$ is given by
		\begin{align*}
			\mathfrak{H}_{[n]}(\theta)
			=
			\
			&
			2 \Deltaslash \Deltaslash \big( h^3_{[n]}(\theta_{[n]}) \big)
			-
			2 \big( \Deltaslash \Deltaslash h^3_{[n]} \big) (\theta_{[n]})
			+
			\frac{2}{r^2} \Big( 3 - \frac{8M_f}{r} \Big) 
			\Big(
			\Deltaslash \big( h^3_{[n]}(\theta_{[n]}) \big)
			-
			\big( \Deltaslash h^3_{[n]} \big) (\theta_{[n]})
			\Big)
			\\
			&
			-
			\frac{2\Omega_{\circ,M_f}^2}{r^2}
			\Big(
			\Deltaslash \big( h^4_{[n]}(\theta_{[n]}) \big)
			-
			\big( \Deltaslash h^4_{[n]} \big) (\theta_{[n]})
			\Big).
		\end{align*}
		Note that
		\[
			\sum_{k=0}^1
			\Vert (r\nablaslash)^k r^4 \mathfrak{H}_{[n]} \Vert_{S_{\hat{u}_f+\delta,v_{\infty}}}
			\lesssim
			\varepsilon
			\sup_{u_{-1} \leq u \leq \hat{u}_f+\delta}
			\sum_{\vert \gamma \vert \leq 5}
			\Vert \mathfrak{D}^{\gamma} f_{[n]} - \mathfrak{D}^{\gamma} f_{[n-1]} \Vert_{S_{u,v_{\infty}}},
		\]
		by Lemma \ref{lem:mvt} (in particular the estimates \eqref{eq:diffiscloseiteratesdifferencethree}, \eqref{eq:diffeomorphismsconverging} and \eqref{eq:diffeomorphismsconverging2}).
		An elliptic estimate, together with the fact that
		\[
			r^3
			\Vert
			\mubar^{\dagger}_{[n]}(x)
			-
			\mubar^{\dagger}(x)
			\Vert_{S_{\hat{u}_f+\delta,v_{\infty}}}
			\lesssim
			\hat{u}_f v_{\infty}^{-1}
			\sup_{u_{-1} \leq u \leq \hat{u}_f+\delta} r^p \Vert \Phi^{[n]}_p(x) - \Phi_p(x) \Vert_{S_{u,v_{\infty}}},
		\]
		then gives, for $k \leq 3$,
		\begin{multline*}
			r^2
			\Vert
			(r\nablaslash)^k (\Omega \tr \chi - \Omega \tr \chi_{\circ,M_f})^{[n]}_{\ell\geq 2}(\hat{u}_f+\delta,v_{\infty})
			\Vert_{S_{\hat{u}_f+\delta,v_{\infty}}}
			\lesssim
			\varepsilon \delta \hat{u}_f^{-1}
			+
			\vert M_{[n-1]} - M_f \vert^2
			\\
			+
			\sum_{l=n-1}^n
			\Big[
			\sum_{\vert \gamma \vert \leq 5}
			\varepsilon
			\Vert \mathfrak{D}^{\gamma} f_{[l]} \Vert_{S_{\hat{u}_f+\delta,v_{\infty}}}
			+
			\varepsilon
			\sup_{u_{-1} \leq u \leq \hat{u}_f+\delta} r^p \Vert \Phi^{[l]}_p(x) - \Phi_p(x) \Vert_{S_{u,v_{\infty}}}
			\Big].
		\end{multline*}
		Note that this estimate exploits the fact that each term of the form $\Pi_{\newsph{S}} \Phi \cdot \mathfrak{D}^{\gamma_1} f$ appearing in $\mathfrak{E}^1$, see \eqref{eq:IgaugecalEerrorsdef}, has the property that
		\[
			r^4 \sum_{\vert \gamma \vert \leq 3}
			\big\Vert \mathfrak{D}^{\gamma} \big( \Pi_{\newsph{S}} \Phi \cdot \mathfrak{D}^{\gamma_1} f \big) \big\Vert_{S_{\hat{u}_f+\delta,v_{\infty}}}
			\lesssim
			\sum_{\vert \gamma \vert \leq 5} r^p \Vert \mathfrak{D}^{\gamma} \Phi_{p} \Vert_{S_{\hat{u}_f+\delta,v_{\infty}}}
			\sum_{\vert \gamma \vert \leq 5} \Vert \mathfrak{D}^{\gamma} f \Vert_{S_{\hat{u}_f+\delta,v_{\infty}}},
		\]
		where the norm of $\mathfrak{D}^{\gamma} f$ is defined in \eqref{eq:iteratesdiffeonorms}.
		Equation \eqref{eq:nIgauge2} gives a similar estimate for $(\Omega^{-1} \tr \chibar - \Omega^{-1} \tr \chibar_{\circ,M_f})^{[n]}_{\ell\geq 1}(\hat{u}_f+\delta,v_{\infty})$, and equation \eqref{eq:nIgauge3} similarly gives  
		\begin{multline*}
			r^2
			\vert
			(\Omega^{-1} \tr \chibar - \Omega^{-1} \tr \chibar_{\circ,M_f})^{[n]}_{\ell =0}(\hat{u}_f+\delta,v_{\infty})
			\vert
			\lesssim
			\varepsilon \delta \hat{u}_f^{-1}
			+
			\vert M_{[n-1]} - M_f \vert
			\\
			+
			\sum_{l=n-1}^n
			\Big[
			\sum_{\vert \gamma \vert \leq 5}
			\varepsilon
			\Vert \mathfrak{D}^{\gamma} f_{[l]} \Vert_{S_{\hat{u}_f+\delta,v_{\infty}}}
			+
			\varepsilon
			\sup_{u_{-1} \leq u \leq \hat{u}_f+\delta} r^p \Vert \Phi^{[l]}_p(x) - \Phi_p(x) \Vert_{S_{u,v_{\infty}}}
			\Big],
		\end{multline*}
		and equation \eqref{eq:nIgauge1a} gives
		\begin{multline*}
			r^5
			\Vert
			(\divslash \Omega \beta)^{[n]}_{\ell=1}(\hat{u}_f+\delta,v_{\infty},\cdot)
			\Vert_{S_{\hat{u}_f+\delta,v_{\infty}}}
			\lesssim
			\varepsilon \delta \hat{u}_f^{-1}
			+
			\vert M_{[n-1]} - M_f \vert^2
			\\
			+
			\sum_{l=n-1}^n
			\Big[
			\sum_{\vert \gamma \vert \leq 5}
			\varepsilon
			\Vert \mathfrak{D}^{\gamma} f_{[l]} \Vert_{S_{\hat{u}_f+\delta,v_{\infty}}}
			+
			\varepsilon
			\sup_{u_{-1} \leq u \leq \hat{u}_f+\delta} r^p \Vert \Phi^{[l]}_p(x) - \Phi_p(x) \Vert_{S_{u,v_{\infty}}}
			\Big],
		\end{multline*}
		exploiting similar structure in the terms of the form $\Pi_{\newsph{S}} \Phi \cdot \mathfrak{D}^{\gamma_1} f$ appearing in $\mathfrak{E}^2$, $\mathfrak{E}^3$ and $\mathfrak{E}^4$ respectively.  The estimates for $(\Omega^{-1} \tr \chibar - \Omega^{-1} \tr \chibar_{\circ,M_f})^{[n]}$ rely on the fact that the change of gauge relation (cf.\@ Proposition \ref{prop:Riccirelations}) takes the form
		\begin{multline*}
			\left( \widetilde{\Omega^{-1} \tr \chibar} - \widetilde{(\Omega^{-1} \tr \chibar)}_{\circ,\widetilde{M}} \right)
			-
			\left( \Omega^{-1} \tr \chibar - (\Omega^{-1} \tr \chibar)_{\circ,M} \right)
			\\
			=
			(\Omega^{-1} \tr \chibar)_{\circ,M}
			-
			\widetilde{(\Omega^{-1} \tr \chibar)}_{\circ,\widetilde{M}}
			+
			2 \widetilde{\Deltaslash} f^4
			-
			\Omega^{-1} \tr \chibar \partial_{\widetilde{v}} f^4
			-
			2 \widetilde{\nablaslash} \partial_{\widetilde{u}} f^3 \cdot \widetilde{\nablaslash} f^4
			+
			\Omega^{-1} \tr \chibar \partial_{\widetilde{u}} f^3 \partial_{\widetilde{v}} f^4
			+
			\mathcal{E}^{2,0}_{\mathfrak{D} \fsc,2},
		\end{multline*}
		so that all of the nonlinear terms which have the worst $r$ behaviour are nonlinear in the diffeomorphism functions.
		Since $(\Omega \tr \chi - \Omega \tr \chi_{\circ,M_f})^{[n]}_{\ell =0}(\hat{u}_f+\delta,v_{\infty}) = 0$ and
		\begin{multline*}
			\sum_{k \leq 3}
			\Big[
			r^2
			\Vert
			(r\nablaslash)^k (\Omega \tr \chi - \Omega \tr \chi_{\circ,M_f})_{\ell \neq 1}
			\Vert_{S_{\hat{u}_f+\delta,v_{\infty}}}
			+
			r^2
			\Vert
			(r\nablaslash)^k (\Omega^{-1} \tr \chibar - \Omega^{-1} \tr \chibar_{\circ,M_f})
			\Vert_{S_{\hat{u}_f+\delta,v_{\infty}}}
			\Big]
			\\
			+
			r^5
			\Vert
			(\divslash \Omega \beta)_{\ell =1}
			\Vert_{S_{\hat{u}_f+\delta,v_{\infty}}}
			\lesssim
			\frac{\varepsilon \delta}{\hat{u}_f},
		\end{multline*}
		it then follows that
		\begin{equation} \label{eq:Ifestimate5}
			\mathcal{S}_{[n]}
			\lesssim
			\varepsilon \delta \hat{u}_f^{-1}
			+
			\vert M_{[n-1]} - M_f \vert
			+
			\sum_{l=n-1}^n
			\Big[
			\sum_{\vert \gamma \vert \leq 5}
			\varepsilon
			\Vert \mathfrak{D}^{\gamma} f_{[l]} \Vert_S
			+
			\varepsilon
			\sup_{u_{-1} \leq u \leq \hat{u}_f+\delta} r^p \Vert \Phi^{[l]}_p(x) - \Phi_p(x) \Vert_{S_{u,v_{\infty}}}
			\Big].
		\end{equation}

\noindent \underline{\textbf{The completion of the proof:}}  Provided $C_1$ is chosen to be sufficiently large, the estimates \eqref{eq:Ifestimate1}, \eqref{eq:Ifestimate2}, \eqref{eq:Ifestimate3}, \eqref{eq:Ifestimate3a}, \eqref{eq:Ifestimate3b}, \eqref{eq:Ifestimate5} now combine to give
		\[
			\sup_{u_{-1} \leq u \leq \hat{u}_f+\delta}
			\sum_{\vert \gamma \vert \leq 5} \Vert \mathfrak{D}^{\gamma} f_{[n]} \Vert_{S_{u,v_{\infty}}}
			+
			\vert M_{[n]} - M_f \vert
			\leq
			\frac{C_1 \varepsilon \delta}{\hat{u}_f},
		\]
		provided $\delta_0$ and $\hat{\varepsilon}_0$ are sufficiently small, thus completing the induction.

\end{proof}

The next part of the proof involves showing that the map which takes the $n$-th sphere to $n+1$-th sphere is a contraction.

\begin{lemma}[Estimates for differences of iterates] \label{lem:fIexistencediff}
	Provided $\hat{\varepsilon}_0$ is sufficiently small (independent of $\hat{u}_f$) and $\delta_0$ is sufficiently small (with respect to $\hat{u}_f$), the diffeomorphisms $f_{[n]}$ and masses $M_{[n]}$ satisfy, for all $n \geq 1$ and all $u_{-1} \leq u \leq \hat{u}_f+\delta$, the estimates
	\[
		\vert M_{[n+1]} - M_{[n]} \vert
		+
		\sum_{\vert \gamma \vert \leq 5}
		\Vert \mathfrak{D}^{\gamma} f_{[n+1]} - \mathfrak{D}^{\gamma} f_{[n]} \Vert_{S_{u,v_{\infty}}}
		\leq
		2^{-n}
		,
	\]
	where $v_{\infty} = v_{\infty}(\hat{u}_f+\delta)$ and the norm $\sum_{\vert \gamma \vert \leq 5} \Vert \mathfrak{D}^{\gamma} f_{[n+1]} - \mathfrak{D}^{\gamma} f_{[n]} \Vert_{S_{u,v_{\infty}}}$ is as in \eqref{eq:iteratesdiffeonorms}.  In particular, $h_{[n]}^3$, $h_{[n]}^4$ satisfy
	\[
		\sum_{k\leq 5}
		\big(
		\Vert (r\nablaslash)^k h_{[n+1]}^3 - (r\nablaslash)^k h_{[n]}^3 \Vert_{S_{u,v_{\infty}}}
		+
		\Vert (r\nablaslash)^k h_{[n+1]}^4 - (r\nablaslash)^k h_{[n]}^4 \Vert_{S_{u,v_{\infty}}}
		\big)
		\leq
		2^{-n}
		.
	\]
\end{lemma}

\begin{proof}
	The proof is similar to the proof of Lemma \ref{lem:fnI}.  The proof is divided into several steps, which follow closely the corresponding steps of Lemma \ref{lem:fnI}.  Define now, for $n \geq 1$,
	\begin{align*}
		&
		\mathfrak{s}_{[n]}
		=
		\sum_{k \leq 3}
		\Big[
		r^2
		\Vert
		\big[
		(r\nablaslash)^k (\Omega \tr \chi - \Omega \tr \chi_{\circ,M_{[n]}})_{[n]}
		-
		(r\nablaslash)^k(\Omega \tr \chi - \Omega \tr \chi_{\circ,M_{[n-1]}})_{[n-1]}
		\big] (\hat{u}_f+\delta,v_{\infty})
		\Vert_{S_{\hat{u}_f+\delta,v_{\infty}}}
		\\
		&
		\quad
		+
		r^2
		\Vert
		\big[
		(r\nablaslash)^k(\Omega^{-1} \tr \chibar - \Omega^{-1} \tr \chibar_{\circ,M_{[n]}})_{[n]}
		-
		(r\nablaslash)^k(\Omega^{-1} \tr \chibar - \Omega^{-1} \tr \chibar_{\circ,M_{[n-1]}})_{[n-1]}
		\big] (\hat{u}_f+\delta,v_{\infty})
		\Vert_{S_{\hat{u}_f+\delta,v_{\infty}}}
		\Big]
		\\
		&
		\quad
		+
		r^5
		\Vert
		\big[
		(\divslash \Omega \beta)_{[n]}
		-
		(\divslash \Omega \beta)_{[n-1]}
		\big] (\hat{u}_f+\delta,v_{\infty})
		\Vert_{S_{\hat{u}_f+\delta,v_{\infty}}}.
	\end{align*}

	\noindent \underline{\textbf{Estimates for differences of Ricci coefficients $\Phi_{[n+1]} - \Phi_{[n]}$:}}  Considering the equations satisfied by the differences $\Phi_{[n+1]}(x) - \Phi_{[n]}(x)$ and revisiting the estimates of Chapter \ref{chap:Iestimates}, it follows from Lemma \ref{lem:fnI} that
			\begin{align} \label{eq:Ifdiffestimate1}
				&
				\sum_{\vert \gamma \vert \leq 3} r^p \Vert \mathfrak{D}^{\gamma} \Phi_p^{[n+1]}(x) - \mathfrak{D}^{\gamma} \Phi_p^{[n]}(x) \Vert_{S_{u,v_{\infty}}}
				+
				\Vert
				(r^p \Phi_p^{[n+1]})_{\ell=0}(x) - (r^p \Phi_p^{[n]})_{\ell=0}(x)
				\Vert_{L^1(\Cbar_{v_{\infty}})}
				\\
				&
				\qquad \qquad
				\lesssim
				\mathfrak{s}_{[n+1]}
				\nonumber
				+
				\varepsilon \frac{\hat{u}_f}{v_{\infty}} \vert M_{[n+1]} - M_{[n]} \vert
				+
				\sum_{\vert \gamma \vert \leq 2}
				\Vert r^5 \mathfrak{D}^{\gamma} \nablaslash_3 \alpha_{[n+1]}(x) - r^5 \mathfrak{D}^{\gamma} \nablaslash_3 \alpha_{[n]}(x) \Vert_{S, \Cbar_{v_{\infty}}}
				\\
				&
				\qquad \qquad \quad
				+
				\sum_{\substack{
				\vert \gamma \vert + k \leq 3
				\\
				k \leq 2
				}}
				\Vert  \mathfrak{D}^{\gamma} (r^2 \nablaslash_4)^k r \alphabar_{[n+1]}(x)
				-
				\mathfrak{D}^{\gamma} (r^2 \nablaslash_4)^k r \alphabar_{[n]} (x) \Vert_{S, \Cbar_{v_{\infty}}},
				\nonumber
			\end{align}
			where the norms $\Vert \cdot \Vert_{L^1(\Cbar_{v_{\infty}})}$ and $\Vert \cdot \Vert_{S, \Cbar_{v_{\infty}}}$ are as in the proof of Lemma \ref{lem:fnI}.  The Gauss equation \eqref{eq:Gauss} implies that,
		\begin{equation} \label{eq:IgaugeGausscontractionproof}
			\big\Vert
			(r^{2} K)_{[n+1]}(\hat{u}_f + \delta,v_{\infty},\cdot)
			-
			(r^{2} K)_{[n]}(\hat{u}_f + \delta,v_{\infty},\cdot)
			\big\Vert_{\mathbb{S}^2}
			\lesssim
			\sum_{\vert \gamma \vert \leq 3} r^p \Vert \mathfrak{D}^{\gamma} \Phi_p^{[n+1]}(x) - \mathfrak{D}^{\gamma} \Phi_p^{[n]}(x) \Vert_{S_{\hat{u}_f + \delta,v_{\infty}}}.
		\end{equation}
		The difference between the metrics, $\gslash(x) - \gslash_{[n]}(x)$, are similarly be estimated.  The estimate \eqref{theestimatesforourspherediffdifferences} implies that
		\[
			\vert (r^{-2} \gslash)_{[n+1]} (\hat{u}_f+\delta,v_{\infty},\theta)
			-
			(r^{-2} \gslash)_{[n]} (\hat{u}_f + \delta,v_{\infty},\theta)
			\vert_{\gamma}
			\lesssim
			\big\Vert
			(r^{2} K)_{[n+1]}(\hat{u}_f + \delta,v_{\infty},\cdot)
			-
			(r^{2} K)_{[n]}(\hat{u}_f + \delta,v_{\infty},\cdot)
			\big\Vert_{\mathbb{S}^2},
		\]
		from which it follows that
		\begin{multline*}
			\vert \gslash_{[n+1]} (\hat{u}_f + \delta,v_{\infty}(\hat{u}_f + \delta),\theta)
			-
			\gslash_{[n]}(\hat{u}_f + \delta,v_{\infty}(\hat{u}_f + \delta),\theta)
			\vert_{\gslash}
			\lesssim
			\vert M_{[n+1]} - M_{[n]} \vert
			\\
			+
			r
			\big\Vert
			K_{[n+1]}(\hat{u}_f + \delta,v_{\infty},\cdot)
			-
			K_{[n]}(\hat{u}_f + \delta,v_{\infty},\cdot)
			\big\Vert_{\mathbb{S}^2}
			,
		\end{multline*}
		with $v_{\infty} = v_{\infty}(\hat{u}_f+\delta)$.  The difference $\gslash_{[n+1]}(x) - \gslash_{[n]}(x)$ is then estimated as in Chapter \ref{chap:Iestimates}, using this estimate in place of where the estimate \eqref{theestimatesforourspherediff} is used (see Section \ref{subsec:finalspheremetricestimate}).

	\noindent \underline{\textbf{Estimates for differences $\alpha_{[n+1]} - \alpha_{[n]}$ and $\alphabar_{[n+1]} - \alphabar_{[n]}$:}} Consider now the differences $\alpha_{[n+1]}(x) - \alpha_{[n]}(x)$ and $\alphabar_{[n+1]}(x) - \alphabar_{[n]}(x)$.  The relation \eqref{eq:curvaturecomp1} implies that, schematically,
	\begin{multline*}
		\alphabar_{[n+1]}(x) - \alphabar_{[n]}(x)
		=
		\Pi_{S_{[n+1]}} \alphabar(x) - \Pi_{S_{[n]}} \alphabar(x)
		\\
		+
		\Pi_{S_{[n+1]}} \Phi \cdot \mathfrak{D} f_{[n+1]}(x)
		-
		\Pi_{S_{[n]}} \Phi \cdot \mathfrak{D} f_{[n]}(x)
		+
		\mathfrak{D} f_{[n+1]} \cdot \mathfrak{D}  f_{[n+1]} (x)
		-
		\mathfrak{D} f_{[n]}\cdot \mathfrak{D} f_{[n]}(x)
		.
	\end{multline*}
	For each $\Phi$, using Lemma \ref{lem:mvt} (in particular \eqref{eq:diffiscloseiteratesdifferencetwo}),
	\[
		\big\vert
		\Pi_{S_{[n+1]}} \Phi (x)
		-
		\Pi_{S_{[n]}} \Phi (x)
		\big\vert
		\lesssim
		\sup_{\theta} \big( \vert \Phi \vert + \vert \mathfrak{D} \Phi \vert \big)
		\sup_{u_{-1} \leq u \leq \hat{u}_f+\delta}
		\sum_{\vert \gamma \vert \leq 3}
		\Vert \mathfrak{D}^{\gamma} f_{[n+1]} - \mathfrak{D}^{\gamma} f_{[n]} \Vert_{S_{u,v_{\infty}}}
		,
	\]
	and so it follows that
	\begin{align*} 
		\sum_{\vert \gamma \vert\leq 3}
		\Vert  \mathfrak{D}^{\gamma} r \alphabar_{[n+1]}(x)
		-
		\mathfrak{D}^{\gamma} r \alphabar_{[n]}(x) \Vert_{S, \Cbar_{v_{\infty}}}
		\lesssim
		\varepsilon
		\sup_{u_{-1} \leq u \leq \hat{u}_f+\delta}
		\sum_{\vert \gamma \vert \leq5}
		\Vert \mathfrak{D}^{\gamma} f_{[n+1]}(x) - \mathfrak{D}^{\gamma} f_{[n]}(x) \Vert_{S_{u,v_{\infty}}},
	\end{align*}
	where the norm $\sum_{\vert \gamma \vert \leq 5} \Vert \mathfrak{D}^{\gamma} f_{[n+1]} - \mathfrak{D}^{\gamma} f_{[n]} \Vert_{S_{u,v_{\infty}}}$ is as in \eqref{eq:iteratesdiffeonorms}.  Similarly for $(r^2 \nablaslash_4)^k r \alphabar_{[n+1]}(x) - (r^2 \nablaslash_4)^k r \alphabar_{[n]}(x)$, for $k=1,2$, and for $\alpha_{[n+1]}(x) - \alpha_{[n]}(x)$, as in the proof of Lemma \ref{lem:fnI}, and so
	\begin{align} 
			&
			\sum_{\vert \gamma \vert \leq 2}
			\Vert r^5 \mathfrak{D}^{\gamma} \nablaslash_3 \alpha_{[n+1]}(x) - r^5 \mathfrak{D}^{\gamma} \nablaslash_3 \alpha_{[n]}(x) \Vert_{S, \Cbar_{v_{\infty}}}
			+
			\sum_{\substack{
			\vert \gamma \vert + k \leq 3
			\\
			k \leq 2
			}}
			\Vert  \mathfrak{D}^{\gamma} (r^2 \nablaslash_4)^k (r \alphabar)_{[n+1]}(x)
			-
			\mathfrak{D}^{\gamma} (r^2 \nablaslash_4)^k (r \alphabar)_{[n]}(x) \Vert_{S, \Cbar_{v_{\infty}}}
			\nonumber
			\\
			&
			\qquad \qquad \qquad \qquad \qquad \qquad
			\lesssim
			\varepsilon
			\sup_{u_{-1} \leq u \leq \hat{u}_f+\delta}
			\sum_{\vert \gamma \vert \leq 5}
			\Vert \mathfrak{D}^{\gamma} f_{[n+1]}(x) - \mathfrak{D}^{\gamma} f_{[n]}(x) \Vert_{S_{u,v_{\infty}}}.
			\label{eq:Ifdiffestimate2}
	\end{align}

\noindent \underline{\textbf{Estimates for differences of diffeomorphisms $f_{[n+1]}(x) - f_{[n]}(x)$:}}  The next step involves estimating the differences $\mathfrak{D}^k f_{[n+1]} - \mathfrak{D}^k f_{[n]}$.  Considering the system \eqref{eq:metriccomp1}--\eqref{eq:curvaturecomp6} which, recall, takes the schematic form \eqref{eq:ellipticfIschematic}, it follows that, schematically,
			\begin{align*}
				\mathfrak{D}^2 f_{[n+1]}(x)
				-
				\mathfrak{D}^2 f_{[n]}(x)
				=
				\
				&
				\Phi_{[n+1]}(x) - \Phi_{[n]}(x)
				+
				\Pi_{S_{[n+1]}} \Phi(x) - \Pi_{S_{[n]}} \Phi(x)
				\\
				&
				+
				\sum_{1 \leq \vert \gamma_1\vert, \vert \gamma_2 \vert \leq 2}
				\big(
				\Pi_{S_{[n+1]}} \Phi \cdot \mathfrak{D}^{\gamma_1} f_{[n+1]}(x)
				-
				\Pi_{S_{[n]}} \Phi \cdot \mathfrak{D}^{\gamma_1} f_{[n]}(x)
				\\
				&
				+
				\mathfrak{D}^{\gamma_1} f_{[n+1]} \cdot \mathfrak{D}^{\gamma_2} f_{[n+1]} (x)
				-
				\mathfrak{D}^{\gamma_1} f_{[n]}\cdot \mathfrak{D}^{\gamma_2} f_{[n]}(x)
				\big).
			\end{align*}
			Each $\Phi_p$ satisfies,
		\begin{multline*}
			r^p \Vert \Pi_{S_{[n+1]}} \mathfrak{D}^{\gamma} \Phi_p(x) - \Pi_{S_{[n]}} \mathfrak{D}^{\gamma} \Phi_p(x) \Vert_{S_{u,v_{\infty}}}
			\\
			\lesssim
			\sum_{\vert \tilde{\gamma} \vert \leq \vert \gamma \vert +1}
			\sup_{S_{u,v_{\infty}}}
			\vert \mathfrak{D}^{\tilde{\gamma}} \Phi_p \vert
			\sup_{u_{-1} \leq u \leq \hat{u}_f+\delta}
			\sum_{\vert \gamma \vert \leq 3}
			\Vert \mathfrak{D}^{\gamma} f_{[n+1]} - \mathfrak{D}^{\gamma} f_{[n]} \Vert_{S_{u,v_{\infty}}},
		\end{multline*}
		by Lemma \ref{lem:mvt}
			and so, again using the reductive structure of Proposition \ref{prop:dIdiff}, if $\varepsilon$ is sufficiently small, for $u_{-1} \leq u \leq \hat{u}_f+\delta$,
			\begin{multline} \label{eq:Ifdiffestimate3}
				\sum_{\vert \gamma \vert \leq 5} \big\Vert \big(\mathfrak{D}^{\gamma} f_{[n+1]}(x)
				-
				\mathfrak{D}^{\gamma} f_{[n]}(x) \big)_{\ell \geq 1} \big\Vert_{S_{u,v_{\infty}}}
				\lesssim
				\sum_{\vert \gamma \vert \leq 3}
				r^p
				\Vert \mathfrak{D}^{\gamma} \Phi_p^{[n+1]}(x) - \mathfrak{D}^{\gamma} \Phi_p^{[n]}(x) \Vert_{S_{u,v_{\infty}}}
				\\
				+
				\sum_{\vert \gamma \vert \leq 3}
				(
				r^p \Vert \mathfrak{D}^{\gamma} \Phi_p \Vert_{S_{u,v_{\infty}}}
				+
				\Vert \mathfrak{D}^{\gamma} f_{[n+1]} \Vert_{S_{u,v_{\infty}}}
				+
				\Vert \mathfrak{D}^{\gamma} f_{[n]} \Vert_{S_{u,v_{\infty}}}
				)
				\sum_{\vert \gamma \vert \leq 5}
				\Vert \mathfrak{D}^{\gamma} f_{[n+1]}(x)
				-
				\mathfrak{D}^{\gamma} f_{[n]}(x) \Vert_{S_{u,v_{\infty}}},
			\end{multline}
			where the norms are again as in \eqref{eq:iteratesdiffeonorms} and \eqref{eq:IgaugeGausscontractionproof} has been used.
			Similarly, as in the proof of Lemma \ref{lem:fnI},
			\begin{align} \label{eq:Ifdiffestimate4}
				&
				\vert (f^3_{[n+1]} - f^4_{[n+1]} - f^3_{[n]} + f^4_{[n]})_{\ell=0} \vert
				+
				\sum_{1 \leq \vert \gamma \vert \leq 2} \vert (\mathfrak{D}^{\gamma} f_{[n+1]} - \mathfrak{D}^{\gamma} f_{[n]})_{\ell =0} \vert
				\nonumber
				\lesssim
				r^p \vert (\Phi_p^{[n+1]}(x) - \Phi_p^{[n]}(x))_{\ell =0} \vert
				\\
				&
				\qquad \quad
				+
				\sum_{\vert \gamma \vert \leq 2}
				(
				r^p \vert \mathfrak{D}^{\gamma} \Phi_p \vert
				+
				\vert \mathfrak{D}^{\gamma} f_{[n+1]} \vert
				+
				\vert \mathfrak{D}^{\gamma} f_{[n]} \vert
				)
				\sum_{\vert \gamma \vert \leq 2}
				\vert \mathfrak{D}^{\gamma} f_{[n+1]}(x)
				-
				\mathfrak{D}^{\gamma} f_{[n]}(x) \vert
				,
			\end{align}
			\begin{align} \label{eq:Ifdiffestimate5}
				&
				\vert M_{[n+1]} - M_{[n]} \vert
				\lesssim
				\frac{1}{r} r^p \vert (\Phi_p^{[n+1]}(x) - \Phi_p^{[n]}(x))_{\ell =0} \vert
				\\
				&
				\qquad \quad
				+
				\sum_{\vert \gamma \vert \leq 2}
				(
				r^p \vert \mathfrak{D}^{\gamma} \Phi_p \vert
				+
				\vert \mathfrak{D}^{\gamma} f_{[n+1]} \vert
				+
				\vert \mathfrak{D}^{\gamma} f_{[n]} \vert
				)
				\sum_{\vert \gamma \vert \leq 2}
				\vert \mathfrak{D}^{\gamma} f_{[n+1]}(x)
				-
				\mathfrak{D}^{\gamma} f_{[n]}(x) \vert
				,
				\nonumber
			\end{align}
			and moreover, equation \eqref{eq:nIgauge4} implies
			\begin{equation} \label{eq:Ifdiffestimate6}
				\vert (f^3_{[n+1]} - f^3_{[n]})_{\ell=0}(u,v_{\infty}) \vert
				\leq
				\int_{u_{-1}}^{u}
				\big\vert \big(
				\partial_u \big( f^3_{\ell=0} \big)_{[n+1]} 
				-
				\partial_u \big( f^3_{\ell=0} \big)_{[n]}
				\big) (u',v_{\infty}) \big\vert
				du'
				.
			\end{equation}

	\noindent \underline{\textbf{Estimate for $\mathfrak{s}_{[n+1]}$:}}  As in the proof of Lemma \ref{lem:fnI}, using now the fact that
		\[
			r^3
			\Vert
			\mubar^{\dagger}_{[n+1]}(x)
			-
			\mubar^{\dagger}_{[n]}(x)
			\Vert_{S_{\hat{u}_f+\delta,v_{\infty}}}
			\lesssim
			\hat{u}_f v_{\infty}^{-1}
			\sup_{u_{-1} \leq u \leq \hat{u}_f+\delta} r^p \Vert \Phi^{[n+1]}_p(x) - \Phi^{[n]}_p(x) \Vert_{S_{u,v_{\infty}}},
		\]
		it follows that,
		\begin{multline}
			\label{eq:Ifdiffestimate7}
			\mathfrak{s}_{[n+1]}
			\lesssim
			\vert M_{[n]} - M_{[n-1]} \vert
			+
			\sum_{l=n-1}^n
			\Big[
			\varepsilon
			\sum_{\vert \gamma \vert \leq 5}
			\Vert \mathfrak{D}^{\gamma} f_{[l+1]}(x) - \mathfrak{D}^{\gamma} f_{[l]}(x) \Vert_{S_{\hat{u}_f+\delta,v_{\infty}}}
			\\
			+
			\varepsilon
			\sup_{u_{-1} \leq u \leq \hat{u}_f+\delta} r^p \Vert \Phi^{[l+1]}_p(x) - \Phi^{[l]}_p(x) \Vert_{S_{u,v_{\infty}}}
			\Big].
		\end{multline}

	\noindent \underline{\textbf{The completion of the proof:}}  The result then follows from the estimates \eqref{eq:Ifdiffestimate1}--\eqref{eq:Ifdiffestimate7} if $\hat{\varepsilon}_0$ is sufficiently small (independent of $\hat{u}_f$).
	
\end{proof}

It follows from Lemma \ref{lem:fIexistencediff} that the sequences $\{h_{[n]}^3\}$ and $\{h_{[n]}^4\}$ converges to limits $h^3$ and $h^4$ respectively, sequence of masses $\{M_{[n]}\}$ converges to some limit $M_{\hat{u}_f+\delta}$, and, moreover, the diffeomorphisms $\{\slashed{H}_{[n]}\}$ converge (for example, in given coordinate charts for $\mathbb{S}^2$) to a limiting diffeomorphism $\slashed{H}$ (see \eqref{eq:diffeomorphismsconverging} and Lemma \ref{lem:fIexistencediff}).
The limit is the unique solution of the system \eqref{eq:newIgauge1}--\eqref{eq:newIgauge5}.  The expressions \eqref{eq:Riccicomp1}, \eqref{eq:Riccicomp2}, \eqref{eq:Riccicomp3}, \eqref{eq:Riccicomp4}, \eqref{eq:Riccicomp7}, \eqref{eq:Riccicomp8}, \eqref{eq:curvaturecomp5} then imply that the spacetime double null foliation defined by the sphere \eqref{eq:Igaugeexistencenewsphere} and Proposition \ref{prop:Iconesfoliations} (with $s^3 = h^3 \circ \slashed{H}^{-1}$, $s^4 = h^4 \circ \slashed{H}^{-1}$) is the desired gauge. The smoothness of the gauge follows from the smoothness of the ambient spacetime, the smoothness of the $\hat{u}_f$ normalised $\I$ gauge and thus the smoothness of $h^3$, $h^4$ and $\slashed{H}$ solving \eqref{eq:newIgauge1}--\eqref{eq:newIgauge5}.

The functions
\[
	M_f \colon [\hat{u}_f,\hat{u}_f+\delta_0] \times \mathfrak{R}(\hat{u}_f) \to \mathbb{R},
	\qquad
	{\bf J} \colon [\hat{u}_f,\hat{u}_f+\delta_0] \times \mathfrak{R}(\hat{u}_f) \to \mathbb{R}^3,
\]
are defined by fixing $\lambda\in \mathfrak{R}(\hat{u}_f)$ as above and setting, for $0 \leq \delta \leq \delta_0$,
\[
	M_f(\hat{u}_f+\delta,\lambda) := M_{\hat{u}_f+\delta},
	\qquad
	{\bf J}(\hat{u}_f+\delta,\lambda) := (J^{-1}_{\I}(\hat{u}_f + \delta,\lambda),J^0_{\I}(\hat{u}_f + \delta,\lambda),J^1_{\I}(\hat{u}_f + \delta,\lambda)),
\]
where one recalls (see Proposition~\ref{prop:oneformdecomp}) that $\Omega \beta$ can be decomposed as
\[
	\Omega \beta (u,v,\theta) 
	=
	r \nablaslash h_{1,\Omega \beta}(u,v,\theta)
	+
	r {}^* \nablaslash h_{2,\Omega \beta}(u,v,\theta),
\]
for two functions $h_{1,\Omega \beta}$, $h_{2,\Omega \beta}$, and $J^m_{\I}(\hat{u}_f + \delta,\lambda)$ are defined, for $m=-1,0,1$, by the relation
\[
	(r^3 h_{2,\Omega \beta})_{\ell=1}(\hat{u}_f + \delta,v_{\infty}(\hat{u}_f + \delta),\theta)
	=
	3 \Omega_{\circ,M_f}^2(\hat{u}_f + \delta,v_\infty(\hat{u}_f + \delta))
	\sum_{m=-1}^1
	J^m_{\I}(\hat{u}_f + \delta,\lambda)
	Y^{1}_m(\hat{u}_f + \delta,v_{\infty}(\hat{u}_f + \delta),\theta).
\]
The continuity of $M_f$ and ${\bf J}$, in both $\delta$ and $\lambda$, follows softly from Cauchy stability for the system \eqref{eq:newIgauge1}--\eqref{eq:newIgauge5}.  Similarly for the continuity in $\delta$ of the energies $\mathbb{E}^{N-2}_{\hat{u}_f+\delta}[P_{\I},\check{\Pbar}_{\I}]$, $\mathbb{E}^{N}_{\hat{u}_f+\delta}[\alpha_{\I},\alphabar_{\I}]$, $\mathbb{E}^N_{\hat{u}_f+\delta,\I}$ of the geometric quantities of the $\hat{u}_f + \delta$ normalised $\I$ gauge with respect to $M_f(\hat{u}_f+\delta,\lambda)$, and the associated diffeomorphism energies $\mathbb{E}_{\hat{u}_f+\delta}[f_{d,\I}]$.  The estimate \eqref{eq:Igaugeexistencemass} follows from Lemma \ref{lem:fnI}.  The estimate \eqref{eq:MMinit} follows from the improved estimate \eqref{improvement1} and continuity, provided $\delta_0$ is sufficiently small.

The fact that each $\hat{u}_f + \delta$ normalised $\I$ gauge can be extended to a $\hat{u}_f + \delta$ normalised $\I$ gauge
\[
	i_{\delta} \colon \mathcal{Z}_{\I}(\hat{u}_f + \delta, \delta_1) \to \mathcal{M},
\]
in the sense of Definition \ref{def:Igaugeextension}, follows as in the proof of Theorem \ref{thm:extendedgauges} if $\delta_1$ is sufficiently small.
The estimates \eqref{eq:diffeodeltadeltaprime1}, \eqref{eq:diffeodeltadeltaprime2} for the diffeomorphisms when $\delta'=0$ follow from Lemma \ref{lem:fnI} (for \eqref{eq:diffeodeltadeltaprime2} see \eqref{eq:diffiscloseiteratescoords}).  For $\delta' >0$ the estimates are obtained similarly, following the proof of Lemma \ref{lem:fnI} and considering now the differences of the geometric quantities in the $\hat{u}_f+\delta$ normalised $\I$ gauge with those of the $\hat{u}_f+\delta'$ normalised $\I$ gauge.

\end{proof}

\begin{remark} \label{rmk:linearIgaugesphere}
 From the proof of Theorem \ref{thm:Igaugeexistence}, one can again distill arguments to the linear setting
 (where matters simplify considerably) 
 so as to complete the proof of Proposition \ref{proplinIpgauge} as follows.  Recall again the notation of~\cite{holzstabofschw}.
	
	Recall that the linearised Schwarzschild solution satisfies
	\[
		\rlin^{\mathscr{K}_{{\mathfrak m},0,0,0}}
		=
		- \frac{2M_f}{r^3} \cdot \mathfrak{m},
		\qquad
		2 \Olin^{\mathscr{K}_{{\mathfrak m},0,0,0}}
		=
		-
		\mathfrak{m},
		\qquad
		\otx^{\mathscr{K}_{{\mathfrak m},0,0,0}}
		=
		\otxb^{\mathscr{K}_{{\mathfrak m},0,0,0}}
		=
		0.
	\] 
	
	Now, any two functions $h^3,h^4 \colon \mathbb{S}^2 \to \mathbb{R}$ give rise to functions $f^3(u,\theta) = h^3(\theta)$ for all $u$ and $f^4(v,\theta) = h^4(\theta)$ for all $v$, which generate a pure gauge solution, denoted $\mathscr{G}_{\mathrm{sphere}}$.  In view of the linearised analogues of the relations \eqref{eq:metriccomp6}, \eqref{eq:Riccicomp4}, \eqref{eq:Riccicomp7}, \eqref{eq:Riccicomp8}, \eqref{eq:curvaturecomp3}, \eqref{eq:curvaturecomp5}, and in anticipation of the linearised analogues of the gauge normalisations, in order to realise the linearised analogues of the gauge normalisations \eqref{eq:Iextragauge1}--\eqref{eq:Iextragauge4} one considers the solution $\mathscr{S} + \mathscr{K}_{{\mathfrak m},0,0,0} +\mathscr{G}_{\mathrm{sphere}}$ arising from $h^3$, $h^4$, $\mathfrak{m}$ which satisfy the linearised analogues of equations \eqref{eq:newIgauge1}--\eqref{eq:newIgauge3}, \eqref{eq:newIgauge5}, namely the elliptic system
	\begin{align*}
		&
		2 \Deltaslash \Deltaslash h^3_{\ell \geq 2}
		+
		\frac{2}{r^2} \Big( 3 - \frac{8M_f}{r} \Big) \Deltaslash h^3_{\ell \geq 2}
		-
		\frac{2\Omega^2}{r^2} \Deltaslash h^4_{\ell \geq 2}
		+
		\frac{4\Omega^2}{r^4} (h^3_{\ell \geq 2} - h^4_{\ell \geq 2})
		\\
		&
		\qquad \qquad \qquad
		=
		-
		\big(
		\Omega^{-2}
		\Deltaslash \otx
		+
		r^{-2} \otx
		+
		\frac{2}{r}
		\divslash \eblin
		+
		\frac{2}{r}
		\rlin
		+
		\frac{1}{r^2} \otxb
		+
		\frac{2\Omega^2}{r^3} 2 \Olin
		\big)_{\ell \geq 2}^{\mathscr{S}}(u_f,v_{\infty},\cdot),
		\\
		&
		\frac{6M_f}{r^3} \Big( 1 - \frac{2M_f}{r} \Big) \Deltaslash h^3_{\ell =1}
		=
		(\divslash \Omega \blin)_{\ell =1}^{\mathscr{S}}(u_f,v_{\infty},\cdot),
		\\
		&
		2 \Deltaslash \Deltaslash h^4_{\ell \geq 1}
		+
		\frac{4}{r^2} \Big( 1 - \frac{3M_f}{r} \Big) \Deltaslash h^4_{\ell \geq 1}
		+
		\frac{12M_f \Omega^2}{r^5} (h^3_{\ell \geq 1} - h^4_{\ell \geq 1})
		\\
		&
		\qquad \qquad \qquad
		=
		-
		\big(
		\Omega^{-2} \Deltaslash \otxb
		+
		\frac{2}{r} \Deltaslash 2 \Olin
		+
		\frac{2}{r}
		\divslash \eblin
		+
		\frac{2}{r}
		\rlin
		+
		\frac{2}{r^2} \otxb
		+
		\frac{4\Omega^2}{r^3} 2 \Olin
		\big)_{\ell \geq 1}^{\mathscr{S}}(u_f,v_{\infty},\cdot),
		\\
		&
		-
		\frac{4}{r^2} \Big( 1 - \frac{3M_f}{r} \Big) (h^3_{\ell = 0} - h^4_{\ell = 0})
		-
		\frac{2\mathfrak{m}}{r}
		=
		-
		\big(
		\Omega^{-2} \otxb
		+
		\frac{2}{r} 2 \Olin
		-
		\Omega^{-2} \otx
		\big)_{\ell = 0}^{\mathscr{S}}(u_f,v_{\infty}),
		\\
		&
		\frac{2M_f}{r^3} \mathfrak{m}
		+
		\frac{6M_f\Omega^2}{r^4} (h^3_{\ell=0} - h^4_{\ell=0})
		= 
		\rlin_{\ell=0}^{\mathscr{S}}(u_f,v_{\infty}).
	\end{align*}
	(Note that the linearised analogue of the condition
	$
		\left(\Omega^{-1} \tr \chibar - (\Omega^{-1} \tr \chibar)_{\circ} \right)(u_f,v_{\infty},\cdot)
		=
		0
	$
	is the condition
	$
		\big(
		\Omega^{-2} \otxb
		+
		\frac{2}{r} 2 \Olin
		\big)(u_f,v_{\infty},\cdot)
		=
		0.
	$)
	The proof of Theorem \ref{thm:Igaugeexistence} in this linearised setting reduces to showing the existence of $h^3$, $h^4$, $\mathfrak{m}$ satisfying this elliptic system.  Note that the solution is not unique in view of the fact that, for any solution $h^3$, $h^4$, $\mathfrak{m}$, the functions $h^3 + \lambda$, $h^4+\lambda$ together with $\mathfrak{m}$ are also a solution for any $\lambda \in \mathbb{R}$ (this non-uniqueness could be broken by imposing a linearised analogue of the anchoring condition \eqref{eq:f3f4Iba}, cf.\@ equation \eqref{eq:newIgauge4}).  Given such a solution $h^3$, $h^4$, $\mathfrak{m}$, the solution $\mathscr{S} + \mathscr{K}_{{\mathfrak m},0,0,0} + \mathscr{G}_{\mathrm{sphere}}$ of the linearised equations satisfies
	\begin{align}
		\big(
		\Omega^{-2}
		\Deltaslash \otx
		+
		r^{-2} \otx
		+
		\frac{2}{r}
		\divslash \eblin
		+
		\frac{2}{r}
		\rlin
		+
		\frac{1}{r^2} \otxb
		+
		\frac{2\Omega^2}{r^3} 2 \Olin
		\big)_{\ell \geq 2}^{\mathscr{S} + \mathscr{K}_{{\mathfrak m},0,0,0} + \mathscr{G}_{\mathrm{sphere}}}(u_f,v_{\infty},\cdot)
		&
		=
		0,
		\label{eq:h3h4frakmlinearisedIgauge1}
		\\
		\big(
		\Omega^{-2} \Deltaslash \otxb
		+
		\frac{2}{r} \Deltaslash 2 \Olin
		+
		\frac{2}{r}
		\divslash \eblin
		+
		\frac{2}{r}
		\rlin
		+
		\frac{2}{r^2} \otxb
		+
		\frac{4\Omega^2}{r^3} 2 \Olin
		\big)_{\ell \geq 1}^{\mathscr{S} + \mathscr{K}_{{\mathfrak m},0,0,0} + \mathscr{G}_{\mathrm{sphere}}}(u_f,v_{\infty},\cdot)
		&
		=
		0,
		\label{eq:h3h4frakmlinearisedIgauge3}
	\end{align}
	and
	\begin{align}
		(\divslash \Omega \blin)_{\ell =1}^{\mathscr{S} + \mathscr{K}_{{\mathfrak m},0,0,0} + \mathscr{G}_{\mathrm{sphere}}}(u_f,v_{\infty},\cdot)
		&
		=
		0,
		\label{eq:h3h4frakmlinearisedIgauge2}
		\\
		\big(
		\Omega^{-2} \otxb
		+
		\frac{2}{r} 2 \Olin
		-
		\Omega^{-2} \otx
		\big)_{\ell = 0}^{\mathscr{S} + \mathscr{K}_{{\mathfrak m},0,0,0} + \mathscr{G}_{\mathrm{sphere}}}(u_f,v_{\infty})
		&
		=
		0,
		\label{eq:h3h4frakmlinearisedIgauge4}
		\\
		\rlin_{\ell=0}^{\mathscr{S} + \mathscr{K}_{{\mathfrak m},0,0,0} + \mathscr{G}_{\mathrm{sphere}}}(u_f,v_{\infty})
		&
		=
		0.
		\label{eq:h3h4frakmlinearisedIgauge5}
	\end{align}
	
	Recall now Remark \ref{rmk:linearIgaugefoliations}.  Adding the pure gauge solution $\mathscr{G}_{\mathrm{foliations}}$ arising from $\tilde{\mathscr{S}} = \mathscr{S} + \mathscr{K}_{{\mathfrak m},0,0,0} + \mathscr{G}_{\mathrm{sphere}}$, the solution $\mathscr{S} + \mathscr{K}_{{\mathfrak m},0,0,0} +  \mathscr{G}_{\mathrm{sphere}} + \mathscr{G}_{\mathrm{foliations}}$ then, in view of the fact that $f^3(u_f,\cdot) = f^4(v_{\infty},\cdot) = 0$ for the functions generating this pure gauge solution, still satisfies the relations \eqref{eq:h3h4frakmlinearisedIgauge1}--\eqref{eq:h3h4frakmlinearisedIgauge5} and, in view of the fact that this solution satisfies the linearised analogues of \eqref{eq:Ifoliation1}--\eqref{eq:Ifoliation5}, thus moreover achieves the linearised analogues of the gauge normalisations \eqref{eq:Iextragauge1}--\eqref{eq:Iextragauge4} (the linear analogue of \eqref{eq:Iextragauge4} being the condition $\rlin_{\ell=0}(u_f,v_{\infty}) = 0$).
	
	Finally, the linearised analogue of the gauge condition \eqref{eq:Iextragauge5} can be achieved by adding an appropriate pure gauge solution $\mathscr{G}_{\mathrm{theta}}$ generated by functions $q_1(u,\theta)$ and $q_2(u,\theta)$ (see Lemma 6.\@1.\@3 of~\cite{holzstabofschw} and note that the difference in the position of the torsion means that here $q_1$ and $q_2$ are indeed functions of $(u,\theta)$, rather than $(v,\theta)$ as in~\cite{holzstabofschw}) satisfying
	\[
		-
		r^2 \nablaslash \partial_u q_1(u,\theta)
		+
		r^2 {}^* \nablaslash \partial_u q_2(u,\theta)
		=
		-
		b^{\mathscr{S} + \mathscr{K}_{{\mathfrak m},0,0,0} +  \mathscr{G}_{\mathrm{sphere}} + \mathscr{G}_{\mathrm{foliations}}}(u,v_{\infty},\theta),
	\]
	for all $u$ and $\theta \in \mathbb{S}^2$ (along with $f^3 \equiv 0$, $f^4\equiv 0$).  The solution
	\[
		\mathscr{S}
		+
		\mathscr{K}_{{\mathfrak m},0,0,0}
		+
		\mathscr{G}_{\mathrm{sphere}}
		+
		\mathscr{G}_{\mathrm{foliations}}
		+
		\mathscr{G}_{\mathrm{theta}},
	\]
	is then normalised as desired.
	
	Note again that $\mathscr{G} = \mathscr{G}_{\mathrm{sphere}} + \mathscr{G}_{\mathrm{foliations}} + \mathscr{G}_{\mathrm{theta}}$ is not the unique pure gauge solution which achieves these normalisations in view of the absence of any linearised analogues of the anchoring condition \eqref{eq:f3f4Iba} and the anchoring condition of Definition \ref{anchoringdef} concerning the sphere diffeomorphism of the $\I$ gauge in Proposition \ref{proplinIpgauge}.
\end{remark}

\begin{remark} \label{rmk:uf0Igaugeexistence2}
	Recall the double null parametrisation \eqref{herethenewone} of Theorem \ref{thm:localEF}, and let $u_f^0$ be as defined in Section~\ref{compediumparameterssec}.
	The proof of Theorem \ref{thm:Igaugeexistence} given in this section can easily be adapted to give a proof of the existence of a $u_f^0$ normalised $\I$ gauge satisfying the anchoring condition \eqref{eq:f3f4Iba} and the anchoring condition of Definition \ref{anchoringdef} concerning the sphere diffeomorphism of the $\I$ gauge (cf.\@ Theorem \ref{existenceofanchoredgaugethe}).  Indeed, the existence of such a gauge reduces to finding smooth functions $h^3, h^4\colon \mathbb{S}^2 \to \mathbb{R}$, a diffeomorphism $\slashed{H} \colon \mathbb{S}^2 \to \mathbb{S}^2$ and a mass $M_f$ satisfying the system \eqref{eq:newIgauge1}--\eqref{eq:newIgauge5} with $\hat{u}_f+\delta$ replaced by $u_f^0$, the geometric quantities of the extended $\hat{u}_f$ normalised $\I$ gauge replaced by the corresponding geometric quantities of the $i_{\mathcal{EF}}$ gauge of Theorem \ref{thm:localEF}, and the diffeomorphism functions $f$ replaced by the diffeomorphism functions relating the new double null foliation arising from Remark \ref{rmk:uf0Igaugeexistence1} with the sphere \eqref{eq:uf0Igaugeexistencesphere} defined by $s^3 = h^3 \circ \slashed{H}^{-1}$, $s^4 = h^4 \circ \slashed{H}^{-1}$, together with the condition that
	\[
		\psi \circ i_{\mathcal{EF}} (u_f^0 + h^3(\theta), v_{\infty}(u_f^0) + h^4(\theta),\slashed{H}(\theta)) = \theta,
	\]
	where $\psi$ is the canonical diffeomorphism of Proposition \ref{determiningthesphere} and $p$, $v$ chosen so that \eqref{eq:Iextragauge6a} and \eqref{eq:Iextragauge6} hold.  The existence of such $h^3$, $h^4$, $\slashed{H}$ and $M_f$ can easily be established by defining a suitable sequence of iterates and following the proof of Theorem \ref{thm:Igaugeexistence}.
\end{remark}

This section is ended with the following estimate for the differences between the $\ell = 1$ modes of the $\hat{u}_f + \delta$ normalised $\I$ gauge, which will be used in the proof of Theorem \ref{thm:lambda} in Section \ref{section:monotonicityR} below.

\begin{proposition}[Estimate for difference between $\ell =1$ modes of $\hat{u}_f+\delta'$ and $\hat{u}_f+\delta$ normalised $\I$ gauges] \label{prop:YYtildedifference}
	Let $\lambda \in \mathfrak{R}(\hat{u}_f)$ be fixed and consider some $0\leq \delta' < \delta \leq \delta_0$.  Let $(Y_m^1)^{\delta}$ and $(Y_m^1)^{\delta'}$ denote the $\ell =1$ modes of the $\hat{u}_f+\delta$ and $\hat{u}_f+\delta'$ normalised $\I$ gauges of Theorem \ref{thm:Igaugeexistence} respectively.  One can view $(Y_m^1)^{\delta}(\hat{u}_f+\delta',v_{\infty}(\hat{u}_f+\delta'),\cdot)$ and $(Y_m^1)^{\delta'}(\hat{u}_f+\delta',v_{\infty}(\hat{u}_f+\delta'),\cdot)$ as functions on $\mathbb{S}^2$ via the identifications $i_{\delta}$ and $i_{\delta'}$ respectively.  The differences satisfy, for $m=-1,0,1$,
	\begin{multline*}
		\Vert (Y_m^1)^{\delta} (\hat{u}_f+\delta',v_{\infty}(\hat{u}_f+\delta'),\cdot) - (Y_m^1)^{\delta'} (\hat{u}_f+\delta',v_{\infty}(\hat{u}_f+\delta'),\cdot ) \Vert_{S_{u,v}}
		\\
		+
		\Vert (r^2 \Deltaslash Y_m^1)^{\delta}(\hat{u}_f+\delta',v_{\infty}(\hat{u}_f+\delta'),\cdot) - (r^2 \Deltaslash Y_m^1)^{\delta'} (\hat{u}_f+\delta' ,v_{\infty}(\hat{u}_f+\delta'),\cdot) \Vert_{S_{u,v}}
		\lesssim
		\frac{\varepsilon (\delta - \delta')}{\hat{u}_f+\delta'}
		.
	\end{multline*}
\end{proposition}

\begin{proof}
	Consider first the case that $\delta' = 0$.  Define $Y_m^1 := (Y_m^1)^0$ and $\widetilde{Y}_m^1 := (Y_m^1)^{\delta'}$, and similarly let $\gslash$ and $\widetilde{\gslash}$ denote the induced metrics of the extended $\hat{u}_f$ normalised $\I$ gauge and the $\hat{u}_f+\delta$ normalised $\I$ gauge respectively.
	Via the identifications $i_0$ and $i_{\delta}$, one can view $\gslash(\hat{u}_f ,v_{\infty}(\hat{u}_f),\cdot)$ and $\widetilde{\gslash}(\hat{u}_f ,v_{\infty}(\hat{u}_f),\cdot)$ as two metrics on the sphere $\mathbb{S}^2$.
	Given a function $h$ on $\mathbb{S}^2$, let $h_{\ell=0}$, $h_{\ell =1}$ and $h_{\ell \geq 2}$ denote the mode projections with respect to the metric $\gslash$, and let $h_{\widetilde{\ell = 0}}$, $h_{\widetilde{\ell = 1}}$ and $h_{\widetilde{\ell \geq 2}}$ denote the mode projections with respect to the metric $\widetilde{\gslash}$.  Let also $\Vert \cdot \Vert_{S_{u,v}}$ and $\Vert \cdot \Vert_{\widetilde{S}_{u,v}}$ denote the $L^2$ norms on $\mathbb{S}^2$ with respect to $\gslash(u,v)$ and $\widetilde{\gslash}(u,v)$ respectively.
	
	Recall now that, for $m = -1,0,1$, 
	$Y^1_m$ is defined in terms of 
	$\breve{Y}^1_m = \Pi_{\mathcal{Y}^1} \mathring{Y}^1_m$ (see Section \ref{basisforprojspace}).  It follows that $\breve{Y}^1_m = (\mathring{Y}^1_m)_{\ell = 1}$, and similarly $\widetilde{\breve{Y}}{}^1_m = (\mathring{Y}{}^1_m)_{\widetilde{\ell = 1}}$, and so
	\[
		(\breve{Y}^1_m)_{\widetilde{\ell = 1}}
		-
		(\widetilde{\breve{Y}}{}^1_m)_{\ell = 1}
		=
		\big( \big( \mathring{Y}^1_m - \mathring{Y}{}^1_m \big)_{\ell =1} \big)_{\widetilde{\ell = 1}}
		=
		0.
	\]
	Hence, for $\breve{Y}^1_m = \breve{Y}^1_m(u,v,\theta)$ and $\widetilde{\breve{Y}}{}^1_m = \widetilde{\breve{Y}}{}^1_m(u,v,\theta)$,
	\begin{align*}
		\breve{Y}^1_m - \widetilde{\breve{Y}}{}^1_m
		&
		=
		(\widetilde{\breve{Y}}{}^1_m)_{\ell =0}
		+
		(\widetilde{\breve{Y}}{}^1_m)_{\ell \geq 2}
		-
		(\breve{Y}{}^1_m)_{\widetilde{\ell =0}}
		-
		(\breve{Y}{}^1_m)_{\widetilde{\ell \geq 2}}
		.
	\end{align*}
	Now
	\[
		(\breve{Y}^1_m)_{\widetilde{\ell =0}}(u,v)
		=
		(\int \sqrt{\det \widetilde{\gslash}(u,v,\theta)} d\theta^1 d \theta^2)^{-1}
		\int
		\breve{Y}^1_m(u,v,\theta)
		\sqrt{\det \widetilde{\gslash}(u,v,\theta)} d\theta^1 d \theta^2,
	\]
	and
	\[
		\int
		\breve{Y}^1_m(u,v,\theta)
		\sqrt{\det \gslash(u,v,\theta)} d\theta^1 d \theta^2
		=
		0,
	\]
	hence
	\[
		\vert (\breve{Y}^1_m)_{\widetilde{\ell =0}} (u,v) \vert
		\lesssim
		\sup_{\theta \in \mathbb{S}^2}
		\vert \widetilde{\gslash}(u,v,\theta) - \gslash(u,v,\theta) \vert
		.
	\]
	Moreover, by Proposition \ref{prop:ellipticestimates},
	\begin{align}
		\sum_{k=0}^2 \Vert \widetilde{\nablaslash}{}^k (\breve{Y}^1_m)_{\widetilde{\ell \geq 2}} \Vert_{\widetilde{S}_{u,v}}
		&
		\lesssim
		\Vert \widetilde{\Dslash}{}_2^* \widetilde{\nablaslash} \breve{Y}^1_m \Vert_{\widetilde{S}_{u,v}}
		=
		\Vert (\widetilde{\Dslash}{}_2^* \widetilde{\nablaslash} - \Dslash{}_2^* \nablaslash) \breve{Y}^1_m \Vert_{\widetilde{S}_{u,v}}
		\nonumber
		\\
		&
		\lesssim
		\Vert \gslash(u,v,\cdot) - \widetilde{\gslash}(u,v,\cdot) \Vert_{\widetilde{S}_{u,v}}
		+
		\Vert \Gammaslash(u,v,\cdot) - \widetilde{\Gammaslash}(u,v,\cdot) \Vert_{\widetilde{S}_{u,v}},
		\label{eq:tildeYgeq2}
	\end{align}
	since
	\[
		(\widetilde{\Dslash}{}_2^* \widetilde{\nablaslash} - \Dslash{}_2^* \nablaslash) \breve{Y}^1_{m}
		=
		(\Gammaslash - \widetilde{\Gammaslash}) \cdot \nablaslash \breve{Y}^1_{m}
		+
		\frac{1}{2} (\Deltaslash - \widetilde{\Deltaslash}) \breve{Y}^1_{m} \cdot \widetilde{\gslash}
		+
		\frac{1}{2} (\gslash - \widetilde{\gslash}) \Deltaslash \breve{Y}^1_{m}.
	\]
	The terms $\Vert (\widetilde{\breve{Y}}{}^1_m)_{\ell =0} \Vert_{\widetilde{S}_{u,v}}$ and $\Vert (\widetilde{\breve{Y}}{}^1_m)_{\ell \geq 2} \Vert_{\widetilde{S}_{u,v}}$ can be estimated similarly.  Hence
	\begin{align*}
		\Vert \breve{Y}^1_m - \widetilde{\breve{Y}}{}^1_m \Vert_{\widetilde{S}_{u,v}}
		\lesssim
		\sup_{\theta \in \mathbb{S}^2}
		\vert \widetilde{\gslash}(u,v,\theta) - \gslash(u,v,\theta) \vert
		+
		\Vert \Gammaslash(u,v,\cdot) - \widetilde{\Gammaslash}(u,v,\cdot) \Vert_{\widetilde{S}_{u,v}}.
	\end{align*}
	Now,
	\[
		Y^1_{-1}
		=
		\Vert \breve{Y}^1_{-1} \Vert_{S}^{-1} \breve{Y}^1_{-1},
		\qquad
		\widetilde{Y}^1_{-1}
		=
		\Vert \widetilde{\breve{Y}}{}^1_{-1} \Vert_{\widetilde{S}}^{-1} \widetilde{\breve{Y}}{}^1_{-1},
	\]
	and so
	\[
		\Vert \widetilde{\breve{Y}}{}_{-1}^1 (u,v,\cdot) - \breve{Y}_{-1}^1 (u ,v,\cdot ) \Vert_{\widetilde{S}_{u,v}}
		\lesssim
		\Vert \breve{Y}^1_{-1} - \widetilde{\breve{Y}}{}^1_{-1} \Vert_{\widetilde{S}_{u,v}}.
	\]
	Similarly for $Y_0^1 - \widetilde{Y}_0^1$ and $Y_1^1 - \widetilde{Y}_1^1$.
	The proof of the estimate for $\widetilde{Y}_m^1 - Y_m^1$ then follows from estimates for $\gslash(u,v,\theta) - \widetilde{\gslash}(u,v,\theta)$ and $\Gammaslash(u,v,\theta) - \widetilde{\Gammaslash}(u,v,\theta)$ which arise from Lemma \ref{lem:fnI} (see, for example \eqref{eq:gslashgslashnufdeltaestimate}).
	The estimate for $\Deltaslash Y{}^1_m - \widetilde{\Deltaslash} \widetilde{Y}{}^1_m$ follows from the fact that
	\[
		\Deltaslash Y{}^1_m - \widetilde{\Deltaslash} \widetilde{Y}{}^1_m
		=
		(\Deltaslash - \widetilde{\Deltaslash}) Y{}^1_m
		+
		\widetilde{\Deltaslash} \big(
		(Y{}^1_m - \widetilde{Y}{}^1_m)_{\widetilde{\ell=1}}
		+
		(Y{}^1_m)_{\widetilde{\ell \geq 2}}
		\big),
	\]
	and so
	\[
		\big\Vert \Deltaslash Y{}^1_m - \widetilde{\Deltaslash} \widetilde{Y}{}^1_m \big\Vert_{\widetilde{S}_{u,v}}
		\lesssim
		\Vert \gslash - \widetilde{\gslash} \Vert_{\widetilde{S}_{u,v}}
		+
		\Vert \Gammaslash - \widetilde{\Gammaslash} \Vert_{\widetilde{S}_{u,v}}
		+
		\Vert Y{}^1_m - \widetilde{Y}{}^1_m \Vert_{\widetilde{S}_{u,v}},
	\]
	by \eqref{eq:tildeYgeq2}.
	
	The proof when $\delta'>0$ is similar.  Defining $\gslash^{\delta}$ and $\gslash^{\delta'}$ to be the induced metrics of the $\hat{u}_f+\delta$ and $\hat{u}_f+\delta'$ normalised $\I$ gauges respectively, one uses now appropriate estimates for $\gslash^{\delta}(\hat{u}_f+\delta',v_{\infty}, \theta) - \gslash^{\delta'}(\hat{u}_f+\delta',v_{\infty}, \theta)$, which follow as in the proof of Lemma \ref{lem:fnI} by considering now the differences of the geometric quantities in the $\hat{u}_f+\delta$ normalised $\I$ gauge with those of the $\hat{u}_f+\delta'$ normalised $\I$ gauge.
\end{proof}

\subsection{Existence of $\hat{u}_f+\delta$ normalised $\Hp$ gauge}
\label{offing2}

This section concerns the proof of the following theorem, which establishes the existence of $\hat{u}_f+\delta$ normalised $\Hp$ gauges.  Recall the function $M_f \colon [\hat{u}_f,\hat{u}_f+\delta_0] \times \mathfrak{R}(\hat{u}_f) \to \mathbb{R}$ from Theorem \ref{thm:Igaugeexistence}.

\begin{theorem}[Existence of $\hat{u}_f+\delta$ normalised $\Hp$ gauge] \label{thm:Hgaugeexistence}
	There exists $\delta_0>0$ such that, for all $\delta \in [0,\delta_0]$ and all $\lambda\in \mathfrak{R}(\hat{u}_f)$, if $\hat{\varepsilon}_0$ is sufficiently small, there exists a smooth $\hat{u}_f + \delta$ normalised $\Hp$ gauge with respect to $M_f(\hat{u}_f+\delta,\lambda)$ satisfying the anchoring conditions of Definition \ref{anchoringdef} with the $\hat{u}_f+\delta$ normalised $\I$ gauge of Theorem \ref{thm:Igaugeexistence}.  Moreover the energies $\mathbb{E}^{N-2}_{\hat{u}_f+\delta}[P_{\Hp},\Pbar_{\Hp}]$, $\mathbb{E}^{N}_{\hat{u}_f+\delta}[\alpha_{\Hp},\alphabar_{\Hp}]$, $\mathbb{E}^N_{\hat{u}_f+\delta,\Hp}$ of the geometric quantities of the $\hat{u}_f + \delta$ normalised $\Hp$ gauge with respect to $M_f(\hat{u}_f+\delta,\lambda)$, and the associated diffeomorphism energies $\mathbb{E}^{N+2}_{\hat{u}_f+\delta}[f_{\Hp,\I}]$ and $\mathbb{E}_{\hat{u}_f+\delta}[f_{d,\Hp}]$, depend continuously on $\delta$.
\end{theorem}

The fact that the energies of the geometric quantities and the relevant diffeomorphism functions of the $\hat{u}_f + \delta$ normalised gauges all depend continuously on $\delta$, as stated in Theorem \ref{thm:Igaugeexistence} and Theorem \ref{thm:Hgaugeexistence}, ensures that, provided $\delta_0$ is suitably small, the estimates  \eqref{eq:bamain}, \eqref{eq:badiffeo} and \eqref{lowerorderpointwise} hold with $u_f:=\hat{u}_f+\delta$.  Theorem \ref{thm:Igaugeexistence} and Theorem \ref{thm:Hgaugeexistence} thus complete the proof of Theorem \ref{thm:newgauge}.

\begin{remark}
	It is again manifest from the proof of Theorem \ref{thm:Hgaugeexistence} that, for fixed $\lambda\in \mathfrak{R}(\hat{u}_f)$ and fixed $0 \leq \delta \leq \delta_0$, the $\hat{u}_f + \delta$ normalised $\Hp$ gauge with respect to $M_f(\hat{u}_f+\delta,\lambda)$ is unique amongst all such nearby gauges which satisfy the anchoring conditions \eqref{eq:anchoringdefcommoncone} and \eqref{eq:anchoringdefaffixingsphere}.
\end{remark}

\begin{remark}
	For each $0 \leq \delta' < \delta \leq \delta_0$, the diffeomorphisms relating the $\hat{u}_f+\delta$ normalised and the (appropriately extended) $\hat{u}_f+\delta'$ normalised $\Hp$ gauges satisfy appropriate estimates, similar to the estimates \eqref{eq:diffeodeltadeltaprime1} and \eqref{eq:diffeodeltadeltaprime2} relating the different $\I$ gauges.  Since such estimates are not used later (unlike \eqref{eq:diffeodeltadeltaprime1} and \eqref{eq:diffeodeltadeltaprime2}) they are not stated explicitly as part of Theorem \ref{thm:Hgaugeexistence}.
\end{remark}

\begin{remark}
\label{sameasbeforebutforthehorizongauge}
	For any $u_f$, the existence of the analogue of a $u_f$ normalised $\Hp$ gauge in linear theory, namely the proof of Proposition \ref{proplinHpgauge}, can again be inferred from (a considerable simplification of) the proof of Theorem \ref{thm:Hgaugeexistence}.  We will explicitly distill such a proof in Remarks \ref{rmk:linearHgaugefoliations} and \ref{rmk:linearHgaugesphere} below.
\end{remark}

\begin{remark}
\label{theotheronexiststoo}
	Recall the double null parametrisation $i_{\mathcal{K}}$ of Theorem \ref{maxCauchythe} and Theorem \ref{thm:localKrus}, and let $u_f^0$ be as defined in Section~\ref{compediumparameterssec}.
	The proof of Theorem \ref{thm:Hgaugeexistence} can readily be adapted to give the existence of a $u_f^0$ normalised $\Hp$ gauge (cf.\@ the proof of Theorem \ref{existenceofanchoredgaugethe}), anchored (in the sense of Definition \ref{anchoringdef}) with the $u_f^0$ normalised $\I$ gauge discussed in Remark \ref{rmk:uf0Igaugeexistence0}, by replacing the role of the $\delta_1$ extended $\hat{u}_f$ normalised $\Hp$ gauge of Theorem \ref{thm:extendedgauges} with the double null parametrisation $i_{\mathcal{K}}$, and replacing any invocation of the smallness of $\delta_0$ with the smallness of $\hat{\varepsilon}_0$.  Again, the existence of $u_f^0$ normalised $\Hp$ gauge is much simpler than the proof of Theorem \ref{thm:Hgaugeexistence} in view of the fact that $\hat{\varepsilon}_0$ can be chosen to be small with respect to $u_f^0$.  It is again implicit from the construction that this gauge is unique amongst all such nearby gauges.  See also the discussion in Remarks \ref{rmk:uf0Igaugeexistence1} and \ref{rmk:uf0Igaugeexistence2} concerning the $u_f^0$ normalised $\I$ gauge.
\end{remark}

Throughout this section $\lambda \in \mathfrak{R}(\hat{u}_f)$ is considered fixed, and hence the dependence of quantities on $\lambda$ is suppressed, and
\[
	M_{\hat{u}_f+\delta} = M_f (\hat{u}_f+\delta,\lambda).
\]

The proof of Theorem \ref{thm:Hgaugeexistence} is divided into two parts.  The first part involves showing that, for any sphere suitably close to the sphere $S_{\hat{u}_f+\delta, v_{-1}}$ of the extended $\hat{u}_f$ normalised $\Hp$ gauge, there exist certain foliations of the corresponding null hypersurfaces defined by the sphere.  The foliations attain a subset of the defining conditions of the $\hat{u}_f+\delta$ normalised $\Hp$ gauge with mass $M_{\hat{u}_f+\delta}$.  The existence of such foliations is addressed in {\bf Section~\ref{subsec:Hfoliationsexistence}}.  See Proposition \ref{prop:Hconesfoliations}.  The second part of the proof of Theorem \ref{thm:Hgaugeexistence} involves finding an appropriate sphere with the property that, in the double null foliation arising from the foliations of the resulting cones of Proposition \ref{prop:Hconesfoliations}, the remaining defining conditions of the $\hat{u}_f+\delta$ normalised $\Hp$ gauge are attained.  The existence of such a sphere is addressed in {\bf Section~\ref{subsec:Hsphereexistence}}.

\subsubsection{Foliations of null hypersurfaces}
\label{subsec:Hfoliationsexistence}

Throughout this section $\lambda\in \mathfrak{R}(\hat{u}_f)$ is fixed and the dependence of quantities on $\lambda$ is again suppressed.
Suppose $\mathring{S}$ is any sphere suitably close to $S_{\hat{u}_f+\delta, v_{-1}}$.  Let $\newsph{\underline{C}}_{v_{-1}}$ denote the incoming component of the boundary of the causal past of $\mathring{S}$, and let $\newsph{C}_{\hat{u}_f+\delta}$ denote the outgoing component of the boundary of the causal future of $\newsph{S}$.  Any foliations of $\newsph{\underline{C}}_{v_{-1}}$ and $\newsph{C}_{\hat{u}_f+\delta}$ by spheres define a spacetime double null foliation by considering the outgoing components of the boundaries of the causal future of the spheres of the foliation of $\newsph{\underline{C}}_{v_{-1}}$, and the incoming components of the boundaries of the causal past of the spheres of the foliation of $\newsph{C}_{\hat{u}_f+\delta}$.  Under suitable smallness conditions, this spacetime double null foliation will be regular in the image of the domain of the extended $\hat{u}_f$ normalised $\Hp$ gauge, $i_{\Hp}(\mathcal{Z}_{\Hp}(\hat{u}_f,\delta_1))$.

Following the proof of Theorem \ref{thm:Igaugeexistence}, the first step in the proof of Theorem \ref{thm:Hgaugeexistence} is to show that, given any such sphere suitably close to $S_{\hat{u}_f+\delta, v_{-1}}$, there exist foliations of the two corresponding null hypersurfaces defined by this sphere such that, in the corresponding spacetime double null foliation, a subset of the defining properties of the $\hat{u}_f+\delta$ normalised $\Hp$ gauge are satisfied.  This statement is the content of Proposition \ref{prop:Hconesfoliations}.  Recall the quantity
\[
	\mu^*
	=
	\divslash \eta + (\rho - \rho_{\circ}) - \frac{3}{2r} \left( \Omega \tr \chi - (\Omega \tr \chi)_{\circ}\right)
\]

\begin{proposition}[Foliation of cones corresponding to any sphere close to $S_{\hat{u}_f+\delta, v_{-1}}$] \label{prop:Hconesfoliations}
	Let $\mathring{S}$ be a sphere defined, in the $(u_{\Hp}, v_{\Hp},\theta_{\Hp})$ coordinate system of the extended $\hat{u}_f$ normalised $\Hp$ gauge, by smooth functions $s^3$, $s^4$,
	\begin{equation} \label{eq:newsphereHh}
		\mathring{S} = \{ i_{\Hp} (\hat{u}_f+\delta + s^3(\theta), v_{-1} + s^4(\theta),\theta) \mid \theta \in \mathbb{S}^2 \}.
	\end{equation}
	Suppose that $s^3$ and $s^4$ satisfy
	\[
		\sum_{k \leq 5}
		\big(
		\Vert (r \nablaslash)^k s^3 \Vert_{\mathring{S}}
		+
		\Vert (r \nablaslash)^k s^4 \Vert_{\mathring{S}}
		\big)
		\leq
		\tau,
	\]
	for some $\tau$ sufficiently small, so that $\mathring{S}$ is appropriately close to the sphere $S_{\hat{u}_f+\delta, v_{-1}}$.  Then, provided $\tau$ and $\delta_0$ are suitably small, there exists a smooth foliation of the outgoing cone, $\newsph{C}_{\hat{u}_f+\delta}$, of $\mathring{S}$ and a smooth foliation of the incoming cone, $\newsph{\underline{C}}_{v_{-1}}$, of the sphere $\mathring{S}$ such that the geometric quantities of the associated spacetime double null foliation satisfy,
	\begin{align}
		\Omega^2
		-
		\Omega_{\circ,M_{\hat{u}_f+\delta}}^2
		&
		= 
		0
		\quad
		\text{on } \newsph{C}_{\hat{u}_f+\delta}^{\Hp} \cap \{ r_{\I} \leq R_2 \};
		\label{eq:Hfoliation1}
		\\
		\partial_u \left( r^3 (\divslash \eta)_{\ell\geq 1} + r^3 \rho_{\ell \geq 1} \right) (u,v_0,\theta) 
		&
		= 
		0
		\quad
		\text{for all }
		u_0 \leq u \leq \hat{u}_f, \theta \in \mathbb{S}^2;
		\label{eq:Hfoliation2}
		\\
		\Omega(u,v_0)_{\ell=0}^2 
		-
		\Omega_{\circ,M_{\hat{u}_f+\delta}} (u,v_0)^2
		&
		=
		0
		\quad
		\text{for all }
		u_0 \leq u \leq \hat{u}_f;
		\label{eq:Hfoliation3}
		\\
		\mu^*_{\ell \geq 1} (\hat{u}_f+\delta,v(R,\hat{u}_f+\delta),\theta) 
		&
		= 
		0
		\quad
		\text{for all }
		\theta \in \mathbb{S}^2;
		\label{eq:Hfoliation4}
		\\
		\big(\Omega \tr \chibar - (\Omega \tr \chibar)_{\circ,M_{\hat{u}_f+\delta}} \big)_{\ell = 0} (\hat{u}_f+\delta, v_{-1})
		&
		=
		0.
		\label{eq:Hfoliation5}
	\end{align}
\end{proposition}

Given a sphere $\mathring{S}$ as in \eqref{eq:newsphereHh}, let $\newsph{C}_{\hat{u}_f+\delta}$ denote the outgoing component of the boundary of the causal future of $\mathring{S}$, and let $\newsph{\Cbar}_{v_{-1}}$ denote the incoming component of the boundary of the causal past of $\mathring{S}$.  If $\tau$ is suitably small, the sphere $\mathring{S}$ is contained in the region covered by the extended $\hat{u}_f$ normalised $\Hp$ gauge, $\mathring{S} \subset i_{\Hp}(\mathcal{Z}_{\Hp}(\hat{u}_f,\delta_1))$, and $\newsph{C}_{\hat{u}_f+\delta}$ and $\newsph{\Cbar}_{v_{-1}}$ are regular null hypersurfaces in $i_{\Hp}(\mathcal{Z}_{\Hp}(\hat{u}_f,\delta_1))$.  The cones $\newsph{C}_{\hat{u}_f+\delta}$ and $\newsph{\Cbar}_{v_{-1}}$ will each be foliated in such a way that the relevant regions of the cones also live in the region $i_{\Hp}(\mathcal{Z}_{\Hp}(\hat{u}_f,\delta_1))$ covered by the extended $\hat{u}_f$ normalised $\Hp$ gauge.

\subsubsection*{Foliations of the outgoing null hypersurface $\newsph{C}_{\hat{u}_f+\delta}$}

The following proposition relates certain geometric quantities associated to two different foliations of the outgoing null hypersurface $\newsph{C}_{\hat{u}_f+\delta}$.

\begin{proposition}[Change of foliation relations on outgoing cones] \label{prop:outHconesrelations}
	For any sphere $\mathring{S}$ as in \eqref{eq:newsphereHh}, let $v$ denote the restriction to $\newsph{C}_{\hat{u}_f+\delta}$ of the $v$ coordinate of the extended $\hat{u}_f$ normalised gauge.  Consider another foliation of $\newsph{C}_{\hat{u}_f+\delta}$ by spheres, described by the level hypersurfaces of a function $\widetilde{v}$.  Suppose $v$ and $\widetilde{v}$ are related by
	\[
		v = \widetilde{v} + f^4(\widetilde{v},\theta^1,\theta^2),
	\]
	for some function $f^4$.  Then the corresponding $\Omega \omegahat$ are related by
	\[
		\frac{1}{2} \partial_{\widetilde{v}} \log(1+\partial_{\widetilde{v}} f^4)(\widetilde{v},\theta)
		=
		\widetilde{\Omega \omegahat}(\widetilde{v},\theta)
		-
		(1+\partial_{\widetilde{v}} f^4) \Omega \omegahat(\widetilde{v} + f^4,\theta).
	\]
\end{proposition}

\begin{proof}
	The proof follows exactly as that of Proposition \ref{prop:outIconesrelations}.
\end{proof}

As in Section \ref{subsec:Ifoliationsexistence}, rather than foliate the outgoing cone $\newsph{C}_{\hat{u}_f+\delta}$ directly by the condition that $(\Omega^2 - \Omega_{\circ}^2) = 0$, it is first shown that the incoming cone can be foliated by certain conditions which involve only $\Omega \omegahat$.

In order to motivate the foliations constructed in Proposition \ref{prop:foliationHoutgoingcone} below, note first that, in a given spacetime double null foliation, equation \eqref{eq:DlogOmega} implies that,
\begin{equation*}
	\Omega_{\circ}^{-1} \Omega (\hat{u}_f+\delta,v,\theta)
	=
	\Omega_{\circ}^{-1} \Omega (\hat{u}_f+\delta,v_{-1},\theta)
	e^{\int_{v_{-1}}^v (\Omega \omegahat - \Omega \omegahat_{\circ})(\hat{u}_f+\delta,v',\theta) dv'}.
\end{equation*}

\begin{proposition}[Foliations of outgoing cones] \label{prop:foliationHoutgoingcone}
	Let $\mathring{S}$ be a sphere defined, in the $(u_{\Hp}, v_{\Hp},\theta_{\Hp})$ coordinate system of the extended $\hat{u}_f$ normalised gauge, by smooth functions $s^3$, $s^4$,
	\[
		\mathring{S} = \{ i_{\Hp} (\hat{u}_f+\delta + s^3(\theta), v_{-1} + s^4(\theta),\theta) \mid \theta \in \mathbb{S}^2\}.
	\]
	such that $s^3$ and $s^4$ satisfy
	\[
		\sum_{k \leq 5}
		\big(
		\Vert (r \nablaslash)^k s^3 \Vert_{\mathring{S}}
		+
		\Vert (r \nablaslash)^k s^4 \Vert_{\mathring{S}}
		\big)
		\leq
		\tau.
	\]
	Consider also a smooth function $p^4 : \mathring{S} \to \mathbb{R}$ satisfying
	\[
		\sup_{\theta \in \mathbb{S}^2} \vert p^4(\theta) \vert
		\leq
		\tau.
	\]
	Then, provided $\tau$ and $\delta_0$ are sufficiently small, there exists a smooth foliation of the outgoing cone, $\newsph{C}_{\hat{u}_f+\delta}$, of $\mathring{S}$ such that, on $\newsph{C}_{\hat{u}_f+\delta}$,
	\[
		(\Omega \omegahat - \Omega \omegahat_{\circ,M_{\hat{u}_f+\delta}})(\widetilde{v}, \theta)
		=
		0,
	\]
	for all $\widetilde{v} \in [v_{-1}, v(R_2, \hat{u}_f+\delta)]$, $\theta \in \mathbb{S}^2$, and,
	\[
		\partial_{\widetilde{v}} f^4 (v_{-1},\theta)
		=
		p^4(\theta),
	\]
	where $f^4$ is the diffeomorphism which relates this foliation of $\newsph{C}_{\hat{u}_f+\delta}$ to the foliation of $\newsph{C}_{\hat{u}_f+\delta}$ obtained by intersecting with the incoming cones $\Cbar_v$, for $v_{-1} \leq u \leq v(R_2,\hat{u}_f + \delta_1)$, of the extended $\hat{u}_f$ normalised $\Hp$ gauge.
\end{proposition}

\begin{proof}
	The proof is similar to that of Proposition \ref{prop:foliationIoutgoingcone}.  By Proposition \ref{prop:outHconesrelations}, the goal is to construct a function $f^4$ such that
	\[
		\frac{1}{2} \partial_{\widetilde{v}} \log(1+\partial_{\widetilde{v}} f^4)(\widetilde{v},\theta)
		=
		\Omega \omegahat_{\circ,M_{\hat{u}_f+\delta}} (\widetilde{v},\theta)
		-
		(1+\partial_{\widetilde{v}} f^4) \Omega \omegahat(\widetilde{v} + f^4,\theta),
	\]
	along with the initial conditions
	\[
		f^4(v_{-1},\cdot) = s^4,
		\qquad
		\partial_{\widetilde{v}} f^4 (v_{-1},\cdot)
		=
		p^4,
	\]
	where $\Omega \omegahat$ denotes the quantity of the foliation of $\newsph{C}_{\hat{u}_f+\delta}$ obtained by intersecting with the incoming hypersurfaces of the extended $\hat{u}_f$ normalised gauge.  Again, by continuity and compactness, this quantity is sufficiently close to its value in the extended $\hat{u}_f$ normalised gauge if $\tau$ and $\delta$ are sufficiently small.
	Define therefore the iterates $f^4_{[0]} = 0$ and, for $n \geq 1$, define $f_{[n]}^4$ to be the unique solution of the linear equation
	\begin{equation} \label{eq:foliationHoutgoingcone}
		\frac{1}{2} \partial_{\widetilde{v}} \log(1+\partial_{\widetilde{v}} f^4_{[n]})(\widetilde{v},\theta)
		=
		\Omega \omegahat_{\circ,M_{\hat{u}_f+\delta}} (\widetilde{v},\theta)
		-
		(1+\partial_{\widetilde{v}} f^4_{[n-1]}) \Omega \omegahat(\widetilde{v} + f^4_{[n-1]},\theta),
	\end{equation}
	with the initial conditions
	\[
		f^4(v_{-1},\cdot) = s^4,
		\qquad
		\partial_{\widetilde{v}} f^4 (v_{-1},\cdot)
		=
		p^4.
	\]
	As in the proof of Proposition \ref{prop:foliationIoutgoingcone} it follows inductively that each $f^4_{[n]}$ satisfies the estimates
	\[
		\sum_{k\leq 2} 
		\sup_{v\in[v_{-1},V], \theta\in \mathbb{S}^2}
		\vert \partial_v^k f^4_{[n]}(v,\theta) \vert \lesssim \tau,
	\]
	on the interval $[v_{-1}, V]$, provided $V- v_{-1}$ is sufficiently small.  It moreover follows that,
	\[
		\sum_{k\leq 2} 
		\sup_{v\in[v_{-1},V], \theta\in \mathbb{S}^2}
		\vert \partial_v^k ( f^4_{[n+1]} - f^4_{[n]})(v,\theta) \vert
		\lesssim
		\vert V - v_{-1} \vert \sum_{k\leq 2} 
		\sup_{v\in[v_{-1},V], \theta\in \mathbb{S}^2}
		\vert \partial_v^k ( f^4_{[n]} - f^4_{[n-1]})(v,\theta) \vert,
	\]
	and so, provided $V- v_{-1}$ is sufficiently small the sequence $\{f^4_{[n]}\}$ converges to a limit $f^4$ on the interval $[v_{-1},V]$, which satisfies the equation \eqref{eq:foliationHoutgoingcone}.  As in the proof of Proposition \ref{prop:foliationIoutgoingcone}, the proof follows from dividing the interval $[v_{-1},v(R_2,\hat{u}_f+\delta)]$ into sub-intervals of size $V- v_{-1}$ and repeating the proof.
\end{proof}

\subsubsection*{Foliations of the incoming null hypersurface $\newsph{\Cbar}_{v_{-1}}$}

The following proposition relates certain geometric quantities associated to two different foliations of the incoming null hypersurface $\newsph{\underline{C}}_{v_{-1}}$.

\begin{proposition}[Change of foliation relations on incoming cones] \label{prop:inHconesrelations}
	Let $u$ denote the restriction to $\newsph{\underline{C}}_{v_{-1}}$ of the $u$ coordinate of the extended $\hat{u}_f$ normalised gauge.  Consider another foliation of $\newsph{\underline{C}}_{v_{-1}}$ by spheres, described by the level hypersurfaces of a function $\widetilde{u}$.  Suppose $u$ and $\widetilde{u}$ are related by
	\begin{equation} 
		u = \widetilde{u} + f^3(\widetilde{u},\theta^1,\theta^2),
	\end{equation}
	for some function $f^3$.  Then the corresponding geometric quantities $\partial_u(r^3 \divslash \eta + r^3 \rho)$, $\Omega \tr \chibar$ and $\Omega \omegabarhat$ are related by
	\[
		\widetilde{r}^3
		\partial_{\widetilde{u}} \widetilde{\Deltaslash} \log (1+\partial_{\widetilde{u}} f^3)
		=
		\partial_{\widetilde{u}}(\widetilde{r}^3 \widetilde{\divslash \eta} + \widetilde{r}^3 \widetilde{\rho})(\widetilde{u},\theta)
		-
		\partial_u(r^3 \divslash \eta + r^3 \rho)(\widetilde{u} + f^3,\theta)
		+
		F (\widetilde{\nablaslash} f^3, \widetilde{\nablaslash}{}^2 f^3,\partial_{\widetilde{u}} f^3, \widetilde{\nablaslash} \partial_{\widetilde{u}} f^3),
	\]
	for a function $F$ which, provided $\vert \widetilde{r \nablaslash} f^3 \vert + \vert (\widetilde{r \nablaslash})^2 f^3 \vert+\vert \partial_{\widetilde{u}} f^3 \vert + \vert \widetilde{r \nablaslash} \partial_{\widetilde{u}} f^3 \vert \lesssim 1$, satisfies
	\[
		\vert F (\widetilde{\nablaslash} f^3, \widetilde{\nablaslash}{}^2 f^3,\partial_{\widetilde{u}} f^3, \widetilde{\nablaslash} \partial_{\widetilde{u}} f^3) \vert
		\lesssim
		\vert \widetilde{r \nablaslash} f^3 \vert
		+
		\vert (\widetilde{r \nablaslash})^2 f^3 \vert
		+
		\vert \partial_{\widetilde{u}} f^3 \vert
		+
		\vert \widetilde{\nablaslash} \partial_{\widetilde{u}} f^3 \vert
		,
	\]
	and
	\[
		\frac{1}{2} \partial_{\widetilde{u}} \log(1+\partial_{\widetilde{u}} f^3)(\widetilde{u},\theta)
		=
		\widetilde{\Omega \omegabarhat}(\widetilde{u},\theta)
		-
		(1+\partial_{\widetilde{u}} f^3(\widetilde{u},\theta)) \Omega \omegabarhat(\widetilde{u} + f^3,\theta).
	\]
\end{proposition}

\begin{proof}
	The proof follows exactly as that of Proposition \ref{prop:inIconesrelations}, using now the fact that
	\begin{align*}
		\partial_{\widetilde{u}} \widetilde{\divslash \eta}
		-
		\partial_u \divslash \eta
		=
		\partial_{\widetilde{u}} \widetilde{\Deltaslash} \log (1+\partial_{\widetilde{u}} f^3)
		+
		\partial_{\widetilde{u}} f^3 \divslash \eta
		+
		\partial_{\widetilde{u}}
		\Big(
		\widetilde{\nablaslash} f^3 \cdot \Omega \nablaslash_{3} \eta
		-
		\frac{1}{2}
		\widetilde{\divslash} \big(
		\Omega \chibar \cdot \widetilde{\nablaslash} f^3 \big)
		+
		2 \widetilde{\divslash} \big( \Omega \omegabarhat \widetilde{\nablaslash} f^3 \big)
		\Big),
	\end{align*}
	and,
	\[
		\partial_{\widetilde{u}} \widetilde{\rho}
		-
		\partial_{u} \rho
		=
		\partial_{\widetilde{u}}f^3 \rho
		+
		\partial_{\widetilde{u}}
		\Big(
		\widetilde{\nablaslash} f^3 \otimes \widetilde{\nablaslash} f^3 \cdot \Omega^2 \alphabar
		-
		2 \widetilde{\nablaslash} f^3 \cdot \Omega \betabar
		\Big),
	\]
	and the fact that $\partial_u (r^3 \rho_{\circ}) = \partial_{\widetilde{u}}(\widetilde{r}^3 \widetilde{\rho_{\circ}}) = 0$.
\end{proof}

Again, rather than foliate the incoming cone $\newsph{\Cbar}_{v_{-1}}$ directly by the condition that $(\Omega^2 - \Omega_{\circ}^2)_{\ell=0} = 0$, it is first shown that the incoming cone can be foliated by certain conditions which involve only the above quantities $\Omega \omegabarhat$ and $\Omega \tr \chibar$.

Recall that, in a given spacetime double null foliation,
\[
	\partial_u \big( \Omega_{\circ}^{-1} \Omega_{\ell=0} - 1 \big)
	=
	\Omega \omegabarhat_{\ell=0}
	-
	\Omega \omegabarhat_{\circ}
	-
	\underline{\mathcal{F}},
\]
where
\[
	\underline{\mathcal{F}}
	=
	-
	\big( \Omega_{\circ}^{-1} \Omega_{\ell=0} - 1 \big)
	\big( \Omega \omegabarhat_{\ell=0}
	-
	\Omega \omegabarhat_{\circ} \big)
	-
	\big(
	\Omega_{\circ}^{-1} \Omega_{\ell \geq 1} (\Omega \omegabarhat)_{\ell \geq 1}
	\big)_{\ell =0}
	-
	\big( \Omega_{\circ}^{-1} \Omega (\Omega \tr \chibar)_{\ell \geq 1} \big)_{\ell=0},
\]
and, since
\begin{equation*}
	\Omega_{\circ}^{-1} \Omega (u,v_{-1},\theta)
	=
	-
	\Omega_{\circ}^{-1} \Omega (\hat{u}_f+\delta,v_{-1},\theta)
	e^{\int_{u}^{\hat{u}_f+\delta} (\Omega \omegabarhat - \Omega \omegabarhat_{\circ})(u',v_{-1},\theta) du'},
\end{equation*}
$\underline{\mathcal{F}}$ can be viewed as a function of $u$, for any given $\Omega \omegabarhat - \Omega \omegabarhat_{\circ}$, $\Omega \tr \chibar - \Omega \tr \chibar_{\circ}$, and $\Omega_{\circ}^{-1} \Omega (u_{\hat{u}_f+\delta},v_{-1},\cdot)$, i.\@e.\@, abusing notation,
\[
	\underline{\mathcal{F}}
	=
	\underline{\mathcal{F}}
	( \Omega \omegabarhat
	-
	\Omega \omegabarhat_{\circ}
	,
	\Omega \tr \chibar - \Omega \tr \chibar_{\circ}
	,
	\Omega_{\circ}^{-1} \Omega (\hat{u}_f+\delta,v_{-1},\cdot))
	: [u_{-1}, \hat{u}_f+\delta] \to \mathbb{R}
	.
\]

In Proposition \ref{prop:Hconesfoliations} a gauge will be constructed in which $\Omega_{\circ}^{-1} \Omega (\hat{u}_f+\delta,v_{-1},\cdot) \equiv 1$, and so the following proposition uses this value of $1$ for $\Omega_{\circ}^{-1} \Omega (\hat{u}_f+\delta,v_{-1},\cdot)$.

\begin{proposition}[Foliations of incoming cones] \label{prop:foliationHincomingcone}
	Let $\mathring{S}$ be a sphere defined, in the $(u_{\Hp}, v_{\Hp},\theta_{\Hp})$ coordinate system of the extended $\hat{u}_f$ normalised $\Hp$ gauge, by smooth functions $s^3$, $s^4$,
	\[
		\mathring{S} = \{ i_{\Hp} (\hat{u}_f+\delta + s^3(\theta), v_{-1} + s^4(\theta),\theta) \mid \theta \in \mathbb{S}^2\}.
	\]
	such that $s^3$ and $s^4$ satisfy
	\[
		\sum_{k \leq 5}
		\big(
		\Vert (r \nablaslash)^k s^3 \Vert_{\mathring{S}}
		+
		\Vert (r \nablaslash)^k s^4 \Vert_{\mathring{S}}
		\big)
		\leq
		\tau.
	\]
	Consider also a smooth function $p^3 : \mathring{S} \to \mathbb{R}$ satisfying
	\[
		\sup_{\theta \in \mathbb{S}^2} \vert p^3(\theta) \vert
		\leq
		\tau.
	\]
	Then, provided $\tau$ and $\delta_0$ are sufficiently small, there exists a smooth foliation of the incoming cone, $\newsph{\underline{C}}_{v_{-1}}$, of $\mathring{S}$ such that, on $\newsph{\underline{C}}_{v_{-1}}$,
	\[
		\partial_u(r^3 \divslash \eta_{\ell \geq 1} + r^3 \rho_{\ell \geq 1})(\widetilde{u},\theta) = 0,
		\qquad
		(\Omega \omegabarhat - \Omega \omegabarhat_{\circ,M_{\hat{u}_f+\delta}})_{\ell = 0}(\widetilde{u})
		=
		\underline{\mathcal{F}}
		[
		\Omega \omegabarhat - \Omega \omegabarhat_{\circ,M_{\hat{u}_f+\delta}}
		,
		\Omega \tr \chibar - \Omega \tr \chibar_{\circ,M_{\hat{u}_f+\delta}}
		,
		1
		]
		(\widetilde{u}),
	\]
	for all $\widetilde{u} \in [u_{-1}, \hat{u}_f+\delta]$, $\theta \in \mathbb{S}^2$, and,
	\[
		\partial_{\widetilde{u}} f^3 (\hat{u}_f+\delta, \theta)
		=
		p^3(\theta),
	\]
	where $f^3$ is the diffeomorphism which relates this foliation of $\newsph{\underline{C}}_{v_{-1}}$ to the foliation of $\newsph{\underline{C}}_{v_{-1}}$ obtained by intersecting with the outgoing cones $C_u$, for $u_{-1} \leq u \leq \hat{u}_f + \delta_1$, of the extended $\hat{u}_f$ normalised gauge.
\end{proposition}

\begin{proof}
	The proof is again similar to that of Propositions \ref{prop:foliationIincomingcone}, \ref{prop:foliationIoutgoingcone} and \ref{prop:foliationHoutgoingcone}.  By Proposition \ref{prop:inHconesrelations}, the goal is to construct a function $f^3:[u_{-1},\hat{u}_f+\delta] \times \mathbb{S}^2 \to \mathbb{R}$ satisfying
	\begin{equation} \label{eq:foliationHincomingcone1}
		\Big[
		\widetilde{r}^3
		\partial_{\widetilde{u}} \widetilde{\Deltaslash} \log (1+\partial_{\widetilde{u}} f^3)
		\Big]_{\widetilde{\ell \geq 1}}
		=
		\Big[
		-
		\partial_u(r^3 \divslash \eta + r^3 \rho)(\widetilde{u} + f^3,\theta)
		+
		F'
		\Big]_{\widetilde{\ell \geq 1}},
	\end{equation}
	and
	\begin{equation} \label{eq:foliationHincomingcone2}
		\frac{1}{2}
		\big[
		\partial_{\widetilde{u}} \log(1+\partial_{\widetilde{u}} f^3)(\widetilde{u},\cdot)
		\big]_{\widetilde{\ell =0}}
		=
		\mathcal{\underline{F}}(\widetilde{u})
		+
		\big[
		\Omega \omegabarhat_{\circ}(\widetilde{u})
		-
		(1+\partial_{\widetilde{u}} f^3(\widetilde{u},\cdot)) \Omega \omegabarhat(\widetilde{u} + f^3(\widetilde{u},\cdot),\cdot)
		\big]_{\widetilde{\ell =0}},
	\end{equation}
	along with the \emph{final} conditions,
	\begin{equation} \label{eq:foliationHincomingcone3}
		f^3(\hat{u}_f+\delta,\cdot) = s^3,
		\qquad
		\partial_{\widetilde{u}} f^3 (\hat{u}_f+\delta,\cdot)
		=
		p^3,
	\end{equation}
	where
	\begin{align*}
		F'
		:=
		F(\widetilde{\nablaslash} f^3, \widetilde{\nablaslash}{}^2 f^3,\partial_{\widetilde{u}} f^3, \widetilde{\nablaslash} \partial_{\widetilde{u}} f^3)
		+
		(\partial_{\widetilde{u}}(\widetilde{r}^3 \widetilde{\divslash \eta} + \widetilde{r}^3 \widetilde{\rho}))_{\widetilde{\ell \geq 1}}
		-
		\partial_{\widetilde{u}}(\widetilde{r}^3 \widetilde{\divslash \eta} + \widetilde{r}^3 \widetilde{\rho})_{\widetilde{\ell \geq 1}}
		,
	\end{align*}
	and $F$ is as in Proposition \ref{prop:inHconesrelations}.  Note that, by Proposition \ref{prop:modesdifference} and Proposition \ref{prop:com0},
	for any function $h$,
	\[
		\vert
		(\partial_{\widetilde{u}}h)_{\widetilde{\ell \geq 1}}
		-
		\partial_{\widetilde{u}}(h_{\widetilde{\ell \geq 1}})
		\vert
		\lesssim
		\sup_{\theta\in \mathbb{S}^2}
		\vert
		h
		\vert
		\cdot
		\big(
		\vert \widetilde{\Omega \betabar} \vert
		+
		\vert \widetilde{\eta} \vert
		+
		\sum_{k \leq 1}
		(
		\vert (\widetilde{r\nablaslash})^k \widetilde{\Omega \hat{\chibar}} \vert
		+
		\vert (\widetilde{r\nablaslash})^k (\widetilde{\Omega \tr \chibar} - \widetilde{\Omega \tr \chibar}_{\circ}) \vert
		)
		\big),
	\]
	and so
	\[
		\vert
		(\partial_{\widetilde{u}}(\widetilde{r}^3 \widetilde{\divslash \eta} + \widetilde{r}^3 \widetilde{\rho}))_{\widetilde{\ell \geq 1}}
		-
		\partial_{\widetilde{u}}(\widetilde{r}^3 \widetilde{\divslash \eta} + \widetilde{r}^3 \widetilde{\rho})_{\widetilde{\ell \geq 1}}
		\vert
		\lesssim
		\varepsilon \big(
		\vert \widetilde{r \nablaslash} f^3 \vert
		+
		\vert (\widetilde{r \nablaslash})^2 f^3 \vert
		+
		\vert \partial_{\widetilde{u}} f^3 \vert
		+
		\vert \widetilde{\nablaslash} \partial_{\widetilde{u}} f^3 \vert \big),
	\]
	by relating $\widetilde{\Omega \betabar}$ etc.\@ to their values in the extended $\hat{u}_f$ normalised gauge (see Proposition \ref{prop:inIconesrelations} and Proposition \ref{prop:inHconesrelations}).  As in the proofs of Propositions \ref{prop:foliationIincomingcone}, \ref{prop:foliationIoutgoingcone} and \ref{prop:foliationHoutgoingcone}, the proof proceeds by considering the iterates $f^3_{[0]} = 0$, and inductively defining $f^3_{[n]}$ to be the solution of the linear system
	\[
		\Big[
		r_{[n-1]}^3
		\partial_{\widetilde{u}} \Deltaslash_{[n-1]} \log (1+\partial_{\widetilde{u}} f^3_{[n]})
		\Big]_{(\ell \geq 1)_{[n-1]}}
		=
		\Big[
		-
		\partial_u(r^3 \divslash \eta + r^3 \rho)(\widetilde{u} + f^3_{[n-1]},\theta)
		+
		F_{[n-1]}'
		\Big]_{(\ell \geq 1)_{[n-1]}},
	\]
	and
	\[
		\frac{1}{2}
		\big[
		\partial_{\widetilde{u}} \log(1+\partial_{\widetilde{u}} f^3_{[n]})(\widetilde{u},\cdot)
		\big]_{(\ell =0)_{[n-1]}}
		=
		\mathcal{\underline{F}}_{[n-1]}(\widetilde{u})
		+
		\big[
		\Omega \omegabarhat_{\circ}(\widetilde{u})
		-
		(1+\partial_{\widetilde{u}} f^3_{[n-1]}(\widetilde{u},\cdot)) \Omega \omegabarhat(\widetilde{u} + f^3_{[n-1]}(\widetilde{u},\cdot),\cdot)
		\big]_{(\ell =0)_{[n-1]}},
	\]
	along with the final conditions,
	\[
		f^3(\hat{u}_f+\delta,\cdot) = s^3,
		\qquad
		\partial_{\widetilde{u}} f^3 (\hat{u}_f+\delta,\cdot)
		=
		p^3,
	\]
	where 
	\[
		\underline{\mathcal{F}}_{[n-1]}
		=
		\underline{\mathcal{F}}
		( 
		\Omega \omegabarhat_{\circ}
		+
		\frac{1}{2} \partial_{u} \log(1+\partial_{u} f^3_{[n-1]})
		+
		(1+\partial_{u} f^3_{[n-1]})\Omega \omegabarhat_{f^3_{[n-1]}}
		,
		\Omega \tr \chibar_{\circ}
		+
		(1+\partial_{u} f^3_{[n-1]})
		\Omega \tr \chibar_{f^3_{[n-1]}}
		,
		q)
		,
	\]
	and
	\begin{align*}
		F_{[n-1]}'
		=
		\
		&
		F(\nablaslash f^3_{[n-1]}, \nablaslash^2 f^3_{[n-1]},\partial_{u} f^3_{[n-1]}, \nablaslash \partial_{u} f^3_{[n-1]})
		\\
		&
		+
		(\partial_{u}( (r^3 \divslash \eta + r^3 \rho)_{[n-1]}))_{(\ell \geq 1)_{[n-1]}}
		-
		\partial_{u}( ((r^3 \divslash \eta + r^3 \rho)_{[n-1]})_{(\ell \geq 1)_{[n-1]}}).
	\end{align*}
	As in the proofs of Propositions \ref{prop:foliationIincomingcone}, \ref{prop:foliationIoutgoingcone} and \ref{prop:foliationHoutgoingcone}, one then shows that the sequence of iterates $\{f^3_{[n]}\}$ converges to a limit $f^3$ which solves \eqref{eq:foliationHincomingcone1}--\eqref{eq:foliationHincomingcone3} as desired.
\end{proof}

\subsubsection*{The proof of Proposition \ref{prop:Hconesfoliations}: foliations of cones of any sphere close to $S_{\hat{u}_f+\delta,v_{-1}}$}

The proof of Proposition \ref{prop:Hconesfoliations} can now be given.

\begin{proof}[Proof of Proposition \ref{prop:Hconesfoliations}]
	\underline{\textbf{Equations:}}	
	The goal is to choose functions $p^3$, $p^4$ appropriately so that, in the spacetime double null foliation arising from the foliations of $\newsph{C}_{\hat{u}_f+\delta}$ and $\newsph{\underline{C}}_{v_{-1}}$ of Propositions \ref{prop:foliationHoutgoingcone} and \ref{prop:foliationHincomingcone} respectively,
	\begin{align}
		\Omega(\hat{u}_f+\delta,v_{-1},\cdot)^2
		-
		\Omega_{\circ,M_{\hat{u}_f+\delta}}(\hat{u}_f+\delta,v_{-1})^2
		&
		= 
		0,
		\label{eq:Hfoliationconstruction1}
		\\
		\mu^*_{\ell \geq 1} (\hat{u}_f+\delta,v(R,\hat{u}_f+\delta),\cdot) 
		&
		= 
		0,
		\\
		\big(\Omega \tr \chibar - (\Omega \tr \chibar)_{\circ,M_{\hat{u}_f+\delta}} \big)_{\ell = 0} (\hat{u}_f+\delta, v_{-1})
		&
		=
		0.
	\end{align}
	The function $q$ will then be chosen so that, in the resulting spacetime double null foliation, the function $\underline{\mathcal{F}}$ of Proposition \ref{prop:foliationHincomingcone} satisfies
	\begin{equation*}
		\underline{\mathcal{F}}
		=
		\partial_u \big( \big( \Omega^2 - \Omega_{\circ,M_{\hat{u}_f+\delta}}^{2} \big)_{\ell=0}\big)
		-
		(\Omega \omegabarhat - \Omega \omegabarhat_{\circ,M_{\hat{u}_f+\delta}})_{\ell=0}.
	\end{equation*}

	Suppose that, for some given $p^4$, the cone $\newsph{C}_{\hat{u}_f+\delta}$ is foliated according to Proposition \ref{prop:foliationHoutgoingcone} and, for some given $p^3$, $q$, the cone $\newsph{\underline{C}}_{v_{-1}}$, is foliated according to Proposition \ref{prop:foliationHincomingcone}.  Suppose that these foliations are described as the level hypersurfaces of functions $\newsph{v}$ and $\newsph{u}$ respectively.  The resulting spacetime double null foliation, obtained by considering the outgoing null cones associated to the spheres of constant $\newsph{u}$ in $\newsph{\Cbar}_{v_{-1}}$ and the incoming null cones associated to the spheres of constant $\newsph{v}$ in $\newsph{C}_{\hat{u}_f+\delta}$, is related to the extended $\hat{u}_f$ normalised double null foliation by functions $f^3$, $f^4$ satisfying
	\begin{equation*}
		u = \newsph{u} + f^3(\newsph{u},\newsph{v},\newsph{\theta}),
		\qquad
		v = \newsph{v} + f^4(\newsph{u},\newsph{v},\newsph{\theta}),
		\qquad
		\theta^A = \newsph{\theta}^A + f^A(\newsph{u},\newsph{v},\newsph{\theta}),
	\end{equation*}
	where $f^3(\cdot,v_{-1},\cdot)$ agrees with $f^3$ of Proposition \ref{prop:foliationHincomingcone}, $f^4(\hat{u}_f+\delta,\cdot,\cdot)$ agrees with $f^4$ of Proposition \ref{prop:foliationHoutgoingcone},
	and $f^1$, $f^2$ are defined so that
	\[
		f^A (\hat{u}_f+\delta,v_{-1},\cdot) \equiv 0,
		\qquad
		A = 1,2,
	\]
	and the spacetime metric takes the standard double null form with $\newsph{b} \equiv 0$ on the cone $\newsph{C}_{\hat{u}_f+\delta}$.  In particular,
	\[
		 f^3(\hat{u}_f+\delta,v_{-1},\cdot)
		 =
		 s^3,
		\qquad
		f^4(\hat{u}_f+\delta,v_{-1},\cdot)
		=
		s^4.
	\]
	The change of spacetime gauge relations \eqref{eq:metriccomp6}, \eqref{eq:Riccicomp3}, \eqref{eq:Riccicomp7}, \eqref{eq:Riccicomp8}, \eqref{eq:curvaturecomp5} then imply that
	\[
		( \newsph{\Omega}_{\circ,M}^{-2} \newsph{\Omega}^2 - 1)(\hat{u}_f+\delta,v_{-1})
		=
		( \Omega_{\circ,M}^{-1} \Omega^2 - 1)(\hat{u}_f+\delta+s^3,v_{-1}+s^4)
		+
		p^3 + p^4
		-
		\frac{2M}{r^2} (s^3-s^4)
		+
		\mathfrak{E}^1,
	\]
	\begin{multline*}
		\big( \newsph{\mu}^* + 3\newsph{\Omega}^2\newsph{r}^{-2} ( \newsph{\Omega}_{\circ,M}^{-2} \newsph{\Omega}^2 - 1) \big)_{\ell \geq 1}(\hat{u}_f+\delta,v_{-1})
		=
		\big( \mu^* + 3\Omega^2r^{-2} ( \Omega_{\circ,M}^{-2} \Omega^2 - 1) \big)(\hat{u}_f+\delta + s^3,v_{-1}+s^4)_{\ell \geq 1}
		\\
		+
		\Deltaslash p^3_{\ell \geq 1}
		+
		\frac{3\Omega^2}{r^2} p^3_{\ell \geq 1}
		-
		\frac{2}{r} \left( 1 - \frac{M}{r} \right) \Deltaslash s^3_{\ell \geq 1}
		+
		\frac{\Omega^2}{r} \Deltaslash s^4_{\ell \geq 1}
		-
		\frac{3\Omega^2}{r^3} (s^3 - s^4)_{\ell \geq 1}
		+
		\mathfrak{E}^2,
	\end{multline*}
	and
	\begin{multline*}
		\newsph{\Omega}_{\circ,M}^{-2} (\newsph{\Omega \tr \chibar} - \newsph{\Omega \tr \chibar}_{\circ,M})_{\ell = 0}(\hat{u}_f+\delta,v_{-1})
		=
		\Omega_{\circ}^{-2} (\Omega \tr \chibar - \Omega \tr \chibar_{\circ,M})(\hat{u}_f+\delta + s^3,v_{-1}+s^4)_{\ell = 0}
		\\
		-
		\frac{2}{r} p^3_{\ell = 0}
		-
		\frac{2}{r^2} \left( 1 - \frac{4M}{r} \right) (s^3-s^4)_{\ell = 0}
		+
		\mathfrak{E}^3,
	\end{multline*}
	where $M = M_{\hat{u}_f+\delta}$ and $\mathfrak{E}^1$, $\mathfrak{E}^2$, $\mathfrak{E}^3$ are nonlinearities, as determined by the relations of Proposition \ref{prop:metricrelations} and Proposition \ref{prop:Riccirelations}, which take the schematic form 
	\[
		\mathfrak{E}^1
		=
		\sum_{\vert \gamma_1 \vert, \vert \gamma_2 \vert \leq 1}
		\Pi_{\mathring{S}} \Phi \cdot \newsph{\mathfrak{D}}^{\gamma_1} f(x_{-1})
		+
		\newsph{\mathfrak{D}}^{\gamma_1} f \cdot \newsph{\mathfrak{D}}^{\gamma_2} f(x_{-1}),
	\]
	\[
		\mathfrak{E}^2
		=
		\sum_{k \leq2 }
		\sum_{\vert \gamma_1 \vert, \vert \gamma_2 \vert \leq 1}
		\Big(
		\Pi_{\mathring{S}} \Phi \cdot (r\newsph{\nablaslash})^k \newsph{\mathfrak{D}}^{\gamma_1} f(x_{-1})
		+
		\newsph{\mathfrak{D}}^{\gamma_1} f \cdot (r\newsph{\nablaslash})^k \newsph{\mathfrak{D}}^{\gamma_2} f(x_{-1}) \Big)_{\ell\geq 1},
	\]
	\[
		\mathfrak{E}^3
		=
		\sum_{k \leq1}
		\sum_{\vert \gamma_1 \vert, \vert \gamma_2 \vert \leq 1}
		\Big(
		\Pi_{\mathring{S}} \Phi \cdot (r\newsph{\nablaslash})^k \newsph{\mathfrak{D}}^{\gamma_1} f(x_{-1})
		+
		\newsph{\mathfrak{D}}^{\gamma_1} f \cdot (r\newsph{\nablaslash})^k \newsph{\mathfrak{D}}^{\gamma_2} f(x_{-1}) \Big)_{\ell=0},
	\]
	with $x_{-1} = (\hat{u}_f+\delta,v_{-1},\cdot)$, and $\newsph{\Omega}$ and $\newsph{\Omega \tr \chi}$ denote the values of $\Omega$ and $\Omega \tr \chi$ in the resulting spacetime double null foliation.
	
	Under the assumption that the gauge condition \eqref{eq:Hfoliationconstruction1} is satisfied, Proposition \ref{prop:nabla4mustaruf} in particular implies that, after relating (see Proposition \ref{prop:outIconesrelations} or Proposition \ref{prop:outHconesrelations}) the quantities $\newsph{\Omega \tr \chi} - \newsph{\Omega \tr \chi}_{\circ}$, $\newsph{\Omega \hat{\chi}}$, $\eta$, $\newsph{\Omega^{-1} \hat{\chibar}}$, $\newsph{\rho}$, $\newsph{\Omega \beta}$, $\newsph{\Omega^2 \alpha}$ to the corresponding quantities in the extended $\hat{u}_f$ normalised gauge and $f^4$ and its derivatives,
	\[
		\mu^*_{\ell \geq 1} (\hat{u}_f + \delta,v_{-1})
		=
		\mu^*_{\ell \geq 1} (\hat{u}_f + \delta,v(R,\hat{u}_f + \delta))
		+
		\mathcal{G}(
		f^4, \nablaslash f^4, \nablaslash^2 f^4,\partial_v f^4, \nablaslash \partial_v f^4, \nablaslash^2 \partial_v f^4)
		,
	\]
	for some appropriate $\mathcal{G}$, which depends on $\Omega \tr \chi - \Omega \tr \chi_{\circ}$, $\eta$, $\Omega^2 \alpha$, etc.\@ in the extended $\hat{u}_f$ normalised gauge, and satisfies
	\[
		\Vert \mathcal{G} \Vert_S
		\lesssim
		\varepsilon^2
		+
		\hat{u}_f
		\sup_{v_{-1} \leq v \leq v(R,\hat{u}_f+\delta)}
		\sum_{k \leq 2}
		(\Vert (r \nablaslash)^k f^4 \Vert_{S_{\hat{u}_f+\delta,v}}^2 + \Vert (r \nablaslash)^k \partial_v f^4 \Vert_{S_{\hat{u}_f+\delta,v}}^2)
		.
	\]

	The goal then is to find $p^3$ and $p^4$ such that
	\begin{equation} \label{eq:Hgaugepdef1}
		0
		=
		( \Omega_{\circ}^{-1} \Omega^2 - 1)(\hat{u}_f+\delta+s^3,v_{-1}+s^4)
		+
		p^3 + p^4
		-
		\frac{2M}{r^2} (s^3-s^4)
		+
		\mathfrak{E}^1,
	\end{equation}
	\begin{multline} \label{eq:Hgaugepdef2}
		\mathcal{G}(
		f^4, \nablaslash f^4, \nablaslash^2 f^4,\partial_v f^4, \nablaslash \partial_v f^4, \nablaslash^2 \partial_v f^4)
		=
		\big( \mu^* + 3r^{-2} (\Omega^2 - \Omega_{\circ}^2) \big)(\hat{u}_f+\delta + s^3,v_{-1}+s^4)_{\ell \geq 1}
		\\
		+
		\Deltaslash p^3_{\ell \geq 1}
		+
		\frac{3\Omega^2_{\circ}}{r^2} p^3_{\ell \geq 1}
		-
		\frac{2}{r} \left( 1 - \frac{M}{r} \right) \Deltaslash s^3_{\ell \geq 1}
		+
		\frac{\Omega^2_{\circ}}{r} \Deltaslash s^4_{\ell \geq 1}
		-
		\frac{3\Omega^2_{\circ}}{r^3} (s^3 - s^4)_{\ell \geq 1}
		+
		\mathfrak{E}^2,
	\end{multline}
	and
	\begin{equation} \label{eq:Hgaugepdef3}
		0
		=
		\Omega_{\circ}^{-2} (\Omega \tr \chibar - \Omega \tr \chibar_{\circ})(\hat{u}_f+\delta + s^3,v_{-1}+s^4)_{\ell = 0}
		-
		\frac{2}{r} p^3_{\ell = 0}
		-
		\frac{2}{r^2} \left( 1 - \frac{4M}{r} \right) (s^3-s^4)_{\ell = 0}
		+
		\mathfrak{E}^3,
	\end{equation}
	where $f^4$ denotes the change of foliation of the cone arising from Proposition \ref{prop:foliationHoutgoingcone} with $\partial_v f^4 (\hat{u}_f+\delta,v_{-1}) = p^4$ and $f^3$ denotes the change of foliation of the cone arising from Proposition \ref{prop:foliationHincomingcone} with $\partial_u f^3 (\hat{u}_f+\delta,v_{-1}) = p^3$.

	\noindent \underline{\textbf{Definition of iterates:}}
	Define first
	\[
		p^3_{[1]} = p^4_{[1]} = 0,
	\]
	and, for $n \geq 2$ inductively define $p^3_{[n]}$ and $p^4_{[n]}$ as solutions of the system
	\begin{multline} \label{eq:Hgaugep1}
		\big[ \Deltaslash_{[n-1]} p^3_{[n]}
		+
		3\Omega^2_{\circ}r^{-2} p^3_{[n]} \big]_{(\ell \geq 1)_{[n-1]}}
		=
		\big[
		\mathcal{G}_{[n-1]}
		-
		\big( \mu^* + 3r^{-2} (\Omega^2 - \Omega_{\circ}^2) \big)(\hat{u}_f+\delta + s^3,v_{-1}+s^4)
		\\
		+
		\frac{2}{r} \left( 1 - \frac{M}{r} \right) \Deltaslash s^3
		-
		\frac{\Omega^2_{\circ}}{r} \Deltaslash s^4
		+
		\frac{3\Omega^2_{\circ}}{r^3} (s^3 - s^4)
		-
		\mathfrak{E}^2_{[n-1]}
		\big]_{(\ell \geq 1)_{[n-1]}}
	\end{multline}
	\begin{equation} \label{eq:Hgaugep2}
		\big[ 2 r^{-1} p^3_{[n]} \big]_{(\ell =0)_{[n-1]}}
		=
		\big[
		\Omega_{\circ}^{-2} (\Omega \tr \chibar - \Omega \tr \chibar_{\circ})(\hat{u}_f+\delta + s^3,v_{-1}+s^4)
		-
		\frac{2}{r^2} \left( 1 - \frac{4M}{r} \right) (s^3-s^4)
		+
		\mathfrak{E}^3_{[n-1]}
		\big]_{(\ell =0)_{[n-1]}},
	\end{equation}
	and
	\begin{equation} \label{eq:Hgaugep3}
		p^4_{[n]}
		=
		-
		p^3_{[n]}
		-
		( \Omega_{\circ}^{-1} \Omega^2 - 1)(\hat{u}_f+\delta+s^3,v_{-1}+s^4)
		+
		\frac{2M}{r^2} (s^3-s^4)
		-
		\mathfrak{E}^1_{[n-1]},
	\end{equation}
	where
	\[
		\mathcal{G}_{[n-1]}
		=
		\mathcal{G}(
		f^4_{[n-1]}, \nablaslash f^4_{[n-1]}, \nablaslash^2 f^4_{[n-1]},\partial_v f^4_{[n-1]}, \nablaslash \partial_v f^4_{[n-1]}, \nablaslash^2 \partial_v f^4_{[n-1]}),
	\]
	and the errors $\mathfrak{E}^1_{[n-1]}$, $\mathfrak{E}^2_{[n-1]}$ and $\mathfrak{E}^3_{[n-1]}$ are defined by replacing $f$ with $f_{[n-1]}$ in the definition of $\mathfrak{E}^1$, $\mathfrak{E}^2$ and $\mathfrak{E}^3$ respectively.  Schematically,
	\[
		\mathfrak{E}^1_{[n-1]}
		=
		\sum_{\vert \gamma_1 \vert, \vert \gamma_2 \vert \leq 1}
		\Pi_{\mathring{S}} \Phi \cdot \mathfrak{D}^{\gamma_1} f_{[n-1]}(x_{-1})
		+
		\mathfrak{D}^{\gamma_1} f_{[n-1]} \cdot \mathfrak{D}^{\gamma_2} f_{[n-1]}(x_{-1}),
	\]
	and similarly for $\mathfrak{E}^2_{[n-1]}$ and $\mathfrak{E}^3_{[n-1]}$.  The diffeomorphisms $f^3_{[n]}$ and $f^4_{[n]}$ are inductively defined as follows.  Suppose $n$ is such that $p^3_{[n]}$ and $p^4_{[n]}$ are defined.  Consider the foliation of $\newsph{\Cbar}_{v_{\infty}}$ arising from Proposition \ref{prop:foliationHincomingcone} with $p^3 = p^3_{[n]}$.  This foliation of $\newsph{\Cbar}_{v_{\infty}}$, described by the level hypersurfaces of a function $u_{[n]}$, is related to $u$ by a function $f^3_{[n]}$ by
	\[
		u = u_{[n]} + f^3_{[n]}(u_{[n]},\theta).
	\]
	Consider the foliation of $C_{u_{-1}}$ arising from Proposition \ref{prop:foliationHoutgoingcone} with $p^4 = p^4_{[n]}$.  This foliation of $C_{u_{-1}}$, described by the level hypersurfaces of a function $v_{[n]}$, is related to $v$ by a function $f^4_{[n]}$ by
	\[
		v = v_{[n]} + f^4_{[n]}(v_{[n]},\theta).
	\]

	\noindent \underline{\textbf{Estimates and existence of limits:}}
	It follows from Proposition \ref{prop:outHconesrelations} that
	\[
		\frac{1}{2} \partial_{\widetilde{v}} \log(1+\partial_{\widetilde{v}} f^4_{[n]})(\widetilde{v},\theta)
		=
		\Omega \omegahat_{\circ,M_{\hat{u}_f+\delta}}(\widetilde{v})
		-
		(1+\partial_{\widetilde{v}} f^4_{[n]}) \Omega \omegahat(\widetilde{v} + f^4_{[n]},\theta),
	\]
	for all $n$, and so
	\[
		\sup_{v_{-1} \leq v \leq v(R,\hat{u}_f+\delta)}
		\sum_{k_1 \leq 4}
		\sum_{k_2 \leq 3}
		\Vert (r\nablaslash)^{k_1} \partial_v^{k_2} f^4_{[n]} \Vert_{S_{\hat{u}_f+\delta,v}}
		\lesssim
		C(\hat{u}_f)(\delta + \tau + \sum_{k \leq 4} \Vert (r\nablaslash)^kp^4_{[n]} \Vert_{S_{\hat{u}_f+\delta,v_{-1}}}),
	\]
	for a constant $C(\hat{u}_f)$ which depends on $\hat{u}_f$.  The relations \eqref{eq:metriccomp1}--\eqref{eq:metriccomp6} moreover imply that, if $\tau$ and $\delta_0$ are sufficiently small,
	\begin{align*}
		&
		\sum_{k \leq 4}
		\big(
		\Vert (r \nablaslash)^k \partial_u f^3_{[n]} \Vert_{S_{\hat{u}_f + \delta,v_{-1}}}
		+
		\Vert (r \nablaslash)^k \partial_v f^3_{[n]} \Vert_{S_{\hat{u}_f + \delta,v_{-1}}}
		+
		\Vert (r \nablaslash)^k \partial_u f^4_{[n]} \Vert_{S_{\hat{u}_f + \delta,v_{-1}}}
		+
		\Vert (r \nablaslash)^k \partial_v f^4_{[n]} \Vert_{S_{\hat{u}_f + \delta,v_{-1}}}
		\big)
		\nonumber
		\\
		&
		\qquad
		\lesssim
		\sum_{k \leq 5}
		\big(
		\Vert (r \nablaslash)^k s^3 \Vert_{S_{\hat{u}_f + \delta,v_{-1}}}
		+
		\Vert (r \nablaslash)^k s^4 \Vert_{S_{\hat{u}_f + \delta,v_{-1}}}
		\big)
		+
		\sum_{k \leq 4}
		\big(
		\Vert (r \nablaslash)^k p^3_{[n]} \Vert_{S_{\hat{u}_f + \delta,v_{-1}}}
		+
		\Vert (r \nablaslash)^k p^4_{[n]} \Vert_{S_{\hat{u}_f + \delta,v_{-1}}}
		\big)
		,
	\end{align*}
	since $f^A_{[n]}(u_{-1},v_{\infty},\cdot) = 0$ and $b_{[n]}(u_{-1},v_{\infty},\cdot) = 0$.
	
	It then follows, using a simple induction argument and an elliptic estimate for the system \eqref{eq:Hgaugep1}--\eqref{eq:Hgaugep3}, that, if $\delta$ and $\tau$ are sufficiently small,
	\[
		\sum_{k \leq 4}
		(
		\Vert (r\nablaslash)^k p^3_{[n]} \Vert_{S_{\hat{u}_f+\delta,v_{-1}}}
		+
		\Vert (r\nablaslash)^k p^4_{[n]} \Vert_{S_{\hat{u}_f+\delta,v_{-1}}}
		)
		\lesssim
		\tau + \delta + \varepsilon,
	\]
	for all $n$.  Similarly, by considering differences of the system \eqref{eq:Hgaugep1}--\eqref{eq:Hgaugep3}, if $\tau$, $\varepsilon$ and $\delta$ are sufficiently small,
	\begin{multline*}
		\sum_{k \leq 4}
		(
		\Vert (r\nablaslash)^k (p^3_{[n+1]} - p^3_{[n]}) \Vert_{S_{\hat{u}_f+\delta,v_{-1}}}
		+
		\Vert (r\nablaslash)^k (p^4_{[n+1]} - p^4_{[n]}) \Vert_{S_{\hat{u}_f+\delta,v_{-1}}}
		)
		\\
		\leq
		\frac{1}{2}
		\sum_{k \leq 4}
		(
		\Vert (r\nablaslash)^k (p^3_{[n]} - p^3_{[n-1]}) \Vert_{S_{\hat{u}_f+\delta,v_{-1}}}
		+
		\Vert (r\nablaslash)^k (p^4_{[n]} - p^4_{[n-1]}) \Vert_{S_{\hat{u}_f+\delta,v_{-1}}}
		).
	\end{multline*}
	It follows that the sequences $\{p^3_{[n]} \}$, $\{p^4_{[n]} \}$ converge to limits $p^3$, $p^4$ respectively which solve the system \eqref{eq:Hgaugepdef1}--\eqref{eq:Hgaugepdef3} as desired.

\end{proof}

\begin{remark} \label{rmk:linearHgaugefoliations}
	Note again (cf.\@ Remark \ref{rmk:linearIgaugefoliations} concerning the linearised $\I$ gauge) that in the linear setting of Proposition \ref{proplinHpgauge},  one can distill from the proof of Proposition \ref{prop:Hconesfoliations} the analogous linear statement, namely that, for any
	solution $\tilde{\mathscr{S}}$ of the linearised equations around Schwarzschild of mass $M_f$,
	there exist functions $f^3(u,\theta)$, $f^4(v,\theta)$, with $f^3(u_f,\cdot) =0$, $f^4(v_{\infty},\cdot)= 0$, which generate a pure gauge solution $\mathscr{G}_{\mathrm{foliations}}$, such that the solution $\tilde{\mathscr{S}}+\mathscr{G}_{\mathrm{foliations}}$ 
	of the linearised equations
	satisfies the linearised analogues of \eqref{eq:Hfoliation1}--\eqref{eq:Hfoliation5}.  Indeed, in the notation of~\cite{holzstabofschw}, by the linearised analogues of the relations of Proposition \ref{prop:outHconesrelations}, Proposition \ref{prop:inHconesrelations} and the the relations \eqref{eq:metriccomp6}, \eqref{eq:Riccicomp3}, \eqref{eq:Riccicomp7}, and \eqref{eq:curvaturecomp5}, it suffices to find $f^3(u,\theta)$ satisfying the equations
	\begin{align*}
		\partial_u
		\Big(
		r^3 \Deltaslash \partial_u f^3_{\ell \geq 1}
		+
		\big( 1 - \frac{4M_f}{r} \big) r^2 \Deltaslash f^3_{\ell \geq 1}
		+
		r^2 \Omega^2 \Deltaslash f^4_{\ell \geq 1}
		-
		\frac{6M_f \Omega^2}{r} (f^3_{\ell \geq 1} - f^4_{\ell \geq 1})
		\Big)
		&
		=
		-
		\big(
		\partial_u(r^3
		\divslash \elin
		+
		r^3\rlin)
		\big)_{\ell \geq 1}^{\tilde{\mathscr{S}}} (\cdot, v_{-1},\cdot),
		\\
		\frac{1}{2} \partial_u^2 f^3_{\ell =0}
		-
		\frac{M_f}{r^2} \partial_u f^3_{\ell = 0}
		-
		\frac{2M_f\Omega_{\circ}^2}{r^3} f^3_{\ell = 0}
		&
		=
		-
		\olinb_{\ell =0}^{\tilde{\mathscr{S}}} (\cdot,v_{-1})
		,
	\end{align*}
	and $f^4(v,\theta)$ satisfying the equation
	\begin{align*}
		\frac{1}{2} \partial_v^2 f^4
		+
		\frac{M_f}{r^2} \partial_v f^4
		-
		\frac{2M_f\Omega_{\circ}^2}{r^3} f^4
		=
		-
		\olin^{\tilde{\mathscr{S}}} (u_f,\cdot,\cdot),
	\end{align*}
	along with the final conditions
	\begin{align*}
		f^3(u_f,\cdot) = f^4(v_{-1},\cdot)
		&
		= 0,
	\\
		\partial_u f^3 (u_f,\cdot) + \partial_v f^4(v_{-1},\cdot)
		&
		=
		-
		2 \Olin^{\tilde{\mathscr{S}}} (u_f,v_{-1},\cdot),
		\\
		\Deltaslash \partial_u f^3_{\ell\geq 1} (u_f,\cdot)
		+
		\frac{3\Omega^2}{r^2} \partial_u f^3_{\ell\geq 1} (u_f,\cdot)
		&
		=
		-
		\big(
		\divslash \elin
		+
		\rlin
		-
		\frac{3}{2r} \otx
		+
		\frac{3\Omega^2}{r^2} 2 \Olin
		\big)_{\ell \geq 1}^{\tilde{\mathscr{S}}} 
		(u_f,v_{\infty},\cdot),
		\\
		\partial_u f^3_{\ell=0} (u_f)
		&
		=
		\frac{r}{2\Omega^2} \otx_{\ell = 0}^{\tilde{\mathscr{S}}} (u_f,v_{\infty}),
	\end{align*}
	where the latter three are the linear analogues of \eqref{eq:Hgaugepdef1}, \eqref{eq:Hgaugepdef2} and \eqref{eq:Hgaugepdef3}.  In this simplified linear setting, the proofs of Propositions \ref{prop:foliationHincomingcone}, \ref{prop:foliationHoutgoingcone}, and \ref{prop:Hconesfoliations} reduce to showing the existence of functions $f^3(u,\theta)$ and $f^4(v,\theta)$ satisfying these linear equations.
\end{remark}

Recall the quantity
\[
		\Upsilon
		=
		\left(1 - \frac{3M_f}{r} \right) (\rho - \rho_{\circ})
		+
		\frac{3M_f}{2r^2} \left( \Omega \tr \chi - (\Omega \tr \chi)_{\circ} \right)
		-
		\frac{3M_f\Omega_{\circ}^2}{2r^2} \Omega^{-2} \left( \Omega \tr \chibar - (\Omega \tr \chibar)_{\circ} \right).
	\]
The following lemma expresses $\big( \Omega \tr \chi - (\Omega \tr \chi)_{\circ} \big)_{\ell=0}$ in terms of $\big( \Omega \tr \chibar - (\Omega \tr \chibar)_{\circ} \big)_{\ell=0}$, $\Upsilon_{\ell=0}$ and nonlinear terms for any sphere as in Proposition \ref{prop:Hconesfoliations}, in the double null foliation of Proposition \ref{prop:Hconesfoliations}.  This expression will be used in the proof of Theorem \ref{thm:Hgaugeexistence}.

\begin{lemma}[Expression for $\tr \chi$ in gauge of Proposition \ref{prop:Hconesfoliations}]
	\label{lem:trchipartialHgauge}
	Given any sphere, as in Proposition \ref{prop:Hconesfoliations}, in the double null foliation of Proposition \ref{prop:Hconesfoliations},
	\begin{equation} \label{eq:trchipartialHgauge1}
		\big( \Omega \tr \chi - (\Omega \tr \chi)_{\circ} \big)_{\ell=0}(\hat{u}_f+\delta,v(R))
		=
		A
		\big( \Omega \tr \chibar - (\Omega \tr \chibar)_{\circ} \big)_{\ell=0}(\hat{u}_f+\delta,v(R))
		+
		B
		\Upsilon_{\ell=0}(\hat{u}_f+\delta,v(R))
		+
		\mathfrak{E},
	\end{equation}
	for some constants $A$, $B$ (where $A \neq 0$ and $A \neq 1$) where $v(R) = v(R,\hat{u}_f+\delta)$
	and $\mathfrak{E}$ is a nonlinear error which satisfies
	\[
		\vert \mathfrak{E} \vert
		\lesssim
		\sum_{\Phi} \int_{v_{-1}}^{v(R)} \vert \Phi(\hat{u}_f+\delta,v) \vert^2 dv.
	\]
	Moreover, in the extended $\hat{u}_f$ normalised $\Hp$ gauge,
	\begin{equation} \label{eq:trchipartialHgauge2}
		\big( \Omega \tr \chi - (\Omega \tr \chi)_{\circ} \big)_{\ell=0}(\hat{u}_f+\delta,v(R))
		=
		A
		\big( \Omega \tr \chibar - (\Omega \tr \chibar)_{\circ} \big)_{\ell=0}(\hat{u}_f+\delta,v(R))
		+
		B
		\Upsilon_{\ell=0}(\hat{u}_f+\delta,v(R))
		+
		\mathfrak{F},
	\end{equation}
	where
	\begin{multline*}
		\vert \mathfrak{F} \vert
		\lesssim
		\vert \big( \Omega \tr \chibar - (\Omega \tr \chibar)_{\circ} \big)_{\ell=0}(\hat{u}_f+\delta,v_{-1}) \vert
		+
		\int_{v_{-1}}^{v(R)} \Omega^2 (\vert (1 - \Omega^{-2}\Omega_{\circ}^2)_{\ell=0} \vert
		+
		\vert (\Omega \omegahat - \Omega \omegahat_{\circ})_{\ell=0} \vert
		)(\hat{u}_f+\delta,v) dv
		\\
		+
		\sum_{\Phi} \int_{v_{-1}}^{v(R)} \vert \Phi(\hat{u}_f+\delta,v) \vert^2 dv.
	\end{multline*}
\end{lemma}

\begin{proof}
	Consider first \eqref{eq:trchipartialHgauge1}.  Note that, on $u=\hat{u}_f+\delta$,
	\[
		\partial_v \Big( \frac{r^2}{\Omega^2}\big( \Omega \tr \chi - (\Omega \tr \chi)_{\circ} \big)_{\ell=0} \Big)
		=
		\Omega^{-2} \mathcal{E},
		\qquad
		\partial_v(r^3 \mu^*_{\ell=0}) = \mathcal{E},
	\]
	and
	\begin{align*}
		\partial_v
		\left( r
		\left(
		\Omega \tr \chibar - (\Omega \tr \chibar)_{\circ}
		\right)_{\ell=0}
		\right)
		&
		=
		4 \Omega^2
		\left(
		\Omega \tr \chi - (\Omega \tr \chi)_{\circ}
		\right)_{\ell=0}
		+
		2 \Omega^2 r\mu^*_{\ell=0}
		+
		\Omega^2 \mathcal{E}
		\\
		&
		=
		\frac{4\Omega^4}{r^2}
		\frac{r^2}{\Omega^2}
		\left(
		\Omega \tr \chi - (\Omega \tr \chi)_{\circ}
		\right)_{\ell=0}
		+
		\frac{2\Omega^2}{r^2} r^3 \mu^*_{\ell=0}
		+
		\Omega^2 \mathcal{E}.
	\end{align*}
	Since $\left( \Omega \tr \chibar - (\Omega \tr \chibar)_{\circ} \right)_{\ell=0}(v_{-1}) = 0$, it follows that
	\[
		\left(
		\Omega \tr \chibar - (\Omega \tr \chibar)_{\circ}
		\right)_{\ell=0}(v(R))
		=
		\int_{v_{-1}}^{v(R)}\frac{4\Omega^4}{r^2} dv
		\frac{R}{\Omega^2}
		\left(
		\Omega \tr \chi - (\Omega \tr \chi)_{\circ}
		\right)_{\ell=0}(v(R))
		+
		\int_{v_{-1}}^{v(R)} \frac{2\Omega^2}{r^2} dv R^2 \mu^*_{\ell=0}(v(R))
		+
		\mathcal{E}.
	\]
	The proof then follows from rewriting $\mu^*_{\ell=0}(v(R))$ in terms of $\Omega \tr \chi - (\Omega \tr \chi)_{\circ}$, $\Omega \tr \chibar - (\Omega \tr \chibar)_{\circ}$ and $\Upsilon$.  The constants $A$, $B$ take the explict form,
	\[
		A = C^{-1} R \Big( \Omega(v(R))^{-2} I_1 - \frac{3}{2} \Big( 1- \frac{3M_f}{R} \Big)^{-1} I_2 \Big),
		\qquad
		B = C^{-1} R^2 \Big( 1- \frac{3M_f}{R} \Big)^{-1} I_2,
	\]
	\[
		C= 1 - \frac{3M_f}{2} \Big( 1- \frac{3M_f}{R} \Big)^{-1} I_2,
		\qquad
		I_1 = \int_{v_{-1}}^{v(R)} \frac{4 \Omega^4}{r^2} dv,
		\qquad
		I_2 = \int_{v_{-1}}^{v(R)} \frac{2 \Omega^2}{r^2} dv,
	\]
	where $v(R) = v(R,\hat{u}_f+\delta)$.
	
	The equality \eqref{eq:trchipartialHgauge2} follows similarly.  The fact that the defining conditions of the $\Hp$ gauge hold for $u=\hat{u}_f$, rather than $\hat{u}_f+\delta$, means there are extra contributions to the error $\mathfrak{F}$.
\end{proof}

\subsubsection{Existence of a new good sphere: the proof of Theorem \ref{thm:Hgaugeexistence}}
\label{subsec:Hsphereexistence}

The remaining defining conditions of the $\hat{u}_f+\delta$ normalised $\Hp$ gauge, namely,
\begin{align}
	\left( \Omega \tr \chibar - (\Omega \tr \chibar)_{\circ,M_{\hat{u}_f+\delta}} \right)_{\ell \geq 1}(\hat{u}_f + \delta,v_{-1})
	&
	=
	0
	\text{ for all } \theta \in \mathbb S^2;
	\label{eq:Hextragauge1}
	\\
	\left( \Omega \tr \chi - (\Omega \tr \chi)_{\circ,M_{\hat{u}_f+\delta}} \right)_{\ell=0}(\hat{u}_f+\delta,v(R,\hat{u}_f+\delta))
	&
	=
	0;
	\label{eq:Hextragauge2}
	\\
	b(\hat{u}_f+\delta,v,\theta)
	&
	=
	0
	\text{ for all } v \geq v_{-1}, \theta \in \mathbb S^2;
	\label{eq:Hextragauge3}
\end{align}
will be achieved by iteratively solving for a suitable sphere $\mathring{S}$ and considering the resulting spacetime double null foliation generated by Proposition \ref{prop:Hconesfoliations}.
Moreover, letting $\newsph{f}_{\I,\Hp}$ denote the diffeomorphism functions which relate the $\hat{u}_f + \delta$ normalised $\Hp$ gauge to the $\hat{u}_f+\delta$ normalised $\I$ gauge,
\[
	\newsph{u}_{\I} = \newsph{u}_{\Hp} + \newsph{f}^3_{\I,\Hp}(\newsph{u}_{\Hp}, \newsph{v}_{\Hp}, \newsph{\theta}^1_{\Hp}, \newsph{\theta}^2_{\Hp}),
	\quad
	\newsph{v}_{\I} = \newsph{v}_{\Hp} + \newsph{f}^4_{\I,\Hp}(\newsph{u}_{\Hp}, \newsph{v}_{\Hp}, \newsph{\theta}^1_{\Hp}, \newsph{\theta}^2_{\Hp}),
\]
\[
	\newsph{\theta}^1_{\I} = \newsph{\theta}^1_{\Hp} + \newsph{f}^1_{\I,\Hp}(\newsph{u}_{\Hp}, \newsph{v}_{\Hp}, \newsph{\theta}^1_{\I}, \newsph{\theta}^2_{\Hp}),
	\quad
	\newsph{\theta}^2_{\I} = \newsph{\theta}^2_{\Hp} + \newsph{f}^2_{\I,\Hp}(\newsph{u}_{\Hp}, \newsph{v}_{\Hp}, \newsph{\theta}^1_{\Hp}, \newsph{\theta}^2_{\Hp}),
\]
the anchoring conditions
\begin{align}
	\newsph{f}^3_{\I,\Hp}(\hat{u}_f+\delta, v(R,\hat{u}_f+\delta), \theta)
	&
	=
	0
	\text{ for all } \theta \in \mathbb S^2;
	\label{eq:Hextragauge4}
	\\
	\newsph{f}^A_{\I,\Hp}(\hat{u}_f+\delta, v(R,\hat{u}_f+\delta), \theta)
	&
	=
	0
	\text{ for all } \theta \in \mathbb S^2,
	\qquad
	A=1,2,
	\label{eq:Hextragauge5}
\end{align}
will also be achieved.  The remaining relevant anchoring conditions of Definition \ref{anchoringdef} (in particular the inclusion \eqref{eq:overlap3}) are trivially satisfied, provided $\delta_0$ is sufficiently small, by continuity and the improved bootstrap assumptions of Theorem \ref{havetoimprovethebootstrap}.

Define $\hat{f}$ to be the diffeomorphism functions which relate the extended $\hat{u}_f$ normalised $\Hp$ gauge to the $\hat{u}_f+\delta$ normalised $\I$ gauge,
\begin{equation} \label{eq:oldHnewIdiffeo1}
	\newsph{u}_{\I} = u_{\Hp} + \hat{f}^3(u_{\Hp}, v_{\Hp}, \theta_{\Hp}),
	\quad
	\newsph{v}_{\I} = v_{\Hp} + \hat{f}^4(u_{\Hp}, v_{\Hp}, \theta_{\Hp}),
\end{equation}
\begin{equation} \label{eq:oldHnewIdiffeo2}
	\newsph{\theta}_{\I}^A
	=
	\hat{\slashed{F}}^A(u_{\Hp}, v_{\Hp}, \theta_{\Hp})
	=
	\theta^A_{\Hp} + \hat{f}^A(u_{\Hp}, v_{\Hp}, \theta_{\Hp}),
	\qquad
	A = 1,2,
\end{equation}
defined on an appropriate subset of $\mathbb{R}^2 \times \mathbb{S}^2$.  The $\mathbb{S}^2$ part of this diffeomorphism in particular has the property that
\begin{equation} \label{eq:Fslashhat1}
		|h(\hat{u}_f+\delta,v,\theta) - h(\hat{u}_f+\delta,v,\hat{\slashed{F}}(\hat{u}_f+\delta, v(R,\hat{u}_f+\delta), \theta))|
		\leq
		C(\hat{u}_f)\delta
		\sup_{S_{\hat{u}_f+\delta,v}} |r\nablaslash h |_{\gslash_{[n]}}
		,
	\end{equation}
	for all functions $h$,
	\begin{equation} \label{eq:Fslashhat2}
		\big\vert \big( \omega - (\hat{\slashed{F}}(\hat{u}_f+\delta, v(R,\hat{u}_f+\delta), \cdot))^*\omega \big) (\hat{u}_f+\delta,v(R,\hat{u}_f+\delta),\theta) \big\vert_{\gslash}
		\leq
		C(\hat{u}_f)\delta
		\sup_{S_{\hat{u}_f+\delta,v}}(|\omega|_{\gslash}
		+
		|r\nablaslash\omega|_{\gslash}),
	\end{equation}
	for any $(0,k)$ $S$ tensor $\omega$, and
\begin{equation} \label{eq:Fslashhat3}
	\sum_{k \leq 3} \Vert (r\nablaslash)^{k+2} \hat{\slashed{F}}(\hat{u}_f+\delta, v(R,\hat{u}_f+\delta), \cdot) \Vert_{\mathbb{S}^2}
	\leq
	C(\hat{u}_f) \delta,
\end{equation}
for some constant $C(\hat{u}_f)$ which depends on $\hat{u}_f$.  The estimates \eqref{eq:Fslashhat1} and \eqref{eq:Fslashhat2} can be viewed as appropriate estimates for $\hat{\slashed{F}}(\hat{u}_f+\delta, v(R,\hat{u}_f+\delta), \cdot)$ and $\nablaslash \hat{\slashed{F}}(\hat{u}_f+\delta, v(R,\hat{u}_f+\delta), \cdot)$.

	Let $\newsph{C}_{\hat{u}_f+\delta}^{\Hp}$ denote the continuation of the cone $\newsph{C}_{\hat{u}_f+\delta}^{\I}$ of the $\hat{u}_f+\delta$ normalised gauge of Theorem \ref{thm:Igaugeexistence}.  By continuity, $\newsph{C}_{\hat{u}_f+\delta}^{\Hp}$ is a regular cone if $\delta$ is sufficiently small and, moreover, in the $(u_{\Hp}, v_{\Hp}, \theta_{\Hp})$ coordinates of the extended $\hat{u}_f$ normalised $\Hp$ gauge,
	\[
		\{ i_{\Hp} (\hat{u}_f+\delta + K(v,\theta), v, \theta) \mid (v, \theta) \in [v_{-1} , v(R_2,\hat{u}_f+\delta)] \times \mathbb{S}^2 \}
		\subset
		\newsph{C}_{\hat{u}_f+\delta}^{\Hp},
	\]
	for some smooth function $K \colon [v_{-1} , v(R_2,\hat{u}_f+\delta)] \times \mathbb{S}^2 \to \mathbb{R}$ satisfying
	\begin{equation} \label{eq:Hnewconebound}
		\sup_{v_{-1} \leq v \leq v(R_2,\hat{u}_f+\delta)} \sum_{k \leq 5} \Vert (r\nablaslash)^k K (v,\cdot) \Vert_{\mathbb{S}^2} \leq C(\hat{u}_f) \delta,
	\end{equation}
	for some constant $C(\hat{u}_f)$ which depends on $\hat{u}_f$.
	
Recall the domain \eqref{WH+}.  Given a smooth function $h^4 \colon \mathbb{S}^2 \to \mathbb{R}$ and a diffeomorphism $\slashed{H} \colon \mathbb{S}^2 \to \mathbb{S}^2$, provided $K\circ \slashed{H}^{-1}$ and $h^4\circ \slashed{H}^{-1}$ are sufficiently small, the sphere
\[
	\newsph{S} = \{i_{\Hp} (\hat{u}_f+\delta + K(v_{-1},\slashed{H}(\theta)), v_{-1}+h^4(\theta), \slashed{H}(\theta)) \mid \theta \in \mathbb{S}^2\},
\]
defines, by Proposition \ref{prop:Hconesfoliations} (with $s^3 = K$, $s^4 = h^4\circ \slashed{H}^{-1}$), a unique smooth double null parametrisation
\[
	\newsph{i} \colon \mathcal{Z}_{\Hp}(\hat{u}_f + \delta) \to \mathcal{M},
\]
such that the metric takes the double null form \eqref{doublenulllongform} in the associated coordinates, the associated geometric quantities satisfy the gauge conditions of Proposition \ref{prop:Hconesfoliations}, \eqref{eq:Hextragauge3} holds, the image of $\mathbb{S}^2$ under $\newsph{i} (\hat{u}_f+\delta, v(R,\hat{u}_f+\delta), \cdot)$ is the sphere $\newsph{S}$, and
\[
	\pi_{\mathbb{S}^2} \circ i^{-1} \circ \newsph{i} (\hat{u}_f+\delta, v_{-1}, \theta)
	=
	\slashed{H}(\theta),
\]
where $i$ is the parametrisation of the extended $\hat{u}_f$ normalised $\Hp$ gauge and $\pi_{\mathbb{S}^2} \colon \mathbb{R}^2 \times \mathbb{S}^2 \to \mathbb{S}^2$ is the projection onto the $\mathbb{S}^2$ argument.  
	Note that, if $\delta$, $K$ and $h^4\circ \slashed{H}^{-1}$ are sufficiently small then $\newsph{i}(\mathcal{Z}_{\Hp}(\hat{u}_f + \delta)) \subset i(\mathcal{Z}_{\mathcal{H}^+}(\hat{u}_f,\delta_1))$ (recalling \eqref{actualHpdomainext} and Theorem \ref{thm:extendedgauges}) where $i$ (see \eqref{localformparam}) is the local parametrisation of the extended $\hat{u}_f$ normalised $\Hp$ gauge, and the parametrisation $\newsph{i}$ can be made arbitrarily close to the parametrisation $i$.  It follows that the inclusion \eqref{impoverlapext3} implies that the cones $\newsph{\Cbar}_{v}^{\Hp}$ of the parametrisation $\newsph{i}$ satisfy
	\begin{align*}
		\bigcup_{v_{-1} \leq v \leq v_2} \newsph{\Cbar}_{v}^{\Hp}
		\subset
		\mathcal{D}^{\mathcal{K}}(V_3) \cap \{ V_{-1} - 3C\varepsilon \leq V_{data} \leq V_{-1}+ 3C \varepsilon \}
			\cap \{U_0 -3C\varepsilon\leq U_{data} \leq 3C\varepsilon \} ,
	\end{align*}
	In particular, $U_{5}$ is large enough so that the region covered by the initial data gauge $\mathcal{D}^{\mathcal{K}}(V_3)$ is not exhausted and there is room to extend in $u$.  Moreover the inclusions \eqref{impoverlapext1}, \eqref{impoverlapext2} guarantee that the diffeomophism functions appearing in \eqref{eq:Hextragauge4}, \eqref{eq:Hextragauge5} are well defined.

In the proof of Theorem \ref{thm:Hgaugeexistence} below, equations will be solved for $f^4_{\ell = 0}(\hat{u}_f+\delta, v (R,\hat{u}_f+\delta))$ and $f^4_{\ell \geq 1}(\hat{u}_f+\delta, v_{-1},\cdot)$. The following lemma guarantees that there indeed exists a double null parametrisation corresponding to such a prescription.

\begin{lemma}[Existence of a double null parametrisation defined by $f^4_{\ell = 0}(\hat{u}_f+\delta, v (R,\hat{u}_f+\delta))$ and $f^4_{\ell \geq 1}(\hat{u}_f+\delta, v_{-1},\cdot)$] \label{lem:h4j4sphere}
	Consider $\tau >0$.  Let $h^4 \colon \mathbb{S}^2 \to \mathbb{R}$ be a smooth function and let $j^4$ be a constant such that
	\begin{equation} \label{eq:lemh4j4sphereassumption}
		\sum_{k \leq 5} \Vert (r\nablaslash)^k h^4 \Vert_{S_{\hat{u}_f+\delta,v_{-1}}} + \vert j^4 \vert \leq \tau.
	\end{equation}
	If $\tau$, $\hat{\varepsilon}_0$ and $\delta_0$ are sufficiently small, there exists a constant $\lambda$ satisfying
	\[
		\vert \lambda \vert \lesssim C(\hat{u}_f)\delta + \tau,
	\]
	and a diffeomorphism $\slashed{H} \colon \mathbb{S}^2 \to \mathbb{S}^2$ such that, in the above double null parametrisation arising from Proposition \ref{prop:Hconesfoliations} and the sphere, in the $(u_{\Hp}, v_{\Hp}, \theta_{\Hp})$ coordinates of the extended $\hat{u}_f$ normalised $\Hp$ gauge,
	\begin{equation} \label{eq:h4j4newsphere}
		\newsph{S} = \{i_{\Hp} (\hat{u}_f+\delta + K(v_{-1},\slashed{H}(\theta)), v_{-1}+h^4(\theta) + \lambda, \slashed{H}(\theta)) \mid \theta \in \mathbb{S}^2\},
	\end{equation}
	the diffeomorphisms which relate the above double null parametrisation to the extended $\hat{u}_f$ normalised $\Hp$ gauge satisfy
	\begin{equation} \label{eq:h4j4spherediffeo}
		f^4_{\ell = 0}(\hat{u}_f+\delta, v (R,\hat{u}_f+\delta)) = j^4,
		\qquad
		\slashed{F}(\hat{u}_f+\delta, v (R,\hat{u}_f+\delta),\theta) = \hat{\slashed{F}}(\hat{u}_f+\delta,v(R,\hat{u}_f+\delta),\theta),
	\end{equation}
	for all $\theta \in \mathbb{S}^2$, where $\hat{\slashed{F}}$ is defined in \eqref{eq:oldHnewIdiffeo1}, \eqref{eq:oldHnewIdiffeo2}.
\end{lemma}

\begin{proof}
		First note that any constant $\lambda_{[n]}$ and diffoemorphism $\slashed{H}_{[n-1]} \colon \mathbb{S}^2 \to \mathbb{S}^2$ define a sphere, in the coordinates of the extended $\hat{u}_f$ normalised $\Hp$ gauge,
		\begin{equation} \label{eq:lambdanhnminusonesphere}
			S_{[n]}
			=
			\{i_{\Hp} (\hat{u}_f+\delta + K(v_{-1},\slashed{H}_{[n-1]}(\theta)), v_{-1}+h^4(\theta) + \lambda_{[n]}, \slashed{H}_{[n-1]}(\theta)) \mid \theta \in \mathbb{S}^2\}.
		\end{equation}
		Provided $K$, $h^4 \circ \slashed{H}^{-1}_{[n-1]}$, and $\lambda_{[n]}$ are suitably small, Proposition \ref{prop:Hconesfoliations} defines a unique double null parametrisation
		\[
			i_{[n]} \colon \mathcal{Z}_{\Hp}(\hat{u}_f + \delta) \to \mathcal{M},
		\]
		such that the metric in the associated coordinates takes the double null form \eqref{doublenulllongform}, the associated double null foliation is that of Proposition \ref{prop:Hconesfoliations}, the condition \eqref{eq:Hextragauge3} holds, and
		\[
			\pi_{\mathbb{S}^2} \circ i^{-1} \circ i_{[n]}(\hat{u}_f+\delta, v(R,\hat{u}_f+\delta), \theta)
			=
			\hat{\slashed{F}}(\hat{u}_f+\delta,v(R,\hat{u}_f+\delta),\theta),
		\]
		for all $\theta \in \mathbb{S}^2$, where $i$ is the parametrisation of the extended $\hat{u}_f$ normalised $\Hp$ gauge. Let $f_{[n]}$ denote the diffeomorphism functions which relate this double null parametrisation to the extended $\hat{u}_f$ normalised $\Hp$ double null parametrisation, so that
		\begin{equation*}
			u_{\Hp} = u_{[n]} + f_{[n]}^3(u_{[n]}, v_{[n]}, \theta_{[n]}),
			\quad
			v_{\Hp} = v_{[n]} + f_{[n]}^4(u_{[n]}, v_{[n]}, \theta_{[n]}),
		\end{equation*}
		\begin{equation*} 
			\theta_{\Hp}^A
			=
			\slashed{F}_{[n]}^A(u_{[n]}, v_{[n]}, \theta_{[n]})
			=
			\theta^A_{[n]} + f_{[n]}^A(u_{[n]}, v_{[n]}, \theta_{[n]}),
			\qquad
			A = 1,2,
		\end{equation*}
		and let $\Phi_{[n]}$ denote the appropriate geometric quantities of the this double null parametrisation.  Note that
		\[
			f^3_{[n]}(\hat{u}_f+\delta,v,\theta) = K(v,\slashed{F}_{[n]}(\hat{u}_f+\delta,v,\theta)),
			\qquad
			\slashed{F}_{[n]}(\hat{u}_f+\delta,v(R),\theta)
			=
			\hat{\slashed{F}}(\hat{u}_f+\delta,v(R),\theta),
		\]
		for all $v_{-1} \leq v \leq v(R_2,\hat{u}_f+\delta)$, $\theta \in \mathbb{S}^2$, where $v(R) = v(R,\hat{u}_f+\delta)$.
		In such a double null parametrisation, $(f^4_{[n]})_{\ell = 0}(\hat{u}_f+\delta,v(R))$ and $(f^4_{[n]})_{\ell = 0}(\hat{u}_f+\delta,v_{-1})$ are related as follows.
		It follows from the relations \eqref{eq:metriccomp6}, \eqref{eq:Riccicomp8}, and \eqref{eq:curvaturecomp5} that, on $u_{[n]} = \hat{u}_f +\delta$, using the gauge condition \eqref{eq:Hfoliation1},
		\begin{align}
			&
			\frac{2}{r} \partial_v (f^4_{[n]})_{\ell=0}(x)
			=
			\Omega^{-2}_{[n]} (\Omega \tr \chibar - \Omega \tr \chibar_{\circ})^{[n]}_{\ell =0}(x)
			-
			\Omega^{-2} (\Omega \tr \chibar - \Omega \tr \chibar_{\circ})_{\ell =0}(x+f_{[n]}(x))
			\nonumber
			\\
			&
			\qquad \qquad
			-
			\frac{2}{r} \Big( \frac{\Omega^2}{\Omega_{\circ}^2} - 1 \Big)_{\ell =0}(x+f_{[n]}(x))
			-
			\frac{r^2}{3M_f} (\rho - \rho_{\circ})^{[n]}_{\ell = 0}(x)
			+
			\frac{r^2}{3M_f} (\rho - \rho_{\circ})_{\ell = 0}(x+f(x))
			+
			\mathcal{X}_0[f_{[n]}](x),
			\label{eq:lemh4j4sphere1}
		\end{align}
		where $\mathcal{X}_0[f_{[n]}]$ is an error satisfying $\vert \mathcal{X}_0[f_{[n]}]\vert \lesssim \mathcal{X}[f_{[n]}]$, and 
		\[
			\mathcal{X}[f_{[n]}] =
			\big(
			\vert M_f - M_{\hat{u}_f+\delta} \vert 
			+
			\sum_{\Phi} \vert \Phi \vert 
			\big)
			\sum_{\vert \gamma \vert \leq 2} \vert \mathfrak{D}^{\gamma} f_{[n]}\vert
			+
			\sum_{\vert \gamma \vert \leq 2} \vert \mathfrak{D}^{\gamma} f_{[n]}\vert^2,
		\]
		where, for any $k$,
		\[
			\sum_{\vert \gamma \vert \leq k} \vert \mathfrak{D}^{\gamma} f_{[n]} \vert
			:=
			\sum_{\vert \gamma \vert \leq k}
			\Big(
			\vert \mathfrak{D}^{\gamma} \Omega^2 f^3_{[n]} \vert
			+
			\vert \mathfrak{D}^{\gamma} f^4_{[n]} \vert
			\Big)
			+
			\sum_{\vert \gamma \vert \leq k-1}
			\Big(
			\vert \mathfrak{D}^{\gamma} \partial_u \slashed{f}_{[n]} \vert
			+
			\vert \mathfrak{D}^{\gamma} \partial_v \slashed{f}_{[n]} \vert
			\Big).
		\]
		Now, as in the proof of Lemma \ref{lem:trchipartialHgauge},
		\begin{align*}
			&
			\int_{v_{-1}}^{v(R)}
			\frac{r}{2} \Omega_{[n]}^{-2} (\Omega \tr \chibar - \Omega \tr \chibar_{\circ})^{[n]}_{\ell =0}(\hat{u}_f+\delta,v)
			-
			\frac{r^3}{6M_f} (\rho - \rho_{\circ})^{[n]}_{\ell = 0}(\hat{u}_f+\delta,v)
			dv
			\\
			&
			\qquad \qquad
			=
			\tilde{c}_1
			(\Omega \tr \chi - \Omega \tr \chi_{\circ})^{[n]}_{\ell =0}(\hat{u}_f+\delta,v(R))
			+
			\tilde{c}_2
			\Upsilon^{[n]}_{\ell =0}(\hat{u}_f+\delta,v(R))
			+
			\mathcal{E},
		\end{align*}
		for an appropriate nonlinearity $\mathcal{E}$ and constants $\tilde{c}_1$, $\tilde{c}_2$, where the gauge conditions \eqref{eq:Hfoliation1}, \eqref{eq:Hfoliation5} have been used.  It then follows, from the change of gauge formula \eqref{eq:Riccicomp7} for $(\Omega \tr \chi - \Omega \tr \chi_{\circ})_{\ell =0}$, together with the gauge condition $(\Omega \tr \chi - \Omega \tr \chi_{\circ})_{\ell =0}(\hat{u}_f,v(R)) = 0$, that
		\begin{align*}
			&
			\int_{v_{-1}}^{v(R)}
			\frac{r}{2}
			\Omega^{-2}_{[n]} (\Omega \tr \chibar - \Omega \tr \chibar_{\circ})^{[n]}_{\ell =0}(x)
			-
			\frac{r}{2} \Omega^{-2} (\Omega \tr \chibar - \Omega \tr \chibar_{\circ})_{\ell =0}(x+f_{[n]}(x))
			-
			\frac{r^3}{6M_f} (\rho - \rho_{\circ})^{[n]}_{\ell = 0}(x)
			\\
			&
			+
			\frac{r^2}{3M_f} (\rho - \rho_{\circ})_{\ell = 0}(x+f_{[n]}(x))
			dv
			\\
			&
			=
			c_1 (f^4_{[n]})_{\ell =0} (\hat{u}_f+\delta,v(R))
			-
			c_1 K(v(R),\slashed{H}_{[n]}(\cdot))_{\ell = 0}
			+
			c_2
			\big(
			\Upsilon^{[n]}_{\ell =0}(\hat{u}_f+\delta,v(R))
			-
			\Upsilon_{\ell =0}(\hat{u}_f+\delta,v(R))
			\big)
			+
			\mathcal{E}_{[n]},
		\end{align*}
		for some constants $c_1$, $c_2$ with $c_1 \neq 1$, where $x = (\hat{u}_f+\delta,v)$ and $\mathcal{E}_{[n]}$ is a nonlinear error which takes the schematic form
		\begin{align*}
			\mathcal{E}_{[n]}
			=
			&
			\int_{v_{-1}}^{v(R)}
			\Phi_{[n]} \cdot \Phi_{[n]} (x)
			-
			\Phi \cdot \Phi(x+f_{[n]}(x))
			dv'
			+
			\int_{v_{-1}}^{v(R)}
			\big(
			\Omega_{\circ}^{-2} \Omega^2(x+f_{[n]}(x)) - 1
			\big)_{\ell = 0}
			dv'
			\\
			&
			+
			\sum_{\vert \gamma \vert \leq 2}
			\Pi_{S_{[n]}} \Phi \cdot \mathfrak{D}^{\gamma} f_{[n]}(v(R))
			+
			\sum_{\vert \gamma_1 \vert + \vert \gamma_2 \vert \leq 2}
			\mathfrak{D}^{\gamma_1} f_{[n]} \cdot \mathfrak{D}^{\gamma_2} f_{[n]} (v(R)).
		\end{align*}
		From \eqref{eq:lemh4j4sphere1} it then follows that
		\begin{align*}
			&
			(f^4_{[n]})_{\ell = 0}(\hat{u}_f+\delta,v(R)) - (f^4_{[n]})_{\ell = 0}(\hat{u}_f+\delta,v_{-1})
			=
			c_1 (f^4_{[n]})_{\ell =0} (\hat{u}_f+\delta,v(R))
			-
			c_1 K(v(R),\slashed{H}_{[n]}(\cdot))_{\ell = 0}
			\\
			&
			\qquad
			+
			c_2
			\big(
			\Upsilon^{[n]}_{\ell =0}(\hat{u}_f+\delta,v(R))
			-
			\Upsilon_{\ell =0}(\hat{u}_f+\delta,v(R))
			\big)
			+
			\mathcal{E}_{[n]}
			+
			\int_{v_{-1}}^{v(R)} \mathcal{X}_0[f_{[n]}] -  \Big( \frac{\Omega^2}{\Omega_{\circ}^2} - 1 \Big)_{\ell =0}(x+f_{[n]}(x)) dv,
		\end{align*}
		and so
		\begin{align*}
			(h^4 + \lambda_{[n]})_{\ell =0}
			=
			&
			(1- c_1) (f^4_{[n]})_{\ell =0} (\hat{u}_f+\delta,v(R))
			+
			\int_{v_{-1}}^{v(R)} \Big( \frac{\Omega^2}{\Omega_{\circ}^2} - 1 \Big)_{\ell =0}(x+f_{[n]}(x)) - \mathcal{X}_0[f_{[n]}] dv
			\\
			&
			+
			c_1 K(v(R),\slashed{H}_{[n]}(\cdot))_{\ell = 0}
			-
			c_2
			\big(
			\Upsilon^{[n]}_{\ell =0}(\hat{u}_f+\delta,v(R))
			-
			\Upsilon_{\ell =0}(\hat{u}_f+\delta,v(R))
			\big)
			-
			\mathcal{E}_{[n]}
			.
		\end{align*}
		Define therefore $\lambda_{[0]} = 0$, $\slashed{H}_{[0]}=Id$ and, for $n \geq 1$, define $\lambda_{[n]}$ inductively by
		\begin{align} \label{eq:lemh4j4spherelambdadef}
			\lambda_{[n]}
			=
			&
			(1-c_1) j^4
			-
			h^4_{\ell=0}
			+
			\int_{v_{-1}}^{v(R)}
			\Big( \frac{\Omega^2}{\Omega_{\circ}^2} - 1 \Big)_{\ell =0}(x+f_{[n-1]}(x))
			-
			\mathcal{X}_0[f_{[n-1]}](x) dv
			\\
			&
			+
			c_1 K(v(R),\slashed{H}_{[n-1]}(\cdot))_{\ell = 0}
			-
			c_2
			\big(
			\Upsilon^{[n-1]}_{\ell =0}(\hat{u}_f+\delta,v(R))
			-
			\Upsilon_{\ell =0}(\hat{u}_f+\delta,v(R))
			\big)
			-
			\mathcal{E}_{[n-1]},
			\nonumber
		\end{align}
		where $x = (\hat{u}_f+\delta,v)$.  
		Recall that, provided $h^4 \circ \slashed{H}^{-1}_{[n-1]}$ and $\lambda_{[n]}$ are suitably small, $\lambda_{[n]}$ and $\slashed{H}_{[n-1]}$ define a sphere \eqref{eq:lambdanhnminusonesphere} and hence a double null parametrisation.  Define $\slashed{H}_{[n]}$ in terms of the diffeomorphisms relating this double null parametrisation to the extended $\hat{u}_f$ normalised double null parametrisation by
		\[
			\slashed{H}_{[n]}(\theta) = \slashed{F}_{[n]}(\hat{u}_f+\delta, v_{-1},\theta).
		\]
		
		The proof will follow from showing that the sequences $\{\lambda_{[n]}\}$ and $\{\slashed{H}_{[n]}\}$ converge to appropriate limits.  Many of the steps are very similar to those of the proof of Theorem \ref{thm:Hgaugeexistence} below and so are only sketched.  Following the proof of Lemma \ref{lem:fnH} below, one sees that, if $\varepsilon$ and $\delta$ are sufficiently small,
		\begin{multline} \label{eq:h4j4lambdaestimate}
			\sup_{v_{-1} \leq v \leq v(R,\hat{u}_f+\delta)}
			\Big(
			\sum_{\vert \gamma \vert \leq 5}
			\Vert \mathfrak{D}^{\gamma} f_{[n]} \Vert_{S_{\hat{u}_f+\delta,v}}
			+
			\sum_{\vert \gamma \vert \leq 3}
			\sup_{v_{-1} \leq v \leq v(R)} \Vert \mathfrak{D}^{\gamma} \Phi_{[n]}(x) - \mathfrak{D}^{\gamma} \Phi(x) \Vert_{S_{\hat{u}_f+\delta,v}}
			\Big)
			\\
			\lesssim
			C(\hat{u}_f) \delta
			+
			\sum_{k \leq 5} \Vert (r\nablaslash)^k h^4 \Vert_{S_{\hat{u}_f+\delta,v_{-1}}}
			+
			\vert (f^4_{[n]})_{\ell=0}(\hat{u}_f+\delta,v(R)) \vert,
		\end{multline}
		where the norm of the diffeomorphism functions is defined below in \eqref{eq:iteratesdiffeonormsH}, for some constant $C(\hat{u}_f)$ which depends on $\hat{u}_f$.
		
		Recall that
		\begin{align*}
			\lambda_{[n]}
			=
			&
			(1-c_1) (f^4_{[n]})_{\ell=0}(\hat{u}_f+\delta,v(R))
			-
			h^4_{\ell=0}
			+
			\int_{v_{-1}}^{v(R)}
			\Big( \frac{\Omega^2}{\Omega_{\circ}^2} - 1 \Big)_{\ell =0}(x+f_{[n]}(x))
			-
			\mathcal{X}_0[f_{[n]}](x) dv
			\\
			&
			+
			c_1 K(v(R),\slashed{H}_{[n]}(\cdot))_{\ell = 0}
			-
			c_2
			\big(
			\Upsilon^{[n]}_{\ell =0}(\hat{u}_f+\delta,v(R))
			-
			\Upsilon_{\ell =0}(\hat{u}_f+\delta,v(R))
			\big)
			-
			\mathcal{E}_{[n]},
		\end{align*}
		and so, subtracting from the definition \eqref{eq:lemh4j4spherelambdadef}, it follows, using the fact that, for all $v_{-1} \leq v \leq v(R_2,\hat{u}_f+\delta)$, $\theta\in \mathbb{S}^2$,
		\[
			\sum_{\vert \gamma \vert \leq 2}
			\vert \mathfrak{D}^{\gamma} ( \Omega_{\circ}^{-2} \Omega^2 - 1) (\hat{u}_f+\delta,v, \theta) \vert
			\lesssim
			\delta,
		\]
		the assumption \eqref{eq:lemh4j4sphereassumption}, and the estimate \eqref{eq:h4j4lambdaestimate},
		that, if $\delta$ is sufficiently small,
		\[
			\vert (f^4_{[n]})_{\ell=0}(\hat{u}_f+\delta,v(R)) \vert
			\lesssim
			C(\hat{u}_f) \delta
			+
			\tau.
		\]
		It then follows from the definition \eqref{eq:lemh4j4spherelambdadef} of $\lambda_{[n]}$ and the estimate \eqref{eq:h4j4lambdaestimate} that the sequence $\{ \lambda_{[n]}\}$ is uniformly bounded.  It follows that there exists a subsequence converging to a limit $\lambda$, which one easily checks is as desired.  Moreover, in any given coordinate chart for $\mathbb{S}^2$, for any $k \leq 4$, $A_1,\ldots,A_k = 1,2$,
		\[
			\vert \partial_{A_1} \ldots \partial_{A_k} f^B_{[n]}(\hat{u}_f+\delta,v,\theta) \vert
			\lesssim
			C(\hat{u}_f) \delta
			+
			\int_v^{v(R)}
			\vert \partial_{A_1} \ldots \partial_{A_k} \partial_v f^B_{[n]}(\hat{u}_f+\delta,v',\theta) \vert
			dv'
			\lesssim
			C(\hat{u}_f) \delta
			+
			\tau,
		\]
		by the estimates \eqref{eq:Fslashhat1}--\eqref{eq:Fslashhat3} for $\hat{\slashed{F}}(\hat{u}_f+\delta,v(R),\cdot)$, and so one similarly sees that the diffeomorphisms $\slashed{H}_{[n]}$ converge to a limiting diffeomorphism $\slashed{H} \colon \mathbb{S}^2 \to \mathbb{S}^2$, which is as desired.
\end{proof}

Note that, for any $h^4$ and $j^4$ as in Lemma \ref{lem:h4j4sphere}, the double null parametrisation of Lemma \ref{lem:h4j4sphere} clearly satisfies the anchoring condition \eqref{eq:Hextragauge4}.  The latter of \eqref{eq:h4j4spherediffeo} moreover implies that the anchoring condition \eqref{eq:Hextragauge5} also holds.

The proof of Theorem \ref{thm:Hgaugeexistence} can now be given.

\begin{proof}[Proof of Theorem \ref{thm:Hgaugeexistence}]
	Let $\lambda\in \mathfrak{R}(\hat{u}_f)$ be fixed.
	The goal is to find an appropriate function $h^4 \colon \mathbb{S}^2 \to \mathbb{R}$ and a constant $j^4$ so that double null parametrisation of Lemma \ref{lem:h4j4sphere}, which, recall, satisfies the gauge conditions of Proposition \ref{prop:Hconesfoliations} and the anchoring conditions \eqref{eq:Hextragauge4}, \eqref{eq:Hextragauge5}, moreover satisfies \eqref{eq:Hextragauge1} and \eqref{eq:Hextragauge2}.

	Recall the quantity
	\[
		\Upsilon
		=
		\left(1 - \frac{3M_f}{r} \right) (\rho - \rho_{\circ})
		+
		\frac{3M_f}{2r^2} \left( \Omega \tr \chi - (\Omega \tr \chi)_{\circ} \right)
		-
		\frac{3M_f\Omega_{\circ}^2}{2r^2} \Omega^{-2} \left( \Omega \tr \chibar - (\Omega \tr \chibar)_{\circ} \right).
	\]
	Using the expressions \eqref{eq:metriccomp6}, \eqref{eq:Riccicomp7}, \eqref{eq:Riccicomp8}, \eqref{eq:curvaturecomp5}, for how geometric quantities change under a change of gauge, along with Lemma \ref{lem:trchipartialHgauge}, it follows that the goal is to find a function $h^4(\theta)$ and a constant $j^4$ such that
	\begin{align}
		&
		\Big[
		2 \newsph{\Deltaslash} \newsph{\Deltaslash} h^4(\theta)
		+
		\frac{2}{r^2} \left( 5 - \frac{12 M}{r} \right) \newsph{\Deltaslash} h^4(\theta)
		+
		\mathfrak{a}^1(r,\newsph{r})(x_{-1})
		\Big]_{\ell \geq 1}
		\label{eq:h4Hgaugeeqn1}
		\\
		&
		\qquad \qquad
		=
		\Big[
		\frac{6\Omega^2_{\circ}}{r^2}
		\newsph{\Deltaslash} h^3(\theta)
		-
		\Deltaslash ( \Omega^{-2} (\Omega \tr \chibar - \Omega \tr \chibar_{\circ})) (x_{-1}+f(x_{-1}))
		-
		\frac{3}{r^2}(\Omega \tr \chibar - \Omega \tr \chibar_{\circ}) (x_{-1}+f(x_{-1}))
		\nonumber
		\\
		&
		\qquad \qquad \qquad
		-
		\frac{6}{r^3} (\Omega^2 - \Omega_{\circ}^2)(x_{-1}+f(x_{-1}))
		+
		\frac{2}{r} \mathfrak{m}(x_{-1})
		+
		\mathcal{E}^1(x_{-1})
		\Big]_{\ell \geq 1}
		,
		\nonumber
		\\
		&
		\mathfrak{a}^2(r,\newsph{r})(x_R)_{\ell =0}
		=
		\Big[
		(1 - A^{-1})\Omega^{-2} (\Omega \tr \chi - \Omega \tr \chi_{\circ})(x_R+f(x_R))
		\label{eq:h4Hgaugeeqn2}
		\\
		&
		\qquad \qquad \qquad
		-
		BA^{-1} \Omega^{-2} \newsph{\Upsilon}(x_R)
		+
		BA^{-1} \Omega^{-2} \Upsilon(x_R+f(x_R))
		-
		\frac{2}{r\Omega_{\circ}^{2}}(\Omega^2 - \Omega_{\circ}^2)(x_R+f(x_R))
		+
		\mathcal{E}^2(x_R)
		\Big]_{\ell =0},
		\nonumber
		\\
		&
		j^4 = f^4_{\ell = 0}(\hat{u}_f+\delta, v(R,\hat{u}_f+\delta)),
		\label{eq:h4Hgaugeeqn3}
	\end{align}
	where $x_{-1} = x_{-1}(\theta) = (\hat{u}_f+\delta, v_{-1},\theta)$, $x_R = x_R(\theta) = (\hat{u}_f+\delta, v(R,\hat{u}_f+\delta),\theta)$.  Here the functions $f^1$, $f^2$, $f^3$, $f^4$ denote the diffeomorphisms relating the double null foliation arising from $h^4$, $j^4$ and Lemma \ref{lem:h4j4sphere} (whose double null coordinates are denoted $(\newsph{u}, \newsph{v}, \newsph{\theta})$), to the extended $\hat{u}_f$ normalised $\Hp$ double null parametrisation, so that
	\begin{equation*}
		u = \newsph{u} + f^3(\newsph{u}, \newsph{v}, \newsph{\theta}),
		\quad
		v = \newsph{v} + f^4(\newsph{u}, \newsph{v}, \newsph{\theta}),
		\quad
		\theta^A 
		=
		\slashed{F}^A(\newsph{u}, \newsph{v},\newsph{\theta})
		= \newsph{\theta}^A + f^A(\newsph{u}, \newsph{v}, \newsph{\theta}).
	\end{equation*}
	The Laplacian $\newsph{\Deltaslash}$ is the Laplacian of the new sphere $\newsph{S}_{\hat{u}_f+\delta, v_{-1}} = \{ \newsph{u} = \hat{u}_f+\delta \} \cap \{ \newsph{v} = v_{-1} \}$ (or, equivalently, the sphere \eqref{eq:h4j4newsphere} arising from Lemma \ref{lem:h4j4sphere}), the mode projection in \eqref{eq:h4Hgaugeeqn1} is that associated to the new sphere $\newsph{S}_{\hat{u}_f+\delta, v_{-1}}$ and the mode projection in \eqref{eq:h4Hgaugeeqn2} is that of the sphere $\newsph{S}_{\hat{u}_f+\delta, v(R)} = \{ \newsph{u} = \hat{u}_f+\delta \} \cap \{ \newsph{v} = v(R,\hat{u}_f+\delta) \}$.
	The nonlinear error $\mathcal{E}^1$ has the schematic form
	\[
		\mathcal{E}^1(x_{-1})
		=
		\sum_{\vert \gamma_1 \vert + \vert \gamma_2 \vert \leq 4}
		\Pi_{\newsph{S}} \Phi \cdot \mathfrak{D}^{\gamma_1} f(x_{-1})
		+
		\mathfrak{D}^{\gamma_1} f \cdot \mathfrak{D}^{\gamma_2} f(x_{-1}),
	\]
	and $\mathcal{E}^2$ takes the form,
	\[
		\mathcal{E}^2(x_{R})
		=
		\sum_{\vert \gamma_1 \vert + \vert \gamma_2 \vert \leq 2}
		\Pi_{\newsph{S}}\Phi \cdot \mathfrak{D}^{\gamma_1} f(x_{R})
		+
		\mathfrak{D}^{\gamma_1} f \cdot \mathfrak{D}^{\gamma_2} f(x_{R})
		+
		A^{-1} \Omega^{-2} \mathfrak{F}
		-
		A^{-1} \Omega_{\circ}^{-2} \mathfrak{E},
	\]
	where $A$, $B$, $\mathfrak{E}$ and $\mathfrak{F}$ are as in Lemma \ref{lem:trchipartialHgauge}.  For each $\theta \in \mathbb{S}^2$, $x_{-1} + f(x_{-1})$ and $x_R + f(x_R)$ denote the points, in the coordinate system of the extended $\hat{u}_f$ normalised $\Hp$ gauge,
	\begin{align}
		x_{-1} + f(x_{-1})
		&
		=
		(\hat{u}_f+\delta + f^3(\hat{u}_f+\delta, v_{-1},\theta), v_{-1} + f^4(\hat{u}_f+\delta, v_{-1},\theta),\slashed{F}(\hat{u}_f+\delta, v_{-1},\theta)),
		\label{eq:Hgaugexminusone}
		\\
		x_R + f(x_R)
		&
		=
		(\hat{u}_f+\delta + f^3(\hat{u}_f+\delta,v(R),\theta)
		,
		v(R) + f^4(\hat{u}_f+\delta, v(R),\theta)
		,
		\slashed{F}(\hat{u}_f+\delta, v(R),\theta))
		,
	\end{align}
	with $v(R) = v(R,\hat{u}_f+\delta)$.
	The function $\mathfrak{m}$ is defined by
	\begin{align*}
		\mathfrak{m}
		&
		=
		\newsph{\Deltaslash} \partial_u f^3
		-
		\frac{3\Omega^2}{r^2} \partial_v f^4
		-
		\frac{2}{r} \left( 1 - \frac{M_f}{r} \right) \newsph{\Deltaslash} h^3
		+
		\frac{\Omega^2}{r} \newsph{\Deltaslash} h^4
		+
		\mathfrak{a}(r,\newsph{r}),
	\end{align*}
	and $\mathfrak{a}^1,\mathfrak{a}^2,\mathfrak{a}$ are smooth functions, determined from the relations \eqref{eq:metriccomp6}, \eqref{eq:Riccicomp7}, \eqref{eq:Riccicomp8}, \eqref{eq:curvaturecomp5}, which satisfy $\mathfrak{a}(r,r) = \mathfrak{a}^i(r,r) = 0$ and
	\begin{align*}
		\vert
		\mathfrak{a}^1(r,\newsph{r})(x_{-1})
		-
		\frac{12 \Omega_{\circ}^4}{r^4}
		(f^4(x_{-1}) - f^3(x_{-1}))
		\vert
		&
		\lesssim
		\vert f^3(x_{-1}) \vert^2
		+
		\vert f^4(x_{-1}) \vert^2
		+
		\vert M_f - M_{\hat{u}_f+\delta} \vert^2,
		\\
		\vert
		\mathfrak{a}^2(r,\newsph{r})(x_{R})
		-
		\frac{4}{R^2} \left( 1 - \frac{3M}{R} \right) (f^4(x_{R}) - f^3(x_{R}))
		\vert
		&
		\lesssim
		\vert f^3(x_{R}) \vert^2
		+
		\vert f^4(x_{R}) \vert^2
		+
		\vert M_f - M_{\hat{u}_f+\delta} \vert,
		\\
		\vert
		\mathfrak{a}(r,\newsph{r})(x_{-1})
		+
		\frac{3\Omega^4}{r^3} (f^4(x_{-1}) - f^3(x_{-1}))
		\vert
		&
		\lesssim
		\vert f^3(x_{-1}) \vert^2
		+
		\vert f^4(x_{-1}) \vert^2
		+
		\vert M_f - M_{\hat{u}_f+\delta} \vert^2.
	\end{align*}
	The dependence of $\mathfrak{a}^i$ on $M_f$ and $M_{\hat{u}_f+\delta}$ is suppressed since both $M_f$ and $M_{\hat{u}_f+\delta}$ are now considered fixed.  Recall that, by Theorem \ref{thm:Igaugeexistence},
	\[
		\vert M_f - M_{\hat{u}_f+\delta} \vert
		\lesssim
		\delta.
	\]
	Note that $\newsph{\mu}^*(x) = \mu^*(x+f(x)) + \mathfrak{m}(x) + \mathcal{O}(\varepsilon^2)$.  The function $\newsph{r}$ is the $r_M$ (see \eqref{EFrdef}) associated to the $\newsph{i}$ double null parametrisation and mass $M=M_{\hat{u}_f+\delta}$, and $r$ is the $r_M$ associated to the extended $\hat{u}_f$ normalised $\Hp$ double null gauge, with mass $M=M_f$.  In particular, in the coordinates associated to the $\newsph{i}$ double null parametrisation,
	\[
		\newsph{r}(u,v,\theta) = r_{M_{\hat{u}_f+\delta}}(u,v),
		\qquad
		r(u,v,\theta) = r_{M_f}(u+f^3(u,v,\theta),v+f^4(u,v,\theta)).
	\]
	The quantities $\Deltaslash \Omega^{-2} (\Omega \tr \chibar - \Omega \tr \chibar_{\circ})$, $\mu^*$, $\Phi$ etc.\@ refer to the quantities in the extended $\hat{u}_f$ normalised gauge.
	
	Note that, for such $h^4$, $j^4$, the diffeomorphism functions satisfy
	\[
		f^4_{\ell \geq 1}(\hat{u}_f+\delta,v_{-1},\theta) = h^4_{\ell \geq 1} (\theta),
		\qquad
		f^4_{\ell =0}(\hat{u}_f+\delta,v(R,\hat{u}_f+\delta)) = j^4,
	\]
	and
	\[
		f^3(\hat{u}_f+\delta,v,\theta)
		=
		K(v,\slashed{F}(\hat{u}_f+\delta,v,\theta)),
	\]
	for all $v_{-1} \leq v \leq v(R,\hat{u}_f+\delta)$ and $\theta \in \mathbb{S}^2$.
	
	Equation \eqref{eq:h4Hgaugeeqn2} arises from Lemma \ref{lem:trchipartialHgauge} and the fact that the diffeomorphisms relating two given double null gauges satisfy (see the relations \eqref{eq:metriccomp1}, \eqref{eq:Riccicomp7}, \eqref{eq:Riccicomp8})
	\begin{multline} \label{eq:trchitrchibarOmegachangeofgauge}
		\frac{4\Omega^2}{r^2} \left( 1 - \frac{3M}{r} \right) (f^3-f^4)_{\ell=0}
		=
		(\widetilde{\Omega \tr \chi} - \widetilde{\Omega \tr \chi}_{\circ})_{\ell=0}
		-
		(\Omega \tr \chi - \Omega \tr \chi_{\circ})_{\ell=0}
		\\
		-
		\big[
		(\widetilde{\Omega \tr \chibar} - \widetilde{\Omega \tr \chibar}_{\circ})_{\ell=0}
		-
		(\Omega \tr \chibar - \Omega \tr \chibar_{\circ})_{\ell=0}
		\big]
		-
		\frac{2}{r}
		\big[
		(\widetilde{\Omega}^2 - \widetilde{\Omega}^2_{\circ})_{\ell=0}
		-
		(\Omega^2 - \Omega^2_{\circ})_{\ell=0}
		\big]
		+
		\mathcal{E}.
	\end{multline}
	The reader is also referred to Remark \ref{rmk:linearHgaugesphere}, concerning the proof of Proposition \ref{proplinHpgauge} on the existence of a linearised $\Hp$ gauge.
	
	Given a solution $h^4$, $j^4$ of \eqref{eq:h4Hgaugeeqn1}, \eqref{eq:h4Hgaugeeqn2}, the double null parametrisation arising from Lemma \ref{lem:h4j4sphere} will then be as desired.

	The solution $h^4$, $j^4$ of \eqref{eq:h4Hgaugeeqn1}, \eqref{eq:h4Hgaugeeqn2} will be constructed as a limit of a sequence of iterates.  Given suitably small iterates $h^4_{[n]}$ and $j^4_{[n]}$, consider the double null parametrisation
	\[
		i_{[n]} \colon \mathcal{Z}_{\Hp}(\hat{u}_f + \delta) \to \mathcal{M},
	\]
	arising from Lemma \ref{lem:h4j4sphere}, whose coordinates are denoted $(u_{[n]},v_{[n]},\theta_{[n]})$.  Let $F_{[n]}:= i^{-1} \circ i_{[n]}$, where $i$ is the parametrisation of the extended $\hat{u}_f$ normalised $\Hp$ double null gauge, so that
	\[
		(u,v,\theta)
		=
		F_{[n]}(u_{[n]}, v_{[n]}, \theta_{[n]}),
	\]
	and let $f_{[n]}$ denote the corresponding diffeomorphism functions,
	\[
		u = u_{[n]} + f^3_{[n]} (u_{[n]}, v_{[n]}, \theta_{[n]}),
		\quad
		v = v_{[n]} + f^4_{[n]}(u_{[n]}, v_{[n]}, \theta_{[n]}),
	\]
	\[
		\theta^A
		=
		\slashed{F}^A_{[n]} (u_{[n]}, v_{[n]}, \theta_{[n]})
		=
		\theta^A_{[n]} + f^A_{[n]}(u_{[n]}, v_{[n]}, \theta_{[n]}).
	\]
	Define $\rho_{[n]}$ etc.\@ to be the geometric quantities of this ``$n$-th'' double null foliation (schematically denoted $\Phi_{[n]}$).

	Define now $h_{[0]}^4 = 0$, $j_{[0]}^4 = 0$ and, for $n \geq 1$, define $h_{[n]}^4\colon \mathbb{S}^2 \to \mathbb{R}$ and the constant $j_{[n]}^4$ inductively as the solutions of
	\begin{align} \label{eq:nHgauge1}
		&
		2 \Deltaslash_{[n-1]} \Deltaslash_{[n-1]} h^4_{[n]}(\theta)
		+
		\frac{2}{r^2_{[n-1]}} \left( 5 - \frac{12 M_f}{r_{[n-1]}} \right) \Deltaslash_{[n-1]} h^4_{[n]}(\theta)
		+
		\frac{12 \Omega_{\circ,[n-1]}^4}{r^4_{[n-1]}} h^4_{[n]}(\theta)
		\\
		&
		\qquad
		=
		\Big[
		\frac{6\Omega^2_{\circ,[n-1]}}{r^2_{[n-1]}}
		\big(
		\Deltaslash_{[n-1]} h^3_{[n-1]}(\theta)
		+
		\frac{2 \Omega_{\circ[n-1]}^2}{r^2_{[n-1]}} h^3_{[n-1]}(\theta)
		\big)
		+
		\mathcal{G}^1_{[n-1]}
		+
		A^1_{[n-1]}
		+
		\mathcal{E}^1_{[n-1]}
		\Big]_{(\ell \geq 1)_{[n-1]}}
		,
		\nonumber
		\\
		&
		\frac{4}{R^2} \left( 1 - \frac{3M_f}{R} \right) j^4_{[n]}
		=
		\Big[
		\frac{4}{R^2} \left( 1 - \frac{3M_f}{R} \right) j^3
		\label{eq:nHgauge2}
		+
		\mathcal{G}^1_{[n-1]}
		+
		A^2_{[n-1]}
		+
		\mathcal{E}^2_{[n-1]}
		\Big]_{(\ell = 0)_{[n-1]}},
	\end{align}
	where
	\[
		h^3_{[n-1]}(\theta) = K(v_{-1},\slashed{F}_{[n-1]}(\hat{u}_f+\delta,v_{-1},\theta)),
		\qquad
		j^3(\theta) = K(v(R,\hat{u}_f+\delta),\hat{\slashed{F}}(\hat{u}_f+\delta,v(R,\hat{u}_f+\delta),\theta)),
	\]
	and
	\begin{align*}
		\mathcal{G}^1_{[n-1]}
		=
		&
		-
		\Deltaslash ( \Omega^{-2} (\Omega \tr \chibar - \Omega \tr \chibar_{\circ})) (x+f_{[n-1]}(x))
		-
		\frac{3}{r^2}(\Omega \tr \chibar - \Omega \tr \chibar_{\circ}) (x+f_{[n-1]}(x))
		\\
		&-
		\frac{6}{r^3} (\Omega^2 - \Omega_{\circ}^2)(x+f_{[n-1]}(x))
		+
		\frac{2}{r} \mathfrak{m}_{[n-1]},
		\\
		\mathcal{G}^2_{[n-1]}
		=
		&
		(1 - A^{-1})\Omega^{-2} (\Omega \tr \chi - \Omega \tr \chi_{\circ})(x_R+f_{[n-1]}(x_R))
		-
		BA^{-1} \Omega^{-2} \Upsilon_{[n-1]}(x_R)
		\\
		&
		+
		BA^{-1} \Omega^{-2} \Upsilon(x_R+f_{[n-1]}(x_R))
		-
		\frac{2}{r\Omega_{\circ}^{2}}(\Omega^2 - \Omega_{\circ}^2)(x_R+f_{[n-1]}(x_R)),
	\end{align*}
	\begin{align*}
		A^1_{[n-1]}
		=
		a^1(r,r_{[n-1]})
		-
		\frac{12 \Omega_{\circ}^4}{r^4}
		(h^4_{[n-1]} - h^3_{[n-1]}),
		\quad
		A^2_{[n-1]}
		=
		a^2(r,r_{[n-1]})
		-
		\frac{4}{R^2} \left( 1 - \frac{3M}{R} \right) (j^4_{[n-1]} - j^3_{[n-1]}),
	\end{align*}
	\begin{align*}
		\mathfrak{m}_{[n-1]}
		=
		\
		&
		\Deltaslash_{[n-1]} \partial_u f^3_{[n-1]}
		-
		\frac{3\Omega^2}{r^2} \partial_v f^4_{[n-1]}
		-
		\frac{2}{r} \left( 1 - \frac{M_f}{r} \right) \Deltaslash_{[n-1]} h^3_{[n-1]}
		+
		\frac{\Omega^2}{r} \Deltaslash_{[n-1]} h^4_{[n-1]}
		+
		a(r,r_{[n-1]}),
	\end{align*}
	and the errors $\mathcal{E}^1_{[n-1]}$ and $\mathcal{E}^2_{[n-1]}$ are defined by replacing $f$ with $f_{[n-1]}$ in $\mathcal{E}^1$ and $\mathcal{E}^2$.  Schematically,
	\begin{align*}
		\mathcal{E}^1_{[n-1]}
		=
		\
		&
		\sum_{\vert \gamma_1 \vert + \vert \gamma_2 \vert \leq 4}
		\Pi_{S_{[n-1]}} \Phi \cdot \mathfrak{D}^{\gamma_1} f_{[n-1]}(x_{-1})
		+
		\mathfrak{D}^{\gamma_1} f_{[n-1]} \cdot \mathfrak{D}^{\gamma_2} f_{[n-1]}(x_{-1}),
		\\
		\mathcal{E}^2_{[n-1]}
		=
		\
		&
		\sum_{\vert \gamma_1 \vert + \vert \gamma_2 \vert \leq 2}
		\Pi_{S_{[n-1]}} \Phi \cdot \mathfrak{D}^{\gamma_1} f_{[n-1]}(x_R)
		+
		\mathfrak{D}^{\gamma_1} f_{[n-1]} \cdot \mathfrak{D}^{\gamma_2} f_{[n-1]}(x_R)
		+
		A^{-1} \Omega^{-2} \mathfrak{F}
		-
		A^{-1} \Omega_{\circ}^{-2} \mathfrak{E}_{[n-1]}.
	\end{align*}
	The Laplacian $\Deltaslash_{[n-1]}$ is defined to be Laplacian associated to the sphere $S^{[n-1]}_{\hat{u}_f+\delta, v_{-1}} = \{ u_{[n-1]} = \hat{u}_f+\delta \} \cap \{ v_{[n-1]} = v_{-1} \}$ (or, equivalently, the sphere \eqref{eq:h4j4newsphere} arising from Lemma \ref{lem:h4j4sphere} with $h_{[n-1]}^4$ and $j_{[n-1]}^4$).  The mode projection in \eqref{eq:nHgauge1} is that of the associated to the new sphere $S^{[n-1]}_{\hat{u}_f+\delta, v_{-1}}$ and the mode projection in \eqref{eq:nHgauge2} is that of the sphere $S^{[n-1]}_{\hat{u}_f+\delta, v(R)} = \{ u_{[n-1]} = \hat{u}_f+\delta \} \cap \{ v_{[n-1]} = v(R,\hat{u}_f+\delta) \}$.
	Finally, $r_{[n-1]}$ is the $r_M$ (see \eqref{EFrdef}) associated to the $i_{[n-1]}$ double null parametrisation and mass $M=M_{\hat{u}_f+\delta}$, and $r$ is the $r_M$ associated to the extended $\hat{u}_f$ normalised $\Hp$ double null gauge, with mass $M=M_f$, so that, in the coordinates associated to the $i_{[n-1]}$ double null parametrisation,
\begin{align*}
	r_{[n-1]}(\hat{u}_f+\delta,v,\theta)
	&
	=
	r_{M_{\hat{u}_f+\delta}}(\hat{u}_f+\delta,v),
	\\
	r(\hat{u}_f+\delta,v,\theta)
	&
	=
	r_{M_f}(\hat{u}_f+\delta+f^3_{[n-1]}(\hat{u}_f+\delta,v,\theta),v+f^4_{[n-1]}(\hat{u}_f+\delta,v,\theta))
	,
\end{align*}
and
\[
	\Omega_{\circ,[n-1]}^2
	=
	1 - \frac{2 M_{\hat{u}_f+\delta}}{r_{[n-1]}}.
\]

	Note that the unique solution $h^4_{[n]}$ of \eqref{eq:nHgauge1} is supported on $(\ell \geq 1)_{[n-1]}$, and the unique solution $j^4_{[n]}$ of \eqref{eq:nHgauge2} is supported on $(\ell =0)_{[n-1]}$, and that
	\begin{equation} \label{eq:h4nj4nf3nFslashed1}
		h^4_{[n]}(\theta)
		=
		(f^4_{[n]})_{(\ell \geq 1)_{[n-1]}}(\hat{u}_f+\delta,v_{-1},\theta),
		\qquad
		j^4_{[n]}
		=
		(f^4_{[n]})_{(\ell =0)_{[n-1]}}(\hat{u}_f+\delta,v(R,\hat{u}_f+\delta)),
	\end{equation}
	and
	\begin{align} \label{eq:h4nj4nf3nFslashed2}
		f^3_{[n]}(\hat{u}_f+\delta,v,\theta)
		&
		=
		K(v,\slashed{F}_{[n]}(\hat{u}_f+\delta,v,\theta)),
	\\
	\label{eq:h4nj4nf3nFslashed3}
		\slashed{F}_{[n]}(\hat{u}_f+\delta,v(R,\hat{u}_f+\delta),\theta)
		&
		=
		\hat{\slashed{F}}(\hat{u}_f+\delta,v(R,\hat{u}_f+\delta),\theta),
	\end{align}
	for all $v_{-1} \leq v \leq v(R_2,\hat{u}_f+\delta)$ and $\theta \in \mathbb{S}^2$.

	In order to show that the sequences $\{h^4_{[n]}\}$ and $\{j^4_{[n]}\}$ converge it is first necessary to estimate the iterates $f_{[n]}$ and their derivatives.
	
	In what follows $x$ will be used to denote a value $x = (u,v,\theta)$.  As in the proof of Theorem \ref{thm:Igaugeexistence}, for fixed $(u,v)$ the geometric quantities $\Phi(u,v)$ and $\Phi_{[n]}(u,v)$ can be viewed, via the identifications \eqref{localformparam} and \eqref{localformparamtwo}, as tensor fields on the sphere $\mathbb{S}^2$.  When the quantities $\Phi$ and $\Phi_{[n]}$ are identified using the values of their respective coordinate functions in this way, we write $\Phi(x) - \Phi_{[n]}(x)$.  When they are considered at the same spacetime point, as in the relations of Propositions \ref{prop:metricrelations}, \ref{prop:Riccirelations} and \ref{prop:curvaturerelations}, we write $\Pi_{S_{[n]}}\Phi - \Phi_{[n]}$, where the projection $\Pi_{S_{[n]}}$ is as in \eqref{eq:tildeprojection}.  For example, as in the proof of Theorem \ref{thm:Igaugeexistence},
\begin{align*}
	(\Pi_{S_{[n]}} \rho - \rho_{[n]} )(u,v,\theta)
	&
	=
	\rho(F_{[n]}(u,v,\theta))
	-
	\rho_{[n]}(u,v,\theta)
	=
	\rho(x+f_{[n]}(x))
	-
	\rho_{[n]}(x),
\\
	(\Pi_{S_{[n]}} \alpha - \alpha_{[n]} )(u,v,\theta)
	&
	=
	\Big(
	\alpha(F_{[n]}(u,v,\theta))_{CD} \partial_{\theta_{[n]}^A} F^C_{[n]} \partial_{\theta_{[n]}^B} F^D_{[n]}
	-
	\alpha_{[n]}(u,v,\theta)_{AB}
	\Big)
	d\theta_{[n]}^A d \theta_{[n]}^B,
\\
	\alpha(u,v,\theta) - \alpha_{[n]}(u,v,\theta)
	&
	=
	\Big( \alpha(u,v,\theta)_{AB} - \alpha_{[n]}(u,v,\theta)_{AB} \Big) d\theta_{[n]}^A d \theta_{[n]}^B,
\\
	\alpha(u,v,\theta) - \Pi_{S_{[n]}} \alpha(u,v,\theta)
	&
	=
	\Big( \alpha(u,v,\theta)_{AB} - 
	\alpha(F_{[n]}(u,v,\theta))_{CD} \partial_{\theta_{[n]}^A} F^C_{[n]} \partial_{\theta_{[n]}^B} F^D_{[n]}
	\Big) d\theta_{[n]}^A d \theta_{[n]}^B.
\end{align*}

For some given $v_{-1} \leq v \leq v(R,\hat{u}_f+\delta)$ and $k \geq 1$, define the norm on diffeomorphisms $f$,
\begin{align} \label{eq:iteratesdiffeonormsH}
	\sum_{\vert \gamma \vert \leq k} \Vert \mathfrak{D}^{\gamma} f \Vert_{S_{\hat{u}_f+\delta,v}}
	=
	\sum_{\vert \gamma \vert \leq k}
	\Big(
	\Vert \mathfrak{D}^{\gamma} \Omega^2 f^3 \Vert_{S_{\hat{u}_f+\delta,v}}
	+
	\Vert \mathfrak{D}^{\gamma} f^4 \Vert_{S_{\hat{u}_f+\delta,v}}
	\Big)
	+
	\sum_{\vert \gamma \vert \leq k-1}
	\Big(
	\Vert \mathfrak{D}^{\gamma} \partial_u \slashed{f} \Vert_{S_{\hat{u}_f+\delta,v}}
	+
	\Vert \mathfrak{D}^{\gamma} \partial_v \slashed{f} \Vert_{S_{\hat{u}_f+\delta,v}}
	\Big),
\end{align}
with, for example,
\[
	\Vert \mathfrak{D}^{\gamma} \Omega^2 f^3 \Vert_{S_{\hat{u}_f+\delta,v}}^2
	=
	\int_{\mathbb{S}^2} \vert \mathfrak{D}^{\gamma} (\Omega^2 f^3) (\hat{u}_f+\delta,v,\theta) \vert^2_{r^2 \gamma} \sqrt{\det \gamma(\theta)} d \theta.
\]
Define also $\sum_{\vert \gamma \vert \leq k} \Vert \mathfrak{D}^{\gamma} f_{\ell \geq 1} \Vert_{S_{u,v_{\infty}}}$ by replacing $f^3$ and $f^4$ with $f^3_{\ell \geq 1}$ and $f^4_{\ell \geq 1}$ respectively in \eqref{eq:iteratesdiffeonorms}, and
\begin{align} \label{eq:iteratesdiffeonormsHell0}
	\sum_{\vert \gamma \vert \leq k} \Vert \mathfrak{D}^{\gamma} f_{\ell=0} \Vert_{S_{\hat{u}_f+\delta,v}}
	=
	\sum_{\vert \gamma \vert \leq k}
	\Big(
	\Vert \mathfrak{D}^{\gamma} \Omega^2 f^3_{\ell=0} \Vert_{S_{\hat{u}_f+\delta,v}}
	+
	\Vert \mathfrak{D}^{\gamma} f^4_{\ell = 0} \Vert_{S_{\hat{u}_f+\delta,v}}
	\Big).
\end{align}

	The following lemma will be used in the proof of Lemma \ref{lem:fnH}.
	
	\begin{lemma}[Estimates for differences between $S$ tensors with pullback by $\slashed{F}_{[n]}$] \label{lem:mvtH}
	For any $v_{-1} \leq v \leq v(R,\hat{u}_f+\delta)$,
	\begin{equation*}
		|h(\hat{u}_f+\delta,v,\theta) - h(\hat{u}_f+\delta,v,\slashed{F}_{[n]}(\hat{u}_f+\delta,v,\theta))|
		\lesssim
		\sup_{S_{\hat{u}_f+\delta,v}} |r\nablaslash h |_{\gslash_{[n]}}
		\Big(
		C(\hat{u}_f)\delta
		+
		\int_v^{v(R,\hat{u}_f+\delta)} \vert \partial_v \slashed{f}_{[n]} (\hat{u}_f+\delta,v',\theta) \vert dv'
		\Big)
		,
	\end{equation*}
	for all functions $h$, and
	\begin{multline} \label{eq:diffiscloseiteratesHgaugetwo}
		\big\vert \big( \omega - (\slashed{F}_{[n]}(\hat{u}_f+\delta,v,\cdot))^*\omega \big) (\hat{u}_f+\delta,v,\theta) \big\vert_{\gslash_{[n]}}
		\\
		\lesssim
		\sup_{S_{\hat{u}_f+\delta,v}}(|\omega|_{\gslash_{[n]}}
		+
		|r\nablaslash\omega|_{\gslash_{[n]}})
		\Big(
		C(\hat{u}_f)\delta
		+
		\sum_{k=0}^1
		\int_v^{v(R,\hat{u}_f+\delta)}
		\vert (r\nablaslash)^k \partial_v \slashed{f}_{[n]} (\hat{u}_f+\delta,v',\theta) \vert
		dv'
		\Big).
	\end{multline}
	for any $(0,k)$ $S$ tensor $\omega$.  Similarly,
	\begin{multline*}
		|
		h(\hat{u}_f+\delta,v_{\infty},\slashed{F}_{[n+1]}(\hat{u}_f+\delta,v_{\infty},\theta))
		-
		h(u,v_{\infty},\slashed{F}_{[n]}(u,v_{\infty},\theta))
		|
		\\
		\lesssim
		\sup_{S_{u,v_{\infty}}} |r\nablaslash h |_{\gslash_{[n]}}
		\int_v^{v(R,\hat{u}_f+\delta)} \vert (\partial_v \slashed{f}_{[n+1]} - \partial_v \slashed{f}_{[n]}) (\hat{u}_f+\delta,v',\theta) \vert dv'
		,
	\end{multline*}
	for all functions $h$, and
	\begin{multline} \label{eq:diffiscloseiteratesdifferenceHgaugetwo}
		\big\vert \big( (\slashed{F}_{[n+1]}(u,v_{\infty},\cdot))^* \omega - (\slashed{F}_{[n]}(u,v_{\infty},\cdot))^*\omega \big) (u,v_{\infty},\theta) \big\vert_{\gslash_{[n]}}
		\\
		\lesssim
		\sup_{S_{u,v_{\infty}}}(|\omega|_{\gslash_{[n]}}
		+
		|r\nablaslash\omega|_{\gslash_{[n]}})
		\sum_{k=0}^1
		\int_v^{v(R,\hat{u}_f+\delta)}
		\vert (r\nablaslash)^k (\partial_v \slashed{f}_{[n+1]} - \partial_v \slashed{f}_{[n]}) (\hat{u}_f+\delta,v',\theta) \vert
		dv'.
	\end{multline}
	for any $(0,k)$ $S$ tensor $\omega$.
\end{lemma}

\begin{proof}
	The proof is similar to that of Lemma \ref{lem:mvt}, using now the estimates \eqref{eq:Fslashhat1}, \eqref{eq:Fslashhat2} in place of Proposition \ref{determiningthesphere}.
\end{proof}

	The following estimate for the diffeomorphisms in terms of the differences $\Phi_{[n]}(x) - \Pi_{S_{[n]}} \Phi(x)$ will be also used in Lemma \ref{lem:fnH}.

	\begin{lemma}[Estimates for diffeomorphisms by differences of geometric quantities] \label{lem:fHnPhiest}
		The diffeomorphism functions $f_{[n]}$ satisfy the estimates
		\begin{multline*}
			\sup_{v_{-1} \leq v \leq v(R)}
			\sum_{\vert \gamma \vert \leq 5}
			\Vert \mathfrak{D}^{\gamma} (f_{[n]})_{\ell \geq 1} \Vert_{S_{\hat{u}_f+\delta,v}}
			\lesssim
			\sup_{v_{-1} \leq v \leq v(R)} \Big[
			\sum_{ \vert \gamma \vert \leq 3}
			\Vert \mathfrak{D}^{\gamma} \Phi_{[n]}(x) - \Pi_{S_{[n]}} \mathfrak{D}^{\gamma} \Phi(x) \Vert_{S_{\hat{u}_f+\delta,v}}
			\\
			+
			\sum_{k\leq 5}
			\Vert (r \nablaslash)^k \Omega^2 f_{[n]}^3 \Vert_{S_{\hat{u}_f+\delta,v}}
			+
			\sum_{\vert \gamma \vert \leq 5}
			\Vert \mathfrak{D}^{\gamma} f_{[n]} \Vert_{S_{\hat{u}_f+\delta,v}}^2
			\Big],
		\end{multline*}
		and
		\begin{multline*}
			\sup_{v_{-1} \leq v \leq v(R)}
			\sum_{\vert \gamma \vert \leq 5}
			\Vert \mathfrak{D}^{\gamma} (f_{[n]})_{\ell =0} \Vert_{S_{\hat{u}_f+\delta,v}}
			\lesssim
			\sum_{\substack{
			\vert \gamma \vert \leq 3
			\\
			\vert \widetilde{\gamma} \vert \leq 5}}
			\Big[
			\Vert \mathfrak{D}^{\gamma} \Phi^{[n]}_{\ell = 0}(x) - \Pi_{S_{[n]}} \mathfrak{D}^{\gamma} \Phi_{\ell = 0}(x) \Vert_{L^1(C_{\hat{u}_f+\delta})}
			\\
			+
			\sup_{v_{-1} \leq v \leq v(R)}
			\big(
			\Vert \mathfrak{D}^{\gamma} \Phi_{[n]}(x) - \Pi_{S_{[n]}} \mathfrak{D}^{\gamma} \Phi(x) \Vert_{S_{\hat{u}_f+\delta,v}}
			+
			\Vert \Omega^2 f_{[n]}^3 \Vert_{S_{\hat{u}_f+\delta,v}}
			+
			v(R)
			(
			\vert M_f - M_{\hat{u}_f+\delta}\vert
			+
			\Vert \mathfrak{D}^{\widetilde{\gamma}} f_{[n]} \Vert_{S_{\hat{u}_f+\delta,v}}^2
			)\big)\Big],
		\end{multline*}
		where $v(R) = v(R,\hat{u}_f+\delta)$ and the norms of the diffeomorphism functions are defined in \eqref{eq:iteratesdiffeonormsH} and \eqref{eq:iteratesdiffeonormsHell0}, provided $\varepsilon$ is sufficiently small (independent of $\hat{u}_f$).
	\end{lemma}
	
\begin{proof}
	Throughout this proof $f = f_{[n]}$.  The subscript is omitted for brevity.
	Define
	\[
		\mathcal{X}
		=
		\big(
		\vert M_f - M_{\hat{u}_f+\delta} \vert 
		+
		\sum_{\Phi} \vert \Phi \vert 
		\big)
		\sum_{\vert \gamma \vert \leq 2} \vert \mathfrak{D}^{\gamma} f\vert
		+
		\sum_{\vert \gamma \vert \leq 2} \vert \mathfrak{D}^{\gamma} f\vert^2,
	\]
	where $\sum_{\vert \gamma \vert \leq 2} \vert \mathfrak{D}^{\gamma} f\vert$ is defined as in \eqref{eq:iteratesdiffeonormsH}.
	
	Consider any $v_{-1} \leq v \leq v(R,\hat{u}_f+\delta)$ and the sphere $S=S_{\hat{u}_f+\delta,v}$.  First note that equations \eqref{eq:Riccicomp2} and \eqref{eq:curvaturecomp4} imply that
	\[
		\Vert (r\nablaslash)^2 f^4_{\ell \geq 2} \Vert_S
		\lesssim
		\Vert \Omega^{-1} \hat{\chibar}_{[n]}(x) - \Pi_{S_{[n]}} \Omega^{-1} \hat{\chibar}(x) \Vert_S
		+
		\Vert
		\mathcal{X}
		\Vert_S,
		\qquad
		\vert f^4_{\ell =1} \vert
		\lesssim
		\vert \Omega^{-1} \beta^{[n]}_{\ell =1}(x) - \Pi_{S_{[n]}} \Omega^{-1} \beta_{\ell =1}(x) \vert
		+
		\mathcal{X}.
	\]
	The relation \eqref{eq:Riccicomp3} implies that
	\[
		\Vert r \nablaslash \Omega^{-2} \partial_u (\Omega^2 f^3)_{\ell \geq 1} \Vert_S
		\lesssim
		\Vert \eta_{[n]}(x) - \Pi_{S_{[n]}} \eta(x) \Vert_S
		+
		\Vert r\nablaslash \Omega^2 f^3 \Vert_S
		+
		\Vert r \nablaslash f^4 \Vert_S
		+
		\Vert \mathcal{X} \Vert_S.
	\]
	The relation \eqref{eq:Riccicomp4} implies that
	\[
		\Vert r \nablaslash \partial_v f^4_{\ell \geq 1} \Vert_S
		\lesssim
		\Vert \etabar_{[n]}(x) - \Pi_{S_{[n]}} \etabar(x) \Vert_S
		+
		\Vert r\nablaslash \Omega^2 f^3 \Vert_S
		+
		\Vert r \nablaslash f^4 \Vert_S
		+
		\Vert \mathcal{X} \Vert_S.
	\]
	The metric relations \eqref{eq:metriccomp1} and \eqref{eq:metriccomp2} give
	\[
		\Vert \Omega^{-2} \partial_u f^4 \Vert_S
		+
		\Vert \partial_v (\Omega^2 f^3) \Vert_S
		\lesssim
		\Vert \mathcal{X} \Vert_S.
	\]
	The metric relations \eqref{eq:metriccomp3} and \eqref{eq:metriccomp4} give
	\[
		\Vert \Omega^{-2} \partial_{\widetilde{u}} \slashed{f}^{\flat} \Vert_S
		\lesssim
		\Vert r\nablaslash f^4 \Vert_S
		+
		\Vert \mathcal{X} \Vert_S,
		\qquad
		\Vert \partial_{v} \slashed{f}^{\flat} \Vert_S
		\lesssim
		\Vert r\nablaslash \Omega^2 f^3 \Vert_S
		+
		\Vert \mathcal{X} \Vert_S.
	\]
	For the $\ell = 0$ modes, first note that
	\begin{align*}
		&
		\frac{2}{r} \partial_v f^4_{\ell=0}
		=
		\Omega^{-2} (\Omega \tr \chibar - \Omega \tr \chibar_{\circ})^{[n]}_{\ell =0}
		-
		\Omega^{-2} (\Omega \tr \chibar - \Omega \tr \chibar_{\circ})_{\ell =0}
		\\
		&
		\qquad \qquad
		+
		\frac{2}{r} \Big( \frac{\Omega^2}{\Omega_{\circ}^2} - 1 \Big)^{[n]}_{\ell =0}
		-
		\frac{2}{r} \Big( \frac{\Omega^2}{\Omega_{\circ}^2} - 1 \Big)_{\ell =0}
		-
		\frac{r^2}{3M_f} (\rho - \rho_{\circ})^{[n]}_{\ell = 0}
		+
		\frac{r^2}{3M_f} (\rho - \rho_{\circ})_{\ell = 0}
		+
		\mathcal{X}_0,
	\end{align*}
	where $\mathcal{X}_0$ is an error satisfying $\vert \mathcal{X}_0\vert \lesssim \mathcal{X}$, and so
	\begin{align*}
		&
		\vert \partial_v f^4_{\ell=0} \vert
		\lesssim
		\vert
		\Omega^{-2} (\Omega \tr \chibar - \Omega \tr \chibar_{\circ})^{[n]}_{\ell =0}(x)
		-
		\Omega^{-2} (\Omega \tr \chibar - \Omega \tr \chibar_{\circ})_{\ell =0}(x+f(x))
		\vert
		\\
		&
		\qquad
		\qquad
		+
		\Big\vert
		\Big( \frac{\Omega^2}{\Omega_{\circ}^2} - 1 \Big)^{[n]}_{\ell =0}(x)
		-
		\Big( \frac{\Omega^2}{\Omega_{\circ}^2} - 1 \Big)_{\ell =0}(x+f(x))
		\Big\vert
		+
		\vert
		(\rho - \rho_{\circ})^{[n]}_{\ell = 0}(x)
		-
		(\rho - \rho_{\circ})_{\ell = 0}(x+f(x))
		\vert
		+
		\mathcal{X}.
	\end{align*}
	Integrating backwards from $r=R$ gives an estimate for $f^4_{\ell=0}$
	\begin{align*}
		\vert f^4_{\ell=0} (v) \vert
		\lesssim
		&
		\int_v^{v(R)} \vert \partial_v f^4_{\ell=0} \vert dv'
		+
		\vert f^4_{\ell=0} (v(R)) \vert
		+
		\mathcal{X}
		\\
		\lesssim
		&
		\int_v^{v(R)} \vert \partial_v f^4_{\ell=0} \vert dv'
		+
		\vert f^3_{\ell=0} (v(R)) \vert
		+
		\vert (\rho - \rho_{\circ})^{[n]}_{\ell=0} (v(R)) - (\rho - \rho_{\circ})_{\ell=0}(v(R) + f) \vert
		+
		\mathcal{X}.
	\end{align*}
	Next, note that
	\[
		\Omega^{-2} \partial_u (\Omega^2 f^3)
		+
		\partial_v f^4
		+
		\frac{2M_f}{r^2} f^4
		=
		\Big( \frac{\Omega^2}{\Omega_{\circ}^2} - 1 \Big)_{[n]}
		-
		\Big( \frac{\Omega^2}{\Omega_{\circ}^2} - 1 \Big)
		+
		\mathcal{X}_0,
	\]
	where $\mathcal{X}_0$ is an error satisfying $\vert \mathcal{X}_0\vert \lesssim \mathcal{X}$, and so
	\[
		\vert \Omega^{-2} \partial_u (\Omega^2 f^3)_{\ell=0} \vert
		\lesssim
		\Big\vert
		\Big( \frac{\Omega^2}{\Omega_{\circ}^2} - 1 \Big)^{[n]}_{\ell =0}(x)
		-
		\Big( \frac{\Omega^2}{\Omega_{\circ}^2} - 1 \Big)_{\ell =0}(x+f(x))
		\Big\vert
		+
		\vert \partial_v f^4_{\ell=0} \vert
		+
		\vert f^4_{\ell=0} \vert
		+
		\mathcal{X}.
	\]
	Finally, the relations \eqref{eq:Riccicomp5}, \eqref{eq:Riccicomp6} give appropriate estimates for $(\Omega^{-2} \partial_u)^2 \Omega^2 f^3$ and $\partial_v^2 f^4$.  Estimates for higher order derivatives follow similarly.  The proof then follows after inserting the estimates for $\Phi$ and taking $\varepsilon$ sufficiently small (independent of $\hat{u}_f$).
\end{proof}
	
	The following Lemma gives estimates for the diffeomorphism functions $f^3_{[n]}$, which take the form \eqref{eq:h4nj4nf3nFslashed2}, using the estimates \eqref{eq:Fslashhat1}--\eqref{eq:Fslashhat3} and \eqref{eq:Hnewconebound}.
	
	\begin{lemma}[Estimates for $f^3_{[n]}$] \label{lem:f3Hnest}
		For each $n$, the diffeomorphism function $f_{[n]}^3$ satisfies the estimates, for all $v_{-1} \leq v \leq v(R,\hat{u}_f+\delta)$,
		\begin{equation} \label{eq:f3Hnest1}
			\sum_{k\leq 5}
			\Vert (r \nablaslash)^k f_{[n]}^3 \Vert_{S_{\hat{u}_f+\delta,v}}
			\lesssim
			C(\hat{u}_f) \delta
			\Big(
			1
			+
			\int_v^{v(R,\hat{u}_f+\delta)} 
			\sum_{k \leq 5} \Vert (r\nablaslash)^k \partial_v \slashed{f}_{[n]} \Vert_{S_{\hat{u}_f+\delta,v'}}
			dv'
			\Big),
		\end{equation}
		where $C(\hat{u}_f)$ is as constant which depends on $\hat{u}_f$.  Moreover,
		\begin{equation} \label{eq:f3Hnest2}
			\sum_{k\leq 5}
			\Vert (r \nablaslash)^k (f_{[n+1]}^3 - f_{[n]}^3) \Vert_{S_{\hat{u}_f+\delta,v}}
			\lesssim
			C(\hat{u}_f) \delta
			\int_v^{v(R,\hat{u}_f+\delta)} 
			\sum_{k \leq 5} \Vert (r\nablaslash)^k (\partial_v \slashed{f}_{[n+1]} - \partial_v \slashed{f}_{[n]}) \Vert_{S_{\hat{u}_f+\delta,v'}}
			dv'.
		\end{equation}
	\end{lemma}
	
	\begin{proof}
		Consider some coordinate chart $(\theta^1,\theta^2)$ of $\mathbb{S}^2$.  Recall (see \eqref{eq:h4nj4nf3nFslashed2}) that,
		\[
			f^3_{[n]}(\hat{u}_f+\delta,v,\theta^A)
			=
			K(v,\theta^A + f_{[n]}^A(\hat{u}_f+\delta,v,\theta)),
		\]
		and so
		\[
			\partial_{\theta^B} f^3_{[n]}(\hat{u}_f+\delta,v,\theta^A)
			=
			(\partial_{\theta^C} K)(v,\theta^A + f_{[n]}^A(\hat{u}_f+\delta,v,\theta))
			\frac{\partial f^C_{[n]}}{\partial \theta^B} (\hat{u}_f+\delta,v,\theta).
		\]
		The estimate \eqref{eq:f3Hnest1} for $f^3_{[n]}$ is clear.  The estimate \eqref{eq:f3Hnest1} for $r\nablaslash f^3_{[n]}$ follows from the fact that
		\[
			\frac{\partial f^C_{[n]}}{\partial \theta^B} (\hat{u}_f+\delta,v,\theta)
			=
			\frac{\partial f^C_{[n]}}{\partial \theta^B} (\hat{u}_f+\delta,v(R),\theta)
			-
			\int^{v(R)}_v
			\frac{\partial^2 f^C_{[n]}}{\partial \theta^B \partial v} (\hat{u}_f+\delta,v',\theta)
			dv',
		\]
		together with \eqref{eq:h4nj4nf3nFslashed3}, the estimate \eqref{eq:Fslashhat2}, which in particular implies that
		\[
			\Big\vert \frac{\partial f^C_{[n]}}{\partial \theta^B} (\hat{u}_f+\delta,v(R),\theta) \Big\vert
			\lesssim
			C(\hat{u}_f) \delta,
		\]
		and \eqref{eq:Hnewconebound}.  Similarly for $(r\nablaslash)^k f^3_{[n]}$ for $k =2,3,4,5$, using now \eqref{eq:Fslashhat3}.  The estimate \eqref{eq:f3Hnest2} follows similarly.
	\end{proof}

	The iterates, $f_{[n]}$, can now be estimated.
	
	\begin{lemma}[Estimates for diffeomorphisms] \label{lem:fnH}
		Provided $\hat{\varepsilon}_0$ is sufficiently small (independent of $\hat{u}_f$), and $\delta_0$ is sufficiently small (with respect to $\hat{u}_f$), for all $n \geq 1$ the diffeomorphisms $f_{[n]}$ satisfy, for all $v_{-1} \leq v \leq v(R,\hat{u}_f+\delta)$, the estimates
		\[
			\sum_{k \leq 5} \Vert \mathfrak{D}^k f_{[n]} \Vert_{S_{\hat{u}_f+\delta,v}}
			\leq
			C(\hat{u}_f) \delta,
		\]
		where the norm is defined in \eqref{eq:iteratesdiffeonormsH} and $C(\hat{u}_f)$ is a constant which depends on $\hat{u}_f$.
	\end{lemma}
	
	\begin{proof}[Proof]
		The proof follows that of Lemma \ref{lem:fnI}.  The diffeomorphisms and their derivatives, $\mathfrak{D}^{\gamma} f_{[n]}$, are estimated inductively.  Suppose $n\geq1$ is such that
		\[
			\sup_{v_{-1} \leq v \leq v(R,\hat{u}_f+\delta)}
			\sum_{\vert \gamma \vert \leq 5} \Vert \mathfrak{D}^{\gamma} f_{[n-1]} \Vert_{S_{\hat{u}_f+\delta,v}}
			\leq
			C_1(\hat{u}_f) \delta,
		\]
		for some appropriate $C_1(\hat{u}_f)$, to be chosen, with the norm defined in \eqref{eq:iteratesdiffeonormsH}.  Under this assumption, the estimates for $f_{[n]}$ are divided into several steps.
		
		Throughout this proof the notation $\underline{S} := \Omega^{-2}(\Omega \tr \chibar - \Omega \tr \chibar_{\circ,M_{\hat{u}_f+\delta}})$, $T := \Omega \tr \chi - \Omega \tr \chi_{\circ,M_{\hat{u}_f+\delta}}$ is used.  Define
		\begin{align*}
					&
					\mathcal{S}_{[n]}
					=
					\sum_{k \leq 3}
					\Vert
					(r\nablaslash)^k \Omega \hat{\chi}_{[n]}(x_R)
					-
					(r\nablaslash)^k \Omega \hat{\chi}(x_R)
					\Vert_{S_{\hat{u}_f+\delta,v(R)}}
					+
					\vert
					(
					\divslash \Omega \beta_{[n]}
					-
					\divslash \Omega \beta
					)_{\ell=1} (x_R)
					\vert
					\\
					&
					\qquad
					+
					\vert
					(
					\Upsilon_{[n]}
					-
					\Upsilon
					)_{\ell=0} (x_R)
					\vert
					+
					\sum_{k \leq 3}
					\Vert
					\big[
					(r\nablaslash)^k \underline{S}_{[n]}
					-
					(r\nablaslash)^k \underline{S}
					\big]_{\ell \geq 1} (x_{-1})
					\Vert_{S_{\hat{u}_f+\delta,v_{-1}}}
					+
					\vert
					(
					T_{[n]}
					-
					T
					)_{\ell=0} (x_R)
					\vert,
		\end{align*}
		to be the sum of the differences between $\Omega \hat{\chi}_{[n]}(x_R)$ and $\Omega \hat{\chi}(x_R)$, and $\divslash \Omega \beta^{[n]}_{\ell=1} (x_R)$ and $\divslash \Omega \beta_{\ell=1} (x_R)$ etc.\@, where $x_R = x_R(\theta) = (\hat{u}_f+\delta,v(R,\hat{u}_f+\delta),\theta)$ and $x_{-1} = x_{-1}(\theta) = (\hat{u}_f+\delta,v_{-1},\theta)$.  Throughout this proof $v(R) = v(R,\hat{u}_f+\delta)$.

		\noindent \underline{\textbf{Estimates for Ricci coefficients of iterates $\Phi_{[n]}$:}}
		The first step is to obtain the following estimates for the Ricci coefficients of the $n$-th gauge, schematically denoted $\Phi_{[n]}$, along with non-sharp estimates for the diffeomorphisms $f_{[n]}$ and their derivatives:
		\begin{equation} \label{eq:PhinfnHnonsharp}
			\sup_{v_{-1} \leq v \leq v(R,\hat{u}_f+\delta)}
			\Big(
			\sum_{\vert \gamma \vert \leq 4}
			\Vert \mathfrak{D}^{\gamma} \Phi_{[n]} \Vert_{S_{\hat{u}_f+\delta,v}}
			+
			\sum_{\vert \gamma \vert \leq 6}
			\Vert \mathfrak{D}^{\gamma} f_{[n]} \Vert_{S_{\hat{u}_f+\delta,v}}
			\Big)
			\lesssim
			\varepsilon,
		\end{equation}
		for all $n \geq 1$, where the norm of the diffeomorphism functions is defined in \eqref{eq:iteratesdiffeonormsH} .  Since this is similar to, but less involved than, the main part of the proof, the estimates are only sketched.  The estimates for $f_{[n]}$ are non-sharp, and much easier to obtain than the sharp estimates, as they do not exploit the fact that the differences $\Phi_{[n]}(x) - \Phi(x)$ can be estimated by $\delta$, but only that $\Phi_{[n]}(x)$ and $\Phi(x)$ can individually be estimated by $\varepsilon$.

		First one estimates $\Omega^2 \alpha_{[n]}(\hat{u}_f+\delta,v)$ and $\Omega^{-2} \alphabar_{[n]}(\hat{u}_f+\delta,v)$ and their derivatives up to order $6$, for $v_{-1} \leq v \leq v(R,\hat{u}_f+\delta)$.  The estimates are obtained, as in the proof of Proposition \ref{thm:gidataestimates}, in terms of nonlinearities involving $\Phi$ and the diffeomorphisms $f_{[n]}$, and linear terms involving $\Pi_{S_{[n]}} \Omega^2 \alpha$ and $\Pi_{S_{[n]}} \Omega^{-2} \alphabar$ and their derivatives up to order $6$ (see for example equations \eqref{eq:curvaturecomp1} and \eqref{eq:curvaturecomp2}).  One then uses the fact that, for example,
		\begin{multline*}
			\sum_{k \leq 4} \Vert \Pi_{S_{[n]}} (r\nablaslash)^k \Omega^{-2} \alphabar \Vert_{S_{\hat{u}_f+\delta,v}}
			\\
			\lesssim
			\varepsilon \big(
			1
			+
			\sup_{v_{-1} \leq v \leq v(R,\hat{u}_f+\delta)}
			\sup_{\theta \in \mathbb{S}^2}
			(\vert f_{[n]}^3 (\hat{u}_f+\delta,v,\theta) \vert
			+
			\vert f_{[n]}^4 (\hat{u}_f+\delta,v,\theta) \vert
			+
			\sum_{k=0}^1
			\vert (r\nablaslash)^k \partial_v \slashed{f}_{[n]} (\hat{u}_f+\delta,v,\theta) \vert
			) \big),
		\end{multline*}
		using the estimates on $\alphabar$ and its derivatives up to order $6$ and Lemma \ref{lem:mvtH}.
		
		Next one estimates, see Lemma \ref{lem:fHnPhiest} and Lemma \ref{lem:f3Hnest}, the diffeomorphisms $f_{[n]}(u,v_{\infty}(\hat{u}_f+\delta))$ and their derivatives in terms of $\Phi_{[n]}$, $\Phi$ and their derivatives.  One then revisits the estimates of Chapter \ref{chap:Hestimates} (or rather the part of Chapter \ref{chap:Hestimates} involving the estimates on the hypersurface $u = u_f$) and estimates $\Phi_{[n]}(\hat{u}_f+\delta,v)$ and their derivatives up to order $4$ in terms of $\Omega^2 \alpha_{[n]}(\hat{u}_f+\delta,v)$, $\Omega^{-2} \alphabar_{[n]}(\hat{u}_f+\delta,v)$, and their derivatives, and
		\[
			\sum_{k \leq 4}
			\Vert
			(r\nablaslash)^k \Omega \hat{\chi}_{[n]}
			\Vert_{S_{\hat{u}_f+\delta,v(R)}}
			+
			\sum_{k \leq 6}
			\Vert
			(r\nablaslash)^k \underline{S}^{[n]}_{\ell \geq 1}
			\Vert_{S_{\hat{u}_f+\delta,v_{-1}}}
			+
			\vert
			\divslash \Omega \beta^{[n]}_{\ell=1} (x_R)
			\vert
			+
			\vert
			\Upsilon^{[n]}_{\ell=0} (x_R)
			\vert
			+
			\vert
			T^{[n]}_{\ell=0} (x_R)
			\vert.
		\]
		A simple induction argument, using the change of gauge relations \eqref{eq:metriccomp6}, \eqref{eq:Riccicomp1}, \eqref{eq:Riccicomp3}, \eqref{eq:Riccicomp7}, \eqref{eq:Riccicomp8}, \eqref{eq:curvaturecomp3}, \eqref{eq:curvaturecomp5}, the equations \eqref{eq:nHgauge1}, \eqref{eq:nHgauge2}, the estimates \eqref{eq:Fslashhat1}--\eqref{eq:Fslashhat3}, \eqref{eq:Hnewconebound}, and the relations \eqref{eq:h4j4spherediffeo}, \eqref{eq:h4nj4nf3nFslashed2} then completes the proof of the \eqref{eq:PhinfnHnonsharp} after taking $\delta$ sufficiently small.

		\noindent \underline{\textbf{Estimates for differences of Ricci coefficients $\Phi_{[n]} - \Phi$:}}
		The differences between the $n$-th Ricci coefficients and curvature components and those of the extended $\hat{u}_f$ normalised $\Hp$ gauge, $\Phi_{[n]}(x) - \Phi(x)$, \emph{identified by the values of their respective coordinate functions}, are estimated.  Revisiting the estimates of Chapter \ref{chap:Hestimates}, it follows that
		\begin{align}
			&
			\sum_{\vert \gamma \vert \leq 3}
			\sup_{v_{-1} \leq v \leq v(R)} \Vert \mathfrak{D}^{\gamma} \Phi_{[n]}(x) - \mathfrak{D}^{\gamma} \Phi(x) \Vert_{S_{\hat{u}_f+\delta,v}}
			+
			\Vert \mathfrak{D}^{\gamma} \Phi^{[n]}_{\ell = 0}(x) - \mathfrak{D}^{\gamma} \Phi_{\ell = 0}(x) \Vert_{L^1(C_{\hat{u}_f+\delta})}
			\nonumber
			\lesssim
			\mathcal{S}_{[n]}
			+
			\varepsilon \delta
			\\
			&
			\label{eq:Hfestimate1}
			+
			\sum_{ \vert \gamma \vert \leq 3}
			\Big[
			\Vert \mathfrak{D}^{\gamma} \Omega^2 \alpha_{[n]}(x) - \mathfrak{D}^{\gamma} \Omega^2 \alpha(x) \Vert_{S, C_{\hat{u}_f+\delta}}
			+
			\Vert \mathfrak{D}^{\gamma} \Omega^{-2} \alphabar_{[n]}(x) - \mathfrak{D}^{\gamma} \Omega^{-2} \alphabar(x) \Vert_{S, C_{\hat{u}_f+\delta}}
			\Big],
		\end{align}
		where
		\[
			\Vert F \Vert_{L^1(C_{\hat{u}_f+\delta})}
			=
			\int_{v_{-1}}^{v(R,\hat{u}_f+\delta)} \int_{S_{\hat{u}_f+\delta,v}} \vert F \vert d \theta dv,
			\qquad
			\Vert F \Vert_{S, C_{\hat{u}_f+\delta}}
			:=
			\sup_{v_{-1} \leq v \leq v(R,\hat{u}_f+\delta)}
			\Vert F \Vert_{S_{\hat{u}_f+\delta,v}}
			+
			\Vert F \Vert_{C_{\hat{u}_f+\delta}}.
		\]
		The difference between the metrics, $\gslash(x) - \gslash_{[n]}(x)$, are similarly be estimated.  The change of gauge relation \eqref{eq:metricid6}, Lemma \ref{lem:mvtH}, the estimates \eqref{eq:Fslashhat1}, \eqref{eq:Fslashhat2}, \eqref{eq:Hnewconebound}, and the relations \eqref{eq:h4nj4nf3nFslashed2}, \eqref{eq:h4nj4nf3nFslashed3}, imply that
		\[
			\vert \gslash (\hat{u}_f+\delta ,v(R),\theta)
			-
			\gslash_{[n]} (\hat{u}_f + \delta,v(R),\theta)
			\vert_{\gslash}
			\lesssim
			C(\hat{u}_f)\delta
			+
			\sum_{k \leq 1}
			\vert (r\nablaslash)^k f^4_{[n]}(\hat{u}_f + \delta,v(R),\theta) \vert.
		\]
		The difference $\gslash(x) - \gslash_{[n]}(x)$ is then estimated in terms of $\Phi(x) - \Phi_{[n]}(x)$ as in Proposition \ref{prop:gslashH}.

		\noindent \underline{\textbf{Estimates for differences $\Omega^2 \alpha_{[n]} - \Omega^2 \alpha$ and $\Omega^{-2} \alphabar_{[n]} - \Omega^{-2} \alphabar$:}}
		Next, the difference $\Omega^2 \alpha_{[n]}(x) - \Omega^2 \alpha(x)$ is estimated.  The relation \eqref{eq:curvaturecomp1} implies that, schematically,
		\[
			\Omega^2 \alpha_{[n]}(x) - \Pi_{S_{[n]}} \Omega^2 \alpha(x)
			=
			\Pi_{S_{[n]}} \Phi \cdot \mathfrak{D} f_{[n]}(x)
			+
			(\mathfrak{D} f_{[n]}(x))^2,
		\]
		and so
		\[
			\Vert \Omega^2 \alpha_{[n]}(x) - \Pi_{S_{[n]}} \Omega^2 \alpha(x) \Vert_{S, C_{\hat{u}_f+\delta}}
			\lesssim
			\varepsilon \sup_{v_{-1} \leq v \leq v(R)} \Vert \mathfrak{D} f_{[n]} \Vert_{S_{\hat{u}_f+\delta,v}}
			+
			\Vert (\mathfrak{D} f_{[n]})^2 \Vert_{S, C_{\hat{u}_f+\delta}}.
		\]
		Moreover, by Lemma \ref{lem:mvtH} (in particular \eqref{eq:diffiscloseiteratesHgaugetwo}),
		\begin{multline*}
			\Vert \Pi_{S_{[n]}} \Omega^2 \alpha(x) - \Omega^2 \alpha(x) \Vert_{S, C_{\hat{u}_f+\delta}}
			\lesssim
			C(\hat{u}_f) \delta
			\\
			+
			\varepsilon
			\sup_{v_{-1} \leq v \leq v(R,\hat{u}_f+\delta)}
			\sup_{\theta \in \mathbb{S}^2}
			(\vert f_{[n]}^3 (\hat{u}_f+\delta,v,\theta) \vert
			+
			\vert f_{[n]}^4 (\hat{u}_f+\delta,v,\theta) \vert
			+
			\sum_{k=0}^1
			\vert (r\nablaslash)^k \partial_v \slashed{f}_{[n]} (\hat{u}_f+\delta,v,\theta) \vert
			).
		\end{multline*}
		Similarly for higher order derivatives, and for $\Omega^{-2} \alphabar$.  It therefore follows that
		\begin{multline} \label{eq:Hfestimate2}
			\sum_{\vert \gamma \vert \leq 3}
			\Big[
			\Vert \mathfrak{D}^{\gamma} \Omega^2 \alpha_{[n]}(x) - \mathfrak{D}^{\gamma} \Omega^2 \alpha(x) \Vert_{S, C_{\hat{u}_f+\delta}}
			+
			\Vert \mathfrak{D}^{\gamma} \Omega^{-2} \alphabar_{[n]}(x) - \mathfrak{D}^{\gamma} \Omega^{-2} \alphabar(x) \Vert_{S, C_{\hat{u}_f+\delta}}
			\Big]
			\\
			\lesssim
			C(\hat{u}_f) \delta
			+
			\sum_{\vert \gamma \vert \leq 5}
			\Big[
			\varepsilon 
			\sup_{v_{-1} \leq v \leq v(R)}
			\Vert \mathfrak{D}^{\gamma} f_{[n]} \Vert_{S_{\hat{u}_f+\delta,v}}
			+
			\Vert \mathfrak{D}^{\gamma} f_{[n]} \Vert_{S, C_{\hat{u}_f+\delta}}^2
			\Big].
		\end{multline}

		\noindent \underline{\textbf{Estimates for diffeomorphisms of iterates $f_{[n]}$:}}
		By Lemma \ref{lem:fHnPhiest}, using also the fact that, for each $\Phi$ and $\vert \gamma \vert \leq 3$,
		\begin{multline*}
			\Vert \mathfrak{D}^{\gamma} \Phi(x) - \Pi_{S_{[n]}} \mathfrak{D}^{\gamma} \Phi(x) \Vert_{S_{\hat{u}_f+\delta,v}}
			\lesssim
			C(\hat{u}_f) \delta
			\\
			+
			\varepsilon
			\sup_{v_{-1} \leq v \leq v(R,\hat{u}_f+\delta)}
			\sup_{\theta \in \mathbb{S}^2}
			(\vert f_{[n]}^3 (\hat{u}_f+\delta,v,\theta) \vert
			+
			\vert f_{[n]}^4 (\hat{u}_f+\delta,v,\theta) \vert
			+
			\sum_{k=0}^1
			\vert (r\nablaslash)^k \partial_v \slashed{f}_{[n]} (\hat{u}_f+\delta,v,\theta) \vert
			),
		\end{multline*}
		by Lemma \ref{lem:mvtH}, it follows that, for $\vert \gamma \vert \leq 5$,
		\begin{multline} \label{eq:Hfestimate3}
			\sup_{v_{-1} \leq v \leq v(R,u)}
			\Vert \mathfrak{D}^{\gamma} f_{[n]} \Vert_{S_{\hat{u}_f+\delta,v}}
			\lesssim
			C(\hat{u}_f) \delta
			+
			\sup_{v_{-1} \leq v \leq v(R,\hat{u}_f+\delta)}
			\Big[
			\sum_{ \vert \tilde{\gamma} \vert \leq 3}
			\Vert \mathfrak{D}^{\tilde{\gamma}} \Phi_{[n]}(x) - \mathfrak{D}^{\tilde{\gamma}} \Phi(x) \Vert_{S_{\hat{u}_f+\delta,v}}
			\\
			+
			\sum_{ \vert \tilde{\gamma} \vert \leq 3}
			\Vert \mathfrak{D}^{\tilde{\gamma}} \Phi^{[n]}_{\ell = 0}(x) - \mathfrak{D}^{\tilde{\gamma}} \Phi_{\ell = 0}(x) \Vert_{L^1(C_{\hat{u}_f+\delta})}
			+
			C(\hat{u}_f)
			\sum_{ \vert \tilde{\gamma} \vert \leq 5}
			\Vert \mathfrak{D}^{\tilde{\gamma}} f_{[n]} \Vert_{S_{\hat{u}_f+\delta,v}}^2
			\Big],
		\end{multline}

		\noindent \underline{\textbf{Estimate for $\mathcal{S}_{[n]}$:}}
		Consider now $\mathcal{S}_{[n]}$.  The change of gauge relations \eqref{eq:metriccomp6}, \eqref{eq:Riccicomp3}, \eqref{eq:Riccicomp7}, \eqref{eq:Riccicomp8}, \eqref{eq:curvaturecomp5}, the relations \eqref{eq:h4nj4nf3nFslashed1} and \eqref{eq:h4nj4nf3nFslashed2}, and equation \eqref{eq:nHgauge1} imply that
		\begin{align*}
			&
			\Big[
			\Deltaslash \underline{S}_{[n]} (x_{-1})
			+
			\frac{3}{r^2}\Omega^2 \underline{S}_{[n]} (x_{-1})
			\Big]_{(\ell \geq 1)_{[n-1]}}
			=
			\Big[
			\Deltaslash \underline{S} (x_{-1}+f_{[n]}(x_{-1}))
			+
			\frac{3}{r^2} \Omega^2 \underline{S} (x_{f_{[n]}})
			\\
			&
			\qquad \qquad \quad
			+
			2 \Deltaslash \Deltaslash h^4_{[n]}(\theta)
			+
			\frac{2}{r^2} \left( 5 - \frac{12 M_f}{r} \right) \Deltaslash h^4_{[n]}(\theta)
			+
			\frac{12 \Omega_{\circ}^4}{r^4} h^4_{[n]}(\theta)
			-
			\mathfrak{K}_{[n]}(\theta)
			\\
			&
			\qquad \qquad \quad
			+
			\frac{6}{r^3} (\Omega^2 - \Omega_{\circ}^2)(x_{f_{[n]}})
			-
			\frac{2}{r} \big( \mu_{[n]}^*(x_{-1}) - \mu^*(x_{f_{[n]}}) \big)
			-
			A^1_{[n]}(x_{-1})
			-
			\mathcal{E}^1_{[n]}(x_{-1})\Big]_{(\ell \geq 1)_{[n-1]}}
			\\
			&
			\qquad \qquad
			=
			\Big[
			\Deltaslash \underline{S} (x_{f_{[n]}})
			-
			\Deltaslash \underline{S} (x_{f_{[n-1]}})
			+
			\frac{3}{r^2} \Omega^2 \underline{S} (x_{f_{[n]}})
			-
			\frac{3}{r^2} \Omega^2 \underline{S} (x_{f_{[n-1]}})
			-
			\mathfrak{K}_{[n]}(\theta)
			+
			\mathfrak{K}_{[n-1]}(\theta)
			\\
			&
			\qquad \qquad \quad
			+
			\frac{6}{r^3} (\Omega^2 - \Omega_{\circ}^2)(x_{f_{[n]}})
			-
			\frac{6}{r^3} (\Omega^2 - \Omega_{\circ}^2)(x_{f_{[n-1]}})
			+
			A^1_{[n-1]}(x_{-1})
			-
			A^1_{[n]}(x_{-1})
			\\
			&
			\qquad \qquad \quad
			+
			\frac{2}{r} \big( \mu_{[n-1]}^*(x_{-1}) - \mu^*(x_{f_{[n-1]}}) \big)
			-
			\frac{2}{r} \big( \mu_{[n]}^*(x_{-1}) - \mu^*(x_{f_{[n]}}) \big)
			+
			\mathcal{E}_{[n-1]}^1(x_{-1})
			-
			\mathcal{E}_{[n]}^1(x_{-1})\Big]_{(\ell \geq 1)_{[n-1]}} ,
		\end{align*}
		where $x_{-1} = (\hat{u}_f+\delta,v_{-1},\theta)$, $x_{f_{[n]}} = x_{-1}+f_{[n]}(x_{-1})$ is as in \eqref{eq:Hgaugexminusone}, and 
		\begin{align*}
			\mathfrak{K}_{[n]}(\theta)
			=
			\frac{12 \Omega_{\circ}^4}{r^4}
			K(v_{-1},\slashed{F}_{[n]}(x_{-1}))
			+
			\frac{6\Omega^2_{\circ}}{r^2}
			\Deltaslash \big(K(v_{-1},\slashed{F}_{[n]}(x_{-1})) \big).
		\end{align*}
		Note that Lemma \ref{lem:fHnPhiest} and Lemma \ref{lem:f3Hnest} imply that $\mathfrak{K}_{[n]}$, $A_{[n]}^1$ and $\mathcal{E}_{[n]}^1$ satisfy, if $\delta$ is sufficiently small,
		\[
			\Vert \mathfrak{K}_{[n]} - \mathfrak{K}_{[n-1]}\Vert_{S_{\hat{u}_f+\delta,v_{-1}}}
			+
			\Vert A_{[n]}^1 - A_{[n-1]}^1 \Vert_{S_{\hat{u}_f+\delta,v_{-1}}}
			+
			\Vert \mathcal{E}_{[n]}^1 - \mathcal{E}_{[n-1]}^1 \Vert_{S_{\hat{u}_f+\delta,v_{-1}}}
			\lesssim
			\varepsilon
			\sum_{\vert \gamma \vert \leq 5}
			\Vert \mathfrak{D}^{\gamma} (f_{[n]} - f_{[n-1]}) \Vert_{S_{\hat{u}_f+\delta,v_{-1}}}
			.
		\]
		An elliptic estimate, together with the fact that
		\[
			\Vert
			\mu_{[n]}^*(x)
			-
			\mubar^*(x)
			\Vert_{S_{\hat{u}_f+\delta,v_{-1}}}
			\lesssim
			\varepsilon \delta
			+
			\varepsilon
			\sup_{v_{-1} \leq v \leq v(R)} \Vert \Phi_{[n]}(x) - \Phi(x) \Vert_{S_{\hat{u}_f+\delta,v}},
		\]
		then gives, for $k \leq 3$,
		\begin{multline*}
			\Vert
			(r\nablaslash)^k \underline{S}^{[n]}_{\ell\geq 1}(\hat{u}_f+\delta,v_{-1})
			\Vert_{S_{\hat{u}_f+\delta,v_{-1}}}
			\lesssim
			\varepsilon \delta
			+
			\sum_{l=n-1}^n
			\Big[
			\varepsilon
			\sum_{\vert \gamma \vert \leq 3}
			\sup_{v_{-1} \leq v \leq v(R)}
			\Vert \mathfrak{D}^{\gamma} \Phi_{[l]}(x) - \mathfrak{D}^{\gamma} \Phi(x) \Vert_{S_{\hat{u}_f+\delta,v}}
			\\
			+
			\sum_{\vert \gamma \vert \leq 5}
			\big(
			\varepsilon
			\Vert \mathfrak{D}^{\gamma} f_{[l]} \Vert_{S_{\hat{u}_f+\delta,v_{-1}}}
			+
			\Vert \mathfrak{D}^{\gamma} f_{[l]} \Vert_{S_{\hat{u}_f+\delta,v_{-1}}}^2
			\big)
			\Big].
		\end{multline*}
		Similarly for $T^{[n]}_{\ell=0}$, Lemma \ref{lem:trchipartialHgauge} and equation \eqref{eq:nHgauge2} imply that
		\begin{align*}
			&
			(1-A^{-1}) \Omega^{-2} T^{[n]}_{\ell=0}(x_R)
			=
			(1-A^{-1}) \Omega^{-2} T_{\ell = 0} (x_R + f_{[n]}(x_R))
			-
			\frac{2}{r\Omega_{\circ}^{2}}(\Omega^2 - \Omega_{\circ}^2)_{\ell=0}(x_R+f_{[n]}(x_R))
			\\
			&
			\quad
			-
			BA^{-1} \Big( \Omega^{-2} \Upsilon^{[n]}_{\ell=0}(x_R) - \Omega^{-2} \Upsilon_{\ell=0}(x_R + f_{[n]}(x_R)) \Big)
			+
			\frac{4}{R^2} \left( 1 - \frac{3M_f}{R} \right) (j^3 - j^4_{[n]})_{\ell=0}
			+
			A_{[n]}^2
			+
			\mathcal{E}_{[n]}^2
			\\
			&
			=
			\mathcal{E}_{[n]}^2
			-
			\mathcal{E}_{[n-1]}^2
			+
			A_{[n]}^2
			-
			A_{[n-1]}^2
			+
			(1-A^{-1})
			\big(
			\Omega^{-2} 
			T_{\ell = 0} (x_R + f_{[n]}(x_R))
			-
			\Omega^{-2} 
			T_{\ell = 0} (x_R + f_{[n-1]}(x_R))
			\big)
			\\
			&
			\quad
			-
			\Big(
			\frac{2}{r\Omega_{\circ}^{2}} (\Omega^2 - \Omega_{\circ}^2)_{\ell=0}(x_R+f_{[n]}(x_R))
			-
			\frac{2}{r\Omega_{\circ}^{2}} (\Omega^2 - \Omega_{\circ}^2)_{\ell=0}(x_R+f_{[n-1]}(x_R))
			\Big)
			\\
			&
			\quad
			-
			\frac{B}{A}
			\Big(
			\Omega^{-2} \Upsilon^{[n]}_{\ell=0}(x_R)
			-
			\Omega^{-2} \Upsilon^{[n-1]}_{\ell=0}(x_R)
			-
			\Omega^{-2} \Upsilon_{\ell=0}(x_R + f_{[n]}(x_R))
			+
			\Omega^{-2} \Upsilon_{\ell=0}(x_R + f_{[n-1]}(x_R))
			\Big),
		\end{align*}
		and so it similarly follows that
		\begin{multline*}
			\vert T^{[n]}_{\ell=0}(x_R) \vert
			\lesssim
			\sum_{l=n-1}^n
			\vert
			\Upsilon^{[l]}_{\ell=0}(x_R)
			-
			\Upsilon_{\ell=0}(x_R + f_{[l]}(x_R))
			\vert
			\\
			+
			\sum_{l=n-1}^n
			\Big[
			\varepsilon
			\sum_{\vert \gamma \vert \leq 3}
			\sup_{v_{-1} \leq v \leq v(R)}
			\Vert \mathfrak{D}^{\gamma} \Phi_{[l]}(x) - \mathfrak{D}^{\gamma} \Phi(x) \Vert_{S_{\hat{u}_f+\delta,v(R)}}
			+
			\sum_{\vert \gamma \vert \leq 2}
			\big(
			\varepsilon
			\Vert \mathfrak{D}^{\gamma} f_{[l]} \Vert_{S_{\hat{u}_f+\delta,v(R)}}
			+
			\Vert \mathfrak{D}^{\gamma} f_{[l]} \Vert_{S_{\hat{u}_f+\delta,v(R)}}^2
			\big)
			\Big].
		\end{multline*}
		
		Consider now the remaining terms in $\mathcal{S}_{[n]}$.  First, the relation \eqref{eq:Riccicomp1} implies that, schematically,
		\[
			\Omega \hat{\chi}_{[n]}(x_R)
			=
			\Pi_{S_{[n]}} \Omega \hat{\chi}(x_R)
			-
			2 \Omega^2 \Dslash_2^* \nablaslash  f_{[n]}^3 (x_R)
			+
			\sum_{1 \leq \vert \gamma_1\vert, \vert \gamma_2 \vert \leq 2}
			\Pi_{S_{[n]}} \Phi \cdot \mathfrak{D}^{\gamma_1} f_{[n]}(x_R)
			+
			\mathfrak{D}^{\gamma_1} f_{[n]} \cdot \mathfrak{D}^{\gamma_2} f_{[n]}(x_R),
		\]
		and so it follows from Lemma \ref{lem:mvtH} and Lemma \ref{lem:f3Hnest}, if $\delta$ is sufficiently small, that
		\begin{align*}
			&
			\sum_{k \leq 3}
			\Vert (r\nablaslash)^k \Omega \hat{\chi}_{[n]}(x_R) - (r\nablaslash)^k \Omega \hat{\chi}(x_R) \Vert_{S_{\hat{u}_f+\delta,v(R)}}
			\lesssim
			C(\hat{u}_f) \delta
			+
			\sum_{\vert \gamma \vert \leq 5}
			\Big(
			\varepsilon
			\Vert \mathfrak{D}^{\gamma} f_{[n]} \Vert_{S_{\hat{u}_f+\delta,v(R)}}
			+
			\Vert \mathfrak{D}^{\gamma} f_{[n]} \Vert_{S_{\hat{u}_f+\delta,v(R)}}^2
			\Big).
		\end{align*}
		Similarly for $\divslash \beta_{\ell = 1}$ and $\Upsilon_{\ell=0}$.
		
		It then follows that
		\begin{multline} \label{eq:Hfestimate5}
			\mathcal{S}_{[n]}
			\lesssim
			C(\hat{u}_f) \delta
			+
			\sup_{v_{-1} \leq v \leq v(R)}
			\sum_{l=n-1}^n
			\Big[
			\sum_{\vert \gamma \vert \leq 5}
			\big(
			\varepsilon
			\Vert \mathfrak{D}^{\gamma} f_{[l]} \Vert_{S_{\hat{u}_f+\delta,v}}
			+
			\Vert \mathfrak{D}^{\gamma} f_{[l]} \Vert_{S_{\hat{u}_f+\delta,v}}^2
			\big)
			\\
			+
			\varepsilon
			\sum_{\vert \gamma \vert \leq 3}
			\Vert \mathfrak{D}^{\gamma} \Phi_{[l]}(x) - \mathfrak{D}^{\gamma} \Phi(x) \Vert_{S_{\hat{u}_f+\delta,v}}
			\Big].
		\end{multline}

		\noindent \underline{\textbf{The completion of the proof:}}
		The estimates \eqref{eq:Hfestimate1}, \eqref{eq:Hfestimate2}, \eqref{eq:Hfestimate3}, \eqref{eq:Hfestimate5} now combine to give
		\[
			\sup_{v_{-1} \leq v \leq v(R)}
			\sum_{k \leq 5} \Vert \mathfrak{D}^k f_{[n]} \Vert_{S_{\hat{u}_f+\delta,v}}
			\lesssim
			C(\hat{u}_f) \delta
			+
			\varepsilon
			\sup_{v_{-1} \leq v \leq v(R)}
			\sum_{k \leq 5} \Vert \mathfrak{D}^k f_{[n-1]} \Vert_{S_{\hat{u}_f+\delta,v}},
		\]
		provided $\varepsilon$ is sufficiently small (independent of $\hat{u}_f$) and $\delta$ is sufficiently small with respect to $\hat{u}_f$. The proof then follows if $C_1(\hat{u}_f)$ is chosen to be sufficiently large.

\end{proof}

	The next part of the proof involves showing that the map which takes the $n$-th sphere to $n+1$-th sphere is a contraction.
	
	\begin{lemma}[Estimates for differences of iterates] \label{lem:fHexistencediff}
		Provided $\hat{\varepsilon}_0$ is sufficiently small (independent of $\hat{u}_f$), and $\delta_0$ is sufficiently small (with respect to $\hat{u}_f$), the diffeomorphisms $f_{[n]}$ satisfy, for all $n \geq 1$ and all $v_{-1} \leq v \leq v(R,\hat{u}_f+\delta)$, the estimates
		\[
			\sum_{k\leq 5}
			\Vert \mathfrak{D}^k f_{[n+1]} - \mathfrak{D}^k f_{[n]} \Vert_{S_{\hat{u}_f+\delta,v}}
			\leq
			2^{-n}
			,
		\]
		where the norm is as in \eqref{eq:iteratesdiffeonormsH}, and, in particular, $h_{[n]}^4$ and $j_{[n]}^4$ satisfy
		\[
			\vert j^4_{[n+1]} - j^4_{[n]} \vert
			+
			\sum_{k\leq 5}
			\Vert (r\nablaslash)^k (h^4_{[n+1]} - h^4_{[n]}) \Vert_{S_{\hat{u}_f+\delta,v}}
			\leq
			2^{-n}
			.
		\]
	\end{lemma}
	
	\begin{proof}
		The proof is similar to the proof of Lemma \ref{lem:fnH} and is divided into several steps, which follow closely the corresponding steps of Lemma \ref{lem:fnH}.  See also Lemma \ref{lem:fIexistencediff}.  Again, throughout this proof the notation $\underline{S} := \Omega^{-2}(\Omega \tr \chibar - \Omega \tr \chibar_{\circ,M_{\hat{u}_f+\delta}})$, $T := \Omega \tr \chi - \Omega \tr \chi_{\circ,M_{\hat{u}_f+\delta}}$ is used.  Define now, for $n \geq 1$,
				\begin{align*}
					&
					\mathfrak{s}_{[n]}
					=
					\sum_{k \leq 3}
					\Vert
					(r\nablaslash)^k \Omega \hat{\chi}_{[n]}(x_R)
					-
					(r\nablaslash)^k \Omega \hat{\chi}_{[n-1]}(x_R)
					\Vert_{S_{\hat{u}_f+\delta,v(R)}}
					+
					\vert
					(
					\divslash \Omega \beta_{[n]}
					-
					\divslash \Omega \beta_{[n-1]}
					)_{\ell=1} (x_R)
					\vert
					\\
					&
					\
					+
					\vert
					(
					\Upsilon_{[n]}
					-
					\Upsilon_{[n-1]}
					)_{\ell=0} (x_R)
					\vert
					+
					\sum_{k \leq 3}
					\Vert
					\big[
					(r\nablaslash)^k \underline{S}_{[n]}
					-
					(r\nablaslash)^k \underline{S}_{[n-1]}
					\big]_{\ell \geq 1} (x_{-1})
					\Vert_{S_{\hat{u}_f+\delta,v_{-1}}}
					+
					\vert
					(
					T_{[n]}
					-
					T_{[n-1]}
					)_{\ell=0} (x_R)
					\vert,
		\end{align*}
		(where quantities are identified by the values of their respective coordinates) where again $x_R = (\hat{u}_f+\delta,v(R,\hat{u}_f+ \delta),\theta)$ and $v(R) = v(R,\hat{u}_f+\delta)$.

		\noindent \underline{\textbf{Estimates for differences of Ricci coefficients $\Phi_{[n+1]} - \Phi_{[n]}$:}}		
		Considering the equations satisfied by the differences $\Phi_{[n+1]}(x) - \Phi_{[n]}(x)$ and revisiting the estimates of Chapter \ref{chap:Hestimates}, it follows from Lemma \ref{lem:fnH} that
				\begin{align}
					&
					\sum_{\vert \gamma \vert \leq 3}
					\Big[
					\sup_{v_{-1} \leq v \leq v(R)}
					\Vert \mathfrak{D}^{\gamma} \Phi_{[n+1]}(x) - \mathfrak{D}^{\gamma} \Phi_{[n]}(x) \Vert_{S_{\hat{u}_f+\delta,v}}
					+
					\Vert \mathfrak{D}^{\gamma} \Phi^{[n+1]}_{\ell = 0}(x) - \mathfrak{D}^{\gamma} \Phi^{[n]}_{\ell = 0}(x) \Vert_{L^1(C_{\hat{u}_f+\delta})}
					\Big]
					\label{eq:Hfdiffestimate3}
					\\
					&
					\lesssim
					\mathfrak{s}_{[n+1]}
					+
					\sum_{ \vert \gamma \vert \leq 3}
					\Big[
					\Vert \mathfrak{D}^{\gamma} \Omega^2 \alpha_{[n+1]}(x) - \mathfrak{D}^{\gamma} \Omega^2 \alpha_{[n]}(x) \Vert_{S, C_{\hat{u}_f+\delta}}
					+
					\Vert \mathfrak{D}^{\gamma} \Omega^{-2} \alphabar_{[n+1]}(x) - \mathfrak{D}^{\gamma} \Omega^{-2} \alphabar_{[n]}(x) \Vert_{S, C_{\hat{u}_f+\delta}}
					\Big],
					\nonumber
				\end{align}
				where the norms $\Vert \cdot \Vert_{L^1(C_{\hat{u}_f+\delta})}$ and $\Vert \cdot \Vert_{S, C_{\hat{u}_f+\delta}}$ are as in the proof of Lemma \ref{lem:fnH}.
				The metric differences $\gslash_{[n+1]}(x) - \gslash_{[n]}(x)$ are similarly be estimated, as in Proposition \ref{prop:gslashH}, using now the fact that
				\[
					\vert \gslash_{[n+1]} (\hat{u}_f+\delta ,v(R),\theta)
					-
					\gslash_{[n]} (\hat{u}_f + \delta,v(R),\theta)
					\vert_{\gslash}
					\lesssim
					\sum_{k \leq 1}
					\vert (r\nablaslash)^k (f^4_{[n+1]} - f^4_{[n]})(\hat{u}_f + \delta,v(R),\theta) \vert,
				\]
				which follows from the change of gauge relation \eqref{eq:metricid6}, Lemma \ref{lem:mvtH}, the estimate \eqref{eq:Hnewconebound}, and the relations \eqref{eq:h4nj4nf3nFslashed2}, \eqref{eq:h4nj4nf3nFslashed3}.
		
		\noindent \underline{\textbf{Estimates for differences $\alpha_{[n+1]} - \alpha_{[n]}$ and $\alphabar_{[n+1]} - \alphabar_{[n]}$:}}
		Consider now the differences $\Omega^2 \alpha_{[n+1]}(x) - \Omega^2 \alpha_{[n]}(x)$ and $\Omega^{-2} \alphabar_{[n+1]}(x) - \Omega^{-2} \alphabar_{[n]}(x)$.  The relation \eqref{eq:curvaturecomp1} implies that, schematically,
		\begin{multline*}
					\Omega^2 \alpha_{[n+1]}(x) - \Omega^2 \alpha_{[n]}(x)
					=
					\Pi_{S_{[n+1]}} \Omega^2 \alpha(x) - \Pi_{S_{[n]}} \Omega^2 \alpha(x)
					\\
					+
					\Pi_{S_{[n+1]}} \Phi \cdot \mathfrak{D}^{2} f_{[n+1]}(x)
					-
					\Pi_{S_{[n]}} \Phi \cdot \mathfrak{D}^{2} f_{[n]}(x)
					+
					(\mathfrak{D}^{2} f_{[n+1]}(x))^2
					-
					(\mathfrak{D}^{2} f_{[n]}(x))^2,
		\end{multline*}
		and so, using Lemma \ref{lem:mvtH} (in particular \eqref{eq:diffiscloseiteratesdifferenceHgaugetwo}),
		\begin{equation} \label{eq:Hfdiffestimate4}
					\sum_{\vert \gamma \vert \leq 3}
					\Vert \mathfrak{D}^{\gamma} \Omega^2 \alpha_{[n+1]}(x) - \mathfrak{D}^{\gamma} \Omega^2 \alpha_{[n]}(x) \Vert_{S, C_{\hat{u}_f+\delta}}
					\lesssim
					\varepsilon
					\sup_{v_{-1} \leq v \leq v(R,\hat{u}_f+\delta)}
					\sum_{\vert \gamma \vert \leq 5}
					\Vert \mathfrak{D}^{\gamma} f_{[n+1]}(x) - \mathfrak{D}^{\gamma} f_{[n]}(x) \Vert_{S_{\hat{u}_f+\delta,v}}.
		\end{equation}
		Similarly for $\Omega^{-2} \alphabar_{[n+1]}(x) - \Omega^{-2} \alphabar_{[n]}(x)$.

		\noindent \underline{\textbf{Estimates for differences of diffeomorphisms $f_{[n+1]}(x) - f_{[n]}(x)$:}}
		The next step involves estimating the differences $\mathfrak{D}^{\gamma} f_{[n+1]} - \mathfrak{D}^{\gamma} f_{[n]}$.  Considering differences of the system \eqref{eq:metriccomp1}--\eqref{eq:curvaturecomp6} for $f_{[n+1]}$ and $f_{[n]}$ and following again Lemma \ref{lem:fHnPhiest}, it follows that, for $\vert \gamma \vert \leq 5$, if $\varepsilon$ is sufficiently small,	
				\begin{multline*}
					\sup_{v_{-1} \leq v \leq v(R)}
					\Vert \mathfrak{D}^{\gamma} f_{[n+1]}(x)
					-
					\mathfrak{D}^{\gamma} f_{[n]}(x) \Vert_{S_{\hat{u}_f+\delta,v}}
					\\
					\lesssim
					\sup_{v_{-1} \leq v \leq v(R)}
					\Big[
					\sum_{ \vert \tilde{\gamma} \vert \leq 3}
					\Vert \mathfrak{D}^{\tilde{\gamma}} \Phi_{[n+1]}(x) - \mathfrak{D}^{\tilde{\gamma}} \Phi_{[n]}(x) \Vert_{S_{\hat{u}_f+\delta,v}}
					+
					\sum_{ \vert \tilde{\gamma} \vert \leq 3}
					\Vert \mathfrak{D}^{\tilde{\gamma}} \Phi^{[n+1]}_{\ell = 0}(x) - \mathfrak{D}^{\tilde{\gamma}} \Phi^{[n]}_{\ell = 0}(x) \Vert_{L^1(C_{\hat{u}_f+\delta})}
					\\
					+
					C(\hat{u}_f)
					\sum_{ \vert \gamma_1 \vert, \vert \gamma_2 \vert \leq 5}
					(
					\Vert \mathfrak{D}^{\gamma_1} f_{[n+1]} \Vert_{S_{\hat{u}_f+\delta,v}}
					+
					\Vert \mathfrak{D}^{\gamma_1} f_{[n]} \Vert_{S_{\hat{u}_f+\delta,v}}
					)
					\Vert \mathfrak{D}^{\gamma_2} f_{[n+1]} - \mathfrak{D}^{\gamma_2} f_{[n]} \Vert_{S_{\hat{u}_f+\delta,v}}
					\Big],
				\end{multline*}
				where Lemma \ref{lem:mvtH} (in particular \eqref{eq:diffiscloseiteratesdifferenceHgaugetwo}) has again been used.
		
		\noindent \underline{\textbf{Estimate for $\mathfrak{s}_{[n+1]}$:}}
		As in the proof of Lemma \ref{lem:fnH}, using now the fact that
		\[
			\Vert
			\mu_{[n+1]}^*(x_{-1})
			-
			\mu_{[n]}^*(x_{-1})
			\Vert_{S_{\hat{u}_f+\delta,v_{-1}}}
			\lesssim
			\varepsilon
			\sup_{v_{-1} \leq v \leq v(R)} \Vert \Phi_{[n+1]}(x) - \Phi_{[n]}(x) \Vert_{S_{\hat{u}_f+\delta,v}},
		\]
		it follows that, if $\delta$ is sufficiently small,
				\begin{align}
					\mathfrak{s}_{[n+1]}
					\lesssim
					\
					&
					\label{eq:Hfdiffestimate5}
					\varepsilon
					\sum_{l=n-1}^n
					\Big[
					\sum_{\vert \gamma \vert \leq 3}
					\sup_{v_{-1} \leq v \leq v(R)}
					\Vert \mathfrak{D}^{\gamma} \Phi_{[l]}(x) - \mathfrak{D}^{\gamma} \Phi_{[l-1]}(x) \Vert_{S_{\hat{u}_f+\delta,v}}
					+
					\sum_{\vert \gamma \vert \leq 5}
					\Vert \mathfrak{D}^{\gamma} f_{[l]} - \mathfrak{D}^{\gamma} f_{[l-1]} \Vert_{S_{\hat{u}_f+\delta,v}}
					\Big].
				\end{align}

		\noindent \underline{\textbf{The completion of the proof:}}
		The result then follows from the estimates \eqref{eq:Hfdiffestimate3}--\eqref{eq:Hfdiffestimate5} if $\hat{\varepsilon}_0$ is sufficiently small (independent of $\hat{u}_f$) and $\delta_0$ is sufficiently small with respect to $\hat{u}_f$.

	\end{proof}

	It follows from Lemma \ref{lem:fHexistencediff} that the sequences $\{h^4_{[n]}\}$ and $\{j^4_{[n]}\}$ converge to limits $h^4$ and $j^4$ respectively with $h^4_{\ell =0} = j^4_{\ell \geq 1} =0$.  Moreover, the limits $h^4$ and $j^4$ are the unique solution of the system \eqref{eq:h4Hgaugeeqn1}--\eqref{eq:h4Hgaugeeqn2}.  The function $h^4$ is smooth in view of the smoothness of the ambient spacetime, the smoothness of the $\hat{u}_f$ normalised $\I$ gauge, and the fact that $h^4$ solves \eqref{eq:h4Hgaugeeqn1}--\eqref{eq:h4Hgaugeeqn2}.   The expressions \eqref{eq:metriccomp6}, \eqref{eq:Riccicomp7}, \eqref{eq:Riccicomp8}, \eqref{eq:curvaturecomp5} then imply that the spacetime double null foliation defined by the sphere arising from Lemma \ref{lem:h4j4sphere} and Proposition \ref{prop:Hconesfoliations} is the desired gauge.  The continuity in $\delta$ of the energies $\mathbb{E}^{N-2}_{\hat{u}_f+\delta}[P_{\Hp},\Pbar_{\Hp}]$, $\mathbb{E}^{N}_{\hat{u}_f+\delta}[\alpha_{\Hp},\alphabar_{\Hp}]$, $\mathbb{E}^N_{\hat{u}_f+\delta,\Hp}$ of the geometric quantities of the $\hat{u}_f + \delta$ normalised $\Hp$ gauge with respect to $M_f(\hat{u}_f+\delta,\lambda)$, and the associated diffeomorphism energies $\mathbb{E}^{N+2}_{\hat{u}_f+\delta}[f_{\Hp,\I}]$ and $\mathbb{E}_{\hat{u}_f+\delta}[f_{d,\Hp}]$, follows softly from continuity.

\end{proof}

\begin{remark} \label{rmk:linearHgaugesphere}
	Again, the proof of Theorem \ref{thm:Hgaugeexistence} can be adapted to the linearised setting of Proposition \ref{proplinHpgauge}, where it simplifies considerably, to complete the proof of Proposition \ref{proplinHpgauge} as follows.  Recall again the notation of~\cite{holzstabofschw}.
	Define
	\[
		\accentset{\scalebox{.6}{\mbox{\tiny (1)}}}{\Upsilon}
		=
		\Big( 1 - \frac{3M_f}{r} \Big) \rlin
		+
		\frac{3M_f}{2r^2} (\otx - \otxb)
		-
		\frac{3M_f \Omega^2}{r^3} 2\Olin,
	\]
	and recall $A$ and $B$ from Lemma \ref{lem:trchipartialHgauge}.
	Any function $h^4 \colon \mathbb{S}^2 \to \mathbb{R}$ gives rise to a function $f^4(v,\theta) = h^4(\theta)$ for all $v$, which generates a pure gauge solution (together with $f^3 \equiv 0$), denoted $\mathscr{G}_{\mathrm{sphere}}$.  In view of the linearised analogues of the relations \eqref{eq:metriccomp6}, \eqref{eq:Riccicomp3}, \eqref{eq:Riccicomp7}, \eqref{eq:Riccicomp8}, \eqref{eq:curvaturecomp5}, and in anticipation of the linearised analogues of the gauge normalisations \eqref{eq:Hfoliation1}--\eqref{eq:Hfoliation5}, to achieve the linearised analogues of the gauge normalisations \eqref{eq:Hextragauge1}--\eqref{eq:Hextragauge2} one considers the solution $\mathscr{S} +\mathscr{G}_{\mathrm{sphere}}$ arising from $h^4$ satisfying the linearised analogues of equations \eqref{eq:h4Hgaugeeqn1}--\eqref{eq:h4Hgaugeeqn3}, namely the elliptic equations
	\begin{align*}
		&
		2 \Deltaslash \Deltaslash h^4_{\ell \geq 1}
		+
		\frac{2}{r^2} \Big( 5 - \frac{12 M_f}{r} \Big) \Deltaslash h^4_{\ell \geq 1}
		+
		\frac{12 \Omega^4}{r^4} h^4_{\ell \geq 1}
		\\
		&
		\qquad \qquad \qquad
		=
		-
		\big(
		\Omega^{-2} \Deltaslash \otxb
		+
		\frac{3}{r^2} \otxb
		+
		\frac{6 \Omega^2}{r^3} 2 \Olin
		+
		\frac{2}{r}
		\divslash \elin
		+
		\frac{2}{r}
		\rlin
		-
		\frac{3}{r^2} \otx
		\big)_{\ell \geq 1}^{\mathscr{S}}(u_f,v_{-1},\cdot),
		\\
		&
		\frac{4\Omega^2}{r^2} \Big( 1 - \frac{3M_f}{r} \Big) h^4_{\ell = 0}
		=
		-
		\big(
		\otx
		-
		\otxb
		-
		\frac{2\Omega^2}{r} 2 \Olin
		-
		\frac{B}{A}
		\accentset{\scalebox{.6}{\mbox{\tiny (1)}}}{\Upsilon}
		\big)_{\ell = 0}^{\mathscr{S}}(u_f,v(R,u_f)).
	\end{align*}
	The proof of Theorem \ref{thm:Hgaugeexistence} in this linearised setting reduces to showing the existence of a function $h^4$ solving these equations.  Indeed, given such a function $h^4$, it follows from the linearised analogues of the relations \eqref{eq:metriccomp6}, \eqref{eq:Riccicomp3}, \eqref{eq:Riccicomp7}, \eqref{eq:Riccicomp8}, \eqref{eq:curvaturecomp5} that
	\begin{equation} \label{eq:Hlinearexistenceafterspherespgs1}
		\big(
		\Omega^{-2} \Deltaslash \otxb
		+
		\frac{3}{r^2} \otxb
		+
		\frac{6 \Omega^2}{r^3} 2 \Olin
		+
		\frac{2}{r}
		\divslash \elin
		+
		\frac{2}{r}
		\rlin
		-
		\frac{3}{r^2} \otx
		\big)_{\ell \geq 1}^{\mathscr{S}+\mathscr{G}_{\mathrm{sphere}}}(u_f,v_{-1},\cdot)
		=
		0,
	\end{equation}
	and (recalling that $A\neq 0$)
	\begin{equation} \label{eq:Hlinearexistenceafterspherespgs2}
		\big(
		\otx
		-
		\otxb
		-
		\frac{2\Omega^2}{r} 2 \Olin
		-
		\frac{B}{A}
		\accentset{\scalebox{.6}{\mbox{\tiny (1)}}}{\Upsilon}
		\big)_{\ell = 0}^{\mathscr{S} + \mathscr{G}_{\mathrm{sphere}} }(u_f,v(R,u_f))
		=
		0,
	\end{equation}
	(see also the transformation law \eqref{eq:trchitrchibarOmegachangeofgauge} and note that $\accentset{\scalebox{.6}{\mbox{\tiny (1)}}}{\Upsilon}_{\ell=0}$ is gauge invariant).
	
	Recall now Remark \ref{rmk:linearHgaugefoliations}.  Adding the pure gauge solution $\mathscr{G}_{\mathrm{foliations}}$ arising from $\tilde{\mathscr{S}} = \mathscr{S} + \mathscr{G}_{\mathrm{sphere}}$, the solution $\mathscr{S} +  \mathscr{G}_{\mathrm{sphere}} + \mathscr{G}_{\mathrm{foliations}}$ then achieves the linearised analogues of the gauge normalisations \eqref{eq:Hfoliation1}--\eqref{eq:Hfoliation5} and, in view of the fact that $f^3(u_f,\cdot) = f^4(v_{-1},\cdot) = 0$ for the functions generating this pure gauge solution, \eqref{eq:Hlinearexistenceafterspherespgs1} and \eqref{eq:Hlinearexistenceafterspherespgs2} are still satisfied.  Moreover, in this gauge the relation (see Lemma \ref{lem:trchipartialHgauge})
	\[
		\big(
		\otx
		-
		A \otxb
		-
		B \accentset{\scalebox{.6}{\mbox{\tiny (1)}}}{\Upsilon}
		\big)_{\ell = 0}^{\mathscr{S}+  \mathscr{G}_{\mathrm{sphere}} + \mathscr{G}_{\mathrm{foliations}}}(u_f,v(R,u_f))
		=
		0
	\]
	holds, and so it follows from \eqref{eq:Hlinearexistenceafterspherespgs2}, together with the fact that $\Olin^{\mathscr{S} + \mathscr{G}_{\mathrm{sphere}} + \mathscr{G}_{\mathrm{foliations}}}(u_f,\cdot,\cdot) = 0$, that
	\[
		(1 - A^{-1})\otx_{\ell = 0}^{\mathscr{S}+  \mathscr{G}_{\mathrm{sphere}} + \mathscr{G}_{\mathrm{foliations}}}(u_f,v(R,u_f)) = 0.
	\]
	It similarly follows from the relation \eqref{eq:Hlinearexistenceafterspherespgs1} that
	\[
		\otxb_{\ell \geq 1}^{\mathscr{S}+\mathscr{G}_{\mathrm{sphere}}+ \mathscr{G}_{\mathrm{foliations}}}(u_f,v_{-1},\cdot)
		=
		0.
	\]
	Hence (recalling that $A\neq 1$) the linearised analogues of \eqref{eq:Hextragauge1}--\eqref{eq:Hextragauge2} are also satisfied by $\mathscr{S} +  \mathscr{G}_{\mathrm{sphere}} + \mathscr{G}_{\mathrm{foliations}}$.
	
	Finally, the linearised analogue of the gauge condition \eqref{eq:Hextragauge3} can be achieved by adding an appropriate pure gauge solution $\mathscr{G}_{\mathrm{theta}}$ generated by functions $q_1(v,\theta)$ and $q_2(v,\theta)$ (see Lemma 6.\@1.\@3 of~\cite{holzstabofschw}) satisfying
	\[
		-
		r^2 \nablaslash \partial_u q_1(v,\theta)
		+
		r^2 {}^* \nablaslash \partial_u q_2(v,\theta)
		=
		-
		b^{\mathscr{S} +  \mathscr{G}_{\mathrm{sphere}} + \mathscr{G}_{\mathrm{foliations}}}(u_f,v,\theta),
	\]
	for all $v$ and $\theta \in \mathbb{S}^2$ (along with $f^3 \equiv 0$, $f^4\equiv 0$).  The solution
	\[
		\mathscr{S}
		+
		\mathscr{G}_{\mathrm{spheres}}
		+
		\mathscr{G}_{\mathrm{foliations}}
		+
		\mathscr{G}_{\mathrm{theta}},
	\]
	is then normalised as desired.
	
	Again, note that $\mathscr{G} = \mathscr{G}_{\mathrm{spheres}} + \mathscr{G}_{\mathrm{foliations}} + \mathscr{G}_{\mathrm{theta}}$ is not the unique pure gauge solution which achieves these normalisations in view of the absence of any linearised analogues of the anchoring conditions \eqref{eq:anchoringdefcommoncone} and \eqref{eq:anchoringdefaffixingsphere} in Proposition \ref{proplinHpgauge}.
\end{remark}

\section{Monotonicity properties of \texorpdfstring{$\mathfrak{R}(\hat{u}_f+\delta)$}{TEXT}: the proof of Theorem \ref{thm:lambda}}
\label{section:monotonicityR}

We restate Theorem~\ref{thm:lambda}.

\lambdaherehere*

\begin{proof}
Since $\hat{u}_f\in \mathcal{B}$, 
the map $(\ref{restrictiondegree})$ is degree $1$ for $\delta=0$.
It follows that there exists
a $\lambda^0\in \mathfrak{R}(u_f^0)$ such that ${\bf J}_{\hat{u}_f}(\lambda^0)=0$.
(See for instance Theorem~1.1.1 of~\cite{nirenberg}.)
But now this can only happen if $u'_f(\hat{u}_f, \lambda^0)=\hat{u}_f$, because if $u'_f(\hat{u}_f, \lambda^0)<\hat{u}_f$,
by continuity of the map ${\bf J}(u'_f, \lambda)$ in $u'_f$ it would follow that
$|{\bf J} (u'_f,\lambda^0)|\ge \varepsilon_0/u'_f(\hat{u}_f, \lambda^0) \ne 0 $.
 Thus, we have that for the $\lambda^0$ above, $\lambda^0\in \mathfrak{R}(\hat{u}_f)$ and 
 ${\bf J}(\lambda^0, \hat{u}_f)=0$.
 
By continuity of ${\bf J}(\hat{u}_f+\delta, \lambda)$ in $\lambda$ and $\delta$, 
it follows that for sufficiently small $\delta_0>0$ and $0\le\delta\le \delta_0$, 
$\lambda^0\in \mathfrak{R}(\hat{u}_f+\delta)$, and thus, in particular,
the set $\mathfrak{R}(\hat{u}_f+\delta)$ is indeed nonempty, i.e.~$(\ref{indeednonemptyhere})$ holds.

To infer the inclusions $(\ref{inclusionstuff})$ we show that, for $0 \leq \delta' < \delta \leq \delta_0$,
\begin{equation} \label{eq:inclusionstufffollows}
	\vert {\bf J}(\hat{u}_f + \delta', \lambda) \vert
	\geq
	\frac{\varepsilon_0}{\hat{u}_f + \delta'}
	\qquad
	\text{implies}
	\qquad
	\vert {\bf J}(\hat{u}_f + \delta, \lambda) \vert
	>
	\frac{\varepsilon_0}{\hat{u}_f + \delta}.
\end{equation}
Let $\lambda \in \mathfrak{R}(\hat{u}_f)$ be fixed.  For each $0\leq \delta \leq \delta_0$, geometric quantities in the $\hat{u}_f+\delta$ normalised $\I$ gauge with respect to $M_f(\hat{u}_f,\lambda)$ are denoted with a $\delta$ superscript.  Recall the definition \eqref{definofboldv} of the angular momentum parameters
\[
	{\bf J} (u_f, \lambda)
	=
	(J^{-1} (u_f, \lambda),J^0 (u_f,\lambda),J^1 (u_f,\lambda))
	.
\]
where $J^{m} (u_f,\lambda)$ is defined in Definition \ref{assocKerparIplus}.
Equations \eqref{eq:beta4}, \eqref{eq:beta3}, \eqref{eq:curletacurletabar}, \eqref{eq:Codazzi} imply that $\curlslash \Omega^{-1} \beta$ satisfies,
\begin{align*}
	\partial_u \big( r^5 \curlslash \Omega^{-1} \beta \big)
	=
	\
	&
	- 
	r^6 \Big( \Deltaslash + \frac{2}{r^2} \Big) \curlslash \Omega^{-1} \beta
	-
	r^6 \Omega_{\circ}^{-2} ( \Deltaslash - 3 \rho_{\circ}) \curlslash \divslash \Omega \hat{\chi}
	+
	\mathcal{E}_{\partial_u}
	\\
	\partial_v \big( r^5 \curlslash \Omega^{-1} \beta \big)
	=
	\
	&
	r^5 \curlslash \divslash \alpha
	+
	\mathcal{E}_{\partial_v}
	,
\end{align*}
where
\begin{align} \label{eq:partialuangularmomentumerror}
	\mathcal{E}_{\partial_u}
	=
	\
	&
	-
	\frac{r^5}{2} \Deltaslash (\hat{\chi} \wedge \hat{\chibar})
	+
	2 r^5 \curlslash (\hat{\chi}^{\sharp} \cdot \betabar)
	\\
	&
	+
	r^5 \curlslash \Big(
	3(\rho - \rho_{\circ}) \eta + 3 {}^* \eta \sigma
	-
	(\Omega \tr \chibar - \Omega \tr \chibar_{\circ}) \Omega^{-1} \beta
	-
	2(\Omega \omegabarhat - \Omega \omegabarhat_{\circ}) \Omega^{-1} \beta
	\Big)
	+
	[\Omega \nablaslash_3, r \curlslash] r^4 \Omega^{-1} \beta
	\nonumber
	\\
	&
	+
	r^6(\Deltaslash - 3 \rho_{\circ}) 
	\Big(
	(1 - \Omega_{\circ}^{-2} \Omega^2) \curlslash(\Omega^{-1} \beta)
	-
	({}^*\eta + {}^* \etabar) \cdot \Omega \beta
	+
	\Omega_{\circ}^{-2} \curlslash(\Omega \hat{\chi}^{\sharp} \cdot \etabar - \frac{1}{2} (\Omega \tr \chi - \Omega \tr \chi_{\circ}) \etabar)
	\Big),
	\nonumber
	\\
	\mathcal{E}_{\partial_v}
	=
	\
	&
	[\Omega \nablaslash_4, r \curlslash] r^4 \Omega^{-1} \beta
	+
	r^5 \curlslash \big(
	\eta^{\sharp} \cdot \alpha
	-
	2(\Omega \tr \chi - \Omega \tr \chi_{\circ}) \Omega^{-1} \beta
	\big)
	.
	\nonumber
\end{align}
For each $0\leq \delta \leq \delta_0$, after recalling that the geometric quantities in the $\hat{u}_f + \delta$ normalised $\I$ gauge with respect to $M_f(\hat{u}_f + \delta,\lambda)$ satisfy the bootstrap estimate \eqref{eq:bamain} with $u_f = \hat{u}_f+\delta$ (see Theorem \ref{thm:newgauge}),
Proposition \ref{prop:almostevalue} and Proposition \ref{prop:com0}, together with the fact that $\curlslash \divslash \alpha_{\ell=1} = \curlslash \divslash \Omega \hat{\chi}_{\ell=1}= 0$, imply that,
\begin{equation} \label{eq:monotonicitycurlbetaestimates}
	\big\vert \partial_u \big( r^5 \curlslash \Omega^{-1} \beta_{\ell = 1}^{\delta} \big) \big\vert
	\lesssim
	\frac{\varepsilon^2}{u^2},
	\qquad
	\big\vert \partial_v \big( r^5 \curlslash \Omega^{-1} \beta_{\ell = 1}^{\delta} \big) \big\vert
	\lesssim
	\frac{\varepsilon^2}{r^2},
\end{equation}
using also Proposition \ref{prop:divcurlmodes}.

Fix now some $0 \leq \delta' < \delta \leq \delta_0$.  It follows, recalling Definition \ref{assocKerparIplus}, that
\begin{equation*}
	\Big\vert
	3 \sum_{m=-1}^1 J^m (\hat{u}_f + \delta,\lambda) (r^2 \Deltaslash Y_m^1)^{\delta}(\hat{u}_f+ \delta,v_{\infty}(\hat{u}_f+ \delta),\theta)
	+
	r^5 \curlslash \Omega^{-1} \beta_{\ell = 1}(\hat{u}_f + \delta',v_{\infty}(\hat{u}_f + \delta'),\theta)
	\Big\vert
	\lesssim
	\frac{\varepsilon^2 (\delta - \delta')}{(\hat{u}_f+\delta')^2}.
\end{equation*}
It moreover follows from Proposition \ref{prop:modesdifference} that
\[
	\vert
	(r^2 \Deltaslash Y_m^1)^{\delta} (\hat{u}_f+ \delta,v_{\infty}(\hat{u}_f+ \delta),\theta)
	-
	(r^2 \Deltaslash Y_m^1)^{\delta} (\hat{u}_f + \delta',v_{\infty}(\hat{u}_f + \delta'),\theta)
	\vert
	\lesssim
	\frac{\varepsilon (\delta- \delta')}{v_{\infty}(\hat{u}_f+\delta') \hat{u}_f+\delta'},
\]
and so
\begin{equation} \label{eq:monotonicityestimate1}
	\Big\vert
	3 \sum_{m=-1}^1 J^m (\hat{u}_f + \delta,\lambda) (r^2 \Deltaslash Y_m^1)^{\delta} (\hat{u}_f,v_{\infty}(\hat{u}_f),\theta)
	+
	r^5 \curlslash \Omega^{-1} \beta_{\ell = 1}^{\delta}(\hat{u}_f+ \delta',v_{\infty}(\hat{u}_f+ \delta'),\theta)
	\Big\vert
	\lesssim
	\frac{\varepsilon^2 (\delta - \delta')}{(\hat{u}_f+\delta')^2}
	.
\end{equation}
Revisiting the proof of \eqref{eq:curvaturecomp3}, one sees that, in fact,
\begin{equation} \label{eq:betachangeofgaugeinfact}
	\Omega \beta^{\delta}
	-
	\Pi_{S^{\delta}} \Omega \beta^{\delta'}
	=
	3 (\Omega^2 \rho)^{\delta} \nablaslash f^3_{\delta',\delta}
	+
	\frac{1}{2} (1 + \partial_{\widetilde{v}} f^4_{\delta',\delta})^2
	\Pi_{S^{\delta}} \big(\partial_{\widetilde{u}} \slashed{f}_{\delta',\delta} \cdot \alpha^{\delta'} \big)
	+
	\mathcal{E}^{1,0}_{\mathfrak{D} \fsc,4}
	+
	\mathcal{E}^{1,0}_{\mathfrak{D} \fsc,5},
\end{equation}
where $f_{\delta',\delta}$ denote the diffeomorphism functions relating the $\hat{u}_f+\delta$ and $\hat{u}_f+\delta'$ gauges (see Theorem \ref{thm:Igaugeexistence}) and the term $\mathcal{E}^{1,0}_{\mathfrak{D} \fsc,4}$ does not involve $\alpha$, $\beta$ or $\Omega\omegahat - \Omega \omegahat_{\circ}$.  Since $\vert \alpha \vert \lesssim \varepsilon r^{-\frac{9}{2}}$, $r^p \vert \Phi_p \vert \lesssim \varepsilon u^{-1}$ for all $\Phi_p \neq \alpha$, $\beta$ or $\Omega\omegahat - \Omega \omegahat_{\circ}$, it follows from the estimate \eqref{eq:diffeodeltadeltaprime1} and the fact that $u_f^2 \leq v_{\infty}(u_f)$ that
\begin{equation*}
	\big\vert
	\big(
	(r^5 \curlslash \Omega^{-1} \beta_{\ell = 1})^{\delta}
	-
	\Pi_{S^{\delta}}(r^5 \curlslash \Omega^{-1} \beta_{\ell = 1})^{\delta'}
	\big)(\hat{u}_f +\delta',v_{\infty}(\hat{u}_f +\delta'),\theta)
	\big\vert
	\lesssim
	\frac{\varepsilon^2 (\delta - \delta')}{(\hat{u}_f+\delta')^2},
\end{equation*}
where, for any function $h$,
\[
	\Pi_{S^{\delta}} h (\hat{u}_f +\delta',v_{\infty},\theta)
	=
	h(\hat{u}_f +\delta' + f^3_{\delta',\delta}(\hat{u}_f +\delta',v_{\infty},\theta)
	,
	v_{\infty} + f^4_{\delta',\delta}(\hat{u}_f +\delta',v_{\infty},\theta)
	,
	\slashed{F}_{\delta',\delta}(\hat{u}_f +\delta',v_{\infty},\theta)).
\]
The estimates \eqref{eq:diffeodeltadeltaprime1}, \eqref{eq:diffeodeltadeltaprime2} for $f^3_{\delta',\delta}$, $f^4_{\delta',\delta}$ and $\slashed{F}_{\delta',\delta}$ then imply that
\begin{equation} \label{eq:monotonicityestimate2}
	\big\vert
	(r^5 \curlslash \Omega^{-1} \beta_{\ell = 1})^{\delta}(\hat{u}_f +\delta',v_{\infty}(\hat{u}_f +\delta'),\theta)
	-
	(r^5 \curlslash \Omega^{-1} \beta_{\ell = 1})^{\delta'}(\hat{u}_f +\delta',v_{\infty}(\hat{u}_f +\delta'),\theta)
	\big\vert
	\lesssim
	\frac{\varepsilon^2 (\delta - \delta')}{(\hat{u}_f+\delta')^2}.
\end{equation}
Now, for any $J = (J^{-1},J^0,J^1)$, $\widetilde{J} = (\widetilde{J}^{-1},\widetilde{J}^0,\widetilde{J}^1) \in \mathbb{R}^3$,
\begin{align*}
	&
	\sum_{m=-1}^1
	\Big(
	J^m (r^2 \Deltaslash Y_m^1)^{\delta}(\hat{u}_f + \delta',v_{\infty},\theta)
	-
	\widetilde{J}^m (r^2 \Deltaslash Y_m^1)^{\delta'} (\hat{u}_f+ \delta',v_{\infty},\theta)
	\Big)
	\\
	&
	=
	\sum_{m=-1}^1
	\Big(
	-2 (J^m - \widetilde{J}^m) (Y_m^1)^{\delta'}(\hat{u}_f+ \delta',v_{\infty},\theta)
	+
	2 (J^m - \widetilde{J}^m) ((Y_m^1)^{\delta'}(\hat{u}_f+ \delta',v_{\infty},\theta)- (Y_m^1)^{\delta}(\hat{u}_f+ \delta',v_{\infty},\theta))
	\\
	&
	+
	(J^m - \widetilde{J}^m) (r^2 \Deltaslash Y_m^1 + 2 Y_m^1)^{\delta}(\hat{u}_f+ \delta',v_{\infty},\theta)
	-
	\widetilde{J}^m ((r^2 \Deltaslash Y_m^1)^{\delta'}(\hat{u}_f+ \delta',v_{\infty},\theta) - (r^2 \Deltaslash Y_m^1)^{\delta}(\hat{u}_f+ \delta',v_{\infty},\theta))
	\Big),
\end{align*}
for $v_{\infty} = v_{\infty}(\hat{u}_f+ \delta')$.
Setting $J= {\bf J} (\hat{u}_f + \delta,\lambda)$, $\widetilde{J} = {\bf J} (\hat{u}_f + \delta',\lambda)$ and using the fact that $(Y_m^1)^{\delta'}$ are orthonormal, for $m=-1,0,1$, it follows from Proposition \ref{prop:almostevalue} and Proposition \ref{prop:YYtildedifference} that, if $\hat\varepsilon_0$ is sufficiently small,
\begin{align*}
	&
	\vert {\bf J} (\hat{u}_f + \delta,\lambda) - {\bf J} (\hat{u}_f+\delta',\lambda) \vert
	\lesssim
	\vert {\bf J} (\hat{u}_f+\delta',\lambda) \vert
	\sum_{m=-1}^1
	\Vert
	(r^2 \Deltaslash Y_m^1)^{\delta'} (\hat{u}_f+\delta',v_{\infty},\cdot)
	-
	(r^2 \Deltaslash Y_m^1)^{\delta}(\hat{u}_f+\delta',v_{\infty},\cdot)
	\Vert_{S^{\delta'}}
	\\
	&
	\qquad \qquad
	+
	\Big\Vert 
	\sum_{m=-1}^1
	\Big(
	J^m (\hat{u}_f+\delta,\lambda) (r^2 \Deltaslash Y_m^1)^{\delta}(\hat{u}_f+\delta',v_{\infty},\cdot)
	-
	J^m (\hat{u}_f+\delta',\lambda) (r^2 \Deltaslash Y_m^1)^{\delta'}(\hat{u}_f+\delta',v_{\infty},\cdot)
	\Big)
	\Big\Vert_{S^{\delta'}}
	,
\end{align*}
with $S^{\delta'} = S^{\delta'}_{\hat{u}_f+\delta',v_{\infty}(\hat{u}_f+\delta')}$.  Since 
\[
	(r^5 \curlslash \Omega^{-1} \beta_{\ell = 1})^{\delta'}(\hat{u}_f+\delta', v_{\infty}(\hat{u}_f+\delta'),\theta)
	=
	-
	3
	\sum_{m=-1}^1
	J^m (\hat{u}_f+\delta',\lambda) (r^2 \Deltaslash Y_m^1)^{\delta'} (\hat{u}_f+\delta', v_{\infty}(\hat{u}_f+\delta'),\theta),
\]
the estimates \eqref{eq:monotonicityestimate1} and \eqref{eq:monotonicityestimate2}, along with Proposition \ref{prop:YYtildedifference} then imply that
\begin{equation} \label{eq:monotonicityestimate3}
	\vert {\bf J} (\hat{u}_f + \delta,\lambda) - {\bf J} (\hat{u}_f+\delta',\lambda) \vert
	\lesssim
	\frac{\varepsilon^2 (\delta - \delta')}{(\hat{u}_f+\delta')^2}.
\end{equation}
Suppose now that 
$\vert {\bf J} (\hat{u}_f,\lambda) \vert \geq \frac{\varepsilon_0}{\hat{u}_f}$.  The estimate \eqref{eq:monotonicityestimate3} implies that, for some constant $C$,
\[
	\vert {\bf J} (\hat{u}_f + \delta,\lambda) \vert
	\geq
	\frac{\varepsilon_0}{\hat{u}_f+\delta'}
	-
	\frac{C\varepsilon^2_0 (\delta- \delta')}{(\hat{u}_f +\delta')^2}
	>
	\frac{\varepsilon_0}{\hat{u}_f + \delta},
\]
where the latter inequality holds if $\varepsilon_0$ is sufficiently small.  This concludes the proof of \eqref{eq:inclusionstufffollows}.

\begin{remark}
\label{pointwheredecay}
	The strong $r^{-\frac{9}{2}}$ decay for $\alpha$ and appropriate derivatives imposed by~\eqref{smallnessofdata} (see Remark \ref{commentaboutpeeling}) was convenient in obtaining \eqref{eq:monotonicityestimate3}, as it allowed us to avoid
	capturing decay in $u$ faster than $u^{-1}$ for any of the geometric quantities of the $\I$ gauge. Indeed, this is one of the main places where this assumption on data is used.  See the first two terms in the error \eqref{eq:partialuangularmomentumerror} and the second term on the right hand side of the change of gauge relation \eqref{eq:betachangeofgaugeinfact}.
\end{remark}

The continuity of the map ${\bf J}_{\hat{u}_f+\delta}$ follows from the continuity of~\eqref{asamaphere} 
in $\lambda$
and the continuity of the original map ${\bf J}_{\hat{u}_f}$ as follows:

Since
\[
{\bf J}_{u_f} (\lambda) = {\bf J}(u'_f(\lambda, u_f),\lambda) 
\]
where 
\begin{equation}
\label{thismapinlambda}
u'_f(\lambda, u_f):=\sup \{u^0_f\le u'_f \le u_f : \lambda \in \mathfrak{R}(u'_f)\},
\end{equation}
it suffices to note that the map $u'_f(u_f,\lambda)$ 
defined by~\eqref{thismapinlambda}
is continuous, for fixed $u_f=\hat{u}_f+\delta$, in $\lambda$ as a map on $\mathfrak{R}(u^0)$.

For this, fix $u_f=\hat{u}_f+\delta$, and consider a sequence $\lambda_i \to \lambda_0$ with $\lambda_i\in
\mathfrak{R}(u^0)$. 

If $u''_f<u'_f(u_f, \lambda_0)$, then by the monotonicity properties~\eqref{monotonofparam} of $\mathfrak{R}(u''_f)$,
it follows that $\lambda_0\in {\rm int}(\mathfrak{R}(u''_f))$. But since ${\rm int}(\mathfrak{R}(u''_f))$ is open, it follows that
$\lambda_i \in \mathfrak{R}(u''_f)$ for sufficiently large $i$, and thus, for such $i$, $u'_f(\lambda_i, u_f)\ge u''_f$.
It follows that that 
\begin{equation}
\label{aboutliminf}
\lim\inf u'_f(u_f,\lambda_i) \ge u'_f(u_f,\lambda_0).
\end{equation}

On the other hand, suppose that 
\begin{equation}
\label{aboutlimsupcontra}
\lim \sup u'_f(u_f,\lambda_i) > u'_f(u_f,\lambda_0).
\end{equation}
It follows again by the first monotonicity property~\eqref{monotonofparam} of $\mathfrak{R}(u''_f)$
that there exists $u''_f>u'_f(u_f,\lambda_0)$ and a sequence of $i_k\to \infty$ such
that $\lambda_{i_k}\in \mathcal{R}(u''_f)$ for all $i$. But since $\mathfrak{R}(u''_f)$ is closed,
it follows that $\lambda_0 \in \mathcal{R}(u''_f)$ but this contradicts the fact that 
 $u''_f>u'_f(u_f,\lambda_0)$.
Thus, we have in fact
\begin{equation}
\label{aboutlimsup}
\lim \sup u'_f(u_f,\lambda_i) \le u'_f(u_f,\lambda_0),
\end{equation}
and inequalities~\eqref{aboutliminf} and~\eqref{aboutlimsup} together give continuity
of the map~\eqref{thismapinlambda} at $\lambda_0$, as desired.

 We note that the  identity $(\ref{restrictiondegree})$
continues to hold for ${\bf J}_{\hat{u}_f+\delta}$ and thus the statement about its degree.
\end{proof}

\section{Higher order estimates: the proof of Theorem~\ref{thehigherordertheorem}}
\label{proofofhigherorderestimatessec}

We restate Theorem~\ref{thehigherordertheorem}:

\thehigherordertheoremhere*

\begin{proof}
Typically, higher order estimates like the above, without additional smallness
at higher order, follow since the equations are linear after sufficient differentiation. This is similar here although slightly complicated by the teleological normalisation of our gauges. For the reader's convenience, we provide some of the details.

We begin with the proof of (\ref{nonlocalisedstatement}). The proof is an induction on $k$. We have already proven the above estimate for $k=N=12$ (base case). Assume now the estimate holds for some fixed $k\geq N$. The estimate at order $k+1$ is now obtained through a sequence of four steps: \\

\noindent {\bf Step 1.} We first consider the wave equations for the quantities of the almost gauge invariant hierarchy, i.e. $\left(\alpha, \psi, P\right)$ and $\left(\underline{\alpha}, \underline{\psi}, \underline{P}\right)$ and repeat the analysis of Chapters  \ref{chapter:psiandpsibar} and \ref{moreherechapter}   for the higher commuted equations.

The key observation is that upon commutation with $k$ derivatives for $\alpha$ ($k-1$ for $\psi$ and $k-2$ for $P$) the (finite) number of top order non-linear error terms on the right hand side (i.e.~those involving $k+1$ derivatives of curvature or Ricci coefficients and one exceptional term involving $k+2$ derivatives of $\Omega tr \chi-\Omega tr \chi_\circ$ in the $\underline{\alpha}$ equation, see (\ref{eq:Teukolskybartilde})) is independent of $k$ and moreover each of the finite terms has the structure\footnote{We do not keep track of the $r$-weights here as it is clear that they don't provide any additional difficulty in closing the higher order estimates compared with closing them at lower order.} $h \Phi \mathfrak{D}^k \Phi^\prime$, where $h$ is a bounded function and both $h$ and the $\Phi, \Phi^\prime$ that appear are independent of $k$. This observation provides the estimate
\begin{align} \label{altk}
\mathbb{E}^{k+1}_{u_f} [\alpha_{\Hp}, \alpha_{\I}] + \mathbb{E}^{k+1}_{u_f} [\underline{\alpha}_{\Hp}, \underline{\alpha}_{\I}] 
\lesssim  &C \varepsilon  \left( \mathbb{E}^{k+1}_{u_f, \Hp}  +  \mathbb{E}^{k+1}_{u_f, \I} \right)  \\
+&C_k \varepsilon \left( \mathbb{E}^k_{u_f, \Hp}  +  \mathbb{E}^k_{u_f, \I}+ \mathbb{E}^{k}_{u_f} [\alpha_{\Hp}, \alpha_{\I}] + \mathbb{E}^{k}_{u_f} [\underline{\alpha}_{\Hp}, \underline{\alpha}_{\I}]   \right) \nonumber \\
+ &C_k \varepsilon \left( \mathbb{E}^{k-6}[f_{\Hp,\I}] + \mathbb{E}_{u_f}[f_{d,\Hp}] + \mathbb{E}_{u_f}[f_{d,\I}]\right) + C_k \mathbb E^{k+1}_{0}[\mathcal{S}(\lambda)] \, , \nonumber
\end{align}
where the constant $C_k$ may depend on $k$ but the constant $C$ does not.
Indeed, the first two lines on the right hand side follow directly from the observation. The last line follows from the fact that we need to compare the top order energies on the timelike hypersurface and estimate the initial energies in the horizon and the infinity gauge from the data. However, all these estimates involve only lower order energies of the diffeomorphisms and -- again for a finite $k$-independent number of terms -- the top order energy of the Ricci coefficients. Recall Sections \ref{subsubsec:cancelT}--\ref{Heretheproofofgida} for the precise statements. \\

\noindent {\bf Step 2.} We next repeat the analysis of Chapters \ref{chap:Iestimates} and \ref{chap:Hestimates} to estimate the Ricci-coefficients in their individual gauges after $k+1$ commutations. Again the equations are linear at top order leading to the estimates
\begin{align}
  \mathbb{E}^{k+1}_{u_f, \I} \lesssim C \left( \mathbb{E}^{k+1}_{u_f} [\alpha_{\Hp}, \alpha_{\I}] + \mathbb{E}^{k+1}_{u_f} [\underline{\alpha}_{\Hp}, \underline{\alpha}_{\I}] \right) + C_k \left( \mathbb{E}^{k}_{u_f} [\alpha_{\Hp}, \alpha_{\I}] + \mathbb{E}^{k}_{u_f} [\underline{\alpha}_{\Hp}, \underline{\alpha}_{\I}] \right) \, , 
\end{align}
\begin{align}
 \mathbb{E}^{k+1}_{u_f, \Hp}   \lesssim &C \left( \mathbb{E}^{k+1}_{u_f} [\alpha_{\Hp}, \alpha_{\I}] + \mathbb{E}^{k+1}_{u_f} [\underline{\alpha}_{\Hp}, \underline{\alpha}_{\I}] \right) + C_k \left( \mathbb{E}^{k}_{u_f} [\alpha_{\Hp}, \alpha_{\I}] + \mathbb{E}^{k}_{u_f} [\underline{\alpha}_{\Hp}, \underline{\alpha}_{\I}] \right) \nonumber \\
 &+C \varepsilon \mathbb{E}^{k+1}_{u_f}[f_{\Hp,\I}] + C_k  \mathbb{E}^{k}_{u_f}[f_{\Hp,\I}] \, .
\end{align}
Note that in the horizon gauge there are extra terms involving the higher order energies of the diffeomorphisms because on the final hypersurface $u=u_f$, some horizon quantities are obtained from the infinity quantities and the estimates on the diffeomorphisms; see Section \ref{proofofinheritingthem}. \\

\noindent {\bf Step 3.} Finally, the higher order diffeomorphisms are estimated from the Ricci coefficients as in 
Chapter~\ref{chap:comparing}. Once more, we observe that the number of top order terms is $k$-independent and we have
\begin{align}
\mathbb{E}^{k+1}_{u_f}[f_{\Hp,\I}]  \lesssim  C \left(  \mathbb{E}^{k+1}_{u_f, \I} +  \mathbb{E}^{k+1}_{u_f, \Hp}   \right) + C_k\left(  \mathbb{E}^k_{u_f, \I} +  \mathbb{E}^k_{u_f, \Hp}   \right) \, .
\end{align}

\noindent {\bf Step 4.} Coupling all the estimates together yields in view of $C\varepsilon \sim \varepsilon$ the estimate
\begin{align}
\nonumber
\mathbb{E}^{k+1}_{u_f} [\alpha_{\Hp}, \alpha_{\I}] + \mathbb{E}^{k+1}_{u_f} [\underline{\alpha}_{\Hp}, \underline{\alpha}_{\I}] +
\mathbb E_{u_f, \mathcal{H}^+}^{k+1}
+ \mathbb E_{u_f, \mathcal{I}^+}^{k+1} + \mathbb{E}_{u_f}^{k+3}[f_{\Hp,\I}]  \le C_k \mathbb E^{k+1}_{0}[\mathcal{S}(\lambda)].
\end{align}
One finally observes that the left hand side controls also $\mathbb{P}_{u_f}^{k-4}[\Phi^{\Hp}]+\mathbb{P}_{u_f}^{k-4}[\Phi^{\I}]$ as in Section \ref{section:proofofpointwiseestimates}. This is the estimate (\ref{nonlocalisedstatement}) at order $k+1$ and hence the induction is complete. \\

We next turn to the proof of  (\ref{localisedstatementhere}). Here we can follow exactly the same steps as above, except that the estimate (\ref{altk}) is now replaced by
\begin{align} \label{altk2}
\mathbb{E}^{k+1}_{u_f} [\alpha_{\Hp}, \alpha_{\I}] + \mathbb{E}^{k+1}_{u_f} [\underline{\alpha}_{\Hp}, \underline{\alpha}_{\I}] 
\lesssim  &C \varepsilon  \left( \mathbb{E}^{k+1}_{u_f, \Hp}  +  \mathbb{E}^{k+1}_{u_f, \I} \right)  \\
+&C_k \varepsilon \left( \mathbb{E}^k_{u_f, \Hp}  +  \mathbb{E}^k_{u_f, \I}+ \mathbb{E}^{k}_{u_f} [\alpha_{\Hp}, \alpha_{\I}] + \mathbb{E}^{k}_{u_f} [\underline{\alpha}_{\Hp}, \underline{\alpha}_{\I}]   \right) \nonumber \\
+ &C_k \varepsilon \left( \mathbb{E}^{k-6}[f_{\Hp,\I}] + \mathbb{E}_{u_f}[f_{d,\Hp}] + \mathbb{E}_{u_f}[f_{d,\I}]\right) \nonumber \\
+& \mathbb E^{N}_{0}[\mathcal{S}(\lambda)]+ C_k (u_f) \mathbb E^{k+1}_{0,v_\infty}[\mathcal{S}(\lambda)] \, , \nonumber
\end{align}
the only difference being that the energy in the last line is now localised at the expense of a constant depending on $u_f$.
This is because we proceed slightly differently when estimating the initial energies of $(\alpha, \psi, P)$ and $(\underline{\alpha}, \check{\underline{\psi}}, \check{\underline{P}})$
in the  $\I$ gauge. Specifically, we now obtain estimates for these initial energies by comparing the almost gauge invariant quantities in the $\I$ gauge directly with those in the preliminary gauge $i^0_{\mathcal{EF}}$ of part (a) of 
Theorem~\ref{thm:localEF} (as opposed to the renormalised gauge~\eqref{herethenewone} of part (b)). Appealing to the 
gauge $i^0_{\mathcal{EF}}$ allows us to exploit the domain of dependence property and localise the region where one does the comparison of energies.\footnote{Note that we cannot do this for the  renormalised initial data gauge~\eqref{herethenewone} in view of its ``teleological'' normalisations~\eqref{somenormalisations} at $\mathcal{I}^+$: The quantities expressed in this gauge in a compact region depend on the full seed energy $\mathbb E^{k+1}_{0}[\mathcal{S}(\lambda)]$.} However, this comes at the expense of constants potentially growing in $u_f$ since the estimates obtainable for the diffeomorphisms relating the $\mathcal{I}^+$ gauge and the preliminary gauge $i^0_{\mathcal{EF}}$ exhibit worse behaviour in $r$. The domain of dependence property and the compactness of the region under consideration, however, allows one to absorb this behaviour into the constant $C(u_f)$. This establishes~\eqref{altk2}, and repeating the steps above now yields~\eqref{localisedstatementhere}.
\end{proof}

\chapter{Conclusions}
\label{conclusionsection}

In this last Chapter, we deduce properties of the limiting foliations and complete the
proof of the statement of Theorem~\ref{thm:main}. We also prove Corollary~\ref{axicorollary}.

\minitoc

We first interpret in {\bf Section~\ref{rescaledandgravrad}} the ``laws of gravitational radiation'' as identities
on null infinity $\mathcal{I}^+$ 
and then in  {\bf Section~\ref{eventsmydear}}, we shall understand the regularity and properties of the event 
horizon $\mathcal{H}^+$.
In {\bf Section~\ref{recapsecti}}, we shall explicitly recap all statements 
appearing in the statement of Theorems~\ref{limitgaugetwo} and~\ref{limitgaugethree},
completing thus the proof of Theorem~\ref{thm:main}.
Finally, in {\bf Section~\ref{axisymmtheorem}} we  give
a proof of Corollary~\ref{axicorollary}  concerning axisymmetric initial data
 with vanishing angular momentum.

\vskip1pc
\emph{Section~\ref{axisymmtheorem} is independent of Sections~\ref{rescaledandgravrad}--\ref{recapsecti}. The reader may wish to refer also to Chapter 17 of~\cite{CK} and to~\cite{Christmem} for background.}

\section{Null infinity $\mathcal{I}^+$ and the laws of gravitational radiation}
\label{rescaledandgravrad}
In this section we shall present the laws of gravitational radiation as a set of relations of
limiting quantities defined on null infinity $\mathcal{I}^+$.

We first define null infinity as a set $\mathcal{I}^+$ in {\bf Section~\ref{asasetnullinf}}
and discuss in what sense it is Bondi normalised. We remark on the completeness of null infinity
in {\bf Section~\ref{thecompletenesssec}}. 
We shall then define the rescaled quantities in {\bf Section~\ref{rescaledquantsec}}
and formulate
the ``laws of gravitational radiation'' in {\bf Section~\ref{lawisthelaw}}, 
following~\cite{CK}. We shall discuss in {\bf Section~\ref{BMSandallthat}} how the BMS group acts on asymptotic
quantities and how our construction has broken BMS symmetry in normalising uniquely
the asymptotic gauge. Finally, we shall give decay statements for the asymptotic quantities
in {\bf Section~\ref{decayalonginfinitysec}}.

\subsection{Null infinity $\mathcal{I}^+$ as a set and Bondi normalisation}
\label{asasetnullinf}

We have  remarked in Section~\ref{proofofisec}
that the existence of a $u_f$-normalised $\mathcal{I}^+$  gauge for all $u_f\ge u_f^0$
already yields the completeness of null infinity, in the sense of Christodoulou~\cite{Chrmil}.

We can describe this completeness in another way by introducing $\mathcal{I}^+$ as an actual set.
We define
\[
\mathcal{I}^+ := [u_{-1},\infty) \times \mathbb S^2.
\]
If desired, one may naturally attach $\mathcal{I}^+$ to the spacetime $(\mathcal{M},g)$ by extending
the embedding~\eqref{limitingembeddinginf},
identifying
$\mathcal{I}^+$ with the image of a limiting $\{\infty\}\times [u_0,\infty)\times \mathbb S^2$ hypersurface
which can be naturally attached to the domain $\mathcal{W}_{\mathcal{I}^+}(\infty, M_{\rm final})\times \mathbb S^2$
of~\eqref{limitingembeddinginf}.

With respect to the limiting $\mathcal{I}^+$ gauge, we note that our bounds~\eqref{asymptpointwisestatementintheorem}
in particular immediately yield
\begin{equation}
\label{Omegaoneatinfinity}
\lim_{v\to\infty} \Omega^2(u,v,\theta)=1.
\end{equation}
One could rescale the metric to obtain a conformal compactification,
but we shall refrain from doing so here. 
Let it suffice here to note that~\eqref{asymptpointwisestatementintheorem}
similarly imply
\begin{equation}
\label{rtrchilimitis2}
\lim_{v\to\infty} r{\tr \chi}(u,v,\theta) = 2
\end{equation}
\begin{equation}
\lim_{v\to\infty} r{\tr \underline\chi}(u,v,\theta) = -2
\end{equation}
and, defining $\mathring{\gamma}$ to be the standard metric
on the $\mathbb S^2$, then we have that 
\begin{equation}
\label{roundlimit}
\lim_{v\to\infty} r^{-2}\slashed{g}_{AB} (u,v,\theta)  =  \mathring \gamma_{AB} (u,\theta).
\end{equation}

Note that 
\begin{equation}
\label{rsquaredclosetoarea}
|r^{-2} {\rm Area} (u,v) - 4\pi | \lesssim \frac{\varepsilon_0}r
\end{equation}
so in all the above $(\ref{rtrchilimitis2})$--$(\ref{roundlimit})$ we could replace $r$ with the geometric area-radius function.

We interpret $(\ref{Omegaoneatinfinity})$ and $(\ref{roundlimit})$
to be the statement that our limiting $\mathcal{I}^+$ normalised  gauge is in an appropriate
sense \emph{Bondi normalised}.

\subsection{The completeness of $\mathcal{I}^+$}
\label{thecompletenesssec}

Given the above concrete realisation of null infinity $\mathcal{I}^+$ as a set, we can view the ``completeness
of null infinity''
as a relation on $\mathcal{I}^+$ itself. Indeed, 
since the range of the asymptotic $u$ coordinate is $[u_{-1},\infty)$ and
the normalisation $(\ref{Omegaoneatinfinity})$ yield
\begin{equation}
\label{tiscompletenullinfinity}
\int_{u_{-1}}^\infty \lim_{v\to\infty} \Omega^2(u,v ,\theta^1,\theta^2) du =\int_{u_{-1}}^\infty 1 du =\infty.
\end{equation}
The relation $(\ref{tiscompletenullinfinity})$ 
is then another manifestation of the ``completeness of null infinity''.

\subsection{Definition of the rescaled quantities}
\label{rescaledquantsec}

We note that the following quantities, thought of as tensors on null infinity $\mathcal{I}^+$,
may now be defined by the following pointwise limits in the $\mathcal{I}^+$ gauge\index{null infinity!$\Xi$, Bondi news}\index{null infinity! $\Sigma$, asymptotic shear}\index{null infinity! ${\bf A}$, rescaled curvature component $\alphabar$}\index{null infinity! ${\bf B}$, rescaled curvature component $\betabar$}\index{null infinity! ${\bf P}$, rescaled curvature component $\rho$}\index{null infinity! ${\bf Q}$, rescaled curvature component $\sigma$}\index{null infinity! $H$, rescaled quantity at null infinity}\index{null infinity! $\underline{N}$, rescaled quantity at null infinity}\index{null infinity! $Z$, rescaled quantity at null infinity}
\[
\Xi_{AB}(u,\theta) :=\lim_{v\to\infty}r^{-1}\underline{\hat\chi}_{AB} (u,v,\theta),
\]
\[
\Sigma_{AB}(u, \theta) :=  \lim_{v\to \infty}\hat\chi_{AB} (u,v, \theta),
\]
\[
H(u,\theta) :=  \lim_{v\to\infty} r^2 \left(\Omega tr \chi - (\Omega tr \chi)_\circ \right)(u,v,\theta),
\]
\[
\underline{N}(u,\theta): =\left( \frac{1}{2} r^3 \slashed{div} \left(\eta - \underline{\eta}\right) + \frac{1}{2} r^3 \hat{\chi} \cdot \underline{\hat{\chi}} - r^3 \rho\right) (u,v,\theta),
\]
\[
Z_A(u,\theta): =\lim_{v\to\infty} \frac12 r \left(\eta_A - \underline{\eta}_A\right)(u,v,\theta),
\]
\[
{\bf A}_{AB}(u, \theta) := \lim_{v\to \infty} r^{-1}\underline{\alpha}_{AB}(u,v,\theta),
\]
\[
{\bf B}_{A} (u,\theta) :=\lim_{v\to\infty} r\underline{\beta}_A (u,v,\theta),
\]
\[
{\bf P}(u,\theta):= \lim_{v\to \infty} r^3\rho (u,v,\theta),
\]
\[
{\bf Q}(u,\theta): = \lim_{v\to \infty} r^3\sigma( u,v,\theta).
\]

We define also the limit of the Hawking mass:\index{null infinity! $M(u)$, Bondi mass}
\begin{equation}
\label{definitionofBondimass}
M(u) = \lim_{v\to \infty} \sqrt{\frac{{\rm Area}(S_{u,v})}{16\pi}} \left(1+\frac{1}{16\pi} \int_{S_{u,v}}{\rm tr}\chi {\rm tr}\underline\chi dA
\right).
\end{equation}
This is indeed the \emph{Bondi mass}, due to $(\ref{roundlimit})$.
We note that in view of $(\ref{rsquaredclosetoarea})$, we may replace ${\rm Area}(S_{u,v})$ with 
$4\pi r^2$ in $(\ref{definitionofBondimass})$.

The reader can easily check that the above pointwise limits are indeed well defined
in view of our estimates~\eqref{asymptpointwisestatementintheorem}.

We will also make use of the ``final'' values of the quantities:\index{null infinity! ${\bf P}^+$, final value of
${\bf P}$  (related to centre of mass)}\index{null infinity! ${\bf Q}^+$, final value of
${\bf Q}$}
\[
\Sigma^+(\theta) := \lim_{u\to \infty}\Sigma(u, \theta)
\]
\[
{\bf P}^+(\theta) = \lim_{u\to \infty} {\bf P} (u,\theta) , \qquad
{\bf Q}^+(\theta)= \lim_{u\to \infty} {\bf Q} (u,\theta).
\]
We will see in Section~\ref{BMSandallthat} that these limits are well defined
and particularly simple in our gauge.

\subsection{The laws of gravitational radiation}
\label{lawisthelaw}

The laws of gravitational radiation (cf.~Chapter~17  of~\cite{CK}) hold,
in particular, the \emph{Bondi  mass loss} formula
\begin{equation}
\label{masslossformula}
\frac{d}{du} M(u) = -\frac1{8\pi} \int_{\mathbb S^2(u)} \left | \Xi(u,\theta)\right|^2 d\mu_{\mathring \gamma}
\end{equation}
and the relations
\[
\frac{\partial \Sigma}{\partial u} = -\Xi,\qquad \frac{\partial \Xi}{\partial u} =-\frac14{\bf A}
\]
and
\[
\frac{\partial H}{\partial u}(u,\theta)=0.
\]

Again, the  validity of these pointwise limiting identities is an easy consequence of~\eqref{boundanddecayasym}
and~\eqref{asymptpointwisestatementintheorem}.

\subsection{BMS symmetry and the normalisation of the asymptotic $\mathcal{I}^+$ gauge}
\label{BMSandallthat}

To understand the limiting behaviour of the quantities of Section~\ref{rescaledquantsec}, 
it is useful, for greater context,
to introduce the notion of a \emph{Bondi normalised double null  gauge}.
This is any double null gauge covering the region $[u_{-1},\infty)\times [v_0,\infty)$ of 
the $\mathcal{I}^+$ gauge
such that the asymptotic relations of Section~\ref{rescaledquantsec} are satisfied.  

We may construct other such gauges starting from our $\mathcal{I}^+$ normalised gauge 
by applying a diffeomorphism   satisfying  appropriate  bounds on $f^3$, $f^4$ and $f^A$.
Given such a gauge, we may again define the asymptotic quantities of
Section~\ref{rescaledquantsec}.
Their transformation properties only depend on the $r\to\infty$ limits of  $f^A$ and $f^3$ 
and the part of $f^4$ growing linearly
in $r$ (which is itself determined by its $\ell=1$ mode, $f^3$ and $f^A$.)
We note that given two Bondi normalised double null gauges, the quantities
$\Sigma$, $\Xi$, ${\bf P}$, ${\bf Q}$, ${\bf A}$, ${\bf B}$ transform 
by the action of the well-known BMS group~\cite{penrose1984spinors}.

Our asymptotic normalised $\mathcal{I}^+$ 
gauge induces a Bondi frame uniquely by the conditions
\begin{equation}
\label{finalPell1}
{\bf P}^+_{\ell =1} =0 
\end{equation}
and\index{null infinity!$\Sigma_+$, final asymptotic shear} 
\begin{equation}
\label{Sigmapluszero}
\Sigma_+=0.
\end{equation}
The condition $(\ref{finalPell1})$ says that our gauge is normalised so as to be a centre of
mass frame for the final black hole that arises while the condition $(\ref{Sigmapluszero})$ chooses
the cuts of null infinity.

Our gauge satisfies the following additional relations:
\begin{equation}
\label{interp}
M_{\rm final} = \lim_{u\to \infty} M(u) =\inf_u M(u),
\end{equation}
\[
{\bf P}_{\ell =0}^+=-2M_{\rm final},
\]
\[
{\bf P}_{\ell\ge 2}^+ =0,
\]
\[
\underline{N}^+_{\ell=0} = \frac1{4\pi} M_{\rm final}, \qquad \underline{N}^+_{\ell\ge 1} =0, \qquad Z^+=0,
\]
\[
{\bf Q}^+ =0,
\]
\[
H=0.
\]

The relation $(\ref{interp})$ can be interpreted as the statement that the final Bondi mass
measured at infinity coincides with the final Schwarzschild parameter of the metric to which the 
spacetime converges at finite points.

\subsection{Decay along null infinity $\mathcal{I}^+$}
\label{decayalonginfinitysec}
We infer from the Bondi mass law formula~\eqref{masslossformula} and
the boundedness of the (second line of the) energy expression~\eqref{estart}
(computed in the asymptotic $\mathcal{I}^+$ gauge, i.e.~as it appears in the energy~\eqref{boundanddecayasym})
the following decay bound for the Bondi mass:
\begin{equation}
\label{decayofmassfinal}
M (u) - M_{\rm final} \lesssim   \varepsilon_0^2 u^{-2}.
\end{equation}
In particular we have the orbital stability statement
\[
|M(u)-M(u_{-1})|\lesssim \varepsilon^2.
\]

Finally, we infer immediately from~\eqref{asymptpointwisestatementintheorem} the following 
pointwise decay bounds along $\mathcal{I}^+$:
\begin{equation}
\label{decayofSigma}
|\Sigma(u,\theta)| \lesssim \varepsilon_0 u^{-1},
\end{equation}
\begin{equation}
\label{decayofXi}
|\Xi(u,\theta)| \lesssim  \varepsilon_0 u^{-1},
\end{equation}
\begin{equation}
\label{decayofA}
|{\bf A}(u,\theta)| \lesssim \varepsilon_0 u^{-1},
\end{equation}
\[
|{\bf B}(u,\theta)| \lesssim \varepsilon_0 u^{-1},
\]
\[
|{\bf P}(u,\theta) +2 M_{\rm final}|\lesssim \varepsilon_0 u^{-1},
\]
\[
|{\bf Q}(u,\theta) |\lesssim \varepsilon_0 u^{-1}.
\]

These can all be viewed as asymptotic stability statements at $\mathcal{I}^+$.

The tensor
\[
\Sigma^+(\theta)-\Sigma(u_{-1},\theta) 
\]
is related to the infinitesimal total displacement of test masses in gravitational wave experiments~\cite{Christmem}.
Its boundedness
\begin{equation}
\label{farfaraway}
|\Sigma^+-\Sigma(u_{-1},\cdot)|\lesssim \varepsilon_0
\end{equation}
follows thus from  the above estimates, and
can be viewed as an additional (orbital) stability statement.

Defining\index{null infinity! $F(\theta)$, total radiation flux to null infinity per unit solid
angle}
\[
F(\theta) :=  \frac{1}{8\pi} \int_{u_{-1}}^{\infty} |\Xi (\theta) |^2du
\]
we have in particular,
that
\begin{equation}
\label{fluxoflinmom}
|F_{\ell = 1}| \lesssim \varepsilon_0^2  
\end{equation}
which is the statement that the \emph{linear momentum radiated} to infinity is bounded.

A similar statement can be made for
the angular momentum, modulo the well-known ambiguities in its general definition along
null infinity. This is related to the bound on $\lambda^{\rm final}$. See Remark~\ref{sizeoflambdafinal}.

\section{The event horizon $\mathcal{H}^+$ and the characterization of the black hole region $\mathcal{M}\setminus
J^-(\mathcal{I}^+)$}
\label{eventsmydear}

In this section, we shall define the event horizon $\mathcal{H}^+$ and infer its basic regularity and completeness properties,
characterizing it moreover as the past boundary of the black hole region $\mathcal{M}\setminus J^-(\mathcal{I}^+)$.

The definition of $\mathcal{H}^+$, with its basic regularity and completeness properties,
will be given in {\bf Section~\ref{definitionofhorizon}}. 
We shall infer polynomial decay statements  along $\mathcal{H}^+$
in {\bf Section~\ref{polyalsoonH}}. We characterize the horizon $\mathcal{H}^+$ as the past boundary of 
$\mathcal{M} \setminus J^-(\mathcal{I}^+)$ in {\bf Section~\ref{whatistheblackholeregion}}, 
connecting with the usual notion 
of black hole event horizon. Finally, we infer higher order
regularity of $\mathcal{H}^+$, under suitable additional assumptions on initial data, in 
{\bf Section~\ref{higherorderregulofhor}}.

\subsection{Definition of $\mathcal{H}^+$, regularity and completeness}
\label{definitionofhorizon}
To define the event horizon, let us
first note that $J^-(\mathcal{D}^{\I}_{\infty})\cap \mathcal{D}^{\mathcal{K}}$ has a non-empty future boundary $B\subset
\mathcal{D}^{\mathcal{K}}$. 
Considering the sets
 $\{v_{\mathcal{H}^+}=v_{-1}\}\cap \{u_{\mathcal{H}^+}=u_n\}\subset  \mathcal{D}^{\Hp}_\infty$,
 then since these are also contained
 in $\mathcal{D}^{\mathcal{K}}$, one easily sees that
$\{v_{\mathcal{H}^+}=v_{-1}\}\cap \{u_{\mathcal{H}^+}=u_n\}$ all lie in a fixed compact subset of $\mathcal{D}^{\mathcal{K}}$.
By examining the transition functions from  $\mathcal{D}^{\Hp}_\infty$ to $\mathcal{D}^{\mathcal{K}}$,
one easily sees that the Kruskalised frame satisfies uniform $C^k$ bounds with respect to the frame
corresponding to $\mathcal{D}^{\mathcal{K}}$ for a $k\ge3$.
On the other hand, with respect to this frame, we have uniform $C^k$ bounds of all geometric quantities
associated to $\{v_{\mathcal{H}^+}=v_{-1}\}\cap \{u_{\mathcal{H}^+}=u_n\}$. Since these are all contained in a compact
subset of $\mathcal{D}^{\mathcal{K}}$, we have that a subsequence converges in $C^{k-1}$ to some
$C^k$ sphere $S\subset B$. 

Now let us define the event horizon $\mathcal{H}^+$ as the future boundary of the set $J^-(\mathcal{D}^{\I}_{\infty})$ in $\mathcal{M}$.
Note that $S\subset \mathcal{H}^+$. General Lorentzian geometry, our bounds on $S$ and Cauchy stability now tell us that
$\mathcal{H}^+\cap  \mathcal{D}^{\mathcal{K}}$ is a regular $C^k$ null hypersurface admitting $S$ as a spatial section.

Now given any $v_n$, we similarly have uniform bounds on the geometry of the sequence of hypersurfaces
$\{v_{-1}\le v_{\Hp}\le v_n \}\cap \{u_{\Hp}=u_n\}$ in the Kruskalised frame. As in the proof of the closedness,
we may extract a convergent limit which is a $C^k$ null hypersurface, denote $\mathcal{H}^+_{v_n, {\rm new}}$, attach this to spacetime, noting that it will be consistent with the geometry of
$\mathcal{H}^+$. We may now solve a characteristic initial value problem (now we have
to lose derivatives) to obtain an extension spacetime. By the $C^{\tilde{k}}$ uniqueness statement of
Theorem~\ref{maxCauchythe}, it follows that this extension embeds into $\mathcal{M}$ in the first place. 
It follows that $\mathcal{H}^+_{v_n, {\rm new}}\subset \mathcal{H}^+$ for all $n$, under the obvious identification
and thus that
$\mathcal{H}^+= \cup_{n}\mathcal{H}^+_{v_n, {\rm new}}\cup (\mathcal{H}^+\cap \mathcal{D}^{\mathcal{K}})$.
It follows that that $\mathcal{H}^+$ is a $C^k$ null hypersurface.

The future completeness of $\mathcal{H}^+$ follows easily from our estimates on $\Omega^2_{\Hp}$ when
Kruskalised and
the range of the $v_{\Hp}$ coordinates.

\subsection{Polynomial decay along $\mathcal{H}^+$}
\label{polyalsoonH}

The limiting hypersurface $\mathcal{H}^+$ inherits the estimates of  the asymptotically
normalised $\mathcal{H}^+$ gauge~\eqref{limitingembeddinghor}. 
Thus, from~\eqref{asymptpointwisestatementintheorem}, we easily see the estimate
that 
\begin{equation}
\label{alphadecayonhorizoninproof}
|\Omega^2\alpha_{\mathcal{H}^+}(\infty, v, \cdot)| \lesssim \varepsilon_0 v^{-1},
 \end{equation}
 where $\Omega^2\alpha_{\mathcal{H}^+}$ is defined as the $u\to\infty$ limit of $\Omega^2\alpha_{\Hp}(u,v,\theta)$
 defined with respect to the gauge~\eqref{limitingembeddinghor}.
Note that  $\Omega^2\alpha_{\mathcal{H}^+}$ is generically a nonvanishing quantity as can be
seen by passing to the Kruskalised gauge.

\subsection{The black hole region as $\mathcal{M}\setminus J^-(\mathcal{I}^+)$}
\label{whatistheblackholeregion}

It is traditional to attach $\mathcal{I}^+$ as a ``conformal boundary'' of spacetime by introducing
a conformally rescaled metric, and using this, ascribe meaning to $J^-(\mathcal{I}^+)$. We prefer to 
define $J^-(\mathcal{I}^+)$ directly as the set 
\[
J^-(\mathcal{I}^+):=\{ p\in \mathcal{M} :\gamma:[0,1)\to \mathcal{M} {\rm\ future\ directed\ causal}, \gamma(0)=p, \lim_{t\to1} \gamma(t) \in \mathcal{I}^+\}
\]
where by $ \lim_{t\to1} \gamma(t) \in \mathcal{I}^+$ we mean that $\gamma$ is eventually in the 
range $\mathcal{D}^{\I}_{\infty}$ of the $\mathcal{I}^+$ asymptotic gauge and, writing $\gamma$ as $(u(t),v(t),\theta(t))$,
then
$v(t)\to \infty$ while $u(t)$, $\theta(t)$ have finite limits.

Under this definition, we manifestly see that $J^-(\mathcal{I}^+)\subset J^-(\mathcal{D}^{\I}_\infty)$, and in fact 
\[
J^-(\mathcal{I}^+)=
 J^-(\mathcal{D}^{\I}_\infty)=
\mathcal{D}^{\Hp}_{\infty}\cup \mathcal{D}^{\I}_{\infty}\cup( \mathcal{D}^{\mathcal{EF}}\cup
\mathcal{D}^{\mathcal{K}} \cap J^-(\mathcal{D}^{\I}_\infty)) .
\]

\subsection{Higher order regularity of $\mathcal{H}^+$}
\label{higherorderregulofhor}
In view of our construction of $\mathcal{H}^+$, the higher order regularity statement
of Theorem~\ref{limitgaugetwo} follows directly from     the statement in Theorem~\ref{limitgaugeone}
on higher regularity of the embedding~\eqref{limitingembeddinginf}.
In particular, if~\eqref{datanormtobesmall} is finite for all $k$, then the event horizon $\mathcal{H}^+$ is
a $C^\infty$ hypersurface in $\mathcal{M}$.

\section{Proof of Theorems~\ref{limitgaugetwo} and~\ref{limitgaugethree}}
\label{recapsecti}

Let us explicitly note that we have indeed obtained all statements in Theorems~\ref{limitgaugetwo} 
and~\ref{limitgaugethree}.

We restate first Theorem~\ref{limitgaugetwo}.
\limitgaugetwohere*

\begin{proof}
The statement concerning $J^-(\mathcal{I}^+)$ has been obtained in Section~\ref{whatistheblackholeregion}, 
whereas the completeness of null infinity was obtained as $(\ref{tiscompletenullinfinity})$ in 
Section~\ref{tiscompletenullinfinity}. 
The laws of gravitational radiation are formulated in Section~\ref{lawisthelaw}, and the statements
on the final shear and centre of mass, i.e.~$(\ref{normalisationsinstatement})$, were obtained in
$(\ref{Sigmapluszero})$ and $(\ref{finalPell1})$ above. 
The statement regarding the completeness and basic regularity properties of the horizon $\mathcal{H}^+$ 
were obtained in Section~\ref{definitionofhorizon} whereas the higher order regularity statement
is given in Section~\ref{higherorderregulofhor}.
\end{proof}

We now restate Theorem~\ref{limitgaugethree}.
\limitgaugethreehere*

\begin{proof}
The decay properties on $\mathcal{H}^+$
were obtained in Section~\ref{eventsmydear}, specifically, in $(\ref{alphadecayonhorizoninproof})$,
while the decay properties on $\mathcal{I}^+$
were obtained in Section~\ref{decayalonginfinitysec},
specifically, in $(\ref{decayofmassfinal})$, $(\ref{decayofSigma})$, $(\ref{decayofXi})$ and $(\ref{decayofA})$.
\end{proof}

\section{Axisymmetric solutions with vanishing angular momentum: the proof of Corollary~\ref{axicorollary}}
\label{axisymmtheorem}

In the notation of Section~\ref{threeparamsection},
let us consider an initial  data set $\mathcal{S}\in \mathfrak{M}(M_{\rm init},c^2\varepsilon_0)$ 
which is moreover assumed to be axisymmetric and represented
such that the initial axisymmetric vector field   is $\partial_{\mathring{\phi}}$, 
where $(\mathring\theta,\mathring\phi)$
represent the 
standard spherical coordinates~\eqref{sphcorddef} of the initial data cones.

It follows that the maximal Cauchy development $(\mathcal{M},g)$ corresponding
to $\mathcal{S}$ inherits a globally defined Killing field $\tilde{Z}$.

We now remark that the renormalised initial Eddington--Finkelstein gauge~\eqref{herethenewone} 
of part (b) of Theorem~\ref{thm:localEF}, in view 
of its geometric characterization and its anchoring~\eqref{polesshouldbethesame} to the
(manifestly symmetric) preliminary gauge (a), also 
inherits the axisymmetry and in fact
$\partial_{\mathring \phi}=h\tilde{Z}$ for $h$ some constant with $|h-1|\lesssim \varepsilon_0$,
where  now $(\mathring\theta,\mathring\phi)$
represent the 
standard spherical coordinates~\eqref{sphcorddef}  of the parametrisation~\eqref{herethenewone}.
It is natural in fact to define ``the'' axisymmetric Killing field as the extension
of $Z:=\partial_{\mathring\phi}$ to $\mathcal{M}$, as
this is the correctly normalised vector at infinity.

Let us assume moreover that the initial 
angular momentum vanishes. This is the statement that
the so-called Komar integral~\cite{Komar, Wald}\index{angular momentum!$\mathscr{J}(S)$, Komar angular momentum associated to $S$}
\begin{equation}
\label{Komarintegral}
\mathscr{J}(S)=
-\frac{1}{16\pi} \int_{S} \nabla^\mu Z^\nu  \epsilon_{\mu\nu\lambda\rho}
\end{equation}
vanishes when computed
for instance on any standard section of the initial cones. 
(We note  that if the data are polarised, then in particular the angular momentum indeed vanishes
in the above sense.)

By the well known conservation relation for the Komar integral~\cite{Wald}, it follows that~\eqref{Komarintegral}
vanishes
when computed
on any spacelike surface $S$ in $(\mathcal{M},g)$ homologous to a standard section of the initial cones.

Note that  $\mathcal{S}$ as above might not satisfy  
$(\eta_{\mathcal{EF}})_{\ell=1}=0$. It will satisfy, however,
$|(\eta_{\mathcal{EF}})_{\ell=1}|\lesssim \varepsilon_0^2$. 
We may thus write $\mathcal{S}=\mathcal{S}_0(\lambda_0)$ for some $|\lambda_0|\lesssim \varepsilon_0^2$
and some $\mathcal{S}_0\in \mathfrak{M}_0(M_{\rm init},c^2\varepsilon_0)$. Note that it follows
that $\lambda_0 \in \mathfrak{R}(u_f^0)$.

For all $u_f\ge u_f^0$, consider the sets $\mathfrak{R}(u_f)$ given in the proof of Theorem~\ref{thm:main}.
For all $u_f\ge u_f^0$ such that
$\lambda_0\in \mathfrak{R}(u_f)$, consider the $u_f$, $M_f$ normalised $\mathcal{I}^+$ gauge~\eqref{fortheIplusgauge} in $(\mathcal{M},g)$ corresponding to $\mathcal{S}=\mathcal{S}(\lambda_0)$.
Since this gauge
is itself again geometrically defined and anchored only to 
the initial Eddington--Finkelstein gauge through~\eqref{choiceofpandv}, it follows, in view also 
of the equivariance of the construction of Proposition~\ref{determiningthesphere} (cf.~Remark~\ref{equivarianceremark}), that $Z=h_{u_f}\partial_{\mathring{\phi}}$ 
for some constant $|h_{u_f}-1|\lesssim \varepsilon_0$, where
$(\mathring\theta,\mathring\phi)$ are now the standard spherical coordinates
corresponding to the
$\mathbb S^2$ appearing in~\eqref{fortheIplusgauge}.

An easy computation  (see~\cite{gallomoreschi} for the analogous computation in the Newman--Penrose formalism
and Bondi framework) shows that, in view of our estimates in the $\mathcal{I}^+$ gauge and the
relation between $u_f$ and $v_\infty$, 
we have the relation:
\begin{equation}
\label{lastoneinpaper}
\big|\,|\mathscr{J}(S^{\mathcal{I}^+}_{u_f,v_\infty})|-|{\bf J}(u_f, \lambda_0)|\,\big|\lesssim \frac{\varepsilon_0^2}{u_f}.
\end{equation}
Since $\mathscr{J}(S^{\mathcal{I}^+}_{u_f,v_\infty})=0$ for this solution, it 
follows then that for no $u_f$ can ${\bf J}(u_f, \lambda_0)$ saturate the bound~\eqref{eq:curletalambda},
i.e.~for no $u_f$ can we have
$|{\bf J}(u_f, \lambda_0)|=\frac{\varepsilon_0}{u_f}$.  From the construction of the sets $\mathfrak{R}(u_f)$, it
then  follows
that in fact we have 
$\lambda_0\in \mathfrak{R}(u_f)$ for all $u_f\ge u_f^0$. We may thus take $\lambda^{\rm final}=\lambda_0$,
and
the corollary  follows.

\begin{remark}
\label{alternativeremark}
An alternative proof of Corollary~\ref{axicorollary}
would of course be to dispense completely with the  $3$-parameter family $\mathcal{L}_{\mathcal{S}_0}$ and simply
use~\eqref{lastoneinpaper} directly on the solution corresponding to the axisymmetric solution $\mathcal{S}=\mathcal{S}(\lambda_0)$ as an a priori estimate for $|{\bf J}(u_f, \lambda_0)|$,
namely $|{\bf J}(u_f, \lambda_0)|\lesssim \frac{\varepsilon_0^2}{u_f}$, since for this solution we have $|\mathscr{J}(S^{\mathcal{I}^+}_{u_f,v_\infty})|=0$.
Given this estimate, one may remove all references to  $\mathcal{L}_{\mathcal{S}_0}$  and
$\mathfrak{R}(u_f)$ from the setup of Theorem~\ref{thm:main}, 
as the modulation was only necessary
to bound $|{\bf J}(u_f,\lambda)|$.  This leads to a number of important simplifications throughout, for instance
one does not require the monotonicity properties of Section~\ref{section:monotonicityR}, which were the main reason for
our precise assumptions on data (cf.~Remark~\ref{pointwheredecay}).
Thus, in this case, one can see relatively easily that the decay assumptions
on the data can indeed be further relaxed (cf.~Remark~\ref{commentaboutpeeling}),
to allow for instance for $\beta\sim r^{-4}\log |r|$ as in~\cite{christorome}.
\end{remark}

\addcontentsline{toc}{chapter}{Index}

\printindex

\addcontentsline{toc}{chapter}{Bibliography}
\bibliographystyle{DHRalpha}
\bibliography{CodimStabrefs}

\end{document}